\documentclass[12pt]{report}
\usepackage[colorlinks,citecolor=blue,urlcolor=blue,breaklinks,pdftitle={xtz},pdfauthor={Benjamin Nachman}]{hyperref}
\usepackage{suthesis-2e}
\usepackage{amsmath, amsthm, amssymb, amsfonts}
\usepackage{cite}
\usepackage{subfig}
\usepackage{graphicx}
\renewcommand{\arraystretch}{1.25}
\usepackage{lscape}
\usepackage[makeroom]{cancel}
\usepackage{multicol}
\usepackage{comment}

\newcommand{\ttbar}{\ensuremath{t\bar{t}}}
\newcommand{\tone}{\ensuremath{{\tilde{t}^{}_{1}}}}

\def\CLs{\ensuremath{\mathrm{CL}_s}}

\def\ltri{\mbox{\begin{picture}(7,10)
\put(1,0){\line(1,0){5}}
\put(6,0){\line(-1,2){5}}
\put(1,0){\line(0,1){10}}
\end{picture}
}}

\makeatletter
\@addtoreset{chapter}{part}
\makeatother  

\usepackage{titletoc}
\titlecontents{chapter}
[2.65em]
{\addvspace{10pt}\bfseries}
{\contentslabel{2.3em}}
{\hspace*{-2.3em}}
{\space\hfill}

\titlecontents{part}
[2.65em]
{\addvspace{10pt}\bfseries}
{\contentslabel{3.3em}}
{\hspace*{-2.3em}}
{\space\hfill}

\usepackage{overpic}
\usepackage{multirow}
\usepackage{adjustbox}

\usepackage{tikz}
\usetikzlibrary{arrows,shapes}
\usetikzlibrary{snakes}
\usetikzlibrary{decorations.pathmorphing}
\usetikzlibrary{decorations.markings}
\usetikzlibrary{backgrounds}

\usepackage{booktabs,siunitx}
\usepackage{fancyvrb}

\tikzset{
	    vector/.style={decorate, decoration={snake}, draw},
	provector/.style={decorate, decoration={snake,amplitude=2.5pt}, draw},
	antivector/.style={decorate, decoration={snake,amplitude=-2.5pt}, draw},
    fermion/.style={draw=black, postaction={decorate},
        decoration={markings,mark=at position .55 with {\arrow[draw=black]{>}}}},
    fermionbar/.style={draw=black, postaction={decorate},
        decoration={markings,mark=at position .55 with {\arrow[draw=black]{<}}}},
    fermionnoarrow/.style={draw=black},
    gluon/.style={decorate, draw=black,
        decoration={coil,amplitude=2pt, segment length=5pt}},
    gluon2/.style={decorate, draw=black,
        decoration={coil,amplitude=5pt, segment length=6pt}},        
    scalar/.style={dashed,draw=black, postaction={decorate},
        decoration={markings,mark=at position .55 with {\arrow[draw=black]{>}}}},
    scalarbar/.style={dashed,draw=black, postaction={decorate},
        decoration={markings,mark=at position .55 with {\arrow[draw=black]{<}}}},
    scalarnoarrow/.style={dashed,draw=black},
    electron/.style={draw=black, postaction={decorate},
        decoration={markings,mark=at position .55 with {\arrow[draw=black]{>}}}},
	bigvector/.style={decorate, decoration={snake,amplitude=4pt}, draw},
}

\usepackage{euler}
\usepackage{verbdef}
\verbdef{\vtext}{generate p p > t t~ a / t t~}
\usepackage{textcomp}

\usepackage[T1]{fontenc}
\usepackage{enumitem}
\usepackage{mathtools}
\newcommand\myeqa{\stackrel{\mathclap{\normalfont\mbox{\tiny(1)}}}{\implies}}
\newcommand\myeqb{\stackrel{\mathclap{\normalfont\mbox{\tiny(2)}}}{\implies}}
\newcommand\myeqc{\stackrel{\mathclap{\normalfont\mbox{\tiny(3)}}}{\implies}}
\newcommand\myeqd{\stackrel{\mathclap{\normalfont\mbox{\tiny(4)}}}{\implies}}

\makeatletter
     \renewcommand*\l@figure{\@dottedtocline{1}{1em}{3.2em}}
\makeatother

\newcommand\blfootnote[1]{  \begingroup
  \renewcommand\thefootnote{}\footnote{#1}  \addtocounter{footnote}{-1}  \endgroup
}

    \title{Investigating the Quantum Properties of Jets and the Search for a Supersymmetric Top Quark Partner with the ATLAS Detector}
    \author{Benjamin Philip Nachman}
    \dept{Physics}
    \principaladviser{Su Dong}
    \coprincipaladviser{Ariel Schwartzman}
    \secondreader{Michael Peskin}
    \thirdreader{Lauren Tompkins}

\begin{document}

    \beforepreface

\prefacesection{Abstract}
Quarks and gluons are the fundamental building blocks of matter responsible for most of the visible energy density in the universe.  However, they cannot be directly observed due to the confining nature of the strong force.  The Large Hadron Collider (LHC) uses proton-proton collisions to probe the highest energy reactions involving quarks and gluons happening at the smallest distance scales ever studied in a terrestrial laboratory.  The observable consequence of quark and gluon production in these reactions is the emergent phenomenon known as the {\it jet}: a collimated stream of particles traveling at nearly the speed of light.  The quantum properties of the initiating quarks and gluons are encoded in the distribution of energy inside and around jets.   These {\it quantum properties of jets} can be used to study the high energy nature of the strong force and provide a way to tag the hadronic decays of heavy boosted particles.  The ATLAS detector at the LHC is well-suited to perform measurements of the internal structure of high energy jets.  A variety of novel techniques utilizing the unique capabilities of the ATLAS calorimeter and tracking detectors are introduced in order to probe the experimental and theoretical limits of the quantum properties of jets.

Studying quarks and gluons may also be the key to understanding the fundamental problems with the Standard Model (SM) of particle physics.  In particular, the top quark has a unique relationship with the newly discovered Higgs boson and as such could be a portal to discovering new particles and new forces.  In many extensions of the SM, the top quark has a partner with similar relationships to other SM particles.   For example, a scalar top partner (stop) in Supersymmetry (SUSY) could solve the Higgs boson mass hierarchy problem.  Miraculously, a SUSY neutralino could also account for the dark matter observed in the universe and may be copiously produced in stop decays.   High-energy top quarks from stop decays result in jets with a rich structure that can be identified using the techniques developed in the study of the quantum properties of jets.  While there is no significant evidence for stop production at the LHC, the stringent limits established by this search have important implications for SUSY and other models.  
 
\clearpage
 \vspace*{\stretch{1}}
\begin{flushright}
\vspace{10mm}
{\it In memory of Isabella Threlkeld}
\end{flushright}
\vspace{\stretch{3}}
\clearpage

    \newpage

 \prefacesection{Acknowledgements}
 
 	Like any good story, a thesis describes a grand adventure with many twists and turns.  I leave the quality of this story for you to judge, but I cannot proceed without thanking the many characters behind the scenes who have supported my adventure and contributed to my ongoing development as a physicist.  My first high energy physics experiment was part of the Cosmic Ray Observatory Project (CROP) when I was in High School.  Thanks to the guidance of John Rogers and the support of Dan Claes and Greg Snow, I had fun building and operating muon detectors on the roof of my school and my house.  In college, I had many inspiring teachers including Csaba Csaki, Yuval Grossman, Flip Tanedo, Camil Muscalu, and Ravi Ramakrishna.  I had the pleasure of working with and learning from Itai Cohen about fluid dynamics and from Keith Dennis about group theory.  I am indebted to Jim Alexander for showing me how to think like an experimental particle physicist.  I learned a lot from and with Adam Dishaw and Nathan Mirman about statistics and top quark physics.  Across the ocean, I learned from Christopher Lester that science really does happen on the back of an envelope.  Studying with Ben Allanach,  I built a foundation in Supersymmetry theory.  
	
	Since formally starting at SLAC and with ATLAS, I have been blessed to work with many kind, passionate, and knowledgable physicists.  There are simply too many wonderful people to thank all of them here - I apologize!   First of all, I would like to thank all of the (sub)group conveners, editorial board (chairs), and members at large for their insightful comments on my work.  It was hard to make small list, but I'm grateful to these people in particular for advice, opportunities, and extensive feedback: Alison Lister, Mark Owen, Tancredi Carli, Andreas Hocker, Jan Kretzschmar, Bogdan Malaescu, Monica D'Onofrio, Jamie Boyd, Tomasso Lari, Iacopo Vivarelli, Michael Begel, David Miller, Cigdem Issever, Judith Katzy, Ian Hinchliffe, Luciano Mandelli, Tony Doyle, and Mathieu Benoit.  Till Eifert has been an amazing mentor, colleague, and friend.  I have thoroughly enjoyed a camaraderie over jets with Max Swiatlowski, David Lopez Mateos, and Nurfikri Norjoharuddeen.  Thank you to Matt Schwartz, Jesse Thaler, and Andrew Larkoski for teaching me jet phenomenology. The SLAC/Stanford ATLAS group has been extremely supportive, especially Charlie Young, Philippe Grenier, Rainer Bartoldus, Lauren Tompkins, Pascal Nef, Michael Kagan, Francesco Rubbo, Qi Zeng, Zihao Jiang, and Aviv Cukierman.  I am very grateful to Michael Peskin for many fun conversations about particle physics - I have never met someone so knowledgable, clear, and enthusiastic.  All of my high energy physics adventures would not have been possible without generous funding from the National Science Foundation, the Department of Energy, and the Stanford Graduate Fellowship.  This funding would have been useless without the strong support, guidance, and mentorship from my fantastic advisers Su Dong and Ariel Schwartzman.  Thank you Lester Mackey for many fun discussions about connecting machine learning and particle physics and for chairing my thesis defense.  Thank you Su Dong, Ariel, Michael P., Lauren, and Till for providing feedback on this thesis (and for your patience with its length).

   	My parents, Beth (to the muons, moon and back!) and Gary, and brothers, Marty and Lev, have always supported me in all my endeavors. No finite amount of text could describe how much I am indebted to them. This thesis is a discussion about the smallest distance scales ever studied (on Earth), so my enormous gratitude for them must wait until I see them in person.  I am additionally grateful for my extended Bay Area family: Bart, Carrie, Nathan, Brynna, Cole, Ali, Oli, Felix, and Maceo.
	
	Miracles do happen.  Three days after defending this thesis, I married my soulmate Hannah Joo.  Not only is she the most beautiful, wise, caring, and careful person I have ever met, she has been my anchor through the ups and downs of graduate student life.  There are no gloomy days when my sunshine is forever nearby. 
    
    \clearpage

    \afterpreface

\section{Preface}

The following sections summarize useful nomenclature and background information.

\subsection{Units}

All physical results can be presented in any unit system, but not all systems are equivalently useful. The familiar meters-kilograms-seconds (SI) unit system will be used to express dimensions of the detector.  Most other discussions will use {\it natural units} in which the rulers are not distance, mass, and time, but instead speed, angular momentum, and energy.  The rulers of time have length $c$, the rulers of angular momentum have length $\hbar$ and the rulers of energy have length giga-electron-volts (GeV). In SI, an object has `length 1' if it is one meter long.  Equivalently, in natural units an object has `speed 1' if it is going at the speed of light.  This nomenclature is used throughout - the $c$ and $\hbar$ will be henceforth implied and not stated explicitly for all dimensionful quantities.  For example, masses, momenta, and energies are all given in units of GeV (the $1/c^2$ and $1/c$ for mass and momentum are implied) and lengths and time are given in units of 1/GeV (the $\hbar c$ and $\hbar$ are implied).  Table~\ref{tab:units} gives a representative set of useful units for high energy physics and their abbreviations.

\vspace{6mm}

\begin{table}[h!]
\centering
\noindent\adjustbox{max width=\textwidth}{
\begin{tabular}{ccccc}
\hline 
\hline
 Quantity & Abbreviated Units & Full Units & SI (approximate) & Comment\\
 \hline
\hline
Speed & 1 & $c$ & $3\times 10^8$ m/s& \\
Angular Momentum & 1 & $\hbar$ &$10^{-34}$ m${}^2$ kg/s&\\
Energy & GeV & GeV &$1.6\times 10^{-10}$ J&\\
Momentum & GeV & GeV/$c$&$10^{-19}$ kg $\cdot$ m/s& \\
Mass & GeV & GeV/$c^2$&$1.8\times 10^{-27}$ kg& \\
Time & 1/GeV & $\hbar$/GeV &$6.6\times 10^{-25}$s&\\
Length & 1/GeV &  $\hbar c$/GeV &$2\times 10^{-16}$ m&\\
Charge & 1 & $e/\sqrt{4\pi\alpha}$ &$5.3\times 10^{-19}$ C& $e=1.6\times 10^{-19}$ C\\
Magnetic Field & (GeV)${}^2$& GeV${}^2$/($\hbar c^2$) &$5\times 10^{16}$ T&$\text{T}=\text{(kg)}/(\text{C}\cdot\text{s})$ \\ 
\hline
\hline
\end{tabular}}
\caption{Natural units.  There are multiple ways to define the electric charge.  In these natural units, $e=\sqrt{4\pi\alpha}$ so that one unit represents $\sim 0.3$ of an elementary charge.}
\label{tab:units}
\end{table}

\subsection{Coordinates}
\label{coords}

\begin{figure}[h!]
\begin{center}
\includegraphics[width=0.5\textwidth]{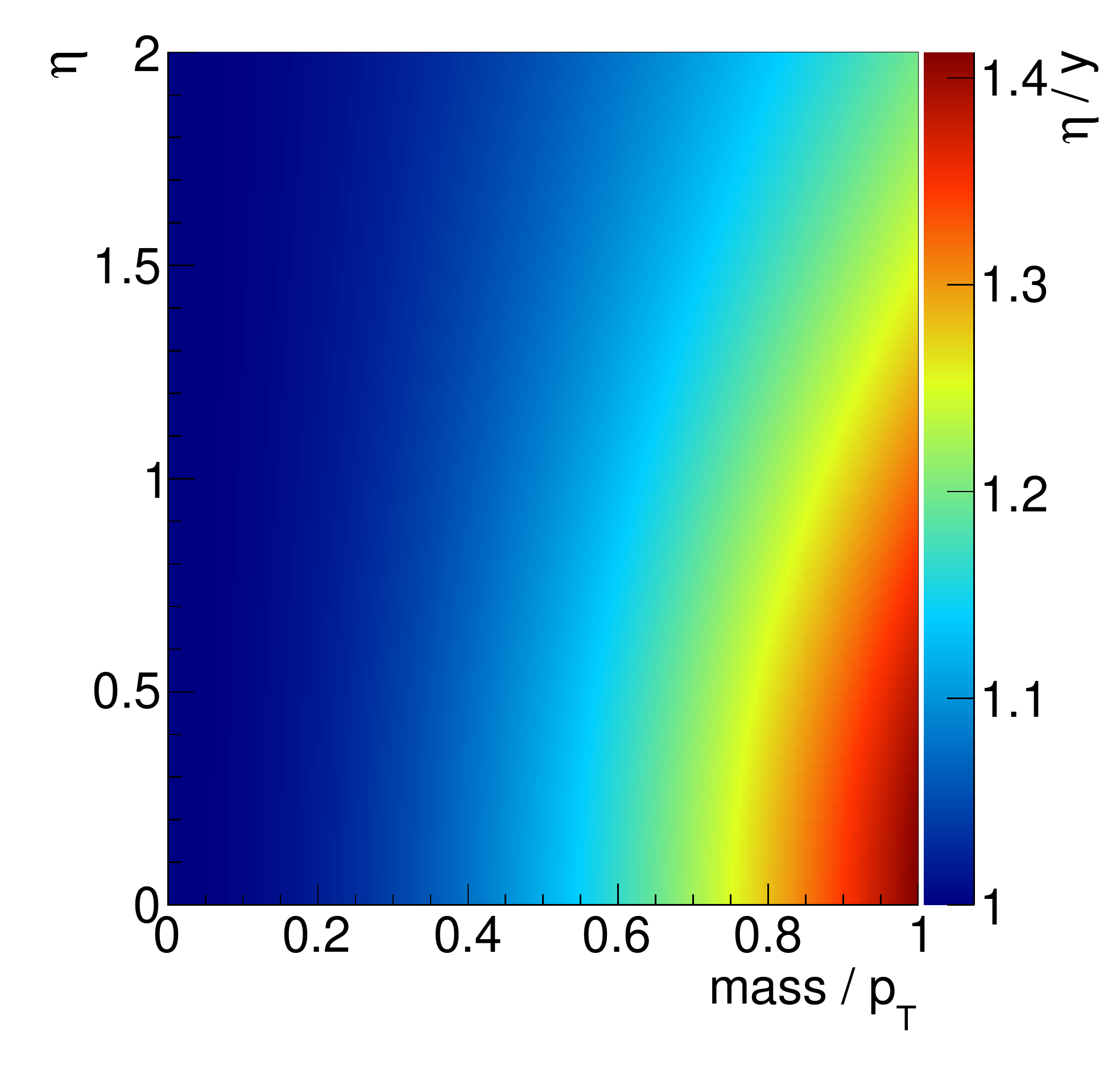}
 \caption{A comparison of $\eta$ and $y$ for a massive particle with $\phi=0$.}
 \label{fig:rapdiff}
  \end{center}
\end{figure}

Two sets of coordinates will be used interchangeably: $(p_x,p_y,p_z)$ and $(\eta,\phi,p_\text{T})$, where the $z$-axis is along the beam (longitudinal) direction, $\phi$ is the azimuthal angle, $p_\text{T}^2=p_x^2+p_y^2$ is the {\it transverse momentum}, and $\eta$ is the {\it pseudo-rapidity}:

	\begin{align}
	\label{rapditiy}
	\eta = -\text{ln}\left(\tan\left(\frac{\theta}{2}\right)\right)=\frac{1}{2}\text{ln}\left(\frac{|\vec{p}|+p_z}{|\vec{p}|-p_z}\right)=\tanh^{-1}\left(\frac{p_z}{|\vec{p}|}\right),
	\end{align}
	
	\noindent where $\theta$ is the angle between the $z$-axis and the transverse plane.  Particles with $\eta=0$ point in the transverse plane and $\eta=\pm\infty$ are moving along the $z$-axis.  These coordinates are particularly useful because after a Lorentz boost along the $z$-axis with magnitude $\beta$, a massless particles with $(\eta,\phi,p_\text{T})$ is described by $(\eta+\tanh^{-1}(\beta),\phi,p_\text{T})$.  In particular, the difference $\Delta\eta$ between two massless particles is invariant under a boost along $z$.  This motivates the distance metric $\Delta R^2=\Delta\eta^2+\Delta \phi^2$, which is invariant under longitudinal boosts for massless particles.  At a hadron collider, the partonic $p_z$ of a collision is in general not known, so the invariance of $\eta$ is crucial.  For massive particles, the generalization of $\eta$ is the {\it rapidity} ($y$), which is defined using the second or third equality Eq.~\ref{rapditiy}, but replacing $|\vec{p}|$ with $E$.  Rapidity is not determined solely by geometry (no equivalent to the first equality in Eq.~\ref{rapditiy}), but does transform additively under a boost along $z$.  Figure~\ref{fig:rapdiff} compares $\eta$ and $y$.  For a particle with $(p_\text{T},m)\approx (200,100)$ GeV at $\eta \sim 1$, the difference is about $10\%$.

\subsection{Statistics}

The distribution $\mathcal{D}(\theta)$ of a random variable $X$ will be denoted $X\sim \mathcal{D}(\theta)$ for some parameters $\theta$.  For example, $X\sim\mathcal{N}(\mu,\sigma^2)$ means that $X$ follows a normal distribution with mean $\mu$ and variance $\sigma^2$.  The probability distribution $f_X(x)$ is related to the cumulative distribution $F_X(x)=\Pr(X \leq x)$ by $f_X(x)=\partial_x F_X(x)$.   If $y=g(x)$, then $f_Y(y)=f_X(g^{-1}(y))|\partial_y g^{-1}(y)|$.  The quantity $\int_{-\infty}^\infty dx x f_X(x)$ will be interchangeably called the expected value, mean, or average and is denoted $E[X]$ or $\langle X\rangle$.  The square root of the variance $\langle X^2\rangle-\langle X\rangle^2$ uses the symbol $\sigma(X)$ and is referred to as the standard deviation.  A distribution's mean and standard deviation are sensitive to outliers and in general do not carry any probability content.  Therefore, frequently used alternatives are the median $m$, defined by $\int_{-\infty}^m dx f_X(x)=\int_{m}^\infty dx f_X(x)$, and the inter-quantile range, which is a symmetric interval around $m$ that contains a specified fraction of the distribution $f_X(x)$.

A technique that is used extensively to numerically estimate the uncertainty in the measured statistics of $X$ is the {\it bootstrap}~\cite{efron1979}.  Let $x_1,...,x_n$ be independent and identically distributed measurements from a random variable $X$.  A bootstrap dataset $x_1',...,x_n'$ is generated by picking $j_1,...,j_n$ with $j_i\sim \text{Uniform}(1,...,n)$ and settings $x_i'=x_{j_{i}}$.  Note that the same measurement $x_i$ may appear multiple times in the bootstrap dataset.  Many such datasets are generated and then the uncertainty on a statistic is estimated by computing moments or percentiles of the distribution of the statistic over the ensemble of bootstrap datasets.  For proofs about the bootstrap, see Ref.~\cite{Horowitz20013159} and references therein.

Additional statistical tools and techniques are introduced when needed in later sections and in Appendix~\ref{additionastats}.

\clearpage

\part[The Theory of Experimental Particle Physics]{The Theory of Experimental Particle Physics\\[4ex]\makebox[0pt]{\includegraphics[width=0.6\paperwidth]{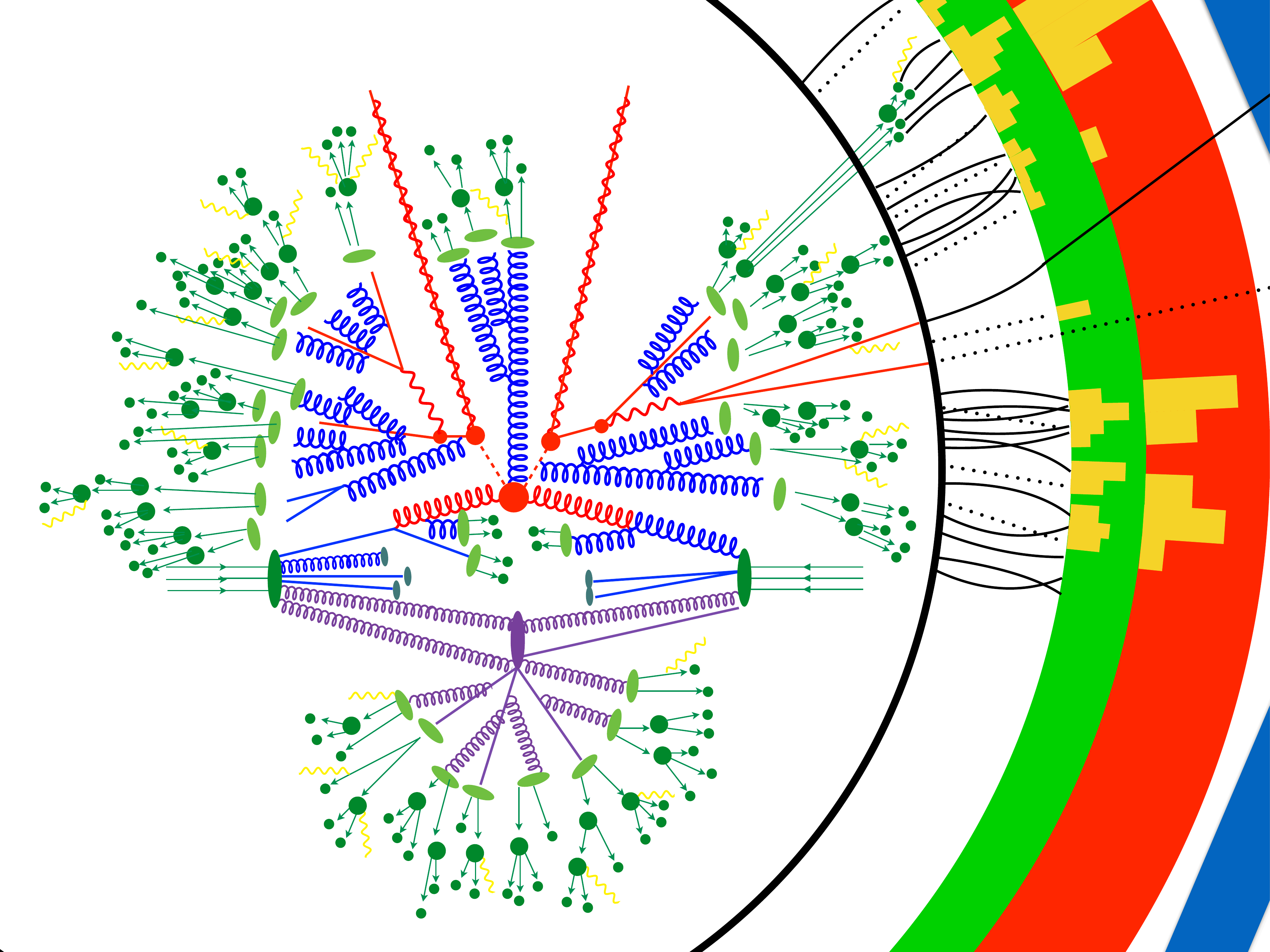}}\\\blfootnote{A schematic diagram illustrating the production and measurement of stop quark pair production, from $10^{-19}$ m up to $10^1$ m.  Modeled after Fig. 1 in Ref.~\cite{Gleisberg:2008ta}.}\label{parti}}

The ultimate goal of particle physics is to uncover the equations of motion of the elementary degrees of freedom: the {\it fundamental laws of the universe}.   Due to quantum mechanics, there is inherent randomness in the these fundamental laws.  Therefore, the elementary degrees of freedom are not particles in the classical sense.  Instead, they are described by {\it quantum fields} which are operator functions on spacetime from which one can compute probability distributions.  Fields, and not individual particle wave functions, are elementary in order to rectify quantum mechanics and special relativity: fields preserve causality and allow for particle creation and destruction.  {\it Quantum field theory} (QFT) is the language of modern particle physics.

The definition of {\it elementary} has evolved over time.  Perhaps the first `particle physicists' where the early chemists at the turn of the 19${}^\text{th}$ century who discovered various distinct elements that were postulated to be built from atoms (literally Greek for `uncuttable').  Subatomic particle physics began with the discovery of the electron by J. J. Thomson in 1897.   Arguably the beginning of modern particle physics was the first {\it fixed target} experiment by Geiger, Marsden, and Rutherford using gold foil in $\sim$ 1910 to show that the positive charge inside atoms is concentrated in a point-like center called the nucleus.   The same idea was used to discover the internal structure of the nucleus 50 years later at SLAC with a much more sophisticated apparatus and significantly higher energy probe particles.   The last 100 years of nuclear and elementary particle physics have been filled with many stories of discovery, confusion, prediction, and success.  Some of these will be introduced throughout Part I;  for a more detailed account, there are many excellent references\footnote{A thorough account can be found in Ref.~\cite{3566} up to the mid 1980s.  Reference~\cite{Schweber:1994qa} is a detailed account of QED.  Various dedicated articles on the LEP program at CERN, the discovery of the top quark, and the discovery of the Higgs boson can be found in e.g.~\cite{Assmann:2002th,Campagnari:1996ai,Ellis:2015tba}.}.

Part~\ref{parti} serves as a brief introduction to experimental particle physics, with the goal of explaining all aspects of the schematic diagram from p.\pageref{parti}.  Chapter~\ref{chapter:SM} introduces the Standard Model of particle physics, which describes all known elementary particles and interactions.  The detection of particles, specifically with the ATLAS detector, is described in Chapter~\ref{chapter:ATLAS}.  Simulation and modeling of particle production and detection is documented in Chapter~\ref{chapter:simulation} and Chapter~\ref{chapter:reco} describes how final states are reconstructed with the ATLAS detector.

\clearpage

\chapter{The Standard Model of Particle Physics}
\label{chapter:SM}

		Due to technical advances in the 1950s, there was an explosion of new unstable subatomic particles discovered with lifetimes ranging from $10^{-23}$ to $10^{-6}$ seconds.  Just as the periodic table of elements reduced the number of degrees of freedom from patterns in atomic spectra, the quark model based on $SU(3)$ (gauge) symmetry was postulated to explain the structure of the newly discovered {\it hadrons}~\cite{Zweig:570209,GellMann:1964nj,Fritzsch:1973pi}.   The $SU(3)$ gauge theory built on the success of the slightly earlier foundational work on the quantum theory of electromagnetism (QED) based on a $U(1)$ gauge group\footnote{See Ref.~\cite{Dyson:1949bp} and references therein by Tomonaga, Schwinger, and Feynman.}.  Around the same time as the strong force, a coherent quantum theory of the weak force and electromagnetic forces was assembled involving a $U(1)\times SU(2)$ gauge theory~\cite{Glashow:1961tr,Weinberg:1967tq,Salam:1968rm} that contained a mechanism for generating masses for SM particles~\cite{Englert:1964et,Higgs:1964pj,Higgs:1964ia,Guralnik:1964eu}.
		
		The Standard Model (SM)\footnote{There are many excellent books on QFT that describe the technicalities of the SM.  See for example, Ref.~\cite{Peskin:1995ev} and Ref.~\cite{Ticciati:1999qp} (based on S. Coleman's lectures).  There are also many dedicated books on the SM or various parts of the SM such as Ref.~\cite{Ellis:1991qj} (QCD).} is a composite theory combining the strong and electroweak forces.  This chapter summarizes the particle content and interactions of the SM (Sec.~\ref{sec:particlesandforces}).  Section~\ref{SMproblems} provides an overview of the success of the SM as well as its limitations, which have lead to an extensive literature on extensions of the SM.  One important class of models is Supersymmetry (SUSY), which is the focus of Part~\ref{part:susy}.
	
		\clearpage
	
		\section{Particles and Forces}
		\label{sec:particlesandforces}
		
			The SM describes three elementary forces: electromagnetic, weak, and strong.  The electromagnetic interaction at the subatomic level is the same long range force that is familiar at everyday distance scales; it is responsible for most aspects of daily life (protein structure, batteries, friction, etc.).  Note that the SM does not describe the other force relevant to daily life: gravity.  This is irrelevant for terrestrial particle physics probed thus far because the strength of gravity is incredibly weak (but is revisited in Sec.~\ref{SMproblems}).  The weak and strong forces are mostly unfamiliar to daily life because they are not long range forces.  At very short distance scales, the weak force is actually {\it stronger} than the electromagnetic force.  However, the weak force analogue to electromagnetism's photon is about 100 times as massive as the proton ($1$ GeV).  For distances comparable to or larger than this mass $\hbar c/(100\text{ GeV})\sim 10^{-18}$ m, the weak force is highly suppressed.  Despite this small distance scale, the weak force is responsible for radioactive decay.  The large mass of the mediator allows many unstable particles to survive for macroscopic times before decaying.  In contrast, the strong force is mediated by a massless particle called the gluon.  The reason the strong force is not long range is because it is {\it too strong}. Unlike electromagnetism, the strong force strengthens with distance; this means that the energy stored in the field of two objects participating in the strong force will be sufficient by $E=mc^2$ to make more particles.  These additional particles {\it screen} the original force.  Beyond about $10^{-15}$ m, the primary strong force is highly suppressed, but there is a residual {\it strong nuclear force} that is responsible for binding protons and neutrons inside the nucleus.  After a few femtometers, the strong force is negligible compared with electromagnetism.
			
			All matter particles participate in the weak interaction\footnote{Right-handed neutrinos are ignored.  If they exist, they do not interact within the SM.}, while only {\it quarks} feel the strong force, and both quarks and {\it charged leptons} interact via the electromagnetic force.  Neutrinos are matter particles that only interact via the weak force.  Each force is mediated by the exchange of force-carrying particles.  Both the electromagnetic and strong forces are mediated by massless spin $1$ bosons (photon and gluon) while the weak force is carried by three massive particles called the $W^\pm$ and $Z$ bosons.   The spin and mass of a particle are associated with their representation of the Poincar\'{e} group, which is the symmetry group of spacetime (rotations, translations, and boosts). In the context of QFT, the electromagnetic, weak, and strong forces are described by {\it internal} (in contrast to spacetime) symmetries of the SM Lagrangian.  Electromagnetism is the result of a dimension one unitary group $U(1)$ symmetry while the weak force and the strong force are described by the special unitary groups $SU(2)$ and $SU(3)$, respectively.  The SM Lagrangian is given by:
			
		\begin{align}
		\label{eq:SM}
		\mathcal{L}=-\frac{1}{4}(F_{\mu\nu}^a)^2+\bar{\psi}(i/\hspace{-2.5mm}D)\psi+y_{ij}\bar{\psi}_i\psi_j\phi+|D_\mu\phi|^2+\mu^2\phi^\dag\phi-\lambda(\phi^\dag\phi)^2,
		\end{align}
		
		\noindent where $F_{\mu\nu}^a=\partial_\mu A_\nu^a-\partial_\nu A_\mu^a+g f^{abc} A_\mu^b A_\nu^c$ and $D_\mu=\partial_\mu-ig A_\mu^a t_r^a$ for fermion fields $\psi$, scalar field $\phi$, gauge fields $A$, representation matrices $t_r^a$, and structure constants $f^{abc}$.  Using the Feynman slash notation, $\hspace{0mm}/\hspace{-2.5mm}D=\gamma^\mu D_\mu$, where $\gamma^\mu$ form a matrix representation of the Clifford algebra.  The {\it Yukawa} couplings $y_{ij}$ and the parameters $\lambda$ and $\mu$ describe the interactions of the Higgs boson with itself and with the fermions.  Equation~\ref{eq:SM} has an implicit sum over gauge groups, fermion types, group indices, and implicit Hermitian conjugates.  Expanding the Lagrangian in Eq.~\ref{eq:SM} would fill multiple pages; however, there are already some interesting observations one can make: (1) $f^{abc}\neq 0$ for non-Abelian groups and therefore the bosons of the weak and strong forces {\it interact with themselves}, i.e. participate in the force they mediate.  In contrast, the structure constant for electromagnetism is zero so photons do not directly interact with other photons via electromagnetism\footnote{This is only true at tree-level in perturbation theory; there are virtual corrections that lead to non-trivial photon-photon interactions.  Due to the smallness of the electromagnetic coupling, these interactions are highly suppressed, which is why at everyday energy scales this is negligible.}.  (2) There are no explicit mass terms in Eq.~\ref{eq:SM}.  The fields participating in the weak force are {\it Weyl} fermions with definite (left-handed) chirality.  Therefore, mass terms which link left- and right-handed fermions such as $m\bar{\psi}_L\psi_R+\text{h.c.}$ are not allowed by symmetry.  This is a serious problem because fermions certainly have mass.  The Lagrangian above is written {\it before electroweak symmetry breaking}, which is the process by which fermions and the bosons acquire a mass\footnote{In fact, the $U(1)$ in Eq.~\ref{eq:SM} is not exactly electromagnetism.  After electroweak symmetry breaking, the $B$ field combines with the neutral $W$ field to form {\it both} the photon and $Z$ boson.} (described below).  Note that there is no problem writing down a mass term for the {\it Higgs field} $\phi$ since $\phi^\dag\phi$ is invariant under an $SU(2)$ rotation.  (3) The parameters $y_{ij}$ are dimensionless.  Integrating the Lagrange density from Eq.~\ref{eq:SM} must be dimensionless so $[y]+[\phi]+2[\psi]-4=0$, where $[*]$ is the mass dimension of $*$ and $0=[\int d^4x\mathcal{L}]=[\mathcal{L}]-4$.  The mass term for a scalar is $m^2\phi^2$ and for Dirac fermion is $m\bar{\psi}\psi$ so $2[\phi]-2=0$ and $2[\phi]-3=0$.  Solving these three equations gives $[y]=0$.  Naively, one might expect these Yukawa couplings to be $\mathcal{O}(1)$; it is therefore a surprise that they span $6$ orders of magnitude.
		
Table~\ref{tab:SMfields} lists all of the fields and their representations as part of Eq.~\ref{eq:SM} (prior to electroweak symmetry breaking).  There are three families each of right-handed up-type and down-type quarks.  The left-handed up- and down-type quarks are grouped into an $SU(2)$ doublet.  Similarly, there are three generations of right-handed charged leptons and three generations of left-handed charged leptons grouped with neutrinos as $SU(2)$ doublets.  One additional $SU(2)$ doublet $\phi$ is occupied by two complex scalar fields.  This field will play a critical role in electroweak symmetry breaking.  The other SM fields are the gauge boson for $U(1)$ called the $B$,  the $W^i, i=1,2,3$ bosons for $SU(2)$, and the gluons for for $SU(3)$.  There are eight gluon fields, one for each generator of the $SU(3)$ Lie algebra (the Gell-Mann matrices).  Likewise, there are three quarks for each entry in Table~\ref{tab:SMfields}, one for each dimension of the fundamental representation of $SU(3)$.  To build an analogy to QED where particles have electrical charge, the three possible $SU(3)$ options for each quark are called {\it color charge} and labeled red, green, and blue.  These have nothing to do with actual color, but are useful because like visible light, a triple of quarks covering all three colors (red, green, blue) acts as if it were colorless.  The eight gluons can be considered as having one color charge and one anti-color charge (the `anti-' refers to the electric charge, since gluons are electrically neutral). Color charge in $SU(3)$ is discussed in more detail in Part~\ref{part:qpj}.  	
\begin{table}[h]
\begin{center}
\noindent\adjustbox{max width=\textwidth}{
\begin{tabular}{|c|c|c|c|c|c|c|}
\hline
Field &  Content &  Spin & $U(1)$ & $SU(2)$ & $SU(3)$ & Comment\\
 \hline  
 $Q_i$ & $(u_L\hspace{2mm}d_L)$ & $\frac{1}{2}$ & $\frac{1}{6}$&  ${\bf 2}$& ${\bf 3}$  & 3 generations\\
  $u_{R,i}$ & $u_R$ & $\frac{1}{2}$ & $\frac{2}{3}$ & ${\bf 1}$  & ${\bf \overline{3}}$  & 3 generations\\
    $d_{R,i}$ & $d_R$ & $\frac{1}{2}$ & $-\frac{1}{3}$ & ${\bf 1}$  & ${\bf \overline{3}}$  & 3 generations\\
   $L_i$ & $(e_L\hspace{2mm}\nu_L)$ & $\frac{1}{2}$ &  $\frac{1}{2}$&  ${\bf 2}$& ${\bf 1}$  & 3 generations\\
  $e_{R,i}$ & $e_R$ & $\frac{1}{2}$ & $-1$ & ${\bf 1}$  & ${\bf 1}$  & 3 generations\\  
    $\phi$ & $(\phi^+\hspace{2mm}\phi^0)$ & $0$ &  $\frac{1}{2}$&  ${\bf 2}$& ${\bf 1}$  & \\ 
 \hline
     $B$ & $B$ & $1$ &  $0$&  ${\bf 1}$& ${\bf 1}$  & \\         
     $W$ & $(W_1\hspace{2mm}W_2\hspace{2mm}W_3)$ & $1$ &  $0$&  ${\bf 3}$& ${\bf 1}$  & \\
     $g$ & $g$ & $1$ &  $0$&  ${\bf 1}$& ${\bf 8}$  & \\           
 \hline                                  
\end{tabular}}
\caption{The particle content of the SM prior to electroweak symmetry breaking.  The values under $U(1)$ are the Abelian charge (the actual representation is one-dimensional) whereas the entries under $SU(2)$ and $SU(3)$ are the representation of the field in the first column.  For example a bold eight denotes the octet (adjoint) representation of $SU(3)$.  }
\label{tab:SMfields}
\end{center}
\end{table}	
	
\begin{figure}[h!]
\begin{center}
\includegraphics[width=0.9\textwidth]{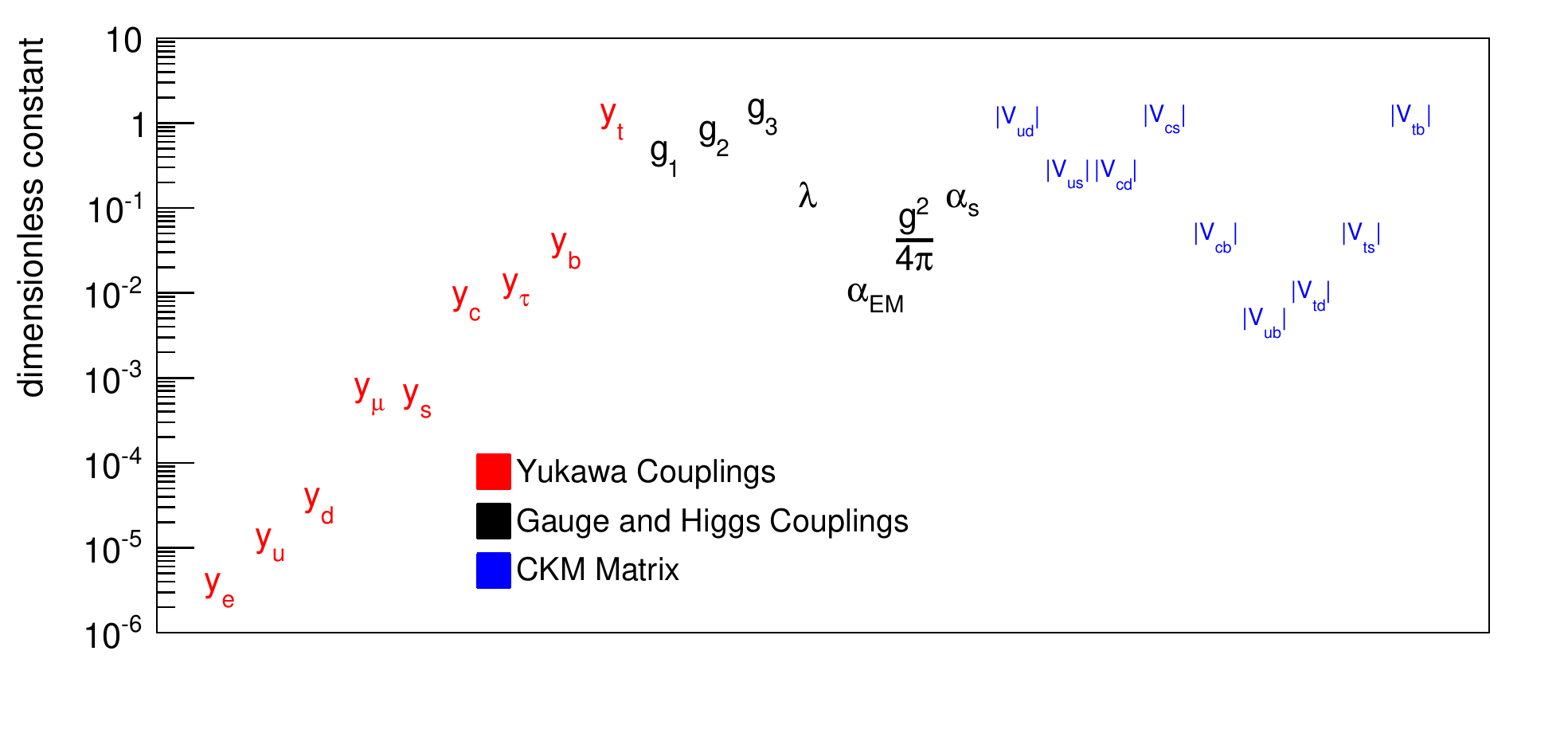}
 \caption{A comparison of all dimensionless SM parameters.  The nine Yukawa couplings are on the left in red and are organized by generation.  The three gauge couplings $g_i$ are $g_1=e/\cos(\theta_W), g_2=e/\sin(\theta_W), g_3=\sqrt{4\pi\alpha_s}$, for the weak mixing angle $\theta_W$, $e$ is the electron charge $e=\sqrt{4\pi\alpha_\text{EM}}$, and $\alpha_s\sim 0.1$ is the strong coupling constant.  The magnitude of all elements of the CKM matrix are in blue on the right. Parameter values are from Ref.~\cite{pdg}.}
 \label{fig:SM:overview}
  \end{center}
\end{figure}

The idea of electroweak symmetry breaking is that the potential for the Higgs field $V(\phi)=-\mu^2\phi^\dag\phi+\lambda(\phi^\dag\phi)^2$ can have a classical nonzero minimum if $\mu^2,\lambda >0$.  The field $\phi$ has four real degrees of freedom, but one can write the minimum as $\frac{1}{\sqrt{2}}(0\hspace{2mm} v)$, with $v=\sqrt{\frac{\mu^2}{\lambda}}$ and then all other points in the minimum are related to this one by $SU(2)$ transformations.  One can re-write the field $\phi$ as a fluctuation about this minimum: $\phi=\frac{1}{\sqrt{2}}(0\hspace{2mm} v+h(x))$, where $h$ is a real-valued scalar field\footnote{This choice is the {\it unitary gauge}.}.  With this formulation of $\phi$, expanding Eq.~\ref{eq:SM} gives rise to terms of the form $\mathcal{L}_{\psi\bar{\psi}v}=-(yv/\sqrt{2})\psi\bar{\psi}$, resulting in masses for the fermions $m=yv/\sqrt{2}$, and

\begin{align}
\label{gaugebosonmasses}
\mathcal{L}_{\text{gauge-boson-}v}=\frac{1}{2}\frac{v^2}{4}\left[g_2^2 (W_\mu^1)^2+g_2^2(W_\mu^2)^2+\left(-g_2W_\mu^3+\frac{1}{2}g_1 B_\mu\right)^2\right],
\end{align}

\noindent which are the mass terms for the electroweak Gauge bosons.  There are three vector boson mass eigenstates from Eq.~\ref{gaugebosonmasses}: $W_\mu^\pm=\frac{1}{\sqrt{2}}(W_\mu^1\pm i W_\mu^2)$ with mass $m_W=gv/2$, $Z_\mu=\cos(\theta_W)W_\mu^3-\sin(\theta_W)B_\mu$ with mass $m_Z=m_W/\cos(\theta_W)$ and the massless photon field $A_\mu=\sin(\theta_W)W_\mu^3+\cos(\theta_W)B_\mu$.  The {\it weak mixing angle} is $\theta_W=\cos^{-1}(g_2/\sqrt{g_1^2+g_2^2})$.  Three of the four real degrees of freedom from the field $\phi$ have been absorbed by the massive gauge bosons, adding a longitudinal polarization and allowing them to be massive.  The fourth real degree of freedom is $h(x)$, known as the Higgs boson.  This field couples to all massive particles and was the last particle of the SM to be discovered~\cite{Aad:2012tfa,Chatrchyan:2012xdj}.  Table~\ref{tab:SMfields2} summarizes the mass eigenstates of the SM after electroweak symmetry breaking.  There are six quark fields and six lepton fields, organized into three families of increasing mass.  The Yukawa couplings are $y_f=\sqrt{s}m_f/v$ where $m_f$ is the fermion mass and $v$ is the Higgs field vacuum expectation value $\sim 250$ GeV.   In total, there are $19$ free parameters of the SM\footnote{This depends on how one counts.  For example, the number of families could be a free parameter.  The electric charge could be viewed as a free parameter, but it is basically fixed by the coupling structure of the SM (including the anomaly cancellation - see Sec.~\ref{sec:MSSM}).  There are also several terms which are allowed by symmetry but are so close to zero that they are neglected (see Sec.~\ref{SMproblems}).}, including nine Yukawa couplings (fermion masses), one Higgs mass parameter, three gauge couplings, one Higgs self coupling, and one $3\times 3$ matrix $V$ (the Cabibbo-Kobayashi-Maskawa (CKM) matrix~\cite{Cabibbo:1963yz,Kobayashi:1973fv}) to describe transitions between quark types from weak decay.  This last item is the result of defining the quark fields in Table~\ref{tab:SMfields2} as the mass eigenstates: this induces off-diagonal components in the electroweak basis.  The matrix $V$ has four independent real number degrees of freedom.  Except for the Higgs mass parameter, all other SM parameters are dimensionless.  Some of the numerical values are given in Table~\ref{tab:SMfields2} and Figure~\ref{fig:SM:overview} graphically compares all of the dimensionless values.  The Yukawa couplings (and therefore the SM masses) span six orders of magnitude.   In contrast, the range for the gauge couplings is less than one order of magnitude.  The CKM matrix is nearly diagonal.	
		
\begin{table}[h]
\begin{center}
\noindent\adjustbox{max width=\textwidth}{
\begin{tabular}{|c|c|c|c|c|c|}
\hline
Field  &$Q$ & $SU(3)$ & Yukawa Coupling & Other Couplings\\
 \hline  
 $u, c, t$  & $\frac{2}{3}$&   ${\bf 3}$  &$10^{-5}, 7\times 10^{-3},1$& --\\  
  $d, s, b$  & $-\frac{1}{3}$&  ${\bf 3}$  &$3\times 10^{-5},5\times 10^{-4},0.03$& --\\  
   $e,\mu,\tau$ &$-1$&   ${\bf 1}$  &$3\times 10^{-6},6\times 10^{-4},0.01$& --\\
  $\nu_e,\nu_\mu,\nu_\tau$&$0$&  ${\bf 1}$  &--& --\\  
    $h$  &$0$&  ${\bf 1}$  &--&$\mu=90$ GeV, $\lambda=0.1$ \\ 
 \hline
     $\gamma$ &$0$&   ${\bf 1}$  &--& $\alpha=1/127$\\   
      $Z$ & 0&  ${\bf 1}$&--& $\sin(\theta_W)=0.5$ \\         
     $W^\pm$&  $\pm 1$&   ${\bf 1}$  &--& $V$ \\      $g$ & 0&  ${\bf 8}$  &--& $\alpha_s=0.1$ \\           
 \hline                                  
\end{tabular}}
\caption{The particle content of the SM after electroweak symmetry breaking.  All couplings are given to one significant figure at the scale $m_Z$.  Parameter values are from Ref.~\cite{pdg}.}
\label{tab:SMfields2}
\end{center}
\end{table}			

\begin{figure}[h!]
\begin{center}
\includegraphics[width=0.9\textwidth]{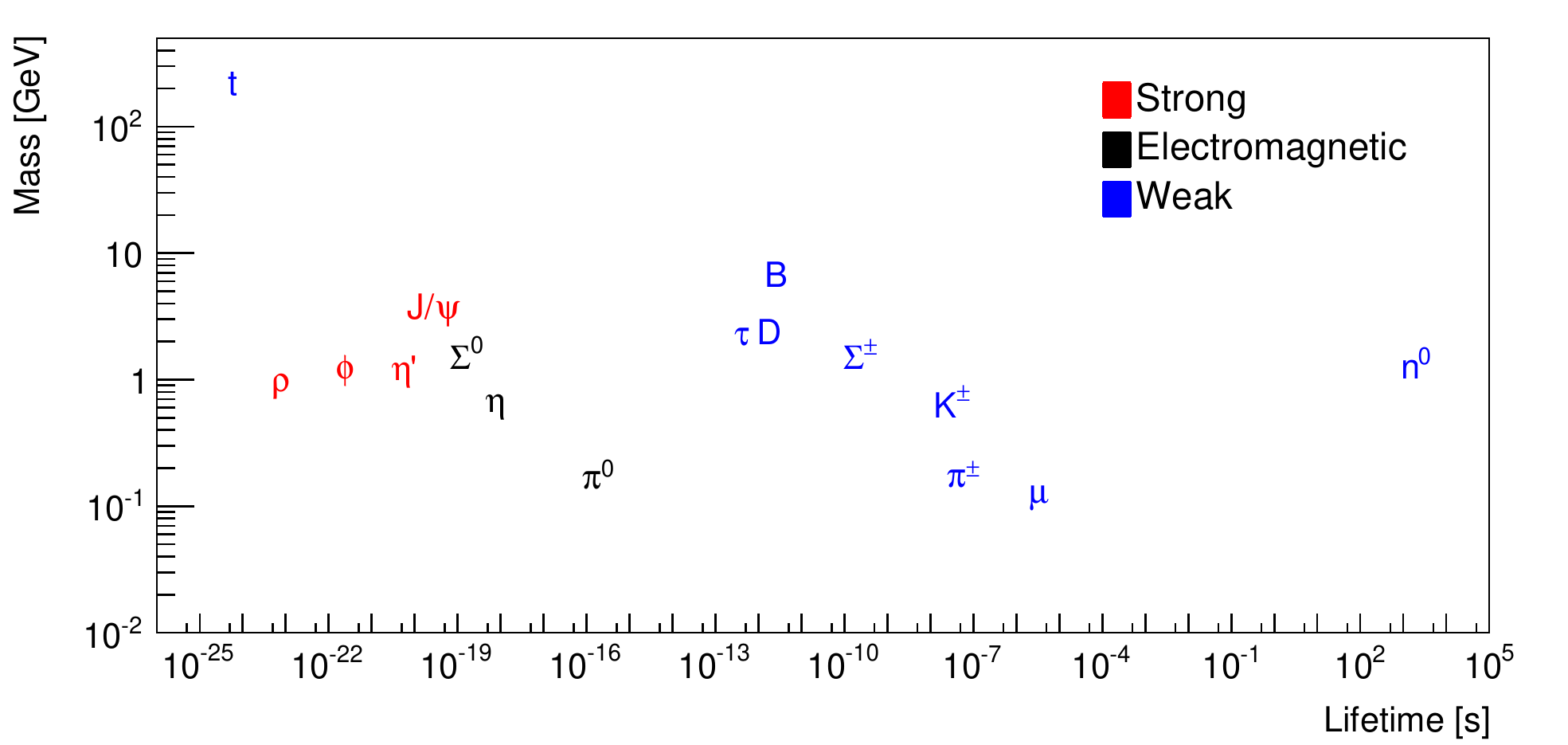}
 \caption{An illustration of the strength of the three forces through the lifetimes of various elementary and composite particles.  The color of the labels is determined by the dominant decay mode.  Values of the mass and lifetimes are from Ref.~\cite{pdg}.  The lifetime is defined as the inverse of the decay width (for the strong force resonances, this is what is measured).}
 \label{fig:lifetimes:overview}
  \end{center}
\end{figure}

Another, `practical' way of visualizing the strength of the three forces is through the decay times of various particles.  Figure~\ref{fig:lifetimes:overview} shows the lifetimes and masses of various elementary and composite particles.  The decay rate $\Gamma$ scales with $g^2$, for gauge coupling $g$.  For charged current weak decays, there is also a factor of $|V_{ij}|^2$ (CKM matrix), which is near unity when the transition is near the diagonal of the CKM matrix.  The decays mediated by the strong force are the fastest, with typical lifetimes $\sim1/(\text{1 GeV})\sim 10^{-24}$ s.  Admixture of electromagnetic decays and phase space factors can increase these lifetimes.  In contrast, the electromagnetic decays are much slower, with $\alpha_\text{EM}<\alpha_s$ reducing $\Gamma$.  Slower still are the weak decays, which can persist for macroscopic timescales.  Even though $g_2>g_1$ (from Fig.~\ref{fig:SM:overview}), the weak decays are highly suppressed because the matrix element squared scales as $g^4/m_W^4$ (often called the Fermi constant, $G_F$ up to an $\mathcal{O}(1)$ constant) when $m\ll m_W$.  The lifetime of the heavier $b$- and $c$-mesons ($B$ and $D$) as well as the $\tau$ are less suppressed than for the lighter mesons and baryons.  In particular, the mass splitting between the proton and the neutron is so small ($\mathcal{O}(0.001)$ GeV) that a free neutron survives for about $15$ minutes on average.  However, Fig.~\ref{fig:lifetimes:overview} also reflects the fact that $g_2$ is not inherently small; the top quark decays via the weak interaction and has a lifetime comparable to the strong force resonances.  This is because $m_\text{top}>m_W$ so there is little phase space suppression.
		
\begin{figure}[h!]
\begin{center}
\includegraphics[width=0.5\textwidth]{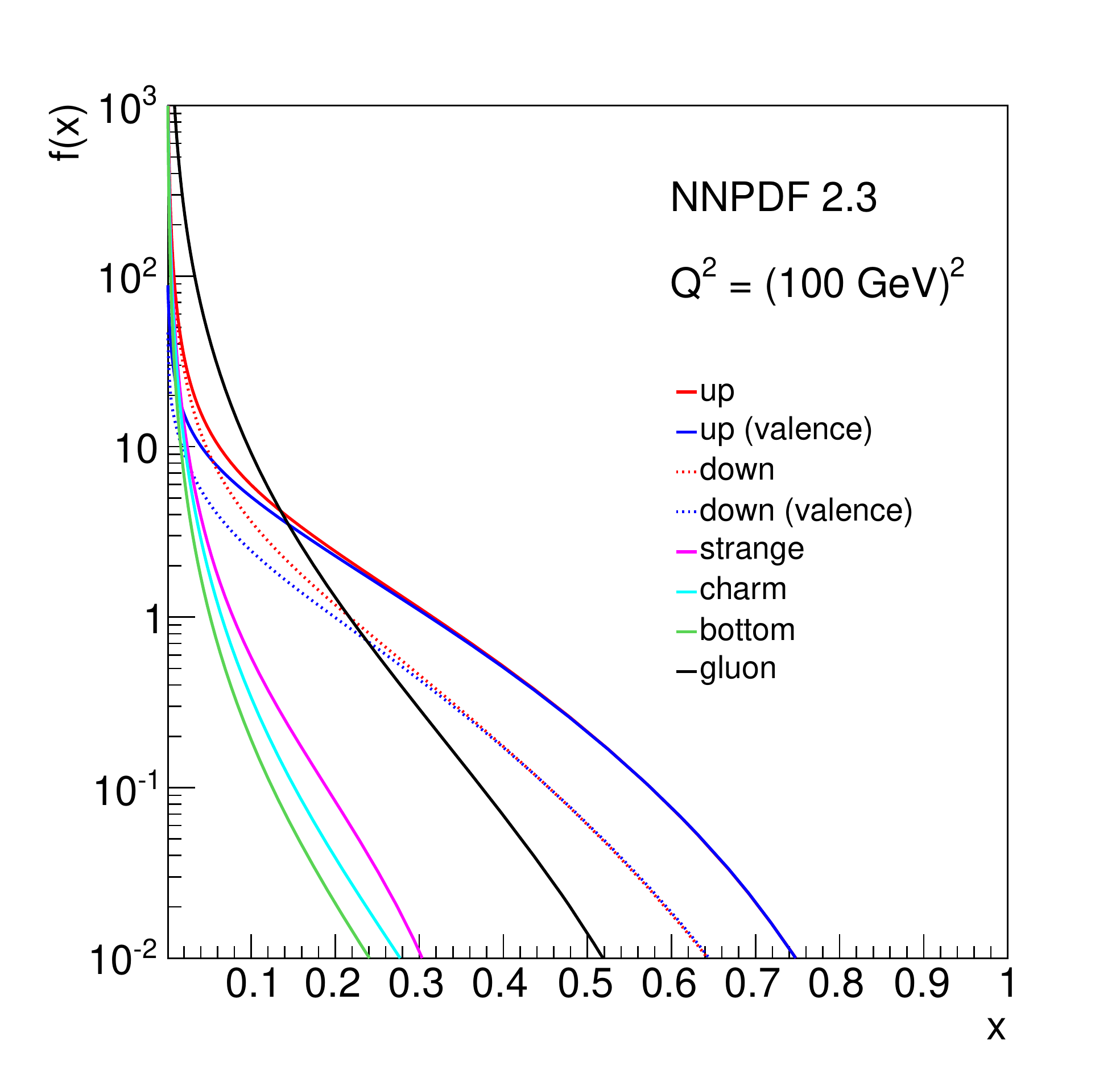}\includegraphics[width=0.5\textwidth]{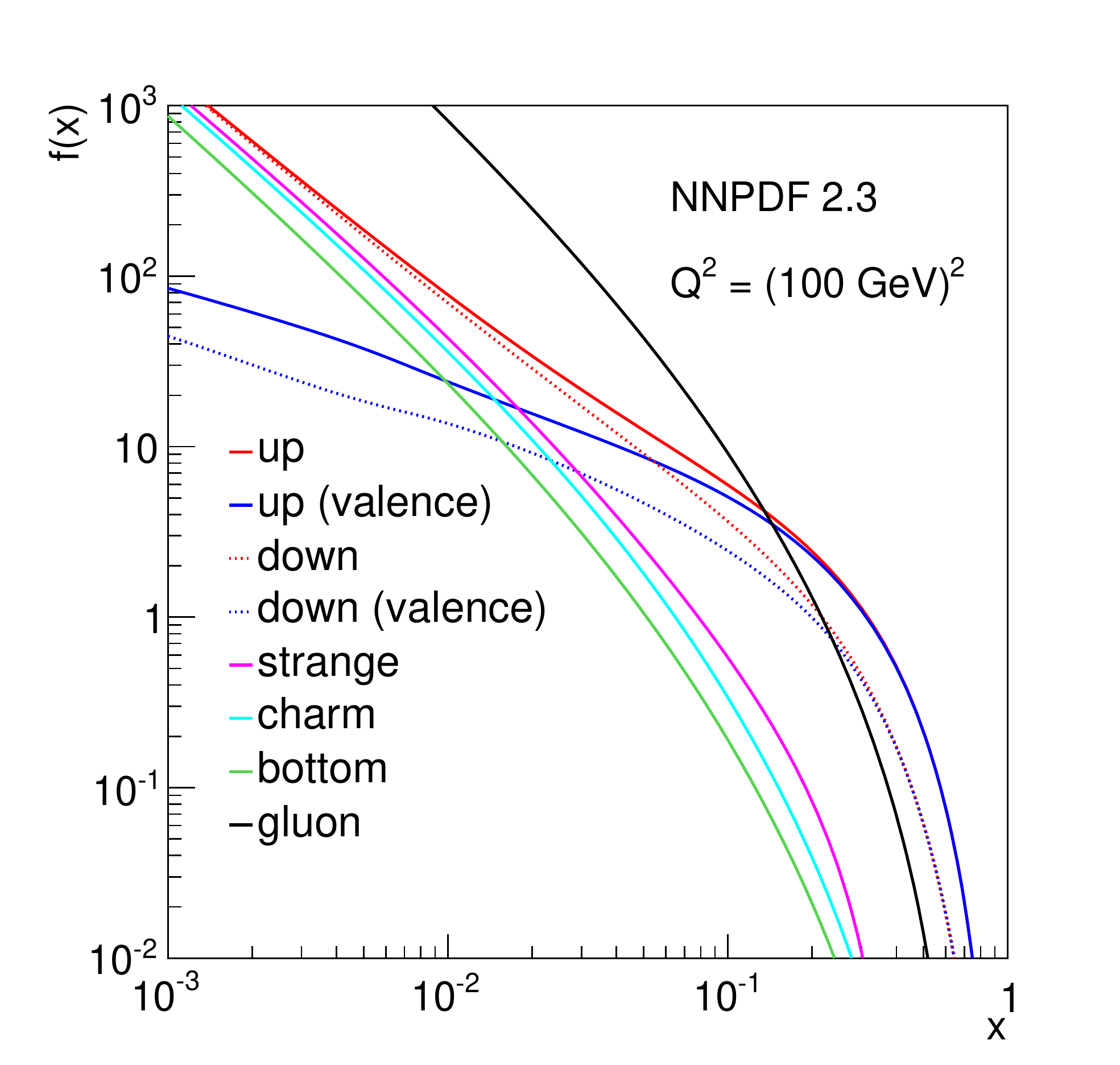}
 \caption{PDF sets from the NNPDF collaboration~\cite{Ball:2012cx} at $Q^2=(100\text{ GeV})^2$ extracted using the HepData~\cite{hepdata} interface. The left and right plots are the same aside from the scale on the horizontal axis.}
 \label{fig:pdfsets}
  \end{center}
\end{figure}

	Aside from the top quark, the other five quarks are not present in Fig.~\ref{fig:lifetimes:overview}.  This is because the strong force exhibits {\it asymptotic freedom}~\cite{Gross:1973id,Politzer:1973fx} at high energy and {\it confinement} at low energy.   The effective coupling for the three forces are energy-scale dependent, as governed by the Callan-Symanzik equations~\cite{Callan:1970yg,Symanzik:1970rt}: 
	
	\begin{align}
	E\frac{\partial g}{\partial E}=\beta(g),
	\end{align}
	
	\noindent where the $\beta$-function on the right-hand side is computed in perturbation theory.  For QCD\footnote{The general equation for the three gauge couplings is given in Sec.~\ref{sec:hiearchyproblem}.  The general form for $SU(N)$ given in e.g. Sec. 16.7 of Ref.~\cite{Peskin:1995ev} only applies to QCD because the other gauge couplings (before electroweak symmetry breaking) involve a scalar field.  The factor $7=11-\frac{2}{3}n_f$ assumes there are only $6$ quarks.}, $\beta(g_3)=-7g^3/(4\pi)^2<0$.  This means that the coupling strength of the strong force is stronger at lower energies and weaker at higher energies.  The top quark lifetime is sufficiently short that it decays before the strong force confines.  In contrast, the other quarks live long enough so that after $\sim 10^{-24}$ s, the strong force is so strong that quarks and gluons are created from the potential energy to surround the bare quarks in color neutral configurations (hadrons).  Aside from the $\mu,\tau$, and top quark, all of the particles in Fig.~\ref{fig:lifetimes:overview} are hadrons.  Hadrons built from a three-quark configuration are called {\it baryons} while those constructed from two-quark configurations are called {\it mesons}.  One well-known baryon is the proton, which is composed of two up-quarks and one down-quark.  The mass of the proton is about $1$~GeV even though its constituent quarks have masses in the MeV range.  This is due to the binding energy from the strong force\footnote{Ironically, even though the Higgs boson has been called the `God particle' that gives all particles their mass, most of the mass around you is due to the binding energy from the strong force and not the Higgs mechanism.  More appropriately, the Higgs mechanism gives rise to radioactivity through bestowing an electroweak scale mass to the $W$ and $Z$ bosons.} that is realized by continuous exchange of soft gluons between the quarks.  These gluons are also `in' the proton and due to quantum fluctuations from gluon splitting, an entire {\it sea} of quarks and gluons are also `in' the proton. The precise statement that a proton is composed of two up quarks and a down quark (called valence quarks) is

	\begin{align}
	\int_0^1 (f_u(x)-f_{\bar{u}}(x))dx=2\hspace{5mm}\int_0^1 (f_d(x)-f_{\bar{d}}(x))dx=1\hspace{5mm}\int_0^1 (f_q(x)-f_{\bar{q}}(x))dx=0,
	\end{align}
	
	\noindent where $q\in\{s,c,b,t\}$ and $f_q(x)$ are {\it parton distribution functions} (PDF) that describe the probability density for a parton of type $q$ to carry a momentum fraction $x$ of the proton.   These functions also depend on the energy scale $|Q|$ at which the proton is probed: $f_q(x)=f_q(x,Q^2)$.  The timescale for the dynamics of parton creation and destruction inside the proton is bounded by $\sim 1/(1\text{ GeV})\sim 10^{-24}$ seconds.  However, when two protons collide at the LHC with energies of $\sim 10$ TeV, the protons pass through each other on the scale of $\sim 1/(10\text{ TeV})\sim 10^{-29}$ seconds.  With this separation of scales, the proton-proton collision is really parton-parton scattering where all of the relevant non-perturbative information about the proton dynamics are neatly bundled into the PDFs.  Figure~\ref{fig:pdfsets} shows representative PDF sets at a hard-scatter scale of $Q^2=(100\text{ GeV})^2$.  The valence $u$ and $d$ PDFs (defined as $f_q-f_{\bar{q}}$) dominate at high $x$ and then all the sea quark PDFs approach each other (and diverge) at low $x$.  The gluon PDF dominates below $x\sim 0.2$.

\vspace{13mm}	
	
\vspace{13mm}

	As they are inherently non-perturbative, PDFs cannot be calculated from an $\alpha_s$ expansion in QCD, but the {\it energy-dependence} of PDFs can be calculated and is an essential ingredient to cross-section predictions at a hadron collider.   The energy ($Q^2$) dependence is governed by the Dokshitzer-Gribov-Lipatov-Altarelli-Parisi (DGLAP)~\cite{Dokshitzer:1977sg,Gribov:1972ri,Altarelli:1977zs} equations:
	
	\begin{align}\nonumber
	\label{dglap}
	\mu\frac{d}{d\mu}f_q(x,Q^2)&=\frac{\alpha_s(Q^2)}{\pi}\int_x^1 \frac{dz}{z}\left[P_{q\leftarrow q}\left(z\right)f_q\left(\frac{x}{z},Q^2\right)+P_{q\leftarrow g}\left(z\right)f_g\left(\frac{x}{z},Q^2\right)\right]\\\nonumber
	\mu\frac{d}{d\mu}f_{\bar{q}}(x,Q^2)&=\frac{\alpha_s(Q^2)}{\pi}\int_x^1 \frac{dz}{z}\left[P_{q\leftarrow q}\left(z\right)f_{\bar{q}}\left(\frac{x}{z},Q^2\right)+P_{g\leftarrow q}\left(z\right)f_g\left(\frac{x}{z},Q^2\right)\right]\\\nonumber
	\mu\frac{d}{d\mu}f_g(x,Q^2)&=\frac{\alpha_s(Q^2)}{\pi}\int_x^1 \frac{dz}{z}\left[P_{g\leftarrow g}\left(z\right)f_{g}\left(\frac{x}{z},Q^2\right)\right.\\
	&\hspace{10mm}+\left. P_{g\leftarrow q}\left(z\right)\sum_{q'} \left(f_{q'}\left(\frac{x}{z},Q^2\right)+f_{\bar{q}'}\left(\frac{x}{z},Q^2\right)\right)\right],
	\end{align}
	
	\noindent where $\mu$ is the running scale and the functions $P_{p_2\leftarrow p_1}$ are the Altarelli-Paressi splitting functions that encode the probability for a parton $p_1$ to split or radiate parton $p_2$.  At leading order in $\alpha_s$, the splitting functions are given by
	
	\begin{align}
	P_{q\leftarrow q}(x)&=C_F \left[\frac{1+x^2}{1-x}\right]_+\\
	P_{g\leftarrow q}(x)&=C_F \frac{1+(1-x)^2}{x}\\
	P_{q\leftarrow g}(x)&=T_R\left[x^2+(1-x)^2\right]\\
	P_{g\leftarrow g}(x)&=2C_A\left[\frac{x}{(1-x)_+}+\frac{1-x}{x}+x(1-x)\right]+\delta(1-x)\frac{11C_A-4n_fT_R}{6},
	\end{align}	
	
	\noindent where the plus-notation $[f(x)]_+$ is defined in the context of an integral equation $\int_0^1 dx g(x)[f(x)]_+=\int_0^1 dx(g(x)-g(1))f(x)$.  The `color factors' $C_A=3,C_F=4/3$ and $T_F=1/2$ are properties of the $SU(3)$ QCD algebra.
	
	The intuition for e.g. the first line of Eq.~\ref{dglap} is as follows: a quark of type $q$ with momentum fraction $x$ could be due to either a quark of the same type with proton momentum fraction $x'\geq x$ that has radiated a gluon or due to a gluon with momentum fraction $x'\geq x$ that has split into a $q\bar{q}$ pair.  The energy fraction of the initial quark or gluon carried by the final quark or gluon is $z=x/x'$ which means that the proton momentum fraction of the initial quark or gluon is $x'=x/z$.  The probability for the final quark to be from an initial quark is (heuristically) $\Pr(\text{$q$ from $q$})=\Pr(\text{final $q$}|\text{initial $q$})\Pr(\text{initial $q$})=P_{q\leftarrow q}(z)f_q(x/z)$.  Similarly for the gluon term, $\Pr(\text{$q$ from $g$})=\Pr(\text{final $q$}|\text{initial $g$})\Pr(\text{initial $g$})=P_{q\leftarrow g}(z)f_g(x/z)$.  The integral in the first line in Eq.~\ref{dglap} is over all emissions from the initial quark or gluon and the $dz/z$ is the phase space for these emissions (see Sec.~\ref{sec:mass:theory}).  The DGLAP equations  and the splitting functions will be revisited in more detail in Part~\ref{part:qpj}.

	Even though proton-proton collisions are well-described by parton-parton scattering, no out-going parton has ever been directly observed.  As the out-going partons travel away from the interaction point, the same processes that fill protons with a sea of quarks and gluons generate a shower of partons described in the soft and collinear limits by the Altarelli-Paressi splitting functions.  Once this {\it parton shower} has cooled to an energy $\lesssim 1$ GeV, the partons hadronize due to confinement.  The resulting collimated spray of hadrons is known as a {\it jet}.  Jets are ubiquitous at the LHC because of the prevalence of quark and gluon radiation from both the initial partons in the proton (initial-state radiation) as well from out-going quarks and gluons participating in the hard-scatter process (final-state radiation).  Information about the initiating quark or gluon is embedded in the complex radiation pattern within jets.  Part~\ref{part:qpj} is dedicated to study of this radiation.

		\clearpage
	
		\section{Successes and Limitations}
		\label{SMproblems}

\begin{figure}[h!]
\begin{center}
\includegraphics[width=0.6\textwidth]{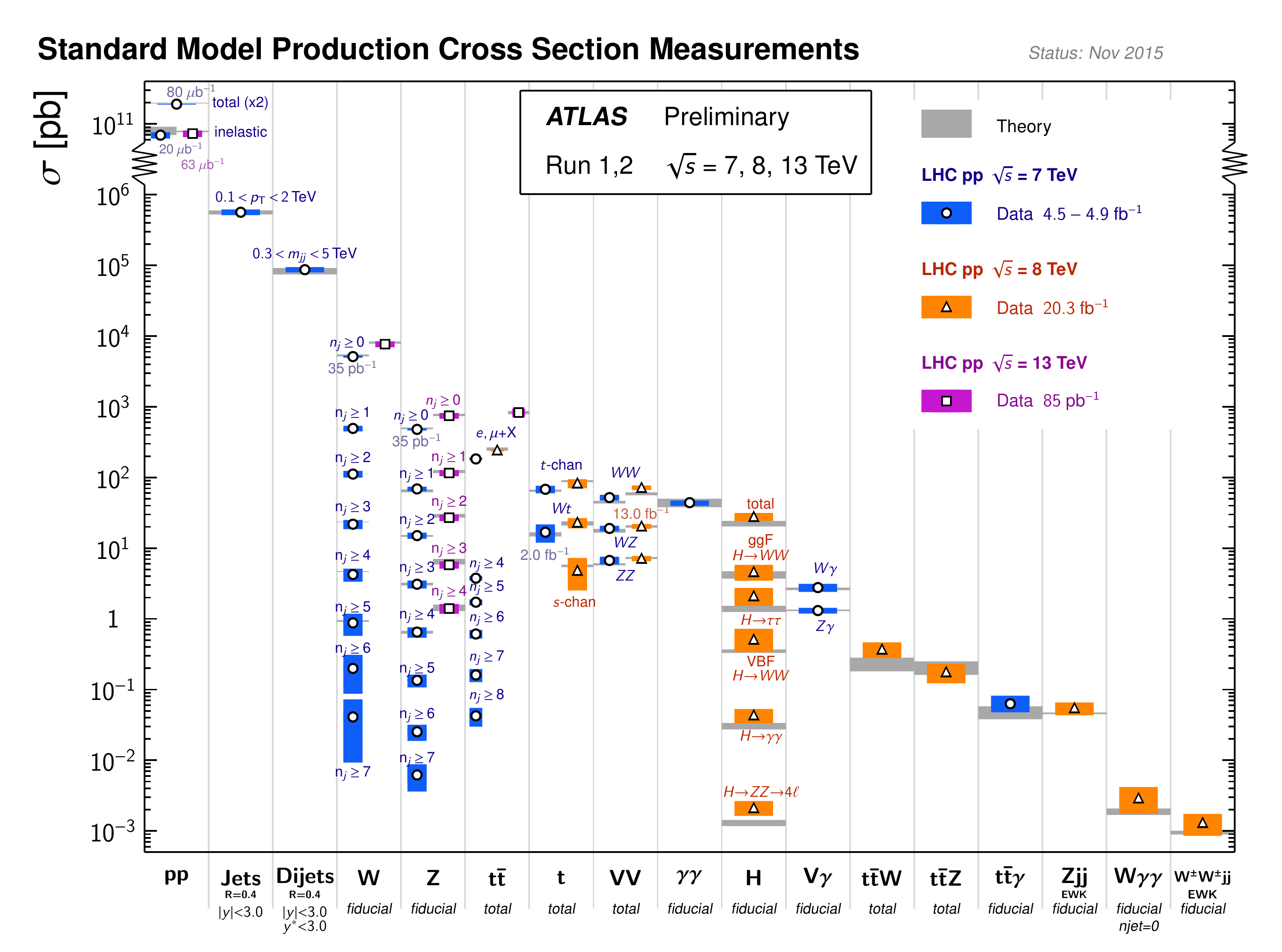}
 \caption{Cross section measurements at $\sqrt{s}=7, 8, 13$ TeV from the ATLAS experiment compared to theoretical measurements.  From Ref.~\cite{SMresults}.}
 \label{fig:lifetimes:overviewSM}
  \end{center}
\end{figure}

The Standard Model is incredibly successful.  Increasingly precise calculations in the context of perturbation theory have accurately predicted and matched cross section measurements over $10$ orders of magnitude at the LHC alone - see Fig.~\ref{fig:lifetimes:overviewSM}.  These calculations and measurements span a wide range of processes probing all three fundamental forces.  In addition to the LHC measurements shown in Fig.~\ref{fig:lifetimes:overviewSM}, there are numerous collider- and non-collider-based experiments that probe various aspects of the SM.   One of the most impressive single measurement is the anomalous magnetic moment of the muon, $g_\mu$.  This quantity can be calculated and measured to nine significant figures.  The value of $g-2$ is dominated by QED, but the accuracy is such that there are non-negligible contributions from electroweak and hadronic processes in loops\footnote{For a nice overview, see Ref.~\cite{Hoecker:2010qn}.  There is some tension between $g-2$ and the SM prediction, but the agreement spans many orders of magnitude.}.  A related success story is the use of complementary measurements across colliders and energies to indirectly determine parameters that are not measured directly (global fits)~\cite{Baak:2012kk}.  Without $m_h$, the $m_\text{top}$ uncertainty is $\sim 4$ GeV (assuming there is only the SM) when the direct measurement has $\sim 1$ GeV uncertainty. 

Despite its great success, there are significant issues as well.  Some of these problems are due to an inability of the SM to describe known phenomena while others are mostly aesthetic problems, but suggestive of the SM's incompleteness.  There are no known logical inconsistencies with the theory.  One important physical sector not described by the SM is gravity.  Currently, there is no coherent quantum theory of gravity.  For most physical phenomena that are known or will be studied in the near future, the SM augmented with general relativity is an accurate description of nature.  However, there are extreme phenomena where quantum gravity effects are non-negligible (see Sec.~\ref{sec:hiearchyproblem} about the Planck scale).  New approaches to QFT such as string theory are a promising direction, but there are currently no unique testable predictions from such models (see e.g. Ref.~\cite{Schwarz:1998ny}).  

Another aspect of nature not described by the SM is dark matter.  There is overwhelming evidence\footnote{See for instance the evidence from the velocity profile of galaxies~\cite{1985ApJ305V} and from the analysis of colliding galaxy clusters~\cite{Clowe:2003tk,Markevitch:2003at}.} that most of the gravitationally interacting matter in the universe is not composed of SM particles.  There is a small component of the non-luminous matter due to neutrinos, but they explain less than one percent of the total dark matter relic density (See e.g. Ref.~\cite{PDG2}).  Massive weakly interacting particles (WIMP) are an excellent dark matter candidate and these particles are a natural aspect of supersymmetric (SUSY) extensions of the SM.  This will be revisited in more detail with Sec.~\ref{sec:hiearchyproblem}.

Another phenomenon\footnote{There are others, including dark energy (related to gravity) and the imbalance between matter and anti-matter.} not explained by the SM is the neutrino mass.  In the SM introduced in Sec.~\ref{sec:particlesandforces}, neutrinos are massless, but it is now known that neutrinos have a nonzero mass~\cite{Fukuda:1998mi}.  One could readily accommodate neutrino masses by adding a Yukawa coupling $y_\nu$ to Eq.~\ref{eq:SM}.  However, the neutrino masses are known to be less than about $0.3$ eV~\cite{PhysRevLett.112.051303}, so there is an enormous hierarchy $y_{\nu_\tau}/y_\tau\sim 10^{-10}$ that is unexplained.  Also, the off-diagonal elements of the corresponding mixing matrix (the CKM matrix analogue) are experimentally constrained to be much smaller than the off-diagonal elements of the CKM matrix.  There is a plethora of theories to extend the SM to naturally explain the smallness of the neutrino mass - see Ref.~\cite{King:2003jb} for a recent review.

The problem of neutrino masses is mostly aesthetic because there is a mechanism in the SM for generating neutrino masses, but the associated parameters ($y_\nu$) are small.  Another issue with the SM of this type is called the {\it strong CP problem}.   In principle, there could be a term in Eq.~\ref{eq:SM} of the form $\theta F_{\mu\nu}^a\epsilon_{\mu\nu\rho\sigma}F^{\rho\sigma a}$, where $\theta$ is a dimensionless parameter and $F$ is the gluon field strength tensor (see e.g. Ref.~\cite{Dine:2000cj}).  The parameter $\theta$ is experimentally constrained by neutron electric dipole moment measurements~\cite{Baker:2006ts,Baluni:1978rf,axionwitten} to be $\theta<10^{-10}$.  A set of popular theories to naturally explain why $\theta$ is so small is to augment the SM with particles called axions~\cite{Peccei:1977hh,Peccei:1977ur,Weinberg:1977ma,Wilczek:1977pj}.  

Arguably the issue with the SM that has received the most theoretical and experimental attention is the size of the Higgs boson mass.  Unlike the fermions and gauge bosons, there is no symmetry principle which protects the mass of the Higgs boson from quantum corrections.  These corrections make the mass sensitive to particles and forces at the highest energy scales.  As a result, there is an enormous unnatural hierarchy between the measured Higgs boson mass and the {\it Planck scale} ($10^{19}$ GeV) where quantum gravity must be important.  This {\it hierarchy problem} is one of the main motivations for SUSY and therefore an entire section is devoted to describe it properly (Sec.~\ref{sec:hiearchyproblem}).

The SM has other (minor) aesthetic and practical problems that may suggest it is incomplete.  For example, there is a large unexplained mass hierarchy for the known SM particles,  $y_e/y_t\sim 10^{-6}$.  In fact, there is no reason within the SM for any of the $18$ dimensionless parameters.  As a result of $g_1\ll g_3$, perturbation theory results in extremely precise predictions for electrodynamic processes but low energy hadron spectra are incalculable in perturbation theory (and difficult to calculate with lattice techniques - see Ref.~\cite{Gupta:1997nd}).  In addition, the non-perturbative nature of low energy QCD requires the introduction of phenomenological models for e.g. hadronization that have many new (non-fundamental) parameters.  

The sensitivity to precise SM measurements to non-perturbative modeling will be revisited in Part~\ref{part:qpj} and all of Part~\ref{part:susy} is dedicated to a search for new particles {\it beyond the SM} (BSM).

\clearpage

\chapter{Experimental Apparatus}
\label{chapter:ATLAS}

\begin{figure}[h!]
\begin{center}
\includegraphics[width=0.6\textwidth]{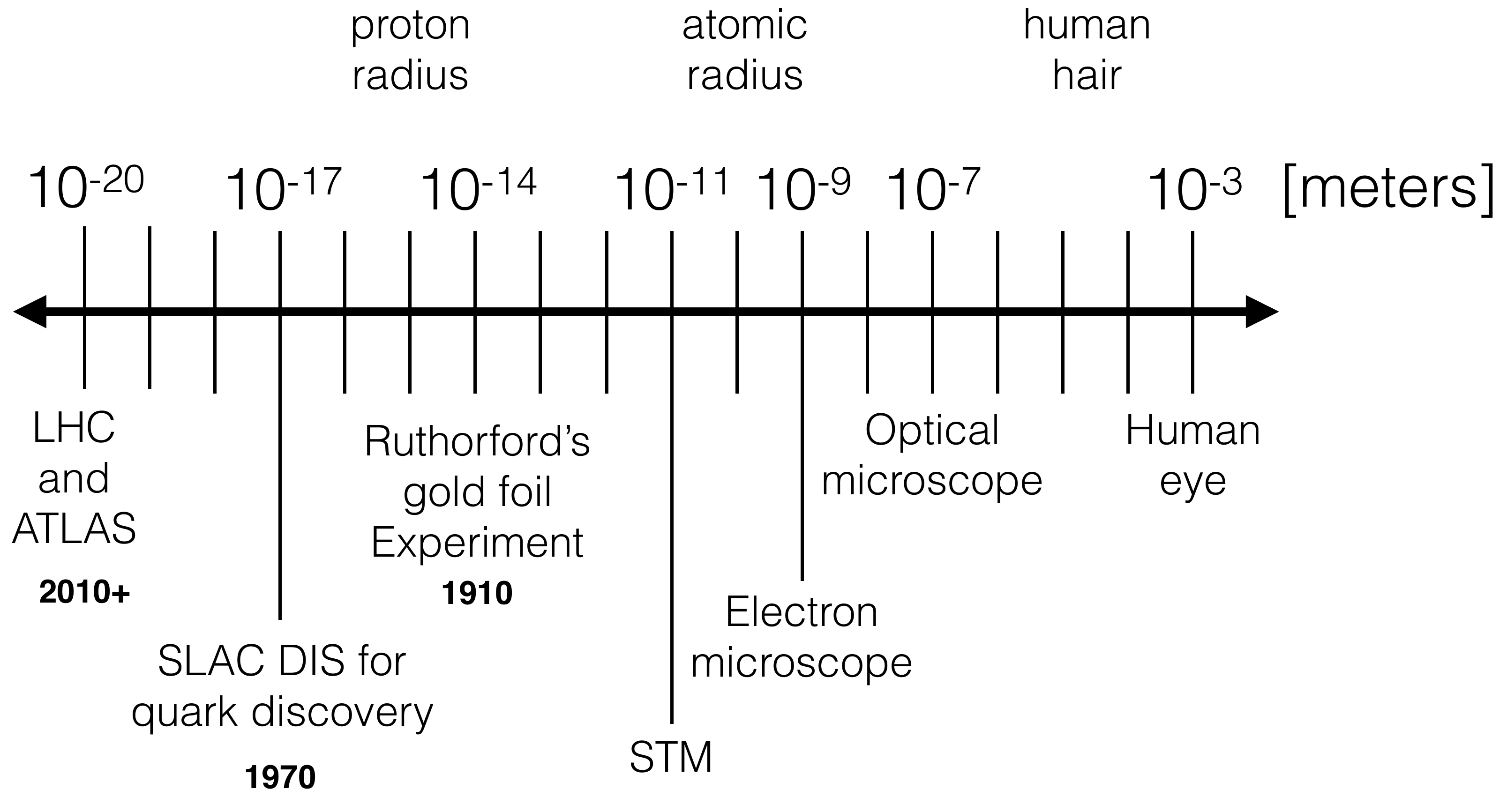}
 \caption{Diagram illustrating the length scales probes by various `microscopes.'}
 \label{fig:microscopes}
  \end{center}
\end{figure}

		Measuring the properties of nature at the smallest distance scales ever recorded with a terrestrial apparatus requires the highest energy particles accelerator ever built.  The size of a structure is related to the probe energy via the de Broglie relation $\lambda = 1/p$.  The same principle governs optical microscopes, limiting their resolution to hundreds of nanometers.  The Large Hadron Collider (LHC) produces protons with energies up to $\sqrt{s}=13$ TeV.  At this energy, {\bf A} {\bf T}oroidal {\bf L}HC {\bf A}pparatu{\bf S} (ATLAS) is able to capture the byproducts of the proton-proton collisions to probe distance scales as small as $10^{-20}$ m.  For comparison, Fig.~\ref{fig:microscopes} shows various technologies that have been used to measure increasingly smaller distance scales.  This chapter explores how the LHC works and how particles are measured by the ATLAS detector.  
		
		\clearpage

		\section{The Large Hadron Collider}

\begin{figure}[h!]
\begin{center}
\includegraphics[width=0.5\textwidth]{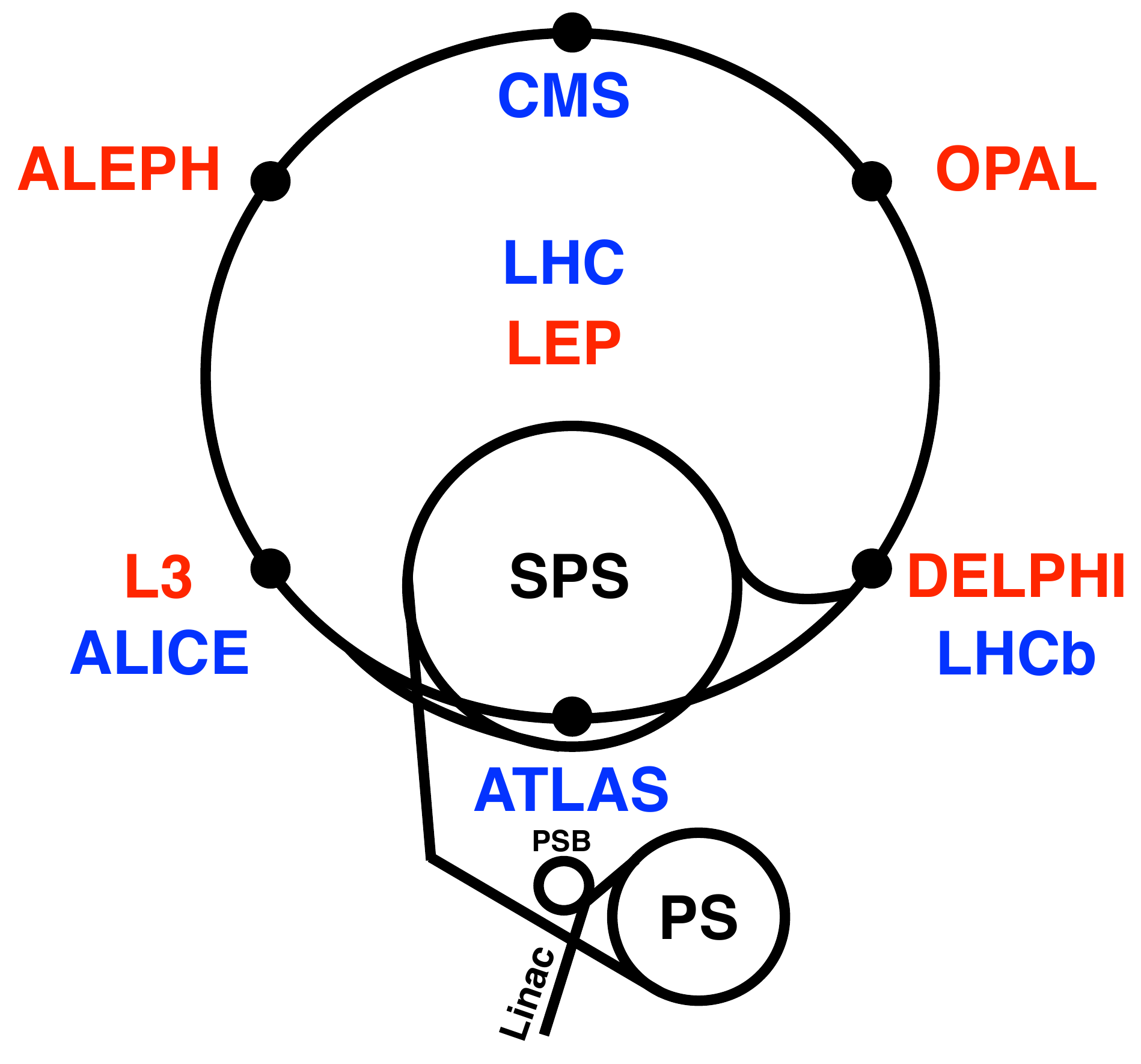}
 \caption{An overview of the CERN accelerator complex.  The LEP experimental areas are indicated in red while the four LHC experiments are labeled in blue.  }
 \label{fig:CERNcomplex}
  \end{center}
\end{figure}

		The life of a proton at the LHC begins as hydrogen.  After being stripped of its electron, the remaining protons proceed through a series of accelerators to successively increase their speed.  To begin, a linear accelerator increases the proton energy to $50$ MeV ($\beta\approx 5\%$).  Then, a small ($25$ m in radius) circular accelerator called the Proton Synchrotron (PS) Booster increases the energy to $1.4$ GeV ($\beta\approx 80\%$) after which a larger synchrotron ($100$ m in radius), the PS, increases the energy to $25$ GeV ($\beta\approx 99.9\%$).  While never itself used as a particle collider, the PS has a rich history~\cite{Plass:1443556} providing a variety of beams to other experiments such as a neutrino beam to the Gargamelle bubble chamber where weak neutral currents were discovered in 1974~\cite{Hasert:1973ff}.  Following the PS, protons are accelerated to $450$ GeV in the $7$ km (in circumference) Super Proton Synchrotron (SPS).   Like the PS, the SPS provides beams for a variety of experiments.  The SPS has also played an important role as a collider in its own right, such as facilitating the discovery of the $W$ and $Z$ bosons by UA1~\cite{Arnison:1983rp,Arnison:1983mk} and UA2~\cite{Banner:1983jy,Bagnaia:1983zx}.  The SPS directly injects into the $27$ km LHC where the energy is ramped up to $\sqrt{s}=8$ TeV (Run 1) or $\sqrt{s}=13$ TeV (Run 2). There are a series of crossing points where the beams collide.  An overview of the CERN accelerator complex is shown in Fig.~\ref{fig:CERNcomplex}. Prior to its use as a proton-proton collider, the LHC tunnel was filled with an $e^+ e^-$ accelerator called the Large Electron Positron (LEP) collider, which had four experiments.  The LHC ring also has four collision points, with two multipurpose experiments ATLAS and the Compact Muon Solenoid (CMS) as well as two special detectors ALICE and LHCb.  				
		
		Particle acceleration occurs via radio frequency (RF, 400 MHz) cavities driven by high-power klystrons.  Each cavity in the LHC provides a gradient of 5 MV/m.  There are eight such cavities, each supplying 2 MV for a total of $16$ MeV added per beam per revolution.  The RF must be an integer multiple of the revolution frequency (asynchronous protons slow down/speed up accordingly) which induces a {\it bucket structure} whereby there are about $36,000$ possible locations along the beam for packets of protons.  For protons traveling at nearly the speed of light, this means that the buckets are separated by about $2.5$ ns.  Only $10\%$ of the possible buckets are filled resulting in a nominal collision rate of 40 MHz.  The buckets that are filled with {\it bunches} contain about $100$ billion protons.   Protons are steered around the LHC ring using dipole magnets (and focused using quadruple magnets).  The relationship between the (dipole) magnet strength $B$, the radius of the accelerator $R$, and the energy of the protons $E$ is given by\footnote{This is $qvB = mv^2/R$ from equating the magnetic and centripetal forces, but accounting for the relativistic factor $\gamma$ so that $p=\gamma mv = qBR$.  For protons, $E\sim p$, and the elementary charge is $e=\sqrt{4\pi\alpha}$.  See Table~\ref{tab:units} for the unit conversions.}
	 
	 \begin{align}
	 \label{eq:magnet}
	E=\sqrt{4\pi\alpha}\times\left(\frac{B}{\text{T}}\right)\left(\frac{R}{\text{m}}\right) \approx 800\times\left(\frac{B}{\text{T}}\right).
	 \end{align}
	 
	 \noindent Therefore, at $\sqrt{s}=8$ TeV, dipole magnets at $\sim 5$ T are required and at $\sqrt{s}=13$, the magnets need to be powered to about $8$ T.  To achieve such high field magnets, super conducting Niobium-Titanium coils must be cooled down to $1.9$ K using superfluid helium-4.  From Eq.~\ref{eq:magnet}, for a fixed radius accelerator, the collision energy is set by the strength of the dipole magnetic field.  The availability of robust high (enough) temperature superconductors was a limiting factor to the design energy of the LHC.  There are promising alternatives to Ni-Ti such as Ni${}_3$-Sn, but this is still an area of active research.  One of the main limitations in the number of protons per bunch is from the cooling of the magnets.  Accelerating charged particles radiate, resulting in an emitted power per proton (synchrotron radiation) given by\footnote{For a derivation, see e.g. Chapter 8 in Ref.~\cite{marian} or Chapter 14 in Ref.~\cite{jackson}.  The full formula is $P=e^4\gamma^2B^2/(6\pi\epsilon_0m^2c)$.  See Table~\ref{tab:units} for unit conversions.}
	 
	 \begin{align}
	 \label{power}
	 P \approx (4\times 10^{-15}\text{ W})\left(\frac{B}{\text{T}}\right)^2\left(\frac{E}{\text{TeV}}\right)^2\left(\frac{m}{\text{GeV}}\right)^{-4},
	 \end{align}

	 \noindent where $m\sim 1$ GeV for protons.  Due to high power of $m$ in the denominator of Eq.~\ref{power}, this is a severe limitation for electron beams.  For proton beams, the synchrotron radiation is highly suppressed but if there are $N_b\sim 10^{11}$ protons per bunch, then the total power at $\sqrt{s}=13$ TeV per bunch is $N_bP\sim 1$ W\footnote{The synchrotron radiation may become dominant in the not-to-distant future if $\sqrt{s}$ is increased by a factor of $10$ at a future collider~\cite{Keil:327302}.}.   There are about $3,000$ bunches (just below $10\%$ of the available buckets) so the total power per beam per meter is about $0.1$ W/m.  This is one of the main challenges to the cryogenics~\cite{Lebrun:411139} and leads to an important justification for keeping $N_b\lesssim 10^{11}$.  Another factor is the collision rate, which is set by the size of the bunches and the number of protons per bunch.  The transverse size of a bunch $\sigma$ is given by $\sqrt{\beta\epsilon}$ where $\epsilon$ is the area of the beam in phase space ({\it beam emittance}) and $\beta$ ({\it betatron function}) captures the changes in the beam due to focusing magnets.  By Louisville's theorem, $\epsilon$ does not depend on the position along the LHC.  In a region without a magnetic field, the betatron function has the form~\cite{wiedmann}
	 
	 \begin{align}
	 \label{minbeta}
	 \beta(z-z_0)=\beta(z_0)+\frac{(z-z_0)^2}{\beta(z_0)}.
	 \end{align}
	 
	 \noindent When $z_0$ is the collision point, $\beta(z_0)$ is called $\beta^*$ and because $\epsilon$ is constant along the beam, by Eq.~\ref{minbeta}, $\beta^*$ measures the distance from the collision point at which the transverse size $\sigma$ doubles.  At the LHC, $\beta^*\sim 0.5$ m and the {\it normalized emittance} $\epsilon_N=\beta\gamma\epsilon\sim 3\mu m$~\cite{1742-6596-455-1-012001}.  The eminence itself is actually not conserved along the beam for nonzero acceleration, but the normalized emittance does obey Louisville's theorem.  At $\sqrt{s}=13$ TeV, $\beta\gamma\sim 13000$ so $\epsilon\sim 2\times 10^{-10}$ m resulting in the physical beam size in the lab frame $\sigma\sim 10$ $\mu$m.  Note that this is significantly larger than the `size' of the proton, which is about one femtometer $\sigma_{p}\sim 10^{-15}$ m.  The probability for one proton-proton collision could be estimated by $p\sim \sigma_p^2/\sigma^2\ll 1$.  For $N_b\sim10^{11}$ protons per bunch, the average number of collisions is $p\sim N_b^2\sigma_p^2/\sigma^2\sim100$, which is another reason to keep $N_b\sim 10^{11}$\footnote{Not all of these collisions result in interesting {\it inelastic} scattering in which the protons dissociate (about $50\%$).  The actual number of collisions per bunch crossing will be called $\mu$ and is discussed in the context of {\it pileup} in Sec.~\ref{atlasdetector}.}.  This quick calculation demonstrates that the average number of events from a particular process in a given bunch crossing can be calculated as the product of a process-dependent {\it cross section} and a quantity related to the rate of collisions.  The later quantity will be called the {\it instantaneous luminosity} and is given in full by
	 
	 \begin{align}
	 \mathcal{L} = \frac{N_b^2fn_\text{bunches}F}{4\pi\epsilon\beta^*},
	 \end{align}

	 \noindent where $f$ is the revolution frequency ($40$ MHz) and $F$ is a $\mathcal{O}(1)$ geometric factor to correct for an off-axis crossing angle.  Van Der Meer~\cite{vanderMeer:296752} scans are combined with a variety of techniques for measuring $\mathcal{L}$ in-situ~\cite{Aad:2013ucp}.  For the data collected so far at the LHC, $\mathcal{L}\sim 10^{34}$-$10^{35}$ cm${}^{-2}$s${}^{-1}$.
	 
	 The {\it integrated luminosity}, $\int\mathcal{L}dt$ is used to quantify the amount of collected data.  Units of the integrated luminosity are {\it inverse barns} (b${}^{-1}\approx 10^{28}/\text{m}^2$); the full $\sqrt{s}=8$ TeV dataset was about 20 fb${}^{-1}$. For any process $pp\rightarrow X$, the average number of predicted events for that process is given by $\int dt\mathcal{L}\sigma_{pp\rightarrow X}$, for $\sigma_{pp\rightarrow X}$ calculated\footnote{From QFT and corrected for various detector effects, discussed in Parts~\ref{part:qpj} and~\ref{part:susy}.} in barns.

		\clearpage

		\section{Interactions of Particles with Matter}

\begin{figure}[h!]
\begin{center}
\includegraphics[width=0.5\textwidth]{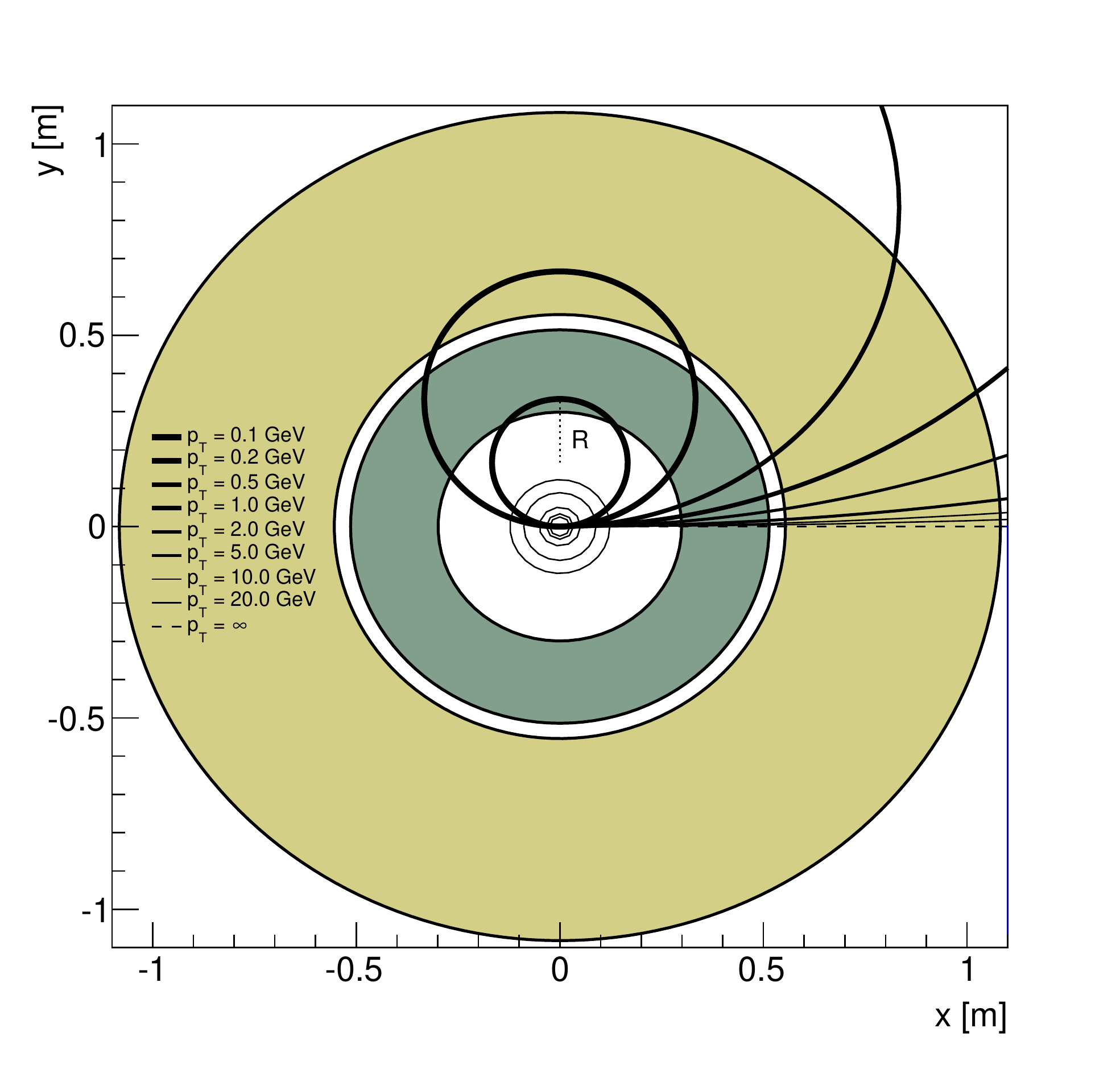}\includegraphics[width=0.5\textwidth]{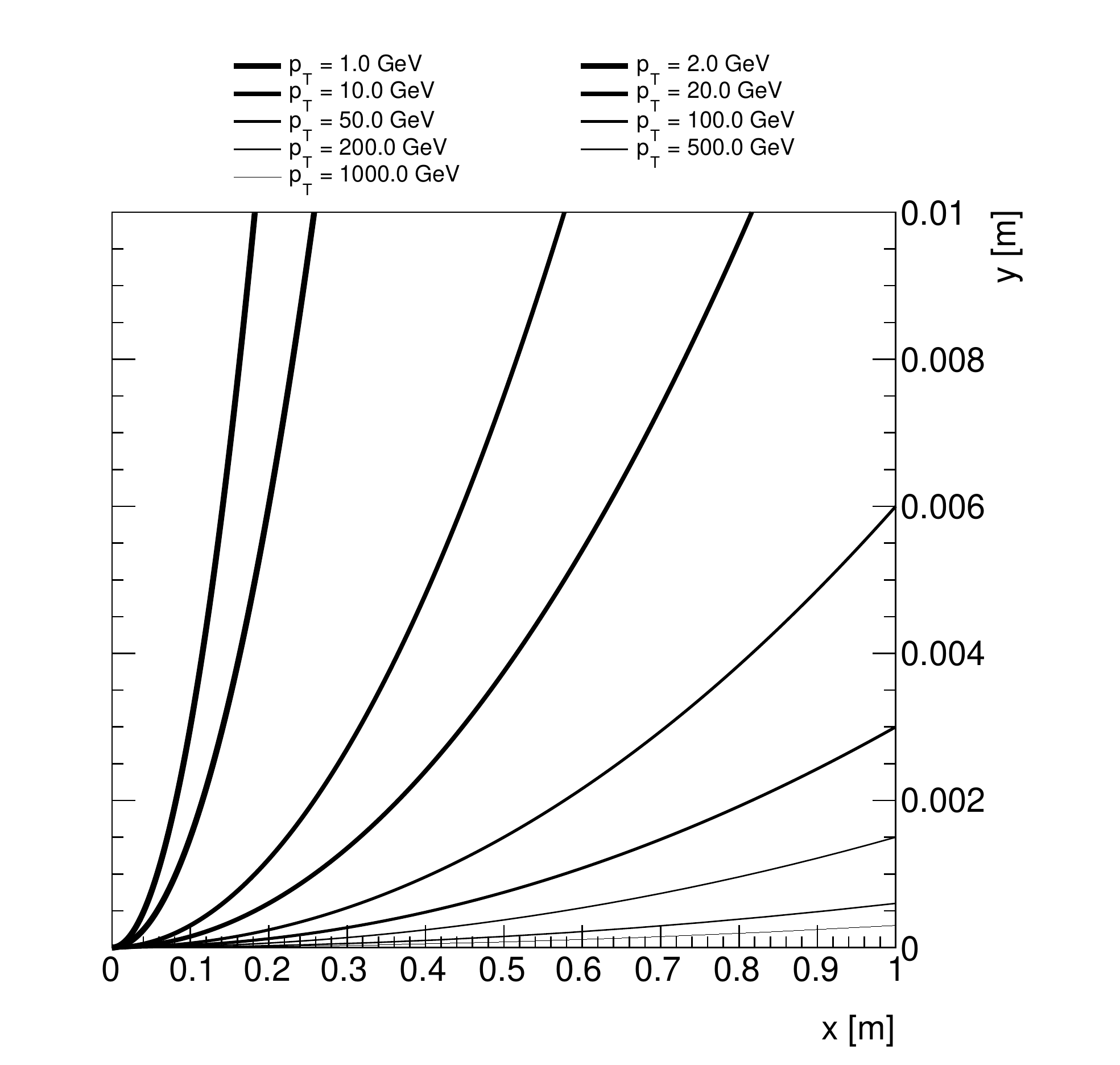}
 \caption{Diagrams illustrating the trajectory of charged particles in a magnetic field.  Both views are cross sections of the detector with the beam axis and $B$-field perpendicular to the page.  Circles on the left plot indicate the locations of various tracking detector elements for the ATLAS detector, discussed in Sec.~\ref{atlasdetector}.  A 2 T solenoid magnet is used to determine $R$.}
 \label{fig:tracktrajectories}
  \end{center}
\end{figure}

\begin{figure}[h!]
\begin{center}
\includegraphics[width=0.5\textwidth]{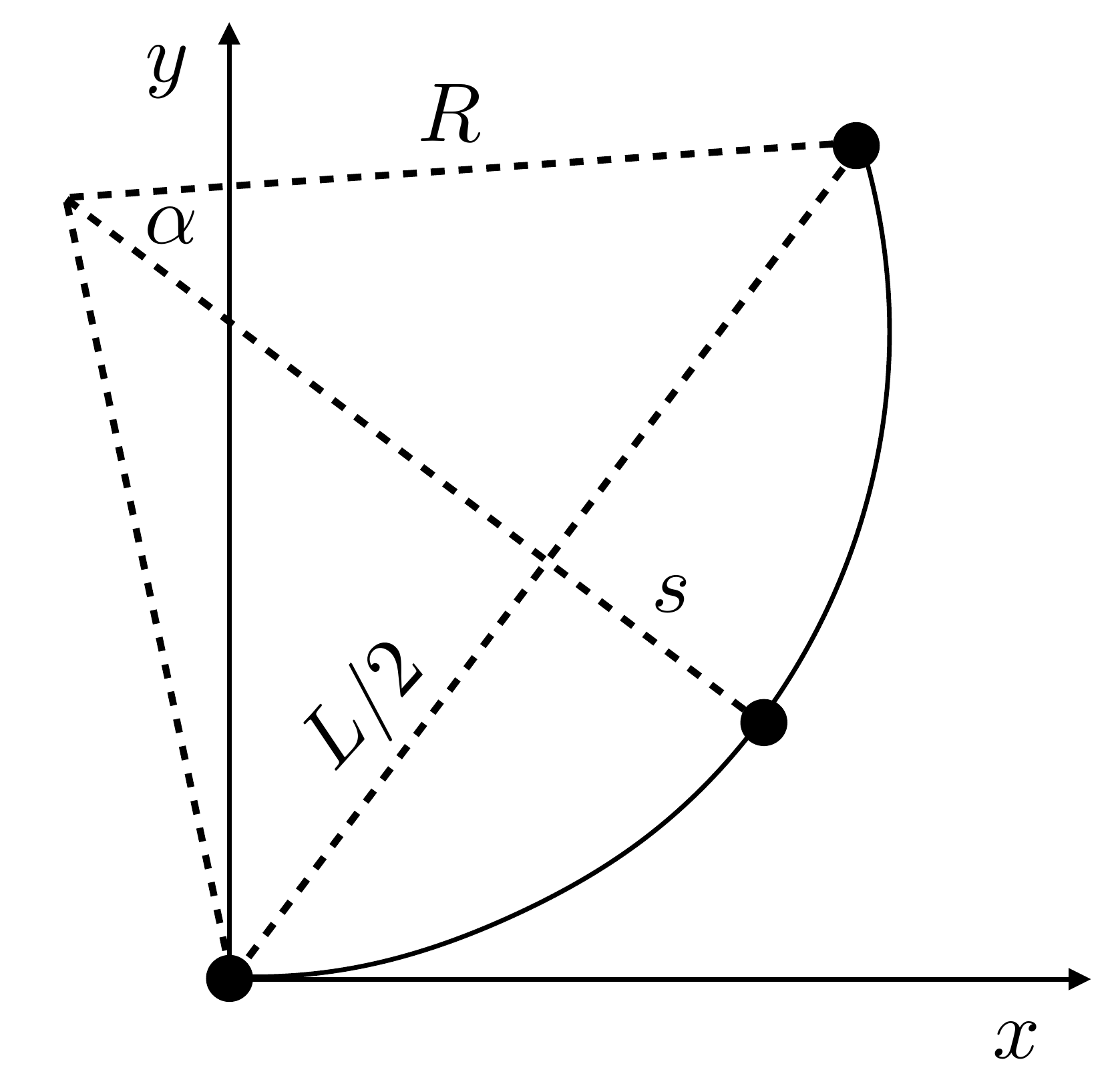}
 \caption{A schematic diagram used to derive the momentum resolution of a charged particle track measurement.  The length $L$ ({\it lever arm}) is the distance over which measurement of the trajectory can be made.  The three dots indicate discrete samplings (measurements) of the trajectory. }
 \label{fig:trackingresolution}
  \end{center}
\end{figure}

		Typical proton-proton collisions delivered by the LHC result in hundreds of particles scattering away from the interaction point.  There are two ways to measure the properties of these particles\footnote{This section will briefly introduce some of the main concepts of particle detection techniques.  There are many books on this subject; see e.g. Ref.~\cite{Grupen:2012zpa,Tavernier:2010zz} and the PDG review~\cite{pdg}.}.  One possibility is to passively observe secondary particle production without disturbing the trajectory of the primary particle.  These techniques are available for charged particles, which can interact electromagnetically with a detector without loosing a significant fraction of their energy.  The distribution of secondary particles contains information about the momentum and type of the original particle.  If a series of such measurements are made along the trajectory of the particle, a fit can reconstruct the particle trajectory with high precision.  The transverse momentum of a particle is related to the curvature of its trajectory in a magnetic field perpendicular to its motion by\footnote{This is the same formula used to derive Eq.~\ref{eq:magnet}, only now the magnetic field is parallel to the beam and perpendicular to the particle trajectory.} 
				
		\begin{align}
		\label{pTbfield}
		\frac{p_\text{T}}{\text{GeV}} \approx 0.3\left(\frac{B}{\text{T}}\right)\left(\frac{R}{\text{m}}\right) ,
		\end{align}
		
		\noindent where one unit of the electric charge is about $0.3$ in natural units and $Tm\sim 10^{-1}$ (see Table~\ref{tab:units}).  In particular, for particles with $p_\text{T} \lesssim 0.3$ GeV in a 2 T magnetic field, they will never travel further than one meter.  Figure~\ref{fig:tracktrajectories} shows the trajectory of charged particles in a solenoidal magnetic field with the same setup as the ATLAS detector, discussed in Sec.~\ref{atlasdetector}.  Note that charged particles of the opposite charge would bend {\it down} instead of up in Fig.~\ref{fig:tracktrajectories}. The momentum resolution from track fitting is determined by how well the {\it sagitta} can be measured.  Figure~\ref{fig:trackingresolution} shows an example charged particle trajectory in the same coordinates as Fig.~\ref{fig:tracktrajectories} where three measurements (dots) of the {\it track} have been measured.  The sagitta $s$ is related to the radius $R$ by $s=R(1-\cos\alpha)$.  When $\alpha\ll 1$, $s\approx \frac{1}{R}\alpha^2$.  Also in this approximation, $\alpha\approx \frac{1}{2}L/R$.  Using Eq.~\ref{pTbfield}, this gives the result $s\approx \frac{1}{8}\frac{L^2eB}{p_\text{T}}$. Linear propagation of errors shows that $\sigma_{p_\text{T}}/p_\text{T}\approx \sigma_s/s$.  Therefore, $\sigma_{p_\text{T}}/p_\text{T}\propto \sigma_s p_\text{T}/L^2B$; the resolution is worse at high $p_\text{T}$ and can be improved with a longer lever arm $L$ and a higher magnetic field $B$.  The resolution $\sigma_s$ is independent of $L$ and $B$ and scales as $\sigma_s\propto 1/\sqrt{N}$ for enough hit measurements $N$\footnote{The exact formula is derived in Ref.~\cite{Gluckstern:1963ng}, which also has an interesting discussion about the optimal spacing of measurements.}.  In addition to measurement uncertainty, there is a contribution to the resolution from {\it multiple scattering} of the primary particle in the detector material.  This term is approximately independent of momentum and scales as~\cite{Grupen:2012zpa,Tavernier:2010zz} $\sigma_{p_\text{T}}/p_\text{T}\propto \frac{1}{BL\beta}\sqrt{L/X_0}$, where $X_0$ is the {\it radiation length} of the detector material\footnote{The radiation of a material is the characteristic length for energy loss via Bremsstrahlung; quantitatively, $dE/dx = E/X_0$.  After $X_0$, the particle has only $1/e$ of its original energy.}. 
		
		A second possibility for particle detection is to stop the primary particle and measure the heat deposited in the material ({\it calorimeter})\footnote{A detailed description of calorimetery is in the dedicated textbook Ref.~\cite{Wigmans:2000vf}.}.  Both charged and neutral particles can be measured this way and the location of the energy deposition provides information about the particle's momentum direction.  For electrons, energy loss at high energy is dominated by the same Bremsstrahlung that is a nuisance for tracking detectors.  Photons undergo pair production, which is also set by the radiation length $X_0$\footnote{Though the exact dependence for photons is different than for electrons: the probability for a photon to survive a distance $x$ before pair production is $e^{-\frac{7}{9}\frac{x}{X_0}}$.}.  The radiation length for muons is much larger than for electrons.  Muons deposit energy via ionization (governed by Bethe-Bloch), but are not stopped by reasonably sized calorimeters.  Hadrons loose energy by a combination of ionization (charged particles) and nuclear (i.e. via the strong force) interactions.  Electromagnetically decaying hadrons subsequently loose energy via the processes listed above for leptons and photons.  These interactions are characterized by the {\it hadronic interaction length} $\lambda$.  For a given material $\lambda $ is often much larger than $X_0$; for example in liquid argon (used by the ATLAS calorimeters), $\lambda \sim 6X_0$.  
		
		Energy in a calorimeter is lost via a cascade of collisions.  Radiative and hadronic processes result in significant energy loss until ionization or other low energy phenomena dominate and the remaining particles slowly lose energy and are absorbed.  The crossover energy is called the {\it critical energy} $E_c$.  If each collision occurs after time $\delta t$ and results in a reduction of the primary particle energy by $\frac{1}{2}$, then the timespan of a particle shower in the calorimeter is proportional to $\log_2(E/E_c)$.  In general, the depth of a shower scales {\it logarithmically} with the energy.  For this reason, calorimeters of a fixed depth can detect particles over many decades in energy.  Showers initiated by electromagnetic particles are shallower than those started by nuclear processes.  For this reason, calorimeters specifically optimized for detecting electromagnetic showers are closer to the interaction point than {\it thicker} calorimeters aimed at stopping nuclear showers from hadrons.   The transverse size of a shower is also significantly larger for hadronic showers compared with electromagnetic showers, which scale with $\lambda$ and $X_0$ (called the Moli\`{e}re radius), respectively.  
		
		In contrast to tracking detectors, the resolution of a calorimeter {\it decreases} with energy.  The energy in a calorimeter is related to the number of particles produced in the shower; as such, the energy follows a Poisson distribution: $\sigma_E/E\propto 1/\sqrt{E}$.   Estimating the proportionality constant is complicated because the hadronic/electromagnetic composition plays a significant role in determining the resolution.  As with tracking detectors, calorimeters also have a constant term due to a variety of sources, such as differences in behavior for electromagnetic and hadronic showers~\cite{Grupen:2012zpa}.  Additional sources of (e.g. electronic) noise result in a constant energy resolution that is independent of the primary particles.  This results in a term $\sigma_E/E\propto 1/E$.  

\begin{figure}[h!]
\begin{center}
\includegraphics[width=0.5\textwidth]{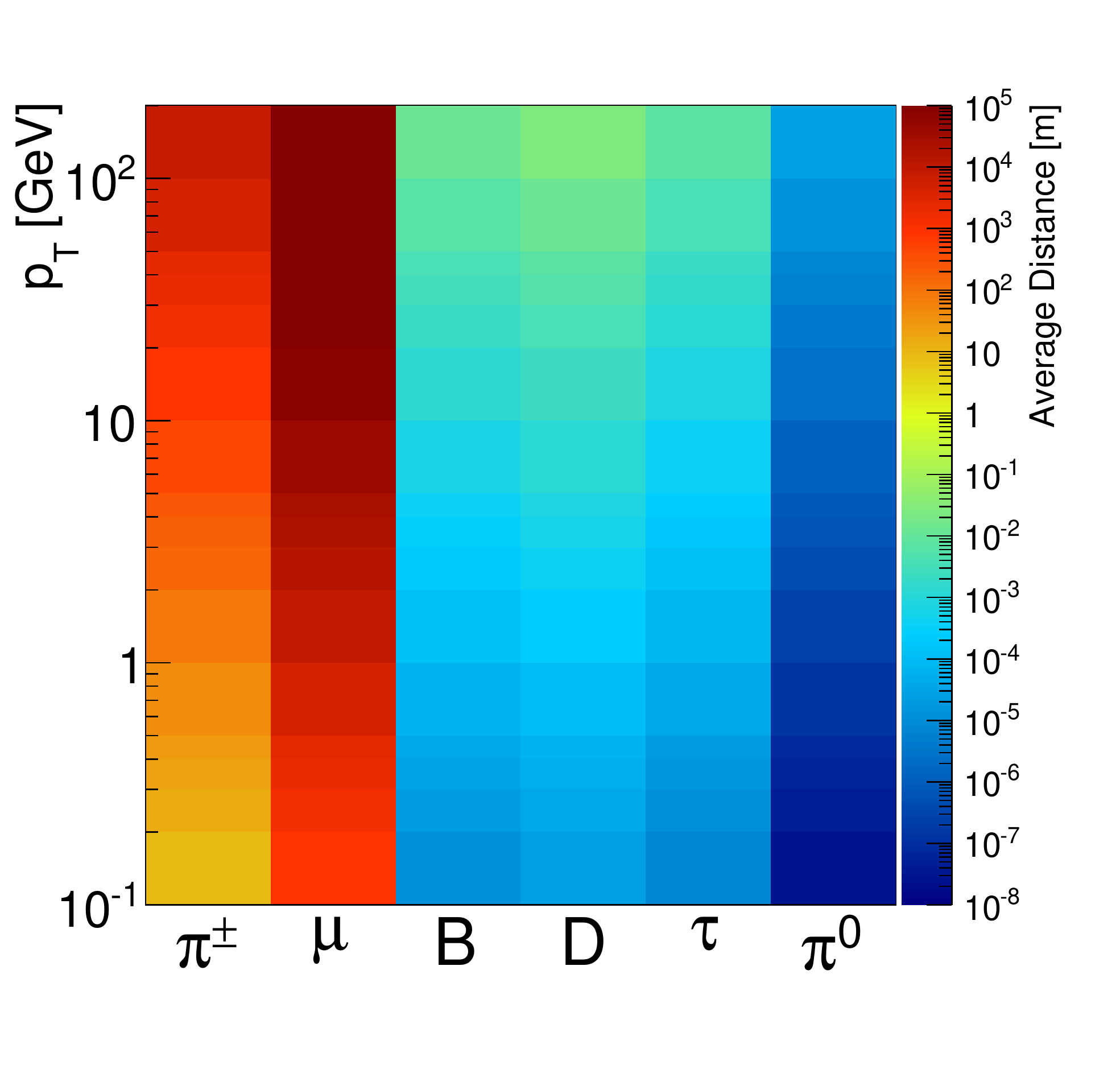}
 \caption{The average transverse distance traveled by various particles labeled on the horizontal axis as a function of their $p_\text{T}$.}
 \label{fig:particletypedd}
  \end{center}
\end{figure}

		In addition to measuring the momentum or energy of a particle, detectors can be used to infer the particle type.  Figure~\ref{fig:particletypedd} shows the average distance that various particles travel in the lab frame before decaying, $\beta\gamma\tau$, as a function of $p_\text{T}$.  Charged pions and muons travel well past any detector element before decaying.  For pions, this is largely irrelevant because of nuclear interactions that stops them in the calorimeters.  Muons loose only a small amount of energy in the tracking detectors and calorimeters.  Therefore, one can identify muons by placing an additional set of tracking detectors {\it beyond} the calorimeters.  Except for occasional {\it punch-through} hadrons and low energy sources of radiation around the detector, particles measured in these outer tracking chambers can be identified as muons.  On the other end of the spectrum, neutral pions decay nearly immediately after production into two photons.  The angular distribution between the two photons scales as $2m_\pi/p_\text{T}$ (see Chapter~\ref{cha:bosonjets}).  One of the reasons that the electromagnetic calorimeter needs to be very finely segmented is to separate high $p_\text{T}$ isolated photons from two photons produced collinearly from a pion decay.  There are a class of particles in Fig.~\ref{fig:particletypedd} that can be produced at the primary collision, but decay after macroscopic distances in the detector.  The existence and properties of the secondary decay vertices for these particles are powerful observables for separating $b$-quark jets, $c$-quark jets, and hadronically decaying $\tau$-lepton jets from light(er) quark and gluon jets.  Since top quark jets decay nearly $100\%$ of the time to $b$-quarks, $b$-quark jet tagging will be a critical aspect of the measurements and search presented in Parts~\ref{part:qpj} and~\ref{part:susy}.  Reference~\cite{Aad:2015ydr} describes $b$-quark jet tagging with the ATLAS detector, which is also discussed when used in subsequent chapters.
		
Various other particle type identification techniques exist that require specialty detector elements or work only in a limited kinematic region. For example, as part of the ATLAS tracking detector, there is a special subdetector for transition radiation emitted by electrons when they traverse different materials.  This is useful for separating charged pion tracks from electron tracks.  The tracking detector can also measure the amount of energy deposited per unit distance, $dE/dx$, which depends on the mass of the primary particle.  This information can be used to separate proton, pion, kaon, and possibly new massive particle tracks from each other as long as $\beta\gamma \lesssim 1$~\cite{ATLAS-CONF-2011-016}.  

\begin{figure}[h!]
\begin{center}
\includegraphics[width=0.5\textwidth]{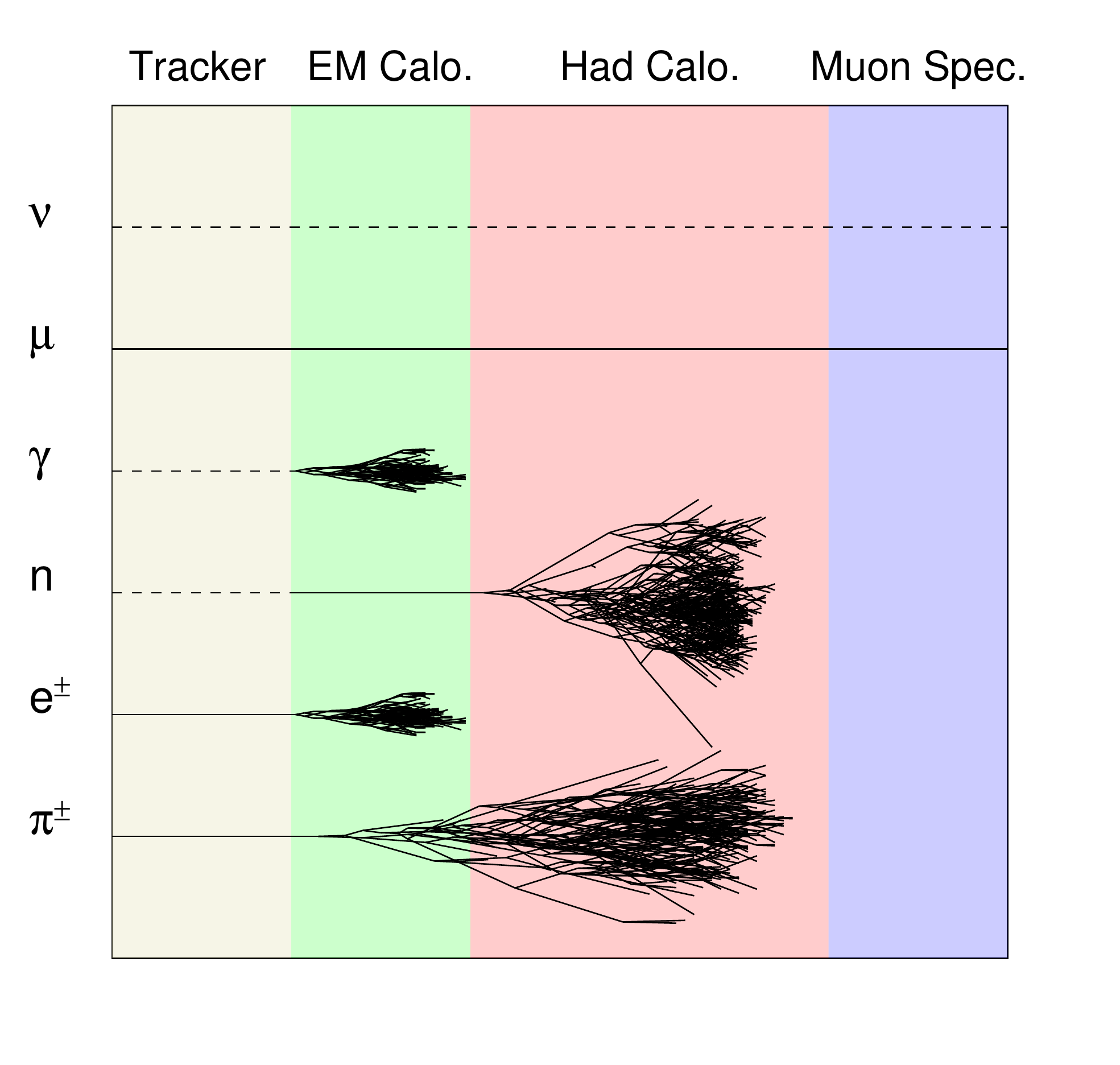}
 \caption{An overview of the measurement pattern for various particle types in the ATLAS detector.  A dashed line means that the given particle leaves no trace in the sub-detector.  A solid horizontal line indicates insignificant energy loss (ionization and small levels of multiple scattering for electrons).}
 \label{fig:particletypeabc}
  \end{center}
          \end{figure}

Figure~\ref{fig:particletypeabc} presents a schematic overview for the detector signature of various particle classes.  The majority of hadrons in jets are pions because they are the lighest hadron.  The mass of the light hadrons is insignificant compared with their momenta when $p_\text{T}\gg 1$ GeV and so there is no distinction between various charged hadron types.  Section~\ref{atlasdetector} presents an overview of all the ATLAS detector elements, which will follow the pattern in Fig.~\ref{fig:particletypeabc}.

		\clearpage

	\section{The ATLAS Detector}
	\label{atlasdetector}

\begin{figure}[h!]
\begin{center}
\includegraphics[width=0.7\textwidth]{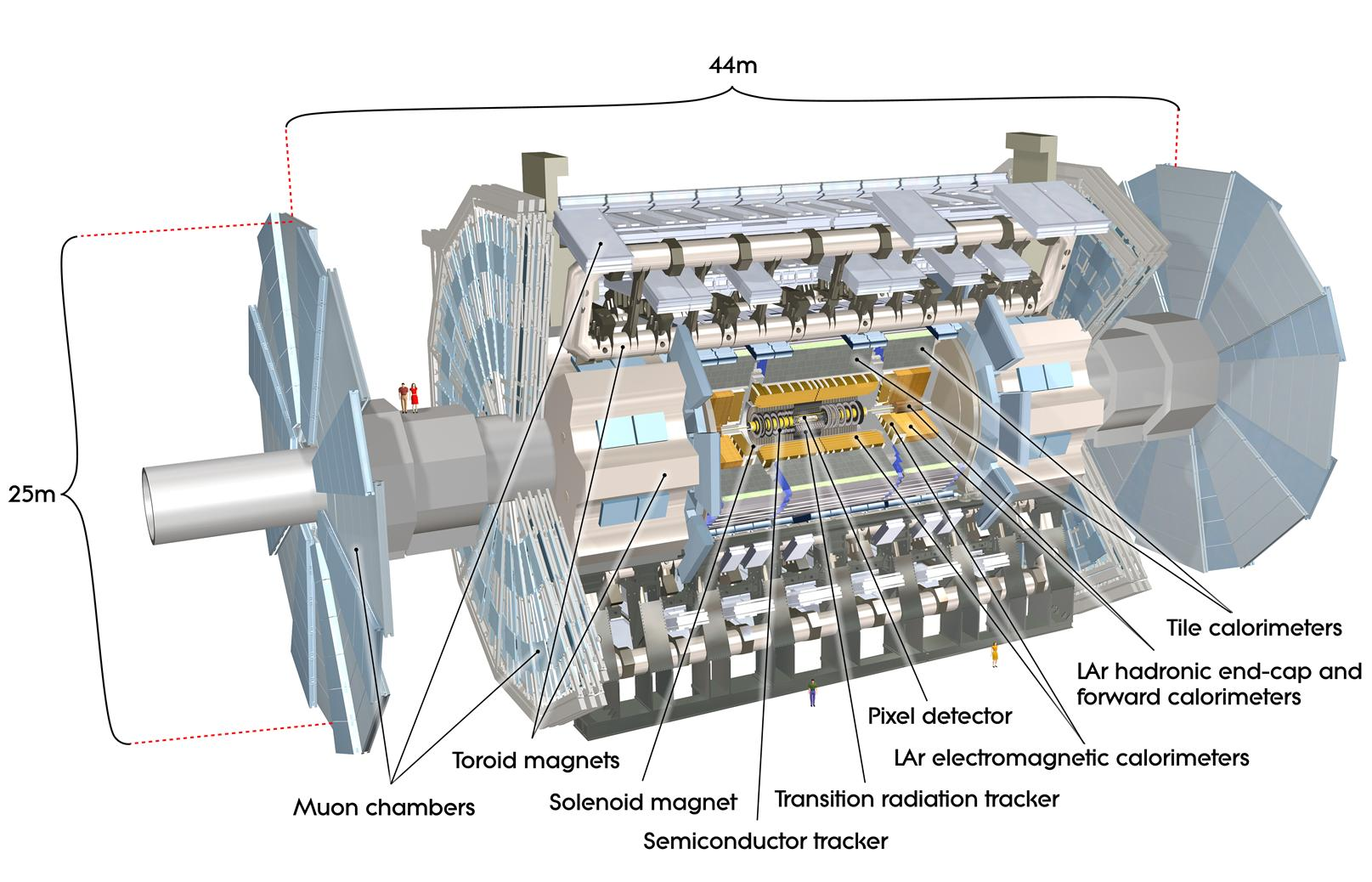}
 \caption{The ATLAS detector and its subsystems (to scale).  Image from Ref.~\cite{Aad:2010mr}.}
 \label{fig:atlasoverview}
  \end{center}
\end{figure}

ATLAS is a general-purpose detector designed to measure the properties of particles produced in high-energy $pp$ collisions with nearly a full $4\pi$ coverage in solid angle\footnote{This section is intended to be a brief overview - for many more details, see Ref.~\cite{Aad:2008zzm}.}.   In order to provide shielding from cosmic rays (and reduce costs), the LHC and the cavern containing the ATLAS detector are about $100$ m below ground.   The innermost subsystem of the detector is a series of tracking devices used to measure charged-particle trajectories bent in a 2~T axial field provided by a solenoid whose axis is parallel with the beam direction.   This inner detector (ID) consists of a silicon pixel detector surrounded by a semiconductor microstrip detector (SCT) and a straw-tube tracker that can detect electron transition radiation (TRT) (Sec.~\ref{innerdetector}).  Surrounding the ID are electromagnetic and hadronic calorimeters that use liquid argon and scintillating tile as active media (Sec.~\ref{calo}).  Beyond the calorimeters is a 4 T toroidal magnetic field and a multi-component tracking system for muon detection (Sec.~\ref{muon}).  Section~\ref{trig} discusses the data acquisition including the {\it trigger}.  A diagram of the subsystems of ATLAS is shown in Fig.~\ref{fig:atlasoverview}.  For scale, people are shown walking on the cavern floor as well as between the muon chamber wheels.
	
	\clearpage
	\subsection{Inner Detector}
	\label{innerdetector}

\begin{figure}[h!]
\begin{center}
\includegraphics[width=0.5\textwidth]{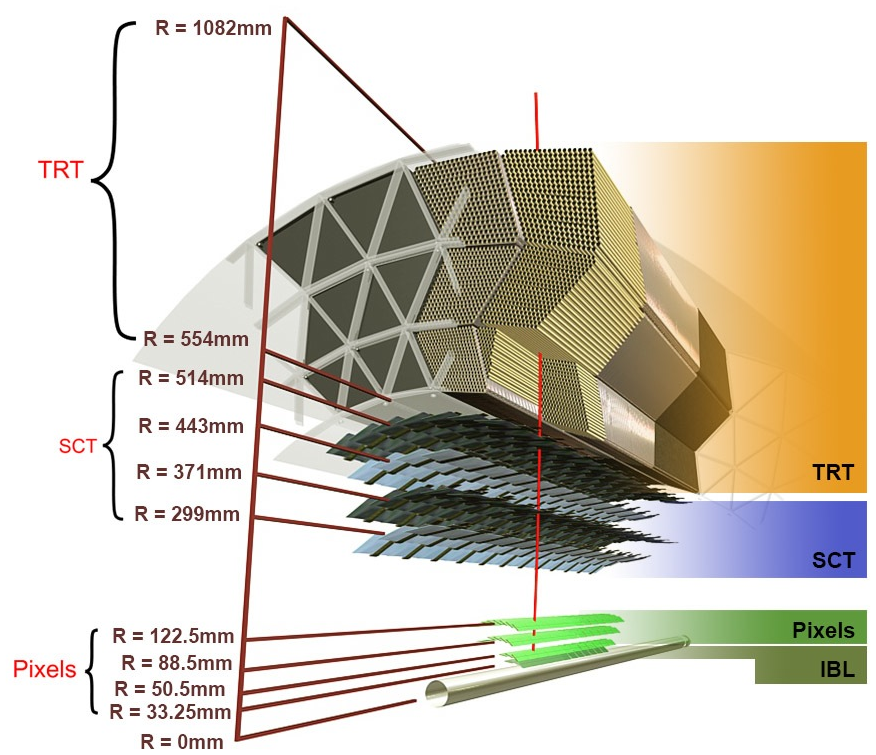}\includegraphics[width=0.5\textwidth]{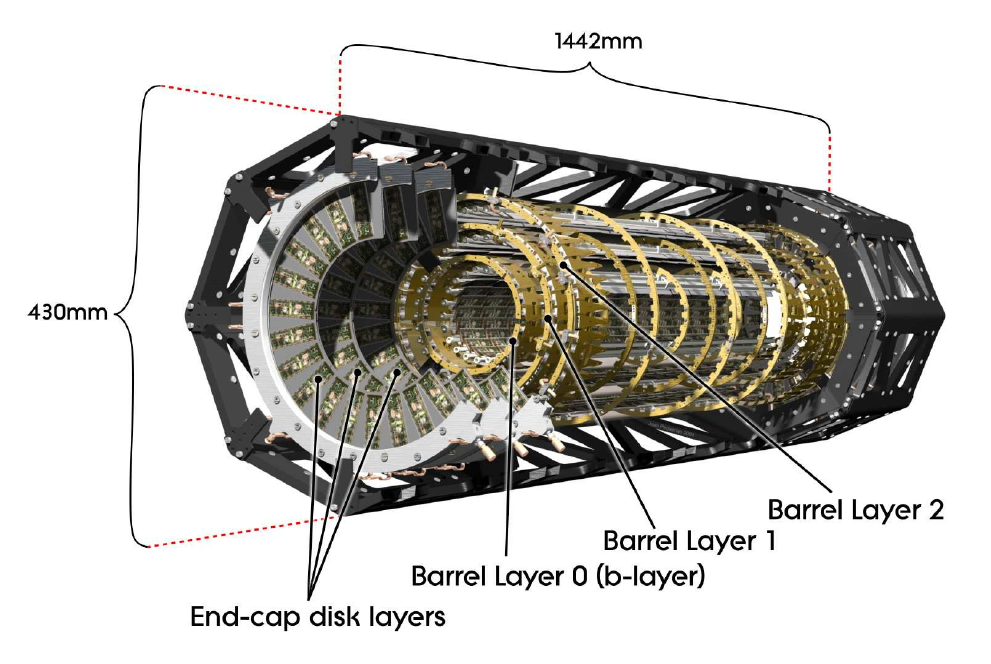}
 \caption{Left: A schematic view of the subsystems of the ATLAS inner detector~\cite{ATL-PHYS-PUB-2015-018}.  The IBL was added between Runs 1 and 2.  The red line indicates the trajectory of a hypothetical particle with $p_\text{T}=10$ GeV at $\eta=0.3$.  Right: an enlarged view of the pixel detector prior to the insertion of the IBL~\cite{Aad:2008zz}.}
 \label{fig:pixel1}
  \end{center}
\end{figure}

\begin{figure}[h!]
\begin{center}
\includegraphics[width=0.95\textwidth]{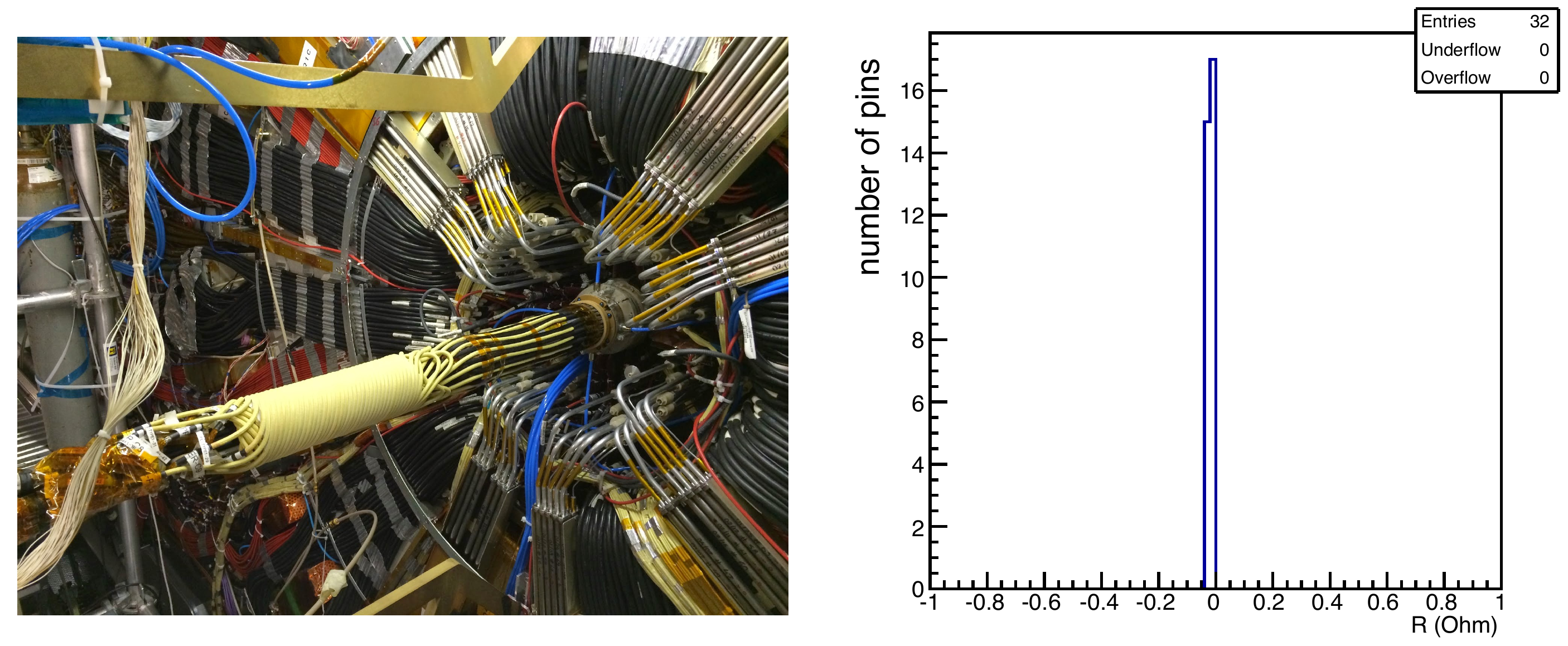}
 \caption{Left: A picture of the IBL just after being inserted into the detector.  The yellow cables are wrapped in order to pass through to the far side.  These cables host data transmission, high/low voltage, and many other services.  Right: The difference in the measured resistance in the data pins before and after a practice wrapping.}
 \label{fig:pixel2}
  \end{center}
\end{figure}

The innermost layer of ATLAS is a series of tracking detectors with three different technologies, illustrated by Fig.~\ref{fig:pixel1}.  Closest to the beam pipe is a pixel detector, which is composed of 3 (4) layers in Run 1 (2).  A new pixel layer (insertable $b$-layer, or IBL) was inserted closer to the collision point between Runs 1 and 2.  This was a significant technical challenge because the beampipe had to be removed and replaced with a smaller radius pipe upon which the IBL was mounted and inserted with all of its services into the small space inside the original pixel detector.  The left image of Fig~\ref{fig:pixel2} shows the just-inserted IBL before the service cables were unwound.  This winding was needed in order to connect tables to the side of the IBL opposite the insertion.  These service cables (assembled at SLAC) were thoroughly tested at every stage of processing, including before and after mechanical stress tests such as a practice winding.  The right plot of Fig.~\ref{fig:pixel2} shows the difference in resistances on all the pins of one data cable before and after a practice winding.  As desired, the resistance is unchanged.

The original three pixel layers are composed of 250 $\mu m$ thick planar sensors most with a $50\times 400$ $\mu$m${}^2$ surface area.  In order to cope with a higher radiation dose, the IBL sensors are smaller $50\times 250$ $\mu$m${}^2$ (also thinner) planar sensors in the central region and 3D sensors~\cite{Parker1997328} at high $|\eta|$ with charges drifting perpendicular to the sensor depth instead of parallel.  Beyond the pixel detector are four layers of silicon microstrips (SCT).  In order to provide a (crude) measurement along the $z$ direction, each SCT module has two sensors that are rotated by $\pm 20$ mrad with respect to each other.  Each module provides about $20\mu$m resolution in the azimuthal direction and about 600 $\mu$m resolution along $z$.  Surrounding both the silicon-based detectors is an annulus between about $50$ cm to $1$ m filled with $2$ mm radius drift tubes.  The region around the tubes is filled with a material that enhances the electron transition radiation.  Charged particles leave ionization energy in an average of $36$ tubes of this transition radiation tracker (TRT).  Dedicated low and high thresholds are used to measure minimum ionizing particles and energy from X-ray photons due electron transition radiation, respectively.

\clearpage
\subsection{Electromagnetic and Hadronic Calorimeters}
\label{calo}

\begin{figure}[h!]
\begin{center}
\includegraphics[width=0.55\textwidth]{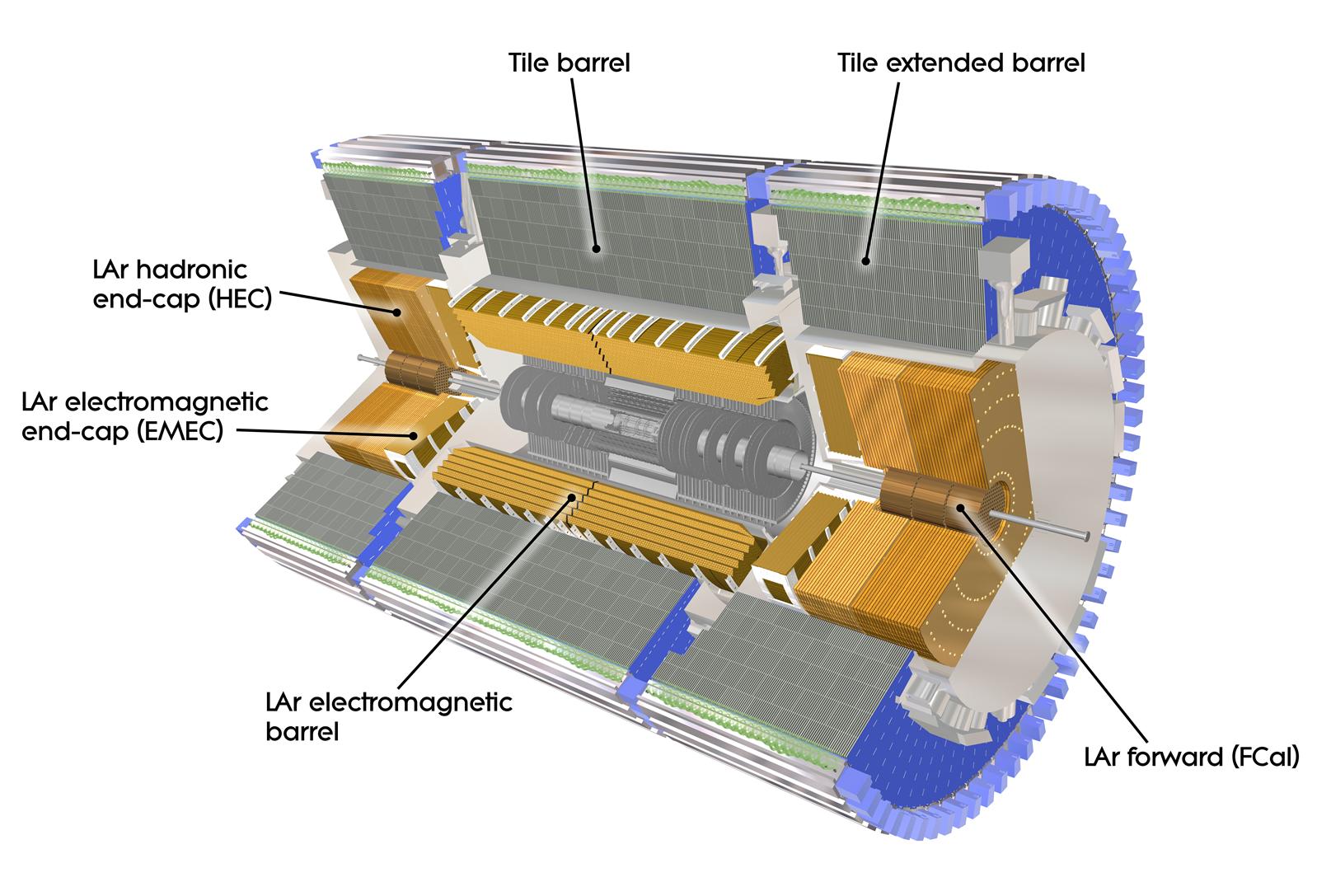}
 \caption{A schematic diagram of the ATLAS calorimeters. Image from Ref.~\cite{Aad:2016upy}.}
     \label{fig:calo}
  \end{center}
\end{figure}

Surrounding the ID and solenoid are electromagnetic and hadronic calorimeters to measure showers from charged and neutral particles.   A high-granularity lead/liquid-argon (LAr) sampling electromagnetic calorimeter is located just beyond the solenoid and spans the range $|\eta| < 3.2$.  Beyond the electromagnetic calorimeter is a two-component hadronic calorimeter that uses steel absorbers and scintillator-tile sampling technology in the range $|\eta| < 1.7$ and copper/LAr sampling technology for $1.5<|\eta| < 3.2$.  Additional calorimetry is provided up to $|\eta|=4.9$ using copper (tungsten)/LAr in the electromagnetic (hadronic) sections.  Figure~\ref{fig:calo} is a schematic diagram of the various calorimeter components.

The bulk of energy deposited in the electromagnetic calorimeter is deposited in the second layer which contains about $17X_0$ (out of about $23X_0$) with a granularity of $0.025\times 0.025$ in $\Delta\eta\times\Delta\phi$.  In contrast, most of the hadronic energy is deposited in the first two layers of the hadronic calorimeter with about $5.5$ (out of about $7.5$) hadronic interaction lengths $\lambda$ with a granularity of  $0.1\times 0.1$ in $\Delta\eta\times\Delta\phi$.  The total detector thickness is about $10$ hadronic interaction lengths at $\eta=0$.  The inner detector material accounts for about $0.5X_0$ and $0.2\lambda$ at $\eta=0$ and about $2X_0$ and $0.7\lambda$ just beyond the edge of the ID ($\eta\sim1.5$).  Most of this material is in the form of support structures, coolant, electronics, and cables.

\clearpage
\subsection{Muon Spectrometer}
\label{muon}

Surrounding the calorimeters is a muon spectrometer with trigger and precision chambers.  Four different detector technologies are used for these purposes.   Monitored Drift Tubes (MDT) provide precision tracking in the central region (except for a small gap at $|\eta|\approx 0$ for services) and Cathode Strip Chambers (CSC) cover the forward region $2<|\eta|<2.7$.  The MDTs have a $35$ $\mu$m resolution along $z$ while the CSCs measure both the $z$ and radial position with $40$ $\mu$m and $5$ mm resolutions, respectively.  The main reason for using CSCs in the forward region is the higher particle flux so the second coordinate measurement is important for resolving track ambiguities.  A long drift time in the MDT ($\mathcal{O}(100)$ ns) makes them unusable for triggering (25 ns crossings).  Therefore, two additional detectors are dedicated to triggering: Resistive Plate Chambers (RPC) in the central region ($|\eta|<1.05$) and Thin Cap Chambers (TGC) up to $|\eta|=2.4$.  RPCs are parallel plate capacitors filled with gas and separated radially for a crude but fast momentum measurement.  TGCs are multi-wire proportional chambers with a finer granularity than RPCs in order to cope with the higher multiplicity and reduced track bending (for a fixed $p_\text{T}$) in the forward region.

\begin{figure}[h!]
\begin{center}
\includegraphics[width=0.55\textwidth]{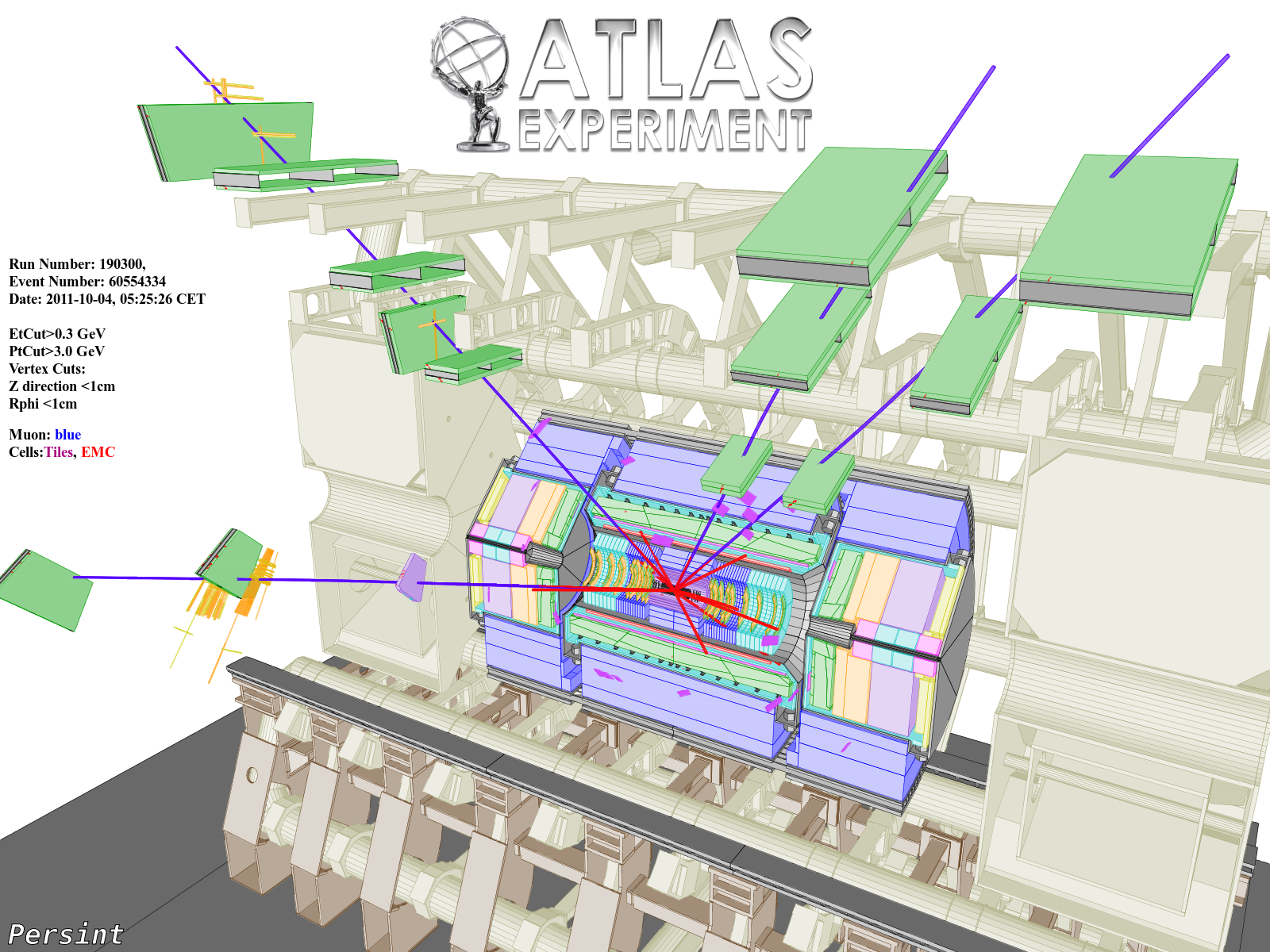}
 \caption{A candidate $H\rightarrow WW^*\rightarrow 4\mu$ event.  The long purple lines indicate reconstructed muon tracks and each muon tracking chamber contributing to those tracks is highlighted in green (MDT) or purple (CSC).  The invariant mass of the four muons is $145.8$ GeV (inconsistent with the now known $m_H\sim 125$ GeV).}
 \label{fig:particletype}
  \end{center}
\end{figure}

\clearpage
\subsection{Trigger System}
\label{trig}

Due to the large event rate, not every collision can be recorded for processing offline.  Events are selected using a three- (two-)level trigger system that is hardware-based at the first level and software-based for the (two) following level(s) in Run 1 (2)~\cite{Aad:2012xs,ATL-DAQ-PUB-2016-001}.  The nominal interaction rate is 40 MHz (=1/25 ns).  It is physically not feasible and undesirable to read out all detector elements at this rate.  First of all, the cross-section for interesting hard-scatter events is significantly below the total cross-section.  For example, the $W$+jets cross-section (highest rate non-QCD process) is about $100$ nb~\cite{Aad:2011dm} while the total inelastic (any collision other than $pp\rightarrow pp$) cross-section is about $70$ mb and the total cross-section is about $95$ mb~\cite{Aad:2014dca}.  With $\mathcal{O}(10)$ simultaneous $pp$ collisions (pileup) per bunch crossing,  this means that only one bunch crossing in $10^5$ produces anything potentially interesting; the rate for $t\bar{t}$, Higgs, etc. is even lower.   Another problem is storage space: an entire event is $\mathcal{O}(1)$ MB.  A readout of $40$ MHz would result in about $100$ TB/s.  These challenges are circumvented by quickly deciding if an event should be saved or discarded.  An event must satisfy all three (two) trigger levels to be recorded for further processing.  The hardware-based trigger system has about $2.5$ $\mu$s to make this decision and reduces the rate from $40$ MHz to $70$ (Run 1) or $100$ (Run 2) kHz.  The total software based trigger operates on an $\mathcal{O}(1)$~s timescale with an output of $400$ (Run 1) or $1000$ (Run 2) Hz readout with a high efficiency for physics processes of interest.  The highest level (software) trigger uses offline-like algorithms while the hardware-based trigger uses crude approximations to object reconstruction in order to increase the speed.   There is redundancy built into the trigger system in order to use one trigger to study another and ensure a high efficiency for processes of interest.  An event can fire multiple triggers, though only one is required to record the event.   As many interesting processes (and rarely any uninteresting processes) contain leptons in the final state, the largest trigger bandwidth is dedicated to single lepton triggers ($\gtrsim 20\%$).  Some triggers are {\it pre-scaled} in order to artificially reduce the rate.  A pre-scale is implemented by randomly keeping only a fraction of events that pass a given trigger.  The pre-scale rates are known, so the luminosity of the data can be corrected to account for the rate reduction (see Sec.~\ref{sec:jettriggers} for more detail).

\clearpage 
     
\chapter{Monte Carlo Simulation}
\label{chapter:simulation}	
	
		In order to interpret the ATLAS data in the context of the SM or any other theory, precise predictions for the detector output are required.  This output depends on physical processes occurring on length scales spanning $10^{-20}$ m up to the macroscopic size of the detector at $\mathcal{O}(10)$ m.  High precision simulation of this entire process is possible because the physical laws {\it factorize}: in order to understand the behavior at one length scale, it is only necessary to know what happened previously at one length scale smaller.  Therefore, each length scale is simulated in series.   At the core of each step is a Monte Carlo (MC) integration.  The basic idea of a MC simulation is that the expected value of a function can be approximated by computing the average value of the function when sampling many times from the underlying probability distribution.  Factorization is realized by using {\it Markov Chain} MC in which the stochastic evolution of a simulated event at one stage only depends on the previous stage.  Some aspects of a simulated event are unphysical, these parts are often called the {\it MC truth} because they are unknowable in reality.  At the stage when the simulated events represent the same information that is present in a real data event, the simulated event is treated exactly as if it were a real event when reconstructing the final state in terms of high level objects (see Chapter~\ref{chapter:reco}).  The only difference is that one simulated event often represents far fewer than one real event; in order for the averaging to be useful (have small uncertainty), the number of simulated events needs to (greatly) exceed the number of real events.  
		
		The following sections briefly introduce the various stages of simulation.  At the smallest distance scales, perturbative calculations are combined with MC techniques to generate the {\it hard-scatter} process, which describes the collision of partons to produce the process of interest (Sec.~\ref{sec:ME}).  Next, perturbative scale evolution takes the outgoing colored particles through radiation down to $\mathcal{O}(1)$ GeV where QCD is no longer well-described by perturbation theory (electromagnetic effects are also included).  Phenomenological models  are then used to convert the quarks and gluons into hadrons as well as describe the (relatively) soft processes related with additional radiation in the event ({\it underlying event} and {\it multiple parton interactions}) (Sec.~\ref{sec:fragmentation}).  Any particle with $\tau \lesssim 30$ ps is decayed before modeling the interaction of the remaining particles with the various detector elements, including inactive components.  The last step in the simulation chain is to model the detector response by converting the energy deposited into digital signals, including the effect of noise (Sec.~\ref{sec:detectorsim}).   
	
		\section{Matrix Elements}
		\label{sec:ME}
		
		Matrix element (ME) calculations describe the hard-scatter process of interest and are computed at fixed order in $\alpha_s$.  In order for such a calculation to be useful for later stages of the simulation, these calculations must be interfaced with another generator that simulates the parton shower (PS) described at the end of Sec.~\ref{sec:particlesandforces}.  The fundamental challenge of the combined ME+PS simulation is how to treat the overlapping soft and collinear regions of phase space.  At lowest order in $\alpha_s$, this problem is manifest when additional quarks and gluons are included in the ME calculation.  This is solved by {\it merging} ME calculations with a PS simulation.  There are several approaches to merging (see Ref.~\cite{Alwall:2007fs} for a comparative review), but the idea used in all of them is to veto emissions in the PS that overlap the ME phase space and then apply event weights based on the probability of the ordering of ME emissions ({\it Sudakov form factors}) involving the splitting functions from Sec.~\ref{sec:particlesandforces}.  The ME emissions are preferred to the PS ones because they better describe hard and wide angle radiation.  At next-to-leading-order (NLO) in $\alpha_s$, there is a phase space overlap between the first real emission with the radiation from the PS.  There are two common schemes for subtracting the overlap from the ME calculation (MC@NLO~\cite{Frixione:2002ik}) or from the PS (POWHEG~\cite{Nason:2004rx}) while still maintaining NLO accuracy in the {\it matched} simulation.  See Ref.~\cite{Nason:2012pr} for a review of these methods.  Algorithms have also recently been developed and (partially) automated to simultaneously match and merge with extra out-going partons in the ME at NLO.  Three actively developed approaches are UNLOPS~\cite{Lonnblad:2012ix}, FxFx~\cite{Frederix:2012ps}, and MEPS@NLO~\cite{Gehrmann:2012yg,Hoeche:2012yf}.

		\section{Fragmentation and the Underlying Event}
		\label{sec:fragmentation}
		
			The role of a PS simulation is to evolve outgoing colored partons from a starting scale $t_\text{hard}\sim\mathcal{O}(10)$-$\mathcal{O}(100)$ GeV down to a cutoff $t_\text{cutoff}$ at which QCD perturbation theory is no longer valid.  The most common MC implementations of the PS are based on a Markov Chain of $1\rightarrow 2$ splittings from scale $t_i$ to $t_{i+1}$ with no-emission probabilities between these two scales given by exponentiating the leading order splitting functions.  This is a {\it leading logarithm} (LL) approximation which numerically accounts for the resummation of logs of the opening angle $\rho$ of the radiation $(\alpha_s\log^2\rho)^n$ to all orders in perturbation theory (see Sec.~\ref{sec:mass:theory} for more detail).  The shower is produced in the limit that there is an infinite number of color charges $(N_c=\infty)$ to avoid complicated non-local effects.  Corrections to this picture are suppressed by $1/N_c^2\sim 1/10$.  Many modern PS generators include effects beyond LL and also beyond leading color.  The two most widely used PS generators\footnote{{\sc Sherpa}~\cite{Gleisberg:2003xi} is also a widely used generator, but its PS and hadronization models are conceptually similar to {\sc Herwig}.} are {\sc Pythia}~\cite{Sjostrand:2006za} and {\sc Herwig}~\cite{Corcella:2002jc}, which are distinguished by their choice of $t$.  {\sc Pythia} uses a $p_\text{T}$-ordered shower~\cite{Sjostrand:2004ef} while {\sc Herwig++} uses angular ordering~\cite{Gieseke:2003rz} in order to explicitly account for coherence effects (see Sec.~\ref{sec:colorcoherence}).  Electromagnetic radiation is also included in the modeling of fragmentation (sometimes with dedicated generators like PHOTOS~\cite{Golonka:2005pn}), but it is suppressed by $\alpha/\alpha_s\sim 1/10$.
			
			After the PS, the remaining partons are combined into color-neutral hadrons.  There is no first-principles model of hadronization\footnote{The words {\it fragmentation} and {\it hadronization} are often used interchangeably, but can also mean different processes depending on the context.  In this document, hadronization will refer to the transition between the end of the parton shower and the formation of hadrons, whereas fragmentation includes both the parton shower and hadronization.}, so {\sc Pythia} and {\sc Herwig} implement physically-inspired phenomenological models with various tunable parameters that can be adjusted to match data.  A model based on color strings with a tension to represent the non-perturbative strong force (Lund string model~\cite{string}) is used by {\sc Pythia} while a cluster model is used by {\sc Herwig}~\cite{Webber:1983if}.  The two generators also differ in other aspects of non-perturbative modeling such as for the underlying-event~\cite{Sjostrand:2004pf,Bahr:2008dy}, which is the production of radiation from the same $pp$ collision as the hard-scatter process, but not directly involving the two scattering partons.  Phenomenological models designed to describe non-perturbative physical effects have many parameters which are tuned to data.  Specific sets of parameter tunes are described in later sections. During the hadronization process, unstable particles are decayed, including $B$-hadrons and $\tau$-leptons, often with decay tables from dedicated programs like EvtGen~\cite{EvtGen} and Tauola~\cite{Jadach:1993hs}, respectively. 
			
			Multiple simultaneous interactions (pileup) are modeled by overlaying independent {\it minimum bias} (inelastic) events on top of the hard-scatter event.  The number of such collisions is stochastic and modeled to match the pileup level in data.  This only accounts for the {\it in-time pileup}: radiation resulting from collisions that occurred in the same bunch crossing as the primary hard-scatter event.  {\it Out-of-time} pileup from bunch crossings before or after the primary one are modeled in the same way, but are offset in time in the simulation to allow for an accurate model of the signal processing that can take $>25$ ns~\cite{Marshall:2014mza}.
		
			Each simulation setup in the subsequent chapters will be specified by the ME and PS generators as well as the various perturbative and non-pertubative tunable parameters.

		\section{Material Interactions and Detector Simulation}
		\label{sec:detectorsim}
		
			Up to this stage, all the steps of the event generation only depend on the beam type ($pp$) and beam energy ($\sqrt{s}=8$ or $13$ TeV).  After fragmentation, hadrons begin to interact with the detector material and so all subsequent stages are tailored to the ATLAS detector composition and geometry.  A detailed model of each detector element, including inactive material, is constructed and imported into the {\sc Geant4} generator~\cite{Agostinelli:2002hh}.  Particles produced from fragmentation are propagated through each subdetector and the nuclear interactions are modeled using a variety of physically-inspired models\footnote{Analyses that are not particularly sensitive to local fluctuations in the energy deposited in the calorimeter use a parameterized description that significantly speeds up the simulation time~\cite{ATLAS:1300517}. }.  Custom algorithms for each subdetector then transform the energy deposited into an analogue and/or digital signal and model the readout~\cite{Aad:2010ah}.  For example, when a high energy pion traverses the doped silicon inside a planar pixel detector, {\sc Geant4} stochastically calculates the energy deposited along the path length.  This energy is assigned to low energy electrons or holes\footnote{Holes are gaps in the electron Fermi distribution that propagate as if they were a positive charged particle.} that are propagated (including thermal diffusion) to the collecting electrodes.  The collected charge is converted into a time over threshold (TOT), which is digitized into $4$ (IBL) or $8$ bits.  The various voltage and tuning parameters of the sensor and readout are part of the simulation.  One important condition that is not currently part of the pixel simulation is the radiation level, which can degrade charge collection.  The Run~1 and early Run~2 dosages are likely not sufficient for a significant degradation in performance, but this will be an important phenomenon to model in the future.  See Appendix~\ref{raddamage} for further details about modeling radiation damage.
		
\clearpage

\chapter{Event Reconstruction}
\label{chapter:reco}		
	
		Once the data are collected (or simulated events are generated), pattern recognition algorithms are employed to reconstruct basic physical objects.  The first step in this process is to build low-level objects representing individual particles.  In the inner detector, tracks are constructed from space point hits (Sec.~\ref{tracks}) and in the calorimeter, calorimeter-cell clusters are formed (Sec.~\ref{clusters}).  In order to reject tracks that do not originate from particles produced in the primary collision, various {\it quality criteria} are imposed for tracks used in subsequent analysis.  The energy of calorimeter-cell clusters is corrected ({\it calibrated}) based on shower properties so that it is an unbiased measurement of the initiating particle energy. From tracks and calorimeter-cell clusters, electrons, photons, muons, taus (Sec.~\ref{sec:leptons}), and jets (Sec.~\ref{sec:jets}) are constructed.  Object properties are used to construct {\it particle identification} schemes intended to reject objects of one type mis-identified as another type.  Various corrections are applied to ensure that the energy or momentum of the reconstructed objects are calibrated.  All of the aforementioned objects are then used to construct the missing transverse momentum, which is a measure of the momentum carried away by undetected particles such as neutrinos (Sec.~\ref{sec:MET}).  
	
		\clearpage
	
		\section{Calorimeter-cell Clusters}
		\label{clusters}
		
		Energy deposits in the calorimeter that are likely to have originated from a single hadron shower are grouped into calorimeter-cell clusters called {\it topo-clusters}~\cite{Aad:2016upy}.  First, cells with energy exceeding $4\sigma$ above the noise are labeled as seeds\footnote{There is a subtle point that the absolute value of the energy is used.  Due to the shaping function in the LAr calorimeter, out of time pileup from previous bunch-crossings can result in negative energy.  Retaining these clusters can be useful for canceling positive energy fluctuations from pileup.}.   Any neighboring cells (or neighbors of the neighbors) with energy exceeding the noise by $2\sigma$ are added to the seeds.  This second step is repeated, but with a lower threshold of zero energy.  The remaining {\it topologically connected} clusters with local maxima are split into multiple pieces, resulting in the final topo-clusters.  Calorimeter noise is highly $\eta$ dependent as a result of changes in detector technology and at $\mu=30$ ranges from about 70 MeV per layer of the Tile calorimeter for $|\eta|<1.5$ to $1$-$10$ GeV in the forward calorimeter at $|\eta|>4.5$.

		A local cluster weighting (LCW) scheme is used to correct for biases in the energy assigned to each topo-cluster~\cite{Aad:2016upy}.  Corrections are applied to each cluster to correct for energy in the calorimeter but outside the cluster, for energy lost in inactive material, and for the different response to the EM and hadronic components of the shower.

		\section{Charged Particle Tracks}
		\label{tracks}

\begin{figure}[h!]
\begin{center}
\includegraphics[width=0.75\textwidth]{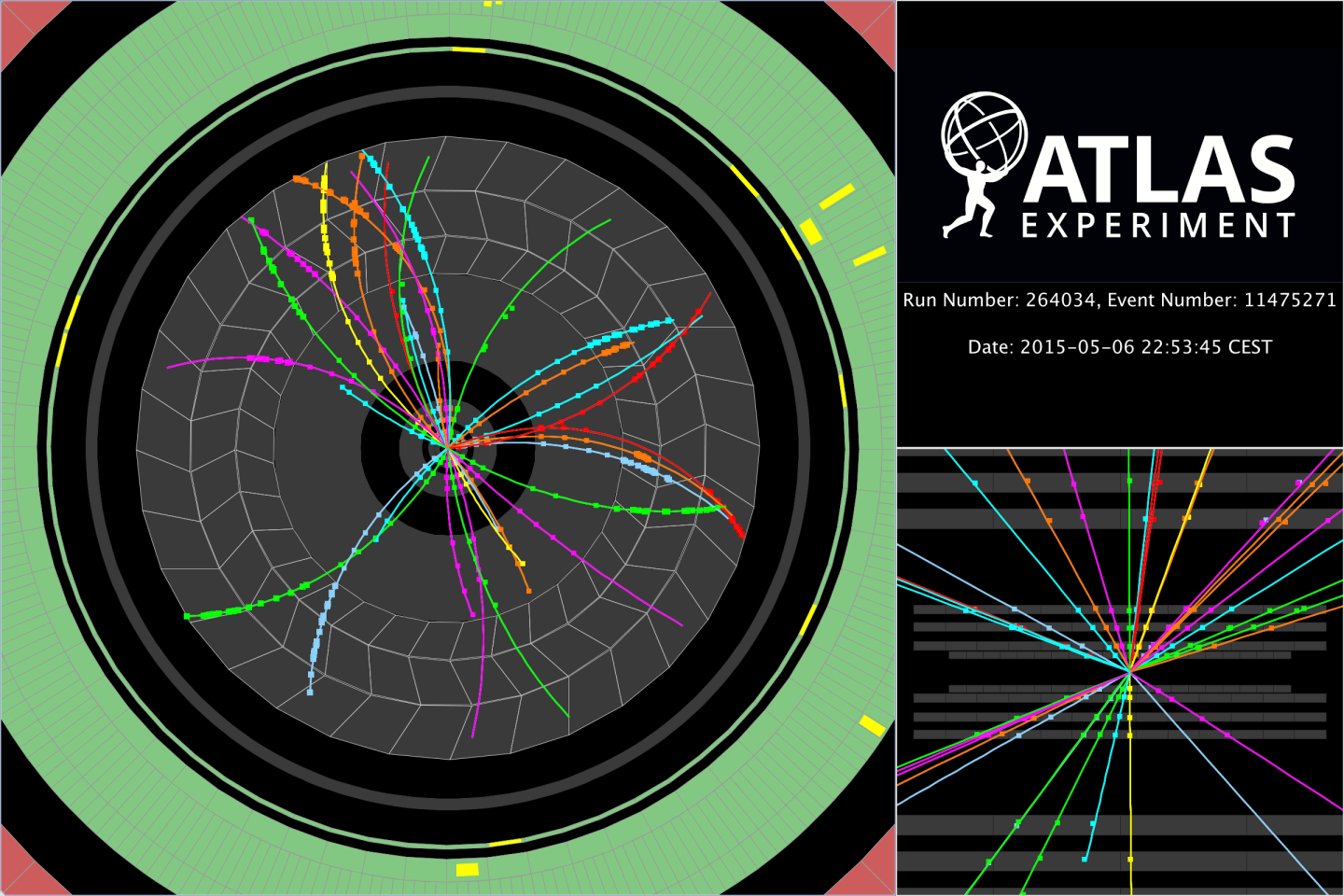}
 \caption{An event display in the early Run 2 data showing charged particle tracks reconstructed from the ID.  The fitted track trajectory is shown as a continuous colored line and the hits in the three sub-detectors are shown as points.  From Ref.~\cite{Collaboration:2014666}.}
 \label{fig:trackingeventdisplay}
  \end{center}
\end{figure}

  Charged-particle tracks are reconstructed from all three inner detector components, providing measurements of the transverse momentum of tracks with a resolution $\sigma_{p_\text{T}}/p_\text{T} \approx 0.05\%\times p_\text{T}/\text{ GeV} \oplus 1\%$, where $\oplus$ indicates a sum in quadrature.  The track reconstruction algorithm fits five track parameters: $d_0$, $z_0$, $\phi$, $\theta,$ and $q/p$, where $d_0$ and $z_0$ are the transverse and longitudinal impact parameters, respectively, $q$ is the track charge and $p$ is the track momentum.  Reference~\cite{Cornelissen:1020106} provides a detailed explanation of the various algorithms used to build tracks and Fig.~\ref{fig:trackingeventdisplay} shows an example event display from the early Run 2 data where tracks are reconstructed from all three ID subdetectors including the IBL. Excellent spatial precision is required to maintain a well-performing track reconstruction out to and exceeding charged-particle $p_\text{T}$ of 1 TeV, where track sagittas are $\lesssim 0.2$~mm.  The large particle density in the core of high $p_\text{T}$ jets is a challenge for track reconstruction.  At low $p_\text{T}$, {\it fake tracks} that are due to combinations of hits from several particles can be suppressed by reducing the number of tracks with shared hits in the pixel detector.  However, at high $p_\text{T}$, real tracks can also have shared hits on the pixel detector.  At $\eta\approx 0$, the innermost (non-IBL) pixel layer has a size of approximately $0.001\times 0.008$ in $\Delta\phi\times\Delta\eta$.  If there are $n\sim\mathcal{O}(10)$ particles in the $\Delta R<0.02$ core of a jet, then multiple particles can deposit hits in the same pixel and one of the resulting tracks can be lost.  Despite this challenge\footnote{Between Runs 1 and 2, a new method for resolving ambiguities in the assignment of hits in the pixel detector to tracks significantly improved the track reconstruction in jets~\cite{Aad:2014yva}.  This may help to improve the resolution of the tracks-in-jets based algorithms presented in Part~\ref{part:qpj}.}, tracking inside jets will play an important role in Part~\ref{part:qpj}.  Methods to measure the reconstruction efficiency in jet cores are presented in Sec.~\ref{sec:uncerts}.

\vspace{10mm}				
						
		\clearpage
	
		\section{Jets}
		\label{sec:jets}

There is no unique way to construct a jet: jets are {\it defined by a jet clustering algorithm}\footnote{The connection between jet clustering and unsupervised machine learning is discussed in Sec.~\ref{sec:HEPML}.}.  For a jet algorithm to be useful experimentally and theoretically it must be {\it IRC safe}.  Let $\vec{\rho}=(y,\phi)$, where $y$ is the rapidity\footnote{Then $|\vec{\rho}|^2$ is the same as $\Delta R^2$ for massless particles, but a different notation is used to distinguish them in the massive case.  Jet masses are often quite small (see Sec.~\ref{sec:jetmass}), but there can be small differences in the jet $p_\text{T}$ when using $y$ or $\eta$ for clustering.  As a  geometric coordinate $\eta$ is useful for relating to rigid detector boundaries, but the full rapidity has the desired Lorentz covariance and is thus used for jet clustering (see Sec.~\ref{coords}).}.  For an algorithm to be IRC safe:

\begin{enumerate}
\item Infrared safe (IR): if a particle $i$ is added with $|p_T|\rightarrow 0$, the jets are unaffected.
\item Collinear safe (C): if a particle $i$ with momentum $p_i$ is replaced with two particles $j$ and $k$ with momenta $p_j+p_k=p_i$ such that $|\vec{\rho}_i-\vec{\rho}_j|= 0$, then the jets are unaffected.
\end{enumerate}

\noindent The most widely used algorithms are categorized by {\it sequential recombination}~\cite{Ellis:1993tq}.  These IRC safe schemes require metrics $d$ on momenta $d_{ij}=d(p_i,p_j):(p_i,p_j)\rightarrow\mathbb{R}^+,d_{iB}=d(p_i):p_i \rightarrow \mathbb{R}^+$ and proceed as follows:

\begin{enumerate}
\item Assign each particle as a proto-jet.
\item Repeat until there are no proto-jets left: Let $(k,\ell)=\text{argmin}_{i,j}d(p_i,p_j)$.  If $d_{mB} < d_{k\ell}$ for $m=\text{argmin}_i d(p_i)$, then declare proto-jet $m$ a jet and remove it from the list.  Otherwise, combine proto-jets $k$ and $\ell$ into a new proto-jet with momentum $p_\text{new}=p_\ell+p_k$.
\end{enumerate}

\begin{figure}[h!]
\begin{center}
\includegraphics[width=0.33\textwidth]{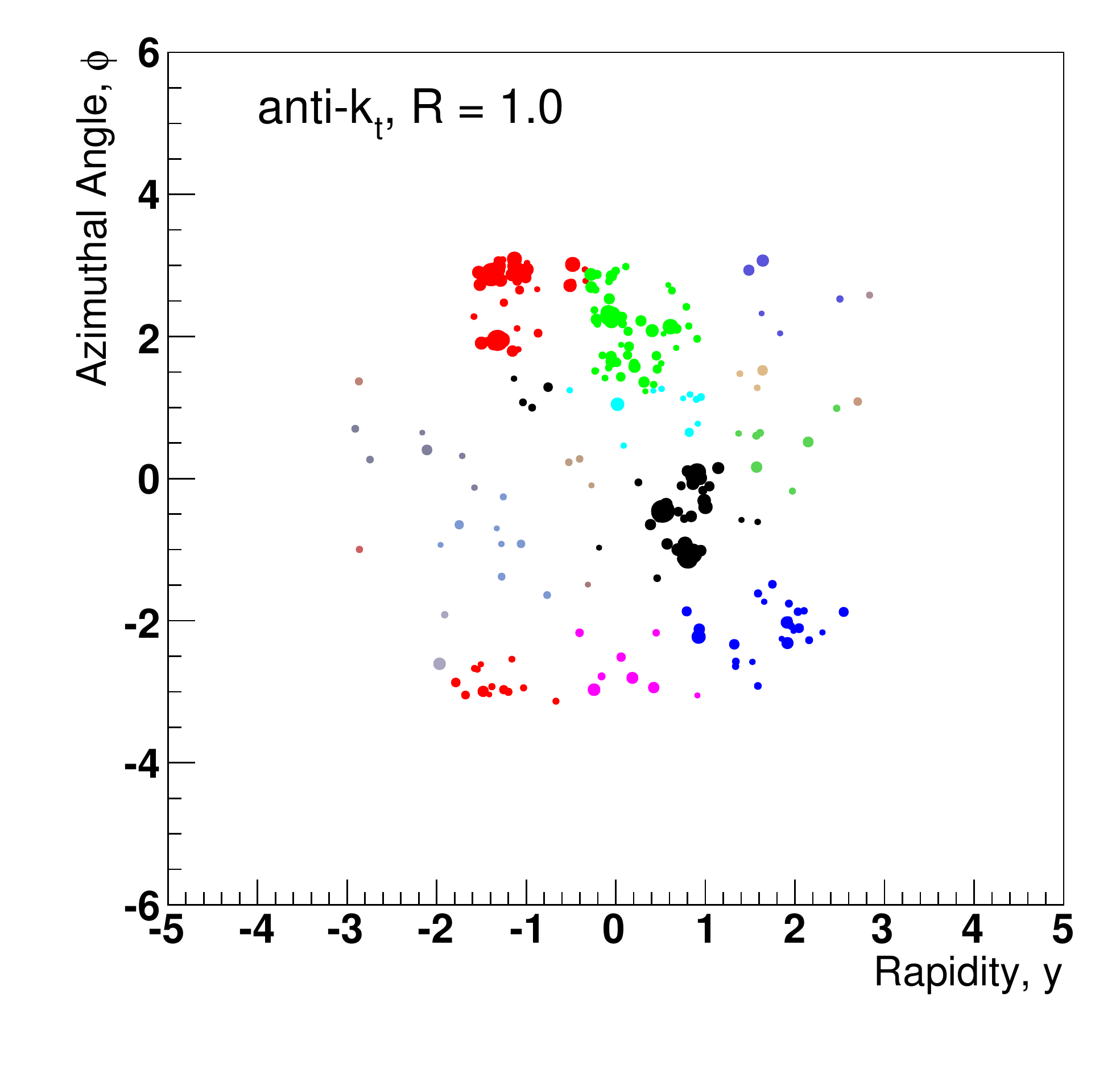}\includegraphics[width=0.33\textwidth]{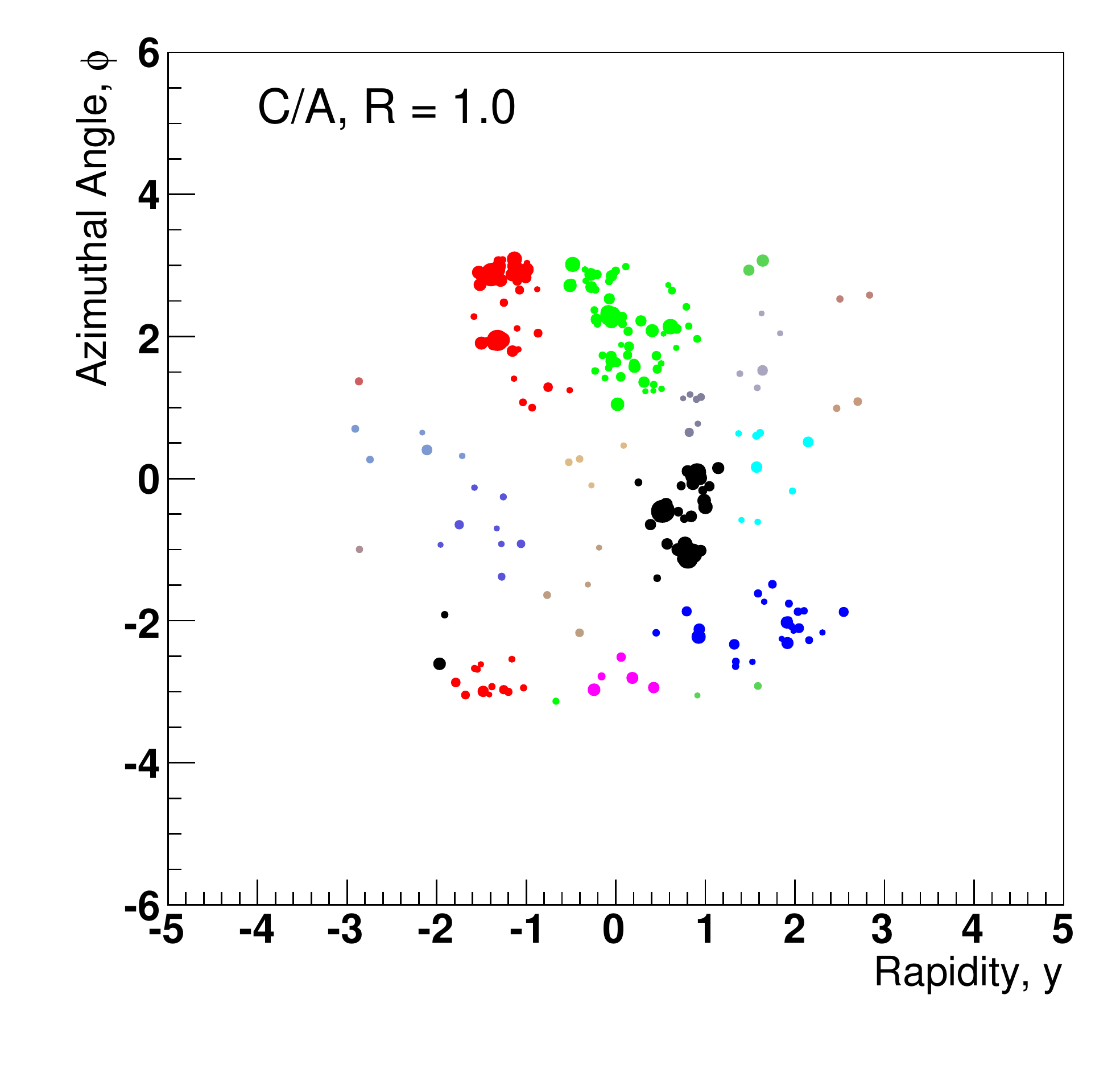}\includegraphics[width=0.33\textwidth]{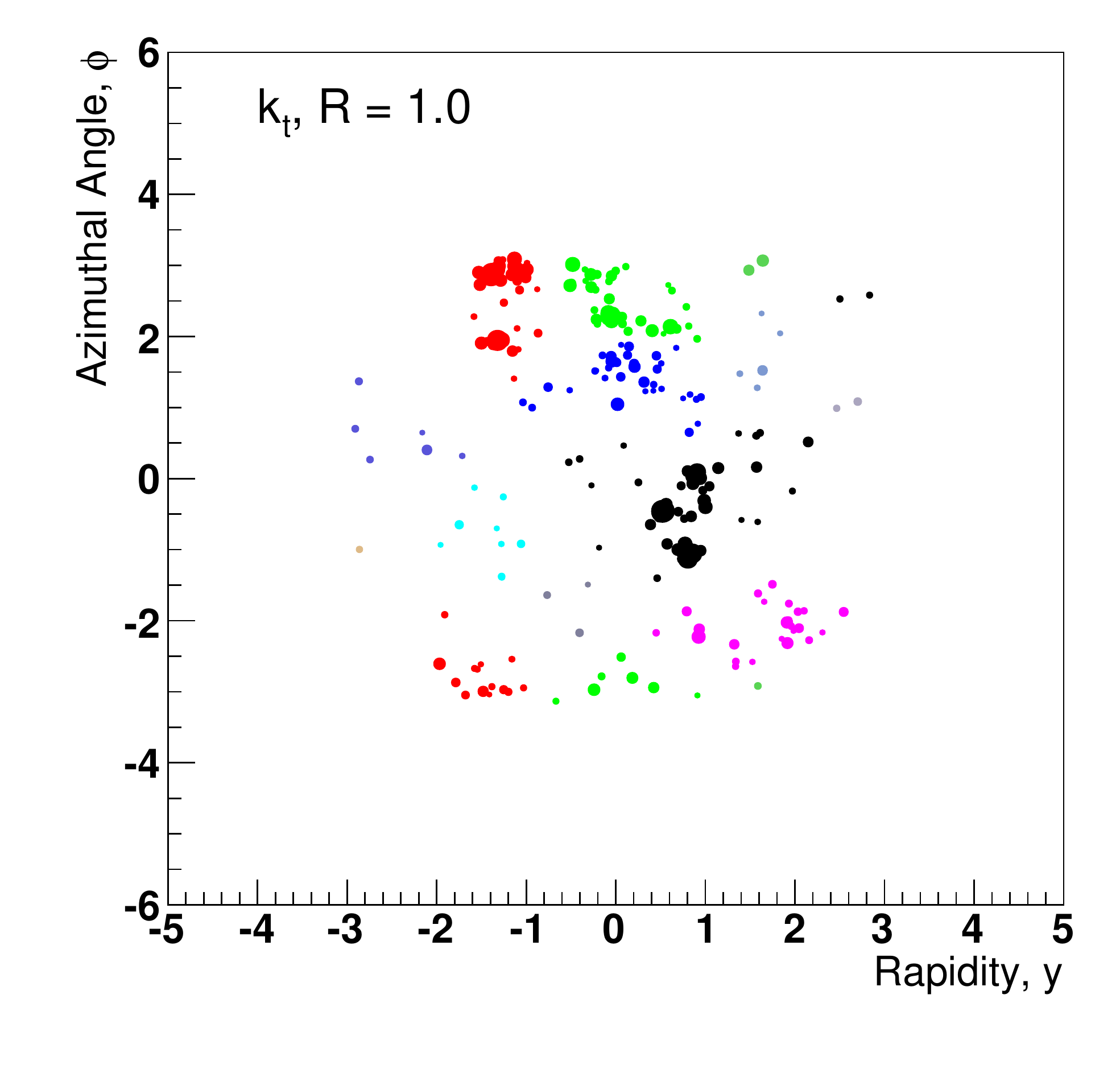}\\
\includegraphics[width=0.33\textwidth]{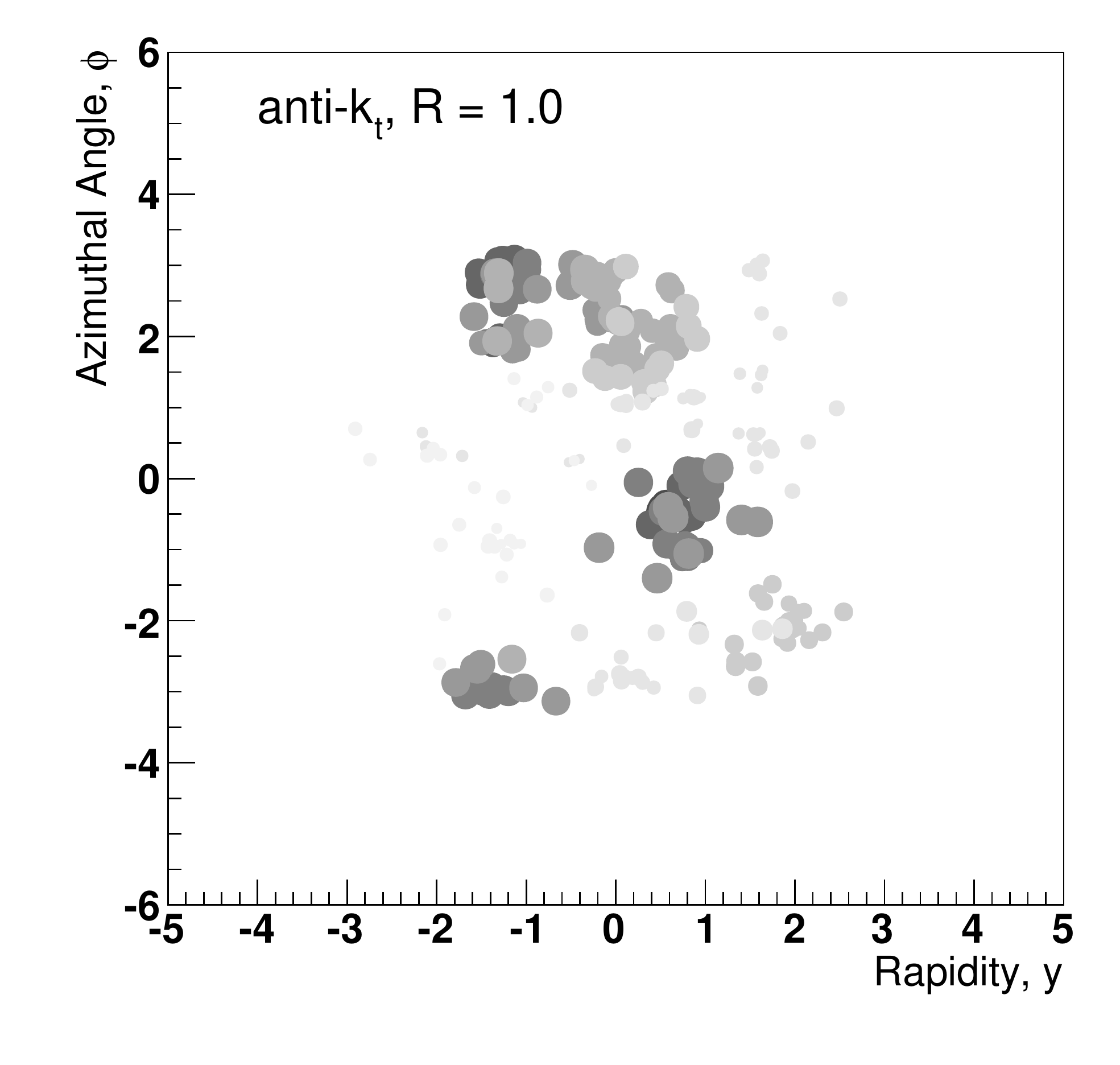}\includegraphics[width=0.33\textwidth]{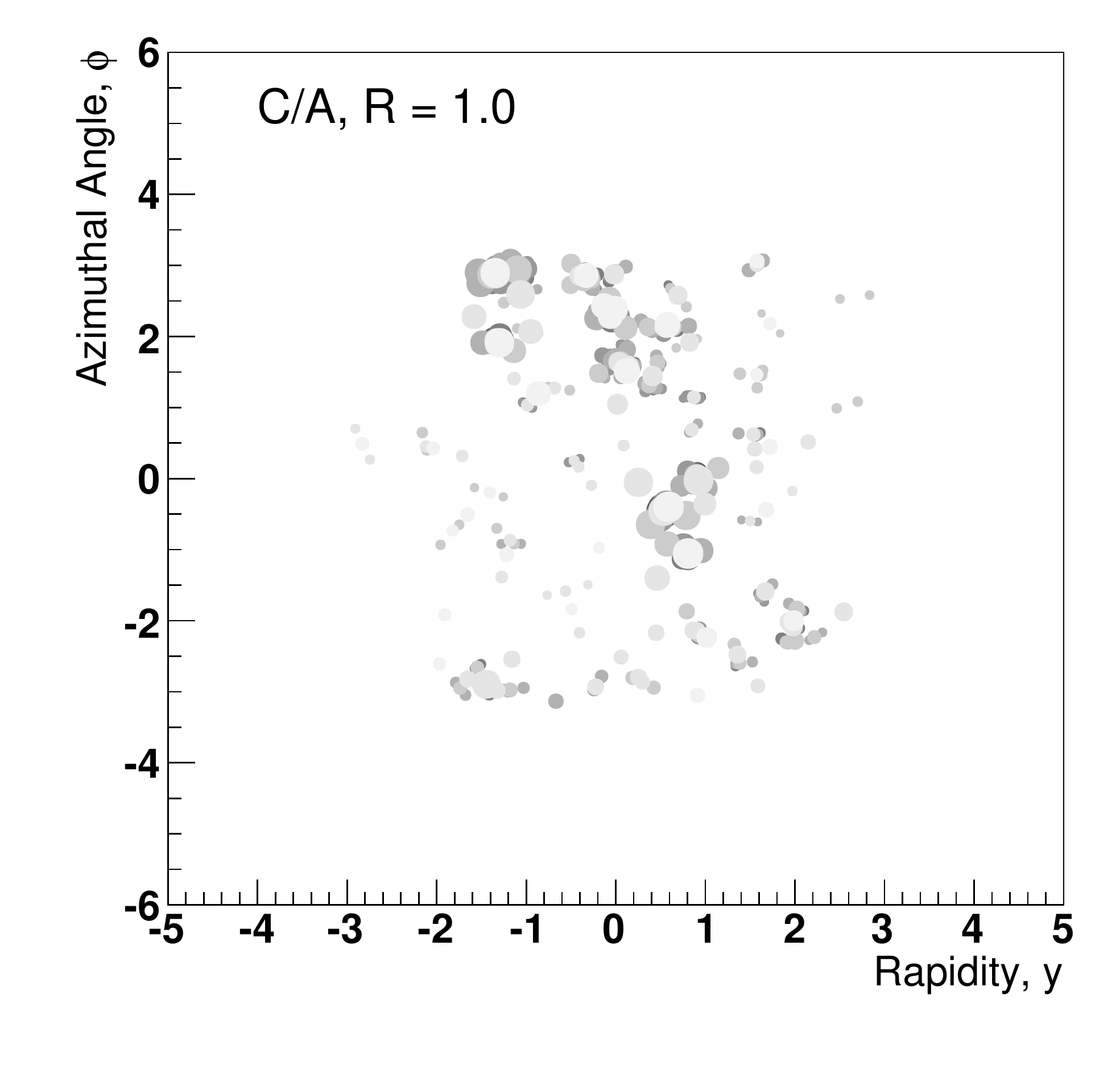}\includegraphics[width=0.33\textwidth]{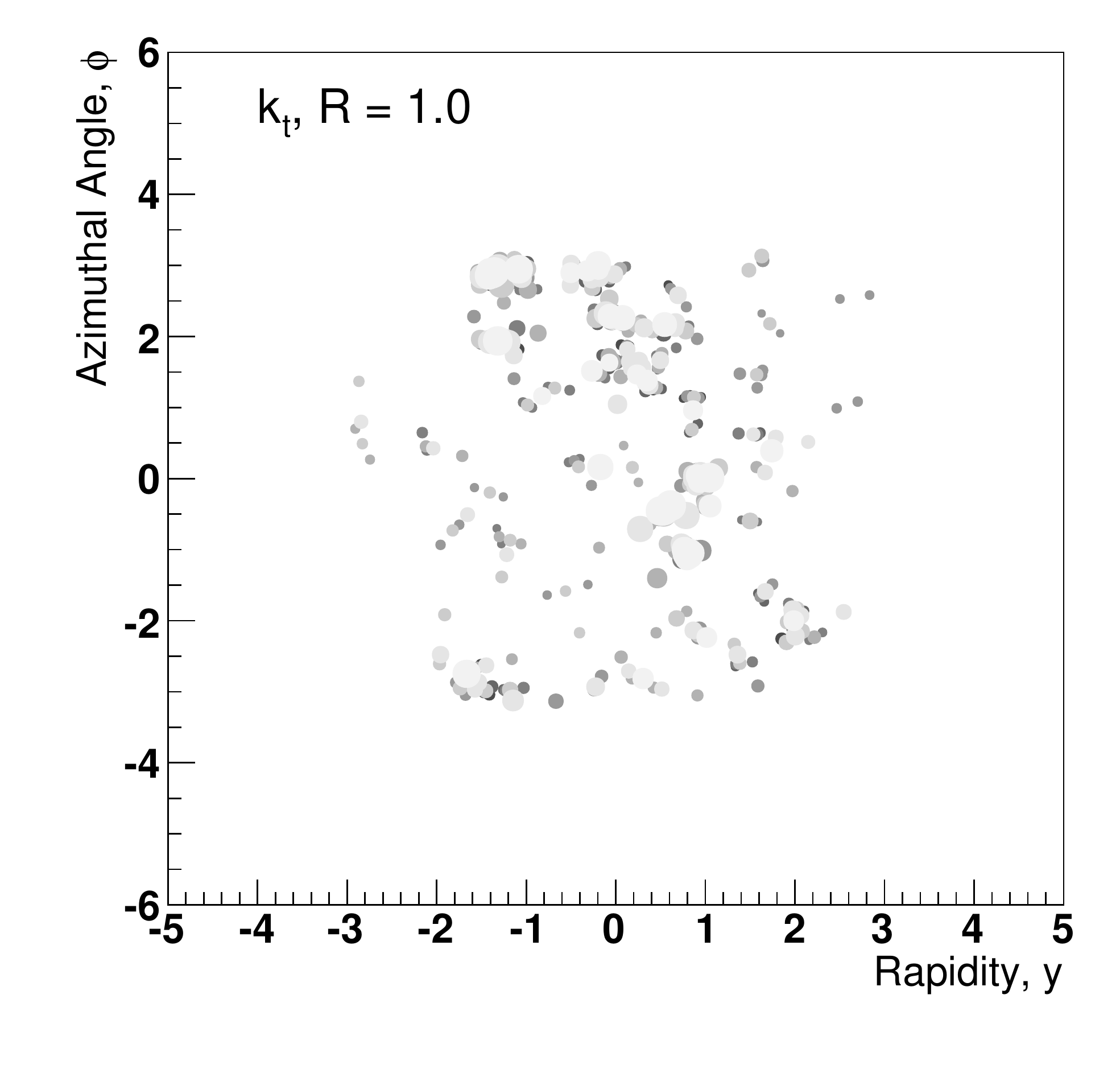}
 \caption{A simulated $Z'\rightarrow t\bar{t}$ event clustered with anti-$k_t$ (left), C/A (middle) and $k_t$ (right).  Particles are colored according to the jet they are clustered in (highest $p_\text{T}$ jet is black, then red, then green).  The particle size is proportional to $\log(10\times p_\text{T}/\text{GeV})$.  The bottom panel shows the clustering history.  Each proto jet merger is recorded with the size of each pre-merger proto jet proportional to $\log(10\times p_\text{T}/\text{GeV})$, where $p_\text{T}$ is the momentum of the merged proto jet.  Mergers are colored from dark (earlier) to light (later).  Note that $\phi$ is $2\pi$ periodic so the particles at $-\pi$ are geometrically close to the particles at $+\pi$.}
 \label{fig:jetclusteringexample}
  \end{center}
\end{figure}

\begin{figure}[h!]
\begin{center}
\includegraphics[width=0.33\textwidth]{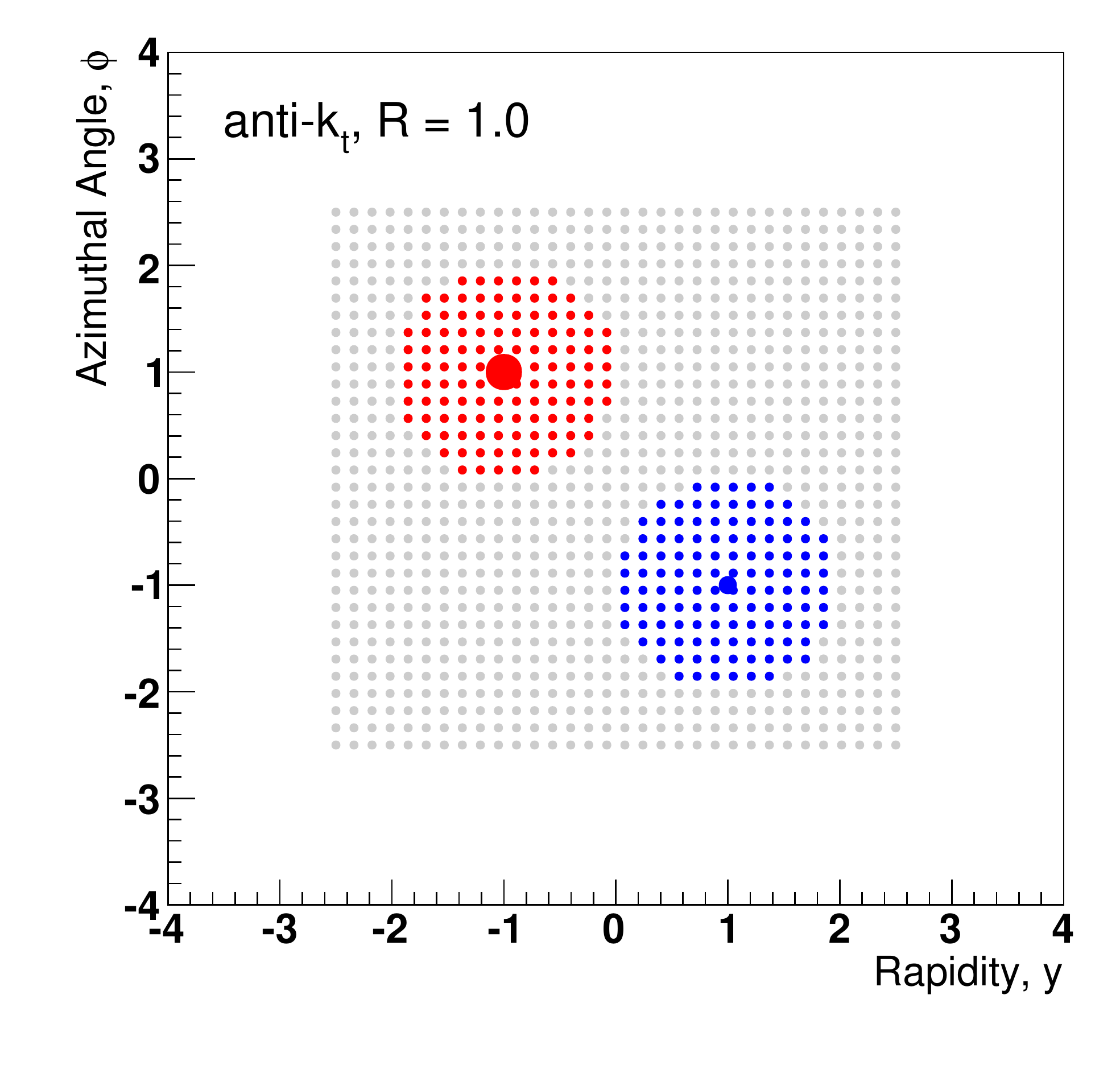}\includegraphics[width=0.33\textwidth]{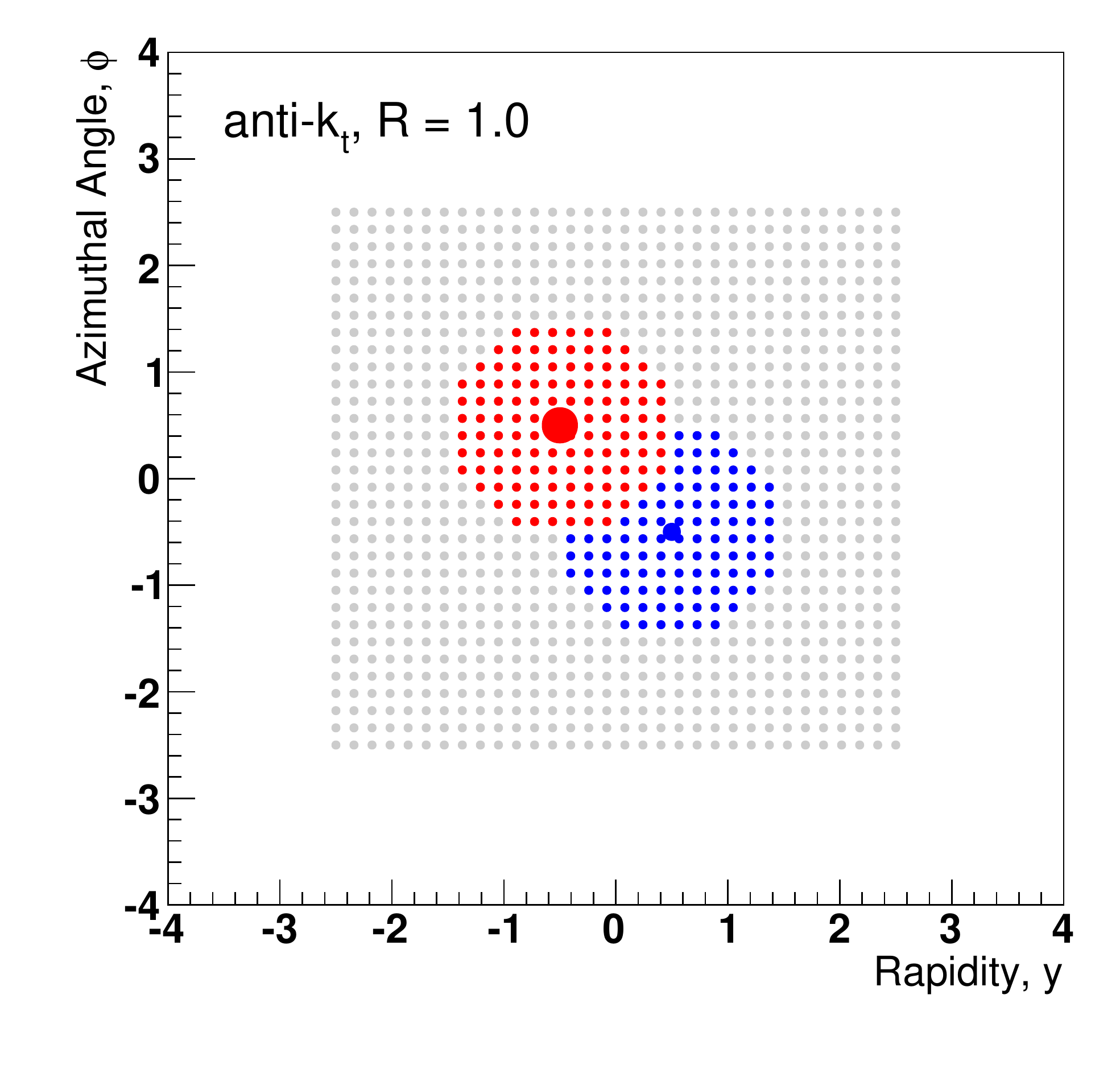}\includegraphics[width=0.33\textwidth]{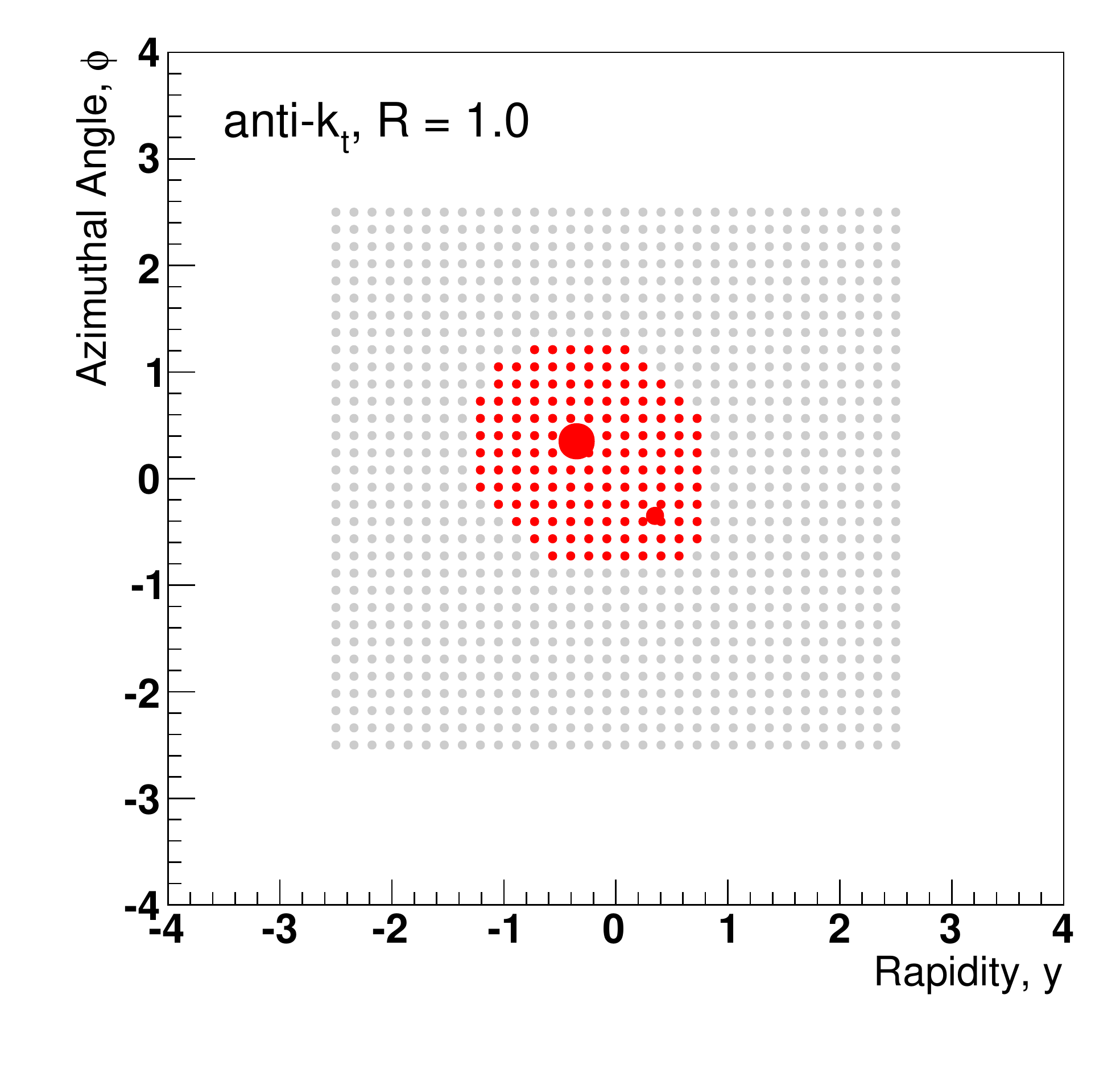}
 \caption{Jet clustering with two high $p_\text{T}$ particles and a grid of ghost particles.  Particles clustered in the higher $p_\text{T}$ jet are colored red and those in the lower $p_\text{T}$ jet are colored blue.  All other ghosts are gray.  The ghosts marker size is arbitrary and the radius of the two high $p_\text{T}$ particle markers $\propto p_\text{T}$.  When the two particles are far away (left), the jet areas are $\pi R^2$, while if they are within $|\Delta\vec{\rho}|<2R$, then one is circular and the other is a crescent.  If $|\Delta\vec{\rho}|<R$ (right), than the jets merge. }
   \label{fig:jetclusteringexampleareas}
  \end{center}
\end{figure}

\noindent One common widely used set of algorithms use the $k_t$ family of metrics, $d_{ij}(k)=\min(p_{T,i}^{2k},p_{T,j}^{2k})|\vec{\rho}_i-\vec{\rho}_j|^2/R^2$ and $d_{iB}(k)=p_{T,i}^{2k}$.  The parameter $R$ is roughly the size of the jet in $(y,\phi)$.  When $k=0$, the clustering procedure is called the Cambridge-Aachen (C/A) algorithm~\cite{Dokshitzer:1997in,Wobisch:1998wt} and the distance metric is independent of $p_\text{T}$.  By far, the most ubiquitous jet algorithm used at the LHC is the anti-$k_t$ algorithm~\cite{Cacciari:2008gp} with $k=-1$ (the $k_t$ algorithm has $k=+1$).  Figure~\ref{fig:jetclusteringexample} shows an example $Z'\rightarrow t\bar{t}$ simulated event clustered with the $k_t$, C/A and anti-$k_t$ algorithms.  The core of the highest $p_\text{T}$ jets is the same for all three algorithms\footnote{This statement can be quantified: in perturbation theory, there is no difference between algorithms in the $k_t$ family at leading order\cite{Dasgupta:2007wa,Salam:2009jx}.}.  However, the soft radiation on the outside of the jets varies between the three approaches.  One way to visualize the origin of these differences is in the bottom plots of Fig.~\ref{fig:jetclusteringexample} which shows the clustering history.  Because of the negative power of $p_\text{T}$, the anti-$k_t$ algorithm clusters higher $p_\text{T}$ particles first.  The $k_t$ algorithm clusters the softest particles first and the C/A algorithm clusters the closest particles first, independent of $p_\text{T}$.  As a result, anti-$k_t$ jets have the most regular catchment area which makes them easier to calibrate.   This is quantified with the notion of the {\it jet area}~\cite{ghost}, which is defined by\footnote{This is called the {\it active area}; there are other less used possibilities such as the Voronoi area~\cite{ghost}.}

\begin{align}
A_J = \lim_{a_g\rightarrow 0}\lim_{p_\text{T,g}\rightarrow 0}\sum_{g\in G}a_g\mathbb{I}(g\in J),
\end{align}

\noindent where $G$ is a set of {\it ghost particles} uniformly spread over $|\eta|<\eta_\text{max}$ and $\phi$, and $\mathbb{I}$ is the indicator function that is $1$ when its argument is true and $0$ otherwise.   The area of a ghost $g$ is $a_g=4\eta_\text{max}\pi/|G|$ (so the limit $a_g\rightarrow 0$ is the same as $|G|\rightarrow\infty$). Each ghost particle has a small but finite $p_\text{T,$g$}$; for an IRC safe algorithm, as $p_\text{T,$g$}\rightarrow 0$, the ghost particles do not influence the clustering.   A ghost particle $g\in J$ when after running jet clustering with the ghost particles, $g$ is clustered in the jet $J$ (which coincides with the jets clustered without ghosts by IRC safety).  In practice, the ghost particle transverse momenta are set to a small number and the number of ghosts is fixed, but large enough so that edge effects are negligible.  The catchment area of an anti-$k_t$ jet is a circle with area $\pi R^2$, except when two anti-$k_t$ jets are within $|\Delta\vec{\rho}|<2R$ in which case the higher $p_\text{T}$ one is a circle and the other is a crescent.  This is demonstrated by Fig.~\ref{fig:jetclusteringexampleareas}.  The regular shape of anti-$k_t$ jets makes the their calibration (defined below) more universal, i.e. less dependent on the event topology.

With the ATLAS detector, jets are built from calorimeter-cell clusters.  If locally calibrated calorimeter-cell clusters are used for the jet clustering, then the resulting jets are at the LCW-scale and otherwise are at the EM-scale.  Jets are calibrated so that on average they have the same energy as a {\it particle-level} jet clustered from all detector-stable particles prior to reaching the detector, excluding muons and weakly interacting particles such as neutrinos~\cite{Aad:2014bia,ATLAS-CONF-2015-037}.  The default clustering scheme is the anti-$k_t$ $R=0.4$ algorithm.  In the first step of the calibration\footnote{At the end of Run 1, an {\it origin correction} was added as a first step to improve the angular resolution.  See Sec.~\ref{origincorrection}.}, the average amount of pileup energy is subtracted from each jet using the jet areas method~\cite{Cacciari:2007fd,Aad:2015ina}.  As a diffuse source of noise, the amount of pileup energy contributing to a jet is proportional to the jet area.  The jet-by-jet correction is given by $p_\text{T,J}\mapsto p_\text{T,J}-\rho A_J$, where $\rho$ is the median pileup density\footnote{$R=0.4$ $k_t$ jets are used to calculate $\rho$.  Hard-scatter jets have little impact on this median.} $\rho=\text{median}_J(p_\text{T,J}/A_J)$.  Additional corrections based on $\mu$ and the number of reconstructed vertices (NPV) remove the residual pileup dependence\footnote{NPV is only sensitive to in-time pileup whereas $\mu$ additionally reflects the out-of-time pileup.}.  The next step is the core calibration, which corrects the energy and $\eta$ of the jets using numerical inversion (See Appendix~\ref{numericalinversion}) based on simulation.  At this stage, the reconstructed jet energy in simulation is inclusively unbiased.  At the end of Run 1, an additional MC-based calibration was introduced ({\it global sequential calibration}~\cite{ATLAS-CONF-2015-002}) to reduce residual biases depending on the jet flavor and energy leaked beyond the hadronic calorimeter (see Sec.~\ref{sec:qgtagging}).   A variety of object balancing techniques are used to validate the calibration procedure in data and result in a residual correction applied to the data~\cite{ATLAS-CONF-2015-017,ATLAS-CONF-2015-057,Aad:2014bia,ATLAS-CONF-2015-037}.

		Basic quality criteria are also used to remove jets from pileup and other sources of noise~\cite{Aad:2014bia,Aad:2015ina}.  A quantitative description of jets reconstruction, including systematic uncertainties, is found in Sec.~\ref{experimentaluncerts}.  Jets and their internal structure are the main focus of Part~\ref{part:qpj} and will therefore be discussed in much more detail in subsequent chapters.

		\clearpage
		
		\section{Electrons, Photons, Muons, and Taus}
		\label{sec:leptons}
		
		Lepton and photon reconstruction all rely on inner detector tracks.   Electrons are built from single tracks matched to a cluster of electromagnetic calorimeter-cells~\cite{Aad:2014nim,Aad:2014fxa,Aad:2011mk}.  Photons are either matched to two tracks if a {\it conversion} happens in the ID\footnote{There is about $0.5  X_0$ for $|\eta|< 0.8$ and about $1.5X_0$ for 0.8<$|\eta<1.5$.} or zero tracks if no conversion happens before the calorimeter~\cite{Aad:2014nim}.   Tracking for particles from secondary vertices (such as conversion electrons) extends as far as 80 cm into the ID.  In addition to stricter matching requirements between tracks and clusters (including energy/momentum and $\Delta R$), further background rejection is achieved by requiring that the electromagnetic shower and the amount of transition radiation from the TRT be consistent with expectations for electrons and photons.  
		
		Muons are constructed from tracks in the ID matched to tracks in the MS~\cite{Aad:2014rra,Aad:2014zya,Aad:2016jkr}.  Additional muons beyond the ID acceptance are built entirely out of MS tracks.  Furthermore, the efficiency for muons is recovered for $|\eta|<0.1$, where the MS is only partially instrumented due to calorimeter and ID services, by using ID tracks matched to either a calorimeter energy deposit consistent with a minimum ionizing particle or a track segment in the MS.  
		
		Hadronically decaying tau leptons are constructed from jets~\cite{Aad:2014rga,Aad:2015unr}.  Tau leptons decay hadronically about 2/3 of the time and of those, about 80\% have one charged pion ({\it one-prong}) while about 20\% have three charged pions ({\it three-prong}).  A series of calorimeter and tracking observables such as the fraction of EM energy, the width in $\Delta R$, and the jet mass are used to discriminate taus from electrons and jets.  
		
		The reconstruction efficiency and energy scale of leptons and photons are calibrated using simulation and corrected based on in-situ studies.  Low mass resonances and $Z$ bosons are used for both the energy calibration and the efficiency measurement.  The latter uses a {\it tag-and-probe} method where one object $o_1$ passes a strict selection ({\it tag}) and another object $o_2$ with $m_{o_1o_2}$ near the resonance mass is probed to see if it passes the particle identification.  A quantitative comparison of the reconstruction efficiencies and resolutions, including systematic uncertainties, can be found in Sec.~\ref{experimentaluncerts}.  Specific particle identification requirements are specified when used.

		\clearpage
		
		\section{Missing Transverse Momentum}
		\label{sec:MET}
		
		Particles that only decay weakly such as neutrinos are not measured directly.  However, the sum of the transverse momenta from all such particles can be inferred using conservation of momentum in the transverse plane.  In the absence of a detector, $\vec{p}_\text{T}^\text{miss}=-\sum\vec{p}_\text{T,i}^\text{visible}$ would be equal to $\vec{p}_\text{T}^\text{non-interacting}$.  The reconstructed $\vec{p}_\text{T}^\text{miss}$ in ATLAS is built from all objects described in this chapter.  Each object's unique calibration is used to improve the overall missing momentum resolution.  The measured energy that is not assigned to jets\footnote{All topo-clusters belong to a jet, but jets below a threshold of $\sim 20$ GeV are not calibratable (no correlation between detector-level and particle-level energy).  Measured energy is `not in a jet' if it is in one of these low $p_\text{T}$ jets. }, electrons, photons, etc. is called the {\it soft-term}.  At $\sqrt{s}=8$ TeV, the baseline soft-term was constructed from calibrated calorimeter-cell clusters not assigned to jets or other objects~\cite{ATLAS-CONF-2013-082,Aad:2012re}.  For the early Run 2 data, this default has shifted to a track-based soft-term~\cite{ATL-PHYS-PUB-2015-023,ATL-PHYS-PUB-2015-027}.  Information about neutral particles is lost when only using tracks, but the neutral contribution cancels on average because charge-to-neutral fluctuations are symmetric in azimuth. The main motivation for the track-based term is the robustness to pileup.  Tracks from collisions other than the primary hard-scatter vertex can be readily identified and removed from the soft-term.  By construction, the contribution from the soft-term is subdominant to the contribution from other high $p_\text{T}$ objects in events with real sources of missing particles.
		
		The magnitude of the missing transverse momentum\footnote{This quantity uses an `$E$' because it has historically been called the {\it missing transverse energy}.  This is a misnomer because energy is a scalar, but has been used because it is mostly due to calorimeter energy measurements (as opposed to momentum measurements from tracks).} is called $E_\text{T}^\text{miss}$ and is a powerful discriminating variable for identifying events with neutrinos, such as the pair production of top quarks, $pp\rightarrow t\bar{t}\rightarrow b\bar{b}W^+W^-\rightarrow b\bar{b}l\nu qq'$ or the production of new particles that do not interact with the detector.  The particles targeted by the SUSY search presented in Part~\ref{part:susy} decay via undetectable particles that can have $p_\text{T}\sim \mathcal{O}(100)$ GeV and so the $E_\text{T}^\text{miss}$ will be one of the most important observables.

\part[The Quantum Properties of Jets]{The Quantum Properties of Jets\\[1ex]\makebox[0pt]{\includegraphics[width=0.5\paperwidth]{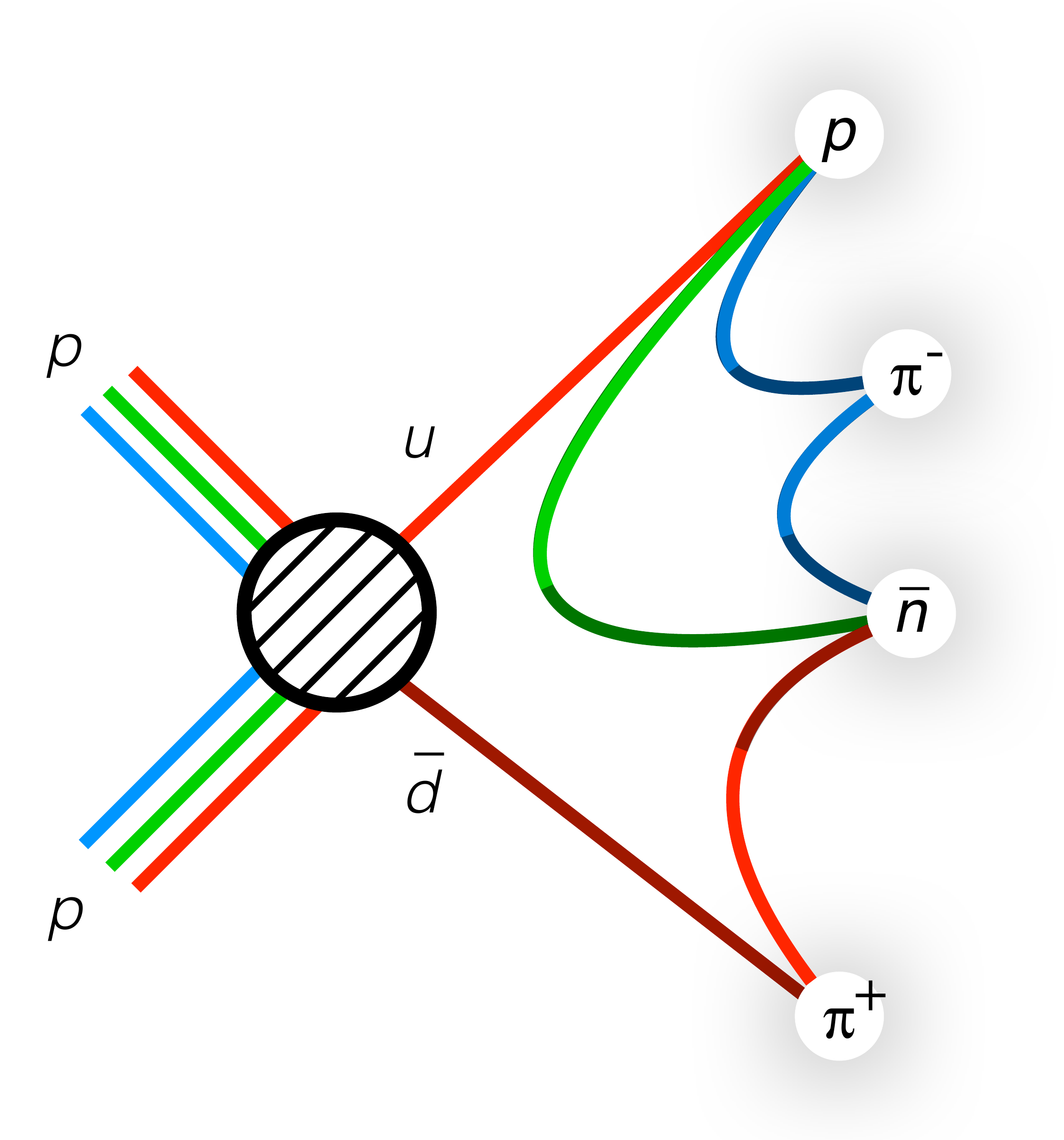}}\\\blfootnote{A schematic (oversimplified) diagram illustrating the transmission of quark charge to the quantum properties of jets. The $u$ and $\bar{d}$ quarks are not directly observable, but their electric and color charge have observable consequences for the pattern of hadrons.}}\label{part:qpj}

The {\it quantum properties of jets} are the observable consequences of the quantum properties of the initiating quarks and gluons.  Quarks are the only elementary particles that are charged under all forces.  Due to confinement, these quantum properties are not directly observable and are instead embedded in the radiation pattern within and around a jet.  The strong coupling constant is sufficiently small that many aspects of this transmission of quark and gluon charge to inter- and intrajet radiation can be understood in the context of perturbation theory.  However, there are important non-perturbative aspects of jet formation.  There are additional theoretical challenges due to the rich structure of QCD; for example the gluon carries color charge unlike the photon in QED.  Measuring the quantum properties of jets also presents a significant experimental challenge.  Differences in radiation patterns between different quark and gluon charges are often subtle and thus require precise measurements of jet constituent energies and locations.  Furthermore, there are several sources of diffuse noise such as pileup that complicate both the measurements and their interpretations.

A parton that initiates a jet is uniquely determined by its charges under all symmetry groups of the Standard Model.  The most basic property is the parton three-momentum, a charge of the Poincar\'{e} group.  For most applications involving jets, this is the only relevant quantum property, as the jet is viewed as a noisy proxy of the parton; the internal structure is a nuisance.  The jet three-momentum also has the least quantum noise of all quantum properties of jets; the average jet $p_\text{T}$ is the same as the average parton $p_\text{T}$ within 5-10\%~\cite{Salam:2009jx}.  A related quantum property is the parton mass.  Jets produced by light quarks and gluons can acquire non-negligible mass resulting from relatively hard wide-angle radiation.  This mass encodes information about color charge of the initiating quark or gluon and is unrelated to the on-shell quark and gluon mass.  On the other hand, jets initiated by the hadronic decays of genuinely heavy particles such as $W/Z$ or Higgs bosons have significant mass that is correlated with the parton mass.  The last charge of the Poincar\'{e} group is the particle spin.  This information is lost for light quark and gluon jets as a result of the hadronization process.  However, the angular distribution of subjets within top quark and heavy boson jets does contain some information about the polarization of the initiating parton.  The quantum properties of jets related to the Poincar\'{e} group symmetries are studied in Chapter~\ref{cha:bosonjets}.

The other charges of the Standard Model are associated with the internal $U(1)\times SU(2)\times SU(3)$ symmetry.  Partons are most likely to fragment into hadrons of the same electric charge.  Therefore the electric charge of the hadrons inside a jet encodes information about the parton electric charge.  This is complicated by the finite acceptance of the detector for low $p_\text{T}$ particles and also the fact that additional charge must flow into quark jets in order to make the net charge an integer.  Higher energy hadrons carry more information about the parton charge and low energy hadrons are subject to threshold effects, so a {\it jet charge} can be constructed by using the energy-weighted charge of hadrons inside a jet as a proxy of the parton charge.  Chapter~\ref{cha:jetcharge} is an extensive study of jet charge, both as a probe of jet formation and for charge tagging.  For example, the energy-dependence of the jet charge is studied for evidence of {\it scale violation}.  Additionally, jet charge in boosted boson jets is studied in Chapter~\ref{cha:bosonjets}, where no additional charge needs to flow into the jet due to the color singlet nature of the initial state.

Analogous to the electric charge for the electroweak force is the color charge for the strong force.  Partons can either be in the singlet (no charge), triplet, or octet representations of $SU(3)$.  Partons in the triplet representation carry one color while partons in the octet representation carry one color and one anti-color.  The radiation pattern from hadronic jets resulting from singlet partons tends to be enhanced within the core of the jet relative to jets from octets that are {\it color-connected} to other partons in the event.  This is particularly important to study because of the applications to jet tagging as a boosted $H\rightarrow b\bar{b}$ jet gives rise to a singlet radiation pattern while the background $g\rightarrow b\bar{b}$ process should resemble the octet radiation pattern.  Chapter~\ref{cha:colorflow} is a study of colorflow in and around boson jets.  In addition to differentiating singlet-induced jets from octet-induced jets, it is essential to study the differences between octet-induced jets (gluons) and triplet-induced jets (quarks).  Quark and gluon tagging is ubiquitous (if only implicit) at the LHC and despite being well-studied, is still an area of active research theoretically and experimentally.  Chapter~\ref{cha:multiplicity} presents a measurement of constituent multiplicity, an observable that is directly proportional to strength of the quark and gluon color charges $C_F$ and $C_A$.

One final quantum property is flavor.  Quark and gluon flavor are uniquely specified by color charge, but there is a further distinction into the various quark types.  Jet charge is sensitive to the up versus down type of the initiating parton, but there are an entire class of observables sensitive to heavy flavor quarks.  The use of $b$-tagging to probe quark flavor is studied in the context of boson jet tagging in Chapter~\ref{cha:bosonjets}.

Part II will explore the {\it substructure} and {\it superstructure}~\cite{Gallicchio:2010sw} of high energy jets in order to understand how quark and gluon quantum charges are realized within the observable pattern of hadrons.  These quantum properties of jets probe the detailed nature of the strong force as well as provide tools for discovering new particles and forces beyond those described by the Standard Model.  Table~\ref{tab:QPJoverview} summarizes quantum properties of jets discussed above as well as which chapters cover them.

\vspace{10mm}

\begin{table}[h!]
\centering
\begin{tabular}{cccc}
\hline 
\hline
 Quantum Property & Charge & Observable & Chapter\\
 \hline
\hline
  Electric Charge & $\pm 2/3,\pm 1/3,0$ & Jet Charge & ~\ref{cha:jetcharge} \\
  Color Charge & $\bf{1},\bf{8}$ & Jet Pull & ~\ref{cha:colorflow} \\
  Color Charge & $\bf{1},\bf{3}$ & Constituent Multiplicity & ~\ref{cha:multiplicity} \\
  Mass & $\alpha_s Rp_\text{T}, m_W,m_Z$ & Jet Mass & ~\ref{cha:bosonjets} \\
  Electric Charge & $\pm 1,0$ & Large-radius Jet Charge & ~\ref{cha:bosonjets} \\  
  Flavor & $b/c/\text{light}$ & $b$-tagging &~\ref{cha:bosonjets} \\  
\hline
\hline
\end{tabular}
\caption{The various quantum properties of jets studied in Part II along with the corresponding chapter.  Even though some of the quantum properties are due to forces other than the strong force, their realization in jets is due to the quantum evolution from partons to hadrons via QCD.}
\label{tab:QPJoverview}
\end{table}
  \chapter{Jet Charge}
 \label{cha:jetcharge}	

Quarks and gluons produced in high-energy particle collisions hadronize before their electric charge can be directly measured.  However, information about the electric charge is embedded in the resulting collimated sprays of hadrons.  One jet observable sensitive to the electric charge of quarks and gluons is the momentum-weighted charge sum constructed from charged-particle tracks in a jet~\cite{Feynman1978}.   
Called the {\it jet charge}, this observable was first used experimentally in deep inelastic scattering studies~\cite{Berge1980,Berge1981,Allen1981,Allen1982,Albanese1984,Barlag1982,Erickson1979} to establish a relationship between the quark model and hadrons.  Since then, jet charge observables have been used in a variety of applications, including tagging the charge of $b$-quark jets~\cite{SLD1995,Tasso1990,Delphi1991,Aleph1991,Opal1992,Opal1994,Delphi1996,CDF1999,Abazov2007,CDF2011,ATLAS2011} and hadronically decaying $W$ bosons~\cite{Barate1998,Abreu:2001rpa,Acciarri:1999kn,Abbiendi:2000ej,ATLAS-CONF-2013-086,CMS-PAS-JME-14-002} as well as distinguishing hadronically decaying $W$ bosons from jets produced in generic quantum chromodynamic (QCD) processes~\cite{Khachatryan:2014vla} and quark jets from gluon jets~\cite{Aad:2014gea,ATLAS-CONF-2013-086,Acton:1992uu,Arnison:1986ja}.

\begin{figure}[h!]
\begin{center}
\includegraphics[width=0.99\textwidth]{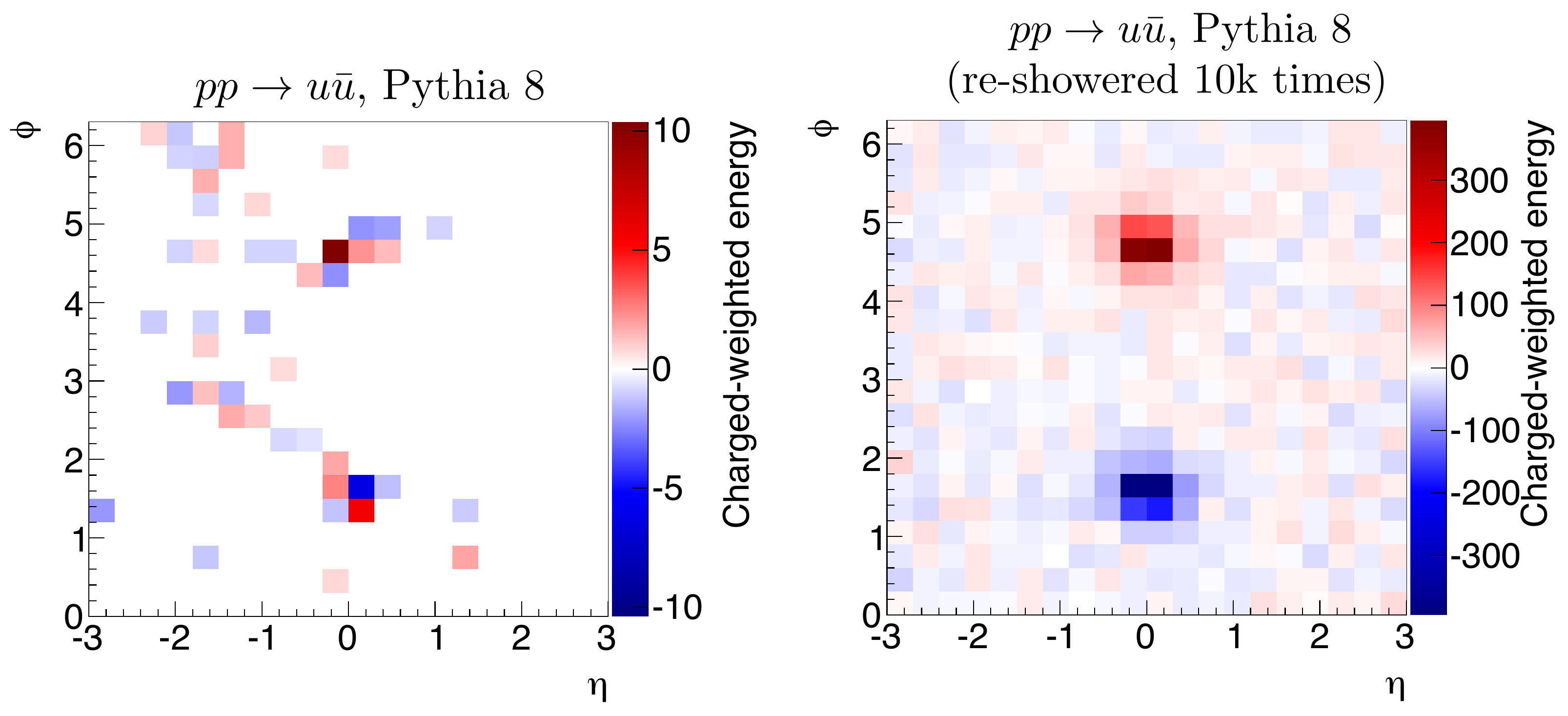}
\end{center}	
\caption{Left: the process $pp\rightarrow u\bar{u}$ is simulated with {\sc Pythia} 8 once; Right: the same hard-scatter process with fragmentation re-simulated $10,000$ times.  Each pixel intensity represents the charge-weighted sum of the energies of all particles produced within the $\phi$ and $\eta$ covered by the pixel area.  }
\label{fig:reshowered}
\end{figure}

As will be a reoccuring theme for the quantum properties of jets, the charge information embedded in the radiation pattern of jets is subtle.  The left plot of Fig.~\ref{fig:reshowered} shows the pixelated energy distribution for a simulated event $pp\rightarrow u\bar{u}$.  The pixel intensity is the charge-weighted sum of the energy of all hadrons resulting from fragmentation, prior to interactions with the detector.  There are clearly two nodes of localized energy deposits (jets), but it is not possible to deduce which jet as initiated by the up quark and which was initiated by the anti-up quark.  In contrast, after re-simulated the parton shower and hadronization $10,000$ times, the right plot of Fig.~\ref{fig:reshowered} clearly shows the upper jet is from the up quark (positive charge) and the bottom jet is from the anti-up quark.  In the data, a given hard-scatter event fragments only once.  The tools developed in this chapter are intended to apply to individual events, but they are most useful when considering an ensemble of events.

This chapter presents\footnote{The performance studies presented in this chapter are published in Ref.~\cite{ATLAS-CONF-2013-086} (with technical help from M. Swiatlowski and manuscript help from A. Arce) and the precision measurement is published in Ref.~\cite{Aad:2015cua} (with help from M. Schwartz on the theory calculation).} performance studies related to the detector reconstruction and charge tagging performance of the jet charge as well as a precision measurement of the jet charge moments as a function of jet $p_\text{T}$ with the ATLAS detector.  The chapter begins in Sec.~\ref{sec:jetcharge:background} with some background information.

\clearpage

\section{Background}
\label{sec:jetcharge:background}

The jet charge is defined in Sec.~\ref{sec:construcing} and its important properties are reviewed in Sec.~\ref{sec:jetcharge:properties}.  Section~\ref{sec:jetchargetheory} describes the theoretical predictions for the jet charge distribution.  The section ends in Sec.~\ref{sec:chargetagging} with some comments about charge tagging.

\subsection{Constructing the jet charge}
\label{sec:construcing}

\begin{figure}[h!]
\begin{center}
\includegraphics[width=0.5\textwidth]{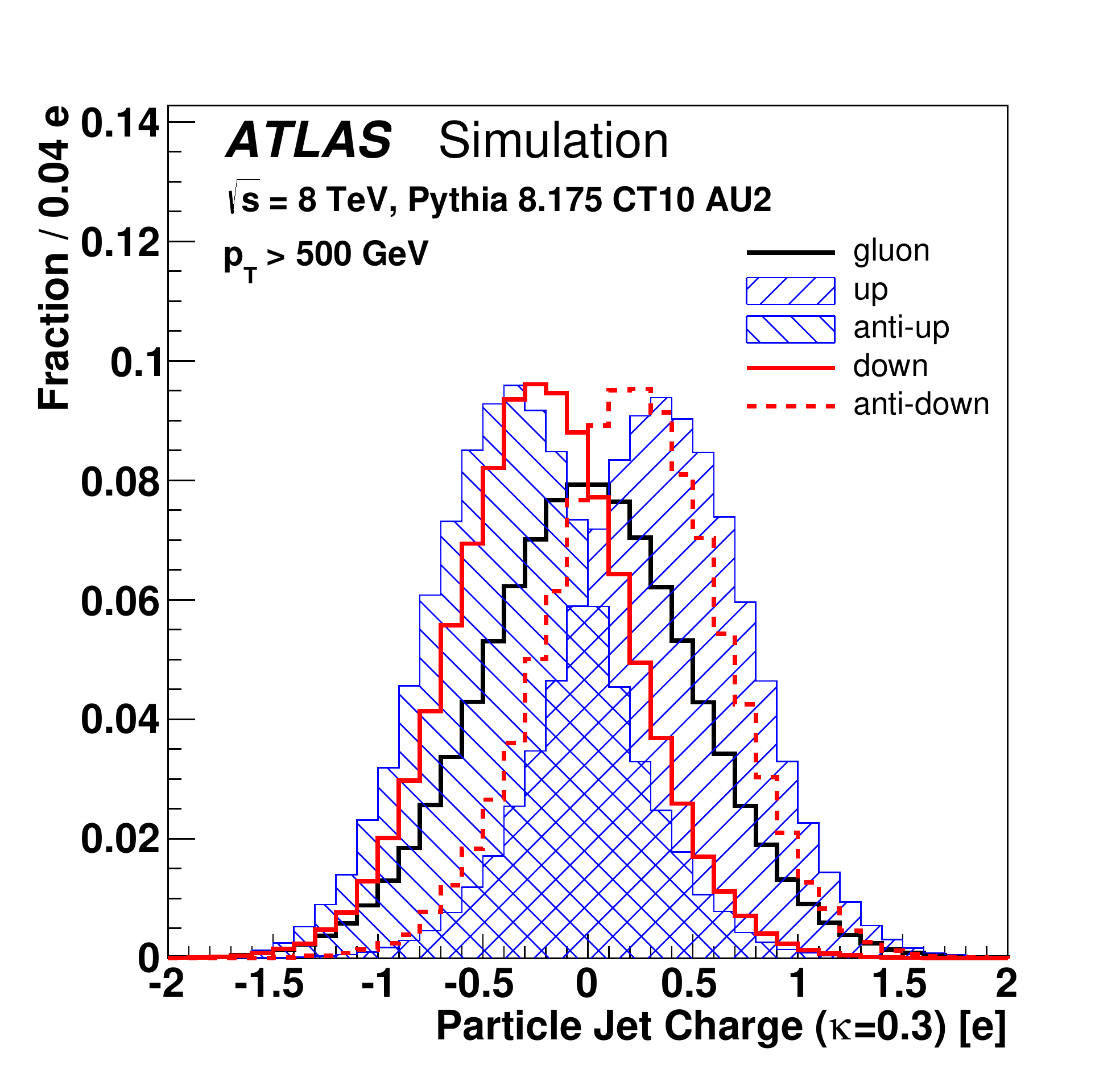}\includegraphics[width=0.5\textwidth]{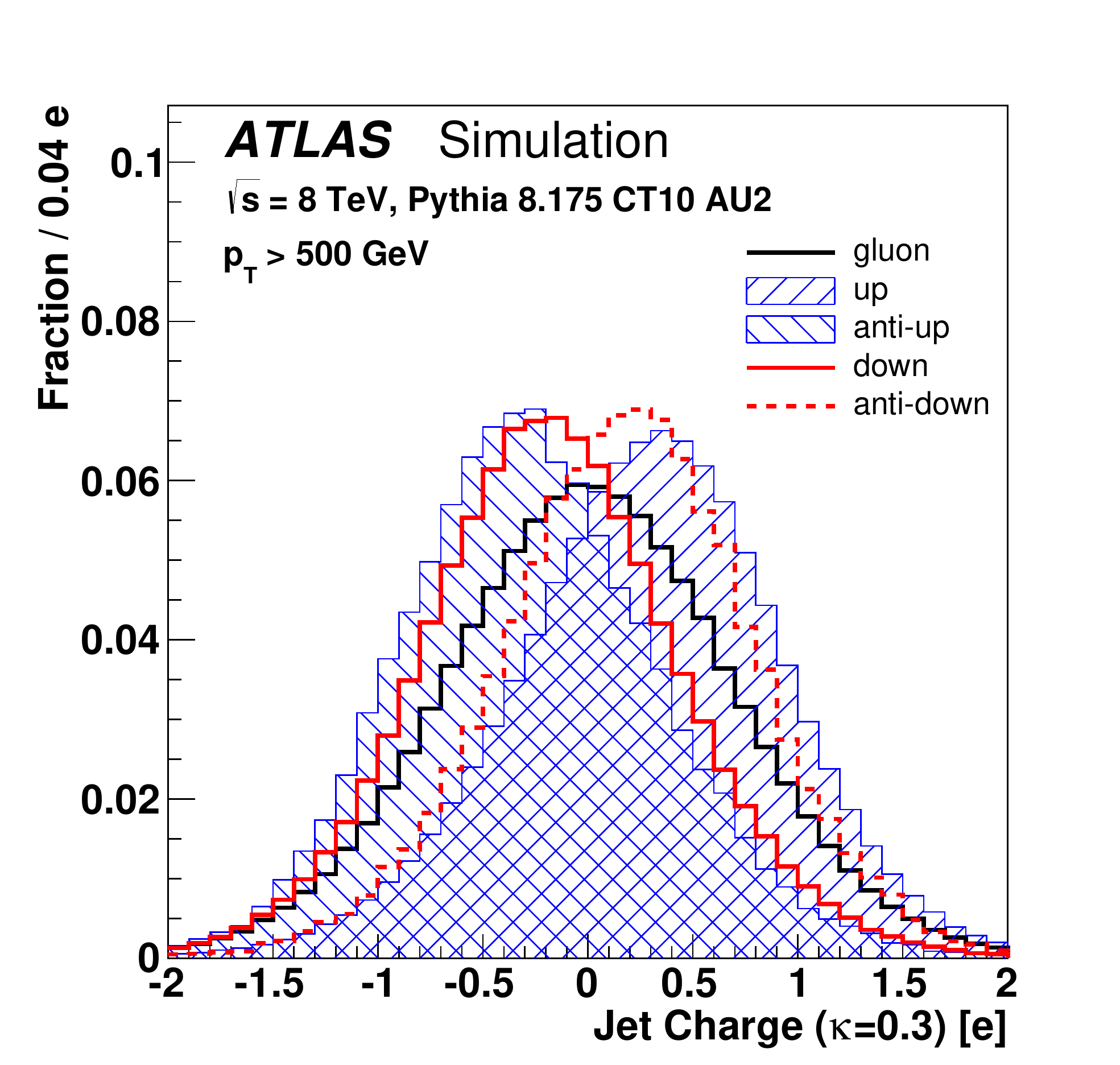}\\
\includegraphics[width=0.5\textwidth]{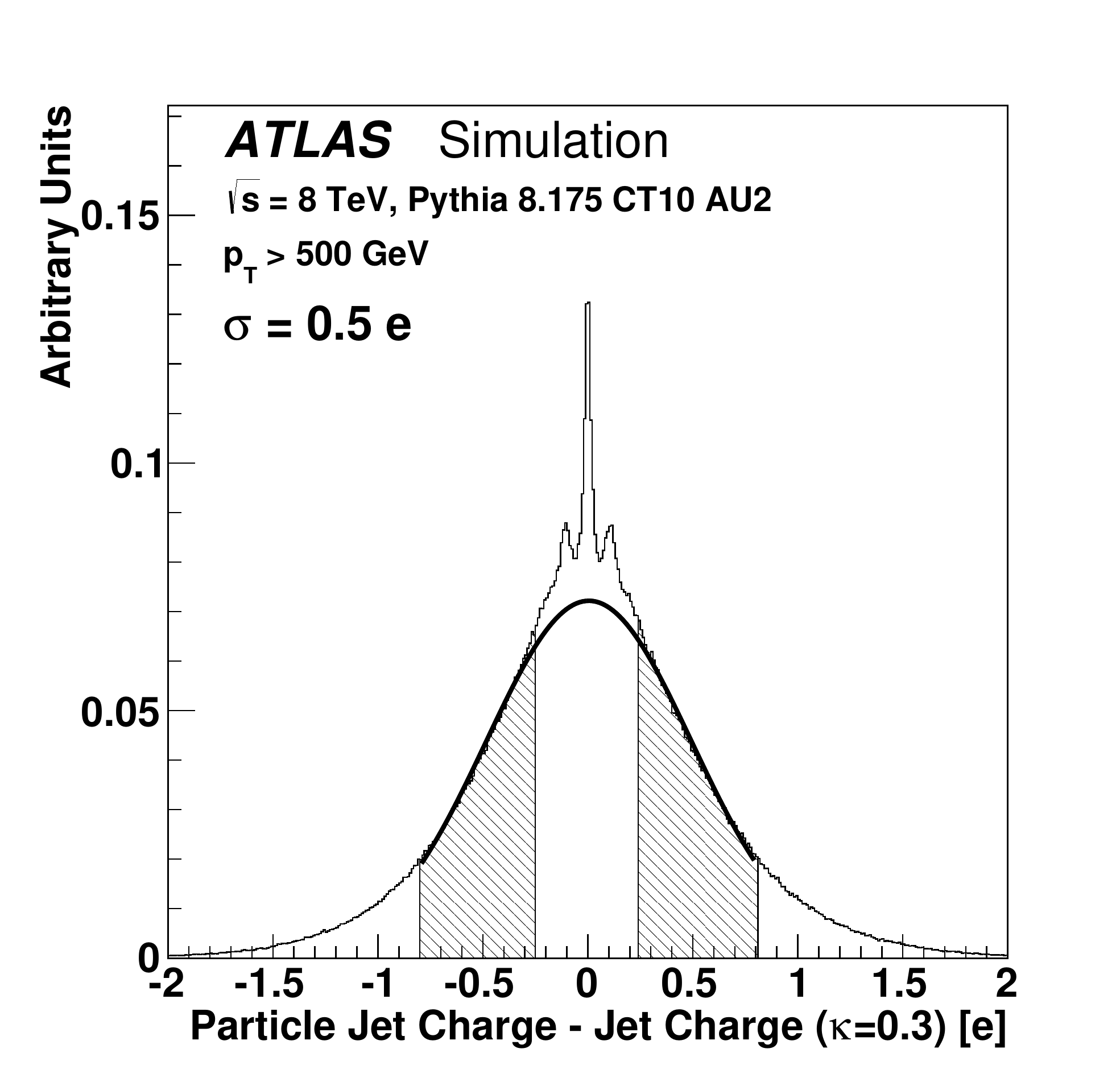}
\end{center}	
\caption{Top left (right): The particle- (detector-)level jet charge distribution for various jet flavors in a sample of jets with $p_\text{T}>500$ GeV~for $\kappa=0.3$. Bottom: the distribution of the jet-by-jet difference between the particle-level and detector-level jet charge distributions.  The shaded region is used to fit a Gaussian function to extract the bulk response resolution, which is $\sigma\sim 0.5$ $e$, where $e$ is the positron charge. See Sec.~\ref{sec:chargesamples} for details about the simulation.}
\label{fig:sortedbypartons}
\end{figure}

There is no unique way to define the jet charge.  The most na\"{i}ve construction is to add the charge of all tracks associated with a jet.   However, this scheme is very sensitive to lost radiation and diffuse soft radiation that contaminates the jet.  Therefore, a weighting scheme is introduced to suppress fluctuations.  The matching of tracks with the calorimeter-based jets is performed via the ghost-association technique~\cite{ghost}: the jet clustering process is repeated with the addition of {\it ghost} versions of measured tracks that have the same direction but infinitesimally small $p_\text{T}$, so that they do not change the properties of the calorimeter jets.  A track is associated with a jet if its ghost version is contained in the jet after reclustering.  Using such tracks, the jet charge $Q_J$ of a jet $J$ is calculated using a transverse-momentum-weighting scheme~\cite{Feynman1978}:

\begin{align}
  \label{chargedefcharge}
  Q_J = \frac{1}{({p_\text{T}}_J)^\kappa}\sum_{i\in \text{\bf Tracks}} q_i\times (p_\text{T,i})^\kappa,  
  \end{align} 

\noindent where $\text{\bf Tracks}$ is the set of tracks associated with jet $J$, $q_i$ is the electric charge of track $i$ in units of the positron charge, $p_{\text{T},i}$ is  transverse momentum of track $i$, $\kappa$ is a free regularization parameter, and ${p_\text{T}}_J$ is the transverse momentum of the calorimeter jet.  The distributions of $Q_J$ for various jet flavors are shown in Fig.~\ref{fig:sortedbypartons} for $\kappa=0.3$.  In the simulation, there is a clear relationship between the jet charge and the initiating parton's charge, as up-quark jets tend to have a higher jet charge than gluon jets.   Furthermore, gluon jets tend to have a higher jet charge than down-quark jets.  However, the jet charge distribution is already broad at particle level and the jet charge response ($Q_\text{particle-level}-Q_\text{detector-level}$) resolution is comparable to the differences in the means of the distributions for different flavors, so one can expect only small changes in the inclusive jet charge distribution for changes in the jet flavor composition.   The three narrow distributions on top of the bulk response distribution in Fig.~\ref{fig:sortedbypartons}(b) are due to cases in which only one or two charged particles dominate the jet charge calculation at particle level.  The two off-center peaks are due to cases in which one of the two high-$p_\text{T}$-fraction tracks is not reconstructed and the widths of the two off-center and central peaks are due to the (single) track and jet $p_\text{T}$ resolutions.  The bulk response is fit to a Gaussian function with standard deviation $\sigma\sim 0.5$ $e$ (units of the positron charge).

\clearpage

\subsection{Jet Charge Properties}
\label{sec:jetcharge:properties}

By using the calorimeter jet $p_\text{T}$ in the denominator of Eq.~\ref{chargedefcharge}, there is some sensitivity to the charge-to-neutral fraction in the jet, which contains useful information about the parton charge.  Alternative definitions using $(\sum_{i\in \text{\bf Tracks}} (p_\text{T,i}))^\kappa$ which lead to a bounded jet charge are studied in Sec.~\ref{sec:altdef}.  Sections~\ref{sec:jetcharge:IRsensitivity} describes how $\kappa$ regulates the sensitivity to soft radiation within a jet and Sec.~\ref{sec:jetcharge:LI} shows how the jet charge transforms under Lorentz boosts.

\subsubsection{Sensitivity to Soft Radiation}
\label{sec:jetcharge:IRsensitivity}

\begin{figure}[h!]
\begin{center}
{\includegraphics[width=0.45\textwidth]{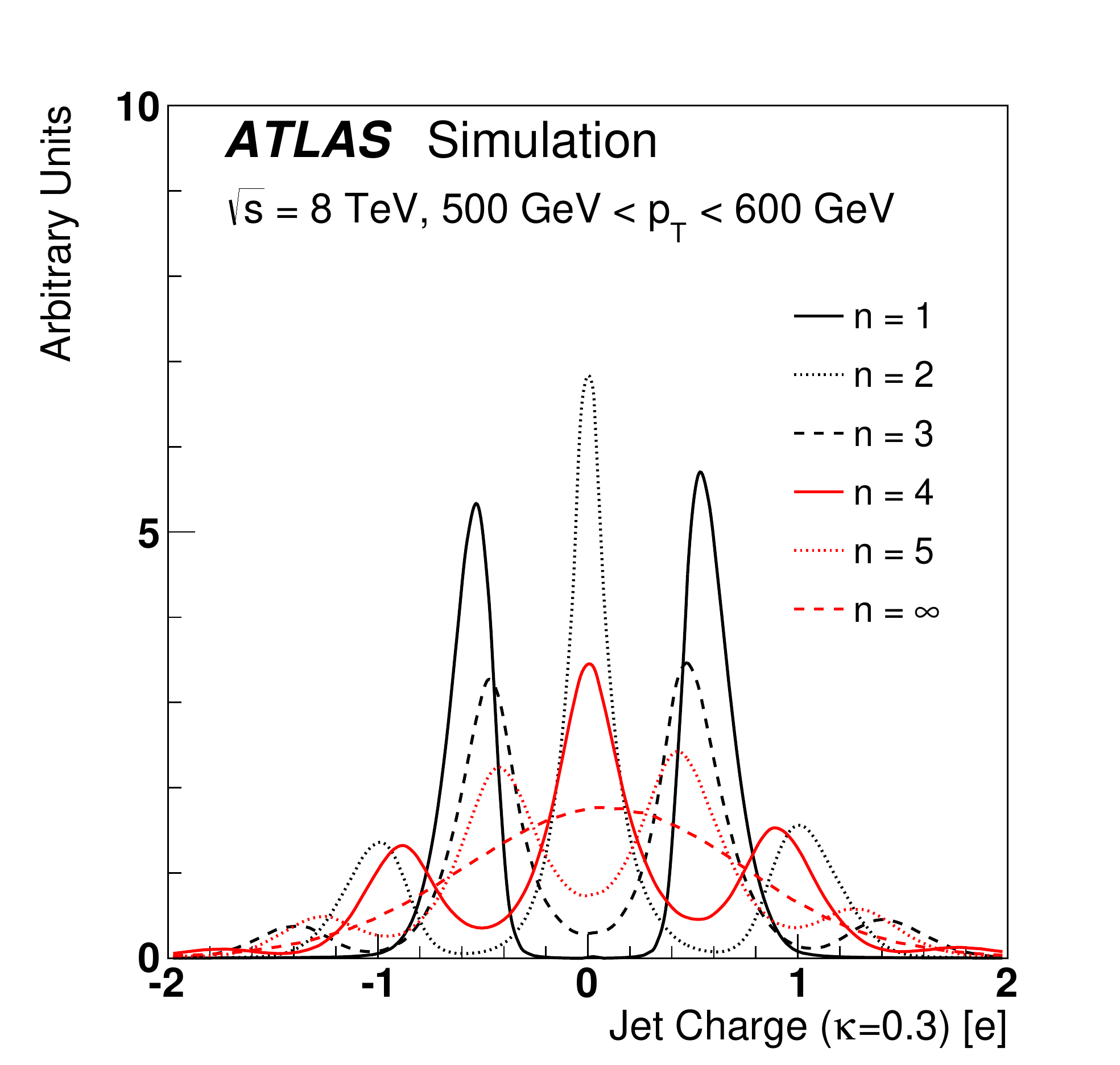}}{\includegraphics[width=0.45\textwidth]{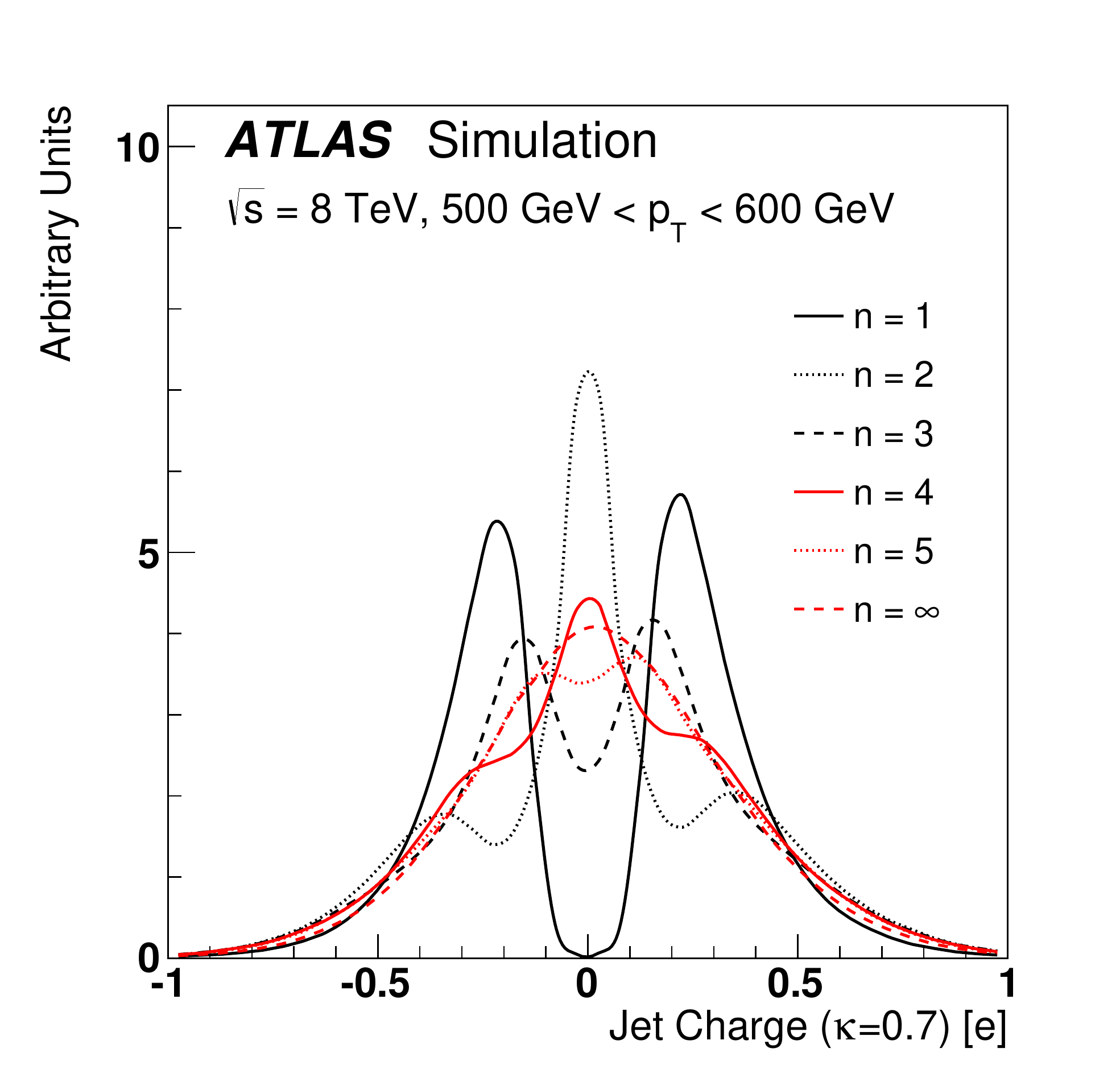}}

 \includegraphics[width=.45\columnwidth]{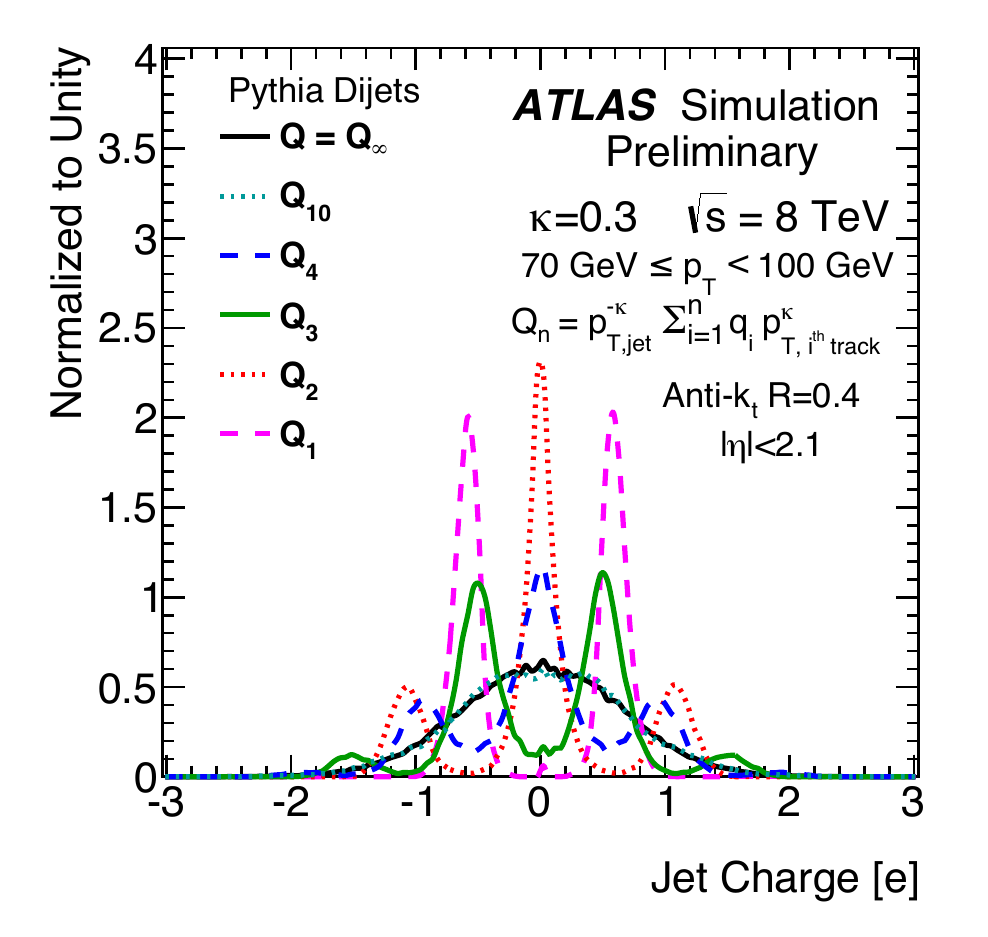}\includegraphics[width=.45\columnwidth]{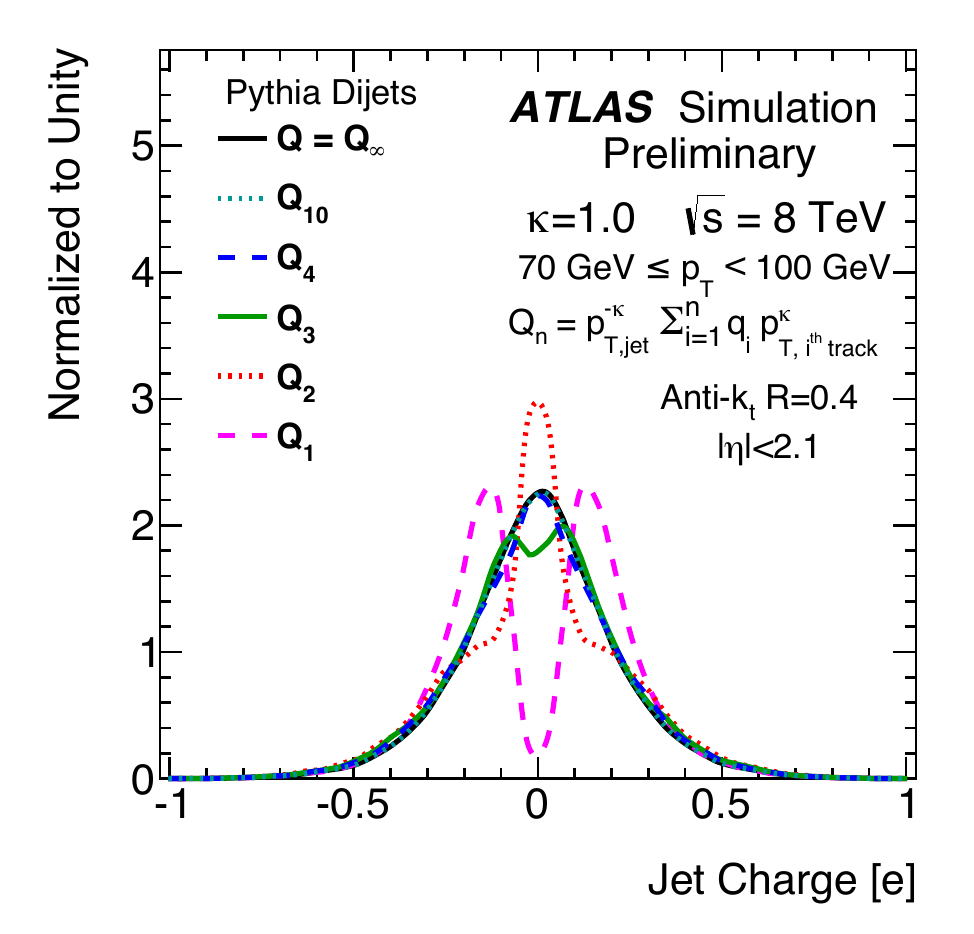} 

\end{center}	
\caption{The distribution of the jet charge built from the leading $n$ tracks ($Q_{J,n}$) for (left) $\kappa=0.3$ and (right) $\kappa=0.7$ or $\kappa=1.0$ for (top) 500 GeV $<p_\text{T}<$ 600 GeV and (bottom) 70 GeV $<p_\text{T}<$ 100 GeV.  In the top (bottom) plots, the mean number of tracks is about 15 (7).  See Sec.~\ref{sec:chargesamples} for details about the simulation.}
\label{fig:jetcharge:qn}
\end{figure}

The parameter $\kappa$ in Eq.~\ref{chargedefcharge} regulates the sensitivity of the jet charge to soft radiation.  Low values of $\kappa$ enhance the contribution to the jet charge from low-$p_\text{T}$ particles while in the $\kappa\rightarrow\infty$ limit, only the highest-$p_\text{T}$ track contributes to the sum in Eq.~\ref{chargedefcharge}.   The dependence on the highest-$p_\text{T}$ tracks is demonstrated using the plots in Fig.~\ref{fig:jetcharge:qn} with the variable $Q_{J,n}$, which is the jet charge in Eq.~\ref{chargedefcharge}, but built from the leading $n$ tracks.  The variable $Q_{J,1}$ is simply the weighted fragmentation function of the leading-track $p_\text{T}$ to the jet $p_\text{T}$ with weight $\kappa$.  The usual $Q_J$ is recovered in the limit $n\rightarrow\infty$.  Figure~\ref{fig:jetcharge:qn} shows the sequence $Q_{J,n}$ for $\kappa=0.3$ and $\kappa=0.7$.  For lower values of $\kappa$, many tracks are required for the sequence of distributions to converge to the full jet charge.  However, for $\kappa\gtrsim 0.7$, the distribution converges quickly, indicating that only the highest-$p_\text{T}$ tracks are contributing.  The peaks in the distributions in Fig.~\ref{fig:jetcharge:qn} are due to the discrete nature of hadron electric charge: if there is only one track, then the peaks are at $\pm \langle p_\text{track}^\kappa/p_\text{jet}^\kappa\rangle $ while if there are two tracks, then a peak at zero develops from the cases in which the two tracks have opposite charge.  If the charge of the tracks are chosen at random, it is twice as likely that the tracks have opposite charge compared with the case that both have positive charge and therefore the peak at zero is taller than the peaks at larger values of $|Q_{J,n}|$.

\begin{figure}[h!]
\begin{center}
\includegraphics[width=0.5\textwidth]{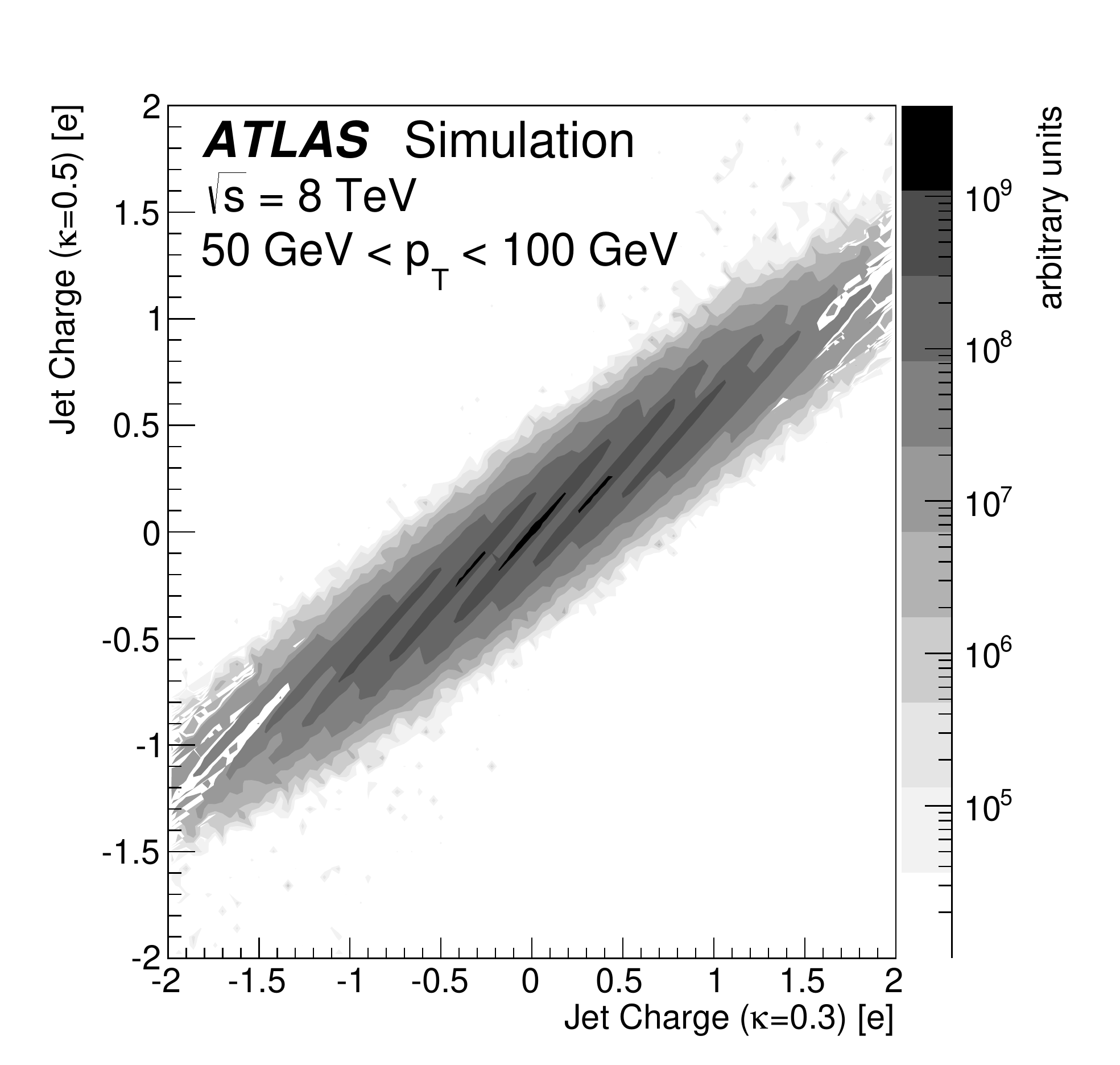}\includegraphics[width=0.5\textwidth]{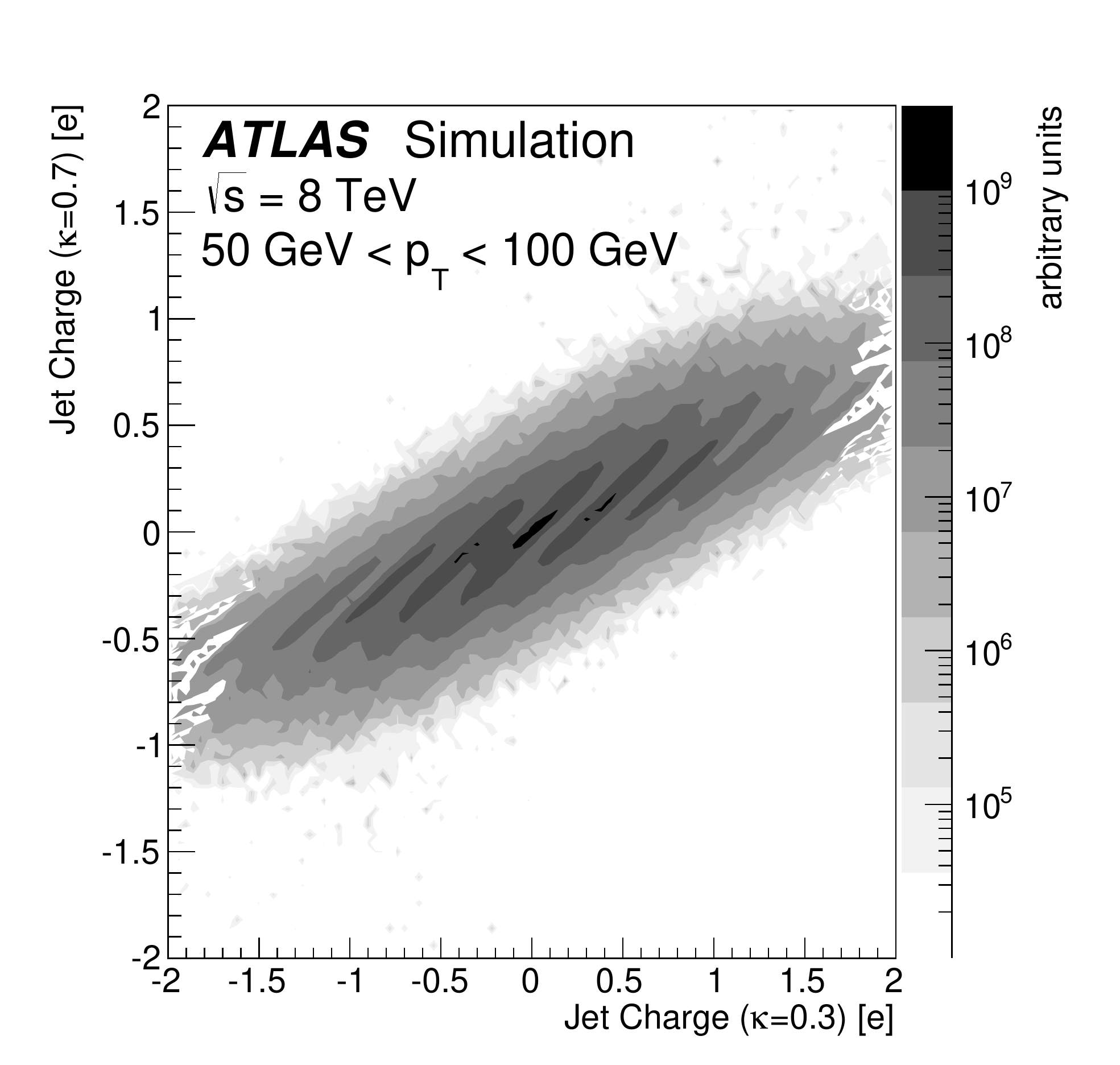}\\
\includegraphics[width=0.5\textwidth]{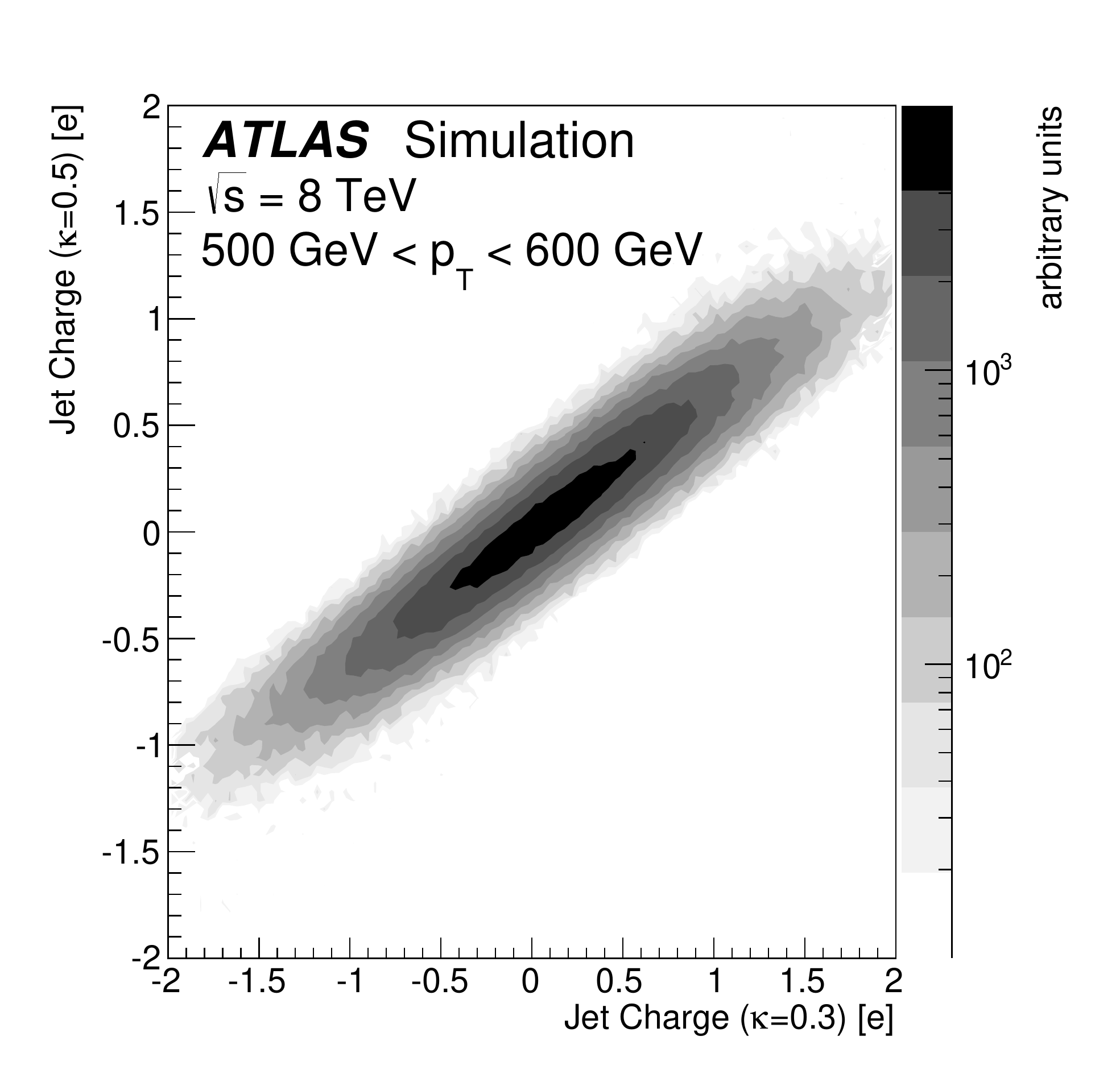}\includegraphics[width=0.5\textwidth]{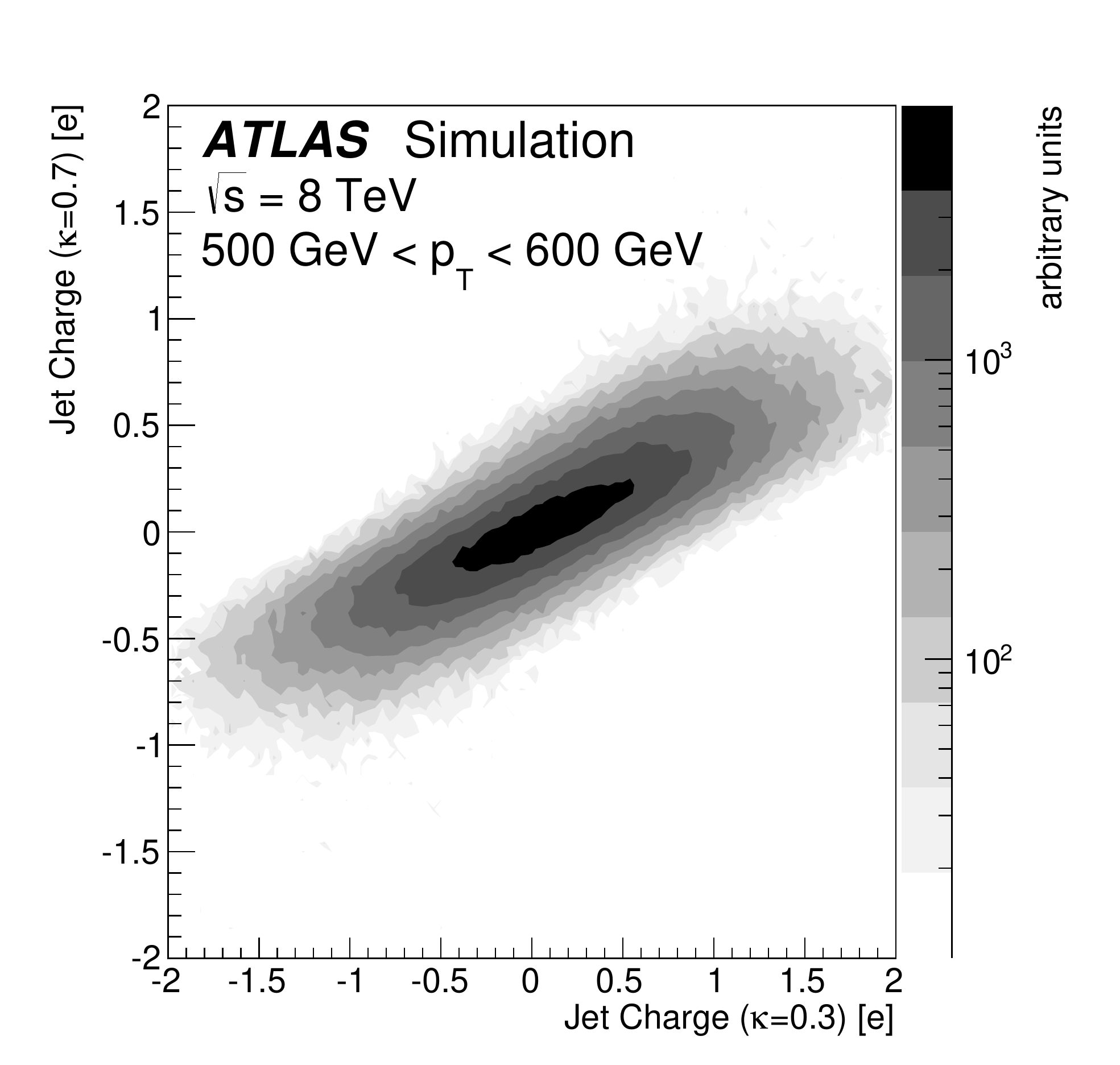}
\caption{The joint distribution of the jet charge defined with different values of the weighting factor $\kappa\in\{0.3,0.5,0.7\}$.  The left plots show the joint distributions of $\kappa=0.3$ and $\kappa=0.5$ while the right plots show the joint distributions of $\kappa=0.3$ and $\kappa=0.7$.  Jets in the top plots have 50 GeV $<p_\text{T}<100$ GeV while those in the bottom plots have 500 GeV $<p_\text{T}<600$ GeV.  See Sec.~\ref{sec:chargesamples} for details about the simulation.}
\label{fig:2Dto1D_recocorr}
\end{center}
\end{figure}

All reconstructed tracks are henceforth used when computing the jet charge, but the plots in Fig.~\ref{fig:jetcharge:qn} give an indication of the contribution of (relatively) high- and low-$p_\text{T}$ tracks.  Figure~\ref{fig:2Dto1D_recocorr} shows the joint distribution of jet charges with different $\kappa$ values.  While the distributions are peaked along the diagonal, there is a significant off-diagonal spread that is bigger when the two $\kappa$ values are further apart.  The stripes in the low $p_\text{T}$ bin are due to cases in which there is only one track; in those cases the jet charge for one value of $\kappa$ is uniquely specified by the jet charge at any other $\kappa\neq 0$ value.  The studies presented in this chapter use a range of $\kappa$ values in order to maintain a broad sensitivity to both hard and soft radiation inside jets.  The impact of low $p_\text{T}$ tracks on the jet charge reconstruction is revisited in Sec.~\ref{sec:JetCharge:CONF}.

\clearpage

\subsubsection{Lorentz Invariance}
\label{sec:jetcharge:LI}

The electric charge of a particle is a Lorentz invariant quantity, but the jet charge is not Lorentz invariant.  It is even possible (sometimes) to flip the sign of the jet charge by performing a suitable Lorentz transformation.  To illustrate the problem, consider a simplified case where $Z\rightarrow e^+ e^-$, as illustrated in the left plot of Fig.~\ref{fig:jetcharge:boost}.  Define the `jet charge' as $Q=\frac{1}{m_Z^\kappa}(p^\kappa-p^\kappa)=0$, where $p=m_Z/2$.  Now, suppose that the $Z$ has some transverse boost with speed $\beta$ along the $\pm x$ direction.  Then,

\begin{align}
Q=\frac{1}{(\gamma m_Z)^\kappa}\Big((\gamma p(1 \pm \beta ))^\kappa -(\gamma p(1 \mp \beta ))^\kappa \Big).
\end{align}

\begin{figure}[h!]
\begin{center}
\begin{tikzpicture}[line width=1.5 pt, scale=0.6]
\draw[->] (0,0) -- (3,0);
\draw[->] (0,0) -- (0,3);
\node at (3.3,0) {$x$};
\node at (0,3.3) {$y$};
\node at (-2,0.7) {$e^+$};
\node at (2,0.7) {$e^-$};
\node at (-2,-0.7) {$-p$};
\node at (2,-0.7) {$p$};
{\color{red}\draw[->] (0,0) -- (2,0);}
{\color{blue}\draw[->] (0,0) -- (-2,0);}
\begin{scope}[shift={(9,0)}]

\draw[->] (0,0) -- (3,0);
\draw[->] (0,0) -- (0,3);
\node at (3.3,0) {$x$};
\node at (0,3.3) {$y$};
\node at (2,2) {$Q>0$};
\node at (-3,0.7) {$e^+$};
\node at (1,0.7) {$e^-$};
{\color{red}\draw[->] (0,0) -- (1,0);}
{\color{blue}\draw[->] (0,0) -- (-3,0);}
\end{scope}

\begin{scope}[shift={(18,0)}]
\draw[->] (0,0) -- (3,0);
\draw[->] (0,0) -- (0,3);
\node at (3.3,0) {$x$};
\node at (0,3.3) {$y$};
\node at (2,2) {$Q<0$};
\node at (-1,0.7) {$e^+$};
\node at (3,0.7) {$e^-$};
{\color{red}\draw[->] (0,0) -- (3,0);}
{\color{blue}\draw[->] (0,0) -- (-1,0);}
\end{scope}
\end{tikzpicture}
\end{center}
\caption{A schematic diagram to illustrate the impact of Lorentz boosts on the `jet' charge.  A $Z$ boson decays at rest at the origin in the lab frame and decays into electrons that are along the $x$ direction (left).  For a boost along the $-x$ direction, the $e^+$ has a higher $p_\text{T}$ than the $e^-$ and vice versa for a boost along $+x$.}
\label{fig:jetcharge:boost}
\end{figure}

\noindent If $\kappa=1$, then $Q=\pm\beta$. If $\kappa\ll 1$, then $Q=\pm\kappa\beta$.  In either case, one can make $Q$ arbitrarily positive or negative depending on the direction of the boost.  Now, slightly more generally consider the decay of a color singlet, like a $W$ boson, which decays into $n$ particles and has speed $\beta$ in the $\hat{r}$ direction in a particular frame.  Then,

\begin{align}
Q &= \frac{1}{(\gamma m_{\text{boson}})^\kappa }\sum_{i=1}^n q_i \gamma^\kappa (E_i-\beta \vec{P}_i\cdot \hat{r})^\kappa.
\end{align}

\noindent When $\kappa=1$,

\begin{align}
\label{eq:jetcharge:generalboost}\nonumber
Q &= Q(\text{rest frame}) - \frac{\beta}{( m_{\text{boson}})}\sum_{i=1}^n q_i \vec{P}_i\cdot \hat{r}\\
&= Q_\text{boson} - \frac{\beta}{m}\sum_{i=1}^n q_i \hat{P}_i\cdot \hat{r}
\end{align}

\noindent For a given event, the second term in Eq.~\ref{eq:jetcharge:generalboost} will not be zero, unless all the tracks are perpendicular to the boost.  For an ensemble of events, the non-closure term (second term in Eq.~\ref{eq:jetcharge:generalboost}) will have average zero since the particles are randomly oriented and so the average jet charge is the boson charge.  However, the standard deviation of the non-closure term  is not zero and so there is an induced smearing of the boson charge due to the distribution of boosts.  This is illustrated with a simulated $W$ boson in Fig.~\ref{fig:jetcharge:LI:change}.  Because the $W$ boson is a color singlet, it is possible to uniquely associate final state hadrons with the decay of the $W$ boson (See Sec.~\ref{sec:colorflow:wcandidateselection}).  The jet charge using all of the $W^+$ decay products in the lab frame is positive, but the sign changes after a large boost in the $-x$ direction.  This change of sign can be understood by the dominance of one high $p_\text{T}$ negatively charged hadron, shown in left plot of Fig.~\ref{fig:jetcharge:LI:change}.    Figure~\ref{fig:jetcharge:LI:change2} shows how the jet charge sign depends on the value of $\kappa$ as well as on the direction and magnitude of the boost.  In the left plot of Fig.~\ref{fig:jetcharge:LI:change2}, $\beta_W = 0.65$.  For $\kappa=0$, $Q=1$ by construction as all $W$ boson decay products are part of the `jet'.  Even though the jet charge is positive for a large range of $\kappa$ values, as $\kappa\rightarrow\infty$ the jet charge is driven negative because the leading charged particle has a negative charge.  The right plot of Fig.~\ref{fig:jetcharge:LI:change2} has $\kappa=1$.  The horizontal axis begins at $p_\text{T}=10$ GeV because the jet charge is $\infty$ as $\beta_W\rightarrow 0$.  The blue region in the center, where the jet charge is negative, corresponds to a boost in the $-x$ direction, resembling a configuration as in the right plot of Fig.~\ref{fig:jetcharge:LI:change}.  

The jet charge is invariant under longitudinal boosts, which is critical at a hadron collider where there is a large range of $p_z$ values that contain much less information about the scale of jet formation compared with $p_\text{T}$.  Even though generic quark and gluon jets do not have a well-defined decay frame, the above example illustrates how the jet charge changes between frames.

\begin{figure}[h!]
\begin{center}
\begin{tikzpicture}[line width=1.5 pt, scale=0.8]
\draw[->] (0,0) -- (3,0);
\draw[->] (0,0) -- (0,3);
\node at (3.2,0) {$x$};
\node at (0,3.3) {$y$};
{\color{black!40!white}\draw[->,dotted,thin] (0,0) -- (0.233000,0.319000);}
{\color{red}\draw[->,thin] (0,0) -- (-0.243000,-0.189500);}
{\color{blue}\draw[->,thin] (0,0) -- (-2.640000,1.191500);}
{\color{red}\draw[->,thin] (0,0) -- (-2.569000,0.819500);}
{\color{blue}\draw[->,thin] (0,0) -- (-2.786500,1.205500);}
{\color{blue}\draw[->,thin] (0,0) -- (1.797000,0.421000);}
{\color{blue}\draw[->,thin] (0,0) -- (-0.146000,-0.134500);}
{\color{blue}\draw[->,thin] (0,0) -- (-0.148500,0.162500);}
{\color{red}\draw[->,thin] (0,0) -- (-0.774500,-0.014500);}
{\color{blue}\draw[->,thin] (0,0) -- (-0.337500,0.073000);}
{\color{red}\draw[->,thin] (0,0) -- (0.029000,-0.117500);}
{\color{black!40!white}\draw[->,dotted,thin] (0,0) -- (-1.114500,0.095500);}
{\color{red}\draw[->,thin] (0,0) -- (-0.138500,0.043500);}
{\color{blue}\draw[->,thin] (0,0) -- (1.458000,0.462000);}
{\color{red}\draw[->,thin] (0,0) -- (4.064000,1.265500);}
{\color{black!40!white}\draw[->,dotted,thin] (0,0) -- (1.003000,0.177500);}
{\color{black!40!white}\draw[->,dotted,thin] (0,0) -- (0.713500,0.175500);}
{\color{black!40!white}\draw[->,dotted,thin] (0,0) -- (0.053000,-0.050000);}
{\color{black!40!white}\draw[->,dotted,thin] (0,0) -- (0.074000,-0.154000);}
{\color{blue}\draw[->,thin] (0,0) -- (-0.188000,-0.380500);}
{\color{red}\draw[->,thin] (0,0) -- (-0.160000,-0.302000);}
{\color{black!40!white}\draw[->,dotted,thin] (0,0) -- (-0.778500,0.452000);}
{\color{black!40!white}\draw[->,dotted,thin] (0,0) -- (-0.023000,0.027000);}
{\color{black!40!white}\draw[->,dotted,thin] (0,0) -- (-0.228000,0.082500);}
{\color{black!40!white}\draw[->,dotted,thin] (0,0) -- (-0.834500,0.171000);}
{\color{black!40!white}\draw[->,dotted,thin] (0,0) -- (1.106500,0.634000);}
{\color{black!40!white}\draw[->,dotted,thin] (0,0) -- (0.114500,0.089500);}
{\color{black!40!white}\draw[->,dotted,thin] (0,0) -- (-0.081000,-0.014000);}
{\color{red}\draw[->,thin] (0,0) -- (-0.008000,-0.017500);}
{\color{blue}\draw[->,thin] (0,0) -- (-0.005000,-0.010000);}
{\color{blue}\draw[->,thin] (0,0) -- (-0.295000,0.124500);}
{\color{red}\draw[->,thin] (0,0) -- (-0.109000,-0.070000);}
{\color{black!40!white}\draw[->,dotted,thin] (0,0) -- (2.156500,0.678000);}
{\color{black!40!white}\draw[->,dotted,thin] (0,0) -- (0.258000,0.099000);}
{\color{black!40!white}\draw[->,dotted,thin] (0,0) -- (-0.044500,-0.072000);}
{\color{black!40!white}\draw[->,dotted,thin] (0,0) -- (-0.508500,-0.464500);}
\begin{scope}[shift={(5,0)}]
\draw[->] (0,0) -- (3,0);
\draw[->] (0,0) -- (0,3);
\node at (3.2,0) {$x$};
\node at (0,3.3) {$y$};
{\color{black!40!white}\draw[->,dotted,thin] (0,0) -- (0.625014,0.319000);}
{\color{red}\draw[->,thin] (0,0) -- (0.236491,-0.189500);}
{\color{blue}\draw[->,thin] (0,0) -- (1.054536,1.191500);}
{\color{red}\draw[->,thin] (0,0) -- (0.793605,0.819500);}
{\color{blue}\draw[->,thin] (0,0) -- (1.238972,1.205500);}
{\color{blue}\draw[->,thin] (0,0) -- (3.143528,0.421000);}
{\color{blue}\draw[->,thin] (0,0) -- (-0.022474,-0.134500);}
{\color{blue}\draw[->,thin] (0,0) -- (0.119025,0.162500);}
{\color{red}\draw[->,thin] (0,0) -- (-0.063842,-0.014500);}
{\color{blue}\draw[->,thin] (0,0) -- (0.207745,0.073000);}
{\color{red}\draw[->,thin] (0,0) -- (0.169449,-0.117500);}
{\color{black!40!white}\draw[->,dotted,thin] (0,0) -- (0.520099,0.095500);}
{\color{red}\draw[->,thin] (0,0) -- (-0.052352,0.043500);}
{\color{blue}\draw[->,thin] (0,0) -- (2.567505,0.462000);}
{\color{red}\draw[->,thin] (0,0) -- (7.153487,1.265500);}
{\color{black!40!white}\draw[->,dotted,thin] (0,0) -- (1.754284,0.177500);}
{\color{black!40!white}\draw[->,dotted,thin] (0,0) -- (1.251373,0.175500);}
{\color{black!40!white}\draw[->,dotted,thin] (0,0) -- (0.108121,-0.050000);}
{\color{black!40!white}\draw[->,dotted,thin] (0,0) -- (0.184092,-0.154000);}
{\color{blue}\draw[->,thin] (0,0) -- (0.067020,-0.380500);}
{\color{red}\draw[->,thin] (0,0) -- (0.088954,-0.302000);}
{\color{black!40!white}\draw[->,dotted,thin] (0,0) -- (0.327985,0.452000);}
{\color{black!40!white}\draw[->,dotted,thin] (0,0) -- (0.014932,0.027000);}
{\color{black!40!white}\draw[->,dotted,thin] (0,0) -- (0.198129,0.082500);}
{\color{black!40!white}\draw[->,dotted,thin] (0,0) -- (0.613009,0.171000);}
{\color{black!40!white}\draw[->,dotted,thin] (0,0) -- (2.016784,0.634000);}
{\color{black!40!white}\draw[->,dotted,thin] (0,0) -- (0.216137,0.089500);}
{\color{black!40!white}\draw[->,dotted,thin] (0,0) -- (0.134606,-0.014000);}
{\color{red}\draw[->,thin] (0,0) -- (0.128333,-0.017500);}
{\color{blue}\draw[->,thin] (0,0) -- (0.072724,-0.010000);}
{\color{blue}\draw[->,thin] (0,0) -- (0.106924,0.124500);}
{\color{red}\draw[->,thin] (0,0) -- (0.124024,-0.070000);}
{\color{black!40!white}\draw[->,dotted,thin] (0,0) -- (3.803824,0.678000);}
{\color{black!40!white}\draw[->,dotted,thin] (0,0) -- (0.457694,0.099000);}
{\color{black!40!white}\draw[->,dotted,thin] (0,0) -- (0.004234,-0.072000);}
{\color{black!40!white}\draw[->,dotted,thin] (0,0) -- (-0.137818,-0.464500);}
\end{scope}
\end{tikzpicture}
\end{center}
\caption{The decay of a simulated $W$ boson event with {\sc Pythia} 8.  The length of the arrow is proportional to the energy of the decay product; red arrow denote negatively charged hadrons, blue arrows mark positively charged hadrons and neutral hadrons and photons are in gray.  In the right plot, the $W$ has received a large boost in the $+x$ direction.}
\label{fig:jetcharge:LI:change}
\end{figure}

\begin{figure}[h!]
\begin{center}
\begin{tikzpicture}[line width=1.5 pt, scale=1.0]
\draw[->] (0,0) -- (3,0);
\draw[->] (0,0) -- (0,3);
\node at (3.2,0) {$\kappa$};
\node at (-0.6,3.3) {$\sum_{i=1}^n q_i (p_\text{T}^i)^\kappa$};
\draw [red, ultra thick] (-0.500000,0.222766) circle [radius=0.05];
\node at (-0.5,-0.2) {\tiny $-0.5$};
\draw [red, ultra thick] (-0.490000,0.196373) circle [radius=0.05];
\draw [red, ultra thick] (-0.480000,0.173364) circle [radius=0.05];
\draw [red, ultra thick] (-0.470000,0.153574) circle [radius=0.05];
\draw [red, ultra thick] (-0.460000,0.136843) circle [radius=0.05];
\draw [red, ultra thick] (-0.450000,0.123022) circle [radius=0.05];
\draw [red, ultra thick] (-0.440000,0.111964) circle [radius=0.05];
\draw [red, ultra thick] (-0.430000,0.103531) circle [radius=0.05];
\draw [red, ultra thick] (-0.420000,0.097592) circle [radius=0.05];
\draw [red, ultra thick] (-0.410000,0.094019) circle [radius=0.05];
\draw [red, ultra thick] (-0.400000,0.092694) circle [radius=0.05];
\draw [red, ultra thick] (-0.390000,0.093500) circle [radius=0.05];
\draw [red, ultra thick] (-0.380000,0.096329) circle [radius=0.05];
\draw [red, ultra thick] (-0.370000,0.101075) circle [radius=0.05];
\draw [red, ultra thick] (-0.360000,0.107639) circle [radius=0.05];
\draw [red, ultra thick] (-0.350000,0.115926) circle [radius=0.05];
\draw [red, ultra thick] (-0.340000,0.125844) circle [radius=0.05];
\draw [red, ultra thick] (-0.330000,0.137307) circle [radius=0.05];
\draw [red, ultra thick] (-0.320000,0.150232) circle [radius=0.05];
\draw [red, ultra thick] (-0.310000,0.164542) circle [radius=0.05];
\draw [red, ultra thick] (-0.300000,0.180160) circle [radius=0.05];
\draw [red, ultra thick] (-0.290000,0.197015) circle [radius=0.05];
\draw [red, ultra thick] (-0.280000,0.215039) circle [radius=0.05];
\draw [red, ultra thick] (-0.270000,0.234168) circle [radius=0.05];
\draw [red, ultra thick] (-0.260000,0.254340) circle [radius=0.05];
\draw [red, ultra thick] (-0.250000,0.275495) circle [radius=0.05];
\draw [red, ultra thick] (-0.240000,0.297579) circle [radius=0.05];
\draw [red, ultra thick] (-0.230000,0.320537) circle [radius=0.05];
\draw [red, ultra thick] (-0.220000,0.344320) circle [radius=0.05];
\draw [red, ultra thick] (-0.210000,0.368879) circle [radius=0.05];
\draw [red, ultra thick] (-0.200000,0.394168) circle [radius=0.05];
\draw [red, ultra thick] (-0.190000,0.420144) circle [radius=0.05];
\draw [red, ultra thick] (-0.180000,0.446766) circle [radius=0.05];
\draw [red, ultra thick] (-0.170000,0.473994) circle [radius=0.05];
\draw [red, ultra thick] (-0.160000,0.501792) circle [radius=0.05];
\draw [red, ultra thick] (-0.150000,0.530123) circle [radius=0.05];
\draw [red, ultra thick] (-0.140000,0.558955) circle [radius=0.05];
\draw [red, ultra thick] (-0.130000,0.588255) circle [radius=0.05];
\draw [red, ultra thick] (-0.120000,0.617994) circle [radius=0.05];
\draw [red, ultra thick] (-0.110000,0.648142) circle [radius=0.05];
\draw [red, ultra thick] (-0.100000,0.678672) circle [radius=0.05];
\draw [red, ultra thick] (-0.090000,0.709560) circle [radius=0.05];
\draw [red, ultra thick] (-0.080000,0.740780) circle [radius=0.05];
\draw [red, ultra thick] (-0.070000,0.772310) circle [radius=0.05];
\draw [red, ultra thick] (-0.060000,0.804128) circle [radius=0.05];
\draw [red, ultra thick] (-0.050000,0.836213) circle [radius=0.05];
\draw [red, ultra thick] (-0.040000,0.868545) circle [radius=0.05];
\draw [red, ultra thick] (-0.030000,0.901107) circle [radius=0.05];
\draw [red, ultra thick] (-0.020000,0.933881) circle [radius=0.05];
\draw [red, ultra thick] (-0.010000,0.966851) circle [radius=0.05];
\draw [red, ultra thick] (0.000000,1.000000) circle [radius=0.05];
\node at (0.0,-0.2) {\tiny $0.0$};
\draw [red, ultra thick] (0.010000,1.033315) circle [radius=0.05];
\draw [red, ultra thick] (0.020000,1.066781) circle [radius=0.05];
\draw [red, ultra thick] (0.030000,1.100386) circle [radius=0.05];
\draw [red, ultra thick] (0.040000,1.134117) circle [radius=0.05];
\draw [red, ultra thick] (0.050000,1.167963) circle [radius=0.05];
\draw [red, ultra thick] (0.060000,1.201912) circle [radius=0.05];
\draw [red, ultra thick] (0.070000,1.235956) circle [radius=0.05];
\draw [red, ultra thick] (0.080000,1.270083) circle [radius=0.05];
\draw [red, ultra thick] (0.090000,1.304285) circle [radius=0.05];
\draw [red, ultra thick] (0.100000,1.338553) circle [radius=0.05];
\draw [red, ultra thick] (0.110000,1.372879) circle [radius=0.05];
\draw [red, ultra thick] (0.120000,1.407256) circle [radius=0.05];
\draw [red, ultra thick] (0.130000,1.441676) circle [radius=0.05];
\draw [red, ultra thick] (0.140000,1.476133) circle [radius=0.05];
\draw [red, ultra thick] (0.150000,1.510621) circle [radius=0.05];
\draw [red, ultra thick] (0.160000,1.545133) circle [radius=0.05];
\draw [red, ultra thick] (0.170000,1.579664) circle [radius=0.05];
\draw [red, ultra thick] (0.180000,1.614209) circle [radius=0.05];
\draw [red, ultra thick] (0.190000,1.648762) circle [radius=0.05];
\draw [red, ultra thick] (0.200000,1.683318) circle [radius=0.05];
\draw [red, ultra thick] (0.210000,1.717875) circle [radius=0.05];
\draw [red, ultra thick] (0.220000,1.752426) circle [radius=0.05];
\draw [red, ultra thick] (0.230000,1.786968) circle [radius=0.05];
\draw [red, ultra thick] (0.240000,1.821498) circle [radius=0.05];
\draw [red, ultra thick] (0.250000,1.856011) circle [radius=0.05];
\draw [red, ultra thick] (0.260000,1.890504) circle [radius=0.05];
\draw [red, ultra thick] (0.270000,1.924974) circle [radius=0.05];
\draw [red, ultra thick] (0.280000,1.959418) circle [radius=0.05];
\draw [red, ultra thick] (0.290000,1.993832) circle [radius=0.05];
\draw [red, ultra thick] (0.300000,2.028213) circle [radius=0.05];
\draw [red, ultra thick] (0.310000,2.062558) circle [radius=0.05];
\draw [red, ultra thick] (0.320000,2.096865) circle [radius=0.05];
\draw [red, ultra thick] (0.330000,2.131131) circle [radius=0.05];
\draw [red, ultra thick] (0.340000,2.165352) circle [radius=0.05];
\draw [red, ultra thick] (0.350000,2.199526) circle [radius=0.05];
\draw [red, ultra thick] (0.360000,2.233651) circle [radius=0.05];
\draw [red, ultra thick] (0.370000,2.267722) circle [radius=0.05];
\draw [red, ultra thick] (0.380000,2.301739) circle [radius=0.05];
\draw [red, ultra thick] (0.390000,2.335697) circle [radius=0.05];
\draw [red, ultra thick] (0.400000,2.369593) circle [radius=0.05];
\draw [red, ultra thick] (0.410000,2.403426) circle [radius=0.05];
\draw [red, ultra thick] (0.420000,2.437191) circle [radius=0.05];
\draw [red, ultra thick] (0.430000,2.470886) circle [radius=0.05];
\draw [red, ultra thick] (0.440000,2.504508) circle [radius=0.05];
\draw [red, ultra thick] (0.450000,2.538053) circle [radius=0.05];
\draw [red, ultra thick] (0.460000,2.571517) circle [radius=0.05];
\draw [red, ultra thick] (0.470000,2.604898) circle [radius=0.05];
\draw [red, ultra thick] (0.480000,2.638191) circle [radius=0.05];
\draw [red, ultra thick] (0.490000,2.671393) circle [radius=0.05];
\draw [red, ultra thick] (0.500000,2.704499) circle [radius=0.05];
\node at (0.5,-0.2) {\tiny $0.5$};
\draw [red, ultra thick] (0.510000,2.737507) circle [radius=0.05];
\draw [red, ultra thick] (0.520000,2.770410) circle [radius=0.05];
\draw [red, ultra thick] (0.530000,2.803205) circle [radius=0.05];
\draw [red, ultra thick] (0.540000,2.835887) circle [radius=0.05];
\draw [red, ultra thick] (0.550000,2.868452) circle [radius=0.05];
\draw [red, ultra thick] (0.560000,2.900893) circle [radius=0.05];
\draw [red, ultra thick] (0.570000,2.933205) circle [radius=0.05];
\draw [red, ultra thick] (0.580000,2.965384) circle [radius=0.05];
\draw [red, ultra thick] (0.590000,2.997423) circle [radius=0.05];
\draw [red, ultra thick] (0.600000,3.029315) circle [radius=0.05];
\draw [red, ultra thick] (0.610000,3.061055) circle [radius=0.05];
\draw [red, ultra thick] (0.620000,3.092636) circle [radius=0.05];
\draw [red, ultra thick] (0.630000,3.124050) circle [radius=0.05];
\draw [red, ultra thick] (0.640000,3.155291) circle [radius=0.05];
\draw [red, ultra thick] (0.650000,3.186351) circle [radius=0.05];
\draw [red, ultra thick] (0.660000,3.217221) circle [radius=0.05];
\draw [red, ultra thick] (0.670000,3.247894) circle [radius=0.05];
\draw [red, ultra thick] (0.680000,3.278361) circle [radius=0.05];
\draw [red, ultra thick] (0.690000,3.308613) circle [radius=0.05];
\draw [red, ultra thick] (0.700000,3.338641) circle [radius=0.05];
\draw [red, ultra thick] (0.710000,3.368434) circle [radius=0.05];
\draw [red, ultra thick] (0.720000,3.397983) circle [radius=0.05];
\draw [red, ultra thick] (0.730000,3.427276) circle [radius=0.05];
\draw [red, ultra thick] (0.740000,3.456304) circle [radius=0.05];
\draw [red, ultra thick] (0.750000,3.485054) circle [radius=0.05];
\draw [red, ultra thick] (0.760000,3.513515) circle [radius=0.05];
\draw [red, ultra thick] (0.770000,3.541675) circle [radius=0.05];
\draw [red, ultra thick] (0.780000,3.569519) circle [radius=0.05];
\draw [red, ultra thick] (0.790000,3.597036) circle [radius=0.05];
\draw [red, ultra thick] (0.800000,3.624211) circle [radius=0.05];
\draw [red, ultra thick] (0.810000,3.651030) circle [radius=0.05];
\draw [red, ultra thick] (0.820000,3.677479) circle [radius=0.05];
\draw [red, ultra thick] (0.830000,3.703541) circle [radius=0.05];
\draw [red, ultra thick] (0.840000,3.729200) circle [radius=0.05];
\draw [red, ultra thick] (0.850000,3.754441) circle [radius=0.05];
\draw [red, ultra thick] (0.860000,3.779246) circle [radius=0.05];
\draw [red, ultra thick] (0.870000,3.803598) circle [radius=0.05];
\draw [red, ultra thick] (0.880000,3.827477) circle [radius=0.05];
\draw [red, ultra thick] (0.890000,3.850865) circle [radius=0.05];
\draw [red, ultra thick] (0.900000,3.873742) circle [radius=0.05];
\draw [red, ultra thick] (0.910000,3.896088) circle [radius=0.05];
\draw [red, ultra thick] (0.920000,3.917882) circle [radius=0.05];
\draw [red, ultra thick] (0.930000,3.939102) circle [radius=0.05];
\draw [red, ultra thick] (0.940000,3.959725) circle [radius=0.05];
\draw [red, ultra thick] (0.950000,3.979729) circle [radius=0.05];
\draw [red, ultra thick] (0.960000,3.999089) circle [radius=0.05];
\draw [red, ultra thick] (0.970000,4.017781) circle [radius=0.05];
\draw [red, ultra thick] (0.980000,4.035779) circle [radius=0.05];
\draw [red, ultra thick] (0.990000,4.053056) circle [radius=0.05];
\draw [red, ultra thick] (1.000000,4.069586) circle [radius=0.05];
\node at (1.0,-0.2) {\tiny $1.0$};
\draw [red, ultra thick] (1.010000,4.085339) circle [radius=0.05];
\draw [red, ultra thick] (1.020000,4.100288) circle [radius=0.05];
\draw [red, ultra thick] (1.030000,4.114402) circle [radius=0.05];
\draw [red, ultra thick] (1.040000,4.127650) circle [radius=0.05];
\draw [red, ultra thick] (1.050000,4.140001) circle [radius=0.05];
\draw [red, ultra thick] (1.060000,4.151421) circle [radius=0.05];
\draw [red, ultra thick] (1.070000,4.161877) circle [radius=0.05];
\draw [red, ultra thick] (1.080000,4.171333) circle [radius=0.05];
\draw [red, ultra thick] (1.090000,4.179755) circle [radius=0.05];
\draw [red, ultra thick] (1.100000,4.187104) circle [radius=0.05];
\draw [red, ultra thick] (1.110000,4.193343) circle [radius=0.05];
\draw [red, ultra thick] (1.120000,4.198432) circle [radius=0.05];
\draw [red, ultra thick] (1.130000,4.202331) circle [radius=0.05];
\draw [red, ultra thick] (1.140000,4.204998) circle [radius=0.05];
\draw [red, ultra thick] (1.150000,4.206390) circle [radius=0.05];
\draw [red, ultra thick] (1.160000,4.206463) circle [radius=0.05];
\draw [red, ultra thick] (1.170000,4.205171) circle [radius=0.05];
\draw [red, ultra thick] (1.180000,4.202468) circle [radius=0.05];
\draw [red, ultra thick] (1.190000,4.198305) circle [radius=0.05];
\draw [red, ultra thick] (1.200000,4.192633) circle [radius=0.05];
\draw [red, ultra thick] (1.210000,4.185401) circle [radius=0.05];
\draw [red, ultra thick] (1.220000,4.176556) circle [radius=0.05];
\draw [red, ultra thick] (1.230000,4.166044) circle [radius=0.05];
\draw [red, ultra thick] (1.240000,4.153810) circle [radius=0.05];
\draw [red, ultra thick] (1.250000,4.139796) circle [radius=0.05];
\draw [red, ultra thick] (1.260000,4.123944) circle [radius=0.05];
\draw [red, ultra thick] (1.270000,4.106193) circle [radius=0.05];
\draw [red, ultra thick] (1.280000,4.086482) circle [radius=0.05];
\draw [red, ultra thick] (1.290000,4.064746) circle [radius=0.05];
\draw [red, ultra thick] (1.300000,4.040920) circle [radius=0.05];
\draw [red, ultra thick] (1.310000,4.014936) circle [radius=0.05];
\draw [red, ultra thick] (1.320000,3.986724) circle [radius=0.05];
\draw [red, ultra thick] (1.330000,3.956215) circle [radius=0.05];
\draw [red, ultra thick] (1.340000,3.923333) circle [radius=0.05];
\draw [red, ultra thick] (1.350000,3.888005) circle [radius=0.05];
\draw [red, ultra thick] (1.360000,3.850153) circle [radius=0.05];
\draw [red, ultra thick] (1.370000,3.809696) circle [radius=0.05];
\draw [red, ultra thick] (1.380000,3.766555) circle [radius=0.05];
\draw [red, ultra thick] (1.390000,3.720644) circle [radius=0.05];
\draw [red, ultra thick] (1.400000,3.671877) circle [radius=0.05];
\draw [red, ultra thick] (1.410000,3.620167) circle [radius=0.05];
\draw [red, ultra thick] (1.420000,3.565422) circle [radius=0.05];
\draw [red, ultra thick] (1.430000,3.507549) circle [radius=0.05];
\draw [red, ultra thick] (1.440000,3.446453) circle [radius=0.05];
\draw [red, ultra thick] (1.450000,3.382034) circle [radius=0.05];
\draw [red, ultra thick] (1.460000,3.314191) circle [radius=0.05];
\draw [red, ultra thick] (1.470000,3.242822) circle [radius=0.05];
\draw [red, ultra thick] (1.480000,3.167819) circle [radius=0.05];
\draw [red, ultra thick] (1.490000,3.089074) circle [radius=0.05];
\draw [red, ultra thick] (1.500000,3.006473) circle [radius=0.05];
\node at (1.5,-0.2) {\tiny $1.5$};
\draw [red, ultra thick] (1.510000,2.919902) circle [radius=0.05];
\draw [red, ultra thick] (1.520000,2.829242) circle [radius=0.05];
\draw [red, ultra thick] (1.530000,2.734373) circle [radius=0.05];
\draw [red, ultra thick] (1.540000,2.635169) circle [radius=0.05];
\draw [red, ultra thick] (1.550000,2.531502) circle [radius=0.05];
\draw [red, ultra thick] (1.560000,2.423242) circle [radius=0.05];
\draw [red, ultra thick] (1.570000,2.310254) circle [radius=0.05];
\draw [red, ultra thick] (1.580000,2.192399) circle [radius=0.05];
\draw [red, ultra thick] (1.590000,2.069537) circle [radius=0.05];
\draw [red, ultra thick] (1.600000,1.941521) circle [radius=0.05];
\draw [red, ultra thick] (1.610000,1.808202) circle [radius=0.05];
\draw [red, ultra thick] (1.620000,1.669427) circle [radius=0.05];
\draw [red, ultra thick] (1.630000,1.525039) circle [radius=0.05];
\draw [red, ultra thick] (1.640000,1.374878) circle [radius=0.05];
\draw [red, ultra thick] (1.650000,1.218777) circle [radius=0.05];
\draw [red, ultra thick] (1.660000,1.056567) circle [radius=0.05];
\draw [red, ultra thick] (1.670000,0.888074) circle [radius=0.05];
\draw [red, ultra thick] (1.680000,0.713121) circle [radius=0.05];
\draw [red, ultra thick] (1.690000,0.531523) circle [radius=0.05];
\draw [red, ultra thick] (1.700000,0.343093) circle [radius=0.05];
\draw [red, ultra thick] (1.710000,0.147640) circle [radius=0.05];
\draw [red, ultra thick] (1.720000,-0.055035) circle [radius=0.05];
\draw [red, ultra thick] (1.730000,-0.265133) circle [radius=0.05];
\draw [red, ultra thick] (1.740000,-0.482863) circle [radius=0.05];
\draw [red, ultra thick] (1.750000,-0.708437) circle [radius=0.05];
\draw [red, ultra thick] (1.760000,-0.942073) circle [radius=0.05];
\draw [red, ultra thick] (1.770000,-1.183995) circle [radius=0.05];
\draw [red, ultra thick] (1.780000,-1.434433) circle [radius=0.05];
\draw [red, ultra thick] (1.790000,-1.693621) circle [radius=0.05];
\node at (-0.3,0.0) {\tiny $0.0$};
\node at (-0.3,0.5) {\tiny $0.5$};
\node at (-0.3,1.0) {\tiny $1.0$};
\node at (-0.3,1.5) {\tiny $1.5$};
\node at (-0.3,2.0) {\tiny $2.0$};

\begin{scope}[shift={(4,0)}]
\node[inner sep=0pt] (russell) at (3,2)
    {\includegraphics[width=0.4\textwidth]{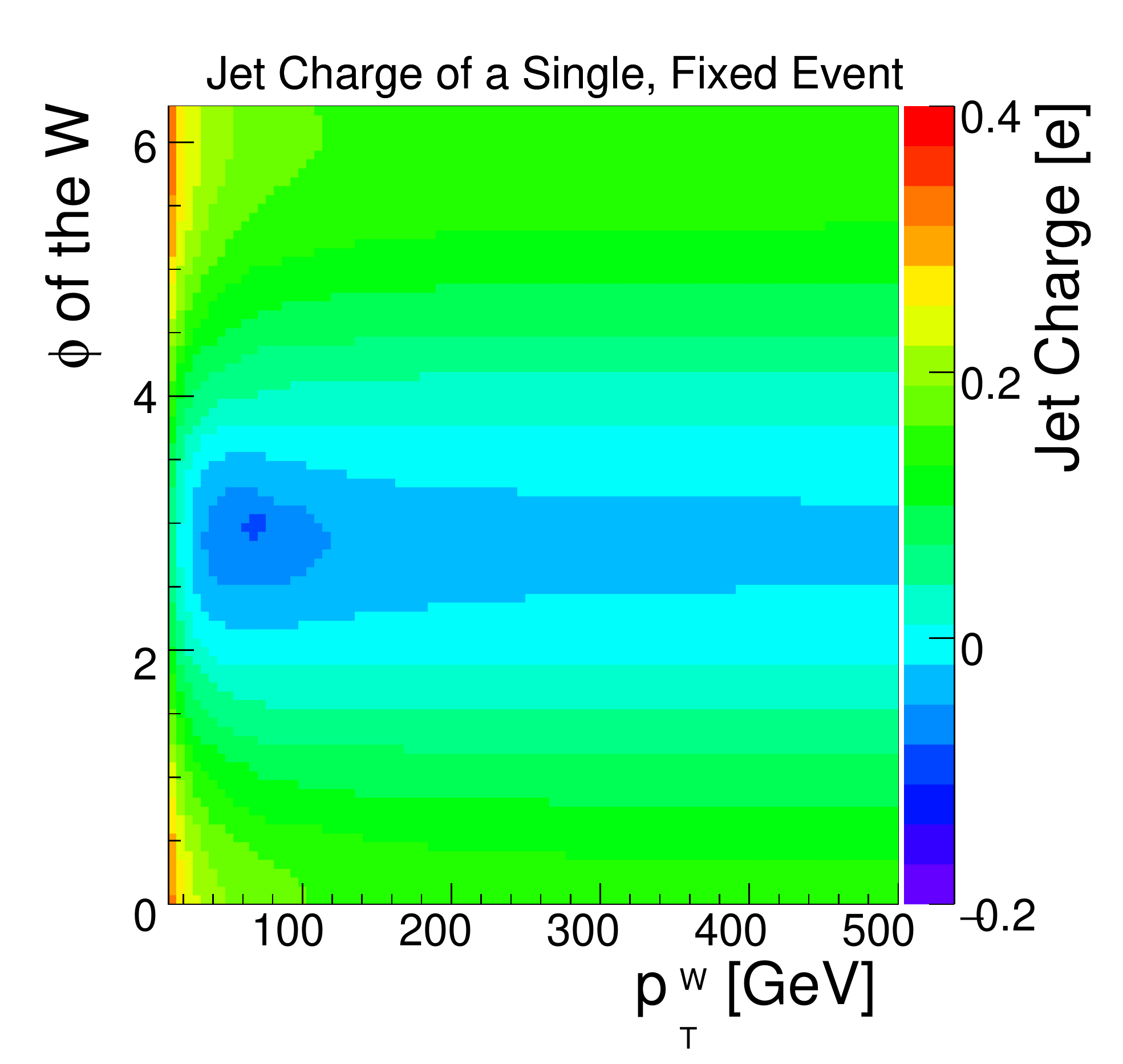}};
\end{scope}

\end{tikzpicture}
\end{center}
\caption{Left: for fixed $\beta_W=0.65$, the dependence of the `jet' ($=$ all $W$ boson decay products) charge on $\kappa$ for the event shown in Fig.~\ref{fig:jetcharge:LI:change}.  Right: for $\kappa=1$, the dependence of the jet charge on the boost direction and magnitude for the same event as in the left plot. }
\label{fig:jetcharge:LI:change2}
\end{figure}

\clearpage

\subsection{From Parton Charge to Jet Charge}
\label{sec:jetchargetheory}

In general, the jet charge is not an infrared and collinear safe observable.  Infrared safety is guaranteed for $\kappa > 0$, since the contribution of an arbitrarily soft particle is suppressed by $p_\text{T,soft}^\kappa$.  However, the charge-weighting in Eq.~\ref{chargedefcharge} spoils collinear safety.  To see this, suppose that a positively charged particle $P$ carries momentum fraction $z_P=p_\text{T,P}/p_\text{T,jet}$ and splits into a charged particle $P_+$ and a neutral particle $P_0$ whose momenta are collinear.  The contribution to the jet charge before the splitting is $z_P^\kappa$, while after the splitting it is $z_{P_+}^\kappa$.  For $z_{P+}<z_p$, these two contributions are not identical.   As a result of collinear sensitivity, hadronization must be included in any reliable description of the jet charge.  This information cannot be described perturbatively within QCD, but the non-perturbative components can be quantified and isolated~\cite{Krohn:2012fg,Waalewijn2012}.  For a parton of type $p$ with energy $E$, the probability for a hadron of type $h$ to carry a fraction $[z,z+dz]$ of the parton's momentum is given by the {\it fragmentation function} $D_p^h(z,E)dz$.   The normalization of $D_p^h(z,E)$ is the average number of hadrons $h$ produced by a jet initiated by a parton of energy $E$, $\langle n_p^h(E)\rangle$.  This can be shown by dividing the interval $[0,1]$ into $N$ pieces so that the probability for multiple hadrons of the same type (e.g. $\pi^+$) to have $z\in[i,i+1]/N$ is small: 

\begin{align}\nonumber
\langle n_p^h(E)\rangle&=\sum_{i=0}^{N-1} \sum_{k=0}^\infty \Pr(\text{$k$ hadrons of type $h$ with $z\in[i,i+1]/N$})\\\nonumber
&=\sum_{i=0}^{N-1} \Pr(\text{one hadron of type $h$ with $z\in[i,i+1]/N$})+\mathcal{O}(1/N^2)\\\nonumber
&=\sum_{i=0}^{N-1} \frac{1}{N}D_i^h(i/N,E)+\mathcal{O}(1/N^2)\\
&\stackrel{N\rightarrow\infty}{=}\int_0^1 dzD_i^h(z,E).
\end{align}

\noindent The average multiplicity will be revisited in Chapter~\ref{cha:multiplicity}.  Ignoring non-strong force processes, conservation of energy requires that the first moment of the fragmentation function summed over all hadron species is equal to one: $\sum_h \int_0^1 dz zD_p^h(z,E)=1$.  The average jet charge follows a related form:

\begin{align}
\label{eq:jetchargenotLO}
\langle Q_p(E,\kappa)\rangle = \sum_h Q_h\int_0^1 dz z^\kappa D_p^h(z,E)\equiv \sum_h Q_h\tilde{D}_p^h(\kappa,E),
\end{align} 

\noindent where $Q_h$ is the charge of hadron $h$ and $\tilde{D}(\kappa,E)$ is the {\it Mellin transform} of $D$ at $\kappa+1$.  One can include perturbative contributions to Eq.~\ref{eq:jetchargenotLO} within the context of Soft Collinear Effective Theory (SCET)~\cite{Bauer:2001yt,Bauer:2001ct,Bauer:2000yr,Bauer:2000ew} with the Fragmenting Jet Function $\mathcal{G}_p^h$~\cite{Procura:2011aq,Jain:2011xz,Procura:2009vm}.  The average jet charge is given by~\cite{Krohn:2012fg,Waalewijn2012}:

\begin{align}
\label{eq:jetchargeLO}
\langle Q_i(E, R,\kappa,\mu)\rangle = \sum_h Q_h\int_0^1 dz z^\kappa \frac{\mathcal{G}_p^h(E,R,z,\mu)}{2(2\pi)^3J_p(E,R,\mu)},
\end{align} 

\noindent where

\begin{align}
\label{eq:FJF}
\mathcal{G}_p^h(E,R,z,\mu)=\sum_{p'}\int_z^1\frac{dz'}{z'}\mathcal{J}_{pp'}(E,R,z',\mu)D_{p'}^h\left(\frac{z}{z'},\mu\right).
\end{align}

\noindent The factors $\mathcal{J}_{pp'}$ are defined as $\mathcal{J}_{pp'}=2(2\pi)^3\delta(1-z)\delta_{pp'}+\mathcal{O}(\alpha_s)$.  Therefore, $\mathcal{G}_p^h=2(2\pi)^3D_p^h+\mathcal{O}(\alpha)$.  Similarly, the {\it jet function} $J_p(E,R,\mu)=1+\mathcal{O}(\alpha_s)$~\cite{Ellis:2010rwa} and so Eq.~\ref{eq:jetchargenotLO} and Eq.~\ref{eq:jetchargeLO} are the same up to $\mathcal{O}(\alpha_s)$ corrections.  These corrections are less than 10\%, but are also not known precisely due to large uncertainties in the fragmentation functions~\cite{Waalewijn2012}.  The fragmenting jet function is the extension of the inclusive fragmentation function in the context of a jet with finite size.  The intuition for Eq.~\ref{eq:FJF} is that the parton $p$ radiates the parton $p'$ which then in turn fragments into hadron $h$.  The parton $p'$ has energy fraction $z' > z$ and the hadron has energy fraction $z/z'$ of this energy which is a fraction $z=z'\times z/z'$ of the initial parton's energy.  The factor $dz'/z'$ is the phase space for parton $p$ to emit $p'$ and the perturbatively calculable functions $\mathcal{J}_{pp'}$ are related to the QCD splitting functions\footnote{See Sec.~\ref{sec:mass:theory} for a discussion of the phase space and the QCD splitting functions.}.  Higher moments of the jet charge distribution can be computed in a similar fashion, but in general depend on additional non-perturbative information encoded in the multi-hadron fragmentation functions~\cite{Krohn:2012fg,Waalewijn2012}.

The jet charge distribution depends on the jet energy due to two related effects.  First, since the jet charge depends on the initiating parton type $p$, the jet charge distribution varies as the parton distribution functions, $f_p(z,\mu)$, change with energy.   Figure~\ref{fig:QCDfeynman} shows a representative set of leading order QCD Feynman diagrams for $2\rightarrow 2$ scattering with an up-quark in the initial state.  In all cases except for the annihilation diagram, the up quark is also an out-going parton.  As discussed earlier, the average gluon jet charge is zero.  The average up quark jet charge is positive, since the probability for an up quark to fragment into a positively charged hadron is larger than the probability for an up quark to fragment into a negatively charged hadron.  If there were only up quarks and gluons, then the average inclusive jet charge would be proportional to the fraction of up quark jets.  As discussed in Sec.~\ref{sec:particlesandforces}, the fraction of up quarks increases with momentum fraction.  The momentum fractions of the two initial partons $x_1, x_2$, and the proton and parton center-of-mass energies $\sqrt{s}$ and $\sqrt{\hat{s}}$ are related by $\sqrt{\hat{s}}=\sqrt{x_1x_2s}$.   In central dijet events, the jet $p_\text{T}\sim\sqrt{\hat{s}}/2$.  Therefore, the fraction of up quark jets increases with jet $p_\text{T}$.  The fraction of down quark jets also increases, but the fraction of up quark jets is expected to be larger.  These considerations predict that the average jet charge should increase with jet $p_\text{T}$\footnote{Taking into account the fact that there are twice as many valence up quarks versus down quarks, there is residual enhancement of the up quark PDF relative to the down quark one at high $x$ because the mass of the spectator valence quarks is larger.}.

\begin{figure}[h!]
\centering
\begin{tikzpicture}[line width=1.5 pt, scale=1.5]
	\draw[fermionbar] (-140:1)--(0,0);
	\draw[fermion] (140:1)--(0,0);
	\draw[gluon] (0:1)--(0,0);
	\node at (-140:1.2) {$\bar{u}$};
	\node at (140:1.2) {$u$};
	\node at (.5,.3) {$g$};	
\begin{scope}[shift={(1,0)}]
	\draw[fermion] (-40:1)--(0,0);
	\draw[fermionbar] (40:1)--(0,0);
	\node at (-40:1.2) {$d$};
	\node at (40:1.2) {$\bar{d}$};	
\end{scope}
\begin{scope}[shift={(4,0)}]
	\draw[gluon] (-140:1)--(0,0);
	\draw[fermion] (140:1)--(0,0);
	\draw[fermionbar] (0:1)--(0,0);
	\node at (-140:1.2) {$g$};
	\node at (140:1.2) {$u$};
	\node at (.5,.3) {$u$};	
\begin{scope}[shift={(1,0)}]
	\draw[fermionbar] (-40:1)--(0,0);
	\draw[gluon] (40:1)--(0,0);
	\node at (-40:1.2) {$u$};
	\node at (40:1.2) {$g$};	
\end{scope}
\end{scope}
\end{tikzpicture}

\vspace{3mm}

\begin{tikzpicture}[line width=1.5 pt, scale=1.5]
	\draw[fermion] (-140:1)--(0.5,-0.65);
	\draw[gluon] (140:1)--(0.5,0.65);
	\draw[fermionbar] (0.5,0.65)--(0.5,-0.65);
	\node at (-140:1.2) {$u$};
	\node at (140:1.2) {$g$};
	\node at (0.2,.0) {$u$};	
\begin{scope}[shift={(1,0)}]
	\draw[gluon] (-40:1)--(-0.5,-0.65);
	\draw[fermionbar] (40:1)--(-0.5,0.65);
	\node at (-40:1.2) {$g$};
	\node at (40:1.2) {$u$};	
\end{scope}
\begin{scope}[shift={(4,0)}]
	\draw[fermion] (-140:1)--(0.5,-0.65);
	\draw[fermion] (140:1)--(0.5,0.65);
	\draw[gluon] (0.5,0.65)--(0.5,-0.65);
	\node at (-140:1.2) {$u$};
	\node at (140:1.2) {$d$};
	\node at (0.2,.0) {$g$};	
\begin{scope}[shift={(1,0)}]
	\draw[fermionbar] (-40:1)--(-0.5,-0.65);
	\draw[fermionbar] (40:1)--(-0.5,0.65);
	\node at (-40:1.2) {$u$};
	\node at (40:1.2) {$d$};	
\end{scope}
\end{scope}
\end{tikzpicture}
\caption{A representative set of leading order QCD Feynman diagrams with an up quark in the initial state.  Every diagram that has a down quark could be replaced with any other parton.  There are also the additional diagrams related to the $t$-channel ones by crossing symmetry.}
\label{fig:QCDfeynman}
\end{figure}

A second effect that makes the jet charge distribution $p_\text{T}$-dependent is the energy-dependence of the fragmentation functions.  The magnitude of the jet charge for a given parton flavor decreases with $p_\text{T}$ because of an increasing contribution from electrically neutral gluons.  The fragmentation functions evolve with energy scale in an analogous manner to the evolution for parton density functions.  In particular, the DGLAP equation also applies\footnote{Note that in some textbooks and papers, there is a factor of two in this equation which depends on if $\mu$ is the energy scale or the virtual mass squared (the factor of two is the Jacobian). }:

\begin{align}
\label{eq:DGLAPforD}
\mu\frac{\partial}{\partial \mu} D_p^h(z,\mu) = \sum_{p'}\int _z^1\frac{dz'}{z'}\frac{\alpha_sP_{p'\leftarrow p}(z')}{\pi}D_{p'}^h\left(\frac{z}{z'},\mu\right)
\end{align}

\noindent The integral of Eq.~\ref{eq:DGLAPforD} gives the evolution equation for the Mellin moment $\tilde{D}_p^h$:

\begin{align}\nonumber
\label{eq:DGLAPforD2}
\mu\frac{\partial}{\partial \mu} \tilde{D}_p^h(\kappa,\mu) &=\frac{\alpha_s}{\pi} \sum_{p'}\int_0^1 dz z^\kappa\int _z^1\frac{dz'}{z'}P_{p'\leftarrow p}(z')D_{p'}^h\left(\frac{z}{z'},\mu\right)\\\nonumber
&= \frac{\alpha_s}{\pi}\sum_{p'}\int _0^1\frac{dz'}{z'} P_{p'\leftarrow p}(z') \int_0^{z'} dz z^\kappa D_{p'}^h\left(\frac{z}{z'},\mu\right)\\\nonumber
&\stackrel{x=z/z'}{=}\frac{\alpha_s}{\pi}\sum_{p'}\int _0^1 dz' (z')^\kappa P_{p'\leftarrow p}(z') \int_0^{1} dx x^\kappa D_{p'}^h\left(x,\mu\right)\\
&=\frac{\alpha_s}{\pi}\sum_{p'}\tilde{P}_{p'\leftarrow p}(\kappa)\tilde{D}_i^h(\kappa,\mu).
\end{align}

\noindent For the average jet charge, the transition $g\leftarrow q$ is irrelevant for quark jets because the net charge from a fragmenting gluon is zero.  The average gluon jet charge is zero by symmetry.  This means that the only relevant term in Eq.~\ref{eq:DGLAPforD2} is for $p=q$ and $p'=q$.  At {\it leading power} (the approximation of narrow jets), the ratio of the jet mass to the jet energy $m/E$ is small and so $p_\text{T} \propto E$ at fixed $\eta$. Therefore Eq.~\ref{eq:DGLAPforD2} and~\ref{eq:jetchargenotLO} can be used to compute the $p_\text{T}$ dependence of the average jet charge for a particular quark type jet:

\begin{align}\nonumber
\label{eq:scaeviolationequation}
\frac{p_\text{T}}{\langle Q_q(\kappa)\rangle}\frac{d\langle Q_q(\kappa)\rangle}{dp_\text{T}}&=\frac{1}{\sum_h\tilde{D}_i^h(\kappa,p_\text{T})}\sum_h p_\text{T}\frac{d}{dp_\text{T}}\tilde{D}_i^h(\kappa,p_\text{T})\\
&=\frac{\alpha_s}{\pi}\tilde{P}_{q\leftarrow q}(\kappa)
\end{align}

\noindent The righthand side of Eq.~\ref{eq:scaeviolationequation} can be computed numerically:

\begin{align}\nonumber
\label{eq:jetcharge:ckappa}
\frac{\alpha_s}{\pi}\tilde{P}_{q\leftarrow q}(\kappa)&=\frac{\alpha_sC_F}{\pi}\int_0^1dz z^\kappa \left[\frac{1+z^2}{1-z}\right]_+\\
&=\frac{\alpha_sC_F}{\pi}\int_0^1 dz(z^\kappa-1)\frac{1+z^2}{1-z}
&\approx
 \begin{cases} -0.024 \pm 0.004 & \kappa = 0.3\\  -0.038 \pm 0.006 & \kappa = 0.5 \\  -0.049 \pm 0.008 & \kappa=0.7 \end{cases},
\end{align}

\noindent where the last form are numerical approximations varying the scale of $\alpha_s$ between 50 and 500 GeV, with the average giving the central value ($\alpha_s(50 \text{ GeV}) = 0.130$ and $\alpha_s(500 \text{ GeV}) = 0.094$).   The solution is of the form $\langle Q_q(\kappa)\rangle\propto p_\text{T}^{c(\kappa)}$, where $c(\kappa)$ is the factor computed in Eq.~\ref{eq:jetcharge:ckappa}.

\begin{figure}[h!]
\begin{center}
\includegraphics[width=0.65\textwidth]{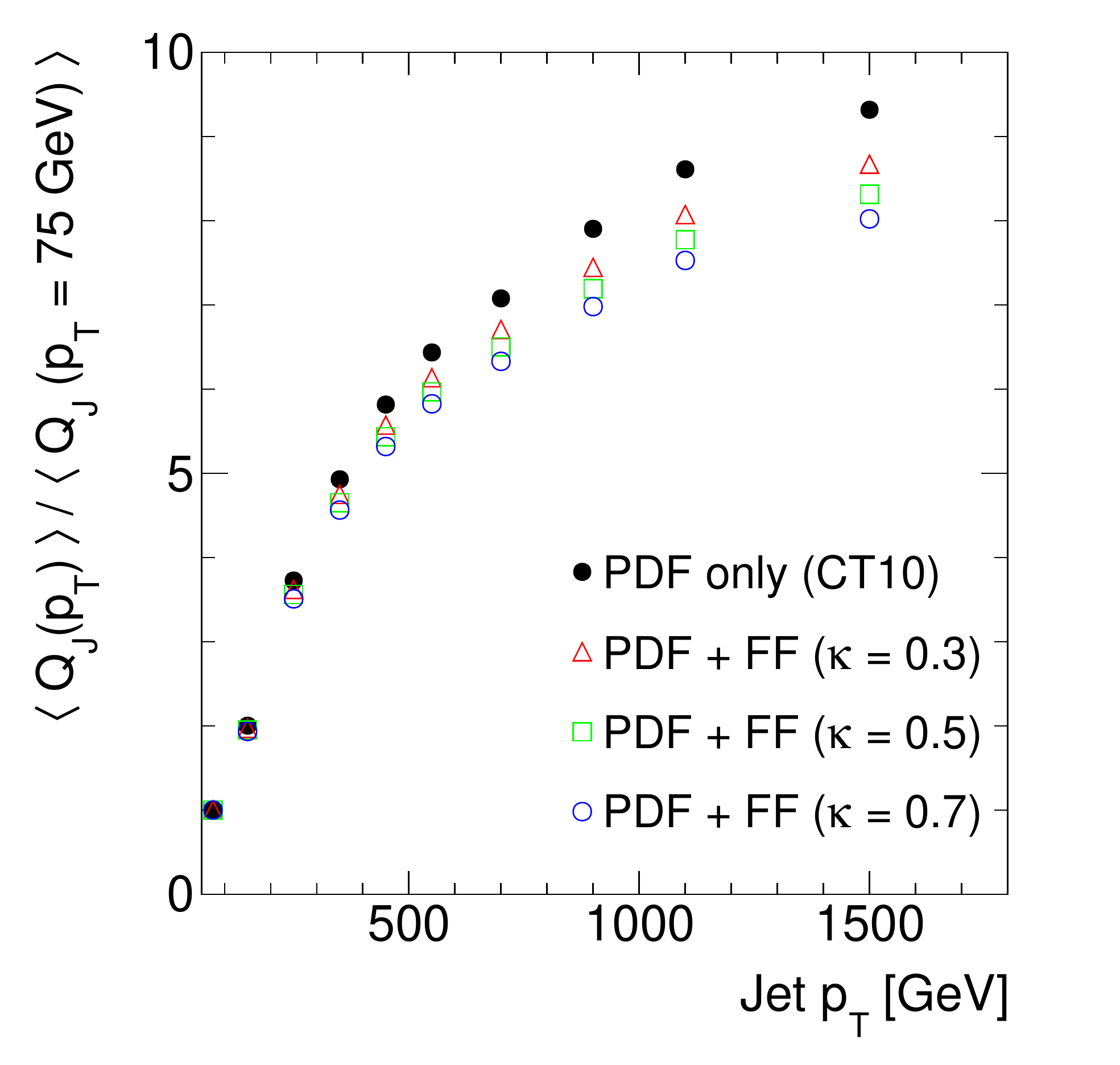}
\end{center}	
\caption{The predicted $p_\text{T}$ dependence of the jet charge with input from the CT10 PDF set and assuming $\langle Q_g\rangle=0,\langle Q_q\rangle=-\langle Q_{\bar{q}}\rangle,\langle Q_u\rangle = -2\langle Q_d\rangle$.  For the red, green, and blue lines, the impact of a $p_\text{T}$-dependent fragmentation function is added on top of the PDF-dependence.  This is the more forward of the two jets in dijet events (see Sec.~\ref{sec:jetcharge:design}) and therefore more likely to be initiated from a quark.  The fragmentation functions (FF) do not depend on $\kappa$, but the energy-dependence of their $\kappa$-moments do depend on $\kappa$.}
\label{fig:twoeffects}
\end{figure}

Figure~\ref{fig:twoeffects} shows the relative size of the two sources of $p_\text{T}$ dependence.   Assuming $\langle Q_g\rangle=0,\langle Q_q\rangle=-\langle Q_{\bar{q}}\rangle,\langle Q_u\rangle = -2\langle Q_d\rangle$, the only free parameter is $\langle Q_u\rangle$, which is removed by normalizing the jet charge at a fixed $p_\text{T}=75$ GeV.  By construction, the relative jet charge is 1 in the first $p_\text{T}$ bin.  The relative change in the average jet charge is a factor of 10 due to PDFs with just over a factor of 10 increase in the jet $p_\text{T}$.  In contrast, the additional impact of $p_\text{T}$-dependent fragmentation functions results in a $\sim10\%$ change in the average jet charge.  

\clearpage

\subsection{Charge Tagging}
\label{sec:chargetagging}

\begin{figure}[h!]
\begin{center}
\begin{overpic}[scale=0.7]{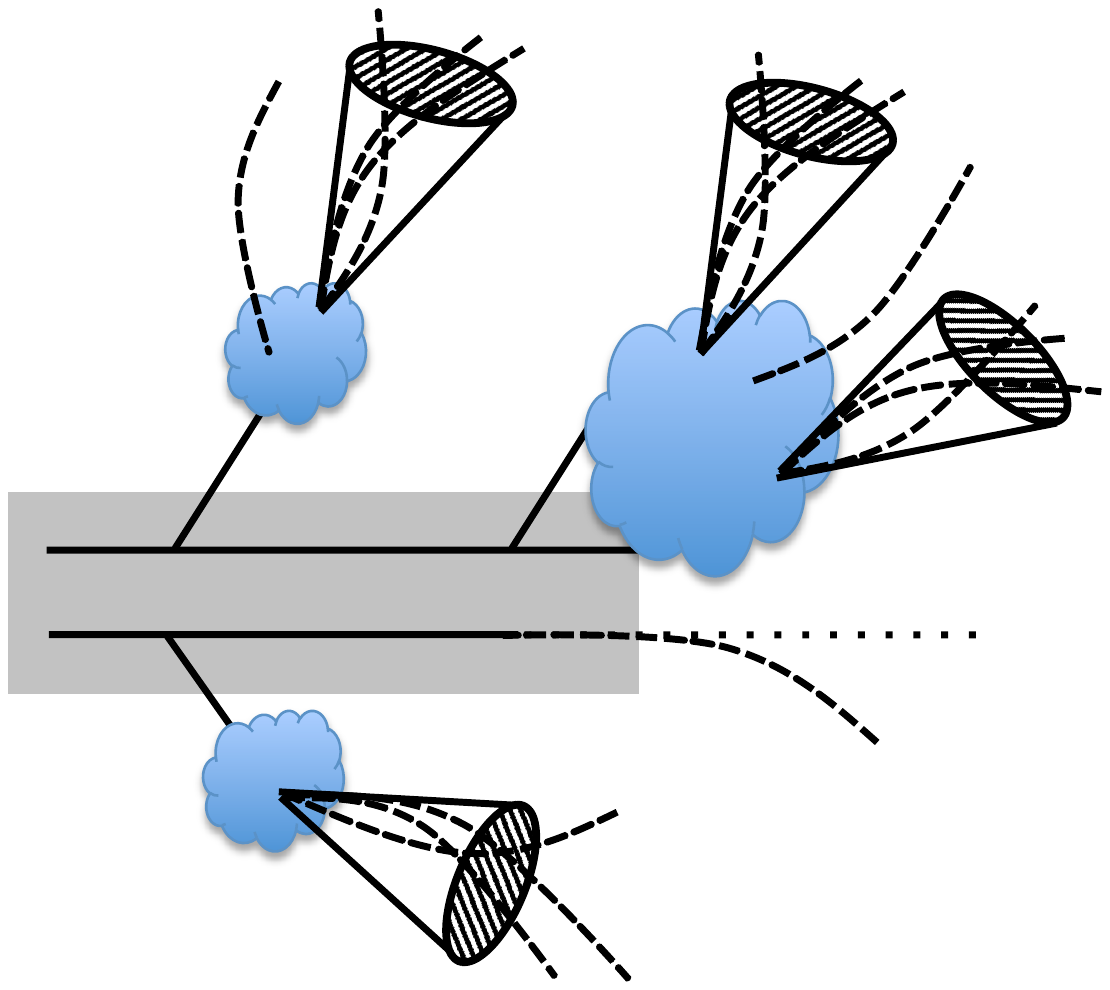}
\put(3,40){$t$}
\put(3,32){$\bar{t}$}
\put(13,29){$\bar{b}$}
\put(13,42){$b$}
\put(30,42){$W^+$}
\put(30,29.3){$W^-$}
\put(43,42.8){$u$}
\put(48,36.2){$\bar{d}$}
\put(68,23){$l^-$}
\put(80,35){$\bar{\nu}$}
\end{overpic}
\caption{A schematic of the $t\bar{t}$ decay topology in the semi-leptonic channel.  The clouds depict the fragmentation process that ultimately lead to observable tracks (dotted lines) and hadronic jets (cones).  The final state is characterized by two $b$-tagged jets, two hadronic jets from the $W$ decay, an isolated lepton, and missing momentum from the undetectable neutrino.  In current kinematic fits, only the momentum of the jets are used and not the charge properties of the tracks.}
\label{fig:JetCharge:Tagging:schematic}
\end{center}
\end{figure}

Section~\ref{sec:jetchargetheory} showed that there are several interesting theoretical aspects of the jet charge that make non-trivial predictions for the $p_\text{T}$-dependence.  This section discusses a practical aspect of studying the jet charge: charge tagging.  In the one-jet-one-parton paradigm, it is often necessary to resolve ambiguities in the matching between partons and jets that could be solved with an additional handle based on the electric charge information.  As an example, consider ambiguity solving in $t\bar{t}$ events in the $t\bar{t}\rightarrow bW(\rightarrow l\nu)bW(\rightarrow qq')$ channel.  Such events can be isolated with high purity due to the leptonically decaying $W$ boson.  However, there are many applications where one needs to directly identify the selected jets with the top decay products.  One example is the measurement presented in Chapter~\ref{cha:colorflow}.  Figure~\ref{fig:JetCharge:Tagging:schematic} schematically illustrates the setup: jets need to be assigned to partons in the top decay topology.  The $W$ boson and top quark masses provide powerful constraints on the jet momenta, but the jet charge could provide additional information.  In particular, the jet charge could help resolve the matching of the $b$-tagged jets with the $b$ or $\bar{b}$ quark\footnote{This idea has now been implemented as a dedicated tagger - see Ref.~\cite{ATL-PHYS-PUB-2015-040} for details.  Vertex charge tagging is also available, when secondary and tertiary charged hadron decay vertices are reconstructed.}.

In addition to jet charge, $b$-quark jets offer additional handles to identify the parton charge from the semi-leptonic decays of $B$ and $D$ hadrons.  These leptons are often too soft to measure\footnote{The threshold used here is $p_\text{T} >4$ GeV.  This requirement comes from the soft-lepton $b$-tagging algorithm used in ATLAS~\cite{Aad:2015ydr}, where there is about a 50\% chance of identifying a soft-lepton.} and even though the lepton charge is highly correlated with the quark charge, there is some contamination from light hadron decays.  Denote the lepton charge as $Q_L$.  In the first paper on jet charge, Field and Feynman~\cite{Feynman1978} describe two criteria for evaluating a charge tagging algorithm:

\begin{itemize}
\item Efficiency ($E$): The percentage of jets to which the algorithm can be applied.  
\item Reliability ($R$): Given that the algorithm is applicable, the probability that the assignment is correct.
\end{itemize}

\noindent For example $R[Q_L]$ is expected to be close to one, but since not all jets have an identified soft lepton ($Q_L\neq 0$), $E[Q_L]<1$. Table~\ref{tab:tabdefs} shows the values of reliability and efficiency for several variations of the jet charge using particle-level simulation with {\sc Pythia}~8.  The algorithm with the best reliability is the lepton charge, but the efficiency to have two reconstructed semileptonic $B$ or $D$ decays is low.  On the other hand, the jet charge  performs well (reliability 66\%) and applies to every jet (efficiency 100\%).  There is not a strong dependence on $\kappa$ for $\kappa\sim 0.5$, but this will be revisited in Sec.~\ref{sec:jetcharge:qcd} with the full ATLAS detector simulation.

\begin{table}
\begin{center}
\begin{tabular}{ c c c c}
   Algorithm & $R$ & $E$ & $E\times R$\\\hline
 $Q_L^{\bar{b}}> Q_L^b$ and $Q_L^{\bar{b}},Q_L^b\neq 0$ & 77\% & 3.7\% & 2.8\% \\
 $Q_L^{\bar{b}}> Q_L^b$  & 42\% &66\% &28\%\\
 $Q_J^{\bar{b}}(\kappa=1) > Q_J^b(\kappa=1)$ & 62\%& 100\% & 62\% \\
 $Q_J^{\bar{b}}(\kappa=0.3) > Q_J^b(\kappa=0.3)$ & 62\%& 100\% & 62\% \\
\end{tabular}
\end{center}
\caption{For each algorithm, the probability that the $b$- and $\bar{b}$-jet assignment is correct (reliability $R$) and the fraction of events to which the algorithm can be applied (efficiency $E$).  Note that the value of R in the second row is less than 50\% of the time, due to cases in which neither jet has an associated lepton ($Q_L=0$).}
\label{tab:tabdefs}
\end{table}

The jet charge reliability is comparable to simple kinematic techniques.  For example, when the initial top quarks are produced with some initial momentum, the resulting $b$ quark and $W$ bosons will tend to be closer in $\Delta R$ than to the anti-top decay products.  Particle-level simulation predicts that a $\Delta R$-based scheme has a similar reliability to the jet charge and is rather uncorrelated; combining the two results in a $\sim 10\%$ increase in reliability.  In addition to $b$-jet charge identification, the jet charge can aid in the assignment of jets to the hadronic $W$ boson decay.  For example, by requiring the dijet charge to be opposite the charge of the lepton, the jet selection based only on the invariant mass of the two jets can be improved by $\sim 15\%$\footnote{There is no unique way to declare an assignment {\it correct}, but the results stated here are nearly the same when using a $\Delta R$ matching between the $W$ boson and the two jets and an energy fraction method, described in Sec.~\ref{sec:colorflow:wcandidateselection}.  The efficiency of this requirement is 60\%.}.  Charge tagging for hadronic $W$ boson decays will be revisited in Sec.~\ref{sec:insitubosoncharge}.

Topological assignments in top quark pair production is only one example where charge tagging could improve the performance of existing methods.  Other examples include high $b$-quark multiplicity final states (e.g. $t\bar{t}H,\tilde{g}\rightarrow t\bar{t}\tilde{\chi}_1^0$, and $T'\rightarrow Ht$) and quark versus gluon tagging. It is therefore important to study the jet charge performance in order to validate and improve the inputs to jet charge-based tagging techniques.   Charge tagging will be revisited in Sec.~\ref{sec:jetcharge:qcd} for small-radius jets and in Sec.~\ref{charge:boosted} for charge tagging large-radius jets.

\clearpage

\section{Analysis Design}
\label{sec:jetcharge:design}

\begin{figure}[h!]
\begin{center}
{\includegraphics[width=0.5\textwidth]{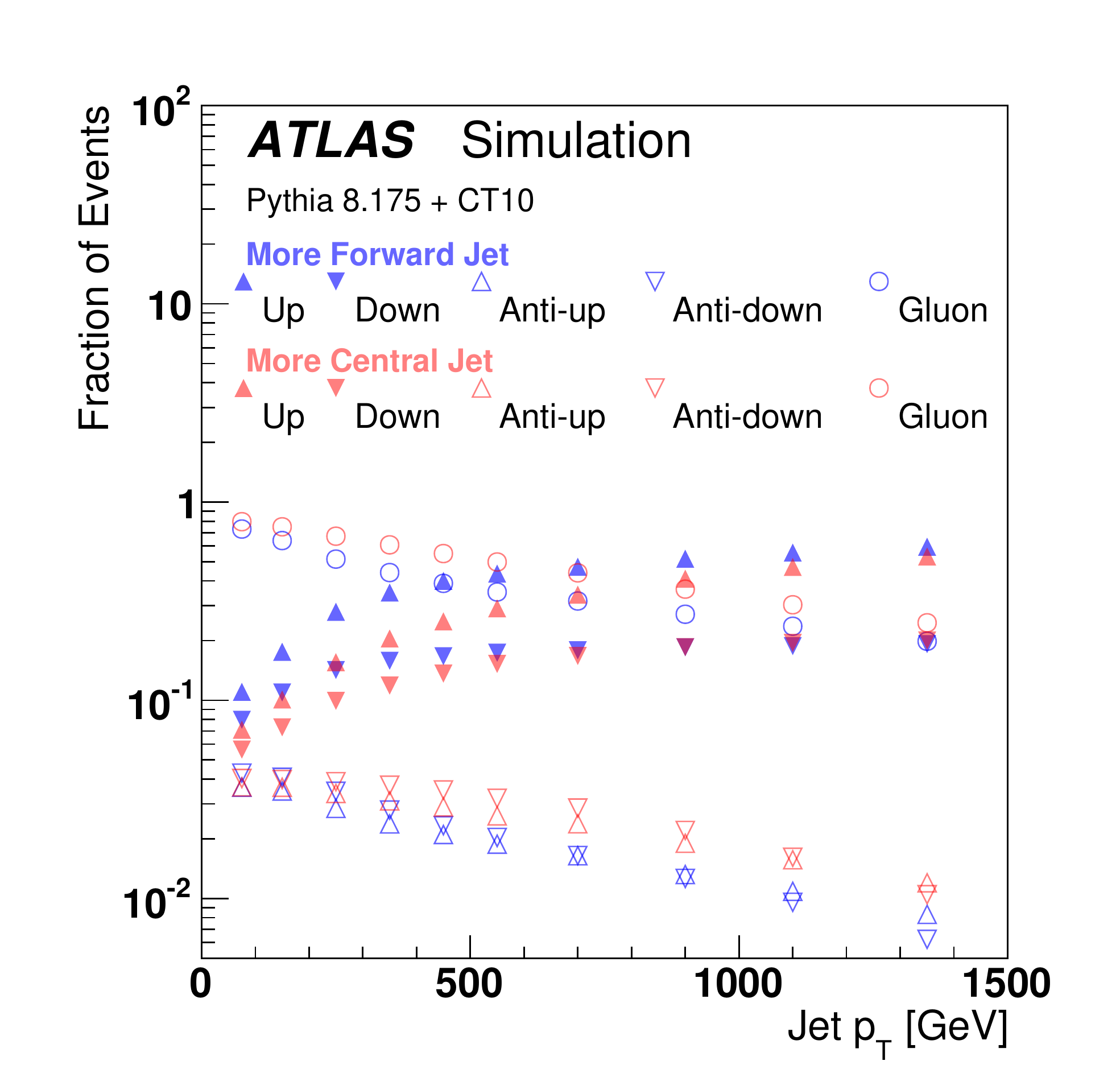}}{\includegraphics[width=0.5\textwidth]{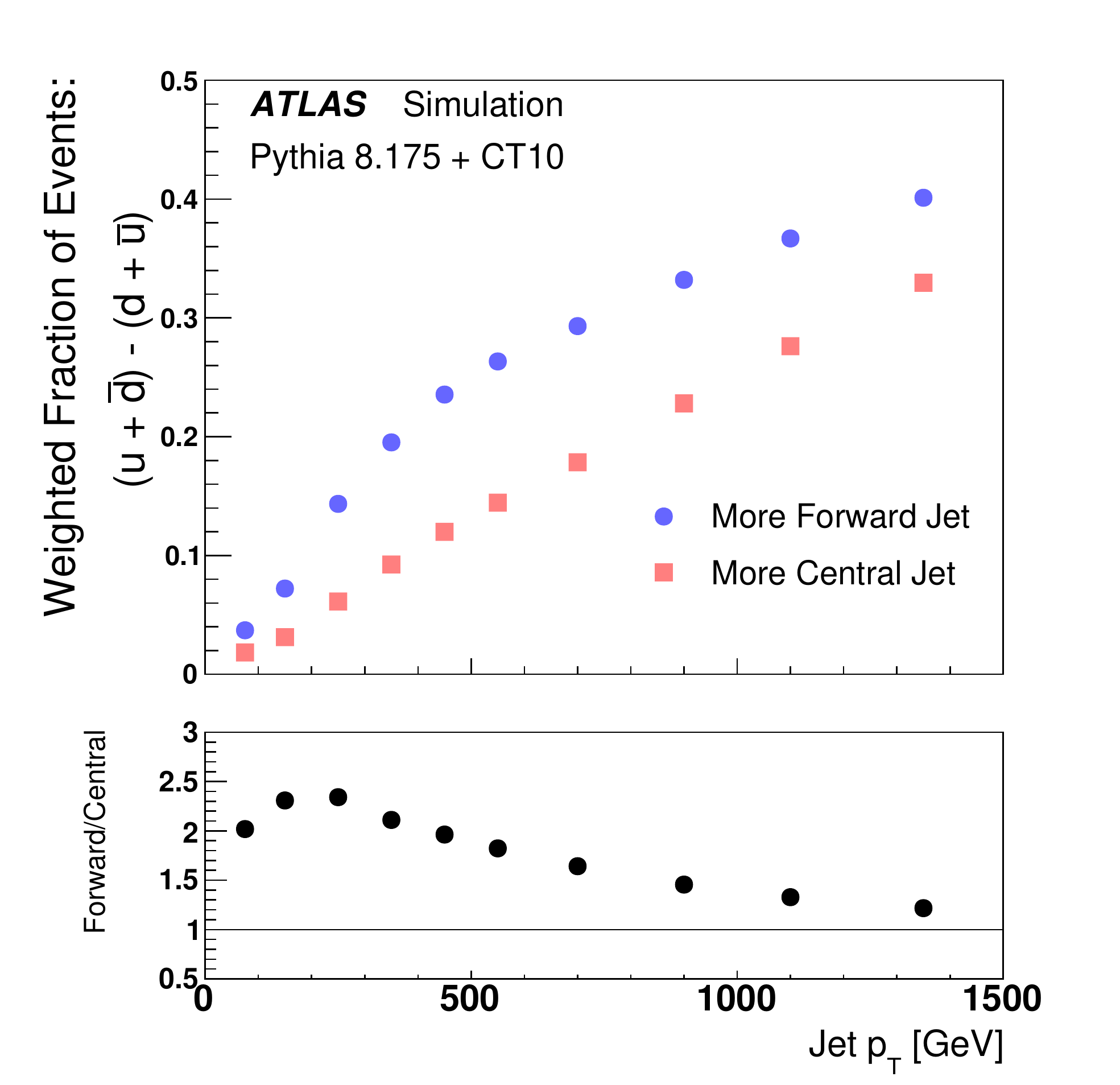}}
\end{center}	
\caption{For a given jet flavor, (a) shows the fraction $f$ of jets with that flavor in events passing the particle-level event selection and (b) shows the $p_\text{T}$ evolution of the flavor fractions weighted by charge-sign: $f_\text{up}+f_\text{anti-down}-f_\text{anti-up}-f_\text{down}$.  The CT10 PDF set is combined with matrix elements from {\sc Pythia} 8.  The forward-central differences between the flavor fractions are largest at low $p_\text{T}$, but the highest quark-jet purity occurs at high jet $p_\text{T}$.  The markers for the more forward and central jets are distinguished by their blue and red colors, respectively.}
\label{fig:flavorfrac}
\end{figure}

The main purpose of this chapter is to present a precision measurement of the jet $p_\text{T}$-dependence of the jet charge distribution.  As part of this analysis, the jet charge reconstruction is studied in order to improve the measurement as well as the understanding of charge tagging.  The jet charge distribution is measured in inclusive dijet events from $pp$ collisions at $\sqrt{s}=8$~TeV.  Inclusive dijet events provide a useful environment for measuring the jet charge as they are an abundant source of gluon-initiated and quark-initiated jets.  There are fewer theoretical ambiguities associated with assigning the jet flavor in events with two jets than in events with higher jet multiplicities.  Furthermore, the transverse momentum ($p_\text{T}$) range accessible in dijet events is broad, $\mathcal{O}(10)$ GeV~up to $\mathcal{O}(1000)$ GeV.  As discussed in Sec.~\ref{sec:jetchargetheory}, the jet charge distribution is expected to change significantly over this kinematic range due to changes in the PDF.  The PDFs are fairly well constrained~\cite{Watt:2012tq,Harland-Lang:2014zoa,Aaron:2009aa,Abramowicz:2015mha,Ball:2014uwa} in the
momentum fraction range relevant for this study, $0.005$---$0.5$.  However, because the jet charge is directly sensitive to the parton flavor, its $p_\text{T}$ dependence can provide a consistency check using new information beyond the jet $p_\text{T}$, which is currently used in PDF fits.  The PDFs are not the only nonperturbative input needed to model the jet charge distribution and its evolution with $\sqrt{\hat{s}}$.  As a momentum-weighted sum over jet constituents, the jet charge is sensitive to the modeling of fragmentation.  Previous studies have shown that there are qualitative differences between the charged-particle track multiplicities of jets in data and as predicted by the leading models of hadron production~\cite{Aad:2014gea}.  Thus, a measurement of the jet charge distribution with a range of quark/gluon compositions can provide a constraint on models of jet formation.  Furthermore, the high energy dataset can be used to probe the sub-leading $p_\text{T}$ dependence of the jet charge due to the $\sqrt{\hat{s}}$-dependence of the fragmentation functions.  This requires new techniques for extracting the jet charge for individual jet flavors. The average jet charge is extracted for both the leading and subleading jet and they are distinguished based on their relative orientation in rapidity.  The more forward of the two jets has a larger energy and is associated with the incoming parton that had a higher momentum fraction of the proton.  As this parton is more likely than the lower momentum fraction parton to be a(n up) quark, the difference in the average jet charge between the more forward and more central jets provides a way to extract the jet charge per jet flavor.  Figure~\ref{fig:flavorfrac}(a) shows the flavor fraction for the more forward and more central particle-level jets that are well-balanced in $p_\text{T}$ (see Sec.~\ref{sec:chargeobjects}).  The fraction of gluon jets decreases with $p_\text{T}$ for both the more forward and the more central jet, but the quark jet purity is higher for the more forward jet.  The $p_\text{T}$ evolution of the sum of the flavor fractions weighted by the sign of the parton charge is shown in Fig.~\ref{fig:flavorfrac}(b).  

Further details about the analysis setup are described in subsequent sections.  The dataset and simulated samples are detailed in Sec.~\ref{sec:chargesamples} and the object reconstruction and event selection are in Sec.~\ref{sec:chargeobjects}.

\clearpage

\subsection{Data and simulated samples}
\label{sec:chargesamples}

This measurement uses the full dataset of $pp$ collisions recorded by the ATLAS detector in 2012, corresponding to an integrated luminosity of 20.3 fb${}^{-1}$ at a center-of-mass energy of $\sqrt{s}=8$ TeV.  Events are only considered if they are collected during stable beam conditions and satisfy all data-quality requirements~\cite{ATLAS-CONF-2010-038}.  To reject noncollision events, there must be a primary vertex reconstructed from at least two tracks each with $p_\text{T}>400$ MeV~\cite{ATLAS-CONF-2010-069}.  Due to the high instantaneous luminosity and the large total inelastic proton-proton cross section, on average there are about $21$ simultaneous ({\it pileup}) collisions in each bunch crossing.

\subsubsection{Jet Triggers}
\label{sec:jettriggers}

\begin{figure}
\begin{center}
\includegraphics[width=0.9\textwidth]{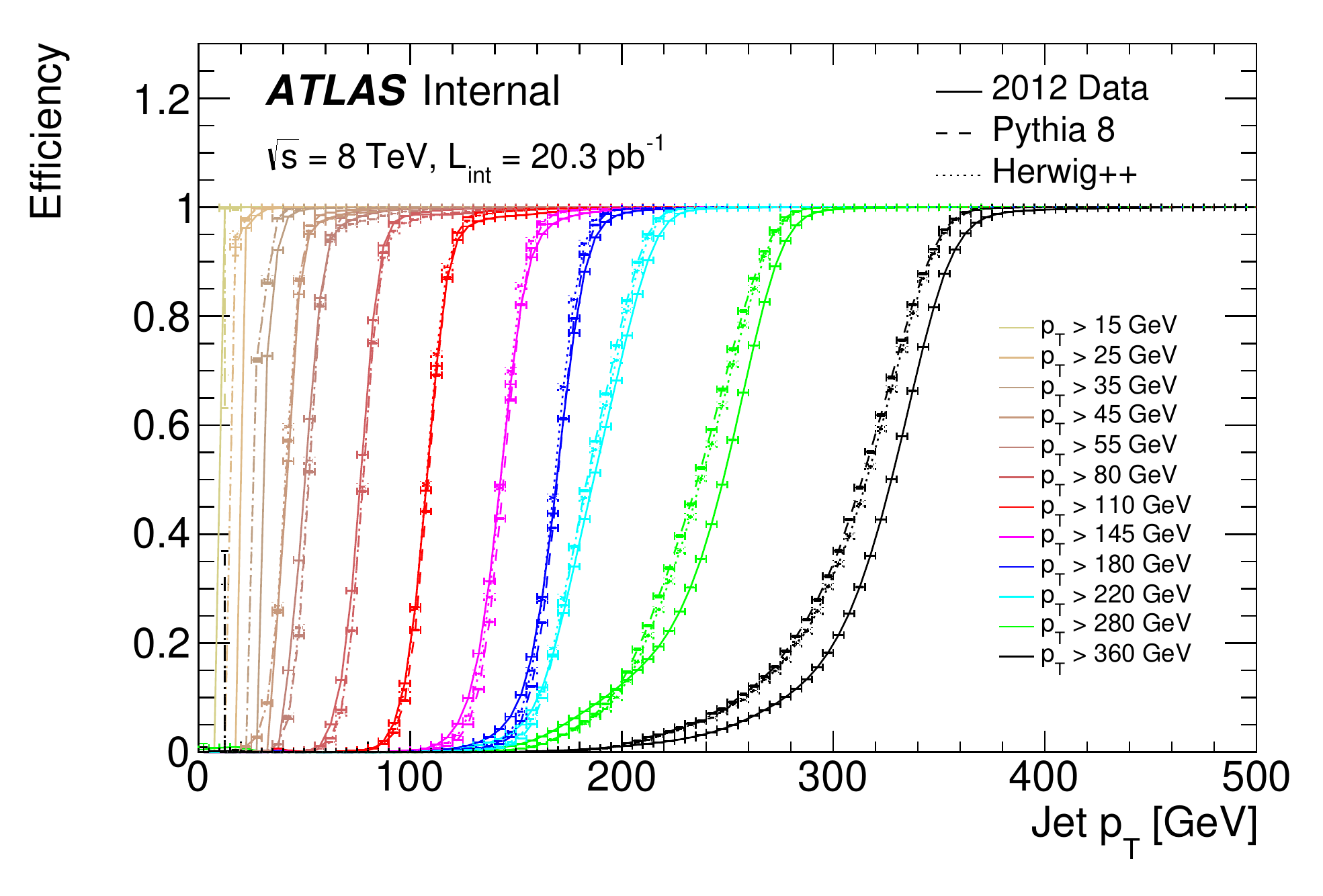}\\
\end{center}
\caption{Trigger Efficiencies for the various single jet triggers used in this analysis.  The numbers in the legend show the point at which the trigger is fully efficient.}
\label{Triggers}
\end{figure}

A set of single-jet triggers is used to collect dijet events with high efficiency.   Due to the large rate for jet production at the LHC and the limited bandwidth, these triggers are {\it pre-scaled}.  For a given trigger $T$, the prescale $\frac{1}{p(T)}=\Pr(\text{save event}|\text{pass $T$})$.  A trigger is not prescaled if $p=1$.  The values $p$ are chosen a priori; the collected data are {\it un-prescaled} by weighting an event by $p(T)$ if the highest $p_\text{T}$ trigger that the event passes is $T$.   A standard method for measuring the trigger efficiency is to use a reference trigger that is fully efficient well below the region of interest and then compute the fraction of events passing the reference trigger that also pass the probe trigger.  The challenge with this method is that by construction the reference trigger will have a larger pre-scale than the probe trigger, and thus a smaller sample size in data.  Another possibility is to {\it emulate} the trigger offline on all collected events.  An event is said to pass the emulated trigger $T$ if the corresponding trigger jet objects all pass the corresponding L1, L2, and Event Filter thresholds.  Figure~\ref{Triggers} shows the trigger efficiency as a function of the offline jet $p_\text{T}$ threshold using the emulation method.  There are some clear differences between data and simulation in the turn-on region of the trigger, but all offline jet thresholds are chosen to avoid this region.  Table~\ref{tab:triggermenu} shows the collected luminosity for each trigger as well as the offline jet $p_\text{T}$ ranges used, chosen such that the trigger is fully efficient.  The highest-$p_\text{T}$ trigger is not prescaled.  The prescale factor is the ratio of total luminosity to the collected luminosity for a given trigger.

\begin{table}[h!]
\centering
\begin{tabular}{ccc}
Trigger threshold [GeV] & Offline Selection [GeV] & Luminosity [fb${}^{-1}$]  \\
\hline 
\hline
25 & [50,100] & 7.84$\times 10^{-5}$ \\
55 & [100, 136] & 4.42$\times 10^{-4}$\\
80 & [136, 190] & 2.32$\times 10^{-3}$ \\
110 & [190, 200] & 9.81$\times 10^{-3}$ \\
145 & [200, 225] & 3.63$\times 10^{-2}$ \\
180 & [225, 250] & 7.88$\times 10^{-2}$ \\
220 & [250, 300] & 2.61$\times 10^{-1}$ \\
280 & [300, 400] & 1.16 \\
360 & $\geq 400$ & 20.3 \\
\hline
\hline
\end{tabular}
\caption{The single-jet trigger menu used to collect dijet events with the 2012 dataset.  The first column is the level-three (Event Filter) jet $p_\text{T}$ threshold and the second column is the offline leading-jet $p_\text{T}$ range corresponding to the given trigger.  The luminosity collected with each trigger is in the last column.  The total 2012 dataset was 20.3 fb${}^{-1}$; the highest-$p_\text{T}$ trigger is not prescaled.}
\label{tab:triggermenu}
\end{table}

Monte Carlo (MC) simulated events are generated in $p_\text{T}$ slices in order to ensure a large number of events over a broad range of reconstructed jet $p_\text{T}$, given constraints on the available computing resources.   The $p_\text{T}$ slices span the interval $0$ to $5$ TeV~in ranges that approximately double with each increasing slice, starting with a range of size $8$ GeV~and ending with a range of size $2240$ GeV.  The baseline sample used for the measurement is generated with {\sc Pythia} 8.175~\cite{Sjostrand:2007gs} with the AU2~\cite{ATL-PHYS-PUB-2012-003} set of tuned parameters (tune) and the next-to-leading-order (NLO) PDF set\footnote{A discussion on the use of NLO PDF sets with LO matrix elements is given in Refs.~\cite{Campbell:2006wx,Sherstnev:2007nd}.} CT10~\cite{Lai:2010vv,Gao:2013xoa}.  Another large sample of events is generated with {\sc Herwig++} 2.63~\cite{Bahr:2008pv,Arnold:2012fq} with tune EE3~\cite{Gieseke:2012ft} and leading-order (LO) PDF set CTEQ6L1~\cite{Pumplin:2002vw} (particle-level samples with CT10 and EE4 are also used for comparisons).  Both {\sc Pythia} and {\sc Herwig++} are LO in perturbative QCD for the ($2\rightarrow 2$) matrix element and resum the leading logarithms (LL) in the parton shower.  However, the ordering of emissions in the MC resummation in the shower differs between these two generators: {\sc Pythia} implements $p_\text{T}$-ordered showers~\cite{Sjostrand:2004ef} whereas {\sc Herwig++} uses angular ordering~\cite{Gieseke:2003rz}.  The phenomenological modeling of the non-pertubative physics also differs between {\sc Pythia} and {\sc Herwig++}.  In addition to different underlying-event models (Ref.~\cite{Sjostrand:2004pf} for {\sc Pythia} and an eikonal model~\cite{Bahr:2008dy} for {\sc Herwig++}) the hadronization models differ between {\sc Pythia} (Lund string model~\cite{string}) and {\sc Herwig++} (cluster model~\cite{Webber:1983if}).  These two schemes are known~\cite{Aad:2014gea} to predict different numbers of charged particles within jets and different distributions of the charged-particle energies within jets, both of which are important for the jet charge.   All tunes of the underlying event that are used with {\sc Pythia} and {\sc Herwig++} in this analysis use LHC data as input.  As discussed in Sec.~\ref{sec:jetcharge:design}, the corrected data are compared to models with various PDF sets; for consistency, each set has a dedicated underlying-event tune constructed in the same way from a fixed set of data inputs (AU2) described in detail in Ref.~\cite{ATL-PHYS-PUB-2012-003}.  The PDF sets include LO sets CTEQ6L1~\cite{Pumplin:2002vw} and MSTW08LO~\cite{Watt:2012tq} as well as NLO sets CT10~\cite{Lai:2010vv,Gao:2013xoa}, NNPDF21 NLO~\cite{Ball:2010de}, and MSTW2008NLO~\cite{Watt:2012tq}.   A sample generated with a NLO matrix element from {\sc Powheg-Box} {\sc r2262}~\cite{Nason:2004rx,Frixione:2007vw,Alioli:2010xd,Frixione:2007nw} (henceforth referred to as {\sc Powheg}) with PDF set CT10 interfaced with {\sc Pythia} 8.175 and the AU2 tune is also used for comparisons.  

\begin{figure}[h!]
\begin{center}
\includegraphics[width=0.5\textwidth]{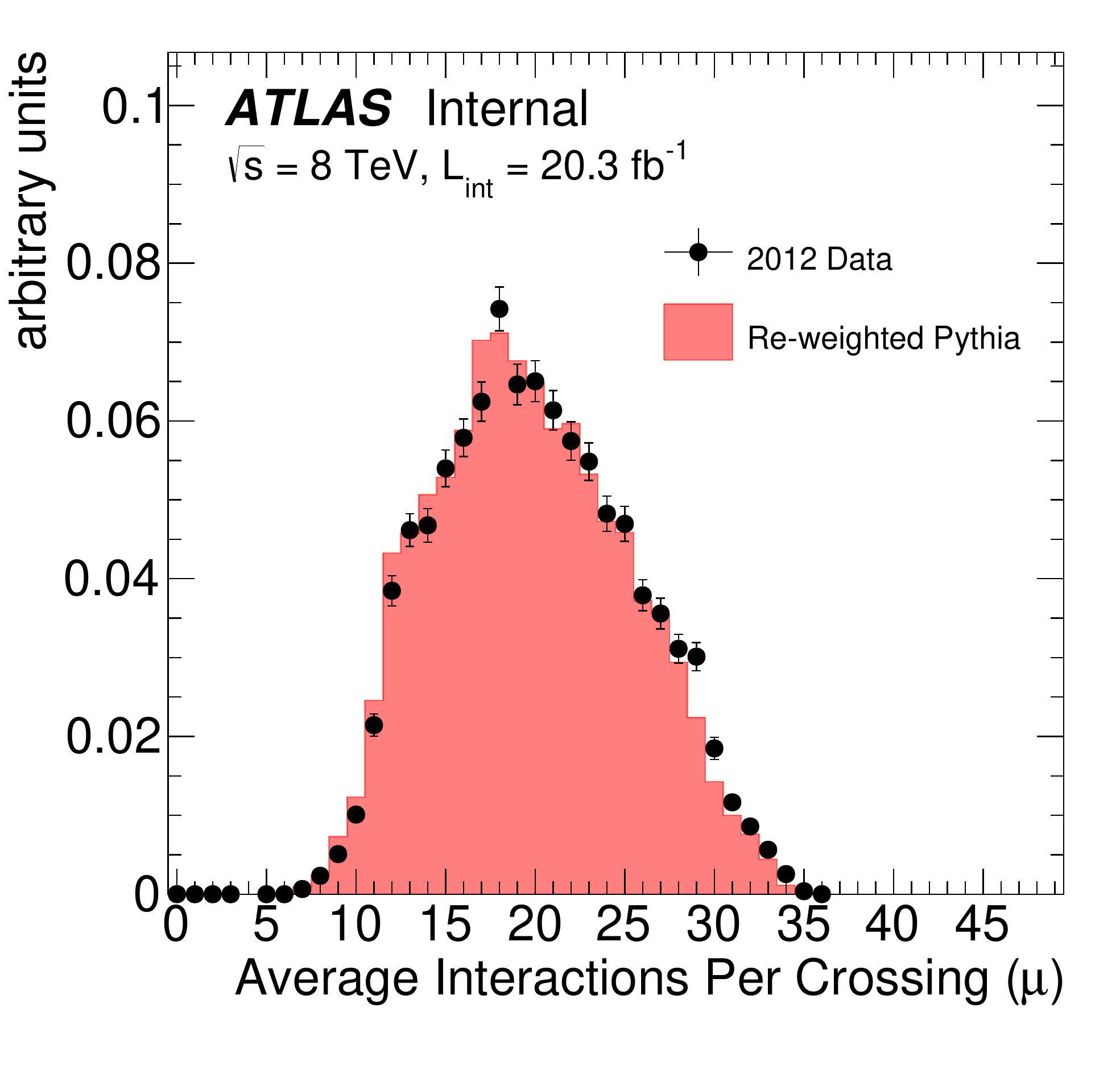}
\end{center}
\caption{The average number of interactions per bunch crossing in data and in Pythia (after re-weighting) after the event selection described in Sec.~\ref{sec:chargeobjects}.}
\label{fig:mu}
\end{figure}

Pileup is simulated by overlaying minimum bias events generated with {\sc Pythia}~8 on top of the hard scatter.  The distribution is re-weighted to match the data as shown in Fig.~\ref{fig:mu}.  All MC samples are processed using the full ATLAS detector simulation~\cite{Aad:2010ah} based on GEANT4~\cite{Agostinelli:2002hh}.

\subsection{Object reconstruction and event selection}
\label{sec:chargeobjects}

The reconstructed objects used for the jet charge as well as for the event selection are described in Sec.~\ref{sec:detectorlevel}.   The fiducial definition of the measurement, unfolded to particle level, is given in Sec.~\ref{sec:particlelevel}.

\subsubsection{Object reconstruction at detector level}
\label{sec:detectorlevel}

Jets are clustered using the anti-$k_t$ jet algorithm~\cite{Cacciari:2008gp} with radius parameter $R=0.4$ implemented in FastJet~\cite{Cacciari:2011ma} from topological calorimeter-cell clusters~\cite{TopoClusters}, calibrated using the local cluster weighting (LCW) algorithm \cite{EndcapTBelectronPion2002,Barillari:2009zza}.  An overall  jet energy calibration accounts for residual detector effects as well as contributions from pileup~\cite{areasATLAS} in order to make the reconstructed jet energy an unbiased measurement of the particle-level jet energy.  Jets are required to be central $(|\eta| < 2.1)$ so that their charged particles are within the $|\eta|<2.5$ coverage of the ID.   

When more than one primary vertex is reconstructed, the one with the highest $\sum p_\text{T}^2$ of tracks is selected as the hard-scatter vertex.  Events are further required to have at least two jets with $p_\text{T}>50$ GeV~and only the leading two jets are considered for the jet charge measurement.  To select dijet topologies, the two leading jets must have $p_\text{T}^\text{lead}/p_\text{T}^\text{sublead} < 1.5$, where $p_\text{T}^\text{lead}$ and $p_\text{T}^\text{sublead} $ are the transverse momenta of the jets with the highest and second-highest $p_\text{T}$, respectively.  The jet with the smaller (larger) absolute pseudorapidity $|\eta|$ is classified as the more central (more forward) jet.  A measurement of the more forward and more central jet charge distributions can exploit the rapidity-dependence of the jet flavor to extract information about the jet charge for a particular flavor.  This is discussed in more detail in Sec.~\ref{sec:particlelevel}.

\begin{figure}[h!]
\begin{center}
\includegraphics[width=0.5\textwidth]{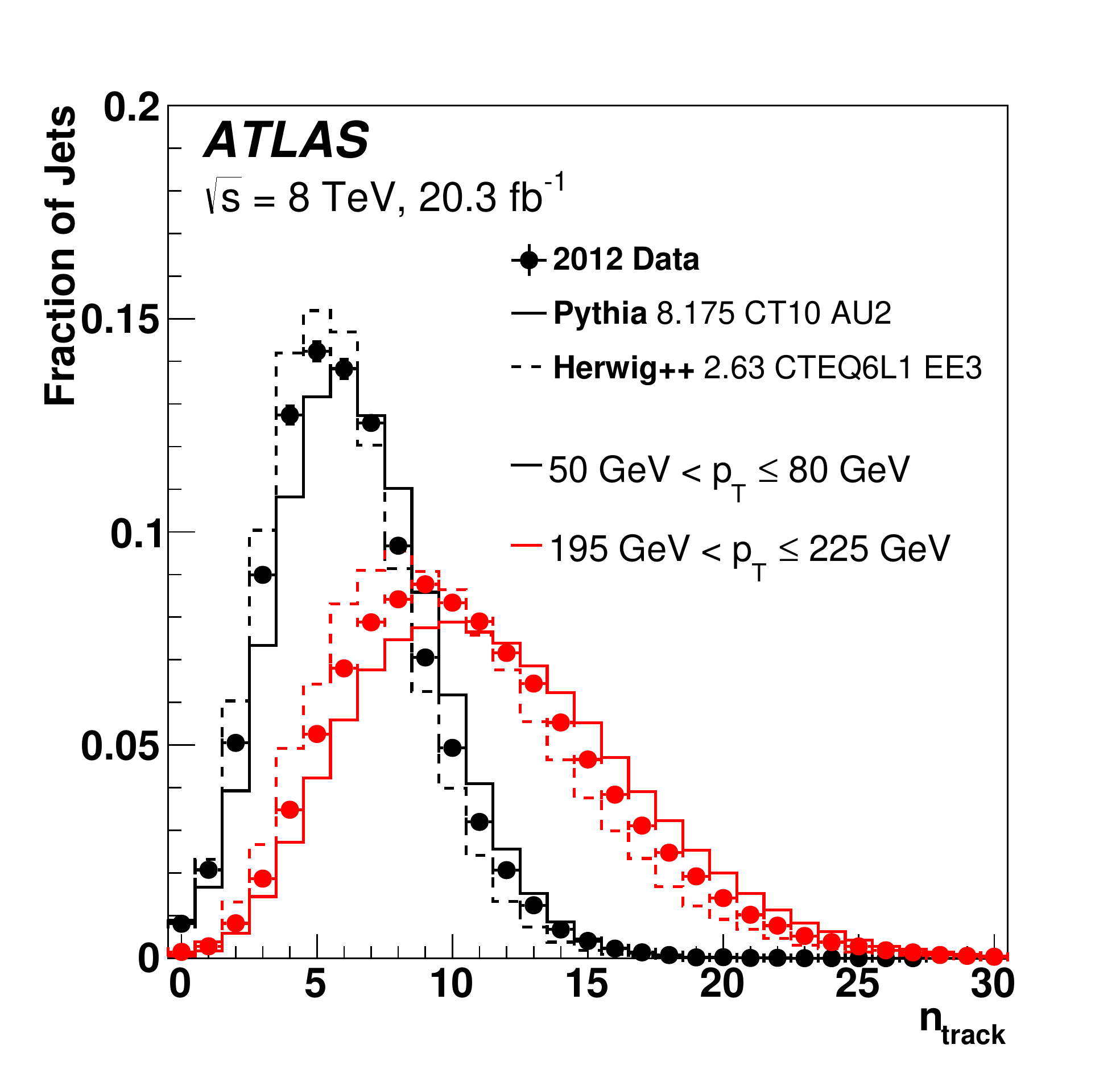}
\end{center}	
\caption{The distribution of the number of tracks associated with a jet in two example jet $p_\text{T}$ ranges.}
\label{fig:tracks}
\end{figure}

Tracks used to calculate the jet charge are required to have $p_\text{T} \geq$ 500 MeV, $|\eta| < 2.5$,~and a $\chi^2$ per degree of freedom (resulting from the track fit) less than 3.0.    Additional quality criteria are applied to select tracks originating from the collision vertex and reject fake tracks reconstructed from random hits in the detector.  In particular, tracks must be well-matched to the hard-scatter vertex with $|z_0\sin(\theta)|<1.5$ mm and $|d_0|< 1$ mm, where $z_0$ and $d_0$ are calculated with respect to the primary vertex.  Tracks must furthermore have at least one hit in the pixel detector and at least six hits in the SCT.   The distribution of the number of tracks in jets in two representative jet $p_\text{T}$ ranges is shown in Fig.~\ref{fig:tracks}.  The number of tracks increases with jet $p_\text{T}$ and the data fall between the predicted distributions of {\sc Pythia} and {\sc Herwig++}.

\subsubsection{Object definitions at particle level}
\label{sec:particlelevel}

\begin{figure}[h!]
\begin{center}
\includegraphics[width=0.5\textwidth]{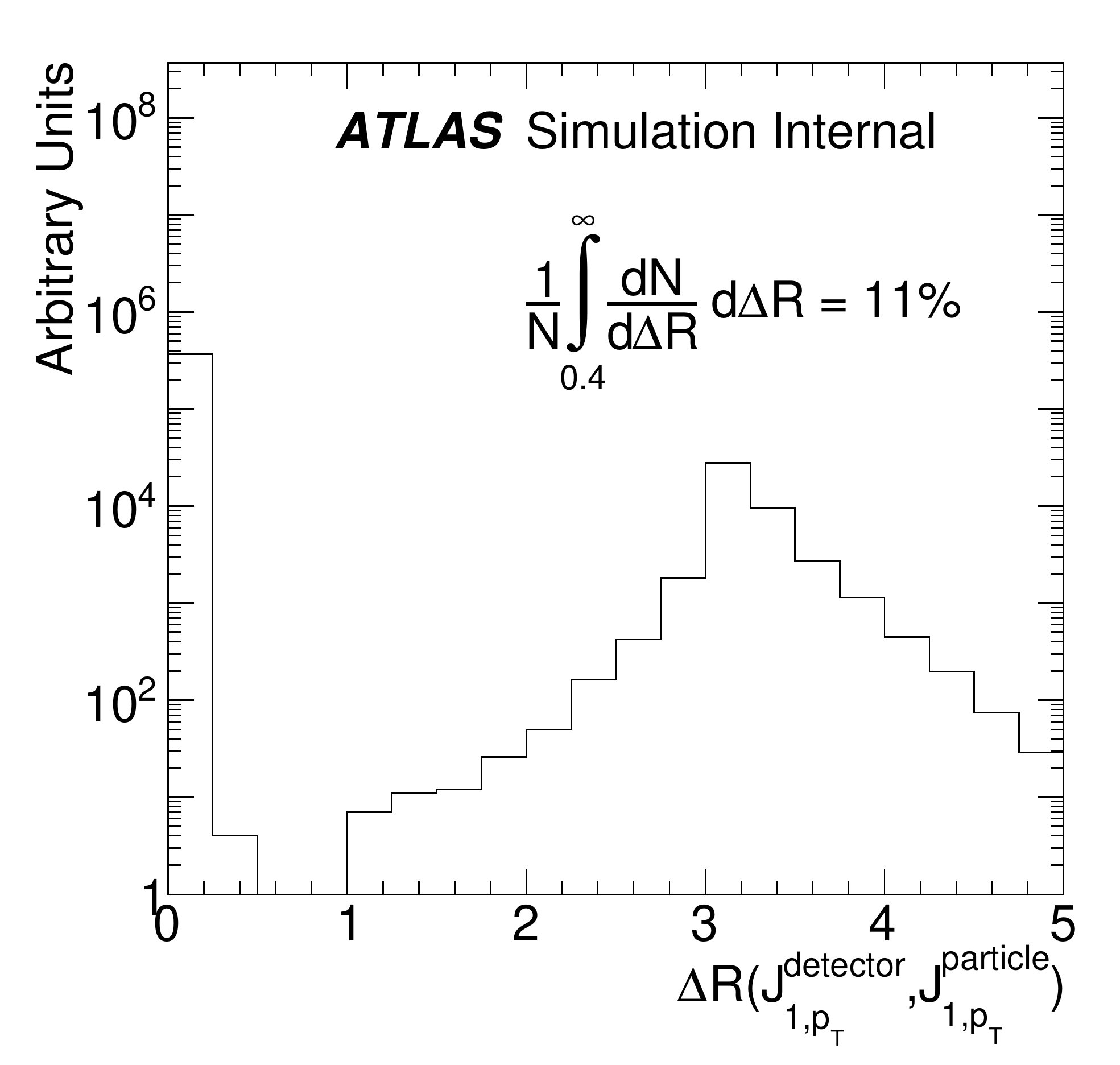}\includegraphics[width=0.5\textwidth]{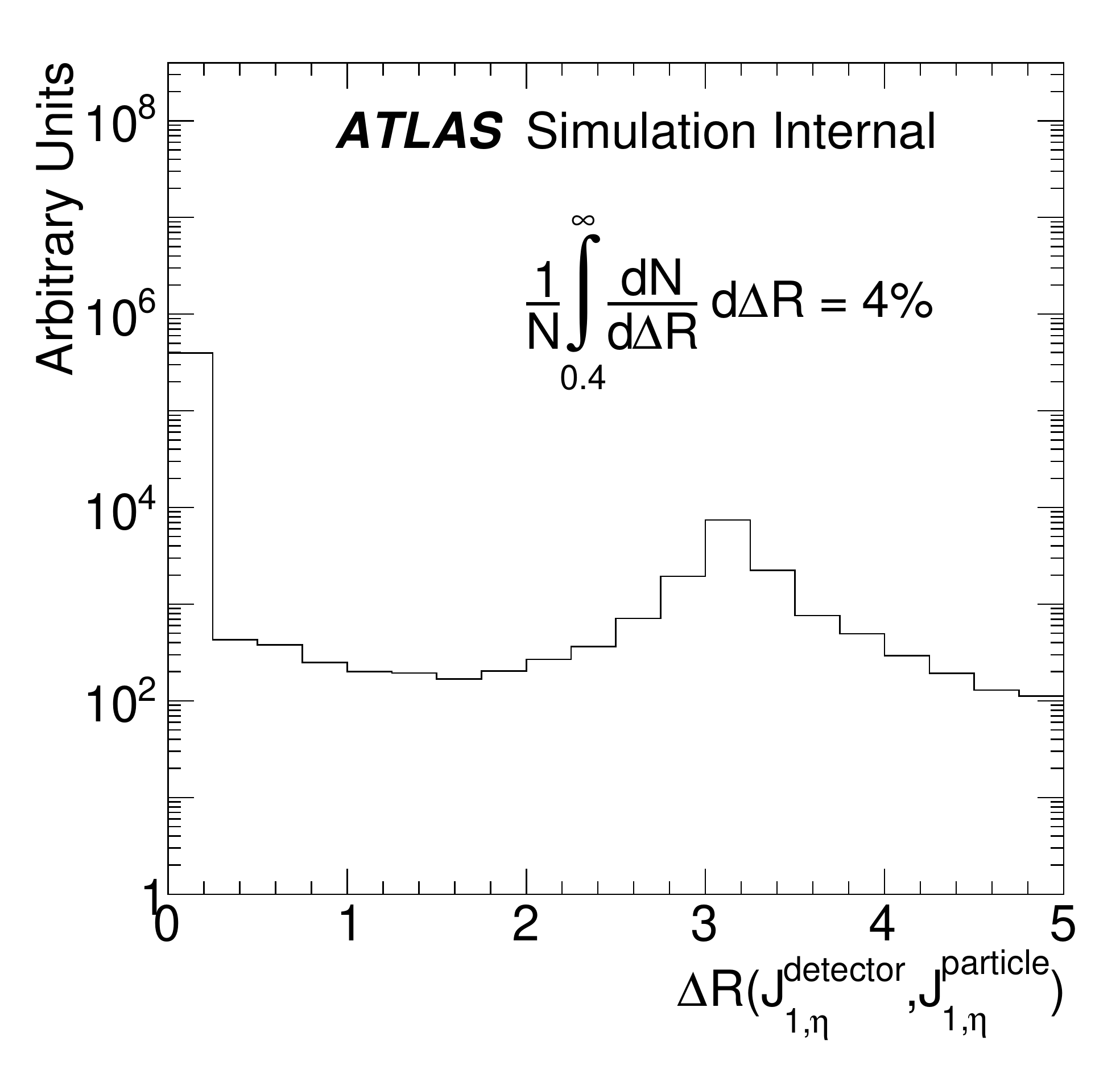}
\end{center}	
\caption{The distribution of the distance $\Delta R$ between the selected detector-level jet and the selected particle-level jet using a momentum scheme (left) and a rapidity scheme (right).}
\label{fig:jetcharge:dRtruthreco}
\end{figure}

\begin{figure}[h!]
\begin{center}
\includegraphics[width=0.45\textwidth]{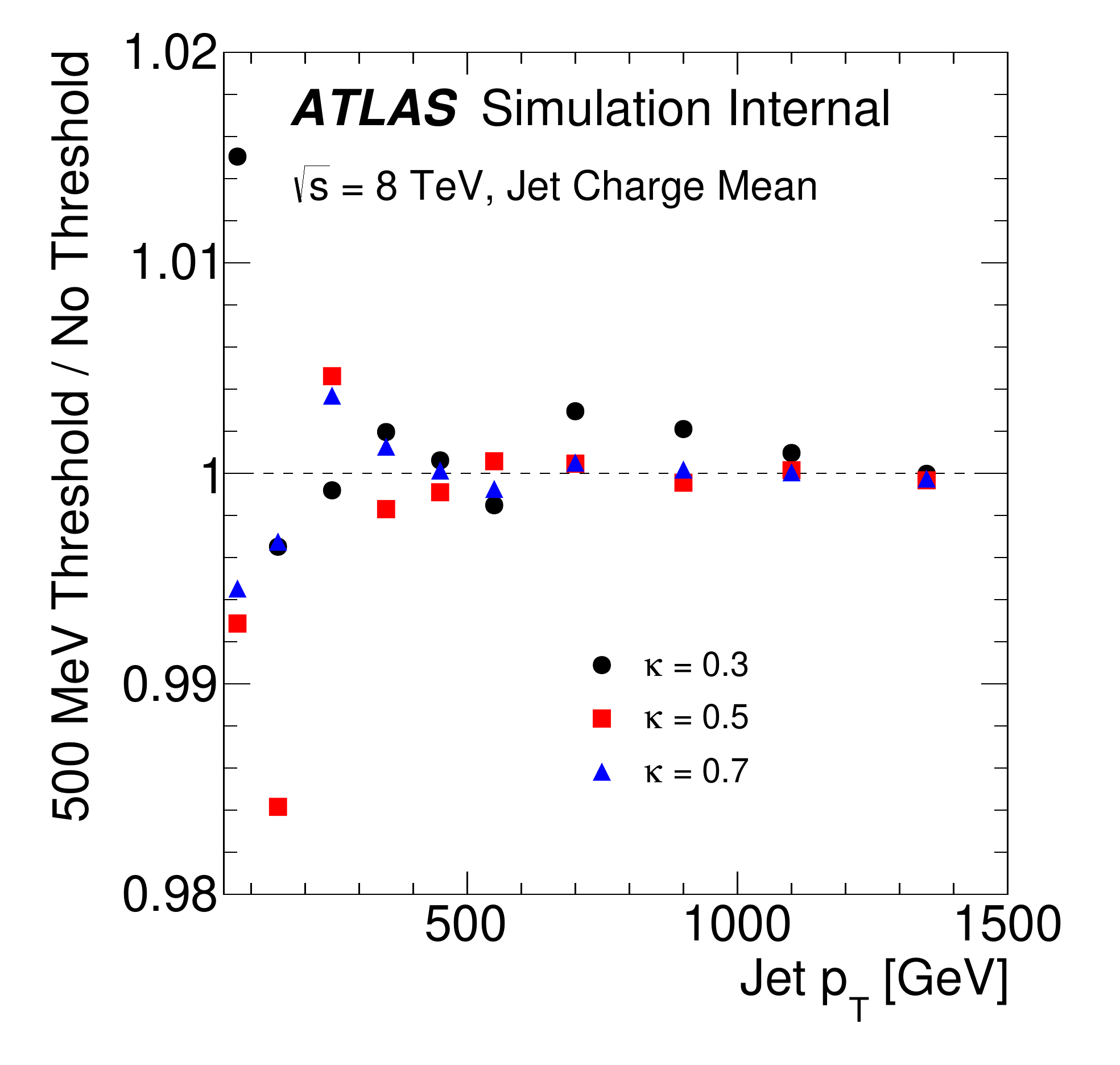}
\includegraphics[width=0.45\textwidth]{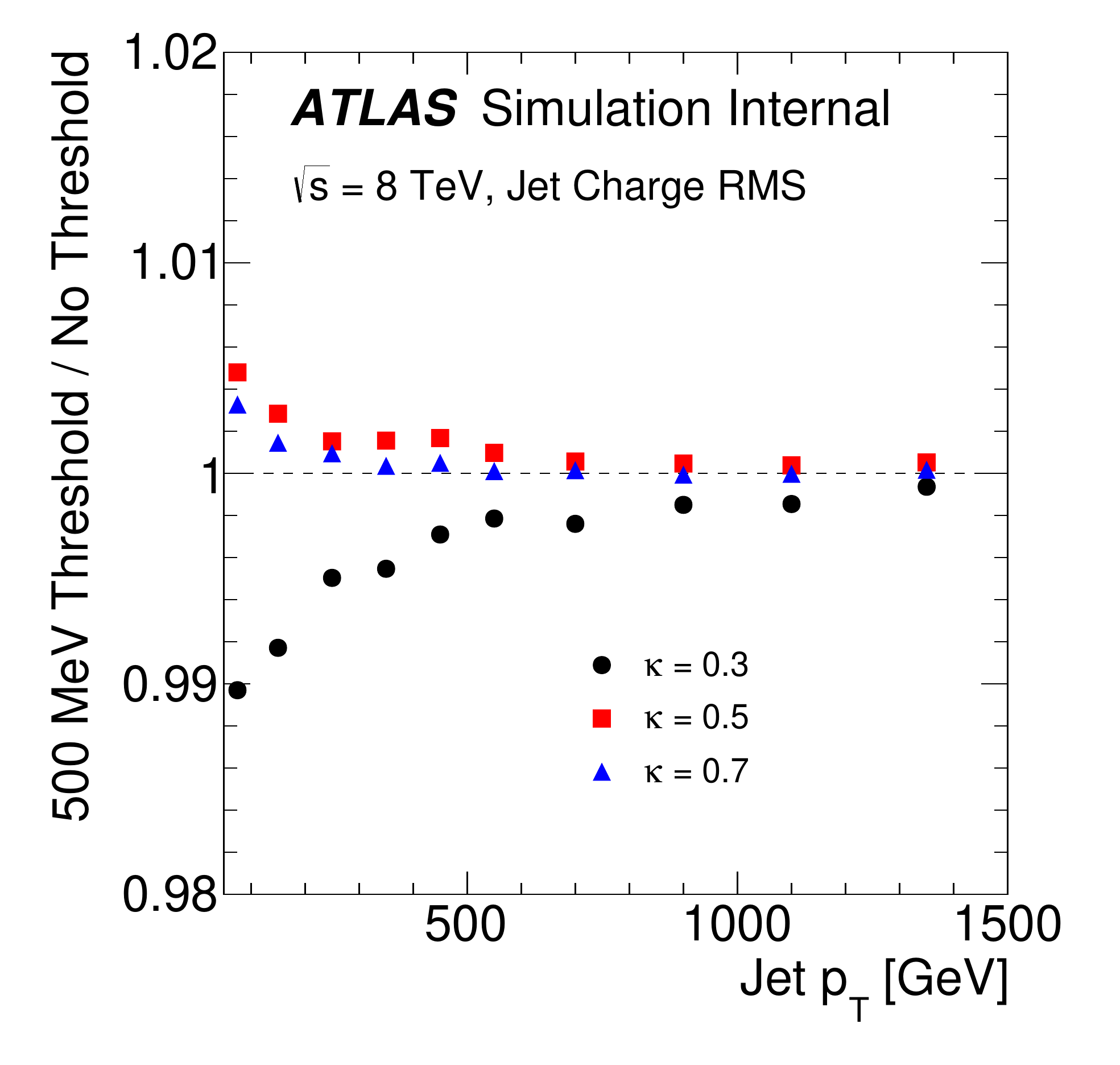}\\
\end{center}
\caption{The impact of adding a particle--level jet $p_\text{T}$ threshold of 500 MeV.  }
\label{fig:truthcut}
\end{figure}

The measurement is carried out within a fiducial volume matching the experimental selection to avoid extrapolation into unmeasured kinematic regions that have additional model-dependence and related uncertainties.  Particle-level (prior to a real or simulated detector) definitions corresponding to the reconstructed objects are chosen to be as close as possible to those described in Sec.~\ref{sec:detectorlevel}.  Particle-level jets are clustered from  generated stable particles with a mean lifetime $\tau>30$~ps,  excluding muons and neutrinos\footnote{Only particles prior to the detector simulation are used in the unfolding.  For example, tracks from photon conversions ($\gamma\rightarrow e^+e^-$) in the inner detector or $K_s\rightarrow \pi^+\pi^-$ decays may be reconstructed as detector-level tracks, but excluded as particle-level tracks.}.  As with the detector-level jets, particle-level jets are clustered with the anti-$k_t$ $R=0.4$ algorithm.  In analogy to the ghost-association of tracks to jets performed at detector level, any charged particle clustered in a particle-level jet is considered for the jet charge calculation\footnote{There is no $p_\text{T}>500$ MeV threshold applied to charged particles.  The impact of applying such a threshold is negligible for all $p_\text{T}$ bins except the first two where effects of up to 1\% are observed in the mean and standard deviation of the jet charge.  See Fig.~\ref{fig:truthcut}.}.  There must be at least two jets with $|\eta|<2.1$ and $p_\text{T}>50$ GeV.  The two highest-$p_\text{T}$ jets must satisfy the same $p_\text{T}$-balance requirement between the leading and subleading jet as at detector level ($p_\text{T}^\text{lead}/p_\text{T}^\text{sublead} < 1.5$).   Due to the high-energy and well-separated nature of the selected jets, the hard-scatter quarks and gluons can be cleanly matched to the outgoing jets.   While it is possible to classify jets as quark- or gluon-initiated beyond leading order in $m_\text{jet}/E_\text{jet}$~\cite{Banfi:2006hf}, the classification is algorithm-dependent and unnecessary for the present considerations (in part because of the large experimental uncertainty).  In this analysis, the flavor of a jet is defined as that of the highest energy parton in simulation within a $\Delta R<0.4$ cone around the particle-jet axis.  The jet flavor depends on rapidity and so the two selected jets are classified as either more forward or more central; the more forward jet tends to be correlated to the higher-$x$ parton and is less likely to be a gluon jet.  Another benefit of this pseudorapidity-based scheme is that the particle-level and detector-level jets are more often the same objects.  This is because by conservation of momentum, the transverse momentum of the two jets in dijet events is similar.  Figure~\ref{fig:jetcharge:dRtruthreco} quantifies this effect; the fraction of events where the selected jets are swapped is half as large under the rapidity scheme compared with the momentum scheme.

\clearpage
\newpage

\section{Performance Studies}
\label{sec:jetchargeperformance}

	For both charge tagging and precision measurements with jet charge, a detailed understanding of the impact of the ATLAS detector on the jet charge reconstruction is critical for improving performance.  Define the {\it jet charge response} as the difference between detector-level jet charge and the particle-level jet charge from jets prior to detector-simulation.   The figures of merit used in this section are the mean and standard deviation of the jet charge response as well as the tradeoff between positive parton type jet efficiency and negative parton type jet efficiency for a given jet charge threshold (charge tagging performance).   The jet charge response is defined as a difference and not a ratio because the jet charge can be close to zero compared with its resolution and thus the ratio with the particle-level jet charge can naturally be  large compared to one.  
			
	Two complementary samples are used to study the response and the charge tagging performance in a variety of settings.  One selection identifies $t\bar{t}$ events to obtain a high-purity sample of hadronically decaying $W$ bosons.  $W$ boson decays are particularly clean because the color singlet $W$ boson is hadronically isolated from the rest of the event.  Additionally, $t\bar{t}$ events in the one lepton final state offer a unique opportunity to use a tag-and-probe technique to study the charge tagging capabilities of the jet charge in-situ.  A second selection targets generic quark and gluon jets in order to probe high $p_\text{T}$ jets and allow for a simulation study of the tagging capabilities of individual quark and gluon jets.

\subsection{Comparisons Between Data and Simulation}

\begin{figure}[h!]
\begin{center}
\includegraphics[width=0.45\textwidth]{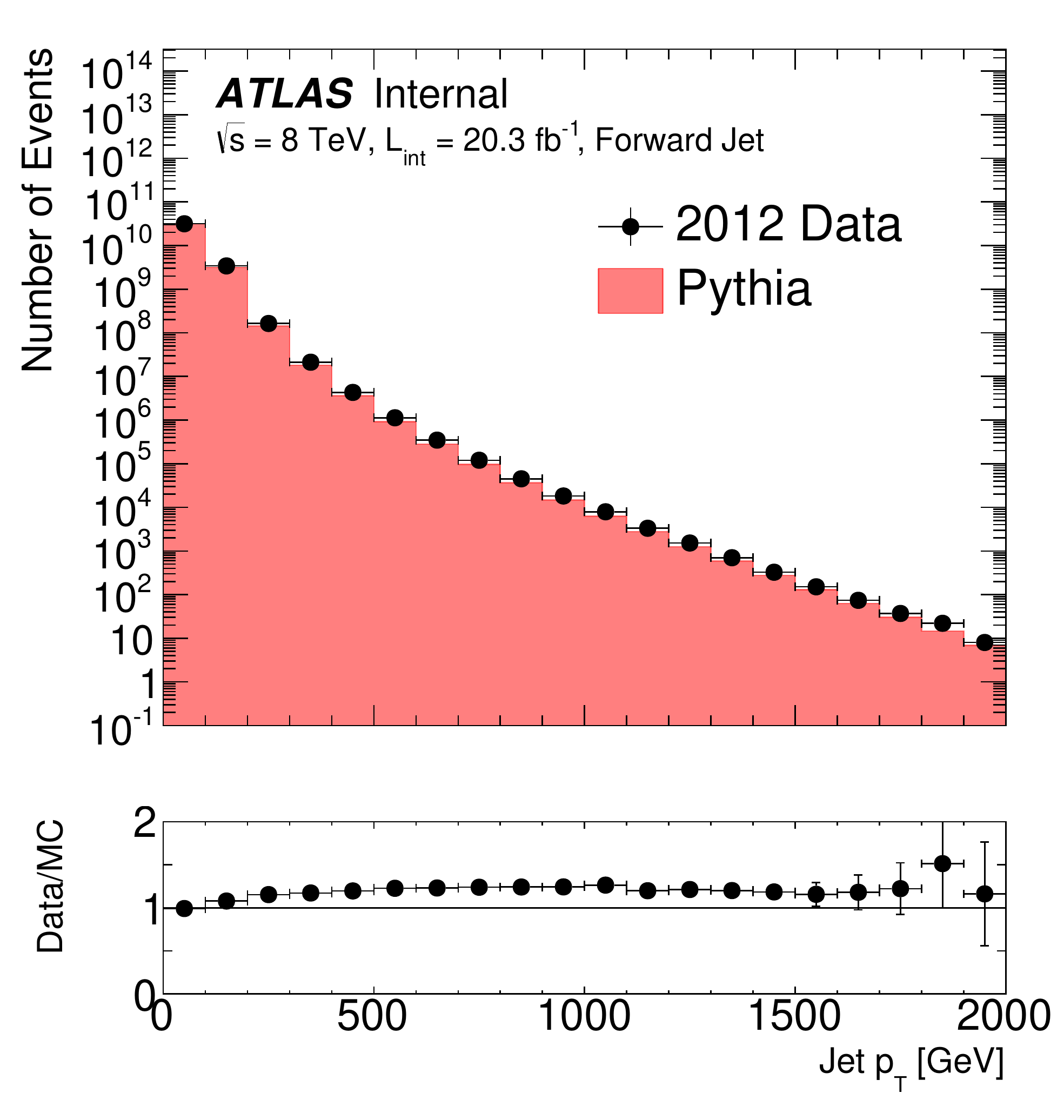}\includegraphics[width=0.45\textwidth]{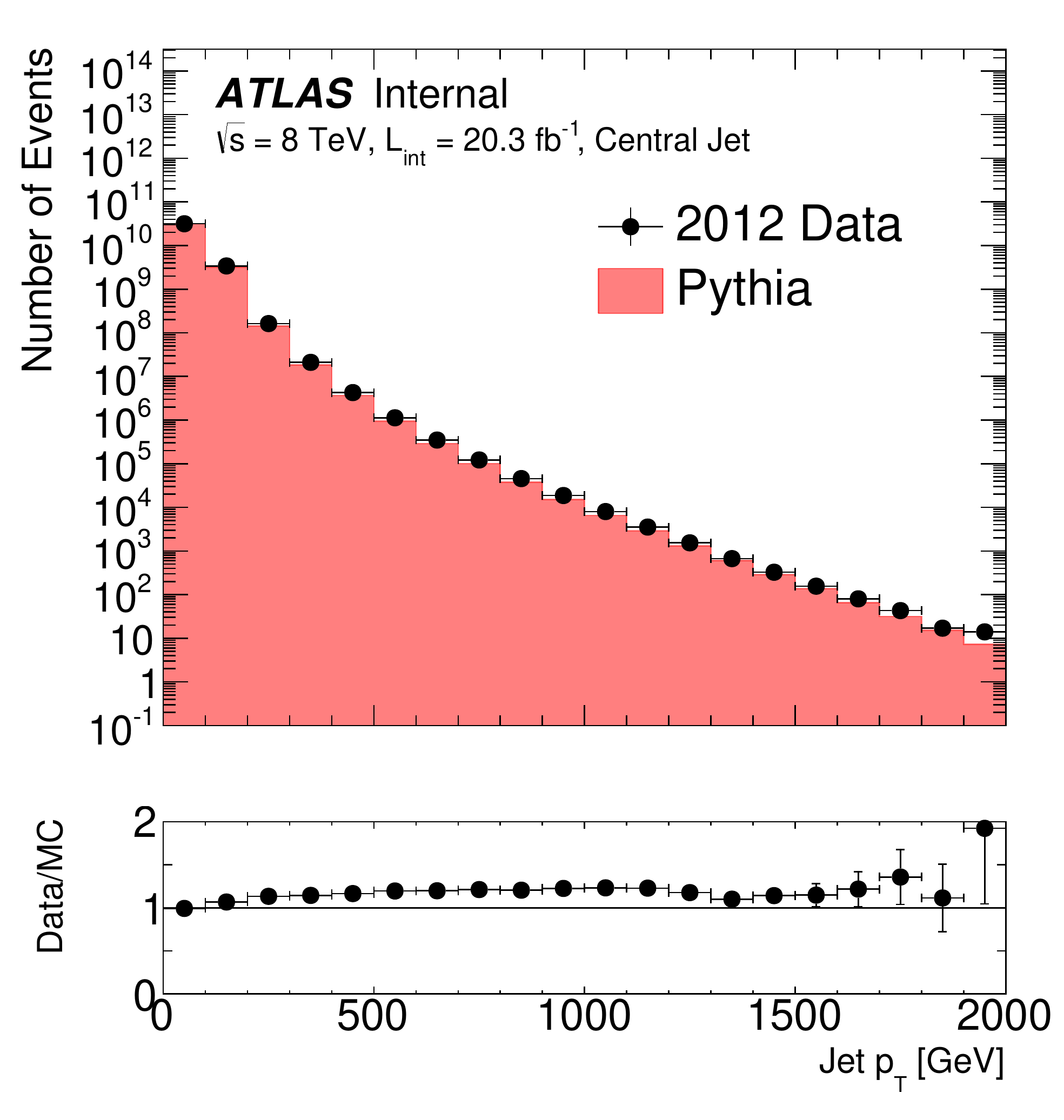}
\end{center}
\caption{Reconstructed jet $p_\text{T}$ spectrum for the more forward jet (left) and the more central jet (right).  Note that the pre-scales are applied to the data to arrive at a smooth and steeply falling distribution of the jet $p_\text{T}$.}
\label{fig:recopT}
\end{figure}

\begin{figure}[h!]
\begin{center}
\includegraphics[width=0.45\textwidth]{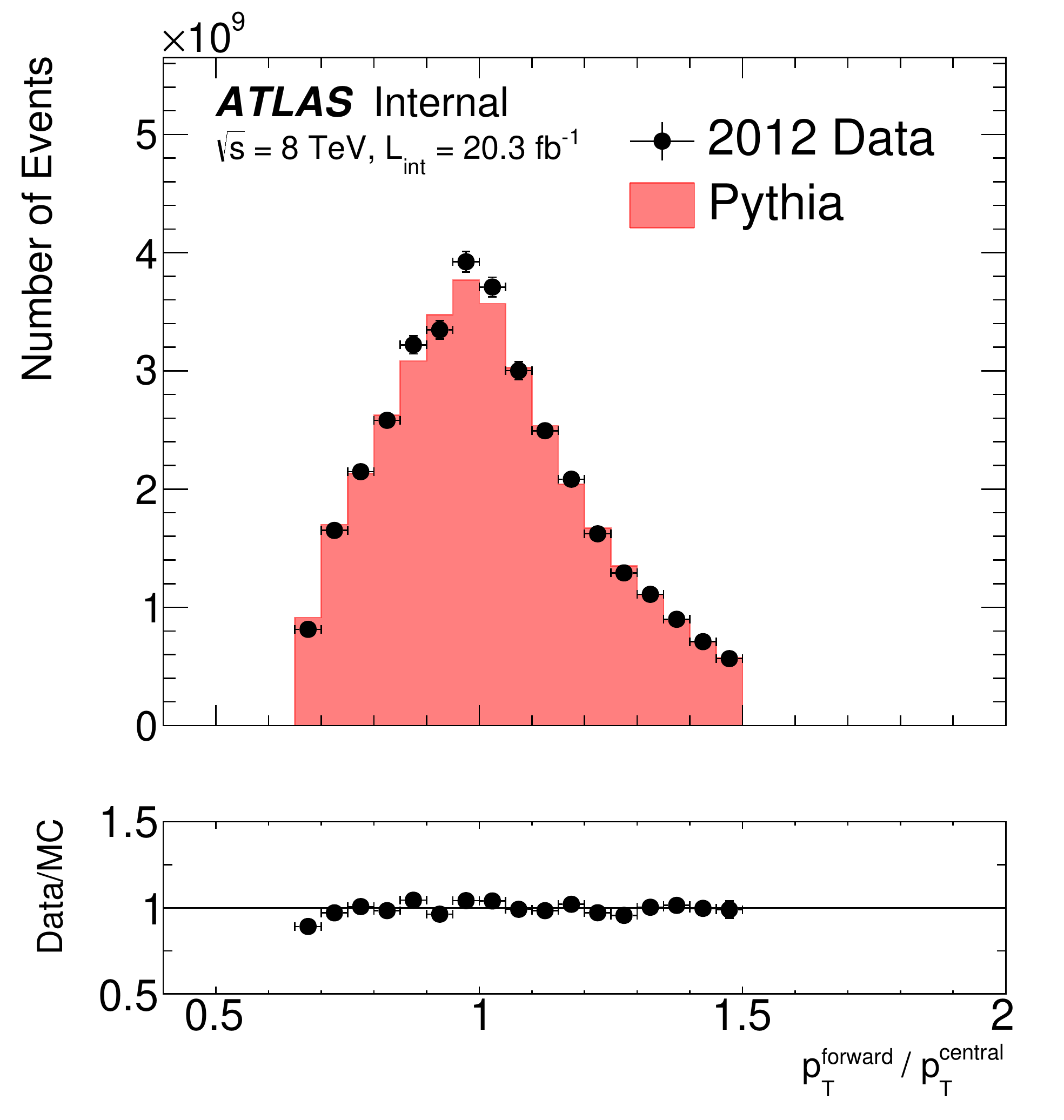}
\end{center}
\caption{The ratio of the more forward to the more central jet $p_\text{T}$ distributions.}
\label{fig:jetratio}
\end{figure}

\begin{figure}[h!]
\begin{center}
\includegraphics[width=0.45\textwidth]{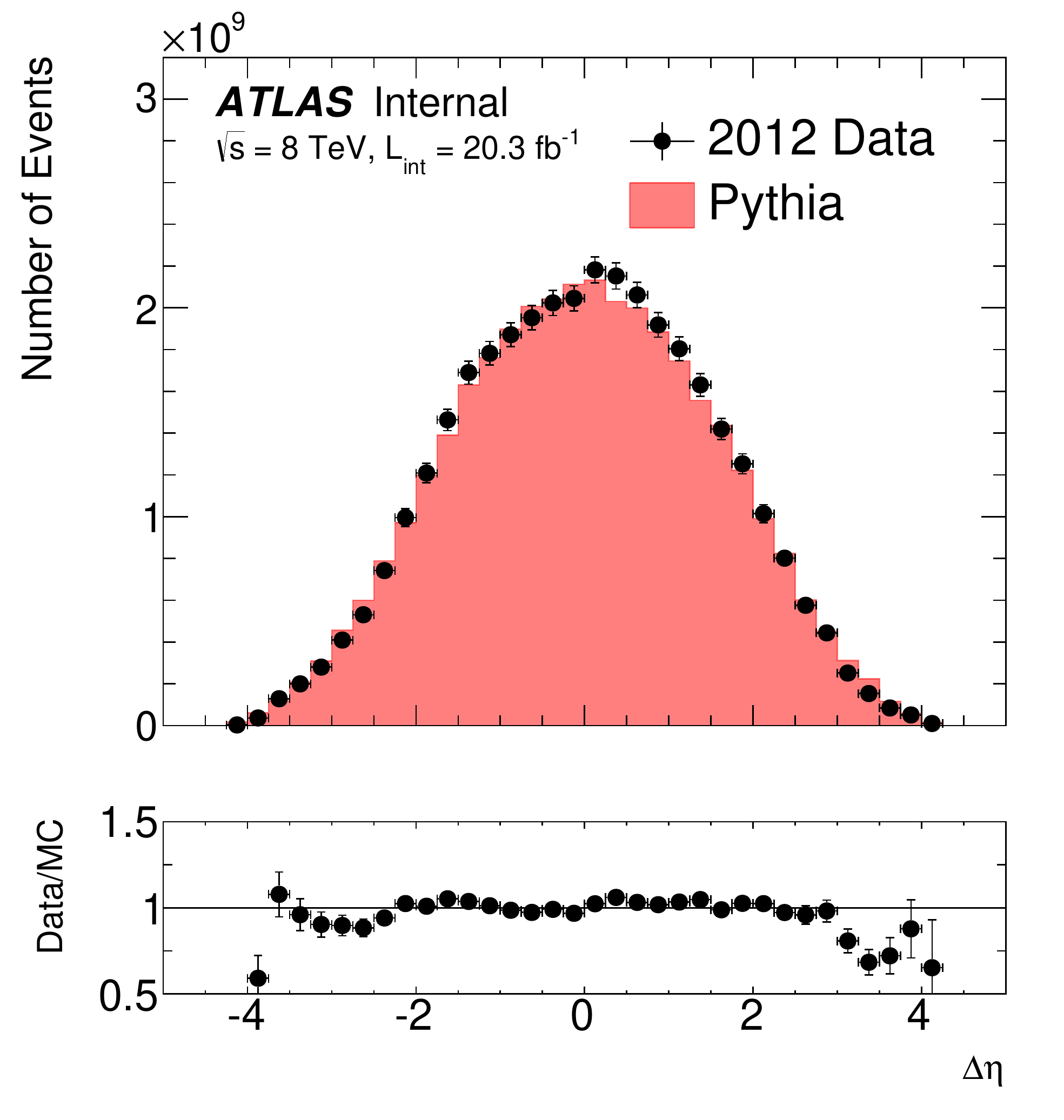}\includegraphics[width=0.45\textwidth]{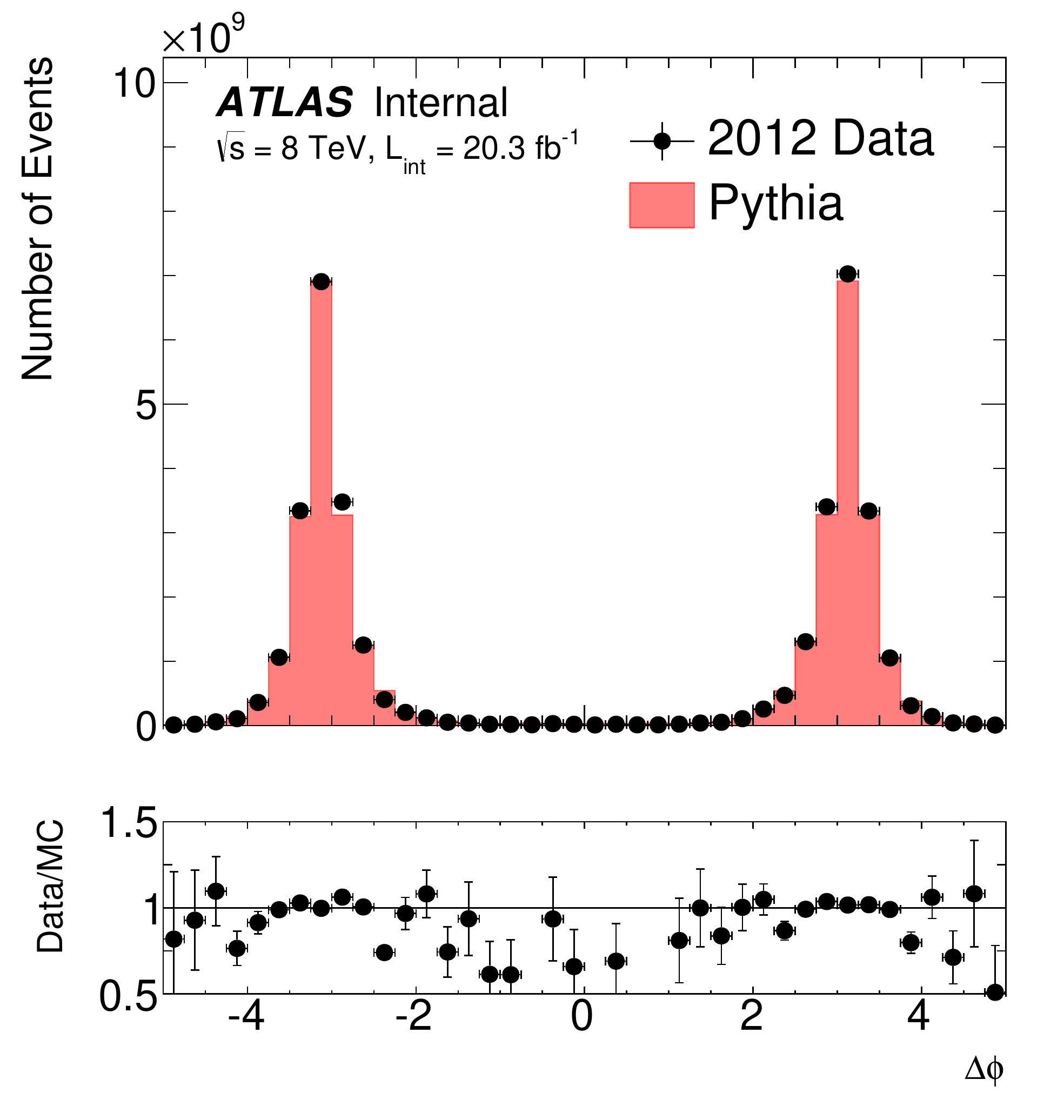}
\end{center}
\caption{The differences in $\eta$ (left) and $\phi$ (right) between the more forward and more central jet.}
\label{fig:jetratioeta}
\end{figure}

\begin{figure}[h!]
\begin{center}
\includegraphics[width=0.45\textwidth]{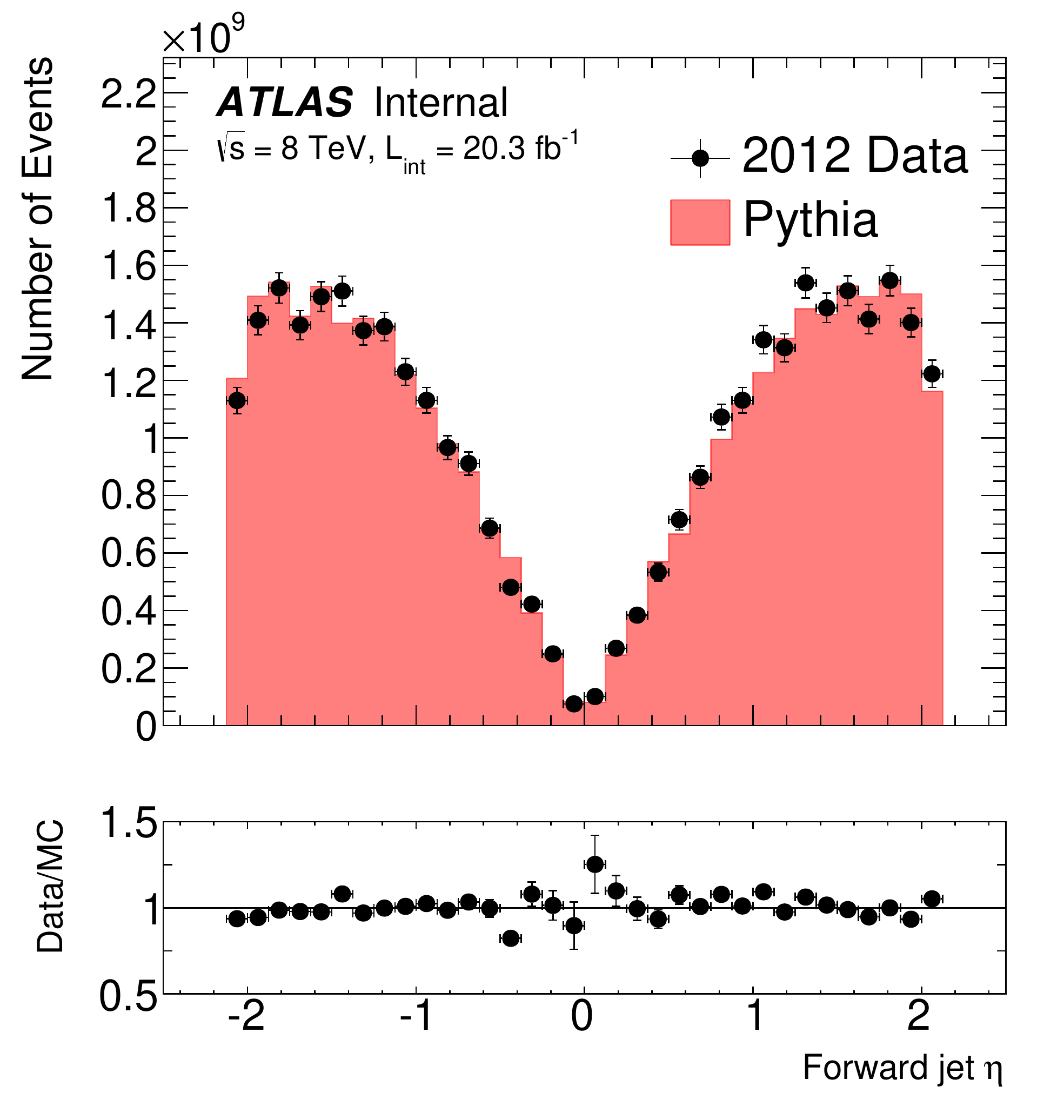}\includegraphics[width=0.45\textwidth]{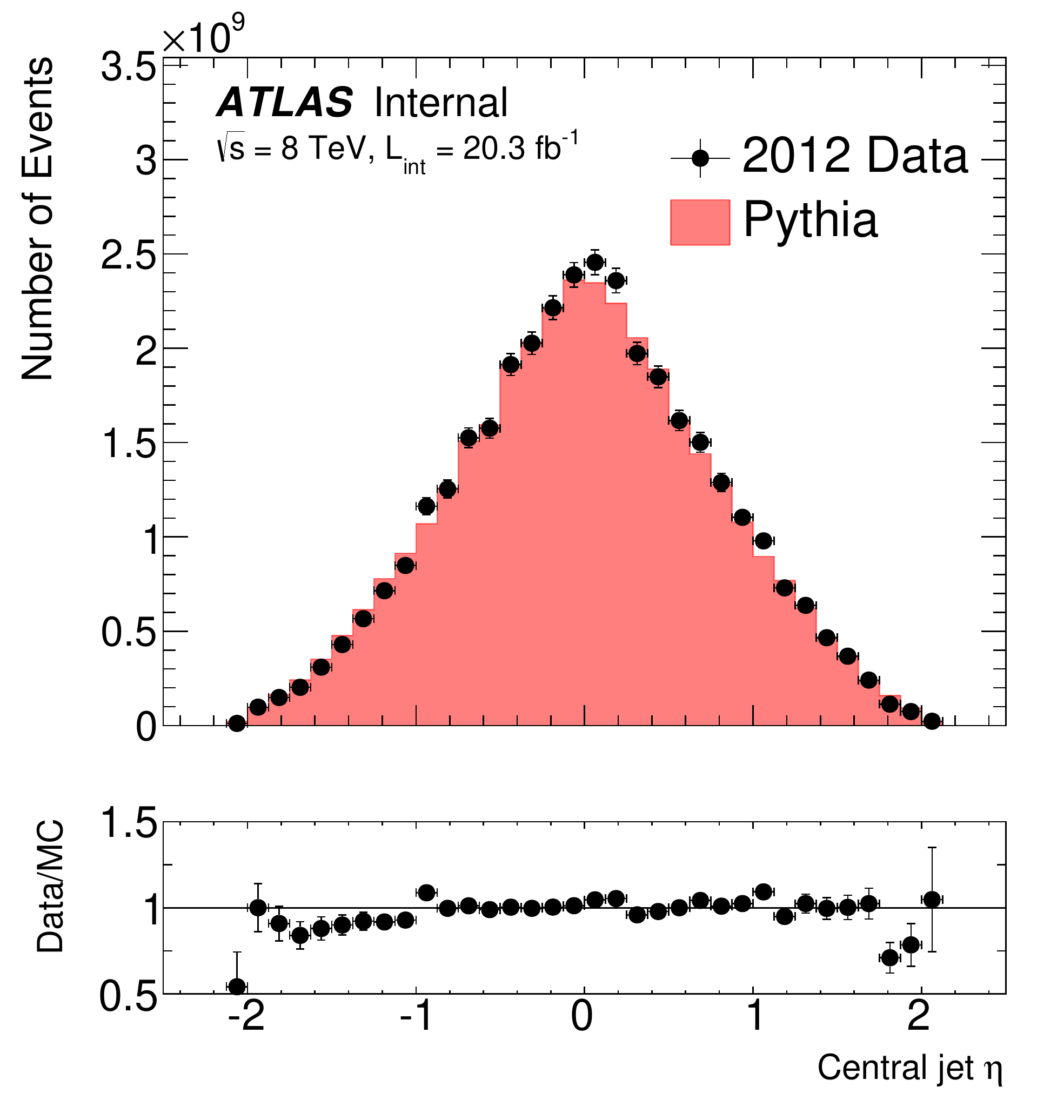}
\end{center}
\caption{The $\eta$ of the more forward (left) and more central (right) jet.}
\label{fig:jetratioeta2}
\end{figure}

\begin{figure}[h!]
\begin{center}
\includegraphics[width=0.45\textwidth]{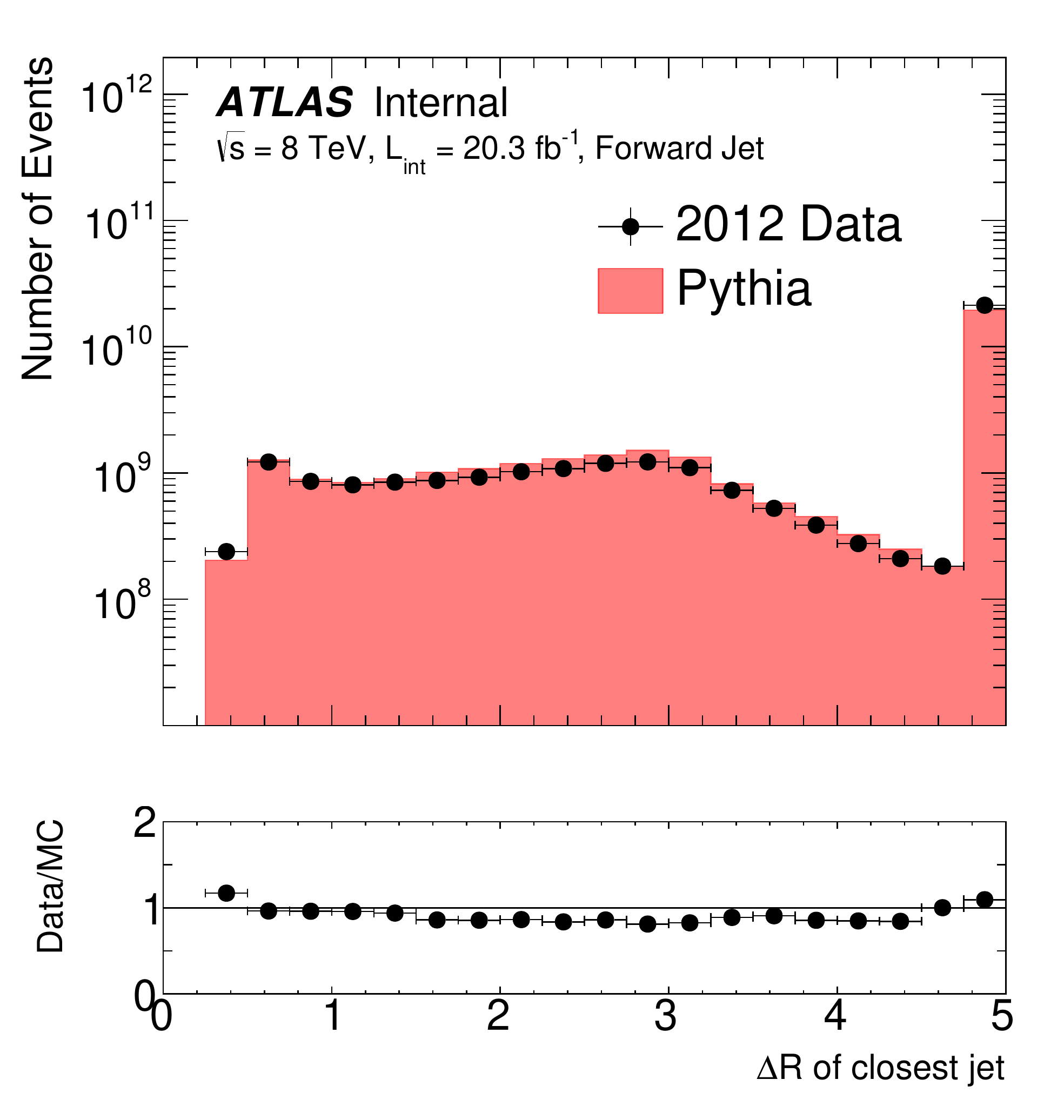}\includegraphics[width=0.45\textwidth]{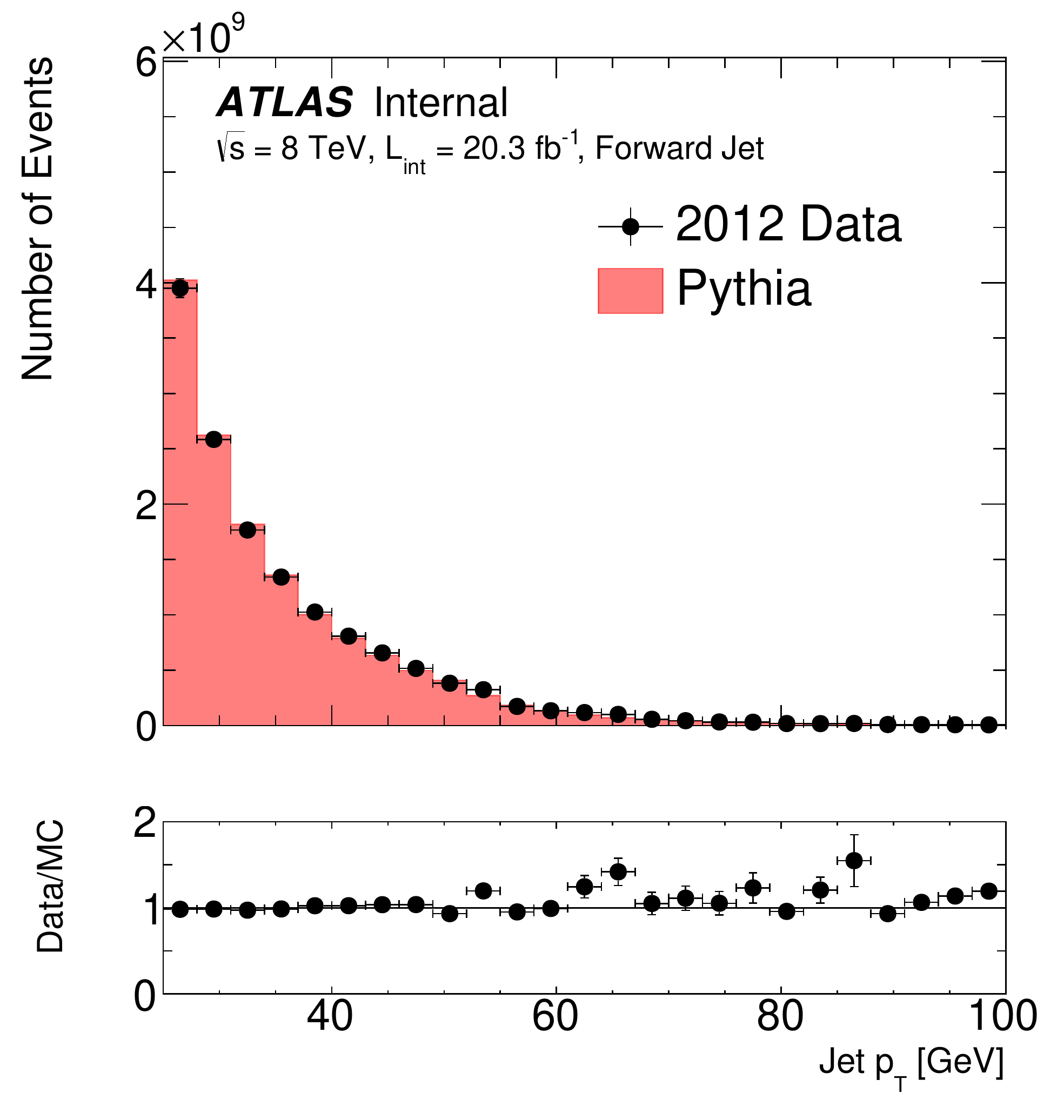}
\end{center}
\caption{The distance in $\Delta R$ to the nearest jet with $p_\text{T}>25$ GeV and the $p_\text{T}$ spectrum of this jet (right).}
\label{fig:deltaR}
\end{figure}

\begin{figure}[h!]
\begin{center}
\includegraphics[width=0.45\textwidth]{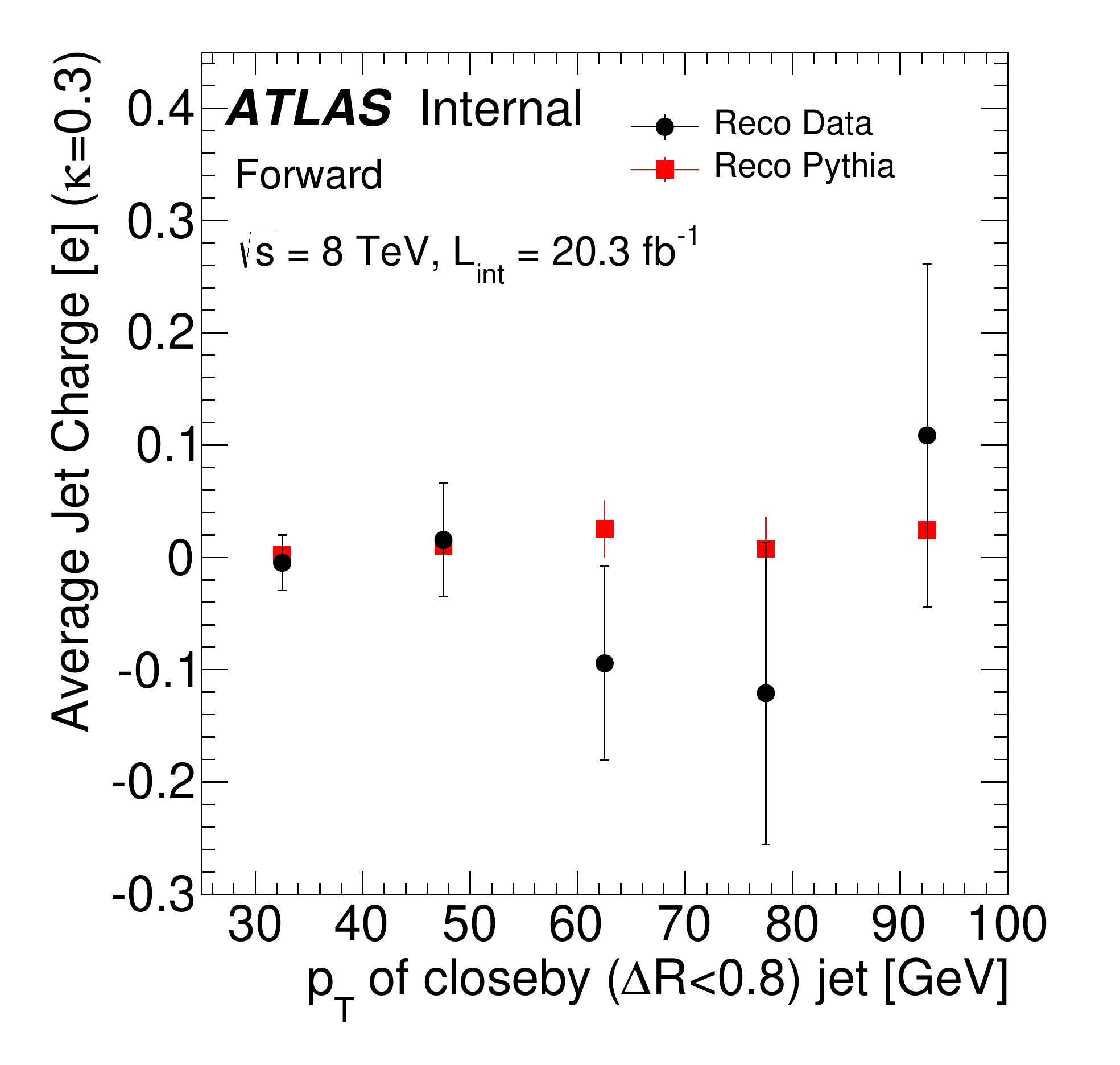}\includegraphics[width=0.45\textwidth]{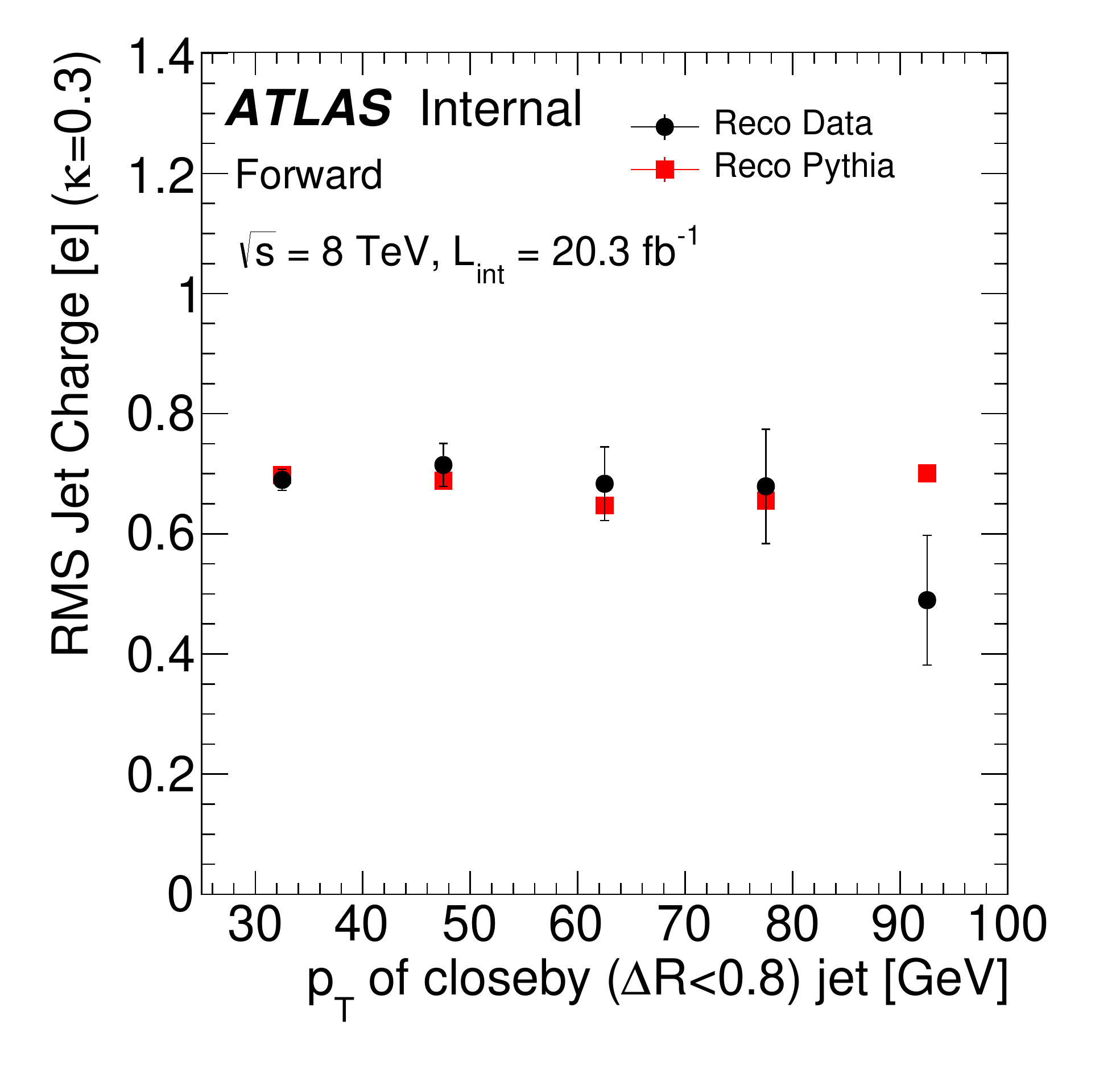}
\end{center}
\caption{The dependance of the average (left) and standard deviation (right) of the more forward jet charge distribution on the $p_\text{T}$ of the nearest jet above 25 GeV.  Events are only plotted if the $\Delta R $ to the nearest such jet is $<0.8$.  Uncertainties are statistical only. }
\label{fig:chargeiso1}
\end{figure}

This section contains various comparisons between the reconstructed MC and the data, using the event selection described in Sec.~\ref{sec:detectorlevel} that targets generic quark and gluon jets.  Figure~\ref{fig:recopT} shows the jet $p_\text{T}$ spectrum for the more forward and the more central of the two leading jets in dijet events.  Over nearly two orders of magnitude in jet $p_\text{T}$, the distribution of events drops by nearly ten orders of magnitude.   The overall shape is well-described by the leading order MC, though there is a small trend at low $p_\text{T}$ in the ratio between data and simulation.   Qualitatively, the left and right plots of Fig.~\ref{fig:recopT} are similar - this is quantified by the ratio between the more forward and more central jets in Fig.~\ref{fig:jetratio}.  As expected, the distribution is peaked at one and is nearly symmetric about the peak (cutoff at $0.5$ and $1.5$ due to the $p_\text{T}$ symmetry requirement).  The difference in $\eta$ and $\phi$ between the more forward and more central jet are shown in Fig.~\ref{fig:jetratioeta}.   The two jets are nearly back-to-back in the transverse plane and are on average close in $\eta$.  Figure~\ref{fig:jetratioeta2} shows the $\eta$ distribution separately for the more forward and the more central jets.  Even though the average $\Delta\eta$ is zero, most of the more central jets are within $|\eta|<1$ and most of the more forward jets have $|\eta|>1$.   There is no explicit isolation requirement, but the requirement for nearly $p_\text{T}$ balanced jets indirectly leads to the two leading jets to be relatively isolated.  The left plot of Fig.~\ref{fig:deltaR} shows the distance in $\Delta R$ to the nearest jet with $p_\text{T} > 25$ GeV, excluding the other selected jet.  A significant fraction of events have only the two selected jets with $p_\text{T}>25$ GeV, which accounts for the spike in the overflow bin.  The $p_\text{T}$ of the closest jet is shown in the right plot of Fig.~\ref{fig:deltaR}, excluding the other selected jet.  For close-by jets, the $p_\text{T}$ spectrum is steeply falling away from 25 GeV.   There is no significant evidence for an impact of these close-by jets on the jet charge distribution.  This is demonstrated by Fig.~\ref{fig:chargeiso1}, which shows the average jet charge and the standard deviation of the jet charge as a function of the $p_\text{T}$ of the close-by jet.  Within the statistical uncertainties of the data and simulation, the jet charge distribution is independent of the $p_\text{T}$ of the close-by jet.  

\begin{figure}[h!]
 \centering
 \includegraphics[width=0.45\columnwidth]{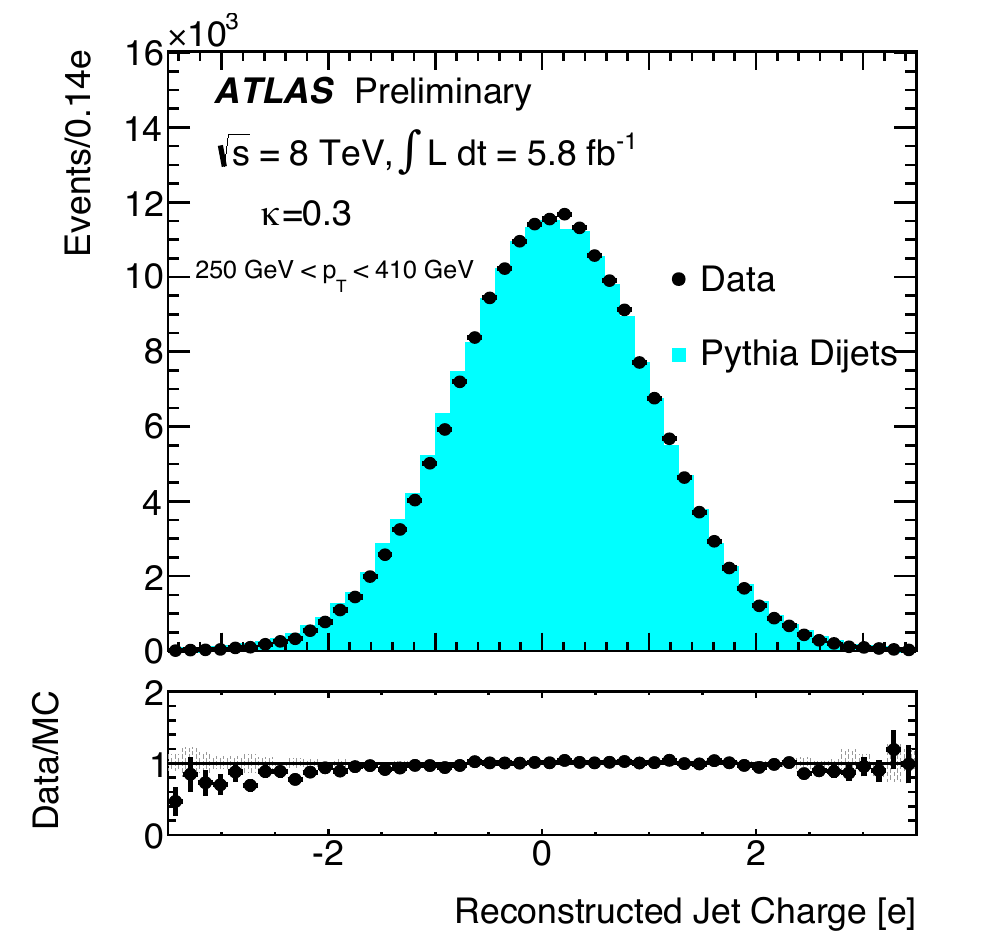}\includegraphics[width=0.45\columnwidth]{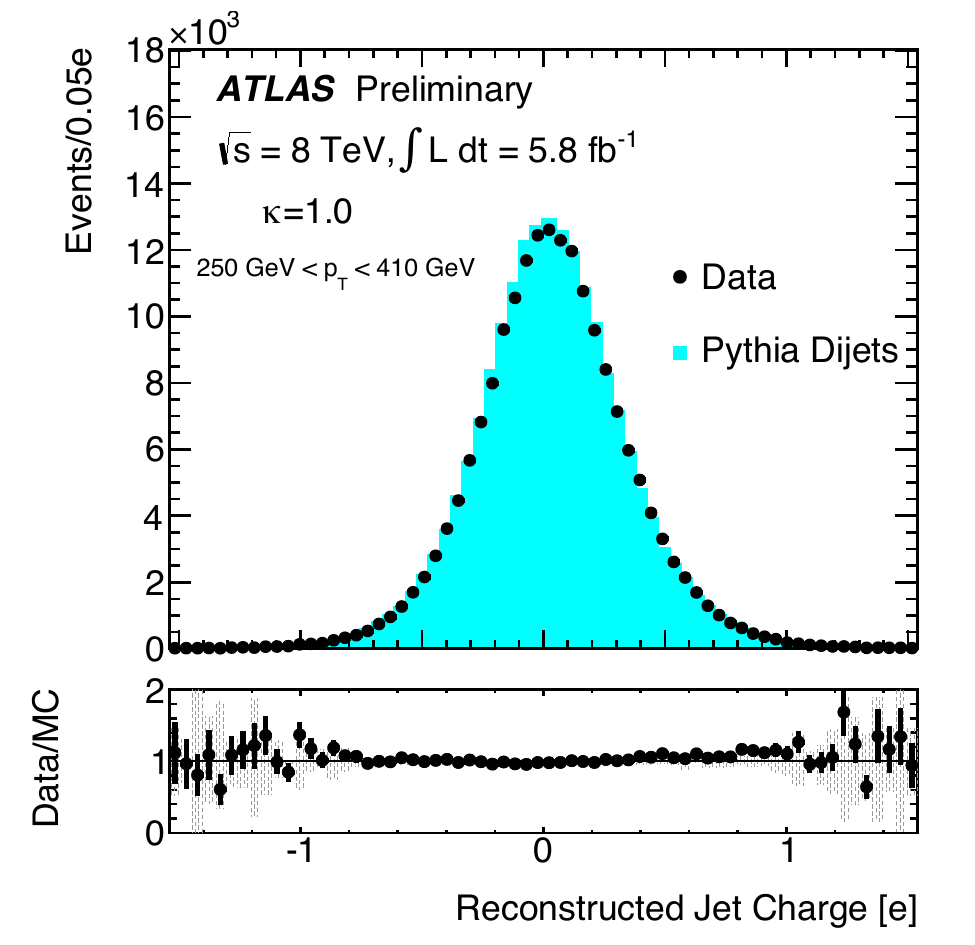} \\
\includegraphics[width=0.45\columnwidth]{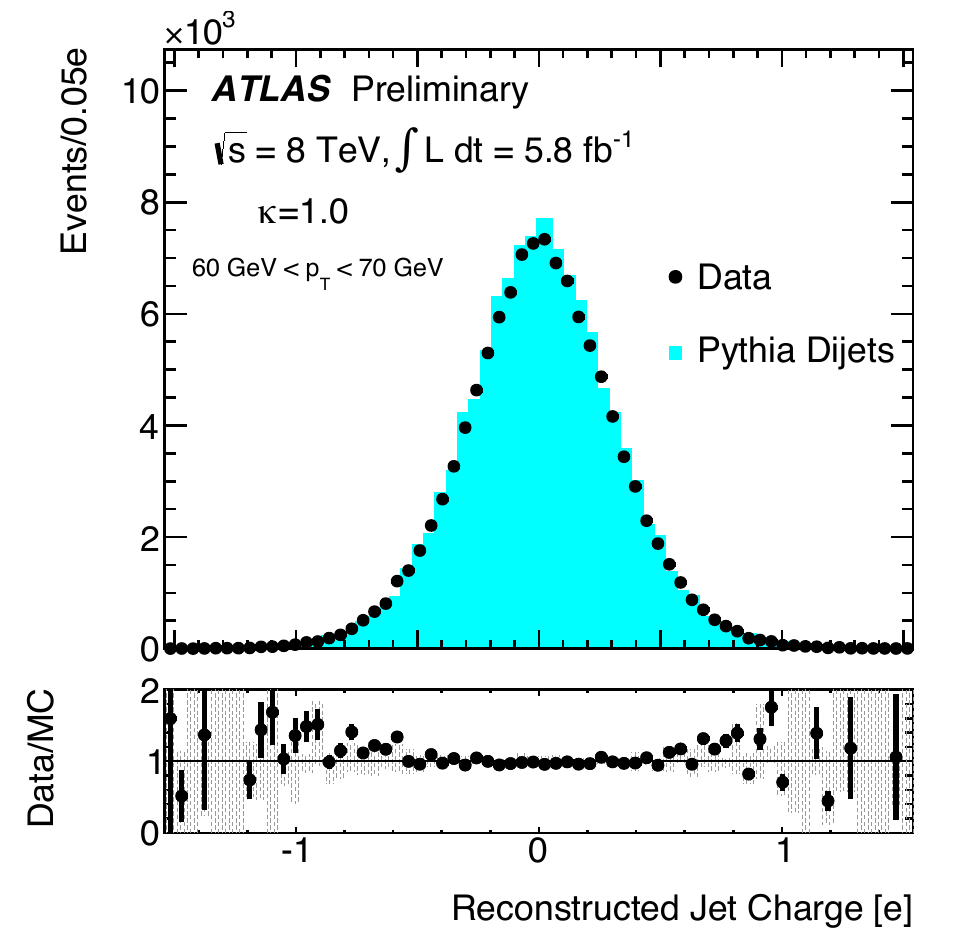}\includegraphics[width=0.45\columnwidth]{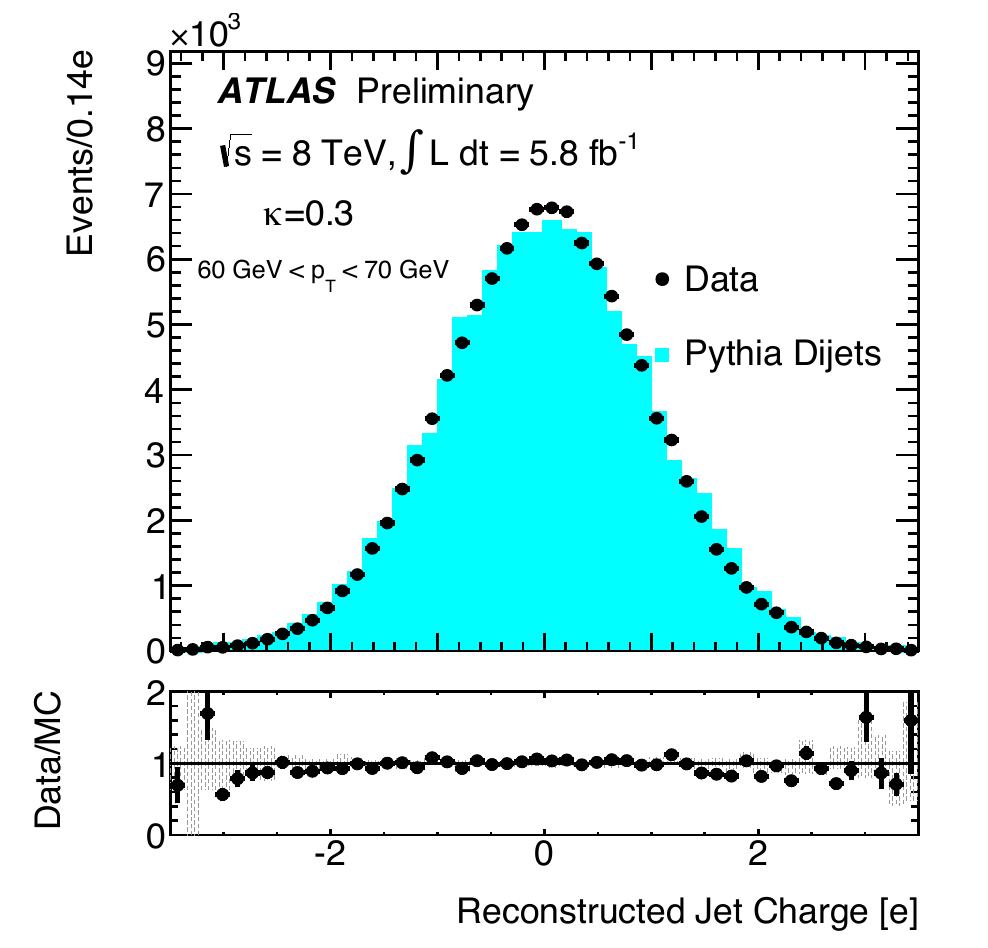} 
 \caption{The top (bottom) row shows the sum of the charges of the two leading jets in dijet events for two bins of the leading jet $p_\text{T}$.
Two values of the $p_{T}$ weighting factor are shown: $\kappa=1.0$ on the left and $\kappa=0.3$ on the right.  The lower panels show the ratios between data and MC.  The gray band in the ratio includes jet $p_\text{T}$ and track reconstruction efficiency uncertainties, described in Sec.~\ref{sec:uncerts}.}
 \label{fig:datamc}
\end{figure}

\begin{figure}
 \centering
 \includegraphics[width=.7\columnwidth]{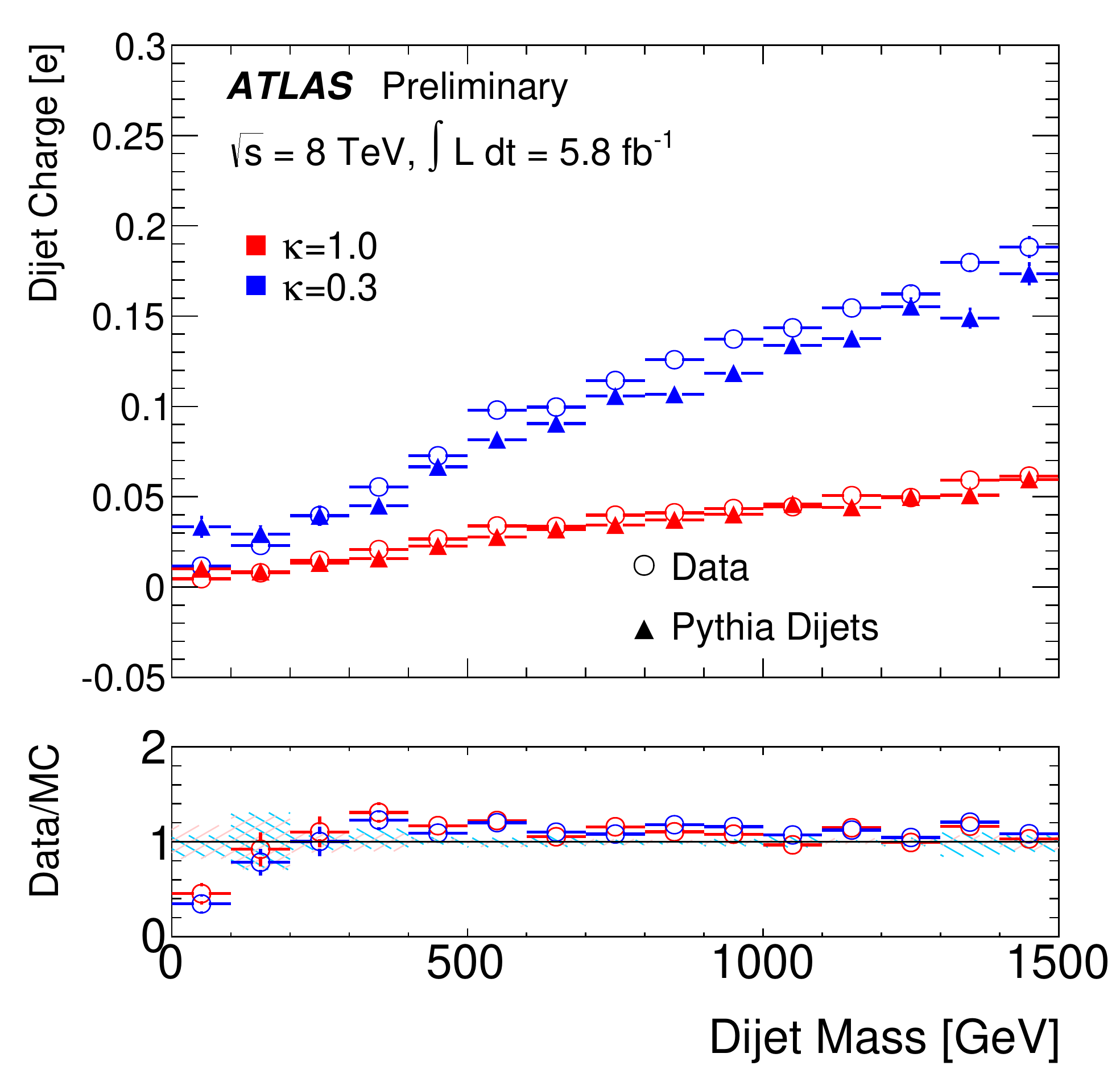}
 \caption{The dependence of the dijet charge on the dijet mass for two different values of $\kappa$ in data and MC for a dijet sample.  The uncertainty band includes preliminary jet $p_\text{T}$ and track isolation uncertainties.}
 \label{fig:dijetMassq}
\end{figure}

The actual jet charge distribution is shown in Fig.~\ref{fig:datamc} for low and high jet $p_\text{T}$ and for $\kappa=0.3$ and $\kappa=0.7$.  Across jet $p_\text{T}$, the jet charge distribution is roughly symmetric around zero, with a small shift to positive values in the higher $p_\text{T}$ bin.  The distribution is wider for $\kappa=0.3$ than for $\kappa=0.7$.  To see this, note that $\partial_\kappa x^\kappa = x^\kappa \text{log}(x)<0$ for $0<x<1$.  Each term in the defining sum for the jet charge has the form $x=p_\text{T,track}/p_\text{T,jet}$ and therefore by decreasing $\kappa$, the absolute value of each contribution to the sum increases.   This effects the mean in addition to the width of the jet charge distribution, as seen by the jet $p_\text{T}$-dependence of the dijet charge in Fig.~\ref{fig:dijetMassq}.  As expected from Sec.~\ref{sec:jetchargetheory}, the average jet charge increases with the energy scale.  There are also qualitative systematic differences between data and simulation.  These observations are revisited in more detail in Sec.~\ref{sec:unfolding}.

\clearpage

\subsection{Modeling and Tagging Performance with QCD Jets}
\label{sec:JetCharge:CONF}

\subsubsection{Single Jet Charge}
\label{sec:jetcharge:qcd}

\begin{figure}
 \centering
 \includegraphics[width=0.45\columnwidth]{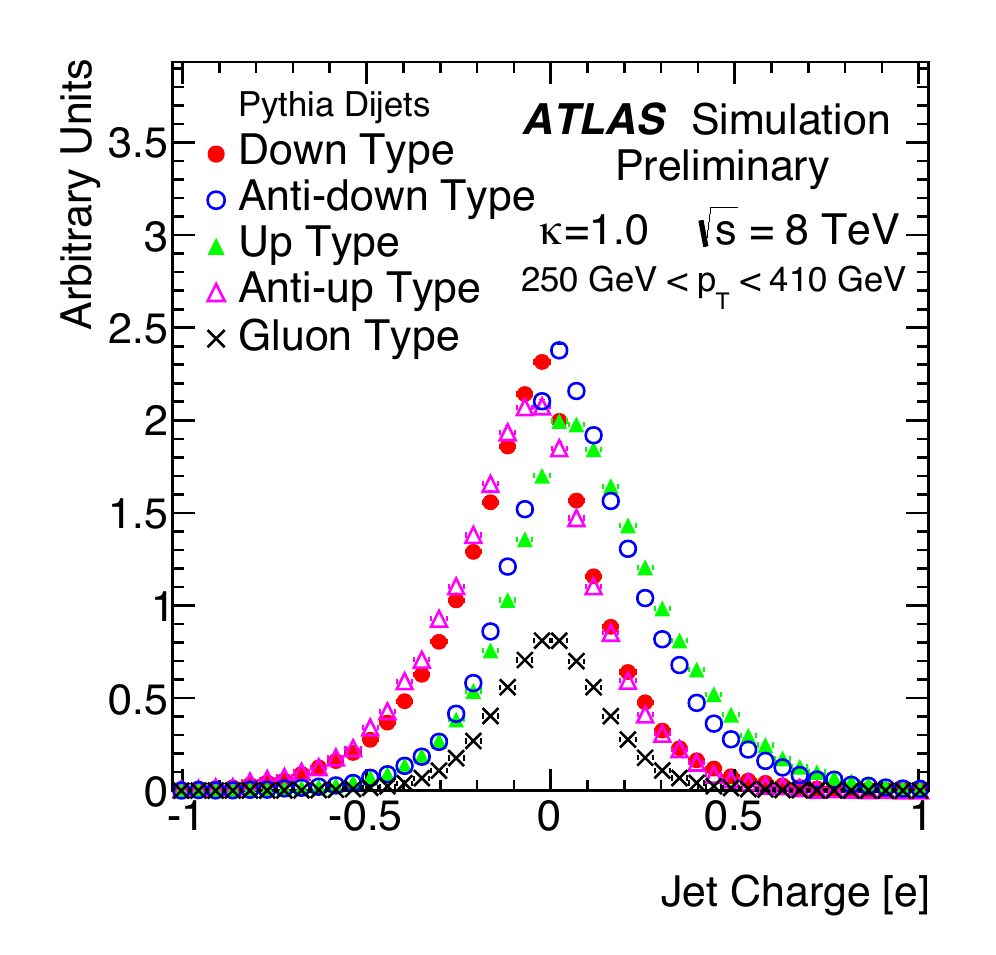} \includegraphics[width=0.45\columnwidth]{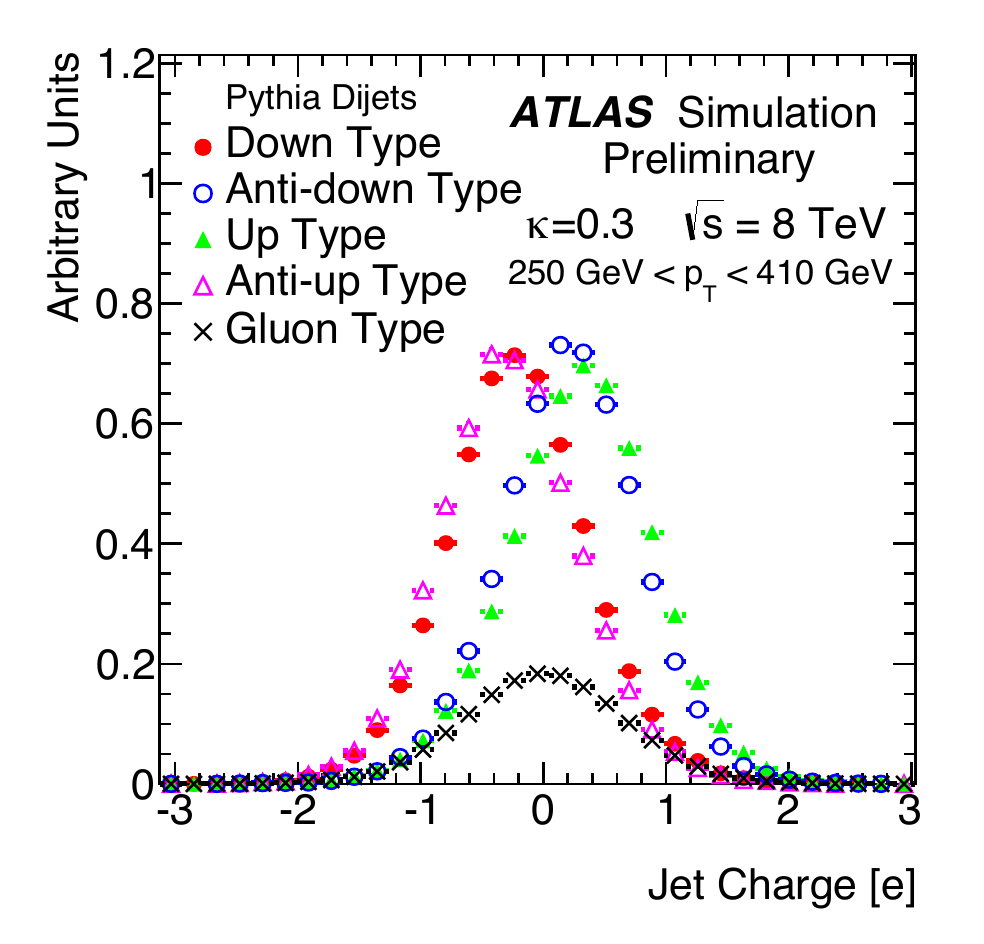}\\
   \includegraphics[width=0.45\columnwidth]{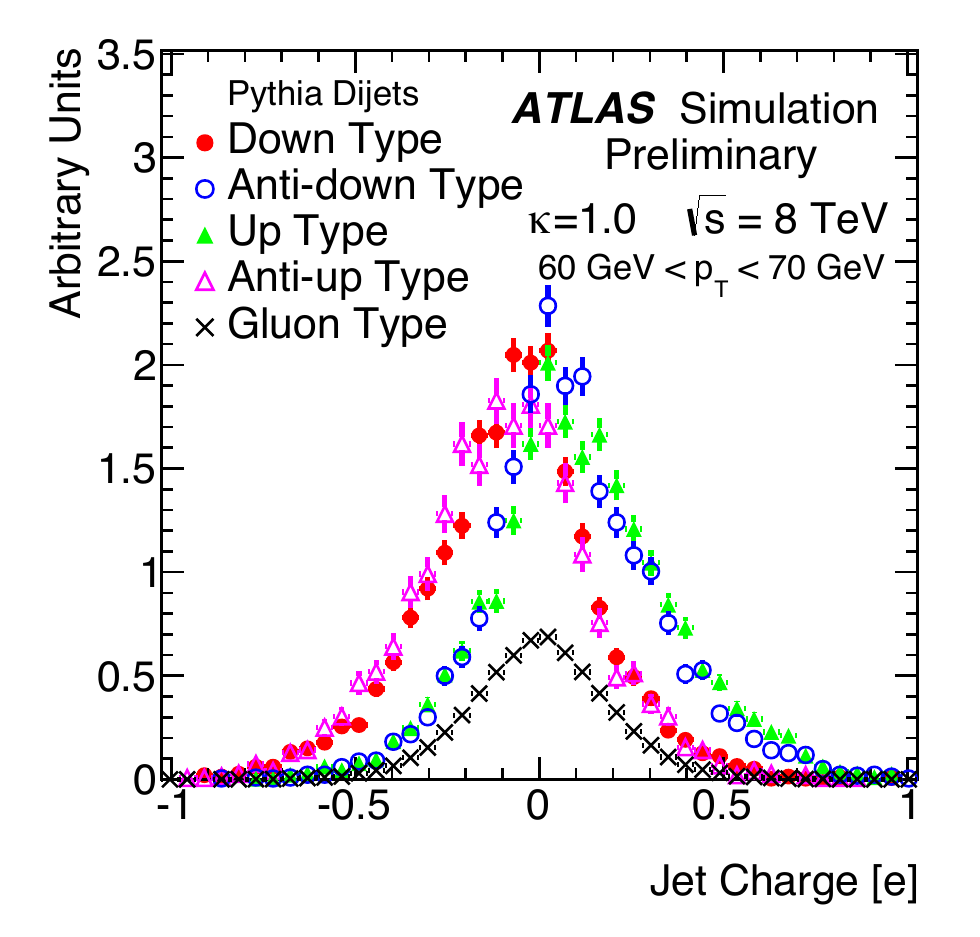}  \includegraphics[width=0.45\columnwidth]{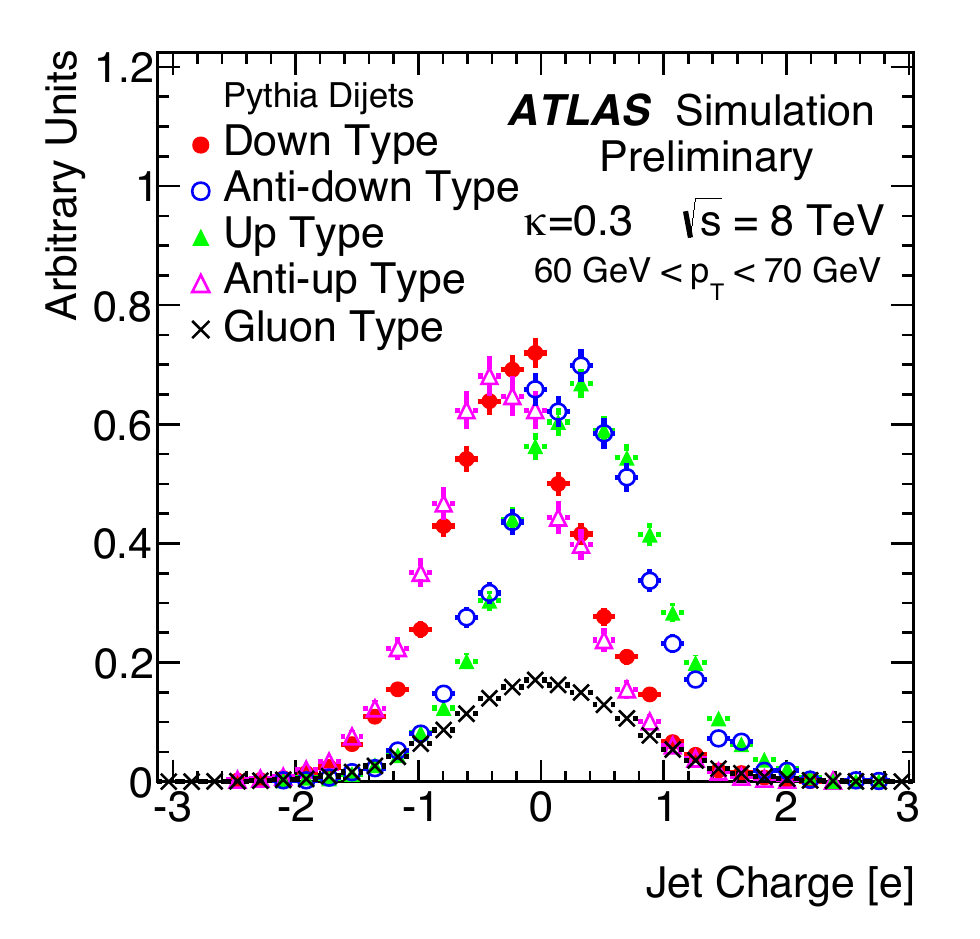}
 \caption{Truth-level jet charge distribution for simulated dijet events for different parton flavours. Each distribution is normalised
to unit area, except the gluon probability distribution function which is normalised to $0.3$ for easier comparison. Distributions are shown in two different $p_\text{T}$ bins for the leading jet
and for two values of the weighting factor $\kappa$. }
 \label{fig:parton}
\end{figure}

\begin{figure}
 \centering
 \includegraphics[width=.5\columnwidth]{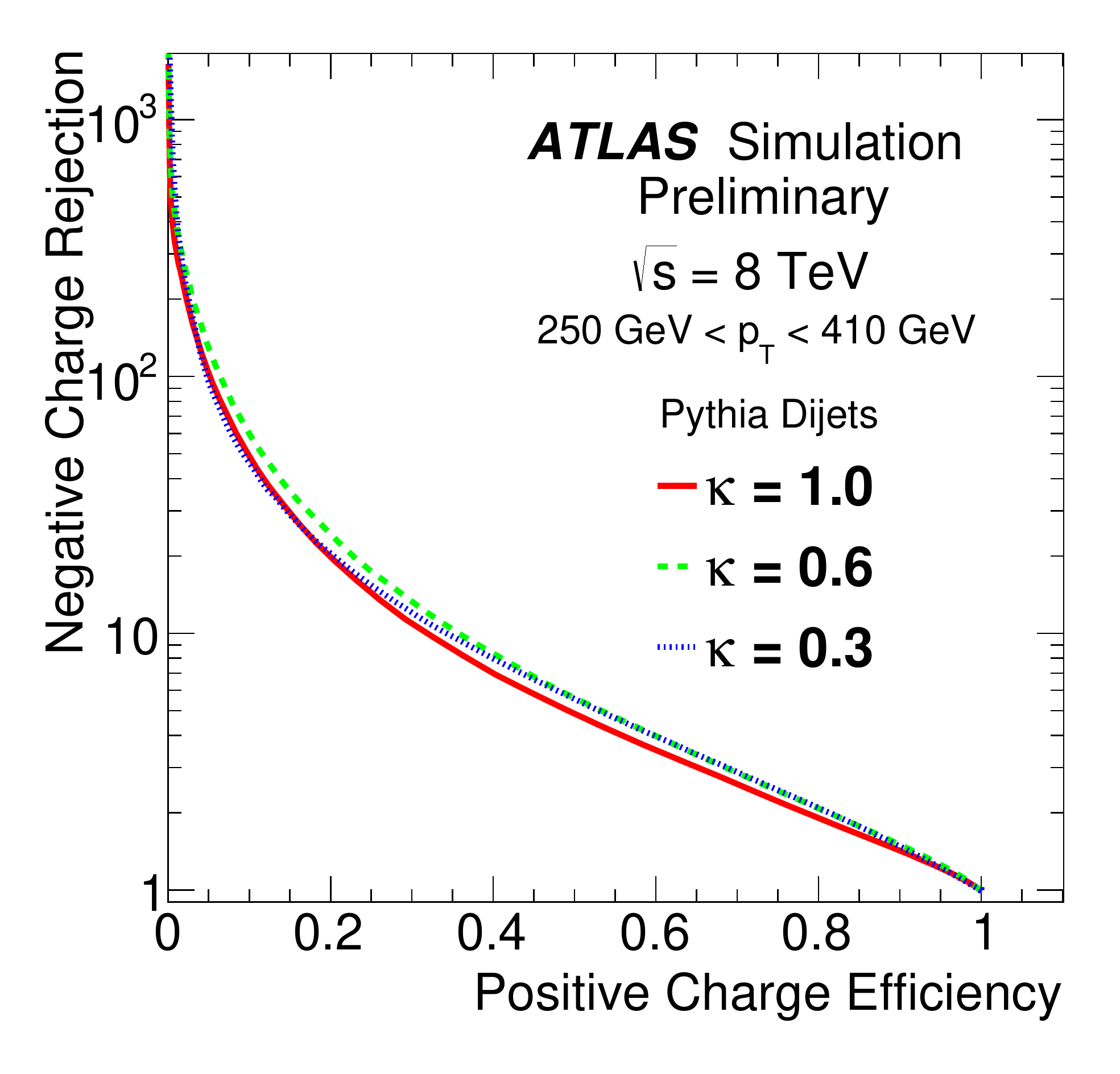}\includegraphics[width=.5\columnwidth]{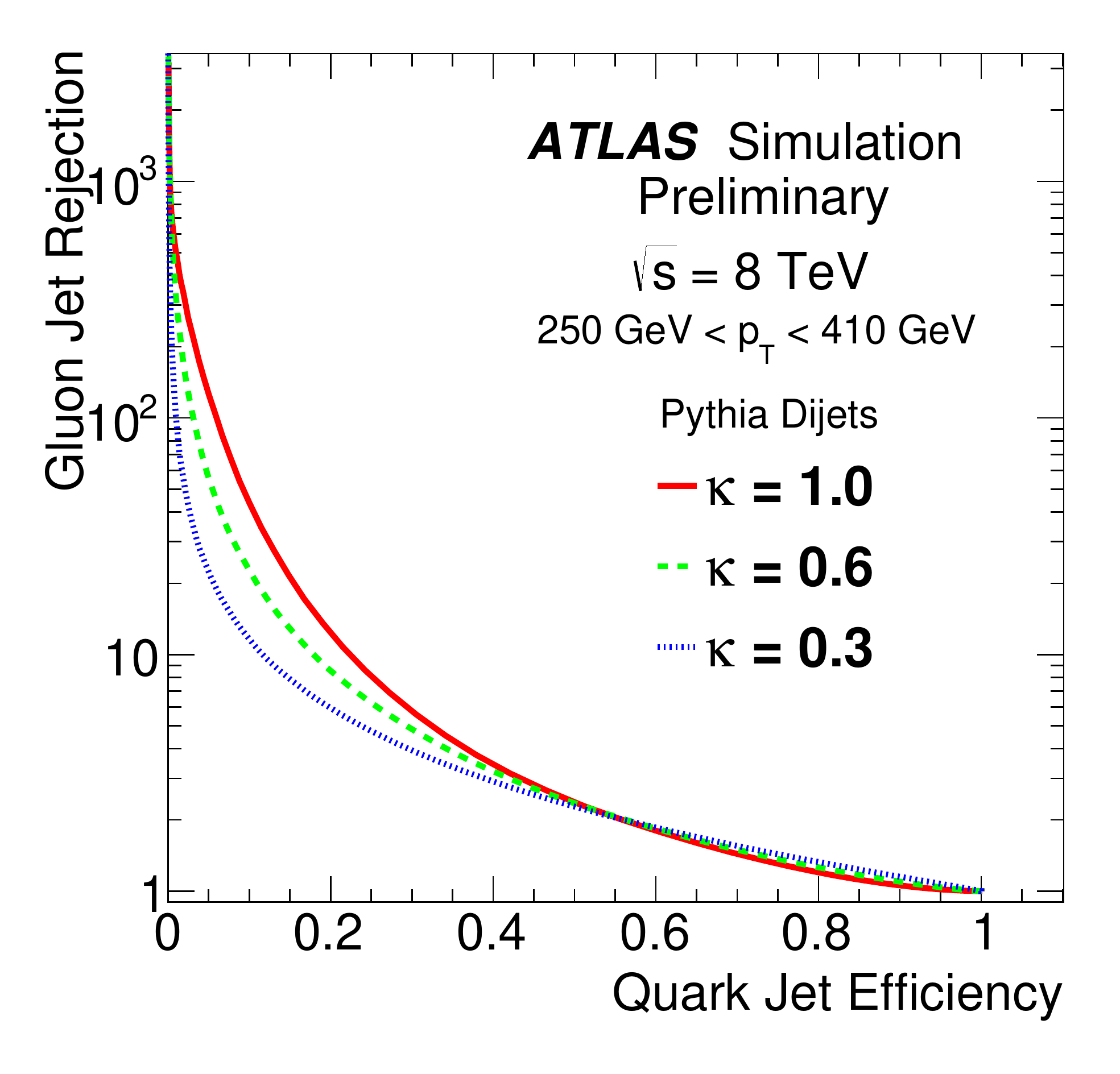}\\
 \caption{Discrimination power of the jet charge to differentiate quark jets of opposite charge (left) and quark from gluon jets (right). Results are obtained from MC truth
information in simulated dijet samples.}
 \label{fig:partons}
\end{figure}

Using the parton-matching scheme described earlier, the truth charge distributions are separated according to
the jet flavour in Fig.~\ref{fig:parton}.  Figure~\ref{fig:parton} is similar to Fig.~\ref{fig:sortedbypartons}, but for multiple $p_\text{T}$ bins and multiple $\kappa$ values.  As observed with the earlier figure, in the simulation there is significant discrimination between the charge sign of the
quark at the generator level, but not between flavour types with the same charge. The rejection of jets with a flavour corresponding to a
negative charged parton as a function of the efficiency for jets with a
flavour corresponding to a positively charged parton is shown in the left plot of Fig.~\ref{fig:partons}.  For an positive quark jet efficiency of about 50\%, there is a rejection($=1/\text{negative quark jet efficiency}$) of about $6$, independent of $\kappa$.  The
discrimination between quark and gluon jets is quantified in the right
plot of Fig.~\ref{fig:partons}.  A rejection of about $3$ against gluon jets is expected for an efficiency of 50\% for quark jets.  While not competitive with dedicated quark/gluon taggers~\cite{qg} on its own, the jet charge could
be used as an additional discriminating variable within a multivariate
approach.

\begin{figure}[hbp]
 \centering
 \includegraphics[width=.45\columnwidth]{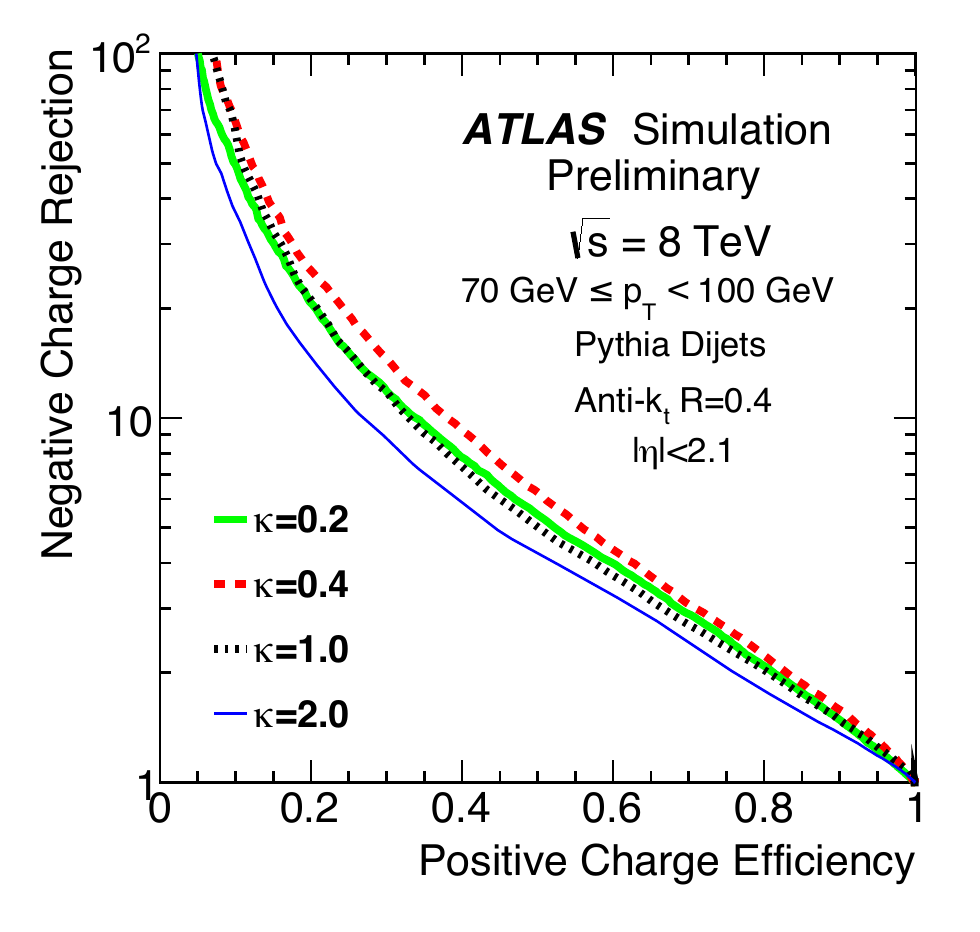}  \includegraphics[width=.45\columnwidth]{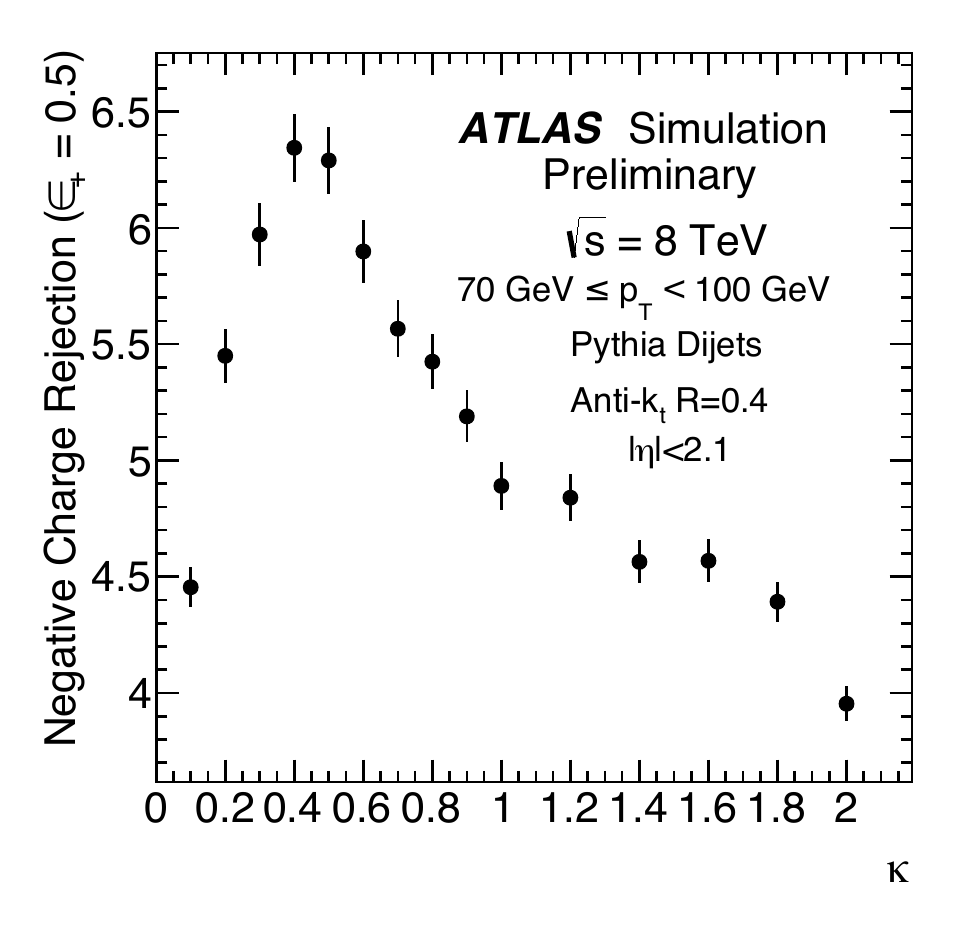} 
 \caption{The parton flavour tagging performance of the jet charge, defined as in Fig.~\ref{fig:partons}.  The left plot shows the positive parton charge jet efficiency versus the negative parton charge rejection (inverse efficiency) for various $\kappa$ values in a fixed $p_\text{T}$ bin.  The right plot fixes the positive charge efficiency at 0.5 and then shows the distribution of the negative charge rejection with $\kappa$. }
 \label{fig:new2}
 \end{figure}

\begin{figure}[hbp]
 \centering
 \includegraphics[width=.45\columnwidth]{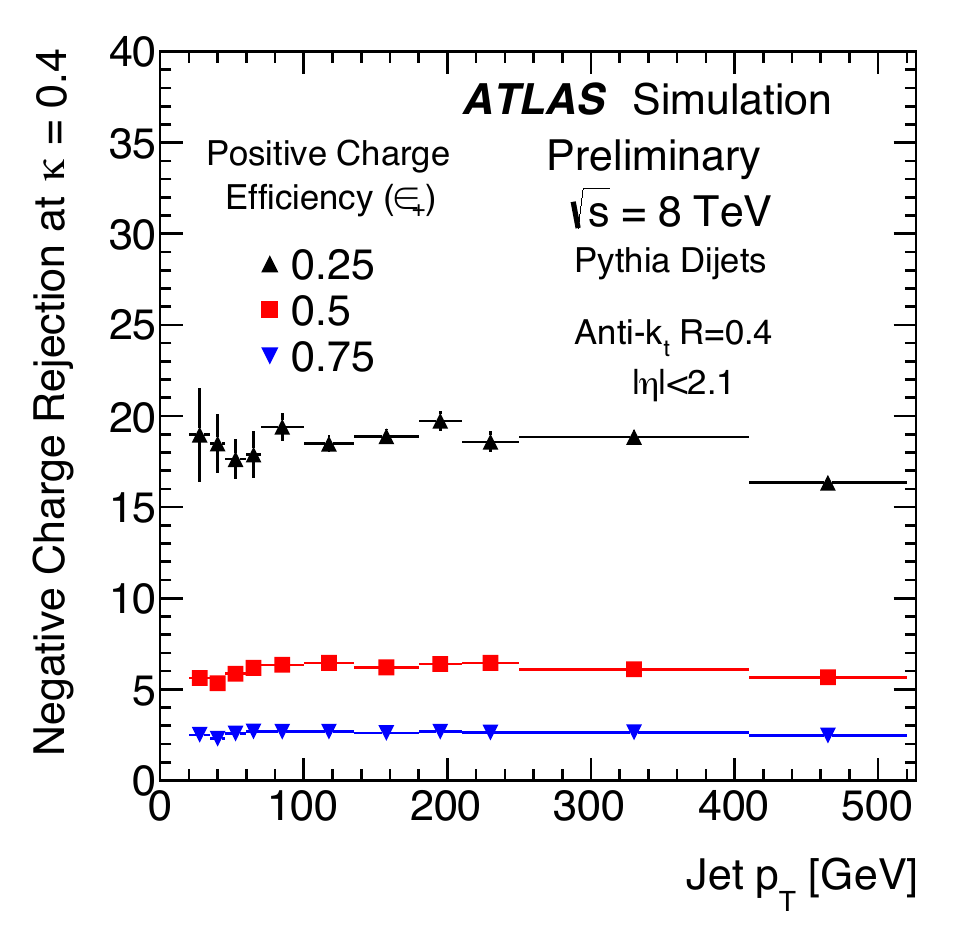}  \includegraphics[width=.45\columnwidth]{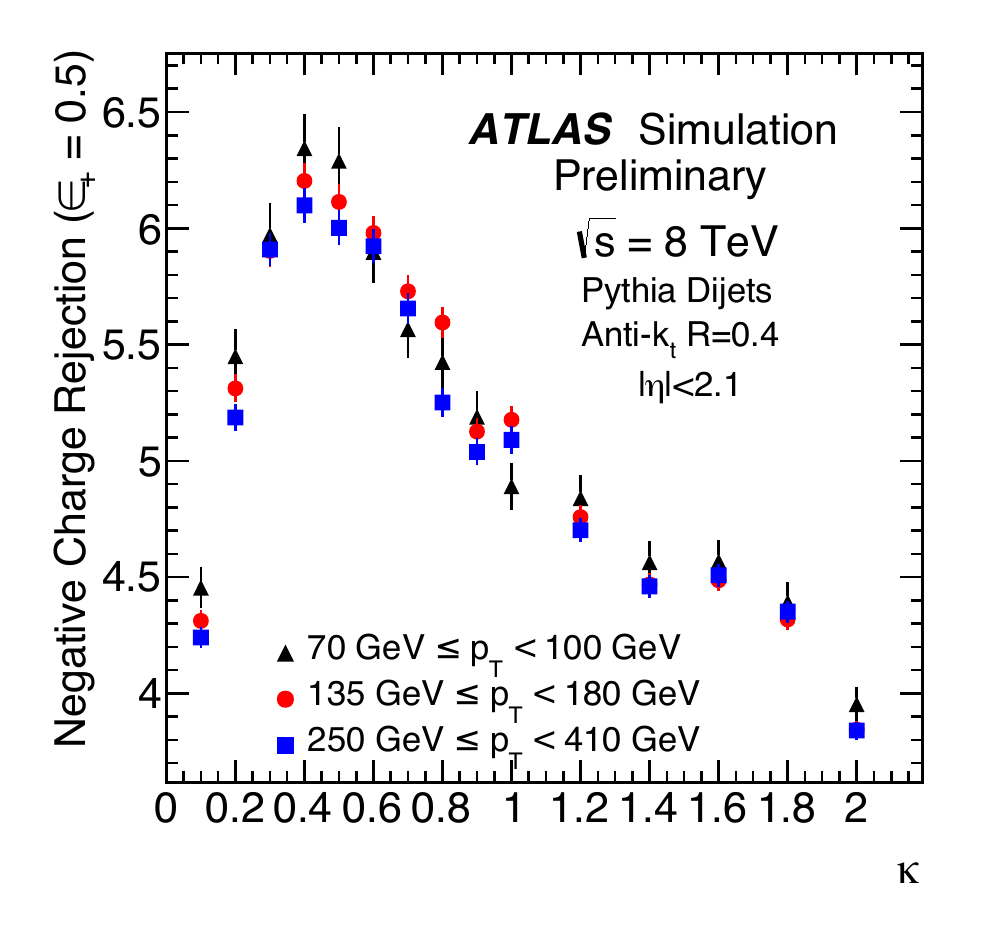} 
 \caption{A summary of the information shown in Fig.~\ref{fig:new2} for many $p_\text{T}$ bins.  The horizontal axis is the jet $p_\text{T}$ and the vertical axis is the maximum negative charge rejection for a fixed positive charge efficiency ($\epsilon_+$) of $0.25, 0.5$ or $0.75$.  For a fixed positive charge efficiency, the optimal $\kappa$ value and the maximum negative charge rejection vary little.  The $p_\text{T}$ bins are chosen based on trigger thresholds.}
  \label{fig:new3}
 \end{figure}

A more extensive scan in $\kappa$ for quark charge tagging performance is shown in Fig.~\ref{fig:new2}.  A value $\kappa\sim 0.5$ is optimal for the chosen $p_\text{T}$ bin; Fig.~\ref{fig:new3} shows that the optimality of this value is nearly independent of $p_\text{T}$.  For $p_\text{T}\lesssim 500$ GeV, the charge tagging performance is also relatively independent of $p_\text{T}$.  For $p_\text{T}\gtrsim 500$ GeV, the performance begins to degrade as the jet charge resolution significantly worsens as discussed below.  

\begin{figure}
 \centering
 \includegraphics[width=0.48\columnwidth]{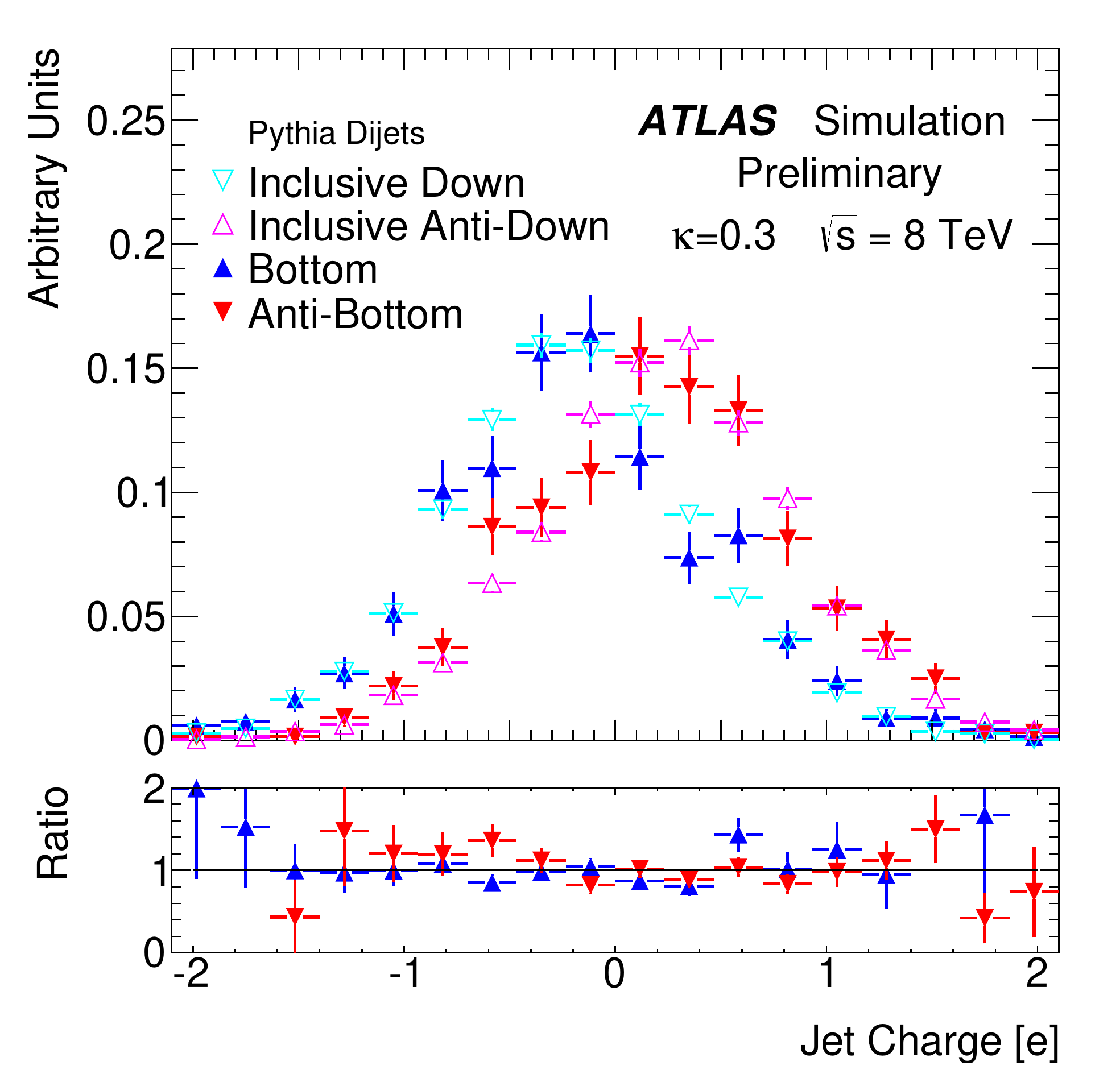}
   \includegraphics[width=0.48\columnwidth]{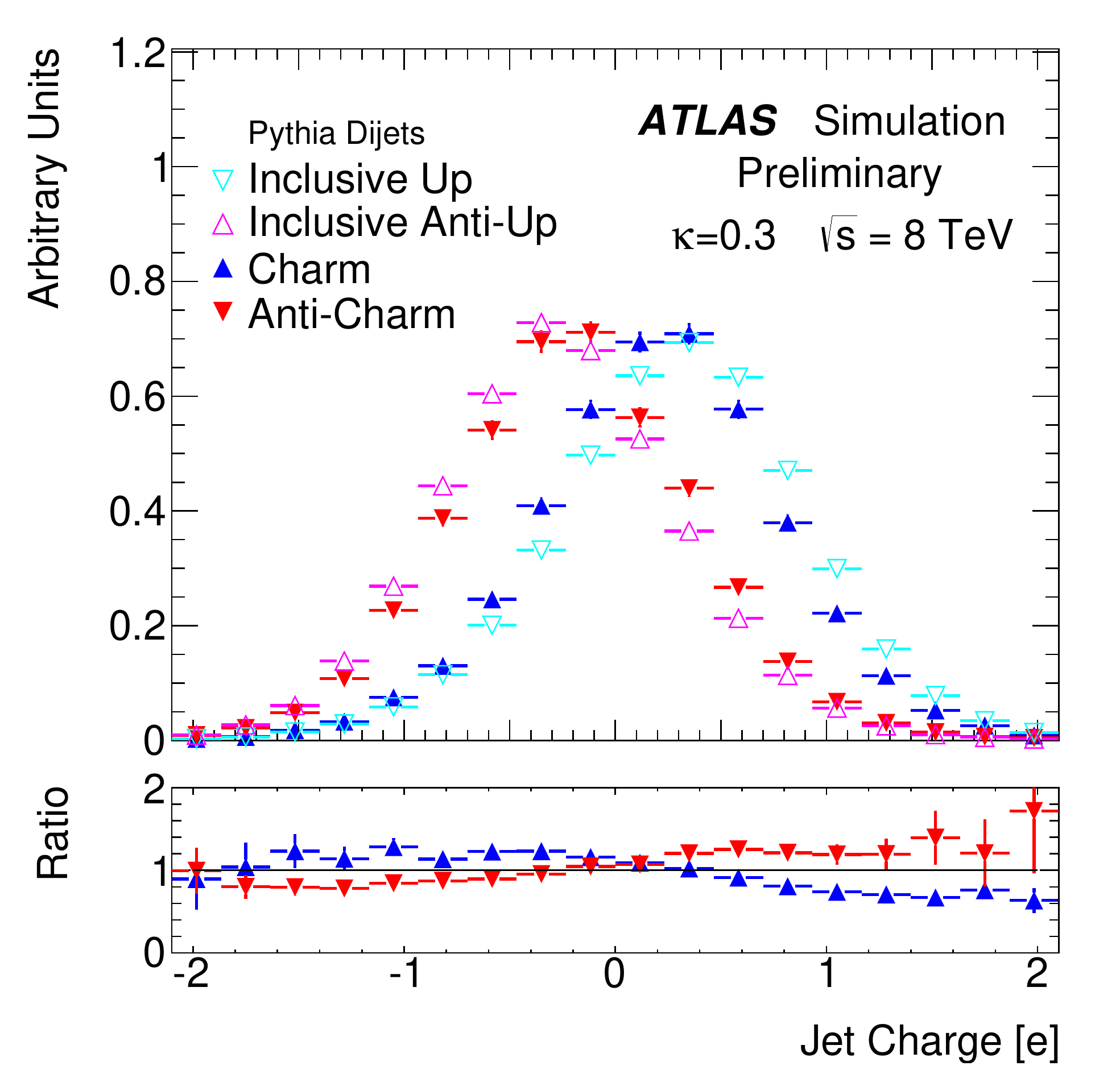}
 \caption{Jet charge distributions for heavy-flavour quarks and inclusive quark types and their ratios (lower panels).
The left plot shows the comparison for bottom type quarks and the right plots for charm type quarks. Results are obtained from MC truth
information in simulated dijet samples. }
 \label{fig:heavy}
\end{figure}

The discrimination is slightly degraded for heavy-flavor jets.  The jet charge distributions for positive and negative charm and bottom quarks is shown in Fig.~\ref{fig:heavy} where the inclusive samples used for comparison are the down type quark jets for the b-quark jets and the up type quark jets for c-quark jet charge.   Both plots of Fig.~\ref{fig:heavy} show that the ratio of heavy-quark jet charge to the inclusive jet charge of the same charge type is low for positive flavor in the positive region (thus high in the negative tail) and vice versa.  This means that the heavy-flavor distributions are shifted towards the center and thus the separation between positive and negative charge is reduced.  This shift is quantified by noting that the difference between the means of the two bottom-flavor distributions is $0.35\pm0.02$ (statistical uncertainty only) while the difference for the inclusive sample is $0.42\pm 0.01$.  Likewise for charm-flavor jets, the difference in means is $0.40\pm 0.02$ while for the inclusive sample the difference is $0.58\pm 0.01$.  This effect cannot be due to the selection on the track vertices, as the $d_0$ requirement is much larger than the
decay length of heavy-flavor mesons. It might be due to differences in the fragmentation, although further investigations are needed
to draw firmer conclusions.  However since the effect is relatively small, the degradation in separation is expected to be small; this may be important for $W^\pm$ discrimination as one of the decay products is a charm quark about 50\% of the time.  The flavor dependence of the jet charge is re-investigated in Sec.~\ref{sec:distinguish} in the context of boosted $W$ and $Z$ boson jets.

\begin{figure}
 \centering
 \includegraphics[width=.48\columnwidth]{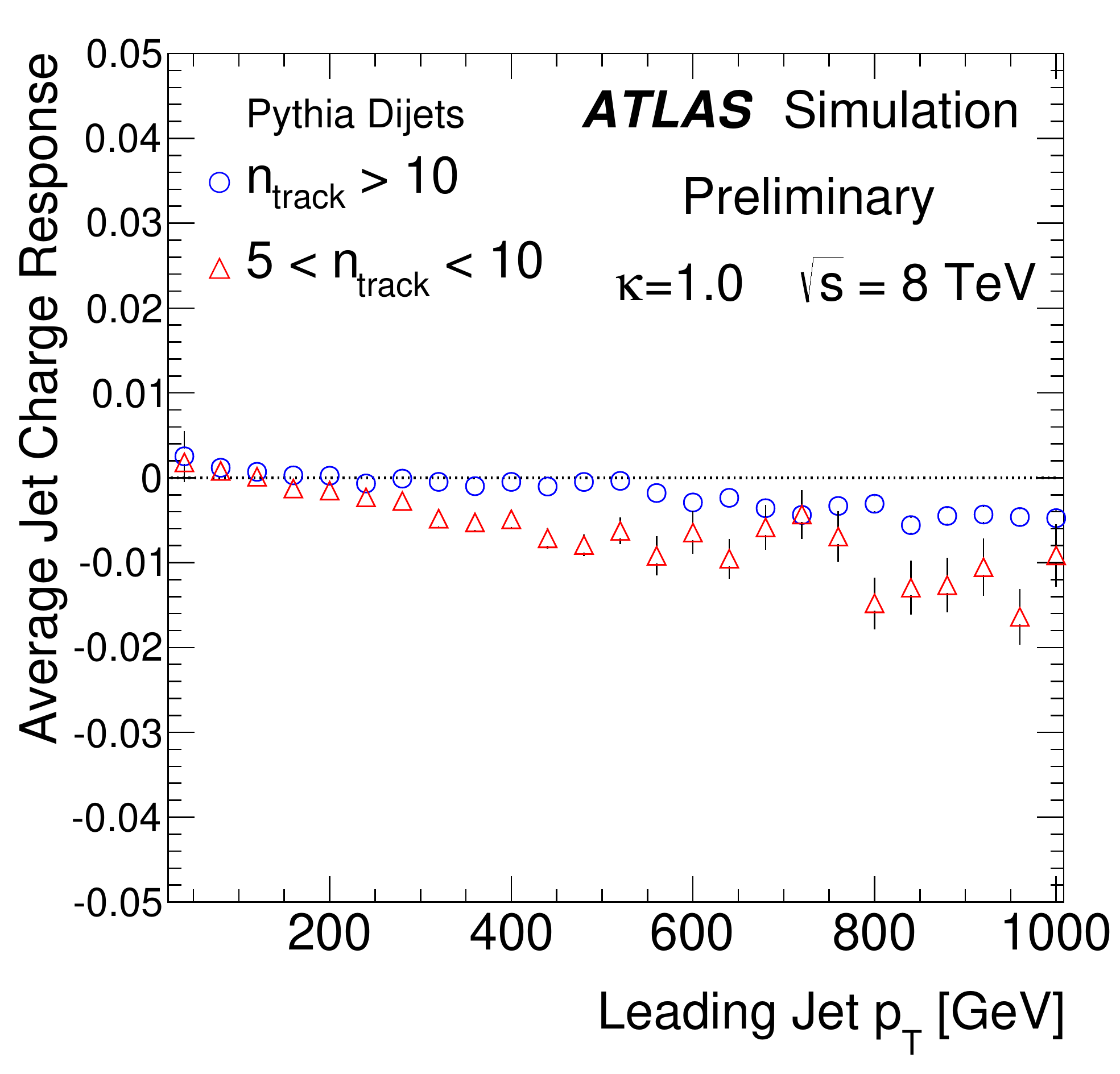}
   \includegraphics[width=.48\columnwidth]{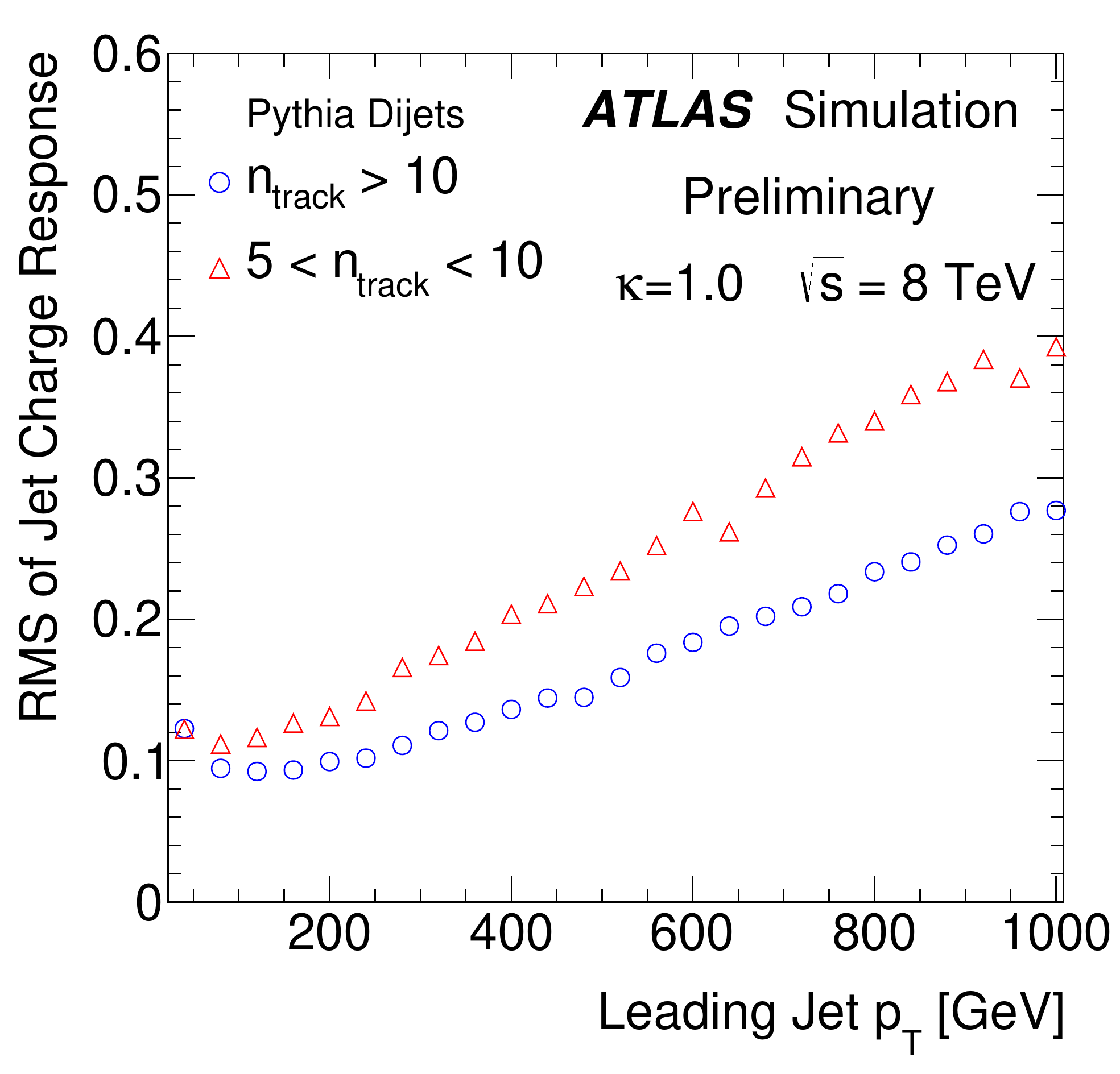}
 \caption{Average (left) and RMS (right) jet charge response ($Q^{\text{reco}}-Q^{\text{truth}}$) as a function of the leading jet $p_\text{T}$ for
a sample of simulated dijet events and for two different bins in track multiplicity. Uncertainties are from the limited size of simulated samples. }
 \label{fig:res_pt}
\end{figure}

\begin{figure}
 \centering
 \includegraphics[width=.48\columnwidth]{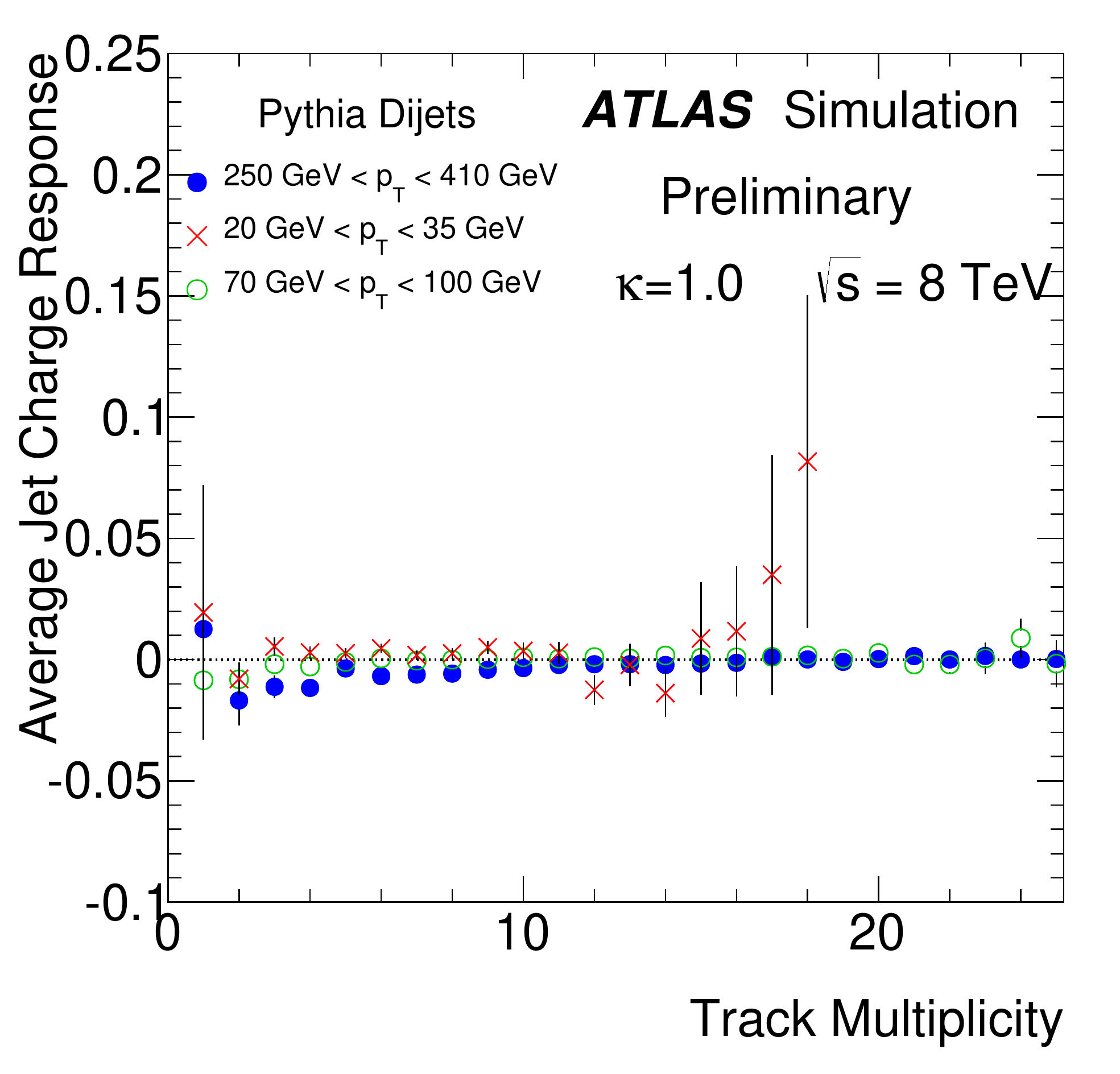}
   \includegraphics[width=.48\columnwidth]{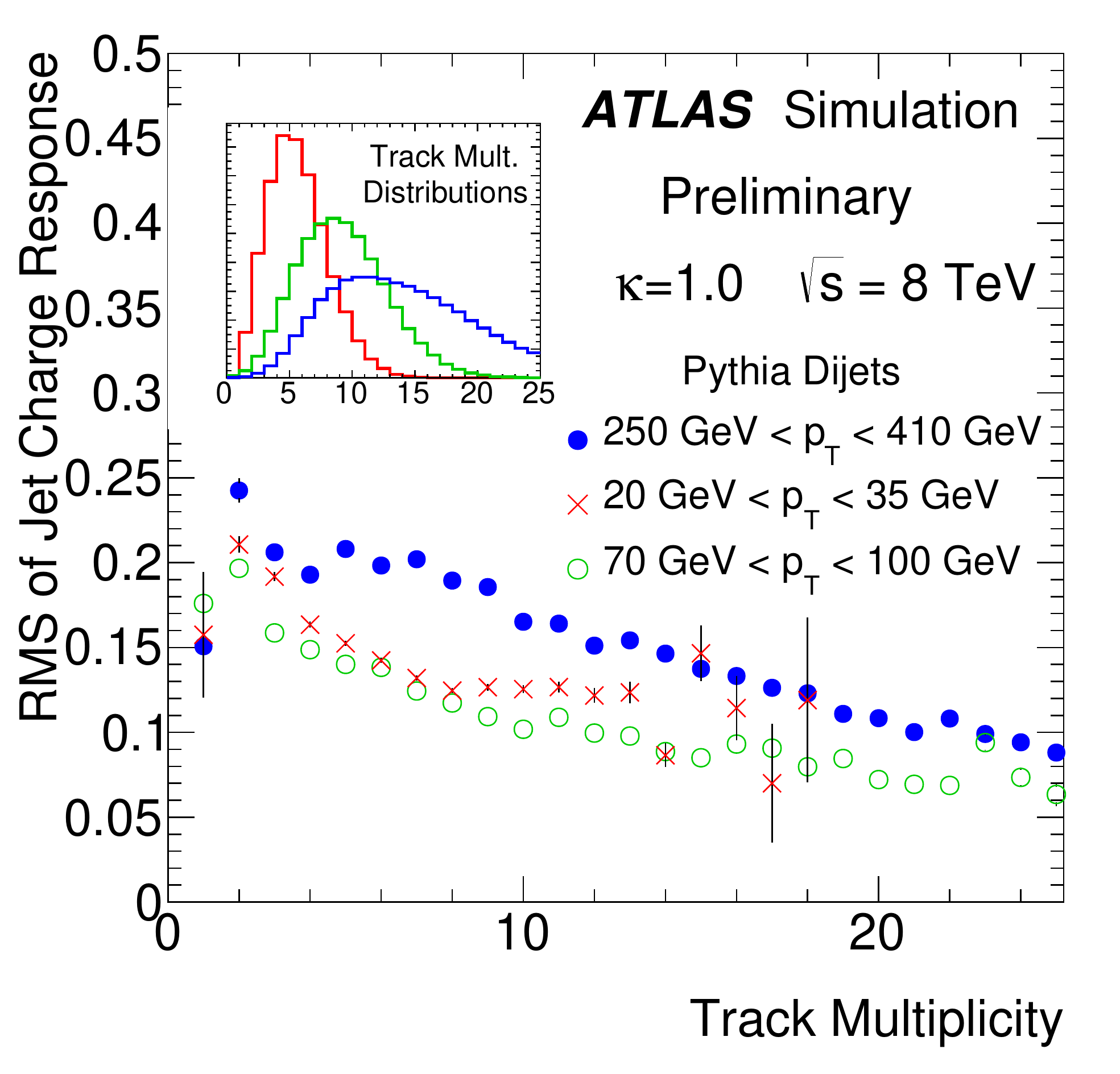}
 \caption{Average (left) and RMS (right) jet charge response ($Q^{\text{reco}}-Q^{\text{truth}}$) as a function of the number of tracks used to compute the charge in different
$p_\text{T}$ bins of the leading jet in dijet simulated events.  Uncertainties are from the limited size of simulated samples. }
 \label{fig:res_tracks}
\end{figure}

The MC can be used to investigate the performance of jet charge reconstruction.  As above, jet charge response is defined as the difference
between the (MC) reconstructed jet charge and the truth jet charge. The
mean of the jet charge response is shown as a function of jet $p_\text{T}$ in
the left plot of Fig.~\ref{fig:res_pt} and as a function of the number
of charged tracks ($n_{\text{track}}$) within the jet in the left plot of
Fig.~\ref{fig:res_tracks}.  The response is nearly
independent of both the $p_\text{T}$ and the number of tracks.  There is a slight decreasing trend in the response with $p_\text{T}$.  A contributing factor to the trend is the loss of tracks in the core of high $p_\text{T}$ jets so that $|Q^{\text{reco}}|<|Q^{\text{true}}|$.  Since the fraction of positive quark jets increases with $p_\text{T}$, this also means that there will be a trend towards $Q^\text{reco}<Q^\text{true}$, which is a negative response. The
spread of the response (as measured by the RMS) does depend both on
the $p_\text{T}$ and number of tracks (right plots in Figs.~\ref{fig:res_pt} and~\ref{fig:res_tracks}).  For low $p_\text{T}$, the RMS of the charge response distribution decreases
with $p_\text{T}$ and for jet $p_\text{T}$ above about 100 GeV, the
response RMS increases with $p_\text{T}$. This high $p_\text{T}$ trend is consistent with the
degradation of the relative momentum resolution as tracks become less
curved and also begin to merge in the dense jet core.  As expected, the RMS tends to decrease with the number of tracks as fluctuations about the mean are suppressed.  However, this trend is less evident at lower $\kappa$ where the individual contribution to the jet charge from any one track is decreased.  There is also a strong correlation between $p_\text{T}$ and number of tracks, which can further weaken the decreasing trend at high track multiplicity. 

\begin{figure}
 \centering
 \includegraphics[width=.48\columnwidth]{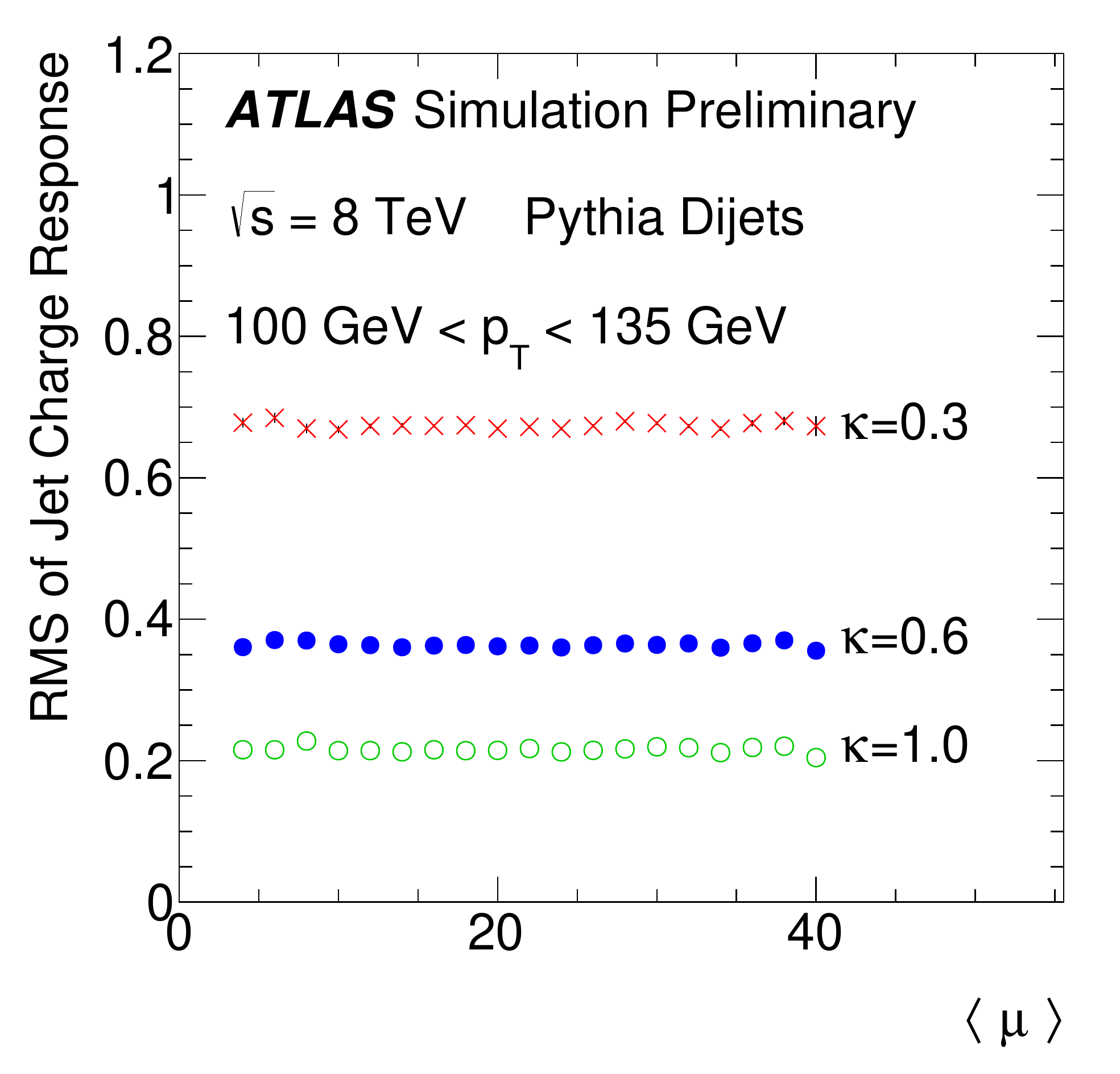}
   \includegraphics[width=.48\columnwidth]{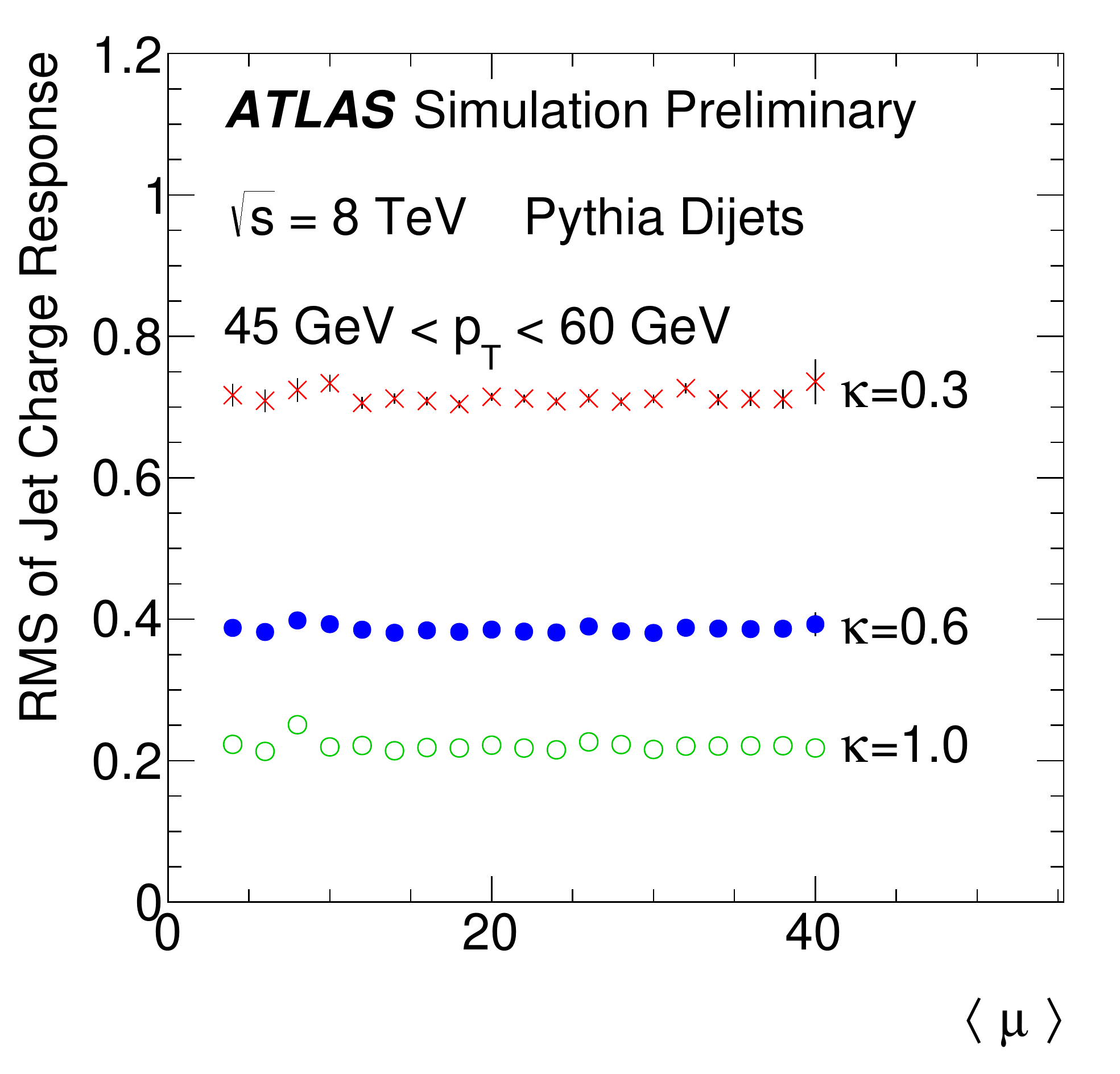}
 \caption{The RMS of the jet charge response ($Q^{\text{reco}}-Q^{\text{truth}}$) as a function of the average number of interactions per crossing for three values of $\kappa$ and in two bins of $p_\text{T}$, as obtained in simulated dijet events.  Uncertainties are from the limited size of simulated samples. }
 \label{fig:res_pileup}
\end{figure}

\begin{figure}[hbp]
 \centering
 \includegraphics[width=.49\columnwidth]{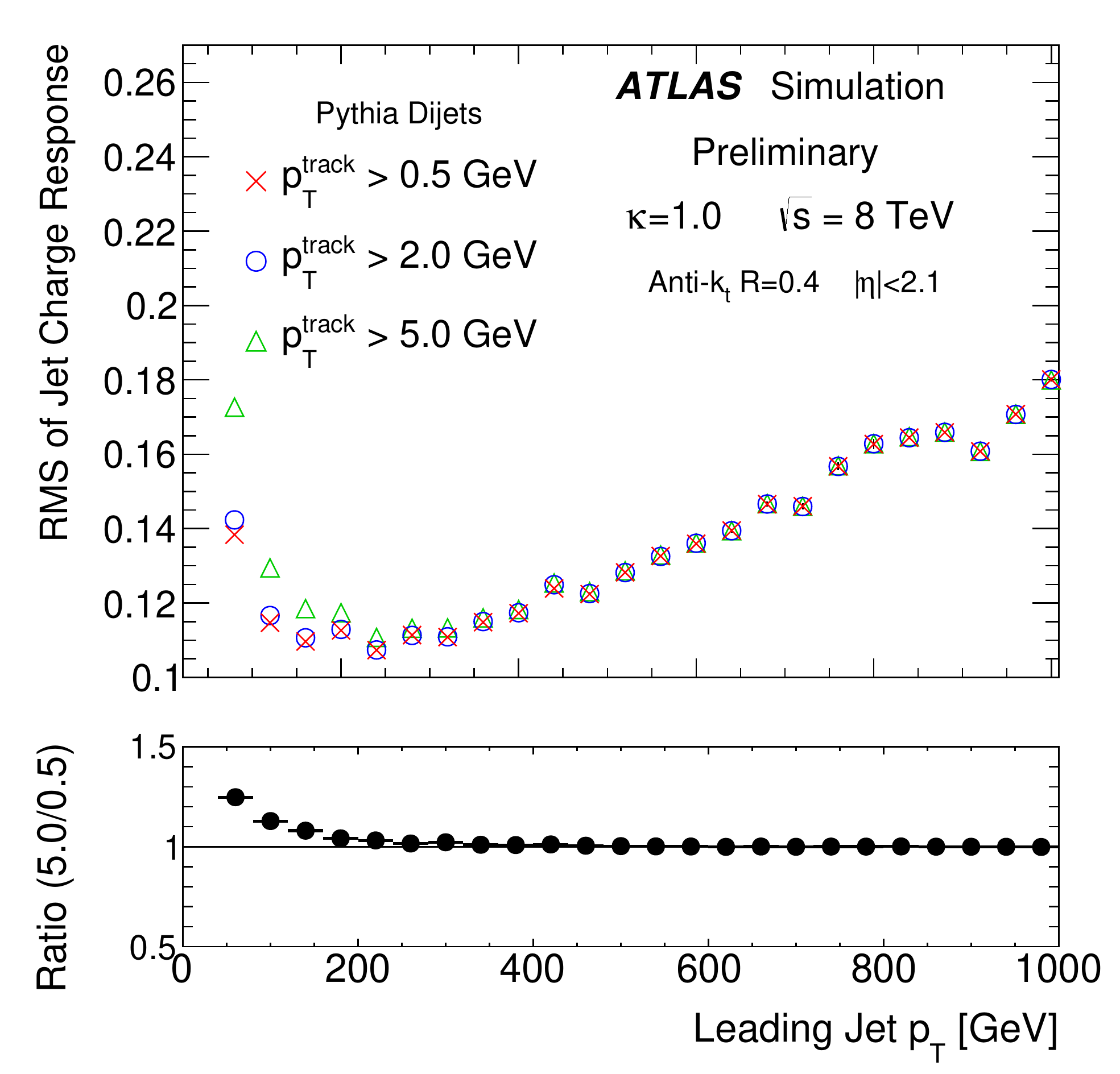} \includegraphics[width=.49\columnwidth]{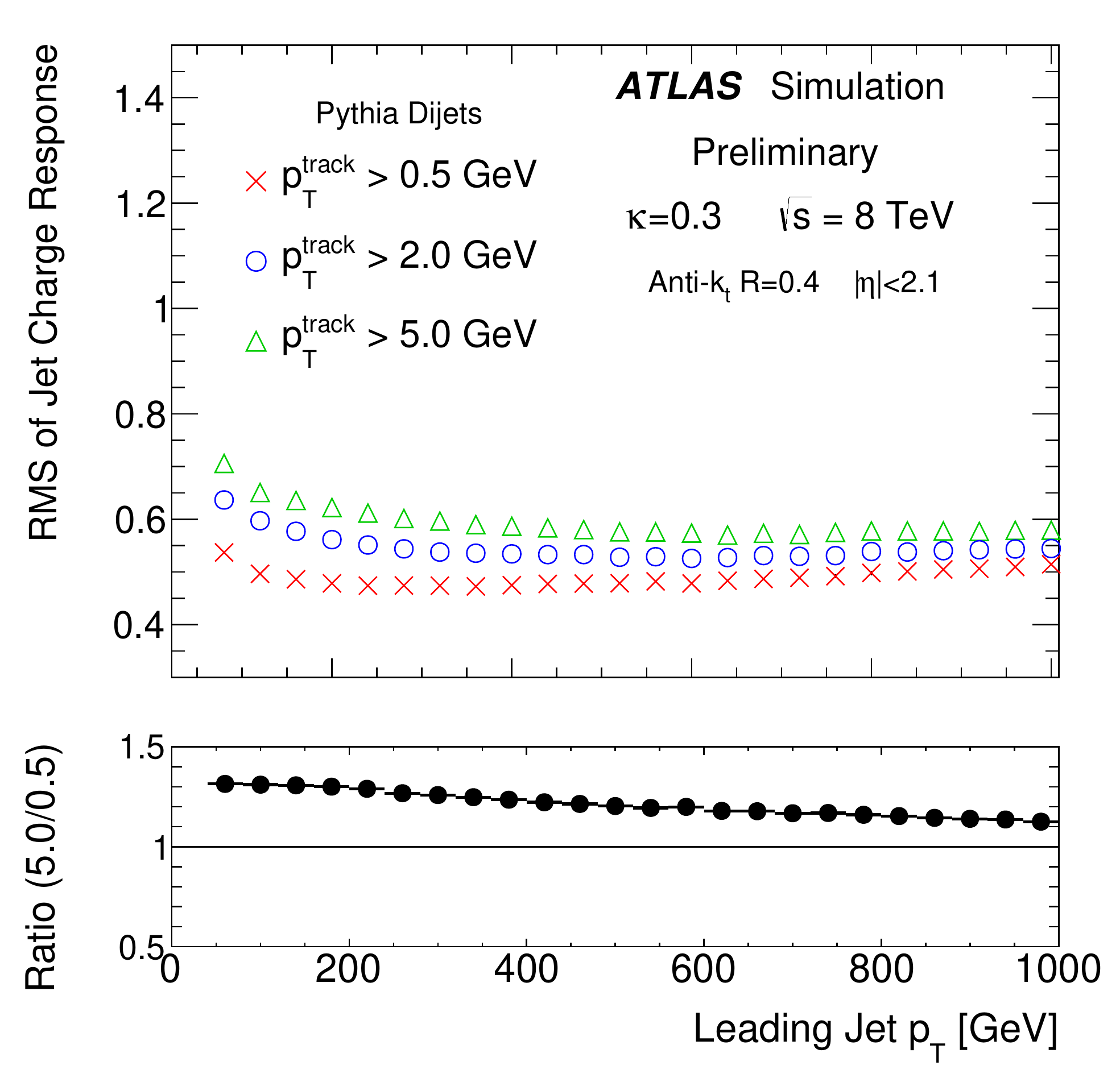}
 \caption{The dependance of the response resolution (measured using RMS) as a function of the jet $p_\text{T}$ for three values of the track $p_\text{T}$ threshold and two $p_\text{T}$ weighting values $\kappa=1$ (left) and $\kappa=0.3$ (right).   One can see that for both $\kappa$ values, there is a big increase in the RMS at low jet $p_\text{T}$ when moving to a track threshold of $5$ GeV.  For the lower value of $\kappa=0.3$, the trend persists for high $p_\text{T}$ since in this case, more weight is given to lower $p_\text{T}$ tracks.  The lower plots show the ratio of the 5 GeV~track threshold distribution with the 500 MeV~track threshold.}
 \label{fig:new1}
\end{figure}

In the 2012 LHC data, pileup has a non-negligible effect on reconstruction.  However, since the jet charge is built mostly of tracks associated to the primary collision vertex, the performance of this variable is expected to be independent of the number of pileup vertices.  This expectation is confirmed in Fig.~\ref{fig:res_pileup} where for two bins of jet $p_\text{T}$ and three values of the $p_\text{T}$-weighting factor $\kappa$, it is shown that the RMS of the jet charge response is independent of the average number of interactions per crossing ($\langle\mu\rangle$).  Related to the dependence of pileup is the choice of track quality criteria used in constructing the charge. The track $p_\text{T}$ threshold (500 MeV) and quality cuts are not expected to have an impact
on the jet charge response, as shown in Fig.~\ref{fig:new1} for the $p_\text{T}$ threshold.  

 \clearpage
 
 \subsubsection{Re-examining the Definition of Jet Charge}
\label{sec:altdef}

\begin{figure}[h!]
\begin{center}
\includegraphics[width=0.9\textwidth]{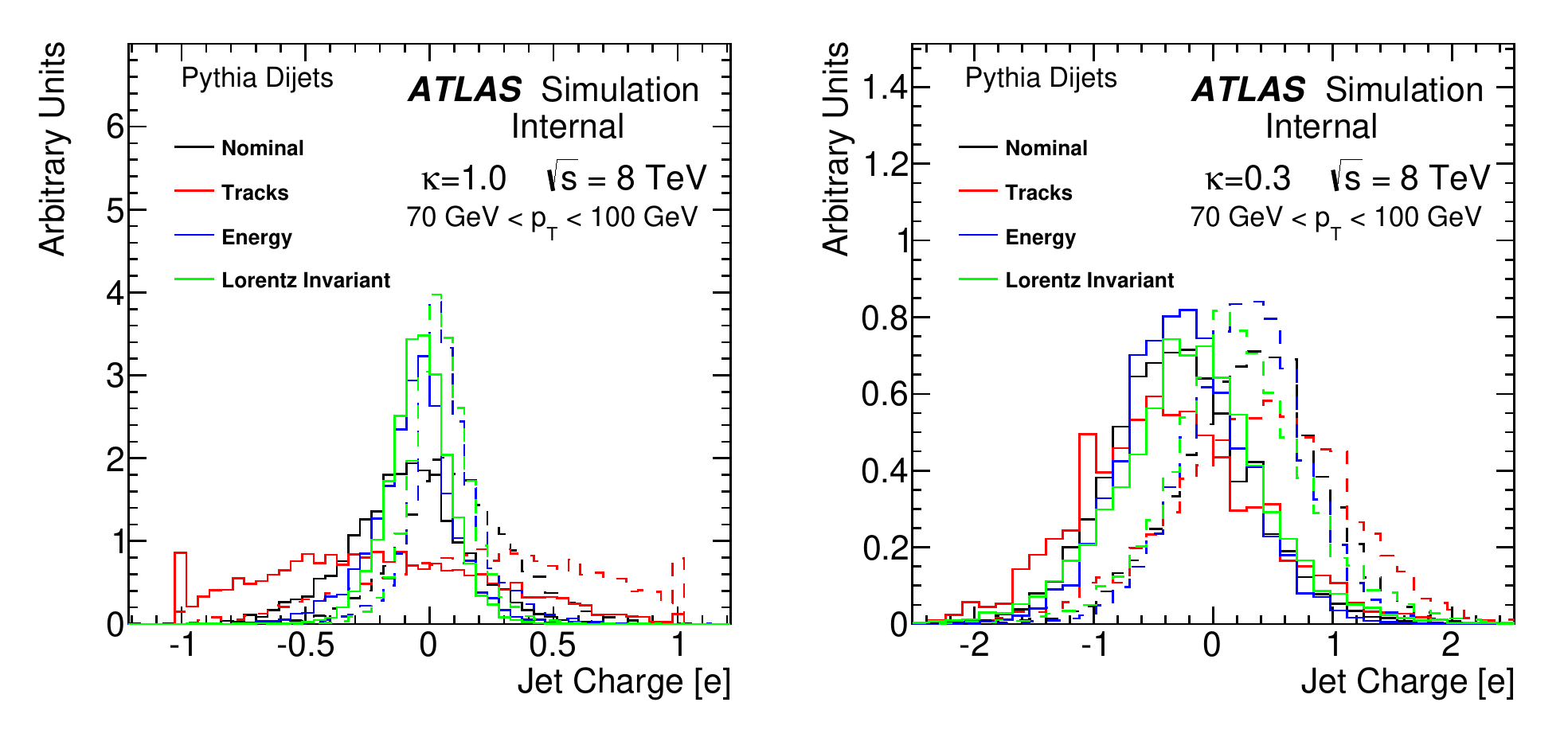}
\end{center}
\caption{Variations on the definition of the jet charge, see the text for details.  The left plot shows $\kappa=0.3$ and the right shows $\kappa=0.7$.  The variants on the colors are for jet originating from a quark with a positive charge versus a negative charge.  The small differences in modes between positive and negative quark initiated jets is likely due to statistical fluctuations from the limited MC sample size.}
\label{fig:variations_CONF}
\end{figure}

\begin{figure}[h!]
\begin{center}
\includegraphics[width=0.9\textwidth]{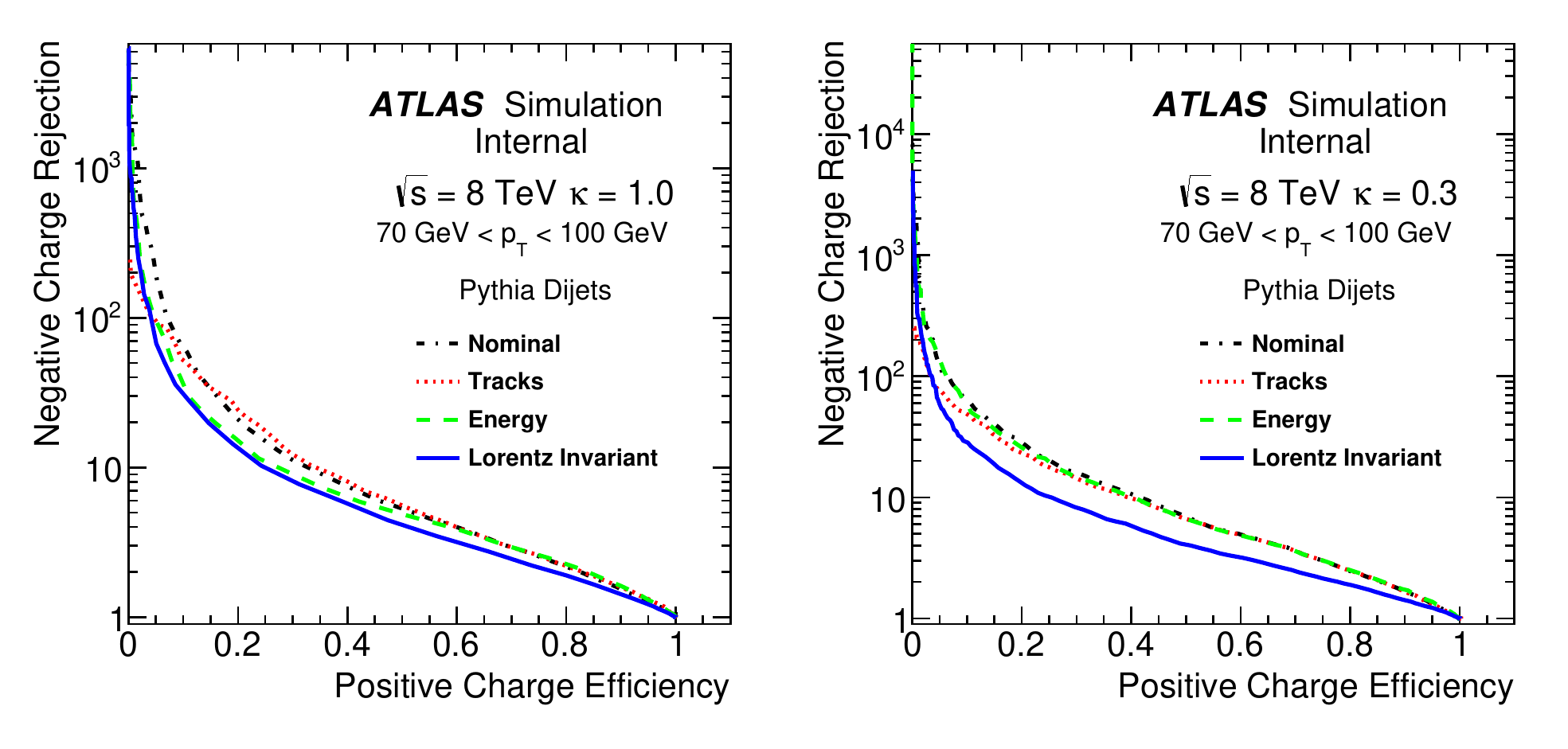}
\end{center}
\caption{ROC curves for the distributions shown in Fig.~\ref{fig:variations_CONF}. The left plot shows $\kappa=0.3$ and the right shows $\kappa=0.7$.}
\label{fig:variations2_CONF}
\end{figure}

This section considers some variations on the definition of jet charge.  As noted in Sec.~\ref{sec:jetcharge:LI}, the jet charge is not Lorentz invariant.  One possible Lorentz invariant definition uses the jet `rest frame' (jets are massive, so this is sensible - see Sec.~\ref{sec:jetmass}).  This variation and others on the jet charge definition are studied in Fig.~\ref{fig:variations_CONF} and the performance is quantified in Fig.~\ref{fig:variations2_CONF}.  One important variation that has been used in some of the analyses mentioned in the chapter introduction is the one labeled `tracks'.  For this definition, the denominator of Eq.~\ref{chargedefcharge} is replaced by the scalar sum of track $p_\text{T}$, raised to the $\kappa$, i.e. $Q=\sum Q_ip_\text{T,i}^\kappa/(\sum p_\text{T,i})^\kappa$.  When $\kappa=1$, this track-only definition is bounded by $1$ and there are spiked in the left plot of Fig.~\ref{fig:variations_CONF} at $\pm 1$ corresponding to cases where there is only one track in the jet.  

The performance of the track-only jet charge in Fig.~\ref{fig:variations2_CONF} is nearly the same as for the `nominal' definition, in which the calorimeter jet $p_\text{T}$ is used instead, except at low efficiency where the calorimeter-based definition is superior.  The Lorentz invariant definition is clearly worse than the other definitions, in part because the mass of generic QCD jets is highly sensitive to diffuse soft radiation.  These soft tracks that may be independent from the initiating parton can have significant momentum in the jet rest frame.  Only the definition from Eq.~\ref{chargedefcharge} is considered in the rest of this chapter. 

 \clearpage
 
\subsection{Simulation and In-situ Studies with $W$ Bosons}
\label{sec:insitubosoncharge}

\subsubsection{Dataset and Simulation Samples}
\label{sec:ttbarsamplesandeventselection}

The studies presented in this section use a subset of the $\sqrt{s}=8$ TeV data from Run~1 corresponding to 5.8~$\mathrm{fb}^{-1}$.  Single lepton triggers are used to select the data.  The MC setup is similar to Sec.~\ref{sec:chargesamples}, except that instead of inclusive dijets as the main process, the target is $t\bar{t}$ production.

Top quark pair production is simulated with two next-to-leading-order (NLO) generators.  When studying $W^\pm$ discrimination in $t\bar{t}$ events, {\sc MC@NLO}~\cite{Frixione:2002ik} is used with the NLO parton density function (PDF) set CT10~\cite{Lai:2010vv,Gao:2013xoa}, and parton showering and underlying event modelled with {\sc Herwig}~\cite{Corcella:2000bw} and {\sc JIMMY}~\cite{JIMMY}, respectively.  For jet charge studies in $W$+jets, $t\bar{t}$ is simulated with {\sc Powheg}~\cite{Nason:2004rx,Frixione:2007vw,Alioli:2010xd} using the PDF set  CT10 and {\sc Pythia} 6.4~\cite{Sjostrand:2006za} for fragmentation and hadronization with the Perugia2011C~\cite{Skands:2010ak} tune that employs the LO CTEQ6L1 PDF set~\cite{Pumplin:2002vw}.  In all $t\bar{t}$ MC events, events are filtered by requiring at least one lepton consistent with the lepton trigger selection used for the measurements in which $t\bar{t}$ is relevant.  Before filtering, the $t\bar{t}$ cross section is $\sigma_{t\bar{t}}= 238^{+22}_{-24}$~pb for a top quark mass of $172.5$ GeV. It has been calculated at next-to-next-to leading-order (NNLO) in QCD including resummation of next-to-next-to-leading logarithmic (NNLL) soft gluon terms with top++2.0~\cite{Cacciari:2011hy,Baernreuther:2012ws,Czakon:2012zr,Czakon:2012pz,Czakon:2013goa,Czakon:2011xx}. The PDF and $\alpha_S$ uncertainties are calculated using the PDF4LHC prescription~\cite{Botje:2011sn} with the MSTW2008 68\% CL NNLO~\cite{Martin:2009iq,Martin:2009bu}, CT10 NNLO and NNPDF2.3 5f FFN~\cite{Ball:2012cx} PDF sets, and added in quadrature to the scale uncertainty.  $W$+ jets production is based on {\sc Alpgen}~\cite{Mangano:2002ea}, with the parton shower modelled with {\sc Pythia} 6.4 and the Perugia2011C tune; for these samples the production of heavy quarks is modelled separately, and overlapping phase space produced in the inclusive samples is removed. 

\begin{figure}[h!]
\begin{center}
\includegraphics[width=0.5\textwidth]{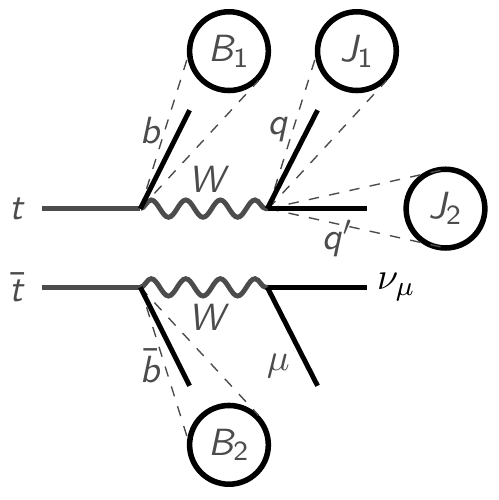}
 \caption{Schematic representation of the object selection.  At least four jets are required: two $b$-tagged jets $B_1,B_2$ and at least two non $b$-tagged jets, labelled $J_1,J_2$.  The jets $J_1$ and $J_2$ are those non b-tagged jets with invariant mass closest to the $W$ boson mass.  The charged lepton is used to trigger and a cut on the missing energy from the neutrino is used to purify the sample in $t\bar{t}$ events.}
 \label{fig:JetCharge:schemeatic}
  \end{center}
\end{figure}

The single top ($s$- and $Wt$-channel) backgrounds are modelled with the same {\sc MC@NLO} setup as $t\bar{t}$ while the $t$-channel is modelled with {\sc AcerMC}~\cite{tchannel} and the CTEQ6L1 PDF set interfaced with {\sc Pythia} using the Perugia2011C tune.  Like $W$+jets, the $Z$+jets backgrounds are modelled with {\sc Alpgen}, {\sc Pythia}~6.4 showering, and the Perugia2011C tune. 
Dibosons are generated with {\sc Herwig}~using the CTEQ6L1 PDF set.  
The $t\bar{t}$ events are selected with exactly one leptonic $W\rightarrow\mu+\nu$ decay to obtain a high-purity source of hadronically-decaying bosons with known charge in~$t\bar{t} \rightarrow (W\rightarrow\ell \nu)(W\rightarrow qq')b\bar{b}$~final-states.  
Candidate events are chosen by requiring a $p_\text{T} > 25$ GeV muon with $|\eta| < 2.5$ and missing transverse momentum $E_\text{T}^\text{miss} > 20$ GeV; in addition, the sum of the missing transverse momentum and the transverse mass\footnote{The transverse mass is defined as $m_{\text{T}}^2=2p_{\text{T}}^{\text{lep}}E_{\text{T}}^{\text{miss}}(1-\cos(\Delta\phi))$, where $\Delta\phi$ is the azimuthal angle between the lepton and the missing transverse momentum direction.} of the $W$ boson reconstructed from the lepton and missing momentum is required to be greater than 60 GeV, as expected for leptonic W decays.   Muons from heavy-flavour decays are suppressed by requiring the muon to be isolated in both the tracker and calorimeter from unclustered objects as well as from jets.  Events must also have at least four jets with $|\eta| < 2.5$ and $p_\text{T} > 25$ GeV.  Exactly two of these jets must be identified as $b$-quark jets using the multivariate discriminant `MV1' \cite{ATLAS-CONF-2012-043} which includes impact parameter and secondary vertex information as inputs.  The chosen MV1 working point corresponds to an average $b$-tagging efficiency of 70\% for $b$-jets in simulated $t\bar{t}$ events.  Among the jets not selected by the $b$-tagger, there must exist a pair each with $|\eta|<2.1$ and a dijet invariant mass within $30$ GeV of the $W$ boson mass.  The two jets with invariant mass closest to the $W$ boson mass are chosen as the $W$ daughter candidates.  This procedure selects a sample that is expected to contain more than $90\%$ $t\bar{t}$ production, as shown in Table~\ref{tab:compcharge} for the positive muon and negative muon channels separately.   Figure~\ref{fig:JetCharge:schemeatic} illustrates the object selection.

\begin{table}[hbp!]
\centering
\begin{tabular}{|l|c|c|} 
\hline
Process                       	& $N_\text{events}$ with $\mu^+$    &$N_\text{events}$ with $\mu^-$        \\\hline
$t\bar{t}$ 	& 3575  $\pm$ 29     &  3522 $\pm$ 20	 \\
Single Top                 	& 126 $\pm$ 3 	&	 97 $\pm$ 3 \\
$W$+jets    	&170 $\pm$ 29   & 	91 $\pm$ 15	   \\
$Z$+jets  	& 23$\pm$ 5	    & 	18 $\pm$ 3	   \\
Dibosons & 3 $\pm$ 0.4 &3 $\pm$ 0.3\\\hline
Total MC             & 3895 $\pm$ 36& 3729 $\pm$ 25\\
2012 Data	&4095  & 3893\\\hline
\end{tabular}\caption{The data and MC signal and background yields after all selections for the 5.8 $\mathrm{fb}^{-1}$ sample, shown separately for $\mu^+$ and $\mu^-$ final states. The MC uncertainties are purely statistical and included solely for the purposes of illustrating the sample composition. }
\label{tab:compcharge}
\end{table}

\begin{figure}
 \centering
 \includegraphics[width=0.48\columnwidth]{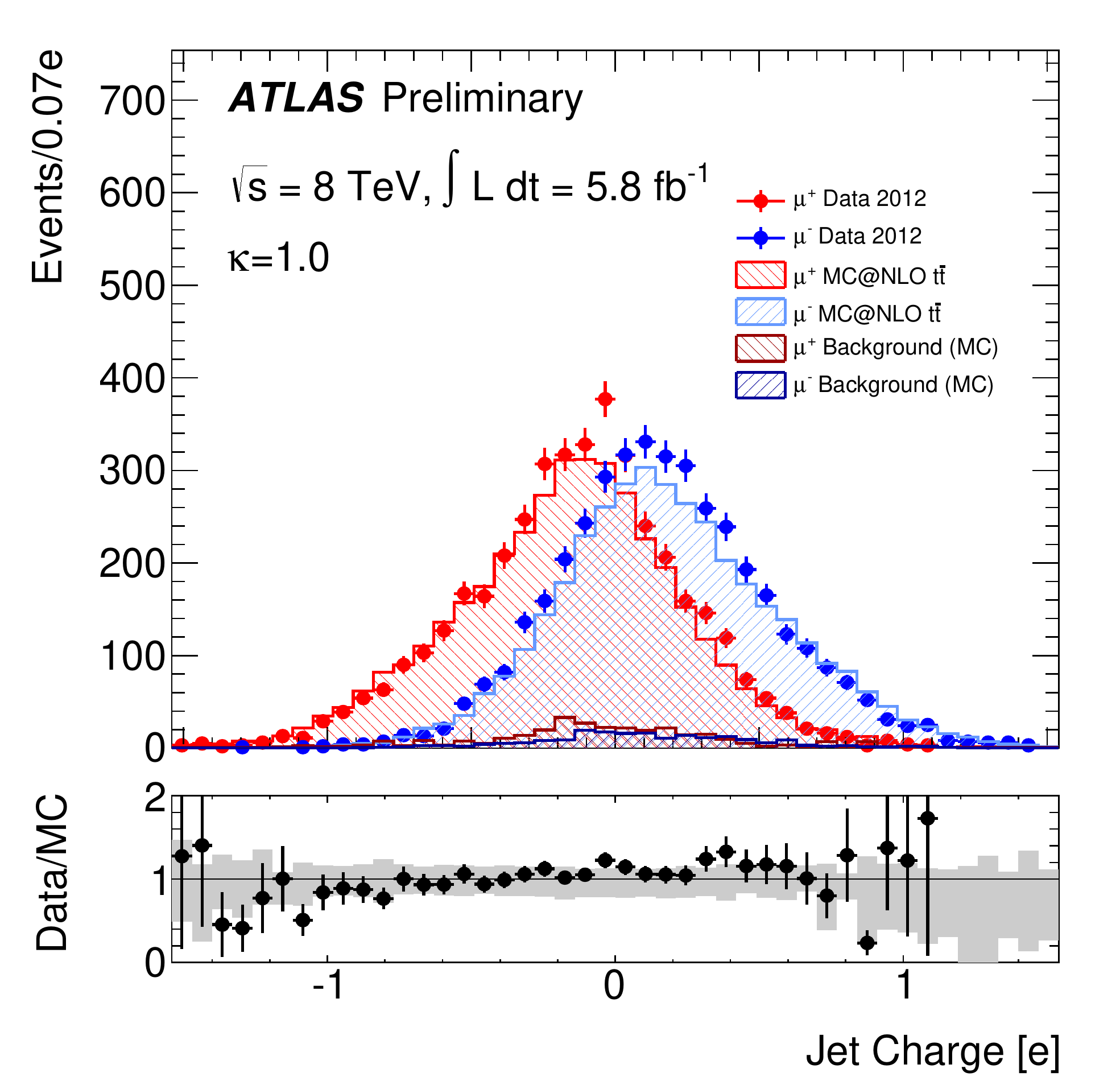}   \includegraphics[width=0.48\columnwidth]{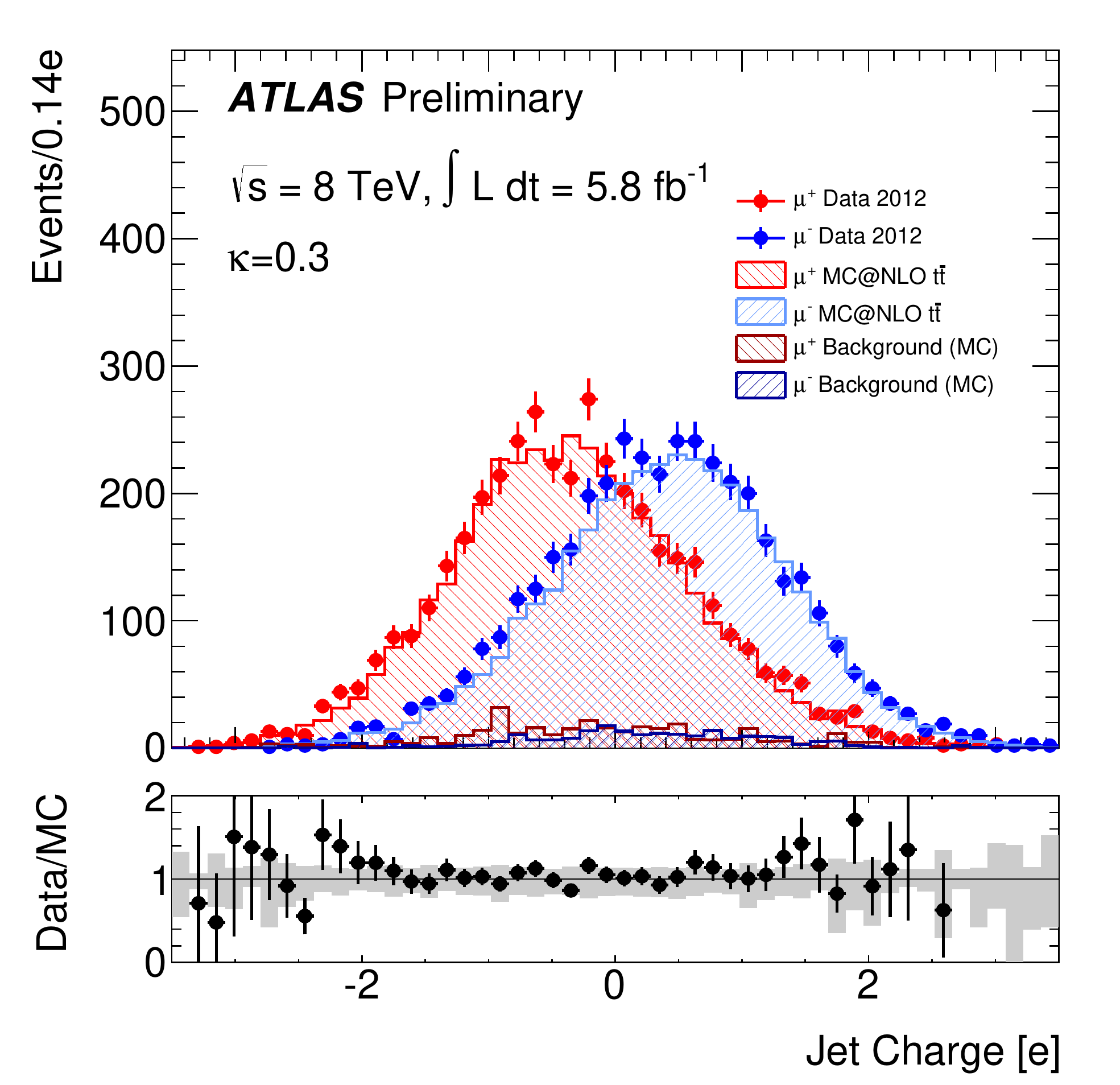} 
 \caption{The distribution of the sum of the jet charges from the two daughter candidates in hadronic W boson decays in semileptonic $t\bar{t}$ events.  The plot on the left is obtained with $p_\text{T}$ weighting factor $\kappa=1.0$ and the right plot with $\kappa=0.3$.  Events with a $\mu^\pm$ correspond to a hadronically-decaying $W^\mp$. The bottom panels show the data/MC ratios with the bands giving the systematic
uncertainties described in the text.}
 \label{fig:distttbar}
\end{figure}

The charge of the hadronically decaying W boson can be inferred from the measured charge
of the selected muon.  Therefore, the discrimination power of the jet charge can be directly determined from data. Figure~\ref{fig:distttbar} shows the distributions of dijet charge from $W^+$ and $W^-$ decays for two different values of the $p_\text{T}$ weighting factor $\kappa$ from Eq.~\ref{chargedefcharge}.  The dijet charge is computed as the sum of the jet charges from the $W$ boson daughter candidates.  In $\mu^+$ events, the dijet charge tends to be negative, while the opposite is true for $\mu^-$ events, indicating that the dijet charge shows correlation with the charge of the hadronically-decaying $W$.  

The ratio between data and MC in the lower panels of Fig.~\ref{fig:distttbar} includes the relevant
systematic uncertainties: the jet energy scale uncertainty (JES), the jet energy resolution uncertainty (JER), tracking efficiency, $b$-tagging related uncertainties, and the uncertainty on the background normalisation.  The JES and its uncertainty are determined from a combination of test-beam data, LHC collision data, and MC simulation~\cite{jes}.   The $\pm 1\sigma $ variations are computed as a function of the $p_\text{T}$ and $\eta$ of each reconstructed jet and are then propagated through to the jet charge distributions.  To model the impact of the uncertainty on the energy resolution, reconstructed jet energies are smeared by a Gaussian function such that the new width incorporates a $+1\sigma$ variation of the JER uncertainty.  The effect on the charge distribution is symmetrized by taking the difference between the nominal and the shifted as $1\sigma$.  The JES and JER uncertainties are comparable and amount to about 20\% of the nominal in the $|Q|<1$ region for $\kappa=1.0$.  Track reconstruction efficiency~\cite{Aad:2010ac} and $b$-tagging uncertainties are much smaller than the JES and JER contributions (less than percent level).  The $b$-tagging only affects acceptance and not the charge itself, unlike JES and JER which contribute to both acceptance and the actual charge via the jet calorimeter energy in the denominator of the jet charge definition.   The uncertainty on the background normalization is taken to be the same as the cross section uncertainty for $t\bar{t}$ stated in Section 2, namely about 6\%.  This is justified because the combinatorial background from $t\bar{t}$ represents more than
90\% of the total background. All uncertainties are added in quadrature for each bin.  

\begin{figure}
 \centering
 \includegraphics[width=.48\columnwidth]{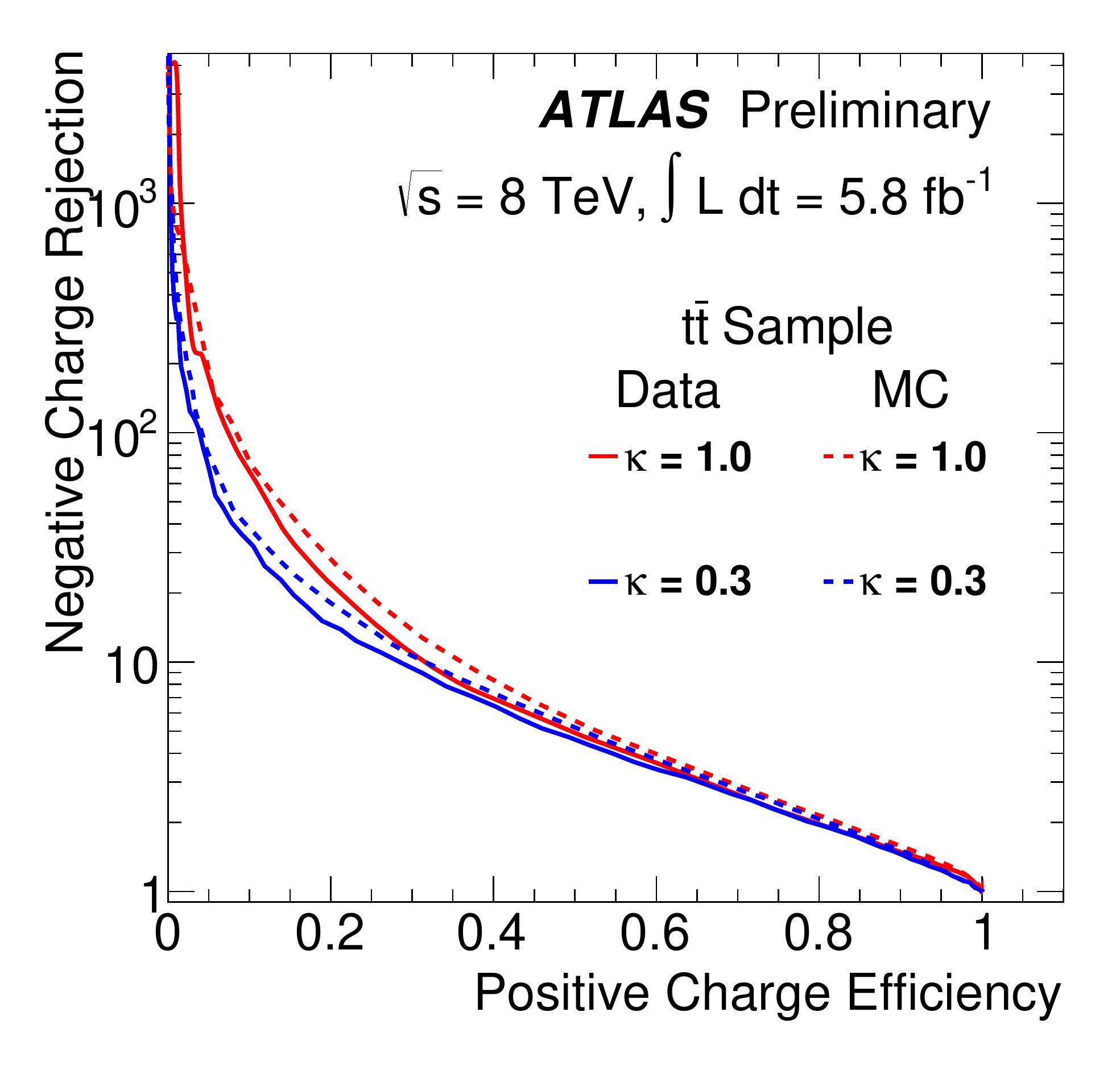} 
  \includegraphics[width=.48\columnwidth]{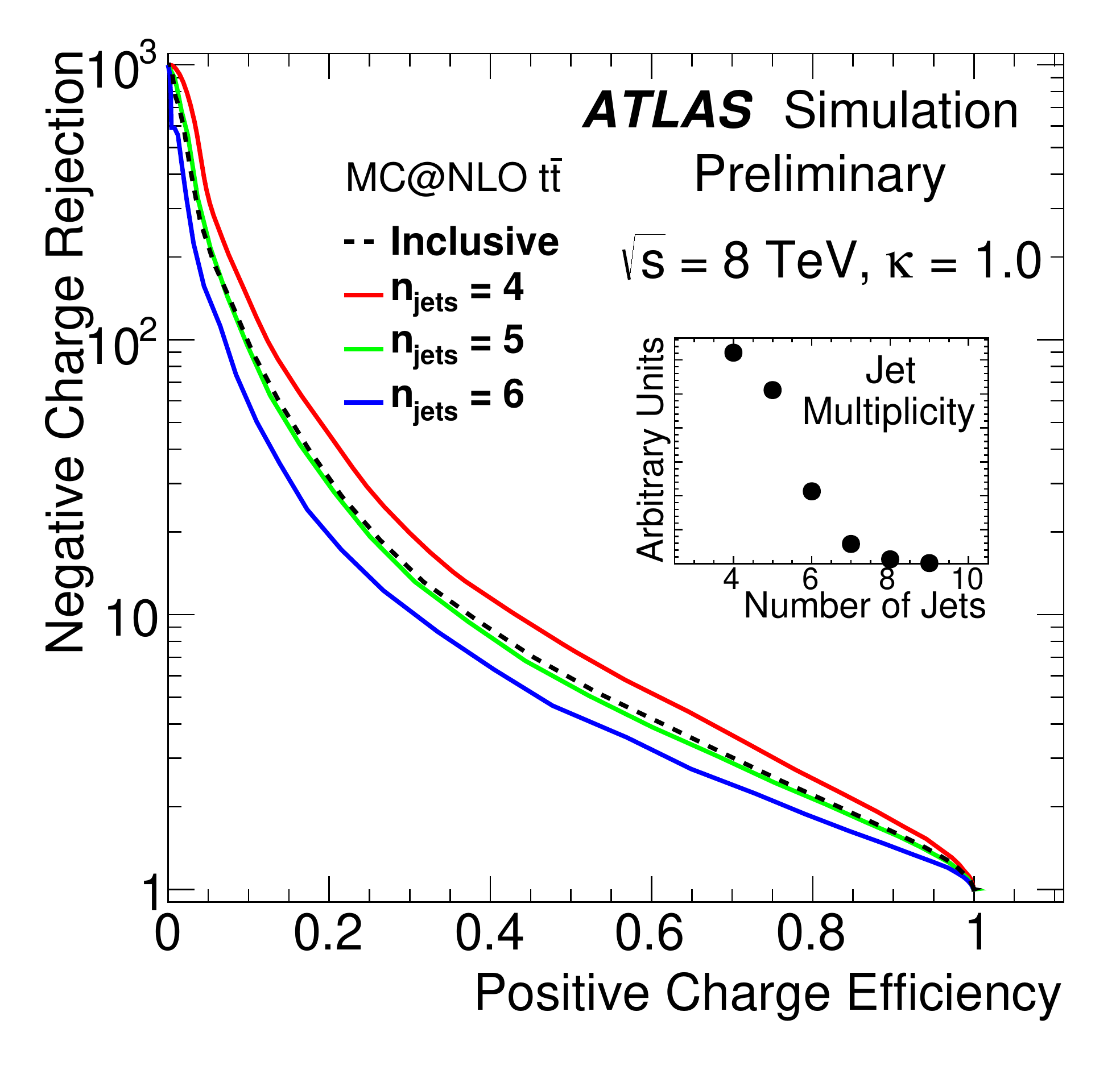} 
   \caption{The power to reject $W^-$ as a function of the efficiency to tag $W^+$ as measured in semileptonic $t\bar{t}$ events.  Each point on the curve corresponds to a cut value on the charge distribution shown in Fig. 1.  The left plot shows results for inclusive jet multiplicity, and the right plot for several
multiplicities and $\kappa=1.0$.}
 \label{fig:rejecttbar}
\end{figure}

To quantify the discriminating power of the dijet charge, the rejection of negatively-charged $W$ bosons is computed against the efficiency for selecting positively-charged $W$ bosons.  This relationship is shown in the left plot of Fig.~\ref{fig:rejecttbar}.  The points along the solid (dashed) lines correspond to cuts on the charge distribution in data (MC) for two values of $\kappa$.    Since $\mu^\pm$ events correspond to hadronic $W^\mp$, the horizontal axis is computed as the fraction of $\mu^-$ events beyond a given cut value with respect to all $\mu^-$ events and the vertical axis values are the reciprocals of the fraction from $\mu^+$ events past the same cut value, with respect to all $\mu^+$ events.  A negative $W$ rejection of about 6 is expected for a positive $W$ boson efficiency of 50\%, almost independent of $\kappa$.  Some degradation of the separation power between positive and negative W bosons is expected to come
from the combinatorial background, i.e. the two $W$ daughter candidates may not have originated from or contain all of the partons associated with the $W$ decay.  Such an effect appears in both data and MC. Its impact can be estimated by selecting a purer sample, which reduces the combinatorial background. The right plot in Fig.~\ref{fig:rejecttbar} shows the positive $W$ efficiency for different numbers of jets in the event.  Jets are required to be above $25$ GeV in $p_\text{T}$ and have $|\eta|<2.5$.  For example, for a fixed positive $W$ efficiency of 50\%, the rejection of negative W increases by 20\% when the
jet multiplicity decreases from six to four.

\subsubsection{Charge Reconstruction Performance}

\begin{figure}
 \centering
 \includegraphics[width=.48\columnwidth]{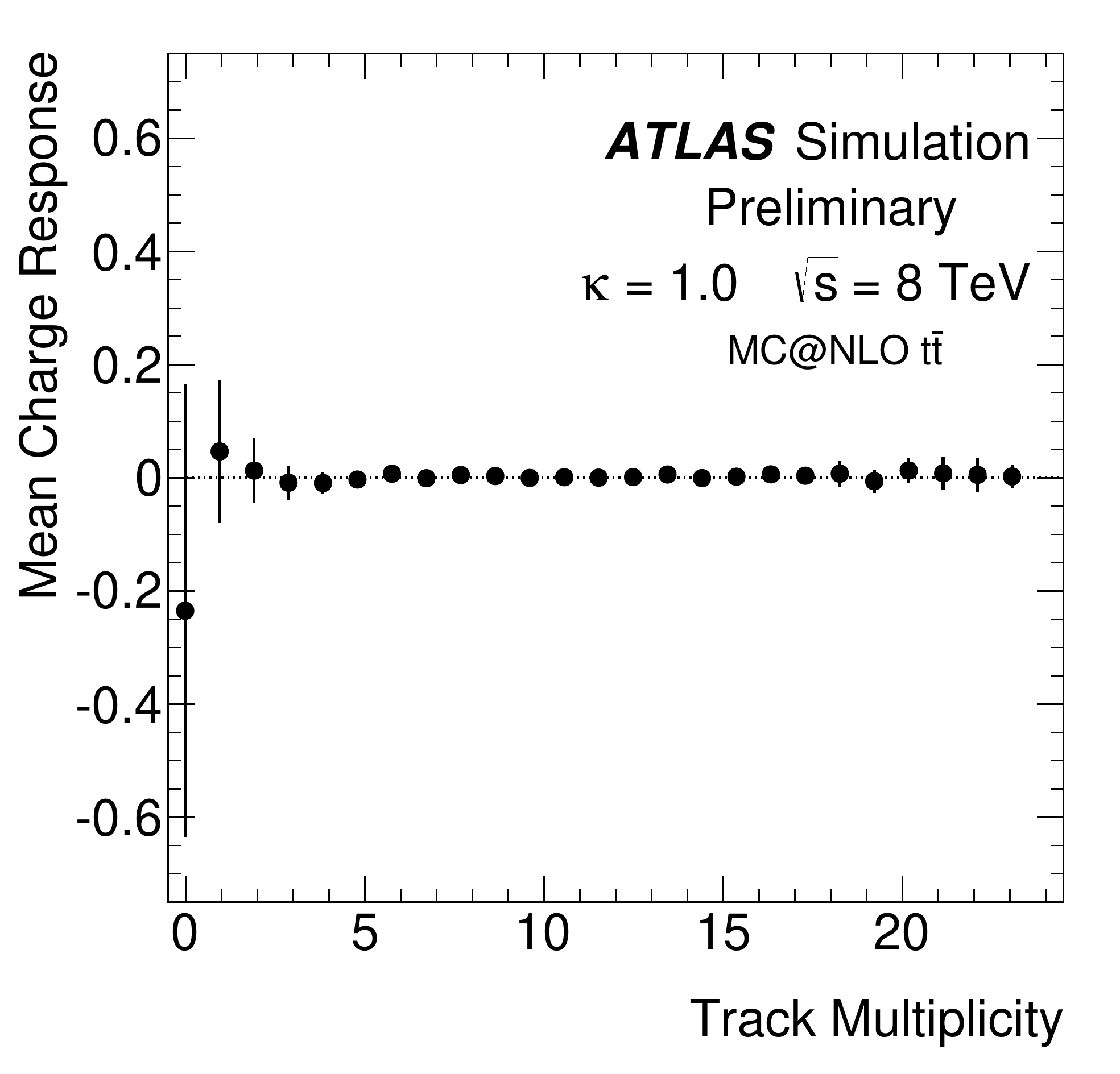}\includegraphics[width=.48\columnwidth]{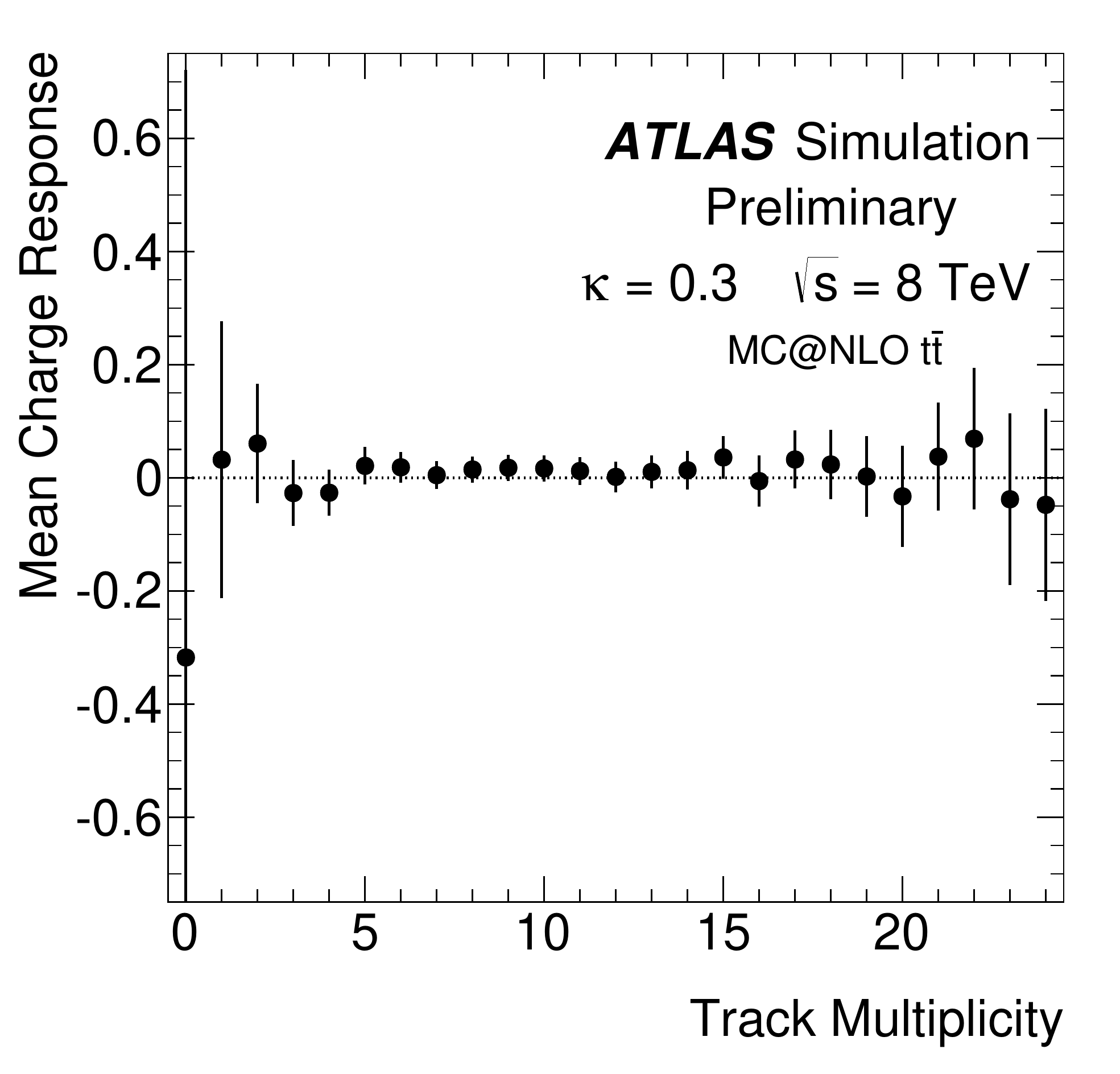}  \\
   \includegraphics[width=.48\columnwidth]{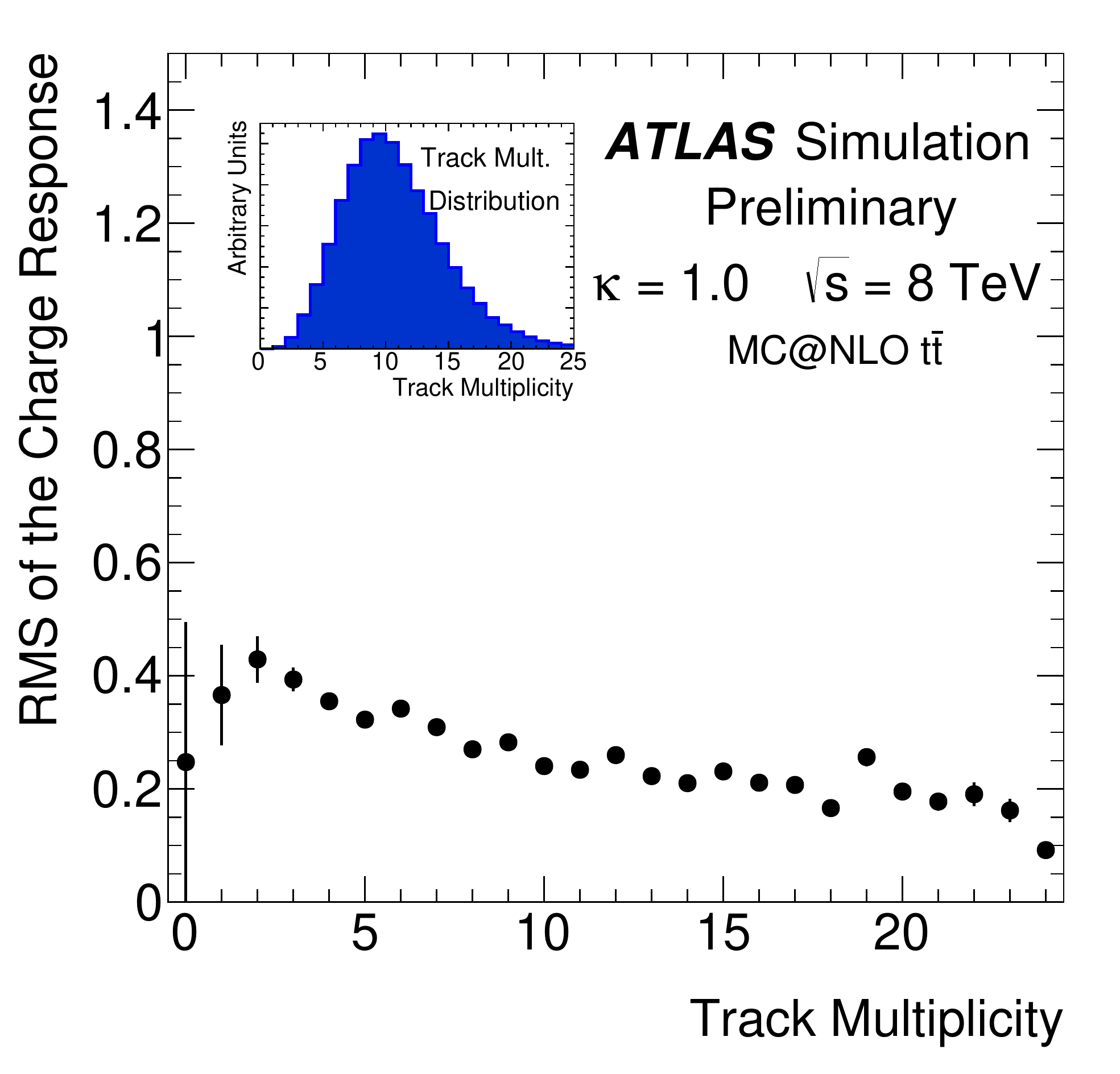}\includegraphics[width=.48\columnwidth]{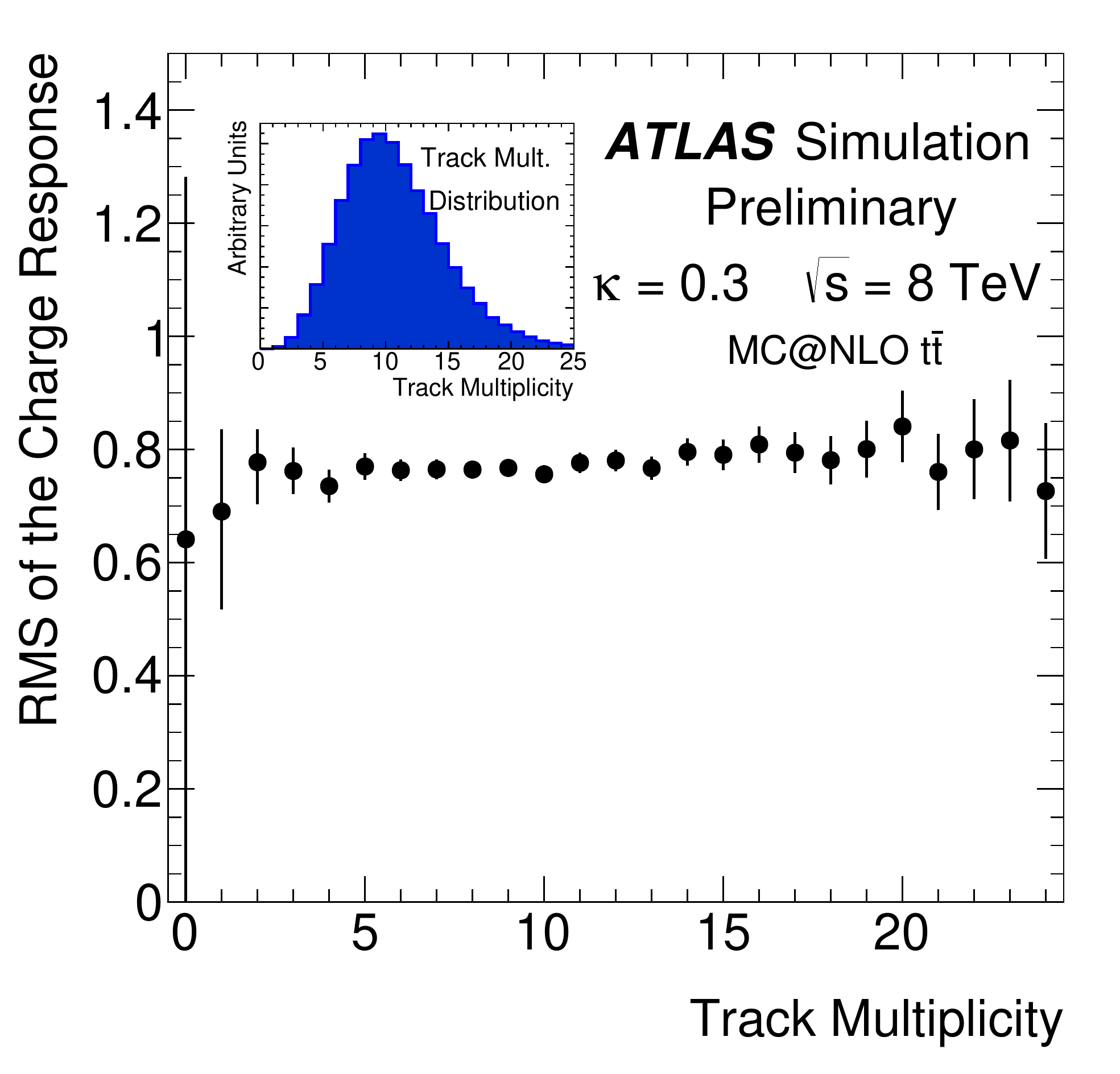}  
   \caption{For $t\bar{t}$ simulated events, the mean (top) and RMS (bottom) distributions of the W daughter dijet charge response as a function of the total number of tracks used to compute the charge for two values of $\kappa$.  The insets show the (arbitrarily normalised) distribution of the number of tracks. The error bars indicate the statistical uncertainties on the number of
MC events.}
 \label{fig:meanvtrack}
\end{figure}

\begin{figure}
 \centering
 \includegraphics[width=.48\columnwidth]{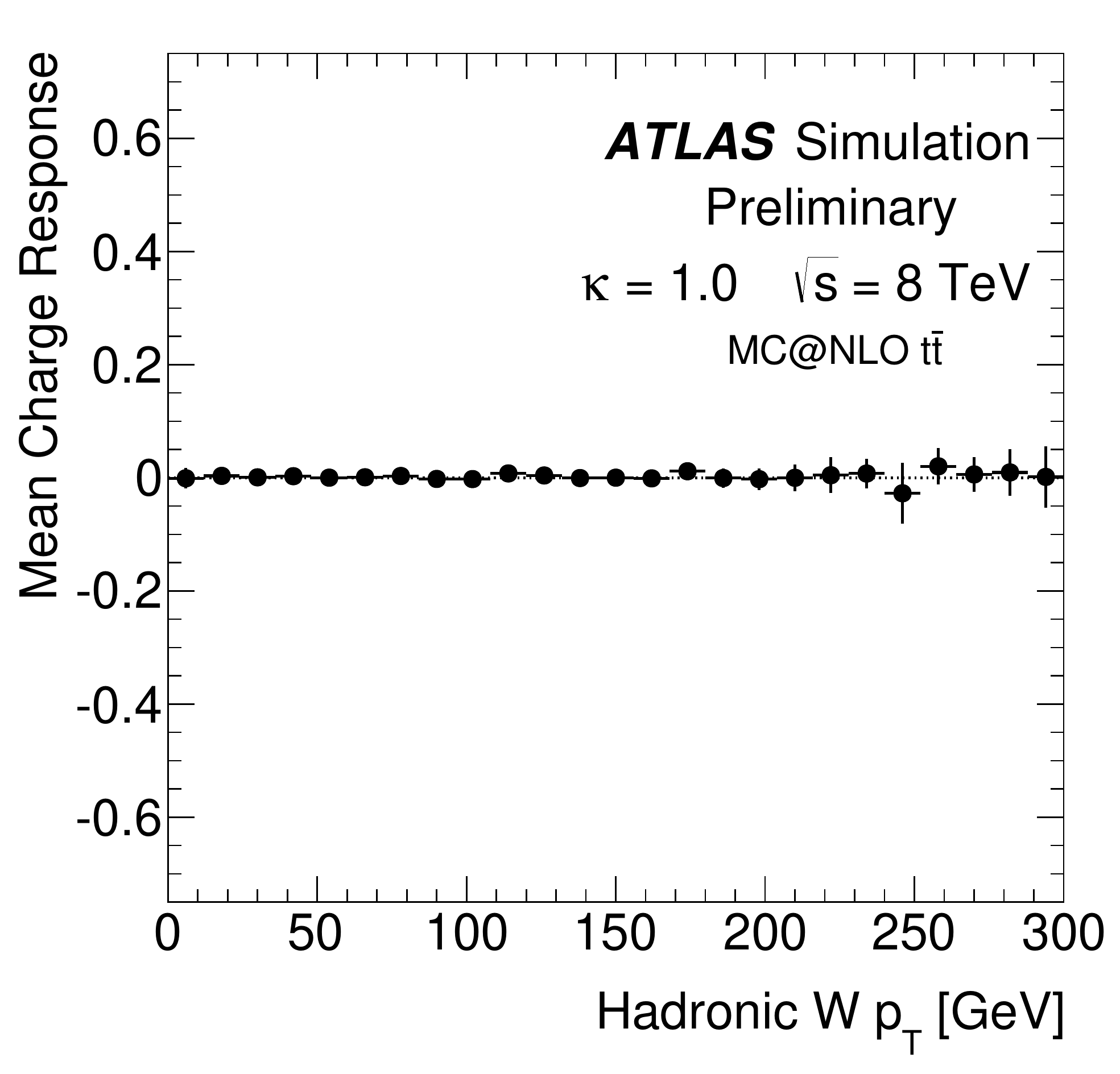} \includegraphics[width=.48\columnwidth]{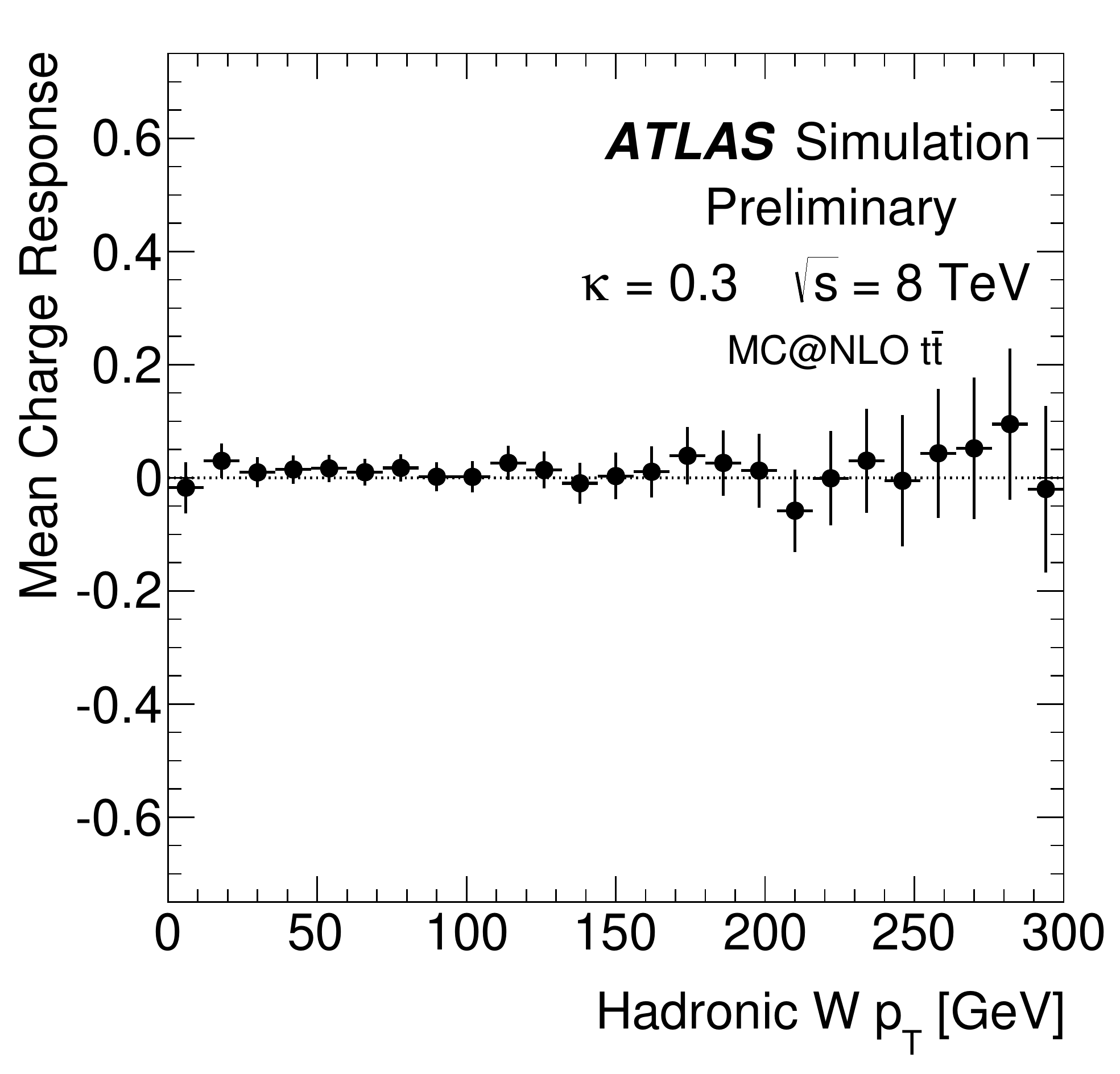}  \\
  \includegraphics[width=.48\columnwidth]{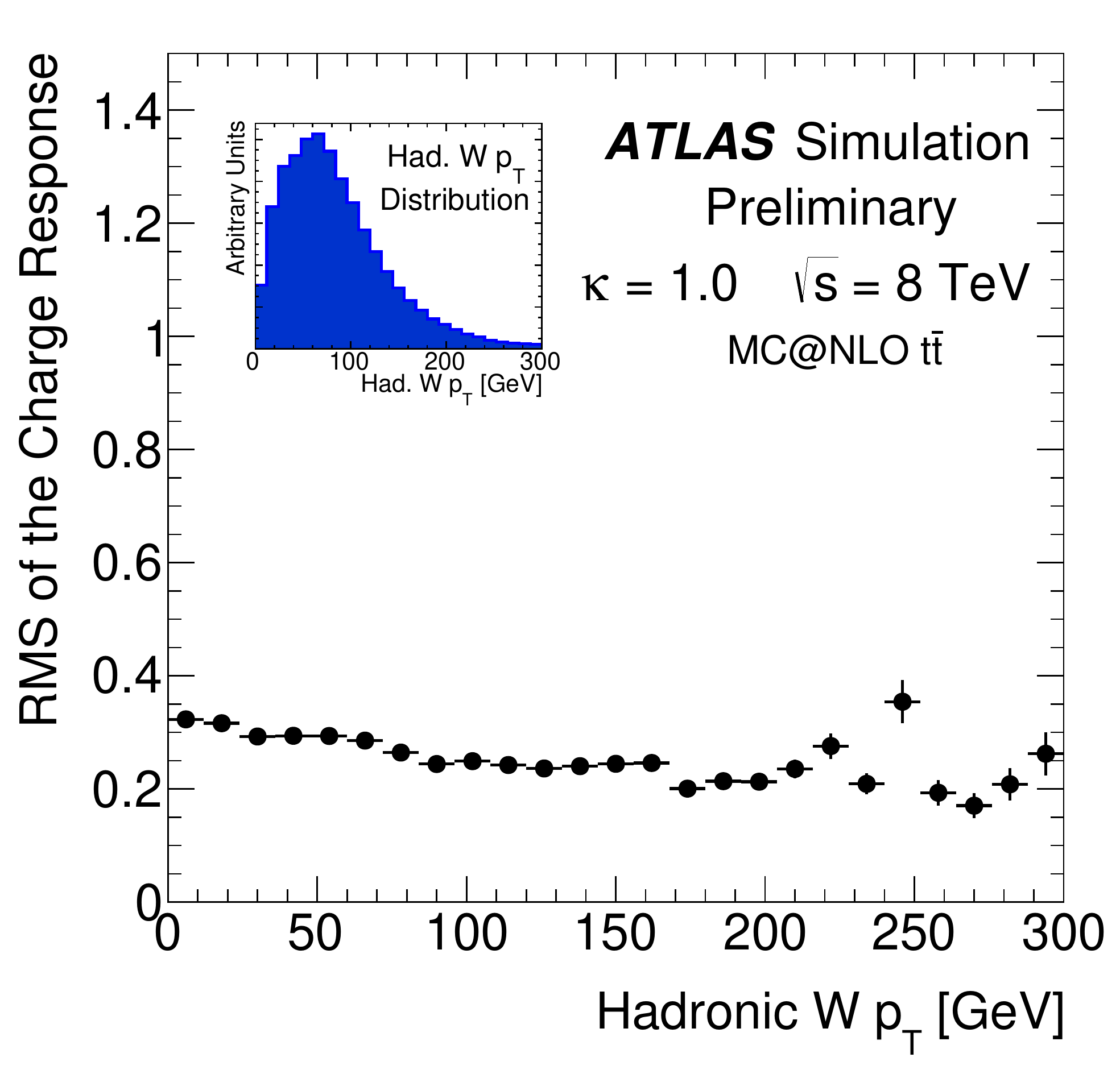} \includegraphics[width=.48\columnwidth]{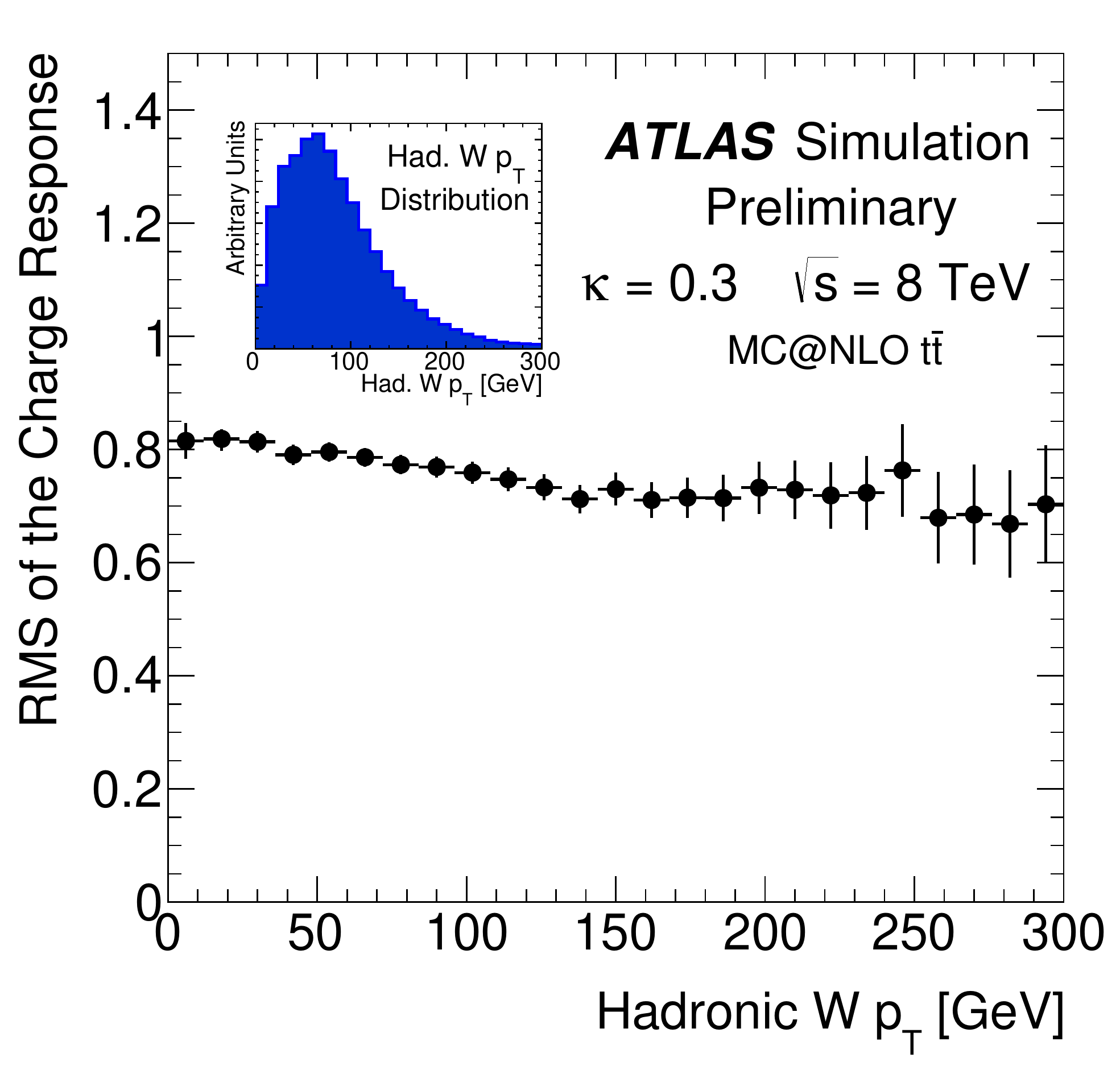}  
   \caption{For $t\bar{t}$ simulated events, the mean (top) and RMS (bottom) distributions of the W daughter dijet charge response as a function of the transverse momentum of the dijet system for two values of $\kappa$. The insets show the (arbitrarily normalised) distribution of the $W$ candidate $p_\text{T}$.  The error bars indicate the statistical uncertainties on the number of
MC events.}
 \label{fig:meanvpt}
\end{figure}

The detector response for jet charge in $W$ events is qualitatively similar to the response studied earlier in generic quark and gluon jets.  The top row of Fig.~\ref{fig:meanvtrack} shows the mean dijet charge response versus the dijet track multiplicity for two values of $\kappa$ in $t\bar{t}$ MC events.  The response is close to zero and constant with respect to the number of tracks.  However, the resolution, parameterized by the distribution RMS, does depend on the number of tracks, as can be seen for the same values of $\kappa$ in the bottom row of Fig.~\ref{fig:meanvtrack}.   The response RMS decreases with the track multiplicity, as in the case of generic quark and gluon jets studied in Sec.~\ref{sec:jetcharge:qcd}.   The top row of Fig.~\ref{fig:meanvpt} shows the mean response as a function of the hadronic $W$ $p_\text{T}$, defined as the transverse momentum of the dijet system formed from the $W$ daughter candidates.  As with track multiplicity, the response is constant around zero (indicating very good agreement between the reconstructed and
the true values), while the RMS (bottom row of Fig.~\ref{fig:meanvpt}) decreases with the W boson $p_\text{T}$.

\clearpage

\subsubsection{Charge Tagging in a Boosted Topology}
\label{charge:boosted}

\begin{figure}[h!]
\begin{center}
 \includegraphics[scale=0.33]{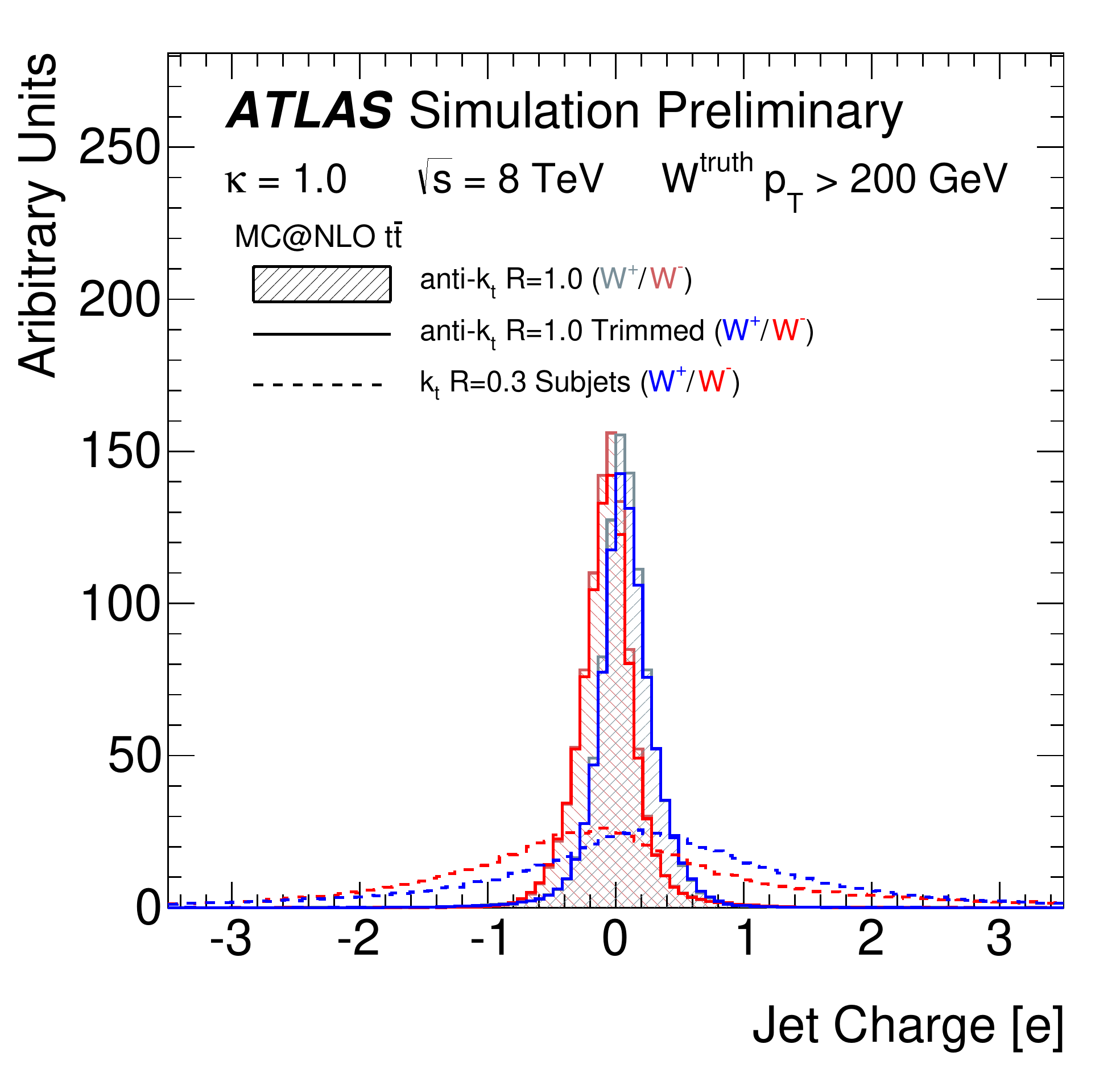}\includegraphics[scale=0.33]{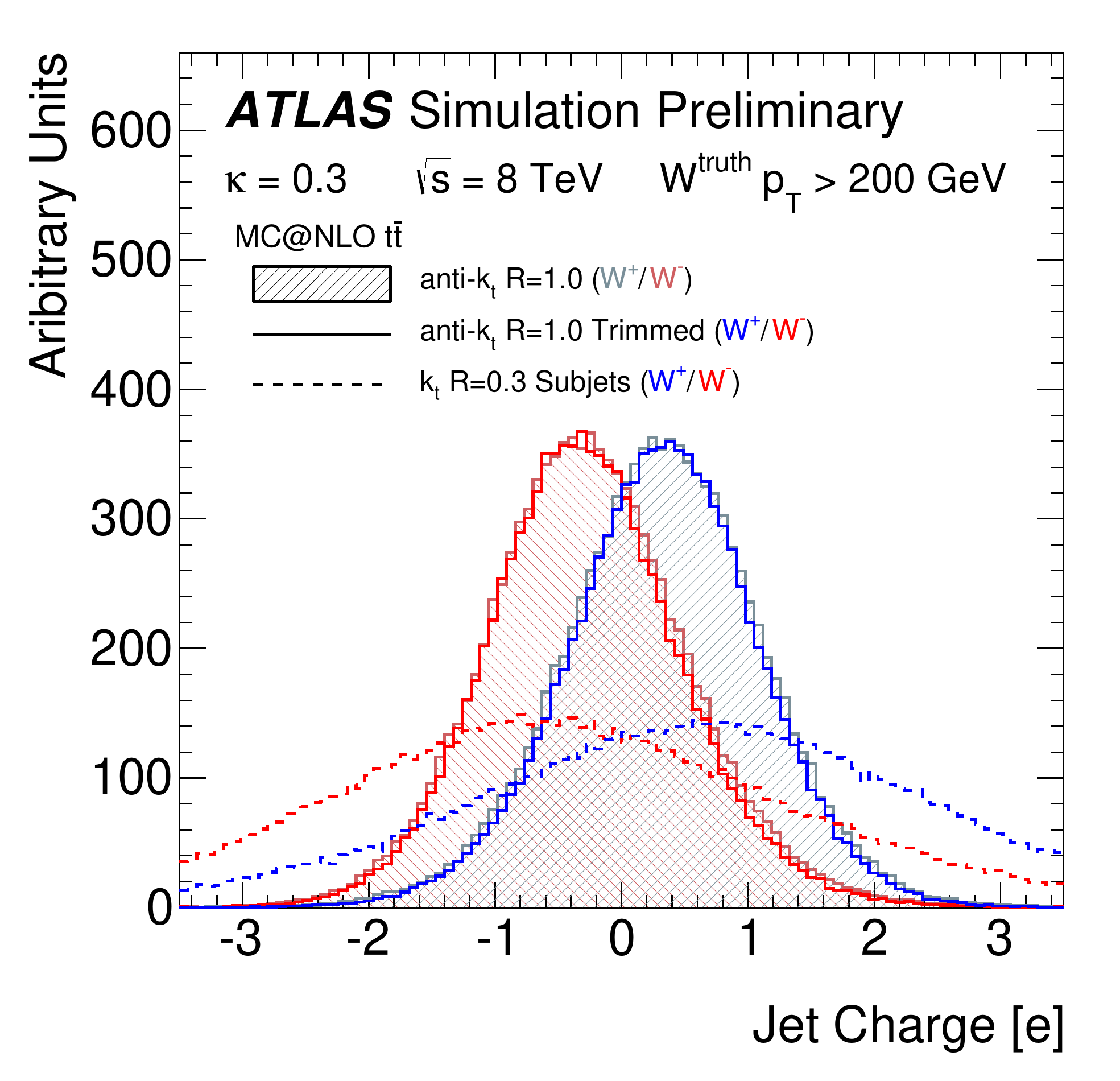}
 \end{center}
 \caption{The charge of a boosted hadronically-decaying $W$ boson in simulated semileptonic $t\bar{t}$ events for $\kappa=1.0$ (left) and $\kappa=0.3$ (right).  The hashed distributions are for the extension of the jet charge definition to large-R jets. The solid line histograms show the distribution of the large-R jet charge after trimming and the dashed lines show the sum of the charge of the two leading $R=0.3$ $k_t$ subjets. Note that the domain of the two jet charge distributions are not the same, but are plotted with the same $x$-axis range, in contrast to Fig.~\ref{fig:distttbar}.}
  \label{large-R}
\end{figure}

In $t\bar{t}$ events, when the hadronic $W$ has a large Lorentz boost, its decay products become merged in the lab frame, obscuring the resolution of the $R=0.4$ jets that are usually associated with the $W$ decay.  In a classical two-body decay of a boosted object,  the separation $\Delta R$ scales as $2m/p_\text{T}$, where $m$ ($p_\text{T}$) is the mass (transverse momentum) of the boosted object; see Chapter~\ref{cha:bosonjets} for more detail.  For a $W$ boson ($m_W$ = 80.4~GeV~\cite{pdg}) with a $p_\text{T}$ of 200~GeV, a $R=1.0$ ({\it large-$R$}) jet often captures most of the hadronic decay products. The jet charge is considered here also in this boosted scenario.  The same {\tt MC@NLO} semileptonic $t\bar{t}$ MC events as described before are used as a source of boosted $W$ bosons.  In each event, the hadronically-decaying $W$ is identified at truth level and its $p_\text{T}$ is required to be above $200$~GeV.  The anti-$k_t$ algorithm is used to cluster the hadronic decay products of the $W$ using an $R=1.0$ radius parameter.   A first definition of jet charge is the simple extension of the procedure described previously: tracks are assigned to the $R=1.0$ jets in the event using ghost association and then Eq.~\ref{chargedefcharge} is used with the large-R (calorimeter) jet $p_\text{T}$ in the denominator.  The distribution of this {\it large-R jet charge} is shown in Fig.~\ref{large-R} for $\kappa=1.0$ and $\kappa=0.3$.  Large-$R$ jets are chosen as the closest $R=1.0$ jet in $\Delta R$ to the truth $p_\text{T}>200$ GeV hadronic $W$ and $\Delta R(\text{jet},W)\leq1.0$ is required.  Since jets are only matched geometrically to the truth $W$ boson, a momentum and mass threshold are imposed: only reconstructed large-R jets with $p_\text{T}>100$ GeV and mass above $30$ GeV are considered.   The jet four-vector ($p$) is corrected for pileup using the area correction~\cite{ghost,areasATLAS} $p\mapsto p-\rho\times A$, where $A$ is the four-vector jet area determined from ghost four-momenta and $\rho$ is the median $p_\text{T}$ density per unit area in $\eta-\phi$ space.

A modification of the large-$R$ jet charge definition can be obtained from {\it trimming}~\cite{Krohn:2009th}.   To form a trimmed large-$R$ jet, first the jet constituent topo-clusters are grouped using the $k_t$ algorithm with a distance parameter of $R=0.3$.  Then, the clusters (and ghosts) of all the subjets that carry less than 5\% of the total jet momentum are removed.  The remaining clusters determine the trimmed jet.  The tracks associated to the trimmed jet are determined by the ghost tracks that remain after subjets removal.  The {\it trimmed large-$R$ jet charge} is defined, as above, by summing over the tracks according to  Eq.~\ref{chargedefcharge}, with the (calorimeter) trimmed jet $p_\text{T}$ in the denominator.  The trimmed large-$R$ jet charge is shown in Fig.~\ref{large-R} for $\kappa=1.0$ and $0.3$ for the same selection as for the untrimmed distribution.  The trimmed and untrimmed jets have similar distributions, with the untrimmed distributions being slightly wider.  The reason why there is not much difference in the charge distributions is that trimming only removes 20\% of tracks, all of which have a low $p_\text{T}^{\text{track}}/p_\text{T}^{\text{jet}}$ weight ($\lesssim \mathcal{O}(1\%)$) and thus do not contribute significantly to the charge.  Trimming removes more than 20\% of calorimeter clusters, but the tracks are required to match to the primary vertex and are thus significantly protected against pileup.

In the process of trimming, $R=0.3$ subjets associated with each large-$R$ jet are clustered together.
This gives rise to a third natural definition of the hadronic $W$ charge: the sum of the charge of the two leading $k_t$ subjets.  Tracks are already matched with subjets from the ghost association to the $R=1.0$ jets.  This {\it subjet charge} is shown in Fig.~\ref{large-R} for $\kappa=1.0$ and $0.3$.  The subjet charge is more spread out than the (trimmed) large-$R$ jet charge.  Part of the stretching is from the definitions.  To see this, consider an example in which the large-$R$ jet transverse momentum $P$ is parallel to the two subjet momenta $p$ and $q$ and assume that $p+q=P$.  Then, $1/p+1/q > 1/(p+q)=1/P$, so the subjet charge will tend to have a larger spread.  

\begin{figure}[h!]
\begin{center}
 \includegraphics[scale=0.33]{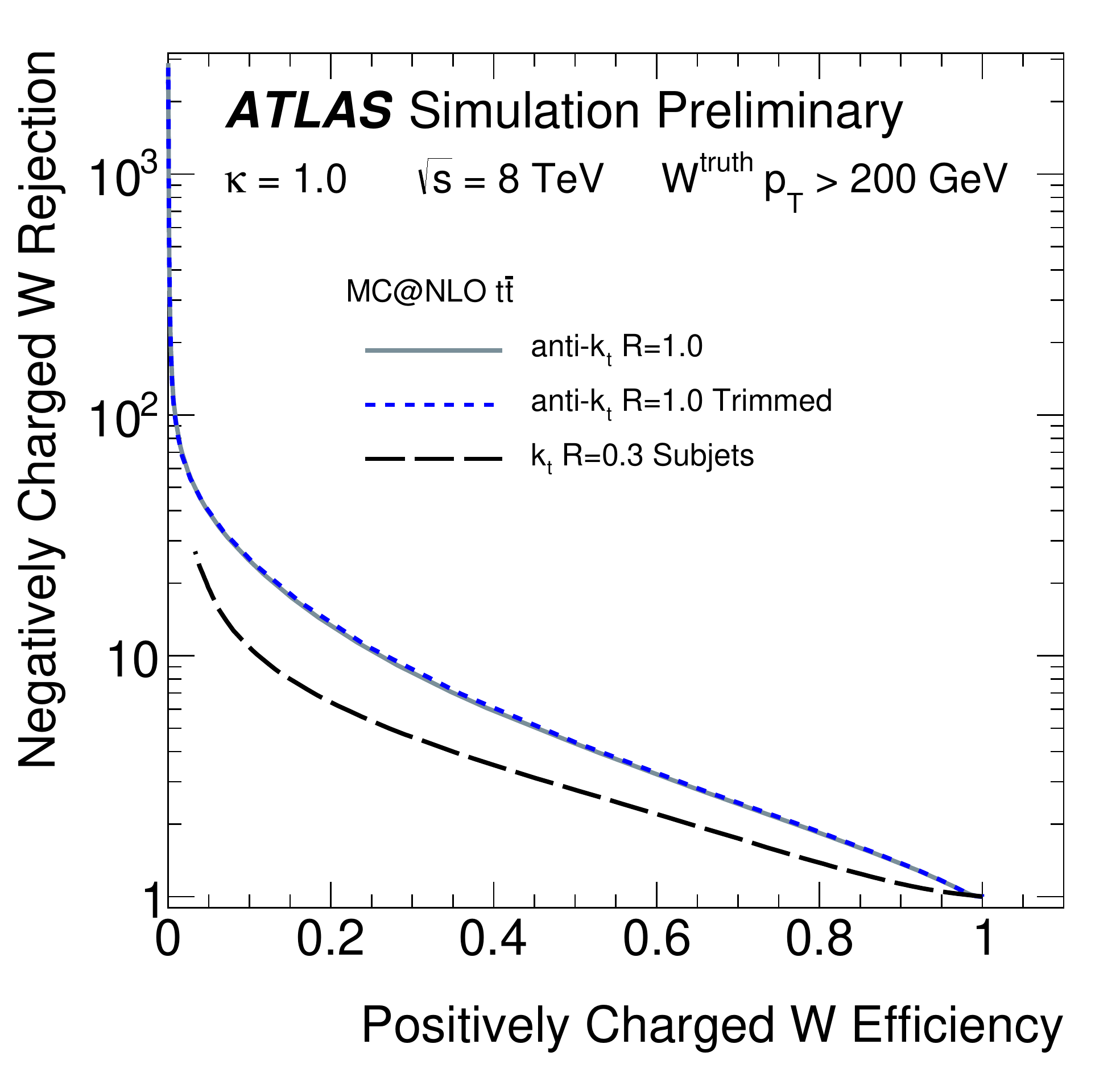}\includegraphics[scale=0.33]{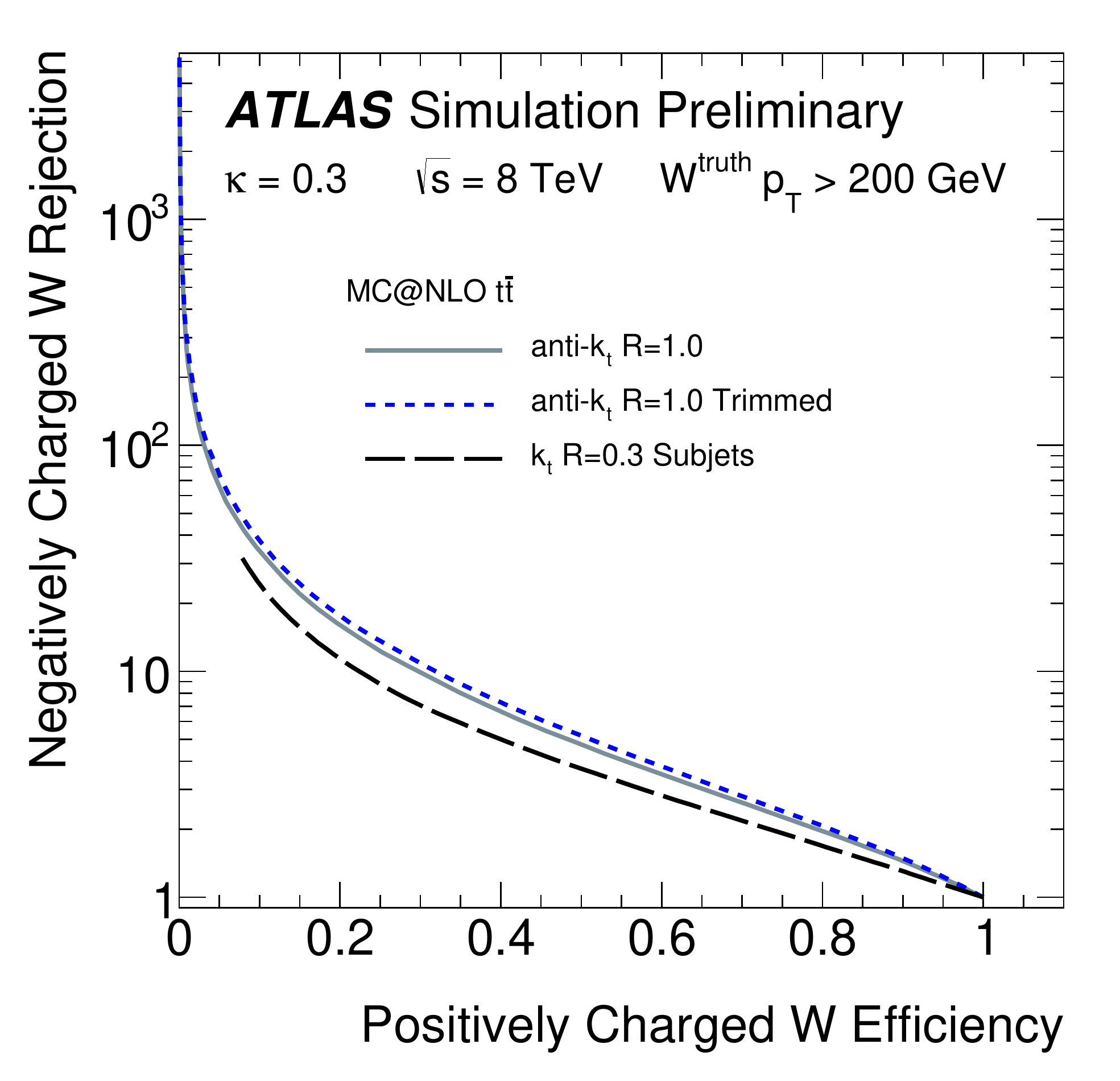}
 \end{center}
  \caption{The inverse efficiency (rejection) of boosted $W^-$ bosons as a function of the efficiency for boosted hadronic $W^+$ bosons for simulated semileptonic $t\bar{t}$ events
and $\kappa=1.0$ (left) and $\kappa=0.3$ (right).  The three curves indicate the performance of three different ways of
measuring the jet charge for boosted $W$ bosons.}
  \label{performlarge-R}
\end{figure}

The performance of hadronic $W$ charge-tagging in the boosted regime is shown in Fig.~\ref{performlarge-R} in terms of  the inverse efficiency (rejection) to identify a $W^-$ as a function of the efficiency to identify a $W^+$.  Since the large-$R$ and trimmed large-$R$ charge distributions are similar, the performance is similar.  For a 50\% $W^+$ efficiency, a factor of four rejection is obtained. The subjet charge performs significantly worse than the (trimmed) large-R jet charge. There are several factors that contribute to the difference in performance.  For example, there are many selected large-$R$ jets with three or more subjets.  In these cases, the jet charge calculation does not include information beyond what is contained in the two leading subjets.   In addition, the decay products of the $W$ may not be fully merged into the $R=1.0$ cone.  The large-$R$ jet charge may take this partial contribution into account, but the subjet charge may miss important information from tracks on the edge of the jet.       

Jet charge for boosted object tagging is revisited in Sec.~\ref{sec:distinguish} in the context of the $W$ versus $Z$ tagger.

\clearpage

\section{Unfolding}
\label{sec:unfolding}

In order to facilitate direct comparisons between the data and particle-level simulations and predictions, it is necessary to remove distortions from detector effects.   Let $h_D$ be the detector-level histogram corresponding to measured values $d_1,...,d_n$ of a particular observable where each event $i=1,...,n$ passed a selection based on detector-level objects.  Furthermore, let $h_T$ be the histogram of particle-level values of another observable $t_1,...,t_m$ for events that pass a particle-level event selection for a particular process.  Unfolding is the name given to an algorithm for estimating $h_T$ given $h_D$.  The name {\it un}folding is used because {\it folding} is a procedure for distorting particle-level measurements to simulate the effects of a detector with finite acceptance and resolution.  Ideally, the detector-level and particle-level definitions of the observables and event selections are close - a notion that will be quantified below.   Intuitively, the stronger the correlation between the detector-level and particle-level object and event selections, the more information the detector-level observations contains about the particle-level quantities and thus the more precise the measurement.  It is useful to focus on measurements of {\it general, well-defined} observables, i.e. that do not depend on the details of the particle-level model or of the particular detector.  For example, parton-level momenta are ill-defined\footnote{This has not stopped many analyses from measuring parton-level quantities, treating fragmentation as a `detector'.  However, the meaning of the measurement depends on the fragmentation model used in the unfolding and thus is not general.} because colored objects are not directly observable and their properties in simulation depend on particular (unphysical) parameter values. When unfolding to events chosen with a non-trivial particle-level selection, the measurement is called a {\it fiducial measurement} and the particle-level selection is called the {\it fiducial volume} of the measurement.  Such measurements are useful from the point of view of unfolding because they require less extrapolation to uninstramented regions of the detector or to unmeasureable kinematic values.  However, fiducial measurements can be a challenge for making theoretical predictions which often require additional assumptions/precision to reduce the calculation to a specified region of phase space.

\clearpage

\noindent In general, unfolding has to correct for many interrelated effects:

\begin{description}
\item[Acceptance and Efficiency] Not every particle produced is measured because of the finite coverage of the detector and even those particles, jets, etc. that are detected are not recorded as such because of quality and identification criteria.  This effect decreases $n$ relative to $m$ (in particular, $n$ and $m$ need not be equal).
\item[Detector Noise] Some of the objects measured in the detector have no particle-level sources.  For instance, tracks can be formed from random hits in the inner detector.  This category also includes jets and tracks from pileup interactions, which do have a particle-level source but not from the hard-scatter particle-level event.  This effect increases $n$ relative to $m$.
\item[Background Processes] A measurement is usually made on a give process and not a particular final state.  For example, one may be interested in a property of $t\bar{t}$ events, so the $W$+jets background needs to be subtracted.
\item[Combinatorics] Objects chosen based on some criteria at detector-level may not correspond to the objects chosen at particle-level based on the same criteria.  For example, the highest $p_\text{T}$ detector-level jet need not originate from the highest $p_\text{T}$ particle-level jet.
\item[Detector Scale] Detector-level quantities are not always unbiased measurements of the corresponding particle-level quantities.  For example, the average jet energy is not exactly the same as the average particle jet energy due to a non-closure in the jet energy scale calibration.
\item[Detector Resolution] The finite resolution of the detector smears out particle-level quantities when measured at detector-level.
\end{description}

\noindent The first two points account for both per-object acceptance and efficiencies as well as the overall (particle- and detector-level) event selection efficiency.

\clearpage

\noindent As a starting point for constructing an unfolding algorithm, consider the {\it folding equation}:

\begin{align}
\label{eq:fold}
h_{D,i} =\sum_{j=1}^m \text{Pr}_{D|T}(i|j) h_{T,j}+h_{F,i},
\end{align}

\noindent where $\Pr_{D|T}(i|j)$ is the probability for an event in bin $h_{T,j}$ to be measured and recorded in bin $i$ of $h_{D}$ and $h_{F}$ is a histogram containing events at detector-level that were not produced by events of the target process passing the particle-level selection.  It is customary to further decompose $h_{F,i}=h_{D,i}f_i+h_{B,i}f_i+h_{B,i}$, where $h_{B}$ is a histogram containing events at detector-level that pass the particle-level selection but originate from a process that is not the target one\footnote{In this chapter, there are no relevant background processes.  However, this will not be true in Chapter~\ref{cha:colorflow}.} and $f_i$ is the {\it fake-factor} that accounts for the fraction of the events at detector-level that do not correspond to particle-level events that pass the selection. 

Letting $R_{ij} =\Pr_{D|T}(i|j)$ and representing the histograms as vectors, the Eq.~\ref{eq:fold} can be written as a matrix equation $\tilde{h}_D=Rh_T$, where $\tilde{h}_D=h_D-h_F$ or equivalently, $\tilde{h}_{D,i}=(1-f_i)(h_{D,i}-h_{B,i}$).  The matrix $R$ is called the {\it response matrix} and is estimated from simulation.  In the matrix form, one may be tempted to solve for $h_T=R^{-1}\tilde{h}_D$.  However, even if $R$ is a square matrix and is invertible, $R^{-1}\tilde{h}_D$ may not be the best estimator for $h_T$ because matrix inversion can enhance statistical fluctuations in both $h_D$, due to a finite dataset, and $R$, due to a finite simulation, when there are significant off-diagonal transition probabilities in $R$.  For example, consider a simple response matrix 

\begin{align}
R=\begin{pmatrix}1-\epsilon & \epsilon \cr \epsilon & 1-\epsilon \end{pmatrix},
\end{align}

\noindent where $0\leq\epsilon < 0.5$ in order to make the matrix invertible by satisfying $\text{Det}(R)=1-2\epsilon>0$.   The problem is that the variance of $R^{-1}\tilde{h}_D$ is proportional to $1/\text{Det}(R)$, which diverges as $\epsilon\rightarrow 0.5$.  Ideally, $\epsilon$ is as small as possible, but there are many cases where it is not small compared to $0.5$ due to a large detector resolution.  As an alternative method\footnote{There are other unfolding techniques that address the matrix inversion challenge with alternative techniques.  One common alternative to the Bayesian method is to apply regularized singular value decomposition (SVD) to the response matrix~\cite{Hocker:1995kb}.  This Bayesian method is used exclusively for the rest of this chapter.}, consider the following Bayesian approach~\cite{D'Agostini:1994zf}.  Using the law of total probability ($T=$ truth, $D=$ detector):

\begin{align}
h_{T,i} = \sum_{j=1}^n \text{Pr}_{T|D}(i|j) \tilde{h}_{D,j}
\end{align}

\noindent Then, the probability $\Pr_{D|T}(i|j)$ can be inverted using Bayes Theorem:

\begin{align}
h_{T,i} = \sum_{j=1}^n \frac{\text{Pr}_{D|T}(i|j)\text{Pr}_T(i)}{\sum_{i'}\text{Pr}_{D|T}(i'|j)\text{Pr}_T(i')} \tilde{h}_{D,j} = \sum_{j=1}^n \frac{R_{ij}\bar{h}_{T,i}^1}{R\bar{h}_T^1} \tilde{h}_{D,j}\equiv B_1\tilde{h}_D,
\end{align}

\noindent where $\bar{h}_T^1$, with $\sum_{i=1}^n\bar{h}_{T,i}^1=1$, is a {\it prior density} for $h_T$.   By construction, when $\bar{h}_T^1\propto h_T$, the solution $B_1\tilde{h}_D$ is unbiased (when it exists, $R^{-1}\tilde{h}_D$ is also unbiased).  In the example above, the matrix $B_1$ is given by

\begin{align}
B_1= \begin{pmatrix}\frac{(1-\epsilon)\bar{h}_{T,1}^1}{(1-\epsilon)\bar{h}_{T,1}^1+\epsilon\bar{h}_{T,2}^1} & \frac{\epsilon\bar{h}_{T,1}^1}{\epsilon\bar{h}_{T,1}^1+(1-\epsilon)\bar{h}_{T,2}^1} \cr \frac{\epsilon \bar{h}_{T,2}^1}{\epsilon\bar{h}_{T,2}^1+(1-\epsilon)\bar{h}_{T,1}^1} & \frac{(1-\epsilon)\bar{h}_{T,2}^1}{(1-\epsilon)\bar{h}_{T,2}^1+\epsilon\bar{h}_{T,1}^1} \end{pmatrix}.
\end{align}

\noindent An important property of $B_1(\epsilon)$ is that its components remain finite as $\epsilon\rightarrow0.5$ and thus the variance of the estimate also remains finite in this limit.  The price paid when the off-diagonal elements of $R$ are large is a dependence on the prior density $\bar{h}_T$.  This can be mitigated by {\it iterating} the above procedure.  Let

\begin{align}
B_k = \sum_{j=1}^n \frac{R_{ij}\bar{h}_{T,i}^{k-1}}{R\bar{h}_T^{k-1}}\hspace{5mm}\text{and}\hspace{5mm} \bar{h}_{T}^{k} = B_{k-1}\tilde{h}_{D}.
\end{align}

\noindent Then the {\it Iterative Bayesian Unfolding Method} (IB) estimates $h_{T} = B_k \tilde{h}_D$.  The free parameter $k$ is the number of iterations and is a regularization parameter that, like $\bar{h}_T^1$, must be chosen ahead of time.  The prior $\bar{h}_{T}^1$ is chosen to be the particle-level distribution in simulation and the number of iterations is chosen to compromise between bias and statistical uncertainty.  One further step is required because there are some events that pass the particle-level selection but are not measured by the detector-level selection.  This is accounted for by an {\it efficiency factor} $\epsilon_i$ that represents the fraction of events in bin $i$ that pass the particle-level selection but not the detector-level selection.  As was the case with the fake factor $f_i$, the efficiency factor $\epsilon_i$ is estimated from simulation.  The unfolding procedure used for the rest of this chapter is summarized as follows:

\begin{enumerate}
\item Using simulation, estimate the response matrix $R$, the fake factors $f$, the efficiency factors $\epsilon$, and the particle-level spectrum $\bar{h}_T^1$ used as the initial prior.
\item Correct the data for background processes and noise via 

$$h_{D,i}\rightarrow\tilde{h}_{D,i}=(1-f_i)(h_{D,i}-h_{B,i}).$$

\item Estimate the particle-level histogram $h_{T}$ by iteratively applying the Bayes method described above:

$$h_{T,i}= \frac{1}{\epsilon_i}\sum_{j=1}^m B_{k,ij}\tilde{h}_{D,j}.$$

\item Estimate the uncertainty based on all the inputs from simulation and the finite statistics of the data.  These are described in more detail in Sec.~\ref{sec:uncerts}.

\end{enumerate}

\clearpage

\subsection{Unfolding for the Jet Charge}
\label{sec:unfoldjetcharge}

The unfolding procedure described in the previous section can be readily extended to histograms of multiple dimensions.  The primary interest for the jet charge measurement is to extract the particle-level dependence of the jet charge spectrum on the particle-level jet $p_\text{T}$.  Since the jet charge and the jet $p_\text{T}$ are not independent, it is important to simultaneously unfold them.  A simple way to extend the IB method for this case is to transform the two-dimensional jet charge and jet $p_\text{T}$ distribution into a one-dimensional histogram.  This is accomplished as follows:

\begin{enumerate}
\item Bin the jet charge and jet $p_\text{T}$ distributions.  The jet charge bin centers are $Q_i,i=1,...,N$ and the jet $p_\text{T}$ bin centers are $P_i,i=1,...,M$.  In general, the jet charge and jet $p_\text{T}$ bins can vary in size and the jet charge bin size can vary as a function of jet $p_\text{T}$.  The number of jet charge bins is the same for all jet $p_\text{T}$ bins.  Binning is described in more detail in Sec.~\ref{sec:JetCharge:binning}.
\item Define the integer map $(i,j)\mapsto z(i,j)=N(i-1)+j$, for jet charge bin $i$ and jet $p_\text{T}$ bin $j$.  Transform the two-dimensional histogram of jet charge and jet $p_\text{T}$ $h_{ij}^\text{2D}$ into a one-dimensional histogram via $h_{ij}^\text{2D}\mapsto h^\text{1D}_{z(i,j)}$.
\item Use the IB unfolding algorithm described in Sec.~\ref{sec:unfolding} to unfold $h^\text{1D}_{z}$.
\item Transform the unfolded one-dimensional histogram back into a two-dimensional histogram via the inverse integer map $i=z\text{ mod } N$ and $j=(z-i)/N+1$ for jet charge bin $i$ and jet $p_\text{T}$ bin $j$.
\end{enumerate}

\begin{figure}[h!]
\begin{center}
\includegraphics[width=0.6\textwidth]{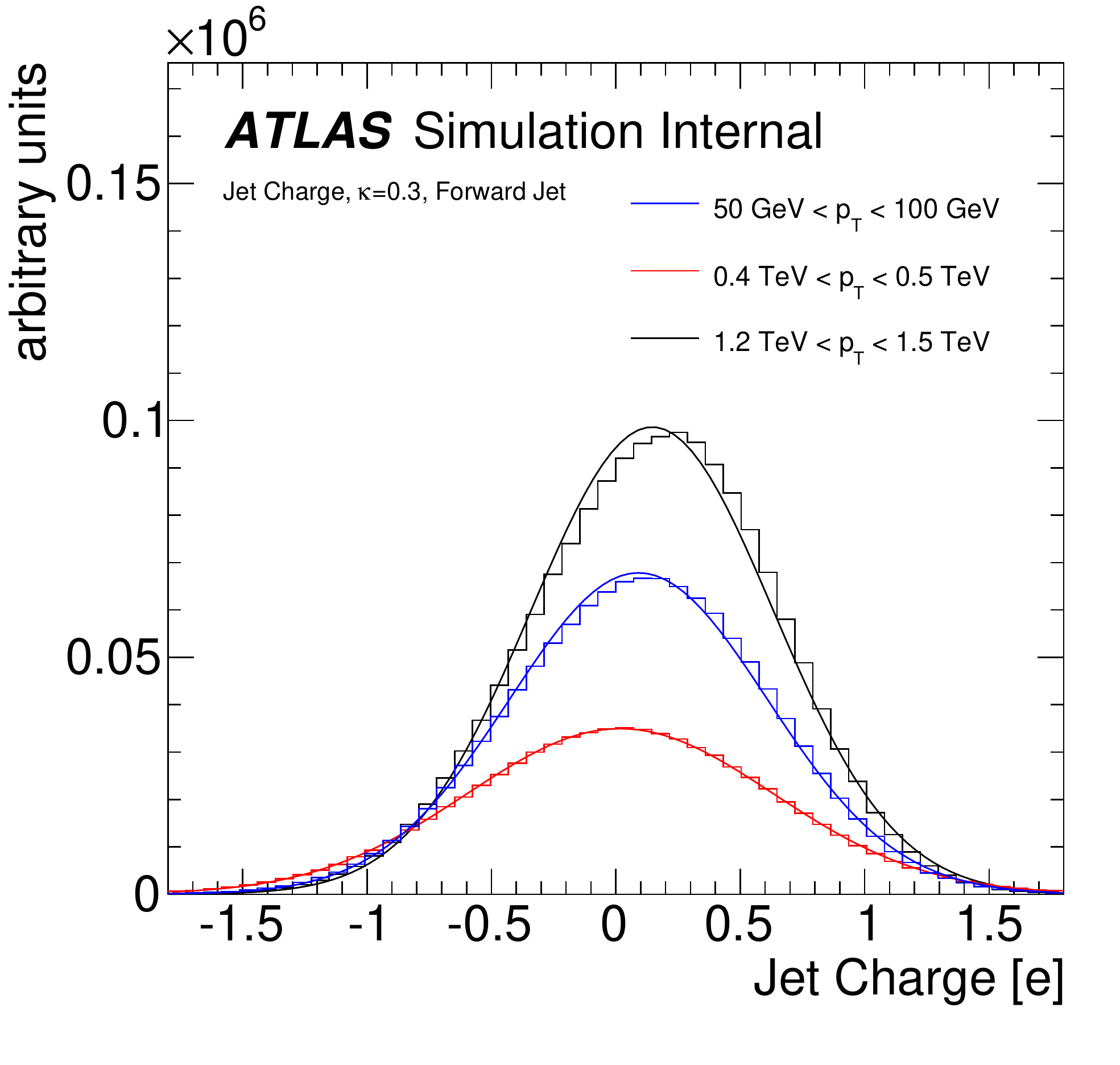}
\caption{A Gaussian fit to the particle-level jet charge distribution in three bins of jet $p_\text{T}$ for the more forward of the two selected jets and $\kappa=0.3$.  The distribution is not exactly Gaussian, which is why the fitted curve is slightly shifted with respect to the histograms.}
\label{fig:JetCharge:Unfold:fittomoments}
\end{center}
\end{figure}

\noindent The jet charge distribution in a fixed jet $p_\text{T}$ bin is nearly Gaussian and so most of the information in the particle-level distribution is contained in the mean and standard deviation.  Figure~\ref{fig:JetCharge:Unfold:fittomoments} shows the particle-level jet charge distribution in three jet $p_\text{T}$ bins along with a $\chi^2$ minimization to Gaussian distributions.  Therefore, instead of measuring the full two-dimensional distribution of the jet charge and the jet $p_\text{T}$, the focus is on the jet $p_\text{T}$ dependence of the jet charge distribution mean and standard deviation. 

\clearpage

\noindent The jet charge distribution average and standard deviation are extracted from the above procedure using Eq.~\ref{eq:JetCharge:extractmoments}:

\begin{align}
\label{eq:JetCharge:extractmoments}
\langle Q_\text{jet}\rangle_i&=\frac{\sum_{j=1}^N n_jQ_j}{\sum_{j=1}^N n_j}\\
\sigma^2_{Q_\text{jet}}&=\frac{\sum_{j=1}^N n_jQ_j^2}{\sum_{j=1}^N n_j}-\langle Q_\text{jet}\rangle_i^2,
\end{align}

\noindent where $i=1,..,M$ is the jet $p_\text{T}$ bin and $n_i$ is the content of jet charge bin $i$ in jet $p_\text{T}$ bin $j$.  One of the important considerations for choosing the binning is to reduce the bias that the discritization procedure introduces in estimating the above moments, i.e.  the difference between $\langle Q_\text{jet}\rangle_i$ at particle level and the true mean in jet $p_\text{T}$ bin $i$.

\clearpage

\subsubsection{Binning}
\label{sec:JetCharge:binning}

\begin{figure}[h!]
\begin{center}
\includegraphics[width=0.5\textwidth]{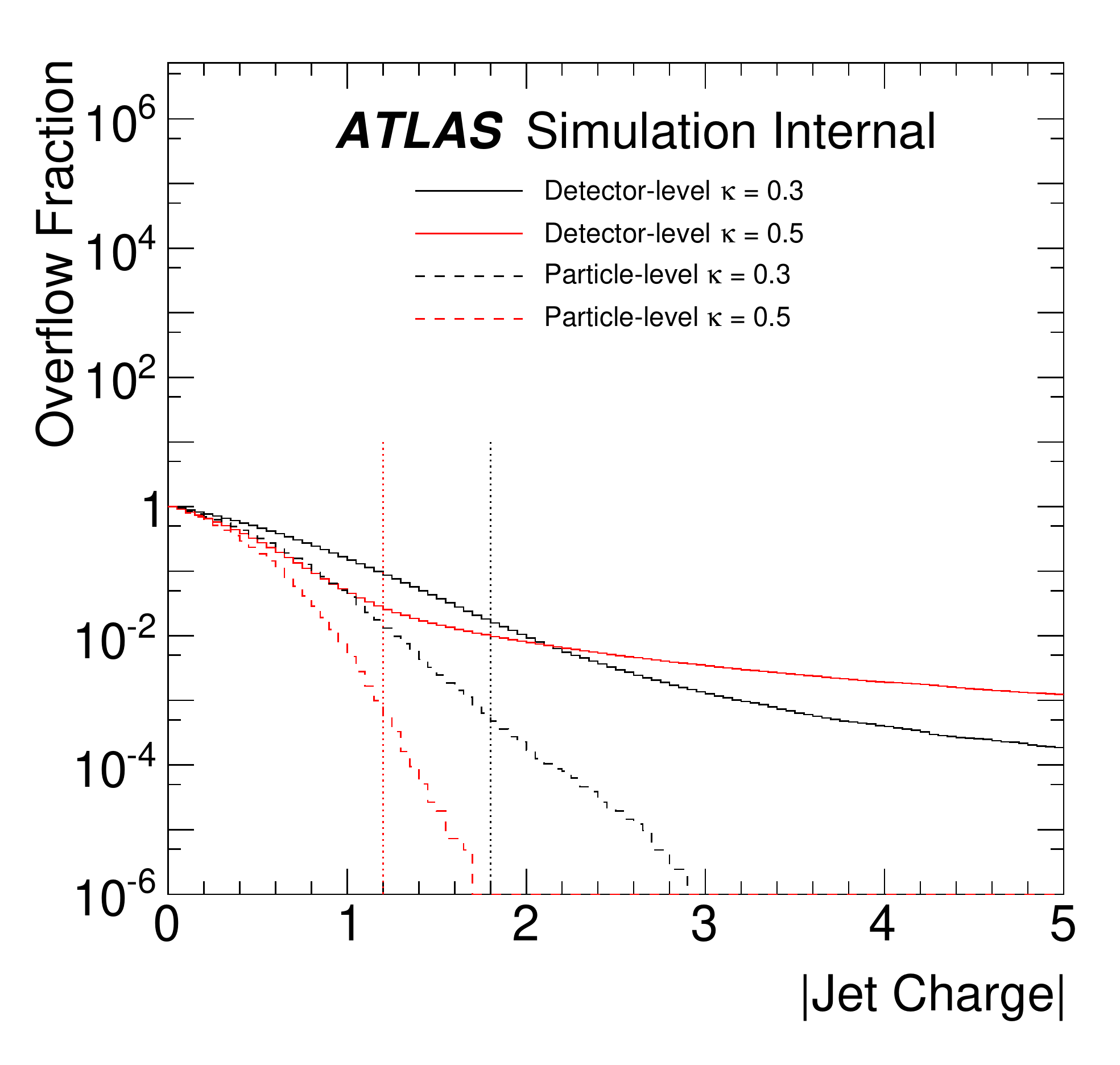}
\caption{Various binning choices for the quantization of jet charge and jet $p_\text{T}$.  The overflow fraction is the fraction of events with a jet charge outside the range set by the $x$-axis.  Vertical lines indicate the values used for $\kappa=0.3$ and $\kappa=0.5$.}  \label{fig:binning0}
\end{center}
\end{figure}

Increasing the number of bins reduces the bias from discretization, at the cost of decreasing the unfolding stability and increasing the uncertainty.  The jet charge and jet $p_\text{T}$ binning are chosen independently.  Due to the falling $p_\text{T}$ spectrum the bins in jet $p_\text{T}$ increase in size as a function of $p_\text{T}$: $[50,100)$, $[100,200)$, $[200,300)$, $[300,400)$, $[400,500)$, $[500,600)$, $[600,800)$, $[800,1000)$, $[1000,1200)$, $[1200,\infty)$ GeV.  For displaying the $p_\text{T}$ dependence of the jet charge distribution moments, the jet $p_\text{T}$ bin mid-point is used to represent the $p_\text{T}$\footnote{Another approach is to simultaneously unfold the $p_\text{T}$ distribution with a finer binning in order to place the measured value at the mean of the $p_\text{T}$ distribution.  This is mostly an aesthetic change and was implemented, but the overhead was sufficiently cumbersome that it was dropped for the simpler presentation.}.  The jet charge is steeply falling away from the mean, but it is important to have fine binning to reduce the discretization bias.  Jet charge bins are equally spaced in the range $|Q_\text{jet}|<1.8$ for $\kappa=0.3$, $|Q_\text{jet}|<1.2$ for $\kappa=0.5$, and $|Q_\text{jet}|<0.9$ for $\kappa=0.7$.  These values are roughly chosen to reduce the overflow fraction to less than about 1\% as shown in Fig.~\ref{fig:binning0}.  Events with a jet charge larger than this upper value are placed in the last bin.

\begin{figure}[h!]
\begin{center}
\includegraphics[width=0.45\textwidth]{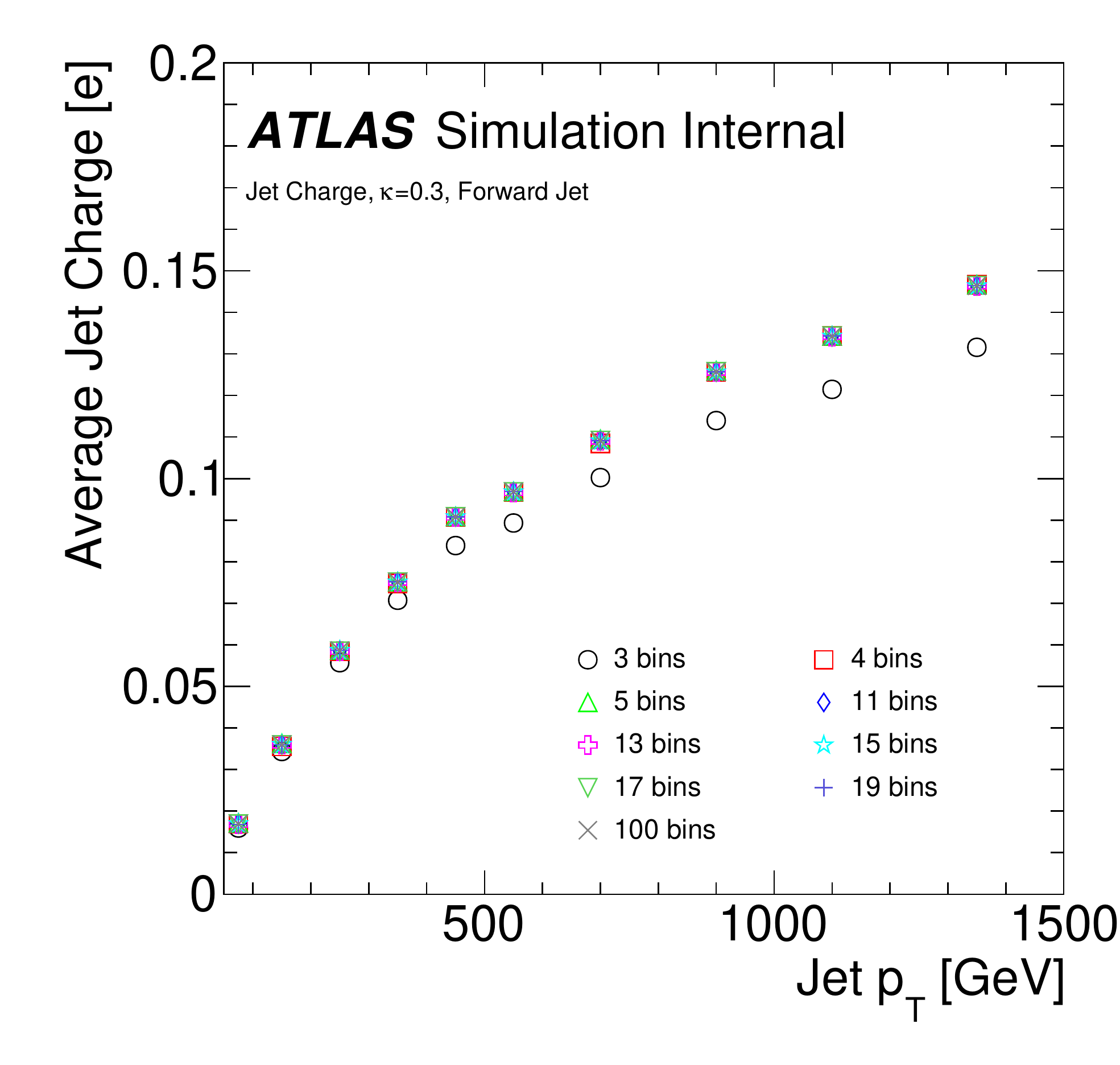}\includegraphics[width=0.45\textwidth]{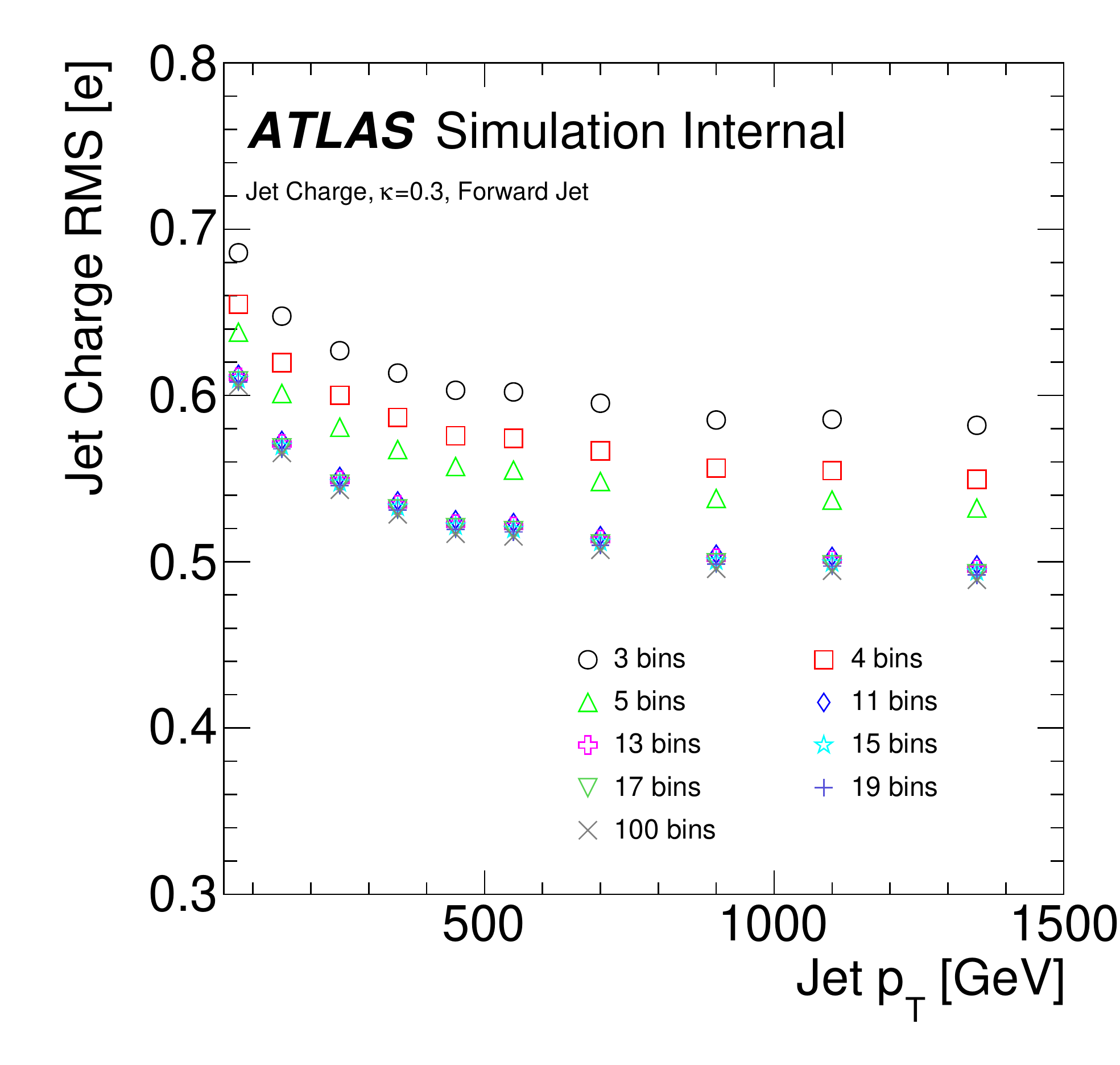}
\caption{Various binning choices for the quantization of jet charge and jet $p_\text{T}$ for the jet charge distribution average (left) and standard deviation (right). For the RMS, there is a significant dependence up to about $10$ iterations, which is why only a few of the small-iteration cases are shown. }
\label{fig:binning1}
\end{center}
\end{figure}

\begin{figure}[h!]
\begin{center}
\includegraphics[width=0.45\textwidth]{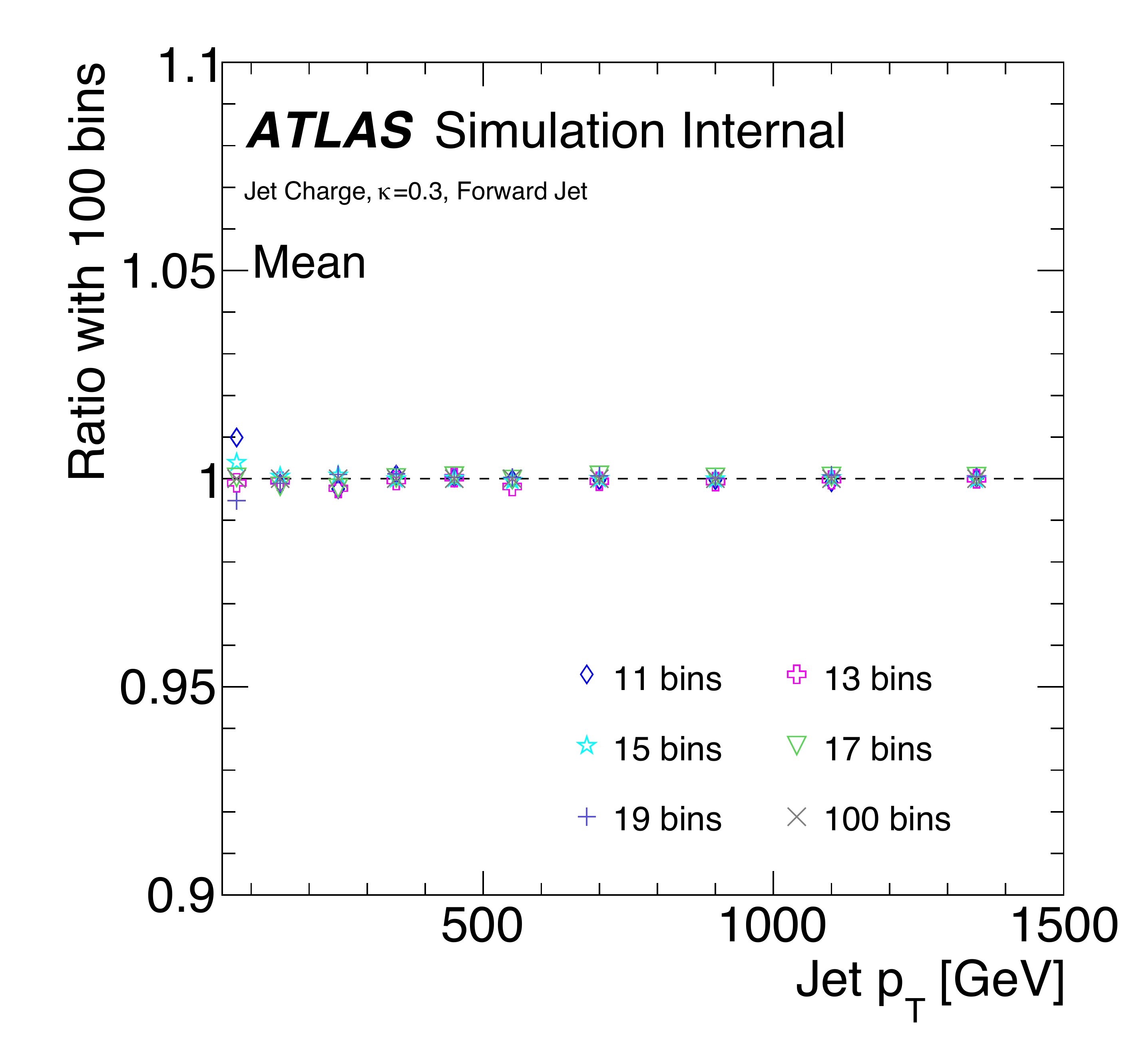}\includegraphics[width=0.45\textwidth]{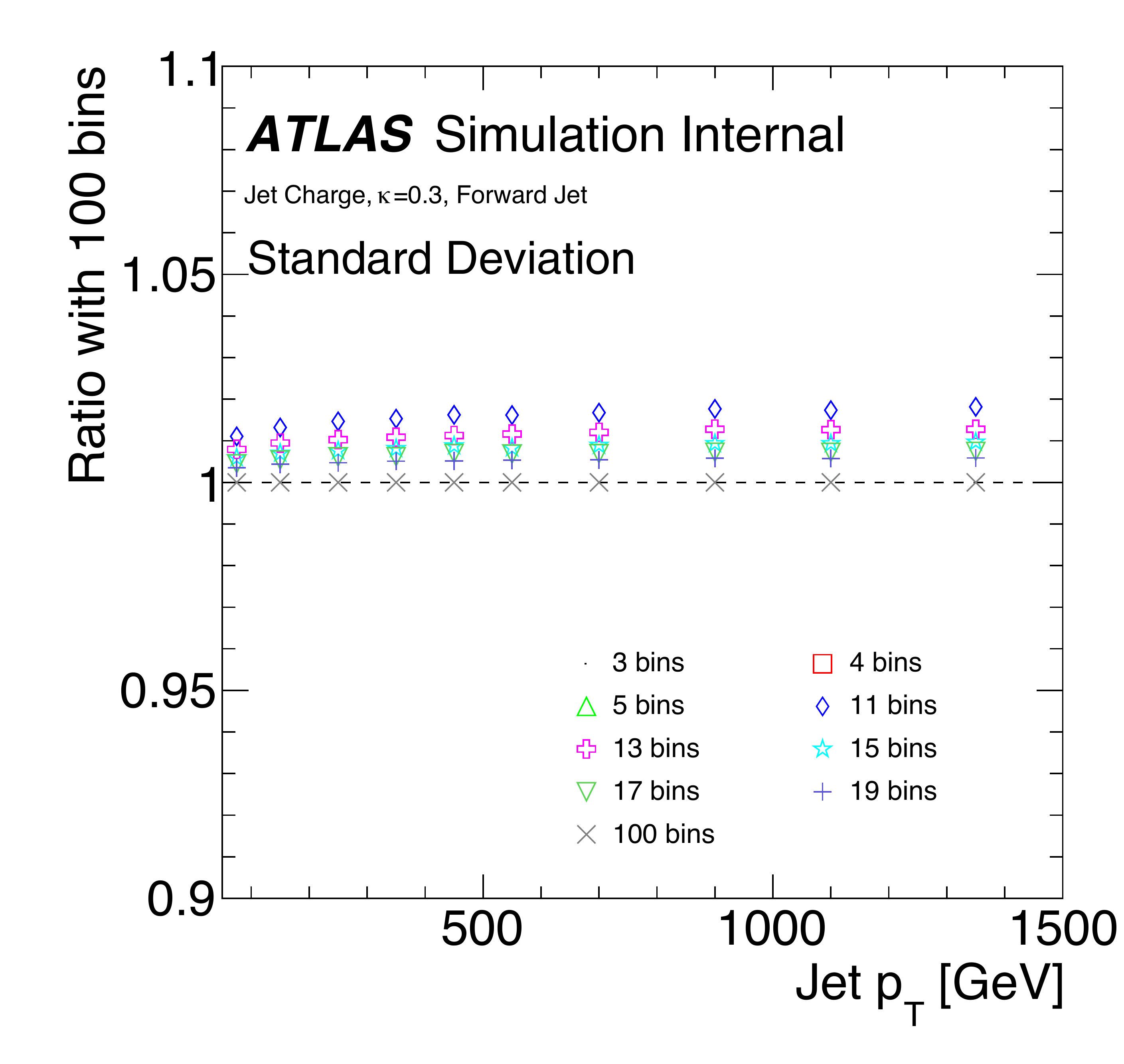}
\caption{The discretization bias for various binning choices for the quantization of jet charge and jet $p_\text{T}$ for the jet charge distribution average (left) and standard deviation (right).}
\label{fig:binning2}
\end{center}
\end{figure}

An optimization for the number of jet charge bins is demonstrated with Fig.~\ref{fig:binning1} and~\ref{fig:binning2}.  The average jet charge is largely insensitive to the number of bins, as long as there are multiple bins on either side of zero.  In contrast, the standard deviation of the jet charge distribution is significantly sensitive to the number of bins\footnote{One can in principle correct for this bias, but the correction depends on the distribution within a bin.  If the bin sizes are small, this renders the impact of any systematic uncertainties on the shape subleading; however if the bins are small the need for a correction is also negligible. }.  For $15$ bins, the discretization bias is $\lesssim 1\%$; this is the value that is used for the remainder of the analysis.

\begin{figure}[h!]
\begin{center}
\includegraphics[width=0.45\textwidth]{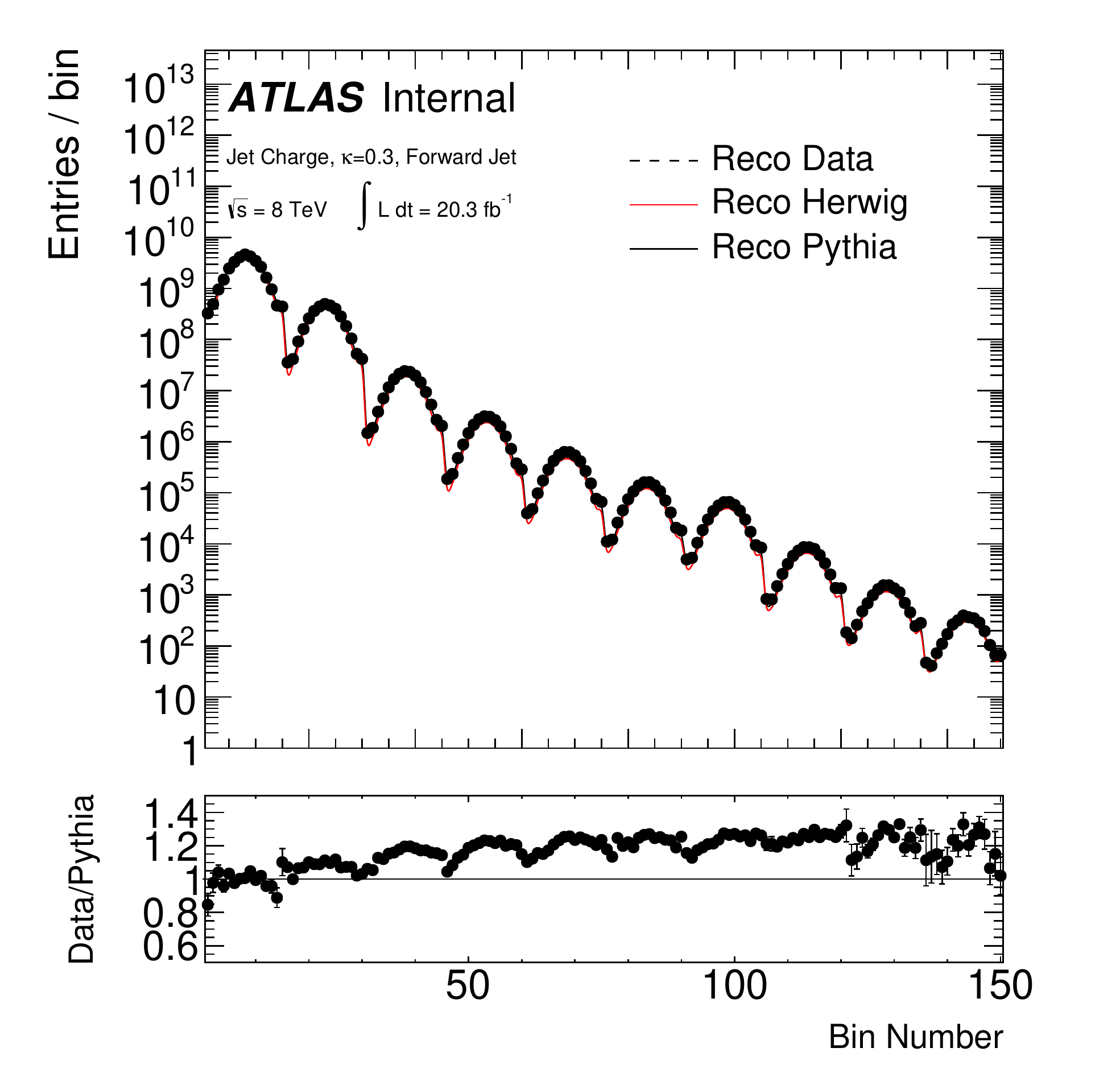}
\caption{The distribution of the one-dimensional transformation of the jet charge and the jet $p_\text{T}$ (see Sec.~\ref{sec:unfoldjetcharge}) for $\kappa=0.5$ for the more forward jet.  All distributions are scaled to have the same normalization.}
\label{fig:2Dto1D_recocharge}
\end{center}
\end{figure}

For the given binning choices, Fig.~\ref{fig:2Dto1D_recocharge} shows the one-dimensional transformation of the joint binned distribution of the jet charge and the jet $p_\text{T}$.  The bumps represent individual jet $p_\text{T}$ bins and the general decreasing trend is due to the steeply falling jet $p_\text{T}$ spectrum.  There are $150$ bins in total which are the input to the unfolding algorithm.

\clearpage

\subsubsection{Correction Factors}
\label{corrrfactors}

\begin{figure}[h!]
\begin{center}
\includegraphics[width=0.45\textwidth]{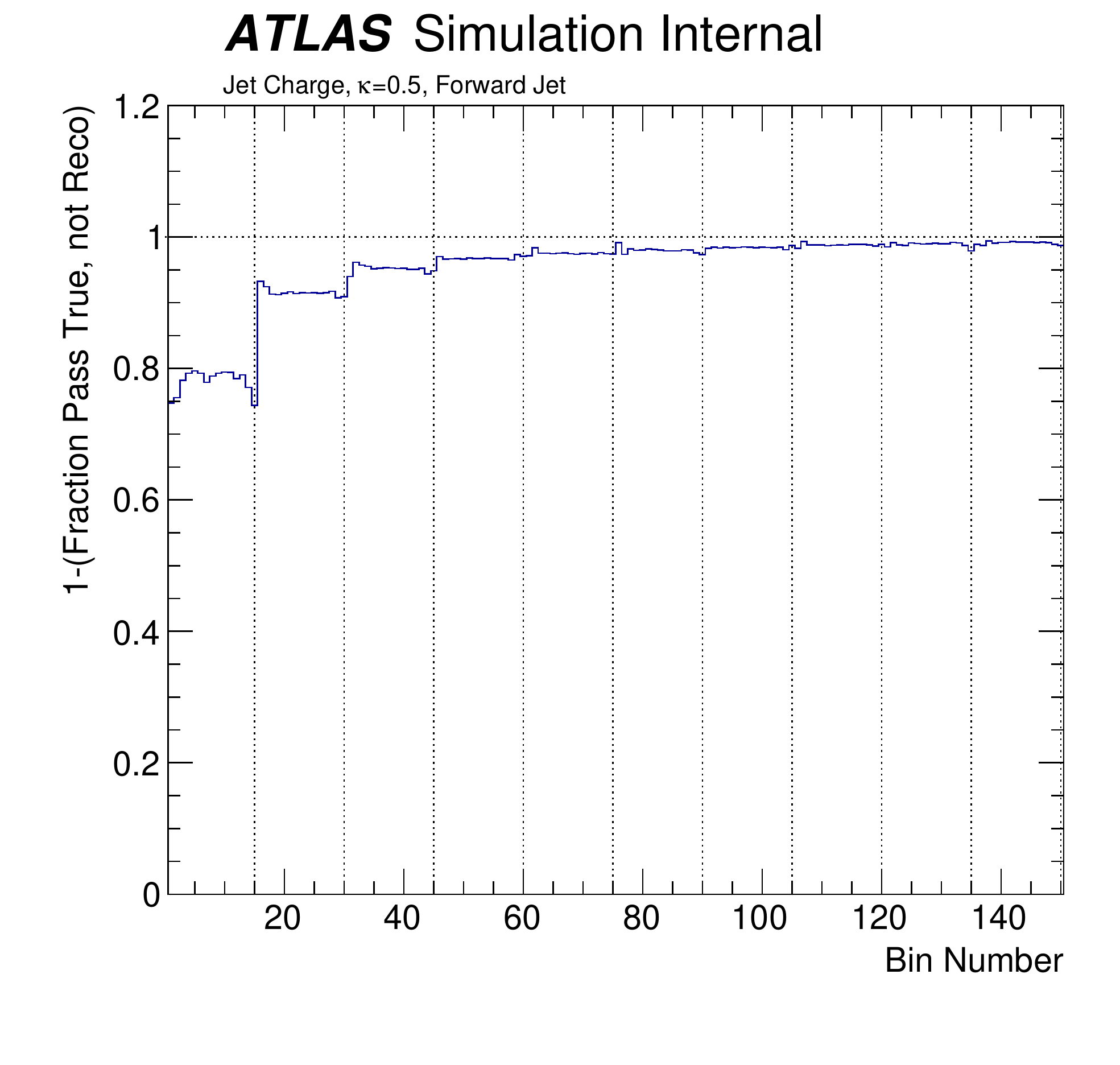}\includegraphics[width=0.45\textwidth]{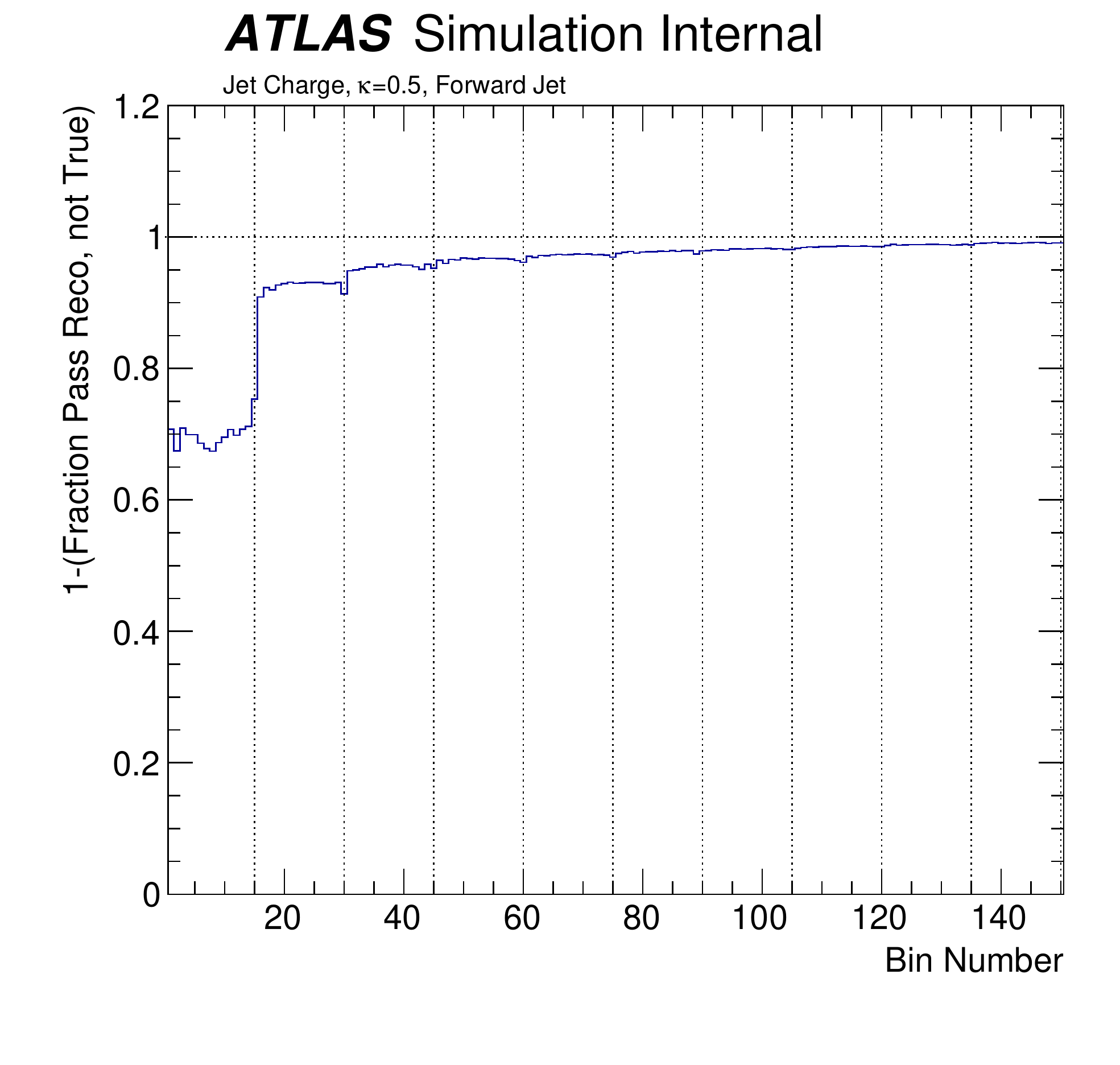}
\caption{The jet $p_\text{T}$ and jet charge distribution is transformed into a one-dimensional variable (see Sec.~\ref{sec:unfoldjetcharge}) and the inefficiency (left) and fake (right) factors are shown as a function of the bin number.  The jet $p_\text{T}$ increases from left to right and the jet $p_\text{T}$ bin edges are marked by vertical dashed lines.  A horizontal line at one indicates that no correction is applied.  The above plots are for $\kappa=0.5$ and the more forward jet, but the distributions for the other cases are qualitatively similar. }
\label{fig:fakes}
\end{center}
\end{figure}

The correction factors, described in the introduction to Fig.~\ref{sec:unfolding} are shown in Fig.~\ref{fig:fakes} as a function of the one-dimensional transformation of the jet $p_\text{T}$ and jet charge described in Sec.~\ref{sec:unfoldjetcharge}.  Both the fake and inefficiency factors are nearly one for high jet $p_\text{T}$.  In the first $p_\text{T}$ bin, there is a significant correction due to threshold effects from the 50 GeV jet $p_\text{T}$ requirement.  Within a given jet $p_\text{T}$ bin, the fake and inefficiency factors are nearly independent of the jet charge.

\begin{figure}[h!]
\begin{center}
\includegraphics[width=0.55\textwidth]{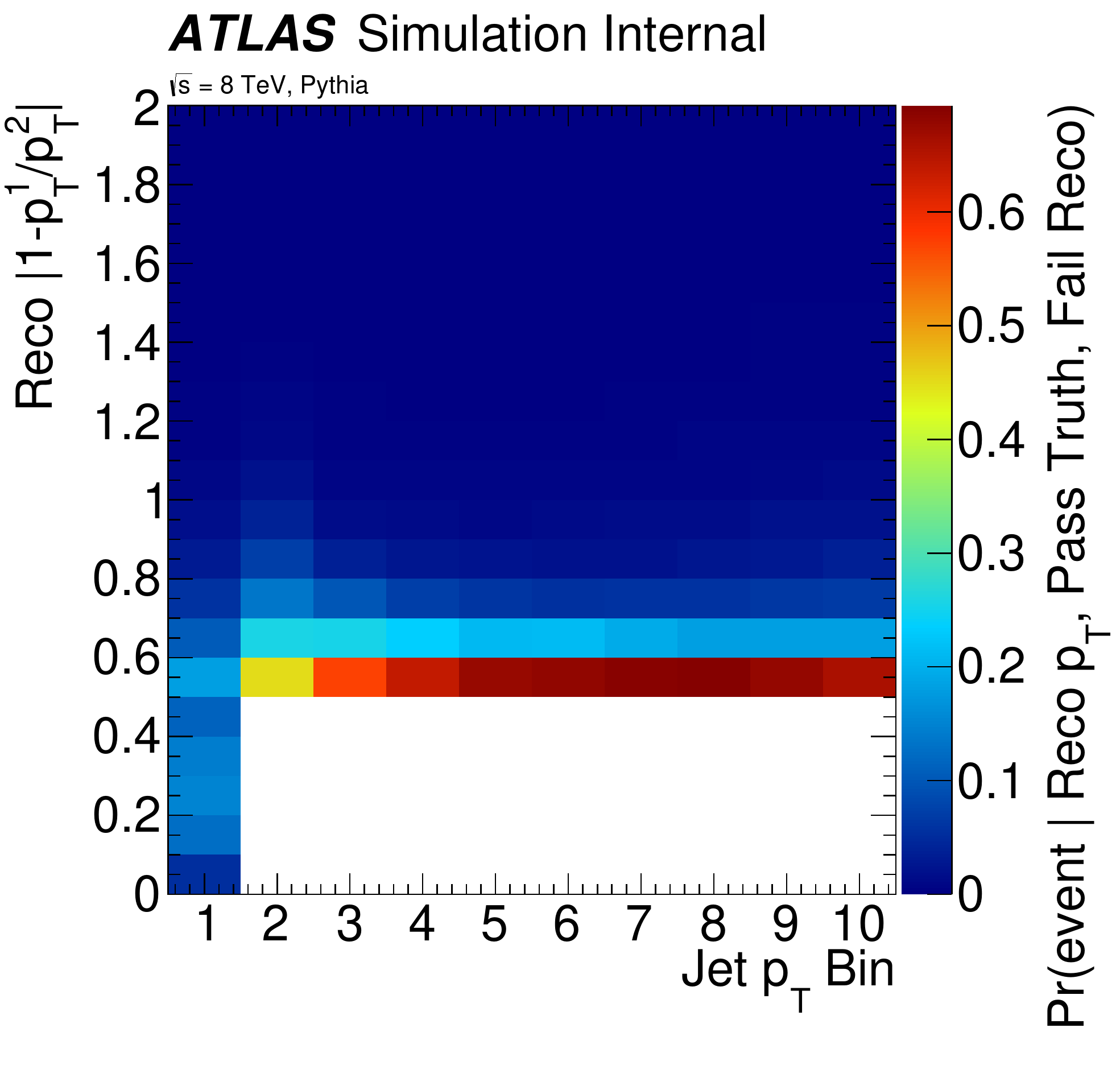}
\caption{For events which pass the truth-based selection but fail the reconstructed event selection, this plot shows $|1-p_\text{T}^1/p_\text{T}^2|$ for the reconstructed jet $p_\text{T}$.  The horizontal axis is the $p_\text{T}$ bin number, from 1-10.  In only the first $p_\text{T}$ bin, there are events which pass the $p_\text{T}$ asymmetry cut but fail the $p_\text{T}$ cut.  In every other bin, the event fails the truth selection due to the asymmetry cut.  The histogram is normalized per $p_\text{T}$ bin.}
\label{fig:fakes_explain}
\end{center}
\end{figure}

The dominant contribution to the corrections outside of the first bin in Fig.~\ref{fig:fakes} is the $p_\text{T}$ symmetry requirement.   This is demonstrated by Fig.~\ref{fig:fakes_explain}.  There are no events in simulation in the second $p_\text{T}$ bin that pass the jet $p_\text{T}$ symmetry requirement but fail the detector-level event selection.  In contrast, there are such events in the first jet $p_\text{T}$ bin.  These jets fail the event selection due to the detector-level jet $p_\text{T}>50$~GeV threshold.

\clearpage

\subsubsection{Response Matrix}

\begin{figure}[h!]
\begin{center}
\includegraphics[width=0.55\textwidth]{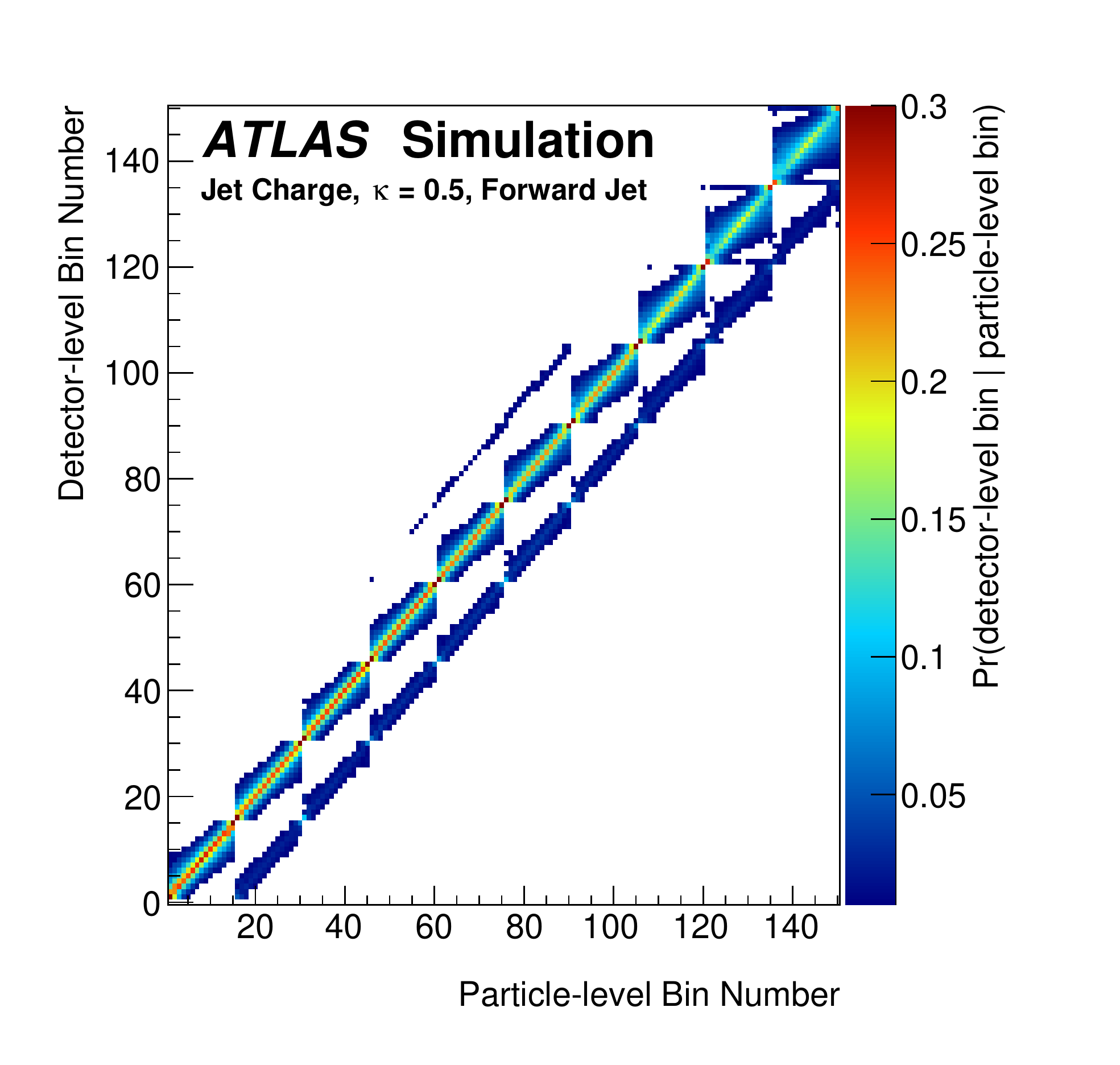}
\caption{The conditional distribution of the detector-level one-dimensional transformation of the jet charge and jet $p_\text{T}$ (see Sec.~\ref{sec:unfoldjetcharge}) in bins of the particle-level analogue (response matrix) for the $\kappa=0.5$ for the more forward jet.  The response matrix for the other variants is qualitatively similar.}
\label{fig:2Dto1D_reco}
\end{center}
\end{figure}

\begin{figure}[h!]
\begin{center}
\includegraphics[width=0.45\textwidth]{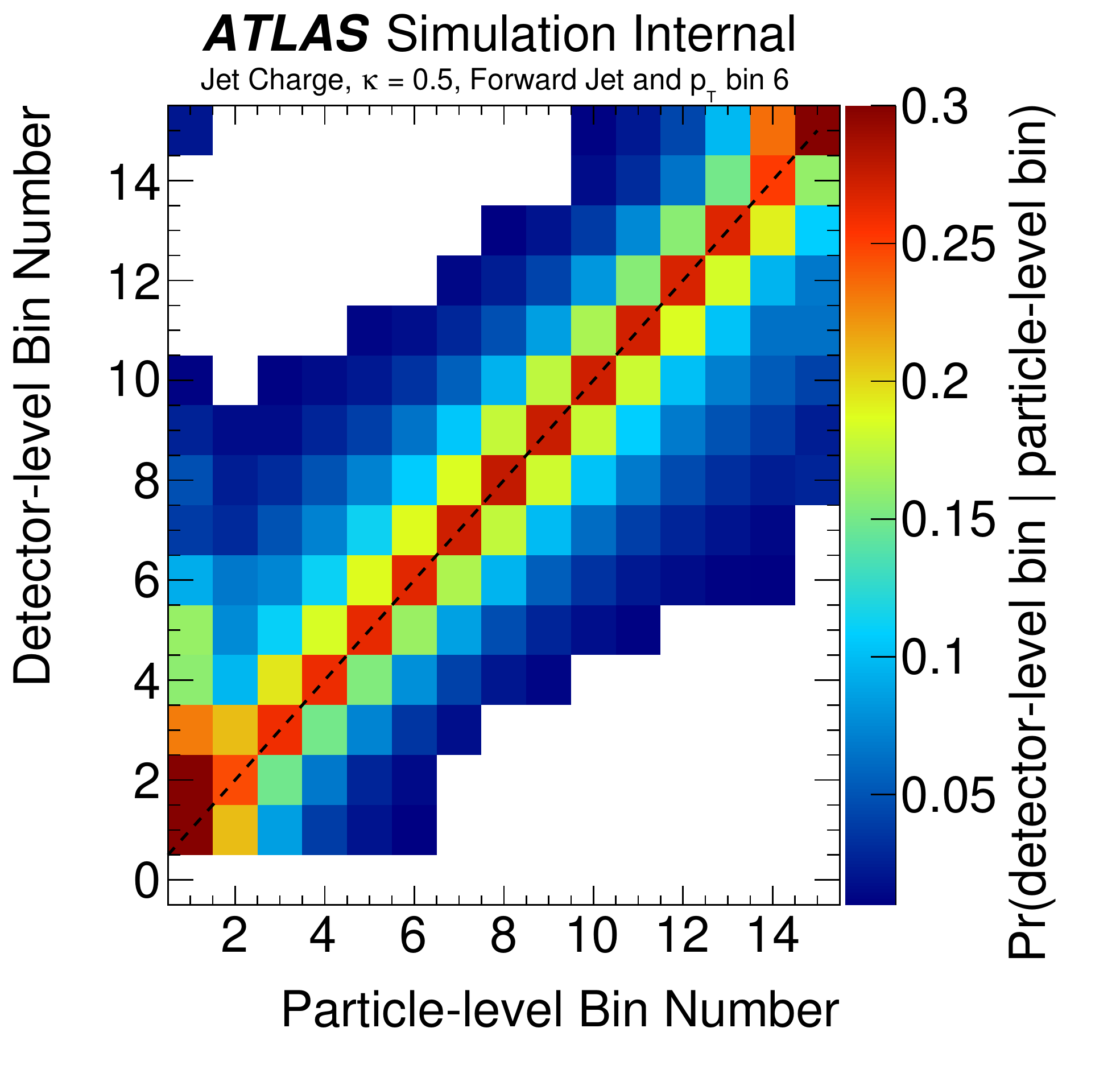}\includegraphics[width=0.45\textwidth]{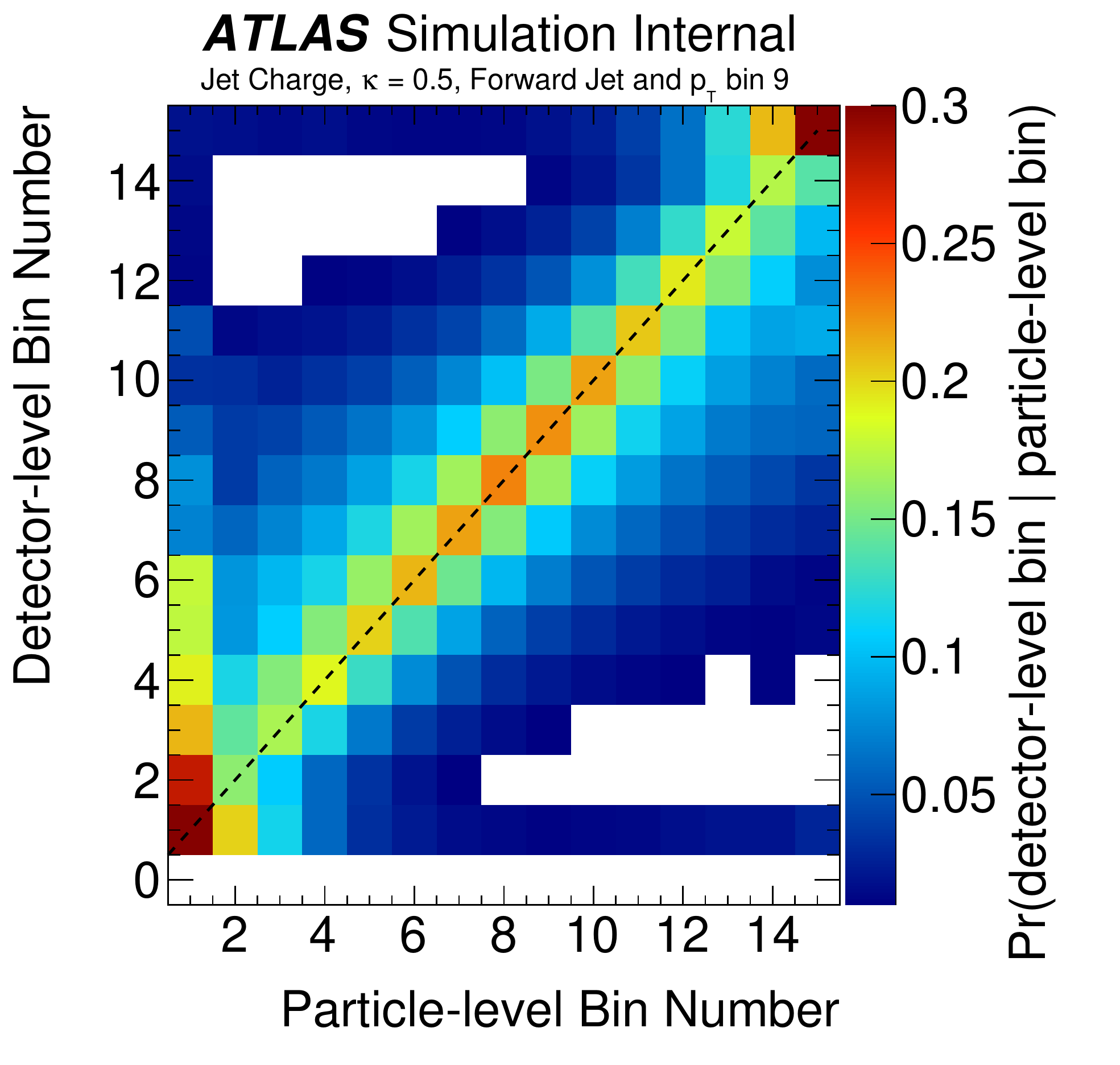}\\
\includegraphics[width=0.45\textwidth]{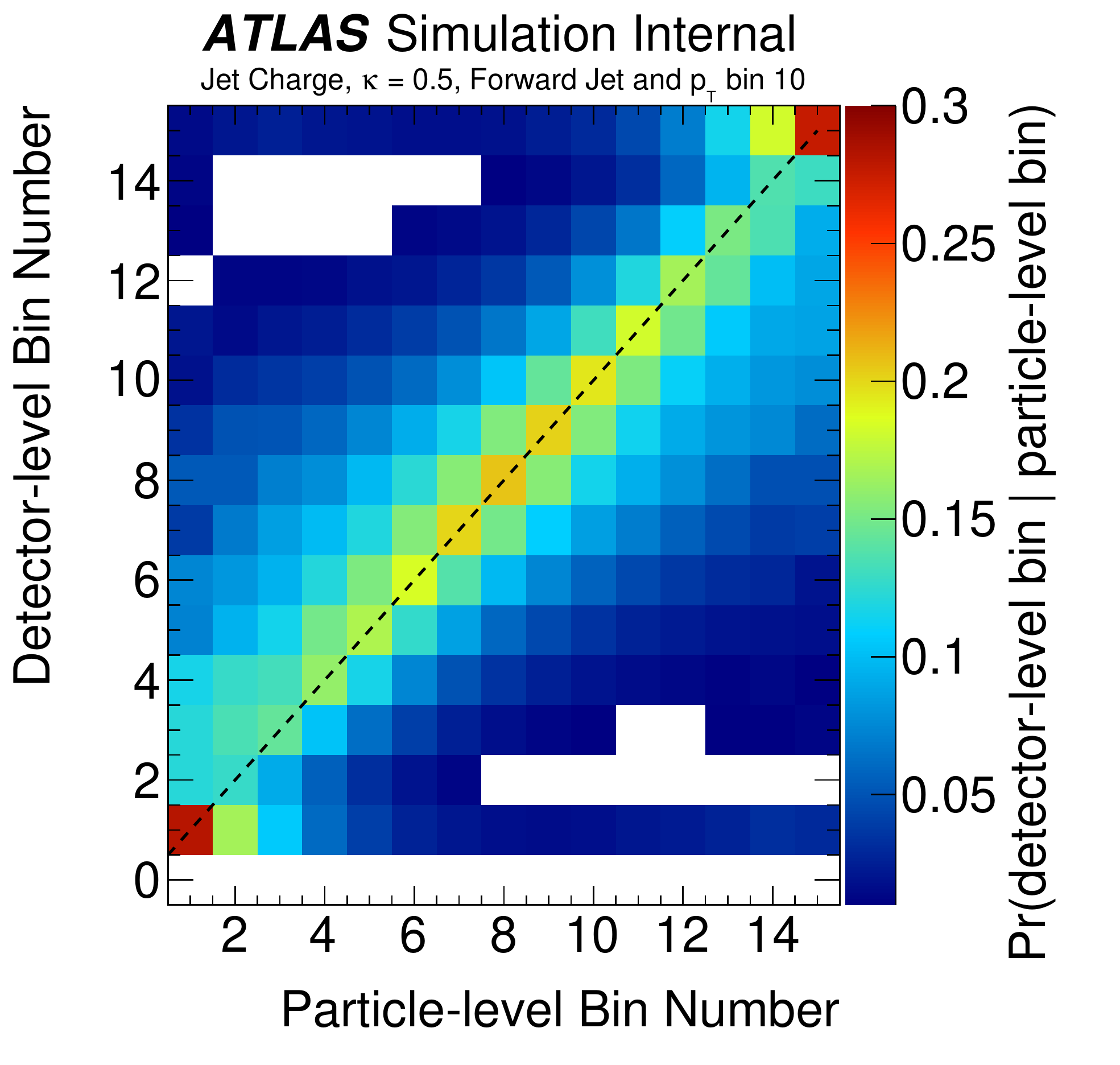}
\end{center}
\caption{The response matrix for $\kappa=0.5$ for the more forward jet in various jet $p_\text{T}$ bins.}
\label{fig:inpTbins1}
\end{figure}

The full response matrix for the one-dimensional transformation of the jet $p_\text{T}$ and the jet charge is shown in Fig.~\ref{fig:2Dto1D_reco}.  There are discrete regions along the diagonal corresponding to the 10 jet $p_\text{T}$ bins.    The large strip just above and below the main diagonal corresponds to events that migrate between jet $p_\text{T}$ bins.  The probability to migrate into a lower jet $p_\text{T}$ bin (below the diagonal) is higher than the probability to migrate to a higher jet $p_\text{T}$ bin because the distribution of jet $p_\text{T}$ within the particle-level bin is steeply falling.   Within one of these regions, the response matrix is peaked along the diagonal, but due to the broad jet charge resolution, the probability for a particle level event to remain in the same bin is $\lesssim 25\%$.  This probability decreases as a function of jet $p_\text{T}$ as the response matrix is more spread out away from the diagonal.   This trend is more evident in Fig~\ref{fig:inpTbins1} which shows the response matrix in a given jet $p_\text{T}$ bin.

\clearpage

\subsubsection{The Number of Iterations}

\begin{figure}[h!]
\begin{center}
\includegraphics[width=0.4\textwidth]{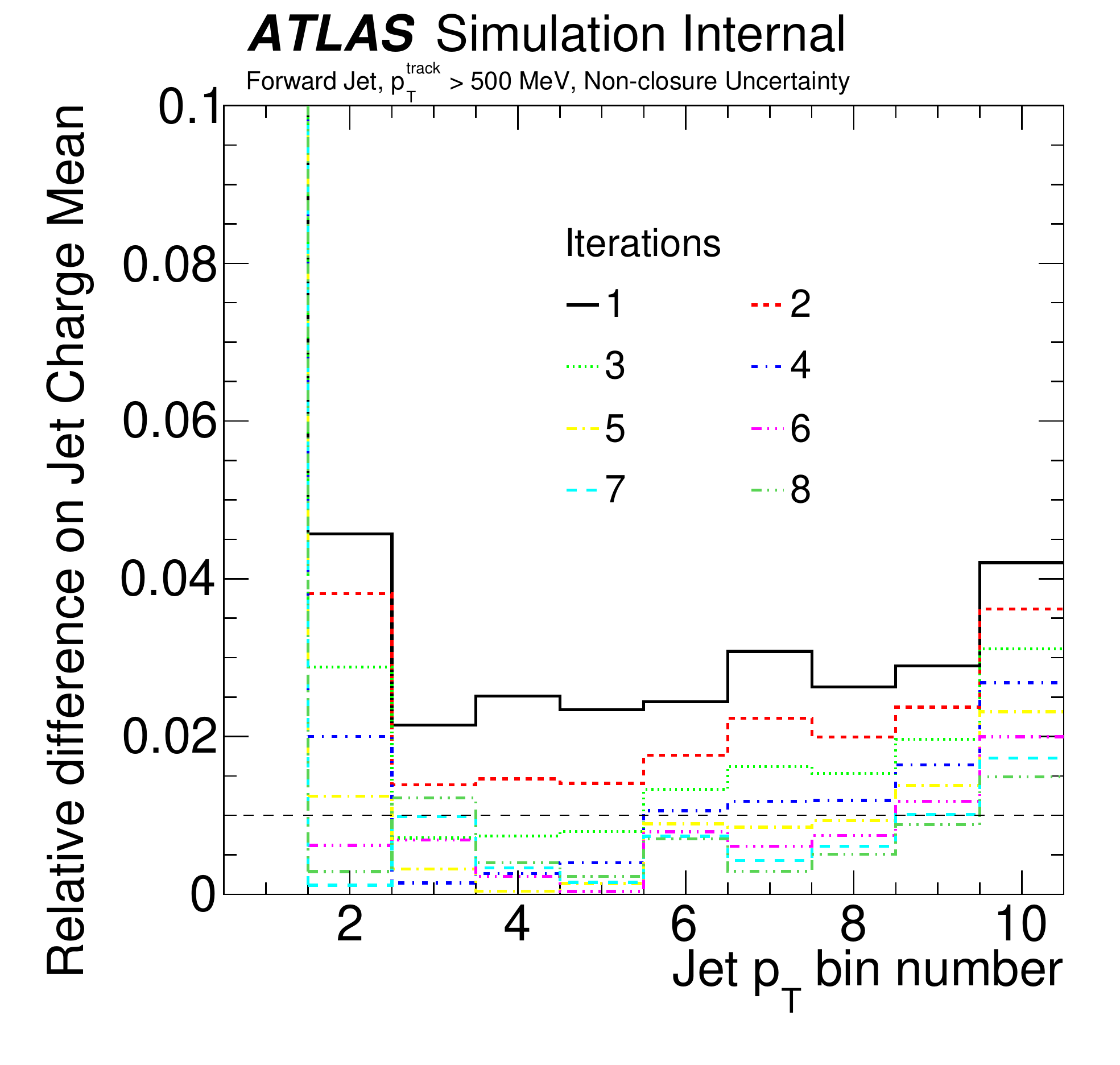}\includegraphics[width=0.4\textwidth]{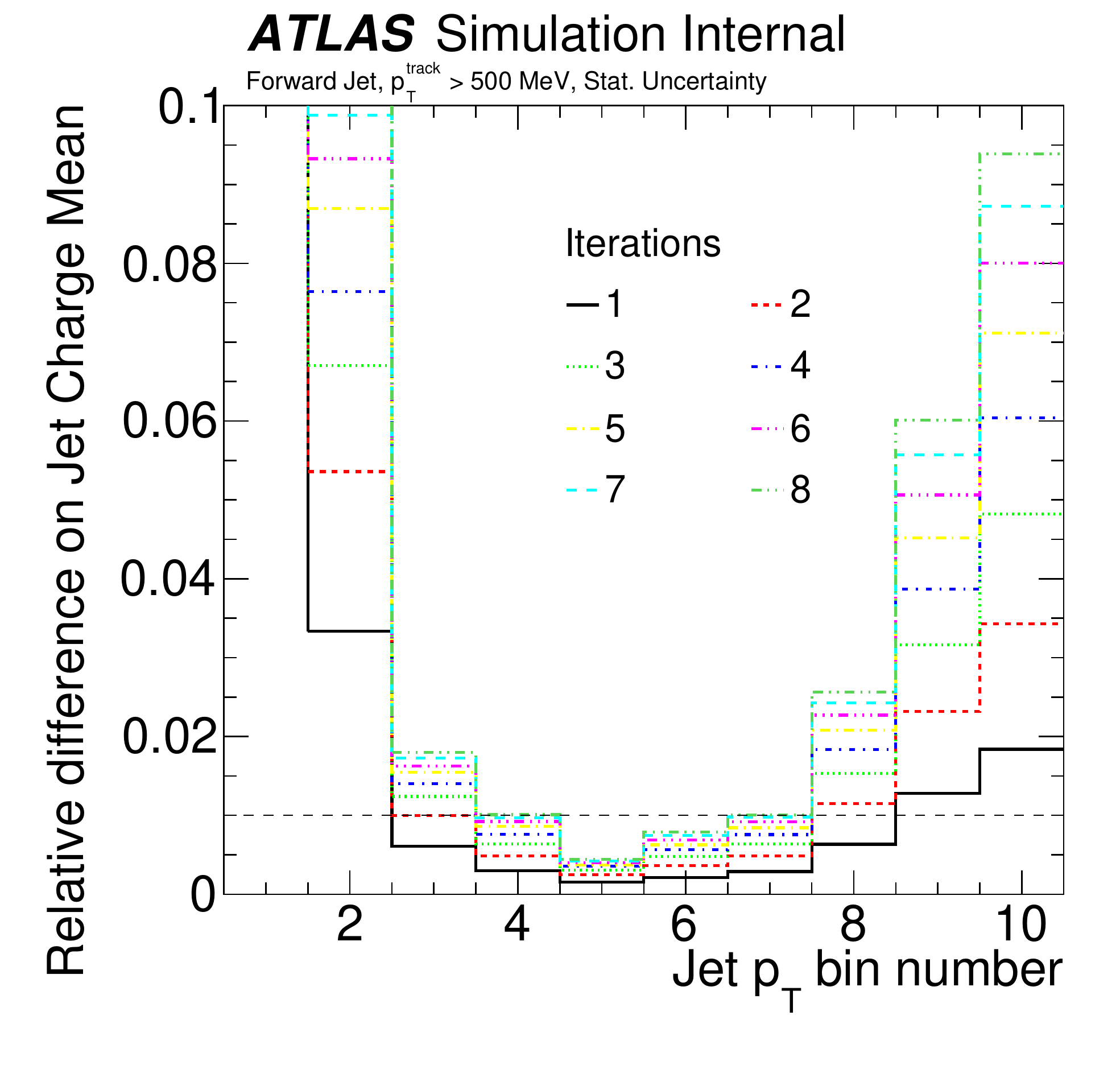}
\includegraphics[width=0.4\textwidth]{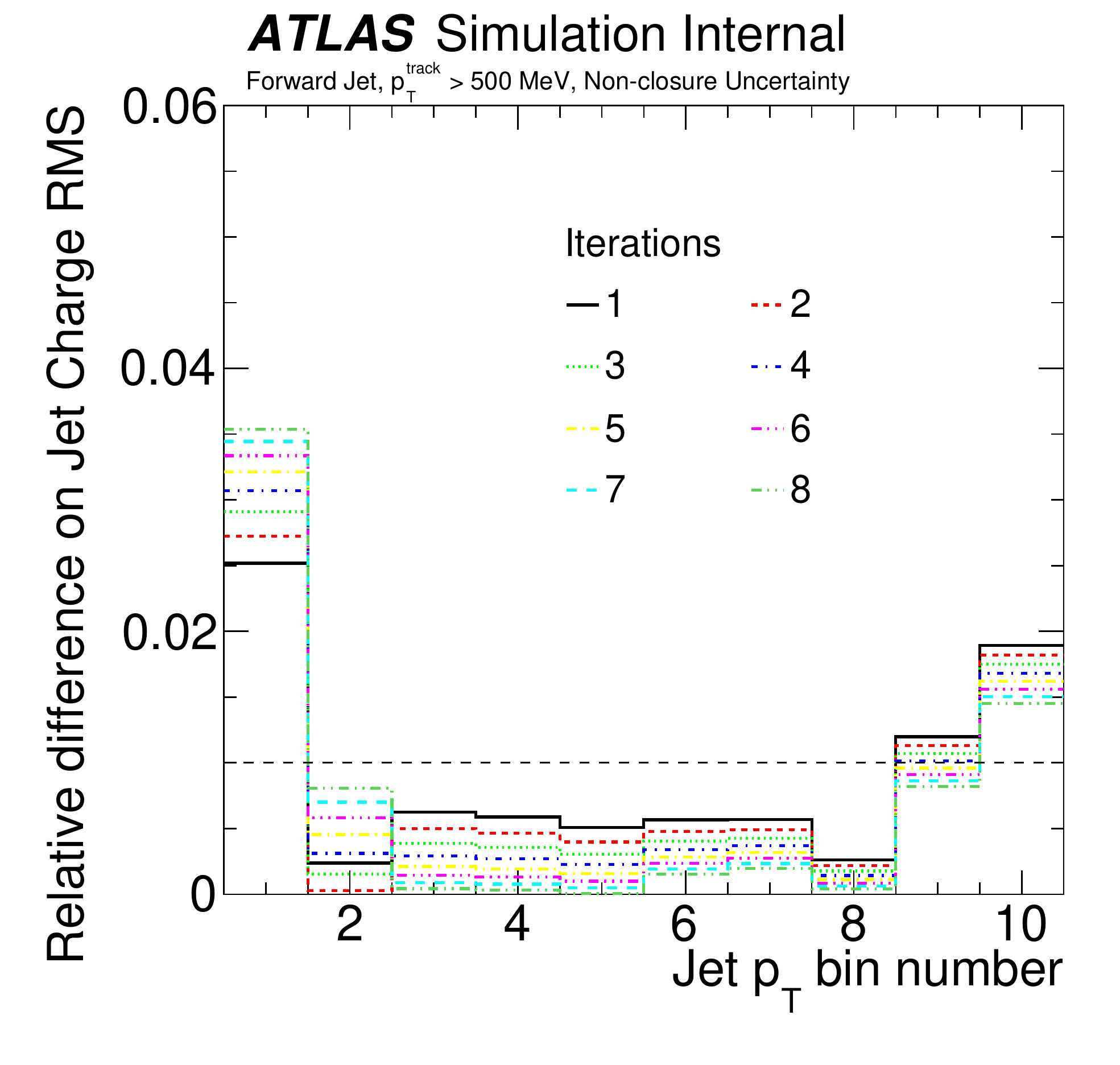}\includegraphics[width=0.4\textwidth]{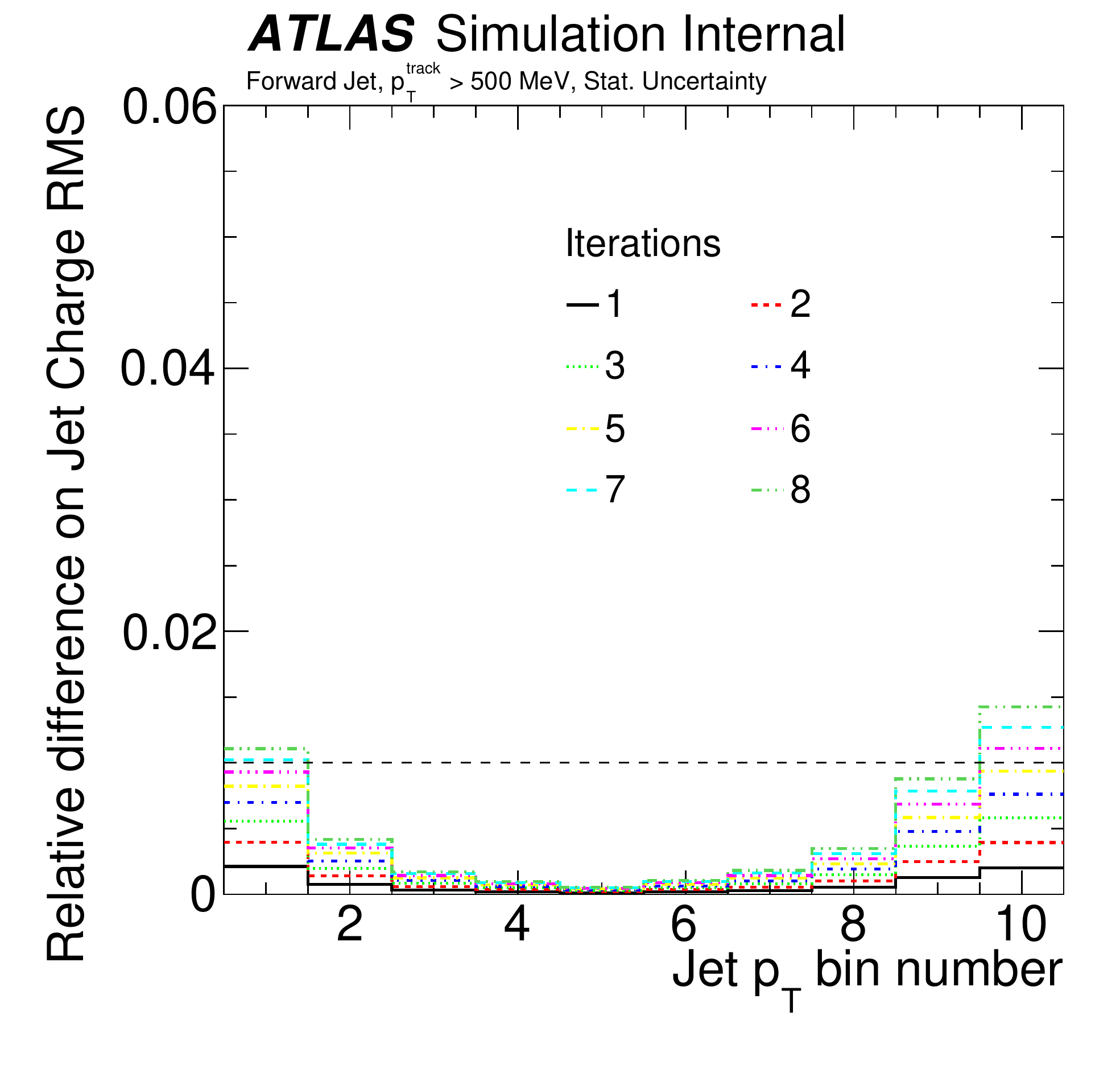}
\caption{Herwig unfolded with Pythia (left) and the statistical uncertainty (right) for various iteration multiplicity settings in the iterative Bayesian unfolding algorithm for the mean (top) and RMS (bottom).}
\label{fig:systs_unfold_iterations}
\end{center}
\end{figure}

The number of iterations is an arbitrary tuning parameter that can be used to tradeoff bias with statistical uncertainty.   Figure~\ref{fig:systs_unfold_iterations} shows the average and RMS of the jet charge distributions as a function of jet $p_\text{T}$. The {\sc Herwig++} MC is treated as data and the response matrix is derived from {\sc Pythia} 8.  About four iterations minimizes the bias and henceforth used as the nominal parameter setting.

\clearpage

\subsubsection{Unfolded Data}

\begin{figure}[h!]
\begin{center}
{\includegraphics[width=0.5\textwidth]{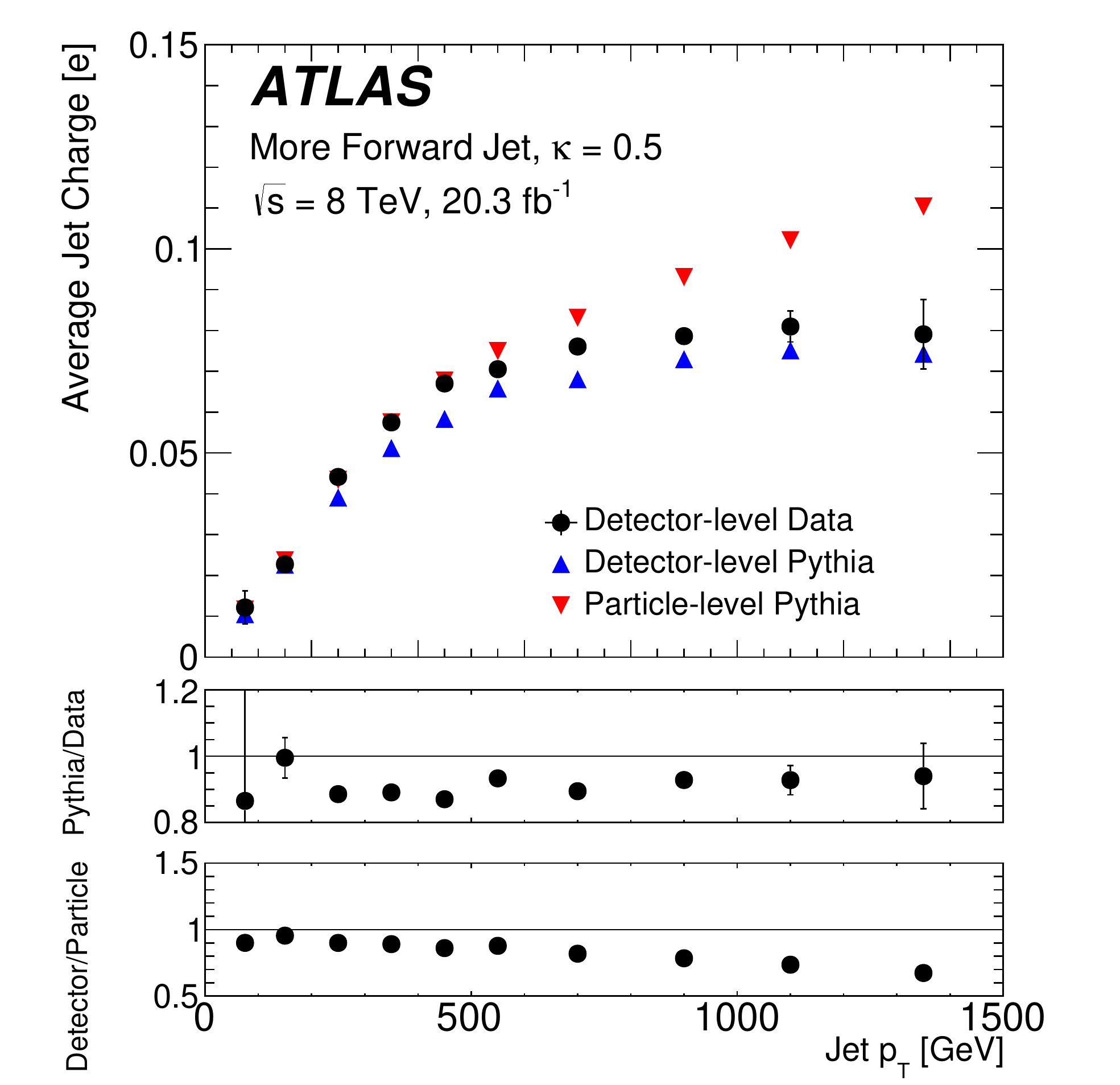}}{\includegraphics[width=0.5\textwidth]{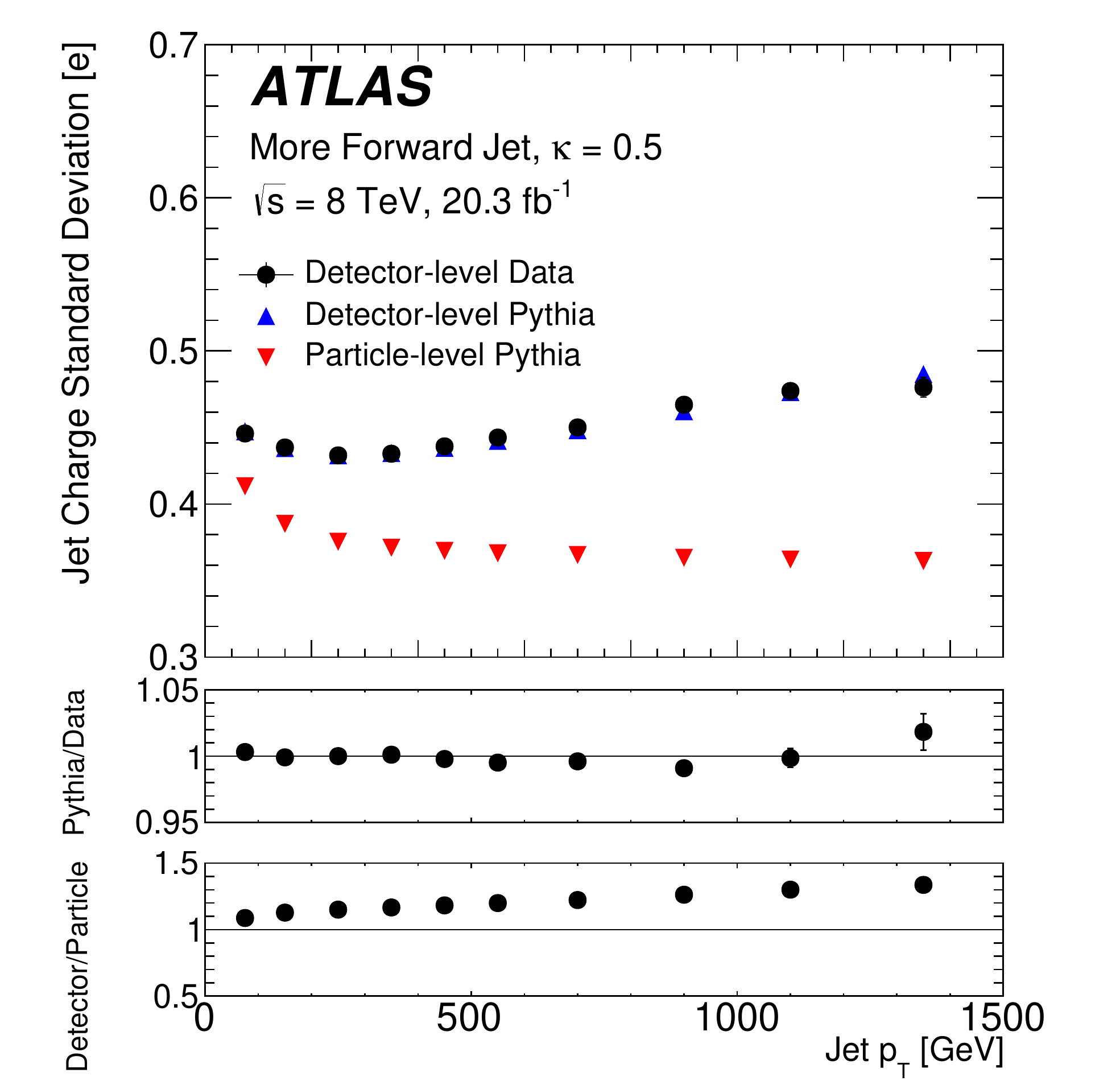}}
\end{center}	
\caption{The detector-level (data and simulation) and particle-level jet charge distribution's (a) average and (b) standard deviation as a function of the jet $p_\text{T}$ for the more forward jet.  The ratios in the bottom panel are constructed from the simulation, and show the prediction of detector-level {\sc Pythia} over the data (top ratio), and detector-level {\sc Pythia} over particle-level {\sc Pythia} (bottom ratio).   Bars on the data markers represent only the statistical uncertainties. For both (a) and (b), $\kappa=0.5$.}
\label{fig:raw}
\end{figure}

Figure~\ref{fig:raw} displays the $p_\text{T}$-dependence of the jet charge distribution's mean and standard deviation for detector-level data and simulation and for particle-level simulation.  The differences between the simulated detector- and particle-level distributions give a indication of the corrections required to account for detector acceptance and resolution effects in the unfolding procedure.  The growing difference between the particle- and detector-level average jet charge is due to the loss of charged-particle momentum inside jets as a result of track merging.  At particle level, the standard deviation of the jet charge distribution decreases with increasing $p_\text{T}$, but at detector level it increases with $p_\text{T}$ due to resolution effects.   

\begin{figure}[h!]
\begin{center}
\includegraphics[width=0.6\textwidth]{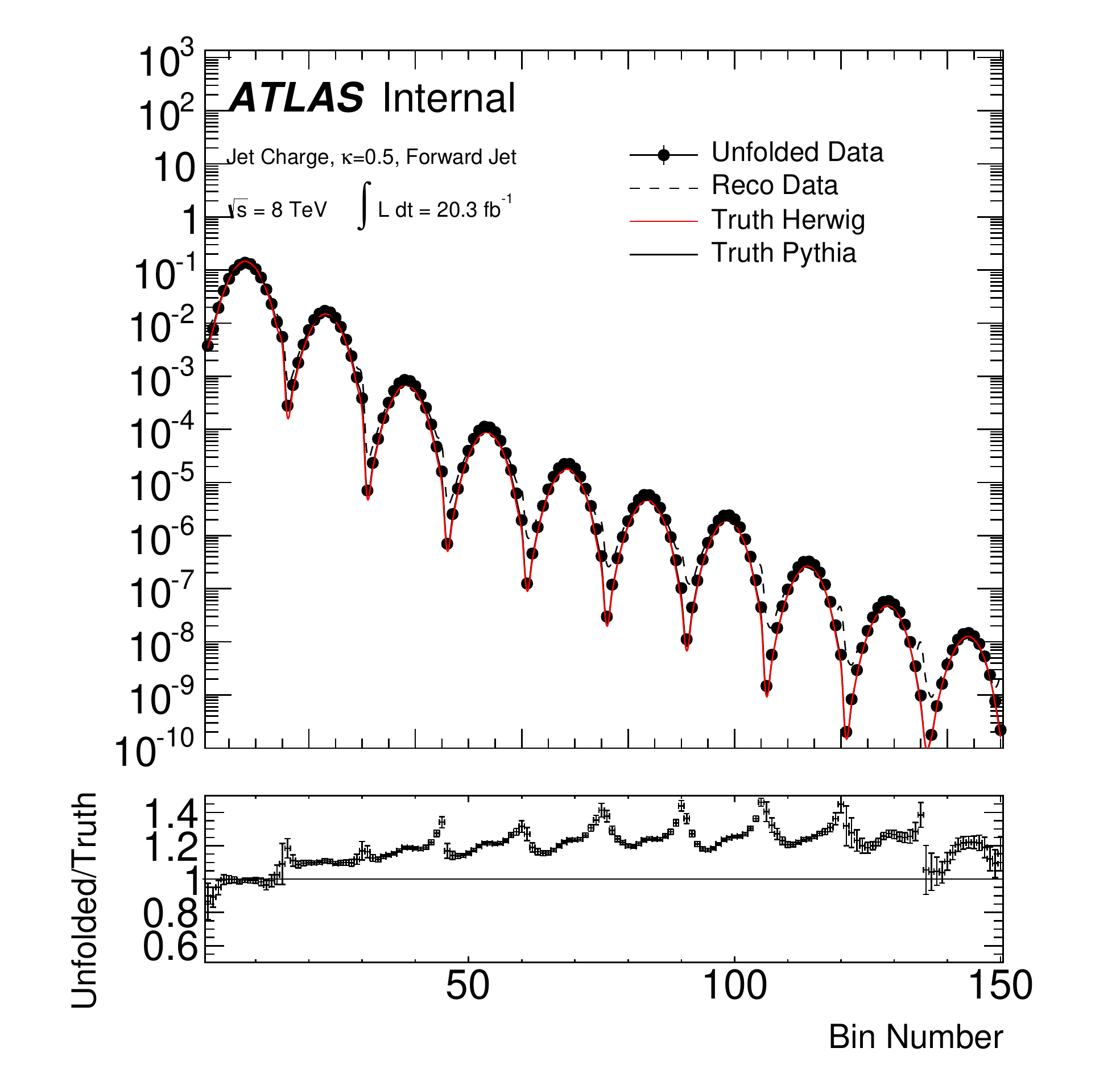}
\caption{The distribution of the 1D transformation of the 2D binned jet charge and jet $p_\text{T}$ distribution for the more forward jet for $\kappa=0.5$.  All distributions are scaled to have the same normalization.}
\label{fig:2Dto1Dunfolded}
\end{center}
\end{figure}

The data are unfolded using the iterative Bayesian technique~\cite{D'Agostini:1994zf}, implemented in the \texttt{RooUnfold} framework~\cite{Adye:2011gm}.  Figure~\ref{fig:2Dto1Dunfolded} shows the unfolded distribution over all bins of the one-dimensional transformation of the jet $p_\text{T}$ and jet charge distributions.  Even though there are $\lesssim 10\%$ differences in the mean and $\lesssim1\%$ differences in the standard deviation between the data and simulation (Fig.~\ref{fig:raw}), there are many bins in Fig.~\ref{fig:2Dto1Dunfolded} with large deviations from unity in the ratio.  This is due to two effects:

\begin{enumerate}
\item The plots all have the same normalization.  Since most events are in the first $p_\text{T}$ bin, there is a compensating offset in the other $p_\text{T}$ bins.  This is an artifact of the normalization.
\item Small changes in the mean and RMS can result in large changes in the ratio of the raw distribution away from zero.  To see this, suppose that the true and unfolded distributions in given $p_\text{T}$ bin are exactly Gaussians with means zero and standard deviations $\sigma_t$ and $\sigma_u$, respectively.  Then, the ratio $r$ will depend on the distance $x$ of the jet charge bin from zero in the following way:

\begin{align}
r=\frac{\sigma_u}{\sigma_t}\exp\left(-\frac{x^2}{2}\left(\frac{1}{\sigma_u^2}-\frac{1}{\sigma_t^2}\right)\right).
\end{align}

\noindent In particular, the ratio will go to zero or blow up to infinity (depending on the ordering of $\sigma_u$ and $\sigma_t$) as $|x|$ becomes large.  For some numerical values, for $\sigma\sim 0.5$ and $|\sigma_u/\sigma_t-1|\sim 2\%$, the ratio will change by $\sim 6\%$ when $x=0.5$ and $\sim 25\%$ when $x=1$, which is consistent with the behavior in Fig.~\ref{fig:raw} and~\ref{fig:2Dto1Dunfolded}.
\end{enumerate}

\clearpage

\section{Systematic uncertainties}
\label{sec:uncerts}

All stages of the jet charge measurement are sensitive to sources of potential bias.  The three stages of the measurement are listed below, with an overview of the systematic uncertainties that impact the results at each stage:

\begin{description}
\item[Correction Factors:] Fake and inefficiency factors are derived from simulation to account for the fraction of events that pass either the detector-level or particle-level fiducial selection, but not both.  These factors are generally between $0.9$ and $1.0$ except in the first $p_\text{T}$ bin, where threshold effects introduce corrections that can be as large as 20\%.  Experimental uncertainties correlated with the detector-level selection acceptance, such as the jet energy scale uncertainty, result in uncertainties in these correction factors.  An additional source of uncertainty on the correction factors is due to the explicit dependence on the particle-level jet charge and jet $p_\text{T}$ spectra.  A comparison of particle-level models ({\sc Pythia} and {\sc Herwig++}) is used to estimate the impact on the correction factors.
\item[Response Matrix:] For events in simulation that pass both the detector-level and particle-level fiducial selections, the response matrix describes migrations between bins when moving between the detector level and the particle level.  The response matrix is taken from simulation and various experimental uncertainties on the jet charge and jet $p_\text{T}$ spectra result in uncertainties in the matrix.  Uncertainties can be divided into two classes: those impacting the calorimeter jet $p_\text{T}$ and those impacting track reconstruction inside jets.
\item[Unfolding Procedure:] A data-driven technique is used to estimate the potential bias from a given choice of prior and number of iterations in the IB method~\cite{Malaescu:2009dm}.  The particle-level spectrum is reweighted using the response matrix so that the simulated detector-level spectrum has improved agreement with data.  The modified detector-level distribution is unfolded with the nominal response matrix and the difference between this and the reweighted particle-level spectrum is an indication of the bias due to the unfolding method.
\end{description}

\begin{figure}[h!]
\begin{center}
{\includegraphics[width=0.5\textwidth]{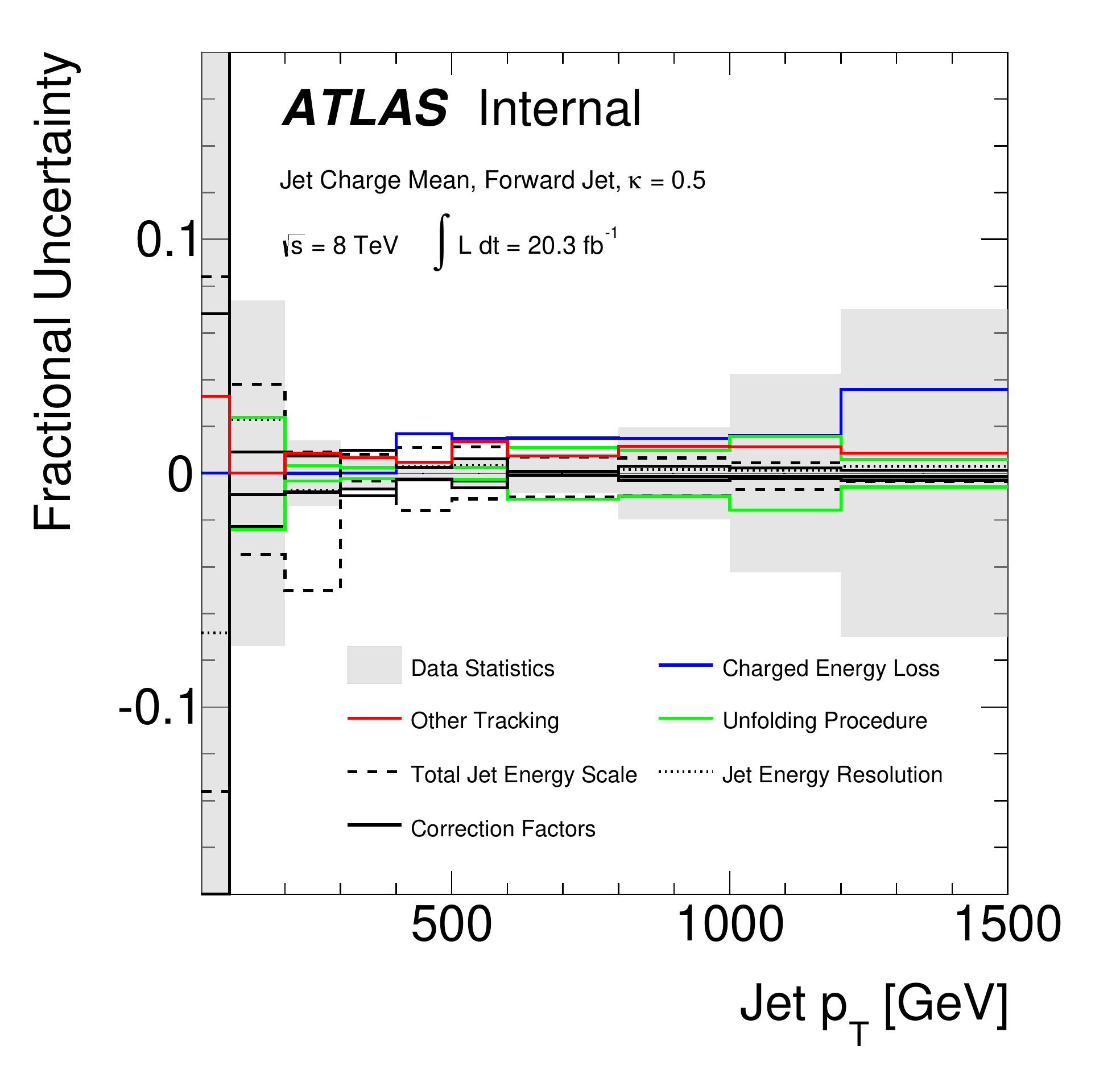}}{\includegraphics[width=0.5\textwidth]{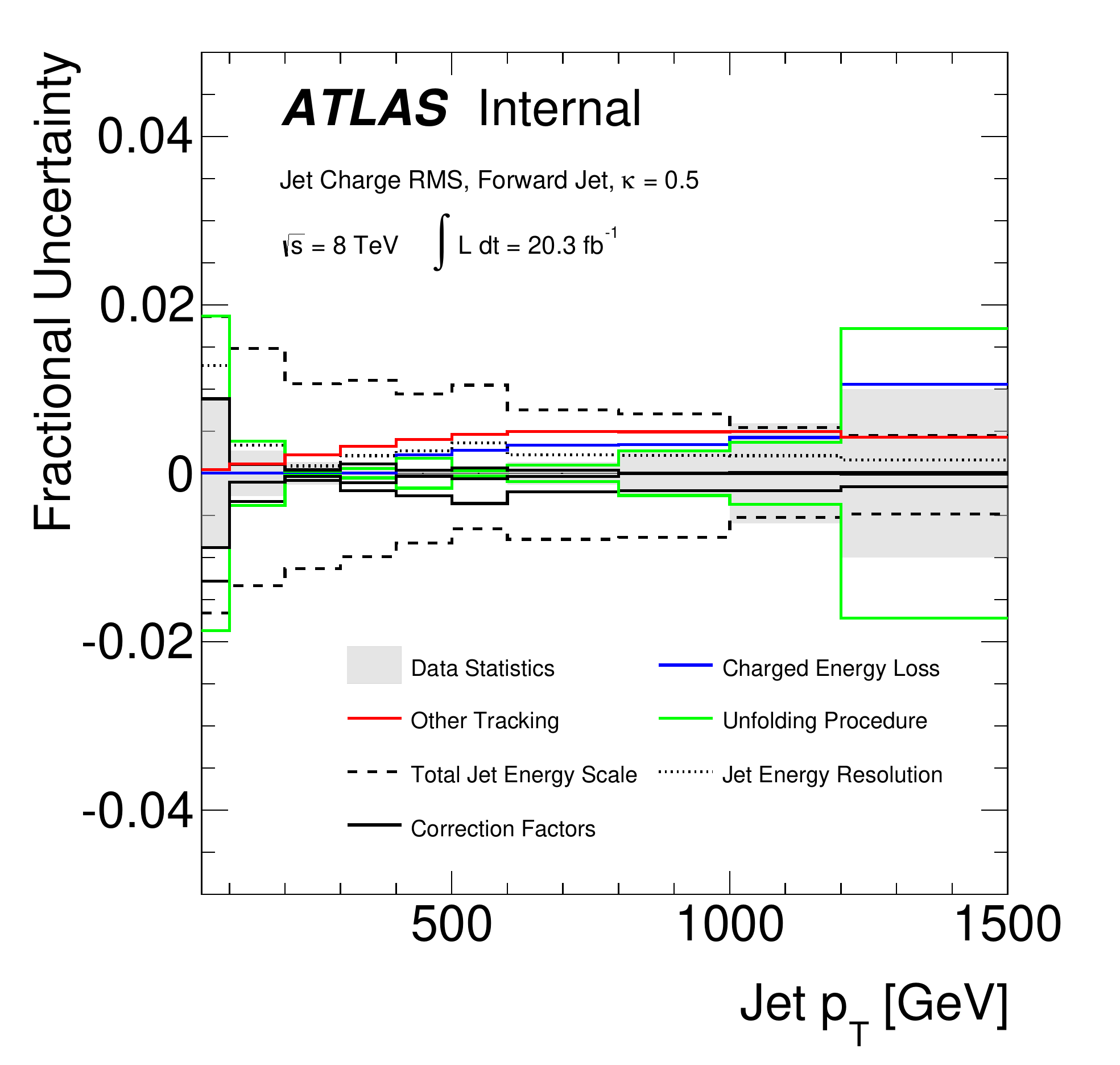}}
\end{center}	
\caption{A visualization of the systematic uncertainties for the jet charge mean (left) and standard deviation (right).  As a result of displaying the uncertainty as a fraction of the mean or RMS, the uncertainty appears artificially large in the first $p_\text{T}$ bin in the left plot where the average jet charge is nearly zero (small compared to the resolution).}\label{fig:JetChargeSystematicOverview}
\end{figure}

The following subsections describe the above uncertainties in more detail.  Uncertainties on the calorimeter jet $p_\text{T}$ are described in Sec.~\ref{sec:calojet} and the uncertainties related to tracking are described in Sec.~\ref{sec:tracking}.   Summaries of the systematic uncertainties for the more forward jet and $\kappa=0.5$ are found in Table~\ref{tab:systs_Mean_all_Forward_5aaa} and Table~\ref{tab:systs_RMS_all_Forward_5aaa} for the average jet charge and the jet charge distribution's standard deviation, respectively\footnote{The uncertainties on the first $p_\text{T}$ bin of the average jet charge are much larger than on the other bins because the mean is small compared to the resolution.}.  The uncertainties for the more central jet are similar.  Figure~\ref{fig:JetChargeSystematicOverview} presents a visualization of the uncertainties in Tables~\ref{tab:systs_Mean_all_Forward_5aaa} and~\ref{tab:systs_RMS_all_Forward_5aaa}.

{\renewcommand{\arraystretch}{1.3}
\setlength{\tabcolsep}{0.15cm}
\begin{table}[h]
\begin{tabular}{ccccccccccc}
 {\bf Average Jet Charge} & \multicolumn{10}{c}{Jet $p_\text{T}$ Range Lower Edge {[}100 GeV{]}} \\
\begin{tabular}[c]{@{}c@{}}Systematic\\ Uncertainty [\%]\end{tabular} & 0.5          & 1             & 2           &3            & 4           & 5             &6            & 8          &10           & 12          \\ \hline
JES & \multicolumn{1}{c}{${}^{+  8.4}_{- 13.6}$}
& \multicolumn{1}{c}{${}^{+  3.8}_{-  3.5}$}
& \multicolumn{1}{c}{${}^{+  0.9}_{-  5.0}$}
& \multicolumn{1}{c}{${}^{+  0.8}_{-  0.3}$}
& \multicolumn{1}{c}{${}^{+  1.1}_{-  1.6}$}
& \multicolumn{1}{c}{${}^{+  1.1}_{-  1.1}$}
& \multicolumn{1}{c}{${}^{+  0.7}_{-  1.0}$}
& \multicolumn{1}{c}{${}^{+  0.7}_{-  0.9}$}
& \multicolumn{1}{c}{${}^{+  0.4}_{-  0.7}$}
& \multicolumn{1}{c}{${}^{+  0.9}_{-  0.3}$}
\\
JER & \multicolumn{1}{c}{${}^{+  6.8}_{-  6.8}$}
& \multicolumn{1}{c}{${}^{+  2.3}_{-  2.3}$}
& \multicolumn{1}{c}{${}^{+  0.7}_{-  0.7}$}
& \multicolumn{1}{c}{${}^{+  0.7}_{-  0.7}$}
& \multicolumn{1}{c}{${}^{+  0.3}_{-  0.3}$}
& \multicolumn{1}{c}{${}^{+  0.3}_{-  0.3}$}
& \multicolumn{1}{c}{${}^{+  0.1}_{-  0.1}$}
& \multicolumn{1}{c}{${}^{+  0.1}_{-  0.1}$}
& \multicolumn{1}{c}{${}^{+  0.1}_{-  0.1}$}
& \multicolumn{1}{c}{${}^{+  0.3}_{-  0.3}$}
\\
Charged Energy Loss & \multicolumn{1}{c}{${}^{+  0.0}_{-  0.0}$}
& \multicolumn{1}{c}{${}^{+  0.0}_{-  0.0}$}
& \multicolumn{1}{c}{${}^{+  0.0}_{-  0.0}$}
& \multicolumn{1}{c}{${}^{+  0.0}_{-  0.0}$}
& \multicolumn{1}{c}{${}^{+  1.7}_{-  0.0}$}
& \multicolumn{1}{c}{${}^{+  1.5}_{-  0.0}$}
& \multicolumn{1}{c}{${}^{+  1.5}_{-  0.0}$}
& \multicolumn{1}{c}{${}^{+  1.5}_{-  0.0}$}
& \multicolumn{1}{c}{${}^{+  1.6}_{-  0.0}$}
& \multicolumn{1}{c}{${}^{+  3.6}_{-  0.0}$}
\\
Other Tracking & \multicolumn{1}{c}{${}^{+  3.3}_{-  1.6}$}
& \multicolumn{1}{c}{${}^{+  0.0}_{-  0.4}$}
& \multicolumn{1}{c}{${}^{+  0.9}_{-  0.2}$}
& \multicolumn{1}{c}{${}^{+  0.7}_{-  0.1}$}
& \multicolumn{1}{c}{${}^{+  0.5}_{-  0.4}$}
& \multicolumn{1}{c}{${}^{+  1.4}_{-  0.6}$}
& \multicolumn{1}{c}{${}^{+  0.7}_{-  0.9}$}
& \multicolumn{1}{c}{${}^{+  1.2}_{-  1.2}$}
& \multicolumn{1}{c}{${}^{+  1.1}_{-  1.3}$}
& \multicolumn{1}{c}{${}^{+  0.9}_{-  1.7}$}
\\
Track Multiplicity & \multicolumn{1}{c}{${}^{+  0.0}_{-  1.5}$}
& \multicolumn{1}{c}{${}^{+  0.1}_{-  0.0}$}
& \multicolumn{1}{c}{${}^{+  0.0}_{-  0.6}$}
& \multicolumn{1}{c}{${}^{+  0.0}_{-  1.1}$}
& \multicolumn{1}{c}{${}^{+  0.0}_{-  0.8}$}
& \multicolumn{1}{c}{${}^{+  0.0}_{-  0.6}$}
& \multicolumn{1}{c}{${}^{+  0.0}_{-  1.2}$}
& \multicolumn{1}{c}{${}^{+  0.0}_{-  1.4}$}
& \multicolumn{1}{c}{${}^{+  0.0}_{-  2.1}$}
& \multicolumn{1}{c}{${}^{+  0.0}_{-  2.9}$}
\\
Correction Factors & \multicolumn{1}{c}{${}^{+   23}_{-   23}$}
& \multicolumn{1}{c}{${}^{+  0.9}_{-  0.9}$}
& \multicolumn{1}{c}{${}^{+  0.8}_{-  0.8}$}
& \multicolumn{1}{c}{${}^{+  1.0}_{-  1.0}$}
& \multicolumn{1}{c}{${}^{+  0.3}_{-  0.3}$}
& \multicolumn{1}{c}{${}^{+  0.6}_{-  0.6}$}
& \multicolumn{1}{c}{${}^{+  0.1}_{-  0.1}$}
& \multicolumn{1}{c}{${}^{+  0.3}_{-  0.3}$}
& \multicolumn{1}{c}{${}^{+  0.2}_{-  0.2}$}
& \multicolumn{1}{c}{${}^{+  0.1}_{-  0.1}$}
\\
Unfolding Procedure& \multicolumn{1}{c}{${}^{+   28}_{-   28}$}
& \multicolumn{1}{c}{${}^{+  2.4}_{-  2.4}$}
& \multicolumn{1}{c}{${}^{+  0.3}_{-  0.3}$}
& \multicolumn{1}{c}{${}^{+  0.2}_{-  0.2}$}
& \multicolumn{1}{c}{${}^{+  0.2}_{-  0.2}$}
& \multicolumn{1}{c}{${}^{+  0.3}_{-  0.3}$}
& \multicolumn{1}{c}{${}^{+  1.1}_{-  1.1}$}
& \multicolumn{1}{c}{${}^{+  1.0}_{-  1.0}$}
& \multicolumn{1}{c}{${}^{+  1.6}_{-  1.6}$}
& \multicolumn{1}{c}{${}^{+  0.6}_{-  0.6}$}
\\
\hline
Total Systematic & \multicolumn{1}{c}{${}^{+   38}_{-   39}$}
& \multicolumn{1}{c}{${}^{+  5.1}_{-  4.9}$}
& \multicolumn{1}{c}{${}^{+  1.7}_{-  5.2}$}
& \multicolumn{1}{c}{${}^{+  1.6}_{-  1.7}$}
& \multicolumn{1}{c}{${}^{+  2.1}_{-  1.9}$}
& \multicolumn{1}{c}{${}^{+  2.4}_{-  1.6}$}
& \multicolumn{1}{c}{${}^{+  2.1}_{-  2.1}$}
& \multicolumn{1}{c}{${}^{+  2.3}_{-  2.3}$}
& \multicolumn{1}{c}{${}^{+  2.6}_{-  3.0}$}
& \multicolumn{1}{c}{${}^{+  3.8}_{-  3.4}$}
\\
Data Statistics & \multicolumn{1}{c}{   28}
& \multicolumn{1}{c}{  7.4}
& \multicolumn{1}{c}{  1.4}
& \multicolumn{1}{c}{  0.7}
& \multicolumn{1}{c}{  0.3}
& \multicolumn{1}{c}{  0.6}
& \multicolumn{1}{c}{  0.9}
& \multicolumn{1}{c}{  2.0}
& \multicolumn{1}{c}{  4.2}
& \multicolumn{1}{c}{  7.0}
\\
Total Uncertainty & \multicolumn{1}{c}{${}^{+   47}_{-   48}$}
& \multicolumn{1}{c}{${}^{+  9.0}_{-  8.9}$}
& \multicolumn{1}{c}{${}^{+  2.2}_{-  5.4}$}
& \multicolumn{1}{c}{${}^{+  1.8}_{-  1.9}$}
& \multicolumn{1}{c}{${}^{+  2.1}_{-  1.9}$}
& \multicolumn{1}{c}{${}^{+  2.5}_{-  1.7}$}
& \multicolumn{1}{c}{${}^{+  2.3}_{-  2.3}$}
& \multicolumn{1}{c}{${}^{+  3.0}_{-  3.0}$}
& \multicolumn{1}{c}{${}^{+  5.0}_{-  5.2}$}
& \multicolumn{1}{c}{${}^{+  8.0}_{-  7.8}$}
\\
\hline\hline
Measured Value [$0.1  e$] & \multicolumn{1}{c}{0.014}
& \multicolumn{1}{c}{0.24}
& \multicolumn{1}{c}{0.49}
& \multicolumn{1}{c}{0.65}
& \multicolumn{1}{c}{0.76}
& \multicolumn{1}{c}{0.82}
& \multicolumn{1}{c}{0.92}
& \multicolumn{1}{c}{1.00}
& \multicolumn{1}{c}{1.08}
& \multicolumn{1}{c}{1.15}
\\
\end{tabular}
\caption{A summary of all the systematic uncertainties and their impact on the mean jet charge for $\kappa=0.5$ and the more forward jet.   The correction factors are the fake and inefficiency corrections applied before/after the response matrix.  The Other Tracking category includes uncertainty on the track reconstruction efficiency, track momentum resolution, charge misidentification, and fake track rate.  All numbers are given in percent.  As a result, the uncertainty appears artificially large in the first $p_\text{T}$ bin where the average jet charge is nearly zero (small compared to the resolution).}
\label{tab:systs_Mean_all_Forward_5aaa}
\end{table} 

{\renewcommand{\arraystretch}{1.3}
\begin{table}[h]
\begin{tabular}{ccccccccccc}
 {\bf Standard Deviation} & \multicolumn{10}{c}{Jet $p_\text{T}$ Range {[}100 GeV{]}} \\
\begin{tabular}[c]{@{}c@{}}Systematic\\ Uncertainty [\%]\end{tabular} & 0.5          &1             & 2           &3            & 4           & 5           &6            & 8            & 10           & 12          \\ \hline
Total Jet Energy Scale & \multicolumn{1}{c}{${}^{+  1.9}_{-  1.7}$}
& \multicolumn{1}{c}{${}^{+  1.5}_{-  1.3}$}
& \multicolumn{1}{c}{${}^{+  1.1}_{-  1.1}$}
& \multicolumn{1}{c}{${}^{+  1.1}_{-  1.0}$}
& \multicolumn{1}{c}{${}^{+  0.9}_{-  0.8}$}
& \multicolumn{1}{c}{${}^{+  1.0}_{-  0.7}$}
& \multicolumn{1}{c}{${}^{+  0.8}_{-  0.8}$}
& \multicolumn{1}{c}{${}^{+  0.7}_{-  0.8}$}
& \multicolumn{1}{c}{${}^{+  0.5}_{-  0.5}$}
& \multicolumn{1}{c}{${}^{+  0.5}_{-  0.5}$}
\\
Jet Energy Resolution & \multicolumn{1}{c}{${}^{+  1.3}_{-  1.3}$}
& \multicolumn{1}{c}{${}^{+  0.3}_{-  0.3}$}
& \multicolumn{1}{c}{${}^{+  0.1}_{-  0.1}$}
& \multicolumn{1}{c}{${}^{+  0.2}_{-  0.2}$}
& \multicolumn{1}{c}{${}^{+  0.3}_{-  0.3}$}
& \multicolumn{1}{c}{${}^{+  0.4}_{-  0.4}$}
& \multicolumn{1}{c}{${}^{+  0.2}_{-  0.2}$}
& \multicolumn{1}{c}{${}^{+  0.2}_{-  0.2}$}
& \multicolumn{1}{c}{${}^{+  0.2}_{-  0.2}$}
& \multicolumn{1}{c}{${}^{+  0.2}_{-  0.2}$}
\\
Charged Energy Loss & \multicolumn{1}{c}{${}^{+  0.0}_{-  0.0}$}
& \multicolumn{1}{c}{${}^{+  0.0}_{-  0.0}$}
& \multicolumn{1}{c}{${}^{+  0.0}_{-  0.0}$}
& \multicolumn{1}{c}{${}^{+  0.0}_{-  0.0}$}
& \multicolumn{1}{c}{${}^{+  0.2}_{-  0.0}$}
& \multicolumn{1}{c}{${}^{+  0.3}_{-  0.0}$}
& \multicolumn{1}{c}{${}^{+  0.3}_{-  0.0}$}
& \multicolumn{1}{c}{${}^{+  0.3}_{-  0.0}$}
& \multicolumn{1}{c}{${}^{+  0.4}_{-  0.0}$}
& \multicolumn{1}{c}{${}^{+  1.1}_{-  0.0}$}
\\
Other Tracking & \multicolumn{1}{c}{${}^{+  0.0}_{-  0.3}$}
& \multicolumn{1}{c}{${}^{+  0.1}_{-  0.3}$}
& \multicolumn{1}{c}{${}^{+  0.2}_{-  0.4}$}
& \multicolumn{1}{c}{${}^{+  0.3}_{-  0.4}$}
& \multicolumn{1}{c}{${}^{+  0.4}_{-  0.5}$}
& \multicolumn{1}{c}{${}^{+  0.5}_{-  0.4}$}
& \multicolumn{1}{c}{${}^{+  0.5}_{-  0.5}$}
& \multicolumn{1}{c}{${}^{+  0.5}_{-  0.5}$}
& \multicolumn{1}{c}{${}^{+  0.5}_{-  0.4}$}
& \multicolumn{1}{c}{${}^{+  0.4}_{-  0.4}$}
\\
Track Multiplicity & \multicolumn{1}{c}{${}^{+  0.0}_{-  0.2}$}
& \multicolumn{1}{c}{${}^{+  0.0}_{-  0.3}$}
& \multicolumn{1}{c}{${}^{+  0.0}_{-  0.2}$}
& \multicolumn{1}{c}{${}^{+  0.0}_{-  0.1}$}
& \multicolumn{1}{c}{${}^{+  0.0}_{-  0.0}$}
& \multicolumn{1}{c}{${}^{+  0.1}_{-  0.0}$}
& \multicolumn{1}{c}{${}^{+  0.2}_{-  0.0}$}
& \multicolumn{1}{c}{${}^{+  0.2}_{-  0.0}$}
& \multicolumn{1}{c}{${}^{+  0.3}_{-  0.0}$}
& \multicolumn{1}{c}{${}^{+  0.2}_{-  0.0}$}
\\
Correction Factors & \multicolumn{1}{c}{${}^{+  0.9}_{-  0.9}$}
& \multicolumn{1}{c}{${}^{+  0.1}_{-  0.1}$}
& \multicolumn{1}{c}{${}^{+  0.0}_{-  0.0}$}
& \multicolumn{1}{c}{${}^{+  0.1}_{-  0.1}$}
& \multicolumn{1}{c}{${}^{+  0.0}_{-  0.0}$}
& \multicolumn{1}{c}{${}^{+  0.1}_{-  0.1}$}
& \multicolumn{1}{c}{${}^{+  0.0}_{-  0.0}$}
& \multicolumn{1}{c}{${}^{+  0.0}_{-  0.0}$}
& \multicolumn{1}{c}{${}^{+  0.0}_{-  0.0}$}
& \multicolumn{1}{c}{${}^{+  0.0}_{-  0.0}$}
\\
Unfolding Procedure& \multicolumn{1}{c}{${}^{+  1.9}_{-  1.9}$}
& \multicolumn{1}{c}{${}^{+  0.4}_{-  0.4}$}
& \multicolumn{1}{c}{${}^{+  0.0}_{-  0.0}$}
& \multicolumn{1}{c}{${}^{+  0.1}_{-  0.1}$}
& \multicolumn{1}{c}{${}^{+  0.2}_{-  0.2}$}
& \multicolumn{1}{c}{${}^{+  0.0}_{-  0.0}$}
& \multicolumn{1}{c}{${}^{+  0.1}_{-  0.1}$}
& \multicolumn{1}{c}{${}^{+  0.3}_{-  0.3}$}
& \multicolumn{1}{c}{${}^{+  0.4}_{-  0.4}$}
& \multicolumn{1}{c}{${}^{+  1.7}_{-  1.7}$}
\\
\hline
Total Systematic & \multicolumn{1}{c}{${}^{+  3.1}_{-  3.0}$}
& \multicolumn{1}{c}{${}^{+  1.6}_{-  1.5}$}
& \multicolumn{1}{c}{${}^{+  1.1}_{-  1.2}$}
& \multicolumn{1}{c}{${}^{+  1.2}_{-  1.1}$}
& \multicolumn{1}{c}{${}^{+  1.1}_{-  1.0}$}
& \multicolumn{1}{c}{${}^{+  1.2}_{-  0.9}$}
& \multicolumn{1}{c}{${}^{+  1.0}_{-  0.9}$}
& \multicolumn{1}{c}{${}^{+  1.0}_{-  1.0}$}
& \multicolumn{1}{c}{${}^{+  1.0}_{-  0.8}$}
& \multicolumn{1}{c}{${}^{+  2.1}_{-  1.8}$}
\\
Data Statistics & \multicolumn{1}{c}{  0.9}
& \multicolumn{1}{c}{  0.3}
& \multicolumn{1}{c}{  0.1}
& \multicolumn{1}{c}{  0.1}
& \multicolumn{1}{c}{  0.0}
& \multicolumn{1}{c}{  0.1}
& \multicolumn{1}{c}{  0.1}
& \multicolumn{1}{c}{  0.3}
& \multicolumn{1}{c}{  0.6}
& \multicolumn{1}{c}{  1.0}
\\
Total Uncertainty & \multicolumn{1}{c}{${}^{+  3.2}_{-  3.1}$}
& \multicolumn{1}{c}{${}^{+  1.6}_{-  1.5}$}
& \multicolumn{1}{c}{${}^{+  1.1}_{-  1.2}$}
& \multicolumn{1}{c}{${}^{+  1.2}_{-  1.1}$}
& \multicolumn{1}{c}{${}^{+  1.1}_{-  1.0}$}
& \multicolumn{1}{c}{${}^{+  1.2}_{-  0.9}$}
& \multicolumn{1}{c}{${}^{+  1.0}_{-  1.0}$}
& \multicolumn{1}{c}{${}^{+  1.1}_{-  1.0}$}
& \multicolumn{1}{c}{${}^{+  1.2}_{-  1.0}$}
& \multicolumn{1}{c}{${}^{+  2.4}_{-  2.1}$}
\\
\hline\hline
Measured Value [$0.1 e$] & \multicolumn{1}{c}{4.10}
& \multicolumn{1}{c}{3.87}
& \multicolumn{1}{c}{3.75}
& \multicolumn{1}{c}{3.72}
& \multicolumn{1}{c}{3.70}
& \multicolumn{1}{c}{3.69}
& \multicolumn{1}{c}{3.68}
& \multicolumn{1}{c}{3.67}
& \multicolumn{1}{c}{3.62}
& \multicolumn{1}{c}{3.55}
\\
\end{tabular}
\caption{A summary of all the systematic uncertainties and their impact on the jet charge distribution's standard deviation for $\kappa=0.5$ and the more forward jet.   The correction factors are the fake and inefficiency corrections applied before/after the response matrix.  The Other Tracking category includes uncertainty on the track reconstruction efficiency, track momentum resolution, charge misidentification, and fake track rate.  All numbers are given in percent.}
\label{tab:systs_RMS_all_Forward_5aaa}
\end{table} 

\clearpage

\subsection{Correction Factors}
\label{sec:stats}

There are two components to the uncertainty in the fake and inefficiency factors described in Sec.~\ref{sec:unfolding}.  Experimental uncertainties are estimated by re-computing the factors coherently with the variations in the response matrix, as described in Sec.~\ref{sec:calojet} and~\ref{sec:tracking}.  The correction factors encode differences between particle-level and detector-level selections.  The experimental systematic uncertainties take into account variations in the detector-level event selection efficiency.  In order to estimate the uncertainty on the particle-level selection efficiency, two particle-level models are compared.  Fixing the response matrix, the fake and inefficiency factors in {\sc Pythia}~8 are re-weighted to match the corresponding factors in {\sc Herwig++}.  The left plots of Fig.~\ref{fig:systs_fake_beforeunfold} and~\ref{fig:systs_fake_beforeunfold2} show the bin-by-bin difference when unfolding the nominal detector-level {\sc Pythia} 8 sample with the nominal {\sc Pythia} 8 response matrix but fake and inefficiency factors from {\sc Herwig}++.  These differences are mostly below 1\% but can be as high as 10\% in the first $p_\text{T}$ bin.  The corresponding differences in the extracted jet charge average and jet charge distribution standard deviation are shown in the middle and right plots of Fig.~\ref{fig:systs_fake_beforeunfold} and~\ref{fig:systs_fake_beforeunfold2}.  In all $p_\text{T}$ bins aside from the first one, the uncertainties are less than 1\%.  For the jet charge distribution standard deviation, these uncertainties are mostly less than 0.1\%.

\begin{figure}[h!]
\begin{center}
\includegraphics[width=0.33\textwidth]{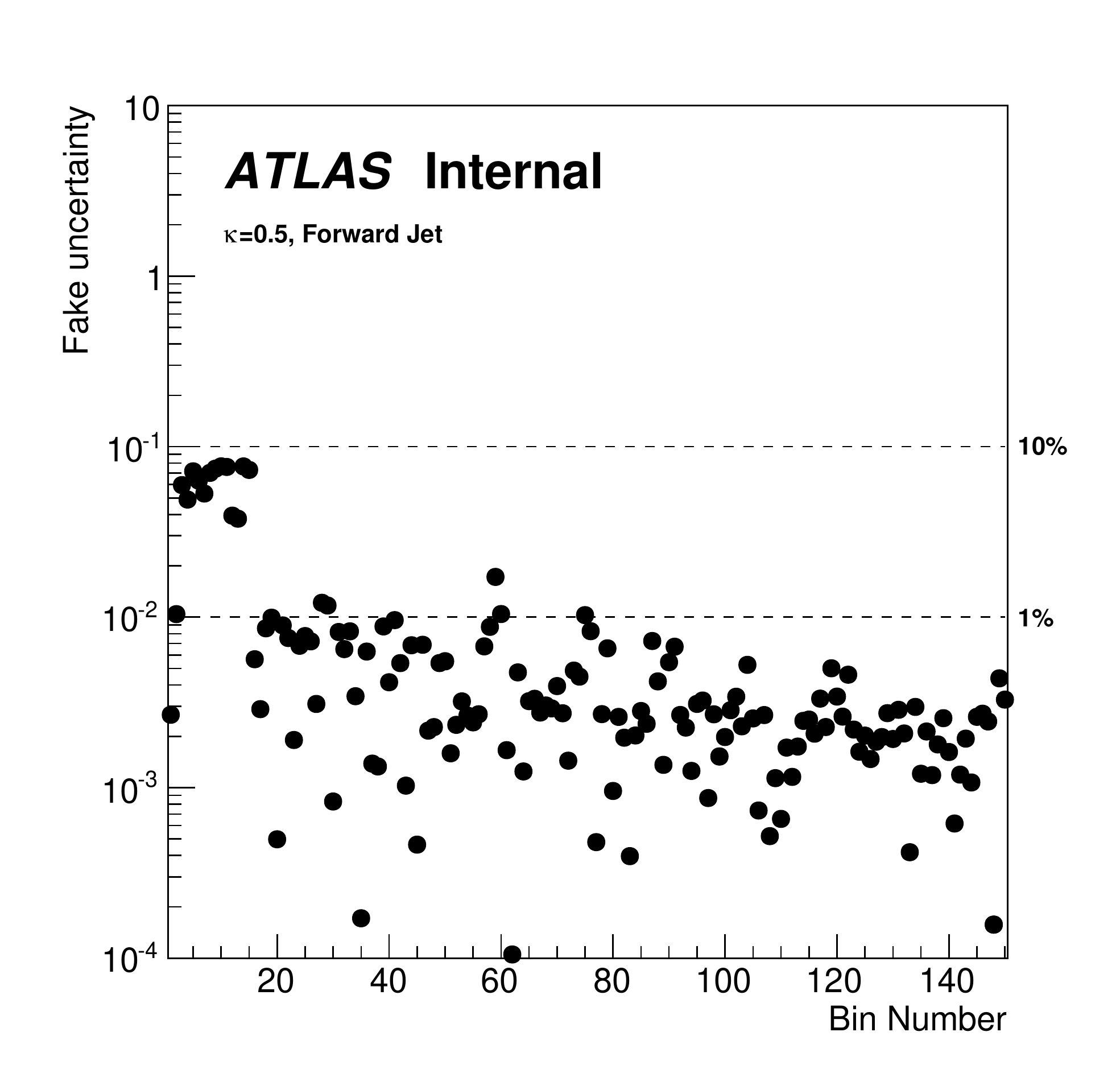}\includegraphics[width=0.33\textwidth]{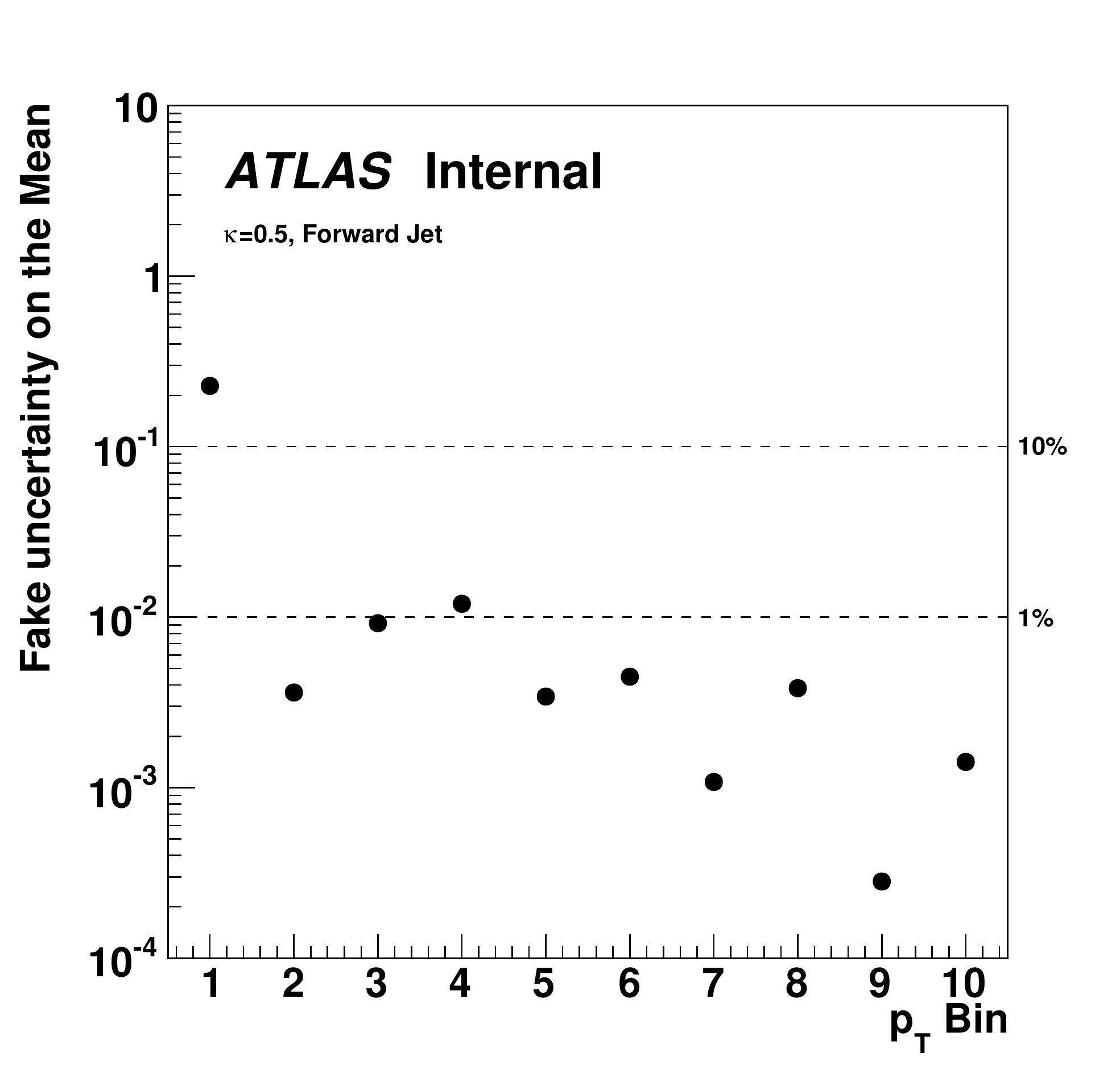}
\includegraphics[width=0.33\textwidth]{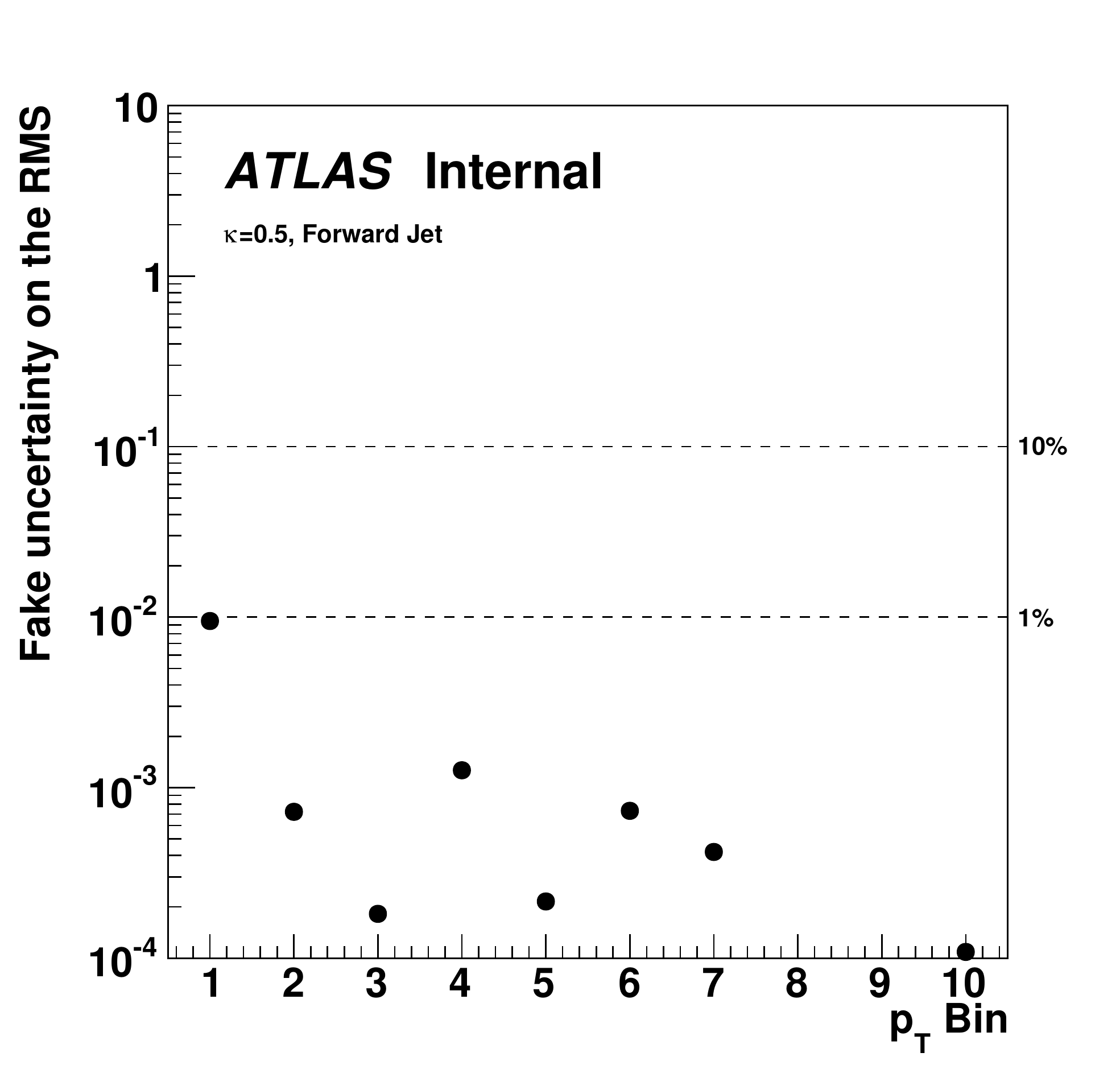}
\caption{The uncertainty on the fake factors by comparing the predictions from {\sc Pythia} 8 and {\sc Herwig} for the more forward jet and $\kappa=0.5$.  The left plot shows the differences in all bins of the combined jet $p_\text{T}$ and jet charge distributions.  The middle and right plots show the uncertainty on the jet charge and jet charge distribution standard deviation, respectively. }
\label{fig:systs_fake_beforeunfold}
\end{center}
\end{figure}

\begin{figure}[h!]
\begin{center}
\includegraphics[width=0.33\textwidth]{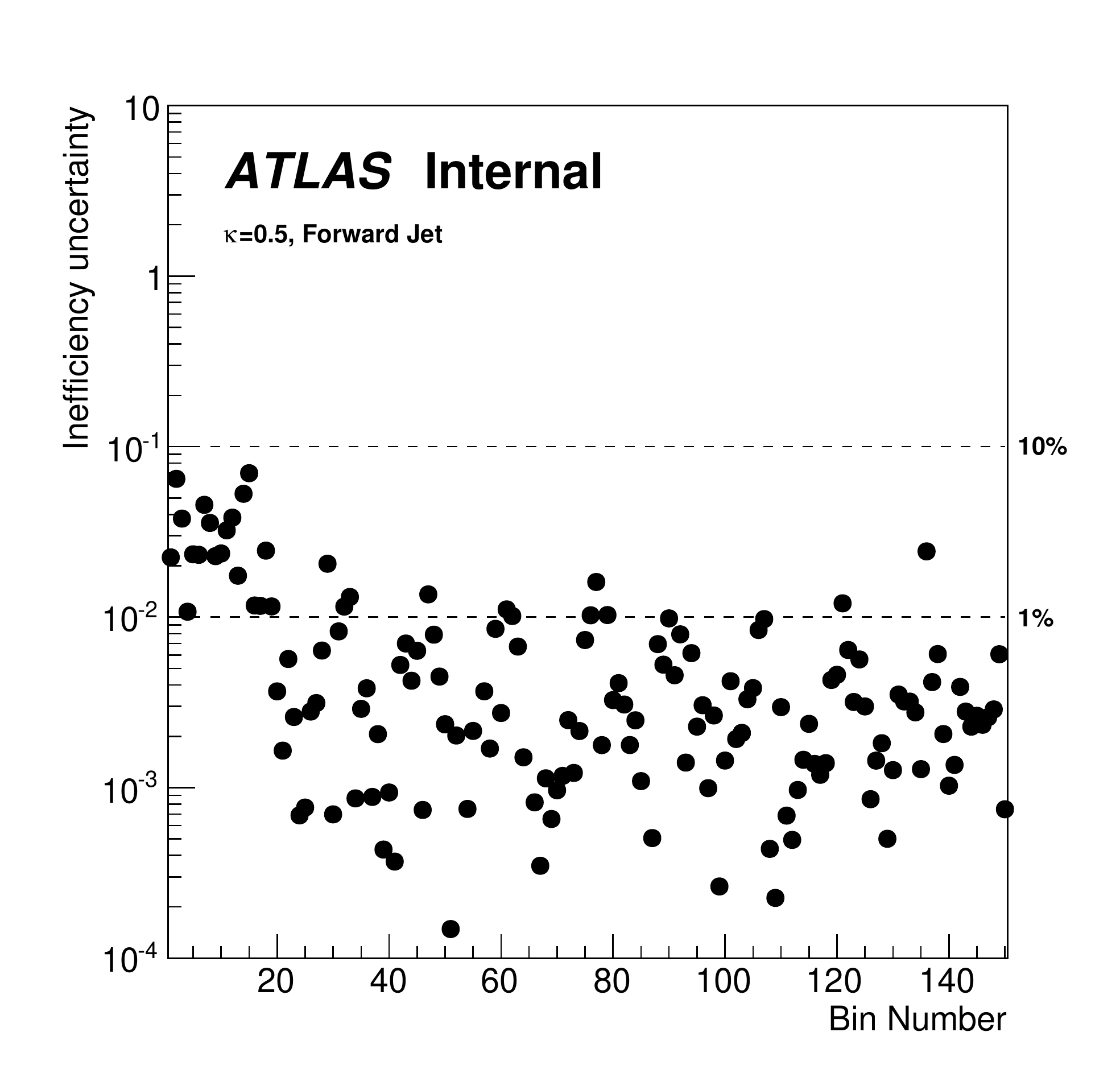}\includegraphics[width=0.33\textwidth]{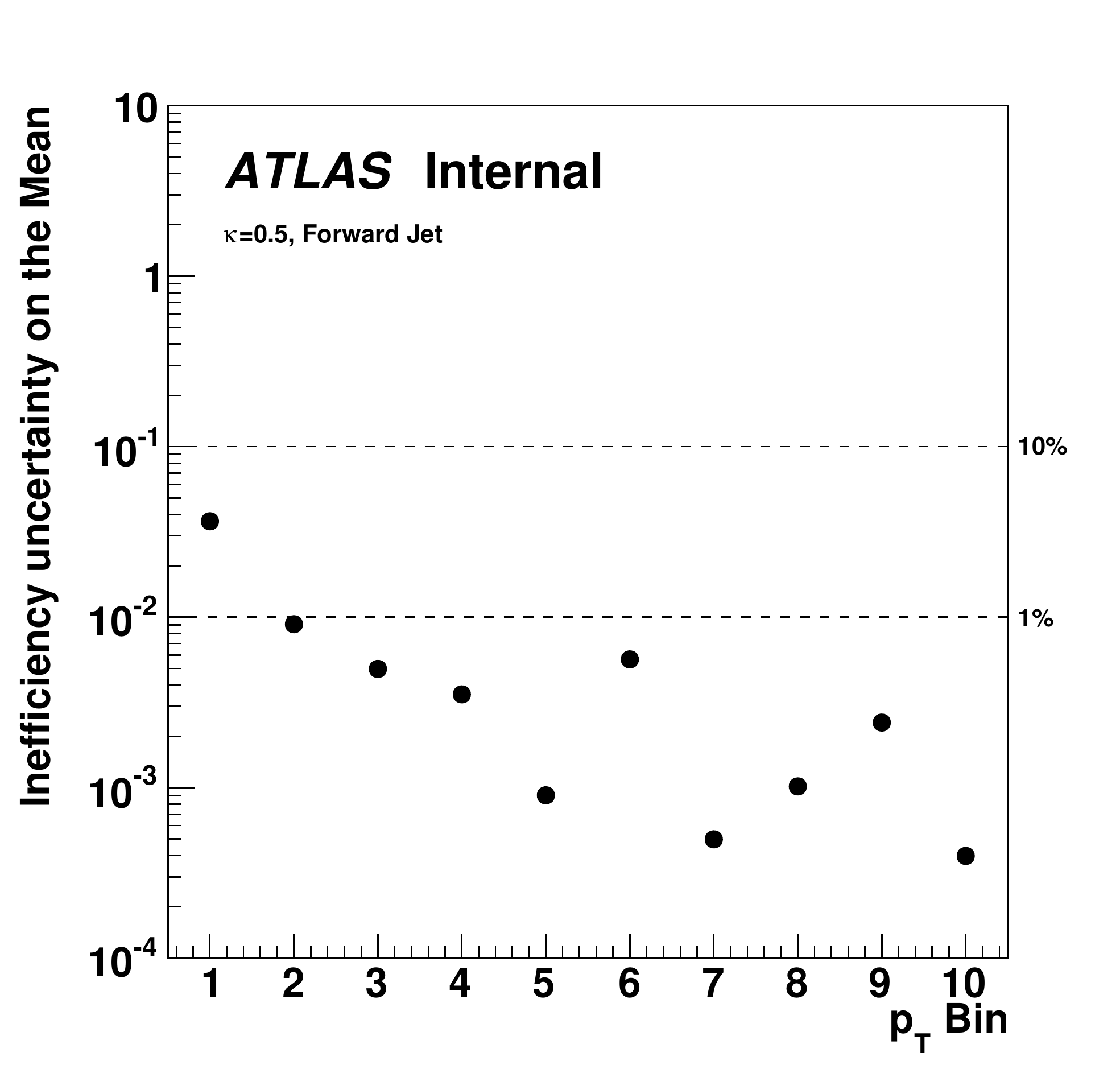}
\includegraphics[width=0.33\textwidth]{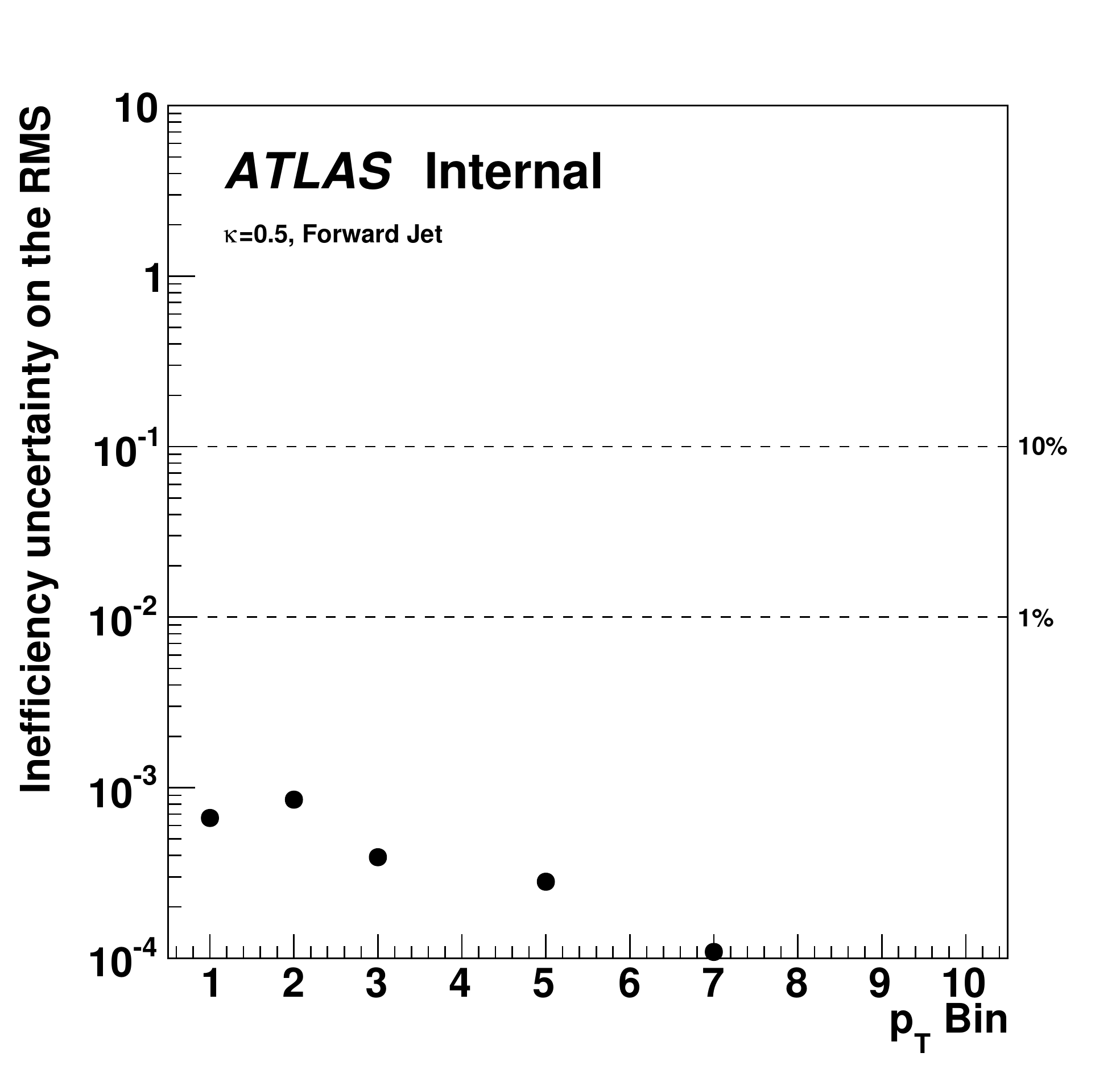}
\caption{The uncertainty on the inefficiency factors by comparing the predictions from {\sc Pythia} 8 and {\sc Herwig} for the more forward jet and $\kappa=0.5$.  The left plot shows the differences in all bins of the combined jet $p_\text{T}$ and jet charge distributions.  The middle and right plots show the uncertainty on the jet charge and jet charge distribution standard deviation, respectively. }
\label{fig:systs_fake_beforeunfold2}
\end{center}
\end{figure}

\clearpage

\subsection{Calorimeter jet uncertainties}
\label{sec:calojet}

Jets are calibrated so that the detector-level $p_\text{T}$ is an unbiased measurement of the particle-level jet $p_\text{T}$ and various data-driven techniques are used to derive {\it in situ} estimates of the difference in this calibration between the data and the simulation.  Uncertainties in the energy scale and resolution of calibrated jets impact the jet charge in the normalization of Eq.~\ref{chargedefcharge} (but preserve the jet charge sign) as well as the binning for the 2D distribution.  Complete details of this source of uncertainty can be found in Ref.~\cite{Aad:2014bia}.  There are many components of the jet energy scale uncertainty.  The {\it in situ} correction is derived from data using the momentum balance in events with $Z$ bosons (low $p_\text{T}$) or photons (moderate $p_\text{T}$) produced in association with jets as well as the balance of multijet (high $p_\text{T}$) and dijet (high $|\eta|$) systems.  Uncertainties on this method stem from the modeling of these processes in simulation.  There is also a contribution from the response to single hadrons~\cite{Aad:2012vm}.  Additional sources of uncertainty are due to the modeling of the in-time and out-of-time pileup corrections to the jet energy scale as well as differences in the response due to the flavor of the jet.  To assess the impact of each component of the jet energy scale uncertainty, the jet energies in simulation are shifted according to the $p_\text{T}$- and $\eta$-dependent $\pm 1\sigma$ variations.  For a fixed variation, the response matrix, and fake and inefficiency factors are recomputed and the unfolding procedure is repeated.  The resulting uncertainty on the jet charge distribution's mean and standard deviation is about 1\% or less for jet $p_\text{T}$ above $200$ GeV.   While subdominant for the average jet charge, the JES uncertainty is dominant for $p_\text{T}\lesssim 1$ TeV for the jet charge distribution RMS.  This is because scaling the jet $p_\text{T}$ by multiplicative factor directly scales the jet charge RMS by the same factor.  The jet charge mean is less effected because of cancellations between positive and negative charges.

The jet energy resolution uncertainty is derived using data-driven techniques in dijet events~\cite{Aad:2012ag}.  To assess the impact of a slightly larger jet energy resolution, jet energies are smeared according to $p_\text{T}$- and $\eta$-dependent factors and propagated through the entire unfolding procedure, as for the jet energy scale uncertainty.  The jet energy resolution uncertainty is subdominant to the jet energy scale uncertainty. 

\clearpage

\subsection{Tracking uncertainties}
\label{sec:tracking}

Uncertainties on tracking are broken down into contributions related to the efficiency of reconstructing tracks and measurements of those tracks that are successfully reconstructed.  In particular, Sec.~\ref{sec:jetcharge:tracksyst:isoeffic} describes the isolated track reconstruction efficiency due to the uncertainty in the inner detector material and Sec.~\ref{sec:jetcharge:tracksyst:TIDE} documents a novel technique for constraining the uncertainty on the modeling of track reconstruction inside the dense hit environment in the core of high $p_\text{T}$ jets.  Then, Sec.~\ref{sec:jetcharge:tracksyst:momentumreso} discusses an estimate of the track momentum resolution due to the modeling of the detector material, magnetic field, and the spatial resolution and alignment of the various detector components.  Additional uncertainties related to the identification of the charge of high $p_\text{T}$ tracks are in Sec.~\ref{sec:jetcharge:tracksyst:chargeid} and the impact of tracks resulting from random combinations of hits is in Sec.~\ref{sec:jetcharge:tracksyst:faketracks}.    Table~\ref{tab:JetCharge:trackingsystoverview} gives an overview of the method and relative size of the various tracking systematic uncertainties.  A common tool for studying and varying the track efficiency and momentum resolution is truth-matching tracks to charged particles, as discussed in Sec.~\ref{sec:jetcharge:tracksyst:truthmatching}.

\vspace{10mm}

\begin{table}[h]
\begin{center}
\begin{tabular}{|c|c|c|c|c|}
\hline
Source & Method & Approximate Size & Section \\ \hline
Isolated Efficiency          & Material Variations & $\lesssim 1\%$ for $|\eta|<2.1$ & ~\ref{sec:jetcharge:tracksyst:isoeffic} \\
Tracking in Jets          & $r_\text{track}$ (Alternate: $\zeta$) & $\lesssim 4\%$ at high $p_\text{T}$ & ~\ref{sec:jetcharge:tracksyst:TIDE} \\
Momentum Resolution          & Resonance Decays & $\sim 2\%$ at high $p_\text{T}$ & ~\ref{sec:jetcharge:tracksyst:momentumreso} \\
Charge Identification          & Resonance Decays & negligible & ~\ref{sec:jetcharge:tracksyst:chargeid} \\
Fake Tracks          & Simulation Variations & $\lesssim 0.5\%$ for $\sigma$(Jet Charge) & ~\ref{sec:jetcharge:tracksyst:faketracks} \\
\hline
\end{tabular}
\end{center}
\caption{An overview of the method and relative size of the various tracking systematic uncertainties.}
\label{tab:JetCharge:trackingsystoverview}
\end{table}

\clearpage

\subsubsection{Truth Matching}
\label{sec:jetcharge:tracksyst:truthmatching}

In the simulation, {\sc Geant4} models the interaction of charged particles with the material of the inner detector.  The deposited energy in each detector element is later digitized and forms the input for the pattern recognition for track reconstruction.  By matching the deposited energy from an individual charged particle with hits on a track, one can associate charged particles to tracks.  This is useful for studying the tracking momentum resolution and charge identification as well as the track reconstruction efficiency.  For each track, define the variable $\Pr_\text{trk}$ as

\begin{align}
\text{Pr}_\text{trk} = \frac{\sum_{i\in\text{ID}} W_i^\text{matched}}{\sum_{i\in\text{ID}} W_i},
\end{align}

\noindent where 

\begin{align}
W_i=\left\{\begin{matrix}w_i & \text{a hit on layer $i$ is part of the track} \cr 0 & \text{else}\end{matrix}\right. ,
\end{align}

\noindent with hit-weight $w_i$ that depends on the detector (defined below) and 

\begin{align}
W_i^\text{matched}=W_i\times \left\{\begin{matrix}1& \text{the matched truth particle deposited energy} \cr 0 & \text{else}\end{matrix}\right. .
\end{align}

\noindent The truth matched particle is the particle in simulation that deposits energy (from {\sc Geant4}) in the most detector elements in common with the track. In other words, if $T$ is the set of truth particles in the simulation, then the matched particle index $i$ is given by

\begin{align}
i = \text{argmax}_{j\in T} \sum_{k\in\text{ID}}\left\{\begin{matrix} 1 & \text{Particle $j$ left energy in $k$ and $k$ is part of the track.} \cr 0 & \text{else}\end{matrix}\right.
\end{align}

\noindent If multiple particles deposited energy in the same pixel, only the one that left the highest energy is considered.  The set $\text{ID}$ contains the various layers of the inner detector (the pixel detector, SCT, and TRT) and $w_i=10$ for the pixel detector, $5$ for the SCT and 1 for the TRT.  The weight for the pixel detector is twice the weight for the SCT because two hits are required in the SCT to give 3D information about the track location, but one hit in the pixel detector already gives this information.  A higher weight is used for the pixel detector over the TRT because even though the TRT is useful for the momentum measurement, many of the important track parameters (such as $d_0,z_0$) are defined in the pixels.  

A track in simulation is considered {\it real} if $\Pr_\text{trk}$ is at least 0.5 and {\it fake} otherwise.  One can remove the resolution of real tracks by replacing their momentum with the matched charged particle $p_\text{T}$.  

\clearpage

\subsubsection{Isolated Track Reconstruction Efficiency}
\label{sec:jetcharge:tracksyst:isoeffic}

The uncertainty on the track reconstruction efficiency is mostly due to the uncertainty in the material in the inner detector.  The material is known to within $\sim 5\%$~\cite{Aad:2011cxa}.  This precise modeling of the material in the ID has led to sub-percent level uncertainties in the track reconstruction efficiency for $|\eta|<2.1$~\cite{Aad:2014xca}.  These uncertainties are estimated as a function of $p_\text{T}$ and $\eta$ by comparing the track reconstruction efficiency in simulated detector geometries with various levels of material in the ID. In the forward region of the tracking acceptance, the material is less constrained and so older and larger uncertainties are still used to set the uncertainty based on the radius dependence of the $K_s^0\rightarrow\pi^+\pi^-$ reconstructed invariant mass and the length of tracks reaching into the SCT~\cite{Aad:2010ac}.  Table~\ref{tab:trackeffic} summarizes the track reconstruction efficiency uncertainties.

\begin{table}[h]
\begin{center}
\begin{tabular}{|c|c|c|c|c|}
\hline
$p_\text{T}$ [GeV] & $|\eta|<1.5$ &  $1.5<|\eta|<2.1$ & $2.1<|\eta|<2.3$  & $2.3<|\eta|<2.5$ \\ \hline
[0.5,1]          & 0.7\%        & 1.2\%         &   4\% & 7\% \\ \hline
$\geq 1$          & 0.5\%       & 1.1\%       &  3.2\%   & 5.6\%    \\ \hline
\end{tabular}
\end{center}
\caption{A summary of the track reconstruction efficiency uncertainties.}
\label{tab:trackeffic}
\end{table}

\noindent In order to estimate the impact of these uncertainties, tracks are randomly removed with $\eta$ and $p_\text{T}$ dependent probabilities as stated in Table~\ref{tab:trackeffic}.  The studies used to determine the inclusive track reconstruction efficiency did not have an explicit track $\chi^2$ requirement.  Since this analysis requires $\chi^2/\text{NDF}\geq 3$, we must also check the data/MC differences of this further selection.  Figure~\ref{fig:track_chi2} shows that the $\chi^2$ cut is very efficient, with a $\geq 99\%$ track efficiency in all jet $p_\text{T}$ bins.  The efficiency is generally higher in the simulation than in the data, with a $\lesssim 10\%$ difference in all $p_\text{T}$ bins and there is no strong evidence for a $p_\text{T}$ dependence in the level of the mis-modeling.  In order to assess the impact of the $\chi^2$ requirement mis-modeling, tracks are randomly removed with a jet $p_\text{T}$-dependent probability.  The $\chi^2/\text{NDF}\geq 3$ requirement efficiency in the simulation (from the left plot of Fig.~\ref{fig:track_chi2}) is well approximated by $1-f(p_\text{T})$ where $f(x)=a+bx+cx^2$ for $a=0.005,b=4\times 10^{-6}/\text{GeV}$, and $c=-6.5\times 10^{-10}/\text{GeV}^2$.  Therefore, tracks are removed randomly with probability given by $10\% \times (1-f(p_\text{T}))$.

\begin{figure}[h!]
\begin{center}
\includegraphics[width=0.45\textwidth]{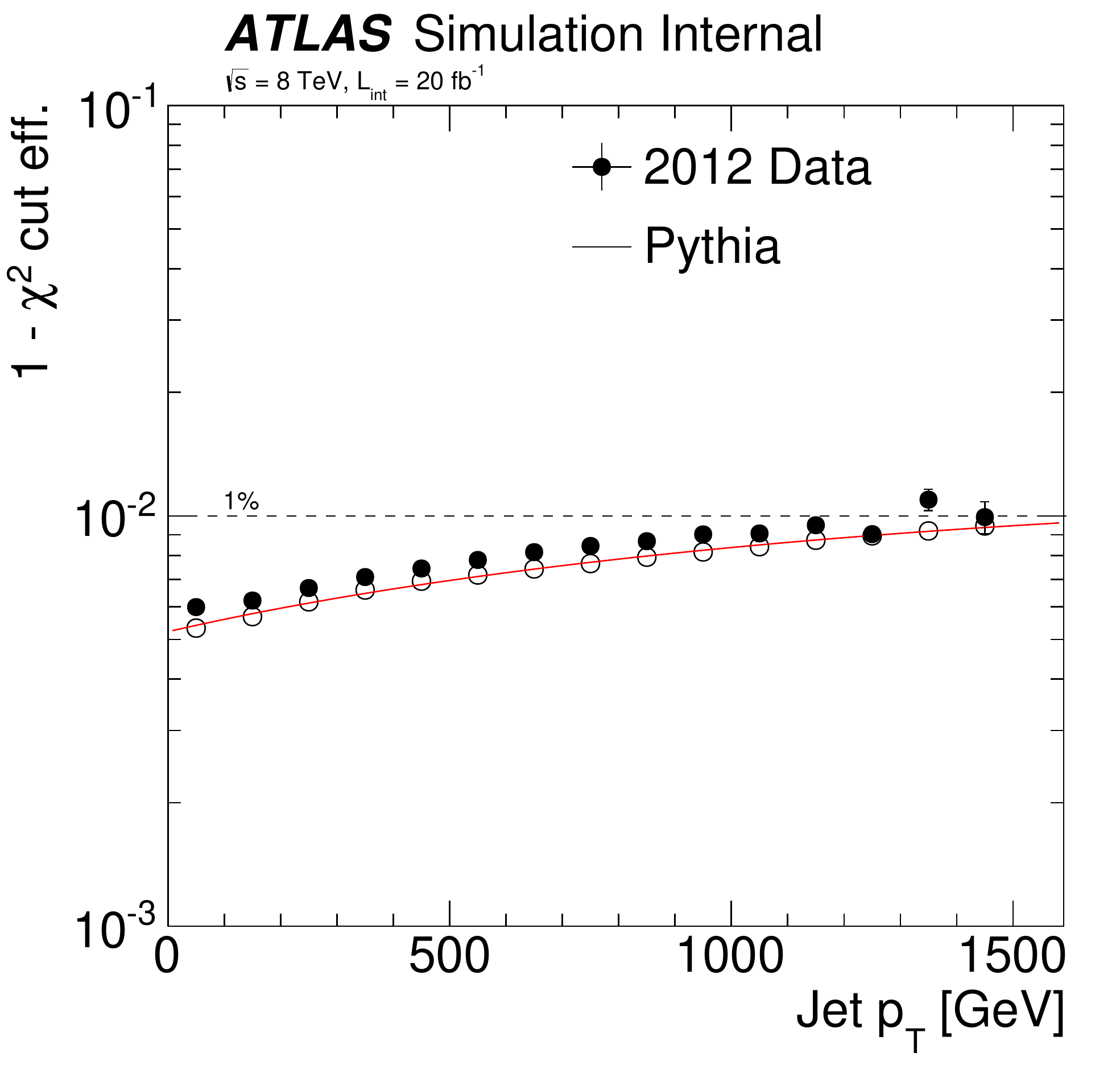}\includegraphics[width=0.45\textwidth]{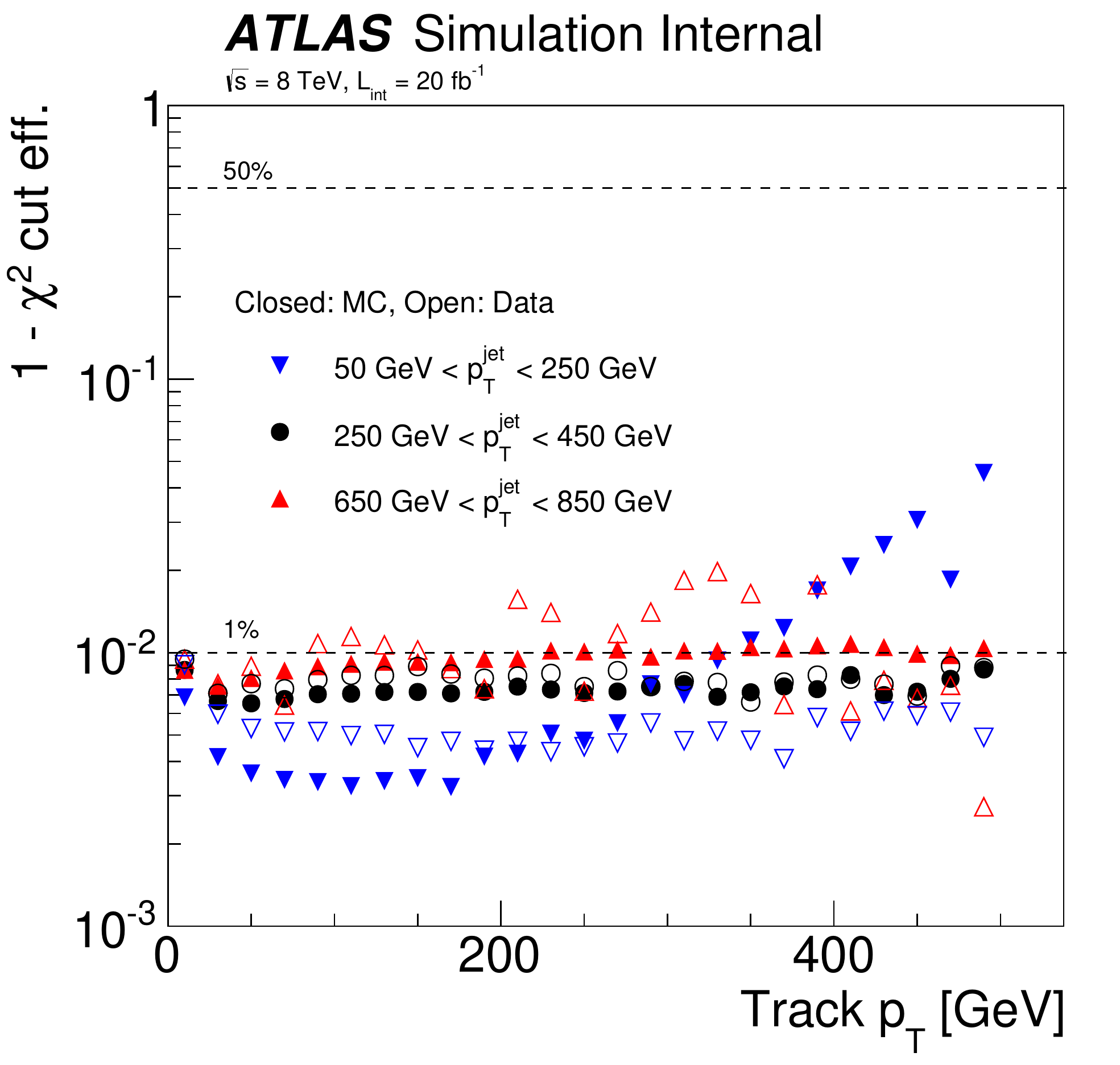}
\caption{Left: The efficiency of the track $\chi^2/\text{NDF}\geq 3$ requirement given all other track requirements as a function of the jet $p_\text{T}$.  The track $p_\text{T}$ dependence for three $p_\text{T}$ bins is shown in the right plot.  Note that no statistical uncertainties are included in the right plot (the size of the bin-to-bin fluctuations indicates the size of these uncertainties).  The low jet $p_\text{T}$ efficiency is poor at high track $p_\text{T}$ due to the large contribution from fakes.}
\label{fig:track_chi2}
\end{center}
\end{figure}

\clearpage

\subsubsection{Track Reconstruction Inside Jets}
\label{sec:jetcharge:tracksyst:TIDE}

In addition to the loss of tracks due to the material in the inner detector, tracks can be lost due to the high hit density inside the core of jets.  A useful variable for quantifying this loss is the charged energy ratio (CER), computed from particle-level jets that are geometrically matched to detector-level jets:

\begin{align}
\text{CER} = \left\langle\frac{\sum p_\text{T}^\text{charged,matched}}{\sum p_\text{T}^\text{charged}}\right\rangle,
\end{align}

\noindent where the denominator runs over all charged particles in the particle-level jet and the numerator runs over all tracks reconstructed inside the detector-level jet.  To remove track resolution effects, the tracks in the numerator are replaced with the matched charged particle (fakes and secondaries\footnote{Tracks from material interactions such as photon conversions, $\gamma\rightarrow e^+e^-$.} are thus excluded).  The CER is plotted in Fig.~\ref{fig:SystematicUncertainties/TIDEfig1} as a function of the jet $p_\text{T}$.  The CER decreases at low jet $p_\text{T}$ due to decreasing importance of losses due to hadronic interactions inside the detector and decreases at high jet $p_\text{T}\gtrsim 500$ GeV due to track merging inside high density jet cores.  The {\it loss} is defined 

\begin{align}
\label{eq:charge:loss}
\text{loss}(\text{jet } p_\text{T}) = \max_{\text{jet } p_\text{T}'}\text{CER}(\text{jet } p_\text{T}')-\text{CER}(\text{jet } p_\text{T}).
\end{align}

\noindent For most analyses using tracks, the uncertainty on the modeling of the loss in Eq.~\ref{eq:charge:loss} is negligible because the jets have $p_\text{T}\lesssim 500$ GeV.  Early Run 1 studies also suggest that in this low $p_\text{T}$ regime, hit sharing is well modeled by the simulation~\cite{Aad:2011sc}.  However, the jet charge measurement is probing a new kinematic regime involving tracks inside jets: the loss is not small and therefore a careful assessment of the systematic uncertainty is critical.  An early Run 2 method uses double peaks in the $dE/dx$ distribution~\cite{dedx}. Such an approach is quite general, but neglects the impact of track $p_\text{T}$, which is important for the jet charge due to the track $p_\text{T}$ weighting in the definition.  The rest of this section describes a new method for constraining the loss modeling with data using a detector-level analogue to the CER.

\begin{figure}[h!]
\begin{center}
\includegraphics[width=0.6\textwidth]{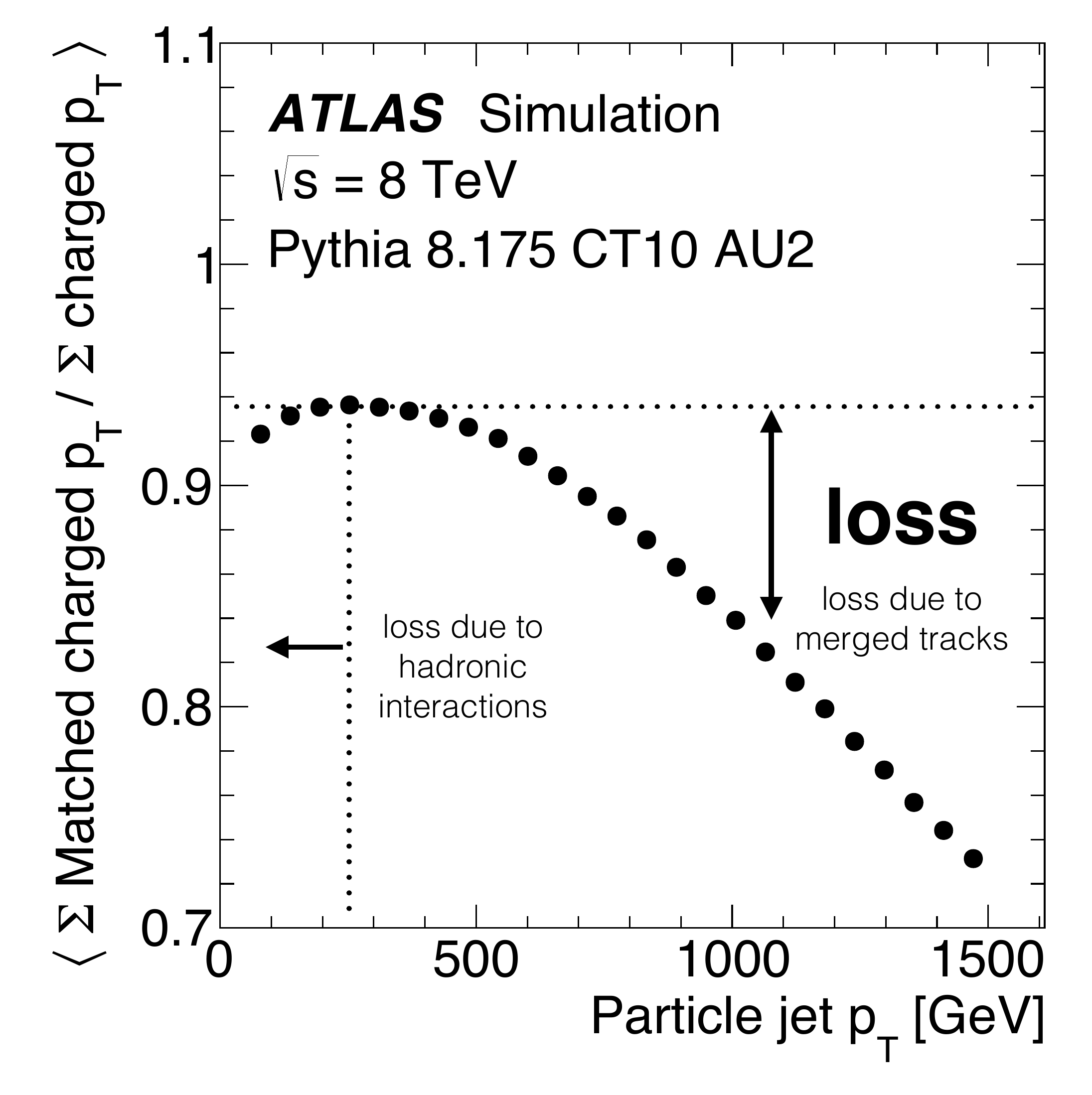}
\caption{The CES as a function of the particle-level jet $p_\text{T}$.}
\label{fig:SystematicUncertainties/TIDEfig1}
\end{center}
\end{figure}

An outline of the new method is as follows:

\begin{enumerate}
\item Demonstrate that the calorimeter loss from the charged-to-total energy ratio (CTER) can be used as a proxy for the loss derived from the CER.
\item Measure the CTER in the data to constrain the loss.
\item Translate the modeling of the loss to an uncertainty on the tracking efficiency.  Consider the impact on the jet charge measurement from all parameterizations of the tracking inefficiency that reproduce the measured loss.
\item Prove that for the (average) jet charge, the tracking inefficiency can be treated as independent per track.
\item The systematic uncertainty on the jet charge is then given by the data/MC difference in the measured loss and is conservatively implemented using an additional tracking inefficiency that has the biggest impact on the jet charge.
\end{enumerate}

The CTER is defined as

\begin{align}
\text{CTER} =\left \langle\frac{\sum p_\text{T}^\text{charged}}{p_\text{T}^\text{jet}}\right\rangle,
\end{align}

\noindent where the denominator is the full (particle- or detector-level jet) and the numerator is the sum over tracks (detector-level) or charged particles (particle-level).  Analogously to the loss, the calorimeter (calo) loss is

\begin{align}
\text{calo loss}(\text{jet } p_\text{T}) = \max_{\text{jet } p_\text{T}'}\text{CTER}(\text{jet } p_\text{T}')-\text{CTER}(\text{jet } p_\text{T}),
\end{align}

\noindent and the fractional calo loss is the calo loss divided by the maximum CTER.  Analogously, the fractional loss is the loss divided by the maximum CER.  The first observation is that in the simulation, the fractional calo loss is similar to the fractional loss.  This makes sense heuristically, since the energy depositions in the calorimeter do not depend on how close the particles are when they reach the calorimeter and so fractional changes in the CTER as a function of jet $p_\text{T}$ should be due to tracking inefficiencies in the core of jets.  Empirical evidence for this similarity is shown in Fig.~\ref{fig:SystematicUncertainties/TIDEfig2}, in which the fractional loss and the fractional calo loss (detector-level) are nearly identical as a function of jet $p_\text{T}$.  Differences between the data and MC in the left plot of Fig.~\ref{fig:SystematicUncertainties/TIDEfig2} indicate that the simulation underestimates the fractional loss by a relative $\sim 10\%$.

\begin{figure}[h!]
\begin{center}
\includegraphics[width=0.5\textwidth]{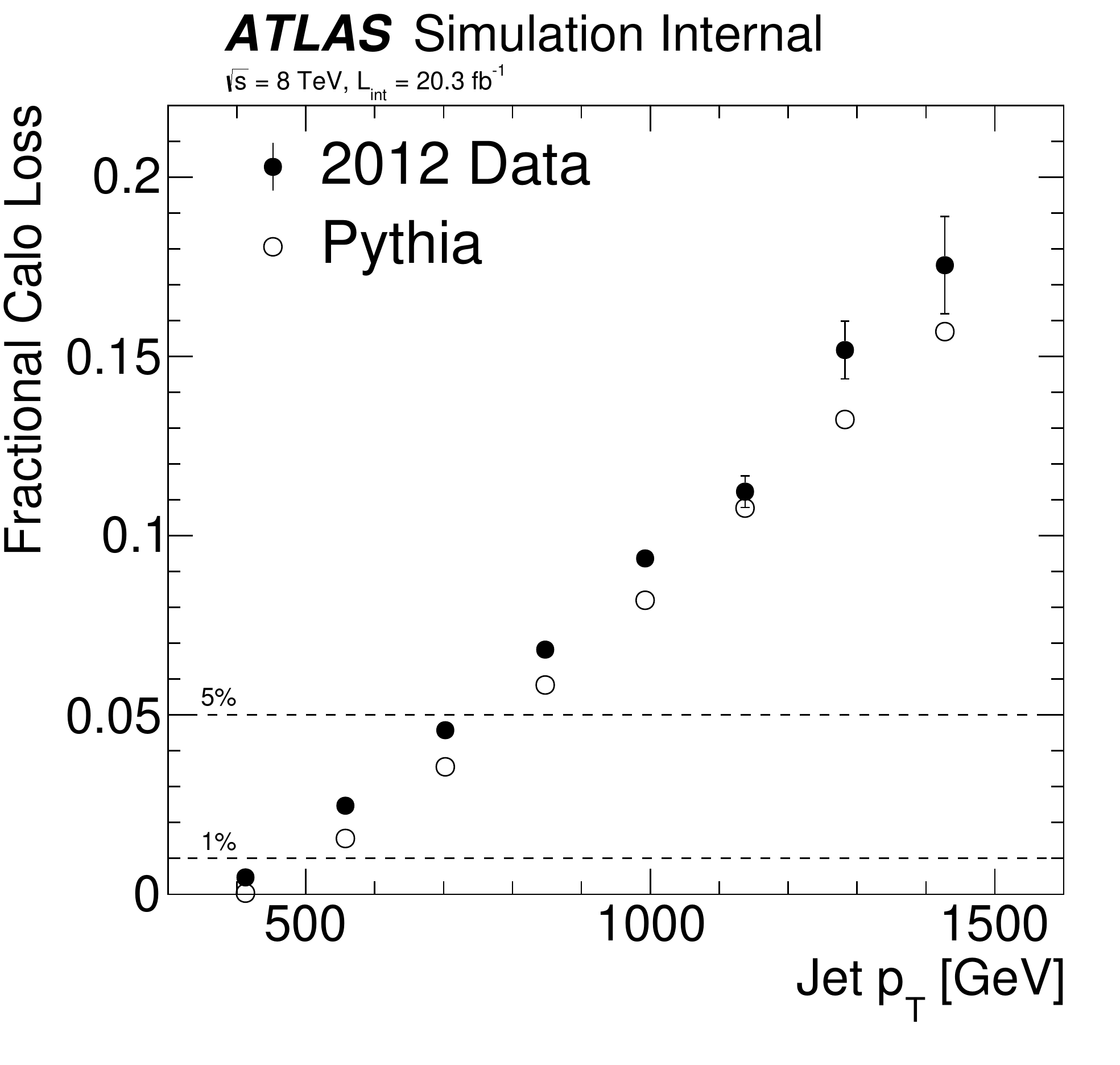}\includegraphics[width=0.5\textwidth]{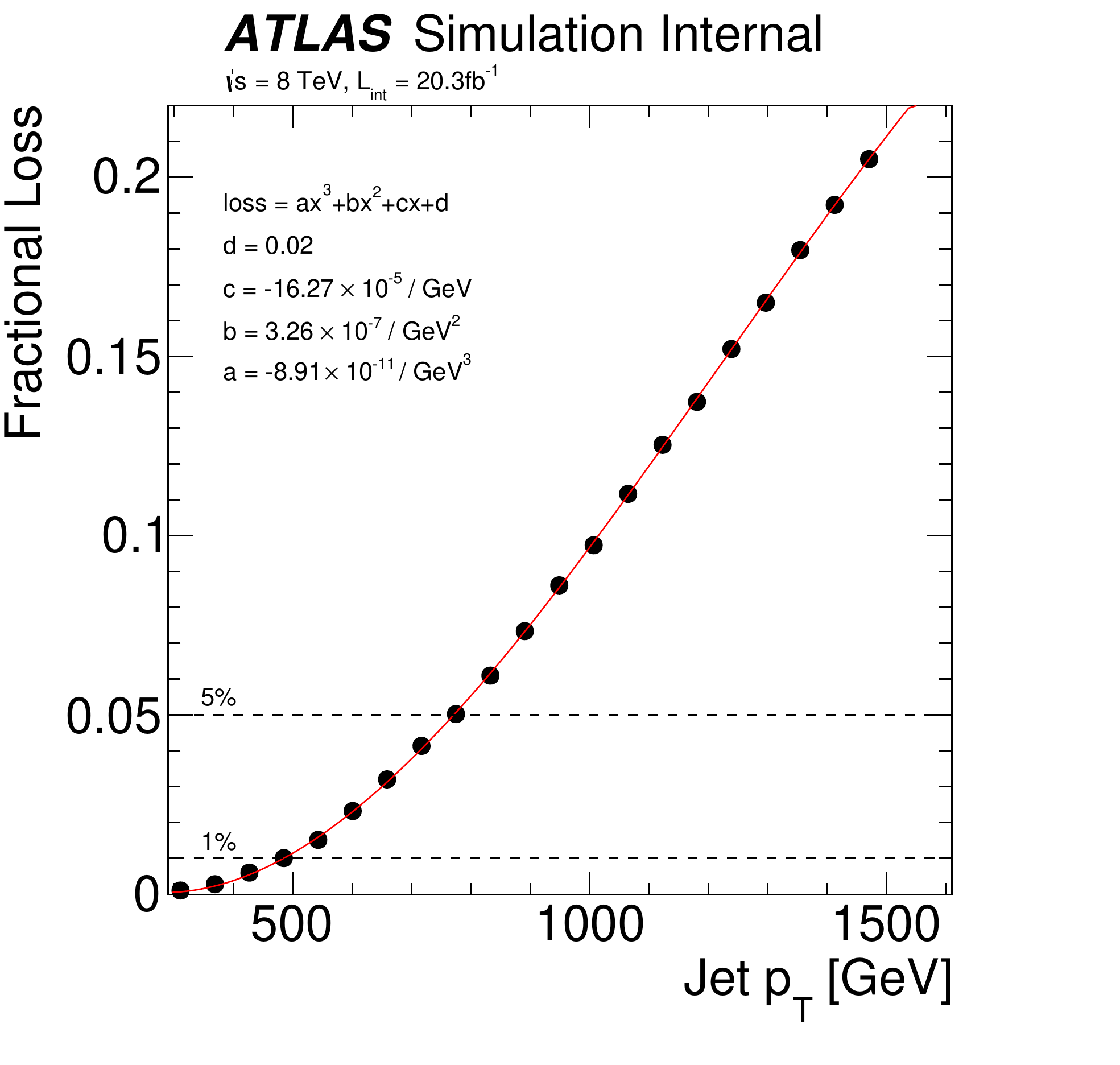}
\caption{Left: The fractional calo loss in data and simulation.  Right: the fractional loss in simulation.}
\label{fig:SystematicUncertainties/TIDEfig2}
\end{center}
\end{figure}

A few more details about the $p_\text{T}$-dependence of the CTER are required before performing a careful measurement of the loss.  First of all, it is important that the fake (see Sec.~\ref{sec:jetcharge:tracksyst:faketracks}) and secondary track (see Fig.~\ref{fig:secondaries}) rates are negligible with the track quality criteria requirements\footnote{It would have been sufficient for these rates to be independent of $p_\text{T}$ for the charged-energy method to work.}.  Next, it is crucial that the particle-level CTER does not depend on $p_\text{T}$.  If it did, then changes in the detector-level CTER may simply be due to changes in the particle-level CTER.  Fortunately, the particle-level CTER is $p_\text{T}$ independent and is nearly 2/3 due to isospin: there are nearly twice as many charged pions inside jets as neutral pions, with small deviations due to presence of heavier hadrons and bremstrahlung photons.  Furthermore, the exact value of the particle-level CTER is largely generator-independent, in part because it is highly constrained by low(er) energy physics.   Figure~\ref{fig:SystematicUncertainties/TIDEfig3} shows the detector- and particle-level CTER as a function of the jet $p_\text{T}$; the difference between {\sc Pythia} 8 and {\sc Herwig++} is at or below the 0.1\% level.

\begin{figure}[h!]
\begin{center}
\includegraphics[width=0.7\textwidth]{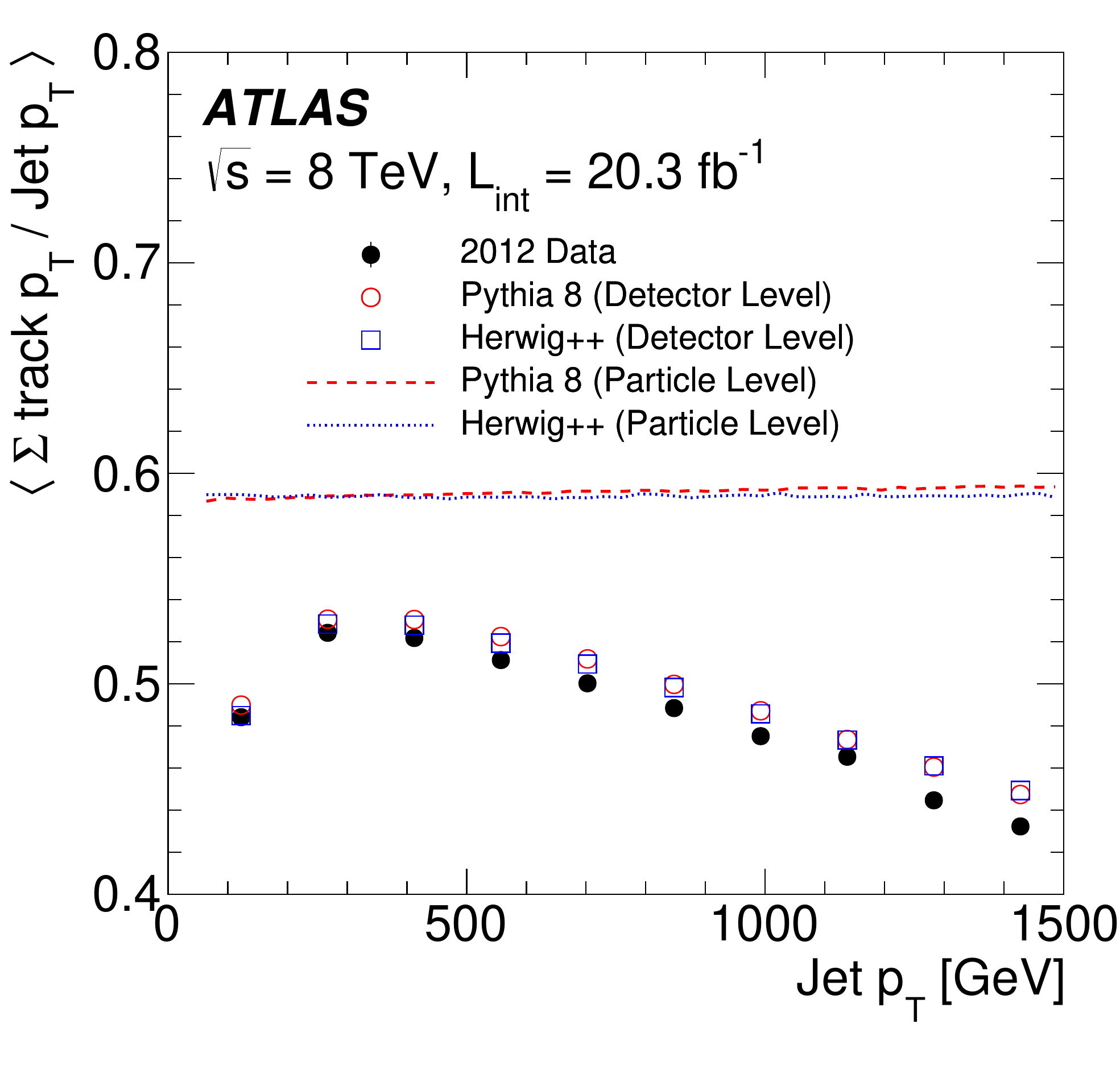}
\caption{The detector- and particle-level CTER as a function of the jet $p_\text{T}$ for Pythia, Herwig, and the data.}
\label{fig:SystematicUncertainties/TIDEfig3}
\end{center}
\end{figure}

The final requirement for the CTER to be a useful proxy for the CES is that the jet $p_\text{T}$ should be an unbiased measurement of the particle-jet $p_\text{T}$.  If there is a $p_\text{T}$-dependent bias, then changes in the CTER as a function of $p_\text{T}$ could be due to the change in the biased measurement of the jet $p_\text{T}$.  This is not exactly satisfied.  The jet energy scale uncertainty is not zero and does depend on $p_\text{T}$.  However, it is small, and one can quantify its influence on the measurement by conservatively adding the JES uncertainty in quadrature to the uncertainty that is determined from the differences between data and MC in describing the calo loss.

All the ingredients are now ready to quantitatively measure the calo loss and by proxy determine an uncertainty on the fractional loss.  The data and MC calo loss distributions have already been shown in Fig.~\ref{fig:SystematicUncertainties/TIDEfig2} and Fig.~\ref{fig:SystematicUncertainties/TIDEfig3}.  The data/MC fractional difference is shown in Fig.~\ref{fig:SystematicUncertainties/TIDEfig4}, also added in quadrature with the data statistical uncertainty and all of the JES uncertainty components.  The total uncertainty is about 1\% until about 1.2 TeV, after which it increases to about 2\%.

\begin{figure}[h!]
\begin{center}
\includegraphics[width=0.5\textwidth]{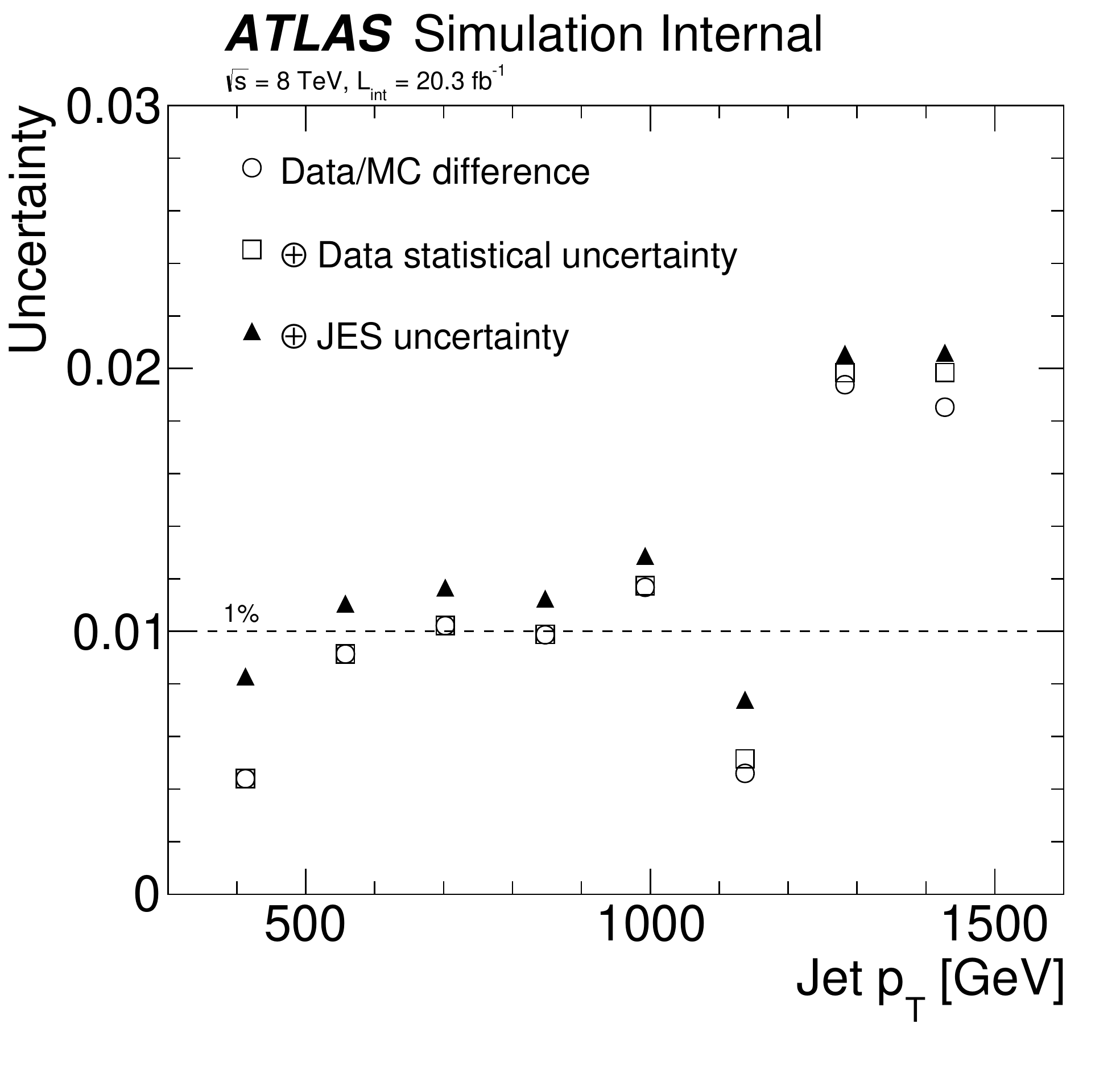}
\caption{The uncertainty in the fractional loss determined from the difference in the data/MC of the fractional calo loss and the JES uncertainties.}
\label{fig:SystematicUncertainties/TIDEfig4}
\end{center}
\end{figure}

The next challenge is to translate the fractional uncertainty in the loss to an uncertainty in the tracking (in)efficiency.  To begin, assume that one can treat the inefficiency as independent per track.  This means that a good model for increasing the loss in the simulation to match the loss in the data is to randomly remove tracks independent of the other tracks in the jet.  This assumption will be justified below.  

\noindent Consider all possible tracking inefficiency uncertainty parameterizations that reproduce the loss:

\begin{align}
\label{eq:charge:epsion}
\left\langle \frac{\sum_i p^\text{kept}_{T,i}}{\sum_i p_{T,i}}  \right\rangle = 1-\epsilon,
\end{align}

\noindent where $\epsilon$ is the uncertainty determined in Fig.~\ref{fig:SystematicUncertainties/TIDEfig4}, the sum runs over all tracks associated to the detector-level jet, and {\it kept} indicates that the track was retained after randomly removing tracks.  Consider a generic parameterization: $\Pr(\text{drop track $i$})=\alpha p_{T,i}^n$, where $n$ is a non-negative integer.  For a fixed parameterization $(n)$, there is one measurement $(\epsilon)$ and one unknown $(\alpha)$.  Due to the form of Eq.~\ref{eq:charge:epsion}, the relationship between $\alpha$ and $\epsilon$ is linear.  Therefore, the solution for $\alpha$ is unique: 

\begin{align}
1-\epsilon = \left\langle \frac{\sum_i p^\text{kept}_{T,i}}{\sum_i p_{T,i}}  \right\rangle &=\frac{\sum_i (1-\alpha p_{T,i}^n)p_{T,i}}{\sum_i p_{T,i}} \implies \alpha = \epsilon\frac{\sum_i p_{T,i}} {\sum_i  p_{T,i}^{n+1}}.
\end{align}

\noindent Any choice of $n$ with the above value of $\alpha$ will exactly reproduce the fractional calo loss observed in the data.  The value of $\epsilon$ is not exactly the the values shown in Fig.~\ref{fig:SystematicUncertainties/TIDEfig4}, which are the absolute difference in fractional loss $x\%$:

\begin{align}
\text{fractional loss (MC)} - \text{fractional loss (data)} = x\%, 
\end{align}

\noindent where $x\sim 1\%$ for $p_\text{T} < 1.2$ TeV and $x\sim 2\%$ for $p_\text{T}>1.2$ TeV.  Let $\text{max}_i =  \text{max}_{\text{jet } p_\text{T}'}\text{CTER}(\text{jet } p_\text{T}')$ for $i\in\{\text{MC,data}\}$.  Then the relationship between $\epsilon$ and $x$ is given by:

\begin{align}\nonumber
\epsilon &= x\%\times\frac{\text{max}_\text{MC}}{\text{CTER}_\text{MC}}+\frac{\text{max}_\text{MC}-\text{CTER}_\text{MC}}{\text{CTER}_\text{MC}}-\frac{\text{max}_\text{MC}}{\text{CTER}_\text{MC}}\left(\frac{\text{max}_\text{data}-\text{data}_\text{MC}}{\text{max}_\text{data}}\right)\\\nonumber
&\sim x\%\times\frac{\text{max}_\text{MC}}{\text{CTER}_\text{MC}}.
\end{align}

\noindent Thus, the value of $\epsilon$ is about 1\% for $p_\text{T} < 1.2$ GeV and about $3\%$ for $p_\text{T} > 1.2$ TeV.  The closure for inefficiency parameterizations for $n=0,1,2,3,10\sim \infty$ are shown in Fig.~\ref{fig:SystematicUncertainties/TIDEfig5}.  For all values of $n$, the fractional calo loss is the same as the data (by construcdtion) and higher than the nominal simulation.

\begin{figure}[h!]
\begin{center}
\includegraphics[width=0.6\textwidth]{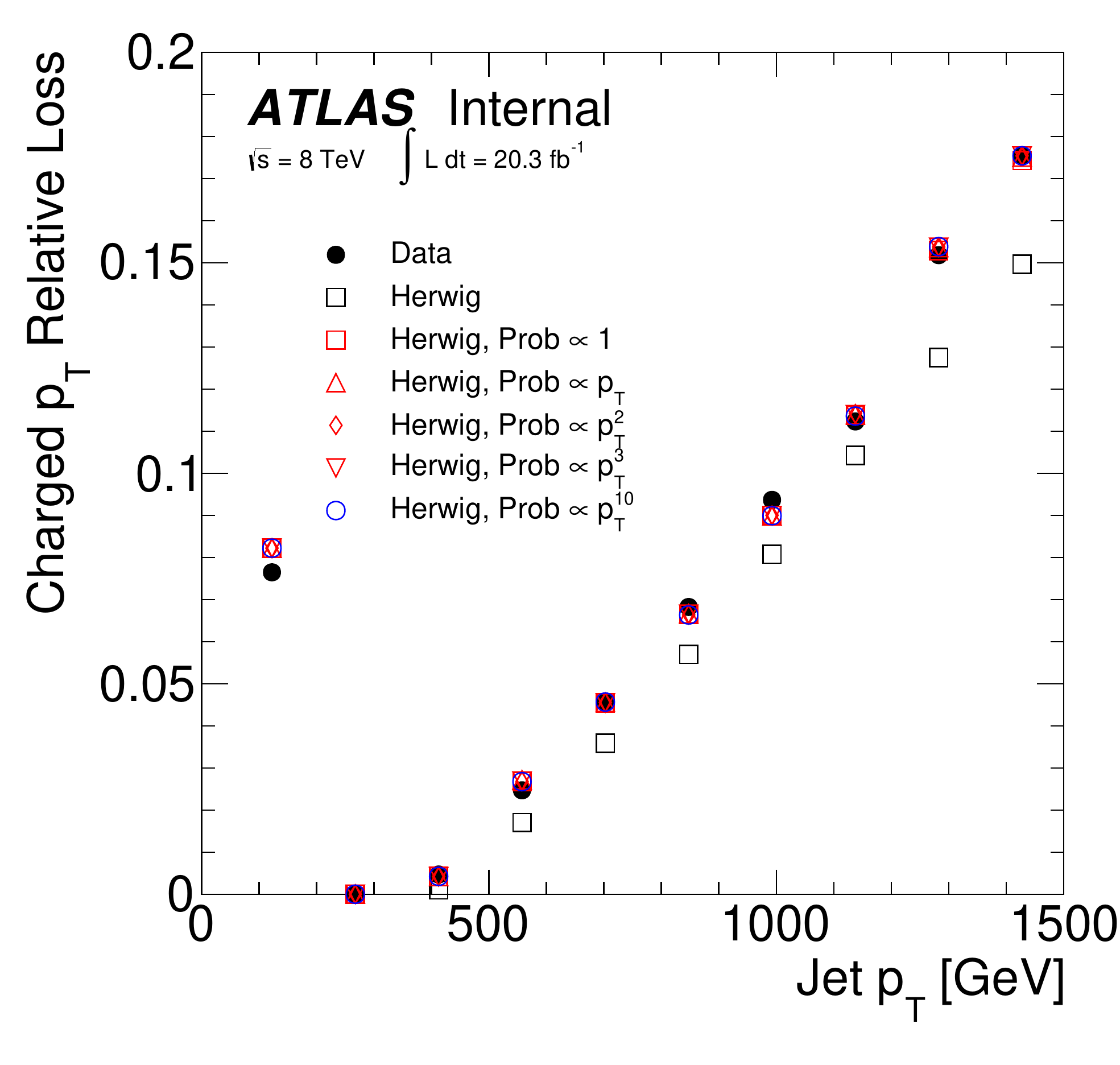}
\caption{The fractional calo loss for the nominal simulation, the data, and various parameterizations of the tracking inefficiency uncertainty.}
\label{fig:SystematicUncertainties/TIDEfig5}
\end{center}
\end{figure}

The next step is to justify the assumption that the tracking inefficiencies can be treated as independent.   In general, the tracking inefficiency will depend on the other tracks present in the jet, measured and unmeasured.  However, the CTER and the average jet charge are special variables which is not sensitive to these effects.  First, for the CTER:

\begin{align}
\label{eq:charge:cterproof}
\text{CTER}&=\left\langle \frac{\sum_{i\in\mathrm{Tr}}p_{T,i}}{p_\text{T}^\text{jet}}\right\rangle =  \frac{\langle\sum_{i\in\mathrm{Tr}}p_{T,i}\rangle }{p_\text{T}^\text{jet}} = \frac{\sum_{j\in J}w_j\sum_{i\in\mathrm{Tr}}p_{T,i}}{p_\text{T}^\text{jet}\sum_{j\in J}w_j}\\
&=\frac{\sum_{p_\text{T}} p_\text{T} n_{p_\text{T}}}{p_\text{T}^\text{jet}\sum_{j\in J}w_j},
\end{align}

\noindent where $J$ is the set of all jets and $w_j$ is the event weight associated with jet $j$.  The first equality is true in a fixed jet $p_\text{T}$ bin and the the last equality is true by exchanging the finite sums.  The quantity $n_{p_\text{T}}=\sum_{j\in J} w_j n_{p_\text{T},j}$, where $n_{p_\text{T},j}$ is the number of tracks in jet $j$ with transverse momentum $p_\text{T}$.  Equation~\ref{eq:charge:cterproof} shows that the CTER only depends on the total number of tracks of a given $p_\text{T}$ in a particular jet $p_\text{T}$ bin and the total weighted number of jets in all of the jet $p_\text{T}$ bins.  This means that a sufficient statistic for the CTER is the joint distribution of track $p_\text{T}$ and jet $p_\text{T}$, i.e. one does not need know the distribution of track $p_\text{T}$ inside each individual jet.   As a closure of the method, one can compute the tracking efficiency as the ratio of the detector-level and particle-level jet and track $p_\text{T}$ joint distributions.  The closure test is then to correct each track and see if the detector-level CTER after the track-by-track correction is given by the particle level CTER\footnote{The tracking efficiency depends also on $\eta$, but since the jet charge is not measured as a function of $\eta$, all distributions are marginalized over $\eta$.}.  The efficiencies are shown in Fig.~\ref{fig:SystematicUncertainties/TIDEfig8b}.  The left plot of Fig.~\ref{fig:SystematicUncertainties/TIDEfig8b} is the track reconstruction efficiency with the impact of jet and track resolutions removed, while the right plot of Fig.~\ref{fig:SystematicUncertainties/TIDEfig8b} is what is applied in practice.  Due to resolution effects, the `efficiency' can exceed one in the right plot of Fig.~\ref{fig:SystematicUncertainties/TIDEfig8b} while the $z$-axis is between $0$ and $1$ by definition in the left plot of Fig.~\ref{fig:SystematicUncertainties/TIDEfig8b}.  The track-by-track correction is applied by replacing $\sum p_{T,i}/p_\text{T}^\text{jet}$ for a given jet by $\sum (p_{T,i}/e(p_{T,i},p_\text{T}^\text{jet}))/p_\text{T}^\text{jet}$, where $e$ is the tracking efficiency, including resolution effects.   Figure~\ref{fig:SystematicUncertainties/TIDEfig10} shows the efficiencies as a function of $\Delta R$ between the track and the jet axis for various $p_\text{T}$ bins.  It is clear that the inefficiency is larger at lower $\Delta R$ (in the jet core) for higher $p_\text{T}$ jets.

The efficiency corrected distribution of $\sum p_{T,i}/p_\text{T}^\text{jet}$ in two jet $p_\text{T}$ bins (before computing CTER as the average) is shown in Fig.~\ref{fig:SystematicUncertainties/TIDEfig9}.  Note that the corrected distributions can be larger than one in order to get the correct average (while the particle-level distributions never exceed one by construction).

\begin{figure}[h!]
\begin{center}
\includegraphics[width=0.5\textwidth]{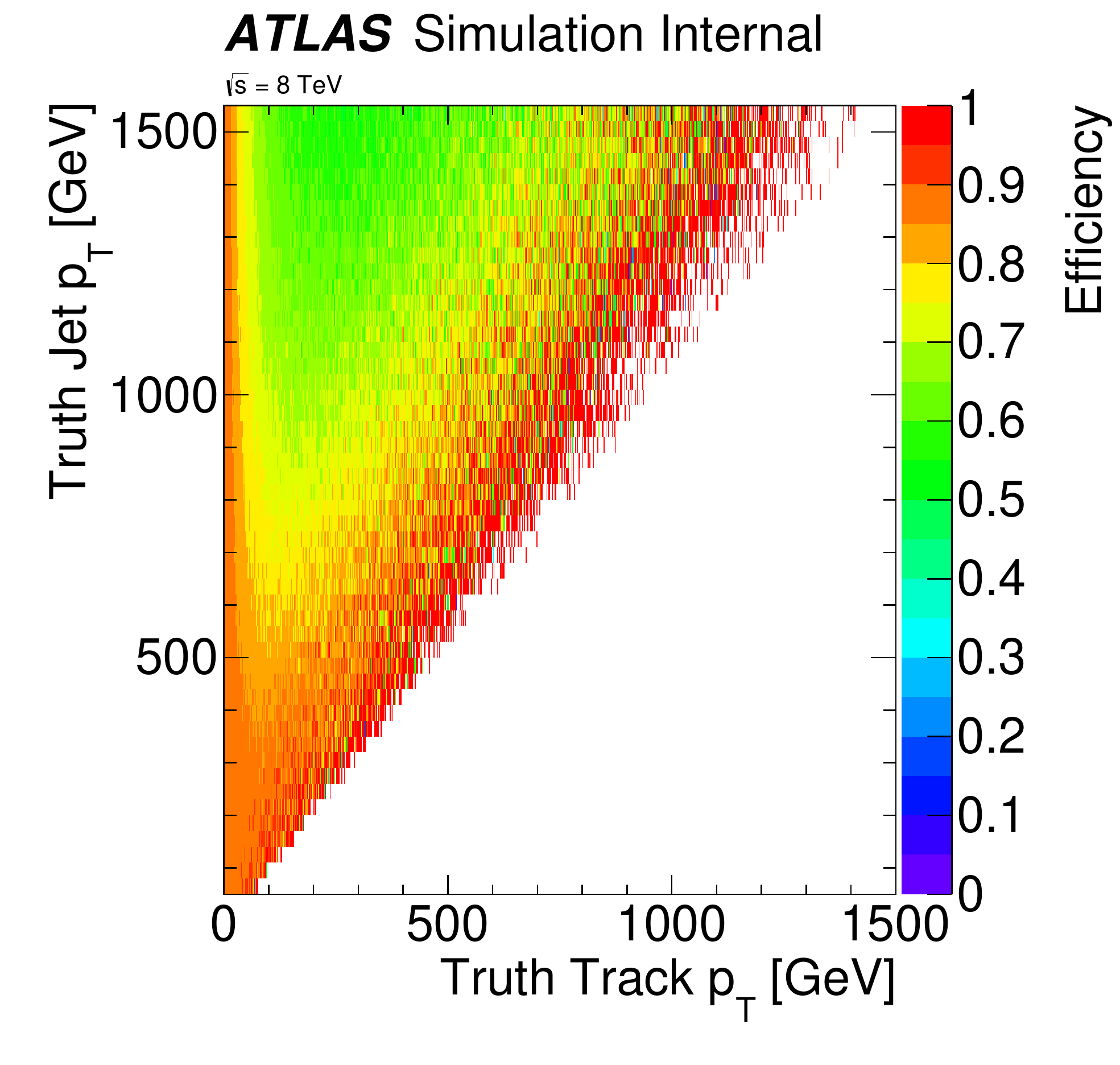}\includegraphics[width=0.5\textwidth]{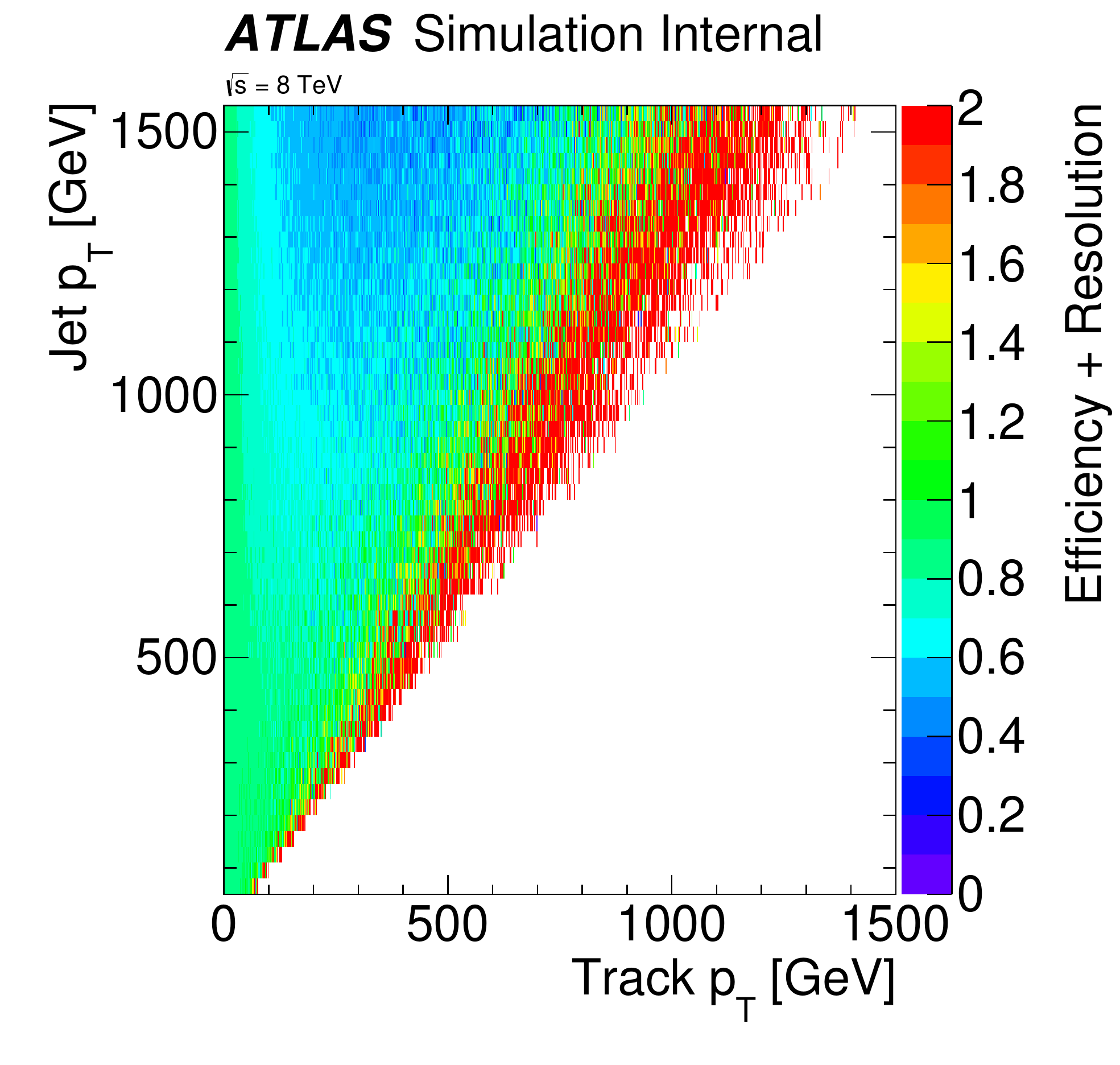}
\caption{Left: The ratio of the number of charged particles matched to reconstructed tracks divided by the total number of charged particles inside particle-level jets as a function of jet $p_\text{T}$ and charged particle $p_\text{T}$.  By construction, this efficiency is between $0$ and $1$.  Right: A similar ratio, but the numerator is replaced with the numerator replaced with all reconstructed tracks.  Due to resolution effects, the `efficiency' in the right plot can exceed unity.}
\label{fig:SystematicUncertainties/TIDEfig8b}
\end{center}
\end{figure}

\begin{figure}[h!]
\begin{center}
\includegraphics[width=0.5\textwidth]{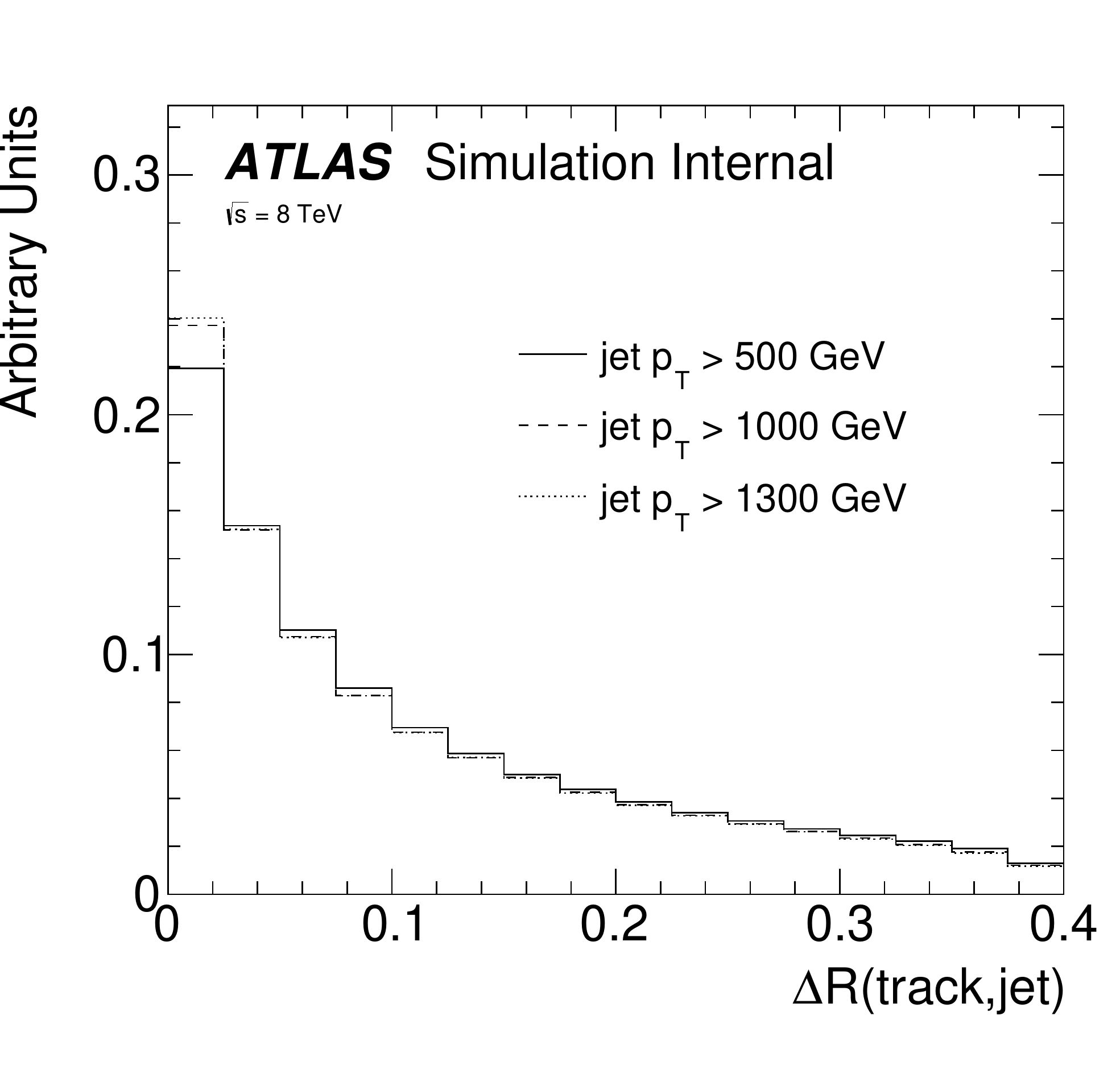}\includegraphics[width=0.5\textwidth]{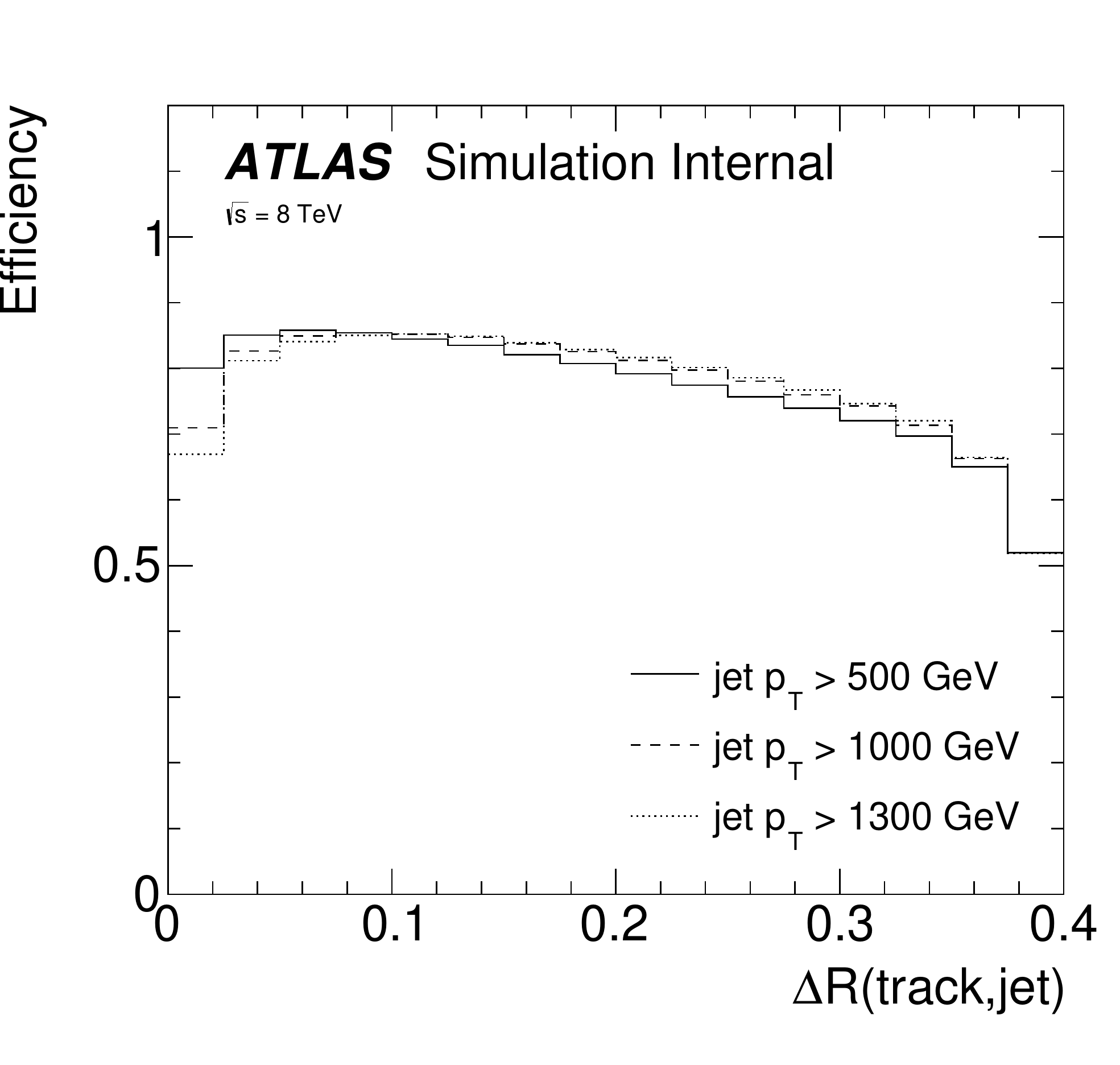}
\caption{The distribution of tracks (left) and tracking efficiency (right) as a function of $\Delta R$ between the track and the jet axis for various $p_\text{T}$ bins and inclusive in track $p_\text{T}>500$ MeV.}
\label{fig:SystematicUncertainties/TIDEfig10}
\end{center}
\end{figure}

\begin{figure}[h!]
\begin{center}
\includegraphics[width=0.45\textwidth]{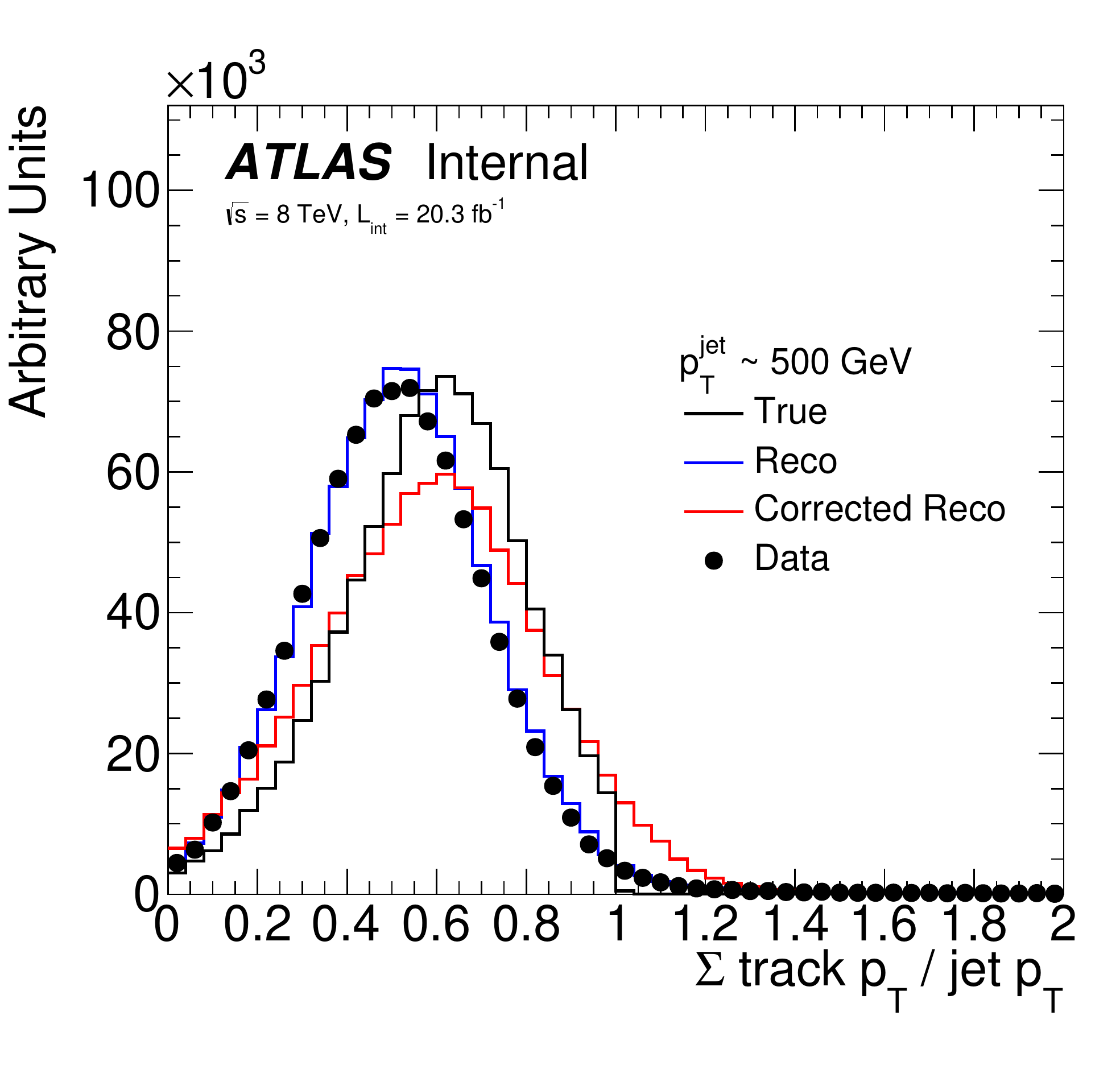}\includegraphics[width=0.45\textwidth]{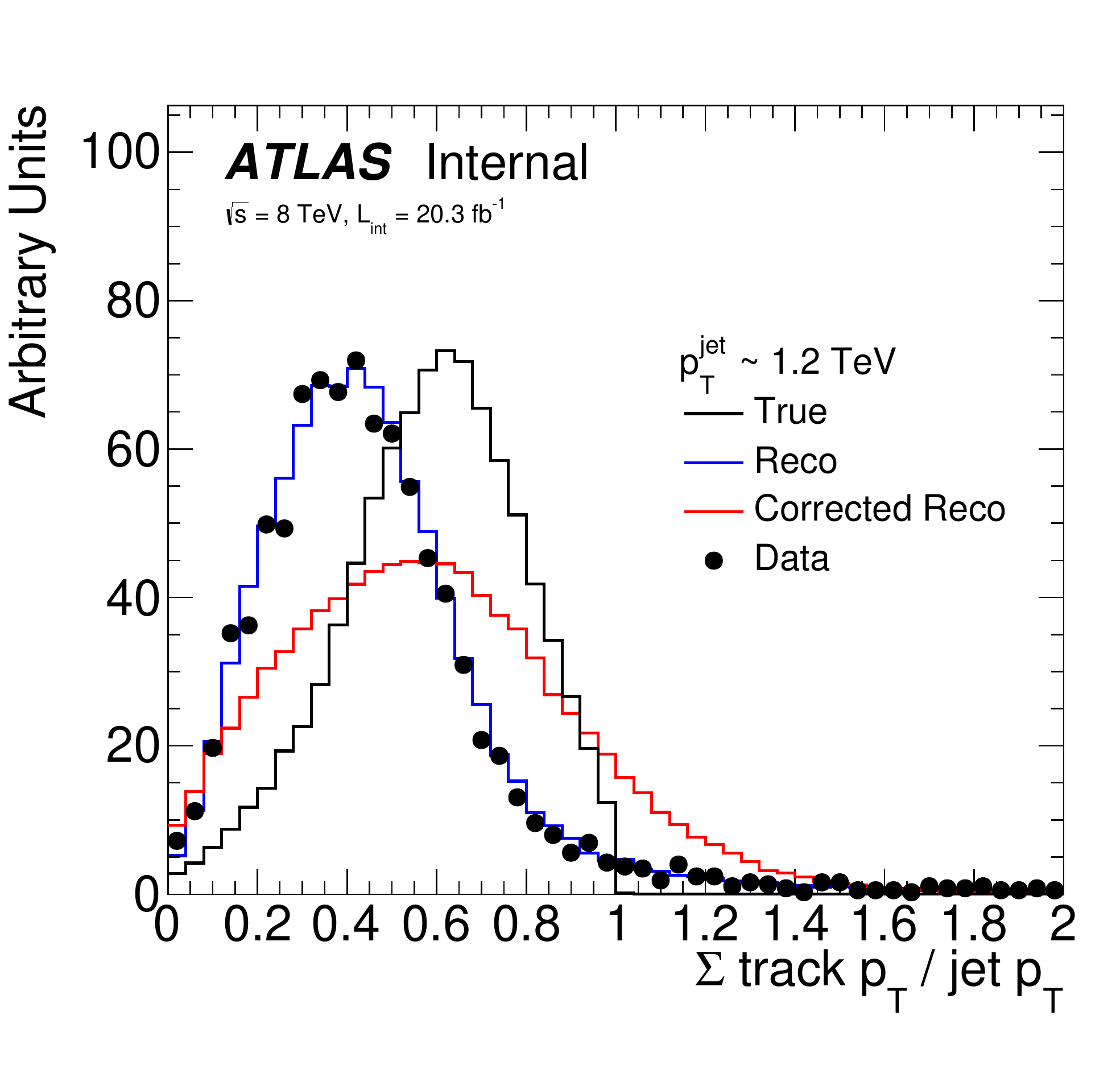}
\caption{The (un)corrected distribution of $\sum p_{T,i}/p_\text{T}^\text{jet}$ in two jet $p_\text{T}$ bins (before computing CTER as the average) for low $p_\text{T}$ jets (left) and high $p_\text{T}$ jets (right).}
\label{fig:SystematicUncertainties/TIDEfig9}
\end{center}
\end{figure}

The actual closure is seen in Fig.~\ref{fig:SystematicUncertainties/TIDEfig11}.  Circles show the particle-level distribution of the CTER, which as already discussed is flat and nearly $2/3$.  Triangles and diamonds show various detector-level versions of the CTER, with(out) fakes, secondaries, and the track resolution.  The crosses are the corrected detector-level CTER values, which nicely fall on top of the circles.  A slight non-closure in the lowest bins is due to the finite binning of the 2D track and jet $p_\text{T}$ distributions.  The various other curves in Fig.~\ref{fig:SystematicUncertainties/TIDEfig11} show the CTER computing using a subset of tracks, indicated by the requirements in the legend.

\begin{figure}[h!]
\begin{center}
\includegraphics[width=0.5\textwidth]{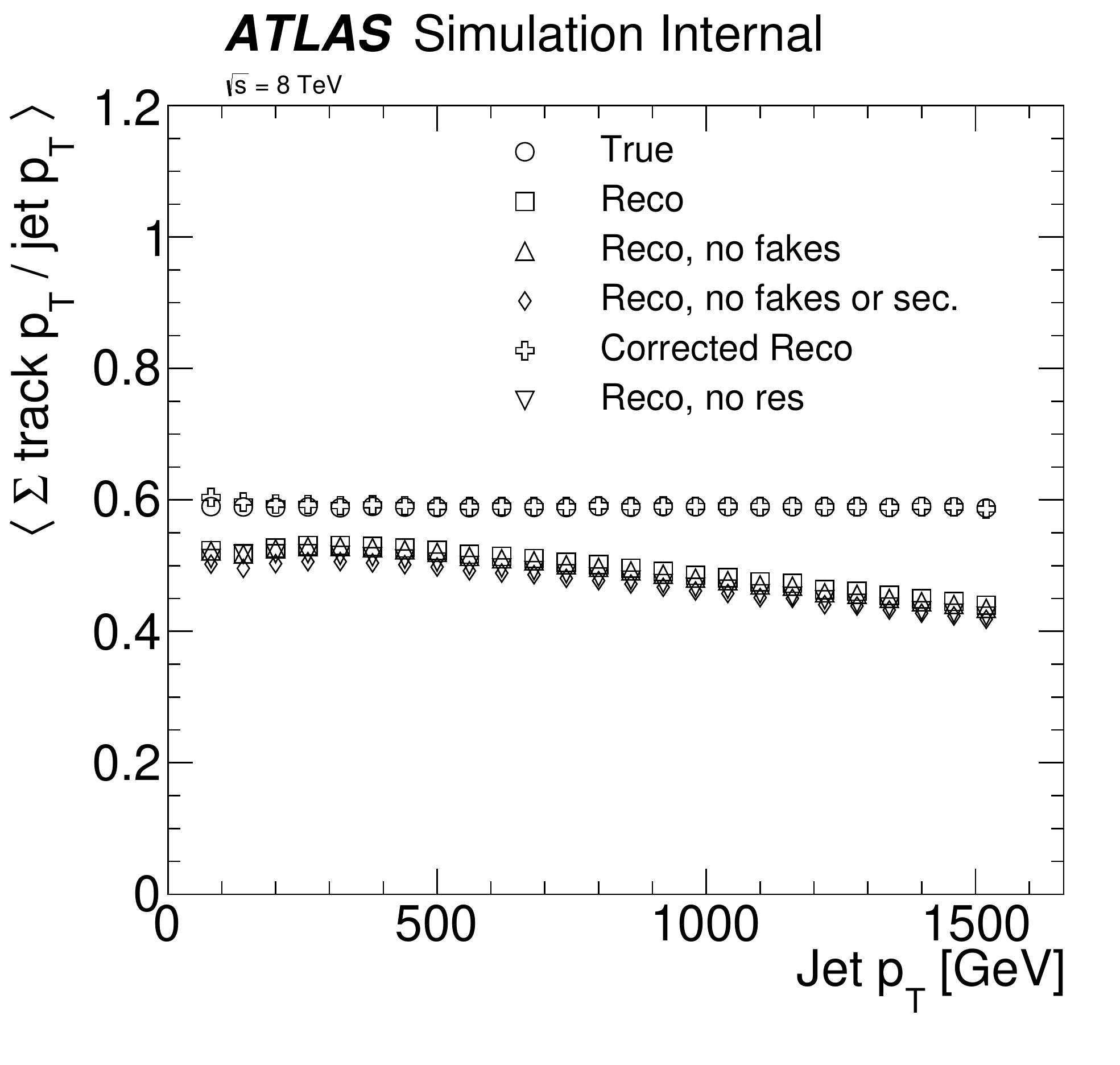}
\caption{The CTER for various particle-level and detector-level definitions (see text for details).}
\label{fig:SystematicUncertainties/TIDEfig11}
\end{center}
\end{figure}

Now that we have shown the method works for the CTER, we quickly prove that the same principle holds for the average jet charge:

\begin{align}
\left\langle \frac{\sum_{i\in\mathrm{Tr}}q_ip_{T,i}^\kappa}{(p_\text{T}^\text{jet})^\kappa}\right\rangle &=  \frac{\langle\sum_{i\in\mathrm{Tr}}q_ip_{T,i}^\kappa\rangle }{(p_\text{T}^\text{jet})^\kappa} = \frac{\sum_{j\in J}w_j\sum_{i\in\mathrm{Tr}}q_ip_{T,i}^\kappa}{(p_\text{T}^\text{jet})^\kappa\sum_{j\in J}w_j}\\
&=\frac{\sum_{p_\text{T}} p_\text{T}^\kappa n_{p_\text{T}}^+}{(p_\text{T}^\text{jet})^\kappa\sum_{j\in J}w_j}-\frac{\sum_{p_\text{T}} p_\text{T}^\kappa n_{p_\text{T}}^-}{(p_\text{T}^\text{jet})^\kappa\sum_{j\in J}w_j},
\end{align}

\noindent where as with the CTER, $J$ is the set of all jets, $w_j$ is the event weight associated with jet $j$, the first inequality is true in a fixed jet $p_\text{T}$ bin, and the the last inequality is true by exchanging the finite sums.  The quantity $n_{p_\text{T}}^\pm=\sum_{j\in J} w_j n_{p_\text{T},j}^\pm$, where $n_{p_\text{T},j}^\pm$ is the number of tracks in jet $j$ with transverse momentum $p_\text{T}$ and charge $\pm$.  So as with the CTER, the average jet charge does not depend on the correlations between the tracking (in)efficiencies of all the constituent tracks\footnote{It is straight-forward to show that the jet charge distribution standard deviation does not share the property of the CTER and the average jet charge - it depends on the first conditional distribution, i.e. depends on every pair (not every track in isolation).  However, the uncertainties on the jet charge standard deviation are much smaller than the average so this subtlety is not considered further and the same prescription for the average is applied for the standard deviation.}.

The last step is to pick a value of $n$.  Representative plots showing the uncertainty for various choices of $n$ are shown in Fig.~\ref{fig:SystematicUncertainties/TIDEfig6a} and Fig.~\ref{fig:SystematicUncertainties/TIDEfig6b}.  The most conservative procedure seems to be $n=10\sim\infty$, which is used for the final prescription.

\begin{figure}[h!]
\begin{center}
\includegraphics[width=0.45\textwidth]{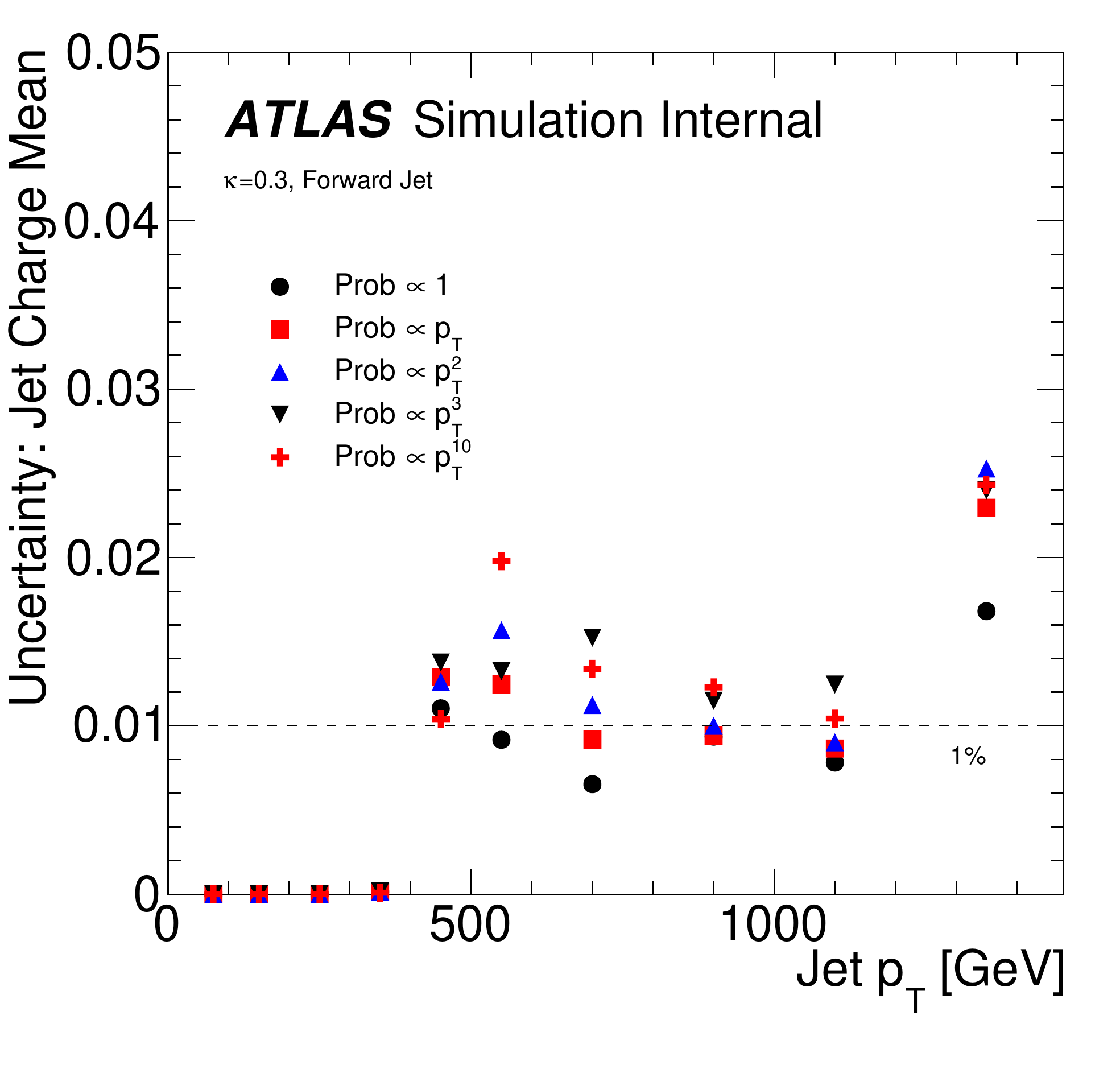}
\includegraphics[width=0.45\textwidth]{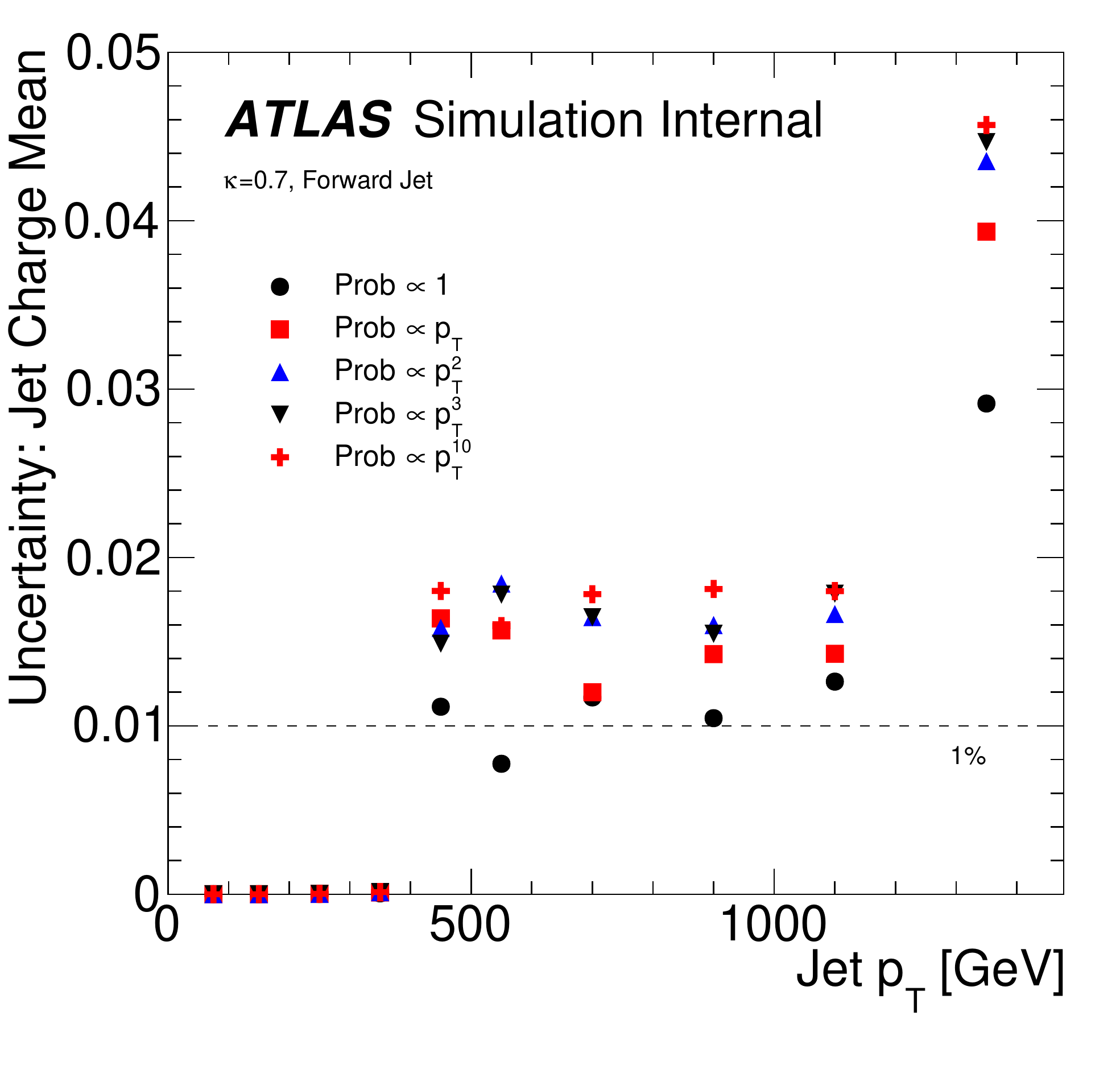}
\caption{The uncertainty on the average jet charge for the more forward (left) and more central (right) jet with $\kappa=0.3$ (top), $\kappa=0.5$ (middle) and $\kappa=0.7$ (bottom).  The uncertainty from tracking in dense environments is assumed to be negligible below $400$ GeV, where nuclear interactions are the dominant source of inefficiency.  }
\label{fig:SystematicUncertainties/TIDEfig6a}
\end{center}
\end{figure}

\begin{figure}[h!]
\begin{center}
\includegraphics[width=0.45\textwidth]{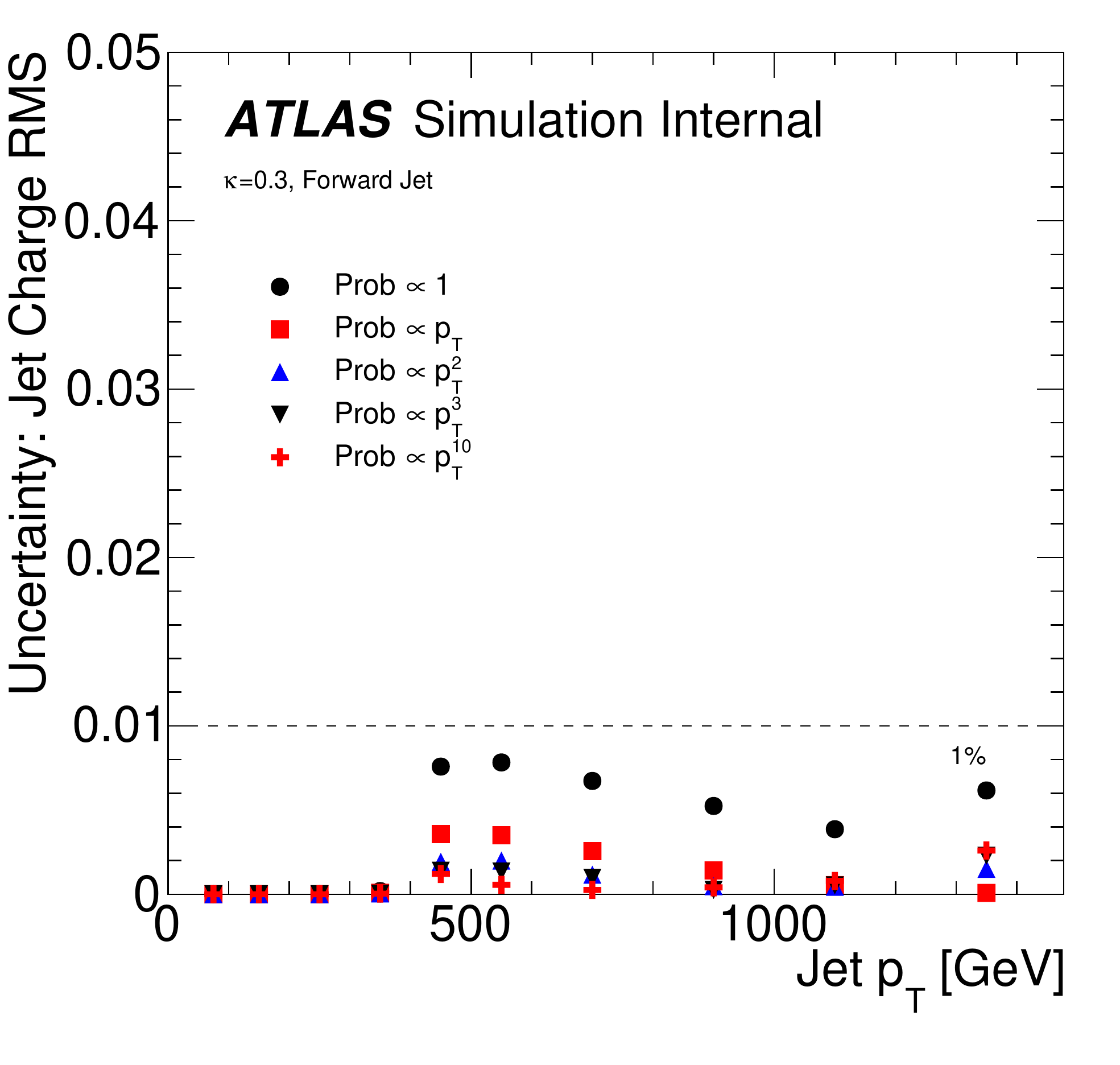}
\includegraphics[width=0.45\textwidth]{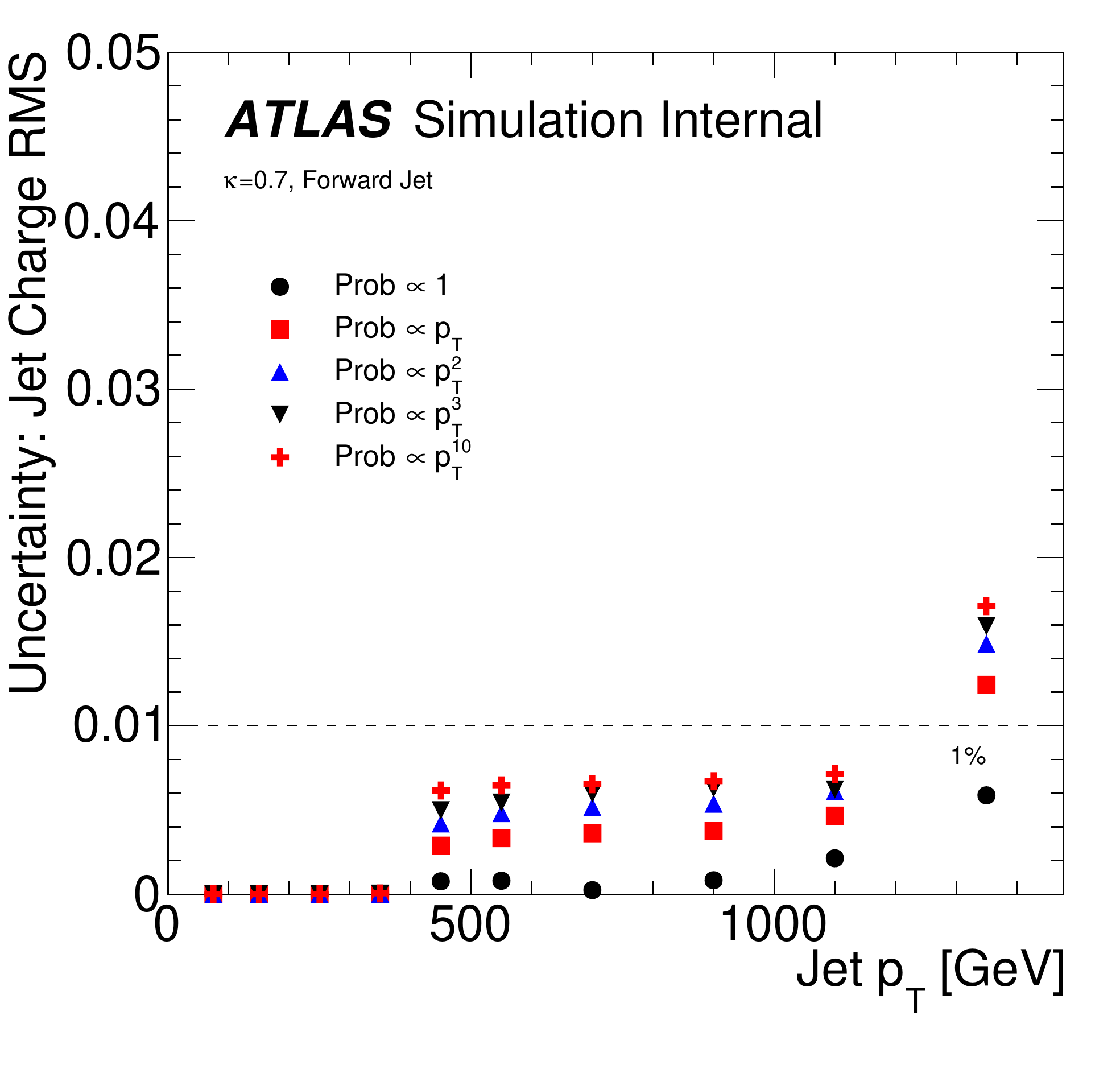}
\caption{The uncertainty on the jet charge distribution standard deviation for the more forward (left) and more central (right) jet with $\kappa=0.3$ (top), $\kappa=0.5$ (middle) and $\kappa=0.7$ (bottom).}
\label{fig:SystematicUncertainties/TIDEfig6b}
\end{center}
\end{figure}
One final note: the uncertainty described in this section is not a generic tracking-in-dense-environments uncertainty.  It can only be applied out-of-the-box to quantities which have the same properties as the CTER and the average jet charge.  It may be possible to use this method in the future to constrain a more general tracking-in-dense-environments uncertainty.

\begin{figure}[h!]
\begin{center}
\includegraphics[width=0.5\textwidth]{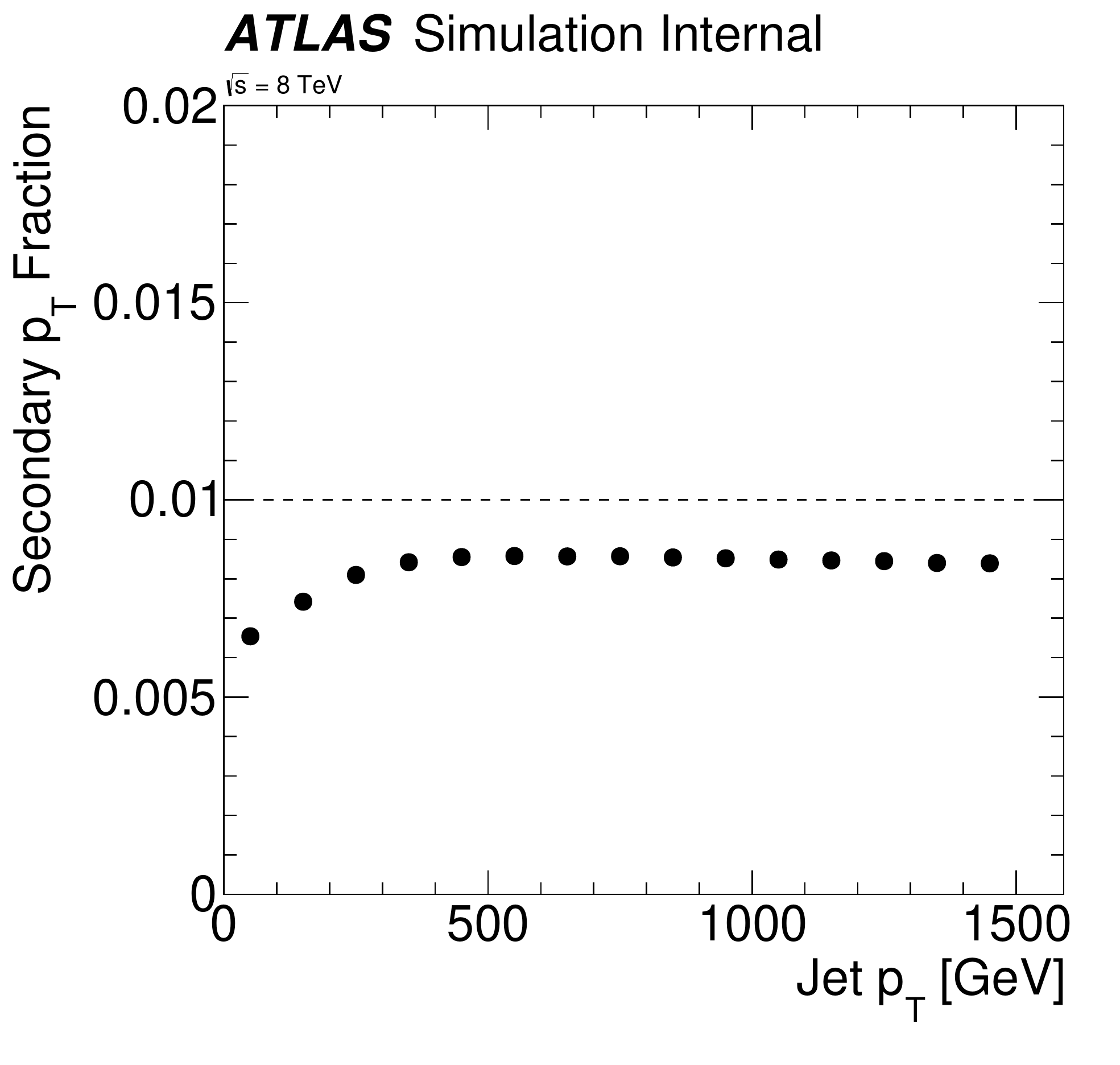}
\caption{The fraction of the sum track $p_\text{T}$ that is due to secondary tracks.}
\label{fig:secondaries}
\end{center}
\end{figure}

\clearpage

\subsubsection{The $\zeta$ Method}

The charged-energy fraction method in Sec.~\ref{sec:jetcharge:tracksyst:TIDE} to determine the systematic uncertainty of charged particle losses inside the core of jets is broadly applicable to observables of the form $\sum_i f_i p_\text{T,i}^\kappa$, where $f_i$ contains information about the track $i$ that does not depend on $p_\text{T,i}$.  This section presents an alternative method\footnote{This method was first introduced by M. Begel, I. Hinchliffe, H. Ma, F. Paige, and M. Shapiro.} based on the asymmetry of the pixels in the inner detector that could be applied for any observable.  The setup is outlined in Fig.~\ref{fig:JetCharge:Syst:TIDE:Zeta:fig0}.  The planar sensors in the pixel detector are about 50 $\mu$m in the $\phi$ direction and $400$ $\mu$m in the $z$ direction.  The outermost pixel layer is at 122.5 mm from the center of ATLAS and so each pixel in that layer covers about $2\pi / [(\pi\times122.5\text{ mm})/40\text{ $\mu$m}]\sim 6.5\times 10^{-4}$ radians in the $\phi$ direction.   Consider two pairs of particles where $\Delta\phi_1=\Delta\eta_2=0$ and $\Delta\eta_1$ and $\Delta\phi_2$ have the same distribution.  Due to the asymmetry in the pixel dimensions, hit merging will occur for higher values of $\Delta R$ in the first pair with respect to the second pair.  More generally, define $\zeta = |\text{atan}(\Delta\phi / \Delta\eta)|$.  Assuming that the distribution of radiation is the same in the $\phi$ and $\eta$ directions, if the pixel dimensions were symmetric, $\zeta\sim\text{Uniform}(0,\pi)$.  However, due to the asymmetry and hit merging due to the high density environment at high $p_\text{T}$, the $\zeta$ distribution is not uniform.

\begin{figure}[h!]
\begin{center}
\includegraphics[width=0.45\textwidth]{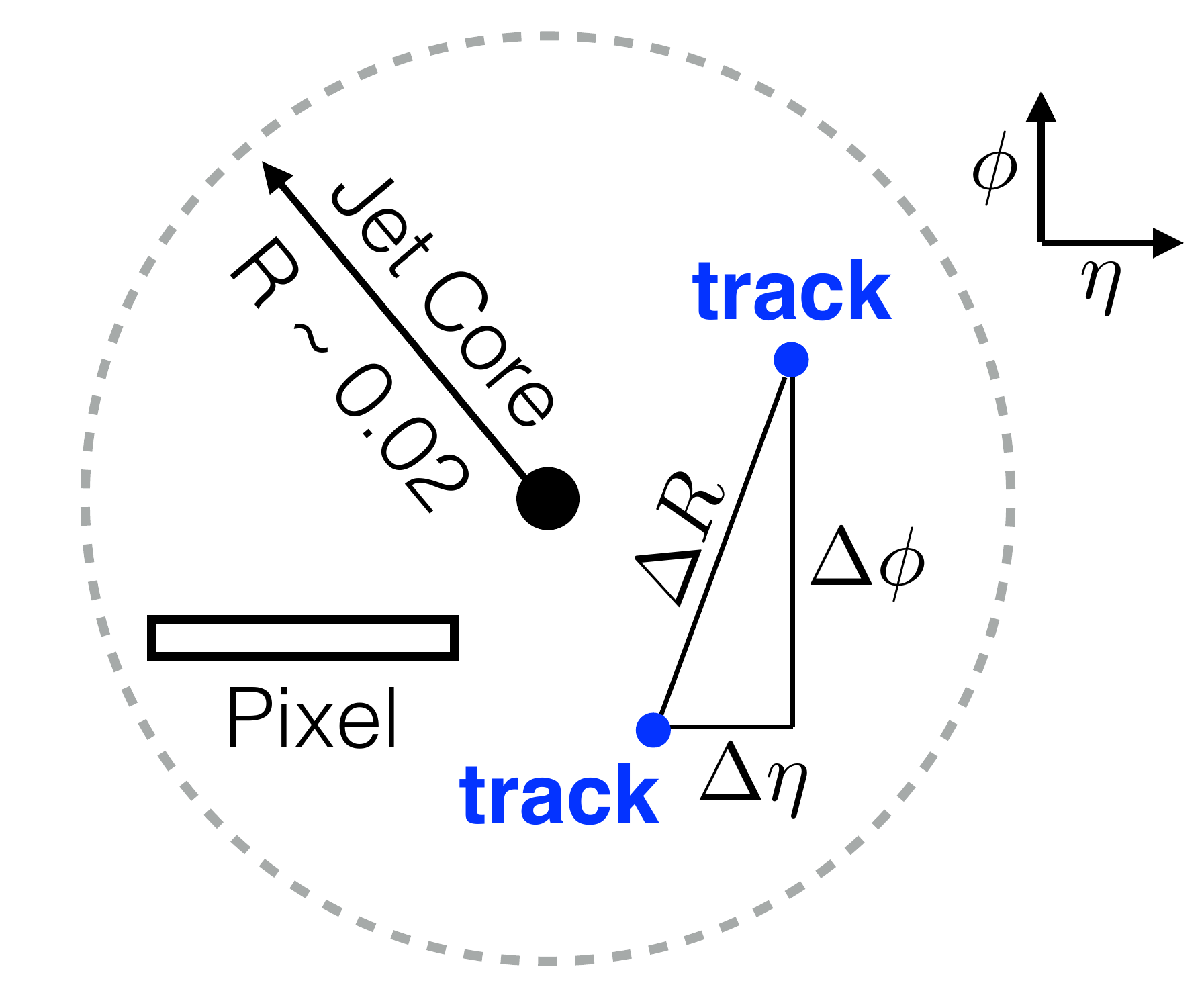}
\caption{A schematic diagram of the jet core, $R\lesssim 0.02$.  The pixel is drawn approximately to scale assuming it is from the third pixel layer in the central region.}
\label{fig:JetCharge:Syst:TIDE:Zeta:fig0}
\end{center}
\end{figure}

Before studying the distribution of $\zeta$ in the data, it is useful to estimate the predicted dependence of $\zeta$ on the loss.  Since the track density is highest in the jet core and the most important impact is on high $p_\text{T}$ tracks, the rest of the section constructs $\zeta$ only using tracks with $p_\text{T}>10$ GeV and with $\Delta R<0.02$ to the calorimeter jet axis.  Define $\phi_0$ to be the characteristic length of a pixel in radians in the $\phi$ direction and let $r$ be the pixel aspect ratio between the $\eta$ and $\phi$ dimensions of the pixel $(r\approx 8)$.  Consider the probability that the track from a particle at exactly the jet center merges with another track with $p_\text{T}>10$ GeV.  Assume that $\eta,\phi\sim\text{Uniform}$ in the jet core.  Then, if there are $n$ tracks with $p_\text{T}>10$ GeV in the jet core,

\begin{align}
\Pr(\text{merger}|\Delta\eta=0)&\sim 1-\prod_{i=1}^n \Pr(\Delta\phi_i>\phi_0)\\
&=1-\left(1-\frac{\phi_0}{\Omega}\right)^n\sim \frac{n\phi_0}{\Omega}\equiv L\text(=Loss),
\end{align}

\noindent where $\Delta\phi$ is between the target track (at the jet center) and another track $i$.  The size of the jet core is $\Omega\sim 0.02$.  A similar calculation shows that $\Pr(\text{merger}|\Delta\phi=0)\sim rL$.  Therefore, the ratio of the distribution of $\zeta$ at $\zeta = 0$ versus $\zeta = 1$ is given by $(1-rL)/(1-L)\sim 1+L(1-r)$.  In other words, the asymmetry of the $\zeta$ distribution depends on the total loss and the aspect ratio.  When the aspect ratio is 1, there is no sensitivity to the loss.  Since $L$ increases with $p_\text{T}$, the asymmetry in the $\zeta$ distribution should also increase with $p_\text{T}$.

\begin{figure}[h!]
\begin{center}
\includegraphics[width=0.4\textwidth]{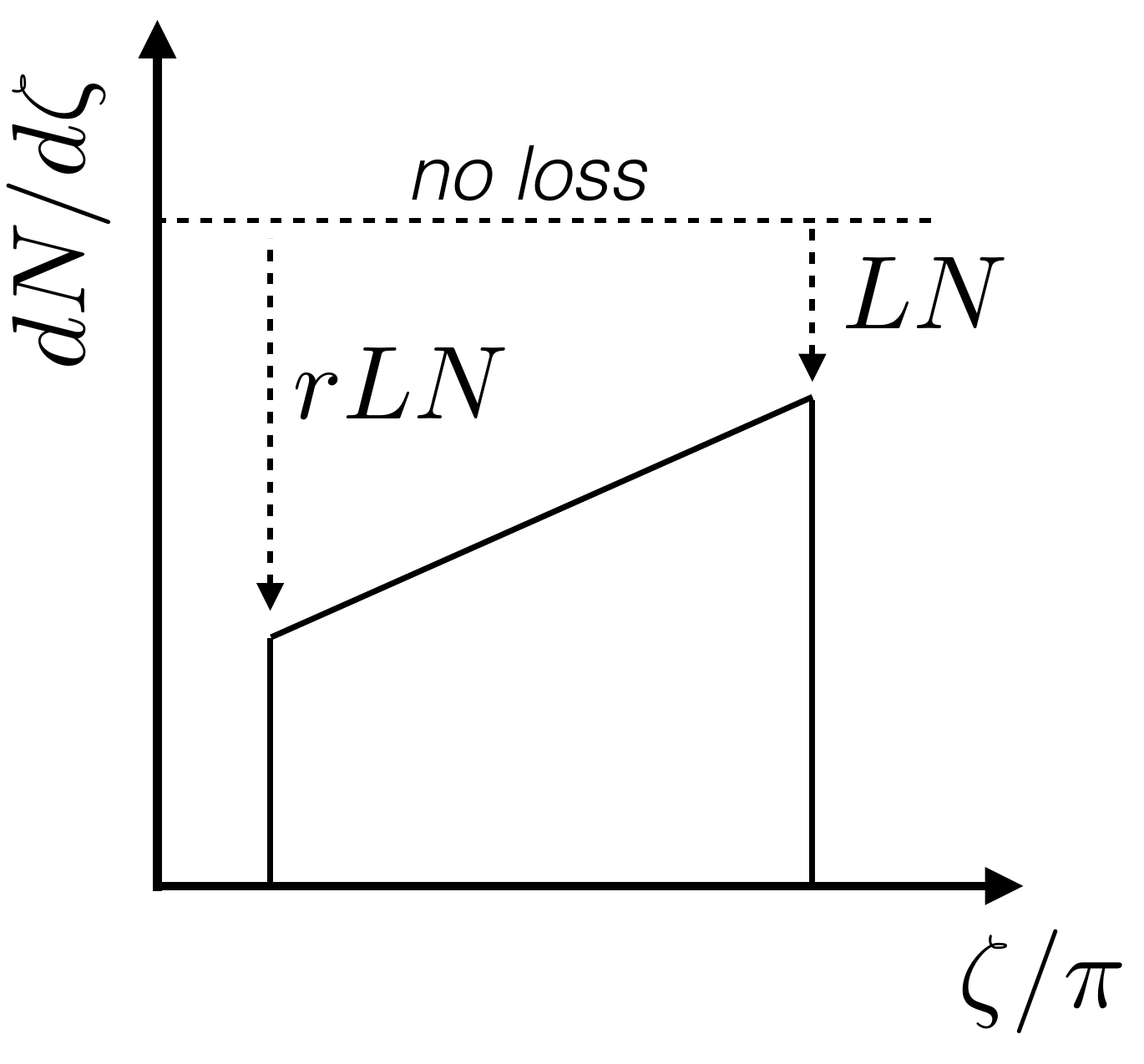}
\caption{A schematic diagram of the distribution of $\zeta$, accounting for merging.}
\label{fig:JetCharge:Syst:TIDE:Zeta:fig0b}
\end{center}
\end{figure}

Figure~\ref{fig:JetCharge:Syst:TIDE:Zeta:fig1} shows the joint distribution of $\zeta$ and the $\Delta R$ between tracks.  Each event contributes multiple tracks to the histograms.  While the distribution of the $\Delta R$ between tracks is qualitatively similar between the two distributions\footnote{If the track locations were uniform in the jet core, one would expect a triangle probability distribution for their $\Delta R$, which resembles the distribution in Fig.~\ref{fig:JetCharge:Syst:TIDE:Zeta:fig1} projected onto the $\Delta R$ axis.}, but there is a clear difference in the distribution of $\zeta$ at low jet $p_\text{T}$ and high jet $p_\text{T}$.  

\begin{figure}[h!]
\begin{center}
\includegraphics[width=0.5\textwidth]{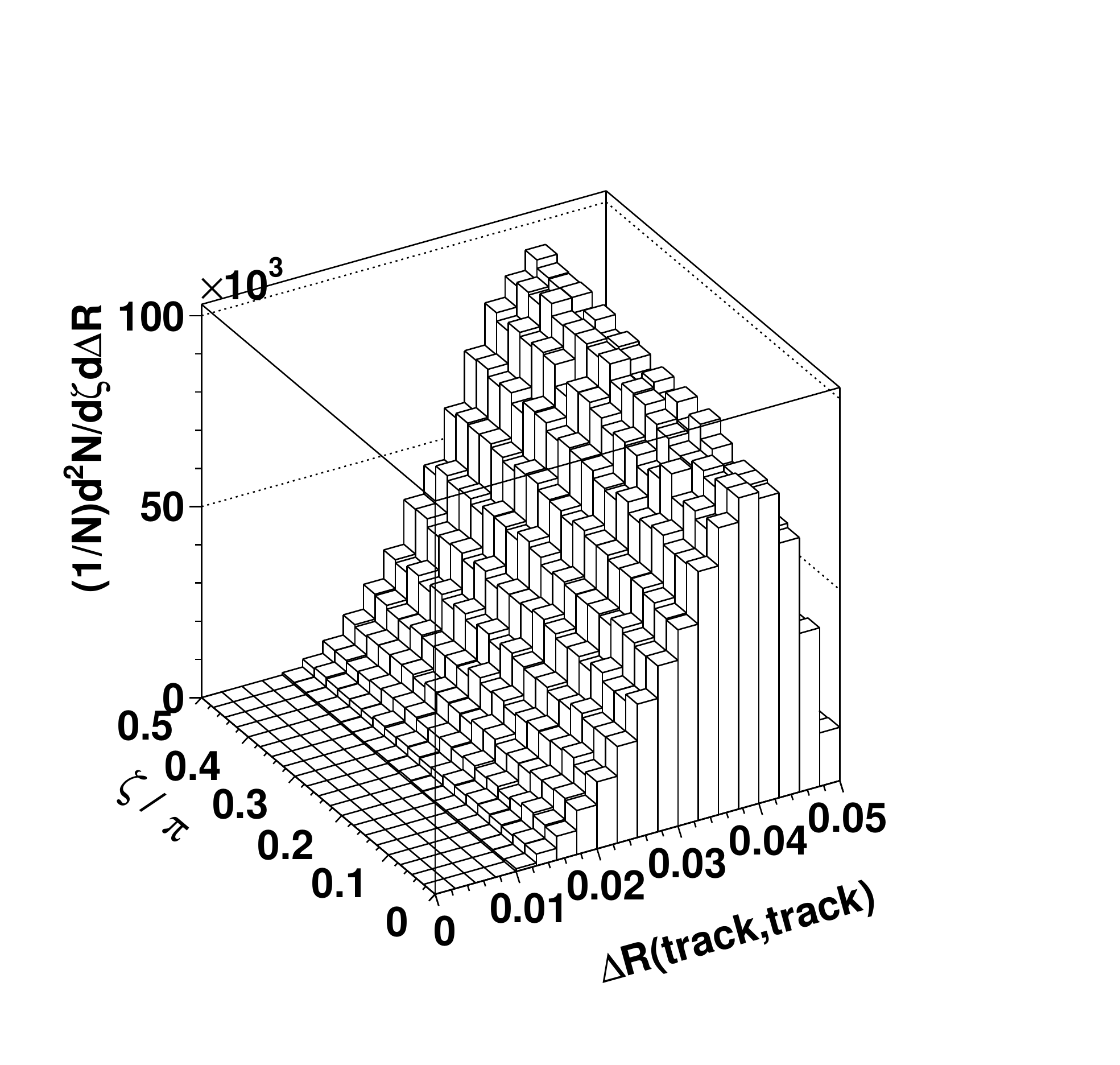}\includegraphics[width=0.5\textwidth]{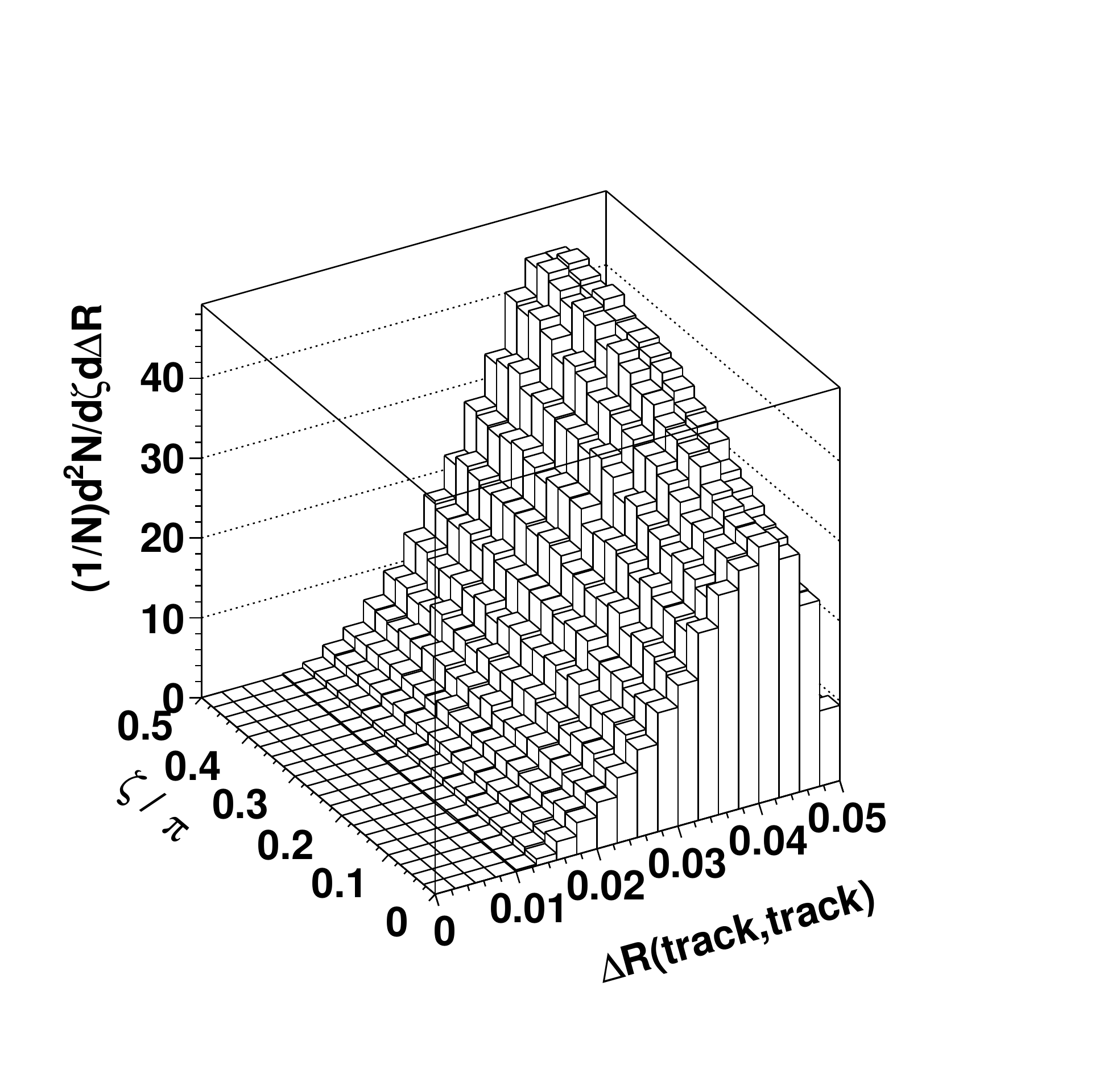}
\caption{The joint distribution of $\zeta$ and the $\Delta R$ between tracks for jet $p_\text{T}\sim 400$~GeV (left) and $p_\text{T}\sim 1.2$~TeV (right).}
\label{fig:JetCharge:Syst:TIDE:Zeta:fig1}
\end{center}
\end{figure}

\noindent The $p_\text{T}$ dependence of the $\zeta$ distribution is quantified in Fig.~\ref{fig:JetCharge:Syst:TIDE:Zeta:fig2}.  The $\zeta$ distribution is nearly uniform for jet $p_\text{T}\lesssim 400$ GeV and the asymmetry in the distribution grows with $p_\text{T}$.  Qualitatively this trend appears for both data and simulation.  Representative slices from Fig.~\ref{fig:JetCharge:Syst:TIDE:Zeta:fig2} are shown in Fig.~\ref{fig:JetCharge:Syst:TIDE:Zeta:fig3}, with the simulation from {\sc Pythia} 8 and {\sc Herwig++} overlaid for comparison.    Since the probability distribution for $\zeta$ in a fixed $p_\text{T}$ bin is approximately linear in $\zeta$, $f(\zeta)\propto \zeta$, a useful statistic of the distribution is the slope, $\partial_\zeta f(\zeta)$.  The heuristic argument above suggests that this slope is proportional to the loss $L$.  The fitted slopes are shown as a function of $p_\text{T}$ in Fig.~\ref{fig:JetCharge:Syst:TIDE:Zeta:fig4}.   As expected from the $p_\text{T}$ dependence of the loss, the slope increases with $p_\text{T}$.  There is a small difference between data and simulation in the slope; the simulation seems to under-predict the loss, in agreement with with the charged-energy fraction in Sec.~\ref{sec:jetcharge:tracksyst:TIDE}.  One way to estimate the relationship between the loss and $\partial_\zeta f(\zeta)$ in simulation is to decrease\footnote{Increasing the loss would be more relevant for matching to the data, but is highly non-trivial because it needs to respect the pixel geometry.} the loss in simulation by adding truth particles without a reconstructed track to the jet.  For a direct comparison, all reconstructed tracks are also replaced with their matched truth particles in order to remove the effect of the detector response.  The difference between the solid squares and open circles in Fig.~\ref{fig:JetCharge:Syst:TIDE:Zeta:fig4} quantifies the impact on the slope when removing these detector distortions.  The other markers in Fig.~\ref{fig:JetCharge:Syst:TIDE:Zeta:fig4} represent various levels of reduced loss ($100\%$ loss means $100\%$ of the loss in the nominal simulation, not $100\%$ of tracks lost).  Slopes of linear fits are given in the legend of Fig.~\ref{fig:JetCharge:Syst:TIDE:Zeta:fig4} and suggest that the slope is quadratically dependent on the loss.  A simple fit yields $\partial_\zeta f(\zeta)\sim 0.8-0.1L+0.0035L^2$, where $L$ is the loss in percent.  Inverting this relationship and using the slopes in the left plot of Fig.~\ref{fig:JetCharge:Syst:TIDE:Zeta:fig4} results in $L_\text{data}/L_\text{MC}\sim 10-15\%$, which is quantitatively similar to the values derived using the charged-energy loss method in Sec.~\ref{sec:jetcharge:tracksyst:TIDE}.  Note that the fit is required because it is non-trivial to simulate more loss, as is the case in data (Fig.~\ref{fig:JetCharge:Syst:TIDE:Zeta:fig4} shows the pattern for a {\it reduced loss} only).

\begin{figure}[h!]
\begin{center}
\includegraphics[width=1\textwidth]{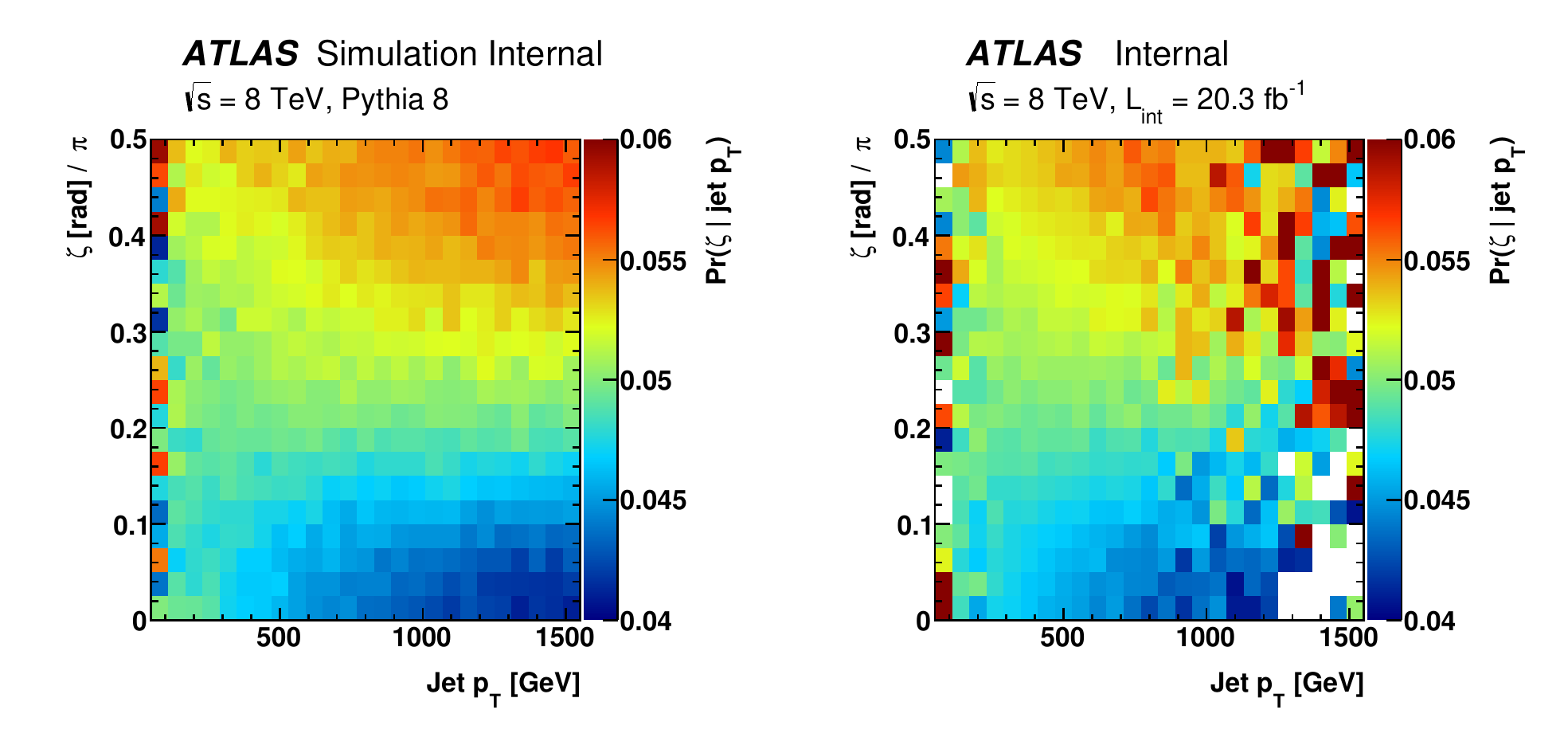}
\caption{The conditional distribution of $\zeta$ given jet $p_\text{T}$ for data (right) and MC (left). }
\label{fig:JetCharge:Syst:TIDE:Zeta:fig2}
\end{center}
\end{figure}

\begin{figure}[h!]
\begin{center}
\includegraphics[width=0.5\textwidth]{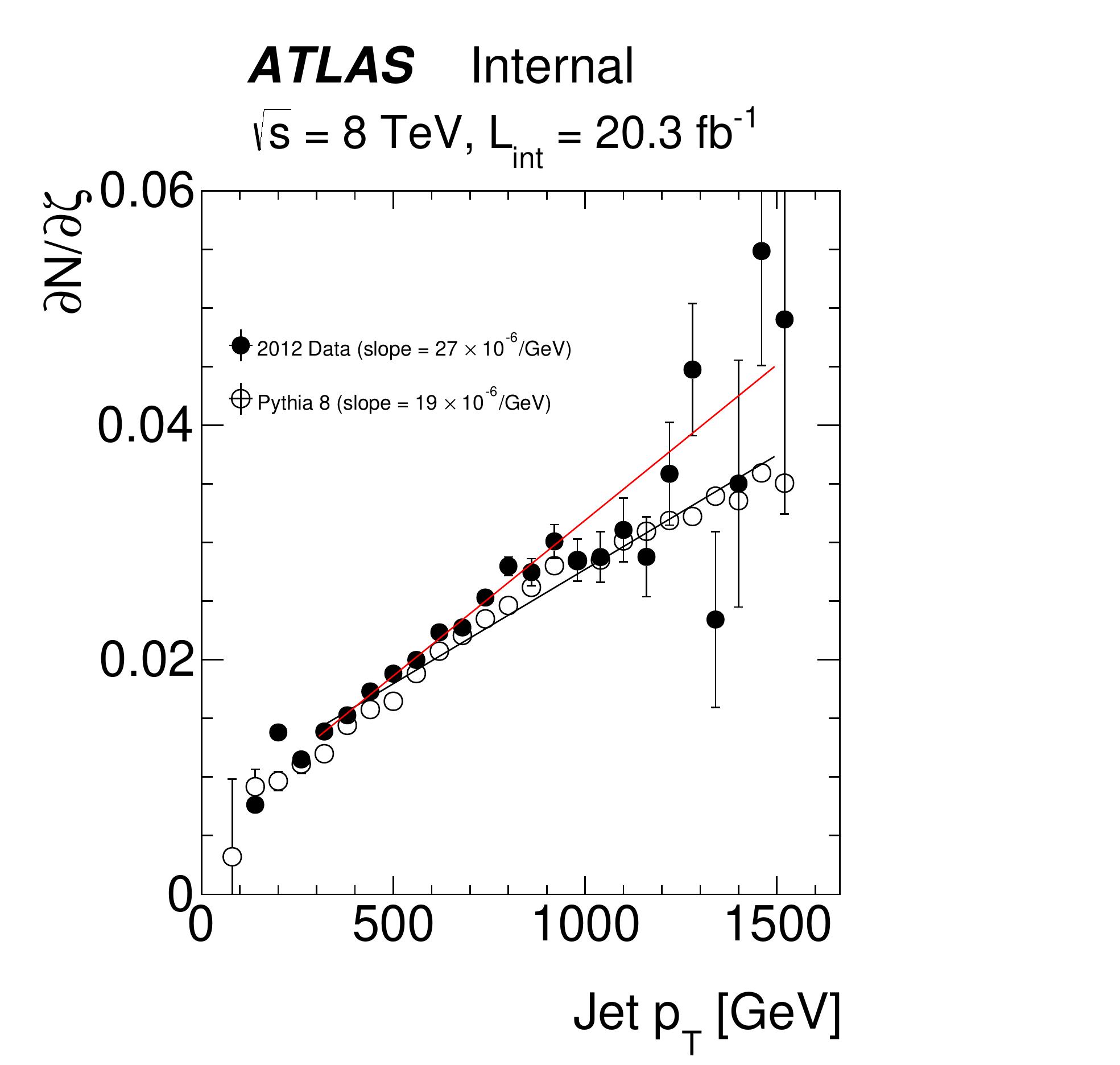}\includegraphics[width=0.5\textwidth]{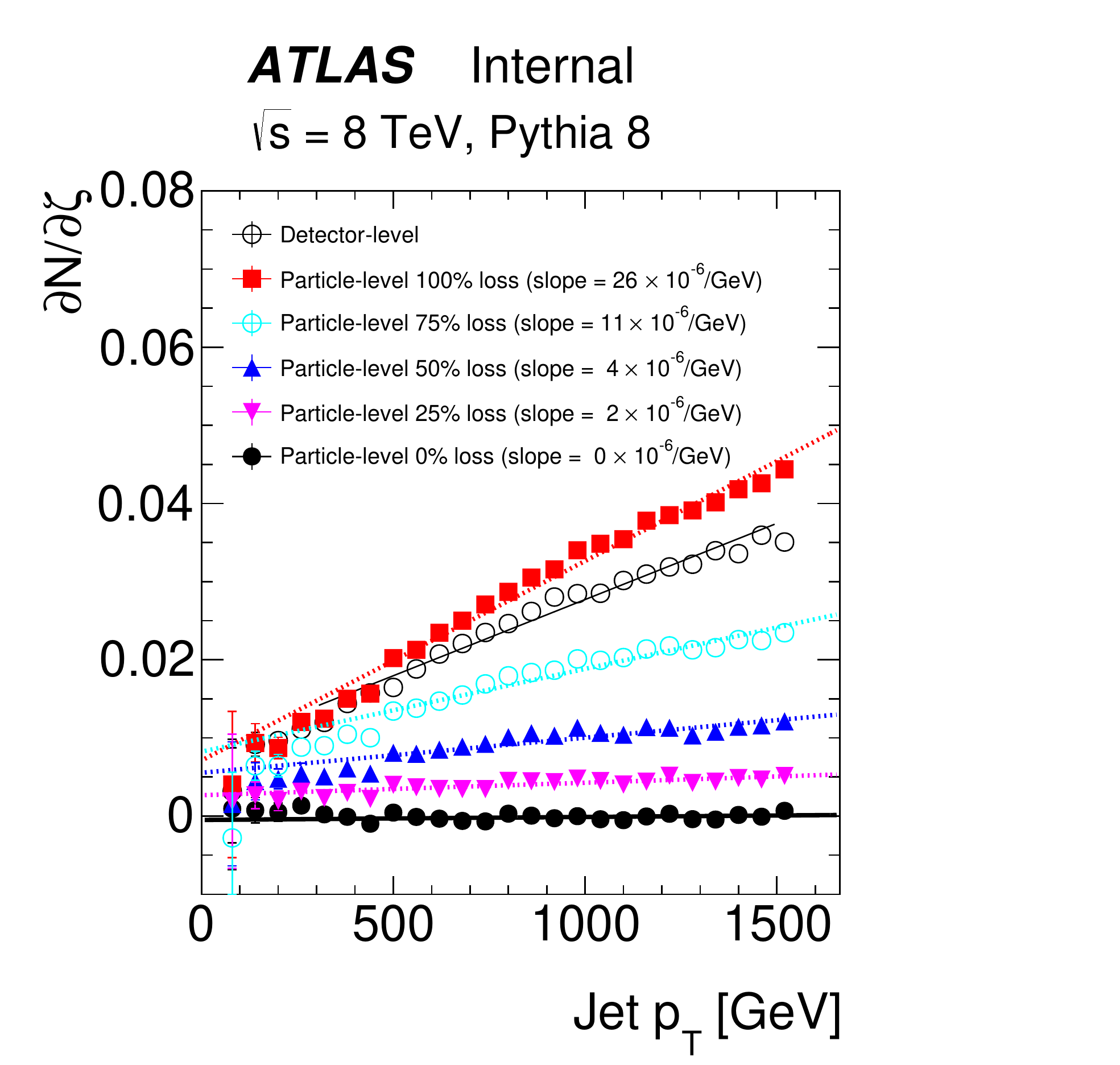}
\caption{Left: the $p_\text{T}$ dependence of $\partial_\zeta f(\zeta)$, for $f(\zeta)$ the probability distribution of $\zeta$.  Right: the impact on the $\zeta$ distribution from reducing the loss in simulation.}
\label{fig:JetCharge:Syst:TIDE:Zeta:fig4}
\end{center}
\end{figure}

\begin{figure}[h!]
\begin{center}
\includegraphics[width=0.45\textwidth]{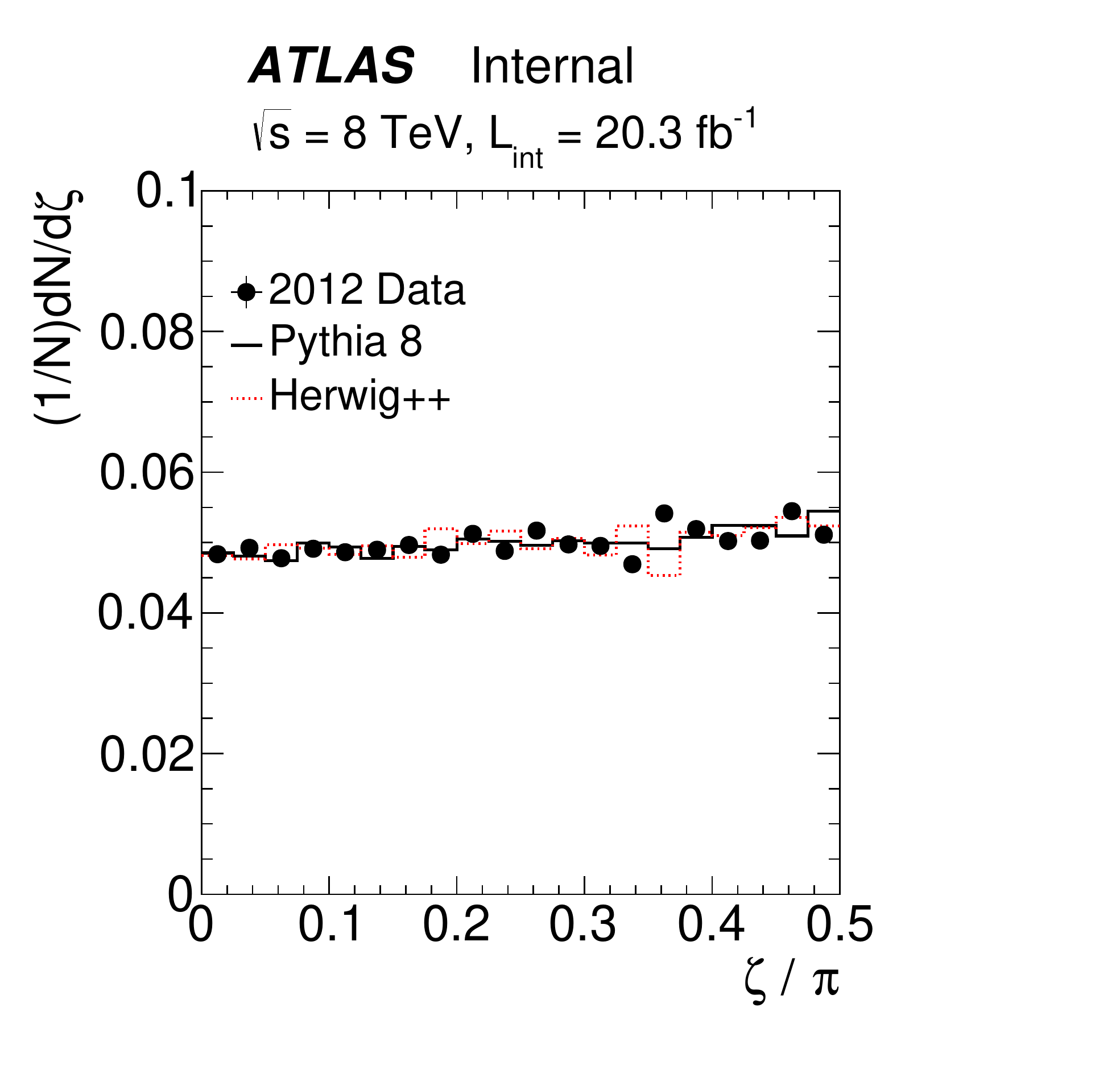}\includegraphics[width=0.45\textwidth]{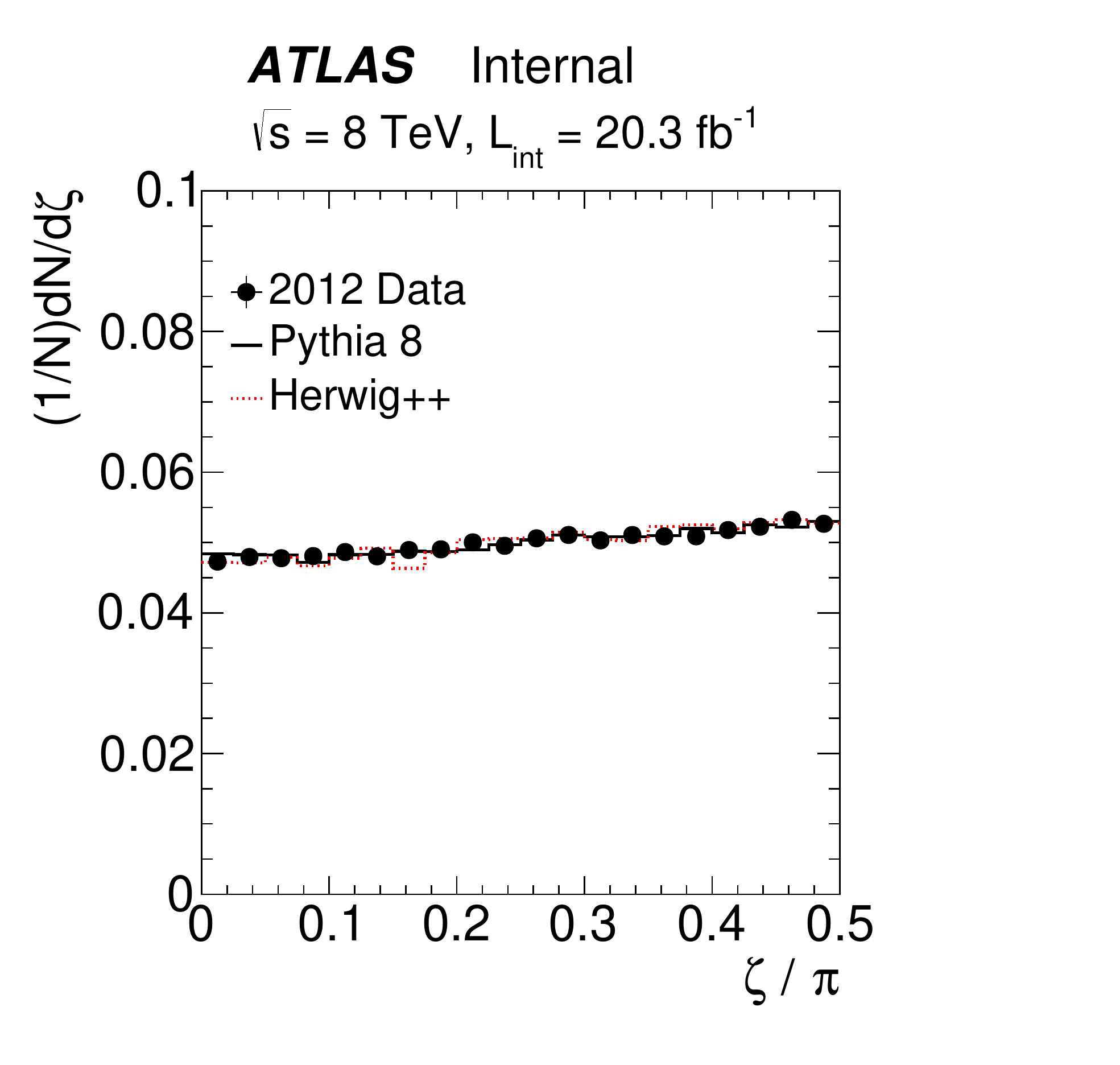} \\
\includegraphics[width=0.45\textwidth]{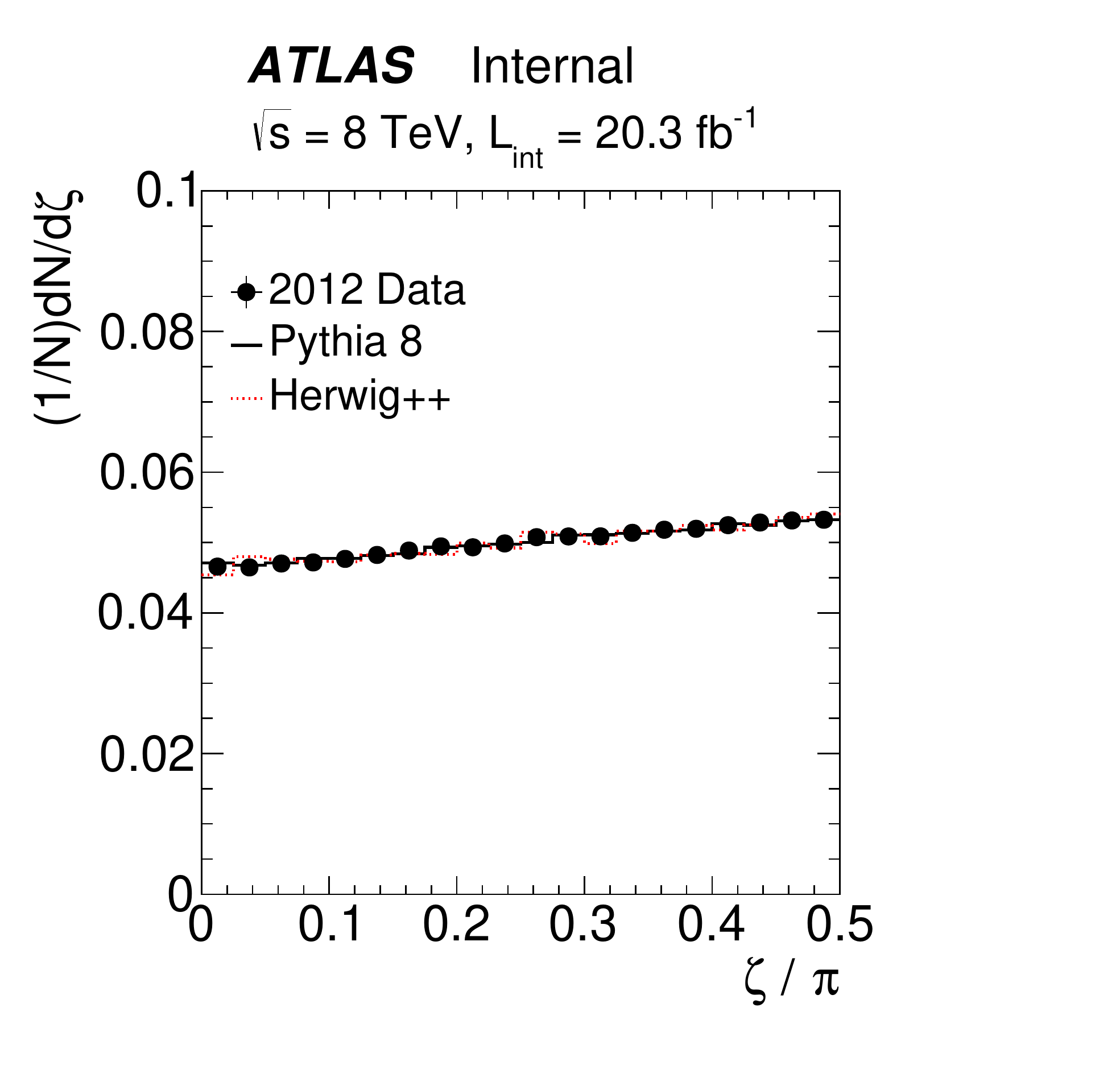}\includegraphics[width=0.45\textwidth]{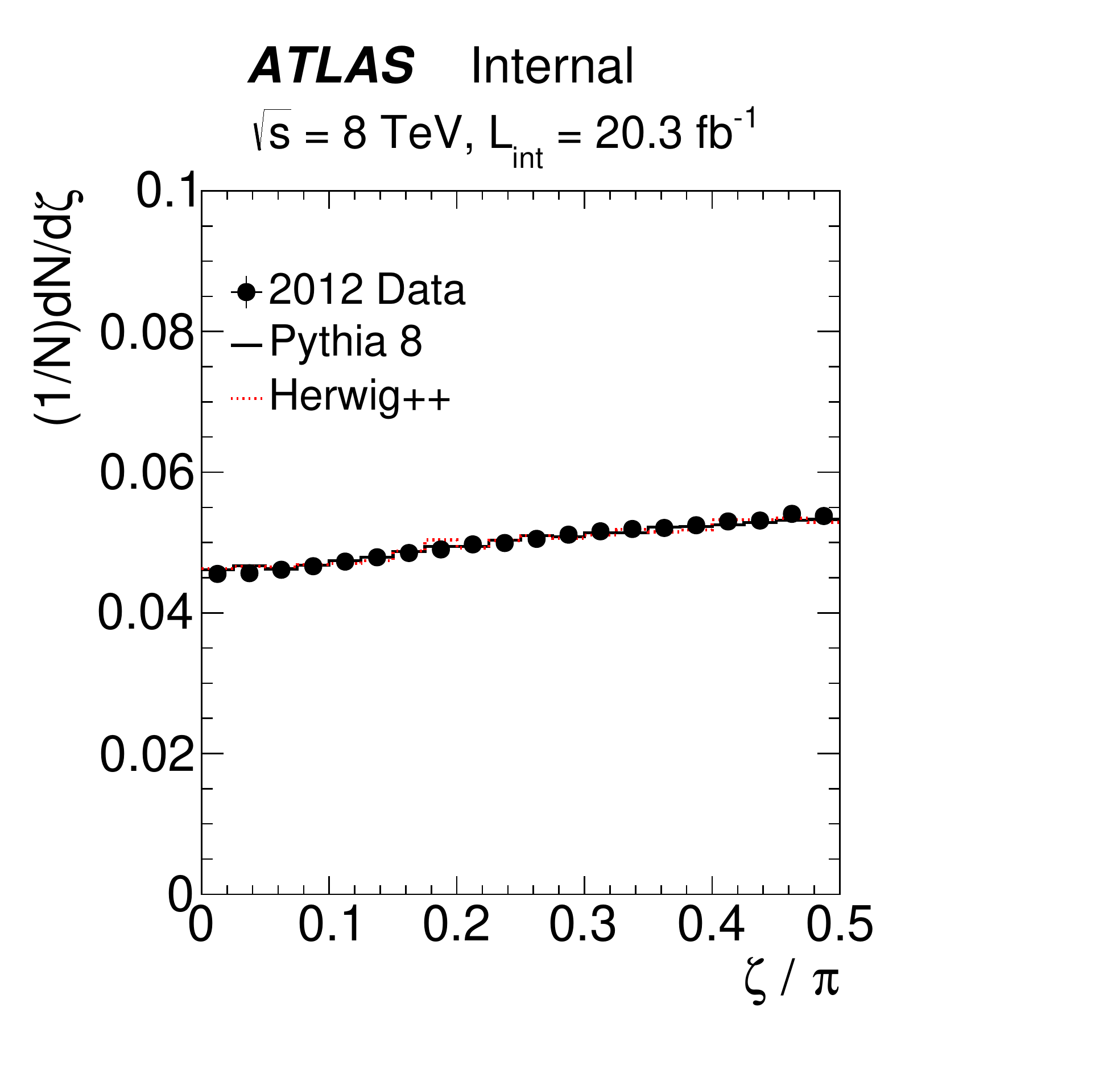}\\
\includegraphics[width=0.45\textwidth]{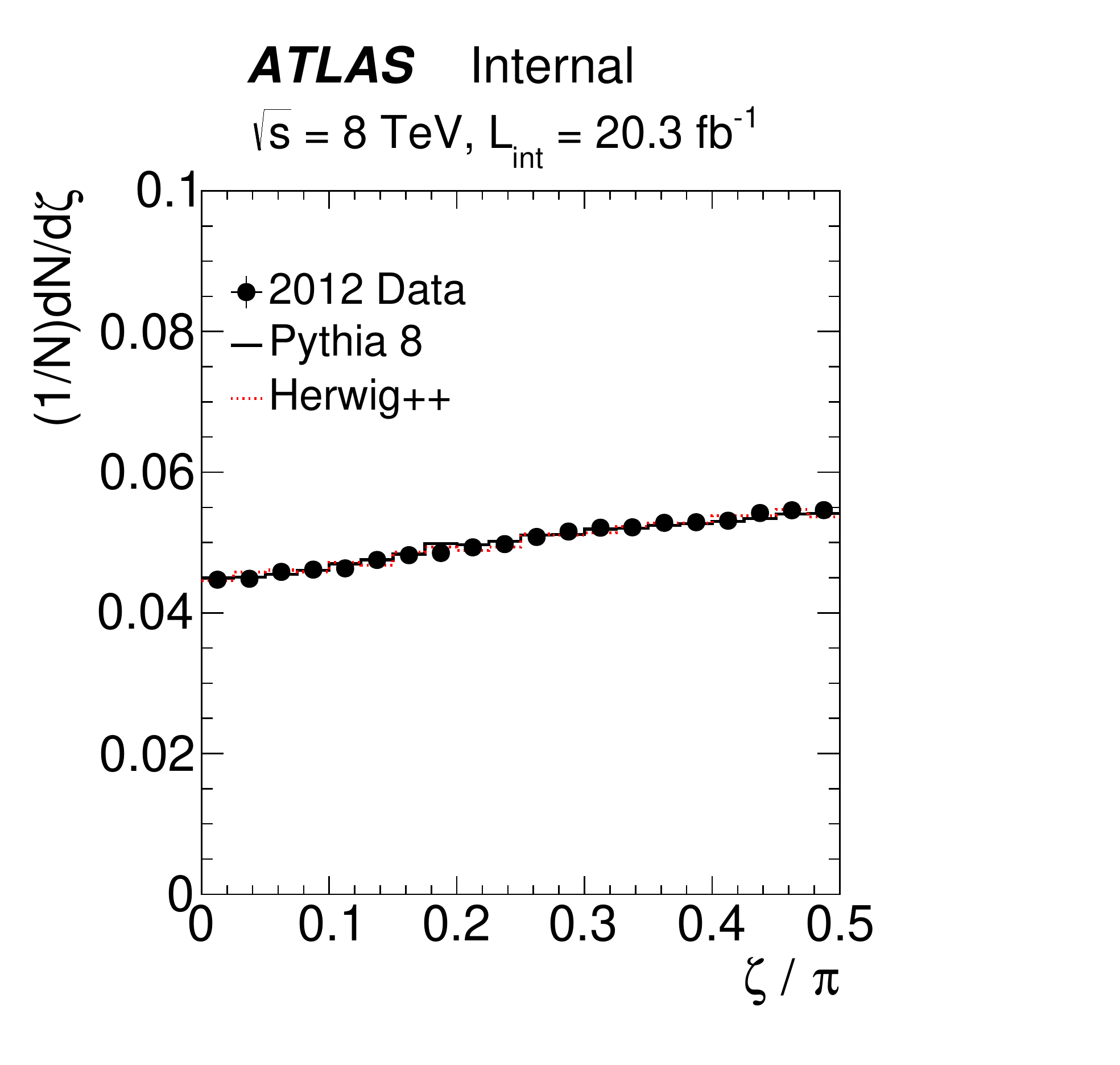}\includegraphics[width=0.45\textwidth]{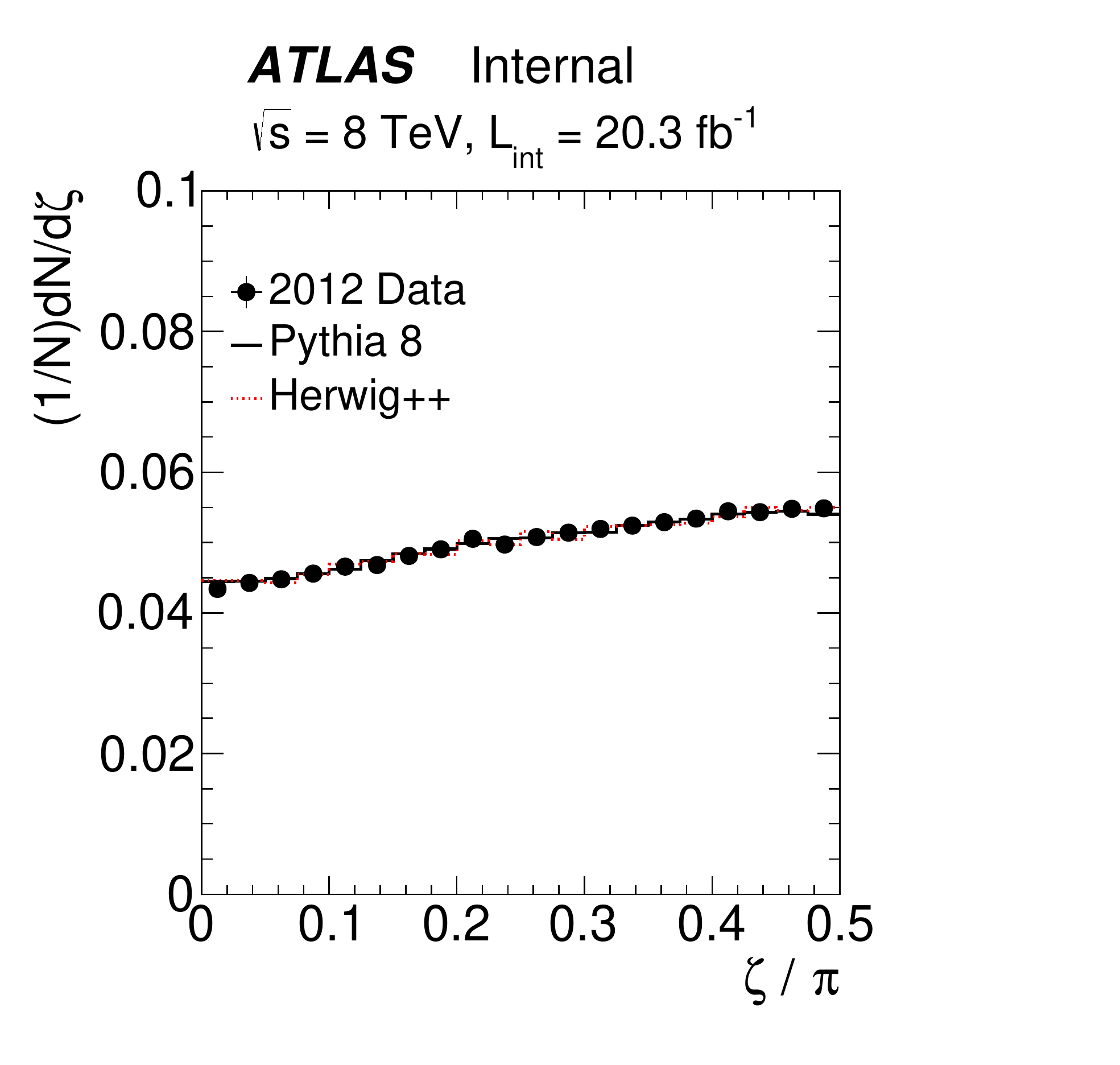}
\caption{The distribution of $\zeta$ in bins of the jet $p_\text{T}$ for the data as well as simulation.  The top left bin corresponds to 140 GeV $<p_\text{T}<200$ GeV and each subsequent plot going from left to right, top to bottom, increases this range by 120 GeV.}
\label{fig:JetCharge:Syst:TIDE:Zeta:fig3}
\end{center}
\end{figure}

\clearpage

\subsubsection{Track Momentum Resolution}
\label{sec:jetcharge:tracksyst:momentumreso}

The momentum resolution of isolated tracks has been well-measured in $J/\psi\rightarrow\mu\mu$ and $Z\rightarrow\mu\mu$ events~\cite{Aad:2014rra}.  In addition to applying this method to muon tracks combining MS and ID information, this technique has been applied to ID-only tracks and is therefore applicable for the jet charge.  The scale and resolution of reconstructed muon candidates are shifted and smeared in the MC to account for differences between the data and the simulation for $m_{\mu\mu}$.  As these corrections are not applied to generic tracks, the correction factors are taken here as the systematic uncertainty on the momentum resolution.  The momentum resolution is parameterized as

\begin{align}
\frac{\sigma(p_\text{T})}{p_\text{T}}=\frac{r_0}{p_\text{T}}\oplus r_1\oplus r_2\cdot p_\text{T},
\end{align}

\noindent where $\oplus$ means `add in quadrature.'  The first term accounts for fluctuations in the energy loss in the detector material, the second term captures effects due to multiple scattering, and the third term accounts for the intrinsic resolution caused by mis-alignment and the finite spatial resolution of hits.  Unlike for muon spectrometer tracks, inner detector tracks do not traverse a significant amount of material and so $r_0$ and its uncertainty are neglected.  The uncertainties on $r_1$, $r_2$ and the momentum scale $s$ are estimated by smearing every track according to

\begin{align}
p_\text{T}^\text{track}\mapsto \frac{p_\text{T}^\text{track}+s\cdot p_\text{T}^\text{track}}{1+\sigma(r_1)\cdot z_1+\sigma(r_2)\cdot p_\text{T}^\text{track}\cdot z_2},
\end{align}

\noindent where $z_i$ are independent random variables that are normally distributed with mean zero and standard deviation 1.  The values of $r_i$ and $s$ as a function of $\eta$ are shown in Table~\ref{tab:trackres}.  A graphical representation of the uncertainties is shown in Fig.~\ref{fig:systs_tracks}.  The impact of this uncertainty is negligible for $p_\text{T}^\text{track}<100$~GeV, but is significant for $p_\text{T}^\text{tracks}\sim 1$~TeV.  

\begin{table}[h]
\begin{center}
\begin{tabular}{|c|c|c|c|}
\hline
 & $\sigma(r_1)$ & $\sigma(r_2)$ [1/TeV] & $s$ \\
\hline
$|\eta|<1.05$ & 0.0068 & 0.146 & $-0.92\times 10^{-3}$\\
$1.05 <|\eta|<2.0 $& 0.0105 & 0.302 & $-0.86\times 10^{-3}$\\
$|\eta|>2.0 $& 0.0069 & 0.088 & $-0.49\times 10^{-3}$\\
\hline
\end{tabular}
\end{center}
\caption{A summary of the momentum (scale and) resolution uncertainties, taken from Ref.~\cite{Aad:2014rra}.}
\label{tab:trackres}
\end{table}

\begin{figure}[h!]
\begin{center}
\includegraphics[width=0.5\textwidth]{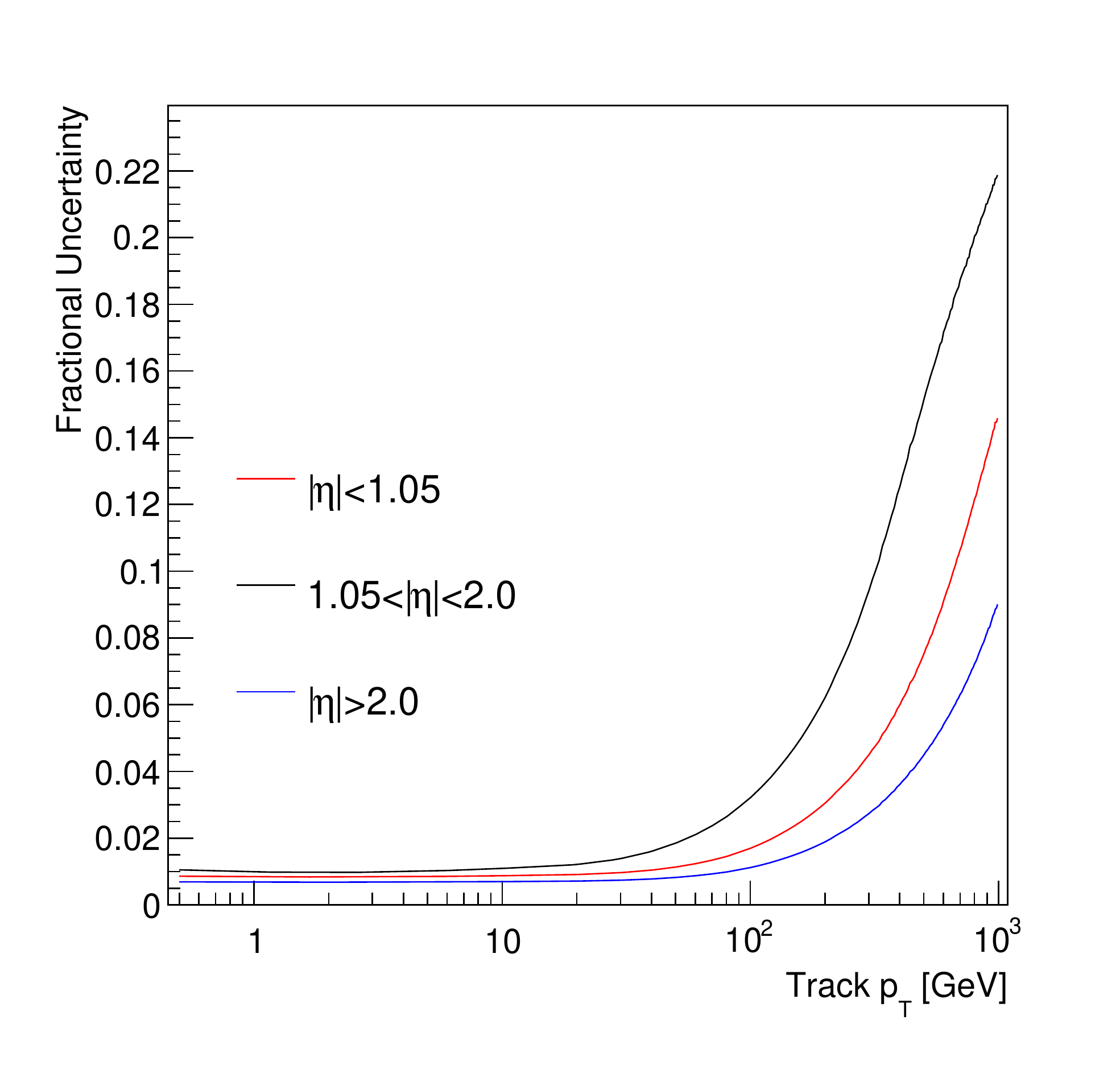}
\caption{For an ensemble of tracks of the same $p_\text{T}$ and $\eta$, the standard deviation of the distribution of the track $p_\text{T}$ divided by the track $p_\text{T}$ by applying the smearing procedure described in the text for the track momentum resolution uncertainty.}
\label{fig:systs_tracks}
\end{center}
\end{figure}

\clearpage

\subsubsection{Track Charge Identification}
\label{sec:jetcharge:tracksyst:chargeid}

Aside from the track $p_\text{T}$, the other track parameter that is relevant for the jet charge is the track charge.  Especially at high $p_\text{T}$ when the tracks are nearly straight, the probability for mis-identifying the track charge increases.  The left plot of Fig.~\ref{fig:qflip} shows the simulation probability for the charge mis-identification as a function of the jet $p_\text{T}$.  The truth charge is the electric charge of the matched truth particle.  At low jet $p_\text{T}$, the charge mis-id rate is less than $0.01\%$ and even in the highest jet $p_\text{T}$ bin, the charge mis-id rate is less than 1\%.   There is a small increase in the mis-id rate as a function of $\eta$, shown in the right plot of Fig.~\ref{fig:qflip}.    The strongest dependence of the mis-id rate is on the track $p_\text{T}$, as shown in Fig.~\ref{fig:qflip2}.  The mis-id rate does not depend strongly on the jet $p_\text{T}$ given the track $p_\text{T}$.  Dedicated charge flipping studies in the data using leptonically decaying resonances suggest that the mis-modelling of the mis-id rate is much less than 50\%~\cite{ATLAS:2014kca}.  Therefore, the charge mis-id uncertainty is conservatively estimated by randomly flipping the charge of tracks at 50\% of the mis-id rate.  The rate extracted from the simulation (Fig.~\ref{fig:qflip2}) is $<0.1\%$ for track $p_\text{T}<100$ GeV, $0.5\%$ for $100$ GeV $<p_\text{T}<200$ GeV,  $1\%$ for $200$ GeV $<p_\text{T}<300$ GeV, $2\%$ for $300$ GeV $<p_\text{T}<400$ GeV and $4\%$ for $p_\text{T}>400$ GeV\footnote{Note that even in the highest $p_\text{T}$ jets that pass the event selection, there are very few with tracks that have $p_\text{T}>400$ GeV.}.  The rate shown in Fig.~\ref{fig:qflip} is likely to be very conservative.  This is because tracks with a truth matching probability of $>50\%$ that are actually fake (see Sec.~\ref{sec:jetcharge:tracksyst:faketracks}) will have the wrong charge $\sim 50\%$ of the time as there is no relation between the track charge and truth particle charge.   As a result, increasing the charge flipping rate has a larger impact (though still negligible) on the jet charge than reducing the mis-id rate.  This is because increasing the mis-id rate mostly impacts correctly classified real tracks while decreasing the mis-id rate effects mostly mis-classified fake tracks.  In other words, a large fraction of the tracks that are classified with a charge mis-id are likely fake tracks, while the majority of tracks with a truth matching probability of $>50\%$ are not fake.  Figure~\ref{fig:qflip22} supports the claim that many of the tracks with charge mis-id are actually fake.  By construction, the tracks have a truth matching probability of $>50\%$, but the distribution of probabilities is not as strongly peaked at one as for tracks with the correct charge.  Furthermore, many tracks with a misclassified charge have a significantly different $p_\text{T}$ than the truth-matched particle.

\begin{figure}[h!]
\begin{center}
\includegraphics[width=0.45\textwidth]{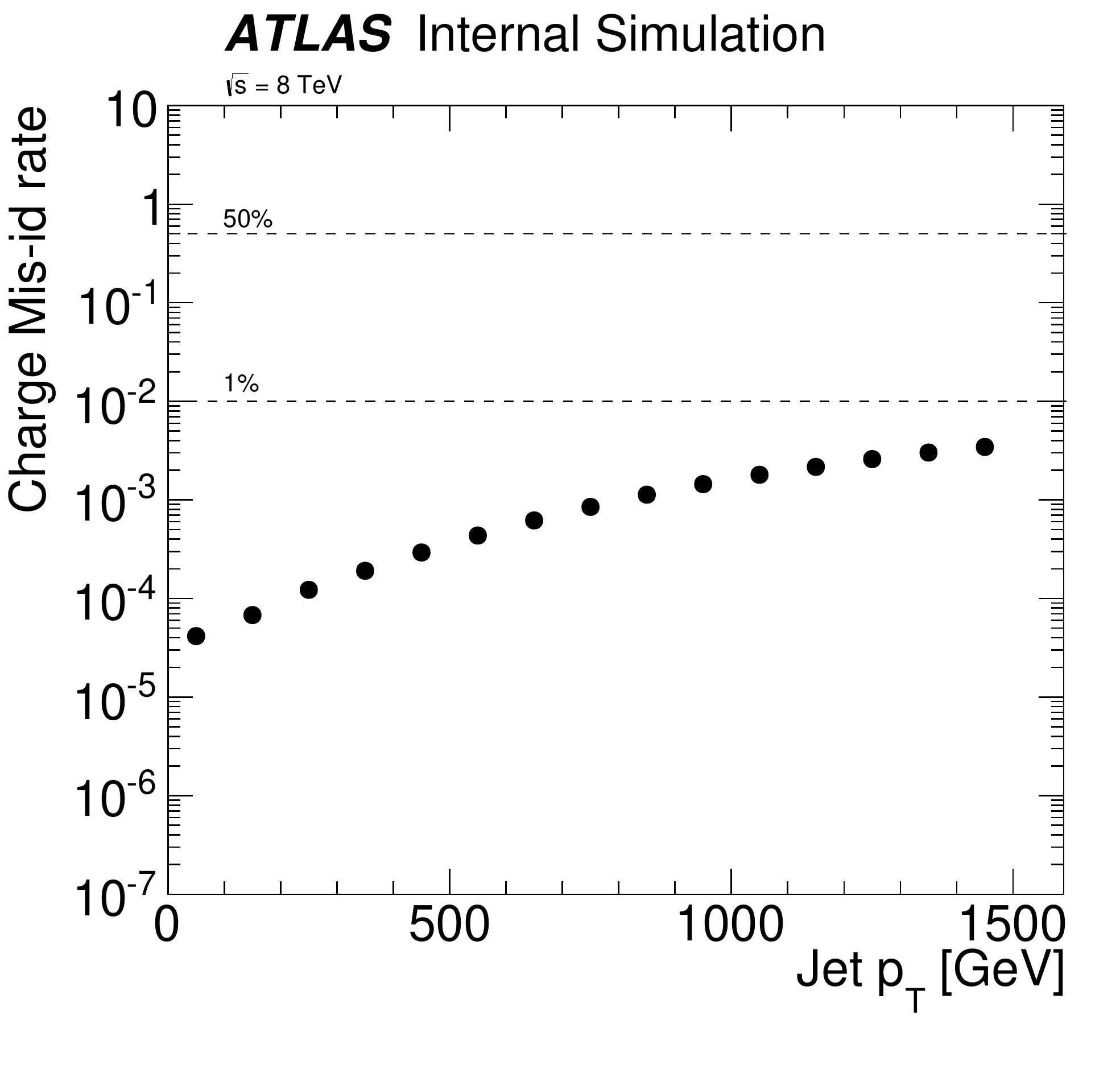}\includegraphics[width=0.45\textwidth]{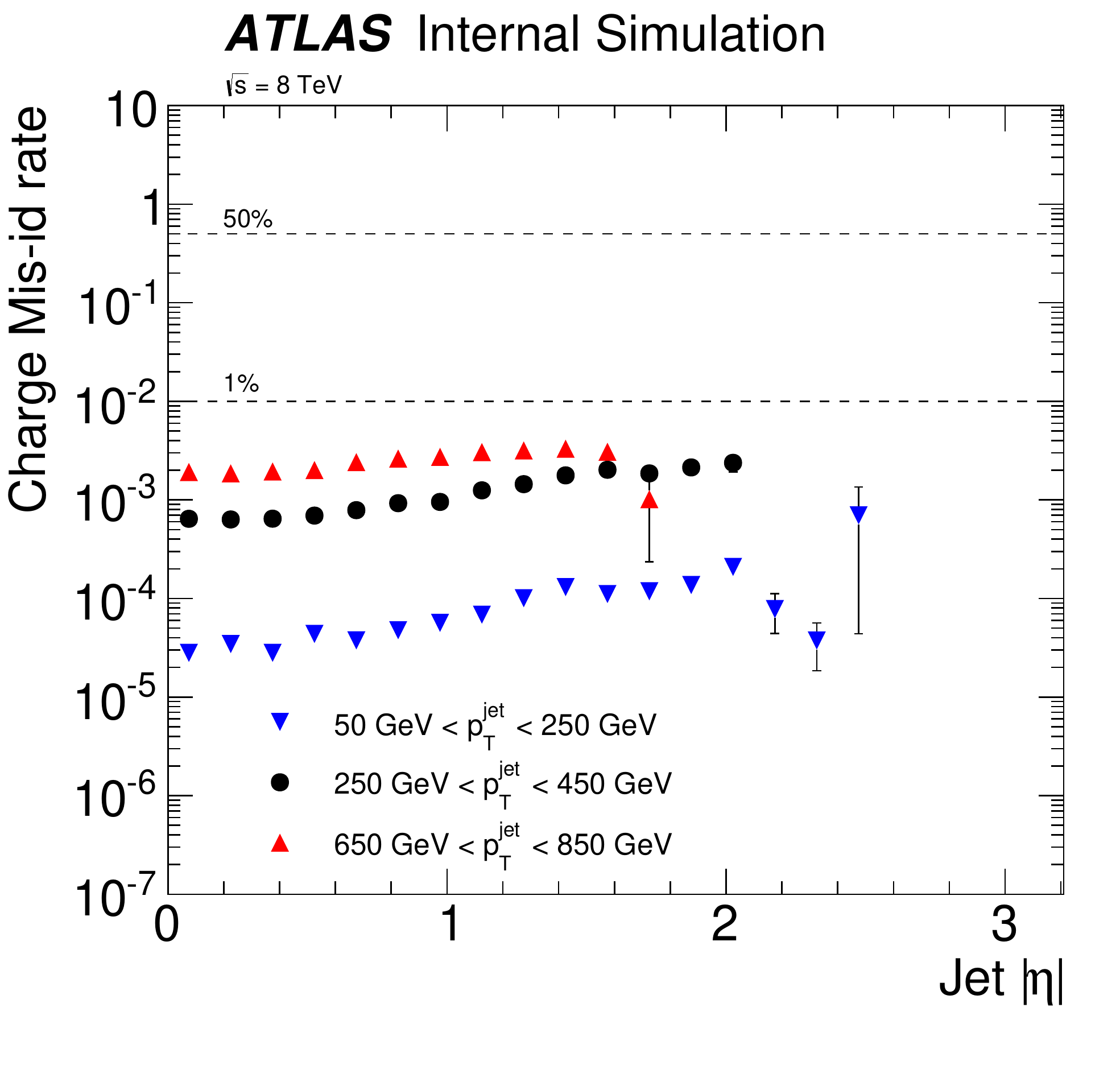}
\caption{The charge mis-id rate as a function of the jet $p_\text{T}$ (left) and $|\eta|$ (right).}
\label{fig:qflip}
\end{center}
\end{figure}

\begin{figure}[h!]
\begin{center}
\includegraphics[width=0.45\textwidth]{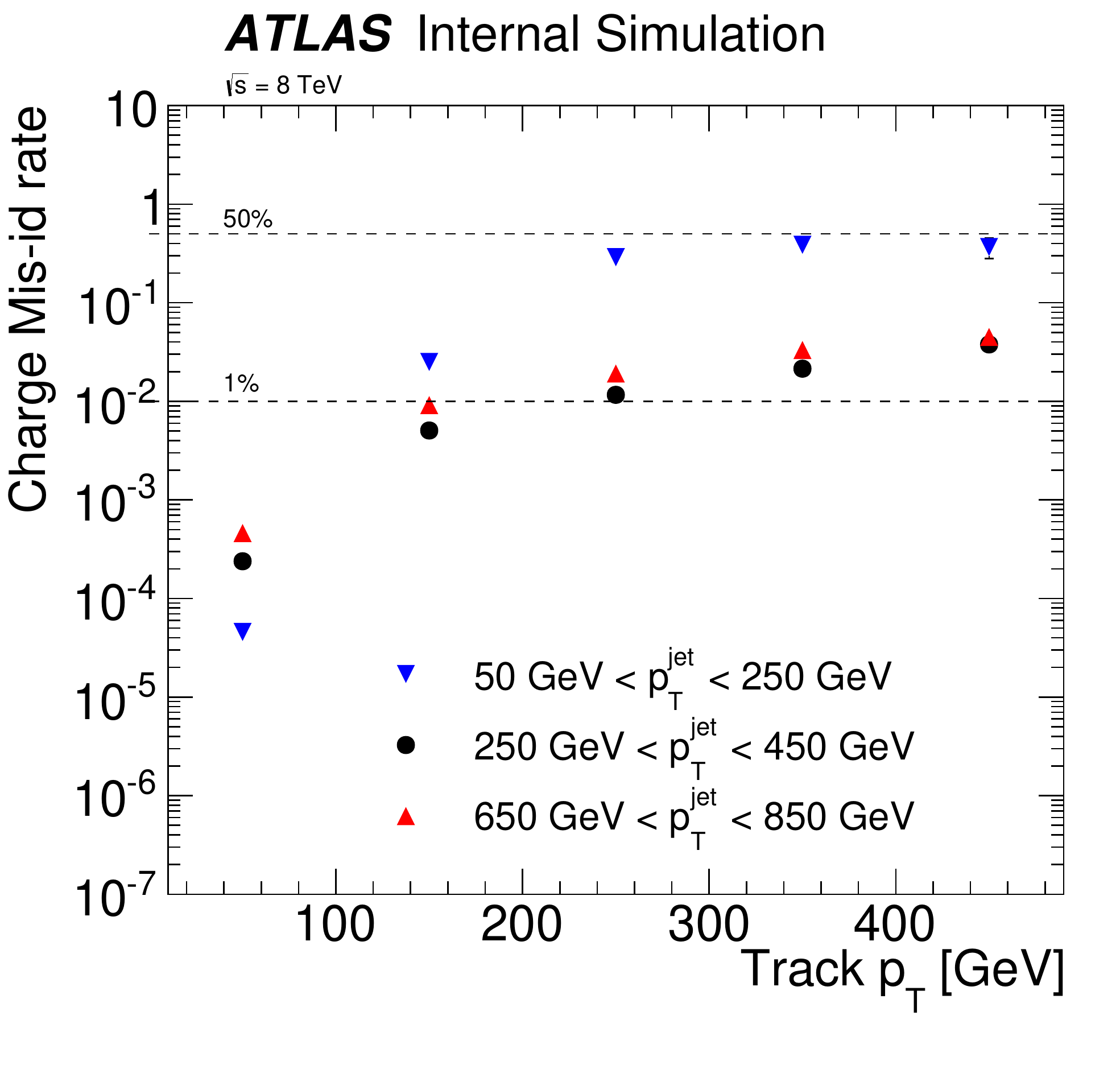}
\caption{The charge mis-id rate as a function of the track $p_\text{T}$.  The charge mis-id rate increases rapidly for the lowest $p_\text{T}$ bin due to fake tracks that happen to have a truth matching probability of $>50\%$ but have a mis-id rate of $\sim 50\%$ because the reconstructed charge is random.}
\label{fig:qflip2}
\end{center}
\end{figure}

\begin{figure}[h!]
\begin{center}
\includegraphics[width=0.5\textwidth]{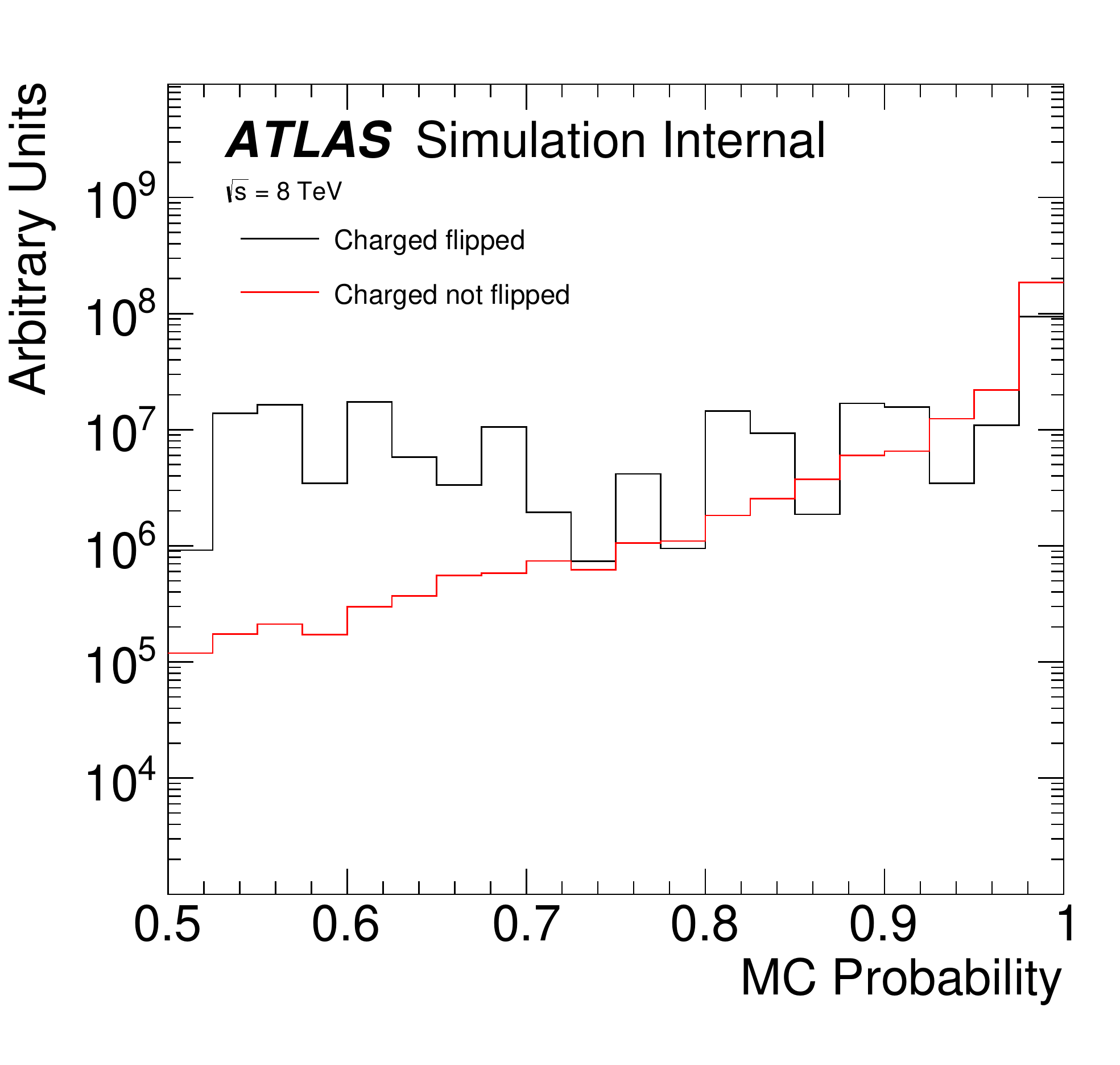}\includegraphics[width=0.5\textwidth]{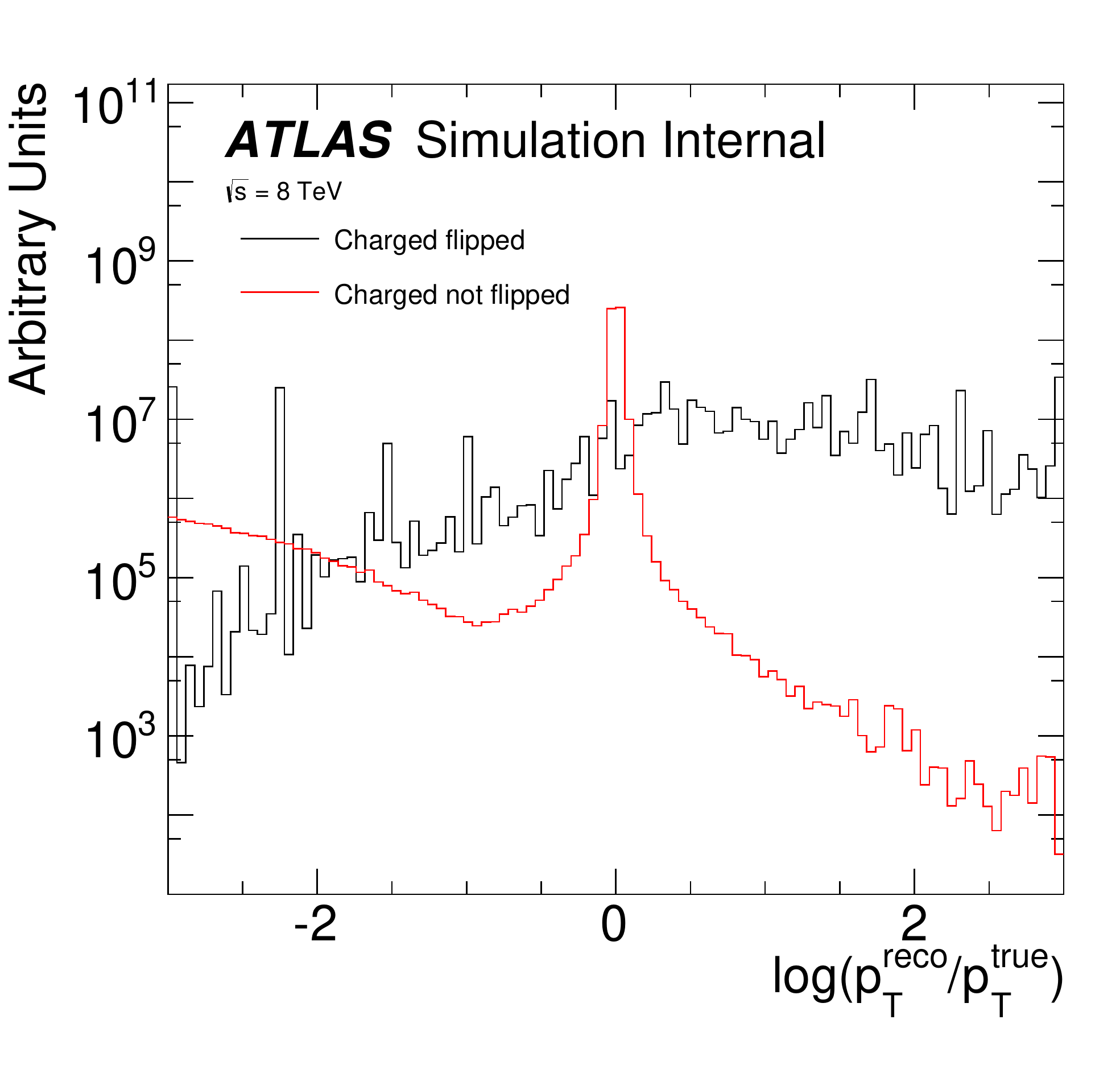}
\caption{The MC matching probability for tracks with a flip and without a flip (left) as well as the $p_\text{T}$ difference for flipped and non-flipped tracks between the `true' and reconstructed track $p_\text{T}$ (right).  The fact that the MC probability is in general lower for tracks with a flipped charge and the $p_\text{T}$ is usually very different from the truth $p_\text{T}$ indicates that many of these tracks are actually fake tracks.}
\label{fig:qflip22}
\end{center}
\end{figure}

\clearpage

\subsubsection{Fake Tracks}
\label{sec:jetcharge:tracksyst:faketracks}

Random combinations of hits in the detector can be combined together to form a reconstructed track.  Tracks resulting in particular from multi-particle trajectories that have kinks can result in a large reconstructed track $p_\text{T}$.  The joint distribution of the fake track $p_\text{T}$ (truth matching less than 50\%) and jet $p_\text{T}$ is shown in the left plot of Fig.~\ref{fig:fake2d}.  Tracks with $p_\text{T}$ larger than the jet $p_\text{T}$ are most likely from fakes and can be used to study the fake rate in data.  The right plot of Fig.~\ref{fig:fake2d} shows the distribution of track $p_\text{T}$ in five jet $p_\text{T}$ bins.  The rate of high $p_\text{T}$ tracks is generally higher in the data than in the simulation, but this is especially relevant beyond the dashed lines where the track $p_\text{T}$ exceeds the jet $p_\text{T}$.  One contribution to the excess of high $p_\text{T}$ tracks is from an underestimation of fake tracks in the simulation.  Figure~\ref{fig:fake2d} suggests that this excess for high $p_\text{T}$ tracks is less than 50\%. 

\begin{figure}[h!]
\begin{center}
\includegraphics[width=0.5\textwidth]{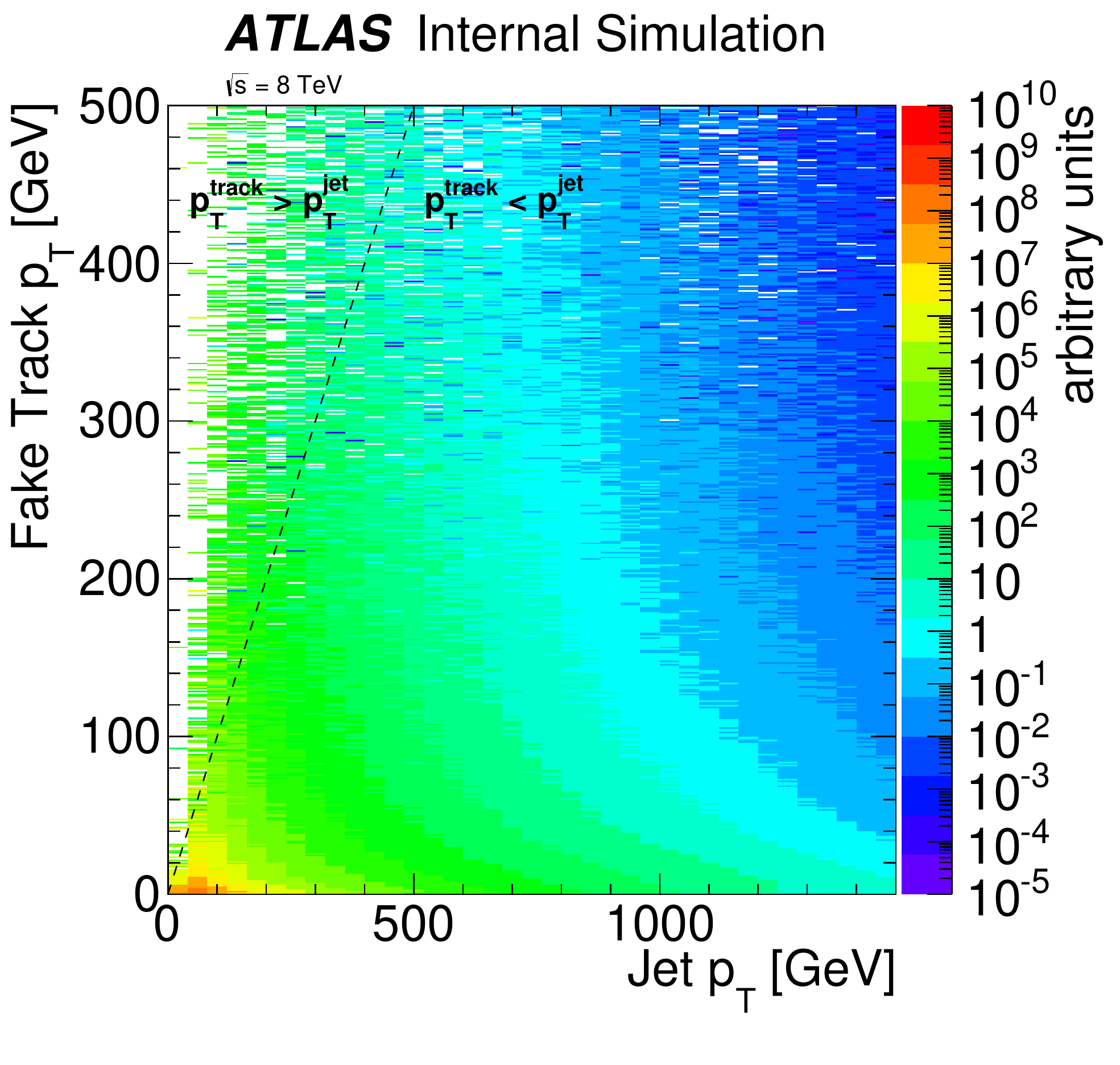}\hspace{4mm}\includegraphics[width=0.43\textwidth]{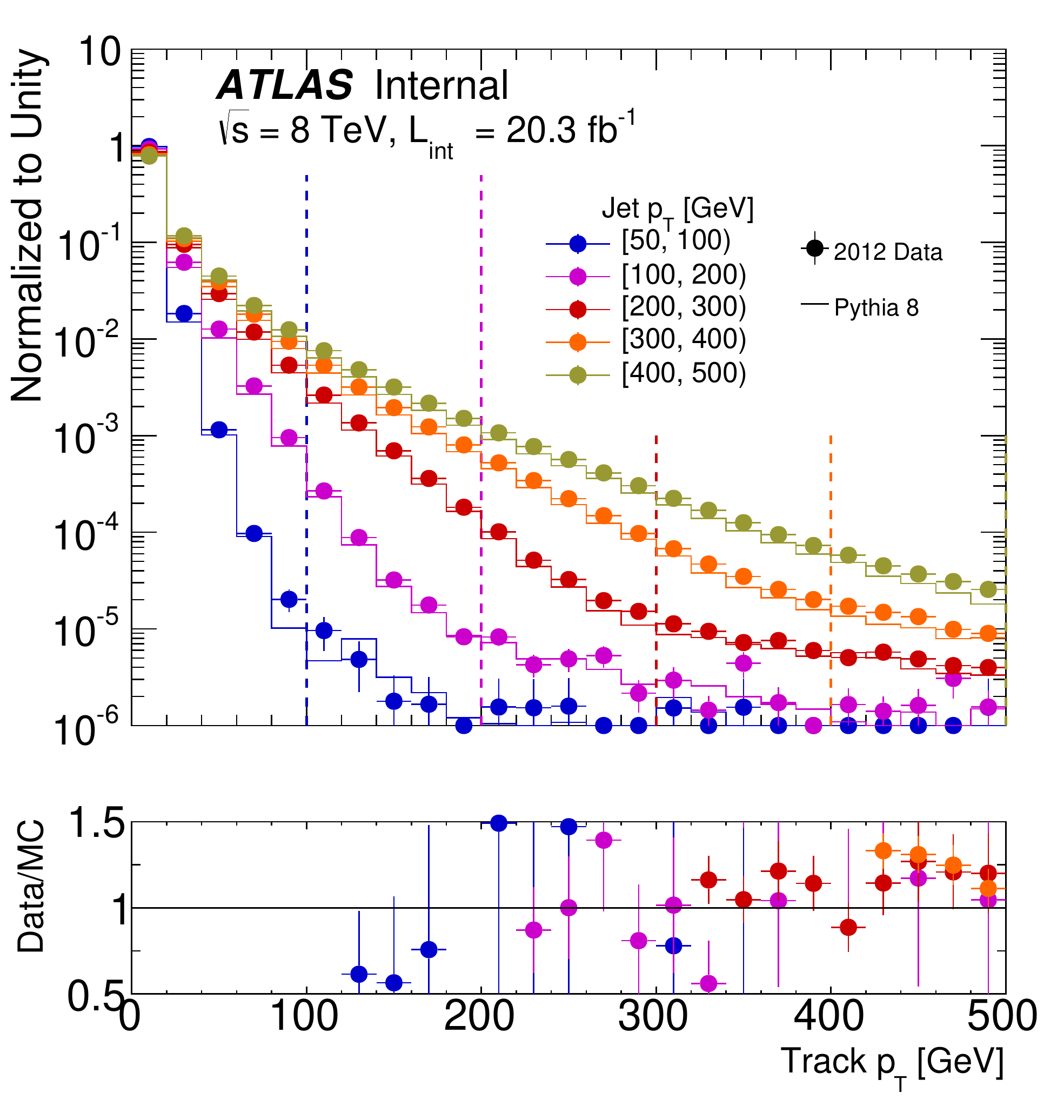}
\caption{Left: The joint distribution of the fake track $p_\text{T}$ and the jet $p_\text{T}$.  The dashed line indicates the $p_\text{T}^\text{track} = p_\text{T}^\text{jet}$ line.  Right: the normalized distribution of the track $p_\text{T}$ in five bins of jet $p_\text{T}$.  The dashed horizontal lines indicate the jet $p_\text{T}$ thresholds.  The ratio only shows points for which the track $p_\text{T}$ exceeds the jet $p_\text{T}$.}
\label{fig:fake2d}
\end{center}
\end{figure}

\noindent To conservatively estimate the impact of fake tracks on the jet charge, fake tracks are randomly removed at a rate that is $\pm 50\%$ of the rate in simulation.  The fraction of fake tracks inside jets, integrating over all track momenta, is shown in Fig.~\ref{fig:fake2d2}.  The fake rate is largely independent of the jet $p_\text{T}$ and is $\lesssim 0.1\%$.

\begin{figure}[h!]
\begin{center}
\includegraphics[width=0.5\textwidth]{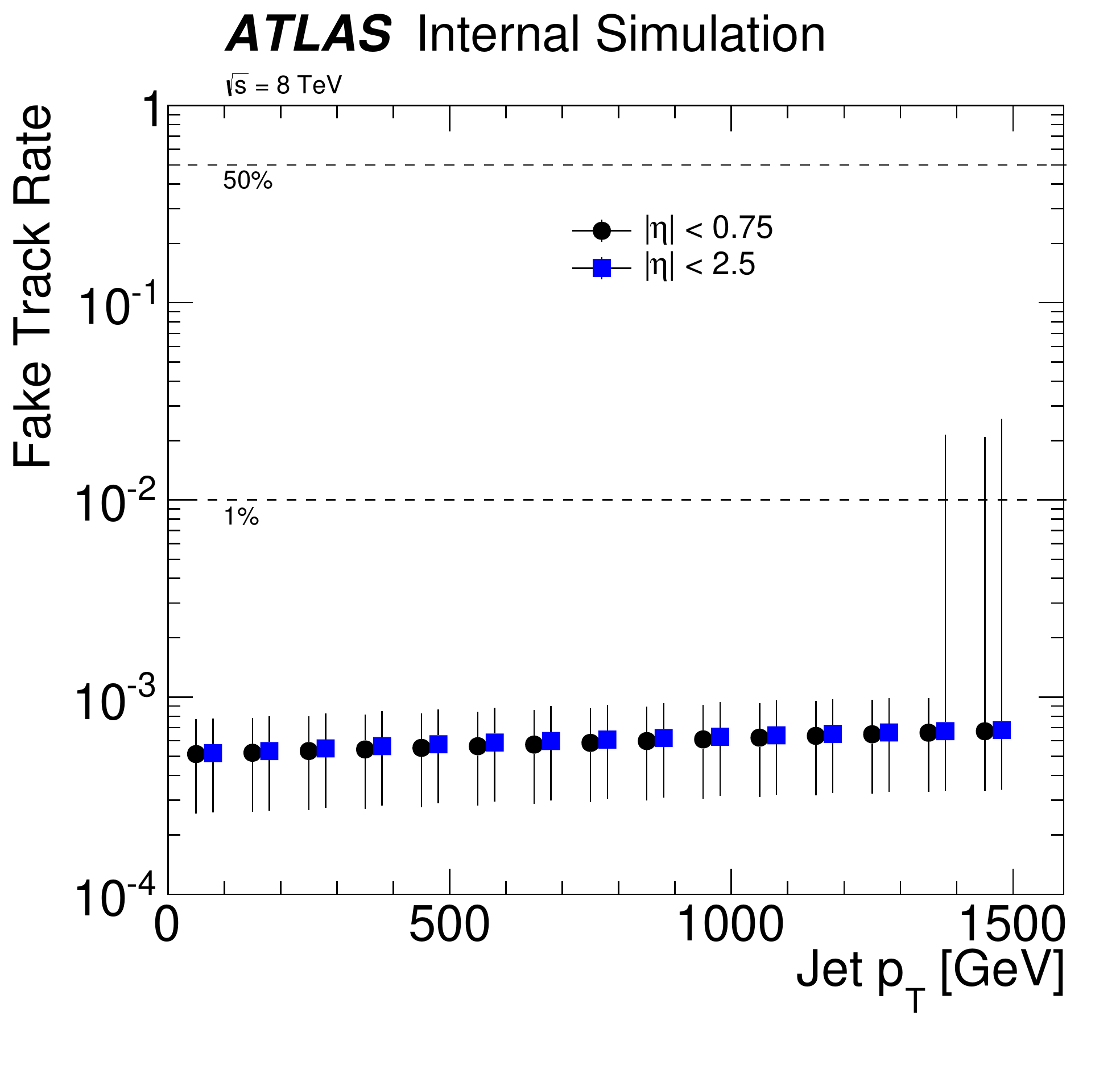}\caption{The fraction of tracks inside a jet that are classified as fake in bins of jet $p_\text{T}$.  The circle markers are for central $|\eta|<0.75$ jets while the squares are for all tracks within the tracker acceptance.  The markers for the squares are offset by 30 GeV, but the actual fake rate is computed with the same $p_\text{T}$ binning as for the circles.  The markers indicate the median of the fake fraction distribution in a given jet $p_\text{T}$ bin and the error bar is the inter-quartile range.}
\label{fig:fake2d2}
\end{center}
\end{figure}

\noindent The distribution of the fake track $p_\text{T}$ conditioned on the jet $p_\text{T}$ is shown in Fig.~\ref{fig:fake2d3}.  As expected, the fake track $p_\text{T}$ spectrum is largely independent of the jet $p_\text{T}$.  There is a small dependence, especially in the lowest $p_\text{T}$ bin, because the hit density and thus fake rate increase monotonically with jet $p_\text{T}$.

\begin{figure}[h!]
\begin{center}
\includegraphics[width=0.45\textwidth]{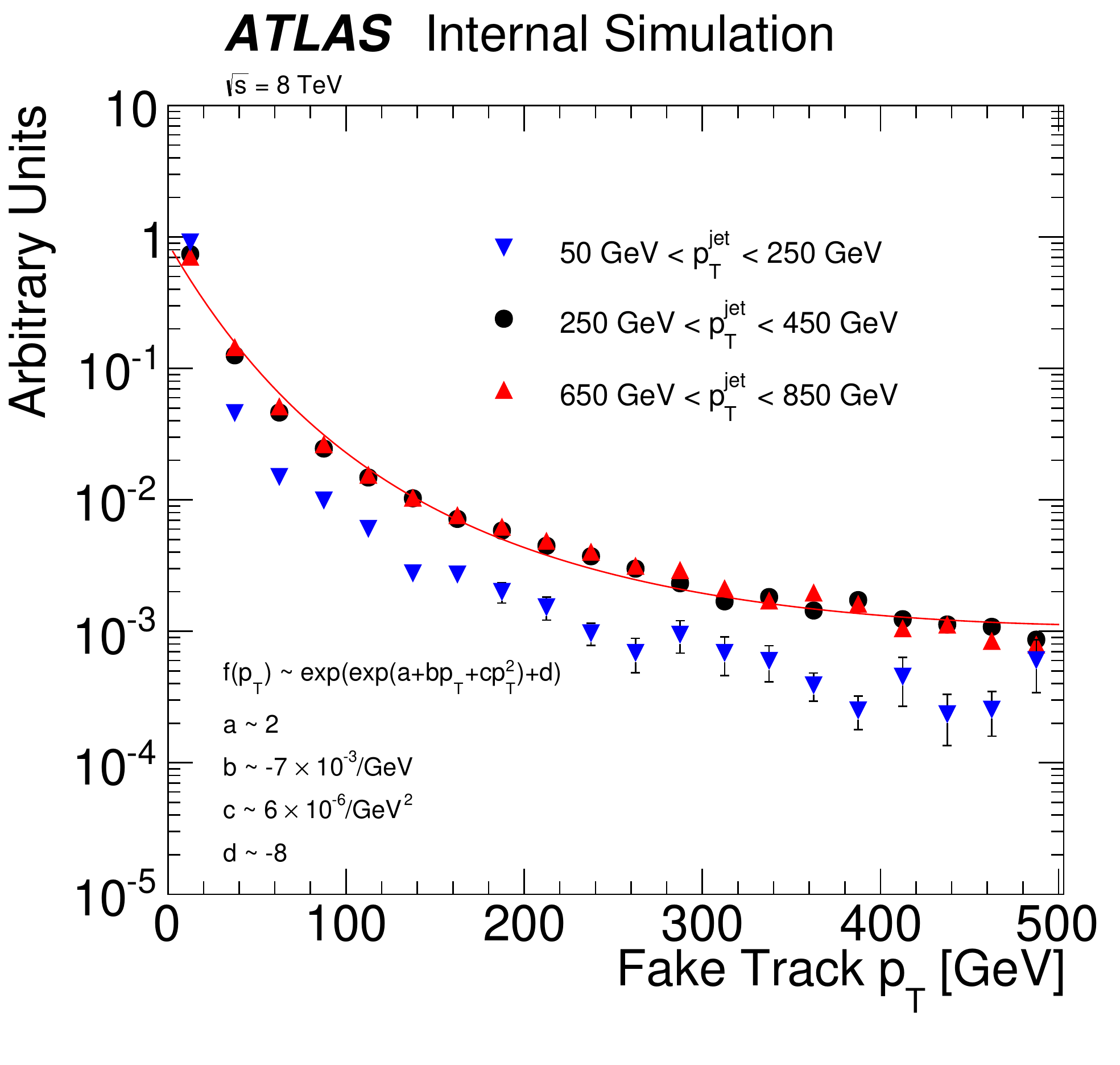}\includegraphics[width=0.45\textwidth]{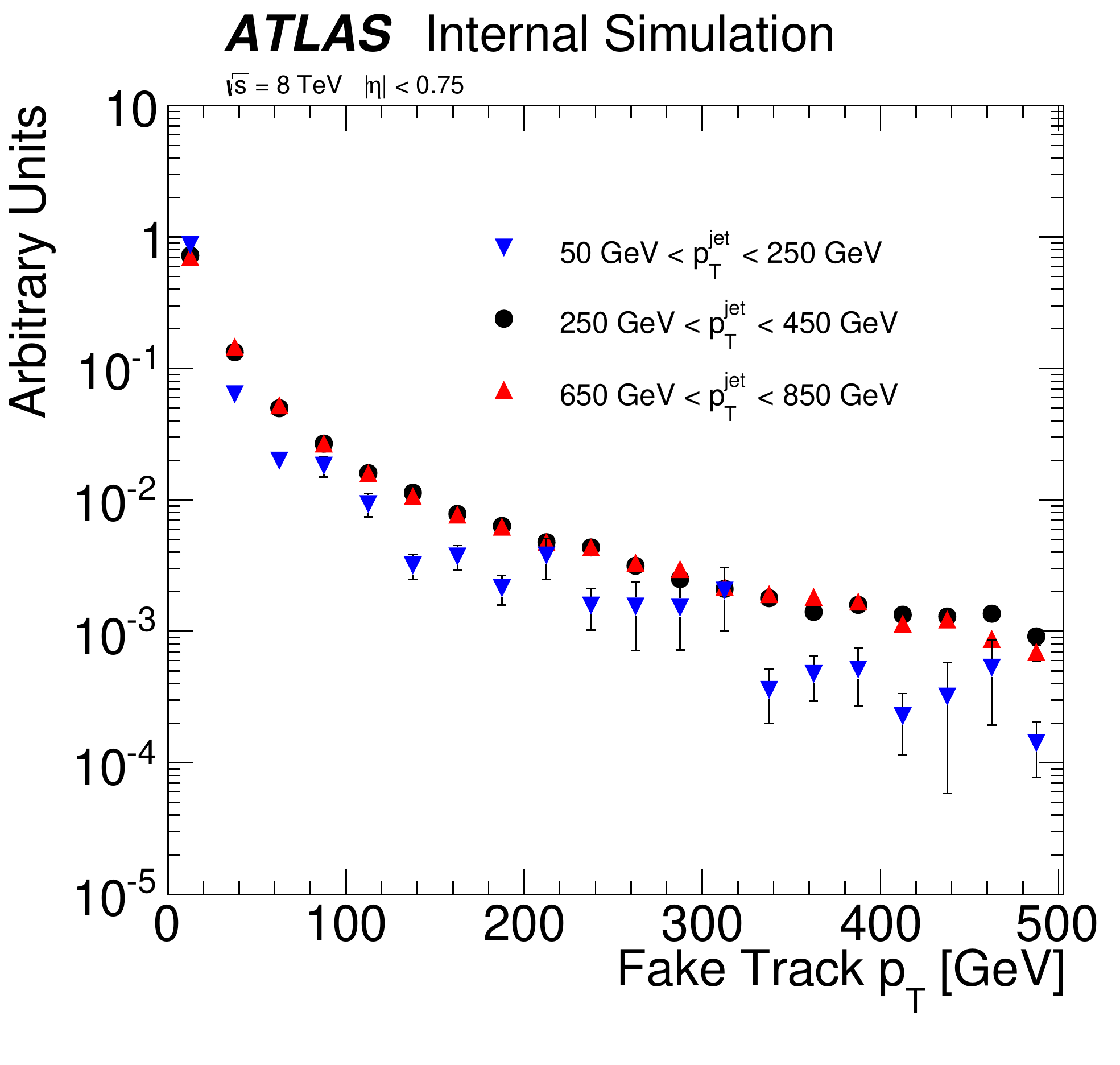}
\caption{The $p_\text{T}$ spectrum of fake tracks inclusively in $\eta$ (left) and for $|\eta|<0.75$ (right).}
\label{fig:fake2d3}
\end{center}
\end{figure}

\clearpage

\subsubsection{Charged Particle Multiplicity}
\label{sec:jetcharge:tracksyst:ncharged}

The tracking uncertainties described so far take into account the resolution and efficiency of the reconstruction of charged-particle momenta.  One last source of systematic uncertainty is the number of charged particles.  The unfolding procedure uncertainty takes into account the uncertainty on the prior due to the charged-particle multiplicity, but the jet charge resolution also changes with the charged-particle multiplicity.  To assess the impact on the response matrix of the mismodeled charged-particle multiplicity, the distribution of $n_\text{track}$ is reweighted in the simulation to match data per jet $p_\text{T}$ bin and the relative difference when unfolding the nominal {\sc Pythia} distribution with the reweighted {\sc Pythia} distribution is taken as a systematic uncertainty\footnote{Since the prior is also changed, this uncertainty at least partially includes the unfolding procedure uncertainty.}.  Fig.~\ref{fig:trackmult:trackmult} shows the track multiplicity in three bins of jet $p_\text{T}$ before any reweighting.  These distributions will be the main focus of Chapter~\ref{cha:multiplicity} and so are not discussed in more detail here.  

\begin{figure}[h!]
\begin{center}
\includegraphics[width=0.33\textwidth]{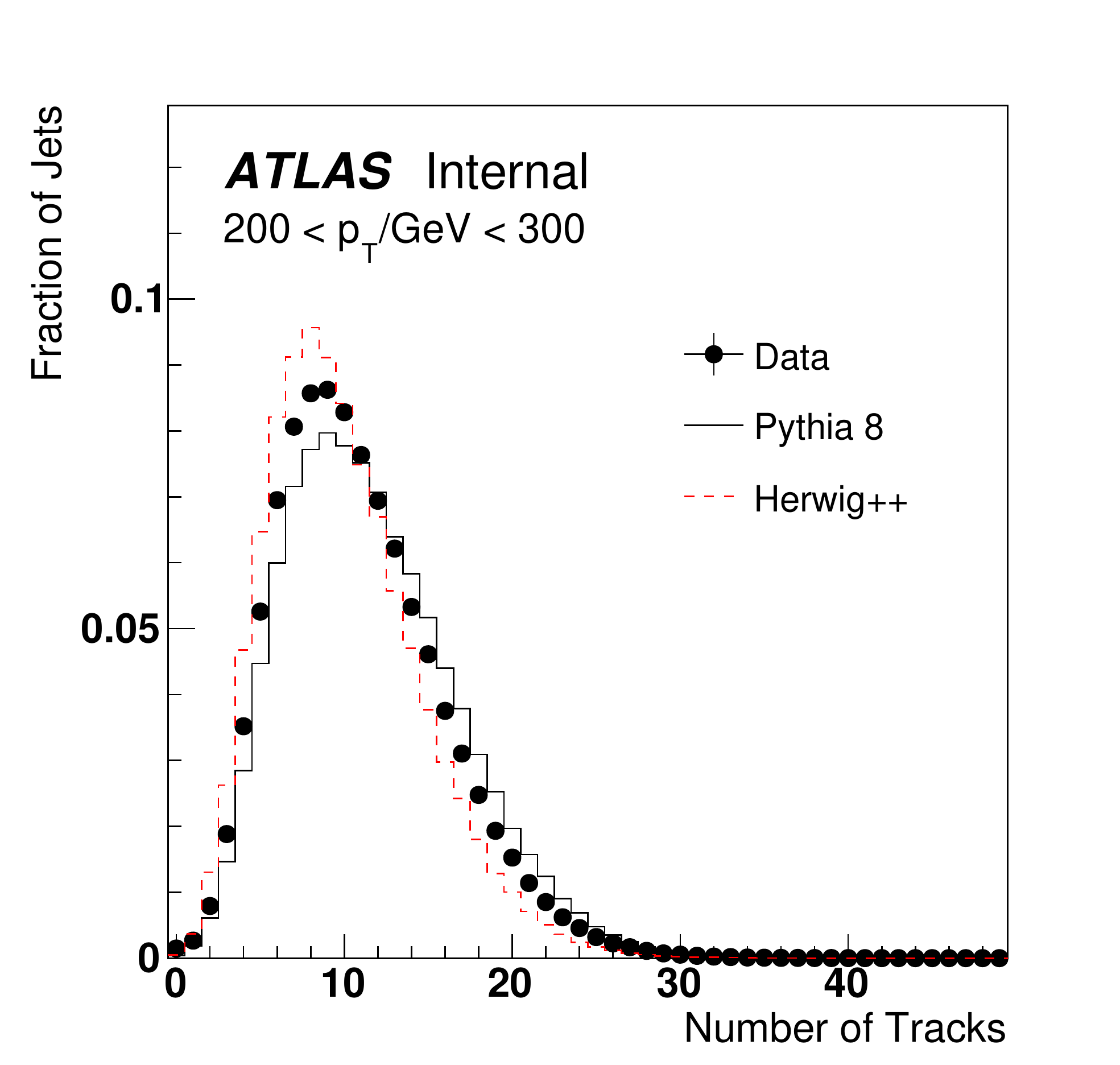}\includegraphics[width=0.33\textwidth]{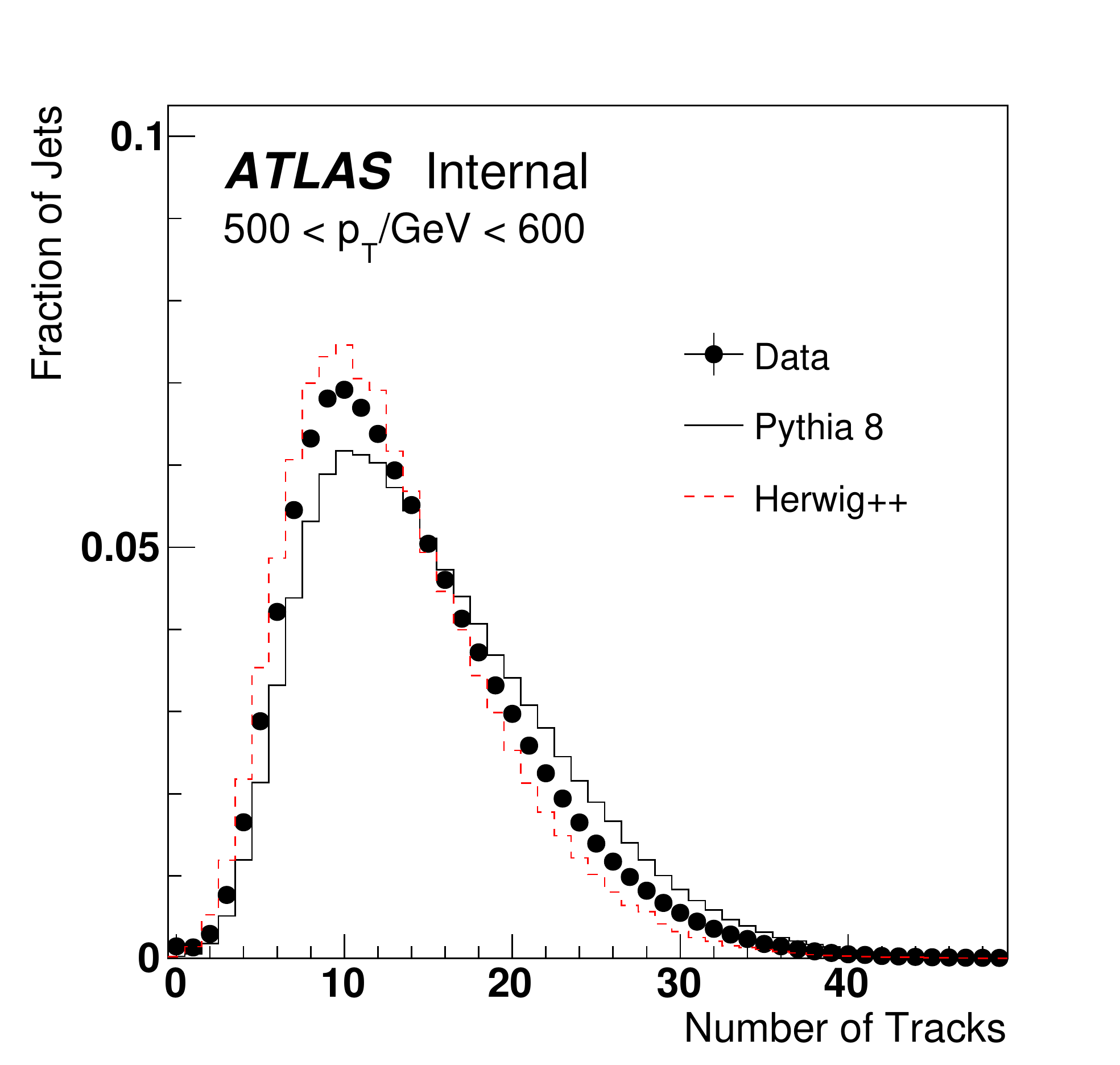}\includegraphics[width=0.33\textwidth]{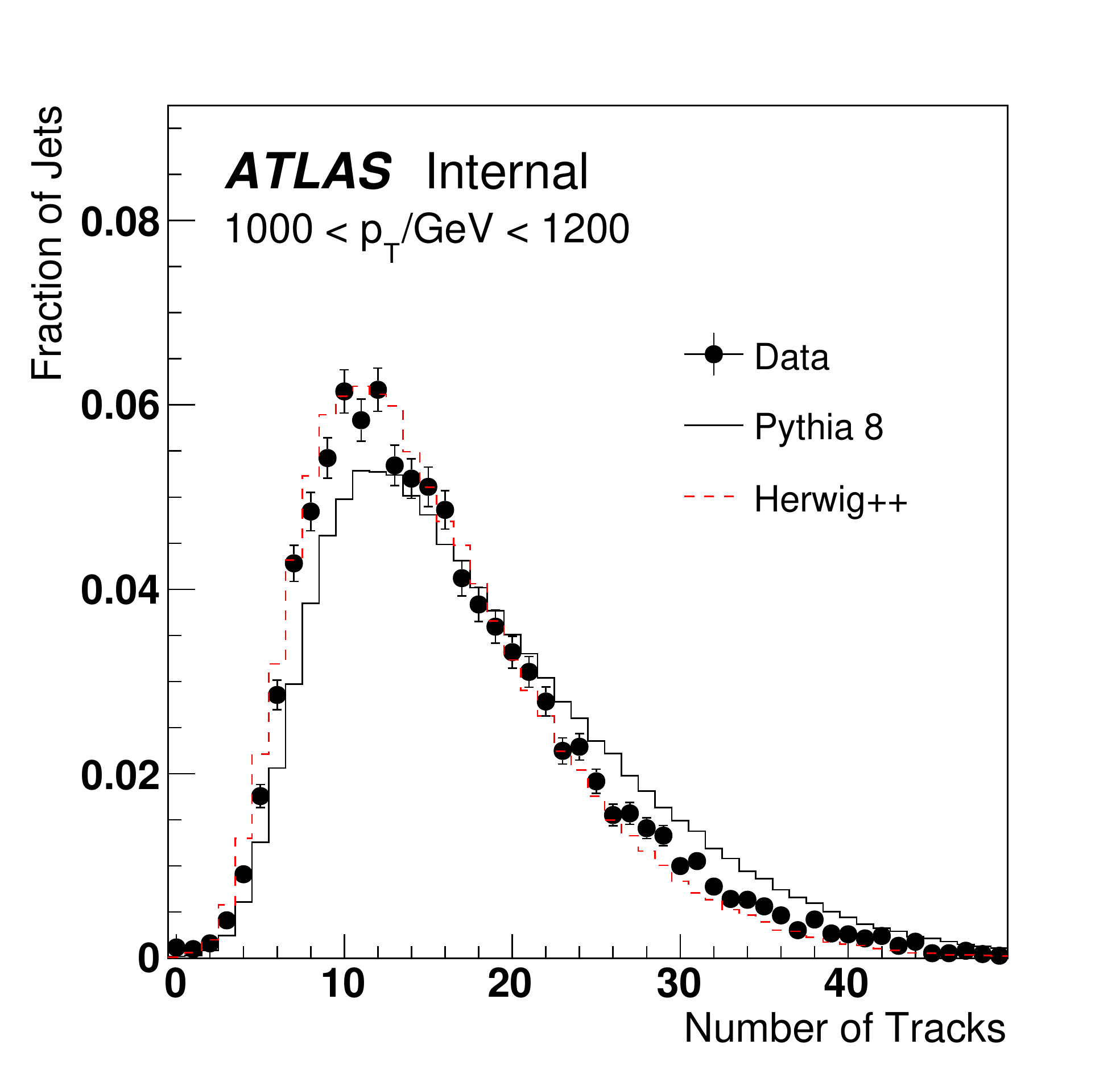}\\
\caption{The track multiplicity for jets in data and in Pythia for various $p_\text{T}$ bins.}
\label{fig:trackmult:trackmult}
\end{center}
\end{figure}

\begin{figure}[h!]
\begin{center}
\includegraphics[width=0.4\textwidth]{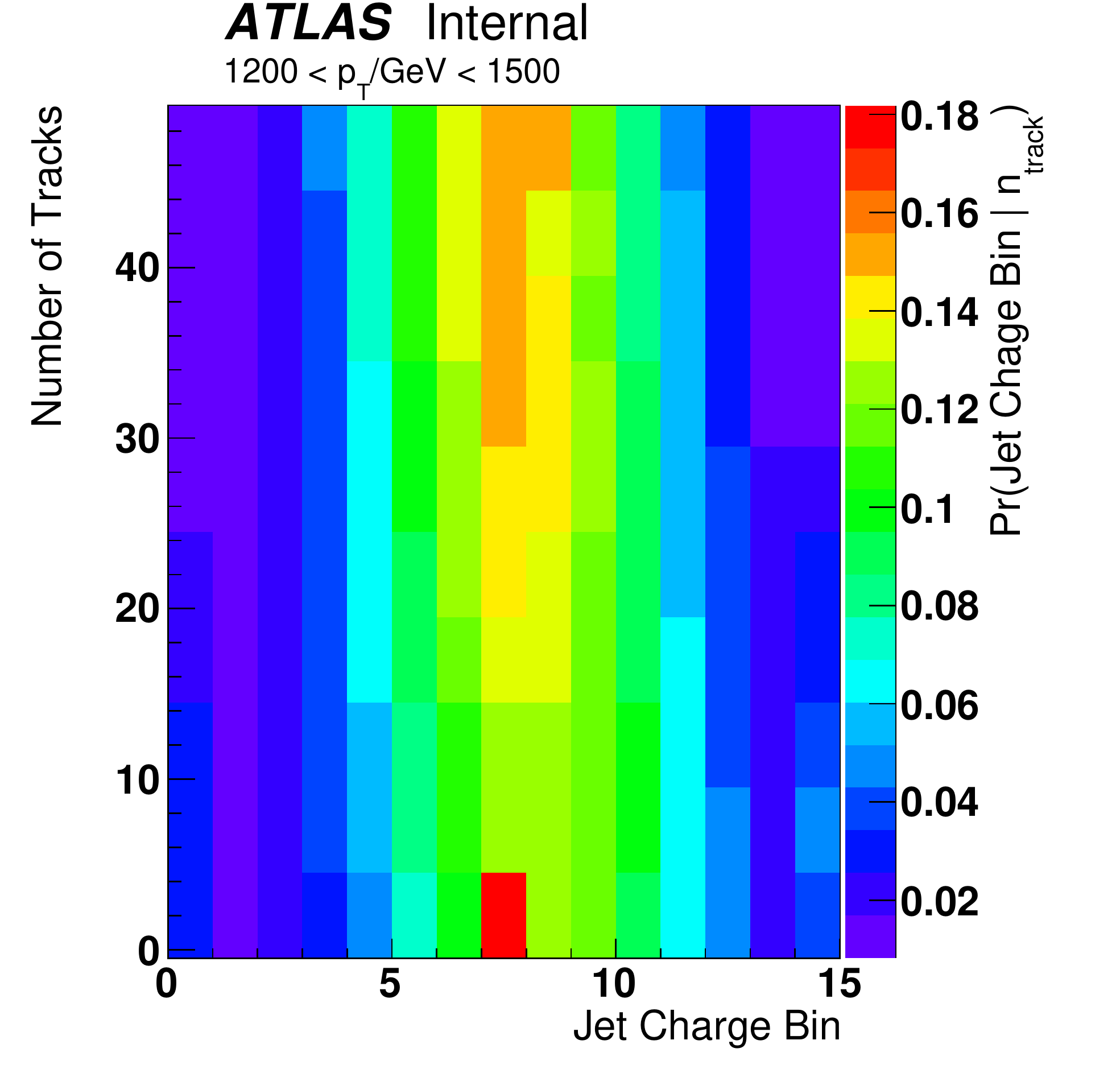}
\caption{The conditional distribution of the jet charge distribution given $n_\text{track}$.  This is for the more forward jet and $\kappa=0.5$.}
\label{fig:trackmult:corr}
\end{center}
\end{figure}

An example conditional distribution of the jet charge given the track multiplicity is shown in Fig.~\ref{fig:trackmult:corr}.  When $n_\text{track}=0$, the jet charge is zero by definition, resulting in a spike in the first row of Fig.~\ref{fig:trackmult:corr}.  The resolution of the jet charge improves with $n_\text{track}$, which is why the higher rows in Fig.~\ref{fig:trackmult:corr} have a jet charge distribution that is more peaked around the mean (see Fig.~\ref{fig:res_tracks}).  The re-weighted jet charge distributions are shown in Fig.~\ref{fig:trackmult:corr2} as a function of jet $p_\text{T}$.  Interestingly, the jet charge distribution average and standard deviation of the re-weighted {\sc Pythia} distribution agree well with the data.  This indicates that a significant contribution to the mis-modeling is from the track multiplicity.   However, the full explanation must be more complicated because the re-weighted {\sc Herwig++} is worse than the un-weighted simulation, though the effect is not as significant as with {\sc Pythia}.

\begin{figure}[h!]
\begin{center}
\includegraphics[width=0.4\textwidth]{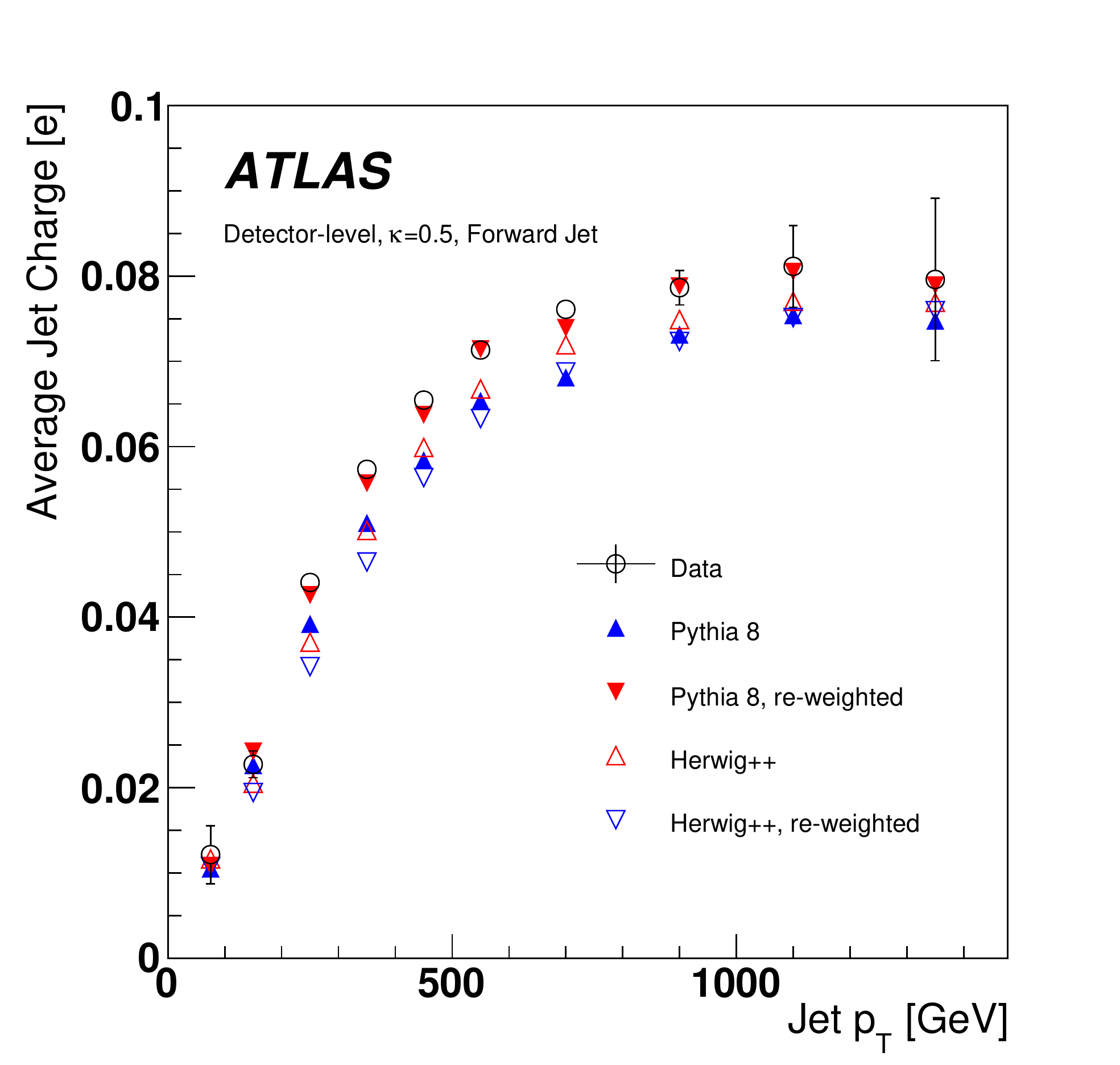}\includegraphics[width=0.4\textwidth]{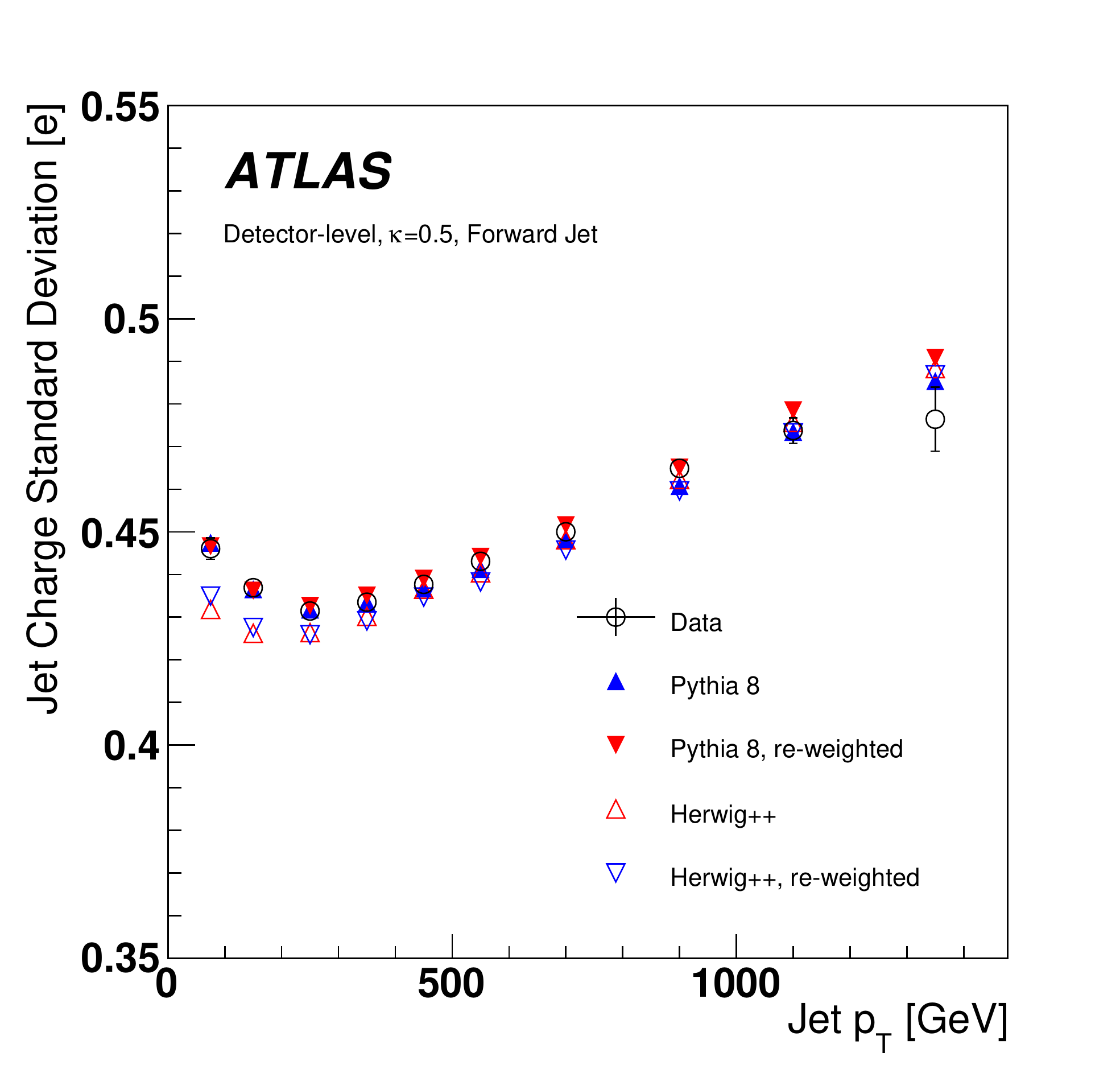}
\caption{The impact of $n_\text{track}$ re-weighting on the jet charge distribution average (left) and standard deviation (right) as a function of jet $p_\text{T}$ for the more forward jet and $\kappa=0.5$.}
\label{fig:trackmult:corr2}
\end{center}
\end{figure}

 The uncertainty associated with the $n_\text{track}$ re-weighting is shown in Fig.~\ref{fig:trackmult:recomean_uncert}. This uncertainty is much smaller than other uncertaitnies for the standard deviation across $p_\text{T}$ and also for the jet charge mean at low to moderate jet $p_\text{T}$.  For the mean jet charge, the largest uncertainty is with the smallest $\kappa$ and for large $p_\text{T}$, where it is $3$--$4\%$ percent in the highest $p_\text{T}$ bin for $\kappa=0.3$ and $\kappa=0.5$. 

\begin{figure}[h!]
\begin{center}
\includegraphics[width=0.4\textwidth]{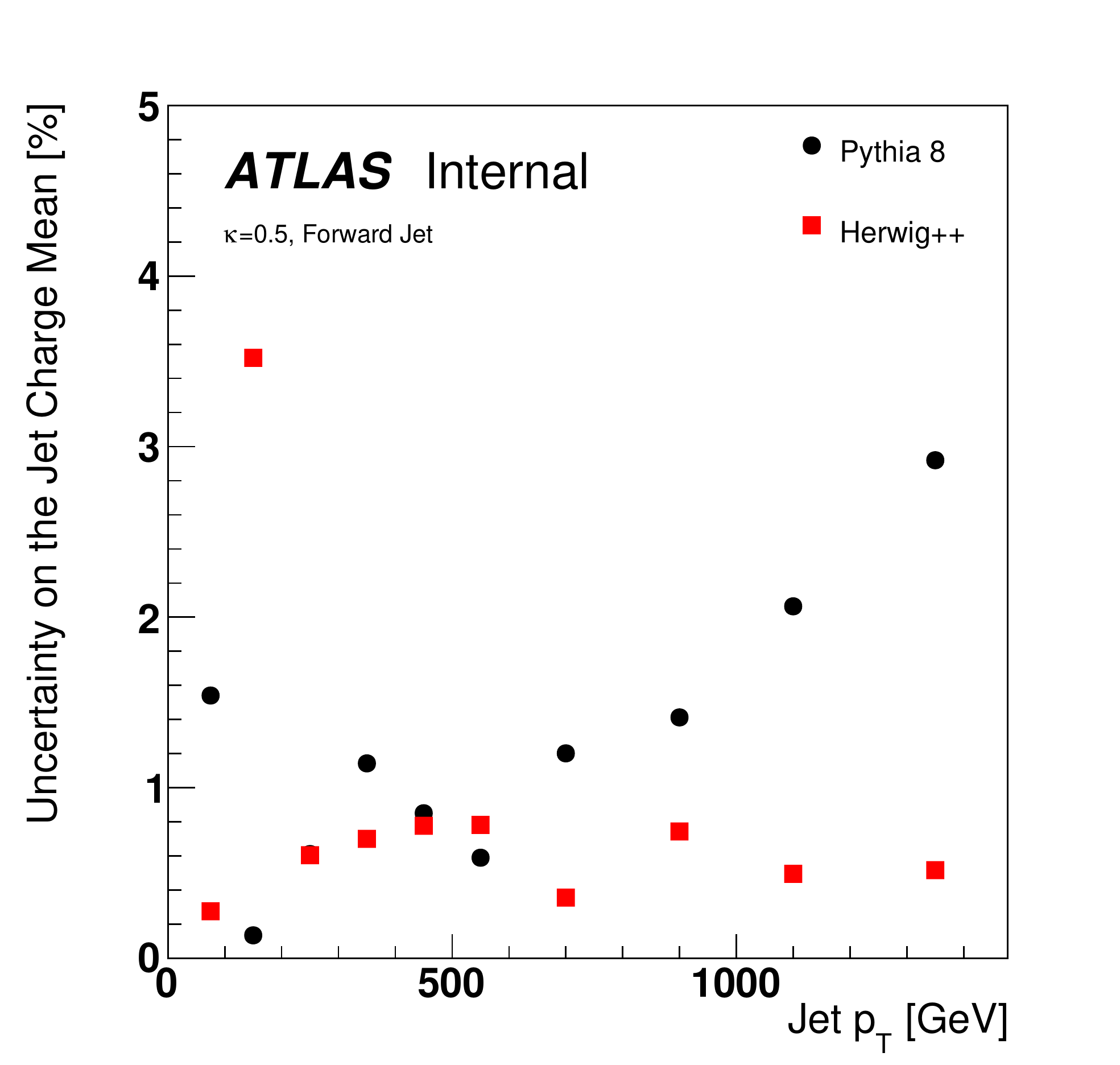}\includegraphics[width=0.4\textwidth]{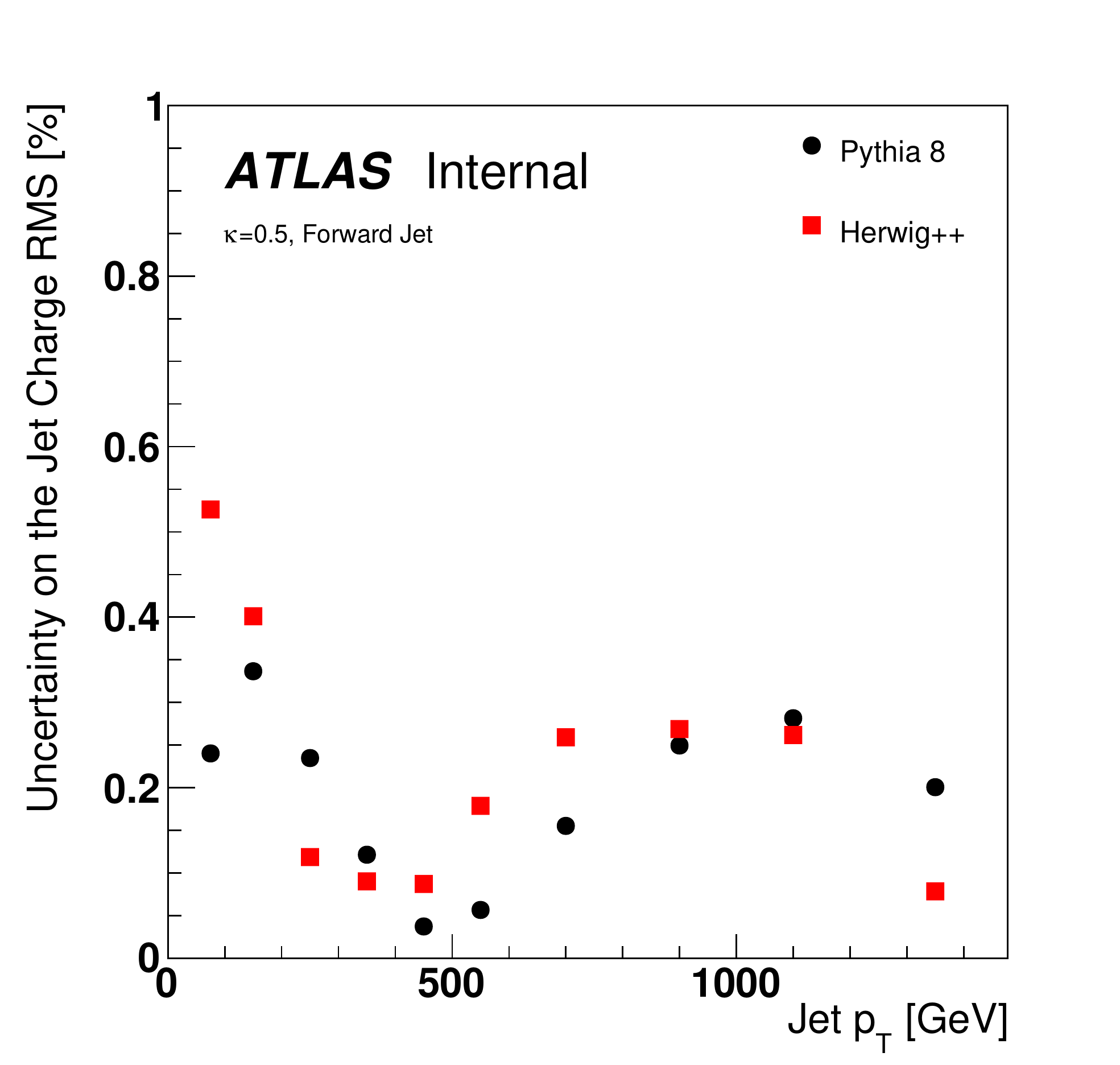}
\caption{The uncertainty on the jet charge distribution average (left) and standard deviation (right) due to the $n_\text{track}$ modeling.  These plots are for  the more forward jet and $\kappa=0.5$.}
\label{fig:trackmult:recomean_uncert}
\end{center}
\end{figure}

In principle, the uncertainty on the track multiplicity is actually part of a larger uncertainty on the full fragmentation.  The remainder of this section explores the impact of the track $p_\text{T}$ spectrum on the jet charge.  The method non-closure includes some aspects of the full fragmentation mis-modeling, but there may be additional sources of uncertainty from variations in the response matrix due to differences in the track $p_\text{T}$ spectrum.  The left plot of Fig.~\ref{fig:trackmult:trackmultpt} shows the track $p_\text{T}$ spectrum inside jets with $200$ GeV $<p_\text{T}<300$ GeV.  Similar to the track multiplicity, the track $p_\text{T}$ distribution from {\sc Pythia} and {\sc Herwig} bracket the data.  A re-weighting procedure analogous to the $n_\text{track}$ reweighting is used to assess the impact of the mis-modeling.  The right plot of Fig.~\ref{fig:trackmult:trackmultpt} shows the conditional distribution of the jet charge given the track $p_\text{T}$.  Every track contributes to the right plot of Fig.~\ref{fig:trackmult:trackmultpt} and since each jet has many tracks, each jet contributes many times.  The fork in the right plot of Fig.~\ref{fig:trackmult:trackmultpt} is due to events with one or a few tracks that carry a significant energy fraction and therefore the jet charge sign is set by the track charge.  Figure~\ref{fig:trackmult:recomean_uncertpt} shows the uncertainty due to the track $p_\text{T}$ after re-weighting to the data.  In all bins, the uncertainty is $<1\%$ and in most bins $\ll 1\%$; therefore it is ignored for the remainder of the analysis.

\begin{figure}[h!]
\begin{center}
\includegraphics[width=0.4\textwidth]{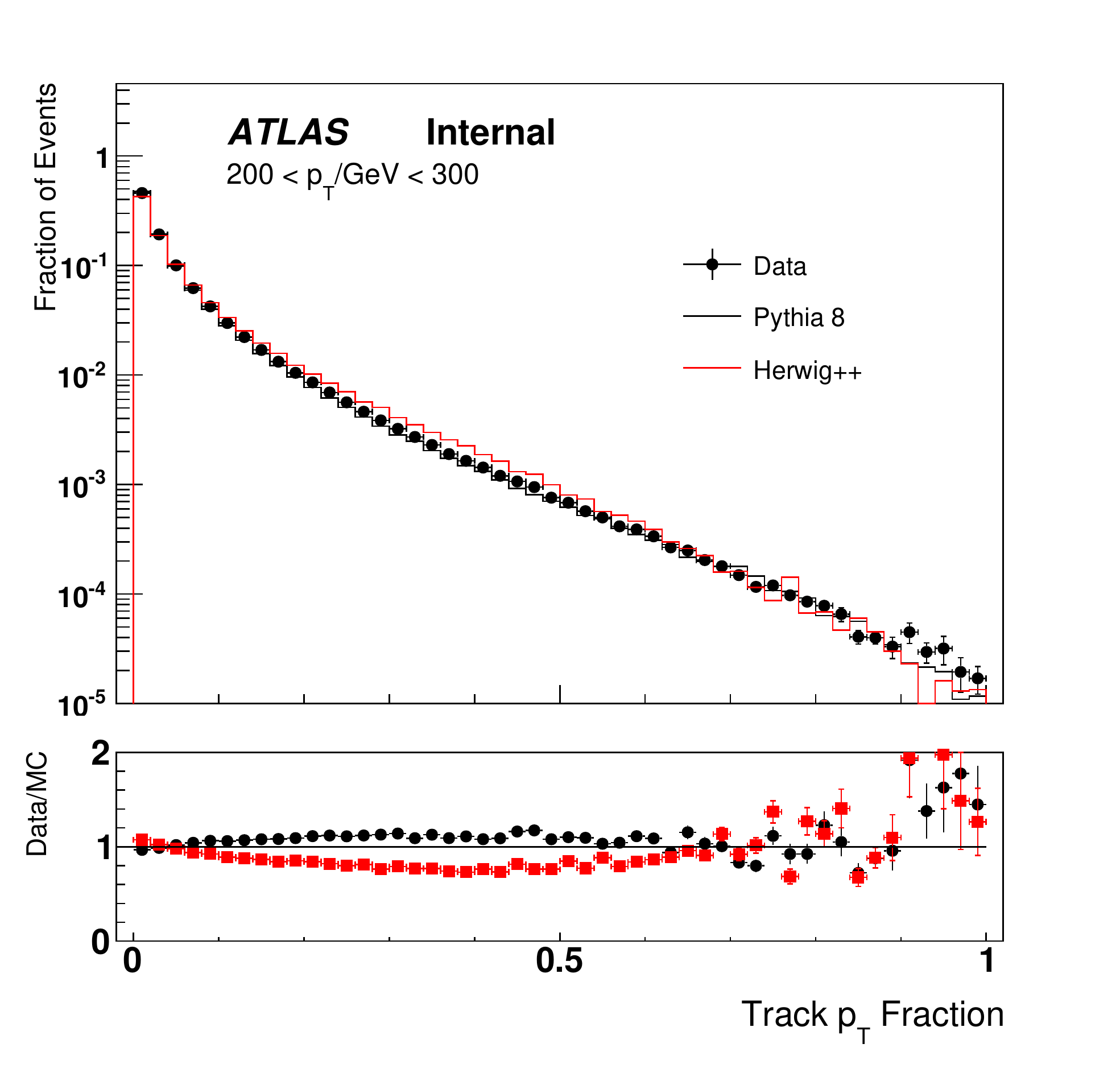}\includegraphics[width=0.4\textwidth]{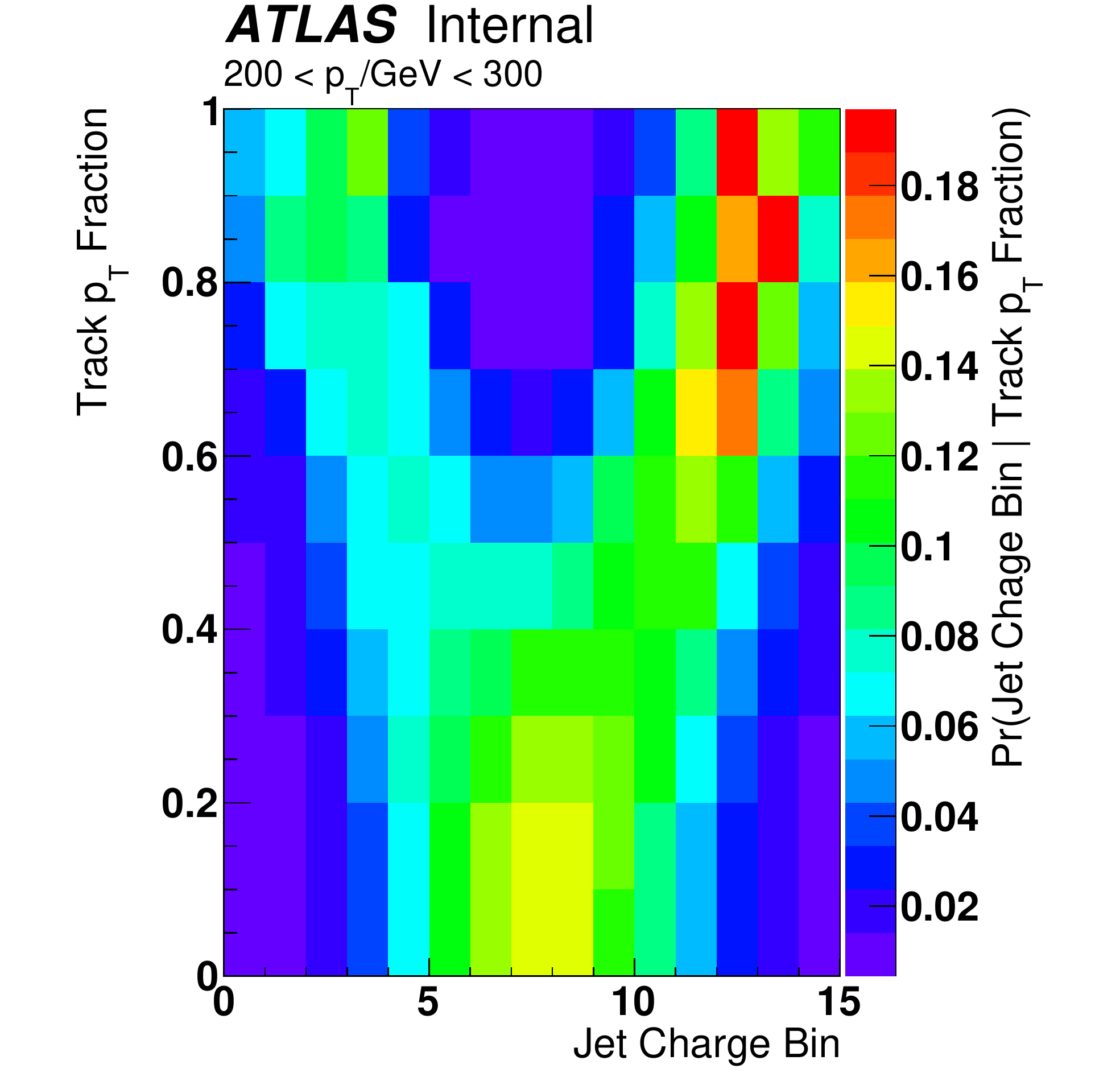}\\
\caption{The track $p_\text{T}$ spectrum in a particular jet $p_\text{T}$ bin (left) and the conditional distribution of the jet charge given the track $p_\text{T}$ (right); see the text for details.  This is for the more forward jet and $\kappa=0.5$.} 
\label{fig:trackmult:trackmultpt}
\end{center}
\end{figure}

\begin{figure}[h!]
\begin{center}
\includegraphics[width=0.4\textwidth]{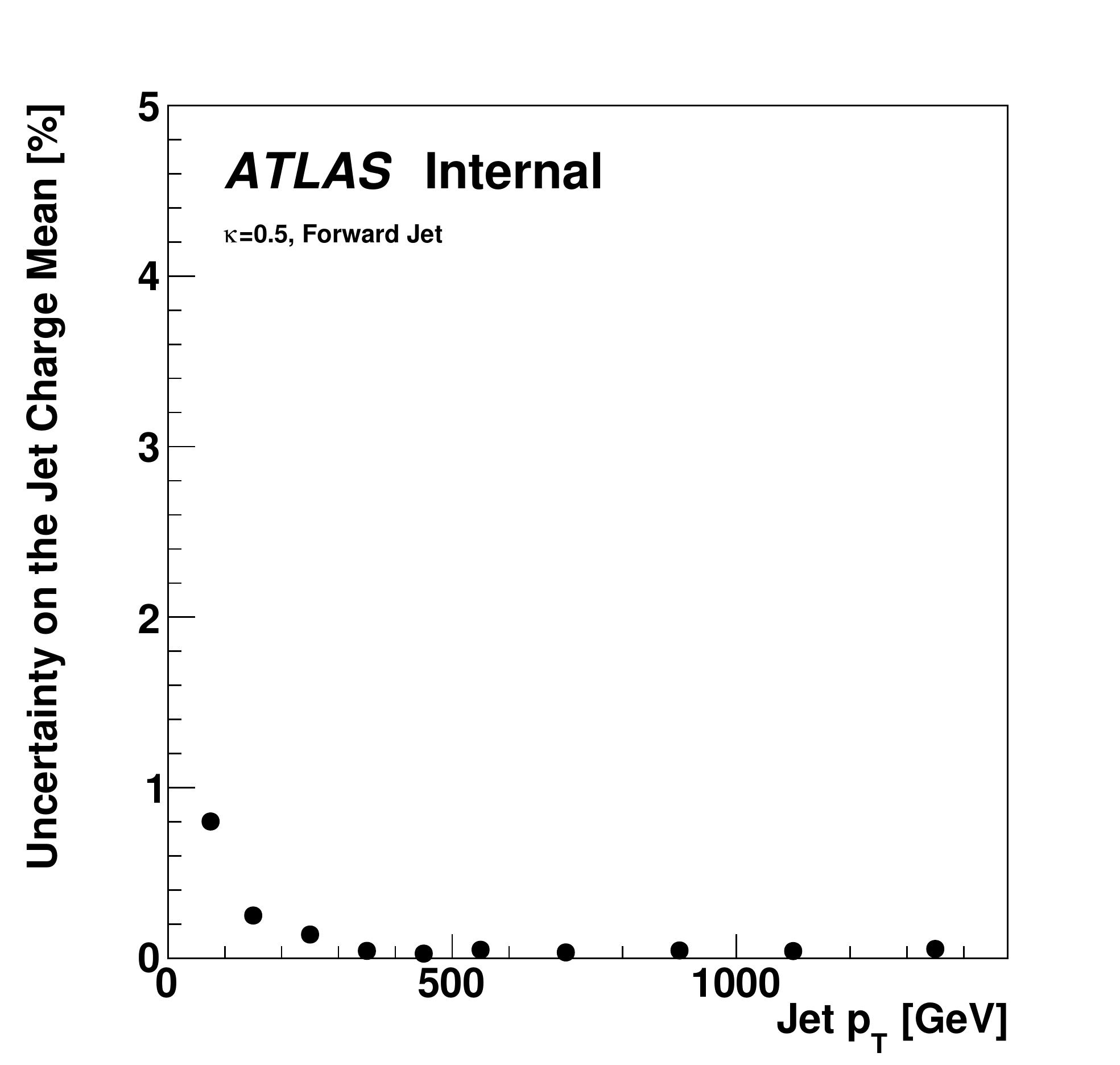}\includegraphics[width=0.4\textwidth]{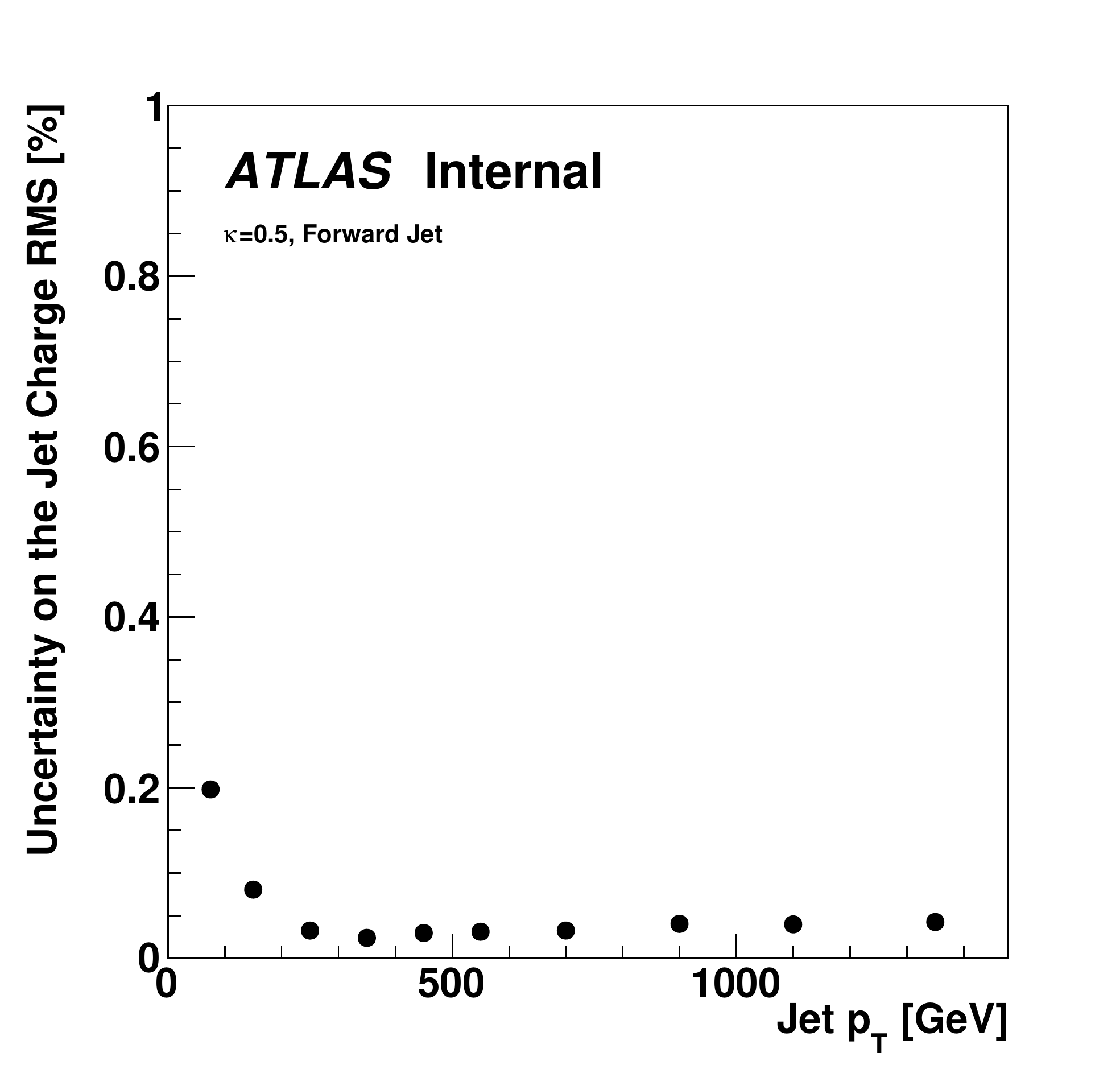}
\caption{The uncertainty due to the track $p_\text{T}$ on the jet charge distribution average (left) and standard deviation (right).}
\label{fig:trackmult:recomean_uncertpt}
\end{center}
\end{figure}

\clearpage

\subsection{Unfolding Non-closure}
\label{sec:statscharge}

A standard method~\cite{Malaescu:2009dm} for evaluating the systematic uncertainty from the procedure is to re-weight the MC to the data and take the difference between the unfolded re-weighted reconstructed MC to the truth MC of the same generator.  The re-weighted truth is a reasonable prior with which one can estimate the bias from the choice of prior in the unfolding method.  Heuristically, let $f(d,p,R)$ be a function that takes as inputs three histograms (data $d$, prior $p$, and the response matrix $R$) and outputs another histogram (the unfolding function).  By construction, $p=f(Rp,p,R)$.  Pick $t$ such that $Rt\sim d$.  Then, the non-closure uncertainty is the difference between $f(Rt,p,R)$ and $t$.  The non-closure is a method uncertainty and not an uncertainty on the prior per se, as the same $p$ is used for $t$ and $f(Rt,p,R)$.  The following is a more detailed and careful description of the non-closure uncertainty, taking note of the proper normalizations for the various histograms and matrices.

Define the following histograms; $x_i$ will interchangeably mean the histogram $x$ and also the content in the $i^{th}$ bin of $x$:

\begin{description}
\item[$d_i$]: The measured spectrum.  There are 150 total bins (10 $p_\text{T}$ bins and 15 jet charge bins) so $i=1,...,150$.
\item[$R_{ij}$]: The unnormalized response matrix; $R_{ij}$ is the number of events in the simulation that fall in the reconstructed bin $i$ and the truth bin $j$.
\item[$t_i$]: $t_i = \sum_j R_{ji}$: the particle-level spectrum for events that pass both particle- and detector-level selections.
\item[$r_i$]: $r_i = \sum_j R_{ij}$: the detector-level spectrum for events that pass both particle- and detector-level selections
\item[$\tilde{R}_{ij}$]: The normalized version of $R_{ij}$ (earlier, this was just called {\it the} response matrix): $r_i = \sum_j \tilde{R}_{ij} t_j$.  Explicitly, $\tilde{R}_{ij} = R_{ij}/\sum_{i'} R_{i'j}$.  The entries of $\tilde{R}_{ij}$ are the conditional probability for a truth event in bin $j$ to be reconstructed in bin $i$.
\end{description}

\noindent The re-weighting procedure can only be applied to simulation events which pass both the particle-level and detector-level event selections and so the first step is to take the data and apply the fake factors bin-by-bin: 

\begin{align}
(1-f_i)=\frac{\text{Pass both reconstructed and truth selections}}{\text{Pass the reconstructed selection}},
\end{align}

\noindent where $i$ is the bin number.   Define $\tilde{d}_i = (1-f_i) d_i$ to be the corrected data histogram.  A reasonable prior $\tilde{t}_i$ is one such that $\tilde{R}_{ij} \tilde{t}_j$ is very close to $\tilde{d}_i$.  Since $\tilde{R}_{ij}$ is not too far from a diagonal matrix, one way of generating (an approximate) $\tilde{t}_i$ is to use weights built from the reconstructed simulation: $w_i = \tilde{d}_i/r_i$.  Define $\tilde{t}_i=w_it_i$.  The left plot of Fig.~\ref{fig:systs_nc_1} shows the distributions of $w_i$.  In order to reduce the sensitivity to statistical fluctuations in the data in generating the weights $w_i$, the histogram of weights is smoothed before generating $\tilde{t}_i$.  A standard median smoothing procedure implemented in ROOT with 20 iterations~\cite{Friedman:695770} is used for this purpose.  There is a clear low-frequency trend in the weight histogram that increases monotonically with the bin number and corresponds to the $p_\text{T}$ spectrum, while the high-frequency trends reflect the fact that the width of the charge distribution in each $p_\text{T}$ bin changes.  The right plot of Fig.~\ref{fig:systs_nc_1} shows that the weights $w_i$ are effective at improving the data/MC agreement of $\tilde{r}_i=\sum_j\tilde{R}_{ij}\tilde{t}_j$ with respect to $r_i$.  In general, the trend is that the data/MC is greatly improved in all but the highest $p_\text{T}$ bins, where the data/MC was already very good to begin with.  Figure~\ref{fig:systs_nc_2} shows the actual non-closure uncertainty for the jet charge and the standard deviation of the jet charge distribution, compared to the raw data/MC differences in the reconstructed version of these quantities.  Except in the first two bins where fractional uncertainties have little meaning due to the small value of the jet charge compared with the uncertainty, the non-closure uncertainty for the jet charge mean is significantly smaller than the raw difference between the data and simulation.  This is also mostly true for the jet charge distribution standard deviation, but the raw differences are already much smaller.  As a comparison, the impact of unfolding the {\sc Pythia} simulation with a {\sc Herwig++} response matrix is shown in Fig.~\ref{fig:systs_hadro}.  The size of the differences shown in Fig.~\ref{fig:systs_hadro} are approximately compatible with those in Fig.~\ref{fig:systs_nc_2}.

\begin{figure}[h!]
\begin{center}
\includegraphics[width=0.45\textwidth]{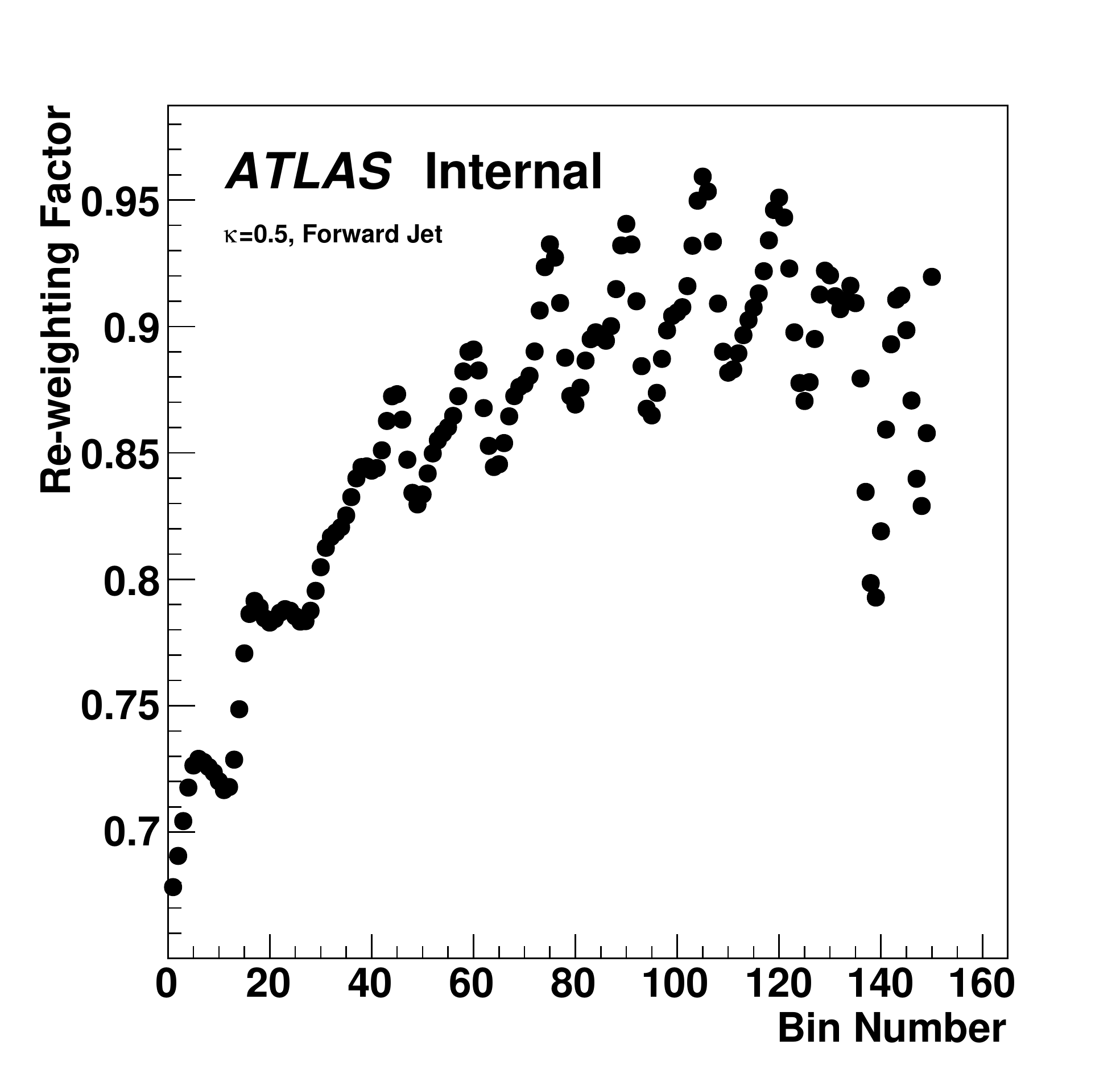}\includegraphics[width=0.45\textwidth]{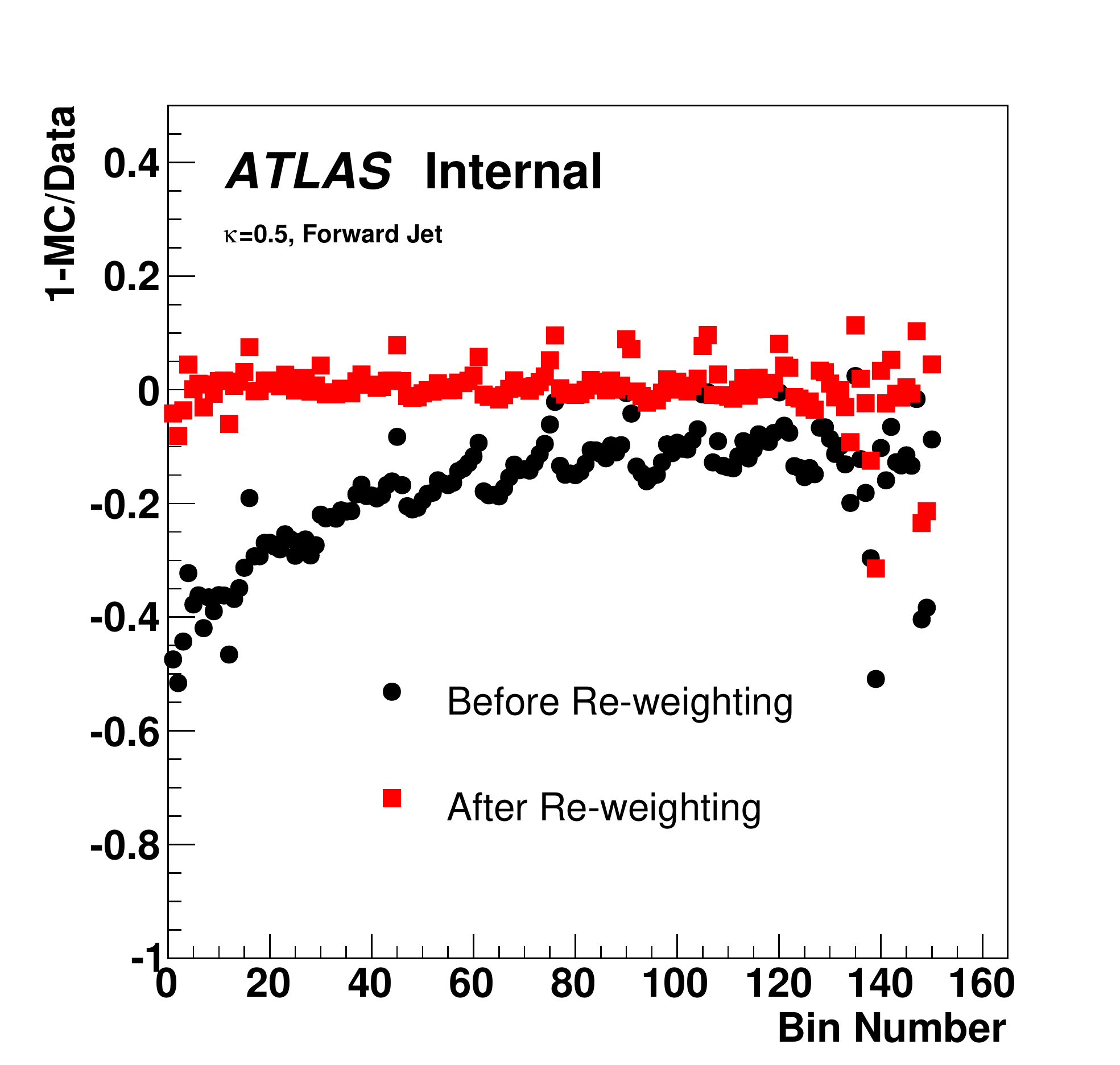}
\caption{The distribution of the weights $w_i$ used to re-weight the MC distribution for the non-closure test (left) and the Data/MC ratio with the re-weighted truth distribution (labeled after) $\tilde{t}_i$ (right) for the more forward jet with $\kappa=0.5$.}
\label{fig:systs_nc_1}
\end{center}
\end{figure}

\begin{figure}[h!]
\begin{center}
\includegraphics[width=0.45\textwidth]{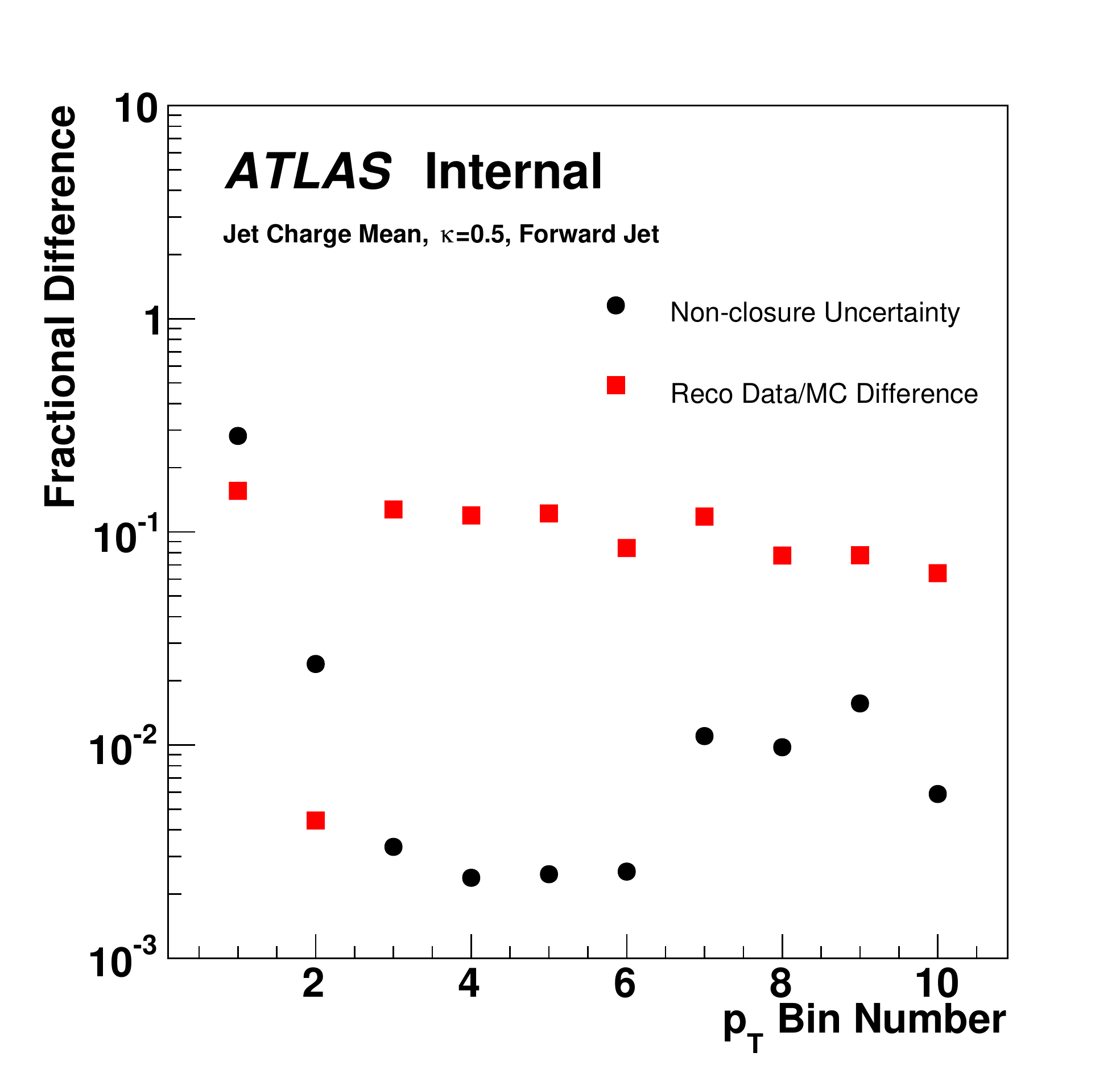}\includegraphics[width=0.45\textwidth]{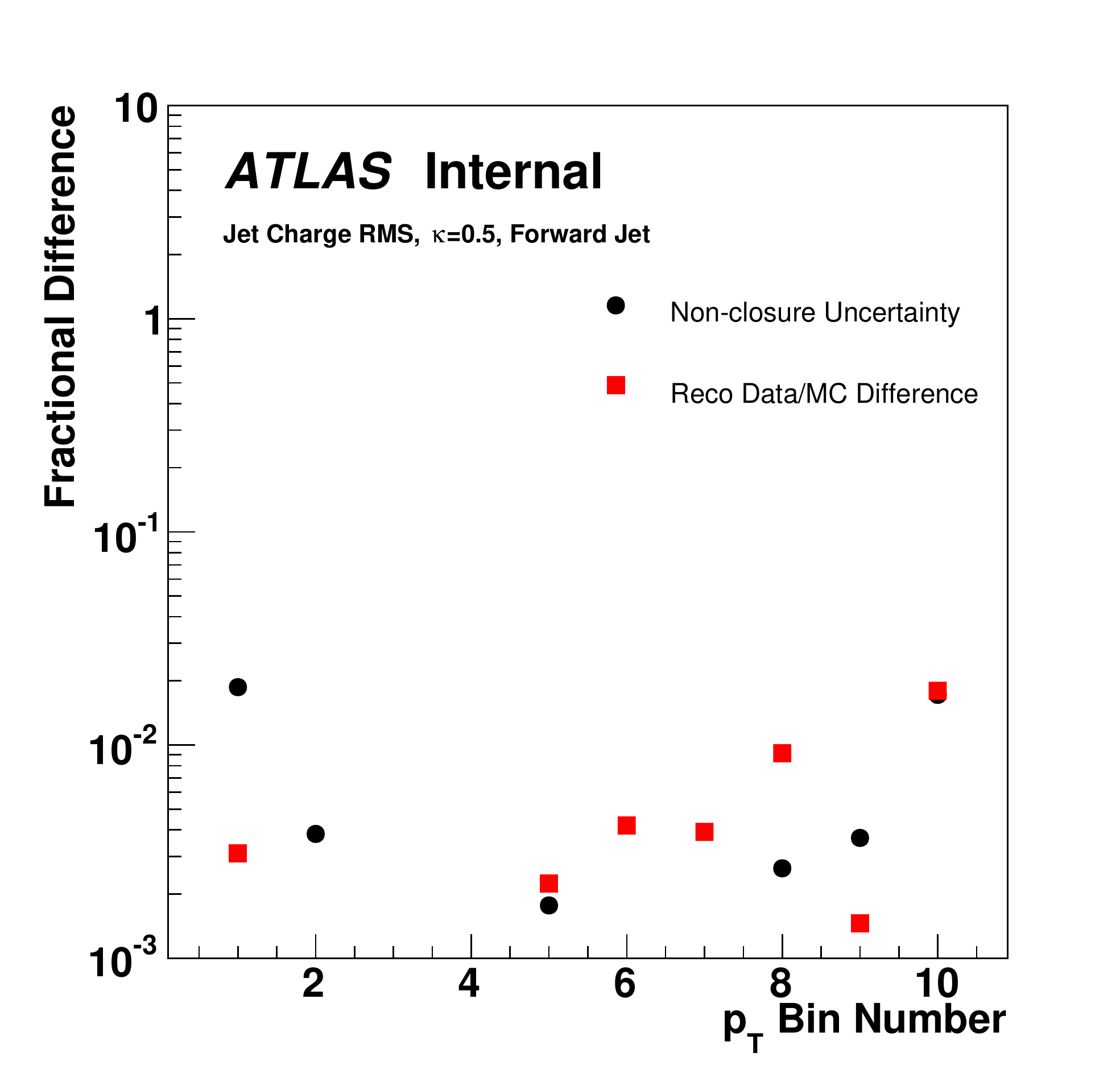}
\caption{The impact of the non-closure uncertainty on the jet charge mean (left) and the jet charge distribution standard deviation (right) for the more forward jet and $\kappa=0.5$.  Bins without a black and red point indicate that one of the two is smaller than $10^{-3}$.}
\label{fig:systs_nc_2}
\end{center}
\end{figure}

\begin{figure}[h!]
\begin{center}
\includegraphics[width=0.45\textwidth]{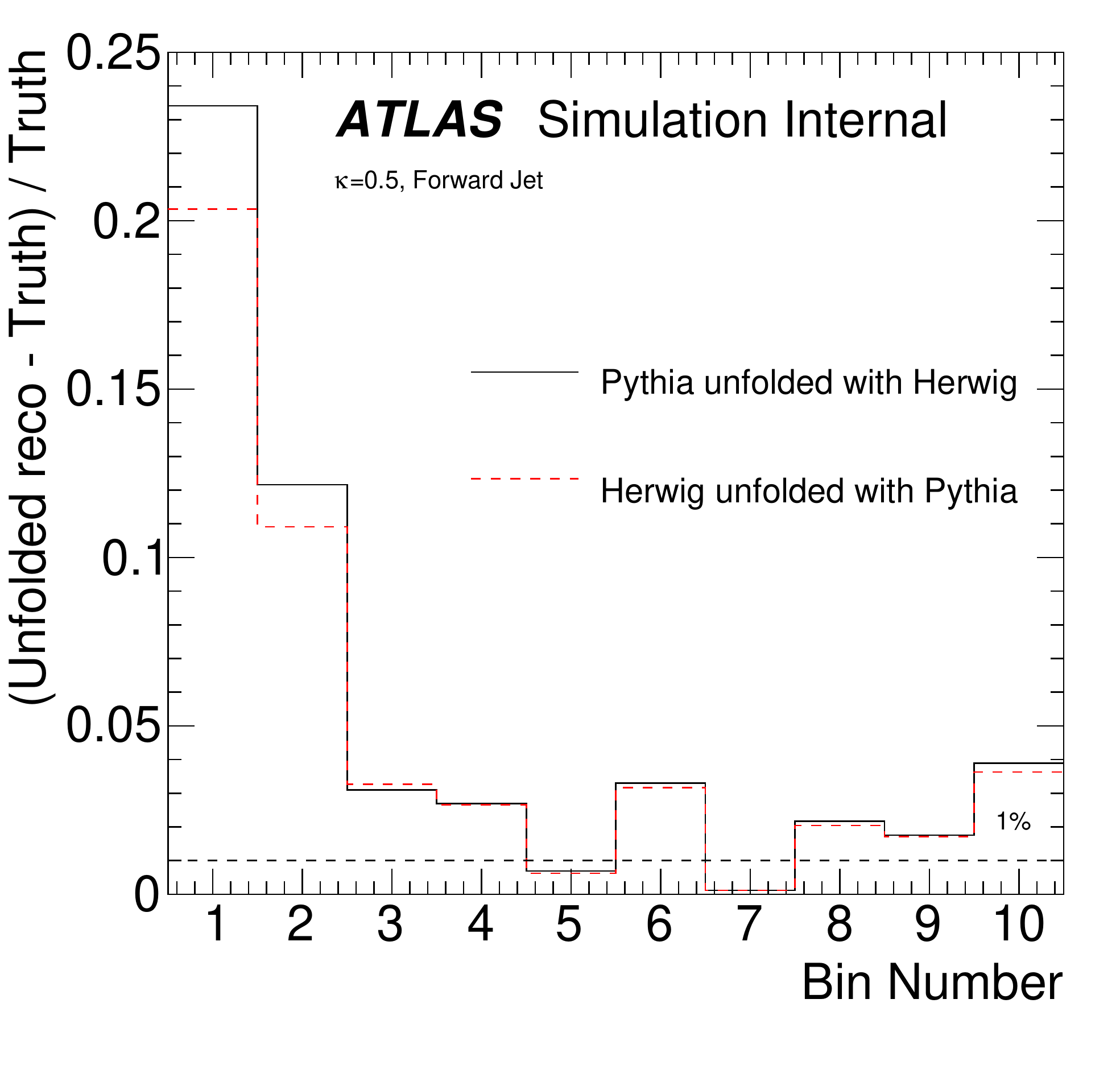}\includegraphics[width=0.45\textwidth]{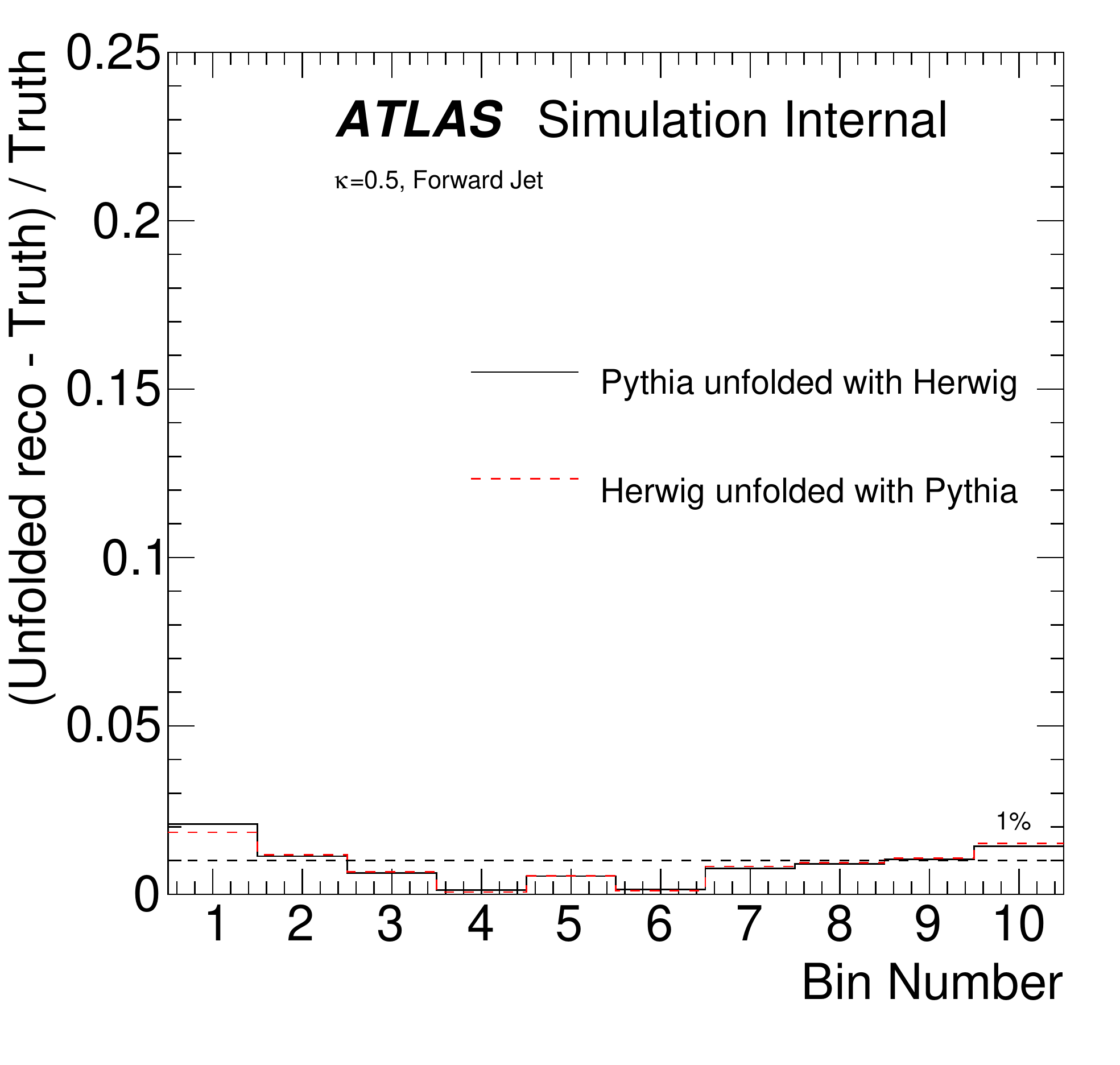}
\caption{The fractional difference on the jet charge mean (left) and standard deviation (right) when unfolding the {\sc Pythia} simulation with a {\sc Herwig++} response matrix and vice versa. }
\label{fig:systs_hadro}
\end{center}
\end{figure}

\clearpage

\section{Results}
\label{sec:results}

The data satisfying the event selection criteria described in Sec.~\ref{sec:objects} are unfolded according to the procedure in Sec.~\ref{sec:unfolding} and the average and standard deviation of the jet charge distribution are computed as a function of the jet $p_\text{T}$.  These results, along with the systematic uncertainties detailed in Sec.~\ref{sec:uncerts}, are discussed in Sec.~\ref{sec:rawunfolded}.  The PDF uncertainty and jet formation uncertainties in the theory predictions are compared to the unfolded data in Secs.~\ref{sec:PDFsensitivity} and~\ref{sec:NPmodeling}, respectively.  Using PDF information as input, the average charge per jet flavor is extracted in Sec.~\ref{sec:updownextract} and its $p_\text{T}$-dependence is studied in Sec.~\ref{sec:scaleviolation}.

\subsection{Unfolded Jet Charge Spectrum}
\label{sec:rawunfolded}

The unfolded jet charge mean is shown as a function of the jet $p_\text{T}$ in the top plots of Fig.~\ref{fig:mean} for $\kappa=0.3$, $0.5$ and $0.7$.  The average charge increases with jet $p_\text{T}$ due to the increase in up-flavor jets from PDF effects.  The average charge increases from $0.01e$ at $p_\text{T}\sim 100$ GeV~to $0.15e$ at $p_\text{T}\sim 1.5$ TeV.  Systematic uncertainties are generally a few percent, except at low jet $p_\text{T}$ where the fractional uncertainty is large because the average jet charge in the denominator is small, and at high $p_\text{T}$ where the tracking uncertainties are not negligible.  The first bin suffers from large statistical uncertainties (up to 170\%), but for the higher $p_\text{T}$ bins the systematic uncertainty is dominant, except at the highest $p_\text{T}$ bin where statistical and systematic uncertainty are of similar size (about 7\%).  The jet charge distributions of the more forward and more central jet differ in shape, in particular at low $p_\text{T}$, due to the different shape of the up/down flavor fractions in those bins as shown in Fig.~\ref{fig:flavorfrac}(b).

Analogous results for the standard deviation of the jet charge distribution are shown in the bottom plots of Fig.~\ref{fig:mean}.   Even though the standard deviation of the reconstructed jet charge distribution increases with jet $p_\text{T}$ (Fig.~\ref{fig:raw}), the particle-level value decreases and approaches an asymptote for $p_\text{T}\gtrsim 300$ GeV.

\begin{figure}[h!]
\begin{center}
\includegraphics[width=0.45\textwidth]{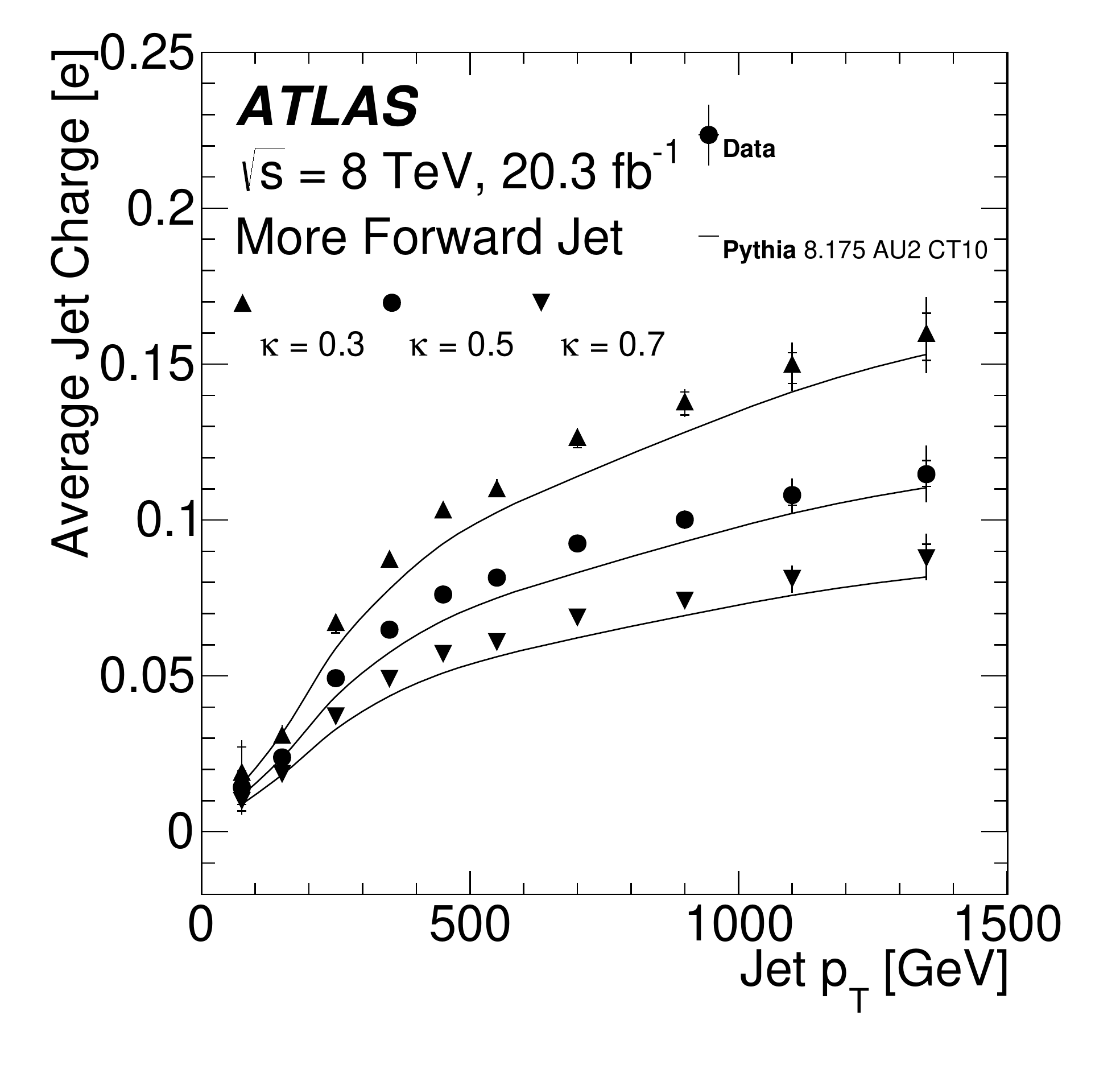}
\includegraphics[width=0.45\textwidth]{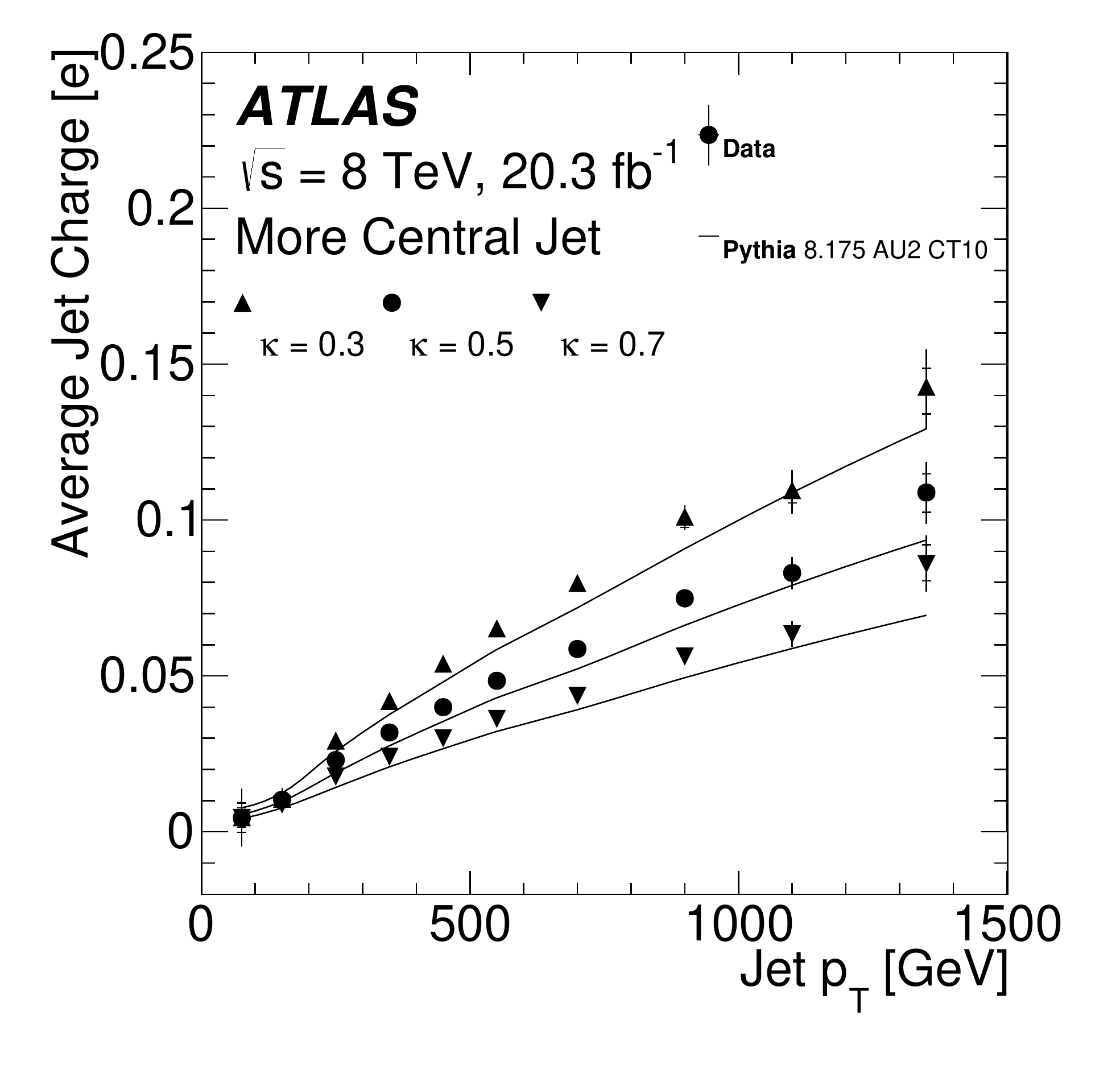}\\
\includegraphics[width=0.45\textwidth]{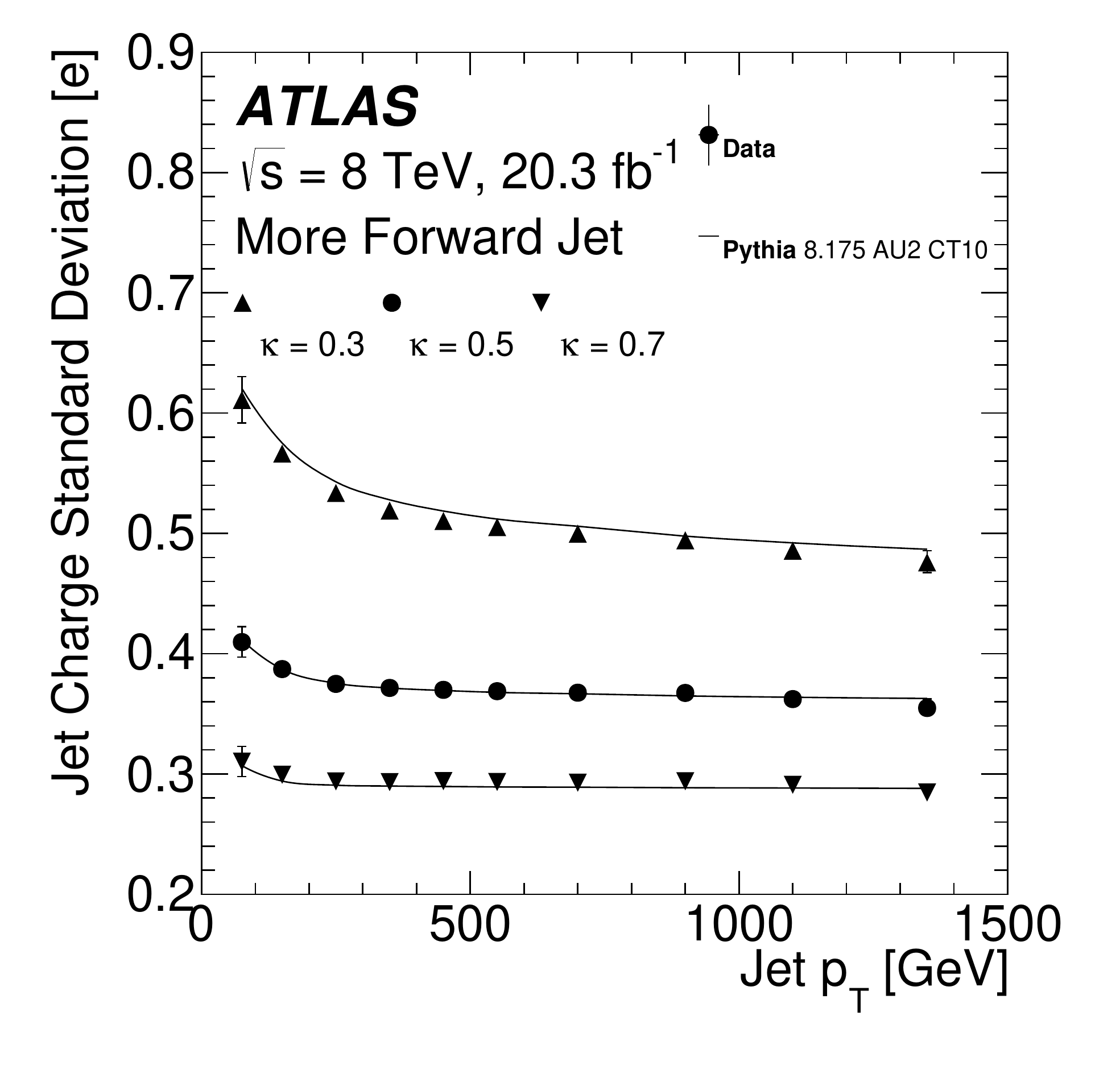}
\includegraphics[width=0.45\textwidth]{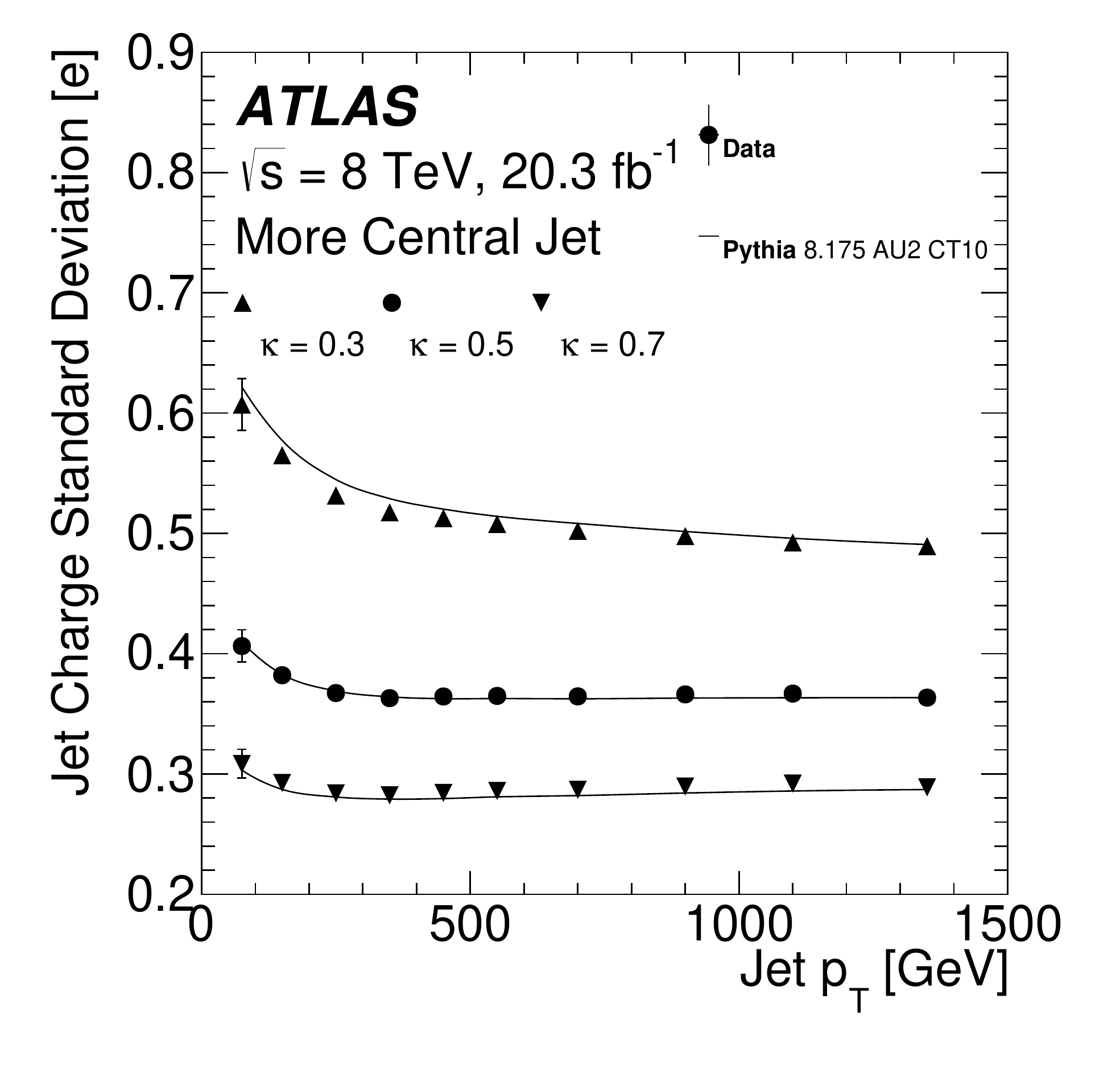}
\end{center}	
\caption{The measured average (standard deviation) of the jet charge distribution on top (bottom) in units of the positron charge as a function of the jet $p_\text{T}$ for $\kappa=0.3, 0.5,$ and $0.7$ for the more forward jet (left) and the more central jet (right).   The crossed lines in the bars on the data indicate the systematic uncertainty and the full extent of the bars is the sum in quadrature of the statistical and systematic uncertainties.  The solid continuous line is a smooth approximation to the {\sc Pythia} prediction.}
\label{fig:mean}
\end{figure}

\clearpage

\subsection{Sensitivity of PDF Modeling}
\label{sec:PDFsensitivity}

Variations in the PDF set impact the relative flavor fractions and thus in turn change the jet charge distribution.  Such changes do not vary much with $\kappa$, since the PDF impacts the jet charge distribution mostly through the flavor fractions.  Figures~\ref{fig:PDFnom} and~\ref{fig:PDFnomrms} compare the unfolded distributions of the jet charge distribution's average and standard deviation with several PDF sets, with tuned predictions for {\sc Pythia} for each PDF, and with the same AU2 family of tunes.   The sampling of PDF sets results in a significant spread for the average jet charge, but has almost no effect on the standard deviation. CTEQ6L1 describes the data best, although the data/MC ratio has a stronger $p_\text{T}$ dependence. In particular, the data/MC differences with CTEQ6L1 are up to 10\% (15\%) at moderate $p_\text{T}$ for the more forward (central) jet.  For high $p_\text{T}$, differences between data and simulation are less significant.  NLO PDFs such as CT10 are consistently below the data by about 10\%-15\%.

\begin{figure}[h!]
\begin{center}
{\includegraphics[width=0.5\textwidth]{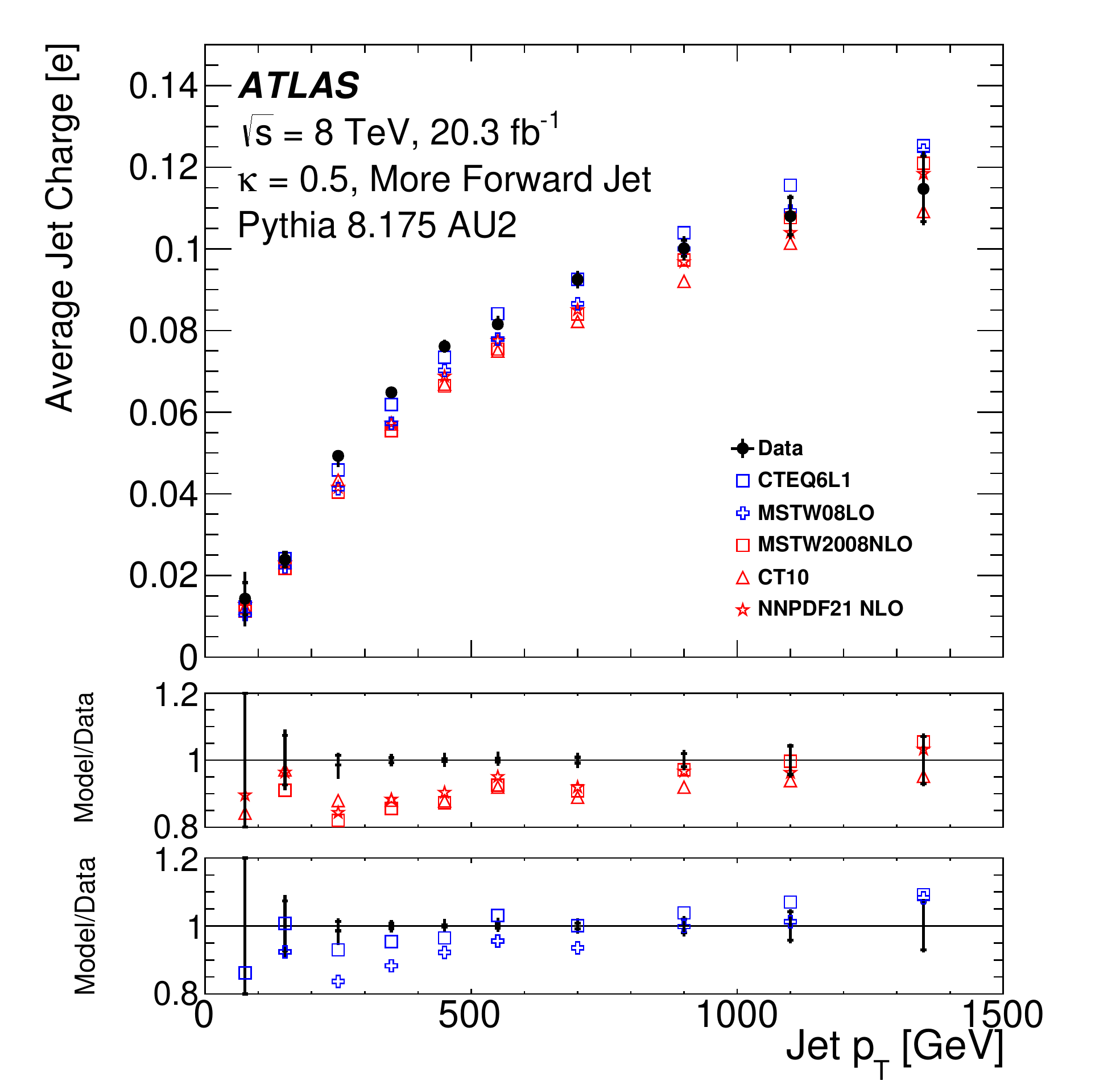}}{\includegraphics[width=0.5\textwidth]{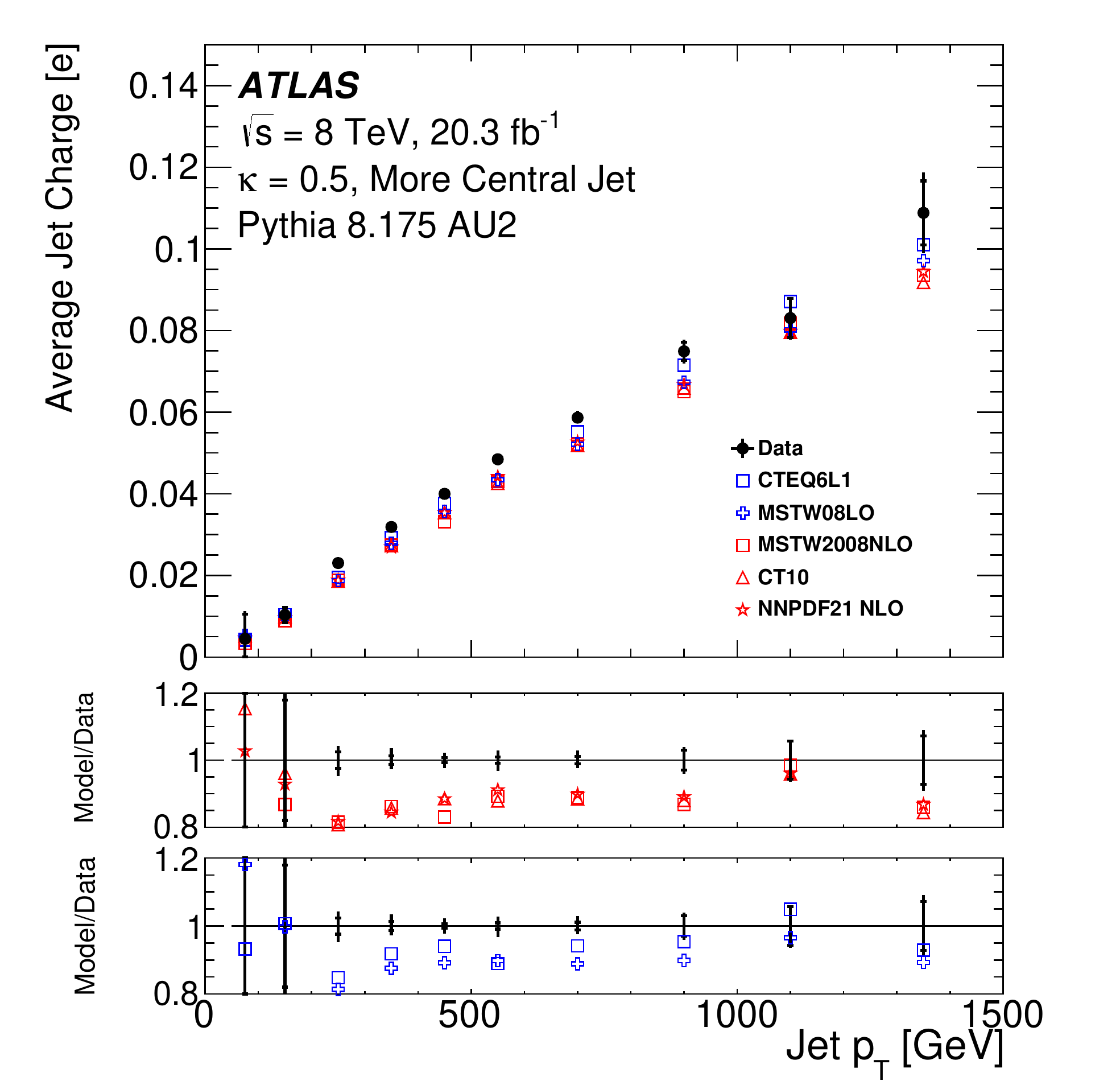}}
\end{center}	
\caption{The average jet charge ($\kappa=0.5$) in units of the positron charge for (a) the more forward jet and (b) the more central jet compared with theory predictions due to various PDF sets. The crossed lines in the bars on the data indicate the statistical uncertainty and the full extent of the bars is the sum in quadrature of the statistical and systematic uncertainties.}
\label{fig:PDFnom}
\end{figure}

\begin{figure}[h!]
\begin{center}
{\includegraphics[width=0.5\textwidth]{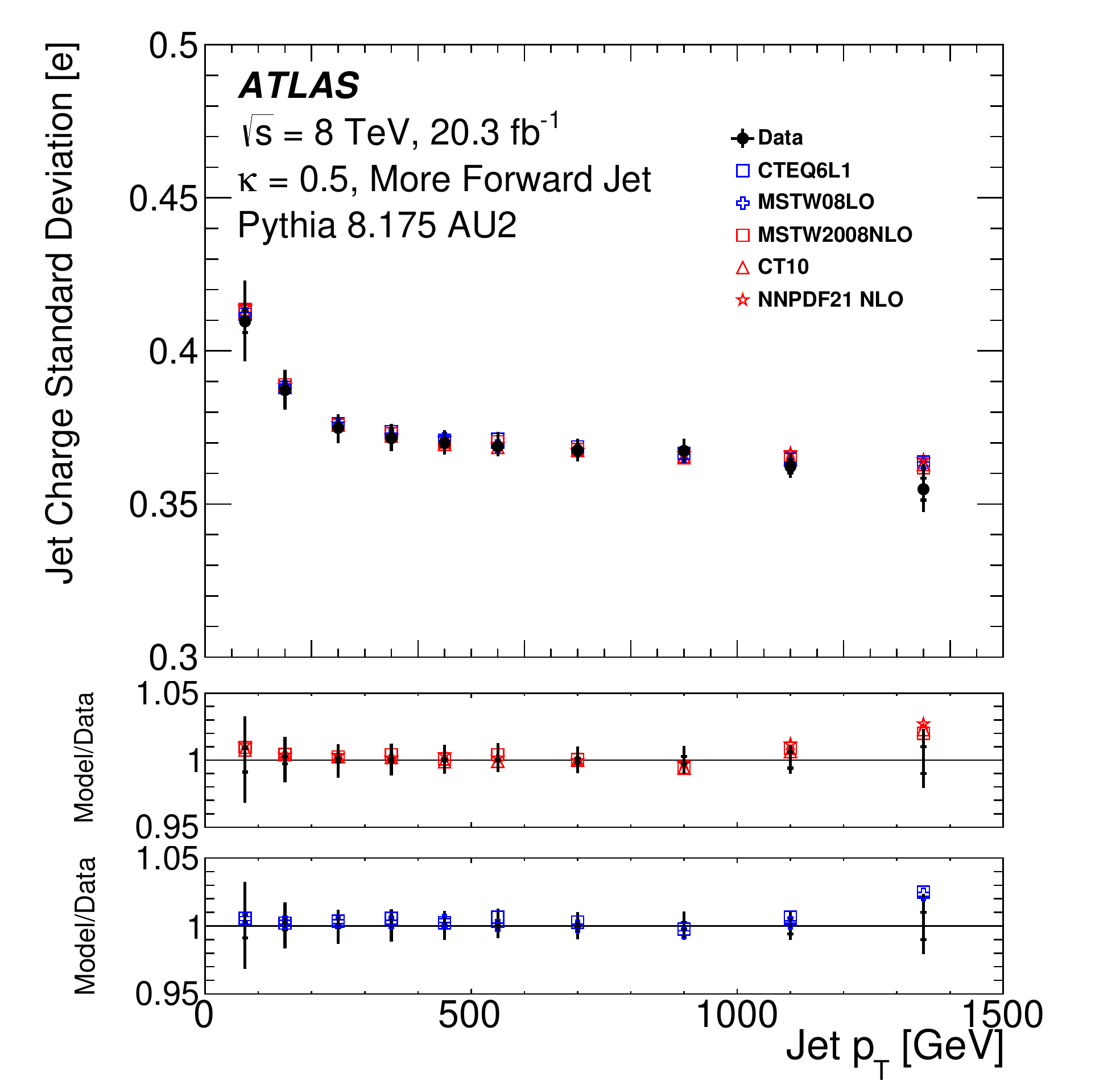}}{\includegraphics[width=0.5\textwidth]{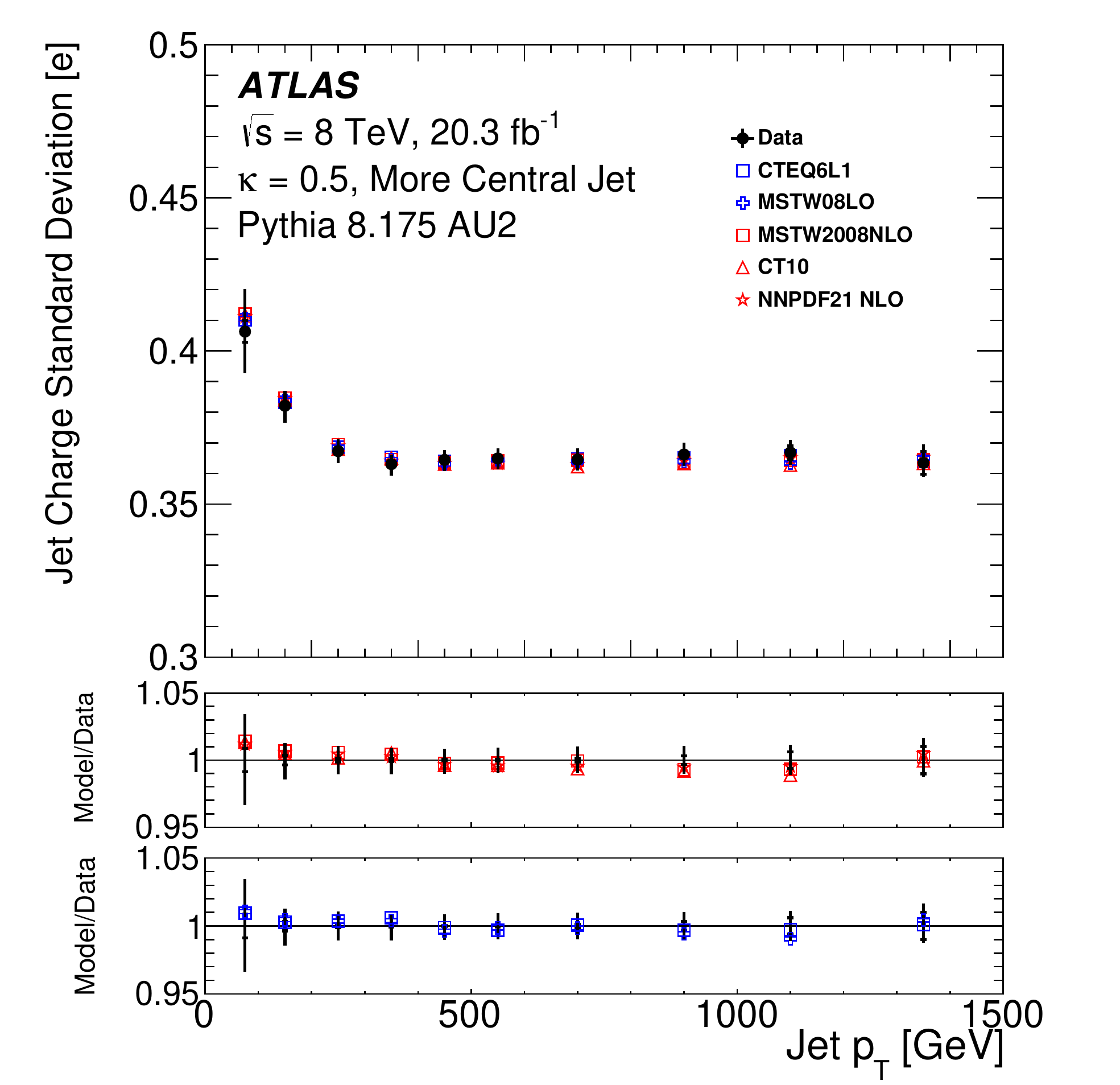}}
\end{center}	
\caption{The standard deviation of the jet charge ($\kappa=0.5$) distribution in units of the positron charge for (a) the more forward jet and (b) the more central jet compared with theory predictions due to various PDF sets. The crossed lines in the bars on the data indicate the statistical uncertainty and the full extent of the bars is the sum in quadrature of the statistical and systematic uncertainties. }
\label{fig:PDFnomrms}
\end{figure}

\subsection{Sensitivity of QCD Models and Tunes}
\label{sec:NPmodeling}

The measurements presented in Sec.~\ref{sec:rawunfolded} show that there are qualitative differences between the data and the MC simulations, and comparisons in Sec.~\ref{sec:PDFsensitivity} suggest that variations in the PDF set cannot fully explain the differences. Differences in Sec.~\ref{sec:rawunfolded} between {\sc Pythia} and {\sc Herwig++} suggest that some aspect of the modeling of fragmentation could lead to the observed differences between the simulation and the data.  One possible source is the hadronization modeling, which differs between {\sc Pythia} (Lund-string fragmentation) and {\sc Herwig++} (cluster fragmentation).  The modeling of final-state radiation (FSR) is expected to have an impact on the jet charge distribution because variations in the radiation lead to different energy flow around the initial parton and hence different fragmentation of the jet.  The plots in Fig.~\ref{fig:nprms} and Fig.~\ref{fig:nprm2} show the measured average jet charge and the jet charge distribution's standard deviation, respectively, for $\kappa=0.3,$ 0.5, and 0.7, compared to various models for a fixed PDF set (CTEQ6L1).   In addition to {\sc Pythia} 8 and {\sc Herwig++} model predictions, Figs.~\ref{fig:nprms} and~\ref{fig:nprm2} contain the predictions from {\sc Pythia} 6 using the Perugia 2012 tune~\cite{Skands:2010ak} and the radHi and radLo Perugia 2012 tune variations.  These Perugia tune variations test the sensitivity to higher/lower amounts of initial- and final-state radiation (via the scaling of $\alpha_\text{s}$), although only variations of the FSR are important for the jet charge distribution. For the mean jet charge, {\sc Pythia} 6 with the P2012 radLo tune is very similar to {\sc Pythia} 8 with the AU2 tune.  The spread in the average jet charge due to the difference between the radHi and radLo tunes increases with $\kappa$, since suppression of soft radiation makes the jet charge distribution more sensitive to the modeling of the energy fraction of the leading emissions.  For the jet charge distribution's standard deviation, the sensitivity to the $\alpha_\text{s}$ scaling is large at both high and low $\kappa$.  However, the sensitivity is inverted: radHi gives a larger standard deviation for $\kappa=0.3$, but a lower standard deviation for $\kappa=0.7$.  Other Perugia 2012 tunes have been studied, testing the sensitivity to color-reconnection and multiple parton interactions, but the differences in the jet charge distribution's mean and standard deviation are small.  The Perugia 2012 tunes may not fully capture the spread in nonperturbative effects, which is also suggested by the increasing difference between {\sc Pythia} 8 and {\sc Herwig++} for decreasing $\kappa$.

\begin{figure}[h!]
\begin{center}
{\includegraphics[width=0.5\textwidth]{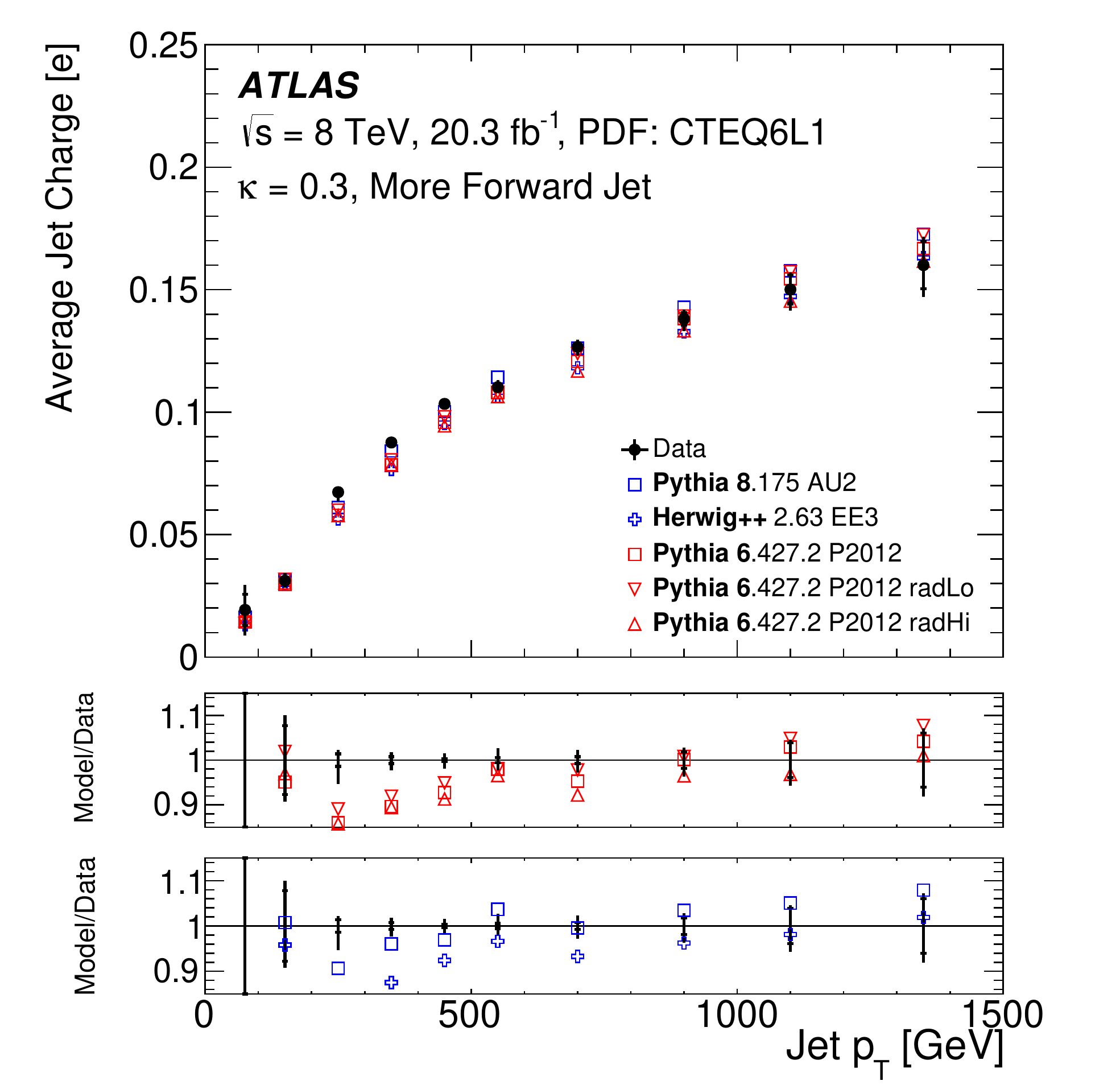}}{\includegraphics[width=0.5\textwidth]{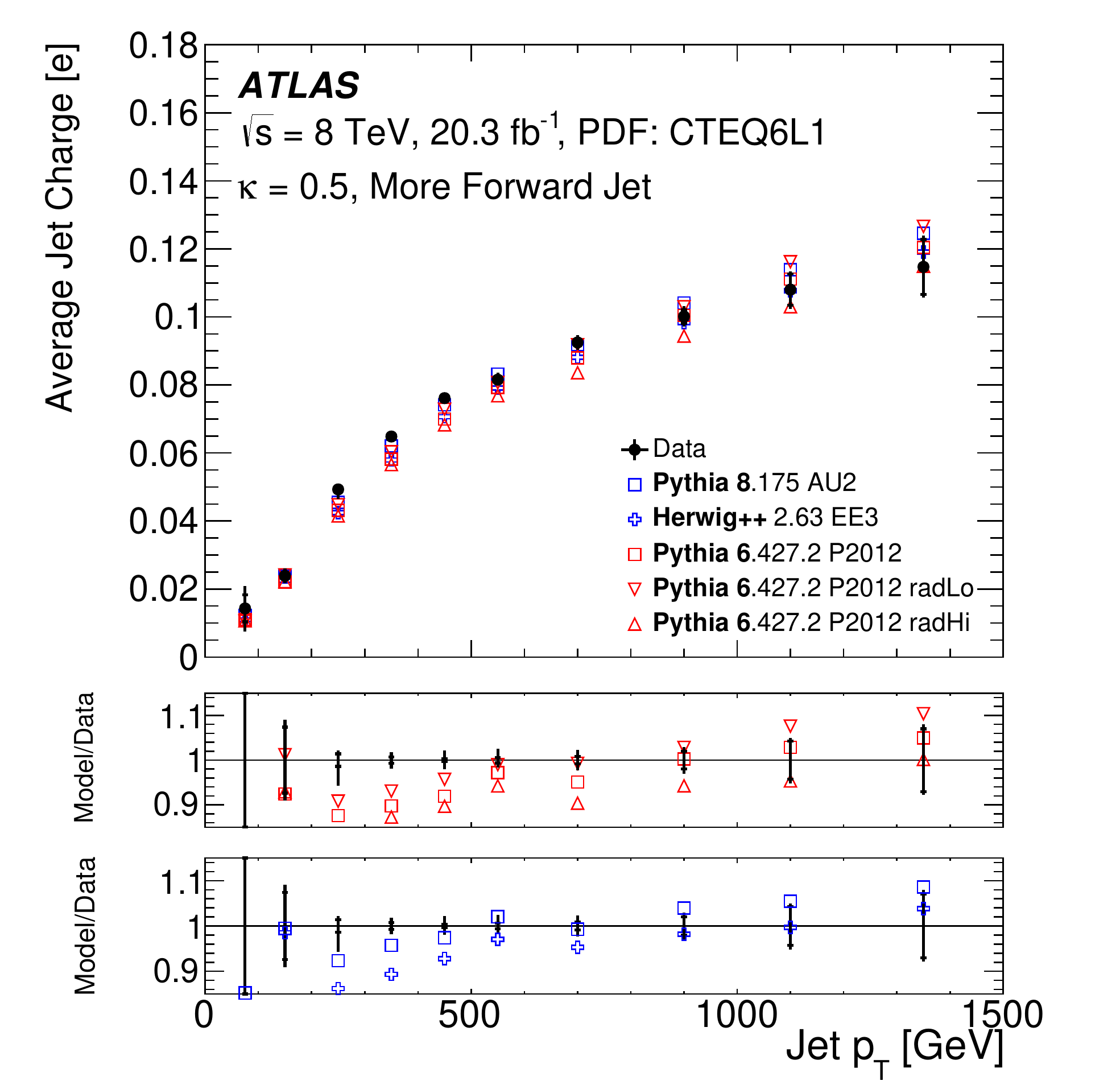}}\\
{\includegraphics[width=0.5\textwidth]{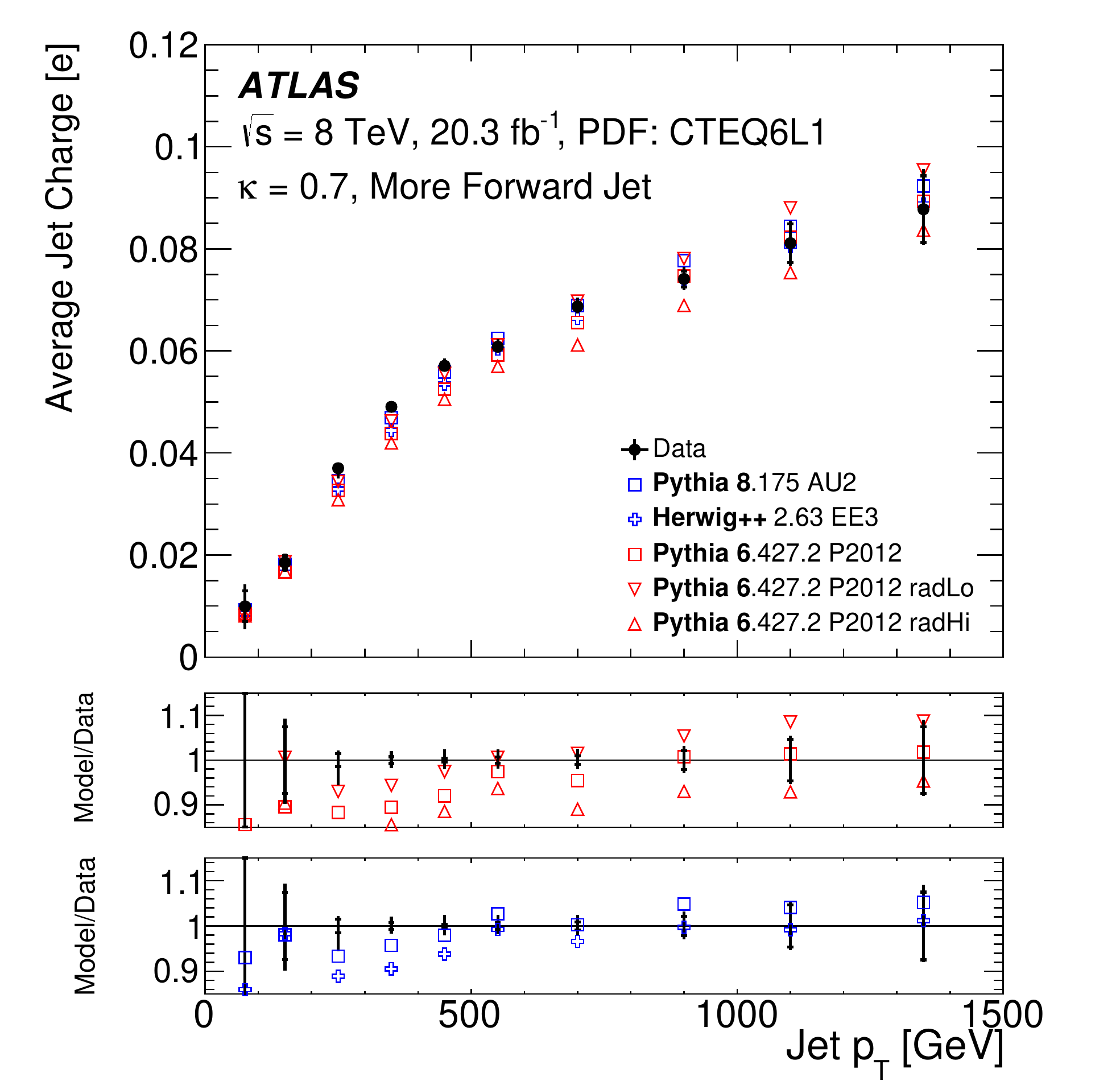}}
\end{center}	
\caption{The average of the jet charge distribution in units of the positron charge for (a) $\kappa=0.3$, (b) 0.5, and (c) 0.7 comparing various QCD MC models and tunes for the more forward jet.  The crossed lines in the bars on the data indicate the statistical uncertainty and the full extent of the bars is the sum in quadrature of the statistical and systematic uncertainties.}
\label{fig:nprms}
\end{figure}

\begin{figure}[h!]
\begin{center}
{\includegraphics[width=0.5\textwidth]{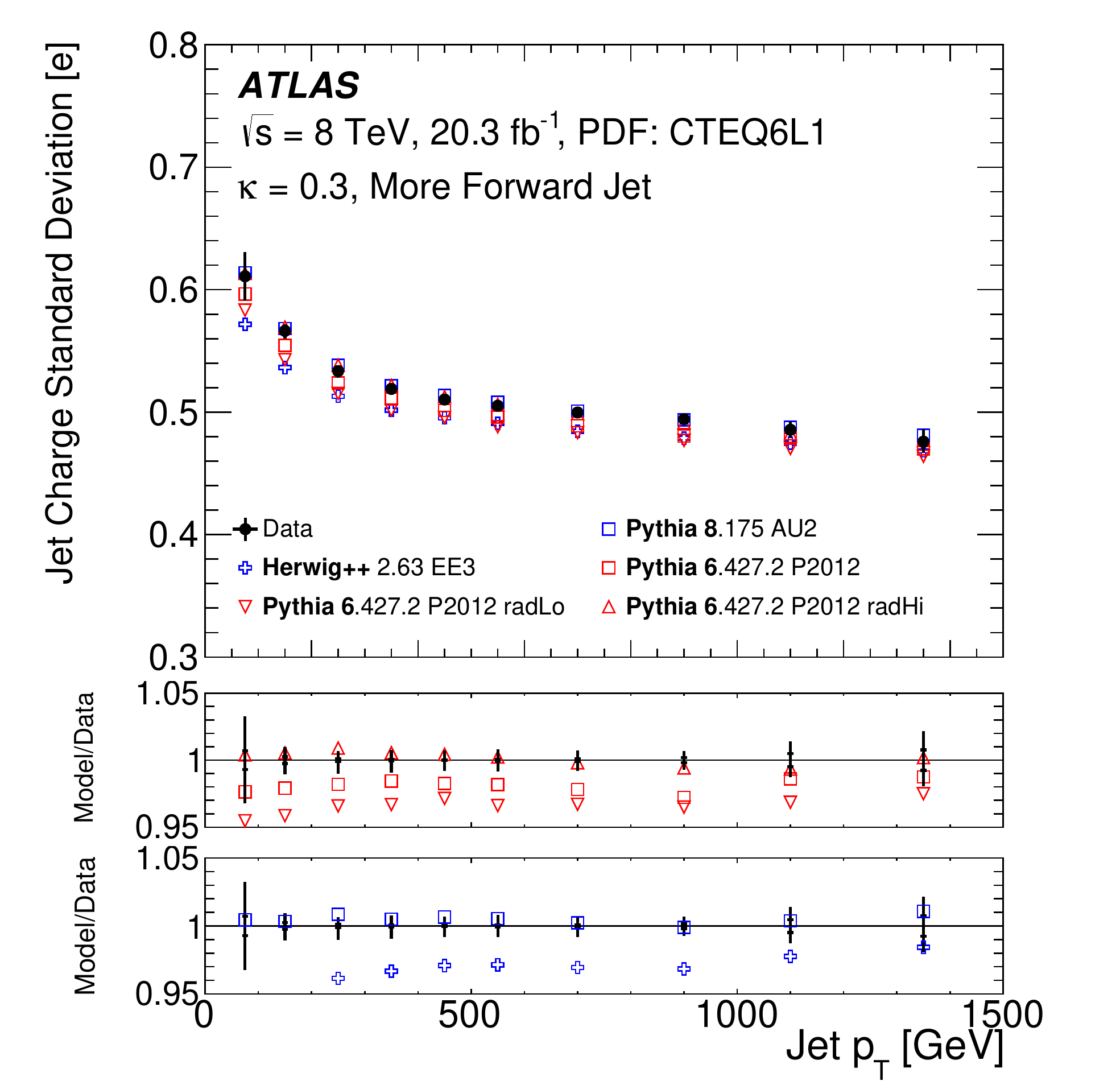}}{\includegraphics[width=0.5\textwidth]{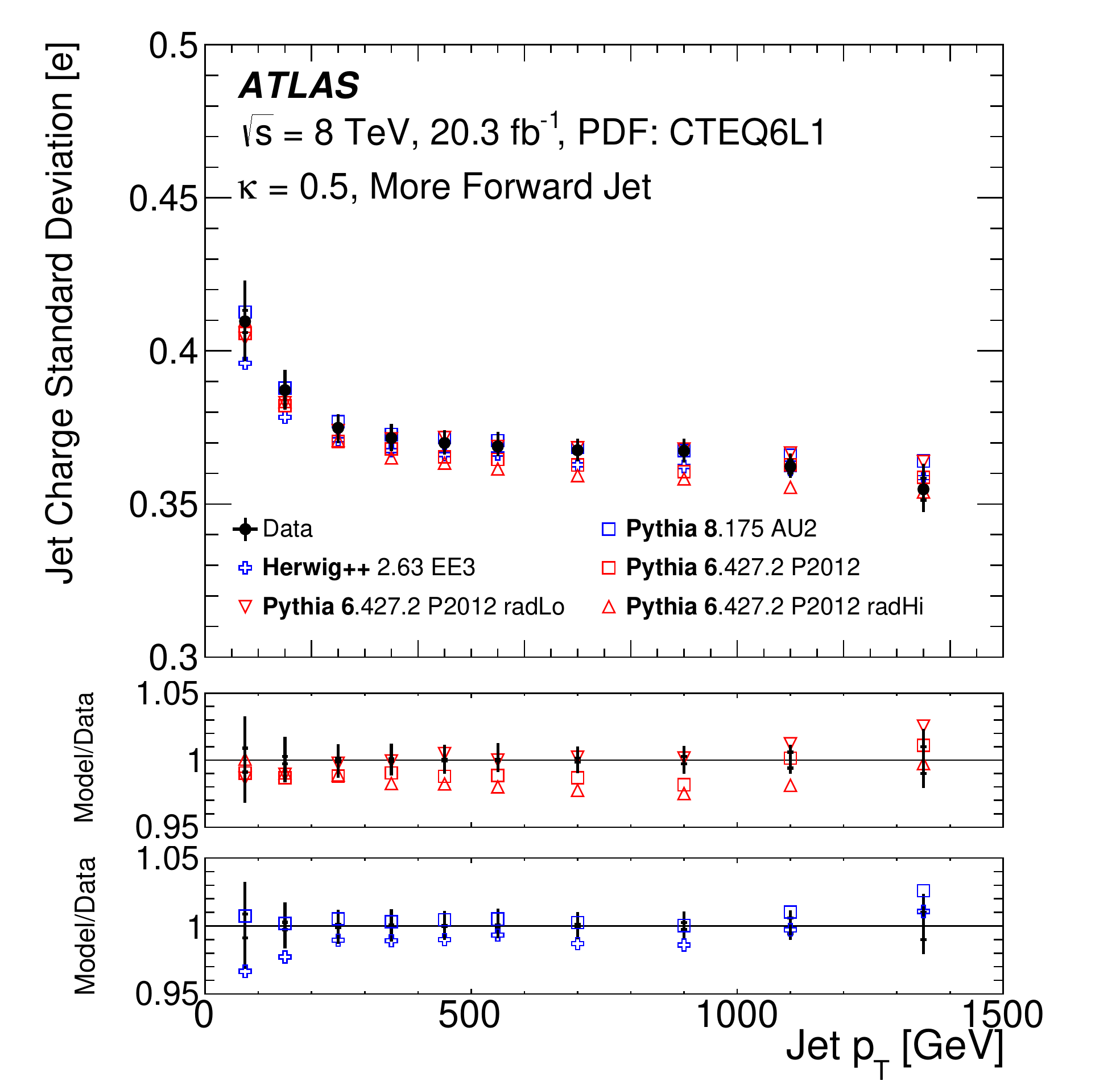}}\\
{\includegraphics[width=0.5\textwidth]{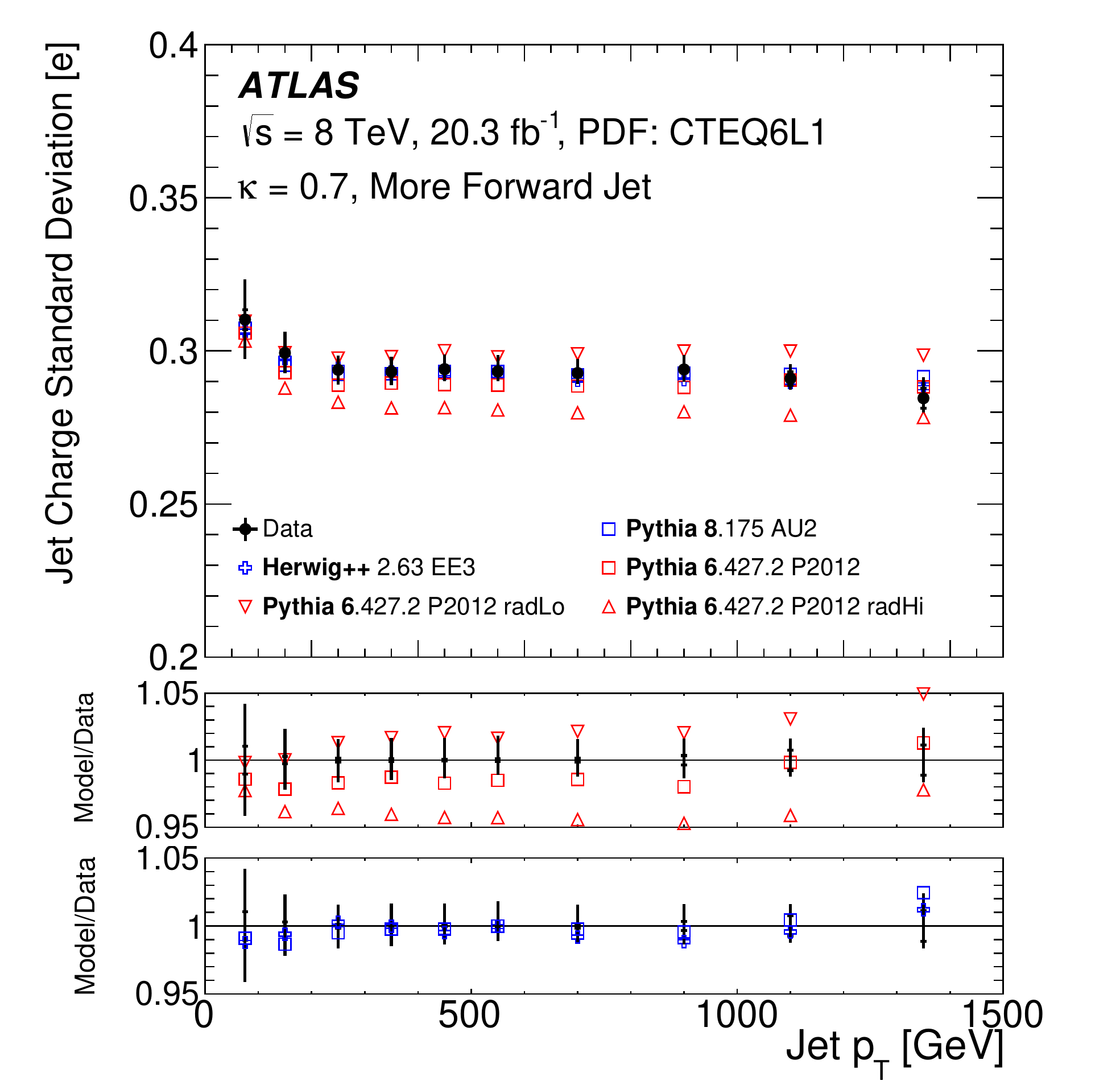}}
\end{center}	
\caption{The standard deviation of the jet charge distribution in units of the positron charge for (a) $\kappa=0.3$, (b) 0.5, and (c) 0.7 comparing various QCD MC models and tunes for the more forward jet.  The crossed lines in the bars on the data indicate the statistical uncertainty and the full extent of the bars is the sum in quadrature of the statistical and systematic uncertainties.}
\label{fig:nprm2}
\end{figure}

\clearpage
\newpage

\subsection{Model comparison overview}

Figures~\ref{fig:meana} and~\ref{fig:mean2} show comparisons of the unfolded jet charge distribution's mean and standard deviation for different QCD simulations using LO and NLO PDF sets. The predictions using the CT10 NLO PDF set as shown in Fig.~\ref{fig:meana} are generally about 10\% below the data.  Consistent with the expectation that the PDF and (nearly collinear) fragmentation are responsible for the jet charge distribution's mean and standard deviation, there does not seem to be an effect from the {\sc Powheg} NLO matrix element.  For the jet charge distribution's standard deviation and $\kappa=0.3$, the data falls between {\sc PYTHIA} (larger standard deviation) and {\sc Herwig++} (smaller standard deviation), but this trend is less evident for larger $\kappa$ values, suggesting a difference due to soft tracks.  As seen in Sec.~\ref{sec:PDFsensitivity}, comparisons with CTEQ6L1 show it be to a better model for the $p_\text{T}$-dependence of the mean jet charge than CT10.  The analogous plots to Fig.~\ref{fig:mean} but using CTEQ6L1 instead of CT10 are shown in Fig.~\ref{fig:mean2}.  Generally, there is agreement between the simulation and the data with only a $\lesssim 5\%$ difference in the lower $p_\text{T}$ bins.

\begin{figure}[h!]
\begin{center}
\includegraphics[width=0.48\textwidth]{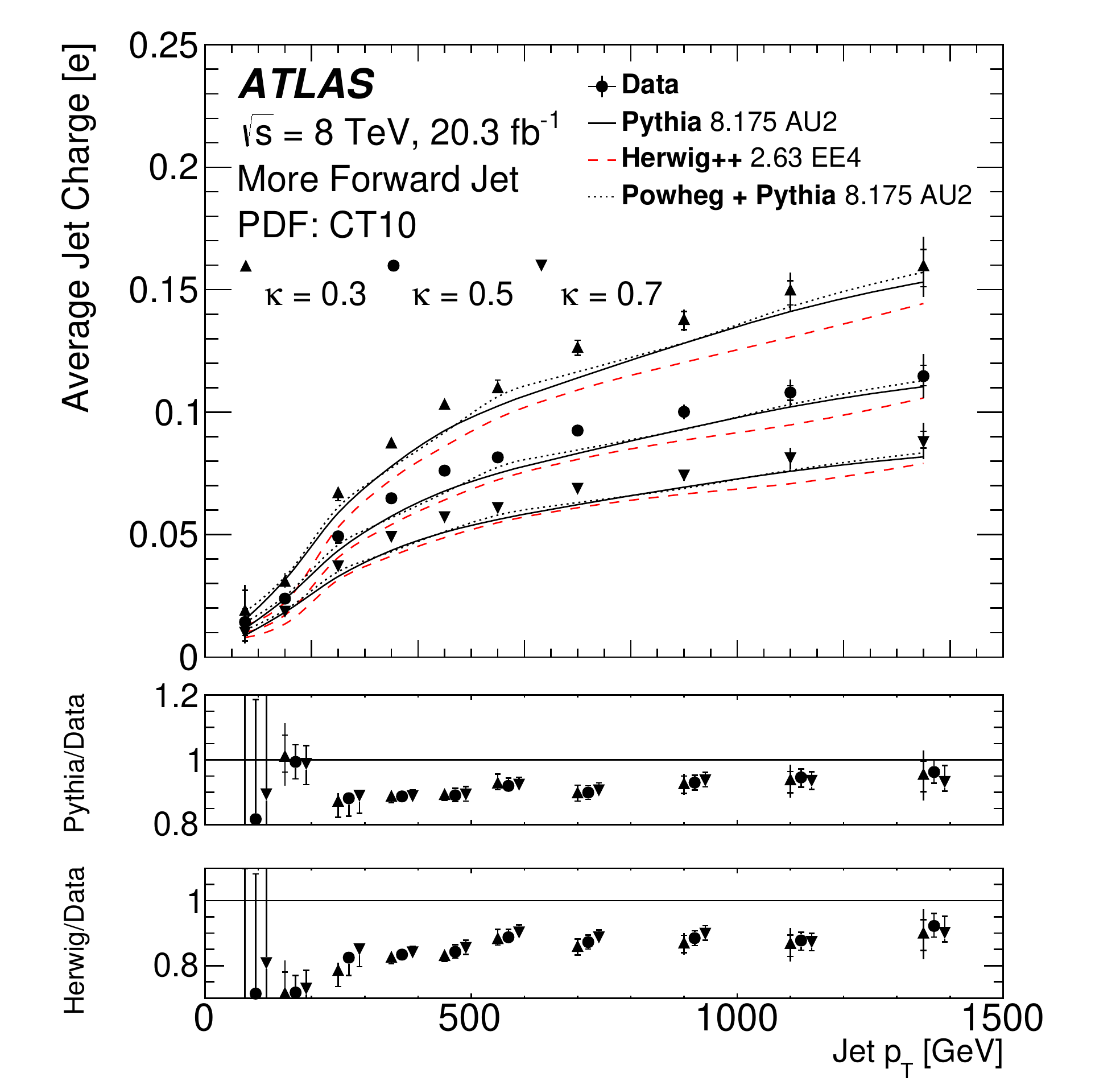}
\includegraphics[width=0.48\textwidth]{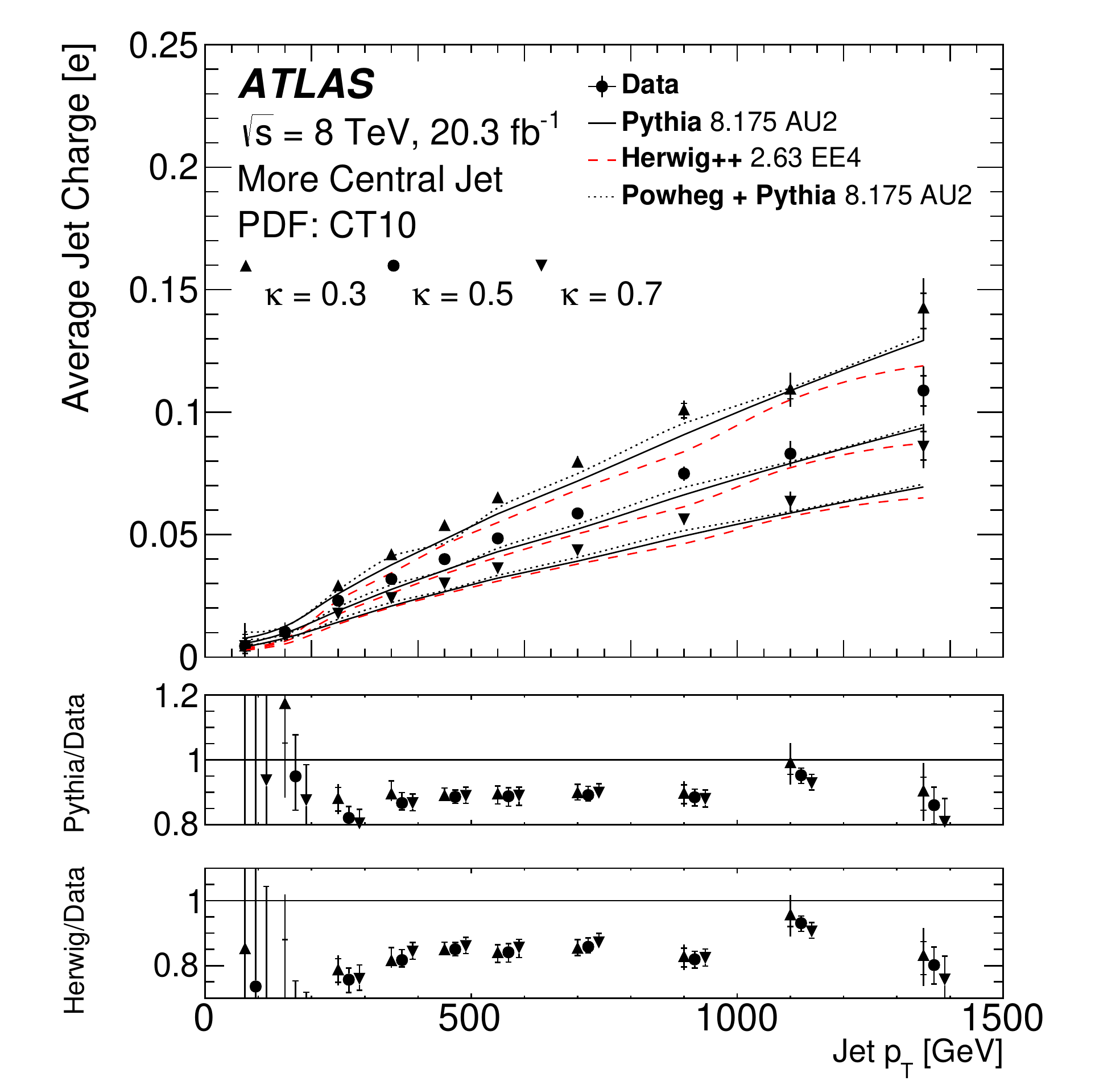}\\
\includegraphics[width=0.48\textwidth]{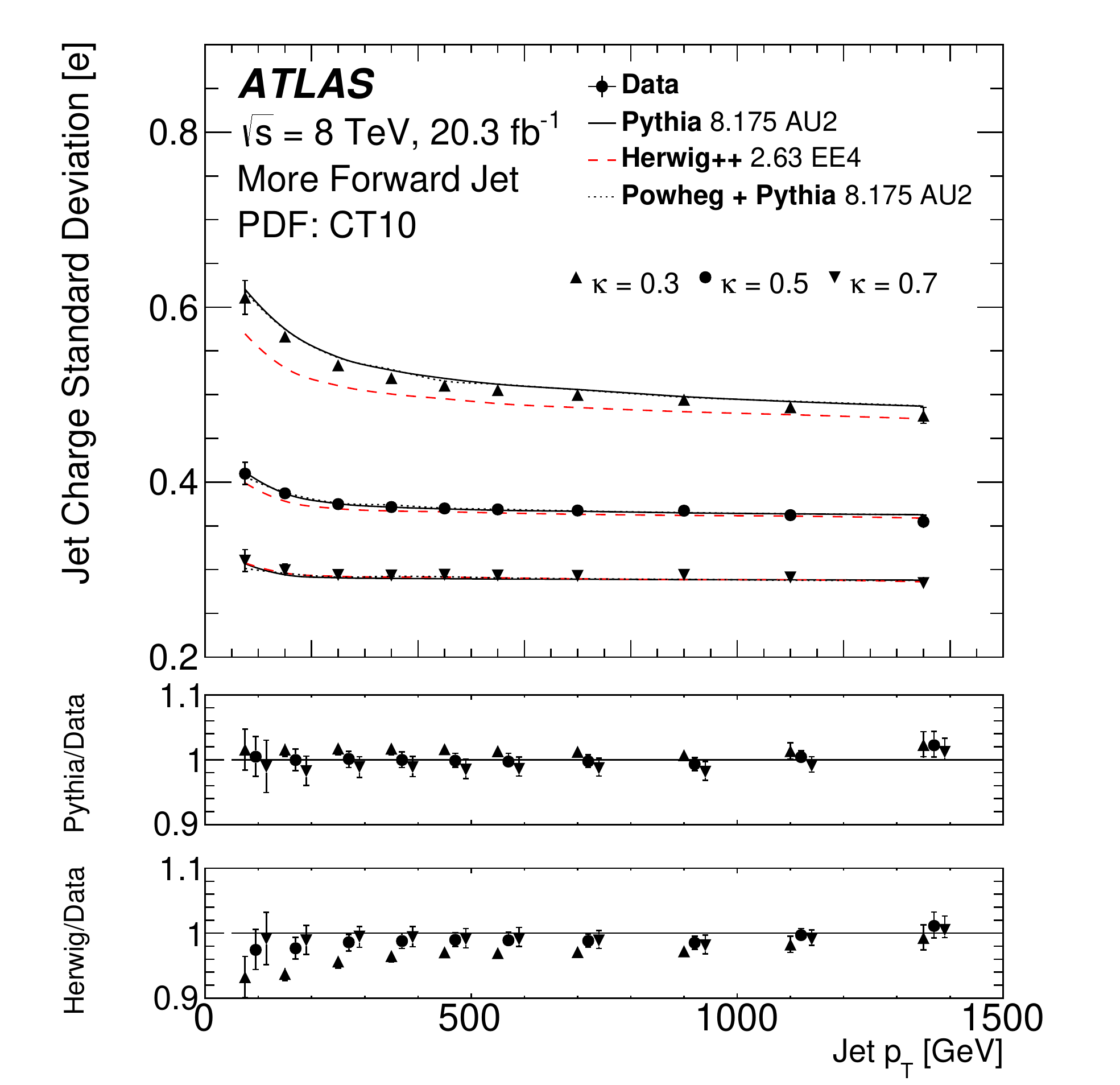}
\includegraphics[width=0.48\textwidth]{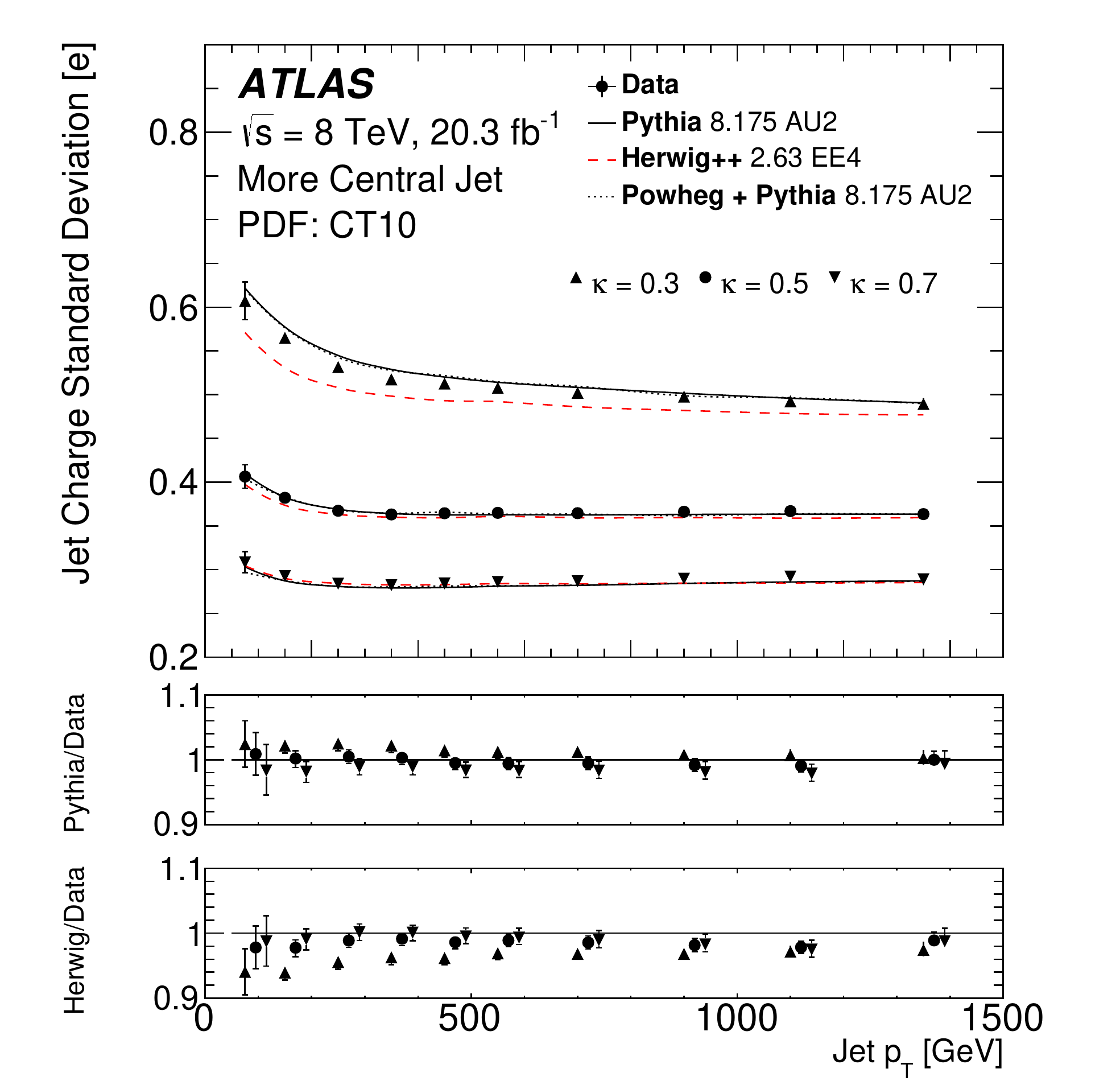}
\end{center}	
\caption{The measured average of the jet charge distribution (top), and the standard deviation (bottom), in units of the positron charge as a function of the jet $p_\text{T}$ for $\kappa=0.3, 0.5,$ and $0.7$ for the more forward jet (left) and the more central jet (right) using CT10 as the PDF set.  The markers in the lower panel are artificially displaced horizontally to make distinguishing the three $\kappa$ values easier.  The {\sc Powheg}+{\sc Pythia} curves are nearly on top of the {\sc Pythia} curves.  The crossed lines in the bars on the data indicate the systematic uncertainty and the full extent of the bars is the sum in quadrature of the statistical and systematic uncertainties.}
\label{fig:meana}
\end{figure}

\begin{figure}[h!]
\begin{center}
\includegraphics[width=0.48\textwidth]{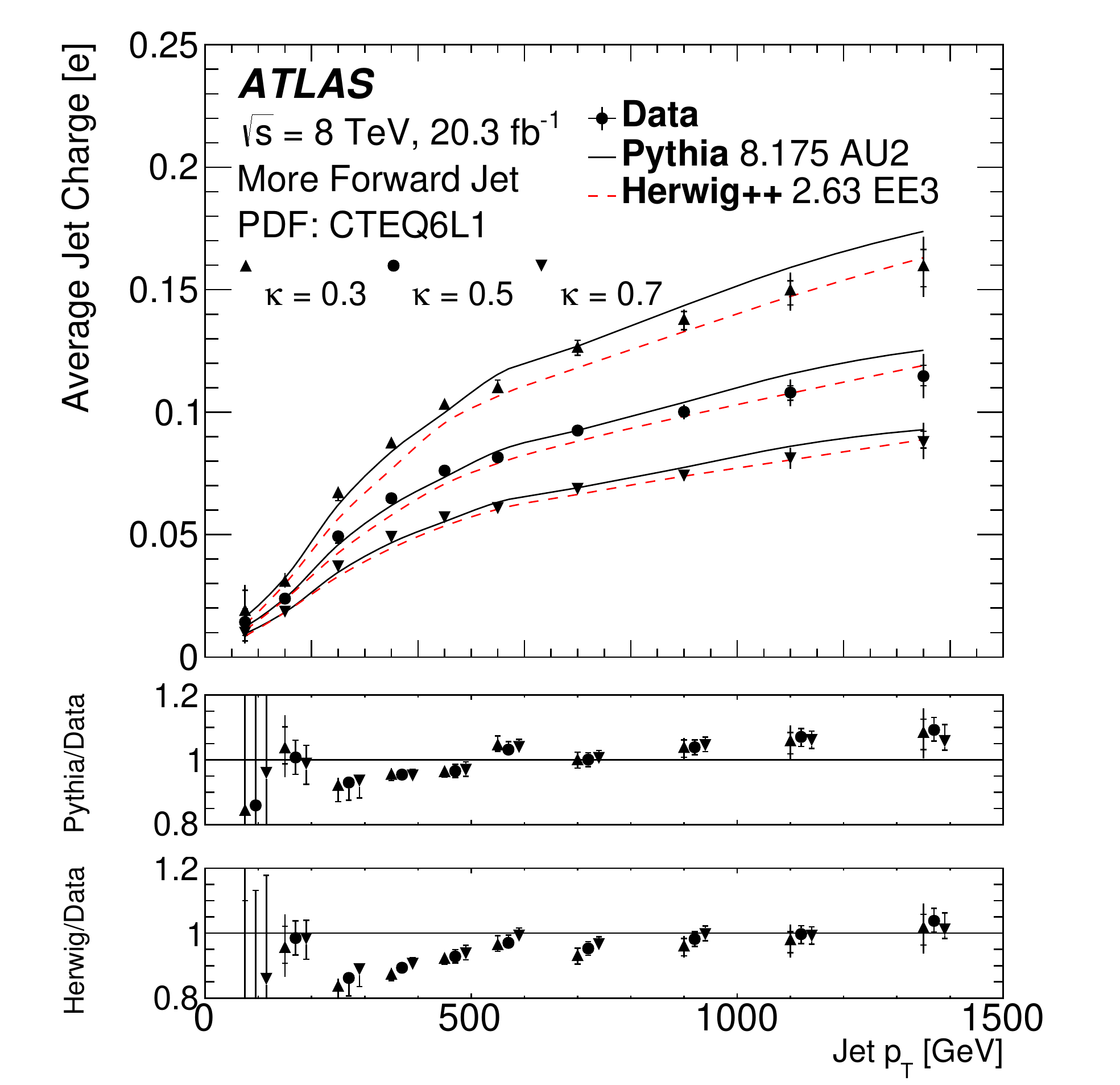}
\includegraphics[width=0.48\textwidth]{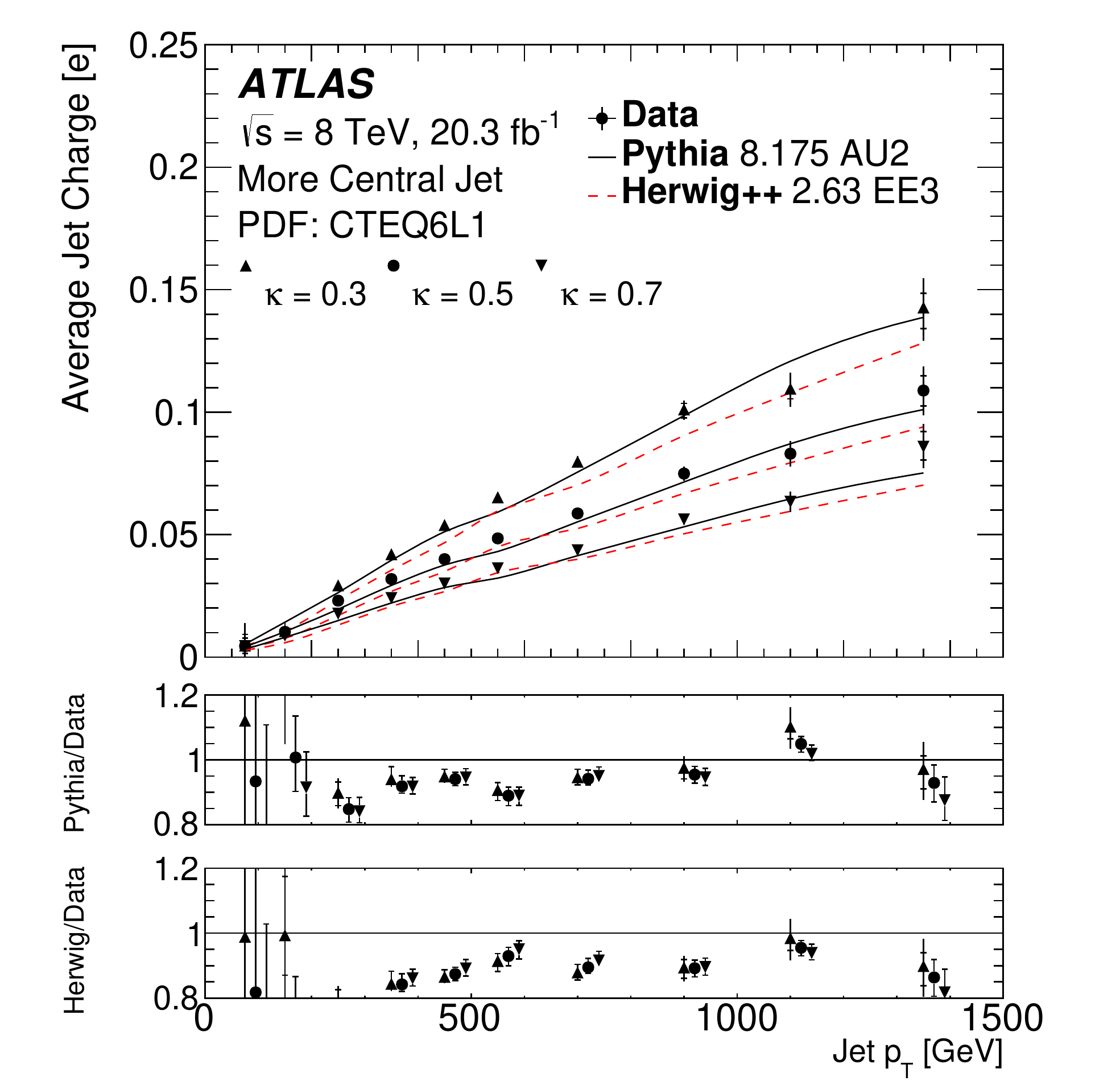}\\
\includegraphics[width=0.48\textwidth]{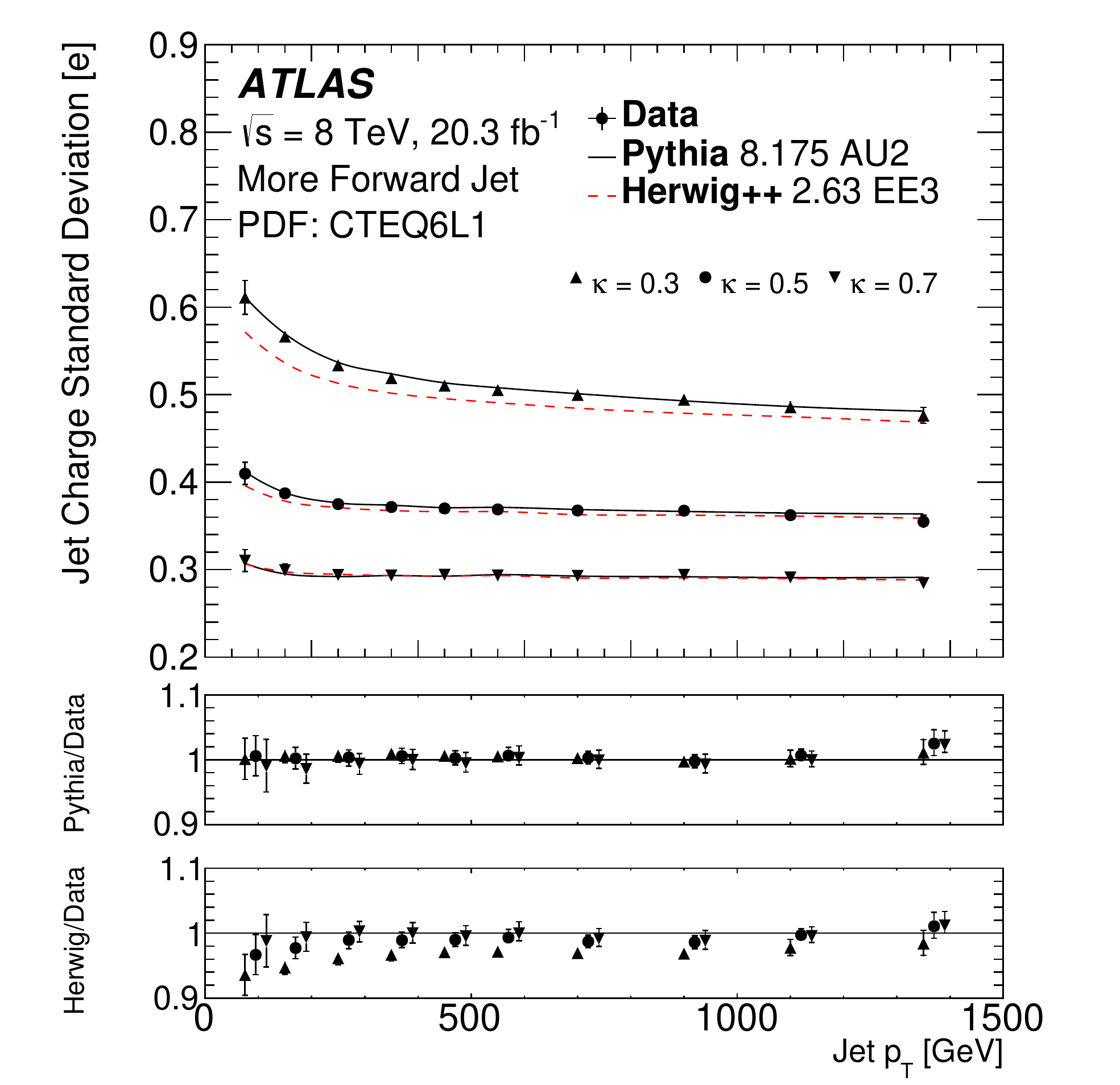}
\includegraphics[width=0.48\textwidth]{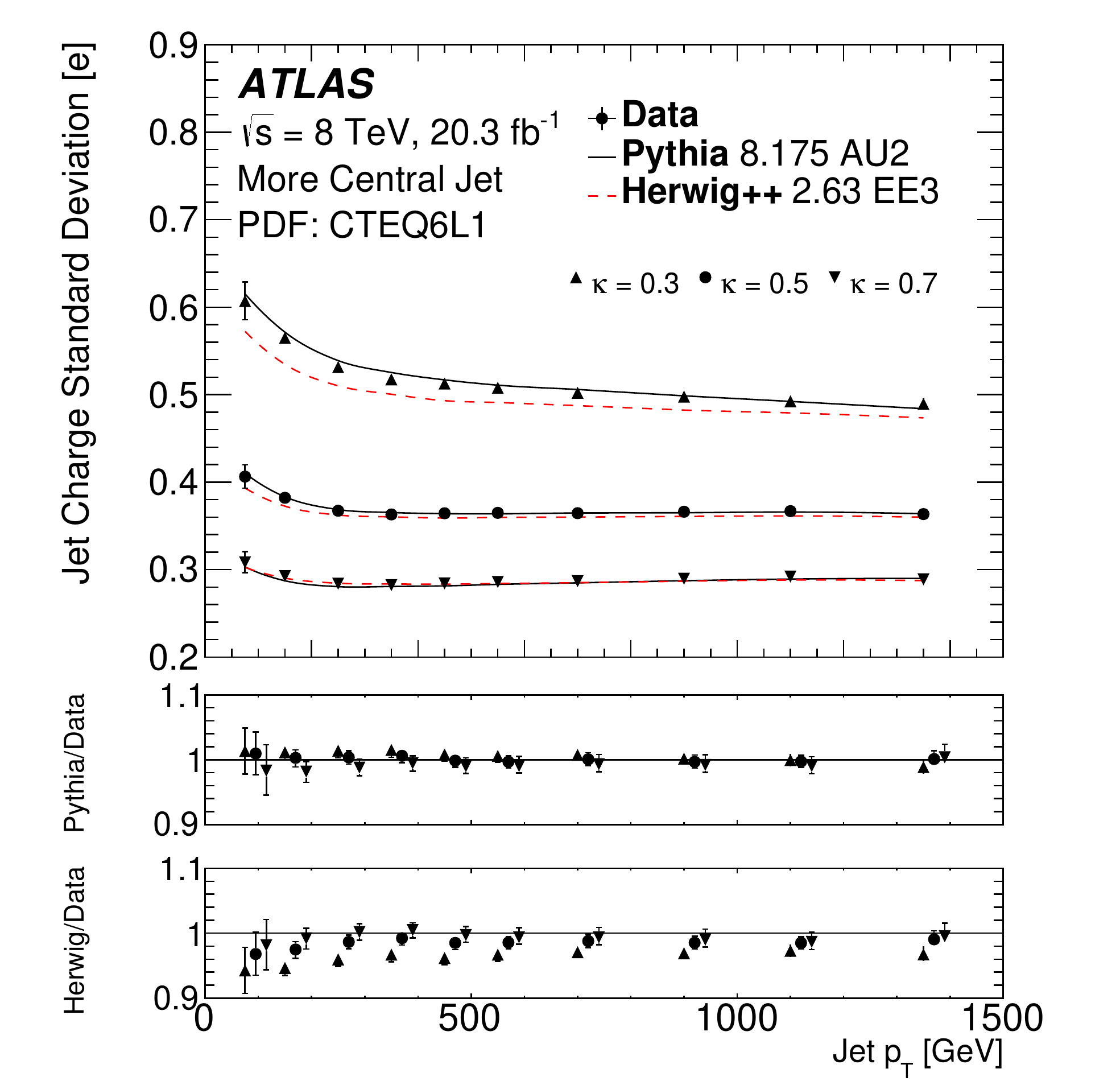}
\end{center}	
\caption{The measured average of the jet charge distribution (top), and the standard deviation (bottom), in units of the positron charge as a function of the jet $p_\text{T}$ for $\kappa=0.3, 0.5,$ and $0.7$ for the more forward jet (left) and the more central jet (right) using CTEQ6L1 as the PDF set.  The markers in the lower panel are artificially displaced horizontally to make distinguishing the three $\kappa$ values easier. The crossed lines in the bars on the data indicate the systematic uncertainty and the full extent of the bars is the sum in quadrature of the statistical and systematic uncertainties.}
\label{fig:mean2}
\end{figure}

\clearpage

\subsection{The average up-quark and down-quark jet charges}
\label{sec:updownextract}

In addition to understanding the trends in the jet charge distribution from PDFs, one can use PDFs to extract information about jets of a particular flavor.  These {\it exclusive} interpretations rely on flavor-fraction information in PDFs and matrix element calculations to extract the jet charge distribution for particular jet (anti-)flavors in each $p_\text{T}$ bin.  The required nonperturbative information is summarized in Fig.~\ref{fig:flavorfrac}(a).  Jets with flavors other than up/down/anti-up/anti-down/gluon are not included in Fig.~\ref{fig:flavorfrac}(a) and give a negligible contribution ($\lesssim 2\%$) in the highest $p_\text{T}$ bins.

One way of extracting the up- and down-flavor average jet charges is to exploit the difference in flavor fractions shown in Fig.~\ref{fig:flavorfrac}(a) between the more forward and the more central jets.  Due to the $p_\text{T}$-balance requirement between the leading and subleading jet in the event selection, to a good approximation, the $p_\text{T}$ spectrum is the same for the more forward and the more central jet.  Assuming that the average jet charge of the sum of flavors that are not up/down/anti-up/anti-down is zero, in each bin $i$ of $p_\text{T}$:

\begin{align}
\label{eq:syst}
\langle Q_J^\text{forward}\rangle_i &= \left(f_\text{up,i}^\text{forward}-f_\text{anti-up,i}^\text{forward}\right)Q_i^\text{up}+(f_\text{down,i}^\text{forward}-f_\text{anti-down,i}^\text{forward})Q_i^\text{down}\\\nonumber
\langle Q_J^\text{central}\rangle_i &= \left(f_\text{up,i}^\text{central}-f_\text{anti-up,i}^\text{central}\right)Q_i^\text{up}+(f_\text{down,i}^\text{central}-f_\text{anti-down,i}^\text{central})Q_i^\text{down},
\end{align}

\noindent where $Q_J$ is the jet charge from Eq.~\ref{chargedef}, $f_{y,i}^x$ is the fraction of flavor $y$ in $p_\text{T}$ bin $i$ for the jet $x\in\{\text{more forward, more central}\}$ and $Q_i^y$ is the average jet charge for such jets (average gluon jet charge is zero).  The values $f_{y,i}^x$ are taken from simulation ({\sc Pythia} with CT10 PDF and AU2 tune), which then allows an extraction of $Q_i^y$ by solving the system of equations in Eq.~\ref{eq:syst}.  This extraction is performed separately in each $p_\text{T}$ bin.  The left plot of Fig.~\ref{fig:extracedupdown} shows the extracted up- and down-flavor jet charges in bins of jet $p_\text{T}$.  At very high jet $p_\text{T}$, the absolute quark flavor fractions are large (Fig.~\ref{fig:flavorfrac}), but the difference between the more forward and more central jets is small and the statistical uncertainty is large.  At low jet $p_\text{T}$, the difference between the more forward and more central jets is large (Fig.~\ref{fig:flavorfrac}), but the absolute quark flavor fraction is small and the statistical uncertainty is once again large because the mean jet charge is close to zero.  In the limit that the flavor fractions are identical for the more forward and more central jet, the equations become degenerate and it is not possible to simultaneously extract the average up- and down-flavor jet charges.  The uncertainties on the flavor fractions and on the measured average jet charges are propagated through the solutions of Eq.~\ref{eq:syst}.  Generally, the uncertainty is larger for the down-flavor jets because the fraction of these jets is smaller than the fraction of up-flavor jets.

The right plot of Fig.~\ref{fig:extracedupdown} compares the extracted up quark and down quark jet charges.  The central value of the up quark jet charge is slightly less than twice the down quark jet charge, though this is not significant beyond one standard deviation for $\kappa=0.5$ and $\kappa=0.7$ and just beyond one standard deviation for $\kappa=0.3$.  

\begin{figure}[h!]
\begin{center}
\includegraphics[width=0.5\textwidth]{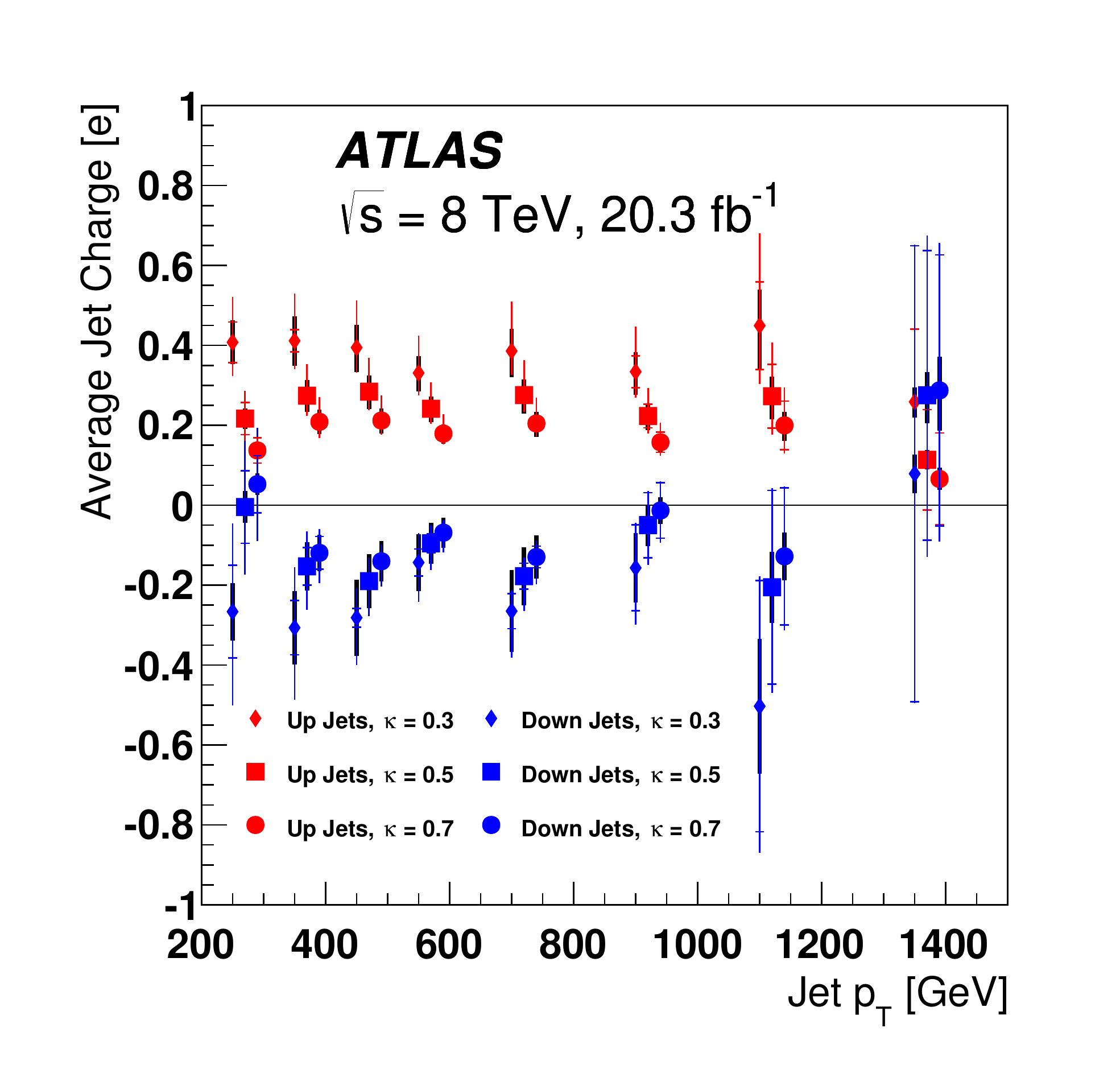}\includegraphics[width=0.5\textwidth]{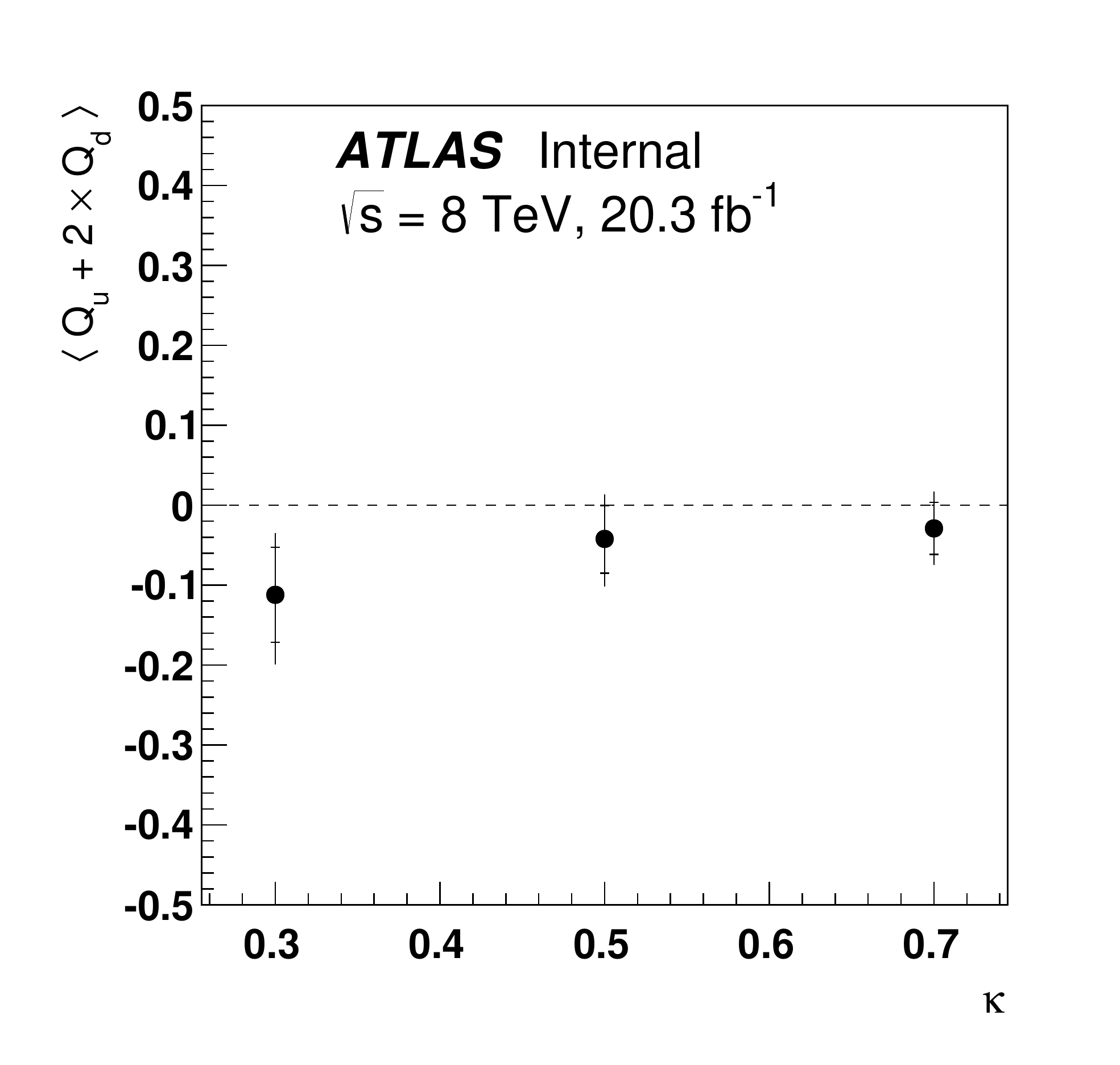}
\end{center}	
\caption{The extracted value of up- and down-quark jet charges in units of the positron charge in bins of jet $p_\text{T}$ for $\kappa=0.3, 0.5,$ and $0.7$.  The error bars include statistical, experimental systematic, and CT10 PDF uncertainties added in quadrature. The thick part of the error bar indicates the PDF contribution to the total uncertainty and the horizontal line on each error bar indicates the contribution from the statistical uncertainty.  The first two $p_\text{T}$ bins in the left plot are excluded due to their very large uncertainties.}
\label{fig:extracedupdown}
\end{figure} 

\subsection{Up- and down-quark jet charge dependence on $p_\text{T}$}
\label{sec:scaleviolation}

Using the methods of Sec.~\ref{sec:updownextract}, one can examine the residual $p_\text{T}$-dependence of the average jet charge {\it after} accounting for PDF effects.  The inclusive jet charge has been shown to increase with $p_\text{T}$ due to a mixing of jet flavors and the following subsection investigates the $p_\text{T}$-dependence of a fixed jet flavor.   Using the theoretical predictions from Sec.~\ref{sec:jetchargetheory}, this section describes how the $p_\text{T}$-dependence is extracted from the data.  Since $c_\kappa\ll 1$ from Eq.~\ref{eq:jetcharge:ckappa}, one can approximate a linear dependence on $c_\kappa$:

\begin{align}
\langle Q_J\rangle(p_\text{T})=\bar{Q} (1+c_\kappa\ln(p_\text{T}/\bar{p}_\text{T}))+\mathcal{O}(c_\kappa^2),
\end{align}

\noindent where $\bar{Q}=\langle Q_J\rangle(\bar{p}_\text{T})$ for some fixed (but arbitrary) transverse momentum, $\bar{p}_\text{T}$.  Therefore, for a fixed $p_\text{T}$ bin $i$, the measured charge is given as a superposition of the average jet charge for various jet flavors:

\begin{align}
\label{eq:scaleviolate}
\langle Q_i\rangle \approx \sum_f \beta_{f,i}\bar{Q}_f(1+c_\kappa\ln(p_{\text{T},i}/\bar{p}_\text{T})),
\end{align}

\noindent where $\beta_{f,i}$ is the fraction of flavor $f$ in bin $i$, $\bar{Q}_f$ is the average jet charge of flavor $f$ and $\bar{p}_\text{T}$ is a fixed transverse momentum.   Fitting the model in Eq.~\ref{eq:scaleviolate} directly to the data to extract $\bar{Q}_f$ is not practical because there are three parameters and only 10 $p_\text{T}$ bins, some of which have very little sensitivity due to low fractions $\beta$ or large uncertainties on $\langle Q_J\rangle$.  One way around this is to extract $\bar{Q}_f$ in one fixed bin of transverse momentum (denoted $\bar{p}_\text{T}$) as described in Sec.~\ref{sec:updownextract}.  Then Eq.~\ref{eq:scaleviolate} is highly constrained, with only one parameter for which each other bin of $p_\text{T}$ gives an estimate.  The systematic uncertainties are propagated through the fit treated as fully correlated between bins and the statistical uncertainty is treated coherently by bootstrapping\footnote{Pseudo-datasets are generated by adding each event in the nominal dataset $j$ times where $j$ is a Poisson random variable with mean $1$.  Since events are coherently added, this respects the correlations in the statistical uncertainty for the more forward and central jet charges.}.  A weighted average is performed across all $p_\text{T}$ bins and for both the more forward and the more central jet.  The procedure is summarized below:

\begin{enumerate}
\item In the bin 600 GeV~$<p_\text{T}<$ 800~GeV,  extract the values $\bar{Q}_\text{up}$ and $\bar{Q}_\text{down}$.  These values can be seen in the fifth $p_\text{T}$ bin of Fig.~\ref{fig:extracedupdown}.
\item With $\bar{Q}_\text{up}$ and $\bar{Q}_\text{down}$ fixed, extract the scale violation parameter estimate $c_{\kappa,i}$ in each $p_\text{T}$ bin $i$ by solving
\begin{align}
\langle Q_i\rangle_\text{measured}=\sum_f \beta_{f,i}\bar{Q}_f(1+c_{\kappa,i}\ln(p_{\text{T},i}/\bar{p}_\text{T}))
\end{align}
\noindent where $\bar{p}_\text{T}=700$~GeV~is the bin center from the previous step.  
\item Repeat the procedure for all systematic variations and for all bootstrap pseudo-datasets to arrive at estimates of the uncertainty $\sigma(c_{\kappa,i})$ for each $p_\text{T}$ bin $i$.  The bin in step 2 is fixed, but the value in the bin varies.
\item The central value for the extracted scale violation parameter is 

$$c_\kappa=\left(\sum_i c_{\kappa,i}/\sigma(c_{\kappa,i})\right)/\sum_i(1/\sigma(c_{\kappa,i})).$$

\item The uncertainty $\sigma(c_\kappa)$ is determined by repeating step (3) with the nominal values $c_{\kappa,i}$ replaced by their systematic varied versions or the bootstrap pseudo-data values for the statistical uncertainty estimate.
\end{enumerate}

\noindent The results are presented in Fig.~\ref{fig:money}.  The data support the prediction that $c_\kappa<0$ and $\partial c_\kappa/\partial\kappa < 0$.  Linear correlations between $\kappa$ values can be determined using the bootstrapped datasets: about $0.9$ between $c_{0.3}$ and $c_{0.5}$ as well as between $c_{0.5}$ and $c_{0.7}$, while the correlation is about 0.7 between $c_{0.3}$ and $c_{0.7}$.  Thus, the three points are quite correlated, but there is additional information from considering more than one $\kappa$ value.
 
\begin{figure}[h!]
\begin{center}
\includegraphics[width=0.8\textwidth]{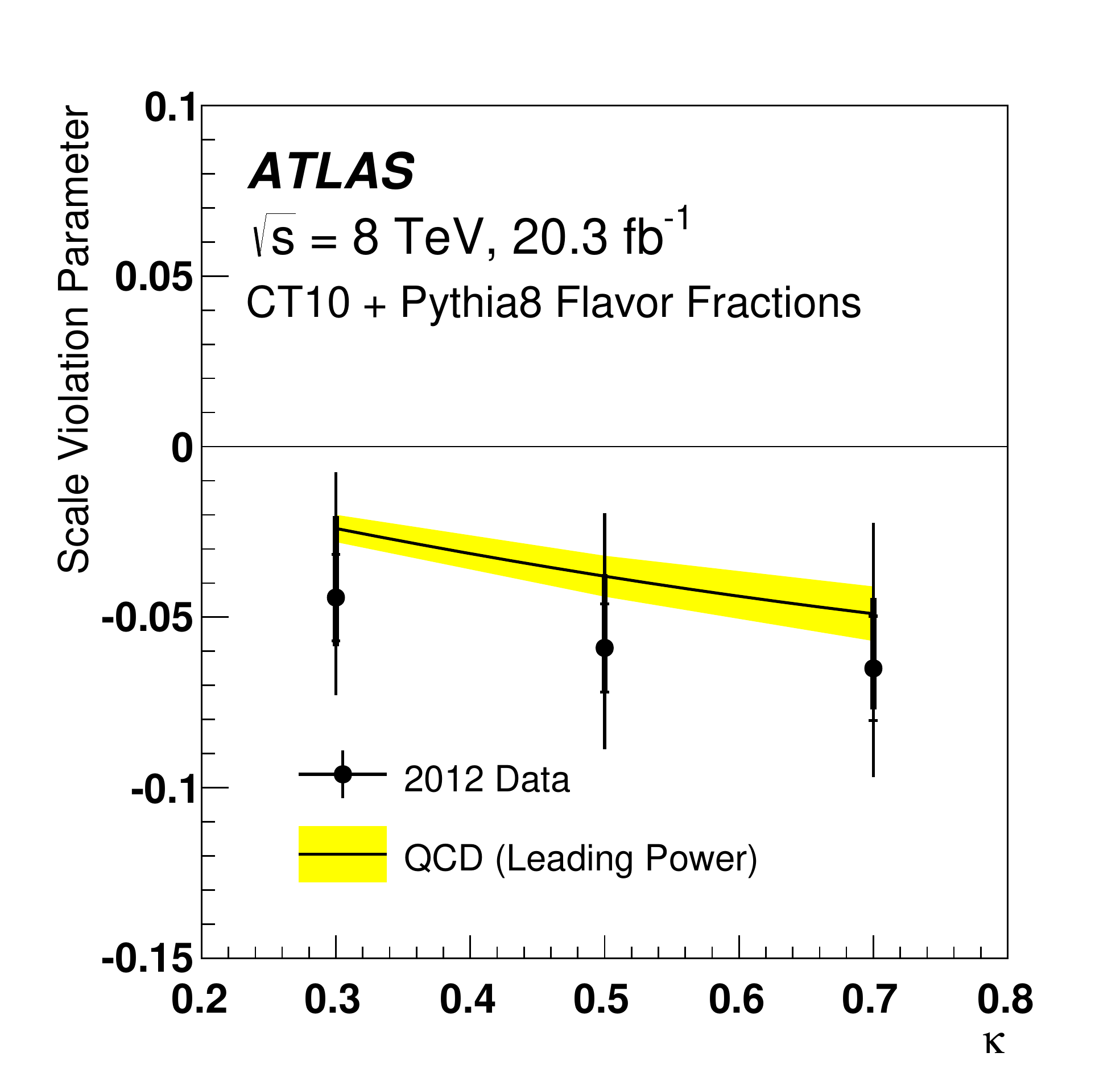}
\end{center}	
\caption{The extracted values of the scale violation parameter $c_\kappa$ from the data compared to theoretical calculations~\cite{Waalewijn2012,Krohn2012}.  The error bars include statistical, experimental systematic, and PDF uncertainties added in quadrature. The thick part of the error bar indicates the PDF contribution to the total uncertainty and the horizontal line on each error bar indicates the contribution from the statistical uncertainty (each shown without adding in quadrature any other source of uncertainty).}
\label{fig:money}
\end{figure} 
 
 \clearpage
 
\section{Summary}
\label{sec:summary}

This chapter presents a measurement of the particle-level $p_\text{T}$-dependence of the jet charge distribution's mean and standard deviation in dijet events from 20.3 fb${}^{-1}$ of $\sqrt{s}=8$ TeV~$pp$ collision data recorded by the ATLAS detector at the LHC.  The measured jet charge distribution is unfolded to correct for the detector acceptance and resolution for direct comparison to particle-level models.  Comparisons are made at particle level between the measured jet charge distribution and various PDF sets and models of jet formation.  Simulations with Pythia 8 using the CTEQ6L1 PDF set describe the average jet charge of the more forward jet within about 5\% and the more central jet within about 10\%. The jet charge distribution's standard deviation is described within 2\%.  {\sc Herwig++} shows a similarly good agreement for $p_\text{T}>500$~GeV~and $\kappa=0.7$.  However, the {\sc Herwig++} predictions decrease systematically for both the average and the standard deviation for decreasing $\kappa$.  Predictions with the CT10 NLO PDF are systematically below the data across jet $p_\text{T}$ for the average jet charge and systematically above for the jet charge distribution's standard deviation.  Taking the PDFs as inputs, the average up- and down-flavor jet charges are extracted as a function of $p_\text{T}$ and are compared with predictions for scale violation.  The data show that the average up- and down-quark jet charges decrease slightly with $p_\text{T}$ and this decrease increases with $\kappa$, as predicted.  The particle-level spectra are publicly available~\cite{hepdata} for further interpretation and can serve as a benchmark for future measurements of the evolution of nonperturbative jet observables to validate QCD MC predictions and tune their free model parameters.
 \chapter{Color flow}	
\label{cha:colorflow}

Due to the confining nature of the strong force, directly measuring the
QCD interactions between quarks and gluons is not possible.
The strength and direction of the strong force depends on the color
charge of the particles involved.
To a good approximation, the radiation pattern in QCD can be described through
a color--connection picture, which consists of color strings connecting quarks and gluons
of one color to quarks and gluons of the corresponding anti--color.
An important question is whether there is evidence of these color connections
(\textit{color flow}) in the observable objects: color--neutral hadrons and the
jets they form.  The study of energy distributions inside and between jets in
various topologies has a long history, dating back to the discovery of gluons
in three--jet events at
PETRA~\cite{Brandelik:1979bd,Barber:1979yr,Berger:1979cj,Bartel:1979ut}.  Color
connections are still a poorly constrained QCD effect,
which motivates the dedicated study presented in this chapter.
If well understood, experiments can exploit color flow to aid Standard Model
measurements and searches for physics beyond the SM.

One of the challenges in studying color flow is the selection of a final state
with a known color composition.  Color--singlet $W$ bosons from $t\bar{t}$
events provide an excellent testing ground because these bosons have a known
initial (colorless) state and such events can be selected with high purity.  As a test that the color flow can be extracted from the observable
final state, the data are compared to models with simulated $W$ bosons that are color--charged or color--neutral.

\clearpage

\section{Introduction}
\label{sec:colorflow:intro}

Information about the color connections of partons participating in the hard-scatter is embedded in the observable final state jets.  This has been demonstrated by studying the energy distribution inside and between jets in events of various topologies.  The first such measurement was by the JADE collaboration in 3-jet events at PETRA~\cite{Bartel:1983ii}.  The JADE collaboration reported that the third leading jet in tri-jet events had a rather diffferent shape than the leading or subleading jet in the same events or the two leading jets in dijet events.  Among other properties, it was found that these third (gluon-like) jets had a broader distribution of energy and particle multiplicity as a function of distance from the jet axis compared to the other (quark-like) jets.  Comparison with the models of the time suggested that this observation was in support of fragmentation along the color axes of the initiating partons.  There are now a variety of three-jet studies aimed at investigating this phenomena of {\it color coherence} performed at PETRA~\cite{Bartel:1983ii,tasso}, PEP~\cite{pep,PhysRevLett.54.270,PhysRevLett.57.945,PhysRevLett.57.1398}, Tevatron~\cite{PhysRevD.50.5562,Abbott:1997bk}, and LHC~\cite{Chatrchyan:2013fha}.

With center of mass energies large enough to produce on-shell $W$ and $Z$ bosons, the highest energy experiments since LEP have studied color flow directly from the decay of heavy color singlets.  For instance, L3~\cite{Achard:2003pe} and DELPHI~\cite{Abdallah:2006uq} studied hadronic diboson $WW$ events in which the energy density between jets associated with the same $W$ decay compared to the density between jets from different $W$ decays was found to be sensitive to the color flow of models used to describe the data.  Additional studies at LEP~\cite{Acciarri:1995hq,Schael:2006ns} and the Tevatron~\cite{Abbott:1999cu} have used the known initial color state of the electroweak bosons to constrain models of color flow in hadronic final states.

All of the studies described thus far have used either the distribution of energy within a jet or the distribution of energy between jets as sensitive observables to constrain models of color flow.   The combination of the orientation and distribution of intrajet and interjet radiation can provide additional discriminating power.  First defined in Ref.~\cite{Gallicchio:2010sw}, the {\it jet pull} is a kinematic variable built from momentum-weighted radial moments of jet constituents ({\it jet substructure}) combined with information from the relative orientations of jets in the event ({\it jet superstructure}) that was designed to be sensitive to the color flow between the initiating partons of jets.  Since its inception, jet pull has been suggested as a discriminating variable to isolate color singlets such as Higgs bosons from color octets (e.g. gluons)~\cite{Gallicchio:2010sw,Gallicchio:2010dq} and has been used for this purpose experimentally in a variety of searches for the Higgs boson~\cite{D0higgs,CMShiggspap,CMShiggspap2}.   However, there has never been significant evidence from the data that this observable is directly sensitive to color flow.

The first experimental measurement of color connection using jet pull was performed in $t\bar{t}$ events with one lepton in the final state at D{\O}~\cite{Abazov:2011vh}.  Such events provide a relatively pure sample of hadronically decaying $W$ bosons.  By fitting the data with MC templates constructed from the jet pull distribution, exotic color flow models can be constrained directly.  However, color flow is subtle and there was not sufficient precision at D{\O} to observe significant differences between the singlet and octet models. 

This chapter describes the first measurement to definitively show that the jet pull angle can differentiate color singlet and color octet dijet resonances\footnote{The jet pull reconstruction studies and precision measurement presented here are published in Ref.~\cite{ATLAS-CONF-2014-048} and Ref.~\cite{Aad:2015lxa}, respectively.  The measurement benefited from fruitful discussions and technical help from T. Neep, K. Joshi, M. Swiatlowski, Y. Peters, D. L. Mateos, and M. Schwartz.}.  The jet pull angle is studied in $t\bar{t}$ at the LHC, where the $t\bar{t}$ cross section and integrated luminosity are much higher than at the Tevatron.  In addition, improved analysis techniques have increased the precision of the measurement.  Furthermore, the jet pull angle distribution is unfolded to correct for distortions from the detector resolution and finite acceptance in order to make the measurement available for MC tuning or testing models of color flow beyond the Standard Model.

 This chapter is organized as follows.  The remainder of Sec~\ref{sec:colorflow:intro} describes color flow in the context of QCD and introduces the jet pull angle.  Details about the analysis design, including the simulation, object reconstruction, and event selection are described in Sec.~\ref{sec:colorflowanalysisdesign}.  The properties of the reconstruction and resolution of the jet pull are in Sec.~\ref{sec:colorflowperformance} in preparation for unfolding the pull angle distribution, described in Sec.~\ref{sec:colorflow:unfolding}.   A detailed description of the systematic uncertainties is documented in Sec.~\ref{sec:colorflow:systematics} and the unfolded results are given in Sec.~\ref{sec:colorflow:results}.  The chapter ends with some concluding remarks in Sec.~\ref{sec:colorflowsummary}.

\clearpage

\subsection{Color flow in QCD}
\label{sec:colorcoherence}

Color flow has important implications for all stages of jet formation.  At the beginning of jet development, the leading effect is due to {\it color coherence}.  A heuristic explanation~\cite{Dokshitzer:1991wu,Ellis:1991qj} for this effect is that soft (long wavelength) gluons cannot resolve individual partons that are close in angle.  The same effect is true for photons.  In electrodynamics, the rate of radiated photons in $\gamma\rightarrow e^+e^-\rightarrow e^+e^-\gamma$ will be suppressed outside of the $e^+e^-$ opening angle.  The soft photons cannot resolve the individual electrons; instead they are sensitive only to the sum, which is neutral.  Symbolically, suppose that the positron has momentum $p$ and the soft photon has momentum $zp$ with $z\ll 1$, as in Fig.~\ref{fig:colorflow:coherence}.  The mass of the virtual positron is 

\begin{align}
m_{e^+}\sim 2zp(1-z)p(1-\cos\theta)\sim 2zp^2\theta_{e\gamma}^2.
\end{align}

\noindent By the uncertainty principle, the virtual $e^+$ can persist for a time $\Delta E\Delta t\sim 1$\footnote{This is a re-writing of the familiar relation $\Delta x\Delta p\gtrsim \hbar/2$.  In natural units, one unit of angular momentum is $\hbar$ and the factor of 2 is absorbed in the $\sim$ sign.}:

\begin{align}
\frac{1}{\Delta t}\sim \sqrt{(m_{e^+})^2+p^2}\sim p\sqrt{1+2z\theta^2}\sim pz\theta^2.
\end{align}

\noindent In this time, the electron and the positron have traveled a distance $\sim \theta_{ee}/pz\theta^2$.   The wavelength of the soft photon in the direction away from the positron is $\lambda\sim 1/p_\text{T}^\gamma\sim 1/pz\theta$.  A soft photons can resolve the separation between the $e^-$ and $e^+$ if

\begin{align}
\frac{1}{pz\theta} < \frac{\theta_{ee}}{pz\theta^2},
\end{align}

\begin{figure}[h!]
\centering
\begin{tikzpicture}[line width=1.5 pt, scale=1.3]
	\draw[vector] (-1,0)--(0,0);
	\draw[fermion] (0,0)--(2,1);
	\draw[fermionbar] (0,0)--(2,-1);
	\draw[vector] (1.5,-0.75)--(2,0);
	\node at (2,-0.65) {$\theta_{e\gamma}$};
	\node at (0.6,0.) {$\theta_{ee}$};
	\node at (0.5,-0.5) {$p$};
	\node at (2.3,0.) {$zp$};
	\node at (2.7,-1.1) {$(1-z)p$};
\end{tikzpicture}
\caption{An illustration of coherence.  Photons emitted at large angles $\theta_{e\gamma}$ are not able to resolve the $e^+e^-$ pair, leading to a suppression of radiation outside of $\theta_{ee}$.}
\label{fig:colorflow:coherence}
\end{figure}

\noindent which is the same as $\theta_{e\gamma} < \theta_{ee}$.  In QCD, the impact is similar, only that the initial gluon in $g\rightarrow q\bar{q}\rightarrow q\bar{q}g$ is colored.  For a parton with no color charge splitting into two quarks with opposite color, the impact of color coherence is the same as in the QED case.  Large angle soft gluon radiation is suppressed because the gluons cannot individually resolve the two quarks and instead are sensitive only to the color of the initial parton, which is zero.   As a consequence of color coherence, the radiation pattern for jets initiated by two quarks originating from color singlets is enhanced in the interjet region relative to jets initiated by color triplet or color octets.  Additionally, the radiation pattern for jets initiated two quarks resulting from a color singlets is suppressed in the intrajet region relative to jets initiated by color triplet or color octets.  These radiation patterns are demonstrated in Fig.~\ref{fig:pull:colorflowimage}.  The same high $p_\text{T}$ Higgs boson undergoes fragmentation many times.  By fixing the hard-scatter parton, the images in Fig.~\ref{fig:pull:colorflowimage} show only the impact of the parton shower and hadronization on the distribution of radiation inside the jet.  The two nodes in the image correspond to the initial quark locations.  The radiation around the two nodes is enhanced in the right plot (octet) with respect to the left plot (singlet).  Figure~\ref{fig:pull:colorflowimage2} shows a quantitative comparison between the two radiation patterns, where the enhancement between the nodes is also apparent for the singlet.

\begin{figure}[htbp]
  \centering
    \includegraphics[width=0.9\textwidth]{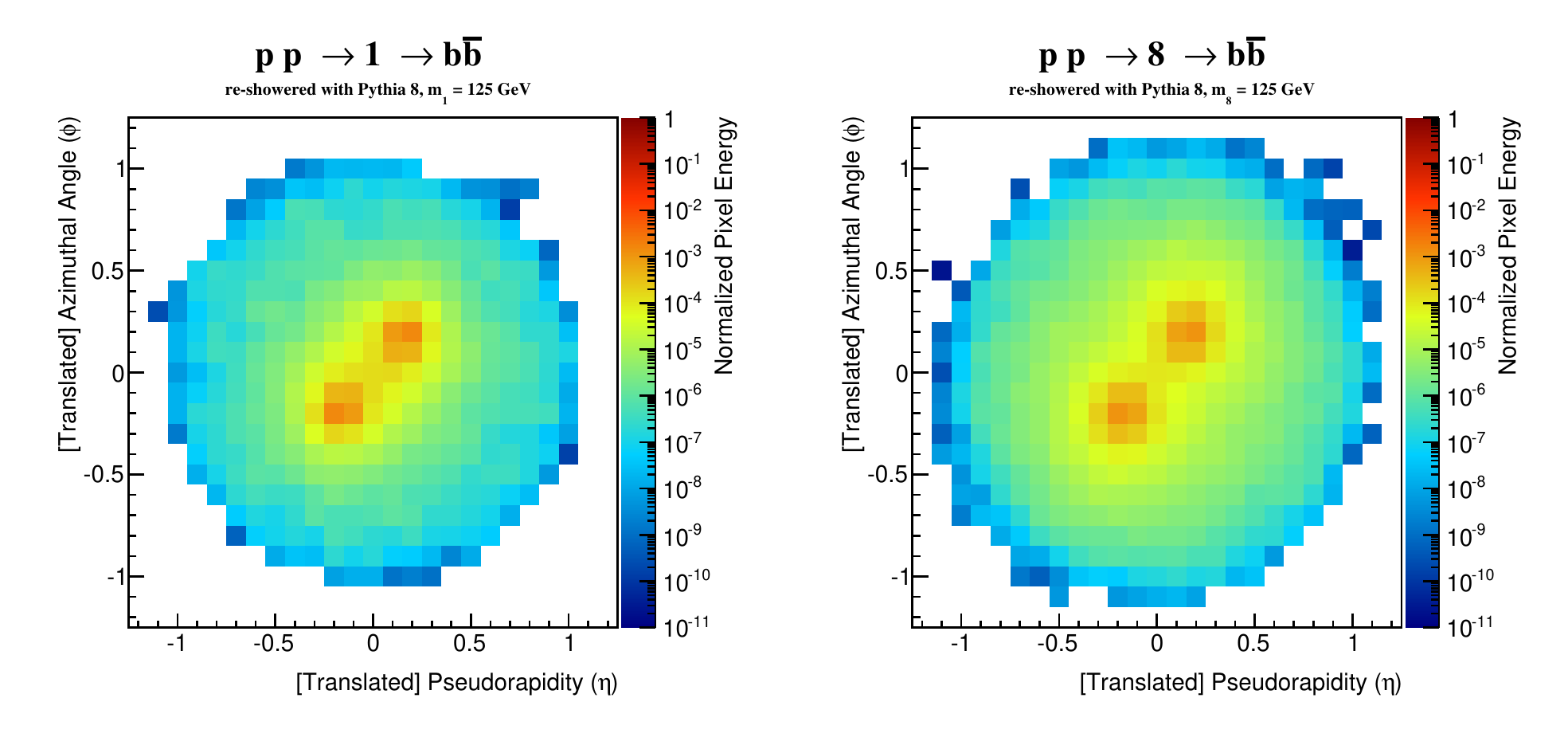}
  \caption{The same high energy parton undergoes fragmentation many times ({\it re-showered}).  Each time the parton fragments, the stable hadrons in the event are clustered into a $R=1.0$ jet and trimmed using $k_t$ subjets with $R_\text{sub}=0.3$ which are removed using the parameter $f_\text{cut}=0.05$.  The high energy parton is a color singlet (left) or color octet (right) Higgs boson $h$ with $m_h=125$ GeV and $p_\text{T}=500$ GeV.  To ensure no other significant radiation in the event, momentum is conserved by balancing the $h$ against a $Z(\rightarrow\nu\bar{\nu})$.  The histograms are the average jet image (see Sec.~\ref{sec:jetimages}) over all re-showers with $p_\text{T}$ intensity and the $L^2$ norm. }
  \label{fig:pull:colorflowimage}
\end{figure}

\begin{figure}[htbp]
  \centering
    \includegraphics[width=0.45\textwidth]{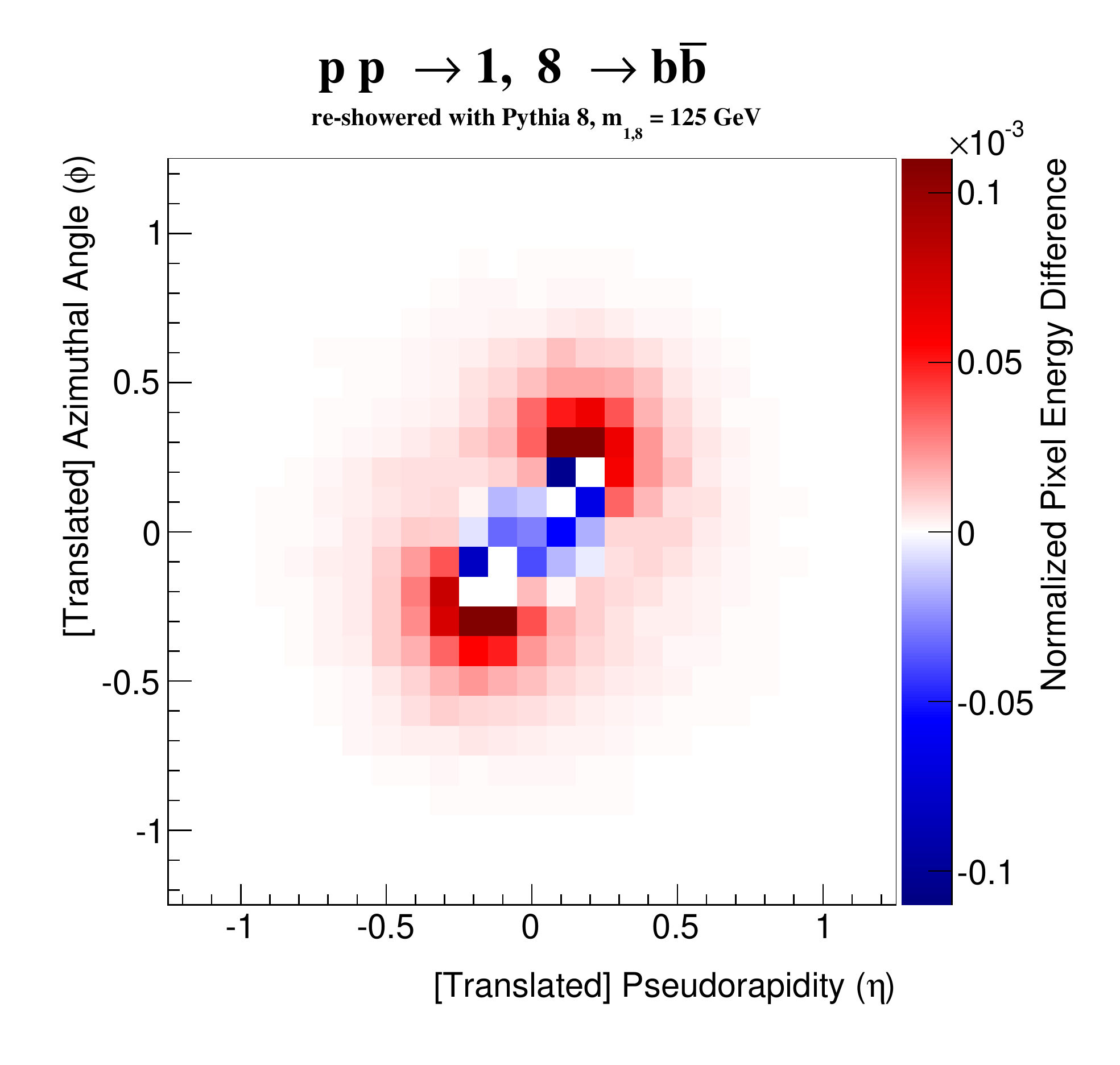}
  \caption{The average difference between the left and right distributions in Fig.~\ref{fig:pull:colorflowimage}.}
  \label{fig:pull:colorflowimage2}
\end{figure}

Color flow also has an impact on hadronization.  While there is no first-principle perturbative description of hadronization, there are a variety of phenomenological QCD-inspired models.  For example, in the popular string model, color connected quarks are bound by a linear confining potential that breaks into hadrons when the potential energy in the `string' is sufficiently large.  The resulting production of hadrons is enhanced between color connected partons. 

The discussion so far has been in the limit $N_c=\infty$.  For a finite number of colors, there are small effects due to {\it color reconnection}.  These effects are suppressed by $1/N_c^2$, which is comparable to $\alpha_s$.

\clearpage

\subsection{Jet Pull}

One observable predicted to contain information about the color representation
of a dijet resonance like the $W$, $Z$,
or Higgs boson, is the \textit{jet pull vector}~\cite{Gallicchio:2010sw}.
The pull vector for a given jet $J$ with transverse momentum, $p_\text{T}^{J}$,
is defined as
\begin{equation}
  \vec{v}^{J}_{p} = \sum_{i\in J} \frac{p_\text{T}^i |r_i|}{p_\text{T}^{J}}\vec{r}_i.
  \label{eq}
\end{equation}
The sum in Eq.~(\ref{eq}) runs over jet constituents with transverse momentum $p_\text{T}^i$ and location
$\vec{r}_i=(\Delta y_i,\Delta\phi_i)$, defined as the vector difference between the constituent and the
jet axis $(y_{J},\phi_{J})$ in rapidity ($y$) - azimuthal angle ($\phi$)
space. Given the pull vector for jet $J_1$, the angle formed between this pull vector and the vector connecting $J_1$ and another jet $J_2$,
$\vec{r}_{J_2}^{J_1}=(y_{J_2}-y_{J_1},\phi_{J_2}-\phi_{J_1})$, is expected to be sensitive to
the underlying color connections between the jets.
This is shown graphically in Fig.~\ref{fig:pull:def}, and the angle is called the
\textit{pull angle}, denoted $\theta_\text{P}(J_1,J_2)$.  Symbolically:

\begin{align}
\cos\theta_\text{P}(J_1,J_2)=(\vec{r}_{J_2}^{J_1}\cdot \vec{v}^{J_1}_{p} ) / (|\vec{r}_{J_2}^{J_1}|| \vec{v}^{J_1}_{p}|).
\end{align}

\noindent The pull angle is symmetric around zero when it takes values between $-\pi$ and
$\pi$ and so henceforth $\theta_\text{P}(J_1,J_2)$ refers to the magnitude of
the angle in $(\Delta y,\Delta\phi)$ space with $0< \theta_\text{P}\leq \pi$.
For jets originating from color--connected
quarks, $\theta_\text{P}\sim 0$ since the radiation is predicted to fall
mostly between the two jets. In other cases, $\theta_\text{P}$ need not be small, so the angle should be useful for determining color connections.

\begin{figure}[htbp]
  \centering
    \includegraphics[width=0.55\textwidth]{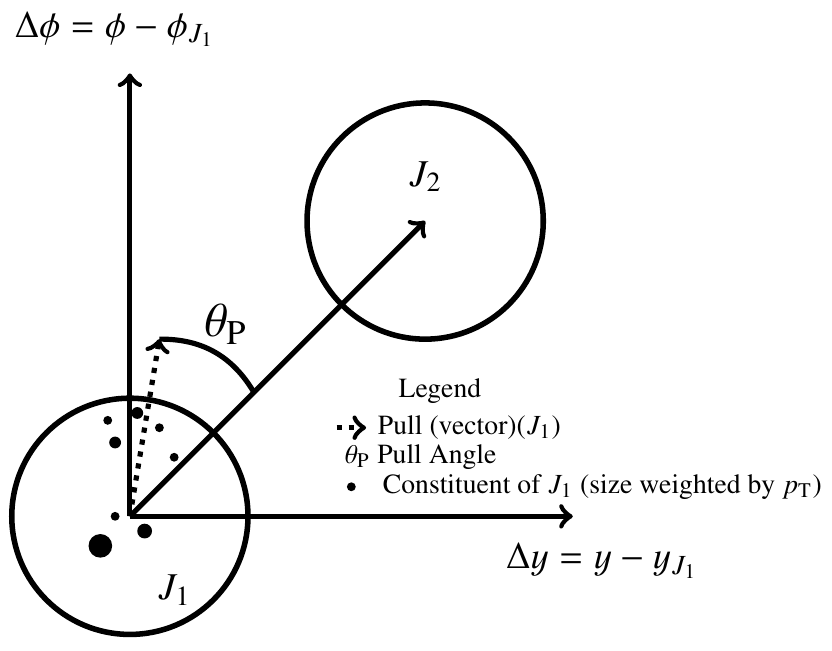}
  \caption{The construction of the jet pull angle for jet $J_1$ with respect to $J_2$.}
  \label{fig:pull:def}
\end{figure}

Due to the angular-weighting in Eq.~\ref{eq}, the contribution of large angle radiation is enhanced with respect to collinear radiation.  For jets of size $R=0.4$, this means that a jet constituent at the edge of a jet with transverse momentum $p_0$ contributes just as much to the jet pull vector as radiation in the jet core $(\Delta R\lesssim0.02)$ with $p_\text{T}\sim p_0(0.4/0.02)^2\sim 400p_0$.  This large radiation is sensitive to color flow, but it is also sensitive to the diffuse uncorrelated radiation in the event due to pileup and the underlying event.  Therefore, the the momentum weighting in Eq.~\ref{eq} is critical to suppress these contributions and formally render the jet pull infrared safe.  The jet pull is also collinear safe: for a particle $P_0\rightarrow P_1P_2$ collinearly, $\vec{r}_{P_0}=\vec{r}_{P_1}=\vec{r}_{P_2}$ and $p_\text{T,$P_0$}=p_\text{T,$P_1$}+p_\text{T,$P_2$}$ so $p_\text{T,$P_1$}|r_{P_1}|\vec{r}_{P_1}+p_\text{T,$P_2$}|r_{P_2}|\vec{r}_{P_2}=p_\text{T,$P_0$}|r_{P_0}|\vec{r}_{P_0}$.

Any jet constituent can be used to construct the jet pull angle, several of which are used in this chapter.  For detector-level jets, the natural constituent choice is the calorimeter-cell cluster.  This results in the {\it calorimeter pull} angle.  The analogous quantity for particle-level jets is the {\it all particles pull} angle using all jet constituents to compute Eq.~\ref{eq}.  Charged particle tracks are not used explicitly in jet reconstruction, but offer superior angular resolution to the calorimeter-cell clusters.  The {\it track pull} angle is built using tracks that are ghost-associated to the calorimeter jet.  An analogous {\it charged-paricles} pull angle is constructed for particle-level jets using only the electrically charged constituents.  The jet axis definition is an implicit input to Eq.~\ref{eq}.  The nominal jet four-vector is used for the particle-level pull angle axis.  A discussion of the axis for detector-level pull angle is postponed until Sec.~\ref{origincorrection}.

There are advantages and disadvantages for the track pull and likewise for the calorimeter pull.  Tracks have a better angular precision than calorimeter clusters and so the charged pull is more precisely measured than the all particles pull angle.  However, by using only charged particles, one is less sensitive to the underlying color flow and may also increase sensitivity to certain modeling uncertainties.  Both constituent inputs are studied in parallel.  

\clearpage

\section{Analysis Design}
\label{sec:colorflowanalysisdesign}

The measurement presented in this chapter demonstrates the ability to extract the color charge of a dijet resonance using the jet pull variable.  As part of this study, various performance aspects of the jet pull are studied in order to increase the precision of the measurement as well as to improve the tagging capabilities of the jet pull angle.  Events enriched in top quark pair production provide a clean environment and a copious source of hadronically decaying $W$ bosons, the model dijet resonance.  The data are unfolded to correct for resolution and acceptance effects.  These unfolded data are compared with various fragmentation models and can be used in the future to constrain models of jet formation.  In order to quantify the sensitivity to the color charge of a dijet resonance, $W$ bosons with an octet color charge are simulated and compared with the unfolded data.  

The dataset and simulated samples are summarized in Sec.~\ref{sec:ColorFlow:simulation}, along with a detailed discussion of the simulation for exotic color flow.  Section~\ref{sec:ColorFlowEventSelection} describes the object reconstruction and event selection.

\subsection{Dataset and Simulation}
\label{sec:ColorFlow:simulation}

The Monte Carlo simulation is similar to setup discussed in Sec~\ref{sec:ttbarsamplesandeventselection}.  Table~\ref{tab:colorflow:mc_samples} contains a summary of the simulation setups used for each SM process.  {\sc Powheg-Box}+{\sc Pythia} 6 is used for the nominal setup to model the $t\bar{t}$ process.  Top quark pair production is derived from the data by subtracting all other background processes, but the $t\bar{t}$ simulation is used to construct the response matrix for unfolding the background-subtracted data.  The additional $t\bar{t}$ samples listed in Table~\ref{tab:colorflow:mc_samples} are used as alternative models to assess systematic uncertainties.  

Aside from the $W$+jets background and multijets backgrounds, all MC samples are normalized to
their theoretical cross--sections, calculated to at least next--to--leading order (NLO) precision in
QCD~\cite{Catani:2009sm,Kidonakis:2010tc,Kidonakis:2010ux,Kidonakis:2011wy,Campbell:1999ah,Campbell:2011bn}.
For the purpose of comparison between between data and
the SM prediction before unfolding, $t\bar{t}$\ events are normalized to
a cross--section of $253\pm 15$ pb, calculated at
next--to-next--to--leading order (NNLO) in QCD
including next--to--next--to--leading logarithmic (NNLL)
soft gluon terms~\cite{Czakon:2011xx},
assuming a top--quark mass of $172.5$ GeV.  The $W$+jets and multijet normalizations are described in Sec.~\ref{sec:colorflow:datadrivenbackground}.

Generated events are processed with a full ATLAS detector
and trigger simulation~\cite{Aad:2010ah} based on
\textsc{Geant4}~\cite{Agostinelli:2002hh} and reconstructed using
the same software as the experimental data.
The effects of pileup are
modelled by adding to the generated hard--scatter events multiple minimum--bias
events simulated with \textsc{Pythia} 8.160 \cite{Sjostrand:2007gs},
the A2 set of tuned MC parameters (tune)~\cite{ATL-PHYS-PUB-2012-003} and the MSTW2008LO
Parton Distribution Function (PDF) set ~\cite{Martin:2009iq}. The distribution of the number of interactions is then weighted to reflect the pileup distribution in the data.

\begin{table}
  \centering
  \noindent\adjustbox{max width=\textwidth}{
  \begin{tabular}{cccccc}
    \toprule
    Process                               & Generator                                        & Type                 & Version & PDF                                  & Tune                          \\
    \midrule
                                                                                                                                                                                                     \\
    \multirow{2}{*}{$t\bar{t}$}               & {\sc Powheg-Box}~\cite{Nason:2004rx,Frixione:2007vw,Alioli:2010xd} & NLO ME               & -       & CT10~\cite{Lai:2010vv,Gao:2013xoa} & -                             \\
                                          & +{\sc Pythia} 6~\cite{Sjostrand:2006za}                          & + PS                 & 6.426.2 & CTEQ6L1~\cite{Pumplin:2002vw}          & Perguia2011c~\cite{Skands:2010ak} \\
                                                                                                                                                                                                     \\
   Single top         & {\sc Powheg-Box}                                          & NLO ME               &         & CT10(4f)                           & DR scheme ($Wt$)~\cite{Frixione:2008yi}                           \\
    ($t$-,$s$-, and $Wt$-channels)                                      & +{\sc Pythia} 6                                         & + PS                 & 6.426.2 & CTEQ6L1                                  & Perguia2011c                      \\
                                                                                                                                                                                                     \\
    $WW,WZ,ZZ$                            & \textsc{Sherpa}~\cite{Gleisberg:2008ta}          & LO multi--leg ME + PS & 1.4.1   & CT10                               & Default                       \\
                                                                                                                                                                                                     \\
    \multirow{2}{*}{$W/Z$+jets}           & \textsc{Alpgen}~\cite{Mangano:2002ea}            & LO multi--leg ME      & 2.1.4   & CTEQ6L1                                & -                             \\
                                          & +{\sc Pythia} 6                                         & + PS                 & 6.426.2 & CTEQ6L1                                  & Perguia2011c                      \\
                                                                                                                                                                                                     \\
    \midrule
                                                                                                                                                                                                     \\
    \multirow{3}{*}{$t\bar{t}^{\,\dagger}$} & {\sc Powheg-Box}                                          & NLO ME               & -       & CT10                               & -                             \\
                                          & +{\sc Herwig}~\cite{Corcella:2000bw}                  & + PS                 & 6.520.2 & CT10                               & AUET2~\cite{AUET2}            \\
                                          & +{\sc Jimmy}~\cite{JIMMY}                             & (MPI)                & 4.31    & -                                    & -                             \\
                                                                                                                                                                                                     \\
    \multirow{3}{*}{$t\bar{t}^{\,\dagger}$} & {\sc MC@NLO}\cite{Frixione:2002ik,Frixione:2003ei}     & NLO ME               & 4.06    & CT10                               & -                             \\
                                          & +{\sc Herwig}                                         & + PS                 & 6.520.2 & CT10                               & AUET2                         \\
                                          & +{\sc Jimmy}                                          & (MPI)                & 4.31    & -                                    & -                             \\
                                                                                                                                                                                                     \\
    \bottomrule
  \end{tabular}}
  \caption{Monte Carlo samples used in this analysis. The abbreviations ME, PS, PDF, 
    MPI, LO and NLO respectively stand for matrix element, parton shower, parton distribution function,
    multiple parton interactions,
    leading order and next--to--leading order in QCD. Tune refers to the used set of 
tunable MC parameters.  Those samples marked
    with a $\dagger$ are used as alternative $t\bar{t}$ samples to evaluate
    uncertainties due to the modeling of $t\bar{t}$ events. The 4-flavor scheme (4f) is used for CT10 only for the $t$-channel single top production. }
  \label{tab:colorflow:mc_samples}
\end{table}

\subsubsection{Data-driven backgrounds}
\label{sec:colorflow:datadrivenbackground}

The event selection is described in Sec.~\ref{sec:ColorFlowEventSelection}, but the $W$+jets and QCD multijet backgrounds are normalized using general data-driven techniques~\cite{Aad:2010ey}.  In order for an event featuring only quarks and gluons to pass the event selection, a non-prompt lepton from a semi-leptonic heavy quark decay or a hadron faking a lepton must be reconstructed as a signal lepton.  This background is estimated from data with the {\it matrix method}. Two data samples -- tight and loose -- are defined based on their observed lepton isolation, where all tight leptons are also loose leptons.  Tight leptons are the ones used in the event selection, described in Sec.~\ref{sec:ColorFlowEventSelection}.  The number of events passing the loose or tight selection can be decomposed as $N^\text{loose} =  N_\text{fake}^\text{loose}+ N_\text{real}^\text{loose}$ and $N^\text{tight}= N_\text{fake}^\text{tight}+ N_\text{real}^\text{tight}$.  Let $\epsilon_X$ be the probability for a $X\in\{\text{real},\text{fake}\}$ event to pass the tight selection given that it passes the loose selection.  Then, since all events that pass the tight selection also passed the loose selection, $N^\text{tight}= \epsilon_\text{fake} N_\text{fake}^\text{loose}+\epsilon_\text{real} N_\text{real}^\text{loose}$.  Solving for number of fake events, $N_\text{fake}^\text{tight}$:

\begin{align}
N^\text{tight}_\text{fake}=\frac{\epsilon^\text{fake}}{\epsilon^\text{real}-\epsilon^\text{fake}}\cdot\left(\epsilon_\text{real}N^\text{loose}-N^\text{tight}\right).
\end{align}

\noindent The probabilities $\epsilon_\text{real}$ and $\epsilon_\text{tight}$ are measured with auxiliary event selections that are enriched in real leptons ($Z\rightarrow l^+l^-$) or fake leptons (low $m_\text{T}$ and low $E_\text{T}^\text{miss}$; see Sec.~\ref{sec:ColorFlowEventSelection}).  Additional details can be found in Ref.~\cite{ATLAS-CONF-2014-058}.  Estimates from the matrix method are often reliable when the number of events is large, but there are known experimental and theoretical challenges\footnote{For a discussion of some of the theoretical challenges, see Ref.~\cite{Gillam:2014xua}; these are of the same flavor as  for the CR method, described in Sec.~\ref{sec:CRmethod}.  Alternative methods similar to the matrix method exist (see e.g. the fake-factor method in Ref.~\cite{ATLAS:2014aga}), but they have their own set of challenges.}.  In this chapter, the multijet background is small and the event selection is inclusive enough that these challenged can be ignored.

The $W$+jets process is normalized by exploiting the asymmetry of $W^+$ and $W^-$ events produced at the LHC due to the charge asymmetric initial state.  There are about $30\%$ more $W^+$ events than $W^-$ events and this ratio slightly increases with jet multiplicity beyond $n_\text{jet}>0$ and the uncertainty on the theoretical uncertainty on the ratio is $\mathcal{O}(\%)$~\cite{Kom:2010mv}.  Since the efficiency for passing the event selection and the distribution of the pull angle are nearly independent of the lepton charge, the total number of $W$+jets events in a given bin can be estimated by

\begin{align}
\label{eq:chargeasymmetry}
N_{W^+}+N_{W^-}=\frac{N_{W^+}^\text{MC}+N_{W^-}^\text{MC}}{N_{W^+}^\text{MC}-N_{W^-}^\text{MC}}\left(N_{W^+}^\text{Data}-N_{W^-}^\text{Data}\right),
\end{align}

\noindent where other backgrounds do not contribute to the parenthetical term as they are charge symmetric.  One large source of uncertainty on the $W$+jets yield is the heavy-flavor composition, as $b$-jets will be part of the event selection.  While $b$-quarks are generated from $t\bar{t}$ production at tree-level, they are generated by higher order processes in $W$ boson production. The charge asymmetry method is extended to include residual scale factors per $W$+jets flavor subprocess through charge-independent corrections for $W$+bb/cc, $W$+c, and $W$+light.  The $W$+cc is grouped with $W$+bb and not with $W$+c because the two processes have a different charge asymmetry.   See Ref.~\cite{Aad:2014zka} for more detail.   The overall scale factors are $1.3\pm 0.03$ for the $W$+bb/cc component, $0.74\pm 0.04$ for the $W$+c component and $0.96\pm 0.02$ for the light component.  

The charge asymmetry method is experimentally and theoretically robust, but the main drawback is the large sample size required for an acceptable precision.  Figure~\ref{fig:chargeasymmetry} shows that almost $1000$ events are required to reach a statistical uncertainty of $10\%$.  In this chapter, the event selection is inclusive enough and the $W$+jets background is sufficiently small that this is unimportant.  However, the search presented in Part~\ref{part:susy} that probes extreme regions of phase space must use a different technique.

\begin{figure}[htbp]
  \centering
    \includegraphics[width=0.5\textwidth]{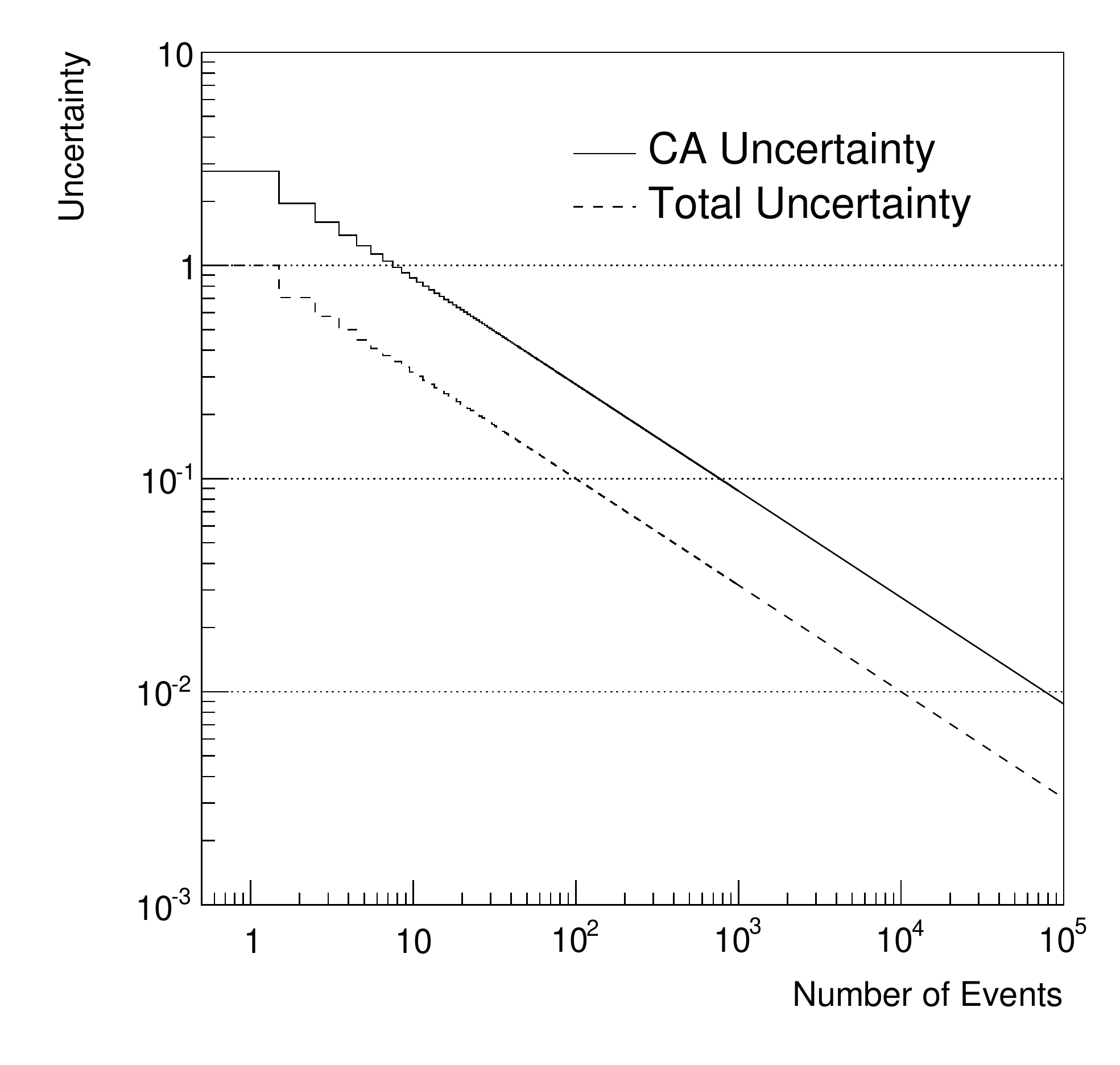}
  \caption{The total statistical uncertainty and the charge-asymmetry statistical uncertainty of a pure sample of $W$+jets events with $30\%$ more $W^+$ than $W^-$.}
  \label{fig:chargeasymmetry}
\end{figure}

\clearpage

\subsubsection{Simulating Exotic Colorflow}
\label{sec:ColorFlowExoticSimulation}

To test the sensitivity of the jet pull angle to the singlet nature of
the $W$ boson, a simulated $t\bar{t}$ is generated with a color--octet $W$ boson.  Using the partons produced with {\sc Powheg-box} recorded in the Les Houches Accord format~\cite{Alwall:2006yp}, the color flow is inverted such that one of the $W$ decay daughters shares a color line with the $b$--quark and the other shares a line with the top quark,
as demonstrated schematically in Fig.~\ref{fig:colorflow}. This sample is referred to as color \textit{flipped} in the rest of this chapter.

\begin{figure}[htbp]
  \centering
    \includegraphics[width=0.6\textwidth]{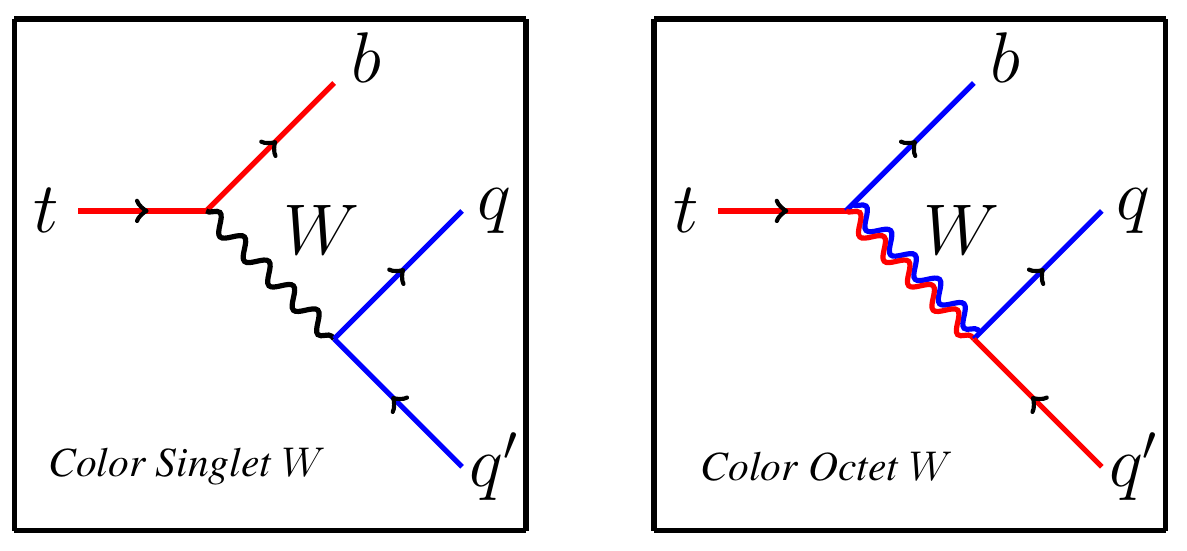}
  \caption{Diagram illustrating the color connections for the nominal sample with a color-singlet $W$ (left) and the flipped sample with a color-octet $W$ (right).}
  \label{fig:colorflow}
\end{figure}

A sample event from a typical $t\bar{t}$ LHE file is shown in the top panel of Fig.~\ref{flip}.  The LHE contains only the hard scatter information, including the incoming protons, the partons participating in the scattering, the outgoing top quarks, and the immediate top quark decay products.  Some events in an LHE file from {\sc Powheg-box} will also contain an additional parton from ISR from real emission in the NLO matrix element.  The second and third columns in Fig.~\ref{flip} give some particle identification information, the fourth-sixth columns describe where in the event record the particles belong and the rest of the columns list the particles quantum numbers.  The most relevant for this section is the column for the `color' quantum number.  All of the MC generators use a large $N_c$ limit and assign a new color-anti color whenever a color charge is created; in this case the color is arbitrarily called 501.  The first of the color columns is the color and the second column is the anti-color.  Color cannot be created or destroyed so the total color must be the same at the end of the hard scatter (status 23) as it was in the beginning (status -21).  The most relevant numbers in the example in Fig.~\ref{flip} are the color and anti-color of the quark and anti-quark decay products of the $W^-$: {\color{red} 502}.  Note that the $W^-$ and the $\bar{t}$ are {\it removed from the event record}.  This is one of the drawbacks from the color-flipping procedure\footnote{An alternative method is to directly simulate a color octet using a UFO in MadGraph.  This setup was found to give qualitatively similar results for the pull angle distribution as the flipped sample.  However, as it is LO, there are significant kinematic differences with compared to {\sc Powheg-box}.}: the top quarks and $W$ bosons participating in the color flipping must be removed\footnote{{\sc Pythia} will produce radiation differently if the top quark is absent from the LHE file.  This is because the virtuality of the top quark is preserved if present and otherwise, the radiation off of the dipole formed from the out-going $b$ and its color-partner in the initial state can produce much larger radiation (as the invariant mass is easily $\mathcal{O}(TeV)$).  This may account for some of the changes in the mass distributions presented in this section.  Thank you to Stefan Prestel for the explanation.}, otherwise the shower generator will identify an unphysical color flow (as $W$ bosons are singlets in the SM).  Just before the $W$ is removed, its decay products are identified and their color strings are flipped with the $\bar{b}$ in the event.  This is seen in the lower panel of Fig.~\ref{flip}, where the quark from the $W$ has color {\color{red}502} while the anti-quark has color {\color{blue}503} (the anti-color from the original $W$ is with the $\bar{b}$).  It is then critical that the particle numbers and mothers are re-aligned since the showering models all have many internal consistency checks.  Some care must be taken for generators that allow for $b$ quarks in the PDF.  This is solved by checking that the parent of the $b$ or $\bar{b}$ is not the proton.

\begin{figure}[h!]
\begin{center}
\begingroup
    \fontsize{6pt}{6pt}\selectfont
\begin{Verbatim}[commandchars=\\\{\},codes={\catcode`$=3\catcode`^=7\catcode`_=8}]
 --------  PYTHIA Event Listing  (hard process)  -----------------------------------------------------------------------------------
 
    no        id   name            status     mothers   daughters     colors      px        py        pz             
     0        90   (system)           -11     0     0     0     0     0     0      0.000      0.000      0.000   8000.000  
     1      2212   (p+)               -12     0     0     3     0     0     0      0.000      0.000   4000.000   4000.000     
     2      2212   (p+)               -12     0     0     4     0     0     0      0.000      0.000  -4000.000   4000.000     
     3        21   (g)                -21     1     0     5     8  \color{green!80!black} 504  \color{blue} 503   \color{black}   0.000      0.000     94.198     94.198    
     4        21   (g)                -21     2     0     5     8  \color{yellow!80!black} 501  \color{green!80!black} 504 \color{black}     0.000      0.000   -466.748    466.748     
     5         6   (t)                -22     3     4     9    10  \color{yellow!80!black} 501 \color{black}    0     81.617    -59.344   -265.703    333.045    
     6        -5   bbar                23     3     4     0     0     0 \color{blue}  503   \color{black}  -0.916    -37.444    -61.245     71.944   
     7         3   s                   23     3     4     0     0   \color{red}502 \color{black}    0    -54.108     10.412     -0.850     55.107      
     8        -4   cbar                23     3     4     0     0     0   \color{red}502\color{black}    -26.593     86.377    -44.751    100.851     
     9        24   (W+)               -22     5     0    11    12     0     0     86.539    -59.228    -98.855    166.058    
    10         5   b                   23     5     0     0     0  \color{yellow!80!black} 501 \color{black}    0     -4.922     -0.116   -166.849    166.988     
    11       -13   mu+                 23     9     0     0     0     0     0     33.200      0.577      6.352     33.807     
    12        14   nu mu               23     9     0     0     0     0     0     53.339    -59.805   -105.207    132.250      
                                   Charge sum:  0.000           Momentum sum:     -0.000      0.000   -372.550    560.947   

 --------  PYTHIA Event Listing  (hard process)  -----------------------------------------------------------------------------------
 
     no        id   name            status     mothers   daughters     colors      px        py        pz             
     0        90   (system)           -11     0     0     0     0     0     0      0.000      0.000      0.000   8000.000  
     1      2212   (p+)               -12     0     0     3     0     0     0      0.000      0.000   4000.000   4000.000     
     2      2212   (p+)               -12     0     0     4     0     0     0      0.000      0.000  -4000.000   4000.000     
     3        21   (g)                -21     1     0     5     8  \color{green!80!black} 504  \color{blue} 503   \color{black}   0.000      0.000     94.198     94.198    
     4        21   (g)                -21     2     0     5     8  \color{yellow!80!black} 501  \color{green!80!black} 504 \color{black}     0.000      0.000   -466.748    466.748     
     5         6   (t)                -22     3     4     9    10  \color{yellow!80!black} 501 \color{black}    0     81.617    -59.344   -265.703    333.045    
     6        -5   bbar                23     3     4     0     0     0 \color{red}  502  \color{black}   -0.916    -37.444    -61.245     71.944   
     7         3   s                   23     3     4     0     0   \color{red}502 \color{black}    0    -54.108     10.412     -0.850     55.107      
     8        -4   cbar                23     3     4     0     0     0   \color{blue}503\color{black}    -26.593     86.377    -44.751    100.851     
     9        24   (W+)               -22     5     0    11    12     0     0     86.539    -59.228    -98.855    166.058    
    10         5   b                   23     5     0     0     0  \color{yellow!80!black} 501 \color{black}    0     -4.922     -0.116   -166.849    166.988     
    11       -13   mu+                 23     9     0     0     0     0     0     33.200      0.577      6.352     33.807     
    12        14   nu mu               23     9     0     0     0     0     0     53.339    -59.805   -105.207    132.250      
                                   Charge sum:  0.000           Momentum sum:     -0.000      0.000   -372.550    560.947   
 
\end{Verbatim}         
\endgroup    
\end{center}
\caption{One LHE event for a nominal (top) and inverted (bottom) color flow.}
\label{flip}
\end{figure}

The remainder of this section shows various distributions at particle-level with the nominal and flipped $t\bar{t}$ simulated samples.  Both samples use {\sc Pythia} 6 with the same settings for the fragmention.  To begin, Fig.~\ref{fig:truth:WMflip_noflip} shows the invariant mass of the leading two non $b$-tagged jets.  Like the pull vector, the dijet mass is sensitive to relatively soft wide angle radiation.  The dijet mass squared is $m_{jj}^2\approx \sum e_ie_j\theta^2$, where $e_i$ is the energy of jet $i$ and $\theta$ is the angle between jets $i$ and $j$.  Because of this angular weighting, the dijet mass distribution is distorted by inverting the color flow.  In particular, the dijet invariant mass is slightly larger for the flipped sample due to the enhancement of radiation around and a reduction of radiation between the two jets with respect to the nominal sample.  Therefore, the jet mass distribution could be used to constrain the color flow, along with the pull angle.  However, this is not used in the subsequent analysis.  This is because if a new resonance were discovered, its mass would not be known a priori and therefore would not contain any useful information about the color flow.  Furthermore, when using the jet pull for tagging jets originating from bosons with a known mass, it is likely that a mass requirement will already be applied.  It is important to ensure that there is more information about the boson color flow aside from the jet mass distribution. This is tested by re-weighting the $m_{jj}$ and $\Delta R$ (between the $W$ daughter jets) distributions, as in the right plot of Fig.~\ref{fig:truth:WMflip_noflip}.   Since they are determined mostly by the hard-scatter, the other kinematic properties of the jets are similar between the simulations.  The individual $\eta$ values of the non $b$-tagged jets are nearly identical, but there are small differences at low $p_\text{T}$ for the momenta, shown in Fig.~\ref{fig:truth:j1ptflip_noflip}.  This is due in part to the higher loss for octet (gluon) jets in the transition from partons to jets compared with triplets (quarks)~\cite{Salam:2009jx}.  The most important distribution is $\theta_\text{P}$, shown in Fig.~\ref{fig:truth:ThetaJ1J2flip_noflip}.  There is a significant difference between the flipped and nominal samples, which is largely invariant to the mass reweighting. This is also true for $|v_\text{P}^J|$, which will be revisited in Sec.~\ref{sec:ColorFlow:UnfoldingParams}.  For comparison, the charged-particles versions of $\theta_\text{P}$ and  $|v_\text{P}^J|$ are shown in Fig.~\ref{fig:truth:ThetaJ1J2qflip_noflip}.

\begin{figure}[h!]
\begin{center}
\includegraphics[width=0.44\textwidth]{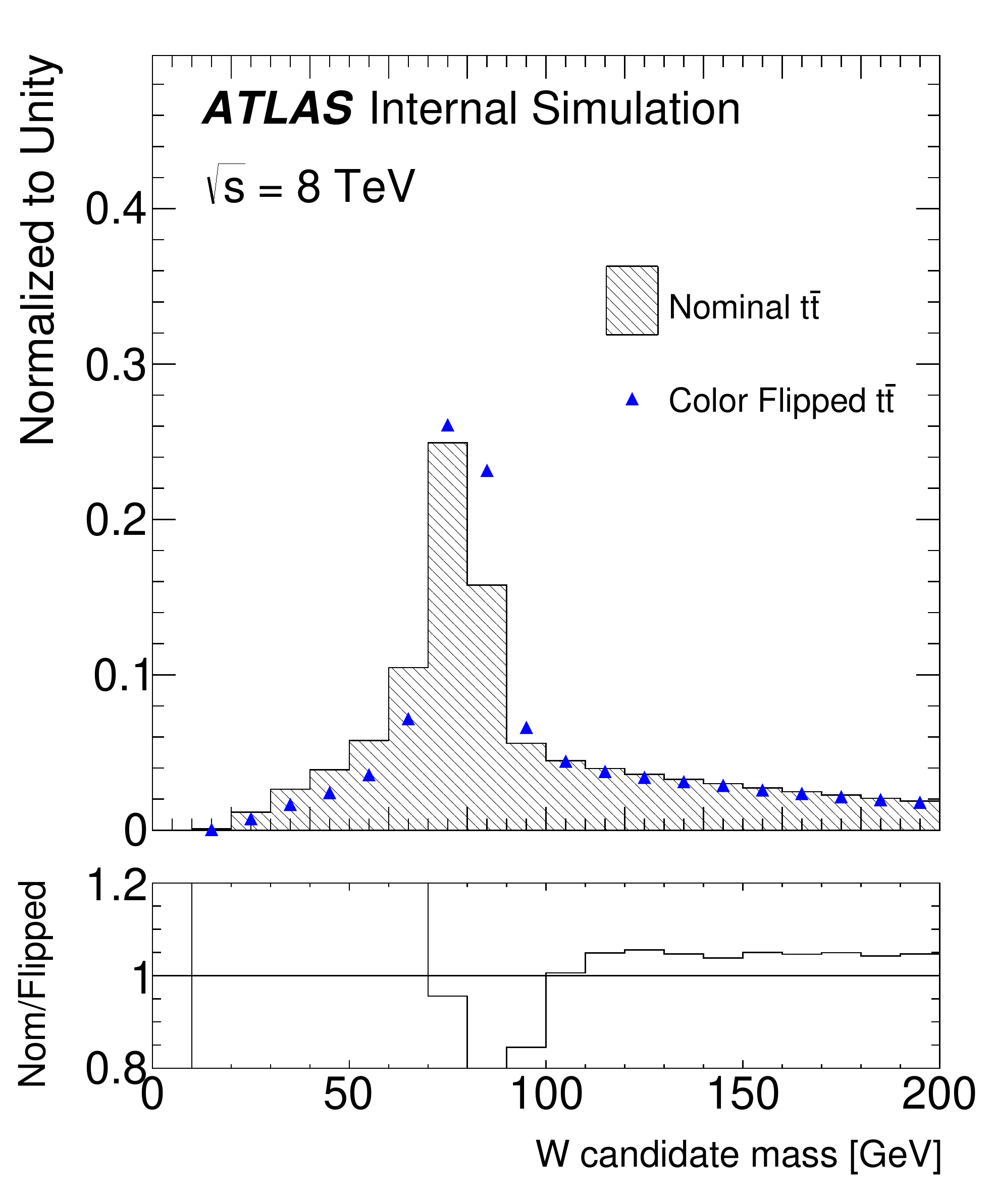}\includegraphics[width=0.44\textwidth]{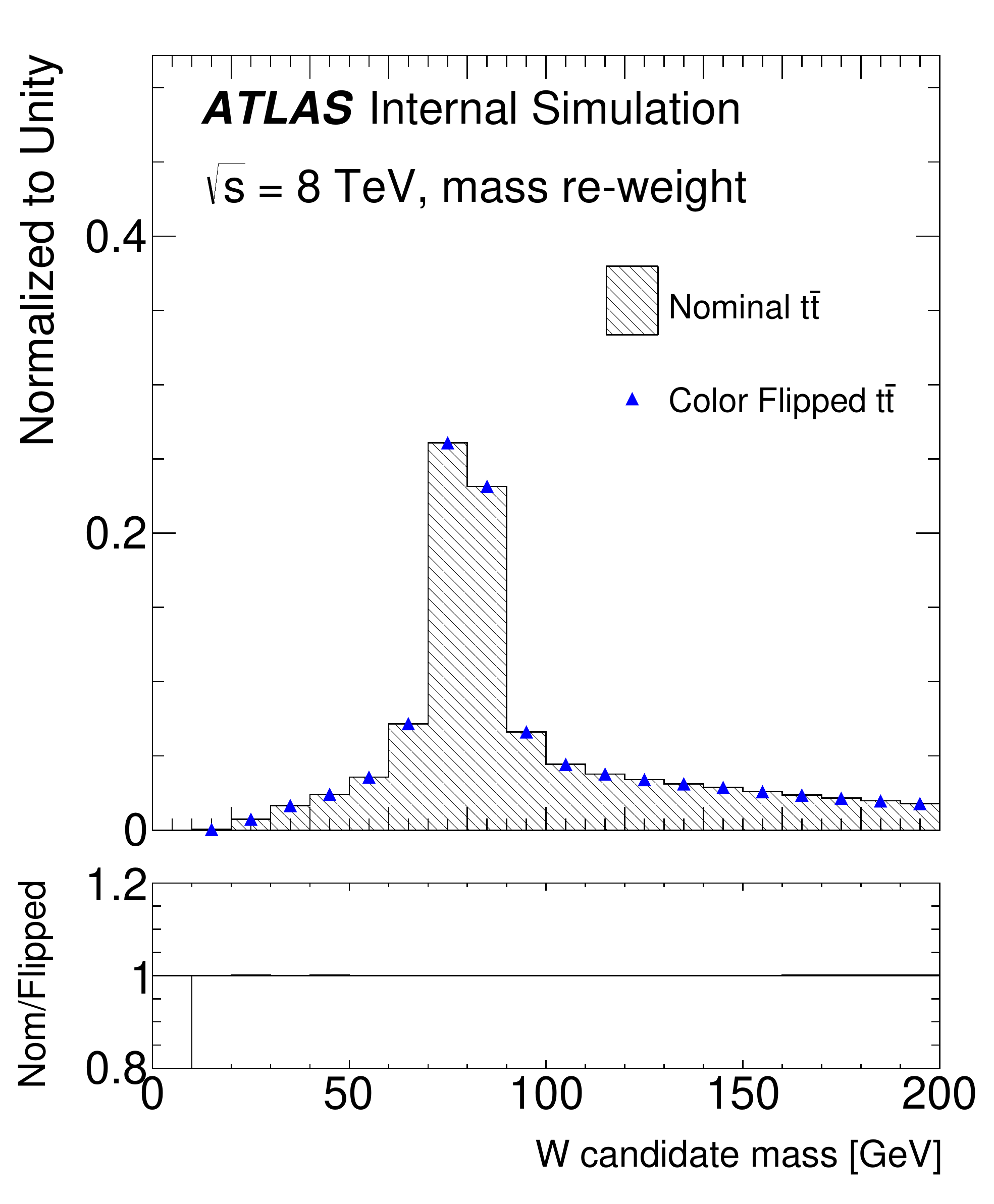}
 \caption{The invariant mass of the leading non $b$-jets with (right) and without (left) a reweighting to the invariant mass spectrum.  }
 \label{fig:truth:WMflip_noflip}
  \end{center}
\end{figure}

\begin{figure}[h!]
\begin{center}
\includegraphics[width=0.44\textwidth]{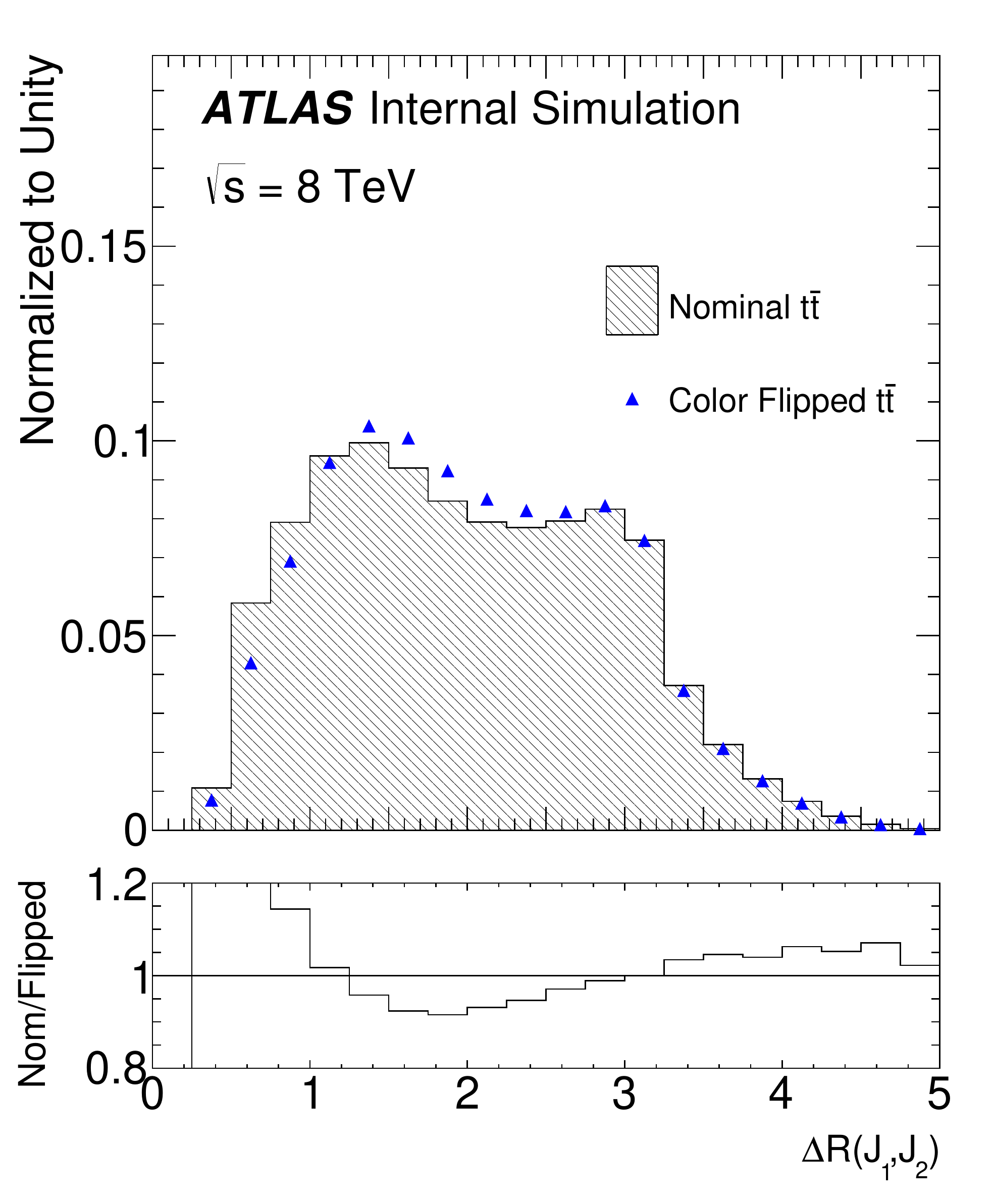}\includegraphics[width=0.44\textwidth]{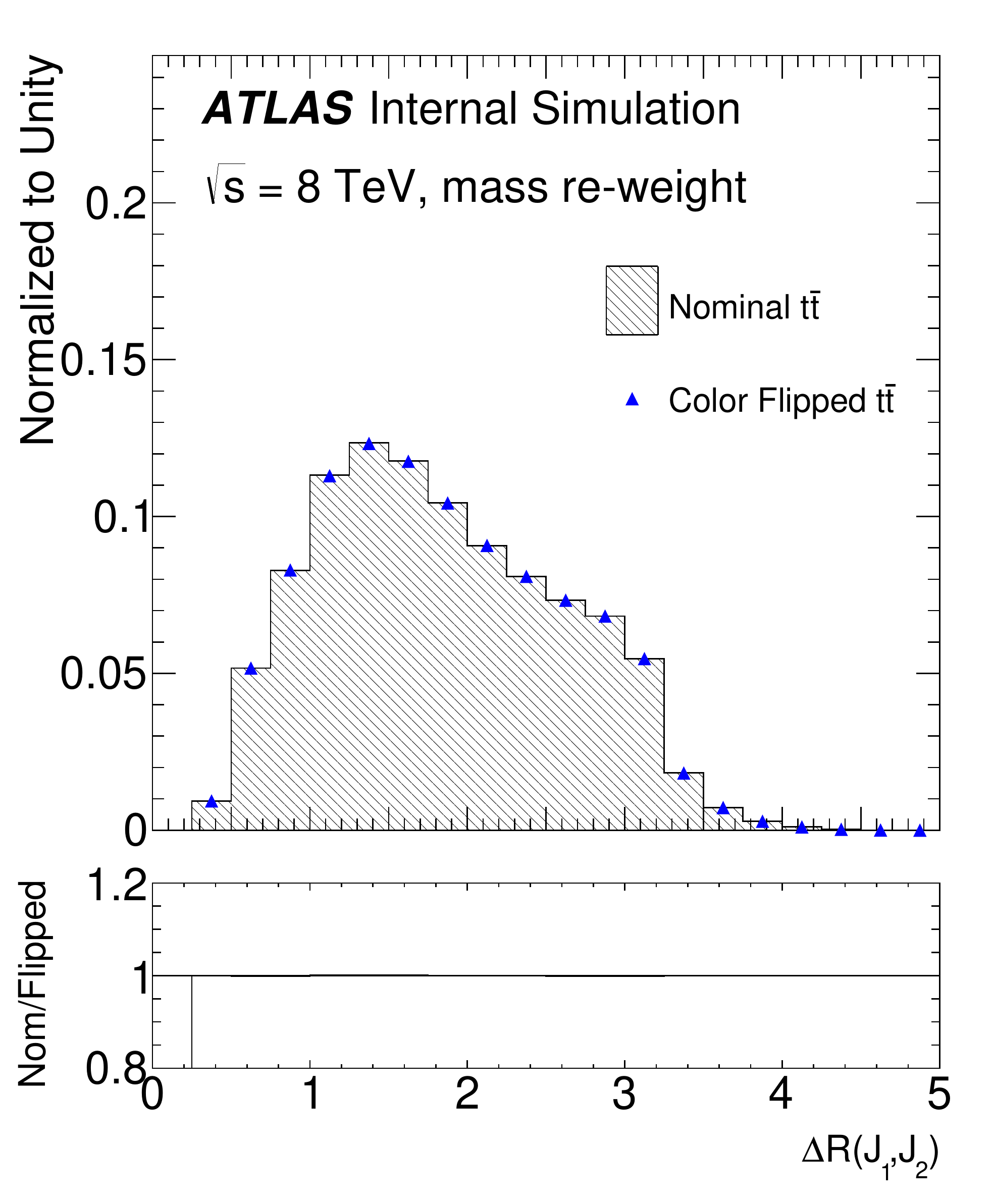}
 \caption{The $\Delta R$ between the non $b$-tagged jets with (right) and without (left) a reweighting to the invariant mass and $\Delta R$ spectra.}
 \label{fig:truth:DRflip_noflip}
  \end{center}
\end{figure}

\begin{figure}[h!]
\begin{center}
\includegraphics[width=0.44\textwidth]{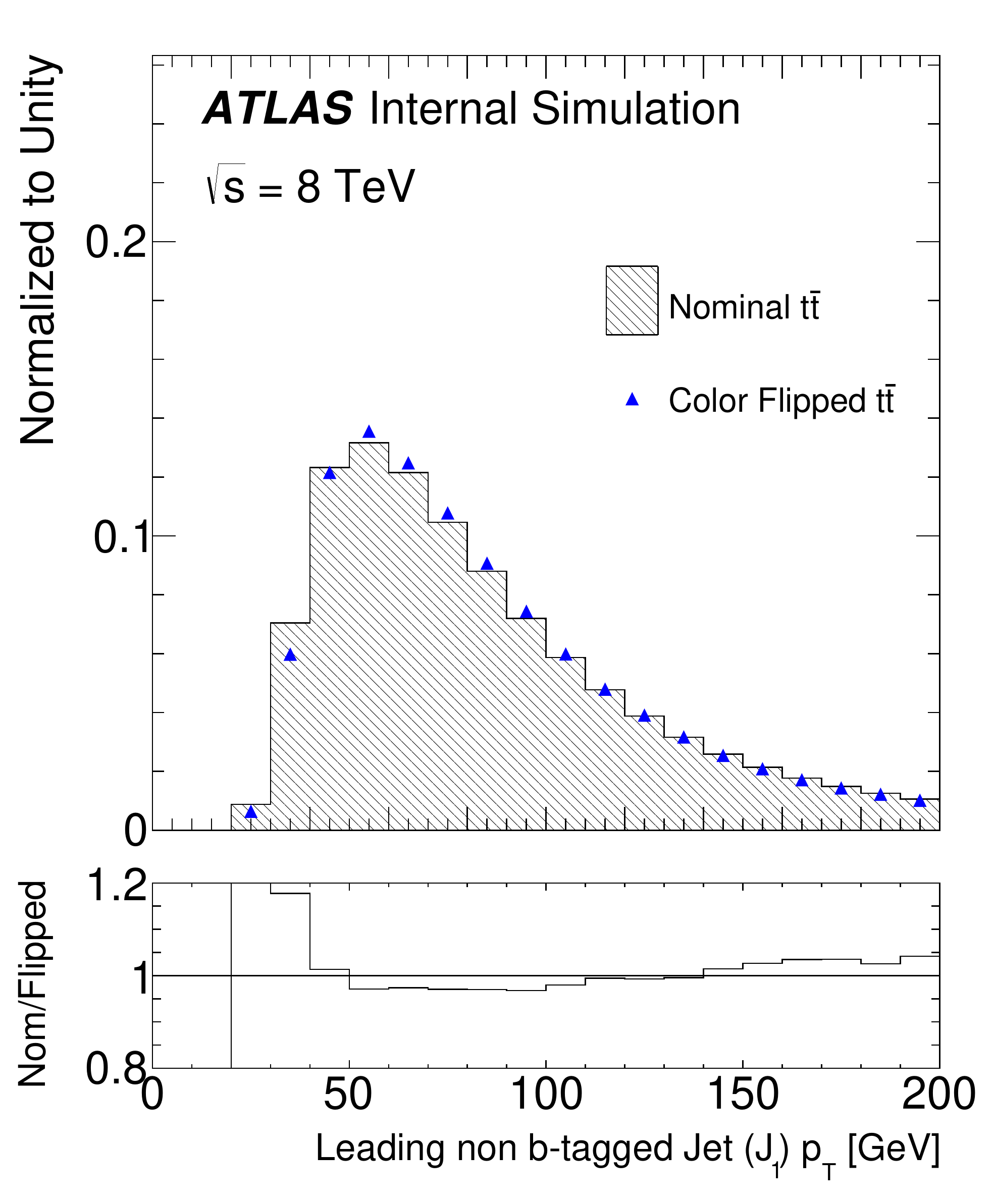}\includegraphics[width=0.44\textwidth]{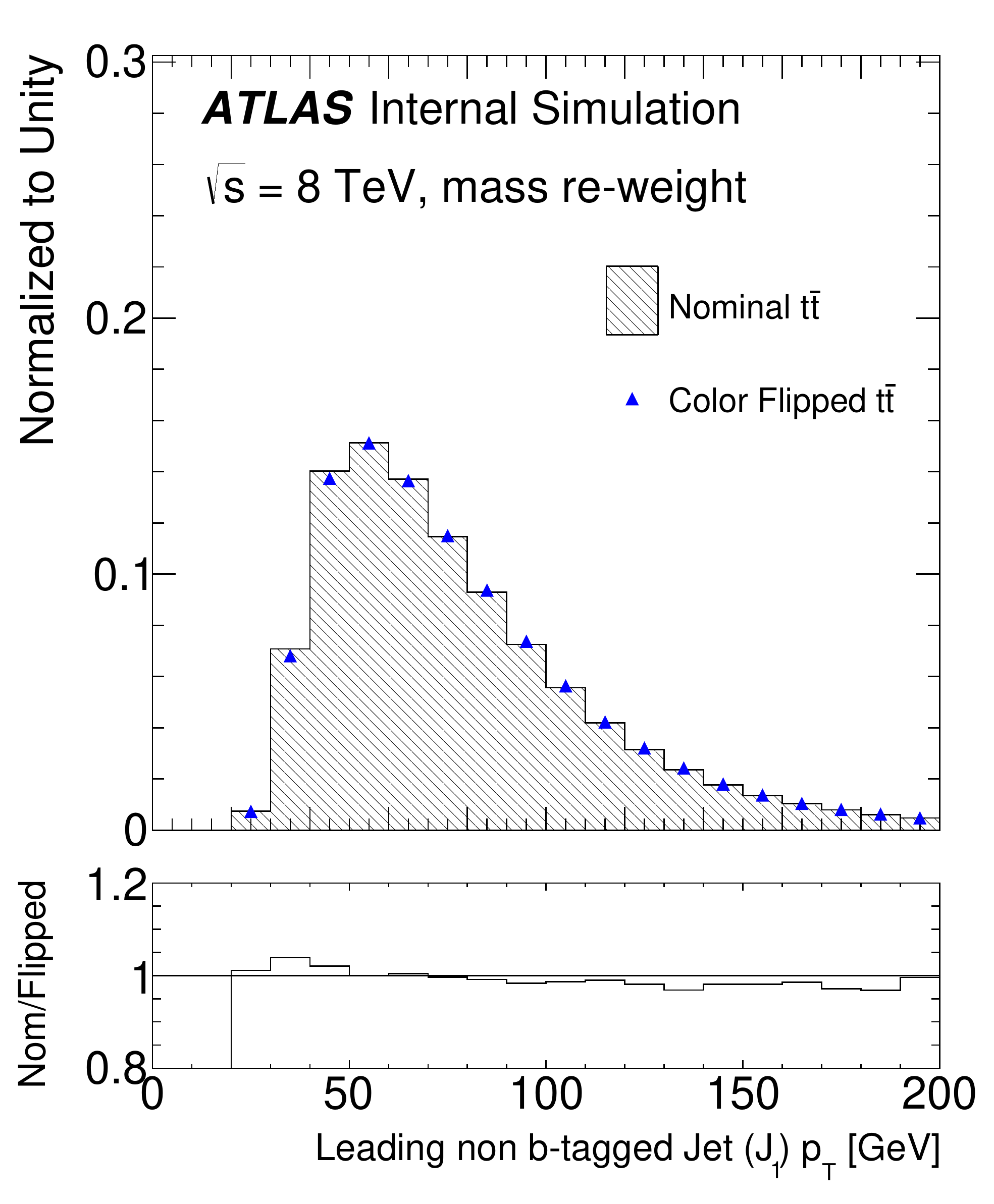}
 \caption{The $p_T$ of the leading non $b$-tagged jet with (right) and without (left) a reweighting to the invariant mass and $\Delta R$ spectra.}
 \label{fig:truth:j1ptflip_noflip}
  \end{center}
\end{figure}

\begin{figure}[h!]
\begin{center}
\includegraphics[width=0.44\textwidth]{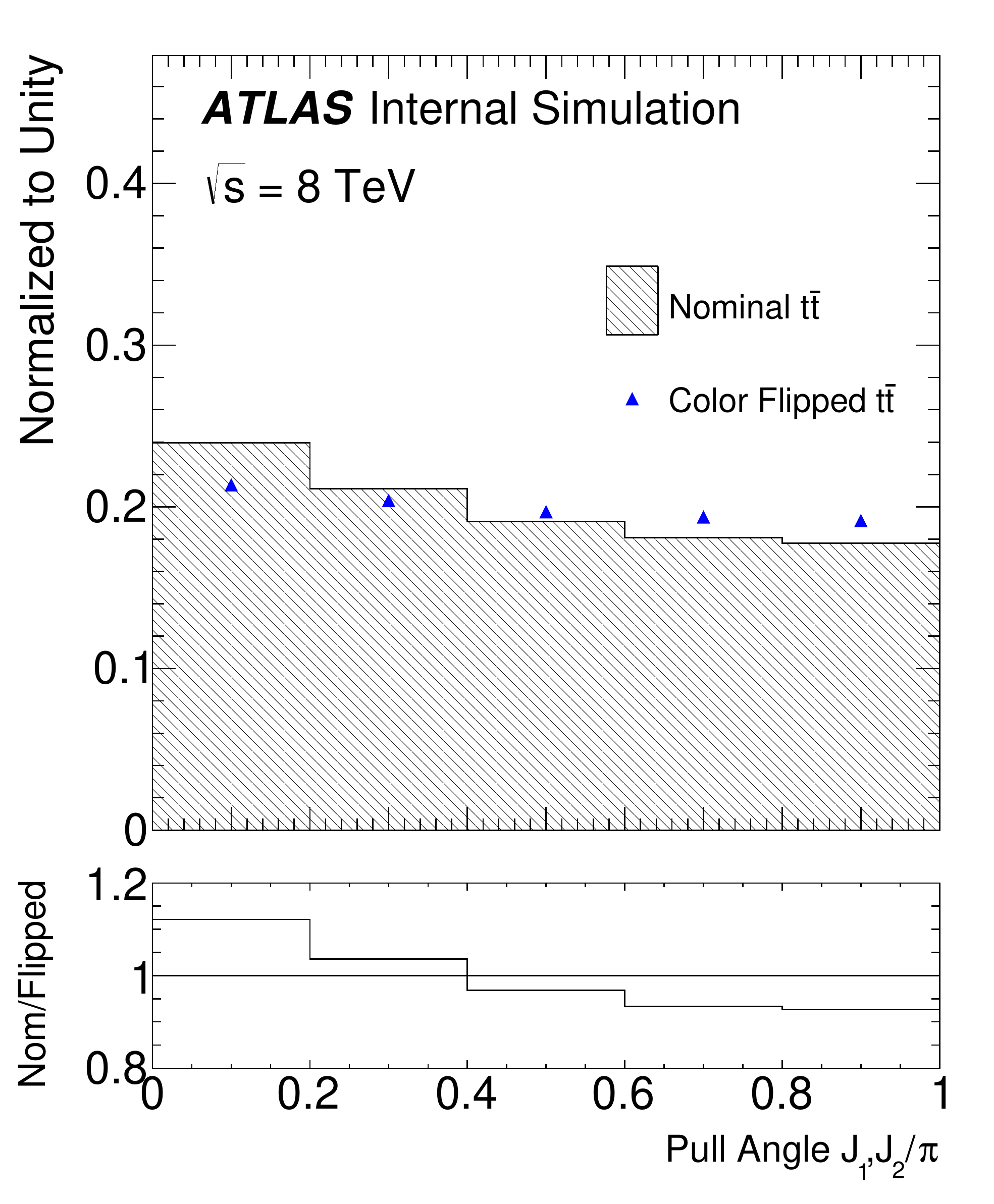}\includegraphics[width=0.44\textwidth]{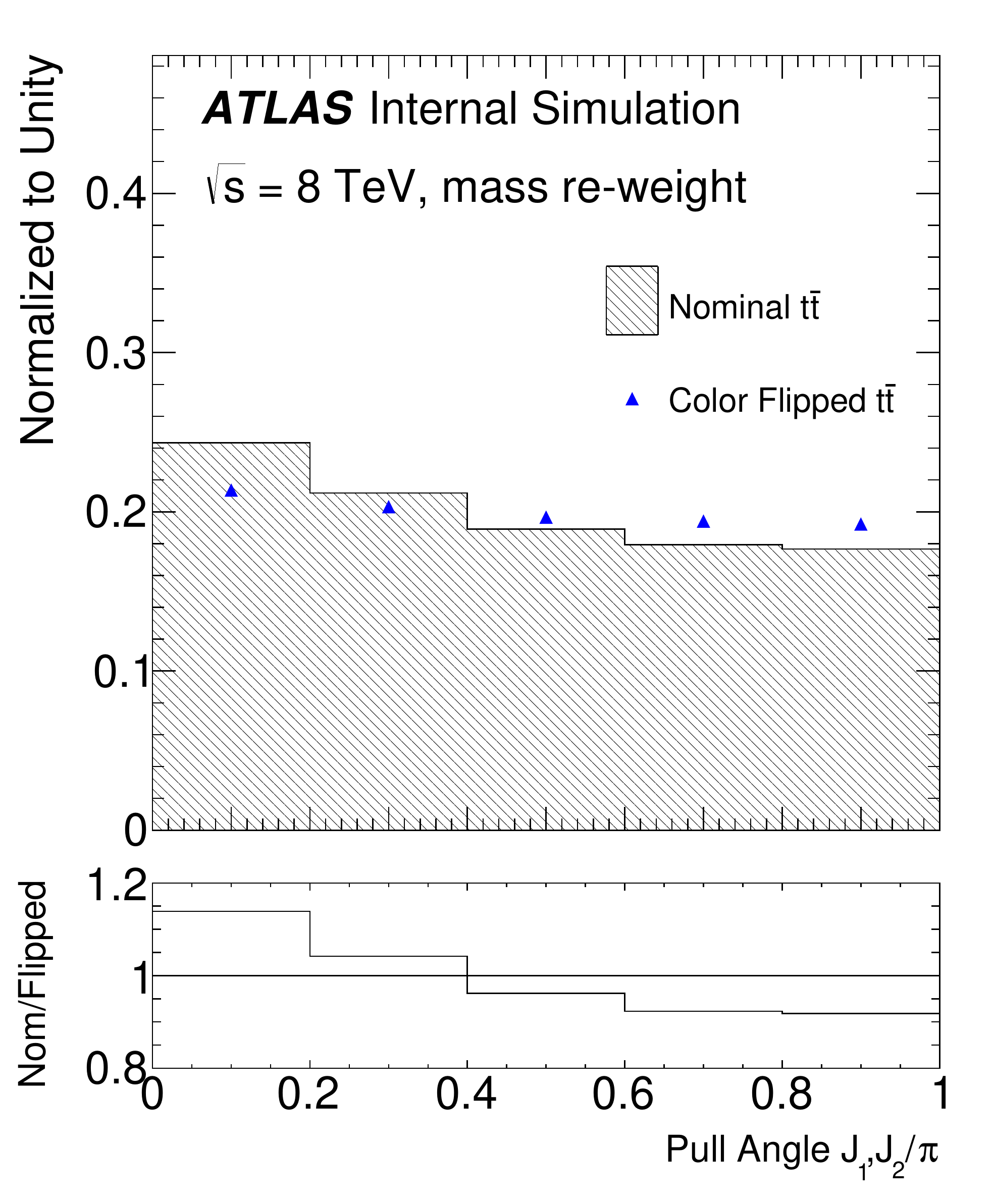}
 \caption{The pull angle with (right) and without (left) a reweighting to the invariant mass and $\Delta R$ spectra.}
 \label{fig:truth:ThetaJ1J2flip_noflip}
  \end{center}
\end{figure}

\begin{figure}[h!]
\begin{center}
\includegraphics[width=0.44\textwidth]{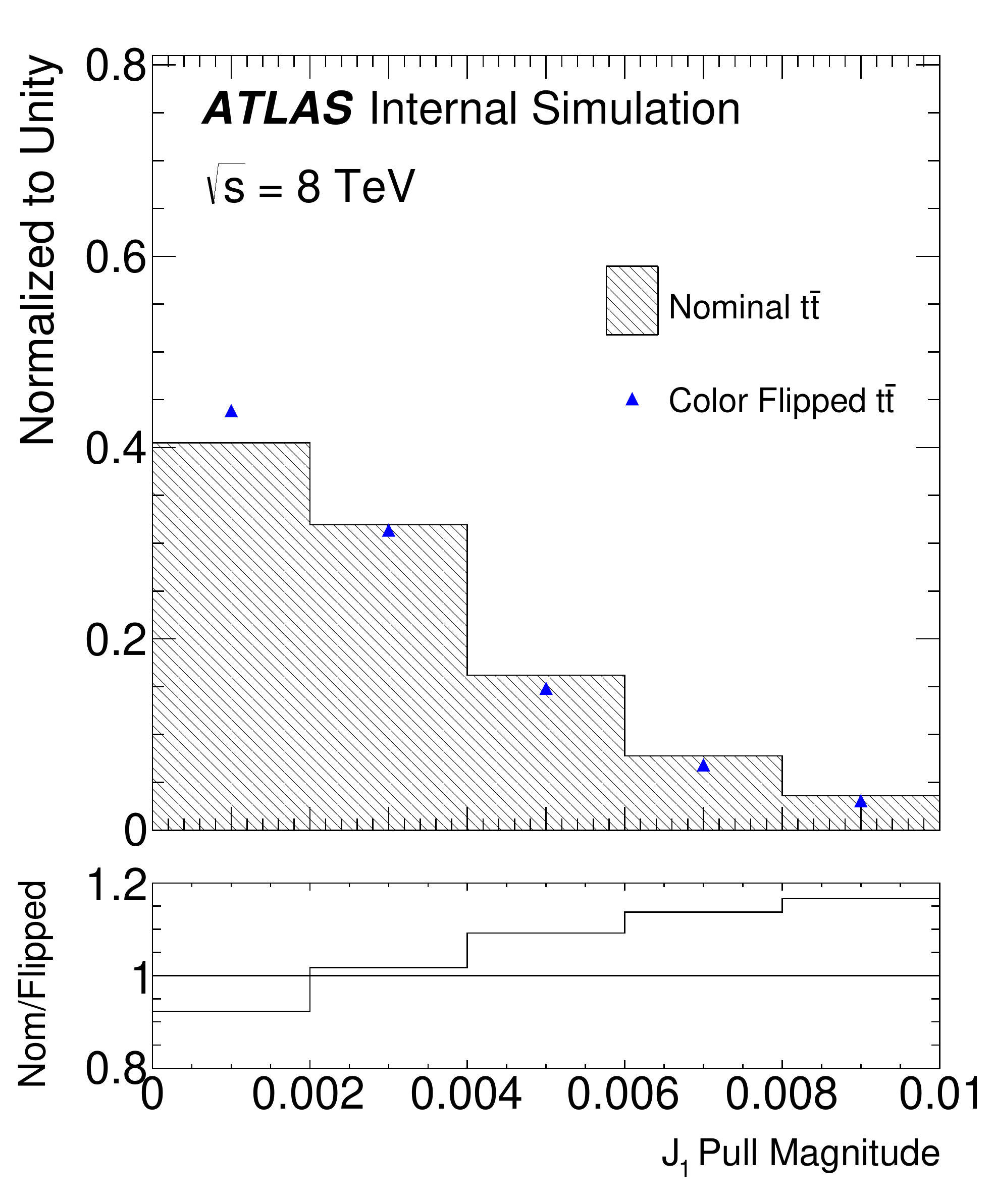}
\includegraphics[width=0.44\textwidth]{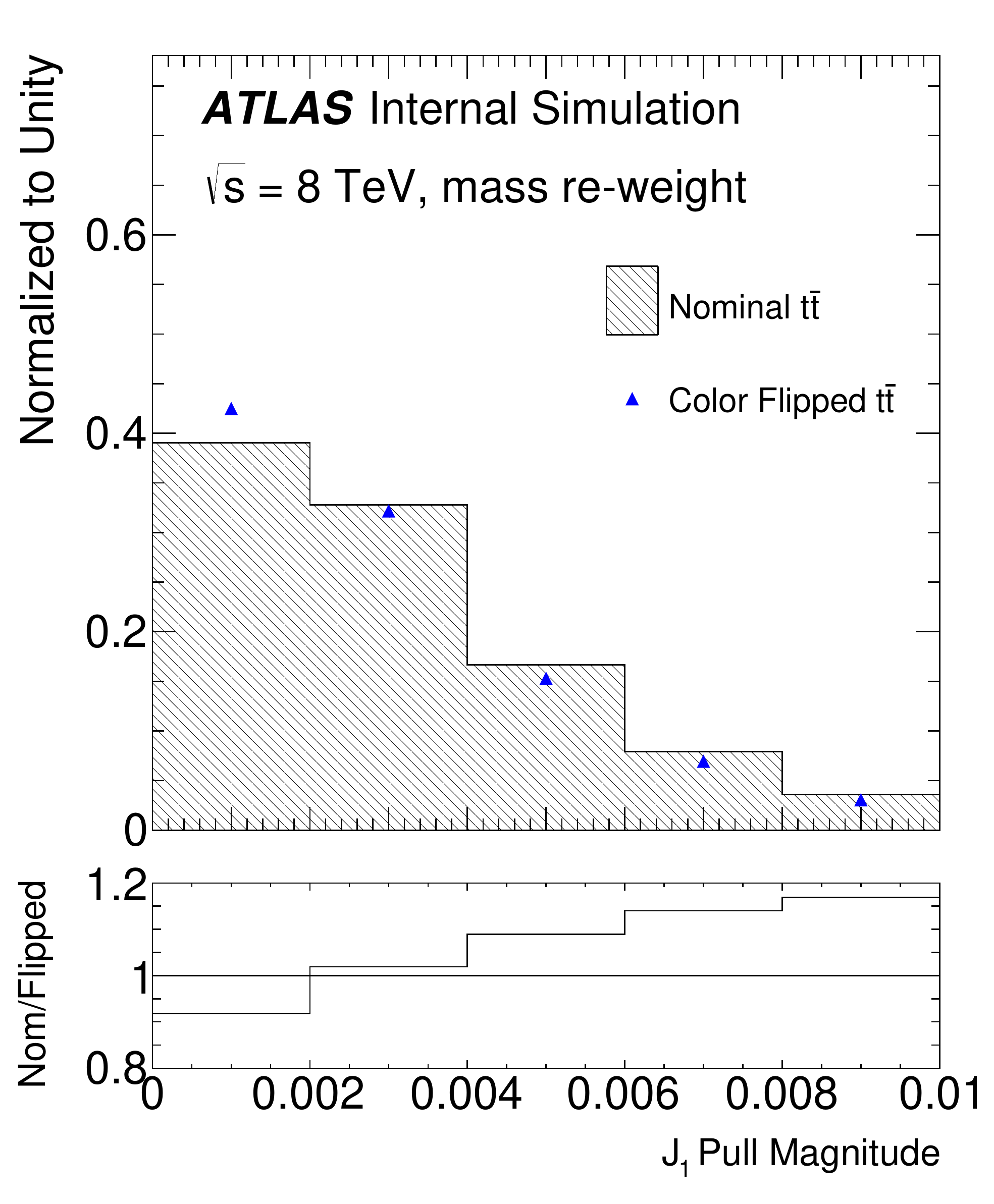}
 \caption{The pull vector magnitude with (right) and without (left) a reweighting to the invariant mass and $\Delta R$ spectra.}
 \label{fig:truth:v1flip_noflip}
  \end{center}
\end{figure}

\begin{figure}[h!]
\begin{center}
\includegraphics[width=0.44\textwidth]{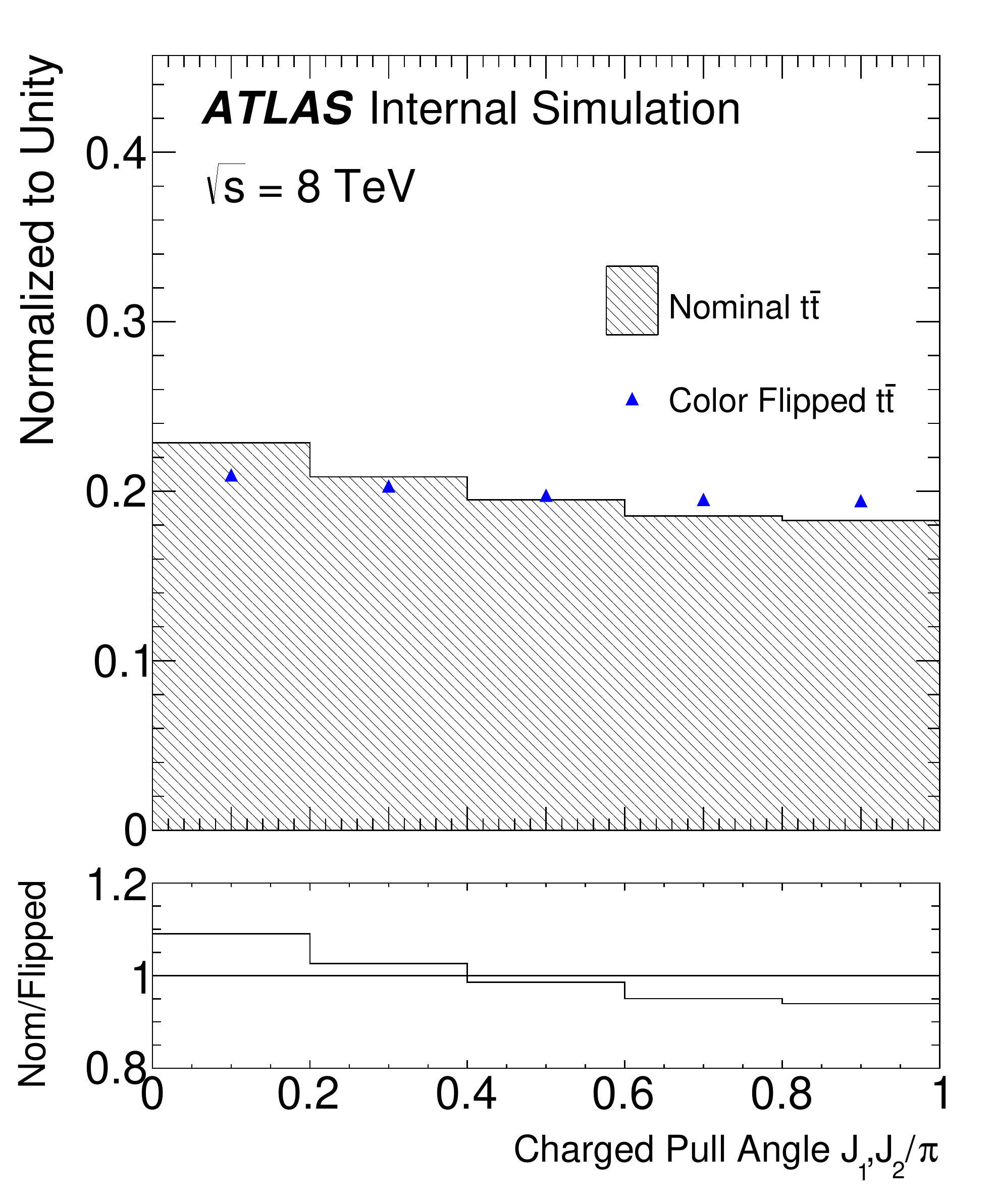}\includegraphics[width=0.44\textwidth]{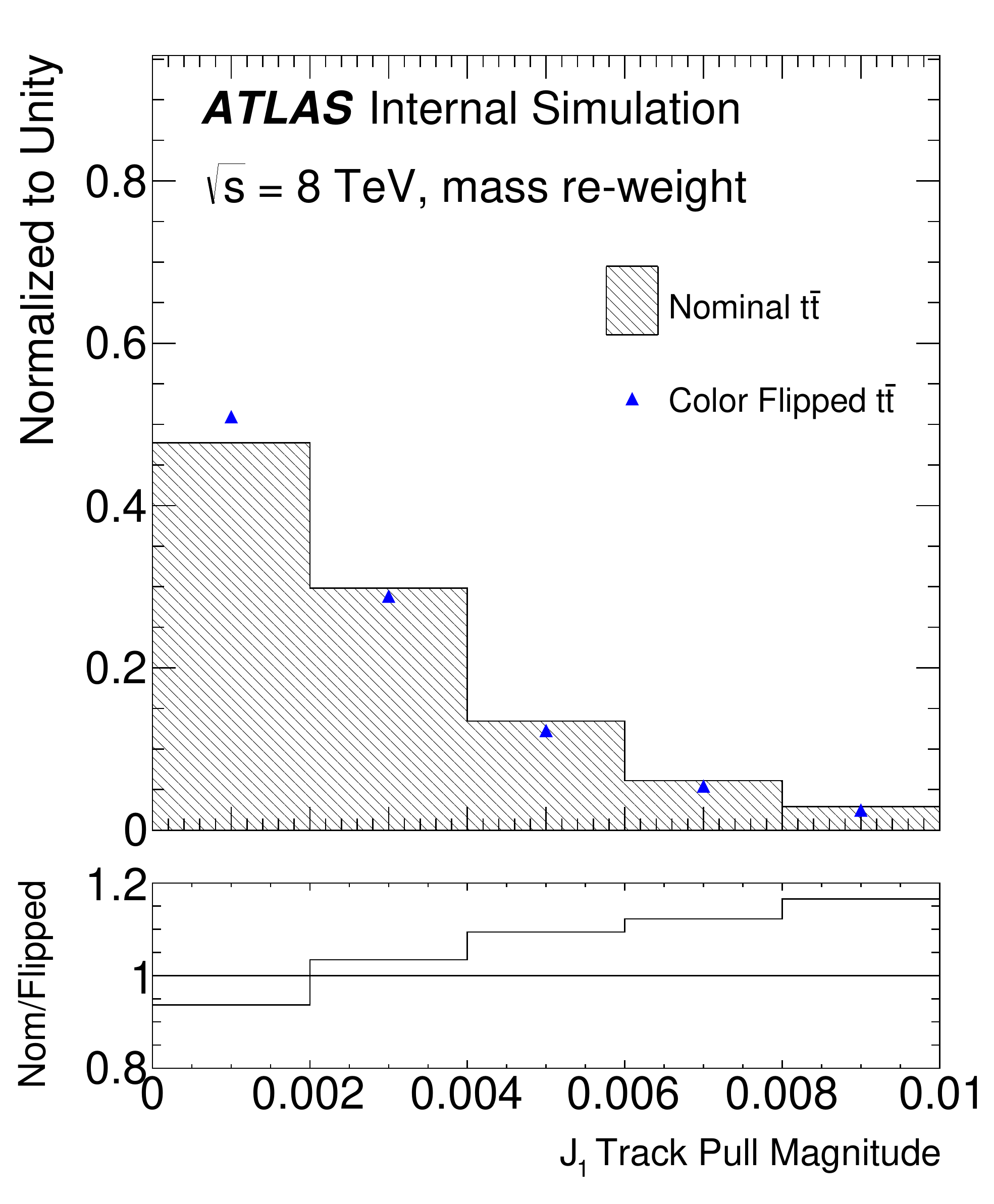}
 \caption{The charged particle pull angle (left) and the charged pull vector magnitude (right).  No reweighting is applied to the invariant mass spectrum.}
 \label{fig:truth:ThetaJ1J2qflip_noflip}
  \end{center}
\end{figure}

\clearpage

\subsection{Object Reconstruction and Event Selection}
\label{sec:ColorFlowEventSelection}

Jet reconstruction and charged particle track association is performed using the algorithms documented in Sec.~\ref{sec:detectorlevel}.  Clusters and jets are corrected to point toward the primary vertex, as motivated and detailed in Sec.~\ref{origincorrection}.  After this correction, the four-vector sum of clusters (treated as massless) is used for the all-particles pull angle jet axis implicit in Eq.~\ref{eq}.  The axis for the charged-particles pull angle is constructed
using the four-momentum sum of all the associated tracks treated as massless.

Aside from the $W$ boson daughter jet selection, the event selection is identical to the one described in Sec~\ref{sec:ttbarsamplesandeventselection}.  In particular, events are selected using single isolated electron and muon triggers and required to have have exactly one reconstructed electron or muon with $p_\text{T}>25$ GeV and $|\eta|<2.5$.  Furthermore, to enrich the selection in $t\bar{t}$ events, $E_{\rm{T}}^{\mathrm{miss}}>20$ GeV and $E_{\rm{T}}^{\mathrm{miss}}+m_\text{T}>60$ GeV.  Events must have $\geq 4$ jets with $p_\text{T}>25$ GeV\,.
At least two of these jets must be tagged using a 70\% target efficiency.  At least two jets must not be $b$--tagged; of these,
the two leading--$p_\text{T}$ jets with $|\eta|<2.1$ are labelled as the jets from the
hadronically decaying $W$ boson, $J_i$ with $p_\text{T}^{J_1}>p_\text{T}^{J_2}$.
The $b$--tagged jets and the jets selected for the pull angle calculation
are required to have $|\eta|<2.1$
so that all constituents are within coverage of the inner detector used for
tracking.  The $W$ daughter jet selection is described in more detail in Sec.~\ref{sec:colorflow:wcandidateselection}.
The event selection produces a sample that is expected to contain approximately
$90\%$ $t\bar{t}$ events.  Table~\ref{tab:comp} shows the predicted composition compared to the data yield.  
\begin{table}
\centering
  \begin{tabular}
    {
 c
 S[table-format=1,
   table-figures-uncertainty=0,group-digits = false,tight-spacing]
}
    Process                         & $\text{Number of Events}$                  \\
\hline
    $t\bar{t}$                      & 95400 \hspace{1mm}$\pm$\hspace{1mm} 14000  \\
    $Wt$--channel single top         & 2730\hspace{1mm} $\pm$\hspace{1mm} 600     \\
    $s$-- and $t$--channel single top & 150\hspace{1mm} $\pm$\hspace{1mm} 10       \\
    $W$+jets                        & 3710\hspace{1mm} $\pm$ \hspace{1mm}120     \\
    $Z$+jets                        & 560 \hspace{1mm}$\pm$ \hspace{1mm}270      \\
    Dibosons                        & 190\hspace{1mm} $\pm$\hspace{1mm} 40       \\
    Multijets                       & 2500\hspace{1mm} $\pm$ \hspace{1mm}910     \\
\hline
    Total SM                        & 105000 \hspace{1mm}$\pm$ \hspace{1mm}14000 \\
    Data                            & 102987                                     \\
\hline
  \end{tabular}
  \caption{Estimated composition of the selected event sample.  The uncertainties are the sum in quadrature of the statistical uncertainties and either the uncertainties of the normalisation method (for the data driven W+jet and multi-jet estimates) or the uncertainties of the cross-section estimates. }
  \label{tab:comp}
\end{table}

\subsection{Particle-level Event Selection}
\label{particlelevelcolorflow}

Particle-level objects and a particle-level event selection are constructed to be as close as possible to the detector-level objects.  The particle-level objects are the target when unfolding the data in order to make direct comparisons with various theoretical models and these objects are also used to study the reconstruction performance.  Particle-level jets are constructed the same way as for the jet charge measurement, described in Sec.~\ref{sec:particlelevel}.  The particle-level inputs to the all-particles pull angle are all of the charged and neutral particles clustered within particle-level jets.
Only the charged particles clustered within the particle-level jets are used
for the charged-particles pull angle.  Particle-level electrons, muons, photons, and neutrinos are only considered if their parent in the MC ancestry is not a hadron or a tau which came from a hadron decay.   Electrons and muons are {\it dressed} with photons by defined the lepton 4-vector as the sum of the electron or muon particle 4-vector and the sum of all photon 4-vectors within $\Delta R<0.1$.  Dressed leptons are a better approximation to the measured leptons than bare leptons because (nearly) collinear radiation cannot be resolved in the detector.  The particle-level $E_\text{T}^\text{miss}$ is the magnitude of the vector sum over all particle-level neutrinos.  Note that particle-level electrons and photons assigned to electrons or muons through dressing are not used for jet clustering.  A particle-level jet is $b$-tagged if a $B$ hadron from the MC event record with $p_\text{T}>5$ GeV is ghost-associated to the jet.  Additional information
about the particle-level object definitions
can be found in Ref.~\cite{Aad:2015eia}.

The particle-level event selection is analogous to the detector-level
selection described in Sec.~\ref{sec:ColorFlowEventSelection} with detector-level
objects replaced with particle--level objects.
Exactly one electron or muon and at least four jets are required,
each with $p_\text{T} > 25$ GeV and $|\eta|<2.5$. The particle-level $E_{\rm{T}}^{\mathrm{miss}}>20$ GeV and the sum of
$E_{\rm{T}}^{\mathrm{miss}}+m_\text{T}>60$ GeV. At least two of the selected
jets are required to be identified as $b$-jets using the same definition
as that found in Ref.~\cite{Aad:2015eia}.
As with the detector-level calculation of the pull angle,
the two leading-$p_\text{T}$ particle-level non $b$--jets
with $|\eta|<2.1$ are labelled as the jets from the
hadronically decaying $W$ boson.  About $80\%$ ($70\%$) of the time, the (sub)leading particle-level jet is within $\Delta R<0.4$ of the detector-level jet.  Since this is not $100\%$, there is non-negligible contribution to the unfolding from combinatorics in addition to per-object resolutions (in Sec.~\ref{sec:particlelevel}, the leading jet contamination is $\sim 4\%$). 

\clearpage

\subsubsection{$W$ Boson Candidate Selection}
\label{sec:colorflow:wcandidateselection}

The $W$ boson candidate is built from the two highest $p_\text{T}$ jets that are not $b$-tagged.  A common alternative $W$ boson identification technique is to use the non $b$-tagged jets whose invariant mass is closest to the $W$ boson mass.  The reason for not using the invariant mass constraint is because the mass can bias the colorflow, as shown in Sec.~\ref{sec:ColorFlowExoticSimulation}.  This section explores the identification efficiency of the the baseline method and the alternative mass-based method.  As a result of fragmentation, it is not possible to uniquely associate jets with quarks and so one must {\it define} a metric for assessing the fidelity of the hadronic $W$ boson reconstruction.  A common scheme is to use $\Delta R$ between the selected jets and the $W$ boson in simulation.  However, such a scheme is not useful when the $W$ boson is produced at low $p_\text{T}$ and also removes most of the information about which jets contain the majority of the $W$ boson energy.  Let $T$ be the set of all truth particles in a simulated event and used for particle-level jet clustering (stable hadrons).  Define $\mathcal{I}(i)=1$ if particle $i\in T$ is a descendent from a $W$ boson and $0$ otherwise.  The function $\mathcal{I}$ is well-defined in a leading log parton shower where particle histories are recoverable from the succession of $1\rightarrow 2$ splittings.  Since the $W$ boson is a color singlet, when no kinematic requirements are placed on the particles entering jet clustering, $\sum_{i\in T}\mathcal{I}(i) e_i = e_W$, where $e_i$ is the energy of particle $i$ and $e_W$ is the energy of the hadronically decaying $W$ boson.  A useful metric for comparing $W$ boson reconstruction algorithms is the fraction of the $W$ boson energy contained in the selected jets, $f_W^\text{jet}=\sum_{i\in\text{jet}} \mathcal{I}(i)e_i/e_W$.  Another useful quantity is the fraction of a jet's energy originated from the $W$ boson, $f_\text{jet}^W=\sum_{i\in\text{jet}} \mathcal{I}(i)e_i/\sum_{i\in\text{jet}}e_i$.  Ideally, $\sum_{i=1}^2 f_W^\text{jet,i}\approx 1$ and each of the two jets is mostly built from $W$ boson radiation, $f_\text{jet,i}^W\approx 1$ for $i=1,2$.  The left plot of Fig.~\ref{fig:ColorFlowMethods1} shows the distribution of $\sum_{i=1}^2 f_W^\text{jet,i}$ for three methods: the baseline method, the alternative method, and the {\it best} method in which the two jets are chosen with the highest $f_W^\text{jet,i}$.  Events are simulated using {\sc Powheg+Pythia} 8 with a simple particle-level event selection that mimics the detector-level selection discussed in Sec.~\ref{sec:ColorFlowEventSelection}.  In particular, jets are clustered with the anti-$k_t$ $R=0.4$  algorithm implemented in {\sc FastJet} using all stable particles from {\sc Pythia} that are not leptons as input.  Jets are considered if they have $p_\text{T}>25$ GeV and are tagged as $b$-jets if there is a $B$-hadron from the event record within $\Delta R<0.3$ of the jet axis.  Events are required to have at least two $b$-tagged jets and at least two non $b$-tagged jets.  Figure~\ref{fig:ColorFlowMethods1} has several interesting features, including the spike at zero from selected jets resulting from ISR or other sources of non $W$ jets and the fact that in the best case, the fraction of the $W$ boson energy contained in the selected jets is often much less than unity.   As expected, baseline method has a lower fraction than the alternative method.  The right plot of Fig.~\ref{fig:ColorFlowMethods1} shows the difference between $\sum_{i=1}^2 f_W^\text{jet,i}$ for the various methods.  The spike at zero corresponds to cases in which the methods select the same jets.  About 70\% of the time, the baseline and best methods are the same and about 80\% of the time, the best and alternative methods are identical.  The alternative is typically better than the baseline, but not always, as indicated by the tail of the red histogram at positive values of the difference.   Since the difference between the alternative and baseline methods is small, and the alternative method has a potential for bias, the baseline method is used excluslvely for the rest of the chapter.

\begin{figure}
\begin{center}
\includegraphics[width=0.45\textwidth]{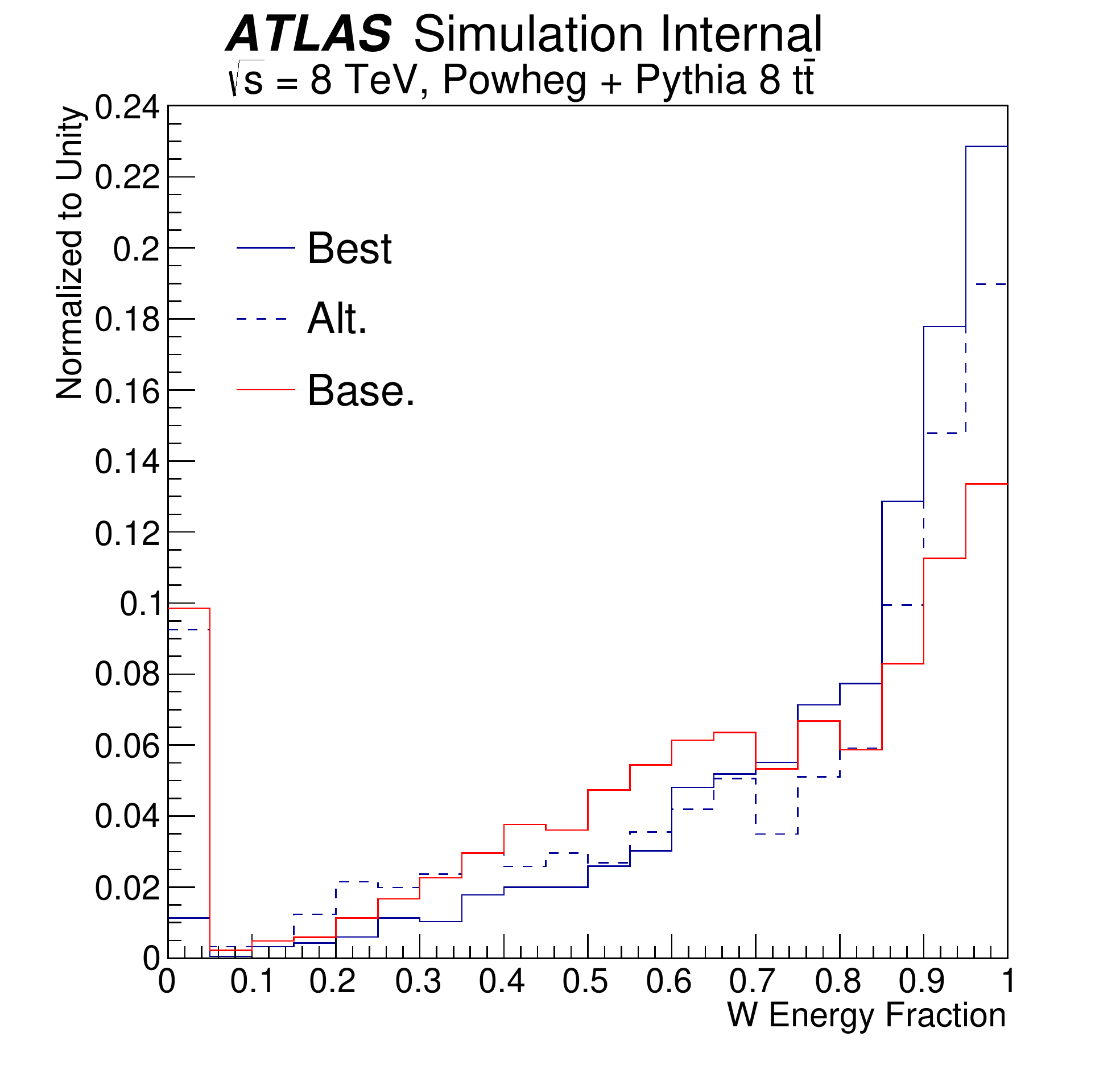}\includegraphics[width=0.45\textwidth]{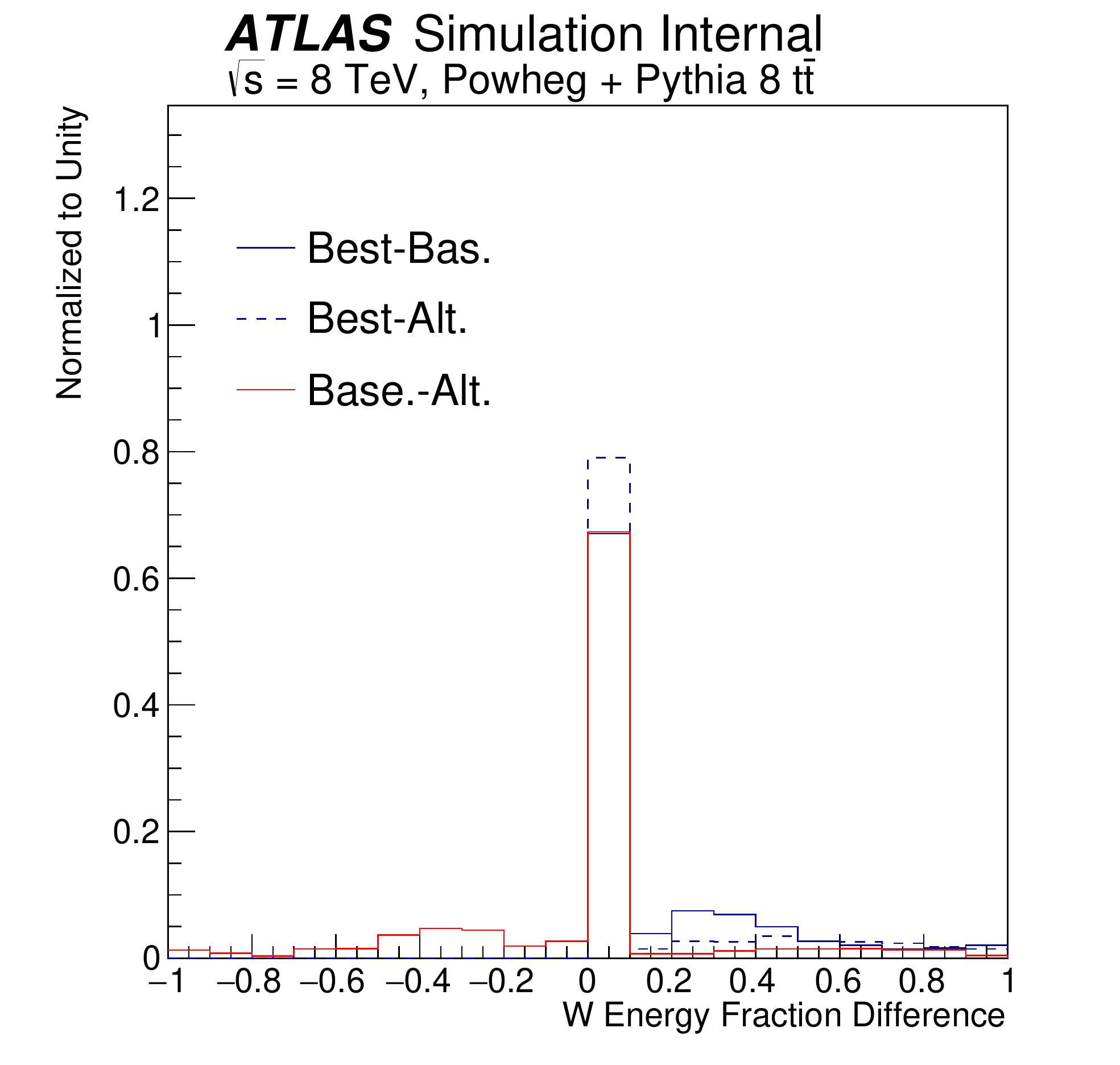}
 \caption{Left: The fraction of the $W$ boson energy carried by the jets selected with three methods: Best, Baseline (`Base.'), and Alternative ('Alt.').  The baseline method uses the leading non-$b$ tagged jets, the alternative method uses an invariant mass constraint, and the best method uses truth information; see the text for details.  Right: The difference in fractions between the various methods.}
 \label{fig:ColorFlowMethods1}
  \end{center}
\end{figure}

Figure~\ref{fig:ColorFlowMethods2} shows how the $W$ energy fraction differs between the leading and sub-leading $W$ candidate daughter jets.  As expected, the higher $p_\text{T}$ jet has more of the $W$ boson energy on average compared to the lower $p_\text{T}$ jet. Many of the two-jet pairs have a roughly symmetric fraction of the $W$ boson energy, but the width of the distribution in Fig.~\ref{fig:ColorFlowMethods2} is broad compared to the range.  While it is possible to form two jet pull angles $\theta_\text{P}(J_1,J_2)$ and $\theta_\text{P}(J_2,J_1)$, only the former is measured in part because it contains more information (radiation) from the hadronic $W$ boson decay\footnote{As will be discussed in Sec.~\ref{jetpullcombo}, the two pull angles are largely uncorrelated so a statistical combination would improve the measurement.  However, the systematic uncertainties are fully correlated and at present, the measurement is limited by systematic and not statistical uncertainty.}.  

\begin{figure}
\begin{center}
\includegraphics[width=0.45\textwidth]{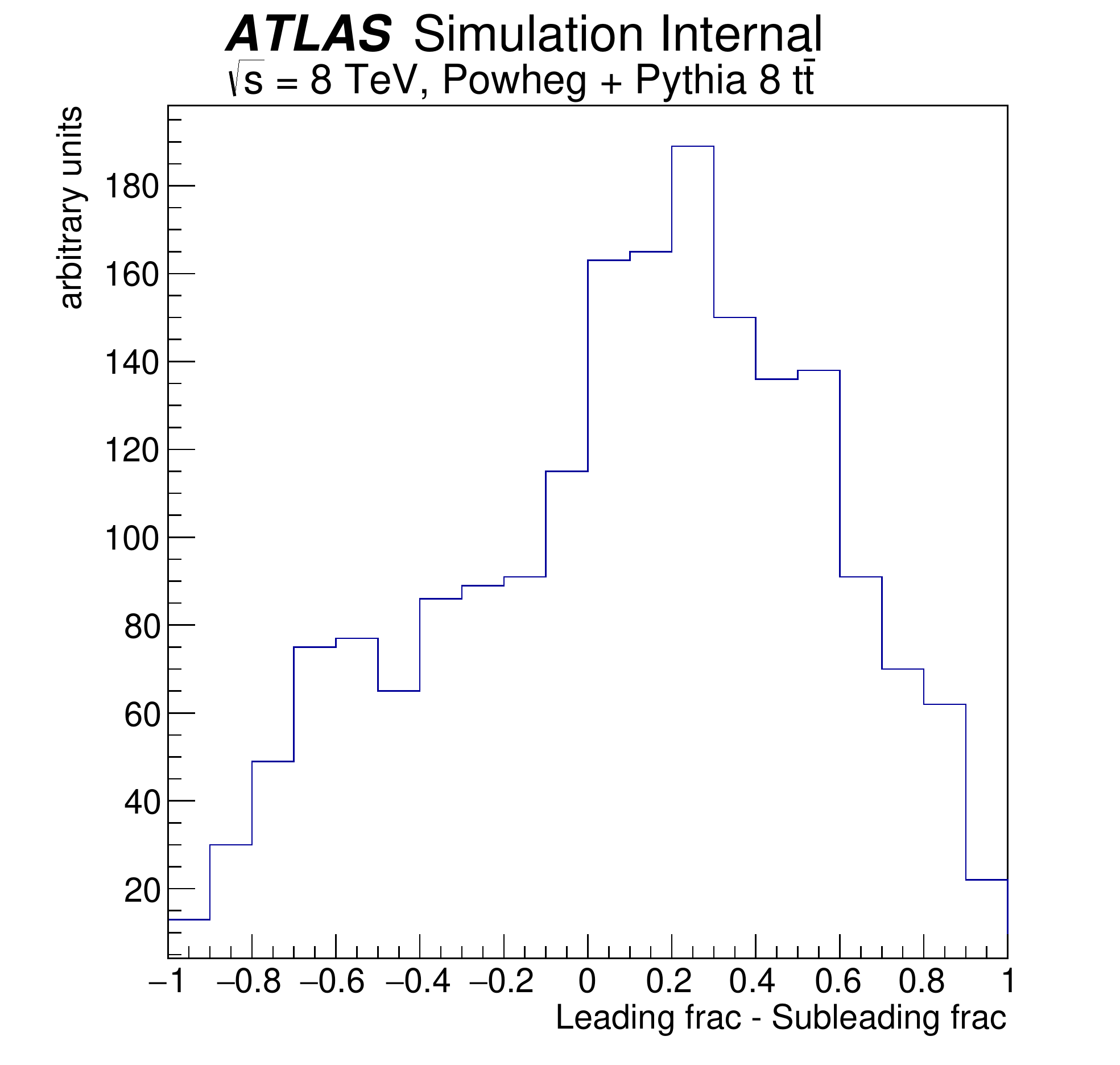}
 \caption{The difference in the $W$ boson energy fractions between the leading (in $p_\text{T}$) and subleading jets, $f_\text{jet,i}^W(\text{lead})-f_\text{jet,i}^W(\text{sublead})$.}
 \label{fig:ColorFlowMethods2}
  \end{center}
\end{figure}

More visualizations of the distribution of the $W$ boson energy inside the selected jets are shown in Fig.~\ref{fig:ColorFlowMethods3}.  The top plots of Fig.~\ref{fig:ColorFlowMethods3} show the two-dimensional distribution of $f_W^\text{jet}$ and $f_\text{jet}^W$ for the leading (left) and sub-leading (right) jets.  When some fraction of the jet energy is from the $W$ boson, the fraction is nearly 100\%.  Consistent with Fig.~\ref{fig:ColorFlowMethods2}, the top right plot of Fig.~\ref{fig:ColorFlowMethods3} is shifted to the left with respect to the top left plot by construction.  The lower left plot in Fig.~\ref{fig:ColorFlowMethods3} shows that many events are along the diagonal, where most of the $W$ energy is captured by the two selected jets.  However, there is a large spread in the bulk where more than two jets are need to capture the full $W$ energy.  The lines at $0$ correspond to ISR jets which have nothing to do with the $W$ boson.  The lower right plot in Fig.~\ref{fig:ColorFlowMethods3} shows that the energy fraction of a selected jet is either dominated by the $W$ energy or is nearly zero.  The fraction of events in which at least one jet has a $W$ energy fraction above 80\% is 90\%, while the fraction of events in which the leading jet has at least 80\% of its energy from the $W$ is 62\%.

In conclusion, the $W$ boson often produces more than two jets and one must be careful when assessing the performance of any matching scheme.

\begin{figure}
\begin{center}
\includegraphics[width=0.41\textwidth]{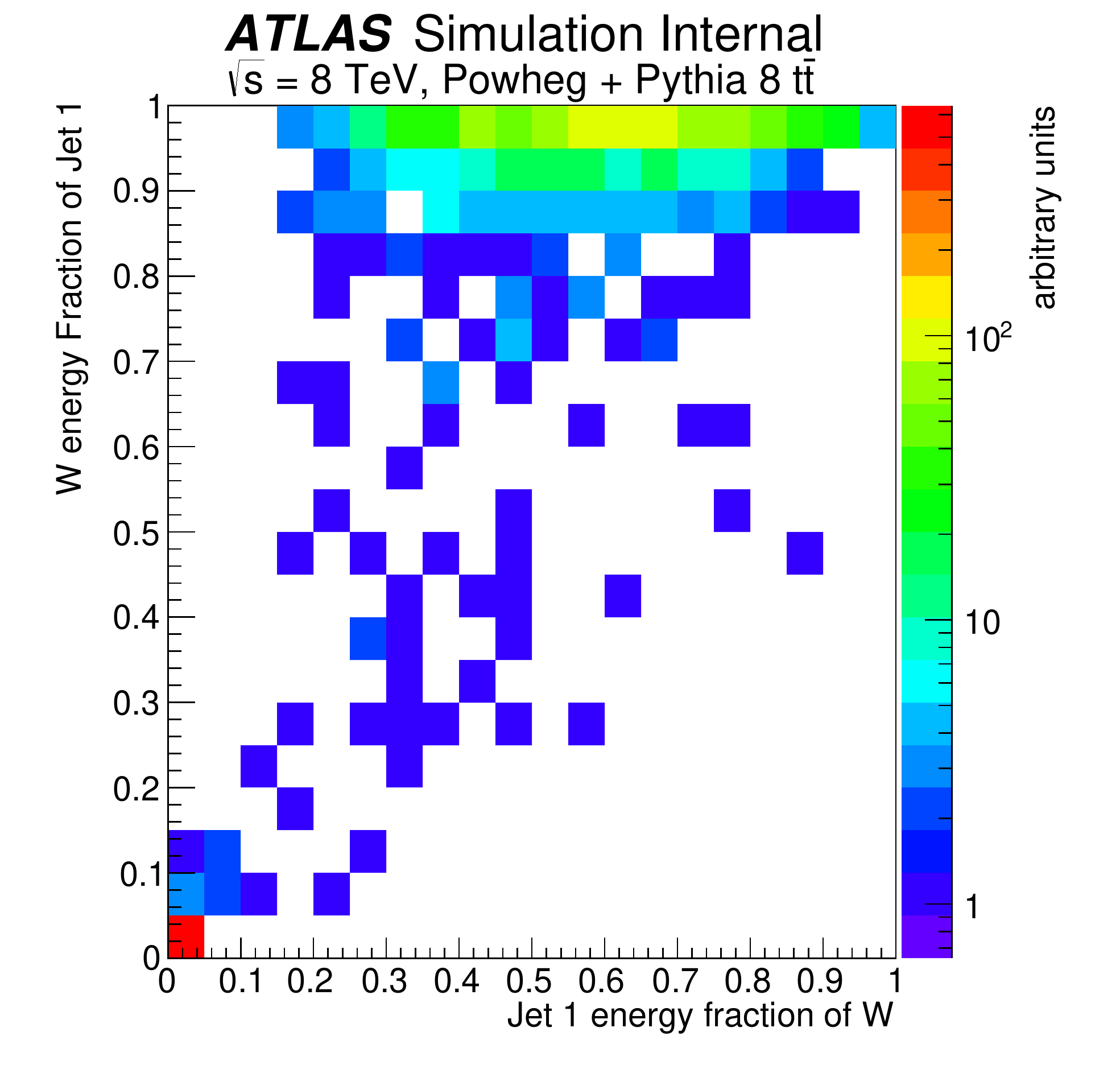}\includegraphics[width=0.41\textwidth]{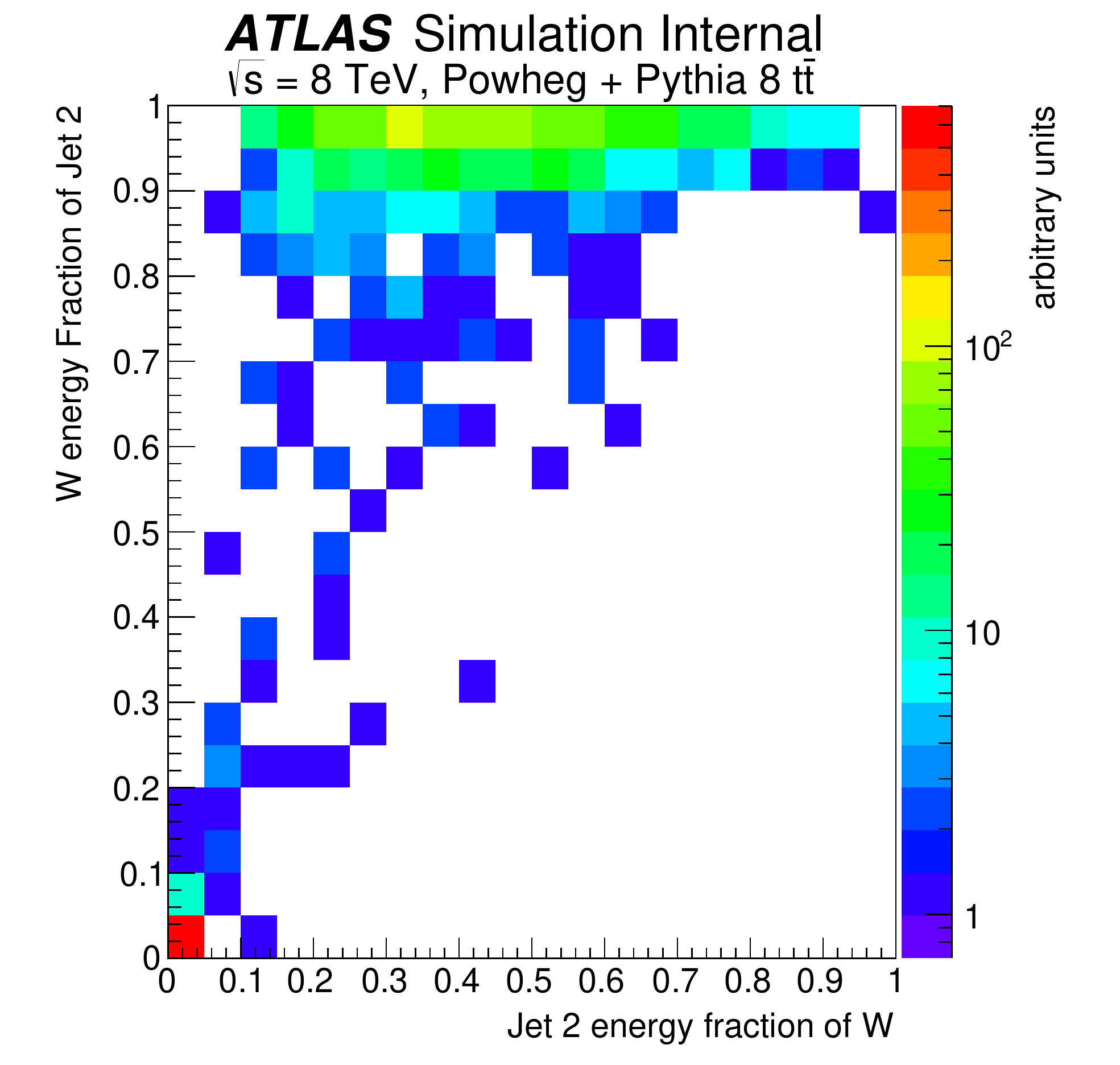}
\includegraphics[width=0.41\textwidth]{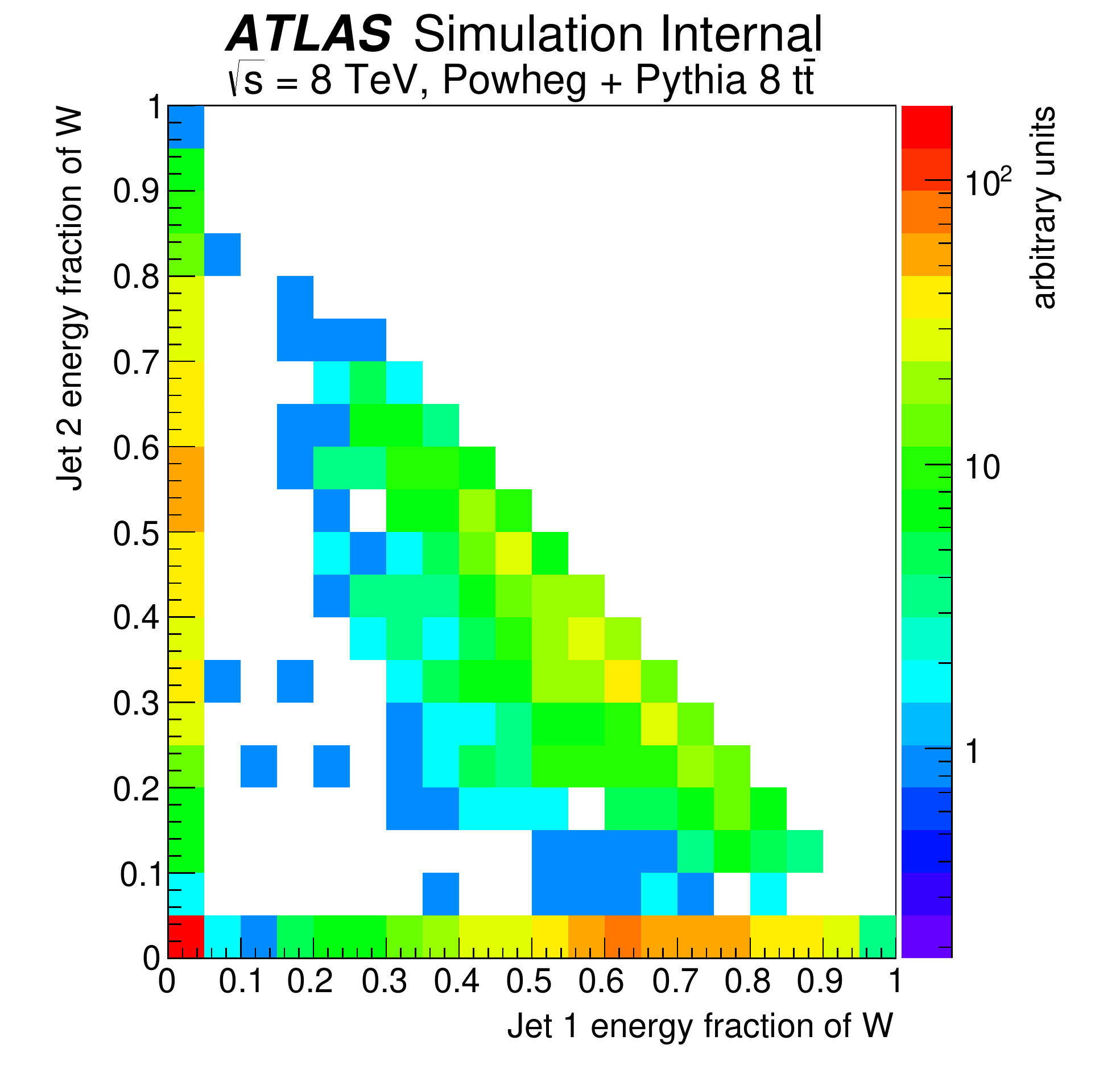}\includegraphics[width=0.41\textwidth]{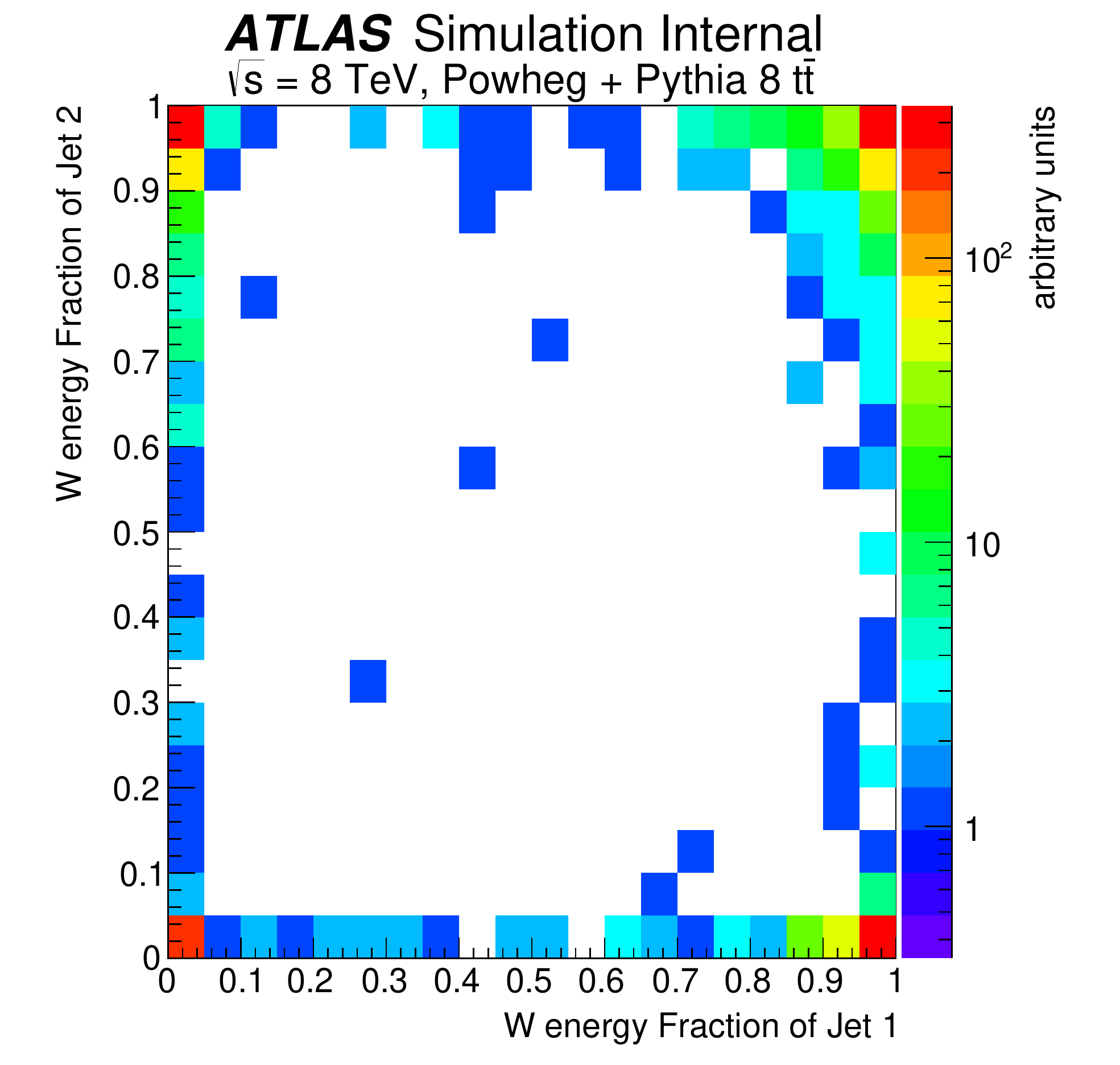}
 \caption{Top: The distribution of $f_W^\text{jet}$ (Jet energy fraction of W) and $f_\text{jet}^W$ (W energy fraction of jet) for the leading (left) and sub-leading (right) jets.  Bottom: the distribution between $f_W^\text{jet,1}$ and $f_W^\text{jet,2}$ (left) and between $f_\text{jet,1}^W$ and $f_\text{jet,2}^W$ (right).}
 \label{fig:ColorFlowMethods3}
  \end{center}
\end{figure}

\clearpage

\subsubsection{Comparisons Between Data and Simulation}

This section briefly describes the modeling of important kinematic distributions related to the event selection described in Sec.~\ref{sec:ColorFlowEventSelection}.  Figure~\ref{fig:pts_sub} shows the individual $p_\text{T}$ of the selected $W$ daughter jets and Fig.~\ref{fig:wpt} shows the dijet $p_\text{T}$.  All of these quantities have a slight slope in the data to MC ratio, which is discussed in more detail in Sec.~\ref{syst:toppt}.   The $\eta$ distributions are shown in Fig.~\ref{fig:etas_sub} and the angular distance between the jets is shown in the left plot of Fig.~\ref{fig:wwmass}.  As the $p_\text{T}$ of the hadronically decaying $W$ bosons is generally $\lesssim 200$ GeV, the two selected jets are generally $\Delta R\gtrsim 1$ (see Chapter~\ref{cha:bosonjets}).   The invariant mass of the two selected $W$ daughter jets is shown in the right plot of Fig.~\ref{fig:wwmass}.  As expected, $m_{jj}$ is peaked near $m_W$, though there is a broad tail from combinatorics and initial and final state radiation.

\begin{figure}[h!]
\begin{center}
\includegraphics[width=0.5\textwidth]{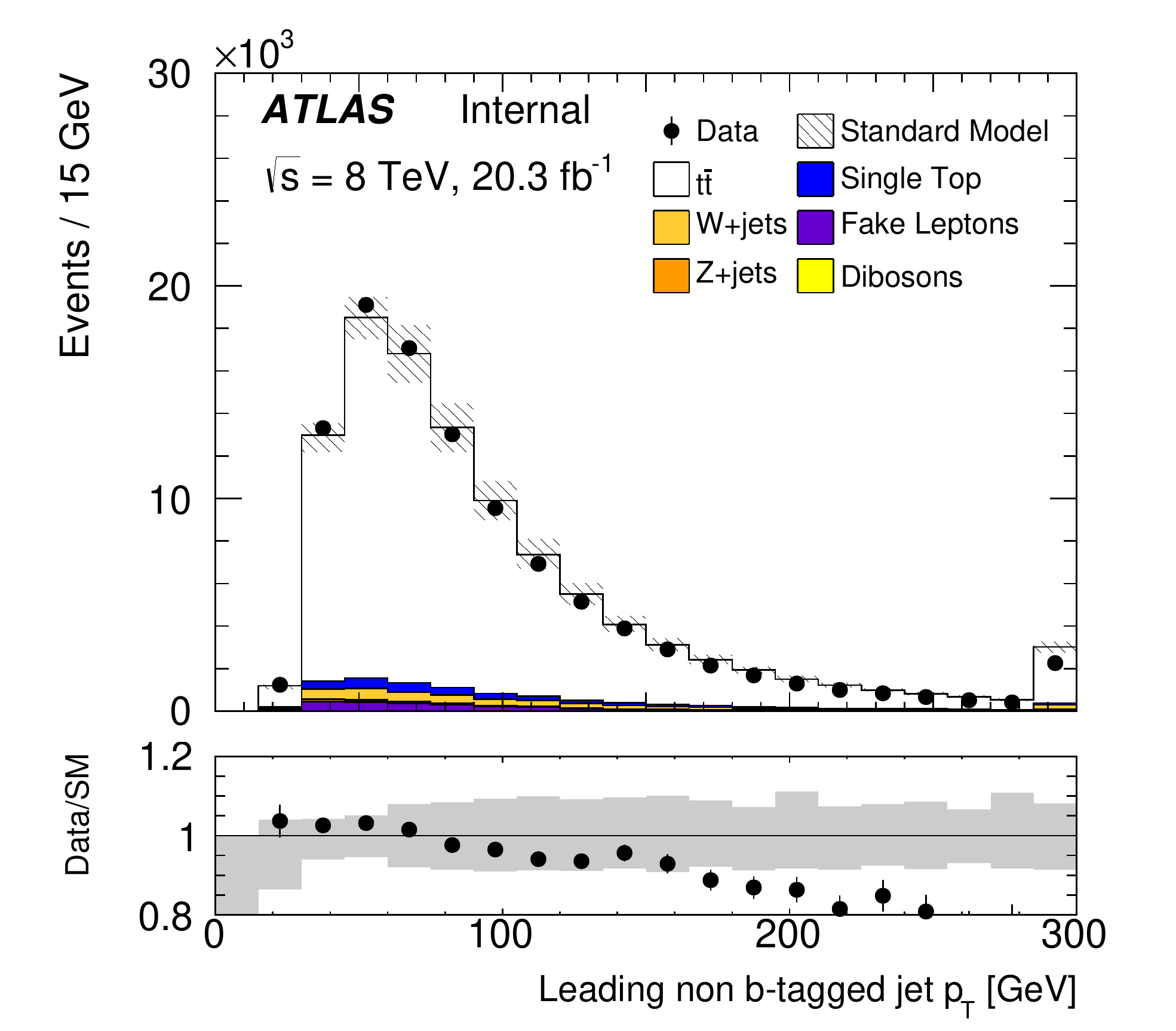}\includegraphics[width=0.5\textwidth]{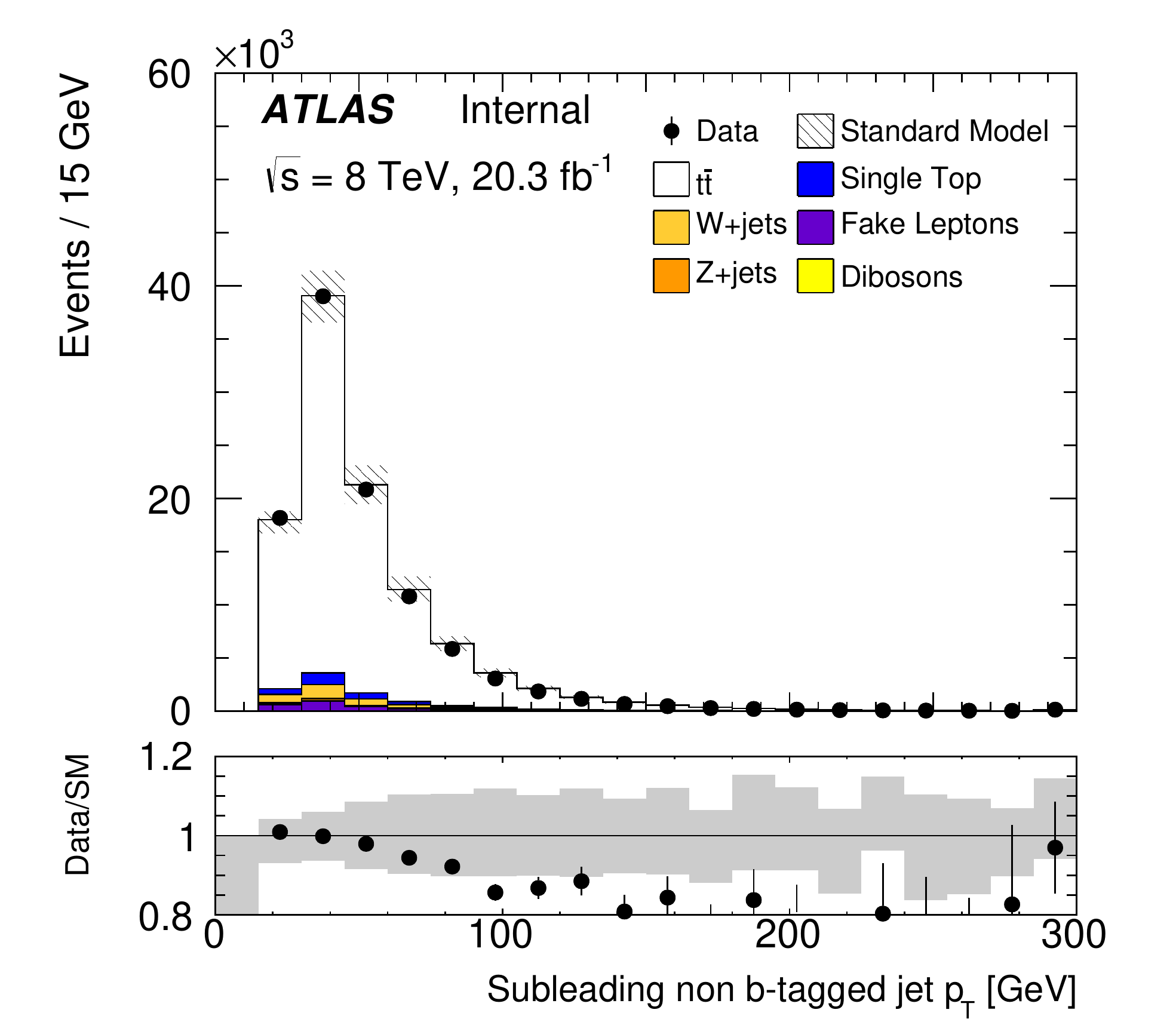}
 \caption{Left (right): The leading (sub-leading) non $b$-tagged jet $p_\text{T}$.  The uncertainty band includes the detector-related experimental uncertainties described in Sec.~\ref{sec:colorflow:systematics}.  The final bin includes overflow.}
 \label{fig:pts_sub}
  \end{center}
\end{figure}

\begin{figure}[h!]
\begin{center}
\includegraphics[width=0.5\textwidth]{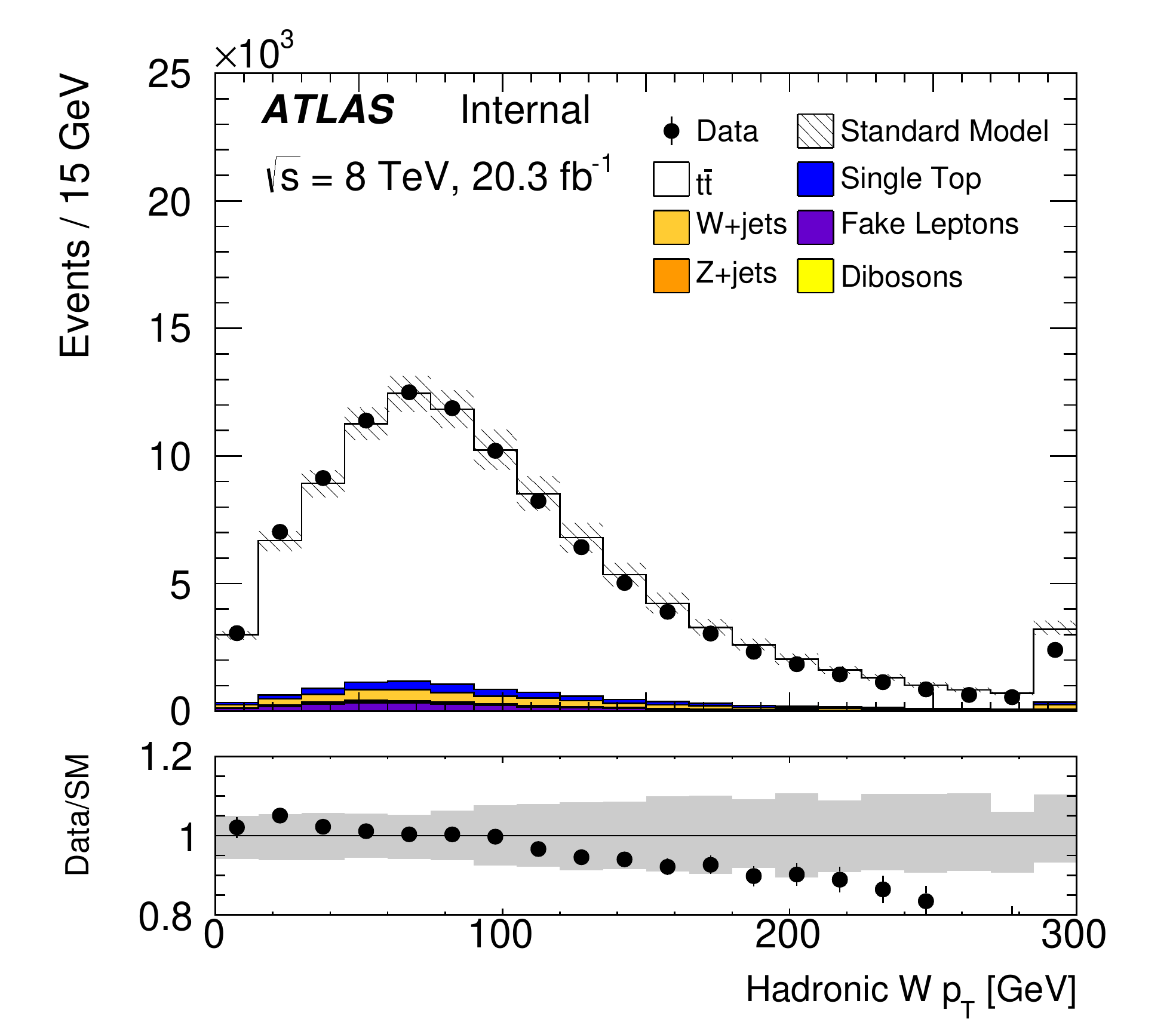}
 \caption{The distribution of the dijet $p_\text{T}$ constructed from the leading two non $b$-tagged jets (hadronically decaying $W$ boson candidate).  The uncertainty band includes the detector-related experimental uncertainties described in Sec.~\ref{sec:colorflow:systematics}.  The final bin includes overflow.}
 \label{fig:wpt}
  \end{center}
\end{figure}

\begin{figure}[h!]
\begin{center}
\includegraphics[width=0.5\textwidth]{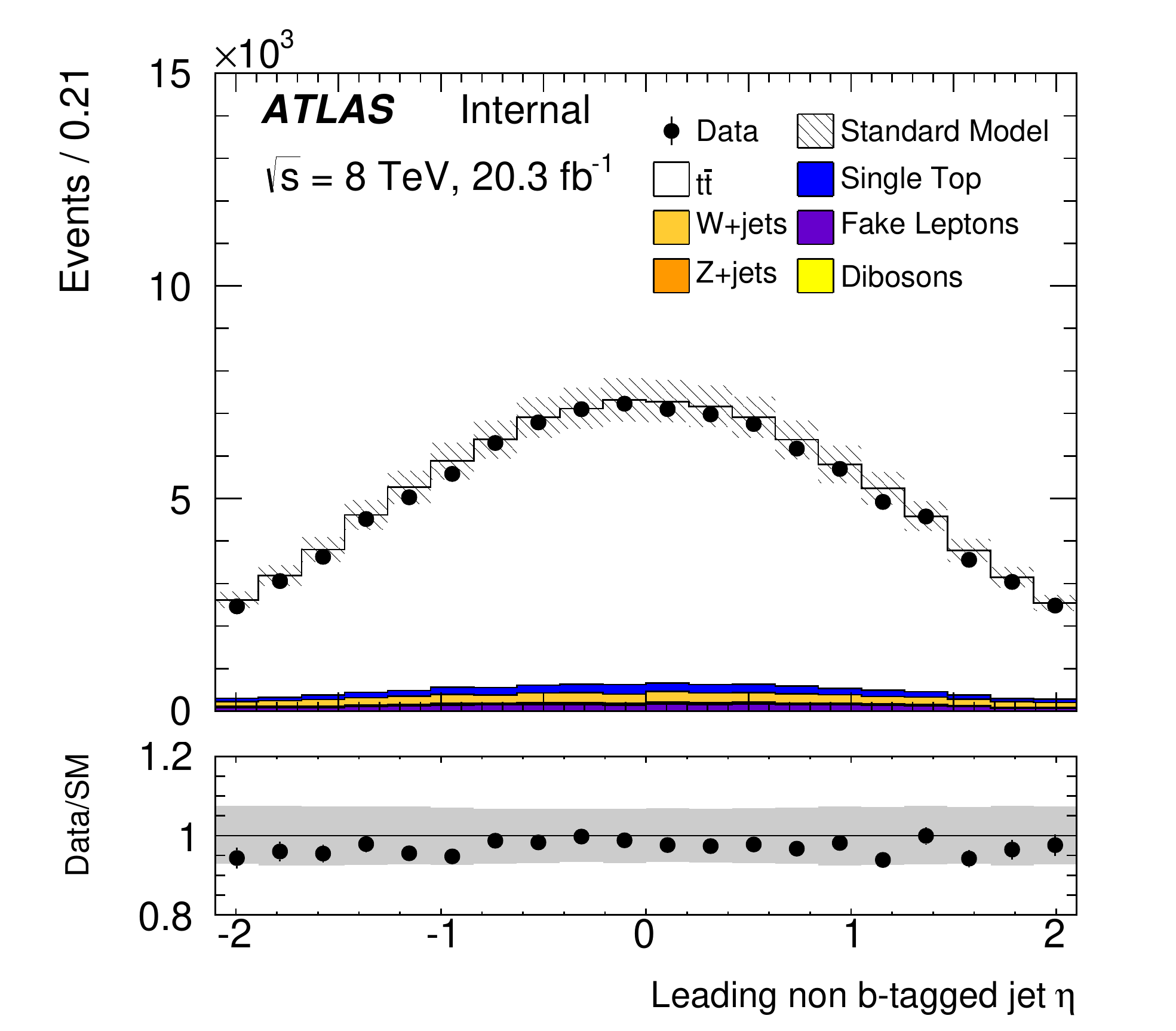}\includegraphics[width=0.5\textwidth]{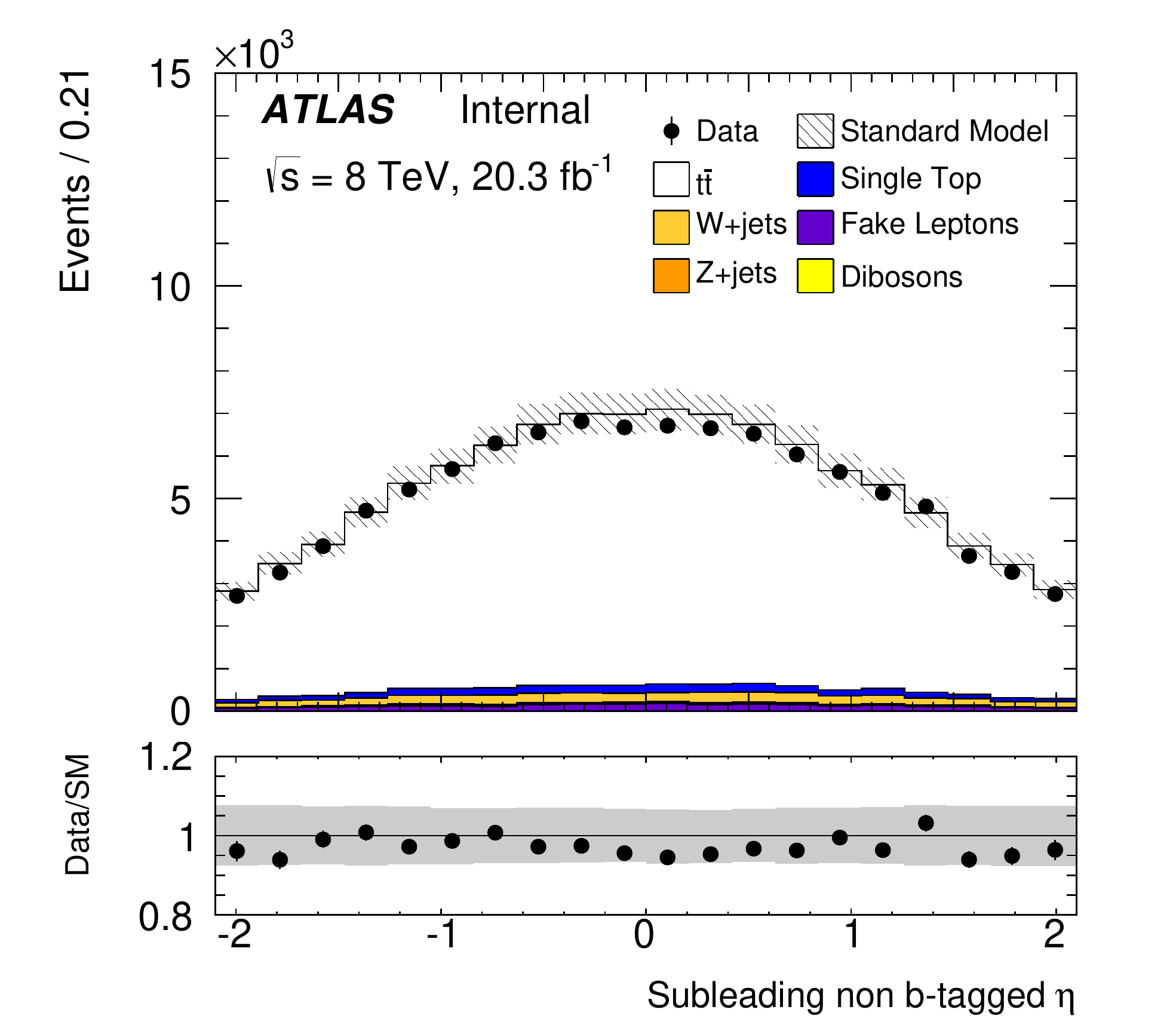}
 \caption{Left (right): The leading (sub-leading) non $b$-tagged jet $\eta$.  The uncertainty band includes the detector-related experimental uncertainties described in Sec.~\ref{sec:colorflow:systematics}.  The final bin includes overflow.}
 \label{fig:etas_sub}
  \end{center}
\end{figure}

\begin{figure}[h!]
\begin{center}
\includegraphics[width=0.5\textwidth]{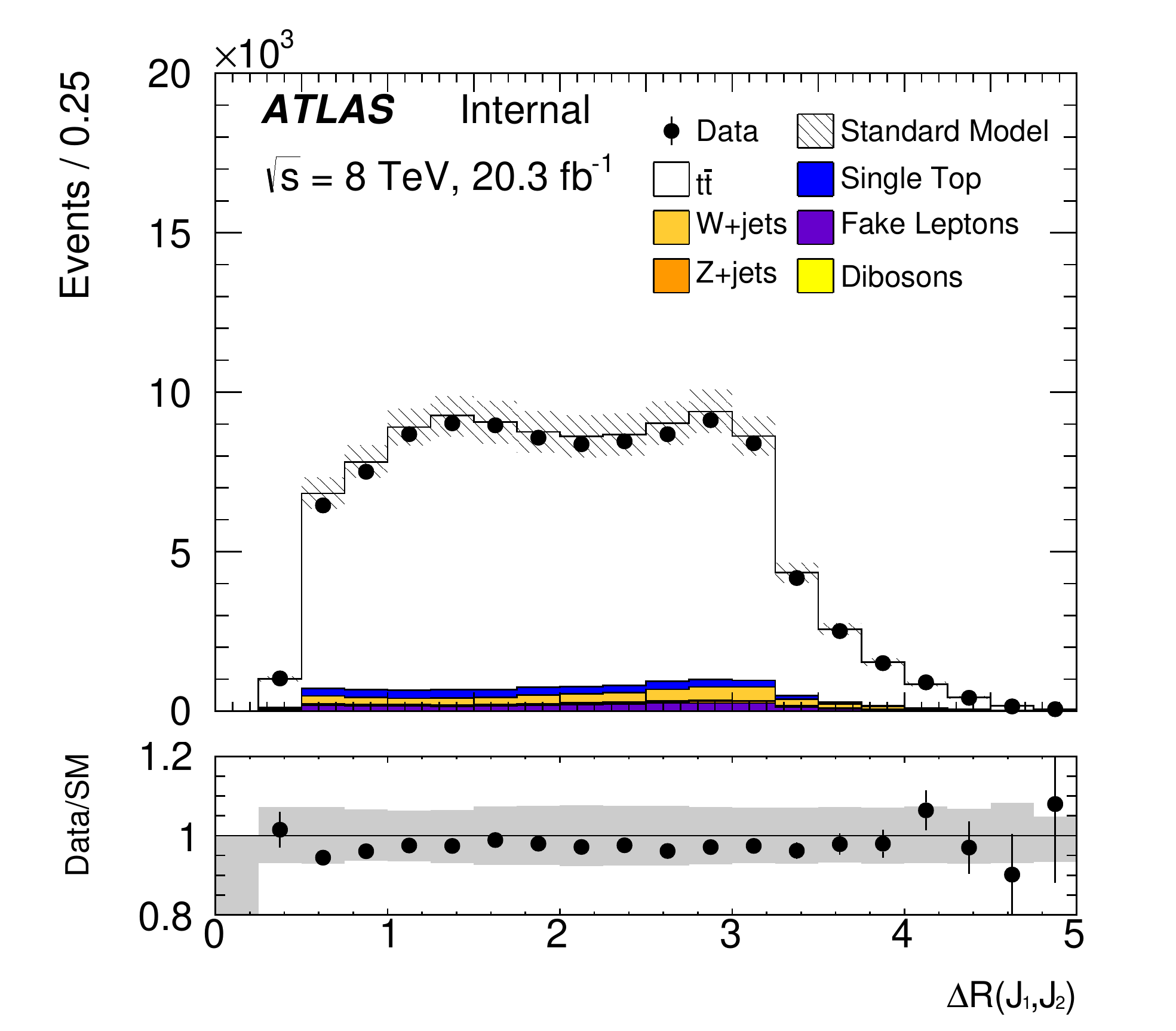}\includegraphics[width=0.5\textwidth]{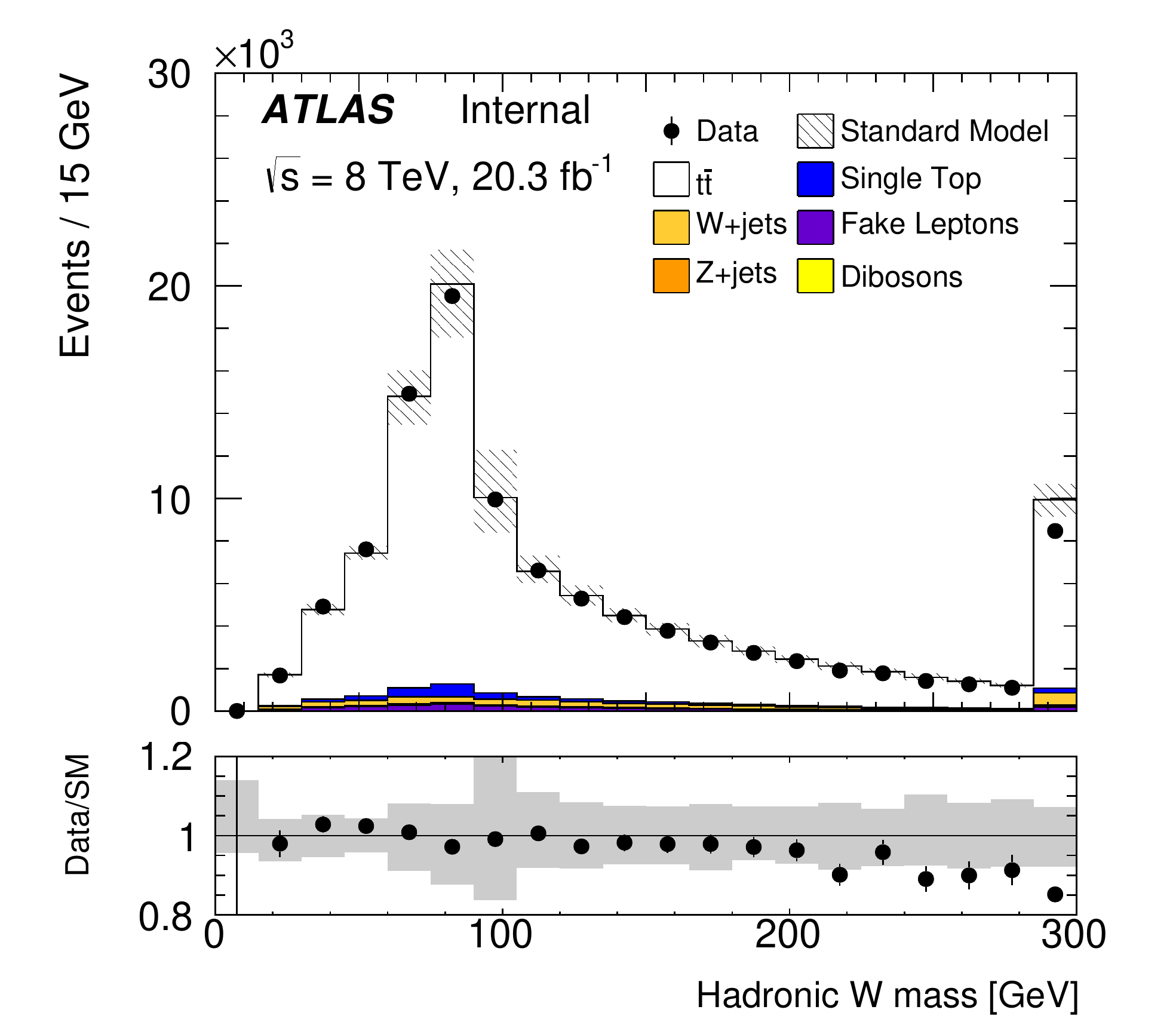}
 \caption{Left (right): The $\Delta R$ (invariant mass) distribution of the leading to non $b$-tagged jets.  The uncertainty band includes the detector-related experimental uncertainties described in Sec.~\ref{sec:colorflow:systematics}.  The final bin includes overflow.}
 \label{fig:wwmass}
  \end{center}
\end{figure}

\clearpage

\section{Jet Pull Reconstruction Performance}
\label{sec:colorflowperformance}

\subsection{Detector effects}

This section uses the particle-level jets described in Sec.~\ref{particlelevelcolorflow}.   Detector-level jets are matched to particle-level jets using a $\Delta R<0.3$ criteria in order to understand how the detector response distorts the particle-level distributions.  The output of the event selection in Sec.~\ref{particlelevelcolorflow} is a set of four jets labeled $B_1,B_2,J_1$ and $J_2$ for every event.  Since the jet pull angle $\theta_P(X,Y)$ requires two jets $X$ and $Y$ as input, there are $12$ possible jet pull angles.  In general $\theta_P(X,Y)\neq \theta_P(Y,X)$ since the former uses the substructure properties of $X$ while the latter uses the substructure properties of $Y$.  Figures~\ref{fig:9}(a)-\ref{fig:9}(f) show the pull angle distributions\footnote{The calorimeter jet axis is used for the detector-level pull angles.  One conclusion of Sec.~\ref{sec:colorflowperformance} will be that instead the constituent axis should be used (see Sec.~\ref{origincorrection}).  As a result, the pull angle distributions in Sec.~\ref{sec:ColorFlow:UnfoldingParams} and and subsequent sections look qualitatively different than the ones shown here.} for all cases that involve the $W$ daughter jets and the leading $b$-jet $B_1$.  The particle-level distributions are consistent with the corresponding particle level studies in the literature, where a peak at zero corresponds to jets which are `color-connected' (e.g. the daughters of the color singlet $W$ boson) and a uniform distribution corresponds to jets without such a connection~\cite{Gallicchio:2010sw}.  
Even though the particle-level distributions in Figures~\ref{fig:9}(c)-\ref{fig:9}(f) are nearly flat, all of the reconstructed shapes are non-uniform.  However, there are clear trends: the track pull has a peak at $\pi/2$ and the calorimeter pull is peaked at zero\footnote{An exception is Fig.~\ref{fig:9}(f) for which the peaks are slightly shifted.  This is due to the dependence of the pull angle on the jet $p_\text{T}$; with a higher $p_\text{T}$ threshold, Fig.~\ref{fig:9}(f) resembles Fig.~\ref{fig:9}(d).}.  Therefore, to understand the detector response for the jet pull in $t\bar{t}$, it suffices to study the truth to reconstructed jet pull angle detector response in Fig.~\ref{fig:9}(a) and Fig.~\ref{fig:9}(d) which are representative of the possible shapes and distortions in Fig.~\ref{fig:9}.  To minimize the dependence on the physics processes creating the peak at zero in Fig.~\ref{fig:9}(a), most of the discussion in this section will be focused on Fig.~\ref{fig:9}(d) where any departure from a uniform distribution provides insight into detector effects.

\begin{figure}[h!]
\centering
\begin{overpic}[width=.4\columnwidth]{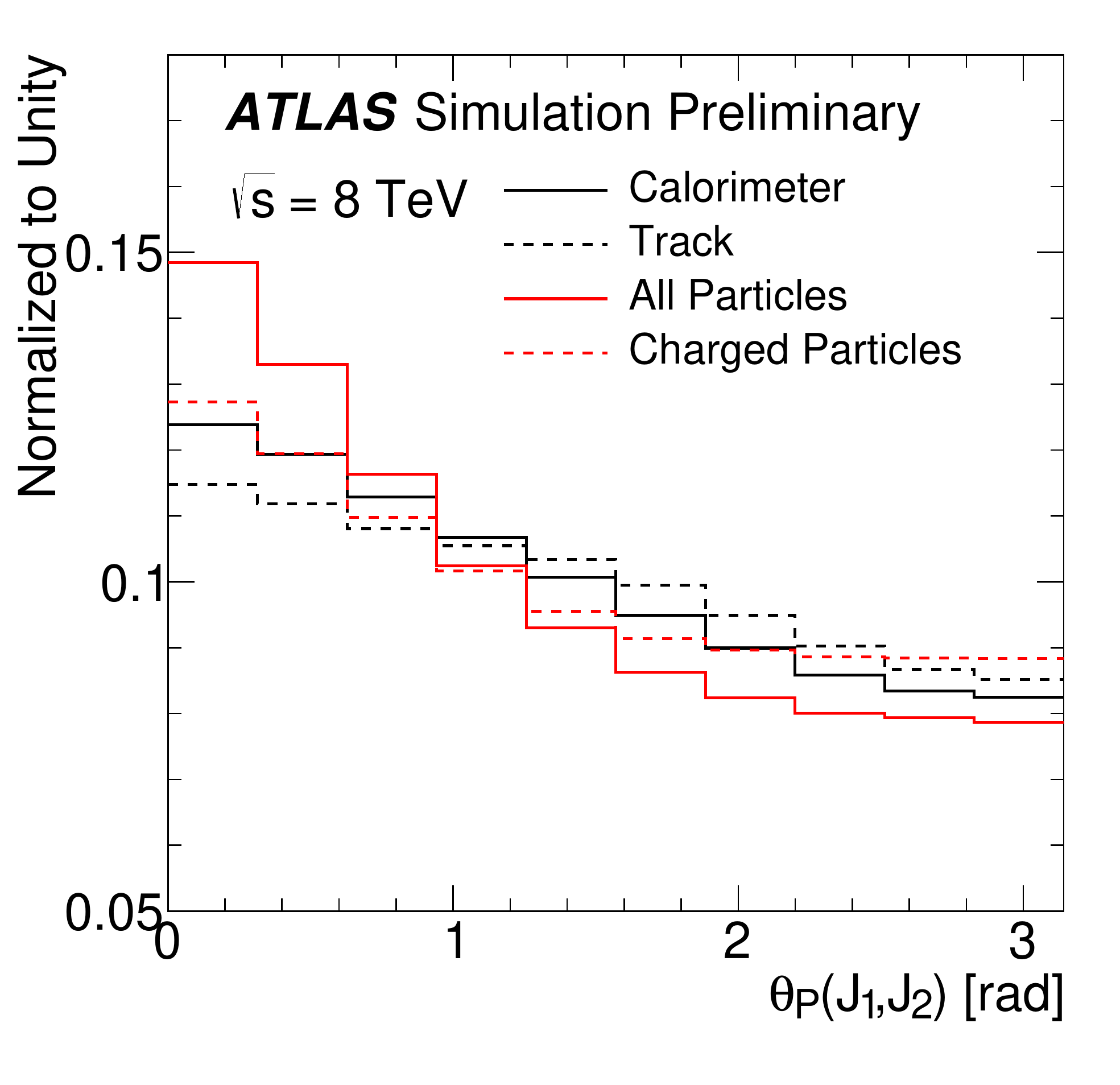}\put(25,25){(a)}
\end{overpic}\begin{overpic}[width=.4\columnwidth]{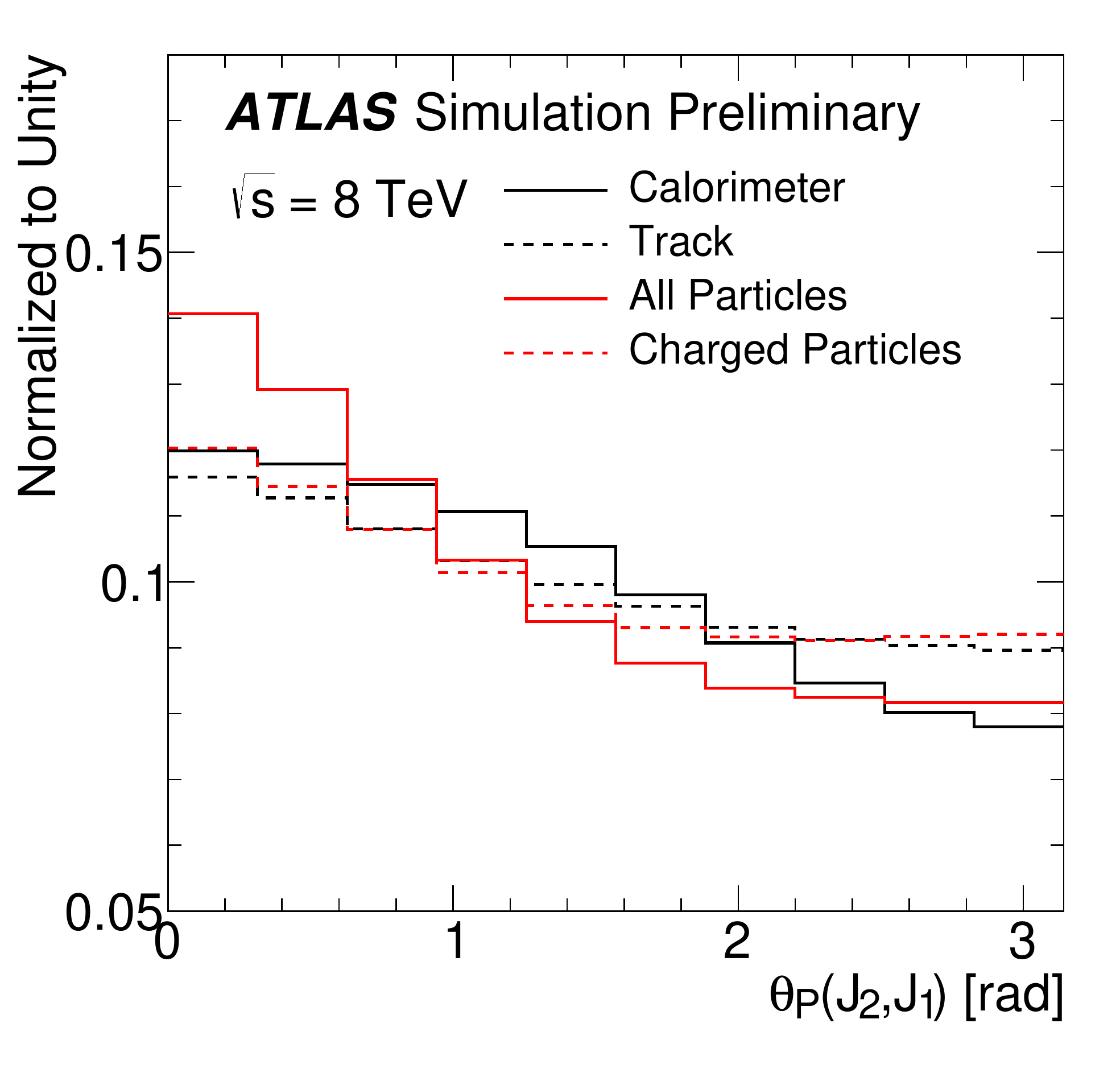}\put(25,25){(b)}\end{overpic}
\begin{overpic}[width=.4\columnwidth]{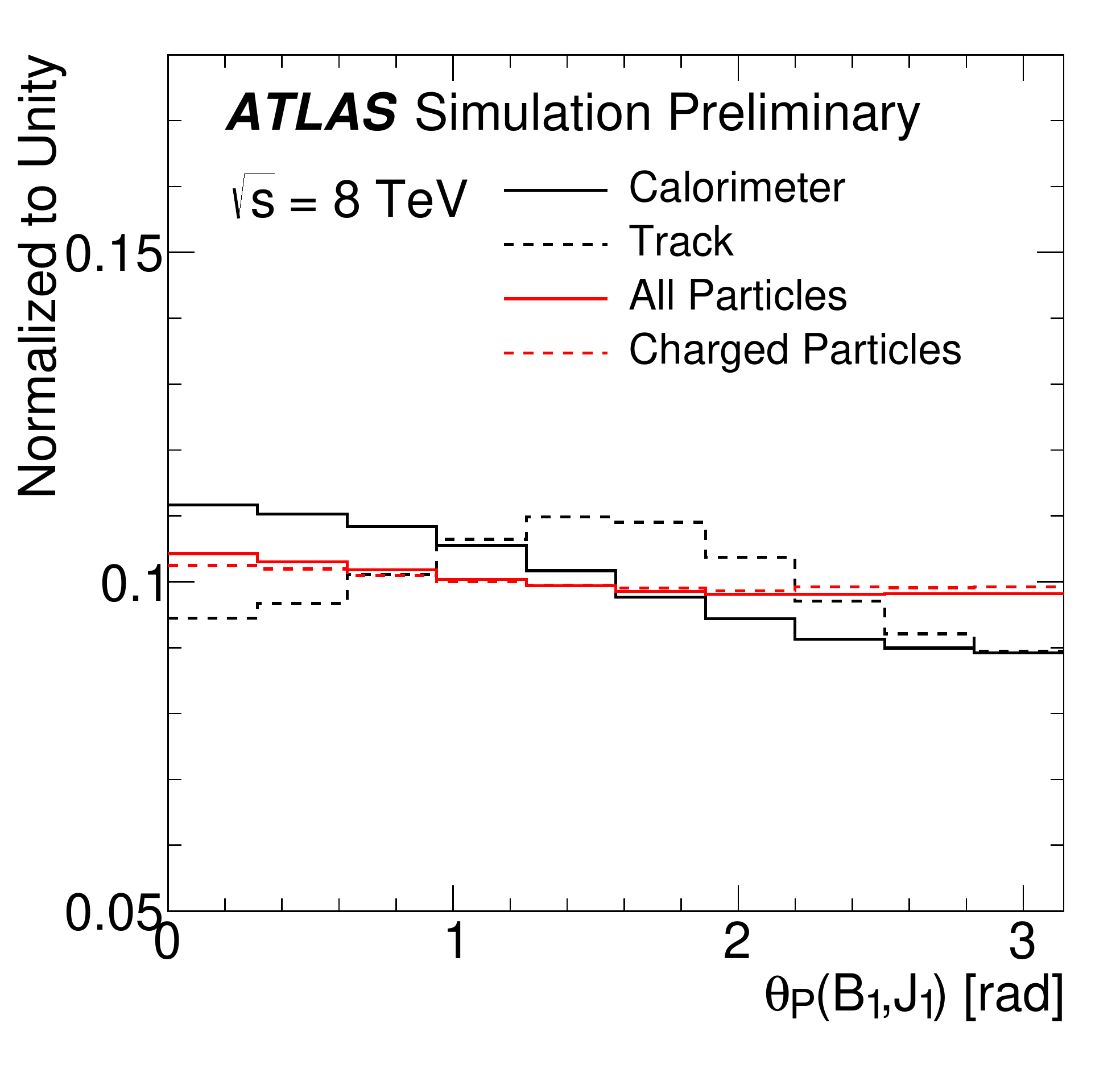}\put(25,25){(c)}\end{overpic}
\begin{overpic}[width=.4\columnwidth]{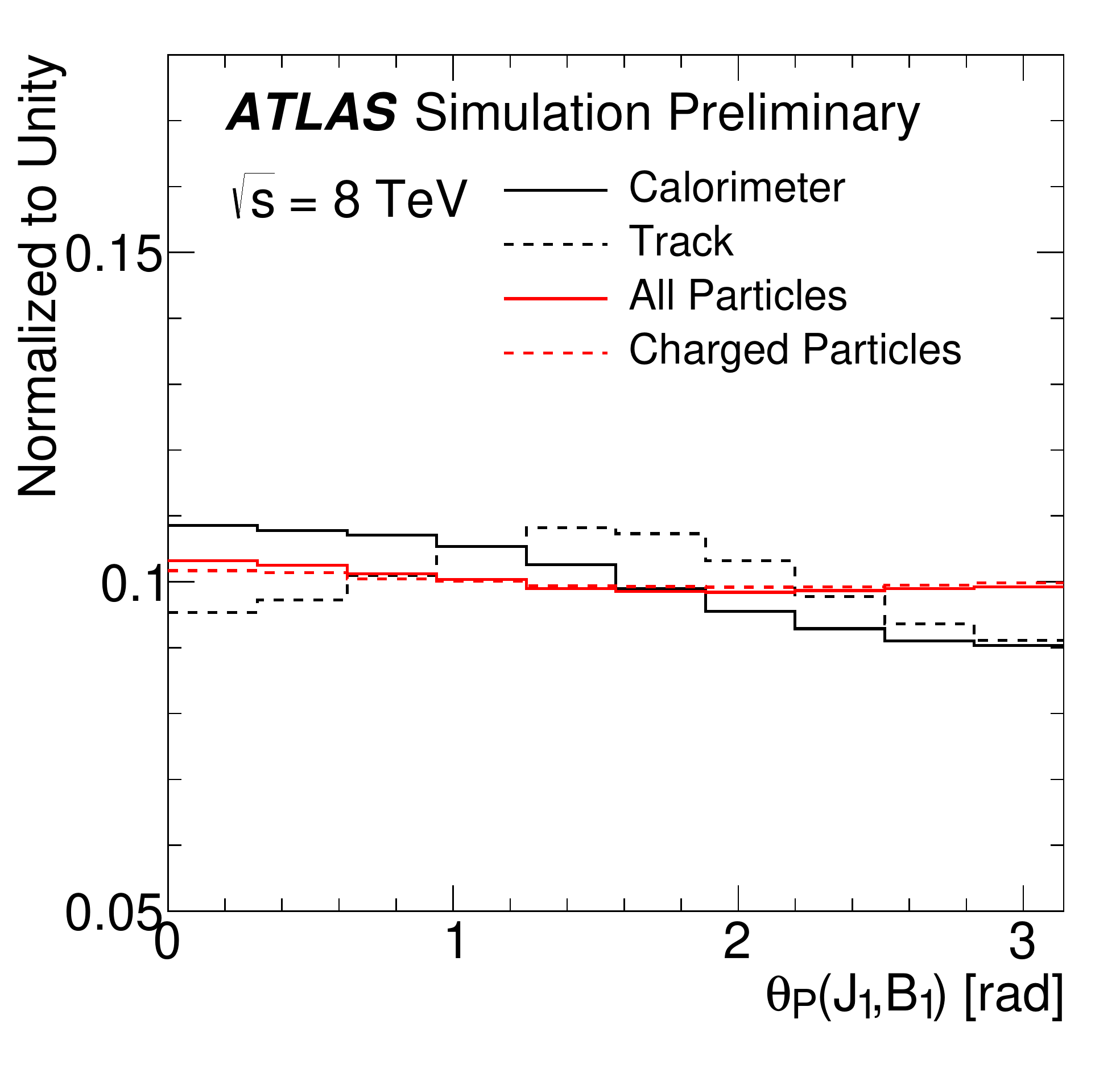}\put(25,25){(d)}\end{overpic}
\begin{overpic}[width=.4\columnwidth]{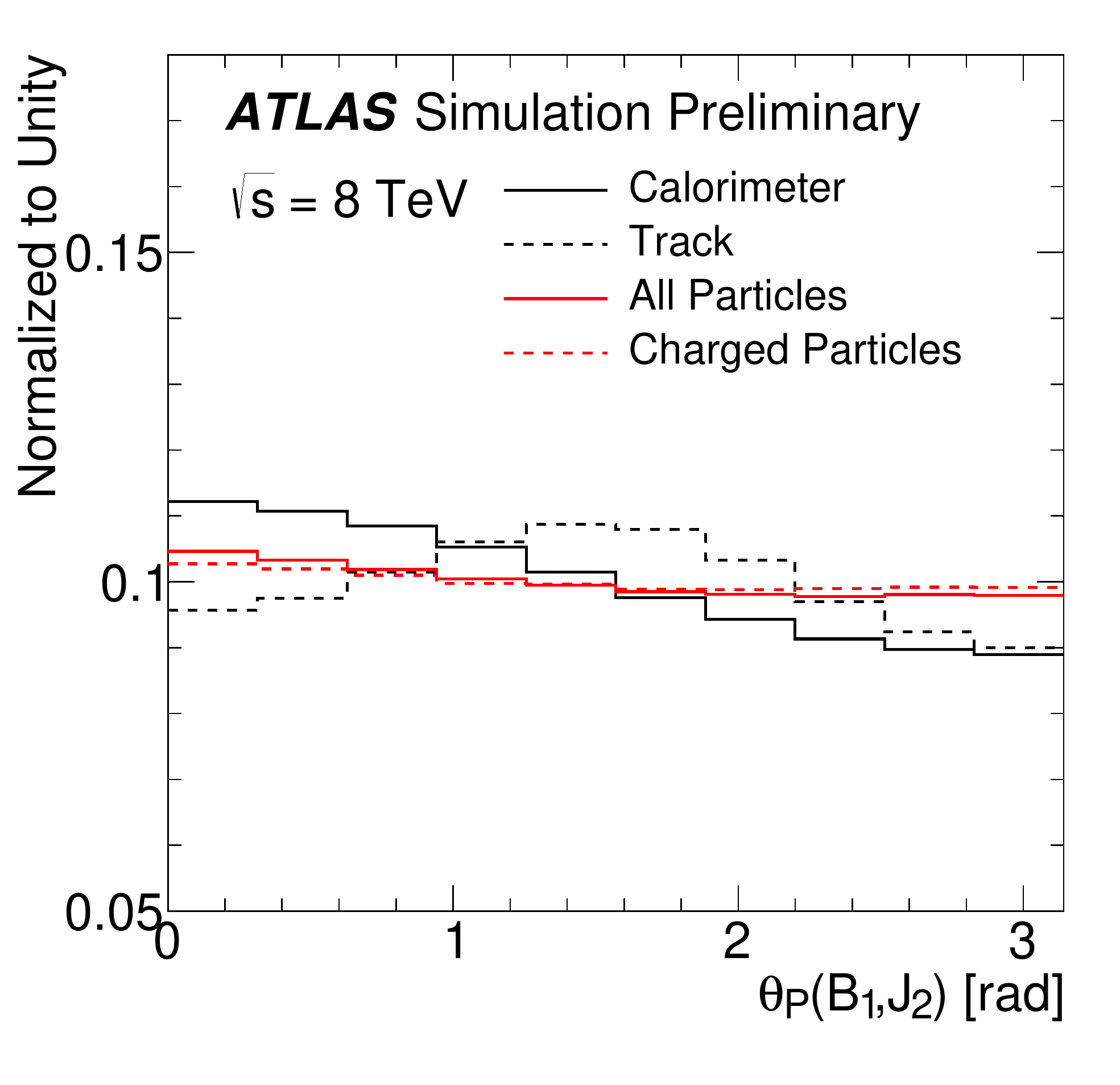}\put(25,25){(e)}\end{overpic}
\begin{overpic}[width=.4\columnwidth]{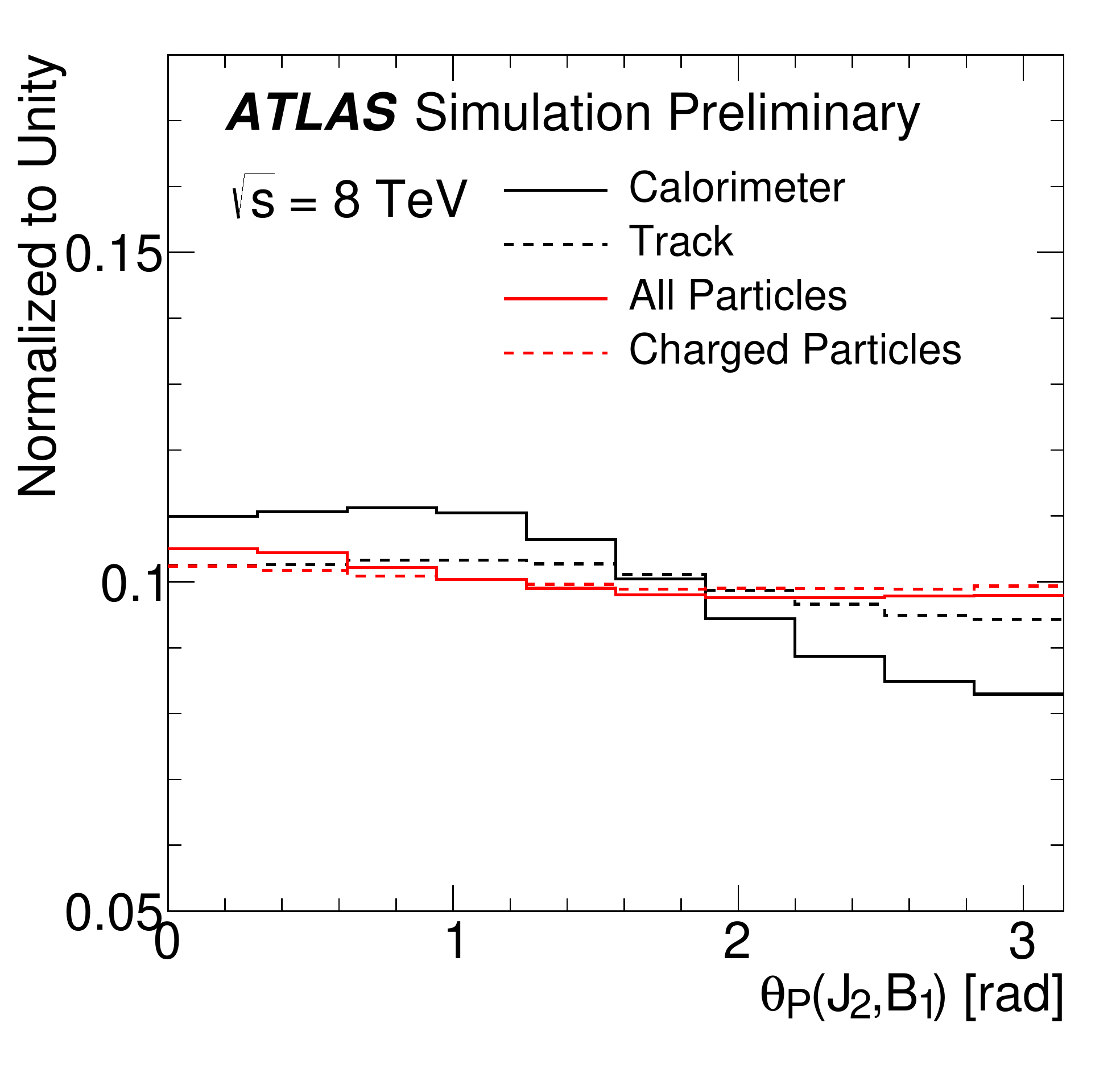}\put(25,25){(f)}\end{overpic}
 \caption{The jet pull angle $\theta_P(X,Y)$ distribution for various choices of $X$ and $Y$ for particle-level jets and also for detector-level jets matched to the particle-level jets.  }
 \label{fig:9}
\end{figure}

\clearpage

\subsection{Jet Pull Angle Response}
\label{sec:pullangleresponse}

The transition between particle-level and detector-level distributions is characterized by the {\it jet pull angle response}, $R(\theta_P)$ -- the difference between the detector-level jet pull angle and the particle-level jet pull angle.   The calorimeter/all-particles pull angle is calculated from clusters for detector-level jets and all constituents for particle-level jets.  The track/charged-particles pull angle uses tracks ghost-associated to the jet for reconstructed jets and charged constituents for particle-level jets.  The resolution of the jet pull angle is significantly different depending on the type of constituent used in the definition.  Figure~\ref{fig:11} shows the inclusive jet pull angle response for both the track/charged particle and calorimeter/all particles pull angles.  It is evident from the different widths of the two sets of distributions in Fig.~\ref{fig:11} that the track pull angle is measured more precisely than the calorimeter pull angle.  In terms of the RMS of the jet pull angle response, this corresponds to about a $20\%$ improved resolution of the track pull angle over the calorimeter pull angle resolution.  The numbers in Fig.~\ref{fig:11} also indicate small biases in the jet pull angle distributions.  These are expected from Fig.~\ref{fig:9}, which show asymmetric shape deformations between the particle-level and detector-level distributions.  

\begin{figure}[h!]
 \centering
\includegraphics[width=.49\columnwidth]{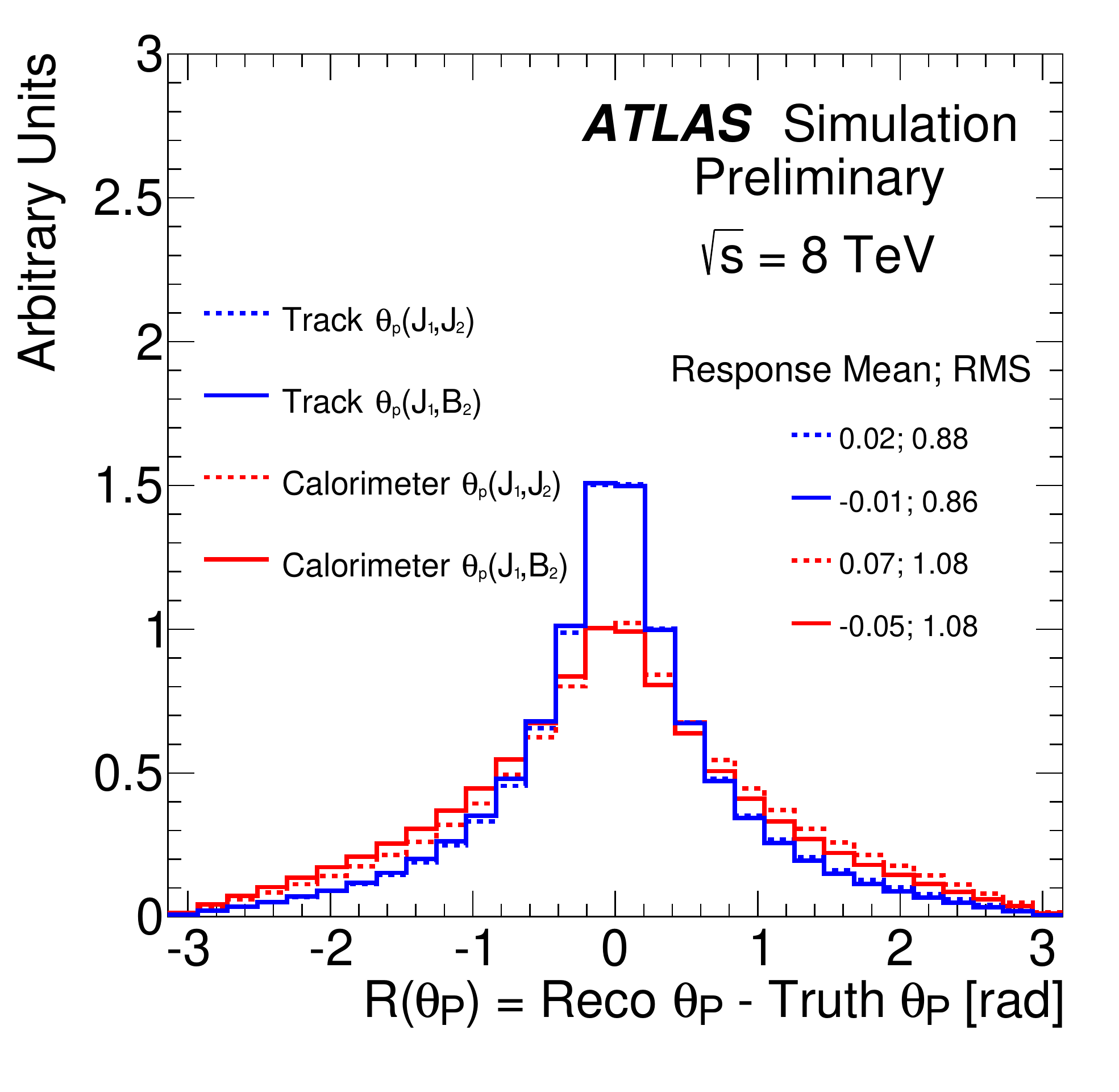}
 \caption{The distribution of the jet pull angle response, $R(\theta_P)$, for both $\theta_P(J_1,J_2)$ and $\theta_P(J_1,B_1)$ as well as for the calorimeter pull angle and the track pull angle. Statistical uncertainties on the mean and RMS are an order of magnitude less than the values shown.}
 \label{fig:11}
\end{figure}

In order to fully understand the transition in shapes between particle-level and detector-level in Fig.~\ref{fig:9} more information is needed beyond the inclusive jet pull angle response from Fig.~\ref{fig:11}.  There are three sources contributing to the resolution of the jet pull angle\footnote{For the track-based pull, the definition also introduces some resolution.  For instance, $K_s$ decays and photon conversions that occur before/inside the pixel detector contribute to reconstructed tracks, but are not in the list of stable MC charged particles.  Also, the $p_T>500$ track threshold is not applied to the MC particles.  All three of these effects have been studied and found to have a very small impact on the resolution and a negligible impact on the pull angle distribution shape.} $\theta_P(X,Y)$ response:  the jet constituent angular resolution and momentum resolution with respect to $X$, the angular resolution of $X$, and the angular resolution of $Y$.  For both the all-particles and charged-particles pull angles, all angles are computed with respect to the calorimeter (or all-particles) jet axes, independent of the constituents used in the calculation of the jet pull angle.  The considerations so far have treated all the resolutions inclusively.  It is difficult to systematically remove the resolution from the jet constituents, but it is straightforward to study the effect of the jet angular resolution on the jet pull angle.  

One measure of the jet angular resolution is $\sigma^\text{match}$: the $\Delta R$ between reconstructed jets and matched particle-level jets\footnote{There are at least two contributions to $\sigma^\text{match}$: 1) the angular distortions in momentum when particles become calorimeter clusters and 2) the set of particles associated with the measured calorimeter clusters may not be the same as the particles in the matched truth jet.  The former effect can be studied by systematically smearing the truth jet axis and this dominates $\sigma^\text{match}$.  The impact of increased distortions of the truth axis is discussed in the context of Fig.~\ref{fig:13}.}.  Figure~\ref{fig:14} shows the impact setting $\sigma^\text{match}=0$ by systematically replacing detector-level jet axes with the corresponding matched particle-level jet axes.  For both the calorimeter and track pull angles $\theta_P(J_1,B_1)$, setting $\sigma^\text{match}=0$ of the $b$-jet has essentially no influence on the jet pull angle distribution due to the large lever-arm spanned by the vector connecting $B_1$ and $J_1$.  However, setting $\sigma^\text{match}=0$ of the $J_1$ axis has a dramatic impact on the pull distribution shape.  For the calorimeter pull, setting $\sigma^\text{match}=0$ of the $J_1$ axis shifts the peak of the distribution to $\pi/2$ instead of at $0$.  Since the track angular resolution is much better than the calorimeter cluster angular resolution, the track pull angle resolution is dominated by the calorimeter jet angular resolution.  By setting $\sigma^\text{match}=0$, the pull angle response RMS decreases and the right plot of Fig.~\ref{fig:14} shows that the jet pull angle distribution is nearly the same as the truth distribution.

\begin{figure}[h!]
 \centering
\includegraphics[width=.49\columnwidth]{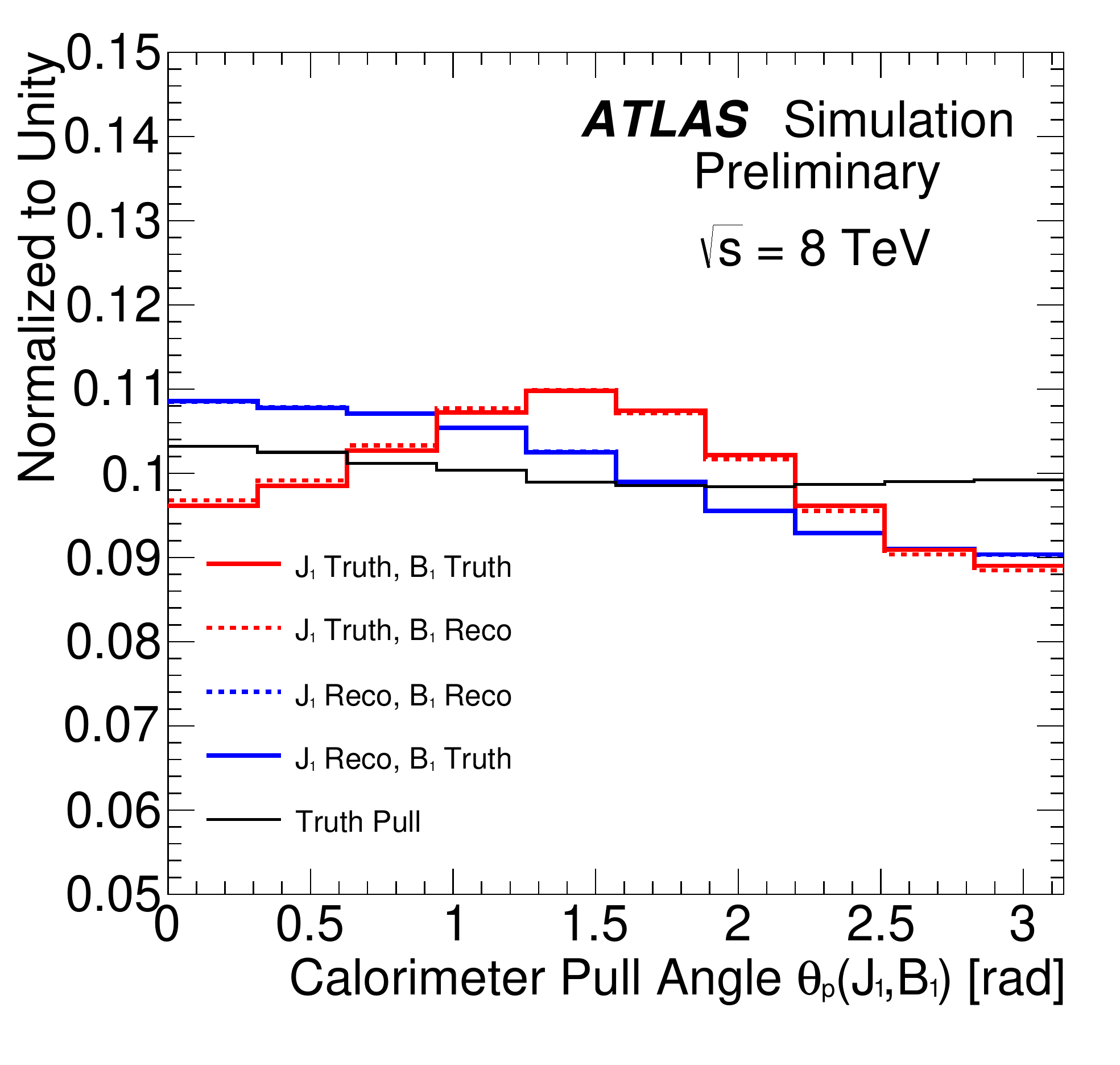}
\includegraphics[width=.49\columnwidth]{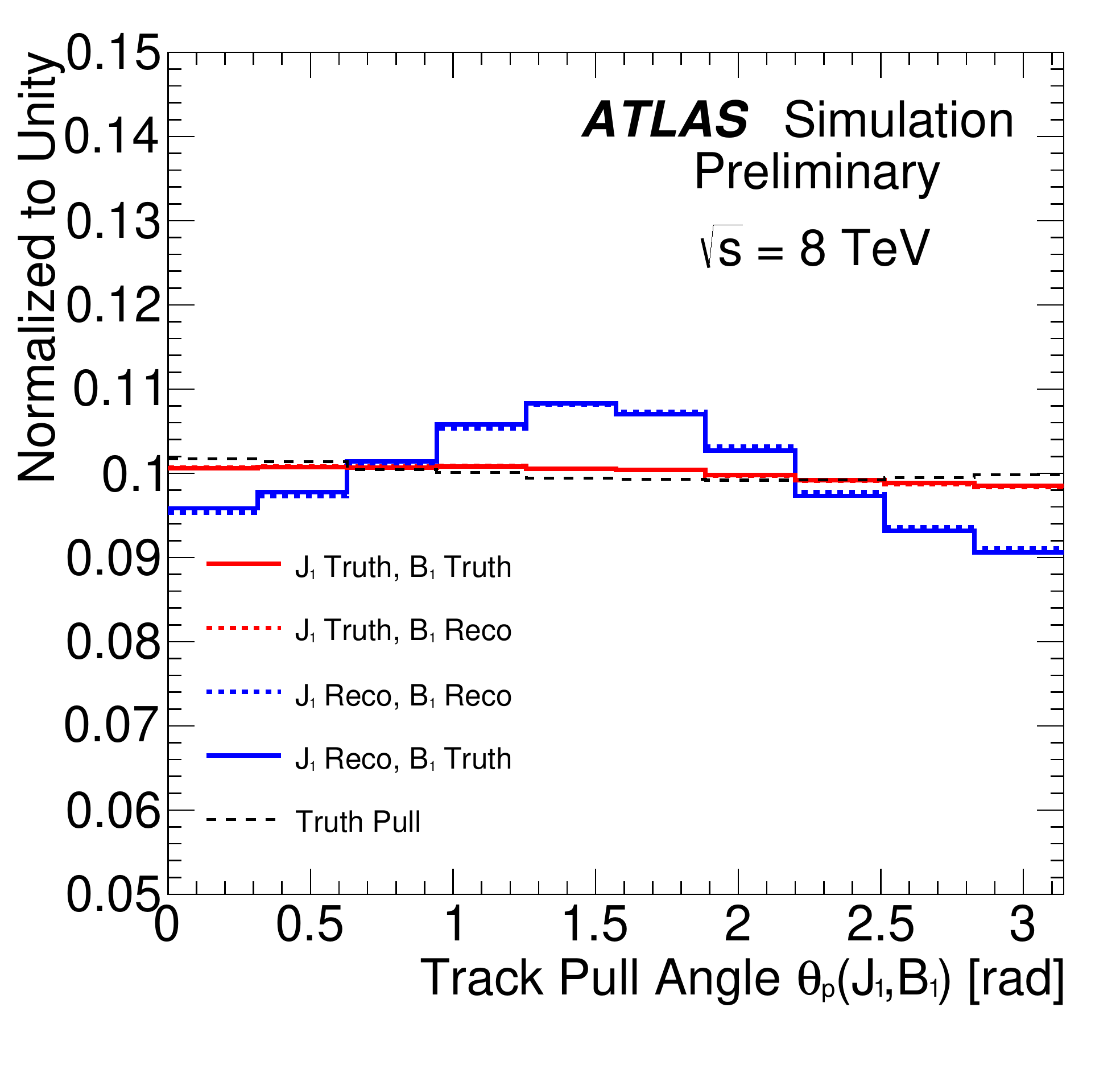}
 \caption{The $\theta_P(J_1,B_1)$ distribution for the calorimeter (left) and track (right) pull angle after replacing detector-level jet axes by particle-level jet axes.  }
 \label{fig:14}
\end{figure}

The right plot of Fig.~\ref{fig:14} suggests a simple model for building intuition for the peak at $\pi/2$.  Consider a {\it pseudo MC} model with $N$ massless particles generated randomly from the decay of a single scalar particle whose mass and boost are tuned so that the lab frame $p_\text{T}$ of the sum of the decay products is specified and such that all the decay products fall within $\Delta R<R$ of the `jet' axis, defined by the vector sum of all the decay products.  The dashed line in figure~\ref{fig:13} shows the jet pull angle distribution for such a model in which two such jets are generated randomly, and with $N=10,p_\text{T}=80$ GeV, and $R=0.4$.  As expected for the undistorted distribution, the jet pull angle is uniform on $[0,\pi)$.  To model the resolution, the constituents are fixed and the jet axis is smeared according to a bivariate normal distribution with zero correlation and $\sigma_\phi^\text{nom},\sigma_y^\text{nom}$ taken from the ATLAS detector simulation: $\sigma_\phi^\text{nom}\approx 0.025\approx \sigma_y^\text{nom}/1.5$.  The resolution used in the simulation is given by $\sigma_\phi=r\times \sigma_\phi^\text{nom},\sigma_y=r\times a\times \sigma_y^\text{nom}$, where $r$ is a multiplicative factor and $a$ is an asymmetry.  The left plot of Fig.~\ref{fig:13} for $r=1$ shows that the peak at $\pi/2$ is a prediction of this simple model.  By tuning the model parameters, one learns that this feature can be explained if the resolution in $y$ and resolution in $\phi$ are not the same in ATLAS; the right plot of Fig.~\ref{fig:13} does not peak at $\pi/2$.   The peak at $\pi/2$ comes from two facts: (1) in $(\Delta y,\Delta\phi)$, the pull vector tends to be stretched towards the $\pm \Delta y$ axis and (2) the distribution of $\Delta y(J_1,B_1)$ is peaked at zero and thus in $(\Delta y,\Delta \phi)$ coordinates, $B_1$ lies on the $\Delta\phi$ axis.  As $r$ is increased so that the asymmetry is no longer relevant, the peak at $\pi/2$ disappears in all cases.  In fact, it is possible to use these observations to {\it measure} the jet angular resolution with the jet pull angle.  Figure~\ref{fig:13333} shows a $\chi^2$ fit between $20$ bins of the pull angle from simulation and templates formed from the toy MC.  The minimum $\chi^2$ is at $a=1.5, r=1$ as desired, though the fit is much more sensitive to $a$ than to $r$\footnote{As a result, after the origin correction (see Sec.~\ref{origincorrection}) this method looses precision.}.

A similar model can be created for the calorimeter pull angle, but the interpretation is less straight-forward.  In particular, using the same pseudo MC model for the particle-level selection, the calorimeter pull angle resolution can be modeled by smearing all particles and then additionally recomputing the jet axis, since the cluster angular resolution need not be small compared to the jet angular resolution as was the case for tracks.  Such a model can generically predict peaks at $0,\pi$ and with angular resolution asymmetry, $\pi/2$, but since there is not a one-to-one matching between particles and clusters, it is not possible to map these models onto a realistic description of the detector.

\begin{figure}[h!]
 \centering
\includegraphics[width=.49\columnwidth]{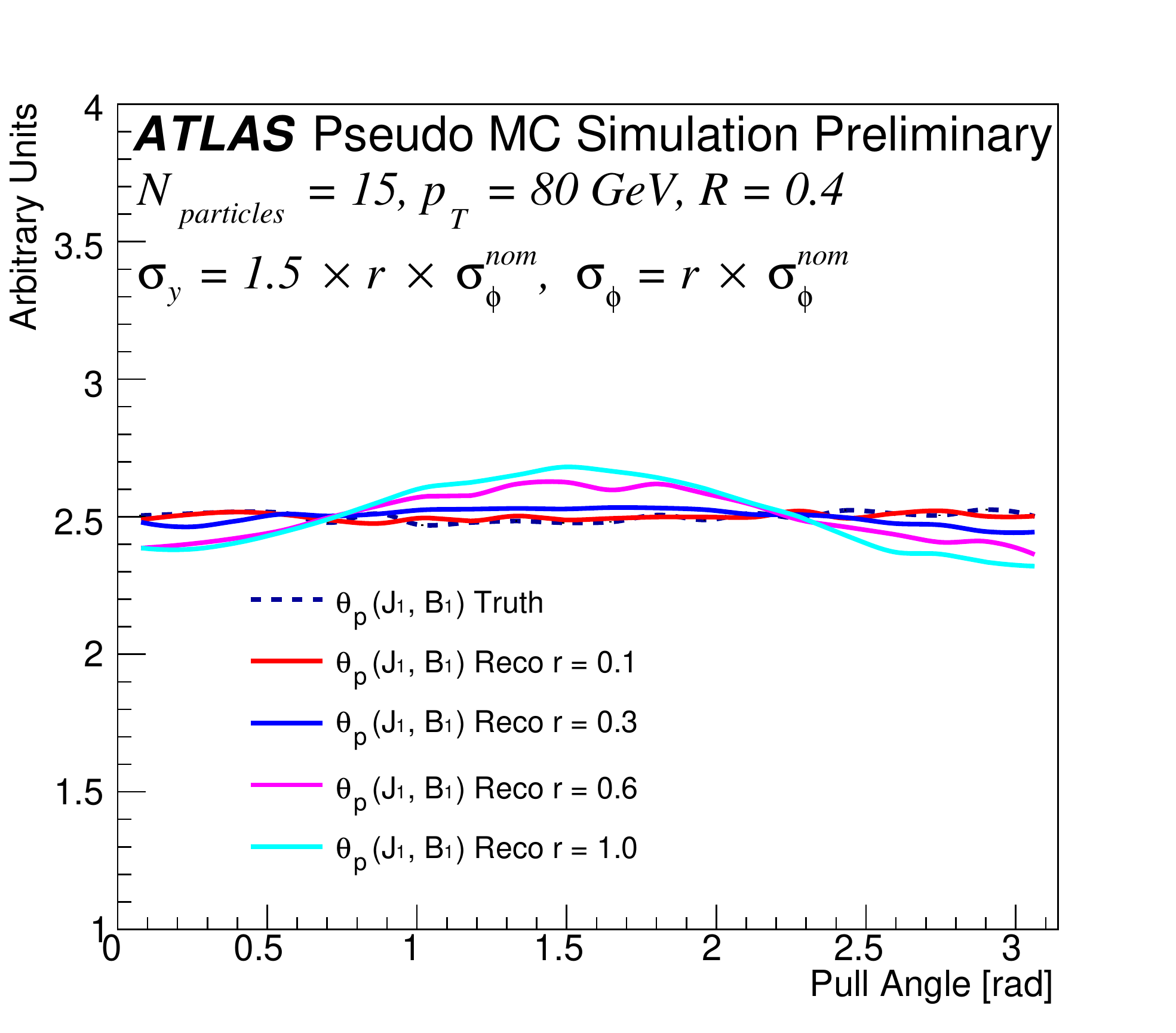}
\includegraphics[width=.49\columnwidth]{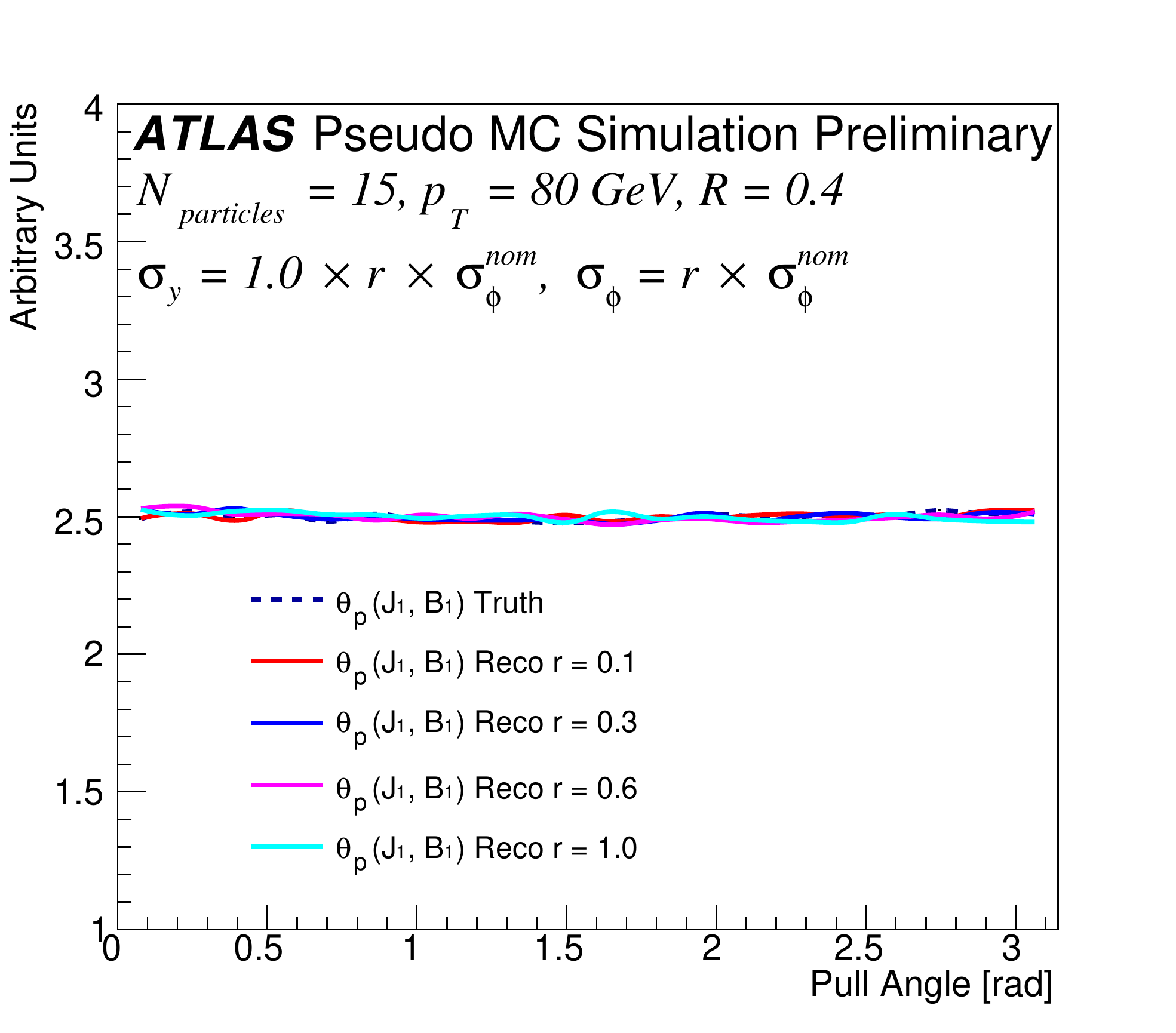}
 \caption{The jet pull angle constructed in a pseudo MC with various resolution settings for the jet constituents.}
 \label{fig:13}
\end{figure}

\begin{figure}[h!]
 \centering
\includegraphics[width=.5\columnwidth]{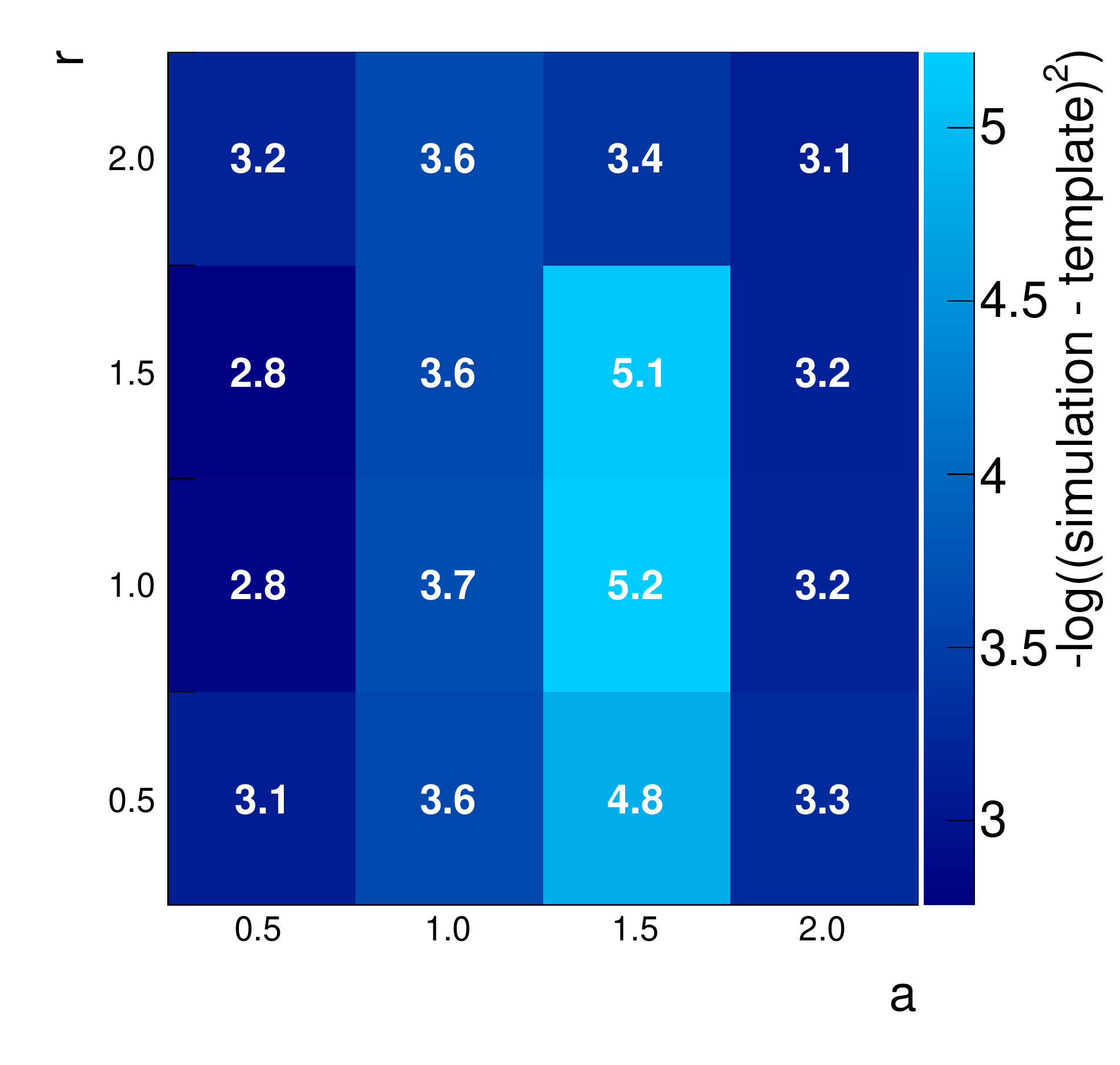}
 \caption{A $\chi^2$-like scan between the simulated charged-particles pull angle distribution and templates using toy MC that are with varying asymmetry $a$ and resolution $r$.  The minimum $\chi^2$ is for $a=1.5$ and $r=1.0$.}
 \label{fig:13333}
\end{figure}

\clearpage

\subsection{Jet Kinematics and the Jet Pull Angle Response}
\label{colorflowkins}

Unlike other jet substructure variables, the jet pull angle depends not only on the orientation of constituents within a jet, but also the placement of jets within an event, hence the term {\it jet superstructure}.  Thus, even at particle-level, the jet pull angle can depend on the relative orientations of jets in a given event.  The right plot of Fig.~\ref{fig:6} shows the relationship between the particle-level jet pull angle $\theta_P(J_1,J_2)$ and the relative distance between jets, $\Delta R(J_1,J_2)$.  The particle-level distribution shows a strong dependence on $\Delta R$, with smaller values of $\Delta R$ corresponding to a larger peak at zero.  In fact, it is mostly through $\Delta R$ that the particle-level distribution of $\theta_P(J_1,J_2)$ depends on the $p_\text{T}$ of $J_1$, as described below. The particle-level pull angle distributions that involve one of the $b$-jets are nearly independent of $\Delta R$ (and $p_\text{T}$).  The right plot of Fig.~\ref{fig:6} shows the RMS of the jet pull angle response as a function of $\Delta R(J_1,B_1)$, which is used because there is no $\Delta R$ dependence at particle-level.

\begin{figure}[h!]
 \centering
\includegraphics[width=.45\columnwidth]{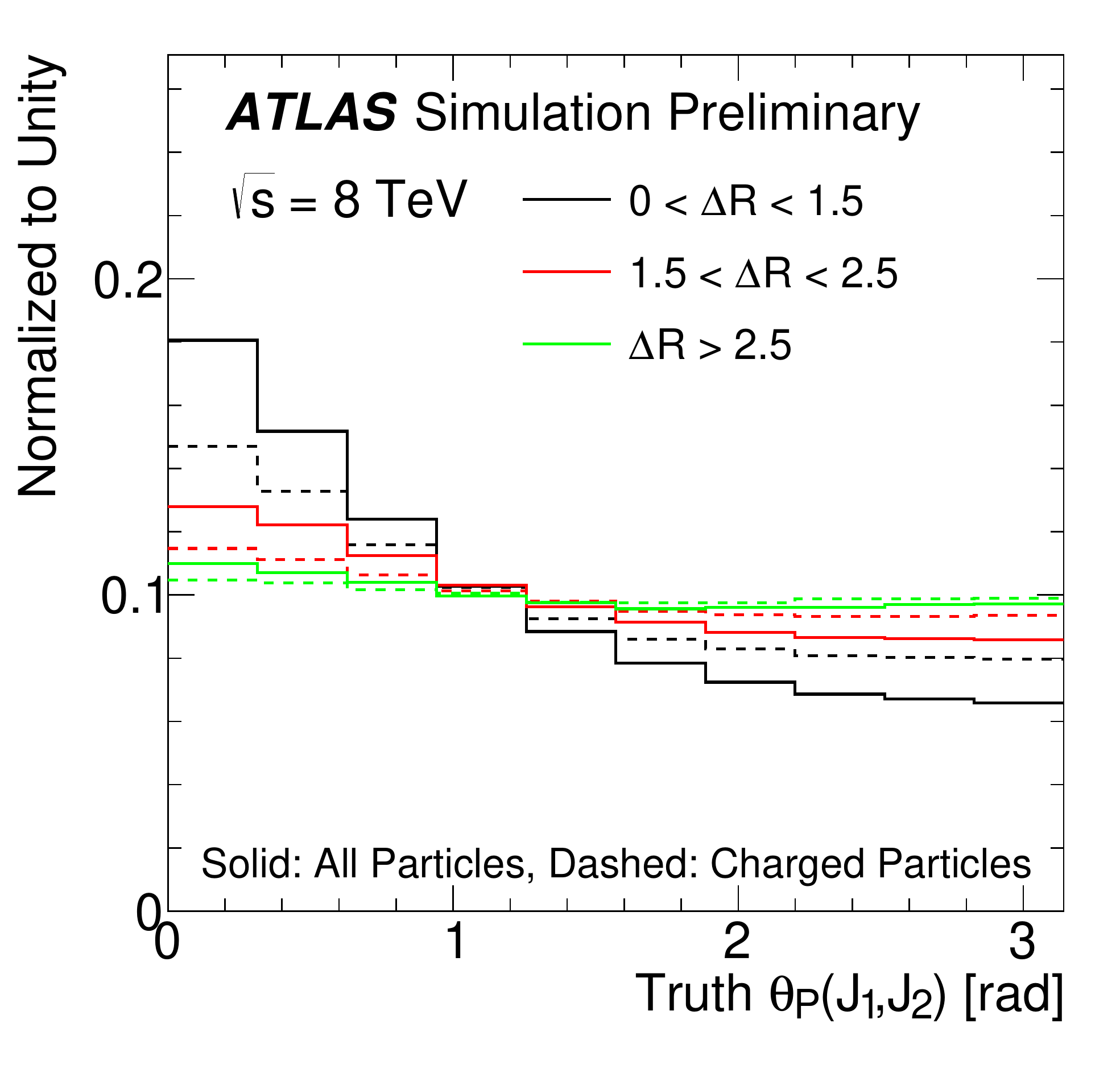}\hspace{2mm}
\includegraphics[width=.45\columnwidth]{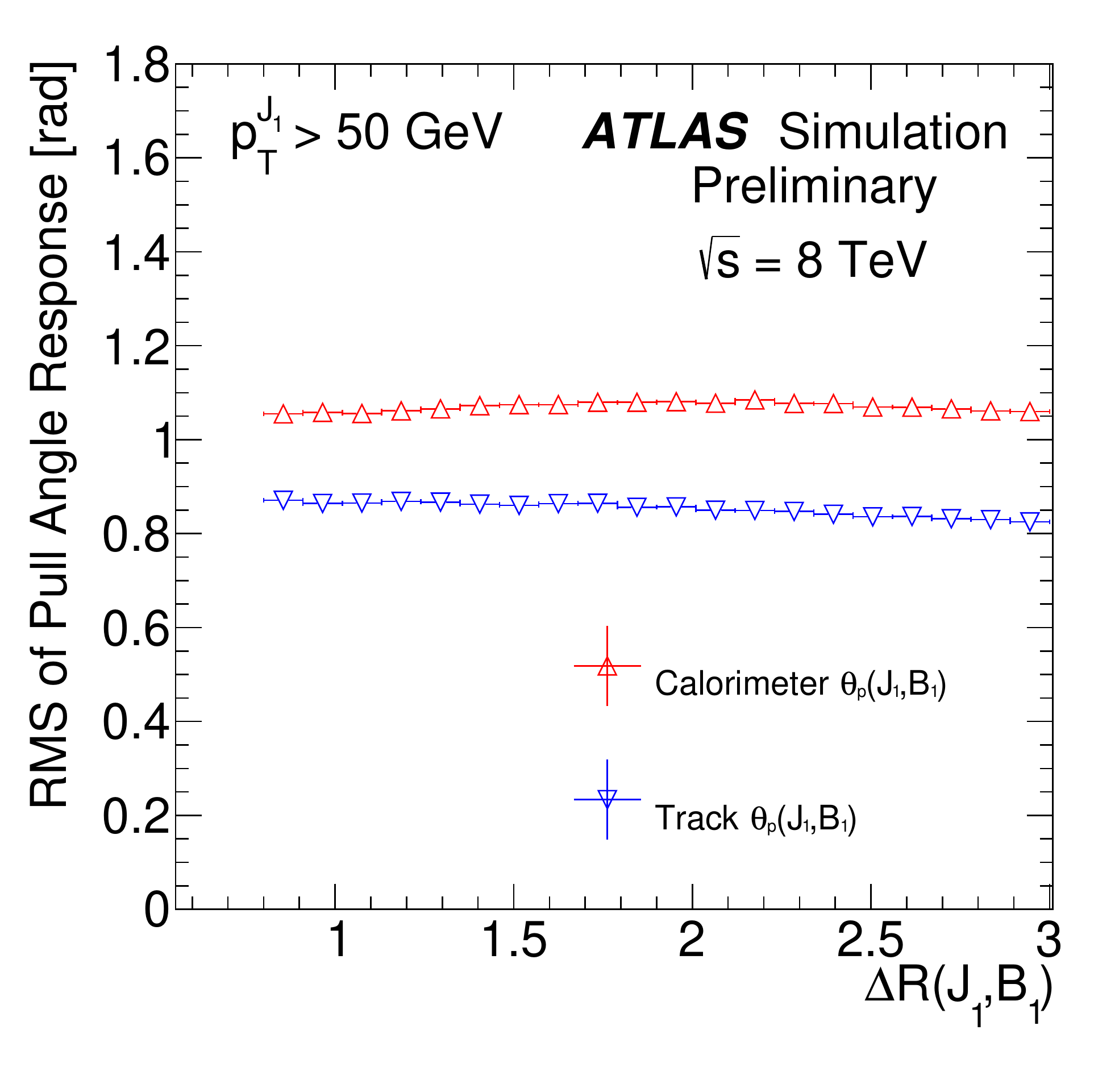}
 \caption{Relationship between the $\theta_\text{P}$ (response) and $\Delta R$ between the jets.}
 \label{fig:6}
\end{figure}

The left plot of Fig.~\ref{fig:66} shows the relationship between the jet pull angle $\theta_P(J_1,J_2)$ and the $p_\text{T}$ of $J_1$.  There seems to be a clear relationship between $p_\text{T}^{J_1}$ and $\theta_P(J_1,J_2)$.  However, this is inconsistent with the truth distributions in Figures~\ref{fig:9}(a) and~\ref{fig:9}(b); these distributions are nearly identical and yet the underlying $p_\text{T}$ distribution for Fig.~\ref{fig:9}(b) must be softer than that of Fig.~\ref{fig:9}(a).  The resolution is that the pull angle distribution depends on $p_\text{T}$ only through $\Delta R$.  At high $J_1$ $p_\text{T}$, $J_1$ and $J_2$ have smaller angular separation since the boost of the $W$ boson in this case is larger (see Chapter~\ref{cha:bosonjets}).  The right plot of Fig.~\ref{fig:66} shows the distribution of $\theta_P(J_1,J_2)$ in bins of the $p_\text{T}$ of $J_1$ for a fixed $\Delta R(J_1,J_2)$.  The $p_\text{T}$ dependence compared to the left plot of Fig.~\ref{fig:66} is significantly reduced.  
\begin{figure}[h!]
 \centering
\includegraphics[width=.45\columnwidth]{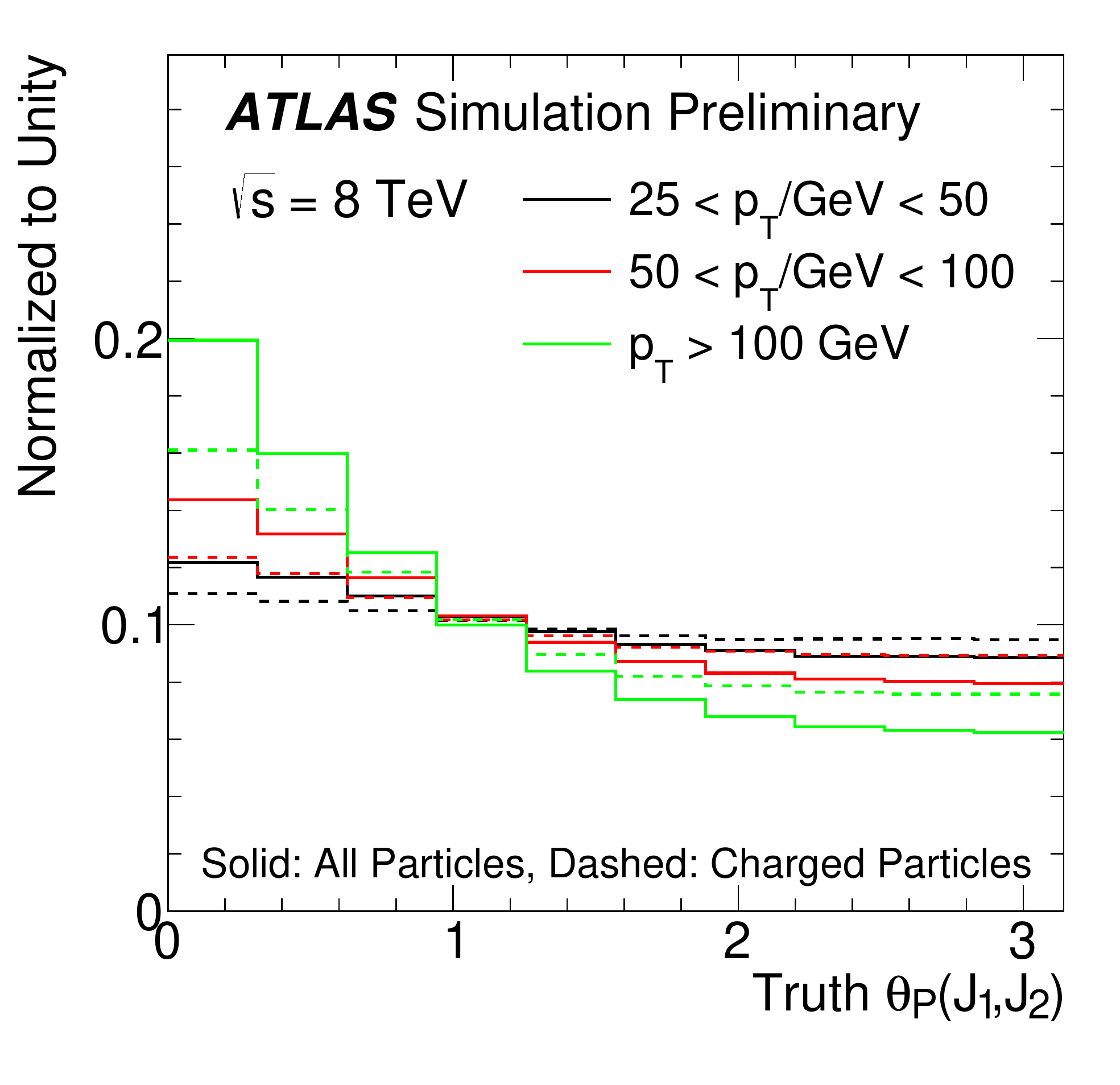}\hspace{2mm}
\includegraphics[width=.45\columnwidth]{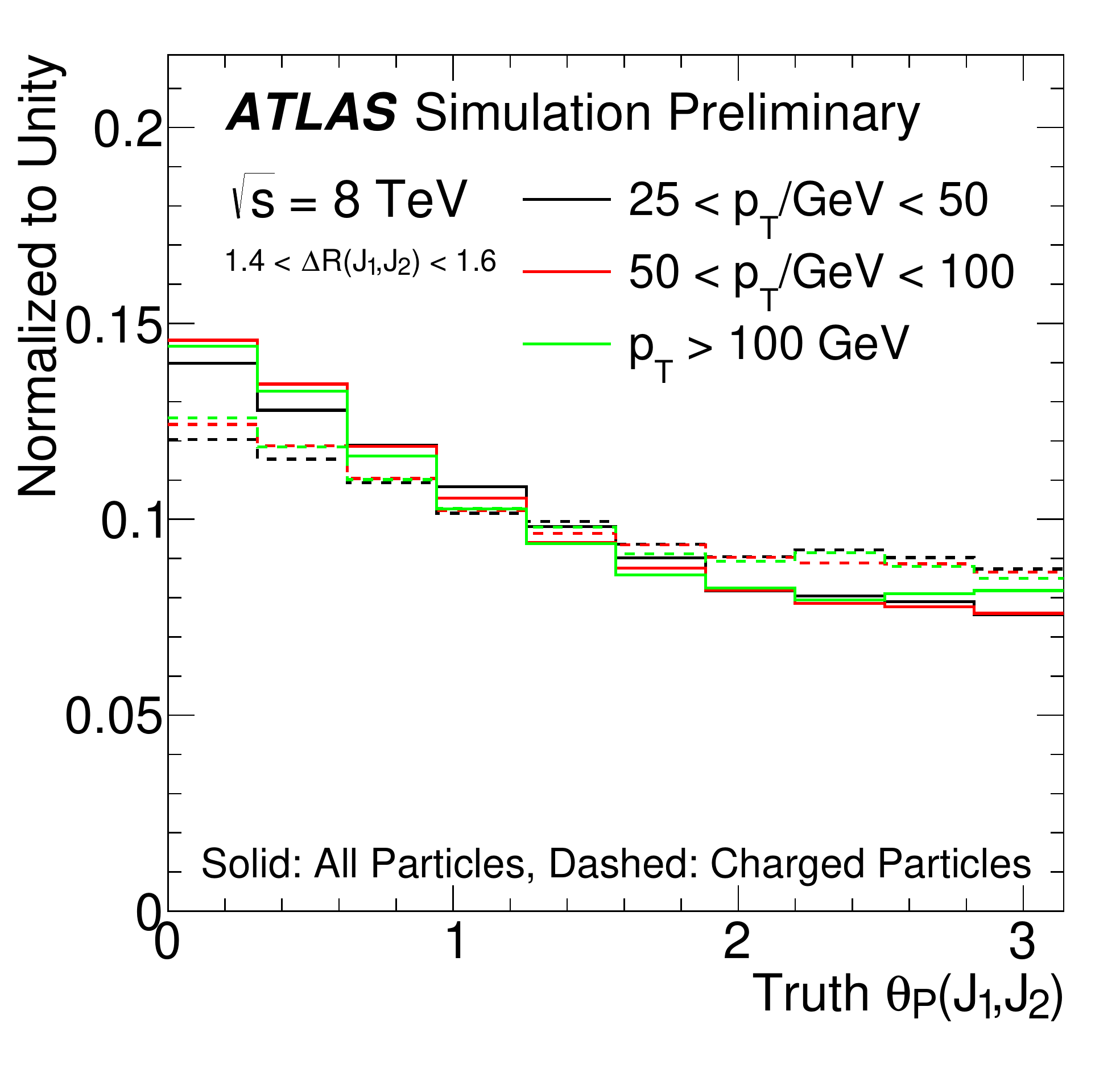}
 \caption{The jet pull angle distribution in three bins of $p_\text{T}$ of $J_1$.}
 \label{fig:66}
\end{figure}

Even though the jet pull angle is relatively independent of $p_\text{T}$, the response RMS does scale with $p_\text{T}$.  Figure~\ref{fig:15} shows the RMS of the jet pull angle response as a function of the $p_\text{T}$ of the leading $W$ daughter jet. The RMS improves with increasing $p_\text{T}$ as the relative jet energy resolution improves with energy.

\begin{figure}[h!]
 \centering
\includegraphics[width=.5\columnwidth]{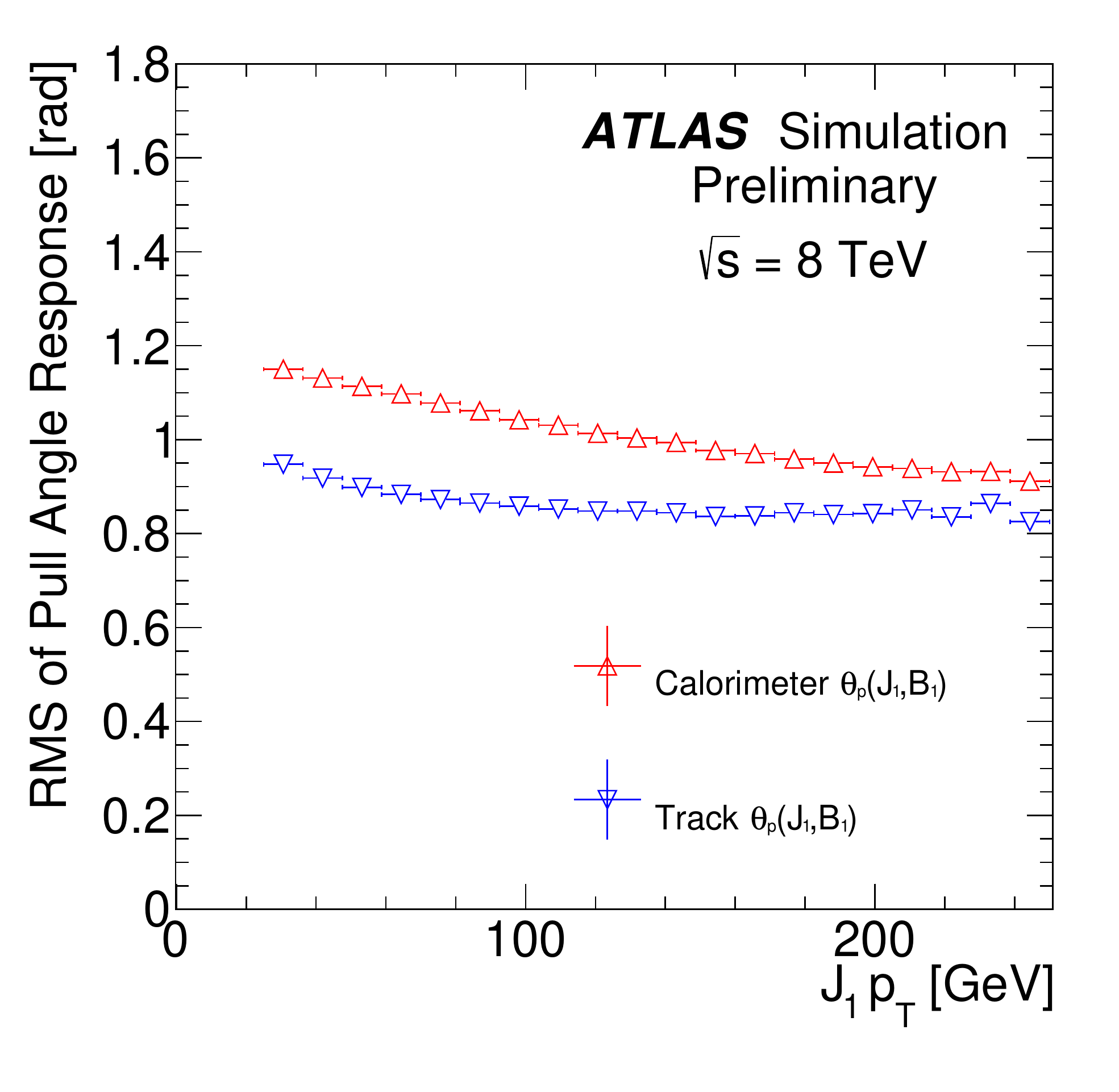}  \caption{The RMS of the jet pull angle response as a function of the jet $p_\text{T}$.  The $\Delta R$ between jets is restricted between $1.4$ and $1.6$ in the right plot.  The track-baed pull angle response is nearly flat by $p_\text{T}\sim 100$ GeV.}
 \label{fig:15}
\end{figure}

\clearpage

\subsection{Relationship Between the Jet Pull Angle Response and Jet Constituents}
\label{sec:ColorFlow:Performance:Constits}

As the pull vector is determined from the constituents inside a jet, the jet pull angle response could depend on the number and orientation of the constituents of $J_1$.   There are many substructure variables which capture various properties of the orientation of constituents within a jet.  One such property is the pull vector magnitude,

\begin{align}
|v_p(J)|=\left|\sum_{i\in J} \frac{p_\text{T}^i |r_i|}{p_\text{T}^{J}}\vec{r}_i\right|.
\end{align}

\noindent In dedicated phenomenological studies, it was shown that the pull magnitude is not useful in discriminating octet from singlet color states~\cite{Gallicchio:2010dq}.  However, this section will show that it is a useful handle on the jet pull angle resolution.  The jet pull vector magnitude can be considered a radial moment, with the radial distance $(\Delta R)^2$ from the jet axis weighted by the fractional constituent $p_\text{T}$.   The distribution of the magnitude for $J_1$ is shown in the left plot of Fig.~\ref{fig:16}.  Events with a reconstructed pull vector magnitude of zero for the track pull, corresponding to cases in which there are no tracks ghost-associated to the jet, are not shown.

The right plot of Fig.~\ref{fig:16} shows the relationship between the jet pull angle response RMS and the pull vector magnitude.   A small pull vector magnitude corresponds to a worse resolution, in some cases because of a small lever arm.  Since the pull vector magnitude can be measured, the right plot of Fig.~\ref{fig:16} suggests that it can be used as an in-situ tool for improving precision.  The left plot of Figure~\ref{fig:17} shows the RMS of the pull response as a function of the efficiency for a threshold requirement on the pull vector magnitude.  For instance, one can achieve a $\sim10\%$ reduction in the RMS of the jet pull angle response while maintaining a $90\%$ selection efficiency.

\begin{figure}[h!]
 \centering
\includegraphics[width=.45\columnwidth]{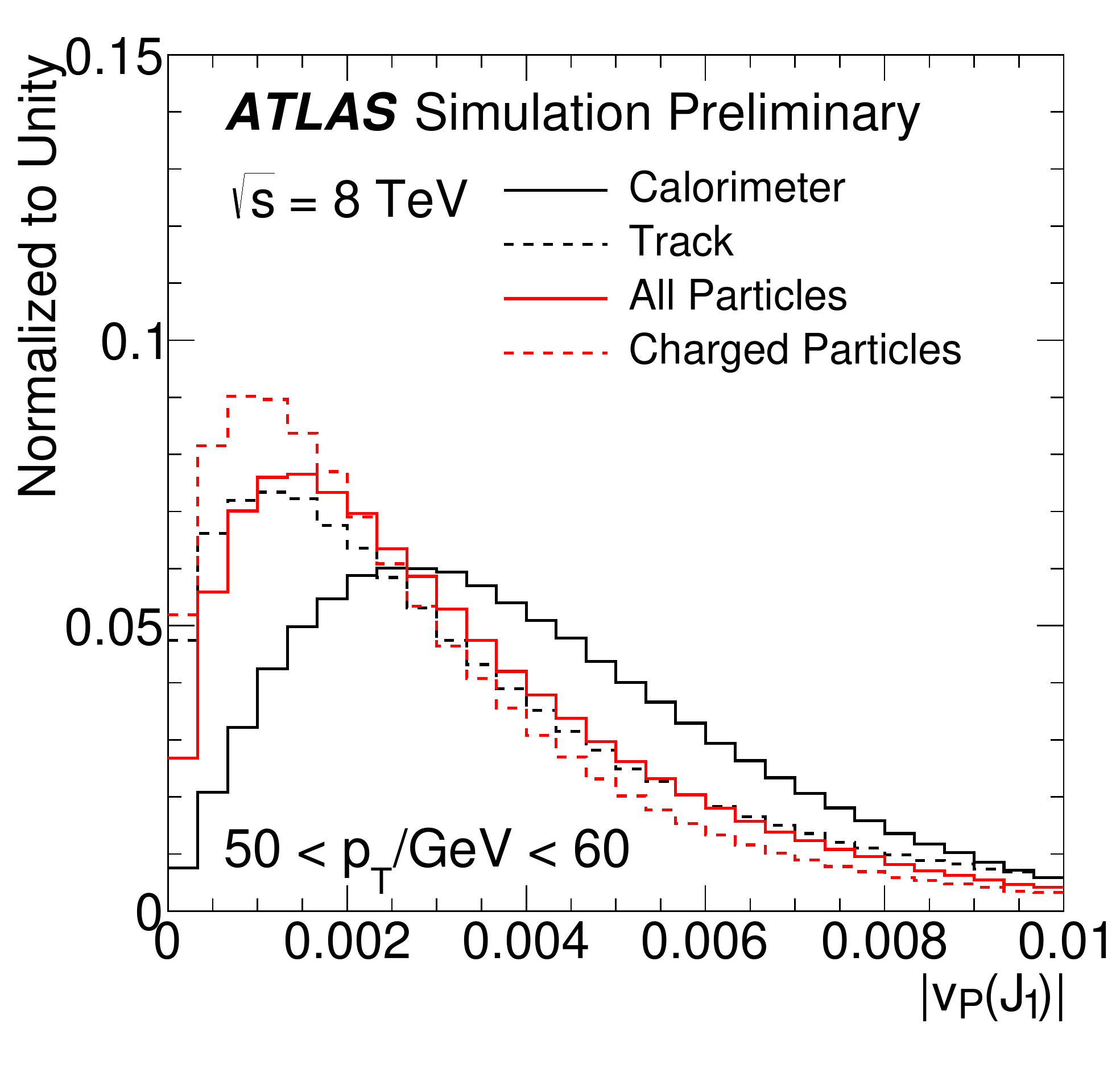}
\includegraphics[width=.45\columnwidth]{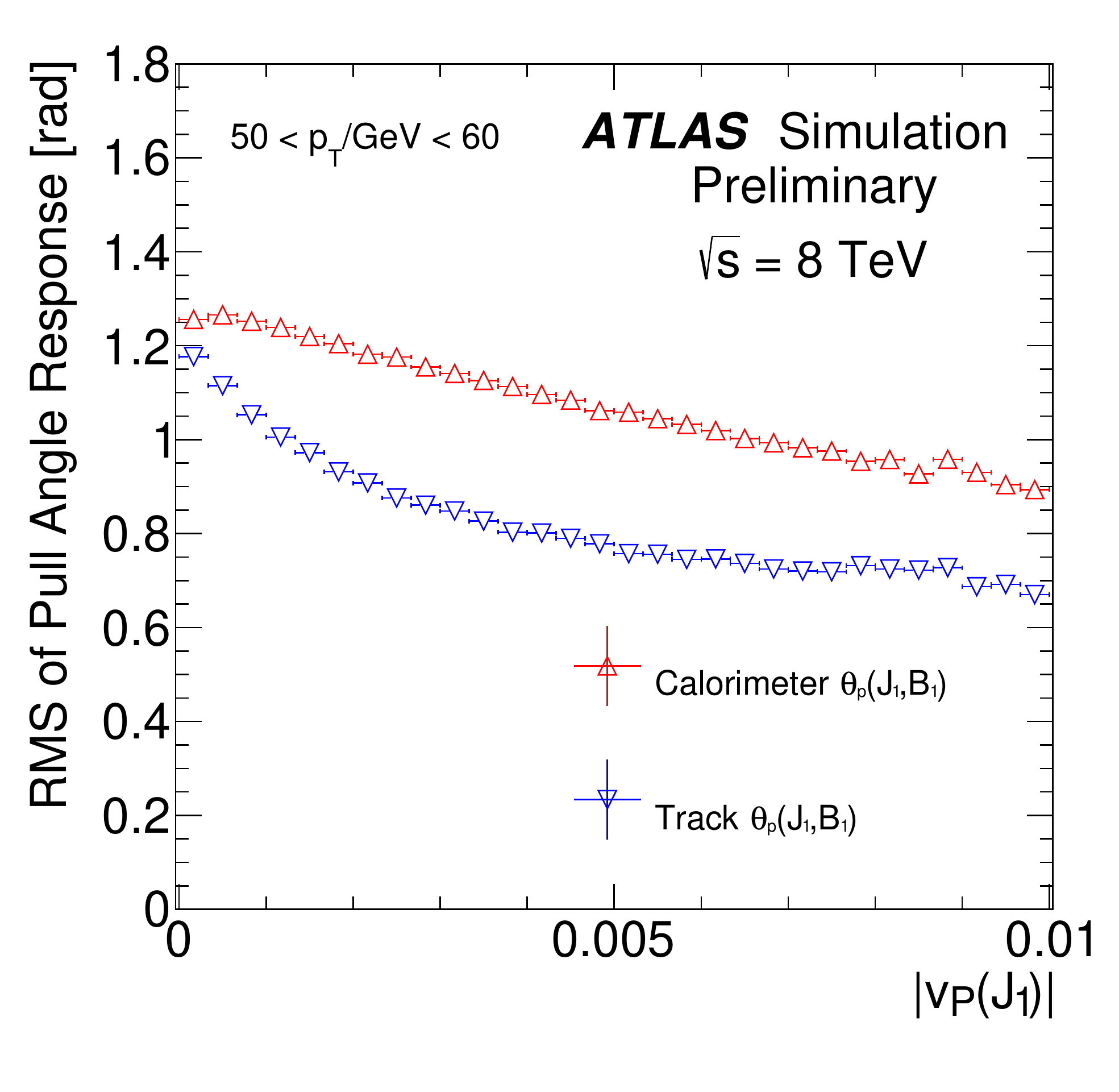}

 \caption{The pull vector magnitude (left) and the relationship (right) between the jet pull angle response and $|v_P(J_1)|$ in a particular bin of jet $p_\text{T}$.  For the track (charged particle) pull magnitude in the left plot, at least one track (charged particle) is required.}
 \label{fig:16}
\end{figure}

\begin{figure}[h!]
 \centering
\includegraphics[width=.45\columnwidth]{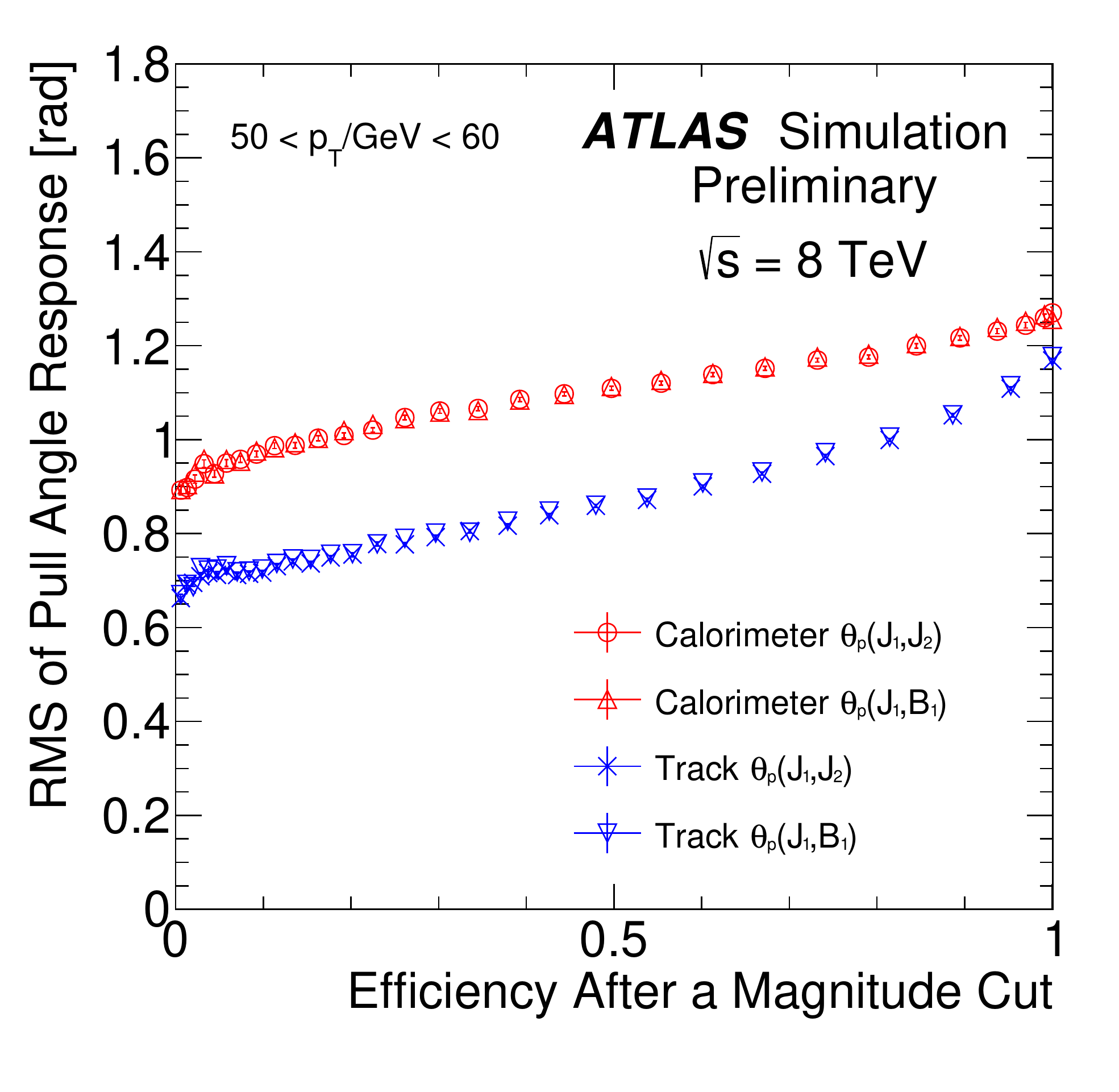}
\includegraphics[width=.45\columnwidth]{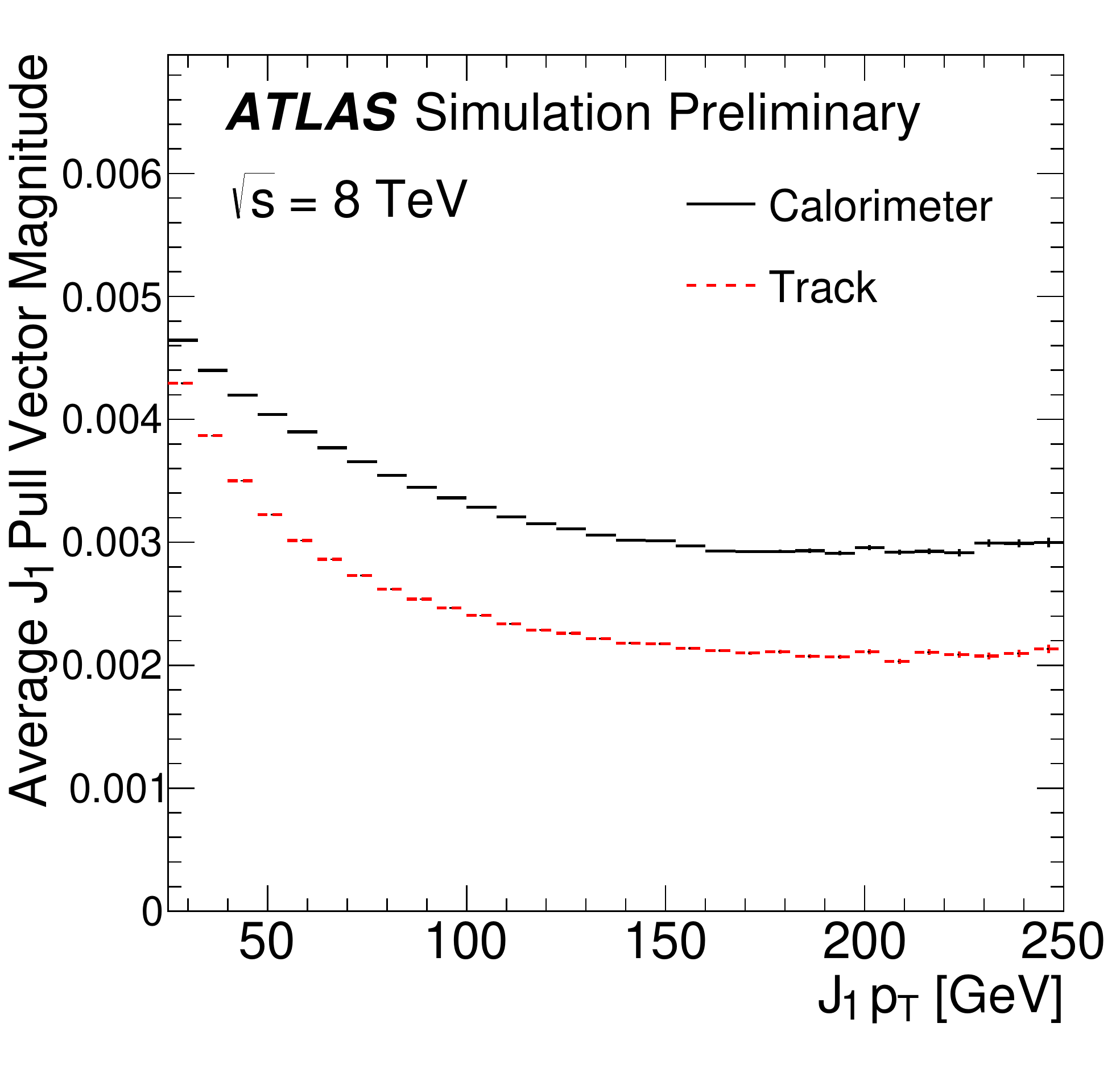}

 \caption{The RMS of the jet pull angle response as a function of the fraction of events that pass a cut on the pull vector magnitude (left) and the $p_\text{T}$ dependence of the average pull vector magnitude (right).}
 
 \label{fig:17}
\end{figure}

One undesirable property of the pull vector magnitude in terms of constraining the resolution is that it is anti-correlated with the jet $p_\text{T}$ as shown in the right plot of Fig.~\ref{fig:17}.    As the jet becomes more collimated, the constituents have a smaller $\Delta R$ with respect to the jet axis and so the pull vector magnitude decreases.  Accordingly, an optimal threshold on the pull vector magnitude would be $p_\text{T}$ dependent.   

Another substructure observable that is correlated with the jet pull angle response is the number of constituents.   The pull angle resolution decreases with the number of constituents at low constituent multiplicity as shown in Fig.~\ref{fig:19}.  The calorimeter pull angle and the track pull angle each require at least one cluster or track, respectively.

\begin{figure}[h!]
 \centering
\includegraphics[width=.45\columnwidth]{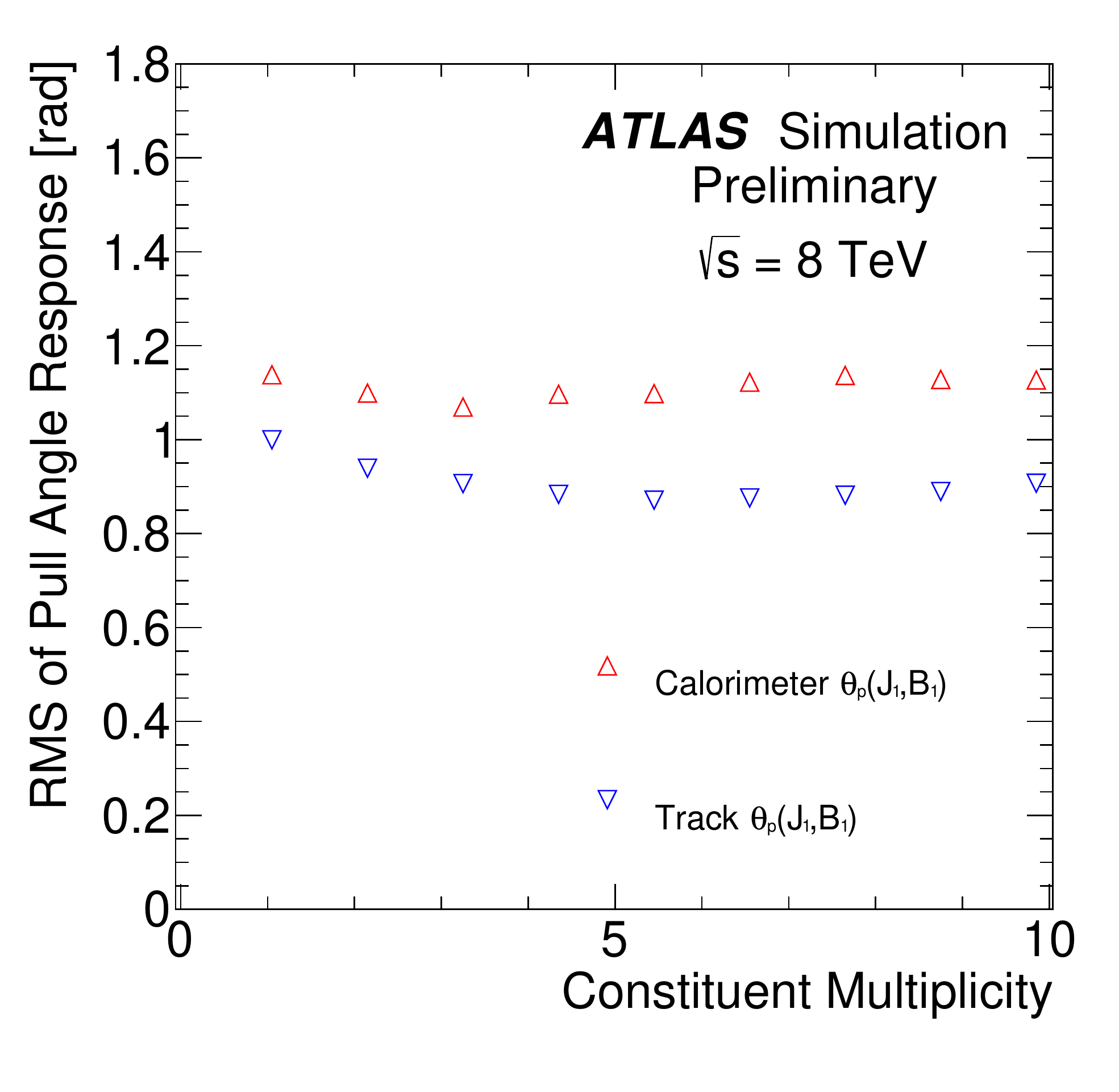}
 \caption{The jet pull angle response as a function of the number of jet constituents for $J_1$.}
 \label{fig:19}
\end{figure}

\subsection{Jet Pull Angle and Event Properties}

\subsubsection{Jet Labeling in $t\bar{t}$ Events}
\label{jetpullcombo}

For a given jet pull angle $\theta_P(X,Y)$, there is the complimentary angle $\theta_P(Y,X)$ which uses different substructure information.  Figure~\ref{fig:cf8} shows that this information is largely uncorrelated.  Furthermore, it is apparent from Figures~\ref{fig:9} that there is a relationship between the shapes of the jet pull angle distributions and the assignment of the jets in the $t\bar{t}$ topology.  For example, one can investigate the frequency with which the $b$-tag and dijet invariant mass assignment of $J_1,J_2,B_1$ and $B_2$ described in Sec. 3 aligns with the observed property that $\theta_P(J_{1},J_{2})$ and $\theta_P(J_{2},J_{1})$ tend to be smaller than $\theta_P(J_{i},B_{j})$, $\theta_P(B_{i},J_j)$ or $\theta_P(B_{i},B_j)$.  

An event is called {\it matched} if $\theta_P(J_a,J_b)<\theta_P(J_a,B_1)$ for any $a,b\in\{1,2\}$.  Conversely, if $\theta_P(J_a,J_b)\geq \theta_P(J_a,B_1)$, an event is called {\it un-matched}.  Figure~\ref{fig:cf7} shows the tradeoff between matched and un-matched event efficiencies for a threshold on the jet pull angle using particle-level jets\footnote{Particle-level jets are used here to illustrate the maximal achievable performance in the absence of selection biases and detector resolution effects.}.  In other words, consider the $\theta_P(J_a,J_b)$ distribution as `signal' and the $\theta_P(J_a,B_1)$ distribution as `background'.  Then, Fig.~\ref{fig:cf7} shows the relationship between signal and background efficiency as a function of the threshold on $\theta_P$.  Also plotted in Fig.~\ref{fig:cf7} is the combined performance curve from both variables ($\theta_P(X,Y)$ and $\theta_P(Y,X)$), which is significantly better than either curve separately.  In absolute units, the overall discrimination is poor -- pull is not intended to be used as a stand-alone tagger.  Since the jet pull angles with $b$-jets are independent of $\Delta R(X,Y)$ but $\theta_P(J_1,J_2)$ becomes more pronounced at smaller $\Delta R$, there is a slight improvement in the efficiency curve, which is shown in the right plot of Fig~\ref{fig:cf7}.

\begin{figure}[h!]
 \centering
 \includegraphics[width=.45\columnwidth]{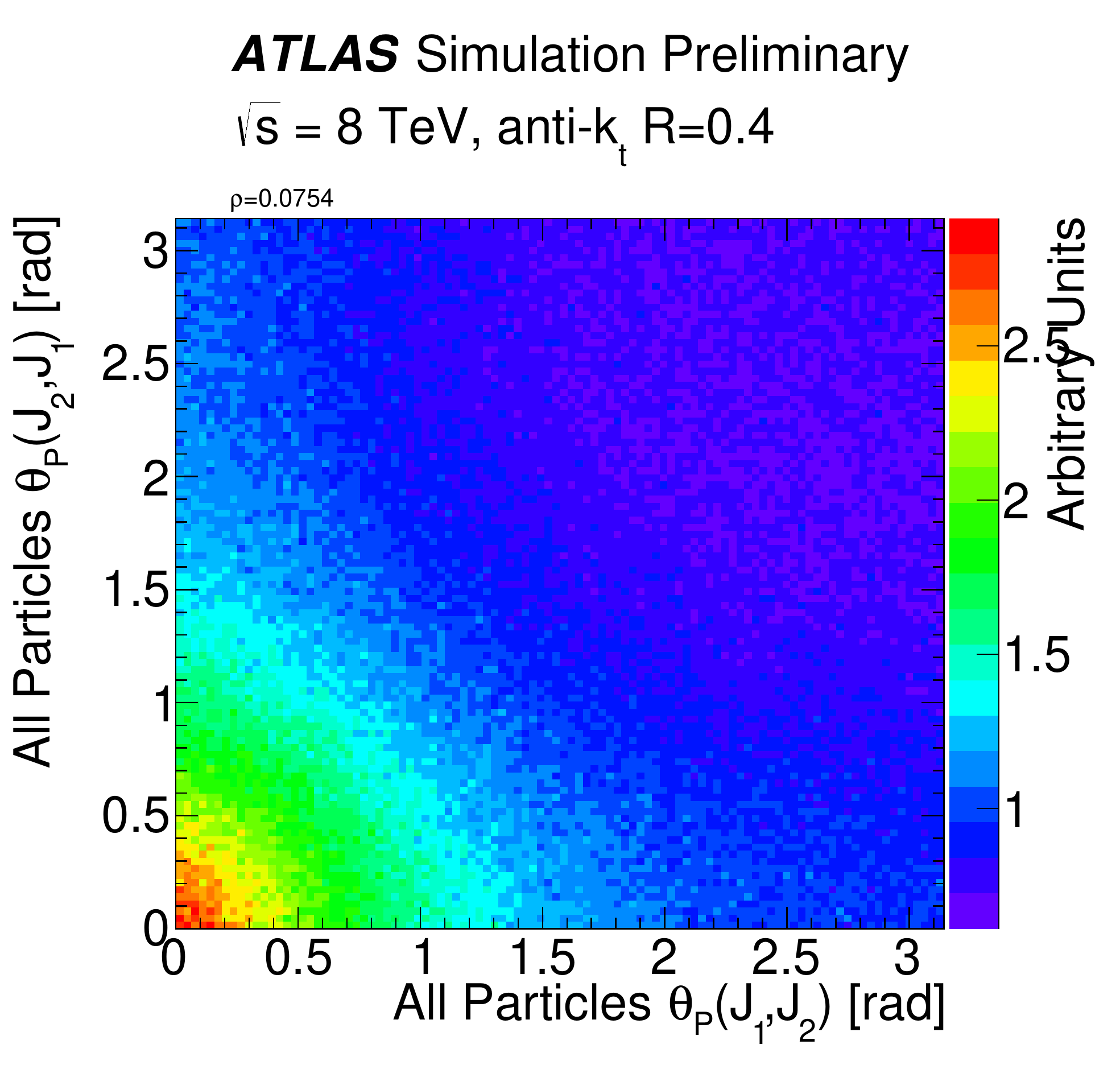}
 \includegraphics[width=.45\columnwidth]{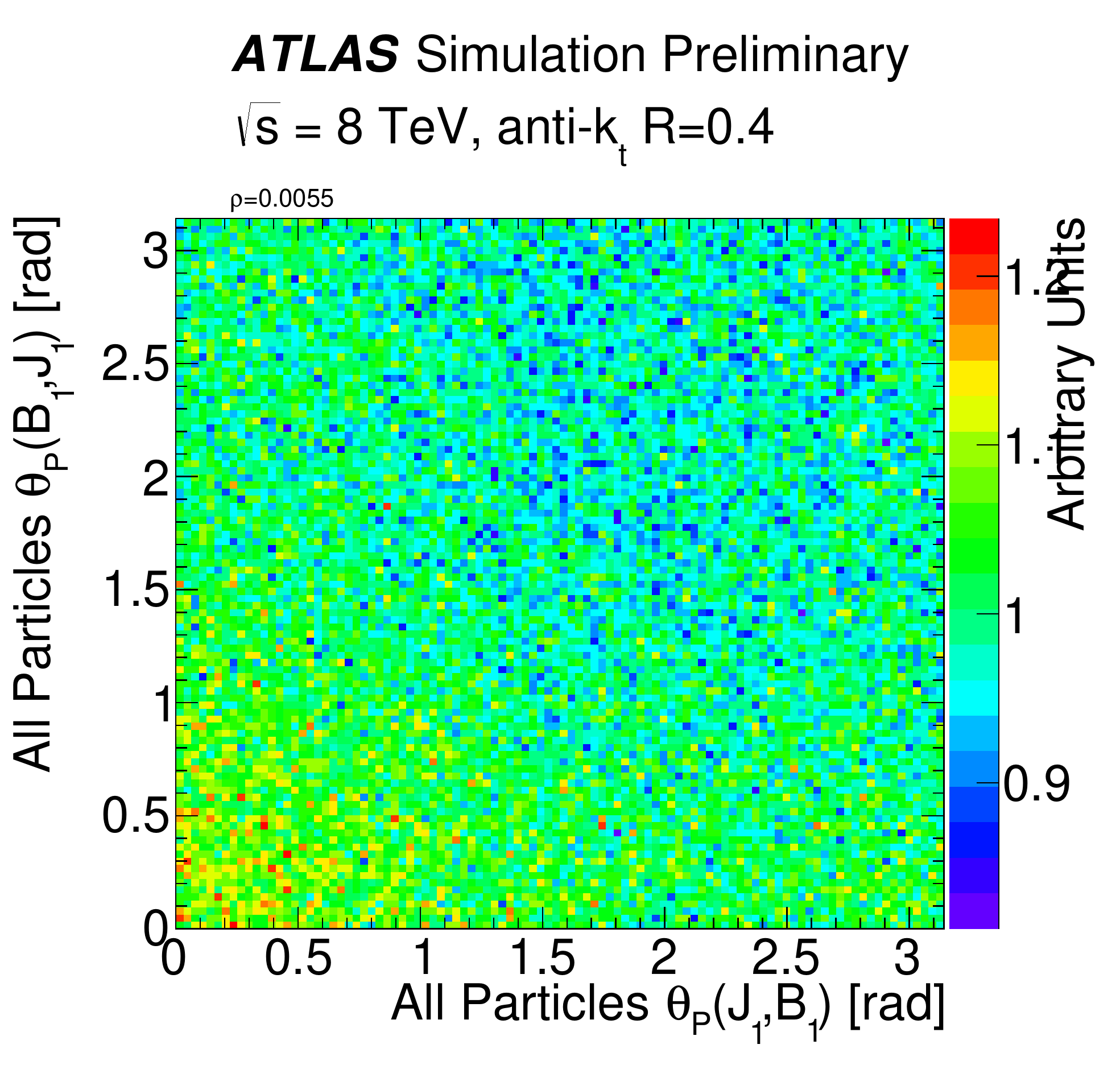}

 \caption{Pairwise all-particles pull angle correlations using particle-level jets.}
 \label{fig:cf8}
\end{figure}

\begin{figure}[h!]
 \centering
\includegraphics[width=.45\columnwidth]{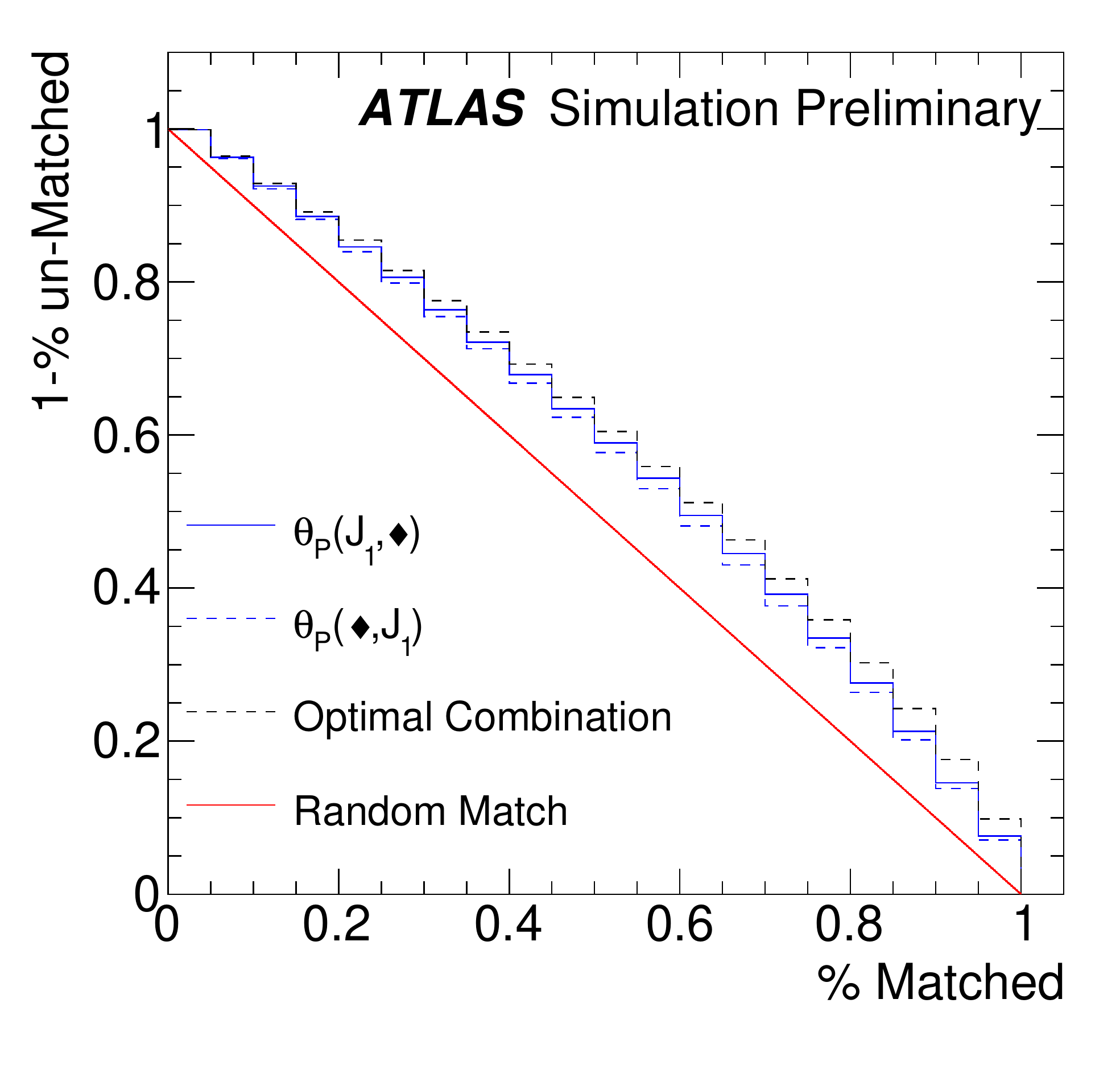}
\includegraphics[width=.45\columnwidth]{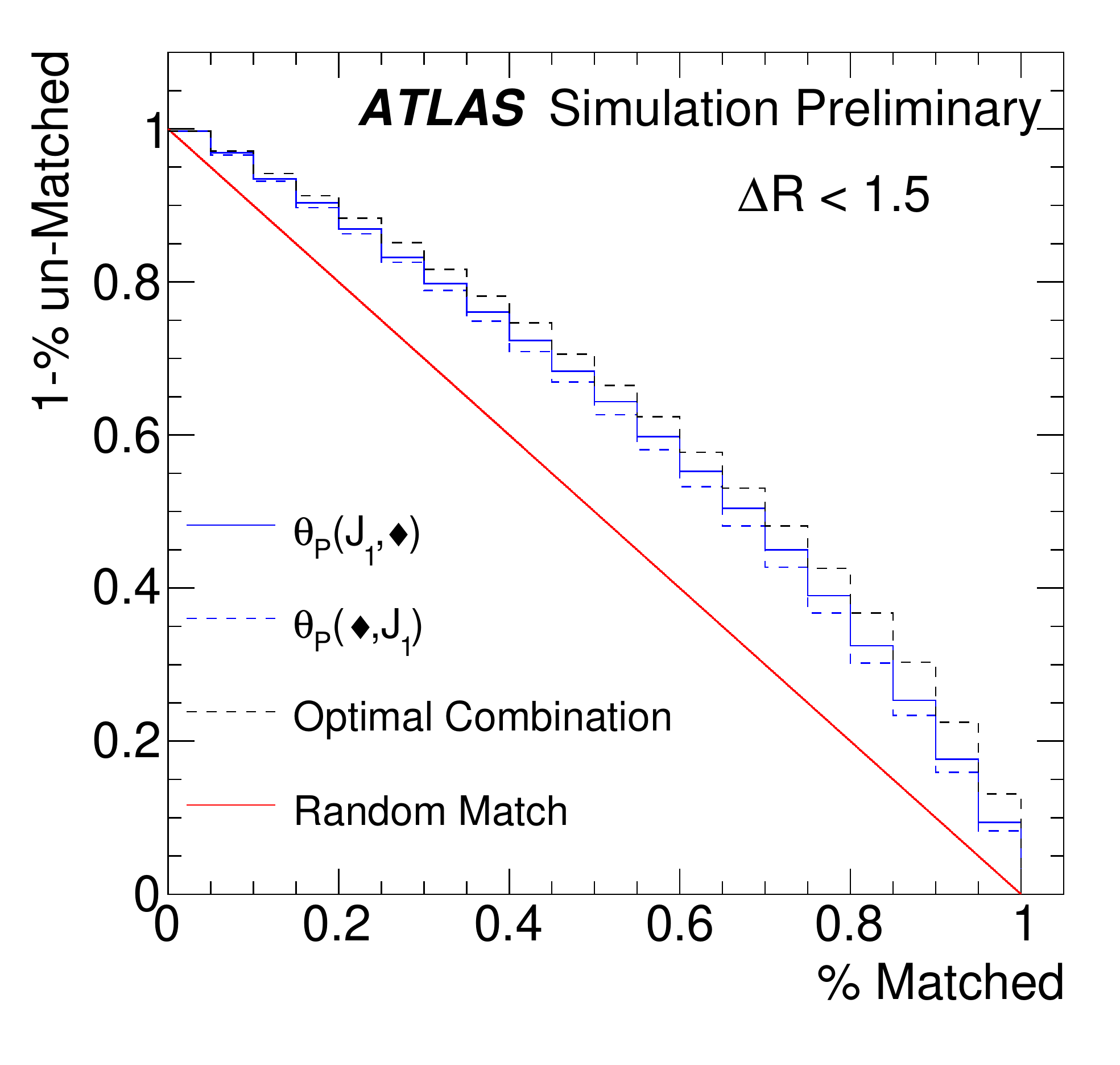}
   \caption{Un-matched event (treated as a background) rejection versus the matched (treated as a signal) efficiency.  The optimal combination is constructed from the full joint likelihood.}

 \label{fig:cf7}
\end{figure}

\subsubsection{Pileup}

An important event property from the point of view of the pull angle RMS is $\mu$ - the average number of additional $pp$ interactions per bunch crossing at the LHC.  The dependence of the RMS of the jet pull angle response is shown as a function of $\mu$ in Fig.~\ref{fig:21}.  The RMS of the jet pull angle response is only weakly dependent on the pileup activity.  For example, a linear fit to the data in Fig.~\ref{fig:21} results in a slope of about $(1.6\pm 0.1)\times 10^{-3}$ rad/interaction for the calorimeter pull angle response RMS and $(1.5\pm 0.1)\times 10^{-3}$ rad/interaction for the track pull angle response RMS in the range $50$ GeV $<p_\text{T}^{J_1}<60$ GeV .  This trend does not vary greatly with $p_\text{T}$.

\begin{figure}[h!]
 \centering
 \includegraphics[width=.45\columnwidth]{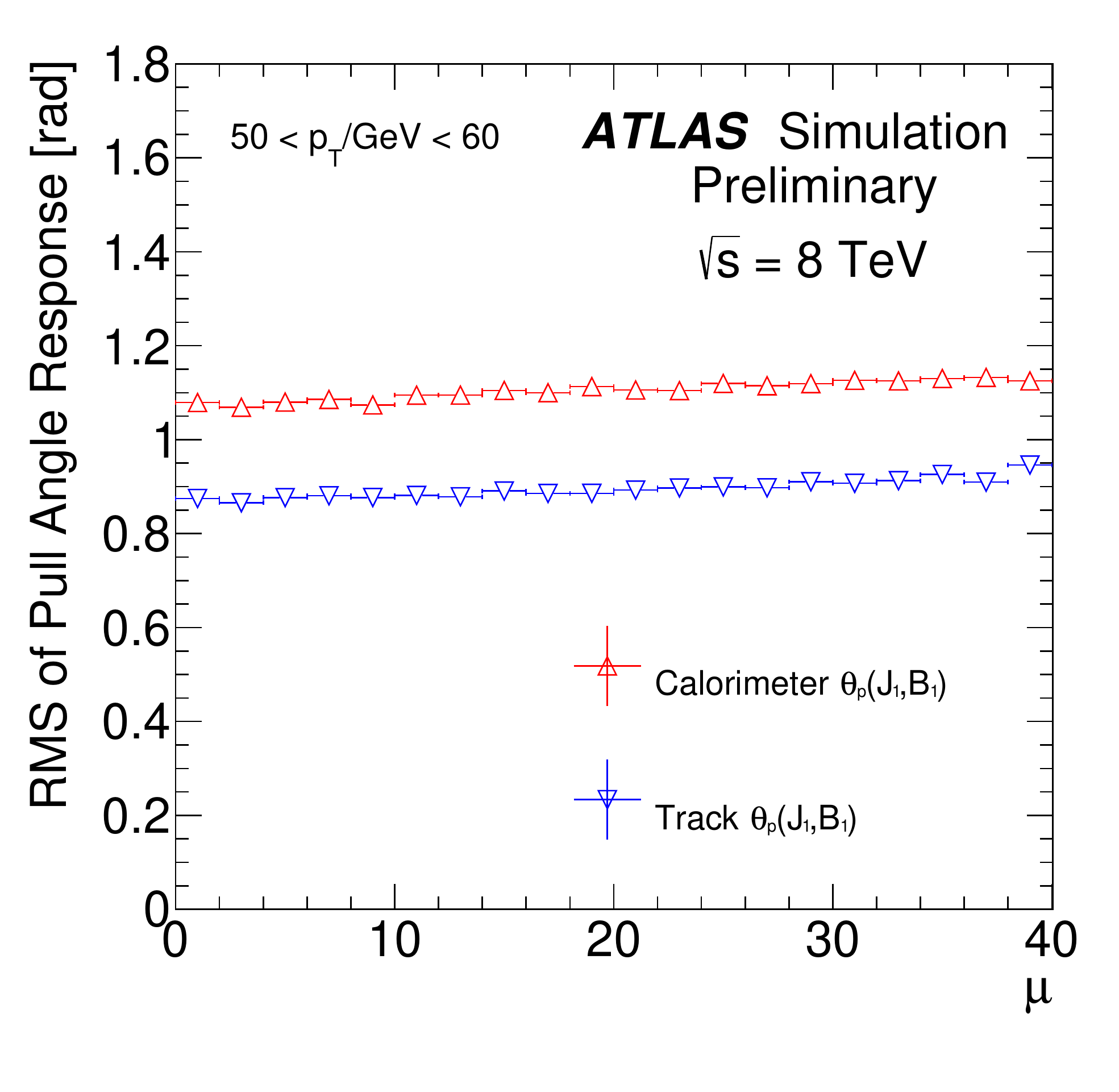}
 \caption{The RMS of the jet pull angle response as a function of $\mu$ for $50$ GeV $<p_\text{T}^{J_1}<60$ GeV.}
 \label{fig:21}
\end{figure}

\clearpage

\subsection{Comparisons between Data and Simulation}

The purpose of this section is to qualitatively compare the pull vector in simulation with data.  A quantitative comparison that disentangles detector-level and particle-level effects through unfolding is in Sec.~\ref{sec:colorflow:unfolding}.  The MC is normalized by area to the data in all the following distributions.  The uncertainty bands on the data/MC ratios include the experimental uncertainty on the tracking efficiency, the jet energy scale and the jet energy resolution in addition to a $\pm6\%$ relative cross-section uncertainty on the $t\bar{t}$ component~\cite{Cacciari:2011hy,Baernreuther:2012ws,Czakon:2012zr,Czakon:2012pz,Czakon:2013goa,Czakon:2011xx}.  For the pull angle, the average uncertainty across all bins is plotted to remove fluctuations due to the small dependence of the pull angle on the jet energy scale and resolution uncertainties.   Uncertainties on the cluster energy scale and angular resolution are not included in this section.

The pull vector magnitude is shown in Figure~\ref{fig:1ab} for both {\sc MC@NLO}+{\sc Herwig} and {\sc Powheg-Box}+{\sc Pythia} 6.  For the track pull angle, at least two tracks are required in order to remove the portion of the resolution curve in Fig.~\ref{fig:19} where the response RMS decreases at low constituent multiplicity.  Both the calorimeter- and track-based distributions are within $10\%$ of the data over nearly the entire range.  There seems to be a minor overall slope in the data/MC ratio for the calorimeter-based pull angle that is due in part to the modeling of the angular resolution (see Sec.~\ref{origincorrection}).  Interestingly, the {\sc Pythia} and {\sc Herwig}\footnote{Sec.~\ref{syst:colorflow:ME} shows that the ME generator is unimportant for the pull angle distribution.} mis-modeling at low pull vector magnitude are in opposite directions of the data; this is likely due to the mis-modeling of the track multiplicity, as discussed in Chapter~\ref{cha:multiplicity}.

\begin{figure}[h!]
 \centering
\includegraphics[width=.45\columnwidth]{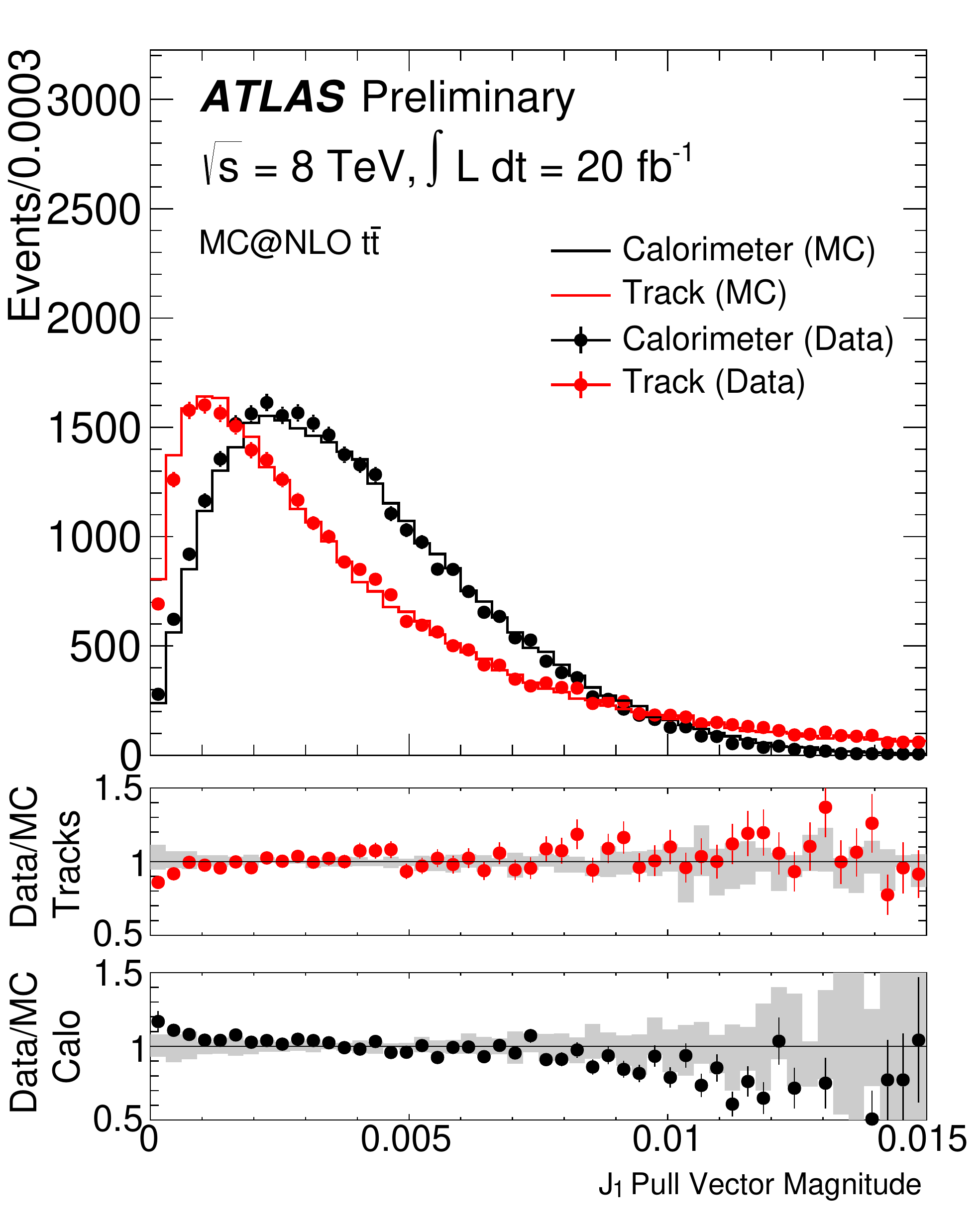}\includegraphics[width=.45\columnwidth]{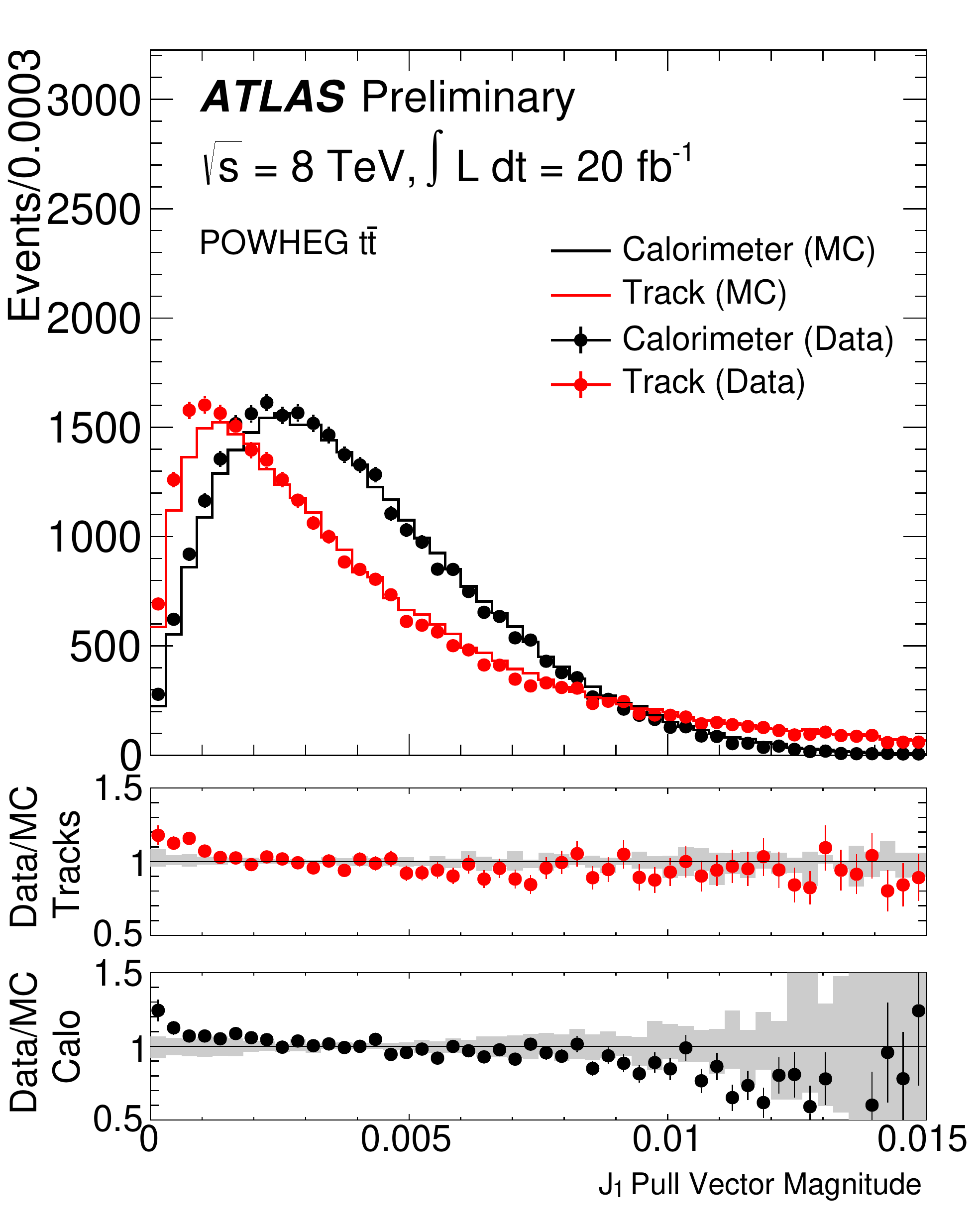}
 \caption{The pull vector magnitude for both calorimeter pull and track pull.  For the track pull, at least two tracks are required.  Uncertainty bands include uncertainties on the jet energy scale and uncertainty as well as on the $t\bar{t}$ component of the MC.  An uncertainty on the tracking efficiency is added for the track pull.  No uncertainty is included for individual calorimeter clusters or for jet angular resolutions.}
 \label{fig:1ab}
\end{figure}

The distribution of the jet pull angle in the data is shown in Fig.~\ref{fig:2ab} for $\theta_P(J_1,J_2)$.  The resolution features at $\pi/2$ for the track-based pull angle and at zero for the calorimeter-based pull angle are both present and well described.  The bias toward zero in the particle-level distribution (Fig.~\ref{fig:9}) that is also present in the particle-level selection (Fig.~\ref{fig:9}(a)) is reduced in Fig.~\ref{fig:2ab} due to a selection bias: in a given event, the particle-level and detector-level assignment of jet labels can differ.  This selection bias decreases with the increasing $p_\text{T}$ of the jets, as is seen in the right plot of Fig.~\ref{fig:2ab}, where the peak at zero for the track pull dominates the resolution peak at $\pi/2$ in the MC.  The size of the peak at zero also increases with $p_\text{T}$ as discussed in Sec.~\ref{colorflowkins}.  Figure~\ref{fig:2ab222} shows the jet pull angle distribution between the leading $W$ daughter jet and the leading $b$-jet.  Based on the studies summarized in Fig.~\ref{fig:13}, the slight parabolic trend in the track-based pull angle ratio in the left plot of Fig.~\ref{fig:2ab222} (and the left plot of Fig.~\ref{fig:2ab}) suggests that the scale or asymmetry parameter of the jet angular resolution may be over-estimated, though quantifying this statement is beyond the scope of this section.

\begin{figure}[h!]
 \centering
\includegraphics[width=.42\columnwidth]{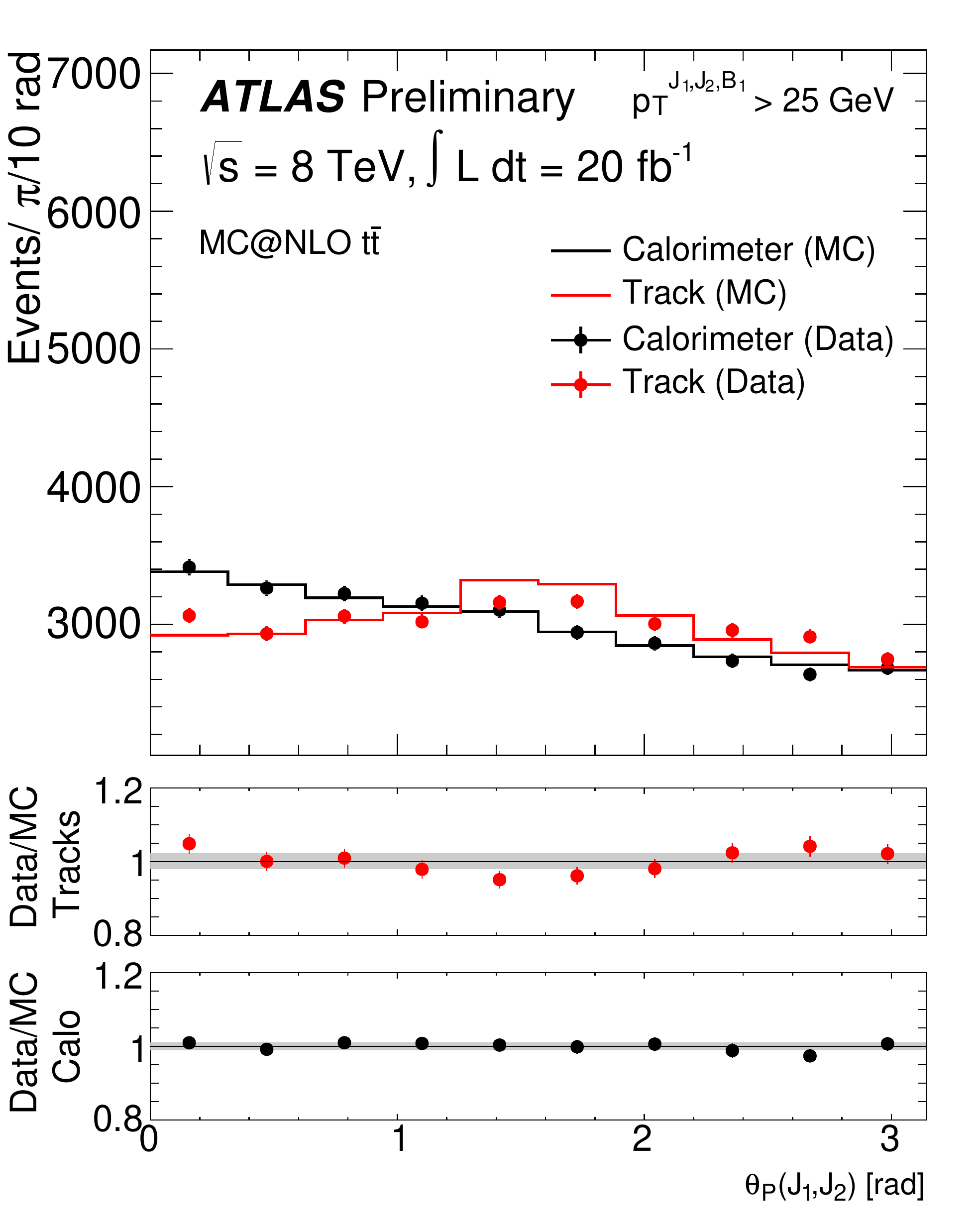}\includegraphics[width=.42\columnwidth]{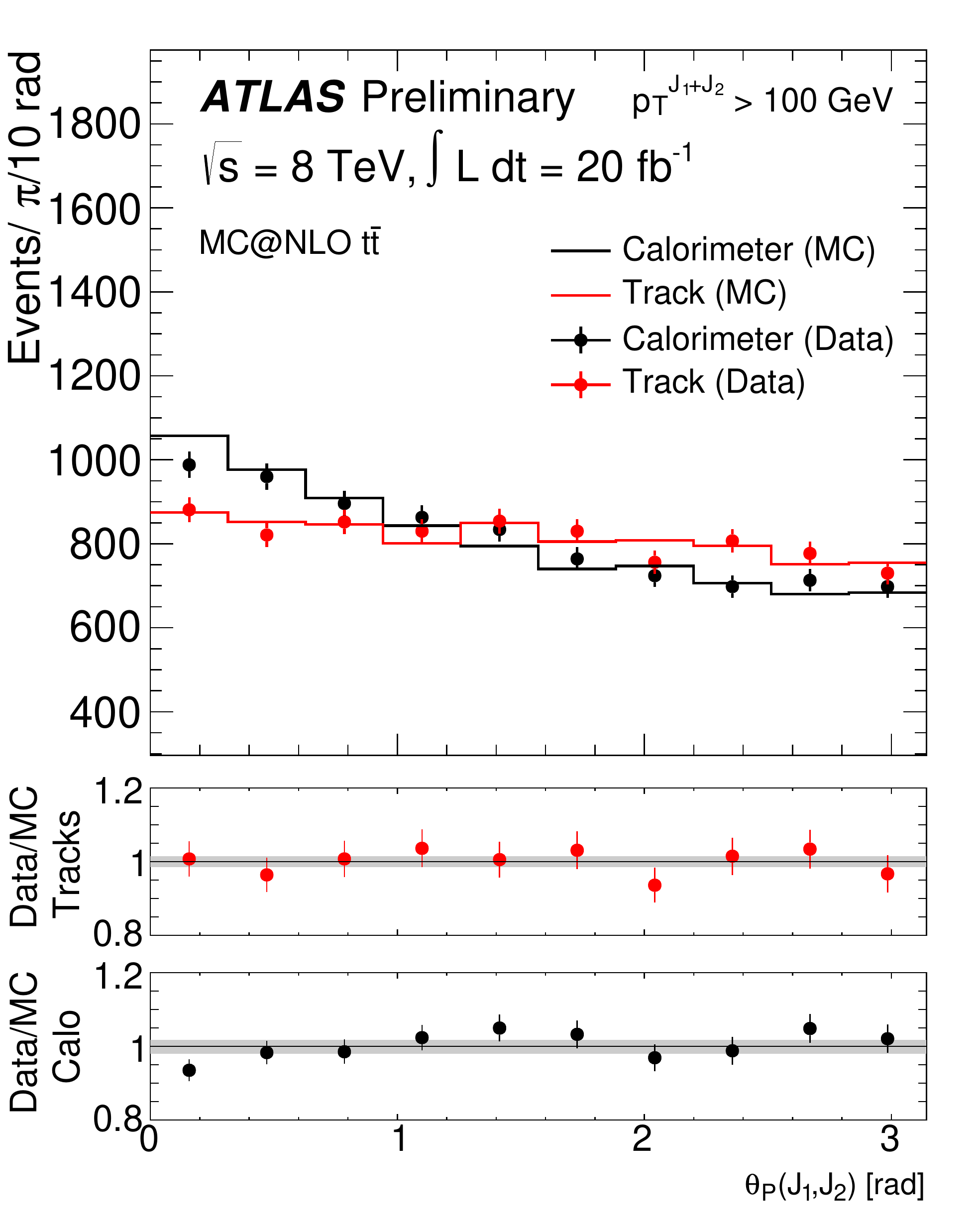}

 \caption{The distribution of the jet pull angle $\theta_P(J_1,J_2)$ for both calorimeter cluster constituents and track constituents in both data and MC.  The left plot has a 25 GeV requirement for the jets while the right plot has a tight threshold placed on the $p_\text{T}$ of the dijet system. }
 \label{fig:2ab}
\end{figure}

\begin{figure}[h!]
 \centering
 
  \includegraphics[width=.42\columnwidth]{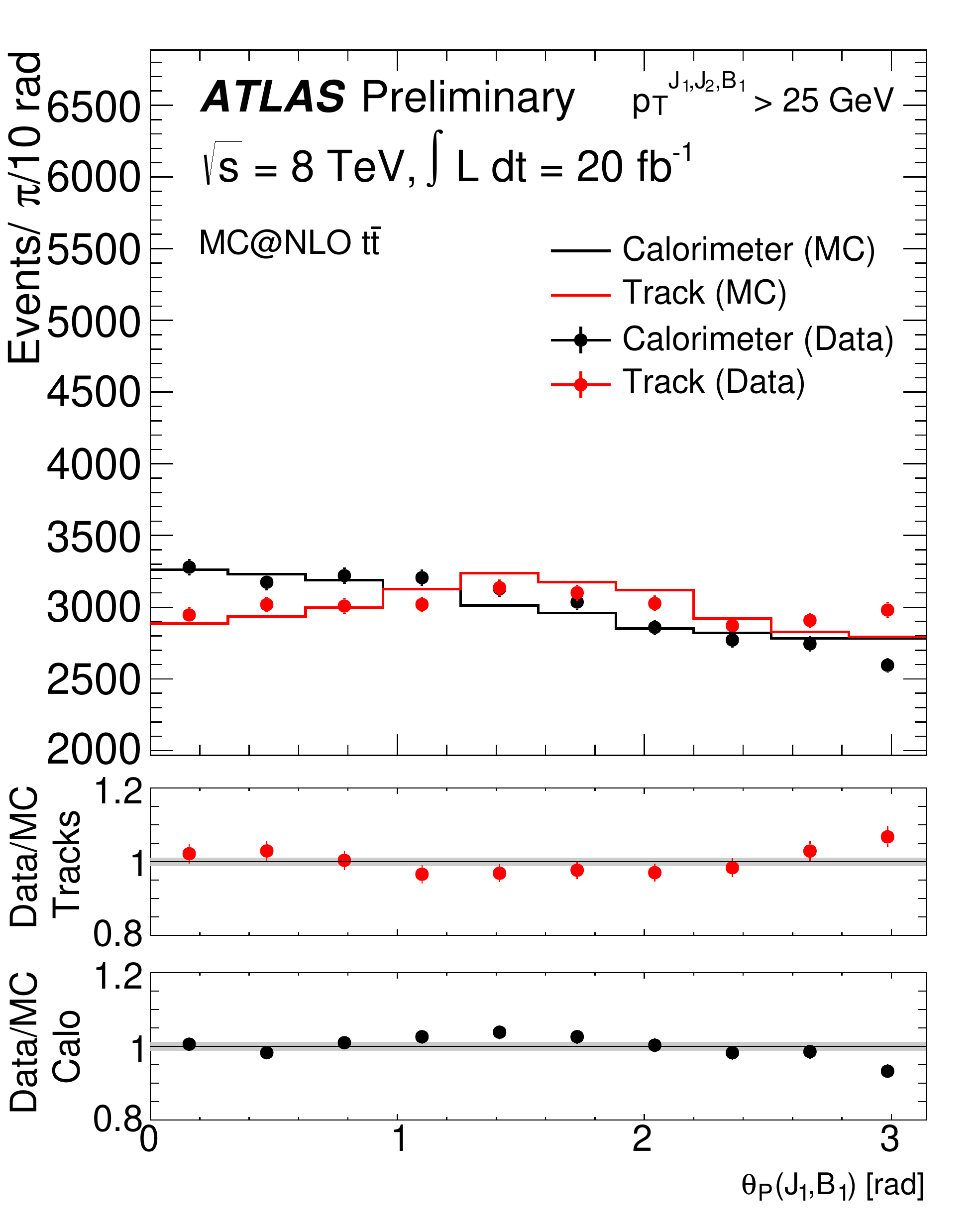}\includegraphics[width=.42\columnwidth]{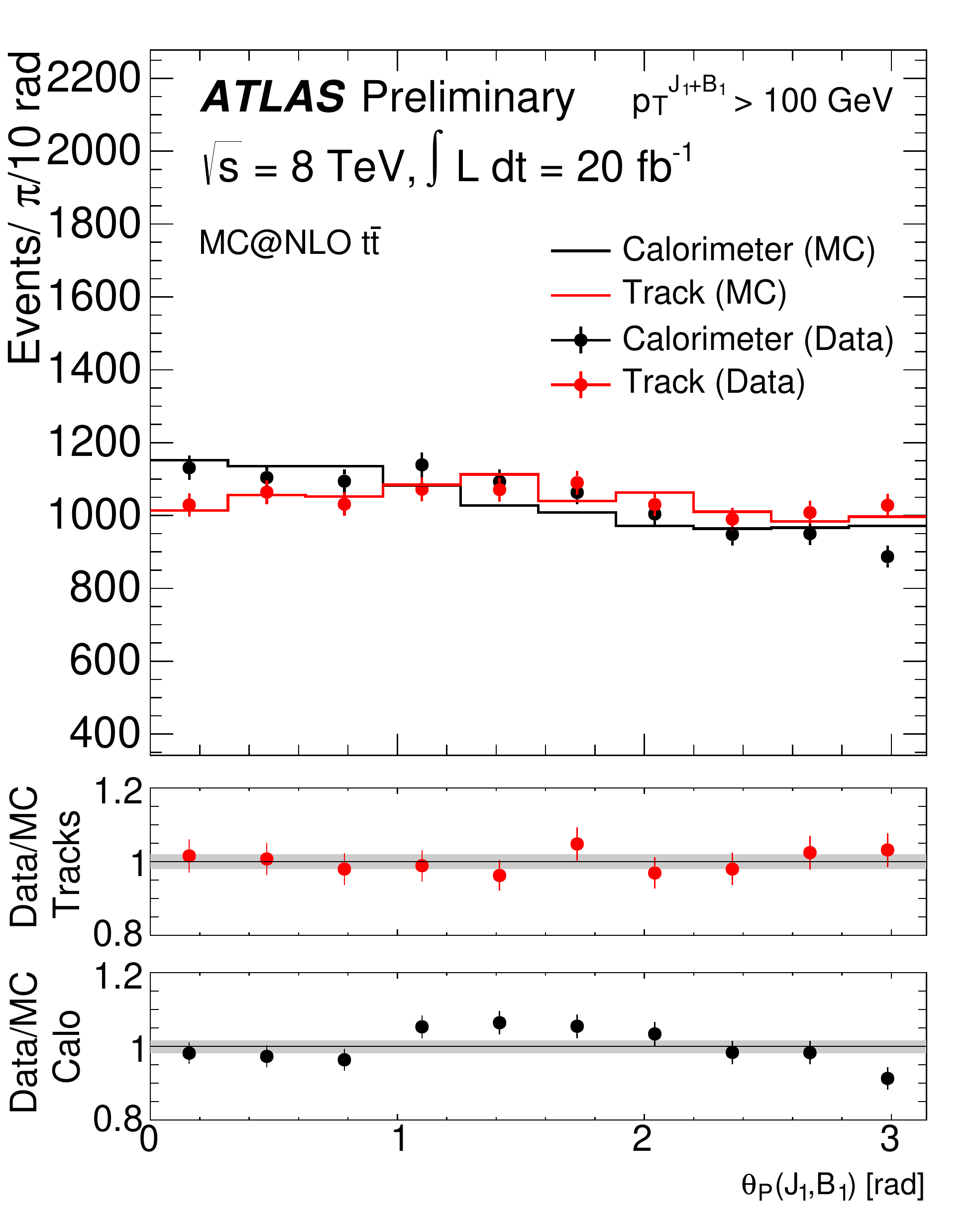}
 
 \caption{Same as Fig.~\ref{fig:2ab} only with $\theta_P(J_1,B_1)$ instead of $\theta_P(J_1,J_2)$.}
 \label{fig:2ab222}
\end{figure}

\clearpage

\subsubsection{Cluster and Jet Origin Corrections}
\label{origincorrection}

Part of the jet calibration procedure is to correct the $\eta$ of jets so that the detector-level pseudo-rapidity is an unbiased measurement of the corresponding particle-level quantity (see Sec.~\ref{sec:jets}).  During the LHC shutdown between Runs 1 and 2, the jet calibration procedure was improved to reduce the resolution in the $\eta$ direction by correcting $\eta$ event-by-event such that the jet axis is radially aligned with the primary vertex (defined by $\sum_\text{tracks}p_\text{T}^2$).  A beamspot with finite size smears out the $\eta$ resolution because of event-by-event distortions in the $\eta$ value from a primary vertex that is not at the geometric center of ATLAS.  Figure~\ref{origin_0} illustrates the geometry of this distortion.  The beamspot is $\mathcal{O}(10)$ cm in the $z$ direction and the calorimeter is about $1$ m away from the primary interaction.  For $\eta=0.5$, this is a correction of $\Delta\eta\sim 0.4$ for a primary vertex of $z\sim 50$ cm and $\Delta\eta\sim 0.1$ for $z\sim 10$ cm\footnote{One can write $\theta_\text{detector}=2\arctan(e^{-\eta_\text{detector}})$ and then the $z$ distance in detector coordinates is $z_\text{detector}\approx \text{$1$ m}/\tan(\theta_\text{detector})$.  The `physics' position $z_\text{physics}=z_\text{detector}-z_\text{PV}$.  The physics angle is then approximately $\theta_\text{physics}\approx \arctan(\text{$1$ m}/z_\text{physics})$.}.

\begin{figure}[h!]
 \centering
		\begin{tikzpicture}[line width=1.5 pt, scale=1.5]
						\draw[->] (0,0) -- (3,0);
			\draw[->] (0,0) -- (0,3);
			\node at (3.2,-0.0) {$z$};
			\node at (-0.0,3.3) {$x$};	
			\draw[->] (0,0) -- (2,3);
			\draw[->] (-1.5,0) -- (2,3);
			\draw[dotted] (-1.5,0) -- (0,0);			
			
			\node [rotate=0] at (2.,3.2) { jet or cluster};
			\node [rotate=0] at (-1.5,-0.2) { PV};
			\node [rotate=0] at (0.7,0.2) { $\theta^\text{detector}$};
			\node [rotate=0] at (-0.7,0.2) { $\theta^\text{physics}$ };
			\fill [black, ultra thick] (-1.5,.0) circle [radius=0.05];
								
		\end{tikzpicture}
\caption{A schematic diagram of the origin correction.  The quantity $\theta^\text{detector}$ is the angle in the $zx$ plane measured with respect to the geometric center of the detector, whereas the `true' $\theta^\text{physics}$ is offset and based from the primary vertex (PV).}
 \label{origin_0}
\end{figure}

\begin{figure}[h!]
\begin{center}
\includegraphics[width=0.95\textwidth]{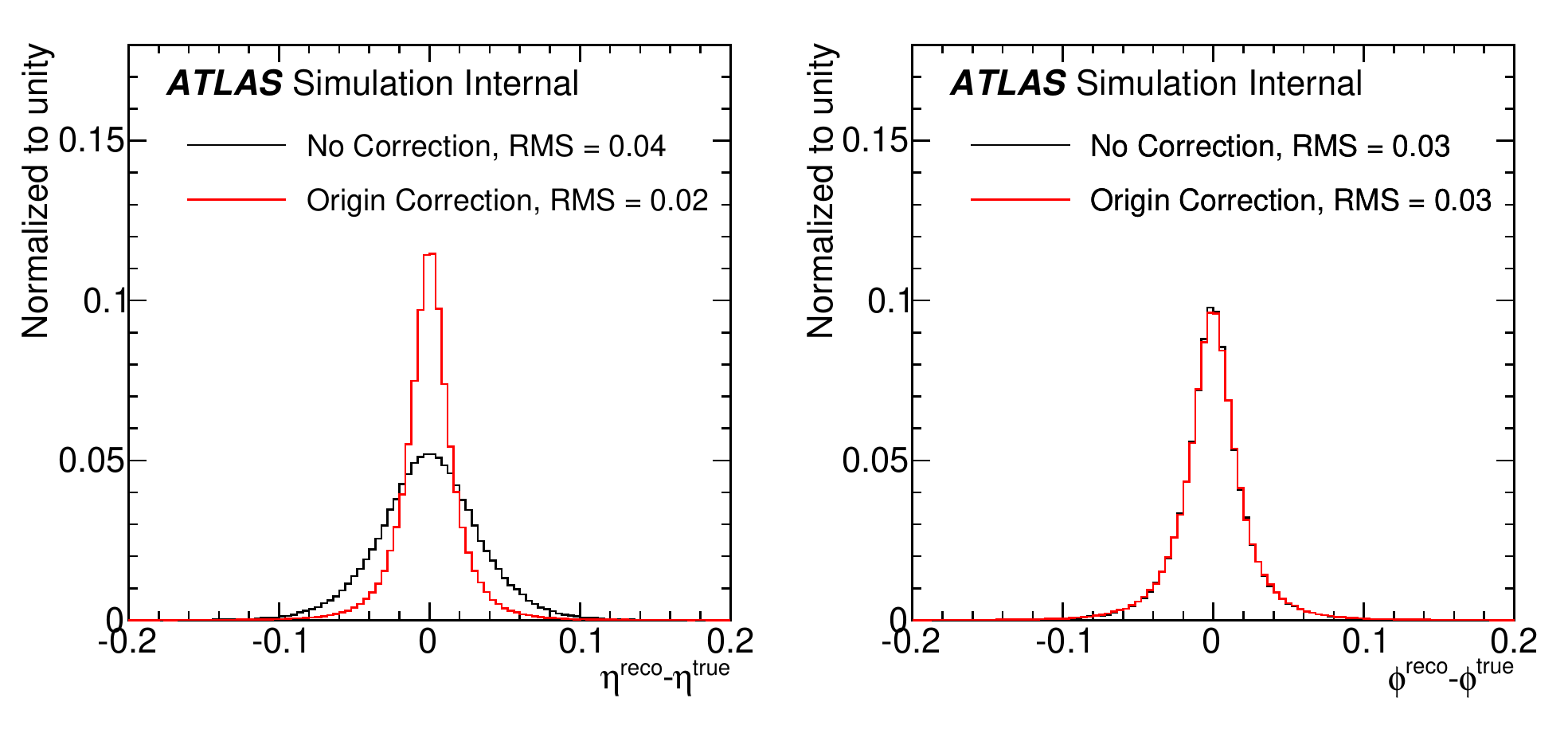}
\end{center}
\caption{The $\eta$ (left) and $\phi$ (right) jet angular response before and after the jet origin correction.  The response is defined as the RMS of the $\Delta\eta$ or $\Delta\phi$ between the reconstructed jet and the $\Delta R < 0.4$ matched particle-level jets.}
\label{origin_1}
\end{figure}

The origin correction significantly improves the jet $\eta$ resolution.  Figure~\ref{origin_1} shows the $\eta$ and $\phi$ angular resolution before and after the origin correction.  The correction has no effect on the $\phi$ resolution, but reduces the width of the $\eta$ resolution by a factor of two.  Since the pull angle resolution significantly depends on the jet axis resolution, the origin correction could significantly improve the pull angle resolution.  However, it is not sufficient to correct the jet axis {\it without correcting the cluster positions as well}.  The jet origin correction is performed {\it after} jet clustering so the jet constituents are unchanged.  For jet substructure variables that depend on the jet axis (such as the jet pull), this introduces a mis-match that can be avoided by coherently origin correcting the constituent calorimeter-cell clusters.  The cluster origin correction is performed as follows.  Let $R_i$ be the calorimeter-cell energy weighted center of the calorimeter-cell cluster $i$ in detector coordinates.  The transverse radius is defined by $R_{T,i}=R_i/\cosh(\eta_i)$.  The new $\eta$ position of the cluster $i$ is 

\begin{align}\nonumber
\eta_\text{physics} &= \text{asinh}\left(\frac{z_\text{physics}}{R_{T,i}}\right)=\text{asinh}\left(\frac{1}{R_{T,i}}(z_\text{detector}-z_\text{PV})\right)\\
&=\text{asinh}\left(\text{sinh}(\eta_\text{detector})-\frac{z_\text{PV}}{R_{T,i}}\right).
\end{align}

\noindent In order to preserve the total energy, the cluster transverse momentum becomes $p_\text{T,physics}=p_\text{T,detector}\text{cosh}(\eta_\text{detector})/\text{cosh}(\eta_\text{physics})$.  Figure~\ref{fig:reco_dists} shows the improvement in the pull angle resolution from coherently applying the cluster origin correction.  The reduction in the width of the pull angle response distribution ($\sim 5\%$) is modest, but there is a significant improvement in the modeling of the pull vector.  Figure~\ref{fig:reco_dists2} compares the simulation to the data before and after coherently applying the origin correction.  When neither the jet or cluster axes are origin corrected, the pull vector is well-modeled (Fig.~\ref{fig:1ab}) even if the resolution with respect to the particle-level quantity is worse because both axes are at the same angular `scale'.   This is also true after both axes are corrected (left plot of Fig.~\ref{fig:reco_dists2}).  However, if only the jet axis is corrected, then the pull vector is maximally sensitive to the modeling of the beamspot because the cluster locations are distributed about the jet axis (which is independent of the PV) according to the width of the beamspot.  The mis-modeling without coherent origin corrections is shown in the right plot of Fig.~\ref{fig:reco_dists2}.  Even though the uncertainty band is large, there is a clear systematic trend in the data/MC ratio and the distribution itself is stretched to higher values due to the offset between axes.  Henceforth, both axes are coherently corrected\footnote{The actual axis used for the jet pull angle is the four-vector sum of the origin-corrected calorimeter-cell clusters.  This is the nearly the same as the origin-corrected jet axis, but the cluster-based systematic uncertainties described in Sec.~\ref{sec:ColorFlowclusteruncerts} are allowed to coherently vary the axis location.}.  The left plot of Fig.~\ref{fig:improvement} confirms that the origin corrections improve over the uncorrected case.  Most of the improvement in the resolution is from the jet axis correction and as already noted by Fig.~\ref{fig:reco_dists}, the resolution improvement from the cluster origin correction is smaller.  However, the pull angle distribution qualitatively changes after each step of the correction, as shown by the right plot of Fig.~\ref{fig:improvement}\footnote{Note the slightly different definition due to the absolute value, $|*|$.  This decreases the response width as events with e.g. pull vectors of $\pi$ and $-\pi$ have a response of zero.  However, this strategy is used for the unfolding described in Sec.~\ref{sec:colorflow:unfolding} in order to reduce the number of bins.  A scheme with an addition bin to account for the extreme migrations does not significantly reduce uncertainties.}.  Interestingly, when the jet axis resolution is significantly reduced with the jet origin correction, the pull angle distribution resembles the uncorrected track-based pull angle (i.e. a resolution peak at $\pi/2$).

\begin{figure}
  \centering
  \includegraphics[width=0.5\textwidth]{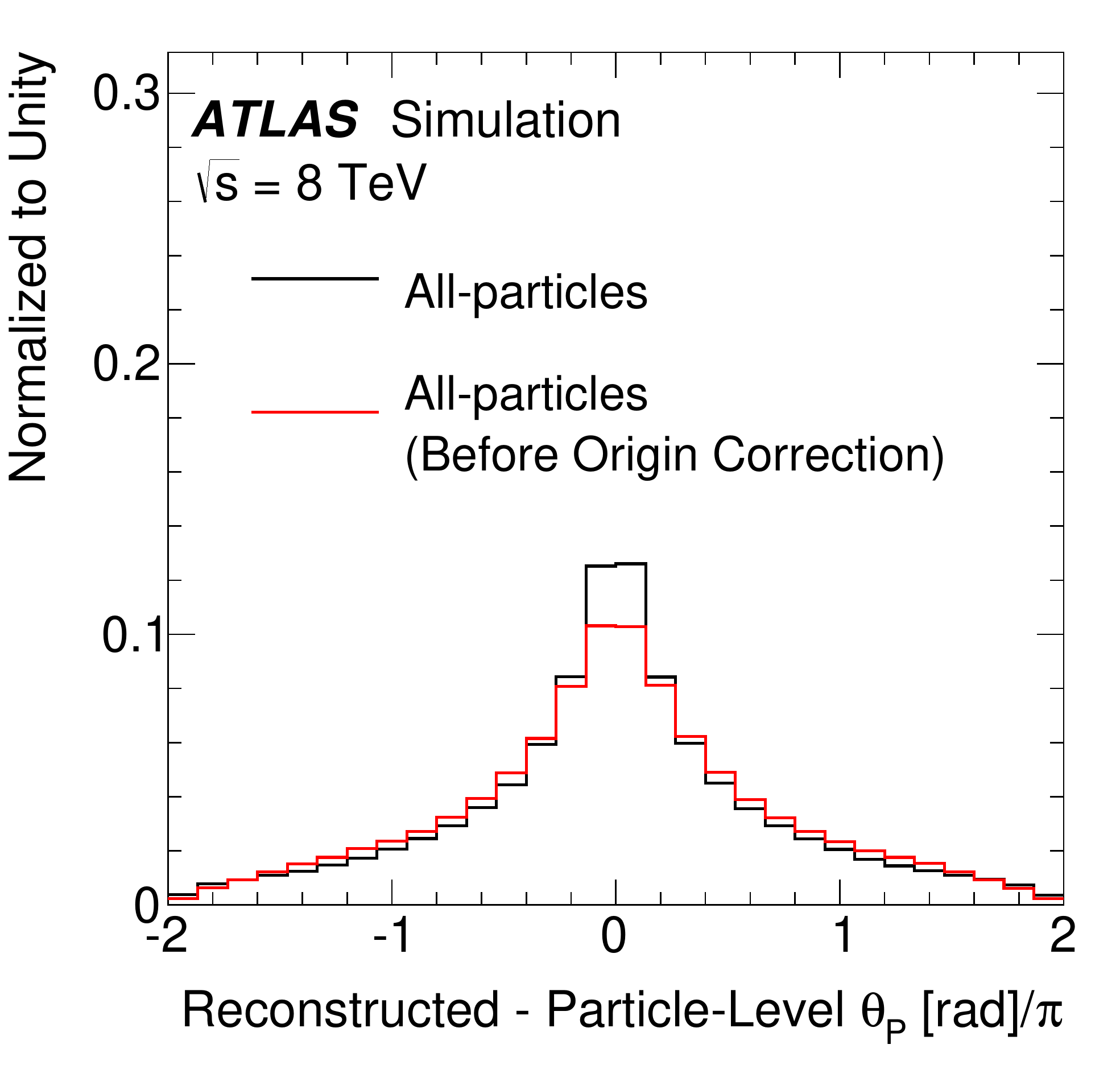}
  \label{fig:reco_charged_particles_pull_angle}
  \caption{The pull angle response before and after applying the cluster origin correction.  The jet origin correction is applied in both cases.}
  \label{fig:reco_dists}
\end{figure}

\begin{figure}
  \centering
  \includegraphics[width=0.5\textwidth]{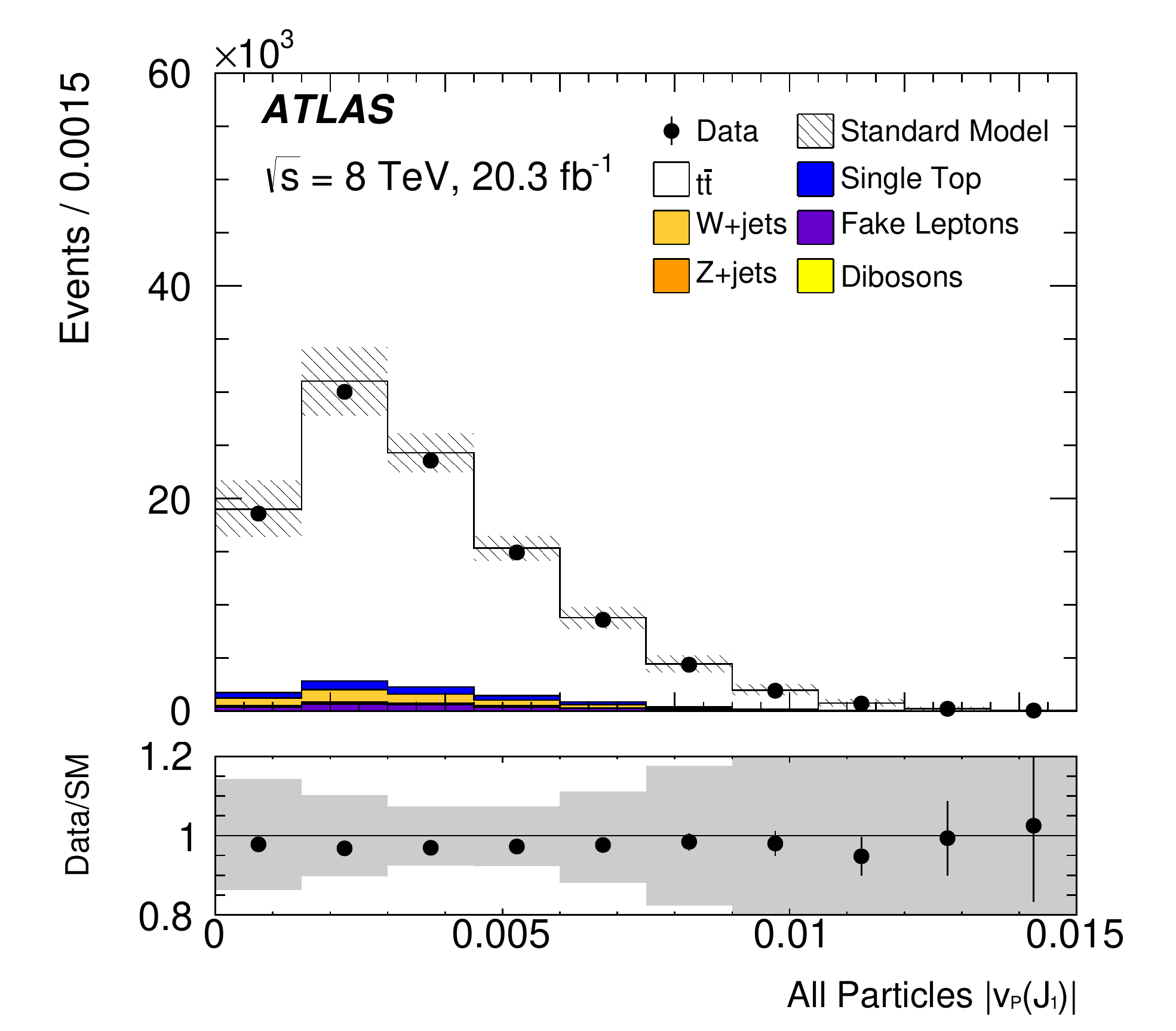}\includegraphics[width=0.5\textwidth]{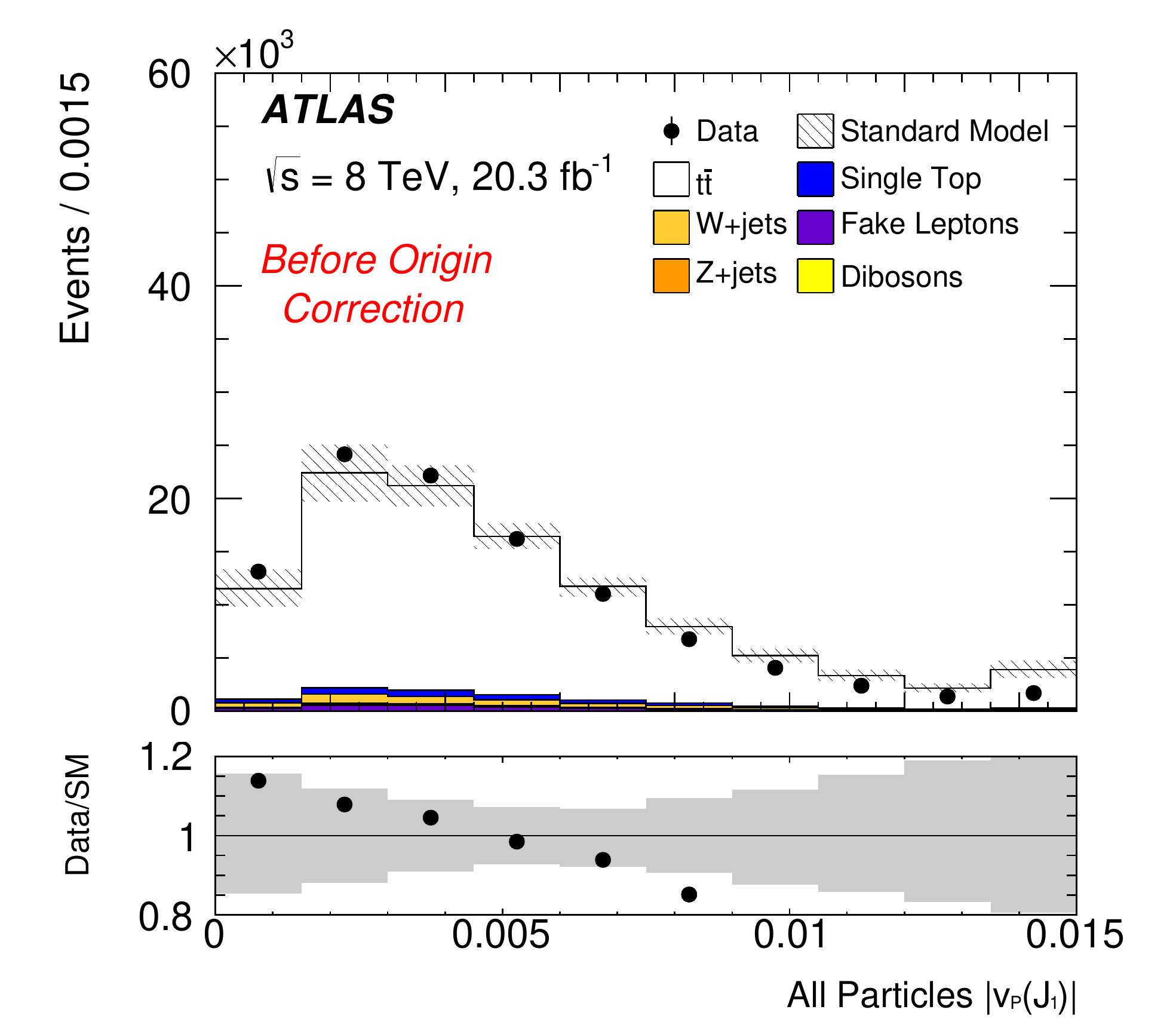}
  \label{fig:reco_charged_particles_pull_angle}
  \caption{The distribution of the all-particles pull vector magnitude with both the jet axis and the cluster axes origin corrected (left) and only the jet axis corrected (right).}
  \label{fig:reco_dists2}
\end{figure}

\begin{figure}
  \centering
  \includegraphics[width=0.5\textwidth]{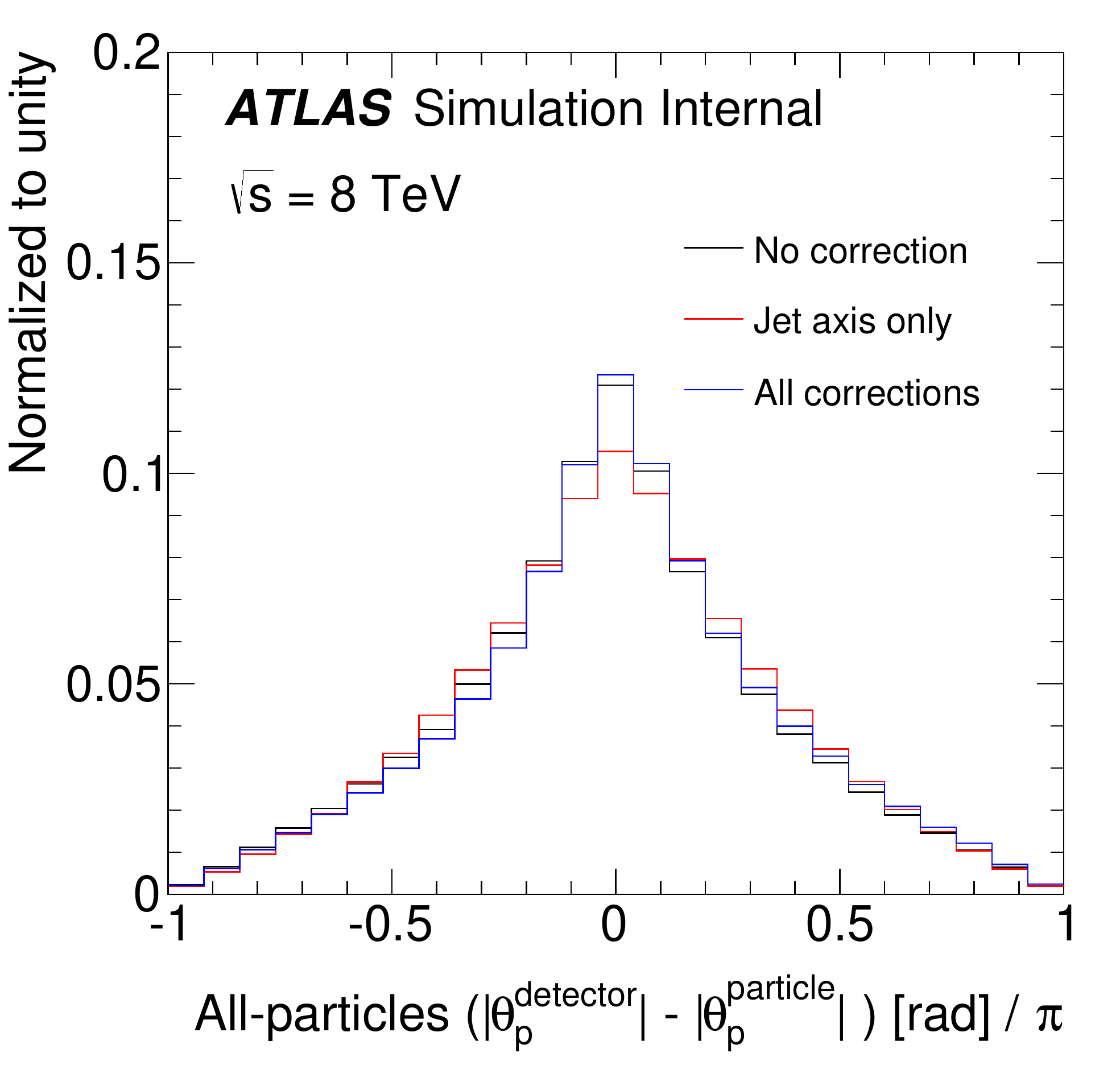}\includegraphics[width=0.5\textwidth]{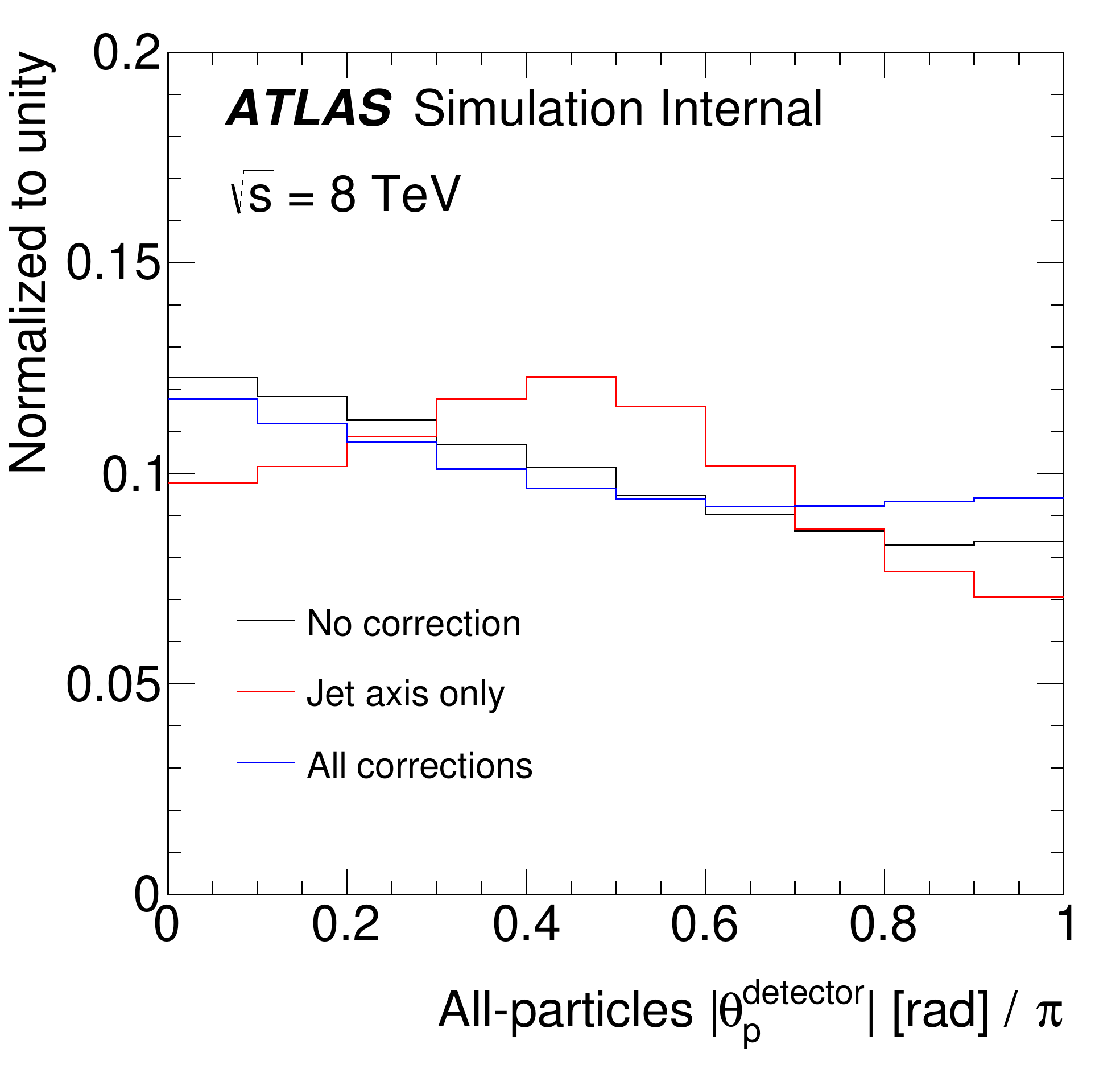}
  \label{fig:reco_charged_particles_pull_angle}
  \caption{Left: All-particles pull angle response.  Right: the all-particles pull angle distribution after various origin corrections.}
  \label{fig:improvement}
\end{figure}

Tracks are already `origin corrected' by construction and so no further correction is required beyond the jet origin correction.  However, it is possible to further improve the performance by using the {\it track-axis} formed from the four-vector sum of the tracks instead of the origin corrected axis.  Figure~\ref{fig:reco_diststrackaxis} shows the axis angular response for the calorimeter jet axis and the track-axis.  They have a similar resolution, but by using the track-axis, the charged-particles pull angle is nearly incentive to the calorimeter angular resolution.  The pull angle response is shown in the left plot of Fig.~\ref{fig:reco_diststrack} for the various jet axis definitions.  As the origin corrected jet axis has a similar resolution to the track axis, the pull angle resolution is similar for these two choices of axis and both are improved with respect to the starting axis.  As with the all-particles pull angle, the reduction in the angular resolution qualitatively changes the pull angle distribution shape (right plot of Fig.~\ref{fig:reco_diststrack}).  As expected from Sec.~\ref{sec:pullangleresponse}, the reduction in the axis resolution removes the resolution peak at $\pi/2$; now the track-based pull angle between the two $W$ daughter jets peaks at zero as is also the case for the all-particles pull angle.  In all subsequent studies, the track axis is used for the charged-particles pull vector.  The pull modeling of the track-based pull vector magnitude is shown in Fig.~\ref{fig:reco_dists4}.  The magnitude is generally shifted toward lower values than the all-particles pull vector due to the smaller constituent multiplicity.

\begin{figure}
  \centering
  \includegraphics[width=0.8\textwidth]{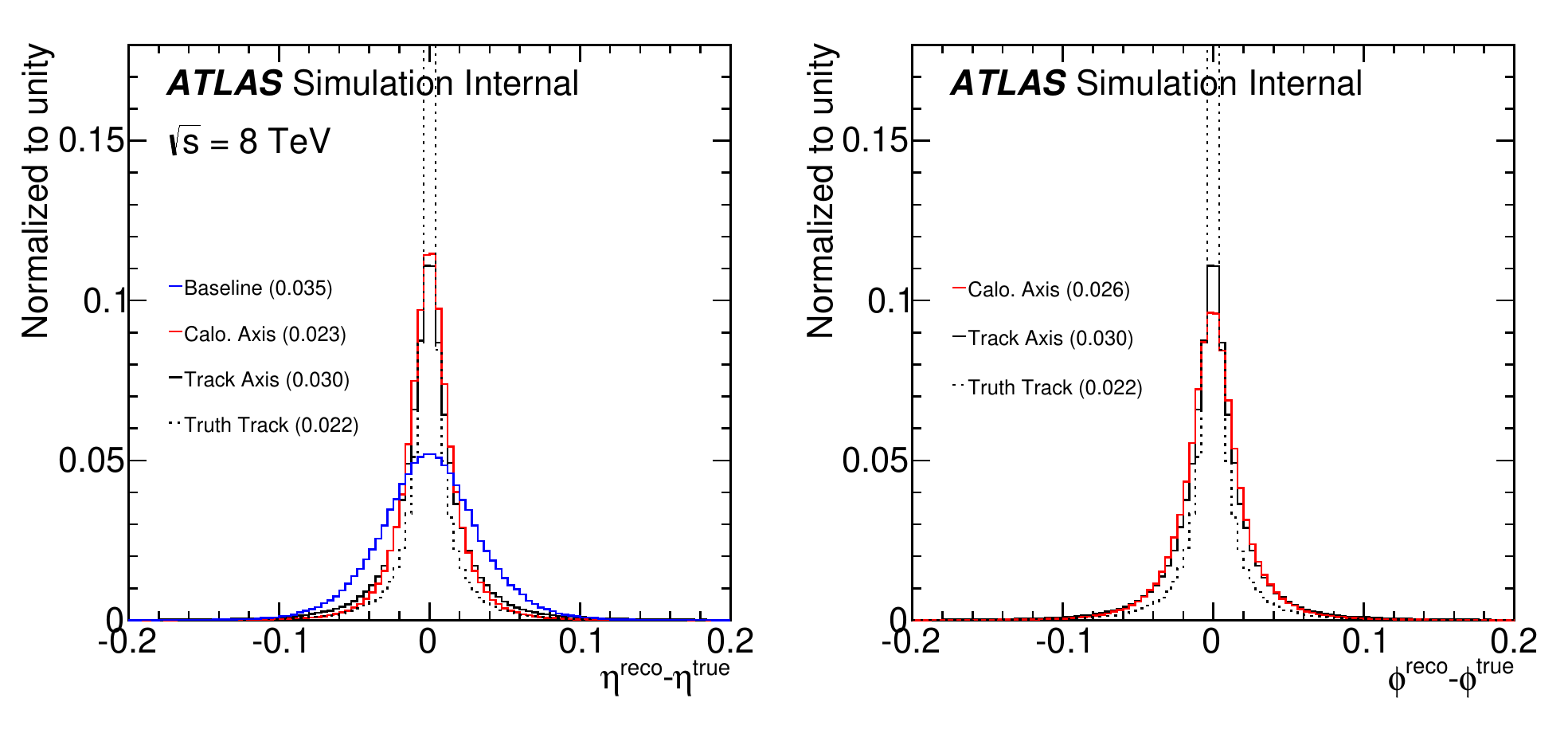}
  \label{fig:reco_charged_particles_pull_angle}
  \caption{The $\phi$ (left) and $\eta$ (right) angular response for various choices of the jet axis.  The `baseline' configuration uses the calorimeter jet axis without the origin correction.  The `calo axis' uses the origin correction and the `track axis' and `truth track' usee the four-vector sum of tracks.  In all cases except the last one, the particle-level reference object is the full particle-level jet axis while in the last case, the reference is the four-vector sum of the charged particles only.  The number in parenthesis is the RMS.}
  \label{fig:reco_diststrackaxis}
\end{figure}

\begin{figure}
  \centering
  \includegraphics[width=0.5\textwidth]{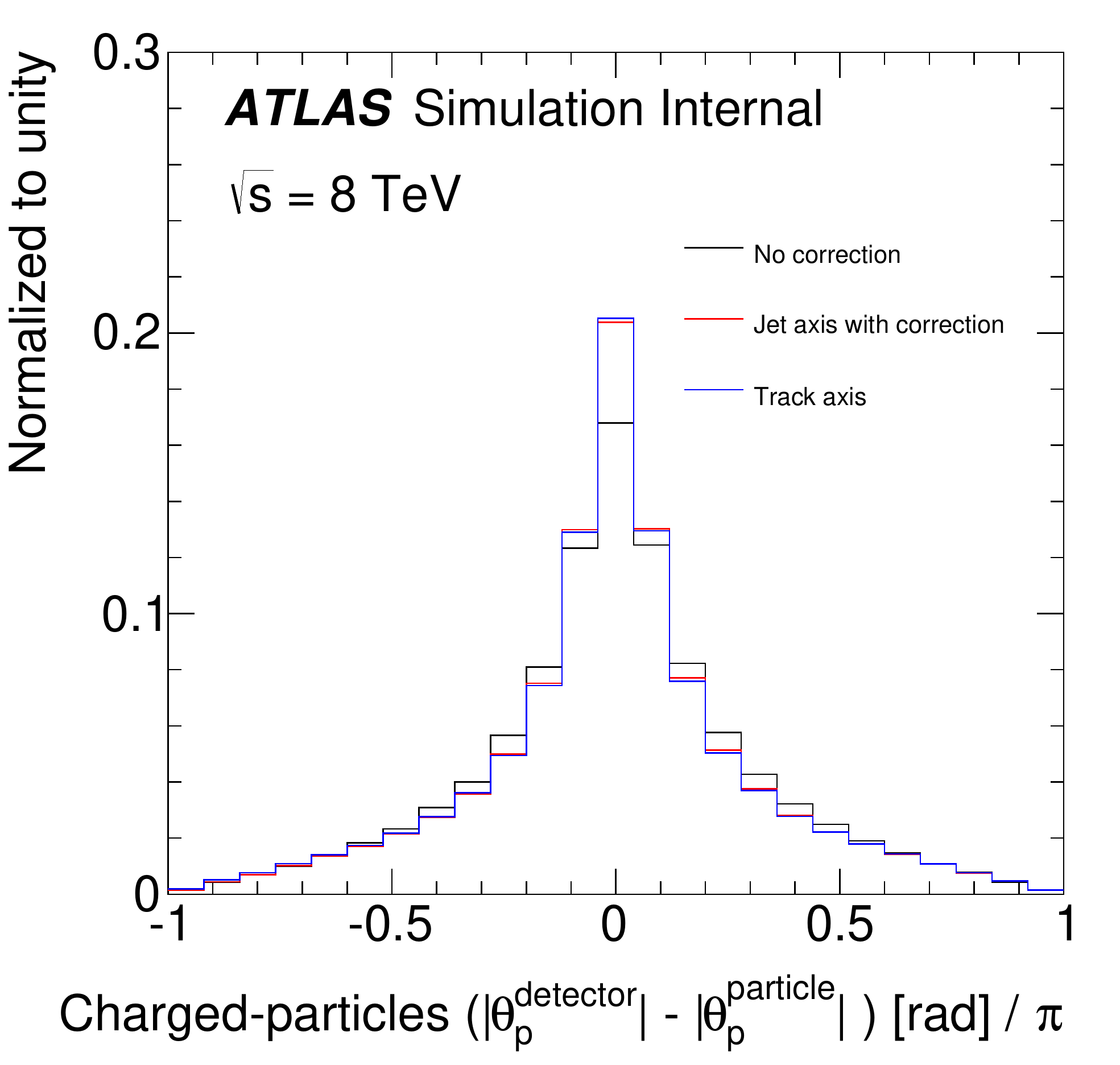}\includegraphics[width=0.5\textwidth]{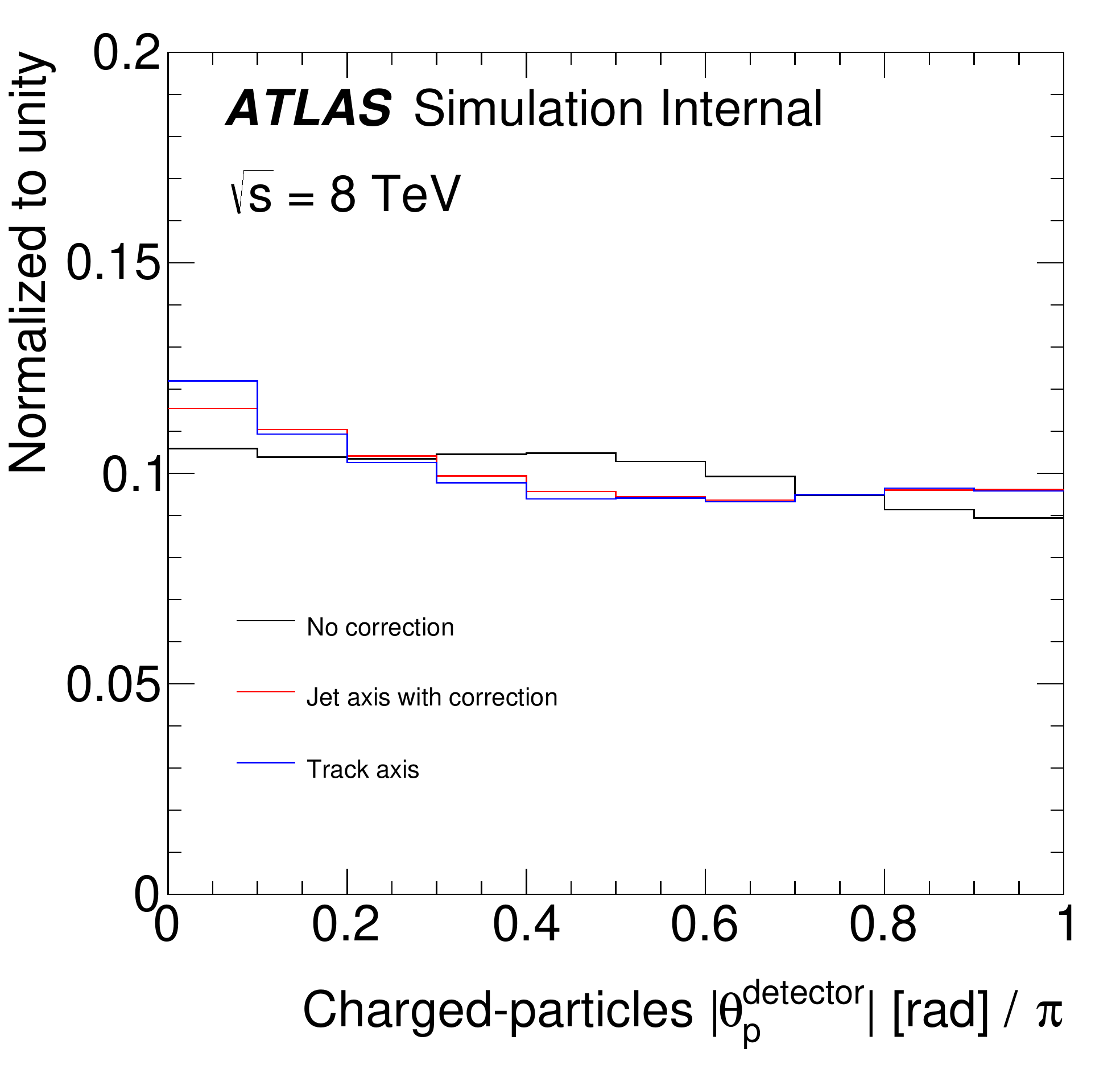}
  \label{fig:reco_charged_particles_pull_angle}
  \caption{Left: Charged-particles pull angle response.  Right: the charged-particles pull angle distribution after various origin corrections.}
  \label{fig:reco_diststrack}
\end{figure}

\begin{figure}
  \centering
  \includegraphics[width=0.48\textwidth]{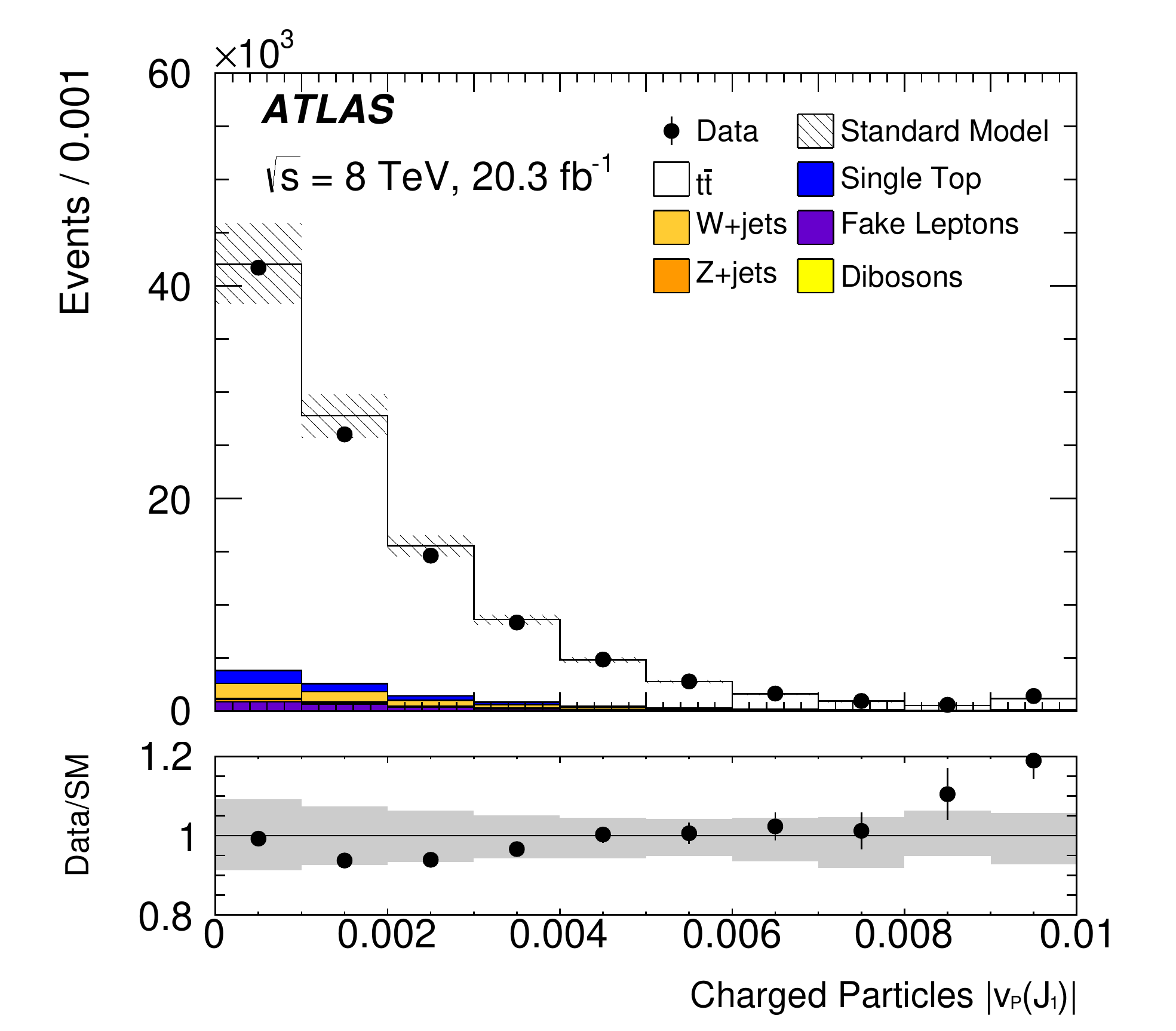} 
  \label{fig:reco_charged_particles_pull_angle}
  \caption{The charged-particles pull vector magnitude using the track four-vector sum for the jet axis.}
  \label{fig:reco_dists4}
\end{figure}

As a summary, the particle-level, detector-level, and response for the all-particles and charged-particles pull angles are shown in Fig.~\ref{fig:reco_dists333}.  With the various axis modifications described above, all the detector-level distributions peak at zero just like the particle-level distributions.  In addition to the SM pull angle distributions, Fig.~\ref{fig:reco_dists333} also shows the flipped $W$ bosons for which the $W$ decay products are not color connected.  The pull angle distribution is more uniform for the octet than for the singlet; the remainder of Chapter~\ref{cha:colorflow} is aimed at studying how well these distributions can be distinguished with the ATLAS data.  Figure~\ref{fig:reco_distsdatapull} shows the all-particles and charged-particles pull angles in data at detector-level for all the axis modifications described in this section.  Removing distortions from detector effects for a direct comparison with the particle-level models is described in Sec.~\ref{sec:colorflow:unfolding}.

\begin{figure}
  \centering
  \includegraphics[width=0.33\textwidth]{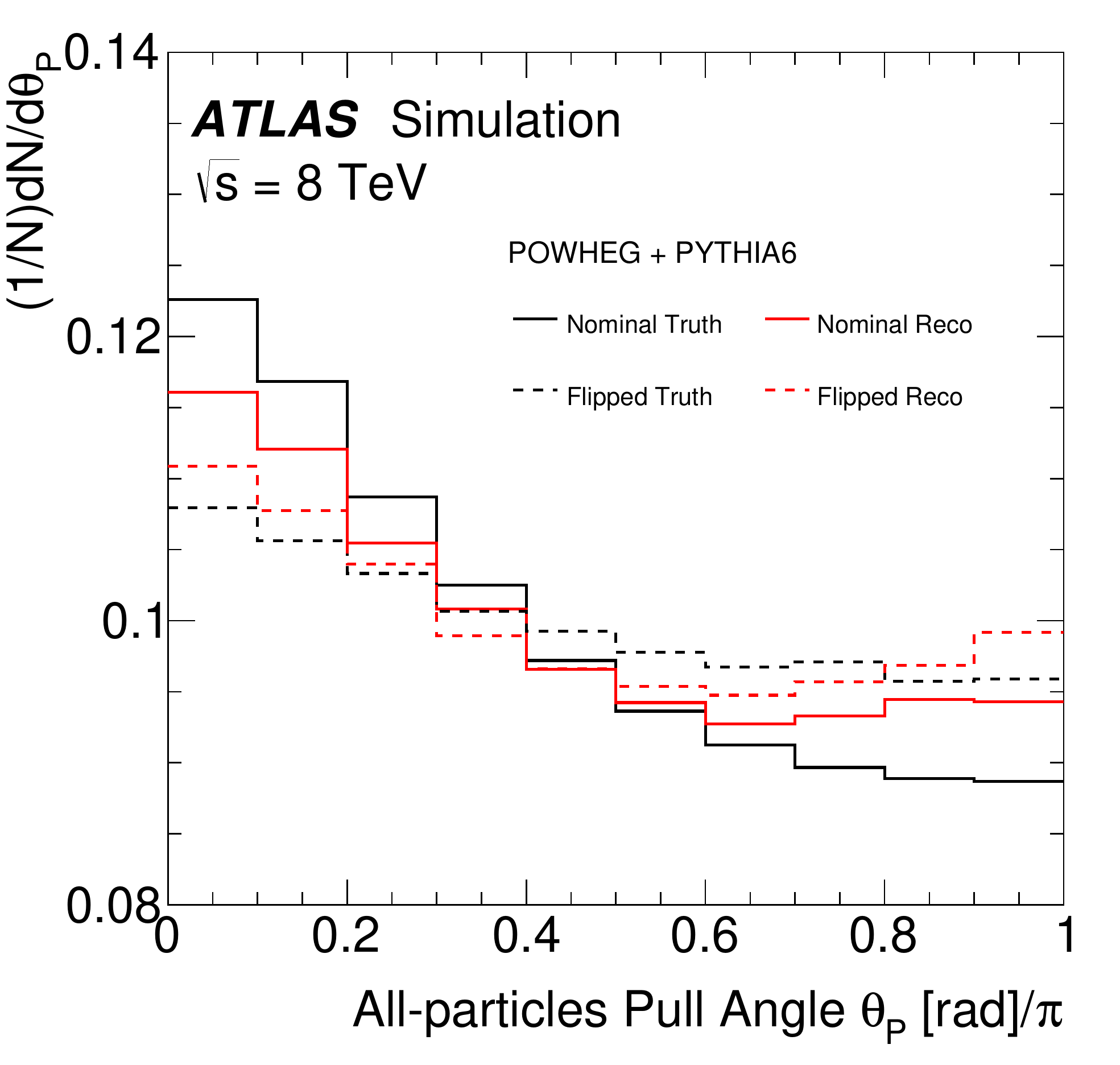}\includegraphics[width=0.33\textwidth]{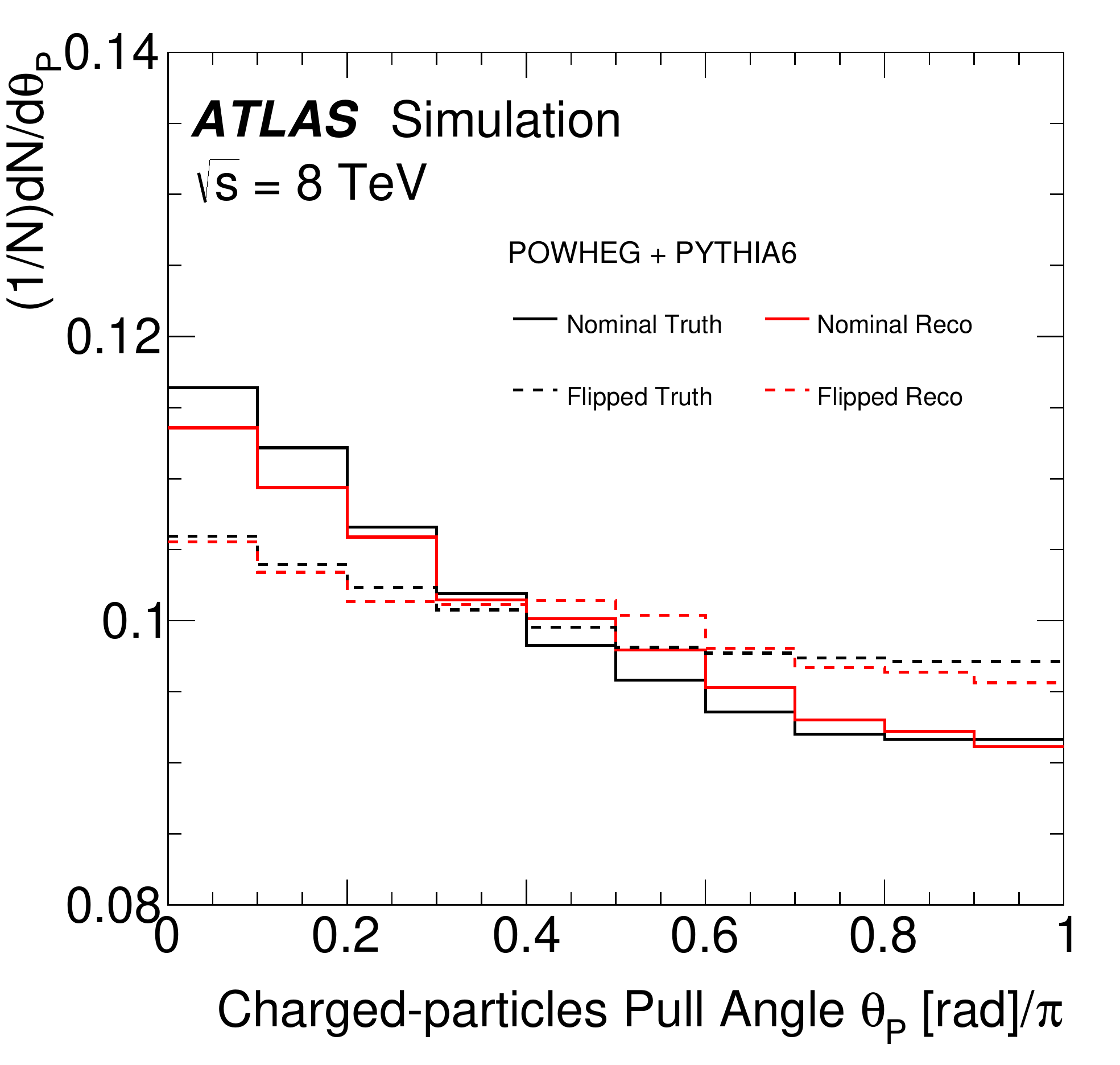} \includegraphics[width=0.33\textwidth]{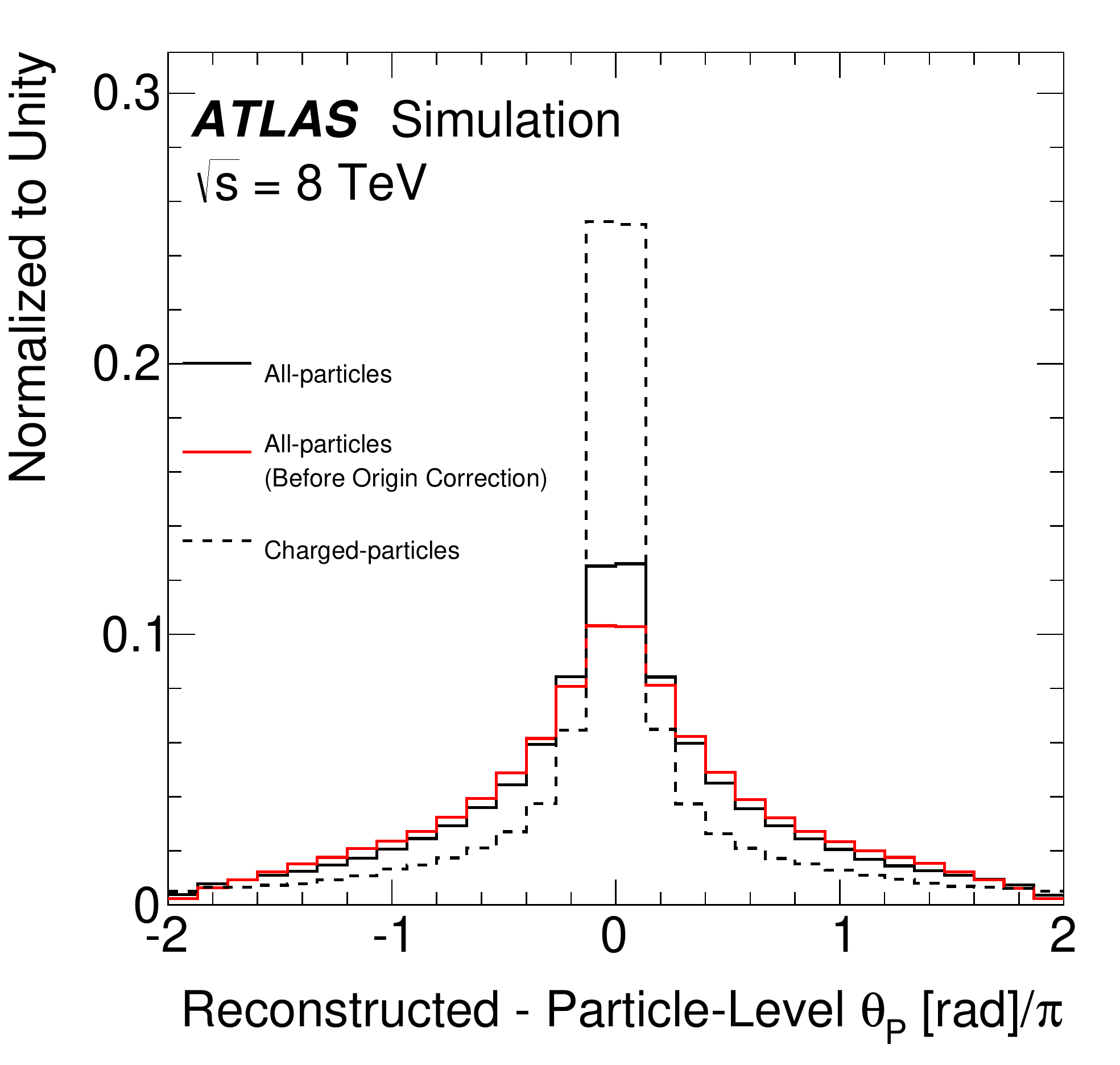}
  \label{fig:reco_charged_particles_pull_angle}
  \caption{The all-particles (left), charged-particles (middle), and pull angle response (right) in simulation using the nominal color singlet $t\bar{t}$ model and additionally with the color octet model (left and middle only).}
  \label{fig:reco_dists333}
\end{figure}

\begin{figure}
  \centering
  \includegraphics[width=0.5\textwidth]{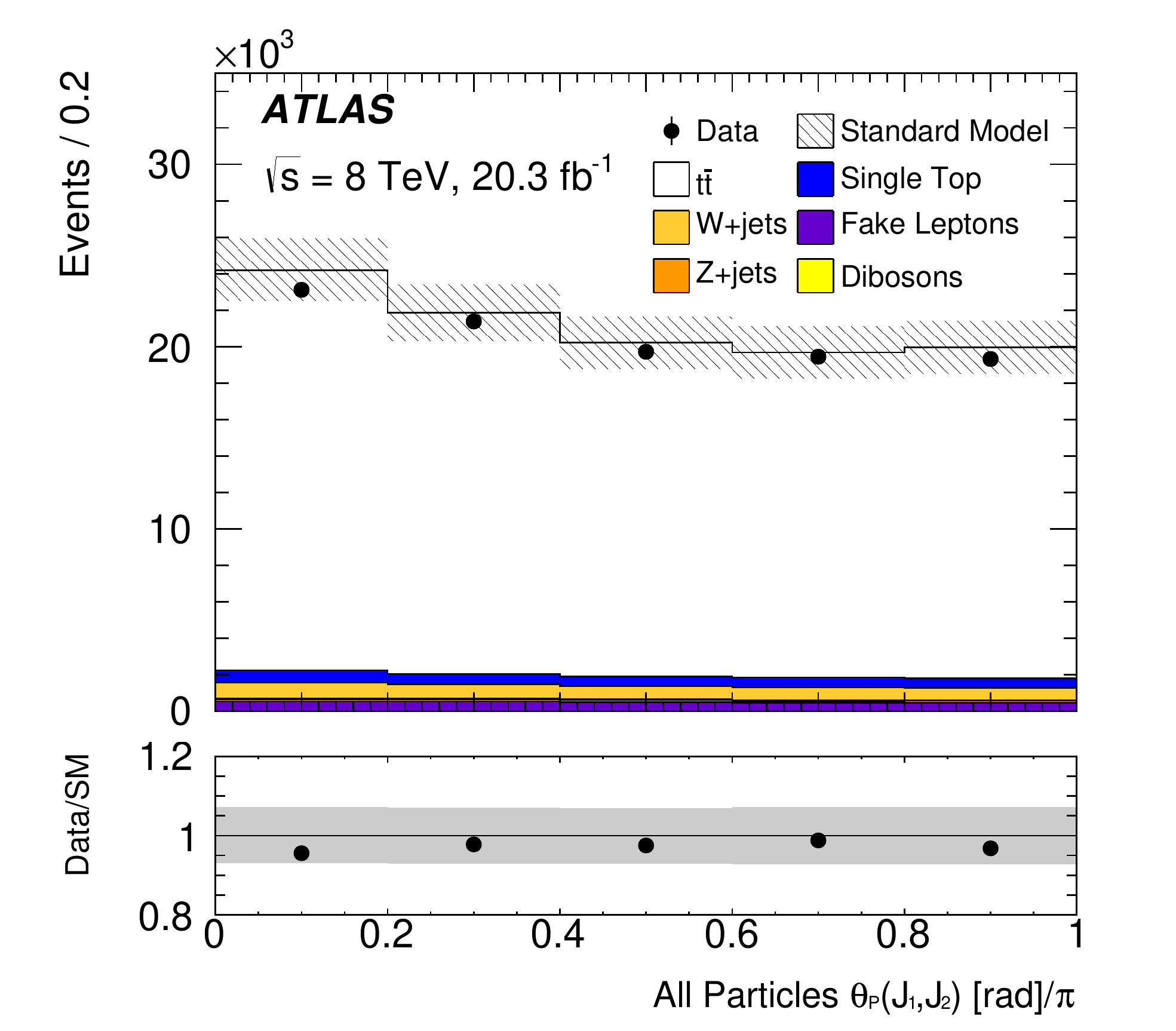}\includegraphics[width=0.5\textwidth]{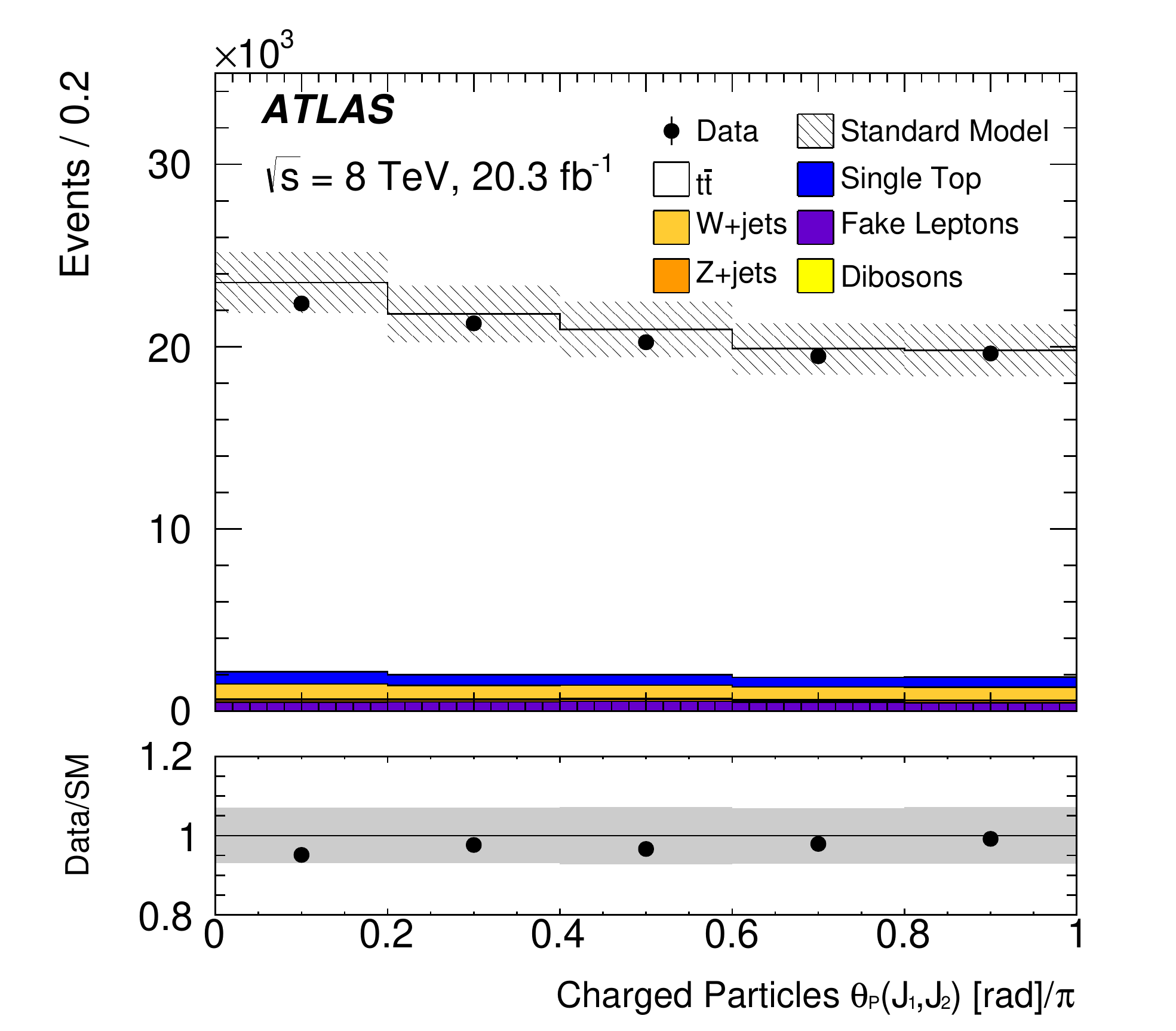}
  \label{fig:reco_charged_particles_pull_angle}
  \caption{The detector-level all-particles (left)
    and charged-particles pull (right)
    angles in data and in simulation.
    The uncertainty band includes only the experimental uncertainties on the
    inputs to the event selection and the jet pull calculation (see Sec.~\ref{sec:colorflow:systematics}).
   A large part of the uncertainty displayed here affects the overall normalization and is correlated between the individual bins.  This component of the uncertainty is cancelled in the unfolded measurement of the unit-normalized pull angle distribution (see Sec.~\ref{sec:colorflow:unfolding}).}
     \label{fig:reco_distsdatapull}
\end{figure}

\clearpage

\section{Unfolding}
\label{sec:colorflow:unfolding}

The rest of Chapter~\ref{cha:colorflow} is dedicated to the measurement of the pull angle by correcting for detector effects through unfolding.

\subsection{Unfolding Parameters}
\label{sec:ColorFlow:UnfoldingParams}

In addition to the number of iterations in the Iterative Bayesian (IB) unfolding algorithm, various aspects of the unfolding setup are optimized to reduce the total uncertainty.  The optimization is performed separately for the all-particles and charged-particles pull angles because the resolution is significantly worse for the former variable.  Three settings were scanned in the optimization procedure:

\begin{description}
\item[Number of bins] The closer the response matrix is to the identity matrix, the less dependent the results will be on the unfolding procedure and in particular on the prior in the IB algorithm.  Generally, it is desirable for the diagonal elements of the response matrix to satisfy $\Pr(\text{bin i}_\text{truth}\rightarrow\text{bin i}_\text{reco})\gtrsim 50\%$.  Since the pull angle resolution is comparable to the allowable range $\theta_\text{p}\in[0,\pi]$, it is expected that only a few, $\pi/\sigma(\theta_\text{p}^\text{reco}-\theta_\text{p}^\text{true})\sim 3$, bins will be possible.  More bins for the charged-particles pull angle are expected due to the superior resolution compared to the all-particles pull angle (see Fig.~\ref{fig:reco_dists333}).

\item[Number of iterations] Increasing the number of iterations in the IB method reduces the dependence on the prior, chosen to be the particle-level spectrum in simulation.  However, after a certain number of iterations the results saturate.  The point at which the results do not change with more iterations (saturation) depends on the resolution.  Figure~\ref{fig:ColorFlow:iterations:toystudy} uses a simple calculation using a Toy MC to show how the saturation point depends on the resolution.  For the pull angle $\sigma/\text{Range}\sim 1/3$, a saturation occurs at $\sim25$ iterations.  One does not necessarily want to use the number of iterations corresponding to the saturation point; increasing the number of iterations usually reduces the dependance on the truth spectrum in the MC used to construct the response matrix, but the cost is a larger statistical uncertainty.

\item[Pull Magnitude cut] The studies in Sec.~\ref{sec:ColorFlow:Performance:Constits} showed that $\sigma(\theta_P^\text{reco}-\theta_P^\text{true})$ depends on $p_\text{T}$ and on the magnitude of the pull vector.  The tradeoffs for a magnitude requirement are a reduction in statistics and a potential increase in model dependence, as the jet pull angle magnitude contains information about color flow.
\end{description}

\begin{figure}[h!]
\begin{center}
\includegraphics[width=0.55\textwidth]{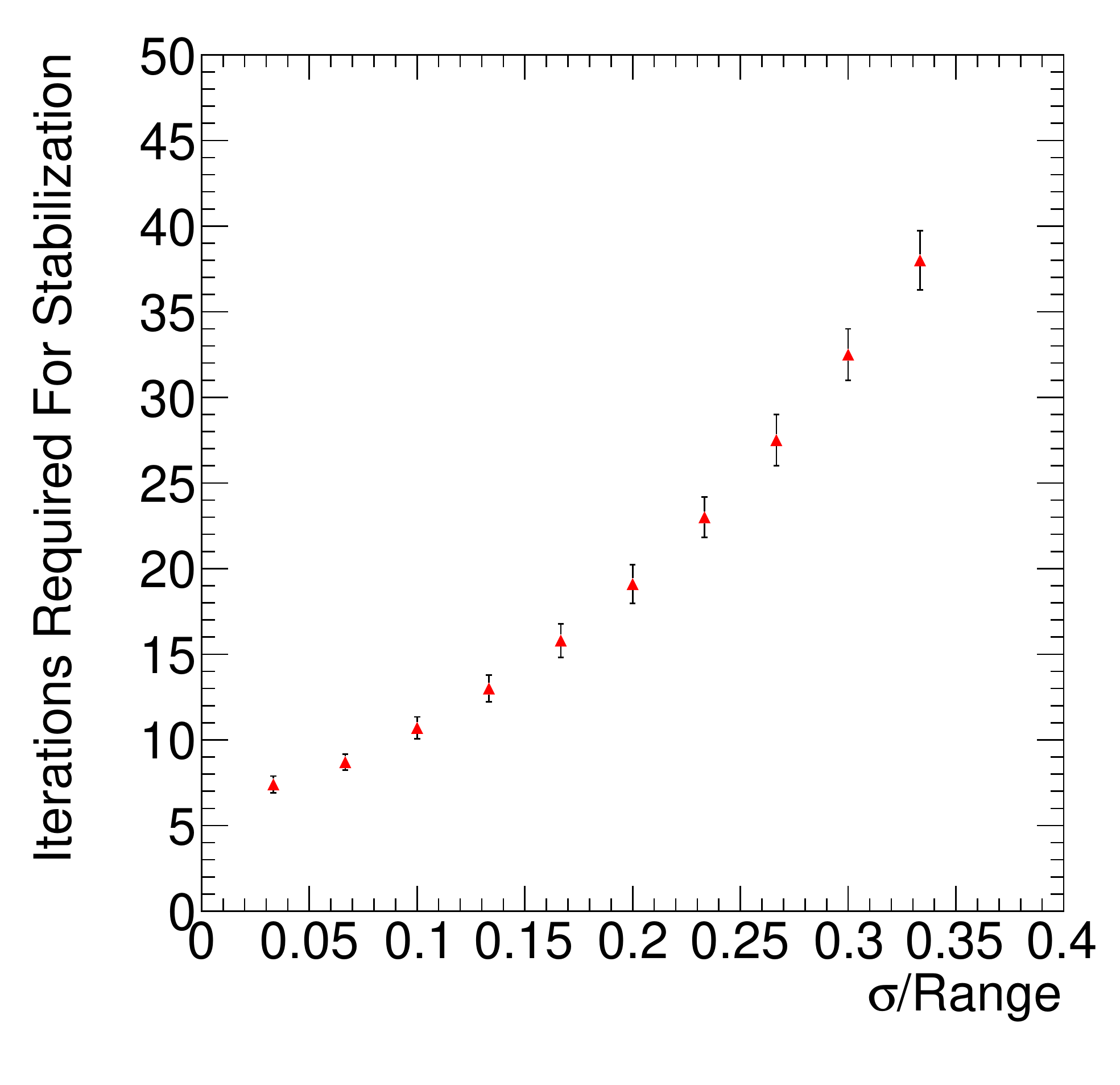}
 \caption{The number of iterations required for the unfolded results to saturate using a Toy MC.  The unfolding is said to saturate if the unfolded bin contents do not change by more than $0.01\%$ between iterations.  In this model, $\theta_\text{p}\sim\text{Uniform}(0,\text{Range})$ and $\theta_P^\text{reco}-\theta_P^\text{true}\sim\mathcal{N}(0,\sigma^2)$, where $\text{Range}=\pi$ and the smearing is done modulo $\pi$.}
 \label{fig:ColorFlow:iterations:toystudy}
  \end{center}
\end{figure}

The parameters described above are interrelated and so the optimization must be performed simultaneously.   The number of equal sized bins was scanned between 3 and 9, the number of iterations was scanned between 1 and 6 (1 and 14 for all-particles) and the pull vector magnitude was scanned between 0 (no requirement) and $3.5\times 10^{-3}$ in steps of $5\times 10^{-4}$ leading to 336 (784) configurations for the charged-particles (all-particles) pull angles.   For each configuration, the data statistical uncertainty was combined with the dominant systematic uncertainties, including the color flow model, fragmentation model, and the data-driven non-closure\footnote{The optimization was performed without data, so the statistical uncertainty is based on the expected yields from the simulation, and the non-closure uncertainty used the data only indirectly.}.  All of these uncertainties require running the unfolding algorithm with the given parameters at least once and are described in Sec.~\ref{sec:colorflow:systematics}.

Using the bin-averaged uncertainty as a metric and allowing for some slight post-hoc modifications with unequal bin sizes, the parameters for the unfolding are as follows:

\begin{description}
\item[All-particles]: 3 bins with ranges $[0,0.275, 0.6375,1.0]\times\pi$, 15 iterations, and no pull vector magnitude requirement.
\item[Charged-particles]: 4 bins with ranges $[0,0.2,0.5,0.8,1.0]\times\pi$, 3 iterations, and no pull vector magnitude requirement.
\end{description}
 
\noindent The optimization procedure suggested that a small requirement on the pull vector magnitude for both the all-particles and charged-particles pull angles could reduce the overall uncertainty.  However, since such a gain is below 1\% in the bin-averaged uncertainty and would introduce a new source of model dependence, the requirement is not used for the final configuration.  Figure~\ref{fig:ColorFlow:numberofiterations} shows the impact of the number of iterations on the uncertainty in each bin using only the color flow model and statistical uncertainties for illustration.   For both the all-particles and charged-particles pull angles, the statistical uncertainty increases monotonically with the number of iterations.  The combined uncertainties for all bins have a minimum in the plotted range except the second bin of the all-particles pull angle, due to the fact that the distributions are normalized before computing the uncertainties.  For nearly all iteration choices, the color flow uncertainty is larger than the statistical uncertainty.  Unlike the statistical uncertainty, the color flow uncertainty decreases with the number of iterations and then increases again as the unfolding overcorrects the simulation.  

\begin{figure}[h!]
\begin{center}
\includegraphics[width=0.45\textwidth]{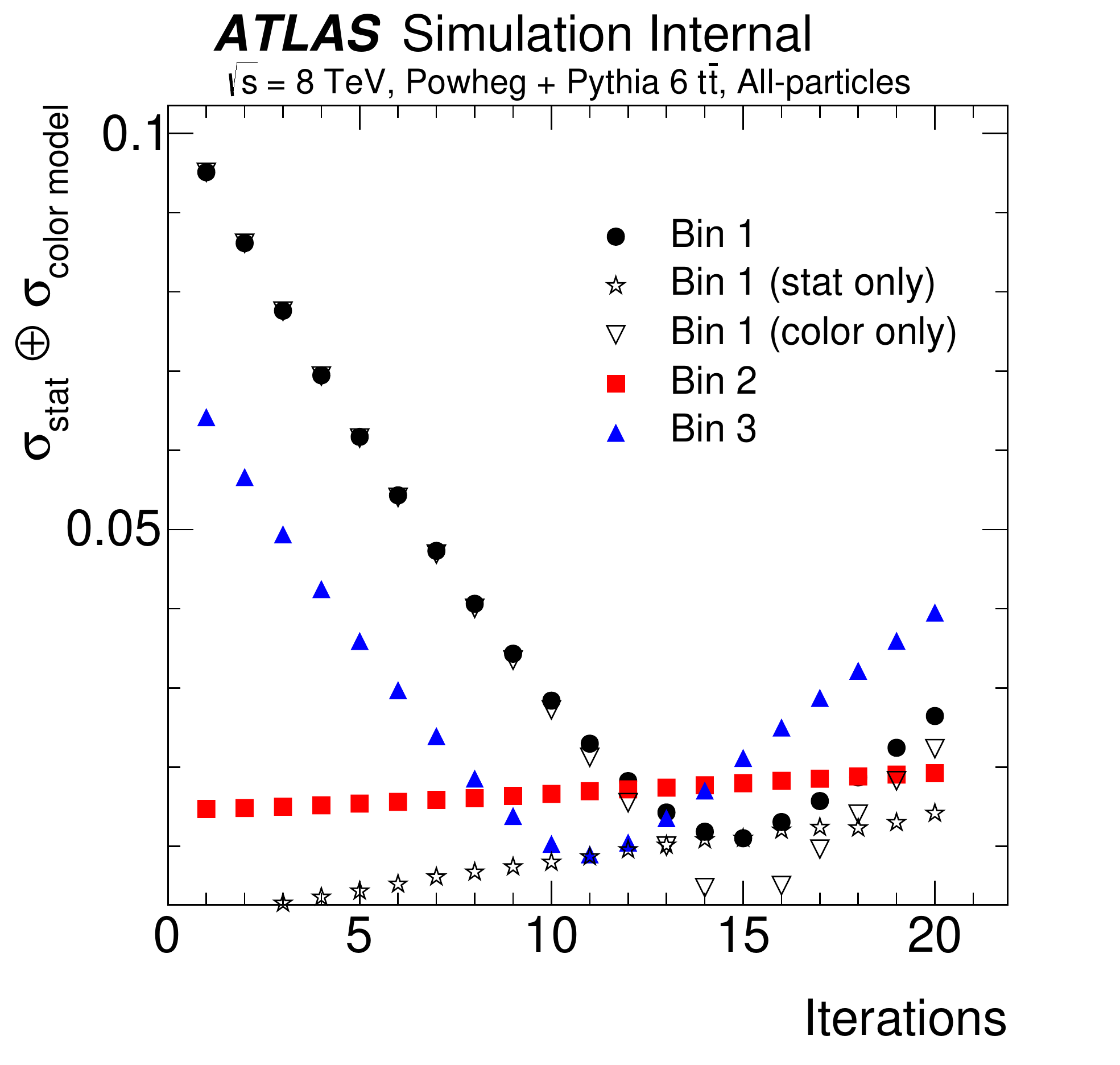}\includegraphics[width=0.45\textwidth]{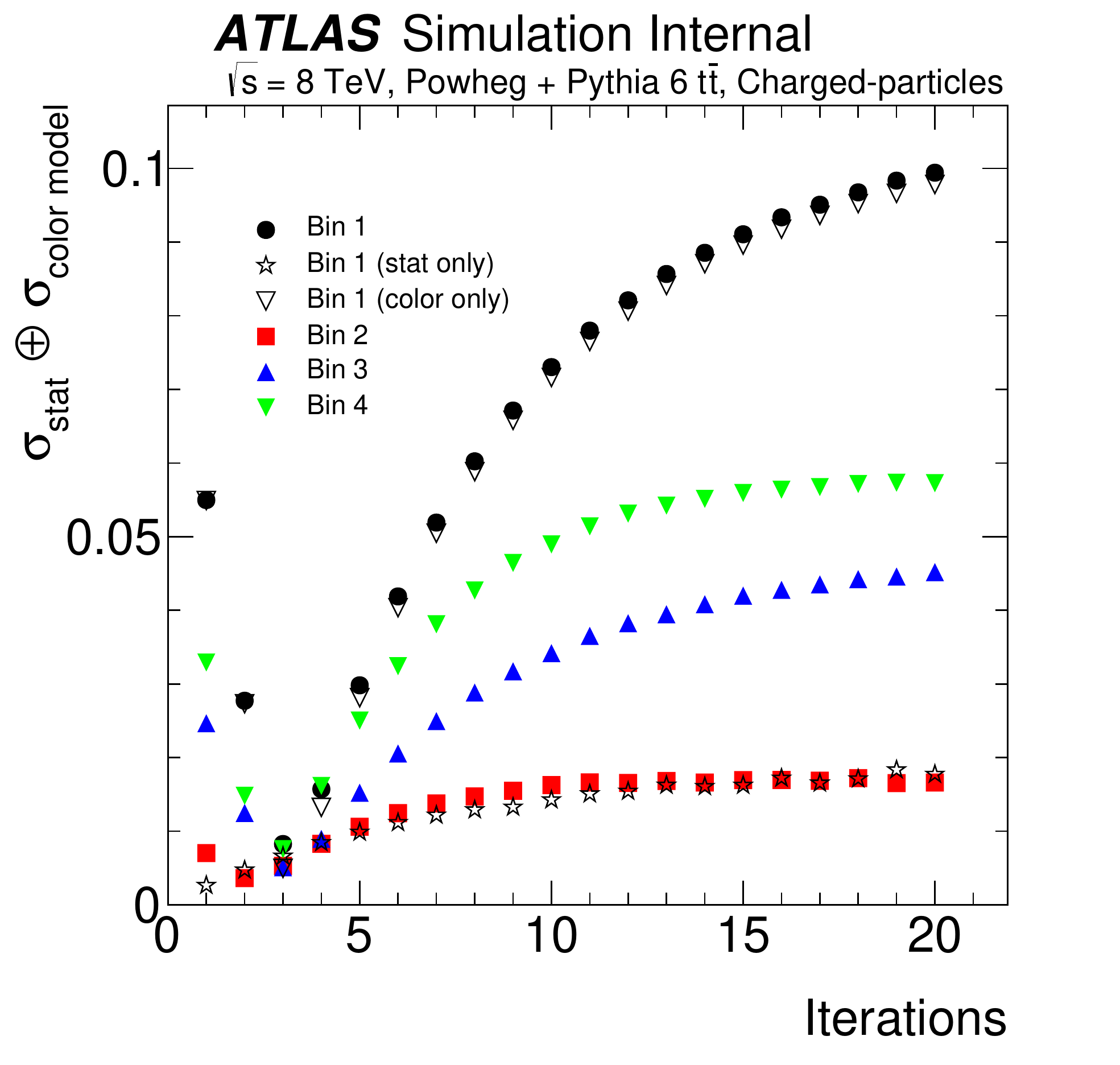}
 \caption{The sum in quadrature of the data statistical uncertainty and the color flow model uncertainty (See Sec.~\ref{sec:colorflow:systematics}) as a function of the number of iterations in the IB unfolding method for the three bins of the all-particles pull angle (left) and for the four bins of the charged-particles pull angle (right).  For the first bin, the open markers show how the total uncertainty is broken down into the two components.}
 \label{fig:ColorFlow:numberofiterations}
  \end{center}
\end{figure}

The remainder of this section describes in more detail the interplay between the resolution and the number of iterations required to reduce the model-dependence uncertainties.  A first observation is that since the jet pull angle is a bounded variable, there is an induced correlation between the pull angle response and the pull angle itself.  This is illustrated schematically in Fig.~\ref{fig:ColorFlow:optimizationtoyreason}.   If the (normalized by $\pi$) particle-level jet pull angle is 0.5, then the difference between the detector-level and the particle-level values can be at most 0.5\footnote{One important subtlety is about when the absolute value is taken when computing the response and the pull angle.  A pull angle of $\pi$ and a pull angle of $-\pi$ have the same probability under a given color flow model, but experimentally, $\pi=\theta_\text{p}^\text{true}\rightarrow\theta_\text{p}^\text{reco}=-\pi$ is a (maximal) mis-measurement.  One way around this is to introduce another bin in the response matrix to account for negative value.  This was tested and did not improve the uncertainty because increasing the number of bins resulted in lower transition probabilities in the response matrix.}.  However, if the particle-level pull angle is 0 or 1, than the difference between the detector-level and the particle-level values can be as large as 1.  This correlation is important because it means the response matrix depends on the particle-level pull angle distribution, which is most relevant for the color flow model uncertainty (which by construction has quite a different pull angle distribution).   

\begin{figure}[h!]
\begin{center}
\includegraphics[width=0.5\textwidth]{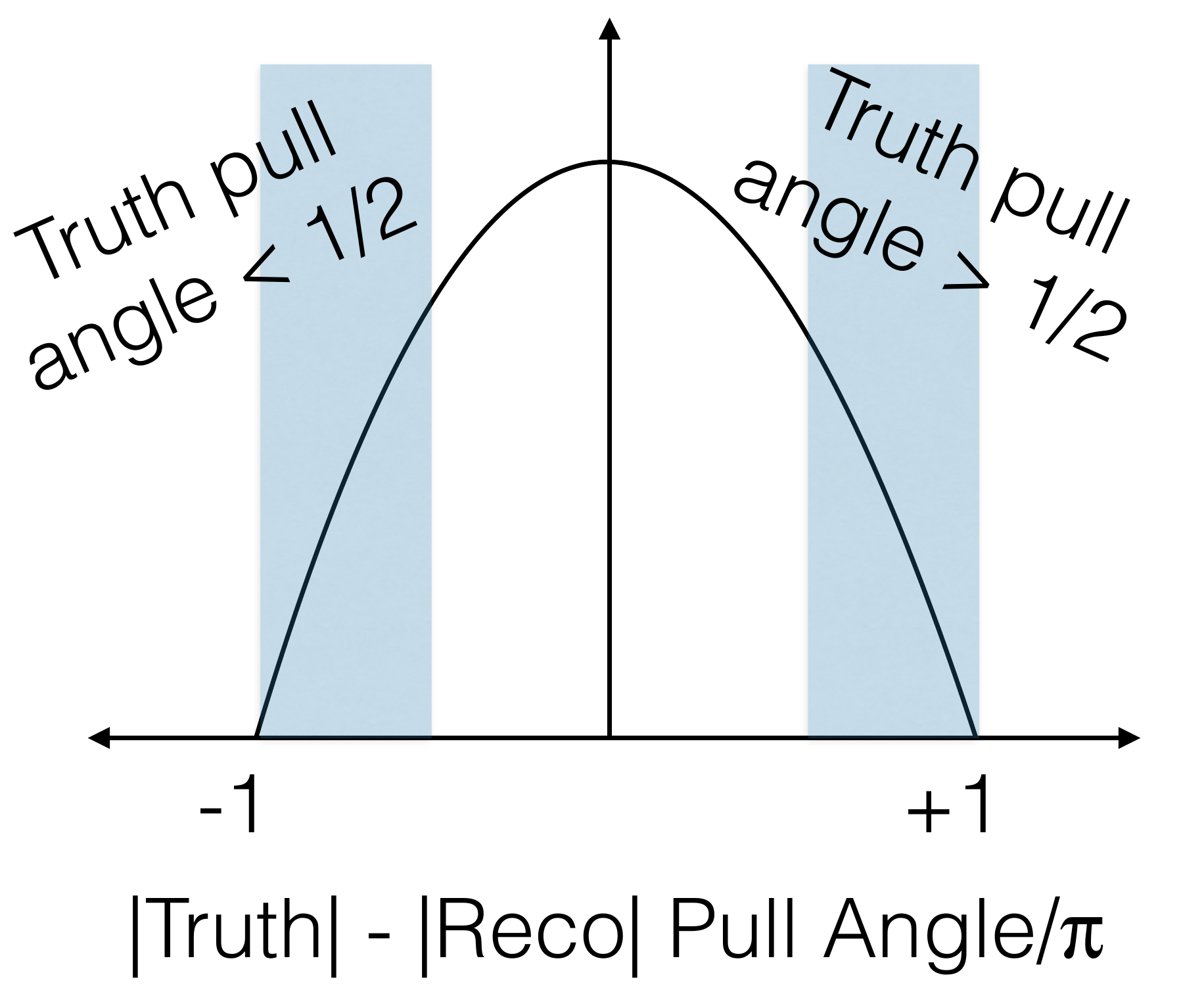}
 \caption{A schematic diagram of the jet pull angle response which illustrates that the pull angle resolution is strongly correlated with the pull angle itself.  For example, if you know that the difference between the particle-level and detector-level (normalized by $\pi$) pull angles is greater than 0.5 (blue shaded box on the right), then the truth pull angle must have been greater than 0.5.}
 \label{fig:ColorFlow:optimizationtoyreason}
  \end{center}
\end{figure}

A Toy MC is constructed to quantify this dependence.  The particle-level spectrum is constructed as a one parameter family of distributions with varying peak heights at zero, emulating the important difference in the pull angle distribution between the singlet and octet color flow models.   Angles are generated uniformly at random between $0$ and $2\pi$ and are then smeared with a Gaussian (modulo $\pi$) that has mean zero and standard deviation $\sigma$.  The `measurement' is performed with the absolute value of the angle divided by $\pi$ so that the range is between 0 and 1.  Each event is then re-weighted such that the truth spectrum probability distribution function is a right triangle with base length $X$ and height set by normalization.  As $X\rightarrow\infty$, the distribution between $0$ and $1$ is uniform and as $X\rightarrow 0$, the distribution is a $\delta$-function at $0$.  Figure~\ref{fig:optimizationtoysetup} shows the distribution corresponding to various values of $X$.  Figure~\ref{fig:optimizationtoy} shows the results of unfolding the measured (i.e. smeared) toy data.  The $z$-axis is the bin normalized fractional uncertainty, defined as the difference between the truth distribution and the unfolded toy data.  The toy truth and toy data both have $X=5$ while the response matrix has a variable $X_\text{MC}$ value.  The uncertainty increases as $X_\text{MC}$ moves away from 5 and the size of this uncertainty is bigger for larger angle smearing $\sigma$.  The difference between the left and right plots in Fig.~\ref{fig:optimizationtoy} shows that the size of the uncertainty can be mitigated by increasing the number of iterations.  For reference, the $\sigma$ for the all-particles pull angle is $\sigma\sim0.35$ and $\sigma\sim0.28$\footnote{The standard deviation does not fully capture the large differences between the resolutions - see Fig.~\ref{fig:reco_dists333}.} for the charged-particles pull angle.  In addition to the difference in resolutions, the absolute difference between the singlet and octet charged-particles pull angle distributions is smaller than for the all-particles pull angle because some of the discriminating information is lost in the neutral radiation.

\begin{figure}[h!]
  \begin{center}
    \includegraphics[width=.5\linewidth]{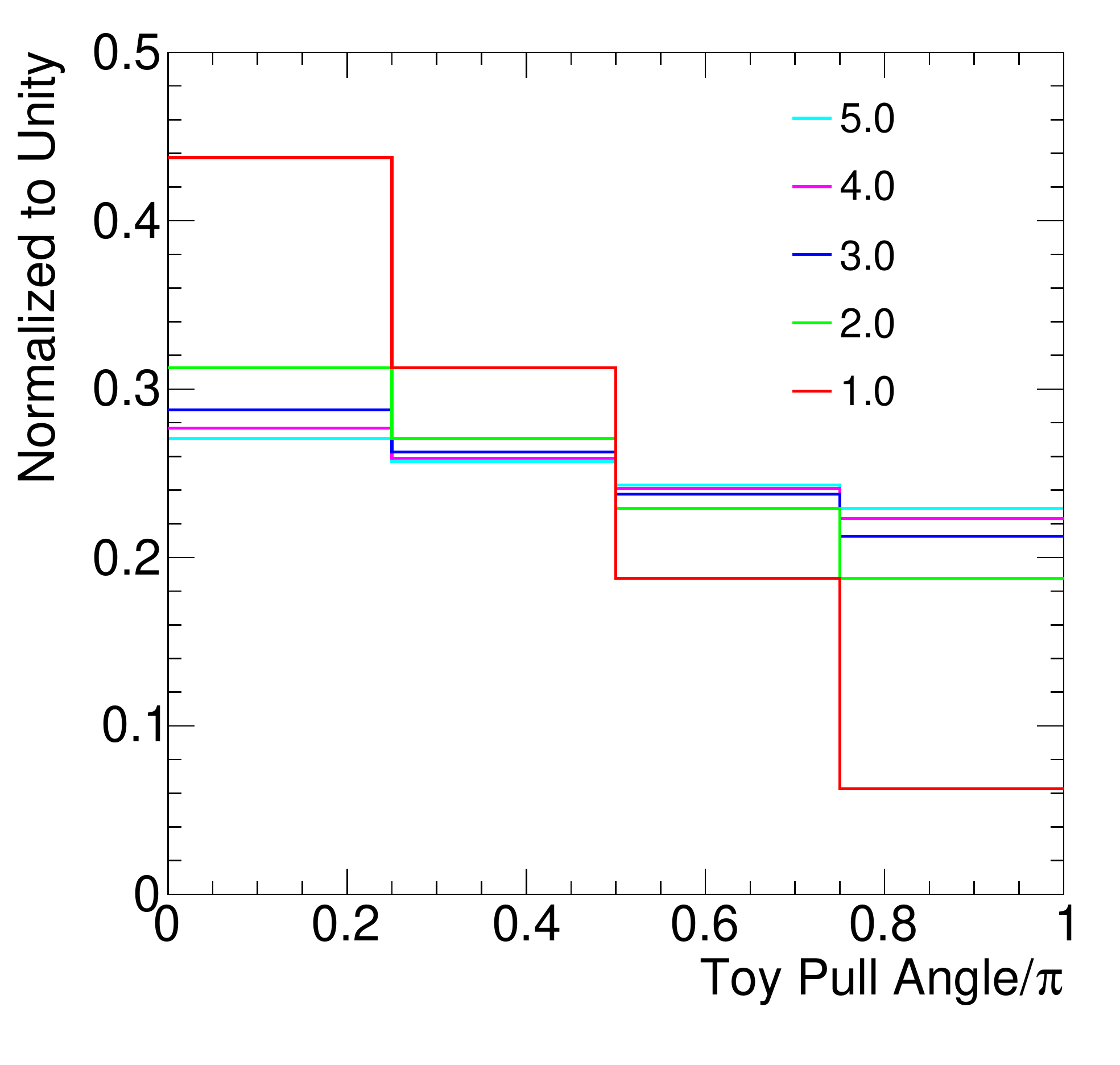}
    \caption{The truth distribution for the toy experiment described in Sec.~\ref{sec:ColorFlow:UnfoldingParams}.  The distributions are indexed by the base of a triangle which varies between $X=1$ and $5$.  When the base length is 5, the distribution is close to uniform and when it is 1, the distribution is strongly peaked at zero.}
    \label{fig:optimizationtoysetup}
  \end{center}
\end{figure}

\begin{figure}[h!]
  \begin{center}
 	\includegraphics[width=.48\linewidth]{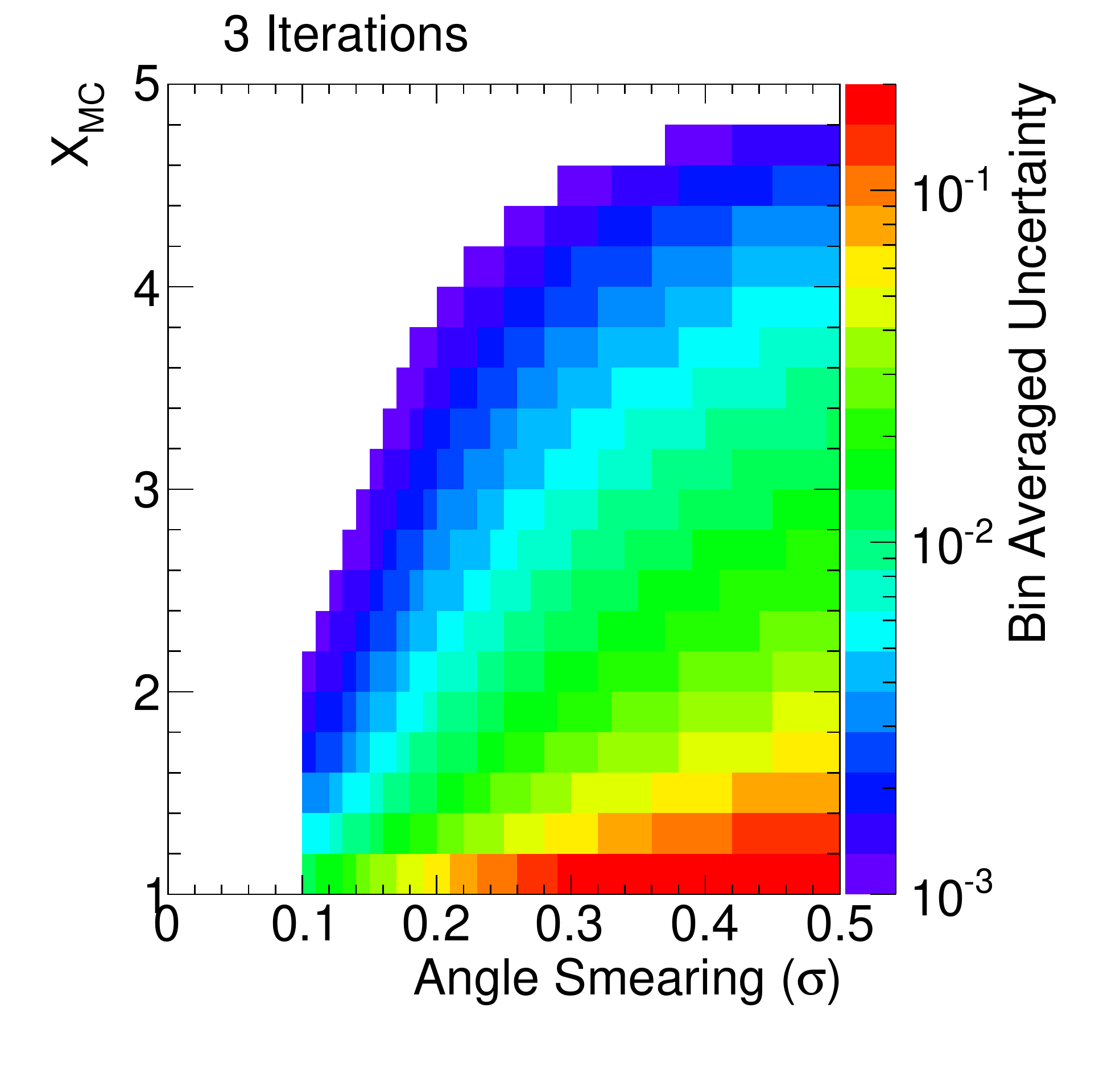}
   	\includegraphics[width=.48\linewidth]{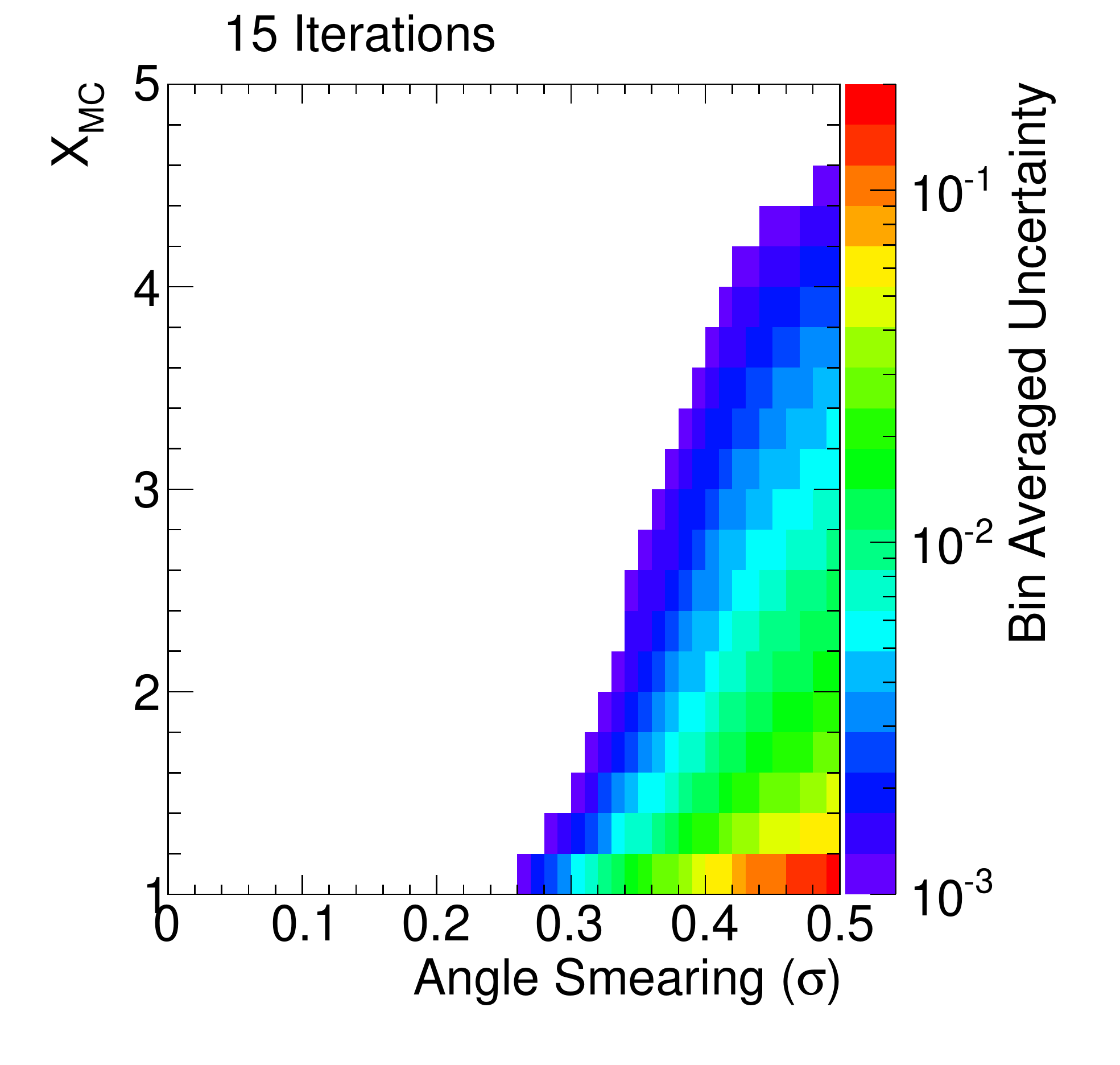}
    \caption{The bin averaged fractional uncertainty from comparing the truth and unfolded toy data ($X=5$) using different response matrices (with $X=X_\text{MC}$).  The value of $\sigma$ is used in both the toy data and the response matrix.  The left plot uses three iterations while the right plot uses 15 iterations.}
    \label{fig:optimizationtoy}
  \end{center}
\end{figure}

\clearpage

\subsection{Correction Factors}
\label{colorflowcorrectionfactors}

With the unfolding setup fixed, the next step in the unfolding procedure is to subtract non-$t\bar{t}$ processes from the data and apply correction factors.  Background estimates (described in Sec.~\ref{sec:ColorFlow:simulation}) are subtracted bin-by-bin in the pull angle distribution.  Even though the expected background composition is about 10\% of the total yield, the background pull angle distributions are nearly independent of the pull angle and therefore this correction has nearly no impact on the {\it normalized} pull angle distribution.  Figure~\ref{fig:ColorFlowBackgroundComp} shows the background composition as a function of the pull angle.  The background distributions vary by less than 3\% across bins.

\begin{figure}[h!]
  \begin{center}
 	\includegraphics[width=.45\linewidth]{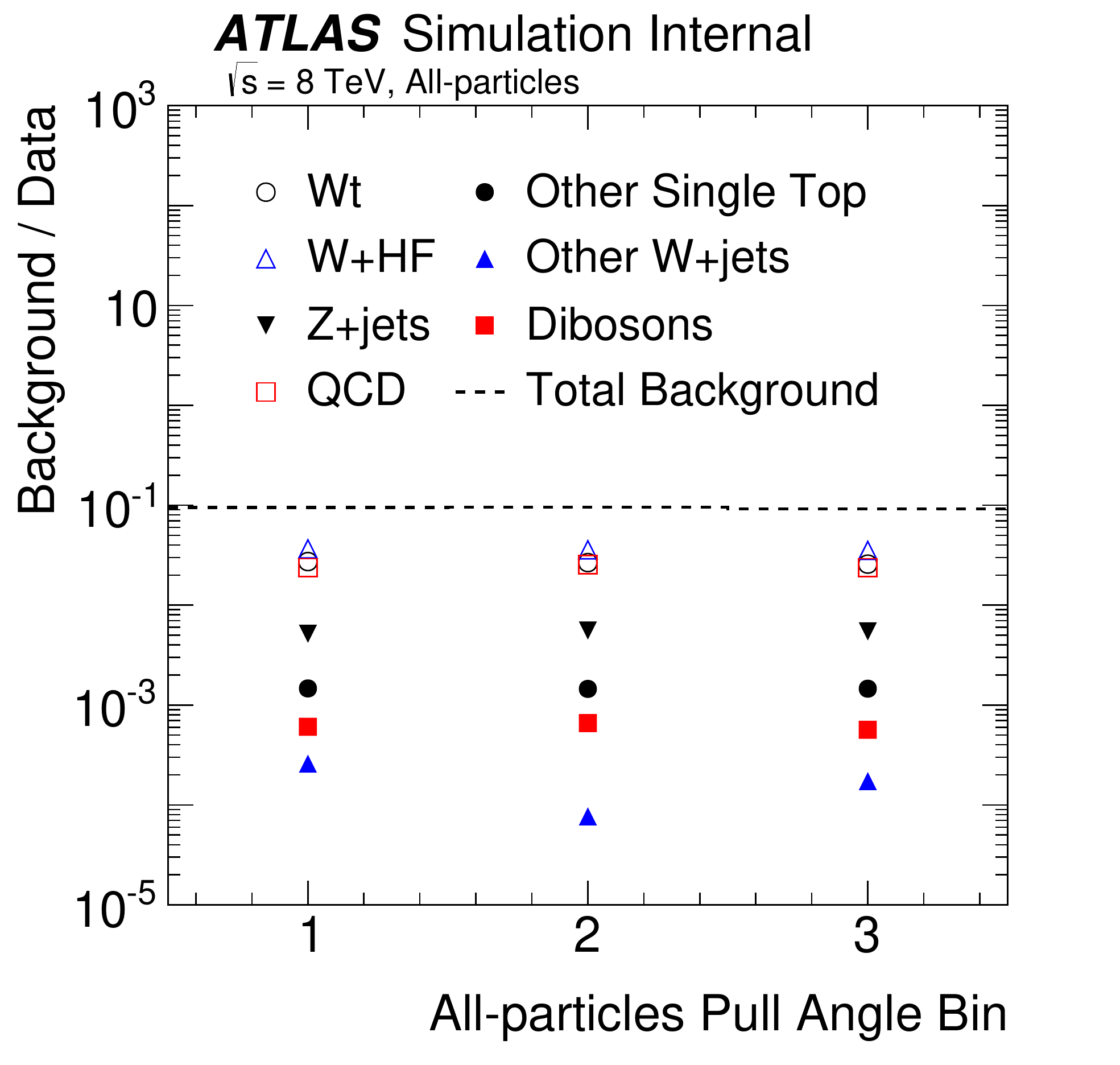}
   	\includegraphics[width=.45\linewidth]{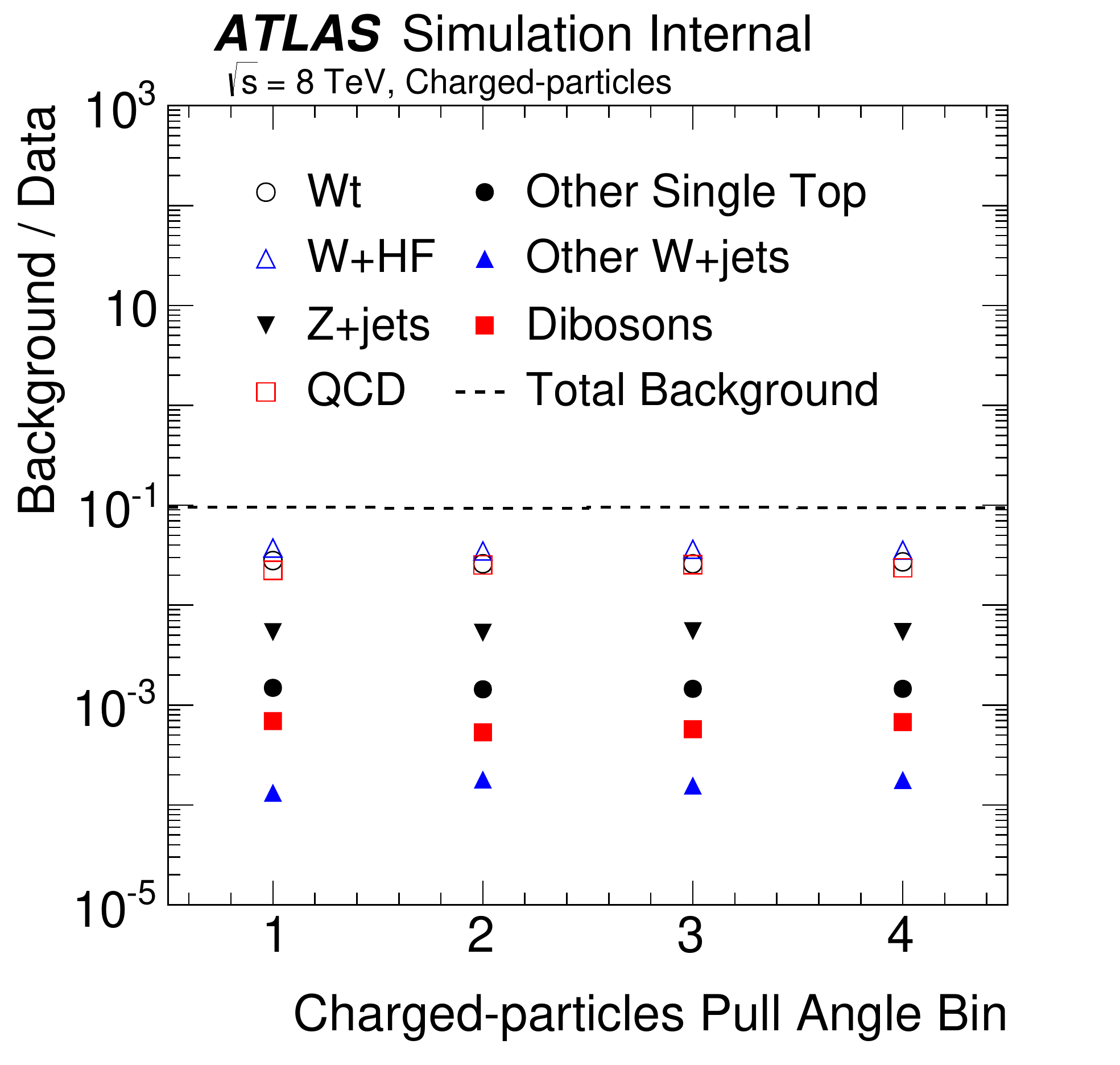}
    \caption{The background composition as a function of the all-particles pull angle (left) and the all-particles pull angle (right).}
    \label{fig:ColorFlowBackgroundComp}
  \end{center}
\end{figure}

One background that requires careful consideration is the single top $Wt$ process.  These events have a hadronically decaying $W$ boson and the pull angle distribution would change depending on the color charge of the $W$ boson.  The nominal procedure is to subtract this component as if it were {\it background}, even though it is expected to behave as the {\it signal}.  To assess the impact of this choice, single top $Wt$ events were replaced by the nominal $t\bar{t}$ events, but scaled to the single top normalization.  This ensemble was compared with an analogous one in which the $Wt$ contribution is replaced with the color octet version of the nominal $t\bar{t}$ sample.  The difference in the unfolded result between these two setups across all bins is much less than the statistical uncertainty and therefore is is ignored for the rest of the analysis (see Sec.~\ref{syst:back:colorflow} for more detail).

After subtracting the non-$t\bar{t}$ backgrounds, the data are corrected to account for events which may pass the detector-level selection but not the particle-level selection.  Unlike for the jet charge measurement, the color flow measurement has a non-trivial event selection with requirements on many reconstructed objects.  Due to the resolution and (in)efficiencies of these objects, there are a large fraction of events that pass one of the particle-level and detector-level selections, but not both.  Figure~\ref{fig:ColorFlowFakeFactor} shows how the ratio of the number of events passing both the particle-level and detector-level event selections to the number of events passing only the detector-level event selection ({\it fake factor} - see Sec.~\ref{corrrfactors}) depends on the pull angle.  The fake factor is about 70\% and is largely independent of the pull angle.  For the same reason as for the background subtraction, this small dependence on the pull angle means that the impact of mis-modeling in the fake factor is suppressed.  

\begin{figure}[h!]
\begin{center}
\includegraphics[width=0.45\textwidth]{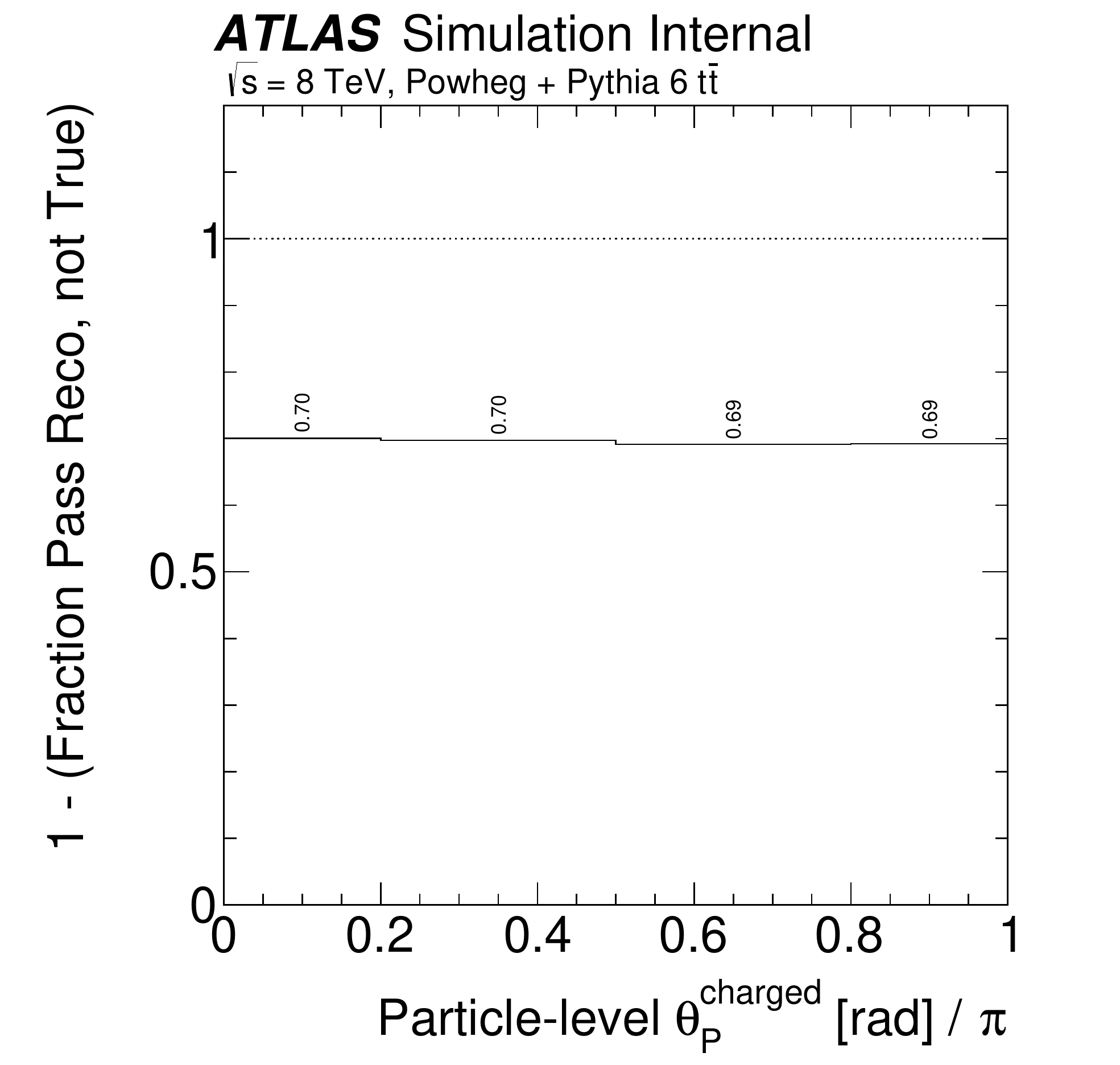}\includegraphics[width=0.45\textwidth]{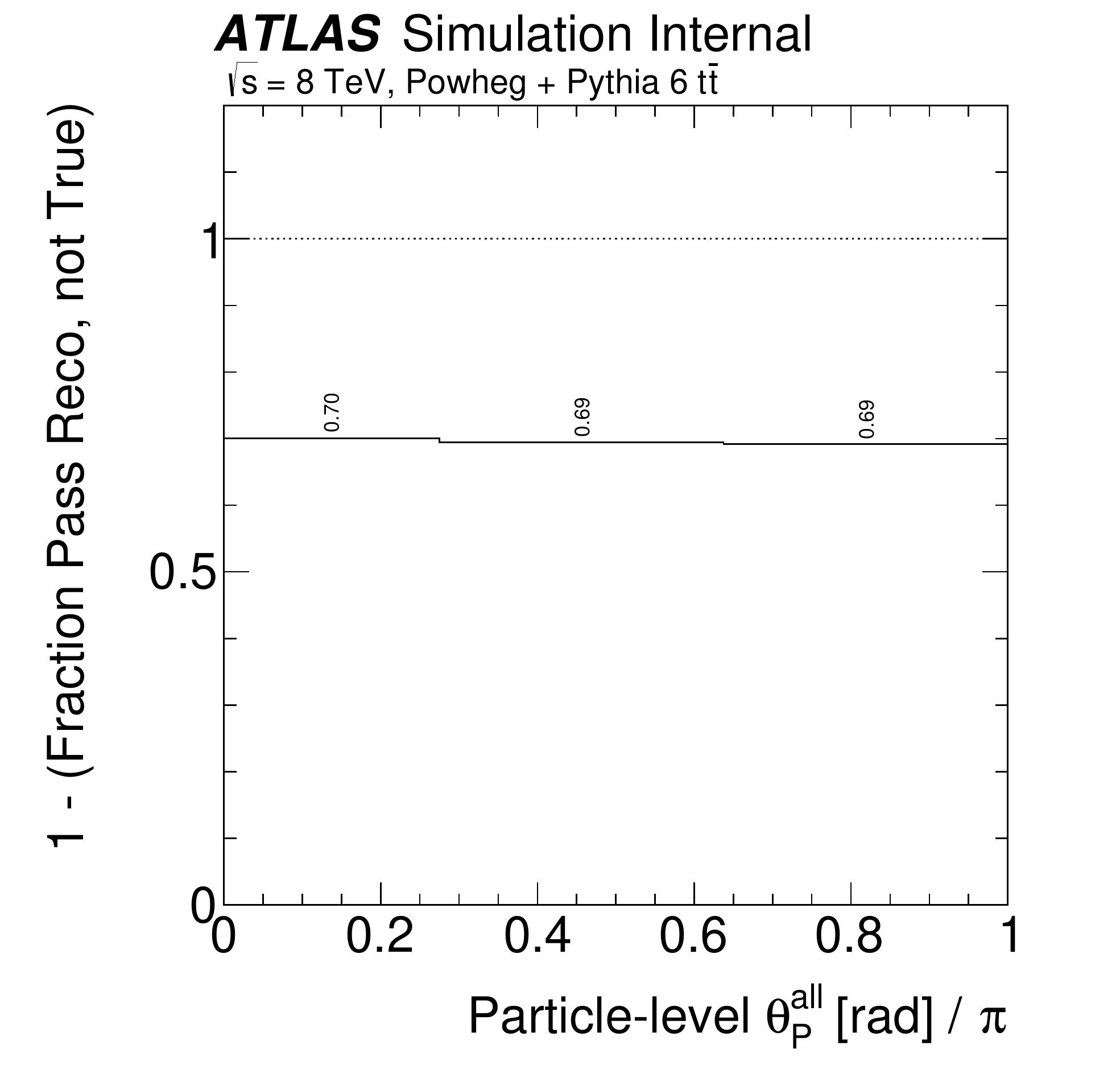}
 \caption{The pull-angle dependence of the fake factors for the all-particles pull angle (left) and for the charged-particles pull angle (right).}
 \label{fig:ColorFlowFakeFactor}
  \end{center}
\end{figure}

After the unfolding with the response matrix, discussed in the next section, {\it inefficiency factors} are applied to account for events in simulation that pass the particle-level selection but not the detector-level selection.  Figure~\ref{fig:ColorFlowInefficiencyFactor} shows the inefficiency factors as a function of the all-particles and charged-particles pull angles.  Due to falling $p_\text{T}$ spectra, the inefficiency factors are much smaller than the fake factors.  However, similar to the fake factors, the inefficiency factors are nearly independent of the pull angle and therefore they have little impact on the final measurement.  

\begin{figure}[h!]
\begin{center}
\includegraphics[width=0.45\textwidth]{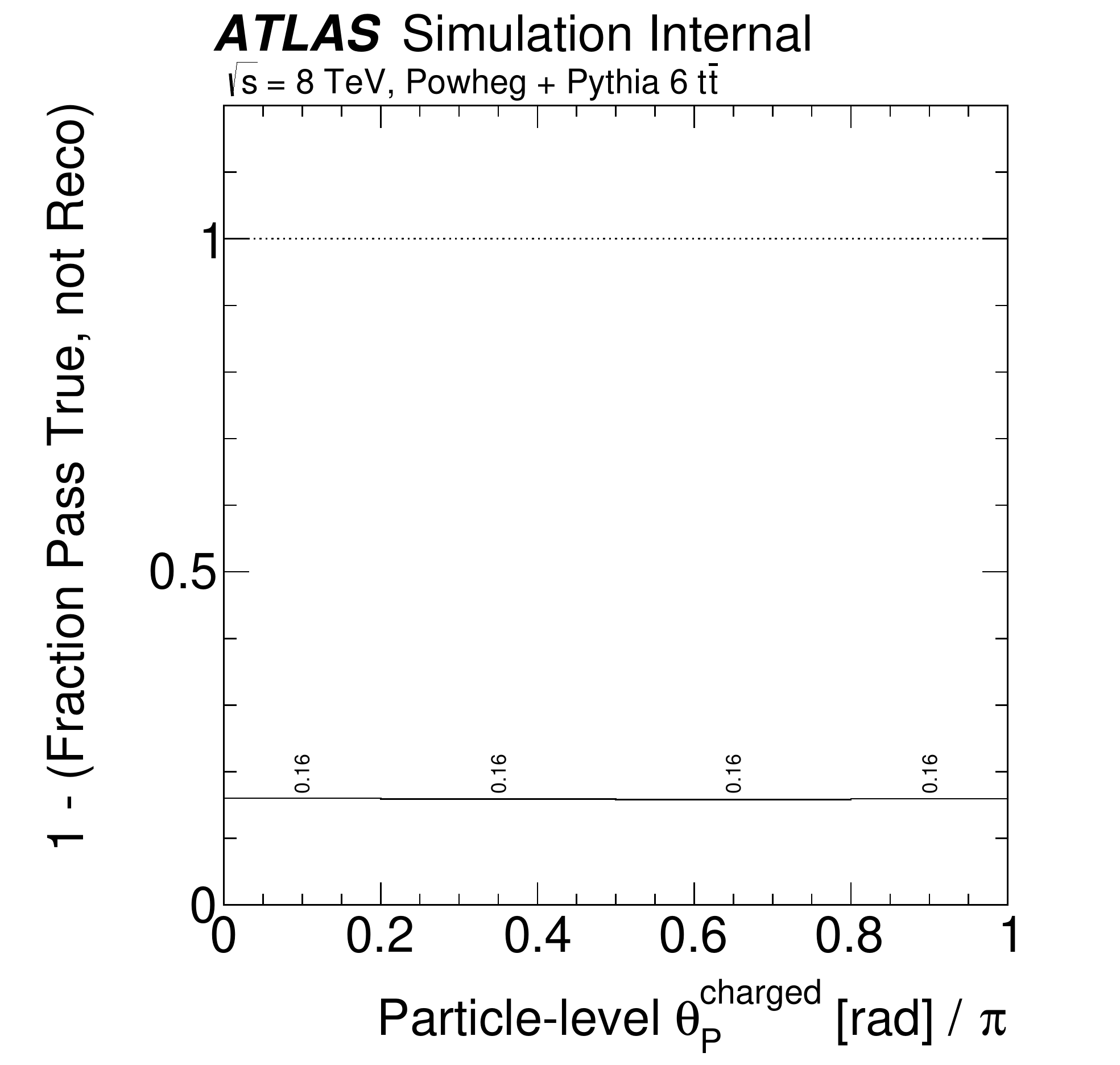}\includegraphics[width=0.45\textwidth]{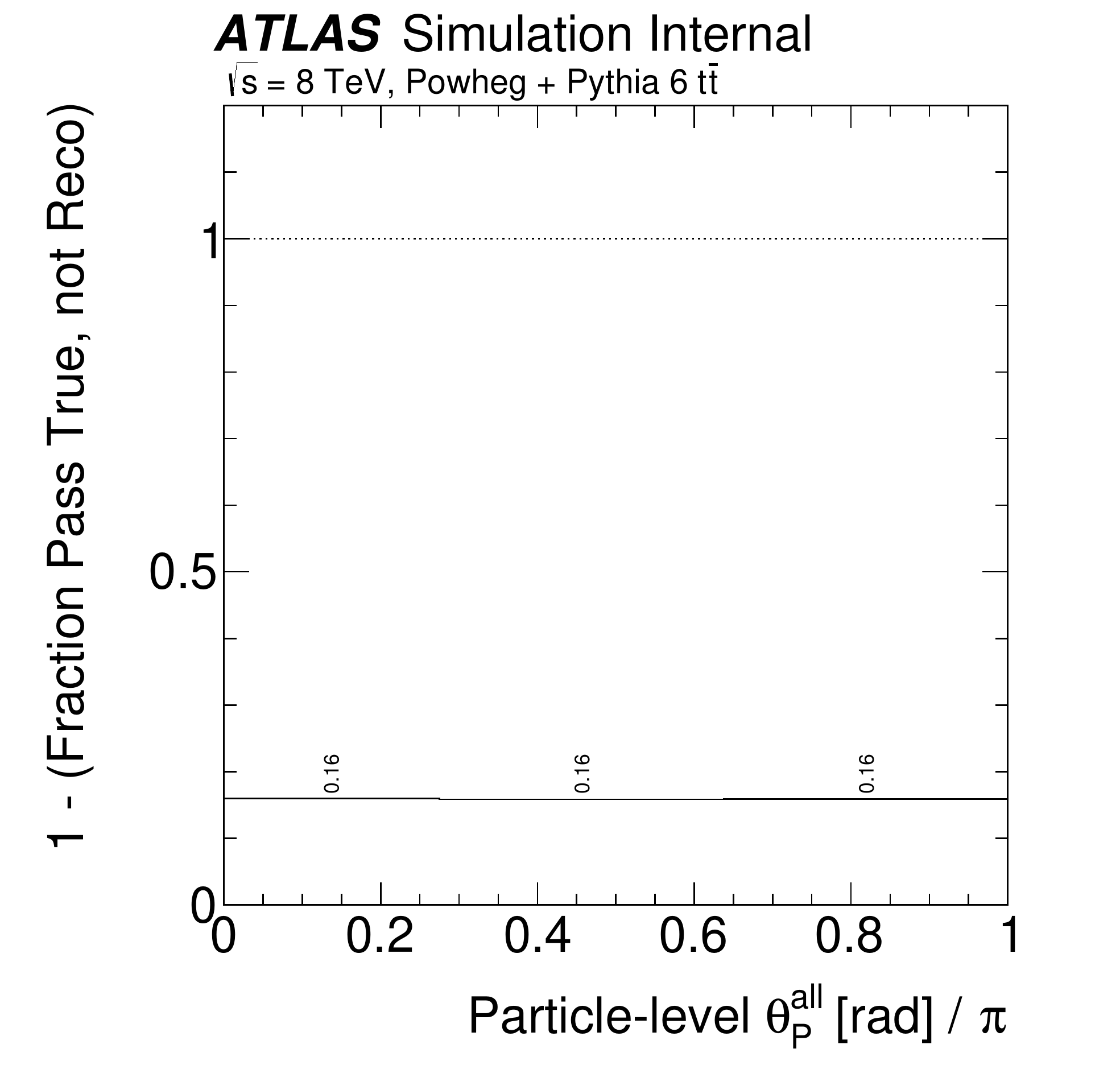}
 \caption{The pull-angle dependence of the inefficiency factors for the all-particles pull angle (left) and for the charged-particles pull angle (right).}
 \label{fig:ColorFlowInefficiencyFactor}
  \end{center}
\end{figure}

The acceptance for electron events is different than for muon events, which is reflected in the difference in fake and inefficiency factors shown in Fig.~\ref{fig:ColorFlowInefficiencyFactorLeptons}.  The fraction of events in simulation that pass the particle-level electron channel and the detector-level muon channel selections (or vice versa) is less than $5\times 10^{-3}\%$ and is ignored for constructing Fig.~\ref{fig:ColorFlowInefficiencyFactorLeptons}.   The fake factors are nearly identical between the two channels while the inefficiency factor is approximately 20\% higher for the muon channel.  This is because the lepton contribution to the fake factor is due mostly to the mis-identification rate (very small) while the lepton contribution to the inefficiency factor is the particle identification efficiency.

\begin{figure}[h!]
\begin{center}
\includegraphics[width=0.45\textwidth]{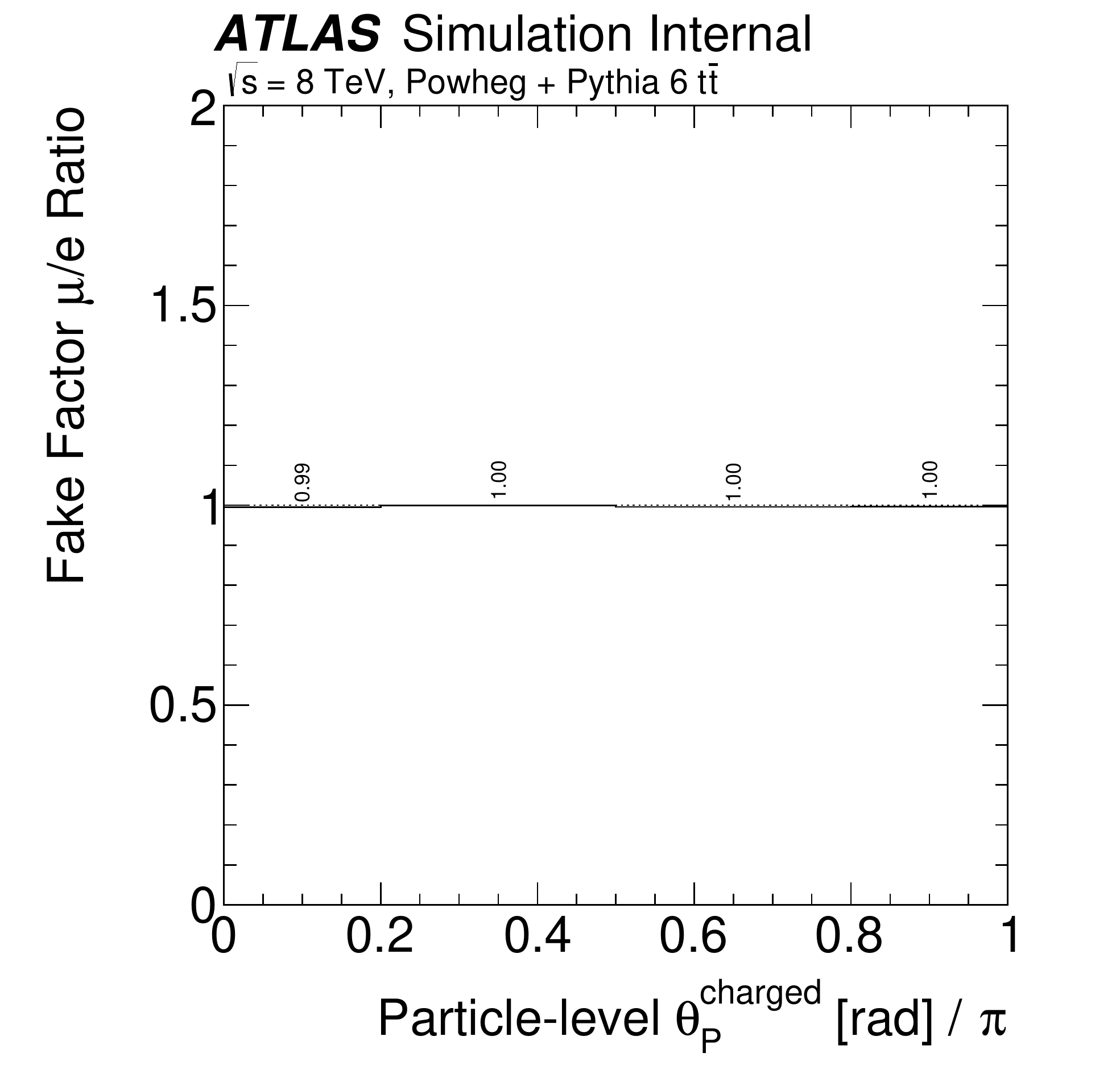}\includegraphics[width=0.45\textwidth]{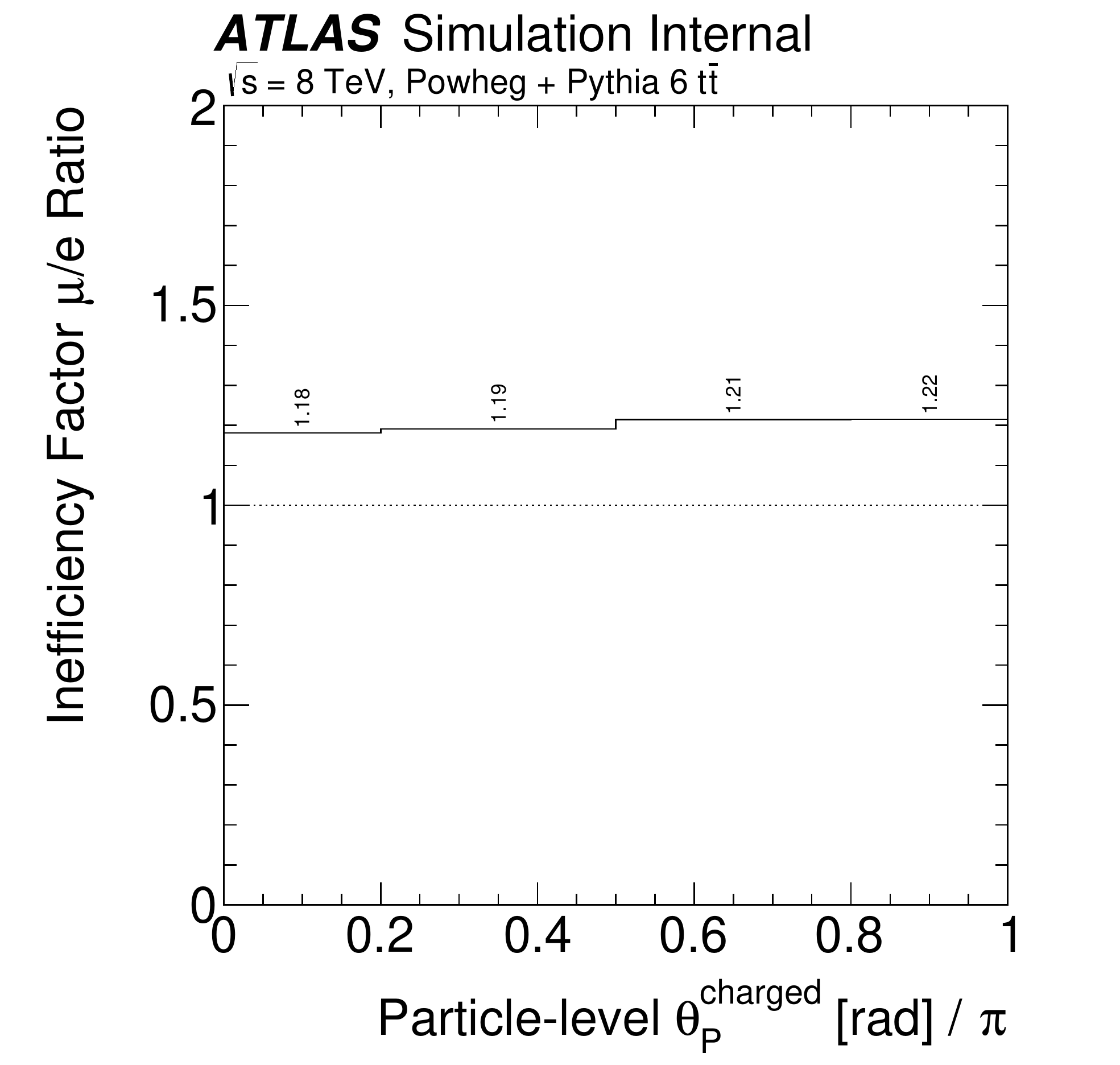}
 \caption{The ratio of the fake (inefficiency) factor for muon events to electron events on the left (right).}
 \label{fig:ColorFlowInefficiencyFactorLeptons}
  \end{center}
\end{figure}

\newpage

\subsection{Response Matrix}

Figure~\ref{fig:colorflowresposematrix} shows the nominal response matrix, constructed from {\sc Powheg-Box}+{\sc Pythia}~6.  Despite the larger bin size, the diagonal entries for the all-particles pull angle are lower than the diagonal entries for the all-particles pull angle.  The binning is chosen roughly so that the diagonal entries are $\gtrsim 50\%$.  Due to the broad resolution, the migration probabilities are significant; as discussed in Sec.~\ref{sec:ColorFlow:UnfoldingParams}, this will have important implications for the theoretical modeling uncertainties described in Sec.~\ref{sec:topquarkmodeling}.

\begin{figure}[h!]
\begin{center}
\includegraphics[width=0.45\textwidth]{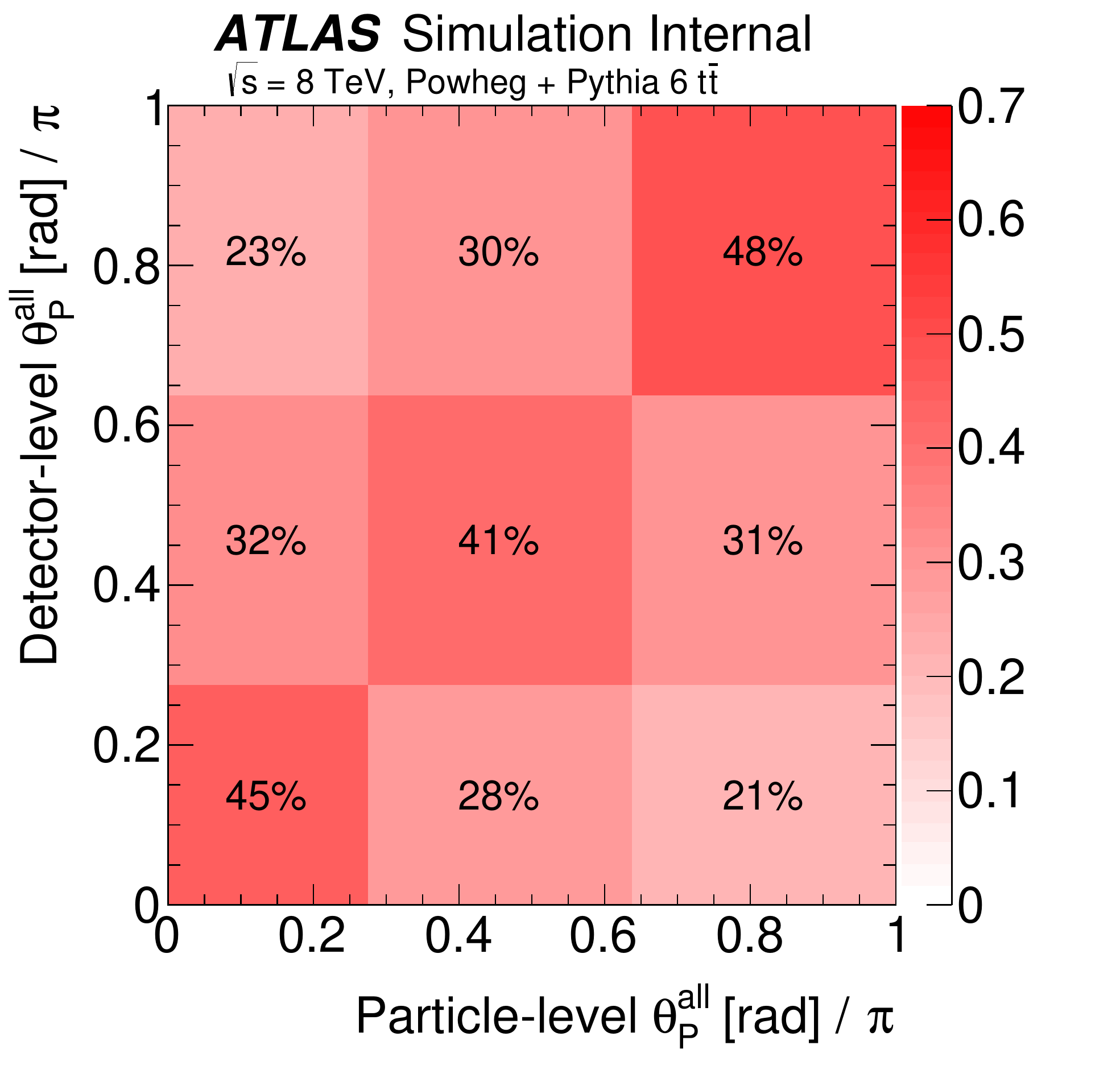}
\includegraphics[width=0.45\textwidth]{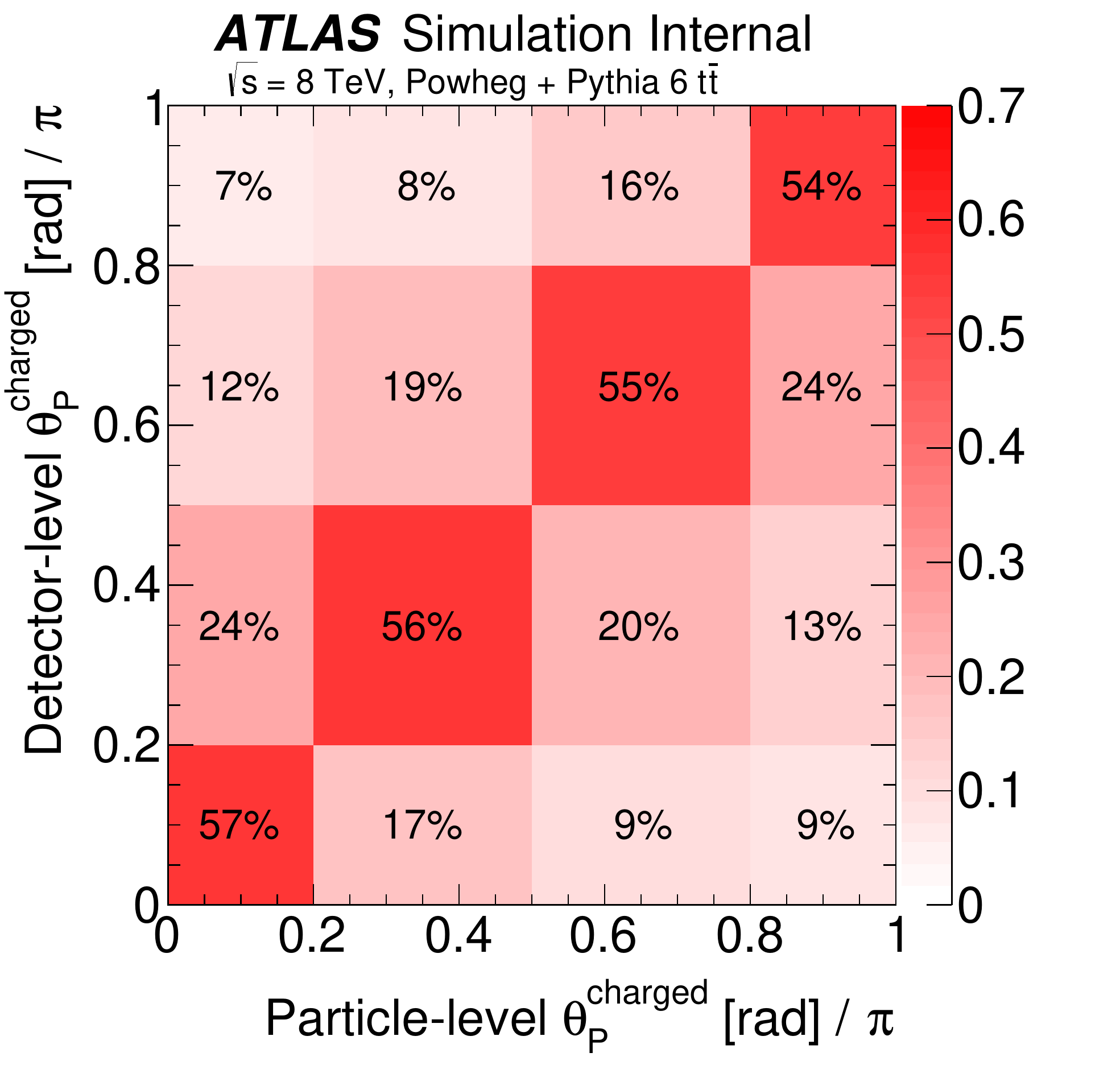} 
\caption{The response matrix for the all-particles pull angle on the left and for the charged-particles pull angle on the right.  The $z$-axis is the probability for an event to be reconstructed in a detector-level bin given that it started in a fixed particle-level bin on the horizontal axis (i.e. the columns are normalized to unity).}
 \label{fig:colorflowresposematrix}
  \end{center}
\end{figure}

Even though the inefficiency factors are slightly different between the electron and muon channels (Sec.~\ref{colorflowcorrectionfactors}), the response matrices are nearly identical.  Figure~\ref{fig:colorflowresponsematrixdifferences} quantifies the difference in the response matrix between the two channels.  Within the simulation statistical uncertainty, they are identical ($\chi^2/\text{NDF}\approx 0.3$).  For all subsequent analysis, the two channels are pooled before unfolding with the response matrix.

\begin{figure}[h!]
\begin{center}
\includegraphics[width=0.5\textwidth]{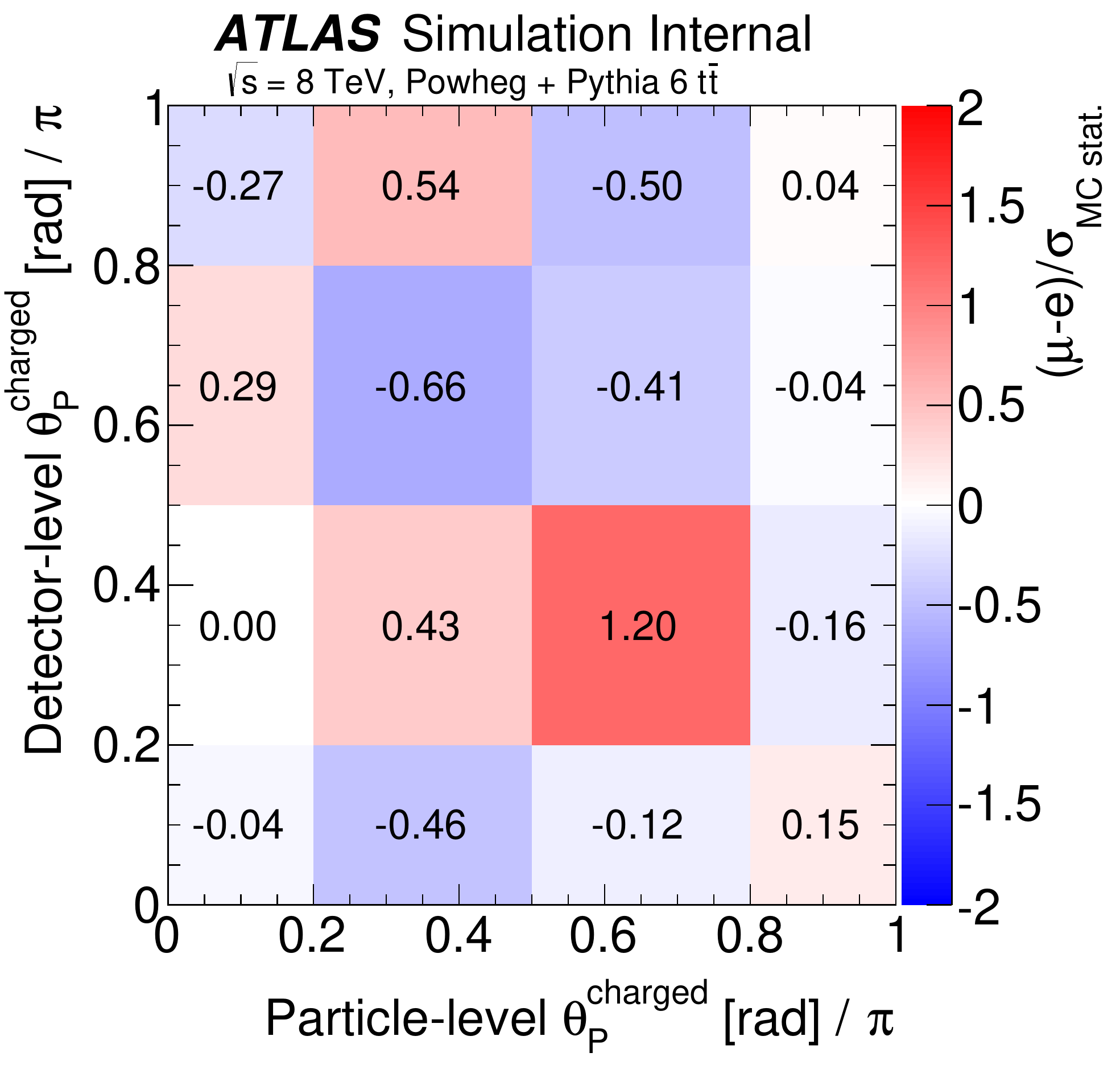}
 \caption{The difference between the response matrices for the muon and electron channels divided by the simulation statistical uncertainty.}
 \label{fig:colorflowresponsematrixdifferences}
  \end{center}
\end{figure}

\clearpage

\section{Systematic uncertainties}
\label{sec:colorflow:systematics}

The sources of uncertainty can be classified into two categories: experimental uncertainties and theoretical modelling uncertainties. In the first category, some uncertainties impact the pull angle directly and the others impact only the acceptance.  As for the jet charge, systematic uncertainties are estimated by varying an aspect of the unfolding procedure, such as the response matrix. The following sections discuss each uncertainty and Sec.~\ref{sec:ColorFlow:syst:summary} contains a summary.

\subsection{Tracking}

The method for evaluating uncertainties related to the track reconstruction are detailed in Sec.~\ref{sec:tracking}.  Unlike for the jet charge, the tracking uncertainties for the jet pull measurement are sub-dominant and so simple but conservative estimates were chosen while many of the detailed prescriptions in Sec.~\ref{sec:tracking} were under development.  For example, the tracking reconstruction efficiency systematic uncertainty is estimated without the final Run I ID material uncertainty constraint and thus tracks are randomly dropped with larger probabilities than are used for the jet charge measurement~\cite{Aad:2010ac}.  The probability in the region $2.3<|\eta|<2.5$ is $7\%$, $1.9<|\eta|<2.3$ corresponds to $4\%$, $1.3<|\eta|<1.9$ is $3\%$, and $0.<|\eta|<1.3$ is $2\%$.  These uncertainties do not explicitly take into account the modeling of the efficiency of the explicit track $\chi^2/\text{NDF}<3$ requirement.  However, the impact of any mis-modeling is subdominant to the already large uncertainties (see Sec.~\ref{sec:jetcharge:tracksyst:isoeffic}) and a comparison of simulation with data of the $\chi^2/\text{NDF}$ distribution in Fig.~\ref{fig:ColorFlow:chisq} confirms that there is no significant mis-modeling.

Most of the jets have $p_\text{T}<400$ GeV where the impact of hit merging is insignificant.  Conservatively, $\sim$50\% of the loss (see Sec.~\ref{sec:jetcharge:tracksyst:TIDE}) is used to determine the rate of dropping tracks based on the jet $p_\text{T}$ for estimating the uncertainty for reconstructing tracks inside high $p_\text{T}$ jets (distinct from the inclusive efficiency described above).  Between 400 and 500 GeV, 0.08\% of tracks are randomly removed, for jets between 500 and 600 GeV,  0.8\% are removed, between 600 and 800 GeV 1.9\% are removed and 3.7\% are removed for $p_\text{T}>800$ GeV.  The impact of a mis-modeling in the track $p_\text{T}$ resolution is conservatively estimated based on early Run I $Z\rightarrow\mu\mu$ studies by smearing track momenta randomly by 10\%~\cite{Aad:2010ac}.  The tracking uncertainties only impact the charged-particle pull angle measurement.  Table~\ref{tab:systs} quantifies the impact of the tracking uncertainties on the measured pull angle distribution.  In all bins, the tracking uncertainties are significantly smaller than the data statistical uncertainty.

\begin{figure}[h!]
\begin{center}
\includegraphics[width=0.45\textwidth]{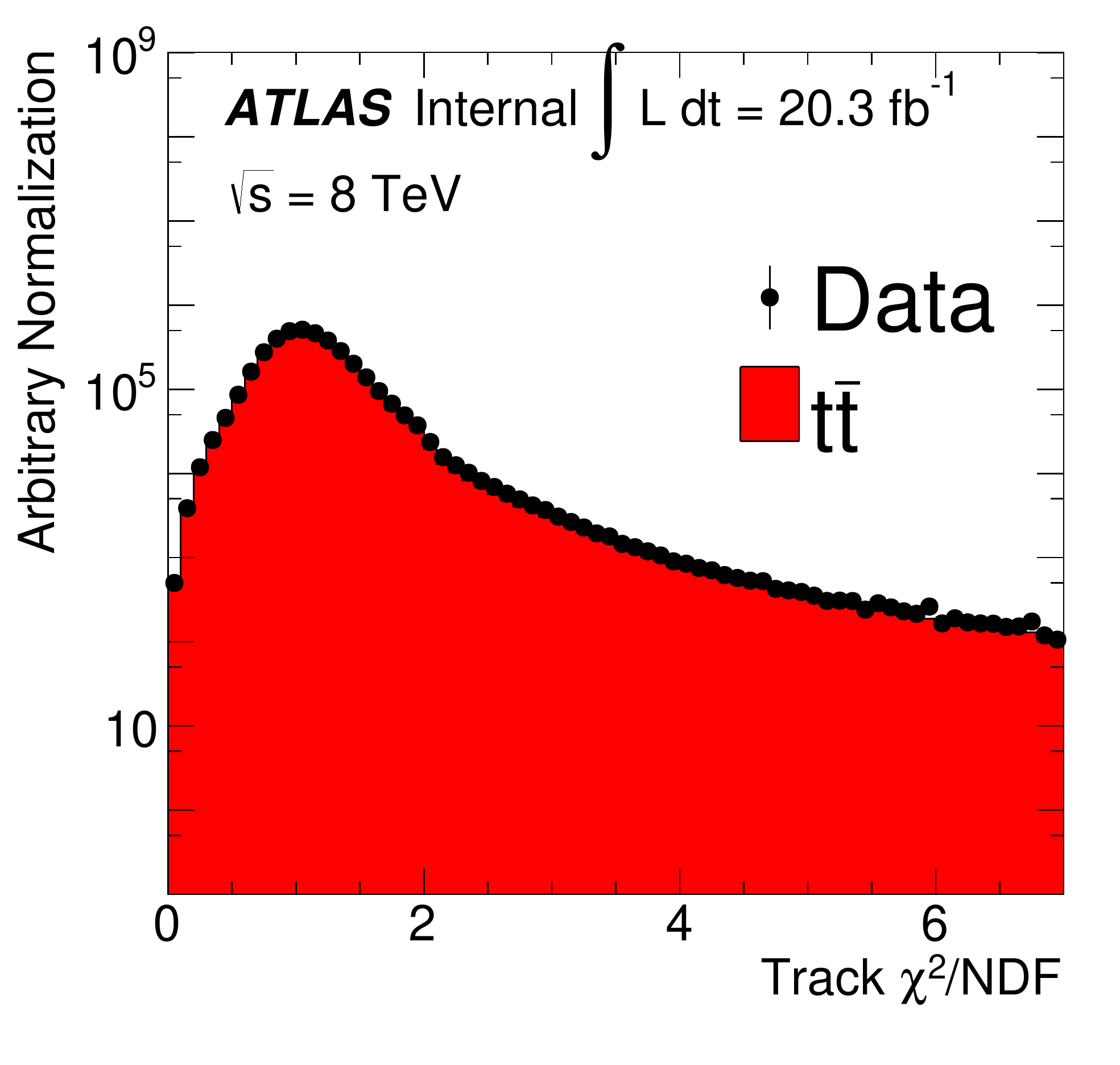}
\end{center}
\caption{The $\chi^2/\text{NDF}$ distribution of tracks before the $\chi^2/\text{NDF}<3$ requirement.}
\label{fig:ColorFlow:chisq}
\end{figure}

\begin{table}[h!]
  \centering
  \vspace{3mm}
  \begin{tabular}{ccccc}
& \multicolumn{4}{c}{Charged-particle $\theta_\text{P}$ Bin} \\
& \multicolumn{4}{c}{(uncertainty in [\%])} \\

    Source              &  1& 2&3& 4 \\
	\hline

    Isolated Efficiency             & 0.17                & 0.14    & 0.05 & 0.11                 \\
    Tracking Inside Jets                & 0.02                & 0.01      & $<$0.01 & 0.03               \\
    Momentum Resolution            & 0.06                & 0.02           & 0.06     & 0.03              \\
	\hline
    Total tracking uncertainty    & 0.18                  & 0.14        & 0.08    & 0.12                \\
    Statistical uncertainty        & 0.68                & 0.47       & 0.48 & 0.74             \\

  \end{tabular}
  \caption{A summary of the tracking systematic uncertainty and their impact on the charged-particle pull angle measurement.  Values are given in percent.  For comparison, the data statistical uncertainty is the last line.}
  \label{tab:systs}
\end{table}

\clearpage

\subsection{Calorimeter Cell Clusters}
\label{sec:ColorFlowclusteruncerts}

Uncertainties on the reconstruction of calorimeter cell clusters are estimated using comparisons between tracks and clusters in data and in simulation.  Earlier versions of these uncertainties based on 2011 data were used in various jet property measurements in early Run 1~\cite{daCosta:2011ni,Aad:2012meb,Aad:2011kq}.  The cluster energy scale and angular resolution uncertainties described in Sec.~\ref{sec:colorflow:ces} and~\ref{sec:colorflow:car}, respectively, are derived for the first time based on the 2012 dataset.  Table~\ref{tab:systs} quantifies the impact of the cluster uncertainties on the measured pull angle distribution.  

\begin{table}[h!]
  \centering
  \vspace{3mm}
  \begin{tabular}{cccc}
& \multicolumn{3}{c}{All-particle $\theta_\text{P}$ Bin} \\
& \multicolumn{3}{c}{(uncertainty in [\%])} \\

    Source              &  1& 2&3 \\
	\hline

    Reconstruction Efficiency             & 0.34                & 0.05    & 0.28               \\
    Energy Scale (Option 1)             & ${}^{+0.28}_{-0.22}$                & ${}^{+0.02}_{-0.66}$      & ${}^{+0.26}_{-0.50}$               \\
    Energy Scale (Option 2)             & 0.04                & 0.18     & 0.24               \\
    Angular Resolution            & 0.28                & 0.05           & 0.34                \\
	\hline
    Total cluster uncertainty    & 0.52                  & 0.66        & 0.67                   \\
    Statistical uncertainty        & 1.14                & 0.58       & 1.19             \\

  \end{tabular}
  \caption{A summary of the cluster systematic uncertainty and their impact on the all-particle pull angle measurement.  See Sec.~\ref{sec:colorflow:ces} for an explanation of the two options for the cluster energy scale uncertainty.  Values are given in percent.  For comparison, the data statistical uncertainty is the last line.}
  \label{tab:systs}
\end{table}

\noindent The calorimeter cell cluster uncertainties described in this section do not fully take into account collective effects on the jet pull angle.  In analogy to the jet energy energy scale uncertainty, it is possible that uncertainties on the jet pull angle from all of the input cluster measurements treated simultaneously may be different than the individual cluster-level approach given in this section.  Developing general {\it bottom-up } cluster-based uncertainties for general jet substructure moments is an area of active research.  Some studies addressing isolation and collective effects are addressed in Sec.~\ref{sec:colorflow:ces} for the cluster energy scale.

\clearpage

\subsubsection{Cluster Reconstruction Efficiency}

Due to the material in and around the ID before the calorimeter, particles may have significant material interaction before reaching the calorimeter that prevent the seeding of calorimeter cell clusters.  Calorimeter cell clusters require seed cells that exceed the noise threshold - if a particle interacts with the material and produces many spread out low energy secondary particles, there may not be sufficient localized energy to seed a cluster.   The rate at which particles do not seed a cluster is studied with tracks that do not match a calorimeter cell cluster within $\Delta R<0.2$ in studies of single hadron response at $\sqrt{s}=900$ GeV~\cite{Aad:2012vm}.  By construction, this technique also includes in the reconstruction efficiency the rate at which particles scatter by a large angle after the ID, but this is a small effect for the choice of $\Delta R<0.2$.  Figure~\ref{syst:eoverp1} shows a comparison between the rate of unmatched tracks in data and simulation as a function of the track momentum.  The rate $P(E=0)\approx \exp(-2E/\text{GeV})$.  To conservatively estimate the uncertainty from mis-modeling the reconstruction efficiency, clusters with $E<2.5$ GeV are randomly dropped 25\% of $P(E=0)$.

\begin{figure}[h!]
\begin{center}
\includegraphics[width=0.6\textwidth]{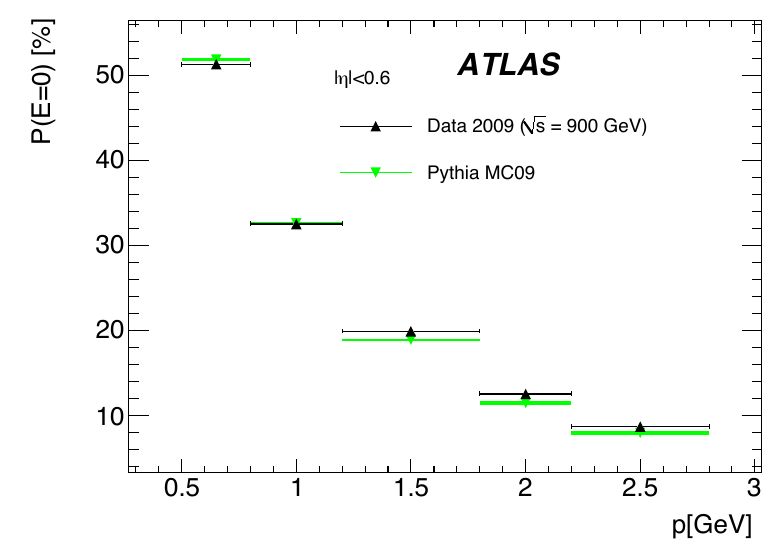}
\end{center}
\caption{The probability for not finding a calorimeter cell cluster matched to a track as a function of the track momentum.  Reproduced from Ref.~\cite{Aad:2012vm}.}
\label{syst:eoverp1}
\end{figure}

\clearpage

\subsubsection{Cluster Energy Scale}
\label{sec:colorflow:ces}

The cluster energy scale (CES) uncertainty is estimated using the $E/p$ measurement based on the 2012 dataset~\cite{ATL-PHYS-PUB-2014-002}.  Tracks are extrapolated to the various layers of the calorimeter and matched to clusters.  Differences in the ratio of the track momentum and the LCW cluster energy between data and simulation are used as an estimate of the uncertainty.  The ratio between data and MC is bounded by the following function:

\begin{align}
\label{eq:ces}
f_\pm(p|\alpha,\beta)= 1\pm \alpha \times\left(1+\frac{\beta\text{ MeV}}{p}\right),
\end{align}

\noindent where $\alpha(\eta)$ and $\beta(\eta)$ are two dimensionless $\eta$-dependent functions and $p$ is the track momentum.  Figure~\ref{syst:eoverp1} show the data and MC used to estimate $\alpha$ and $\beta$ in two bins of $\eta$ and Table~\ref{tab:systs} summarizes the values over all seven $|\eta|$ bins spanning $0<|\eta|<2.3$.

\begin{table}[h!]
  \centering
  \vspace{3mm}
  \begin{tabular}{c|ccccccc}
& \multicolumn{6}{c}{$|\eta|$ bin lower edge} \\
    Coefficient              &  0 & 0.6 & 1.1 & 1.4 & 1.5 & 1.8 & 1.9  \\
	\hline

  $\alpha$           &0.05&0.05&0.07&0.07&0.07&0.04&  0.04             \\
     $\beta$          &500&500&500&0&500&0& 500              \\
\hline
  \end{tabular}
  \caption{A summary of the $\alpha$ and $\beta$ coefficients in Eq.~\ref{eq:ces} used to bound the differences between data and simulation.}
  \label{tab:systs}
\end{table}

\noindent To estimate the impact of the CES uncertainty, the cluster energies inside the jet (after jet-finding) are scaled using the function $f$.  The $E/p$ measurement subtracts out the impact of neutral particles and so is directly applicable only to charged particle induced clusters.  However, given the conservative nature of the prescription described below and that the CES uncertainty is subdominant, the same CES uncertainty is applied to all clusters.

\clearpage

Taking into account the correlations between the CES uncertainty is non-trivial and so two approaches are used, with the more larger one retained per bin of the pull angle.  

\begin{enumerate}
\item For the `up' (`down') uncertainty, multiply the four-vector of all clusters inside the jet by $f_+(p|\alpha,\beta)$ ($f_-(p|\alpha,\beta)$).  The shift will be coherent for all clusters, but the actually scaling will change based on $p$ and $\eta$.
\item There is only one uncertainty: multiply the four-vector of each cluster by a random number with mean one and standard deviation $f_+(p|\alpha,\beta)-1$.  Generate the random numbers for this procedure in strips of $\eta$ (with bin size half that of the $E/p$ measurement bins) to allow for some coherence, but still mostly emulating local fluctuations.
\end{enumerate}

\begin{figure}[h!]
\begin{center}
\includegraphics[width=0.45\textwidth]{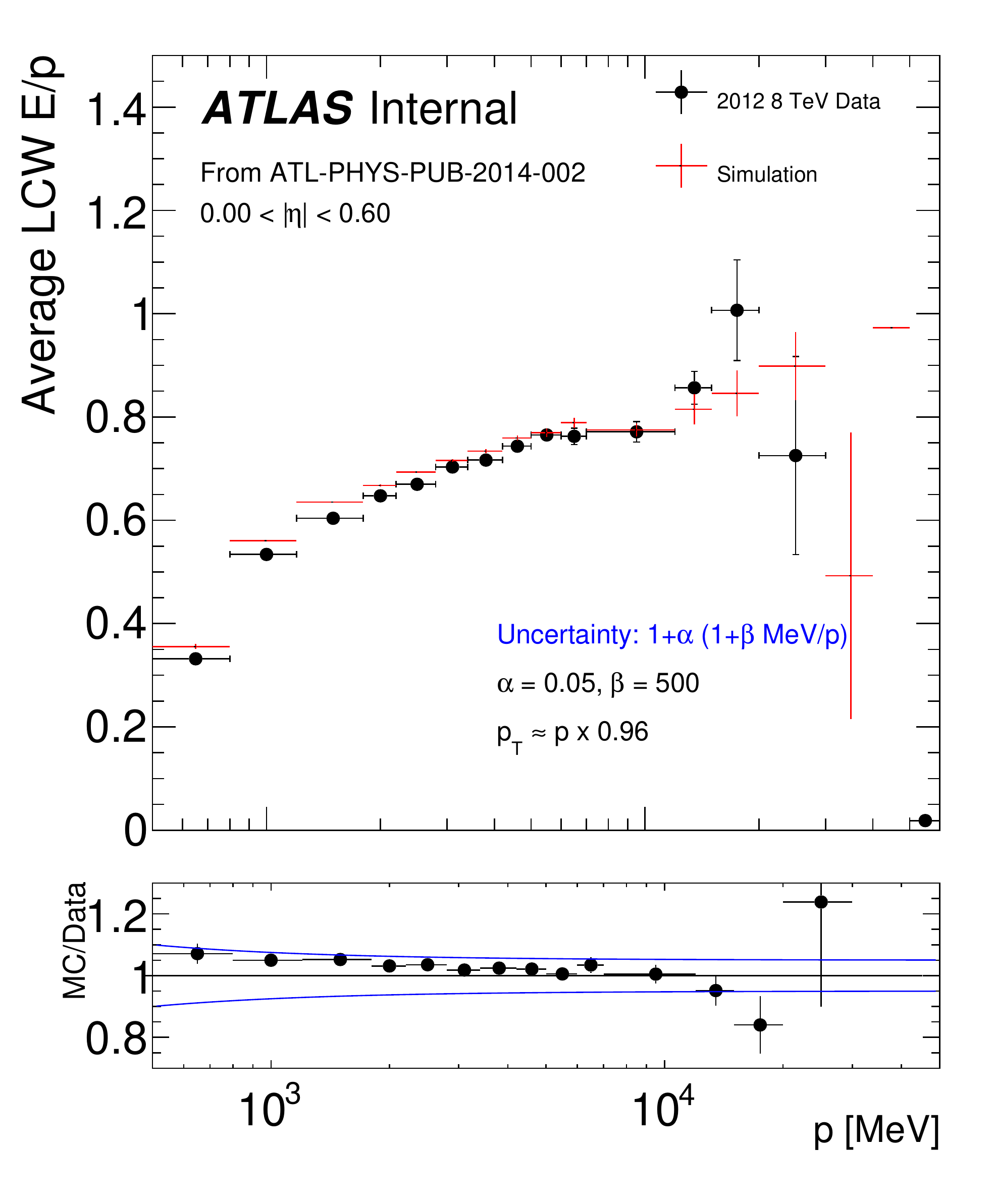}\includegraphics[width=0.45\textwidth]{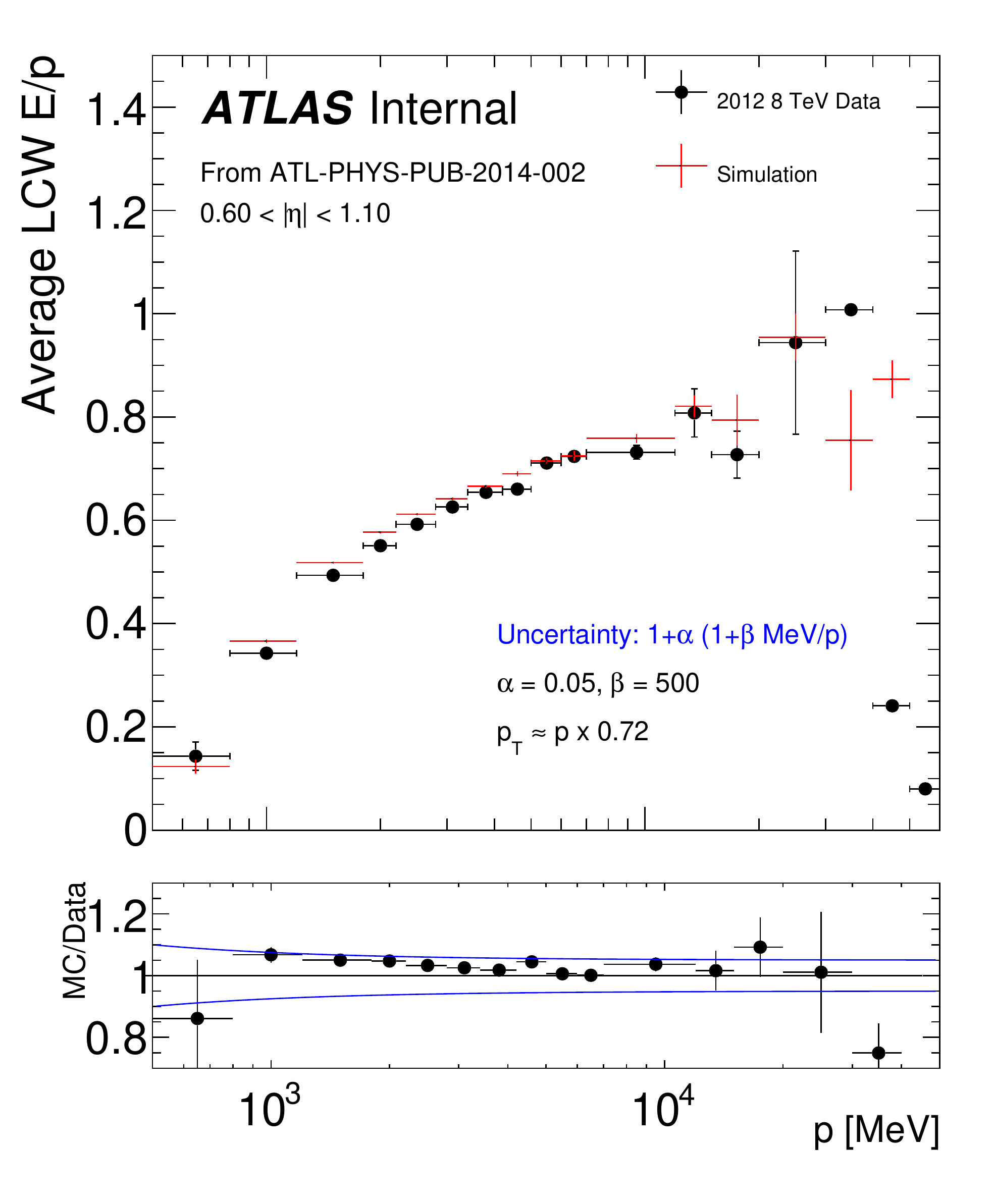}
\end{center}
\caption{The average LCW $E/p$ for $0<|\eta|<0.6$ (left) and $0.6<|\eta|<1.1$ (right), using the same analysis framework as in Ref.~\cite{ATL-PHYS-PUB-2014-002} (but LCW is not in the public note - thank you B. Axen for the inputs).  The blue band in the ratio shows the estimated uncertainty used for the cluster energy scale uncertainty.}
\label{syst:eoverp1}
\end{figure}

By construction, the clusters used in the $E/p$ measurement are isolated.  However, the clusters inside the jets used for the jet pull angle measurement can be non-isolated.   Figure~\ref{fig:iso3} shows the distribution of the cluster energy inside jets in various bins of the cluster isolation ($f_\text{iso}$ in \cite{Aad:2016upy}).   The cluster isolation measures the sampling layer energy-weighted fraction of non-clustered neighbor cells on the outer perimeter of a topocluster.  An isolation of $1$ indicates that the clusters are isolated and an isolation of $0$ indicates that the cluster is non-isolated.  There is no evidence for significant isolation-dependent energy mis-modelling.

\begin{figure}[h!]
\begin{center}
\includegraphics[width=0.4\textwidth]{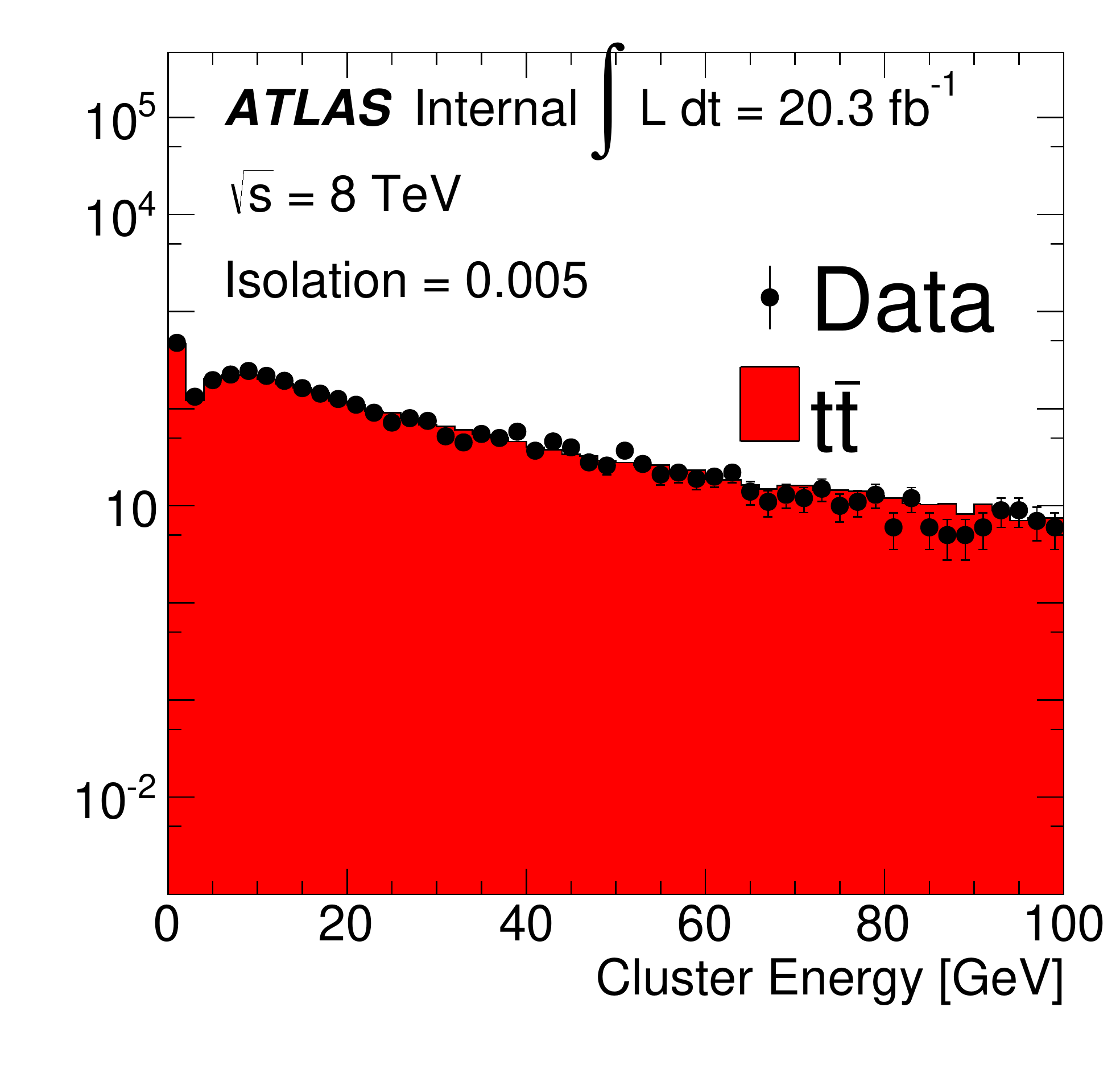}\includegraphics[width=0.4\textwidth]{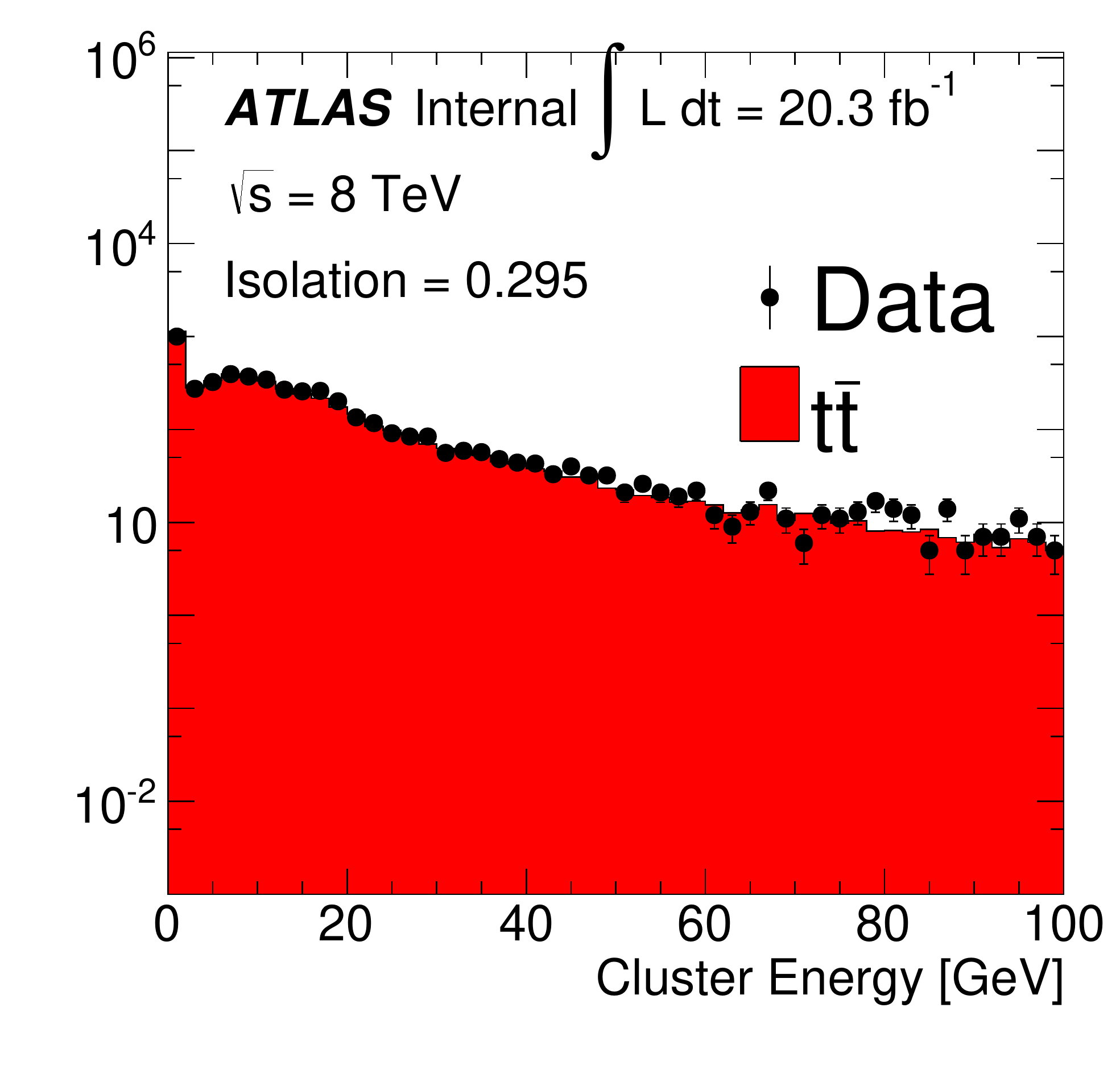}\\
\includegraphics[width=0.4\textwidth]{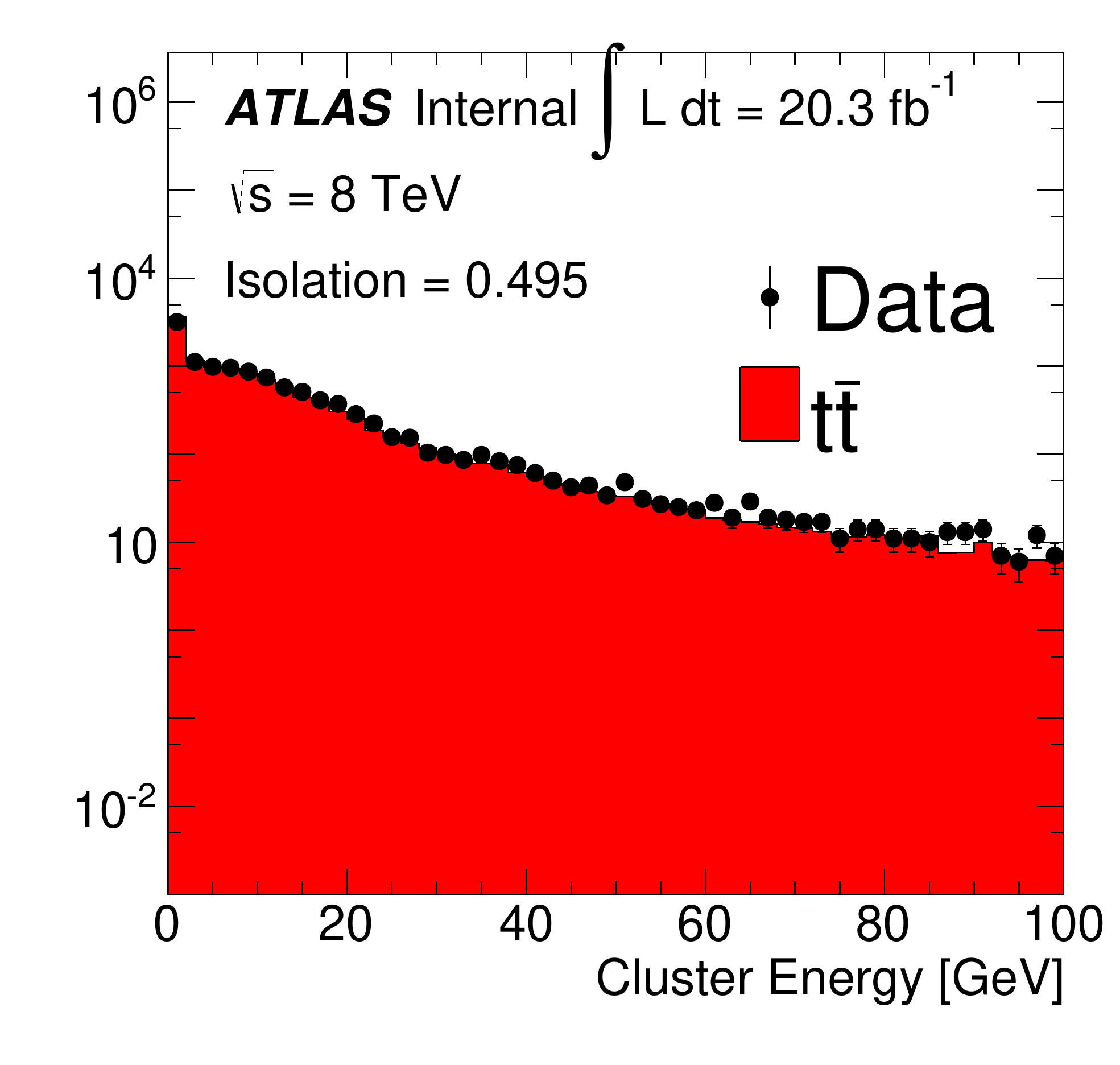}\includegraphics[width=0.4\textwidth]{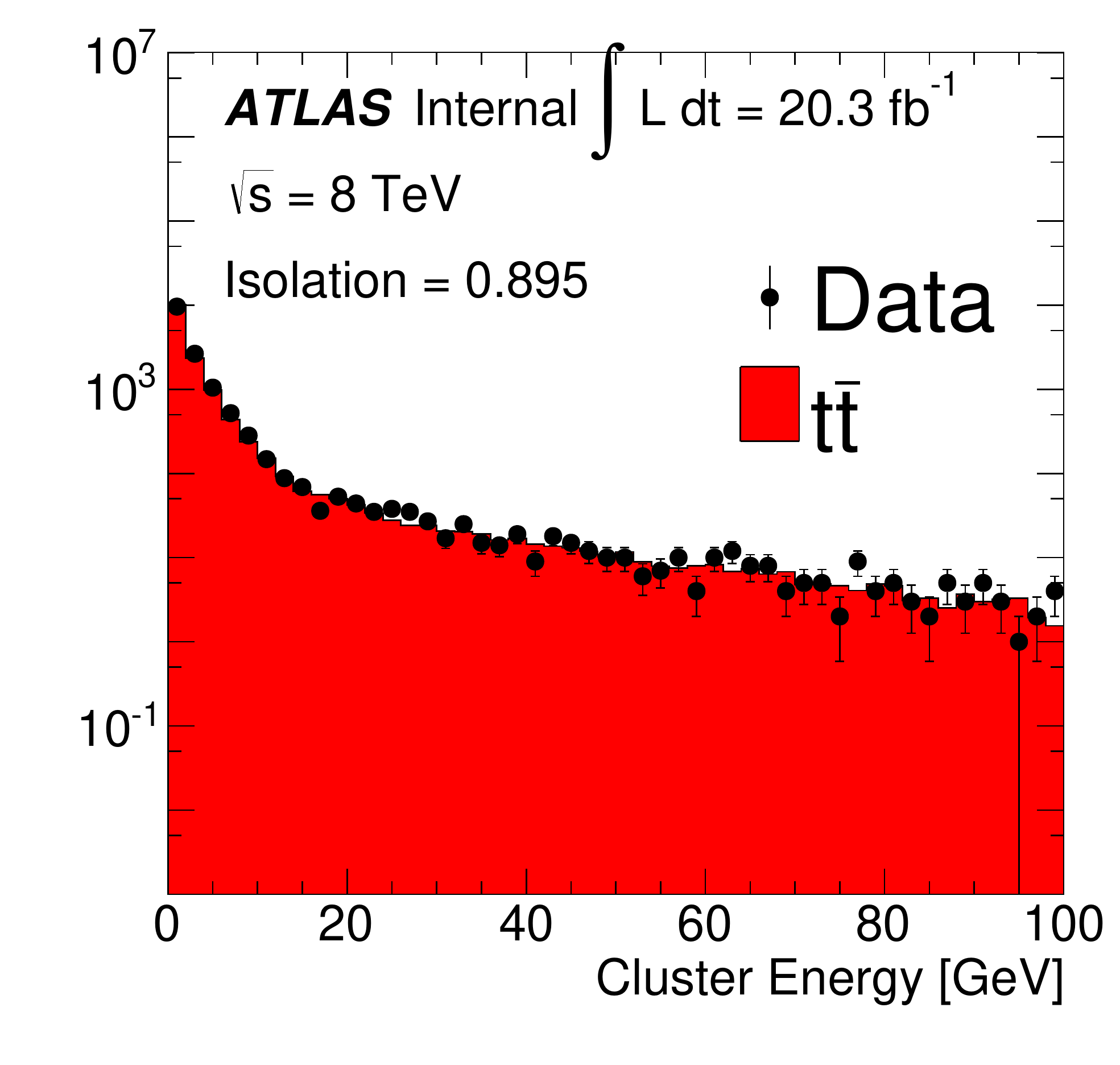}
\end{center}
\caption{The distribution of cluster energy in three bins of the cluster isolation, with less isolated clusters on the left and more isolated cluster on the right.  See the text for the definition of the cluster isolation.}
\label{fig:iso3}
\end{figure}

\clearpage

\subsubsection{Cluster Angular Resolution}
\label{sec:colorflow:car}

Early Run 1 measurements of jet properties at $\sqrt{s}=7$ TeV based on constituent clusters used a cluster angular resolution uncertainty based on differences between data and simulation in the $\Delta\phi$ and $\Delta\eta$ between isolated tracks and clusters.  Cluster positions were smeared by 5 mrad independently in $\eta$ and $\phi$ to account for potential mis-modeling~\cite{Aad:2014pua}.  Similar studies are shown in this section, based on the full $\sqrt{s}=8$ TeV dataset that is about a factor of $4$ larger than the $\sqrt{s}=7$ dataset.

A $Z\rightarrow\mu\mu$ ($p_T^Z> 30 $ GeV) event selection is chosen so that a significant fraction of clusters are isolated (no jet requirement).  Tracks are selected which have a maximum of one cluster within $\Delta R<0.15$ around their position extrapolated to the second layer of the calorimeter, excluding the muon tracks and with no cluster requirements other than $E>0$.   Fig.~\ref{syst:angle1bb} shows the distribution of $\Delta R(\text{track},\text{calo})$ for such tracks in the barrel ($|\eta|<0.6$) and Fig.~\ref{syst:angle2bb} shows the same distribution in the endcap ($2<|\eta|<2.4$).  In all plots, there are clearly two peaks.  The second peak is an artifact of the requirement that there be no additional clusters within $\Delta R<0.15$.  To study the impact of single particles, further analysis is only performed on cases in which $\Delta R(\text{track},\text{calo})<0.075$ to remove the second peak.  The momentum dependence of the $\Delta \phi$ and $\Delta\eta$ between tracks and clusters is tabulated in Fig.~\ref{syst:angle3bb} and~\ref{syst:angle4bb}.  Differences between the data and simulation are generally $\lesssim 1$~mrad.  These differences are significantly smaller than the ones reported in the $\sqrt{s}=7$ TeV analysis (by a factor of 5 in the endcap and 50 in the barrel).  One reason is the restriction to the first peak and thus effectively suppressing the contribution from neutral particles.  A version of Fig.~\ref{syst:angle3bb} including the second peak results in resolutions similar to the early Run 1 numbers.  However, due to the lack of additional studies to probe the full impact of neutral particles in clusters, a (likely) conservative 5 mrad smearing is also adapted at $\sqrt{s}=8$~TeV.

\begin{figure}[h!]
\begin{center}
\includegraphics[width=0.42\textwidth]{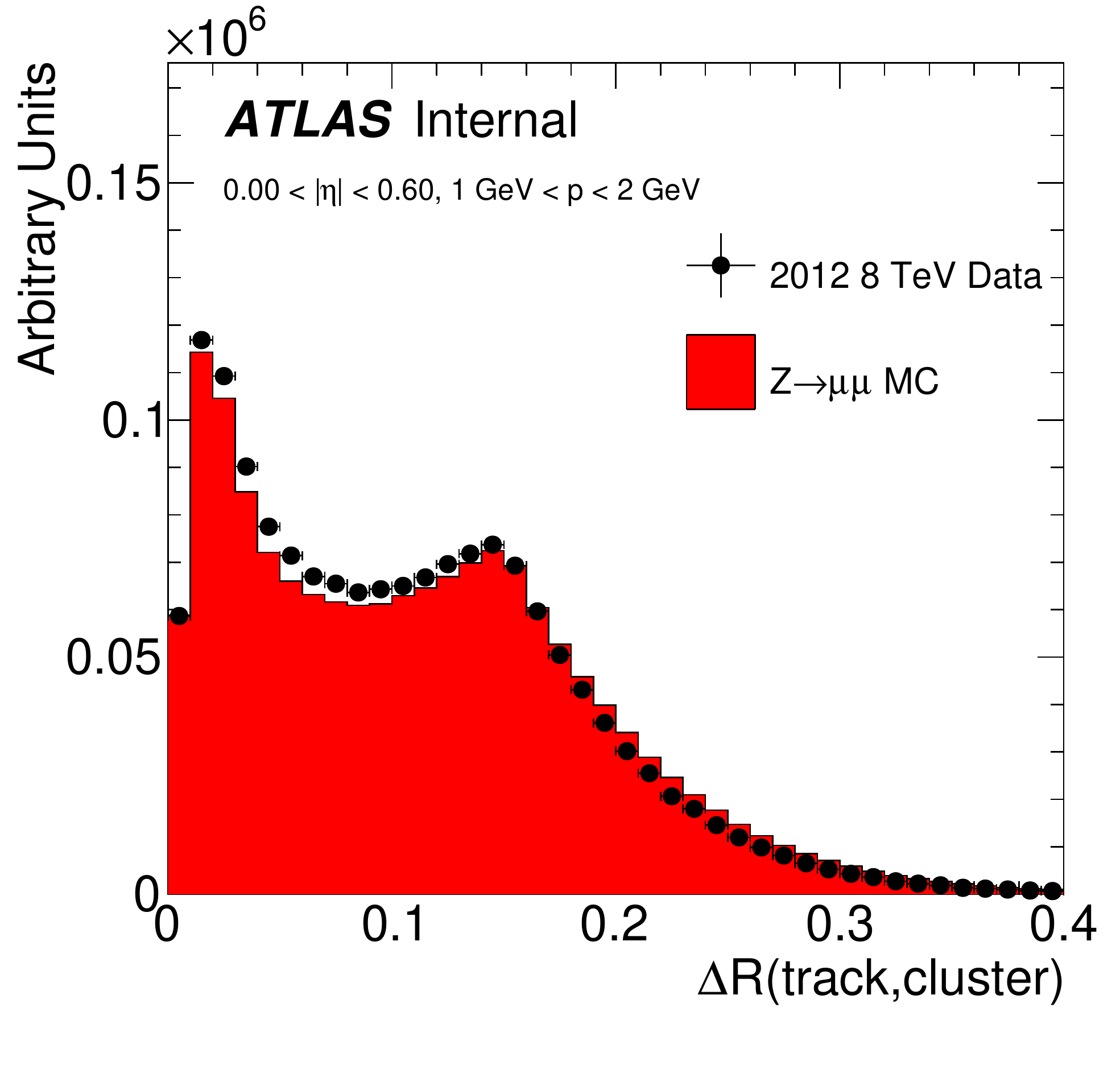}
\includegraphics[width=0.42\textwidth]{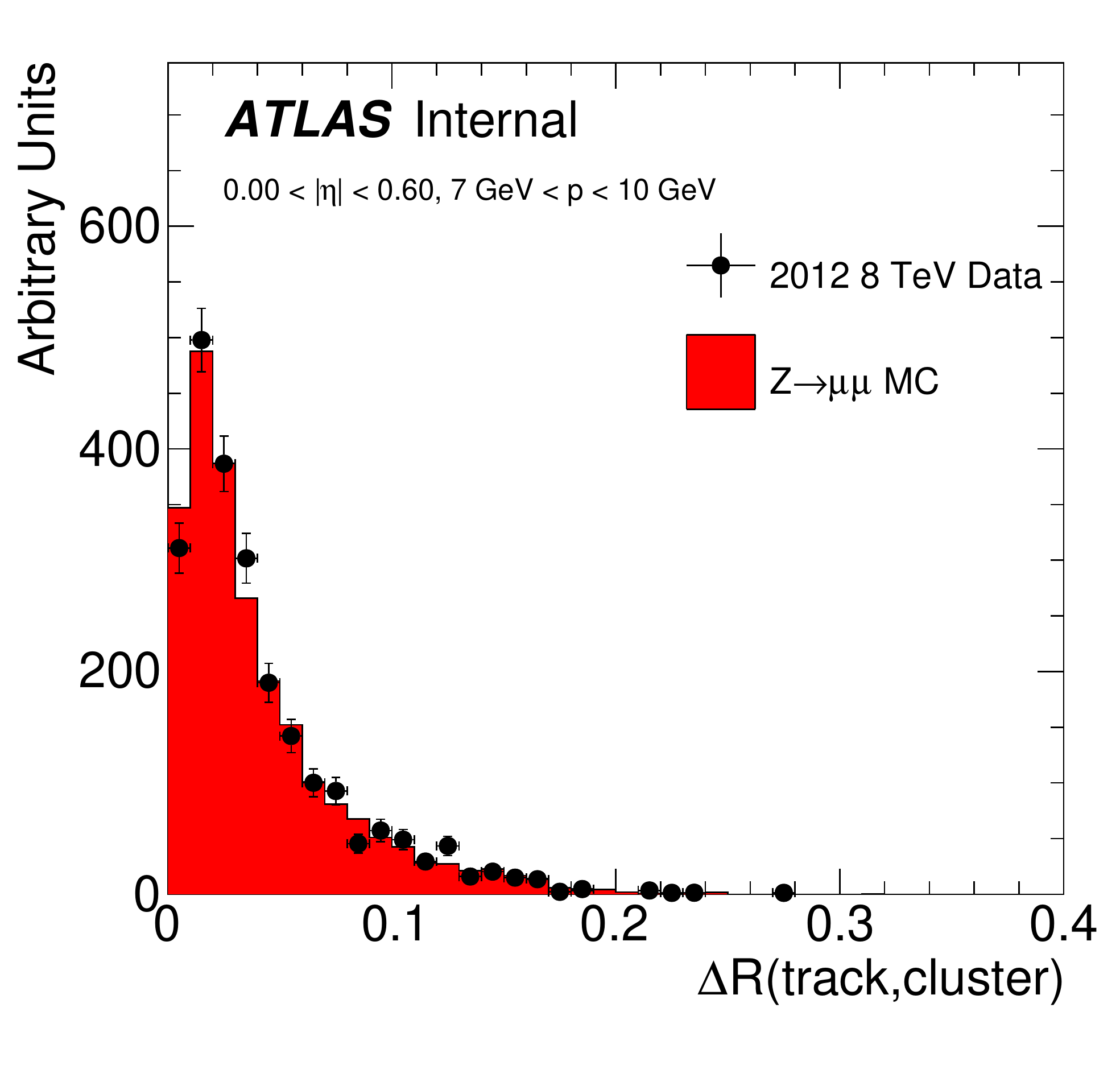}
\end{center}
\caption{Left (Right): The $\Delta R$ between isolated low (high) momentum tracks and clusters in $Z\rightarrow\mu\mu$ events in the barrel of the detector.  Inputs from C. Young.}
\label{syst:angle1bb}
\end{figure}

\begin{figure}[h!]
\begin{center}
\includegraphics[width=0.42\textwidth]{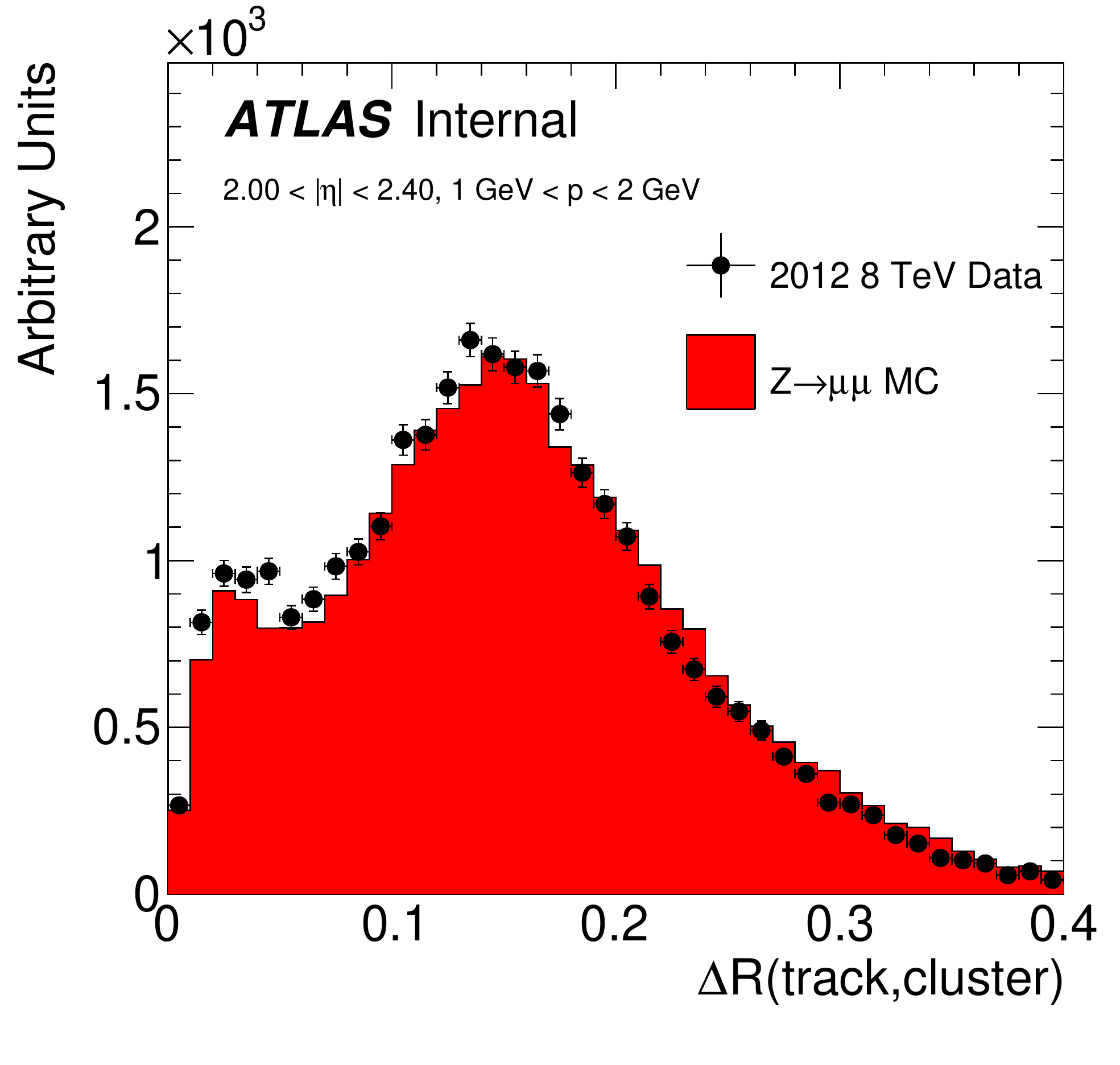}\includegraphics[width=0.42\textwidth]{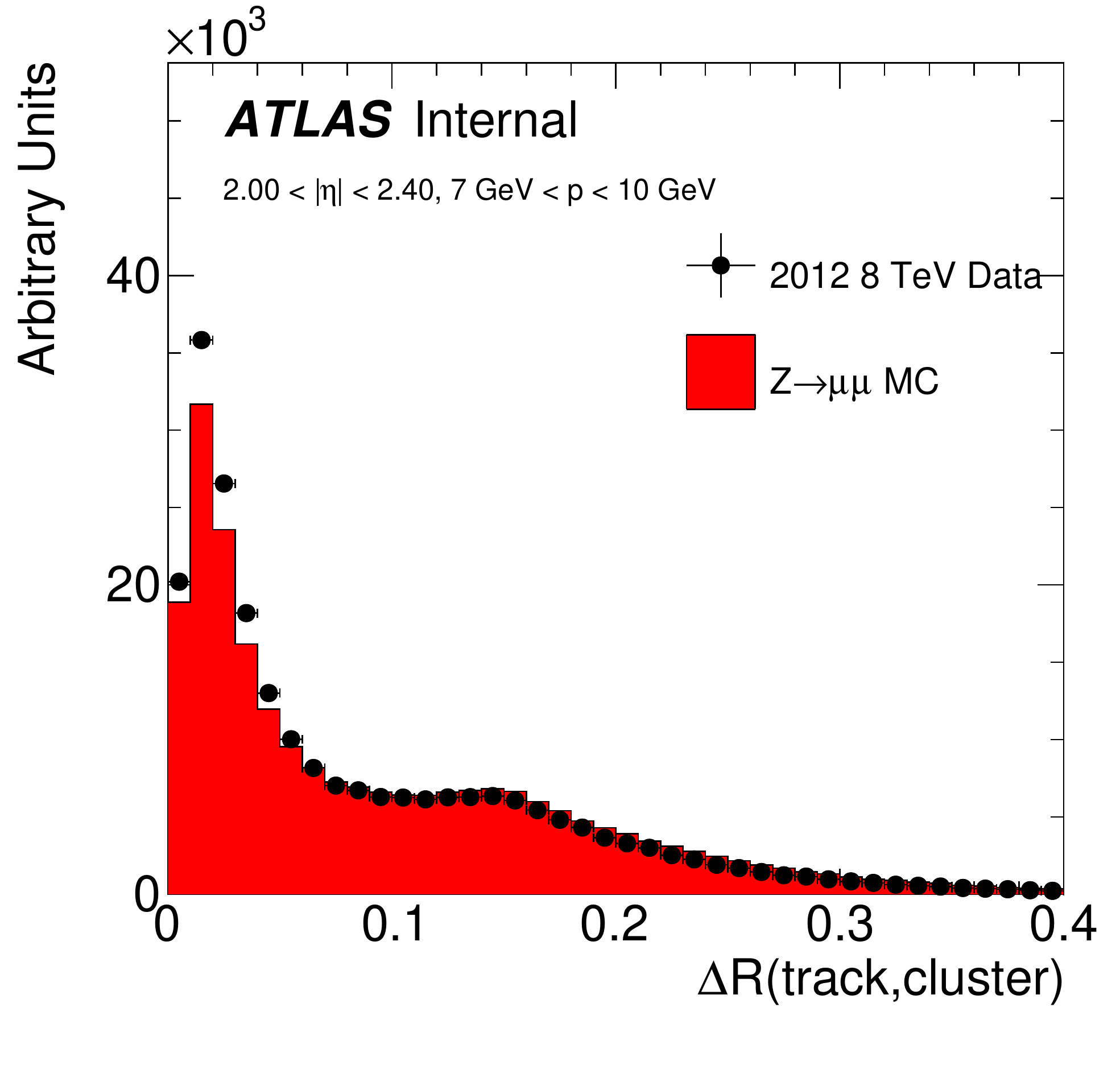}
\end{center}
\caption{Left (Right): The $\Delta R$ between isolated low (high) momentum tracks and clusters in $Z\rightarrow\mu\mu$ events in the endcap of the detector.  Inputs from C. Young.}
\label{syst:angle2bb}
\end{figure}

\begin{figure}[h!]
\begin{center}
\includegraphics[width=0.4\textwidth]{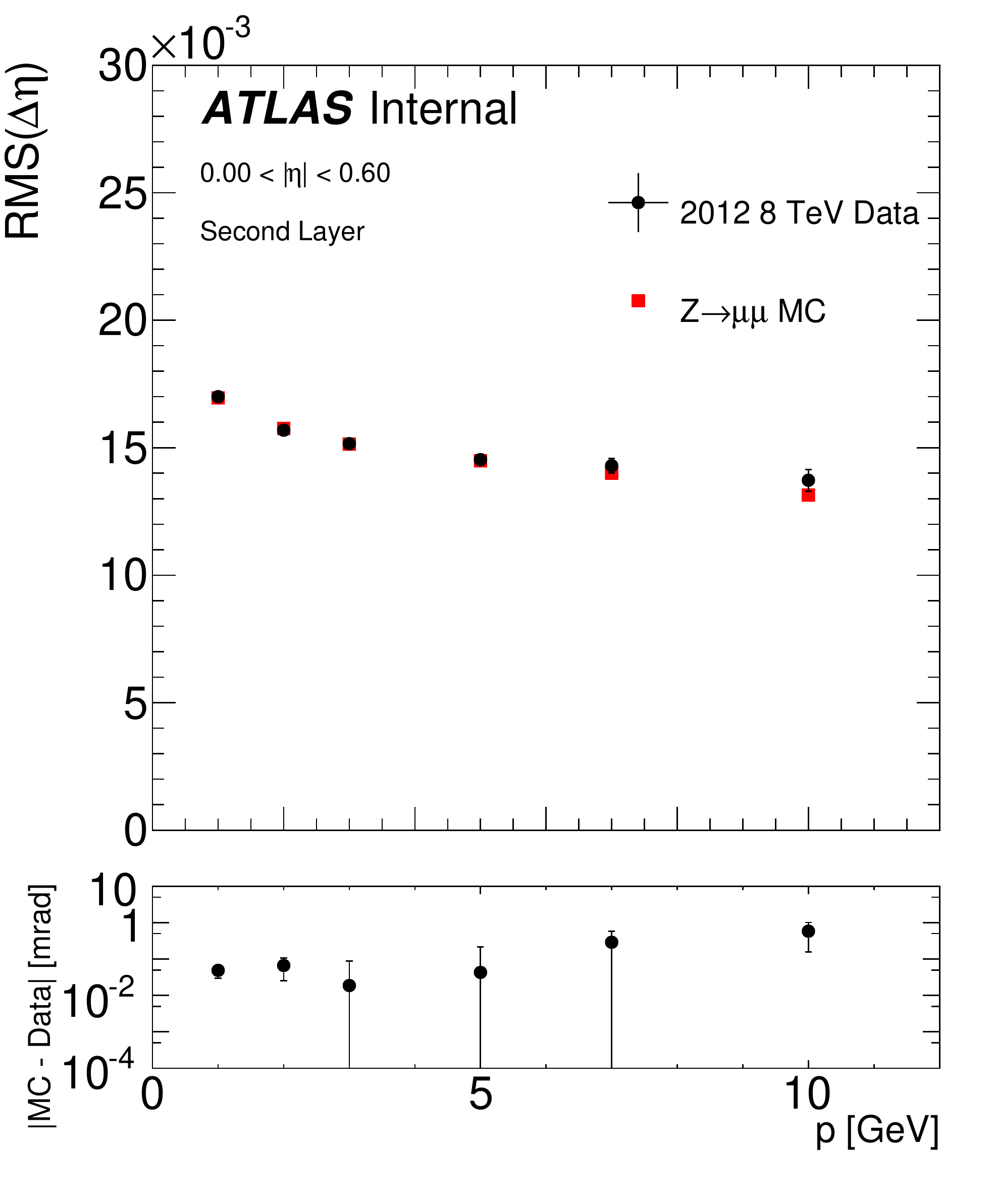}
\includegraphics[width=0.4\textwidth]{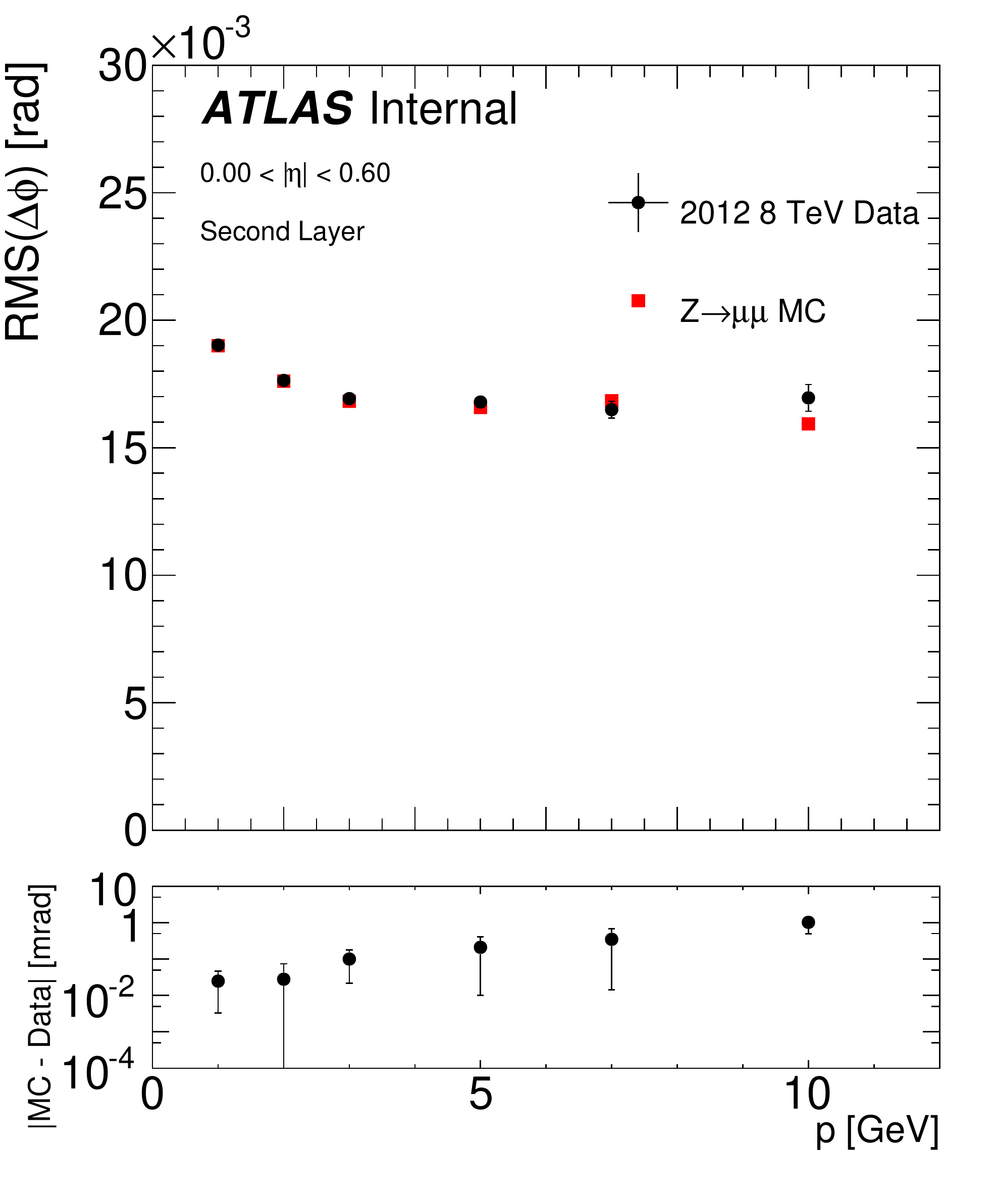}
\end{center}
\caption{Left (Right): The RMS of the $\Delta \eta$ ($\Delta\phi$) between isolated single particle tracks and clusters for tracks extrapolated to the second layer of the calorimeter in the barrel of the detector.}
\label{syst:angle3bb}
\end{figure}

\begin{figure}[h!]
\begin{center}
\includegraphics[width=0.4\textwidth]{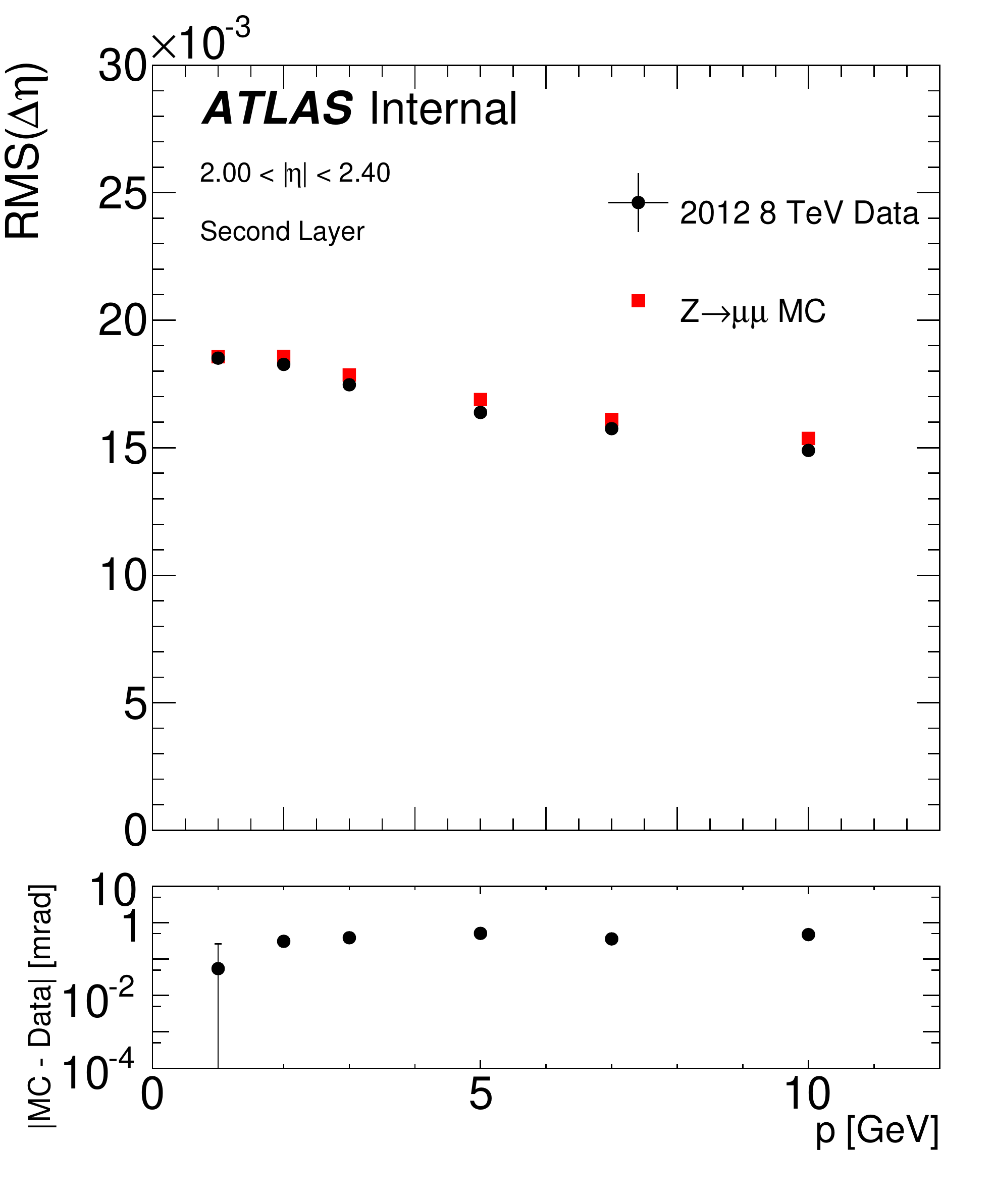}
\includegraphics[width=0.4\textwidth]{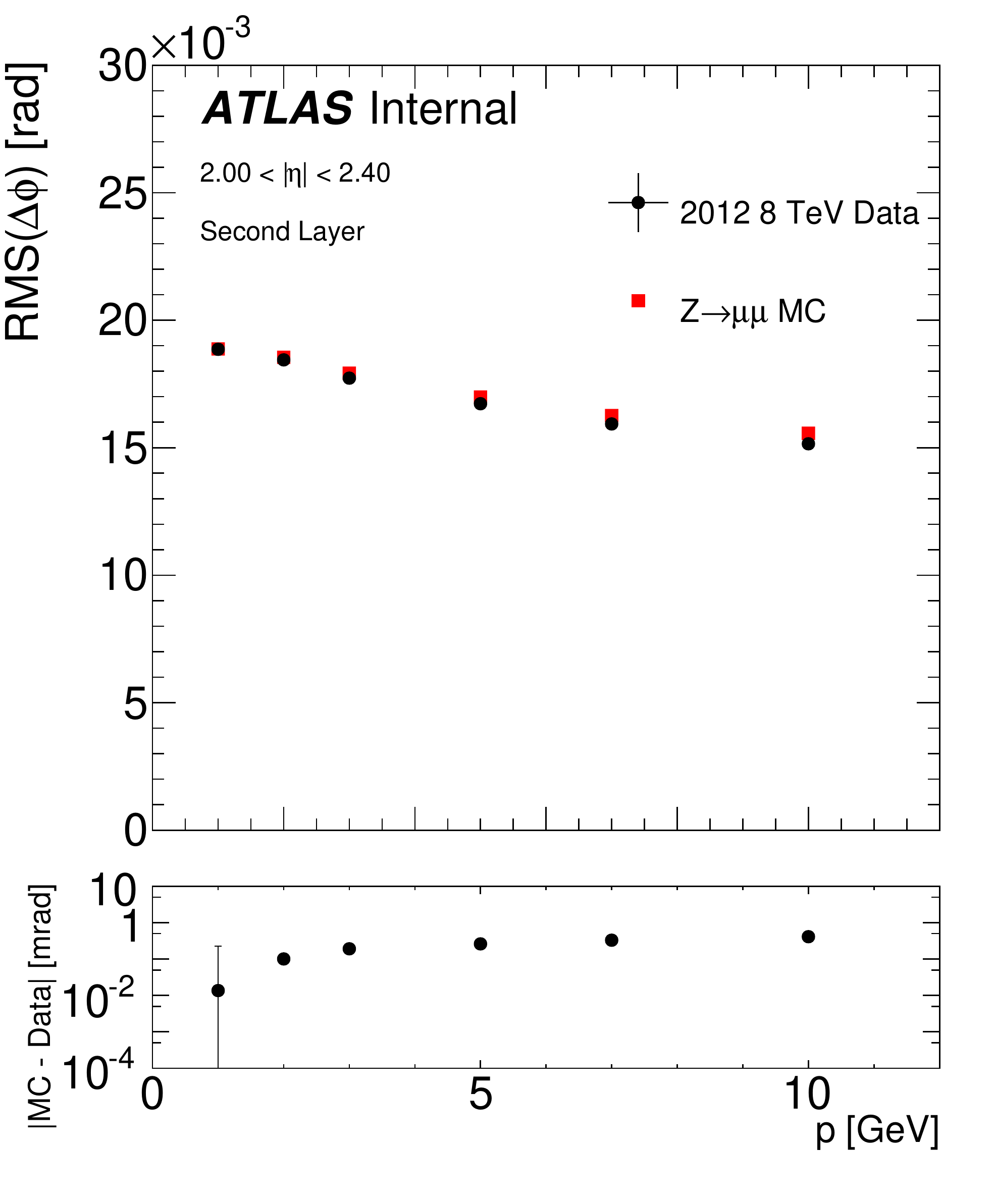}
\end{center}
\caption{Left (Right): The RMS of the $\Delta \eta$ ($\Delta\phi$) between isolated single particle tracks and clusters for tracks extrapolated to the second layer of the calorimeter in the endcap of the detector.}
\label{syst:angle4bb}
\end{figure}

\clearpage

\subsection{Jet Angular Resolution}
\label{jar}

As demonstrated in Sec.~\ref{origincorrection} in the context of the origin correction, the jet pull angle is sensitive to the choice of the jet axis and therefore on the modeling of the jet axis angular resolution (with respect to the particle-level jet axis).  One method to estimate the jet angular resolution (JAR) uncertainty is to use the angular displacement between calorimeter jets and track jets.   Section~\ref{sec:JARalternate} below describes this method in detail, but it is not used as the baseline JAR uncertainty for two reasons:

\begin{enumerate}
\item The uncertainty in the jet angular resolution should be very correlated with the uncertainties on the clusters.  The uncertainty computed with the track jet difference would treat these uncertainties as fully uncorrelated.
\item The uncertainties determined with the track jet method are significantly larger than those determined from propagating cluster uncertainties (which is the baseline method), due at least in part to limited MC statistics in the measurement.   Figure~\ref{syst:angle3} compares the track-jet method of Sec.~\ref{sec:JARalternate} with the JAR induced from the cluster uncertainties (baseline prescription).
\end{enumerate}

\begin{figure}[h!]
\begin{center}
\includegraphics[width=0.45\textwidth]{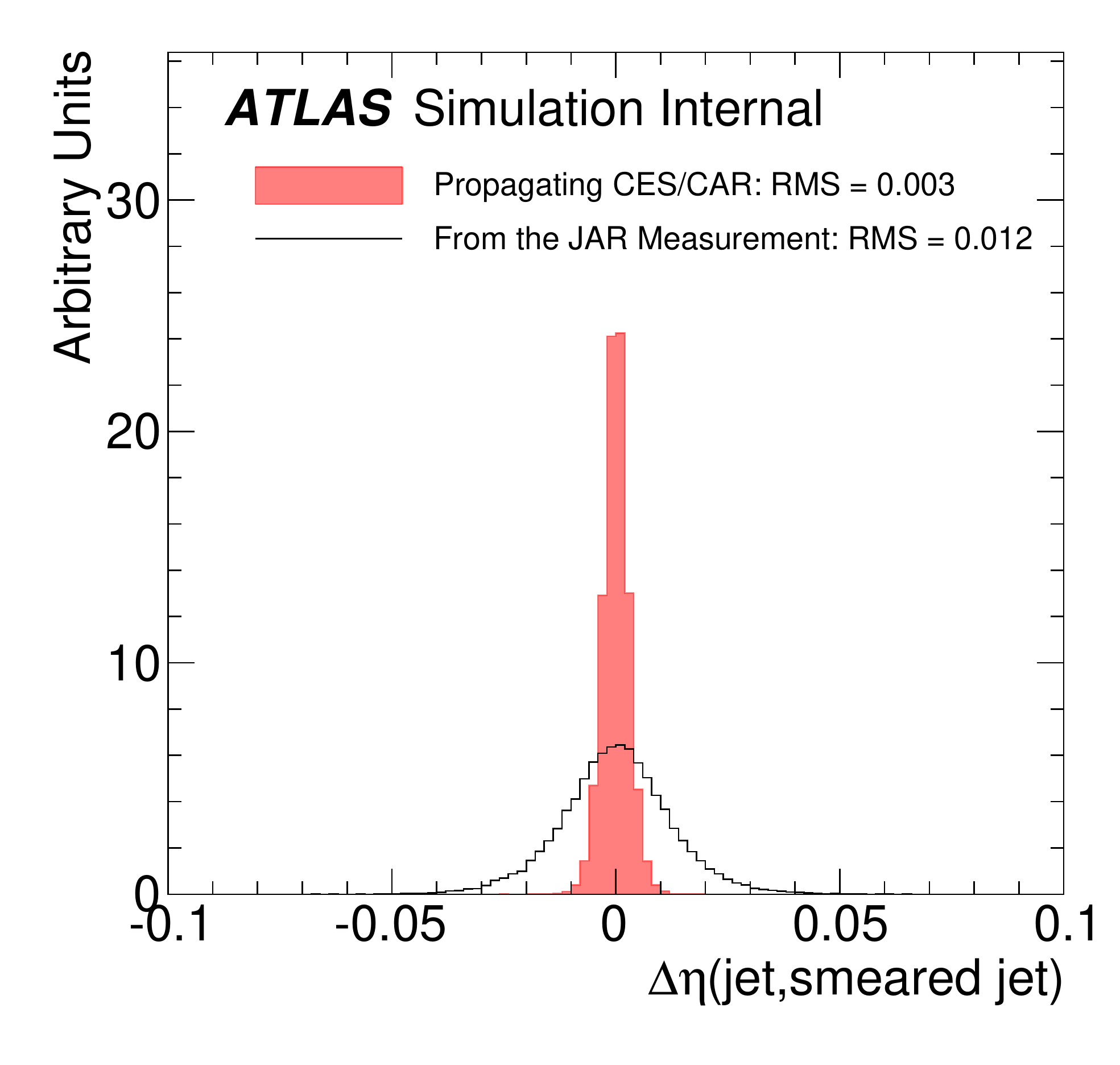}
\includegraphics[width=0.45\textwidth]{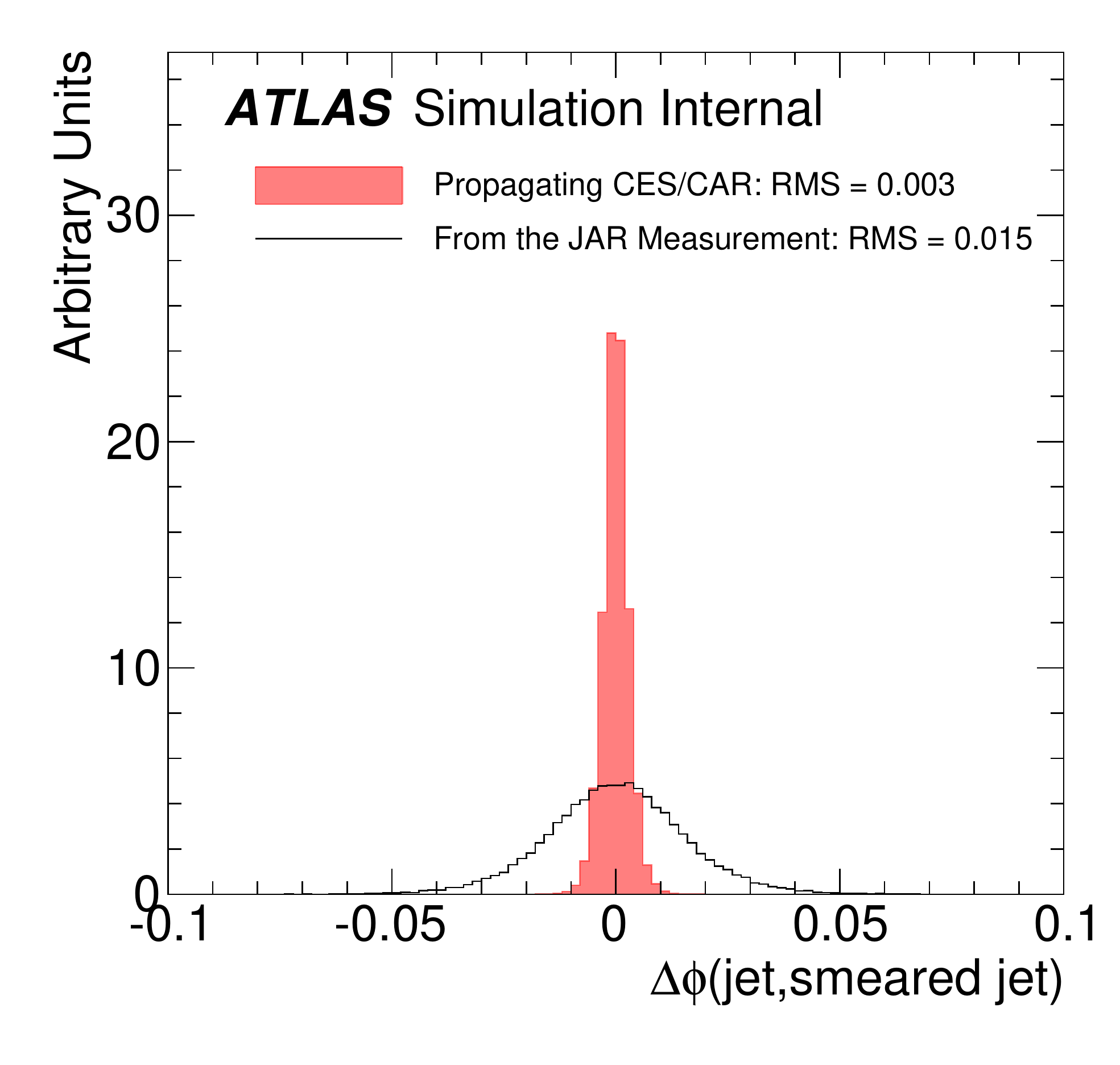}
\end{center}
\caption{A comparison between the track-jet and cluster-induced JAR uncertainties (baseline) for the jet $\eta$ (left) and the jet $\phi$ (right).}
\label{syst:angle3}
\end{figure}

\clearpage

\subsubsection{In-situ method using track jets}
\label{sec:JARalternate}

Track jets are constructed from tracks using the same jet clustering algorithm as for calorimeter jets.  The in-situ JAR uncertainty presented in this section utilizes the excellent angular resolution of these track jet.  Let $\Delta\phi=\phi_\text{track jet}-\phi_\text{calo jet}$ and $\Delta\eta=\eta_\text{track jet}-\eta_\text{calo jet}$.  The resolution of $\Delta x$ for $x\in\{\phi,\eta\}$ is given by

\begin{align}
\label{sec:JAR:eq1}
\sigma_{\Delta x}\sim\sqrt{\sigma_{x_\text{track jet}}^2+\sigma^2_{x_\text{calo jet}}},
\end{align}

\noindent where the resolution for the calorimeter jet $\sigma_{x_\text{calo jet}}$ is with respect to the corresponding particle-level jet with both charged and neutral particles while the resolution of the track jet $\sigma_{x_\text{track jet}}$ is with respect to the corresponding particle level jet with only charged particles.  The resolution of the track jets with respect to particle-level jets using both charged and neutral particles is even worse than the calorimeter jet angular resolution due to charge-to-netural ratio fluctuations that are large compared to the detector-resolution.  Standard error propagation on Eq.~\ref{sec:JAR:eq1} gives an equation involving the uncertainty on the resolution of $x$, $\sigma_{\sigma_x}$:

\begin{align}
\label{eq:angularres}
\sigma_{\sigma_{\Delta x}}^2 \sigma^2_{\Delta x}\sim \sigma^2_{\sigma_{x_\text{track jet}}} \sigma^2_{x_\text{track jet}}+\sigma^2_{\sigma_{x_\text{calo jet}}}\sigma^2_{x_\text{calo jet}}.
\end{align}

\noindent Compared to the calorimeter angular resolution uncertainty, the track jet angular resolution uncertainty should be second order.  Dropping the corresponding terms in Eq.~\ref{eq:angularres} and solving for $\sigma_{\sigma_x}$ gives an estimate for the uncertainty on the resolution of $\sigma_x$:

\begin{align}
\label{JARtrackjets}
\sigma_{\sigma_{x_\text{calo jet}}}(p_T,\eta)\sim \frac{ \sigma_{\Delta\phi}(p_T,\eta)}{\sigma_{x_\text{calo jet}}(p_T,\eta)}\times \sigma_{\sigma_{\Delta x}}(p_T,\eta). 
\end{align}

\noindent The track jet method uses differences between data and simulation in the quantity $\sigma_{\sigma_{\Delta x}}$ to estimate the uncertainty on $\sigma_{\sigma_{x_\text{calo jet}}}$ via a scaling by $ \sigma_{\Delta\phi}/\sigma_{x_\text{calo jet}}$ that is determined from simulation.  The practical implementation of the JAR uncertainty would be to smear the $\phi$ and $\eta$ of each jet by a Gaussian with mean zero and standard deviation $s_x$ that solves the following equation ($\sigma_x\rightarrow\sigma_x+\sigma_{\sigma_x}$): 

\begin{align}
\label{JARtrackjets2}
\sigma_x^2+s_x^2=(\sigma_x+\sigma_{\sigma_x})^2\implies s=\sqrt{2\sigma\sigma_\sigma+\sigma_\sigma^2}.
\end{align}

\noindent Figures~\ref{syst:ColorFlow:JAR_in_ttbar} and~\ref{syst:ColorFlow:JAR_in_ttbar_eta} show the jet $p_\text{T}$ and jet $\eta$ dependence, respecitvely, of the jet $\phi$ and $\eta$ resolutions in simulation.   The resolution decreases with jet $p_\text{T}$, dropping below two mrad at about 100 GeV, and is stable for central $|\eta|$, degrading at high $|\eta|$ due to the worse calorimeter granularity.  

\begin{figure}[h!]
\begin{center}
\includegraphics[width=0.4\textwidth]{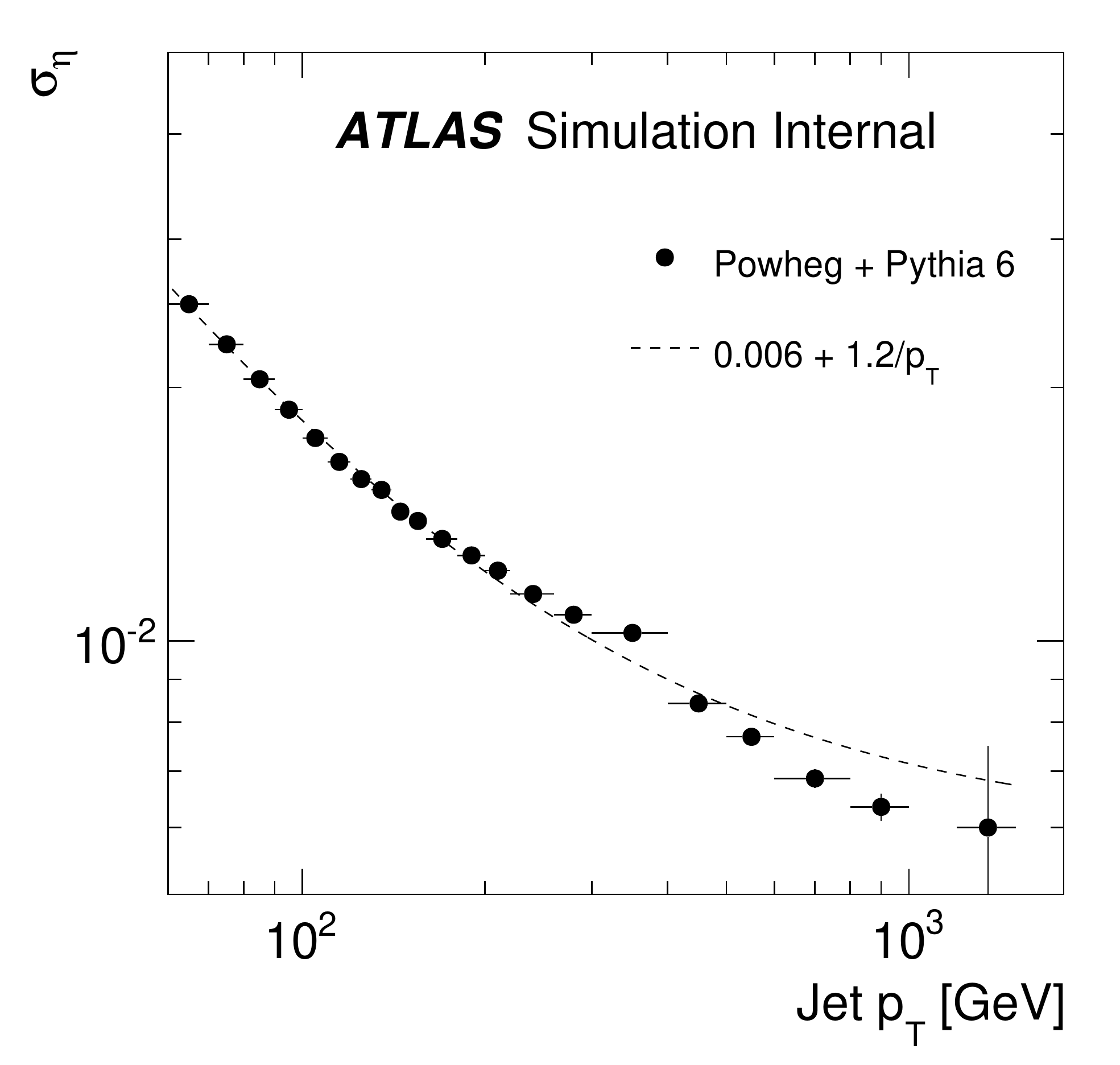}
\includegraphics[width=0.4\textwidth]{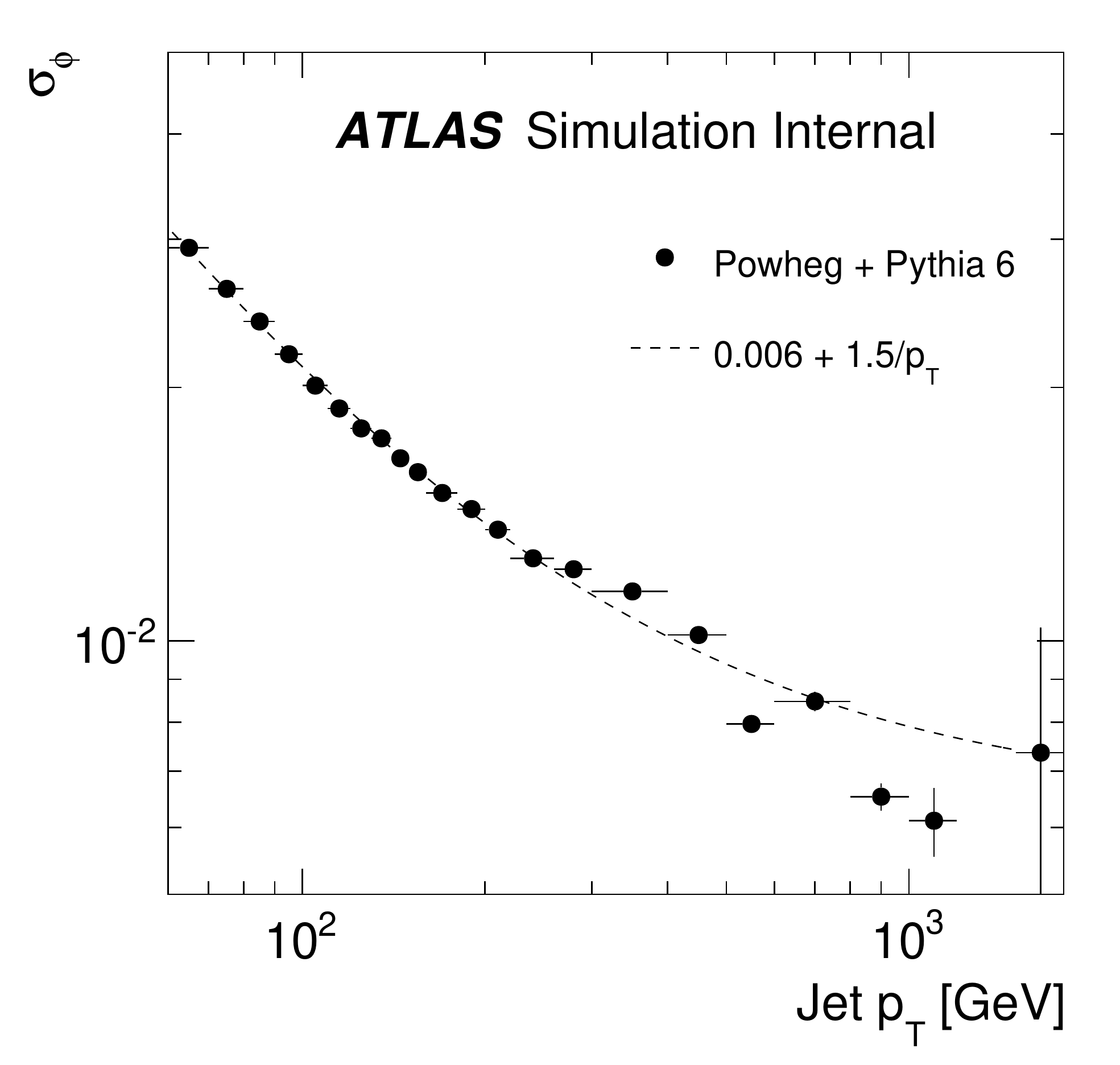}
\end{center}
\caption{The $p_\text{T}$ dependence of the jet angular resolution for $\eta$ (left) and $\phi$ (right) for the leading non $b$-tagged jets in $t\bar{t}$ events  The error bars reflect the statistical uncertainty and the dashed line is a fit to $a+b/p_\text{T}$.}
\label{syst:ColorFlow:JAR_in_ttbar}
\end{figure}

\begin{figure}[h!]
\begin{center}
\includegraphics[width=0.4\textwidth]{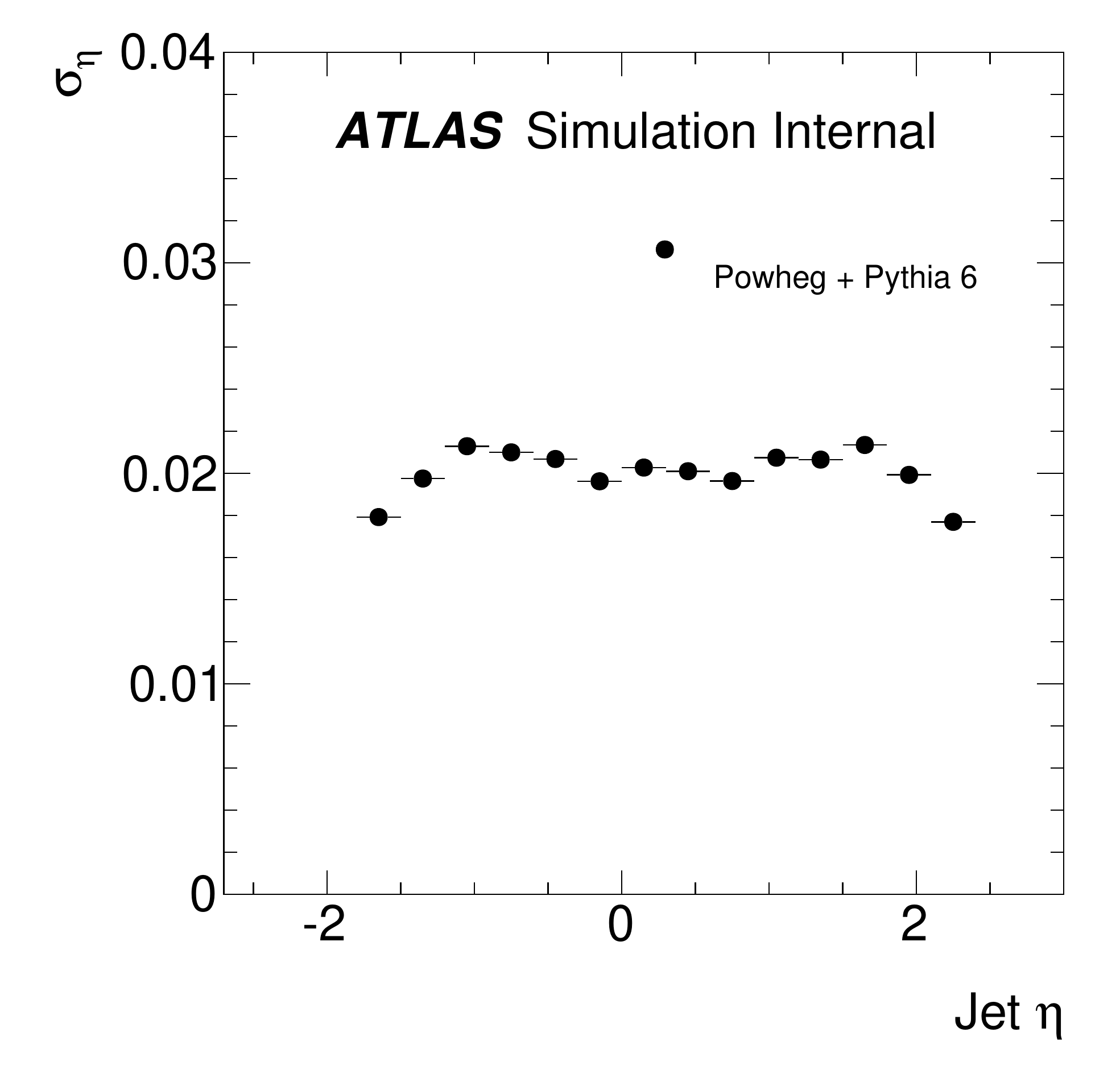}
\includegraphics[width=0.4\textwidth]{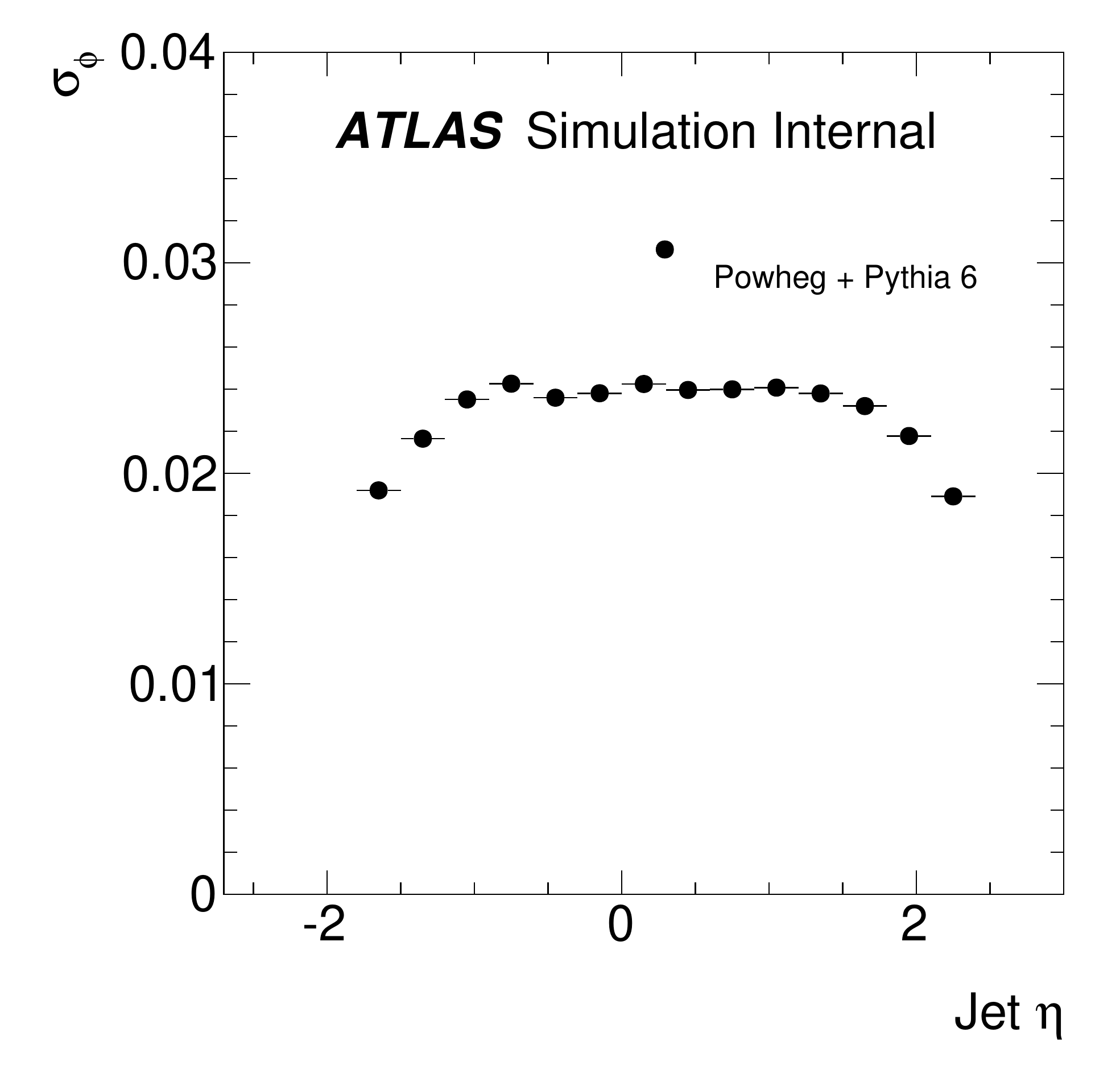}
\end{center}
\caption{The $\eta$ dependence of the jet angular resolution for $\eta$ (left) and $\phi$ (right) for the leading non $b$-tagged jets in $t\bar{t}$ events.}
\label{syst:ColorFlow:JAR_in_ttbar_eta}
\end{figure}

The next input to Eq.~\ref{JARtrackjets} is the resolution $\sigma_{\Delta x}$ and the corresponding uncertainty, $\sigma_{\sigma_{\Delta x}}$.  Figures~\ref{syst:angle3} and~\ref{syst:angle32} show the $p_\text{T}$ and $\eta$ dependence of the track jet - calorimeter jet angular resolution.  As expected, the resolution is larger than the corresponding distributions in Fig.~\ref{syst:ColorFlow:JAR_in_ttbar} and~\ref{syst:ColorFlow:JAR_in_ttbar_eta}.  The uncertainty bands in Fig.~\ref{syst:angle3} and~\ref{syst:angle32} are the result of various simulation variations, including changes in the fragmentation model and comparisons of the amount of inner detector material.  The data (not shown) are consistent with the simulation within these large $\sim 10\%$ uncertainties.  

\begin{figure}[h!]
\begin{center}
\includegraphics[width=0.4\textwidth]{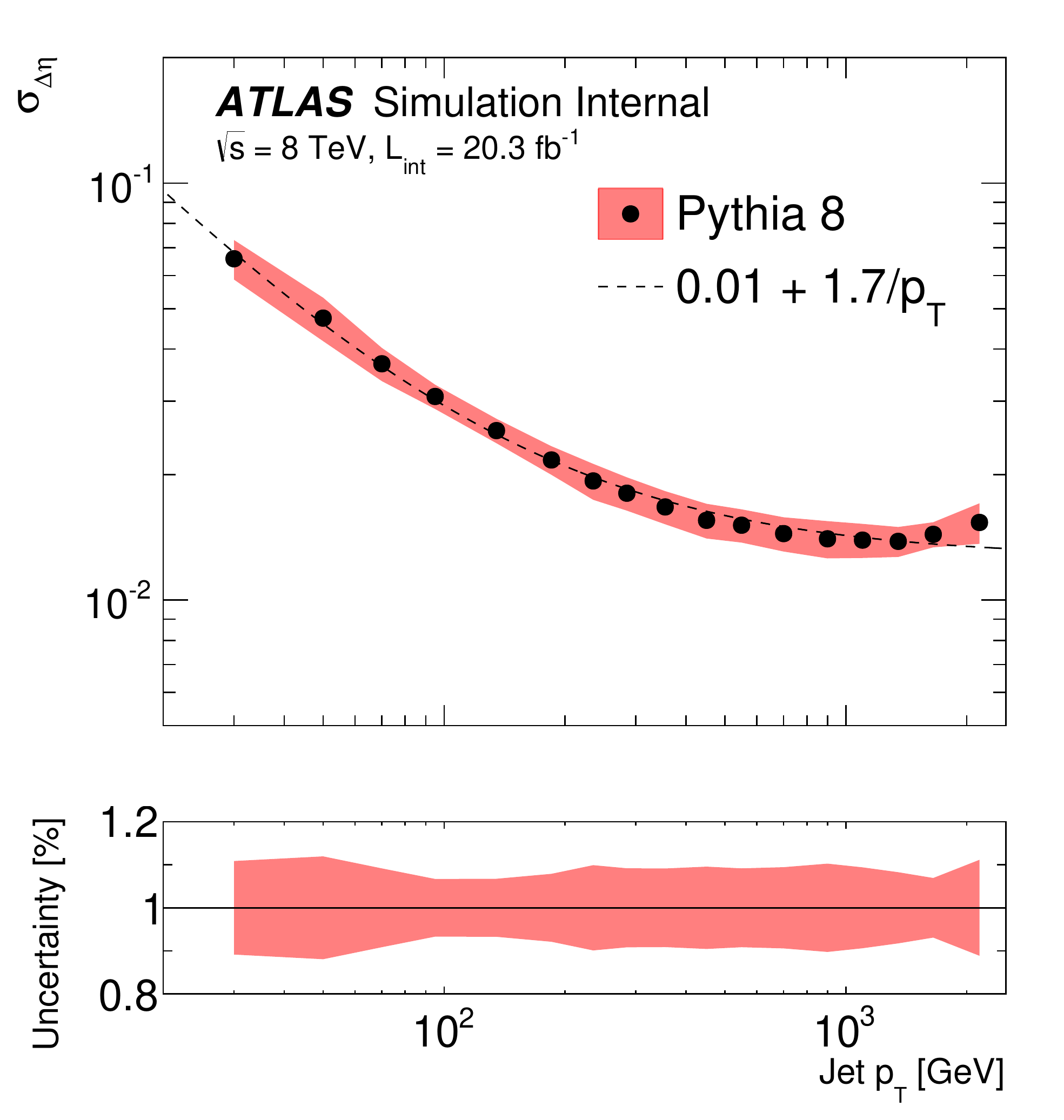}
\includegraphics[width=0.4\textwidth]{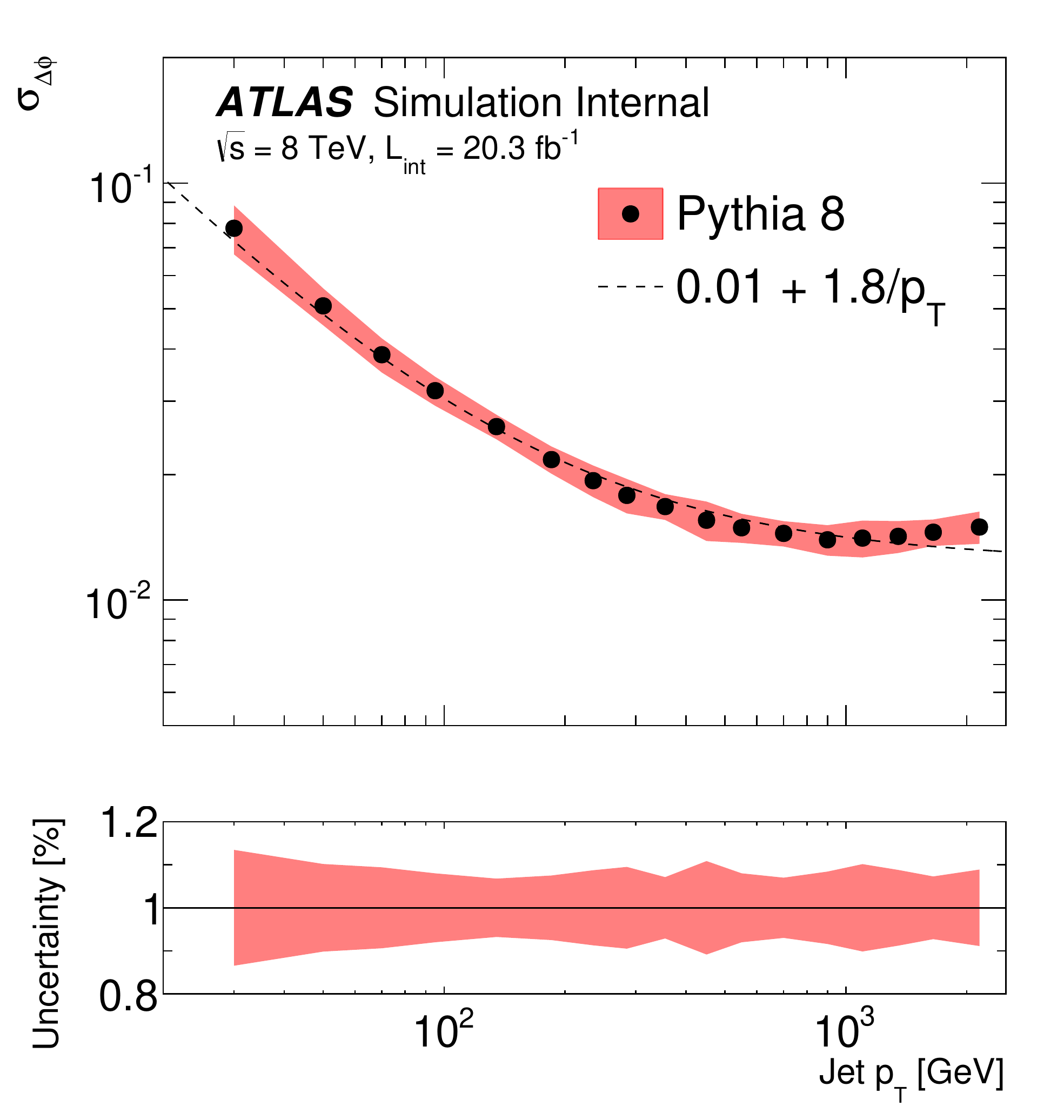}
\end{center}
\caption{The jet $p_\text{T}$ dependence of the resolution of the $\Delta\eta$ (left) and $\Delta\phi$ (right) between track jets and calorimeter jets in simulated dijet events.  The error band is described in the text.  A dashed line is a fit to $a+b/p_\text{T}$.  Inputs from F.  Guescini.}
\label{syst:angle3}
\end{figure}

\begin{figure}[h!]
\begin{center}
\includegraphics[width=0.4\textwidth]{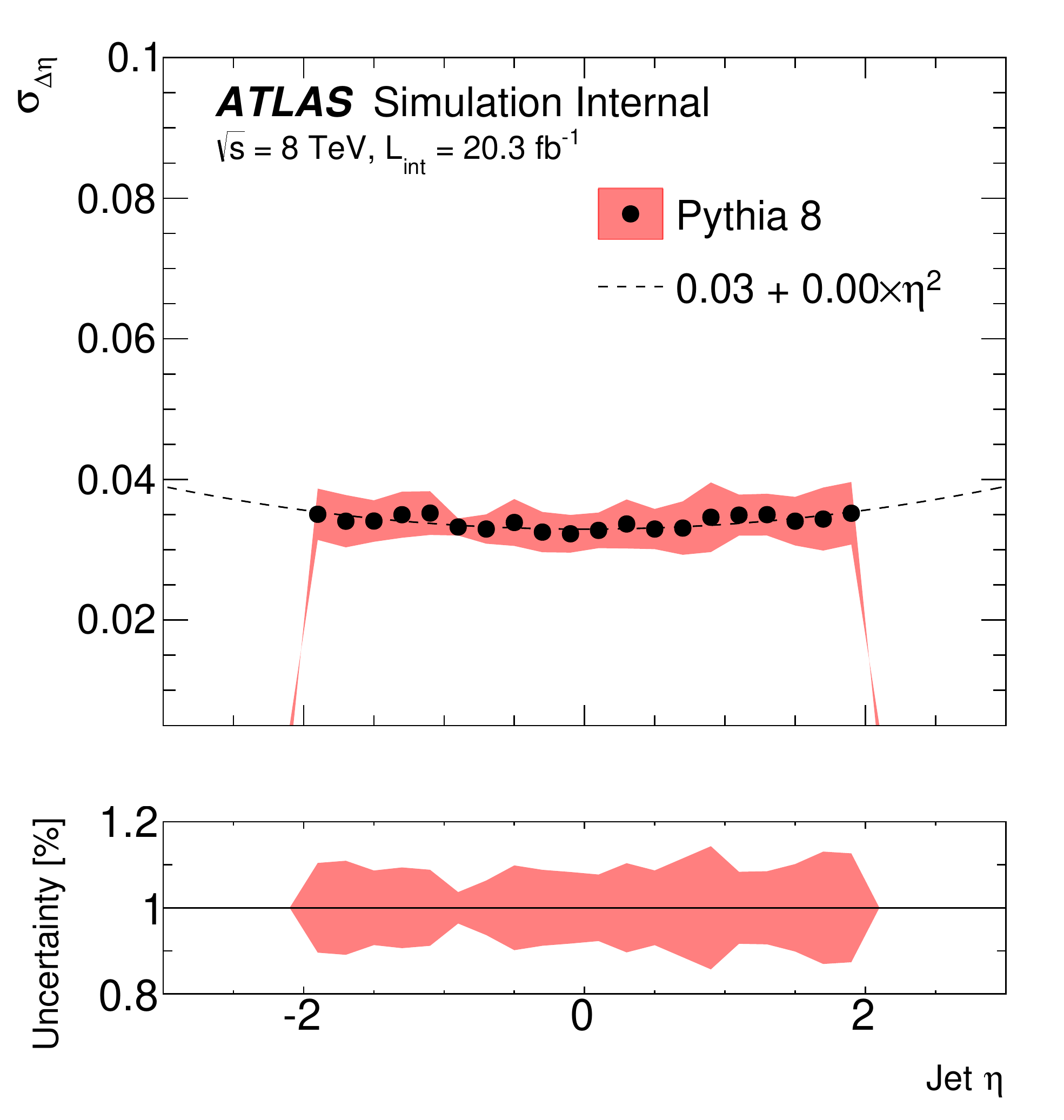}
\includegraphics[width=0.4\textwidth]{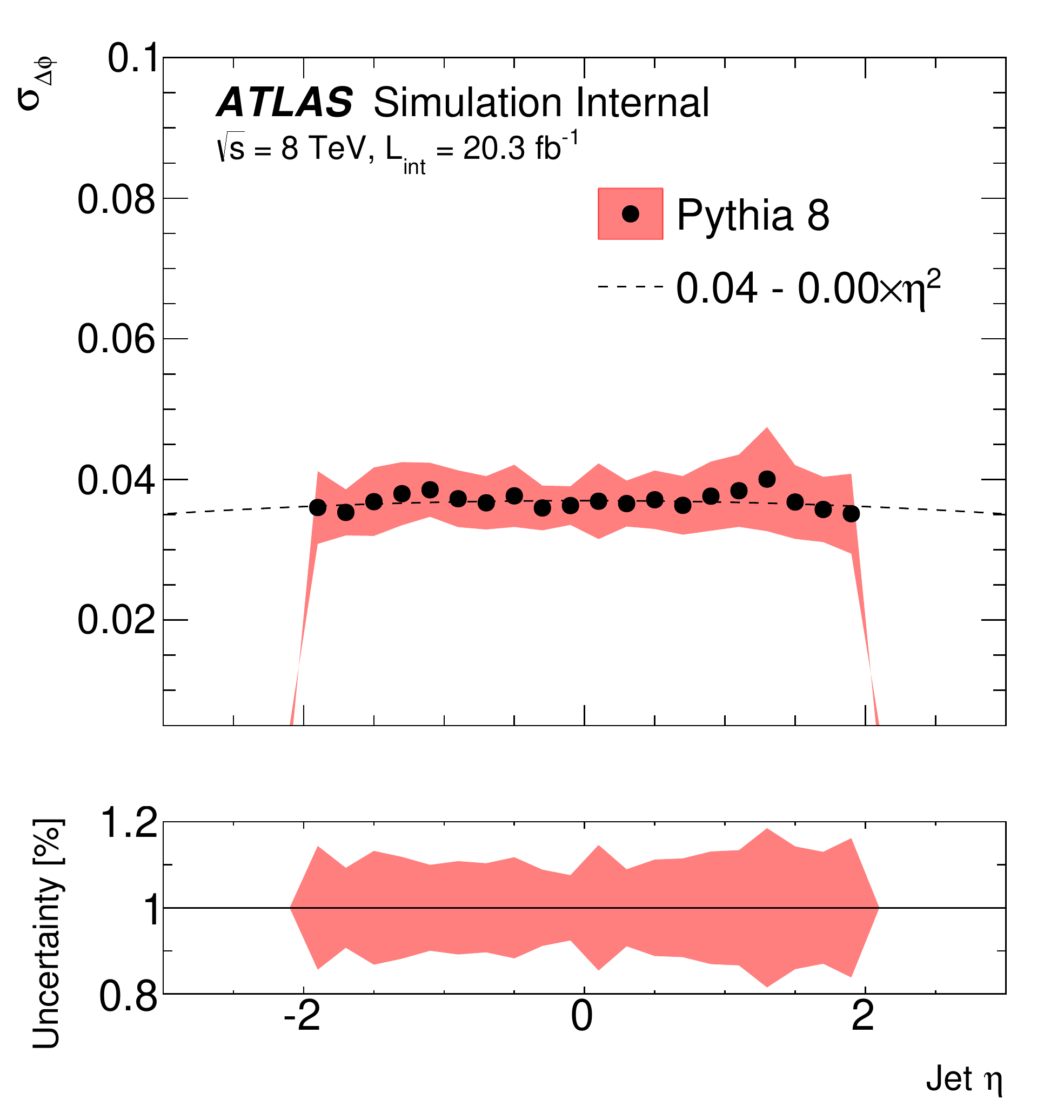}
\end{center}
\caption{The same as Fig.~\ref{syst:angle3}, only with $\eta$ instead of $\phi$.  The dashed line is a fit to $a+b\eta^2$.  Inputs from F.  Guescini.}
\label{syst:angle32}
\end{figure}

Figure~\ref{syst:angle33} shows the total fractional angular resolution and the smearing amount from Eq.~\ref{JARtrackjets2}.  Due to the origin correction, the $\eta$ and $\phi$ resolutions are comparable and the uncertainty on $\eta$ is even smaller than for $\phi$.  However, the uncertainties are large - $15\%$-$20\%$ around $p_\text{T}\sim 50$ GeV.  A significant contribution to this uncertainty is the limited MC sample size\footnote{This is not apparent from the nearly smooth error band in Fig.~\ref{syst:angle3} and~\ref{syst:angle32} because the fluctuations from many variations are summed together.}.  A combination of larger simulations sets and reduced modeling systematic uncertainties will allow this technique to be a competitive validation in the future.

\begin{figure}[h!]
\begin{center}
\includegraphics[width=0.4\textwidth]{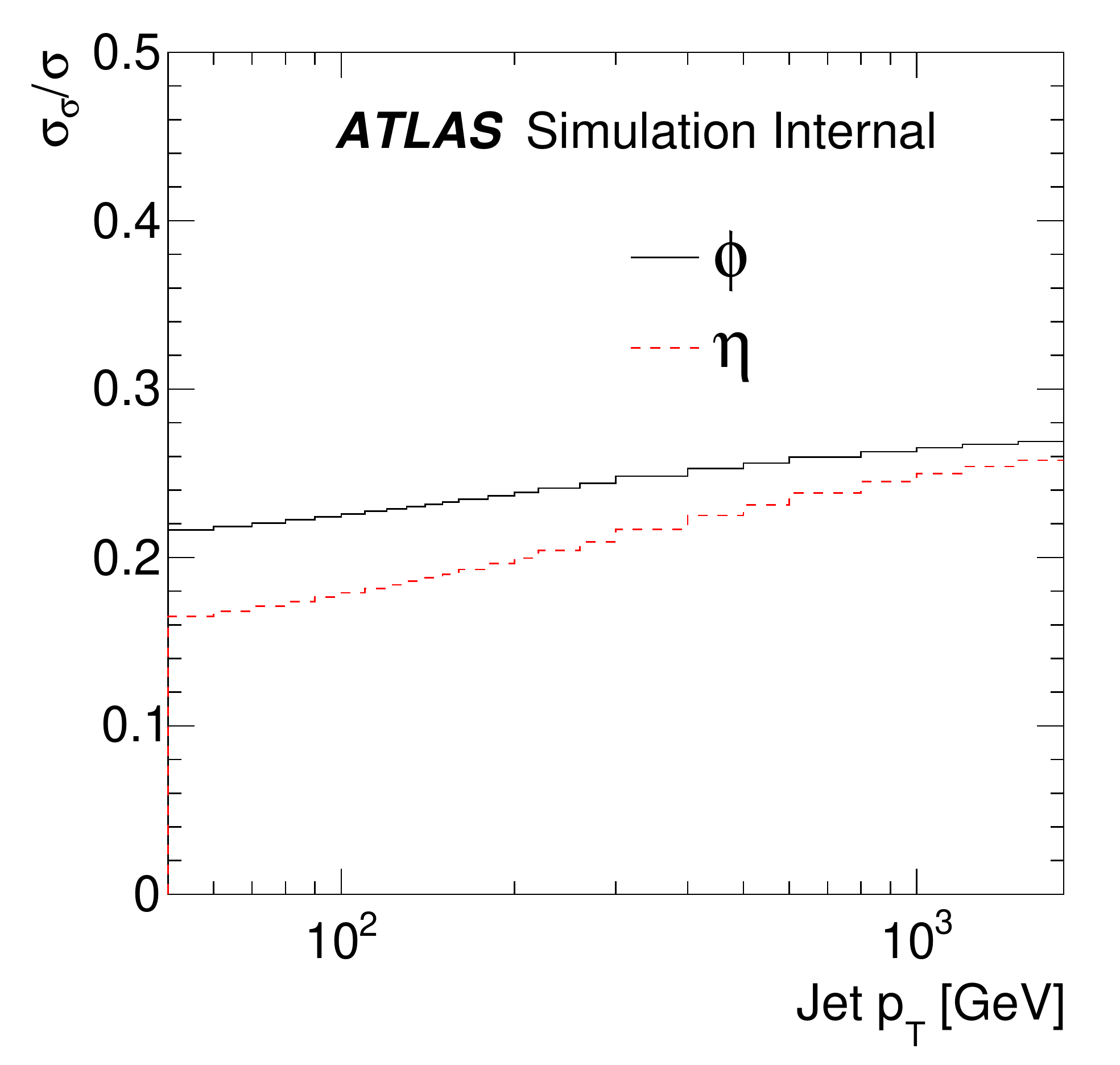}
\includegraphics[width=0.4\textwidth]{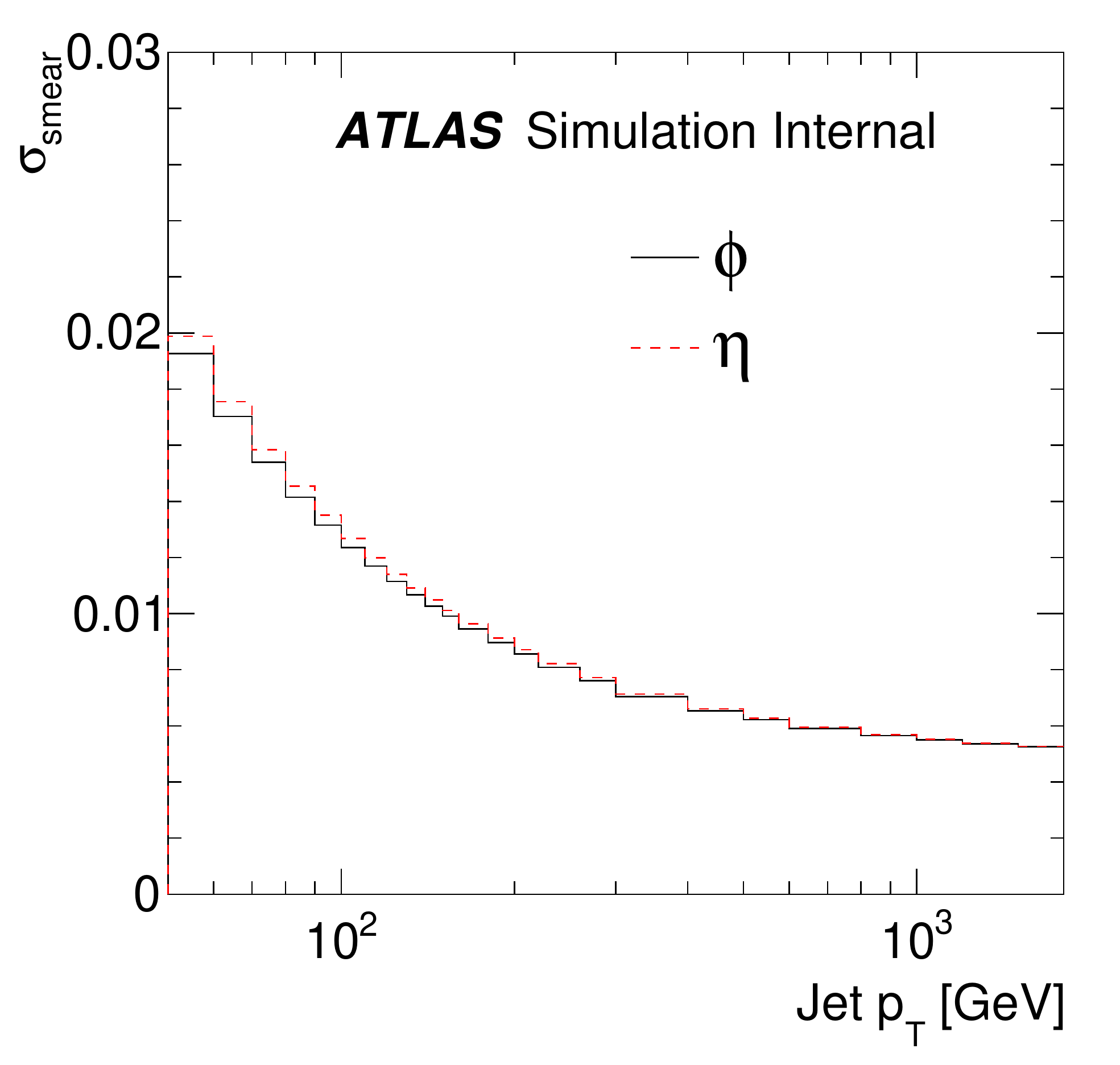}
\end{center}
\caption{The relative jet angular resolution for $\phi$ and $\eta$ (left) and the amount by which the jet angle should be smeared to conservatively cover the uncertainty in the angular resolution (right).}
\label{syst:angle33}
\end{figure}

\clearpage

\subsection{Other Experimental Uncertainties}

As the pull vector definition uses the calorimeter jet $p_\text{T}$,
both the all--particles and charged--particles pull angle are affected by
the uncertainty in the jet energy scale~\cite{Aad:2011he,Aad:2014bia}
and resolution~\cite{Aad:2012ag}.  However, changes in the jet energy scale and resolution do not impact the
pull \textit{angle}, but do impact the results via the acceptance due to
$p_\text{T}$ thresholds (evaluated in the same way as in Sec.~\ref{sec:calojet}).
Similarly, uncertainties in the lepton energy scale, trigger efficiency,
$E_\text{T}^\text{miss}$ resolution and $b$--tagging
efficiencies~\cite{ATLAS-CONF-2012-040,ATLAS-CONF-2014-004,ATLAS-CONF-2012-043,ATLAS-CONF-2012-039}
indirectly affect the results through changes in acceptance.

\subsection{Background Processes}
\label{syst:back:colorflow}

Other (minor) sources of uncertainty on the acceptance, which impact the measurement
through the background subtraction, include those related to
the luminosity~\cite{Aad:2013ucp},
the multijet estimation,
and the normalisation and heavy flavour content of the $W$+jets
background~\cite{Aad:2014zka}. The luminosity uncertainty of $\pm 2.8\%$ only affects those backgrounds that are estimated directly from simulation, including the single top, diboson, and $Z$+jets processes.  Like the $t\bar{t}$ signal, the single production of a top quark in association with a $W$ boson also can have one leptonically decaying $W$ boson and one hadronically decaying $W$ boson ($Wt$).  An uncertainty on the quantum interference of the NLO $Wt$ process with leading order $t\bar{t}$ process is estimated by comparing the DR and DS overlap removal schemes~\cite{Frixione:2008yi} (more detail in Sec.~\ref{sec:singletopuncerts}).  Additionally, there is an uncertainty related to the flipped model.  If the $W$ boson radiation follows an octet pattern, then the contribution from $Wt$ will be more like the flipped model than the SM $t\bar{t}$.  Since the $Wt$ is subtracted along with the other minor backgrounds, this could bias the measurement.  However, the $Wt$ is only about $3\%$ of the total background composition and the difference between the singlet and octet radiation pattern is $\mathcal{O}(\%)$.  A flipped $Wt$ sample is not generated, but the impact can be estimated by replacing the $Wt$ background with the flipped $t\bar{t}$ scaled to the predicted $Wt$ yield.  Figure~\ref{syst:pullwt} shows that such a conservative uncertainty would be much smaller than the statistical uncertainty (already subdominant to the uncertainties in Sec.~\ref{sec:topquarkmodeling}) and is thus not considered for the remainder of the analysis. 

\begin{figure}[h!]
\begin{center}
\includegraphics[width=0.9\textwidth]{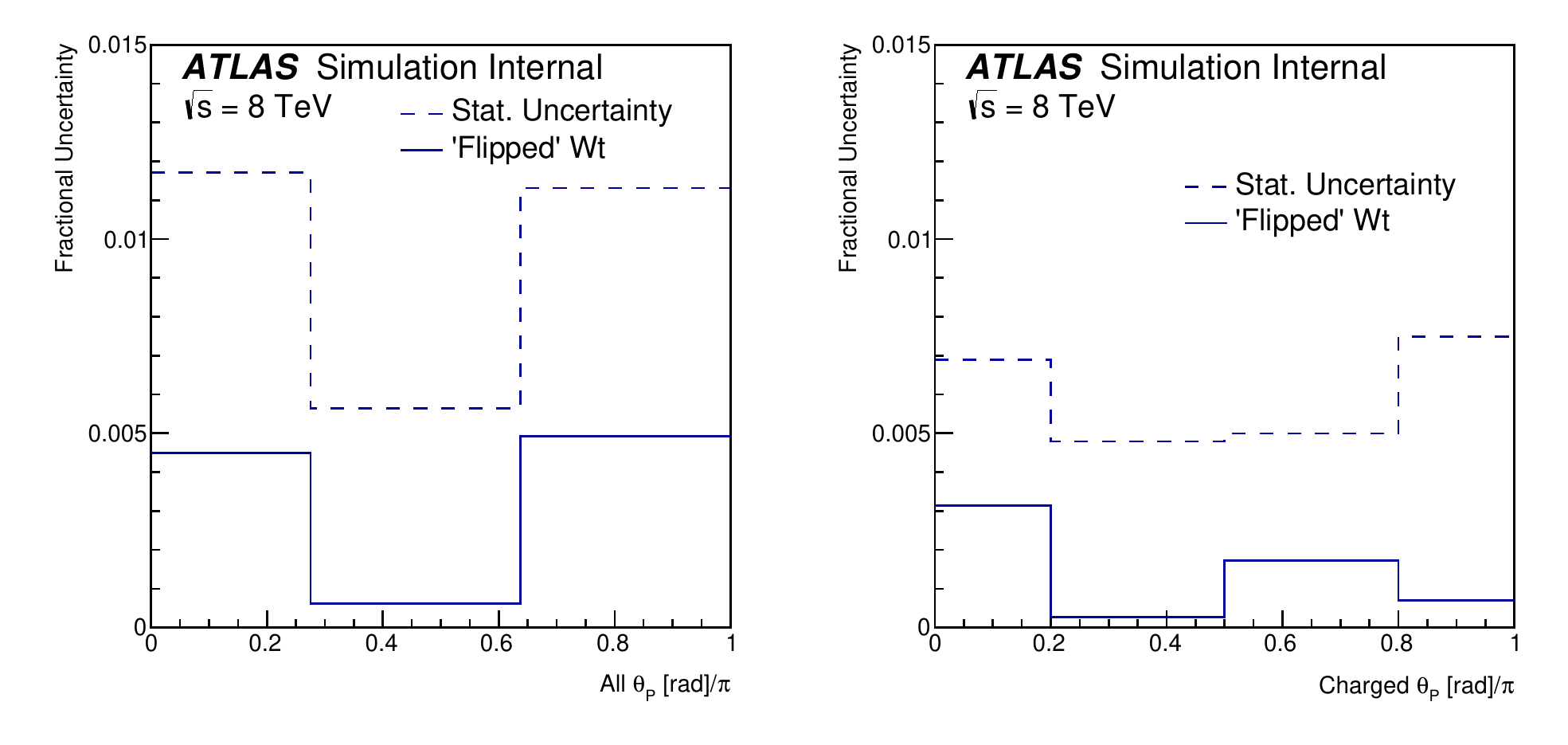}
\end{center}
\caption{The fractional difference in the unfolded result when the $Wt$ is replaced with the flipped $t\bar{t}$ scaled to the $Wt$ yield compared with the data statistical uncertainty for the all particles pull angle (left) and the charged particles pull angle (right).}
\label{syst:pullwt}
\end{figure}

\subsection{Top Quark Pair Production Modeling}
\label{sec:topquarkmodeling}

\subsubsection{ME Generator and Fragmentation Model}
\label{syst:colorflow:ME}

As expected, differences in the the pull angle distribution for a fixed fragmentation model and variable ME generator are small compared to the reverse setup.  This is demonstrated by Fig.~\ref{comparemCs} and~\ref{comparemCs2}.  There are percent-level differences in the pull angle distribution between {\sc Pythia} 6 and {\sc Herwig}.

\begin{figure}[h!]
\begin{center}
\includegraphics[width=0.45\textwidth]{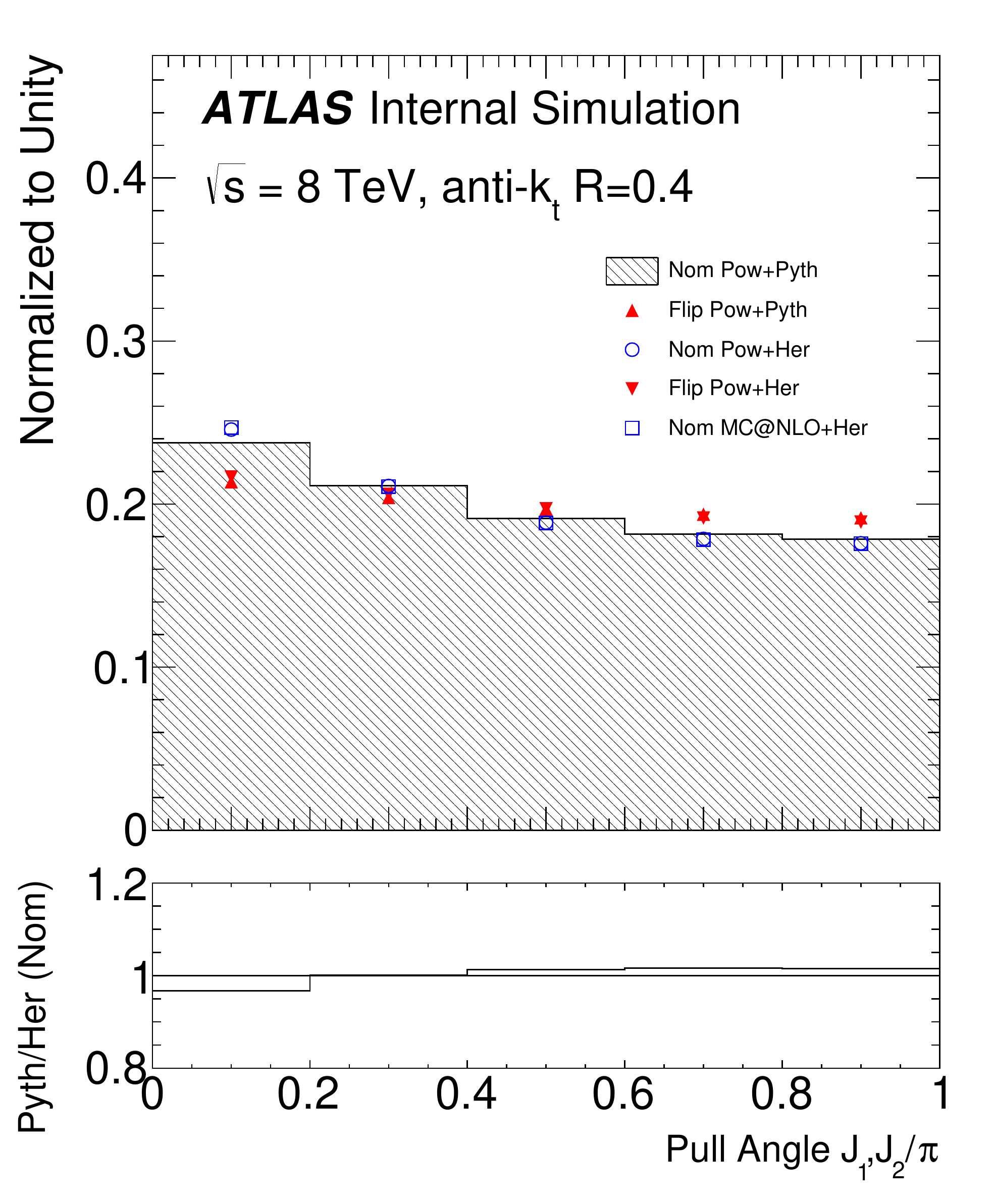}\includegraphics[width=0.45\textwidth]{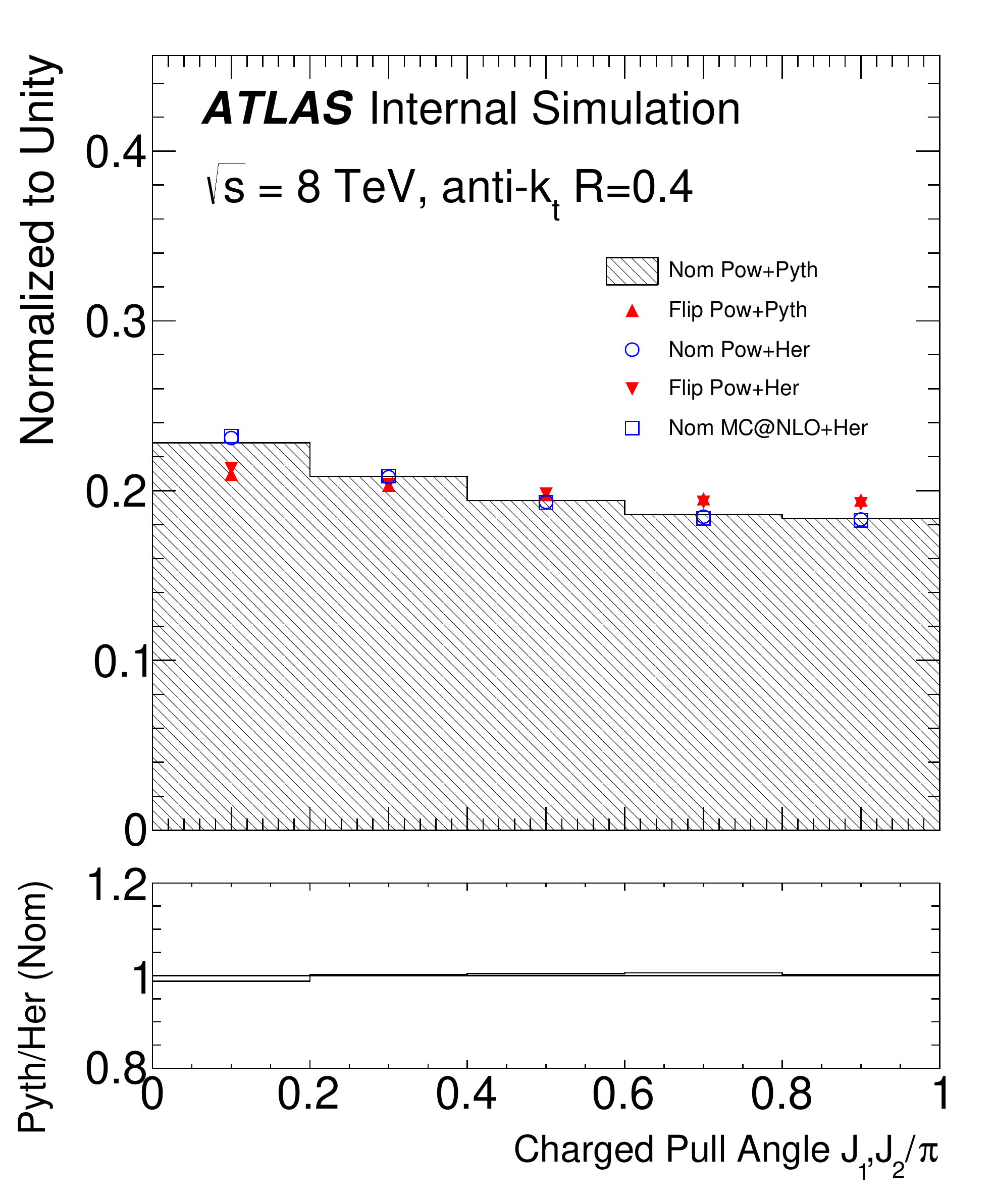}
\caption{Particle-level distributions comparing {\sc Pythia} 6 and {\sc Herwig} for the all particles pull angle (left) and the charged particles pull angle (right).  }
\label{comparemCs}
\end{center}
\end{figure}

\begin{figure}[h!]
\begin{center}
\includegraphics[width=0.45\textwidth]{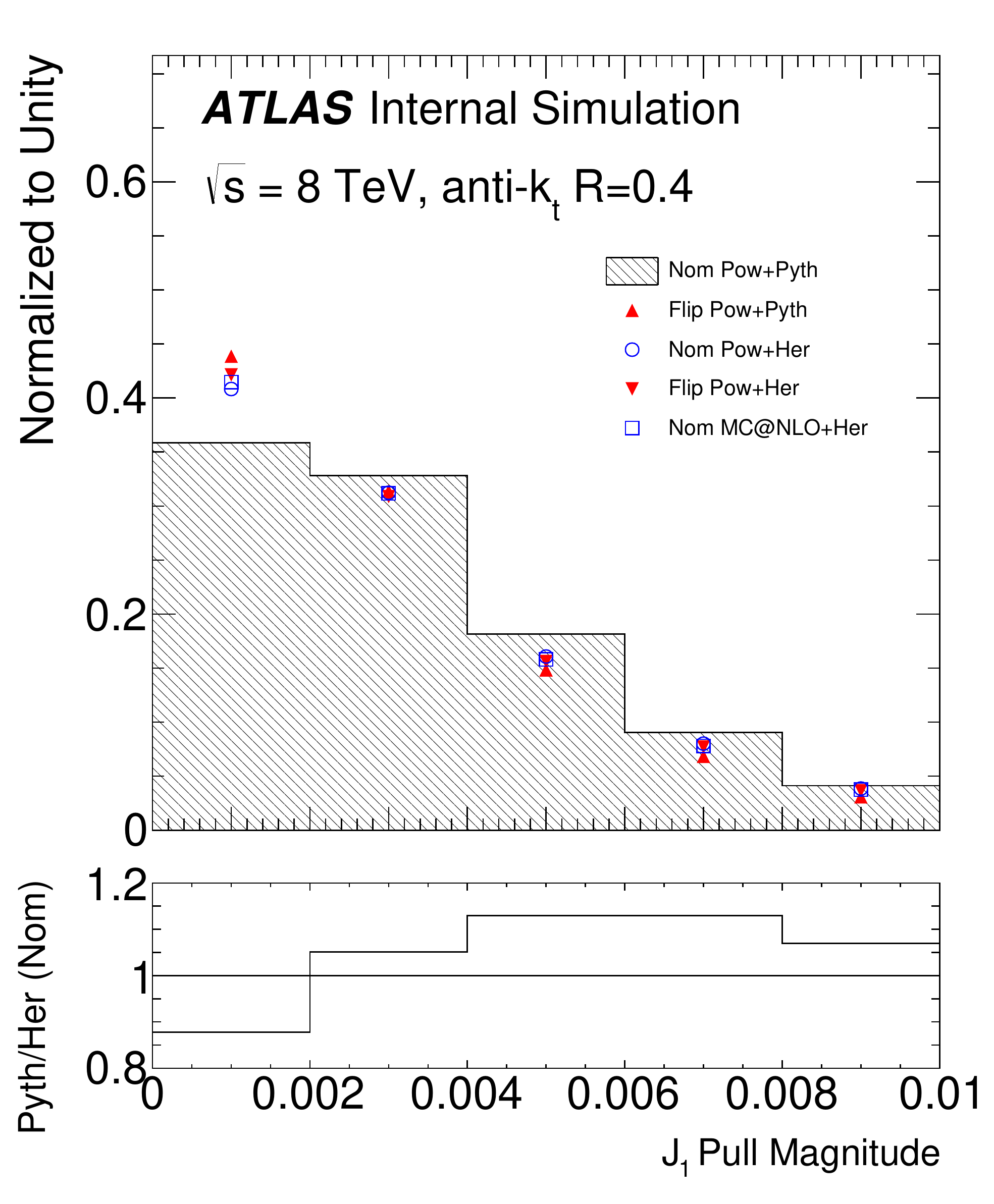}\includegraphics[width=0.45\textwidth]{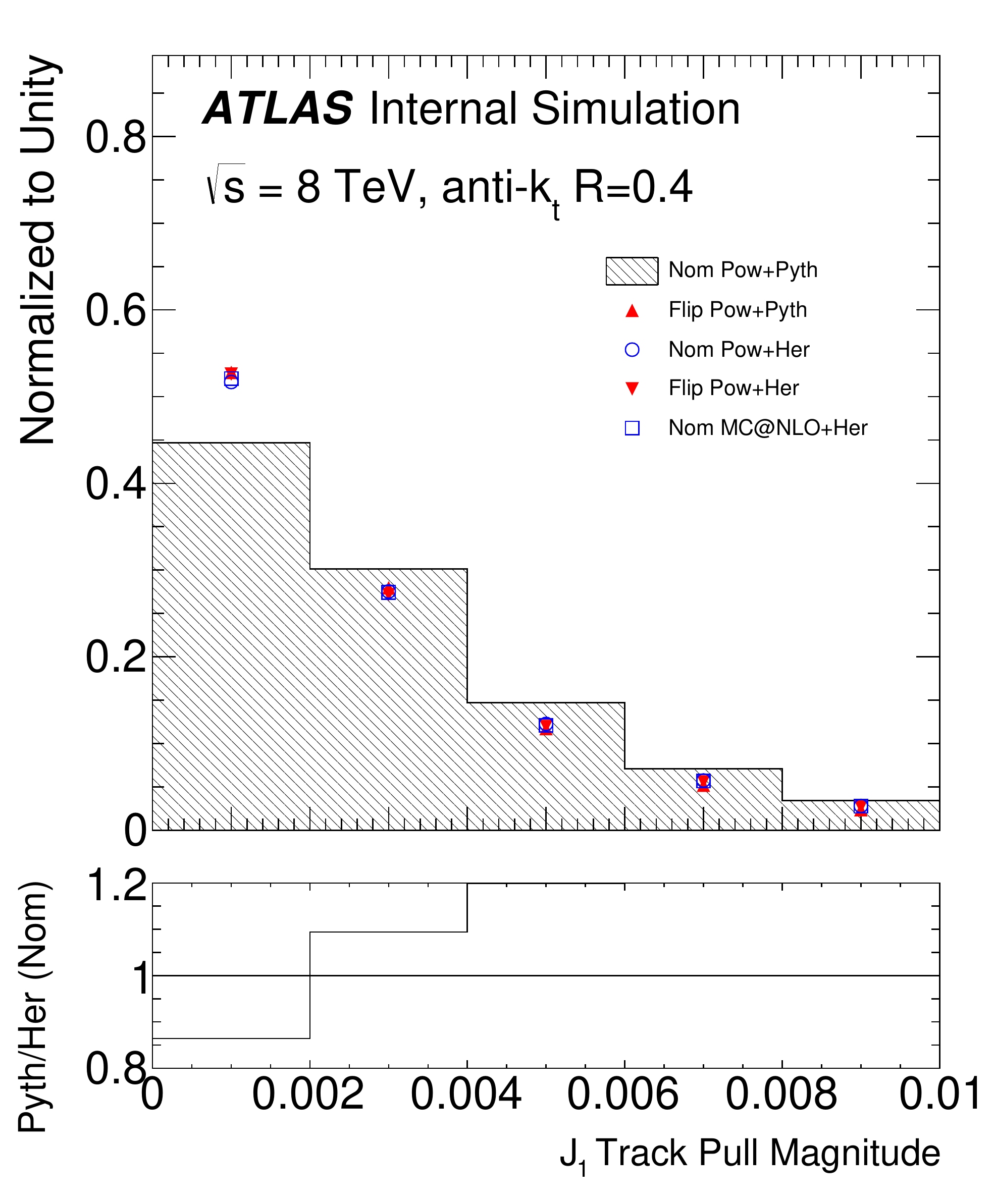}
\caption{Particle-level distributions comparing {\sc Pythia} 6 and {\sc Herwig} for the all particles pull vector magnitude (left) and the charged particles pull vector magnitude (right).  }
\label{comparemCs2}
\end{center}
\end{figure}

\subsubsection{Color Reconnection}
\label{CRuncert}

While the color reconnection in the parton shower is also due to the exchange of color charge, it is expected to not have a large impact on the jet pull distribution as the pull vector should be set by the color flow at the hard scatter.  This is supported by early studies in Ref.~\cite{Altheimer:2013yza}.  Figure~\ref{CR} compares the nominal {\sc Pythia} 6 tune (P2011C) with the 2012 Perugia lowCR tune~\cite{Skands:2010ak} (as well as a tune for higher MPI).  The lowCR Perugia tune differs from the nominal tune in the method and strength for calculating the reconnection probability for colored partons in the PS.  The two parameters which differ are {\tt MSTP(95)} (probability calculation method) and {\tt MSTP(78)} (strength of the connection).   In the lowCR tune, the probability for a string piece to preserve its original connection is given by

$$P_\text{keep}=(1-\zeta\times\text{\tt MSTP(78)})^{n_\text{int}},$$

\noindent where $n_\text{int}$ is the number of parton-parton interactions in the current event.  The parameter $\zeta^{-1}=1+\text{\tt MSTP(77)}^2\times\langle p_T\rangle^2$ is a way to make this $p_\text{T}$ dependent.  In all the Perugia tunes, {\tt MSTP(77)}=1.  The probability in the nominal tune is given by 

$$P_\text{keep}=(1-\zeta\times\text{\tt MSTP(78)})^{\langle n_s\rangle(y_1,y_2)},$$

\noindent where this tries to be more `local' with the function $\langle n_s\rangle(y_1,y_2)$ that counts the number of string pieces (not counting the ones under consideration) between the rapidity endpoints of the piece under consideration $y_1$ and $y_2$.  The loCR tune is set to be consistent with the minimum bias data, with as low a CR setup as possible.  
\vspace{2mm}

Figure~\ref{CR} shows that the impact from varying the CR tune is very small at truth level, $\lesssim 1\%$.  A similar trend is observed for the flipped model.

\begin{figure}[h!]
\begin{center}
\includegraphics[width=0.4\textwidth]{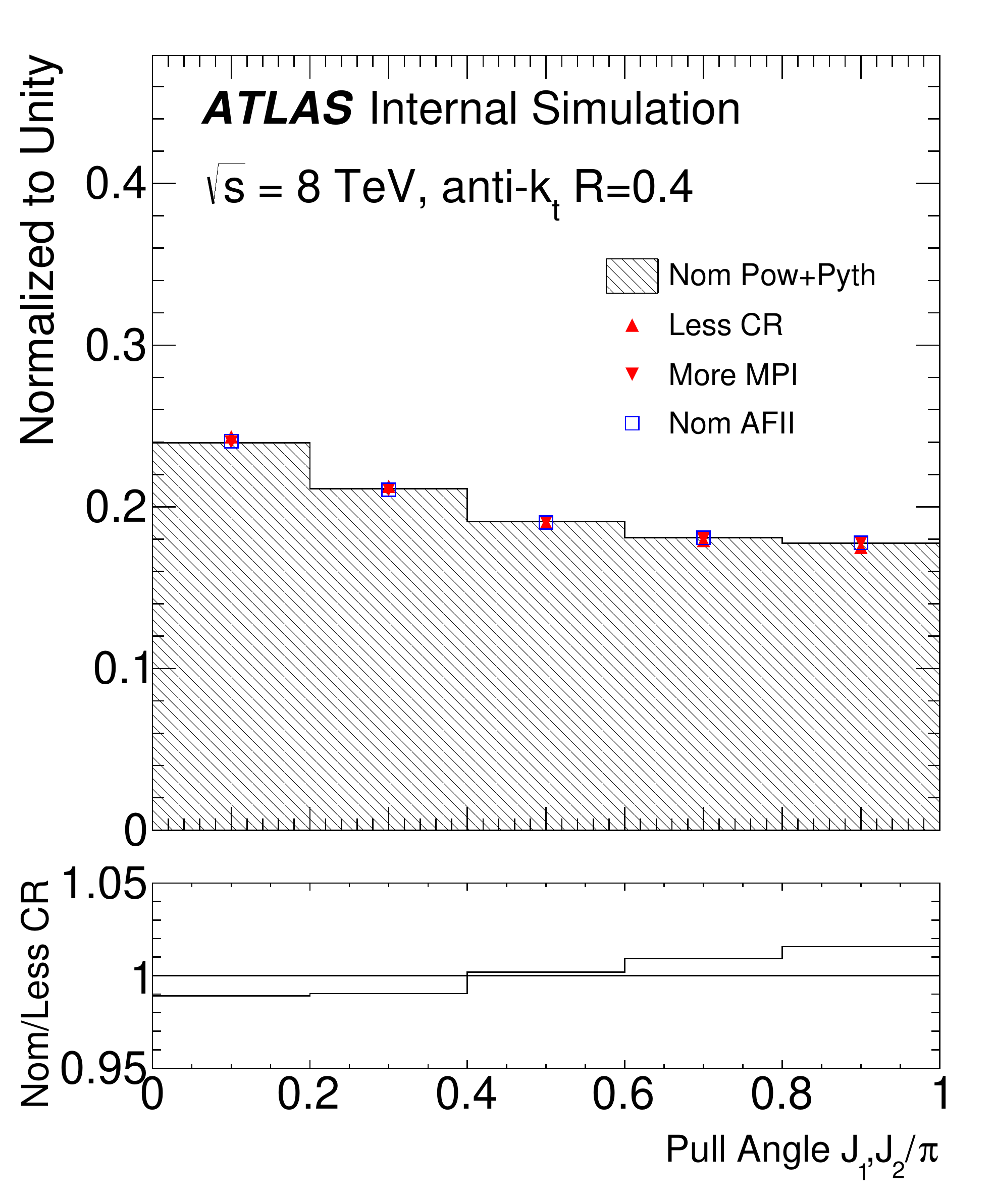}\includegraphics[width=0.4\textwidth]{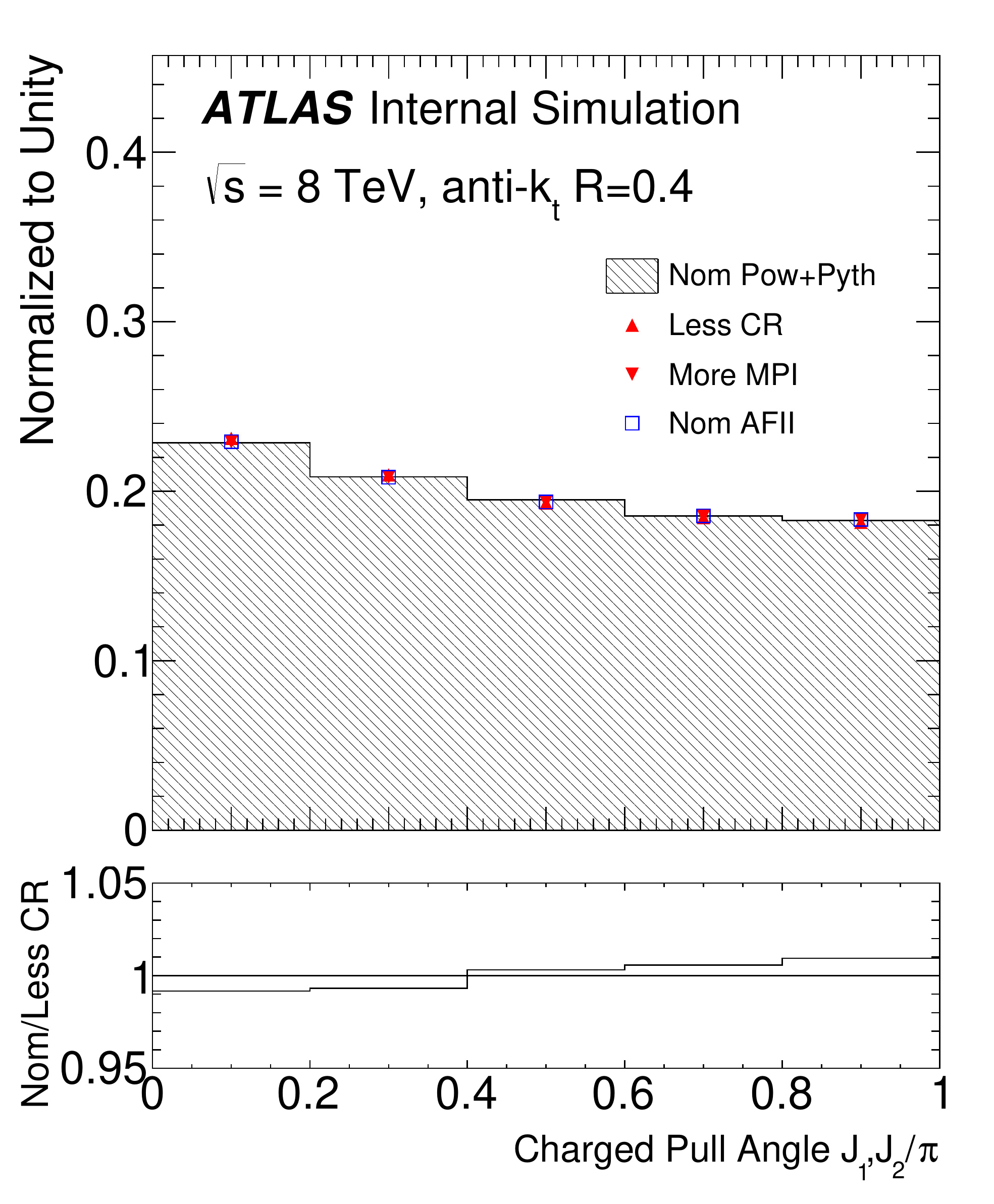}
\caption{Truth level distributions comparing the nominal and low CR tunes of Pythia for the all particles pull angle (left) and the charged particles pull angle (right).  }
\label{CR}
\end{center}
\end{figure}

\subsubsection{Initial and Final State Radiation}
\label{sec:ISRFSRuncert}

Figure~\ref{ISRFSR} shows the impact on the pull angle distribution due to variations in the ISR/FSR modeling from varying the
radiation simulated with \textsc{AcerMC} 3.8~\cite{Kersevan:2004yg} constrained by Ref.~\cite{ATLAS:2012al}.  The ISR/FSR could impact the pull angle either directly by introducing more radiation around the two selected jets or indirectly by changing the event kinematics or by changing the number of jets in the event (and thus impact the combinatorics of which jets are selected).  

\begin{figure}[h!]
\begin{center}
\includegraphics[width=0.45\textwidth]{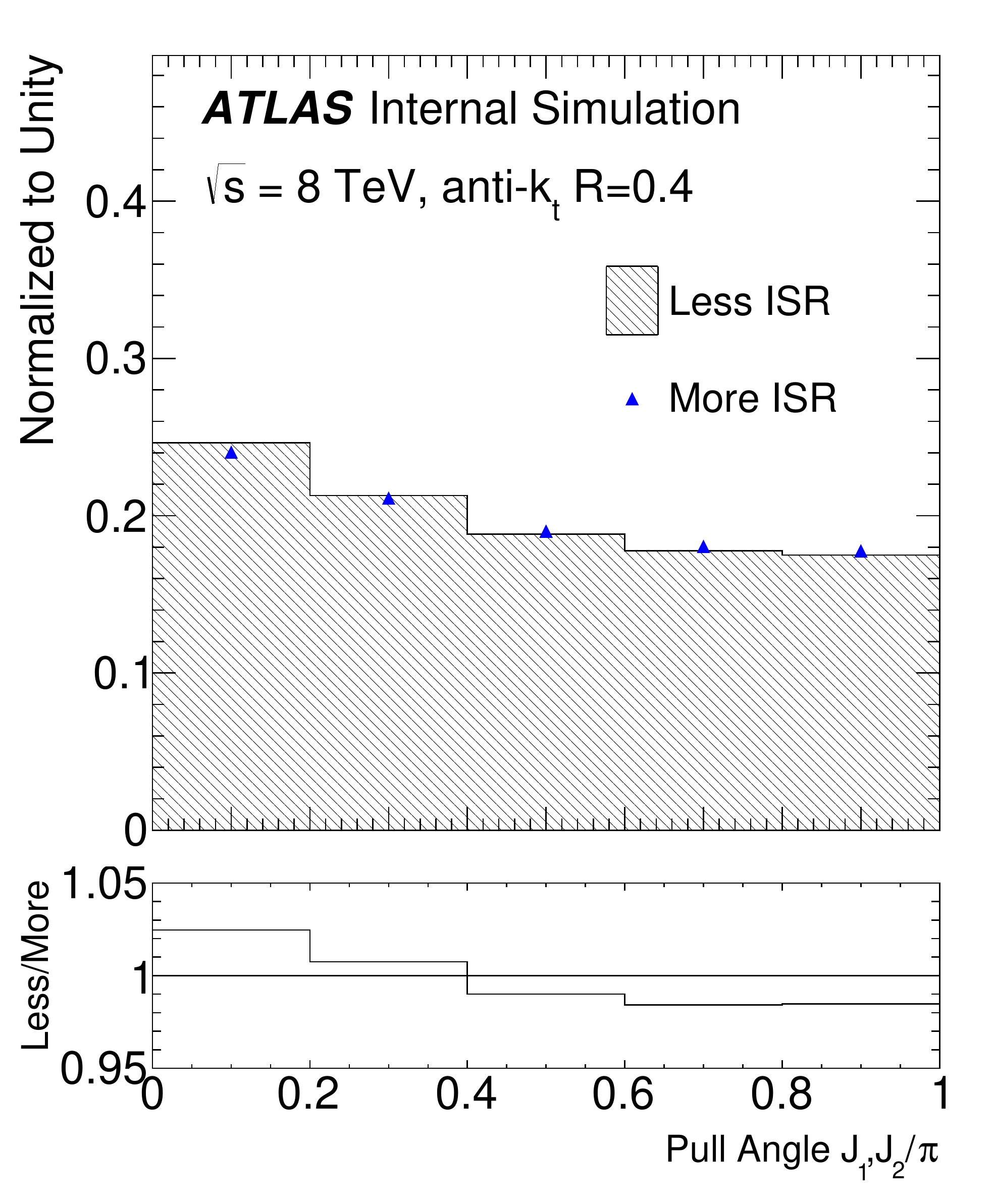}\includegraphics[width=0.45\textwidth]{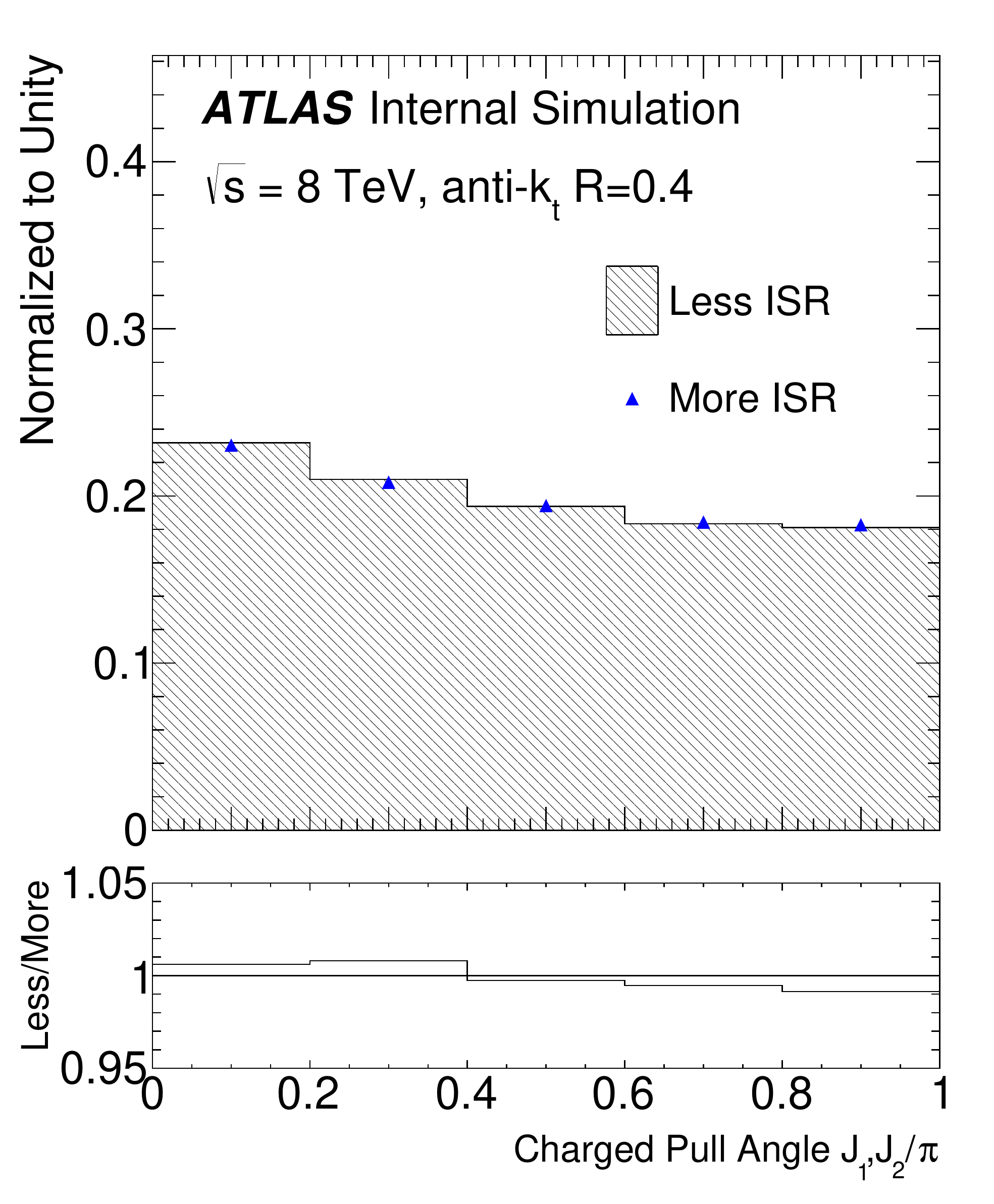}
\caption{The ISR/FSR variations at truth level for the all particle pull (left) and the charged particles pull (right).  Electron and muon channels combined.}
\label{ISRFSR}
\end{center}
\end{figure}

\subsubsection{Top $p_\text{T}$}
\label{syst:toppt}

It is well known~\cite{Aad:2014zka} that the {\sc Powheg-Box} + {\sc Pythia} 6 $t\bar{t}$ simulation at $\sqrt{s}=8$ TeV with  $h_\text{damp}=\infty$ (see Ref.~\cite{ATL-PHYS-PUB-2015-011}) has a slight mis-modeling at high $p_\text{T}$, which is seen clearly in the slope in the ratio plot in Figures~\ref{fig:pts_sub} and~\ref{fig:wpt}.  The $t\bar{t}$ $p_\text{T}$ spectrum only enters the measurement indirectly through the unfolding, since the pull angle distribution and the pull angle resolution depend on the jet $p_\text{T}$ spectrum.   The following procedure is used to assess the impact of this mis-modeling:

\begin{enumerate}[label={(\arabic*)}]
\item Extract a joint distribution of particle-level pull angle distribution and the particle-level leading jet $p_\text{T}$ from the nominal $t\bar{t}$ simulation.

\item Generate random points from (1) and smear the angle according to the Fig~\ref{fig:15}\footnote{The pull angle in Fig~\ref{fig:15} is not exactly the same as the one here because of the origin correction.  For the purpose of this test, the differences are sub-dominant.}.  From this step we get a nominal response matrix and a nominal detector-level distribution.

\item Generate random points from (1) but re-weight the $p_\text{T}$ spectrum (via assign event weights) so that it matches matches the data from Fig.~\ref{fig:pts_sub}.  This produces a shifted response matrix.

\item Unfold the nominal detector-level distribution with the shifted response matrix and compare to the nominal. 
\end{enumerate}

\noindent For this test, five equally spaced bins between $0$ and $\pi$ are used for the pull angle distribution.  The relative change in each bin after doing the comparison in step (4) is shown in Table~\ref{tab:reweight}.  The changes are negligibly small and are ignored for the remainder of the analysis.

\begin{table}[h]
\begin{center}
\begin{tabular}{|c|c|c|}
\hline
Bin Number & All Particles & Charged Particles \\ \hline
1          & 0.01\%        & 0.13\%            \\ \hline
2          & -0.07\%       & -0.08\%           \\ \hline
3          & -0.08\%       & -0.04\%           \\ \hline
4          & 0.05\%        & 0.006\%           \\ \hline
5          &  0.06\%             & -0.02\%           \\ \hline
\end{tabular}
\caption{The impact on the unfolded jet pull angle distribution from re-weighting the jet $p_\text{T}$ spectrum to match the data.}
\label{tab:reweight}
\end{center}
\end{table}

\subsubsection{Color flow Model}
\label{seccolorflowmodeluncert}

For the purpose of comparing the unfolded data with the flipped color model, it is necessary to take into account any potential biases the model has on the unfolding.  One way to estimate this uncertainty is to take the difference in the unfolded result when using the nominal versus flipped model for the response matrix.  This procedure is excessively conservative, because the model dependence is already covered by the non-closure uncertainty described in Sec.~\ref{sec:colorflow:nonclosure}.  Therefore, for any result other than a comparison between the unfolded data and the flipped model, this uncertainty should not be included. Since the pull angle distribution for the flipped sample is significantly different than the nominal one, Sec.~\ref{sec:ColorFlow:UnfoldingParams} suggests that this uncertainty may not be small.  This is confirmed by Fig.~\ref{colorflowmodeluncert}.  As desired, the color fow model uncertainty is significantly smaller than the color flow model difference, but is still $\sim 1\%$ in some bins.

\begin{figure}[h!]
\begin{center}
\includegraphics[width=0.95\textwidth]{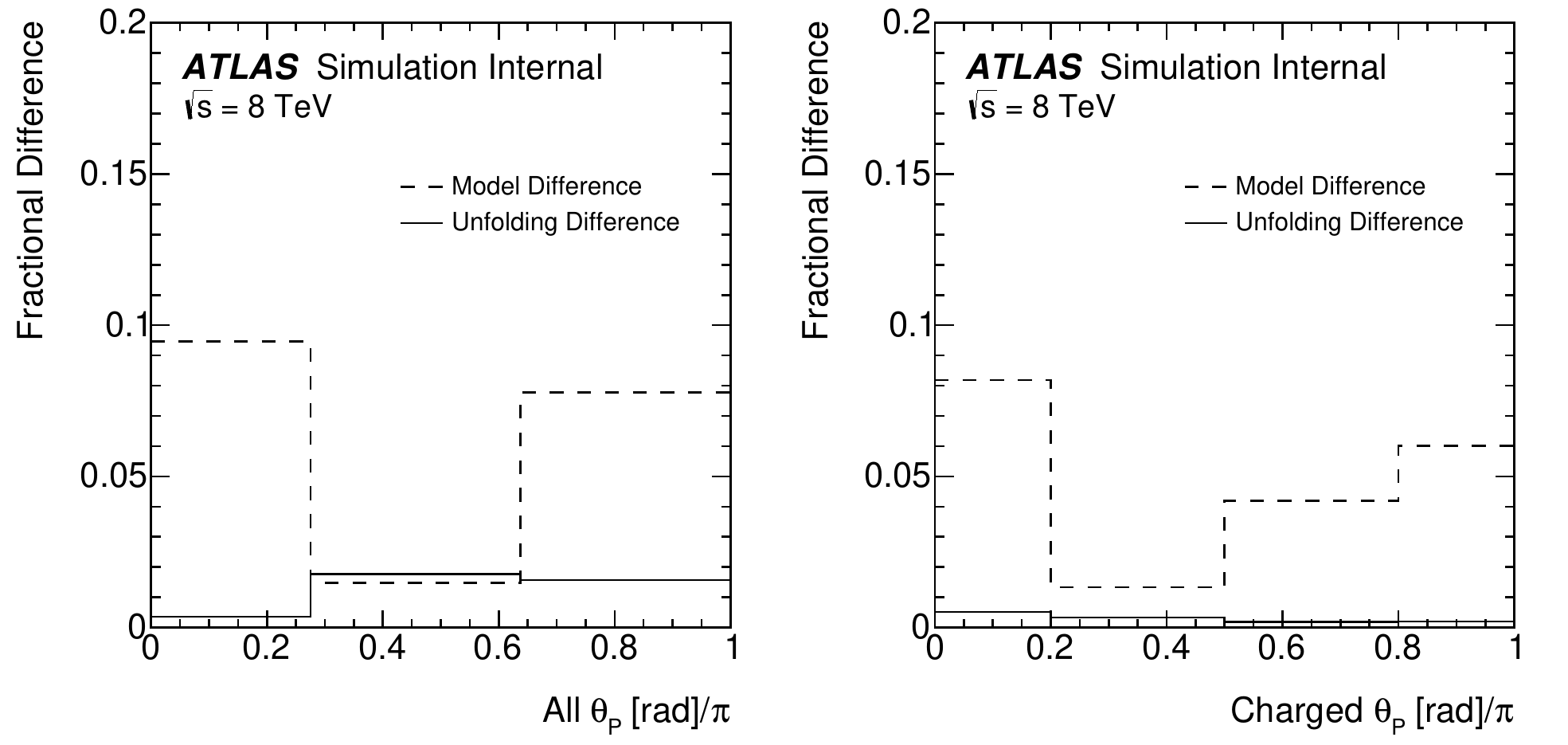}
\caption{The fractional difference between the nominal particle-level distribution and the nominal detector-level distribution unfolded with a response matrix constructed from the flipped sample ({\it unfolding difference}).  The dashed line shows the fractional difference at particle-level between the two models.  All distributions are normalized to unity before computing fractional differences.}
\label{colorflowmodeluncert}
\end{center}
\end{figure}

\subsubsection{Other}

Other sources of uncertainty include the choice of factorization and renormalization scale in the ME calculation and the PDF~\cite{Botje:2011sn}.  As observed in Sec.~\ref{syst:colorflow:ME}, these variations have little impact on the color flow.  Additionally, varying the top quark mass by $\pm 1$ GeV\ has a negligible impact on this measurement.

\subsection{Correction Factors}

Uncertainties in the correction factors from Sec.~\ref{colorflowcorrectionfactors} are accounted for as part of all other uncertainties described thus far.  The fake and inefficiency factors are modified in addition to the response matrix for all the sample variations.  As an example, Fig.~\ref{syst:fake} (\ref{syst:effic}) shows the variation in the fake (inefficiency) factor for the various ME and fragmentation models considered in Sec.~\ref{syst:colorflow:ME}.  The correction factors are largely independent of the pull angle, and the largest uncertainty is on the overall acceptance from the fragmentation model.  {\sc Pythia} and {\sc Herwig} predict $\sim 3\%$ differences in the fake factors and $\sim 15\%$ in the inefficiency factors.  Since the unfolded distributions are normalized to unity for the final result, overall differences in acceptance from the unfolding have no effect on the measurement.

\begin{figure}[h!]
\begin{center}
\includegraphics[width=0.45\textwidth]{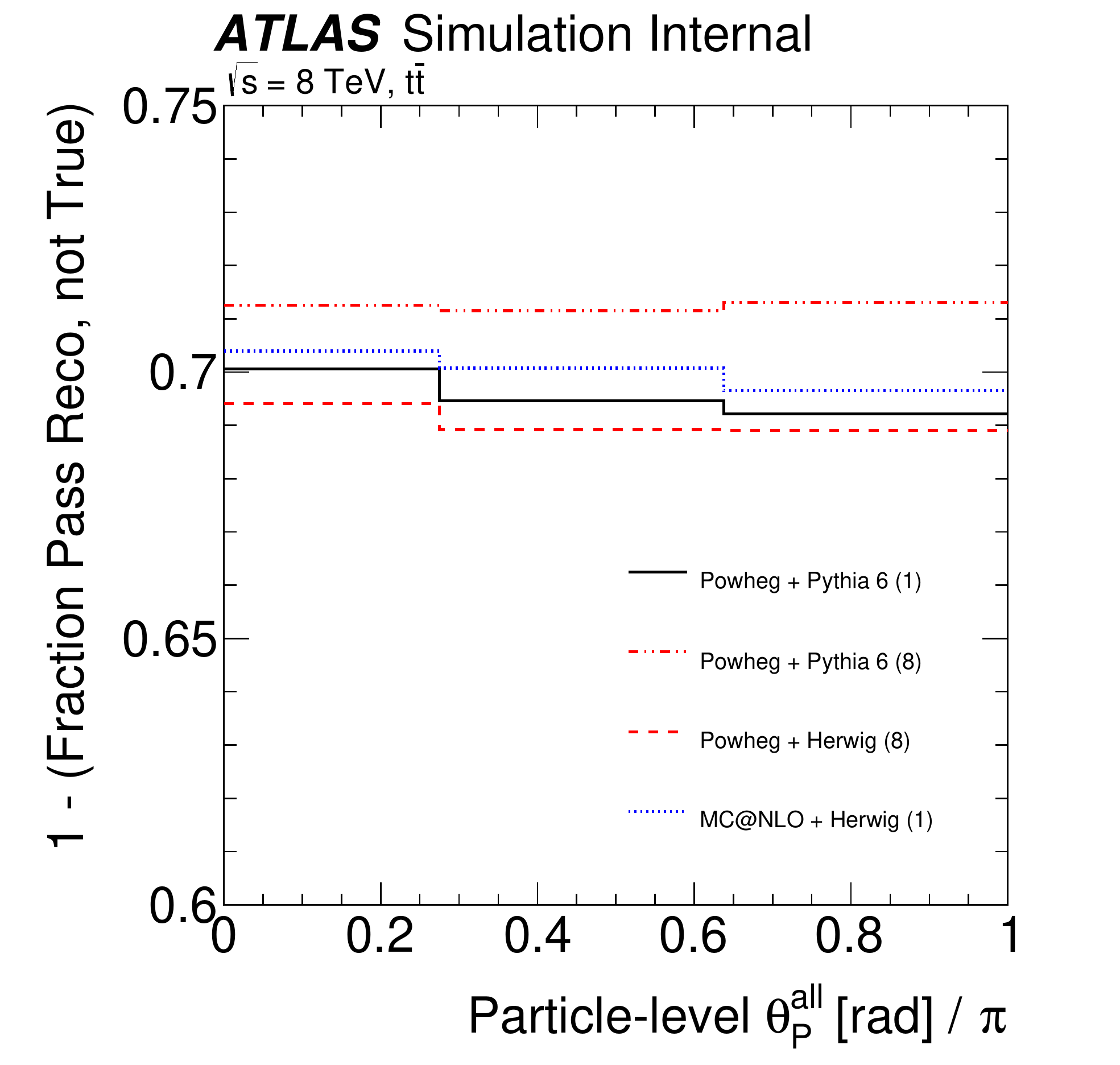}
\includegraphics[width=0.45\textwidth]{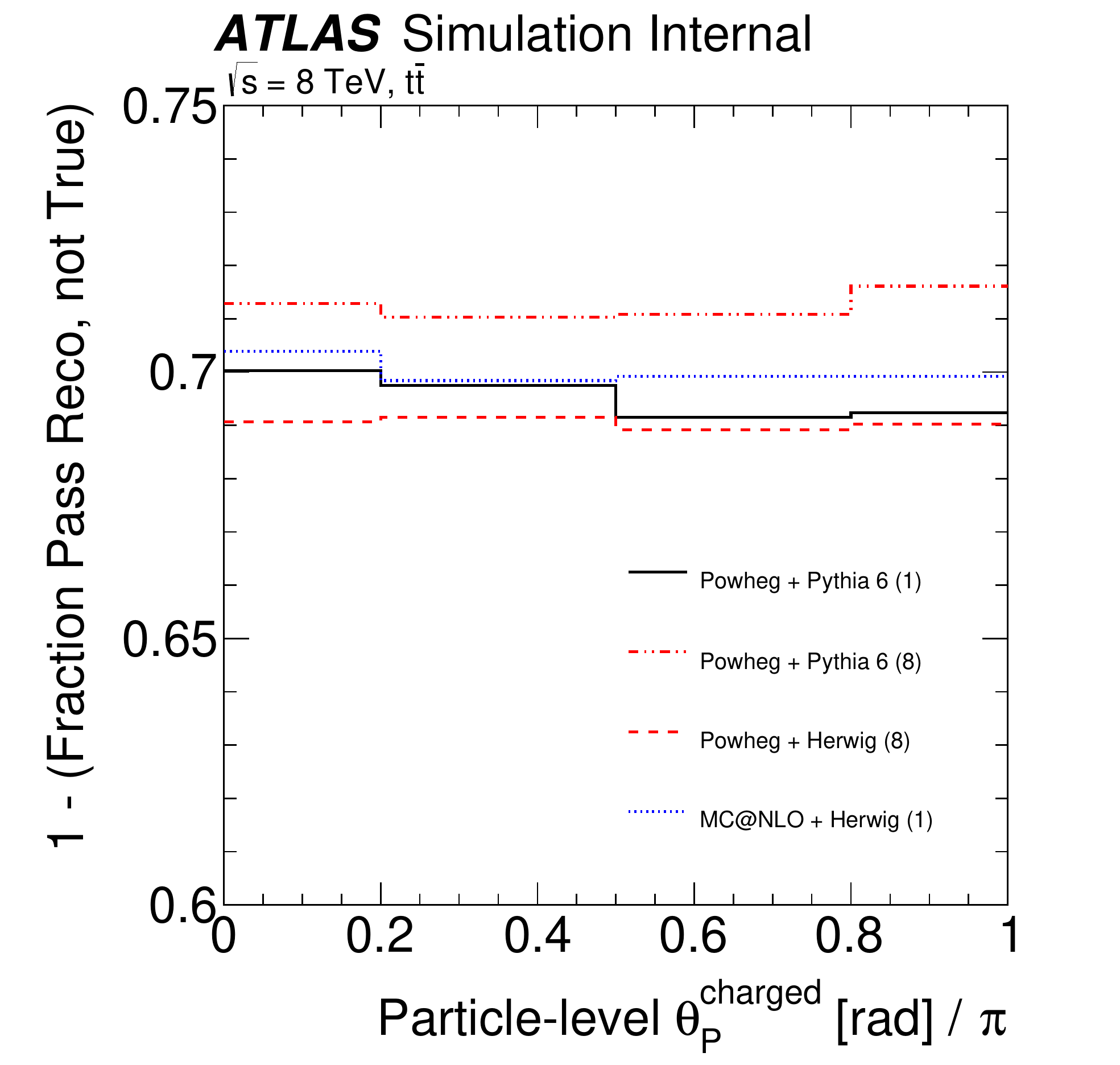}
\end{center}
\caption{The fake factors for the all-particles pull angle (left) and the charged-particles pull angle (right).  The (1) and (8) in the legend refers to the nominal and flipped sample, respectively.}
\label{syst:fake}
\end{figure}

\begin{figure}[h!]
\begin{center}
\includegraphics[width=0.45\textwidth]{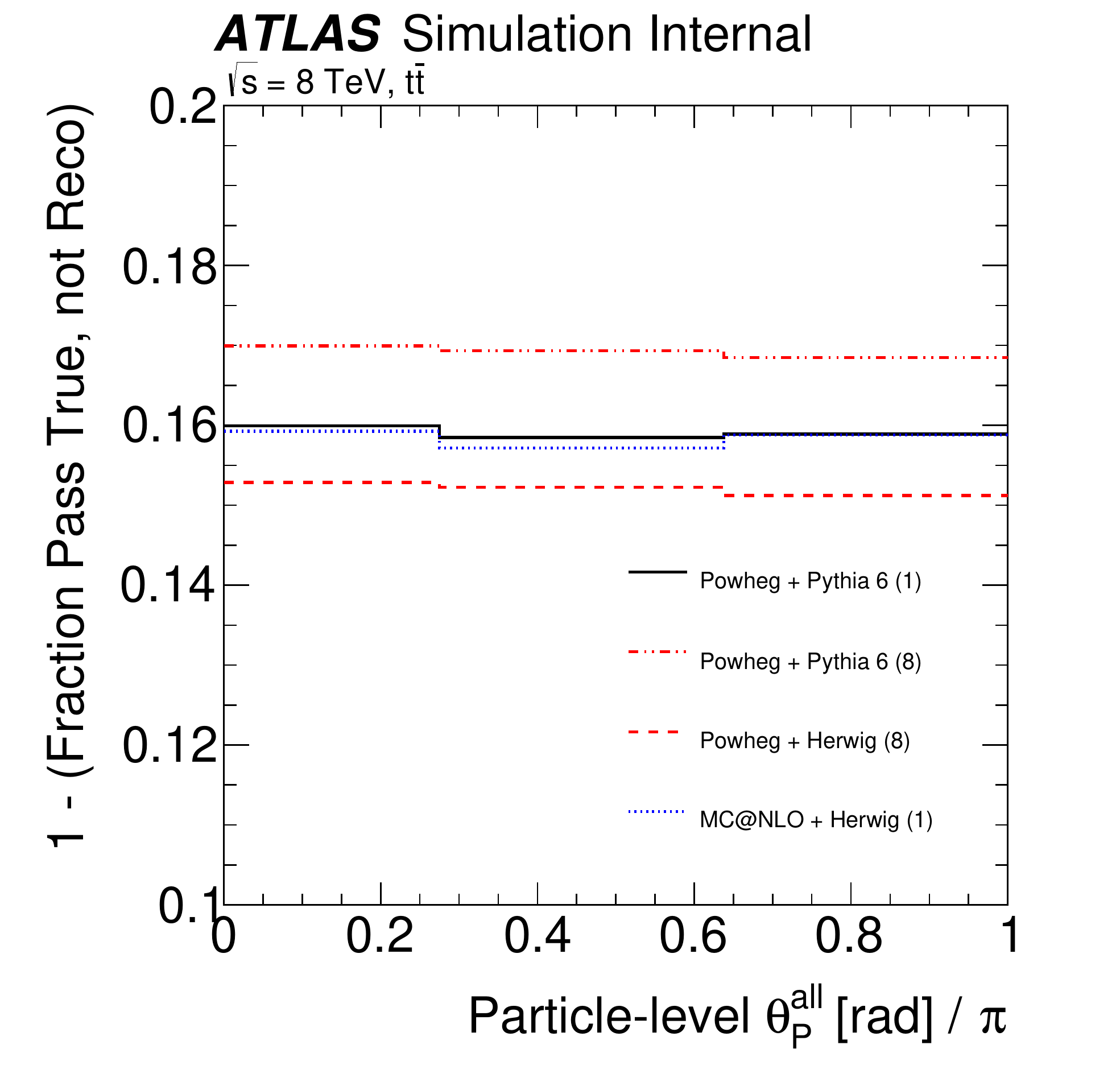}
\includegraphics[width=0.45\textwidth]{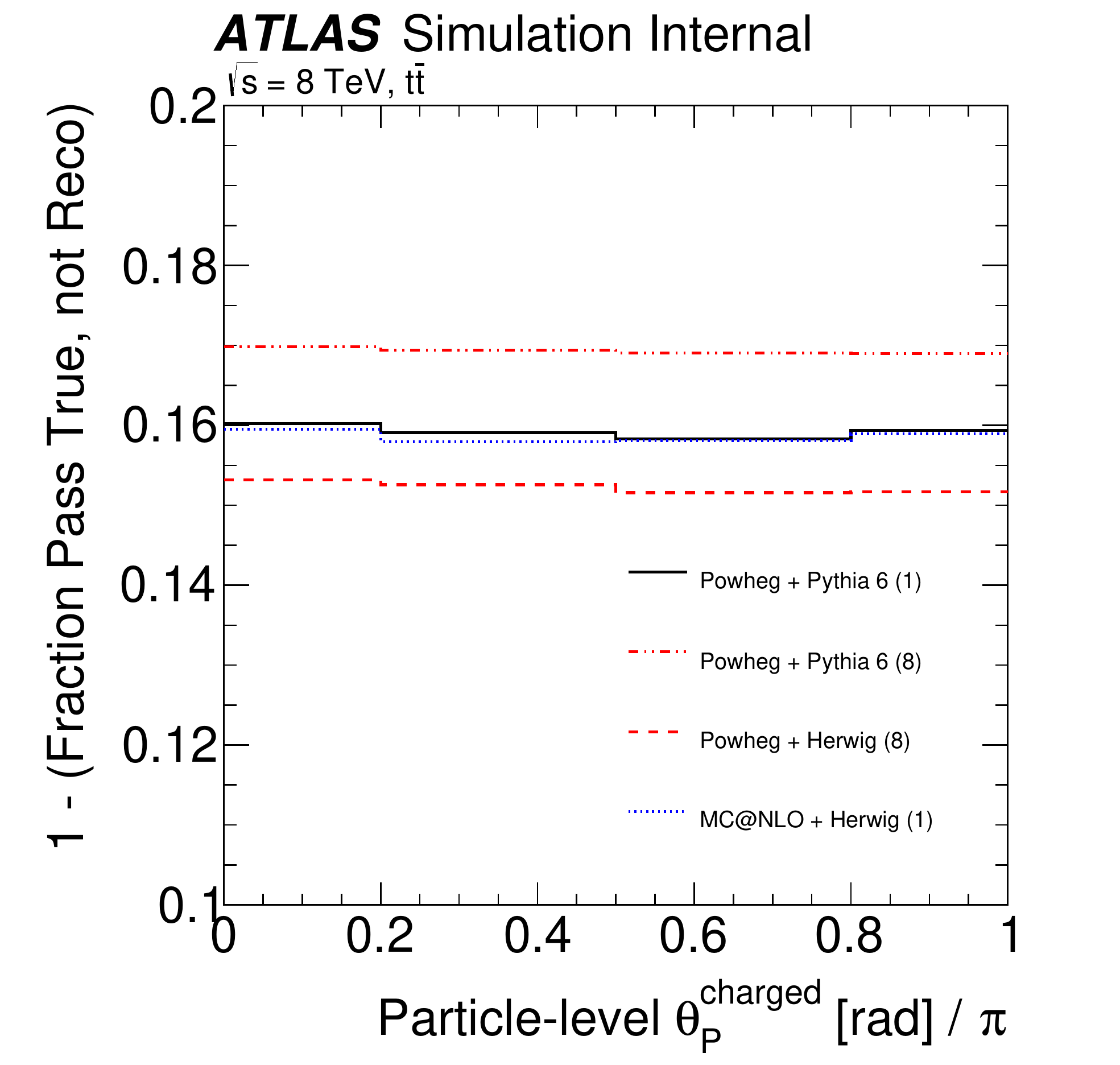}
\end{center}
\caption{The inefficiency factors for the all-particles pull angle (left) and the charged-particles pull angle (right).  The (1) and (8) in the legend refers to the nominal and flipped sample, respectively.}
\label{syst:effic}
\end{figure}

\subsection{Non-closure}
\label{sec:colorflow:nonclosure}

The non-closure uncertainty uses the same data-driven technique that is described in detail in Sec.~\ref{sec:statscharge}.  As the detector-level simulation agrees well with the data (Fig.~\ref{fig:reco_dists}), the amount of reweighting is minimal. 

\clearpage

\subsection{Summary}
\label{sec:ColorFlow:syst:summary}

The systematic uncertainties are estimated by unfolding the data with varied
response matrices or by subtracting varied background predictions from the data.   Table~\ref{tab:unfolding:systematics_all} summarizes the various sources of systematic and statistical uncertainty for both the all-particles and charged-particles pull angles.  The modeling uncertainties dominate and the total uncertainty is about $3\%$ for the all-particles pull angle and about $2\%$ for the charged-particles pull angle.

\begin{table}[h!]
    \centering
      \noindent\adjustbox{max width=\textwidth}{
    \begin{tabular}{ |c|ccc |cccc|}
     \hline
        \multirow{2}{*}{Uncertainty [\%]}&& \multicolumn{1}{c}{$\theta_p^{\mathrm{all}}$ [rad]$/\pi$}&&& \multicolumn{2}{c}{$\theta_p^{\mathrm{charged}}$ [rad]$/\pi$}& \\
        \cline{2-8}
      & 0.0 - 0.275 & 0.275 - 0.6375 & 0.6375 - 1.0  & 0.0 - 0.2 & 0.2 - 0.5 & 0.5 - 0.8 & 0.8-1.0 \\     
     \hline
       \hline
            $t\bar{t}$ NLO generator                   & 1.61        & 0.50           & 1.00       &  0.94      & 0.17      & 0.05      & 1.47   \\
      Fragmentation Model  & 1.61        & 0.98           & 0.48     & 0.52      & 0.31      & 0.46      & 0.56   \\
      ISR/FSR                                   & 1.18        & 0.61           & 0.47      &  0.22      & 0.04      & 0.00      & 0.34 \\      
      Color reconnection                      & 0.54        & 0.37           & 0.92   & 0.40      & 0.29      & 0.16      & 0.23       \\
   MPI                                     & 0.20        & 0.13           & 0.04      &  0.59      & 0.32      & 0.41      & 0.42  \\        
       Color model                             & 1.22        & 1.70           & 0.69      & 1.12      & 0.18      & 0.52      & 0.25  \\  
        \hline   
            Non-closure                             & 0.47        & 0.06           & 0.38     &0.61      & 0.58      & 0.32      & 1.19     \\
      JES                                     & 0.43        & 0.18           & 0.49       & 0.22      & 0.15      & 0.16      & 0.00    \\
      JER                                     & 0.27        & 0.01           & 0.26      &0.03      & 0.12      & 0.17      & 0.49    \\
      Clusters                                & 0.03        & 0.06           & 0.04    &  & \multicolumn{2}{c}{N/A} &       \\
     Tracks                                &         & N/A           &    & 0.04      & 0.02      & 0.05      & 0.00      \\      
      Other                                   & 0.38        & 0.13           & 0.45      &  0.20      & 0.15      & 0.14      & 0.00    \\      
          \hline
            \hline
      Stats.                                  & 1.12        & 0.63           & 1.12     & 0.68      & 0.51      & 0.54      & 0.77      \\
      Total                                   & 3.20        & 2.26           & 2.16     & 1.97      & 1.00      & 1.07      & 2.26     \\
                     \hline
    \end{tabular}
    }
    \caption{Uncertainties in each bin of the all-particle pull angle.
      The ``Other'' category includes uncertainties due to the non-\ttbar\
      backgrounds.}
    \label{tab:unfolding:systematics_all}
\end{table}

Figure~\ref{syst:minorfullcovariance} shows the full systematic uncertainty covariance matrix for the experimental and background normalization uncertainties\footnote{There is some ambiguity on the sign of the off-diagonal terms, especially for the modeling uncertainties for which there is no well-defined notion of `shift up/down'.  For the experimental uncertainties, a natural choice is to take the (signed) uncertainty as nominal - shifted.  The covariance matrix for the modeling uncertainties is omitted here, but is revisited in Sec.~\ref{sec:colorflow:results}.}.  The covariance matrix is the sum of the matrices from individual sources of uncertainty.  A partition of these uncertainties into four components is displayed in Fig.~\ref{syst:minorfullcovariance2}.  By construction, the cluster and tracking uncertainties are only relevant for the all-particles or charged-particles pull angles, but not both.  The significant correlation in the first bin of Fig.~\ref{syst:minorfullcovariance} is from the jet energy resolution, shown in the bottom left matrix in Fig.~\ref{syst:minorfullcovariance2}.  The per-bin uncertainties are dominated by the diagonal components. 

As observed in Fig.~\ref{syst:minorfullcovariance}, the systematic uncertainties induce correlations between bins of the same observable and between the all-particles and charged-particles pull angles.  Correlations between the variables are also present from coherent jet-by-jet statistical fluctuations and correlations between bins of the same variable are induced from the unfolding and from normalization.  The top right plot of Fig.~\ref{syst:minorfullcovariance} shows that the the all-particles and charged-particles pull angles are largely uncorrelated but there is a positive association, as expected ($\rho=0.23$).  The realization of this correlation in the binning used for the measurement is shown in the bottom right plot of Fig.~\ref{syst:minorfullcovariance}.  By construction, the bins of the same variable are independent of each other and correlations are measured by the off-diagonal blocks.  There is a general positive correlation across all off-diagonal bins because the total yield between the two variables is correlated.  This general correlation is removed in the bottom left plot of Fig.~\ref{syst:minorfullcovariance} by normalizing per variable.  For example, the residual correlation between the first bins of both variables is about $15\%$.   The normalization also induces a significant negative correlation between bins of the same variable due to the small number of total bin: if one bin content fluctuates up, the others have to be lower if the total integral is the same.  The amount of the normalization-induced correlation scales with the bin width.  After unfolding and normalization, there is still a significant negative correlation between bins of the same variable, but the magnitudes have changed.  This is in part due to the large correlation between neighboring bins induced by the fact that the pull angle resolution is not small compared to the range, $\pi$.

\begin{figure}[h!]
\begin{center}
\includegraphics[width=0.7\textwidth]{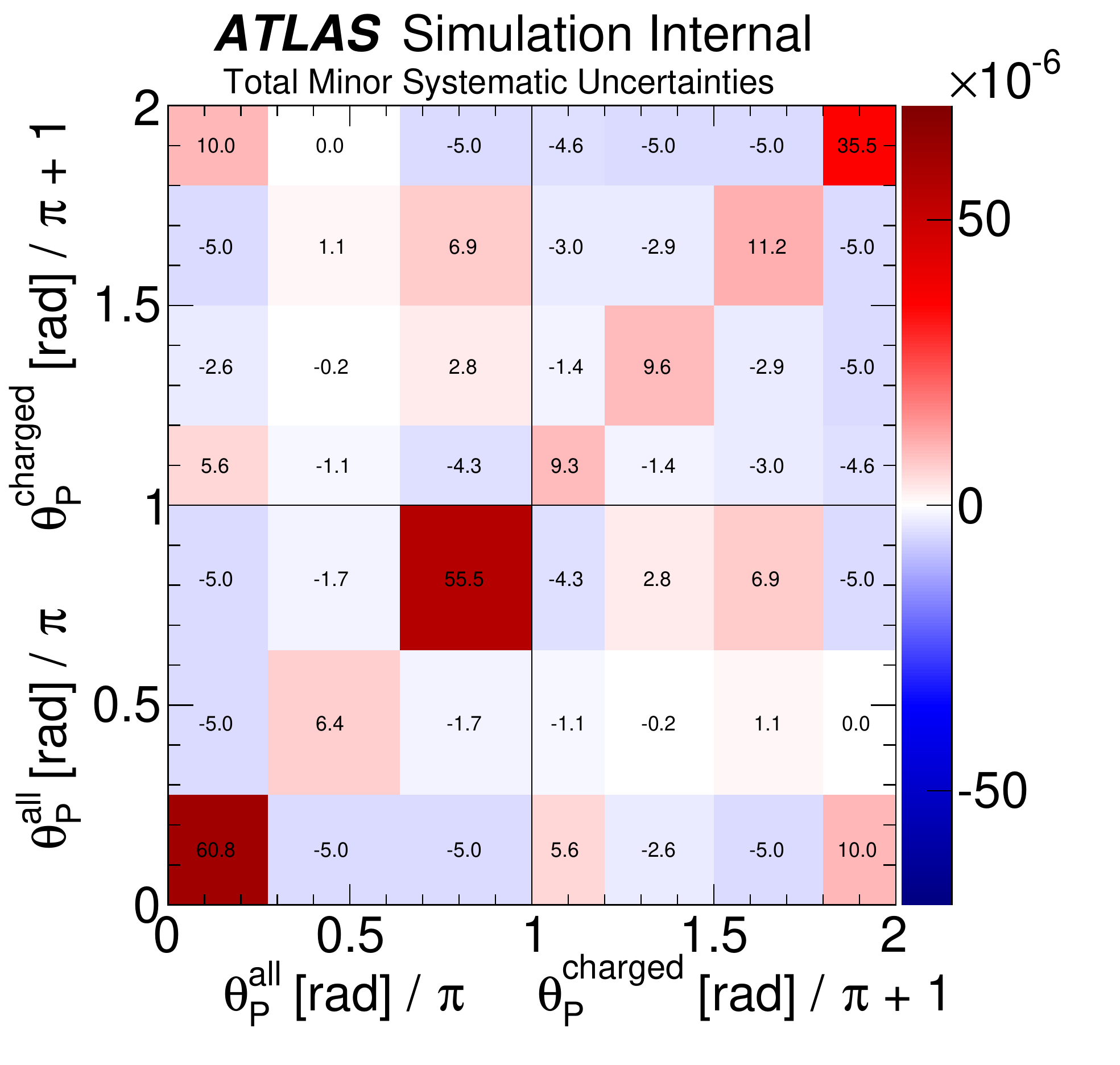}
\end{center}
\caption{The systematic uncertainty covariance matrix for all experimental and background normalization uncertainties.  If the matrix where proportional to the identity matrix, than the systematic uncertainty in bin $i$ of the all-particles pull angle would be $ \sqrt{\Sigma_{ii}}$ and in bin $i$ of the charged-particles pull angle would be $\sqrt{\Sigma_{i+3,i+3}}$ for $\Sigma$ a matrix representing the plot above. The matrix $\Sigma$ is the sum of the matrices from all individual sources of uncertainty.  }
\label{syst:minorfullcovariance}
\end{figure}

\begin{figure}[h!]
\begin{center}
\includegraphics[width=0.5\textwidth]{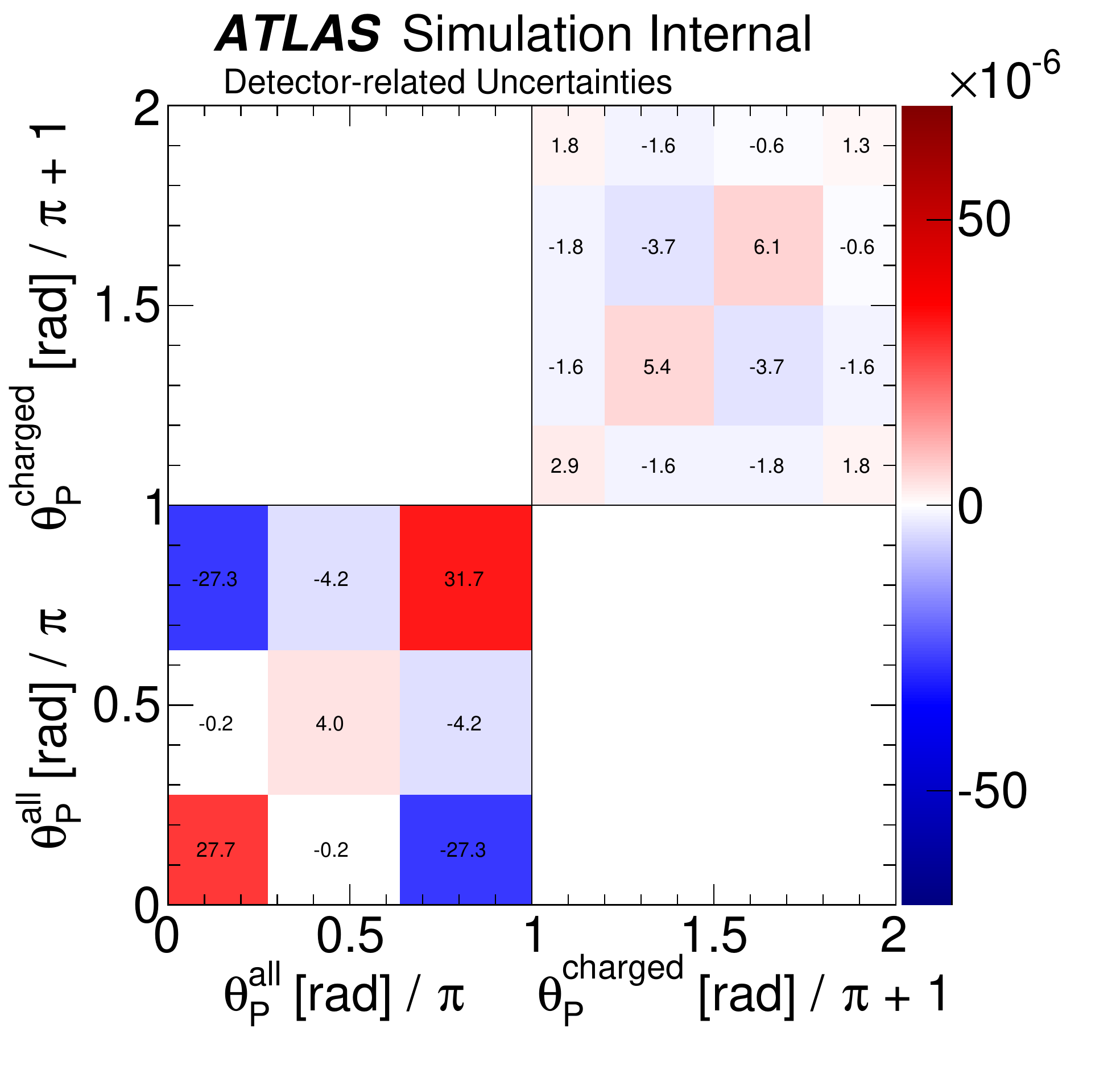}\includegraphics[width=0.5\textwidth]{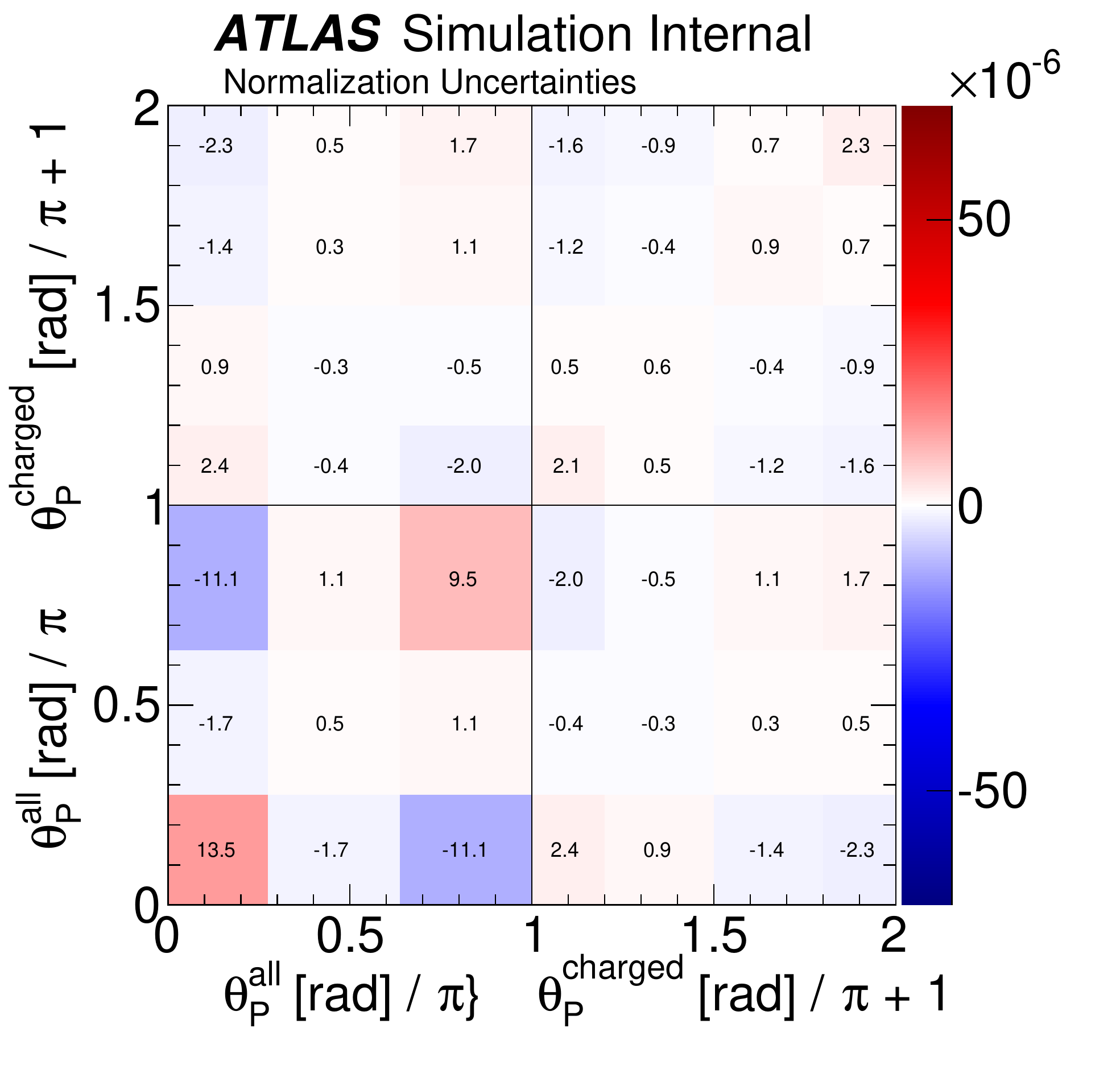}\\
\includegraphics[width=0.5\textwidth]{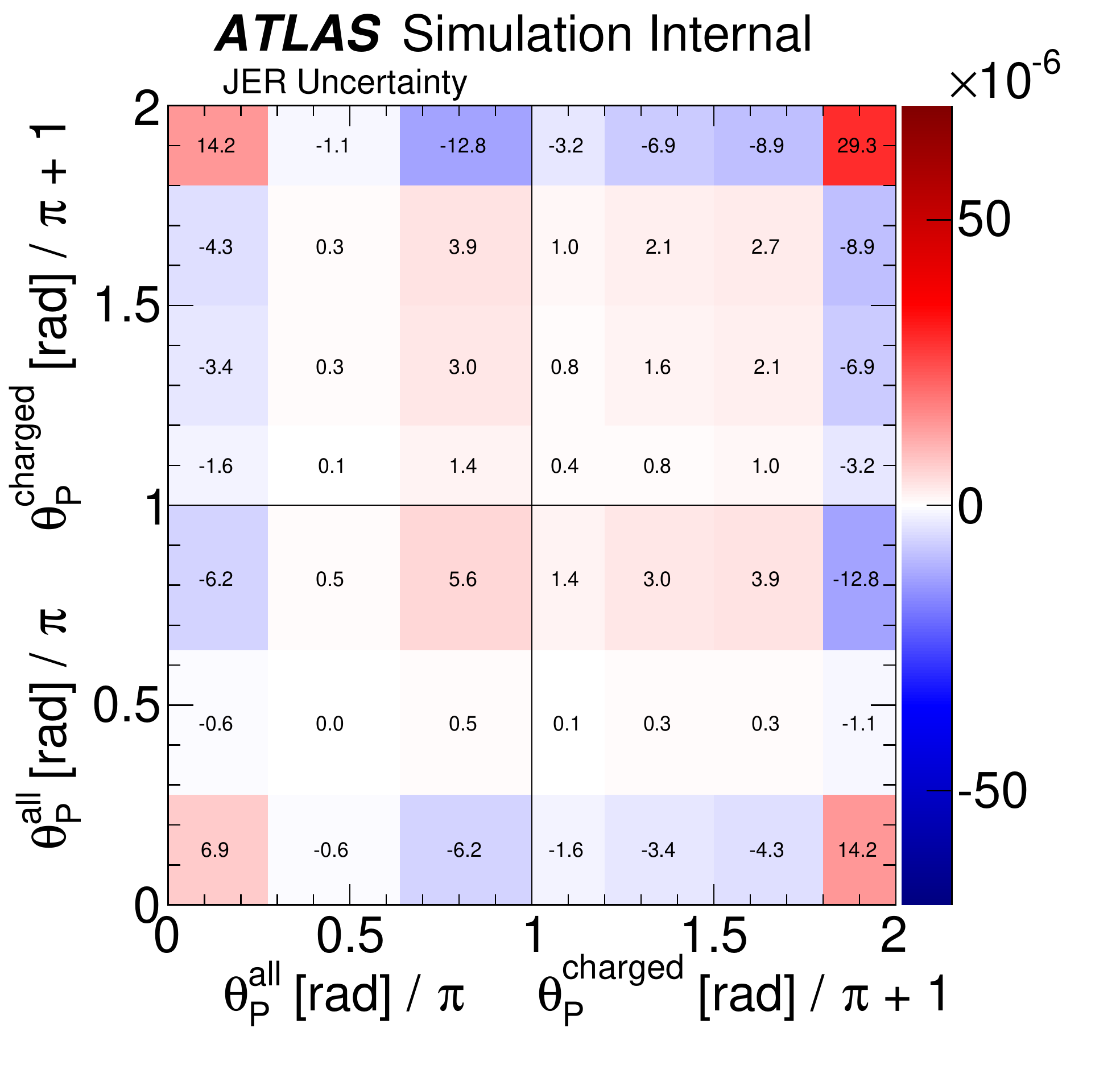}\includegraphics[width=0.5\textwidth]{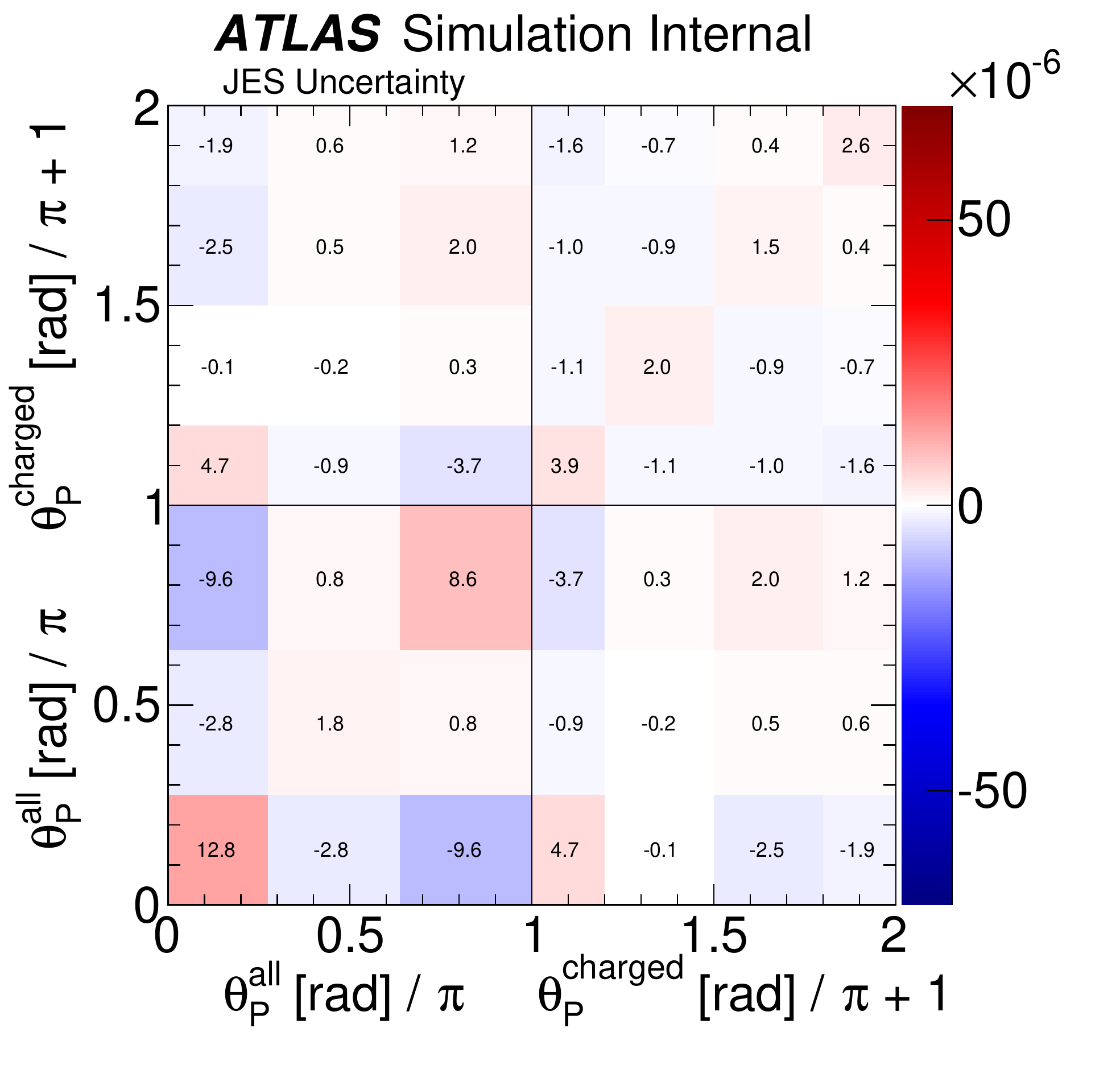}
\end{center}
\caption{The same as Fig.~\ref{syst:minorfullcovariance}, but broken into four categories: cluster and tracking (top left), background normalization (top right), jet energy resolution (bottom left), and jet energy scale (bottom right).}
\label{syst:minorfullcovariance2}
\end{figure}

\begin{figure}[h!]
  \centering
    \includegraphics[width=0.45\textwidth]{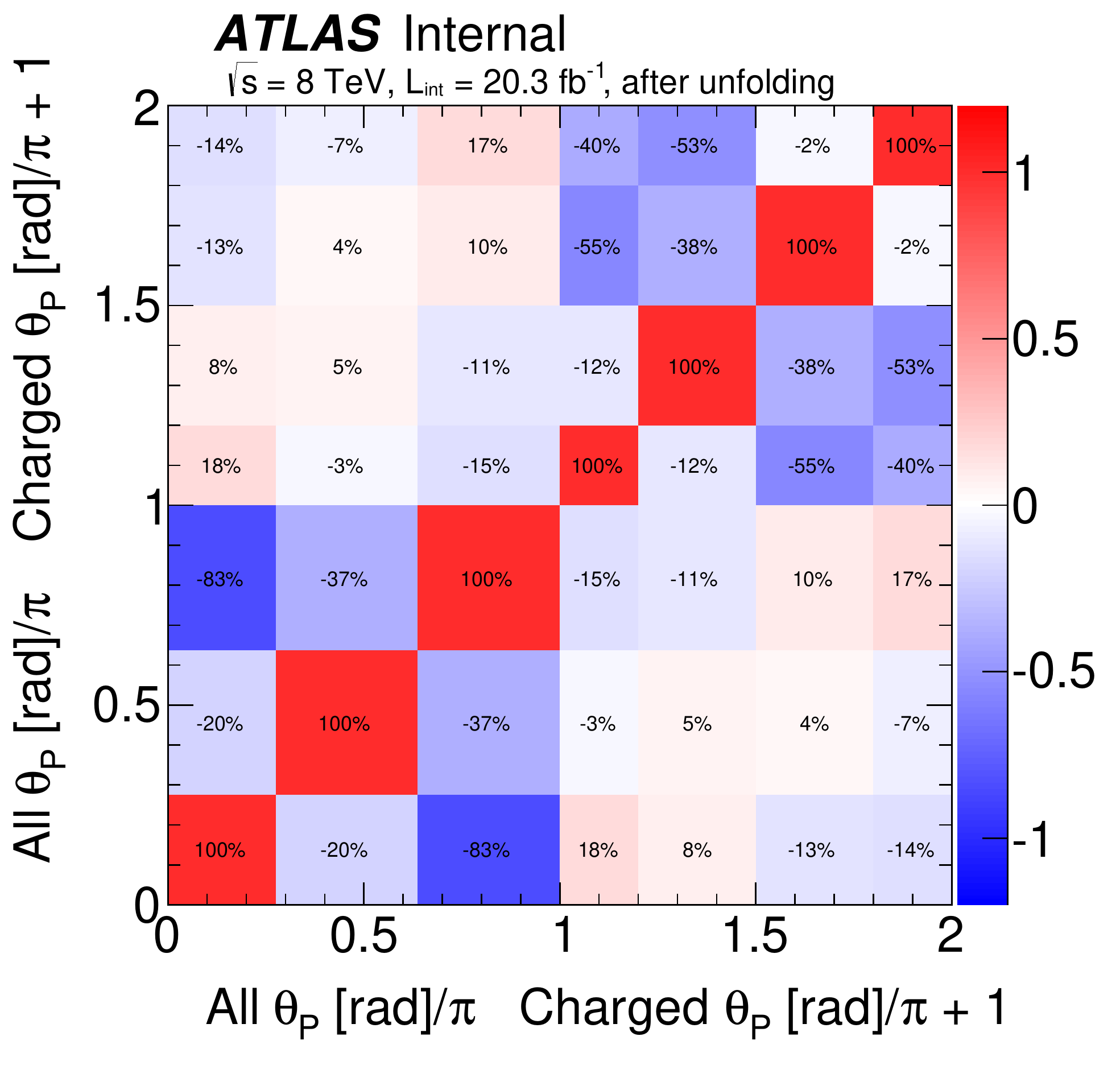}
     \includegraphics[width=0.45\textwidth]{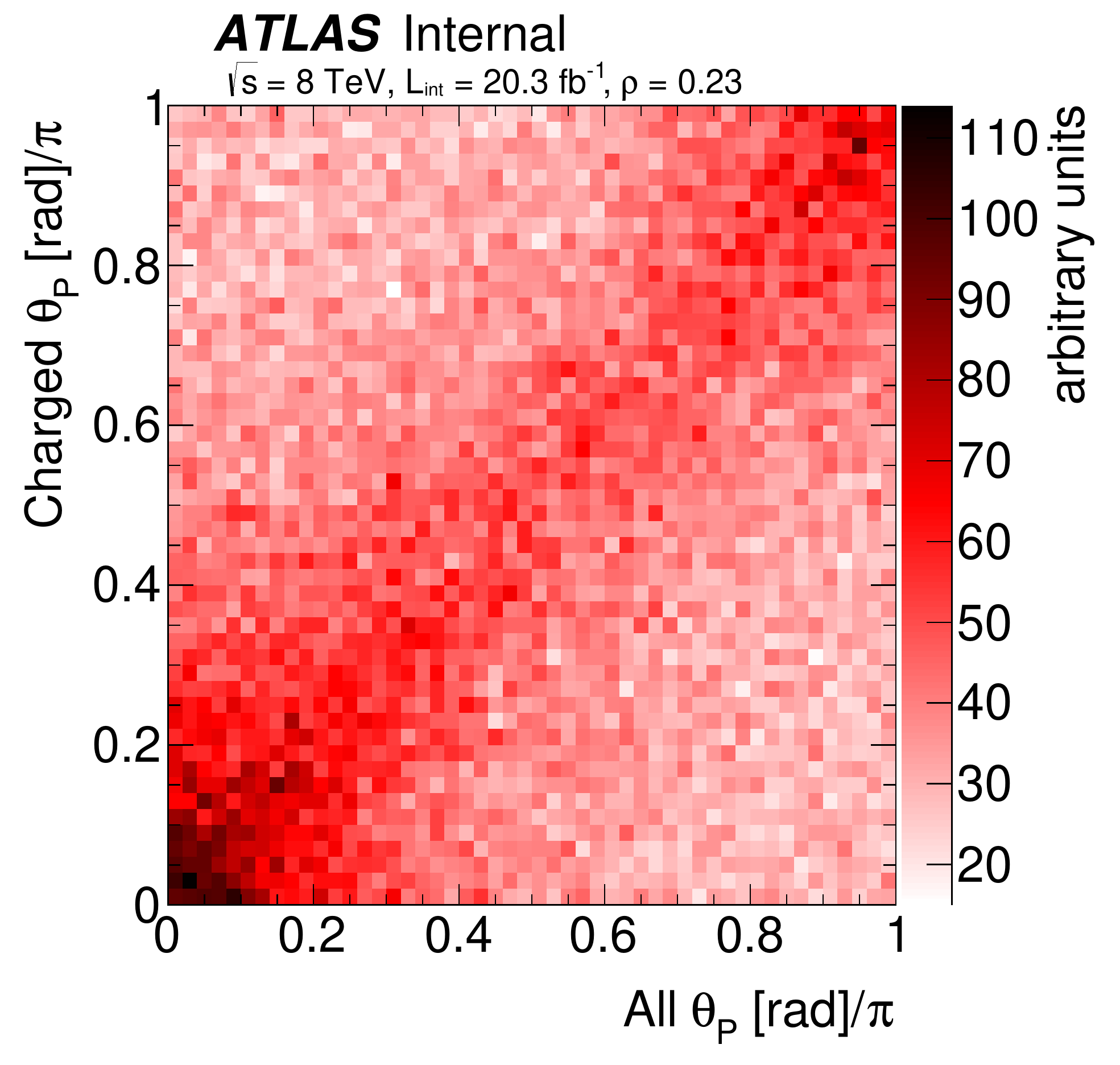}
     \includegraphics[width=0.45\textwidth]{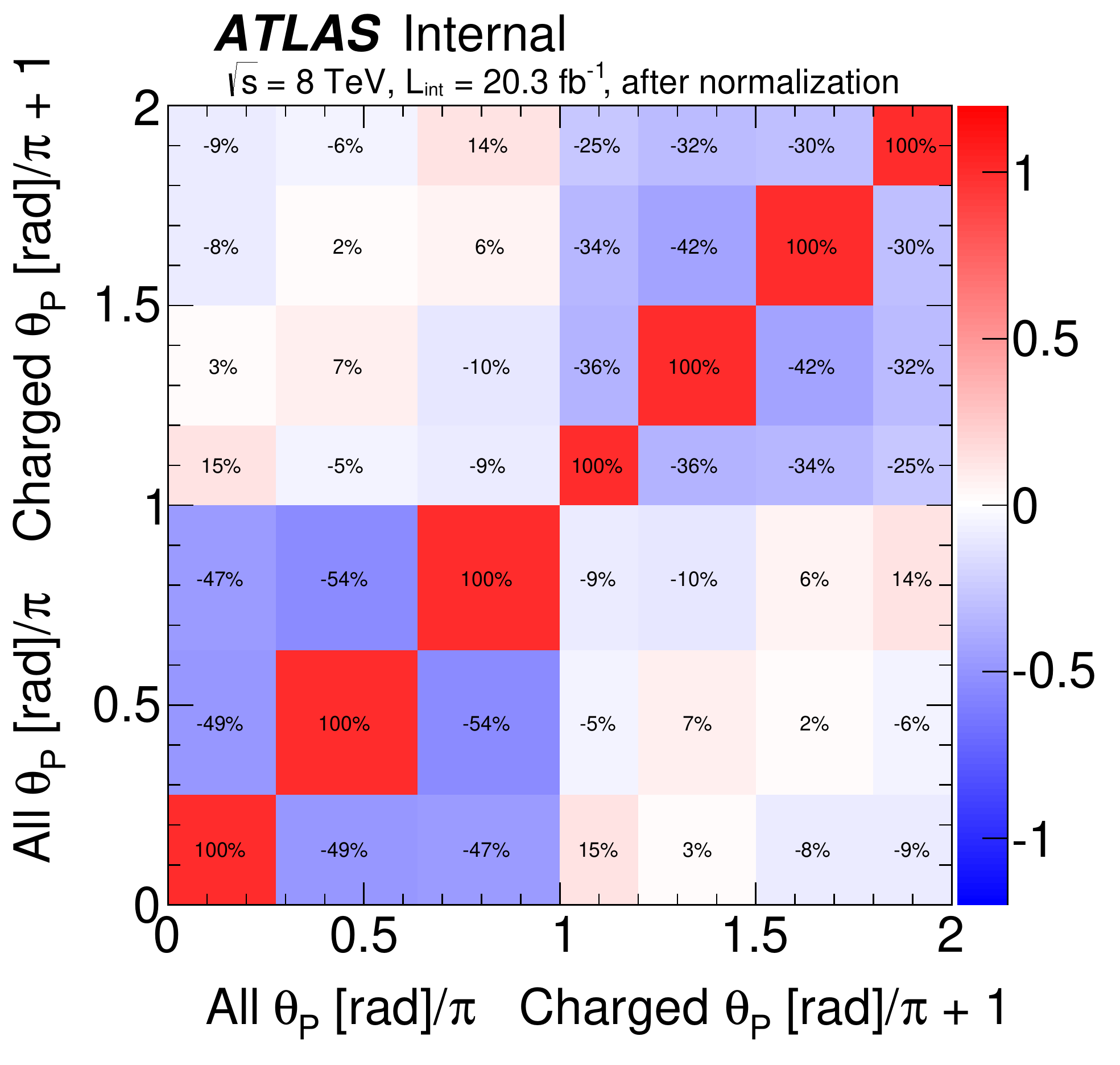}
     \includegraphics[width=0.45\textwidth]{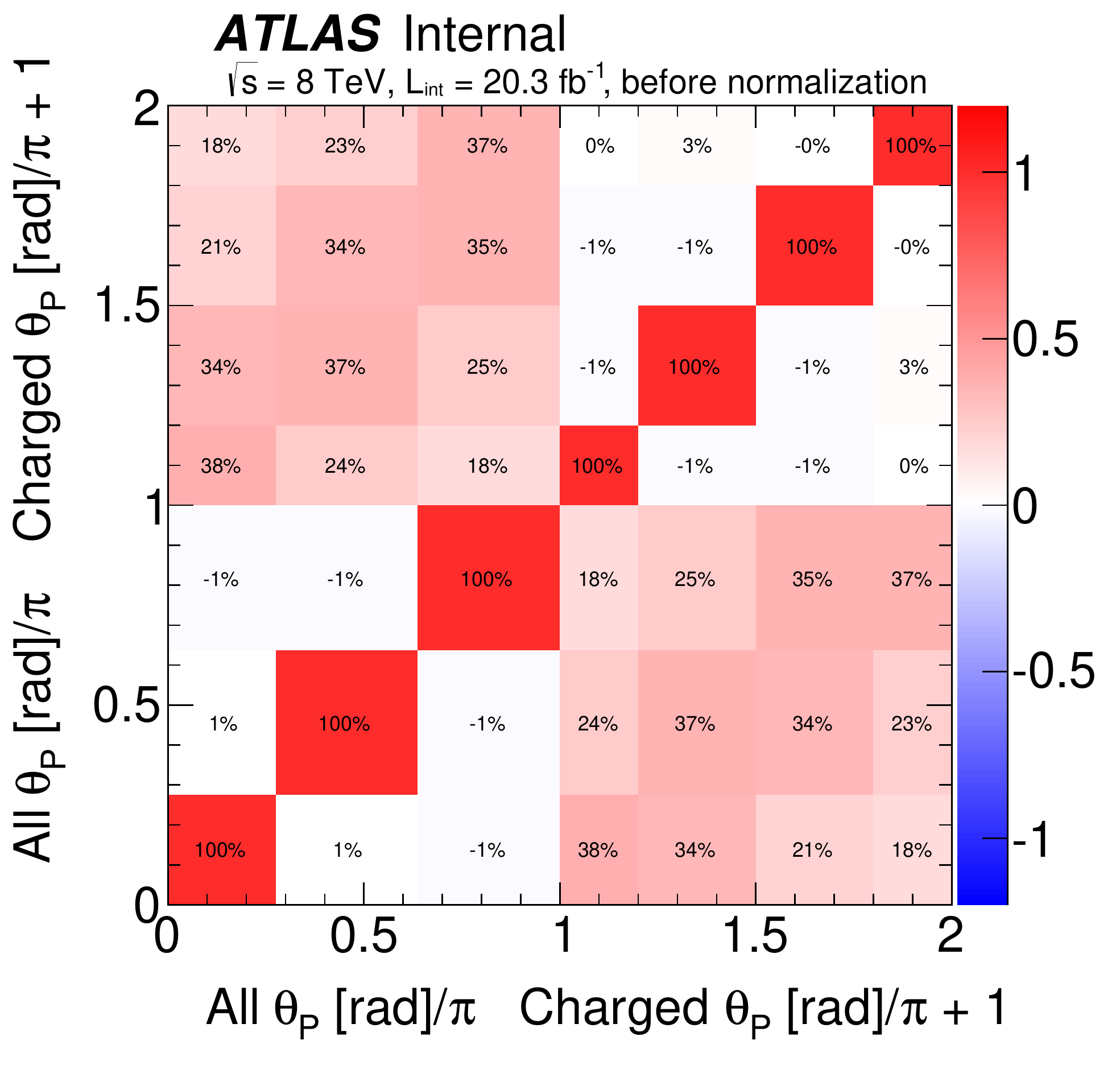}
  \caption{Representations of correlations between and within bins of the all-particles and charged-particles pull angles induced from the unfolding (top left), normalization (bottom left), and jet-by-jet statistical fluctuations (top and bottom right).  The unfolding correlation includes normalization.  These matrices are generated by running the unfolding procedure on the data $10,000$ times, each time generating {\it pseudo-data} $X_{ij}\sim\text{Poisson}(\lambda_{ij})$, where $\lambda_{ij}$ is the number of events measured in bin $(i,j)$.  For each pseudo-dataset, the backgrounds are subtracted before (normalization, unfolding, and) computing the correlation.}
  \label{fig:statuncert}
\end{figure}

\clearpage

\section{Results}
\label{sec:colorflow:results}

The unfolded data are shown in Fig.~\ref{fig:unfolded:pull_all} for both the all-particles and charged-particles pull angles.  The data generally fall between the SM color flow and the flipped model, though the agreement with the SM is significantly better.  Most of the difference between the models is in the first bin and due to the normalization, there is little spread in the distributions for the second bin of both variables.  The flipped model is about $2.3\sigma$ away from the data in the first bin, while it is about $3.3\sigma$ in the first bin for the charged-particles pull angle.  It is possible to quantify the compatibility using all bins by computing the probability distribution of the log likelihood ratio\footnote{The likelihood ratio test is the most {\it powerful} by the Neyman-Pearson lemma.  See Sec.~\ref{sec:susy:stats} for details.  Since the logarithm is a monotonic function, the log-likelihood ratio test is also the most powerful.}.  Assuming the distribution of the pull angle follows a multivariate Gaussian distribution, the log likelihood ratio is (up to constants) simply the difference in $\chi^2$:

\begin{align}
\label{eq:deltachi2}
\log(p_\text{SM}/p_\text{flipped})(\vec{x})=\sum_{i=1}^4\frac{(x_i-x_{i,\text{SM}})^2}{\sigma_i^2}-\sum_{i=1}^4\frac{(x_i-x_{i,\text{flipped}})^2}{\sigma_i^2},
\end{align}

\noindent where $\sigma_i$ is the uncertainty on bin $i$ and $x_{i,\text{M}}$ is the $i^\text{th}$ bin content under model $\text{M}$.  The probability distribution of Eq.~\ref{eq:deltachi2} can be evaluated numerically, taking into account correlation between bins, by generating pseudo-data from the measurement covariance matrix.  As mentioned in Sec.~\ref{sec:ColorFlow:syst:summary}, the covariance matrix is well-defined for the experimental systematic and statistical uncertainties but is not well-defined for the (dominant) modeling uncertainties.  Despite this, one can estimate the mutlibin sensitivity by selecting a convention; in this case, the sign of the covariance matrix is from the varied sample prediction subtracted from the nominal prediction.   Figure~\ref{fig:unfolded:deltachi2} shows the probability distribution for Eq.~\ref{eq:deltachi2} (charged-particles pull angle) under both the SM and flipped hypotheses using ten million pseudo-experiments.  The test statistic for the data is indicated by an arrow.  The data is inconsistent with the flipped model at about $4\sigma$ (observed) while the nominal MC is inconsistent with the flipped model at about $5\sigma$ (expected), for $\sigma = \Phi^{-1}(1-\text{$p$-value})$, for the Gaussian cumulative distribution function $\Phi$.  A similar exercise with the all-particles pull angles results in lower significances such that the full combination of the two variables is dominated by the charged-particles pull angle significance. 

\begin{figure}[htbp]
  \centering
    \includegraphics[width=0.5\textwidth]{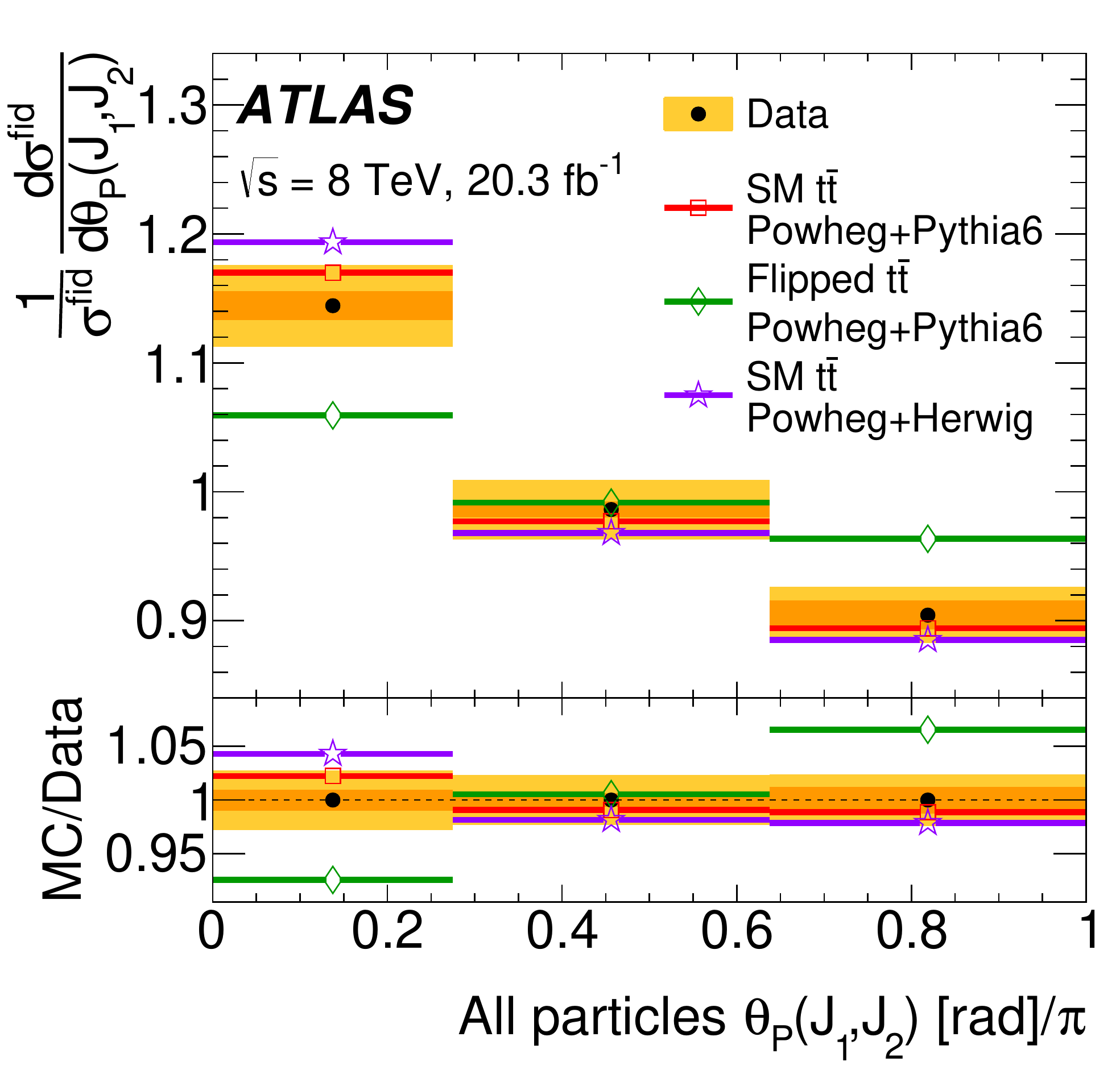}\includegraphics[width=0.5\textwidth]{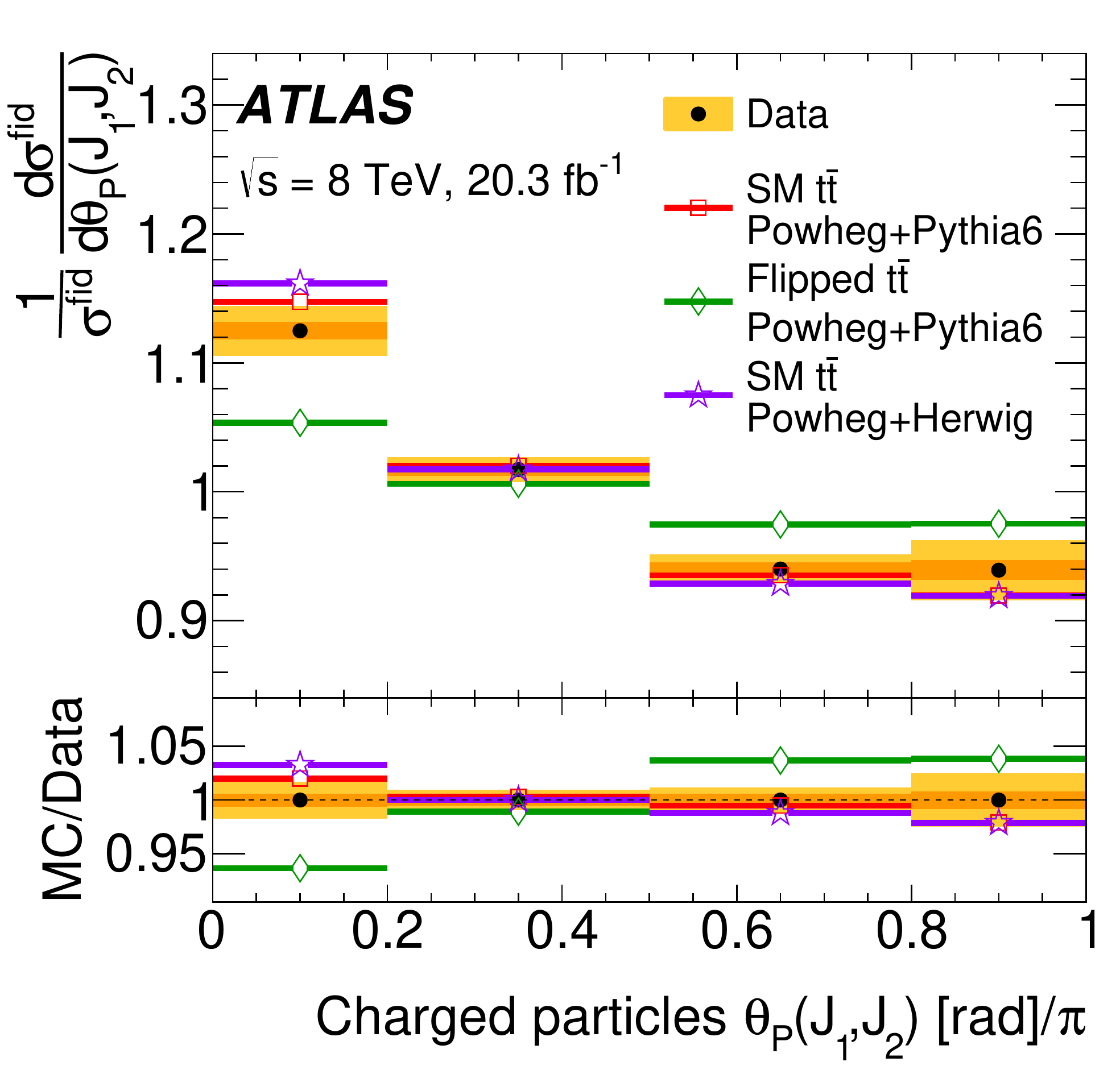}
    \caption{Left (Right): the all-particles (charged-particles) pull angle distribution for the unfolded data and three particle-level simulations.  The orange inner band on the data represents the statistical uncertainty while the yellow band is the sum in quadrature of the statistical and systematic uncertainty.  Final version of this plot is from T. Neep.}
  \label{fig:unfolded:pull_all}
\end{figure}

\begin{figure}[htbp]
  \centering
    \includegraphics[width=0.5\textwidth]{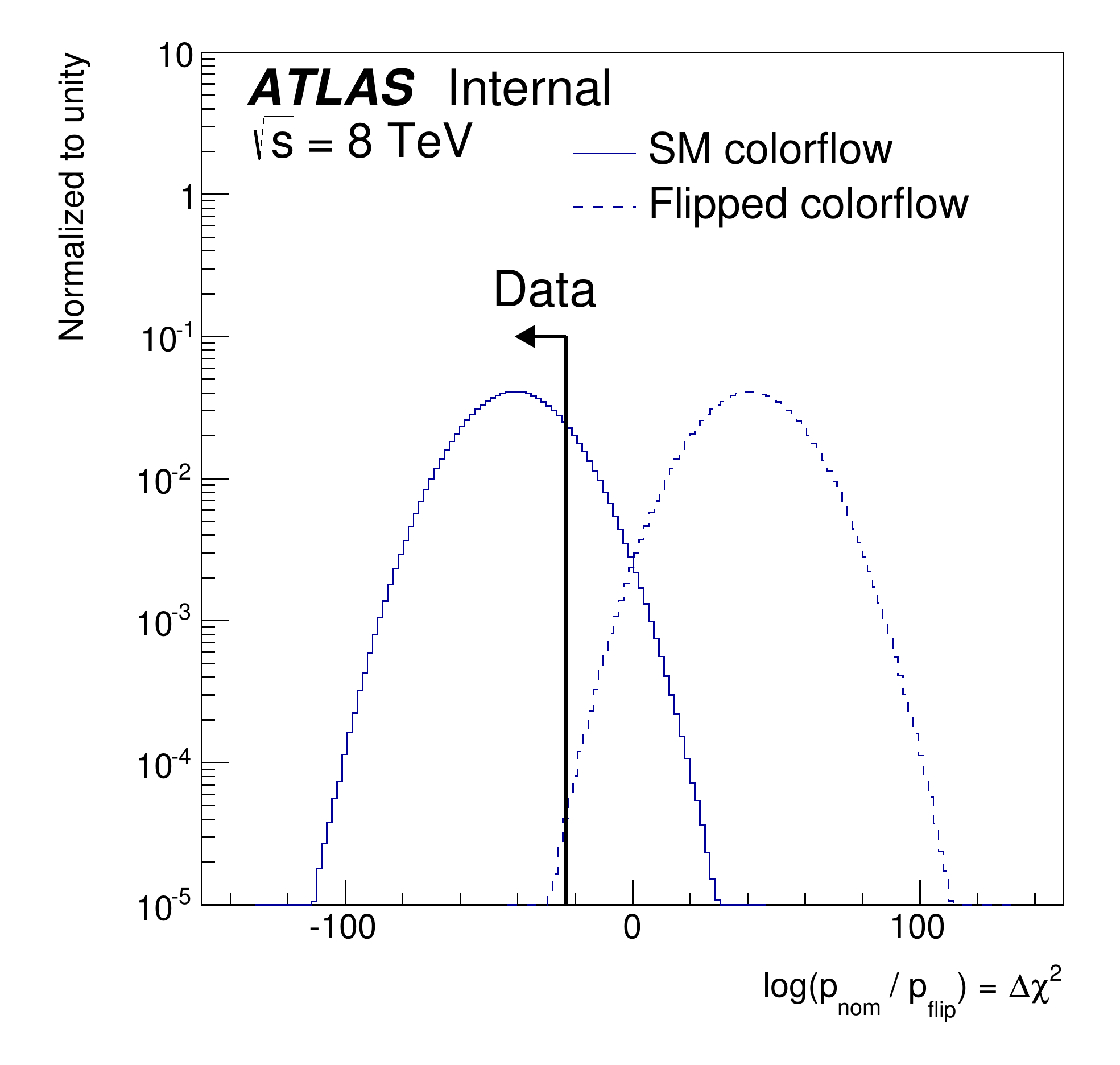}
    \caption{The distribution of the log-likelihood ratio test statistic under the SM and flipped color flow models.  To generate the distributions, the full covariance matrix is resampled ten million times.  In order to impose unitarity of the pseudo-data, only the first $n$ bins are sampled and the $(n-1)^\text{th}$ is fixed by normalization.  Additionally removing this last bin from the log-likelihood ratio has little impact on the approximate significances quoted in the text.}
  \label{fig:unfolded:deltachi2}
\end{figure}

Interestingly, the SM color flow from {\sc Powheg-Box}+{\sc Pythia} 6 is closer to the data than {\sc Powheg-Box}+{\sc Herwig} in Fig.~\ref{fig:unfolded:pull_all}.  With only the first bin, the {\sc Herwig} model is about $2\sigma$ away from the SM\footnote{Using only the charged-particles pull angle and ignoring the explicit color flow model uncertainty - see Sec.~\ref{seccolorflowmodeluncert}.} while the {\sc Pythia} 6 model is $\lesssim 1\sigma$ away.  Both the {\sc Herwig} and {\sc Pythia} 6 model predictions are further away from the flipped model than the data is from the flipped model.  The unfolded data and particle-level analysis code are publicly available~\cite{hepdata,rivet} for further interpretation and can provide useful information for the tuning and model development of color flow.

\clearpage

\section{Summary}
\label{sec:colorflowsummary}

The analysis presented in this chapter describes a measurement of the orientation
of radiation from jets identified as originating from a $W$ boson in $t\bar{t}$\
events. The measurement uses 20.3 fb$^{-1}$ of $\sqrt{s}$ = 8 TeV $pp$ collision
data recorded by the ATLAS detector at the LHC.
To quantify the distribution of energy inside one jet relative to another,
the distribution of the \textit{jet pull angle} is extracted from the data using
information from both the ATLAS calorimeter and tracking detectors.
The jet pull angle is found to correctly characterize the $W$ boson as
a color singlet, with data disfavouring an alternative color--octet model
at greater than $3\sigma$.
This illustrates the potential to use the jet pull angle in future
SM measurements and BSM searches.
The jet pull angle measurement is presented as a normalized
fiducial $t\bar{t}$\ differential cross--section, allowing the results to be
used to constrain implementations of color connection.
 \chapter{Constituent Multplicity}
\label{cha:multiplicity}

As discussed in earlier chapters, quarks and gluons produced in high-energy particle collisions hadronize before they can be observed directly.  However, the properties of the resulting jets depend on the type of parton which initiated them.  One jet observable sensitive to the quark or gluon nature is the number of charged particles inside the jet.  Due to their larger color-charge under the strong force, gluon-initiated jets contain on average more particles than quark-initiated jets and the average (charged) particle multiplicity inside jets increases with jet energy~\cite{Ellis:1991qj}.   These properties were used recently at the Large Hadron Collider (LHC) to differentiate between jets originating from a quark or a gluon~\cite{Aad:2014gea,Aad:2015owa,Aad:2014bia,Khachatryan:2015bnx,Khachatryan:2014dea}. These studies have found significant differences in the charged-particle multiplicity between the available simulations and data.  Improved modelling based on measurements of the number of charged particles inside jets is thus crucial for future studies.

This chapter presents a measurement of the average charged-particle multiplicity inside jets as a function of the jet transverse momentum in dijet events in $pp$ collisions at $\sqrt{s}=8$ TeV~with the ATLAS detector\footnote{This analysis has been published in Ref.~\cite{Aad:2016oit}.}.  The measurement of the charged-particle multiplicity inside jets has a long history from the SPS~\cite{Arnison:167874,BAGNAIA1984291,ua22},  PETRA~\cite{Bartel:1982ub,zphys}, PEP~\cite{Derrick:163229,PhysRevLett.55.1954,PhysRevLett.72.3145,Abe:1996zi}, TRISTAN~\cite{PhysRevLett.63.1772}, CESR~\cite{PhysRevD.56.17}, LEP~\cite{Alexander:1991ce,Acton:1993jm,xyz,Buskulic1995389,Alexander1996659,Buskulic1996353,delphintrack,refId0,Abreu1999383,refId1,Akrawy1990617,opallep2}, and the Tevatron~\cite{Affolder:2001jx}.  At the LHC, both ATLAS~\cite{Aad:2011gn,Aad:2011sc} and CMS~\cite{Chatrchyan:2012mec} have measured the charged-particle multiplicity inside jets at $\sqrt{s}=7$ TeV.  One ATLAS result used jets that are reconstructed with tracks and have transverse momentum less than $40$ GeV.   A second ATLAS analysis~\cite{Aad:2011sc} has measured charged particles inside jets with transverse momenta spanning the range from 50 to 500 GeV with approximately constant 3--4\% uncertainties.  The CMS measurement spans jet transverse momenta between 50 and 800 GeV with 5--10\% uncertainties in the bins of highest transverse momentum.  The analysis presented here uses the full $\sqrt{s}=8$ TeV ATLAS dataset, which allows for a significant improvement in the precision at high transverse momentum up to and beyond 1.5 TeV.

This chapter is organized as follows.  Section~\ref{sec:NCharge:Motivation} describes in more detail the motivation for a measurement of the charged particle multiplicity, including some theoretical considerations from QCD.  The setup of the analysis, the corrections to remove detector distortions, and the systematic uncertainties, which are similar to the techniques used for the jet charge measurement (Chapter~\ref{cha:jetcharge}), are discussed in Sec.~\ref{sec:NCharge:Design}, Sec.~\ref{sec:NCharge:Unfolding}, and Sec.~\ref{sec:NCharge:Systs}, respectively.  The results are in Sec.~\ref{sec:NCharge:Results}, both inclusive and exclusive in jet type.  Section~\ref{sec:NCharge:Summary} ends the chapter with a summary and outlook.

\clearpage

\section{Motivation}
\label{sec:NCharge:Motivation}

Despite being a basic jet quantity, the constituent multiplicity is non-trivial to describe precisely in perturbative QCD due to its sensitivity to very soft energy scales.  Section~\ref{sec:QCDmultiplicity} describes lowest order and state-of-the-art calculations that attempt to recover perturbative predictions for multiplicity.  In addition to providing a basic probe of QCD at the highest energies,  the constituent multiplicity is an important discriminant between quark and gluon initiated jets, as motivated in Sec.~\ref{sec:qgtagging}.

\subsection{QCD Predictions for Multiplicity}
\label{sec:QCDmultiplicity}

The average particle multiplicity inside a jet was calculated in Sec.~\ref{sec:jetchargetheory}:

\begin{align}
\langle n_p(E)\rangle = \sum_h \int_0^1 dz D_p^h(z,E),
\end{align}

\noindent where $D_p^h$ is the fragmentation function describing the probability to find a hadron $h$ with energy fraction $z$ of the parton $p$.  One could try to compute the $p_\text{T}$ dependence of $\langle n_p\rangle$ using similar techniques as for the jet charge for which the multiplicity is related to the $\kappa\rightarrow 0$ limit.  However, $\tilde{P}_{q\leftarrow q}(\kappa\rightarrow 0)\rightarrow\infty$ and so Eq.~\ref{eq:scaeviolationequation} cannot be used.  One way to make a sensible lowest order prediction for the multiplicity is to include the suppression of large angle soft radiation due to color coherence (see Sec.~\ref{sec:colorcoherence}).
Color coherence can be incorporated into the DGLAP equation by using $t=E\theta$ instead of $\mu$ as the evolution variable, where $E$ is the parton energy and $\theta$ is the opening angle of the radiation.  Evolution to smaller values of $\theta$ is the {\it angular ordering} scheme.  In this case, the equation governing the scale-dependence of $\tilde{D}$ is given by

\begin{align}
\label{eq:DGLAPforDcoherence}
t\frac{\partial}{\partial t} D_p^h(z,t) = \sum_{p'}\int _z^1\frac{dz'}{z'}\frac{\alpha_sP_{p'\leftarrow p}(z')}{\pi}D_{p'}^h\left(\frac{z}{z'},z't\right),
\end{align}

\noindent which is identical to Eq.~\ref{eq:DGLAPforD} except that the last term has an explicit dependence on the integrand in its second coordinate.   The corresponding equation for the Mellin moment of $\tilde{D}$ is

\begin{align}
\nonumber
t\frac{\partial}{\partial t} \tilde{D}_p^h(\kappa,t) &=\frac{\alpha_s}{\pi} \sum_{p'}\int_0^1 dzz^\kappa\int _z^1\frac{dz'}{z'}P_{p'\leftarrow p}(z')D_{p'}^h\left(\frac{z}{z'},z't\right)\\\nonumber
&\stackrel{x=z/z'}{=}\frac{\alpha_s}{\pi}\sum_{p'}\int _0^1 dz' (z')^\kappa P_{p'\leftarrow p}(z') \int_0^{1} dx x^\kappa D_{p'}^h\left(x,z't\right)\\\label{eq:DGLAPforDcoherence1}
&=\frac{\alpha_s}{\pi}\sum_{p'}\int _0^1 dz' (z')^\kappa P_{p'\leftarrow p}(z') \tilde{D}_{p'}^h\left(\kappa,z't\right)
\end{align}

\noindent which does not fully factor like Eq.~\ref{eq:DGLAPforD2}.  Nonetheless, one can try a solution of the same form that solves Eq.~\ref{eq:DGLAPforD2}: $\tilde{D}(\kappa,t)\propto t^{\gamma(\kappa)}$ ($\gamma$ is called the anomalous dimension).  With this ansatz, Eq.~\ref{eq:DGLAPforDcoherence1} becomes

\begin{align}
\label{eq:DGLAPforDcoherence}
\gamma(\kappa) &=\frac{\alpha_s}{\pi}\sum_{p'}\int _0^1 dz' (z')^{\kappa+\gamma(\kappa)} P_{p'\leftarrow p}(z').
\end{align}

\noindent The most relevant regime is $z'\ll 1$, where the integral of the splitting function diverges using the original ordering scheme in Sec.~\ref{sec:jetchargetheory}.  In this regime, $P_{p'\leftarrow p}(z')\approx \frac{2 C}{\pi}\frac{1}{z}\delta_{pg}$, where $C=C_F$ for quarks and $C=C_A$ for gluons.   Therefore,

\begin{align}
\label{eq:DGLAPforDcoherence}
\gamma(\kappa) &\approx \frac{2\alpha_sC}{\pi}\int _0^1 dz' (z')^{\kappa+\gamma(\kappa)-1}\\
&=\frac{2\alpha_sC}{\pi}\frac{1}{\kappa+\gamma(\kappa)},
\end{align}

\noindent which is readily solved for $\gamma$:

\begin{align}
\label{sec:mult:anamalousdimension}
\gamma&=-\frac{\kappa}{2}+\sqrt{\frac{\kappa^2}{4}+\frac{2\alpha_s C}{\pi}}.
\end{align}

\noindent As desired, Eq.~\ref{sec:mult:anamalousdimension} is finite as $\kappa\rightarrow 0$.  The difference with the solution in Sec.~\ref{sec:jetchargetheory} is that Eq.~\ref{sec:mult:anamalousdimension} is the start of a series that is in powers of {\it the square root of $\alpha_s$}.  This is not the Taylor series of any function and thus the convergence of the series is not governed in the usual way for a perturbative series in $\alpha_s$.  This $\sqrt{\alpha_s}$ behavior has been observed and catalogued for a variety of related variables~\cite{Larkoski:2015lea,Larkoski:2013paa} ({\it Sudakov safe}).  For comparison, one could expand Eq.~\ref{sec:mult:anamalousdimension} in $\alpha_s$ and compare with the energy ordered calculation from earlier, using the gluon splitting function instead of the quark one:

\begin{align}
\label{eq:reshuffle}
\gamma_\text{angular ordered}&=\text{finite}+\frac{2\alpha_s C}{\pi\kappa}+\mathcal{O}(\alpha^2)\\
\gamma_\text{energy-ordred}&=\tilde{P}_{g\leftarrow p}=\frac{2C\alpha_s}{\pi}\int_0^1 \frac{dz}{z}=\frac{2\alpha_s C}{\pi\kappa}+\mathcal{O}(\alpha_s^2).
\end{align}

\noindent Equation~\ref{eq:reshuffle} shows the importance of the $\sqrt{\alpha_s}$ expansion to recover a finite prediction, which is not achievable with any finite $\alpha_s$ expansion.  Inserting the anomalous dimension from Eq.~\ref{sec:mult:anamalousdimension} into the ansatz $\tilde{D}(\kappa,t)\propto t^{\gamma(\kappa)}$ results in:

\begin{align}
\label{multlo}
\langle n_p(E)\rangle &\propto p_\text{T}^\gamma=\exp\left(\sqrt{\frac{2\alpha_s C}{\pi}}\log(p_\text{T}/\Lambda)\right)\sim \exp\left(\sqrt{C\log(p_\text{T}/\Lambda)}\right),
\end{align}

\noindent where the last line uses\footnote{This can be properly derived by including the running $\alpha_s$ in the ansatz for $\tilde{D}\propto \exp(\gamma \log(t))\propto\exp(\int_{t_0}^t \gamma(\alpha_s(t')) dt'/t')$.  See e.g. Sec. 6.1 in Ref.~\cite{Ellis:1991qj} for details.} $\alpha_s(p_\text{T})\sim 1/\log(p_\text{T}/\Lambda)$.  For quark and gluon jets, the first gluon emission is proportional to $C_F$ and $C_A$, respectively.  However, the subsequent parton shower is dominated in the $z\rightarrow0$ limit by the gluon splitting function $g\rightarrow gg$ because the conversion of gluons back into quarks, $g\rightarrow q\bar{q}$, is suppressed by a factor of $\alpha_s$.  Therefore, 

\begin{align}
\label{multlo2}
\langle n_p(E)\rangle &\propto C_i\exp\left(\sqrt{C_A\log(p_\text{T}/\Lambda)}\right),
\end{align}

\noindent where $i=F$ for $p=$ quark and $i=A$ for $p=$ gluon.  The main features of Eq.~\ref{multlo2} are that the multiplicity increases with $p_\text{T}$ and is larger for gluon jets than for quark jets.  At lowest order, the ratio of quark to gluon multiplicity is a constant $C_F/C_A$.  

The calculation of the anomalous dimension $\gamma$ can be systematically improved as a series in $\sqrt{\alpha_s}$ despite the lack of control in $\alpha_s$.  Currently, the most precise calculation in this context is at next-to-next-to-next-to-leading-order (N${}^3$LO) using pQCD~\cite{Capella:1999ms,Dremin:1999ji}:

\begin{align}
\label{eqscaling}
\langle n_g(y)\rangle&\propto \exp\left(f_\text{LO}\sqrt{y}+f_\text{NLO}\log(y)+f_\text{N${}^2$LO}(y)\frac{1}{\sqrt{y}}+f_\text{N${}^3$LO}(y)\frac{1}{y}\right)\\ \label{eqscaling2}
\langle n_q(y)\rangle &=\frac{\langle n_g(y)\rangle}{r_0(1-r_1\gamma_0-r_2\gamma_0^2-r_3\gamma_0^3)},
\end{align}

\noindent where $y=\log(t/\Lambda)$ and

\begin{align}\nonumber
\label{orders1}
f_\text{LO}&=2C\\\nonumber
f_\text{NLO}&=-a_1C^2\\\nonumber
f_\text{N${}^2$LO}(y)&=C\left(2a_2C^2+\frac{\beta_1}{\beta_0^2}(\log(2y)+1)\right)\\
f_\text{N${}^3$LO}(y)&=C^2\left(a_3C^2-\frac{a_1\beta_1}{\beta_0^2}(\log(2y)+1)\right).
\end{align}

\noindent The values of $r_i$ and $a_i$ are in Table~\ref{tab:numbersmultiplicity}.  At the N${}^3$LO, the ratio of the quark and gluon jet multiplicities does vary with $p_\text{T}$, though the overall $C_F/C_A$ scaling is preserved.

\begin{table}[h!]
\centering
\begin{tabular}{|c|cccc|}
\hline 
\multirow{2}{*}{Coefficient} & \multicolumn{4}{ c| }{Order} \\
& 0 & 1 & 2 & 3  \\
\hline
quark-gluon ratio $r$ &$C_A/C_F=2.25$ & 0.198 & 0510 & -0.041\\
gluon $p_\text{T}$ scaling $a$ & -- & 0.314 & -0.301 & 0.112\\
\hline
\end{tabular}
\caption{The coefficients of various parts of the N${}^3$LO prediction (Ref.~\cite{Capella:1999ms,Dremin:1999ji}) for the quark and gluon constituent multiplicity distributions from Eq.~\ref{eqscaling2} and~\ref{orders1}.}
\label{tab:numbersmultiplicity}
\end{table}

\clearpage

\subsection{Quark and Gluon Tagging}
\label{sec:qgtagging}

In addition to using the charged particle multiplicity in jet to directly test the predictions of QCD, constituent multiplicity is a ubiquitous feature for discriminating quark jets from gluon jets.  For example, the jet energy response in the ATLAS calorimeter differs between quark and gluon jets.  The most recent jet calibration procedure, described in Sec~\ref{sec:jets}, uses a residual correction based on the number of tracks inside jets.  For a fixed jet energy, the jet energy response is lower for higher constituent multiplicity jets due to the non-linear calorimeter response.  The impact of this residual calibration is shown in Fig.~\ref{fig:NCharge:Motivation:GSC}.  The bottom panels show that the constituent multiplicity increases with jet $p_\text{T}$ and the response decreases with the number of tracks.   As a result of the residual calibration, the jet energy scale is less dependent on the composition of quark and gluon jets for a particular analysis since the distribution of the number of tracks inside jets significantly differs between quark and gluon jets.

\begin{figure}[h!]
\begin{center}
\includegraphics[width=0.5\textwidth]{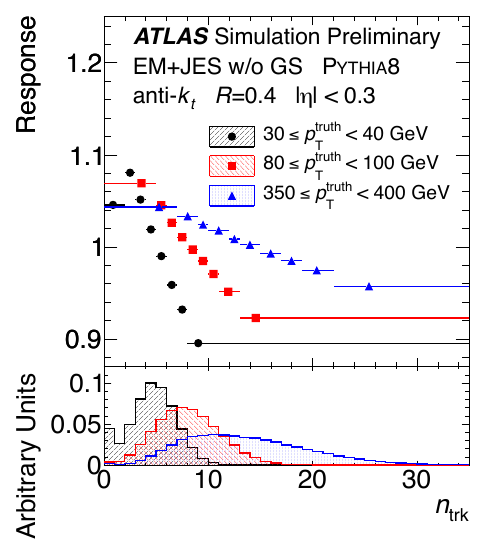}\includegraphics[width=0.5\textwidth]{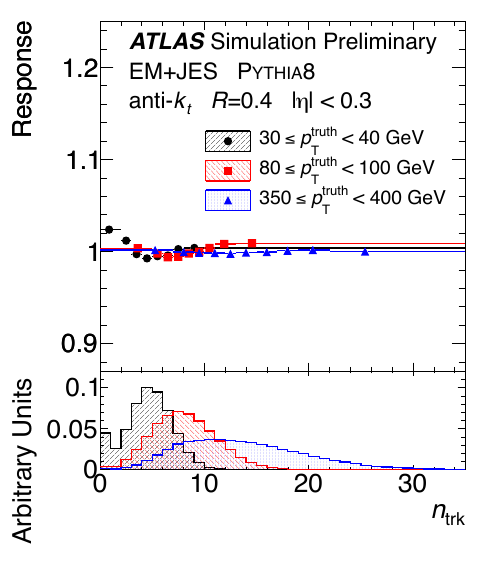}
\end{center}	
\caption{The response after applying an inclusive jet energy calibration (left) and after the residual correction (right).  The lower panel is the distribution of the number of tracks inside jets for three jet $p_\text{T}$ ranges.  Reproduced from Ref.~\cite{ATLAS-CONF-2015-002}.}
\label{fig:NCharge:Motivation:GSC}
\end{figure}

Many SM processes and new physics scenarios of interest are produced with predominately quark jets.  For example, top quark and $W$ bosons decaying hadronically produce mostly quark jets and cascade decays of SUSY squarks or gluinos can result in high multiplicity quark jet final states.  Therefore, it is desirable to have a tool that can differentiate quark jets from gluon jets.  A dedicated performance study using early Run I data showed that $n_\text{track}$ as well as the $p_\text{T}$- and $\Delta R$-weighted sum of tracks (track width) inside jets are good variables for this task.  However, these track-based variables have different distributions in data and simulation.  As a result, the tagger performance in simulation is optimistic.  Figure~\ref{fig:NCharge:Motivation:QGtagger} shows the 2D likelihood ratio used for the tagger.  There are qualitative differences between the two distributions, in particular the large likelihood in the lower left corner in simulation that is not as significant in data.  The implication of this study is that quark/gluon tagging is significantly mis-modeled and one likely source\footnote{This analysis did not assess the systematic uncertainties related to the modeling of the detector response, which could account for some of the differences between data and simulation.} is the modeling of jet fragmentation.  An improved model of the number of particles inside jets is crucial for improved descriptions of quark/gluon tagging in the future.

\begin{figure}[h!]
\begin{center}
\includegraphics[width=0.45\textwidth]{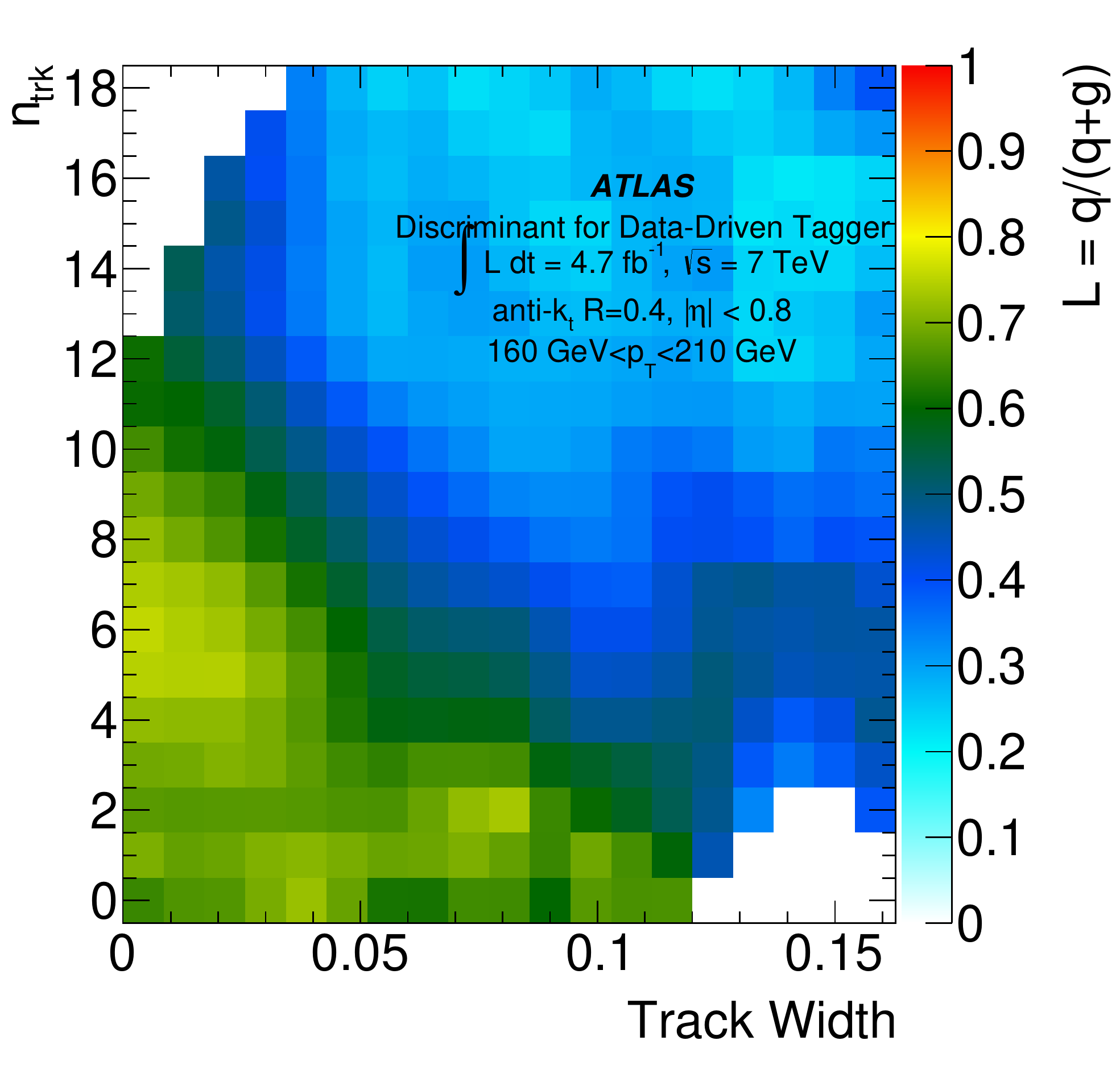}\includegraphics[width=0.45\textwidth]{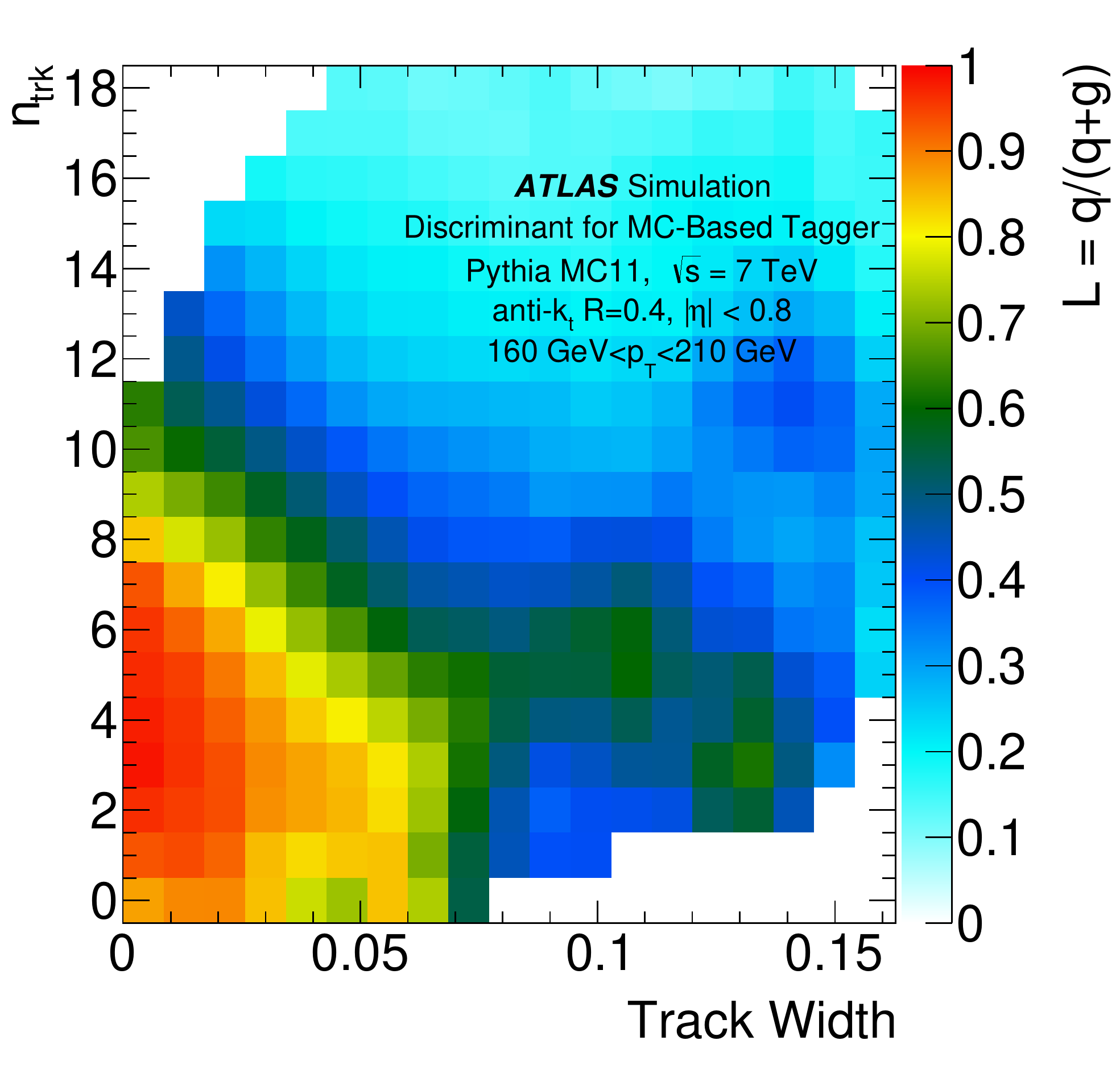}
\end{center}	
\caption{The two-dimensional likelihood ratio for the track multiplicity (vertical axis) and track width (horizontal axis) quark/gluon tagger in data (left) and simulation (right).  Reproduced from Ref.~\cite{Aad:2014gea}.}
\label{fig:NCharge:Motivation:QGtagger}
\end{figure}

Despite the known mis-modeling of the $n_\text{track}$ distribution, both ATLAS and CMS have used $n_\text{track}$-based quark/gluon tagging to search for new physics, taking care to assess the impact of potential sources of systematic bias.  One prominent example is the ATLAS search for all-hadronic diboson resonances.  Figure~\ref{fig:NCharge:Motivation:Dibosons} shows the final dijet invariant mass spectrum before and after applying a requirement on the number of tracks.  The signal to background ratio for the 2 TeV $W'$ model increases because the $W$ and $Z$ bosons from the $W'$ decay predominately produce quark jets.  Interestingly, the excess in data also increases with this requirement.

\begin{figure}[h!]
\begin{center}
\includegraphics[width=0.5\textwidth]{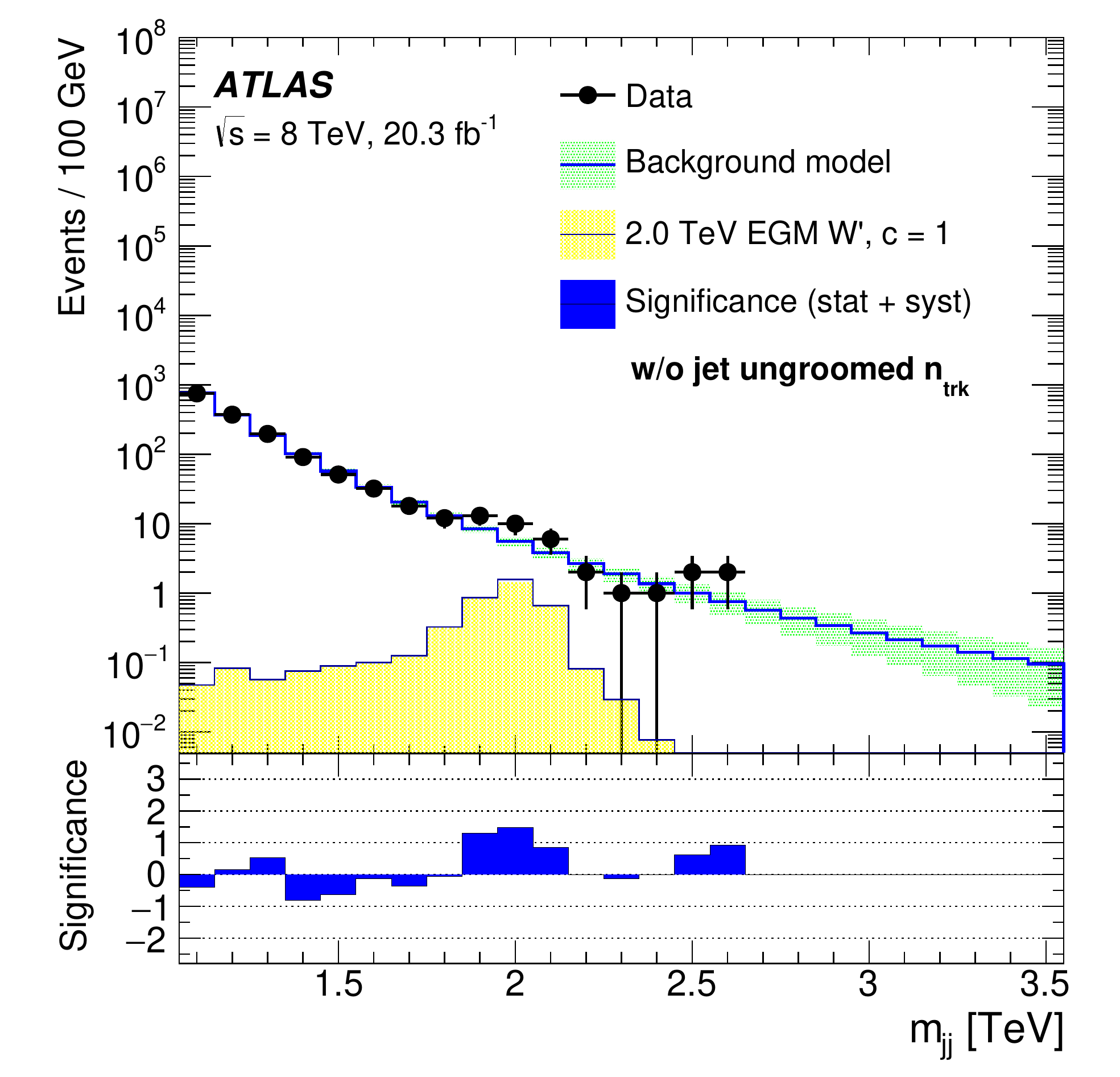}\includegraphics[width=0.5\textwidth]{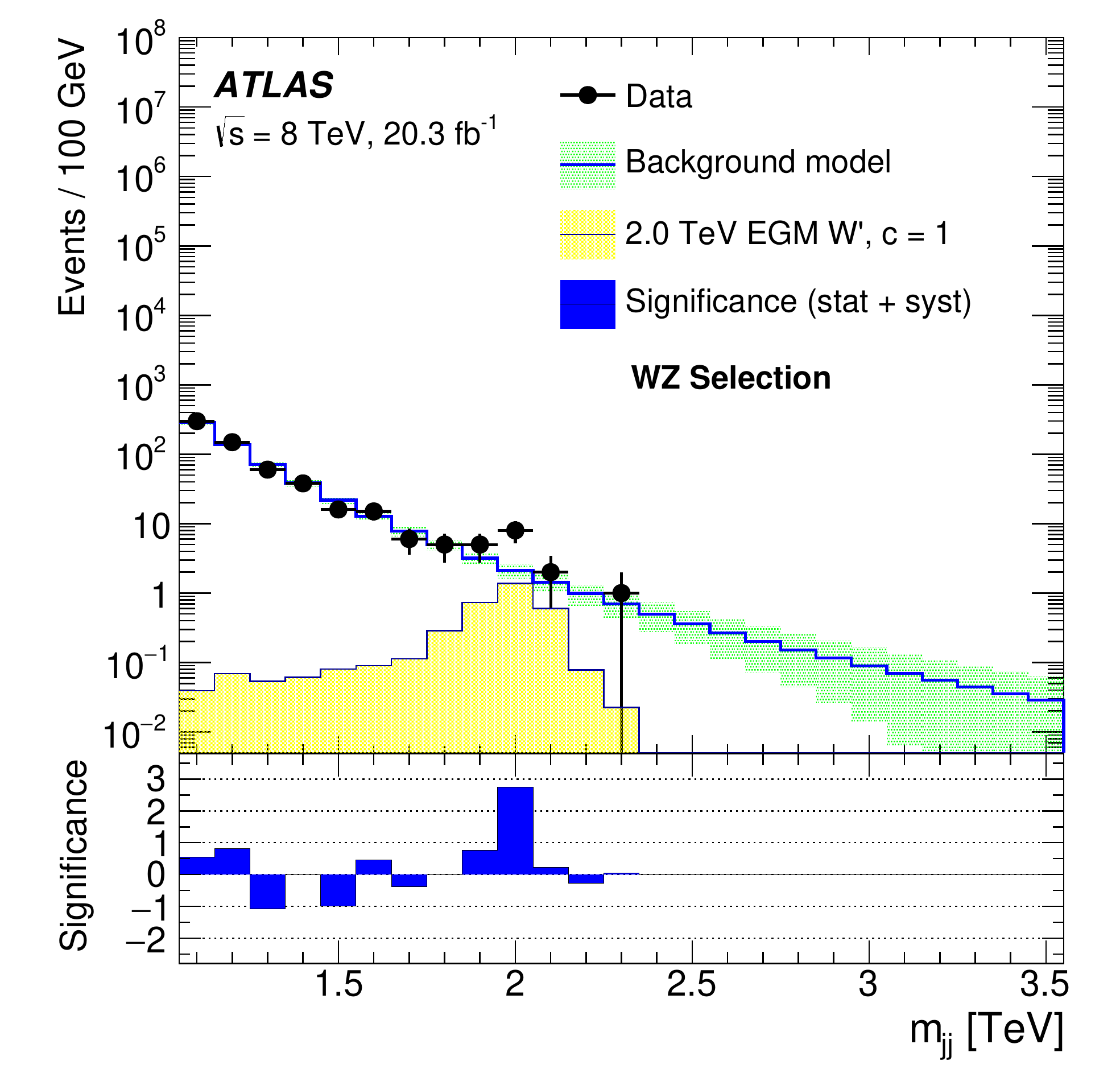}
\end{center}	
\caption{The dijet invariant mass spectrum for the full jet selection without the $n_\text{track}$ requirement (left) and with all requirements (right).  Reproduced from Ref.~\cite{Aad:2015owa}.}
\label{fig:NCharge:Motivation:Dibosons}
\end{figure}

The particle multiplicity inside jets is a powerful tool for probing the high energy behavior of QCD as well as for distinguishing quark jets from gluon jets to improve the significance of other SM measurements and searches for new physics beyond the SM.  The remainder of this chapter describes a measurement of the charged particle multiplicity inside jets, exploiting both aspects of this tool.

\clearpage

\section{Analysis Design}
\label{sec:NCharge:Design}

In a fixed $p_\text{T}$ bin, number of charged particles inside jets is nearly identical to the jet charge with the momentum-weighting factor $\kappa=0$.  As a result, most of the framework for the jet charge measurement can be re-used to perform the measurement of the $p_\text{T}$-dependence of $\langle n_\text{track}\rangle$.  In particular, events are selected using single jet triggers and required to have at least two jets with $p_\text{T}>50$ GeV that are well-balanced in $p_\text{T}$.  One new $p_\text{T}$ bin is added at $p_\text{T}>1.5$ TeV and  the measurement is performed for three track $p_\text{T}$ thresholds (500 MeV, 2 GeV, and 5 GeV) in order to investigate the sensitivity of the modeling to the softness of the radiation.  Figure~\ref{fig:NCharge:tracks} shows the track multiplicity ($p_\text{T}^\text{track}>500$ MeV) in three jet $p_\text{T}$ bins.  As expected, the average $n_\text{track}$ increases with jet $p_\text{T}$.  Interestingly, the data distributions are largely between the predictions from {\sc Pythia 8} with the AU2 tune and {\sc Herwig++} 2.63 with the EE3 tune.

\begin{figure}[h!]
\begin{center}
\includegraphics[width=0.6\textwidth]{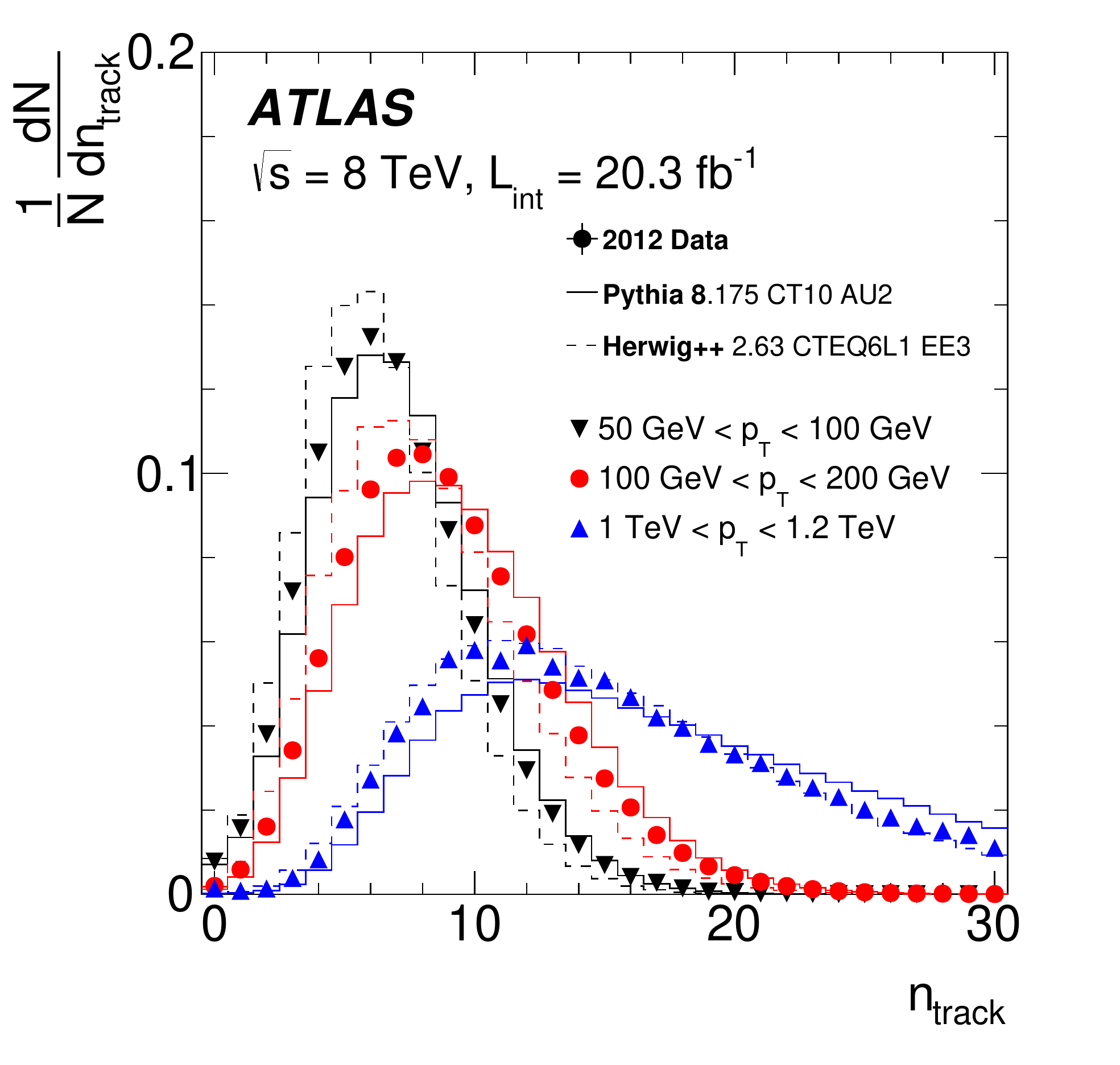}
\end{center}	
\vspace{-5mm}
\caption{The number of reconstructed tracks associated with a jet in three example jet $p_\text{T}$ ranges for data and for {\sc Pythia 8} and {\sc Herwig++} predictions.  The data points have statistical uncertainties which in all bins are smaller than the marker size. }
\label{fig:NCharge:tracks}
\end{figure}

\clearpage

The simulation samples are the same as for the jet charge measurement except for three new particle-level models representing the latest underlying event tunes of {\sc Pythia} 8 and {\sc Herwig++}.  The differences between these models and the older ones will be discussed in the context of the unfolded results in Sec.~\ref{sec:NCharge:Results}.   The details of the samples used are shown in Table~\ref{tab:mc_samples1ntrack}. 

\begin{table}[h!]
\centering
\begin{tabular}{ccccccc}
ME Generator & PDF & Tune \\
\hline 
\hline
 {\sc Pythia} 8.175~\cite{Sjostrand:2007gs} & CT10~\cite{Gao:2013xoa} & AU2~\cite{ATL-PHYS-PUB-2012-003} \\
 {\sc Pythia} 8.186 & NNPDF2.3~\cite{Ball:2012cx} & Monash~\cite{Skands:2014pea} \\
 {\sc Pythia} 8.186 & NNPDF2.3 & A14~\cite{ATL-PHYS-PUB-2014-021}  \\
 {\sc Herwig++} 2.6.3~\cite{Bahr:2008pv,Arnold:2012fq} & CTEQ6L1~\cite{Stump:2003yu} & UE-EE3~\cite{Gieseke:2012ft} \\
{\sc Herwig++} 2.7.1~\cite{Bellm:2013hwb} & CTEQ6L1 & UE-EE5~\cite{Seymour:2013qka} \\
 {\sc Pythia} 6.428~\cite{Sjostrand:2006za} & CTEQ6L1 & P2012~\cite{Skands:2010ak} \\
 {\sc Pythia} 6.428 & CTEQ6L1 & P2012RadLo~\cite{Skands:2010ak} \\
 {\sc Pythia} 6.428 & CTEQ6L1 & P2012RadHi~\cite{Skands:2010ak} \\
\hline
\hline
\end{tabular}
\caption{Monte Carlo samples used for measuring and studying the charged particle multiplicity inside jets.}
\label{tab:mc_samples1ntrack}
\end{table}

\begin{figure}[h!]
\begin{center}
\includegraphics[width=0.8\textwidth]{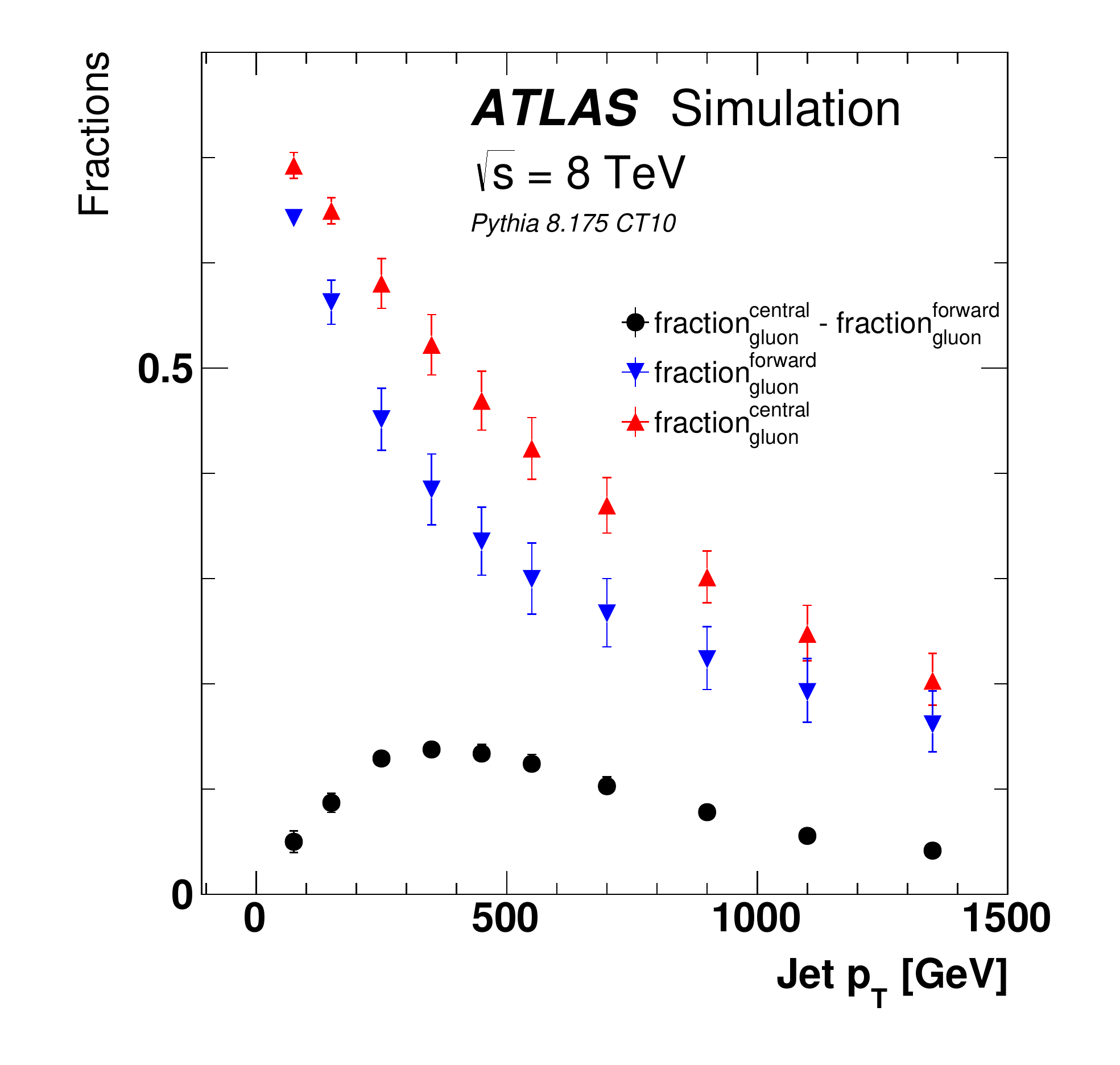}
\end{center}	
\caption{The simulated fraction of jets originating from gluons as a function of jet $p_\text{T}$ for the more forward jet (down triangle), the more central jet (up triangle), and the difference between these two fractions (circle).  The fractions are derived from {\sc Pythia} 8 with the CT10 PDF set and the error bars represent the PDF and matrix element uncertainties, further discussed in Sect.~\ref{sec:results}.  The uncertainties on the fraction difference are computed from propagating the uncertainties on the more forward and more central fractions, treating as fully correlated.}
\label{fig:NCharge:qgfrac}
\end{figure}

In analogy to the procedure for the jet charge, the distribution of the jet $p_\text{T}$ and $n_\text{track}$ are discretized into a two-dimensional histogram.  This histogram is {\it unfolded} to remove detector distortions.  The average $n_\text{track}$ is computed in each $p_\text{T}$ bin, which is compared to a variety of particle-level models.  In addition to studying the inclusive modeling of the $n_\text{track}$ distribution, a novel technique is employed to extract the average charged particle multiplicity separately for quark and gluon jets.  As discussed in the context of the jet charge measurement, the distribution of the jet type depends on rapidity.  The more forward jet in dijet events is more likely to be the quark jet because the higher longitudinal momentum indicates a higher momentum fraction of the colliding proton.  However, the scale of the shower is largely unaffected by the longitudinal momentum and therefore for a fixed jet $p_\text{T}$, the difference in $\langle n_\text{charge}\rangle$ between the more forward and the more central jet is due to the difference in the quark/gluon composition.   Figure~\ref{fig:NCharge:qgfrac} shows the gluon jet fraction of the selected jets in simulation as a function of the jet $p_\text{T}$.  The fraction of gluon jets decreases monotonically as a function of jet $p_\text{T}$ due to the higher fraction of momentum carried on average by quarks in the proton.  However, the {\it difference} in the fractions between the more forward and more central jet peaks around $p_\text{T}\sim 350$ GeV and goes to zero at low and high jet $p_\text{T}$.   Given the quark and gluon fractions $f_{q,g}^{f,c}$ with $f=\text{more forward}$, $c=\text{more central}$, $q=\text{quark}$, $g=\text{gluon}$ and $f_{q}+f_{g}=1$, the average charged-particle multiplicity for quark- and gluon-initiated jets is extracted by solving the system of equations in Eq.~\ref{eq:system}.

\begin{align}
\label{eq:system}
\langle n_\text{charged}^f\rangle&=f_q^f\langle n_\text{charged}^q\rangle+f_g^f\langle n_\text{charged}^g\rangle\\\nonumber
\langle n_\text{charged}^c\rangle&=f_q^c\langle n_\text{charged}^q\rangle+f_g^c\langle n_\text{charged}^g\rangle.
\end{align}

Figure~\ref{fig:NCharge:closure} shows the closure of the extraction method based on Eq.~\ref{eq:system}.   The filled circles show the more forward and more central $\langle n_\text{charge}\rangle$ as a function of the jet $p_\text{T}$, which are nearly identical at low and high $p_\text{T}$ and are maximally different around $p_\text{T}\sim 350$ GeV as expected based on the discussion above and Fig.~\ref{fig:NCharge:qgfrac}.  The $\langle n_\text{charge}\rangle$ for the more forward gluons (quark) and the more central gluons (quarks) are identical with each other (upper ratio) and with the extracted $\langle n_\text{charge}\rangle$ gluon (quark) distribution (lower ratio) within MC statistical uncertainty.  The small non-closure at low and high $p_\text{T}$ is due in part to the effective number of MC events in those regions is very small due to the negligible difference between the more forward and the more central jet $\langle n_\text{charge}\rangle$.   This method has several benefits compared to similar techniques for extracting quark and gluon jet properties.  First, because only one sample is used for the entire extraction (as opposed to using e.g. $\gamma$+jets and dijets), the sample dependent differences between quark and gluon jets are suppressed\footnote{This advantage only holds before comparing to quark and gluon jets from a different topology.}.  Second, as the same events are used for the more forward and the more central jet, many of the experimental uncertainties cancel.  This is also true of the PDF uncertainties: the absolute quark and gluon fractions have a bigger uncertainty than the rapidity-dependent differences between the quark and gluon fractions.

\begin{figure}[h!]
\begin{center}
\includegraphics[width=0.8\textwidth]{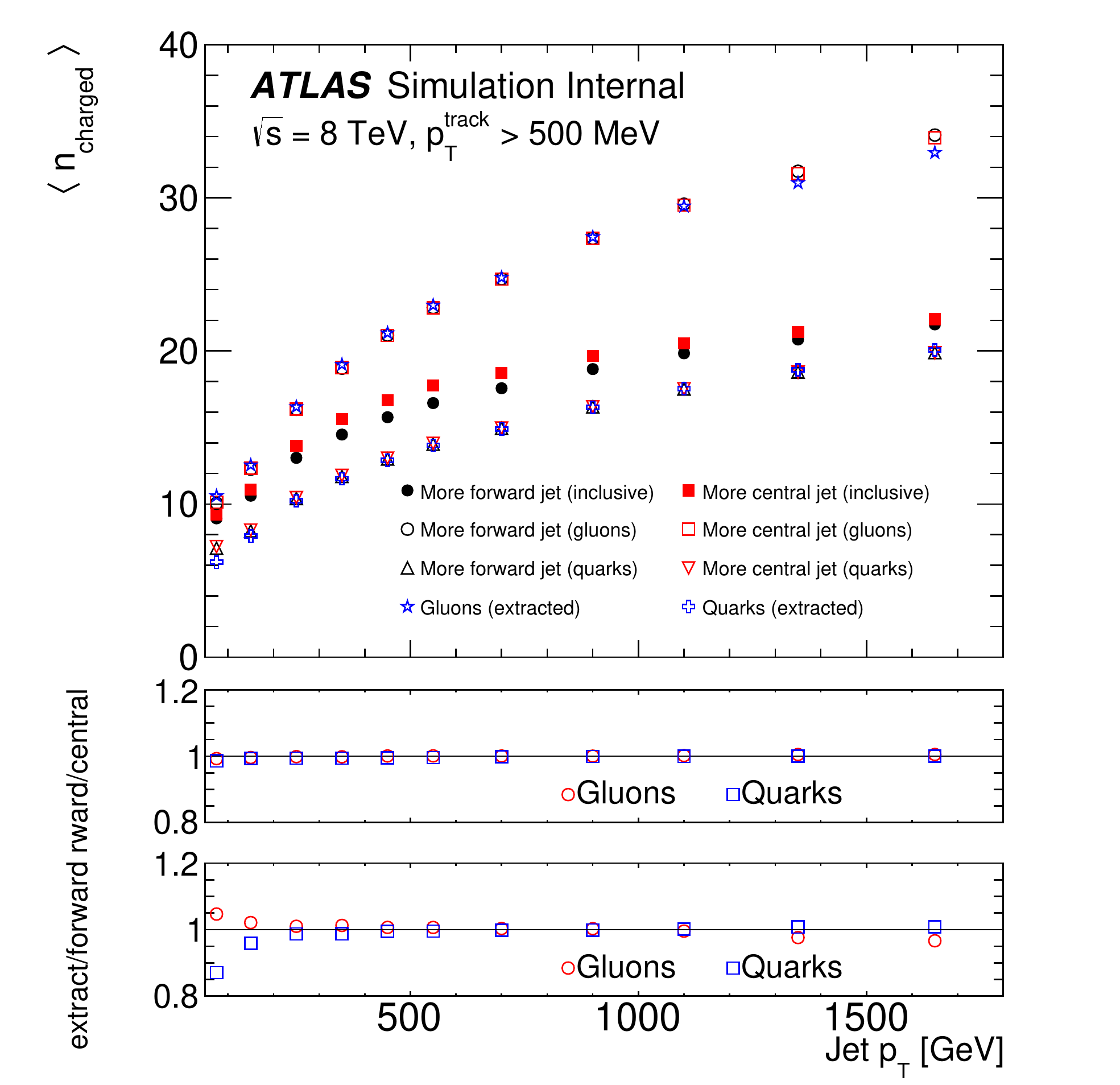}
\caption{The $p_T$ dependence of $\langle n_\text{charged}^f\rangle,\langle n_\text{charged}^c\rangle,\langle n_\text{charged}^g\rangle$, and $\langle n_\text{charged}^q\rangle$ in simulation for {\sc Pythia} 8 AU2 as well as the extracted values of $\langle n_\text{charged}^g\rangle$, and $\langle n_\text{charged}^q\rangle$ using the procedure described in the text.}
\label{fig:NCharge:closure}
\end{center}
\end{figure}

\clearpage

\section{Unfolding}
\label{sec:NCharge:Unfolding}

The procedure for removing detector distortions is the same procedure as was used for the jet charge measurement.   In particular, the measurement is carried out within a fiducial volume matching the experimental selection to avoid extrapolation into unmeasured kinematic regions that have additional model dependence and related uncertainties.  The particle level definitions, described in Sec.~\ref{sec:particlelevel}, are constructed to be as close as possible to the corresponding measured objects.  For the jet charge measurement, the charged particle $p_\text{T}$ threshold was irrelevant because of the $p_\text{T}$-weighting factor $\kappa$.  However, the charged particle multiplicity is maximally infrared-sensitive and so it is crucial to specify a particle-level $p_\text{T}$ threshold on the charged particles.  In this case, the same threshold (500 MeV, 2 GeV, or 5 GeV) that is used for tracks is used for charged particles.  The unfolding is performed over 11 bins in jet $p_\text{T}$: [0.5,1), [1,2), [2,3), [3,4), [4,5), [5,6), [6,8), [8,10), [10,12), [12,15), and [15,$\infty$)$\times 100$~GeV.   For the jet charge, there was no natural binning and the choice of bin size was chosen based on the resolution of the distribution.  For the charged particle multiplicity, there is a natural bin size: one track.  Figure~\ref{fig:NCharge:binning} shows that it is important to use this binning scheme.  For any coarser binning, there is a bias in the average charged particle multiplicity introduced when recovering the mean from the full distribution.  In principle, one can correct for this bias, but since it is easy to remove and the unfolding can handle the large number of bins, a one-track-per-bin scheme is used.  Another aspect of the binning is the total range.  Figure~\ref{fig:NCharge:binning2} shows the fraction of events with more than 60 charged particles.  Even in the highest $p_\text{T}$ bin, this fraction is below 0.1\%, so 60 bins is a conservative range for the measurement. Therefore there are 61 (including no charged particles/tracks) $\times$ 11 = 671 total bins in the measurement.

\begin{figure}[h!]
\begin{center}
\includegraphics[width=0.5\textwidth]{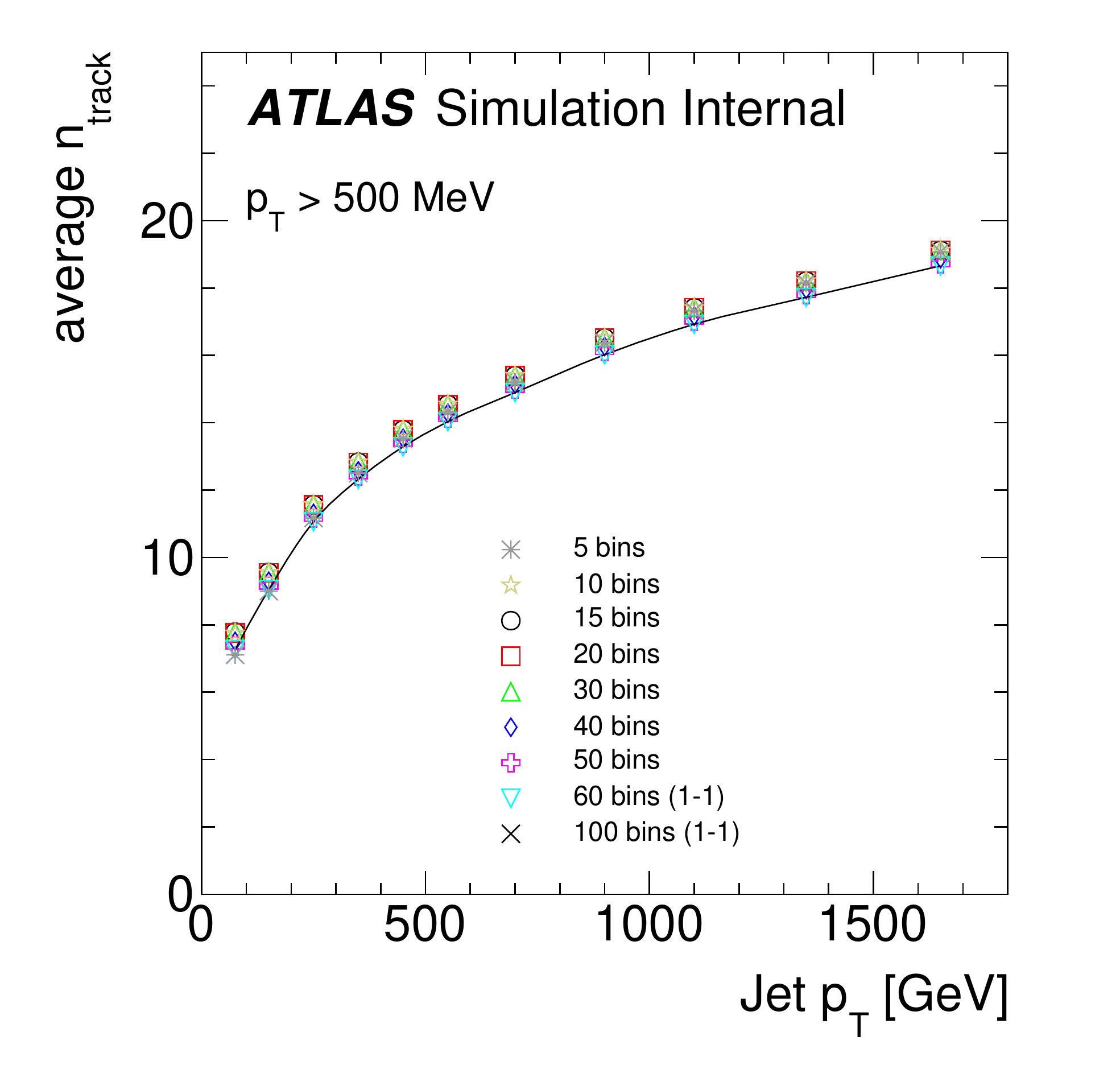}\includegraphics[width=0.5\textwidth]{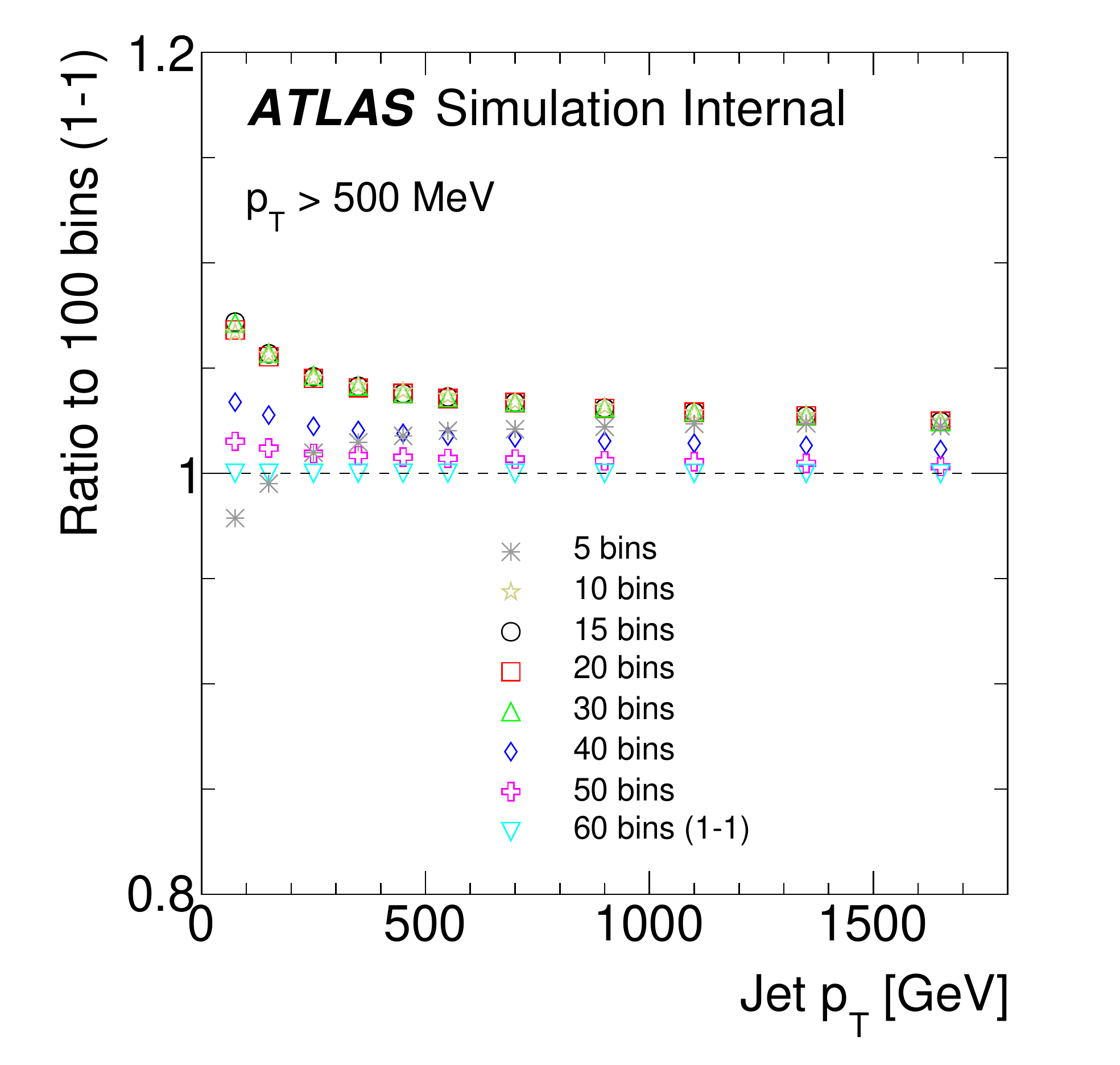}
\caption{The impact of binning the $n_\text{charged}$ distribution at particle-level.  The notation 1-1 means that there is one bin per charged particle multiplicity.}
\label{fig:NCharge:binning}
\end{center}
\end{figure}

\begin{figure}[h!]
\begin{center}
\includegraphics[width=0.5\textwidth]{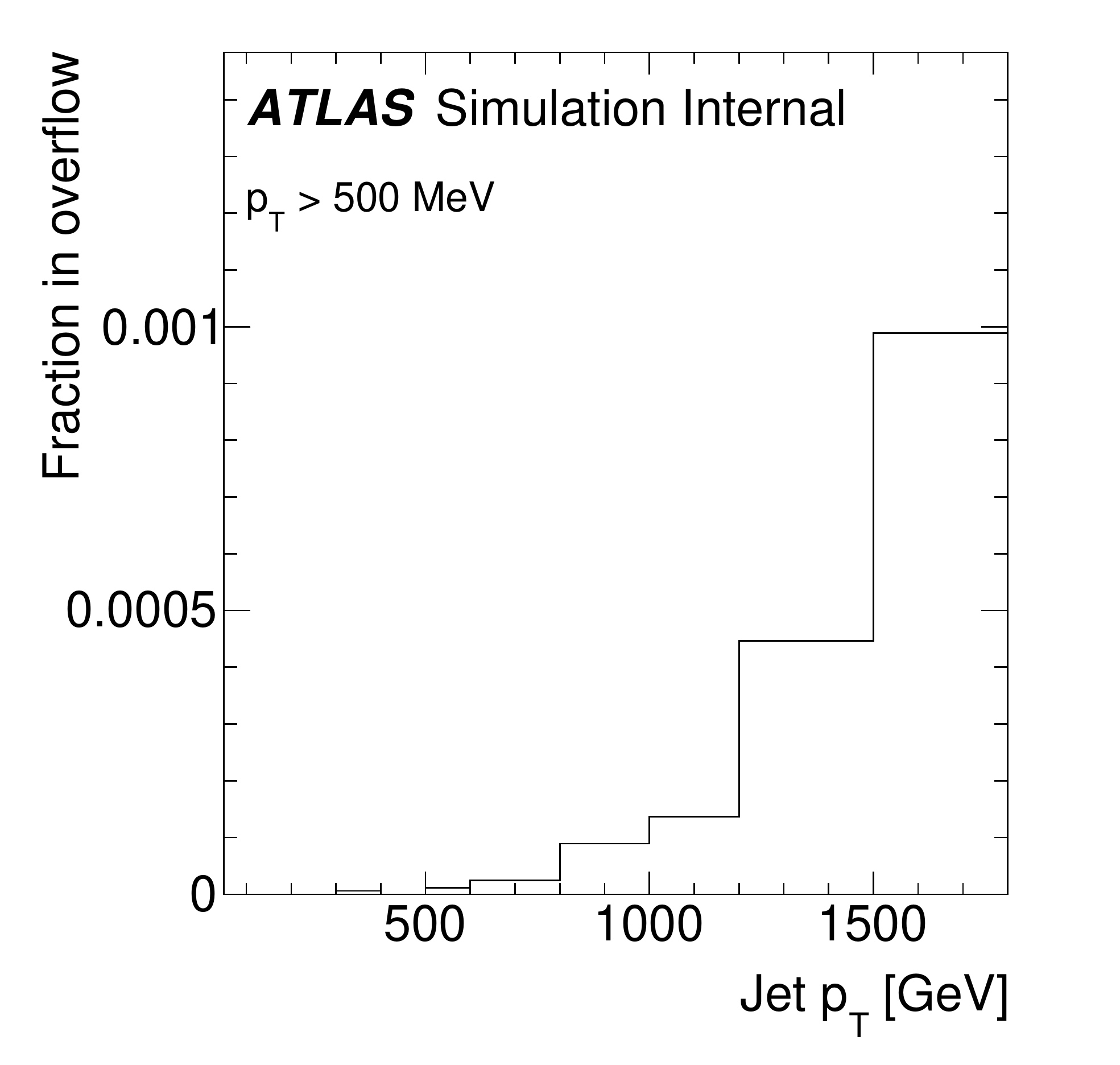}
\caption{The fraction of events with more than 60 charged particles.}
\label{fig:NCharge:binning2}
\end{center}
\end{figure}

Figure~\ref{fig:NCharge:fakeineffic} shows the {\it fake} and {\it inefficiency} factors as a function of the bin number $i=1,...,671$ that are applied in simulation before the response matrix can be used to perform the unfolding. There are some structures that are similar to the analogous figure for the jet charge measurement (Sec.~\ref{corrrfactors}), such as the generally decreasing correction as a function of jet $p_\text{T}$.  However, the within $p_\text{T}$-bin structure is new - for the jet charge the fake and inefficiency factors are largely independent of the jet charge.  These structures are due to the jet calibration - the jets in this measurement do not have the residual track-based correction from the global sequential calibration.  The response is lower for jets with a large number of tracks and so there are cases where an event does not pass the jet $p_\text{T}$ symmetry requirement at particle-level but does at detector-level as a result of the lower response, leading to the cyclic dips in the right plot of Fig~\ref{fig:NCharge:fakeineffic}\footnote{In principle, the left plot in Fig.~\ref{fig:NCharge:fakeineffic} cannot exceed unity; it appears to do so in a few bins due to rounding errors.}.

\begin{figure}[h!]
\begin{center}
\includegraphics[width=0.45\textwidth]{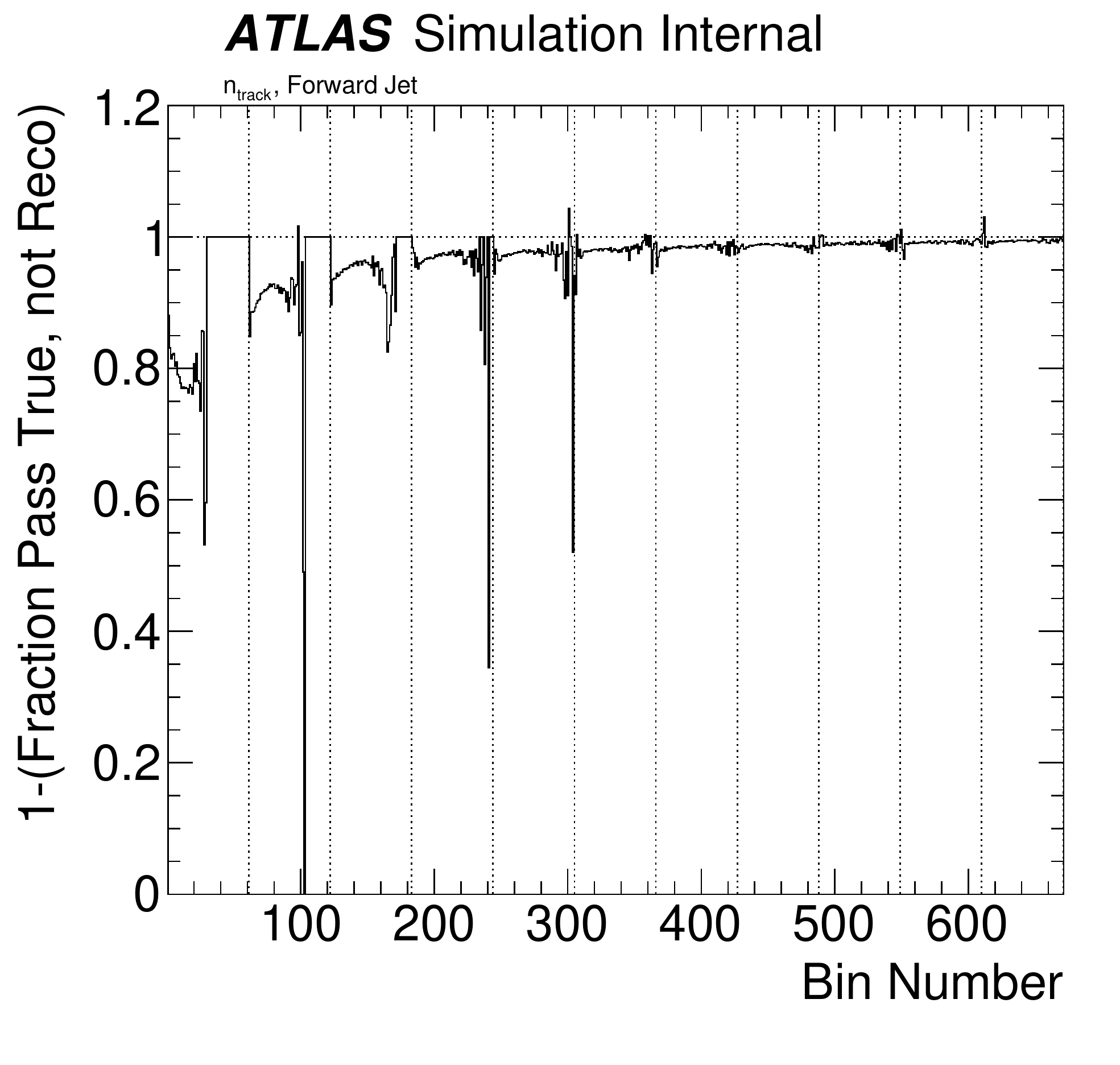}\includegraphics[width=0.45\textwidth]{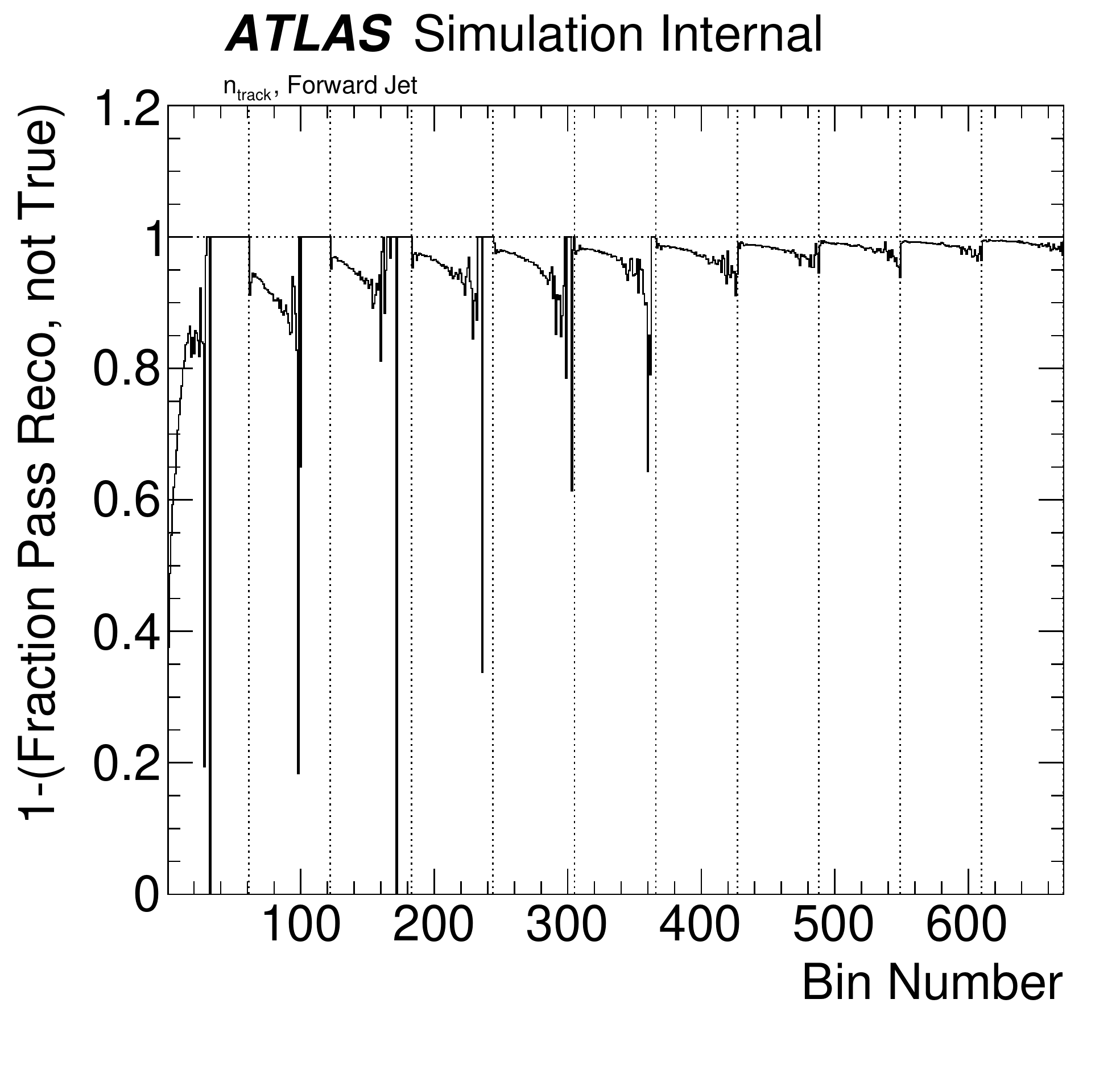}\\
\caption{For each bin of the combined jet $p_\text{T}$ and $n_\text{track}$ distribution, the inefficiency factors (left) and the fake factors (right) for the more forward jet.}
\label{fig:NCharge:fakeineffic}
\end{center}
\end{figure}

After the correction factors are applied, the two-dimensional distribution of the $n_\text{track}$ and jet $p_\text{T}$ is unfolded using the same iterative Bayesian (IB) technique as for the jet charge measurement.  The number of iterations, trading off unfolding bias with statistical fluctuations, is chosen by studying the unfolding bias when unfolding pseudo-data derived from {\sc Herwig++} using a prior distribution and a response matrix derived from {\sc Pythia}. Figure~\ref{fig:NChargeIts} shows the bias induced from a variety of iteration choices.  The improvement from increasing the number of iterations beyond three is marginal, but to be consistent with the jet charge measurement, four iterations are used for all subsequent results.

\begin{figure}[h!]
\begin{center}
\includegraphics[width=0.5\textwidth]{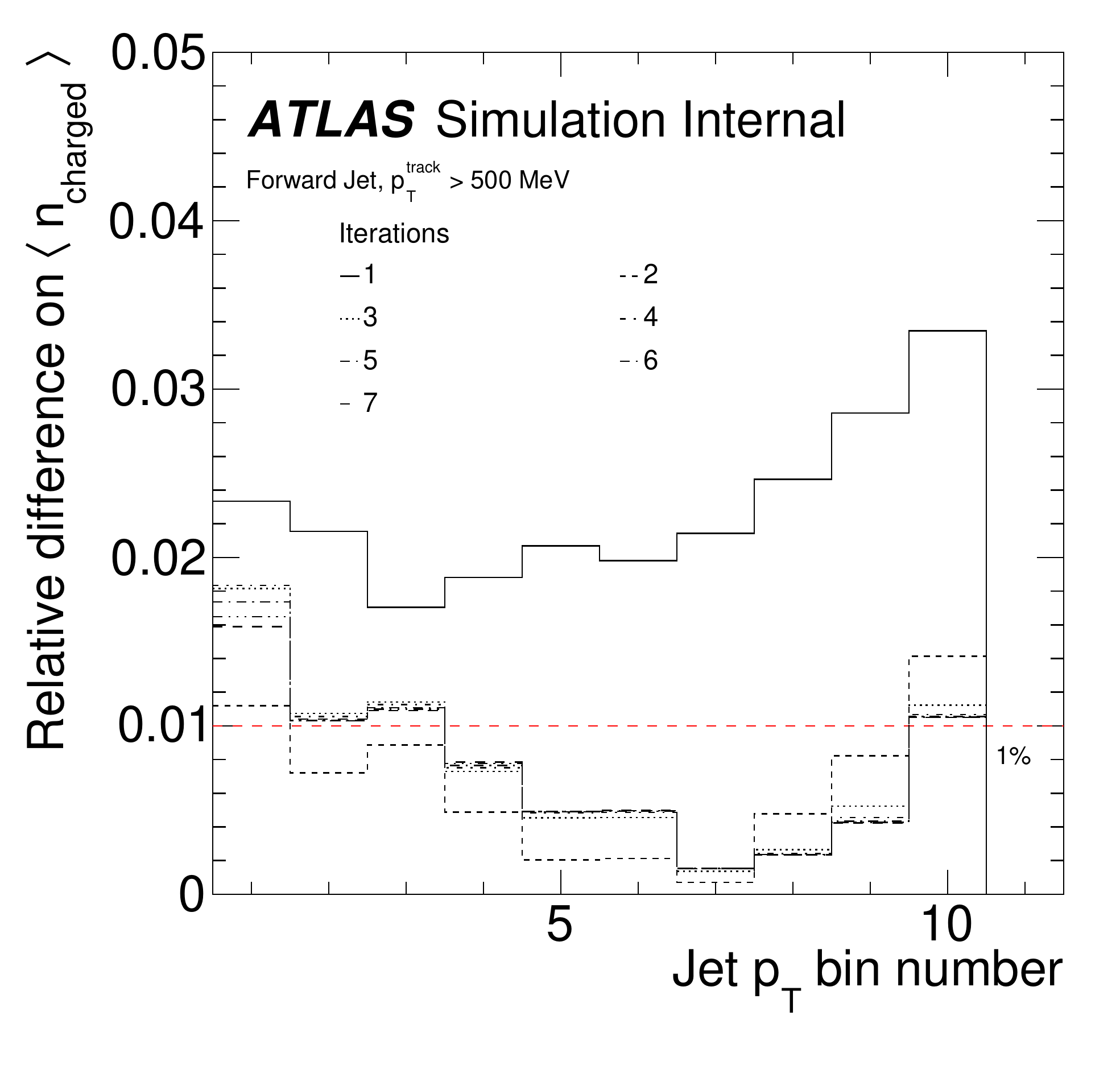}
\end{center}	
\caption{The fractional bias induced when unfolding {\sc Pythia} 8 simulation with a {\sc Herwig++} response matrix for various numbers of iterations in the IB technique.}
\label{fig:NChargeIts}
\end{figure}

The response matrix connects the prior to the posterior distribution in each step of the IB method.  Figure~\ref{fig:NChargeresponsematrix} shows the nominal response matrix from {\sc Pythia} 8.  The matrix is nearly diagonal with several structures due to the nature of the binning.  In particular, the nearly diagonal stripe in the left plot of Fig.~\ref{fig:NChargeresponsematrix} corresponds to events that were in the same particle- and detector-level $p_\text{T}$ bin.  The strip below the diagonal is more prominent than the one above the diagonal because given the jet $p_\text{T}$ distribution is steeply falling and thus migrating to lower detector-level jet $p_\text{T}$ values is more likely than higher jet $p_\text{T}$ values.  This effect diminishes as the size of the $p_\text{T}$ bin goes to zero.  The right plot in Fig.~\ref{fig:NChargeresponsematrix} shows the response matrix over the 61 $n_\text{charged}$ bins averaged over all jet $p_\text{T}$ bins.  The matrix spreads away from the diagonal at high $n_\text{track}$ due in part to the binomial effect\footnote{If every track is lost with probability $p$, then for $n$ charged particles, the average number of reconstructed tracks is $np$ and the standard deviation is $\sqrt{np(1-p)}$.} and there is a bias that the unfolding needs to correct: the average detector-level $n_\text{track} < $ particle-level $n_\text{charged}$.  This bias increases with jet $p_\text{T}$, as shown in Fig.~\ref{fig:NChargeresponsematrix2}.  The lower panel of Fig.~\ref{fig:NChargeresponsematrix2} shows the average detector-level $n_\text{track}$ divided by the particle-level $n_\text{track}$ in each particle-level $n_\text{track}$ bin.  This offset is bigger for the higher jet $p_\text{T}$ bin because of the increased loss of tracks due to hit merging.

\begin{figure}[h!]
\begin{center}
\includegraphics[width=0.42\textwidth]{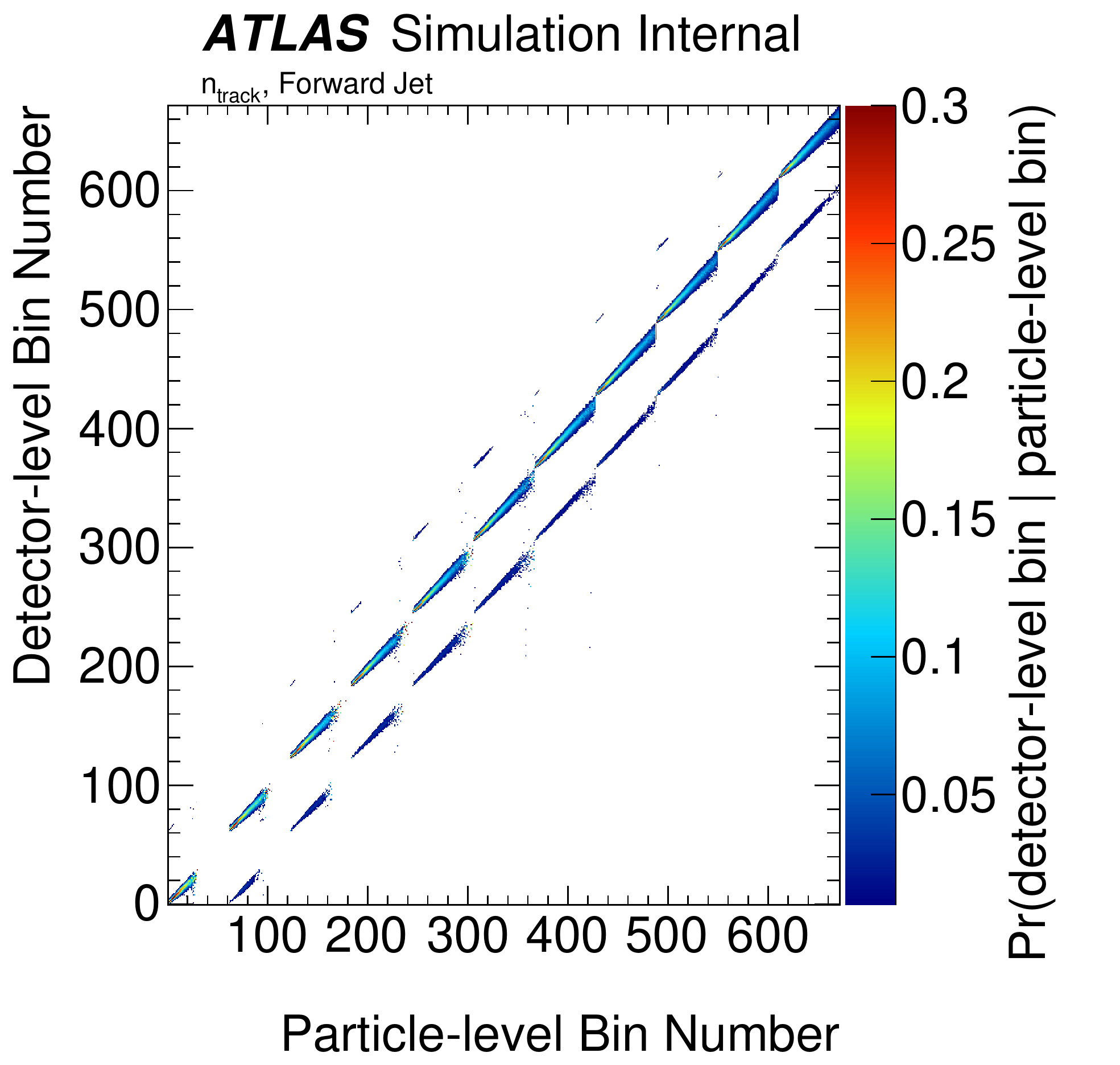}\includegraphics[width=0.42\textwidth]{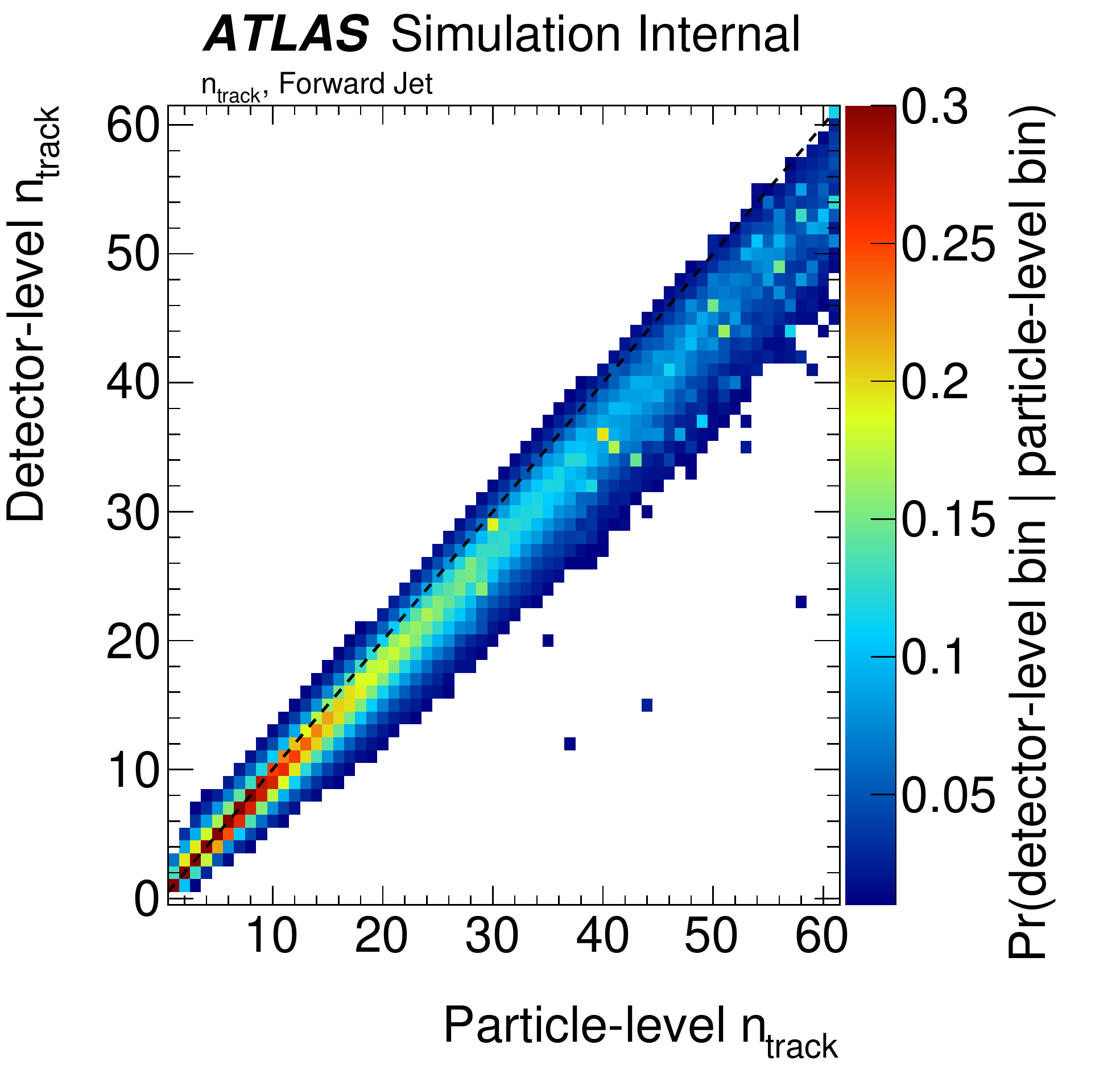}
\caption{The total response matrix (left) and the response matrix for the 61 $n_\text{track}$ bins averaged over the 11 jet $p_\text{T}$ bins.  The z-axis is truncated at 1\%.}
\label{fig:NChargeresponsematrix}
\end{center}
\end{figure}

\begin{figure}[h!]
\begin{center}
\includegraphics[width=0.42\textwidth]{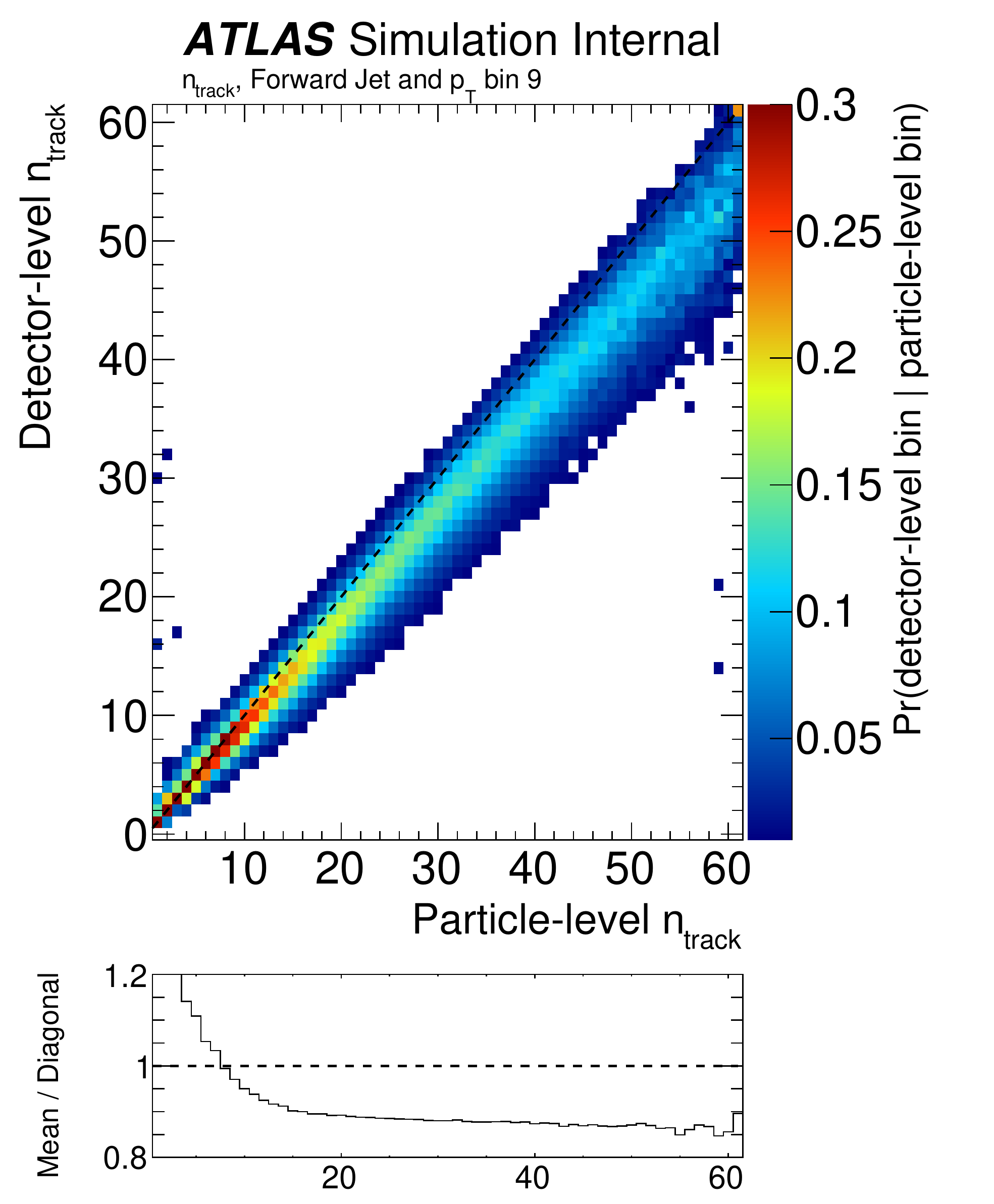}\includegraphics[width=0.42\textwidth]{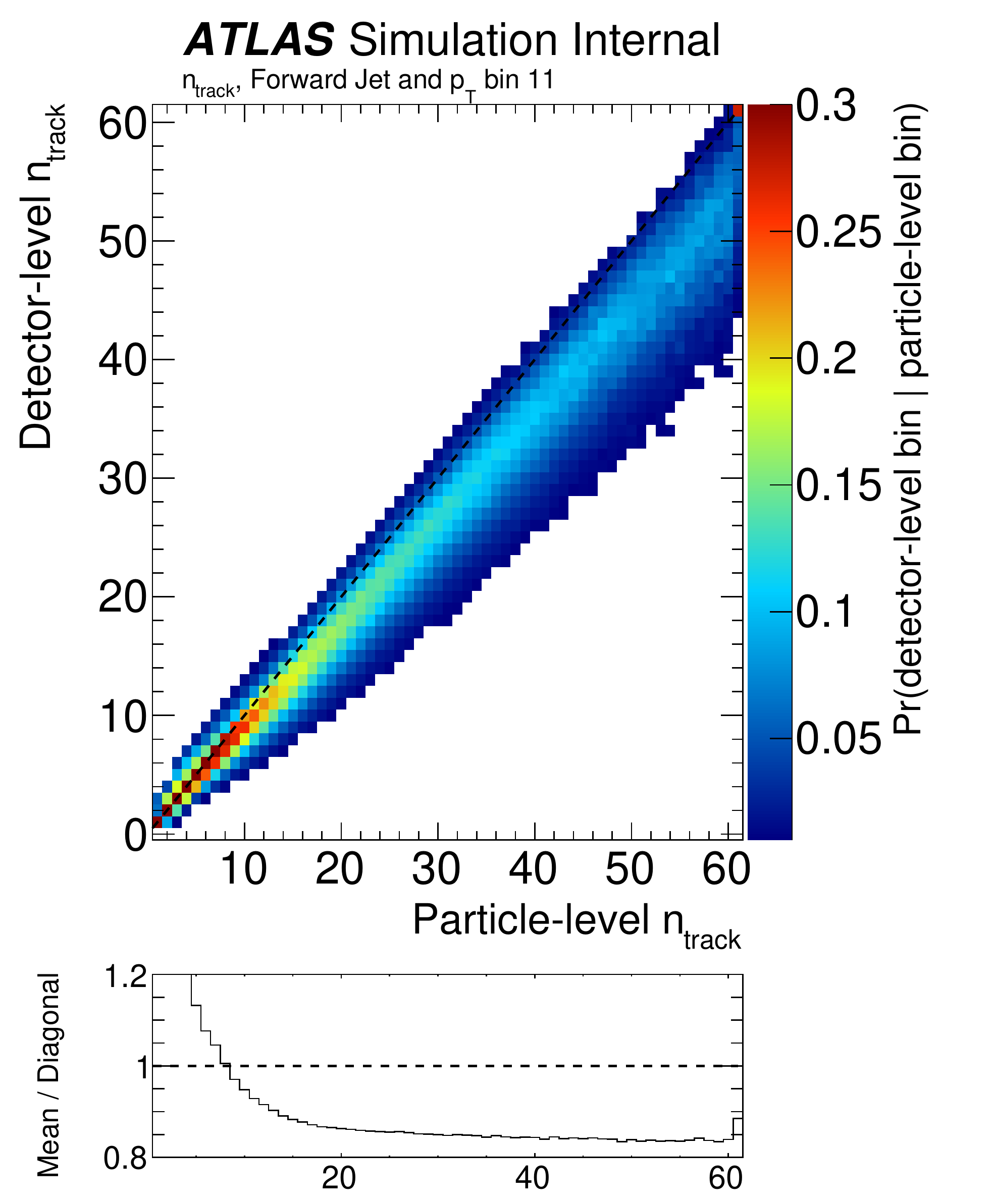}
\caption{The response matrix for the 61 $n_\text{track}$ bins in the ninth jet $p_\text{T}$ bin (1.0 TeV $<p_\text{T}<$ 1.2 TeV) on the left and the last jet $p_\text{T}$ bin ($p_\text{T}>$ 1.5 TeV) on the right.  The lower panel shows the average detector-level $n_\text{track}$ divided by the particle-level $n_\text{track}$ in each particle-level $n_\text{track}$ bin.  The z-axis is truncated at 1\%.}
\label{fig:NChargeresponsematrix2}
\end{center}
\end{figure}

\clearpage

An overview of the unfolding is shown in Fig.~\ref{fig:truthreco}.   The top left plot in Fig.~\ref{fig:truthreco} shows the jet $p_\text{T}$ dependence of $n_\text{track}$ before unfolding for the three track $p_\text{T}$ thresholds.  As observed earlier, the {\sc Pythia} 8 sample with the AU2 over-predicts the number of tracks inside jets.  The relative over-prediction does not largely vary for the three track $p_\text{T}$ bin.  The top right plot of Fig.~\ref{fig:truthreco} shows the particle-level and detector-level simulations to illustrate the size of the bias corrected by the unfolding.  This is quantified in the bottom right plot of Fig.~\ref{fig:truthreco}, which is the ratio of the solid and dotted lines in the top right plot of Fig.~\ref{fig:truthreco}.  Note that this is for illustration purposes - the actual corrections are done over the 671 jet $p_\text{T}$ and $n_\text{track}$ bins and not to the $\langle n_\text{track}\rangle$ itself.  The unfolded data with statistical uncertainty determined by bootstrapping are shown in the bottom left plot of Fig.~\ref{fig:truthreco}.  The next step to determine the quality of the modeling from simulation is to assess sources of systematic uncertainty.

\begin{figure}[h!]
\begin{center}
\includegraphics[width=0.99\textwidth]{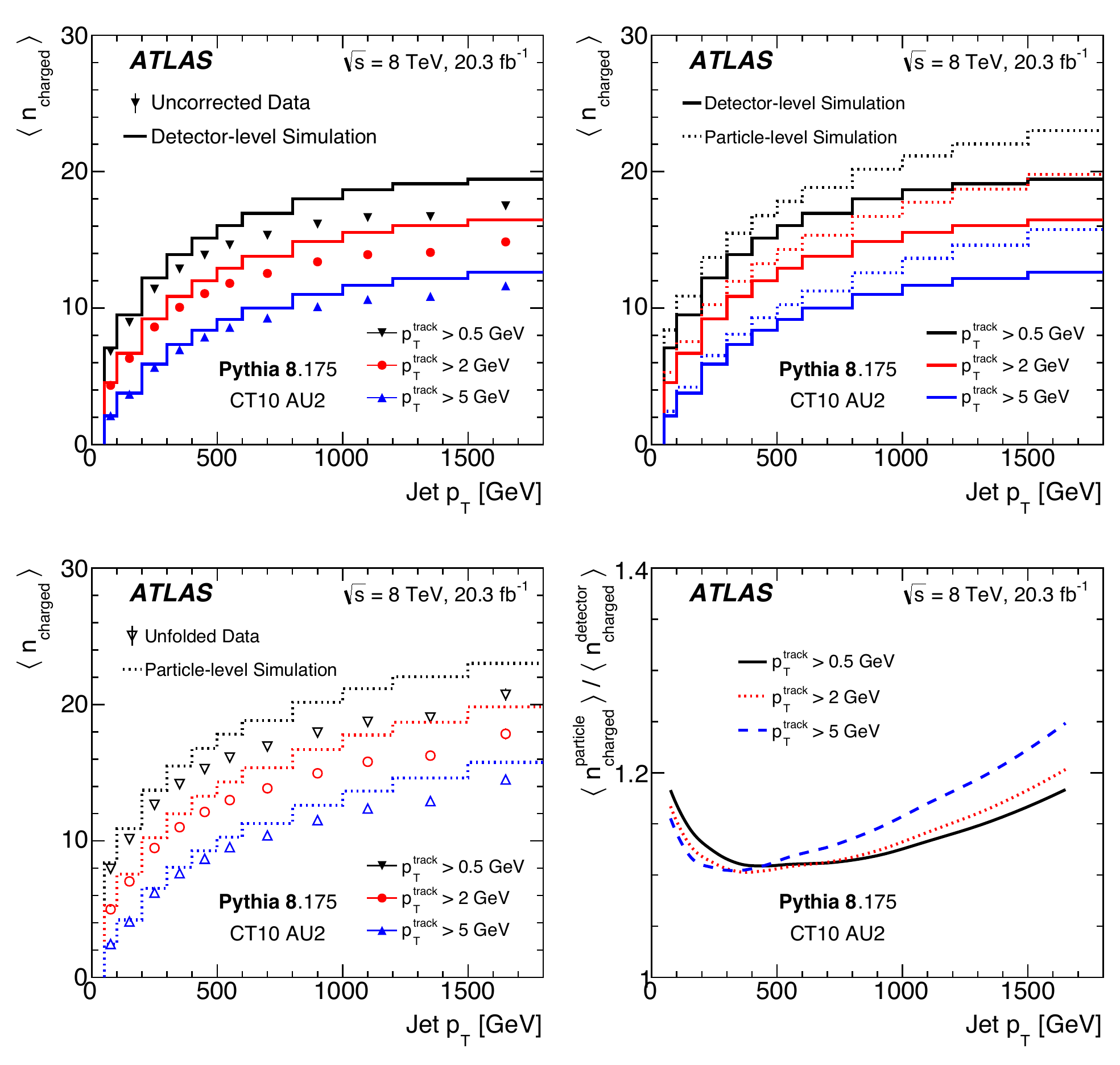}
\end{center}	
\caption{The jet $p_\text{T}$ dependence of (a) the average reconstructed track multiplicity for uncorrected data and detector-level simulation, (b) the average reconstructed track multiplicity for the detector-level simulation and the average charged-particle multiplicity for the particle-level simulation, (c) the average charged-particle multiplicity for the unfolded data and the particle-level simulation, and  (d) the average charged-particle multiplicity divided by the average reconstructed track multiplicity in simulation. For the data, only statistical uncertainties are included in the error bars (which are smaller than the markers for most bins).}
\label{fig:truthreco}
\end{figure}

\clearpage

\section{Systematic uncertainties}
\label{sec:NCharge:Systs}

All stages of the charged-particle multiplicity measurement are sensitive to sources of potential bias.  The method for evaluating the uncertainties is the same as for the jet charge measurement, as described below.

\subsection{Response Matrix}

For events that pass both the detector-level and particle-level fiducial selections, the response matrix describes migrations between bins when moving between the detector level and the particle level.  The response matrix is taken from simulation and various experimental uncertainties in the charged-particle multiplicity and jet $p_\text{T}$ spectra result in uncertainties in the matrix.  These uncertainties can be divided into two classes: those impacting the calorimeter-based jet $p_\text{T}$ and those impacting track reconstruction inside jets.   The dominant uncertainty at high jet $p_\text{T}$ is due to the loss of charged-particle tracks in the jet core due to track merging.  This charged energy loss uncertainty is estimated using the data/MC differences in the ratio of the track-based jet $p_\text{T}$ to the calorimeter-based jet $p_\text{T}$ as was also done for the jet charge.  More charged energy is lost in the data than in the MC and thus this uncertainty is one-sided.  There are other tracking uncertainties in the track momentum scale and resolution, the track reconstruction efficiency, and the rate of tracks formed from random combinations of hits (fake tracks).  The uncertainties related to the calorimeter-based jet are sub-dominant (except in the lowest $p_\text{T}$ bins) and are due to the uncertainty in the jet energy scale and the jet energy resolution.

\subsubsection{Charged-energy loss in the dense core of jets}

The uncertainty on the charged-energy loss in the dense core of jets is estimated from the modeling of $\sum p_{T}^\text{track}/p_T^\text{calo jet}$, as described in Sec~\ref{sec:jetcharge:tracksyst:TIDE}.  The prescription for the uncertainty is to drop tracks randomly with the following probability: $\Pr(\text{drop track $i$})=\alpha p_{T,i}^n$, where $n$ is some non-negative integer.  The value of $\alpha$ is fixed by requiring the data and MC to agree on the average $\sum p_{T}^\text{track}/p_T^\text{calo jet}$.  The only free parameter of the prescription is the power $n$ of the track $p_T$ used to model the uncertainty so one must find the power that has the biggest impact on the observable.   In the jet charge measurement, it was found that a very large power was conservative because at high $p_T$, the largest contribution to the jet charge comes from the highest $p_T$ tracks.  However, for $n_\text{track}$, a low power is conservative because all tracks are treated equally, independent of their momentum (as long as the $p_T$ is large enough to pass the threshold).  This is illustrated in figure~\ref{fig:systs_tide}, where the impact of dropping tracks randomly with $\Pr(\text{drop track $i$})=\alpha p_{T,i}^n$ for various values of $n$.  Therefore, a power of $0$ (i.e. all tracks are dropped with equal probability, regardless of their $p_T$) is adapted.  Note that even for a power of zero the uncertainty increases with $p_T$ because the data/MC difference in $\sum p_{T}^\text{track}/p_T^\text{calo jet}$ increases with $p_T$.   The uncertainty is slightly larger for $n>0$ for a $p_T$ threshold of 5 GeV versus 0.5 GeV, but by construction the uncertainty is independent of the threshold when $n=0$.

\begin{figure}[h!]
\begin{center}
\includegraphics[width=0.5\textwidth]{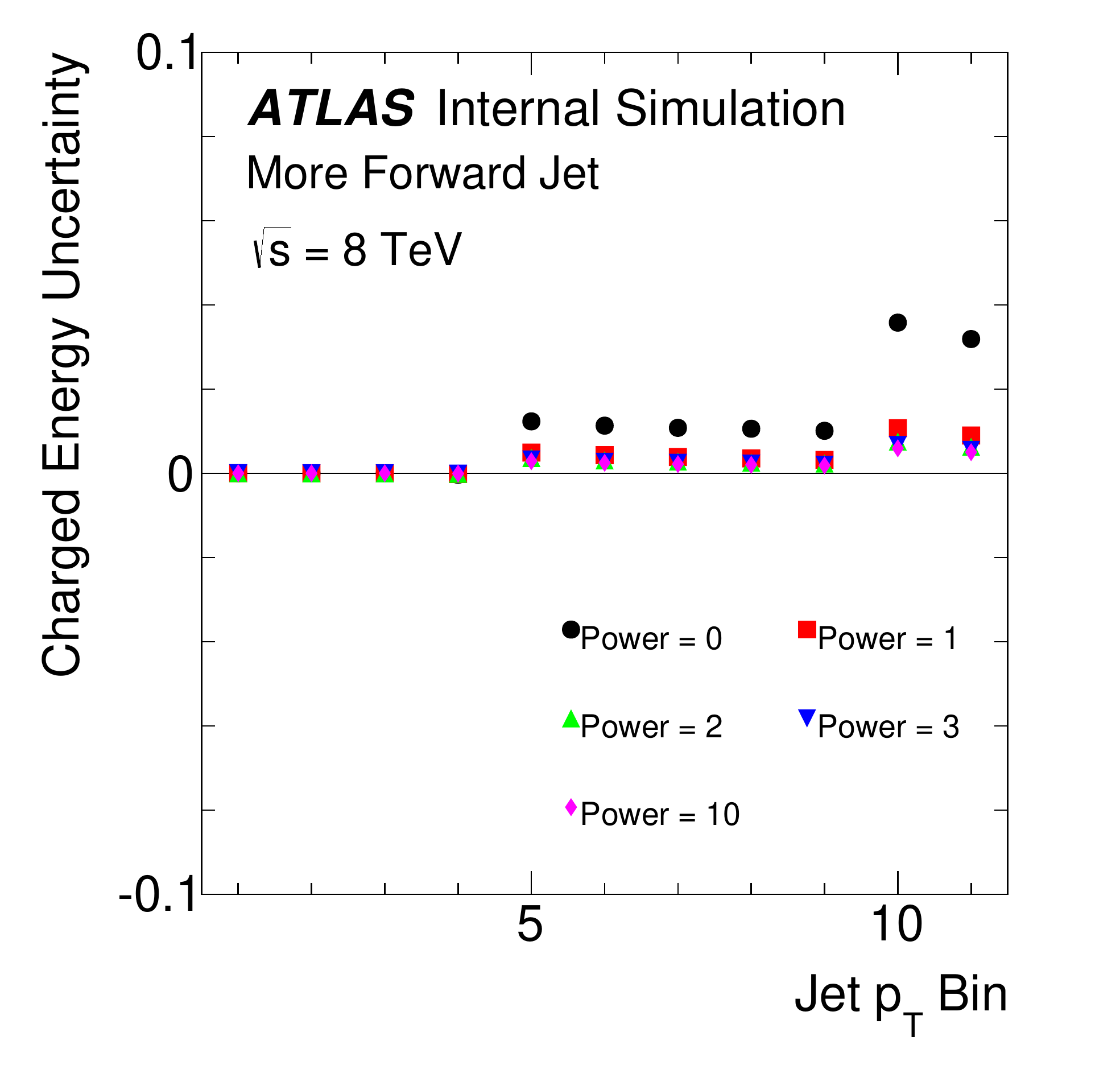}\includegraphics[width=0.5\textwidth]{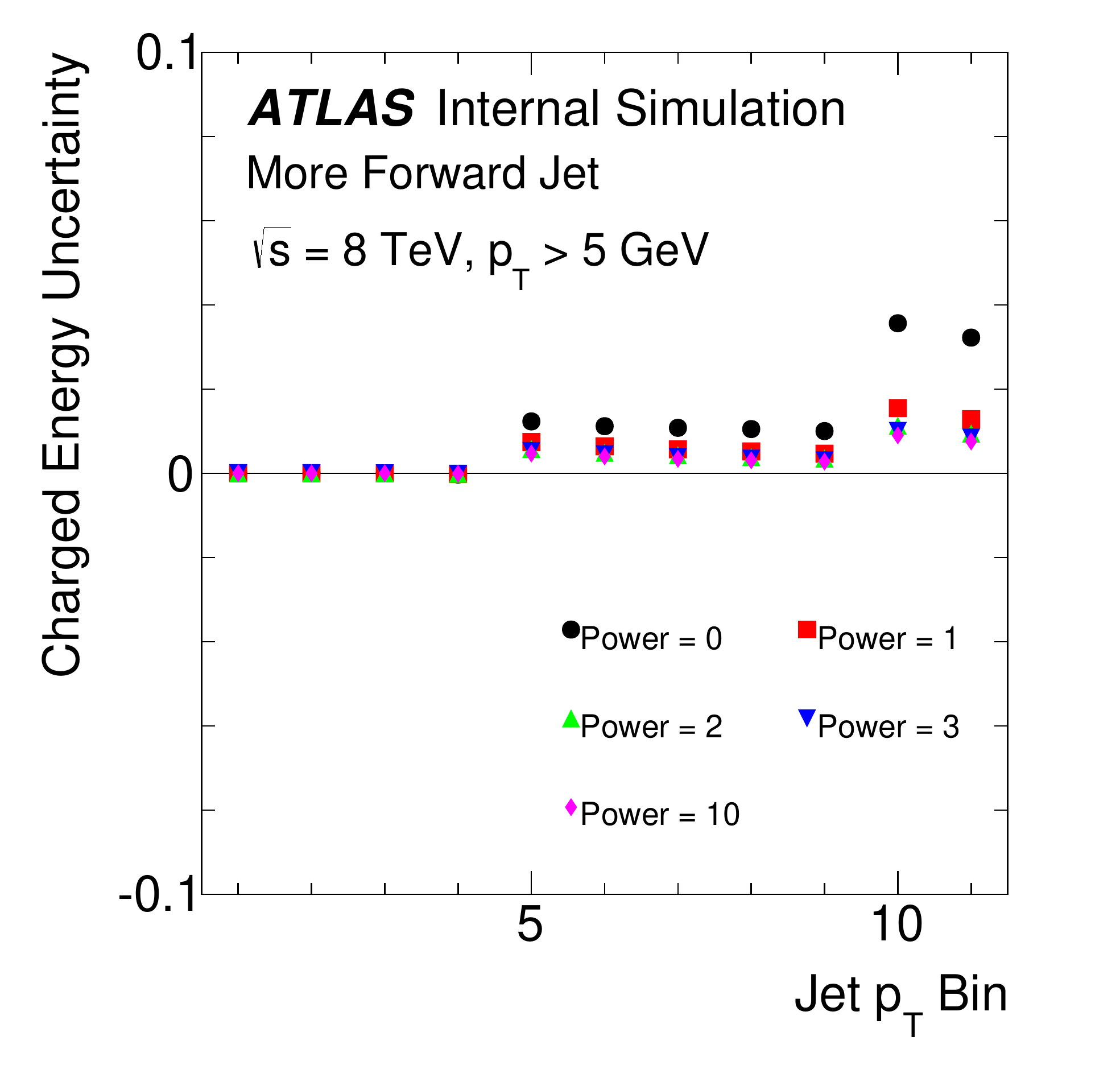}
\caption{The impact of the charged energy loss uncertainty on the average number of charged particles in each $p_T$ bin.  The vertical axis is the relative uncertainty on the average unfolded $n_\text{charged}$.  For each $n$, tracks are dropped randomly with a probability given by $\Pr(\text{drop track $i$})=\alpha p_{T,i}^n$, where $\alpha$ is fixed such that the MC is the same as the data in Fig.~\ref{fig:SystematicUncertainties/TIDEfig3}.  Below $400$ GeV, nuclear interactions dominate the track reconstruction efficiency uncertainty.}
\label{fig:systs_tide}
\end{center}
\end{figure}

\clearpage

\subsection{Correction Factors}
\label{sec:NCharge:corrections}

Fake and inefficiency factors are derived from simulation to account for the fraction of events that pass either the detector-level or particle-level fiducial selection, but not both.  These factors are generally between $0.9$ and $1.0$ except in the first jet-$p_\text{T}$ interval (50~$<p_\text{T}<100$~GeV), where threshold effects cause the correction factors to take values down to 0.8 (see Fig.~\ref{fig:NCharge:fakeineffic}).  Experimental uncertainties correlated with the detector-level selection acceptance, such as the jet energy scale uncertainty, result in uncertainties in these correction factors.  Another source of uncertainty in the correction factors is the explicit dependence on the particle-level multiplicity and jet $p_\text{T}$ spectrum.  A comparison of particle-level models ({\sc {\sc Pythia}} and {\sc {\sc Herwig++}}) is used to estimate the impact on the correction factors.  As was also done for the jet charge, the nominal fake and inefficiency factors from {\sc Pythia} 8 are re-weighted to those from {\sc Herwig++} and the unfolding is performed with the nominal {\sc Pythia}~8 response matrix.  Figure~\ref{fig:NCharge:effic_beforeunfold} shows the impact of the bin-by-bin re-weighting on the two-dimensional jet $p_\text{T}$ and $n_\text{track}$ distributions as well as their impact on the unfolded $\langle n_\text{track}\rangle$ distribution.  In general, these uncertainties are $<0.1\%$.  A similar set of plots with nearly the same conclusion is shown for the fake factor uncertainty in Fig.~\ref{fig:NChargefake_beforeunfold}.

\begin{figure}[h!]
\begin{center}
\includegraphics[width=0.45\textwidth]{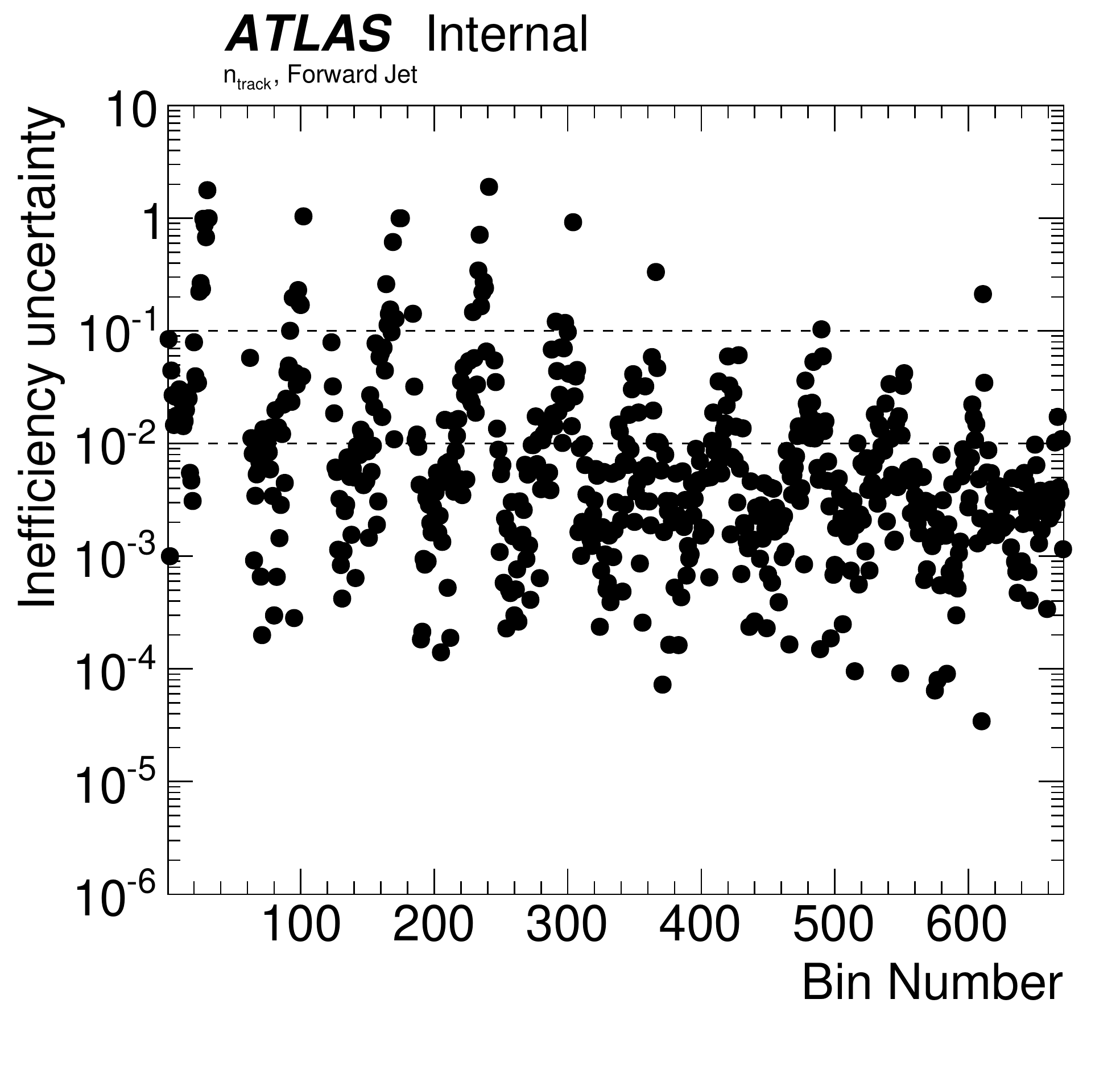}\includegraphics[width=0.45\textwidth]{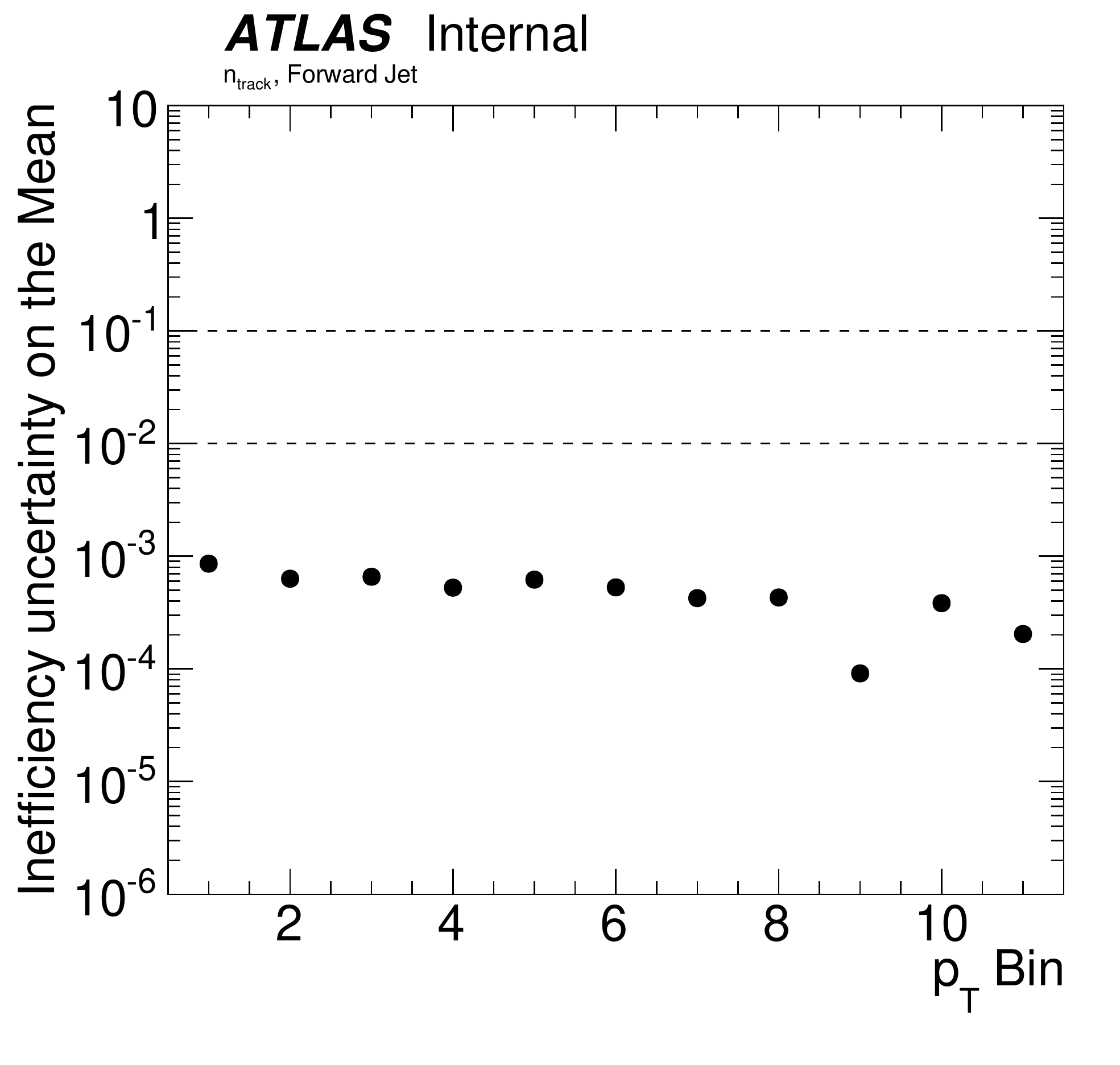}
\caption{The bin-by-bin inefficiency factor fractionl uncertainty before unfolding (left) and the inefficiency factor fractional uncertainty on $\langle n_\text{track}\rangle$ as a function of the jet $p_\text{T}$ bin (right) for track $p_\text{T}>500$ MeV.}
\label{fig:NCharge:effic_beforeunfold}
\end{center}
\end{figure}

\begin{figure}[h!]
\begin{center}
\includegraphics[width=0.45\textwidth]{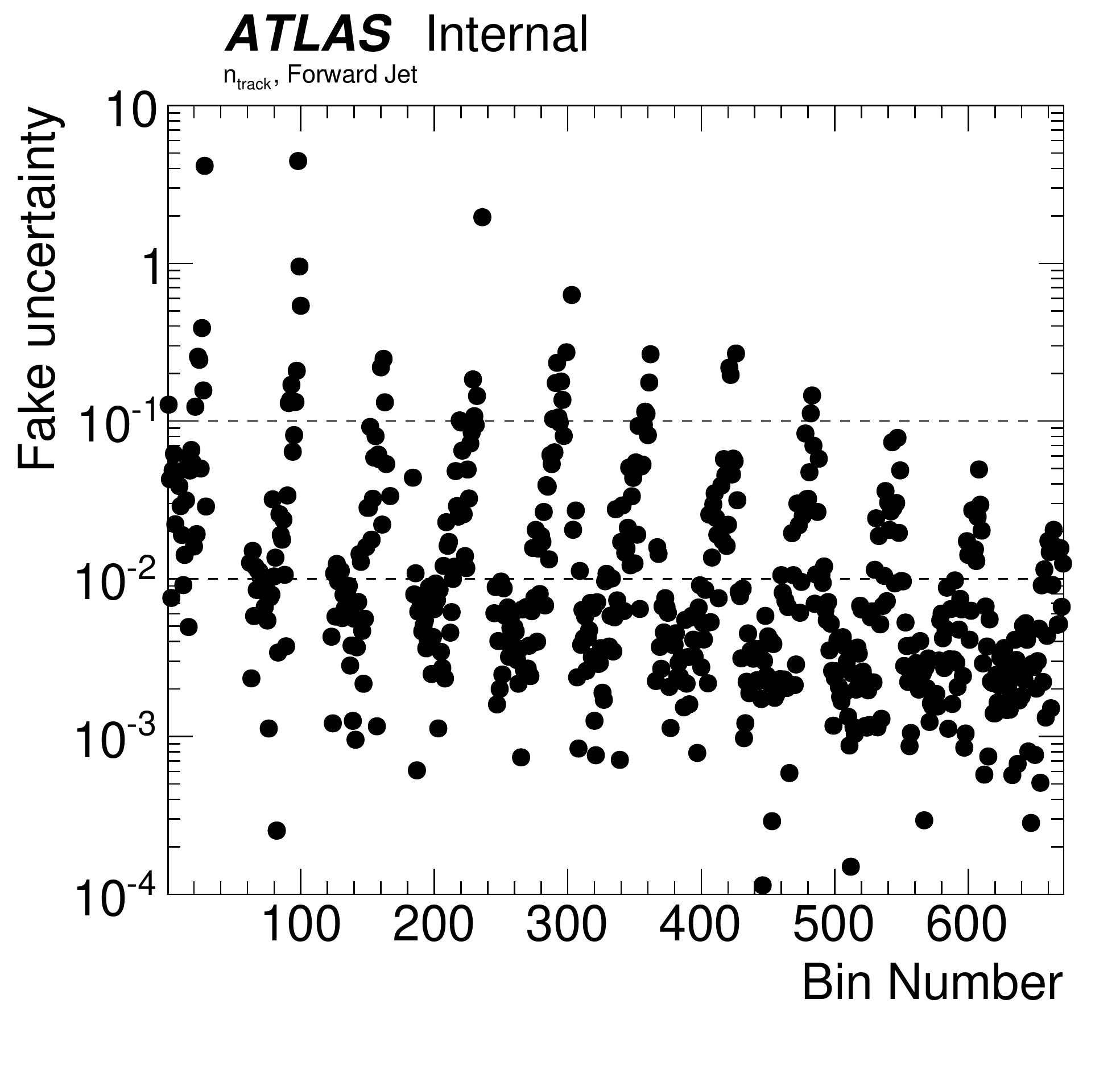}\includegraphics[width=0.45\textwidth]{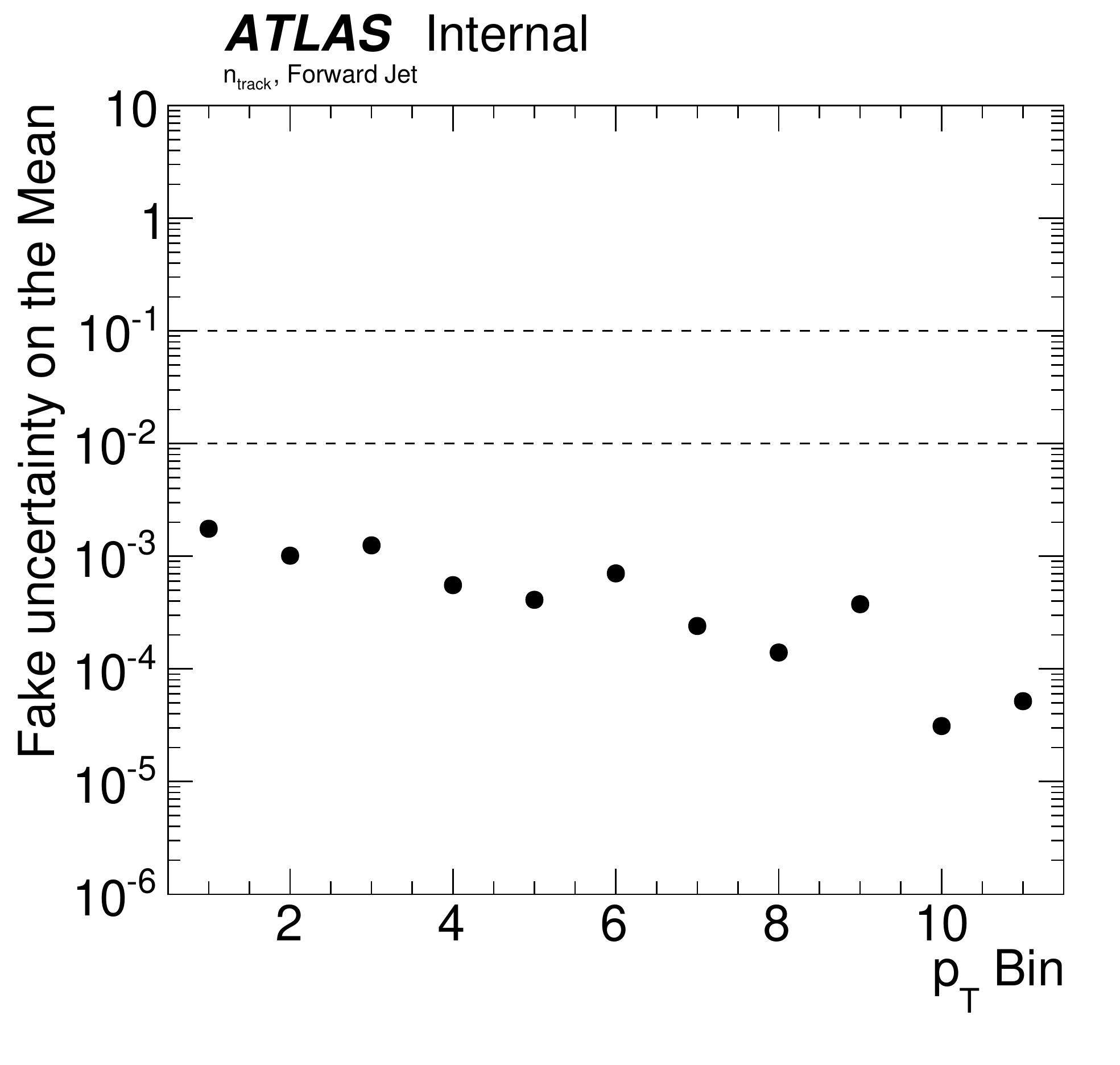}
\caption{The bin-by-bin fake factor fractionl uncertainty before unfolding (left) and the fake factor fractional uncertainty on $\langle n_\text{track}\rangle$ as a function of the jet $p_\text{T}$ bin (right) for track $p_\text{T}>500$ MeV.}
\label{fig:NChargefake_beforeunfold}
\end{center}
\end{figure}

\clearpage

\subsection{Unfolding Procedure} 

The same data-driven non-closure uncertainty technique used for the jet charge measurement is used for $\langle n_\text{charge}\rangle$.  In particular, the particle-level spectrum is reweighted so that the simulated detector-level spectrum, from propagating the reweighted particle-level spectrum through the response matrix, has significantly improved agreement with the uncorrected data.  The modified detector-level distribution is unfolded with the nominal response matrix and the difference between this and the reweighted particle-level spectrum is an indication of the bias due to the unfolding method (in particular, the choice of a prior distribution).  The re-weighting factors are simply determined at detector-level and applied at particle-level.  These factors are shown in the left plot of Fig.~\ref{fig:NCharge:systs_nc_1} and the improvement in the data/MC agreement induced from the particle-level re-weighting is shown in the middle plot of Fig.~\ref{fig:NCharge:systs_nc_1}.  The right plot of Fig.~\ref{fig:NCharge:systs_nc_1} illustrates that the particle-level re-weighting brings the $\langle n_\text{track}\rangle$ distribution into nearly 100\% agreement with the data.

\begin{figure}[h!]
\begin{center}
\includegraphics[width=0.33\textwidth]{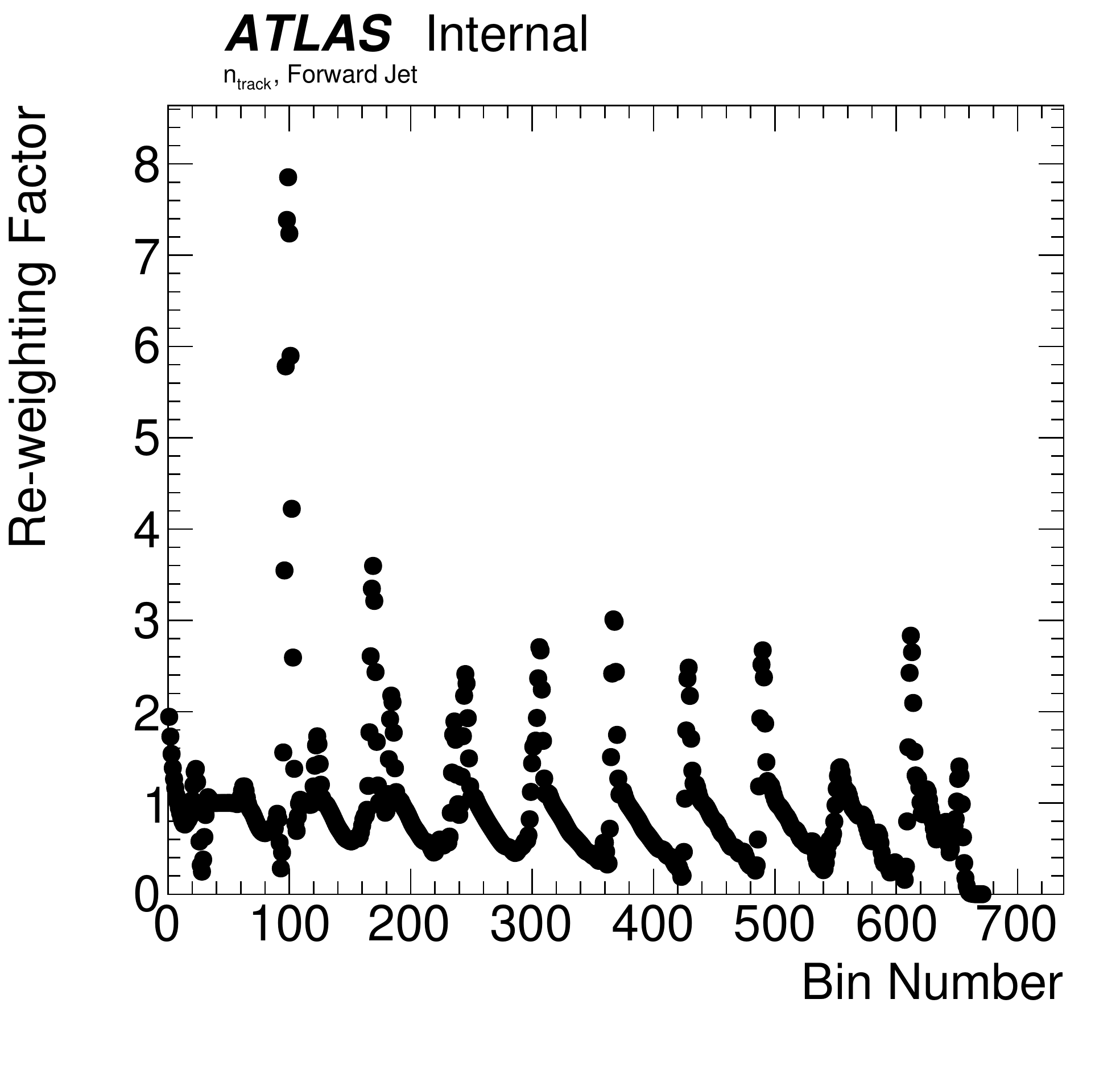}\includegraphics[width=0.33\textwidth]{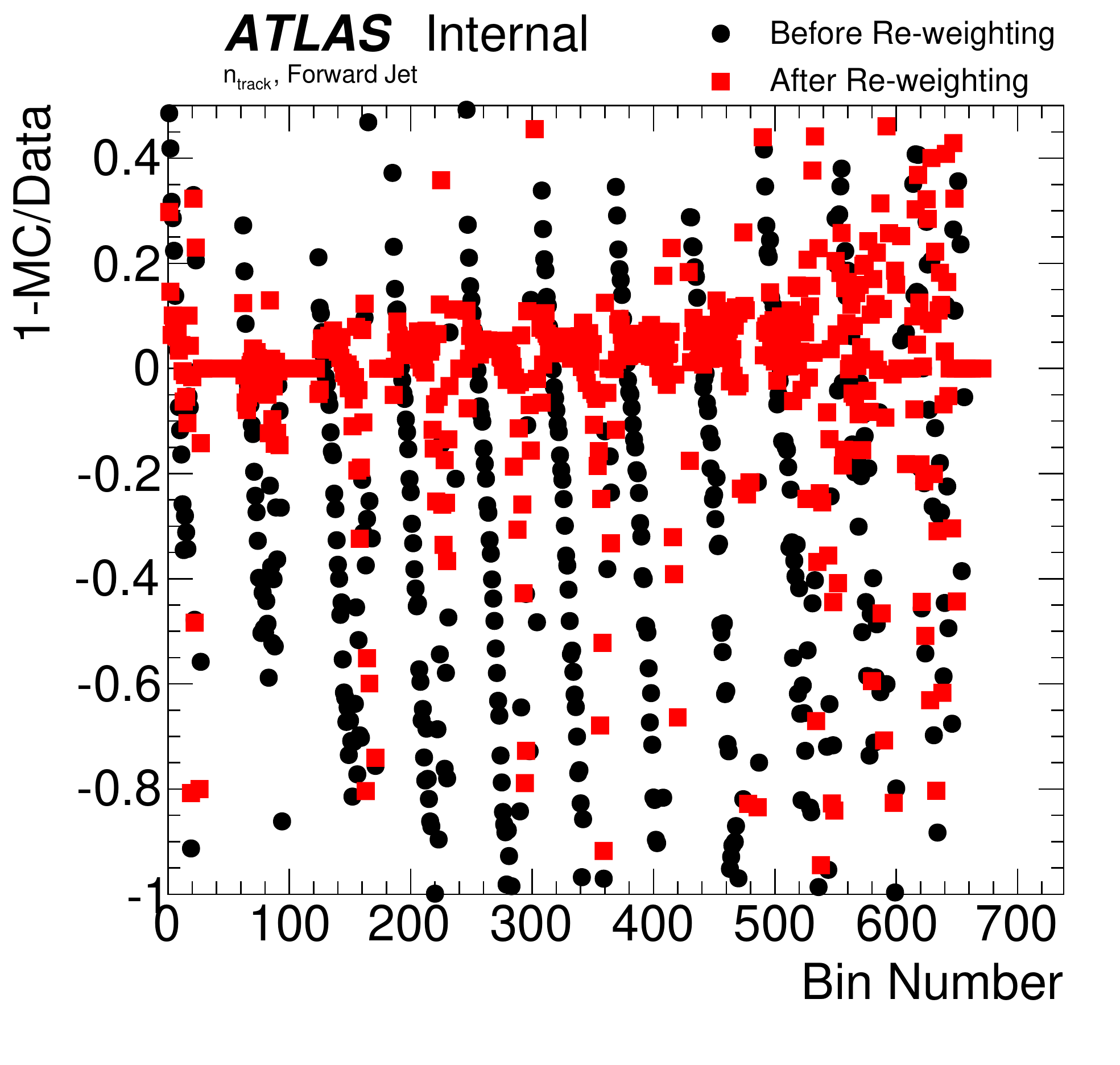}\includegraphics[width=0.33\textwidth]{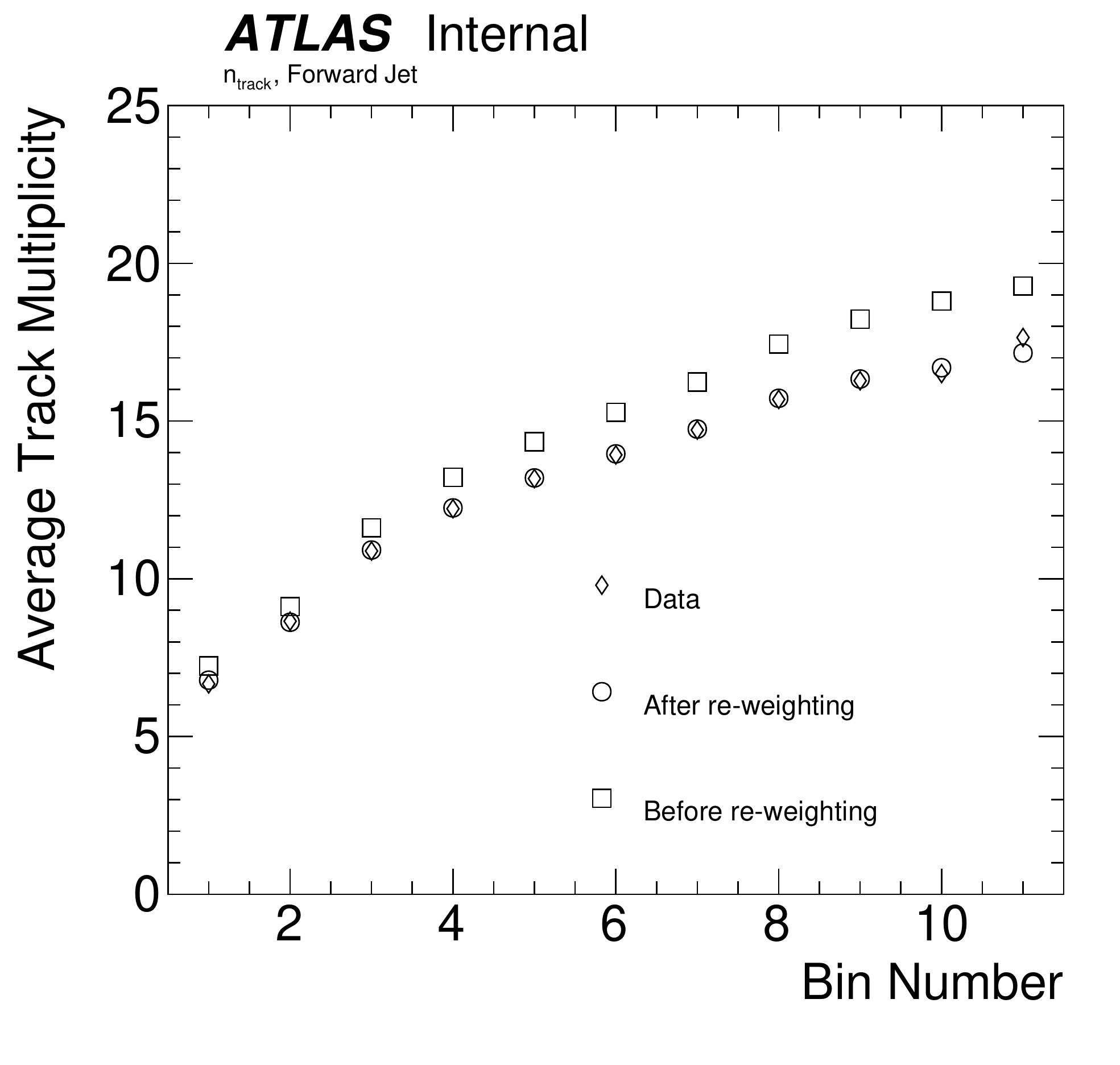}
\caption{The distribution of the weights used to re-weight the MC distribution for the non-closure test (left) and the re-weighted truth distribution (labeled after) (middle) for the more forward jet.  The $\langle n_\text{track}\rangle$ in data and simulation before and after the re-weighting is shown in the right plot.}
\label{fig:NCharge:systs_nc_1}
\end{center}
\end{figure}

The uncertainty on $\langle n_\text{track}\rangle$ due to the data-driven non-closure is shown in the left plot of Fig.~\ref{fig:NCharge:systs_nc_5} and compared with the raw data/MC difference.  Except in the first jet $p_\text{T}$ bin where the track multiplicity is spread out over only a small number of bins, the non-closure uncertainty ($\lesssim 1\%$) is much smaller than the raw data/MC difference ($\sim10\%$).  As a comparison, the relative difference in $\langle n_\text{track}\rangle$ when unfolding {\sc Pythia} 8 with {\sc Herwig++} is shown in the right plot of Fig.~\ref{fig:NCharge:systs_nc_5}.  This difference is not used as an uncertainty as it would over-count the non-closure uncertainty, but it is reassuring that the approximate size of the uncertainty is is comparable to the data-driven technique.  Note that the impact of the difference between {\sc Pythia } 8 and {\sc Herwig++} in the fake/inefficiency factors is already accounted for in Sec.~\ref{sec:NCharge:corrections}.

\begin{figure}[h!]
\begin{center}
\includegraphics[width=0.5\textwidth]{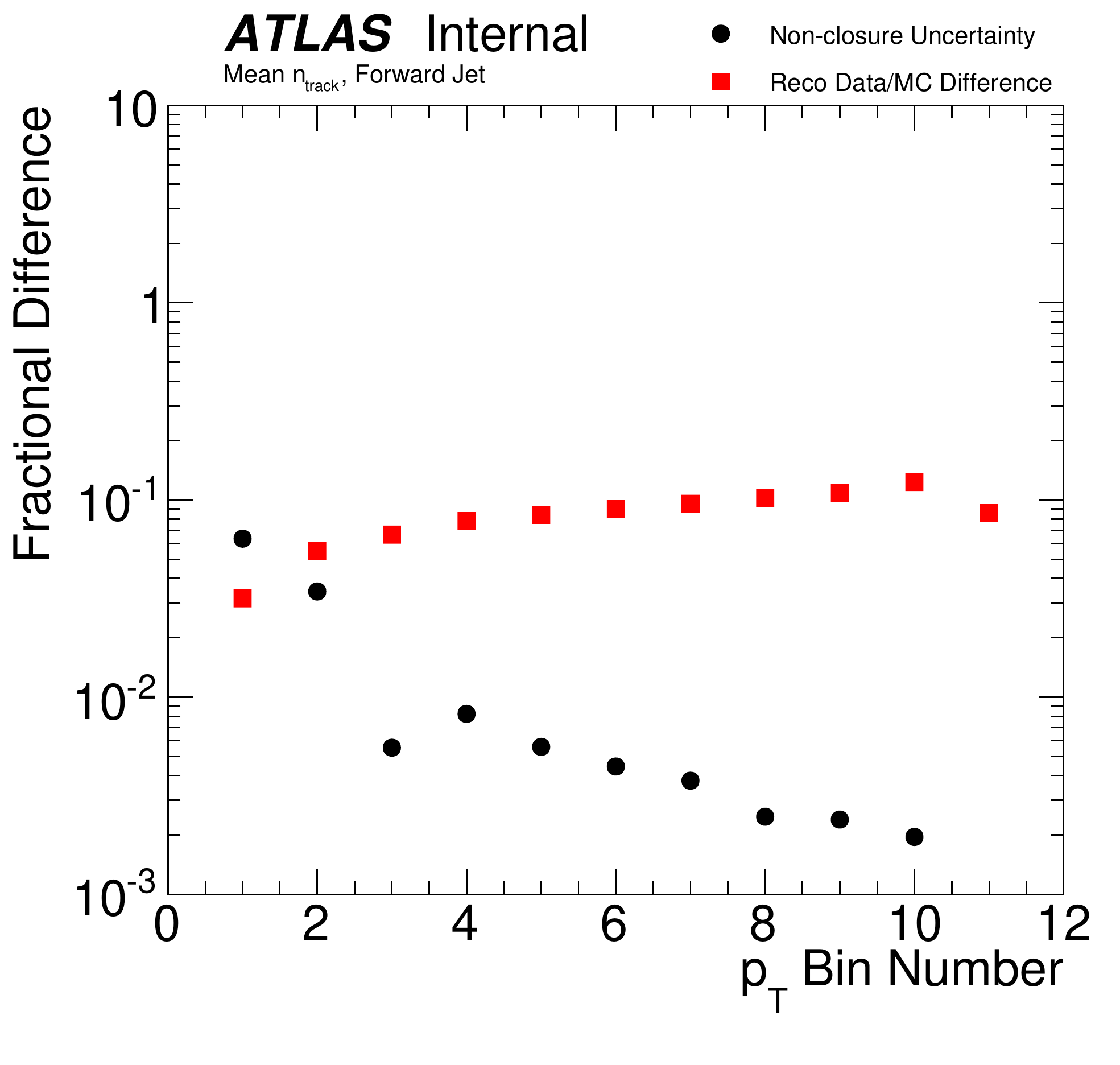}\includegraphics[width=0.5\textwidth]{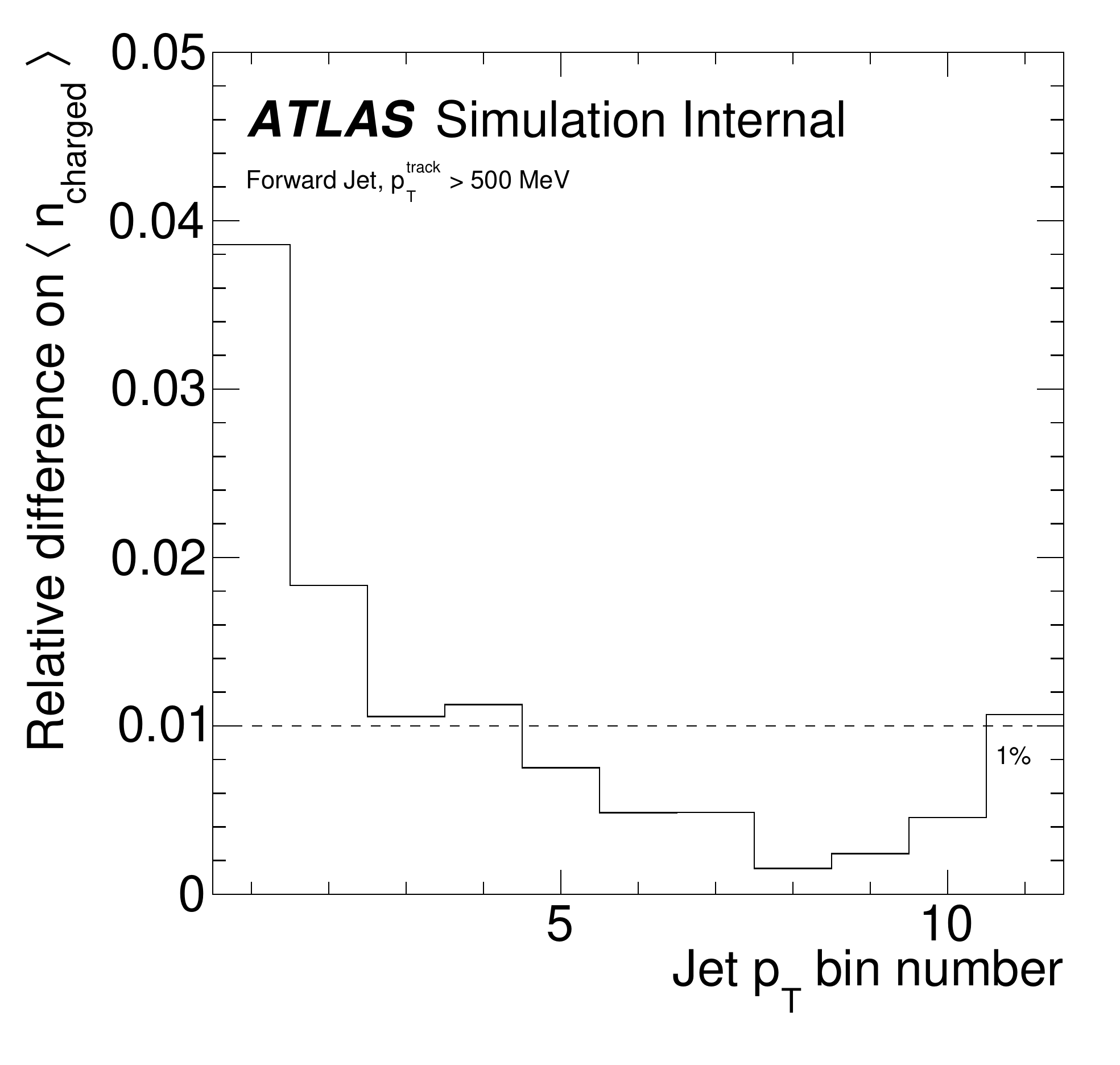}
\caption{The non-closure uncertainty for $\langle n_\text{charged}\rangle $ compared with the raw data/MC difference in the same quantity (left) and the relative difference in $\langle n_\text{charged}\rangle $ when unfolding {\sc Pythia} 8 with itself (= particle-level {\sc Pythia} by construction) and with a {\sc Herwig++} response matrix (right).}
\label{fig:NCharge:systs_nc_5}
\end{center}
\end{figure}

\clearpage

\subsection{Summary}

A summary of the systematic uncertainties can be found in Table~\ref{tab:NCharge:systs_all} and visualized in Fig.~\ref{fig:NCharge:systtotal}.  The relative size of the uncertainties are similar for the three charged particle $p_\text{T}$ thresholds.  Aside from the first jet $p_\text{T}$ bins, the dominant uncertainties are due to the isolated track reconstruction efficiency and the reconstruction efficiency of tracks inside jets.  The statistical and systematic uncertainties are comparable in size ($\sim 4\%$) in the highest jet $p_\text{T}$ bin.

\begin{figure}[h!]
\begin{center}
\includegraphics[width=0.8\textwidth]{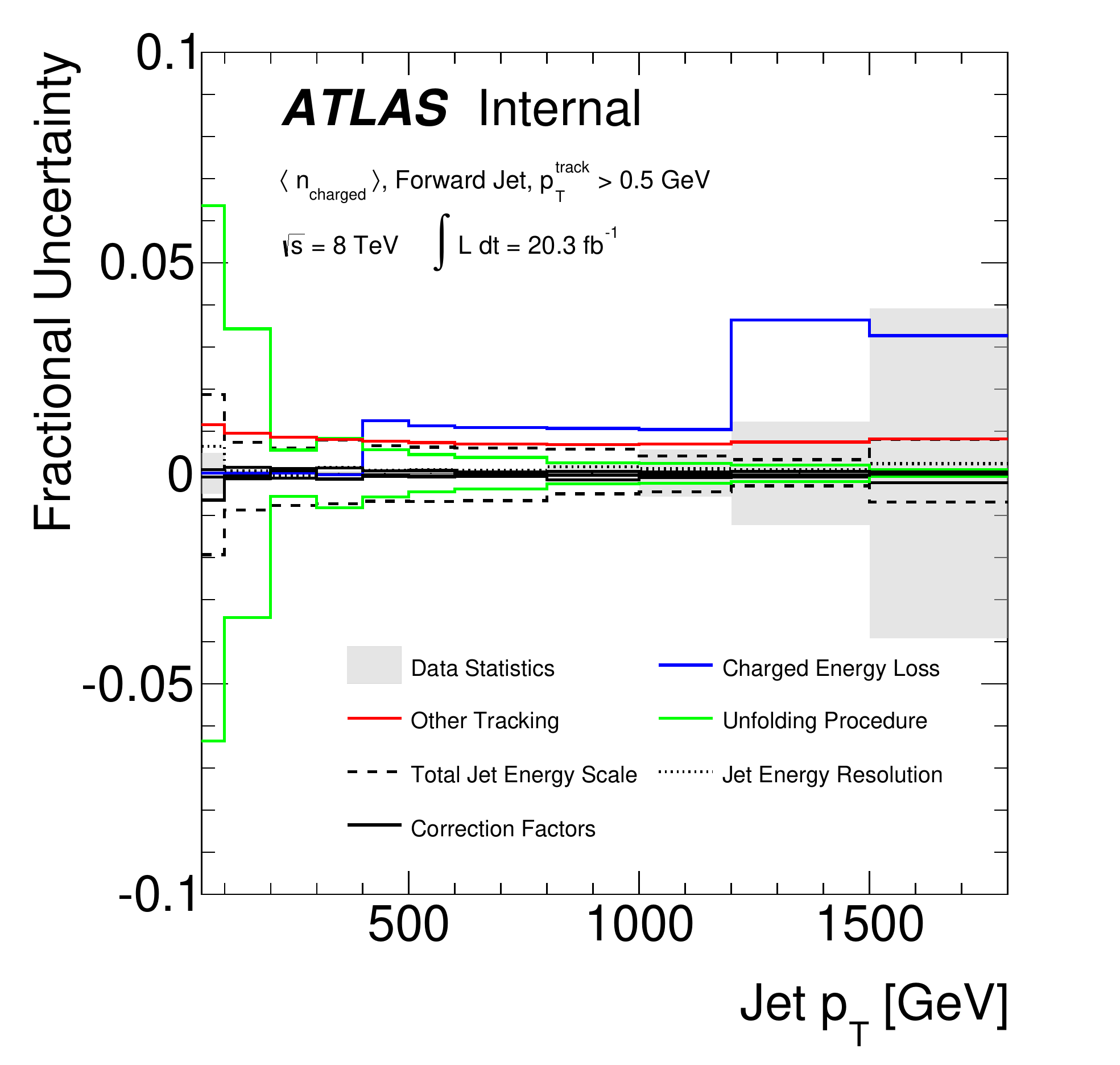}
\caption{A visualization of the systematic uncertainties for the $\langle n_\text{charge}\rangle$ distribution.}
\label{fig:NCharge:systtotal}
\end{center}
\end{figure}

\begin{landscape}
\setlength\tabcolsep{1.1mm}
\begin{table}[h]
\begin{tabular}{lclcllllllll}
 {\bf Average $n_\text{charged}$} & \multicolumn{10}{c}{Jet $p_\text{T}$ Range {[}100 GeV{]}} \\
\begin{tabular}[c]{@{}c@{}}Systematic\\[-3mm]Uncertainty [\%]\end{tabular} & {[}0.5,1{]}          & {[}1,2{]}             & {[}2,3{]}            & {[}3,4{]}             & {[}4,5{]}             & {[}5,6{]}             & {[}6,8{]}             & {[}8,10{]}            & {[}10,12{]}           & {[}12,15{]}       & {[}15,18{]}     \\ \hline
Response Matrix\\\hspace{5mm}Total Jet Energy Scale & \multicolumn{1}{c}{${}^{+  1.9}_{-  1.9}$}
& \multicolumn{1}{c}{${}^{+  0.7}_{-  0.9}$}
& \multicolumn{1}{c}{${}^{+  0.6}_{-  0.8}$}
& \multicolumn{1}{c}{${}^{+  0.8}_{-  0.7}$}
& \multicolumn{1}{c}{${}^{+  0.7}_{-  0.7}$}
& \multicolumn{1}{c}{${}^{+  0.6}_{-  0.7}$}
& \multicolumn{1}{c}{${}^{+  0.6}_{-  0.7}$}
& \multicolumn{1}{c}{${}^{+  0.6}_{-  0.5}$}
& \multicolumn{1}{c}{${}^{+  0.4}_{-  0.4}$}
& \multicolumn{1}{c}{${}^{+  0.3}_{-  0.3}$}
& \multicolumn{1}{c}{${}^{+  0.8}_{-  0.7}$}
\\
\hspace{5mm}Jet Energy Resolution & \multicolumn{1}{c}{${}^{+  0.6}_{-  0.6}$}
& \multicolumn{1}{c}{${}^{+  0.1}_{-  0.1}$}
& \multicolumn{1}{c}{${}^{+  0.1}_{-  0.1}$}
& \multicolumn{1}{c}{${}^{+  0.1}_{-  0.1}$}
& \multicolumn{1}{c}{${}^{+  0.1}_{-  0.1}$}
& \multicolumn{1}{c}{${}^{+  0.1}_{-  0.1}$}
& \multicolumn{1}{c}{${}^{+  0.1}_{-  0.1}$}
& \multicolumn{1}{c}{${}^{+  0.2}_{-  0.2}$}
& \multicolumn{1}{c}{${}^{+  0.1}_{-  0.1}$}
& \multicolumn{1}{c}{${}^{+  0.1}_{-  0.1}$}
& \multicolumn{1}{c}{${}^{+  0.2}_{-  0.2}$}
\\
\hspace{5mm}Charged Energy Loss & \multicolumn{1}{c}{${}^{+  0.0}_{-  0.0}$}
& \multicolumn{1}{c}{${}^{+  0.0}_{-  0.0}$}
& \multicolumn{1}{c}{${}^{+  0.0}_{-  0.0}$}
& \multicolumn{1}{c}{${}^{+  0.0}_{-  0.0}$}
& \multicolumn{1}{c}{${}^{+  1.2}_{-  0.0}$}
& \multicolumn{1}{c}{${}^{+  1.1}_{-  0.0}$}
& \multicolumn{1}{c}{${}^{+  1.1}_{-  0.0}$}
& \multicolumn{1}{c}{${}^{+  1.1}_{-  0.0}$}
& \multicolumn{1}{c}{${}^{+  1.0}_{-  0.0}$}
& \multicolumn{1}{c}{${}^{+  3.6}_{-  0.0}$}
& \multicolumn{1}{c}{${}^{+  3.3}_{-  0.0}$}
\\
\hspace{5mm}Other Tracking & \multicolumn{1}{c}{${}^{+  1.2}_{-  0.0}$}
& \multicolumn{1}{c}{${}^{+  1.0}_{-  0.0}$}
& \multicolumn{1}{c}{${}^{+  0.9}_{-  0.0}$}
& \multicolumn{1}{c}{${}^{+  0.8}_{-  0.0}$}
& \multicolumn{1}{c}{${}^{+  0.8}_{-  0.0}$}
& \multicolumn{1}{c}{${}^{+  0.7}_{-  0.0}$}
& \multicolumn{1}{c}{${}^{+  0.7}_{-  0.0}$}
& \multicolumn{1}{c}{${}^{+  0.7}_{-  0.0}$}
& \multicolumn{1}{c}{${}^{+  0.7}_{-  0.0}$}
& \multicolumn{1}{c}{${}^{+  0.7}_{-  0.0}$}
& \multicolumn{1}{c}{${}^{+  0.8}_{-  0.0}$}
\\
Correction Factors & \multicolumn{1}{c}{${}^{+  0.1}_{-  0.1}$}
& \multicolumn{1}{c}{${}^{+  0.1}_{-  0.1}$}
& \multicolumn{1}{c}{${}^{+  0.1}_{-  0.1}$}
& \multicolumn{1}{c}{${}^{+  0.1}_{-  0.1}$}
& \multicolumn{1}{c}{${}^{+  0.1}_{-  0.1}$}
& \multicolumn{1}{c}{${}^{+  0.1}_{-  0.1}$}
& \multicolumn{1}{c}{${}^{+  0.0}_{-  0.0}$}
& \multicolumn{1}{c}{${}^{+  0.0}_{-  0.0}$}
& \multicolumn{1}{c}{${}^{+  0.0}_{-  0.0}$}
& \multicolumn{1}{c}{${}^{+  0.0}_{-  0.0}$}
& \multicolumn{1}{c}{${}^{+  0.0}_{-  0.0}$}
\\
Unfolding Procedure& \multicolumn{1}{c}{${}^{+  6.4}_{-  6.4}$}
& \multicolumn{1}{c}{${}^{+  3.4}_{-  3.4}$}
& \multicolumn{1}{c}{${}^{+  0.6}_{-  0.6}$}
& \multicolumn{1}{c}{${}^{+  0.8}_{-  0.8}$}
& \multicolumn{1}{c}{${}^{+  0.6}_{-  0.6}$}
& \multicolumn{1}{c}{${}^{+  0.4}_{-  0.4}$}
& \multicolumn{1}{c}{${}^{+  0.4}_{-  0.4}$}
& \multicolumn{1}{c}{${}^{+  0.2}_{-  0.2}$}
& \multicolumn{1}{c}{${}^{+  0.2}_{-  0.2}$}
& \multicolumn{1}{c}{${}^{+  0.2}_{-  0.2}$}
& \multicolumn{1}{c}{${}^{+  0.1}_{-  0.1}$}
\\
\hline
Total Systematic & \multicolumn{1}{c}{${}^{+  6.8}_{-  6.7}$}
& \multicolumn{1}{c}{${}^{+  3.6}_{-  3.5}$}
& \multicolumn{1}{c}{${}^{+  1.2}_{-  1.0}$}
& \multicolumn{1}{c}{${}^{+  1.4}_{-  1.1}$}
& \multicolumn{1}{c}{${}^{+  1.7}_{-  0.9}$}
& \multicolumn{1}{c}{${}^{+  1.5}_{-  0.8}$}
& \multicolumn{1}{c}{${}^{+  1.5}_{-  0.8}$}
& \multicolumn{1}{c}{${}^{+  1.4}_{-  0.6}$}
& \multicolumn{1}{c}{${}^{+  1.3}_{-  0.5}$}
& \multicolumn{1}{c}{${}^{+  3.7}_{-  0.4}$}
& \multicolumn{1}{c}{${}^{+  3.5}_{-  0.7}$}
\\
Data Statistics & \multicolumn{1}{c}{  0.5}
& \multicolumn{1}{c}{  0.2}
& \multicolumn{1}{c}{  0.1}
& \multicolumn{1}{c}{  0.1}
& \multicolumn{1}{c}{  0.0}
& \multicolumn{1}{c}{  0.1}
& \multicolumn{1}{c}{  0.1}
& \multicolumn{1}{c}{  0.3}
& \multicolumn{1}{c}{  0.6}
& \multicolumn{1}{c}{  1.2}
& \multicolumn{1}{c}{  3.9}
\\
Total Uncertainty & \multicolumn{1}{c}{${}^{+  6.8}_{-  6.7}$}
& \multicolumn{1}{c}{${}^{+  3.6}_{-  3.6}$}
& \multicolumn{1}{c}{${}^{+  1.2}_{-  1.0}$}
& \multicolumn{1}{c}{${}^{+  1.4}_{-  1.1}$}
& \multicolumn{1}{c}{${}^{+  1.7}_{-  0.9}$}
& \multicolumn{1}{c}{${}^{+  1.5}_{-  0.8}$}
& \multicolumn{1}{c}{${}^{+  1.5}_{-  0.8}$}
& \multicolumn{1}{c}{${}^{+  1.4}_{-  0.6}$}
& \multicolumn{1}{c}{${}^{+  1.5}_{-  0.8}$}
& \multicolumn{1}{c}{${}^{+  3.9}_{-  1.3}$}
& \multicolumn{1}{c}{${}^{+  5.2}_{-  4.0}$}
\\
\hline\hline
Measured Value& \multicolumn{1}{c}{ 7.87}
& \multicolumn{1}{c}{ 9.87}
& \multicolumn{1}{c}{12.19}
& \multicolumn{1}{c}{13.54}
& \multicolumn{1}{c}{14.59}
& \multicolumn{1}{c}{15.41}
& \multicolumn{1}{c}{16.28}
& \multicolumn{1}{c}{17.41}
& \multicolumn{1}{c}{18.25}
& \multicolumn{1}{c}{18.71}
& \multicolumn{1}{c}{20.78}
\\
\end{tabular}
\caption{A summary of all the systematic uncertainties and their impact on the $n_\text{track}$ mean for $p_\text{T}^\text{track}>0.5$ GeV and the more forward jet.  Uncertainties are given in percent.  The last row is the measured average charged particle multiplicity.  A value of 0.0 is quoted if the uncertainty is below 0.05\%.}
\label{tab:NCharge:systs_all}
\end{table} 
 \end{landscape}

\clearpage
\newpage

\section{Results}
\label{sec:NCharge:Results}

The unfolded average charged-particle multiplicity combining both the more forward and the more central jets is shown in Fig.~\ref{fig:NCharge:result1} for $p_\text{T}^\text{track}>500$ MeV and Fig~\ref{fig:NCharge:result2} for $p_\text{T}^\text{track}>2$ GeV and 5 GeV, compared with various model predictions.  As was already observed for the reconstructed data in Fig.~\ref{fig:tracks}, the average charged-particle multiplicity in data falls between the predictions of {\sc Pythia 8} and {\sc Herwig++}, independently of the underlying-event tunes.  The {\sc Pythia 8} predictions are generally higher than the data and this is more pronounced at higher jet $p_\text{T}$.  The default ATLAS tune in Run 1 (AU2) performs similarly to the Monash tune, but the prediction with A14 (the ATLAS default for the analysis of Run 2 data) is significantly closer to the data.  A previous ATLAS measurement~\cite{Aad:2011gn} of charged-particle multiplicity inside jets was included in the tuning of A14, but the jets in that measurement have $p_\text{T}\lesssim 50$ GeV.  One important difference between A14 and Monash is that the value of $\alpha_\text{s}$ governing the amount of final-state radiation is about 10\% lower in A14 than in Monash.  This parameter has a large impact on the average charged-particle multiplicity, which is shown by the {\sc Pythia 6} lines in Fig.~\ref{fig:NCharge:result1} where the Perugia radHi and radLo tunes are significantly separated from the central P2012 tune.  The $\alpha_\text{s}$ value that regulates final-state radiation is changed by factors of one half and two for these tunes with respect to the nominal Perugia 2012 tune.  The recent (and Run 2 default) EE5 underlying-event tune for {\sc Herwig++} improves the modelling of the average charged-particle multiplicity with respect to the EE3 tune (Run 1 default).  The general differences between data and simulation are similar for the three track $p_\text{T}$ thresholds, but the level of agreement is slightly better for higher thresholds.

\begin{figure}[h!]
\begin{center}
\includegraphics[width=0.8\textwidth]{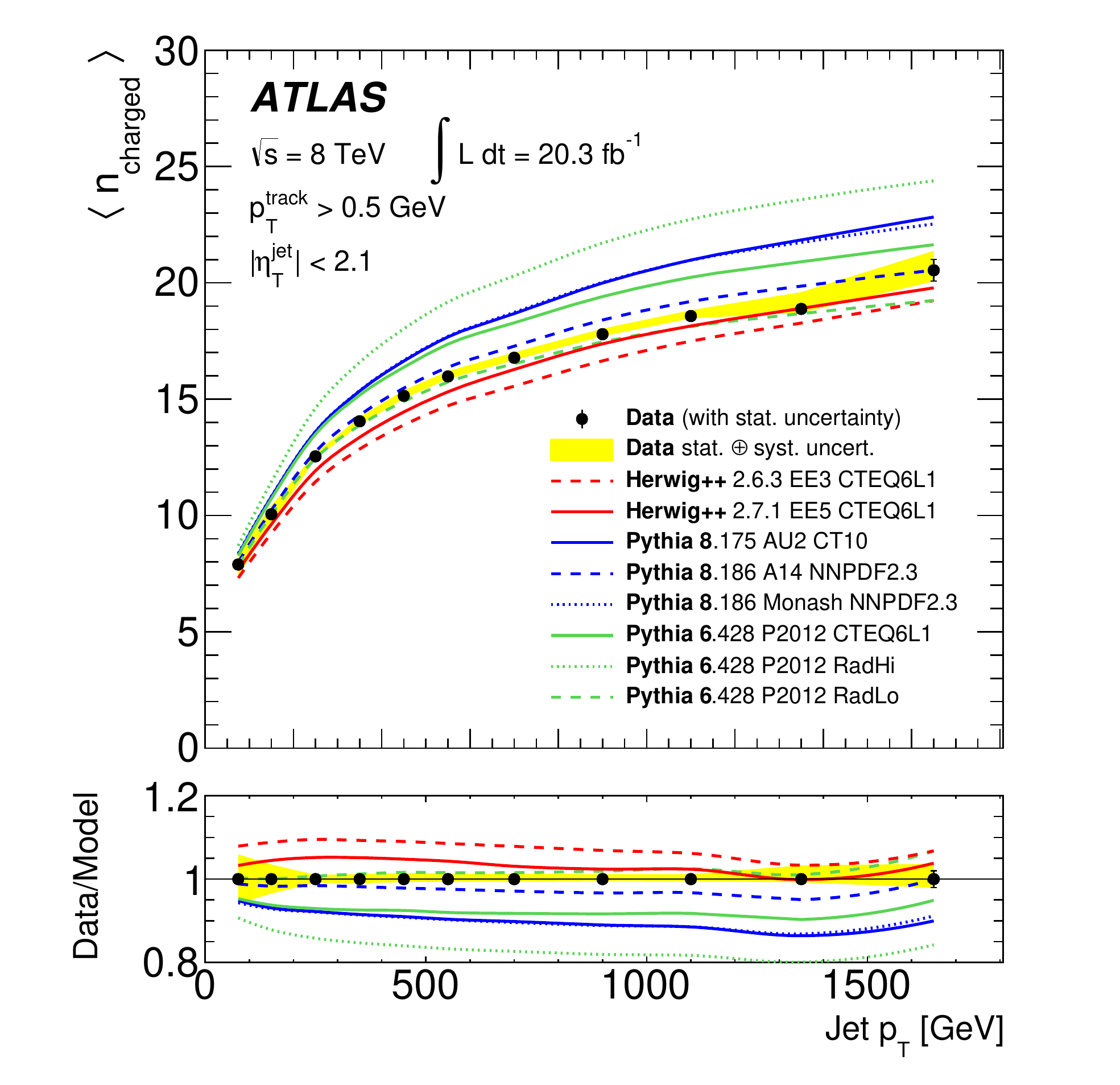}
\end{center}	
\caption{The measured average charged-particle multiplicity as a function of the jet $p_\text{T}$, combining the more forward and the more central jets for $p_\text{T}^\text{track}>0.5$ GeV.  The band around the data is the sum in quadrature of the statistical and systematic uncertainties.  Error bars on the data points represent the statistical uncertainty (which are smaller than the markers for most bins). }
\label{fig:NCharge:result1}
\end{figure}

\begin{figure}[h!]
\begin{center}
\includegraphics[width=0.45\textwidth]{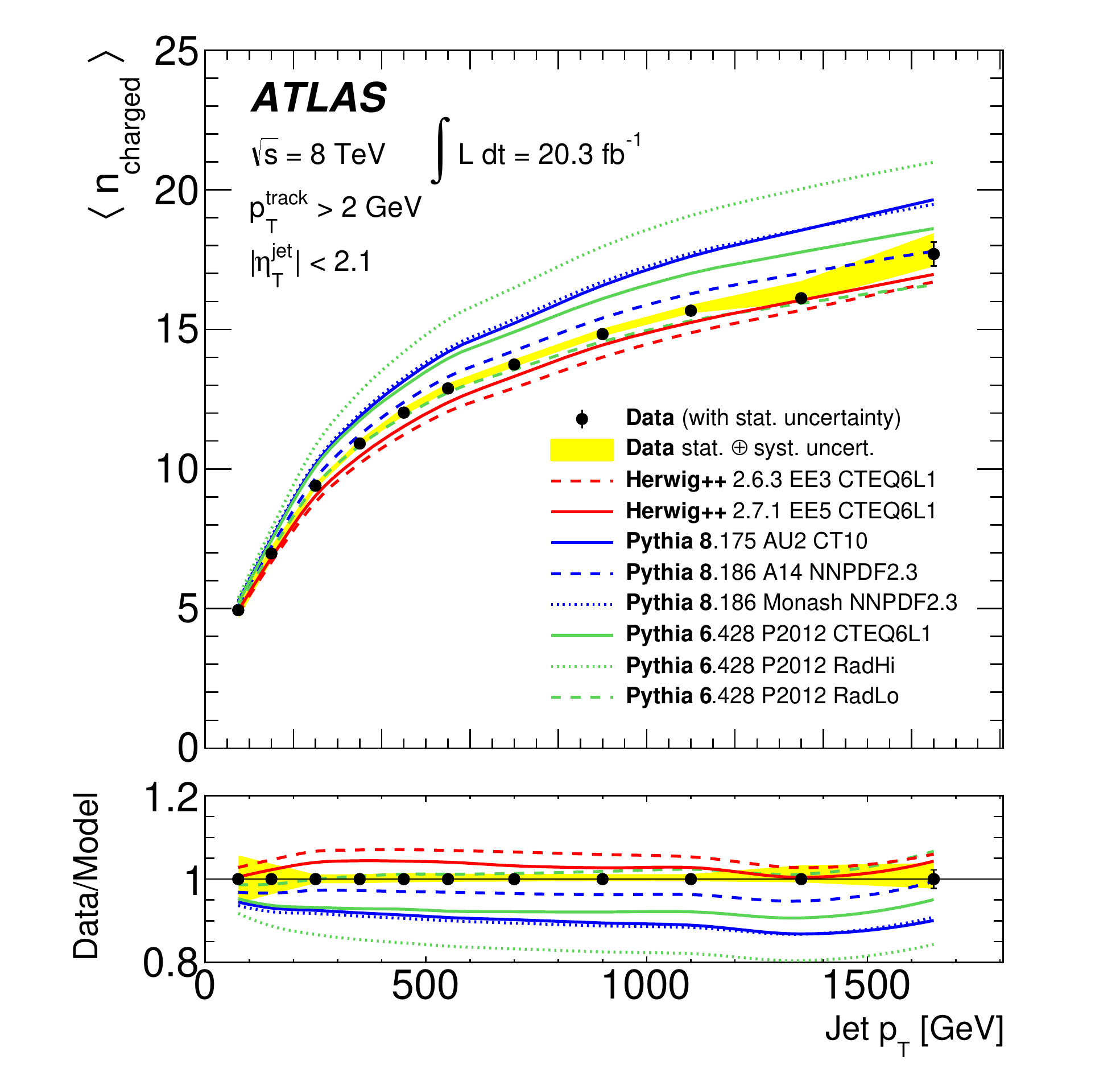}
\includegraphics[width=0.45\textwidth]{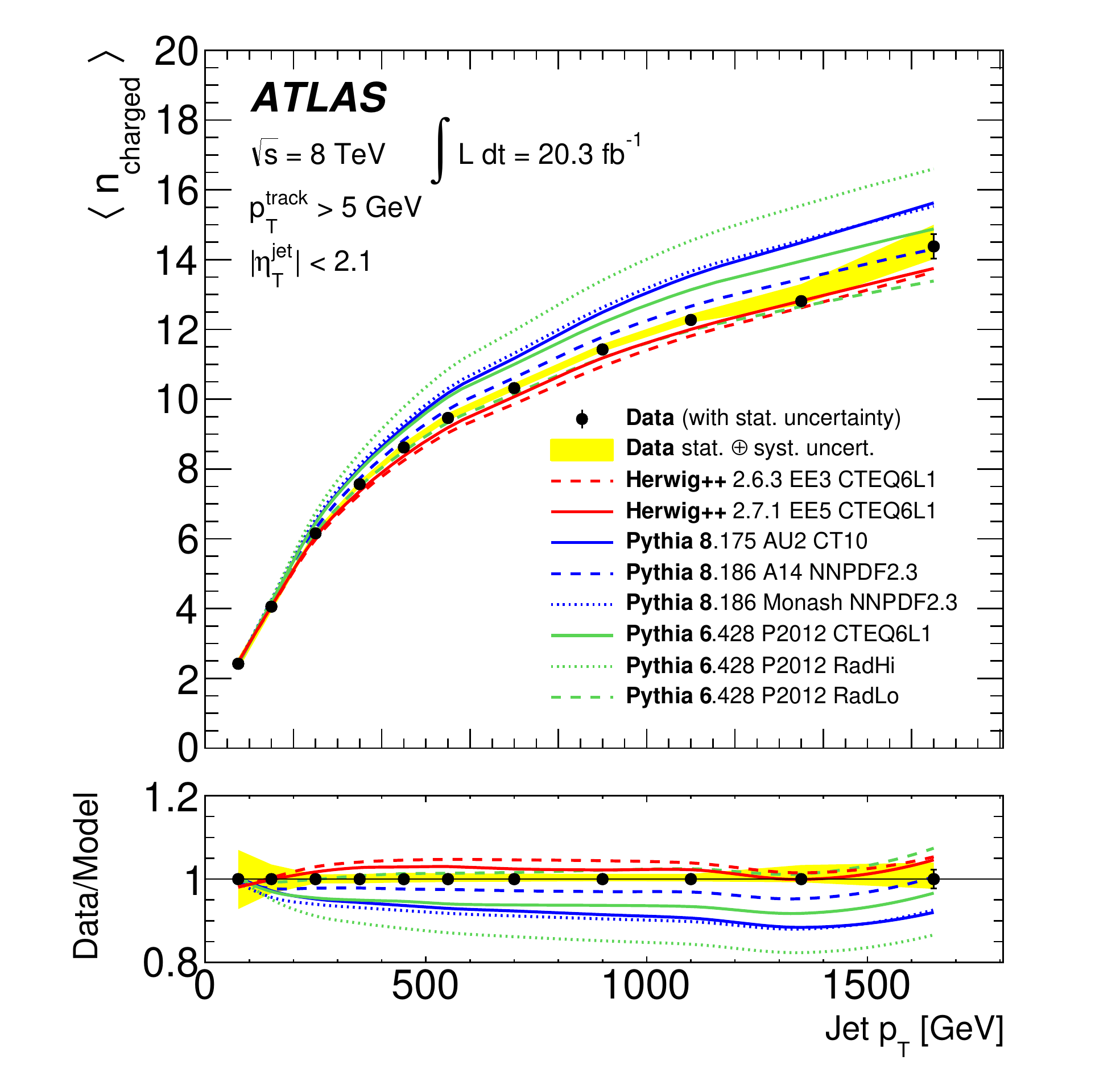}
\end{center}	
\caption{The measured average charged-particle multiplicity as a function of the jet $p_\text{T}$, combining the more forward and the more central jets for $p_\text{T}^\text{track}>2$ GeV (left) and $p_\text{T}^\text{track}>5$ GeV (right).  The band around the data is the sum in quadrature of the statistical and systematic uncertainties.  Error bars on the data points represent the statistical uncertainty (which are smaller than the markers for most bins). }
\label{fig:NCharge:result2}
\end{figure}

\clearpage

\subsection{Quark and Gluon Multiplicity}

As discussed in Sec.~\ref{sec:NCharge:Design}, the difference in the average charged-particle multiplicity between the more forward and the more central jet is sensitive to the difference between quark and gluon constituent multiplicities.   Figure~\ref{fig:Ncharge:diff} shows that the difference is significant for $p_\text{T}\lesssim 1.1$ TeV.  The shape is governed by the difference in the gluon fraction between the more forward and the more central jet\footnote{However, the peak is not in exactly the same location because the multiplicity for quarks and gluons is not the same and depends on $p_\text{T}$: $\langle n^c-n^f\rangle = \langle n^c\rangle - \langle n^f\rangle = (f_g^c n_g + f_q^c n_q) - (f_g^f n_g + f_q^f n_q)$, where $n$ is the charged particle multiplicity for quarks ($q$) or gluons ($g$) and for the more forward ($f$) or more central ($c$) jets.}, which was shown in Fig.~\ref{fig:NCharge:qgfrac} to peak around $p_\text{T}\sim 350$ GeV.   The systematic uncertainties are significantly smaller on the difference than on the pooled (more forward and more central combined) average $n_\text{track}$.  For example, at the peak around $\sim350$ GeV, the systematic uncertainty is about a factor of three smaller for the difference compared with the combination of the more forward and more central jets.

\begin{figure}[h!]
\begin{center}
\includegraphics[width=0.6\textwidth]{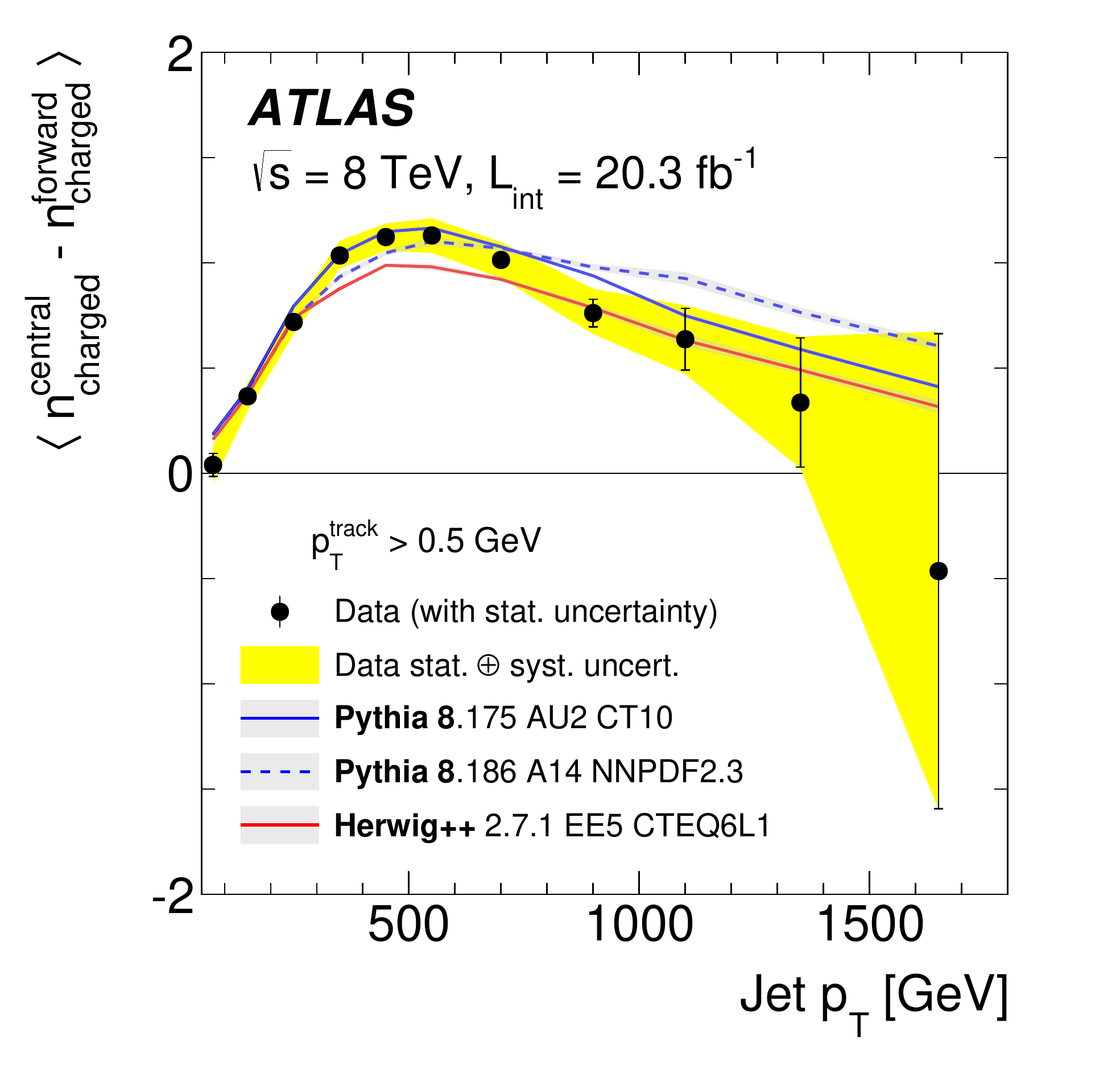}
\end{center}	
\caption{The jet $p_\text{T}$ dependence of the difference in the average charged-particle multiplicity ($p_\text{T}^\text{track}>0.5$ GeV) between the more forward and the more central jet.  The band for the data is the sum in quadrature of the systematic and statistical uncertainties and the error bars on the data points represent the statistical uncertainty.  Bands on the simulation include MC statistical uncertainty. }
\label{fig:Ncharge:diff}
\end{figure}

The average difference, combined with the gluon fraction, can be used to extract the average charged-particle multiplicity for quark- and gluon-initiated jets separately.   The extracted $p_\text{T}$ dependence of the average charged-particle multiplicities for quark- and gluon-initiated jets is shown in Fig.~\ref{fig:Chargeqg}.  {\sc Pythia 8} with the CT10 PDF set is used to determine the gluon fractions.  The experimental uncertainties are propagated through Eq.~\ref{eq:system} by recomputing the quark and gluon average charged-particle multiplicities for each variation accounting for a systematic uncertainty; the more forward and more central jet uncertainties are treated as being fully correlated.  In addition to the experimental uncertainties, the error bands in Fig.~\ref{fig:Chargeqg} include uncertainties in the gluon fractions from both the PDF and matrix element (ME) uncertainties.  The PDF uncertainty is determined using the CT10 eigenvector PDF sets and validated by comparing CT10 and NNPDF.  The ME uncertainty is estimated by comparing the fractions $f_{q,g}^{f,c}$ from {\sc Pythia 8} and {\sc Herwig++} after reweighting the {\sc Pythia 8} sample with CT10 to CTEQ6L1 to match the PDF used for {\sc Herwig++}.  All PDF re-weighting is performed using LHAPDF6~\cite{Buckley:2014ana}.  The PDF and ME uncertainties are comparable in size to the total experimental uncertainty.  As expected, the average multiplicity increases with jet $p_\text{T}$ for both the quark-initiated jets and gluon-initiated jets.  Furthermore, the multiplicity is significantly higher for gluon-initiated jets than for quark-initiated jets.  The average charged-particle multiplicity in {\sc Pythia 8} with the AU2 tune is higher than in the data for both the quark- and gluon-initiated jets.  

\begin{table}[h]
\begin{tabular}{cclclllllll}
 {\bf  $\langle n_\text{charged}\rangle $} & \multicolumn{10}{c}{Jet $p_\text{T}$ Range {[}100 GeV{]}} \\
\begin{tabular}[c]{@{}c@{}}Systematic\\ Uncertainty \end{tabular} & {[}0.5,1{]}          & {[}1,2{]}             & {[}2,3{]}            & {[}3,4{]}             & {[}4,5{]}             & {[}5,6{]}             & {[}6,8{]}             & {[}8,10{]}            & {[}10,12{]}           & {[}12,15{]}            \\ \hline\hline
Total exp. & \multicolumn{1}{c}{${}^{+ 0.44}_{- 0.34}$}
& \multicolumn{1}{c}{${}^{+ 0.29}_{- 0.24}$}
& \multicolumn{1}{c}{${}^{+ 0.15}_{- 0.24}$}
& \multicolumn{1}{c}{${}^{+ 0.24}_{- 0.17}$}
& \multicolumn{1}{c}{${}^{+ 0.21}_{- 0.33}$}
& \multicolumn{1}{c}{${}^{+ 0.37}_{- 0.43}$}
& \multicolumn{1}{c}{${}^{+ 0.48}_{- 0.58}$}
& \multicolumn{1}{c}{${}^{+ 1.01}_{- 1.03}$}
& \multicolumn{1}{c}{${}^{+ 2.20}_{- 2.39}$}
& \multicolumn{1}{c}{${}^{+ 6.09}_{- 6.16}$}
\\
ME & \multicolumn{1}{c}{ 0.04}
& \multicolumn{1}{c}{ 0.06}
& \multicolumn{1}{c}{ 0.05}
& \multicolumn{1}{c}{ 0.12}
& \multicolumn{1}{c}{ 0.14}
& \multicolumn{1}{c}{ 0.16}
& \multicolumn{1}{c}{ 0.06}
& \multicolumn{1}{c}{ 0.01}
& \multicolumn{1}{c}{ 0.05}
& \multicolumn{1}{c}{ 0.22}
\\
PDF & \multicolumn{1}{c}{${}^{+ 0.01}_{- 0.01}$}
& \multicolumn{1}{c}{${}^{+ 0.06}_{- 0.05}$}
& \multicolumn{1}{c}{${}^{+ 0.11}_{- 0.10}$}
& \multicolumn{1}{c}{${}^{+ 0.18}_{- 0.19}$}
& \multicolumn{1}{c}{${}^{+ 0.22}_{- 0.27}$}
& \multicolumn{1}{c}{${}^{+ 0.25}_{- 0.34}$}
& \multicolumn{1}{c}{${}^{+ 0.30}_{- 0.48}$}
& \multicolumn{1}{c}{${}^{+ 0.30}_{- 0.60}$}
& \multicolumn{1}{c}{${}^{+ 0.41}_{- 1.01}$}
& \multicolumn{1}{c}{${}^{+ 0.23}_{- 0.81}$}
\\
PDF II$^{*}$ & \multicolumn{1}{c}{ 0.03}
& \multicolumn{1}{c}{ 0.09}
& \multicolumn{1}{c}{ 0.00}
& \multicolumn{1}{c}{ 0.04}
& \multicolumn{1}{c}{ 0.01}
& \multicolumn{1}{c}{ 0.10}
& \multicolumn{1}{c}{ 0.33}
& \multicolumn{1}{c}{ 0.84}
& \multicolumn{1}{c}{ 1.76}
& \multicolumn{1}{c}{ 1.69}
\\
Half Cone$^{**}$ & \multicolumn{1}{c}{ 0.01}
& \multicolumn{1}{c}{ 0.03}
& \multicolumn{1}{c}{ 0.03}
& \multicolumn{1}{c}{ 0.04}
& \multicolumn{1}{c}{ 0.03}
& \multicolumn{1}{c}{ 0.03}
& \multicolumn{1}{c}{ 0.03}
& \multicolumn{1}{c}{ 0.02}
& \multicolumn{1}{c}{ 0.03}
& \multicolumn{1}{c}{ 0.01}
\\
ME ID$^{***}$ & \multicolumn{1}{c}{ 0.06}
& \multicolumn{1}{c}{ 0.03}
& \multicolumn{1}{c}{ 0.04}
& \multicolumn{1}{c}{ 0.05}
& \multicolumn{1}{c}{ 0.04}
& \multicolumn{1}{c}{ 0.03}
& \multicolumn{1}{c}{ 0.01}
& \multicolumn{1}{c}{ 0.01}
& \multicolumn{1}{c}{ 0.04}
& \multicolumn{1}{c}{ 0.05}
\\
\hline
\hline
\end{tabular}
\caption{A summary of the systematic uncertainties on the average charged multiplicity extraction for gluons.  (*) NNPDF versus CT10, used only as a cross-check. (**) Using a cone size of $\Delta R<0.2$ instead of the nominal 0.4 in the q/g identification.  Used only as a cross-check.  (***) Matching the jets with the outgoing partons in the ME to do the q/g ID.  Used only as a cross-check.  The uncertainties are in units of $n_\text{charged}$.}
\label{tab:guncert}
\end{table} 
\begin{table}[h]
\begin{tabular}{cclclllllll}
 {\bf  $\langle n_\text{charged}\rangle$} & \multicolumn{10}{c}{Jet $p_\text{T}$ Range {[}100 GeV{]}} \\
\begin{tabular}[c]{@{}c@{}}Systematic\\ Uncertainty\end{tabular} & {[}0.5,1{]}          & {[}1,2{]}             & {[}2,3{]}            & {[}3,4{]}             & {[}4,5{]}             & {[}5,6{]}             & {[}6,8{]}             & {[}8,10{]}            & {[}10,12{]}           & {[}12,15{]}            \\ \hline\hline
Total exp. & \multicolumn{1}{c}{${}^{+ 0.82}_{- 1.16}$}
& \multicolumn{1}{c}{${}^{+ 0.36}_{- 0.41}$}
& \multicolumn{1}{c}{${}^{+ 0.26}_{- 0.28}$}
& \multicolumn{1}{c}{${}^{+ 0.22}_{- 0.30}$}
& \multicolumn{1}{c}{${}^{+ 0.25}_{- 0.32}$}
& \multicolumn{1}{c}{${}^{+ 0.30}_{- 0.35}$}
& \multicolumn{1}{c}{${}^{+ 0.32}_{- 0.36}$}
& \multicolumn{1}{c}{${}^{+ 0.41}_{- 0.47}$}
& \multicolumn{1}{c}{${}^{+ 0.69}_{- 0.67}$}
& \multicolumn{1}{c}{${}^{+ 1.42}_{- 1.70}$}
\\
ME & \multicolumn{1}{c}{ 0.06}
& \multicolumn{1}{c}{ 0.23}
& \multicolumn{1}{c}{ 0.19}
& \multicolumn{1}{c}{ 0.23}
& \multicolumn{1}{c}{ 0.22}
& \multicolumn{1}{c}{ 0.25}
& \multicolumn{1}{c}{ 0.26}
& \multicolumn{1}{c}{ 0.22}
& \multicolumn{1}{c}{ 0.23}
& \multicolumn{1}{c}{ 0.16}
\\
PDF & \multicolumn{1}{c}{${}^{+ 0.02}_{- 0.02}$}
& \multicolumn{1}{c}{${}^{+ 0.11}_{- 0.10}$}
& \multicolumn{1}{c}{${}^{+ 0.17}_{- 0.16}$}
& \multicolumn{1}{c}{${}^{+ 0.27}_{- 0.24}$}
& \multicolumn{1}{c}{${}^{+ 0.33}_{- 0.27}$}
& \multicolumn{1}{c}{${}^{+ 0.38}_{- 0.28}$}
& \multicolumn{1}{c}{${}^{+ 0.44}_{- 0.30}$}
& \multicolumn{1}{c}{${}^{+ 0.47}_{- 0.28}$}
& \multicolumn{1}{c}{${}^{+ 0.62}_{- 0.33}$}
& \multicolumn{1}{c}{${}^{+ 0.45}_{- 0.21}$}
\\
PDF II$^{*}$ & \multicolumn{1}{c}{ 0.04}
& \multicolumn{1}{c}{ 0.01}
& \multicolumn{1}{c}{ 0.17}
& \multicolumn{1}{c}{ 0.23}
& \multicolumn{1}{c}{ 0.17}
& \multicolumn{1}{c}{ 0.10}
& \multicolumn{1}{c}{ 0.01}
& \multicolumn{1}{c}{ 0.21}
& \multicolumn{1}{c}{ 0.44}
& \multicolumn{1}{c}{ 0.39}
\\
Half Cone$^{**}$ & \multicolumn{1}{c}{ 0.01}
& \multicolumn{1}{c}{ 0.02}
& \multicolumn{1}{c}{ 0.02}
& \multicolumn{1}{c}{ 0.02}
& \multicolumn{1}{c}{ 0.02}
& \multicolumn{1}{c}{ 0.01}
& \multicolumn{1}{c}{ 0.01}
& \multicolumn{1}{c}{ 0.01}
& \multicolumn{1}{c}{ 0.01}
& \multicolumn{1}{c}{ 0.00}
\\
ME ID$^{***}$ & \multicolumn{1}{c}{ 0.07}
& \multicolumn{1}{c}{ 0.03}
& \multicolumn{1}{c}{ 0.01}
& \multicolumn{1}{c}{ 0.01}
& \multicolumn{1}{c}{ 0.01}
& \multicolumn{1}{c}{ 0.01}
& \multicolumn{1}{c}{ 0.02}
& \multicolumn{1}{c}{ 0.02}
& \multicolumn{1}{c}{ 0.02}
& \multicolumn{1}{c}{ 0.02}
\\
\hline
\hline
\end{tabular}
\caption{A summary of the systematic uncertainties on the average charged multiplicity extraction for quarks.  (*) NNPDF versus CT10, used only as a cross-check.  (**) Using a cone size of $\Delta R<0.2$ instead of the nominal 0.4 in the q/g identification.  Used only as a cross-check.  (***) Matching the jets with the outgoing partons in the ME to do the q/g ID.  Used only as a cross-check. The uncertainties are in units of $n_\text{charged}$.}
\label{tab:quncert}
\end{table}

In addition to predictions from leading-logarithm parton shower simulations, calculations of the scale dependence for the parton multiplicity inside jets have been performed in perturbative quantum chromodynamics (pQCD).  Up to a non-perturbative factor that is constant for the jet $p_\text{T}$ range considered in this analysis\footnote{This factor is found to be about 0.19 for gluon jets and 0.25 for quark-initiated jets.}, these calculations can be interpreted as a prediction for the scale dependence of $\langle n_\text{charged}\rangle$ for quark- and gluon-initiated jets.  There are further caveats to the predictability of such a calculation since $n_\text{charged}$ is not infrared safe or even Sudakov safe~\cite{Larkoski:2015lea}.  Therefore, the formal accuracy of the series expansion in $\sqrt{\alpha_\text{s}}$ is unknown.  Given these caveats, the next-to-next-to-next-to-leading-order (N${}^3$LO) pQCD calculation~\cite{Capella:1999ms,Dremin:1999ji} is overlaid in Fig.~\ref{fig:Chargeqg} with renormalization scale $\mu=Rp_\text{T}$ in the five-flavour scheme and $R=0.4$.  The theoretical error band is calculated by varying $\mu$ by a factor of two.  The prediction cannot give the absolute scale, and therefore the curve is normalized to the data in the second $p_\text{T}$ bin ($100$ GeV $<p_\text{T}<200$ GeV) where the statistical uncertainty is small.  The predicted scale dependence for gluon-initiated jets is consistent with the data within the uncertainty bands while the curve for quark-initiated jets is higher than the data by about one standard deviation.

\begin{figure}[h!]
\begin{center}
\includegraphics[width=0.65\textwidth]{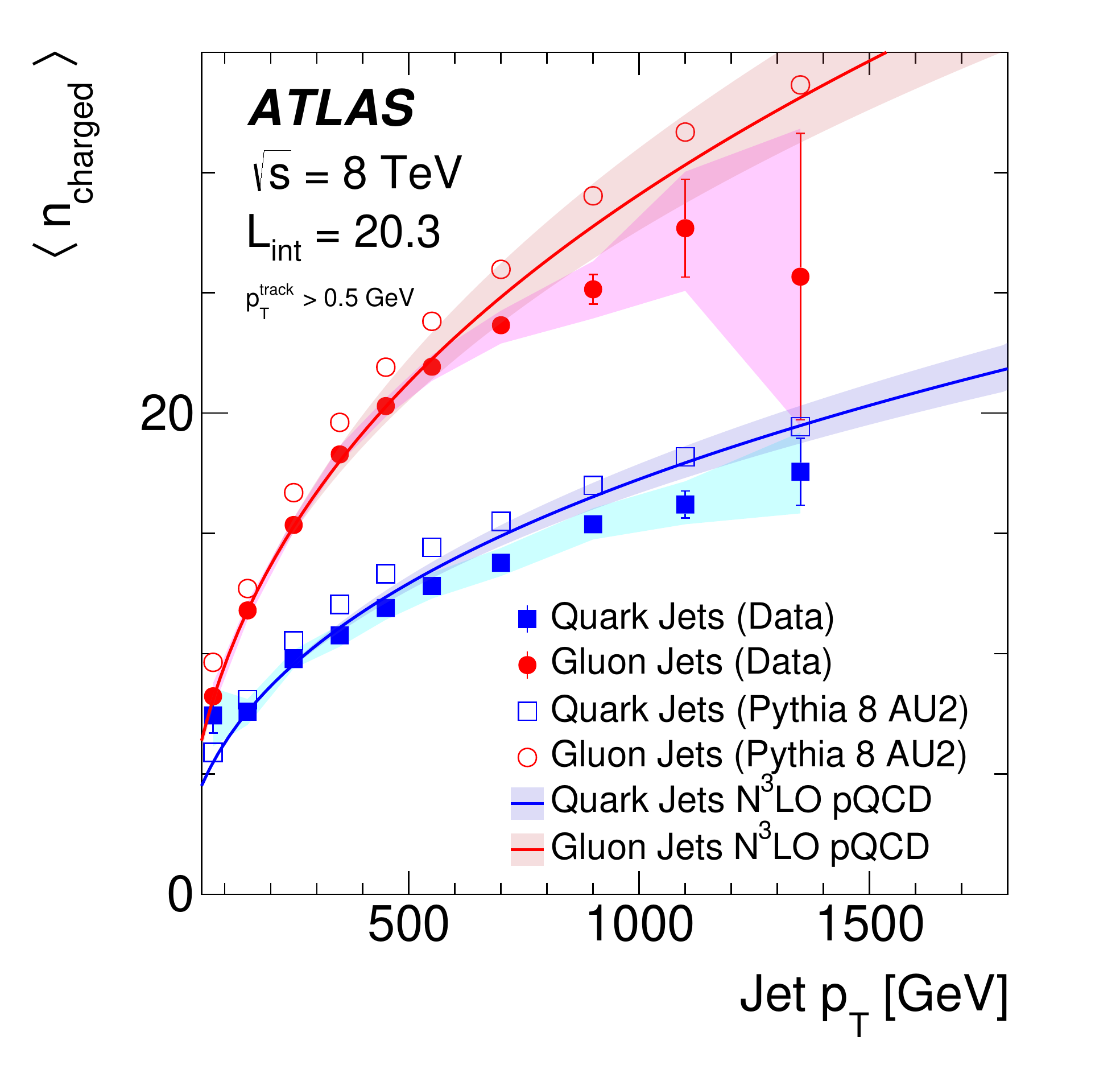}
\end{center}	
\caption{The jet $p_\text{T}$ dependence of the average charged-particle multiplicity ($p_\text{T}^\text{track}>0.5$ GeV) for quark- and gluon-initiated jets, extracted with the gluon fractions from {\sc Pythia} 8.175 with the CT10 PDF.  In addition to the experimental uncertainties, the error bands include uncertainties in the gluon fractions from both the PDF and ME uncertainties.  The MC statistical uncertainties on the open markers are smaller than the markers.  The uncertainty band for the N${}^3$LO pQCD prediction is determined by varying the scale $\mu$ by a factor of two up and down.  The markers are truncated at the penultimate $p_\text{T}$ bin in the right because within statistical uncertainty, the more forward and more central jet constituent charged-particle multiplicities are consistent with each other in the last bin.}
\label{fig:Chargeqg}
\end{figure}

 \clearpage
 
\section{Summary}
\label{sec:NCharge:Summary}

This chapter presents a measurement of the $p_\text{T}$ dependence of the average jet charged-particle multiplicity in dijet events from 20.3 fb${}^{-1}$ of $\sqrt{s}=8$ TeV~$pp$ collision data recorded by the ATLAS detector at the LHC.  The measured charged-particle multiplicity distribution is unfolded to correct for the detector acceptance and resolution to facilitate direct comparison to particle-level models.  Comparisons are made at particle level between the measured average charged-particle multiplicity and various models of jet formation.  Significant differences are observed between the simulations using Run 1 tunes and the data, but the Run 2 tunes for both {\sc Pythia 8} and {\sc Herwig++} significantly improve the modelling of the average $n_\text{charge}$.  Furthermore, quark- and gluon-initiated jet constituent charged-particle multiplicities are extracted and compared with simulations and calculations.  As expected, the extracted gluon-initiated jet constituent charged-particle multiplicity is higher than the corresponding quantity for quark-initiated jets and a calculation of the $p_\text{T}$-dependence accurately models the trend observed in the data.  The particle-level spectra are available~\cite{hepdata} for further interpretation and can serve as a benchmark for future measurements of the evolution of non-perturbative jet observables to validate MC predictions and tune their model parameters.
 \chapter{Boson and Top Quark Jets}
\label{cha:bosonjets}

Processes involving the production and decay of $W$, $Z$, and $H$ bosons as well as top quarks provide benchmarks for testing the Standard Model (SM), as well as probes of physics beyond the SM (BSM).  Since the cross section for the direct strong production of events with multiple jets (QCD multijets) at the LHC is many orders of magnitude larger than for the production of electroweak bosons or top quarks, it is usually the case that leptonic decays must be used to reduce the overwhelming background.   This is an unfortunate limitation because the hadronic branching ratios are larger than the leptonic ones\footnote{There are more {\it active} lepton than active quark types (five quarks since $m_\text{top}>m_{W/Z/H}$ and six total leptons) but since the electroweak bosons are blind to color, there are many more quarks.} and in some BSM theories, new particles similar to the SM electroweak bosons or top quarks do not couple directly to leptons.   However, when the momentum of a boson or top quark is comparable with its mass,  the spatial proximity of the decay products allows for a new set of tools that can be used to distinguish between single jets from hadronic boson decays and jets originating from QCD multijet backgrounds.   Some of these {\it jet substructure} tools have already been introduced in earlier sections including the jet charge in Sec.~\ref{cha:jetcharge}, jet pull in Sec.~\ref{cha:colorflow}, and multiplicity in Sec.~\ref{cha:multiplicity}.  The most powerful tool is related to another quantum property of jets - the {\it jet mass}.  Before discussing this jet observable in detail, it is important to quantify the size of a jet needed to capture most of the decay products of a boosted boson or top quark.  

\begin{figure}[h!]
 \centering
\includegraphics[width=0.6\textwidth]{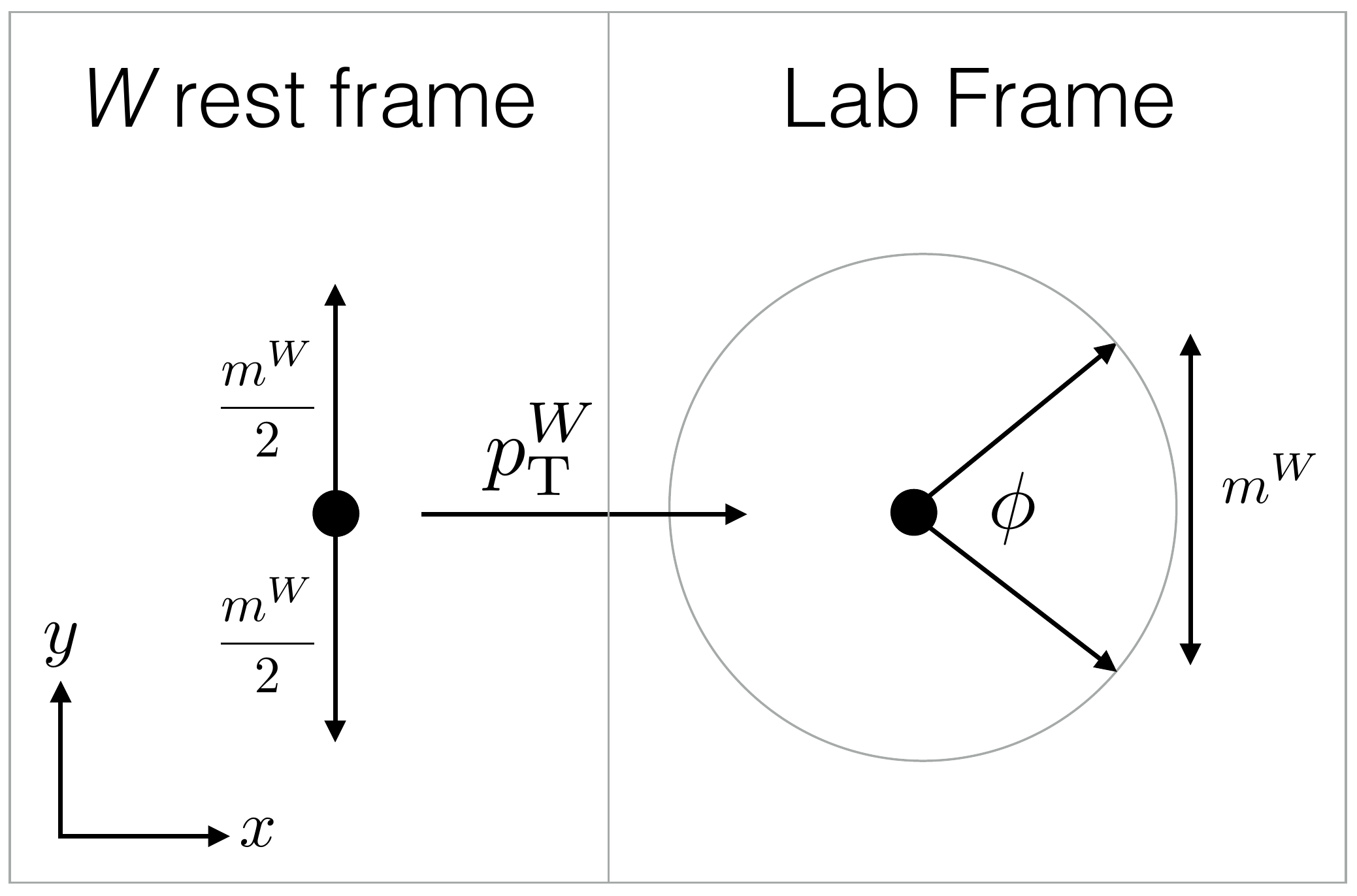}
\caption{A schematic illustration of the setup described in the text to compute the $p_\text{T}$ dependence of $\Delta R$.  The black dot represents the $W$ boson and the arrows from the dot represent the momentum of the quark decay products.}
\label{fig:JMR:bosonintro1}
\end{figure}

To illustrate the scaling of the angular distance $\Delta R$ between decay products, consider a $W$ boson with momentum directed along the $x$-axis in the lab frame with magnitude $p_\text{T}^W$ and assume $W\rightarrow qq'$.   In the $W$ boson rest frame, the two quarks each have energy $m^W/2$ and are back-to-back due to the conservation of energy and momentum.  The angular distance in the lab frame will be maximized when the quark momenta are along the $y$ direction (the $y$ direction is the same in both the lab and $W$ boson frames).  Figure~\ref{fig:JMR:bosonintro1} illustrates this setup.  Ignoring the quark masses, the momentum in the $y$ direction before and after the boost is $m^W/2$ (momenta orthogonal to the boost direction are unchanged) and the $x$ momentum goes from $0$ in the $W$ boson rest frame to $\gamma\beta m^W/2$ in the lab frame.  Therefore, 

\begin{align}
\Delta R=\phi \sim \frac{m^W}{\gamma\beta m^W/2} = \frac{2}{\gamma\beta}=\frac{2m^W}{p_\text{T}^W},
\end{align}

\noindent where $\phi$ is the opening angle between the quarks in the lab frame, $\beta$ is the speed of the $W$ boson in the lab frame $(\beta=p/E)$, $\gamma=\frac{1}{\sqrt{1-\beta^2}}=E/m$ is the usual relativistic enhancement factor and $\gamma\beta=\frac{E}{m}\frac{p}{E}=\frac{p}{m}$.  The $\sim$ represents the small angle approximation.  The full form is given by

\begin{align}
\label{eq:2mptfull}
\Delta R=2\text{arctan}\left(\frac{1}{\gamma\beta}\right) = \frac{2m}{p_\text{T}^W}-\frac{2}{3}\left(\frac{m}{p_\text{T}^W}\right)^3+\mathcal{O}\left(\frac{m^5}{p_\text{T}^5}\right).
\end{align}

\noindent Since the sub-leading term in Eq.~\ref{eq:2mptfull} is negative and the original setup was chosen to maximize $\Delta R$, in general $\Delta R\geq \frac{2m}{p_\text{T}}$ for a particle of mass $m$ and transverse momentum $p_\text{T}$ decaying into two massless particles.  The full joint distribution of $\Delta R$ and $p_\text{T}$ is shown in the right plot of Fig.~\ref{fig:JMR:bosonintro2}, ignoring effects of particle spin.  The $2m/p_\text{T}$ scaling is an excellent approximation for the $W$ decay where the decay products are nearly massless, but there are clear deviations in the case of the top quark where $m_W/m_t\sim 1/2$ is not negligible.  The right plot of Fig.~\ref{fig:JMR:bosonintro2} shows the fraction of events in which the decay products are within $\Delta R<1$ of the parent boosted $W$ boson or top quark in the production of $t\bar{t}$.  At low top quark $p_\text{T}$, the decay products are isotropically distributed.   As the top quark boost increases, the $W$ decay products are close enough to be contained within a cone of size $\Delta R=1$ and then for very large top quark $p_\text{T}$, the $b$-quark is also contained within the cone.  

\begin{figure}[h!]
 \centering
\includegraphics[width=0.45\textwidth]{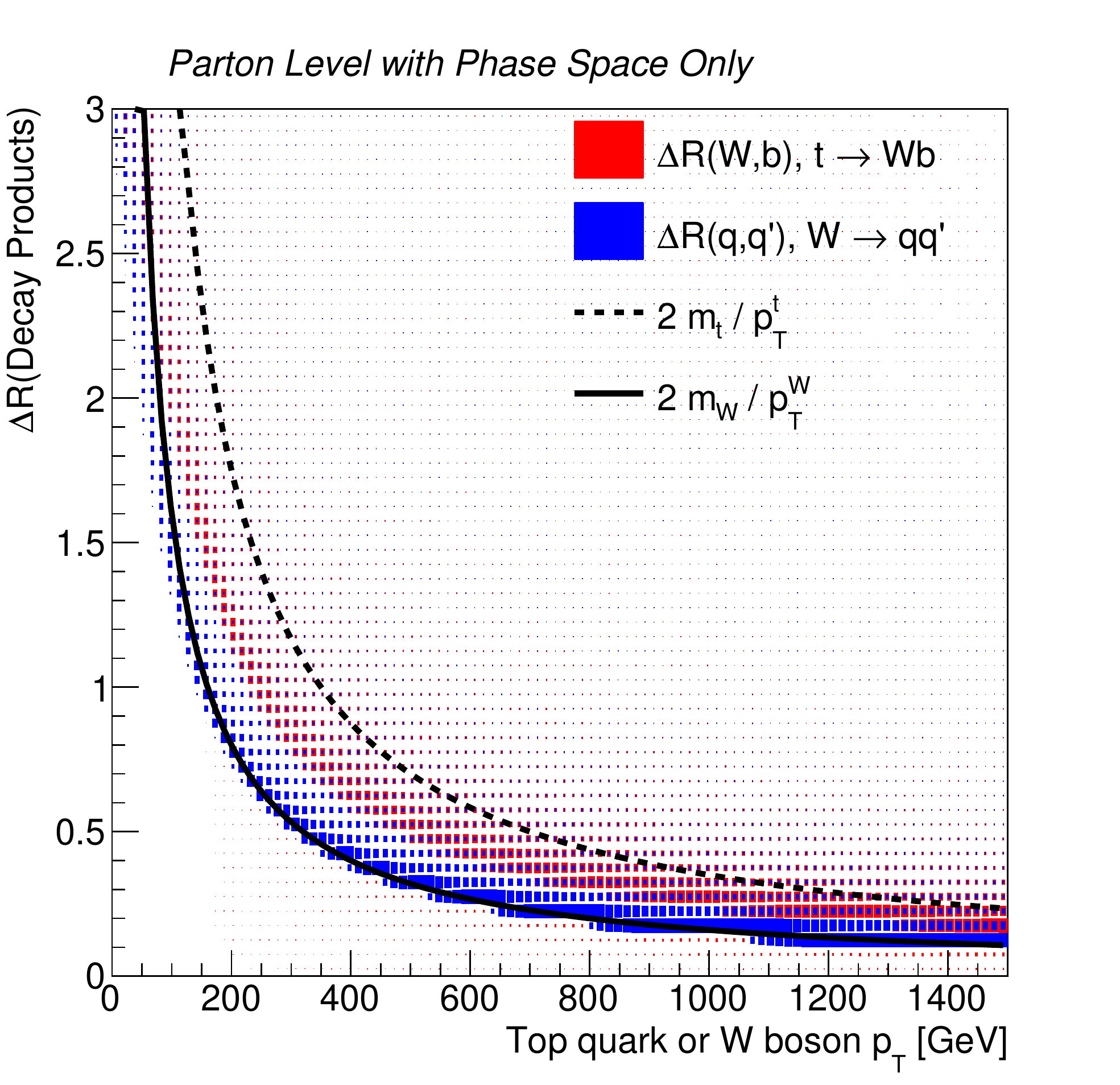}\includegraphics[width=0.45\textwidth]{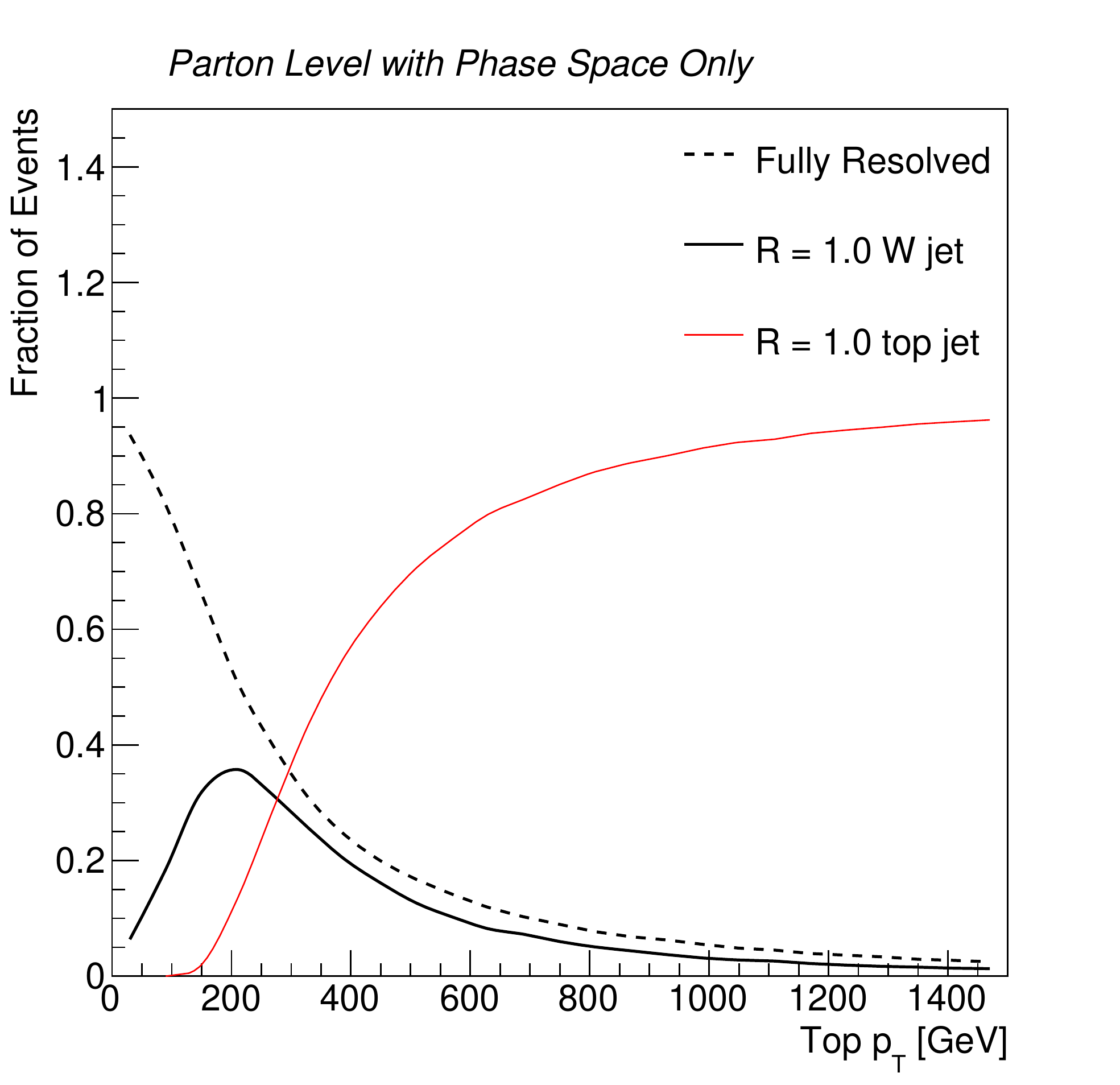}
\caption{Left: the joint distribution of $\Delta R$ and $p_\text{T}$ and Right: the fraction of events in which the decay products ($W\rightarrow qq'$ or $t\rightarrow bqq'$) are within $\Delta R<1$ of the parent particle.}
\label{fig:JMR:bosonintro2}
\end{figure}

Individual anti-$k_t$ R=0.4 jets are an appropriate description of the fragmentation from the well separated daughter quark decay products of low $p_\text{T}$ boson and top quarks.  However, isolating these events from the multijet background is an insurmountable challenge.  Consider the case of $W\rightarrow qq'$, identified from two jets.  If an event has $N$ jets, then there are $N\choose 2$ permutations which could give the $W$ boson daughter jets.  All hadronic top quark pair events produce six jets at leading order and there are often several additional jets from initial and final state radiation.   A powerful discriminant is the invariant mass of the two jets, $m_{jj}$.  One could require $m_{jj}\sim m_W$ in order to pick the two jets, but then $m_{jj}$ is a less useful discriminant because the background will be sculpted.  This is possible because while $m_{jj}\sim m_W$ for the signal, $m_{jj}$ is set by $\sqrt{\hat{s}}$ for the background, which is often near $m_W$ by coincidence.   These challenges are resolved at high boson or top quark $p_\text{T}$.  In that case, the decay products are geometrically close together, so a large radius jet $R\sim 1$ is likely to capture all of the decay products.  Of course, one could always pick $R$ large enough to capture all the decay products of the boson or top quark, but $R\sim 1$ is sufficiently small that jets of this size do not contain significant radiation from other sources.  The dijet invariant mass now becomes the large radius jet mass $m_j$, whose square is defined as the square of the sum of the jet constituent's four-vectors.  Just as in the resolved case, $m_j\sim m_W$ for the signal.  The power of jet mass is that for the multijet background, $m_j\sim \alpha_s Rp_\text{T,J}\ll \sqrt{\hat{s}}$ (see Sec.~\ref{sec:mass:theory}).  This chapter describes the properties of boson and top quark jets in detail.  Section~\ref{sec:jetmass} focuses on the jet mass, including measurements of the calorimeter jet mass resolution and new alternative jet mass definitions.   The jet mass is combined with other jet substructure variables in Sec.~\ref{sec:bosontypetagger} to distinguish boosted hadronically decaying bosons of different types, a natural extension of isolating these boson jets from multijet backgrounds.  The chapter ends in Sec.~\ref{sec:HEPML} with a new paradigm for studying the rich structure of boson and top quark jets in the context of {\it machine learning}.   State-of-the-art classification techniques are adapted to high energy physics for reconstructing and classifying boosted boson and top quark jets.

\clearpage

\section{Jet Mass}
\label{sec:jetmass}

	When a jet is sufficiently large to contain most of the energy from a hadronically decaying boosted boson or top quark, the mass of a jet is approximately the boson or top quark mass.  However, the particle-level and detector-level mass resolutions are both significant.  At particle-level, the mass is obscured due to finite radius effects and sources of diffuse uncorrelated radiation.  The detector-level radiation is affected by both the calorimeter-cell energy and angular resolution.  Various techniques for improving the jet mass resolution and measuring its reconstruction properties using data-driven techniques will be discussed in this section.  First, section~\ref{sec:mass:theory} describes the mechanism by which generic quark and gluon jets acquire mass.  Experimental techniques for calibrating the jet mass are discussed in Sec.~\ref{sec:JMR}.  Alternative jet mass definitions are investigated in Sec.~\ref{sec:ReclusteredJetMass} and~\ref{sec:TAMass}.  Concluding remarks and future outlook are provided in Sec.~\ref{sec:mass:conclusions}.

\subsection{The Mass of Quark and Gluon Jets}
\label{sec:mass:theory}

	While the mass of on-shell quarks and gluons is negligibly small compared with the mass of electroweak boson and top quark jets, this is not always true for the mass of a generic QCD jet.  Quark and gluon jets acquire significant mass as a result of (relatively) hard or wide angle gluon radiation.  Many properties of the jet mass distribution can be well-described within the context of perturbation theory.  The jet mass distribution at a hadron collider is known to approximate next-to-next-to-leading logarithmic order (NNLL${}_p$) in the absence of non-global logarithms (extra-jet radiation that re-emits back into the jet)~\cite{Liu:2014oog} and to even higher order (N${}^3$LL+NLO) for hemisphere mass at lepton colliders~\cite{Chien:2010kc}.  This section uses the lowest order results for illustration.  
	
\begin{figure}[h!]
\centering
\includegraphics[width=0.18\textwidth]{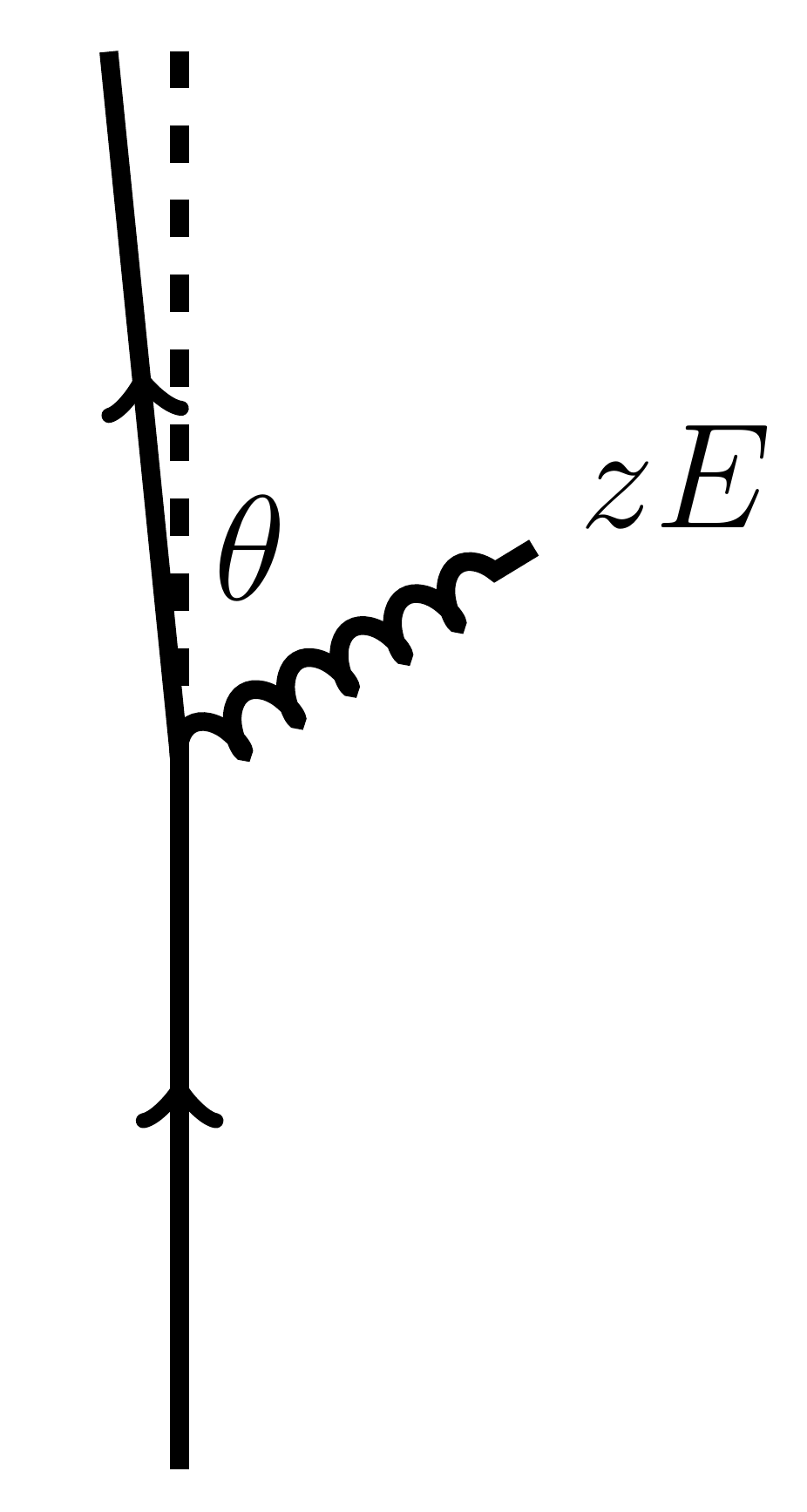}
\caption{A schematic diagram for the emission of one soft and collinear gluon off of a quark.  In this section, $z\ll 1$ and $\theta\ll 1$.}
\label{fig:JetMasQCD0}
\end{figure}
	
	Since the quark masses are small compared with the energy scales relevant at the LHC, the QCD Lagrangian is approximately scale invariant.  Consider a quark or gluon of energy $E$ that radiates a gluon with energy $e$ at an angle $\theta$ relative to the initial parton direction as depicted in Fig.~\ref{fig:JetMasQCD0}.  Define the energy fraction $z=e/E$.  This section will consider the soft ($z\ll 1$) and collinear $(\theta\ll 1)$ region of phase space.  As a result of the approximate scale invariance of the Lagrangian, one expects that the probability distribution of $z$ is the approximately the same on all decades.  More generally, for fixed $0<a<b<1$, $\Pr(a < z < b) = \Pr( ac < z < cb)$ for all $0<c<1/b$.  In particular, taking the derivative shows that the probability distribution function of $z$, $f_z$, has the property $f_z(z)=cf_z(cz)$.  As a result, for all $k=cz$, $f(k)k=f(cz)cz=f(z)z$.  Therefore, $f(z)z$ must be constant.  Let $y=\ln(z)$.  Then, the probability density of $y$, $f_y(y)$, has the form
	
	\begin{align}
	f_y(y) = f_x(x(y))\left|\frac{\partial x(y))}{\partial y}\right| = f_x(x(y))e^{\ln(x(y))}=f(x(y))x(y)=\text{constant},
	\end{align}
	
	\noindent and therefore the logarithm of $z$ is uniformly distributed from $-\infty$ to $0$.  The same argument applies for the logarithm of $\theta$.  To compute the differential cross section, the phase space is multiplied by the coupling factor $\alpha_s/\pi$ and the Altarelli-Parisi splitting functions~\cite{Altarelli:1977zs}, which were briefly introduced in Sec.~\ref{sec:particlesandforces}.  The relevant functions are
	
	\begin{align}
	P_{gq} &= C_F\frac{1+(1-z)^2}{z} \\
	P_{gg}&=2C_A\left[\frac{z}{(1-z)_+}+\frac{1-z}{z}+z(1-z)\right]+\frac{1}{2}\beta_0\delta(1-z),
	\end{align}
	
	\noindent where $\beta_0=(11C_A-4n_fT_F)/3$ is the leading order coefficient of the QCD $\beta$-function, $T_F=1/2$, $C_A=4/3$ and $C_F=3$ are the quark and gluon color factors, and $n_f$ is the number of active quark flavors, which is five.  The function $g(x)_+$ is defined by:
	
	\begin{align}
	\int_0^1 dx f(x)g(x)_+ = \int_0^1 dx (f(x)-f(1))g(x),
	\end{align}
	
	\noindent for some function $f(x)$.  These functions already include the singular behavior $\frac{1}{z}$ and for small $z$, these functions reduce to $P\approx 2C/z$.  A useful space\footnote{This calculation is based on similar discussions in Ref.~\cite{Dasgupta:2013yea,Larkoski:2014wba,Larkoski:2013paa}.} for describing the probability distribution of the soft and collinear gluon emissions is shown in Fig.~\ref{fig:JetMasQCD1}.  Since $\log(z)$ is uniform on $-\infty$ to $0$, $\log(1/z)$ is uniform on $0$ (hard) to $\infty$ (soft).  Likewise, $\log(R/\theta)$ is uniform from $0$ (wide angle) to $\infty$ (collinear).  The invariant mass $m$ of two massless particles with energies $(1-z)E$ and $zE$ is given by $m^2=z(1-z)E^2(1-\cos\theta)\approx zE^2\theta^2$.  In Fig.~\ref{fig:JetMasQCD1}, contours of constant invariant mass squared are thus straight lines given by
	
	\begin{align}
	\log\left(\frac{1}{z}\right) = -2\log\rho-2\log\left(\frac{R}{\theta}\right),
	\end{align}
	
	\noindent where $\rho=m/(ER)$.  At leading order (one real emission), the probability for the jet to have mass squared less than some fixed value $m_0^2$ is $\Pr(m^2\leq m_0^2)=1-\frac{2\alpha_s}{\pi}\ltri$, where $\ltri=\log^2\rho$ is the area of the blue triangle in Fig.~\ref{fig:JetMasQCD1}.  Therefore, the leading order probability distribution for $m^2$ is given by
	
	\begin{align}
	f_\text{LO}(m^2)=-\frac{2\alpha_s C}{\pi}\frac{\partial \ltri}{\partial m^2} = -\frac{\alpha_s C}{2\pi}\frac{\partial}{\partial m^2} \log^2\rho^2 = -\frac{\alpha_sC}{\pi m^2}\log\left(\frac{m^2}{E^2R^2}\right).
	\end{align}
	
	\noindent Changing the variables to $m$ gives $f_\text{LO}(m)=f_\text{LO}(m^2)\frac{\partial m^2}{\partial m}=2mf_\text{LO}(m^2)$, which is
	
	\begin{align}
	f_\text{LO}(m)= -\frac{4\alpha_sC}{\pi m}\log\left(\frac{m}{ER}\right).
	\end{align}

	\noindent The leading order distribution of the mass is not useful because it diverges too quickly\footnote{The divergence at zero can be regulated by considering the virtual corrections, which contribute at exactly $m=0$ by construction (if there is no second particle, then the jet mass is zero).  However, the leading logarithm approach is still more useful for understanding the full distribution of the jet mass, especially at low jet mass.} as $m\rightarrow 0$ (so $\int_0^{ER}f_\text{LO}(m)=\infty$).  Therefore, a different approximation is needed in order to make a sensible prediction of the jet mass distribution.  For the leading order calculation, the soft and collinear regions of phase space are unregulated for one emission.  However, the probability of many significant emissions is non-negligible and therefore another possibility is to consider all possible single gluon emissions.  The initial quark or gluon is treated as a final state object that can radiate an arbitrary number of gluons (the {\it eikonal} approximation) with $z\ll 1$ for each emission.  The leading order calculation showed that each emission has the form $\alpha_s\log^2\rho$ - this approximation is therefore a {\it leading logarithm} approximation in which all double-logarithms $(\alpha_s\log^2\rho)^n$ are summed to all orders.  The beginning of the calculation is the same as for the leading order one - the emission with the highest $z\theta^2$ in Fig.~\ref{fig:JetMasQCD1} will set the jet mass.  Therefore for a fixed $m_0$ and $n$ emissions, one is interested in the probability that all emissions have $z\theta^2<m_0^2$.  To compute this probability, divide the blue triangle in Fig.~\ref{fig:JetMasQCD1} into $N$ little boxed of equal area $a=\ltri/N$.  The size of the boxes is chosen so that the probability of multiple emissions within the box is small.  In this leading logarithm approximation, all emissions are assumed independent of each other.  Therefore,
	
	\begin{align}
	\label{eq:cdsforjetmass}
	\Pr(\text{no emissions in $\ltri$}) &= \prod_\text{$N$ boxes} \Pr(\text{no emission in the box})\\
	&=\prod_\text{$N$ boxes}(1-\Pr(\text{emission in box}))\\
	&=\prod_\text{$N$ boxes}\left(1-\frac{2\alpha_sCa}{\pi}\right)\\
	&=\left(1-\frac{2\alpha_sC\ltri}{\pi N}\right)^N \stackrel{N\rightarrow\infty}{=} e^{-\frac{2\alpha_sC}{\pi }\ltri}
	\end{align}

	\noindent The derivative Eq.~\ref{eq:cdsforjetmass} gives the probability distribution of the jet mass $f_\text{LL}(m^2)$:
	
	\begin{align}
	\label{eq:jetmassatll}
	f_\text{LL}(m) &= 2mf_\text{LL}(m^2) = 2m\frac{\partial}{\partial m^2}\Pr(\text{no emissions in $\ltri$})\\
	&=-2m\frac{\alpha_sC}{\pi m^2}\ln\left(\frac{m^2}{E^2R^2}\right)\exp\left(-\frac{\alpha_s C}{2\pi}\log^2\left(\frac{m^2}{E^2R^2}\right)\right)\\
	&=-\frac{4\alpha_s C}{\pi m}\ln\left(\frac{m}{ER}\right)\exp\left(-\frac{2\alpha_s C}{\pi}\log^2\left(\frac{m}{ER}\right)\right),
	\end{align}
	
	\noindent which is finite (actually zero) as $m\rightarrow 0$.  The exponential suppression factor in Eq.~\ref{eq:jetmassatll} is called a {\it Sudakov factor}.  The left plot of Fig.~\ref{fig:JetMasQCD2} shows the distribution in Eq.~\ref{eq:jetmassatll} plotted for ($E$,$R$) = (200 GeV, 1) and ($E$,$R$) = (400 GeV, 0.4) separately for quark and gluon jets.  In general, the quark jet mass distribution is shifted to lower values of the jet mass.  The energies and radii are chosen to approximately correspond to $2m/E = R$ for a $W$ boson in order to illustrate how the mass distribution compares to $m_W\approx 80$ GeV.  Since $f_\text{LL}(m)$ is bounded and has compact support, it has finite moments.  The average jet mass is given by

	\begin{align}
	\label{eq:averagemassinqcd}
	\langle m \rangle = \alpha_s Rp_\text{T}C\left(4\int_0^1d\rho\log(\rho)\exp\left(-\frac{2\alpha_s C}{\pi}\log^2\rho\right)\right).
	\end{align}
	
	\noindent The expression in parenthesis in Eq.~\ref{eq:averagemassinqcd} is an $\mathcal{O}(1)$ number that is approximately $0.9$ for quark jets and $0.7$ for gluon jets.  The right plot of Fig~\ref{fig:JetMasQCD2} shows the average jet mass as a function of jet $p_\text{T}$, compared with the electroweak boson and top quark masses.  Especially for quark jets, which dominate at high $p_\text{T}$ (see Sec.~\ref{sec:NCharge:Design}), the average mass is significantly less than the mass of the boosted `signal' objects.  Interestingly, at some high $p_\text{T}$ the average QCD jet mass will be the same and even higher than the mass of electroweak bosons and top quarks.  For this reason, analyses using ultra-boosted bosons and top quarks would use a ceiling requirement on the jet mass instead of a lower mass threshold.

\begin{figure}[h!]
\centering
\includegraphics[width=0.5\textwidth]{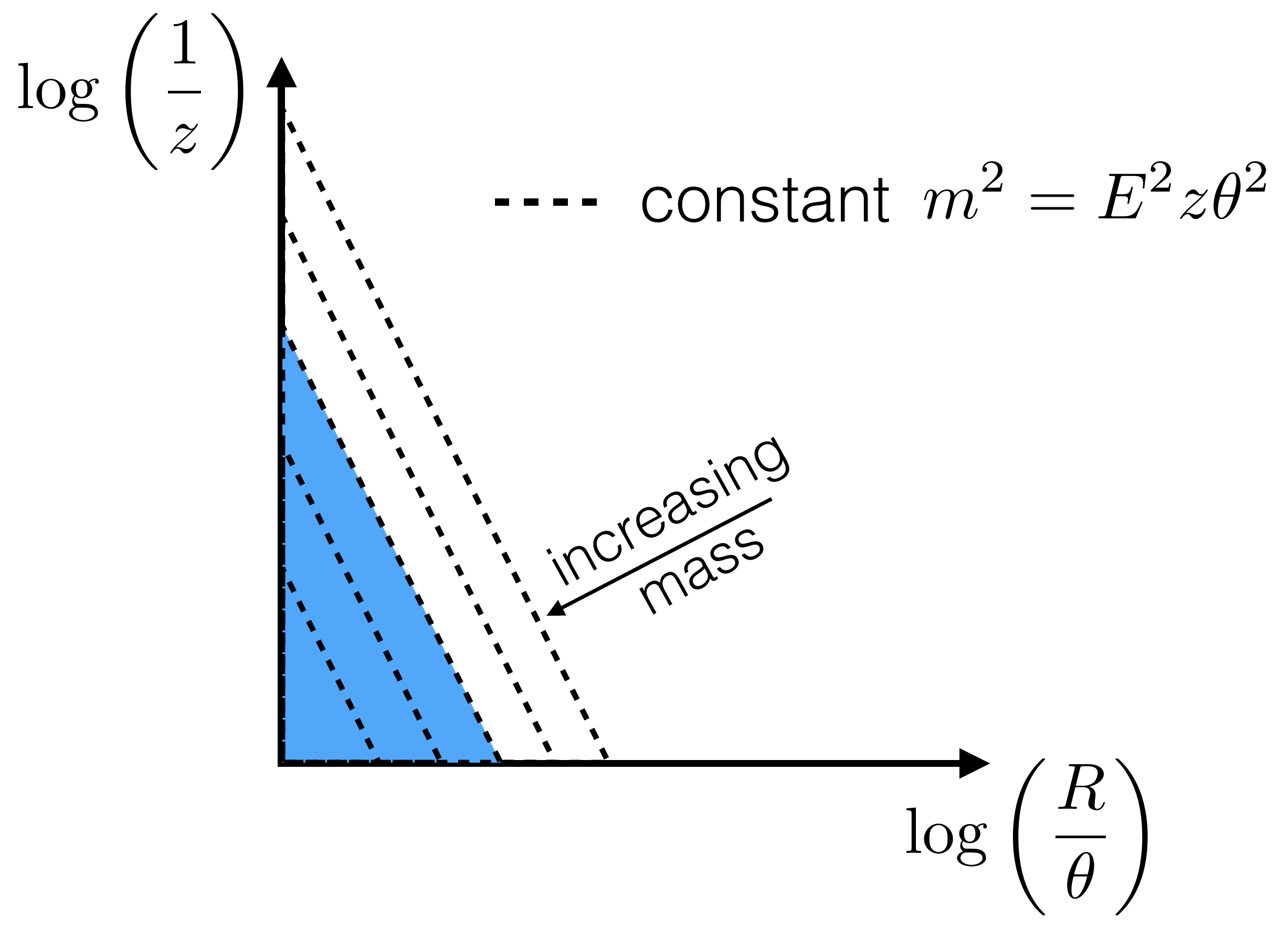}
\caption{A schematic diagram of the $\log(1/z)$ versus $\log(R/\theta)$ plane in which the probability for the emission of a gluon is approximately uniform.  The dashed lines show contours of constant $m^2$, which increase from upper right to lower left. The blue triangle corresponds to mass $m_0^2$.}
\label{fig:JetMasQCD1}
\end{figure}

\begin{figure}[h!]
 \centering
\includegraphics[width=0.45\textwidth]{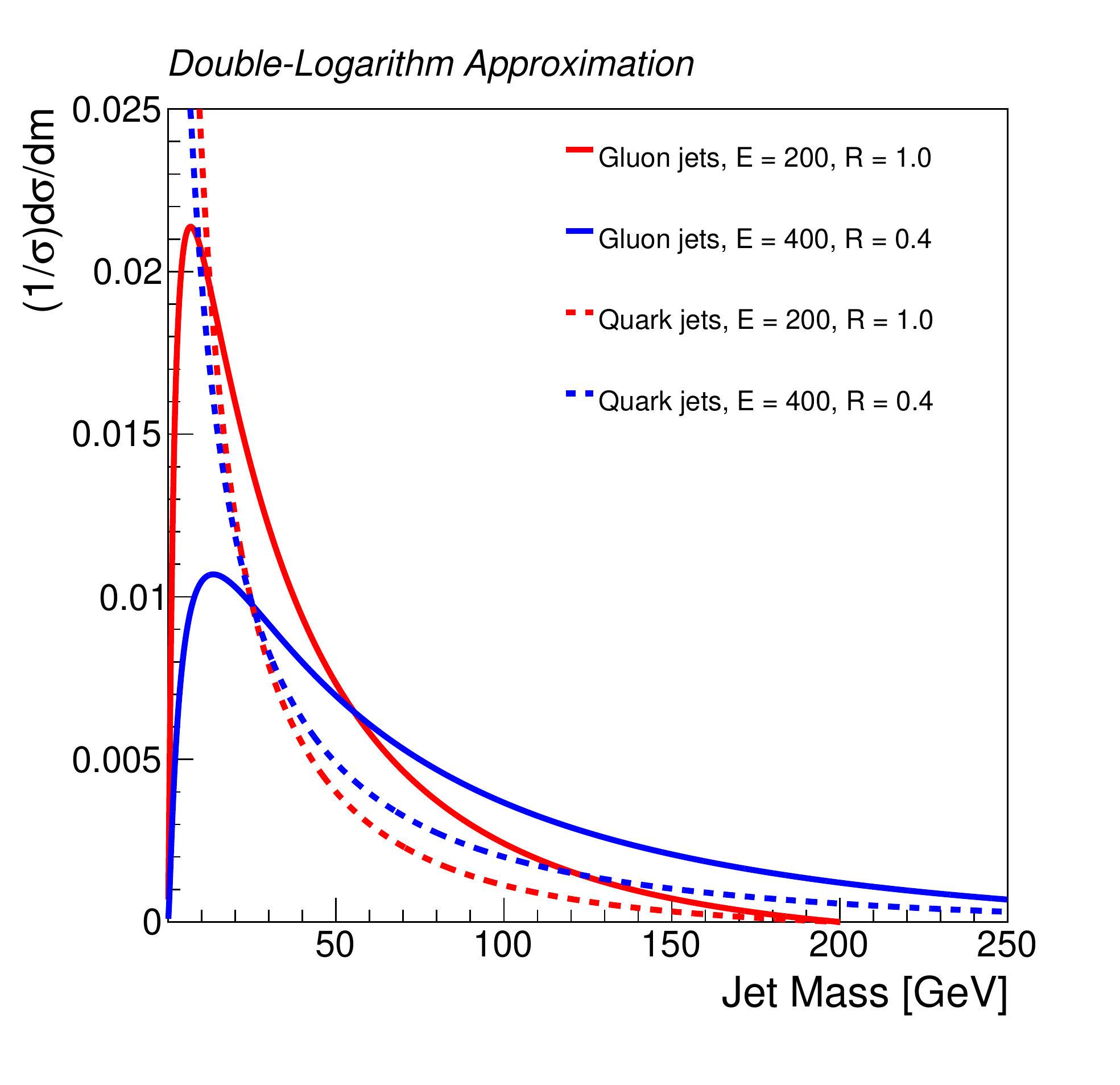}\includegraphics[width=0.45\textwidth]{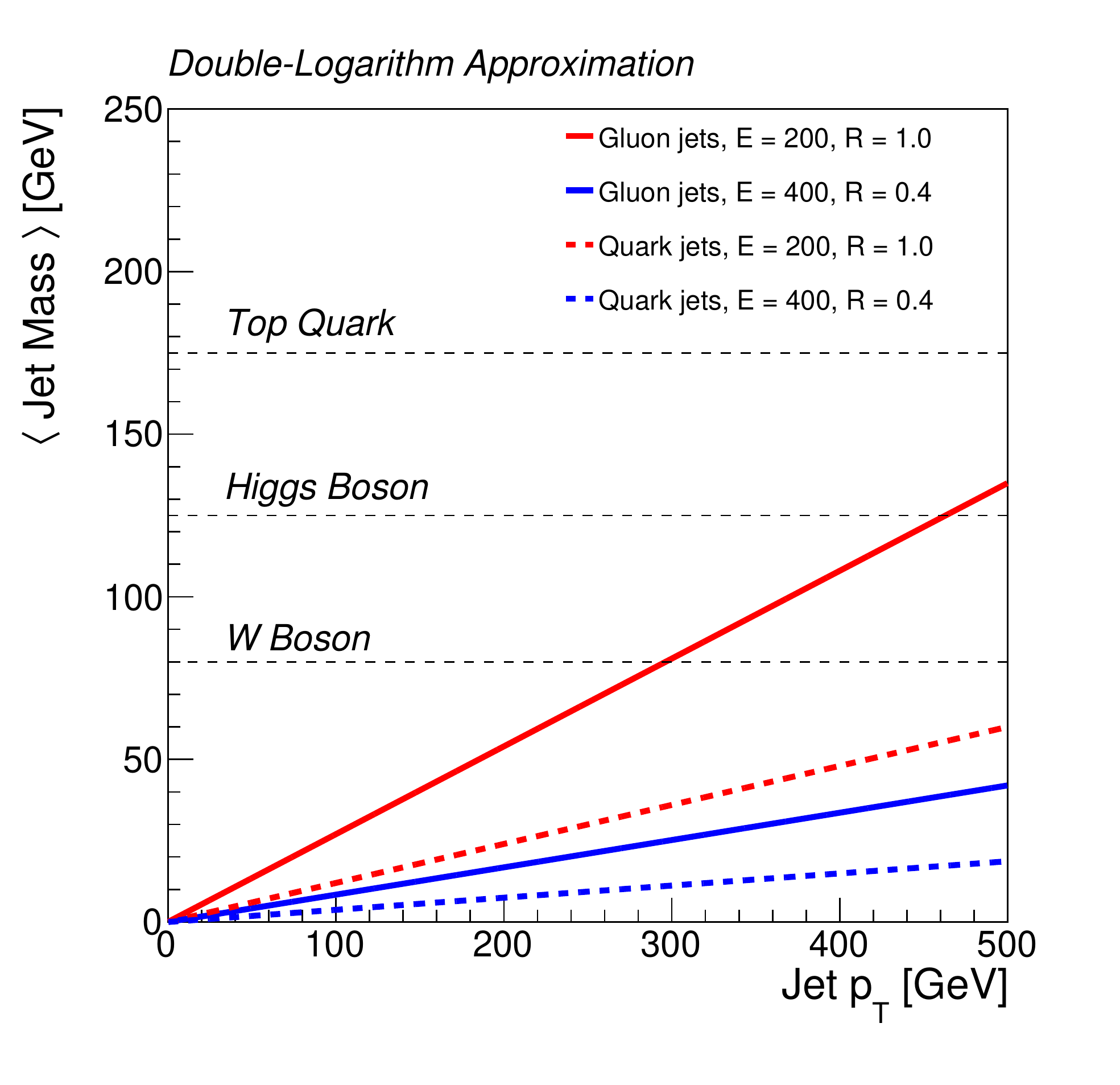}
\caption{Left: The leading logarithm distribution of the jet mass for ($E$,$R$) = (200 GeV, 1) and ($E$,$R$) = (400 GeV, 0.4) separately for quark and gluon jets.  The quark distribution is suppressed at zero, but it increases for finite mass much faster than the gluon distribution.  Right: The average jet mass as a function of jet $p_\text{T}$ for the same four settings as the left plot. Horizontal lines indicate the $W$ boson, Higgs boson and top quark masses. }
\label{fig:JetMasQCD2}
\end{figure}

	Figure~\ref{fig:JetMasQCD3} shows how the average jet mass depends on the jet $p_\text{T}$ for QCD jets in the early Run 2 data compared with simulation.  Jets are clustered with a radius $R=1.0$ and trimmed (see Sec.~\ref{sec:JMR} for details).  As expected, the average jet mass increases monotonically with $p_\text{T}$.  The exact shape in Fig.~\ref{fig:JetMasQCD3} deviates from linear because (a) the composition of quarks and gluons changes as a function of $p_\text{T}$ and (b) the jets are trimmed and so the {\it effective area} of the jet depends on $p_\text{T}$.  The jet mass in the simulation is generated in the parton shower implemented in {\sc Pythia}~8 which is based on the leading logarithm approximation, but includes additional effects such as a running $\alpha_s$ and the full LO quark and gluon splitting functions.  The next sections describe how the jet mass is reconstructed in practice.

\begin{figure}[h!]
 \centering
\includegraphics[width=0.45\textwidth]{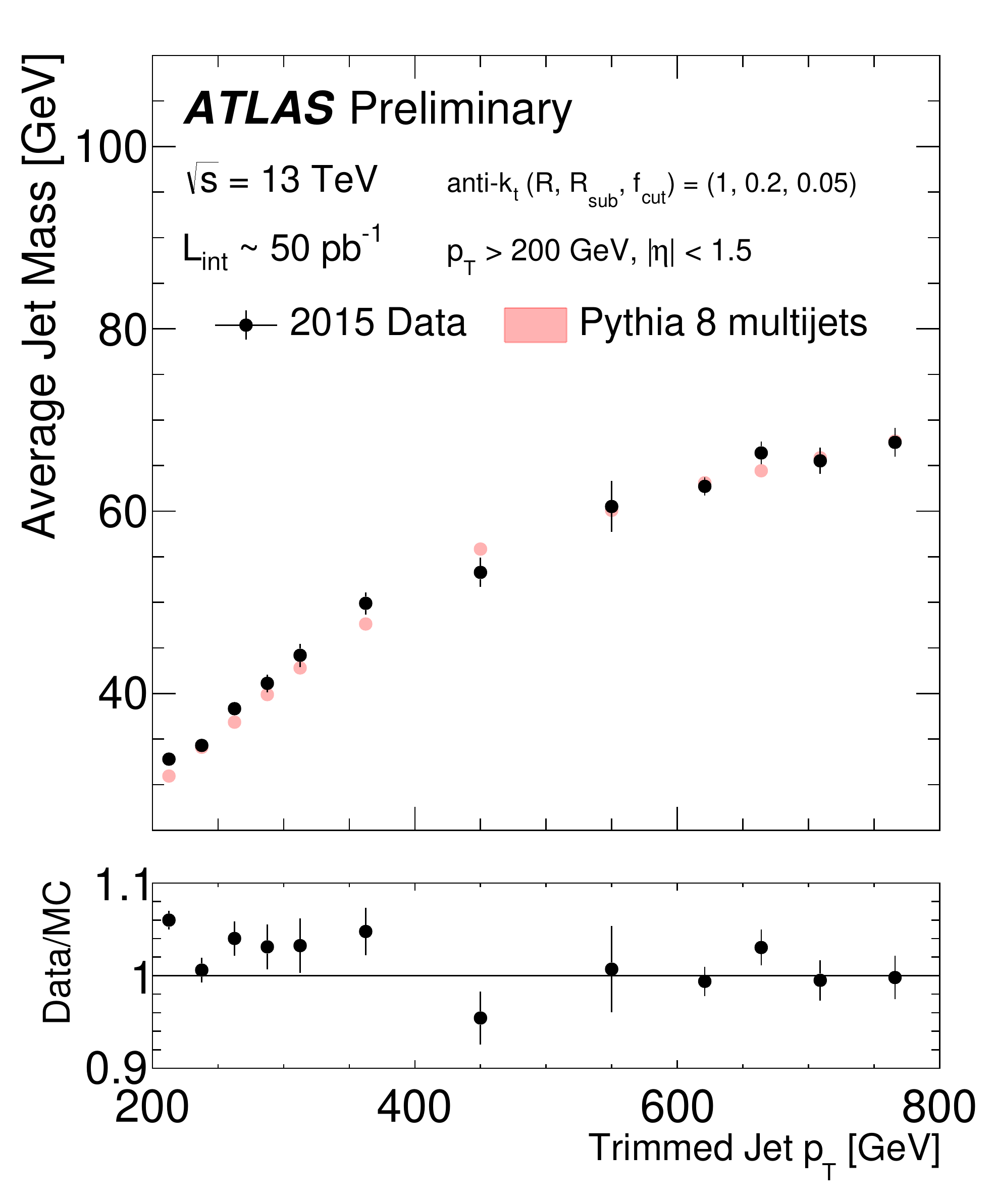}
\caption{The $p_\text{T}$ dependence of the average leading anti-$k_t$ $R=1.0$ trimmed with $f_\mathrm{cut}=0.05$ and $R_\mathrm{sub}=0.2$ jet mass.   See Sec.~\ref{sec:JMR} for details.}
\label{fig:JetMasQCD3}
\end{figure}

\clearpage	
	
	\subsection{Reconstructing the Calorimeter Jet Mass}
	\label{sec:JMR}
	
	Reconstructing the jet mass is an experimental challenge because it requires a precise measurement of both the energy and location of particles inside a jet.  This property of the jet mass is illustrated in Fig.~\ref{fig:JMR:intro} for a hadronically decaying boosted $W$ boson in a MC model.  Particles carrying a small fraction of the jet's $p_\text{T}$ can contribute just as much to the mass as particles carrying a large fraction of the total momentum.   Furthermore, at a hadron collider there is no conservation law that can be used for an in-situ study of the jet mass response.   For the jet momentum, conservation in the plane transverse the beam is a powerful constraint that has no analogue for mass as $\sqrt{\hat{s}}$ is unknown.  Even at a high energy electron-positron collider, for cases of interest for tagging, the jet mass is typically much smaller than the jet energy and therefore a constraint on the total energy is not useful.
	
\begin{figure}[h!]
 \centering
\includegraphics[width=0.55\textwidth]{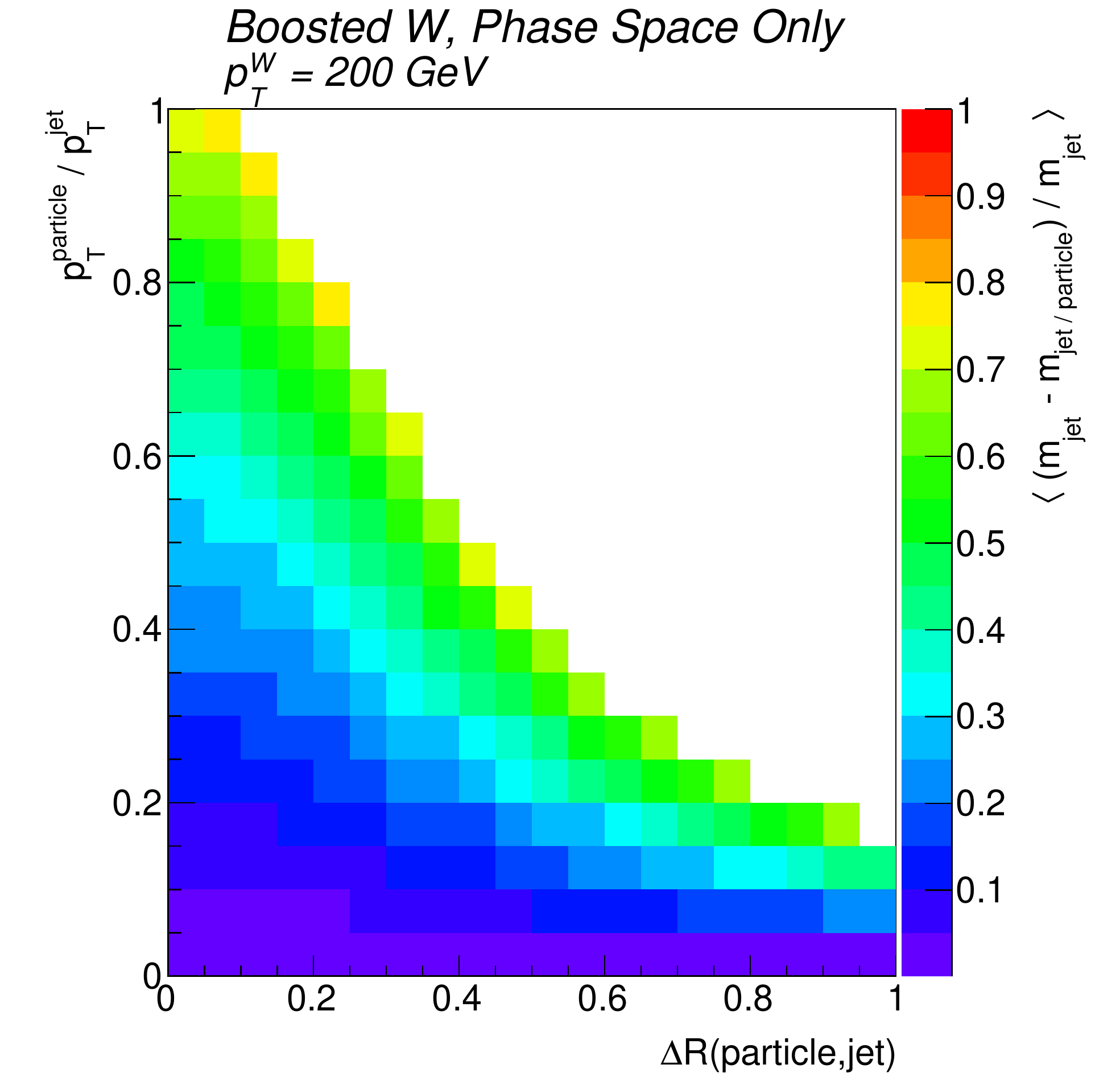}
\caption{The impact of a jet constituent on that jet's mass as a function of $\Delta R$ between the constituent and the jet axis and the $p_\text{T}$ fraction carried by the constituent. $W$ bosons are generated with $p_\text{T}=200$ GeV in a random direction and decay via a scalar two-body phase space into quarks which subsequently decay into 10 massless particles each also with the scalar phase space.  The `jet' is the four-vector sum of all these particles, which has mass 80 GeV and $p_\text{T}=200$ GeV.}
\label{fig:JMR:intro}
\end{figure}

In Run 1 of the LHC, the most used definition of the jet mass takes as input calibrated calorimeter-cell clusters.  Most of this section will be focused on the mass of {\it large-radius} jets clustered with the anti-$k_t$ algorithm using $R=1.0$ and groomed with the trimming procedure~\cite{Krohn:2009th} to reduce the sensitivity of the jet mass to contamination from pileup and the underlying event.  These sources of diffuse energy are detrimental to the jet mass resolution because low-energy wide-angle radiation can have a big impact on the jet mass  as illustrated by Fig.~\ref{fig:JMR:intro}.  For trimming, the jet constituents are re-clustered with the $k_t$ algorithm\footnote{The $k_t$ instead of anti-$k_t$ algorithm is used for subjets because it results in a more balanced distribution of energy - see Ref.~\cite{Krohn:2009th} for more details.} using $R=R_\text{sub}$ and then the constituents of the resulting {\it subjets} with $p_\text{T}^\text{subjet}<f_\text{cut}\times p_\text{T}^\text{jet}$ are removed.  Note that this requirement is applied before any pileup mitigation and therefore the trimming becomes harsher for higher levels pileup.  This is solved naturally by the re-clustering algorithm, described in Sec.~\ref{sec:ReclusteredJetMass}.  As a result of an extensive campaign\cite{Aad:2015rpa,Aad:2013gja} to optimize $R_\text{sub}$ and $f_\text{cut}$, the values $f_\text{cut}=0.05$ and $R_\text{sub}=0.3$ (0.2) are used in Run 1 (Run 2).  The smaller $R_\text{sub}$ value improves the jet mass resolution at high $p_\text{T}$ where the jet constituents of a resonance with fixed mass are closer together.   Figure~\ref{fig:JMR:jetmassresponseED} shows an event display in data illustrating the impact of trimming.  Two high $p_\text{T}$ well-isolated jets are nearly back-to-back in the transverse plane ($\Delta\phi$ mod $\pi\approx 0$).  The isolated anti-$k_t$ jets have a circular catchment area whereas the $k_t$ subjets have irregular areas whose sum is much less than the ungroomed jet area.   Trimming has a small effect on the jet $p_\text{T}$, but a non-trivial impact on the jet masses.  For example, the lower left jet looses less than 2\% of its $p_\text{T}$ after trimming while the jet mass is reduced by over 10\%.  

\begin{figure}[h!]
 \centering
\includegraphics[width=0.7\textwidth]{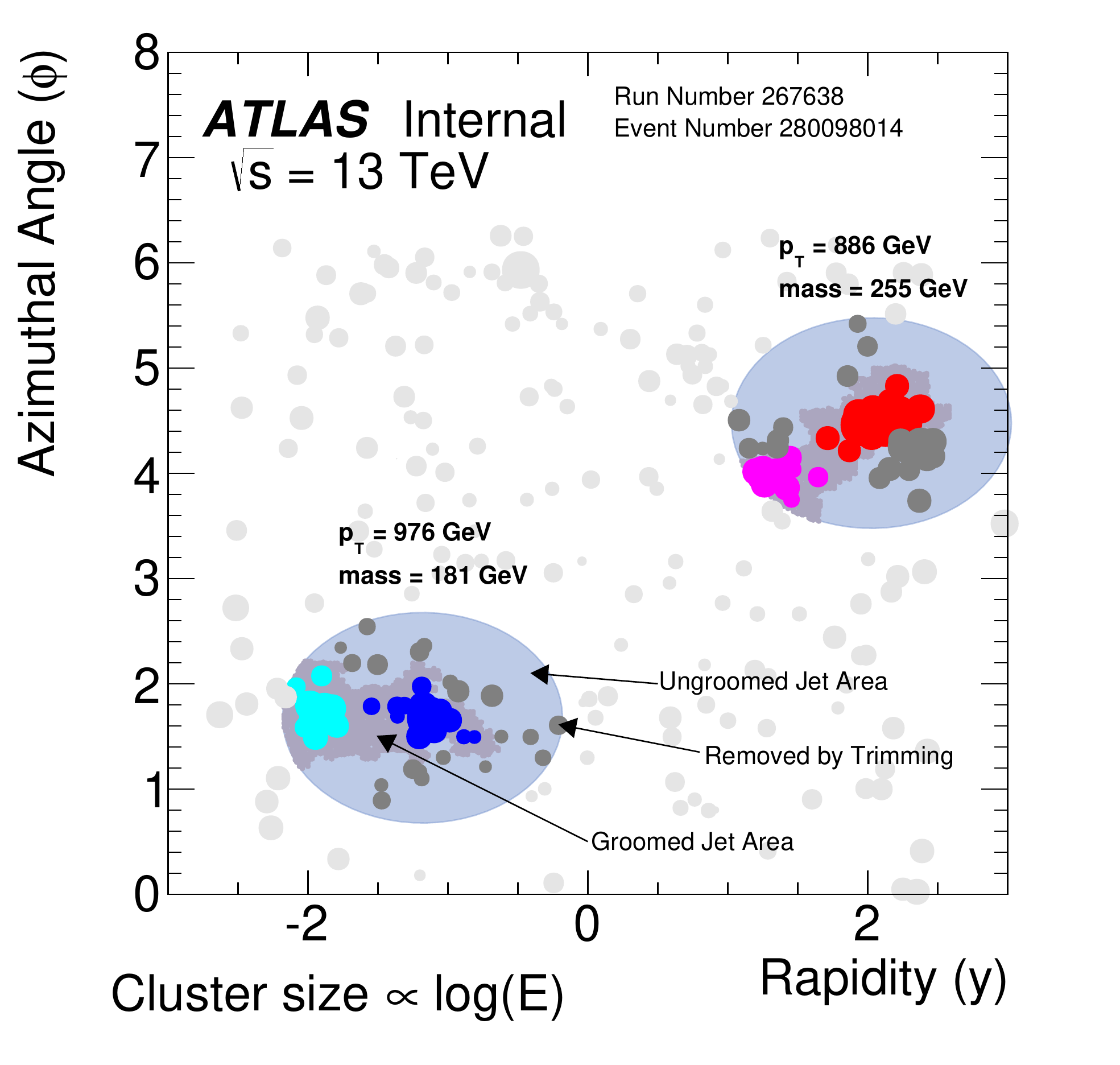}
\caption{An event display of a dijet event in the early Run 2 data.  The gray and colored filled circles correspond to calorimeter cell clusters where the circle radius is proportional to the log of the cluster energy.  Colored circles represent the constituents of the trimmed jets.  The blue-gray circles are the ungroomed anti-$k_t$ R=1.0 jets.  The dark gray circles are the constituent clusters removed by trimming.  The remaining gray area underneath the colored circles is the ghost area of the trimmed jet.}
\label{fig:JMR:jetmassresponseED}
\end{figure}

A jet-level calibration is applied to account for the residual detector response.   This correction is first applied to the jet energy and then to the jet mass.  In particular, the calibrated jet mass $m$ of a jet $J$ reconstructed with $\eta_J$ is given by

\begin{align}
m = c_\text{JMS}\left(c_\text{JES}\left( \sum_{i\in J} E_i,\eta_{J}\right),\eta_{J}\right)  \times \sqrt{\Bigg(\sum_{i\in J} E_i\Bigg)^2-\Bigg(\sum_{i\in J} \vec{p}_i\Bigg)^2},
\end{align}

\noindent where $E_i$ is the LCW calibrated energy of cluster $i$.  Each cluster is treated as massless with three-momentum $\vec{p}_i=(E_i/\cosh\eta)(\cos\phi_i,\sin\phi_i,\sinh\eta)$.  The calibration functions $c_\text{JMS}$ and $c_\text{JES}$ are for the jet mass and jet energy scales determined using numerical inversion.  When generic QCD jets are used to derive the calibration, one needs to also control for the jet size.  Early Run 2 calibrations therefore use $m/p_\text{T}$ as one of the inputs to $c_\text{JMS}$.  After this jet $p_\text{T}$- and jet mass-dependent calibration, the average reconstructed jet mass is the same as the particle-level jet mass in simulation for quark and gluon jets: the calibration {\it closes}.   The response depends on the quark/gluon nature of the jets, so the calibration is only guaranteed to close in a sample of events with the same composition as the one used to derive the calibration.  Also, as a result of the dependence of the response on jet substructure, the calibration may not exactly close for boosted $W/Z/H$ boson or top quark jets.  This is not necessarily a problem for jet tagging, but it can be mitigated by controlling for jet substructure in the calibration or performing the calibration on signal jets.  Alternative jet mass definitions are described in Sec.~\ref{sec:ReclusteredJetMass} and~\ref{sec:TAMass}.  

Large radius jet 4-vector reconstruction performance is quantified by properties of the response $(R)$\footnote{Not to be confused with the jet radius, which is a constant.  Unfortunately, the use of the symbol $R$ is standard for both quantities.}: the ratio of the reconstructed jet mass to the jet mass of the corresponding particle-level jet.  When distinguishing boosted hadronic resonance jets from generic quark and gluon jets, the most important property of $R$ is its width.  Since the distribution of $R$ is not Gaussian, there is no universally accepted definition of the width.  Figure~\ref{fig:JMR:jetmassresponse} shows the distribution of $R$ for boosted hadronically decaying $W$ and $Z$ bosons in four boson $p_\text{T}$ ranges from 200 GeV up to 2 TeV.  For illustration, two different fits are performed and overlaid on the input distributions.  The first fit is an iterative $\chi^2$ fit to a Gaussian that uses the histogram mean and standard deviation as seeds and then subsequently uses the fitted mean and standard deviation to set the fit range.  By focusing on a $\pm1\sigma$ interval about the mean, the fit captures the {\it core} of the distribution of $R$.  Another way to isolate the core and down-weight the heavy tails is to fit a double-Gaussian.  In Fig.~\ref{fig:JMR:jetmassresponse} the core Gaussian is shown in red while the tail Gaussian is shown in blue.  Both normal distributions are constrained to have the same mean and the core Gaussian is seeded with half the histogram standard deviation while the tail Gaussian is seeded with twice the histogram standard deviation.  Similar fits are performed for the jet $p_\text{T}$ response in Fig.~\ref{fig:JMR:jetpTresponse}.  The $p_\text{T}$ dependence of the core Gaussian resolution, along with the histogram standard deviation and 68\% median-centered quantile are shown in Fig.~\ref{fig:JMR:response} for both the jet mass and jet $p_\text{T}$ response.  For all definitions the measure of spread is divided by a measure of the distribution center, which is either the (fitted) mean or the median.  The two fitting methods designed to isolate the core of the response distribution give similar results for both quantities (and for the inter-quantile range).  However, the standard deviation is significantly larger, indicating the presence of non-negligible heavy tails.

\begin{figure}[h!]
 \centering
\includegraphics[width=0.5\textwidth]{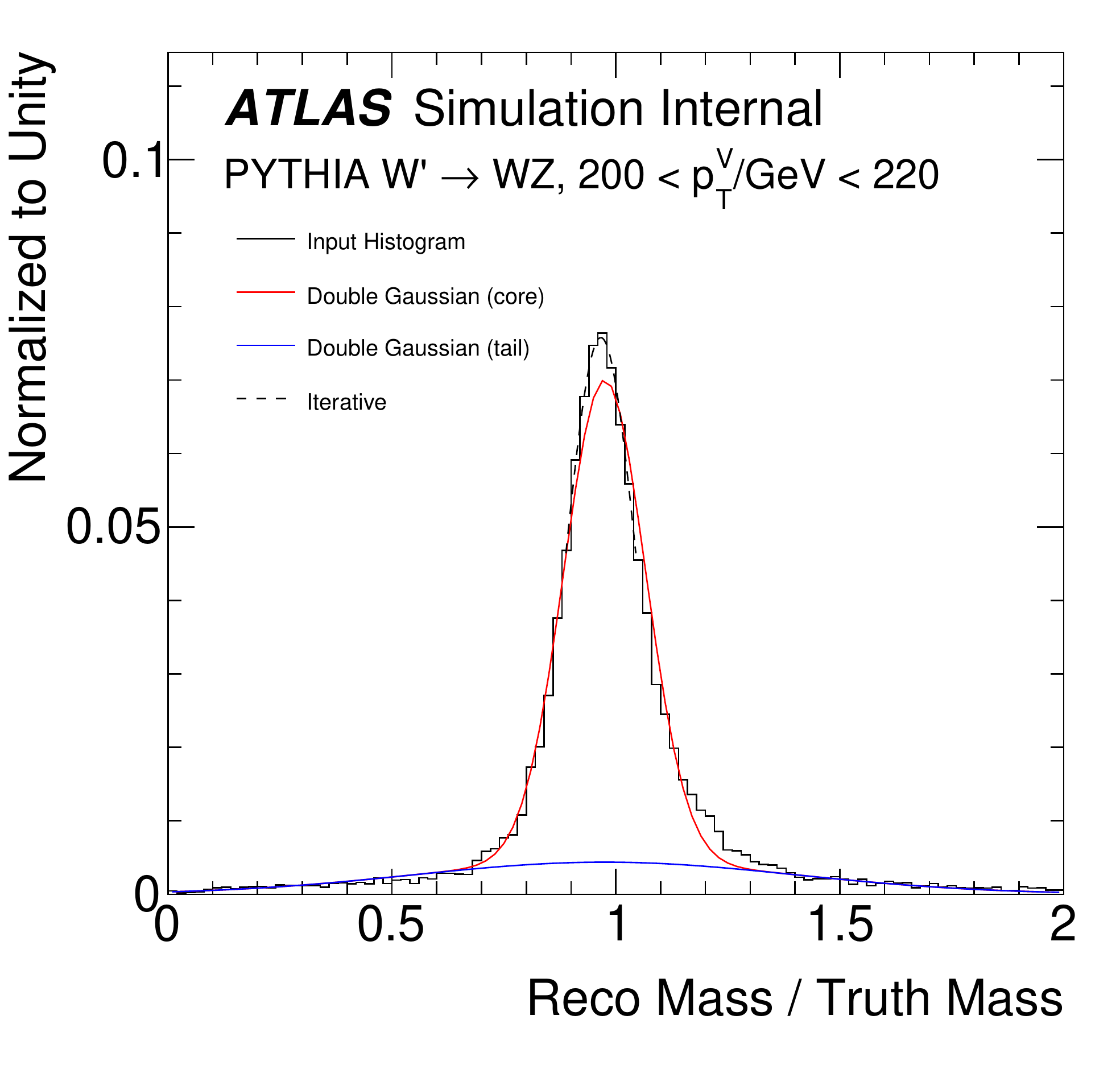}\includegraphics[width=0.5\textwidth]{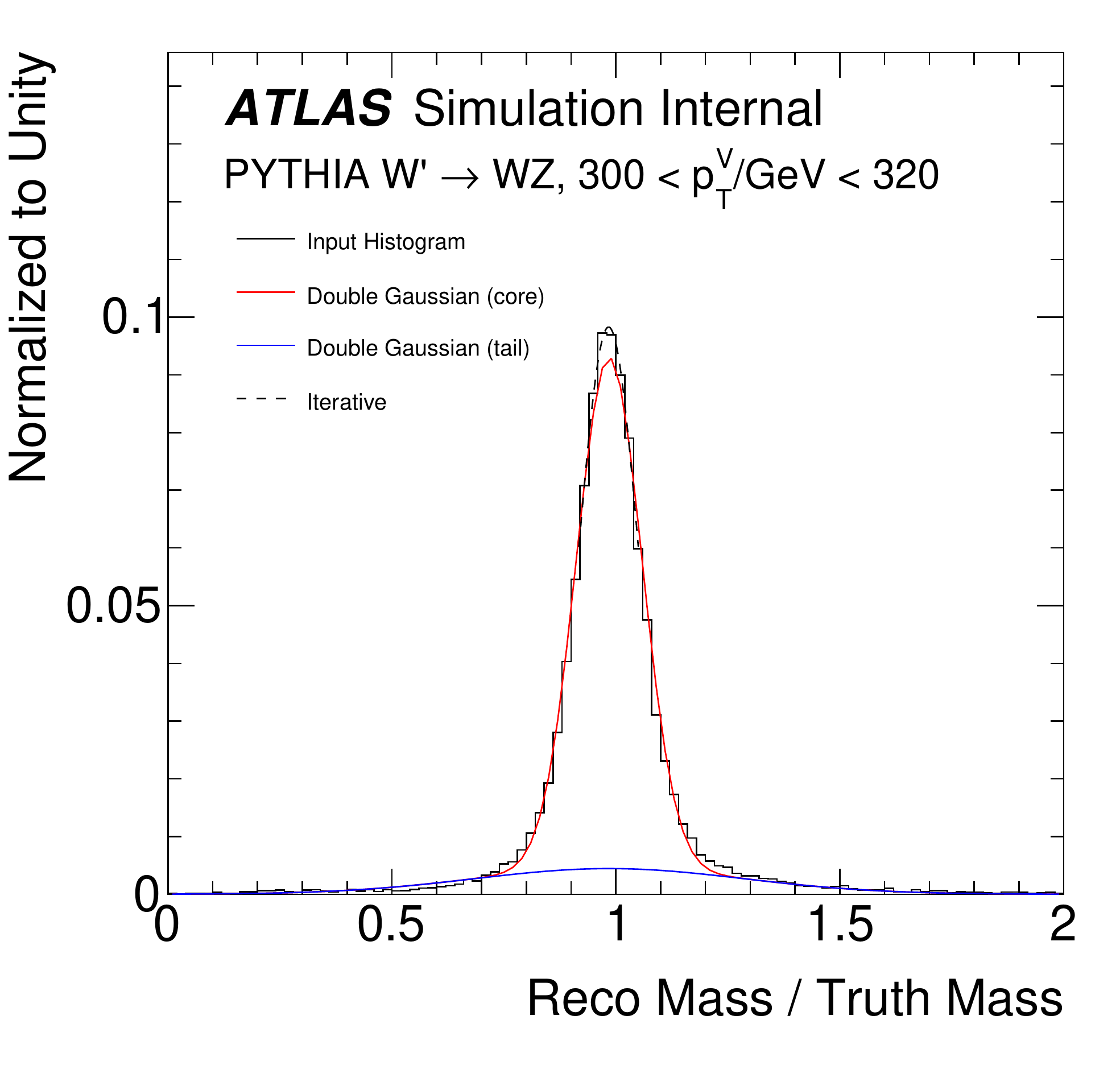}
\includegraphics[width=0.5\textwidth]{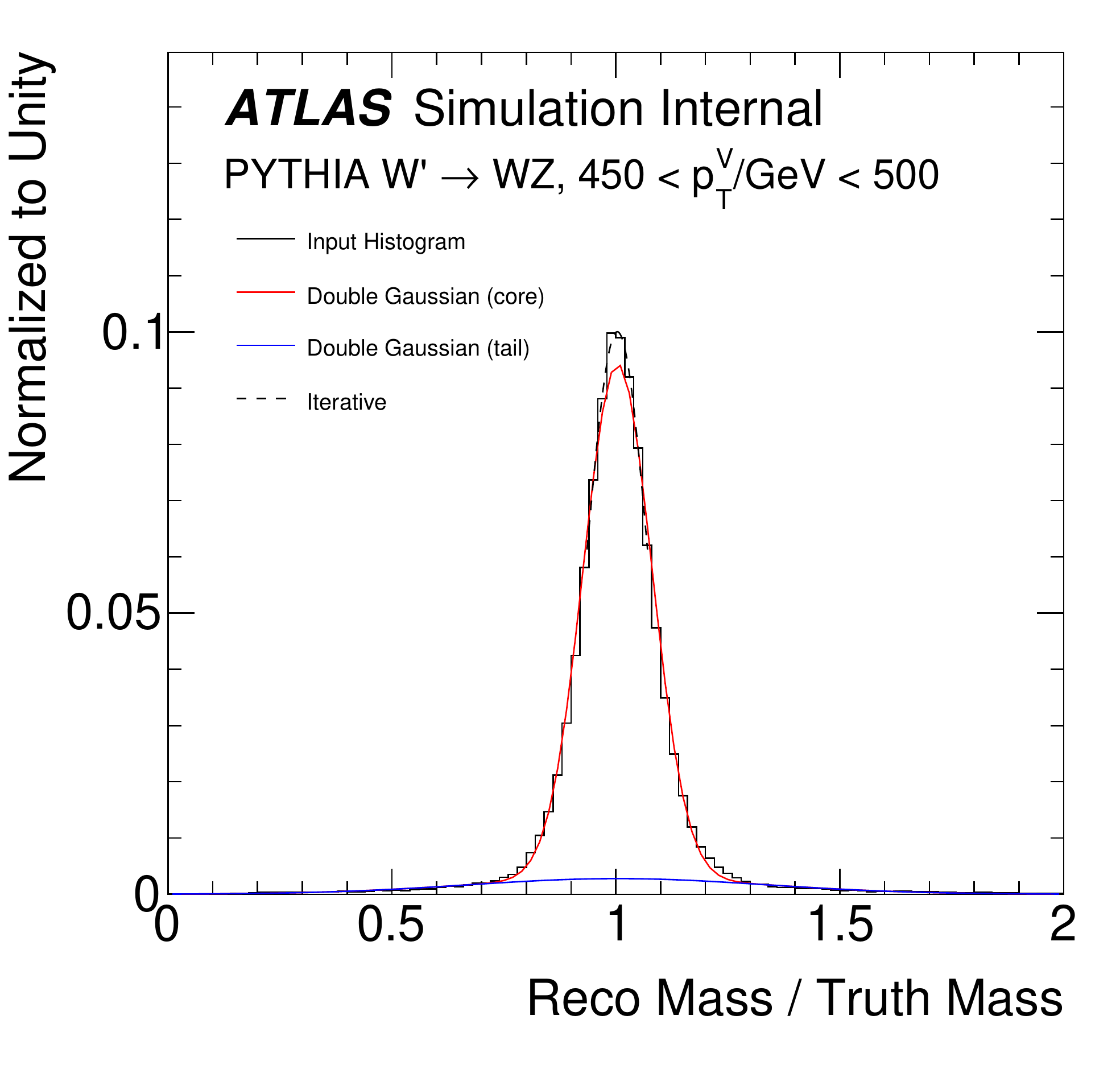}\includegraphics[width=0.5\textwidth]{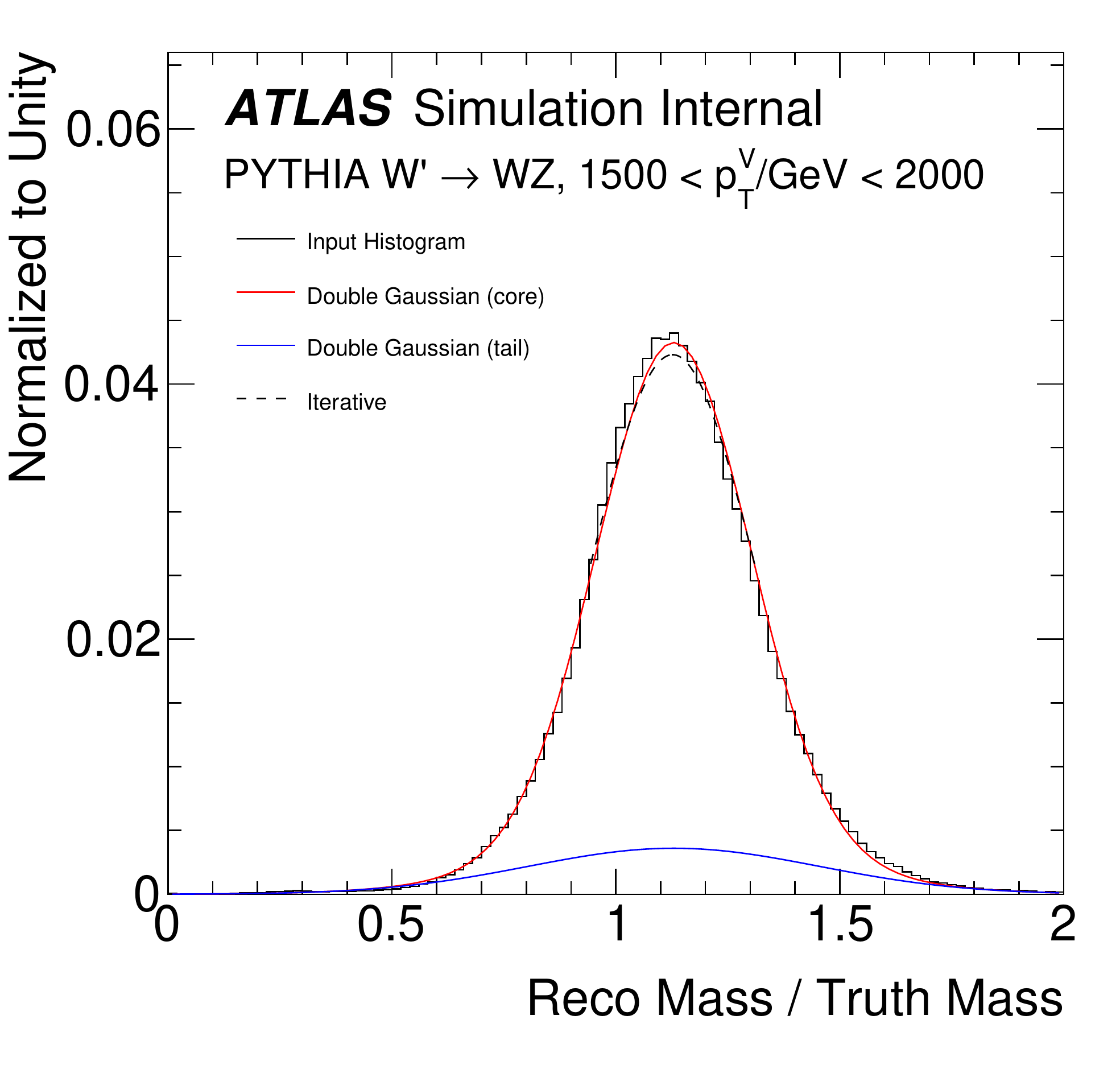}
\caption{The distribution of the jet mass response $(R)$ for boosted hadronically decaying $W$ and $Z$ bosons in four boson $p_\text{T}$ ranges from 200 GeV up to 2 TeV.  See the text for a description of the various fitting methods.}
\label{fig:JMR:jetmassresponse}
\end{figure}

\begin{figure}[h!]
 \centering
\includegraphics[width=0.5\textwidth]{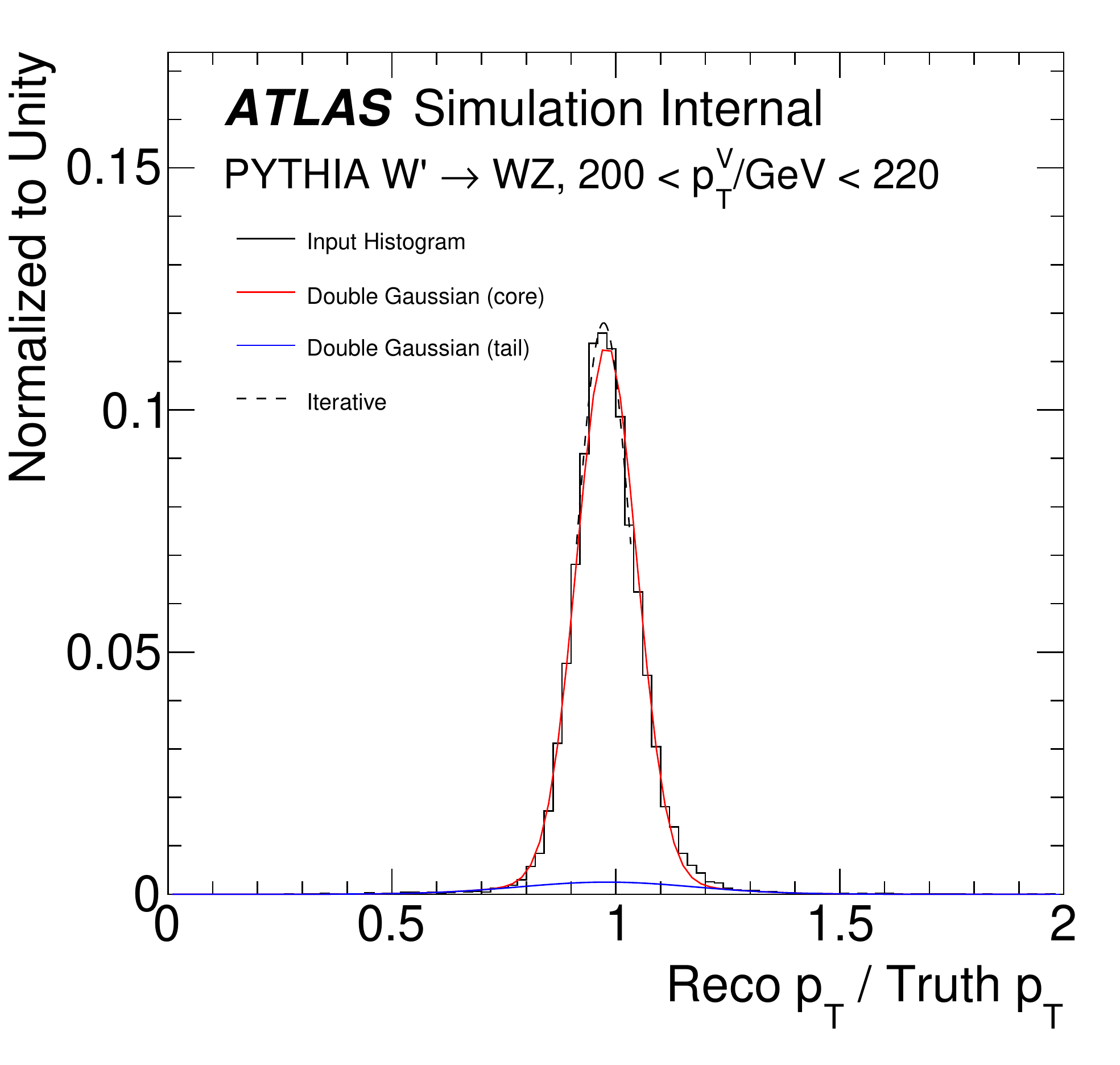}\includegraphics[width=0.5\textwidth]{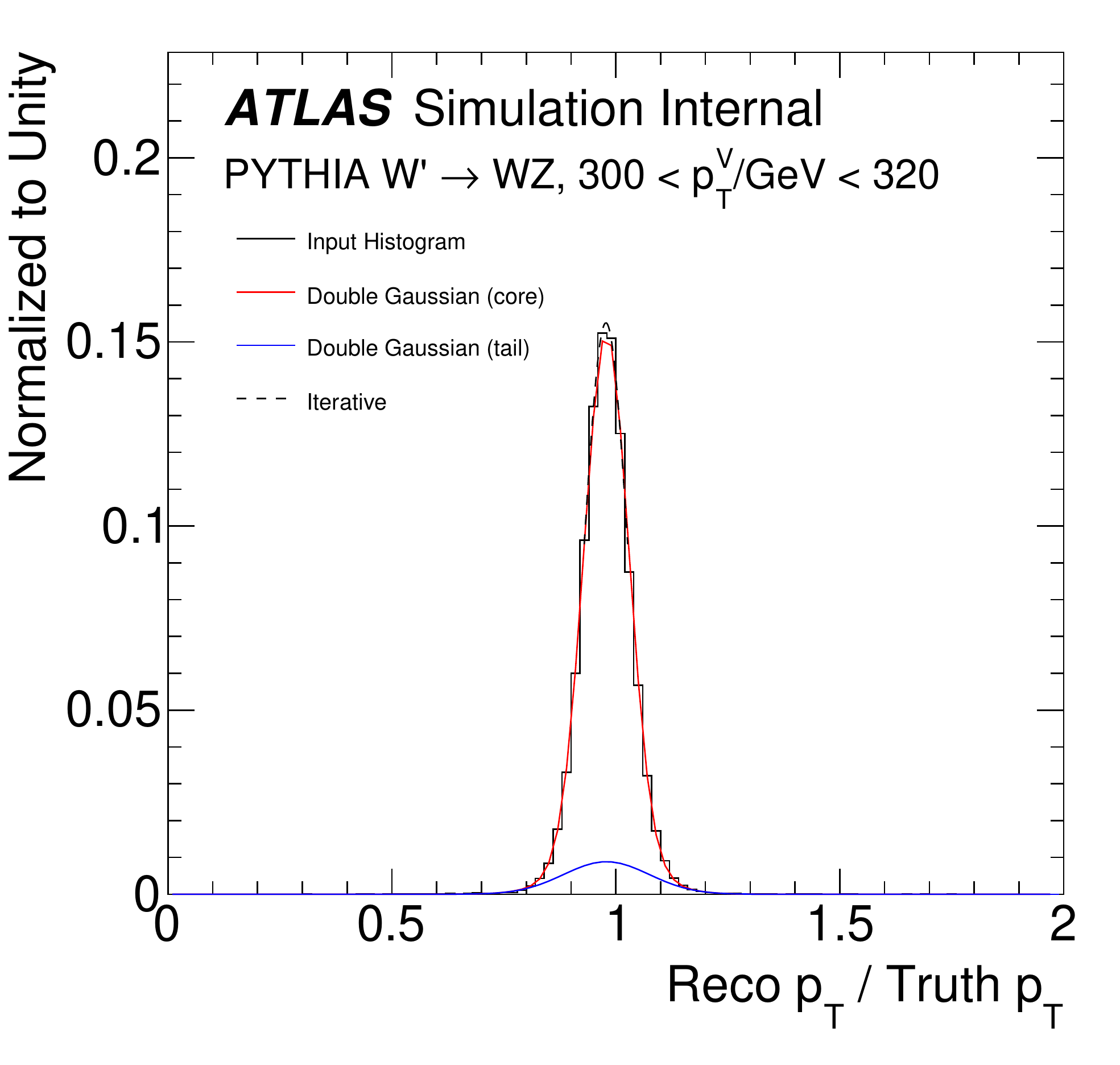}
\includegraphics[width=0.5\textwidth]{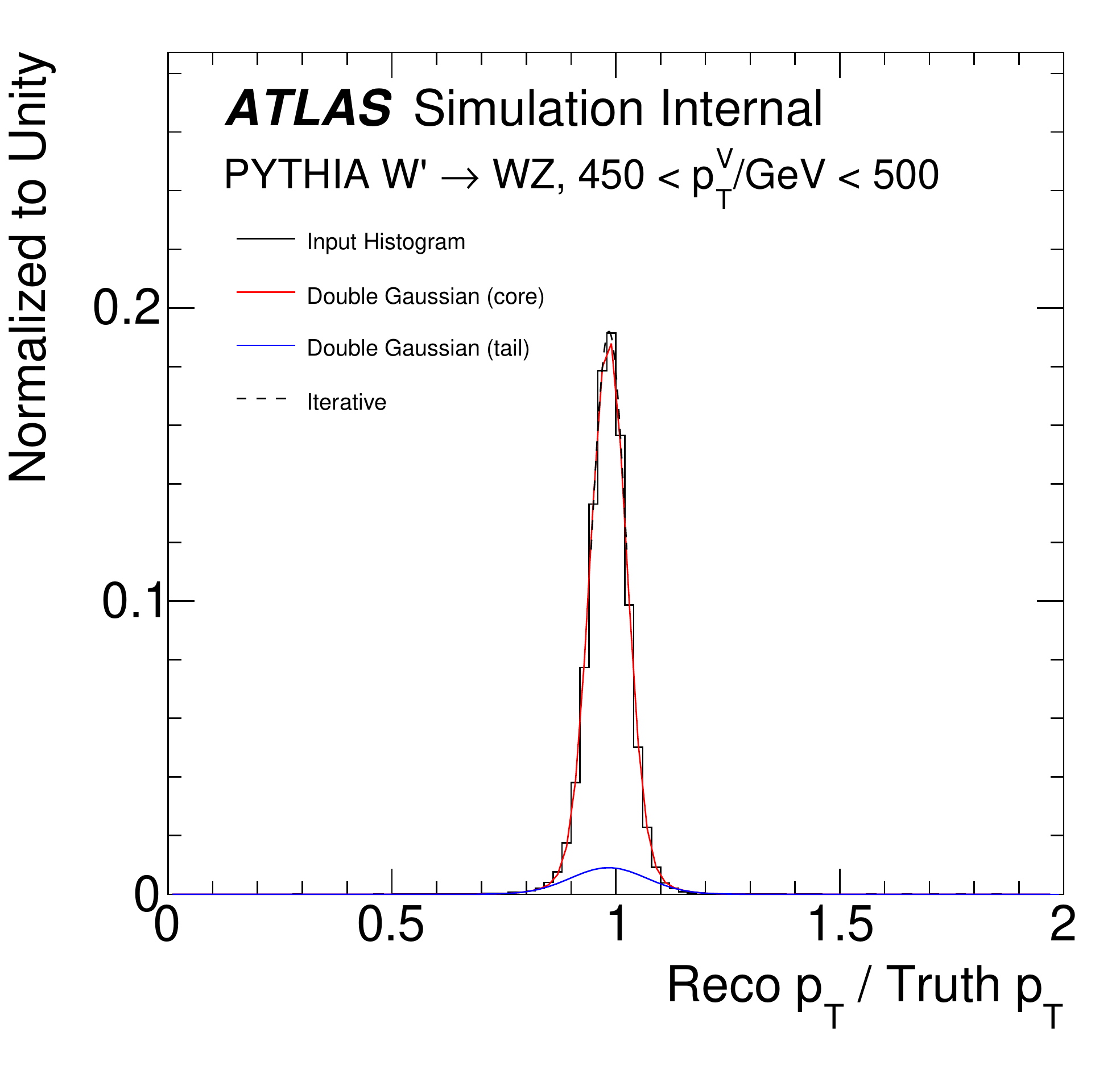}\includegraphics[width=0.5\textwidth]{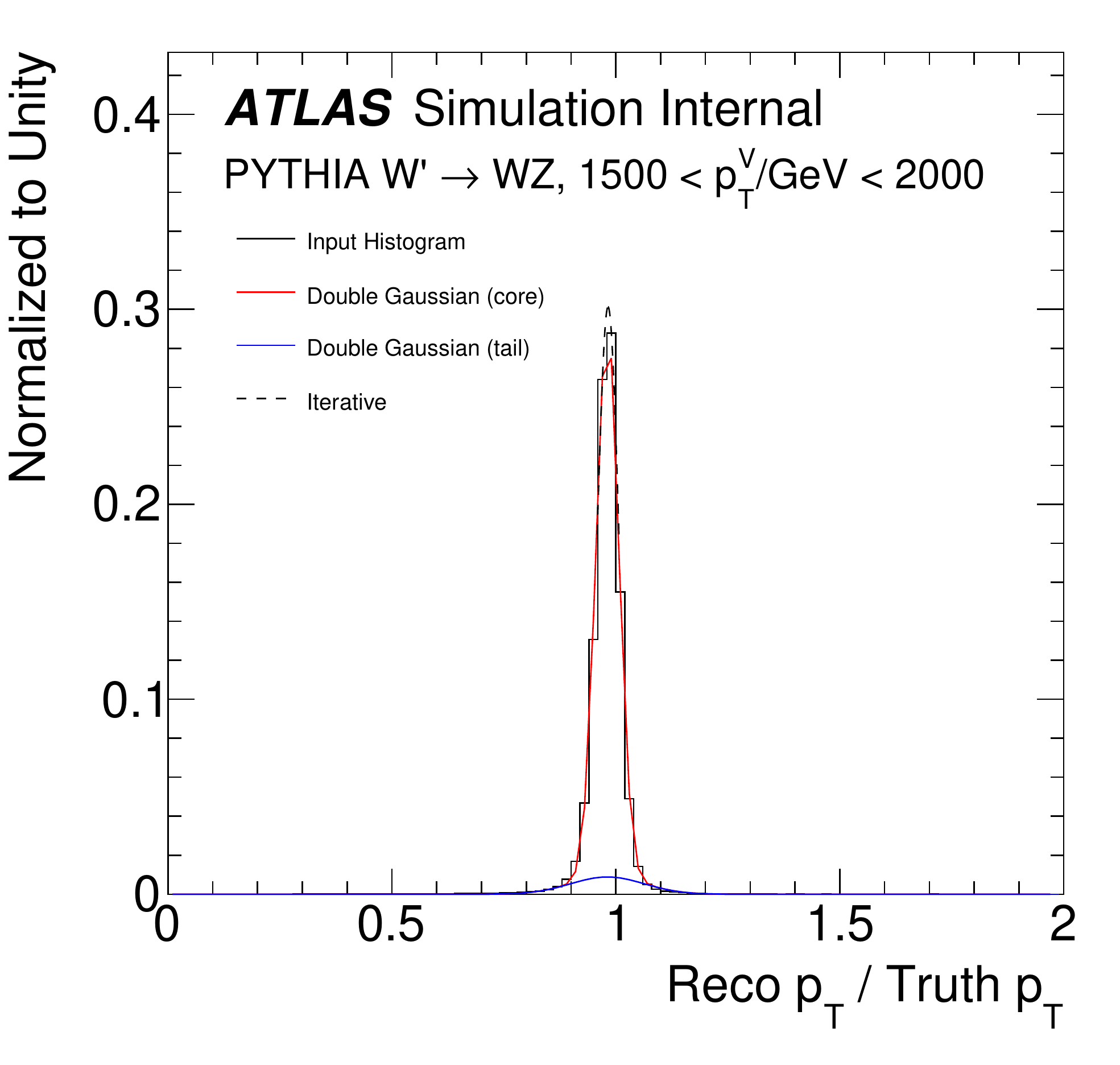}
\caption{The distribution of the jet $p_\text{T}$ response $(R)$ for boosted hadronically decaying $W$ and $Z$ bosons in four boson $p_\text{T}$ ranges from 200 GeV up to 2 TeV.  See the text for a description of the various fitting methods.}
\label{fig:JMR:jetpTresponse}
\end{figure}

\clearpage

\begin{figure}[h!]
 \centering
\includegraphics[width=0.5\textwidth]{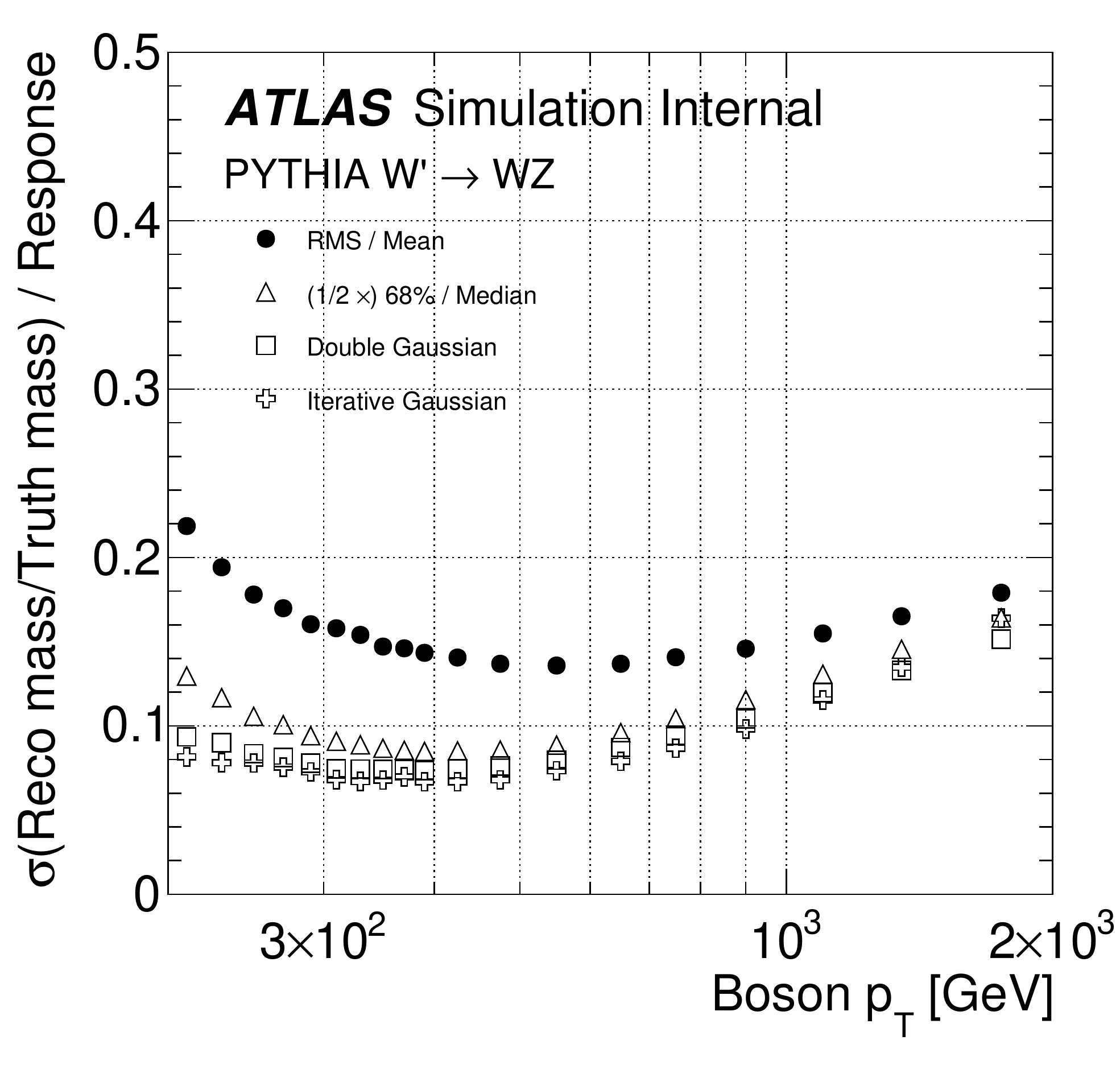}\includegraphics[width=0.5\textwidth]{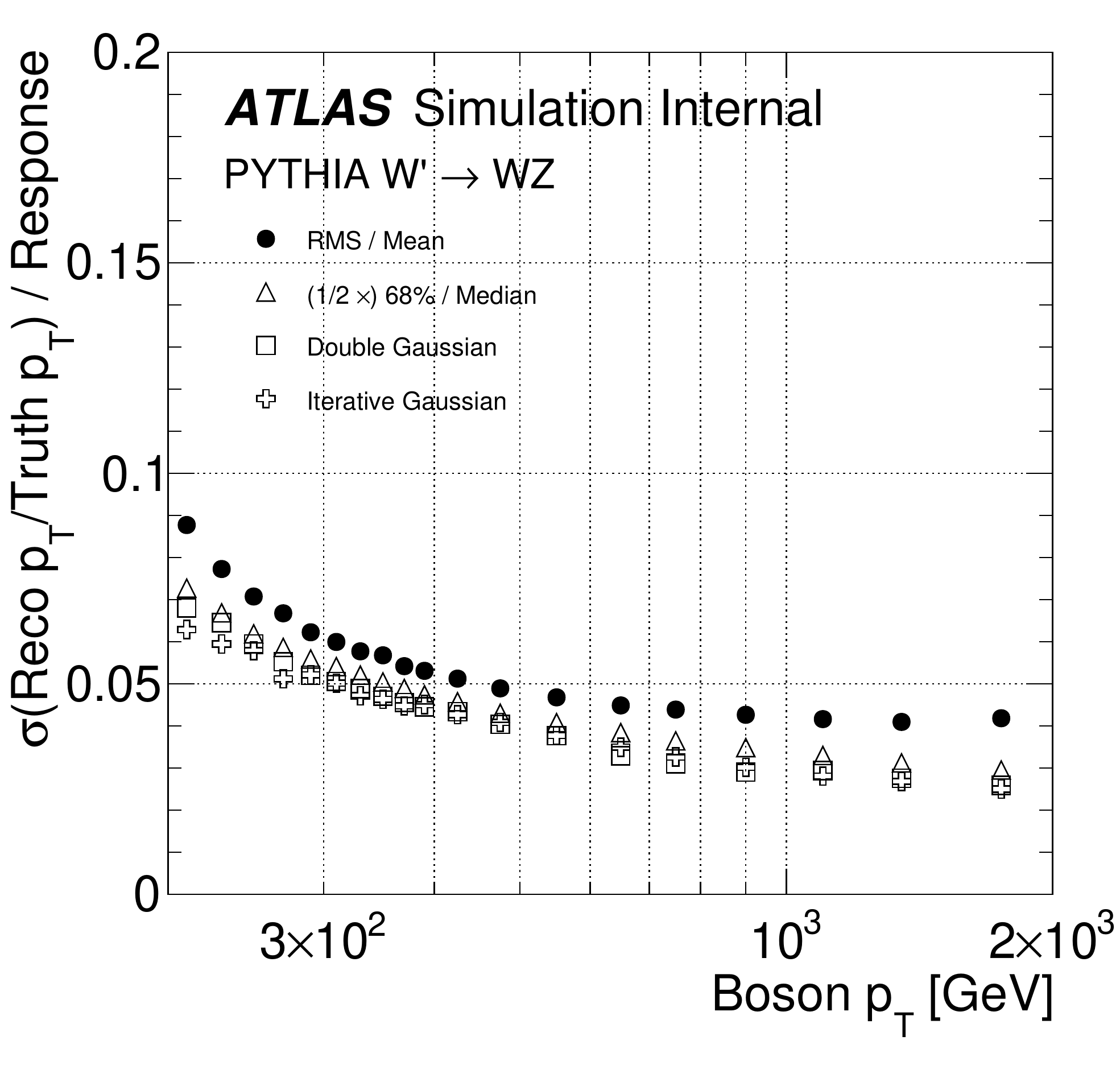}
\caption{A summary of the $p_\text{T}$-dependence of the jet mass (left) and jet $p_\text{T}$ (right) resolution. See the text for a description of the various fitting methods. }
\label{fig:JMR:response}
\end{figure}

As expected from a calorimeter measurement, the jet $p_\text{T}$ resolution monotonically decreases with $p_\text{T}$.  At low $p_\text{T}$, this is also true for the jet mass resolution, but there is a turning point at about 500 GeV where the jet mass resolution degrades with $p_\text{T}$.  This is because for a fixed mass, the particles inside a jet get closer together with increasing $p_\text{T}$.  Due to the finite granularity of the detector, small angular separations cannot be resolved which reduces the jet mass scale and degrades the jet mass resolution.  When the subjets inside the large-radius jet are well-separated (at low $p_\text{T}$), the mass and $p_\text{T}$ resolutions are similar in magnitude because the mass resolution is mostly due to the energy resolution of the isolated subjets.

With an optimized definition of the jet mass and a calibration to remove most of the detector response, the key challenge is to determine the closure of the four-vector calibration and the jet mass and $p_\text{T}$ resolutions in data.  Section~\ref{sec:JMR:trackjet} introduces the most widely used technique for determining the closure of the large-radius jet $p_\text{T}$ and jet mass calibration in data - the track-jet method.  After a brief introduction to a bottom-up method in Sec.~\ref{sec:JMR:bottomup}, the remainder of this section (Sec.~\ref{sec:JMR:resmethod}) focuses on a new technique based on fitting resonance peaks.

\clearpage

\subsubsection{Track-jet Method}
\label{sec:JMR:trackjet}

The baseline method for measuring the closure of the calibration in data uses track jets.  Tracks are clustered into jets using the same algorithm as for the calorimeter jets.  These large-radius track jets are geometrically matched to calorimeter jets and their jet mass provides an independent measurement of the particle-level jet mass.  Track-jets are particularly useful because the typical difference between the reconstructed track jet mass and the jet mass from the particle-level jet using only charged particles is small compared to the calorimeter jet mass resolution.  However, the resolution of the track-jet mass with respect to the full particle-level jet mass is not small compared to the calorimeter jet mass resolution due to the large fluctuations in the charge-to-neutral ratio of particles inside the jet.  Therefore, it is not possible to perform a measurement of the absolute closure of the jet mass calibration using track jets.  Instead, the closure in data is studied relative to the closure in simulation.  Define $r_\text{track}$ as the ratio of the calorimeter jet mass to the matched track jet mass.  Then, 

\begin{align}
\label{eq:rtrack:formula}
r_\text{track} = \frac{m^\text{calorimeter}}{m^\text{particle}}\times \frac{m^\text{particle}}{m^\text{charged-particle}}\times \frac{m^\text{charged-particle}}{m^\text{track}},
\end{align}

\noindent where the first term is the jet mass response $(R)$, the second term is the inverse of the charged ratio of the jet $(f_Q^{-1})$ and the last term is the inverse of the track-jet resolution with respect to the particle-level jet mass using only charged particles $(R_\text{T}^{-1})$.  The statement that track jets are well measured means that $\sigma(R_\text{T})\ll \sigma(R)$.  Due to isospin, $\langle f_Q\rangle\sim 2/3$.  However, $\langle f_Q^{-1}\rangle > 3/2$, as shown by the left plot of Fig.~\ref{fig:JMR:chargedfraction}.  The right plot of Fig.~\ref{fig:JMR:chargedfraction} shows that $\langle f_Q\rangle $ is nearly independent of $p_\text{T}$, a fact that was used in Sec.~\ref{sec:jetcharge:tracksyst:TIDE} to determine tracking uncertainties for the jet charge. Interestingly, there is a slight difference between the the ratio based on the mass and the one based on $p_\text{T}$ due to subtle differences in jet fragmentation to charged and neutral particles.  The standard deviation of the $f_Q$ distribution is also nearly independent of $p_\text{T}$ and is approximately 0.2 which corresponds to $\langle f_Q^{-1}\rangle\sim 2$ based on the left plot of Fig.~\ref{fig:JMR:chargedfraction}.  The value $\sigma(f_Q)$ is smaller than one would expect if all particles carry an equal fraction of the jet's energy (see Fig.~\ref{fig:JMR:chargedfraction}), but is not negligibe compared to $\langle f_Q\rangle$.

\begin{figure}[h!]
 \centering
\includegraphics[width=0.45\textwidth]{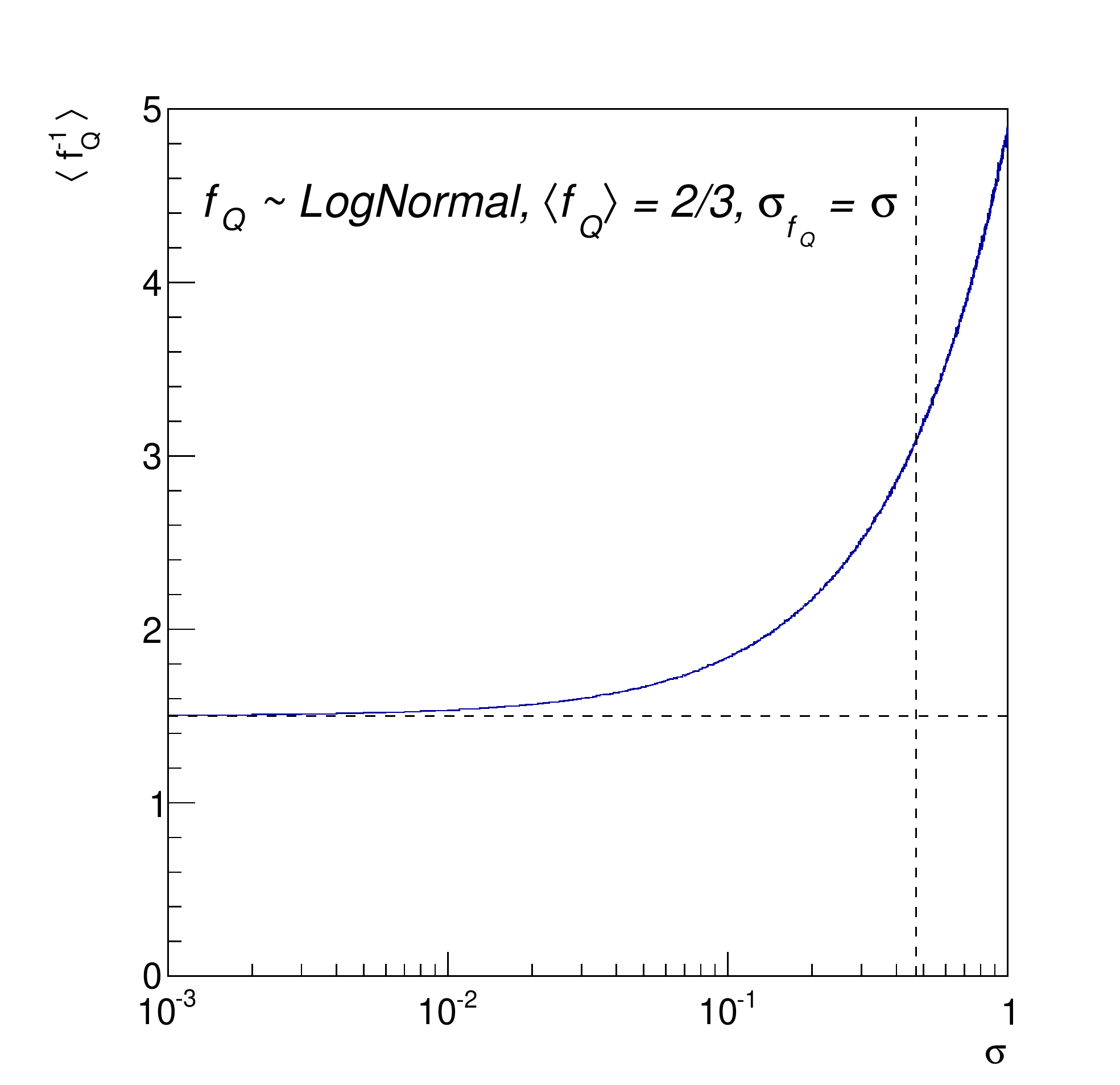}\includegraphics[width=0.45\textwidth]{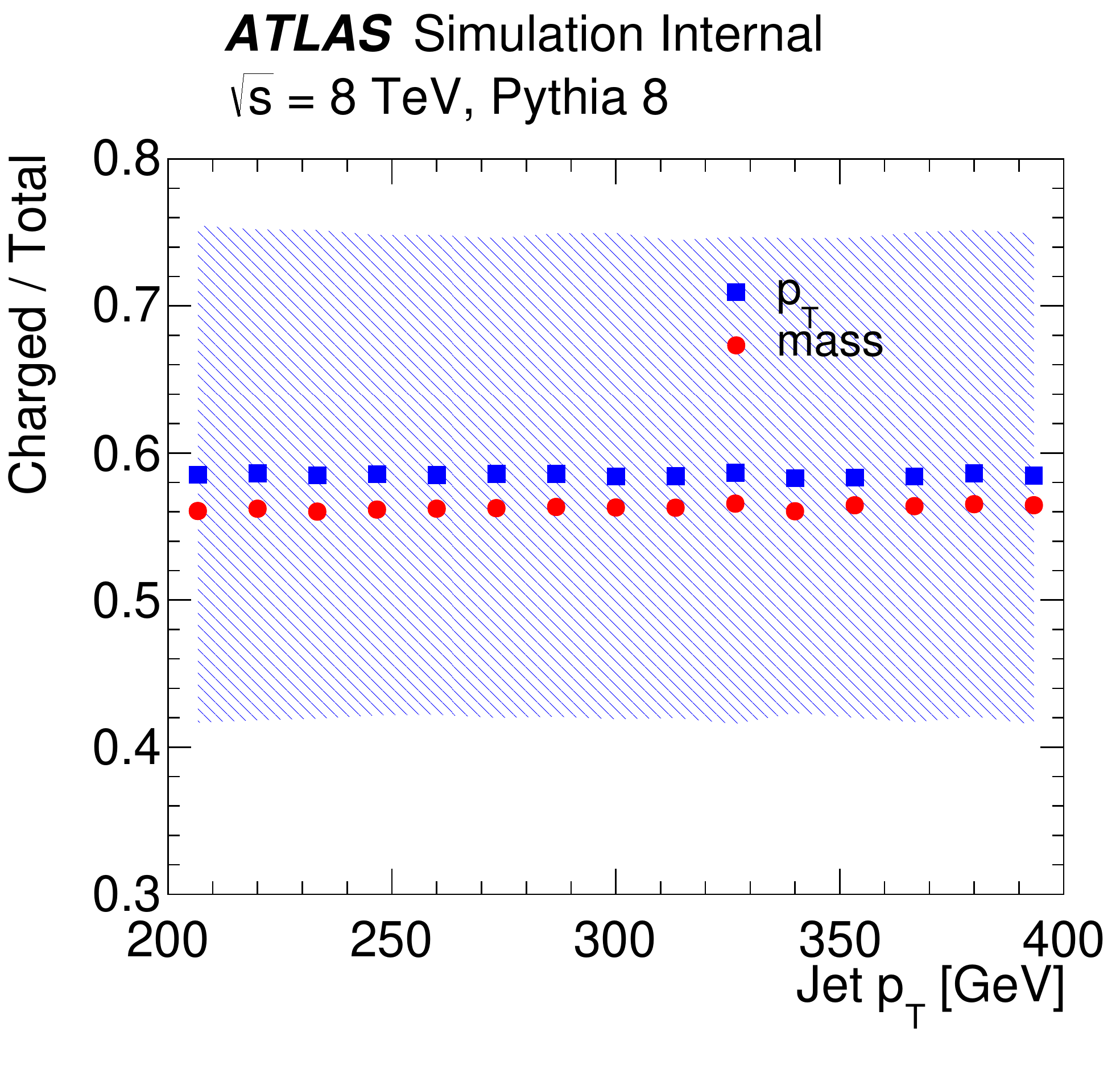}
\caption{Left: The dependence of $\langle f_Q^{-1}\rangle$ on the standard deviation of the charged-to-neutral fluctuations.  The charged fraction $f_Q$ is modeled as a log normal distribution with mean $2/3$ and variable standard deviation $\sigma$.  The horizontal dashed line is at 2/3.  The vertical dashed line is at $\sigma= 0.47\approx\sqrt{\frac{2}{3}(1-\frac{2}{3})}$, which is what one expects if each particle inside a jet carries an equal fraction of the jet energy and has a probability 2/3 of being charged.  Right: the $p_\text{T}$ dependence of $\langle f_Q\rangle $ for both jet mass and jet $p_\text{T}$.  The band is the standard deviation of the $f_Q$ distribution using $p_\text{T}$.}
\label{fig:JMR:chargedfraction}
\end{figure}

When the terms on the righthand side of Eq.~\ref{eq:rtrack:formula} are independent, $\langle r_\text{track}\rangle\propto \langle R\rangle$.  In this case, an estimate for the relative uncertainty in the non-closure is given by

\begin{align}
\vspace{-3mm}
1-\frac{\langle R^\text{data}\rangle}{\langle R^\text{MC}\rangle}=1-\frac{k^\text{MC}}{k^\text{data}}\times\frac{\langle r_\text{track}^\text{data}\rangle}{\langle r_\text{track}^\text{MC}\rangle},
\end{align}

\noindent where $k$ is the constant of proportionality between $\langle r_\text{track}\rangle$ and $\langle R\rangle$.  This measurement is limited by data statistics at high jet $p_\text{T}$ and elsewhere by uncertainties in $k^\text{MC}/k^\text{data}$ due to systematic uncertainties on the reconstruction of charged particle tracks and the modeling of jet fragmentation. 

In principle, the track-jet method can also be used to determine the relative jet mass resolution, $\sigma(R)^\text{data}/\sigma(R)^\text{MC}$.  Further assuming that $R^2$ and $f_Q^{-2}\times R_\text{T}^{-2}$ are independent, one can write $\langle r_\text{track}\rangle=k_1\langle R\rangle$ and $\langle r_\text{track}^2\rangle = k_2\langle R^2\rangle$ for some constants $k_1,k_2$ that are in general different between data and simulation.  Then,

\begin{align}
\sigma^2(R) = \frac{\langle r_\text{track}^2\rangle}{k_2}-\frac{\langle r_\text{track}\rangle^2}{k_1^2}.
\end{align}

\noindent This procedure is not applied in practice because it depends explicitly on the value of $k_i$, whereas for $\langle R\rangle$, only the relative $k^\text{MC}/k^\text{data}$ are required.

Figures~\ref{fig:JMR:rtrackdataQCD} and~\ref{fig:JMR:rtrackdataW} show the distribution of $r_\text{track}$ and $\langle r_\text{track}\rangle(p_\text{T})$ for generic quark and gluon jets and $W$ boson-like jets in both data and simulation.  For both sets of jets, the $r_\text{track}$ distribution peaks just below 2 and the predicted $\langle r_\text{track}\rangle$ is within $\lesssim 5\%$ of the data.  The populations of jets in Fig.~\ref{fig:JMR:rtrackdataQCD} and~\ref{fig:JMR:rtrackdataW} are a representative set for applications of jet mass.  Generic QCD jets are used to calibrate jets and set the most precise uncertainties using $r_\text{track}$ because of their abundance.  The jet mass is mostly used for tagging boosted hadronically decaying bosons and top quarks.  There are no indications for significant biases in this procedure based on the $r_\text{track}$ method.  The total uncertainty from the track-jet method is about 5\%, independent of jet $p_\text{T}$~\cite{ATLAS-CONF-2015-037}.

\begin{figure}[h!]
 \centering
\includegraphics[width=0.4\textwidth]{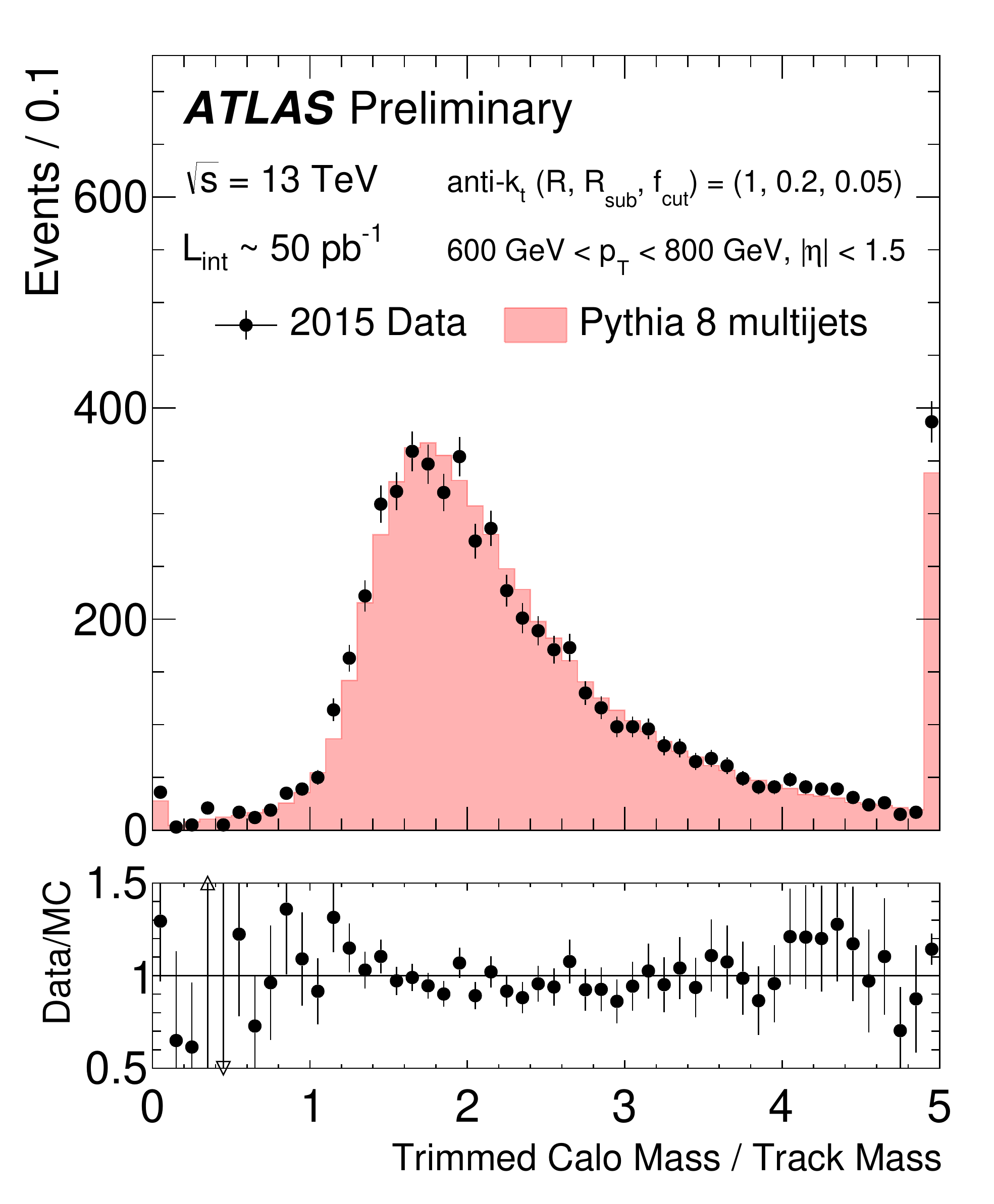}\includegraphics[width=0.4\textwidth]{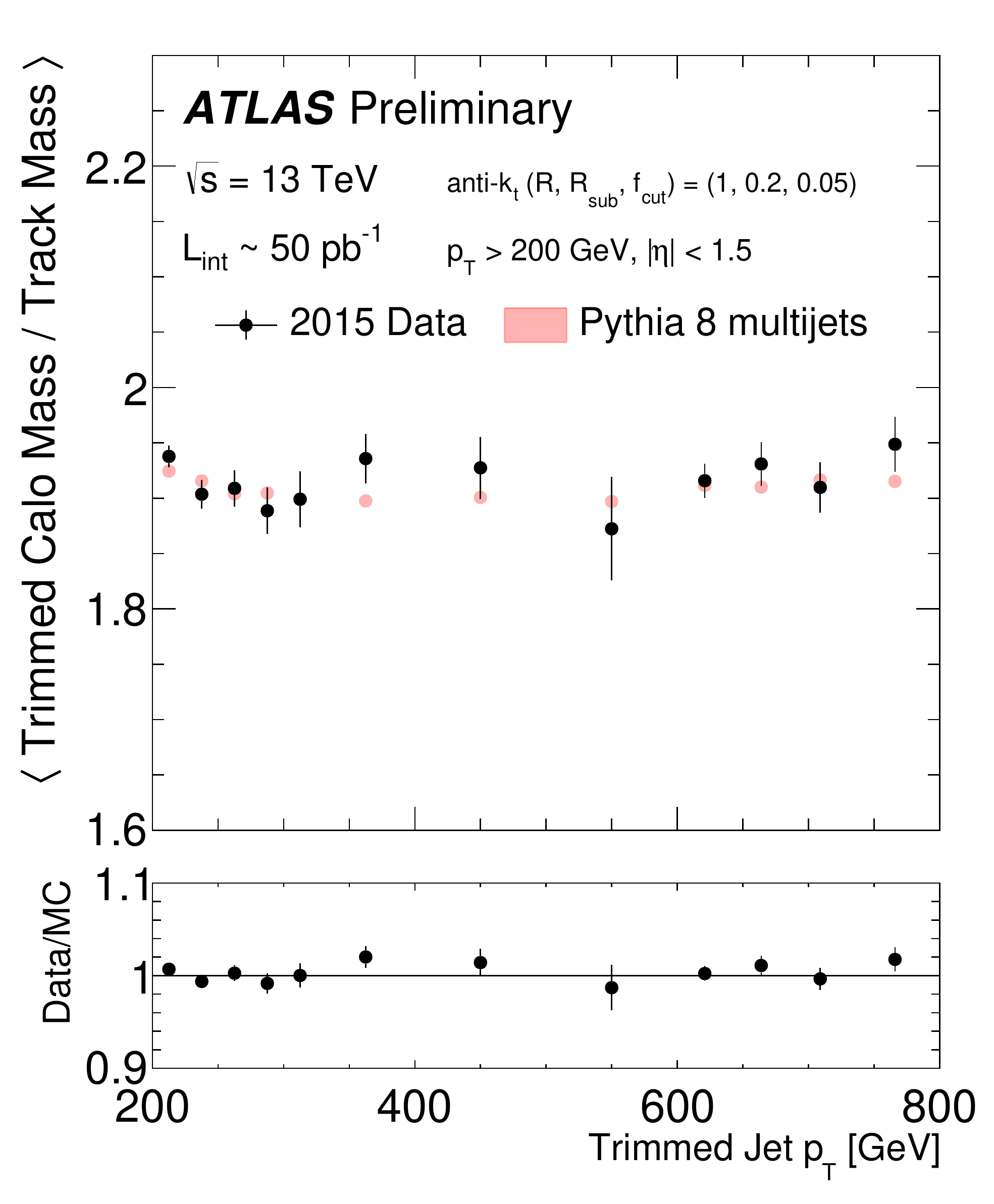}
\caption{The (a) ratio of the calorimeter jet mass to track jet mass and (b) average value of this ratio as a function of the calorimeter jet mass for leading anti-$k_t$ $R=1.0$ trimmed with $f_\mathrm{cut}=0.05$ and $R_\mathrm{sub}=0.2$ jets.  Calorimeter jets which contain only a single cluster have a mass of zero.  MC is normalized to the number of events observed in data.  The last bin includes overflow events.}
\label{fig:JMR:rtrackdataQCD}
\end{figure}

\begin{figure}[h!]
 \centering
\includegraphics[width=0.55\textwidth]{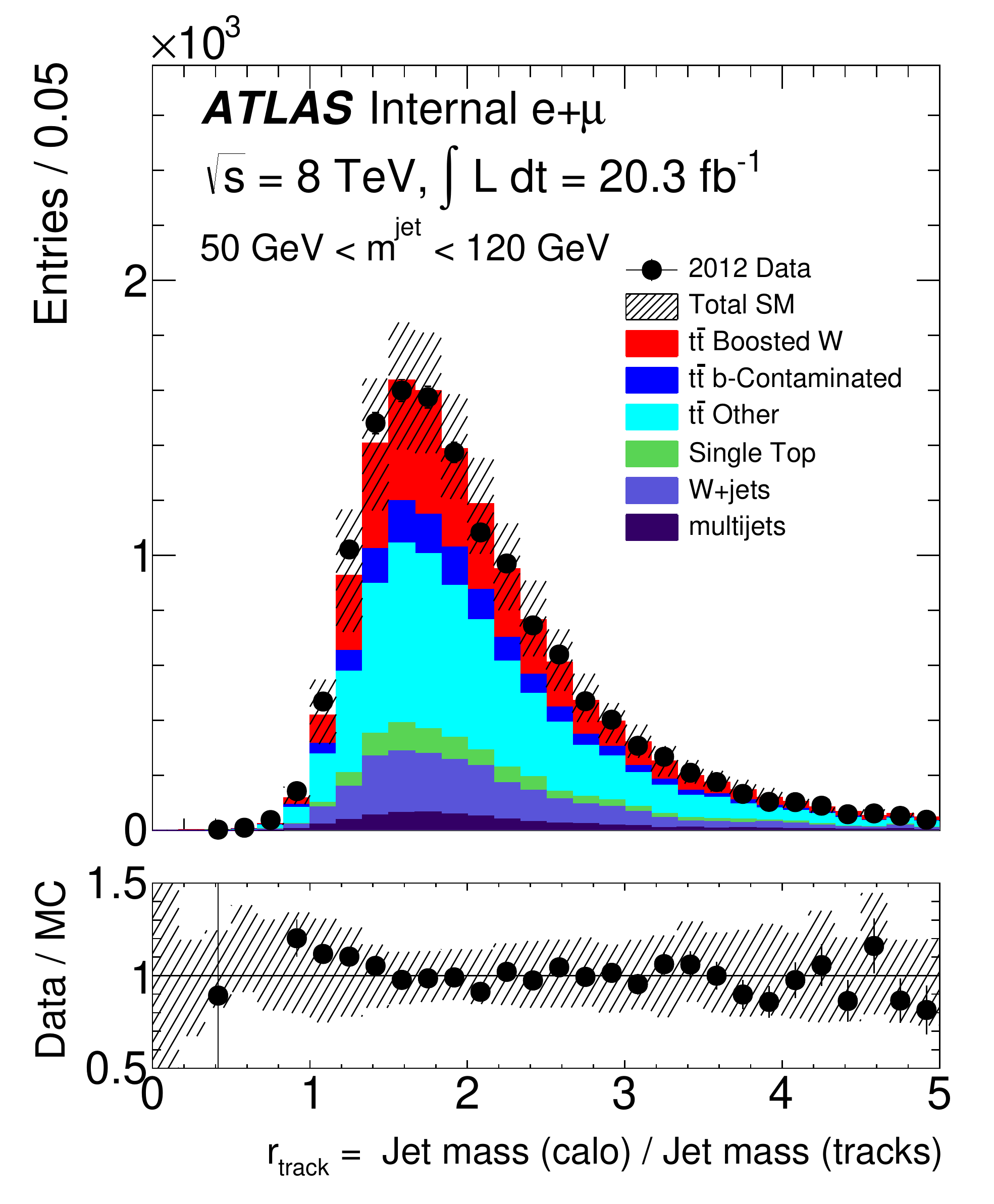}\\
\includegraphics[width=0.45\textwidth]{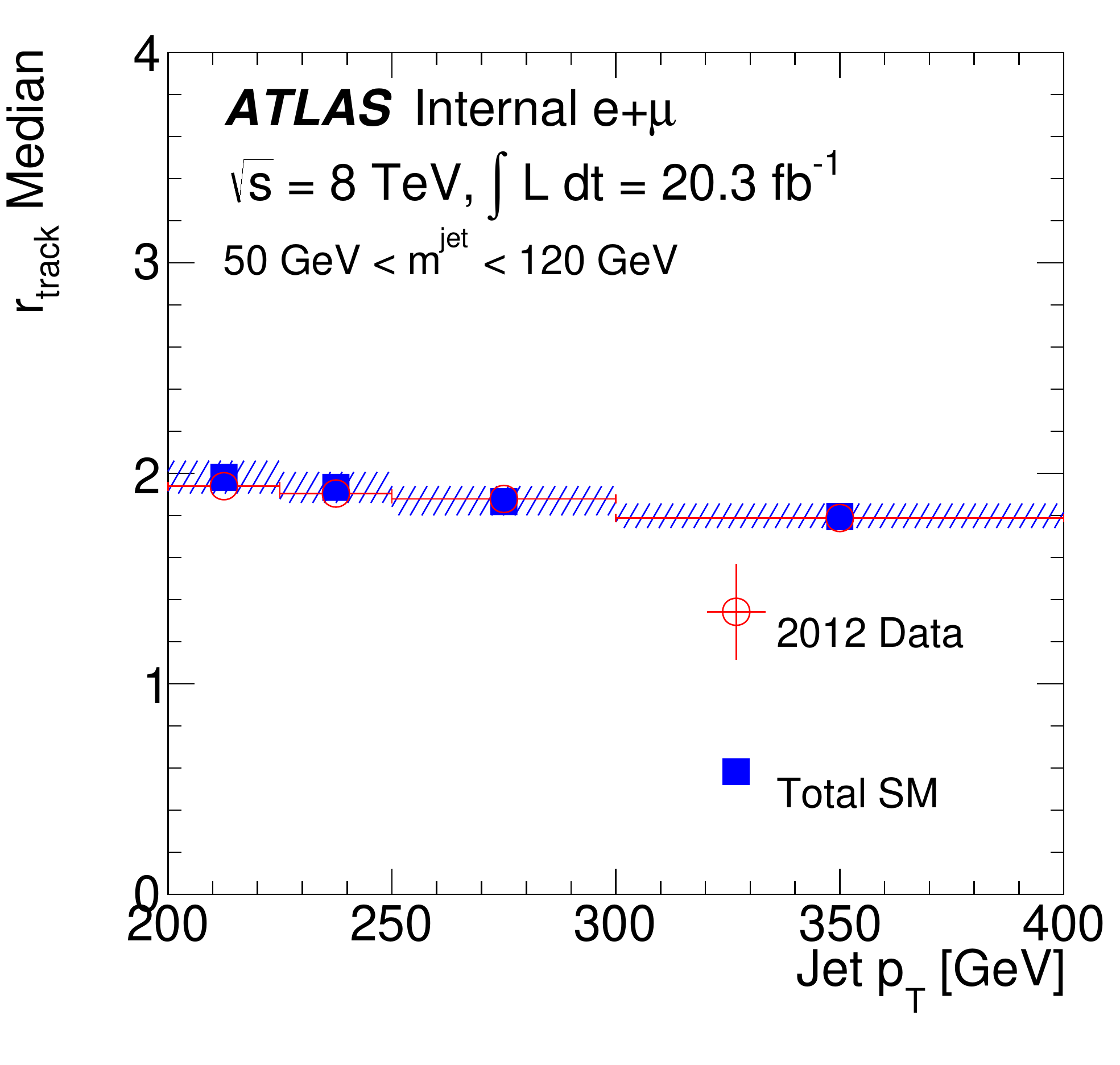}\includegraphics[width=0.45\textwidth]{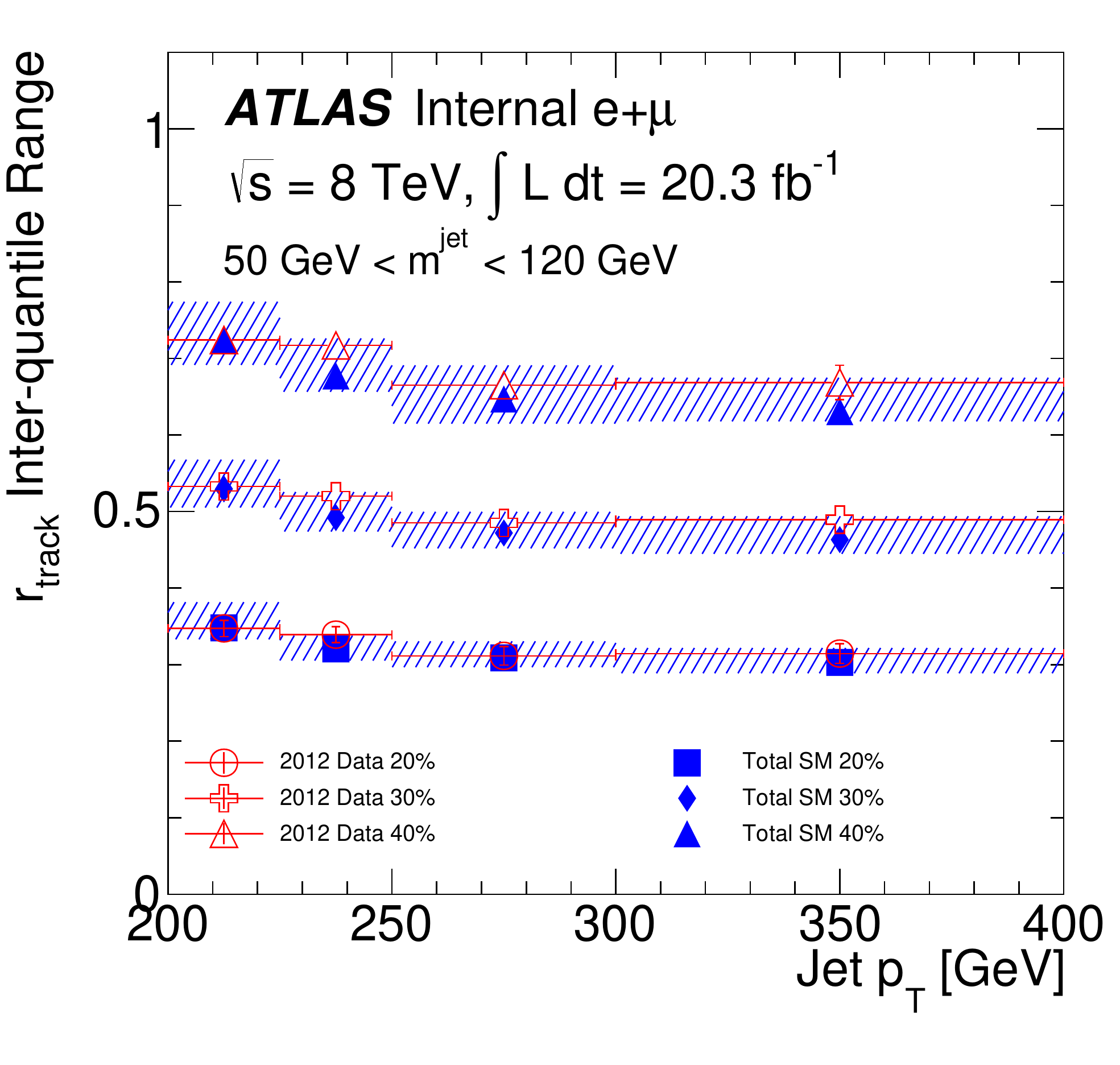}
\caption{(a) The distribution of $r_\text{track}$ in the data for semi-leptonic $t\bar{t}$ events with the selected jet in the range $50$~GeV $<m^\text{jet}<$ $120$~GeV.  (b) The median of the $r_\text{track}$ distribution as a function of the jet $p_\text{T}$.   (c) The inter-quantile range as a measure of the width.  See Sec.~\ref{sec:bosontypetagger} for a description of the event selection.}
\label{fig:JMR:rtrackdataW}
\end{figure}

\clearpage

\paragraph{Interpreting $r_\text{track}$ Uncertainties}\mbox{}\\

While the track-jet method is simple to apply, there is an important caveat when interpreting the results. In general, $f_Q$ and $R$ are {\bf not} independent.  The calorimeter response is different between charged and neutral particles.  The LCW corrects the difference on average, but the classification of individual clusters as EM or hadronic has a non-zero error and the finite calorimeter energy resolution can be non-negligible.   Figure~\ref{fig:JMR:correlation} shows the joint distribution of $R$ and $f_{Q}^{-1}\times R_\text{T}$.  The linear correlation between these two variables is small, but a small correlation is not sufficient for the average of two random variables $X$ and $Y$ to factorize: $\langle XY\rangle=\langle X\rangle\langle Y\rangle$.  

\vspace{5mm}

\begin{figure}[h!]
 \centering
\includegraphics[width=0.55\textwidth]{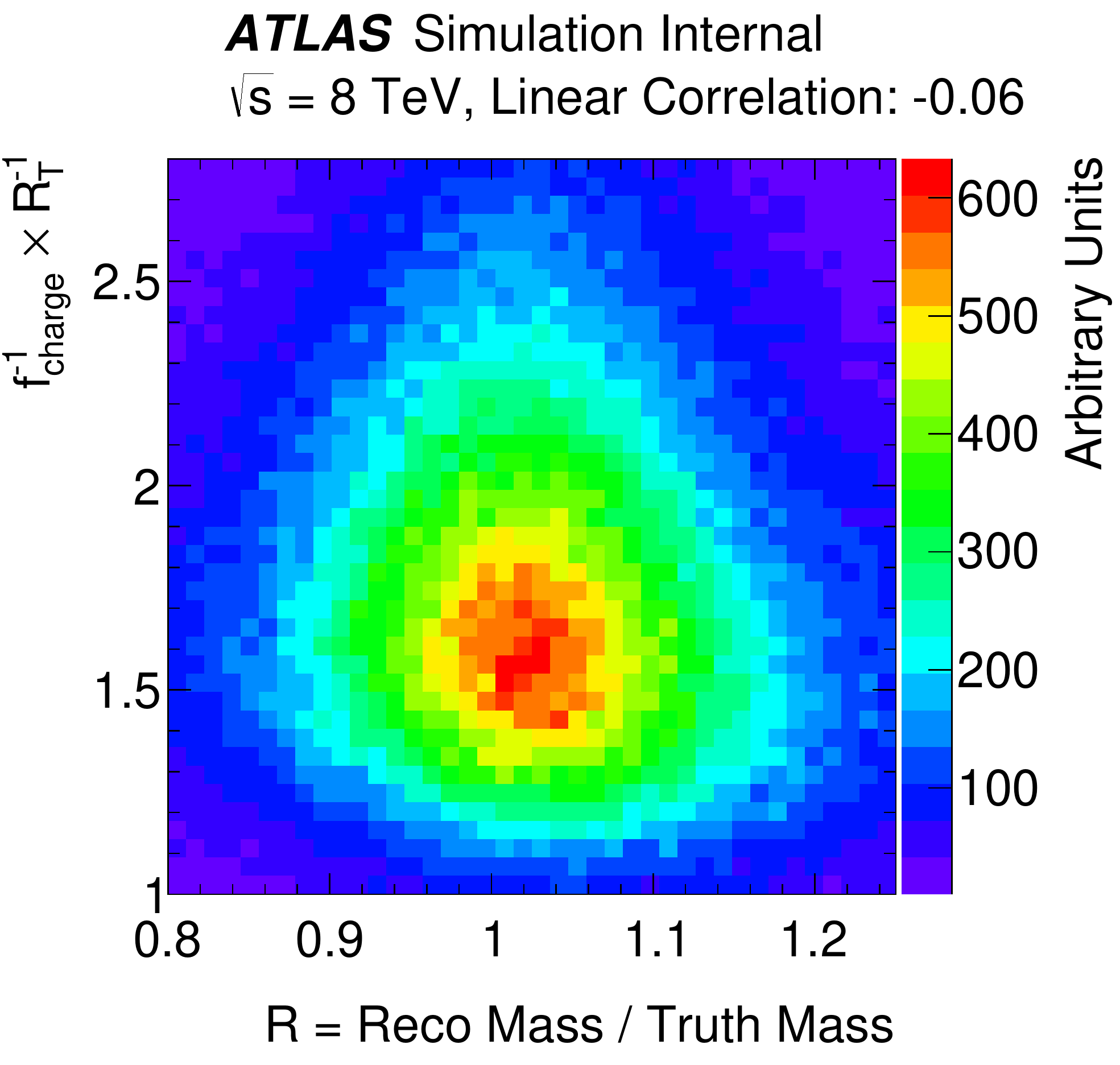}
\caption{The joint distribution of the response $(R)$ and the product of the inverse charged ratio $f_Q^{-1}$ and the inverse track mass response $R_\text{T}^{-1}$ using simulated $W$ boson jets with $p_\text{T}^\text{truth}>200$ GeV from {\sc Powheg-box}+{\sc Pythia} 6 $t\bar{t}$ production.}
\label{fig:JMR:correlation}
\end{figure}

\clearpage

Figure~\ref{fig:JMR:closureofrtrak} shows the gap between $\langle R\rangle \langle f_Q^{-1}R_\text{T}^{-1}\rangle$ and $\langle r_\text{track}\rangle=\langle R f_Q^{-1}R_\text{T}^{-1}\rangle$.  Defining $c$ to be the size of the gap, one can try to estimate if the $r_\text{track}$-based uncertainties are at least conservative, i.e. is $|\sigma| < |\Delta|$ for 

\begin{align}
\sigma = 1-\frac{\langle R^\text{data}\rangle}{\langle R^\text{MC}\rangle} \hspace{5mm}\text{and}\hspace{5mm}\Delta = 1-\frac{\langle r_\text{track}^\text{data}\rangle}{\langle  r_\text{track}^\text{MC}\rangle}.
\end{align}

\noindent In other words, $\sigma$ is the `true' uncertainty and $\Delta$ is the uncertainty one estimates using the track-jet method.  Assuming that the difference in $c$ and $\langle f_{Q}^{-1}R_\text{T}^{-1}\rangle$ between data and simulation is small compared to $\epsilon = c/\langle r_\text{track}^\text{MC}\rangle$, one can compute $\sigma = \Delta(1-\epsilon)+\mathcal{O}(\epsilon^2)$.  Since $\epsilon > 0$, $|\sigma| < |\Delta|$, as desired.  If $\Delta\sim 5\%$ and $\epsilon\sim 30\%$, then the size of the bias could be 1-2\%.  This bias is currently not accounted for when applying the track-jet method, which can be justified if a precision uncertainty is not the goal.

\vspace{5mm}

\begin{figure}[h!]
 \centering
\includegraphics[width=0.6\textwidth]{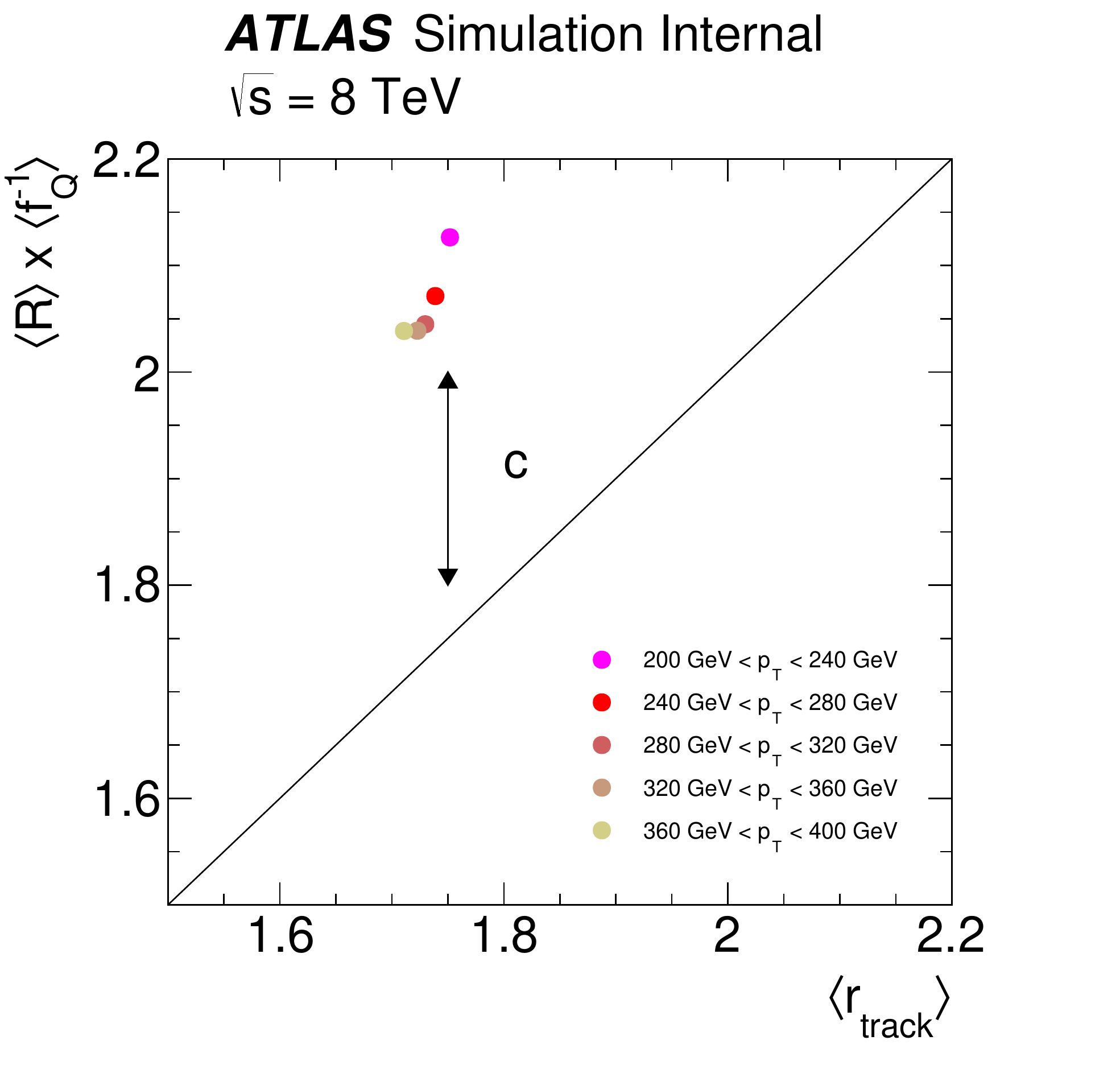}\caption{Left: The gap between $\langle R\rangle \langle f_Q^{-1}R_\text{T}^{-1}\rangle$ and $\langle r_\text{track}\rangle=\langle R f_Q^{-1}R_\text{T}^{-1}\rangle$. }
\label{fig:JMR:closureofrtrak}
\end{figure}

\clearpage

\subsubsection{Bottom-up Method}
\label{sec:JMR:bottomup}

Up to\footnote{This section includes technical input from Z. Marshall.} $p=350$ GeV, the response of individual particles is well-measured using isolated tracks and test-beam experiments.  The idea of the bottom-up method is to model the (average) jet response as the sum of the average jet response for all the constituent particles inside a jet.  Symbolically,

\begin{align}
\label{bottomup}
p^\mu_\text{jet,reco} = \sum_{i\in\text{particle-level jet}} \langle R_i\rangle p_i^\mu,
\end{align}

\noindent where $p_i^\mu$ is the four-vector and $\langle R_i\rangle$ is the average calorimeter energy response of particle $i$.  From $p^\mu_\text{jet,reco} $, one can compute the jet $p_\text{T}$ or jet mass response for a given jet by diving by the appropriate property of the particle-level jet.  Figure~\ref{fig:JMR:bottomup} shows the average $p_\text{T}$ and mass response using this bottom-up approach for boosted hadronically decaying $W$ bosons.  Only particles with energy above $500$ MeV are included.  For electrons and photons, $\langle R_i\rangle =1$ and since muons do not usually deposit significant energy in the calorimeter, $\langle R_i\rangle=0$.  The ratio of charged-particle calorimeter energy to track $p_\text{T}$ (E/p) is used for charged hadrons up to $p_\text{T}=20$ GeV, after which test-beam data~\cite{Abat:2010zza} is used up until $p_\text{T}=350$ GeV.  The test-beam response ranges from $0.65$ at $E<35$ to $0.78$ for $p_\text{T}>125$ GeV (for central $\eta$).   For $p_\text{T}>350$ GeV, $\langle R_i\rangle$ is not constrained by data; in Fig.~\ref{fig:JMR:bottomup} it is set to the highest value from the test-beam: $0.78$.   Protons and pions with $E<10$ GeV can be identified using the amount of energy deposited as a function of distance traversed in the detector ($dE/dx$) and so their response values are individually computed while all other charged hadrons use generic values.  For nearly collinear constituents (and mostly uncorrelated energy fluctuations), Eq.~\ref{bottomup} is a good approximation for the jet $p_\text{T}$ and therefore the bottom-up $p_\text{T}$ response in Fig.~\ref{fig:JMR:bottomup} well-models the full response.  However, jet mass is the result of significant angular splittings and so Eq.~\ref{bottomup} is not a good approximation.  It is therefore not surprising that the two models diverge at high $p_\text{T}$ in the right plot of Fig.~\ref{fig:JMR:bottomup}.

\begin{figure}[h!]
 \centering
\includegraphics[width=0.99\textwidth]{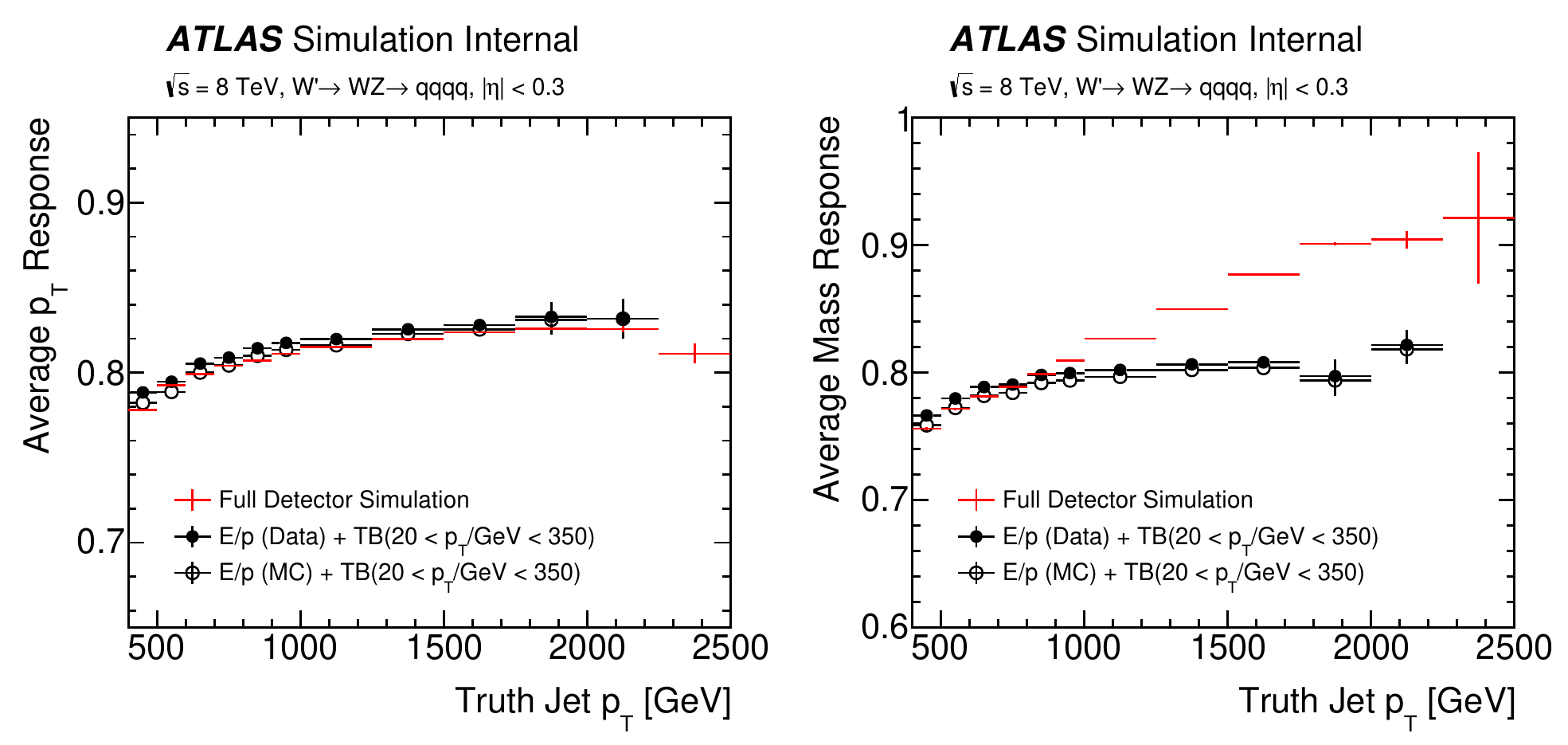}
\caption{The average jet $p_\text{T}$ (left) and mass (right) response as a function of $p_\text{T}$ using simulated detector-level jets and propagating single particle responses via Eq.~\ref{bottomup}.  The points marked data use $E/p$ as measured with the ATLAS detector.  Error bars on the points indicate the MC statistical uncertainty on the mean.}
\label{fig:JMR:bottomup}
\end{figure}

The jet $p_\text{T}$ and mass scale uncertainty in the bottom-up approach arrises due to uncertainty in the values of $\langle R_i\rangle$.  This approach may be a useful technique for extending the jet mass and $p_\text{T}$ scale (uncertainty) to high values beyond the capability of in-situ studies, though there seems to be some challenges for $p_\text{T}\gtrsim 1$ TeV.  Furthermore, it can (approximately) describe some aspects of correlations between variables. However, it is not able to fully describe correlations and does not model fluctuations about the mean.

	\clearpage
	
	\subsubsection{Resonance Method}
	\label{sec:JMR:resmethod}

	Known resonance decays provide a standard reference for in-situ calibration and uncertainty studies.  For example, $Z$ boson, $J/\psi$, and $\Upsilon$ decays are used to measure the scale and resolution of the response function for muons~\cite{Aad:2014rra}, electrons and photons~\cite{Aad:2014nim}, and tau leptons~\cite{Aad:2014rga}.  However, these techniques are not directly applicable to hadronic resonance decays because the parton shower and jet clustering introduce a non-trivial distortion of the resonance's Breit-Wigner mass line-shape.  This is illustrated by the difference between the black and red curves in the left plot of Fig.~\ref{fig:JMR:particlelevel}.  The particle-level distribution depends on pertubative properties of the parton shower as well as non-perturbative effects such as hadronization and the underlying event.  A measurement of the resonance peak will probe the convolution of these particle-level effects and the detector response.  An extraction of the jet mass scale and resolution from a hadronic resonance requires the particle-level spectrum as input and therefore the precision can be limited by the corresponding modeling uncertainties.   The right plot of Fig.~\ref{fig:JMR:particlelevel} shows the impact of varying $\alpha_s$ in the parton shower on the particle-level jet mass spectrum - this uncertainty directly limits the precision of the resonance method.   It is not possible to obtain a pure sample of hadronically decaying $Z$ bosons at a hadron collider.  However, it is possible to select events enriched in hadronic $W$ boson decays from $t\bar{t}$ events where the second $W$ boson is used to tag the event through its leptonic decay.   Low $p_\text{T}$ hadronic $W$ boson decays have been used as a validation of the light quark jet energy scale in early Run 1~\cite{Aad:2014bia}.  The precision of this measurement was limited by the modeling of the parton shower.  
	
	The measurement presented here\footnote{The $\sqrt{s}=8$ TeV analysis presented in this section has been published in Ref.~\cite{ATLAS-CONF-2016-008} and includes technical inputs from J. Veatch.} is the first full estimate of the jet mass scale and resolution on the jet mass directly from boosted $W$ boson jets.   After a brief description of the event selection and simulation in Sec.~\ref{sec:samples}, Sec.~\ref{sec:fitting} describes a new technique for extracing the jet mass scale and resolution from resonance decays called the {\it forward-folding} method.  The impact of systematic uncertainties in the measurement are given in Sec.~\ref{sec:systsforward} and the Run 1 result is summarized in Sec.~\ref{sec:results}.  Finally, Sec.~\ref{sec:JMR13TeV} contains improvements and extensions of the methods as well as results with the early Run 2 data.
	
\begin{figure}[h!]
 \centering
\includegraphics[width=0.45\textwidth]{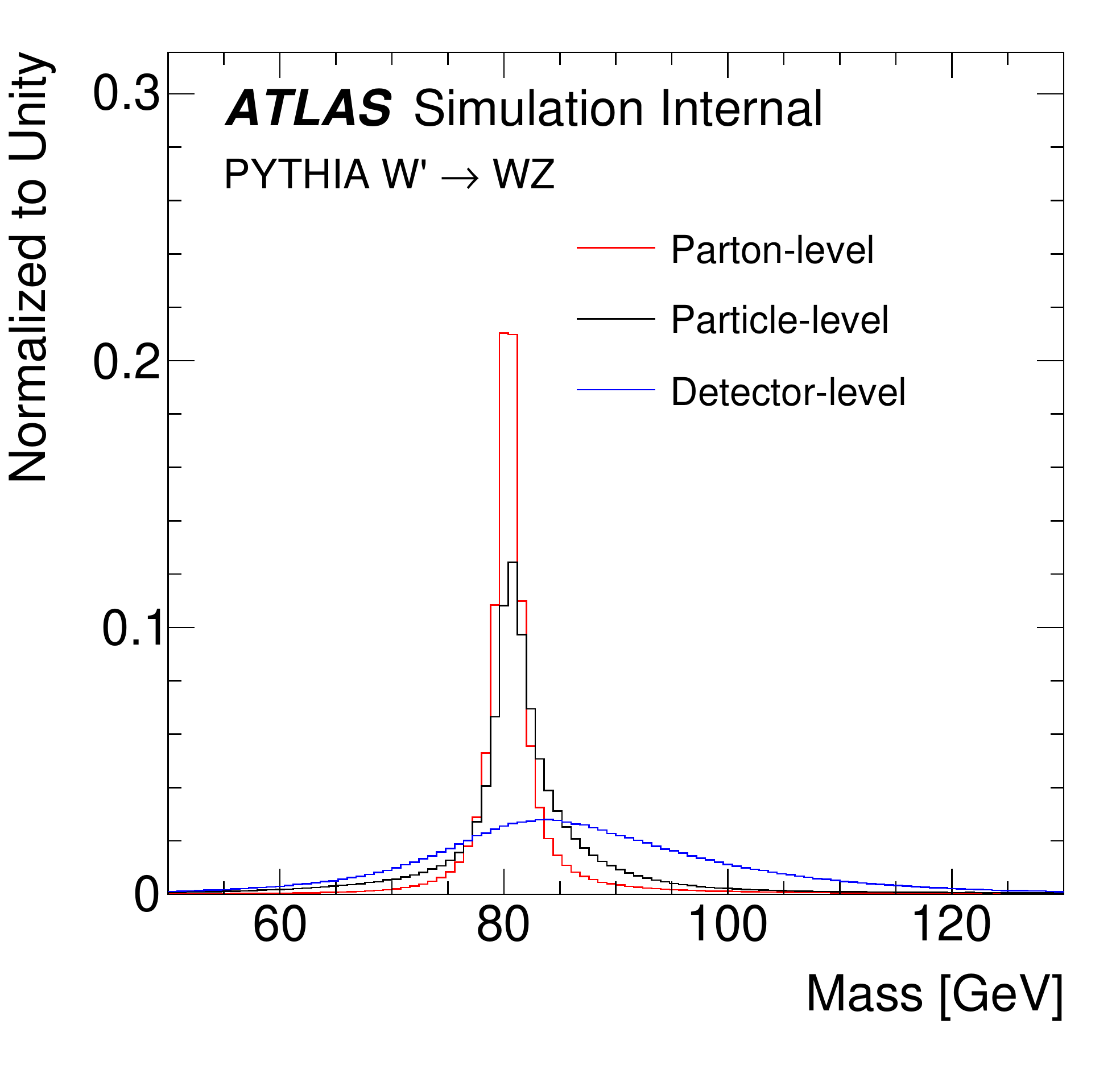}\includegraphics[width=0.45\textwidth]{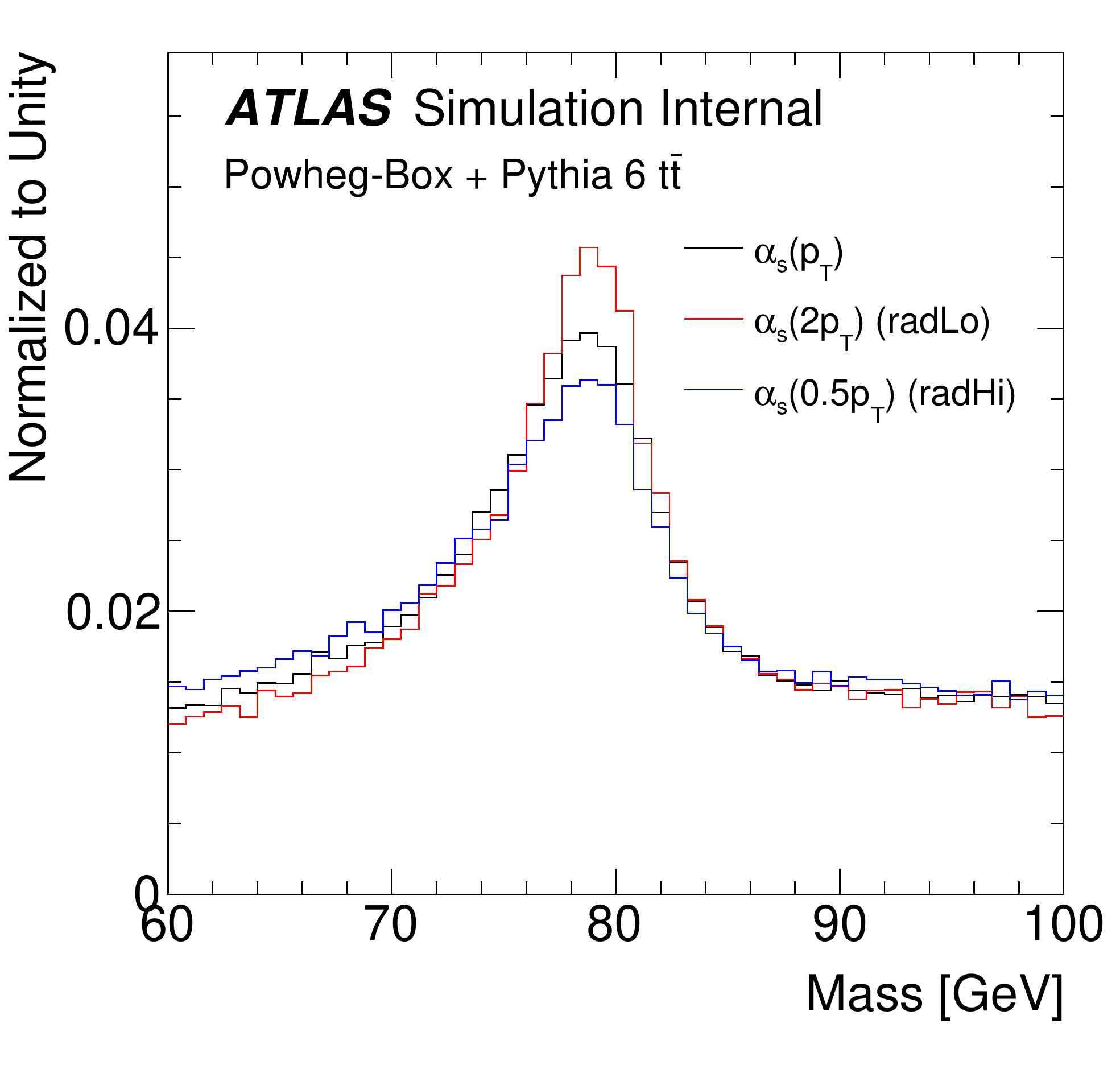}
\caption{Left: the mass distribution of the $W$ boson (parton-level) and boosted and isolated $W$ boson jets at particle-level and the same jet at detector-level.  The jets are required to have $p_\text{T}>200$ GeV.  See Sec.~\ref{sec:ColorFlow:simulation} for details of the simulation.  Right: the particle-level jet mass spectrum for candidate boosted $W$ boson jets from $t\bar{t}$ events using the default Perugia 2012 {\sc Pythia} 6 tune or the radHi/radLo variations.  The one-loop running of the strong coupling-constant is $\alpha_s\propto1/\text{ln}(Q^2/\Lambda^2)$ with $\Lambda=0.26$ in the default Perugia 2012 tune.  This value is double or halved in the radHi/radLo variations.}
\label{fig:JMR:particlelevel}
\end{figure}

\paragraph{Dataset and Event Reconstruction} \mbox{}\\
\label{sec:samples}

The dataset and MC simulations are the same as those used for the color flow measurement, described in Sec.~\ref{sec:ColorFlow:simulation}.  In particular, the data are collected with single electron and muon triggers from the entire 2012 dataset corresponding to n integrated luminosity of $20.3$ fb${}^{-1}$.  {\sc Powheg-box}+{\sc Pythia} 6 is used for modeling the nominal $t\bar{t}$ sample.  The definitions of reconstructed objects, aside from the addition of large-radius jets, are also the same as for the color flow measurement - see Sec.~\ref{sec:ColorFlowEventSelection}.  The only exception is the isolation of electrons.  Just as the size of $W$ boson and top jets decreases with $p_\text{T}$, the leptons from $W$ decays are closer, on average, to the $b$-jets originating from the same parent top quark with increasing $p_\text{T}$.  A relative isolation based on a shrinking cone is straightforward to apply for muons and recovers the efficiency at high $p_\text{T}$.  A non-trivial complication for electrons is that they deposit most of their energy in the calorimeter which can be clustered with the radiation from the $b$-quark to form a single jet.  Therefore, to recover efficiency at high top quark $p_\text{T}$, the electron energy deposit in the calorimeter is removed from the closest jet with $\Delta R<0.4$ before applying a similar relative isolation procedure.

The event and object selections are based on the ATLAS search for $t\bar{t}$ resonances~\cite{Aad:2015fna} and are summarized here for completeness.  Candidate reconstructed $t\bar{t}$ events are chosen by requiring an electron or a muon with $p_\text{T} > 25$ GeV and $|\eta| < 2.5$, as well as a missing transverse momentum $E_\text{T}^\text{miss} > 20 $ GeV.  Events are rejected if there is not exactly one electron or muon.   In addition, the sum of the $E_\text{T}^\text{miss} $ and the transverse mass of the $W$ boson, reconstructed from the lepton momentum and $\vec{p}_\text{T}^\text{miss} $, is required to be greater than 60 GeV.   Events must have at least one $b$-tagged jet (at the 70\% efficiency working point) and have at least one large-radius trimmed jet with $p_\text{T}>200$~GeV and $|\eta|<2$.  Furthermore, there must be a small-radius jet with $p_\text{T}>25$ GeV, and $\Delta R<1.5$ to the selected lepton (targeting the decay chain $t\rightarrow bW(\rightarrow \ell\nu)$).  The candidate $W$ jet used for the measurement is selected as the leading large-radius trimmed jet with $\Delta\phi > 1.5$ from the lepton $\Delta R >1.2$ from the small-radius jet that is matched to the lepton.  To further ensure that the selected jet contains only the decay products of a $W$ boson, it is required to have $\Delta R > 1.0$ to the nearest $b$-tagged small-radius jet.  The jet mass and jet $p_\text{T}$ distributions after the above event selections are shown in Fig.~\ref{fig:forward0} and~\ref{fig:forward0b}.  There are about 35,000 events in data that pass the full selection; about 10,000 events that have $p_\text{T}>300$ GeV; about 3,000 that have $p_\text{T}>400$ GeV, and just over 1000 events with $p_\text{T}>500$ GeV.    The purity of events is about 65\% over the entire mass range and about 80\% for jet masses above 65 GeV.

\begin{figure}[h!]
\centering
\includegraphics[width=0.45\textwidth]{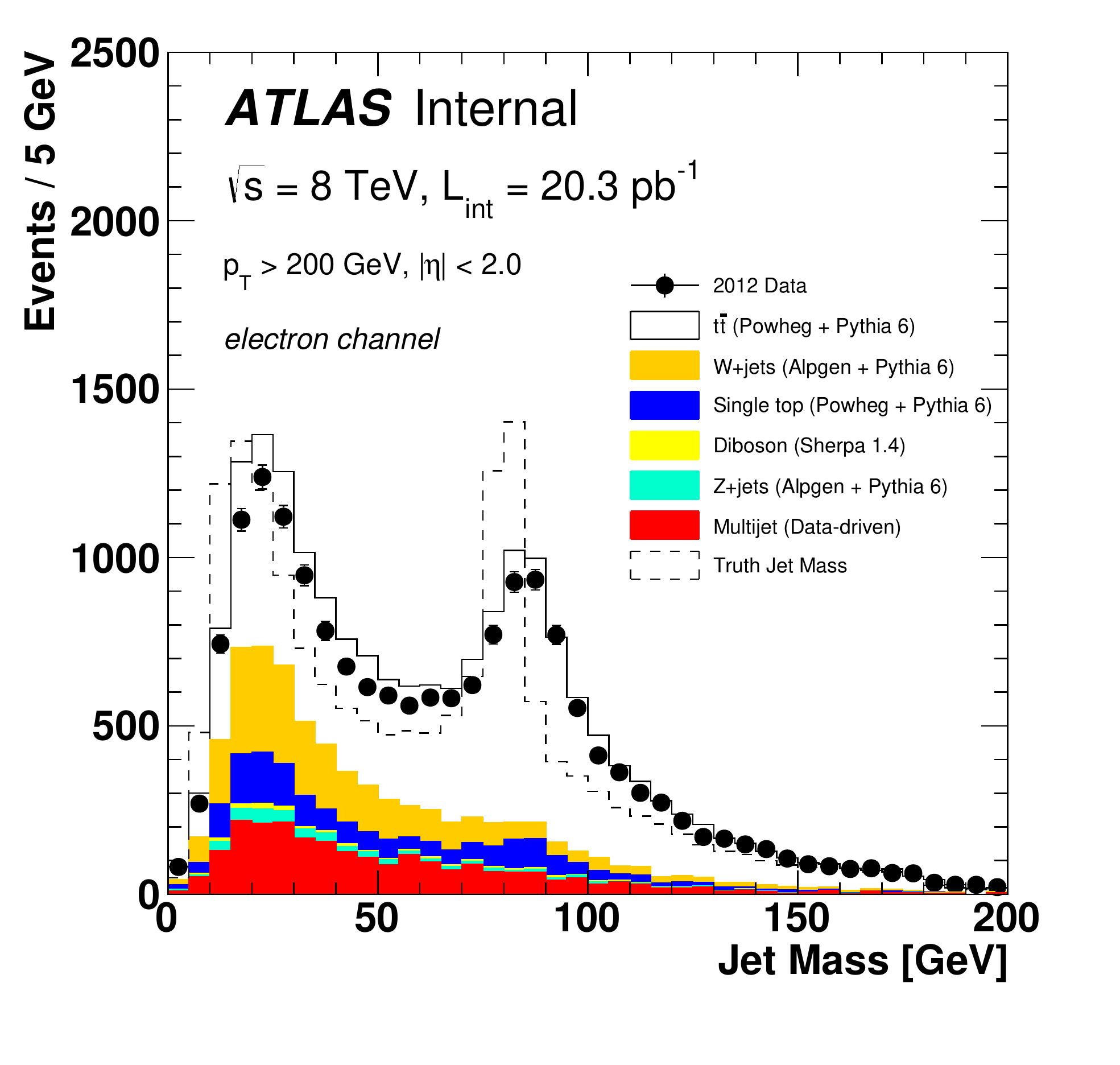}\includegraphics[width=0.45\textwidth]{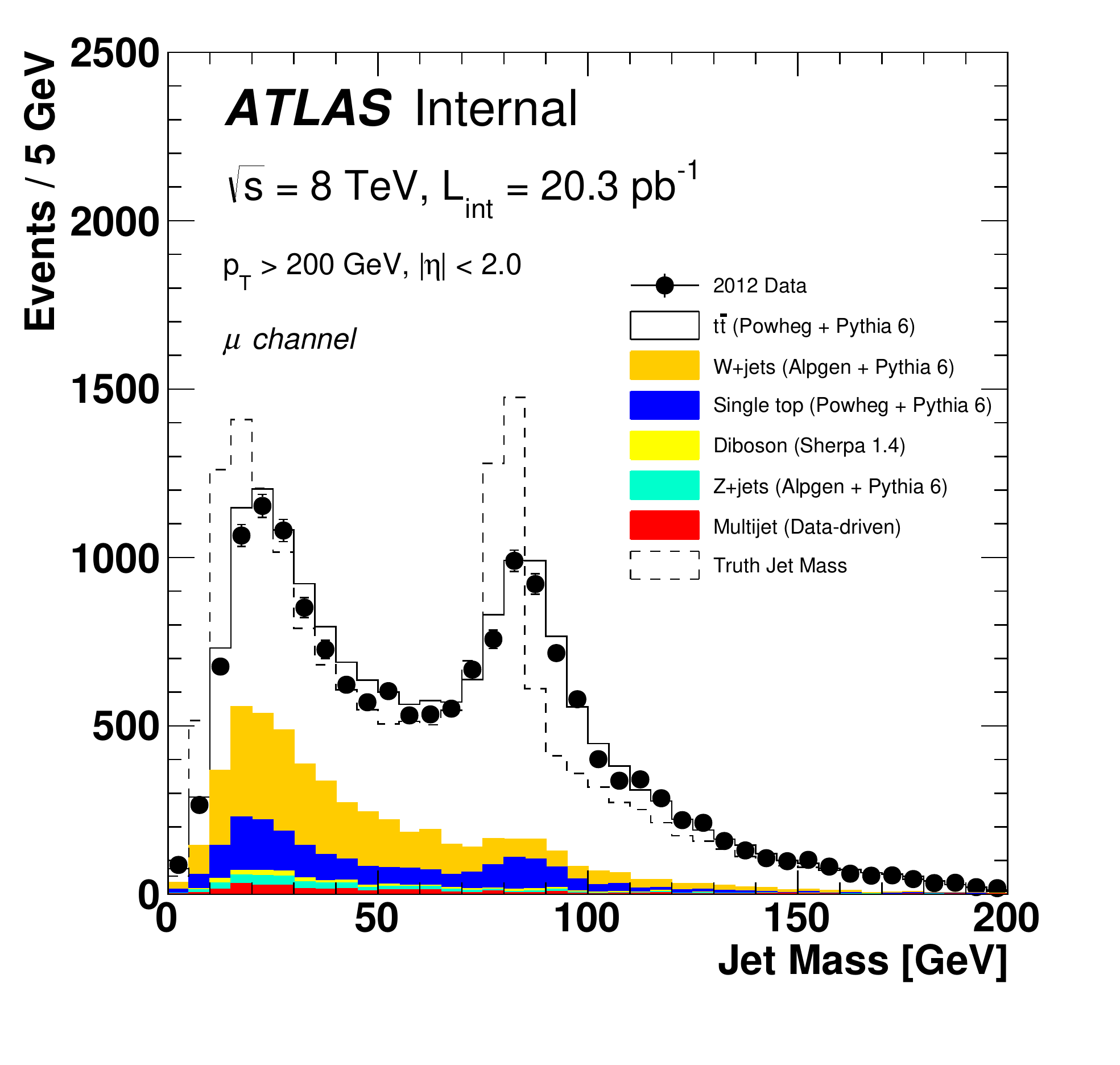}
\caption{The jet mass distribution for events passing the selection described in Sec.~\ref{sec:samples} for electron events (left) and muon events (right).}
\label{fig:forward0}
\end{figure}

\begin{figure}[h!]
\centering
\includegraphics[width=0.45\textwidth]{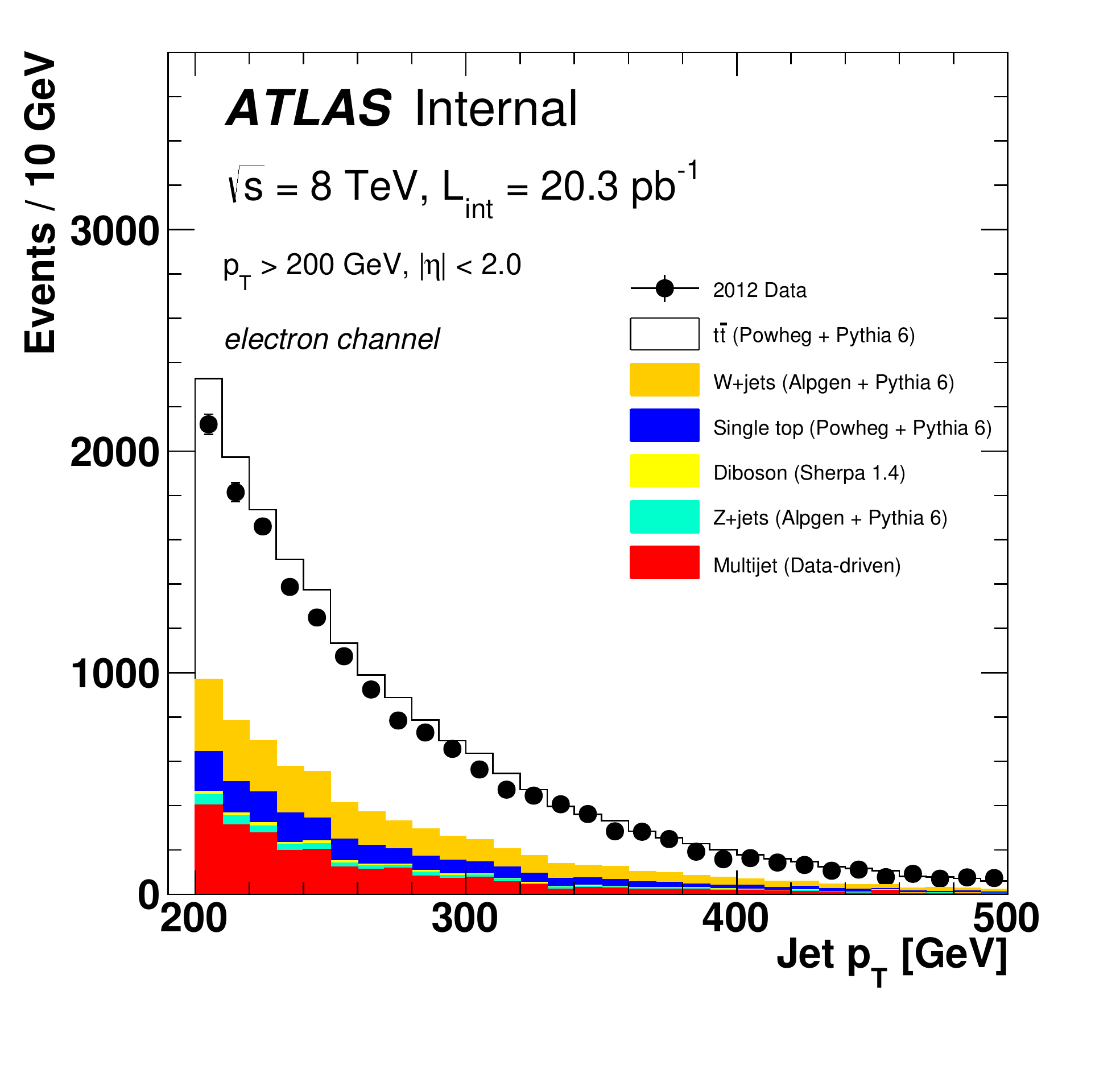}\includegraphics[width=0.45\textwidth]{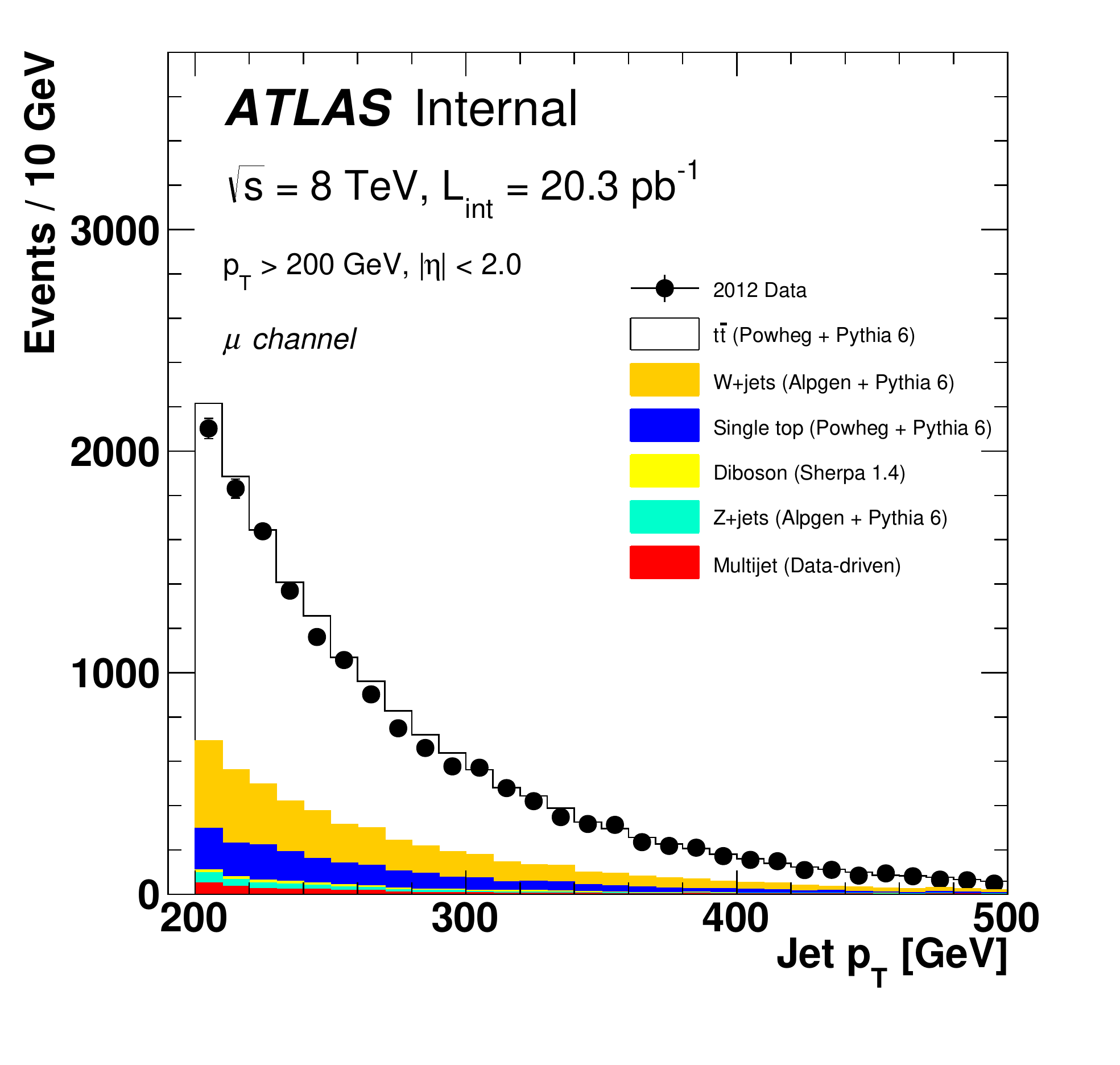}
\caption{The jet $p_\text{T}$ distribution for events passing the selection described in Sec.~\ref{sec:samples} for electron events (left) and muon events (right).}
\label{fig:forward0b}
\end{figure}

\paragraph{Extracting the Jet Mass Scale and Resolution} \mbox{}\\
\label{sec:fitting}

There are two sets of considerations when constructing a procedure for extracting the jet mass response from the measured spectrum.  First, the particle-level spectrum and/or response function can be parameterized or a non-parametric procedure can be used.   In general, parametric forms are useful if the underlying function is known and also to make interpretation easier.  In this case, approximations to the particle-level mass distribution are known (See Sec.~\ref{sec:mass:theory}), but these analytic forms do not capture non-perturbative effects such as hadronization and the underlying event that are accounted for in the state-of-the-art MC generators.  Therefore, the analysis presented in this section uses the non-parametric particle-level mass distribution from the simulation.  The resolution plots in Sec.~\ref{sec:JMR} showed that the jet mass response function is non-Gaussian and depends on the jet $p_\text{T}$.  One possibility is to find a functional form (such as the double-Gaussian) for the non-Gaussian function and then to let the parameters depend on $p_\text{T}$.  The benefit of using a parametric form is that the fitted parameters can be directly interpreted as the scale and resolution of the response function.  However, even though the functions in Sec.~\ref{sec:JMR} worked well to describe the jet mass response, there were still deviations from the empirical distributions from simulation and in general there is a dependence on the jet mass itself in addition to the jet $p_\text{T}$.  Therefore, the method presented in this section is fully non-parametric, using the response function from the simulation.  This response function is stretched and shifted so that when the particle-level mass distribution is {\it forward-folded}, it best matches with the data.  In order to take into account the jet mass and jet $p_\text{T}$ dependence of the response function, the response distribution from the simulation is binned in the jet $p_\text{T}$ and the jet mass.  More details about the forward-folding technique are described below.  

The second consideration for constructing the extraction procedure is the particle-level event selection (fiducial volume).  Ideally, the detector-level event selection would result in a sufficient pure selection.  For generic jets from $t\bar{t}$ events, this is mostly true for jet masses near $m_W$.  However, there are a large fraction of events that originate from top quark pair production, but the selected jet is not a fully contained $W$ boson jet.  This is evident from the large contribution to the particle-level jet mass spectrum in Fig.~\ref{fig:forward0} for jet masses far from $m_W$.  One approach is to subtract the non-resonant $t\bar{t}$ and non-$t\bar{t}$ backgrounds from the detector-level distribution prior to fitting the mass response (See the {\it subtraction method} in Ref.~\cite{ATLAS-CONF-2016-008}).  This makes the definition of the measurement conceptually cleaner, but introduces significant sources of model dependence.  For example, the background jet mass distribution is not uniform under the $W$ boson peak (it is falling) and therefore an uncertainty in the normalization of the backgrounds results in an uncertainty on the measured jet mass distribution shape.  Additionally, the jet mass response of the background must be taken as an input which can bias the measurement of the response in the signal.  For these reasons, the measurement presented in this section is defined only by its detector-level selection - no components are subtracted prior to the measurement.

From the above considerations, a forward-folding method is used to extract the relative differences in the jet mass response between data and simulation.   Let $R(m_\text{true},p_{T}^\text{reco})$ be the distribution of the jet mass response for given values of the particle-level jet mass $m_\text{true}$ and the reconstructed (fully calibrated) jet transverse momentum $p_{T}^\text{reco}$.  In general, $R$ is non-Gaussian and the full non-parametric form is taken from simulation, as well as the distribution of $m_\text{true}$.  For random variable $\rho$ with $\rho\sim R(m_\text{true},p_{T}^\text{reco})$ and fixed $0 <r,s<\infty$ (to be determined below), define the new random variable $m_\text{folded}$ as follows:

\begin{align}
\label{eq:transform}
m_\text{folded}|m_\text{true},p_{T}^\text{reco},r,s&=\left[ s \rho+\left(\rho-\langle \rho\rangle\right)(r-s) \right] m_\text{true}\\
&=( r m_\text{true})\rho +(s-r)\langle\rho\rangle m_\text{true} \label{eq:transofrm:part2}
\end{align}

\noindent The functional form of the transformation in Eq.~\ref{eq:transform} is chosen so that the distribution of $m_\text{folded}|m_\text{true},p_{T}^\text{reco}$ is the same as the distribution of $m_\text{reco}|m_\text{true},p_{T}^\text{reco}$ but with the average response scaled by $s$ and the standard deviation of the response scaled by $r$.  Symbolically:

\begin{align}
\label{eq:forward2}
\left\langle \frac{m_\text{folded}}{m_\text{true}}\Big|m_\text{true},p_{T}^\text{reco}\right\rangle &=  s  \langle\rho\rangle\\
\sigma\left( \frac{m_\text{folded}}{m_\text{true}}\Big|m_\text{true},p_{T}^\text{reco}\right) &= r\sigma(\rho), \label{eq:forward2:part2}
\end{align}

\noindent where $\sigma(X)$ is the standard deviation of the random variable $X$.   Eq.~\ref{eq:forward2} follows by inspection of Eq.~\ref{eq:transform} because the second term has mean zero and Eq.~\ref{eq:forward2:part2} is evident from Eq.~\ref{eq:transofrm:part2} because the second term is not random.  For $r=s=1$, $m_\text{folded}|m_\text{true},p_{T}^\text{reco}$ and $m_\text{reco}|m_\text{true},p_{T}^\text{reco}$ have the same distribution.   The values of $s$ and $r$ are chosen such that the distribution of $m_\text{folded}$ best matches the data.  (Un)folding methods usually need to correct for migrations between the particle-level and detector-level selections, but this is not necessary because the event selection is on the {\it reconstructed} jet $p_\text{T}$ while the fitting is performed on the jet mass\footnote{The jet $p_\text{T}$ spectrum in simulation well reproduces the data (Fig.~\ref{fig:forward0b}) and any residual differences could be removed by re-weighting.} The fit to the detector-level jet mass distribution is performed by minimizing a $\chi^2$ per degrees of freedom:

\begin{align}
\label{eq:jmr:chisq}
r,s = \text{argmin}_{r',s'}\frac{1}{n-1}\left(\sum_{i=1}^n\frac{h_i(m_\text{folded}|r',s')-h_i(m_\text{data})}{\sigma^2_{i,m_\text{reco}}+\sigma^2_{i,m_\text{data}}}\right)^2,
\end{align}

\noindent where $h_i(\cdot)$ is the content of a histogram of the variable $\cdot$ with $n$ bins, $\sigma_{i,m_\text{data}}=\sqrt{h_i(m_\text{data})}$, and $\sigma_{i,m_\text{reco}}$ is the MC statistical uncertainty in bin $i$.  Since $n$ is fixed, the normalization factor is only needed for the visualizations that appear later and do not impact the fitted values of $r$ and $s$. In order to render the fit insensitive to overall changes in the normalization, each template is normalized to have the same integral as the data.  Since the multijet background is derived directly from the data, it is added unchanged to the MC-derived templates for each value of $r$ and $s$.  In order to maximize the sensitivity to $W$ boson-like jets, the fit is only performed in the mass range 50 GeV $<m_\text{reco}<$ 120 GeV.

There are a variety of methods that could be used to generate $h_i(m_\text{folded}|r',s')$.  The most straight-forward but computationally intensive method would be to generate enough MC events to numerically estimate the full $p_\text{T}^\text{reco}$ and $m_\text{true}$ dependence of $R$, sample events from $f(m_\text{true},p_{T}^\text{reco})$, and then subsequently sample events from $ f(m_\text{folded}|r',s',m_\text{true},p_{T}^\text{reco})$.  This method requires a large number of sampled events per $r'$ and $s'$ in addition to a one-time cost of generating enough simulated events to populate the bins of the three-dimensional template for $R$.  The rest of the section uses a much simpler method that takes advantage of the fact that every event that is used for the estimation of the particle-level jet mass distribution also has a detector-level jet mass value.  Each particle level jet mass $m_\text{true}$ is transformed in the following way:

\begin{align}
\label{eq:JMR:closure}
m_\text{true}\mapsto m_\text{true}'|r,s=sm_\text{reco}+(m_\text{reco}-\langle m_\text{reco}|m_\text{true},p_\text{T}^\text{reco}\rangle)(r-s),
\end{align}

\noindent where $m_\text{reco}$ is the detector-level jet mass from {\it the same simulated event} as $m_\text{true}$.   By construction, the distribution of $m_\text{true}'|r,s$ is the same as $m_\text{reco}|r,$, but does not require the distribution of $R$ to be estimated directly, since it is built in automatically to the relationship between $m_\text{true}$ and $m_\text{reco}$.  The only additional input that is required is a two-dimensional template for $\langle m_\text{reco}|m_\text{true},p_\text{T}^\text{reco}\rangle$, an example of which is shown in the left plot of Fig.~\ref{fig:forward11}.  With the transformation in Eq.~\ref{eq:JMR:closure}, the closure of the method is trivial: the $\chi^2$ in Eq.~\ref{eq:jmr:chisq} is exactly zero when $r=s=1$ no matter how many MC events are available.   This closure is illustrated in the right plot of Fig.~\ref{fig:forward11}.

\begin{figure}[h!]
\centering
\includegraphics[width=0.45\textwidth]{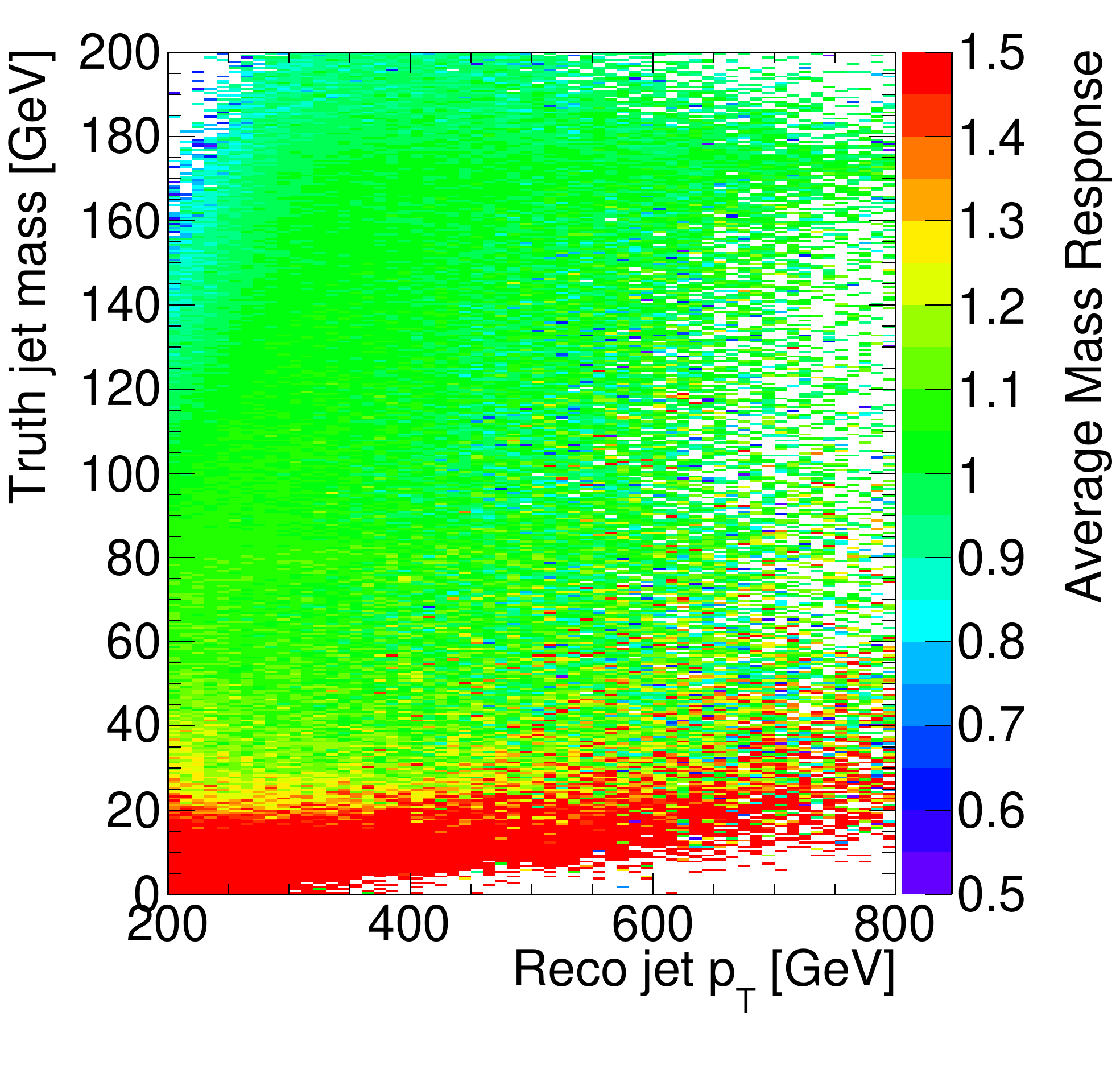}\hspace{5mm}\includegraphics[width=0.45\textwidth]{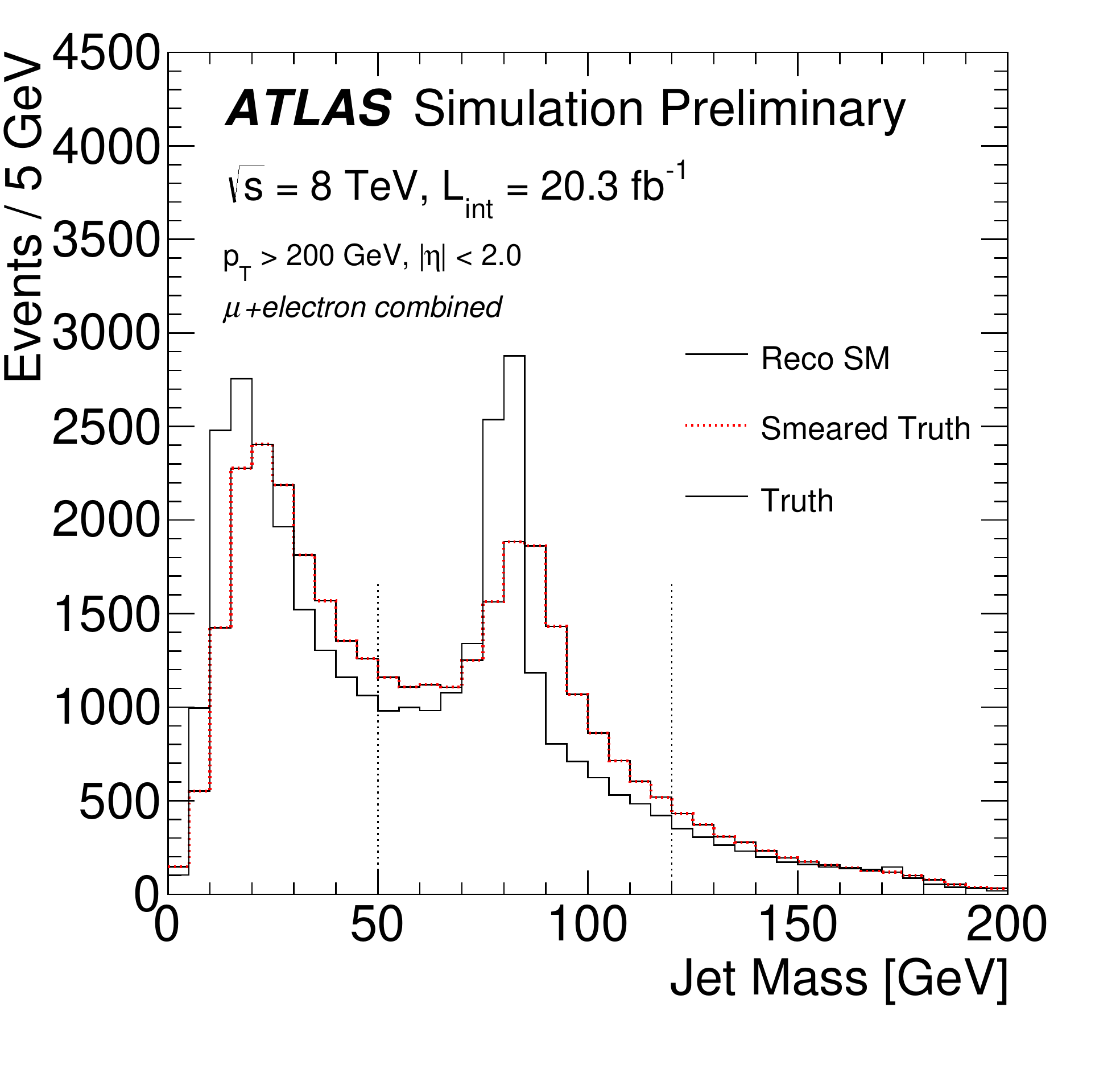}
\caption{Left: dependence of the average response $\langle m_\text{reco}/m_\text{true}|m_\text{true},p_\text{T}^\text{reco}\rangle$ as a function of $p_\text{T}^\text{reco}$ and  $m_\text{true}$.  Right: the fitted distribution when using the detector-level simulation in place of the data.  By construction, the fitted distribution is identical to the input detector-level distribution $(r=s=1)$.}
\label{fig:forward11}
\end{figure}

\paragraph{Systematic Uncertainties}\mbox{}\\
\label{sec:systsforward}

There are two sources of systematic uncertainty in the extraction of the relative jet mass scale and jet mass resolution when using the forward-folding method.  First, there is a theoretical modeling uncertainty on the particle-level jet mass distribution.  Second, there are sources of theoretical or experimental uncertainties that impact the mass response.  The method is not sensitive to sources of uncertainty that only change the overall normalization.  Uncertainties are estimated by varying the simulation and then re-fitting the data.  The difference between the nominal fitted values of $r,s$ and the variation fits is used as the systematic uncertainty.  Sources of theoretical modeling uncertainty include the NLO matching scheme, fragmentation, and initial- and final-state radiation (ISR/FSR).  Any impact on the NLO matching scheme is estimated by replacing the nominal {\sc Powheg-box}+{\sc Pythia} 6 $t\bar{t}$ sample with alternative samples generated with {\sc Powheg-box}+{\sc Herwig} and {\sc MC@NLO}+{\sc Herwig} (all other processes remain unchanged).  The fragmentation uncertainty uses a comparison between {\sc Powheg-box}+{\sc Pythia} 6 with {\sc Powheg-box}+{\sc Herwig} and the ISR/FSR uncertainty is estimated by comparing two variations of {\sc Powheg-box}+{\sc Pythia} 6 with different Perugia 2012 tunes\footnote{The factorization/renormalization scales and the $h_\text{damp}$ parameter are simultaneously varied, but are expected to have a smaller impact on the jet mass (response).  See Fig.~\ref{fig:JMR:particlelevel} for details.}.  The background modeling and experimental uncertainties have a much smaller impact on the relative jet mass scale and resolutions compared with the theoretical modeling uncertainties.  The uncertainty on the $W$+jets background is the most relevant near the $W$ mass peak, which is due to the statistical uncertainty on the charge asymmetry method used to derive the normalization~\cite{Aad:2014zka}.

\paragraph{Results} \mbox{}\\
\label{sec:results}

Figure~\ref{fig:forward1} shows the $\chi^2$ minimization for the relative jet mass scale and the relative jet mass resolution.  Each value on the curve is the $\chi^2$ per degree of freedom when fitting either the simulation or the data with a template from the simulation using a resolution function whose scale or resolution is shifted or stretched by the value indicated on the horizontal axis.  As the fit is performed simultaneously for the relative scale and resolution, the curves in Fig.~\ref{fig:forward1} are the value of the $\chi^2$ per degree of freedom at a given relative jet mass scale or jet mass resolution minimized over the relative jet mass resolution or jet mass scale, respectively.  By construction, the minimum $\chi^2$ for the simulation fit to itself is zero at a relative scale of one.  The relative jet mass scale is 1.001 and the relative jet mass resolution is 0.96.  The range of the vertical axes relative to the range of the horizontal axes indicates that there is significantly more sensitivity to the relative scale than the relative resolution.  The fit improves the $\chi^2$ per degree of freedom by about 10\%.  A comparison between the template from simulation with the fitted parameters and the data is shown in Fig.~\ref{fig:forward2}.  The dashed line is the particle-level jet mass spectrum that is smeared to detector-level before comparing with the data.  The solid line shows the detector-simulation before fitting the relative jet mass scale and resolution and the dotted red line shows the post-fit distribution.  There is only a small decrease in the $\chi^2/\text{NDF}$ from the fit, so the two distributions are similar.  In the ratio plot, the band is the statistical uncertainty from the data while the black and red points are the pre- and post-fit ratios of the simulation with the data.

\begin{figure}[h!]
\centering
\includegraphics[width=0.5\textwidth]{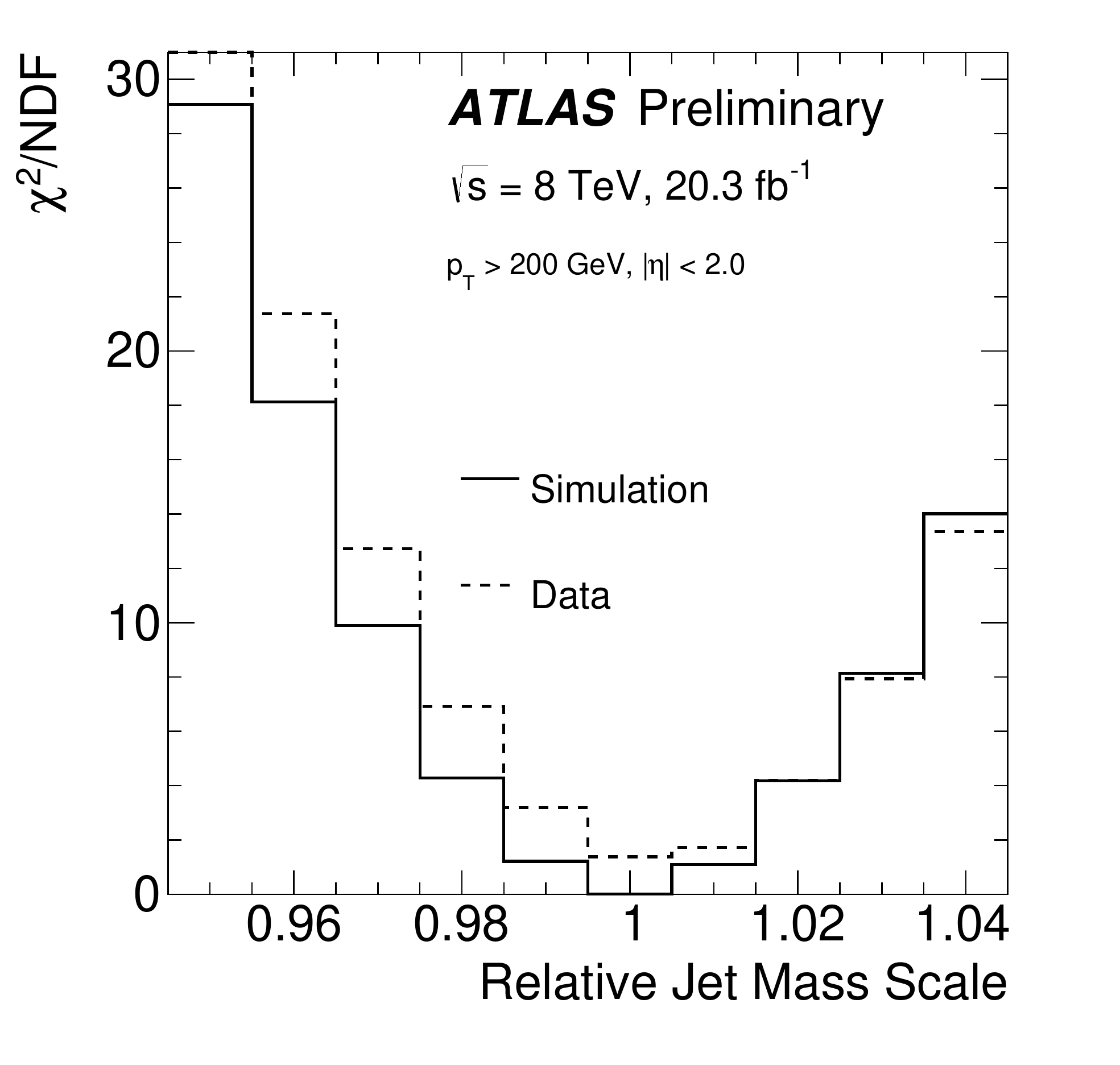}\includegraphics[width=0.5\textwidth]{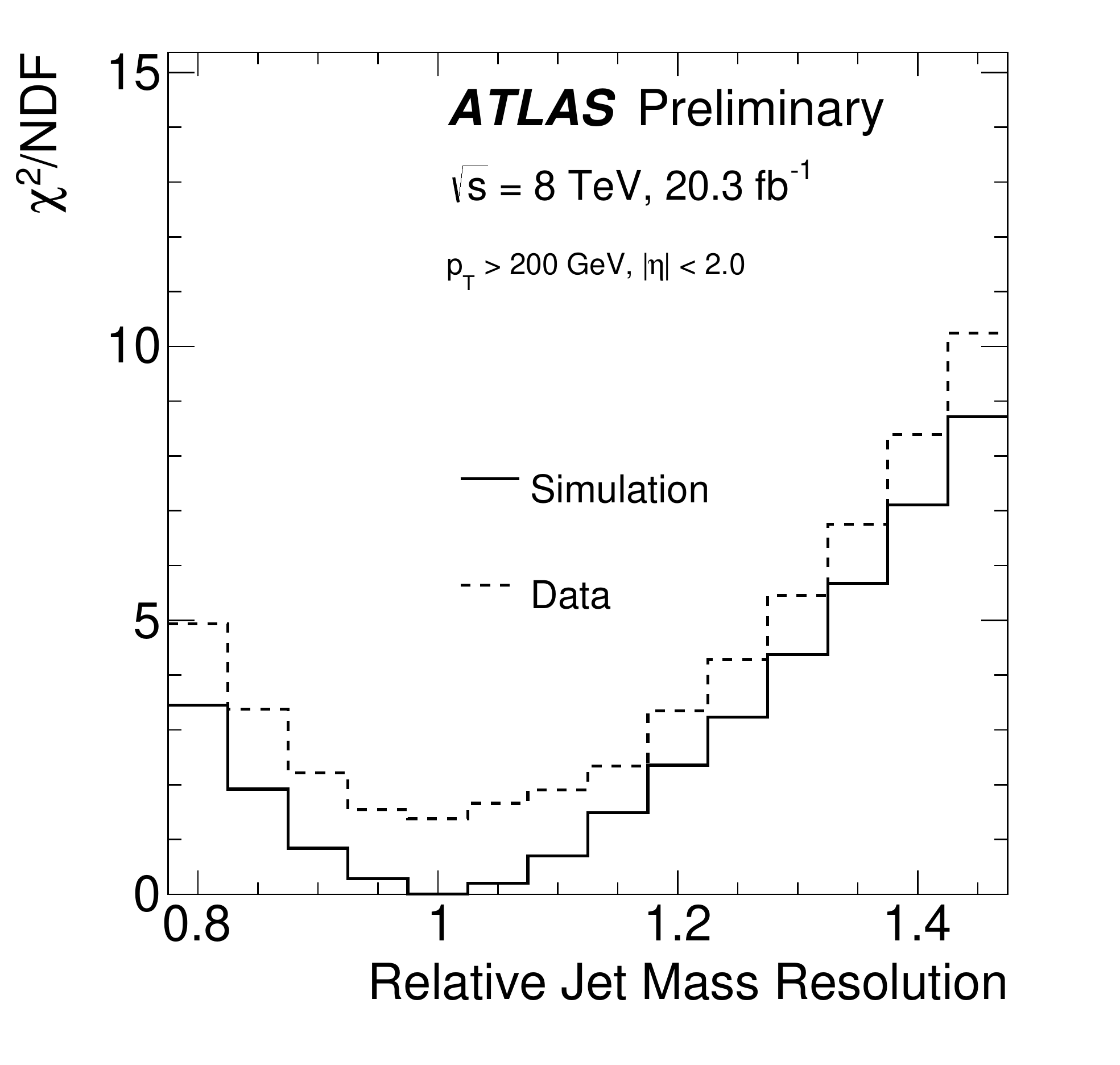}
\caption{The value of the $\chi^2$ per degree of freedom (NDF) at a given relative jet mass scale (left) or jet mass resolution (right) minimized over the variable not shown.}
\label{fig:forward1}
\end{figure}

\begin{figure}[h!]
\centering
\includegraphics[width=0.55\textwidth]{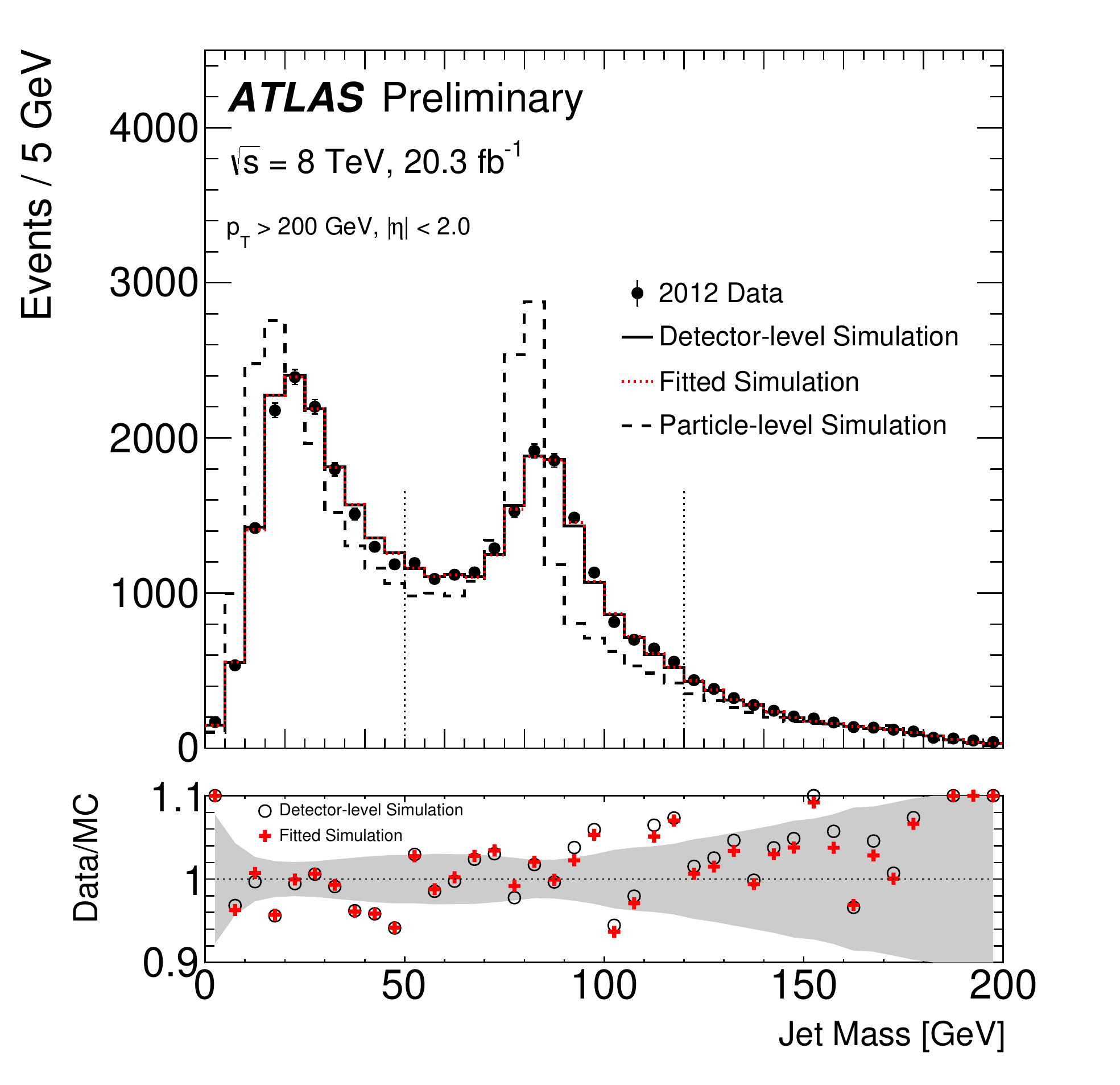}
\caption{A comparison between the post- and pre-fit simulation and the data.  The vertical dotted lines indicate the fit range.}
\label{fig:forward2}
\end{figure}

\begin{figure}[h!]
\centering
\includegraphics[width=0.55\textwidth]{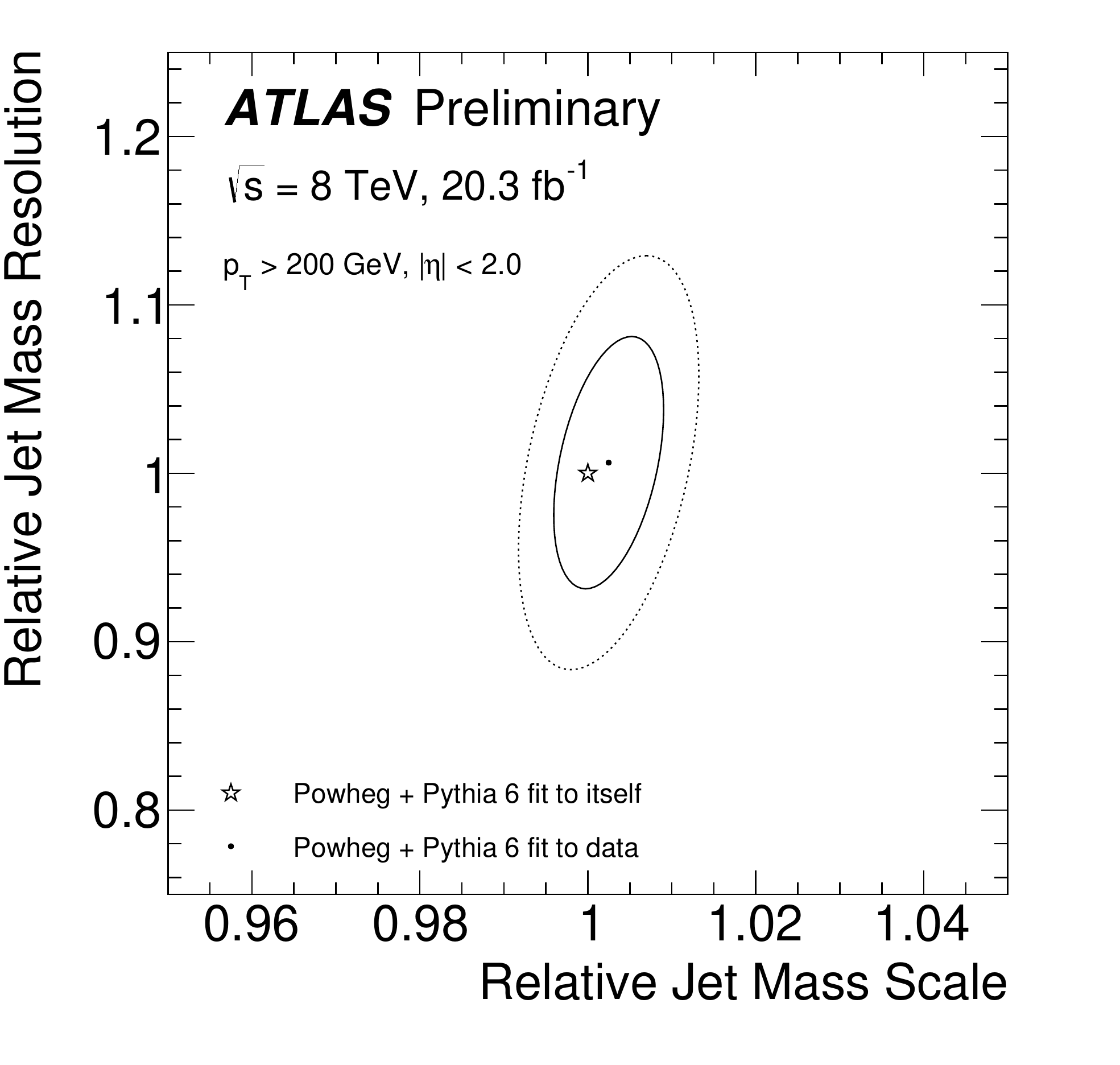}
\caption{The one (solid) and two (dashed) $\sigma$ uncertainty ellipses for the relative jet mass scale and the relative jet mass resolution.  }
\label{fig:forward3}
\end{figure}

One- and two-$\sigma$ statistical uncertainty ellipses are shown in Fig.~\ref{fig:forward3}.  The statistical uncertainty is computed using the bootstrap technique: $N$ pseudo-datasets are generated by (re)sampling from the data with replacement.  Each pseudo-dataset $i$ is then fit with the nominal simulation and the measured values $r_i$ and $s_i$ are recorded.  The circular marker in Fig.~\ref{fig:forward3} represents $(\langle s_i\rangle,\langle r_i\rangle)$ while the the star indicates the result of fitting the simulation to itself, which is at the point (1,1) by construction.  Appendix~\ref{sec:app:uncertaintyellipse} describes how uncertainty ellipses are constructed.  The mean and standard deviation of the joint distribution are estimated using the sample mean and standard deviation over the ensemble of pseudo-datasets. The values $C$ for the $1\sigma$ and $2\sigma$ ellipses are computed by solving

\begin{align}
\frac{1}{2}\int_0^Cdxe^{-\frac{x}{2}}=\frac{1}{\sqrt{2\pi}}\int_{-Z}^Zdxe^{-\frac{x^2}{2}},
\end{align}

\noindent where $Z=1$ for the $1\sigma$ contour and $Z=2$ for the $2\sigma$ contour.  These results are combined with the systematic uncertainties in Table~\ref{tab:forward-fold}.  The amount of (ISR and) FSR is the dominant uncertainty for extracting the relative JMR because the fitted resolution width compensates for changes in the width of the particle-level mass distribution (see Fig.~\ref{fig:JMR:particlelevel}).  The total systematic uncertainty is about $2\%$ for the relative jet mass scale and about $19\%$ for the relative jet mass resolution.

\begin{table}[h!] 
\centering

\begin{tabular}{c|cccc}
Source of Uncertainty &   Jet Mass Scale $(s)$ &  Jet Mass Resolution $(r)$ \\ 
\hline\hline
NLO matching 				 	& 0.017 		& 0.08 \\ 
Fragmentation			   		& 0.018		& 0.05 \\ 
ISR/FSR      					& 0.004		& 0.15 \\ 
Jet Energy Scale     		 		& 0.002		& 0.03 \\ 
Jet Energy Resolution 		 	& 0.001 		& 0.03 \\ 
$b$-tagging 				& $<0.001$			& 0.01 \\
MC Normalization	 		& 0.001		& 0.01 \\
\hline\hline
Total Systematic Uncertainty 	&0.024	&0.18	 \\ 
Data Statistical Uncertainty    	&0.004	&0.05	& \\ 
\hline                                                            
Value               				& 1.001	& 0.96	&\\ 
\hline
\end{tabular}
\caption{A summary of the measured relative jet mass scale and jet mass resolution using both the subtraction and the forward folding methods.  Uncertainties are given as a fraction of the nominal.  The jet energy scale, ISR/FSR, and MC Normalization background uncertainties are treated as asymmetric but the maximum of the two variations are reported in this table. }
\label{tab:forward-fold}
\end{table}

\clearpage

\paragraph{Intermediate Conclusions} \mbox{}\\
\label{sec:conclusions}

This section has reported a measurement of the relative jet mass scale and jet mass resolution using a sample enriched in boosted hadronically decaying $W$ bosons from $t\bar{t}$ events in the $\sqrt{s}=8$ TeV data collected by the ATLAS detector.  A new method called {\it forward folding} uses non-parameteric shapes for both the particle-level distribution and the response function, derived from the simulation.  The relative jet mass scale and jet mass resolution are compatible with unity within the statistical uncertainties at 0.4\% for the jet mass scale and 5\% for the jet mass resolution.  This measurement can be used in the future to set a systematic uncertainty on the jet mass scale and the jet mass resolution for BSM searches.  The scale and resolution are compatible with unity, but the uncertainty on the measurement should be used as uncertainties for analyses that use the large-radius jet mass to identify jets are resulting from boosted heavy particle decays.  However, there are important caveats to this measurement:

\begin{description}
\item[Topology Dependence] The jet mass scale and the jet mass resolution presented in this measurement use boosted hadronically decaying $W$ bosons from $t\bar{t}$ events.  The relative mass scale and resolution may depend on the jet $p_\text{T}$, the jet mass, the number of subjets within the jet, close-by radiation, and the presence of heavy flavor decays inside the jet. 
\item[Particle-level Input] The measured hadronically decaying $W$ boson resonance peak contains information about the convolution of the particle-level spectrum and the resolution function.  In this measurement, the particle-level spectrum is taken as input to extract the resolution function.  Therefore, the relative scale and resolution presented here are not applicable as uncertainties for precision measurements of the particle-level spectrum.
\end{description}

\noindent More, higher energy data will be available in Run 2 that will allow for many of the above challenges to be addressed as a new  frontier is opened for new physics searches and precision measurements at high energies.	
	
\clearpage	
	
\paragraph{Improvements and Prospects with 13 TeV data} \mbox{}\\
\label{sec:JMR13TeV}
	
	The $3.2$ fb${}^{-1}$ of $\sqrt{s}=13$ TeV data collected  in 2015 are used in this section for a preliminary measurement of the relative jet mass scale and resolution, as well as for the introduction of a new technique to measure the jet $p_\text{T}$ scale and resolution with forward-folding\footnote{The results in this section are published in Ref.~\cite{ATLAS-CONF-2016-035} and include input from D. Melini, N. Norjoharuddeen, and M. Vos.}.  Even though the integrated luminosity is significantly lower with the early Run 2 dataset compared with Run 1, the increase in the inclusive $t\bar{t}$ cross-section coupled with a further increase at high $p_\text{T}$ makes the total number of $t\bar{t}$ events roughly comparable.  In particular, the inclusive cross-section increases by about a factor of $3.5$ and Fig.~\ref{fig:forward3ttbar} shows that there is another factor of $\sim 2$ at high $p_\text{T}$.  The ratio of the number of top quarks pairs is therefore about $20.3/(3.2\times 3.5\times 2)\sim 1$.

\begin{figure}[h!]
\centering
\includegraphics[width=0.48\textwidth]{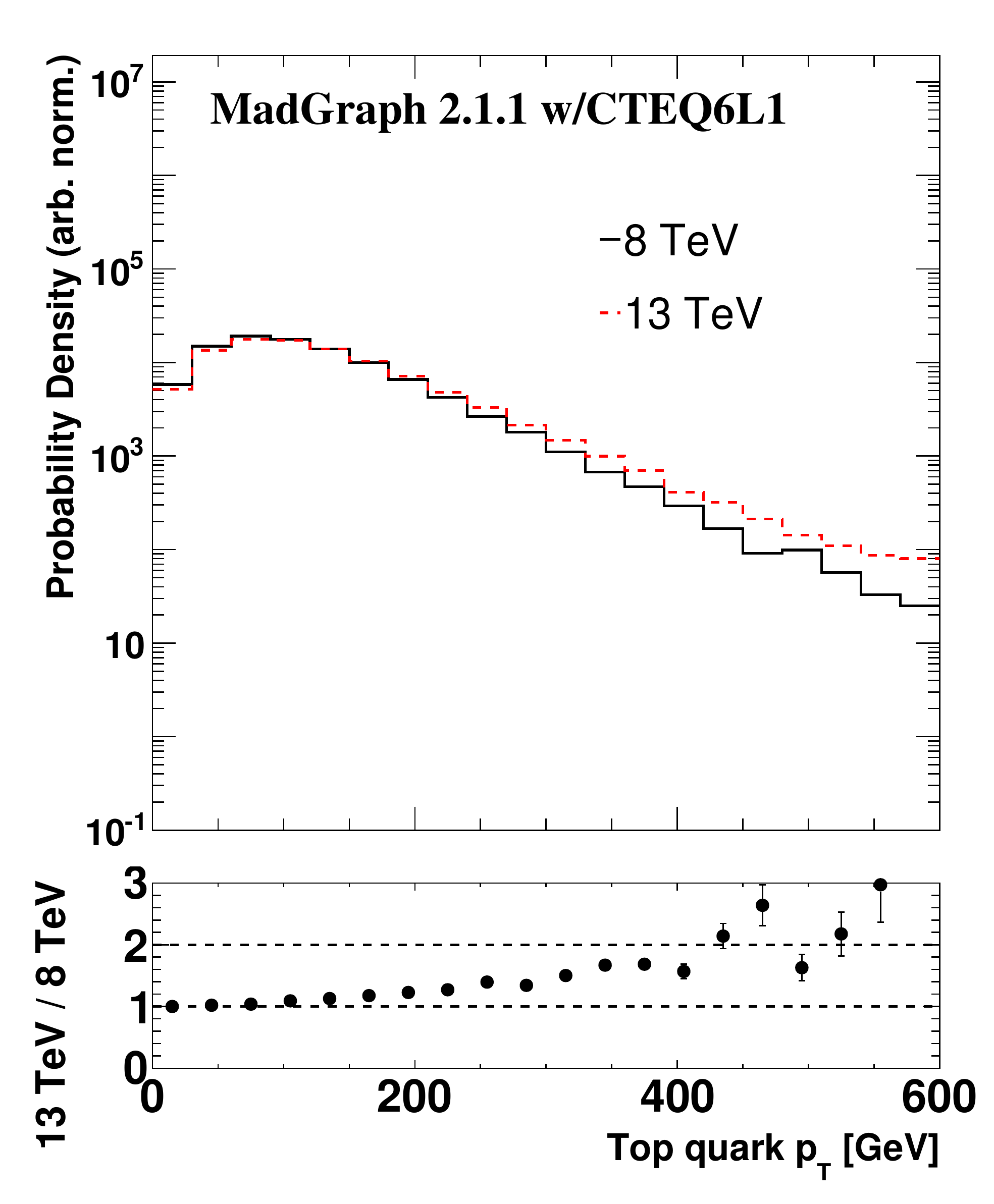}\includegraphics[width=0.48\textwidth]{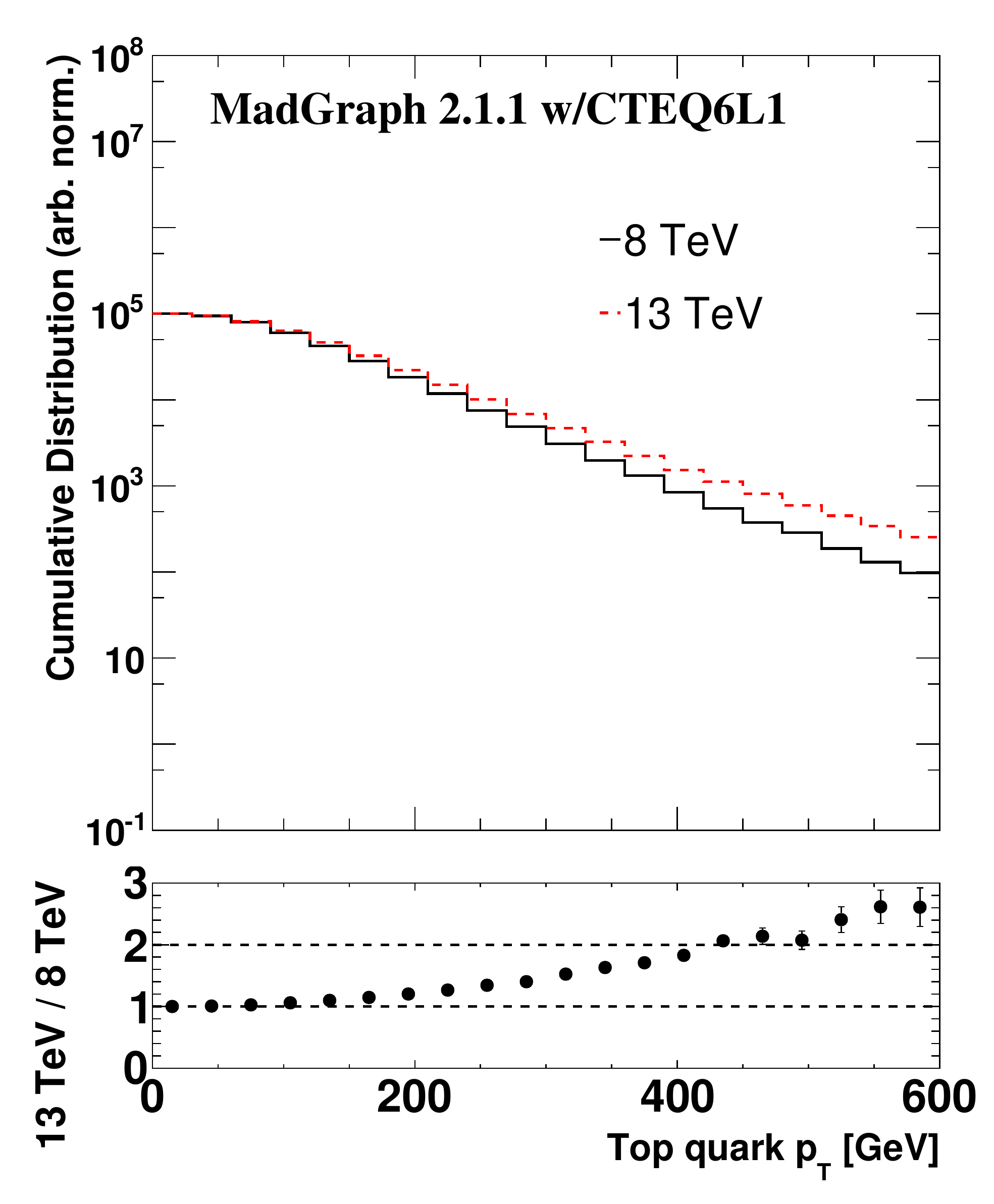}
\caption{The probability distribution (left) and the integral of the probability distribution above a given threshold (strictly speaking, this is one minus the cumulative distribution) on the top quark $p_\text{T}$ for $\sqrt{s}=8$ and $13$ TeV.  The increase in the inclusive $t\bar{t}$ cross-section between these two center-of-mass energies is about $3.5$.}
\label{fig:forward3ttbar}
\end{figure}

Figure~\ref{fig:forwardJETMASS2015} shows the jet mass distribution at $\sqrt{s}=13$ TeV.  The forward-folding method can be applied to any jet mass definition, such as the track-assisted jet mass (Sec.~\ref{sec:TAMass}) shown in the bottom plots of Fig.~\ref{fig:forwardJETMASS2015}.  A scan in the relative jet mass scale and jet mass resolution produce the $\chi^2$ curves in Fig.~\ref{fig:forwardchi22015} that are analogous to the $\sqrt{s}=8$ TeV curves from Fig.~\ref{fig:forward1}.  As was the case at $\sqrt{s}=8$ TeV, there is more sensitivity to the relative JMS than the JMR, evident from the width near the minimum $\chi^2$.

\begin{figure}[h!]
\centering
\includegraphics[width=0.49\textwidth]{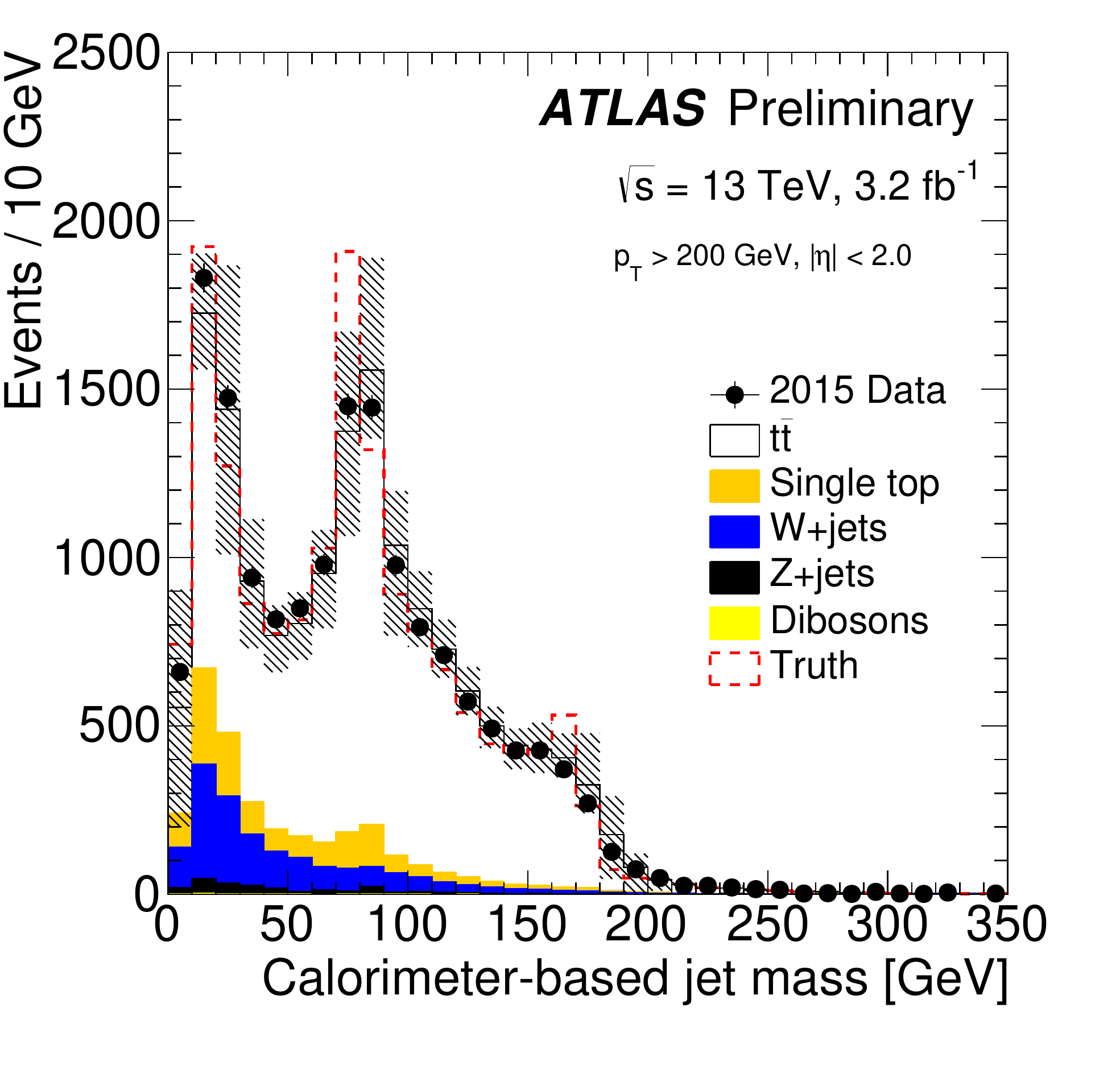}
\includegraphics[width=0.49\textwidth]{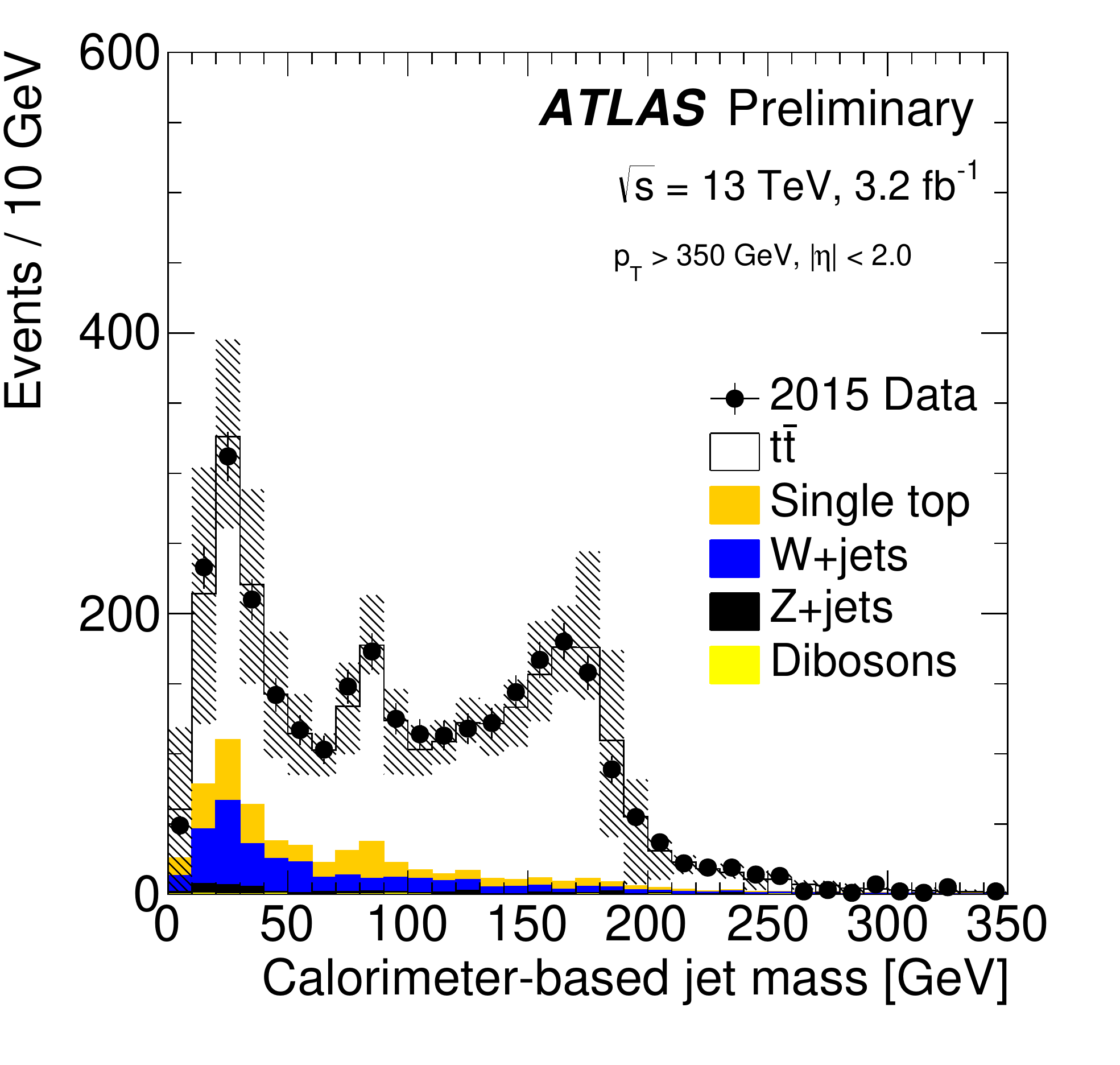}
\includegraphics[width=0.49\textwidth]{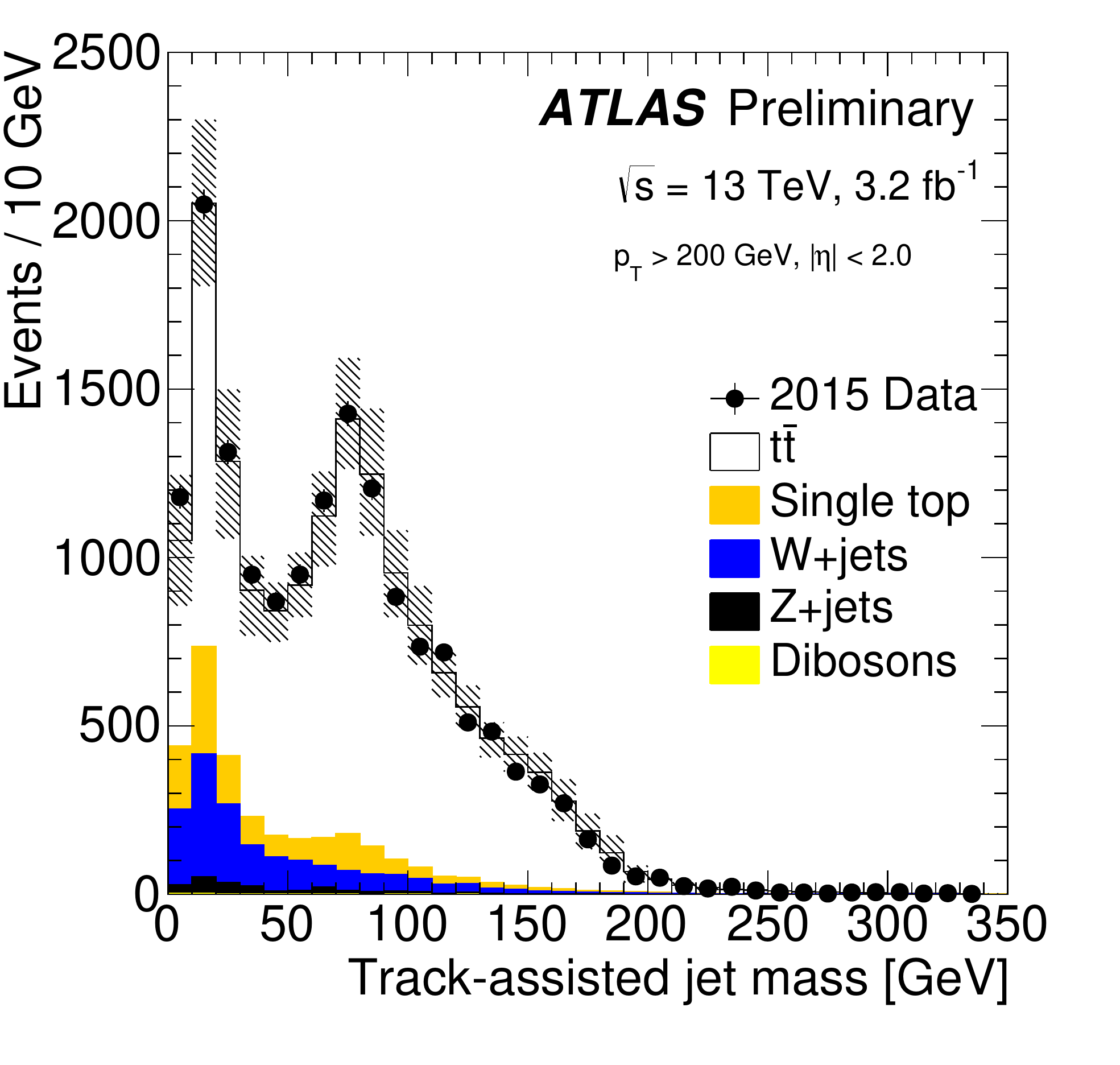}
\includegraphics[width=0.49\textwidth]{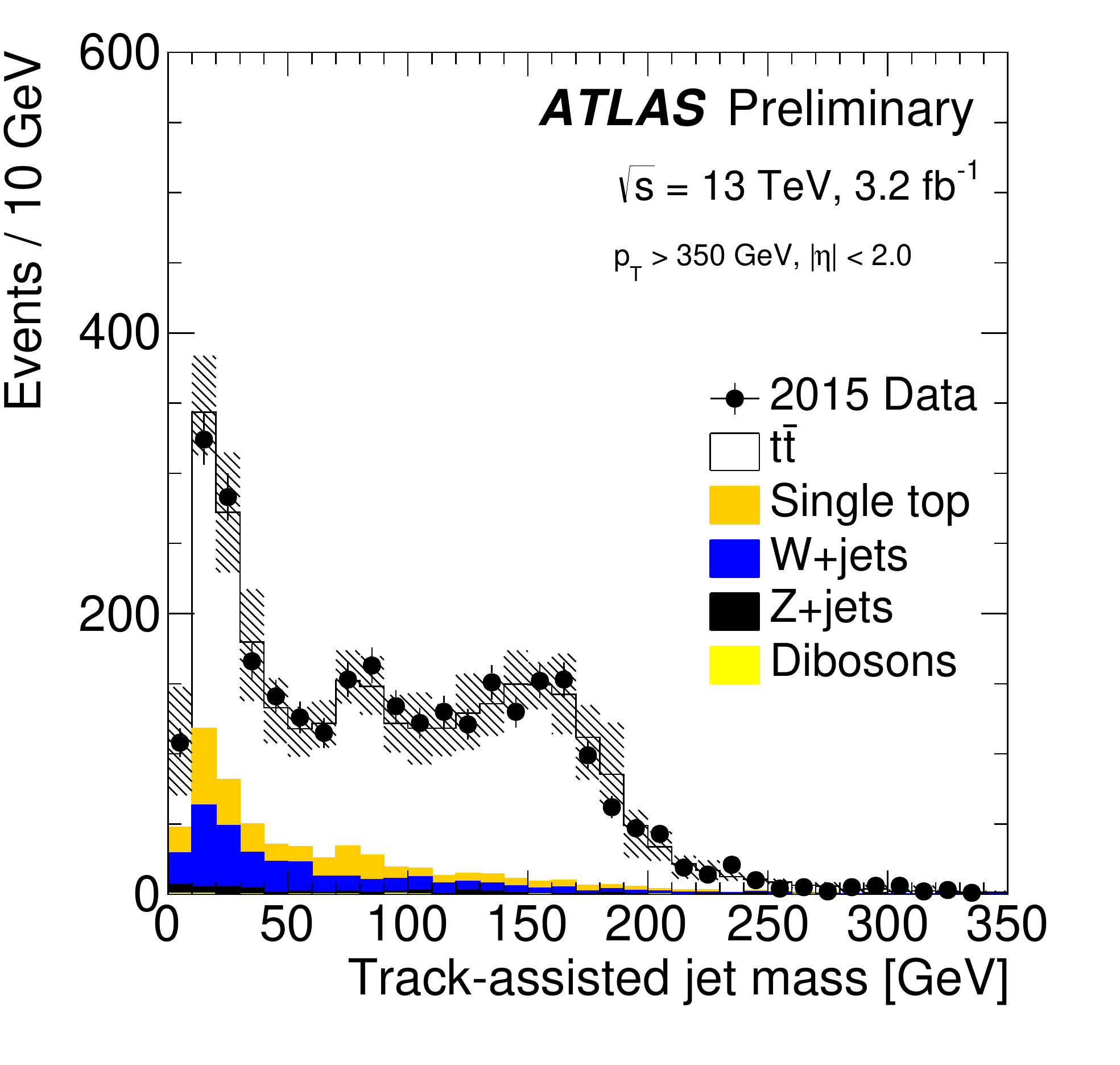}
\caption{The jet mass computed directly from calorimeter-cell clusters (top) and with the track-assisting procedure (bottom) for a lower $p_\text{T}$ (left) and a higher $p_\text{T}$ (right).  The track-assisted jet mass is defined in Sec.~\ref{sec:TAMass}. Note that the ATLAS default trimming parameters for Run 2 are slightly different than Run 1: $R_\text{sub}$ is now $0.2$ instead of $0.3$ in order to improve the resolution at high $p_\text{T}$.  The event selection is identical to the one at $\sqrt{s}=8$ TeV, but only the muon channel is used in this section (negligible multijet contribution).  The bands include detector-level jet and particle-level modeling systematic uncertainties.  For illustration, the top left plot shows also the particle-level distribution.  Note that all distributions are normalized to the data integral.}
\label{fig:forwardJETMASS2015}
\end{figure}

\begin{figure}[h!]
\centering
\includegraphics[width=0.49\textwidth]{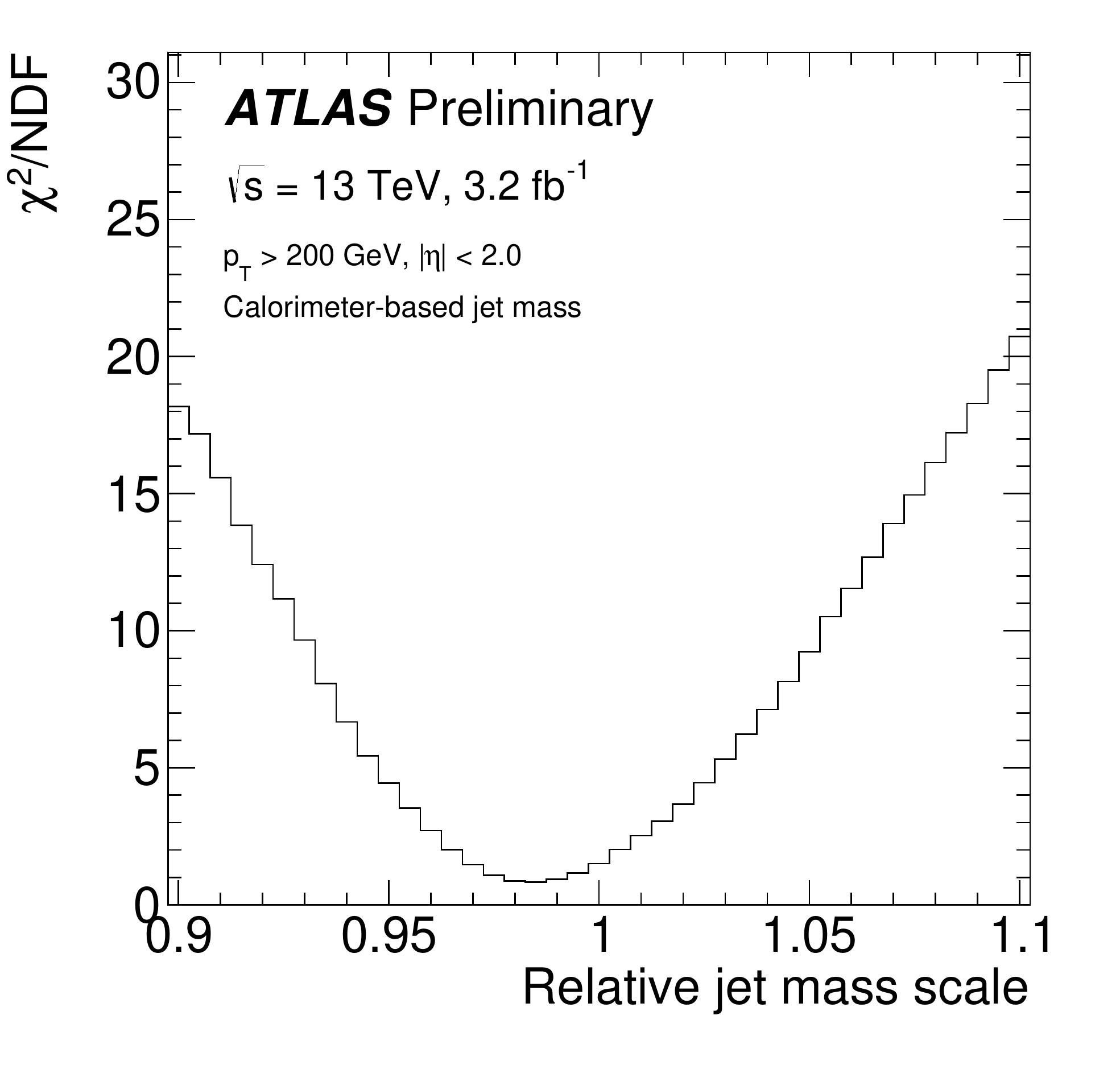}
\includegraphics[width=0.49\textwidth]{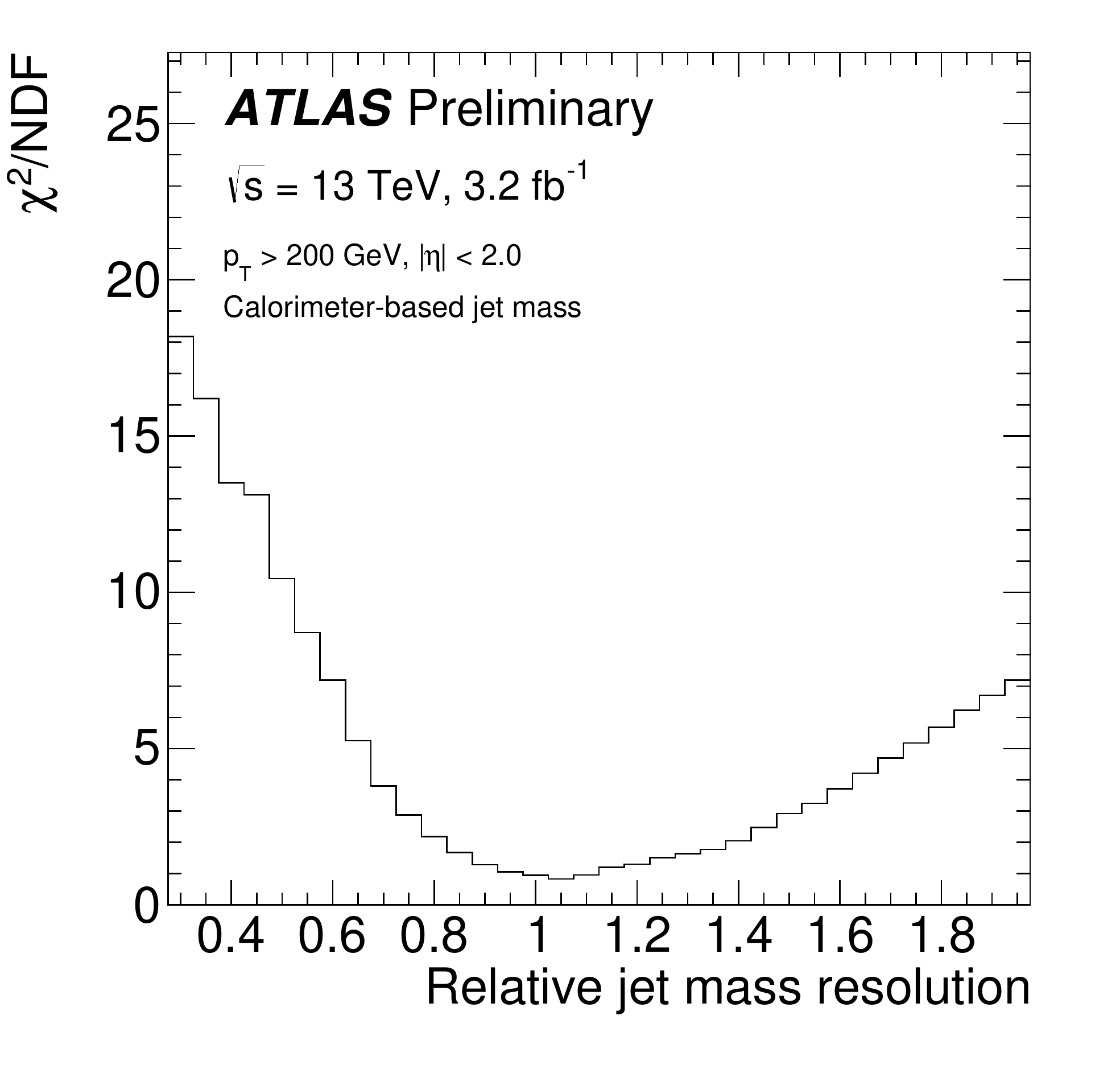}
\caption{The value of the $\chi^2$ per degree of freedom (NDF) at a given relative jet mass scale (left) or jet mass resolution (right) minimized over the variable not shown.}
\label{fig:forwardchi22015}
\end{figure}

A comparison of the Run 1 and early Run 2 measured JMS and JMR is shown in the left plot of Fig.~\ref{fig:forward3comapre813TeV}.  Even though the number of top quark pairs is comparable between the two datasets, the statistical uncertainty ellipse is slightly larger for the $\sqrt{s}=13$ TeV measurement as only the muon channel is used.  As indicated in the caption of Fig.~\ref{fig:forwardJETMASS2015}, the jet mass definition is slightly different between the two datasets: $R_\text{sub}$ is $0.2$ instead of $0.3$ in order to be able to resolve the subjets of ultra boosted $W/Z/H$ bosons and top quarks.  Despite this difference, the relative resolutions are statistically comparable with each other with a significance slightly above $1\sigma$.   The right plot of Fig.~\ref{fig:forward3comapre813TeV} shows the relative JMS and JMR for the two mass reconstruction algorithms in Fig.~\ref{fig:forwardJETMASS2015}.  The values are similar within the statistical uncertainties, but as expected, the systematic uncertainty is smaller for the track-assisted jet mass (more details in Sec.~\ref{sec:TAMass}).

\begin{figure}[h!]
\centering
\includegraphics[width=0.5\textwidth]{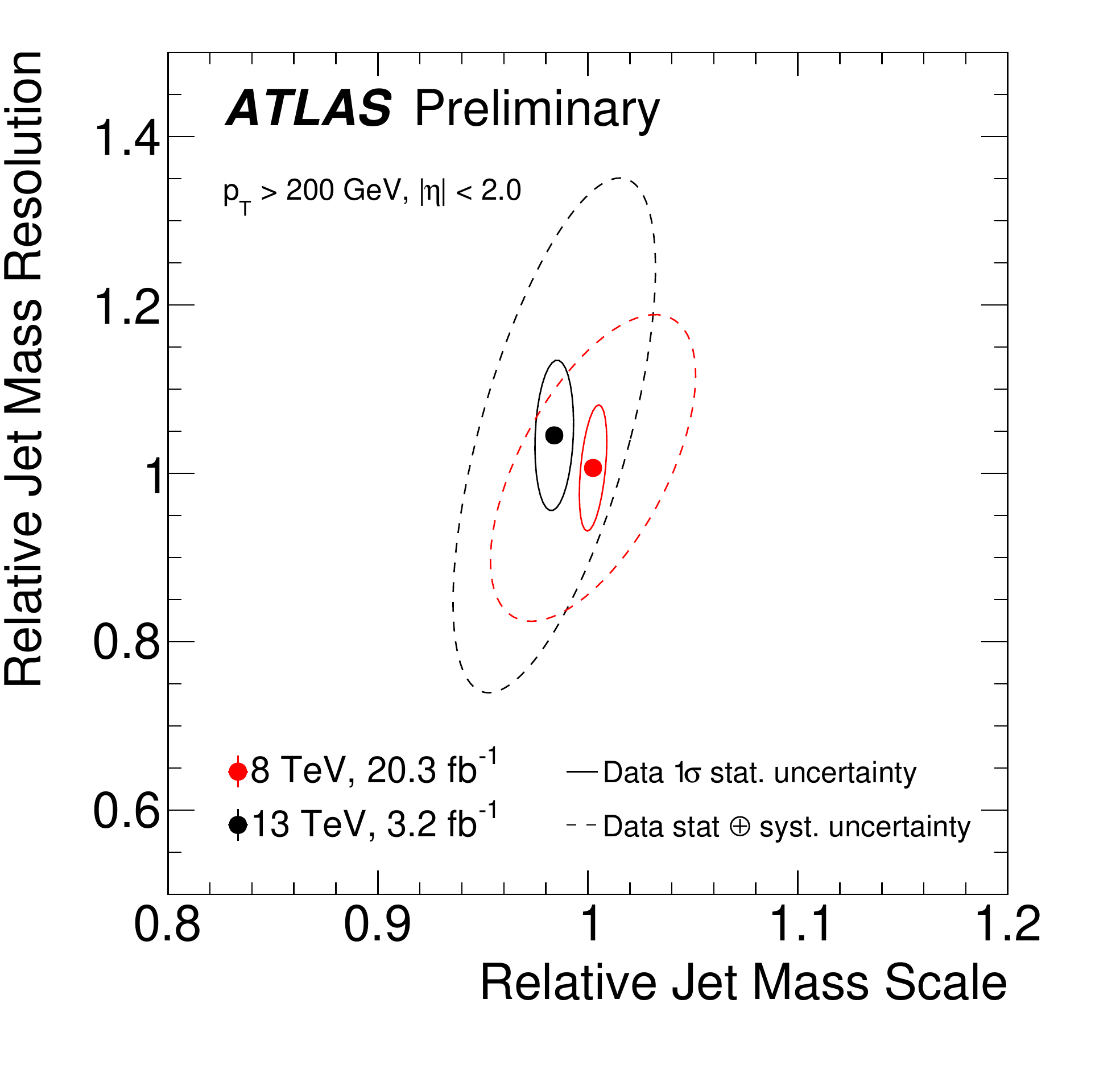}\includegraphics[width=0.5\textwidth]{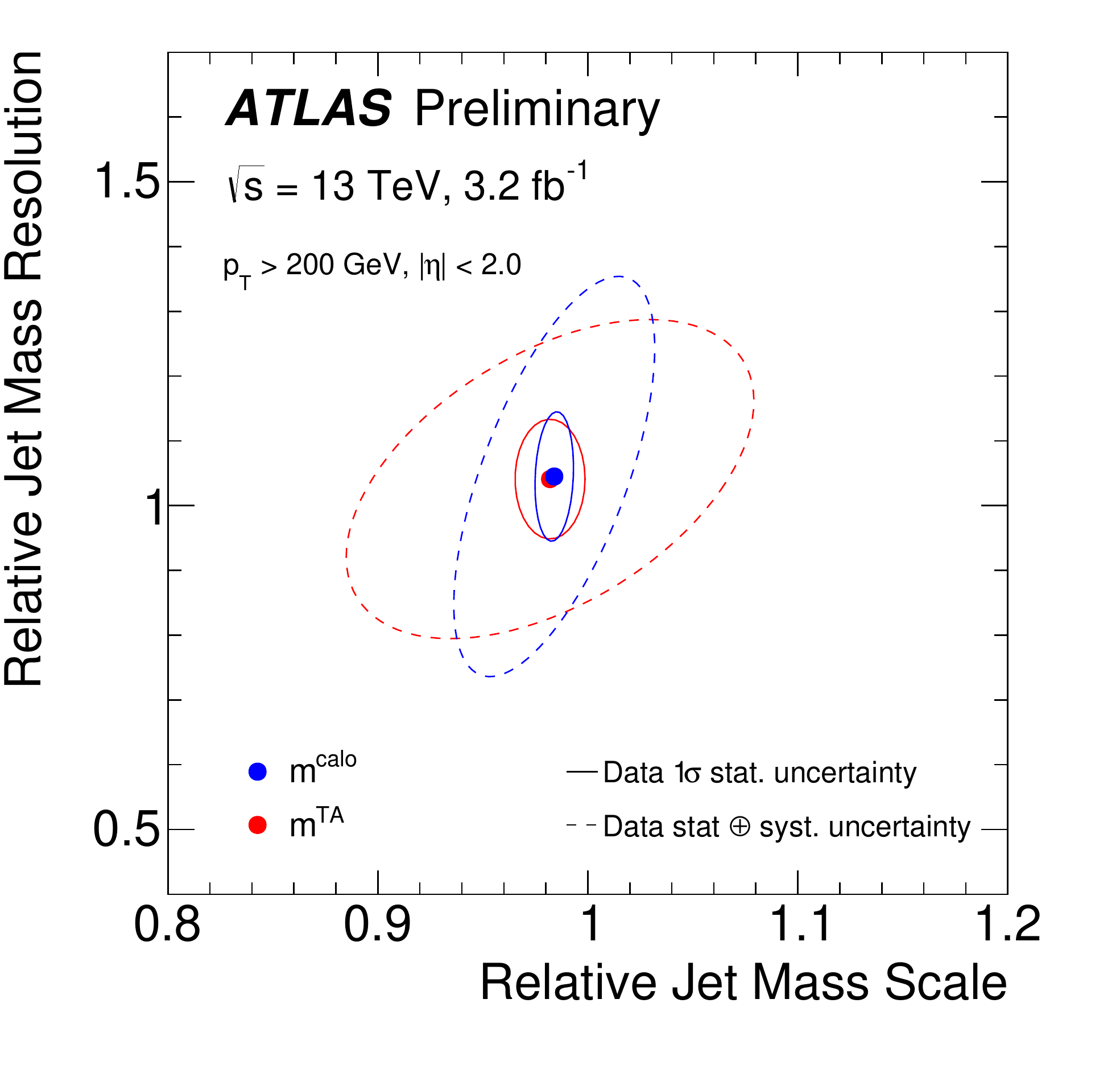}
\caption{Left: The statistical (solid) and total (dashed) 1$\sigma$ uncertainty ellipses for the relative jet mass scale and the relative jet mass resolution at $\sqrt{s}=8$ TeV (red) and $\sqrt{s}=13$ TeV (black).  Right: the relative scale and resolutions for the two mass reconstruction algorithms in Fig.~\ref{fig:forwardJETMASS2015}. }
\label{fig:forward3comapre813TeV}
\end{figure}

An innovation of the early Run 2 analysis is the extension of the forward-folding technique to measure the relative jet $p_\text{T}$ scale and resolution.  The resolution of any quantity can be measured with the forward-folding technique so long as one can identify a detector-level distribution that depends strongly on that resolution.  The two top quarks in $t\bar{t}$ production tend to be produced with a similar $p_\text{T}$.  Therefore, the leptonic top quark can be used as a proxy for the hadronically decaying top quark and thus $p_\text{jet}/p_\text{T}^\text{lep top}$ should be sensitive to the jet $p_\text{T}$ scale and resolution of the hadronically decaying top quark.  The full leptonic top quark $p_\text{T}$ can be reconstructed from the two-vector sum of the lepton momentum, the $\vec{p}_\text{T}^\text{miss}$, and the transverse momentum of the nearby jet (see Sec.~\ref{sec:samples}).  One disadvantage of using the full leptonic top $p_\text{T}$ is that it depends on calorimeter quantities ($\vec{p}_\text{T}^\text{miss}$ and the jet $p_\text{T}$).  One can reduce the calorimeter-dependence by using either just the two-vector sum of the jet $p_\text{T}$ and the lepton $p_\text{T}$ or just the lepton $p_\text{T}$.  The tradeoff for the reduced calorimeter-dependence is the reduced sensitivity to the large-radius jet $p_\text{T}$ resolution.  This is illustrated by the `peakiness' of the three distribution in Fig.~\ref{fig:forwardleppt}.  All three ratios show a peak near one and so would shift if the jet $p_\text{T}$ scale where mis-modeled.   However, the peak is sharpest with the full leptonic top, which suggests that the statistical uncertainty on the JER will be smallest when using this quantity. 

\begin{figure}[h!]
\centering
\includegraphics[width=0.33\textwidth]{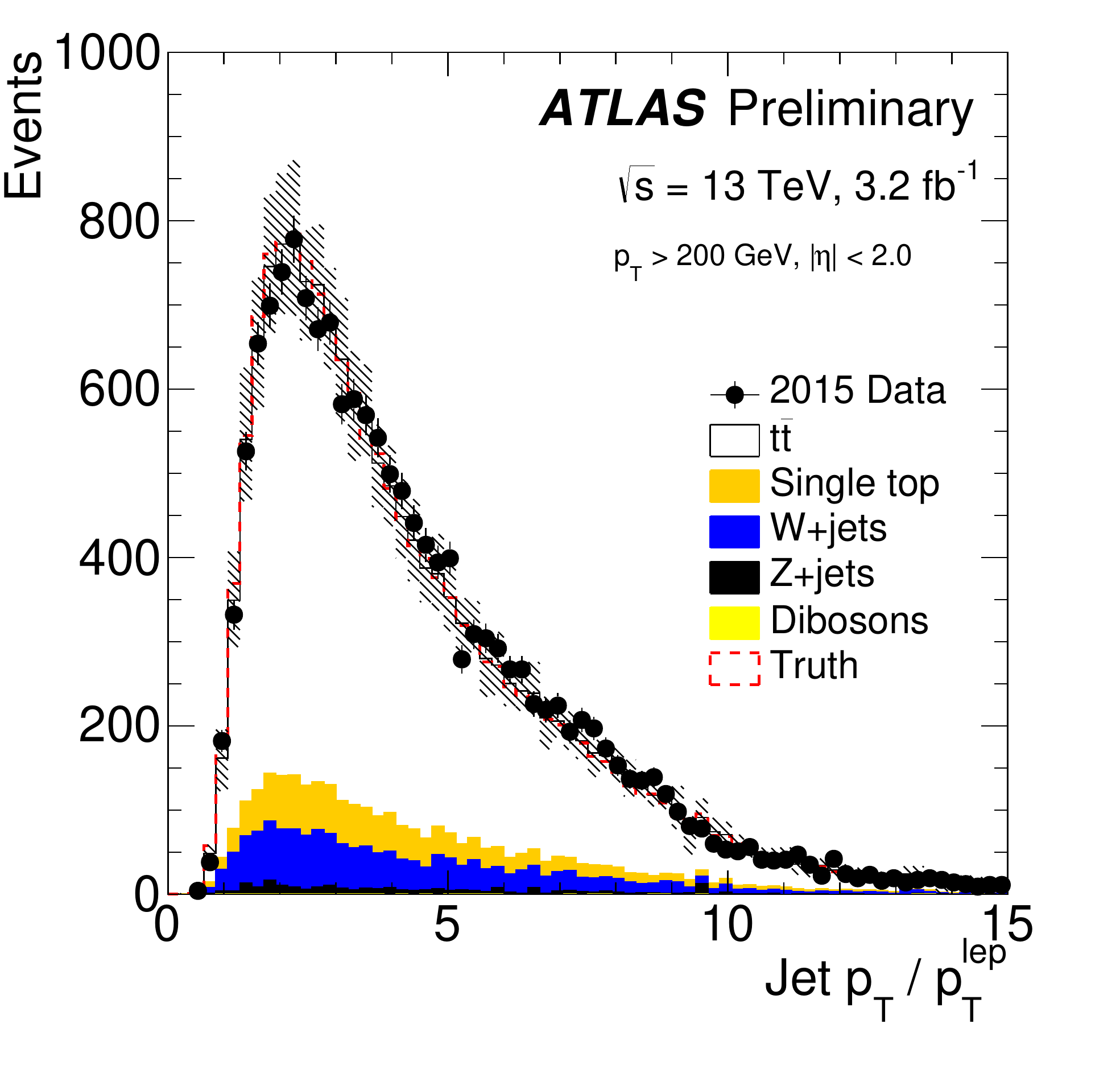}
\includegraphics[width=0.33\textwidth]{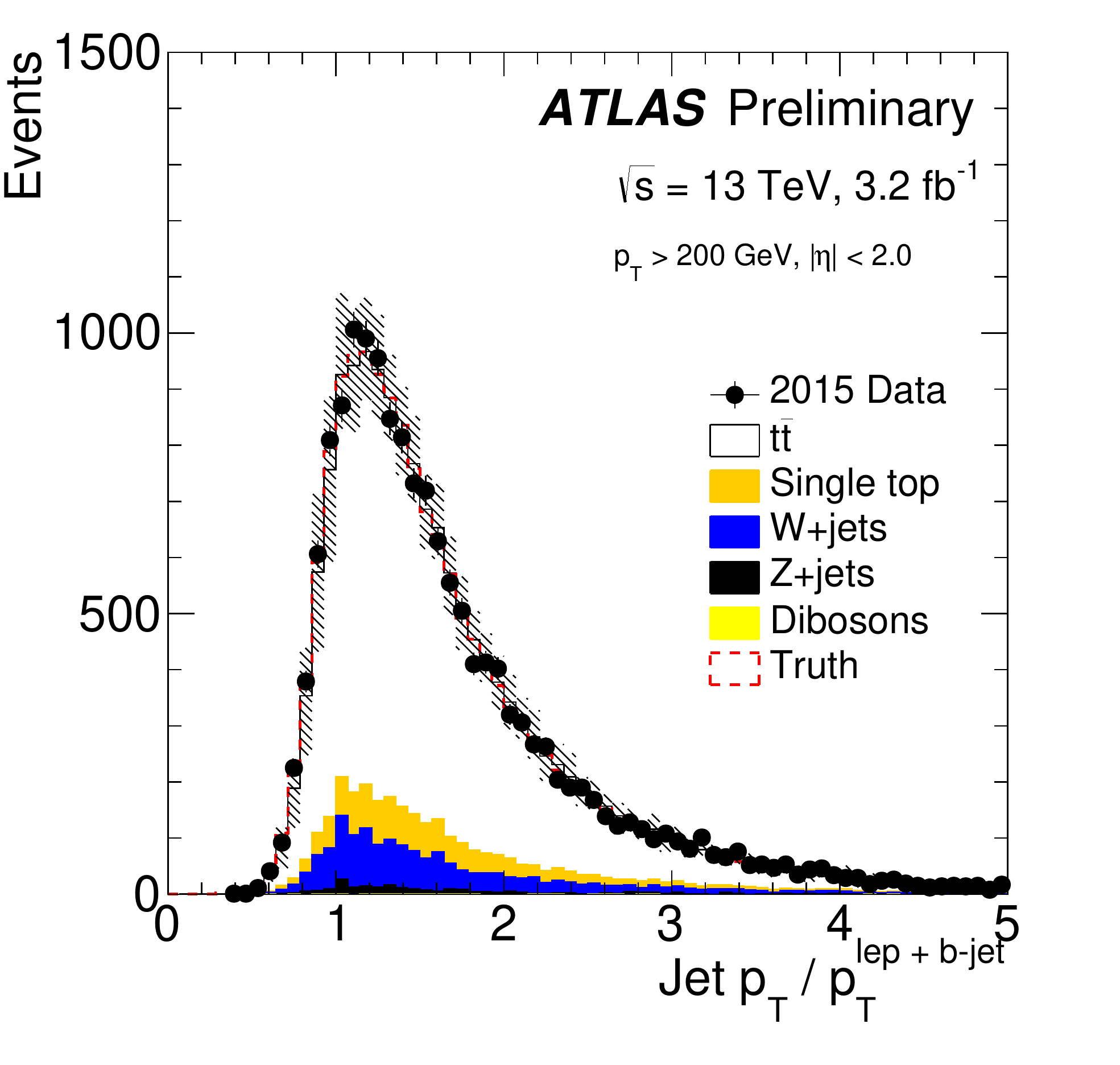}\includegraphics[width=0.33\textwidth]{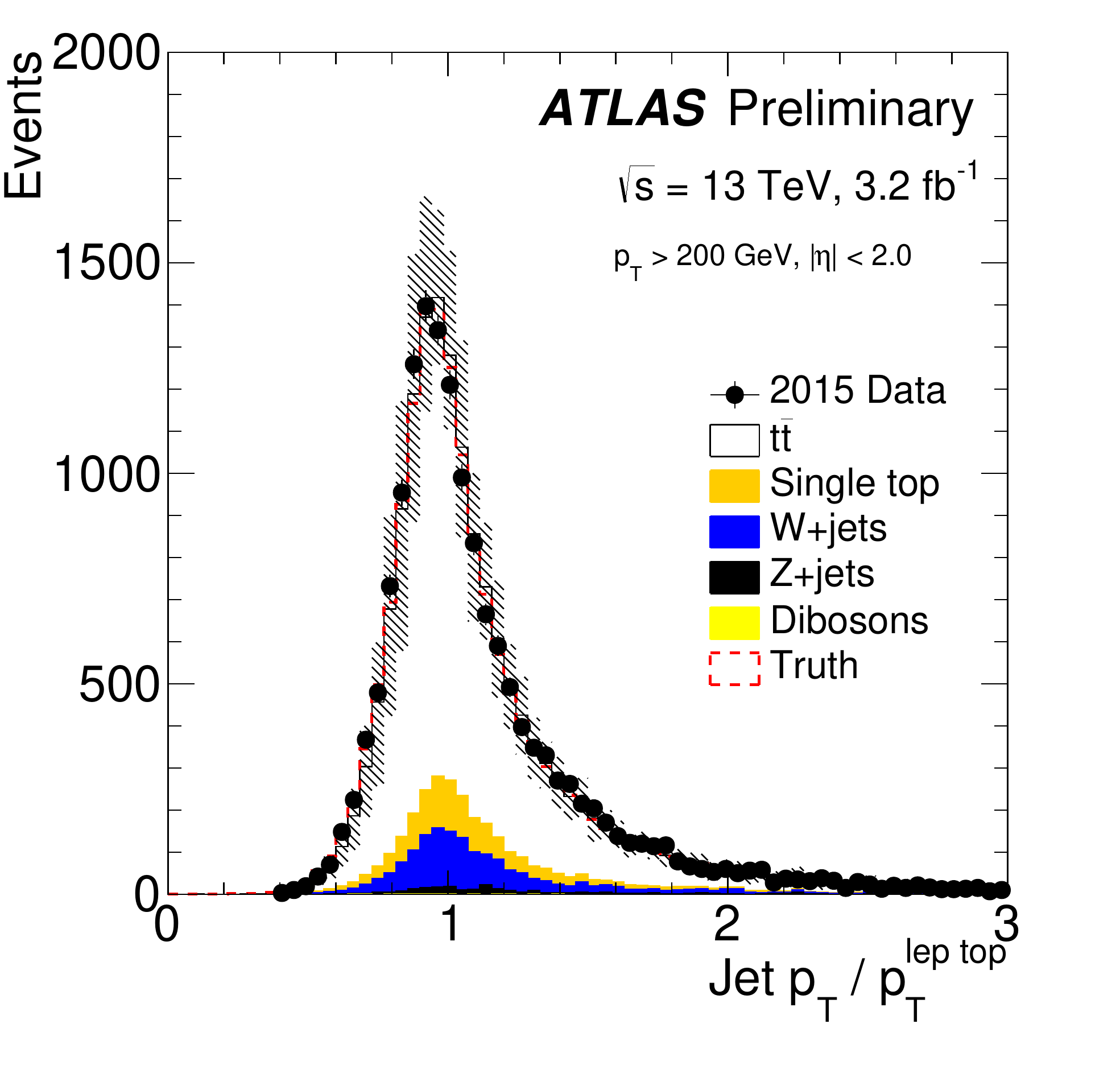}
\caption{The ratio of the large-radius jet $p_\text{T}$ to the lepton $p_\text{T}$ (left), the two-vector sum of the lepton and close-by jet transverse momentum (middle), and the full leptonically decaying top quark using the lepton, the close-by jet, and the $\vec{p}_\text{T}^\text{miss}$ (right).}
\label{fig:forwardleppt}
\end{figure}

A $\chi^2$ fit using the forward-folding method for each of the three observables from Fig.~\ref{fig:forwardleppt} is shown in Fig.~\ref{fig:forward3ptmarginalizeoverres} for the JES and in Fig.~\ref{fig:forward3ptmarginalizeoverscale} for the JER.  The particle-level quantity entering the forward-folding is $p_\text{T}^\text{particle-level jet}/p_\text{T}^\text{ref}$, where $p_\text{T}^\text{ref}$ is one of the {\it detector-level} quantities from Fig.~\ref{fig:forwardleppt}.  As expected from the sharpness of the ratio distributions in Fig.~\ref{fig:forwardleppt}, the ratio with the full leptonic top quark candidate has the deepest $\chi^2$ for the JES.  The $\chi^2$ distribution near the minimum in Fig.~\ref{fig:forward3ptmarginalizeoverscale} is rather flat: there is not much sensitivity to the JER due to the large width of the ratio distributions (large relative to the JER itself).  However, there is currently no in-situ constraint on the large-radius jet JER and so even a crude uncertainty is an important step forward.

\begin{figure}[h!]
\centering
\includegraphics[width=0.33\textwidth]{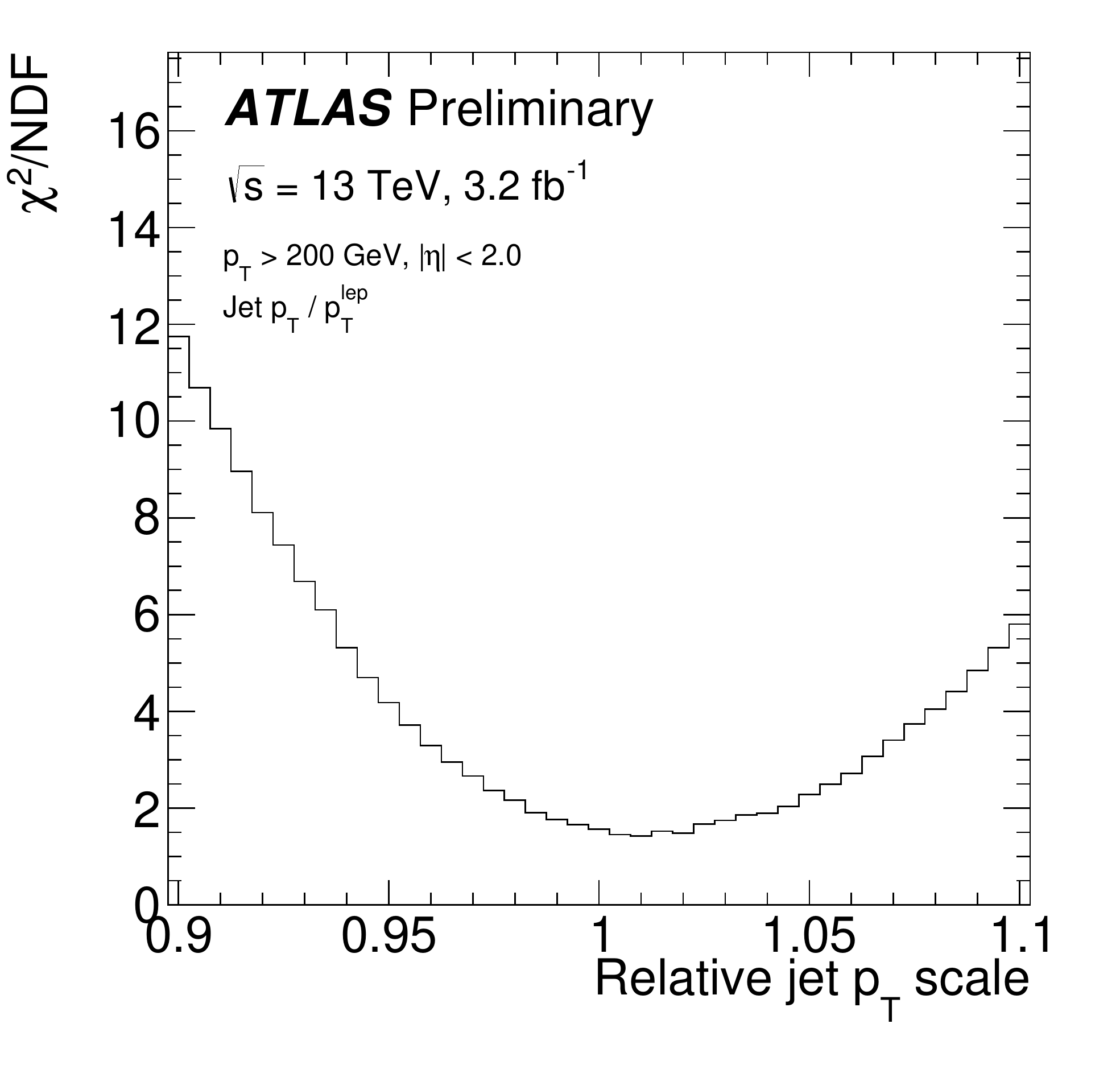}
\includegraphics[width=0.33\textwidth]{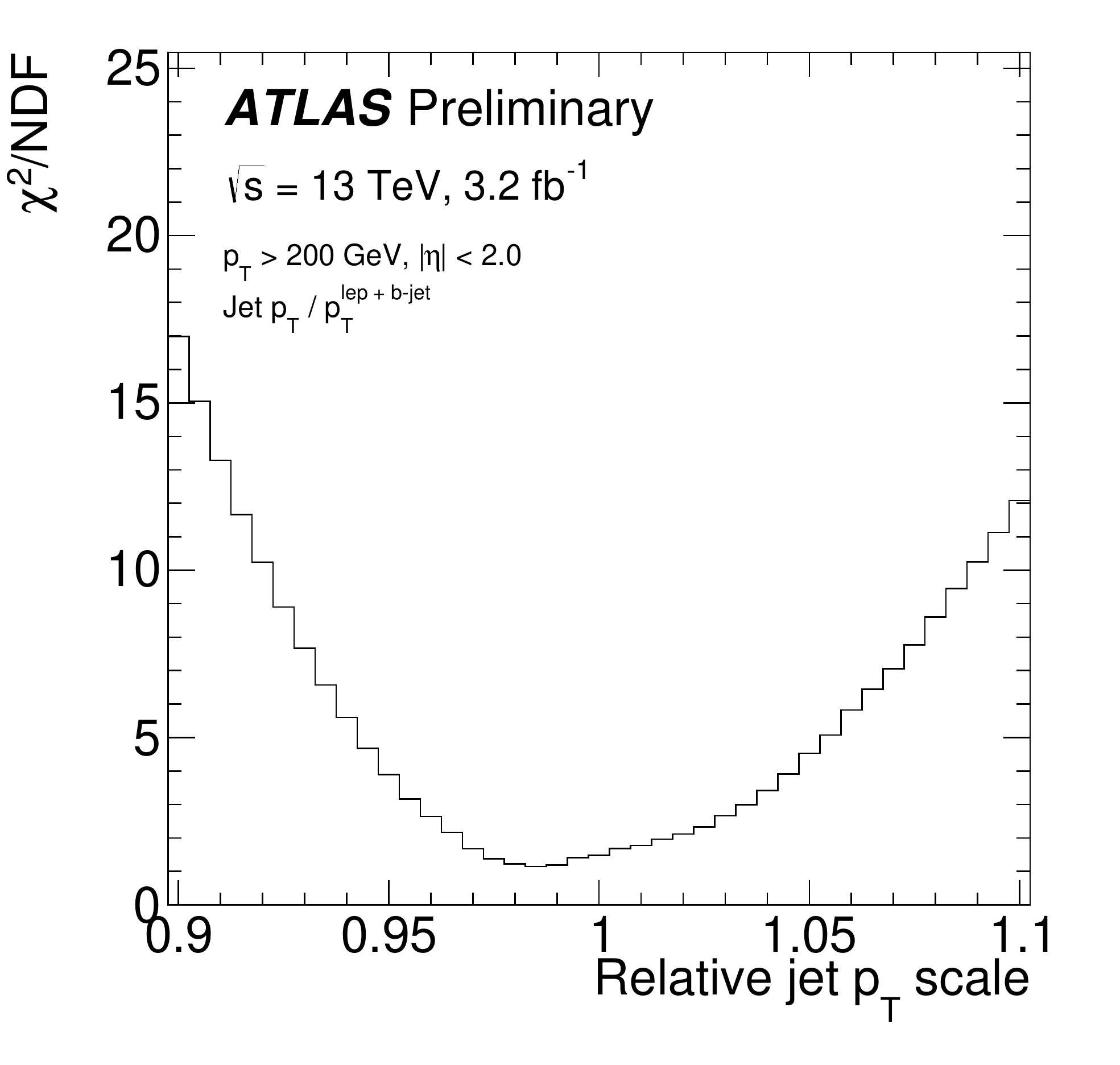}\includegraphics[width=0.33\textwidth]{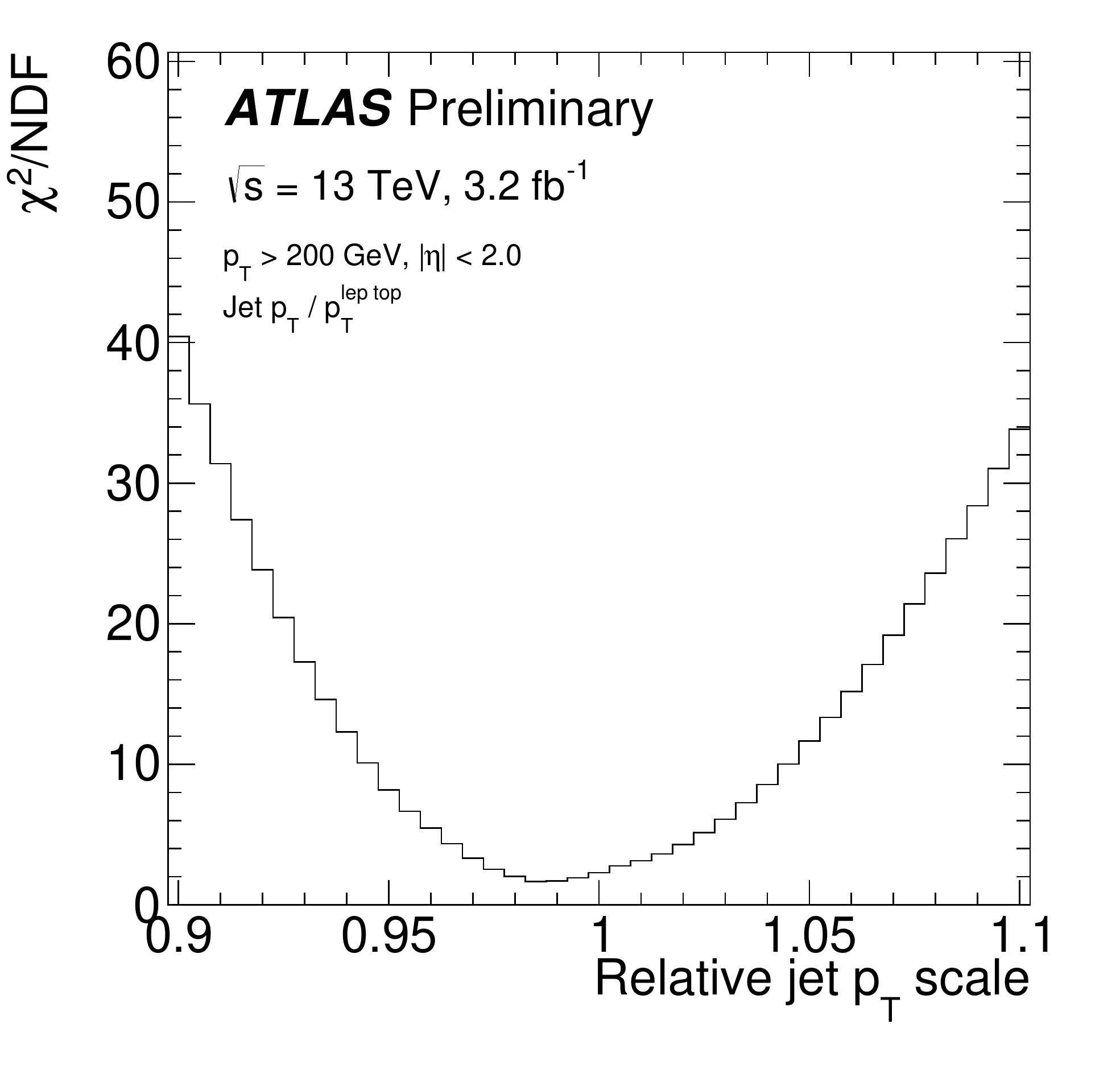}
\caption{The value of the $\chi^2$ per degree of freedom at a given relative jet $p_\text{T}$ scale minimized over the jet $p_\text{T}$ resolution for three different reference objects: the lepton $p_\text{T}$ (left), the two-vector sum of the lepton and close-by jet transverse momentum (middle), and the full leptonically decaying top quark (right).}
\label{fig:forward3ptmarginalizeoverres}
\end{figure}

\begin{figure}[h!]
\centering
\includegraphics[width=0.33\textwidth]{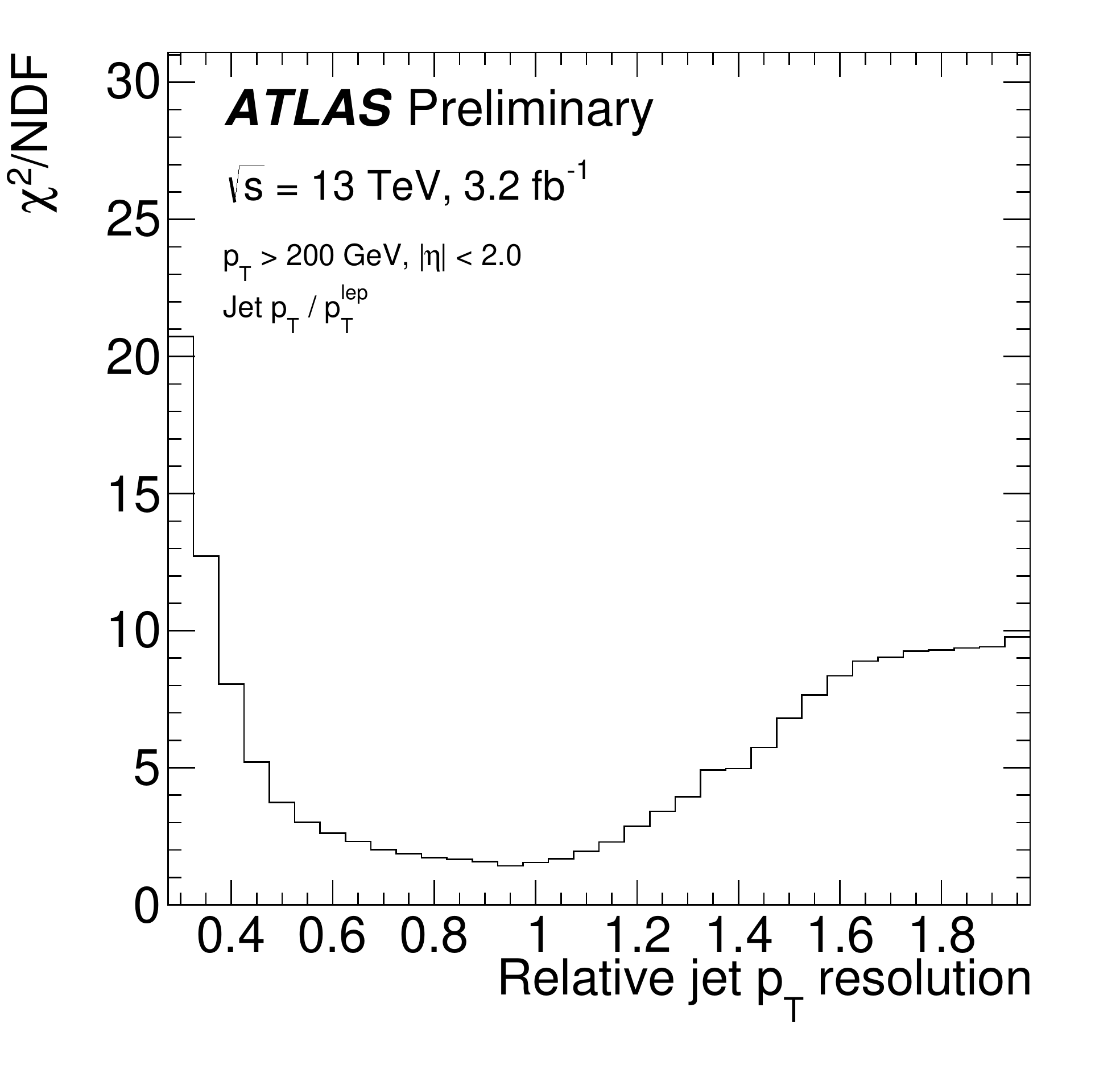}
\includegraphics[width=0.33\textwidth]{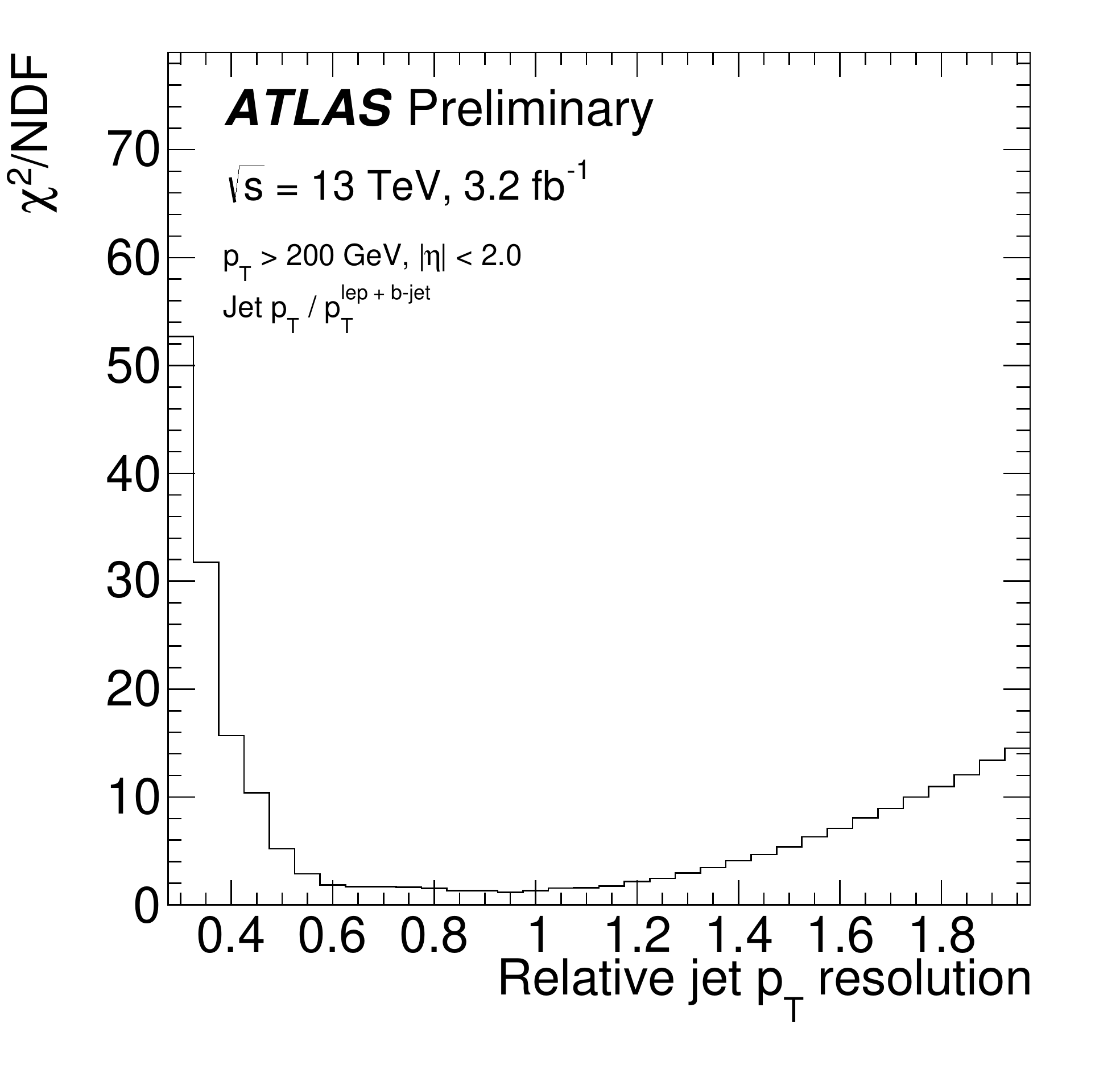}\includegraphics[width=0.33\textwidth]{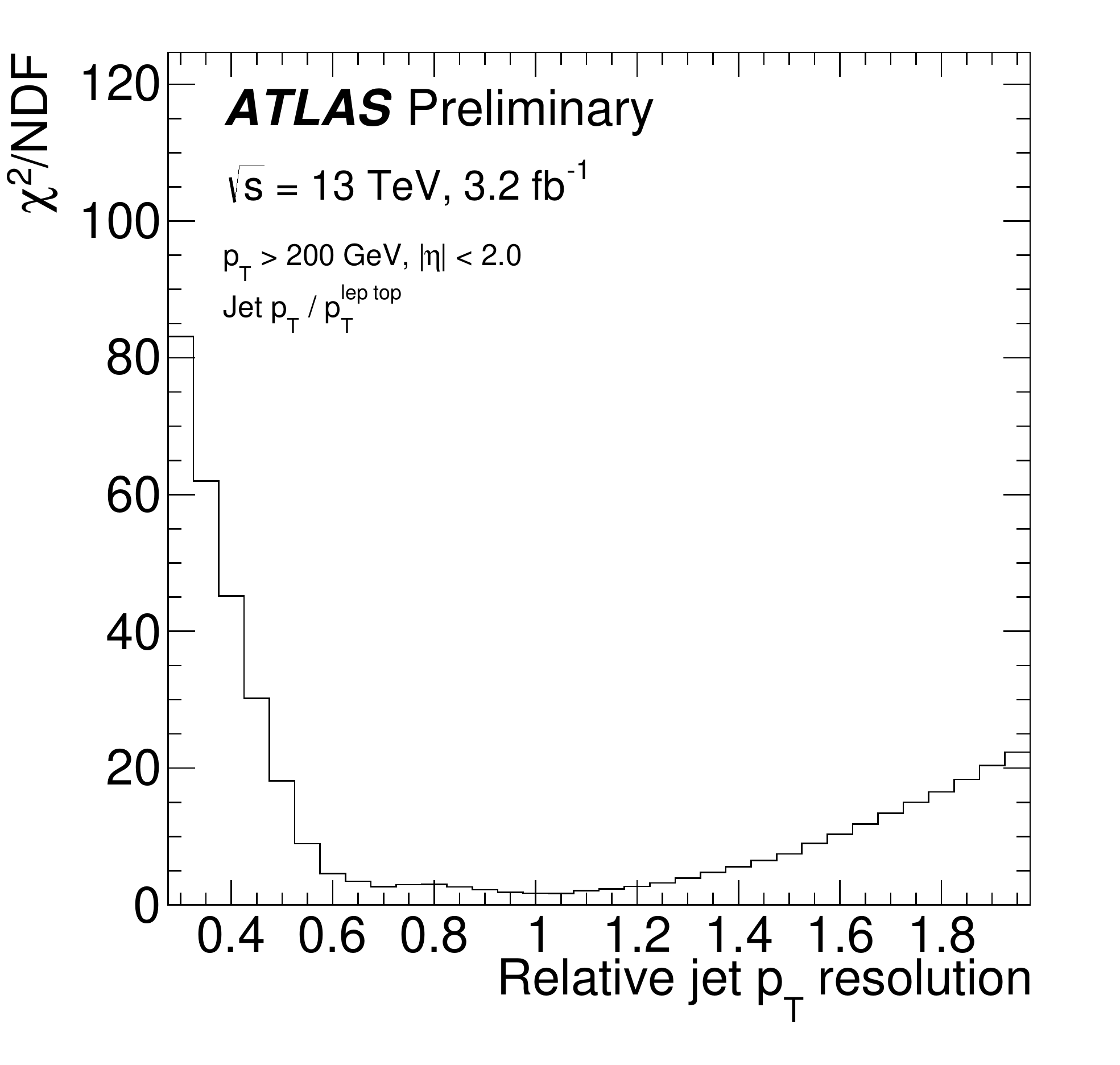}
\caption{The value of the $\chi^2$ per degree of freedom at a given relative jet $p_\text{T}$ resolution minimized over the jet $p_\text{T}$ scale for three different reference objects: the lepton $p_\text{T}$ (left), the two-vector sum of the lepton and close-by jet transverse momentum (middle), and the full leptonically decaying top quark (right).}
\label{fig:forward3ptmarginalizeoverscale}
\end{figure}

Table~\ref{tab:sum} summarizes the fitted values of the relative JES, JER, JMS, and JMR using the early Run 2 dataset for the mass and $p_\text{T}$.  The three jet $p_\text{T}$ scale determinations are not all statistical consistent with each other, but there are significant systematic uncertainties that are not fully correlated between the methods.  Theoretical modeling uncertainties, which are an important ingredient for the particle-level templates dominate dominate over the JES/JMS uncertainty\footnote{The large-R JES (JMR) is not included as an uncertainty when the JES (JMR) is measured.} in most cases.  The more calorimeter information used to extract the $p_\text{T}$ scale and resolution, the larger the experimental uncertainty.  The two methods that do not use the missing momentum have a similar precision for both the JES and JER.

	These early measurements of the relative jet mass and jet $p_\text{T}$ scale and resolution are already dominated by systematic uncertainties.  In the future, it is therefore crucial to perform this measurement differentially in $p_\text{T}$ and jet substructure (such as $n_\text{subjets}$) which will mitigate (some) of the modeling dependence.

\begin{table}[h]
\centering
\begin{tabular}{l|c|c|c|c|c|c}
Quantity &  &Value & Stat. Uncert & Modeling & Jets & Total Syst.\\
\hline\hline

$m^\text{calo}$ & $s_\text{data}^\text{MC}$  & 0.984 & 0.6 \% & 1.7 \% & 1.6 \% & {\bf 2.3 \%}\\
$m^\text{calo}$ & $r_\text{data}^\text{MC}$  & 1.047 & 6.6 \% & 18.1 \% & 7.0 \% & {\bf 19.4 \%}\\
$m^\text{TA}$ & $s_\text{data}^\text{MC}$  & 0.981 & 1.1 \% & 2.4 \% & 4.8 \% & {\bf 5.3 \%}\\
$m^\text{TA}$ & $r_\text{data}^\text{MC}$  & 1.036 & 6.1 \% & 14.6 \% & 5.0 \% & {\bf 15.5 \%}\\
$p_\text{T,jet}/p^\text{lep}_\text{T}$ & $s_\text{data}^\text{MC}$ & 1.011 & 0.7 \% & 1.3 \% & 0.4 \% & {\bf 1.3 \%}\\
$p_\text{T,jet}/p^\text{lep}_\text{T}$ & $r_\text{data}^\text{MC}$ & 0.945 & 4.1 \% & 6.8 \% & 2.7 \% & {\bf 7.3 \%}\\
$p_\text{T,jet}/p^\text{lep + b-jet}_\text{T}$ & $s_\text{data}^\text{MC}$ & 0.985 & 0.4 \% & 0.7 \% & 1.2 \% & {\bf 1.4 \%}\\
$p_\text{T,jet}/p^\text{lep + b-jet}_\text{T}$ & $r_\text{data}^\text{MC}$ & 0.903 & 6.1 \% & 5.5 \% & 4.7 \% & {\bf 7.2 \%}\\
$p_\text{T,jet}/p^\text{lep top}_\text{T}$ & $s_\text{data}^\text{MC}$ & 0.987 & 0.2 \% & 0.3 \% & 2.1 \% & {\bf 2.1 \%}\\
$p_\text{T,jet}/p^\text{lep top}_\text{T}$ & $r_\text{data}^\text{MC}$ & 1.024 & 3.1 \% & 6.2 \% & 6.0 \% & {\bf 8.6 \%}\\
\hline\hline\end{tabular}\caption{Summary of the systematic uncertainties for the relative jet mass or energy scales ($s_\text{data}^\text{MC}$) and resolutions ($r_\text{data}^\text{MC}$). The first column states which observable is used to extract the relative jet mass (first four rows) or jet energy (rows 5-10) scale and resolutions.}\label{tab:sum}\end{table}

	\clearpage

	\subsection{Re-clustered Jet Mass}
	\label{sec:ReclusteredJetMass}

As discussed\footnote{The phenomenological studies presented in this section are published in Ref.~\cite{recluster} and include technical inputs from M. Swiatlowski and P. Nef.} in Sec.~\ref{cha:bosonjets}, the angular separation between decay products of a massive particle $\mathcal{P}$, such as a $W$ or $Z$ boson, scales as $2 m_\mathcal{P}/p_T^\mathcal{P}$.  This suggests that the radius parameter $R$ of jet clustering algorithms aimed at collecting the hadronic decay products of $\mathcal{P}$ should be process dependent and scale with the momentum under consideration.  However, at the LHC, most analyses use one global value of $R$ fixed ahead of time.  In ATLAS, this value is $R=1.0$ for large-radius jets and $R=0.4$ for small-radius jets.  The reason for a fixed jet radius is that every jet configuration, which includes the algorithm, radius, and grooming parameters, must be calibrated to account for unmeasured energy deposits and other experimental effects~\cite{jes,Chatrchyan:2011ds}, even though the inputs to jet clustering are themselves calibrated. The calibration of inputs provides a partial calibration to the jet, but jet energy and mass scale corrections provide a \textit{full} calibration by also correcting for particles that were missed, merged, or below noise thresholds, energy loss in un-instrumented regions of the calorimeter, and additionally takes into account correlations between particles. The dependence on these additional calibrations thus makes it desirable to reconsider the current jet clustering paradigm in favor of a modular structure that allows for a much broader class of algorithms and radius parameters to be selected by analyses.

One solution is to introduce a new angular scale $r<R$, such that jets of radius $r$ can be the inputs to the clustering algorithm of large radius $R$ jets\footnote{Similar ideas have been proposed in the past such as variable $R$ jets~\cite{Krohn:2009zg}.  While these methods address the variability of $R$, they do not address the concerns about calibrations and uncertainties.}.  If chosen appropriately, the fully calibrated small radius jets can make the calibration of the re-clustered large radius jets automatic.  Furthermore, with no additional calibration needed, any large radius $R$, any clustering algorithm, and many grooming strategies can be simultaneously implemented in an analysis.  Using optimal parameters can, for instance, significantly improve the discovery potential of searches for new physics~\cite{Salam:2009jx}.  In particular, every kinematic region of every analysis for every data-taking condition (e.g. level of pileup) can be individually optimized in order to maximize the sensitivity to particular physics scenarios.  Another benefit is that the uncertainties on the re-clustered $p_\text{T}$ and mass are also automatic consequences of propagating the corresponding uncertainties computed for small radius jets.  In this way, the re-clustered jet mass can be viewed as any other kinematic variable, such as di-, tri-, or multi-jet invariant masses that are ubiquitous in measurements and searches for new physics. 
The idea of re-clustering small radius jets is not new.  These objects first appeared in an ATLAS search for supersymmetry in the multijet final state~\cite{Aad:2013wta} and more recently in an ATLAS search for direct stop quark pair production in the all hadronic final state~\cite{Aad:2014bva}.  There are also related techniques which group small radius jets together to form pseudo-jets~\cite{Chatrchyan:2012jx} or mega-jets~\cite{Chatrchyan:2011ek,Chatrchyan:2012uea,Chatrchyan:2014goa}.  This section introduces a new way of thinking about re-clustering.  Instead of viewing jet grouping as a high-level analysis technique, the idea is to consider re-clustered jets as if they were any other jet collection clustered directly from low-level objects.  This is a signifiant paradigm shift because there is an entire class of techniques for using and improving large-radius jets.  For example, re-clustered jets can be groomed and their substructure can be useful for tagging.  

This section is organized as follows.  Section~\ref{sec:reclusteringintro} introduces the technical details and benefits of re-clustering small-radius jets.  Sections~\ref{sec:reluster:particlelevelperformance} an~\ref{sec:recluster:jss} describe performance studies at particle-level for the jet mass and other jet substructure, with a particular emphasis on pileup.  The dependence of jet tagging on re-clustering parameters with the full ATLAS detector simulation is studied in Sec.~\ref{sec:recluster:tagging} and the performance of re-clustered jet mass reconstruction relative to standard large-radius jets is discussed in Sec.~\ref{sec:recluster:calibrations}.  One of the key assumptions of re-clustering is that the impact of close-by jets on the jet energy scale response is small or at least well-modeled by the simulation.  In-situ track jet methods are used to investigate the impact of close-by jets in Sec.~\ref{sec:recluster:closeby}. Section~\ref{sec:reclustering:conclucion} provides an overview and outlook for re-clustering.

\clearpage

\subsubsection{Re-clustering Jets}
\label{sec:reclusteringintro}

The inputs of jet clustering algorithm are typically stable particles (Monte Carlo truth studies), topological clusters (ATLAS), or particle flow objects (CMS).  Re-clustered large radius $R$ jets take as input the output of the  small radius $r$ jet clustering.  Small radius jets have been calibrated with $r$ as small as $0.2$~\cite{ATL-PHYS-PUB-2015-053} and there are no indications of sizable mis-modelling of close-by effects in the jet response for the standard $R=0.4$ jets~\cite{jes} (see Sec.~\ref{sec:recluster:closeby} for more detail).  In general, the algorithm used to cluster the small radius jets can be different than the algorithm used for re-clustering the entire event.  Fig.~\ref{fig:eventdisplay} shows a simple example of an event clustered with  anti-$k_t$ $R=1.0$ and with anti-$k_t$ $R=1.0$ re-clustered $r=0.3$ anti-$k_t$ jets.  Unlike the inputs of  clustering which are e.g. measured in a calorimeter and can be reconstructed and individually calibrated with very low energy, small radius jets can only be reliably fully calibrated for $\gtrsim 15$ GeV~\cite{jes,Chatrchyan:2011ds}, where the actual threshold may depend on $r$.  This minimum $p_\text{T}$ threshold acts as an effective grooming for the re-clustered jets (RC).  This is seen clearly in Fig.~\ref{fig:eventdisplay}, where the blue  large radius jet has many constituents far away from the jet axis (which have low $p_\text{T}$) and are not part of the re-clustered jet.   One could choose a more aggressive threshold to, for instance, remove the impact of additional $pp$ collisions (i.e. pileup) on the jets.  A more dynamic grooming scheme, named {\it re-clustered jet trimming} in analogy to large radius jet trimming~\cite{Krohn:2009th},  sets the $p_\text{T}$ cut on the small radius jets based on the large radius jet $p_\text{T}$ (calculated before any small-$r$ jets are removed).  Specifically, for re-clustered and trimmed jets (RT), the grooming removes any small radius jet constituent $j$ of a large $R$ re-clustered jet $J$ if $p_\text{T}^j<f_\text{cut} \times p_T^J$.  The parameter $f_\text{cut}$ can be optimized for a particular kinematic selection and event topology.  Other grooming schemes are possible, but beyond the scope of this section\footnote{Jet grooming procedures applied to jets-as-inputs have been studied in the past (see for instance Ref.~\cite{Gouzevitch:2013qca}); these and other algorithms can be adopted to the re-clustering paradigm.}.

\begin{figure}[h!]
\begin{center}
\includegraphics[width=.95\textwidth]{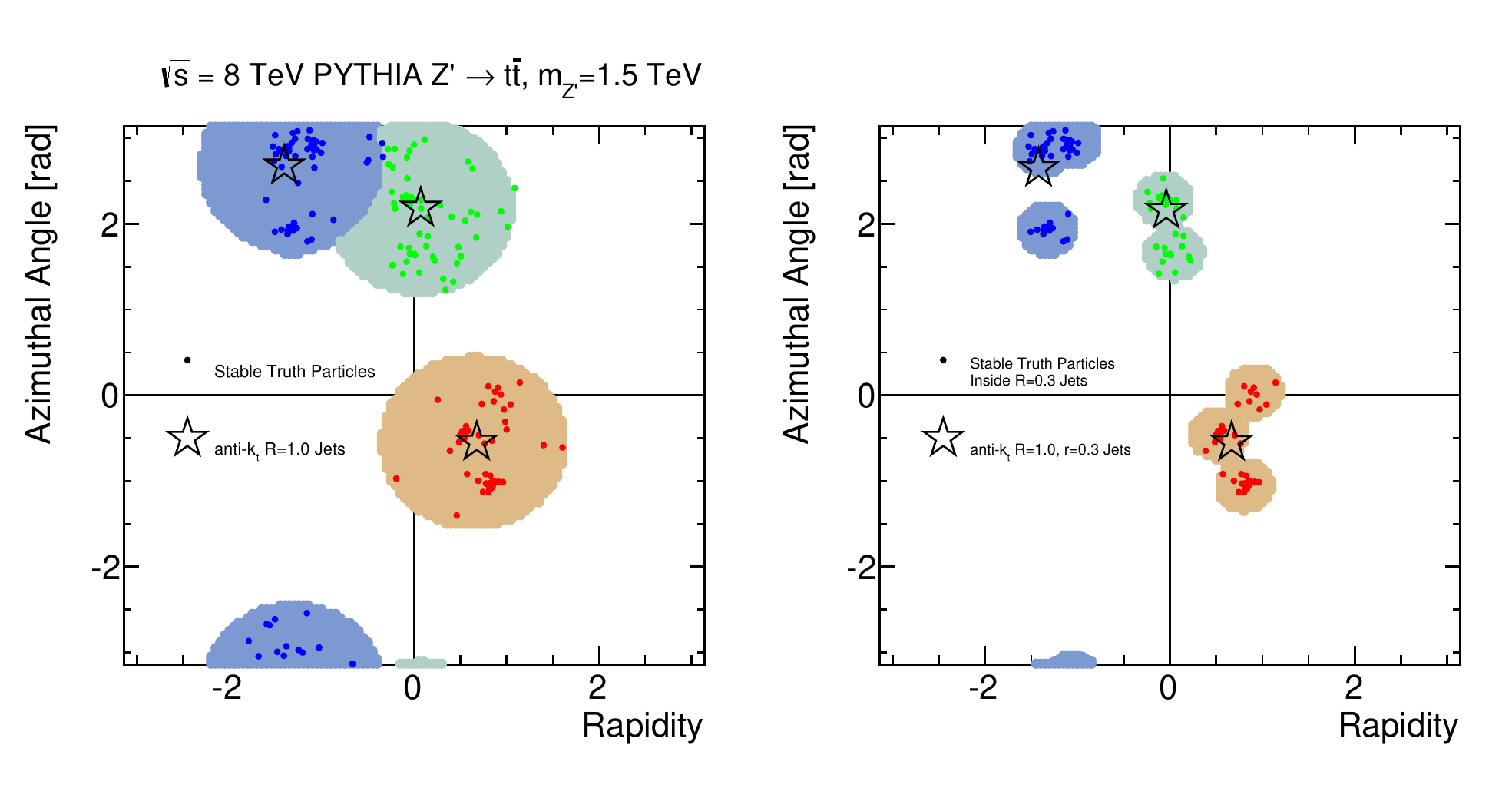}
\end{center}
\caption{An example event which has been clustered using the  anti-$k_t$ $R=1.0$ (left) and with anti-$k_t$ $R=1.0$ re-clustered $r=0.3$ anti-$k_t$ jets (right).  The shaded regions show the jet area determined by clustering ghost particles.  Only large radius jets with $p_\text{T}>50$ GeV are shown and small radius jets are required to have $p_\text{T}>15$ GeV.   As with groomed large radius jets, re-clustered jets can have non-circular shapes.}
\label{fig:eventdisplay}
\end{figure}

Due to the increased catchment area of large radius jets over small radius jets, they are more susceptible to contributions from pileup.  Just as there are pileup correction techniques for large radius jets and their subjets, one can benefit from pileup corrections to the small radius jet inputs that propagate to re-clustered jets.  In particular, one can remove jets from pileup interactions with techniques like JVT~\cite{Aad:2015ina} or pileup jet identification~\cite{CMS:2013wea} and can correct the remaining jets with methods like the four-vector jet areas subtraction.    Another way to mitigate the impact of pileup is to correct jet constituents before clustering~\cite{puppi,constsub,softkill}, which is similar to the $r\rightarrow 0$ limit.  However, applying a jet constituent pileup correction still requires an overall calibration and an intermediate scale $r\sim 0.1-0.5$ is one possibility.

In the growing field of jet substructure, there are many jet observables which depend explicitly on the jet constituents, not just the jet four-vector.  These techniques are still applicable for re-clustered jets.  One possibility is to compute substructure observables using the small-radius jet constituents inside the re-clustered jet.  This approach should be similar to the jet substructure of a traditional groomed large-radius jet.  An alternative {\it bottom-up} approach to jet substructure is to use the radius $r$ jets directly as the inputs to jet substructure.   The advantages and limitations of bottom-up substructure are described in Section~\ref{sec:recluster:jss}.
	
There are other technical benefits to re-clustering.  For example, re-clustering can be much faster than traditional jet clustering.  Jet clustering is an order $N \log{N}$ operation~\cite{Cacciari:2005hq} -- as the number of jet inputs in an event approaches $500$ or more at high pileup conditions, jet clustering can take a significant amount of the full event reconstruction time.  This is particularly relevant if one wants to scan the jet clustering parameters.   However, there are typically 10 or fewer jets above the calibrated $p_\text{T}$ threshold in any given event (though this obviously depends on the $R$ size and threshold). With these typical numbers, creating a re-clustered jet is about 100 times faster than clustering a large-$R$ jet directly. This kind of computational speed-up can allow analysis end-users -- and not just large, central productions -- to produce their own large-$R$ jets, allowing for more creativity in exploring the optimal jet algorithms and parameters for analyses. 
	
	\subsubsection{Particle-level Jet Mass Performance}
	\label{sec:reluster:particlelevelperformance}
	
	Three processes are generated using \textsc{Pythia} 8.170~\cite{Sjostrand:2007gs,Sjostrand:2006za} at $\sqrt{s}=14$ TeV for studying the efficacy of re-clustered jets.  Hadronic $W$ boson and top quarks are used for studying hard 2- and 3-prong type jets.  To simulate high $p_\text{T}$ hadronic $W$ decays, $W'$ bosons are generated which decay exclusively into a $W$ and $Z$ boson which subsequently decay in quarks and leptons, respectively.  The $p_\text{T}$ scale of the hadronically decaying $W$ is set by the mass of the $W'$ which is tuned to $800$ GeV for this study so that the $p_\text{T}^W \lesssim 400$ GeV. In this $p_\text{T}^W$ range, not all of the decay products of the $W$ are expected to merge into a small radius jet of $r\lesssim 0.4$, but should merge within a cone of $R=1.0$.  A sample enriched in 3-prong type jets is generated with $Z'\rightarrow t\bar{t}$, with $m_{Z'}=1.0$ TeV, so that $p_\text{T}^t \gtrsim 350$ GeV.  To study the tradeoff between signal and background jet identification, QCD dijets are generated with a $p_\text{T}$ spectrum similar to the relevant signal process.  Pileup is modeled by overlaying additional independently generated minimum-bias interactions with each signal event.  The number of pileup interactions is between LHC Run 1 conditions, $n_\text{PU}=20$, and the conditions toward the end of the LHC Run 2, $n_\text{PU}=80$.

Jet are re-clustered using \textsc{FastJet}~\cite{Cacciari:2011ma} 3.0.3. While the large radius jets can be defined using any set of parameters, the studies in this section use a fixed large jet algorithm: anti-$k_t$ algorithm with $R=1.0$.  The reference jets are trimmed using  $R_\text{sub}=0.3$ $k_t$ subjets with a $p_\text{T}$ fraction threshold of $f_{cut}=0.1$. Unlike the procedure used by most analyses, all momenta are pileup corrected using the jet areas technique prior to grooming.  This is the natural setup for re-clustering and makes the optimal grooming parameters independent of $n_\text{PU}$.

Re-clustering is investigated with a series of schemes for the small radius jets: anti-$k_t$ radius parameters in $\{0.2,0.3,0.4\}$ grooming $f_\text{cut}=0.1$ and $0.2$ (with a $p_\text{T}=15$ GeV threshold).  This list is not exhaustive, but encompasses a relevant set of parameters.   Radii below $r=0.2$ are not considered due to experimental limitations from calorimeter granularity and theoretical considerations from non-trivial non-perturbative effects.  All small radius jets are required to have $p_\text{T}>15$ GeV.

As the jet mass is the mostly widely used large-radius jet observable, it is used to benchmark various re-clustering schemes.  The jet mass performance is quantified by the average jet mass $\langle m\rangle$, the standard deviation of the jet mass distribution ($\sigma$), and the dependance of these quantities with the amount of pileup.  The averages and  deviations are computed over a fixed mass range: $60$-$100$ GeV.  Another useful metric is the efficiency of a $60<m_\text{jet}/\text{GeV}<100$ requirement.  Figures~\ref{fig:fixed} and~\ref{fig:fixed2} compare RC with two settings of RT.  In the region near the $W$ mass peak, re-clustered trimming with $f_\text{cut}=0.2$ performs the best in terms of the mass distribution standard deviation in the $W$ mass window.  However, there is a sizable peak at low mass where too many jets have been cut out by the aggressive trimming parameter.  The fixed cut of $15$ GeV is too low, especially at very high pileup where the large high mass tail is much bigger for RC than for RT.  The re-clustered trimming using anti-$k_t$ with the same $f_\text{cut}$ as the  trimming has very similar performance, though the peak position is slightly higher.  Figure~\ref{fig:fixed2} shows the performance metrics as a function of NPV for the various grooming schemes.  The average mass for RT is very stable, whereas there is a slight slope for RC.   The mass resolution for RC is slightly worse than for RT, but the efficiency of RC is better because it avoids the peak at low masses well below the $W$ boson mass.

\begin{figure}[h!]
\begin{center}
\includegraphics[width=0.45\textwidth]{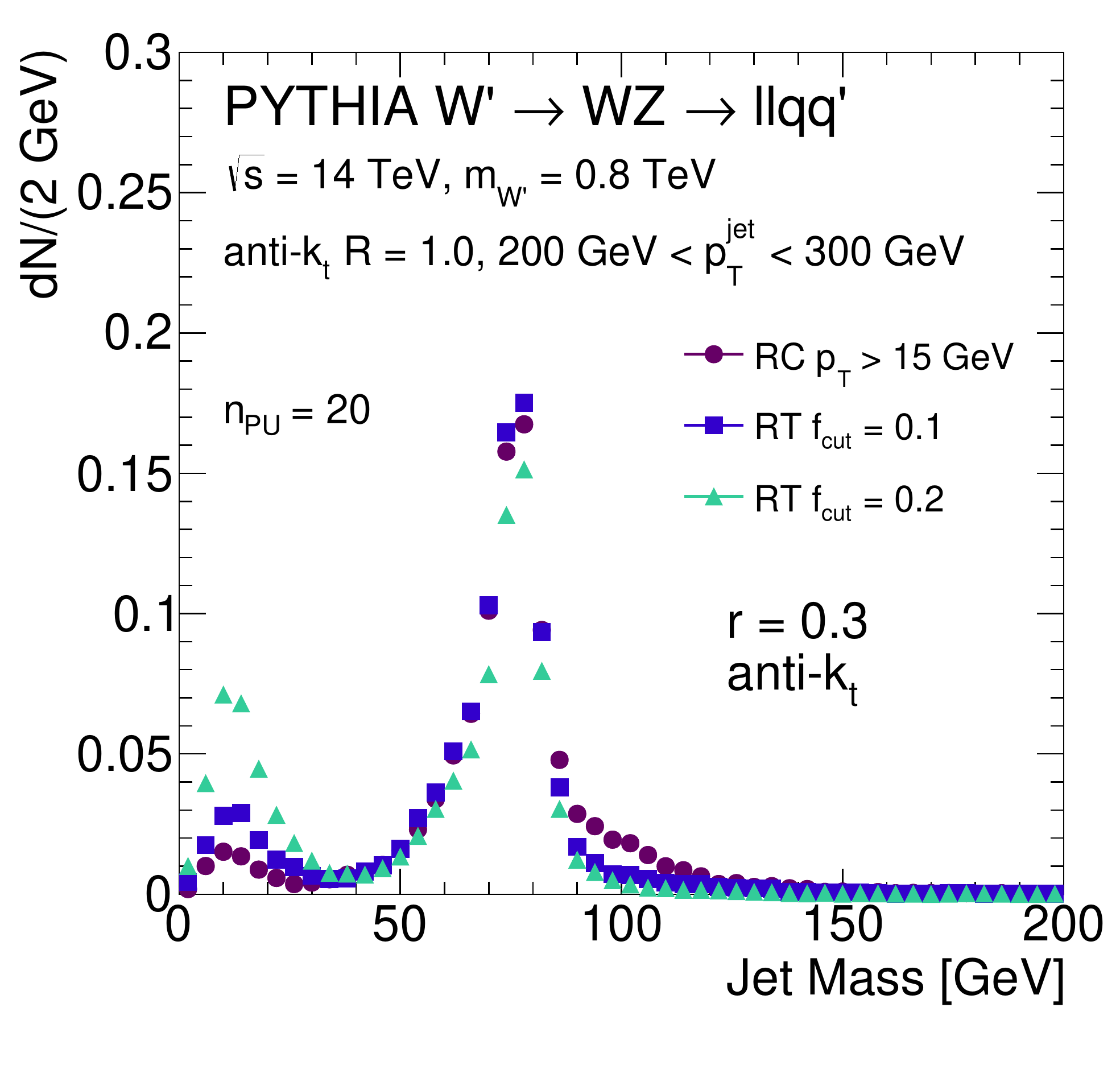}\includegraphics[width=0.45\textwidth]{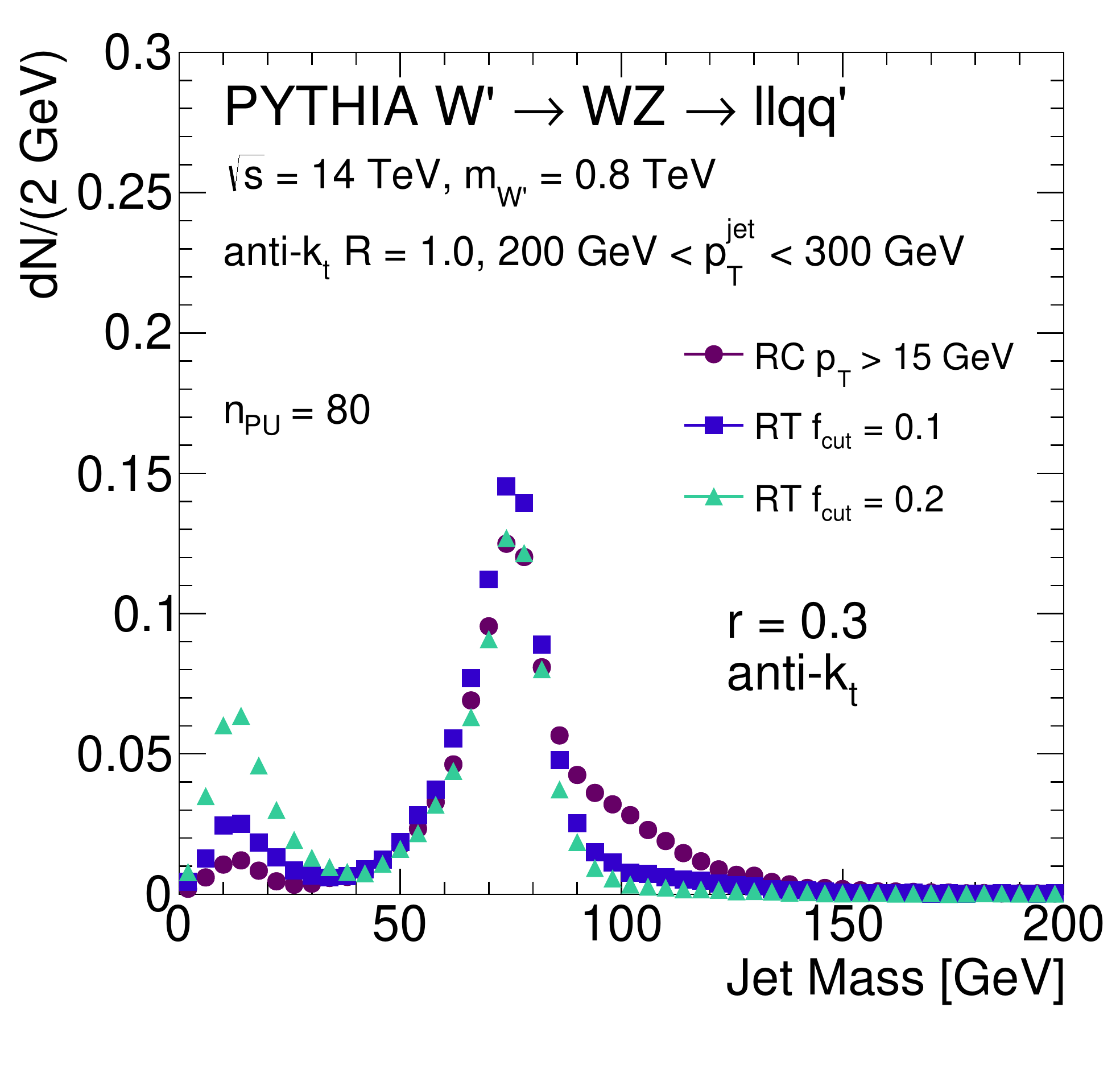}
\end{center}
\caption{Various re-clustered grooming parameters for anti-$k_t$ $r=0.3$ jets for NPV = 20 on the left and NPV = 80 on the right.}
\label{fig:fixed}
\end{figure}

\begin{figure}[h!]
\begin{center}
\includegraphics[width=0.33\textwidth]{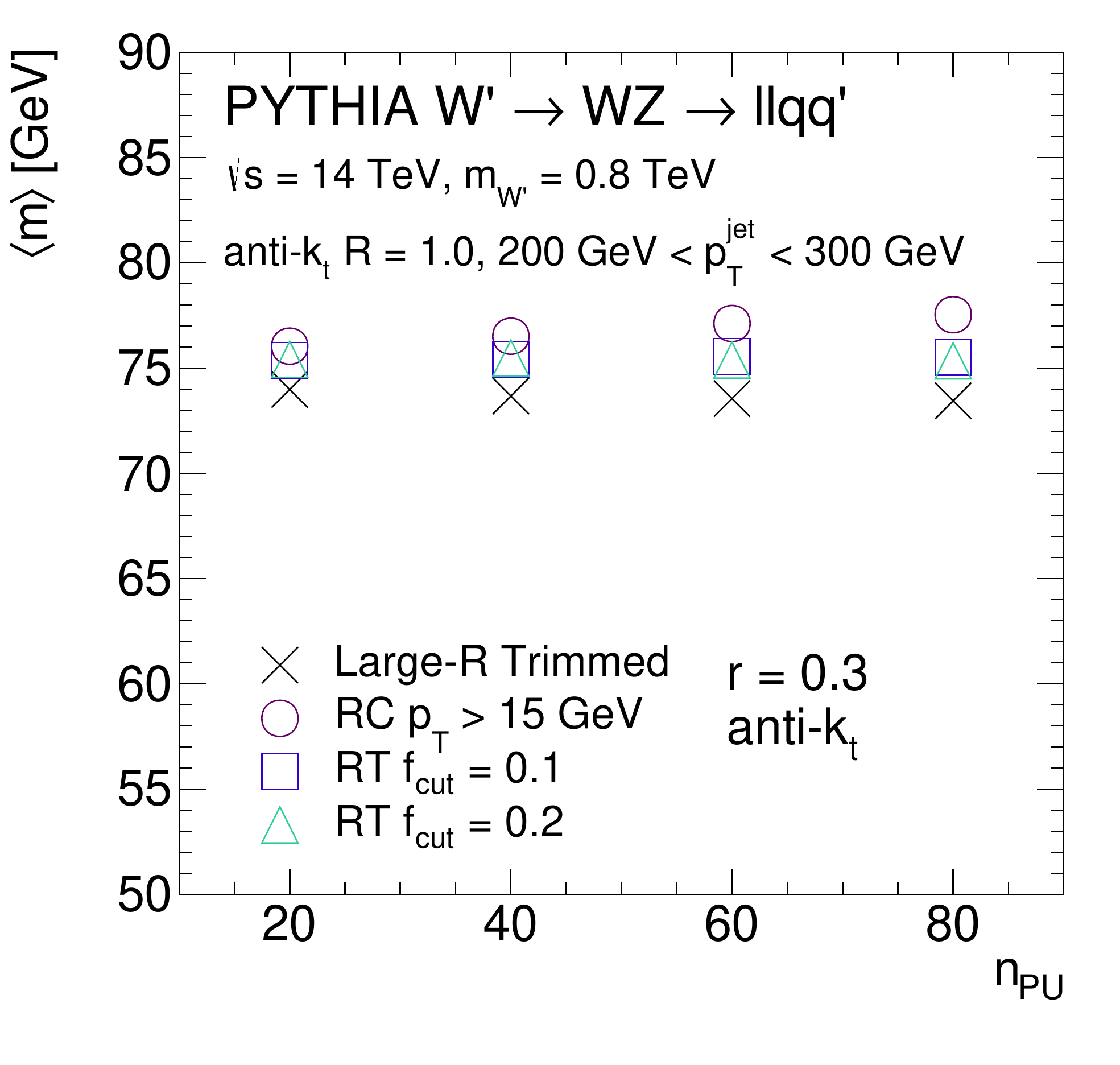}\includegraphics[width=0.33\textwidth]{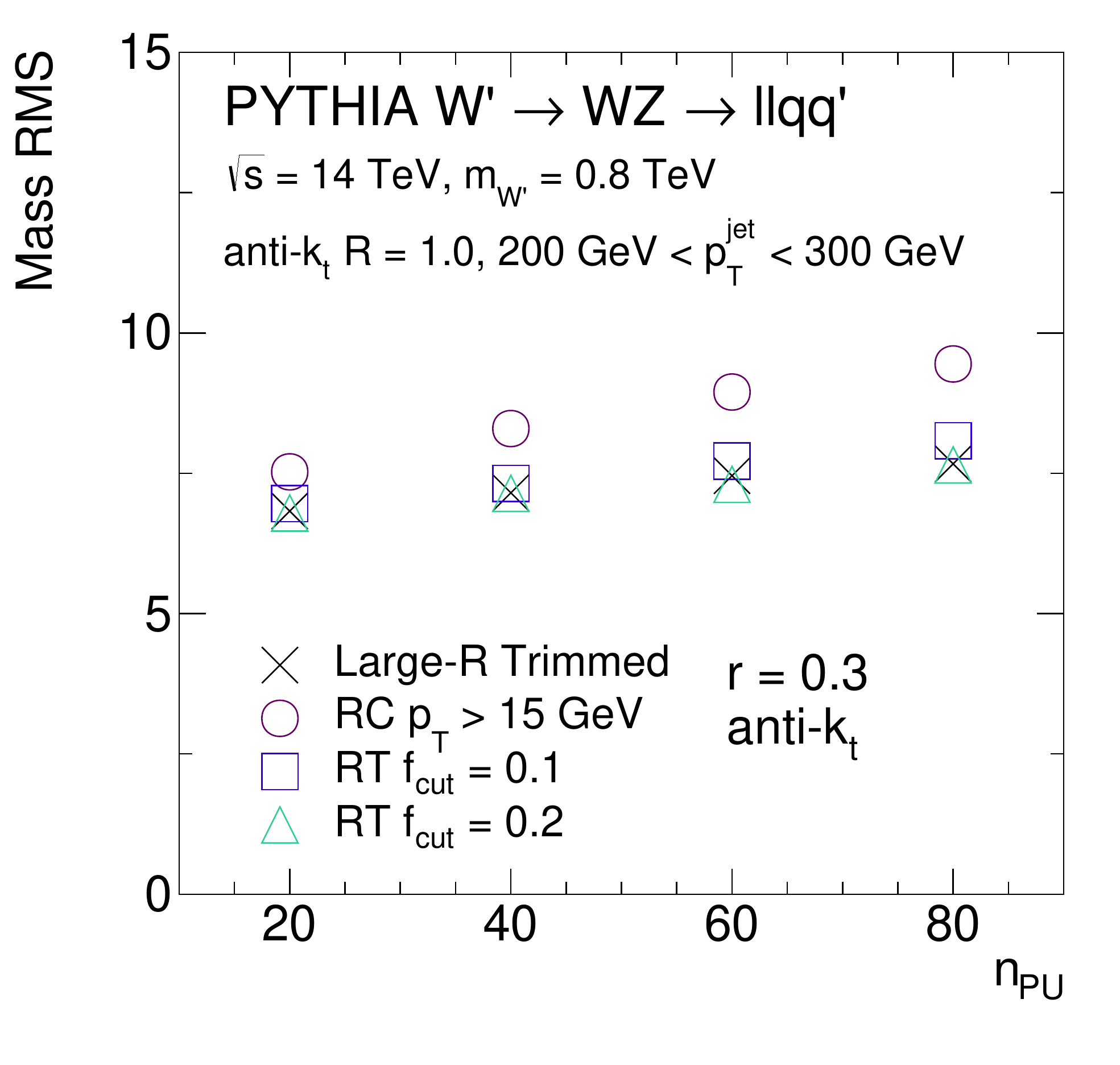}
\includegraphics[width=0.33\textwidth]{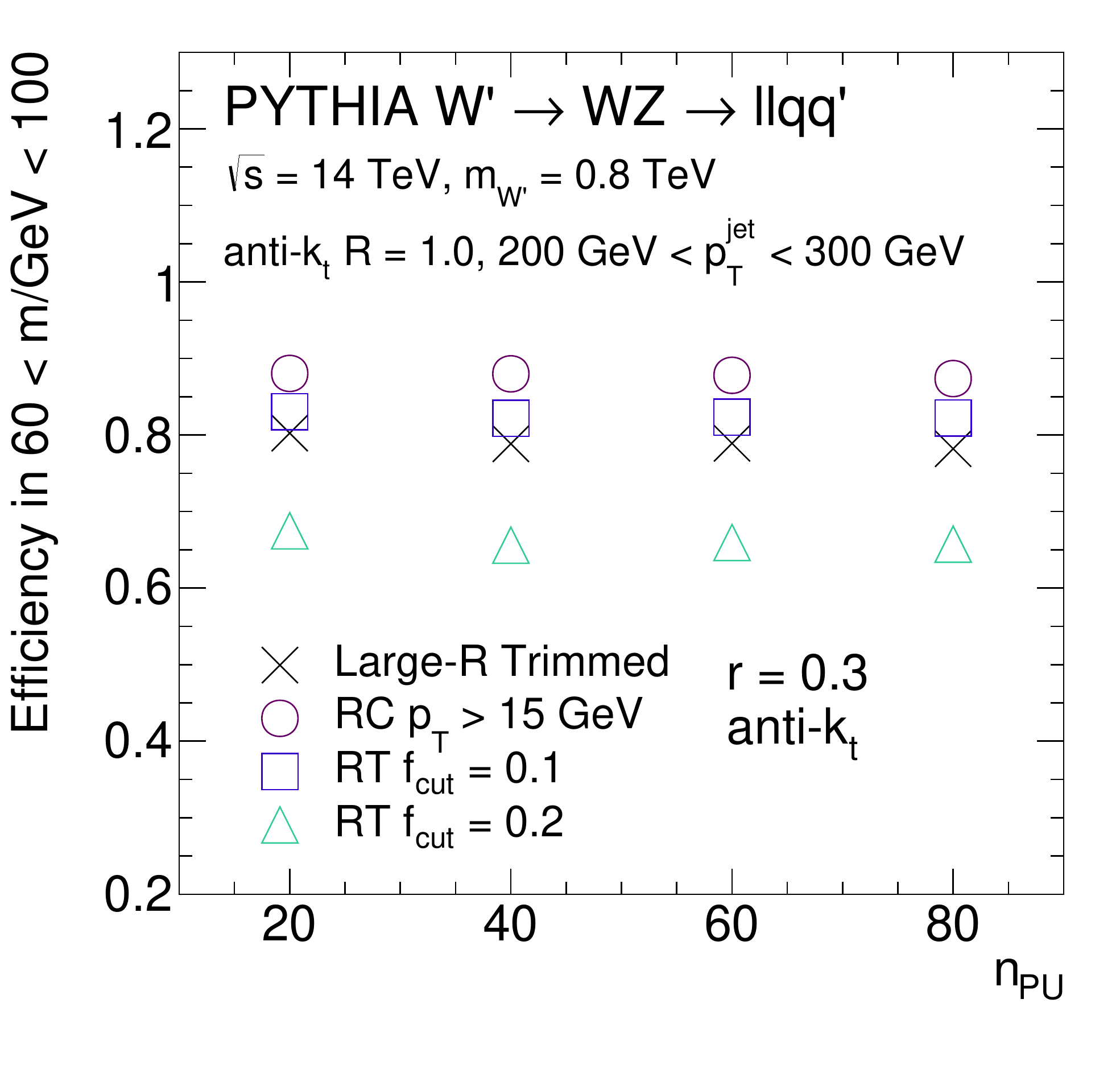}
\end{center}
\caption{Mean, mass resolution, and mass window efficiency of the mass distribution as a function of the number of additional vertices for various re-clustered jet grooming schemes.}
\label{fig:fixed2}
\end{figure}

The re-clustered jet mass distribution for several small radius jet sizes is shown in Figure~\ref{fig:sizes} and the performance metrics are quantified in Fig.~\ref{fig:sizes2}.  For all three considered values of $r$, the minimum $p_\text{T}$ cut is 15 GeV.  In practice, this could be optimized, since smaller radius jets may be calibrated at smaller values of $p_\text{T}$.  An alternative approach is to use {\it iterative re-clustering} by re-clustering $r=0.2$ into $r'=0.4$ and then into $R=1.0$ to further increase the flexibility of the jet algorithms (also this reduces the effective jet area and so the resulting jets would be less susceptible to pileup\footnote{If viewed as a uniform noise in the calorimeter, the contribution of pileup to a given jet scales proportionally to its area.  However, there are local fluctuations that complicate this picture.}).  The right plot of Figure~\ref{fig:sizes} and the top right plot of Fig.~\ref{fig:sizes2} show the $r=0.2$ setting as resulting in the most peaked mass distribution.  

\begin{figure}[h!]
\begin{center}
\includegraphics[width=0.45\textwidth]{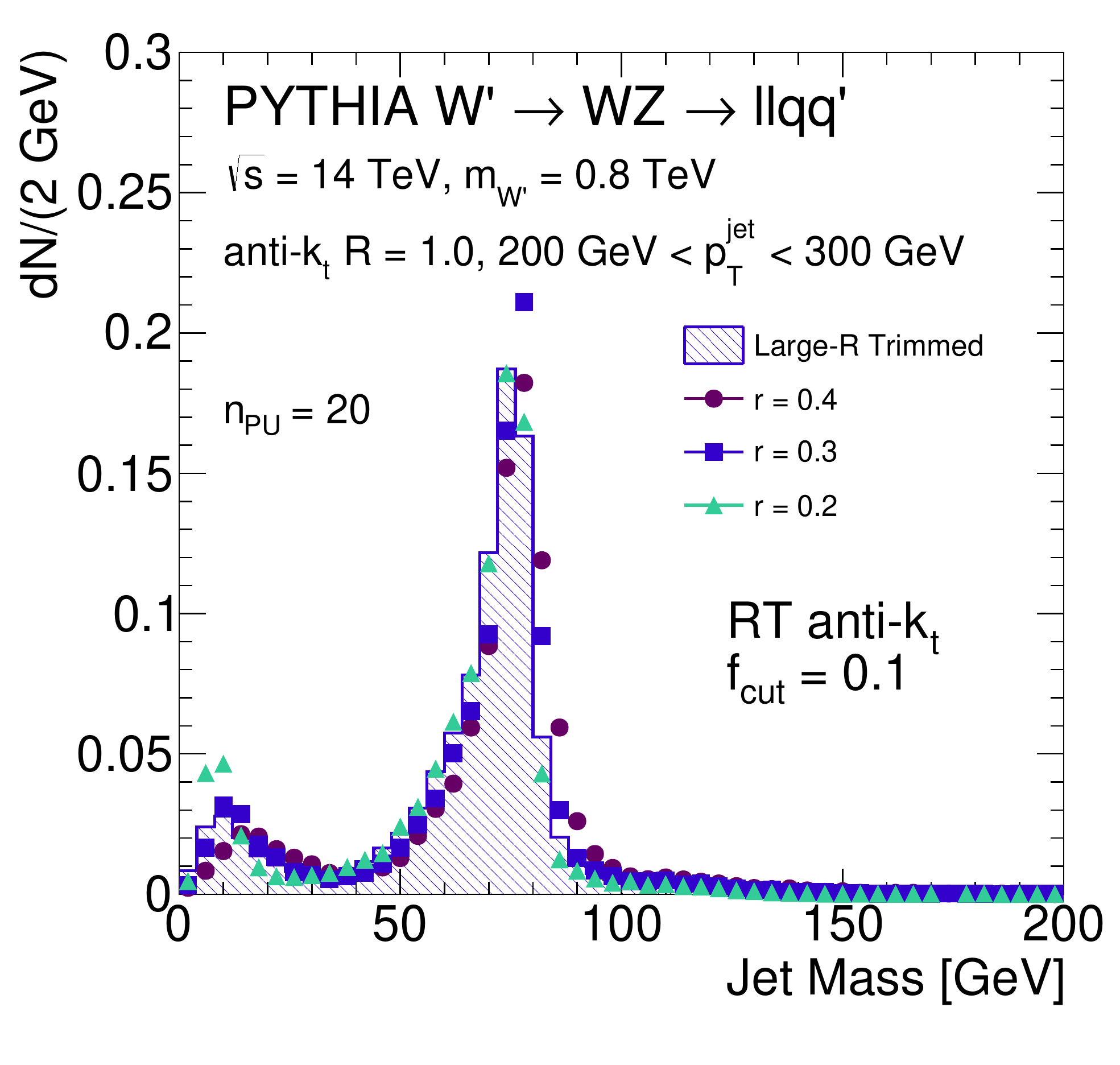}\includegraphics[width=0.45\textwidth]{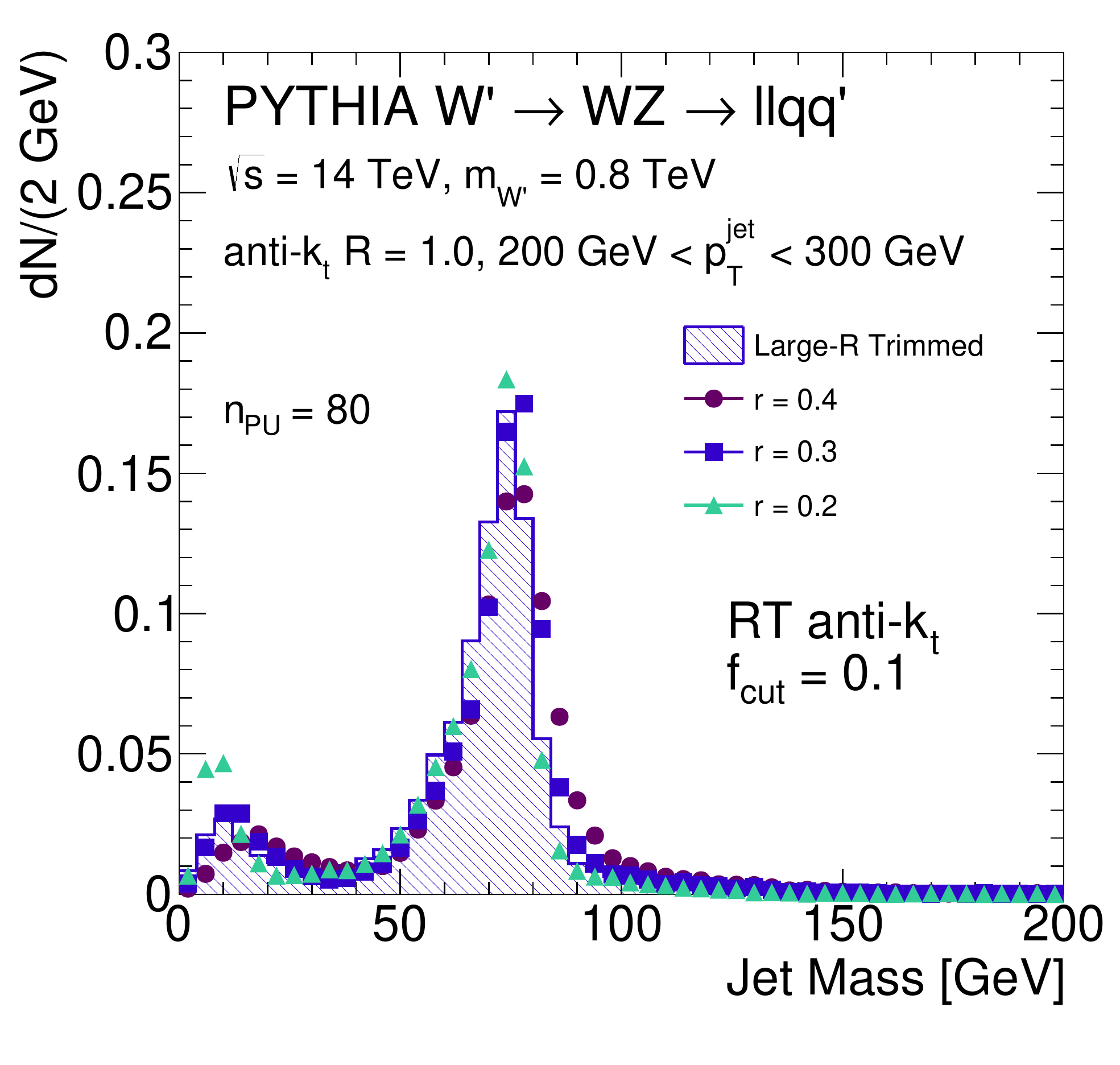}
\end{center}
\caption{Various small radii for a fixed algorithm of anti-$k_t$ for NPV = 20 on the left and NPV = 80 on the right.}
\label{fig:sizes}
\end{figure}

\begin{figure}[h!]
\begin{center}
\includegraphics[width=0.33\textwidth]{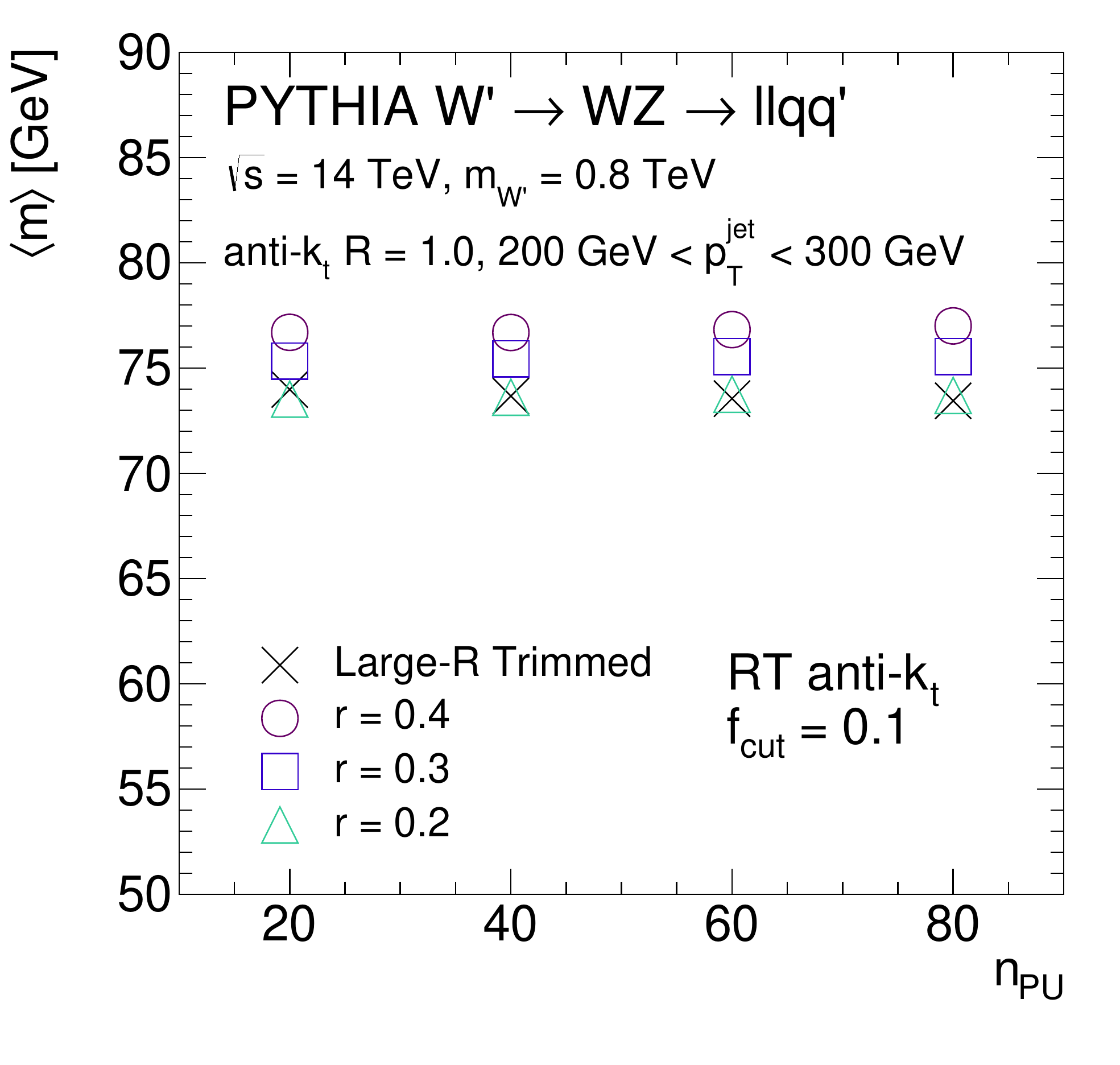}\includegraphics[width=0.33\textwidth]{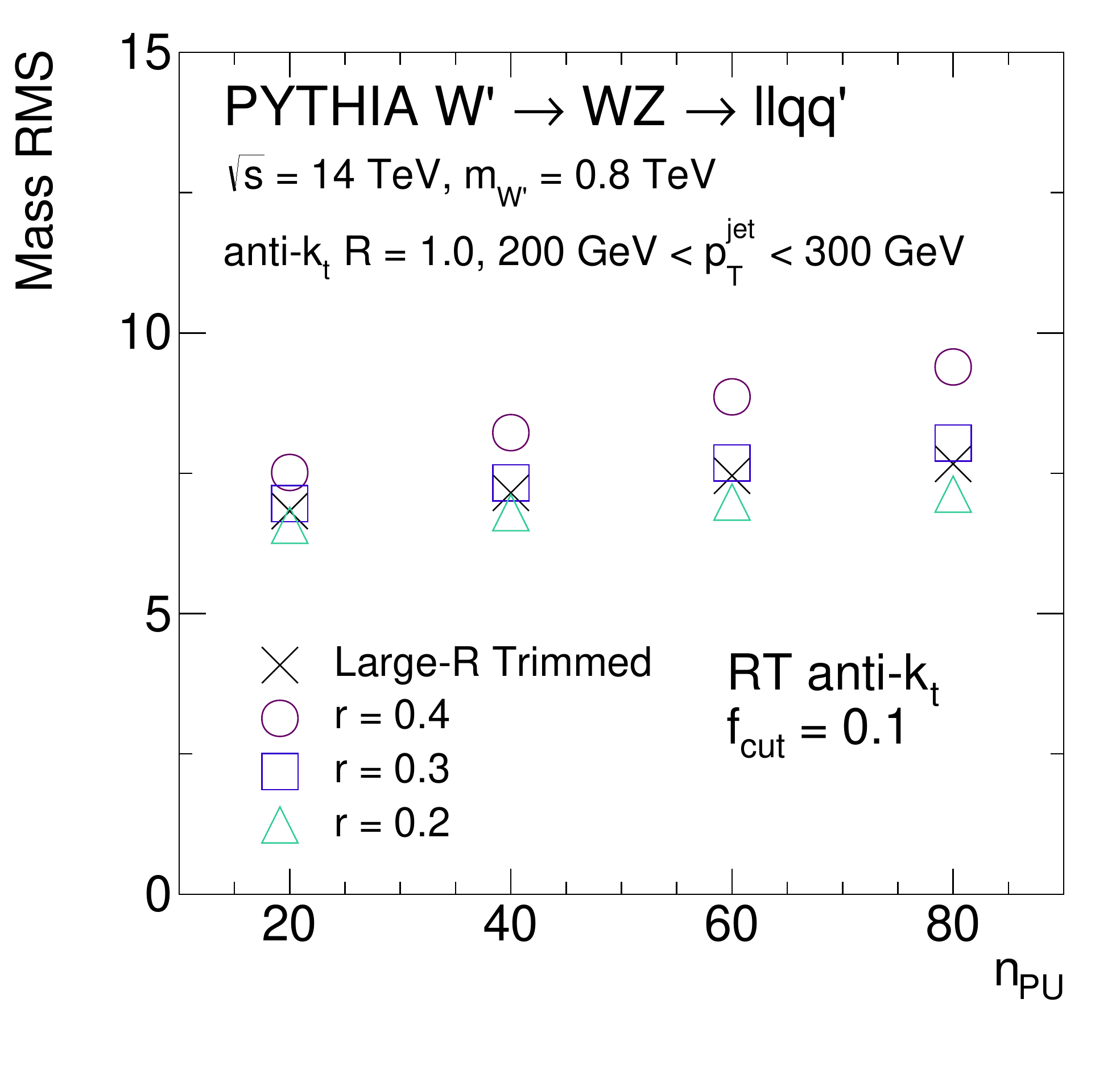}
\includegraphics[width=0.33\textwidth]{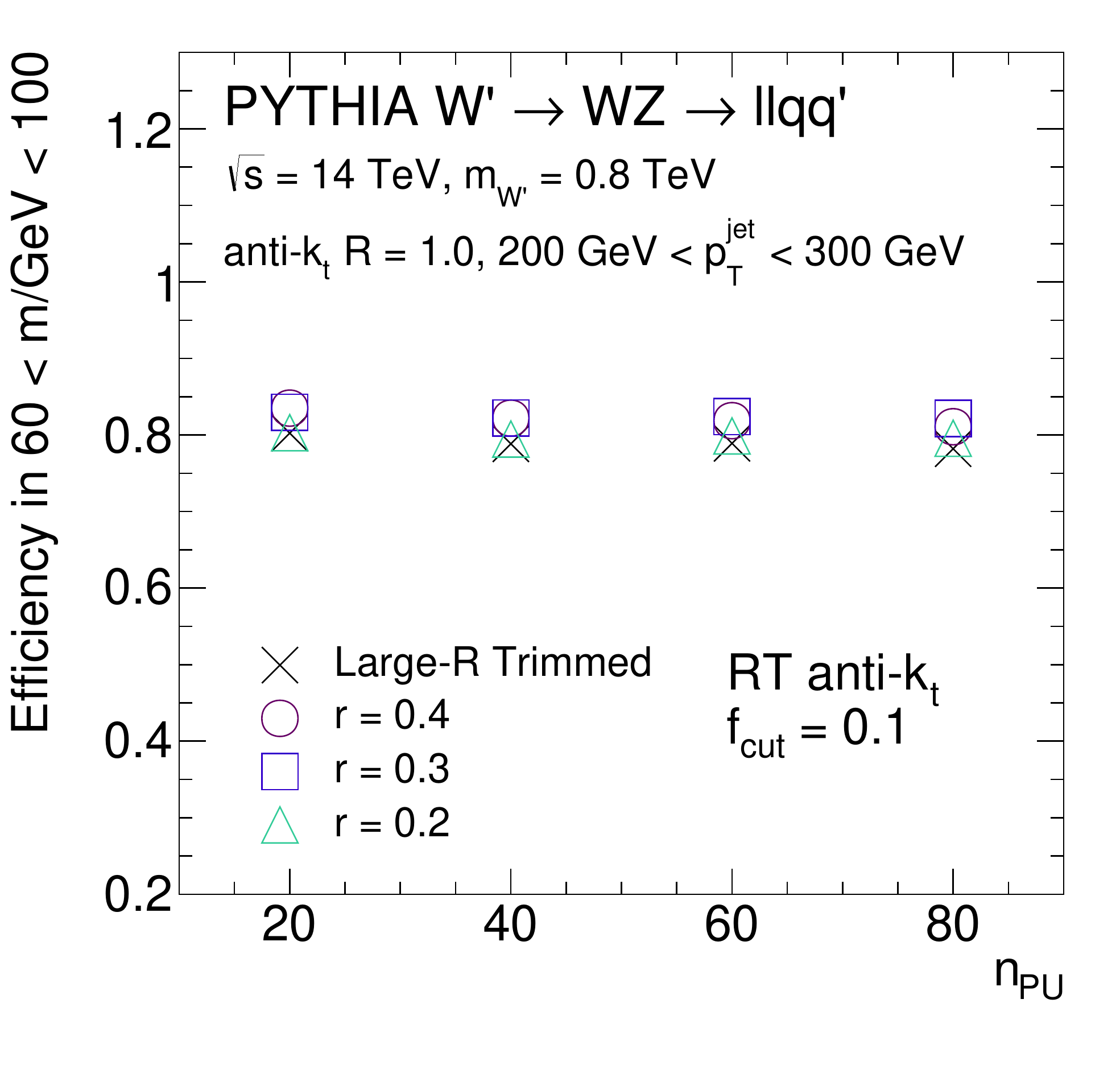}
\end{center}
\caption{Mean, mass resolution, and mass window efficiency of the mass distribution as a function of the number of additional vertices for various small jet radii.}
\label{fig:sizes2}
\end{figure}	
	
	\clearpage
	
	\subsubsection{Re-clustered Jet Substructure}
	\label{sec:recluster:jss}
	
	One natural possibility for computing jet substructure observables for re-clustered jets is to use the constituents of the small-radius jets inside the large-radius jet.  An alternative method is to use the radius $r$ jet momenta directly.  For example, consider the $k_t$ splitting scale\footnote{Computed by re-clustering a jet's constituents using the $k_t$ algorithm and then considering the distance metric of the last $n$ un-clusterings.} $\sqrt{d_{n,n+1}}$, which is sensitive to hard $(n+1)$-prong structure in a jet.  One can use directly the radius $r$ jets inside a radius $R$ re-clustered jet to compute $\sqrt{d_{n,n+1}}$.  If there are only two radius $r$ jets, then $\sqrt{d_{12}}$ is simply the $k_t$ distance between the radius $r$ jets.  The advantage of this approach is that there is a natural prescription for calibrations and systematic uncertainties.  The jet energy scale calibration and its uncertainties directly translate into the calibration of the bottom-up substructure variables.  Furthermore, in this approach one knows how the substructure variable calibrations and uncertainties are correlated with the re-clustered jet calibrations and uncertainties.   This information is available for the first time with this bottom-up procedure.

Figure~\ref{fig:bottomup} compares bottom-up and top-down jet substructure variables in classifying $Z'\rightarrow t\bar{t}$ and QCD multijet events.  For the chosen parameters, the two techniques have comparable performance.  The main drawback of bottom-up substructure is that the relative efficacy depends on $p_\text{T}$ (and $r$).  When $r\gtrsim m/p_T$, or equivalently, when there are not many radius $r$ jets inside the radius $R$ jet, the experimental gains from bottom-up substructure are diminished.  For instance, if there is only one radius $r$ jet, then $\sqrt{d_{12}}=0$.  Thus, in certain kinematic regimes, bottom-up substructure may provide a powerful alternative to standard methods, but in other regimes a more dedicated analysis is required to understand correlations in calibrations and uncertainties (when jet substructure observables are built from the jet constituents).

\begin{figure}[htbp!]
\begin{center}
\includegraphics[width=0.75\textwidth]{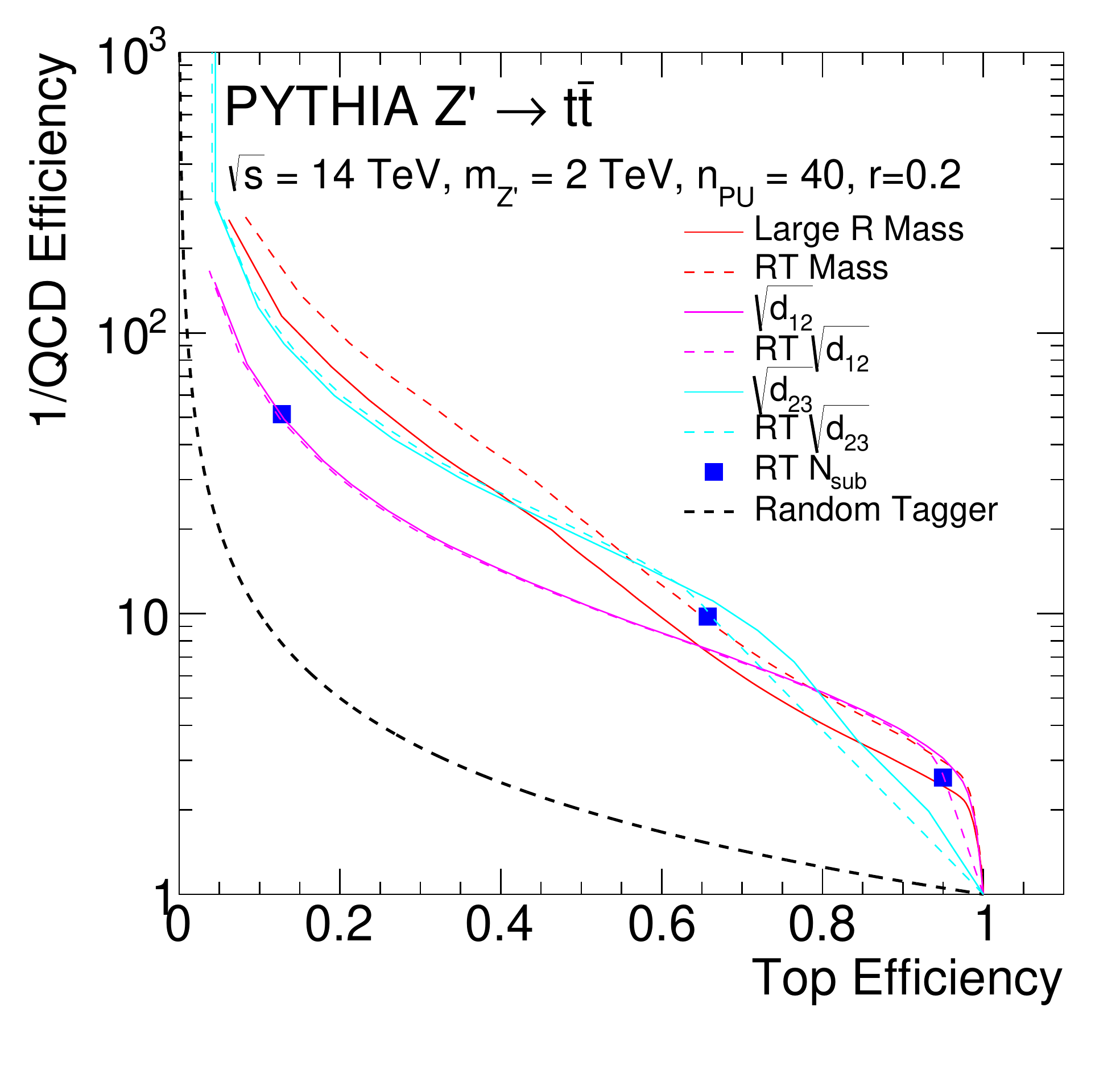}
\end{center}
\caption{The performance of a bottom-up approach to jet substructure where the radius $r$ jets are inputs to substructure variables.  Solid lines show the performance curves for large radius trimmed jets ($R_\text{sub}=0.3, f_\text{cut}=0.05$) and the dashed lines show the analogous re-clustered variable. Random tagger denotes a classifier which picks signal and background with equal probability. The variable $N_\text{sub}$ is the number of re-clustered subjets.  The curves are determined by placing threshold requirements on the variable likelihoods.}
\label{fig:bottomup}
\end{figure}

\clearpage
	
	\subsubsection{Detector-level Jet Tagging}
	\label{sec:recluster:tagging}
	
	The studies\footnote{The results presented in this section include input from M. Solt.  In particular, Solt made the final versions of the plots comparing the various algorithms.} in Sec.~\ref{sec:reluster:particlelevelperformance} are an important first step to quantifying the dependence of re-clustered jet mass performance on one of the most important experimental conditions, pileup.  This section expands upon the study by using the full ATLAS detector-simulation to investigate how the full detector-resolution impacts the re-clustered jet mass performance.  Re-clustering parameters are varied and the resulting jets are compared with traditional large-radius jets clustered directly from calorimeter-cell clusters. The relevant re-clustering parameters are $f_\text{cut}$, $r$, $R$ and $p_\text{T}^\text{cut}$ (the small radius jet $p_\text{T}$ threshold).  Two metrics used for comparing algorithms are the {\it window size}, which is the size of the mass interval which contains at least $68\%$ of the signal, and {\it window efficiency}, which is the fraction of background events which fall in the $68\%$ window.  {\sc Pythia} $W'\rightarrow WZ$ events are signal and {\sc Pythia} QCD dijets are background. The signal $p_\text{T}$ spectrum is re-weighted to match that of the background\footnote{Instead of re-weighting for each jet algorithm, the $p_\text{T}$ spectrum is weighted based on the leading C/A particle-level ungroomed jet with $R=1.2$. This algorithm was chosen because in the signal, it has a high efficiency for capturing all of the $W$ boson energy.}.  The study is decomposed into three $p_\text{T}$ ranges: $p_\text{T}\in [200-350],[350,500],$ and $[500-1000]$ GeV.  The radius $r$ jets used for re-clustering are pileup corrected but not calibrated.  Pruned~\cite{Ellis:2009me,Ellis:2009su} C/A jets with $R=0.8$ are used as a benchmark as they perform well across a wide range of phase space~\cite{Khachatryan:2014vla}.

\paragraph{Low $p_\text{T}$: 200 GeV - 350 GeV}\mbox{}\\
\label{recluster:low}

Figure~\ref{fig:w:low:window2_4} compares anti-$k_t$ $R=1.0$ trimmed jets with $f_\text{cut}=0.05$ and $R_\text{sub}=r$ with the analogous RT jets in the range $200$ GeV $<p_\text{T}<350$ GeV.  The main difference between the RT and traditional large-radius jets is that the small-radius jets for the former are anti-$k_t$ while the $k_t$ algorithm is used for subjet finding for the latter.  As the $W$ decay products are well-resolved by $r=0.2,0.3$, and $r=0.4$ small-radius jets, the mass window is about the same size for all three algorithms.  However, the small-radius jets with a larger size have a worse rejection (higher efficiency) for the background because the large small-radius jets tend to have a higher $p_\text{T}$ and so a second (or third) background jet can survive the trimming and significantly increase the mass.  
\begin{figure}[h!]
\begin{center}
\includegraphics[width=0.5\textwidth]{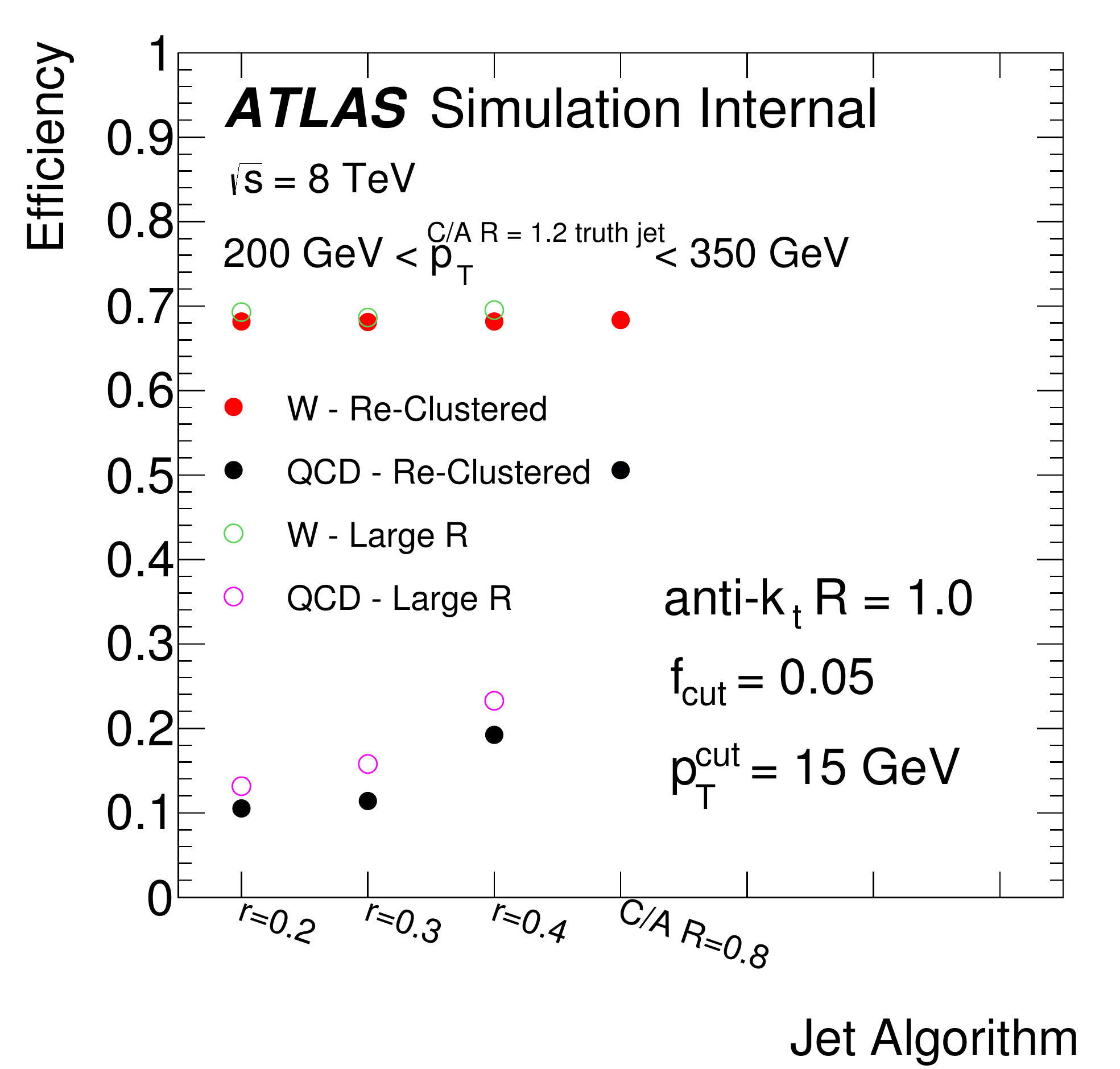}\includegraphics[width=0.5\textwidth]{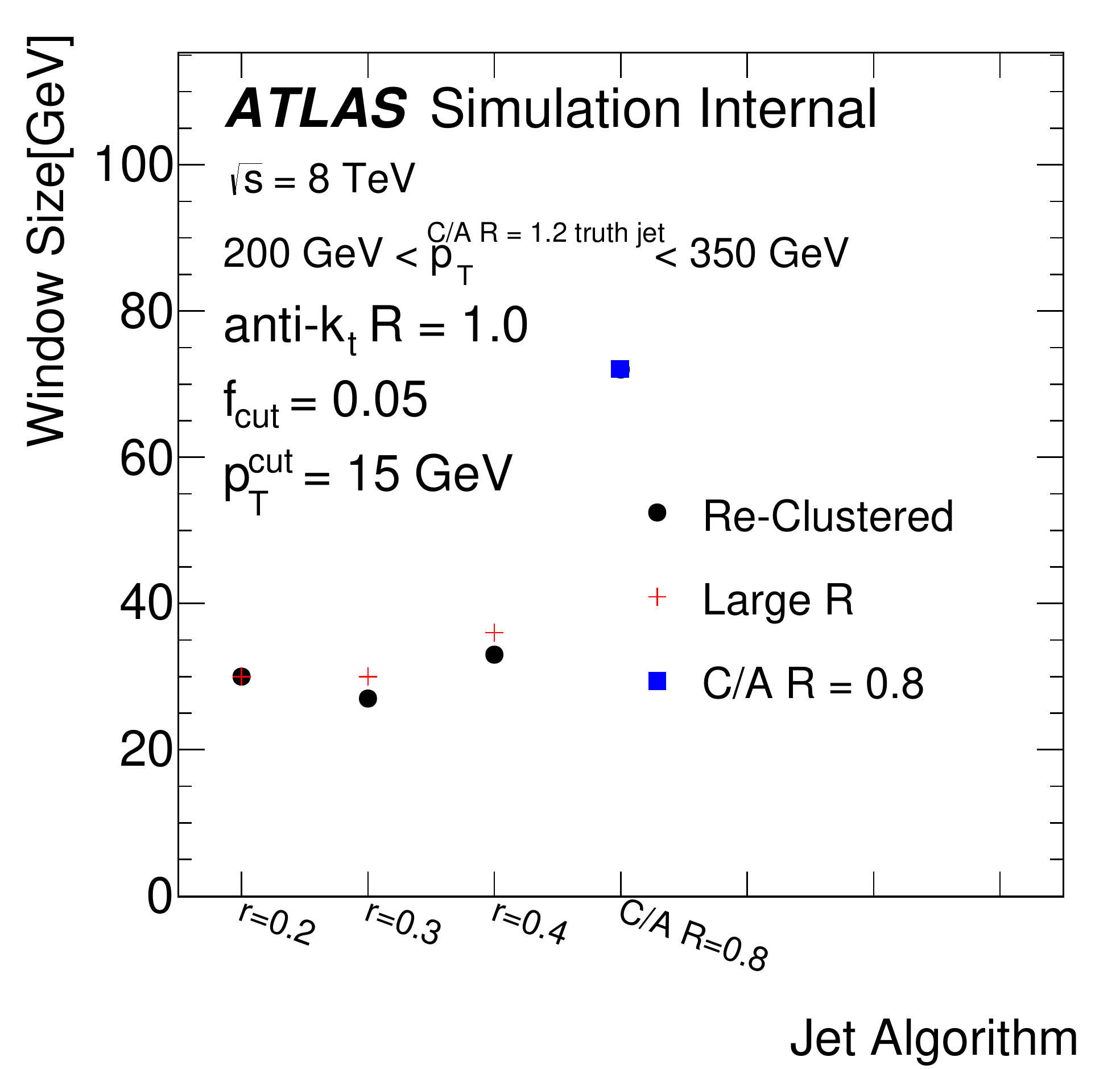}
\end{center}
\vspace{-5mm}
\caption{Left: The efficiency for signal and background jets to be reconstructed in the smallest mass window containing $68\%$ of the signal.  Right: the size of the window from the left plot.  Both plots compare large radius trimmed jets with $R=1.0, f_\text{cut}=0.05$ and $R_\text{sub}=0.05$ with re-clustered jets that have analogous parameters. }
\label{fig:w:low:window2_4}
\end{figure}

At low $p_\text{T}$, the choice of $R$ is particularly important as the boson boost is not yet high enough to capture all of the decay products into one small radius jet.  Therefore, re-clustered jets need to have multiple constituents in order to have a mass compatible with the $W$ boson mass.  Figure~\ref{fig:w:low:NR} shows the small-radius jet constituent multiplicity for several choices of $R$.

Figure~\ref{fig:w:low:NR} shows the $p_\text{T}$ spectrum of the re-clustered jets in the lowest $p_\text{T}$ range along with the number of small radius jets for $r=0.2$ and several values of $R$.  The distributions for $R\gtrsim 0.8$ are similar and collectively are significantly different than the $R=0.6$ case.  The loss of constituents degrades the mass-tagging performance, as illustrated in Fig.~\ref{fig:w:low:windowR}.   The jets with $R\gtrsim 1.0$ have a significantly higher background rejection than re-clustered jets with $R<1.0$.  For $R=0.6$, this is explained by the fact that many re-clustered jets have only one small-radius jet constituent for which the jet mass distribution is nearly identical to the mass distribution for the QCD jet background.  The $R=0.8$ point in the left plot of Fig.~\ref{fig:w:low:windowR} seems inconsistent with Fig.~\ref{fig:w:low:NR}, which suggests that there is only a small difference in the constituent multiplicity between $R=0.8$ and $R=1.0$.  This is explained by the bimodal mass distribution in the left plot of Fig.~\ref{fig:w:low:MassRQCD}. Re-clustered jets with one constituent have a low-mass Sudakov peak while jets with multiple constituents have a jet mass near $m_W$.  When the low mass peak contains more than $(100-68)\%$ of the distribution, the $68\%$ interval is split across the two peaks, which is why the $R=0.8$ point has such a larger window in the right plot of Fig.~\ref{fig:w:low:windowR}.  In contrast, the $R=0.6$ point has at least $68\%$ in the low mass peak alone while the $R>0.8$ points have at least that much probability in the high mass peak.

\begin{figure}[h!]
\begin{center}
\includegraphics[width=0.5\textwidth]{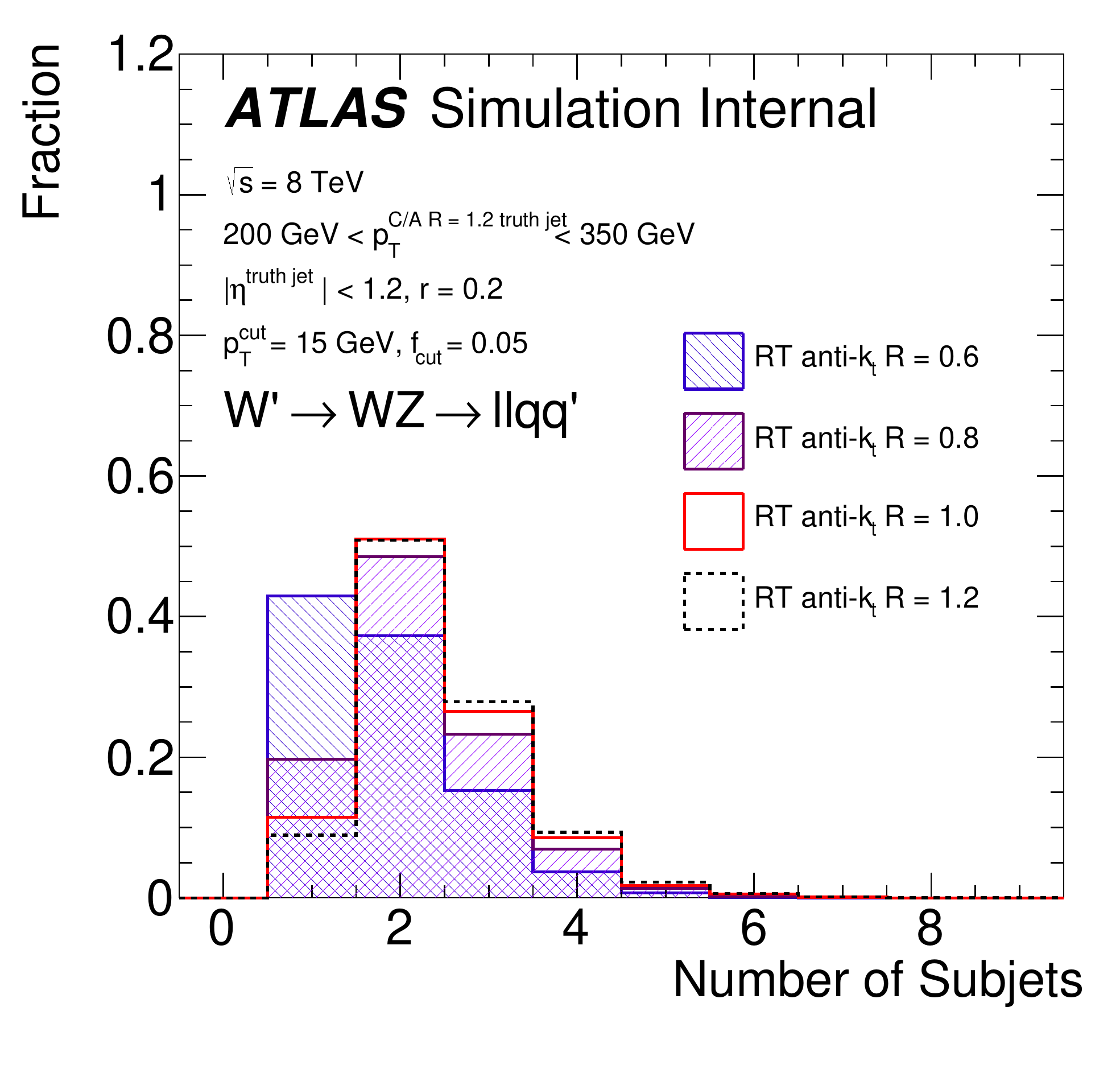}
\end{center}
\vspace{-5mm}
\caption{The small-radius jet constituent multiplicity for several choices of $R$ and a fixed $r=0.2$.}
\label{fig:w:low:NR}
\end{figure}

\begin{figure}[h!]
\begin{center}
\includegraphics[width=0.5\textwidth]{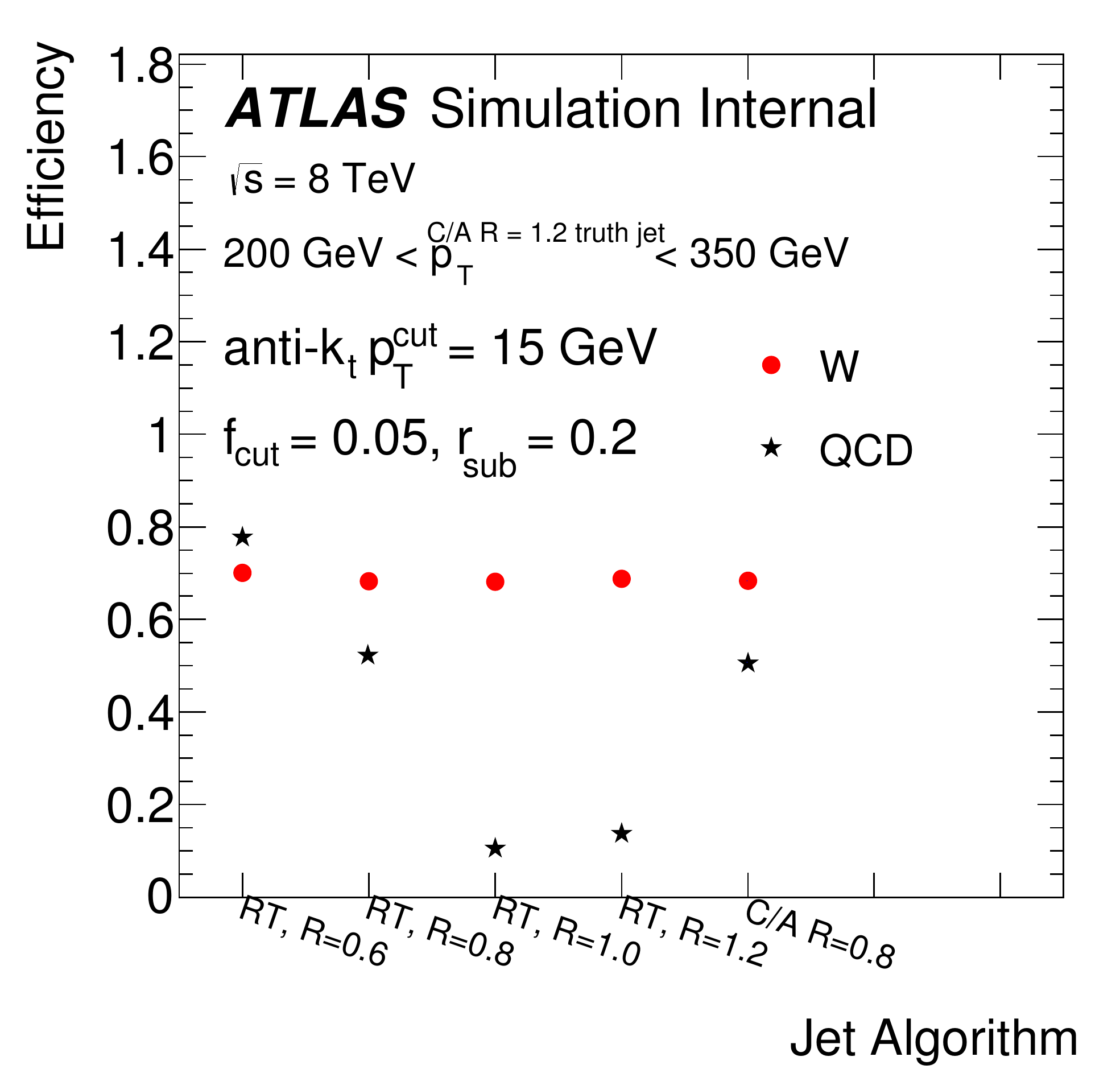}\includegraphics[width=0.5\textwidth]{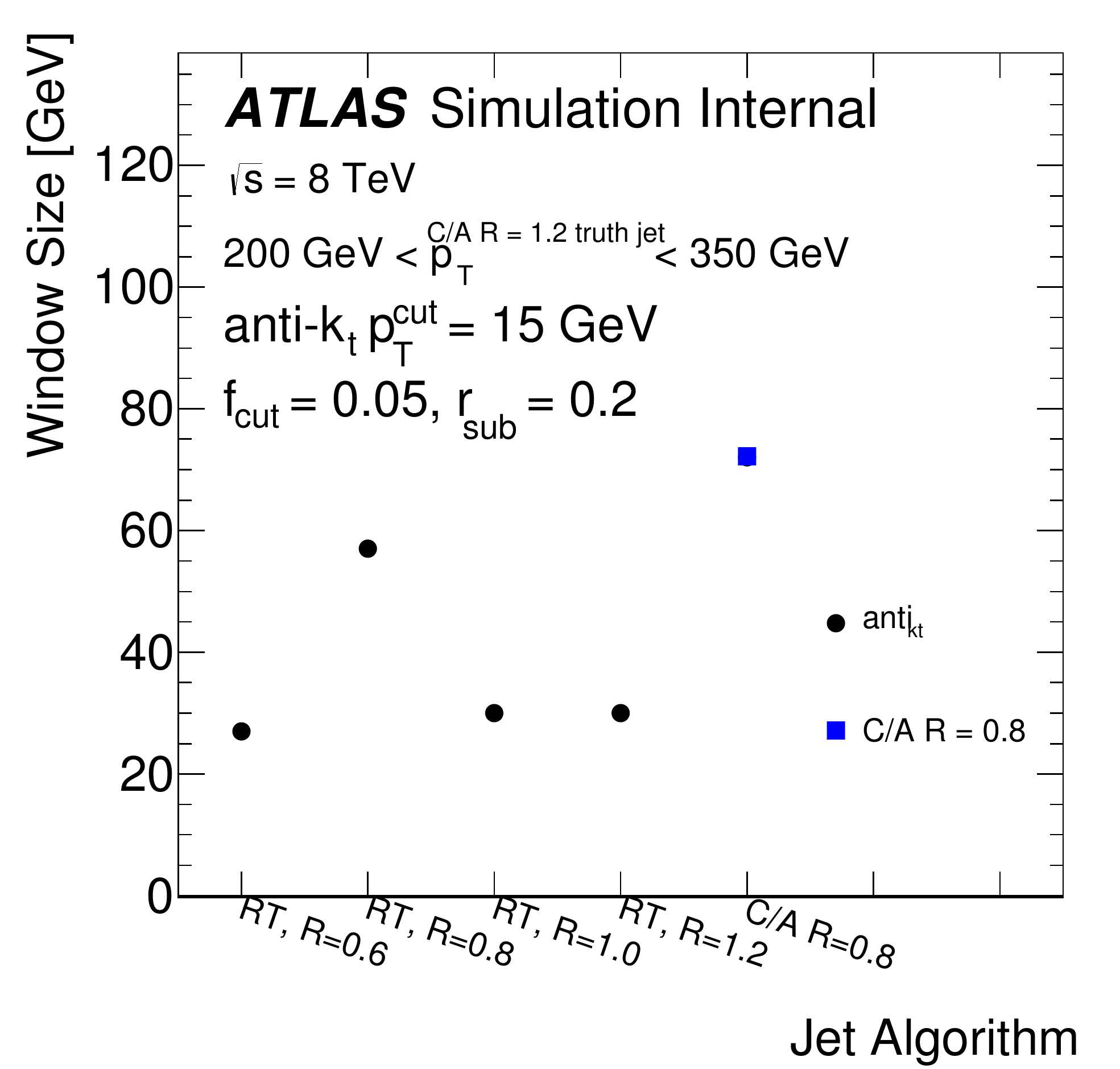}
\end{center}
\vspace{-3mm}
\caption{Left: The efficiency for signal and background jets to be reconstructed in the smallest mass window containing $68\%$ of the signal.  Right: the size of the window from the left plot.  In both plots, $r=0.2$ and $f_\text{cut}=0.05$. }
\label{fig:w:low:windowR}
\end{figure}

\begin{figure}[h!]
\begin{center}
\includegraphics[width=0.5\textwidth]{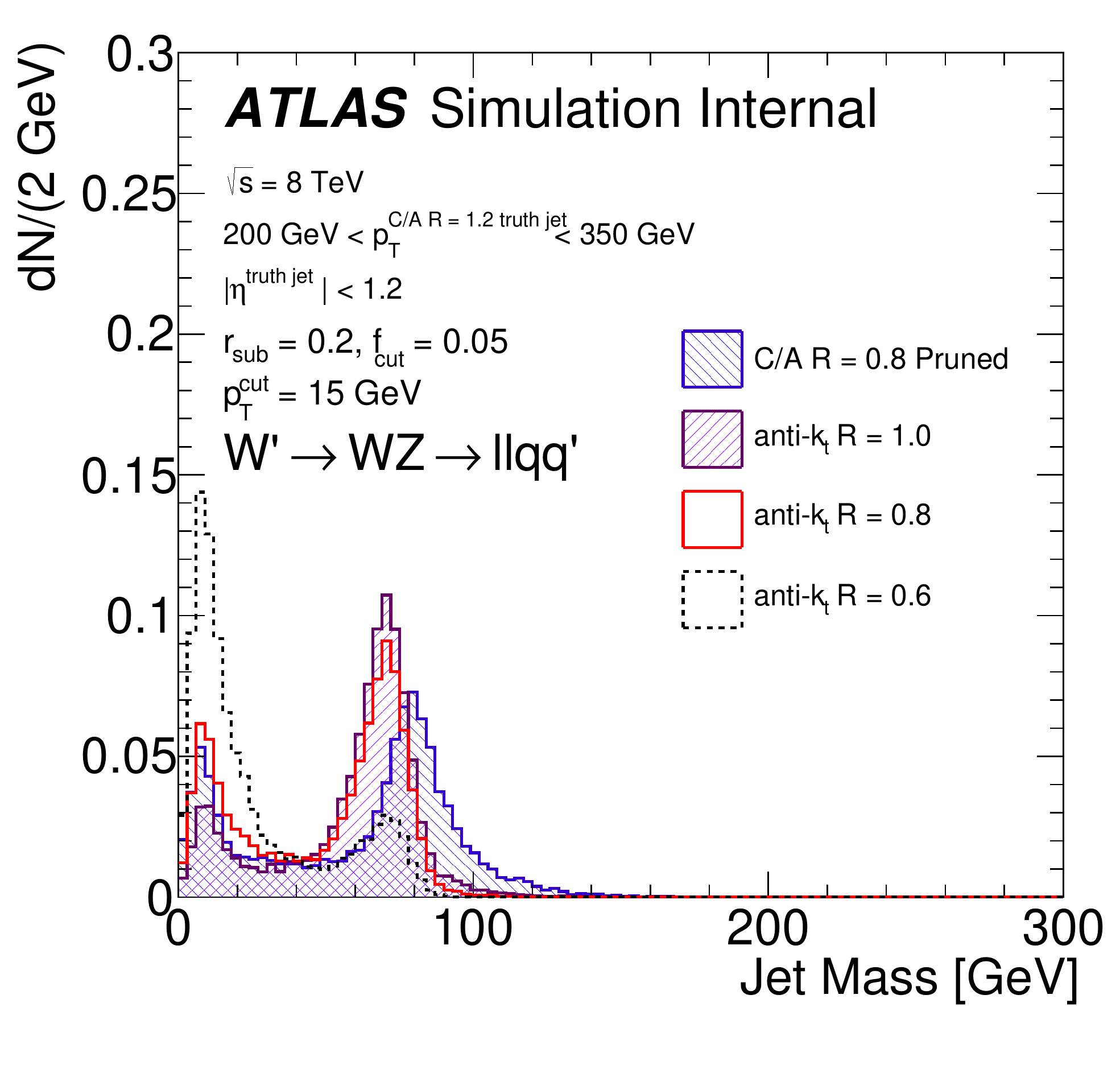}\includegraphics[width=0.5\textwidth]{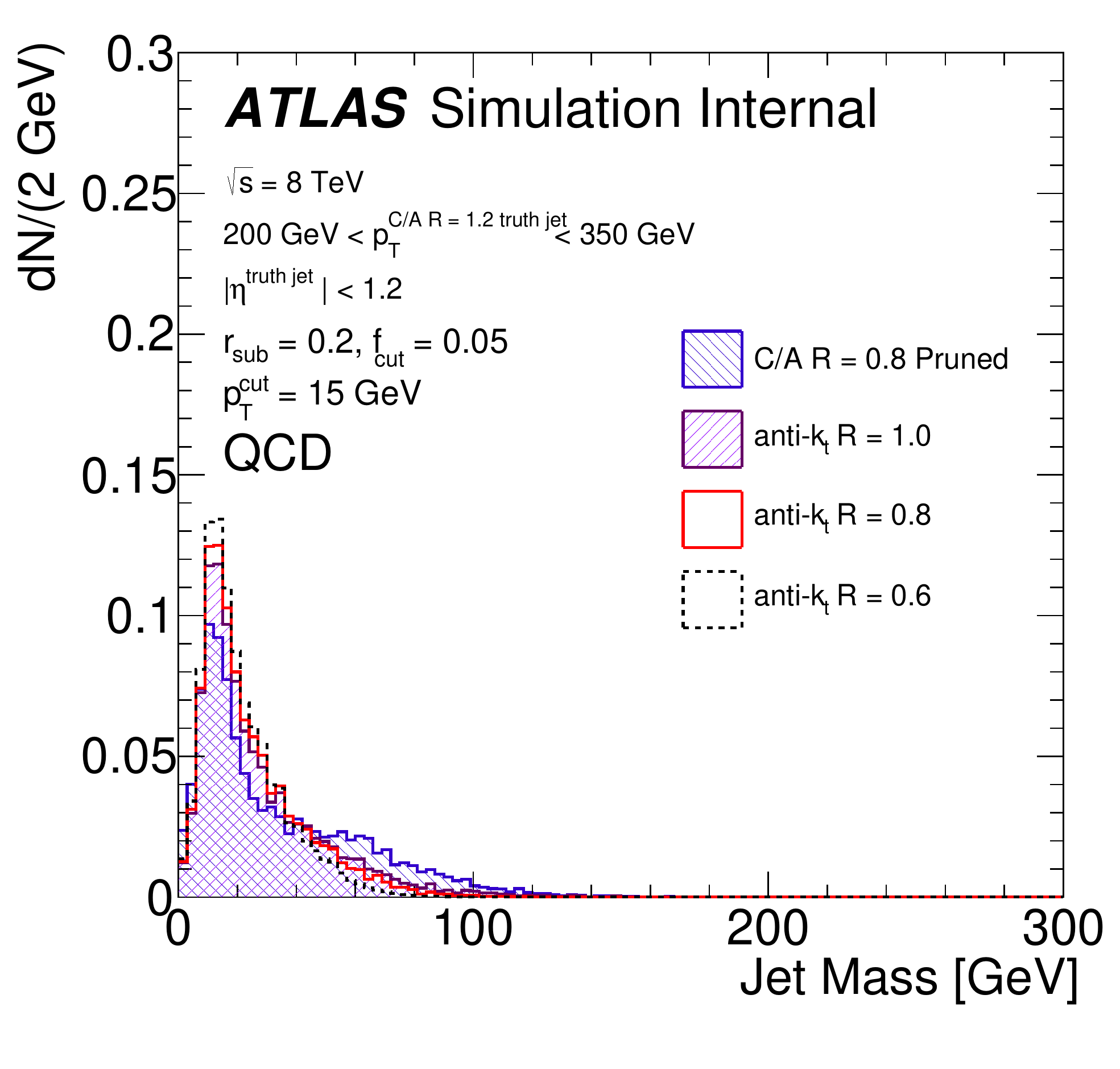}
\end{center}
\caption{The jet mass distribution for the signal (background) for various $R$ values.  In both plots, $r=0.2$ and $f_\text{cut}=0.05$.}
\label{fig:w:low:MassRQCD}
\end{figure}

For a fixed value of $R$, the $f_\text{cut}$ is also an important parameter, as it determines the amount of radiation that is preserved in the clustering.  Figure~\ref{fig:w:low:windowf} shows that so long as $f_\text{cut}$ is large enough to remove unwanted radiation ($f_\text{cut}\sim 0.04$) and is small enough to preserve the hard structure ($f_\text{cut}\sim 0.1$), the performance metrics do not depend strongly on $f_\text{cut}$.  Outside of these regimes, there is a strong dependance as unwanted radiation is preserved or desired radiation is removed.  
\begin{figure}[h!]
\begin{center}
\includegraphics[width=0.5\textwidth]{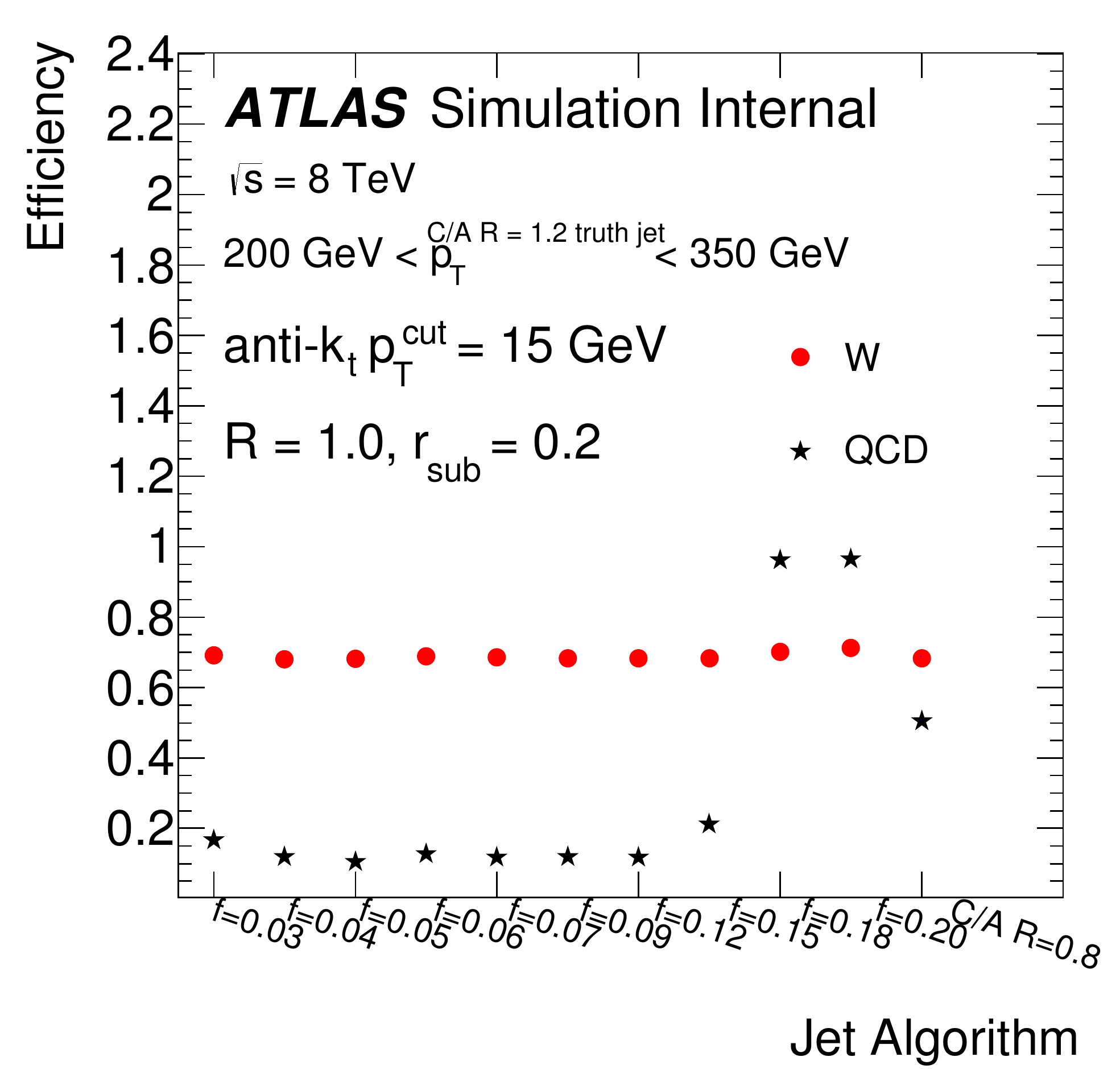}\includegraphics[width=0.5\textwidth]{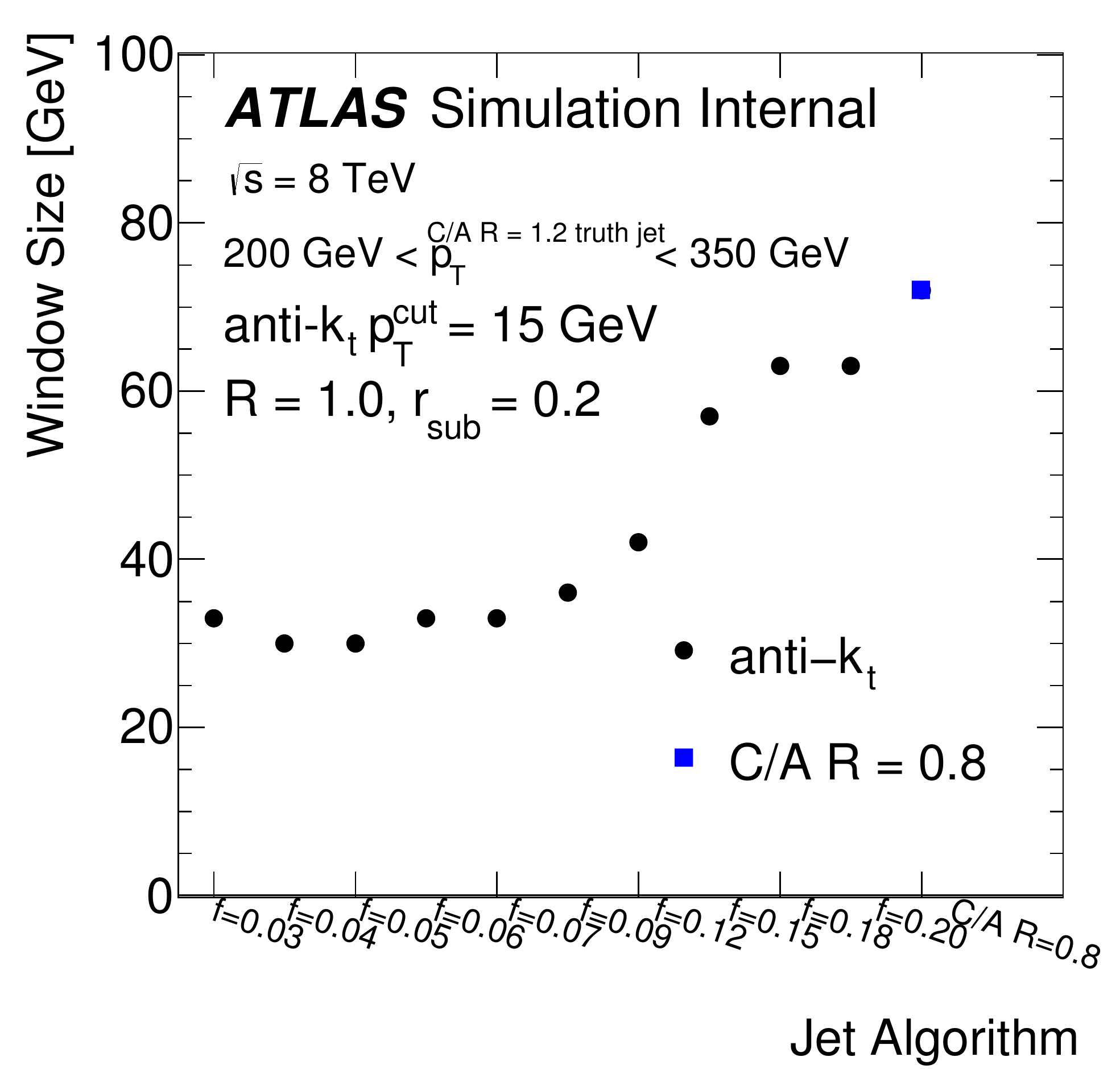}
\end{center}
\caption{Left: The efficiency for signal and background jets to be reconstructed in the smallest mass window containing $68\%$ of the signal.  Right: the size of the window from the left plot.  In both plots, $r=0.2$ and $R=1.0$. }
\label{fig:w:low:windowf}
\end{figure}

\paragraph{High $p_\text{T}$: 500 GeV - 1000 GeV}\mbox{}\\

The trimming is harsher at high $p_\text{T}$ so the $r$-dependence of the QCD rejection is slightly lower in Fig.~\ref{fig:w:high:window2_4} compared with Fig.~\ref{fig:w:low:window2_4}.  As a result, the background rejection for re-clustered jets is nearly independent of $r$ and large-radius trimmed jets have only a small dependence.   In this high $p_\text{T}$ regime, most $r=0.4$ jets have only one constituent (right plot of Fig.~\ref{fig:w:high:windowfNfhigh}) while $r=0.2$ jets still usually have at least two (left plot of Fig.~\ref{fig:w:high:windowfNfhigh}).  As a result, the width of the mass peak is largely insensitive to $f_\text{cut}$ as long as it is not too large that it removes the second jet ($r=0.2$) or to low that it lets in extraneous radiation ($r=0.4$).  In the range $0.04\lesssim f_\text{cut} \lesssim 0.1$, the mass peak has approximately the same size for both algorithms in the signal.  This is a similar range as for the low $p_\text{T}$ bin, though the mass window size itself is significantly smaller for higher $p_\text{T}$ bosons.  The conclusion from this section is that there is a small preference for smaller radii at high $p_\text{T}$, but there is not nearly as much sensitivity as for lower $p_\text{T}$ for the values of $r$ and $R$.  At low $p_\text{T}$, it is non-trivial to pick an appropriate $R$ (which can be compensated to some extend by varying $r$).  From a practical point of view, there is a large incentive for using a smaller radius at high $p_\text{T}$ because the jet mass from two constituents is mostly due to the $p_\text{T}$ of those constituents, which is much better constrained than the mass of a single jet.  The response of small radius jet mass is revisited in the context of close-by jets in Sec.~\ref{sec:recluster:closeby} and for isolated high $p_\text{T}$ jets in Sec.~\ref{sec:TAMass}.

\vspace{15mm}

\begin{figure}[h!]
\begin{center}
\includegraphics[width=0.5\textwidth]{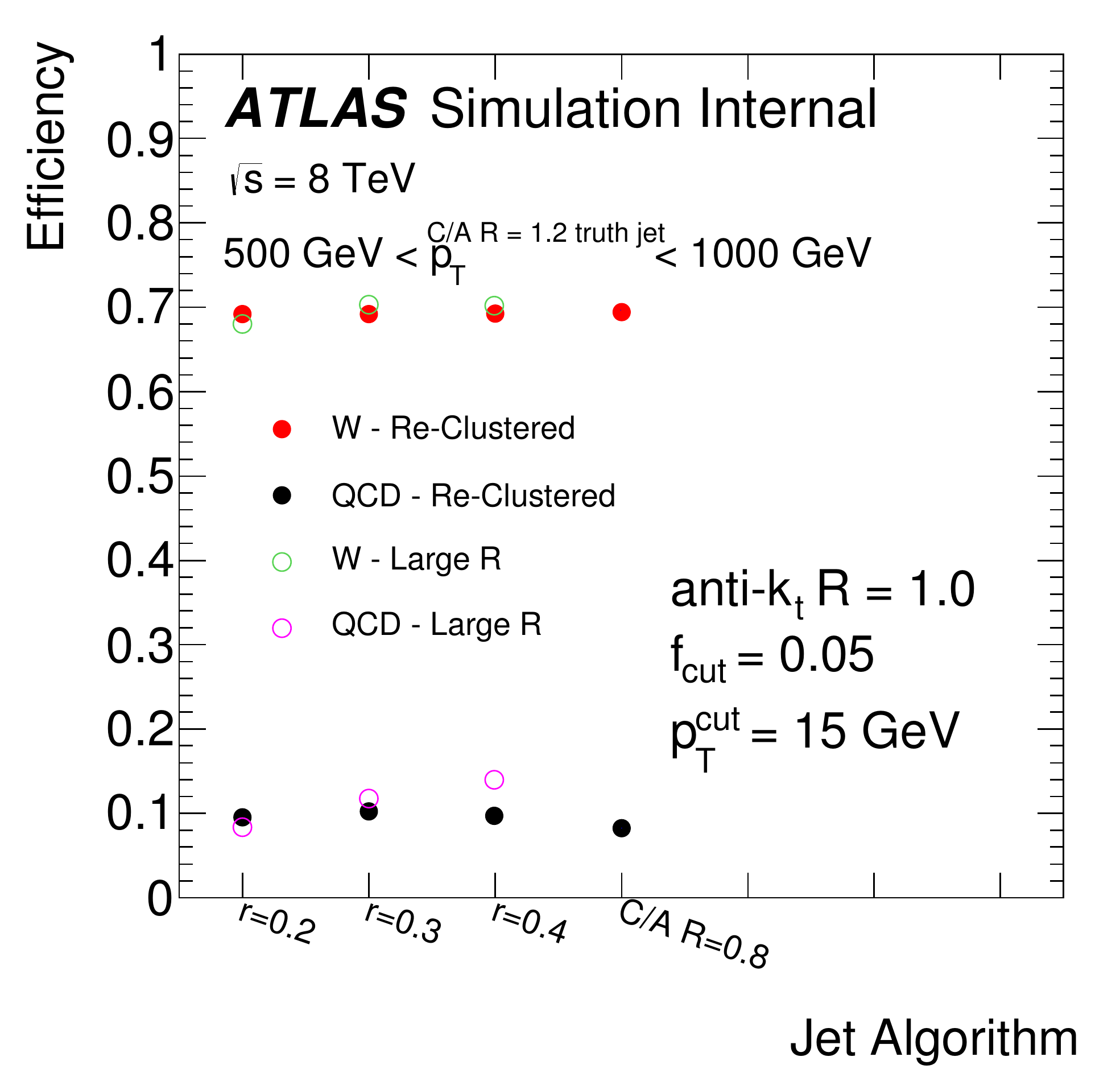}\includegraphics[width=0.5\textwidth]{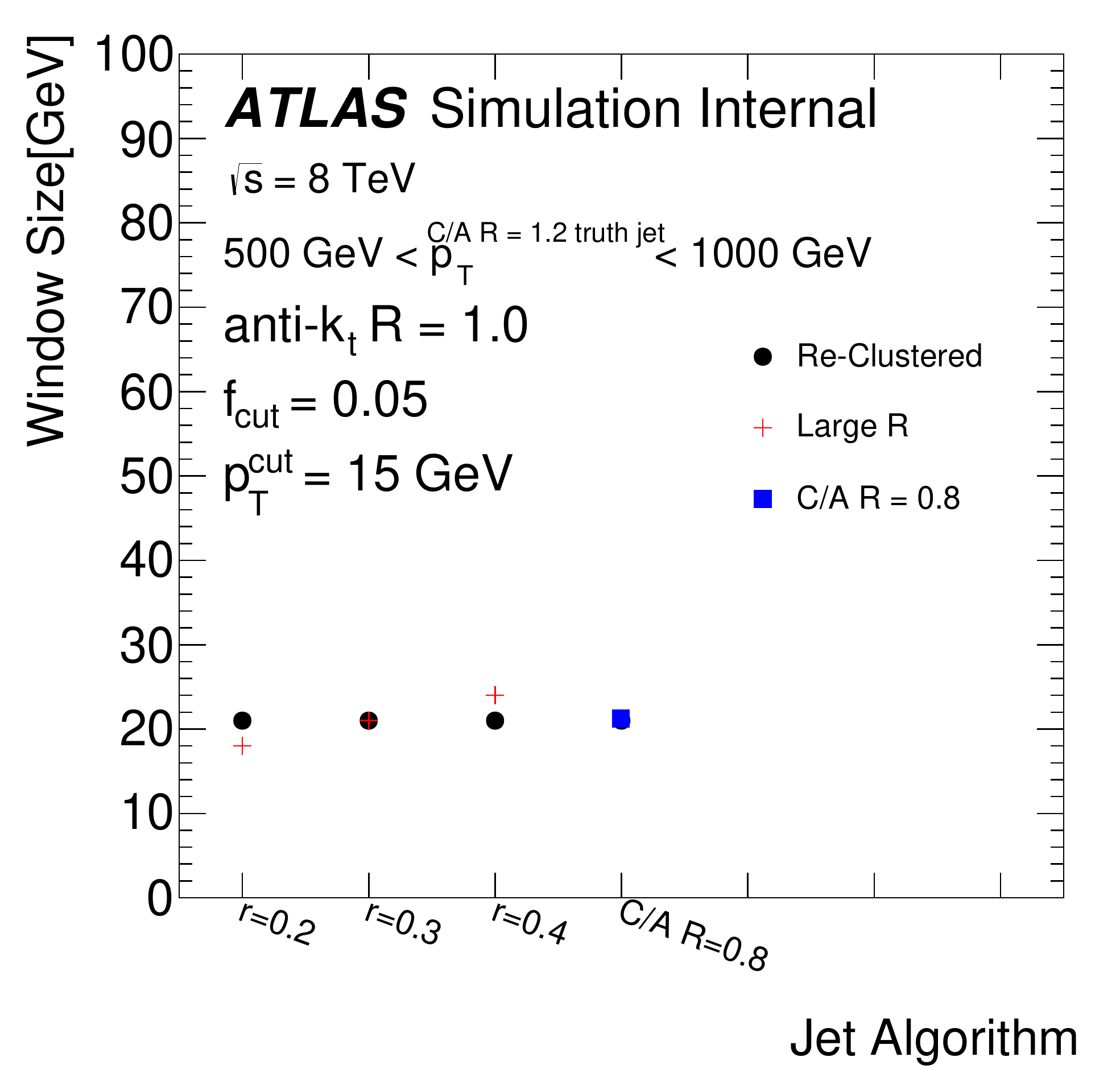}
\end{center}
\caption{Left: Window efficiency for various $r$ values.  Right: Size of the smallest 68\% window.  Both plots compare large radius trimmed jets with $R=1.0, f_\text{cut}=0.05$ and $R_\text{sub}=0.05$ with re-clustered jets that have analogous parameters.}
\label{fig:w:high:window2_4}
\end{figure}

\begin{figure}[h!]
\begin{center}
\includegraphics[width=0.5\textwidth]{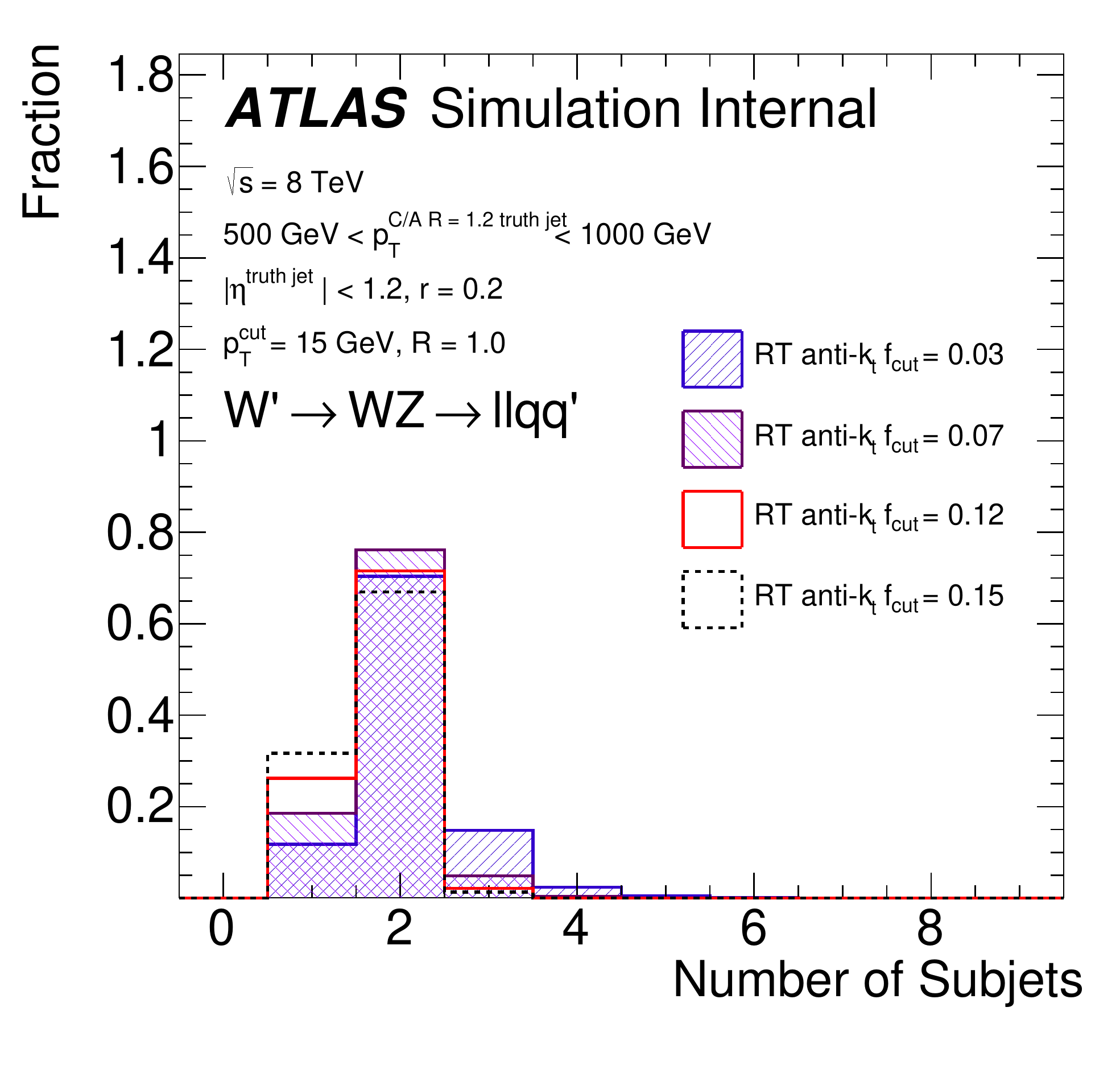}\includegraphics[width=0.5\textwidth]{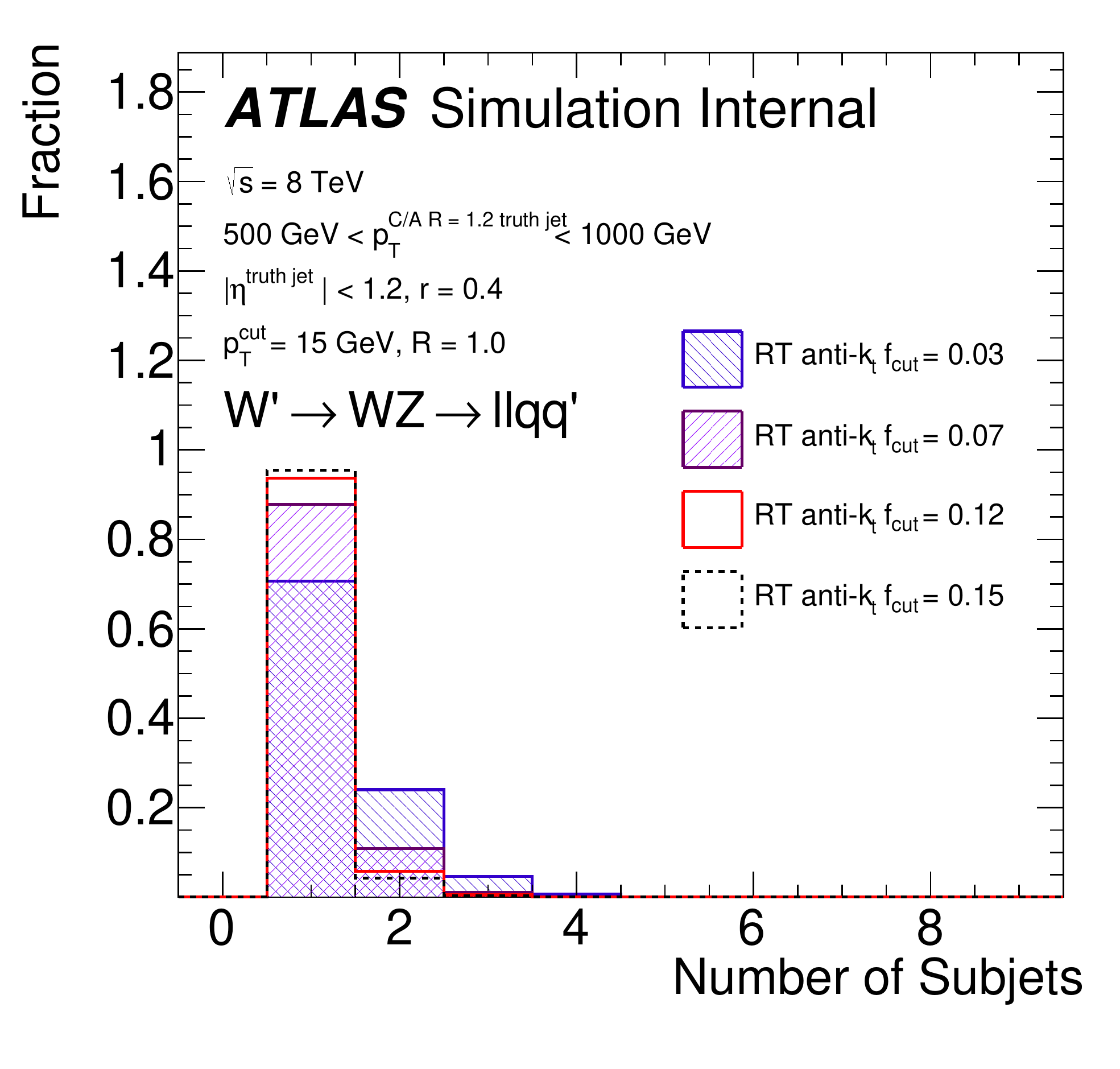}
\end{center}
\caption{The small-radius jet constituent multiplicity for several choices of $f_\text{cut}$ and a fixed $r=0.2$ (left) and $r=0.4$ (right).}
\label{fig:w:high:windowfNfhigh}
\end{figure}

\begin{figure}[h!]
\begin{center}
\includegraphics[width=0.5\textwidth]{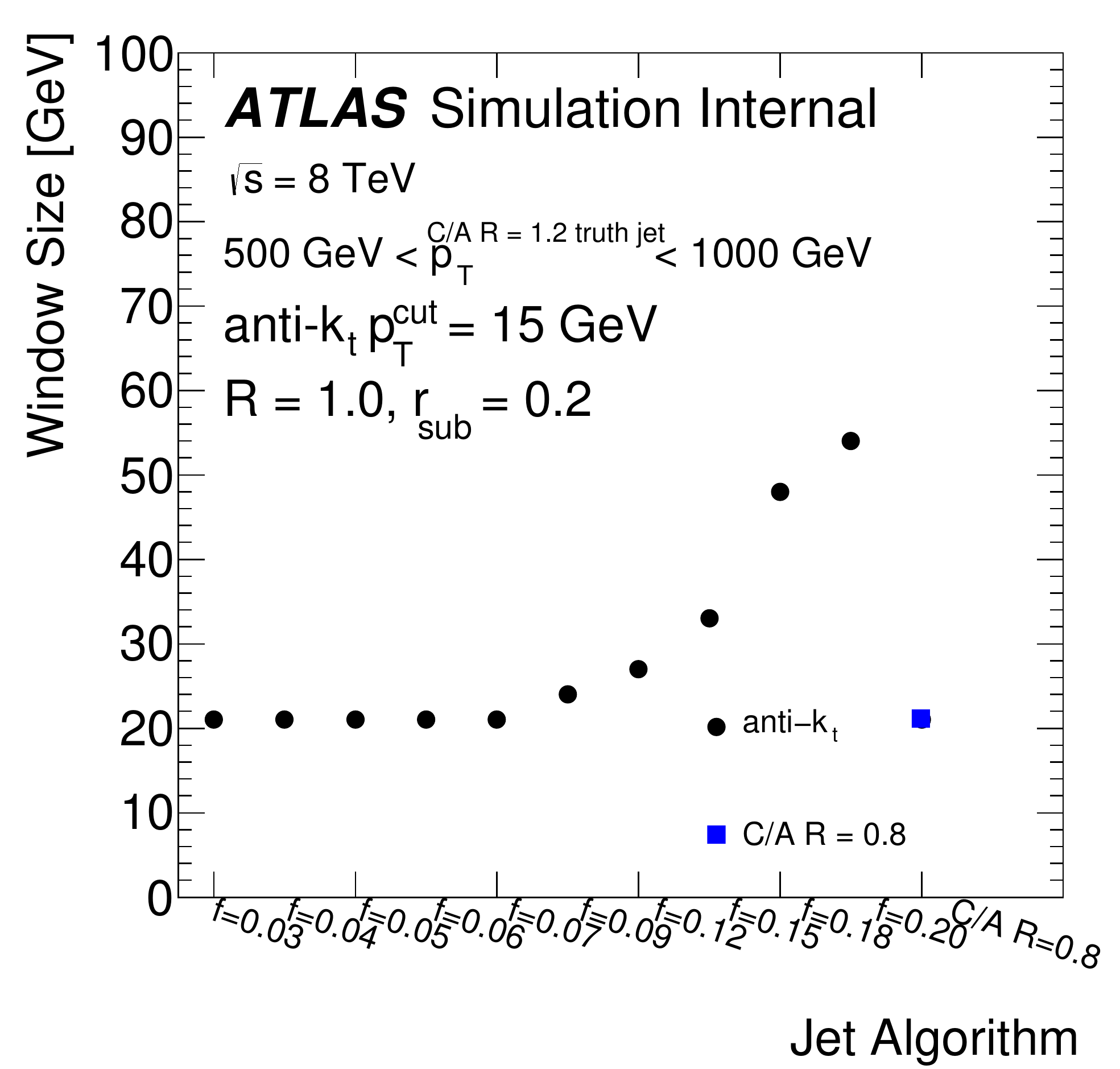}\includegraphics[width=0.5\textwidth]{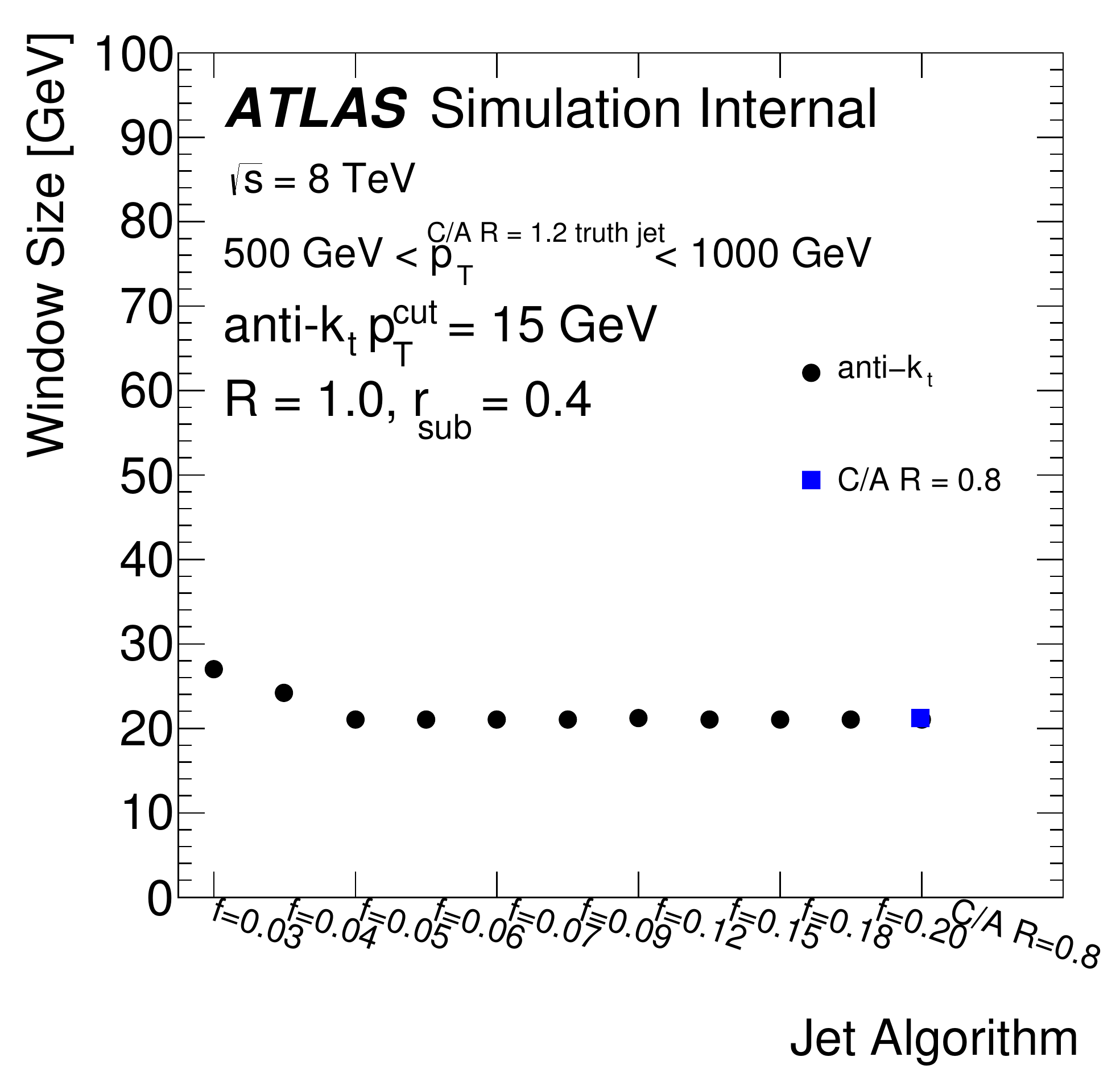}
\end{center}
\caption{The size of the smallest $68\%$ window as a function of $f_\text{cut}$ for $r=0.2$ (left) and $r=0.4$ (right).}
\label{fig:w:high:window2_4}
\end{figure}

\clearpage

	\subsubsection{Re-clustered Jet Resolution}
	\label{sec:recluster:calibrations}

This section augments the signal and background comparisons from the previous section with a study of the four-vector resolution of signal jets.  As they are both built directly or indirectly from calorimeter-cell clusters, one may expect the resolutions of re-clustered jets and large-radius trimmed jets to be similar.   Figure~\ref{fig:mcv:response_600} shows both the jet $p_\text{T}$ and mass response.  For both algorithms, $R=1.0$ and $f_\text{cut}=0.05$.  The large-radius trimmed jets use $k_t$ $R_\text{sub}=0.3$ subjets and the re-clustered jets use anti-$k_t$ $R=0.4$ small-radius jets\footnote{Smaller radius jets are not (yet) calibrated and understood at the same level as $R=0.4$ jets; the previous section does suggest that smaller would be better for the future.}.  This setting will now be default for the rest of the chapter.  For both re-clustered and large-radius trimmed jets, the particle-level reference is defined by running the same algorithm over detector-stable particles.  The core of the response distributions are nearly identical (even slightly better for re-clustered jets in the case of the mass), but there are heavy tails for re-clustered jets.  This can be explained by cases in which the re-clustering procedure picks a different number of jets at detector-level and particle-level.  For nearly symmetric $W$ boson decays, this asymmetry naturally introduces responses as big as $2$ or as small as $1/2$ for $p_\text{T}$ and even smaller for mass ($m^\text{small-radius}\ll m_W$).  One way to quantify this effect is to introduce a new jet collection called {\it re-clustered trimmed truth}, which is constructed by matching each detector-level constituent of a re-clustered jet with a small-radius particle-level jet and then replacing the detector-level jet four-vector by the particle-level jet properties.  When the detector-level and particle-level algorithms choose the same jets, the response of re-clustered trimmed truth jets is identically zero.  The heavy tails of the red dashed lines in Fig.~\ref{fig:mcv:response_600} show how the tails are explained by this mis-match in definition at particle-level and detector-level.  One way to remove the impact of the mis-match is to use an algorithm-independent reference object.  Figure~\ref{fig:mcv:ptVresponse_6002} is the analogue to Fig.~\ref{fig:mcv:response_600}, but using the $W$ boson four-vector as a reference for both re-clustered and large-radius trimmed jets.  The tails as well as the core of both distributions are similar.  One slight disadvantage of this method is that the resolution is now convolved with a non-negligible particle-level resolution from fragmentation that can hide differences between the algorithms in the tails.  A direct way of comparing the resolutions of the two methods is to directly compare the four-vectors jet-by-jet.  The ratio of the re-clustered $p_\text{T}$ to the large-radius trimmed jet $p_\text{T}$ is plotted as a function of boson $p_\text{T}$ in the left plot of Fig.~\ref{fig:mcv:2d_v4}.  For $p_\text{T}\gtrsim 200$ GeV where one $R\sim 1$ jet is expected to capture most of the boson decay products, the ratio is strongly peaked at one.  The two jets share most of the same regions of the calorimeter and so the response fluctuations shown in the middle and right plots of Fig.~\ref{fig:mcv:2d_v4} are highly correlated\footnote{In fact, one could use one of the collections to calibrate or establish uncertainties for the other. }.   One potential difference in the calibration between re-clustered jets and large-radius trimmed jets is the treatment of close-by hadronic activity.  Large-radius jets are calibrated as one unit, integrating over the distribution of energy inside the jet.  However, the calibration of re-clustered jets is factorized, first calibrating small-radius jets.  Figure~\ref{fig:mcv:2d} shows that the simulation does not predict a significant $\Delta R$ dependence of the re-clustered jet $p_\text{T}$ scale.  The next section describes a measurement of close-by effects on the jet mass using the $\sqrt{s}=8$ TeV data.

\clearpage

\begin{figure}[h!]
\begin{center}
\includegraphics[width=0.45\textwidth]{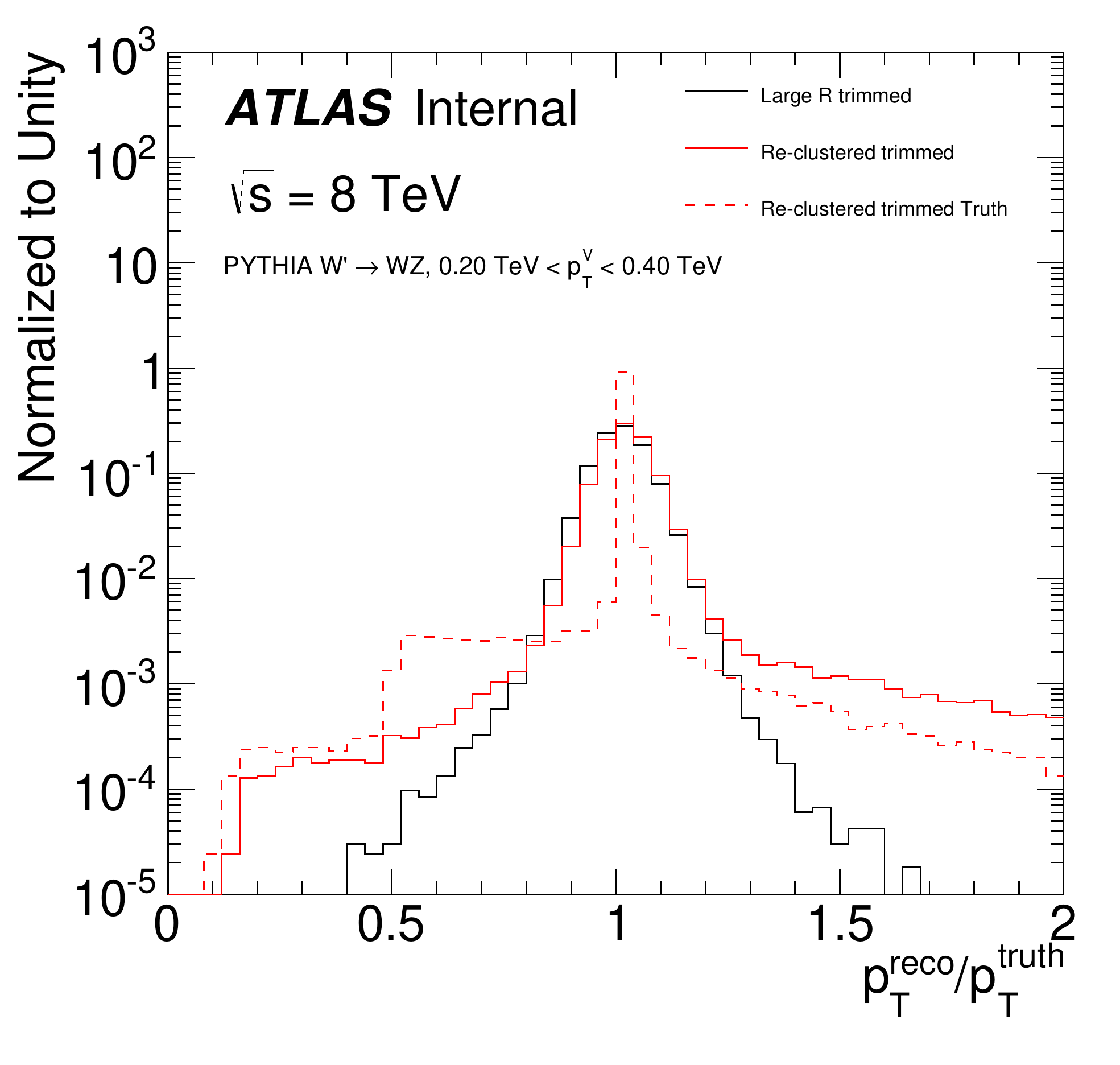}\includegraphics[width=0.45\textwidth]{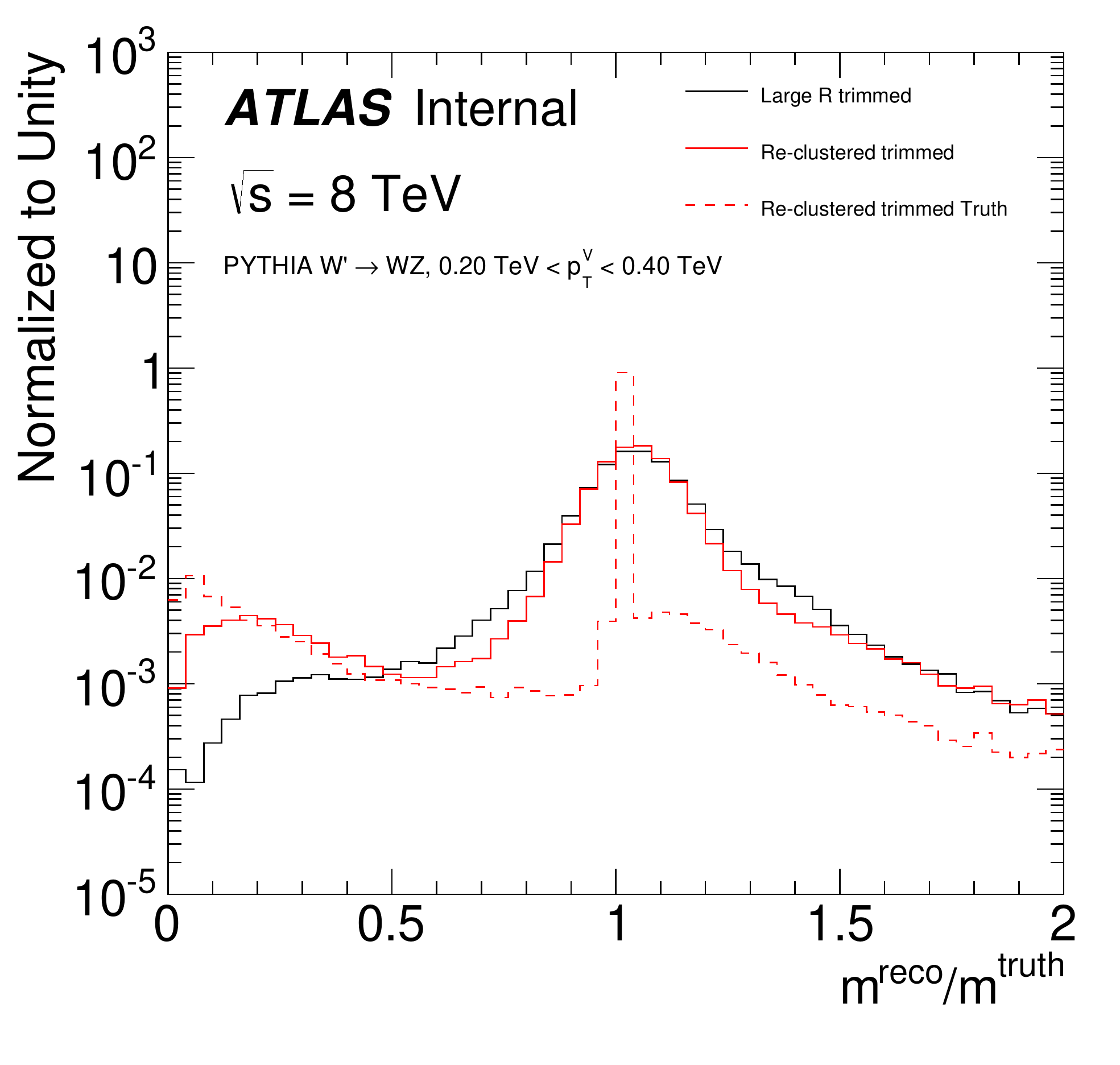}
\end{center}
\caption{The $p_\text{T}$ (left) and mass (right) response of large radius trimmed jets and re-clustered jets for $200$ GeV $<p_\text{T}^V<400$ GeV.  In this $p_\text{T}$ bin, over $80\%$ of the re-clustered jets have at least two constituents. }\label{fig:mcv:response_600}
\end{figure}

\begin{figure}[h!]
\begin{center}
\includegraphics[width=0.45\textwidth]{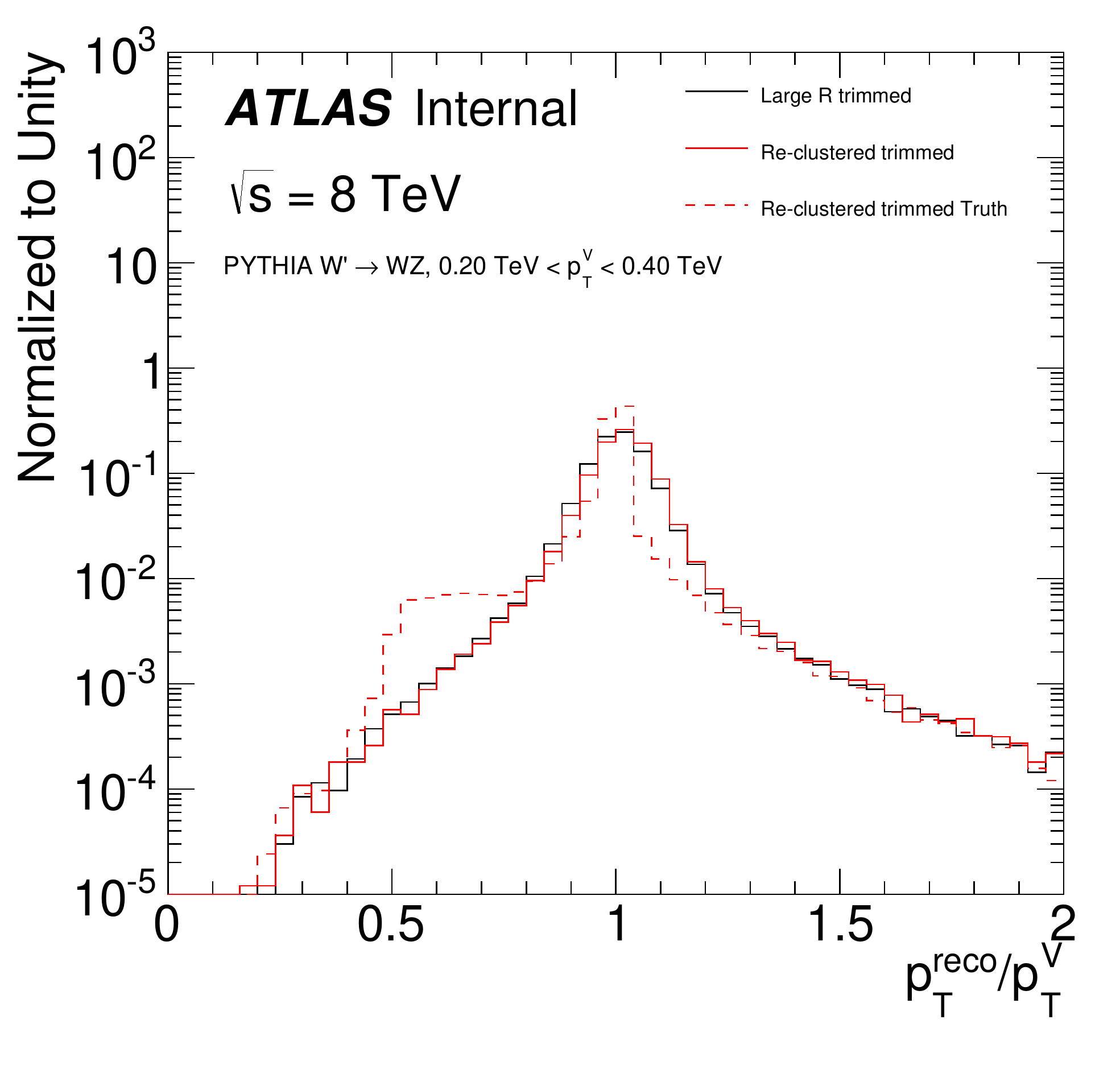}\includegraphics[width=0.45\textwidth]{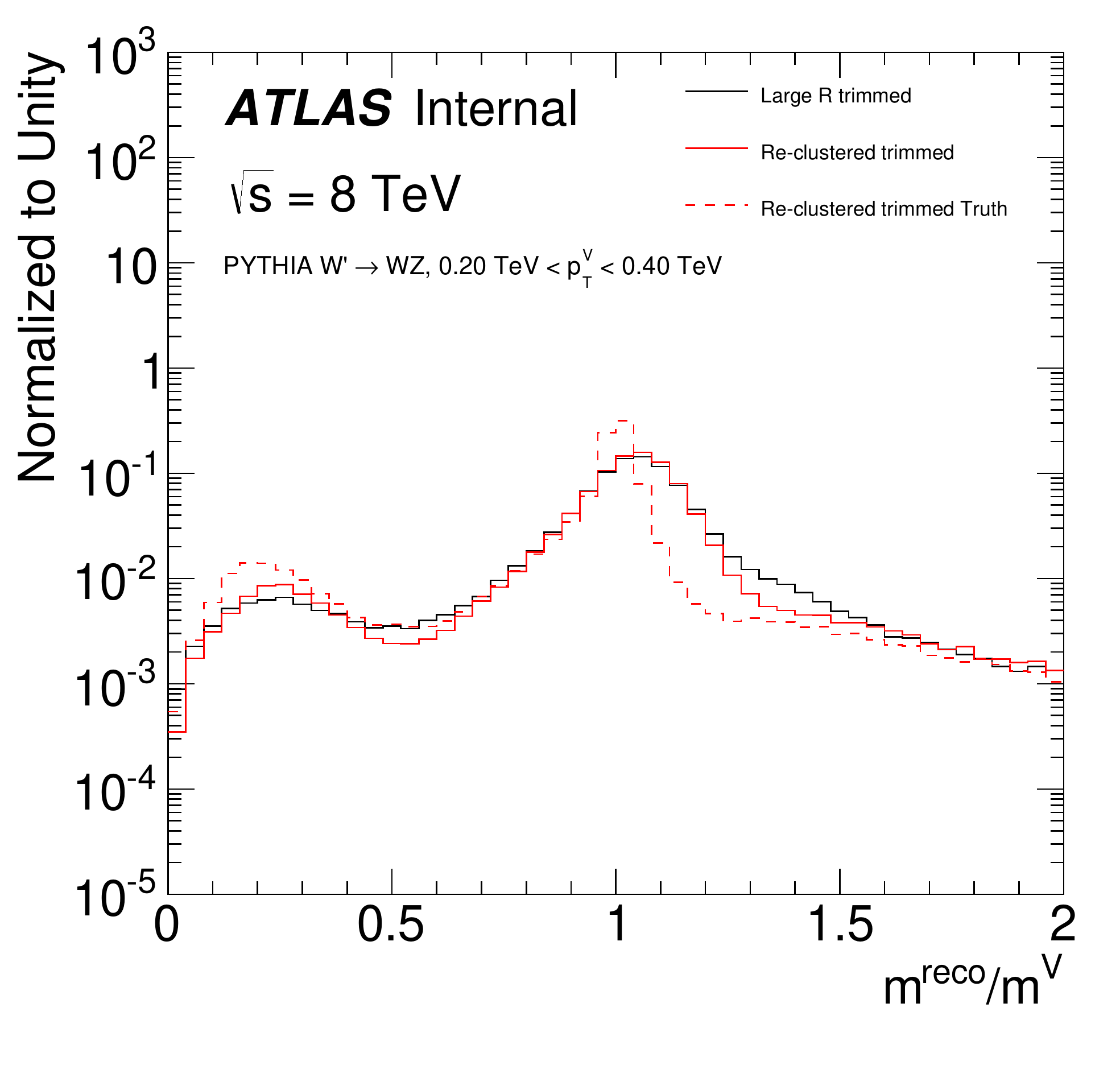}
\end{center}
\caption{The $p_T$ response, with the $p_T^V$ in the denominator instead of the truth jet $p_T$, of large radius trimmed jets and re-clustered jets for various bins of boson $p_T$.}\label{fig:mcv:ptVresponse_6002}
\end{figure}

\begin{figure}[h!]
\begin{center}
\includegraphics[width=0.33\textwidth]{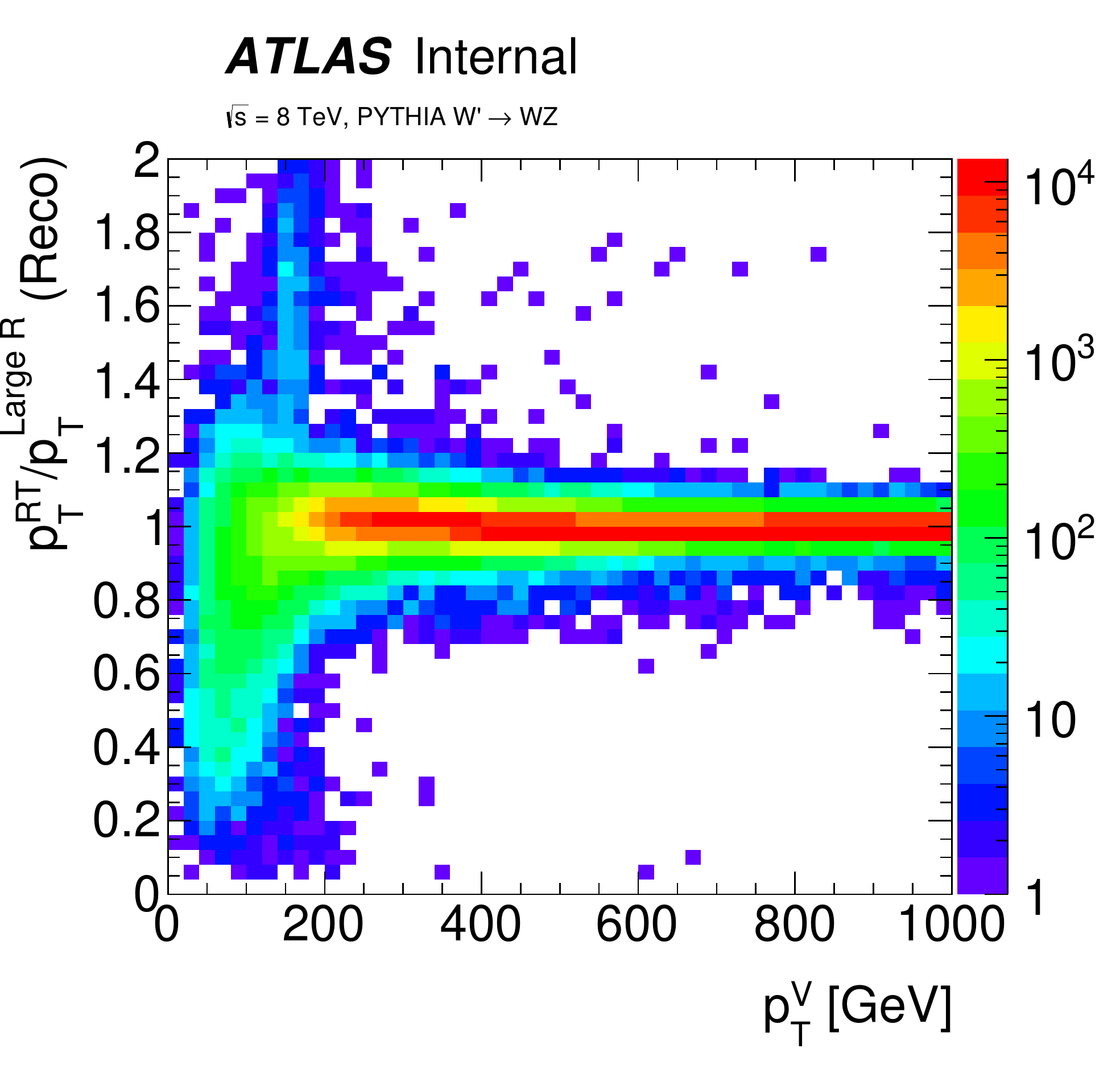}\includegraphics[width=0.33\textwidth]{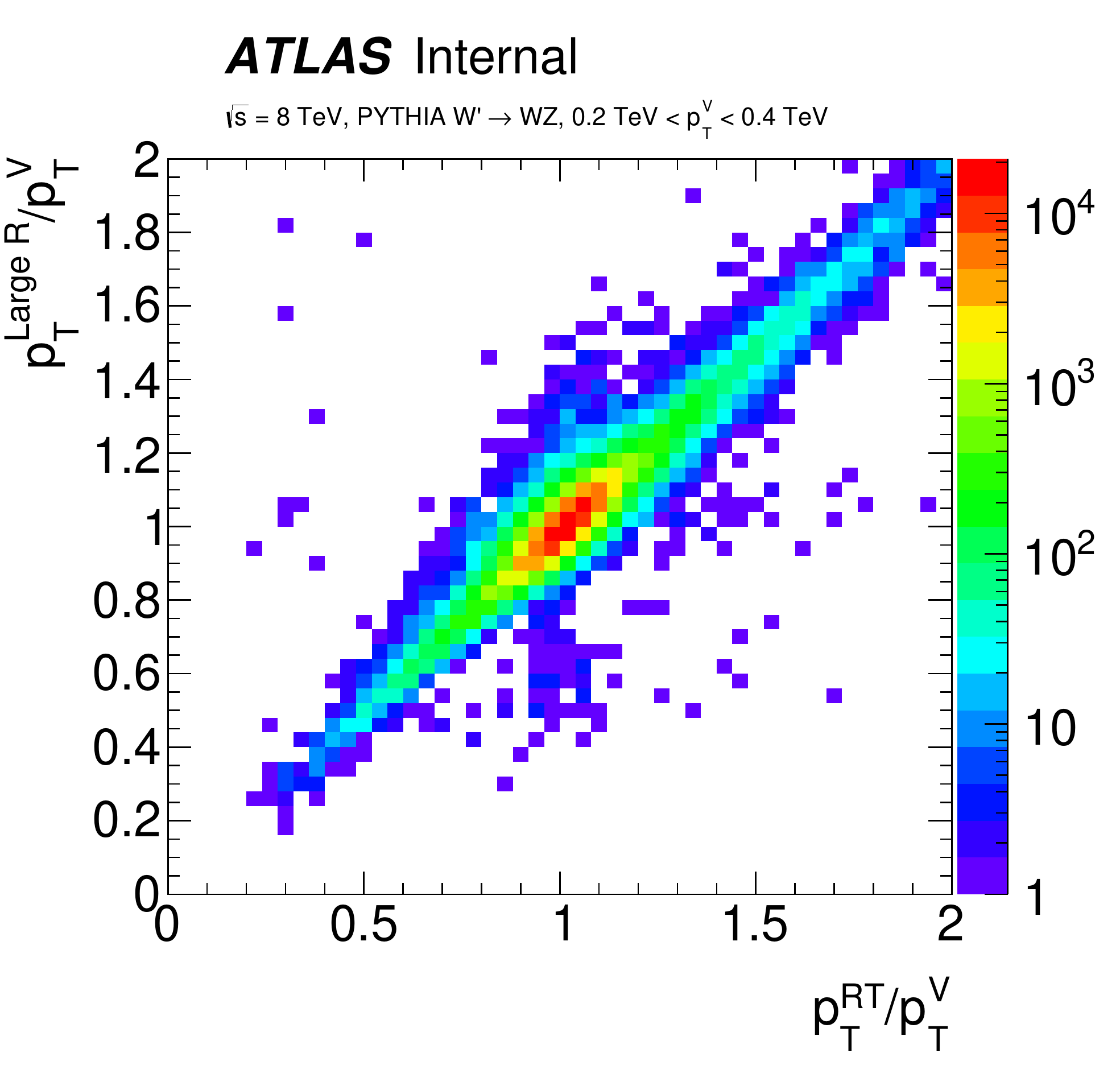}\includegraphics[width=0.33\textwidth]{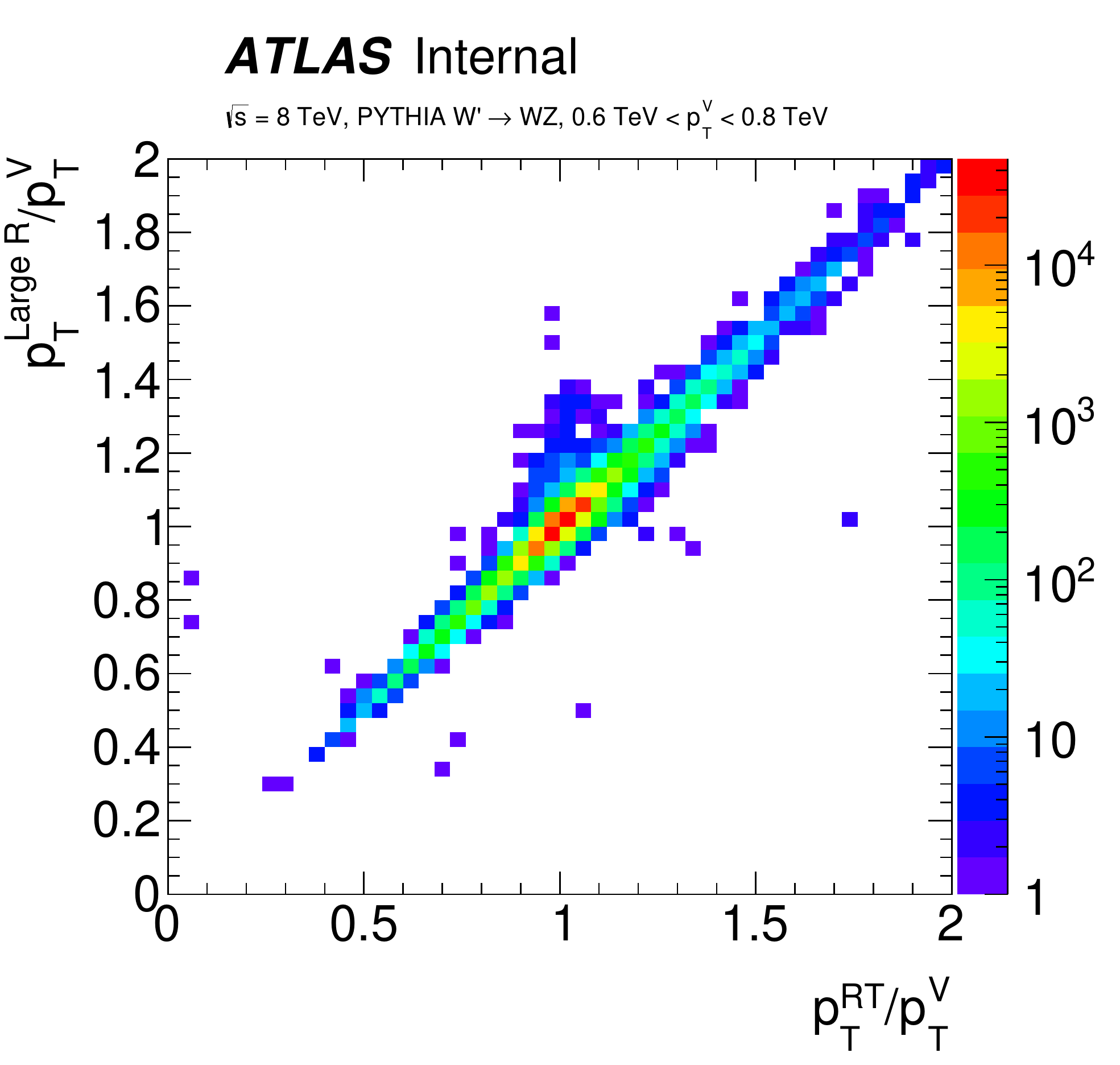}
\end{center}
\caption{Left: the ratio of the re-clustered jet $p_\text{T}$ to the large radius trimmed jet $p_\text{T}$ as a function of the boson $p_\text{T}$.  Middle (right): the joint distribution of the large-radius jet response and the re-clustered jet mass response in a low (high) $p_\text{T}^V$ bin.}\label{fig:mcv:2d_v4}
\end{figure}

\begin{figure}[h!]
\begin{center}
\includegraphics[width=0.5\textwidth]{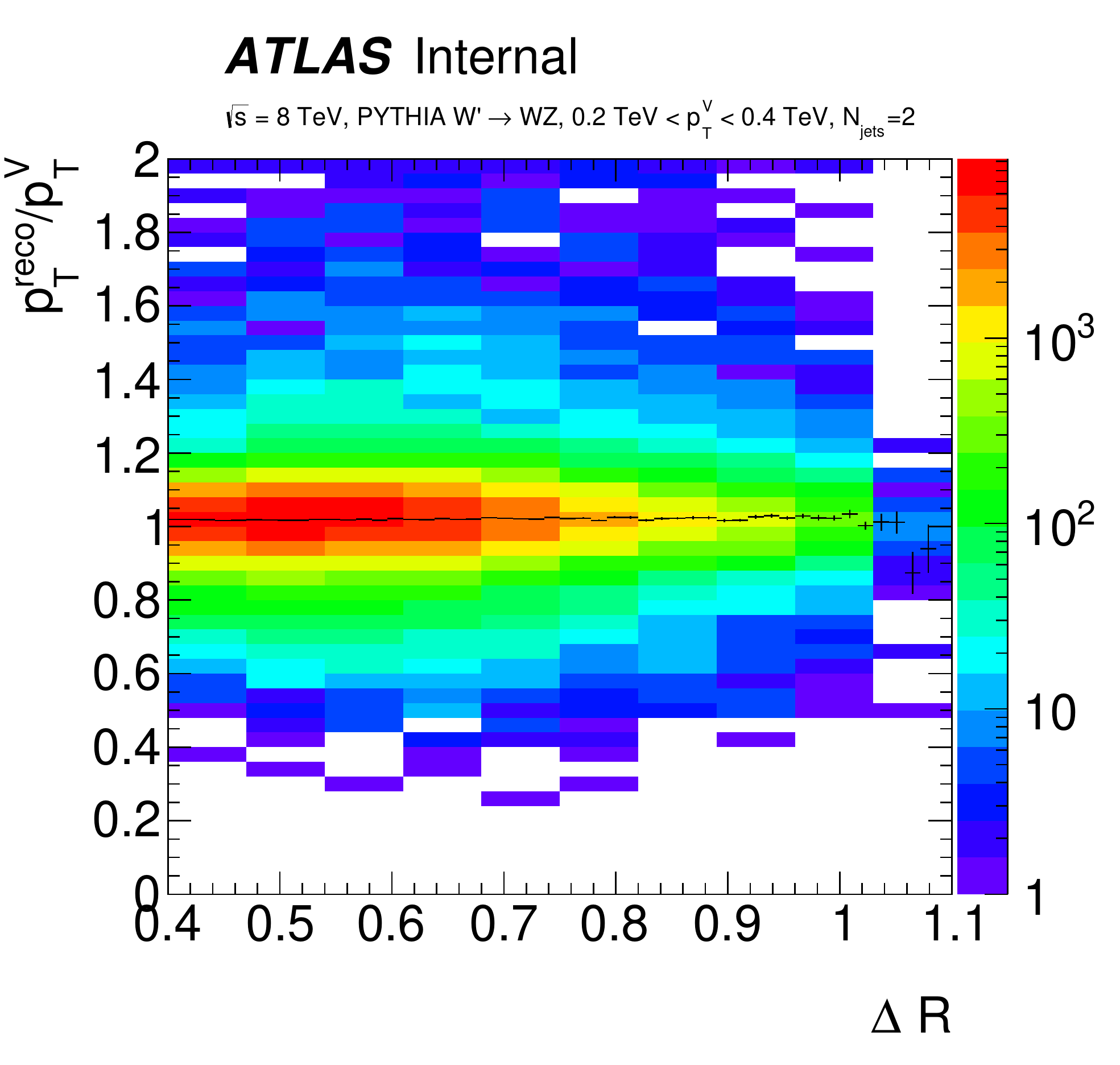}
\end{center}
\caption{The $p_\text{T}$ response as a function of the $\Delta R$ between the re-clustered jet constituents in events where the re-clustered jet has exactly two constituents and $200$ GeV $<p_\text{T}^\text{re-clustered jet}<400$ GeV.  }\label{fig:mcv:2d}
\end{figure}

	\clearpage

	\subsubsection{Close-by Effects}
	\label{sec:recluster:closeby}
	
When two jets are physically close in $\Delta R$, their four-vector response can be different with respect to isolated jets.  For instance, the energy deposits from particles in one particle-level jet may end up clustered into a different reconstructed jet.  In these cases, the energy response will be high for one jet and low for the other.  Even at particle-level, the presence of close-by jets can skew jet shapes due to the properties of the jet clustering algorithm, or changes in the physical distribution of energy due to color flow between jets.  The jet response depends on these jet shapes and so this is another source of bias in the response of non-isolated jets.  Non-isolated jets are common in high multiplicity final states but are always present in re-clustered jets with more than one small-radius jet constituent.	  Jet calibrations and their corresponding systematic uncertainties are derived in simulation from jets that are well-isolated.  These calibrations are applied to all jets, regardless of other close-by hadronic activity.  Studies at $\sqrt{s}=7$ TeV suggested that 2\%-5\% shifts in the jet energy scale are possible due to the presence of close-by jets~\cite{Aad:2011he}.  However, detailed studies with the larger $\sqrt{s}=8$ TeV dataset demonstrated that these shifts are well-modeled by the simulation and therefore no additional uncertainty is applied to the jet $p_\text{T}$ to account for close-by hadronic activity~\cite{ATLAS-CONF-2015-037}.  The right plot of Fig~\ref{fig:JMR:bosonintro2} shows that it is not sufficient for the jet $p_\text{T}$ to be well-modeled - there are important kinematic regimes where small-radius jets with a significant mass are in close proximity to other jets.  For example, this occurs in boosted top quark jets with the $W$ boson decay products are merged inside one small-radius jet distinct from a close-by $b$-quark jet.  Close-by effects on the jet mass response have never studied, but are critical for jet tagging in dense environments.  This section presents the first such measurement using an extension of the track-jet method (see Sec.~\ref{sec:JMR:trackjet}) called the {\it triple-ratio technique}, described in Sec.~\ref{sec:tripleratio}.  Large-radius jets in $t\bar{t}$ events are used to study the jet-area dependence of close-by shifts in the jet mass response because there are not enough small-radius jets with a significant mass and close-by activity.  These results are presented in Sec.~\ref{sec:closebyresults}.

\paragraph{Triple Ratio Technique} \mbox{} \\
\label{sec:tripleratio}

The first step in studying the dependence of the jet mass response on close-by jet activity in the data is to quantify the level of nearby radiation.  One possibility is the quantity $f_\text{closeby}$, defined as

\begin{align}
\label{eq:fcloseby}
f_\text{closeby} = \sum_j \frac{\vec{p}_j\cdot \vec{p}}{|\vec{p}|^2},
\end{align}
\def\dr{\ensuremath{\Delta R_\text{min}}}
\def\fcloseby{\ensuremath{f_\text{closeby}}}

\noindent where the sum runs over all jets above a $p_T$ threshold inside a cone of radius $\Delta R<X$ with respect to the probe jet.  Less isolated jets have a larger value of $f_\text{closeby}$.  There are a few ways to naturally extend the definition of the small-radius jet $f_\text{closeby}$ in Eq.~\ref{eq:fcloseby} to large radius jets.   The most obvious definition is to take all small-radius jets that are in some annulus of the jet axis, as is done with small radius jets.   A disadvantage of this definition is that the trimmed jet area can be rather small so that small-radius jets with $\Delta R\sim 1$ can already be quite far away from the jet center.  An alternative definition uses calorimeter-cell clusters instead of jets in Eq.~\ref{eq:fcloseby}.  The list of clusters could be inclusive or exclude those clusters dropped in the trimming process.   Figure~\ref{fig:largeRprops1} compares three definitions of $f_\text{closeby}$ for large-radius jets.  The definition based on small-radius jet (clusters) uses $1<\Delta R < 2$ ($\Delta R<2$).  By construction, the clusters removed from trimming are relatively soft, so the cluster-based definitions are highly correlated.  If there are no close-by clusters, there will not be close-by jets, but there can also be no close-by jets above $25$ GeV but many close-by clusters.  For the rest of this section, the cluster-based definition, excluding trimmed clusters, is used for making comparisons with data.

\begin{figure}[h!]
\centering
\includegraphics[width=0.45\textwidth]{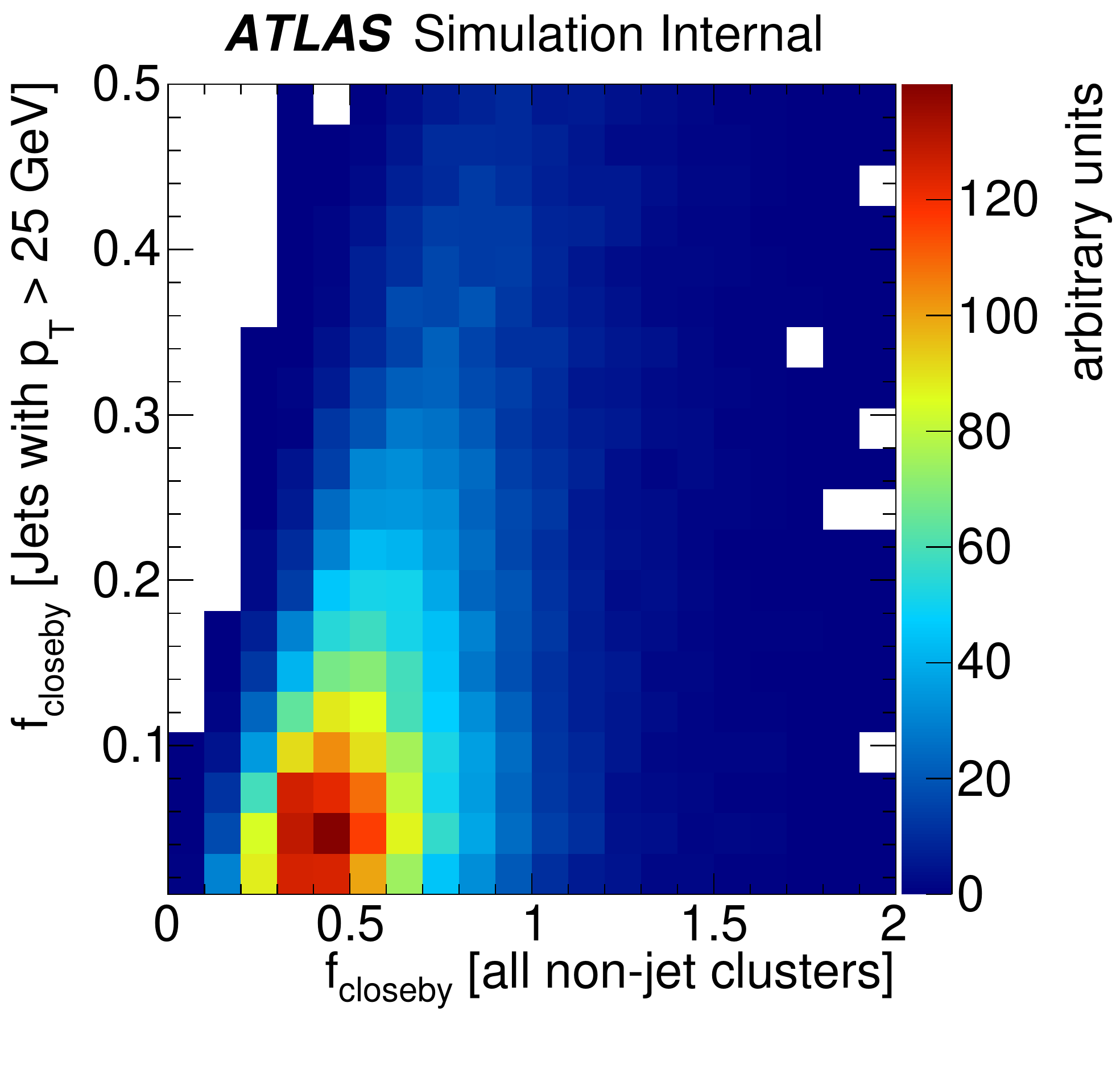}\hspace{4mm}\includegraphics[width=0.45\textwidth]{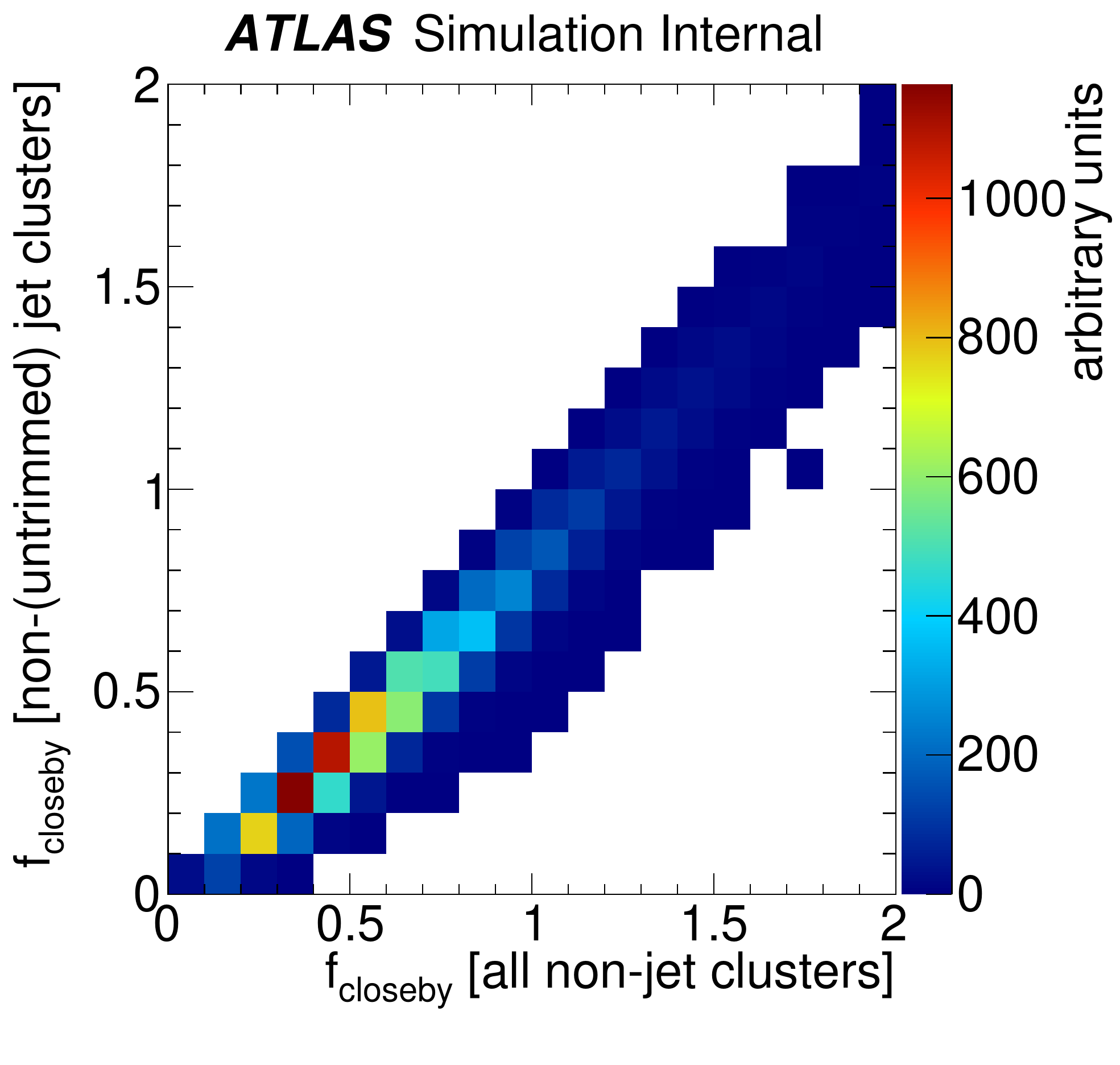}
\caption{The joint distributions of $f_\text{closeby}$ computed with nearby ($1<\Delta R <2$) small-radius jets or all non-jet clusters within $\Delta R<2$ (left) and two cluster-based definition that either include or exclude clusters dropped during trimming (right).  }
\label{fig:largeRprops1}
\end{figure}

In order to probe the impact of close-by effects on $R=0.4$ jets, the jet area dependence is studied in a sample of large-radius trimmed jets with a range of sizes.  Figure~\ref{fig:largeRprops} shows the jet $m/p_\text{T}$ and jet area regions considered in this analysis in addition to the distribution of the cluster $f_\text{closeby}$.   The $f_\text{closeby}$ distribution does not depend strongly on the jet area, but does decrease with $p_\text{T}$ due to the close-by $b$-jet.

\begin{figure}[h!]
\centering
\includegraphics[width=0.45\textwidth]{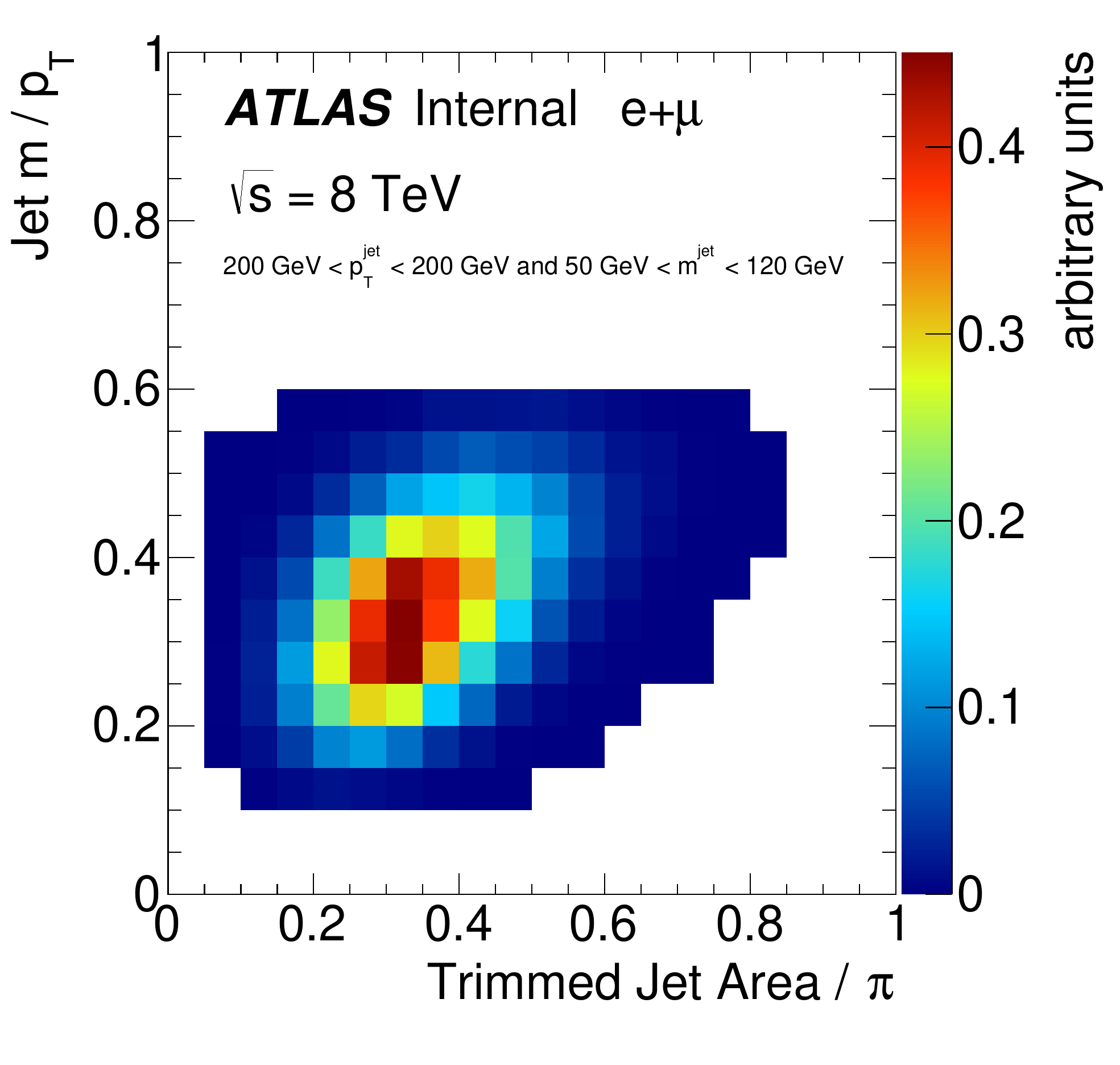}\hspace{4mm}\includegraphics[width=0.45\textwidth]{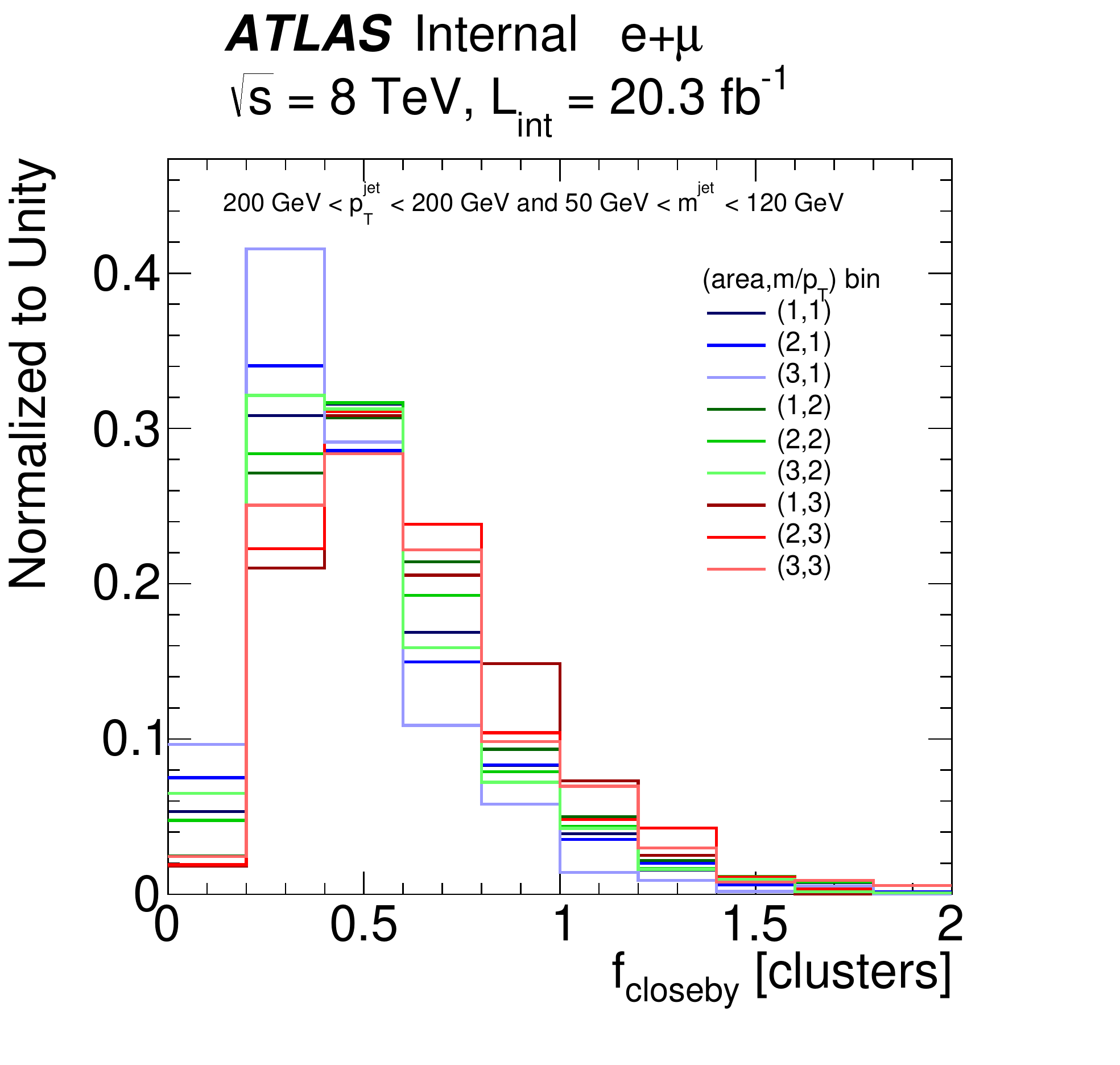}
\caption{Left: the jet $m/p_T$ and trimmed jet area regions used in this analysis.  Right: the distribution of the cluster $f_\text{closeby}$ in all the nine bins of jet $m/p_T$ and trimmed jet area, evenly spaced between $3\times 3$ bins in the range $0.2$ and $0.5$ in $m/p_T$ and in the range $0.2\pi$ and $0.5\pi$ in the trimmed jet area. }
\label{fig:largeRprops}
\end{figure}

The impact of close-by jets is quantified by comparing the response (using $r_\text{track}=m_\text{calo} / m_\text{tracks}$) of isolated jets with non-isolated jets: $R=r_\text{track}^\text{non-iso}/r_\text{track}^\text{iso}$.  A third ratio ($r_\text{track}$ is itself a ratio) is formed to compare data and simulation: $R^\text{data}/R^\text{MC}$. 

\clearpage

\paragraph{In-situ Close-by Results for the Jet Mass} \mbox{} \\
\label{sec:closebyresults}
	
The ratio the median\footnote{The median is less sensitive to outliers than the mean.} $r_\text{track}=m_\text{from calo}/m_\text{from tracks}$ as a function of the jet $m/p_\text{T}$ and the trimmed jet area between low $f_\text{closeby}<0.6$ and $f_\text{closeby}>0.6$ in data and MC are presented in Fig.~\ref{fig:largeRpropsfirstratio}.  The value $f_\text{closeby}=0.6$ is chosen to separate isolated and non-isolated jets because it is approximately the median of the $f_\text{closeby}$ distribution.  A small negative trend in $m/p_\text{T}$ is predicted by the simulation in the right plot of Fig.~\ref{fig:largeRpropsfirstratio}, but this is not in the data distribution.  Instead, there is a small trend in the opposite direction.  This is quantified by the triple ratio in Fig.~\ref{fig:largeRpropstriple} that is the ratio of the left and right plots from Fig.~\ref{fig:largeRpropsfirstratio}.  Except for low $m/p_\text{T}$ and large jet area, the triple ratio is statistical consistent with unity, suggesting that no additional uncertainty is required for re-clustered jets due to the modeling of close-by jets.  There may be a $\lesssim 2\%-5\%$ bias for large jet areas, but this is not the focus of this section.  Figure~\ref{fig:largeRpropsfirstratio} does not include any systematic uncertainties, but many of the experimental sources of bias cancel in one of the three ratios.  For example, Fig.~\ref{fig:largeRpropsJMS} shows the impact of varying the JMS up and down within its uncertainty.  The resulting change in the triple ratio is $\lesssim 1\%$ in all bins.
	
\begin{figure}[h!]
\centering
\includegraphics[width=0.43\textwidth]{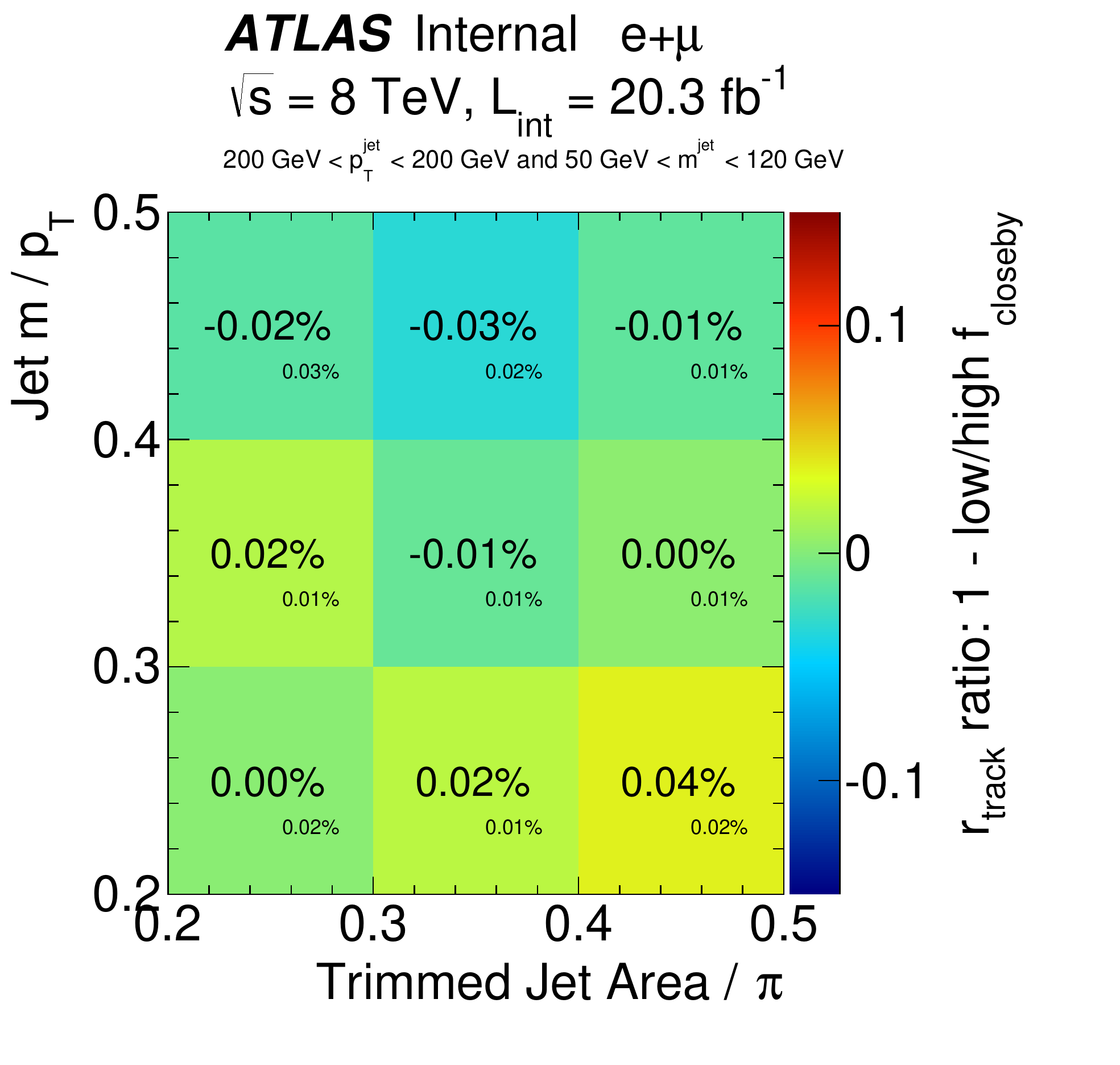}\hspace{4mm}\includegraphics[width=0.43\textwidth]{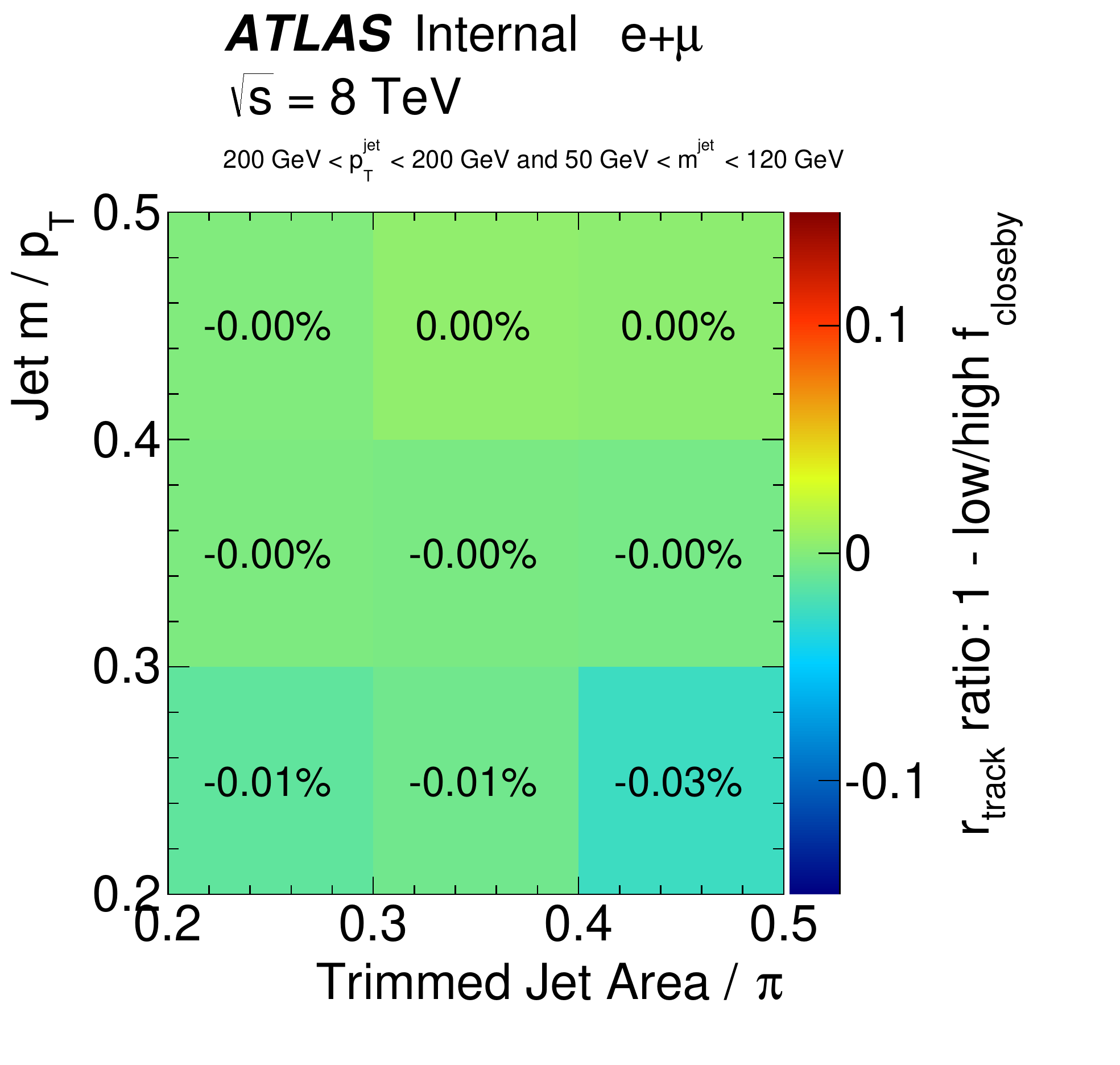}
\caption{The ratio of $r_\text{track}$ between isolated (lose $f_\text{closeby}$) and non-isolated jets (high $f_\text{closeby}$) as a function of the jet area and $m/p_\text{T}$ in data (left) and simulation (right).  The smaller number in the data plot is the statistical uncertainty.}
\label{fig:largeRpropsfirstratio}
\end{figure}	

\begin{figure}[h!]
\centering
\includegraphics[width=0.5\textwidth]{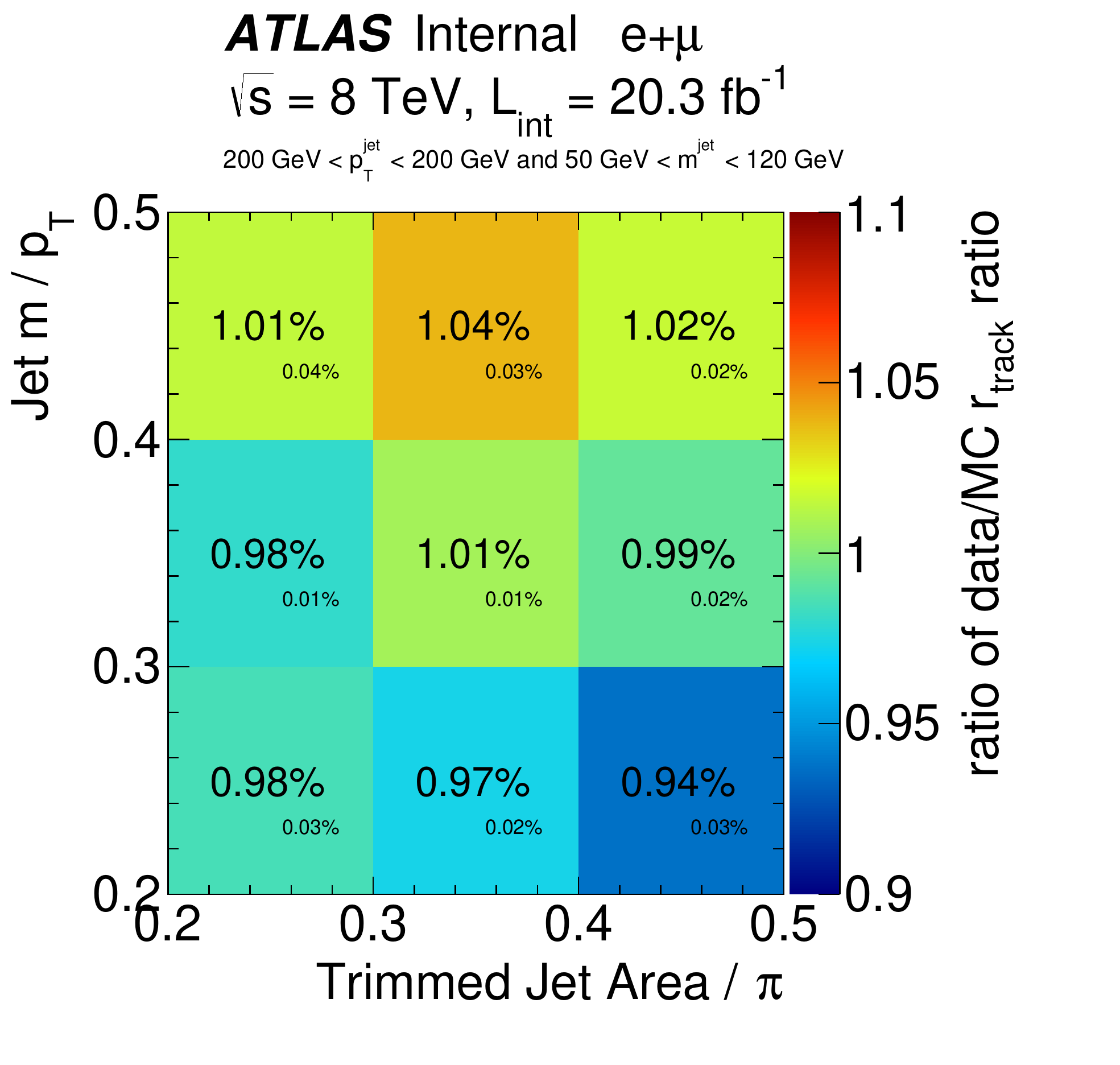}
\caption{The ratio of the left and right plots from Fig.~\ref{fig:largeRpropsfirstratio}.  The smaller number in each bin indicates the data statistical uncertainty.}
\label{fig:largeRpropstriple}
\end{figure}	

\begin{figure}[h!]
\centering
\includegraphics[width=0.43\textwidth]{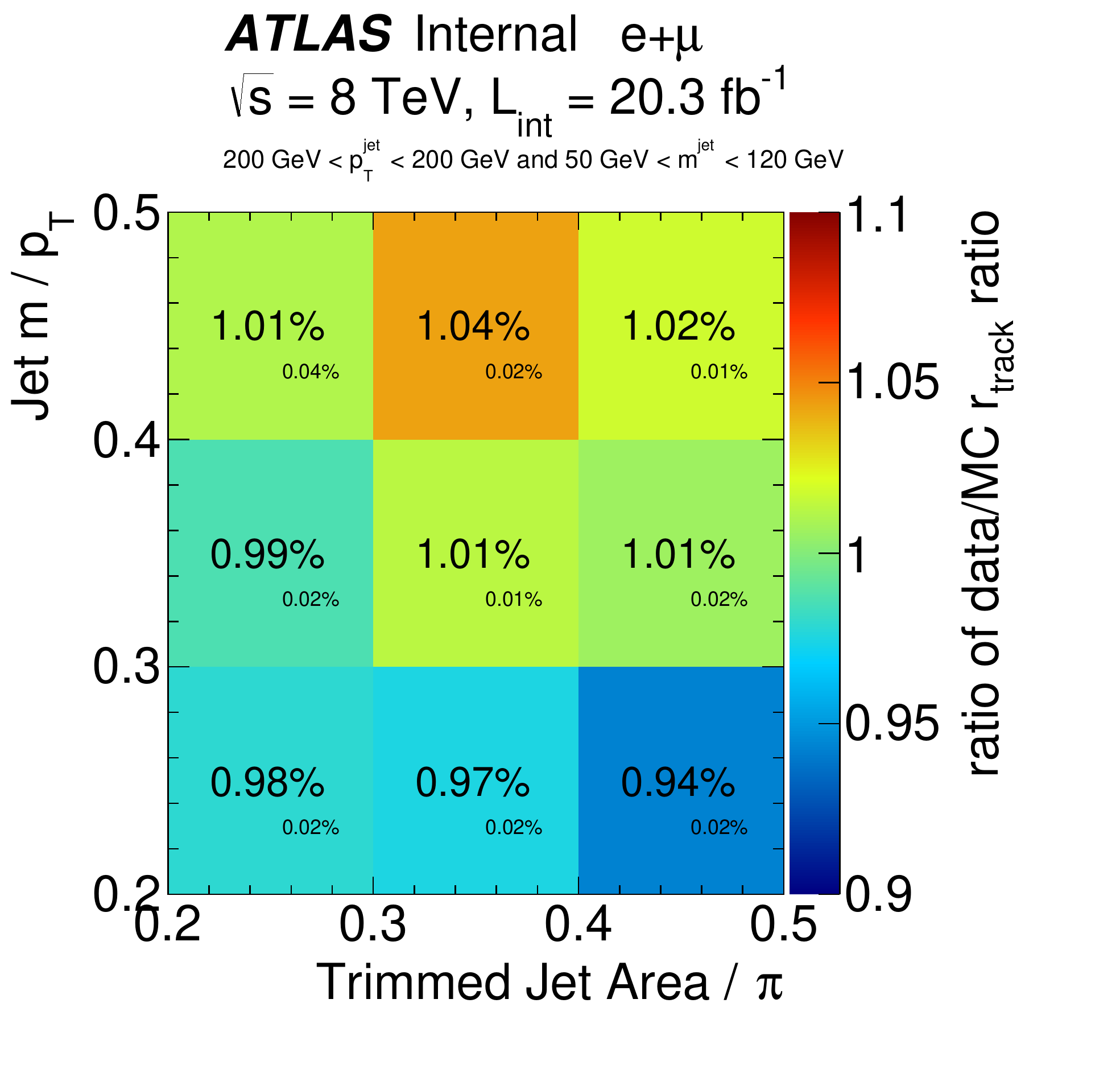}\hspace{4mm}\includegraphics[width=0.43\textwidth]{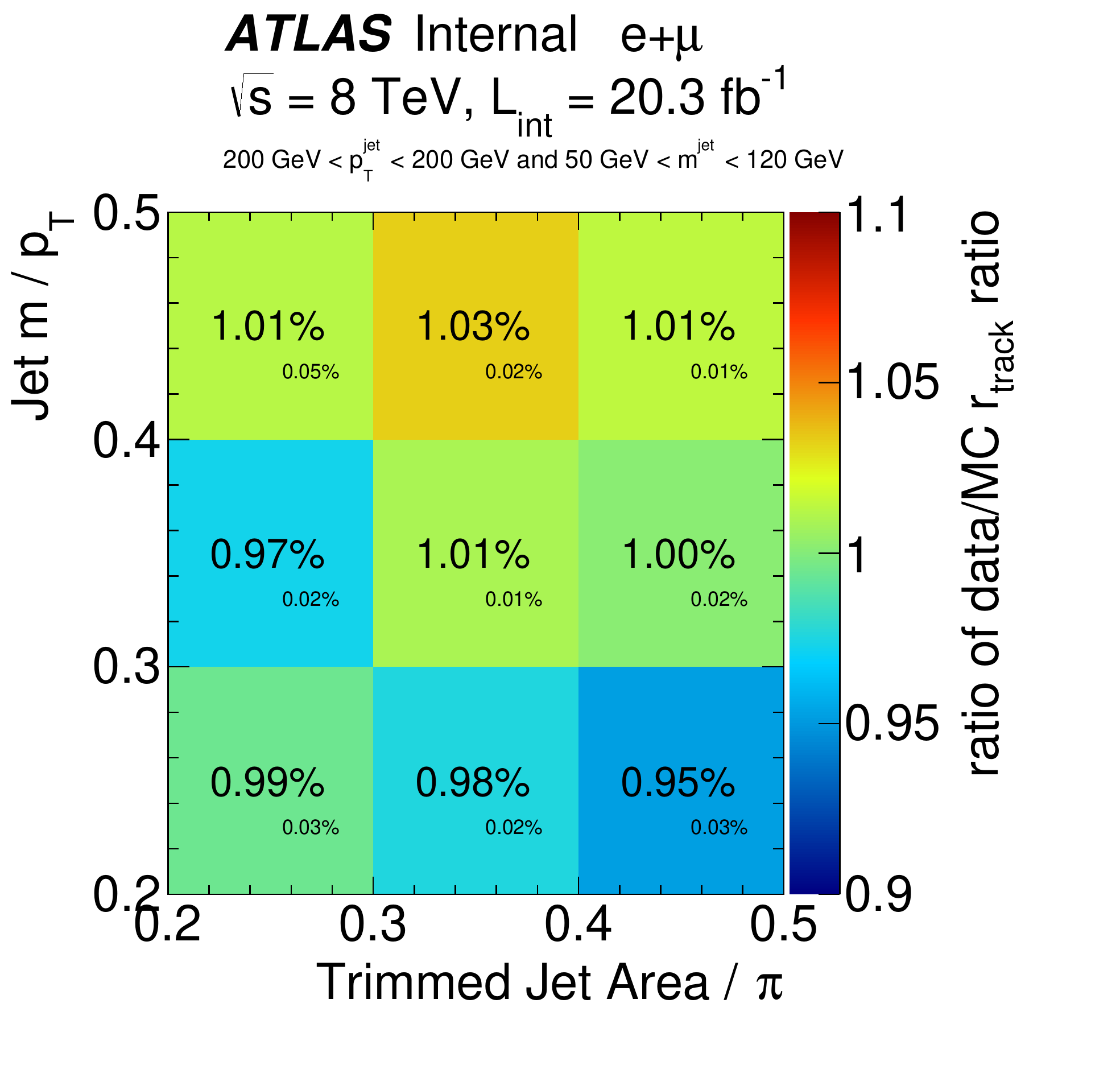}
\caption{The same as Fig.~\ref{fig:largeRpropstriple}, but with the JMS varied down (left) or up (right).}
\label{fig:largeRpropsJMS}
\end{figure}
	
	\clearpage
	
	\subsubsection{Summary and Outlook}
	\label{sec:reclustering:conclucion}
	
	Re-clustering is a modular paradigm for large-radius jet clustering that introduces analysis flexibility and a natural scheme for estimating systematic uncertainties.  This flexibility can increase the discovery potential of the LHC as the large-radius jet parameters can be individually optimized for each analysis.  One last appealing property of re-clustering is that it provides a continuous bridge between the low and high $p_\text{T}$ regimes.  At low $p_\text{T}$, re-clustered jets tend to have several constituents and most of the mass information is from the $p_\text{T}$ of the small-radius jets.  At high $p_\text{T}$, a re-clustered jet is identical to a single small-radius jet (see Fig.~\ref{fig:largeRpropsratiorecluster}).  Calibrations (and uncertainties) of small-radius jet mass is not well-constrained, but this is also true for traditional `large-radius jets' at high $p_\text{T}$ with small groomed area.  The only difference is that the division between the two cases is made explicit in the re-clustering paradigm.  Where re-clustering reduces to a single jet, one can use the {\it track-assisted} jet mass introduced in the next section to obtain calibrations and uncertainties. 
	
	\vspace{5mm}
	
	\begin{figure}[h!]
\centering
\includegraphics[width=0.5\textwidth]{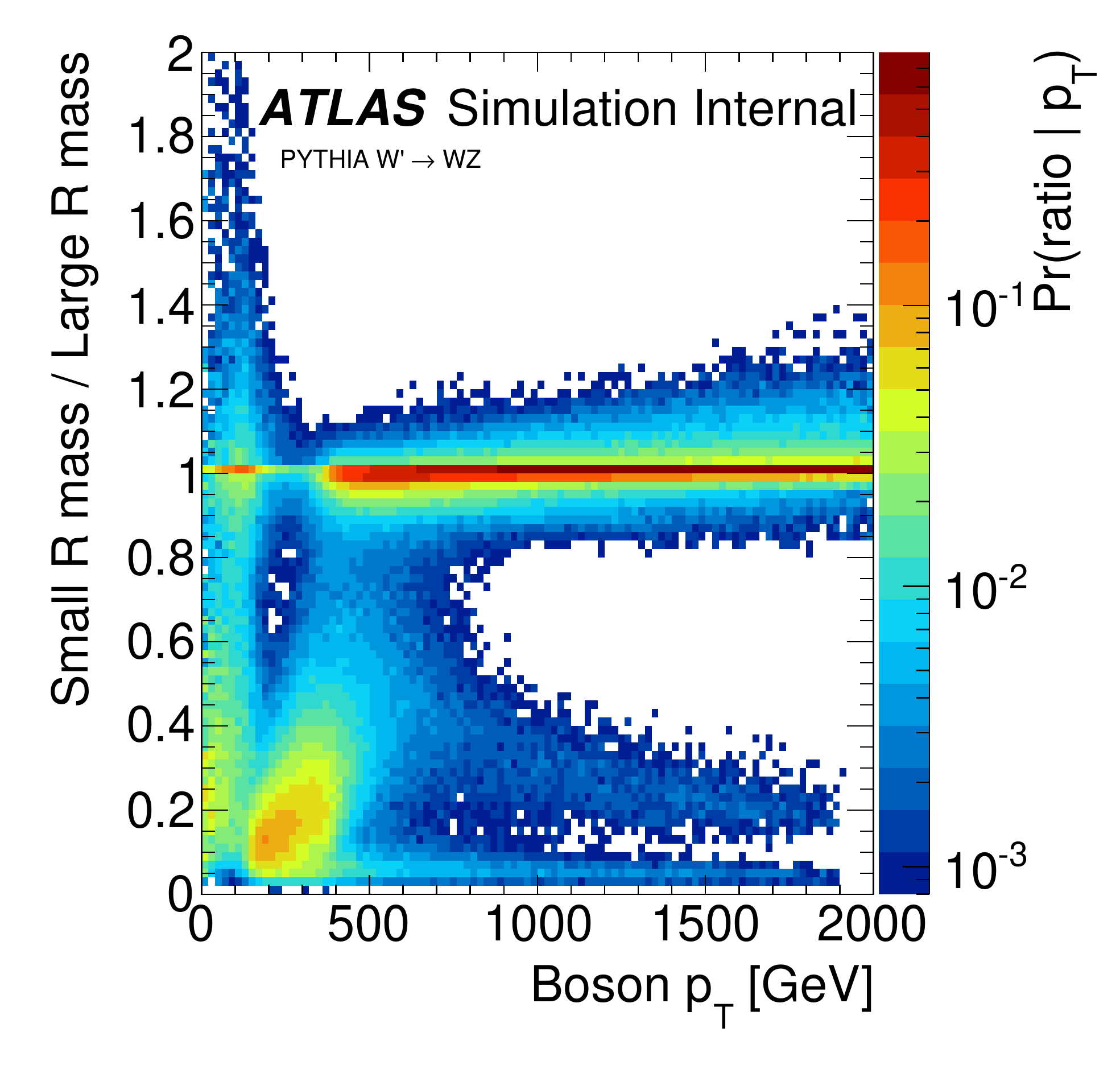}\caption{The distribution of the ratio of the leading large-radius jet mass to small-radius jet mass  as a function of $W$ boson $p_\text{T}$.}
\label{fig:largeRpropsratiorecluster}
\end{figure}
	
	\clearpage
	
	\subsection{Track-assisted Jet Mass}
	\label{sec:TAMass}
	
	The main challenge of re-clustering is at high $p_\text{T}$ where the re-clustered large-radius jets have only one small radius jet constituent.   The mass of small-radius jets is less constrained than large-radius jets from the data due to the lack of ultra high $p_\text{T}$ $W$ boson and top quark jets.  Additionally, the mass resolution for both small- and large-radius jets degrades at high $p_\text{T}$, as discussed in Sec.~\ref{sec:JMR}, as the distance between particles approaches the cluster angular resolution and ultimately the detector granularity.  One strategy to mitigate this degradation in the resolution at high $p_\text{T}$ is to use information from charged particle tracks {\it as part of the jet mass reconstruction}.  The track momentum resolution also degrades with $p_\text{T}$, but the angular resolution is significantly superior to the calorimeter angular resolution.  Track and calorimeter information are already combined as part of particle-flow techniques in CMS~\cite{CMS:2009nxa}, but these procedures suffer at high $p_\text{T}$ from the inability to accurately match tracks and clusters, especially in ATLAS where the magnetic field is weaker than in CMS by a factor of two.  The idea in this section is to use track-based properties of (sub)jets without attempting to match tracks with individual calorimeter clusters.	
	
	The left plot Fig.~\ref{fig:tamass:first} shows the average number of tracks and clusters inside an $R=0.4$ boosted boson jet as a function of the jet $p_\text{T}$.  The particle multiplicity for a $W$ boson jet should be nearly independent of $p_\text{T}$ because the starting scale for the parton shower is set by the quark $p_\text{T}$ in the $W$ boson center-of-mass frame, which is independent of the boost.  The number of tracks is nearly constant up to $p_\text{T}\sim 2$ TeV and then drops by about 1 track over the next 1 TeV.  In contrast, the number of calorimeter clusters decreases significantly with jet $p_\text{T}$ as the particles become more collimated with increasing boost.  This is illustrated with an event display of a particular high $p_\text{T}$ $W$ boson jet in the right plot of Fig.~\ref{fig:tamass:first}.  The $p_\text{T}\sim 3.5$ TeV $W$ boson jet has a particle level mass of about 80 GeV and a calorimeter mass of about 150 GeV.  This large mass is due in part to the very soft radiation at the periphery of the jet (possibly from pileup), while the true mass is mostly contained within the core $\Delta R\sim2m_W/p_\text{T}\sim 0.05$.  There are only eight reconstructed calorimeter clusters while there are 20 reconstructed tracks.  
	
\begin{figure}[h!]
\centering
\includegraphics[width=0.45\textwidth]{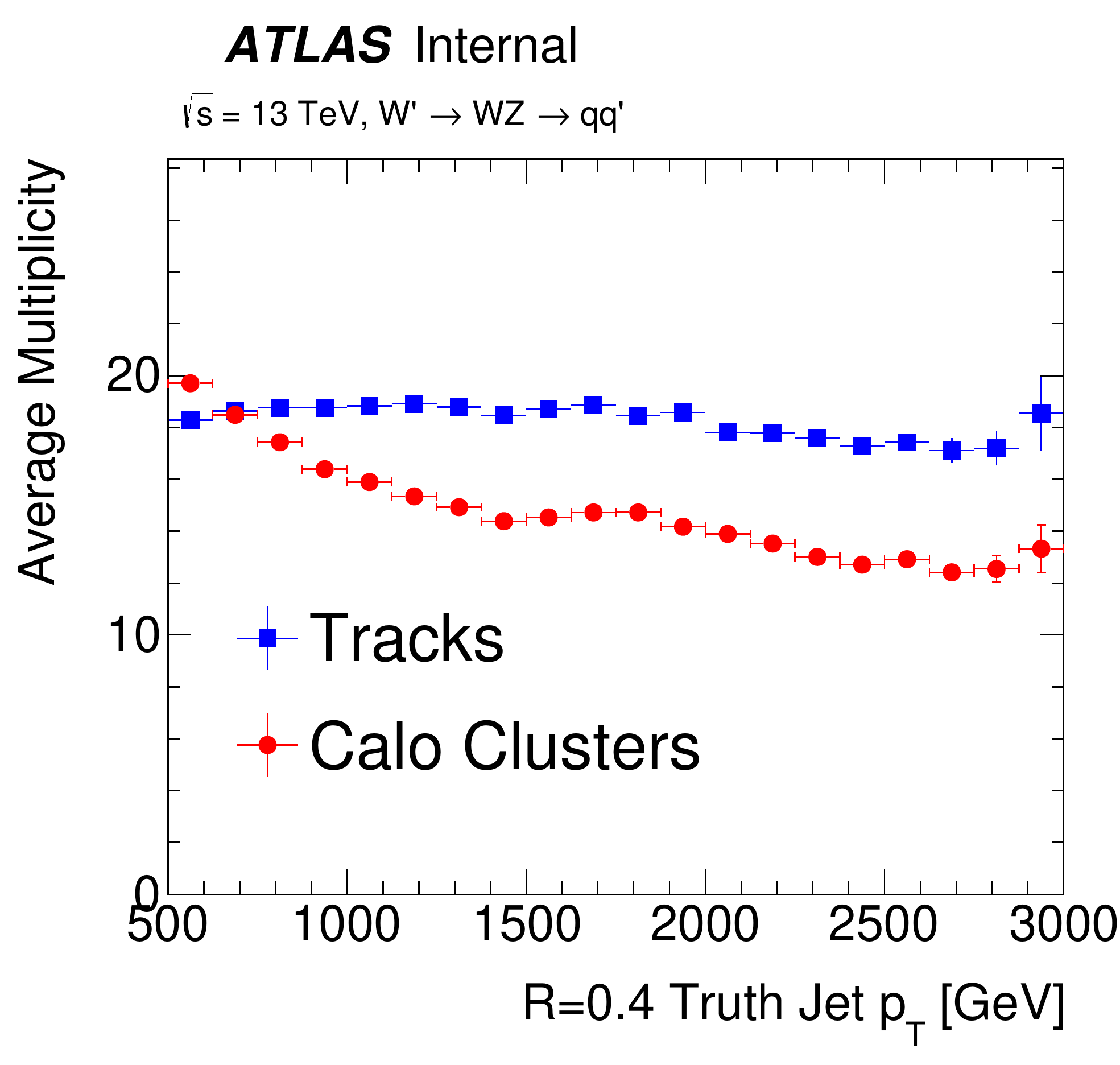}\hspace{5mm}\includegraphics[width=0.45\textwidth]{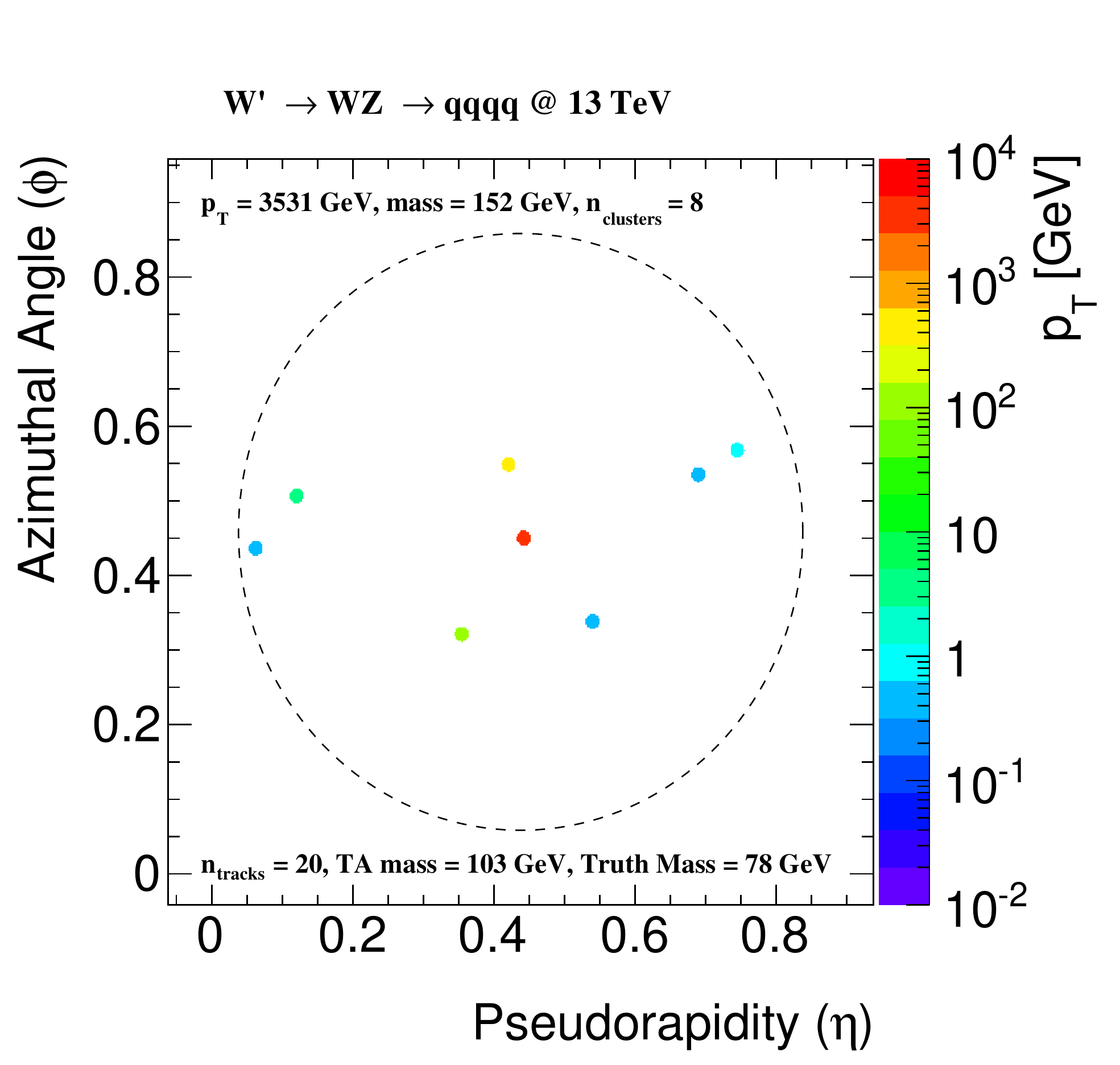}
\caption{Left: the average multiplicity of tracks and clusters in $R=0.4$ boosted $W$ and $Z$ boson jets as a function of the particle-level jet $p_\text{T}$.  Right: an event display of the clusters inside one particular high $p_\text{T}$ boosted boson jet. The TA mass is the track-assisted jet mass (see text for details).}
\label{fig:tamass:first}
\end{figure}		

While tracks are measured precisely, the jet mass built only from tracks is not directly useful.  There are significant charged-to-netural fluctuations that induce a resolution with respect to the particle-level jet mass constructed from all constituents.  This resolution is significantly larger than the calorimeter mass resolution.  Figure~\ref{fig:tamass:second} shows the calorimeter-only and track-only jet mass distributions.  The uncalibrated track-only mass has a much lower average value than the calorimeter jet mass due to the missed neutral energy.  When accounting for this average, the width of the track-mass is substantially broader than the width of the calorimeter mass distribution.

\begin{figure}[h!]
\centering
\includegraphics[width=0.5\textwidth]{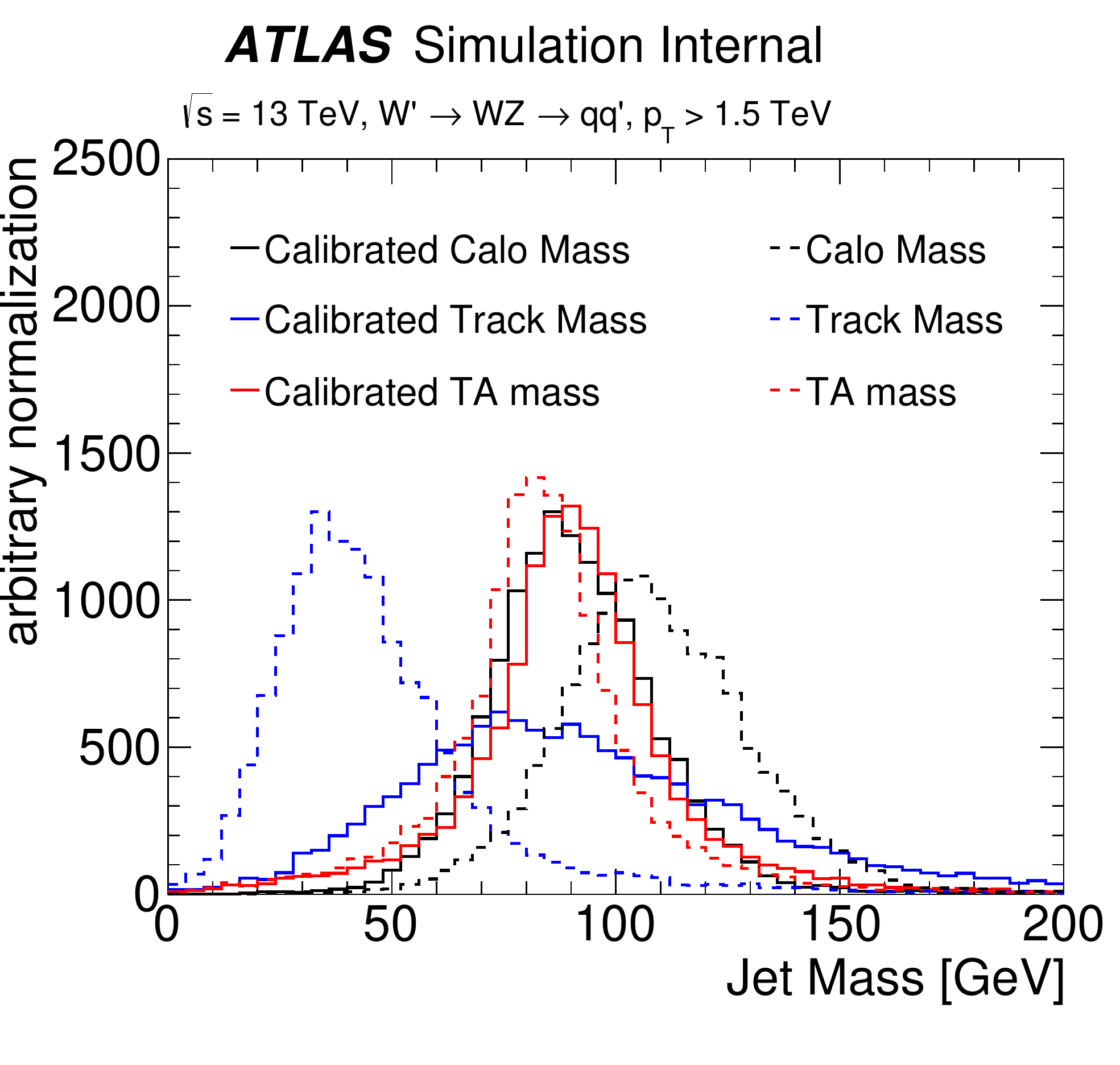}
\caption{The jet mass distribution of reconstructed boosted hadronically decaying $W$ and $Z$ boson jets with $p_\text{T}>1.5$ TeV.  A jet collection is calibrated if the average value of $m_\text{reco jet}/m_\text{truth jet}$ is unity.}
\label{fig:tamass:second}
\end{figure}		

A minimal, but powerful way to improve the track-only mass is to apply a jet-by-jet correction for the charged-to-neutral ratio to form  the {\it track-assisted jet mass}:

\begin{align}
\label{eq:TAmassdef}
m_\text{track-assisted} = m_\text{track}\times \frac{p_\text{T,calo}}{p_\text{T,track}},
\end{align}

\noindent where $p_\text{T,track}$ is the 4-vector sum of the tracks associated to a (trimmed) jet.  The track-assisted jet mass for the example in Fig.~\ref{fig:tamass:first} is closer to the particle-level jet mass, in part because of the larger number of tracks, improved angular resolution, and reduced sensitivity to pileup.  The reduced sensitivity to the charged-to-neutral fluctuations compared with the track-only mass leads to a resonance peak in Fig.~\ref{fig:tamass:second} that has a sharper maximum near the boson mass, comparable to the calorimeter jet mass.

A procedure for correcting the jet mass as in Eq.~\ref{eq:TAmassdef} was first proposed using hadronic calorimetery to correct electromagnetic-only measurements~\cite{Son:2012mb,Katz:2010mr}. The extension to charged particle tracks was introduced in the context of top-quark jet tagging~\cite{Schaetzel:2013vka} using the HEPTopTagger algorithm~\cite{Plehn:2010st,Plehn:2009rk}.  Since that time, there have been phenomenological studies using track-assisted jet mass\footnote{The phenomenological studies have not given a name to the quantity to Eq.~\ref{eq:TAmassdef}, so it is defined here as the track-assisted jet mass.} for ultra boosted ($p_\text{T}\gtrsim \mathcal{O}(10)$ TeV) top quark and boson jets~\cite{Larkoski:2015yqa,Bressler:2015uma}.  This remainder of this section is the first experimental study of the track-assisted jet mass, including a discussion of calibrations and systematic uncertainties.  

Figure~\ref{fig:tamass:third} shows the jet mass distribution for boosted $W$ and $Z$ boson jets clustered using the $R=0.4$ anti-$k_t$ algorithm.  For jets with 600 $<p_\text{T}<800$ GeV, the calorimeter jet mass peak is sharper than for the track-assisted jet mass; the inter-quantile range divided by the median (the quantile analogue to the coefficient of variation) is about 40\% larger for the track-assisted jet mass.  However, at higher momenta, $p_\text{T}> 2$ TeV, the peaks have nearly the same resolution.  Figure~\ref{fig:tamass:fourth} quantifies the $p_\text{T}$ dependence of the inter-quantile range for the jet mass response distribution.  The truth jet mass in the definition of the response is the same for both the track-assisted and calorimeter jet mass.  The calorimeter jet mass response distribution is broader for the track-assisted jet mass for $p_\text{T}\lesssim1.7$ TeV after which the track-assisted jet mass resolution is smaller than the calorimeter jet mass resolution.  These differences are not confined to the core of the response distribution as the trends are quantitatively the same for both the $\pm 10\%, \pm 20\%, $ and $\pm 30\%$ quantiles centered around the median.

\begin{figure}[h!]
\centering
\includegraphics[width=0.45\textwidth]{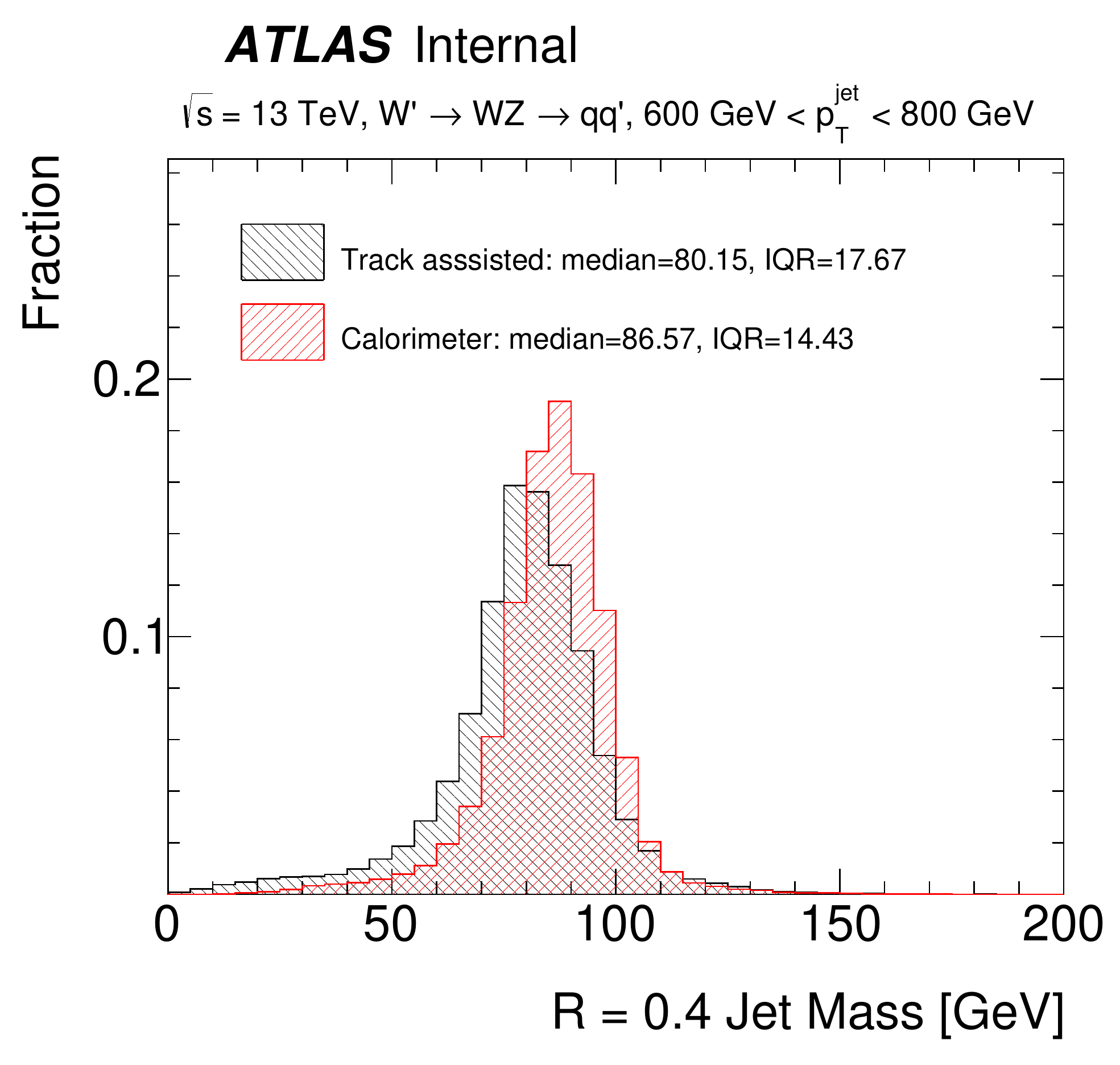}\hspace{5mm}\includegraphics[width=0.45\textwidth]{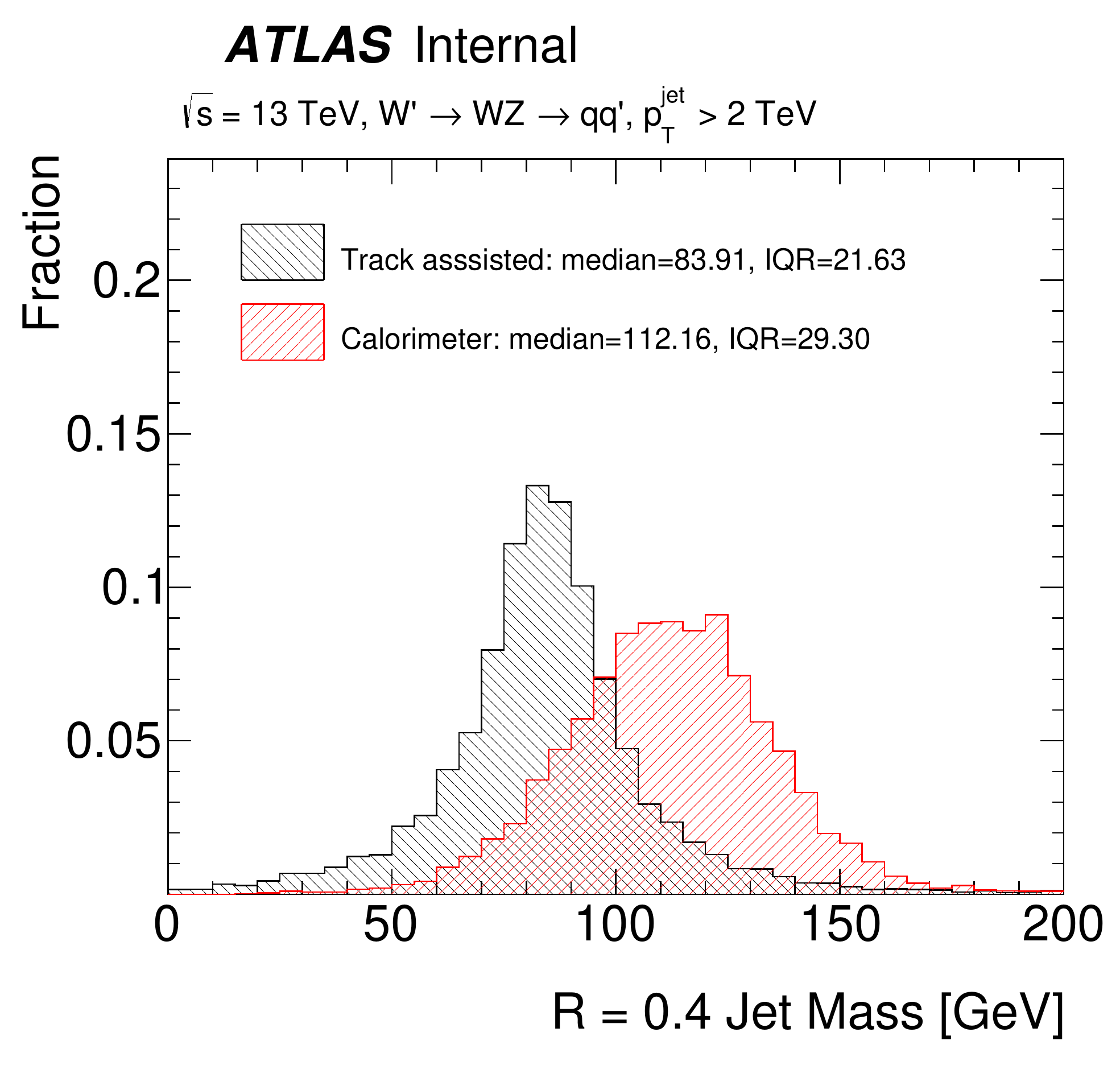}
\caption{The track-assisted and calorimeter jet mass distribution for boosted $W$ and $Z$ boson jets clustered using the $R=0.4$ anti-$k_t$ algorithm for 600 $<p_\text{T}<800$ GeV (left) and $p_\text{T}>2$ TeV (right). }
\label{fig:tamass:third}
\end{figure}	
	
\begin{figure}[h!]
\centering
\includegraphics[width=0.55\textwidth]{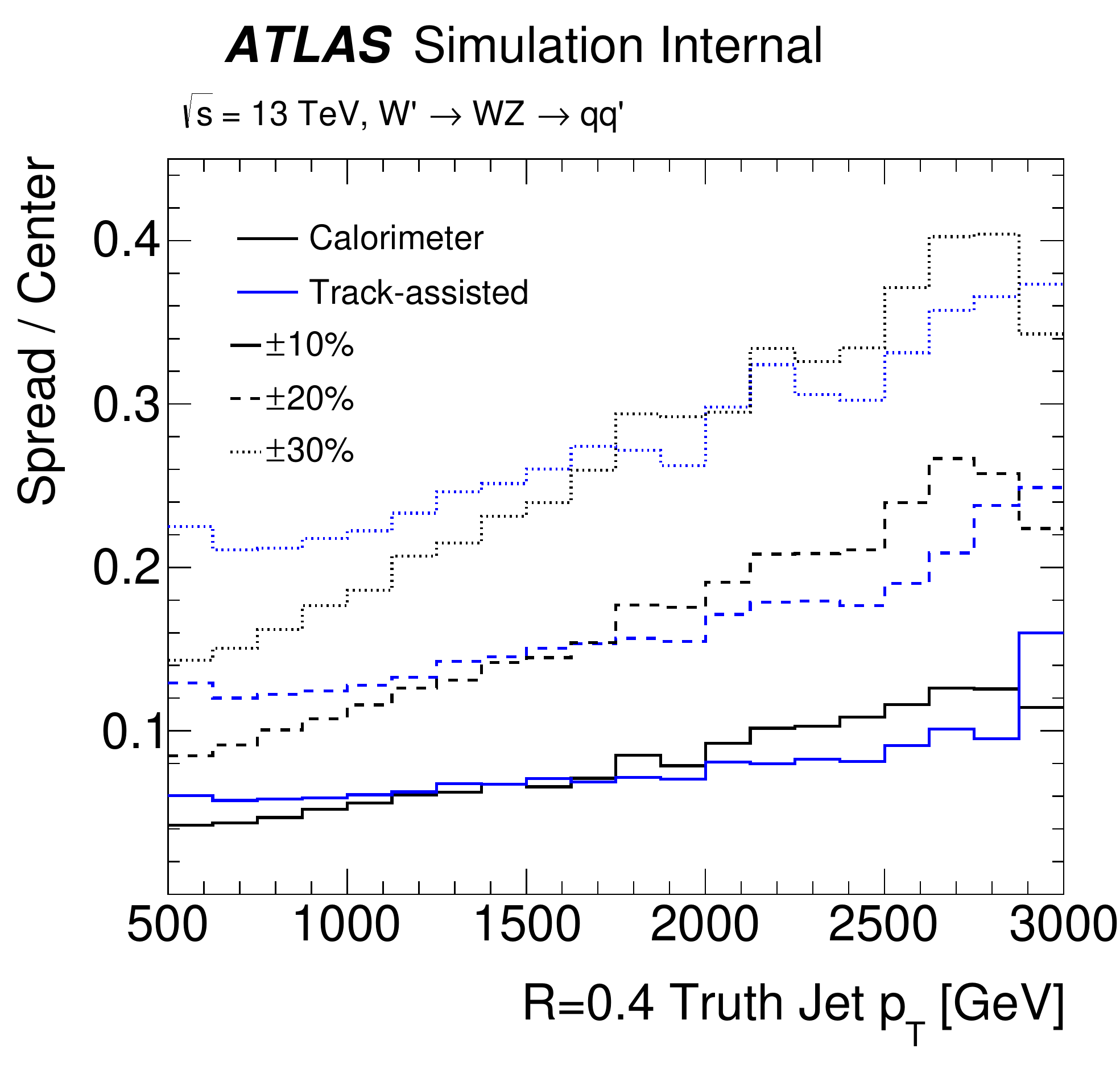}
\caption{The $p_\text{T}$ dependence of the inter-quantile range divided by the median of the track-assisted and calorimeter jet mass distributions.  The 10\%, 20\%, and 30\% inter-quantile ranges are centered around the median.}
\label{fig:tamass:fourth}
\end{figure}	

In addition to improving the jet mass resolution at high jet $p_\text{T}$, there are many experimental benefits to the track-assisted jet mass compared with the traditional calorimeter jet mass.  First of all, if the jet $p_\text{T}$ in Eq.~\ref{eq:TAmassdef} is already calibrated, the track-assisted jet mass is also nearly calibrated without any extra effort.  This is observed in Fig.~\ref{fig:tamass:third}: the masses are constructed with uncalibrated EM-scale jet mass, but EM+JES $p_\text{T}$; the track-assisted jet mass peak is around $m_W$ while the average calorimeter jet mass is much larger.  Additionally, if the jet $p_\text{T}$ is corrected for pileup, the track-assisted jet mass is also corrected for pileup, since pileup tracks can be removed from $m_\text{track}$.  Most importantly, the in situ momentum balancing techniques used to calibrate and estimate systematic uncertainties for the calorimeter jet $p_\text{T}$ apply directly to the track-assisted jet mass.  In particular, the jet mass scale and resolution uncertainty are the convolution of calorimeter $p_\text{T}$ and tracking uncertainties.  Unlike the uncertainty on the calorimeter jet mass, these two components can be well-estimated to the highest accessible jet momenta.  This is true for both small and large radius jets.  For example, Fig.~\ref{fig:tamass:fifth} shows how the small radius jet energy resolution uncertainty translates into an uncertainty on the track-assisted jet mass resolution.  By construction, the resolution uncertainty on the response is nearly the same for the $p_\text{T}$ and the track-assisted jet mass.  For small radius jets, the resolution is measured precisely and so the uncertainty is $\mathcal{O}(1\%)$.  For comparison, the difference between {\sc Pythia} and {\sc Herwig++} is the same order of magnitude.  This last quantity gives a sense of the fragmentation uncertainty, entering through the tracking component of the track-assisted jet mass.
	
\begin{figure}[h!]
\centering
\includegraphics[width=0.6\textwidth]{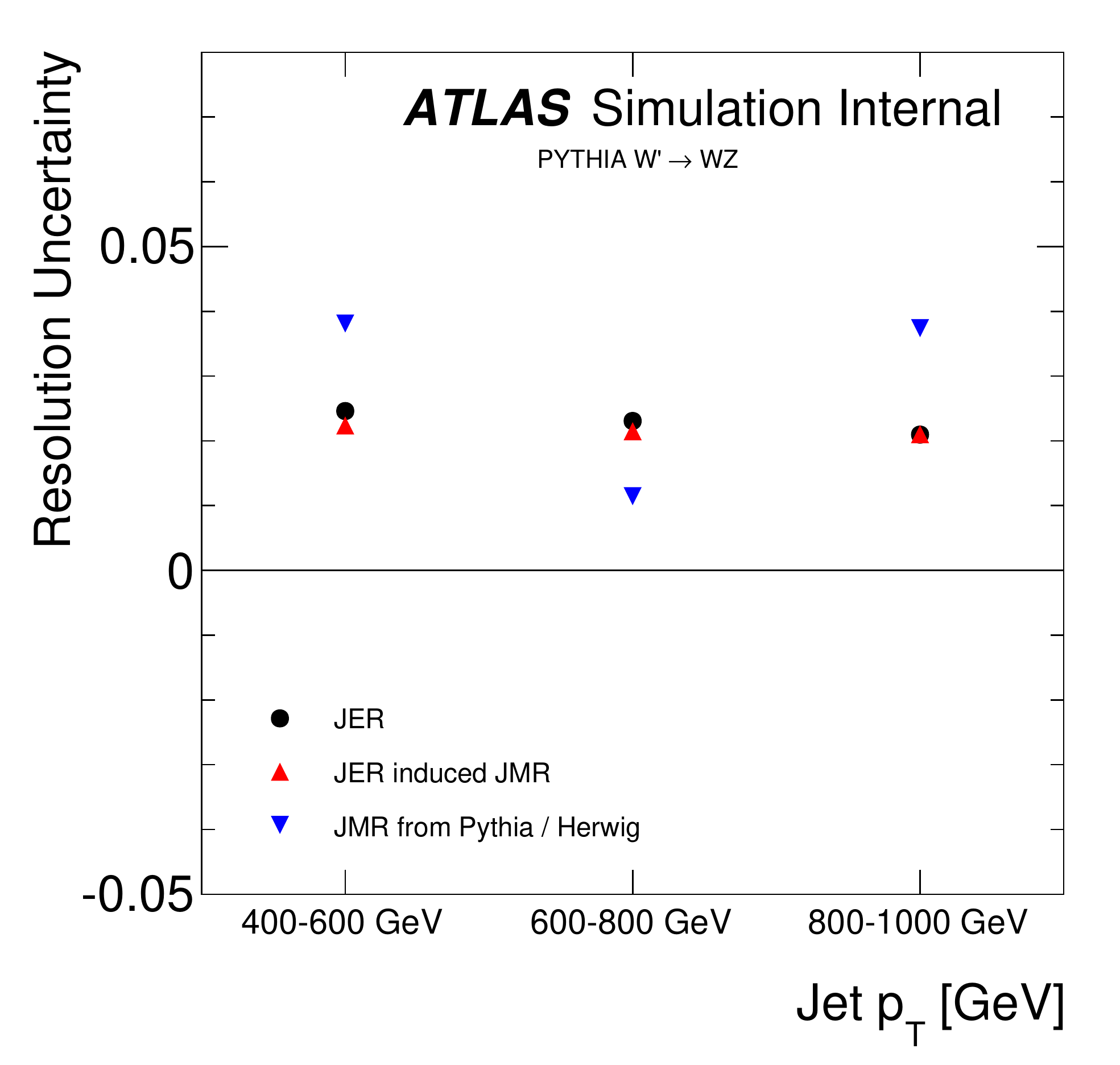}
\caption{The calorimeter jet momentum resolution uncertainty (circle), the jet $p_\text{T}$ resolution-induced track-assisted jet mass resolution uncertainty (up triangle), and the difference in the track-assisted jet mass resolution when comparing {\sc Pythia} and {\sc Herwig++}.  The resolution uncertainty is defined as $\sqrt{v^2-n^2}$, where $n$ is the nominal response resolution and $v$ is the response resolution from the alternative sample.  The alternative sample for the first two cases is created by smearing the jet energy within the resolution uncertainty and the alternative sample for the last case is the {\sc Herwig++} simulation.}
\label{fig:tamass:fifth}
\end{figure}

Since calorimeter jet mass is not used in the construction of the track-assisted jet mass, one may consider combining the two mass definitions to achieve even better performance.  Even though fluctuations in the calorimeter induce correlations between the jet mass and jet $p_\text{T}$ response, the left plot of Fig.~\ref{fig:tamass:sixth} shows that the correlation between the track-assisted jet mass response and the calorimeter jet mass response is negligible.  It is therefore a good approximation to treat the track-assisted jet mass $X$ and the calorimeter jet mass $Y$ as independent when determining the optimal combination.  For a fixed truth mass, $X\sim\mathcal{N}(1,\sigma_1)$ and $Y\sim\mathcal{N}(1,\sigma_2)$.  In this application, $X$ is the calibrated track-assisted jet mass response and $Y$ is the calibrated calorimeter jet mass response.  Let\footnote{The linear combination is optimal over a wider class of functions, but this is beyond the scope of this section.} $Z=\alpha X+\beta Y$.  Assuming $X$ and $Y$ are calibrated, $Z$ is also calibrated if $\alpha+\beta=1$.  The variance of $Z$ is $\sigma^2(Z)\approx\alpha^2\sigma^2(X)+\beta^2\sigma^2(Y)$.  The first order conditions $\partial_\alpha\sigma^2(Z)=0$ and $\partial_\beta\sigma^2(Z)=0$ result in the minimum variance unbiased estimator of the particle-level mass $\hat{Z}$: $\alpha\propto 1/\sigma_1^2, \beta\propto 1/\sigma_2^2$.   The performance of the optimal combination of the calorimeter and track-assisted jet mass is shown in Fig.~\ref{fig:tamass:sixth}.  For $p_\text{T}\lesssim 1$ TeV, the calorimeter jet mass resolution is significantly better than that of the track-assisted jet mass and so the improvement from the combination is negligible.  However, for $p_\text{T}\gtrsim$ 1 TeV, there is a full $\approx 40\%(\approx\sqrt{2}-1)$ improvement in the resolution.  Distributions of the track-assisted jet mass, the calorimeter jet mass, and the optimal combination are shown in a few jet $p_\text{T}$ bins in Fig.~\ref{fig:tamass:seventh}.

\begin{figure}[h!]
\centering
\includegraphics[width=0.45\textwidth]{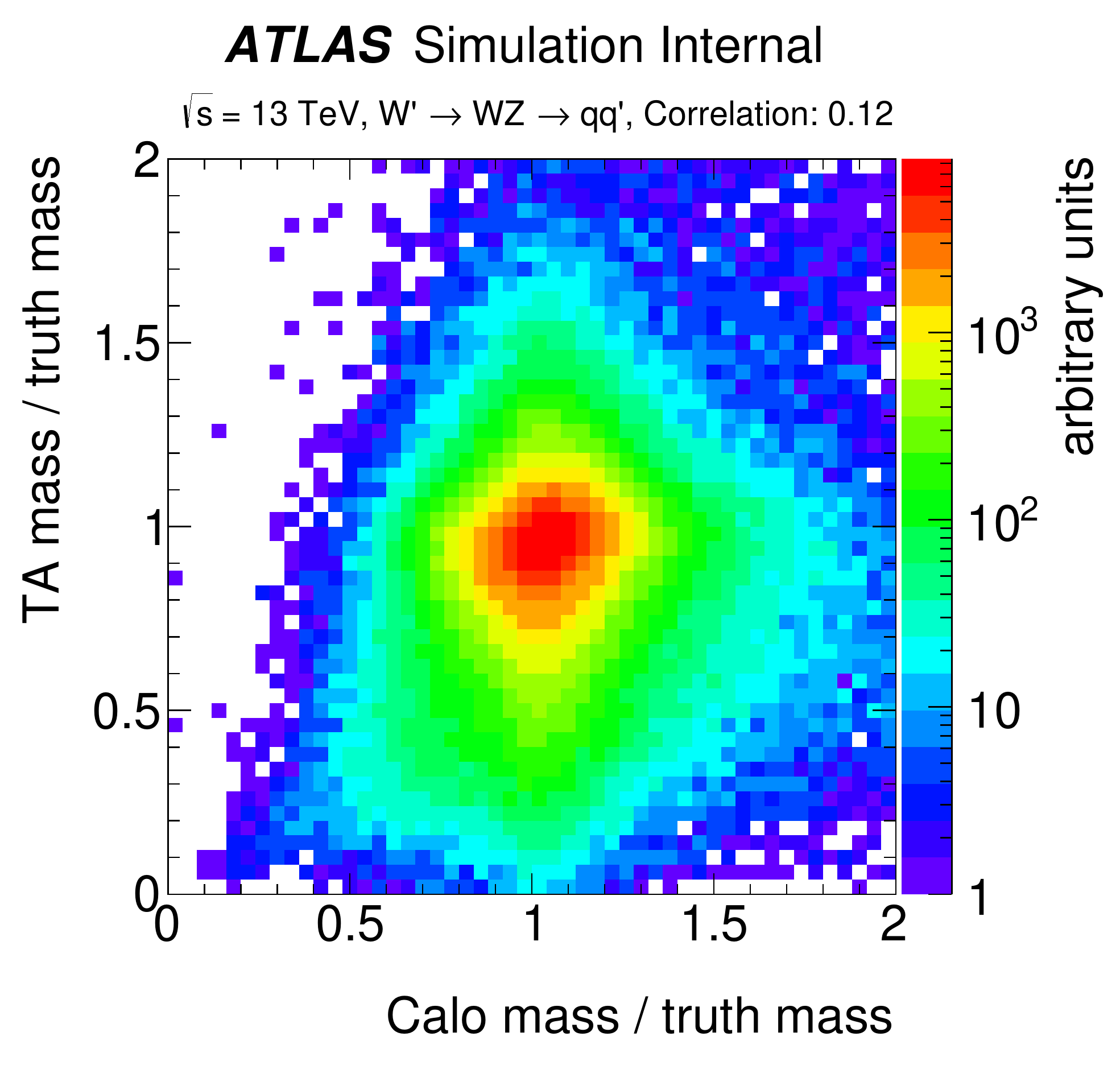}\hspace{5mm}\includegraphics[width=0.45\textwidth]{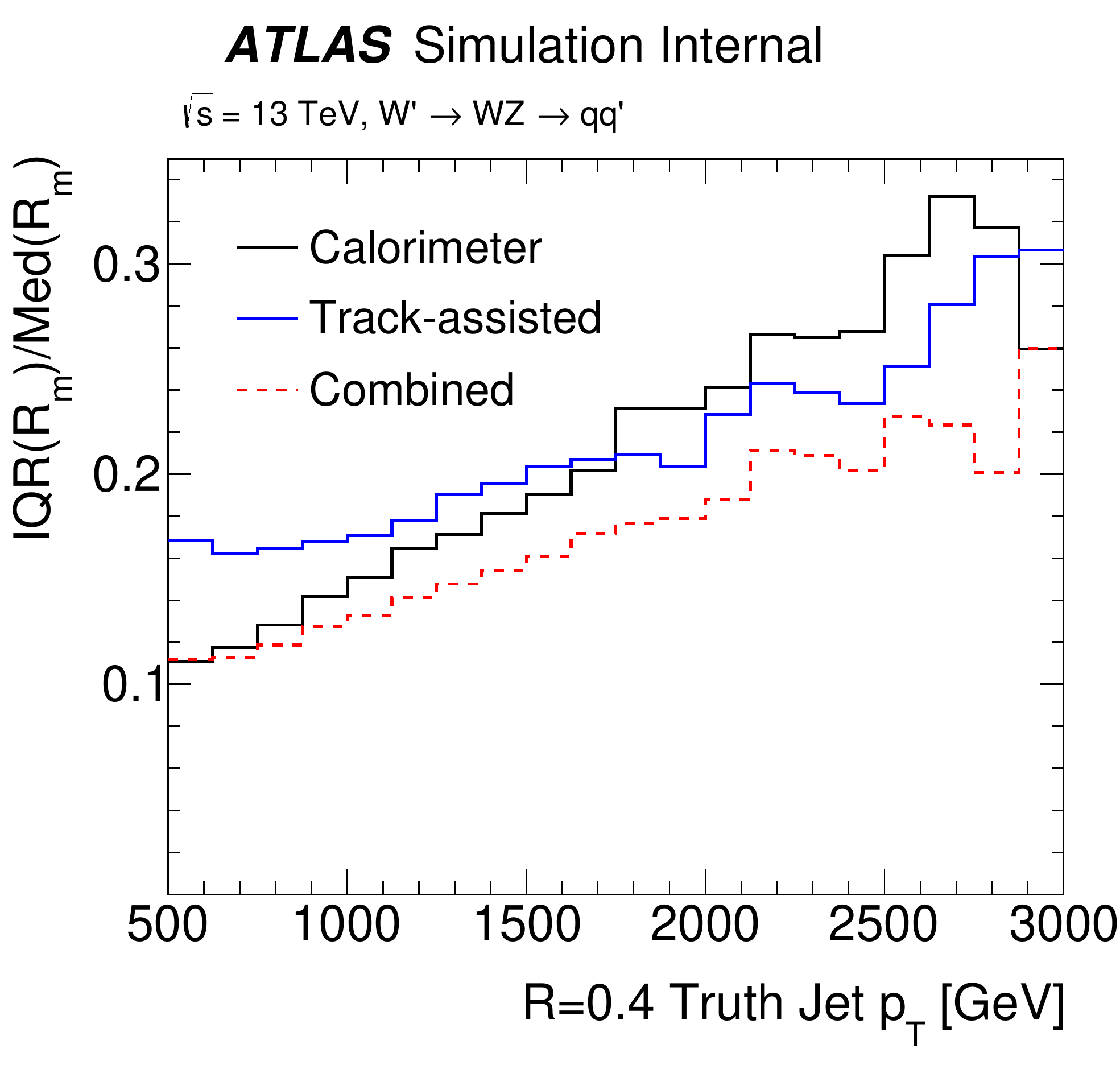}
\caption{Left: the joint distribution of the track-assisted jet mass response and the calorimeter jet mass response.  Right: the $p_\text{T}$ dependence of the normalized jet mass resolution for the track-assisted jet mass, the calorimeter jet mass, and the optimal combination of the two mass definitions.}
\label{fig:tamass:sixth}
\end{figure}

\begin{figure}[h!]
\centering
\includegraphics[width=0.45\textwidth]{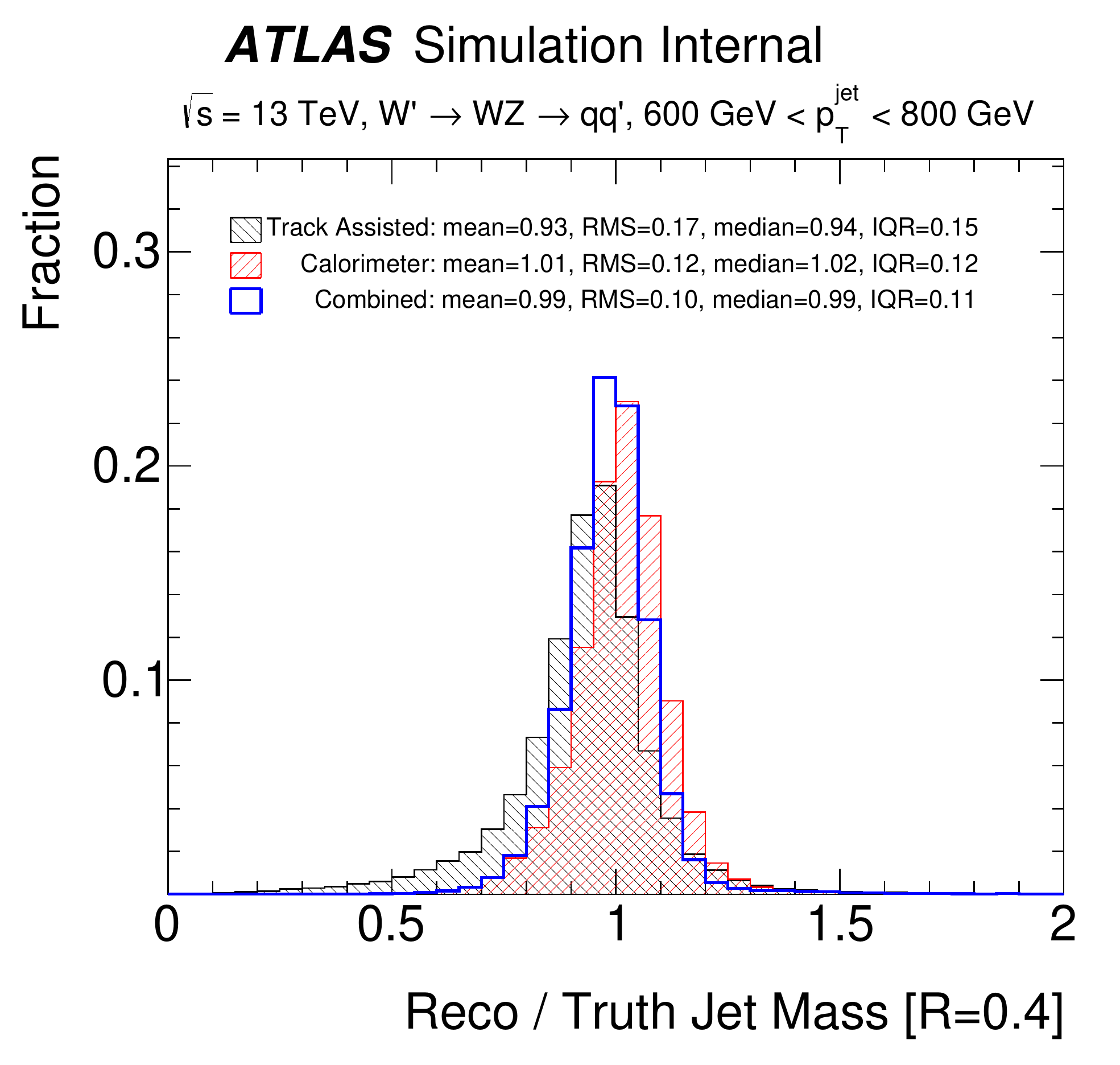}\includegraphics[width=0.45\textwidth]{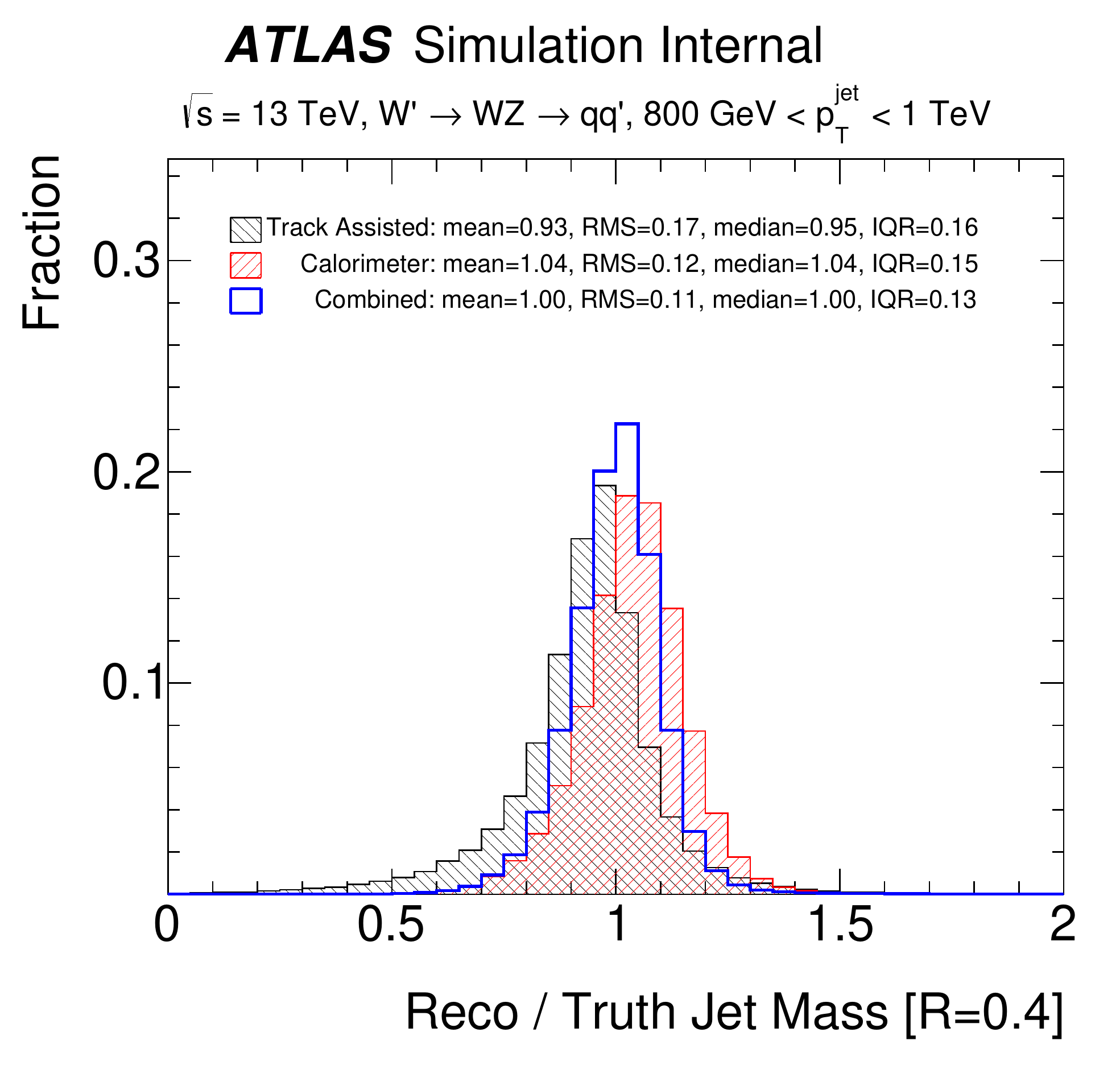}\\
\hspace{5mm}\includegraphics[width=0.45\textwidth]{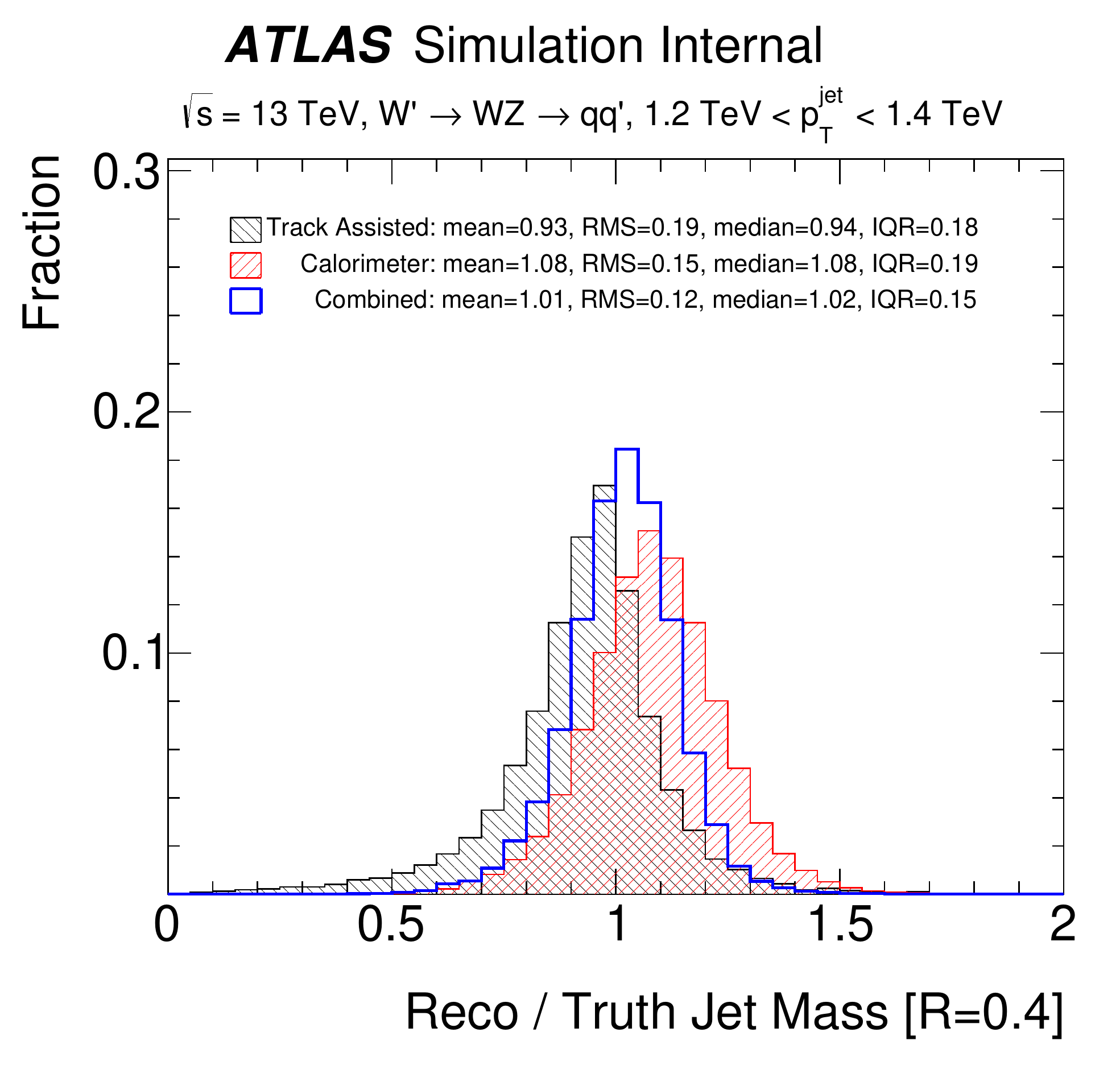}\includegraphics[width=0.45\textwidth]{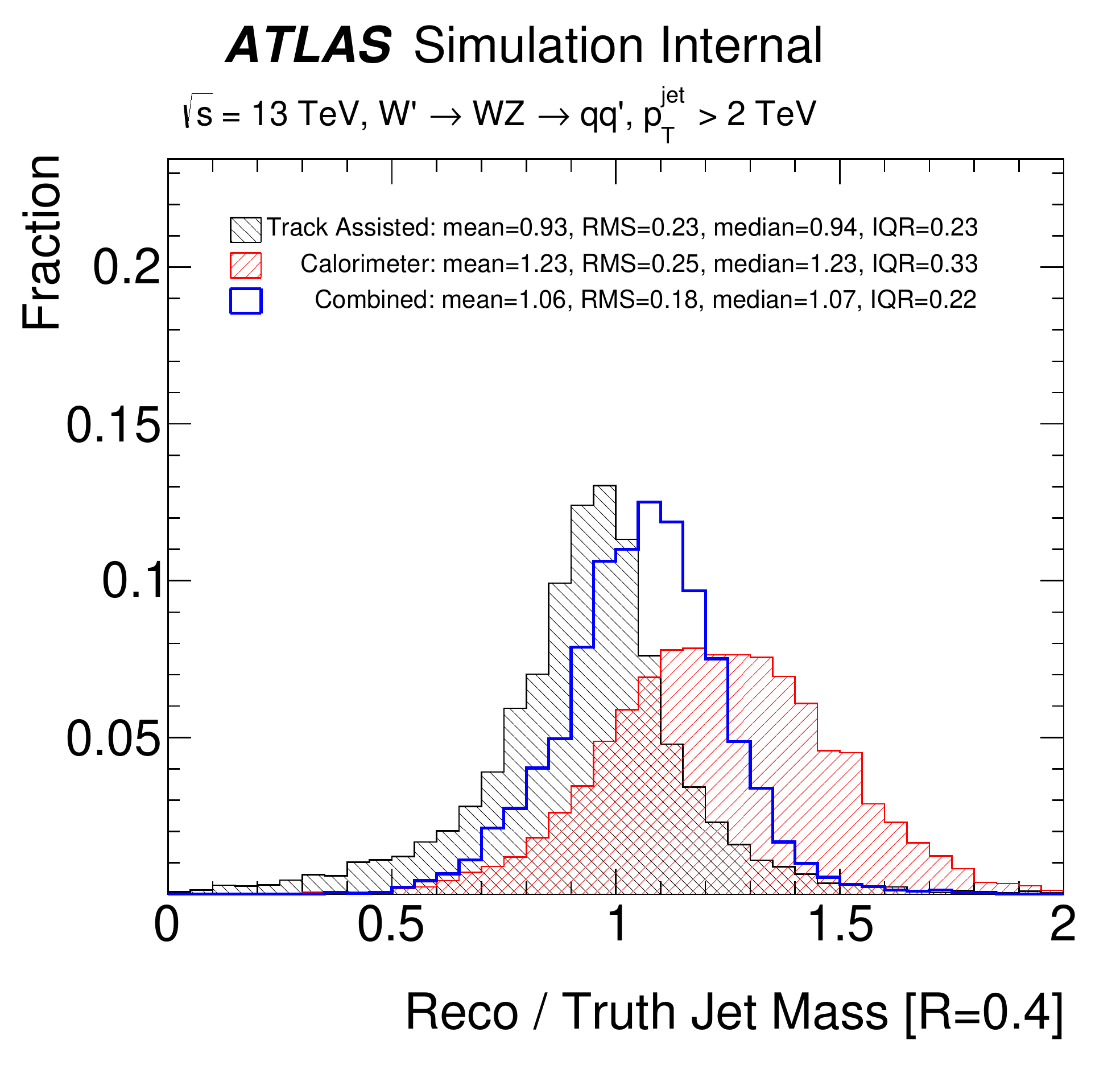}
\caption{Distributions of the track-assisted jet mass, the calorimeter jet mass, and the optimal combination are shown in a few jet $p_\text{T}$ bins.}
\label{fig:tamass:seventh}
\end{figure}

The track-assisted jet mass is a promising technique for jet mass at high $p_\text{T}$.  At moderate and low $p_\text{T}$, there may be modifications of the algorithm to recover performance.  In particular, since the charge-to-neutral ratio has large local fluctuations, subjet corrections may be able to improve the resolution at lower $p_\text{T}$ when the calorimeter can reliably resolve a jet's substructure~\cite{Schaetzel:2013vka}.
	
	\clearpage
	
	\subsection{Conclusions and Future Outlook}
	\label{sec:mass:conclusions}
	
	Section~\ref{sec:jetmass} has presented several new techniques for improving jet four-vector reconstruction and measuring the performance in data.  At low $p_\text{T}$, re-clustering offers a powerful and flexible scheme for tailoring jet clustering that naturally reduces to traditional small-radius jets at high $p_\text{T}$.  The mass of single small-radius jets is not well-constrained due to the lack of a pure sample of massive small radius jets in the data, but one can use the track-assisted mass at high $p_\text{T}$ to improve the performance and retain a natural scheme for calibrations and uncertainties.  As more data at higher $p_\text{T}$ is collected in Run 2, the in-situ resonance method can be used as an independent measurement of the uncertainties for both re-clustering and track-assistance across a wide range of jet $p_\text{T}$.  This toolkit will hopefully be able to improve the sensitivity to boosted bosons and top quarks for a wide range of searches in Run 2 and beyond.  Part~\ref{part:susy} will already show that re-clustering is a useful tool for extending the sensitivity to electroweak scale SUSY using the early $\sqrt{s}=13$ TeV data.
	
	The next section presents a new application of the traditional large-radius trimmed mass to distinguish different types of boson jets.

	\clearpage

\section{Boson Type Tagger}
\label{sec:bosontypetagger}

\subsection{Introduction}
\label{sec:intro}

Jet substructure techniques\footnote{The tool presented in this section has been published in Ref.~\cite{Aad:2015eax}.} developed to distinguish hadronically decaying $W$ and $Z$ bosons from QCD multijet background processes have become increasingly sophisticated.  A recent review is given in Ref.~\cite{Altheimer:2013yza}.  Both ATLAS~\cite{ATL-PHYS-PUB-2014-004} and CMS~\cite{Khachatryan:2014vla} have performed detailed comparisons of the various tagging variables and jet-grooming techniques with the overall conclusion that large QCD multijet suppression factors\footnote{$\mathcal{O}(1\%)$ QCD multijet efficiency at ~50\% signal efficiency.} are possible while maintaining acceptable levels of boson tagging efficiency.   Given a $W/Z$-boson tagger, a natural next step is to distinguish between boson types, e.g. $W$-boson jets from $Z$-boson jets.

There are several important possible applications of a boson-type tagger at the LHC.  First, a type tagger could enhance the SM physics program with $W$ and $Z$ bosons in the final state.  Measurements of this kind include the determination of the cross sections for $V$+jets, $VV$, and $t\bar{t}+V$.  Another important use of a boson-type tagger is in searches for flavor-changing neutral currents (FCNC).  Due to the Glashow--Iliopoulos--Maiani (GIM) mechanism~\cite{PhysRevD.2.1285}, FCNC processes in the SM are highly suppressed.   Many models of new physics predict large enhancements to such processes.  Both ATLAS and CMS have performed searches for FCNC~\cite{Aad:2012ij,Chatrchyan:2013nwa} of the form $t\rightarrow Zq$ in the leptonic channels, but these could be extended by utilizing the hadronic $Z$ decays as well.  FCNC process mediated by a leptophobic $Z'$ may be detected only via hadronic type-tagging methods.   A third use of a boson-type tagger is to categorize the properties of new physics, if discovered at the LHC.  For instance, if a new boson were discovered as a hadronic resonance, a boson-type tagger could potentially distinguish a $W'(\rightarrow qq)$ from a $Z'(\rightarrow qq)$ (where mass alone may not be useful).  This is especially relevant for leptophobic new bosons, which could not be distinguished using leptonic decays.

Labelling jets as originating from a $W$ or $Z$ boson is less ambiguous than quark/gluon labelling.   A $W$ boson can radiate a $Z$ boson, just like a quark can radiate a gluon, but this is heavily suppressed for the former and not for the latter.  The radiation pattern of jets from $W$- and $Z$-bosons is less topology dependent because it is largely independent of the other radiation in the event as $W$ and $Z$ bosons are color singlets.  Aside from the production cross section and subtle differences in differential decay distributions, the only features that distinguish between $W$ and $Z$ bosons are their mass, charge, and branching ratios.  Experimentally, this means that the only variables that are useful in discriminating between hadronic decays of $W$ and $Z$ bosons are those which are sensitive to these properties.  The three variables used in the analysis presented here are {\it jet mass}, sensitive to the boson mass, {\it jet charge}, sensitive to the boson charge, and a {\it b-tagging} discriminant which is sensitive to the heavy-flavor decay branching fractions of the bosons.  The application of a boson-type tagger in practice will be accompanied by the prior use of a boson tagger (to reject QCD multijet processes).  The type-tagger variables are largely independent of typical boson-tagger discriminants like $n$-subjettiness~\cite{Thaler:2010tr}, which rely on the two-prong hard structure of both the $W$ and $Z$ decays.  

Due to the large QCD backgrounds with experimental signatures similar to hadronic electroweak boson production, isolating $W$ and $Z$ bosons at the LHC is challenging.  However, at lepton machines, electroweak boson production is often dominant and can be a background for many other interesting processes (see e.g. Ref.~\cite{Marshall:2012ry}).  Jet tagging in boosted topologies has matured considerably since LEP and so some of the techniques presented here may be applicable to future high energy lepton machines.

This section introduces a new jet tagging method to distinguish between hadronically decay $W$ and $Z$ bosons at the LHC, and documents its performance with the ATLAS detector at $\sqrt{s}=8$ TeV.  The section is organized as follows.  Section~\ref{sec:datasets} describes the simulated datasets used in constructing and evaluating the boson-type tagger.  Following a discussion of the differences between the properties of $W$ and $Z$ bosons in Sec.~\ref{sec:distinguish}, Sec.~\ref{sec:objs} defines the three discriminating variables.  The construction and performance of the tagger are detailed in Sec.~\ref{sec:perfm} and the sensitivity to systematic uncertainties is described in Sec.~\ref{sec:systs}.  The input variables are studied in a dataset enriched in boosted $W$ bosons in Sec.~\ref{sec:data}.   The section ends with a discussion of possible uses of the tagger in Sec.~\ref{sec:outlook} and conclusions in Sec.~\ref{sec:concWZ}.

\clearpage

\subsection{Datasets}
\label{sec:datasets}

Two sets of Monte Carlo (MC) simulations are generated, one to study the tagger's $W$ versus $Z$ performance and the other to compare the tagger inputs for $W$ bosons with the data.  For the tagger performance, it is useful to have a source of isolated high $p_\text{T}$ $W$ and $Z$ bosons.  One physics process that produces such final states is the production of a hypothetical $W'$ boson.  Predicted in models of new physics with an SU(2) gauge group, the $W'$ is analogous to the SM $W$ boson.  For this analysis, a 100\% branching ratio $W'\rightarrow WZ$ is used to generate events with simultaneously boosted $W$ and $Z$ bosons; the $p_\text{T}$ of the SM bosons is set by the mass of the $W'$ boson.  Unfortunately, there is no evidence in real data for $W'$ bosons and it is not possible to measure the tagger performance directly in the data due to the lack of a pure sample of boosted, hadronically decaying $Z$ bosons.  However, the modelling of the tagger inputs can be studied using hadronically decaying $W$ bosons from $t\bar{t}$ events in the data.  The simulation and event selection used for the modelling studies are identical to those from Sec.~\ref{sec:JMR}.

A simulated sample of $W'$ bosons is generated with {\tt PYTHIA}~8 using the leading-order parton distribution function set (PDF) {\tt MSTW2008}~\cite{Lai:2010vv,Gao:2013xoa} and the {\tt AU2}~\cite{ATL-PHYS-PUB-2012-003} set of tunable parameters (tune) for the underlying event.  The baseline samples use {\tt PYTHIA} for the $2\rightarrow 2$ matrix element calculation, as well as $p_\text{T}$-ordered parton showers~\cite{Sjostrand:2004ef} and the Lund string model~\cite{string} for hadronization.  Additional samples are produced with {\tt HERWIG++}~\cite{Bahr:2008pv}, which uses angular ordering of the parton showers~\cite{Gieseke:2003rz}, a cluster model for hadronization~\cite{Webber:1983if}, as well as the {\tt EE3}~\cite{Gieseke:2012ft} underlying-event tune.  In order to remove artifacts in the $p_\text{T}$ distributions of the $W$ and $Z$ bosons due to the generation of $W'$ particles with discrete masses, the $p_\text{T}^V$ spectra are re-weighted to be uniform in the range $200$~GeV$ <p_\text{T}^V<400$~GeV.  As is discussed in Sec.~\ref{sec:intro}, for $p_\text{T}>200$~GeV, a jet with large radius is expected to capture most of the $W$ or $Z$ boson decay products.  The range is truncated to $p_\text{T}<400$~GeV because hadronically decaying $W$ bosons can be probed with data in this $p_\text{T}$ range; there are too few events in the 8 TeV dataset for $p_\text{T}>400$ GeV.  Figure~\ref{fig:WZ:pTspectrum} shows the boson $p_\text{T}$ spectrum before any re-weighting.  The shape of the distribution is set by the available range of $W'$ masses that spans a few hundreds of GeV to 4 TeV.  Spikes are due to kinematic jacobian peaks from individual $W'$ masses.  Since the $W$ and $Z$ mass difference is small compared to the $W'$ masses, the shapes of the $W$ and $Z$ boson $p_\text{T}$ spectrum are nearly identical.

The $W'$ events are processed with a full simulation of the ATLAS detector~\cite{Aad:2010ah} based on the $\mathrm{\tt Geant4}$~\cite{Agostinelli:2002hh} toolkit, and reconstructed using the same software as for the experimental data.  The average number of additional $pp$ collisions per bunch crossing (pileup interactions) was $20.7$ over the full 2012 run.  The effects of pileup are modelled by adding multiple minimum-bias events, which are simulated with Pythia~8.{\tt 160} \cite{Sjostrand:2007gs}, to the generated hard-scatter events. The distribution of the number of interactions is then weighted to reflect the pileup distribution in the 2012 data.  

\begin{figure}[h!]
 \centering
\includegraphics[width=0.55\textwidth]{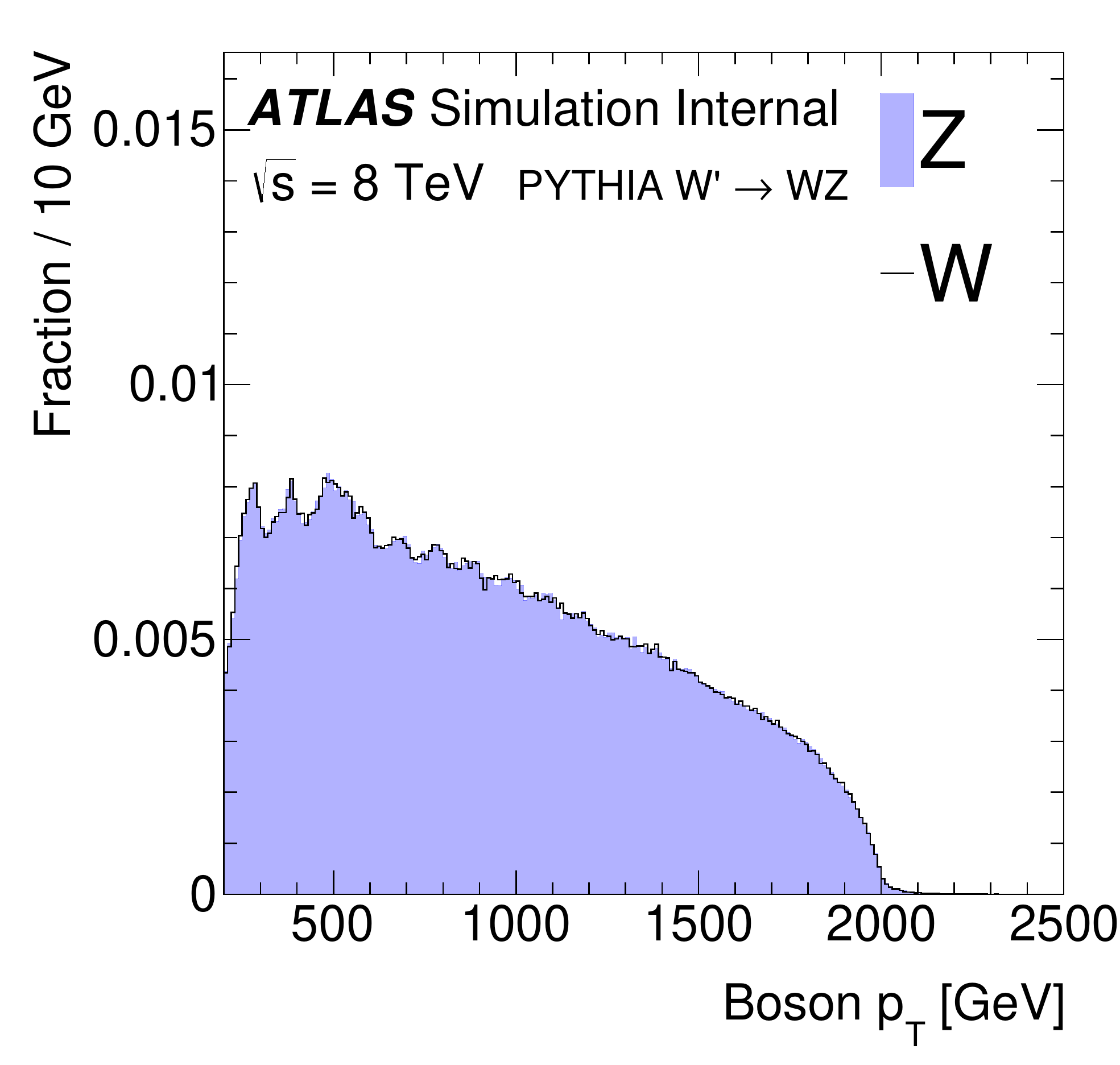}
\caption{The $p_\text{T}$ spectrum of the simulated $W$ and $Z$ bosons from $W'\rightarrow WZ$ decays before applying any $p_\text{T}$ re-weighting. }
\label{fig:WZ:pTspectrum}
\end{figure}

\clearpage

\subsection{Distinguishing a $Z$ boson from a $W$ boson}
\label{sec:distinguish}

Decays of $W$ or $Z$ bosons are characterized by the boson's mass and coupling to fermions.   The mass difference between the $W$ and $Z$ boson is about $10$~GeV and if produced from a hard scatter or the decay of a heavy enough resonance, both bosons are produced nearly on-shell since the width $\Gamma_V=2.1$ ($2.5$)~GeV is much less than the mass $m_V=80.4$ ($91.2$)~GeV for $W$ ($Z$) bosons~\cite{pdg}.  The Breit--Wigner resonance curves for $W$ and $Z$ bosons are shown in Fig.~\ref{fig:mass}(a).  The separation between the curves is a theoretical limit on how well mass-sensitive variables can distinguish between $W$ and $Z$ bosons.  For hadronic boson decays, the mass peaks measured with jets are broader. This is because the jet-clustering algorithm for final-state hadrons loses particles at large angles to the jet axis and includes extra particles from the underlying event and pileup.

The generic coupling of a boson $V$ to fermions is given by $g_\text{V} \gamma_\mu[c_\text{V}-\nobreak c_\text{A}\gamma_5]$, where $g_\text{V}$ is a boson-dependent overall coupling strength, and $c_\text{V}$ and $c_\text{A}$ are the vector and axial-vector couplings, respectively.   The $W$ boson couples only to left-handed fermions so $c_\text{V}=c_\text{A}=1$ with $g_W\propto kN_\text{C}G_\text{F}m_W^3|V_{ij}|^2$, where $G_\text{F}$ is the Fermi coupling constant, $V_{ij}$ is a Cabibbo--Kobayashi--Maskawa (CKM) matrix element~\cite{PhysRevLett.10.531,Kobayashi:1973fv}, $k$ represents higher-order corrections, and $N_\text{C}=3$ for the three colours of quarks and $N_\text{C}=1$ for leptons.  The CKM matrix is nearly diagonal so $W^+\rightarrow u\bar{d}$ and $W^+\rightarrow c\bar{s}$ are the dominant decay modes.  Small off-diagonal elements contribute to the other possible decay modes, and the overall branching ratios assuming hadronic decay are approximately $50\%$ for $W\rightarrow cX$ and $50\%$ for $W\rightarrow \text{light-quark pairs}$.  The $W$ boson has electric charge $\pm 1$ in units of the electron charge, so by conservation of charge, its decay products have the same net charge.   The scalar sum of the charge of all the final-state hadrons originating from a $W$ boson decay is not infrared safe (directly sensitive to the non-zero detection threshold), so there are limits to the performance of charge tagging dictated by the energy threshold placed on charged particles in the event reconstruction.

In contrast to $W$ boson decays, $Z$ bosons decay to both the left- and right-handed fermions.  The partial width for $Z\rightarrow f\bar{f}$ is proportional to $kN_\text{C}G_\text{F}m_Z^3[c_\text{V}^2+c_\text{A}^2]$.  The factors $c_\text{V}$ and $c_\text{A}$ are slightly different for up- and down-type fermions.  The $b\bar{b}$ branching ratio is 22\%, the $c\bar{c}$ branching ratio is $17\%$ and the sum of the remaining branching ratios is $61\%$, assuming a hadronic decay.  $W$ boson decays to $b$-quarks are highly suppressed by the small CKM matrix elements $V_{cb}$ and $V_{ub}$, so that identifying $b$-hadron decays associated with a hadronically decaying boson is a powerful discriminating tool.  Branching ratios are plotted in Fig.~\ref{fig:mass}d for $Z$ decays to light quarks, $c$-quarks, and $b$- quarks, and in Fig.~\ref{fig:mass}(c) for the $W$ boson decays to light quarks and $c$-quarks.

Since the coupling structure is not identical for $W$ and $Z$ bosons, the total decay rates differ, and the angular distributions of the decay products also differ slightly.  However, even at parton level without any combinatoric noise, the differences in the angular distributions are subtle.   This is illustrated in Fig.~\ref{fig:mass}(b) for transversely polarized $W$ and $Z$ bosons (details can be found in Appendix~\ref{sec:app:wzdecay}).  The angular distributions are identical for the two bosons for longitudinal polarization because the distributions for right- and left-handed fermions is the same.  The relative contribution of left- and right-handed components for the $Z$ decays depends on the quark flavor; for up-type quarks the relative contribution from right-handed fermions is $15\%$ while it is only $3\%$ for down-type quarks.  In $t\bar{t}$ decays, the fraction of longitudinally polarized $W$ bosons (ignoring the $b$-quark mass) is $m_t^2/(m_t^2+2m_W^2)\sim 0.7$.  In contrast, the boson is mostly transversely polarized in inclusive $V$+jets events.  See Appendix~\ref{sec:app:wpolarization} for a derivation of these polarization properties.  Any discrimination shown in Fig.~\ref{fig:mass}(b) is diluted by combinatorics (including distinguishing $q$ from $\bar{q}$ jets), non-perturbative effects, and detector reconstruction, so angular distributions are not considered further in this section\footnote{Boson polarization does have an impact on the jet mass distribution and thus on distinguishing boson jets from QCD jets~\cite{Khachatryan:2014vla}.  However, the impact on distinguishing $W$ jets from $Z$ jets is highly suppressed because there are only (small) differences when the bosons are transversely polarized.  Polarizations would be important only if the $W$ and $Z$ were predominately produced with different polarizations, which does not happen in e.g. $V$+jets, $t\bar{t}$, or $W'\rightarrow WZ$ events.}.

\begin{figure}[h!]
\begin{overpic}[width=0.95\textwidth]{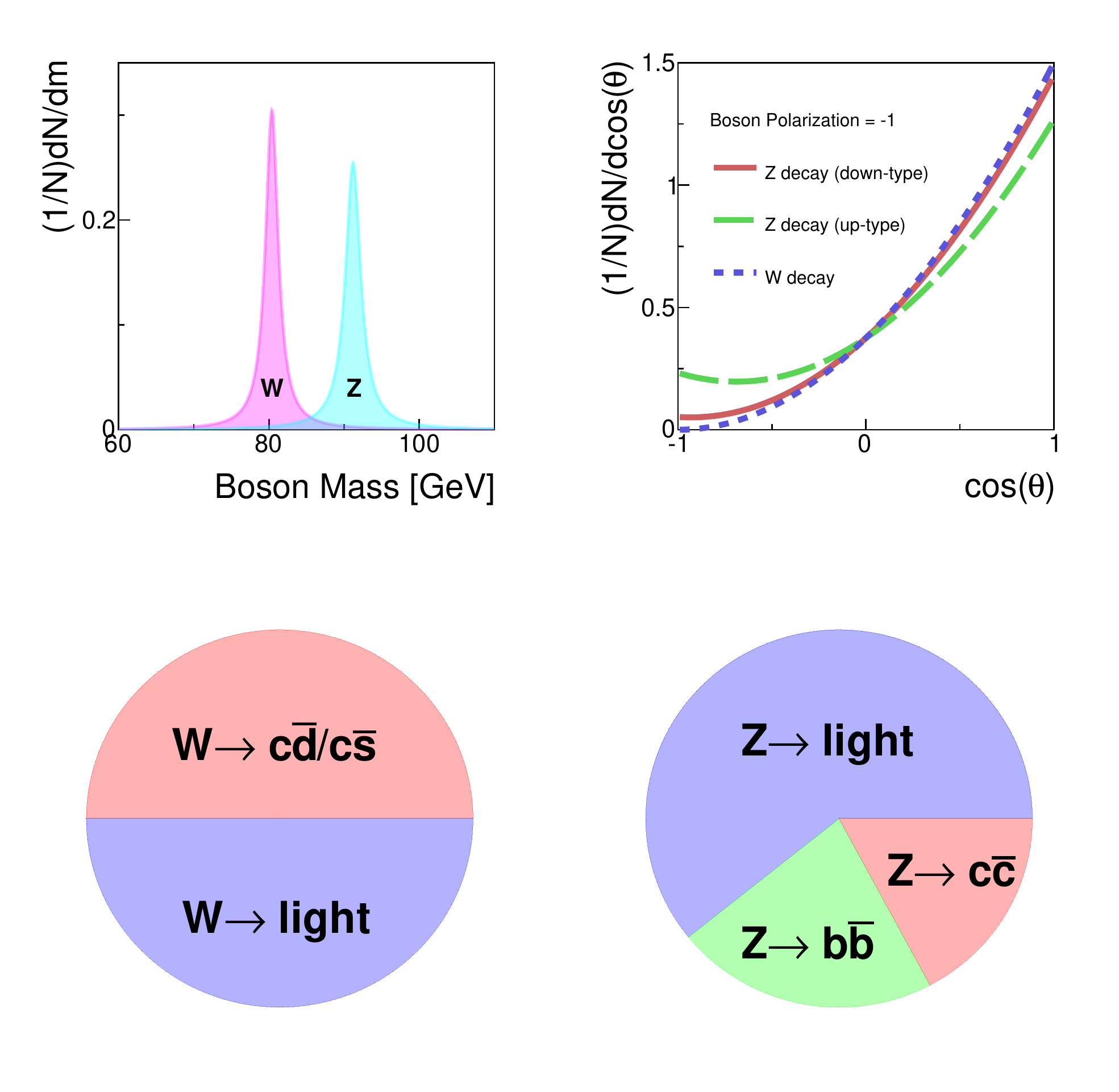}
\put(25,49){(a)}
\put(74,49){(b)}
\put(25,2){(c)}
\put(74,2){(d)}
\end{overpic}
\vspace{3mm}
\caption{(a) Breit--Wigner resonances for the $W$ (red) and $Z$ (blue) bosons, (b) angular distribution of the decay products of transversely polarized $W/Z$ bosons with respect to the spin direction in the boson rest frame, (c) hadronic branching fractions of the $W^+$ boson, and (d) of the $Z$ boson.  In (c) and (d), {\it light} stands for decay modes not involving $c$ and $b$ quarks.}
\label{fig:mass}
\end{figure}

\clearpage
\newpage

\subsection{Definitions of reconstructed objects}
\label{sec:objs}

Jets are formed from clusters using two different jet algorithms.  Small-radius jets are built with the anti-$k_t$ algorithm with jet radius parameter $R=0.4$ and large-radius jets are formed using the anti-$k_t$ algorithm with $R=1.0$ and then trimmed using $k_t$ $R=0.3$ subjets with $f_\text{cut}=0.05$.  Since the $W$ and $Z$ boson masses differ by about $10$~GeV, the jet mass can be used to discriminate between these two particles.  The distributions of the boson masses and jet masses for hadronically decaying $W$ and $Z$ bosons are shown in Fig.~\ref{fig:fat_Jmv0}.  The particle-level (`truth') jet mass is constructed from stable particles in the MC simulation ($c\tau > 10$ mm), excluding neutrinos and muons, clustered with the same jet algorithm as for calorimeter-cell clusters.  The QCD multijet processes that govern the formation of stable particles from the $W$ and $Z$ decay products create a broad distribution of jet masses even without taking into account detector resolution.  Constructing the jet mass from calorimeter-cell clusters further broadens the distribution.  The jet-mass resolution (physical $\oplus$ detector) is large compared to the natural width of the $W$ and $Z$ bosons and comparable to the difference in their masses.  For example, the standard deviation of $p_\text{T}^\text{reco jet}/p_\text{T}^\text{truth jet}$ is approximately 10\%.  The jet-mass variable nevertheless has some discriminating power.  Figure~\ref{fig:aux4} shows that as long as $p_\text{T}\gtrsim 200$ GeV, the jet mass distributions are relatively independent of $p_\text{T}$ for isolated $W$ and $Z$ bosons.

\begin{figure}[h!]
 \centering
\includegraphics[width=0.45\textwidth]{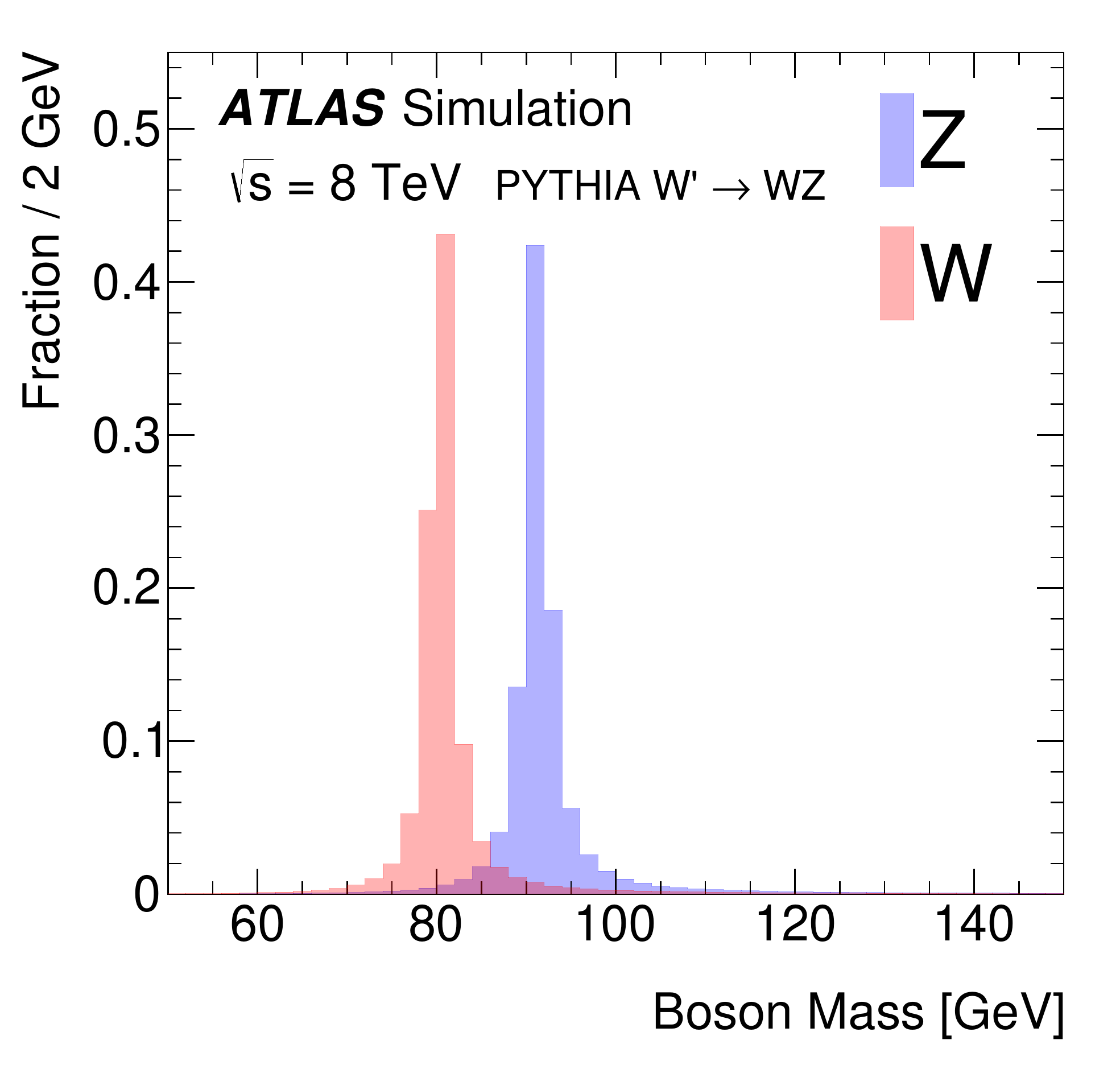}
\includegraphics[width=0.45\textwidth]{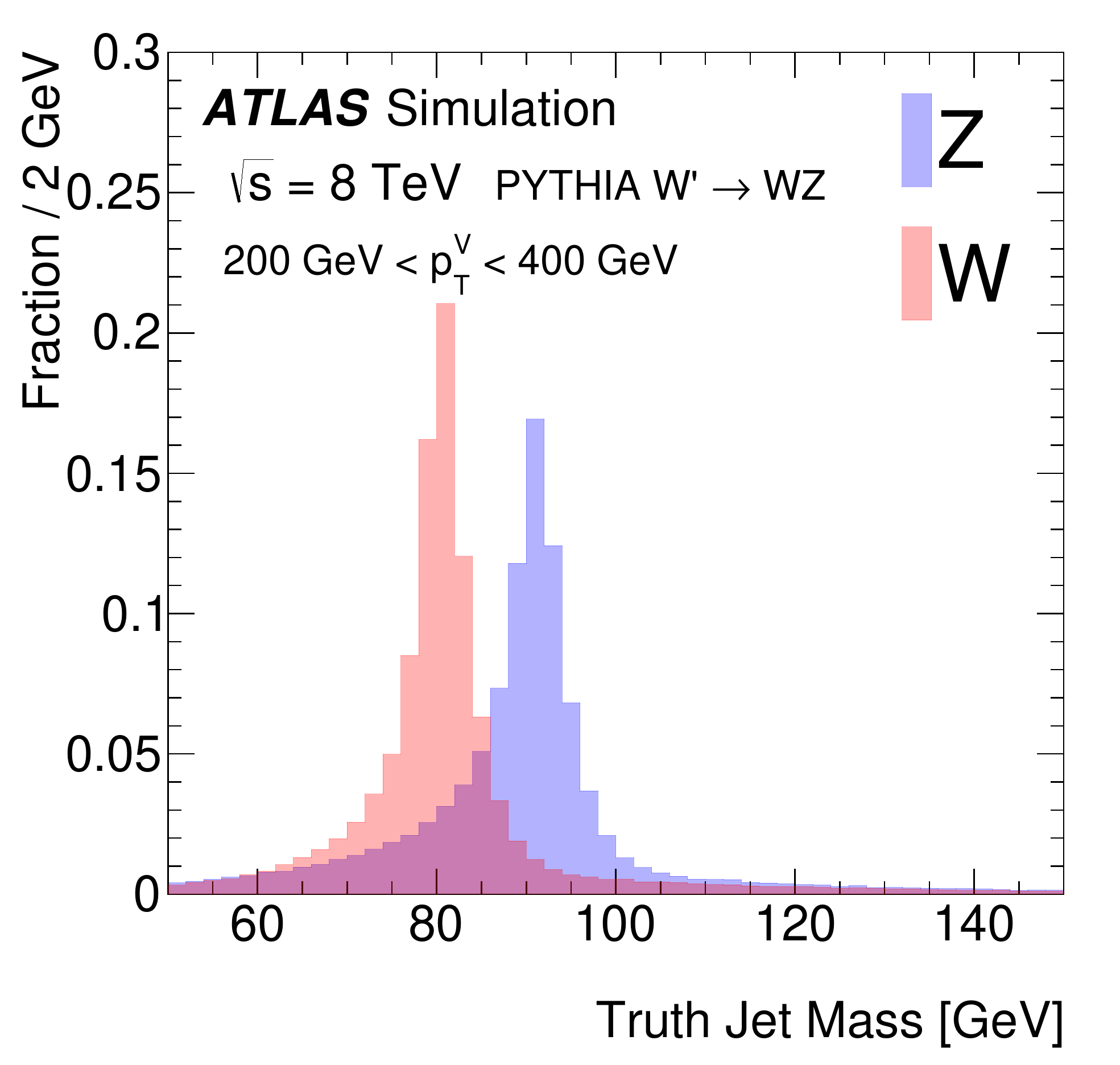}
\includegraphics[width=0.45\textwidth]{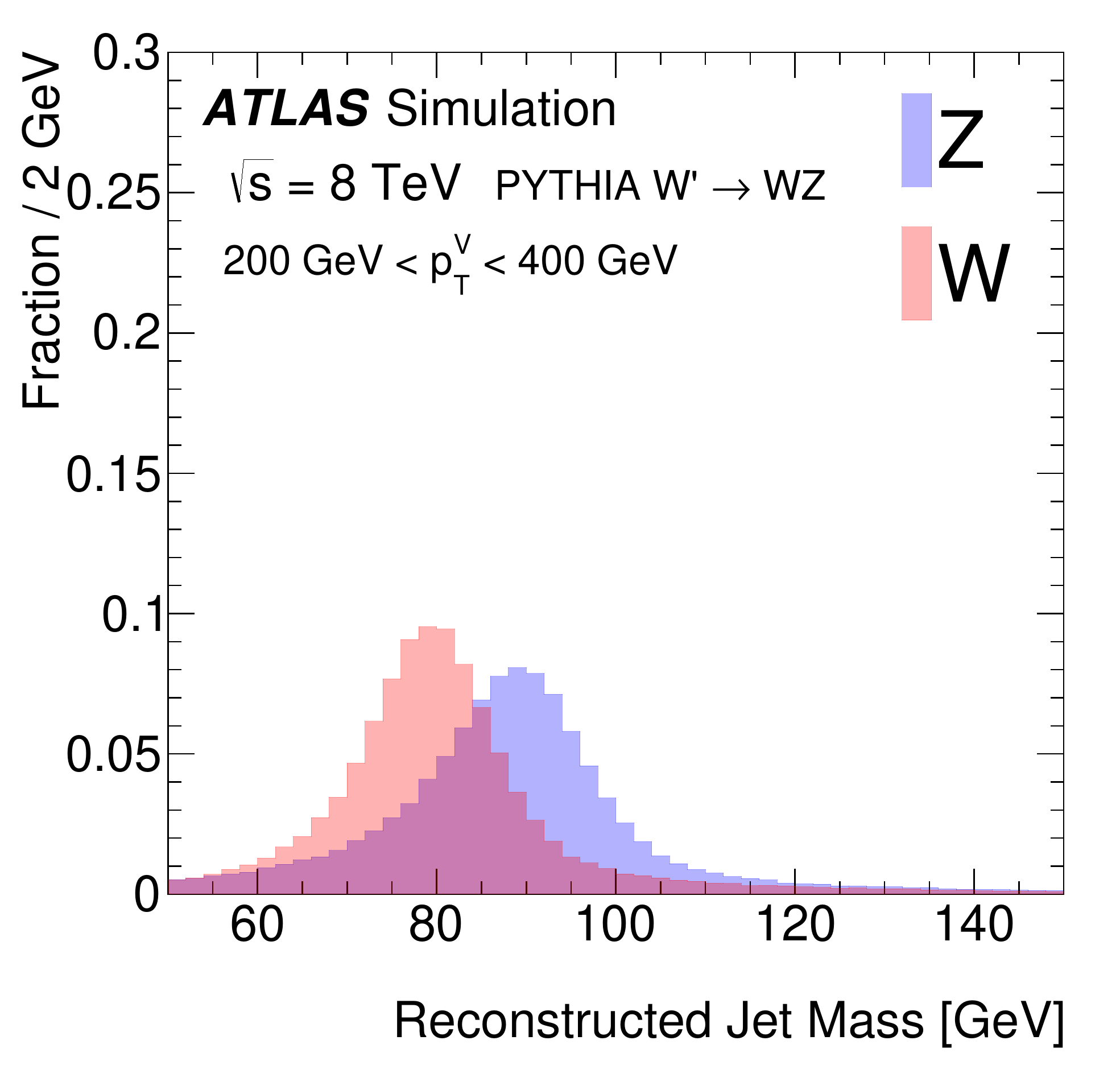}
\includegraphics[width=0.45\textwidth]{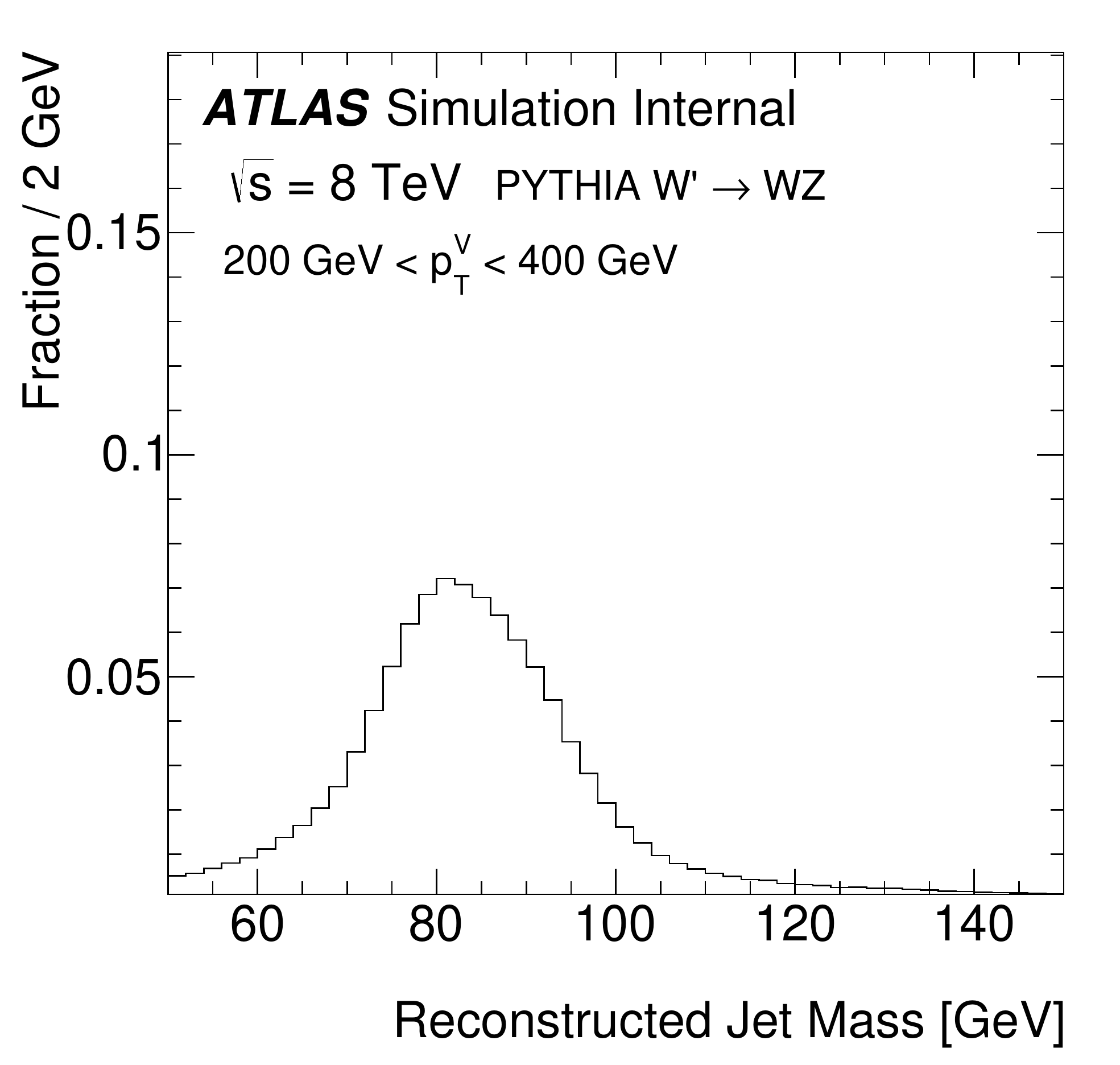}
\caption{The boson mass at generator level (top left), the `truth' jet mass at particle-level after parton fragmentation (top right), and the detector-level jet mass distributions (bottom) for $W$ and $Z$ boson jets separately (left) and for an even admixture of the two jet types (right).  The parton-level plot has a different vertical scale than the other plots and also has no $p_\text{T}$ requirement.}
\label{fig:fat_Jmv0}
\end{figure}

\begin{figure}[h!]
\begin{center}
\includegraphics[width=0.45\textwidth]{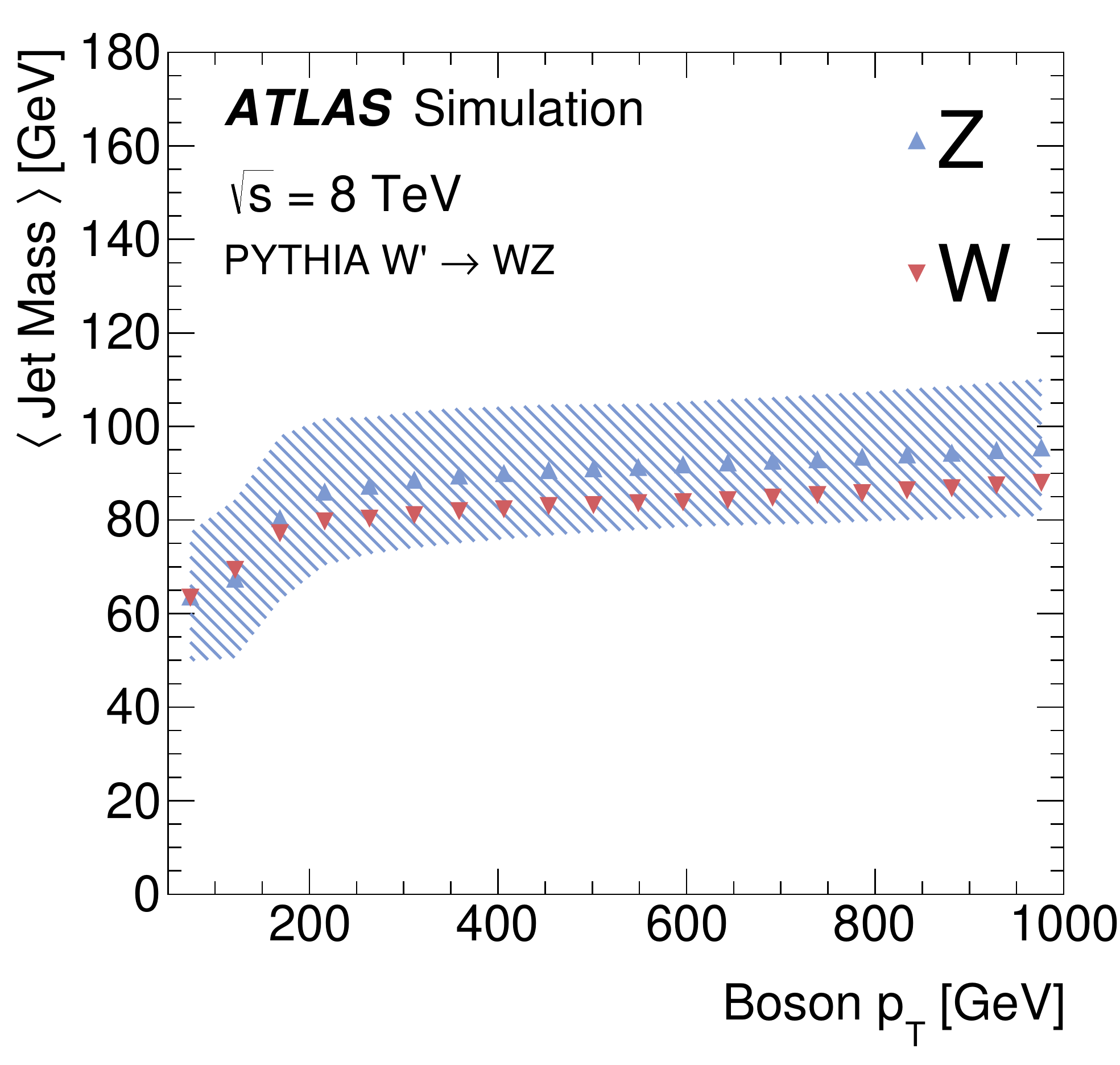}
\caption{The $p_\text{T}$ dependence of the jet mass distribution.  The points are the mean value in a given boson $p_\text{T}$ bin and the shaded region is the standard deviation for the $Z$ boson distribution.}
\label{fig:aux4}
\end{center}
\end{figure}

The momentum and electric charge of particles traversing the detector contain information about the charge of their parent boson.  To suppress the impact of pileup, tracks are required to originate from the primary collision vertex, which is defined as the vertex with the largest $\sum p_\text{T}^2$ computed from associated tracks.  Additionally, tracks must satisfy a very loose quality criterion for the track fit $\chi^2$ per degree of freedom, which must be less than three.  Tracks are associated with jets using ghost association~\cite{ghost}.  The charge of tracks associated with a jet is sensitive to the charge of the initiating parton.  In order to minimize the fluctuations due to low-$p_\text{T}$ particles, the {\it jet charge} is calculated using a $p_\text{T}$-weighting scheme (see Chapter~\ref{cha:jetcharge} for details):

\begin{align}
  \label{chargedef}
  Q_J = \frac{1}{({p_\text{T,J}})^\kappa}\sum_{i\in \text{\bf Tracks}} q_i\times (p_\text{T}^i)^\kappa,    \end{align} 

\noindent where $\text{\bf Tracks}$ is the set of tracks with $p_\text{T}>500$ MeV associated with jet $J$, $q_i$ is the charge (in units of the electron charge) determined from the curvature of track $i$ with associated $p_\text{T}^i$, $\kappa$ is a free parameter, and ${p_\text{T,J}}$ is the transverse momentum of the jet measured in the calorimeter.  The calorimeter energy is used in the denominator to determine $p_\text{T}$ instead of the sum of track momenta to account for the contribution from neutral particles.   Dedicated studies have shown that $\kappa=0.5$ is generally best for determining the charge of partons from the jets they produce (see Sec.~\ref{sec:jetchargeperformance}).   The distributions of the jet charge for jets initiated by $W^+,W^-$ and $Z$ bosons are shown in Fig.~\ref{fig:fat_Jqv0}.  There is an observable separation between positive and negative $W$ bosons, though the width of the jet charge distribution is larger than the separation of means.  Figure~\ref{fig:aux4} shows that the standard deviation is about twice as large as the separation between the jet charge means of $W^+$ boson jets and $Z$ jets.  As with Fig.~\ref{fig:aux4}, the jet charge distribution is relatively stable with $p_\text{T}\gtrsim 200$ GeV, though there is a small increase in the standard deviation due to the degradation in tracking performance at high $p_\text{T}$.

The expected charge composition of a $W$ sample is process dependent.  There are more $W^+$ than $W^-$ bosons in inclusive $W'$ production because of the initial charge asymmetry of quarks in the proton resulting in more $W'{}^{+}$ than $W'{}^{-}$.  The discrimination between $Z$ bosons and a near even mixture of $W^\pm$ is greatly diminished with respect to e.g. $Z$ versus $W^+$.   In that case charge sensitive variables are not very useful for the tagger and so all results are shown also without such variables.  In a variety of physics processes, the charge of the hadronically decaying $W$ boson is known from other information in the event.  For example, in searches for FCNC effects in $t\bar{t}$ events with one leptonically decaying $W$ boson, the charge of the lepton is opposite to the charge of the hadronically decaying $W$ boson.  Henceforth, only $W^+$ bosons are used for constructing the boson-type tagger; the results are the same for $W^-$ bosons.

\begin{figure}[h!]
\begin{center}
\includegraphics[width=0.45\textwidth]{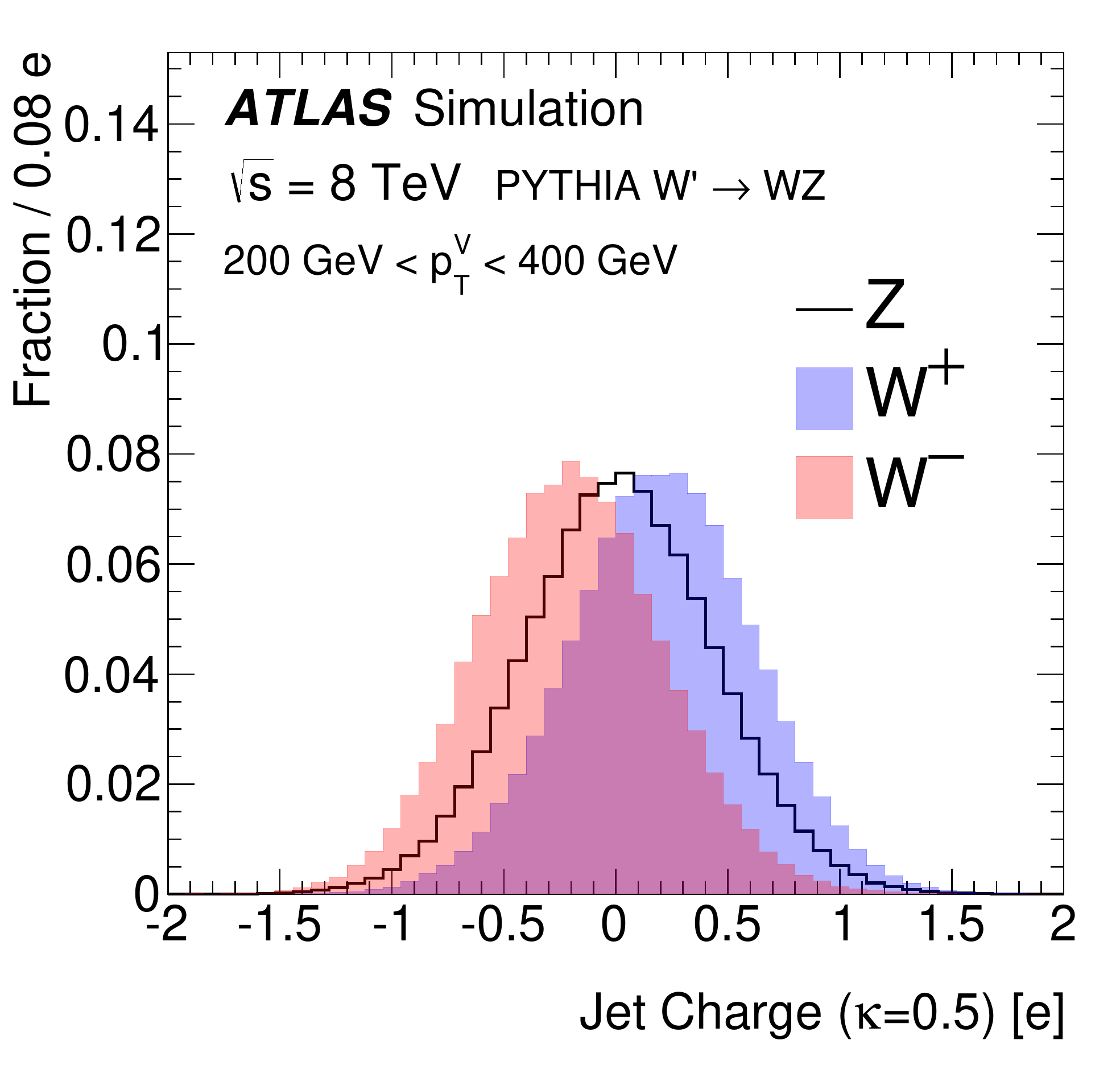}
\caption{The jet charge distribution for jets originating from $W^\pm$ and $Z$ bosons in simulated $W'$ decays.  Each distribution is normalized to unity.  The parameter $\kappa$ controls the $p_\text{T}$-weighting of the tracks in the jet charge sum.}
\label{fig:fat_Jqv0}
\end{center}
\end{figure}

\begin{figure}[h!]
\begin{center}
\includegraphics[width=0.45\textwidth]{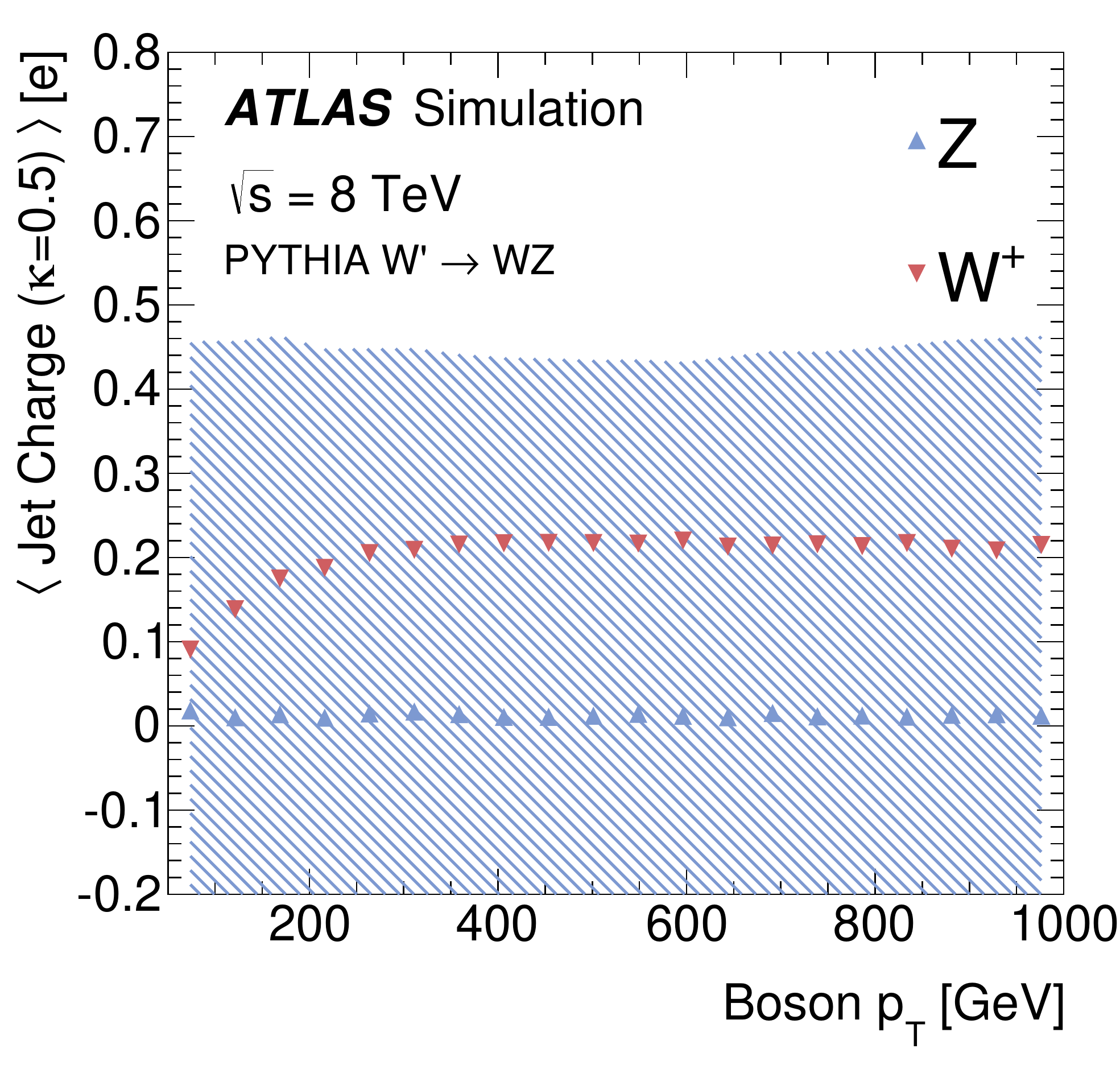}
\caption{The $p_\text{T}$ dependence of the jet charge distribution.  The points are the mean value in a given boson $p_\text{T}$ bin and the shaded region is the standard deviation for the $Z$ boson distribution.}
\label{fig:aux4aa}
\end{center}
\end{figure}

The tracks from charged particles can be used further to identify the decays of certain heavy-flavor quarks inside jets due to the long $b$-hadron lifetime.  This is useful for boson-type tagging because the $Z$ boson couples to $b\bar{b}$ while decays of the $W$ boson to $b$-quarks are highly suppressed and can be neglected.  ATLAS has commissioned a $b$-tagging algorithm called MV1 (defined in Ref.~\cite{Aad:2015ydr}) which combines information about track impact-parameter significance with the explicit reconstruction of displaced $b$- and $c$-hadron decay vertices.  The MV1 distribution is shown in Fig.~\ref{fig:fat_mv1dist} for the leading and sub-leading small-radius jets matched to the leading large-radius jet. The boson-type tagger presented here uses multiple bins of the MV1 distribution simultaneously.  Five bins of MV1 are defined by $b$-tag efficiencies (probability to tag a $b$-quark jet as such) of 0\%--50\%, 50\%--60\%, 60\%--70\%, 70\%--80\%, and 80\%--100\% as determined in simulated $t\bar{t}$ events.  A lower $b$-tag efficiency leads to higher light-quark jet rejection.   The five $b$-tagging efficiency bins are exclusive and MV1 is constructed as a likelihood with values mostly between zero and one (one means more like a $b$-jet).  For example, a $100\%$ $b$-tagging efficiency corresponds to a threshold of MV1 $>0$ and an $80\%$ $b$-tagging efficiency corresponds to a threshold value of MV1 $>z$ for $z\ll1$.  The 80\%--100\% $b$-tag efficiency bin then corresponds to jets with an MV1 value between $0$ and $z$.  Constructed in this way, the fraction of true $b$-jets inside an efficiency bin $x$\%--$y$\% should be $(y-x)\%$.  

Small-radius jets are matched to a large-radius jet by geometric matching\footnote{In the definition of jets, $R$ is the characteristic size in $(y,\phi)$ and the rapidity $y$ is used in the jet clustering procedure, whereas geometrical matching between reconstructed objects is performed using $(\Delta R)^2=(\Delta \phi)^2+(\Delta\eta)^2$, where $\eta$ is the pseudorapidity.} ($\Delta{R}<1.0$).   Of all such small-radius jets, the two leading ones are considered.  There are thus 30 possible bins of combined MV1 when considering the leading and sub-leading matched small-radius jet.  The number of bins is $25$ from the $5\times 5$ efficiency-binned MV1 distributions in addition to five more for the case in which there is no second small-radius jet matched to the large-radius jet.  The distribution for the efficiency-binned MV1 variable for the leading and sub-leading matched small-radius jets is shown for  $W$ and $Z$ bosons in Fig.~\ref{fig:fat_Jbv0}.  The flavor of a small-radius jet is defined as the type of the highest energy parton from the parton shower record within $\Delta R < 0.4$.   As expected, a clear factorization is seen in Fig.~\ref{fig:fat_Jbv0} -- the MV1 value depends on the flavor of the jet and not the process that created it.  This means that, for example, $c$-jets from $W$ decays have the same MV1 distribution as $c$-jets from $Z$ decays.  Jets originating from $b$-hadron decays tend to have a larger value of MV1, which means they fall in a lower efficiency bin.  Jets not originating from $b$- or $c$-decays are called light jets and are strongly peaked in the most efficient bin of MV1.  There is always one small-radius jet matched to the large-radius jet, but about $20\%$ of the time there is no sub-leading jet with $p_\text{T}>25$~GeV matched to the large-radius jets.  These jets are all predicted to originate from light-quark decays of the $W$ and $Z$ bosons. 

\begin{figure}[h!]
\begin{center}
\includegraphics[width=0.4\textwidth]{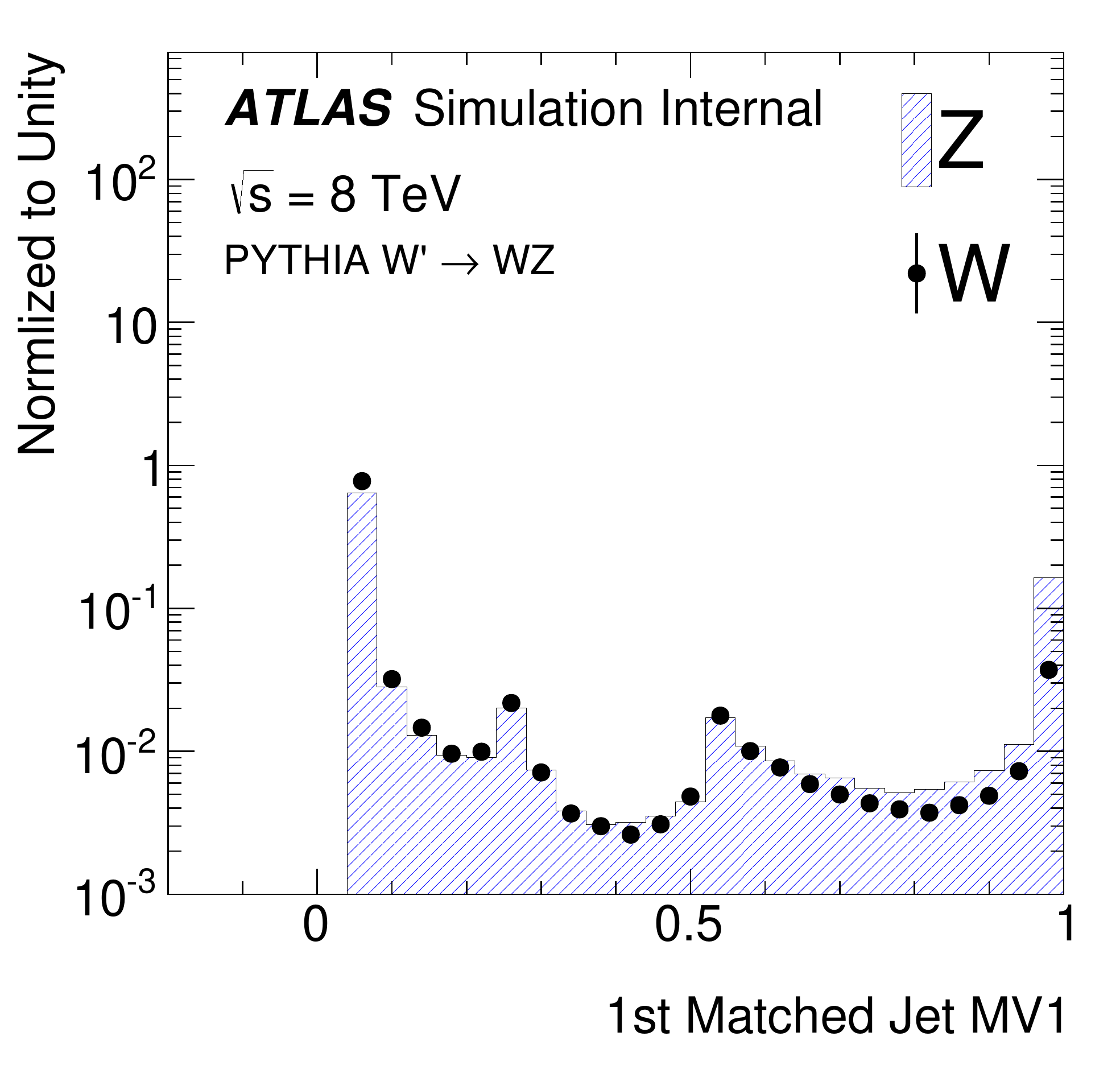}\includegraphics[width=0.4\textwidth]{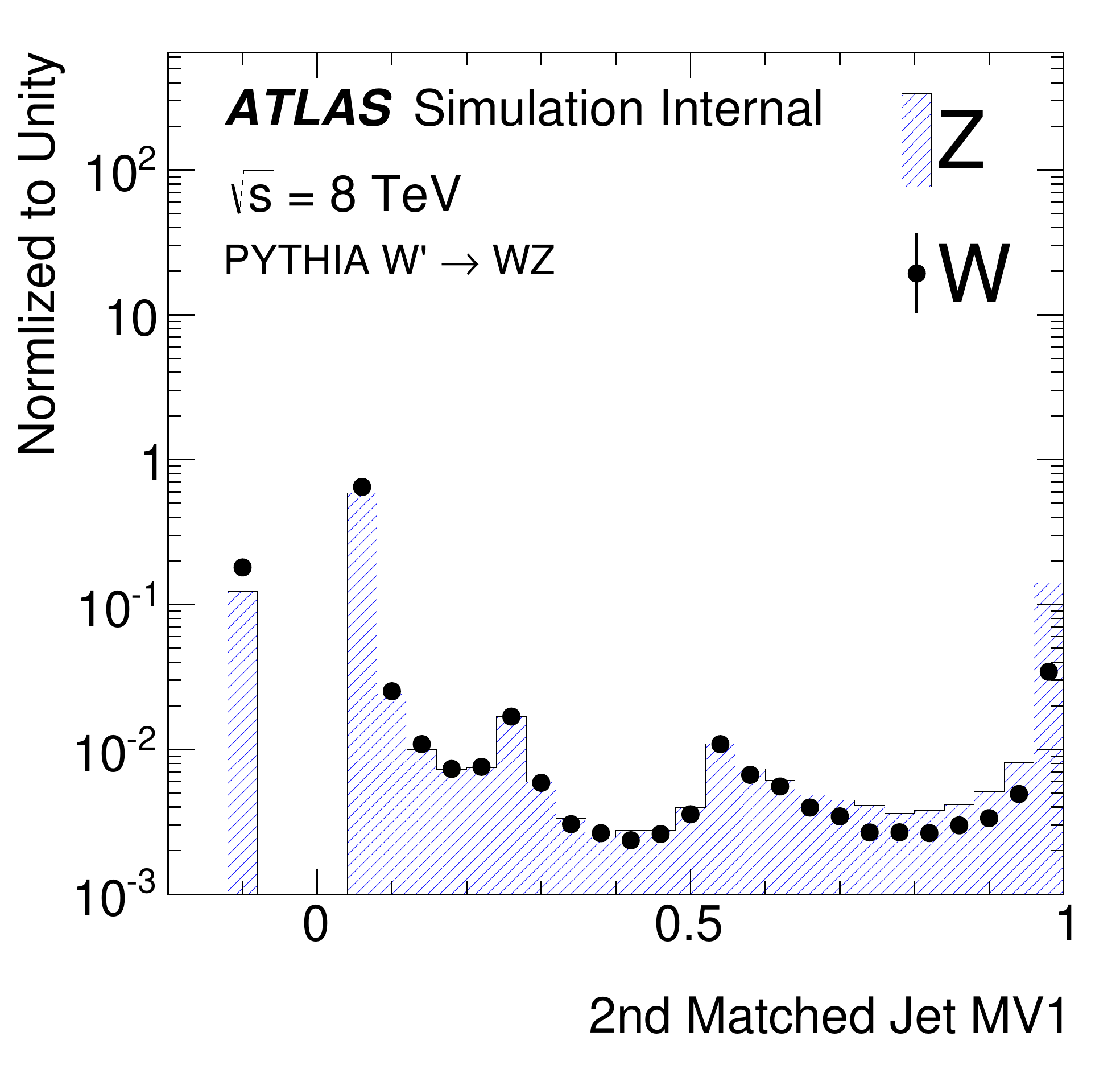}
\caption{The distribution of the MV1 discriminant for the leading (left) and subleading (right) small-radius jets matched to the leading large-radius jet.  The spike at $-1$ in the right plot corresponds to cases in which there is not a second small-radius jet.  The other features in the distribution correspond to transitions in the dominant input algorithms to MV1~\cite{Aad:2015ydr}.}
\label{fig:fat_mv1dist}
\end{center}
\end{figure}

\begin{figure}[h!]
\begin{center}
\includegraphics[width=0.45\textwidth]{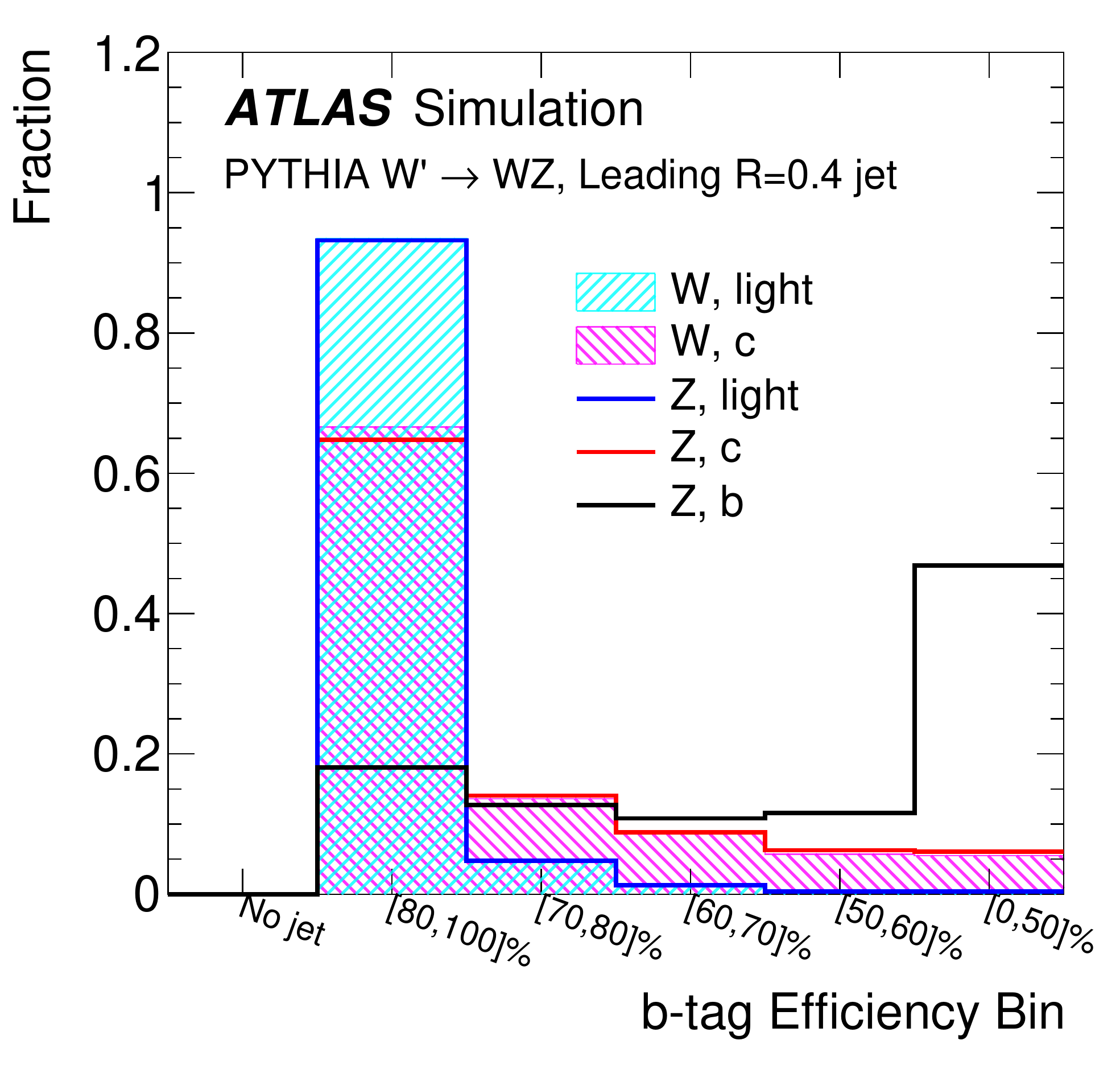}\includegraphics[width=0.45\textwidth]{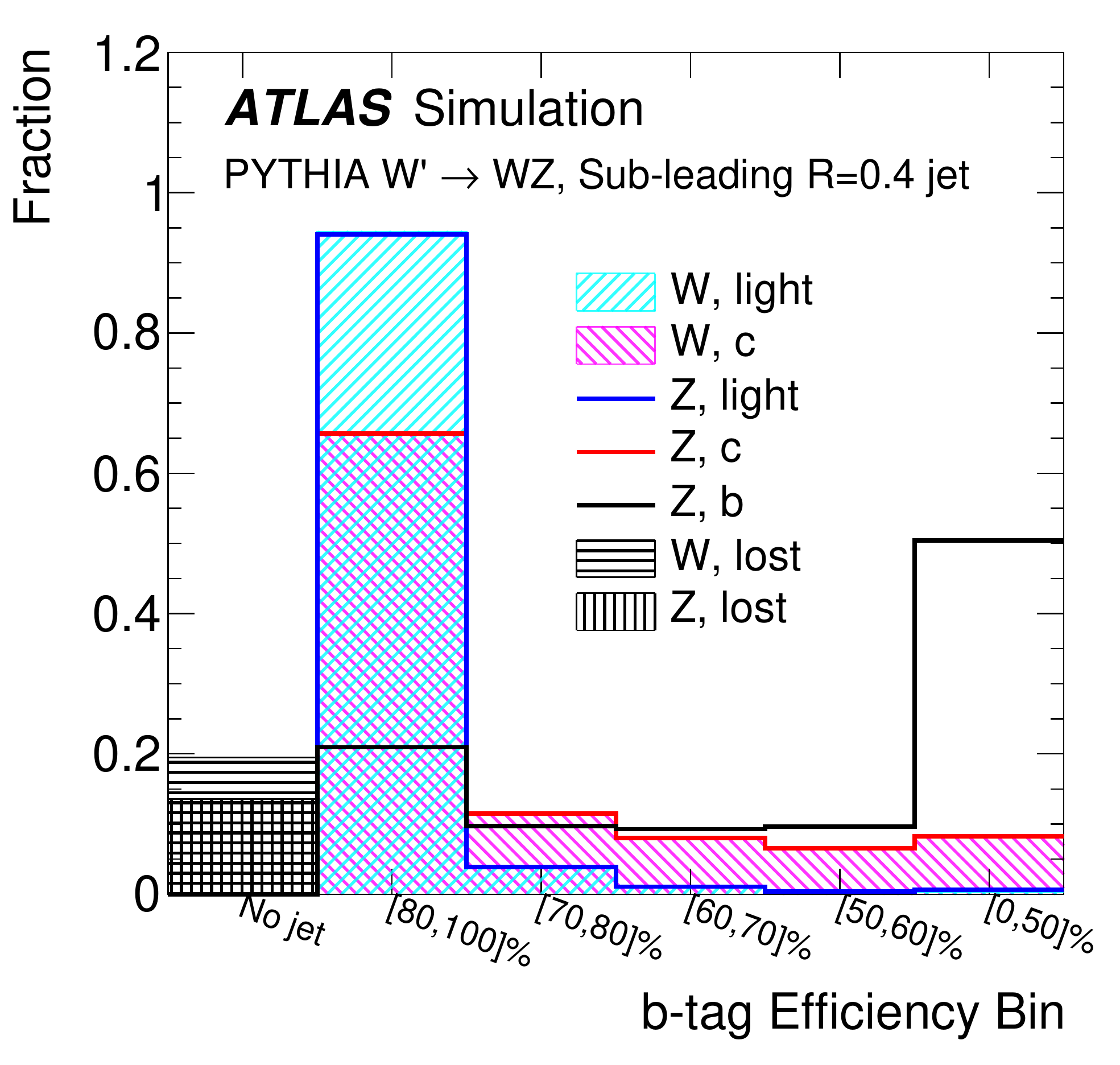}
\caption{The efficiency-binned MV1 distribution for small-radius jets associated with large-radius jets resulting from $W$ and $Z$ boson decays.  The left (right) plot shows the leading (sub-leading) small-radius jet MV1 distribution.  The bins correspond to exclusive regions of $b$-jet efficiency.  As such, the bin content of the black line ($b$-tagging for $b$-jets) should be proportional to the size of the efficiency window: about 50\% for the rightmost bin, 10\% for the three middle bins and 20\% for the second bin.}
\label{fig:fat_Jbv0}
\end{center}
\end{figure}

Figures~\ref{fig:aux4555} illustrates why there is a slightly different fraction of $W$ events that have no second matched small radius jet compared to $Z$ events.  The $W$ and $Z$ transverse momentum spectrum are identical, so the boson mass difference has no effect on the spectrum at the low end, i.e. the probability for the subleading jet to be below threshold is independent of the boson.  However, since $m_W<m_Z$, the angular separation between the two boson decay products is slightly smaller for $W$ bosons and thus at high $p_\text{T}$, the $W$ daughter jets merge into a single small radius jet earlier than for $Z$ jets.

The $p_\text{T}$-dependence of the matched $b$-jet multiplicity is shown in Fig.~\ref{fig:aux4bjet}.  As with the jet mass and jet charge, there is a clear turn on for $p_\text{T}\sim 200$ GeV.  However, there is a second feature at $p_\text{T}\gtrsim 400$ GeV when the two small-radius jets begin to merge into a single small-radius $b$-tagged jet.

\begin{figure}[h!]
\begin{center}
\includegraphics[width=0.45\textwidth]{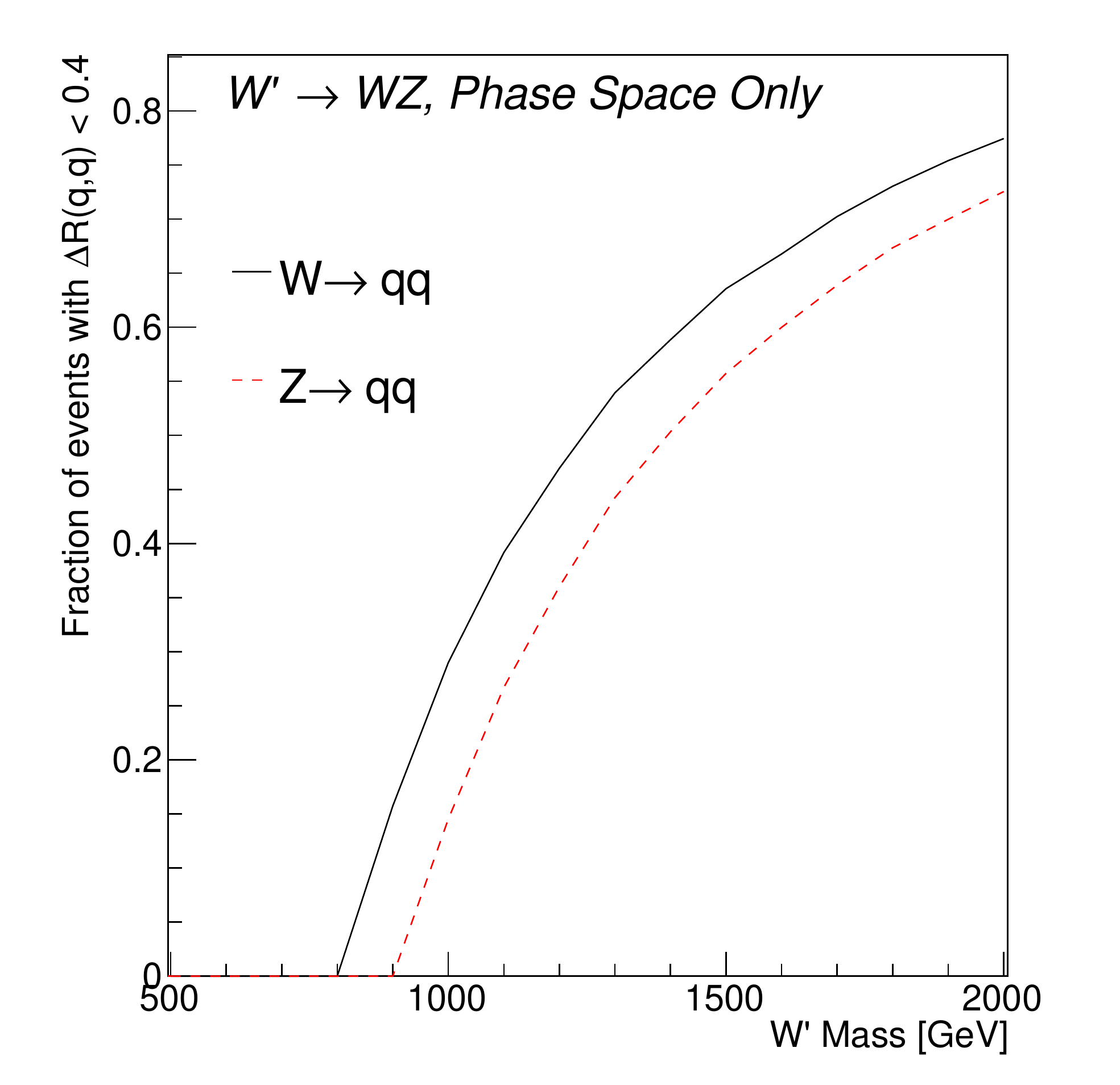}\includegraphics[width=0.45\textwidth]{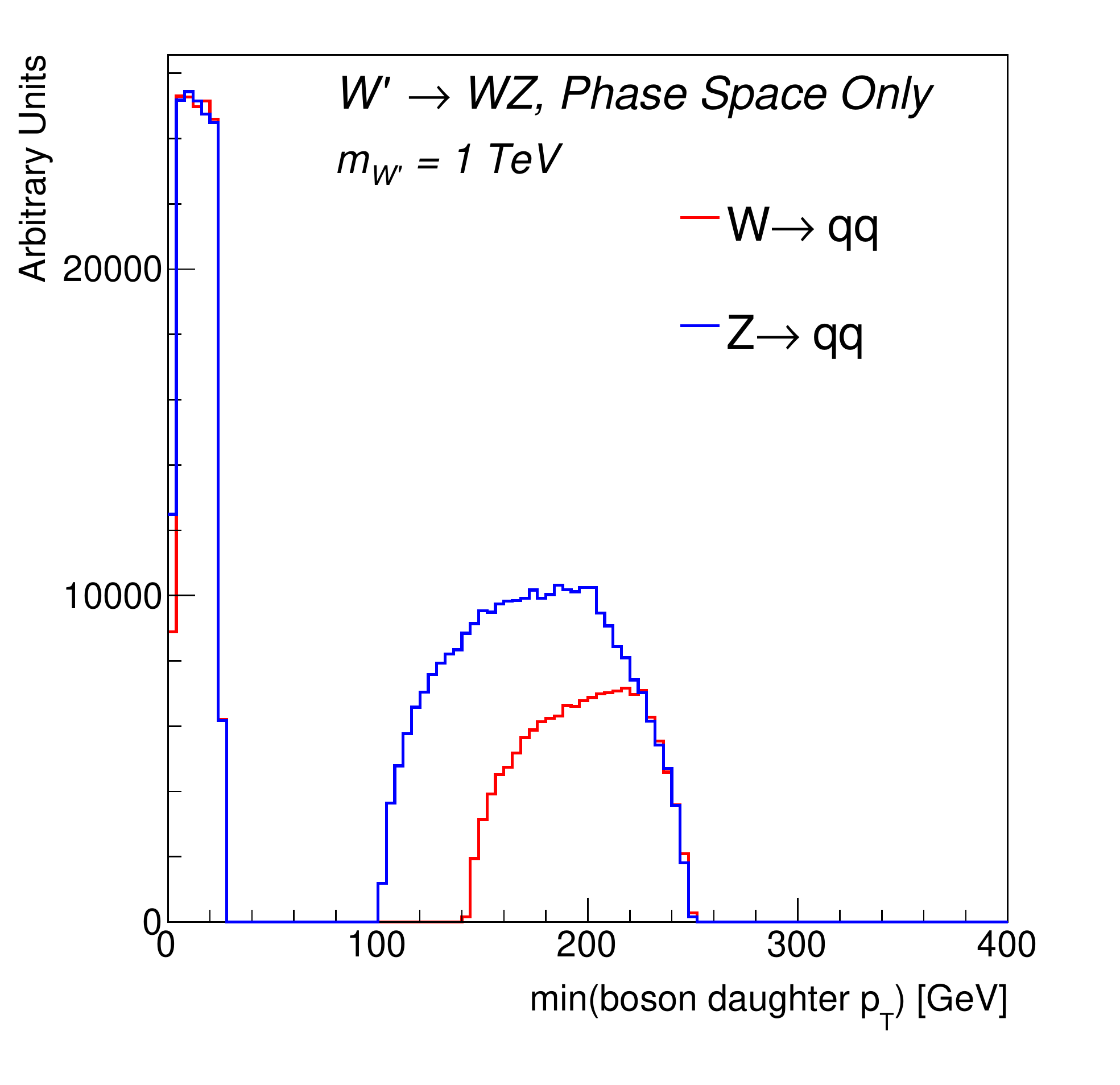}
\caption{Left: The fraction of events with $\Delta R(q,q)<0.4$ in a toy MC simulation as a function of the simulated boson mass. Right: The $p_\text{T}$ spectrum of the softer of the two decay products for a simulated (scalar) boson with mass 1 TeV when one of the decay products is below 25 GeV or the two decay products are within $\Delta R <0.4$.}
\label{fig:aux4555}
\end{center}
\end{figure}

\begin{figure}[h!]
\begin{center}
\includegraphics[width=0.45\textwidth]{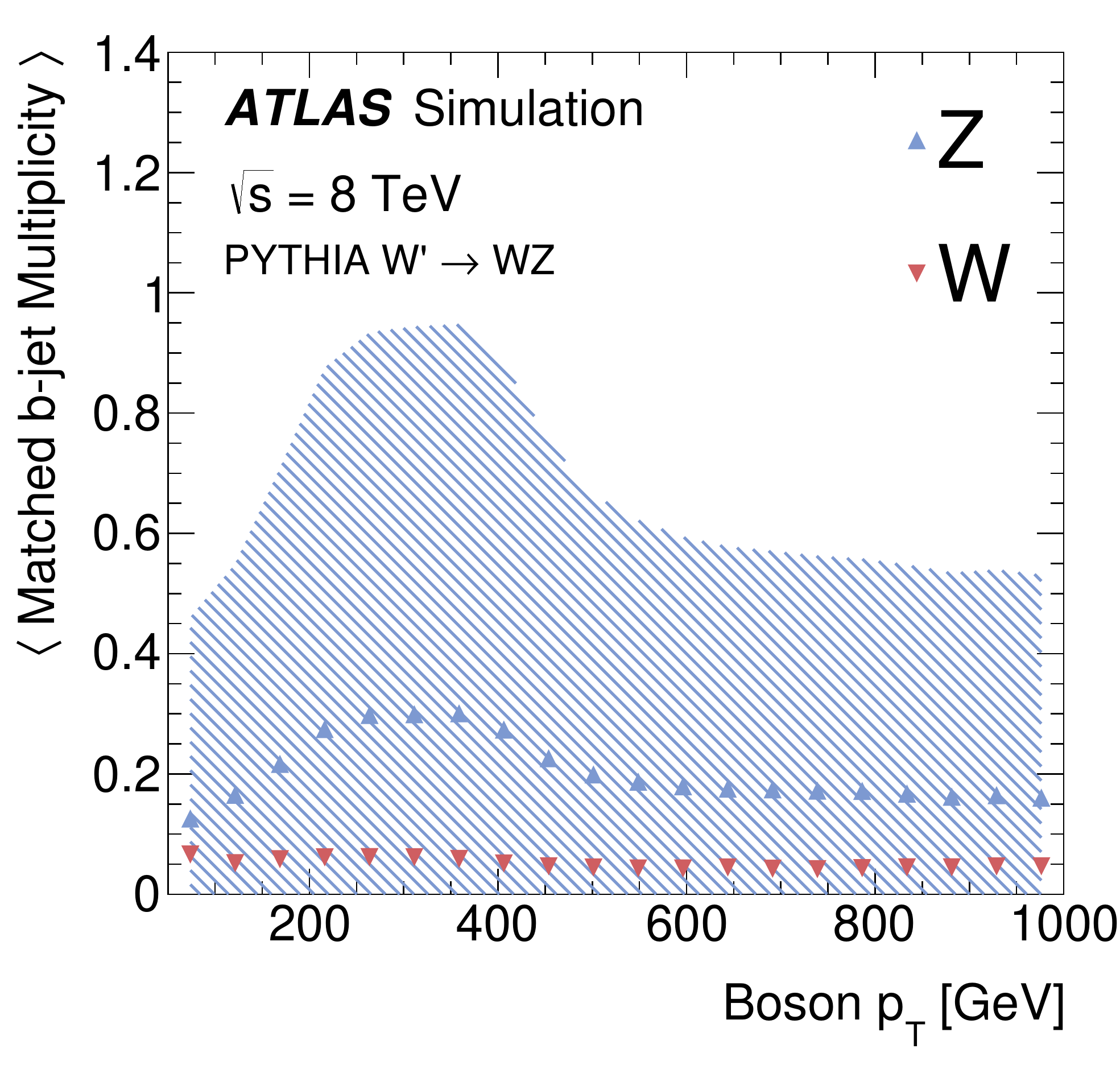}
\caption{The $p_\text{T}$ dependence of the number of small radius $b$-tagged jets (70\% working point) associated to the large radius jet.  The points are the mean value in a given boson $p_\text{T}$ bin and the shaded region is the standard deviation for the $Z$ boson distribution.}
\label{fig:aux4bjet}
\end{center}
\end{figure}

\clearpage

\subsection{Tagger performance}
\label{sec:perfm}

The optimal multivariate tagger combining jet mass, jet charge, and the MV1 of matched small-radius jets is constructed from a three-dimensional (3D) likelihood ratio.  For $N$ bins each of jet mass and jet charge, as well as $30$ combined MV1 bins, the 3D likelihood ratio would have $30\times N^2$ total bins.  Populating all of these bins with sufficient MC events to produce templates for the likelihood ratio requires an unreasonable amount of computing resources, especially for the high-efficiency bins of combined MV1. Estimating the 3D likelihood as the product of the 1D marginal distributions, where all variables but the one under consideration are integrated out, is a poor approximation for jet mass and combined MV1 due to the correlation induced by the presence of semileptonic $b$-decays, which shift the jet mass to lower values due to the presence of unmeasured neutrinos\footnote{The muons from semileptonic decays are added back to the jet using a four-momentum sum.  Adding back the muon has a negligible impact on the inclusive mass distribution due to the semileptonic branching ratios and lepton identification requirements.  For details about the muon reconstruction and selection, see Sec.~\ref{sec:data} (the only difference here is that the isolation is not applied).  Figure~\ref{fig:auxaddmuon} shows the impact of this muon correction on the jet mass.}.  It is still possible to use a simple product by noting that all three tagger inputs are independent when the flavor of the decaying boson has been determined.  Thus, for each possible boson decay channel, templates are built for the jet mass, the jet charge, and the efficiency-binned MV1 distributions.  For a particular decay flavor, the joint distribution is then the product of the individual distributions.  Summing over all hadronic decay channels then gives the full distribution.  To ease notation, the efficiency-binned MV1 is denoted $B=(B_\text{lead},B_\text{sub-lead})$.  The distribution for $B_\text{lead}$ ($B_\text{sub-lead}$) is shown in the left (right) plot in Fig.~\ref{fig:fat_Jbv0}. Symbolically, for decay flavor channel $\mathcal{F}$, mass $M$, charge $Q$, and efficiency-binned MV1 $B$, the likelihood is given by:

\begin{align}
\label{eq:like}
p(M,Q,B|V)=\sum_\mathcal{F} \Pr(\mathcal{F}|V)p(M|\mathcal{F},V)p(Q|\mathcal{F},V)\Pr(B|\mathcal{F},V),
\end{align}

\noindent where\footnote{The symbol $p$ denotes a probability density whereas $\Pr$ denotes a discrete probability distribution.} $V\in\{W,Z\}$ and the sum is over $\mathcal{F}=bb,cc,cs,cd$ and light-quark pairs.  The distribution of $B$ is well approximated as the product of the distributions for $B_\text{lead}$ and $B_\text{sub-lead}$ when the flavors of the leading and sub-leading jets are known.  This is exploited for hadronically decaying $W$ bosons and for the light-quark flavor decays of $Z$ bosons to construct templates for $B$ that have a sufficient number of simulated events for large values of $B$, i.e. $\Pr(B|\mathcal{F},V)=\Pr(B_\text{lead}|\mathcal{F},V)\Pr(B_\text{sub-lead}|\mathcal{F},V)$.  Figure~\ref{fig:btaggingcorr} shows that factorization holds within the statistical uncertainty of the simulation.  The unit-normalized templates for $B$ are shown in Fig.~\ref{fig:fat_Jbv0} and the unit-normalized templates $p(M|\mathcal{F},V)$ and $p(Q|\mathcal{F},V)$ are shown in Fig.~\ref{fig:templates}.  For a given boson type, the jet-charge template is nearly independent of the flavor.  However, there is a dependence of the jet mass on the (heavy) flavor of the boson decay products.  The independence of the jet mass and jet charge distributions is demonstrated in Fig.~\ref{fig:auxcorrjetchargemass}.

\begin{figure}[h!]
\begin{center}
\includegraphics[width=0.45\textwidth]{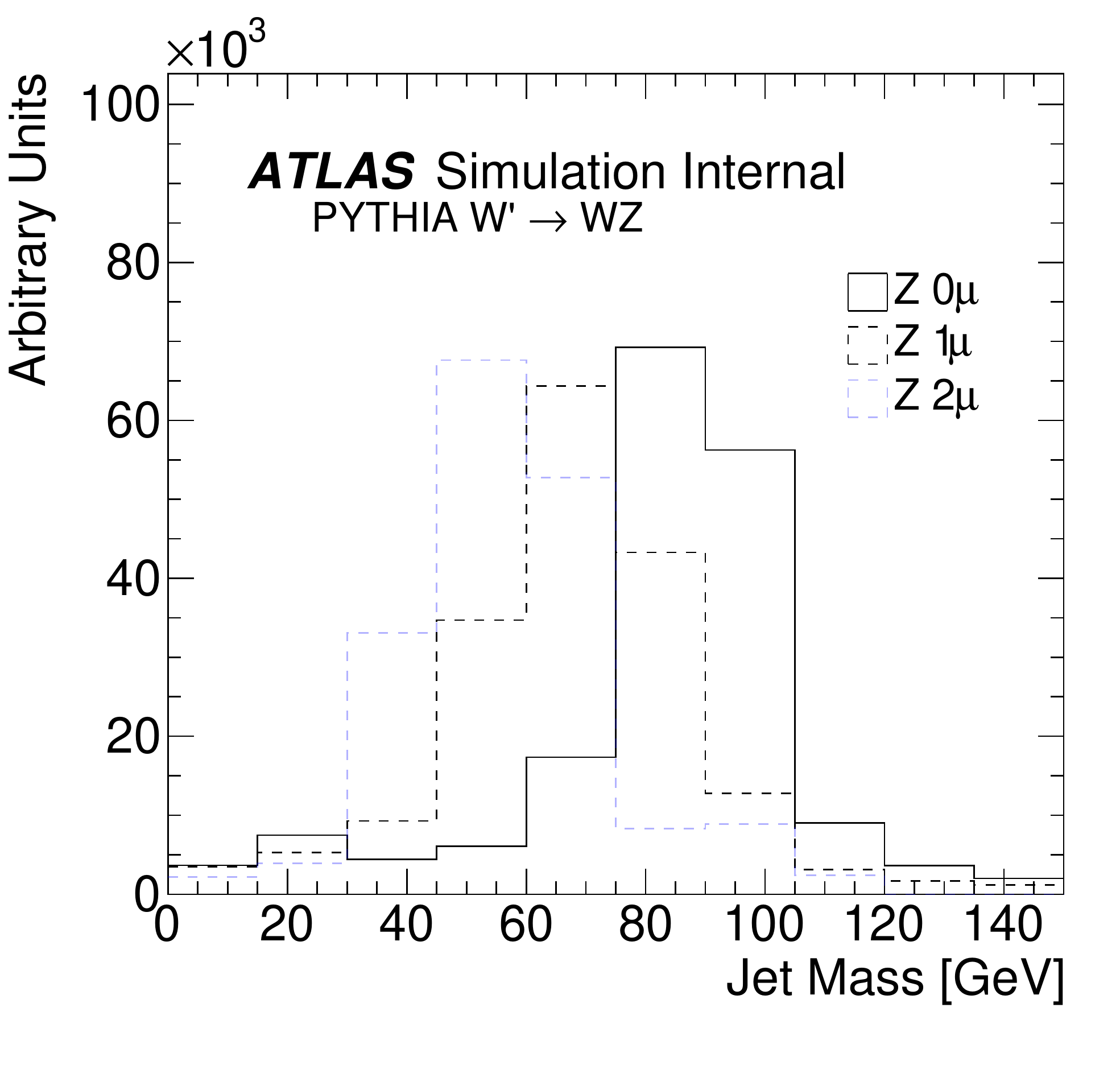}\includegraphics[width=0.45\textwidth]{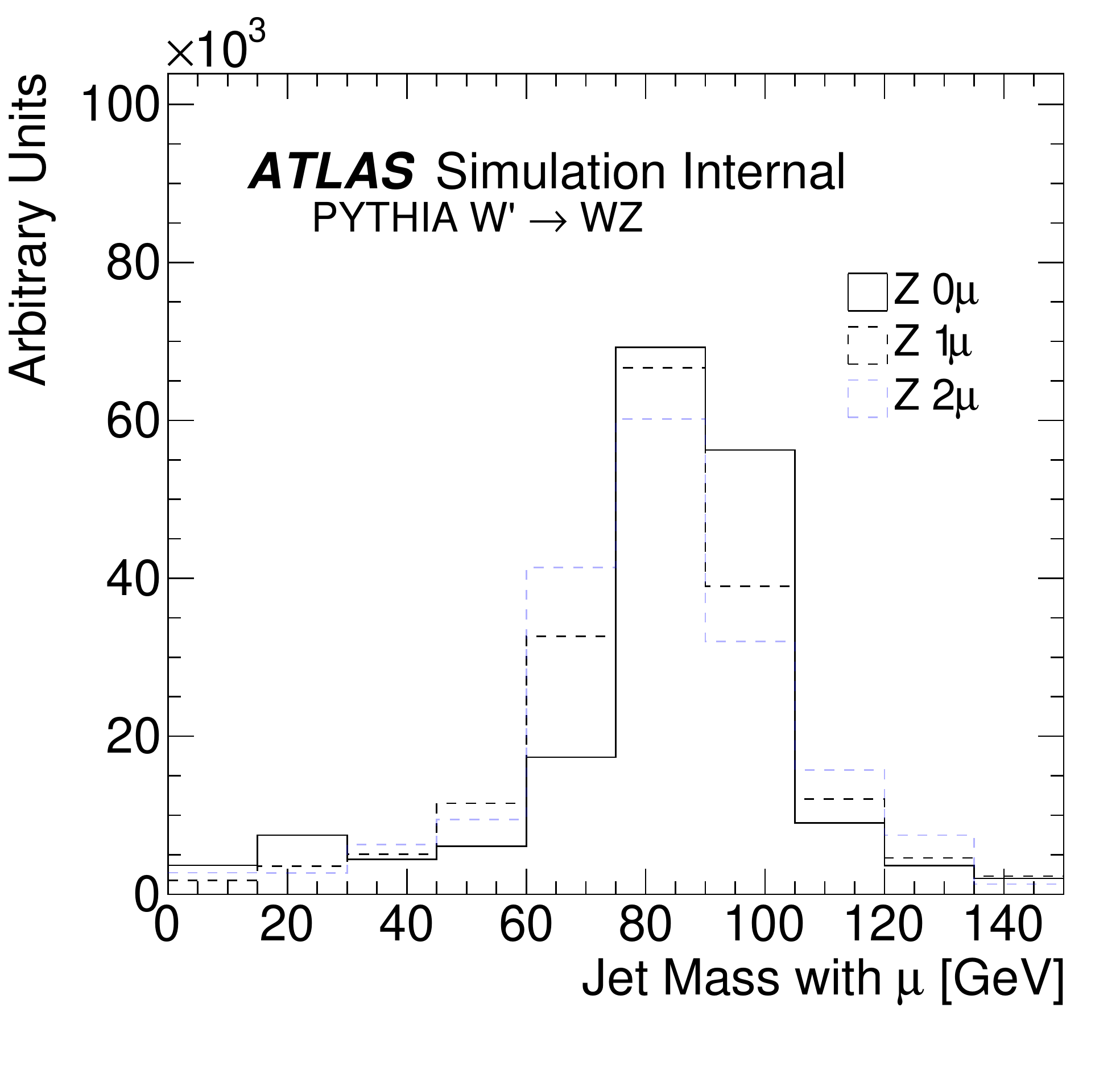}\\
\caption{The distribution of the jet mass for $Z$ boson jets with $0$, $1$, or $2$ matched muons before (left) and after (right) adding the muon to the jet four-vector before computing the mass.}
\label{fig:auxaddmuon}
\end{center}
\end{figure}

\begin{figure}[h!]
\begin{center}
\includegraphics[width=0.45\textwidth]{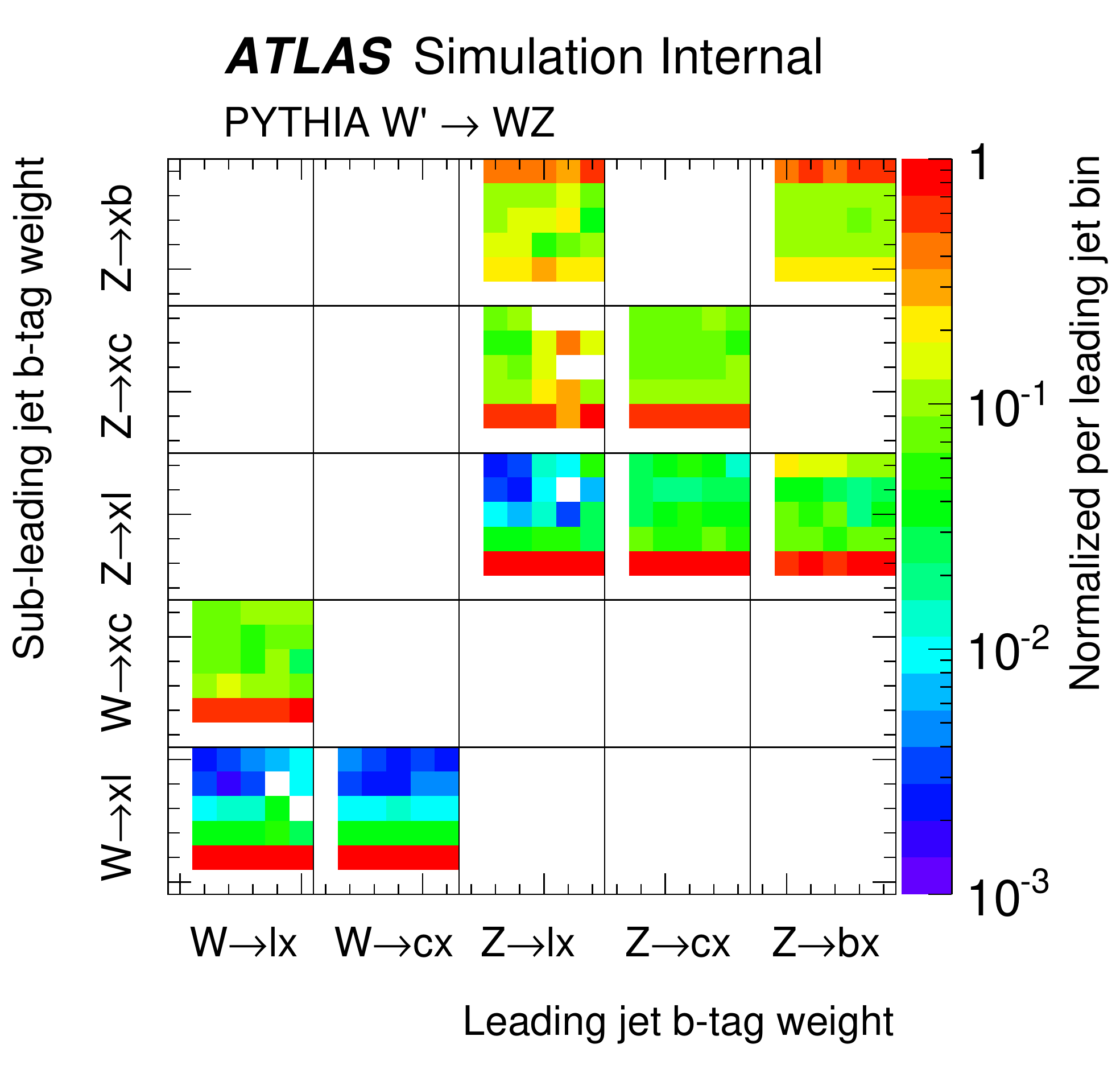}\includegraphics[width=0.45\textwidth]{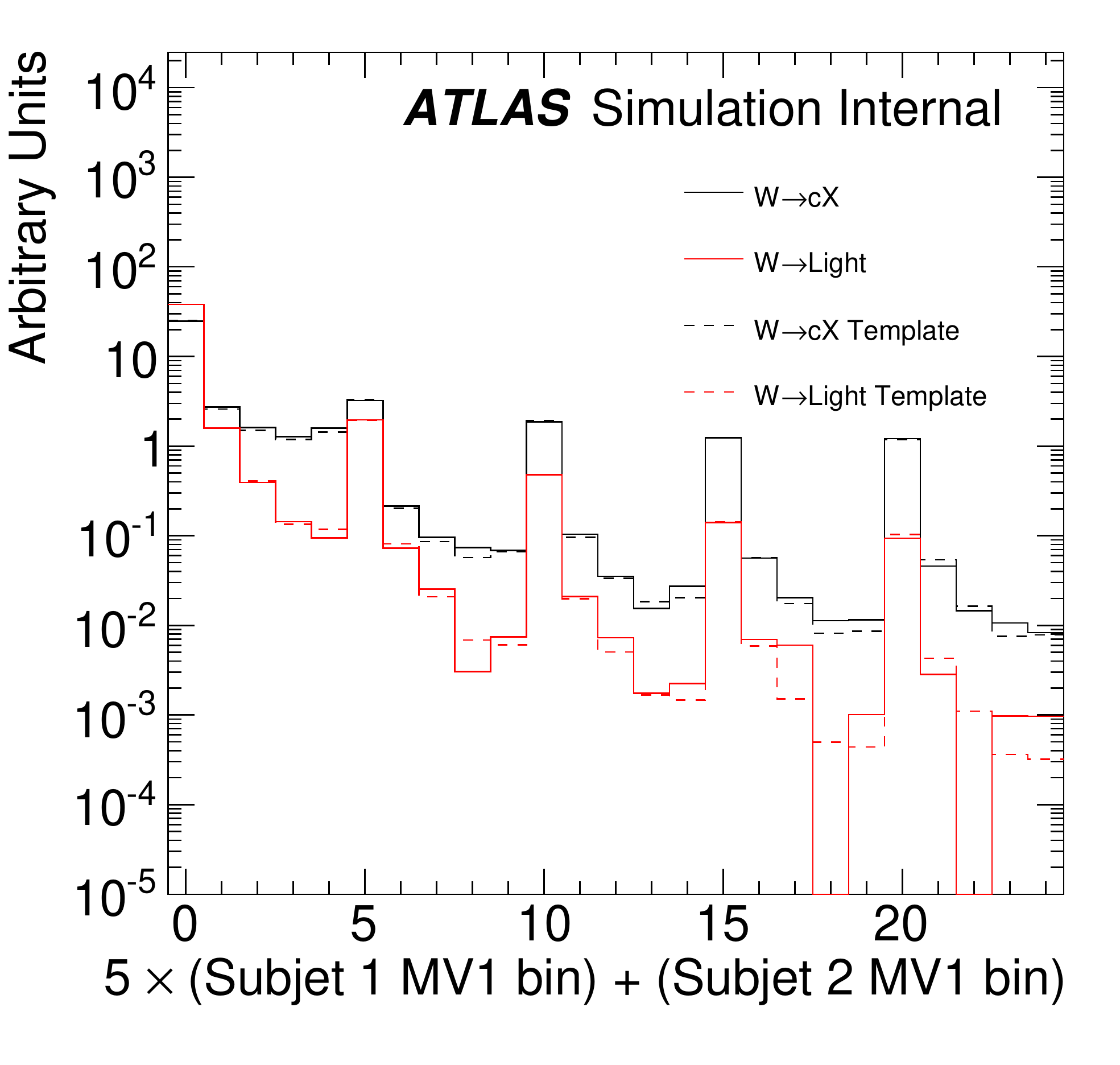}
\caption{Left: The relationship between the leading and sub-leading MV1 distributions. The distributions are normalized per bin of the leading small-radius jet (horizontal axis). The two MV1 values are independent if the distribution in each bin along the vertical axis does not change as a function of the horizontal axis (which is true within the MC statistical uncertainties).  Right: A validation of the templates for the combined binned MV1 $B$ (template is dashed).}
\label{fig:btaggingcorr}
\end{center}
\end{figure}

\begin{figure}[h!]
\begin{center}
 \centering
\includegraphics[width=0.45\textwidth]{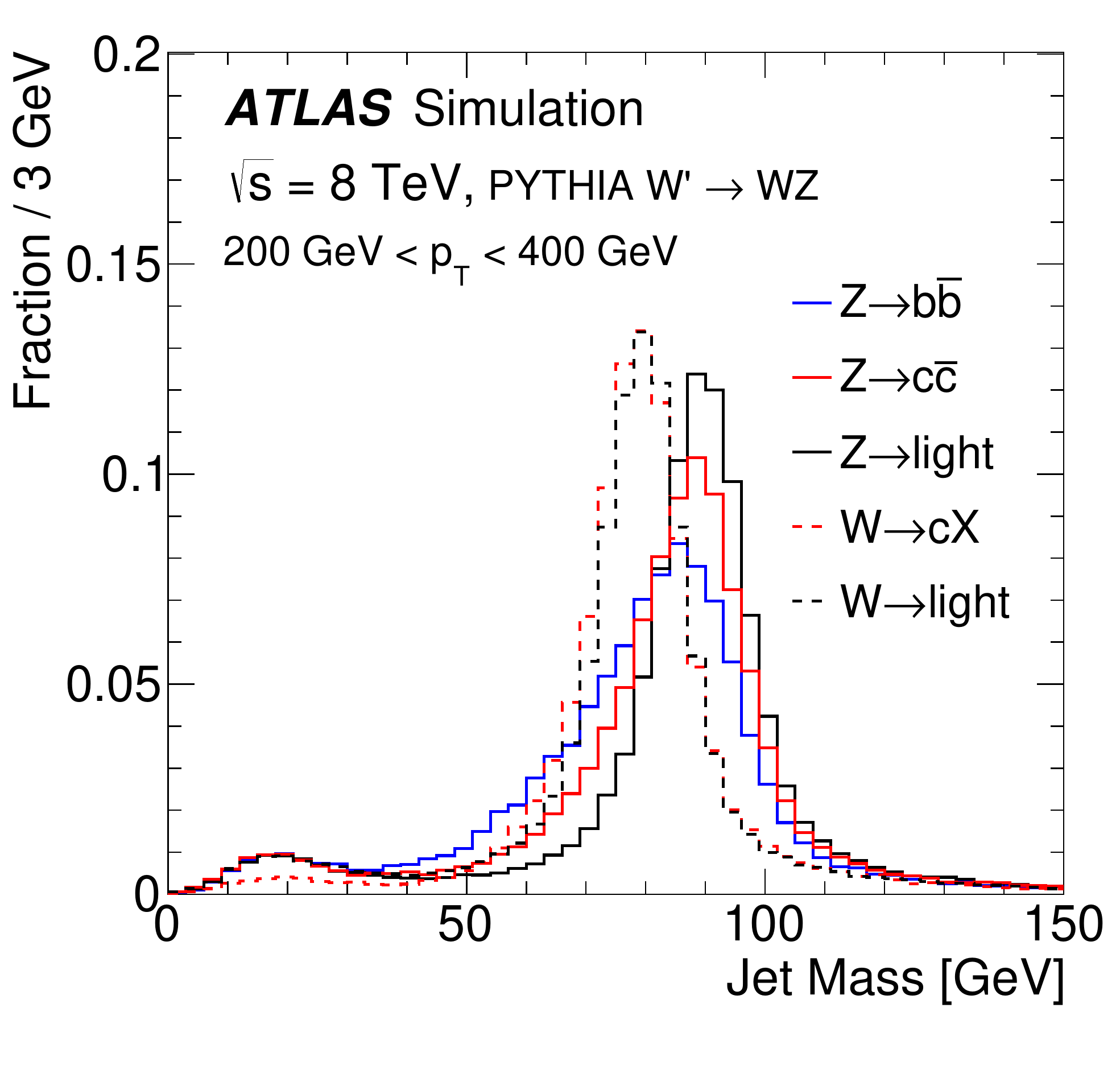}
\includegraphics[width=0.45\textwidth]{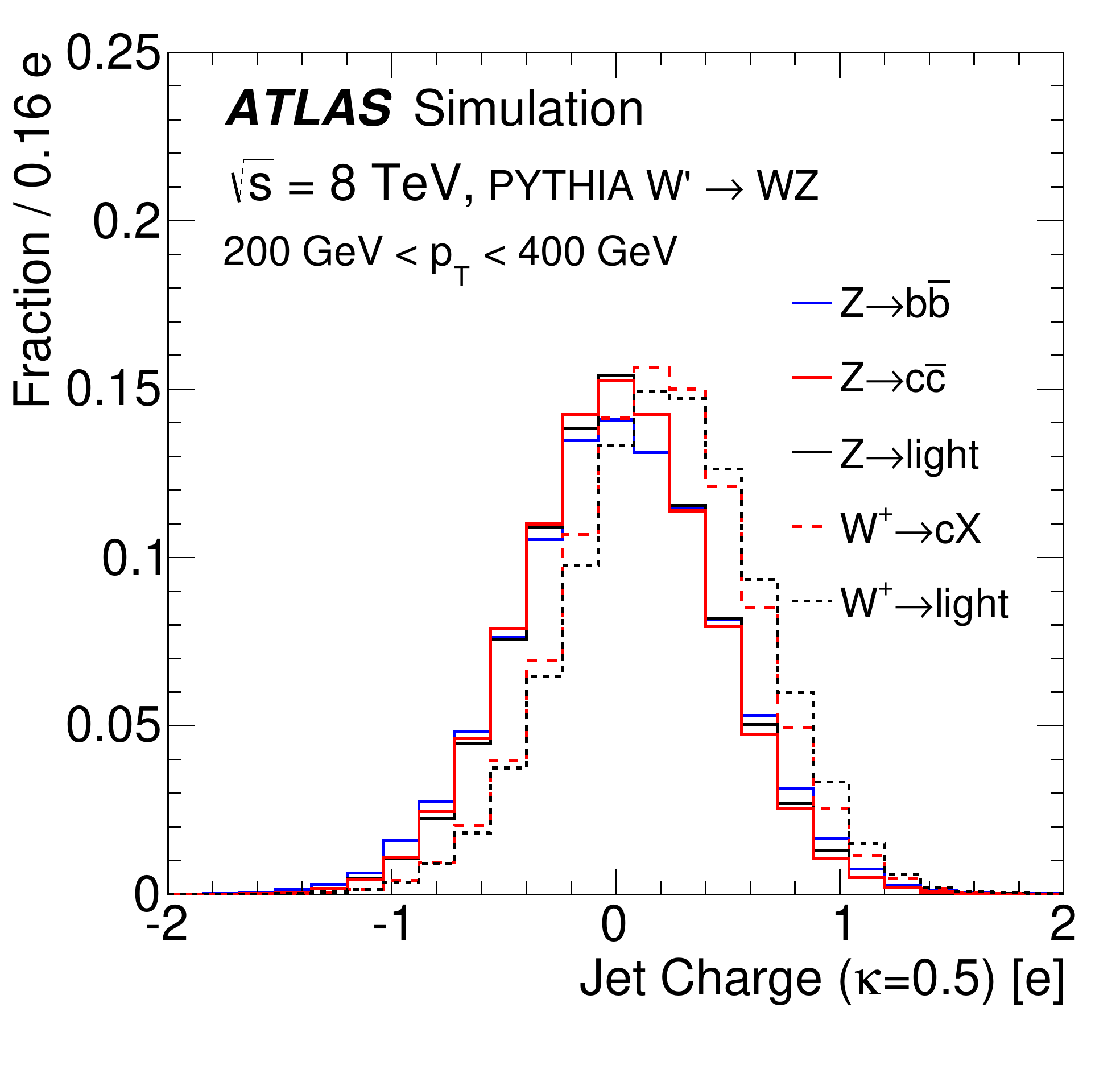}
\caption{(a) The jet mass $p(M|\mathcal{F},V)$ and (b) jet charge $p(Q|\mathcal{F},V)$ templates conditioned on the flavor $\mathcal{F}$ of the boson $V$ decay for jets with $200$ GeV $<p_\text{T}<400$ GeV.   The solid lines are for $Z$ boson decays and the dashed lines are for $W$ boson decays.}
\label{fig:templates}
\end{center}
\end{figure}

\begin{figure}[h!]
\begin{center}
\includegraphics[width=0.95\textwidth]{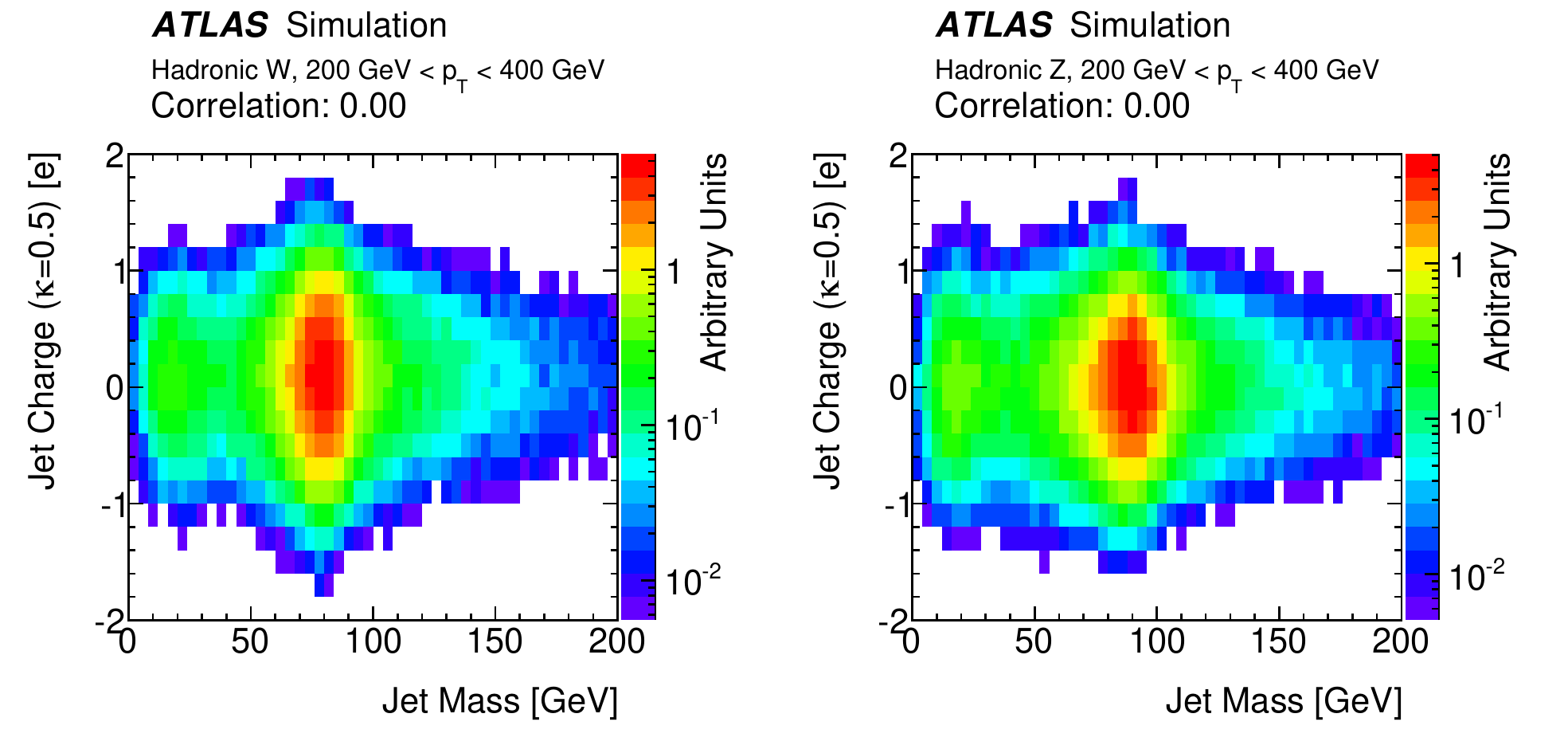}\\
\caption{The joint distribution of the jet mass and jet charge for $W$ boson jets (left) and $Z$ boson jets (right).  The peak of the distribution along the jet mass axis is shifted toward higher values for the $Z$ due to its higher mass.  The linear correlation is less than $1\%$ in both cases and the two distributions are nearly independent.  }
\label{fig:auxcorrjetchargemass}
\end{center}
\end{figure}

The likelihood function is constructed by taking the ratio of the probability distribution functions $p(M,Q,B|V)$, for $V\in\{W,Z\}$, determined from the templates in Eq.~\ref{eq:like}.  Every bin $i$ of the 3D histogram that approximates $p(M,Q,B|V)$ is assigned a pair of numbers $(i,s_i/b_i)$ where $s_i$ is the overall fraction of the signal ($Z$ or $W$) in bin $i$ and $b_i$ is the fraction of the overall background (the other boson flavor) in bin $i$.  Bins are then sorted from largest to smallest $s_i/b_i$, with $f(i)$ defining a map from the old bin index to the new, sorted one. There are then two 1D histograms: for the signal, bin $j$ has bin content $s_{f^{-1}(j)}$ and for the background, bin $j$ has bin content $b_{f^{-1}(j)}$.   The optimal tagging procedure is then to set a threshold on the new 1D histograms.  The full likelihood ratio of the combined tagger is shown in Fig.~\ref{fig:nomlike} where the thresholds required for $90\%$, $50\%$, and $10\%$ $Z$-boson tagging efficiency are marked with shaded regions.

Curves displaying the tagging performance for all possible subsets of $\{M,Q,B\}$ are shown in Fig.~\ref{fig:nomROCs}.  There are $30$ possible values for $B$, which are therefore represented by discrete points.  The jet mass is the best performing single variable for medium to high $Z$-boson efficiencies, with visible improvement for $M$+$B$ and $M$+$Q$.  There is a significant gain from combining all three variables for $Z$-boson tagging efficiency above about $20\%$.  Below $20\%$, the combined tagger is dominated by $B$ where the $Z\rightarrow b\bar{b}$ branching fraction no longer limits $Z$-boson tagging efficiency.   For $Z$-boson efficiencies of about $50\%$, one can achieve $W^+$ rejection factors ($1/\epsilon_{W^+}$) of $3.3$ by using $Q$ or $B$ alone and about 5.0 using mass alone. For $Z$ efficiencies of $\epsilon_Z=90\%$, 50\%, and 10\%, $W^+$ rejection factors of $1.7$, $8.3$, and $1000$, respectively, can be achieved with the combined tagger.  Although most applications of boson-type tagging will target $Z$ bosons as the signal while rejecting $W$ bosons as background, the likelihood constructed in Fig.~\ref{fig:nomlike} can also be used to optimally distinguish $W^{+}$ bosons from $Z$ bosons.  The corresponding performance curves are shown in Fig.~\ref{fig:nomROCs_inverted}.  The locations of the $b$-tagging points are all now shifted to high efficiency with respect to Fig.~\ref{fig:nomROCs} because, for $W^+$ tagging, one wants to operate in the high-efficiency $b$-tagging bins (whereas the opposite is optimal for $Z$ tagging).  At an efficiency of $\epsilon_{W^+}=50\%$, a $Z$-boson rejection factor of $1/\epsilon_Z \approx 6.7$ can be achieved.  

\begin{figure}[h!]
\centering
\includegraphics[width=0.7\textwidth]{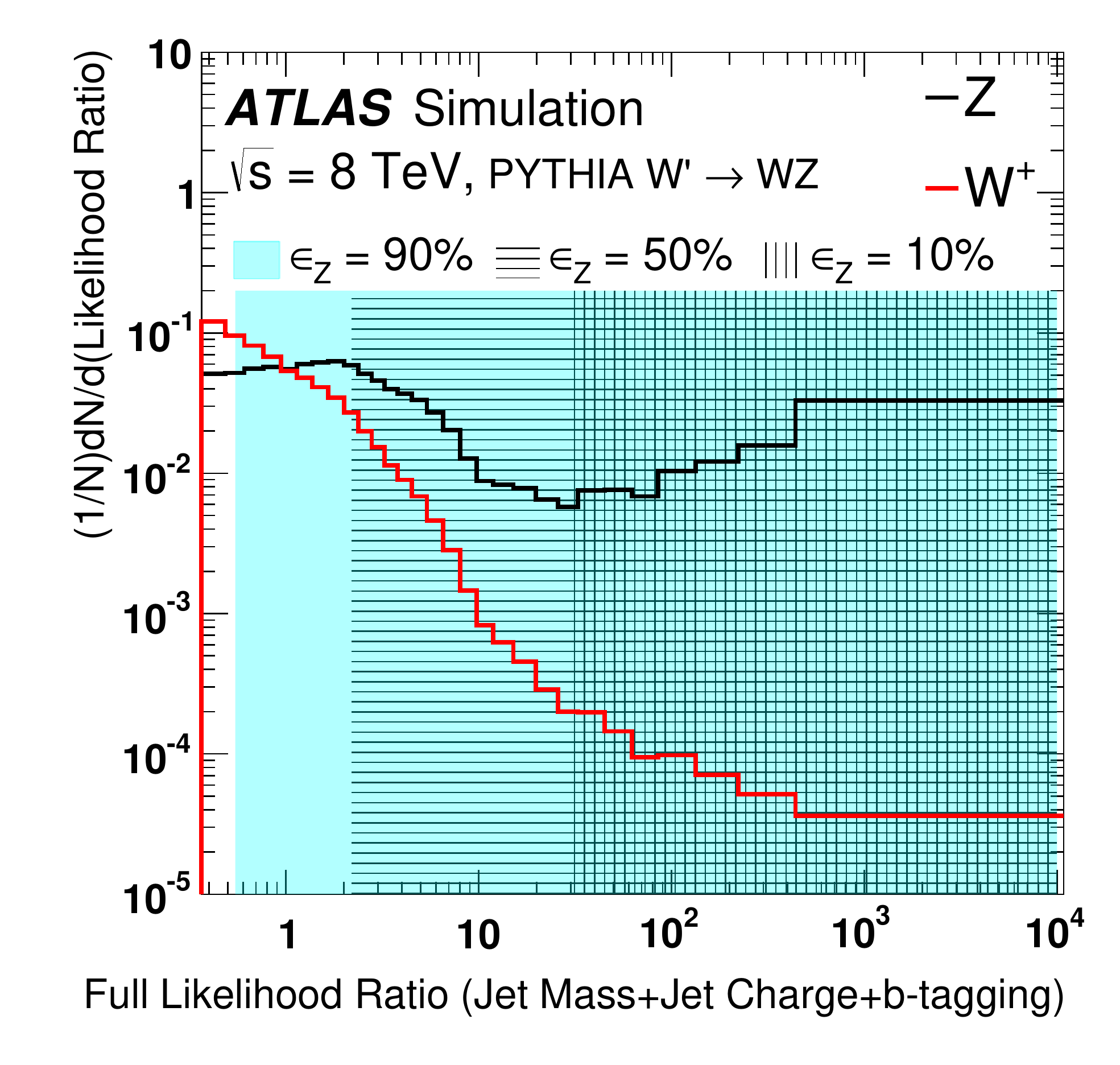}
\caption{The full likelihood ratio for the tagger formed from jet mass, jet charge, and a small-radius jet $b$-tagging discriminant.  The black histogram shows the likelihood ratio for $Z$ bosons and the red histogram is the likelihood ratio for $W^+$ bosons.  The shaded areas show the region of the likelihood ratio corresponding to $90\%$, $50\%,$ and $10\%$ working points of the $Z$-boson tagging efficiency.}
\label{fig:nomlike}
\end{figure}

\begin{figure}[h!]
\centering
\includegraphics[width=0.45\textwidth]{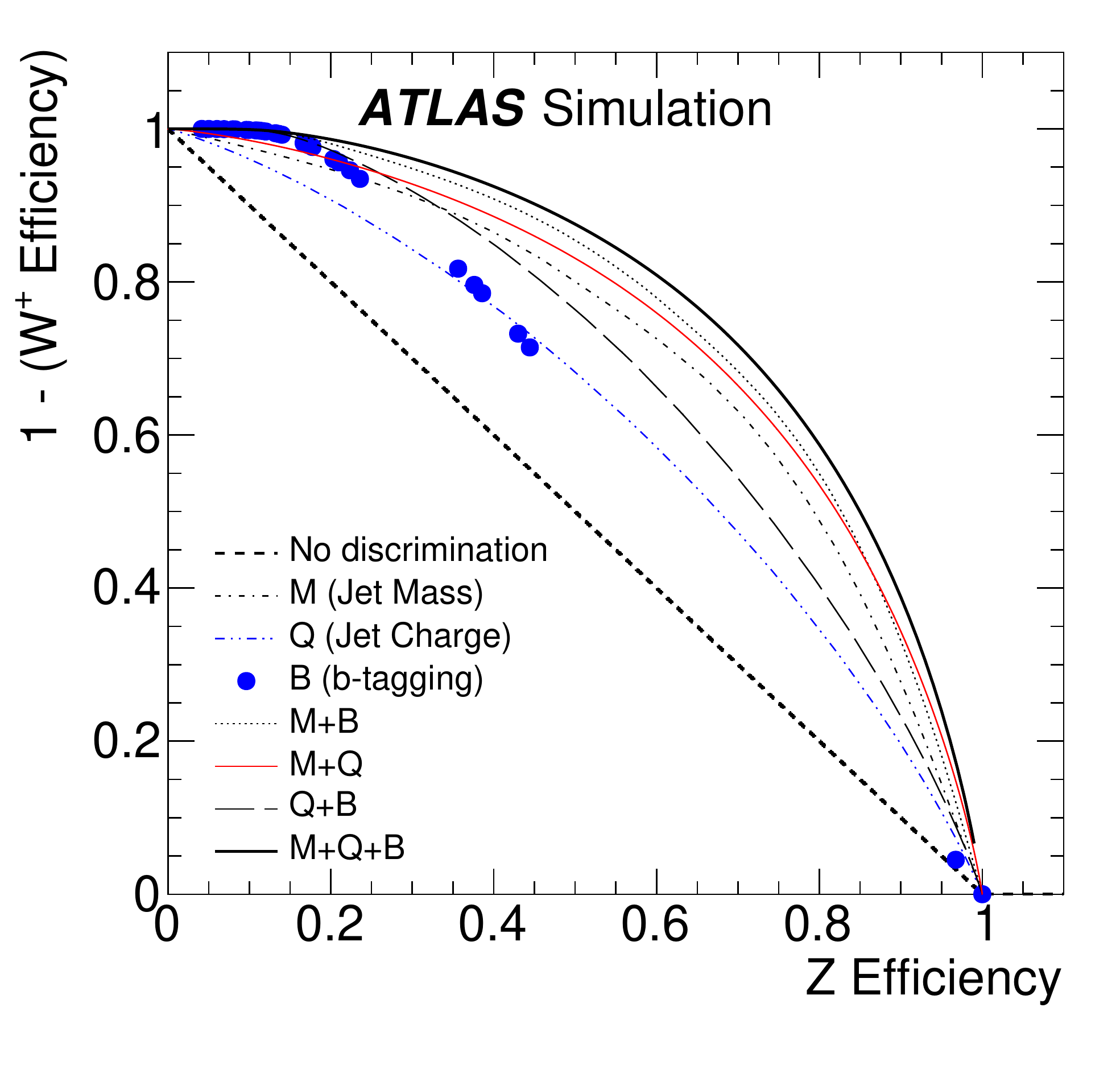}
\includegraphics[width=0.45\textwidth]{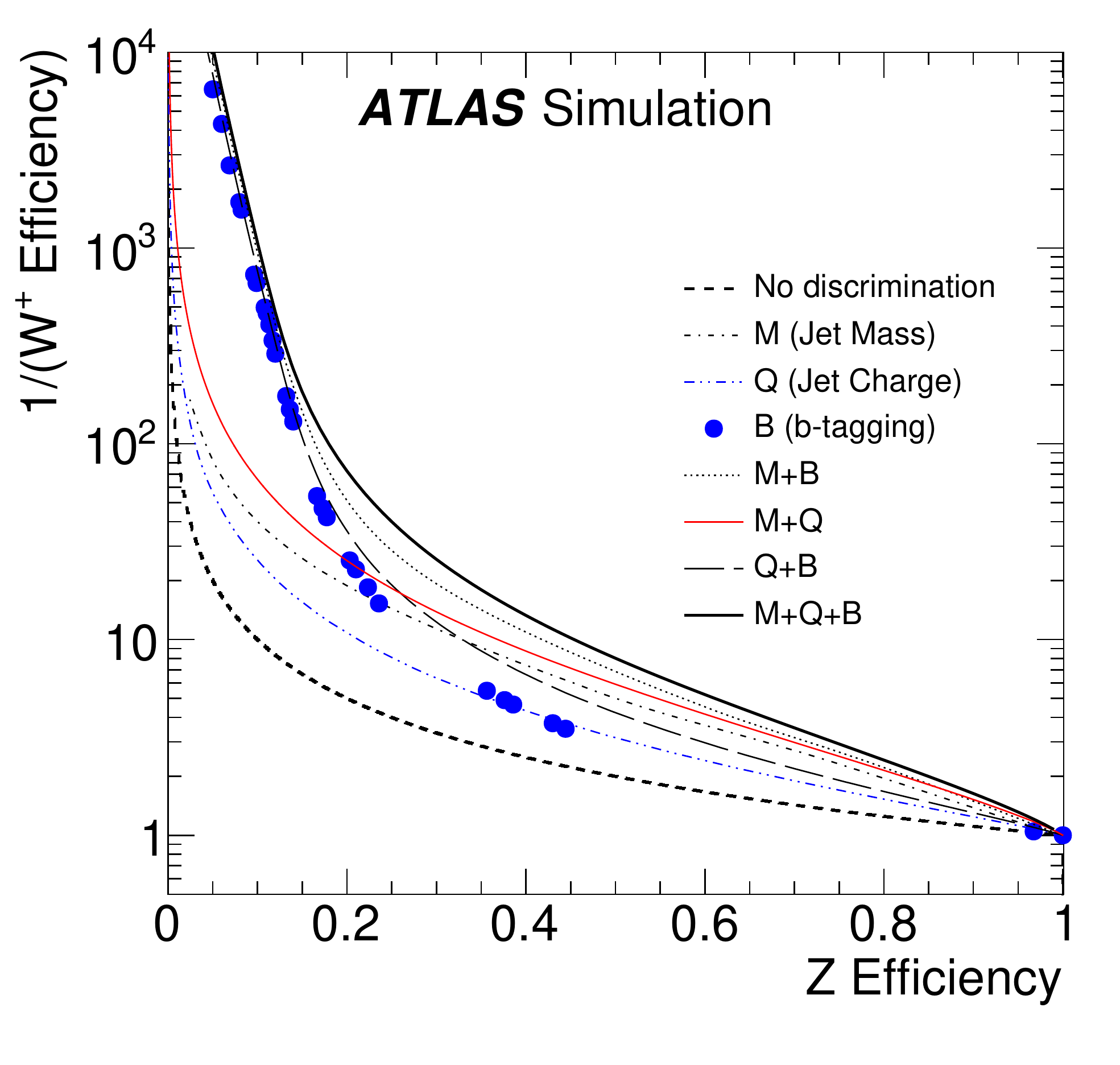}
\caption{The tradeoff between $Z$ efficiency and (a) $1-$ ($W^+$ efficiency) (b) or $1/$($W^+$ efficiency) on (a) a linear scale and (b) a logarithmic scale.  Each curve is constructed by placing thresholds on the likelihood constructed from the inputs indicated in the legend.  Since the $b$-tagging discriminant is binned in efficiency, there are only discrete operating points for the tagger built only from $B$.}
\label{fig:nomROCs}
\end{figure}

\begin{figure}[h!]
\centering
\includegraphics[width=0.45\textwidth]{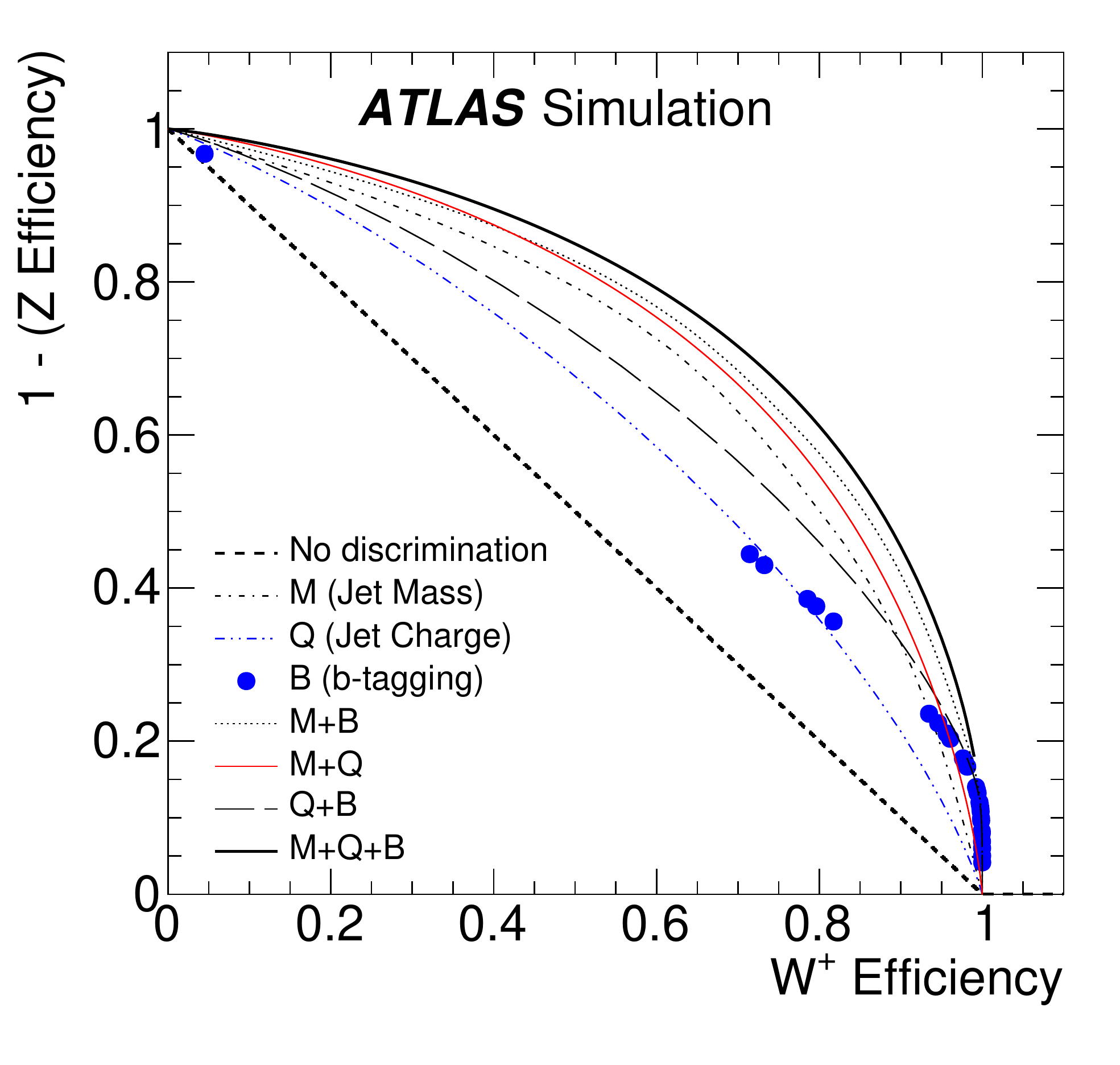}
\includegraphics[width=0.45\textwidth]{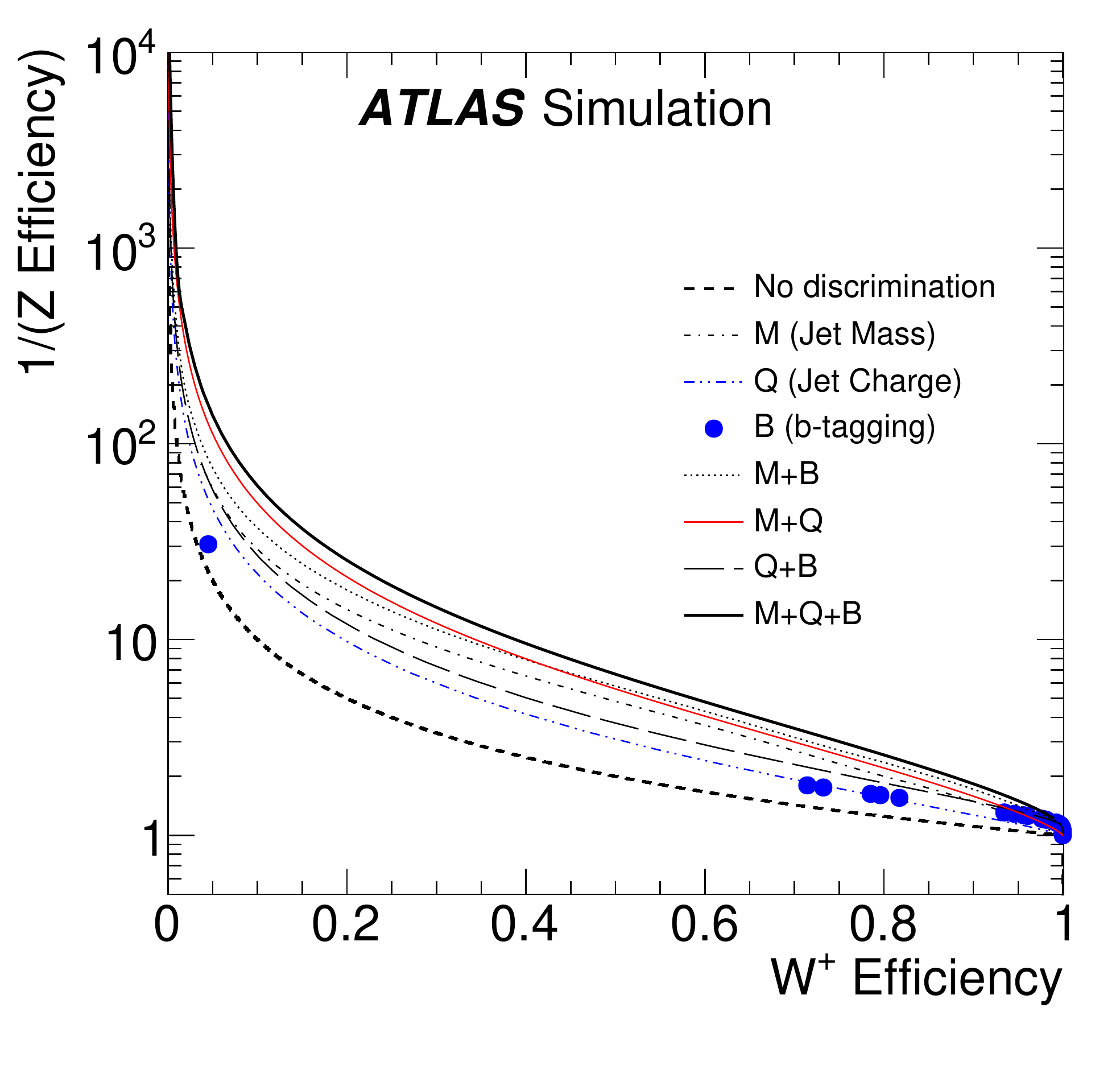}
\caption{The tradeoff between $W^+$ efficiency and (a) $1-$ ($Z$ efficiency) or (b) $1/$($Z$ efficiency) on (a) a linear scale and (b) a logarithmic scale.  Each curve is constructed by placing thresholds on the likelihood constructed from the inputs indicated in the legend.  Since the $b$-tagging discriminant is binned in efficiency, there are only discrete operating points for the tagger built only from $B$.}
\label{fig:nomROCs_inverted}
\end{figure}

\clearpage

\subsection{Systematic uncertainties}
\label{sec:systs}

The performance curves in Figures~\ref{fig:nomROCs} and~\ref{fig:nomROCs_inverted} are based on the nominal modelling parameters of the ATLAS simulation.  Additional studies show how the curves change due to the systematic uncertainties on the inputs to the likelihood function.   Sources of experimental uncertainty include the calibrations of the large- and small-radius jet four-momenta, the $b$-tagging (which incorporates e.g. impact parameter modelling), and the modelling of track reconstruction.  

The uncertainty on the scale of the large-radius jet mass calibration is estimated using the double ratio in data and MC simulation of calorimeter jet mass to track jet mass~\cite{Aad:2013gja}.  Tracks associated with a jet are well measured and provide an independent observable correlated with the jet energy.  Uncertainties on the jet-mass resolution can have a non-negligible impact on the performance of the tagger.  The jet-mass resolution uncertainty is determined from the difference in the widths of the boosted $W$ boson jet-mass peak in semileptonic $t\bar{t}$ simulated and measured data events~\cite{Aad:2013gja} and also from varying the simulation according to its systematic uncertainties~\cite{ATLAS:2012am}.  The resolution is about 5~GeV in the Gaussian core of the mass spectrum and its uncertainty is about 20\%.  The impact of the jet-mass scale and resolution uncertainties on the boson-type tagger built using only the jet mass is shown in Fig.~\ref{fig:nomROCs_syst} for two nominal working points of $50\%$ and $90\%$  $Z$-boson tagging efficiency.   Both the likelihood map $f$ from Sec.~\ref{sec:perfm} and the threshold value are fixed.  Inputs to the tagger are shifted by their uncertainties and the 1D histograms described above are re-populated.  The efficiencies for $W$ and $Z$ bosons are recomputed and shown as markers in Fig.~\ref{fig:nomROCs_syst}(a).  Coherent shifts of the jet masses (JMS) for $W$ and $Z$ bosons result in movement along the nominal performance curve corresponding to $\pm10\%$ changes in the efficiency. However, there are also shifts away from the nominal curve because the optimal jet-mass cut is not a simple threshold.  Variation of the jet-mass resolution (JMR) preserves the scale and so the movement is nearly perpendicular to the original performance curve, at the $\lesssim 5\%$ level, because of the increased overlap in the $Z$ and $W$ mass distributions\footnote{Although such shifts retain optimal use of the tagger (highest rejection for a fixed efficiency), they can degrade the quality of e.g. a cross-section measurement.}.  Shifts along the nominal curve optimally use the input variables (albeit at different efficiencies), while shifts away from the nominal curve are a degradation in the performance.  The impact of the fragmentation is estimated by using input variables from {\tt HERWIG} but with the likelihood map from {\tt PYTHIA}.   {\tt PYTHIA} and {\tt HERWIG} have similar $W/Z$ efficiencies at both the $50\%$ and $90\%$ benchmark points.

The systematic uncertainty on the efficiency of the tracking reconstruction is estimated by removing tracks associated with jets using an $\eta$-dependent probability~\cite{Aad:2010ac}.  The probability in the region $2.3<|\eta|<2.5$ is $7\%$; it is 4\% for $1.9<|\eta|<2.3$, 3\% for $1.3<|\eta|<1.9$, and 2\% for $0<|\eta|<1.3$.  These probabilities are known to be conservative in the most central $\eta$ bins.  There is also an uncertainty on the modelling of track merging for high-$p_\text{T}$ jets, but the loss is expected to be negligible for jets with $p_\text{T}<400$~GeV.   Differences in the modelling of fragmentation can affect the expected performance for all the input variables, especially for the track-dependent observables.  The impact of various uncertainties on the boson-type tagger built using only the jet charge is shown in Fig.~\ref{fig:nomROCs_syst}(b).  Since $W$ and $Z$ boson decays produce on average many tracks (see Sec.~\ref{sec:data}), removing a small number of them does not have a big impact on the jet-charge tagger as a result of the $p_\text{T}$-weighting in the jet charge sum. 
The efficiency to $b$-tag jets of various flavors ($b$, $c$, and light) is measured in data using $t\bar{t}$ events, jets with identified charm hadrons, and multijet events~\cite{Aad:2015ydr}.  The differences between data and MC simulation are typically a few percent and are applied as independent correction factors on a per-jet basis.  The uncertainties on these scale factor measurements are used as estimates of the systematic uncertainty on the $b$-tagging.  The sources of uncertainty are decomposed into many uncorrelated components ($24$ for $b$-jets, $16$ for $c$-jets, and $48$ for light-flavor jets) and the impact on the rejection is added in quadrature for a fixed value of $\epsilon_\text{signal}$.  The $b$-tagging of matched small-radius jets is also affected by uncertainties on the jet-energy scale and resolution.  These quantities are varied within their uncertainties and if the shifted jet has $p_\text{T}<25$~GeV, its MV1 value is not considered.   The impact of various uncertainties on the boson-type tagger built using only the $b$-tagging discriminant for a 10\% nominal $Z$ efficiency is shown in Fig.~\ref{fig:nomROCs_syst2}.  At this efficiency, the full boson-type tagger is dominated by the $b$-tagging inputs, as seen in Fig.~\ref{fig:nomROCs}.   The scale factor uncertainty for $b$-jets has no impact on the $W$ efficiency (no real $b$-jets), but there is approximately a 10\% uncertainty on the $Z$ efficiency.  The uncertainties on the jet-energy scale for small-radius jets are relevant only because of the 25~GeV $p_\text{T}$ threshold.  Since all of the large-radius jets are required to have $p_\text{T}>200$~GeV, the threshold is relevant only in the rare case that one of the $W$ daughters is nearly anti-parallel in the $W$ rest frame to the direction of the $W$ boost vector.  
The $b$-tagging scale factors are only determined up to $p_\text{T}=300$ GeV and then are extrapolated up to $p_\text{T}\gtrsim 500$ GeV using simulation.  The left plot of Fig.~\ref{fig:nomROCs_systbtagg} shows that the fraction of $b$-tagged jets with $p_\text{T}>300$ GeV is negligible in the large-radius jet $p_\text{T}$ range $200<p_\text{T}<400$ GeV.  In principle, the (mis)modeling could depend on the $\Delta R$ between the $b$-jets as the scale factors are extracted for isolated $b$-jets.  However, the studies in Ref.~\cite{ATLAS-CONF-2016-002} based on $g\rightarrow b\bar{b}$ suggest that the systematic uncertainties are small.

\begin{figure}[h!]
\begin{center}
\includegraphics[width=0.4\textwidth]{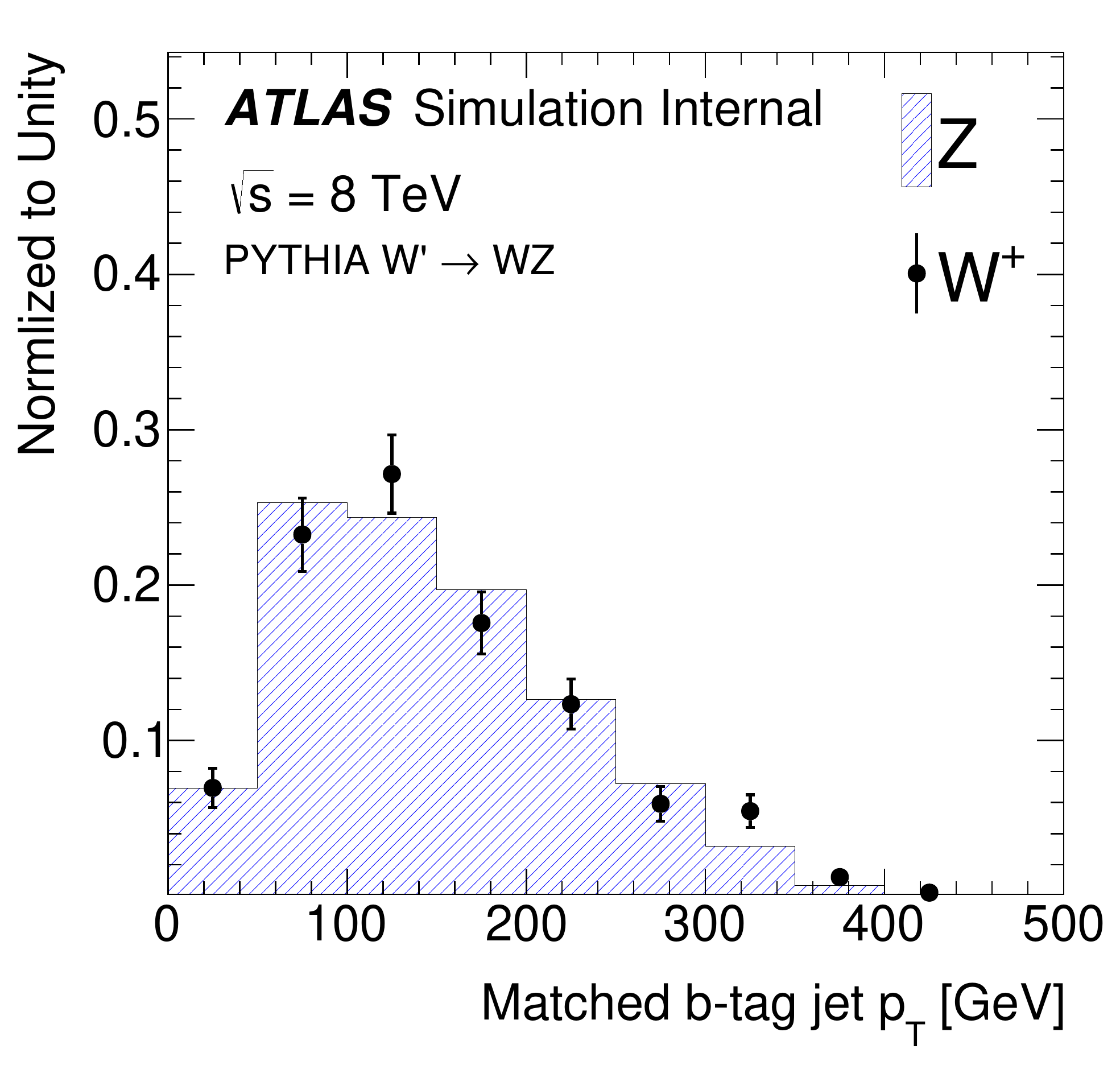}\includegraphics[width=0.4\textwidth]{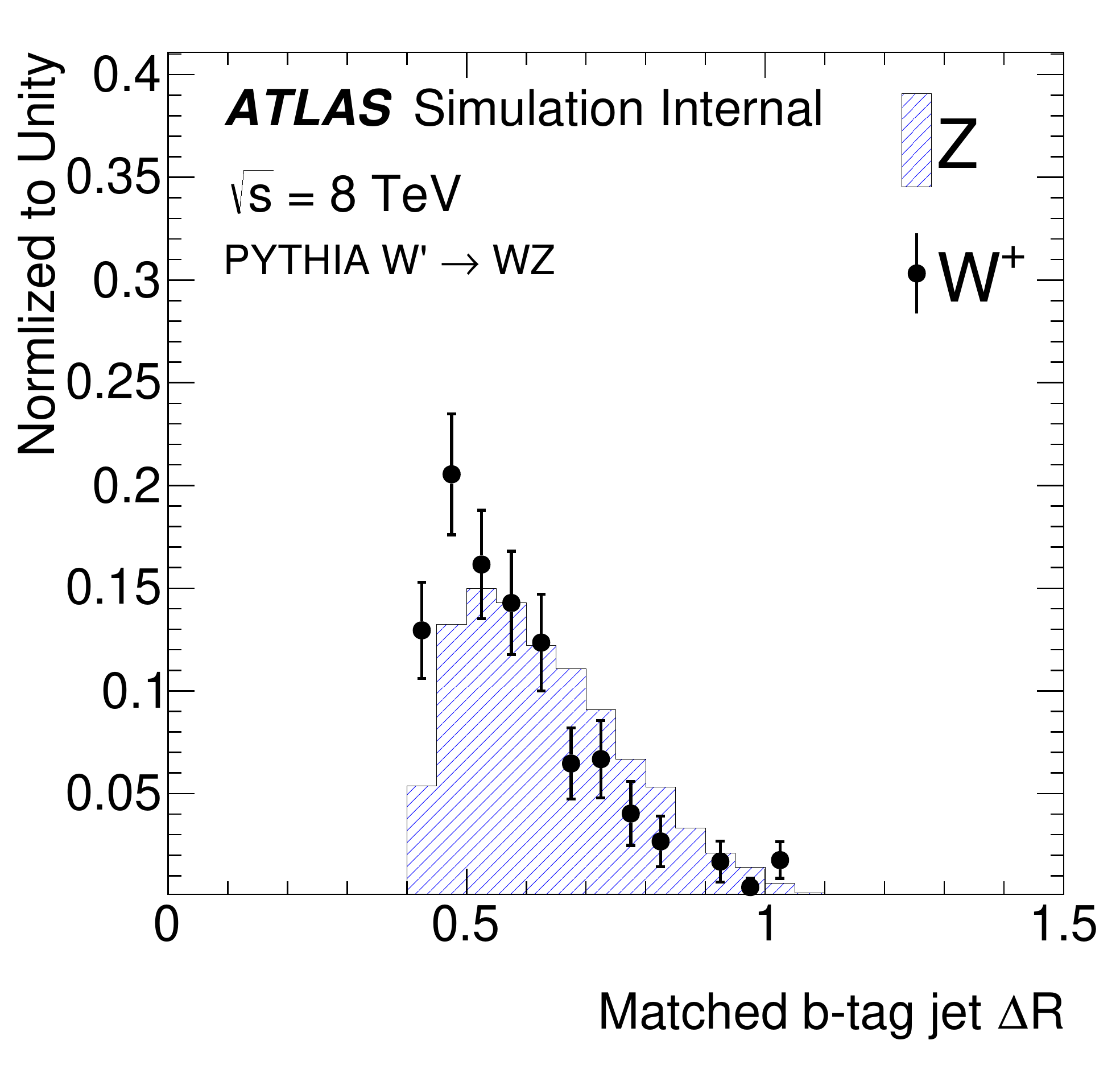}\\
\caption{The $p_\text{T}$ spectrum of the small-radius $b$-jets and the $\Delta R$ between them when there are two.}
\label{fig:nomROCs_systbtagg}
\end{center}
\end{figure}

The impact of the uncertainties on the jet-mass scale and resolution on the boson-type tagger built using all of the inputs (jet mass, jet charge, and $b$-tagging) is shown in Fig.~\ref{fig:nomROCs_sys3Dt}(a).   At very low $Z$-boson tagging efficiency, the tagger is dominated by $b$-tagging, so Fig.~\ref{fig:nomROCs_syst2} is a good representation of the uncertainty on the full tagger's performance.  For higher efficiencies, the tagger is dominated by the jet mass, although the jet charge and $b$-tagging discriminant significantly improve the performance.  The uncertainty on the full tagger's performance at the $50\%$ and $90\%$ $Z$-boson tagging efficiency benchmark points is due mostly to the uncertainty on the jet mass, which is why these uncertainties are shown in Fig.~\ref{fig:nomROCs_sys3Dt}.

\begin{figure}[h!]
\begin{center}
\includegraphics[width=0.45\textwidth]{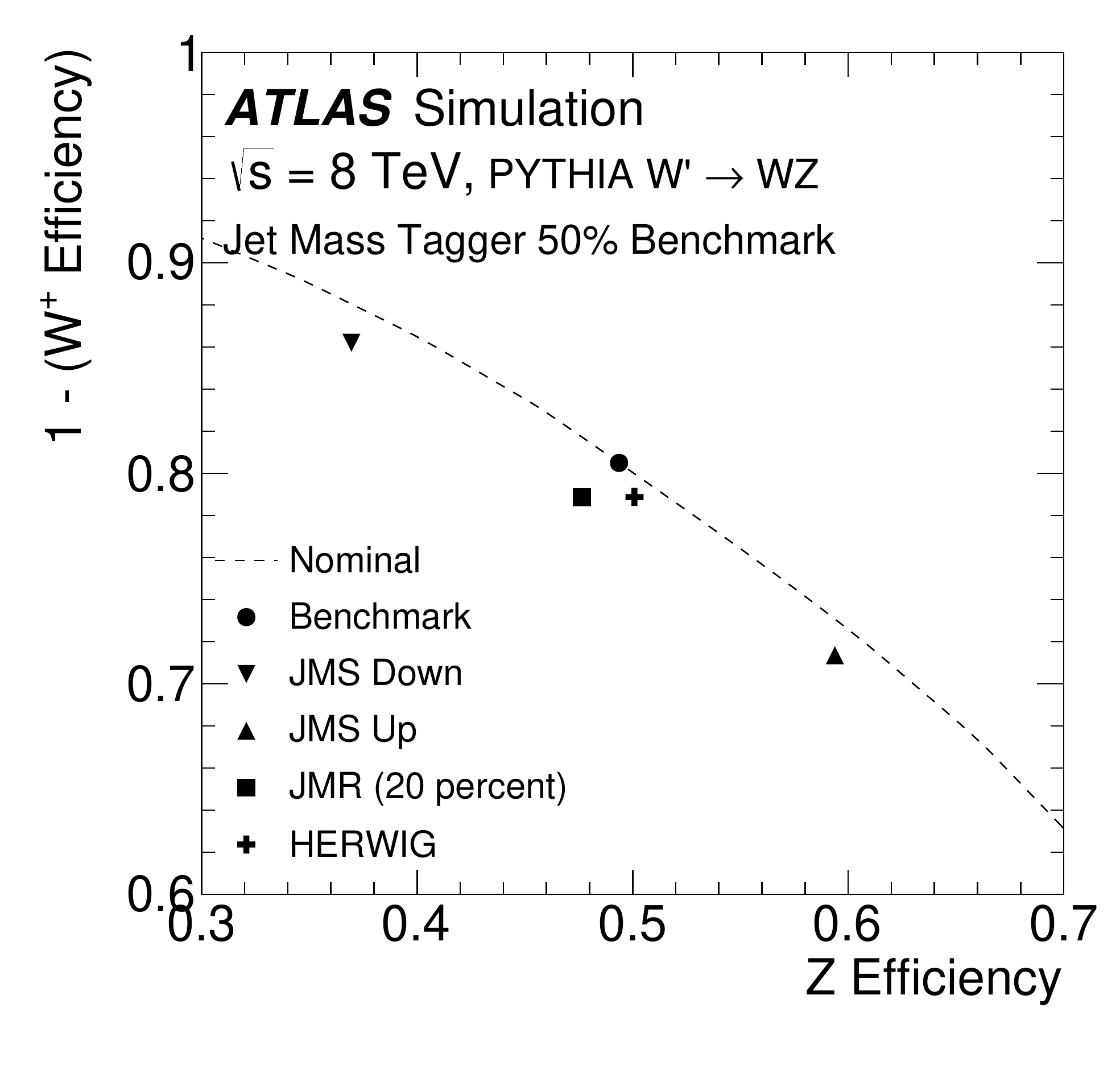}\includegraphics[width=0.45\textwidth]{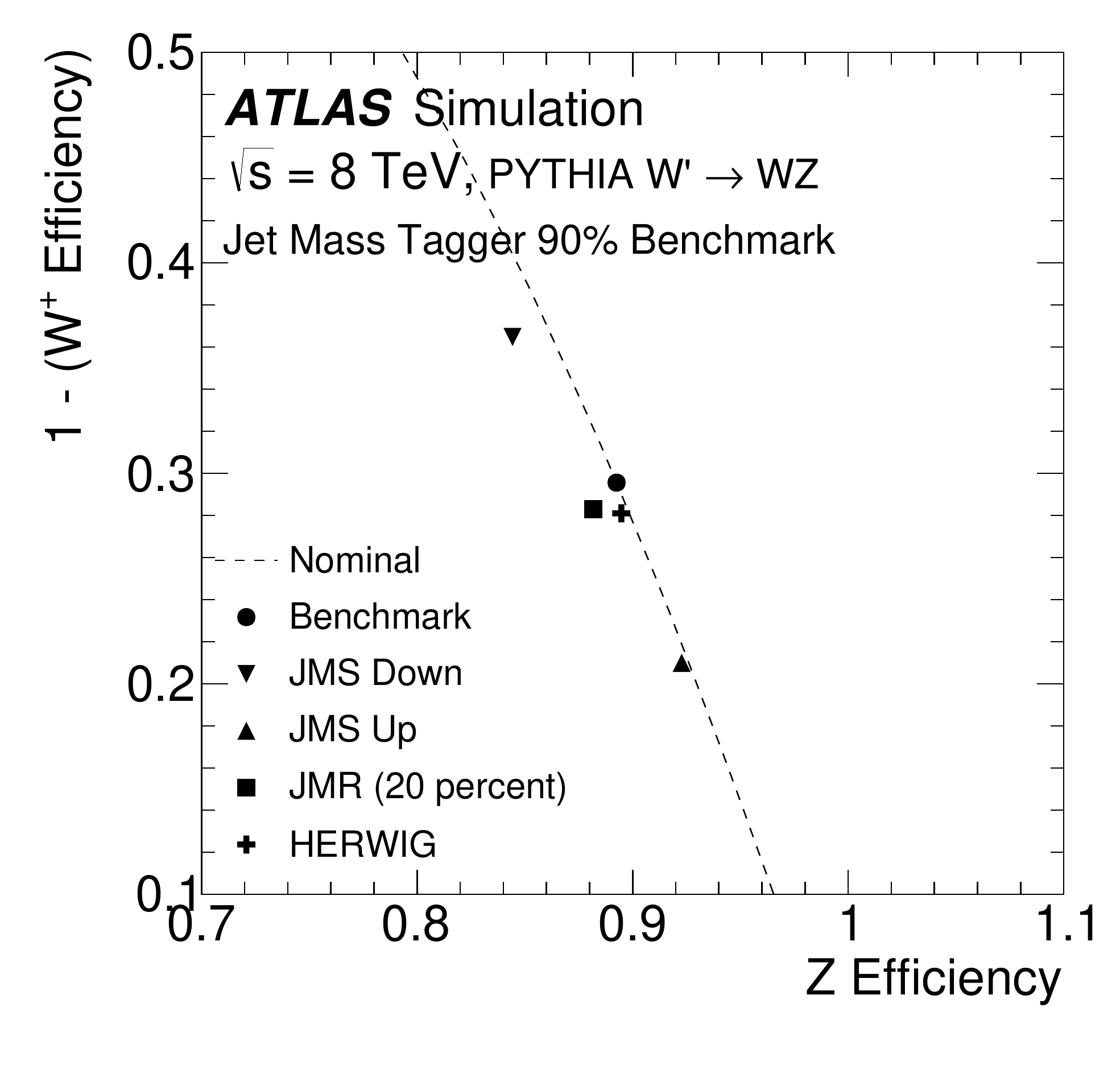}\\
\vspace{-3mm}
\includegraphics[width=0.45\textwidth]{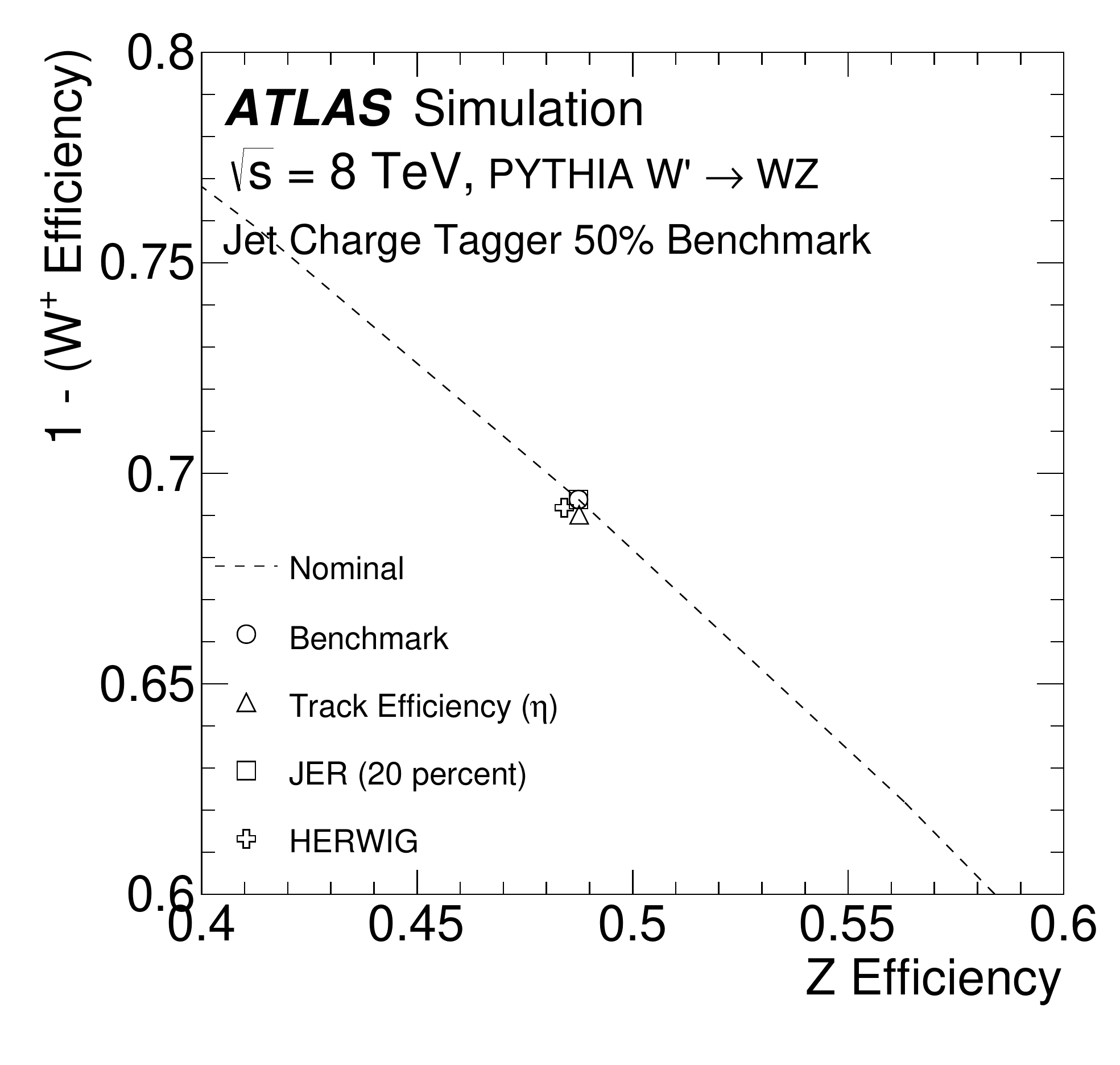}\includegraphics[width=0.45\textwidth]{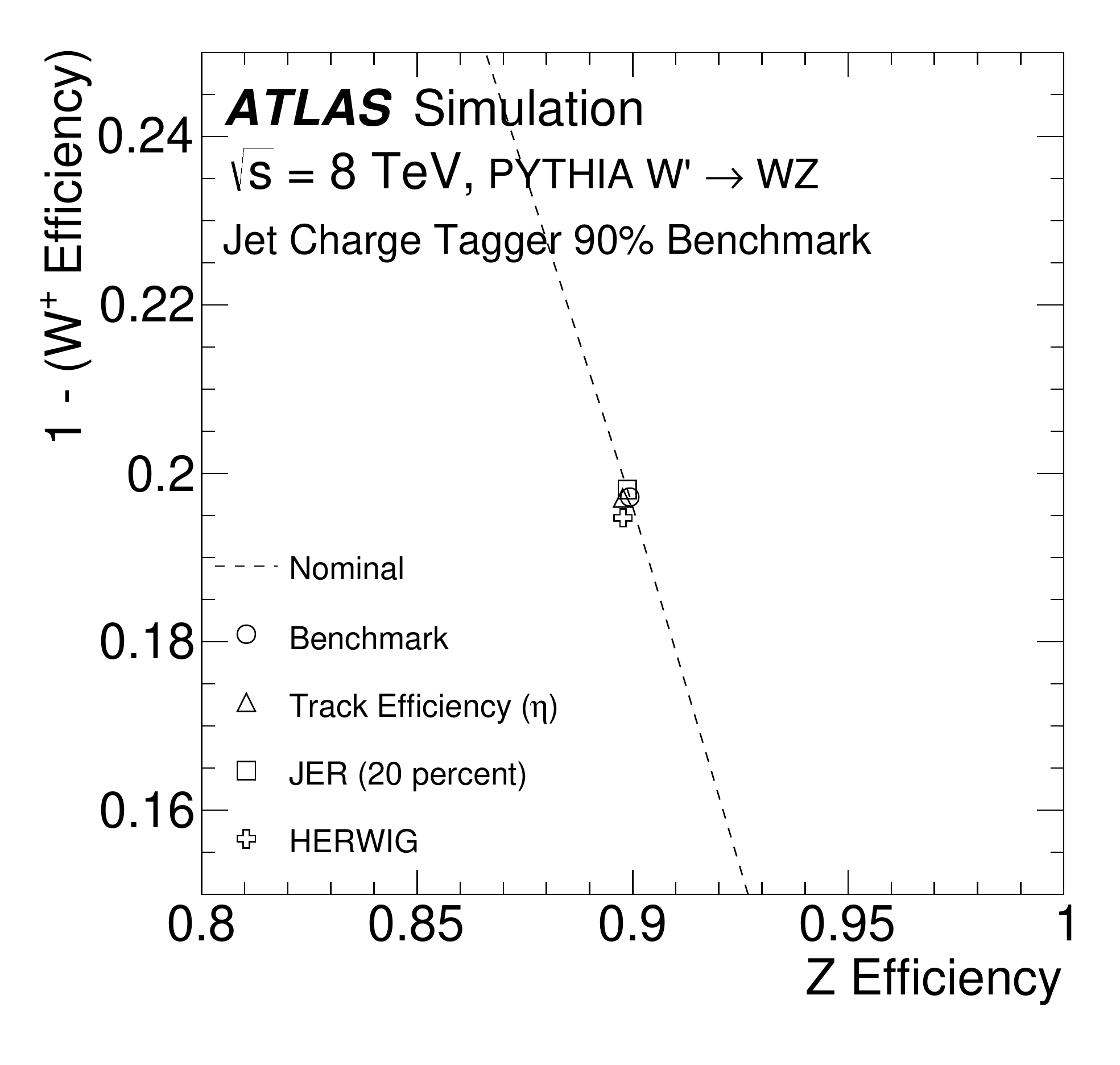}\\
\vspace{-3mm}
\caption{The impact of selected systematic uncertainties on benchmark working points of the boson-type tagger.  (a) a jet-mass-only tagger, for 50\% (left) and 90\% $Z$ efficiency benchmarks.  (b) a jet-charge-only tagger, for 50\% (left) and 90\% $Z$ efficiency benchmarks. The point marked {\tt HERWIG} uses the alternative shower and hadronization model for the simulation, with the likelihood template from {\tt PYTHIA}. See the text for an explanation of the notation in the legend.}
\label{fig:nomROCs_syst}
\end{center}
\end{figure}

\begin{figure}[h!]
\begin{center}
\includegraphics[width=0.65\textwidth]{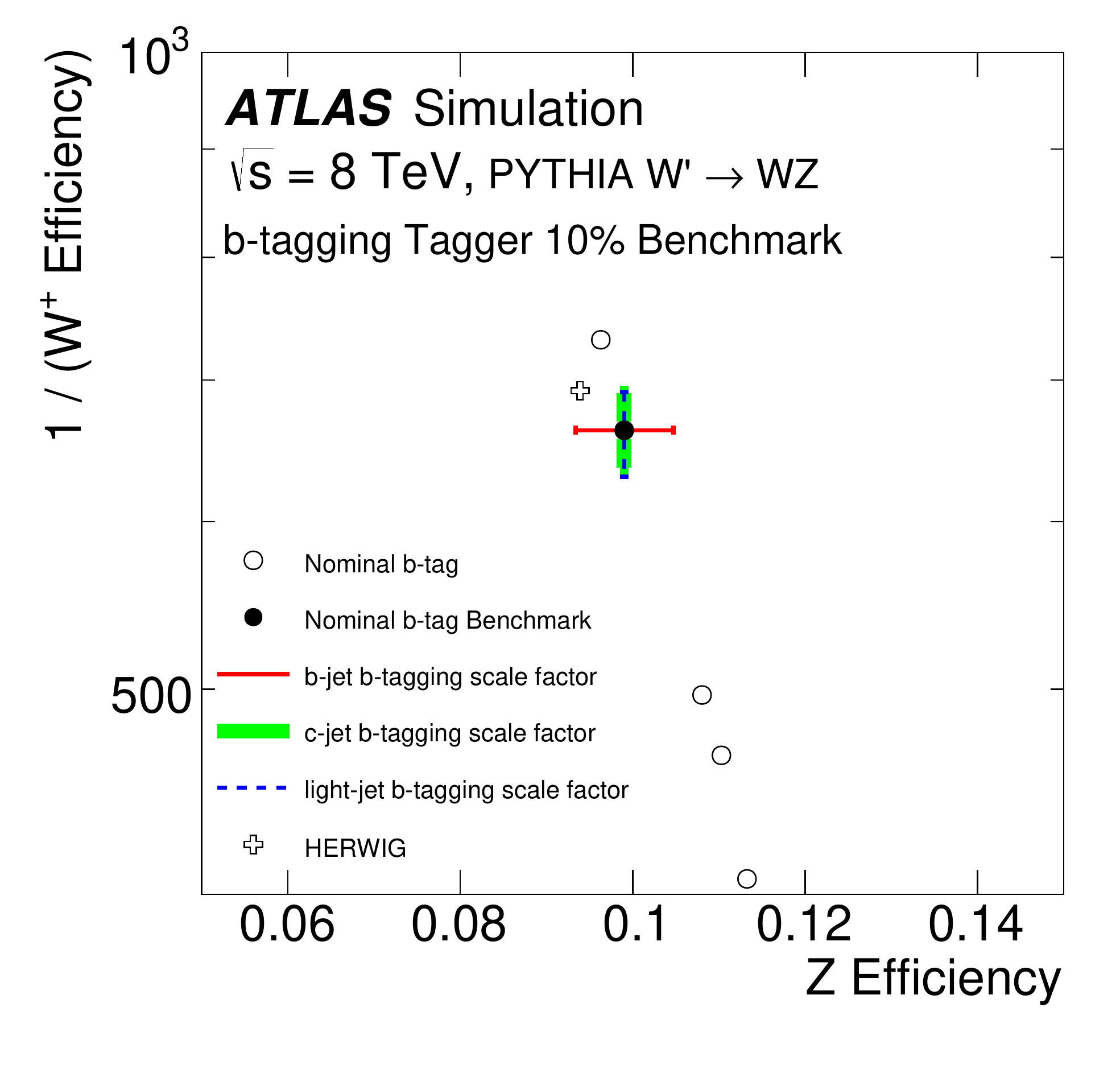}
\caption{The impact of selected systematic uncertainties on benchmark working points of a $b$-tagging-only tagger at a $10\%$ $Z$ efficiency benchmark.  The $b$-tagging discriminant is binned, so there are only discrete operating points.  The point marked {\tt HERWIG} uses the alternative shower and hadronization model for the simulation, with the likelihood template from {\tt PYTHIA}.  The $b$-tagging scale factor uncertainties are determined separately for $b$-, $c$-, and light-quark jets.  Variations are added in quadrature for each `truth' jet flavor.  There is no contribution from the $b$-jet scale factor uncertainties on the $W$ rejection because there are no `truth' $b$-jets.  Conversely, the $c$- and light-jet scale factor uncertainties do not impact the $Z$ bosons because at this low efficiency, all the selected $Z$ bosons decay into $b\bar{b}$.}
\label{fig:nomROCs_syst2}
\end{center}
\end{figure}

\begin{figure}[h!]
\begin{center}
\includegraphics[width=0.45\textwidth]{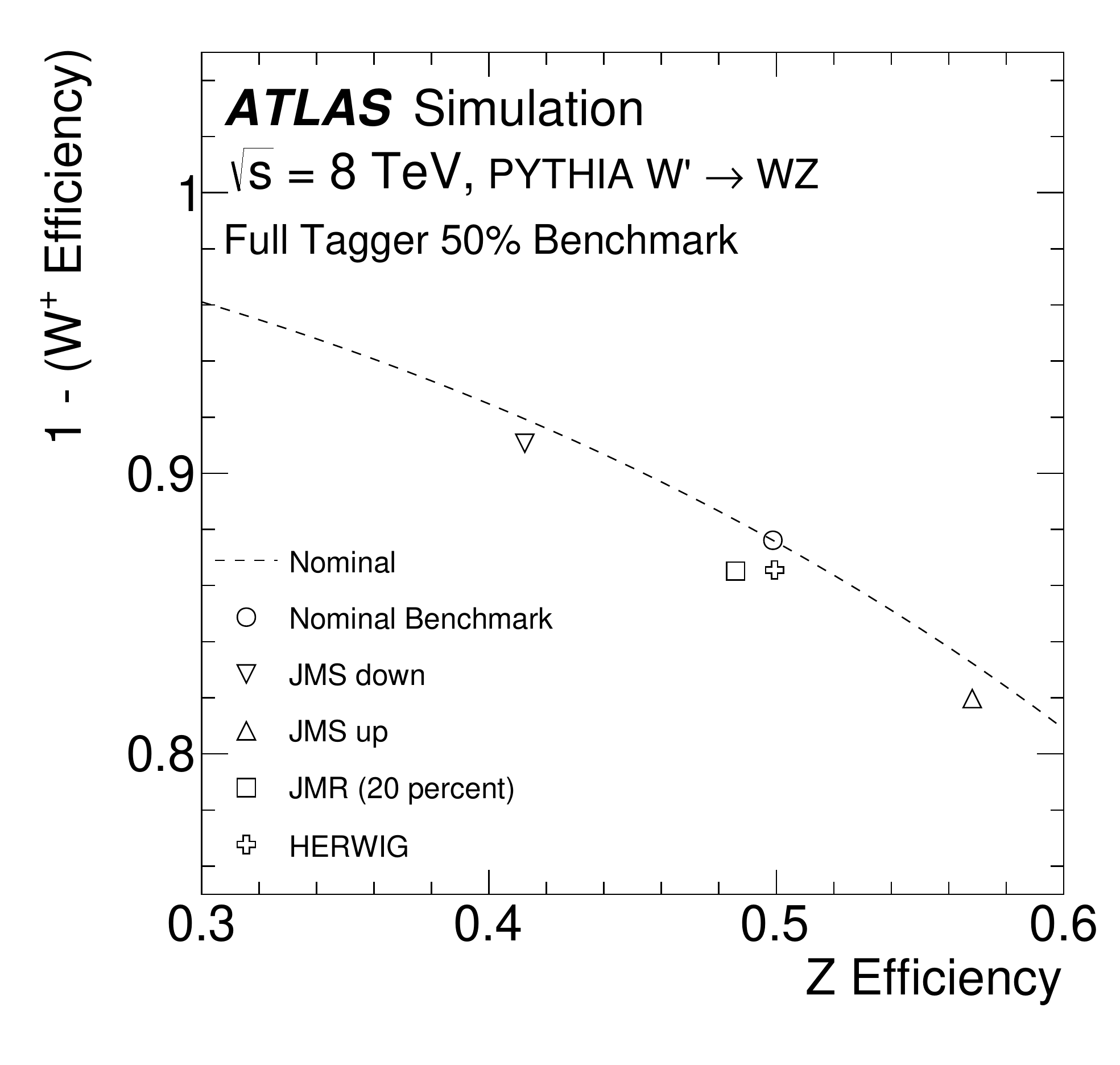}\includegraphics[width=0.45\textwidth]{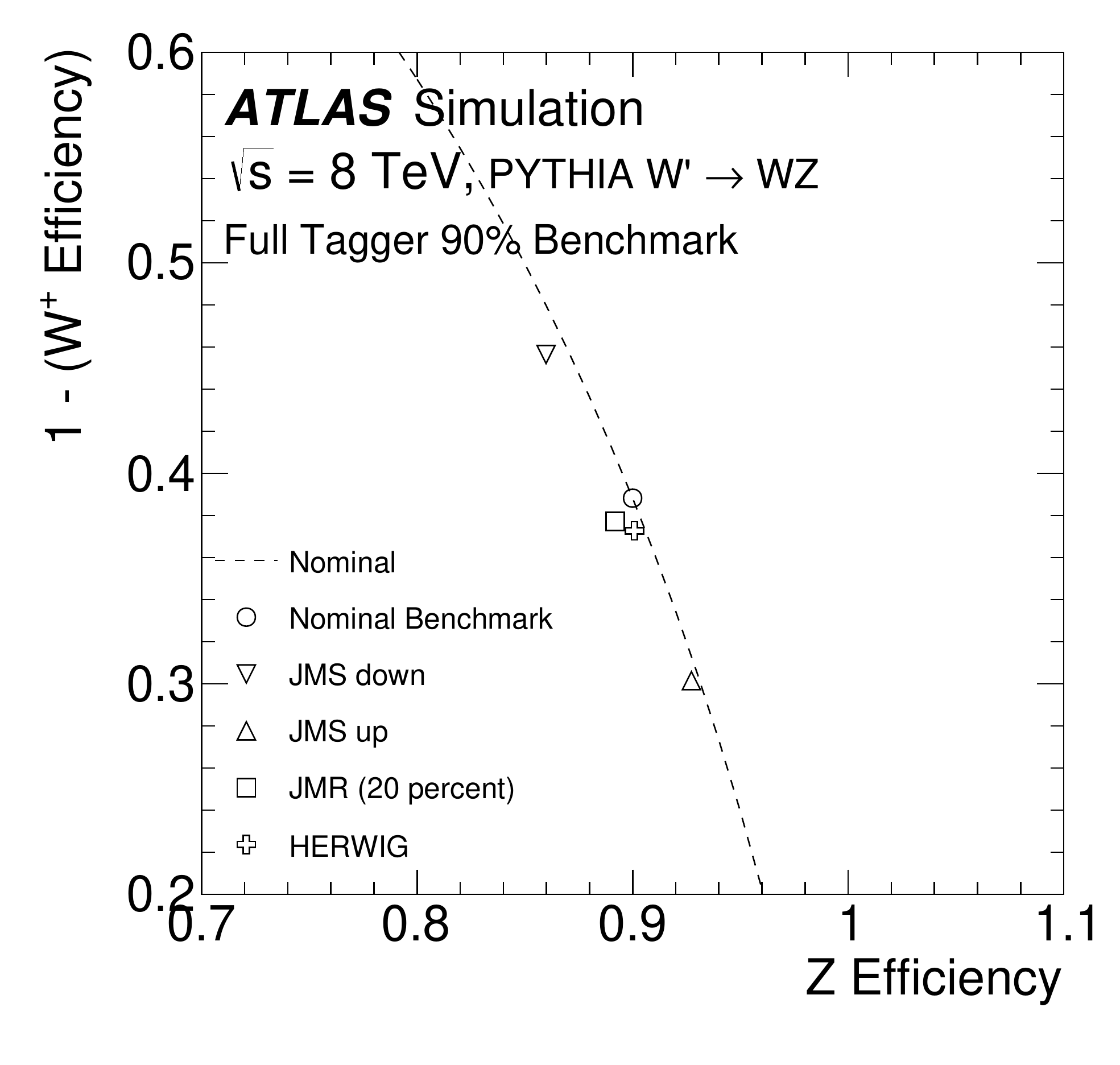}\\
\caption{The impact of uncertainties on the jet-mass scale and resolution for 50\% (a) and 90\% (b) $Z$ efficiency working points of the full boson-type tagger. The point marked {\tt HERWIG} uses the alternative shower and hadronization model for the simulation, with the likelihood template from {\tt PYTHIA}.  }
\label{fig:nomROCs_sys3Dt}
\end{center}
\end{figure}

\clearpage
\newpage

\subsection{Validation of tagging variables using data}
\label{sec:data}

The tagger cannot be fully tested with data because it is not possible to isolate a pure sample of hadronically decaying $Z$ bosons in $pp$ collisions.  However, the modelling of the variables used to design the tagger can be studied with a relatively pure and copious sample of hadronically decaying $W$ bosons in $t\bar{t}$ events which can be tagged by the leptonic decay of the other $W$ boson in the event (semileptonic $t\bar{t}$ events).  Single-lepton triggers are used to reject most of the events from QCD multijet background processes.  Candidate reconstructed $t\bar{t}$ events are chosen by requiring an electron or a muon with $p_\text{T} > 25$ GeV and $|\eta| < 2.5$, as well as a missing transverse momentum $E_\text{T}^\text{miss} > 20$ GeV.  The electrons and muons are required to satisfy a series of quality criteria, including isolation\footnote{Leptons are considered isolated if they are well separated from jets ($\Delta R > 0.4$) and the track/calorimeter energy within a small cone, centred on the lepton direction but excluding the lepton itself, is below a fixed relative value. }.  Events are rejected if there is not exactly one electron or muon.   In addition, the sum of the $E_\text{T}^\text{miss}$ and the transverse mass\footnote{The transverse mass, $m_\text{T}$, is defined as $m_{\text{T}}^2=2p_{\text{T}}^{\text{lep}}E_{\text{T}}^{\text{miss}}(1-\cos(\Delta\phi))$, where $\Delta\phi$ is the azimuthal angle between the lepton and the direction of the missing transverse momentum.} of the $W$ boson, reconstructed from the lepton and $E_\text{T}^\text{miss}$, is required to be greater than 60 GeV.   Events must have at least one $b$-tagged jet (at the 70\% efficiency working point) and have at least one large-radius trimmed jet with $p_\text{T}>200$~GeV and $|\eta|<2$.  Furthermore, there must be a small-radius jet with $p_\text{T}>25$ GeV, and $\Delta R<1.5$ to the selected lepton (targeting the decay chain $t\rightarrow bW(\rightarrow \ell\nu)$).  The other $W$ boson candidate is selected as the leading large-radius trimmed jet with $\Delta R>1.5$ from the small-radius jet that is matched to the lepton.   The $W$+jets and multijet backgrounds are estimated from the data using the charge asymmetry and matrix methods, respectively~\cite{Aad:2014zka}.  The other backgrounds are estimated directly from MC simulation.  Although the resulting event selection is expected to have a high $t\bar{t}$ purity (about 75\%), the events cannot be compared directly to the isolated $W$ bosons from the simulated $W'$ boson decays.  This is because there are several effects that make the typical large-radius jet in semileptonic $t\bar{t}$ events different from isolated $W$ and $Z$ boson jets in typical $W'$ boson events:

\begin{enumerate}
\item The event selection is based on the reconstructed jet $p_\text{T}$, so even if $p_\text{T}^\text{jet}\gtrsim 200$~GeV for an $R=1.0$ jet, the true hadronically decaying $W$ boson in the event may have $p_\text{T}^W<200$~GeV and thus the $W$ boson decay products might not be collimated within $\Delta R <1$.
\item There are more (close-by) jets in semileptonic $t\bar{t}$ events than in $W'$ boson events.  Jets not originating from the $W$ boson can form the leading large-radius jet, or the $b$-jet from the same top-quark as the hadronically decaying $W$ bosons can merge with the $W$ boson decay products to form a large-radius jet.
\end{enumerate}

\begin{figure}[h!]
\centering
\includegraphics[width=0.45\textwidth]{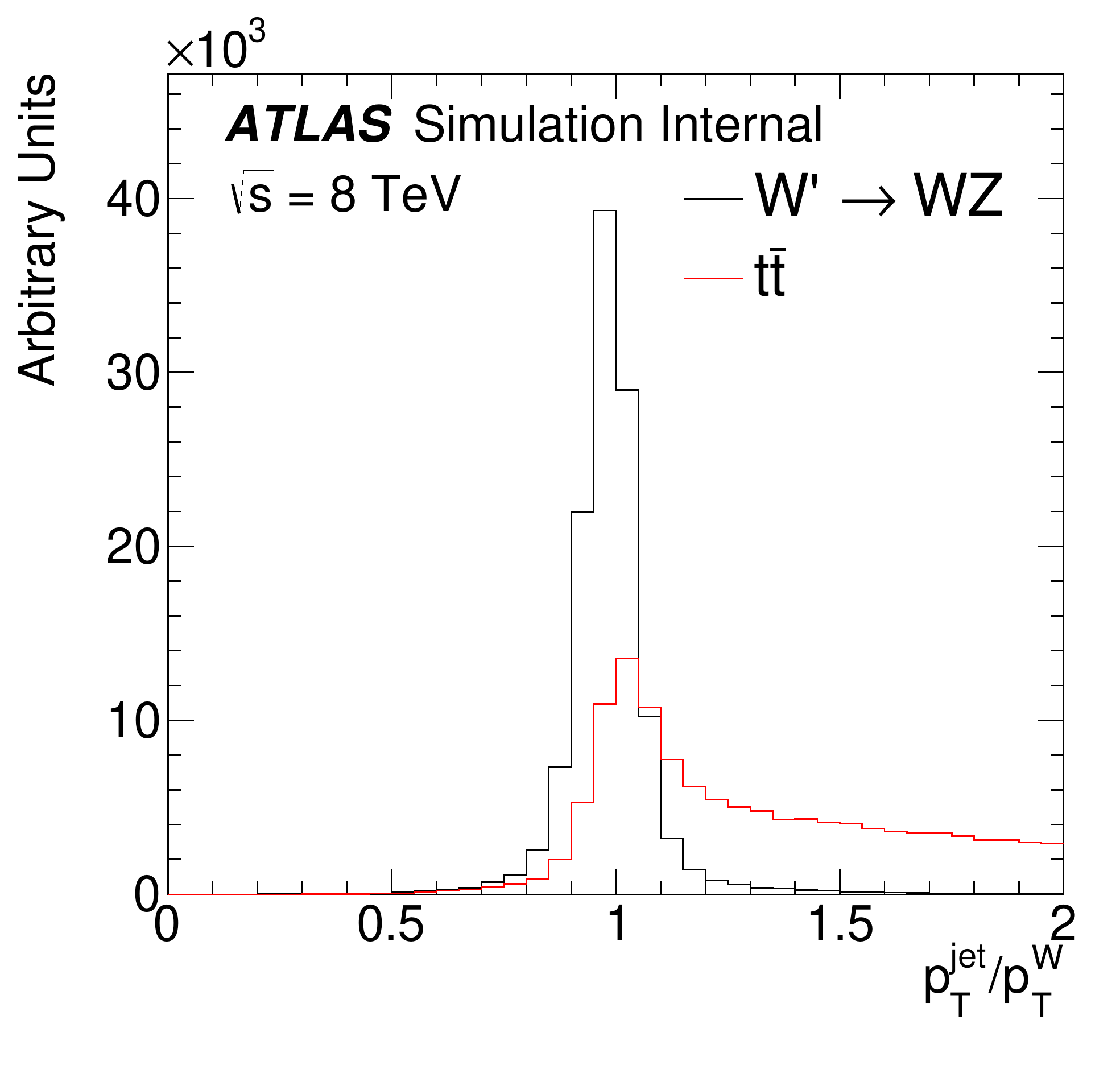}\includegraphics[width=0.45\textwidth]{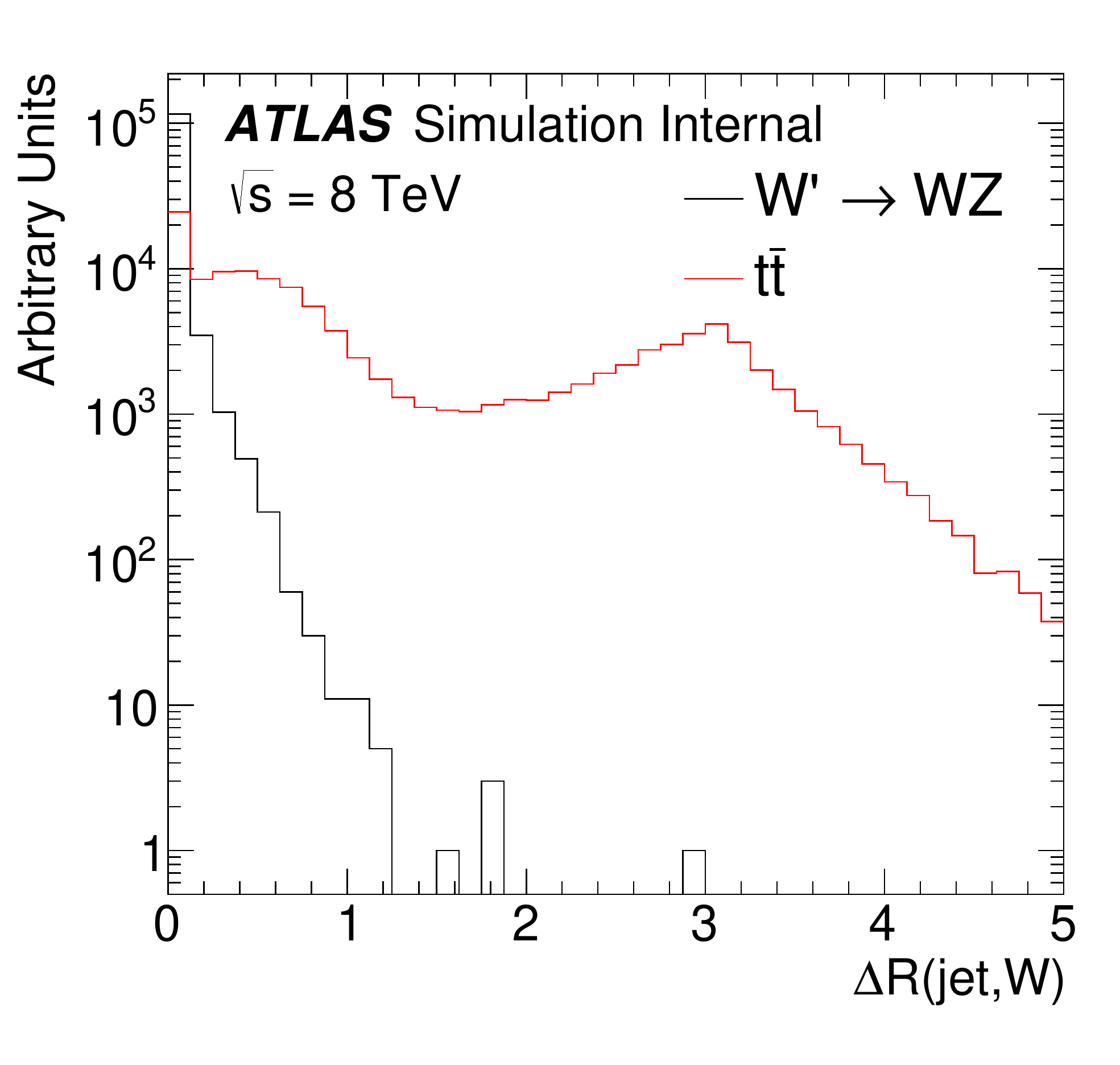}
\caption{The distribution of $p_\text{T}^\text{jet}/p_\text{T}^W$ (left) and $\Delta R(\text{jet},W)$ (right) for $W'$ and $t\bar{t}$ events.  The large tails in $t\bar{t}$ events are due to cases in which more than the $W$ boson hadronic decay products are merged inside the large-radius jet.}
\label{fig:WZlabel1}
\end{figure}

\begin{figure}[h!]
\centering
\includegraphics[width=0.45\textwidth]{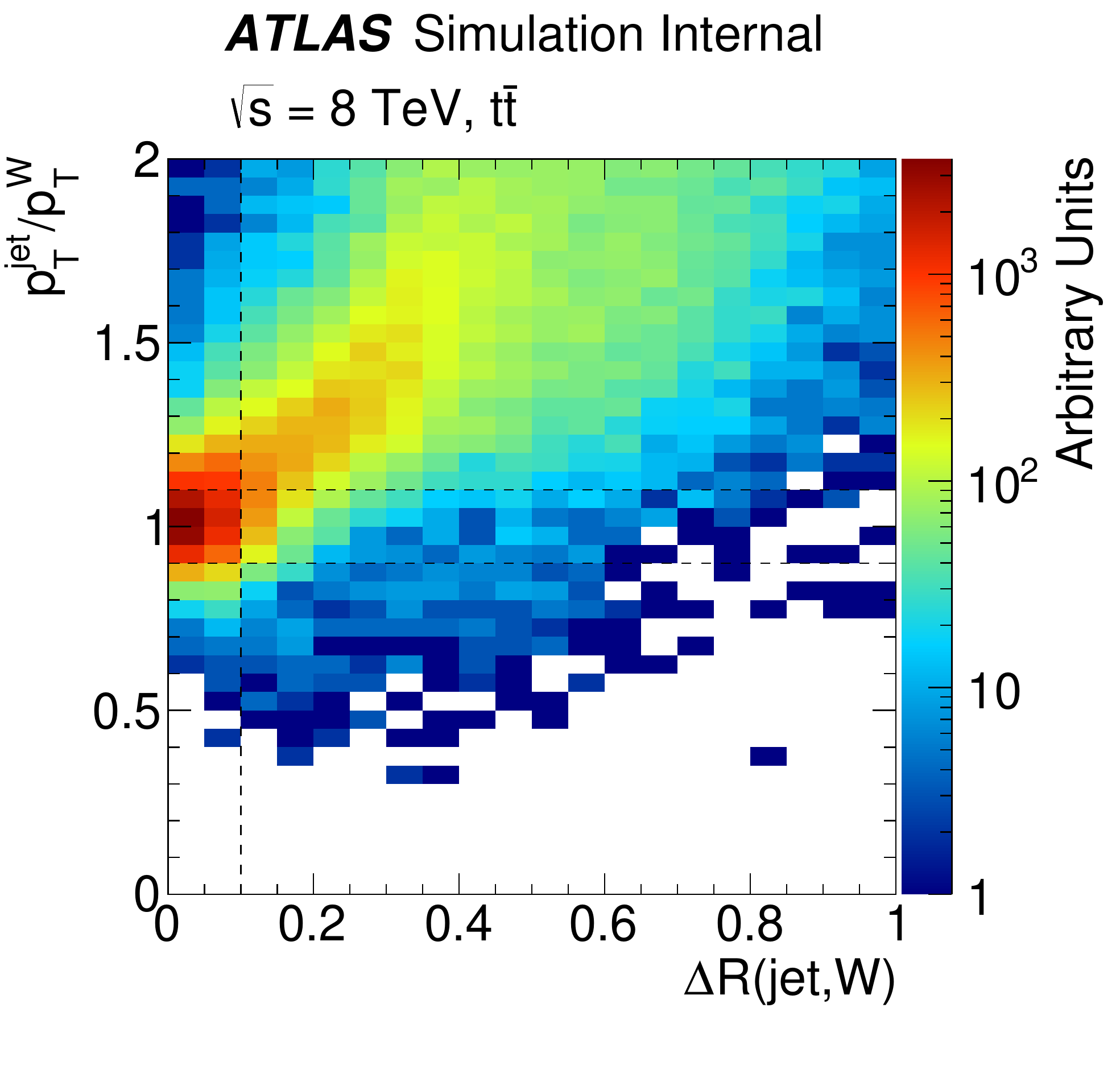}\includegraphics[width=0.45\textwidth]{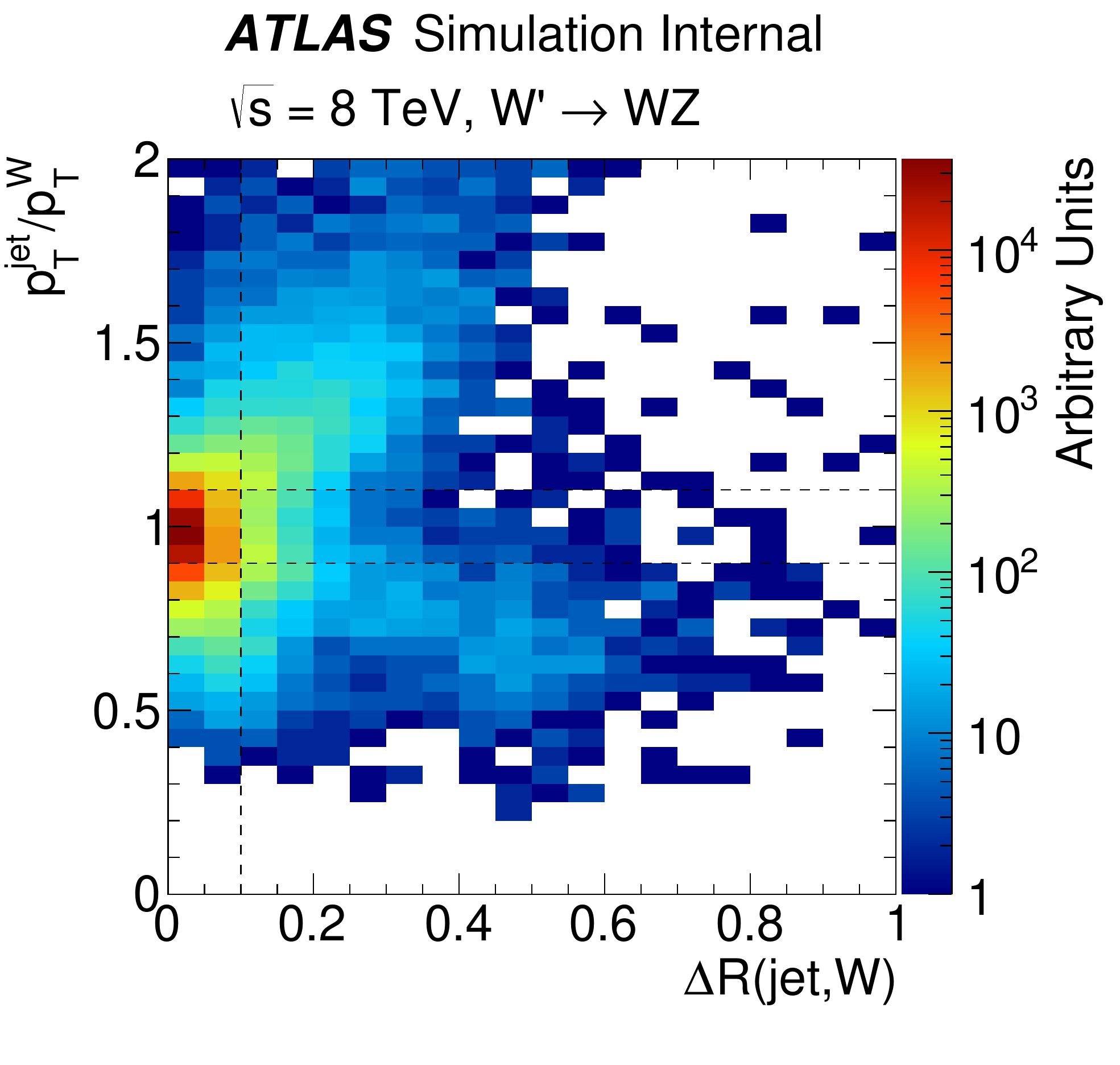}
\caption{The joint distribution of $p_\text{T}^\text{jet}/p_\text{T}^W$ and  $\Delta R(\text{jet},W)$ for $t\bar{t}$ events (left) and $W'$ events (right).  Dashed lines indicate the selection for the {\bf Boosted $W$} category.}
\label{fig:WZlabel2}
\end{figure}

\noindent The variables $p_\text{T}^\text{jet}/p_\text{T}^W$ and $\Delta R(\text{jet},W)$, for the $W$ boson from the MC `truth' record and the selected large-radius jet, are used to classify the various $t\bar{t}$ event sub-topologies.  Events are labelled as having a {\bf Boosted $W$} if $|p_\text{T}^\text{jet}/p_\text{T}^W-1|<0.1$ and $\Delta R(\text{jet},W)<0.1$.  These numbers are based on the distribution for isolated $W$ and $Z$ bosons from the $W'$ simulation.  Figures~\ref{fig:WZlabel1} shows the distributions of $p_\text{T}^\text{jet}/p_\text{T}^W$ and $\Delta R(\text{jet},W)$ in both the $W'$ sample and in the $t\bar{t}$ simulation.  If the $b$-quark from the top-quark decay has an angular distance $\Delta R<1.0$ from the selected large-radius jet, this jet is labelled as {\bf $b$-contaminated}.  All other $t\bar{t}$ events, including events where both $W$ bosons decay into leptons, are labelled as {\bf Other}.  The $p_\text{T}$ spectrum of the jets from the classified events is shown in Fig.~\ref{fig:data}.  In Fig.~\ref{fig:data} and subsequent figures, systematic uncertainties on the simulation include the jet $p_\text{T}$ and jet mass uncertainties described in Sec.~\ref{sec:systs}, but exclude tracking uncertainties, which are sub-dominant.   Events are vetoed if the selected large-radius jet has $p_\text{T}>400$~GeV or if the $\Delta R$ between the selected large-radius jet and a tagged $b$-jet is less than $1.0$.  This suppresses the $b$-contaminated $t\bar{t}$ events.  The effectiveness of the $t\bar{t}$ event classification is most easily seen from the jet mass distribution, shown in Fig.~\ref{fig:datamass}(a).  The mass of the boosted $W$ bosons from $t\bar{t}$ events is peaked around $m_W$, as is a small contribution from the hadronically decaying $W$ bosons in single-top events in the $Wt$ channel.  There is no peak at $m_t$ in the $b$-contaminated spectrum because of the $b$-jet veto, but there is a small non-resonant contribution below the top-quark mass, due to events in which one $W$ daughter is matched with the $b$-jet.  This is akin to the $b$-jet+lepton invariant mass used in other circumstances to measure top-quark properties and naturally has a scale around $150$~GeV~\cite{Chatrchyan:2013boa}. The low-mass peak in $W$+jets and the `other' $t\bar{t}$ events is due to the Sudakov peak from QCD jets, the location of which scales with $R \times p_\text{T}$.  The dependence on $p_\text{T}$ of the $W$-peak position in Fig.~\ref{fig:datamass}(a) is shown in Fig.~\ref{fig:datamass}(b).  Events with the leading jet in a window around the $W$ mass, $50$~GeV $<m^\text{jet}<120$ GeV are selected and the median of the mass distribution is plotted in Fig.~\ref{fig:datamass}(b) as a function of the jet $p_\text{T}$.  The similar trend for the simulation and the data shows that the combination of the reconstructed jet-mass scale and `truth' jet-mass scale is well modelled.  To quantify the spread in the jet mass peak, various inter-quantile ranges are shown as a function of $p_\text{T}$ in Fig.~\ref{fig:datamass}(c).  The inter-quantile range of size $0\%<X<50\%$ is defined as the difference between the $50\%+X\%$ quantile and the $50\%-X\%$ quantile, and is a measure of the spread in the distribution.  The width of the boosted-$W$ mass peak is well modelled within the statistical precision of the 2012 data sample.

\begin{figure}[h!]
\centering
\includegraphics[width=0.5\textwidth]{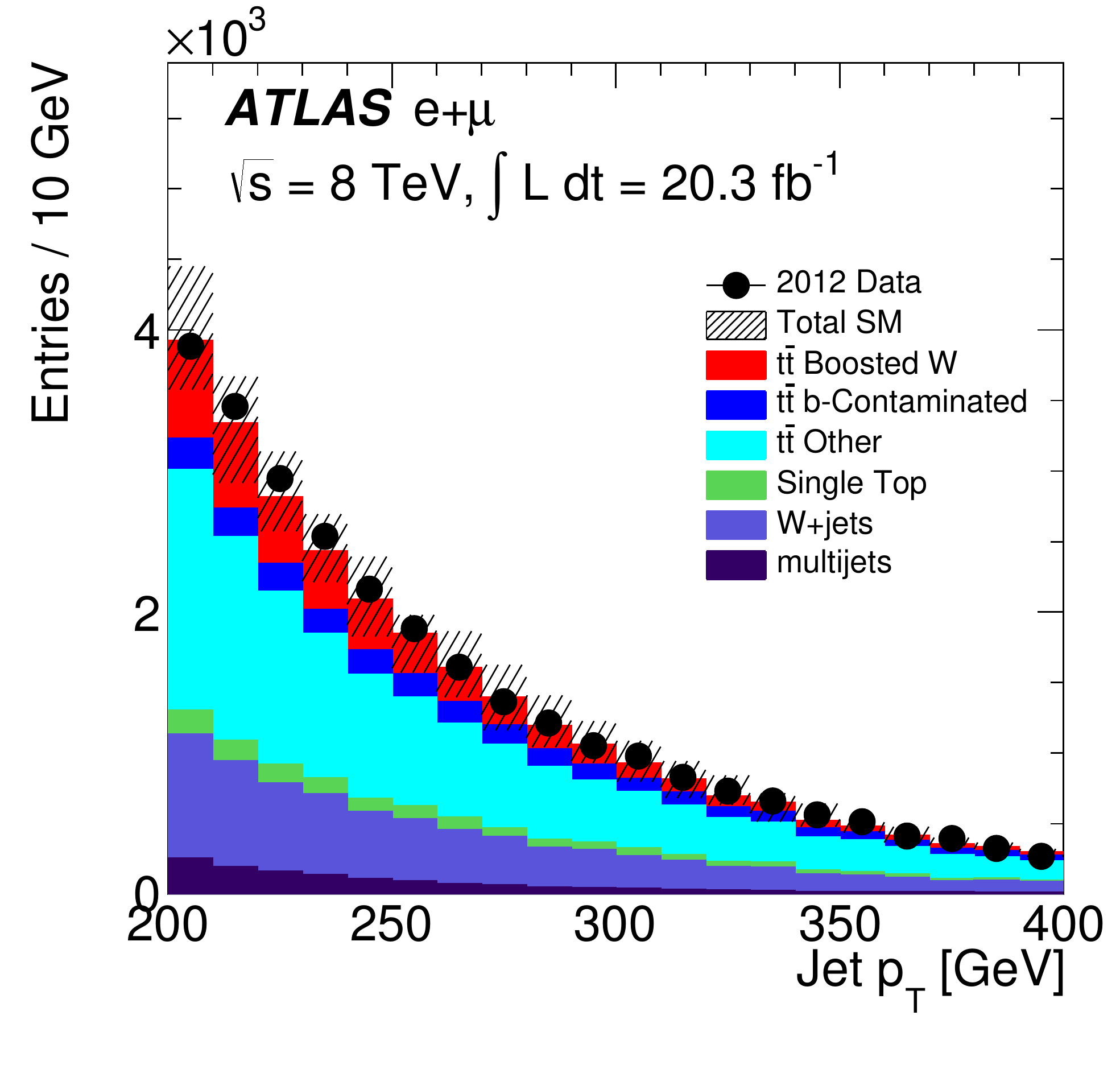}
\caption{The $p_\text{T}$ distribution of the selected large-radius jets.  The uncertainty band includes all the experimental uncertainties on the jet $p_\text{T}$ and jet mass described in Sec.~\ref{sec:systs}.}
\label{fig:data}
\end{figure}

\begin{figure}[h!]
\centering
\includegraphics[width=0.55\textwidth]{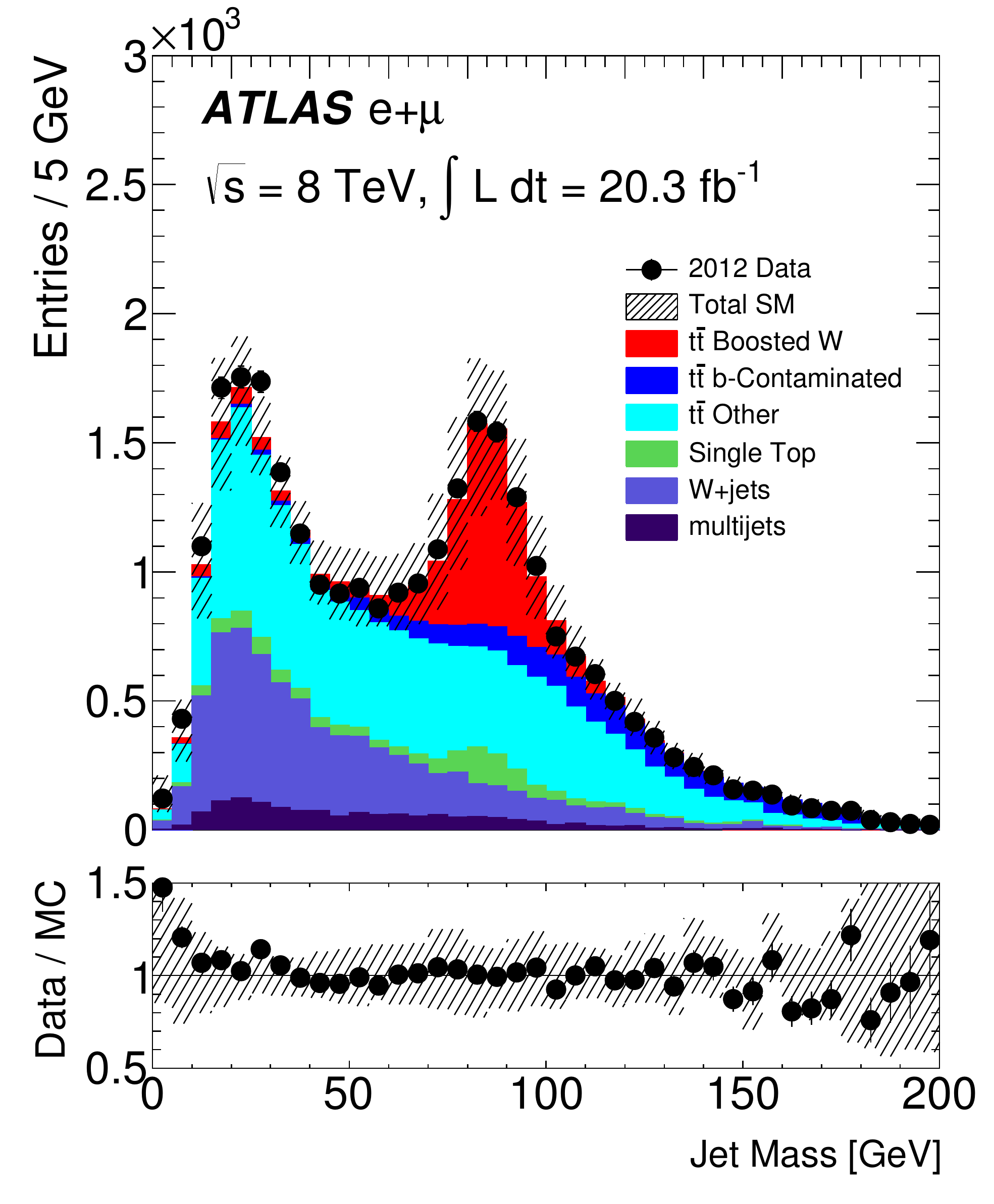}\\\includegraphics[width=0.45\textwidth]{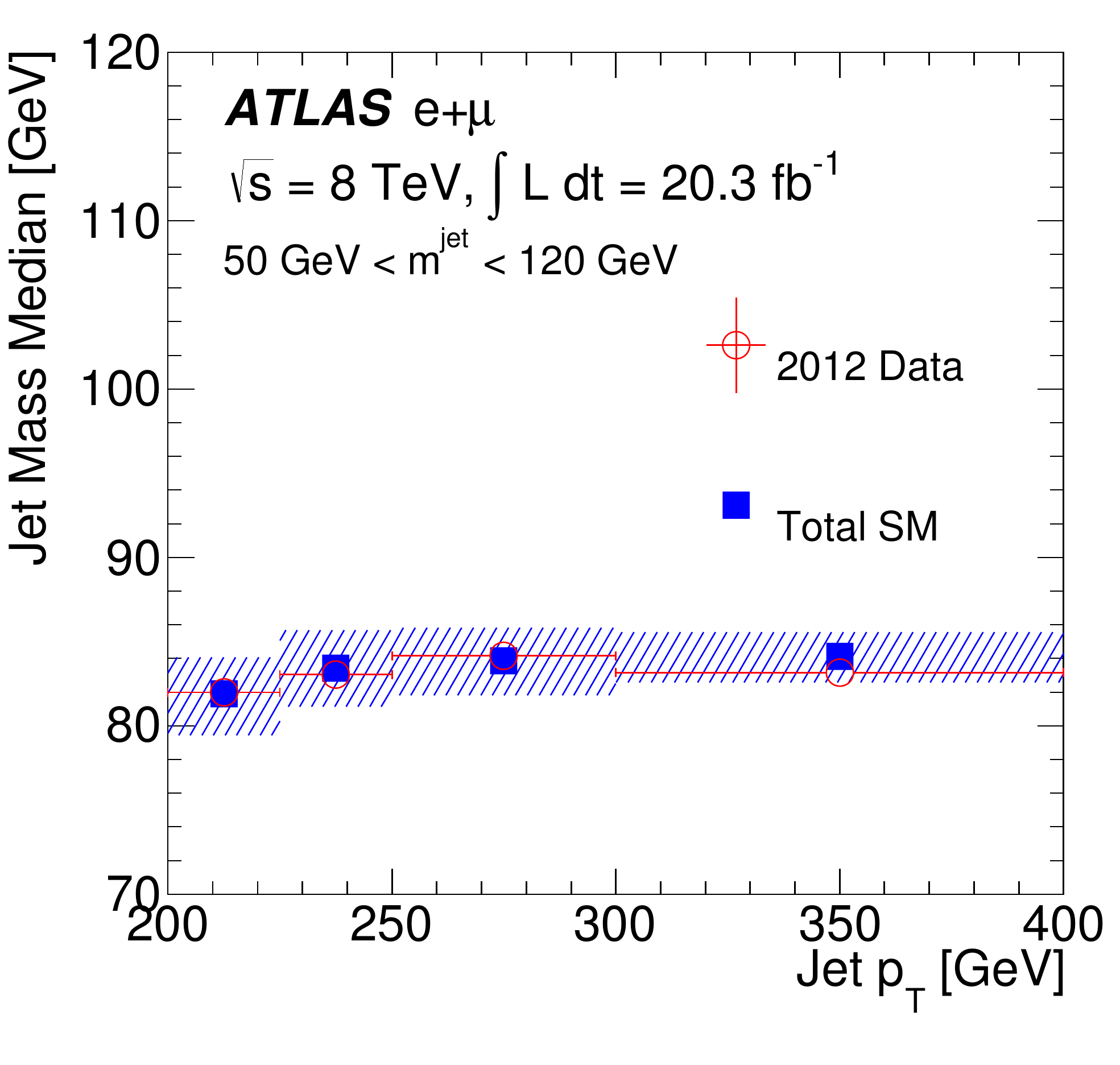}\includegraphics[width=0.45\textwidth]{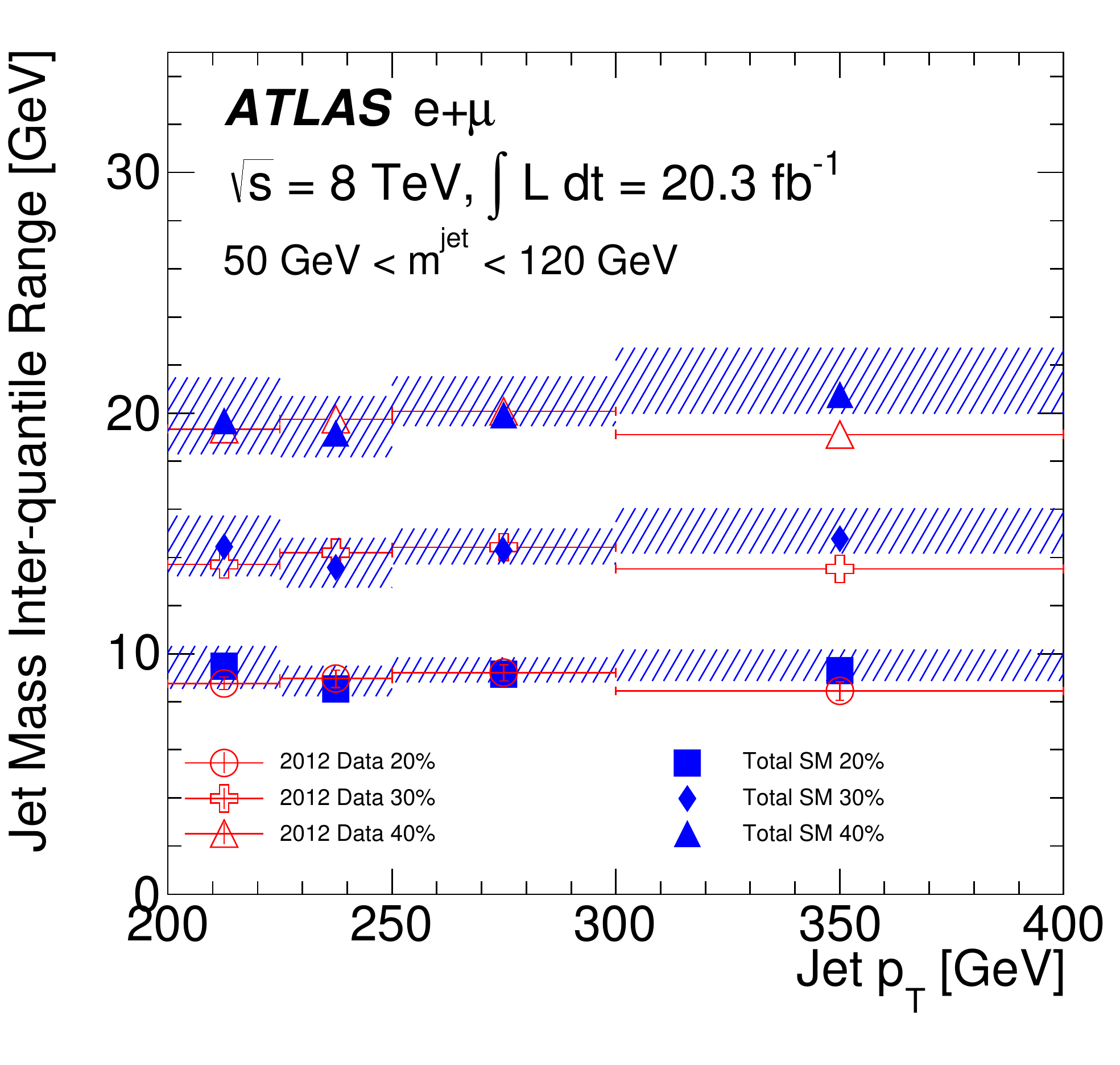}
\caption{(a) The jet-mass distribution of the selected jets in semi-leptonic $t\bar{t}$ events.  (b) The median of the mass distribution as a function of the jet $p_\text{T}$ for events with the selected jet in the range $50$~GeV $<m^\text{jet}<$ $120$~GeV.  This includes the contributions from events which are not classified as Boosted $W$.  (c) For the same events as in (b), the inter-quantile range as a measure of spread.  The quantiles are centred at the median.   The uncertainty band includes all the experimental uncertainties on the jet $p_\text{T}$ and jet mass described in Sec.~\ref{sec:systs}.  The inter-quantile range of size $0\%<X<50\%$ is defined as the difference between the $50\%+X\%$ quantile and the $50\%-X\%$ quantile. Statistical uncertainty bars are included on the data points but are smaller than the markers in many bins.}
\label{fig:datamass}
\end{figure}

The modelling of boosted $W$ bosons can also be studied using the jet-mass scale measured from tracks.  Defining the variable $r_{\mathrm{track}}$ as the ratio of the jet mass determined from tracks to the jet mass determined from the calorimeter, the jet mass scale uncertainty is related to the difference from unity of the ratio of $\langle r_\text{track}\rangle$ in data to $\langle r_\text{track}\rangle$ in MC simulation.  The mass scale uncertainty is calculated using the procedure described above, but with $r_{\mathrm{track}}^{-1}$.  If the jet consists only of pions, the natural scale for $r_\text{track}$ is 2/3, although there are significant physics and detector effects that introduce a large spread of values.  The distribution of $r_\text{track}$ in the $t\bar{t}$--enriched event sample with the same $p_\text{T}$ and $b$-jet veto requirements as in Fig.~\ref{fig:datamass} is shown in Fig.~\ref{fig:datartrak}(a).   Unlike the raw jet-mass distribution, the $r_\text{track}$ distribution is similar for all of the sub-processes, as expected.  The scale and spread of the $r_\text{track}$ distribution are quantified in figures~\ref{fig:datartrak}(b) and~\ref{fig:datartrak}(c) using the $p_\text{T}$ dependence of the median and inter-quantile ranges.   Previous studies have indicated that the track multiplicity, $n_\text{track}$, in quark and gluon jets is not well modelled, especially for gluon jets, where $n_\text{track}$ is lower in the data with respect to {\tt PYTHIA} (see Chapter~\ref{cha:multiplicity}).  The distribution of the track multiplicity for large-$R$ jets in the $t\bar{t}$-enriched event sample is shown in Fig.~\ref{fig:datantrack}.  The boosted $W$ events are peaked at slightly lower values of the number of associated tracks compared to the quark/gluon jets from the other processes.  The (charged) particle multiplicity increases for generic quark and gluon jets as a function of jet energy.  However, the mass-scale of the jets produced from $W$ boson decays is set by $m_W$ so that in the absence of detector reconstruction effects, the track multiplicity distribution should be largely $p_\text{T}$ independent.  The $p_\text{T}$ dependence of the track multiplicity is shown in Fig.~\ref{fig:datantrack}(b) and~\ref{fig:datantrack}(c) in the form of the median and the inter-quantile ranges.  The median does increase because of the large non-$W$ component as well as the finite detector acceptance for charged particles from the boosted $W$ boson decay.  The width is well modelled within the statistical precision of the data.  However, there is disagreement for the median.  Previous studies (including Ref.~\cite{Aad:2014gea}) suggest that this is due to fragmentation modelling and not the modelling of the detector response.

The $p_\text{T}$-weighted distribution of the track charges defines the jet charge, which is shown in Fig.~\ref{fig:datacharge}(a).   The charge of the lepton from the leptonic $W$ boson decay determines the expected charge of the hadronically decaying $W$ boson candidate, allowing for a tag-and-probe study of the capability of charge tagging in hadronic $W$ boson decays~\cite{ATLAS-CONF-2013-086}.  The jet charge for boosted $W$ bosons for positively (negatively) charged leptons is clearly shifted to the left (right) of zero.  There is also some separation between positive and negative $W$ boson decays when the selected large-radius jet does not satisfy the criteria for being a boosted $W$ boson.  This is because the jet still contains some of the $W$ boson decay products, and the jet charge is correlated with the charge of the $W$ boson.  The difference between the inclusive and boosted $W$-boson jets is clearer in the $p_\text{T}$ dependence plot of the median jet charge shown in Fig.~\ref{fig:datacharge}(b).  The medians of the distributions for boosted $W$ jets are nearly twice as far apart as the medians for inclusive jets.  However, in both cases the spread is less than the width of the distribution, shown as the inter-quantile range (inter-quantile range with $X=25\%$) in Fig.~\ref{fig:datacharge}(c).  Even though there is some small disagreement for the median number of tracks, the $p_\text{T}$-weighted sum defining the jet charge is reasonably well modelled.

The remaining input to the boson tagger is the $b$-tagging discriminant for the matched small-radius jets.  The efficiency-binned MV1 distributions are shown in Fig.~\ref{fig:datamv1}(a) and~\ref{fig:datamv1}(b) with the same selection criteria as for the previous figures, except that the $b$-jet veto is removed.  The contamination due to the $b$-jet from the top-quark decay complicates a direct study of the MV1 distribution for boosted $W$ jets; contamination from the $b$-quark decay products is seen clearly in the MV1 distribution at lower values of the efficiency.  Most of the boosted $W$ jets are in the highest efficiency bin because they have no real $b$-hadron decay.

Overall, the simulation models all three input variables well.

\begin{figure}[h!]
\centering
\includegraphics[width=0.55\textwidth]{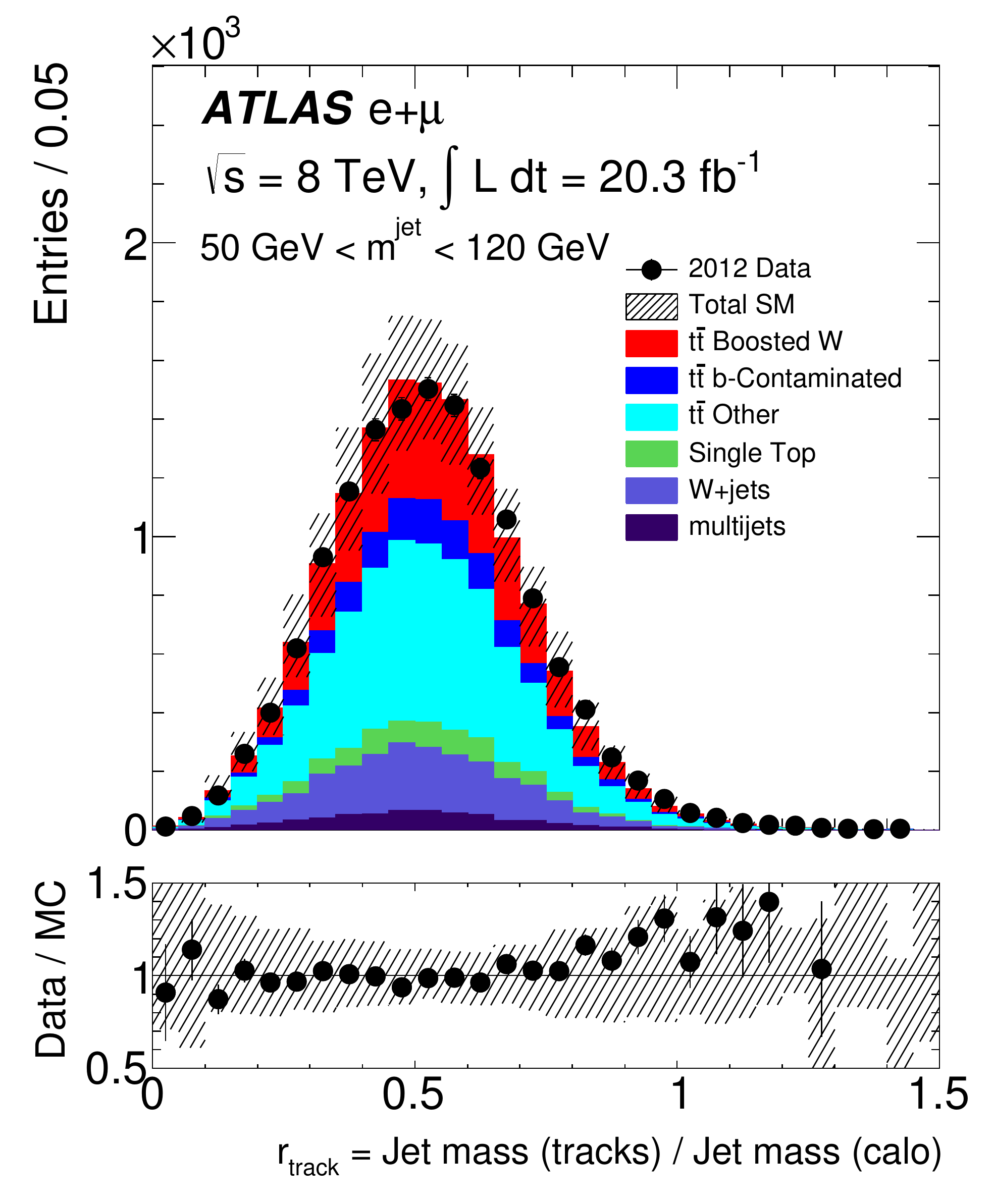}\\\includegraphics[width=0.45\textwidth]{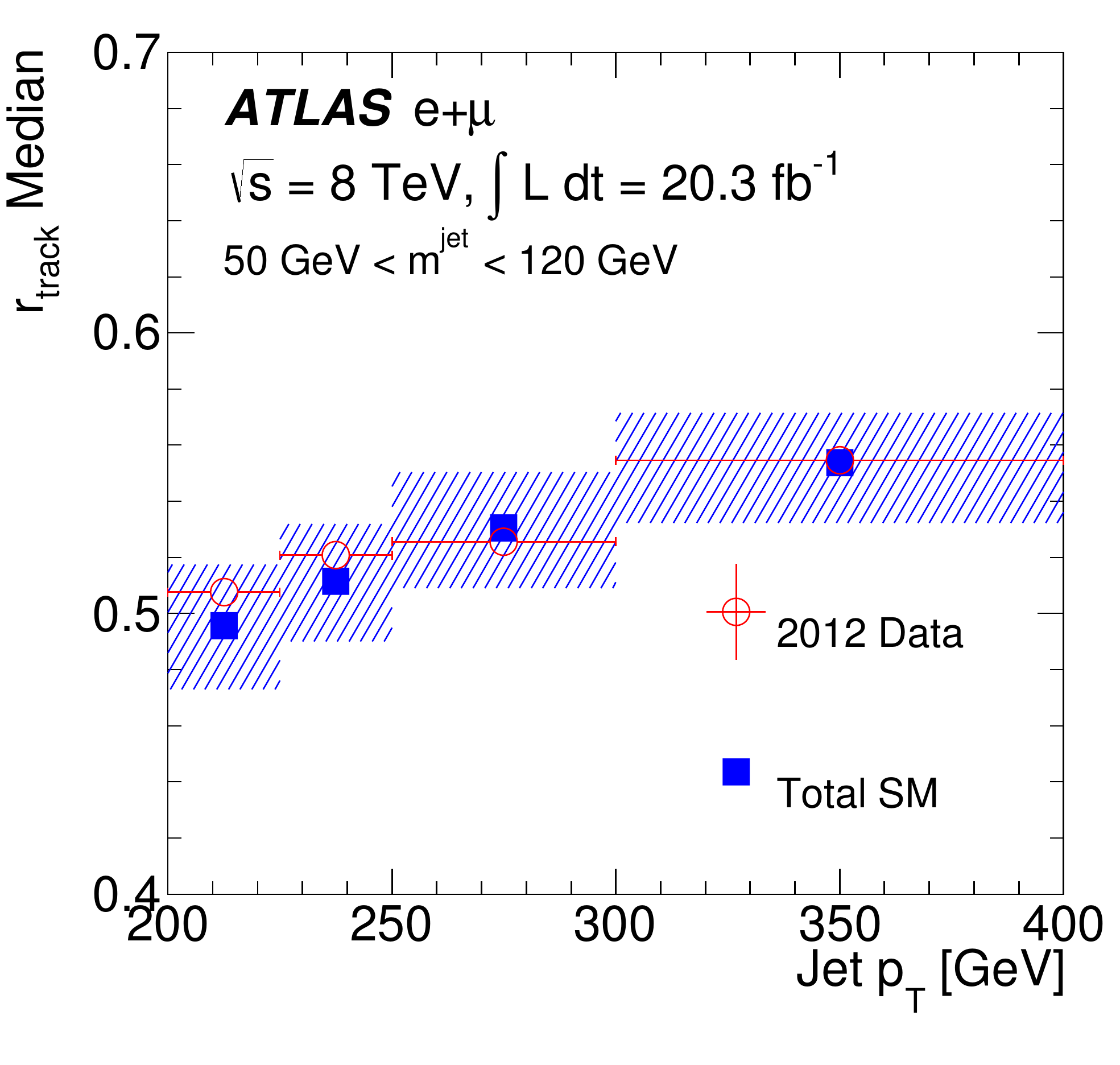}\includegraphics[width=0.45\textwidth]{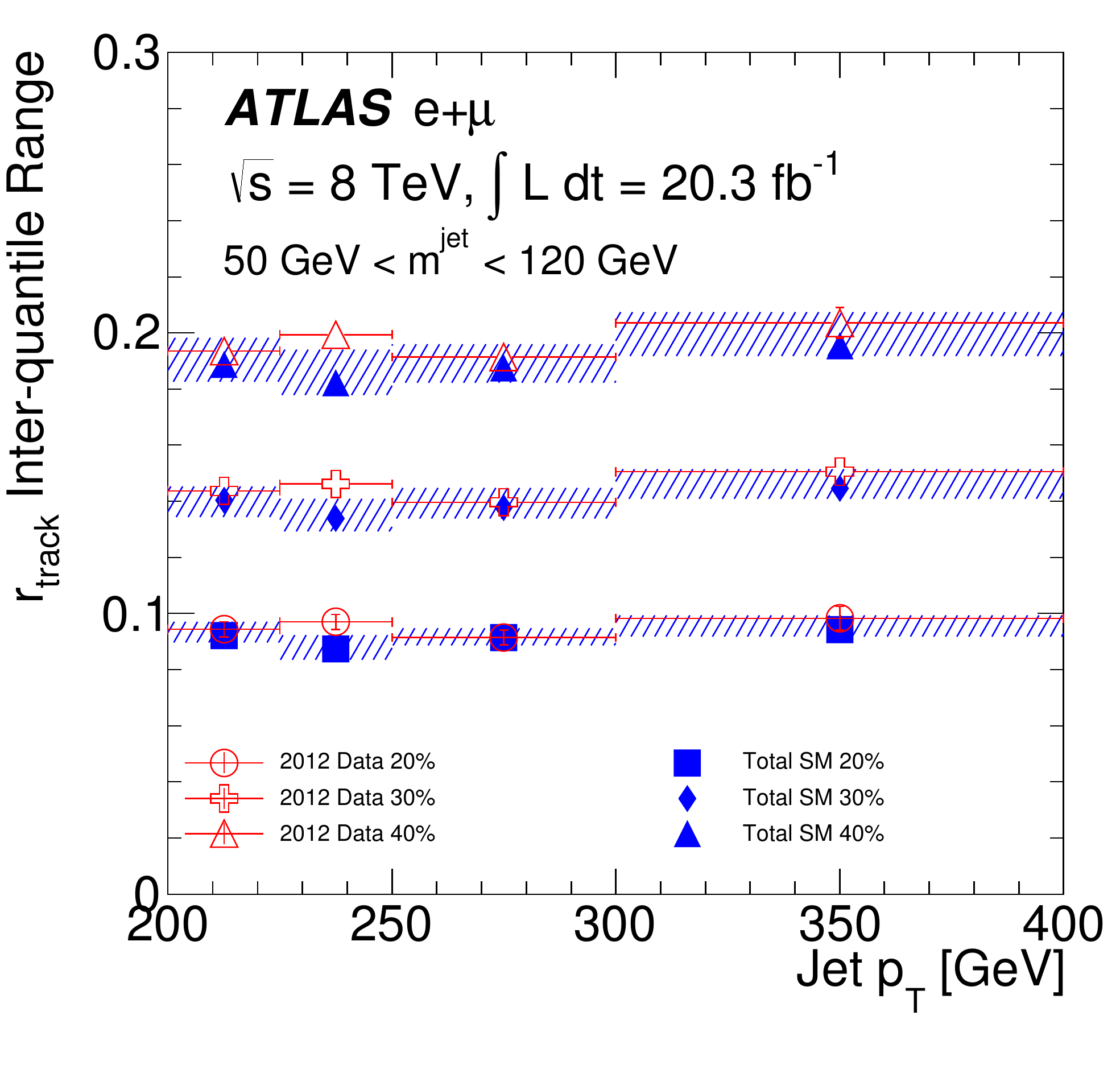}
\caption{(a) The distribution of $r_\text{track}$ in the data for semi-leptonic $t\bar{t}$ events with the selected jet in the range $50$~GeV $<m^\text{jet}<$ $120$~GeV.  (b) The median of the $r_\text{track}$ distribution as a function of the jet $p_\text{T}$.  This includes the contributions from events that are not classified as Boosted $W$.  (c) The inter-quantile range as a measure of the width.  The quantiles are centred at the median.   The uncertainty band includes all the experimental uncertainties on the jet $p_\text{T}$ and jet mass described in Sec.~\ref{sec:systs}.  The inter-quantile range of size $0\%<X<50\%$ is defined as the difference between the $50\%+X\%$ quantile and the $50\%-X\%$ quantile. Statistical uncertainty bars are included on the data points but are smaller than the markers in many bins.}
\label{fig:datartrak}
\end{figure}

\begin{figure}[h!]
\centering
\includegraphics[width=0.55\textwidth]{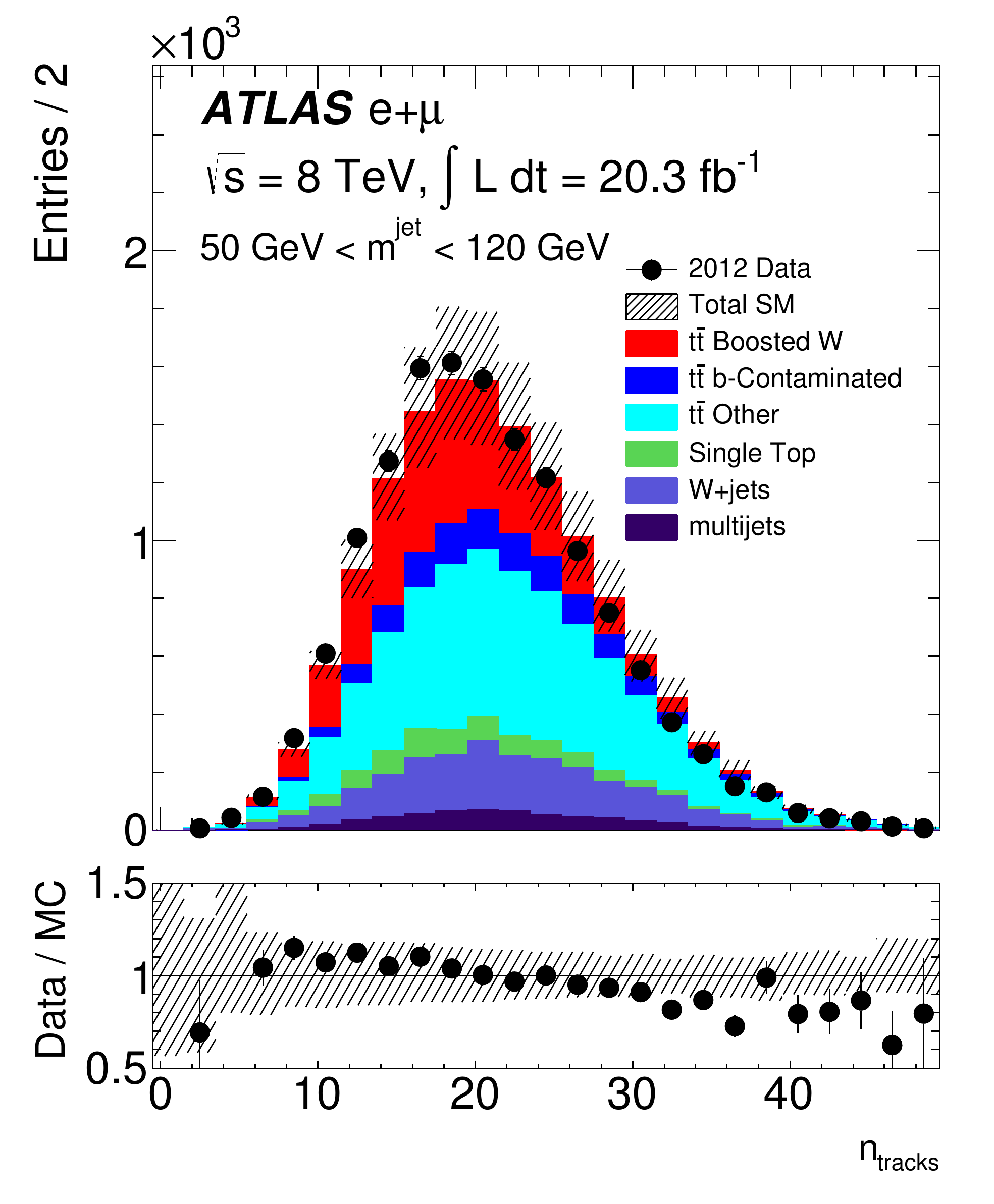}\\\includegraphics[width=0.45\textwidth]{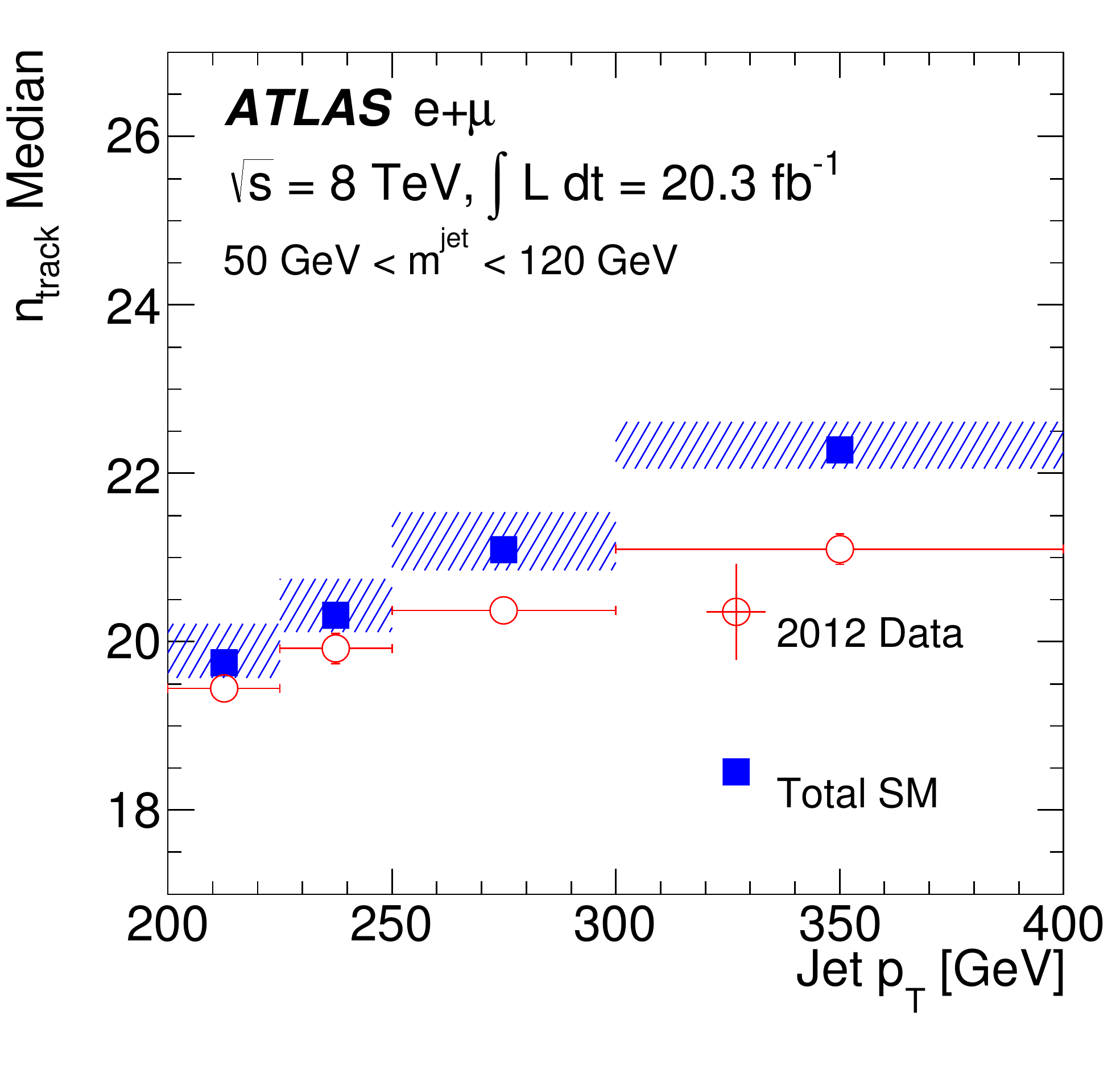}\includegraphics[width=0.45\textwidth]{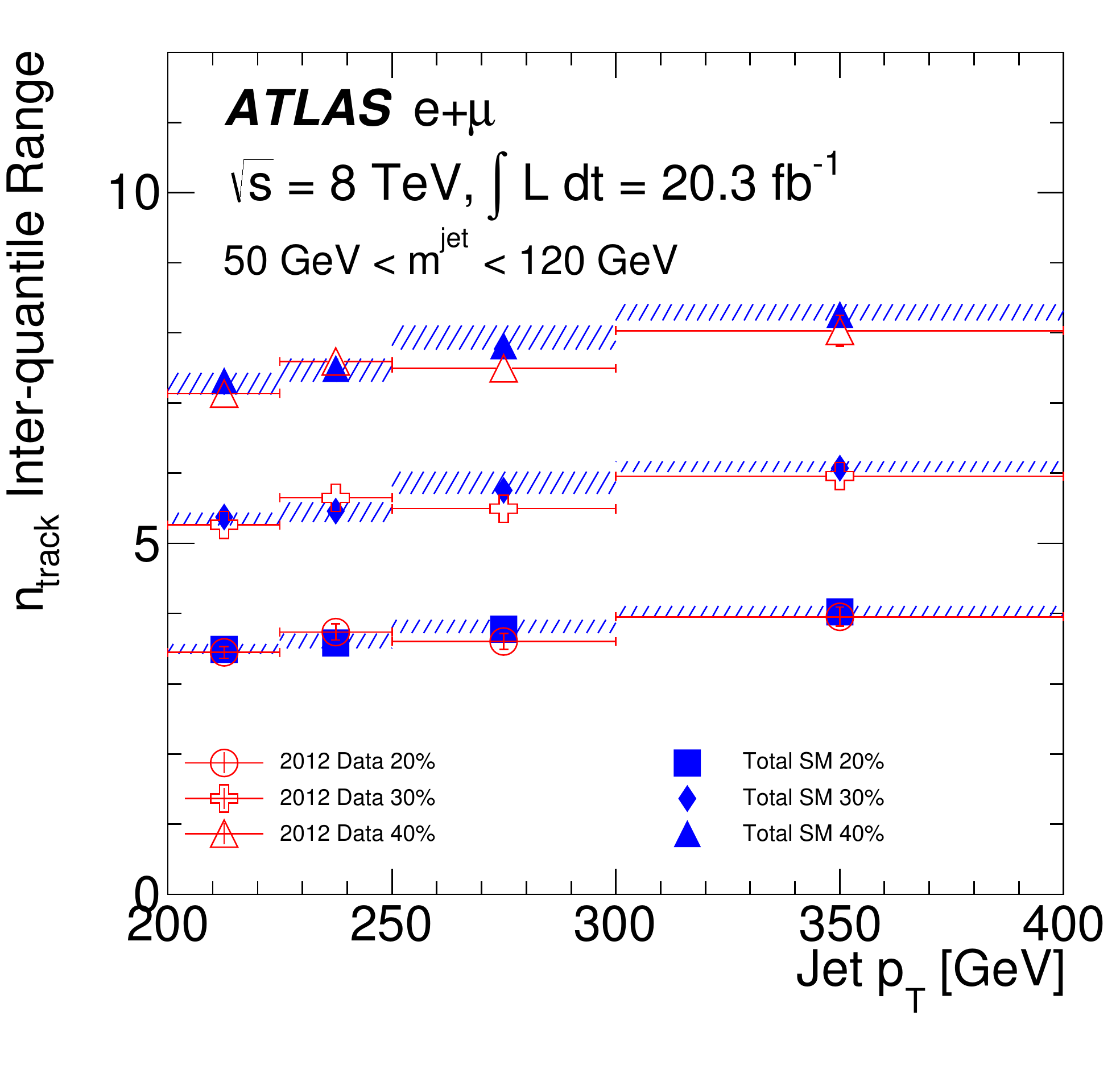}
\caption{(a) The distribution of the number of tracks associated with the selected large-radius jet in the semi-leptonic $t\bar{t}$ data for events with the selected jet in the range $50$~GeV $<m^\text{jet}<$ $120$~GeV.  (b) The median of the distribution of the number of tracks as a function of the jet $p_\text{T}$.  This includes the contributions from events that are not classified as Boosted $W$.  (c) The inter-quantile range as a measure of the width.  The quantiles are centred at the median.  The uncertainty band includes all the experimental uncertainties on the jet $p_\text{T}$ and jet mass described in Sec.~\ref{sec:systs}.  The inter-quantile range of size $0\%<X<50\%$ is defined as the difference between the $50\%+X\%$ quantile and the $50\%-X\%$ quantile. Statistical uncertainty bars are included on the data points but are smaller than the markers in many bins.}
\label{fig:datantrack}
\end{figure}

\begin{figure}[h!]
\centering
\includegraphics[width=0.55\textwidth]{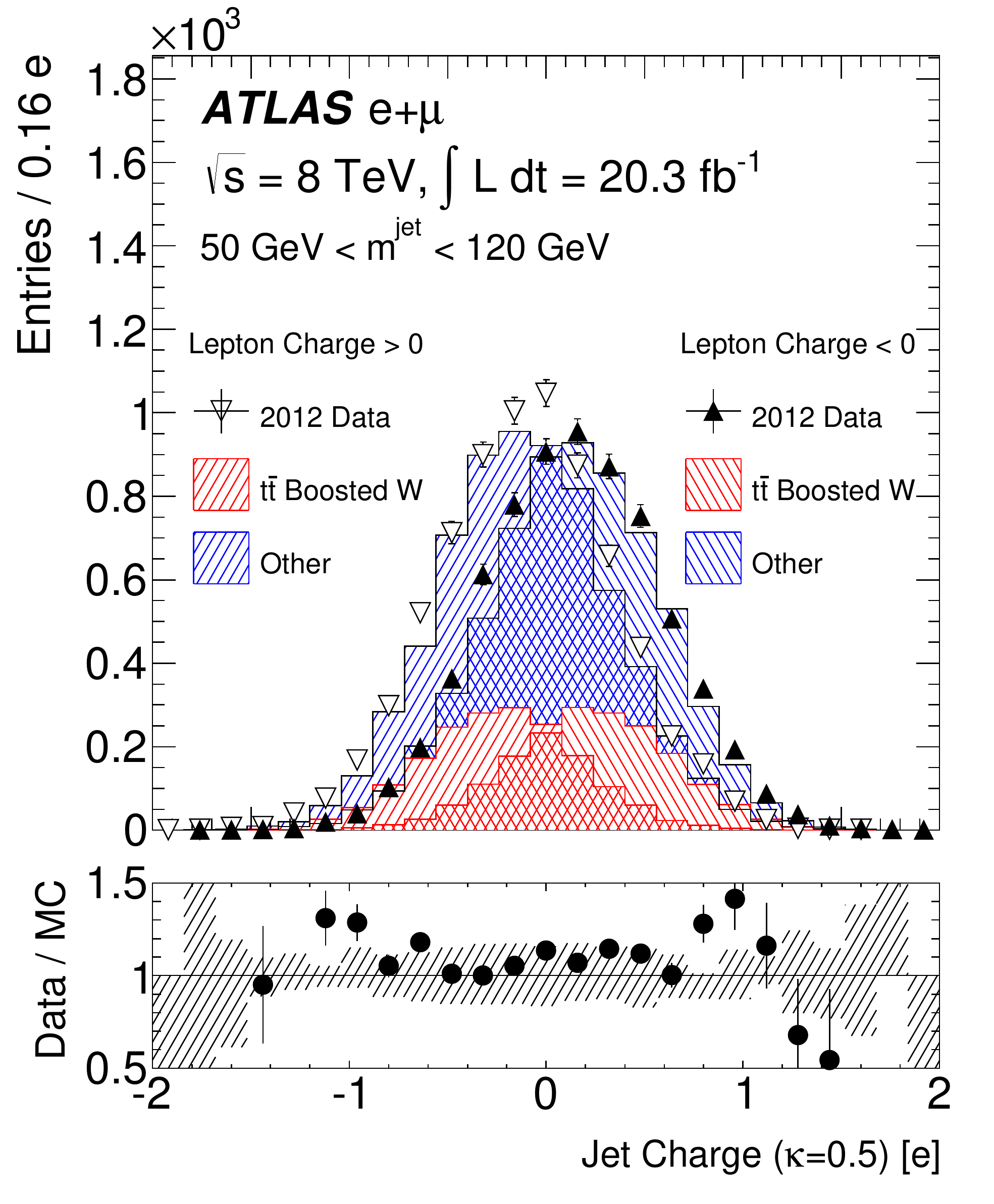}\\\includegraphics[width=0.45\textwidth]{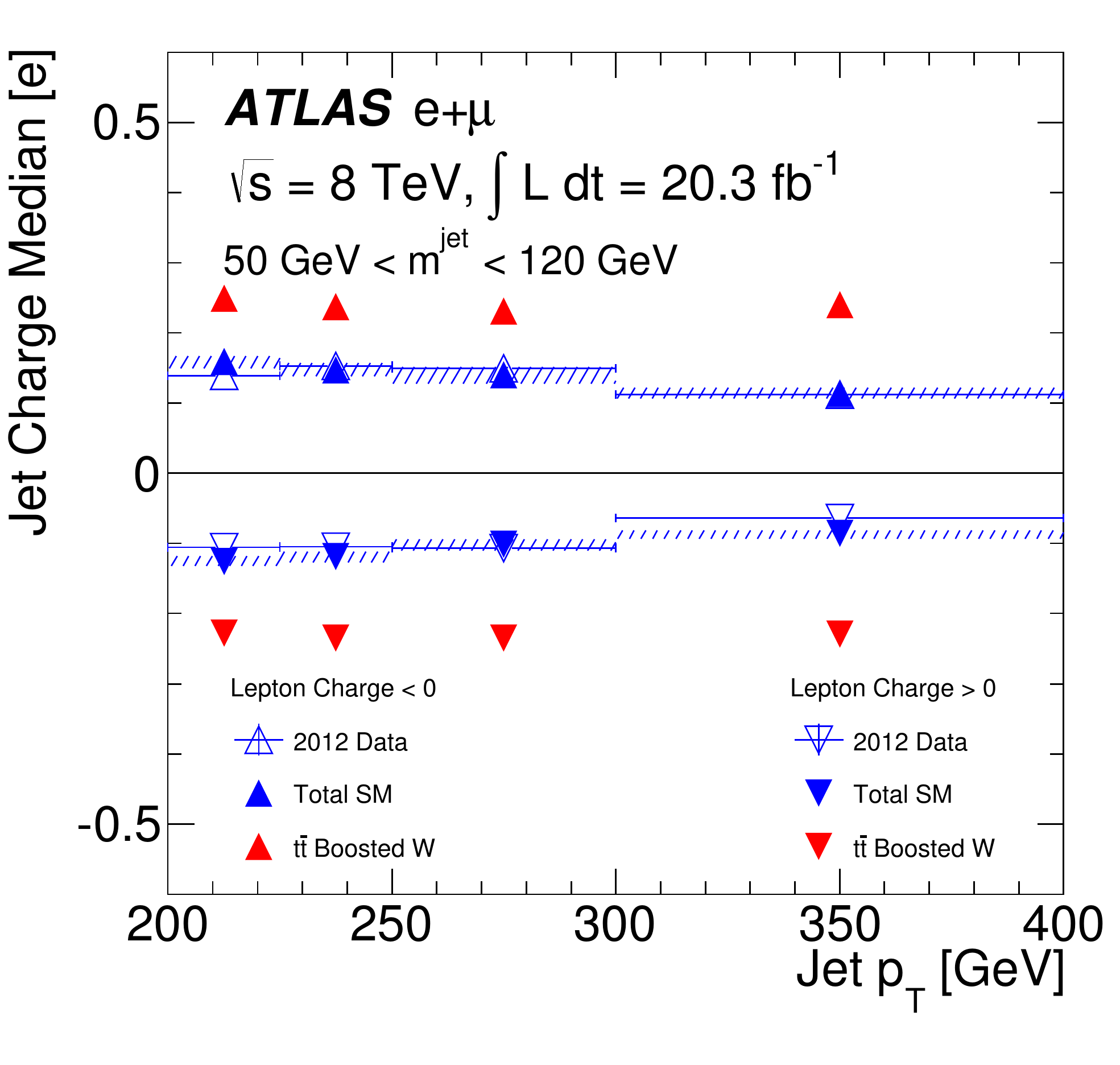}\includegraphics[width=0.45\textwidth]{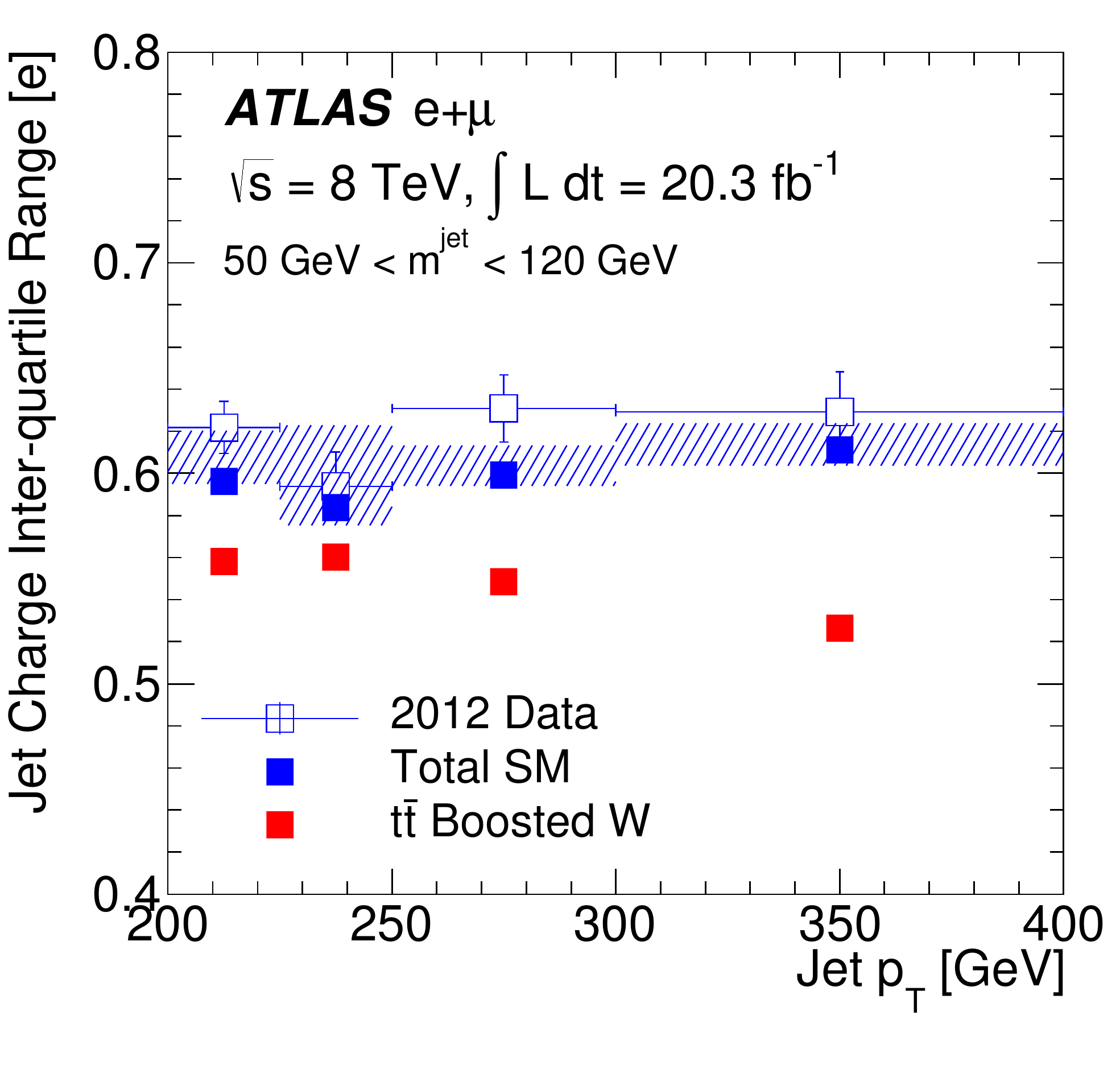}
\caption{(a) The distribution of the jet charge in the data for semi-leptonic $t\bar{t}$ events with the selected jet in the range $50$~GeV $<m^\text{jet}<$ $120$~GeV.  The ratio uses the positive lepton charge.  (b) The median of the jet charge distribution as a function of the jet $p_\text{T}$.  This includes the contributions from events that are not classified as Boosted $W$ (except for the blue triangles, for which only the Boosted $W$ is included).  (c) The inter-quartile range as a measure of the width.  The quantiles are centred at the median.   The uncertainty band includes all the experimental uncertainties on the jet $p_\text{T}$ and jet mass described in Sec.~\ref{sec:systs}.  The inter-quantile range is defined as the difference between the $75\%$ quantile and the $25\%$ quantile.  Statistical uncertainty bars are included on the data points but are smaller than the markers in many bins.}
\label{fig:datacharge}
\end{figure}

\begin{figure}[h!]
\centering
\includegraphics[width=0.5\textwidth]{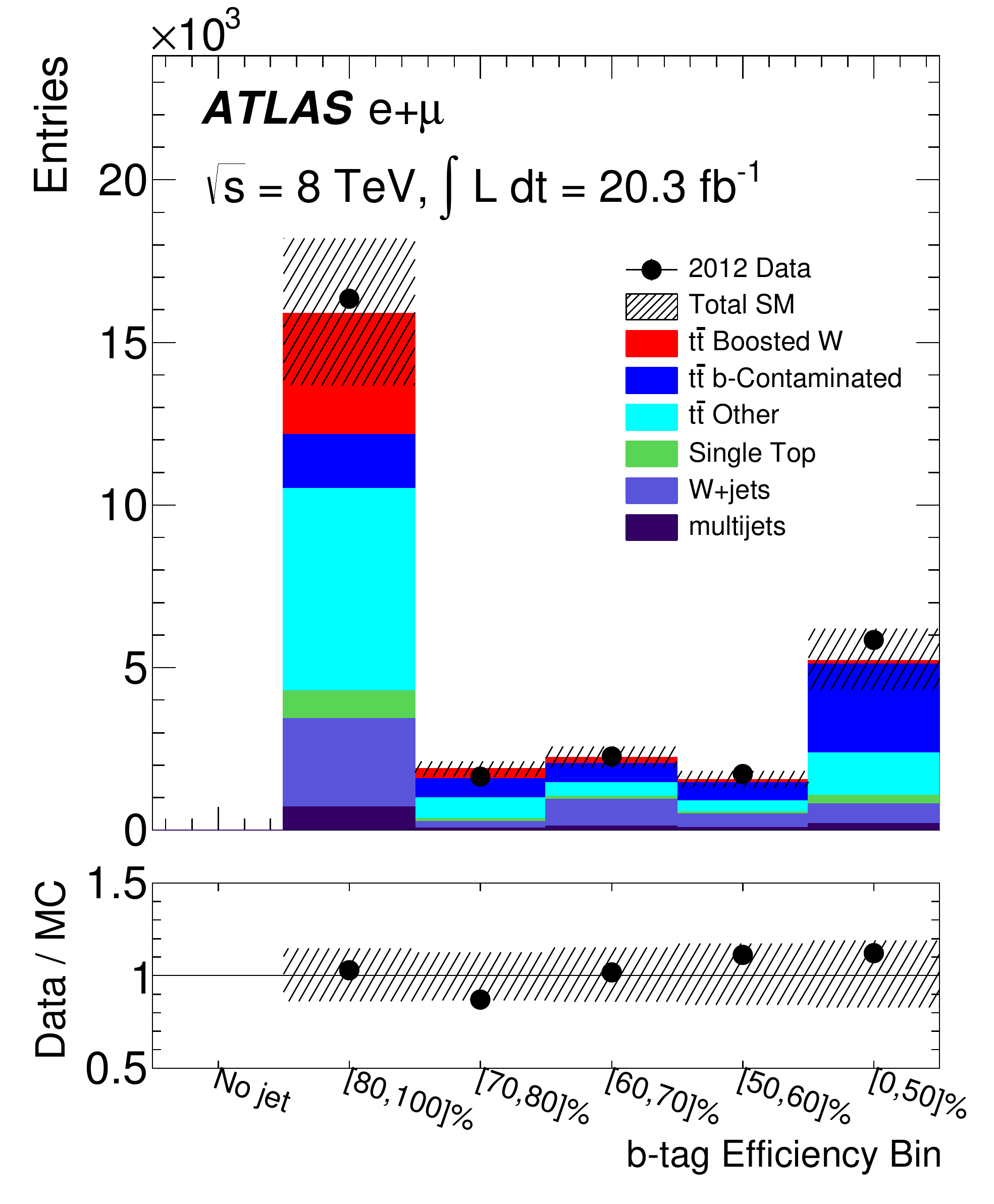}\includegraphics[width=0.5\textwidth]{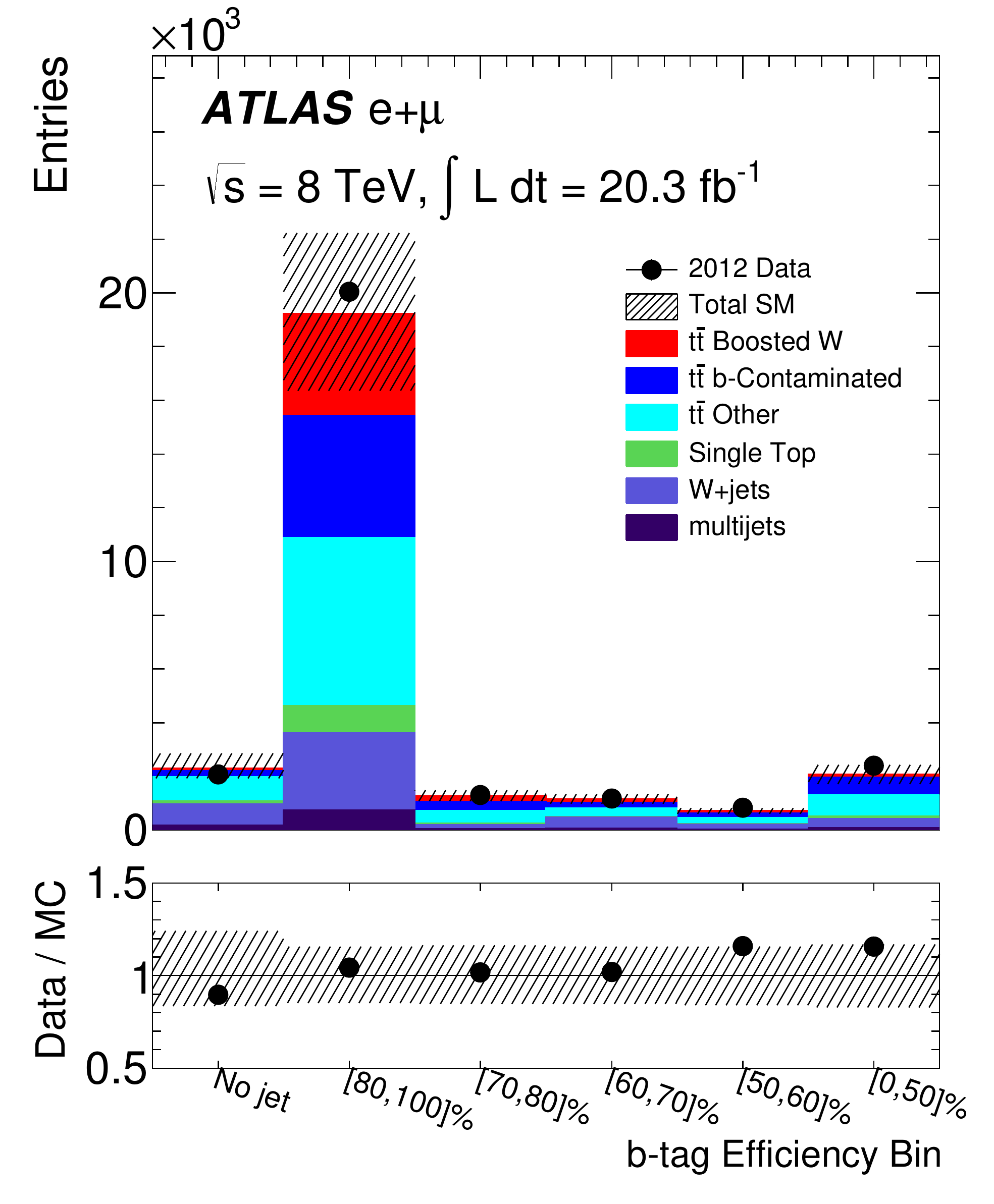}
\caption{The efficiency-binned MV1 distribution for the (a) leading and (b) sub-leading matched small-radius in semi-leptonic $t\bar{t}$ events.  If there is no second small-radius jet with $p_\text{T}>25$~GeV and $\Delta R<1$ to the selected large-radius jet axis, the event is put in the `No jet' category in (b).  The uncertainty band includes all the experimental uncertainties on the jet $p_\text{T}$ and jet mass and those related to the $b$-tagging described in Sec.~\ref{sec:systs}.  Statistical uncertainty bars are included on the data points but are smaller than the markers in many bins.}
\label{fig:datamv1}
\end{figure}

\newpage
\clearpage

\subsection{Outlook}
\label{sec:outlook}

The simulation studies of the boson-type tagger presented in Sec.~\ref{sec:perfm} show that for bosons with $200$~GeV $<p_\text{T}<$ 400~GeV, it is possible to achieve $Z$-boson efficiencies of $\epsilon_Z=90\%$, $50\%$, and $10\%$ with $W^+$ boson rejections of $1.7, 8.3$ and $1000$, respectively.   Putting this into context, with $R(\epsilon_Z)$ defined as the lowest possible $W$-boson tagging efficiency at a fixed $Z$-boson tagging efficiency: 

\begin{itemize}
\item The $WZ$/$WW$ cross-section ratio is $\sim 20\%$~\cite{Aad:2014mda}.  At the 50\% type-tagger working point, one can change the ratio of events to 

\begin{align}
\frac{50\%}{R(50\%)}\times \frac{  \sigma(WZ)}{ \sigma(WW)} = \frac{50\%}{12\%}\times \frac{ \sigma(WZ)}{ \sigma(WW)} = \frac{50}{12}\times 20\% \approx 83\%,
\end{align}

\noindent with the possibility for a high-purity extraction of the $WZ$ cross section in the semileptonic channel ($\ell\nu q\bar{q}$).

\item Diboson resonances are predicted by many models of physics beyond the Standard Model.  The all-hadronic channel provides a significantly higher yield than the leptonic channels.  At the $90\%$ type-tagger working point, one can distinguish $ZZ$ from $WZ$ with a likelihood ratio of $0.9^2/(0.9\times 0.6)\sim 1.5$.  
\item At the 10\% type-tagger working point, a leptophobic flavor-changing neutral current (with decays like in the SM) with a branching ratio of 1\% would have the same number of events as the $t\rightarrow bW$ decay in $t\bar{t}$ production:\footnote{Up to impurities due to the high-occupancy $t\bar{t}$ environment.}

\begin{align}
\frac{10\%}{R(10\%)}\times \frac{\Gamma(t\rightarrow Zc)}{ \Gamma(t\rightarrow Wb)}=\frac{10\%}{0.1\%}\times \frac{ \Gamma(t\rightarrow Zc)}{\Gamma(t\rightarrow Wb)} = 100 \times 1\% = 100\%.
\end{align}

\end{itemize}

Only the range $200$ GeV $<p_\text{T}<400$ GeV was studied thus far due to the availability of $W$ bosons in the data.  MC simulation suggest that the separation between $W$ bosons and $Z$ bosons from jet mass and jet charge is still powerful up to and beyond 1 TeV.   The information from $b$-tagging degrades around 400 GeV as the two decay products from the boson become too close to resolve as two separate jets\footnote{Smaller radius (track) jets can be used to recover the efficiency in this regime~\cite{ATL-PHYS-PUB-2014-013}.}.

\subsection{Conclusions}
\label{sec:concWZ}

A tagger for distinguishing hadronically decaying boosted $Z$ bosons from $W$ bosons using the ATLAS detector has been presented.   It will most likely be used after a boson tagger has rejected most QCD multijet events\footnote{See Fig.~\ref{fig:aux2a} and~\ref{fig:aux3a} for a demonstration the (near-)independence of the jet mass and jet charge with a standard boson-versus-QCD tagging variable, $2$-subjettiness~\cite{Thaler:2010tr}.}.  Three discriminating variables are chosen which are sensitive to the differences in boson mass, charge, and branching ratios to specific quark flavors: large-radius jet mass, large-radius jet charge, and an associated small-radius jet $b$-tagging discriminant.  For moderate and high $Z$-boson tagging efficiencies, the jet mass is the most discriminating of the three variables, but there is significant improvement in discrimination when combining all three inputs into a single tagger.  At low $Z$-boson efficiencies, smaller than the $Z\rightarrow b\bar{b}$ branching ratio, the $b$-tagging discriminant is the most useful for rejecting $W$ bosons.  The full tagger is largely unaffected by many systematic uncertainties on the inputs, with the exception of the uncertainties on the jet-mass scale and resolution.  While it is not possible to measure the tagger efficiencies directly in data due to the lack of a pure sample of boosted, hadronically decaying $Z$ bosons, modelling of the likelihood function using hadronically decaying $W$ bosons has been studied in the data.  Overall, the simulation agrees well with the 20.3 fb${}^{-1}$ of $\sqrt{s}=8$ TeV $pp$ data collected at the LHC.    
\begin{figure}[h!]
\begin{center}
\includegraphics[width=0.99\textwidth]{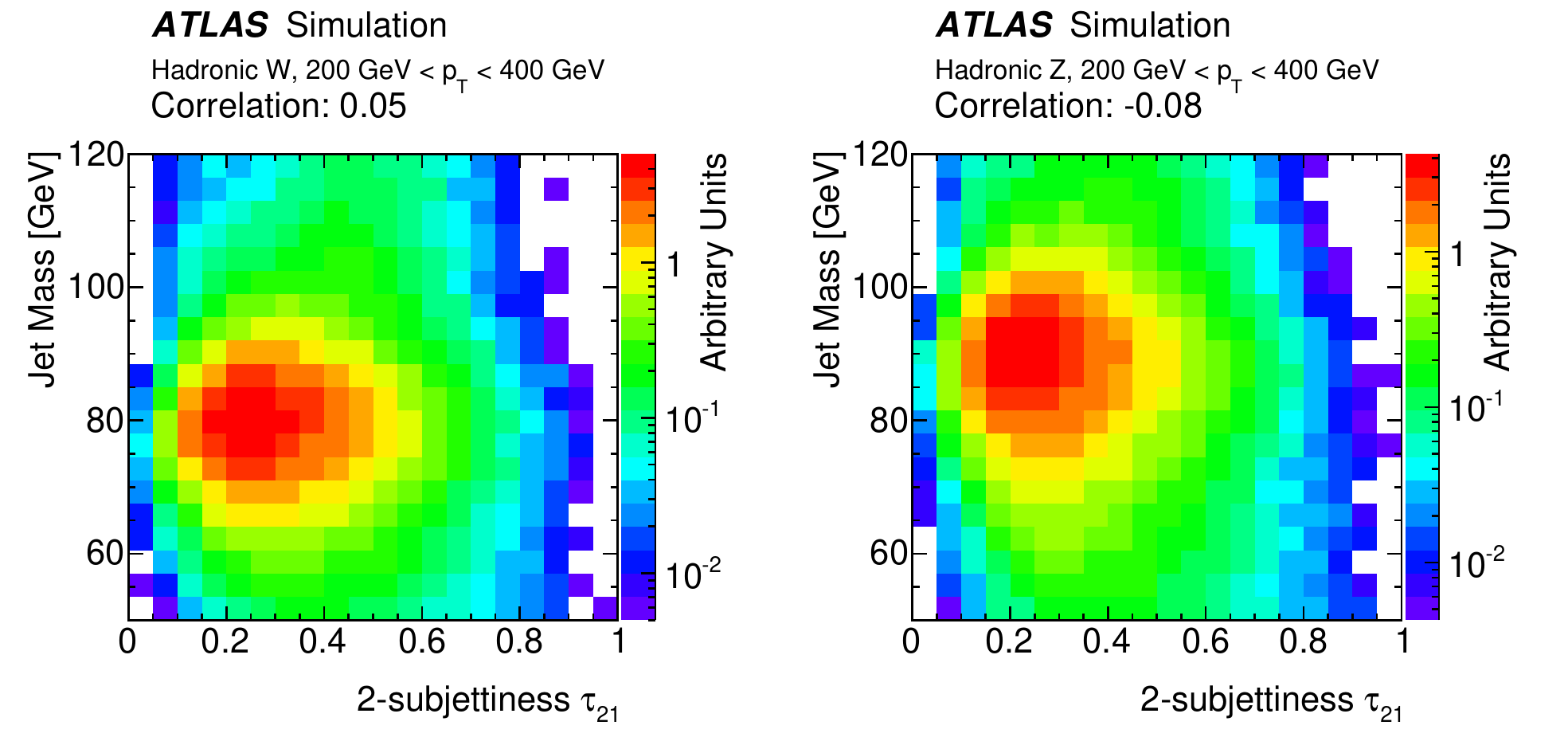}\\
\caption{The joint distribution of the jet mass and $2$-subjettiness for $W$ boson jets (left) and $Z$ boson jets (right).  The peak of the distribution along the jet mass axis is shifted toward higher values for the $Z$ due to its higher mass.  The linear correlation is less than $\pm10\%$ in both cases and the two distributions are nearly independent.}
\label{fig:aux2a}
\end{center}
\end{figure}

\begin{figure}[h!]
\begin{center}
\includegraphics[width=0.99\textwidth]{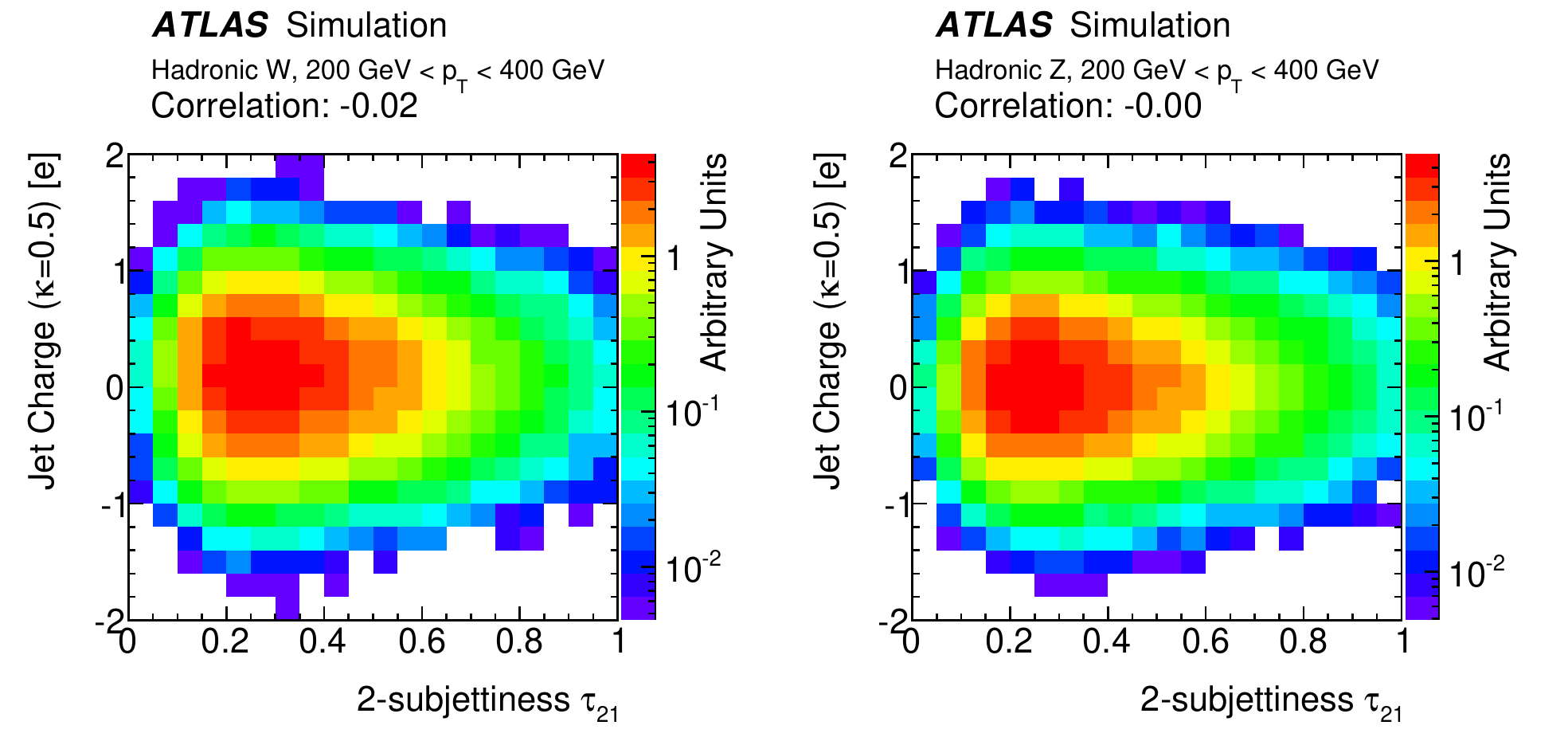}\\
\caption{The joint distribution of the jet charge and $2$-subjettiness for $W$ boson jets (left) and $Z$ boson jets (right). The linear correlation is less than $\pm 2\%$ in both cases and the two distributions are nearly independent.}
\label{fig:aux3a}
\end{center}
\end{figure}

\clearpage

\section{Machine Learning for Jet Tagging}
\label{sec:HEPML}

{\it Machine learning} is a generic term to describe procedures for identifying and classifying structure within a dataset.   As such, most analysis techniques can be described as a form of machine learning.  However, there is a deeper connection between machine learning and jet physics: the fundamental object of study only exists in the context of machine learning.  A jet is defined by a clustering algorithm, which is an example of an {\it unsupervised machine learning} technique.  Unlike the output of most clustering procedures, jets have a {\it physical meaning}.  The earlier sections in this chapter have shown that the quantum properties of jets can be calculated, calibrated, and measured with high precision.  Even though there is an extensive literature on clustering techniques, the most commonly used jet algorithms were established within the high energy physics community.  This is because the physical meaning of a jet only makes sense if the defining algorithm satisfies particular properties such as infrared and collinear safety.  The first half of this section (Sec.~\ref{sec:fuzzyjets}) is dedicated to bridging this gap by minimally modifying one of the most common unsupervised learning techniques for use in jet physics.  A new jet algorithm called {\it fuzzy jets} uses {\it mixture modeling} to cluster jets and is demonstrated on events with the hadronic decays of boosted boson and top quarks.  The parameters of the learned fuzzy jets contain information about the quantum properties of jets, which can be used for jet tagging.

Jet tagging is an example of {\it supervised learning}.  High energy physics is a unique setting for supervised learning because it is possible to generate arbitrarily large high fidelity simulation datasets that are {\it labeled} (have a known type or origin).  This chapter has introduced many jet substructure observables useful for separating jets initiated by different partons or particles.  The optimal tagger is one that uses the likelihood ratio based on the full radiation pattern within the jet.  In practice, it is not possible to compute the full likelihood.  However, many sophisticated supervised learning techniques have been designed to be close approximations to the likelihood and can achieve near-optimal performance.  Section~\ref{sec:jetimages} will demonstrate how state-of-the-art techniques borrowed from computer vision can improve the performance of jet tagging by thinking of the jet radiation pattern as an image.  Machine learning is a tool to guide but not replace physical intuition.  Therefore, one of the main focuses of Sec.~\ref{sec:jetimages} is to visualize what the machine learning algorithms are learning from the radiation pattern in jets.

There is an ever-growing machine learning literature that will aid physics analyses at the LHC to fully exploit the data.  This section ends with a brief discussion of prospects for the future in Sec.~\ref{sec:MLconclusions}.

\subsection{Fuzzy Jets} 
\label{sec:fuzzyjets}

The purpose of this section is to introduce a new paradigm for jet clustering, called {\it fuzzy jets}\footnote{The ideas presented in this section are published in Ref.~\cite{Mackey:2015hwa}.  Many of the studies presented in this section were performed by Conrad Stansbury.  In particular, Stansbury made the final versions of Fig.~\ref{fig:eventdisplay}-\ref{fig:pileup_ed},~\ref{fig:corr_mean_var}, and~\ref{fig:datastatsFuzzy}.}, based on probabilistic mixture modeling and to demonstrate its use in boosted topologies.  Section~\ref{sec:stats} introduces the statistical concept of a mixture model and describes the necessary modification to make the procedure IRC safe (see Sec.~\ref{sec:jets}).  Section~\ref{sec:EMalgorithm} gives one efficient method for clustering fuzzy jets based on the Expectation-Maximization (EM) algorithm.  Section~\ref{sec:tagging} contains several examples comparing fuzzy jets with sequential recombination and Sec.~\ref{fuzzypileup} describes how one might mitigate the impact of overlapping proton-proton collisions (pileup).  Conclusions are presented in Sec.~\ref{sec:conclusions} with some summary remarks and outlook for the future.

\subsubsection{Mixture Model Jets}
\label{sec:stats}

Mixture models~\cite{opac-b1097397} are a statistical tool for clustering which postulate a particular class of probability densities for the data to be clustered.   Generically, for grouping $n$ $m$-dimensional data points into $k$ clusters, the mixture model density is

\begin{align}
\label{eq:mm}
p(x_1,...,x_m|\pi,\theta)=\prod_{i=1}^n\left(\sum_{j=1}^k \pi_j f(x_i|\theta_j)\right),
\end{align}

\noindent where $\pi_j$ is the unknown weight of cluster $j$ such that $\sum_j \pi_j=1$ and $f(x_i|\theta_j)$ is a probability density on $n$-dimensions with unknown parameters $\theta_j$ to be learned from the data.  A common choice for $f$ is the normal density $\Phi$ with $\theta_j=(\mu_j,\Sigma_j)$ for $\mu_j$ the $m$-dimensional mean and $\Sigma_j$ the $m\times m$ covariance matrix.   In the mixture model paradigm, the $\theta_j$ are the cluster properties; in the Gaussian case, $\mu_j$ is the location of cluster $j$ and $\Sigma_j$ describes its shape in the $m$-dimensional space.  When clustering with a finite mixture, the number of clusters $k$ must be specified ahead of time\footnote{There is a wealth of literature on the subject of choosing $k$, for a survey of methods, see~\cite{determiningk}.  The likelihood monotonically increases with $k$; as alternatives to maximum likelihood, one can for instance look for kinks in the likelihood as a function of $k$~\cite{gap}.}, which is dual to the usual use of sequential recombination\footnote{It is similar to the exclusive form of the $k_t$ sequential recombination scheme~\cite{Catani:1993hr}.  The exclusive nature of the algorithm (and the minimization procedure used to find the jets) is similar to the XCone algorithm~\cite{Stewart:2015waa,Thaler:2015xaa}.} in which $k$ is learned and the size of jets is specified ahead of time.  The standard objective in mixture modeling is to select the parameters $\theta_j$ which maximize the likelihood (Eq.~\ref{eq:mm}) of the observed dataset.  Figure~\ref{fig:densitymap} illustrates what the learned event density might look like for $k=3$ and Gaussian $f=\Phi$ in $m=2$ dimensions.  

\begin{figure}[h!]
\begin{center}
\includegraphics[width=0.99\textwidth]{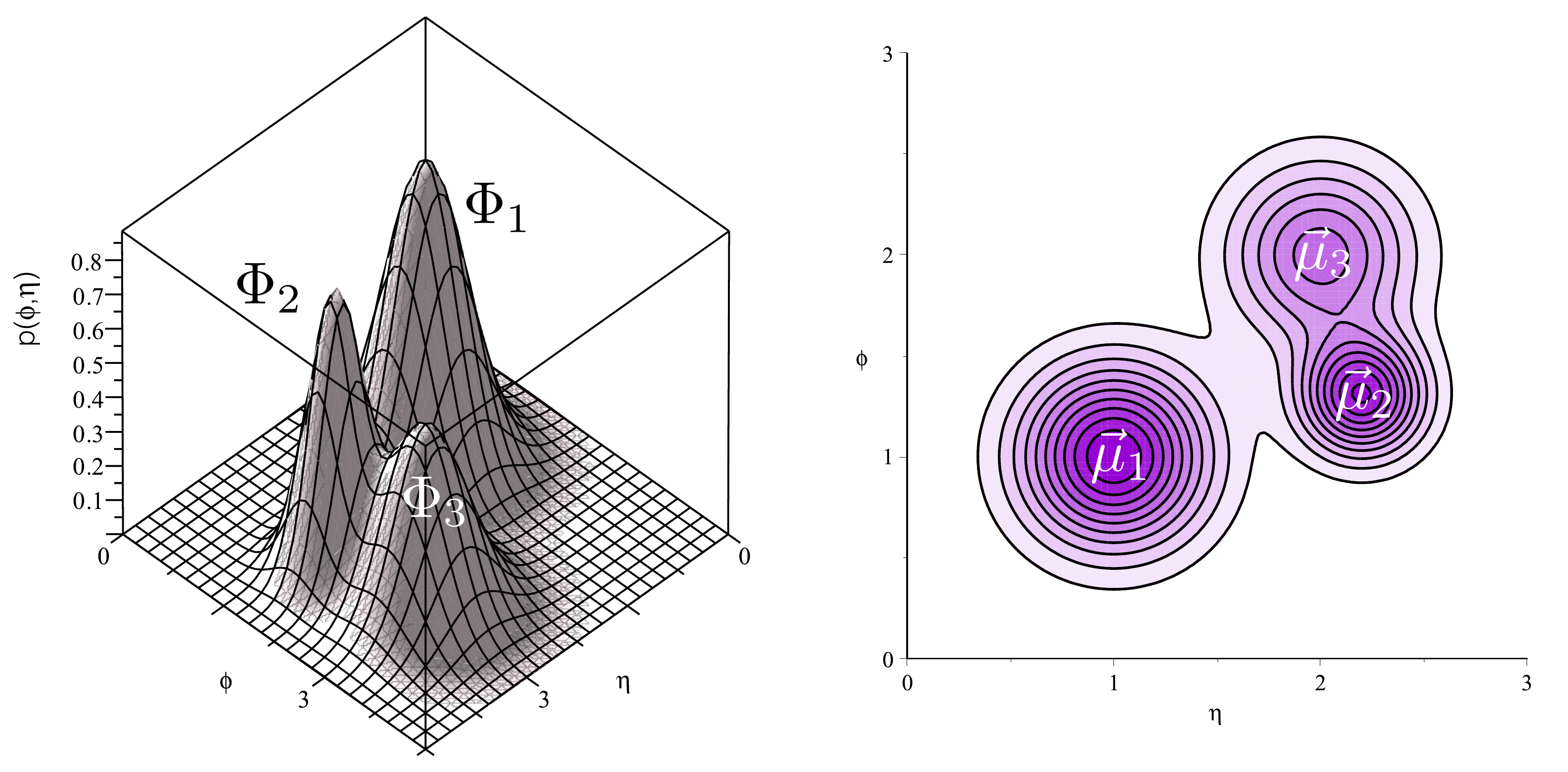}
\caption{An example of the learned per-particle probability density specified in Eq.~\ref{eq:mm} with $k=3$ and Gaussian $f=\Phi$ in $m=2$ dimensions.  One cluster is associated with each component density $\Phi_i=\Phi(\cdot \mid \mu_i,\Sigma_i)$, where the dot $\cdot$ is a placeholder for the function argument.}
\label{fig:densitymap}
\end{center}
\end{figure}

An equivalent way of approaching mixture modeling is to view Eq.~\ref{eq:mm} as the density used to generate the data.  In other words, view the data as having been drawn randomly from the density specified in Eq.~\ref{eq:mm}, with the following setup:

\begin{enumerate}
\item Throw $n$ independent and identical $k$-sided dice with probability $\pi_j$ to land on side $j=1,...,k$ and label the outcomes $\lambda_1,...,\lambda_n$.  \item Independent of the others, data point $i\in\{1,...,n\}$ is drawn randomly from $f(\cdot\mid\theta_{\lambda_i})$.\end{enumerate}

\noindent Once $\theta$ and $\pi$ are learned by minimizing Eq.~\ref{eq:mm}, one can compute $q_{ij} = \Pr(\lambda_i = j \mid x_i)$, the posterior probability that $x_i$ was generated by $f(\cdot\mid\theta_j)$ or, intuitively, the posterior probability that $x_i$ belongs to cluster $j$.  The $q_{ij}$ are the {\it soft assignments} of particles $i$ to jet $j$ and will play an important role in Sec.~\ref{sec:EMalgorithm} when showing how to maximize the likelihood in Eq.~\ref{eq:mm}.  In particular, $q_{ij}=\pi_jf(x_i|\theta_j)/\sum_{j'}\pi_{j'}f(x_i|\theta_{j'})$.  Jets produced with mixture modeling are called {\it fuzzy jets} because of the soft memberships - every particle can belong to every jet with some probability\footnote{Soft assignments for jets during clustering was studied in the context of the ``optimal jet finder''~\cite{Grigoriev:2003tn} which maximizes a function of the soft assignments.}.  This can be seen explicitly in Fig.~\ref{fig:densitymap} where the densities of all three clusters are everywhere nonzero, so $q_{ij}>0$ for all $j$.  The idea of probabilistic membership was recently studied in the context of the Q-jets algorithm~\cite{Ellis:2012sn} in which the same event is interpreted many times by injecting randomness into the clustering procedure.  Unlike Q-jets, fuzzy jets allocates the soft membership functions deterministically throughout the clustering procedure.  However, like Q-jets, there is an ambiguity in how to assign kinematic properties to the clustered jets.  Fuzzy jets are defined by their shape (and location), not their constituents.  This is in contrast to anti-$k_t$ jets, which are defined by their constituents without an explicit shape determined from the clustering procedure.  One simple assignment scheme is to define the momentum of a jet $j$ as

\begin{align}
p_\text{jet $j$}=\sum_{i=1}^m p_{i}\left\{\begin{matrix}1 & j=\text{argmax}_kq_{ik} \cr 0 & \text{else} \end{matrix}\right\} .
\end{align}

\noindent This procedure assigns every particle to its most probable associated jet and will be known as the hard maximum likelihood (HML) scheme, but is not the only possible assignment algorithm.  The dual problem in sequential recombination is the jet area, which must be defined~\cite{areas}, whereas the jet kinematics are the `natural' coordinates.

For the remainder of the section, the likelihood in Eq.~\ref{eq:mm} is specialized to the case of clustering particles into jets at a collider like the LHC.  Consider a mixture model in two dimensions\footnote{\label{second}One must take care in selecting a class of densities appropriate for the angular quantity $\phi$.  For more details on the wrapped Gaussian distribution and motivation for its use in this context, see Appendix~\ref{sec:wrapped}.} with $x_i=\rho_i$.  The resulting mixture model (MM) jets are inherently not IR safe: particle $p_\text{T}$ does not appear in the likelihood and therefore arbitrarily low energy particles can influence the clustering procedure.  Therefore, the log likelihood is slightly modified:

\begin{align}
\label{eq:mm2}
\log\mathcal{L}(\{p_{T,i},\rho_i\}|\theta)=\sum_{i=1}^m p_{T,i}^\alpha \log\left( \sum_{j=1}^k \pi_j f(\rho_i|\theta_j)\right),
\end{align}

\noindent where $\alpha$ is a weighting factor.  Equation~\ref{eq:mm2} is the log of Eq.~\ref{eq:mm} with the term $p_{T,i}^\alpha$  inserted in the outer sum.  For $\alpha > 0$, the resulting {\it modified} mixture model (mMM) jets are IR safe, and when $\alpha=1$, the jets are C safe.  Therefore, for $\alpha=1$, the jets are IRC safe.  Different choices of component densities $f$ in Eq.~\ref{eq:mm2} give rise to different IRC safe MM jet algorithms.  Several possibilities for $f$ have been studied, but for the remainder of this section uses a (wrapped) Gaussian\footnote{When $f$ is a circular step function, the algorithm is related to the Snowmass iterative cone algorithm~\cite{snowmass} via the `Snowmass Potential'~\cite{Ellis:2007ib}.} $f=\Phi$.  The resulting fuzzy jets are called modified Gaussian Mixture Model jets (mGMM) and are parameterized by the locations $\mu_j$, the covariance matrices $\Sigma_i$, and the cluster weights $\pi_j$.  For initialization, $\pi_j=1/k$ and $\Sigma_j = I$.  
Since practical procedures for maximizing the modified likelihood in Eq.~\ref{eq:mm2} may converge to stationary points that are not globally optimal, the output of a fuzzy jet algorithm will depend on an initial setting of the cluster parameters $\theta$ and $\pi$.  One simple procedure, used exclusively for the rest of the section, is to seed fuzzy jets based on the output of a sequential recombination jet algorithm. This guarantees an IRC safe initial condition and therefore the entire procedure is IRC safe. 
\subsubsection{Clustering Fuzzy Jets: the EM Algorithm}
\label{sec:EMalgorithm}

One iterative procedure for maximizing the mixture model likelihood in Eq.~\ref{eq:mm} is the {\it Expectation-Maximization} (EM) algorithm~\cite{em1,em2,em3}.  After initializing the cluster locations and prior density $\pi$, the following two steps are repeated:

\begin{description}   
\item[Expectation] Given the current values of $\theta_j$, compute the fuzzy membership probabilities $q_{ij}=\pi_j\Phi(\vec{\rho}_i|\mu_j,\Sigma_j)/\sum_{j'}\pi_{j'}\Phi(\vec{\rho}_i|\mu_{j'},\Sigma_{j'})$.

\item[Maximization] Given $q_{ij}$, maximize the {\it expected modified complete log likelihood} over the parameters $\pi,\mu,\Sigma$.
\end{description}

\noindent The expected modified complete log likelihood has the form

\begin{align}
\label{eq:cll}
\sum_{i=1}^N\sum_{j=1}^kp_{Ti}^\alpha(q_{ij}\log\Phi(\vec{\rho}_i;\vec{\mu}_j,\Sigma_j)+q_{ij}\log\pi_j).
\end{align}

\noindent Note that the expected modified complete log likelihood is not the same as the expected modified log likelihood, shown in Eq.~\ref{eq:mm2}.  They differ in that the complete log likelihood has the second sum outside the logarithm while Eq.~\ref{eq:mm2} has the sum inside the logarithm.  The power of the EM algorithm is that maximizing the complete log likelihood results in iteration scheme that monotonically improves the original log likelihood.  This desirable property of the EM algorithm is still true when $\alpha>0$; for a proof, see Appendix~\ref{sec:emalgo}.  Many choices for $f$ have closed form maxima for the M step; in the Gaussian $f=\Phi$ case outlined above, the updates are given by

\begin{align}
\label{eq:emupdates}
\mu_j^* = \sum_{i=1}^n \tilde{q}_{ij}x_i \hspace{7mm}\Sigma_j^*=\sum_{i=1}^n  \tilde{q}_{ij}(x_i-\mu_j)(x_i-\mu_j)^\mathsf{T}\hspace{7mm}\pi_j^*=\frac{1}{\sum_{i=1}^np_{Ti}^\alpha}\sum_{i=1}^n p_{Ti}^\alpha  \tilde{q}_{ij},
\end{align}

\noindent where $ \tilde{q}_{ij}=q_{ij}p_{Ti}^\alpha/\sum_{l=1}^nq_{lj}p_{Tl}^\alpha$.  The well-known $k$-means clustering algorithm~\cite{macqueen1967} can be recovered as the limit of expectation-maximization in a Gaussian mixture model with $\Sigma=\sigma^2 I, \sigma^2\rightarrow 0$.  Figure~\ref{fig:em} illustrates GMM clustering using the EM algorithm with $k = 2$ clusters.  The EM algorithm readily accommodates constraints on the model parameters.  One constraint for simplicity that is used throughout the rest of the section is $\Sigma_j =\sigma_j ^2 I$ for all $j$, which requires the curves of constant likelihood in $(y,\phi)$ to be circular.  The learned value of $\sigma_j$ will be useful for distinguishing jets originating from different physics processes.  Note that since the modified complete log likelihood is IRC safe, the EM algorithm does not break the IRC safety of the original log likelihood.

\begin{figure}[h!]

\begin{center}
\begin{overpic}[width=0.2\textwidth]{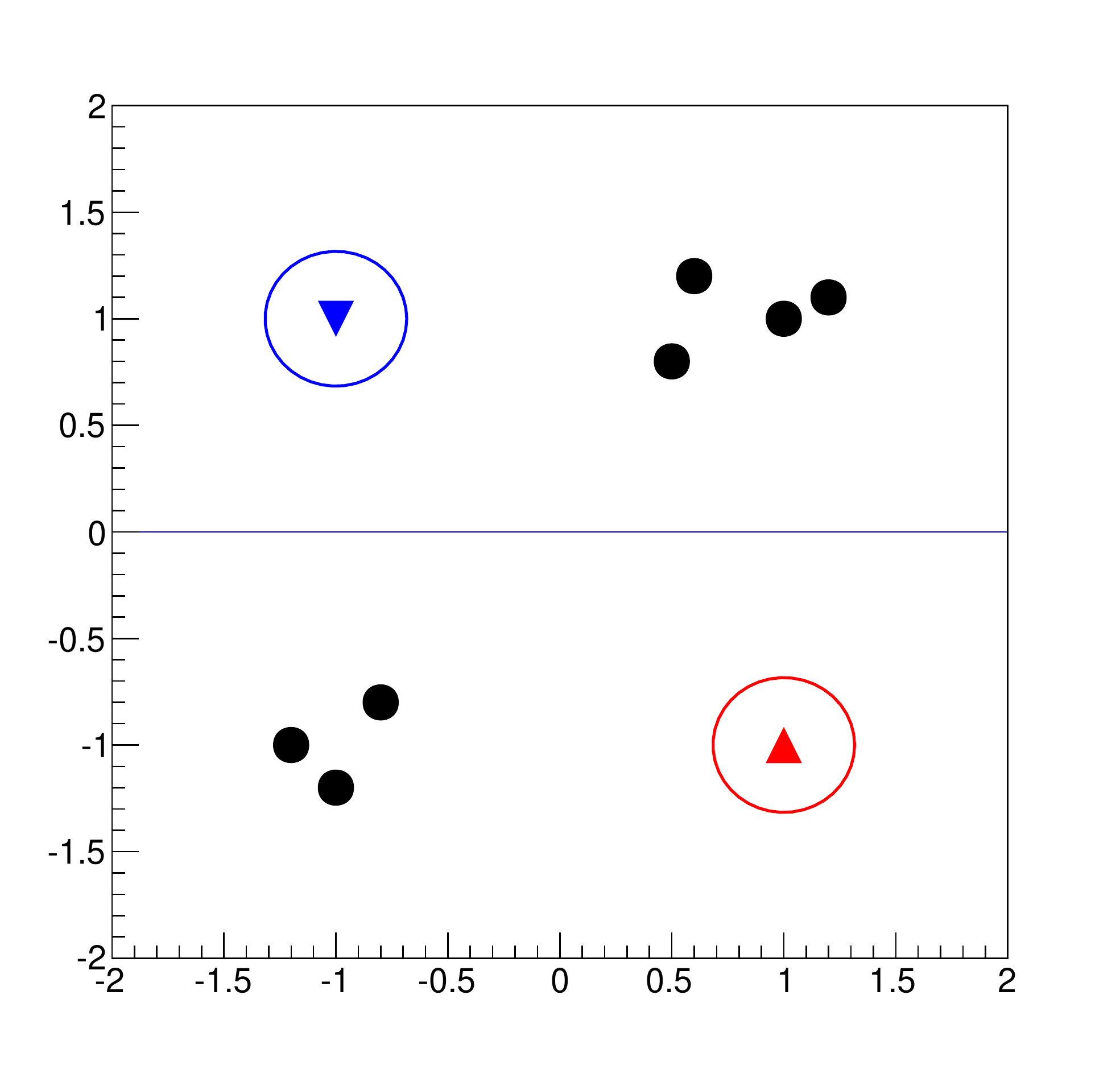}
\put(18,90){\scriptsize Initialization}
\end{overpic}\begin{overpic}[width=0.2\textwidth]{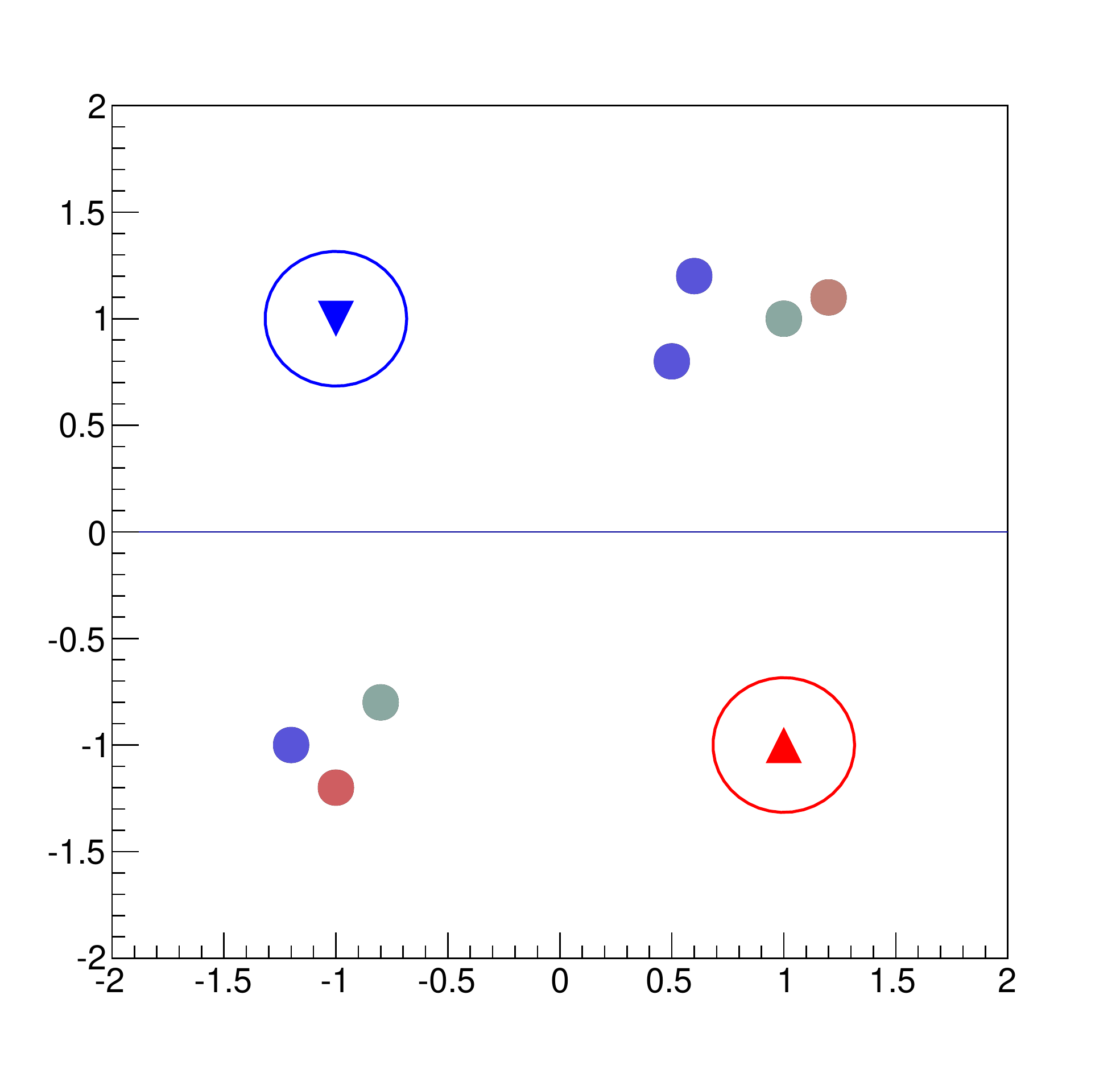}
\put(25,90){\scriptsize $1^\text{st}$ E step}
\end{overpic}\begin{overpic}[width=0.2\textwidth]{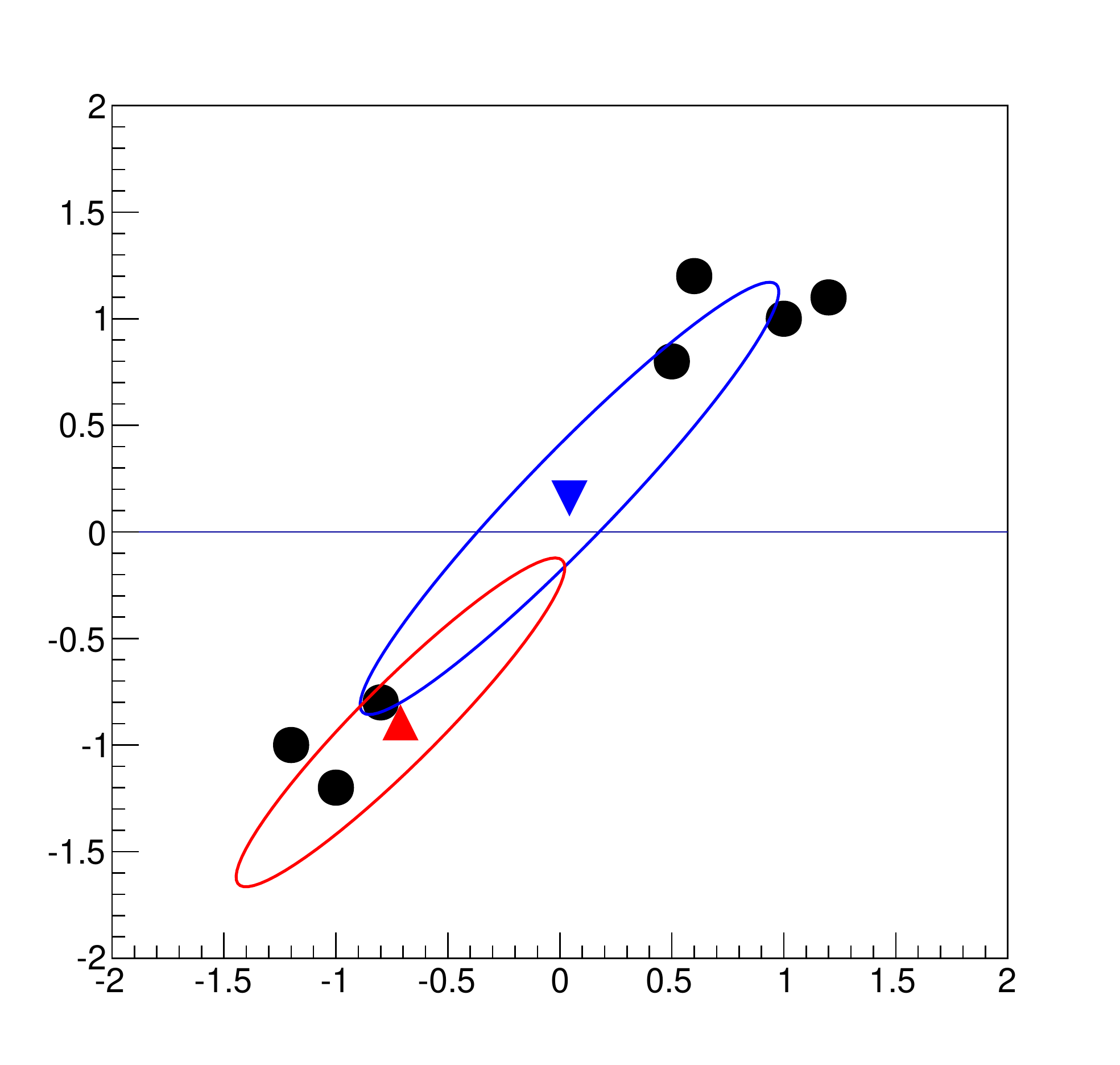}
\put(25,90){\scriptsize $1^\text{st}$ M step}
\end{overpic}\begin{overpic}[width=0.2\textwidth]{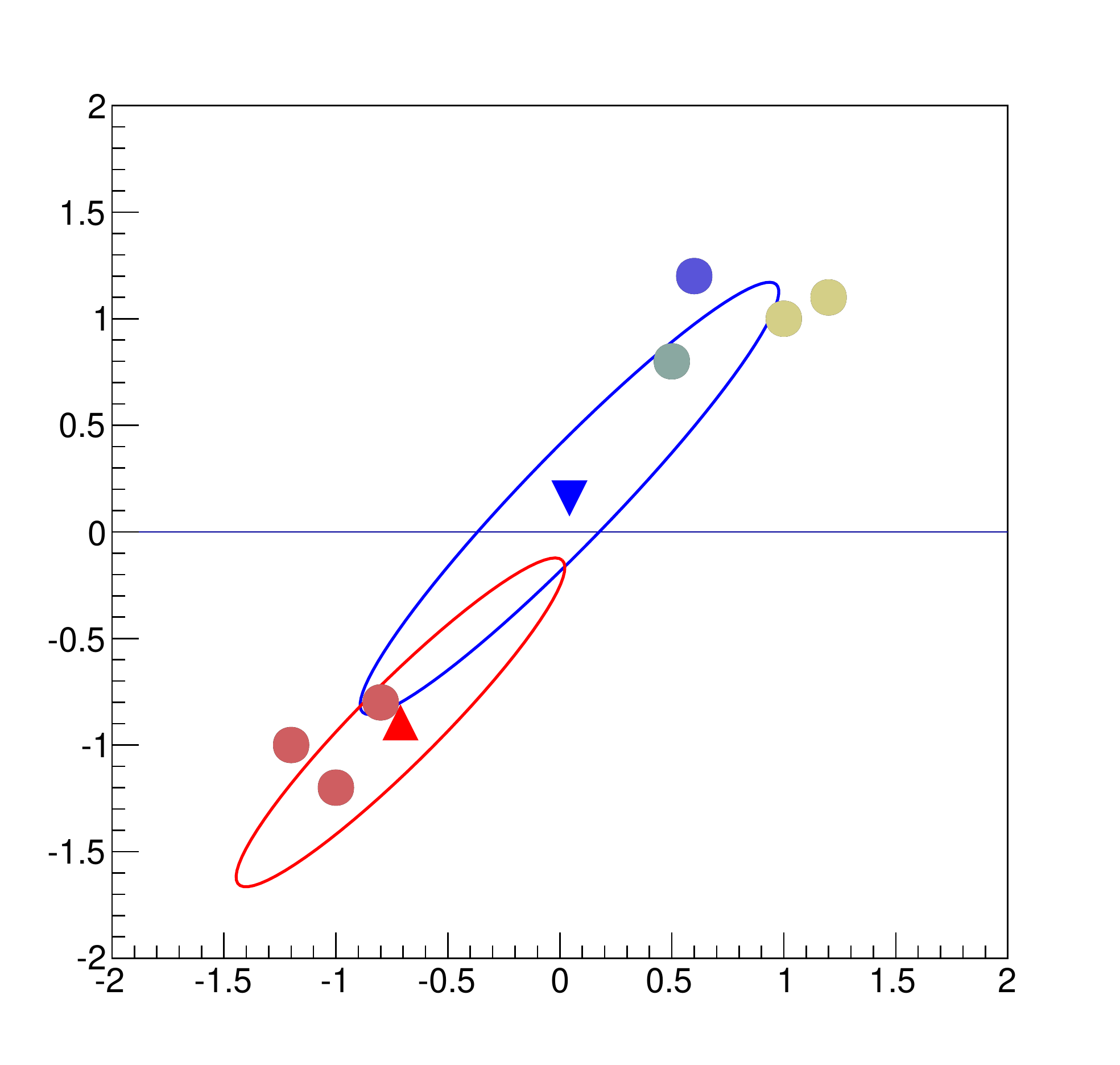}
\put(25,90){\scriptsize $2^\text{nd}$ E step}
\end{overpic}\begin{overpic}[width=0.2\textwidth]{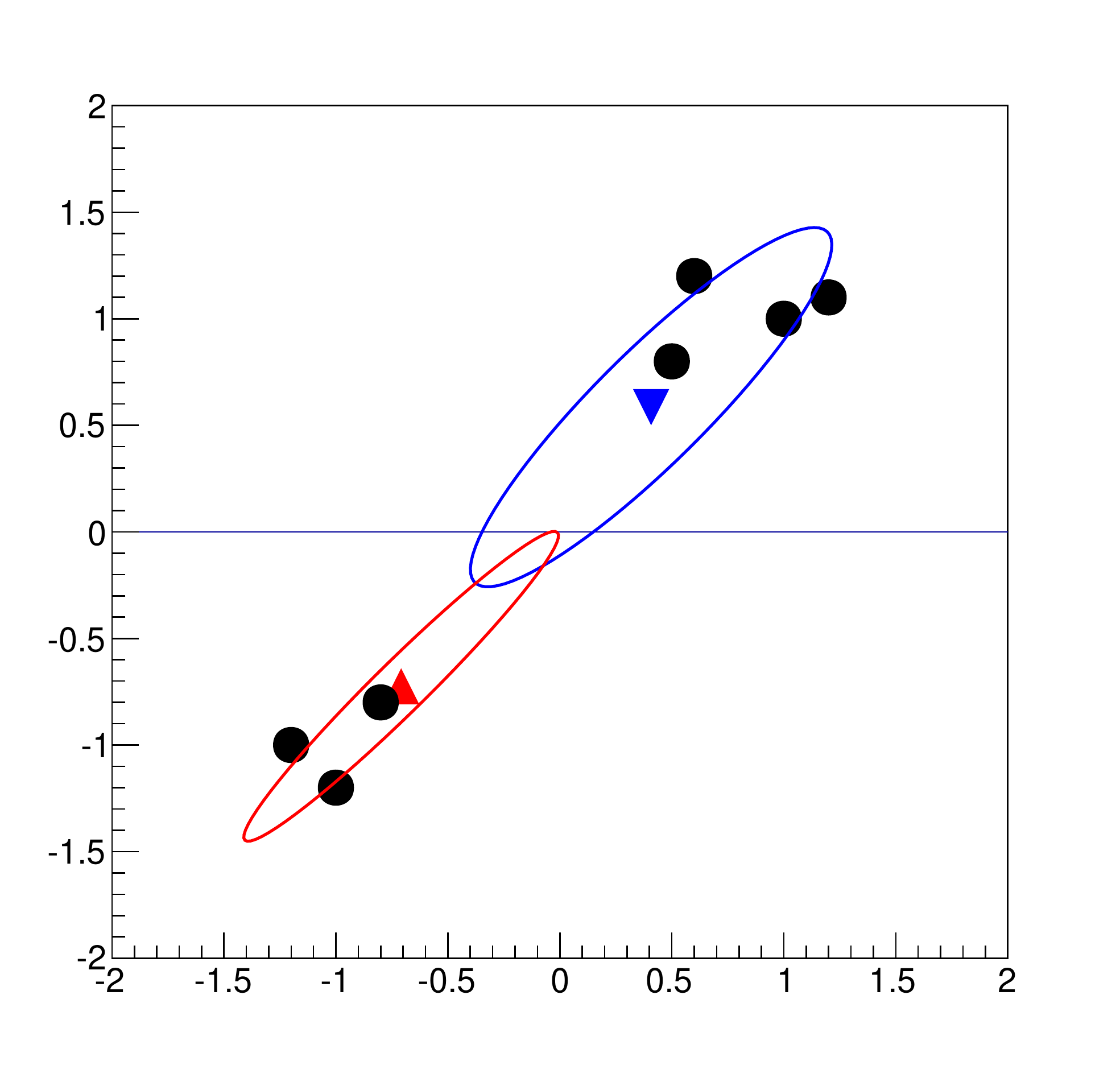}
\put(25,90){\scriptsize $2^\text{nd}$ M step}

\put(-305,0){

\begin{tikzpicture}
  \draw[thick,rounded corners=8pt] (0,1) -- (0,3) -- (6,3) 
   -- (6,0) -- (0,0) -- (0,1);
  \end{tikzpicture}

}

\put(-105,0){

\begin{tikzpicture}
  \draw[thick,rounded corners=8pt] (0,1) -- (0,3) -- (6,3) 
   -- (6,0) -- (0,0) -- (0,1);
  \end{tikzpicture}

}

\end{overpic}\\

\vspace{10mm}

\begin{overpic}[width=0.2\textwidth]{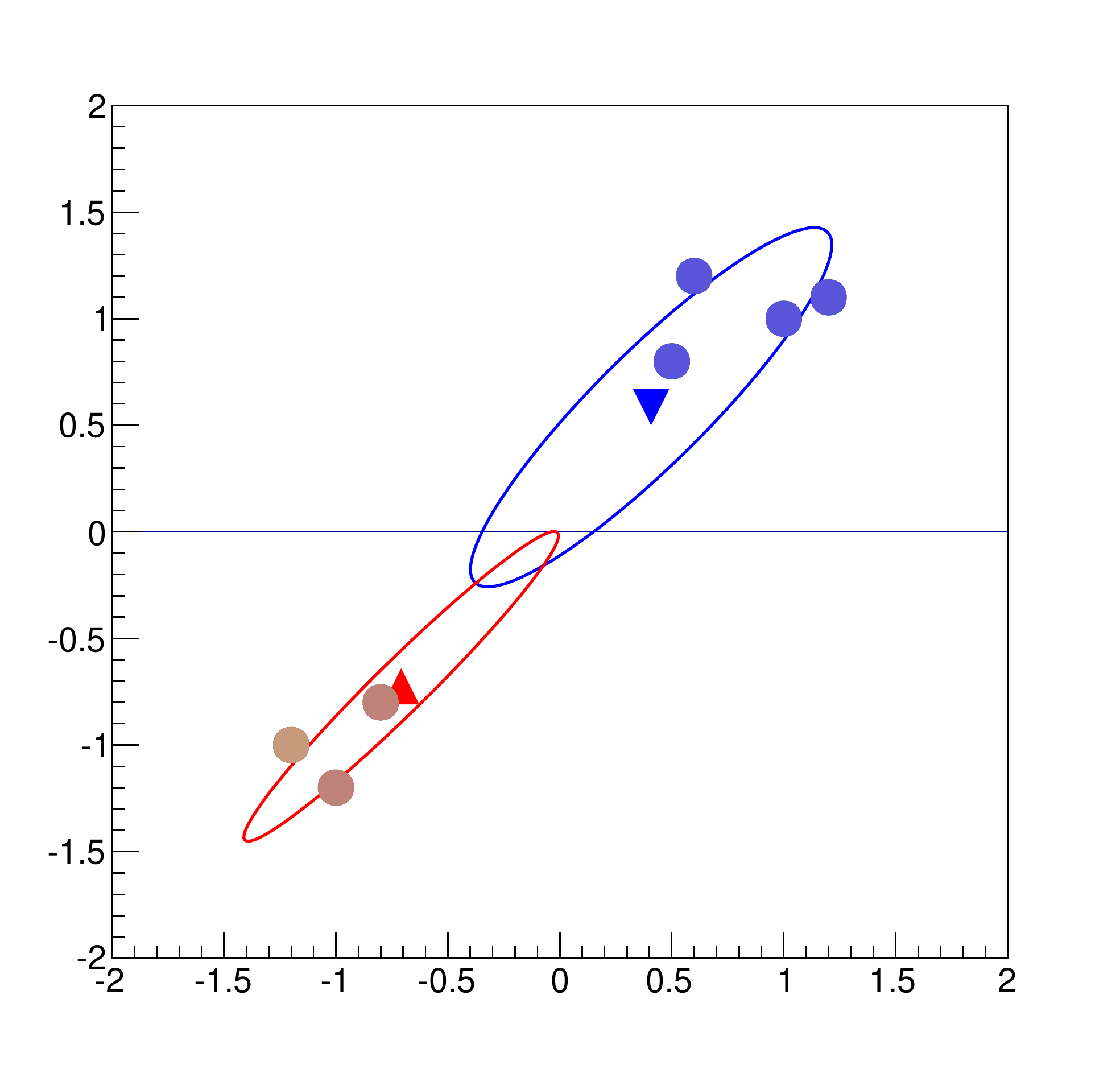}
\put(25,90){\scriptsize $3^\text{rd}$ E step}
\end{overpic}\begin{overpic}[width=0.2\textwidth]{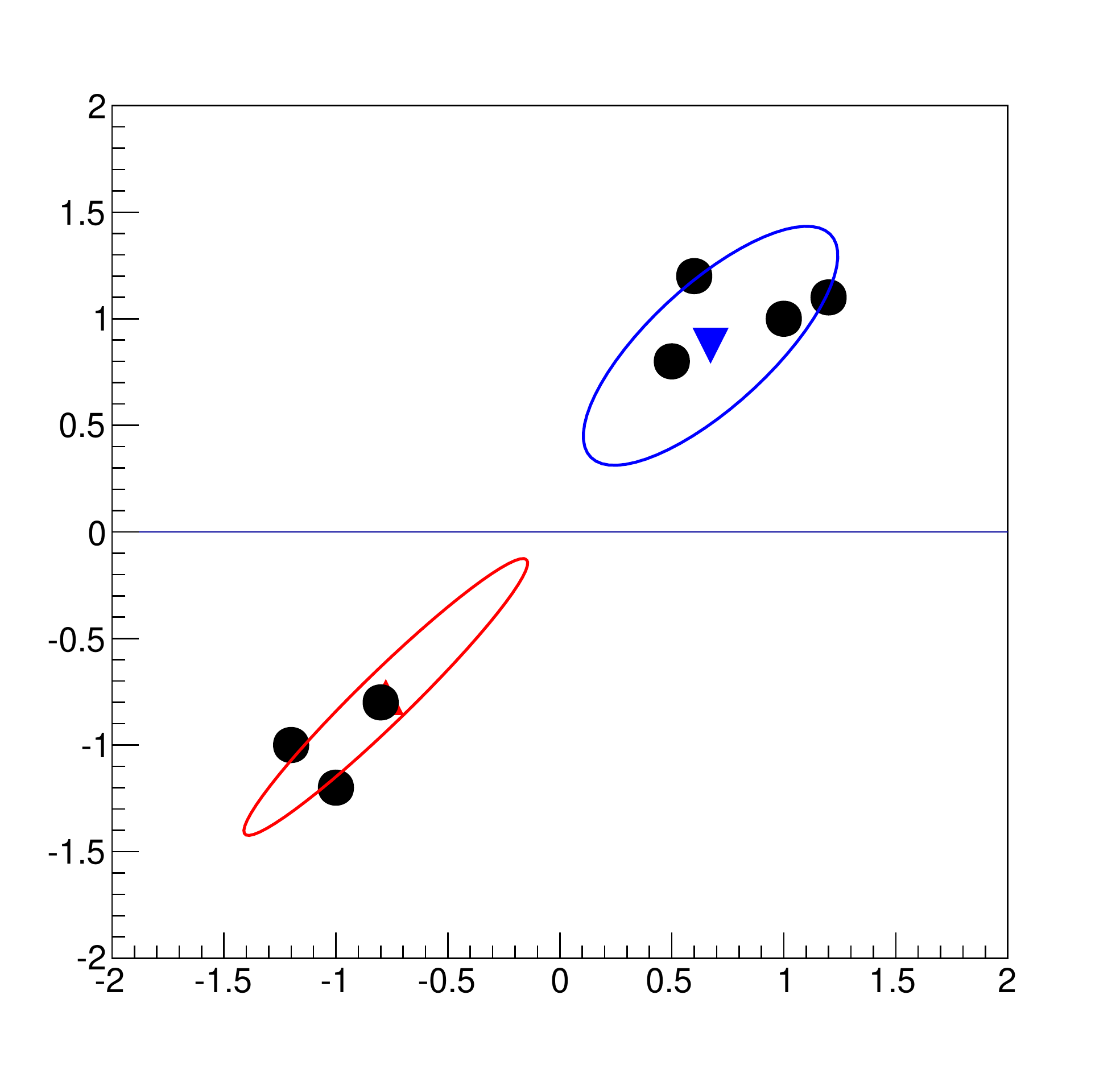}
\put(25,90){\scriptsize $3^\text{rd}$ M step}
\end{overpic}\begin{overpic}[width=0.2\textwidth]{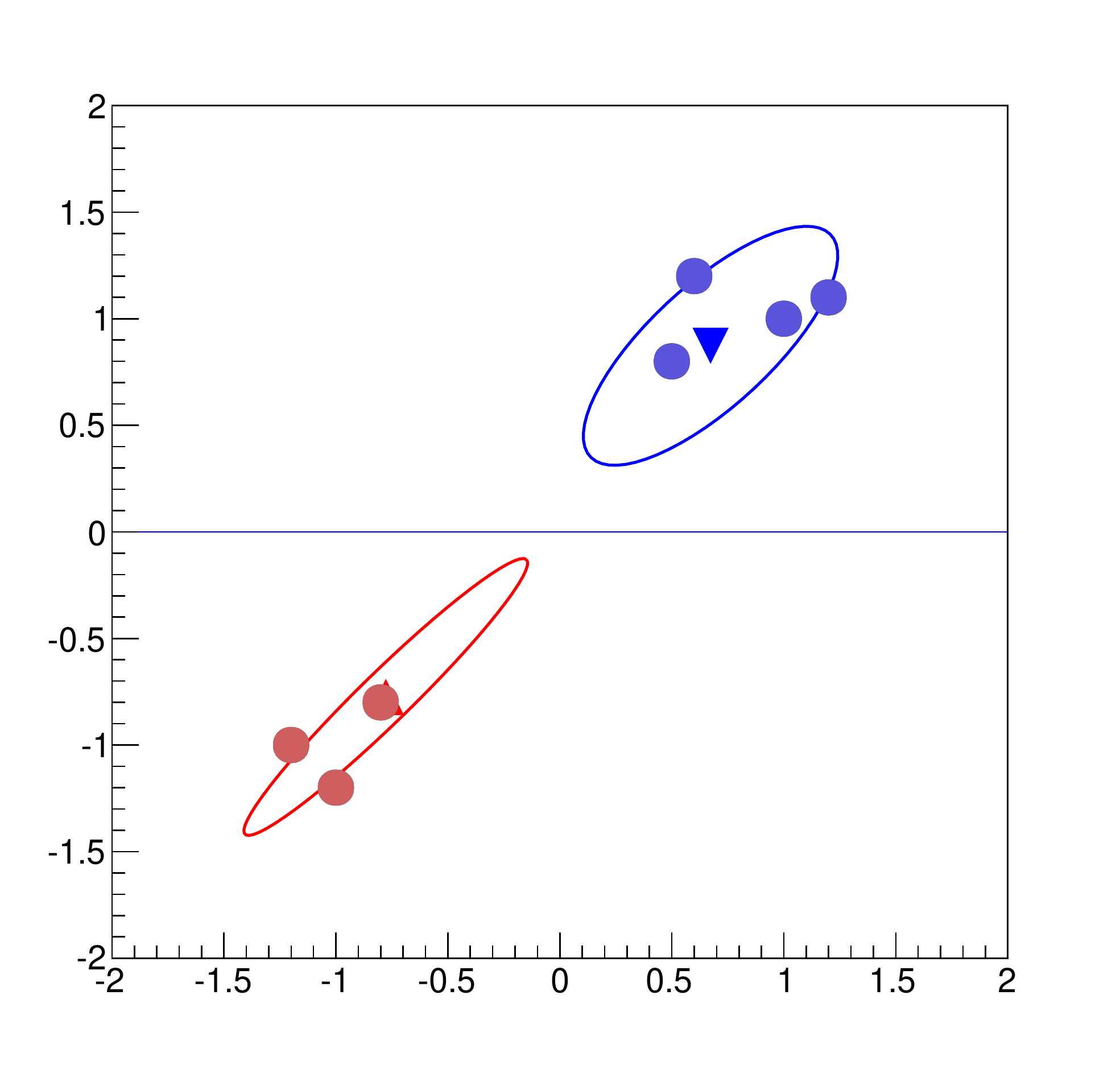}
\put(25,90){\scriptsize $4^\text{th}$ E step}
\end{overpic}\begin{overpic}[width=0.2\textwidth]{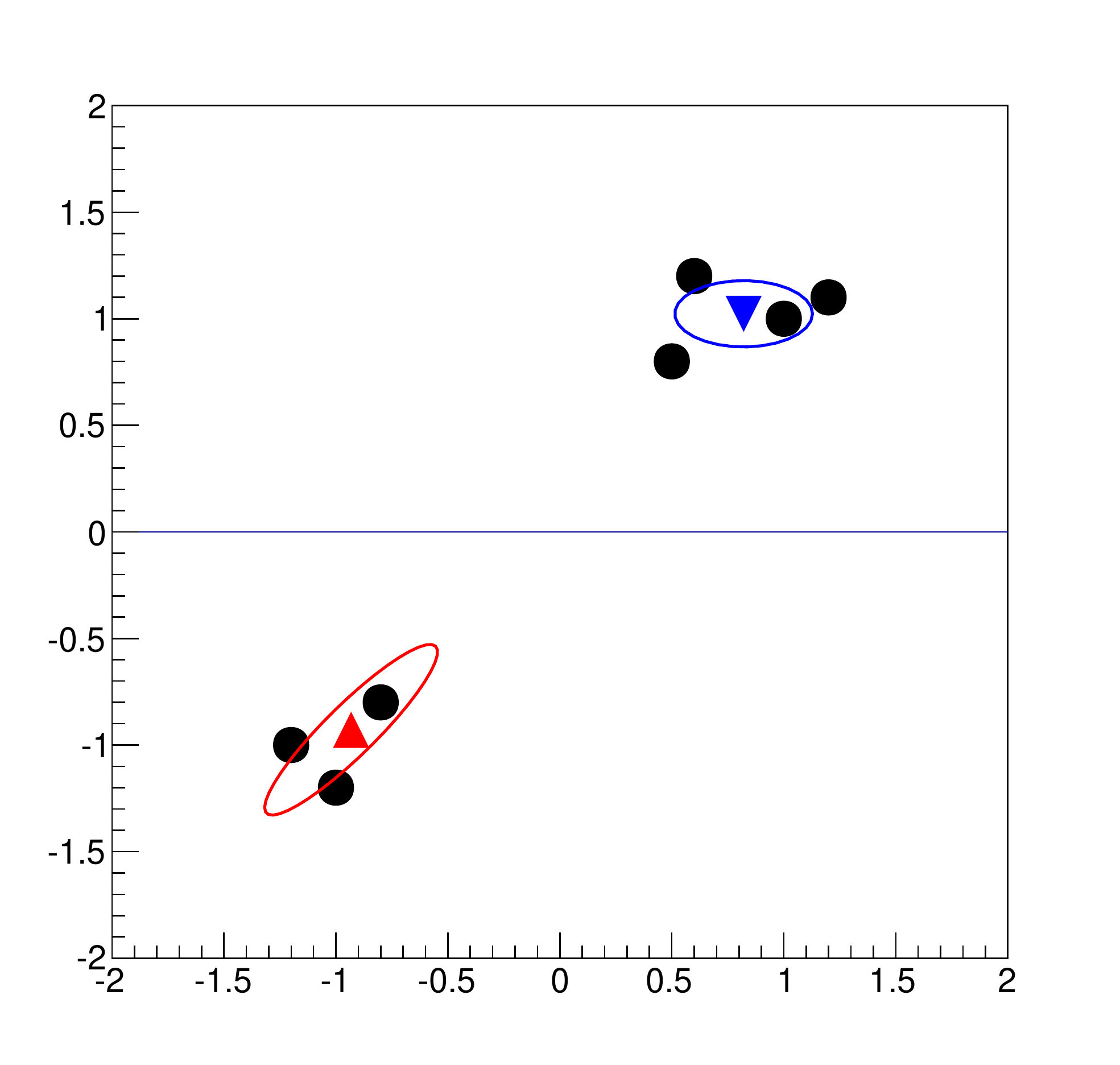}
\put(25,90){\scriptsize $4^\text{th}$ M step}
\end{overpic}\begin{overpic}[width=0.2\textwidth]{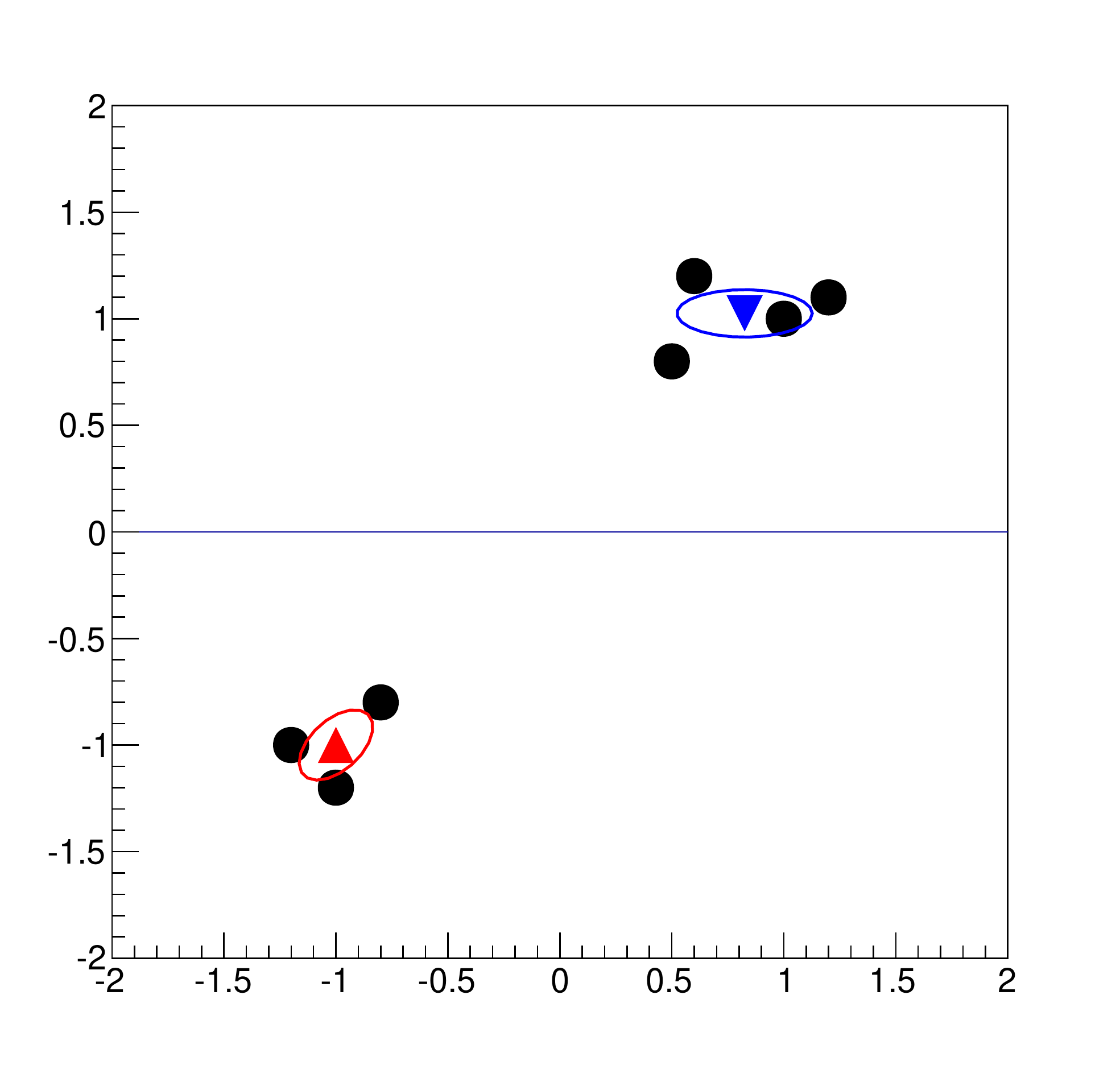}
\put(21,103){\scriptsize $5^\text{st}$ M step:}
\put(22,90){\scriptsize\color{red} Converged}

\put(-405,0){

\begin{tikzpicture}
  \draw[thick,rounded corners=8pt] (0,1) -- (0,3) -- (6,3) 
   -- (6,0) -- (0,0) -- (0,1);
  \end{tikzpicture}

}

\put(-205,0){

\begin{tikzpicture}
  \draw[thick,rounded corners=8pt] (0,1) -- (0,3) -- (6,3) 
   -- (6,0) -- (0,0) -- (0,1);
  \end{tikzpicture}

}

\end{overpic}

\caption{An illustration of of the EM algorithm for $k=2$.  The circles represent data points, the triangles represent the estimated cluster locations $\mu_j$, and the ellipsoids are equidensity contours describing the shapes $\Sigma_j$ of the learned cluster distributions.  In the E-step, bluer colors correspond to higher value of $p_{i,\text{\color{blue} blue jet}}$.}
\label{fig:em}
\end{center}
\end{figure}

\subsubsection{Comparisons with Sequential Recombination and Jet Tagging}

This section describes some numerical comparisons between sequential recombination and fuzzy jets.  Section~\ref{sec:details} summarizes the simulation details with some first event displays showing both fuzzy and sequential recombination jets.  These two approaches to jet clustering are studied over an ensemble of events in Sec.~\ref{sec:compare}.  A third subsection, Sec.~\ref{sec:newinfo},  illustrates that fuzzy jets captures new information about the hadronic final state, and in the fourth section, Sec~\ref{sec:tagging}, it is demonstrated that this new information can be used to classify the jet type.

\paragraph{Details of the Simulation} \mbox{} \\
\label{sec:details}

Simulated $W'$, $Z'$, and QCD multijet events are generated using the same setup as in Sec.~\ref{sec:reluster:particlelevelperformance}.  Large-radius $R=1.0$ anti-$k_t$ trimmed jets with $k_t$ $R=0.3$ subjets groomed with $f_\text{cut}=0.05$ are used as a benchmark.  These jets are also used to seed the fuzzy jets using a threshold of $5$ GeV\footnote{This low threshold guarantees that there are enough seed jets around to capture the radiation from the underlying event.  Another strategy could be to use the {\it event jet} (see Sec.~\ref{fuzzypileup}) even when there is no pileup.}.  The choice of the parameters for the anti-$k_t$ jet seeds is akin to the radius parameter $R$ in the usual sequential recombination paradigm in that they can have a significant impact on the clustered jet properties.  In complete analogy to the choice of $R$, the choice of seed jet parameters will depend on the targeted final state and the initial event conditions (e.g. pileup).  

The EM algorithm for fuzzy jet clustering is terminated when the per iteration increase in the log likelihood is less than $10^{-6}$ for five consecutive iterations, or when a maximum of $100$ iterations is reached. In practice most events converge after a much smaller number of iterations than this bound, with only a small fraction of events stopping for lack of convergence, and then only in high pileup scenarios ($n_\text{PU} > 80$).

To model the discretization and finite acceptance of a real detector, a calorimeter of towers with size $0.1\times 0.1$ in $(y,\phi)$ extends out to $y=5.0$.  The total momentum of the simulated particles incident upon a particular cell are added as scalars and the four-vector $p_j$ of any particular tower $j$ is given by

\begin{align}
\label{eq:calo}
p_j = \sum_{i\text{ incident on $j$}}E_i(\cos\phi_j/\cosh y_j,\sin\phi_j/\cosh y_j,\sinh y_j/\cosh y_j,1).
\end{align}

\noindent Without any corrections, fuzzy jets are significantly sensitive to pileup (see Sec.~\ref{fuzzypileup}).  One simple way to mitigate this sensitivity is to use a local pileup mitigation technique such as charged-hadron-subtraction, by which charged pileup particles (identified by their primary vertex) are subtracted from towers within the acceptance of the tracker $|\eta|<2.5$.  
\begin{figure}[h!]
\vspace{1cm}
\begin{center}
\begin{tabular}{cc}
\begin{overpic}[width=0.5\textwidth]{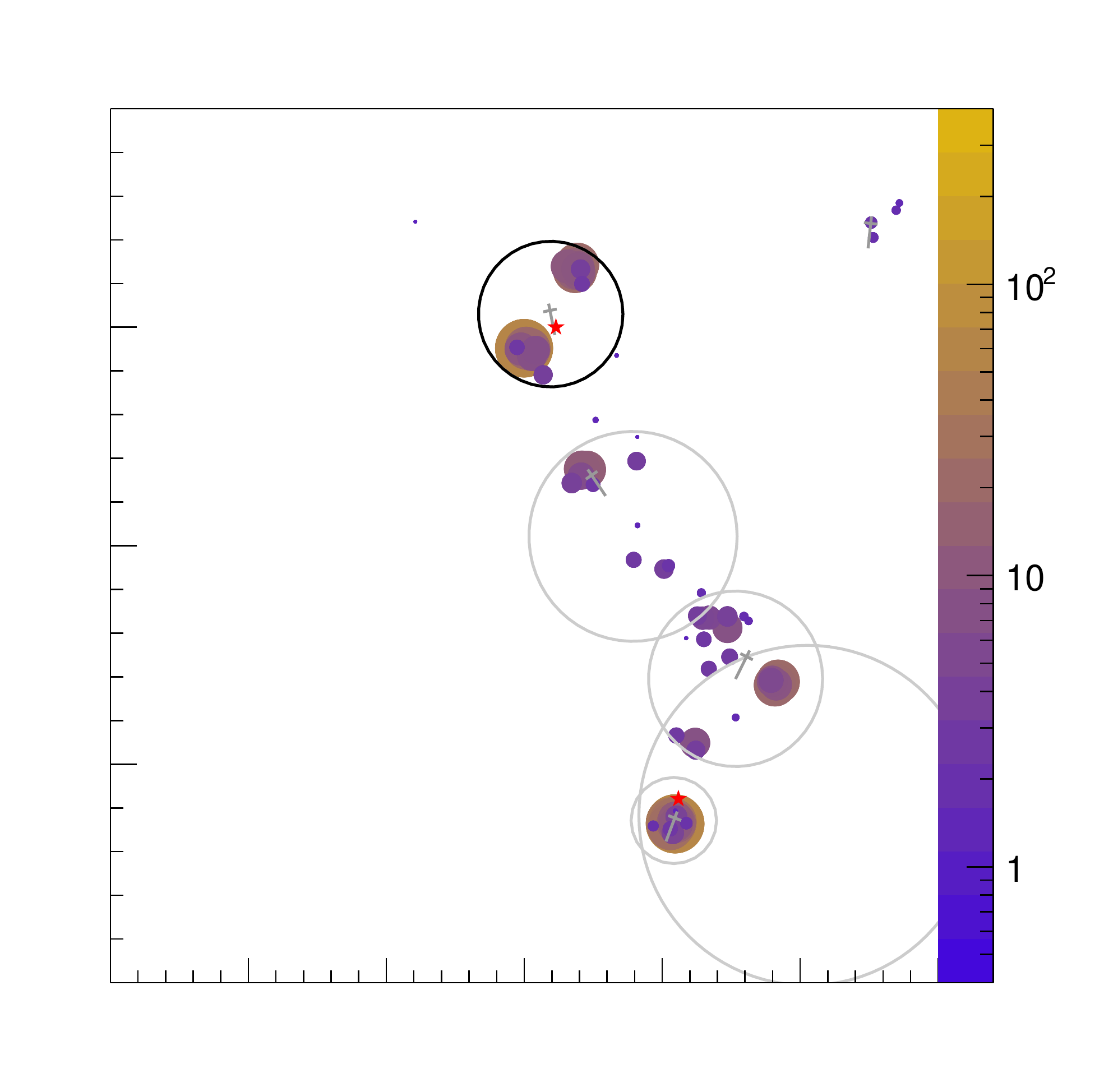}
\put(12, 91){$Z'\rightarrow t\bar{t}$, $\sqrt{s}=8$ TeV, {\sc Pythia} 8}
 
\put(24, -4){ \small Pseudorapidity ($\eta$)}
\put(-4, 18){\rotatebox{90}{ \small Rotated Azimuthal Angle
    ($\phi$)}}
\put(95, 25){\rotatebox{90}{ \small Particle $p_\text{T} \text{ [GeV]}$}}

\put(5, 8){\bfseries \small \sffamily $0$}
\put(4, 28.3){\bfseries \small \sffamily $\frac{\pi}{2}$}
\put(5, 48.3){\bfseries \small \sffamily $\pi$}
\put(4, 68){\bfseries \small \sffamily $\frac{3\pi}{2}$}
\put(4, 88){\bfseries \small \sffamily $2\pi$}

\put(4,  4){\bfseries \small \sffamily $-3$}
\put(17, 4){\bfseries \small \sffamily $-2$}
\put(29.6, 4){\bfseries \small \sffamily $-1$}
\put(46.2, 4){\bfseries \small \sffamily $0$}
\put(58.8, 4){\bfseries \small \sffamily $1$}
\put(71.3, 4){\bfseries \small \sffamily $2$}
\put(83.8, 4){\bfseries \small \sffamily $3$}

\end{overpic} &
\hspace{0.5cm}
\begin{overpic}[width=0.5\textwidth]{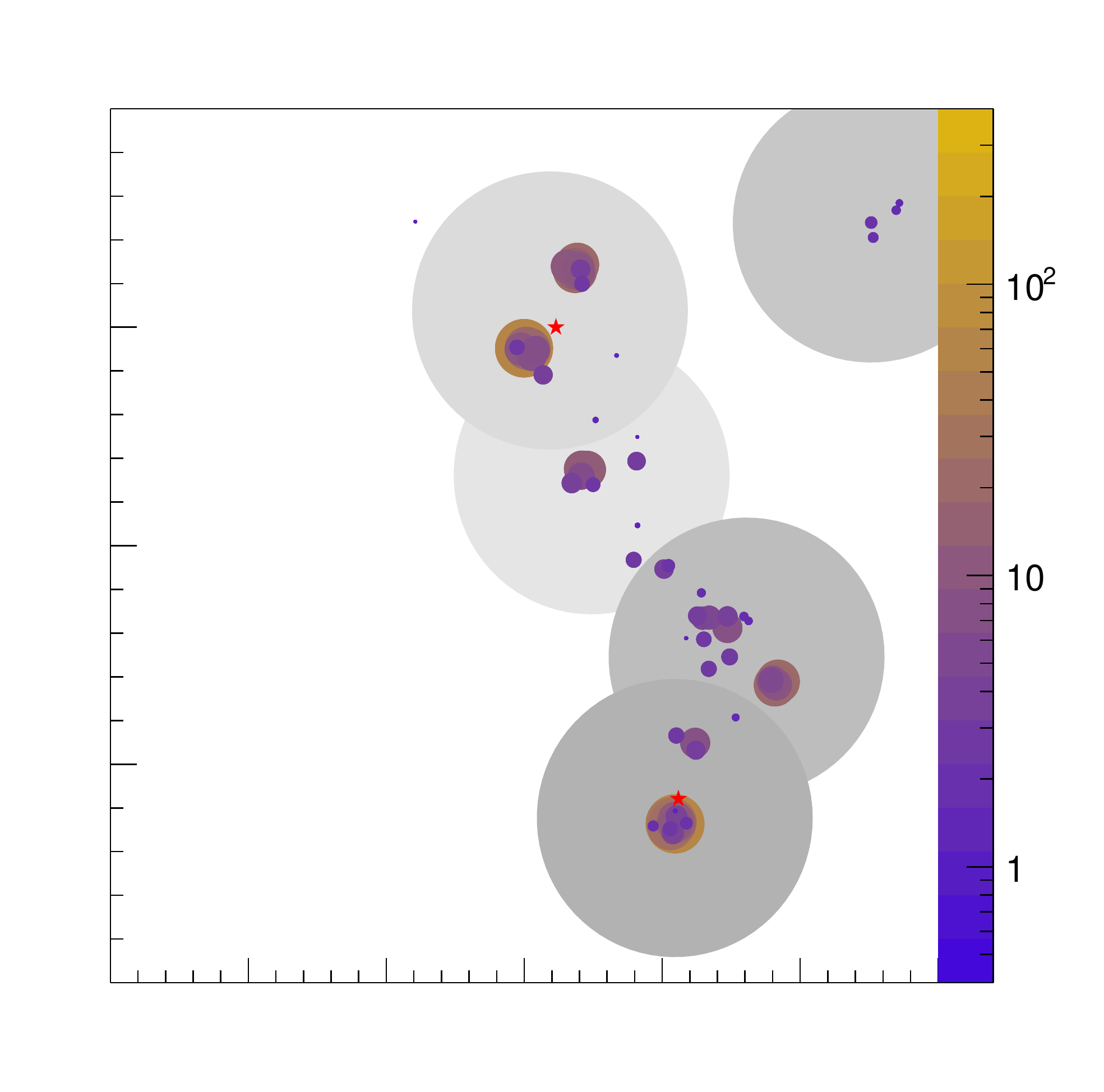}
\put(12, 91){$Z'\rightarrow t\bar{t}$, $\sqrt{s}=8$ TeV, {\sc Pythia} 8}

\put(24, -4){ \small Pseudorapidity ($\eta$)}
\put(-4, 18){\rotatebox{90}{ \small Rotated Azimuthal Angle
    ($\phi$)}}
\put(95, 25){\rotatebox{90}{ \small Particle $p_\text{T} \text{ [GeV]}$}}

\put(5, 8){\bfseries \small \sffamily $0$}
\put(4, 28.3){\bfseries \small \sffamily $\frac{\pi}{2}$}
\put(5, 48.3){\bfseries \small \sffamily $\pi$}
\put(4, 68){\bfseries \small \sffamily $\frac{3\pi}{2}$}
\put(4, 88){\bfseries \small \sffamily $2\pi$}

\put(4,  4){\bfseries \small \sffamily $-3$}
\put(17, 4){\bfseries \small \sffamily $-2$}
\put(29.6, 4){\bfseries \small \sffamily $-1$}
\put(46.2, 4){\bfseries \small \sffamily $0$}
\put(58.8, 4){\bfseries \small \sffamily $1$}
\put(71.3, 4){\bfseries \small \sffamily $2$}
\put(83.8, 4){\bfseries \small \sffamily $3$}
\end{overpic} \vspace{1.6cm}

\\
\end{tabular}

\end{center}
\vspace{-12mm}
\caption{A $Z'\rightarrow t\bar{t}$
  event clustered with fuzzy jets (left) or anti-$k_t$ $R=1$ (right) without pileup ($n_\text{PU}=0$).  The circles indicate the $1\sigma$ contour (fuzzy jets) or the ungroomed jet area (anti-$k_t$).   The small filled colored circles are the particles, with the color and size indicating their energy.  The top quark locations from the generator-record are indicated by red stars.  In the left plot, anti-$k_t$ jet locations are shown with gray crosses where the long tail points towards the mGMM jet
  for which it was a seed.  In the right plot, darker discs correspond to higher $p_\text{T}$ jets.  The highest $p_\text{T}$ fuzzy jet has a black $1\sigma$ contour while all others are shown in gray.}  \label{fig:eventdisplay}
\end{figure}

A representative $Z'\rightarrow t\bar{t}$ event is shown in Figure~\ref{fig:eventdisplay}.  In contrast to the anti-$k_t$ jets, fuzzy jets vary widely in radial size.  The jets centered around the top quark locations did not move far from their anti-$k_t$ seed jets, though the final size is much smaller than one.  The lower $p_\text{T}$ fuzzy jets moved a long distance from the seed jet location and are bigger than $1$ in order to accommodate the diffuse radiation in the event.  Another new feature of fuzzy jets compared to anti-$k_t$ jets is that they can overlap with each other.  Overlapping mGMM jets are an expression of structure inadequately captured with a single Gaussian shape. The ability to learn features at different
scales in the same event without relying on a size parameter like the anti-$k_t$ radius parameter can give mGMM fuzzy jets additional
descriptive power over anti-$k_t$ and other traditional jet algorithms. 
\paragraph{Kinematic Properties of Fuzzy Jets} \mbox{} \\
\label{sec:compare}

Due to the $p_\text{T}$ weighting in the event likelihood, the hard mGMM jets (under HML) have a similar location and total energy as the leading anti-$k_t$ jets.   This is demonstrated by Fig.~\ref{fig:kinematics_pt}, which shows that the $p_\text{T}$ spectrum of the leading mGMM jet is nearly identical to spectrum for the leading anti-$k_t$ jet.    The mGMM algorithm differs from the anti-$k_t$ algorithm in the size and structure of clustered jets.  One important variable sensitive to the distribution of energy within a jet is the jet mass. Figure~\ref{fig:kinematics_mass} shows the jet mass distribution for the same jets as in Fig.~\ref{fig:kinematics_pt}, still using the HML scheme.  Even though the two algorithms learn a similar core, the mass distributions are significantly different.  Both mass distributions show clear peaks near the $W$ boson and top quark masses, but the size and shape of the peaks differs by algorithm.  The $W$ mass peak is higher using fuzzy jets for both the $W'$ and $Z'$ processes.  In $Z'$ events, the fuzzy jets tend to resolve the three-prong structure of top quark jets into two (often overlapping) fuzzy jets.  One of these jets captures the hadronic $W$ decay while the other corresponds to the $b$-jet.  The low mass peak for the $W'$ in the right plot of Fig.~\ref{fig:kinematics_mass} occurs when the fuzzy jets decompose a single boosted $W$ jet into two jets that each have a QCD-like jet mass.  The trend toward lower masses is also observed for the leading jet in QCD multijets.

\begin{figure}[h!]
\begin{center}
\begin{tabular}{cc}
\begin{tikzpicture}
  \node[anchor=south west,inner sep=0] (image) at (0,0)
  {\includegraphics[width=0.43\textwidth]{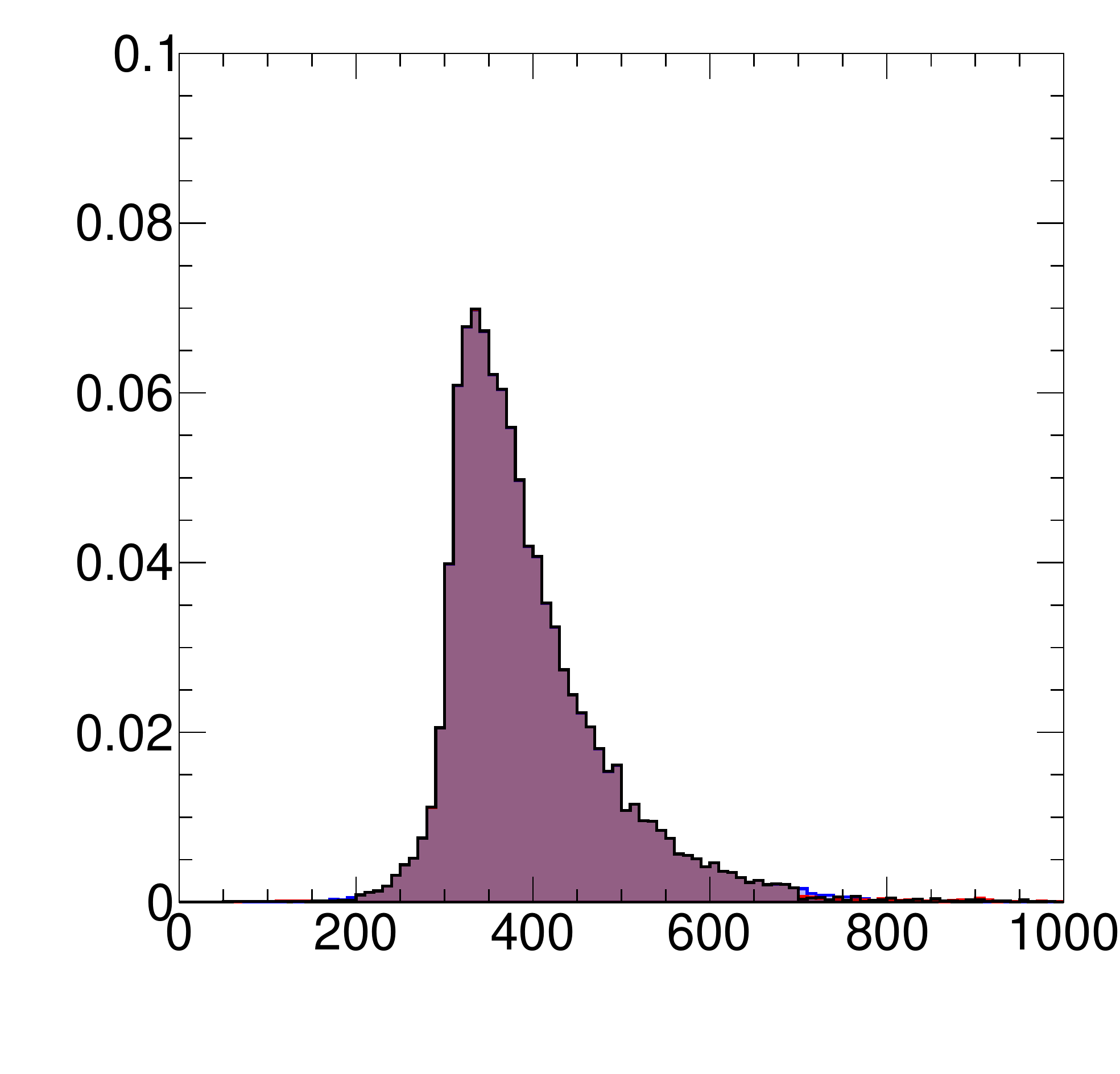}};
  \begin{scope}[x={(image.south east)},y={(image.north west)}]
        \draw[blue,fill=white!72!blue,thick] (0.59,0.85) rectangle
    (0.64,0.9);
    \draw[red,fill=white!72!red,thick] (0.59,0.78) rectangle
    (0.64,0.83);
    \draw[black,fill=white!72!black,thick] (0.59,0.71) rectangle
    (0.64,0.76);
    \node[draw=none, anchor=west] at (0.65, 0.875) { 
      \normalsize $\text{QCD}$};
    \node[draw=none, anchor=west] at (0.65, 0.805) { 
      \normalsize $Z' \rightarrow t\bar{t}$};
    \node[draw=none, anchor=west] at (0.65, 0.73) { 
      \normalsize $W\rightarrow qq'$};

        \node[draw=none] at (0.56,0.065) { \small Leading
      anti-$k_t$ Jet $p_\text{T}$ [GeV]};
    \node[draw=none, rotate=90] at (0.02, 0.55){ \small Arbitrary
    Units};
    \node[draw=none, anchor=west] at (0.17,0.88) { 
      {\sc Pythia} 8};
    \node[draw=none, anchor=west] at (0.17,0.80) { 
      $\sqrt{s} = 8 \text{ TeV}$};
  \end{scope}
\end{tikzpicture} &
\begin{tikzpicture}
  \node[anchor=south west,inner sep=0] (image) at (0,0)
  {\includegraphics[width=0.43\textwidth]{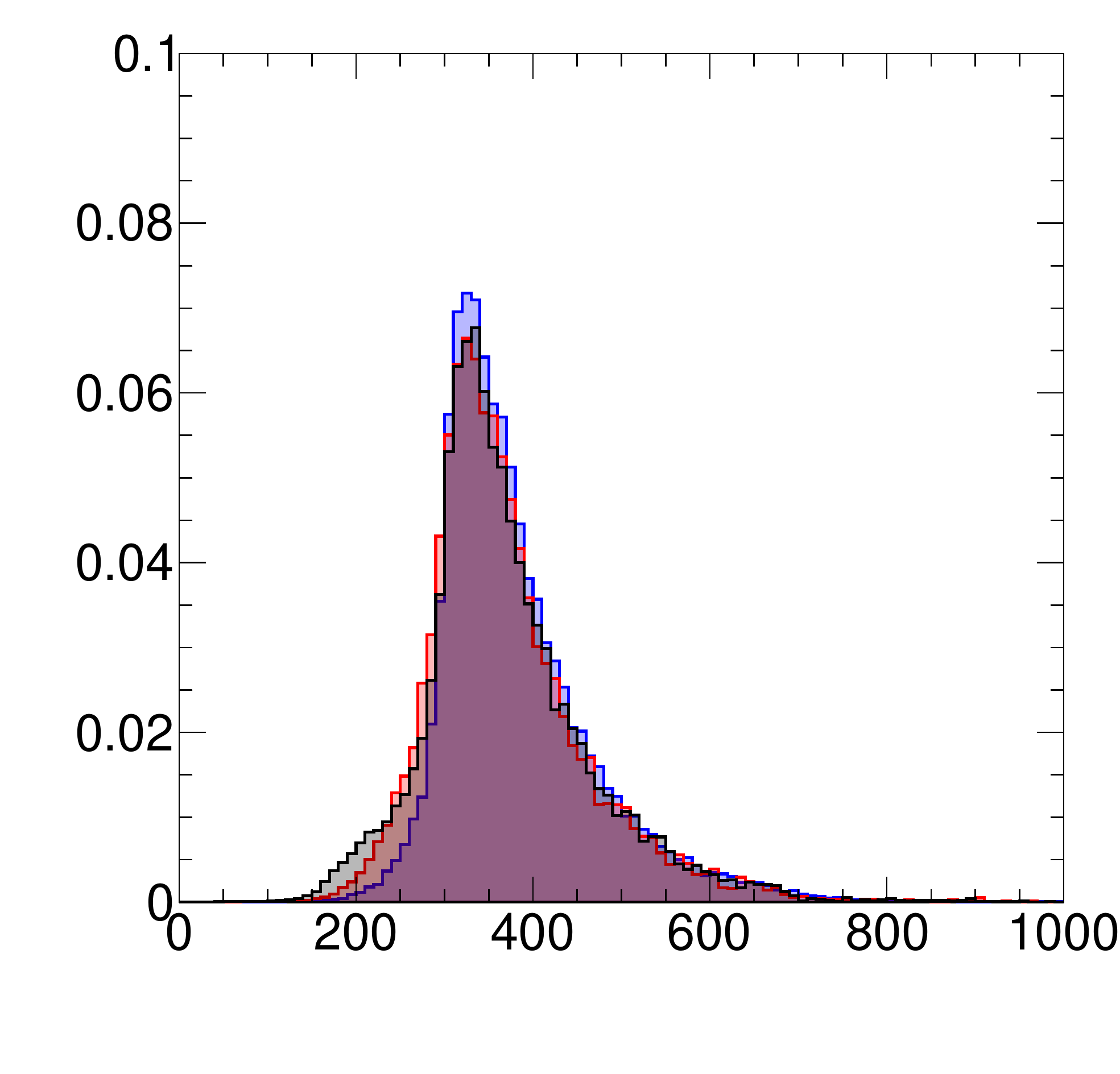}};
  \begin{scope}[x={(image.south east)},y={(image.north west)}]
        \draw[blue,fill=white!72!blue,thick] (0.59,0.85) rectangle
    (0.64,0.9);
    \draw[red,fill=white!72!red,thick] (0.59,0.78) rectangle
    (0.64,0.83);
    \draw[black,fill=white!72!black,thick] (0.59,0.71) rectangle
    (0.64,0.76);
    \node[draw=none, anchor=west] at (0.65, 0.875) { 
      \normalsize $\text{QCD}$};
    \node[draw=none, anchor=west] at (0.65, 0.805) { 
      \normalsize $Z' \rightarrow t\bar{t}$};
    \node[draw=none, anchor=west] at (0.65, 0.73) { 
      \normalsize $W\rightarrow qq'$};

        \node[draw=none] at (0.56,0.065) { \small Leading
      mGMM Jet $p_\text{T}$ [GeV]};
    \node[draw=none, rotate=90] at (0.02, 0.55){ \small Arbitrary
    Units};
    \node[draw=none, anchor=west] at (0.17,0.88) { 
      {\sc Pythia} 8};
    \node[draw=none, anchor=west] at (0.17,0.80) { 
      $\sqrt{s} = 8 \text{ TeV}$};
  \end{scope}
\end{tikzpicture} \\
\end{tabular}
\caption{The jet $p_\text{T}$ for the leading anti-$k_t$ jet (left) and leading fuzzy jet under the HML particle assignment scheme (right).  All the processes are re-weighted so that the anti-$k_t$ $p_\text{T}$ spectra are the same.}
\label{fig:kinematics_pt}
\end{center}
\end{figure}

\begin{figure}[h!]
\begin{center}
\begin{tabular}{cc}
\begin{tikzpicture}
  \node[anchor=south west,inner sep=0] (image) at (0,0)
  {\includegraphics[width=0.43\textwidth]{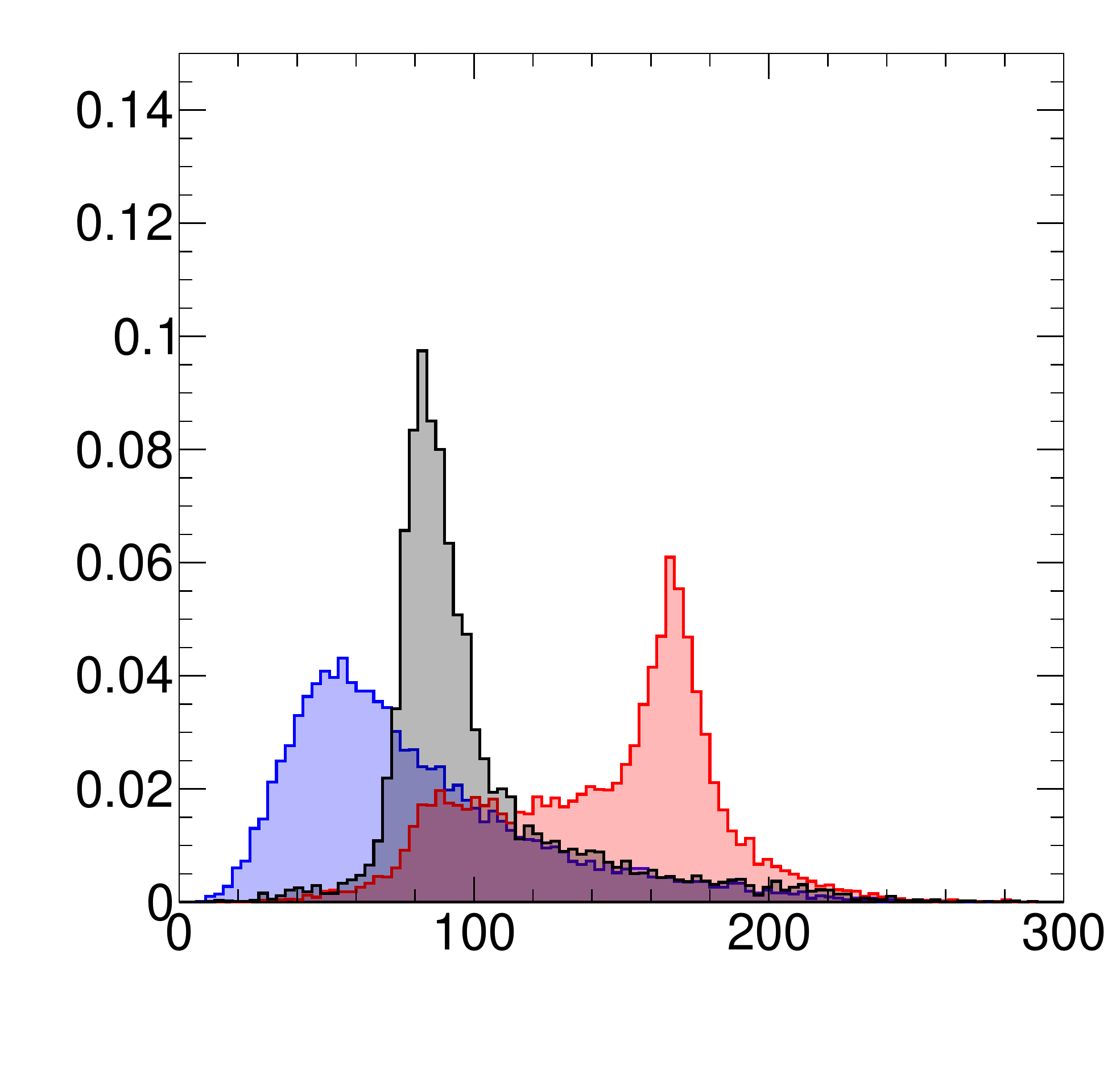}};
  \begin{scope}[x={(image.south east)},y={(image.north west)}]
        \draw[blue,fill=white!72!blue,thick] (0.59,0.85) rectangle
    (0.64,0.9);
    \draw[red,fill=white!72!red,thick] (0.59,0.78) rectangle
    (0.64,0.83);
    \draw[black,fill=white!72!black,thick] (0.59,0.71) rectangle
    (0.64,0.76);
    \node[draw=none, anchor=west] at (0.65, 0.875) { 
      \normalsize $\text{QCD}$};
    \node[draw=none, anchor=west] at (0.65, 0.805) { 
      \normalsize $Z' \rightarrow t\bar{t}$};
    \node[draw=none, anchor=west] at (0.65, 0.73) { 
      \normalsize $W \rightarrow qq'$};

        \draw [dashed, color=white, semithick] (0.3715, 0.1622) -- (0.3715, 0.76);
    \draw [dashed, color=white, semithick] (0.61, 0.1622) -- (0.61, 0.66);

        \node[draw=none] at (0.56,0.065) { \small Leading
      anti-$k_t$ Jet Mass [GeV]};
    \node[draw=none, rotate=90] at (0.02, 0.55){ \small Arbitrary
    Units};
    \node[draw=none, anchor=west] at (0.17,0.89) { 
     {\sc Pythia} 8};
    \node[draw=none, anchor=west] at (0.17,0.82) { 
      $\sqrt{s} = 8 \text{ TeV}$};
    \node[draw=none, anchor=west] at (0.17, 0.76) { \tiny $350 \leq
      p_\text{T}^{\text{Jet}} \leq 450 \text{ GeV}$};
  \end{scope}
\end{tikzpicture} &
\begin{tikzpicture}
  \node[anchor=south west,inner sep=0] (image) at (0,0)
  {\includegraphics[width=0.43\textwidth]{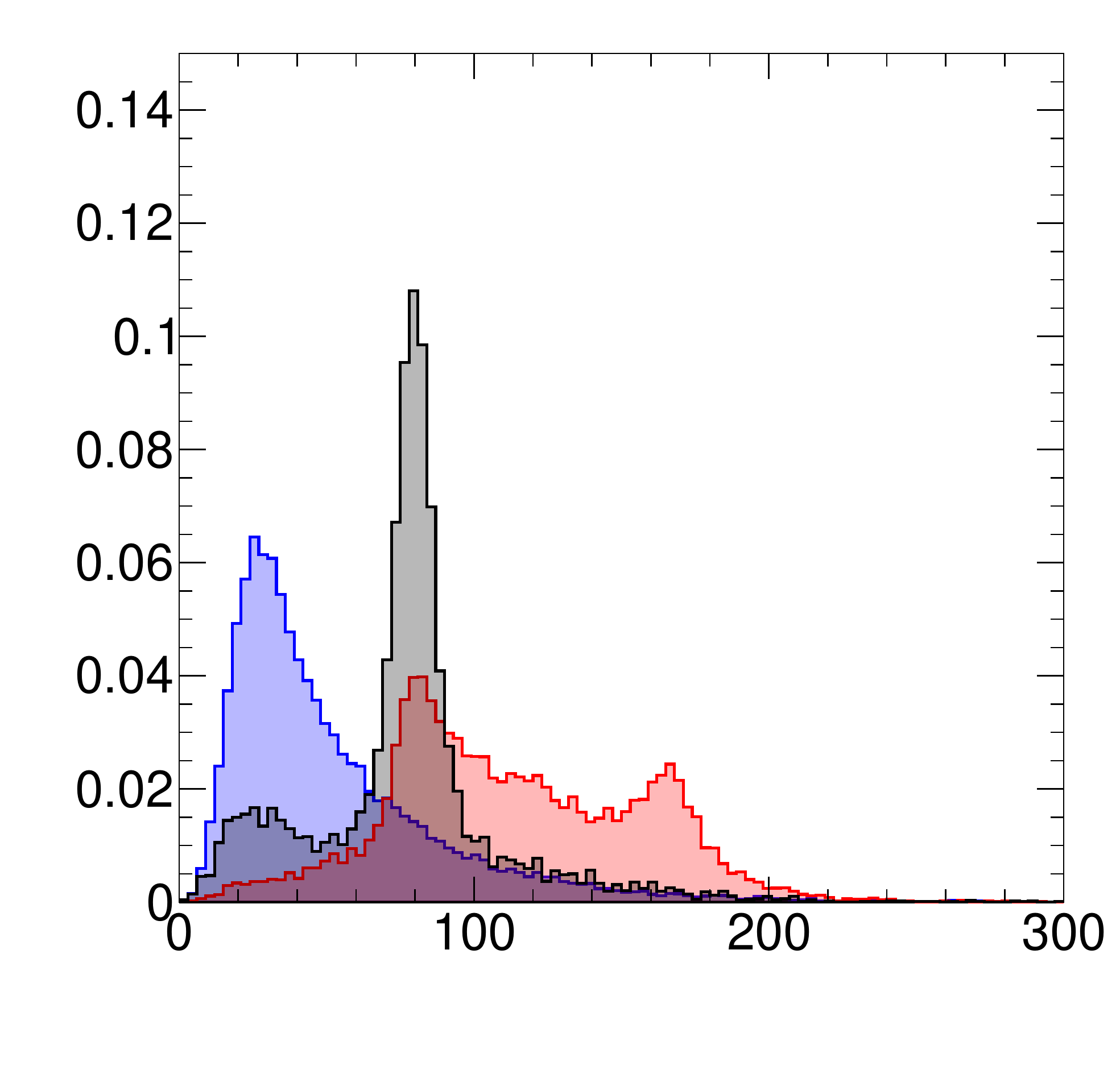}};
  \begin{scope}[x={(image.south east)},y={(image.north west)}]
        \draw[blue,fill=white!72!blue,thick] (0.59,0.85) rectangle
    (0.64,0.9);
    \draw[red,fill=white!72!red,thick] (0.59,0.78) rectangle
    (0.64,0.83);
    \draw[black,fill=white!72!black,thick] (0.59,0.71) rectangle
    (0.64,0.76);
    \node[draw=none, anchor=west] at (0.65, 0.875) { \sffamily
      \normalsize $\text{QCD}$};
    \node[draw=none, anchor=west] at (0.65, 0.805) { \sffamily
      \normalsize $Z' \rightarrow t\bar{t}$};
    \node[draw=none, anchor=west] at (0.65, 0.73) { \sffamily
      \normalsize $W \rightarrow qq'$};

        \draw [dashed, color=white, semithick] (0.3715, 0.1622) -- (0.3715, 0.76);
    \draw [dashed, color=white, semithick] (0.61, 0.1622) -- (0.61, 0.66);

        \node[draw=none] at (0.56,0.065) { \small Leading
      mGMM Jet Mass [GeV]};
    \node[draw=none, rotate=90] at (0.02, 0.55){ \small Arbitrary
    Units};
    \node[draw=none, anchor=west] at (0.17,0.89) { 
     {\sc Pythia} 8};
    \node[draw=none, anchor=west] at (0.17,0.82) { 
      $\sqrt{s} = 8 \text{ TeV}$};
    \node[draw=none, anchor=west] at (0.17, 0.76) { \tiny $350 \leq
      p_\text{T}^{\text{Jet}} \leq 450 \text{ GeV}$};
  \end{scope}
\end{tikzpicture} \\
\end{tabular}
\caption{The jet mass for the leading anti-$k_t$ (left) and
  leading fuzzy jet under the HML particle assignment scheme (right),
  in an anti-$k_t$ leading jet $p_\text{T}$ window of 350 to 450 GeV.
  All the processes are re-weighted so that the anti-$k_t$ $p_\text{T}$
  distributions are the same.  The dashed white lines mark $m_W \approx 80 \text{
    GeV}$ and $m_\text{top} \approx 175 \text{ GeV}$. }
\label{fig:kinematics_mass}
\end{center}
\end{figure}

\clearpage

\paragraph{New Information from Fuzzy Jets}\mbox{} \\
\label{sec:newinfo}

The properties $\Sigma$ of a fuzzy jet can be useful in distinguishing jets resulting from different physics processes.  In the simplest realization of mGMM jets already described above, $\Sigma = \sigma^2 I $, where $\sigma$ is a measure of the size of the core of a jet.  Although $\sigma$ is a simple variable to construct from the
wealth of data available after clustering with the mGMM algorithm, it captures at least some of the
schematic differences in the likelihood for $Z'\rightarrow
t\bar{t}$ and $W'\rightarrow WZ$ relative to a QCD multijet background.   The left plot of Fig.~\ref{fig:sigma_hist} also shows the distribution of $\sigma$ over all fuzzy jets.  The generic jet is nearly independent of the hard-scatter process and tends to be much larger than the usual small-radius jet size ($R=0.4$).  Fuzzy jets capturing the highest $p_\text{T}$ structure in the event tend to be small (as the structure tend to be small), but the rest of the diffuse radiation in the event requires large fuzzy jets spread out over the detector.  The distribution for the leading fuzzy jet $\sigma$ is shown in the right plot of Fig.~\ref{fig:sigma_hist}.  The distribution for the sub-leading jet in signal events is qualitatively similar to the leading jet and is largely uncorrelated.  In background events, the subleading jet is systematically wider than the leading jet.  As expected from the $2m/p_\text{T}$ scaling\footnote{At leading order, there is an exact relationship between $\sigma$ and $m/p_\text{T}$ - See Appendix~\ref{sec:leadingordersigma}.} of the jet size, the right plot of Fig.~\ref{fig:sigma_hist} shows that top quark jets have a larger $\sigma$ than $W$ jets which have a larger $\sigma$ than generic quark and gluon jets.  However, Fig.~\ref{fig:corr_mpt} shows that $\sigma$ is not $100\%$ correlated with $m/p_\text{T}$; the next section will show that $\sigma$ provides additional information for jet tagging beyond $m/p_\text{T}$.   Note that part of the new information in $\sigma$ is resulting from the clustering procedure itself and not just the definition of the observable.  For example, computing $\sigma$ from the constituents of an anti-$k_t$ jet (i.e. running fuzzy jets on these constituents with $k=1$) would result in

\begin{align}
\sigma^2 = \frac{\sum_{i=1}^n p_{\text{T},i}\Delta R^2}{\sum_{i=1}^n p_{\text{T},i}},
\end{align}

\noindent which is nearly the same as $m/p_\text{T}$.

\begin{figure}[h!]
\begin{center}
\begin{tabular}{cc}
\begin{tikzpicture}
  \node[anchor=south west,inner sep=0] (image) at (0,0)
  {\includegraphics[width=0.43\textwidth]{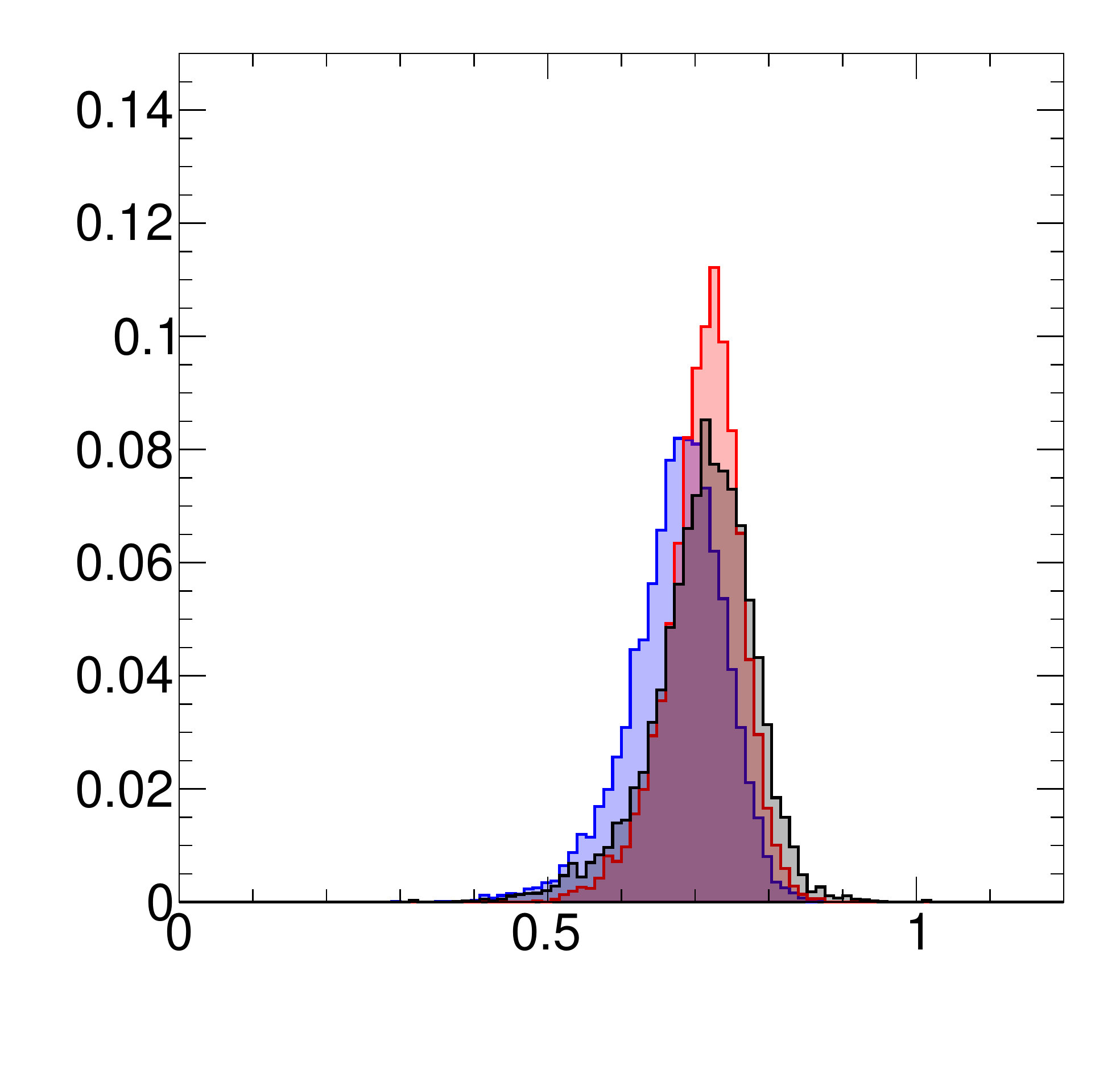}};
  \begin{scope}[x={(image.south east)},y={(image.north west)}]
        \draw[blue,fill=white!72!blue,thick] (0.59-0.38,0.85-0.3) rectangle
    (0.64-0.38,0.9-0.3);
    \draw[red,fill=white!72!red,thick] (0.59-0.38,0.78-0.3) rectangle
    (0.64-0.38,0.83-0.3);
    \draw[black,fill=white!72!black,thick] (0.59-0.38,0.71-0.3) rectangle
    (0.64-0.38,0.76-0.3);
    \node[draw=none, anchor=west] at (0.65-0.38, 0.875-0.3) { 
      \normalsize $\text{QCD}$};
    \node[draw=none, anchor=west] at (0.65-0.38, 0.805-0.3) { 
      \normalsize $Z' \rightarrow t\bar{t}$};
    \node[draw=none, anchor=west] at (0.65-0.38, 0.73-0.3) { 
      \normalsize $W \rightarrow qq'$};

        \node[draw=none] at (0.56,0.065) { \small Leading
      Learned $\sigma$};
    \node[draw=none, rotate=90] at (0.02, 0.55){ \small Arbitrary
    Units};
    \node[draw=none, anchor=west] at (0.17,0.88) { 
       {\sc Pythia} 8};
    \node[draw=none, anchor=west] at (0.17,0.80) { 
      $\sqrt{s} = 8 \text{ TeV}$};
    \node[draw=none, anchor=west] at (0.17, 0.74) { \tiny $350 \leq
      p_\text{T}^{\text{Jet}} \leq 450 \text{ GeV}$};
  \end{scope}
\end{tikzpicture} &

\begin{tikzpicture}
  \node[anchor=south west,inner sep=0] (image) at (0,0)
  {\includegraphics[width=0.43\textwidth]{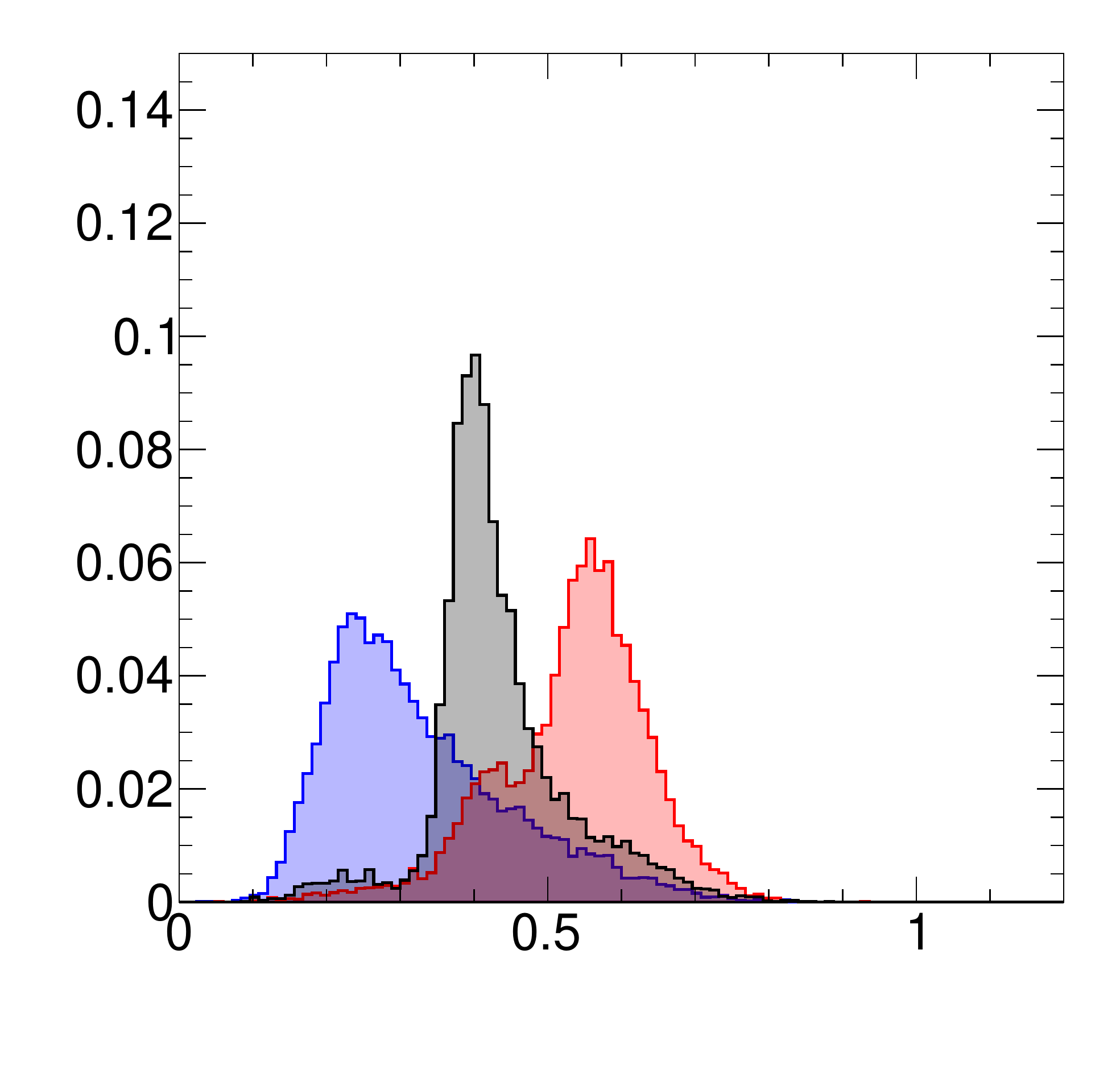}};
  \begin{scope}[x={(image.south east)},y={(image.north west)}]
        \draw[blue,fill=white!72!blue,thick] (0.59,0.85) rectangle
    (0.64,0.9);
    \draw[red,fill=white!72!red,thick] (0.59,0.78) rectangle
    (0.64,0.83);
    \draw[black,fill=white!72!black,thick] (0.59,0.71) rectangle
    (0.64,0.76);
    \node[draw=none, anchor=west] at (0.65, 0.875) { 
      \normalsize $\text{QCD}$};
    \node[draw=none, anchor=west] at (0.65, 0.805) { 
      \normalsize $Z' \rightarrow t\bar{t}$};
    \node[draw=none, anchor=west] at (0.65, 0.73) { 
      \normalsize $W \rightarrow qq'$};

        \node[draw=none] at (0.56,0.065) { \small Leading
      Learned $\sigma$};
    \node[draw=none, rotate=90] at (0.02, 0.55){ \small Arbitrary
    Units};
    \node[draw=none, anchor=west] at (0.17,0.88) { 
       {\sc Pythia} 8};
    \node[draw=none, anchor=west] at (0.17,0.80) { 
      $\sqrt{s} = 8 \text{ TeV}$};
    \node[draw=none, anchor=west] at (0.17, 0.74) { \tiny $350 \leq
      p_\text{T}^{\text{Jet}} \leq 450 \text{ GeV}$};
  \end{scope}
\end{tikzpicture}
 \\
\end{tabular}
\caption{The learned value of $\sigma$ for all fuzzy jets (left) and for the highest $p_\text{T}$ jet under the HML scheme (right).}
\label{fig:sigma_hist}
\end{center}
\end{figure}

\begin{figure}[h!]
\begin{center}
\begin{tabular}{cc}
\begin{overpic}[width=0.43\textwidth]{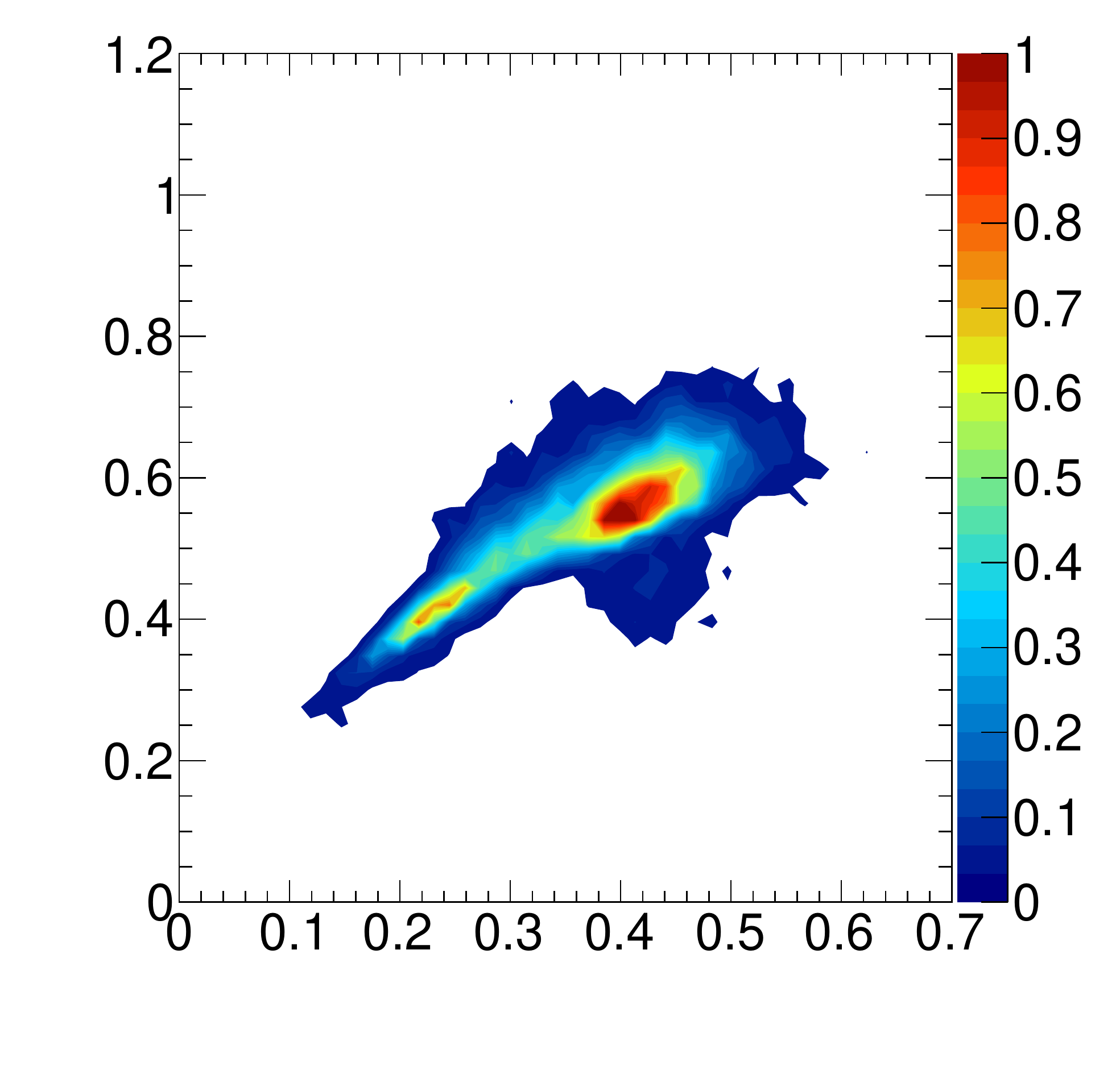}
\put(18, 4){ \small Leading anti-$k_t$ Jet $m/p_\text{T}$}
\put(3, 30){\rotatebox{90}{ \small Leading Learned $\sigma$}}
\put(99, 30){\rotatebox{90}{ \small Arbitrary Units}}
\put(62, 93){  \large $Z'
    \rightarrow \text{t}\bar{\text{t}}$}

\put(20, 82){   {\sc Pythia} 8}
\put(20, 75){  $\sqrt{s} = 8 \text{ TeV}$}
\put(18, 70){ \tiny $350 \leq
  p_T^{\text{Jet}} \leq 450 \text{ GeV}$}

\put(42, 23){\sffamily \small $\rho_{\sigma,\text{m}/p_\text{T}} \approx 0.68$}
\end{overpic} &
\begin{overpic}[width=0.43\textwidth]{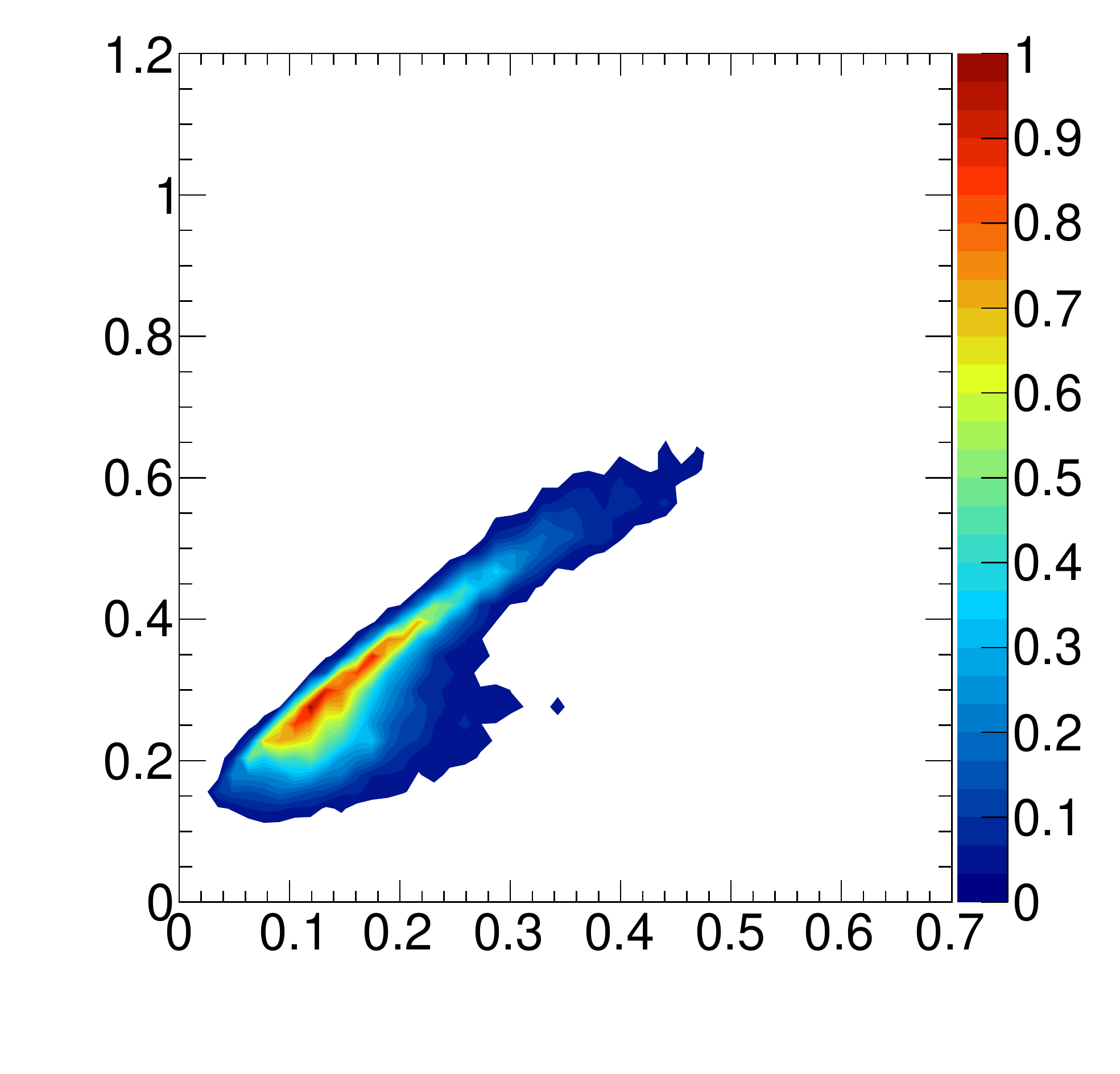}
\put(18, 4){ \small Leading anti-$k_t$ Jet $m/p_\text{T}$}
\put(3, 30){\rotatebox{90}{ \small Leading Learned $\sigma$}}
\put(99, 30){\rotatebox{90}{ \small Arbitrary Units}}
\put(69, 93){  \large $\text{QCD}$}

\put(20, 82){   {\sc Pythia} 8}
\put(20, 75){  $\sqrt{s} = 8 \text{ TeV}$}
\put(18, 70){ \tiny $350 \leq
  p_T^{\text{Jet}} \leq 450 \text{ GeV}$}

\put(42, 23){\sffamily \small $\rho_{\sigma,\text{m}/p_\text{T}} \approx 0.69$}
\end{overpic} \\
\end{tabular}
\caption{The joint distribution of the leading fuzzy jet $\sigma$ and the leading anti-$k_t$ jet $m/p_\text{T}$ for $Z'$ events (left) and QCD multijet events (right).  The Pearson correlation coefficient is shown in the bottom right of both plots. }
\label{fig:corr_mpt}
\end{center}
\end{figure}

\clearpage

\paragraph{Fuzzy Jets for Tagging}\mbox{} \\
\label{sec:tagging}

Many properties of events clustered with fuzzy jets may be useful for jet tagging, but for a brief illustration, Fig.~\ref{fig:tagging_tmva2} shows the performance of a tagger based on $\sigma$.  The $\sigma$-based tagger is significantly better than the random tagger, providing a rejection of $\sim 40$ at a signal efficiency of $50\%$ for both top quark event tagging and $W$ boson event tagging.  The word `event' is used as a reminder that even though $Z'$ events produce boosted top quarks, the fuzzy or anti-$k_t$ jet may only contain the $W$-boson decay products (see Fig.~\ref{fig:kinematics_mass}).  A relevant benchmark variable is the anti-$k_t$ jet $m/p_\text{T}$, which is similarly useful and in contains similar information.  The likelihood (i.e. optimal) combination of $\sigma$ and $m/p_\text{T}$ is significantly better than $\sigma$ or $m/p_\text{T}$ alone, indicating the the information in $\sigma$ that is uncorrected with $m/p_\text{T}$ from Fig.~\ref{fig:corr_mpt} adds useful discriminating information.  

\vspace{5mm}

\begin{figure}[h!]
\begin{center}
\begin{tabular}{cc}
\begin{tikzpicture}
  \node[anchor=south west,inner sep=0] (image) at (0,0)
  {\includegraphics[width=0.43\textwidth]{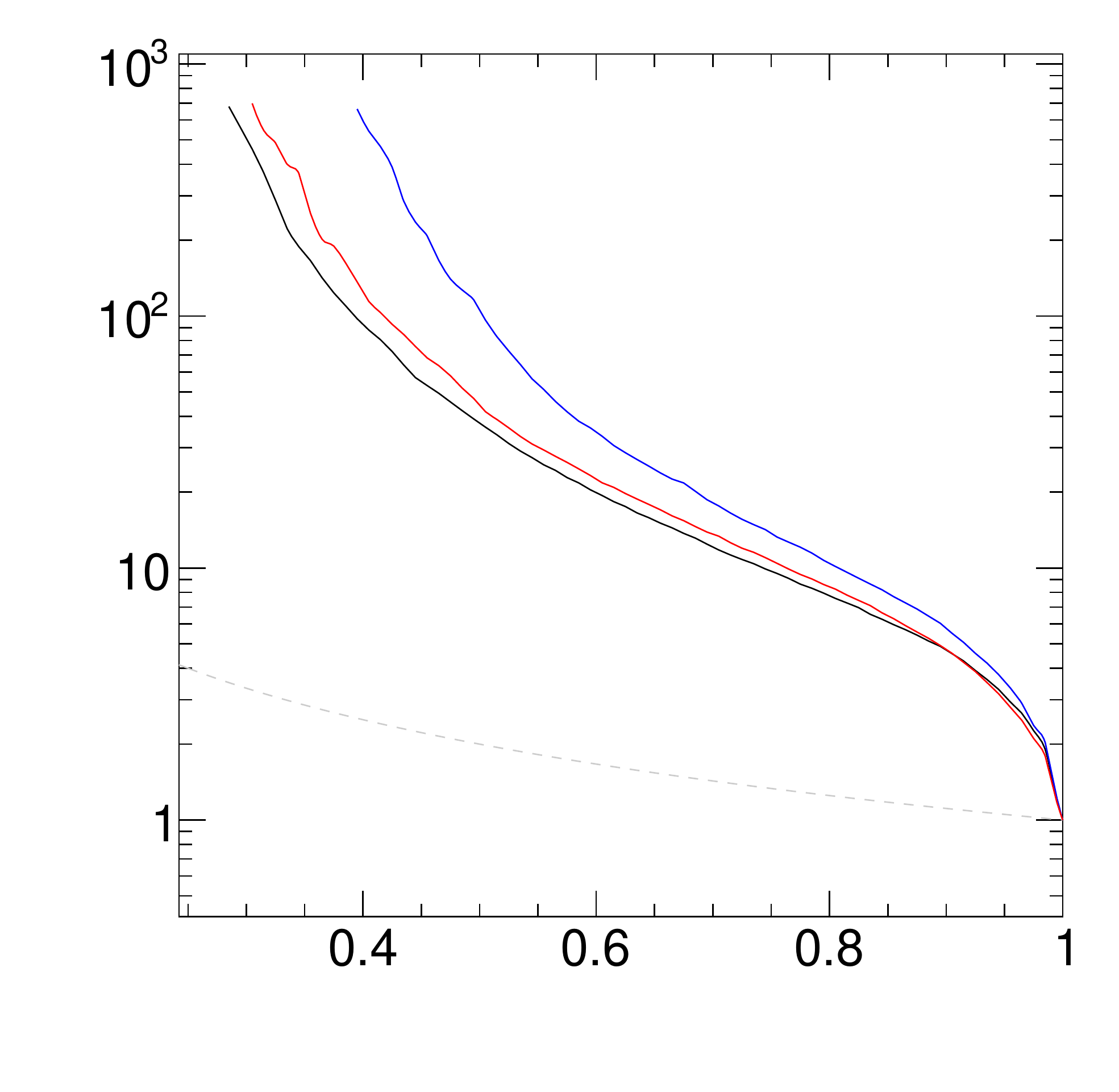}};
  \begin{scope}[x={(image.south east)},y={(image.north west)}]
                    
    \draw[blue,thick] (0.46,0.9) -- (0.53,0.9);
    \draw[red,thick] (0.46,0.83) -- (0.53,0.83);
    \draw[black,thick] (0.46,0.76) -- (0.53,0.76);
    \draw[white!80!black,dash pattern=on 2pt off 2pt] (0.46,0.69) -- (0.53,0.69);

    \node[draw=none, anchor=west] at (0.53, 0.9) {
      \normalsize $m/p_\text{T} \text{ \& } \sigma$};
    \node[draw=none, anchor=west] at (0.53, 0.83) {
      \normalsize $\sigma$};
    \node[draw=none, anchor=west] at (0.53, 0.76) {
      \normalsize $m/p_\text{T}$};
    \node[draw=none, anchor=west] at (0.53, 0.685) {
      \normalsize random};

        \node[draw=none] at (0.56,0.055) { \small Top Quark Efficiency};
    \node[draw=none, rotate=90] at (0.055, 0.55){ \small QCD Rejection};
    \node[draw=none, anchor=west] at (0.15,1.065) { 
       {\sc Pythia} 8};
    \node[draw=none, anchor=west] at (0.15,0.99) { 
      $\sqrt{s} = 8 \text{ TeV}$};
    \node[draw=none, anchor=east] at (0.97,0.99) { 
      $Z' \rightarrow t\bar{t}$};

    \node[draw=none, anchor=east] at (0.97, 1.045) { \tiny $350 \leq
      p_\text{T}^{\text{Jet}} \leq 450 \text{ GeV}$};
    \node[draw=none, anchor=east] at (0.97, 1.1) { \tiny $150 \leq
      m^{\text{Jet}} \leq 200 \text{ GeV}$};
      
  \end{scope}
\end{tikzpicture} &
\begin{tikzpicture}
  \node[anchor=south west,inner sep=0] (image) at (0,0)
  {\includegraphics[width=0.43\textwidth]{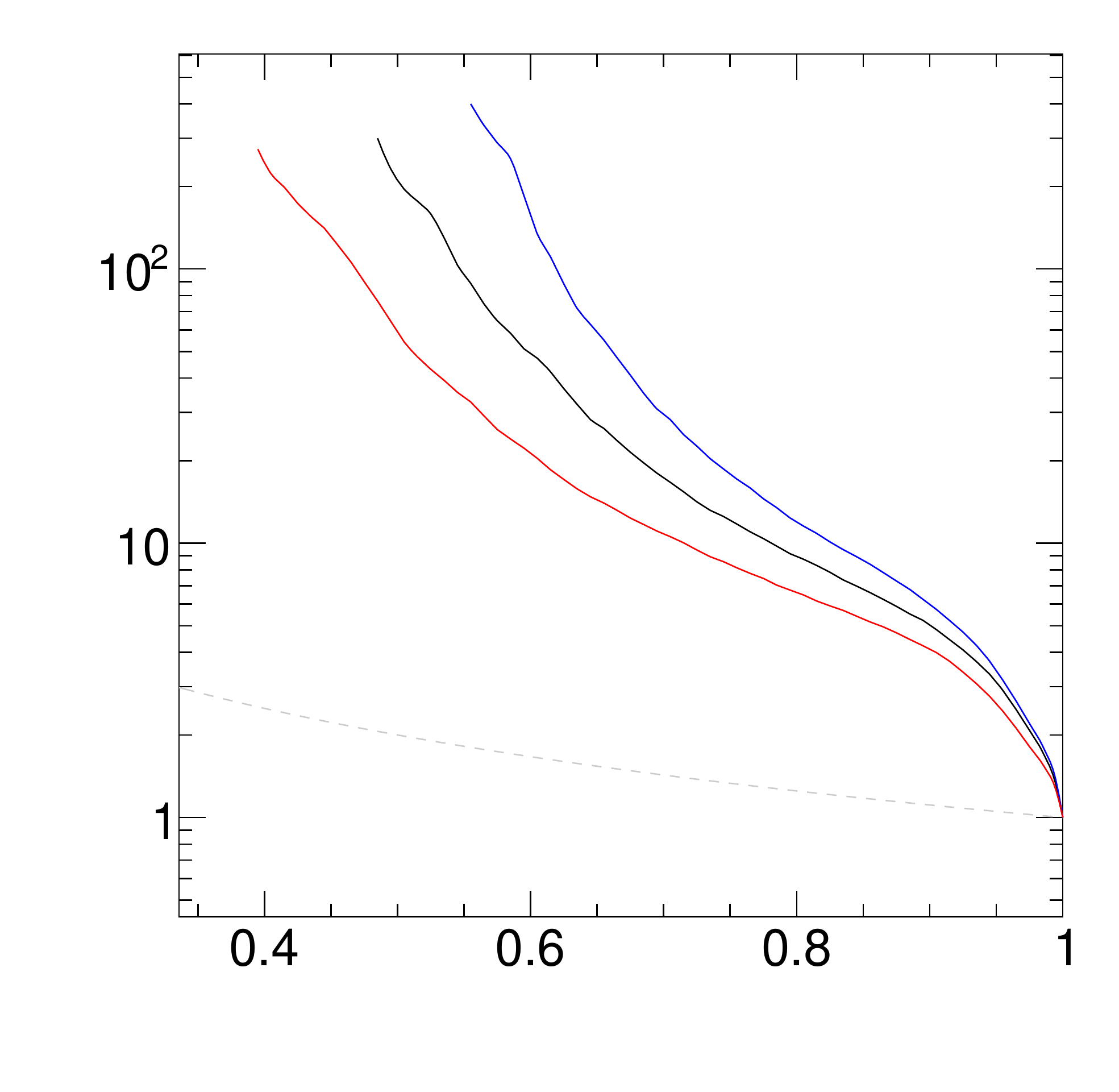}};
  \begin{scope}[x={(image.south east)},y={(image.north west)}]
                    
    \draw[blue,thick] (0.46+0.06,0.9) -- (0.53+0.06,0.9);
    \draw[red,thick] (0.46+0.06,0.83) -- (0.53+0.06,0.83);
    \draw[black,thick] (0.46+0.06,0.76) -- (0.53+0.06,0.76);
  
    \node[draw=none, anchor=west] at (0.6, 0.9) {
      \normalsize $m/p_\text{T} \text{ \& } \sigma$};
    \node[draw=none, anchor=west] at (0.6, 0.83) {
      \normalsize $\sigma$};
    \node[draw=none, anchor=west] at (0.6, 0.76) {
      \normalsize $m/p_\text{T}$};
        
        \node[draw=none] at (0.56,0.055) { \small $W$ Efficiency};
    \node[draw=none, rotate=90] at (0.055, 0.55){ \small QCD Rejection};
    \node[draw=none, anchor=west] at (0.11,1.065) { 
       {\sc Pythia} 8};
    \node[draw=none, anchor=west] at (0.11,0.99) { 
      $\sqrt{s} = 8 \text{ TeV}$};
    \node[draw=none, anchor=east] at (0.98,0.98) { \footnotesize 
      $W' \rightarrow WZ \rightarrow qq'\nu\bar{\nu}$};

    \node[draw=none, anchor=east] at (0.97, 1.04) { \tiny $350 \leq
      p_\text{T}^{\text{Jet}} \leq 450 \text{ GeV}$};
    \node[draw=none, anchor=east] at (0.97, 1.095) { \tiny $60 \leq
      m^{\text{Jet}} \leq 110 \text{ GeV}$};

  \end{scope}
\end{tikzpicture} \\
\end{tabular}
\caption{The tradeoff between signal efficiency versus QCD multijet rejection (=1/efficiency) when the signal process is $Z'\rightarrow t\bar{t}$ (left) or $W'\rightarrow WZ$ (right).  The {\it random} tagger line the curve signal efficiency = background efficiency. }
\label{fig:tagging_tmva2}
\end{center}
\end{figure}

\clearpage

\subsubsection{Underlying Event and Pileup}
\label{fuzzypileup}

One of the interesting features of fuzzy jets is that for densities $\Phi$ with infinite support (such as the Gaussian), the area over which particles can belong to that jet is infinite.  This is in sharp contrast to anti-$k_t$ jets for which the area is bounded by $\pi R^2$.  Hard-scatter anti-$k_t$ jets are unaffected by relatively soft nearby jets.  However, if there are not enough fuzzy jets to capture the diffuse soft radiation in an event, the jets that would otherwise capture the hard-scatter energy must become larger.  This is illustrated in Fig.~\ref{fig:pileup_ed} where both top-quark jets are significantly larger at $n_\text{PU}=40$. 

\begin{figure}[h!]
\vspace{1cm}
\begin{center}
\begin{tabular}{cc}
\begin{overpic}[width=0.43\textwidth]{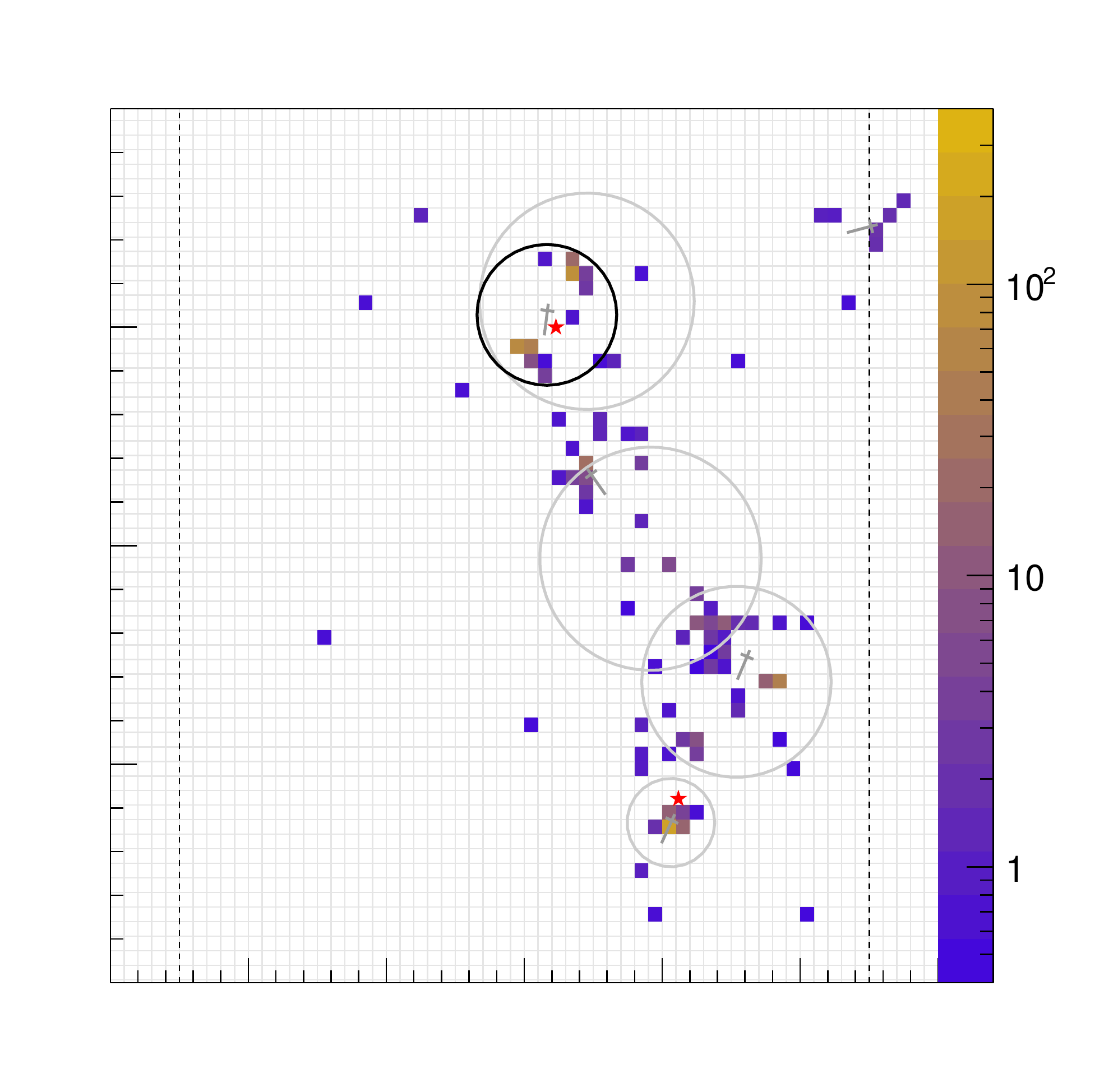}
\put(12, 91){$Z'\rightarrow t\bar{t}$, $n_\text{PU}=0$, {\sc Pythia} 8}
 
\put(24, -4){ \small Pseudorapidity ($\eta$)}
\put(-4, 10){\rotatebox{90}{ \small Rotated Azimuthal Angle
    ($\phi$)}}
\put(95, 27){\rotatebox{90}{ \small Tower $p_\text{T} \text{ [GeV]}$}}

\put(5, 8){\bfseries \small \sffamily $0$}
\put(4, 28.3){\bfseries \small \sffamily $\frac{\pi}{2}$}
\put(5, 48.3){\bfseries \small \sffamily $\pi$}
\put(4, 68){\bfseries \small \sffamily $\frac{3\pi}{2}$}
\put(4, 88){\bfseries \small \sffamily $2\pi$}

\put(4,  4){\bfseries \small \sffamily $-3$}
\put(17, 4){\bfseries \small \sffamily $-2$}
\put(29.6, 4){\bfseries \small \sffamily $-1$}
\put(46.2, 4){\bfseries \small \sffamily $0$}
\put(58.8, 4){\bfseries \small \sffamily $1$}
\put(71.3, 4){\bfseries \small \sffamily $2$}
\put(83.8, 4){\bfseries \small \sffamily $3$}
\end{overpic} &
\begin{overpic}[width=0.43\textwidth]{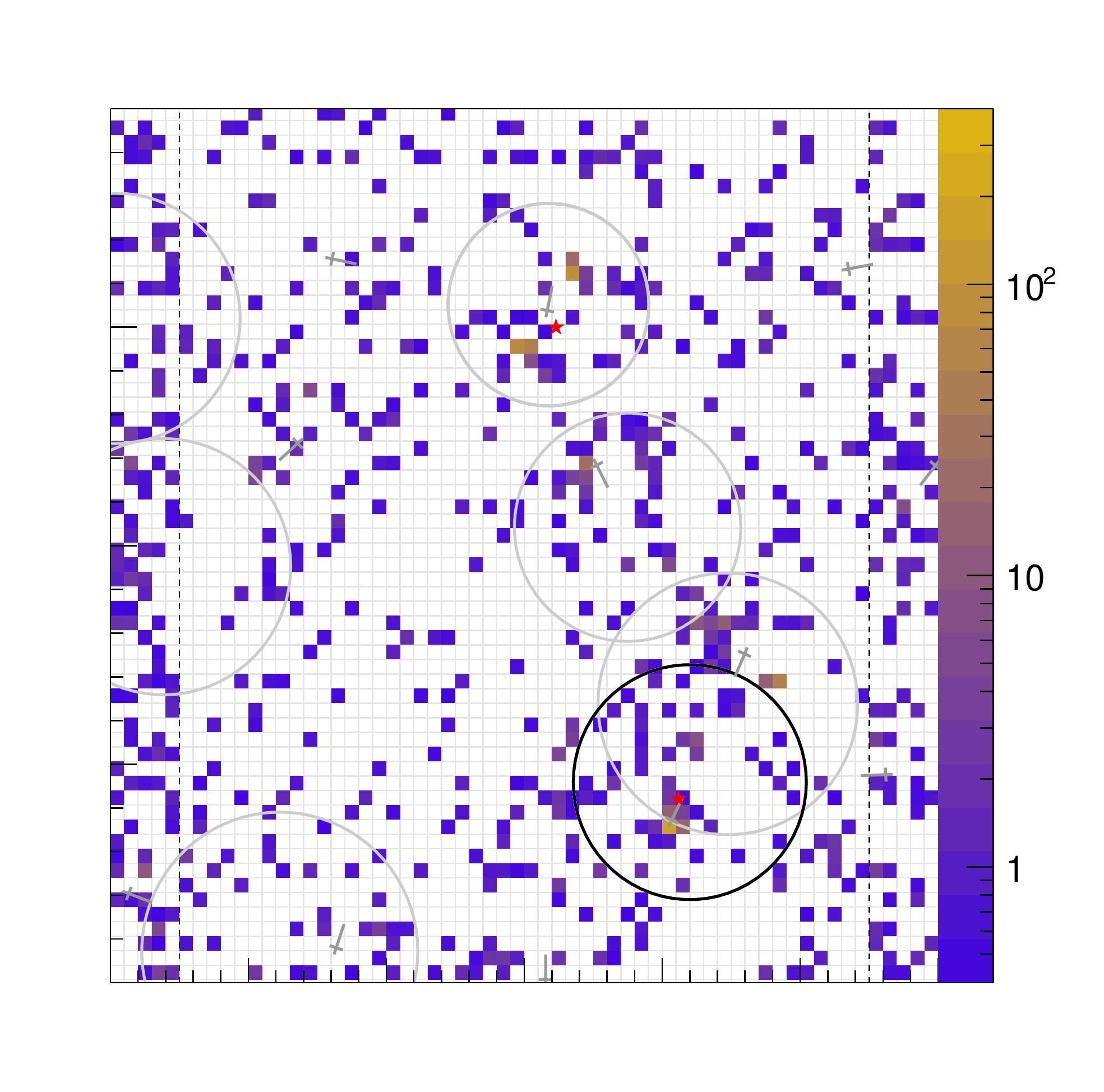}
\put(12, 91){$Z'\rightarrow t\bar{t}$, $n_\text{PU}=40$, {\sc Pythia} 8}
 
\put(24, -4){ \small Pseudorapidity ($\eta$)}
\put(-4, 10){\rotatebox{90}{ \small Rotated Azimuthal Angle
    ($\phi$)}}
\put(95, 27){\rotatebox{90}{ \small Tower $p_\text{T} \text{ [GeV]}$}}

\put(5, 8){\bfseries \small \sffamily $0$}
\put(4, 28.3){\bfseries \small \sffamily $\frac{\pi}{2}$}
\put(5, 48.3){\bfseries \small \sffamily $\pi$}
\put(4, 68){\bfseries \small \sffamily $\frac{3\pi}{2}$}
\put(4, 88){\bfseries \small \sffamily $2\pi$}

\put(4,  4){\bfseries \small \sffamily $-3$}
\put(17, 4){\bfseries \small \sffamily $-2$}
\put(29.6, 4){\bfseries \small \sffamily $-1$}
\put(46.2, 4){\bfseries \small \sffamily $0$}
\put(58.8, 4){\bfseries \small \sffamily $1$}
\put(71.3, 4){\bfseries \small \sffamily $2$}
\put(83.8, 4){\bfseries \small \sffamily $3$}
\end{overpic} \\
\end{tabular}
\end{center}
\caption{The same $Z'$ event with  $n_{\text{PU}} = 0$ (left) and  $n_{\text{PU}} = 40$ (right).  The grid lines show the $0.1\times 0.1$ tower size and the vertical dashed lines show the range over which the charged pileup energy is subtracted from each tower.  The top quark locations from the generator-record are indicated by red stars and anti-$k_t$ jet locations are shown with gray crosses where the long tail points towards the mGMM jet for which it was a seed. The fuzzy jets themselves are represented by their $1\sigma$ contour.}
\label{fig:pileup_ed}
\end{figure}

\begin{figure}[h!]
\begin{center}
\includegraphics[width=0.4\textwidth]{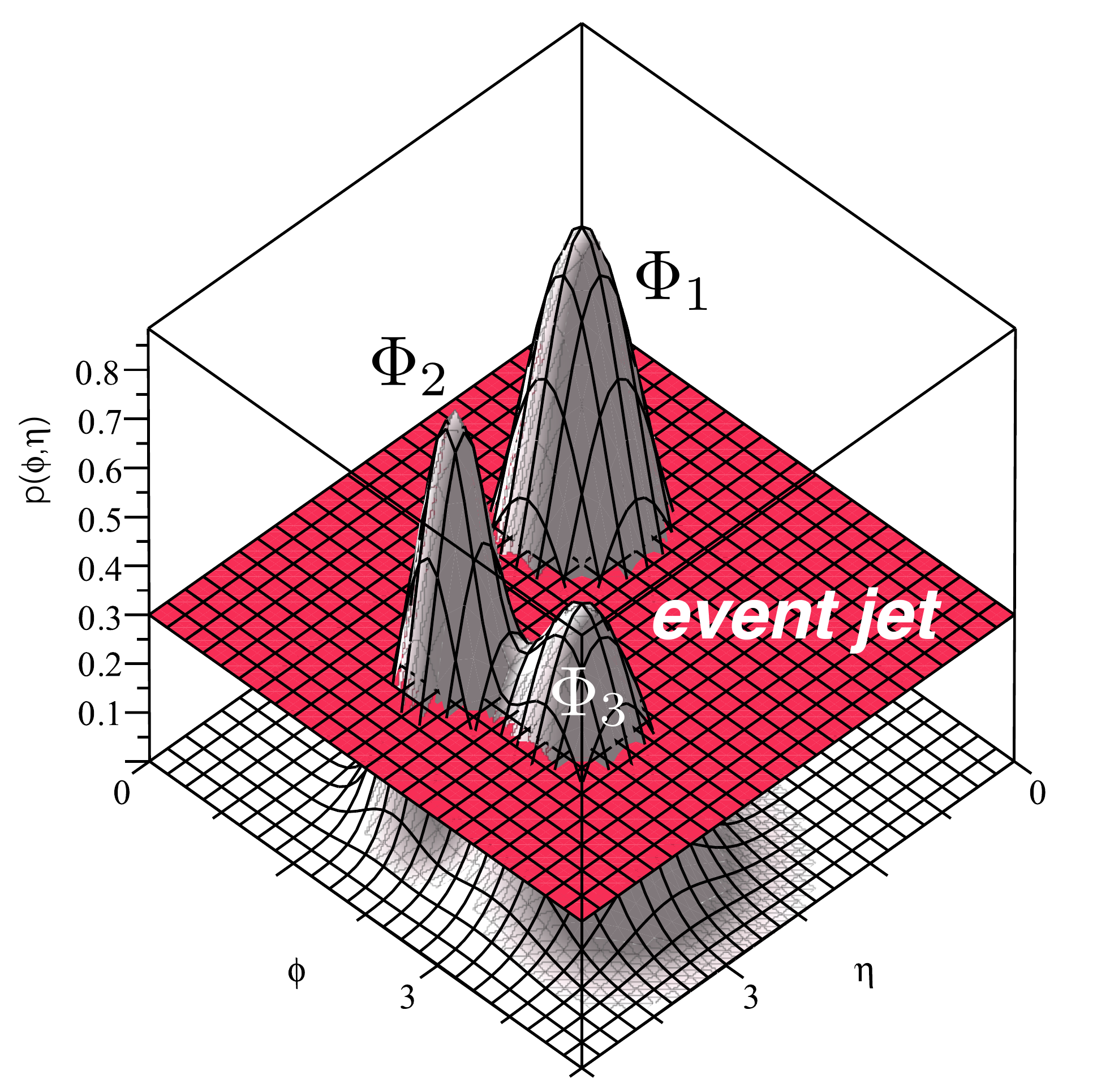}
\caption{The same schematic $k=3$ per-particle probability density from Fig.~\ref{fig:densitymap}, but with a constant likelihood added to represent the event jet.}
\label{fig:densitymap2}
\end{center}
\end{figure}

One can force the fuzzy jets algorithm to focus on the hard-scatter by using $\alpha>1$, but at the cost of losing IRC safety.  Another possibility is to increase the number of seed jets.  A third possibility is to artificially add a jet to the event likelihood that has a uniform constant density over the entire detector.  This {\it event jet} can absorb the diffuse soft radiation and allow the other fuzzy jets to `focus' on the hard-scatter.  The idea of an event jet is illustrated schematically in Fig.~\ref{fig:densitymap2}.  A constant density can provide probability to soft particles far from the hard-scatter jets, which can stabilize the size of the hard-scatter fuzzy jets.  Quantitatively, the algorithm is modified with $q_{ij} \rightarrow \frac{q_{ij}}{\gamma + \sum_k p_{ik}}$, where $\gamma$ is the event jet weight.  In principle, the algorithm could learn $\gamma$, but since it should scale linearly with the median pileup density $\rho$, one could reduce the algorithm complexity by fixing $\gamma=\kappa \rho$.  A value of $\kappa\sim 0.3$ was found to be optimal over a wide range of processes.  Under the HML scheme, a particle is assigned to the event jet if $\max_k p_{ik} < \gamma$.  Studies indicate that when the event-jet is coupled with a simple tower-level pileup subtraction scheme, the resulting properties of the leading fuzzy jet are robust against pileup.  Many complex constituent-based pileup subtraction schemes exist (see e.g. Ref.~\cite{puppi,constsub,softkill}); one simple procedure used here for illustration is $p_\text{T}\mapsto \text{max}\left( p_{T,\text{uncorrected}} - \rho A, 0 \right)$, where $A=0.1^2$.  The same event from Fig.~\ref{fig:pileup_ed} is shown with the above pileup corrections in Fig.~\ref{fig:pileup_corr}.  As desired, the the two leading jets corresponding to the top quarks are nearly the same size for $n_\text{PU}=0$ and $40$.  The sub-leading jets shift as the soft radiation is balanced between between them and the event jet.  The stability of $\sigma$ is quantified in Fig.~\ref{fig:corr_mean_var} where both the mean and standard deviation of the $\sigma$ distribution are nearly independent of $n_\text{PU}$.  Note that the standard deviation of the $\sigma$ distribution {\it decreases} at high $n_\text{PU}$ as all jets are large and nearly all the same size.

\begin{figure}[h!]
\vspace{1cm}
\begin{center}
\begin{tabular}{cc}
\begin{overpic}[width=0.43\textwidth]{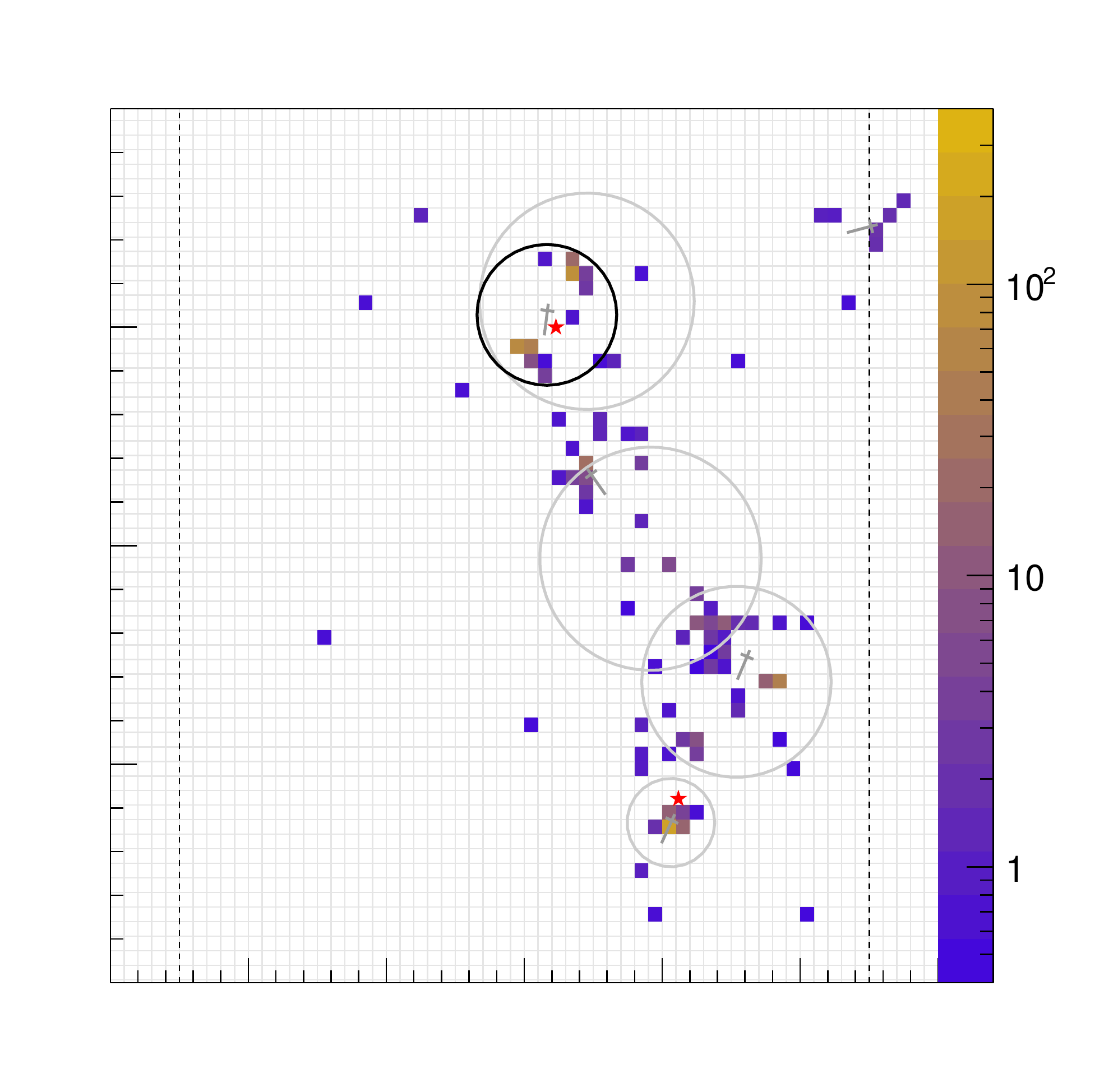}
\put(20, 91){$Z'\rightarrow t\bar{t}$, {\sc Pythia} 8}
\put(1, 100){  \large $n_{\text{PU}} = 0$, pileup-corrected}

\put(24, -4){ \small Pseudorapidity ($\eta$)}
\put(-4, 10){\rotatebox{90}{ \small Rotated Azimuthal Angle
    ($\phi$)}}
\put(95, 27){\rotatebox{90}{ \small Tower $p_\text{T} \text{ [GeV]}$}}

\put(5, 8){\bfseries \small \sffamily $0$}
\put(4, 28.3){\bfseries \small \sffamily $\frac{\pi}{2}$}
\put(5, 48.3){\bfseries \small \sffamily $\pi$}
\put(4, 68){\bfseries \small \sffamily $\frac{3\pi}{2}$}
\put(4, 88){\bfseries \small \sffamily $2\pi$}

\put(4,  4){\bfseries \small \sffamily $-3$}
\put(17, 4){\bfseries \small \sffamily $-2$}
\put(29.6, 4){\bfseries \small \sffamily $-1$}
\put(46.2, 4){\bfseries \small \sffamily $0$}
\put(58.8, 4){\bfseries \small \sffamily $1$}
\put(71.3, 4){\bfseries \small \sffamily $2$}
\put(83.8, 4){\bfseries \small \sffamily $3$}
\end{overpic} &
\begin{overpic}[width=0.43\textwidth]{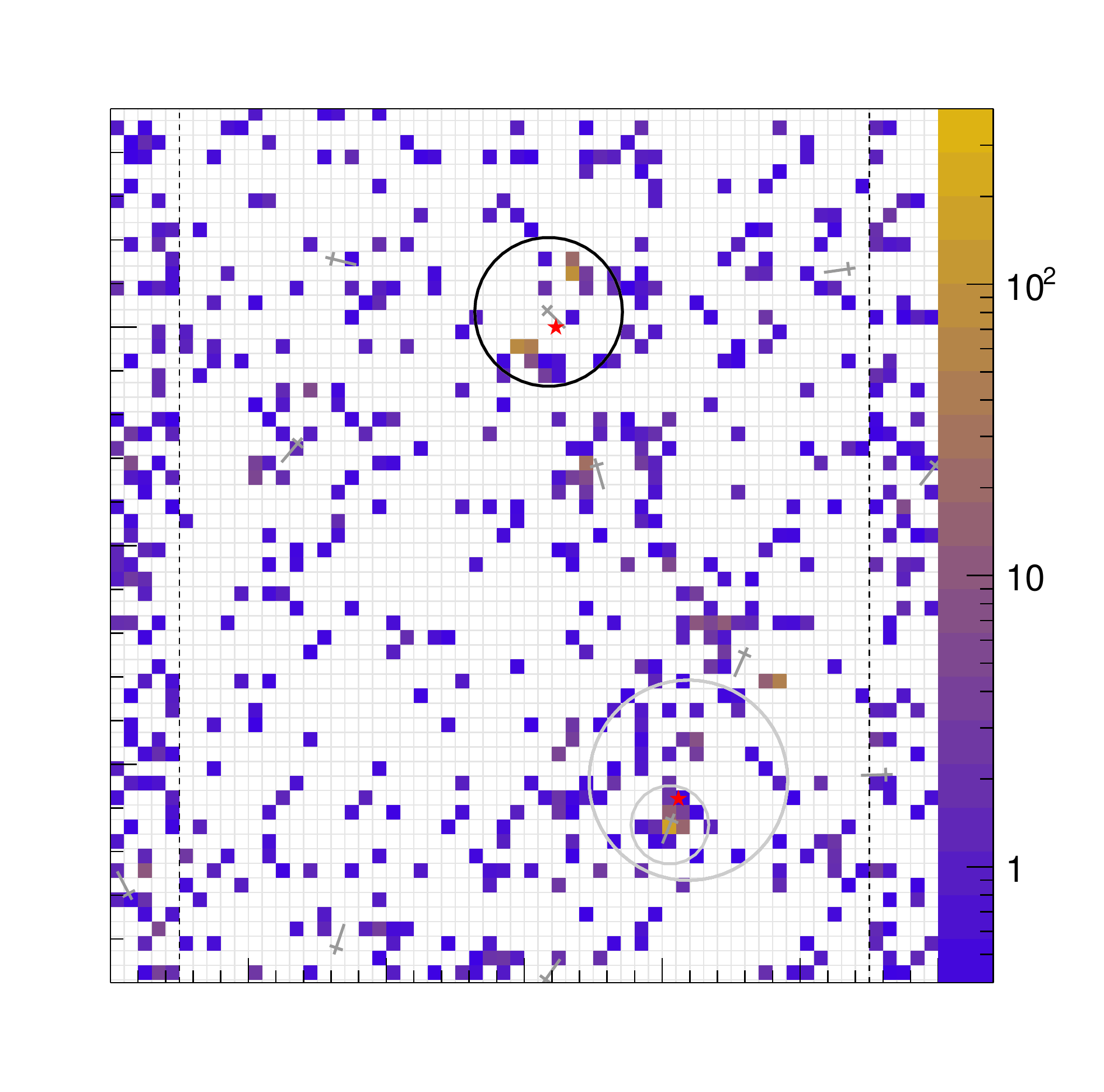}
\put(20, 91){$Z'\rightarrow t\bar{t}$, {\sc Pythia} 8}
\put(1, 100){  \large $n_{\text{PU}} = 40$, pileup-corrected}

\put(24, -4){ \small Pseudorapidity ($\eta$)}
\put(-4, 10){\rotatebox{90}{ \small Rotated Azimuthal Angle
    ($\phi$)}}
\put(95, 27){\rotatebox{90}{ \small Tower $p_\text{T} \text{ [GeV]}$}}

\put(5, 8){\bfseries \small \sffamily $0$}
\put(4, 28.3){\bfseries \small \sffamily $\frac{\pi}{2}$}
\put(5, 48.3){\bfseries \small \sffamily $\pi$}
\put(4, 68){\bfseries \small \sffamily $\frac{3\pi}{2}$}
\put(4, 88){\bfseries \small \sffamily $2\pi$}

\put(4,  4){\bfseries \small \sffamily $-3$}
\put(17, 4){\bfseries \small \sffamily $-2$}
\put(29.6, 4){\bfseries \small \sffamily $-1$}
\put(46.2, 4){\bfseries \small \sffamily $0$}
\put(58.8, 4){\bfseries \small \sffamily $1$}
\put(71.3, 4){\bfseries \small \sffamily $2$}
\put(83.8, 4){\bfseries \small \sffamily $3$}
\end{overpic} \\
\end{tabular}
\end{center}
\caption{The same events as in Fig.~\ref{fig:pileup_ed}, but with event-jet and tower-based pileup subtraction applied.}
\label{fig:pileup_corr}
\end{figure}

\begin{figure}[h!]
\begin{center}
\begin{tikzpicture}
  \node[anchor=south west,inner sep=0] (image) at (0,0)
  {\includegraphics[width=0.43\textwidth]{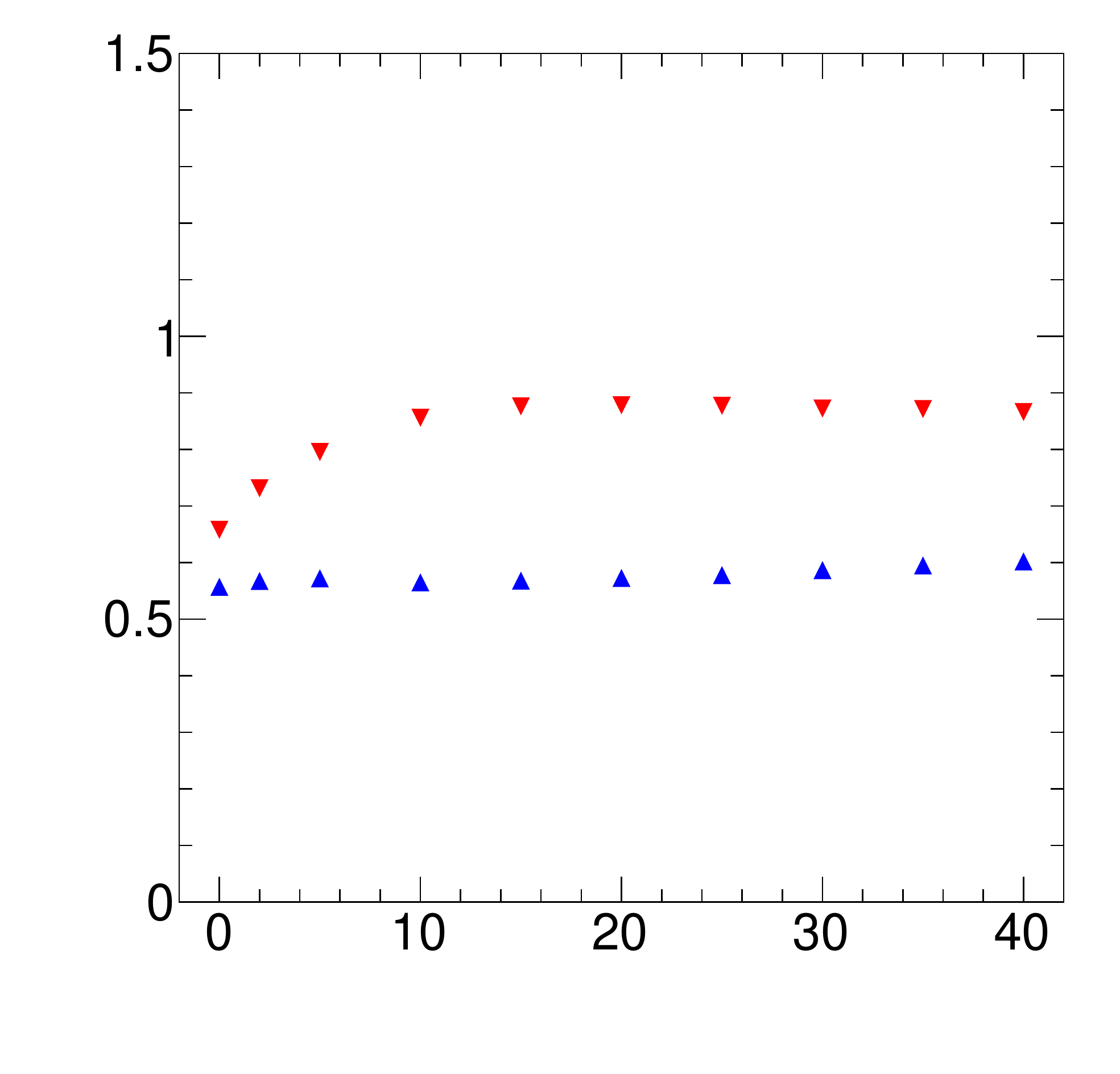}};
  \begin{scope}[x={(image.south east)},y={(image.north west)}]
        \draw[fill=red,draw=none] (0.625-0.02,0.883)--(0.645-0.02,0.883)--(0.635-0.02,0.863)--cycle;
    \draw[fill=blue,draw=none] (0.625-0.02,0.793)--(0.645-0.02,0.793)--(0.635-0.02,0.813)--cycle;
    \node[draw=none, anchor=west] at (0.68, 1) {  \large $Z'
    \rightarrow t\bar{t}$};
    \node[draw=none, anchor=west] at (0.63, 0.875) { 
      \footnotesize Uncorrected};
    \node[draw=none, anchor=west] at (0.63, 0.805) { 
      \footnotesize Corrected};

        \node[draw=none] at (0.56,0.065) { \small $n_{\text{PU}}$};
    \node[draw=none, rotate=90] at (0.05, 0.55){ \small Mean
      of $\sigma$};
    \node[draw=none, anchor=west] at (0.17,0.88) { 
       {\sc Pythia} 8};
    \node[draw=none, anchor=west] at (0.17,0.80) { 
      $\sqrt{s} = 8 \text{ TeV}$};
  \end{scope}
\end{tikzpicture} 
\begin{tikzpicture}
  \node[anchor=south west,inner sep=0] (image) at (0,0)
  {\includegraphics[width=0.43\textwidth]{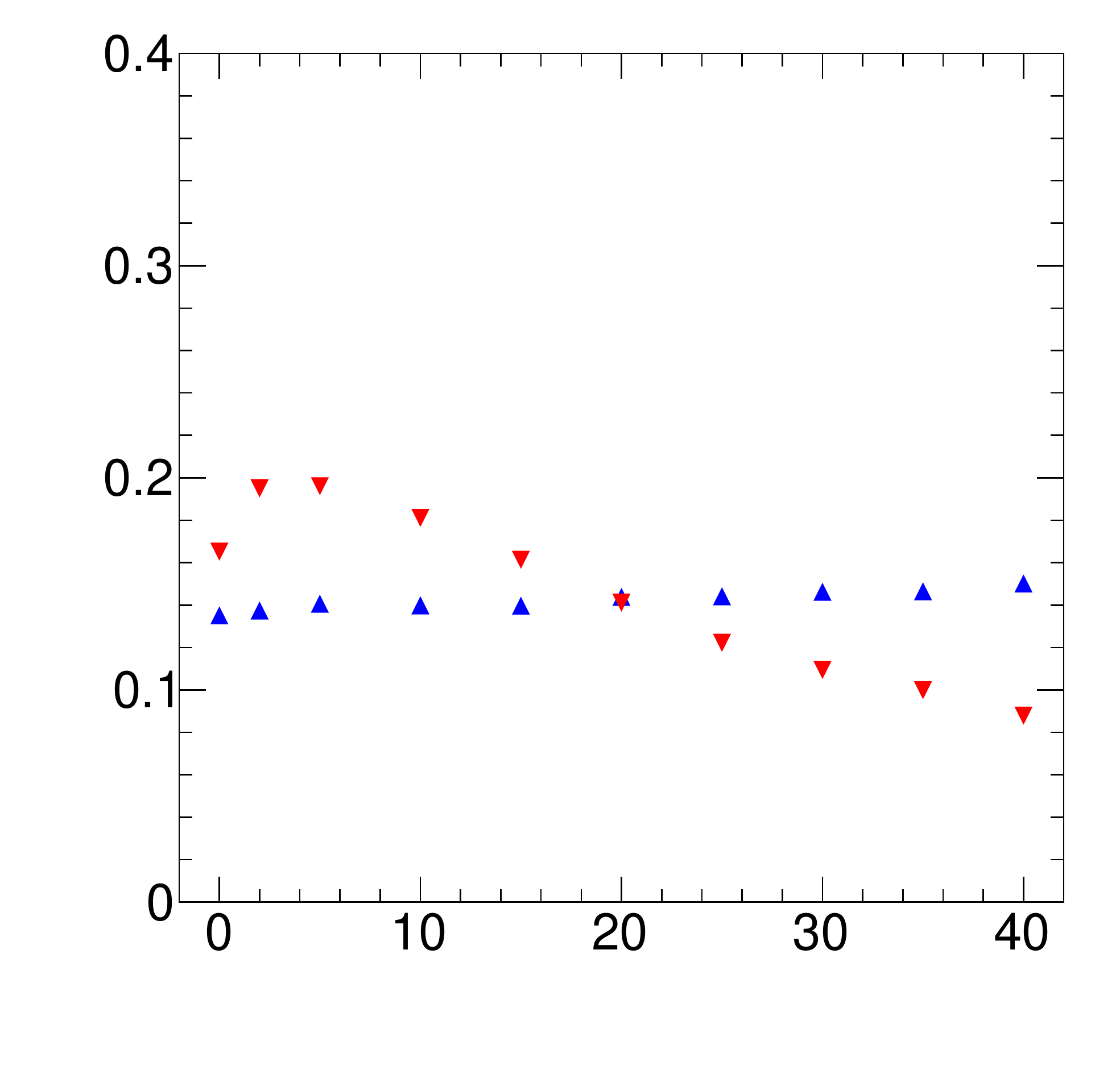}};
  \begin{scope}[x={(image.south east)},y={(image.north west)}]
        \draw[fill=red,draw=none] (0.625-0.02,0.883)--(0.645-0.02,0.883)--(0.635-0.02,0.863)--cycle;
    \draw[fill=blue,draw=none] (0.625-0.02,0.793)--(0.645-0.02,0.793)--(0.635-0.02,0.813)--cycle;
    \node[draw=none, anchor=west] at (0.68, 1) {  \large $Z'
    \rightarrow t\bar{t}$};
    \node[draw=none, anchor=west] at (0.63, 0.875) { 
      \footnotesize Uncorrected};
    \node[draw=none, anchor=west] at (0.63, 0.805) { 
      \footnotesize Corrected};

        \node[draw=none] at (0.56,0.065) { \small $n_{\text{PU}}$};
    \node[draw=none, rotate=90] at (0.05, 0.55){ \small
      Standard Deviation of $\sigma$};
    \node[draw=none, anchor=west] at (0.17,0.88) { 
       {\sc Pythia} 8};
    \node[draw=none, anchor=west] at (0.17,0.80) { 
      $\sqrt{s} = 8 \text{ TeV}$};
  \end{scope}
\end{tikzpicture}
\caption{The mean and standard deviation of the $\sigma$ distribution in $Z'$ events as a function of $n_\text{PU}$ with and without the event jet and tower-based pileup corrections.}
\label{fig:corr_mean_var}
\end{center}
\end{figure}

\clearpage
\newpage

\subsubsection{Conclusions}

The modified mixture model algorithms provide a new way of looking at whole event
structure.  In contrast to the usual uses of hierarchical-agglomerative algorithms like anti-$k_t$, the number of seeds is fixed ahead of time and their properties are learned during the clustering process.  The learned parameters provide a new set of handles for distinguishing jets of different types.  Even simple variables
constructed out of the learned parameters of a mixture of isotropic
Gaussian jets, like $\sigma$, offer complementary information for tagging $W$
boson and top quark jets.  Even though the variable $\sigma$ is sensitive to pileup, small modifications to the fuzzy jets algorithm can mitigate the impact of pileup.

Fuzzy jets provide a new paradigm for jet clustering in high energy physics.  These IRC safe likelihood-based clustering schemes set the stage for many possibilities for future studies related to jet tagging, probabilistic clustering, and pileup suppression.  Figure~\ref{fig:datastatsFuzzy} is the first step to bridge the gap between new machine-learning motivated unsurvised learning algorithms and analysis at the LHC: a first glimpse at the fuzzy jet $\sigma$ with the $\sqrt{s}=8$ TeV ATLAS data.

\begin{figure}[htbp!]
  \begin{center}
        \includegraphics[width=0.45\textwidth]{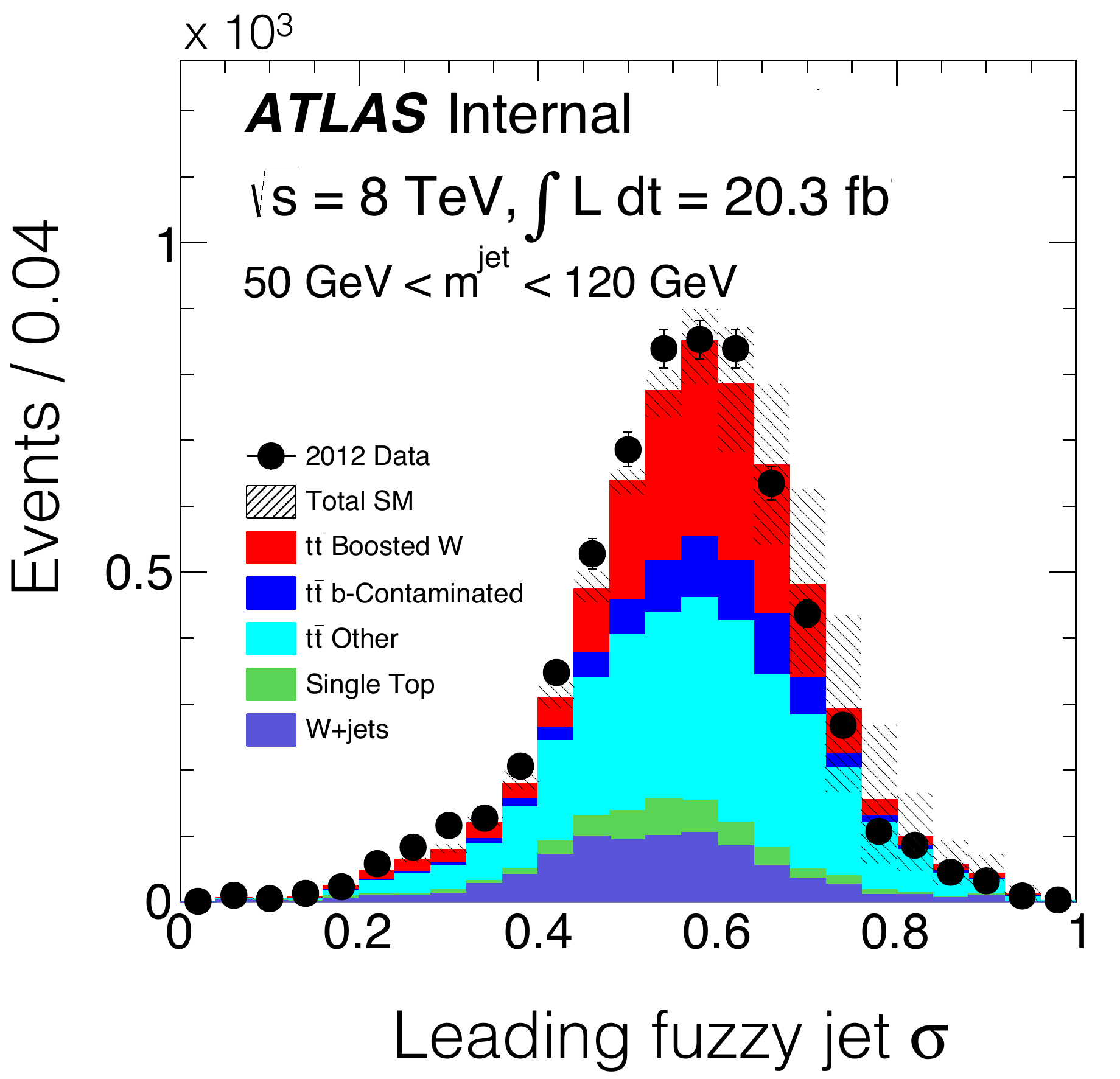} 
      \caption{ The distribution of the leading jet $\sigma$ using the same event selection as in Sec.~\ref{sec:data}.  Only the muon channel is included (negligible QCD multijets contribution).
      \label{fig:datastatsFuzzy} }
    \end{center}
\end{figure}

\clearpage

\subsection{Jet Images}  
\label{sec:jetimages}

One of the most complex and important supervised learning tasks is facial recognition.  The field of computer vision has developed sophisticated tools for performing this task with ever-increasing gains in performance.  The ATLAS calorimeter is analogous to a digital image: it is a scalar field\footnote{A natural extension of these methods is to use vector fields incorporating information from calorimeter segmentation (like RGB images) or even tracks.  These provide interesting challenges as the granularity would vary by component.} in two discrete dimensions where the pixels are calorimeter cells and the intensity is the measured energy.  By using image representations of jets ({\it jet images}~\cite{Cogan:2014oua}), the entire set of computer vision techniques can be directly applied to jet tagging.  Linear-discriminant based tagging with jet images can provide a similar performance as a simple tagger based on jet observables motivated by physical intuition~\cite{Cogan:2014oua}.  With shallow neural networks, the jet images-based tagging performance can out-perform simple jet observables~\cite{Almeida:2015jua}.  This section\footnote{The ideas presented in this section are published in Ref.~\cite{deOliveira:2015xxd}.  Many of the studies presented in this section were performed by Luke de Oliveira.  In particular, de Oliveira developed the network architectures and ran the training.  In addition, M. Kagan helped setup some of the technical framework for the studies.} investigates the use of deep neural networks (DNN) that are the state-of-the-art algorithms in the field of computer vision~\cite{vggnet,maxout:goodfellow,dropout:and:LRN}.  In addition to studying the performance of these algorithms, the focus is on exploring where the networks have extracted discriminating information.  To begin, Sec.~\ref{sec:preprocess} is a detailed description of image pre-processing and the physical impact of each step.  Section~\ref{sec:arch} briefly summarizes describes the neural network architectures explored in Sec.~\ref{sec:studies}.  The section ends with conclusions in Sec.~\ref{sec:conclusion}.

\subsubsection{Pre-processing and the Symmetries of Spacetime}
\label{sec:preprocess}

The setup from Sec.~\ref{sec:reluster:particlelevelperformance} is used to simulate boosted $W$ bosons and QCD multijets and the detector discretization from Sec.~\ref{sec:details} is used to pixelate the energies.  In practice, the detector and jet image granularities can be different, but are set equal here for simplicity.  Large-radius trimmed jets are clustered with $R=1.0$ with $k_t$ $R=0.3$ subjets groomed with $f_\text{cut}=0.05$.  Trimming mitigates the contribution from pileup; a detailed investigation into the performance of the neural network for $n_\text{PU}>0$ is beyond the scope of Sec.~\ref{sec:jetimages}.

Three key jet features for distinguishing between $W$ jets and QCD jets are the {\it jet mass}, {\it n-subjettiness}\footnote{Defined using the winner-takes-all axis that increases the robustness to pileup~\cite{Larkoski:2014uqa}.} and the $\Delta R$ between subjets of the trimmed jet. These observables are used for benchmarking the performance of the neural network in Sec.~\ref{sec:studies}. The distributions of these three discriminating variables are shown in Fig.~\ref{fig:datastats}.  The transverse momentum is also a useful observable for distinguishing signal from background.  However, in practice different techniques may be optimized for individual $p_\text{T}$ bins because most of the input variables have a strong particle-level and/or detector-level $p_\text{T}$ dependence.  To prevent the neural network from learning the jet $p_\text{T}$ as a useful discriminant, the momentum spectrum is re-weighted so that the signal has the same $p_\text{T}$ distribution as the background.

\begin{figure}[htbp!]
  \begin{center}
        \includegraphics[width=0.32\textwidth]{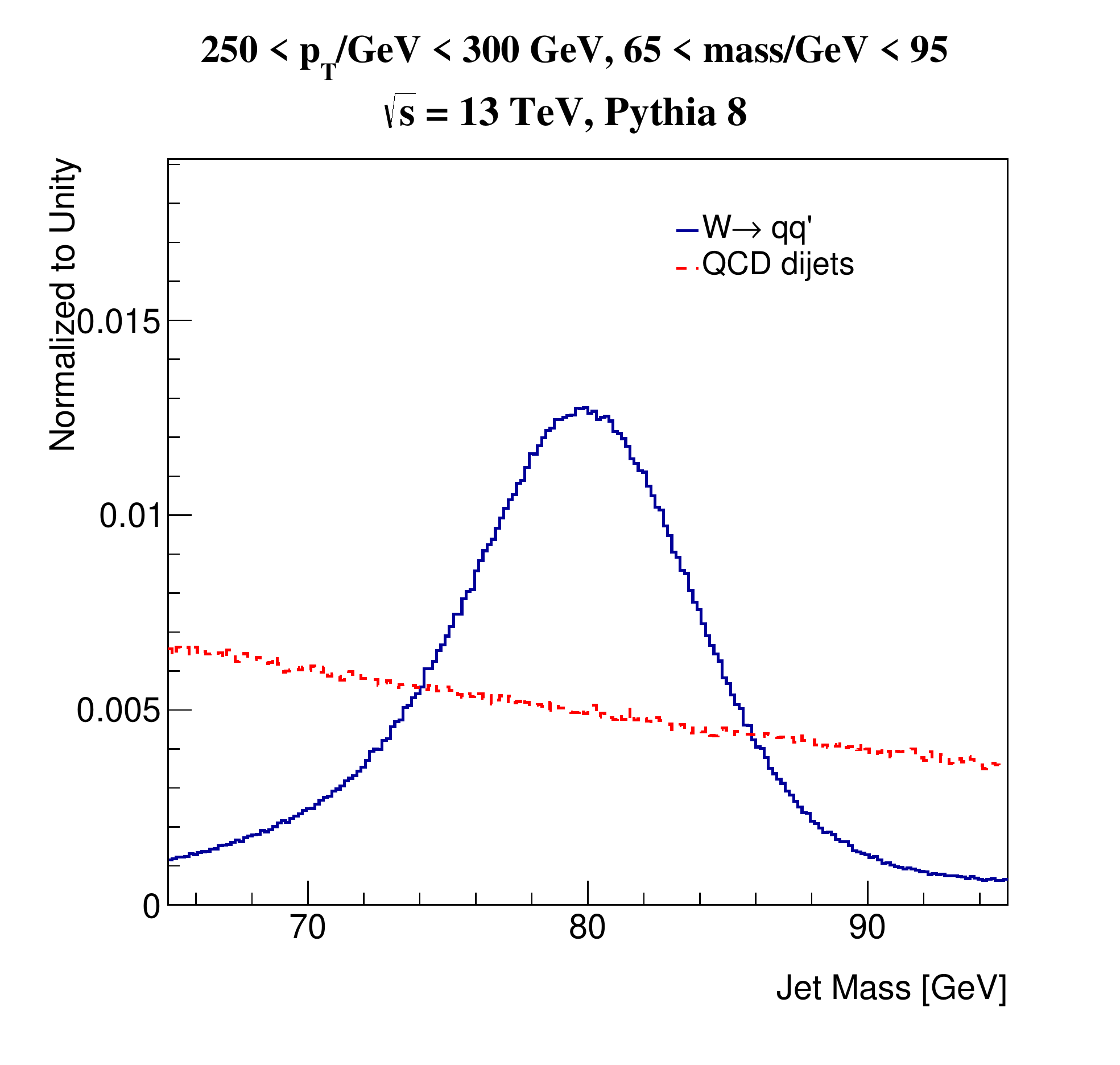} 
        \includegraphics[width=0.32\textwidth]{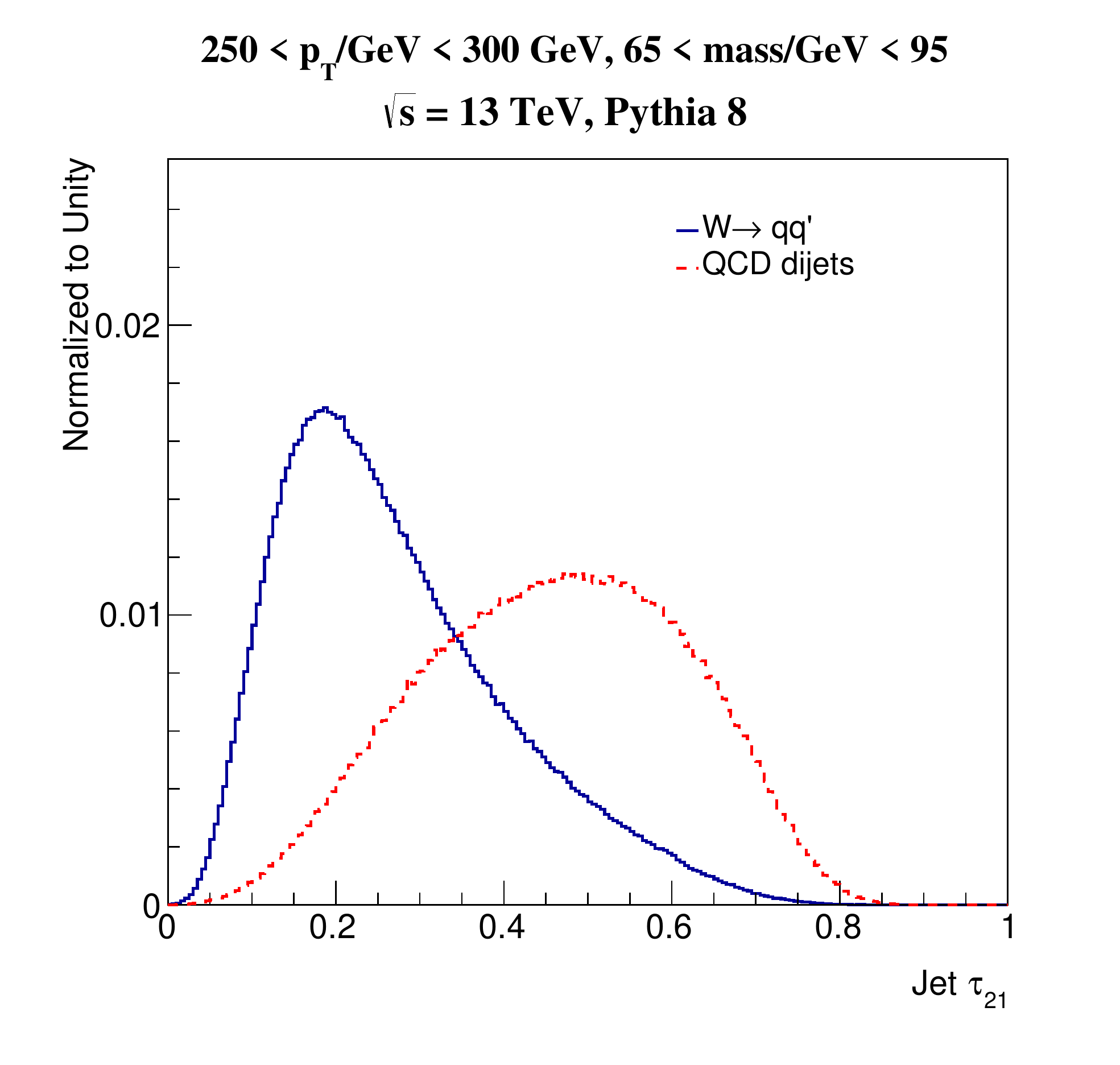} 
        \includegraphics[width=0.32\textwidth]{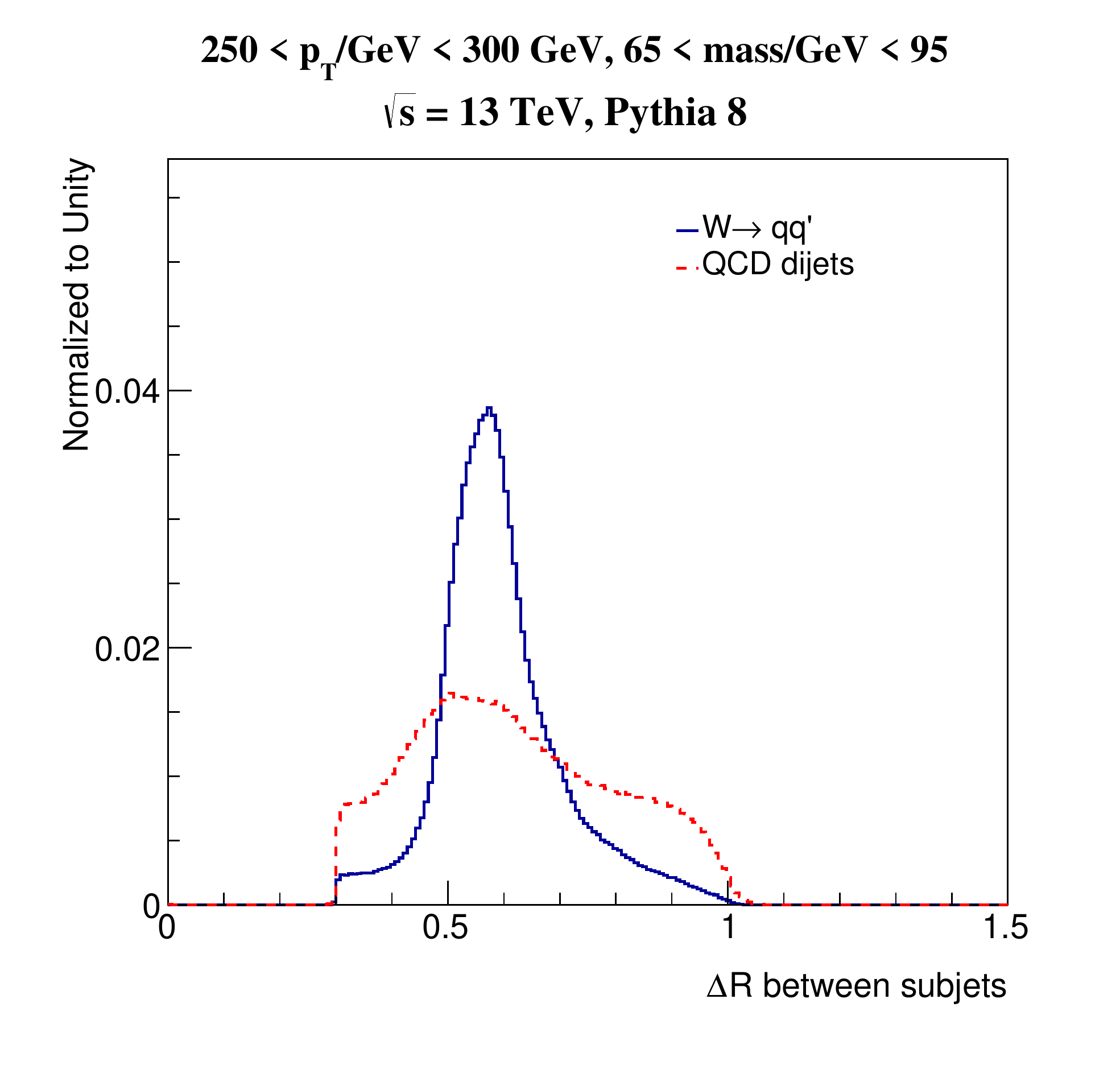}
      \caption{ The distributions of the jet mass (left), $2$-subjettiness ($\tau_{21}$) (middle) and the $\Delta R$ between subjets (right) for signal (blue) and background (red) jets.
      \label{fig:datastats} }
    \end{center}
\end{figure}

 A {\it jet image} is formed by taking the constituents of a jet and discretizing its energy into pixels in ($\eta,\phi$).  In this section, the detector and jet images have the same granularity, so the first step in forming the image is to draw a box of a fixed size ($25\times 25$) around the jet area.
 
In order for the machine learning algorithms to most efficiently learn discriminating features between signal and background and to not learn the symmetries of space-time, the jet images are pre-processed.  This procedure can greatly improve performance and reduce the required size of the sample used for testing.  The pre-processing procedure happens in four steps: translation, rotation, re-pixelation, and inversion.  To begin, the jet images are translated so that the leading subjet is at $(\eta,\phi)=(0,0)$.  Translations in $\phi$ are rotations around the $z$-axis and so the pixel intensity is unchanged by this operation.  On the other hand, translations in $\eta$ are {\it Lorentz boosts} along the $z$-axis, which do not preserve the pixel intensity.  A proper translation in $\eta$ would modify the intensity.  One simple modification of the jet image to circumvent this change is to replace the pixel intensity $E_i$ with the transverse energy $p_{T,i}=E_i/\cosh(\eta_i)$.  This new definition of intensity is invariant under translations in $\eta$ and is used exclusively for the rest of this section.

The second step of pre-processing is to rotate the images around the center of the jet.  If a jet has a second subjet, then the rotation is performed so that the second subjet is at $-\pi/2$.  If no second subjet exists, then the jet image is rotated so that the first principle component of the pixel intensity distribution is aligned along the vertical axis.  Unless the rotation is by an integer multiple of $\pi/4$, the rotated grid will not line up with the original grid.  Therefore, the energy in the rotated grid must be re-distributed amongst the pixels of the original image grid.  A cublic spline interpolation is used in this case - see Ref.~\cite{Cogan:2014oua} for details.  The last step is a parity flip so that the right side of the jet image has the highest sum pixel intensity.  

Figure~\ref{fig:preprocess} shows the average jet image for $W$ boson jets and QCD jets before and after the rotation, re-pixelation, and parity flip steps of the pre-processing.  The more pronounced second-subjet can already be observed in the left plots of Fig.~\ref{fig:preprocess}, where there is a clear annulus for the signal $W$ jets which is nearly absent for the background QCD jets.  However, after the rotation, the second core of energy is well isolated and localized in the images.  The spread of energy around the leading subjet is more diffuse for the QCD background which consists largely of gluon jets that have an octet radiation pattern.  This is compared to the singlet nature of the $W$ jets where the radiation is mostly restricted to the region between the two hard cores (see Chapter~\ref{cha:colorflow}).

\begin{figure}[htbp!]
  \begin{center}
        \includegraphics[width=0.99\textwidth]{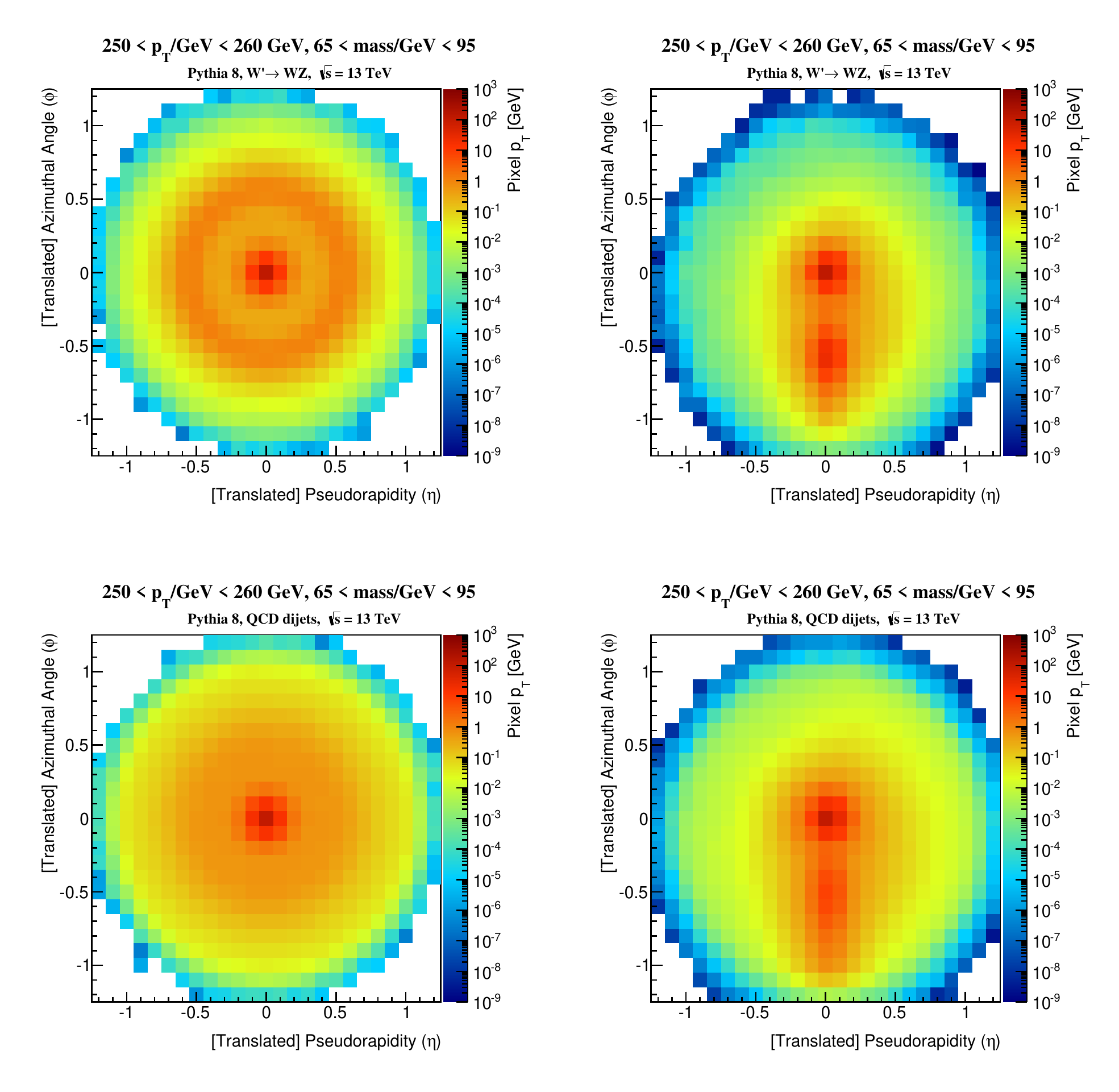}
      \caption{ The average jet image for signal $W$ jets (top) and background QCD jets (bottom) before (left) and after (right) applying the rotation, re-pixelation, and inversion steps of the pre-processing.  The average is taken over images of jets with $240$ GeV $<p_T<$ 260 GeV and 65~GeV~$<$ mass $<$~95~GeV.
      \label{fig:preprocess} }
    \end{center}
\end{figure}

One standard pre-processing step that is often additionally applied in computer vision algorithms is normalization.  A common normalization scheme is the $L^2$ norm such that $\sum I_i^2=1$ where $I_i$ is the intensity of pixel $i$.  This is particularly useful for the jet images where pixel intensities can span many orders of magnitude, and when there is large pixel intensity variations between images.  In this study, the jet transverse momenta are all around 250 GeV, but this can be spread amongst many pixels or concentrated in only a few. The $L^2$ norm helps mitigate the spread and thus makes training easier for the machine learning algorithms.  However, normalization can distort the information contained within the jet image.  Some observables, such as the $\Delta R$ between subjets, is invariant under all of the pre-processing steps as well as normalization.  However, consider the {\it image mass}, 

\begin{align}
m_I^2=\sum_{i<j} E_iE_j(1-\cos(\theta_{ij})),
\end{align}

\noindent where $E_i=I_i/\cosh(\eta_i)$ for pixel intensity $I_i$ and $\theta_{ij}$ is the angle between massless four-vectors with $\eta$ and $\phi$ at the $i$ and $j$ pixel centers.  The image mass is not invariant under all pre-processing steps but does encode key information to identify highly boosted bosons that would ideally be preserved by the pre-processing steps.  As discussed earlier, with the proper choice of pixel intensity, translations preserve the image mass since it is a Lorentz invariant quantity.  However, the rotation pre-processing step does not preserve the image mass.  To understand this effect, consider two four-vectors: $p^\mu=(1,0,0,1)$ and $q^\mu=(0,1,0,1)$.   The invariant mass of these vectors is $\sqrt{2}$.  The vector $p^\mu$ is at the center of the jet image coordinates and the vector $q^\mu$ is located at $\pi/2$ degrees.  If the image is rotated around the jet axis so that the vector $q^\mu$ is at $0$ degrees, akin to rotating the jet image so that the sub-leading subjet goes from $\pi/2$ to $0$, then $p^\mu$ is unchanged but $q^\mu\rightarrow (1,0,\sinh(1),\cosh(1))$.  The new invariant mass of $q^\mu$ and $p^\mu$ is about $1$, which is reduced from its original value of $\sqrt{2}$.  The parity inversion pre-processing step does not impact the image mass, but a $I^2$ normalization does modify the image mass.  The easiest way to see this is to take a series of images with exactly the same image mass but variable $I^2$ norm.  The map $I_i\mapsto I_i/\sum_j I_j^2$ modifies the mass by $m_I\mapsto m_I/\sum_j I_j^2$ and so the variation in the normalizations induces a smearing in the jet-image mass distribution.

The impact of the various stages of pre-processing on the image mass are illustrated in Fig.~\ref{fig:preprocess2}.  The finite segmentation of the simulated detector slightly degrades the jet mass resolution, but the translation and parity inversion (flip) have no impact, by construction, on the jet mass.  The rotation that will have the biggest potential impact on the image mass is when the rotation angle is $\pi/2$ (maximally changing $\eta$ and $\phi$), which does lead to a small change in the mass distribution.  A translation in $\eta$ that uses energy as the intensity instead of $p_\text{T}$ (referred to as the \textit{naive translation}) and the $L^2$ normalization scheme both significantly broaden the mass distribution.  One way to quantify the amount of information in the jet mass that is lost by various pre-processing steps is shown in the Receiver Operator Characteristic (ROC) curve of Fig.~\ref{fig:preprocess3}.  Information about the mass is lost when the ability to use the mass to differentiate signal and background is diminished.  The naive translation and the $I^2$ normalization schemes are significantly worse than the other image mass curves which are themselves similar.

\begin{figure}[h!]
  \begin{center}
        \includegraphics[width=0.5\textwidth]{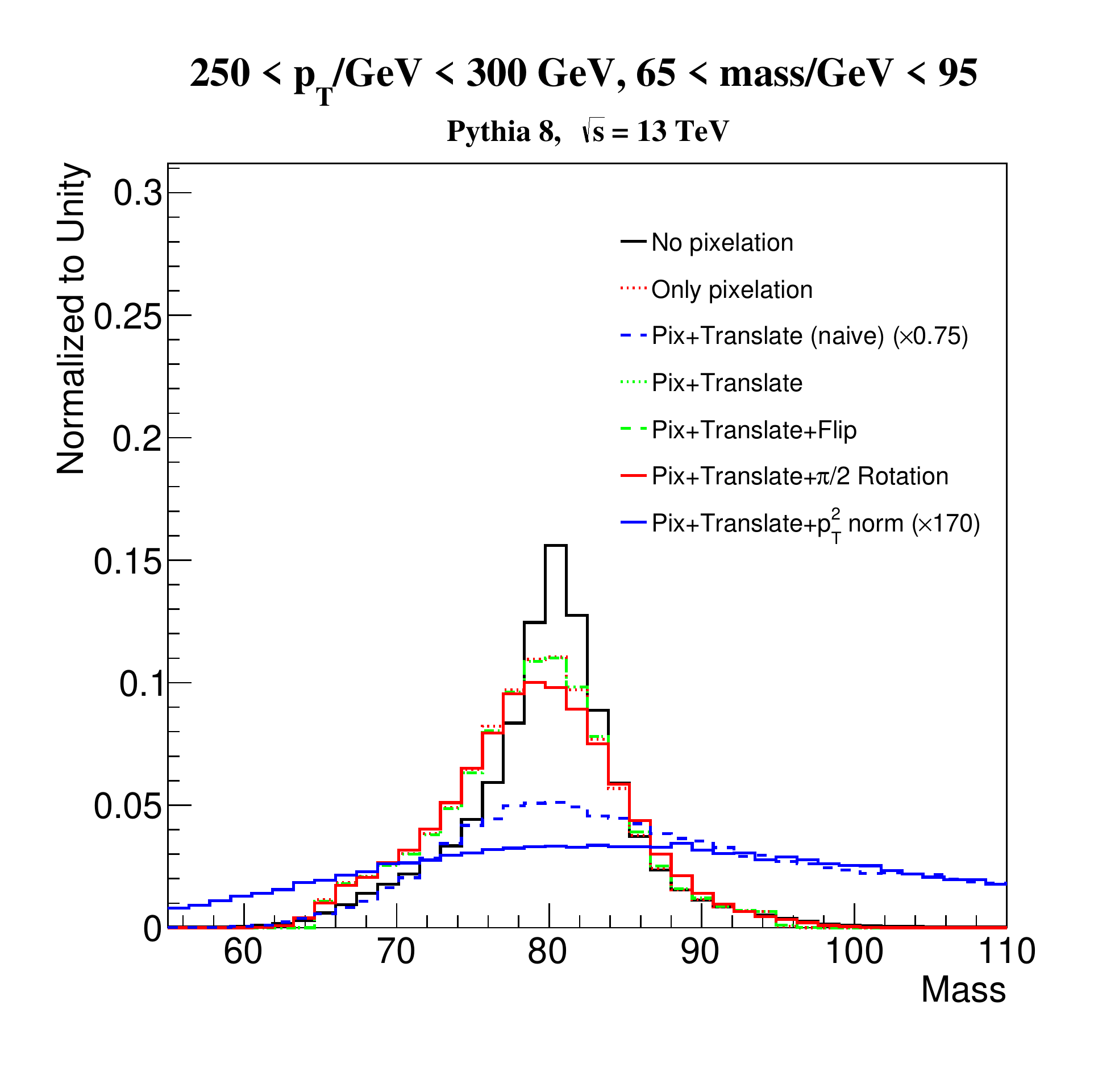}\includegraphics[width=0.5\textwidth]{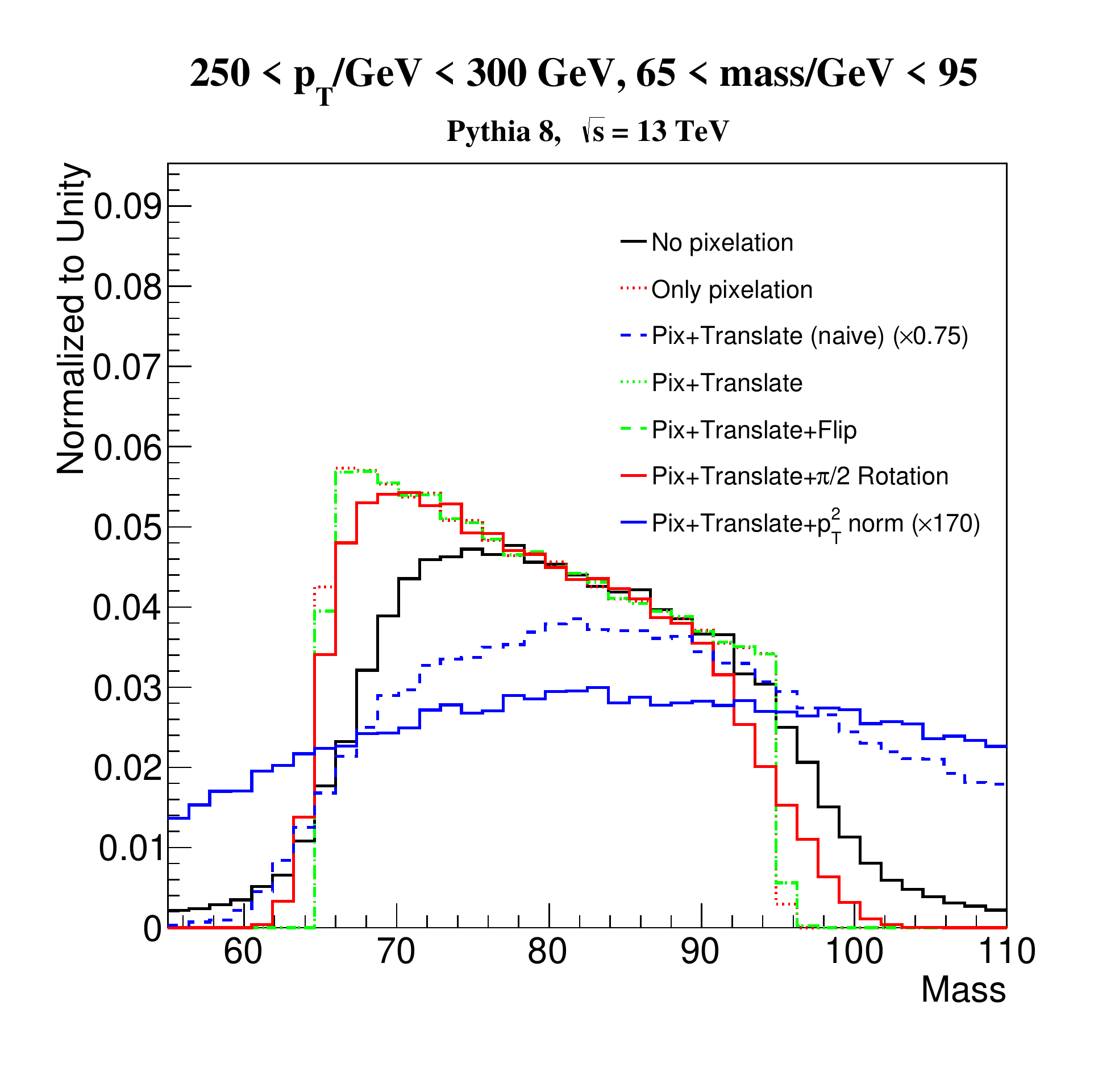}
      \caption{ The distribution of the image mass after various states of pre-processing for signal jets (left) and background jets (right). The naive translation and the $I^2$ normalization image masses are both multiplied by constants so that the centers of the distribution are roughly in the same location as for the other distributions.
      \label{fig:preprocess2} }
    \end{center}
\end{figure}

\begin{figure}[h!]
  \begin{center}
        \includegraphics[width=0.5\textwidth]{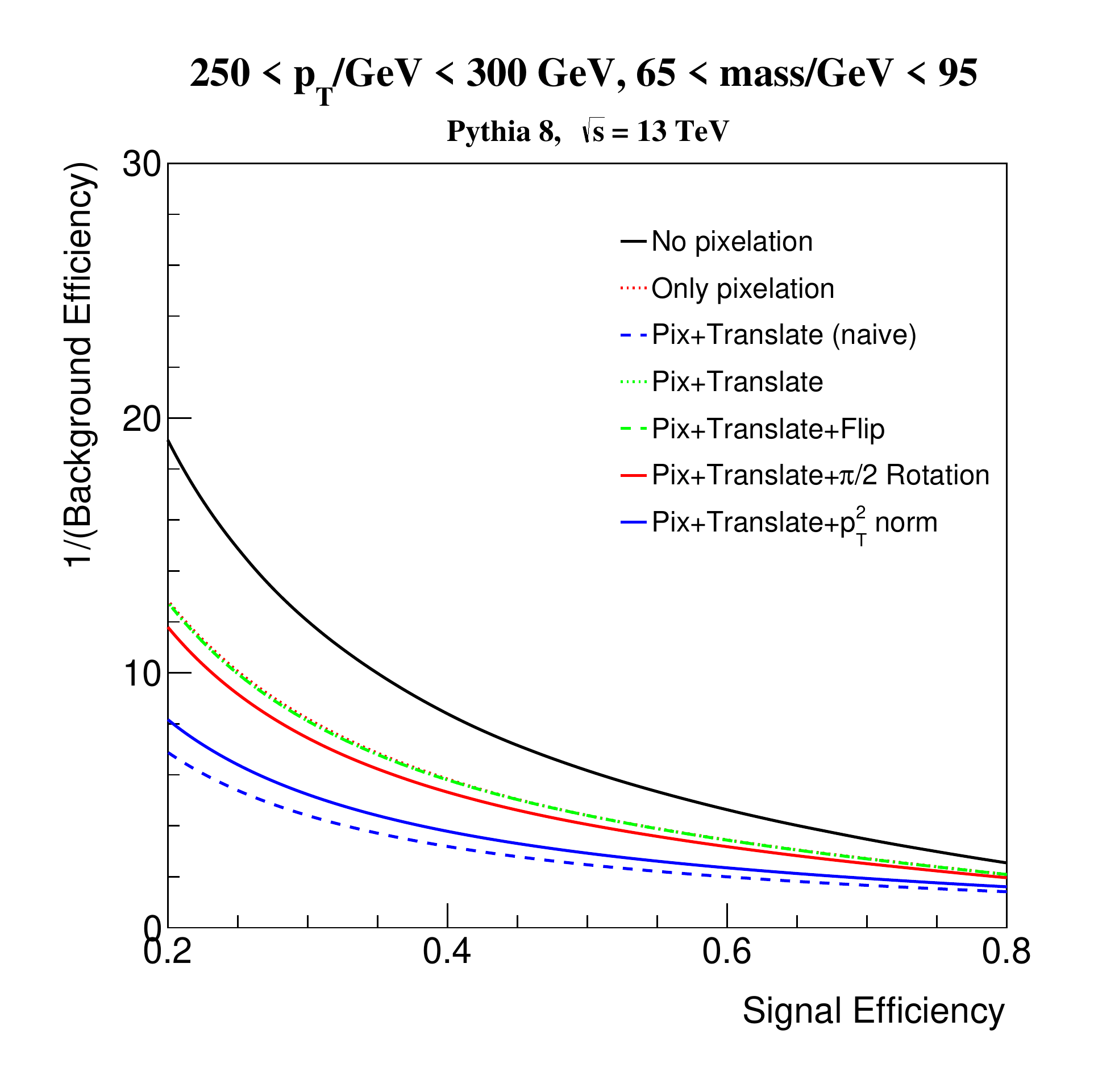}
      \caption{ The tradeoff between $W$ boson (signal) jet efficiency and inverse QCD (background) efficiency for various pre-processing algorithms applied to the jet (images). 
      \label{fig:preprocess3} }
    \end{center}
\end{figure}

\subsubsection{Network Architecture}
\label{sec:arch}

One of the most successful architectures for modern computer vision is the convolution neural network (CNN or Convnet).  A detailed description of the CNN, its components, and related ideas, is beyond the scope of this section.  Traditional (shallow) neural networks are now standard tools (and likely common knowledge), but Ref.~\cite{Goodfellow-et-al-2016-Book} is a thorough textbook on {\it deep} neural networks and Ref.~\cite{888} is a review with many references to current research.  The basic feature of a CNN that distinguishes it from a regular ({\it fully connected}) network is that each node of the output layer is connected to only a small number of nodes (=pixels for the first layer) from the input layer.  The connection from an output node to an input node is the result of a discrete convolution of a filter with a patch of the input.  Convolutional networks work well for detecting features, wherever they may be in the image.  However, unlike images of natural or human-made scenery, jet images are very sparse.   Figure~\ref{fig:occupancy} shows the distribution of the occupancy.  Typically only $5$-$10\%$ of pixels are non-zero and lack edges or other obvious features.  Tests with different filter sizes found that an usually large filter of $11\times 11$ was optimal.  This size is just big enough to capture the only clear jet-by-jet feature: a core separated from a second node of radiation.  To complement the convolutional network, a fully connected network based on the MaxOut activation function~\cite{maxout:goodfellow} is used for comparisons in the next section.  For a complete description of the sequence of activation functions, non-linearities (rectified linear except at the last layer, where a sigmoid is used), and down-samplings see Sec.~4 in Ref.~\cite{deOliveira:2015xxd}.

\begin{figure}[htbp!]
  \begin{center}
        \includegraphics[width=0.5\textwidth]{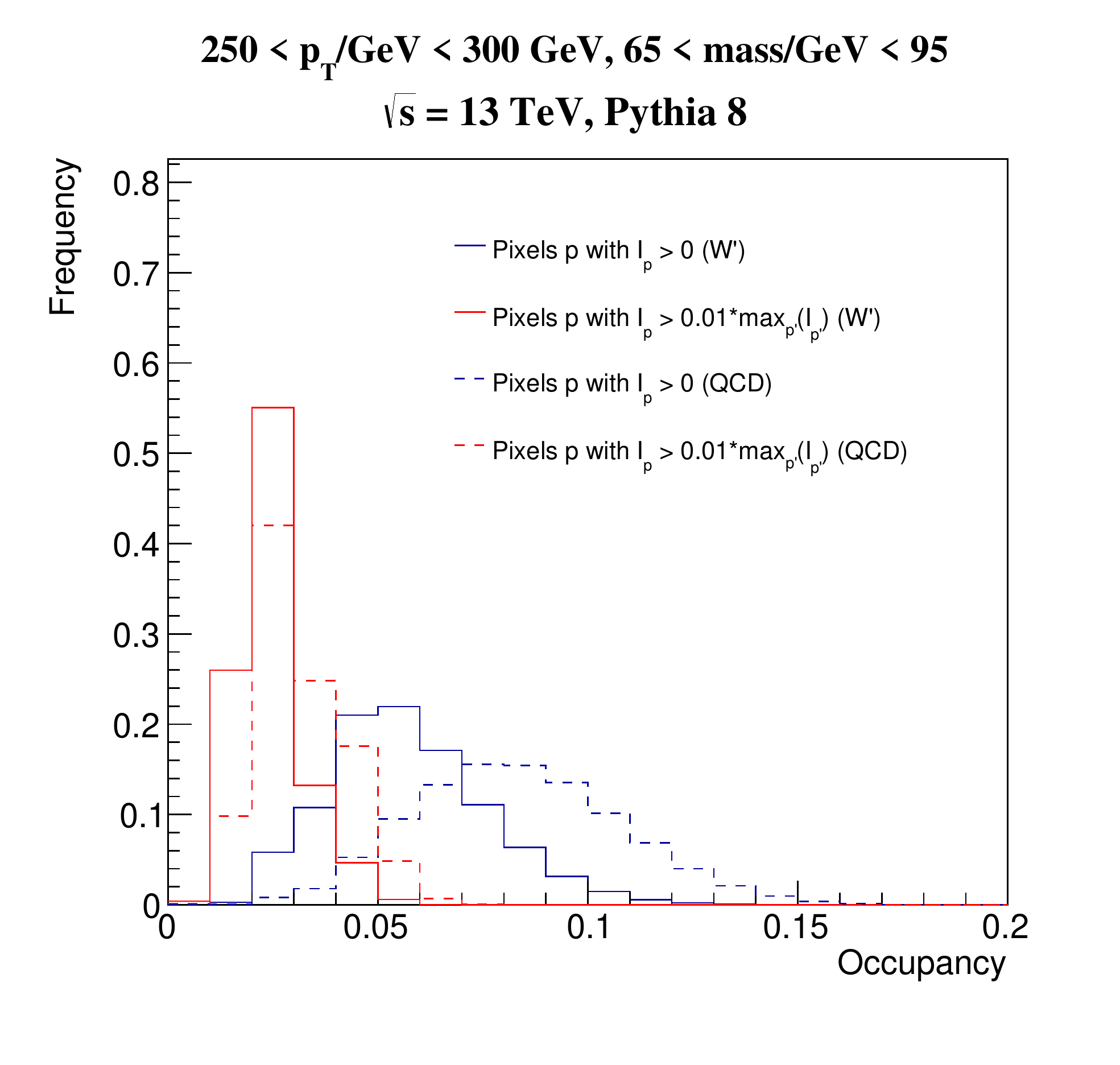}\includegraphics[width=0.5\textwidth]{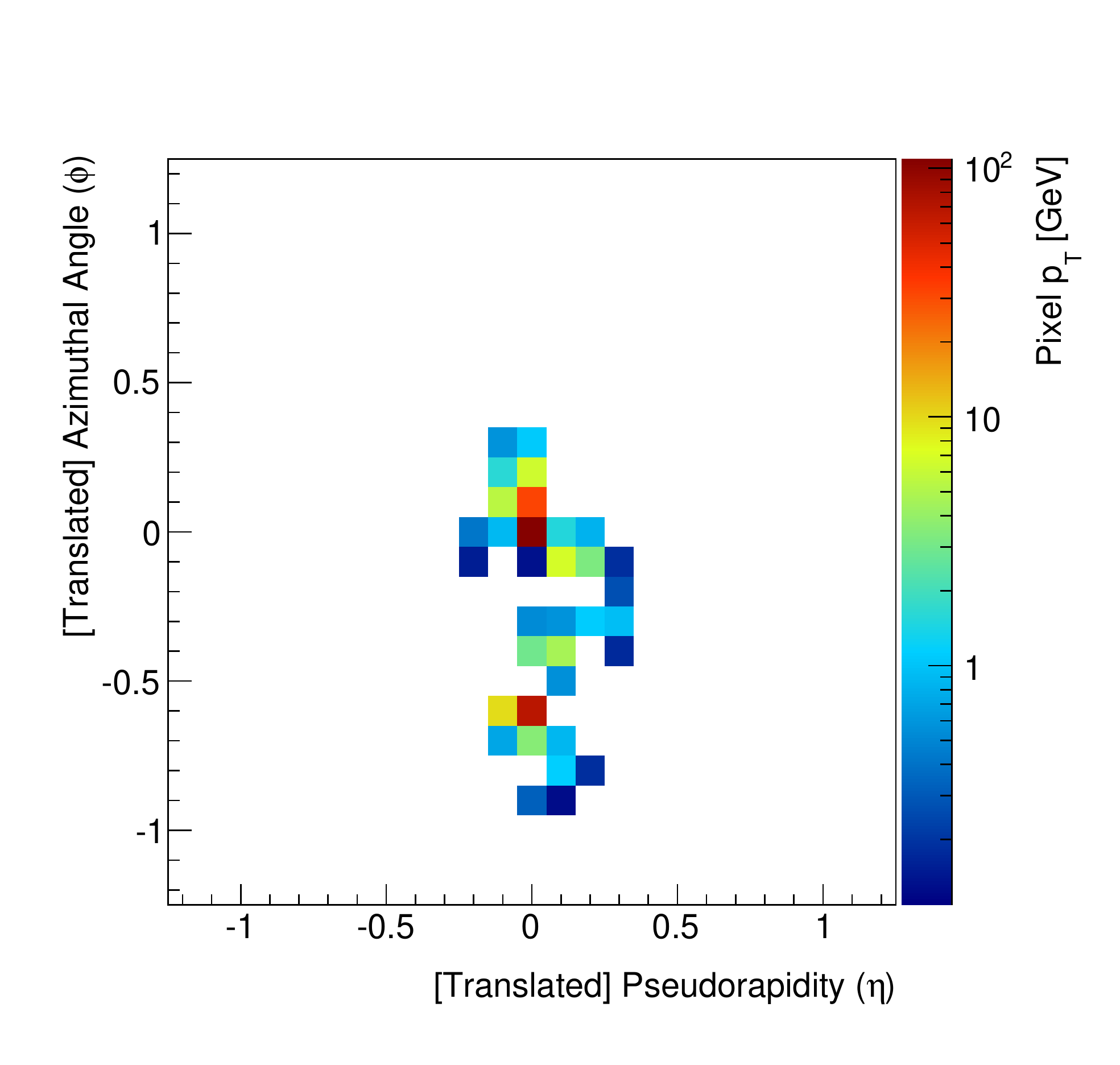}
      \caption{ Left: The distribution of the fraction of pixels (occupancy) that have a nonzero entry (blue) or at least 1\% of the scalar sum of the pixel intensities from all pixels (red).  Right: A typical single $W$ jet image.
      \label{fig:occupancy} }
    \end{center}
\end{figure}

\clearpage

\subsubsection{Performance and Visualization} \label{sec:studies}

Figure~\ref{fig:combinedROC1} shows the $W$ tagging performance of the DNNs compared with the benchmark physically-motivated observables.   Both the CNN and MaxOut networks out-perform the single benchmarks and their pairwise combinations. For example, at a signal efficiency of $30\%$, the best DNN has a $60\%$ larger rejection than the likelihood combination of mass and $\tau_{21}$.  The fully connected network outperforms the CNN and interestingly the CNN with normalized input images outperforms the CNN with unnormalized images.  Section~\ref{sec:preprocess} showed that normalization washes out information about the jet mass, which is the first indication (more below) that the network(s) are not fully learning information about the jet mass.

\begin{figure}[!htbp]
\begin{center}
\includegraphics[width=0.45\textwidth,angle=0]{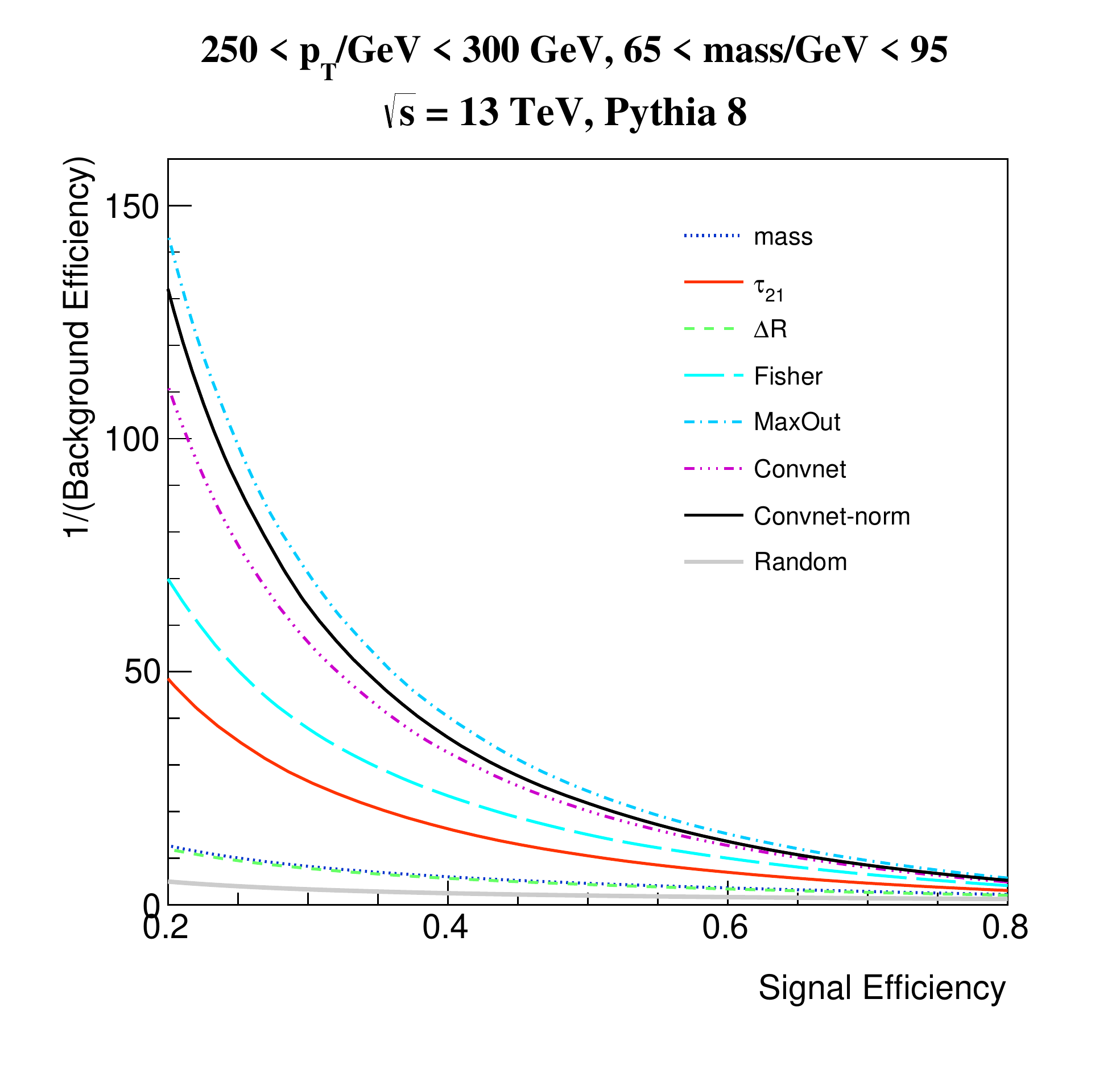}\includegraphics[width=0.45\textwidth,angle=0]{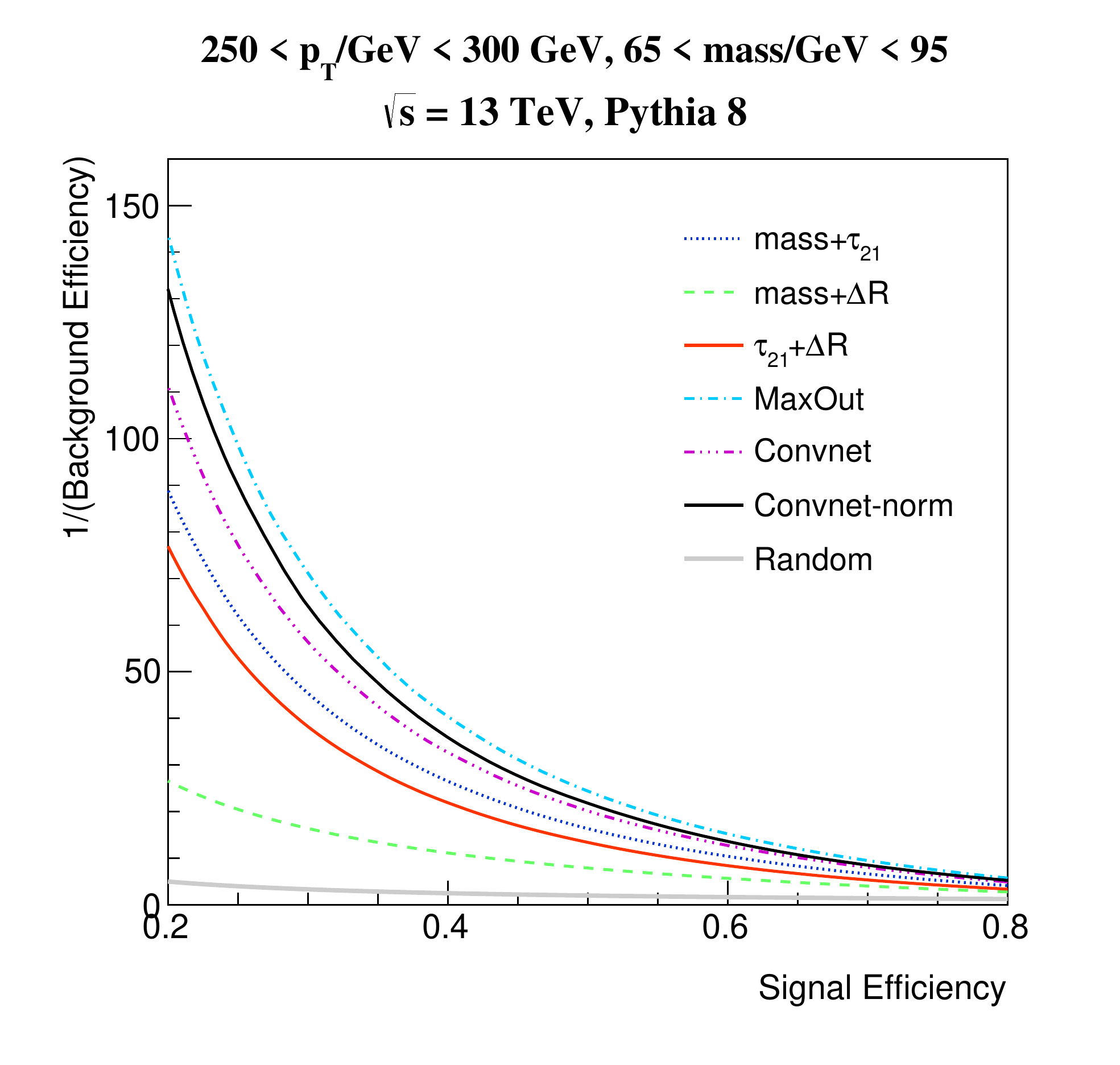}
\end{center}
   \caption{Left: ROC curves for individual physically motivated features as well as three deep neural network discriminants.  Right: the DNNs are compared with pairwise combinations of the physically motivated benchmarks.}
  \label{fig:combinedROC1}
\end{figure}

One way to test if a neural network has learned the discriminating information in a benchmark observable is to assess the performance of a combination of the variable with the DNN output.  The combinations of jet mass, $\tau_{21}$ and $\Delta R$ with the MaxOut network are shown in Fig.~\ref{fig:combinedROC2}.  Combining $\Delta R$ or $\tau_{21}$ with the DNN output does not improve the performance while there is a significant improvement for the mass+DNN combination.  One common feature of $\Delta R$ and $\tau_{21}$ is that they are {\it scale-invariant}, i.e. scaling the jet image by a constant amount (as in normalization) does not change their values.  They encode strictly geometric information about the radiation pattern within the jet.  In contrast, the jet mass depends on both geometric and scale information.  Figure~\ref{fig:combinedROC2} may indicate that scale information is not well-learned by the network.  Corresponding curves for the CNN show the same qualitative features as Fig.~\ref{fig:combinedROC2}.

\begin{figure}[!htbp]
  \begin{center}
			\includegraphics[width=0.45\textwidth,angle=0]{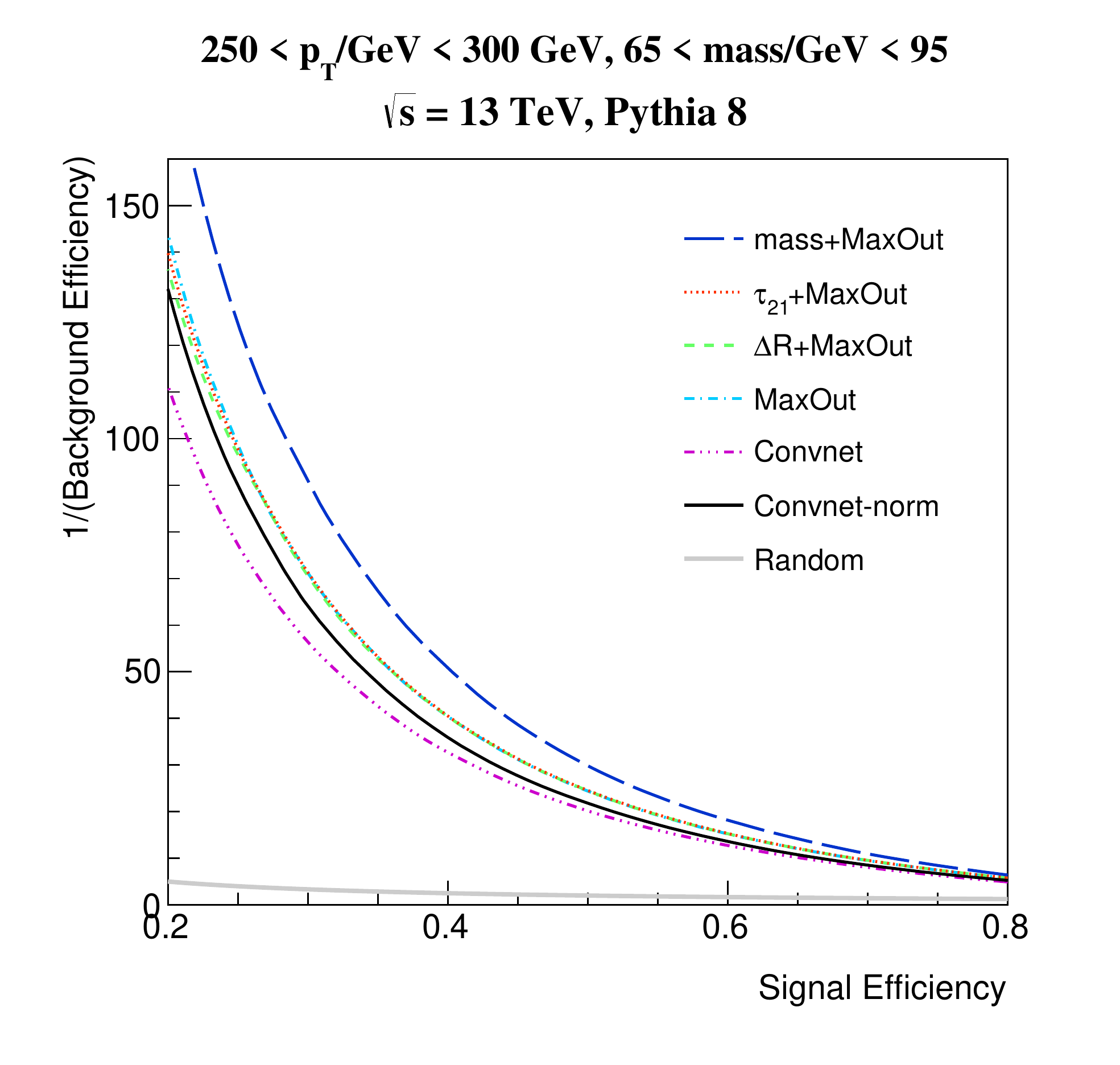}
	\end{center}
  \caption{ROC curves that combined the DNN outputs with physics motivated features MaxOut architecture.}
  \label{fig:combinedROC2}
\end{figure} 

One way to visualize the trends from Fig.~\ref{fig:qcdsculpt} is to consider how the DNN sculpts the distribution of the benchmark observables in background events, i.e. $p(x|\text{DNN})$.  If the background distribution of $x$ for signal-like DNN output is not the same as the signal distribution of $x$, then there is more information in $x$ than is contained in the network output. The left plot of Fig.~\ref{fig:qcdsculpt} shows that the peak of the $\tau_{21}$ distribution is essentially a non-linear function of the DNN output.  For low values of the network output, the $\tau_{21}$ distribution is peaked at high (background-like) values while high DNN output morphs the distribution to be peaked at low (signal-like) values.  A similar trend is observed for $\Delta R$.  When the network output is small (background-like), the distribution of $\Delta R$ is nearly uniform.  However, for high DNN output (signal-like), the $\Delta R$ distribution is peaked around $0.6$ (set by $m$ and $p_\text{T}$) just like the signal distribution in Fig.~\ref{fig:datastats}.  Even though Fig.~\ref{fig:combinedROC2} indicates that not all of the information about the jet mass is learned by the network, the DNN does appropriately sculpt the background distribution for the extreme DNN outputs.  When the DNN output is close to one or close to zero, the jet mass distribution is peaked at $m_W$ or $65$ (i.e. steeply falling) GeV, respectively.  However, the distribution at intermediate values of the DNN is broad much broader than either extreme.

\begin{figure}[htbp!]
  \begin{center}
  
       \includegraphics[width=0.32\textwidth]{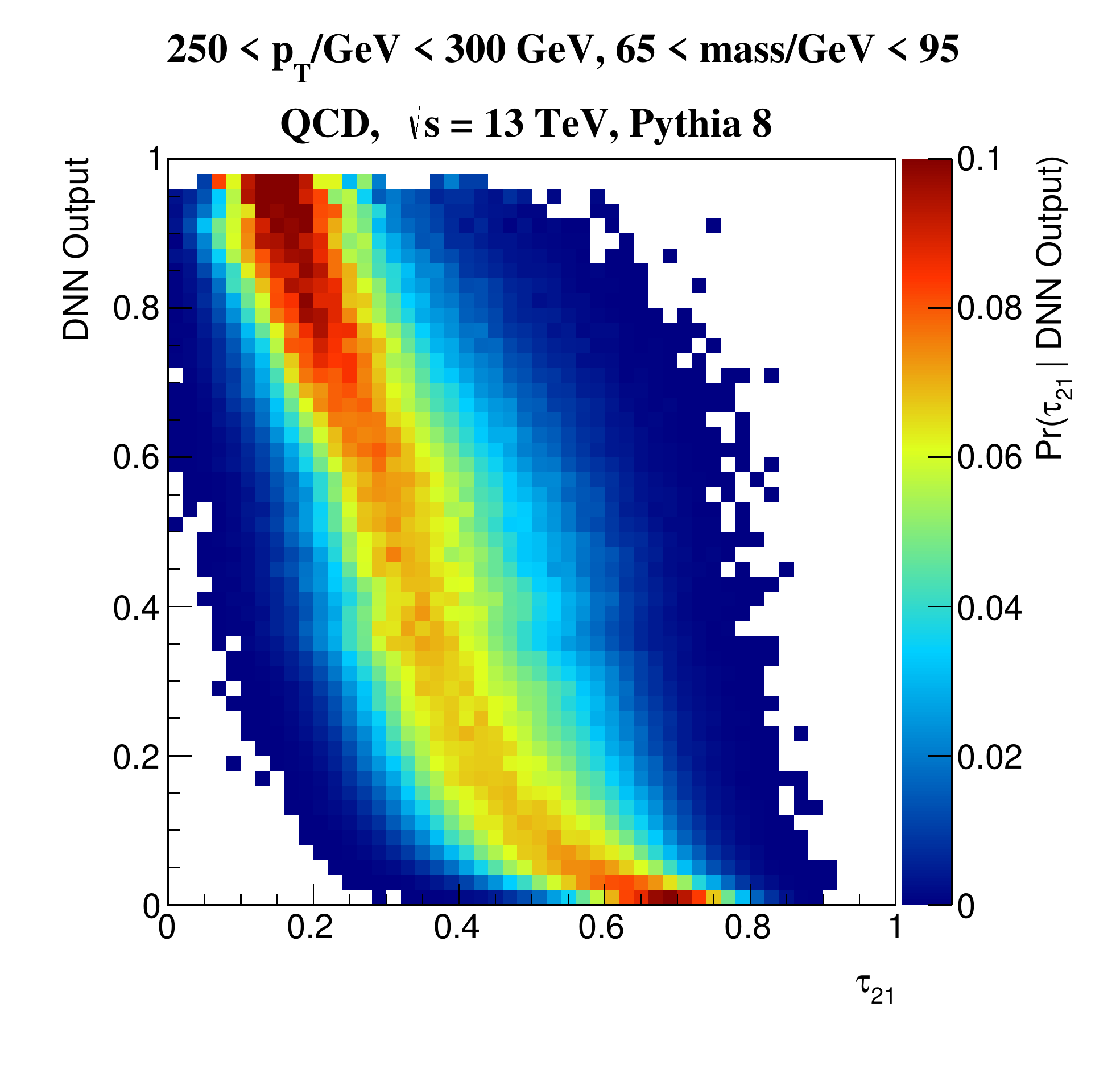}
        \includegraphics[width=0.32\textwidth]{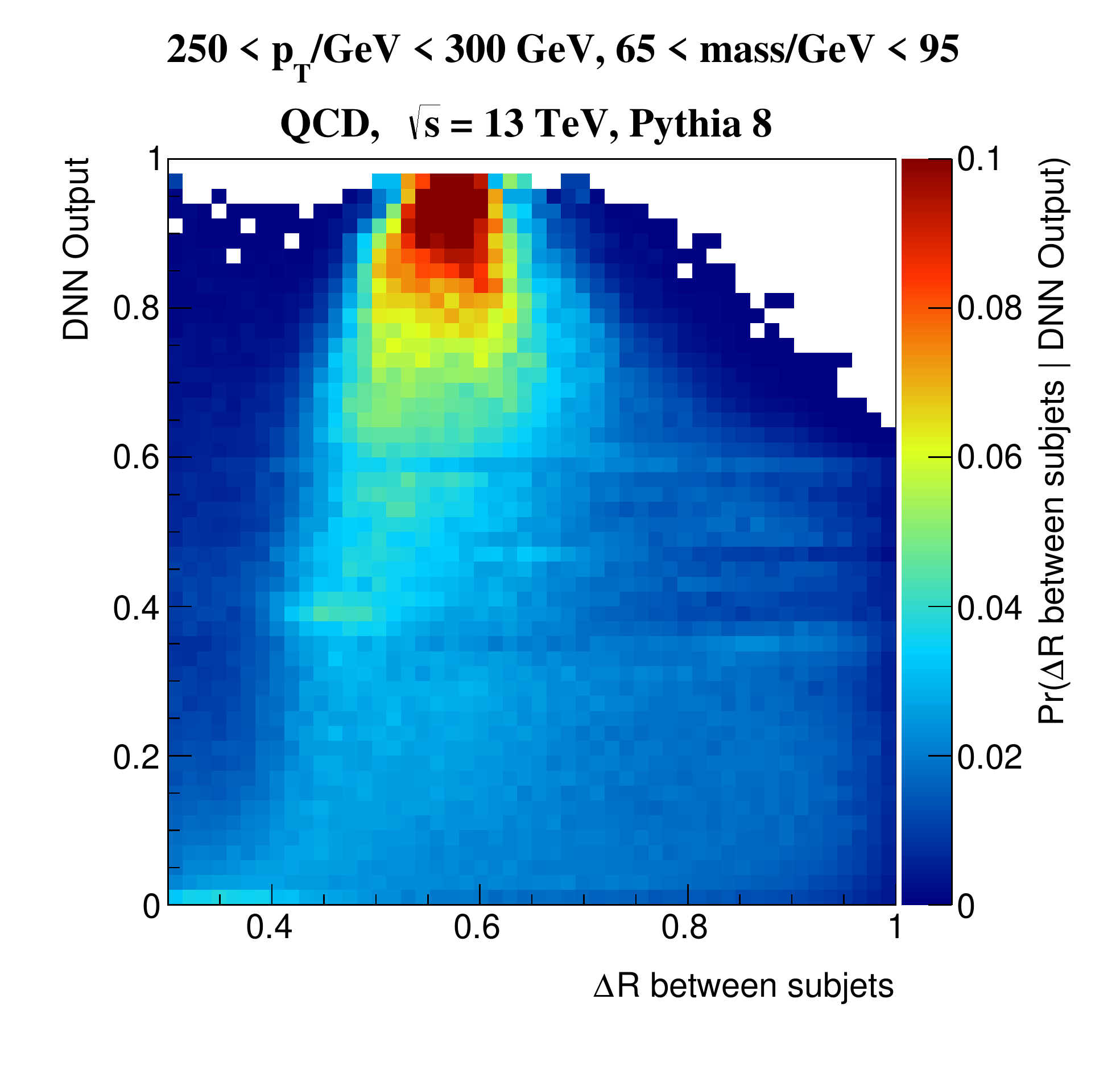} 
        \includegraphics[width=0.32\textwidth]{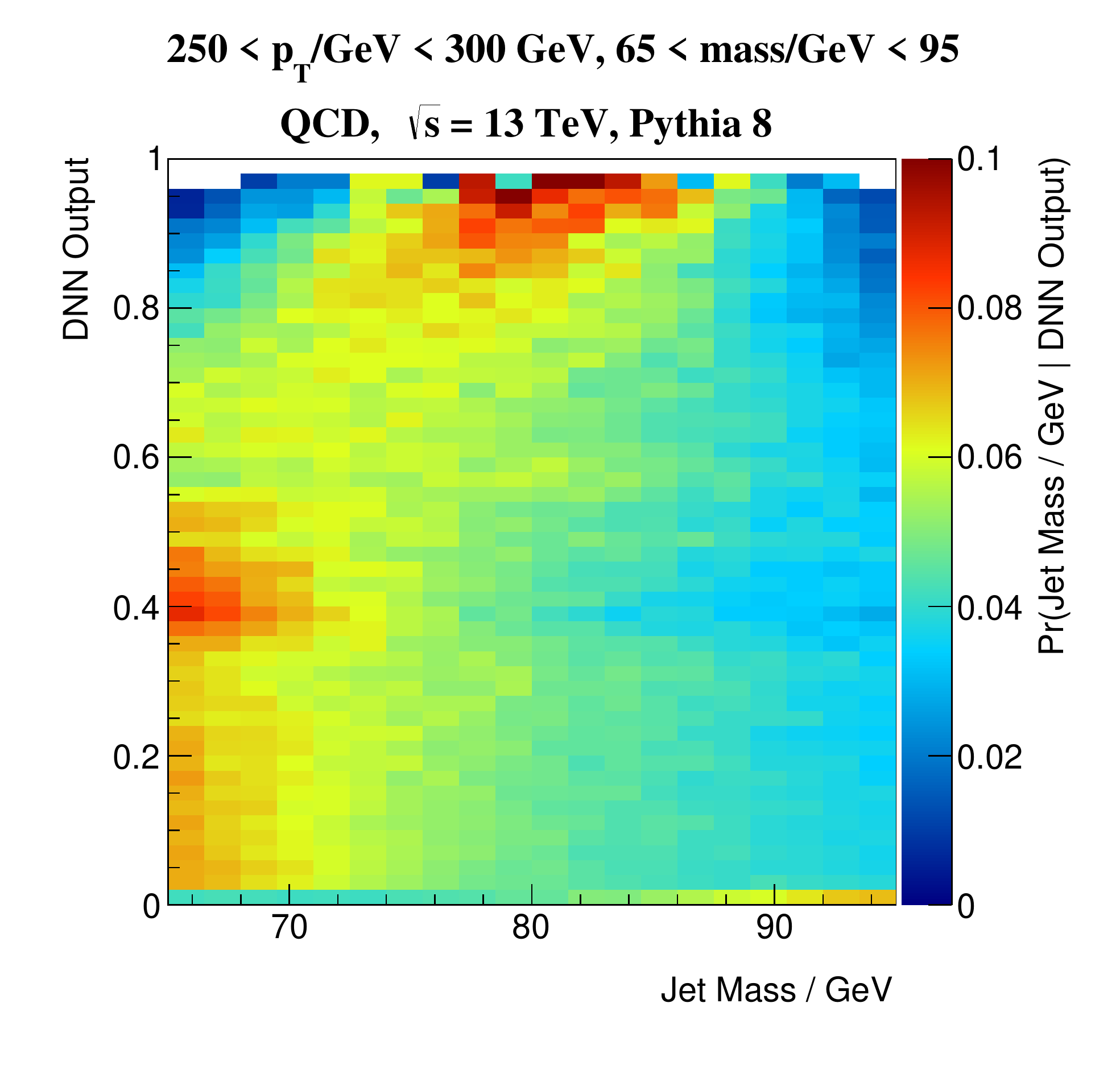}

      \caption{The distribution of $\tau_{21}$ (left), $\Delta R$ (middle) and the jet mass conditioned on the CNN output for background jets.  Distributions for the MaxOut network look qualitatively the same.}
      \label{fig:qcdsculpt}

    \end{center}
\end{figure}

The remainder of this section is dedicated to probing what about the radiation pattern the DNN has learned beyond e.g. $\tau_{21}$ and the jet mass.  A first step is to study {\it what information could be learned} by the network by considering the typical signal and background jet images in a small window of $\tau_{21}$ and jet mass.  Figure~\ref{fig:meanImagesWindow} shows the average jet image in three windows of $\tau_{21}$ for a fixed small window of the jet mass and jet $p_\text{T}$.  As expected, in these small windows the signal and background distributions look nearly identical: at low $\tau_{21}$ the jets have two distinct cores of energy and at high $\tau_{21}$ there are no longer two clear subjets.  The subtle differences between the top and bottom rows of Fig.~\ref{fig:meanImagesWindow} are magnified by taking the image differences, shown in Fig.~\ref{fig:meanImagesWindow2}. In the window with $\tau_{21}\in$[0.19,0.21], there are five features: a localized blue patch in the bottom center, a localized red patch just above that, a red diffuse region between the red patch and the center and then a blue dot just left of center surrounded by a red shell to the right.  Each of these have a physics meaning: the lower two localized patches give information about the orientation of the second subjet ($\Delta R$) which is slightly wider for the QCD jets that need a wider angle to satisfy the mass requirement.  The red diffuse region just above the localized patches is likely an indication of color flow: the $W$ bosons are color singlets compared to the color octet gluon jet background, and thus one expects the radiation pattern to be mostly between the two subjets for the $W$.  One can draw similar conclusions for all the features in each of the plots in Figure~\ref{fig:meanImagesWindow2}.

\begin{figure}[h!]
  \begin{center}
  
        \includegraphics[width=0.99\textwidth]{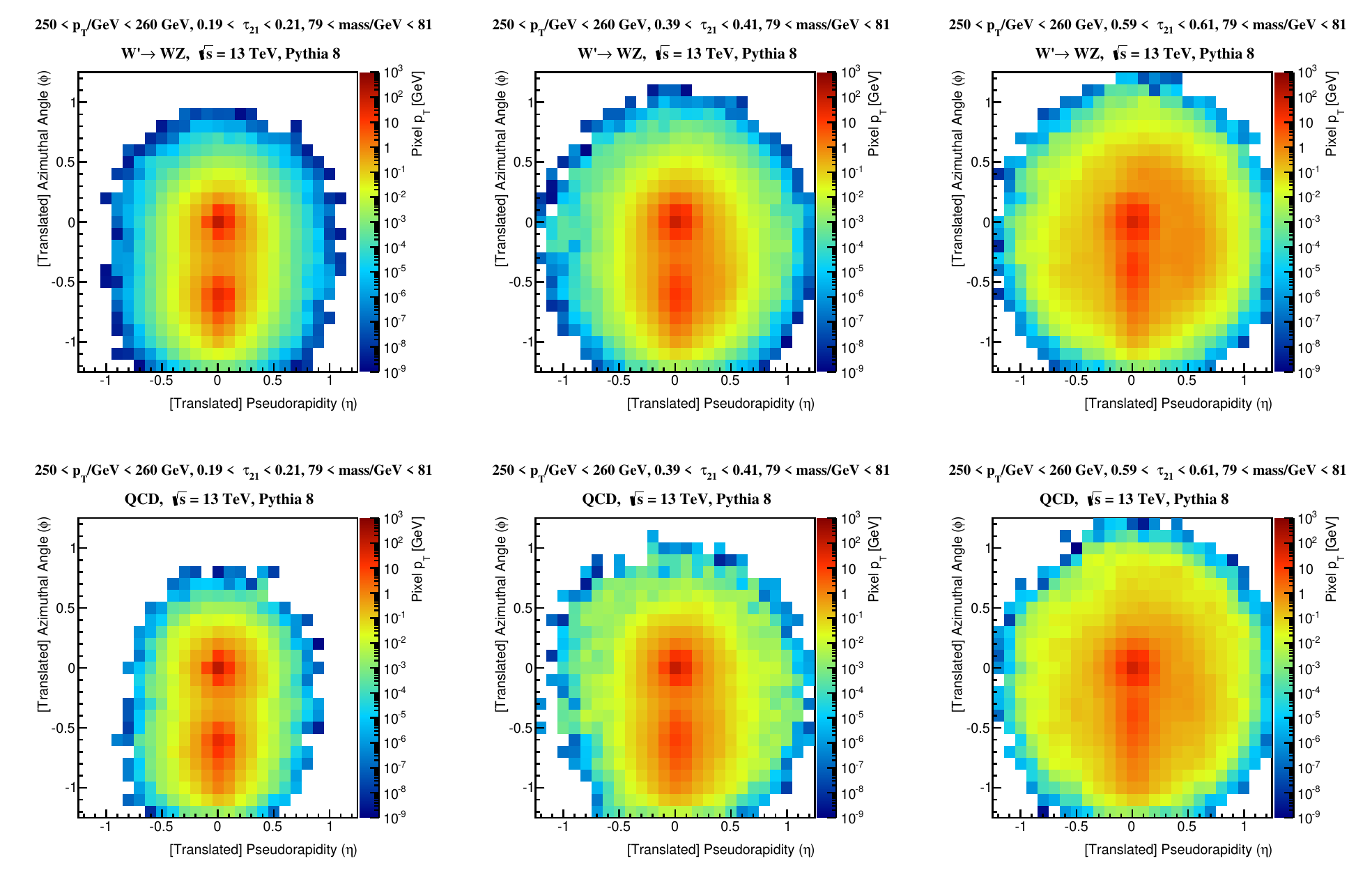}

      \caption{
        $W'\rightarrow WZ$ (top) and QCD (bottom) average jet-images in three small windows of $\tau_{21}$: [0.19, 0.21] (left), [0.39, 0.41] (middle), and [0.59, 0.61] (right).  In all cases, jet mass is restricted to be between 79 GeV and 81 GeV and the jet $p_\text{T}$ is required to be in the interval [250,260] GeV.
        \label{fig:meanImagesWindow} 
      }
    \end{center}
\end{figure}  

\begin{figure}[h!]
  \begin{center}
  
        \includegraphics[width=0.99\textwidth]{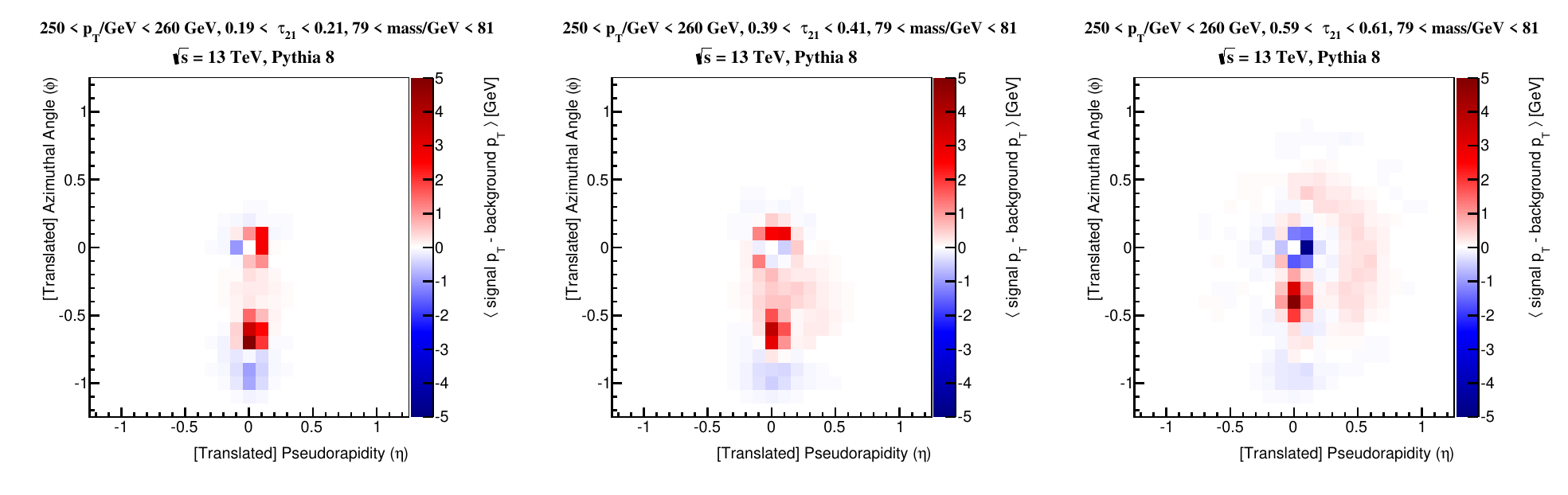}
        
      \caption{
         The average difference between $W'\rightarrow WZ$ jet-images in same small windows of $\tau_{21}$ as Fig.~\ref{fig:meanImagesWindow}.  Red (blue) colors are more signal- (background-)like.
        \label{fig:meanImagesWindow2} 
      }
    \end{center}
\end{figure}  

Figure~\ref{fig:corrWindow} is one way of visualizing if the information available in Fig.~\ref{fig:meanImagesWindow2} is learned by the network.  Each pixel shows the linear correlation with the network output.  The DNN is output is a non-linear function of the inputs, but the distribution of the correlation contains non-linear spatial information about where discrimination information is contained in the jet radiation pattern.  Many of the same features from Fig.~\ref{fig:meanImagesWindow2} appear in these correlation images.  In particular, the radiation between the subjets does seem to be strongly correlated with the DNN - an indication that color flow information is playing a role in the DNN performance.

\begin{figure}[h!]
  \centering
  \includegraphics[width=0.3\textwidth]{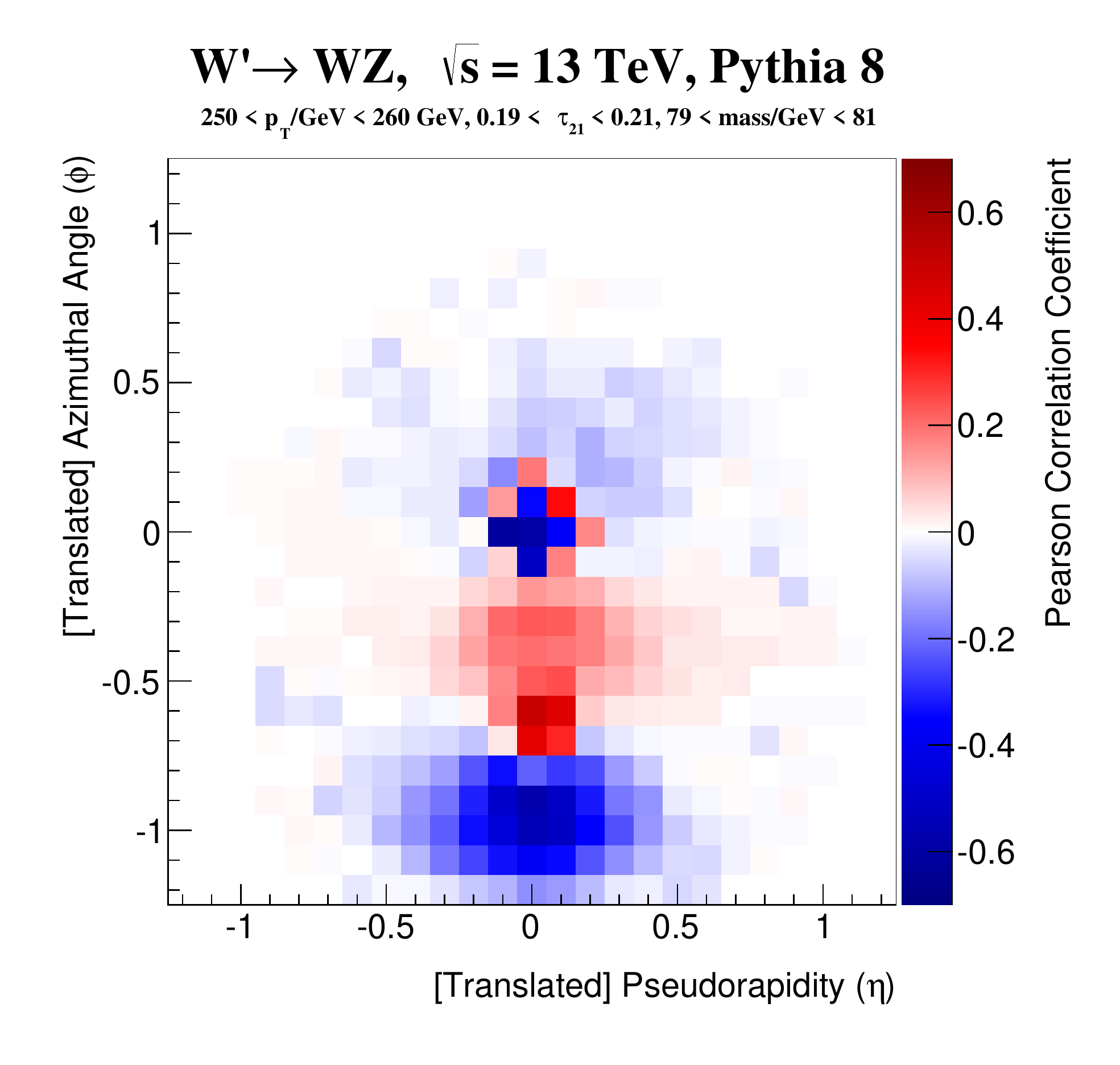}  \includegraphics[width=0.3\textwidth]{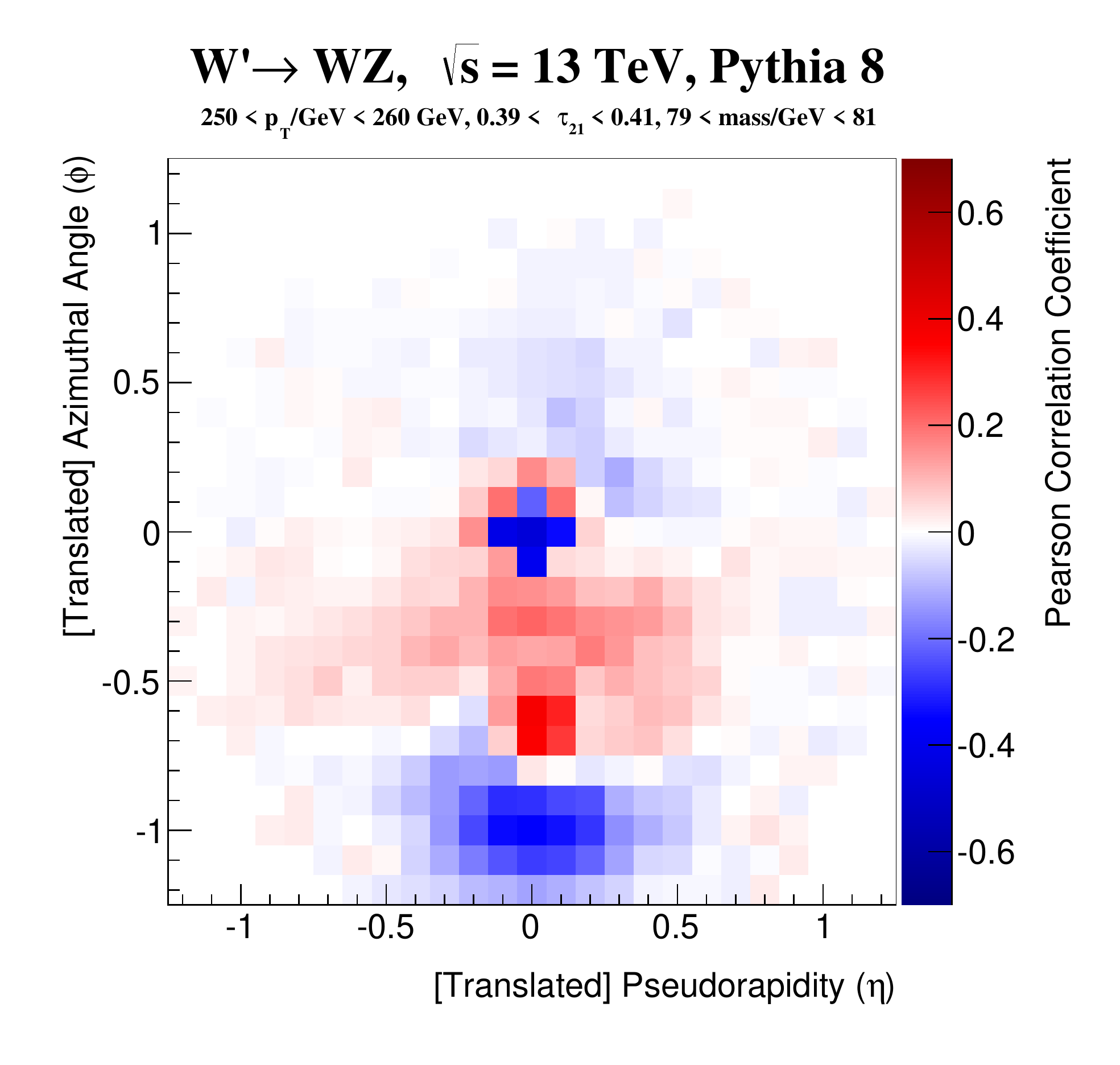}  \includegraphics[width=0.3\textwidth]{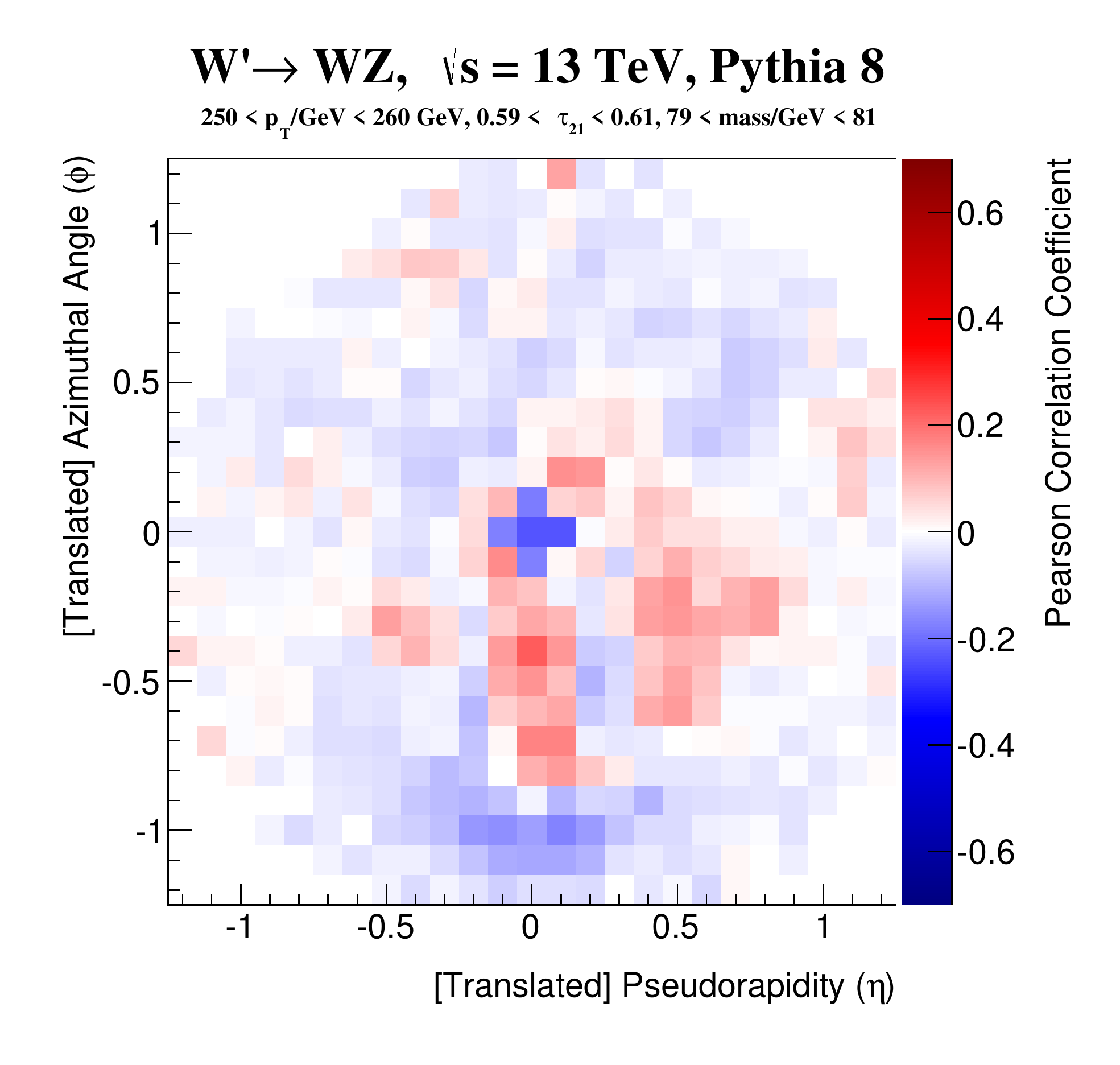}
  \caption{Pearson correlation coefficient for pixel intensity and the CNN output for $W'\rightarrow WZ$ and QCD (combined) in the same small windows of $\tau_{21}$ as Fig.~\ref{fig:meanImagesWindow}.}
  \label{fig:corrWindow}
\end{figure}

Chapter~\ref{cha:colorflow} introduced the jet pull variable and showed that it is sensitive to color flow.  The jet pull angle $\theta_P$ is adapted here for large-radius jets by using subjets instead of resolved jets.  If the leading subjet is labeled $J$ and the subleading subjet is labeled $j$, then there are two pull angles that may contain useful discriminating information related to color flow: $\theta_P(J,j)$ and $\theta_P(j,J)$.  The former pull angle contains substructure information about $J$ and the latter angle uses the substructure of $j$.  Figure~\ref{fig:blahROC} compares the performance of $\theta_P$-based taggers with the other benchmark observables and the DNNs.  In the left plot of Fig.~\ref{fig:blahROC}, the jet mass and $\tau_{21}$ are restricted to a small range as in the previous figures.  By construction, the jet mass and $\tau_{21}$ observable have little discriminating information.  The DNNs are significantly better than the random tagger, but much worse than the inclusive performance from Fig.~\ref{fig:combinedROC1} (jet mass and $\tau_{21}$ are important inputs to the DNN).  The jet pull-based taggers perform significantly better than the random tagger, but are significantly worse than the DNNs.  A similar trend is true for the right plot of Fig.~\ref{fig:blahROC}.  Instead of restricting the phase space, the event weights have been applied in the right plot of Fig.~\ref{fig:blahROC} that make the joint distribution of jet mass and $\tau_{21}$ identical (uniform) for both the signal and background.  All events are used, but by construction the jet mass and $\tau_{21}$ do not contain any useful discriminating information.  The performance of the $\theta_P$-based taggers are a significant fraction of the DNN-based tagger performance.  However, when the DNN is trained with the weighted applied, it significantly out-performs the pull angles.  This suggest that there is possibly more color flow information in the DNN that is not captured by $\theta_P$ and also shows that a significant fraction of the DNN `memory' is dedicated to learning about $\tau_{21}$ and mass.

\begin{figure}[htbp!]
  \begin{center}
        \includegraphics[width=0.45\textwidth]{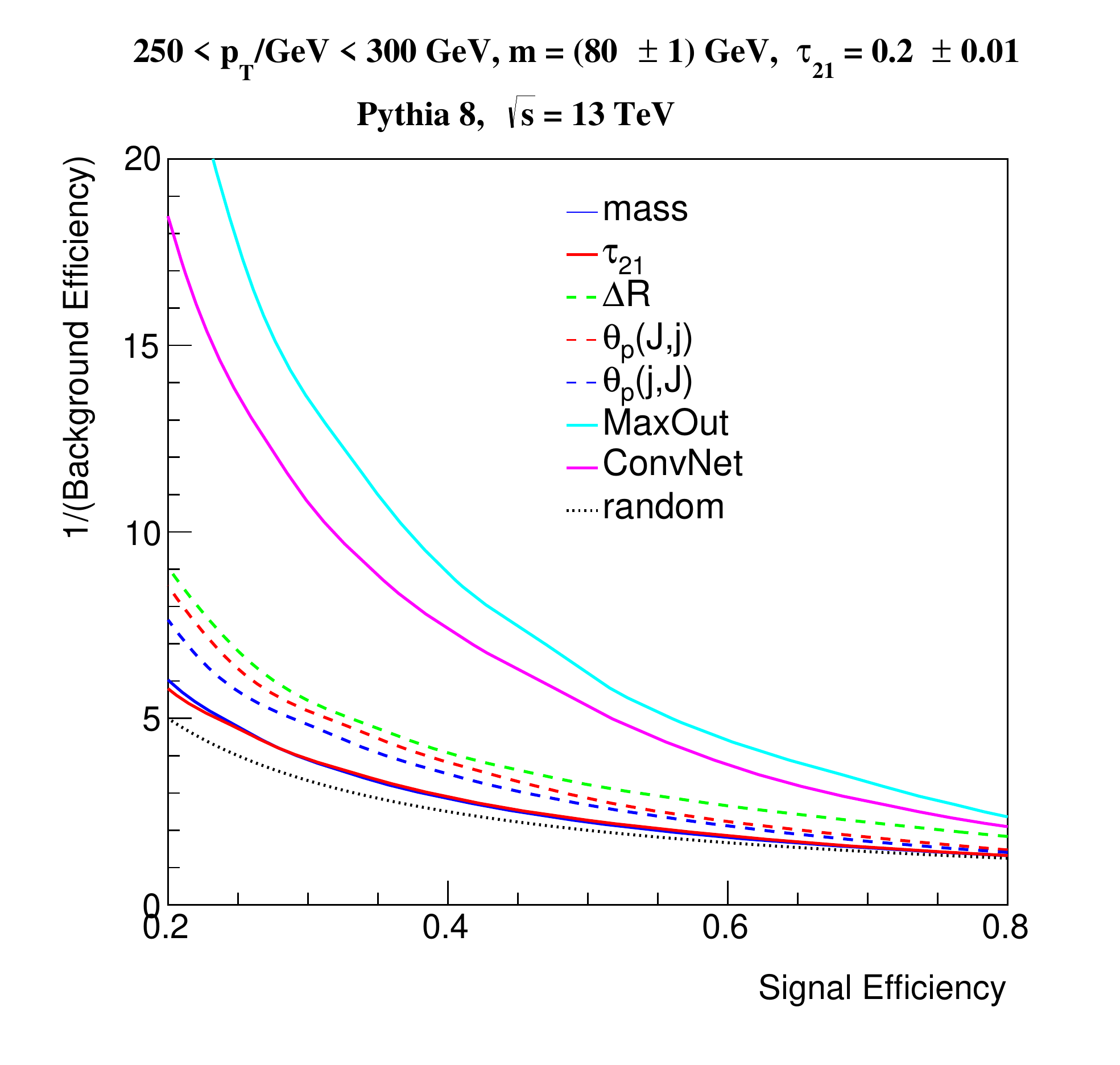}\includegraphics[width=0.45\textwidth]{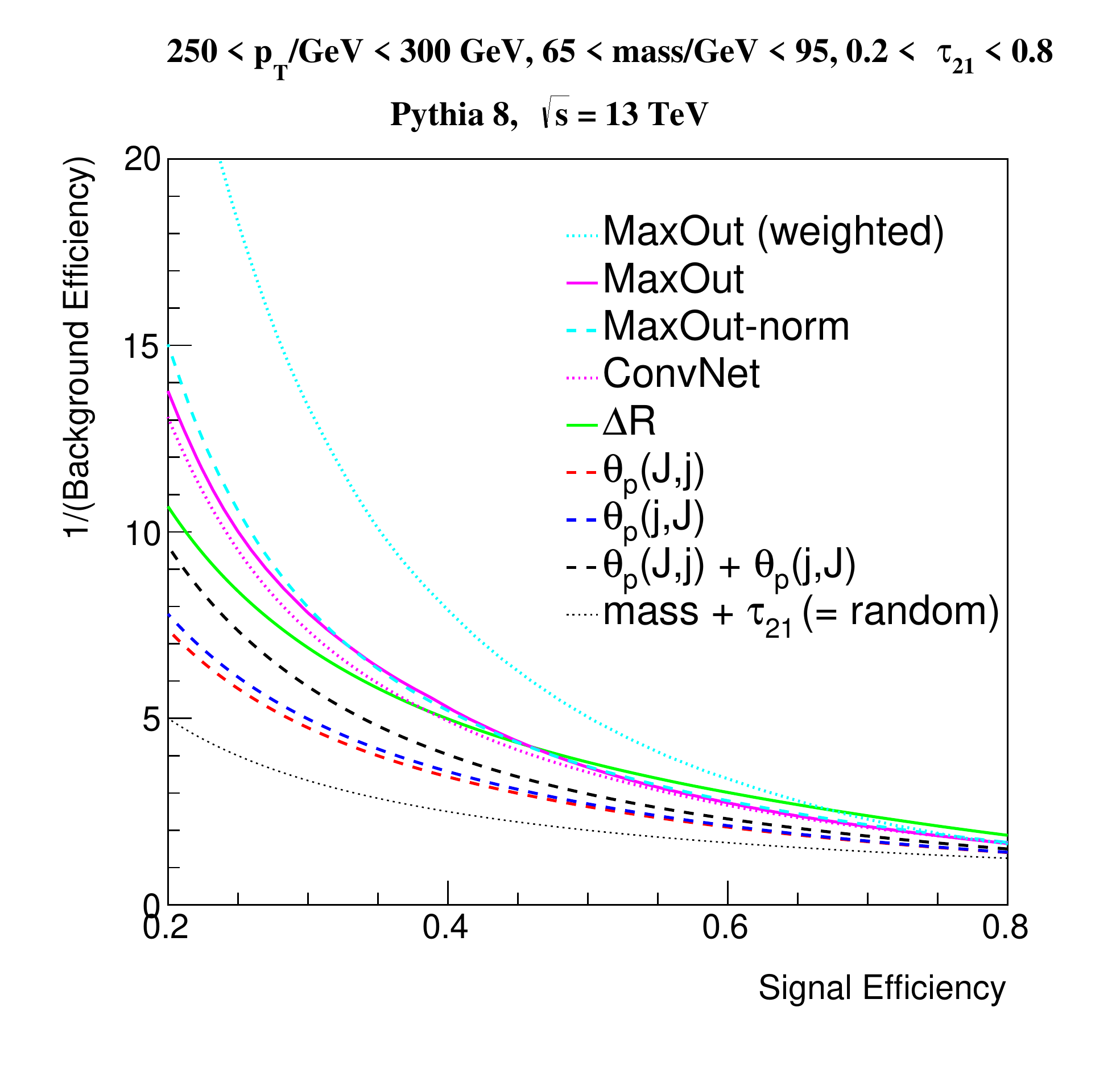}
      \caption{ROC curves including the jet pull angle $\theta_P$ in a restricted phase space (left) and using an inclusive event selection, but with events weighted so that the joint jet mass and $\tau_{21}$ distributions are identical between signal and background. 
      \label{fig:blahROC} }
    \end{center}
\end{figure}

\clearpage

\subsubsection{Outlook and Conclusions}
\label{sec:conclusion}
Jet Images are a powerful paradigm for visualizing and classifying jets.  This section has shown that when applied to jet images, deep neural networks outperform several known and highly discriminating physically-motivated jet observables such as the jet mass and $n$-subjettiness, $\tau_{21}$.  A series of tests have shown that some of these benchmark jet features are learned by the network, but others are not.  In particular, the networks are able to effectively learn geometric information about the radiation pattern, but not scale information as captured in part by the jet mass.  It is an important next step to develop techniques that allow the networks to also learn mass-like features.  The visualization studies in re-weighted or redacted regions of phase space show that some of the residual information learned by the network can be attributed to the differences in color flow between the signal and background.  Chapter~\ref{cha:colorflow} introduced the jet pull variable and demonstrated that it is sensitive to color flow.  In this section, the jet pull has been adapted to large-radius jets using subjets.  While the jet pull angle does carry some discriminating power beyond the jet mass and $\tau_{21}$, it does not contain enough information to fully explain the DNN performance.  Further studies of the visualizations may help to identify a simple feature like the jet pull angle that captures all or most of the color flow information learned by the networks.

The methods presented in this section have built a new link between high energy physics and computer vision. State-of-the-art classification techniques applied to jet images shows that there is a great potential to improve the performance of tagging algorithms using the extensive machine learning literature. In addition to improving the sensitivity of BSM searches, these new techniques may ultimately be able to improve the physical understanding of jets and their complex radiation pattern.

\clearpage

\subsection{Conclusions and Future Outlook}  
\label{sec:MLconclusions} 

Sections~\ref{sec:fuzzyjets} and~\ref{sec:jetimages} have shown two successful applications of adapting machine learning techniques to jet physics.  Domain specific knowledge (IRC safety, the symmetries of spacetime, etc.) have played an important role in specializing these techniques to jet physics and understanding what they have learned.  Three main conclusions from these studies:

\begin{enumerate}
\item State-of-the-art machine learning techniques can significantly improve upon the performance of traditional techniques motivated directly by physical intuition. 
\item Representing the data in new ways can expand physical intuition by highlighting properties that are not readily captured by current methods.
\item Most importantly: advanced machine learning techniques are tools to guide but not replace physical intuition.  An algorithm is most useful if the performance gains can be physically understood and independently validated. 
\end{enumerate}

There are many interesting directions to take this work in the future.  The extensive machine learning literature offers numerous possibilities for studying more complex tagging and reconstruction tasks such as low level tracking/calorimeter-cell clustering and calibration, full event tagging, pileup discrimination, and combining multiple detector elements into a single (multi-`color') jet image.  With the large amount of high energy data to be collected in Run 2 and beyond, it will be important to study these techniques in-situ in order to develop calibrations and systematic uncertainties.  Then, advanced machine learning techniques can be fully utilized to increase the sensitivity of LHC searches and measurement, including the study of rare and or subtle aspects of the SM and beyond.
 	
\part[The Search for a Light Stop Quark]{The Search for a Light Stop Squark\\[3ex]\makebox[0pt]{\includegraphics[width=0.6\paperwidth]{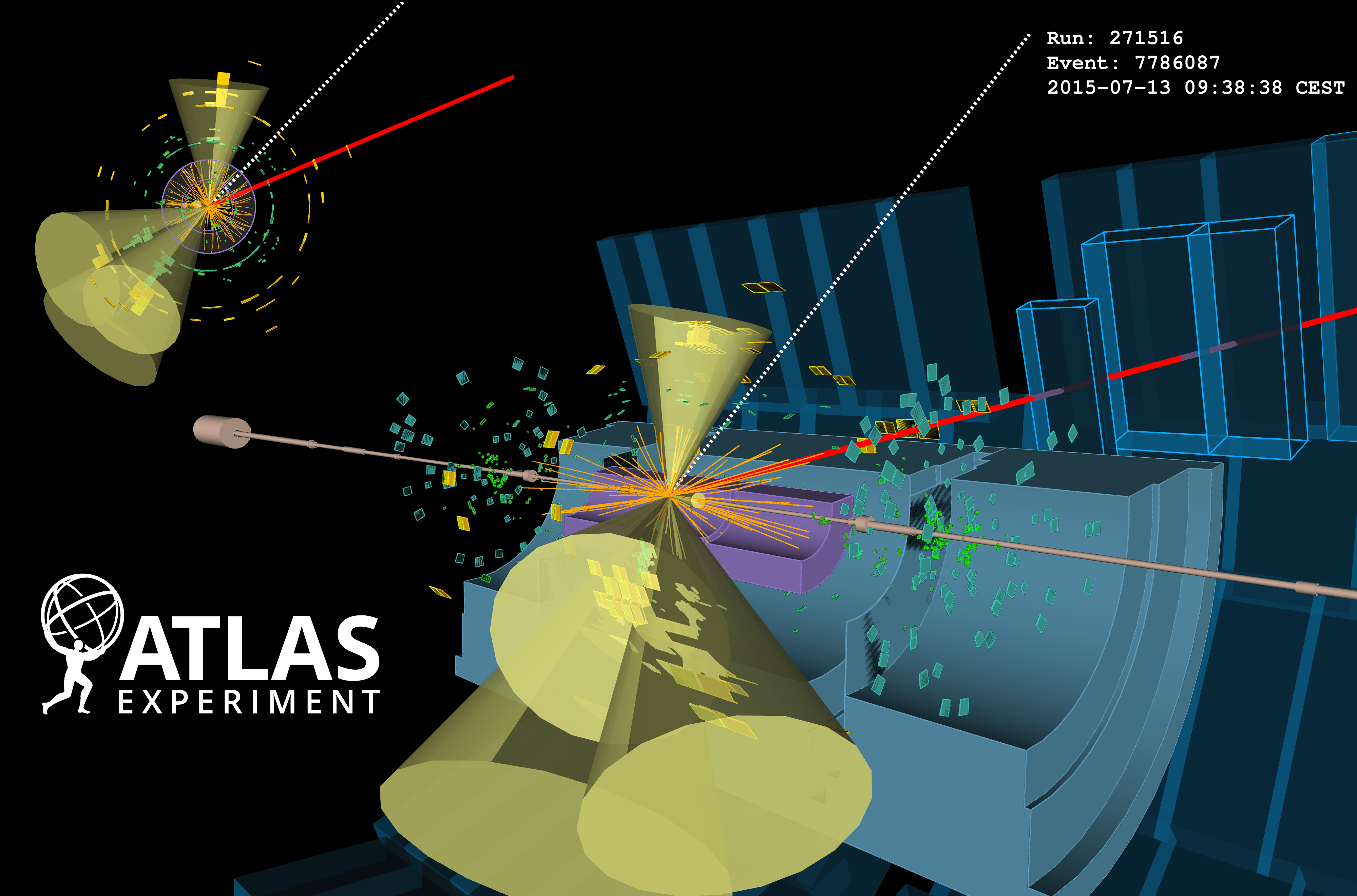}}\\\blfootnote{Display of a candidate boosted top quark pair production event from proton-proton collisions recorded by ATLAS with LHC stable beams at a collision energy of 13 TeV. The red line shows the path of a muon with transverse momentum around 50 GeV through the detector. The dashed line shows the direction of the missing transverse momentum, which has a magnitude of about 470 GeV. The green and yellow bars indicate energy deposits in the liquid argon and scintillating-tile calorimeters, from these deposits 4 small-radius ($R=0.4$) jets are identified with transverse momenta between 70 and 300 GeV. Three of these small-radius jets are re-clustered into the leading large-radius ($R=1.0$) jet (not shown explicitly) with a transverse momentum of about 600 GeV and a jet mass of about 180 GeV, near the top quark mass. One of these three jets in addition to the fourth jet above 70 GeV are identified as having originated from b-quarks. Tracks reconstructed from hits in the inner tracking detector are shown as arcs curving in the solenoidal magnetic field.}}	\label{part:susy}

The studies in Part~\ref{part:qpj} showed that quarks and gluons contain a wealth of information about the structure of the SM; in Part~\ref{part:susy}, they will serve as a window to beyond the SM.  In particular, the top quark holds a special place in the SM.  With a near-unity Yukawa coupling, the top quark is the most massive known elementary particle and has the strongest coupling to the Higgs boson.  As such, many extensions of the SM predict new particles that couple strongly or even exclusively to top quarks.  One of the most compelling such theories is Supersymmetry (SUSY).  Theoretically elegant and practical, SUSY is a powerful paradigm for explaining some of the issues with the SM discussed in Sec.~\ref{SMproblems}.  At the core of weak-scale SUSY is a light top squark (stop), the supersymmetric partner to the top quark.  If sufficiently light, the stop will be copiously produced at the LHC and can result in experimentally rich final states, often via top quarks.  Under mild assumptions, the top quarks from stop decays will always be accompanied by a stable weakly interacting SUSY particle that escapes detection.  This results in a $t\bar{t}+E_\text{T}^\text{miss}$ topology that will be main focus of Part~\ref{part:susy}.   

Chapter~\ref{sec:theory} begins Part~\ref{part:susy} with an introduction and motivation for SUSY and in particular for a relatively light stop.  This introduction is slightly nontraditional, beginning with a purely theoretical motivation instead of the usual practical one associated with the `hierarchy problem', which is discussed in Sec.~\ref{sec:hiearchyproblem}.  This order is chosen to stress that SUSY is a logical model-building extension of the SM, despite being broken below the electroweak scale.   The hierarchy problem and the dark matter relic density (the `WIMP' miracle) motivate the close proximity of the SUSY breaking scale with the electroweak scale.  Light stops are a generic prediction of SUSY models that naturally solve the hierarchy problem and are produced in association with the lightest supersymmetric particle (LSP) that is a dark matter candidate.

Stop pair production produces an experimentally complex and challenging final state.  Chapter~\ref{chapter:susy:analysisstrategy} provides an overview of the analysis strategy that involves both simulation-based and data-driven techniques to estimate and validate background predictions.  A variety of event selections are constructed to target a wide range of phenomenological signatures that are discussed in Sec.~\ref{sec:targetpheno}.  The search presented in Part~\ref{part:susy} spans all of Run 1 of the LHC with $\sqrt{s}=8$ TeV and the beginning of Run 2 at $\sqrt{s}=13$ TeV.  The analysis strategy has evolved over time, increasing in sophistication and sensitivity.  The focus will be on the state-of-the-art, but the early methods are also discussed in order to show the origin of the enhanced sensitivity at each stage.  The search targets the one lepton final state of stop pair production.  Leptons are precisely measured with high efficiency and provide useful handles for differentiating signal from backgrounds.  In particular, the generic production of quark and gluon jets is highly suppressed by requiring at least one reconstructed lepton.  The two-lepton final state offers a particularly clean environment for searching for stop pair production, but the branching ratio is significantly smaller than the zero- and one-lepton final states.  

An extensive toolkit of discriminating variables is constructed specifically for the $t\bar{t}+E_\text{T}^\text{miss}$ topology in the one lepton final state.  Many of the variables utilize the missing momentum vector combined with kinematic properties of the other reconstructed objects.   Chapter~\ref{chapter:susy:variables} describes all of the variables in detail, including new techniques that are used in this search for the first time.  Due to its similarity to the signal signature, SM top quark pair production is one of the most important background processes.  However, $t\bar{t}$ events with a single lepton can be reduced to a negligible level based on kinematic endpoints (Sec.~\ref{sec:mT}).  One of the dominant residual backgrounds is the pair or single production of top quarks resulting in final states with two real leptons.  The construction of variables that can effectively suppress dilepton $t\bar{t}$ events will be a large focus of Chapter~\ref{chapter:susy:variables}.

The discriminating variables from Chapter~\ref{chapter:susy:variables} are combined to form signal-sensitive event selections called {\it signal regions}.  Chapter~\ref{chapter:susy:signalregions} describes the construction of the signal regions, including the optimization procedure for maximizing the sensitivity to stop pair production.  The kinematic properties of the stop decay products depend on the mass of the stop as well as the mass difference between the stop and the LSP.  Higher stop masses and wider mass gaps give rise to harder energy spectra.  However, the stop cross section decreases with mass leading to a tradeoff between acceptance and absolute yield.  Compressed spectra are challenging because the signature is relatively similar to SM top quark pair production.  For low stop masses, the cross-section is sufficiently high to take advantage of subtle differences in the shapes of distributions to increase the sensitivity when the phase space for the LSP is restricted.

In order to reduce the dependence on simulation and the sensitivity to mis-modeling the data, a series of background-enriched event selections are constructed to normalize background yields.  These {\it control regions} are designed to be as close as possible to the signal regions in order to reduce the required simulation-based extrapolation in phase space to the signal region.  Chapter~\ref{chapter:background} documents the background estimation, including the construction of the control regions.  Chapter~\ref{chapter:uncertainites} describes a complete study of potential sources experimental and theoretical bias on the background estimates.   Many sources of uncertainty are reduced by normalizing the predicted yield in the control region to the observed data.  In addition, the total number of predicted events in the signal region is sufficiently small that the data statistical uncertainty dominates any residual systematic uncertainty.  Nonetheless, there are some signal regions for which the systematic uncertainty is significant and plays a major role in setting the sensitivity of the search.

After combining the background predictions with the observations in the signal regions, there is no significant evidence for electroweak scale stops.  As a result, limits are calculated to set bounds on the excluded models.  Chapter~\ref{chapter:results} documents these limits and discusses future directions for the search as well as a retrospective analysis of all Run 1 LHC SUSY searches. 

As a result of the search presented in Part~\ref{part:susy}\footnote{The ATLAS search results from Part~\ref{part:susy} are published in Ref.~\cite{ATLAS-CONF-2012-166,ATLAS-CONF-2013-037,Aad:2014kra,CERN-EP-2016-113} and include technical input and many useful discussions with the entire ATLAS stop one-lepton analysis team, including M. Barisonzi, J. Montejo Berlingen, D. Boerner, T. Eifert, J. Gramling, A. Henrichs, J. Kuechler, P. Pani, S. Pataraia, K. Rosbach, S. Strandberg, M. Ughetto, X. Wang, A. Yiming, and K. Yoshihara.}, simple stop models with $m_\text{stop}$ up to almost $800$ GeV for a wide range of LSP masses are excluded.  This puts a severe constraint on electroweak scale SUSY and many other models that predict light top quark partners.   There are always loopholes and the search will continue to push the limits up to and beyond $m_\text{stop}=1$ TeV as well as fill in gaps at lower masses where more complicated models can allow stops to evade the current limits.  This rich program has a strong foundation in the tools and techniques developed in Part~\ref{part:susy} and will hopefully result in uncovering new aspects of the SM or discovering new particles in the (near) future.

 \chapter{Introduction and Motivation}
\label{sec:theory}			
		
	The most elegant construction of a model is to be as extensive as possible while respecting all known symmetries of nature.  In classical mechanics, this leads to the familiar $\mathcal{L} \propto mv^2$ and in the SM requires and forbids certain terms in the Lagrangian.  Supersymmetry is no exception.  The symmetry group of the SM can be written as $S\simeq \mathfrak{P}\times \mathrm{SU}(3)\times \mathrm{SU}(2)\times \mathrm{U}(1)$, where $\mathfrak{P}$ is the Poincar\'e group that encodes the symmetries of spacetime and the second part of $S$ is the internal symmetry group of the SM.  The famous `no-go' theorem from Coleman and Mandula~\cite{PhysRev.159.1251} in 1967 showed that this structure is maximal: there is no non-trivial (direct product) way to mix the spacetime symmetry group with the internal symmetry group in 3+1 dimensions and retain non-zero scattering amplitudes.  However, there is a unique~\cite{Haag:1974qh} loophole - the symmetry group can be extended if the generators are not {\it bosonic}.  This is quantified in Sec.~\ref{sec:superspace} with construction of the super Poincar\'e group leading to SUSY as a symmetry of super spacetime.  The effect of SUSY is to relate bosons and fermions and as such predicts many new particles.  No SUSY partners have been observed and so SUSY must be broken below (at least) the electroweak scale, as explained by Sec.~\ref{sec:broken}.  A motivation for electroweak scale SUSY breaking is in Sec.~\ref{sec:hiearchyproblem}, where SUSY is shown to be an elegant solution to the hierarchy problem and provides a natural candidate for dark matter.  The minimal supersymmetric SM (MSSM) is introduced in Sec.~\ref{sec:MSSM} along with the properties of a light stop.
		
		\clearpage
		
		\section{Superspace as an Extension of Spacetime}
		\label{sec:superspace}
		
To illustrate\footnote{Part of this introduction is based on the Part III essay in Ref.~\cite{part3}.} the construction of superspace and the action of SUSY without a heavy burden of notation, this section uses a $1+1$ dimensional model that retains most of the key features of the full $3+1$ setting\footnote{Some aspects of this model appear while studying superstring theory~\cite{Superstringtheory}, although in that context it is an auxiliary device.  This section will take an approach which resembles aspects of previous work on $1+1$ super QED~\cite{SUSYQED2D,more2DQED,SUSYQED2Dv2}, but with a slightly different (concrete) angle.  The complications of multidimensional representations can obscure the physical intuition and simplistic motivation for supersymmetry.  The construction in 1+1 removes many complications, such as those associated with the properties of spinors~\cite{Todorov:2011dr}. }.  To begin, consider the $1+1$ representations of the usual spacetime.  Let $\eta=\mathrm{diag}(1,-1)$ be the 1+1 dimensional metric.  As in $3+1$ dimensions,  a Lorentz transformation is a linear map which preserves the Minkowski distance.  If $v$ is a two-vector (the $1+1$ analogue of a four-vector), then the Minkoski distance is given by $v^T\eta v$, where $v^T$ denotes transpose of $v$ and the product represents matrix multiplication.  Then, a Lorentz transformation $\Lambda$ is a matrix such that $v^T \Lambda^T\eta\Lambda v=v^T\eta v$ for all two-vectors $v$.  Therefore, $\Lambda$ is characterized by $\Lambda^T\eta\Lambda=\eta$.  Without loss of generality, let $\Lambda_{12}=-\sinh(y)$ for $y\in\mathbb{R}$ (the rapidity).  Then, $\Lambda^T\eta\Lambda=\eta$ results in three equations:

\begin{align}\nonumber
\Lambda_{11}^2-\sinh(y)^2&=1\\\nonumber
\Lambda_{12}^2-\Lambda_{22}^2=1\\
\Lambda_{11}\Lambda_{21}-\Lambda_{22}\sinh(y)=0.
\end{align}

\noindent Solving these equations leads to the general form of a Lorentz transformation:

\begin{align}
\Lambda(y)=\begin{pmatrix}\cosh(y)&-\sinh(y)\\-\sinh(y)&\cosh(y)\end{pmatrix}=\exp\left(iy J\right)\text{, where $J=\begin{pmatrix}0&i\cr i&0\end{pmatrix}$,}
\end{align}

\noindent A general {\it Poincar\'e transformation} is a combination of a Lorentz transformation and a translation in spacetime: $v\mapsto\Lambda v+w$, for a two-vector $w$.  

\clearpage

\noindent The Poincar\'e transformation can be represented by $3\times 3$ matrix multiplication:
 
 \begin{align}
 \label{3dmatrix}
 \begin{pmatrix}\Lambda & w\cr 0 & 1\end{pmatrix}\hspace{5mm}\text{acting on}\hspace{5mm} \begin{pmatrix}v\cr 1\end{pmatrix}.
 \end{align}

\noindent The representation in Eq.~\ref{3dmatrix} allows for an easy computation of the group laws of the Lie group $\mathfrak{P}$ of Poincar\'e transformations.  The matrix $J$ embedded in the $3\times 3$ matrix is one of the generators.  The full set of generators are given from the Taylor series expansion around the identity matrix:

\begin{align}
M=\begin{pmatrix}0 & i & 0\cr i & 0 & 0 \cr 0 & 0 & 0\end{pmatrix}\hspace{5mm}E=\begin{pmatrix}0 & 0 & -i\cr 0 & 0 & 0 \cr 0 & 0 & 0\end{pmatrix}\hspace{5mm}P=\begin{pmatrix}0 & 0 & 0\cr 0& 0 & -i \cr 0 & 0 & 0\end{pmatrix}.
\end{align}

\noindent Simple matrix multiplication with these explicit representations, shows that the defining commutation relations of the Poincar\'e algebra are $[M,E]=iP$ and $[M,P]=iE$ (boosts do not commute with translations).  In 1+1 dimensions, there is no `spin,' but one can construct the analogy of a spinor representation of the Lorentz subgroup $\mathfrak{L}$ of the Poincar\'e group.  In higher dimensions, $\mathfrak{L}$ has multiple generators with non-trivial commutation relations.  However, in the lower-dimensional case, the Lorentz group is Abelian and as such all irreducible representations are one-dimensional.  Define the lower dimensional analogues of the gamma matrices~\cite{Superstringtheory}:

\begin{align}
\label{gamma}
\gamma^0=\begin{pmatrix}0 & 1\cr 1 & 0\end{pmatrix}\hspace{5mm} \gamma^1=\begin{pmatrix}0 & -1\cr 1 & 0\end{pmatrix}
\end{align}

\noindent The matrices in Eq.~\ref{gamma} satisfy the Clifford algebra $\{\gamma^\mu,\gamma^\nu\}=2\eta^{\mu\nu}I_2$, where $I_2$ is the $2\times 2$ identity matrix.  A representation of the Lorentz group is $K=\frac{i}{4}[\gamma^0,\gamma^1]=\frac{i}{2}\eta$ which is {\it similar} to $J$.  Define a {\it Dirac spinor} $\psi$ as a two component object which transforms as $\psi\mapsto e^{iy K}\psi$, where $e^{iy K}=\mathrm{diag}(\exp(-y/2),\exp(y/2))$.   In order to construct a Lagrangian out of Dirac spinors, they need to be combined to form Lorentz invariant quantities.  As in the 3+1 case, $\psi^\dag\psi$ does not work since $\psi^\dag\psi\mapsto \psi^\dag e^{2iy K} \psi$.  Instead, let $\overline{\psi}=\psi^\dag\gamma^0$, then $\overline{\psi}\psi\mapsto \psi^\dag e^{iy K}\gamma^0 e^{iy K} \psi=\overline{\psi}\psi$.  Simple matrix multiplication shows that $\overline{\psi}\gamma^\mu\psi$ is a Lorentz vector, i.e. transforms by $\exp(iy J)$.  This leads to the the Lorentz invariant Dirac Lagrangian $\mathcal{L}=\overline{\psi}(i\gamma^\mu\partial_\mu-m)\psi$.

As in the 3+1 case, the Dirac spinors are not irreducible representations of the Lorentz group.  In the 1+1 case this is evident because all irreducible representations of an Abelian group are one dimensional.  This is also clear because under the action of the Lorentz group, the two components of $\psi$ transform independently, as $e^{iy K}$ is diagonal.  Let $\psi=(\psi_L\hspace{1mm}\psi_R)^T$ where the $\psi_L$ and $\psi_R$ are called {\it Weyl spinors} and transform as $\psi_{L}\mapsto e^{-y/2}\psi_L$ and $\psi_{R}\mapsto e^{y/2}\psi_R$. A curiosity of 1+1 dimensions is that one can choose $\psi_L$ and $\psi_R$ to be purely real and thus are {\it Majorana-Weyl} spinors~\cite{Todorov:2011dr}. 

Now, Minkowski space is extended to include two new {\it Grassman-valued} degrees of freedom, $\theta_1$ and $\theta_2$.  Unlike {\it bosonic} degrees of freedom (regular commuting numbers), the Grassman-valued degrees of freedom {\it anti}-commute with themselves and each other.  In particular, this means that $\theta_i^2=0$.  Furthermore, these new coordinates do not transform as a vector.  Instead, they transform as Weyl spinors and when combined into $\theta$, transform as a Dirac spinor. The resulting space is known as {\it superspace}.   In addition to extending the space of coordinates, one can extend the group of transformations to include translations in the spinorial degrees of freedom.  In general, let a spinorial translation, $\theta\mapsto\theta+\epsilon$, also affect the vector components of the superspace coordinate.  If the effect is required to be linear, then for $a,b\in\mathbb{R}$, the most general form of a spinor coordinate translation is $(x^\mu,\theta_i)\mapsto (x^\mu+\bar\epsilon\gamma^\mu\theta,\theta+\epsilon)$ where $\bar\epsilon=\epsilon^T\gamma^0$.  To see this, note that the only way to combine $\epsilon_i$ with one of $x,t,\theta_i$ and form a vector-like object is $\epsilon_i\theta_i$, which is both a commuting number and transforms as $\exp(\pm y)$.  Thus, in combinations of $\pm\epsilon_i\theta_i$, one may hope to get the correct transformation law of $x^\mu$.  

\clearpage

\noindent A general transformation then has the form

\begin{align}
\begin{pmatrix}t\cr x\end{pmatrix}\mapsto \begin{pmatrix}t'\cr x'\end{pmatrix}=\begin{pmatrix}t+a\epsilon_1\theta_1+d\epsilon_2\theta_2\cr x+a'\epsilon_1\theta_1+d'\epsilon_2\theta_2\end{pmatrix}.
\end{align}

\noindent One can transform $(x^\mu)'$ using the vector law (lefthand side of Eq.~\ref{transform}) and compare to the transformation of the summands  (righthand side of Eq.~\ref{transform}).

\begin{align}
\label{transform}
\Lambda(\phi)\begin{pmatrix}t'-t\cr x'-x\end{pmatrix}=\begin{pmatrix}ae^{-2\phi/2}\epsilon_1\theta_1+de^{2\phi/2}\epsilon_2\theta_2\cr a'e^{-2\phi/2}\epsilon_1\theta_1+d'\epsilon_2e^{2\phi/2}\theta_2\end{pmatrix}
\end{align}

\noindent Equating terms results in $d=-d',a=a'$.  After renaming constants\footnote{\label{subtle}This is a subtle point.  The spinors are Weyl-Majorana and so are real. Thus, one would not be able to absorb imaginary constants and so a proiri cannot be ruled out.  If one wants $\{Q_1,Q_1\}$ to be real, then a factor of $i$ is required.}, this becomes 

\begin{align}
\begin{pmatrix}t'\cr x'\end{pmatrix}=\begin{pmatrix}t+\epsilon_1\theta_1+\epsilon_2\theta_2\cr x+\epsilon_1\theta_1-\epsilon_2\theta_2\end{pmatrix}=x^\mu+\bar\epsilon\gamma^\mu\theta.
\end{align}

\noindent Note that this is not the most general transformation one could make.  For example, a non-linear transformation of the form $x^\mu\mapsto(1+\epsilon_1\theta_2) x^\mu$ is valid.  In addition, one could try to generalize the affect of a vector translation on the spinorial coordinates, but there is no nontrivial linear transformation.  

Combining the form of a spinor coordinate transformation with the action of the Poincar\'e group, one then can construct the full group of isometries on superspace, called the super-Poincar\'e group. The three dimensional matrix representation from Eq.~\ref{3dmatrix} of the Poincar\'e group can be extended to a five dimensional representation of the super-Poincar\'e group: 

 \begin{equation}
 \\\nonumber
 \text{$\scriptsize\begin{pmatrix}\cosh\phi & -\sinh\phi & \exp(-\phi/2)\epsilon_1 & \exp(\phi/2)\epsilon_2 & w_1\cr -\sinh\phi & \cosh\phi  & \exp(-\phi/2)\epsilon_1 & -\exp(\phi/2)\epsilon_2&w_2\cr 0 & 0 & \exp(-\phi/2) &0 & \epsilon_1\cr 0 & 0 & 0 & \exp(\phi/2) & \epsilon_2\cr 0 & 0 & 0 & 0 & 1\end{pmatrix}$}\hspace{5mm}\text{acting on}\hspace{5mm} \begin{pmatrix}v\cr \theta\cr 1\end{pmatrix},
 \\\nonumber
 \end{equation}
 
\noindent which gives rise to the generators of the super-Poincar\'e group algebra:
 
 \begin{equation}
 \\\nonumber
M=\text{\scriptsize $\begin{pmatrix}0 & i & 0& 0 & 0\cr i & 0 & 0 & 0 & 0\cr 0 & 0 & i/2& 0 & 0\cr 0 & 0 & 0& -i/2 & 0\cr 0 & 0 & 0& 0 & 0\end{pmatrix}$}\hspace{5mm}E=\text{\scriptsize $\begin{pmatrix}0 & 0 &  0 & 0 &-i\cr 0 & 0 & 0& 0 & 0 \cr 0 & 0 & 0& 0 & 0\cr 0 & 0 & 0& 0 & 0\cr 0 & 0 & 0& 0 & 0\end{pmatrix}$}\hspace{5mm}P=\text{\scriptsize $\begin{pmatrix}0 & 0 & 0& 0 & 0\cr 0& 0 & 0 & 0 & -i \cr 0 & 0 & 0& 0 & 0\cr 0 & 0 & 0& 0 & 0\cr 0 & 0 & 0& 0 & 0\end{pmatrix}$}
\end{equation}

 \begin{equation}
 \\\nonumber
Q_1=\text{\scriptsize $\begin{pmatrix}0 & 0 & -i& 0 & 0\cr 0 & 0 & -i & 0 & 0\cr 0 & 0 & 0& 0 & -i\cr 0 & 0 & 0& 0 & 0\cr 0 & 0 & 0& 0 & 0\end{pmatrix}$}\hspace{5mm}Q_2=\text{\scriptsize $\begin{pmatrix}0 & 0 &  0 & -i &0\cr 0 & 0 & 0& i & 0 \cr 0 & 0 & 0& 0 & 0\cr 0 & 0 & 0& 0 & -i\cr 0 & 0 & 0& 0 & 0\end{pmatrix}$}.
\end{equation}

The explicit form of the generators $M,E,P,Q_1$ and $Q_2$ allows for easy computation of the defining relations of the super-Poincar\'e graded algebra.  In particular,  the SUSY translations commute with the space-time translations   $[P,Q_i]=[E,Q_i]=0$ and anticommute with each other $\{Q_1,Q_2\}=0$.  Furthermore, SUSY translation operators have a spinor Lorentz structure: $[M,Q_1]=\frac{i}{2}Q_1$ and $[M,Q_2]=-\frac{i}{2}Q_2$. The most important relation is the self anticommutation of the SUSY translations, which yield $\{Q_1,Q_1\}=-2i (E+P)$ and $\{Q_2,Q_2\}=-2i (E-P)$.  Heuristically, these self anticommutation relations say that a SUSY translation is the `square root' of spacetime translations. 

As a quantum field theory, the fundamental objects in the Standard Model (SM) are the fermionic and bosonic quantum fields.  Likewise, in SUSY, quantum fields are the objects governed by the equations of motion.  The only difference is that the fields in the SM are maps from Minkowski space, while in SUSY, fields are maps from superspace.  The latter are called {\it superfields}.  A scalar superfield $\Psi$ is a map from superspace into $\mathbb{C}$ which is invariant under a super-Poincar\'e transformation\footnote{In this one-dimensional case, the field is real-valued.}.  Since $\theta_i^2=0$, a Taylor expansion of a generic scalar superfield is of the form 

\begin{align}
\label{form}
\Psi(x^\mu,\theta_i)=\phi(x^\mu)+\bar\theta \psi(x^\mu)+\theta_1\theta_2 F(x^\mu),
\end{align}

\noindent where $\phi$ is a scalar, $F$ is a pseudoscalar, and $\psi$ is a Dirac spinor.  The field $F$ is a pseudoscalar because under parity, a Dirac spinor $S$ transforms into $\gamma^0 S$ , which in the chosen basis means that the two components of $\theta$ are interchanged (for a nice discussion, see 8.10 in~\cite{Ticciati:1999qp} or 3.6 in~\cite{Peskin:1995ev}).  Thus, $\theta_1\theta_2\mapsto \theta_2\theta_1=-\theta_1\theta_2$ and since $\Psi$ is a scalar, $\Psi\mapsto\Psi$ under parity.  To compensate, $F\mapsto-F$. Under a SUSY translation of $\epsilon^TQ$ on the coordinates of superspace, $\Psi(x^\mu,\theta,\omega)\mapsto \Psi'(x^\mu,\theta,\omega)=\Psi(x^\mu-\bar\epsilon\gamma^\mu\theta,\theta-\epsilon)$, where $\epsilon$ is a Dirac spinor with components $\epsilon_1$ and $\epsilon_2$.  The minus sign in the expression for $\Psi'$ comes from the fact that a SUSY translation has been defined to shift the coordinates forward and thus the field must compensate by evaluation at a shifted backward location in superspace.  A simple computation shows that one can express $\Psi$ in terms of an operator action on $\Psi$ in the following way:  

\begin{align}
\Psi'(x^\mu,\theta,\omega)=\exp(-\bar\epsilon\gamma^\mu\theta\partial_\mu-\epsilon_i\partial_{\theta_i})\Psi(x^\mu,\theta),
\end{align}

\noindent with the standard notation $\partial_\mu=(\partial_t,\vec{\nabla})$ and $x^\mu=(t,\vec{x})$ so coordinates are initially given raised while derivatives are all positive when lowered\footnote{Following the convention of~\cite{Peskin:1995ev}.}.  Taylor expanding the expression for $\Psi'$ gives the form of an infinitesimal SUSY translation along the $\theta$ direction:

\begin{align}
\delta\Psi\equiv \Psi'(x^\mu,\theta,\omega)-\Psi(x^\mu,\theta,\omega)=\left(-\bar\epsilon\gamma^\mu\theta\partial_\mu-\epsilon_i\partial_{\theta_i}\right)\Psi(x^\mu,\theta).
\end{align}

\noindent Let

\begin{align}
\label{opdef}
\mathcal{Q}_i=i(\gamma^0\gamma^\mu)_{ij}\theta_j\partial_\mu+i\partial_{\theta_i},
\end{align}

\noindent so that $\delta\Psi=i\epsilon^T\mathcal{Q}\Psi$. The factors of $i$ come from the desire to have (anti)Hermitian operators, as is done for the familiar construction of $i\mathcal{P}_\mu=-\partial_\mu$.  Similarly, define $i\mathcal{M}=x\partial_t+t\partial_x+\frac{1}{2}\theta_1\partial_{\theta_1}-\frac{1}{2}\theta_2\partial_{\theta_2}$.   With these identifications, $\mathcal{Q},\mathcal{P},$ and $\mathcal{M}$ form a representation of the super-Poincar\'e group through their action on superfields.  This can be shown by computing the (anti)commutation relations of the various operators.  The momentum operators have the expected Lorentz structure: $[\mathcal{M},\mathcal{P}_0]=i\mathcal{P}_1$ and $[\mathcal{M},\mathcal{P}_1]=i\mathcal{P}_0$ and the index on the operators $\mathcal{Q}_i$ is indeed a spinor index, since $[M,\mathcal{Q}_1]=\frac{i}{2}\mathcal{Q}_1$ and $[M,\mathcal{Q}_2]=-\frac{i}{2}\mathcal{Q}_2$.  As in the case with the matrix representation, $[\mathcal{P}_\mu,\mathcal{Q}_i]=\{\mathcal{Q}_1,\mathcal{Q}_2\}=0$.  The only difference in the defining algebra of the operator versus matrix representation is the self anticomutation relations of the SUSY translations: $\{\mathcal{Q}_1,\mathcal{Q}_1\}=2i\left(\mathcal{P}_0+\mathcal{P}_1\right)$ and $\{\mathcal{Q}_2,\mathcal{Q}_2\}=2i\left(\mathcal{P}_0-\mathcal{P}_1\right)$, which differ by a relative minus sign.  This sign comes from the fact that the fields compensate for a coordinate change in the opposite way that the coordinates themselves shift and thus the SUSY algebra defined by the operators is not identical to the algebra we encountered earlier.  

With the form of a SUSY translation in Equation~\ref{opdef} one can compute the changes in the component fields of $\Psi$ as in $\delta\Psi=\delta\phi+\bar\theta\delta\psi+\theta_1\theta_2\delta F$.  The transformations are

\begin{align}
\label{trans}
\delta\phi&=-\bar\epsilon\psi\\\nonumber
\delta\psi&=\gamma^\mu\epsilon\partial_\mu\phi+\gamma^5\epsilon\\\nonumber
\delta F&=-\partial_\mu\bar\psi\gamma^5\gamma^\mu\epsilon,
\end{align}

\noindent where $\gamma^5\equiv\gamma^0\gamma^1$.  The key feature of Eq.~\ref{trans} and the main result of this section is that the boson $\phi$ transforms into the spinor $\psi$ and the spinor transforms into a (translated) boson.   In this way, SUSY is a symmetry relating bosons and fermions by transforming one into the other.

		\clearpage
		
		\section{Broken Supersymmetry}
		\label{sec:broken}		
		
		Still working in 1+1 dimensions, let $\Psi$ be an irreducible super-Poincar\'e {\it multiplet} containing some bosonic and fermionic degrees of freedom.  Massive Poincar\'e multiplets are identified by their mass and spin.  This means that a single multiplet can contain only bosonic degrees of freedom or fermionic degrees of freedom, but not both.  Consider the operator $N_F$ which is defined by $N_F|\text{boson}\rangle=|\text{boson}\rangle$ and $N_F|\text{fermion}\rangle=-|\text{fermion}\rangle$.   This definition is chosen such that the operator trace $\mathrm{Tr}(N_F)=\sum_\text{boson in $\Psi$}\langle \text{boson} |N_F|\text{boson}\rangle+\sum_\text{fermion in $\Psi$}\langle \text{fermion} |N_F|\text{fermion}\rangle$ is simply the number of bosonic degrees of freedom minus the number fermionic degrees of freedom. The operator $N_F$ anti-commutes with $\mathcal{Q}_i$:
		
		\begin{align}\nonumber
		(N_F\mathcal{Q}_i&+\mathcal{Q}_iN_F)|\text{boson/fermion}\rangle\\\nonumber
		&=N_F|\text{fermon/boson}\rangle+(+/-)\mathcal{Q}_i|\text{boson/fermion}\rangle\\
		&=(-/+) +(+/-)=0.
		\end{align}
		
		\noindent Since the trace is linear and has the cyclic property $\mathrm{Tr}(Q_iN_FQ_i)=\mathrm{Tr}(N_FQ_iQ_i)=\mathrm{Tr}(-Q_iN_FQ_i)$ and therefore, this quantity is zero.  However, $\mathrm{Tr}(N_F\{Q_i,Q_i\})=2(E-P)\mathrm{Tr}(N_F)$. Thus, $\mathrm{Tr}=0$ and the number of fermions and bosons must be the same in the multiplet.  Just as for the  Poincar\'e group, in the super-Poincar\'e group, $E^2-P^2$ commutes with all the generators and so the mass is still characterizes a multiplet.  This means that in SUSY, every boson has a {\it superpartner} fermion with the same mass and vice versa.   Even though this was derived in $1+1$ dimensions, it holds for $3+1$ as well.
		
		While the construction in Sec.~\ref{sec:superspace} is elegant, it cannot be true - no superpartners of the SM particles have been observed.  Therefore, if is a real symmetry of nature, SUSY must be broken below the energy scales currently accessible to experiments.  The next section describes a strong motivation for the SUSY breaking scale to be close to the electroweak energy scale.
		
		\clearpage
		\section{The Hierarchy Problem and Weak-Scale SUSY}
		\label{sec:hiearchyproblem}

			One of the fundamental limitations of the SM is that it does not describe gravity.  This should not be relevant for physics at the electroweak scale, where the strength of classical gravity is negligible compared with the other forces.  However, at energies near the {\it Planck scale}\footnote{This is the energy scale $E$ at which the gravitational potential energy from two objects with mass $E/c^2$ separated by a distance $r$ is the same as a photon with wavelength $r$, i.e. $GE^2/rc^2=\hbar c/r$. At this energy scale, gravitational effects are not small compared to quantum mechanical effects.} $\Lambda\sim 10^{19}$ GeV, gravity will be comparable in strength to the other forces at which point there must be significant contribution from physical laws beyond the SM.  The electroweak scale and the Planck scale are theoretically connected by quantum corrections to particle properties.   For example, the input mass parameter for a particle in the Lagrangian receives corrections from next-to-leading-order effects encoded by Feynman diagrams like the one shown in Fig.~\ref{HiggsRenormalization} for the Higgs boson mass.  If the SM is valid up to $\Lambda$, then the correction from Fig.~\ref{HiggsRenormalization} has the form
			
\begin{align}
\label{label:correction}
\delta m\sim -\frac{m_f^2}{v^2}\int_0^\Lambda\frac{d^4k}{(2\pi)^4}\frac{k\hspace{-2mm}\slash}{k^2}\frac{k\hspace{-2mm}\slash}{k^2}\sim -\frac{m_f^2\Lambda^2}{v^2},
\end{align}

\noindent for the vacuum expectation value $v$ and where the minus sign is for the closed fermion loop and results from the difference between Fermi versus Dirac statistics.  Similar calculations show that other particles are also sensitive to this {\it cutoff} scale $\Lambda$, but not all quadratically (for fermions, it is $\log(\Lambda)$).   For particle masses near the electroweak scale, this seems like an enormous cancellation of $\mathcal{O}(\Lambda/v)$ effects.  However, for fermions and gauge bosons, the impact of the corrections is naturally suppressed by symmetry.  Gauge invariance ensures that the mass of the photon, gluon, $Z$ and $W$ bosons before electroweak symmetry breaking is exactly zero.  Corrections for massless fermions are zero by chiral symmetry\footnote{Invariance under independent transformations of left- and right-handed fermions.}.  Therefore, the corrections for fermions with a small mass must go to zero as the mass goes to zero.  This accounts for all the SM particles except the Higgs boson, which has no symmetry to suppress quantum corrections to the bare mass.  This is further complicated because all of the SM masses are tied to the Higgs boson mass after electroweak symmetry breaking.  The apparent large cancellation giving rise to the physical Higgs boson mass is called the {\it hierarchy problem}.  Before proceeding, it should be stressed that the hierarchy problem is a formal/aesthetic problem and not a logical inconsistency in the theory.  This is in contrast to related problems arising earlier in the history of particle physics such as the non-renormalizability of the Fermi theory of the weak force which had a cutoff at the electroweak scale.  However, the hierarchy problem is intriguing/suggestive and continues to be one of the core drivers of model building in high energy physics research.

\begin{figure}[h!]
\centering
\begin{tikzpicture}[line width=1.5 pt, scale=2]
	\draw[dashed] (0:1)--(0,0);
	 \node at (1.5,0.8) {$f$};
	 \node at (1.5,-0.8) {$\bar{f}$};
	 \node at (.5,.2) {$h$};	
	\draw[fermion] (1,0) arc (180:0:.5);
	\draw[fermion] (2,0) arc (0.:-180:0.5);
	\draw[dashed] (2,0) --(3,0);
	 \node at (2.5,.2) {$h$};	
\end{tikzpicture}
\caption{A one-loop diagram contributing to the Higgs boson self energy at next-to-leading order.}
\label{HiggsRenormalization}
\end{figure}			
			
One elegant method for eliminating the hierarchy problem is to protect the Higgs boson mass using similar strategies as for the fermions or Gauge bosons.  Gauge symmetries do not directly help because there still needs to be a mechanism for generating a non-zero mass.  Suppose that there was a new fermion which shared a mass parameter with the Higgs boson.  Chiral symmetry would protect the fermion from receiving large quantum corrections to its mass and therefore would indirectly suppress corrections for the Higgs boson.  Such a theory was introduced in Sec.~\ref{sec:superspace}: Supersymmetry.  Under SUSY, there is a fermion partner to the Higgs boson (called the {\it Higgsino}) which is in the same multiplet as the Higgs boson with exactly the same mass.  Under exact SUSY, the Higgs boson mass is protected by chiral symmetry.   

However, as discussed in Sec.~\ref{sec:broken}, SUSY is not exact.  There are many mechanisms for breaking SUSY below the electroweak scale such that the SUSY partners to the SM particles are heavier than experimental limits.  Before discussing models with SUSY breaking (see Sec.~\ref{sec:MSSM}), consider the largest contributions to Eq.~\ref{label:correction}.  Since the correction scales with the Yukawa coupling of the fermions, the dominant contribution is from top quark loops.  The corresponding largest corrections from SUSY therefore needs to come from stop loops, shown in Fig.~\ref{HiggsRenormalization2}.  Therefore, if broken SUSY is to provide a solution to the Hierarchy problem, the stop must be relatively light.  This can be quantified by limiting the amount of {\it fine-tuning}~\cite{Barbieri:1987fn,deCarlos:1993yy} required for a SUSY model to reproduce the observed SM spectrum at the electroweak scale.  The SM has a large amount of fine-tuning because the input Higgs boson mass parameter in the Lagrangian and the quantum corrections to the Higgs boson mass, each $\mathcal{O}(10^{19})$ GeV, must cancel at one part in $10^{17}$ to produce the measured 125 GeV Higgs boson mass.  There is no unique way to quantify fine-tuning, but there is general consensus that $\mathcal{O}(1\%)$ tuning (suitably defined) requires $m_\text{stop}\lesssim 1$ TeV\footnote{Early references include Ref.~\cite{Cohen:1996vb,Dimopoulos:1995mi} and this is an area of active research - see for instance Ref.~\cite{Baer:2012uy,Wymant:2012zp,Papucci:2011wy,Kitano:2006gv,Kitano:2005wc,Brust:2011tb}}.  			
\begin{figure}[h!]
\centering
\begin{tikzpicture}[line width=1.5 pt, scale=2]
\begin{scope}[shift={(-1.9,0)}]
	 \node at (1.5,0.8) {$\tilde{f}$};
	 \node at (.5,.2) {$h$};	
	 \node at (2.5,.2) {$h$};		 
	\draw[scalar] (2,0.5) arc (0.:-360:0.5);
	\draw[dashed] (1.5,0) --(3,0);
	\draw[dashed] (0,0) --(1.5,0);
	\end{scope}
\begin{scope}[shift={(1.9,0)}]
	\draw[dashed] (0:1)--(0,0);
	 \node at (1.5,0.8) {$\tilde{f}$};
	 \node at (1.5,-0.8) {$\tilde{\bar{f}}$};
	 \node at (.5,.2) {$h$};	
	 \node at (2.5,.2) {$h$};	
	\draw[scalar] (1,0) arc (180:0:.5);
	\draw[scalar] (2,0) arc (0.:-180:0.5);
	\draw[dashed] (2,0) --(3,0);	
\end{scope}
\end{tikzpicture}
\caption{SUSY NLO corrections to the Higgs boson self energy.  While the right diagram is topologically the same as the leading NLO fermion diagram in Fig.~\ref{HiggsRenormalization}, it is suppressed with respect to the left diagram due to the two scalar propagators $(\sim\int d^4k/k^4)$ compared with one $(\sim\int d^4k/k^2)$.  There are also two powers of the coupling constant for the right diagram, but the Yukawa coupling is nearly one for the top/stop.}
\label{HiggsRenormalization2}
\end{figure}			
			
There is other indirect evidence for weak scale SUSY in addition to solving the Hierarchy problem.  For example, any of the neutral SUSY particles could make a natural dark matter candidate due to the {\it Weakly Interacting Massive Particle (WIMP) miracle}.  Define $\Omega_\chi=\rho_\chi/\rho_\text{critical}$ as the normalized mass density of dark matter particle $\chi$.  The critical density $\rho_\text{critical}$ is derived from the Friedmann metric for a flat, homogeneous, and isotropic universe and is  given by $\rho_\text{critical}=3H^2/(8\pi G)\sim 10 (\text{GeV}/c^2)/\text{m}^3$, where $H\sim 100 (\text{km}/s)/\text{Mpc}$~\cite{Ade:2013sjv} is the Hubble constant.  The equation of motion for the number density of a dark matter particle $\chi$ is given by the Boltzman equation:

\begin{align}
\label{eq:boltzman}
\dot{n}_\chi+3n_\chi\dot{a}/a = -\langle \sigma v\rangle \left(n_\chi^2-n_{\chi,\text{equilibrium}}^2\right),
\end{align}

\noindent where the dot denotes a derivative with respect to time and $a$ is the scale factor of the universe $(H=\dot{a}/a)$.  The lefthand side of Eq.~\ref{eq:boltzman} is the equation for an expanding universe with constant mass; the factor of 3 simply results in the usual $n_\chi\propto 1/a^3$.  The righthand side of Eq.~\ref{eq:boltzman} accounts for creation and annihilation of $\chi$ where $\langle \sigma v\rangle$ is the thermally averaged annihilation cross section multiplied by the relative speed.  In the early universe when the temperature was very high, $k_BT\gg m_\chi c^2$, pairs of dark matter particles were constantly being created and destroyed.  When the temperature dropped below $m_\chi$, the dark matter particles no longer annihilated and so the density was fixed at $n_{\chi,\text{equilibrium}}$, giving rise to the relic density $\Omega_\chi$ observed today.  The solution to the equilibrium number density from Eq.~\ref{eq:boltzman} is given by~\cite{Jungman:1995df}

\begin{align}
n_\chi \sim s_0 10^{-8}\left[\left(\frac{m_\chi}{\text{GeV}}\right)\left(\frac{\langle \sigma v\rangle}{10^{-27}\text{cm}^3/s} \right)\right]^{-1},
\end{align}

\noindent where $s_0$ is the current entropy density of the universe.  For the highly relativistic particles contributing to the entropy density, the only dimensionful number is the temperature $T$ (by definition, the particle masses are irrelevant) and so $s\propto T^3$.    The exact form is $s=2\pi^2g(T)T^3/45$, where $g(T)$ is the number of effective degrees of freedom\footnote{See Ref.~\cite{DanielBaumann} for a pedagogical explanation.}.  The current temperature of the universe is about $3$ K at which basically only photons and neutrinos contribute to $g(T)\sim 4$.  In units of cm${}^3$ (with units in which $k_b=1$), 

\begin{align}
s_0\sim 2T^3\sim 2 (3K)^3\times\left(\frac{1\text{eV}}{10^4K}\right)\times\left(\frac{\frac{1}{\text{eV}}}{2\times 10^{-7}\text{m}}\right)\sim 4000 cm^{-3}.
\end{align}

\noindent Using $\Omega_\chi=m_\chi n_\chi/\rho_\text{critical}$, 

\begin{align}
\label{eq:relic}
\Omega_\chi h^2\sim\left(4\times 10^{-27}\text{cm}^3s^{-1}/\langle \sigma v\rangle\right).
\end{align}

\noindent The total dark matter relic density has been measured to be $\Omega h^2\sim 0.1$~\cite{pdg}.  The ``WIMP miracle'' is that the cross-section for a weak-scale interaction is about $\alpha^2_\text{weak}/m_\text{weak}^2$ and for $\alpha_\text{weak}\sim 0.01$ and $m_\text{weak}\sim\text{100}$ GeV, Eq.~\ref{eq:relic} is the same order of magnitude as the measurement.

Two other related sources of indirect motivation for electroweak SUSY are grand unification and successful electroweak symmetry breaking.  An intriguing curiosity of a minimal SUSY extension of the SM (see Sec.~\ref{sec:MSSM}) is that the three gauge group coupling constants seem to be equal to each other at a high energy and this {\it grand unification} (GUT) scale is near the Plank scale.  As described in Sec.~\ref{sec:particlesandforces}, the running of the coupling constant $g$ is described by solutions to the Callan-Symanzik equation:

\begin{align}
\frac{dg}{d\log(Q/M)}=\beta(g),
\end{align}

\noindent where $M$ is a fixed energy scale (such as $m_Z$) and $Q$ is the running energy scale.  At leading order, $\beta(g)=b_0g^3/(4\pi)^2$, with $b_0=\sum_\text{fields $f$} \kappa_f C(G,r_f)$, where\footnote{Somewhat surprisingly, the full derivation of these factors is not usually presented all at once in the main QFT texts.  Most advanced QFT students will have derived the equations for QED and QCD, but there is a small jump to the general $U(1)$ from QED (also to include complex scalars) - see for instance Chapter 66 in Ref.~\cite{Srednicki:1019751}.  With some careful thought, the inclusion of complex scalars in the non-Abelian case can be extracted using the results of the background field method presented in Chapter 16.6 in Ref.~\cite{Peskin:1995ev}.} $f\in\{\text{gauge},\text{Weyl fermion},\text{scalar}\}$, $\kappa_f=-11/3$ for gauge fields, $2/3$ for Weyl fermions, and $1/3$ for scalars.  The factors $C(G,r_f)$ depend on the gauge group $G$ as well as the representation of the field $r_f$.  In the adjoint representation, $C(G)=C_2(G)$, the quadratic Casimir operator of group that is $N$ for $SU(N)$, $N>1$ and $0$ for $U(1)$.  In the fundamental representation of $SU(N)$, $C(G)=\frac{1}{2}$ and for $U(1)$, $C(G)=Y^2$, where $Y$ is the weak hypercharge.  For $SU(3)$, the Higgs does not contribute and there is no distinction between left and right handed fields, so the Weyl fermion combine to give the familiar equation

\begin{align}
b_0=-\frac{11}{3}N_c+\frac{2}{3}n_f,
\end{align}

\noindent where $N_c$ is the number of colors and $n_f$ is the number of quarks.  For the three gauge couplings of the SM, $g_1=e/\cos(\theta_W),g_2=e/\sin(\theta_W),g_3$ corresponding to the gauge groups $U(1),SU(2)$, and $SU(3)$, the three leading order $\beta$ functions are 

\begin{align}
\label{eq:RGE}\nonumber
b_0^1 &= {\color{red}\frac{2}{3}}\times\left(2{\color{blue}Y_{E_L}^2}+{\color{blue}Y_{e_R}^2}+2N_c{\color{blue}Y_{Q_L}^2}+N_c{\color{blue}Y_{u_R}^2}+N_c{\color{blue}Y_{d_R}^2}\right)\times 3+{\color{red}\frac{1}{3}}\times 2{\color{blue}Y_H^2}=\frac{41}{6}\\\nonumber
b_0^2 &= {\color{red}-\frac{11}{3}}\times {\color{blue}2}+{\color{red}\frac{2}{3}}\times{\color{blue}\frac{1}{2}}\times 12+{\color{red}\frac{1}{3}}\times{\color{blue}\frac{1}{2}}=-\frac{19}{6}\\
b_0^3 &= {\color{red}-\frac{11}{3}}\times {\color{blue}3}+{\color{red}\frac{2}{3}}\times{\color{blue}\frac{1}{2}}\times 12=-7,
\end{align}

\noindent where the terms in red are the $\kappa$ factors, the terms in blue are the $C(G,r)$ factors and the remaining terms count the number of fields.  For instance, there are $12$ Weyl quarks ($=6$ Dirac fermion quarks) and $12$ total left-handed fields contributing to $b_0^2$ (each quark type contributes three times, one for each color).  The weak-hyercharges are $Y_{E_L}=-\frac{1}{2}$, $Y_{e_R}=1$, $Y_{Q_L}=\frac{1}{6}$, $Y_{u_R}=-\frac{2}{3}$, an $Y_{d_R}=\frac{1}{3}$, where $Y=Q-T_3$ ($T_3$ is the weak isospin).  For one copy of SUSY added to the SM, the gauge bosons have fermionic partners in the adjoint representation and the fermions have complex scalar partners in the fundamental representation.  Therefore, instead of $-11/3 C$ in Eq.~\ref{eq:RGE}, in SUSY the contribution is $(-11/3+2/3)C=-3C$.  Likewise for the fermions (and the complex scalars), instead of $2/3C$ (or $1/3$), in SUSY the contribution is $(2/3+1/3)C=C$.  Therefore, the leading order $\beta$ functions become

\begin{align}
\label{eq:RGE2}\nonumber
b_0^1 &= {\color{red}1}\times\left(2{\color{blue}Y_{E_L}^2}+{\color{blue}Y_{e_R}^2}+2N_c{\color{blue}Y_{Q_L}^2}+N_c{\color{blue}Y_{u_R}^2}+N_c{\color{blue}Y_{d_R}^2}\right)\times 3+{\color{red}1}\times 2{\color{blue}Y_H^2}\times 2=11\\\nonumber
b_0^2 &= {\color{red}-3}\times {\color{blue}2}+{\color{red}1}\times{\color{blue}\frac{1}{2}}\times 12+{\color{red}1}\times{\color{blue}{\frac{1}{2}}}\times 2=1\\
b_0^3 &= {\color{red}-3}\times {\color{blue}3}+{\color{red}1}\times{\color{blue}\frac{1}{2}}\times 12=-3,
\end{align}

\noindent where the extra factor of two for the Higgs fields is due to a second Higgs doublet that is required in the minimal SUSY extension of the SM (see Sec.~\ref{sec:MSSM}).  Figure~\ref{fig:GUT} shows the one-loop running of the three gauge group constants, $\alpha_i=g_i^2/4\pi$.  Conveniently, for $t=\log(Q/M)$,

\begin{align}
\label{couplingRGE}
\frac{d\alpha^{-1}(t)}{dt}=-\frac{1}{\alpha^2}\frac{d\alpha}{dt}=\frac{g}{2\pi\alpha^2}\frac{dg}{dt}=\frac{b_0g^4}{2\pi\alpha^2(4\pi)^2}=-\frac{b_0}{2\pi},
\end{align}

\noindent i.e. the inverse coupling depends linearly on $\log(Q/M)$.  The $U(1)$ coupling in Fig.~\ref{fig:GUT}  is scaled by $\sqrt{\frac{5}{3}}$ as predicted by grand unified theories\footnote{Chapter 8.3 in Ref.~\cite{Aitchison:2007fn} has a simple explaination of this factor and a more detailed approach can be found in e.g. Chapter 97 of Ref.~\cite{Srednicki:1019751}.} such as the $SU(5)$ theory of Giorgi and Glashow~\cite{PhysRevLett.32.438}.  The PDG values  of $\alpha_{EM}^{-1}(m_Z)=127.916\pm 0.015$, $\sin^2(\theta_W)(m_Z)=0.23116\pm 0.00013$, and $\alpha_s(m_Z)=0.1184\pm 0.0007$ are used as the initial condition (the error bands are too small to see)~\cite{Agashe:2014kda}.  In the SM, the three couplings do not unify at any scale, but amazingly for the MSSM, there is a point around $Q=10^{16}$ GeV where all three couplings are the same within uncertainties.  There are some changes to this picture by including higher order corrections, but the prospect of unification is unchanged.

\begin{figure}[h!]
\begin{center}
\includegraphics[width=0.55\textwidth]{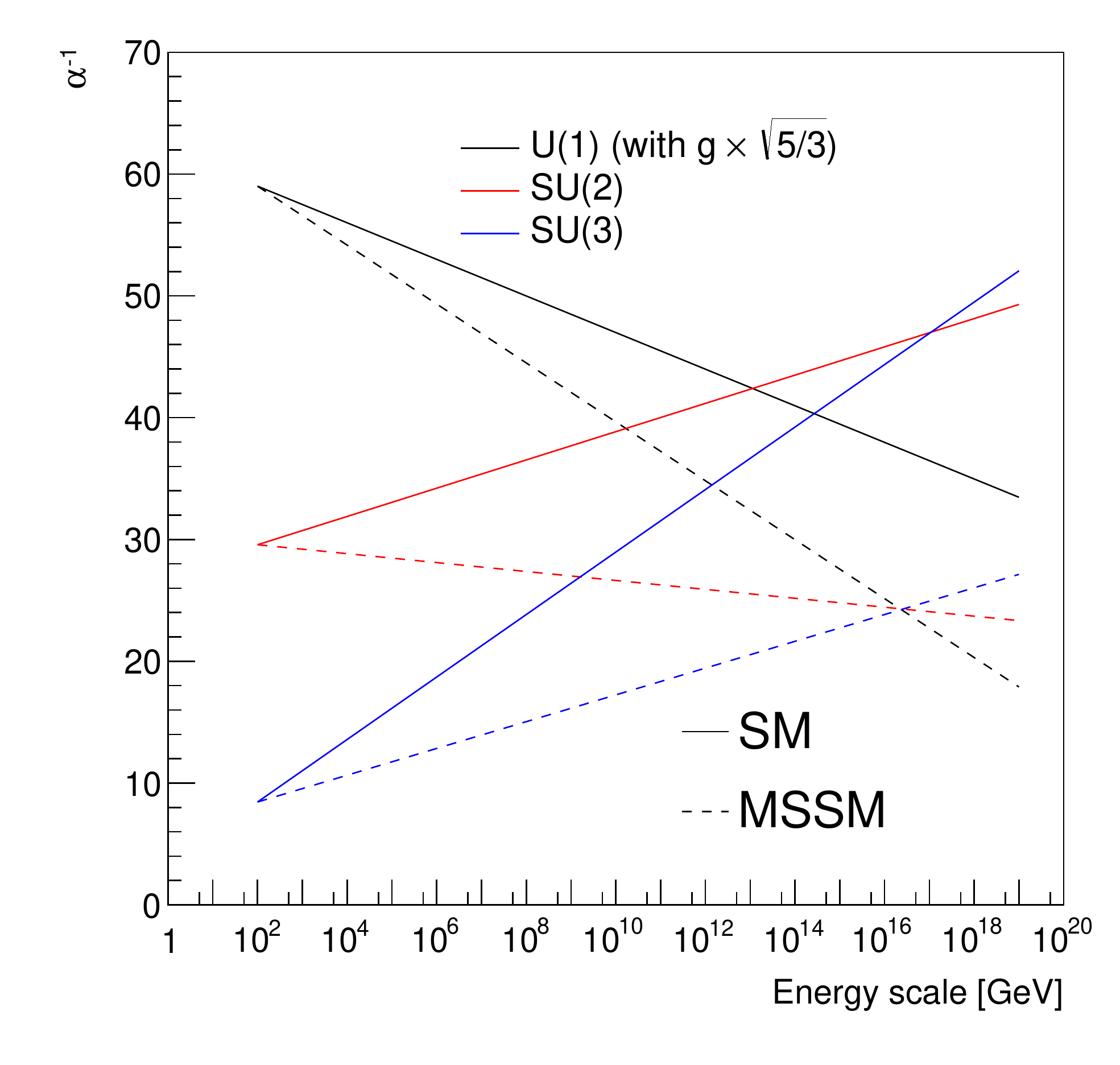}
\end{center}
\caption{The one-loop renormalization group evolution of the inverse couplings as a function of energy.  The $U(1)$ coupling is scaled by $\sqrt{\frac{5}{3}}$, which is the quantity that unifies with the others in grand unified theories.}
\label{fig:GUT}
\end{figure}	
		
Related to supersymmetric grand unification is the successful breaking of electroweak symmetry.   In order for the Higgs potential to have a local minimum and thus a positive vacuum expectation value, the Higgs boson mass squared in the Lagrangian must be negative.  In grand unified SUSY theories where the Higgs mass is set to a positive value at the grand unified scale, the renormalization group flow down to the electroweak scale drives down the Higgs boson mass squared and over a large range of parameter space is negative due to the large top quark Yukawa coupling\footnote{In the minimal SUSY extension of the SM, there are two Higgs boson doublets, one associated with up type quarks and one associated with down type quarks (see Sec.~\ref{sec:MSSM}).  As the top quark is up-type, the associated Higgs mass squared is the one usually driven negative.}.  Therefore, GUT SUSY can {\it explain} why electroweak symmetry is broken~\cite{Ibanez:1981yh,Ibanez:1982fr}.  Another impact of the large top Yukawa coupling for RGE glow in SUSY grand unified theories is that the stop is often the lightest squark near the electroweak scale, even if all the scalar masses are unified at the GUT scale (see Sec.~\ref{sec:MSSM}).

		The next section describes a complete model of weak-scale SUSY called the {\it Minimal Supersymmetric Standard Model} (MSSM) and will be the default SUSY model discussed for the remainder of Part~\ref{part:susy}.	
		
		\clearpage
		
		\section{The Minimal Supersymmetric Standard Model}
		\label{sec:MSSM}
		
			There are many ways to extend the SM with SUSY.  For example, one could augment spacetime with multiple copies of the fermionic dimensions ($N>1$ SUSY) or add additional SUSY multiplets beyond those that match the SM fields (for one additional scalar, this is the NMSSM).  However, the focus of this section and much of the SUSY literature is the {\it minimal} SUSY extension to the SM (MSSM) that has one chiral multiplet\footnote{The procedure for constructing a scalar superfield introduced in Sec.~\ref{sec:superspace}, namely Taylor expanding a field with certain transformation properties can be generalized to form chiral and vector superfield - see e.g. Chapter $4$ in Ref.~\cite{Martin:1997ns}.} for each SM fermion and one vector multiplet for each gauge boson.  One new multiplet in the MSSM with respect to the SM is a second Higgs field.  The main point of SUSY was to add a fermionic partner to the Higgs so that the mass would be protected by chiral symmetry.  However, all electroweakly interacting fermions contribute to the Feynman diagrams in Fig.~\ref{anamoly} which generate an anomaly in the SM\footnote{See for instance Chapter 20.2 in Ref.~\cite{Peskin:1995ev}.}: if this diagram does not exactly vanish, then $U(1)$ symmetry is violated beyond leading order in perturbation theory.  The matrix element from Fig.~\ref{anamoly} is proportional to $Y^3$, which amazingly sums to zero in the SM:			
			
			\begin{align}\nonumber
			\label{eq:anam}
			\mathcal{M}&\propto \sum_\text{left-handed} Y^3-\sum_\text{right-handed}Y^3\\
			&=2{\color{black}Y_{E_L}^3}-{\color{black}Y_{e_R}^3}+2N_c{\color{black}Y_{Q_L}^3}-N_c{\color{black}Y_{u_R}^3}-N_c{\color{black}Y_{d_R}^3}=0
			\end{align}

			\noindent Since the superpartners of the SM particles all have the same hypercharge, the sum in Eq.~\ref{eq:anam} remains zero.  When only one fermionic Higgs partner is added, the anomaly will not vanish.  This is solved by simply adding a second Higgs field with {\it opposite } hypercharge.  Table~\ref{tab:MSSMfields} summarizes the complete field content of the MSSM.  There are $17$ chiral supermultiplets, each containing one fermion and one complex scalar, and three vector supermultiplets, each containing a vector boson and a fermion.   The unbroken MSSM Lagrangian is the same (in form and number of parameters) as the SM case except for the Higgs and the lepton/baryon number violating sectors.  The Higgs part of the Lagrangian is given by
			
\begin{figure}
\centering
\begin{tikzpicture}[line width=1.5 pt, scale=1.2]	 
	\draw[vector] (-1,0) -- (0,0);
	\draw[fermion] (1,1) -- (0,0);
	\draw[fermion] (0,0) -- (1,-1);
	\draw[fermion] (1,-1) -- (1,1);
	\draw[vector] (1,1) -- (2,1);	
	\draw[vector] (1,-1) -- (2,-1);
	\node at (-1.2,0) {$B$};	
	\node at  (2.2,1) {$B$};	
	\node at  (2.2,-1) {$B$};	
\end{tikzpicture}
\caption{The Feynman diagram that is the source of the chiral anomaly in the SM.  Any electroweakly interacting fermion contributes to the loop.}
\label{anamoly}
\end{figure}				
			
			\begin{align}
			\mathcal{L}_\text{Higgs}^\text{MSSM}=(y_U)_{ij}Q_{i}H_uu_j^c+(y_D)_{ij}Q_{i}H_dd_j^c+(y_L)_{ij}L_{i}H_de_j^c+\mu H_u H_d,
			\end{align}  
			
			\noindent where the group indices are suppressed, $y_U,y_D$, and $y_L$ are the SM Yukawa mass matrices and $\mu$ is a new term that is allowed by $SU(2)$ symmetry.  Under electroweak symmetry breaking, the $\mathcal{L}_\text{Higgs}^\text{MSSM}$ behaves similarly to the SM case (ignoring the $\mu$ term\footnote{The $\mu$ term actually introduces a fine-tuning problem - its value is arbitrary, yet needs to be near the electroweak scale.  There is a large literature on this subject - see any papers which cite the earliest ideas: Ref.~\cite{Giudice:1988yz,Kim:1983dt}.}), except masses are generated separately for up-type quarks by $H_u$ and for down-type quarks and leptons by $H_d$\footnote{See e.g. Chapter 10 in Ref.~\cite{Aitchison:2007fn} or Sec. 8.1 in Ref.~\cite{Martin:1997ns} for slightly more information and Ref.~\cite{Gunion:1984yn} for extensive details.}.  In addition to the Higgs sector, there are a set of terms allowed by all of the internal symmetries, but explicitly violate low energy effective symmetries of the SM (lepton and baryon number conservation):

			\begin{align}
			\mathcal{L}_\text{RPV}^\text{MSSM}=\lambda_{ijk}L_i L_je_k^c+\lambda_{ijk}'L_iQ_jd_k^c+\lambda_{ijk}''u_i^cd_j^cd_k^c+\kappa_i L_i H_u,
			\end{align}  
			
			\noindent where the $\lambda$ and $\kappa$ terms are new dimensionless parameters.  While a priori there is no symmetry which forbids $\mathcal{L}_\text{RPV}^\text{MSSM}$, it has significant phenomenological consequences.  Most importantly, if the $\lambda_i\neq 0$\footnote{Technically, two of the $\lambda$ need to be nonzero for proton decay.  However, there are other constraints and issues of naturalness if only one of the $\lambda\neq 0$.}, the proton could rapidly decay even though the experimental lifetime is greater than $10^{33}$ years~\cite{Abe:2014mwa} (see e.g. Sec. 6.2 in Ref.~\cite{Martin:1997ns}).  One process contributing to proton decay is illustrated in Fig.~\ref{RPV}.   The standard assumption to remove $\mathcal{L}_\text{RPV}^\text{MSSM}$ is to impose a new $\mathcal{Z}_2$ symmetry called $R$-parity~\cite{Farrar:1978xj} (RPV = $R$-parity violation) under which the SM particles are neutral and the SUSY partners are charged.  Symbolically, the $R$-charge of a product of fields $F_1F_2\cdots F_n$ is given by
			
			\begin{align}
			R(F_1F_2\cdots F_n) = (-1)^{\sum_i 3B_i+ L_i+2s_i},
			\end{align}
			
			\noindent where $B_i$, $L_i$, and $s_i$ are the baryon number, lepton number, and spin of field $F_i$.  Requiring conservation of $R$-charge has many important phenomenological consequences.  First of all, there is no baryon or lepton number violation at tree level in the MSSM.  Second, at collider experiments with a SM-only initial state, SUSY particles must be produced in pairs.  The lightest SUSY particle (LSP) must be stable because it cannot decay into only SM particles.  When combined with the WIMP miracle, this last property makes the LSP an attractive dark matter candidate particle.  For the remainder of Part~\ref{part:susy}, $R$-parity is assumed conserved.

\begin{figure}[h!]
\centering
\begin{tikzpicture}[line width=1.5 pt, scale=1.3]
	\everymath{\displaystyle}
	\node[] (id1) at (-0.7em,1.9em) {$d$};
	\node[] (id2) at (-0.7em,0em) {$u$};
	\node[] (iu1) at (-0.7em,-1.8em) {$u$};
	\draw[dashed, black!30] (-0.7em,0) ellipse (1.8em and 3em); 	
	\node[circle] (od1) at (8.2em,2.2em) {\Large$e^+$};
	\node[] (ou1) at (8em,0em) {$\bar{u}$};
	\node[] (ou2) at (8em,-1.8em) {$u$};
	
	\node at (-1.4,0) {\Large$p^+$};
	\node at (4.2,-0.35) {\Large$\pi^0$};
	
	\node at (1.1,0.05) {$\lambda''$};
	\node at (1.99,0.05) {$\lambda'$};
	\draw[dashed, black!30] (8em,-0.9em) ellipse (1.2em and 2em);
	\draw[fermion] (0,0.75) -- (1,0.75/2);
	\draw[fermion] (0,0) -- (1,0.75/2);
	\draw[scalar] (1,0.75/2) -- (2,0.75/2);
	\draw[fermion] (2,0.75/2) -- (3,0.75);
	\draw[fermion] (2,0.75/2) -- (3,0);	
	\draw[fermion] (iu1) -- (ou2);
							\end{tikzpicture}
\caption{The Feynman diagram illustrating proton decay with RPV couplings.}
\label{RPV}
\end{figure}

\begin{table}[h]
\begin{center}
\noindent\adjustbox{max width=\textwidth}{
\begin{tabular}{|c|c|c|c|c|c|c|c|c|}
\hline
 & SM  & SM  & SUSY  & Partner  & &  &  & \\
Field &  component &  spin &  partner &  spin & $U(1)$ & $SU(2)$ & $SU(3)$ & Comment\\
 \hline  
 $Q_i$ & $(u_L\hspace{2mm}d_L)$ & $1/2$ & $(\tilde{u}_L\hspace{2mm} \tilde{d}_L)$ & $0$ &  $\frac{1}{6}$&  ${\bf 2}$& ${\bf 3}$  & 3 generations\\
  $u_i^c$ & $u_R^c$ & $1/2$ & $\tilde{u}_R^\dagger$ & $0$ & $\frac{2}{3}$ & ${\bf 1}$  & ${\bf \overline{3}}$  & 3 generations\\
    $d_i^c$ & $d_R^c$ & $1/2$ & $\tilde{d}_R^\dagger$ & $0$ & $-\frac{1}{3}$ & ${\bf 1}$  & ${\bf \overline{3}}$  & 3 generations\\
   $L_i$ & $(e_L\hspace{2mm}\nu_L)$ & $1/2$ & $(\tilde{e}_L\hspace{2mm} \tilde{\nu}_L)$ & $0$ &  $\frac{1}{2}$&  ${\bf 2}$& ${\bf 1}$  & 3 generations\\
  $e_i^c$ & $e_R^c$ & $1/2$ & $\tilde{e}_R^\dagger$ & $0$ & $-1$ & ${\bf 1}$  & ${\bf 1}$  & 3 generations\\  
    $H_u$ & $(H_u^+\hspace{2mm}H_u^0)$ & $0$ & $(\tilde{H}_u^+\hspace{2mm}\tilde{H}_u^0)$ & $1/2$ &  $\frac{1}{2}$&  ${\bf 2}$& ${\bf 1}$  & \\ 
    $H_d$ & $(H_d^0\hspace{2mm}H_d^-)$ & $0$ & $(\tilde{H}_d^0\hspace{2mm}\tilde{H}_d^-)$ & $1/2$ &  $-\frac{1}{2}$&  ${\bf 2}$& ${\bf 1}$  & \\ 
 \hline
     $B$ & $B$ & $1$ & $\tilde{B}$ & $1/2$ &  $0$&  ${\bf 1}$& ${\bf 1}$  & \\         
     $W$ & $W^\pm,W^0$ & $1$ & $\tilde{W}^\pm,\tilde{W}^0$ & $1/2$ &  $0$&  ${\bf 3}$& ${\bf 1}$  & \\
     $G$ & $g$ & $1$ & $\tilde{g}$ & $1/2$ &  $0$&  ${\bf 1}$& ${\bf 8}$  & \\           
 \hline                                  
\end{tabular}}
\caption{A summary of the MSSM field content in terms of electroweak eigenstates before symmetry breaking.  Even though the SUSY partners of the left- and right-handed fermions are scalars, they still carry the $L$ or $R$ subscript to emphasize their relationship to the SM particles.  The superpartners of the fermions are called {\it sfermions} (squarks and sleptons) and the superpartners of the bosons are called {\it bosinos} (bino, wino, gluino, and higgsino).  The right-handed field are specified in terms of the charge conjugate of left-handed fields because chiral superfields only contain left-handed fermions.}
\label{tab:MSSMfields}
\end{center}
\end{table}

There are many ways to break SUSY in the MSSM, but they all involve additional model assumptions.  To avoid making specific model assumptions, consider the Lagrangian of the MSSM augmented with terms that explicitly violate SUSY, $\mathcal{L}^\text{MSSM}_\text{$\cancel{SUSY}$}$.  Terms are only allowed if they do not {\it reintroduce} the hierarchy problem and preserve all other symmetries ({\it softly broken SUSY}).  Since the quadratic divergences giving rise to the hierarchy problem are associated with the dimensionless Yukawa couplings ($y\propto m/v$ as in Eq.~\ref{label:correction}), the Hierarchy problem can be avoided by omitting dimensionless interactions in $\mathcal{L}^\text{MSSM}_\text{$\cancel{SUSY}$}$.  Fermion mass terms are not allowed by $SU(2)$, but mass terms for the complex scalars and fermionic partners of the gauge bosons are allowed\footnote{For a longer explaination, see Chapter 9.2 in Ref.~\cite{Aitchison:2007fn}.}.  The full soft SUSY breaking Lagrangian is given by

\begin{align}
\label{susybreak}
\mathcal{L}^\text{MSSM}_\text{$\cancel{SUSY}$}=\frac{1}{2}M_i G_i^2+m^2_{\tilde{\Phi},ij}\tilde{\Phi}_i^\dag\tilde{\Phi}_j+m^2_{\tilde{\phi},ij}\tilde{\phi}_{R,i}^\dag\tilde{\phi}_{R,j}+A_{ijk}\tilde{\phi}_i\tilde\Phi_j\tilde\Phi_k+\text{h.c.},
\end{align}

\noindent where $G_i\in\{\tilde{g},\tilde{W},\tilde{B}\}$, $\tilde\Phi\in\{\tilde{Q},\tilde{L},\tilde{H}\}$ for e.g. $\tilde{Q}=(\tilde{u}_L\hspace{2mm} \tilde{d}_L)$, $\tilde\phi\in\{\tilde{u},\tilde{d},\tilde{e}\}$, and all group indices are suppressed.  The mass terms in Eq.~\ref{susybreak} allow the SUSY partners to have a mass much higher than and unrelated to the SM fermions and bosons which acquire a mass through EWSB.  In particular, the partners of the left- and right-handed SM fields have different soft SUSY masses and therefore can have significantly different masses.  In total, the full softly broken MSSM (from now on, {\it this} will be called the MSSM) has $105$ new parameters with respect to the SM~\cite{Dimopoulos:1995ju}.  Many of these terms are highly constrained by current experiments.  For example, the off-diagonal terms in the mass matrix induce large neutral flavor changing processes ruled out by flavor physics experiments.  However, the MSSM still has an enormous parameter space.  There is a vast literature of SUSY models that make various predictions for the relationships between parameters.  One well-studied set of models is the Constrained Minimal Supersymmetric Standard Model (CMSSM)~\cite{Fayet:1976et,Fayet:1977yc,Farrar:1978xj,Fayet:1979sa,Dimopoulos:1981zb} in which the particle masse in addition to the gauge couplings unify at a GUT scale.  In particular, at the GUT scale the scalar supersymmetric particles have the same mass $m_0$, the gauge fermion supersymmetric particles have the mass $M_{1/2}$ and the trilinear scalar couplings are given by a new parameter $A_0$ multiplied by the corresponding Standard Model Yukawa matrices.  The only other required input to fully specify the full MSSM is the ratio of the Higgs' vacuum expectation values $\tan\beta$ and the sign of the Higgsino mass term $\mathrm{sign}(\mu)$.  The value of $\mu$ is set by requiring the calculated $Z^0$ mass is equal to the measured value.  Thus, the CMSSM has only five more parameters than the SM, far fewer than the full MSSM. The SUSY particle spectrum at any given scale is then determined by solving the
RGEs with boundary conditions at the three scales: GUT, SUSY breaking\footnote{This intermediate scale is used because the radiative corrections associated with EWSB are smallest~\cite{Casas:1998vh}.}, and electroweak.  The standard is fixed point iteration~\cite{Allanach:2001kg,Baer:1993ae,Porod:2003um,Djouadi:2002ze}.  Figure~\ref{fig:RGEcMSSM} shows an example calculation, running the CMSSM GUT scale parameters down to the electroweak scale.  By construction, the gauge boson masses are equal to $m_{1/2}=500$ GeV at the GUT scale, which is just beyond $10^{16}$ GeV.   Successful EWSB is a {\it prediction} of this model and the stop is generally lighter than the other sfermions at the electroweak scale.   The RGEs for the soft SUSY breaking masses $M_i$, $i=1,2,3$ are similar to the equations for the gauge couplings discusses earlier (Eq.~\ref{couplingRGE}).  In particular, at leading order~\cite{Martin:1997ns}, 

\begin{align}
\frac{dM_i}{dt}=-\frac{b_i}{2\pi}\alpha_i M_i,
\end{align}

\noindent where $b_i$ was defined in Eq.~\ref{eq:RGE2} for the MSSM.  Amazingly, 

\begin{align}
\frac{d}{dt}\left(\frac{M_i}{\alpha_i}\right)=\frac{1}{\alpha_i}\frac{dM_i}{dt}+M_i\frac{d\alpha_i^{-1}}{dt}=0,
\end{align}

\noindent which means that this ratio does not run with scale, at one-loop.  In the CMSSM and in any other model where the gauge masses unify at the GUT scale, this gives a concrete prediction for the mass hierarchy in the MSSM.  Using the input parameters at $m_Z$ from earlier,

\begin{align}
\frac{M_1}{M_2}&=\frac{\alpha_1}{\alpha_2}=\frac{5}{3}\frac{\sin^2\theta_W}{\cos^2\theta_W}\sim 0.5\\
\frac{M_3}{M_2}&=\frac{\alpha_s\sin^2\theta_W}{\alpha_{EM}}\sim 3.5,
\end{align}

\noindent where the factor of $5/3$ is assuming that the gauge unification happens with the GUT scaling mentioned in Sec.~\ref{sec:hiearchyproblem}.  This gives the famous ratio $M_3:M_2:M_1\sim 7:2:1$ at the electroweak scale and the expectation that the gluino mass is higher than the mass of the electroweak superpartners.

Specifying parameters at the GUT scale and at the electroweak scale is a powerful technique for reducing the number of input parameters, but it also creates challenges.  In particular, the five parameters of the CMSSM are actually not enough to uniquely define an electroweak scale SUSY spectrum - there can be multiple solutions to the RGE equations~\cite{Allanach:2013cda,Allanach:2013yua}.  Figure~\ref{fig:RGEcMSSM2} illustrates the presence of these multiple solutions, some of which have significantly different phenomenology.  This loophole may allow CMSSM-type models to evade current limits, though all of the known extra spectra have similar masses to the previous spectra and so inclusive search results should be largely unaffected.

The CMSSM was the main set of models used by experiments at LEPP, the Tevatron, and the early part of LHC Run 1 for designing and interpreting experimental searches.  However, it has largely fallen out of favor because it is {\it too constrained} and many physical parameters such as the measured Higgs mass are not predicted correctly. 

\begin{figure}[h!]
\begin{center}
\includegraphics[width=0.5\textwidth]{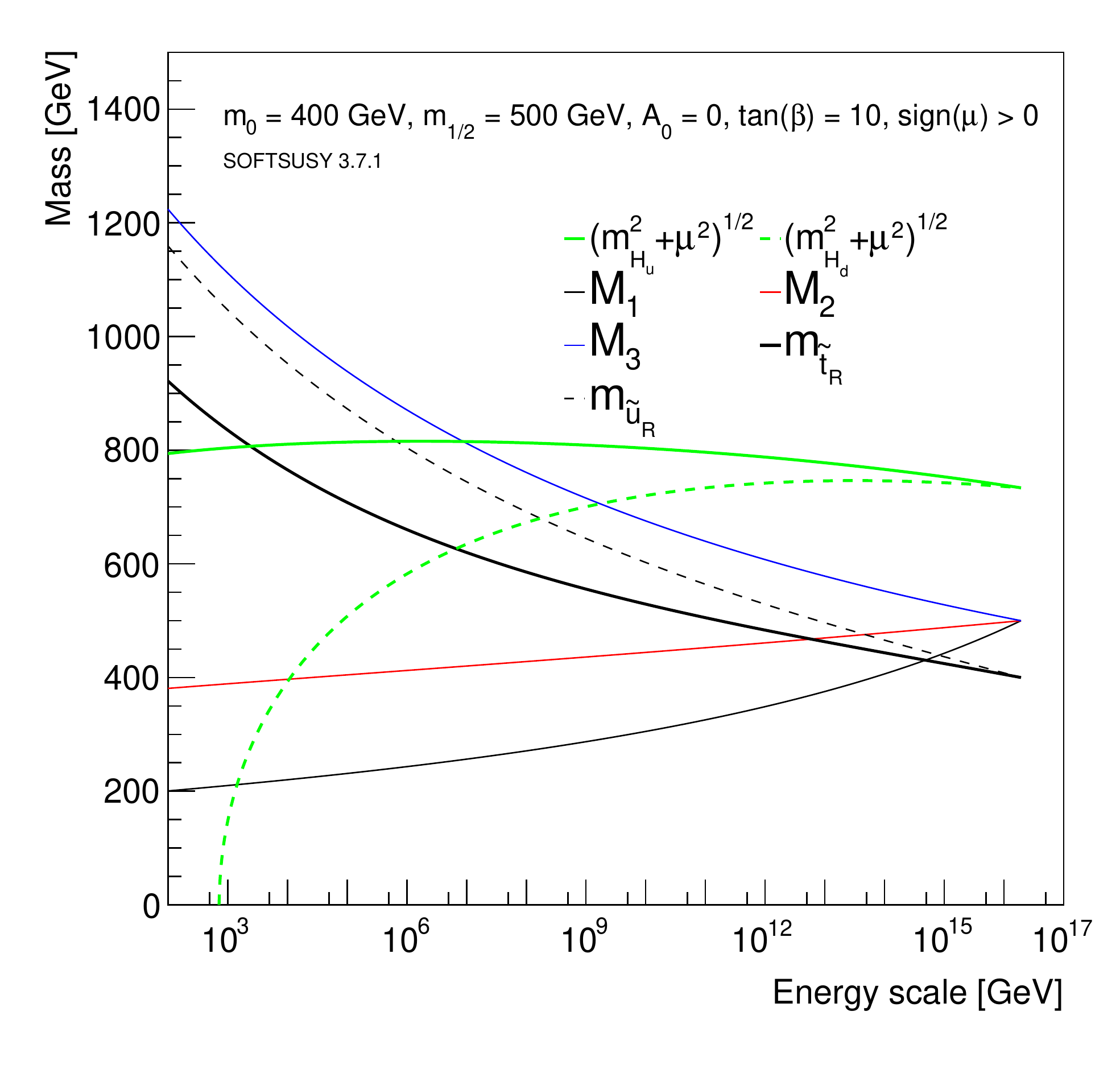}
\end{center}
\caption{An example CMSSM model specified with sfermion mass $m_0=400$ GeV, gaugino mass $m_{1/2}=500$ GeV, zero trilinear couplings, $\tan(\beta)>0$ and a positive $\mu$.  The running of the masses is calculated at NLO using SOFTSUSY 3.7.1~\cite{Allanach:2001kg}.}
\label{fig:RGEcMSSM}
\end{figure}
	
\begin{figure}[h!]
\begin{center}
\includegraphics[width=0.45\textwidth]{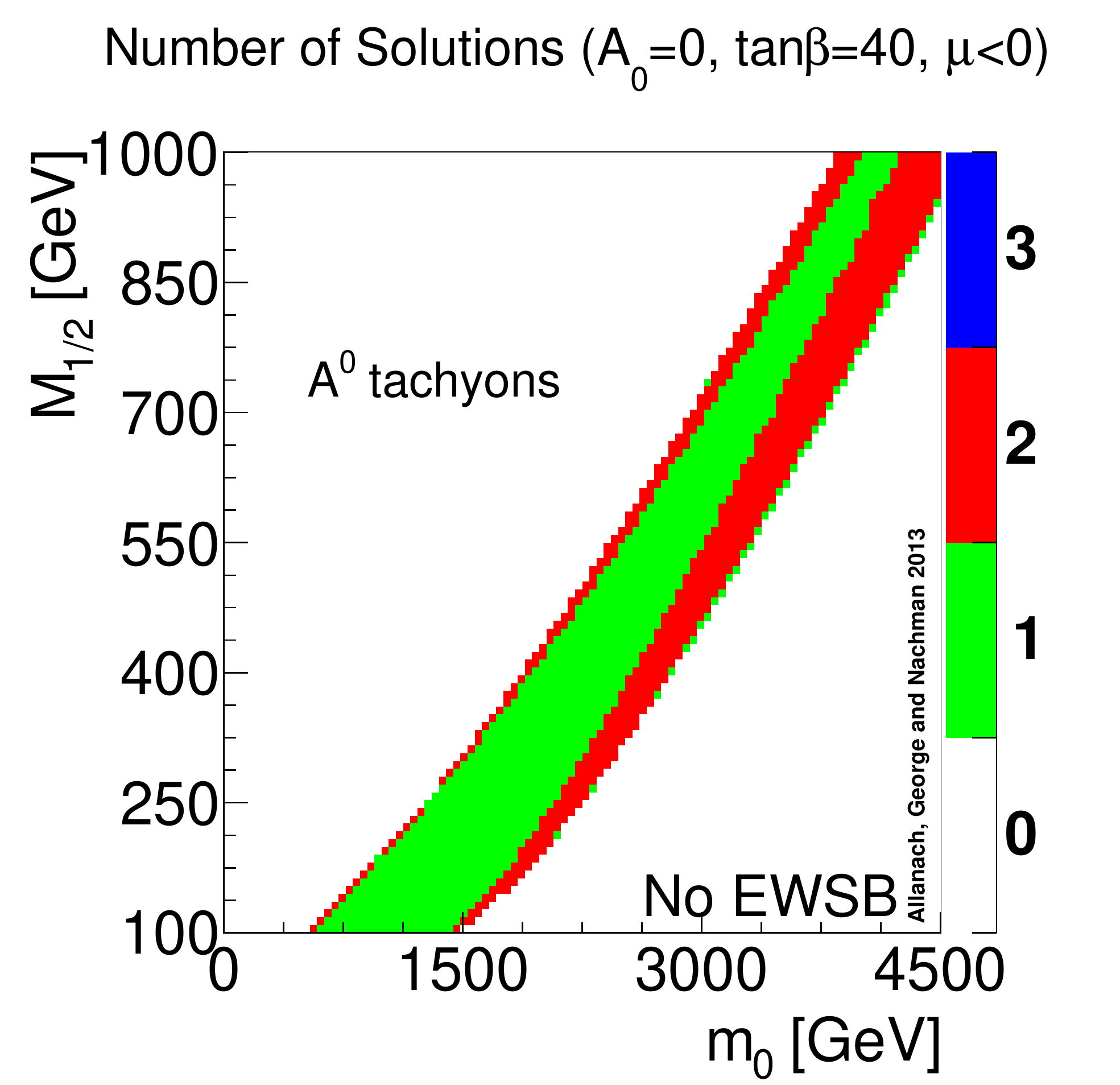}\includegraphics[width=0.53\textwidth]{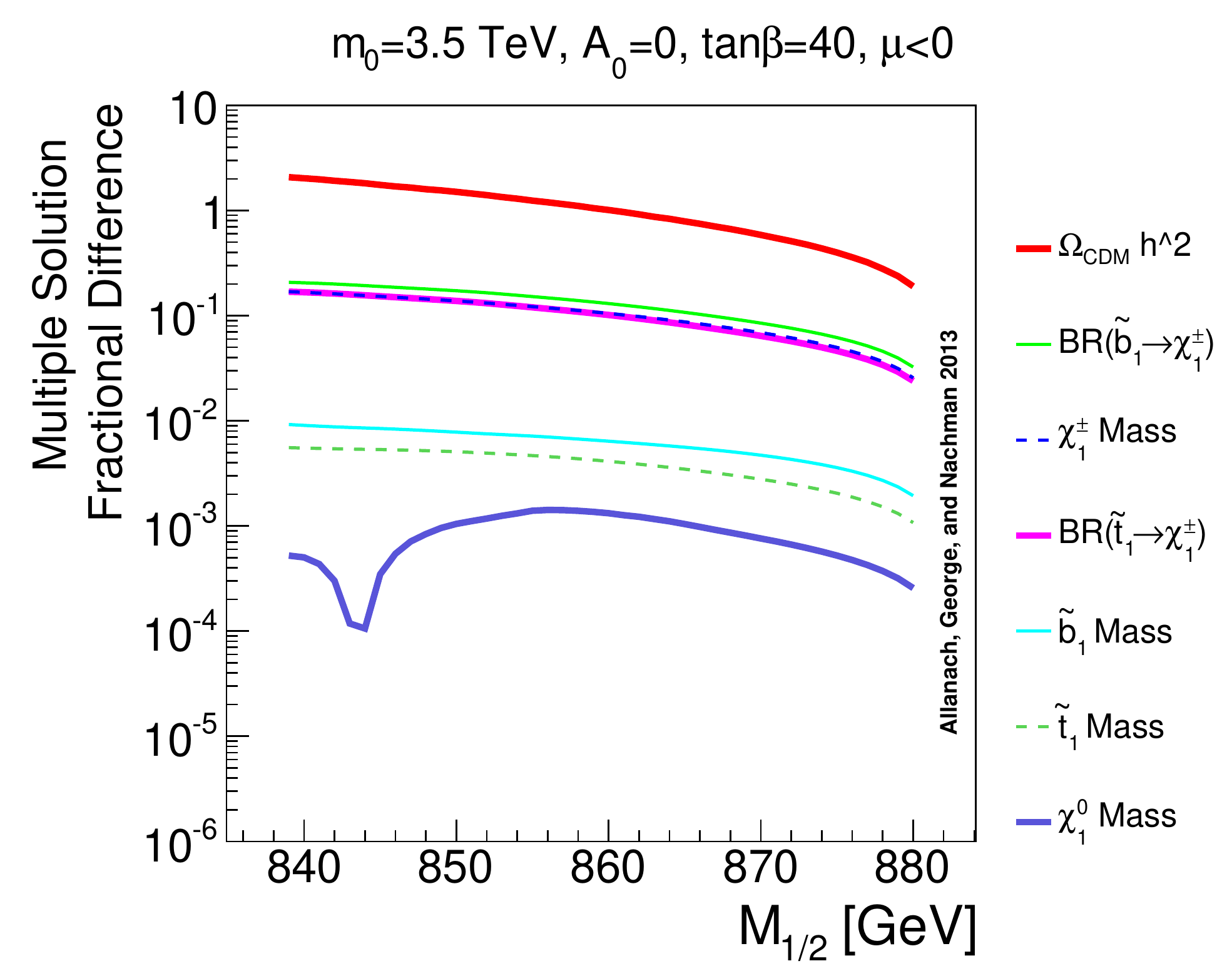}
\end{center}
\caption{Left: the number of electroweak scale spectra consistent with the CMSSM parameters as a function of $m_0$ and $M_{1/2}$ for fixed $A_0=0, \tan\beta<0$ and $\mu < 0$.  Right: the difference in select phenomenological parameters along the strip of two solutions from the left plot just below where $A^0$ is tachyonic.  Sparticle masses are nearly identical between the two spectra, but stop and sbottom branching ratios vary by more than $10\%$ and the predicted dark matter relic density differs by more than $100\%$ for $M_{1/2}\sim 840$ GeV.  See Ref.~\cite{Allanach:2013yua} for more detail.}
\label{fig:RGEcMSSM2}
\end{figure}

The opposite extreme to the CMSSM is an approach where parameters are only specified near the electroweak scale.  One class of such models is called the phenomenological MSSM (pMSSM)~\cite{Djouadi:1998di,Berger:2008cq}, as it reduces the number of MSSM parameters by imposing reasonable phenomenological constraints.  In particular, by requiring pMSSM models to not introduce non-SM sources of CP violation, lack flavor changing neutral currents, and have degenerate first and second generations, the total number of parameter is reduced to $19$.  Various groups have performed scans in (subsets of) this $19$ parameter space to identify regions of the pMSSM that are also consistent with SM measurements and SUSY searches.  For example, Fig.~\ref{fig:pMSSM} shows one part of a pMSSM model with a light stop that is not ruled out by the direct stop searches, but is excluded by searches with a more inclusive scope due to the complexity of the final state.  While there is an inherent bias in these scans due to the choice of parameter priors, they are useful for identifying a class of `realistic' models.
		
\begin{figure}[h!]
\begin{center}
\includegraphics[width=0.5\textwidth]{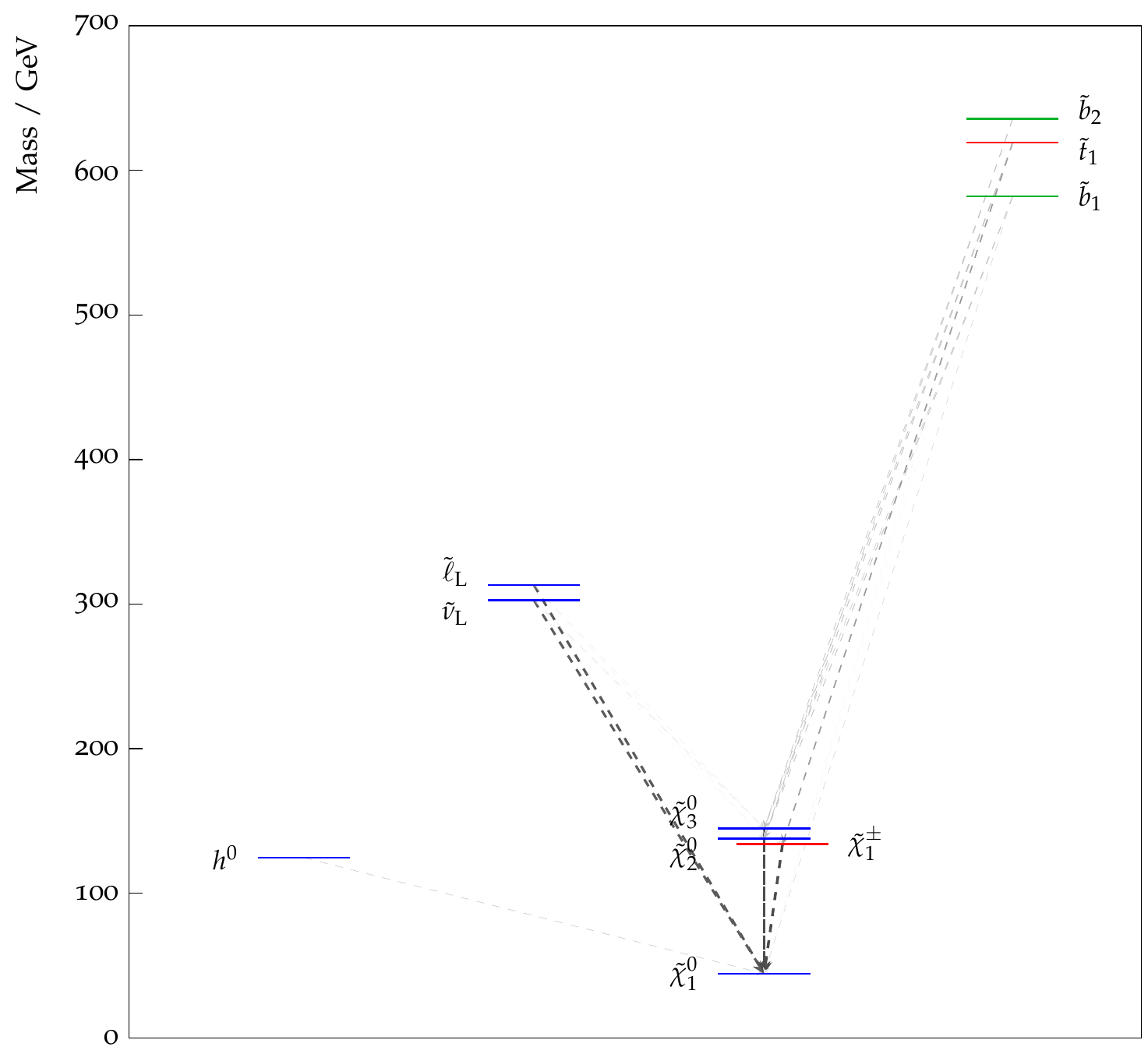}
\end{center}
\caption{One model from a recent ATLAS summary~\cite{Aad:2015baa} of SUSY searches interpreted in the context of a random scan in the pMSSM~\cite{Berger:2008cq,CahillRowley:2012cb,CahillRowley:2012kx,Cahill-Rowley:2014twa}.  This particular model is not excluded by direct stop searches but is ruled out by other searches with a broader scope.  One reason the direct searches do not rule out this model, and one powerful use of the pMSSM, is the model complexity: there are multiple light neutralinos/charginos with cascade decays.   This plot was created with PySLHA~\cite{Buckley:2013jua}.  All sparticles not shown are heavier than 700 GeV.} \label{fig:pMSSM}
\end{figure}

The $19$ parameters of the pMSSM is still too large for most practical purposes.  Currently, the most popular approach is to focus on specific topologies or {\it simplified models} and {\it ignore} the rest of the spectrum, assuming it is largely decoupled or at least factorized from the process of interest~\cite{Alwall:2008ve,Alwall:2008ag,Alves:2011wf}.  Simplified models are useful for organizing searches based on experimental signatures instead of unobservable theoretical parameters.  In addition, searches based on simplified models can easily be reinterpreted in any model that has a simplified model-like component.  To construct simplified models, it is useful to recast the MSSM fields in terms of the mass eignestates instead of the weak eigenbasis (though this is not particular to simplified models).  After electroweak symmetry breaking, the mass matrix for the neutral electroweak superpartners is given at leading order by (e.g. Sec. 8.2 in Ref.~\cite{Martin:1997ns}):

\begin{align}
\label{eq:neutralinos}
M_N=\begin{pmatrix}M_1 & 0 & -c_\beta s_Wm_Z & s_\beta s_Wm_Z \cr 0 & M_2 & c_\beta c_Wm_Z & -s_\beta c_Wm_Z \cr -c_\beta s_Wm_Z &c_\beta c_W m_Z & 0 & -\mu \cr s_\beta s_W m_Z & -s_\beta c_Wm_Z & -\mu & 0 ,\end{pmatrix},
\end{align}

\noindent where $s_x=\sin(x)$ and $c_x=\cos(x)$.  The diagonalization of Eq.~\ref{eq:neutralinos} results in the mass matrix for the four {\it neutralinos} $\tilde{\chi}_i^0$, $i=1,..,4$ with the convention $m_{\tilde{\chi}_i^0}<m_{\tilde{\chi}_{i+1}^0}$.  The lightest neutralino is an excellent dark matter candidate as it is stable if it is the LSP (assumed henceforth) and only interacts via the weak force\footnote{See Ref.~\cite{Ellis:1983ew} for an argument why any of the electrically or color charged particles would not make good dark matter candidates.  The relic abundance of SUSY LSP dark matter depends on the field content of the lightest neutralino - mostly higgsino and wino LSP dark matter tends to overproduce and mostly bino LSP tends to underproduce the measured density (see e.g. the review Ref.~\cite{Jungman:1995df} and references therein).}. When $m_Z \ll |\mu\pm M_i|$, $i=1,2$, the neutralinos are nearly pure {\it bino}, {\it wino}, and {\it higgsino}.   In such a case, one may expect the LSP to be mostly bino-like or higgsino-like (assuming $M_1<M_2$ as in the GUT-inspired scenario).  Similarly, there is a two-by-two matrix for the charged electroweak superpartners that forms the two electrically positive and two negative {\it charginos} $\tilde{\chi}_i^\pm,i=1,2$.   All of the scalar sfermions can also mix to form the mass eigenstates.  The most important is the stop mass matrix\footnote{In principle, there can be mixing between the sfermion families, but this is assumed negligible due to the often unacceptable flavor changing neutral currents.}:

\begin{align}
\label{eq:stopmixing}
M_{\tilde{t}}^2=\begin{pmatrix} m_{\tilde{Q},33}^2+m_\text{top}^2+\left(\frac{1}{2}-\frac{2}{3}s_W^2\right)m_Z^2c_{2\beta}&m_\text{top}(A_{\tilde{t}_R\tilde{Q}_3\tilde{H}_u}-\mu\cot(\beta)) \cr m_\text{top}(A_{\tilde{t}_R\tilde{Q}_3\tilde{H}_u}-\mu\cot(\beta)) & m^2_{q,33}+m_\text{top}^2+ \frac{2}{3}s_W^2m_Z^2c_{2\beta}\end{pmatrix}.
\end{align}

\noindent The stop mixing angle $\theta_t$ is defined as the angle of the rotation matrix required to diagonalize Eq.~\ref{eq:stopmixing}.  After diagonalizing the fields, the two stop mass eigenstates are called $\tilde{t}_1$ and $\tilde{t}_2$ with $m_{\tilde{t}_1}<m_{\tilde{t}_2}$.  In the literature, $X_t= A_{\tilde{t}_R\tilde{Q}_3\tilde{H}_u}-\mu\cot(\beta)$ is often called the effective mixing parameter, as it controls the amount of mixing between the weak eigenstates in Eq.~\ref{eq:stopmixing}. 

The main motivation of electroweak scale SUSY was the cancellation of quantum corrections to the Higgs boson mass.  In the MSSM, the lightest Higgs boson mass is not a free parameter; at tree level, it is given by

\begin{align}
m_h^2=\frac{1}{2}\left(m_{A^0}^2+m_Z^2-\sqrt{(m_{A^0}^2-m_Z^2)^2+4m_Z^2m_{A^0}^2 s_{2\beta}^2}\right),
\end{align}

\noindent where $A^0$ is the pseudoscalar Higgs boson generated after electroweak symmetry breaking by the scalar part of the Higgs field and has mass $m_{A^0}^2=2\mu^2+m^2_{H_u}+m^2_{H_d}$.  It seems that in SUSY, the Higgs mass is actually {\it too well regulated}:

\begin{align}
m_h^2 \leq \lim_{m_{A^0}\rightarrow\infty} m_h^2(m_{A^0}) = m_Z^2(1-s_{2\beta}^2)=m_Z^2\cos^2(2\beta) \leq m_Z^2.
\end{align}

\noindent If the tree-level calculation was (close) to the full answer, the MSSM would be ruled out by the fact that $m_h\approx 125$ GeV $>m_Z$.  Fortunately, the corrections to the Higgs boson mass are not small.  The dominante correction comes from loops with stops\footnote{Due to its importance, there is an extensive literature on this calculation, see e.g. Ref.~\cite{Draper:2016pys} for a review.  This equation is based on Eq. 8.1.24 in Ref.~\cite{Martin:1997ns}, which is a nicer version of Eq. 62 in Ref.~\cite{Draper:2016pys}. }:

\begin{align}
\label{higgsmasscorrection}
\Delta(m_{h}^2)=\frac{3}{4\pi^2}c_\beta^2 y_t^2 m_t^2\left[\ln\left(\frac{m_{\tilde{t}_1}m_{\tilde{t}_2}}{m_t^2}\right)+\frac{\Delta_\text{mix}}{m_t^2}\right]+\mathcal{O}\left(\frac{m_t^2m_{Z}}{m_{A^0}},\frac{m_t^2y_t^4}{(4\pi^2)^2}\right),
\end{align}

\noindent where the term $\Delta_\text{mix}$ goes to zero when $\theta_{\tilde{t}}\rightarrow 0$:

\begin{align}
\label{eq:higgsmass}
\Delta_\text{mix}=c_{\tilde{t}}^2s_{\tilde{t}}^2(m_{\tilde{t}_2}^2-m_{\tilde{t}_1}^2)\ln\left(\frac{m_{\tilde{t}_2}^2}{m_{\tilde{t}_1}^2}\right)+\frac{c_{\tilde{t}}^4s_{\tilde{t}}^4}{m_t^2}\left[(m_{\tilde{t}_2}^2-m_{\tilde{t}_1}^2)^2-\frac{1}{2}(m_{\tilde{t}_2}^4-m_{\tilde{t}_1}^4)\ln\left(\frac{m_{\tilde{t}_2}^2}{m_{\tilde{t}_1}^2}\right)\right].
\end{align}

\noindent Even with the correction in Eq.~\ref{higgsmasscorrection}, it is difficult to accommodate a $125$ GeV Higgs boson.  Figure~\ref{fig:higgsmass} shows the maximum value of Eq.~\ref{higgsmasscorrection} without stop mixing and with mixing that maximizes the Higgs mass correction.  In order for the Higgs mass to be heavy enough, one or both of the stops have to be relatively heavy ($m\gtrsim$ 1 TeV).  In order for one of the stops to be light $(m\lesssim 1$ TeV), there must be significant stop mixing.   

\begin{figure}[h!]
\begin{center}
\begin{overpic}[width=0.45\textwidth]{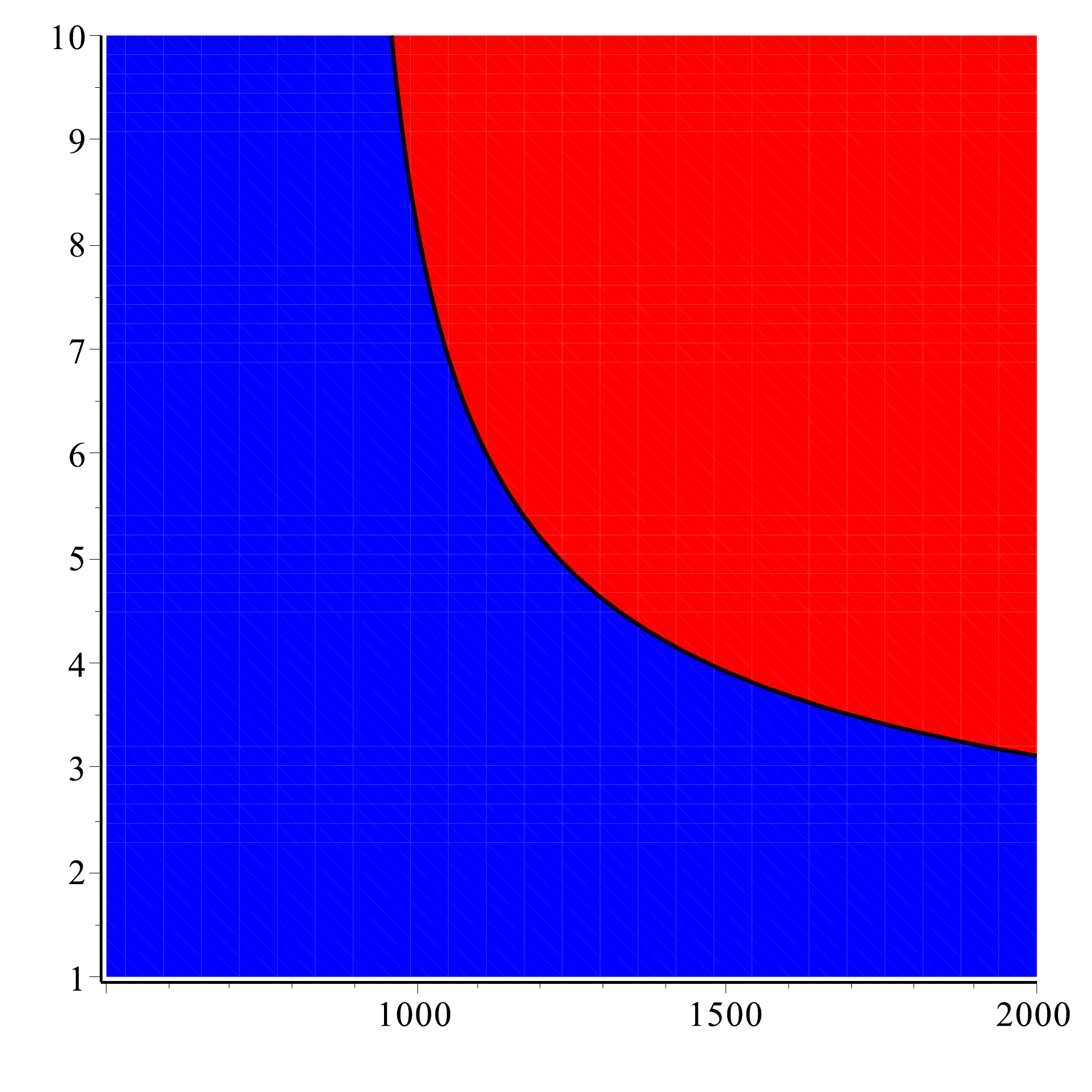}\put(40,-5){$m_{\tilde{t}_2}$ [GeV]}\put(50,60){{\color{white}$m_h > 125$ GeV}}\put(30,25){{\color{white}$m_h < 125$ GeV}}\put(-5,45){\rotatebox{90}{$\tan(\beta)$}}\end{overpic}\hspace{8mm}\begin{overpic}[width=0.45\textwidth]{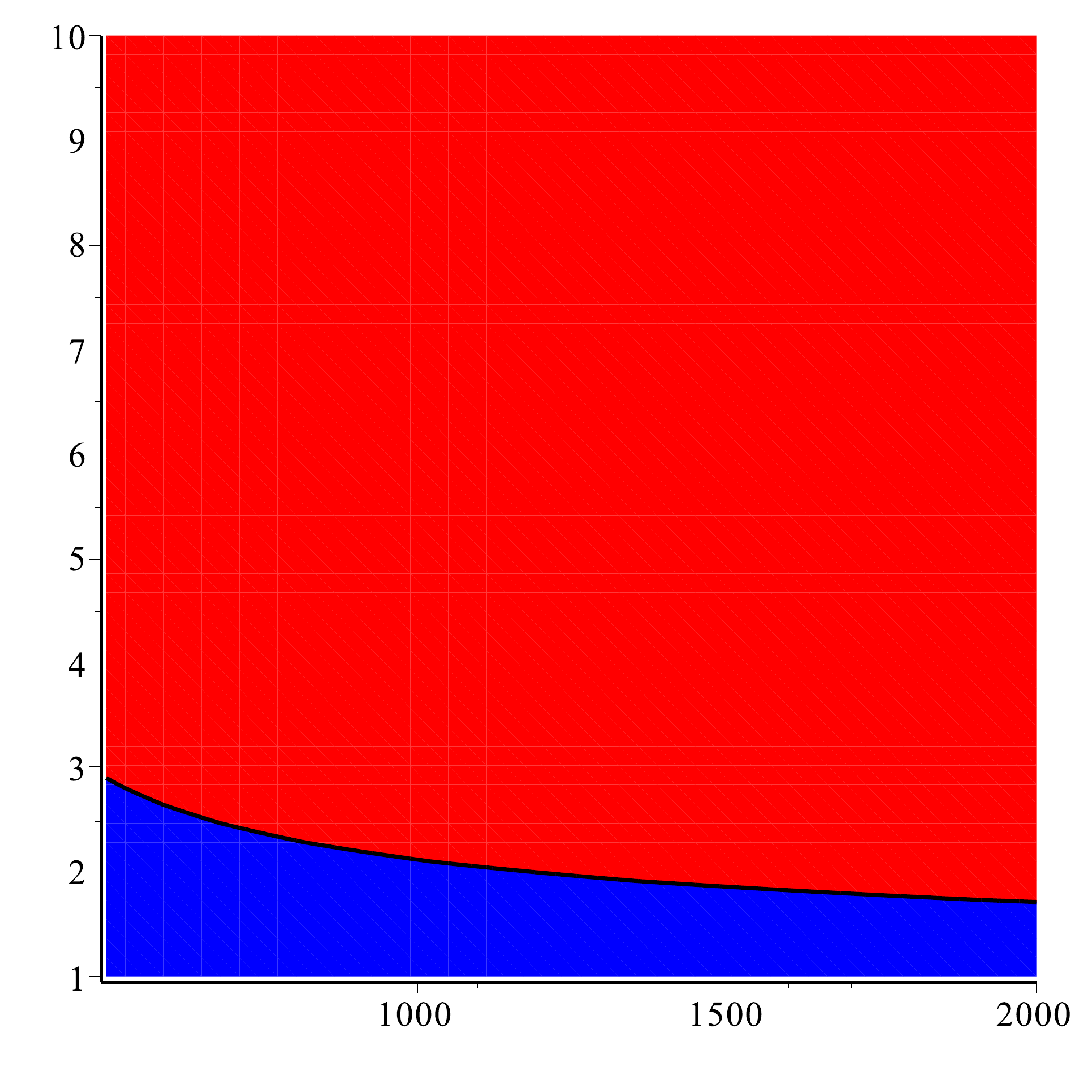}\put(40,-5){$m_{\tilde{t}_2}$ [GeV]}\put(40,25){{\color{white}$m_h > 125$ GeV}}\put(20,13){{\color{white}$m_h < 125$ GeV}}\put(-5,45){\rotatebox{90}{$\tan(\beta)$}}\end{overpic}
\end{center}
\caption{The maximum Higgs mass $\sqrt{m_Z^2c_{2\beta}^2+\Delta(m_h^2)}$ from Eq.~\ref{eq:higgsmass} for no stop mixing $\theta_{\tilde{t}}=0$ (left) and mixing that maximizes the correction (right).}
\label{fig:higgsmass}
\end{figure}

In addition to setting the mass hierarchy, stop mixing has a significant impact on stop decay.  The stop can decay via a neutral current $\tilde{t}\rightarrow t+\tilde{\chi}^0$ or charged current $\tilde{t}\rightarrow b+\tilde{\chi}^\pm$.  If the lightest chargino is heavier than $\tilde{t}_1$, then the neutral current decay dominates unless $m_{\tilde{t}}-m_{\tilde{\chi}_1^0}\ll m_t,m_W$ so that loop-induced processes can compete with the off-shell top/$W$ boson in the decay.  Since the coupling between the stop and the bino is via the weak hypercharge, by Table~\ref{tab:MSSMfields}, the coupling is stronger to $\tilde{t}_R$ than to $\tilde{t}_L$.  Therefore,  $\mathcal{BR}(\tilde{t}_R\rightarrow t+\tilde{\chi}^0_1)/\mathcal{BR}(\tilde{t}_L\rightarrow t+\tilde{\chi}^0_1) > 1$ for a mostly bino LSP.  In fact, if the chargino is mostly a wino and $\theta_{\tilde{t}}\sim\pi/2$ (i.e. $\tilde{t}_1\sim\tilde{t}_R$), then the $\mathcal{BR}(\tilde{t}_1\rightarrow t+\tilde{\chi}^0_1)\approx 100\%$ regardless of the mass of $\tilde{\chi}^\pm_1$ since the superpartner of the right handed top does not couple to $\tilde{W}^\pm$.  In general, the partial widths of the two processes are determined by $\theta_{\tilde{t}}, m_{\tilde{t}}$, $m_{\tilde{\chi}_1^\pm}$, $m_{\tilde{\chi}_1^0}$, and the neutralino/chargino mixing matrices which depend on $\mu,M_1,M_2$, and $\tan(\beta)$ - see Ref.~\cite{Bartl:1996wt} for a full set of formulae at leading order\footnote{The formulae include the branching partial widths to the other three neutralinos and the higher mass charginos.  A derivation in the case of the neutral current decay can be found in e.g. Chapter 12.1 of Ref.~\cite{Aitchison:2007fn}.  Reference~\cite{Belanger:2012tm} presents a clear discussion of the dependence of the $\mathcal{BR}(\tilde{t}_1\rightarrow t+\tilde{\chi}^0_1)$ on $\mu$ and $M_1$, including the case of additional neutralinos.}.  Due to its general importance and unique final state, the reminder of Part~\ref{part:susy} focuses exclusively on $\tilde{t}_1\rightarrow t+\tilde{\chi}^0_1$.

Given that the stop decays via $\tilde{t}_1\rightarrow t+\tilde{\chi}^0_1$, the stop and neutralino mixing parameters determine the polarization of the top quark.  Standard Model $t\bar{t}$ production results in unpolarized top quarks, i.e. equal numbers of left- and right-handed quarks.  However, the production of top quarks via stops can result in significant asymmetry.   The interaction vertex is proportional to~\cite{Perelstein:2008zt}:

\begin{align}
 \tilde{t}_1\tilde{\chi}_1^0\left(\cos(\theta_\text{eff})P_L+\sin(\theta_\text{eff})P_R\right)t,
\end{align}

\noindent where $P_L$ and $P_R$ are the usual spin projection operators $\frac{1}{2}(1\pm\gamma^5)$ and the effective mixing angle is given by

\begin{align}
\label{eq:effectivemixing}
\tan\theta_\text{eff} = \frac{Y_tN_{14}\cos(\theta_{\tilde{t}})-\frac{2\sqrt{2}}{3}g_1 N_{11}\sin(\theta_{\tilde{t}})}{\sqrt{2}\left(\frac{g_2}{2}N_{12}+\frac{g_1}{6}N_{11}\right)\cos(\theta_{\tilde{t}})+Y_t N_{14}\sin(\theta_{\tilde{t}})}.
\end{align}

\noindent The parameter $Y_t=y_t/\sin\beta$, where $y_t$ is the SM top quark Yukawa coupling and the matrix $N$ diagonalizes the mass matrix $M_N$, $N^\dag M_N N^{-1}=\text{Diag}(m_{\tilde{\chi}_i^0})$.  In the case $m_Z\ll |\mu\pm M_i|, i=1,2$, $N_{11}\approx 1$ and $N_{1j}\approx 0$ for $j>0$; then, $-\tan\theta_\text{eff} \sim \tan(\theta_{\tilde{t}})/6$.  The factor of six is due to the asymmetric coupling of the bino to $\tilde{t}_L$ and $\tilde{t}_R$\cite{Belanger:2012tm}.  Note that the effective mixing angle depends on both the stop mixing matrix {\bf and} the neutralino mixing matrix.  Changes in the top quark polarization result in different energy spectra of the final state objects, leading to changes in the efficiency for a given event selection~\cite{Low:2013aza,Belanger:2012tm}.  The phenomenology of stop decay is discussed in more detail in Sec.~\ref{sec:targetpheno}.

		\clearpage

			\section{Related Models}
			\label{relatedmodels}			

	Before describing the analysis strategy for searching for stops, it is important to note that the $t\bar{t}+E_\text{T}^\text{miss}$ signature is an important property of many extensions of the SM.   Another natural source of $t\bar{t}+E_\text{T}^\text{miss}$ within SUSY is the pair production of gluinos where each gluino decays $\tilde{g}\rightarrow t \tilde{t}$ ({\it gluino mediated stop} or GMS).   In a natural SUSY spectrum, the gluino should not be too much heavier than the stop because the stop mass receives large quantum corrections from the gluino just as the Higgs receives large contributions from the stop (see e.g. Ref.~\cite{Brust:2011tb}).  When the stop is significantly heavier than the LSP, GMS models can have fantastic signatures including many top or $b$ quarks.  However, when $m_\text{stop}\sim m_\text{LSP}$ as might be needed to regulate the amount of dark matter~\cite{Profumo:2004at,Edsjo:2003us,Boehm:1999bj,Ellis:2001nx}, the stop decay products can be too soft to measure and therefore the total signature is $t\bar{t}+E_\text{T}^\text{miss}+\text{soft}$, as illustrated by Fig.~\ref{fig:Gtc}.  The properties of these models are discussed more detail in Sec.~\ref{stoprecast}.
	
 There is also a wide range of non-SUSY models that produce $t\bar{t}$ in association with weakly interacting particles.   For example, new particles with both lepton and baryon number ({\it leptoquarks}~\cite{Buchmuller:1986zs}) could decay to a top quark and a neutrino.  These third generation leptoquarks have been recently proposed as a model to explain the $\bar{B}\rightarrow D^{*}\tau\bar{\nu}$ excess~\cite{Freytsis:2015qca}.  Scalar leptoquark production is identical to stop pair production, but there can be differences in kinematic distributions of the decay products due to the spin configurations of the final state objects.  Vector leptoquarks have a significantly higher cross section due to the extra spin states.  Another model that results in an increased cross-section is the case of vector-like quarks~\cite{delAguila:1982fs} $T'$ that are fermions with right-handed charged current interactions.   These spin $1/2$ particles often occur in theories where the Higgs is not fundamental (as a solution to the hierarchy problem) as in the little Higgs~\cite{Perelstein:2005ka,Perelstein:2003wd,ArkaniHamed:2002qy}, top-color assisted technicolour~\cite{Hill:1994hp}, composite Higgs~\cite{Kaplan:1983fs,Kaplan:1983sm,Banks:1984gj,Georgi:1984af,Dugan:1984hq,Georgi:1985hf,Bellazzini:2014yua,Georgi:1984ef} models.  When the $T'\rightarrow tZ$ and $Z\rightarrow\nu\nu$, the final state is similar to the pair production of stops.  Figure~\ref{fig:Gtc} also shows diagrams for the leptoquark and the vector-like quark.  There are also dark matter models with non-resonant production giving rise to $t\bar{t}\chi\chi$ for dark matter particle $\chi$~\cite{Bhattacherjee:2012ch}.  The coupling to mass is a strategy to avoid large flavor changing neutral currents~\cite{Lin:2013sca}.
			
\begin{figure}[h!]
\begin{center}
\includegraphics[width=0.45\textwidth]{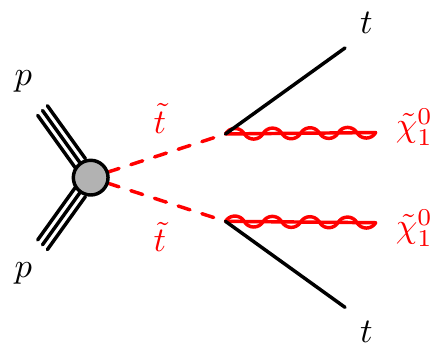}\includegraphics[width=0.45\textwidth]{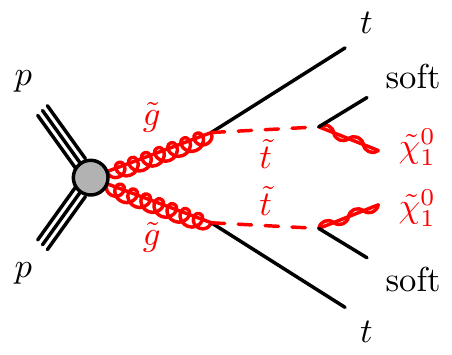}\\
\includegraphics[width=0.45\textwidth]{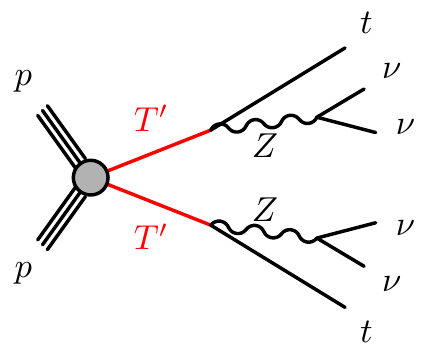}\includegraphics[width=0.45\textwidth]{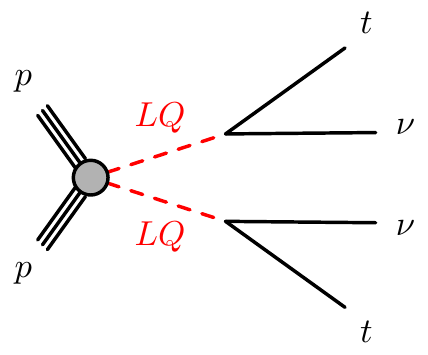}
\end{center}
\caption{Schematic diagrams illustrating models with similar signatures as stop pair production (top left).  Top right: the gluino mediated stop process with nearly mass-degenerate stop and neutralino such that the stop decay products are too soft to be reconstructed.  Bottom left: pair production of vector-like quarks $T'$ decaying into $Z$ bosons that decay into neutrinos that are undetected just like neutralinos.  Bottom right: leptoquarks carry both lepton and baryon numbers and can decay to a top quark and a neutrino.  All of these models are constrained by the search presented in Part~\ref{part:susy}.} 
\label{fig:Gtc}
\end{figure}

 \chapter{Analysis Strategy}
\label{chapter:susy:analysisstrategy}
	
	While stop pair production shares many similarities with other searches for new particles, it also requires a dedicated approach.  For example, high mass stops produce many high $p_\text{T}$ jets and a large $E_\text{T}^\text{miss}$, but without explicitly targeting final states with top quarks, there is a significant loss in sensitivity.  The first searches for the $t\bar{t}+E_\text{T}^\text{miss}$ topology were performed by the CDF collaboration at the Tevatron at $\sqrt{s}\approx2$ TeV using the one-lepton~\cite{Aaltonen:2011rr} and all-hadronic final states~\cite{Aaltonen:2011na}.  However, the stop pair production cross-section is too low for any model to be excluded with $95\%$ confidence (fermionic top quark partners were excluded up to about $400$ GeV).  The first stop search to be sensitive to $t\bar{t}+E_\text{T}^\text{miss}$ was an early $\sqrt{s}=7$ TeV result by ATLAS in the one-lepton channel~\cite{Aad:2011wc} using $1$ fb${}^{-1}$ of data that excludes stops with massless LSP up to about $m_\text{stop}\sim 280$ GeV.  Using the full $\sqrt{s}=7$ TeV dataset, ATLAS was able to exclude simplified stop models with stop masses between 230 GeV and 440 GeV for massless LSPs, and top squark masses around 400 GeV are excluded for LSP masses up to 125 GeV~\cite{Aad:2012xqa}.   These early analyses focused on applying standard tools to relatively low stop mass models.  The remaining natural parameter space is complex and requires a series of dedicated techniques to effectively suppress and estimate backgrounds.  Chapter~\ref{chapter:susy:analysisstrategy} begins with an introduction to stop phenomenology for $m_\text{stop}\lesssim 1$ TeV (Sec.~\ref{sec:targetpheno}).  The background estimation paradigm, called the {\it control region method} is described in Sec.~\ref{sec:CRmethod} and the technical setup of the analysis is documented in Sec.~\ref{sec:datasetandMC}.

	\clearpage
	
	\section{Phenomenology}
		\label{sec:targetpheno}
		
In the approximation that the stop sector decouples from the rest of the MSSM, the cross section for stop pair production depends only on the stop mass.  Like any other non-resonant process with $m\ll \sqrt{s}$, the stop production cross section falls off rapidly as a function of $m_\text{stop}$, with a $1/m^2$ matrix element suppression compounded with a significant PDF suppresion.  Figure~\ref{fig:stopcrosssection} shows the pair-production cross-section at $\sqrt{s}=8$ and $\sqrt{s}=13$ TeV as a function of mass from $100$ GeV $<m_\text{stop} < 2$ TeV.  For $m_\text{stop}\gtrsim 250$ GeV, $\sigma(m_\text{stop})\sim 1/m_\text{stop}^6$.  For $m_\text{stop}\sim m_\text{top}$, the stop cross section is about $15\%$ of the $t\bar{t}$ cross section due to the additional spin states available for the spin $1/2$ top quark.  Around $m_\text{stop}\sim 600$ GeV, the $\tilde{t}\tilde{t}$ cross section is comparable to the irreducible SM $t\bar{t}+Z(\rightarrow\nu\bar{\nu})$ background.  The high mass stop pair production cross section increases more than $t\bar{t}$ between $\sqrt{s}=8$ and $\sqrt{s}=13$ TeV due to the relatively larger gain in parton luminosity at high momentum fraction.  However, event selections targeting stop production will enhance the high $m_{t\bar{t}},p_\text{T,$t\bar{t}$}$ tails, which for the same reason also get a larger increase with energy than the inclusive production.  This is illustrated in the right plot of Fig.~\ref{fig:stopcrosssection}, which shows that even though the inclusive $t\bar{t}$ cross section increases by a factor of about three, after requiring $E_\text{T}^\text{miss}> 300$ GeV the increase is by more than a factor of six.
		
\begin{figure}[h!]
\begin{center}
\includegraphics[width=0.55\textwidth]{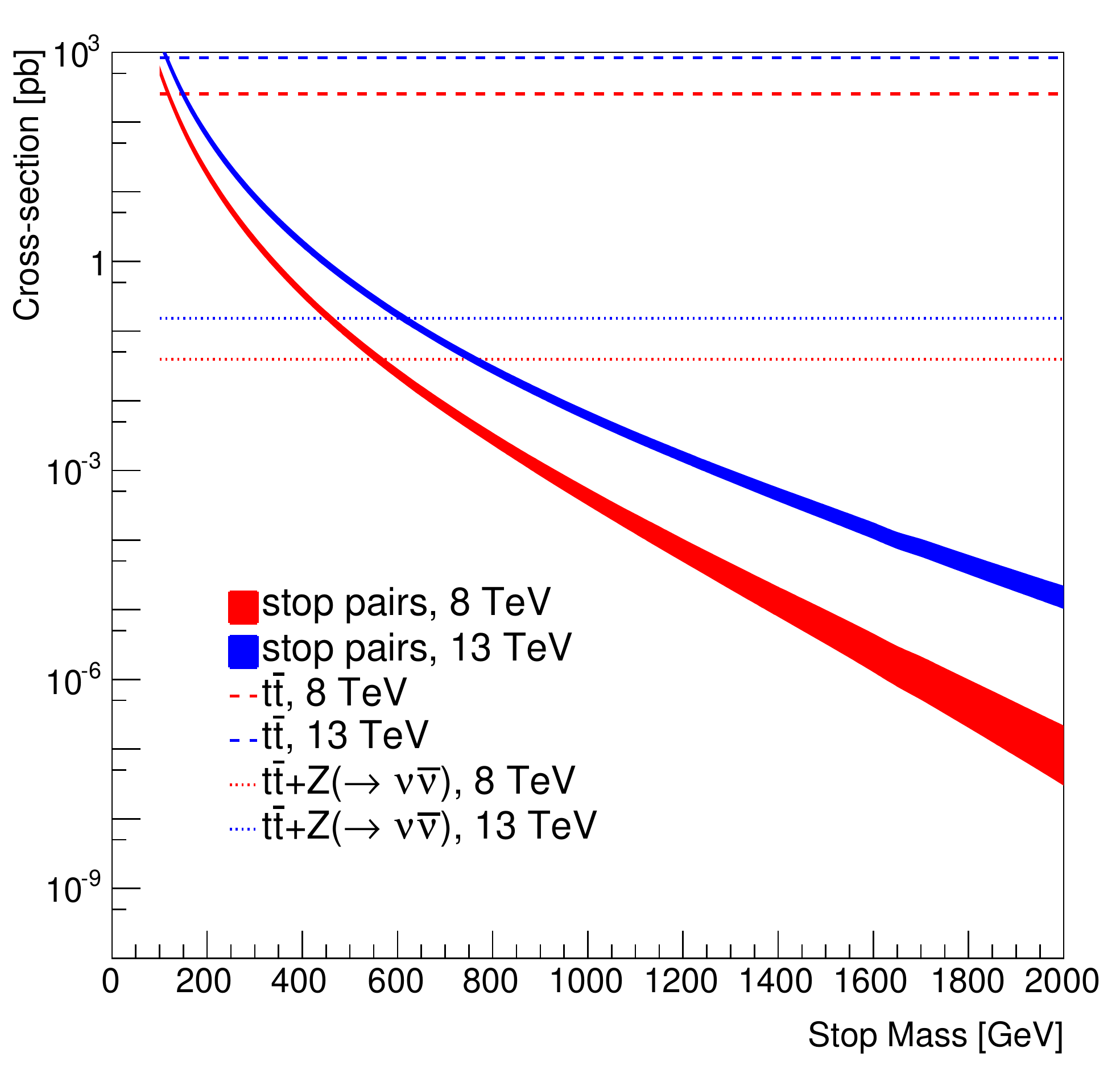}\includegraphics[width=0.41\textwidth]{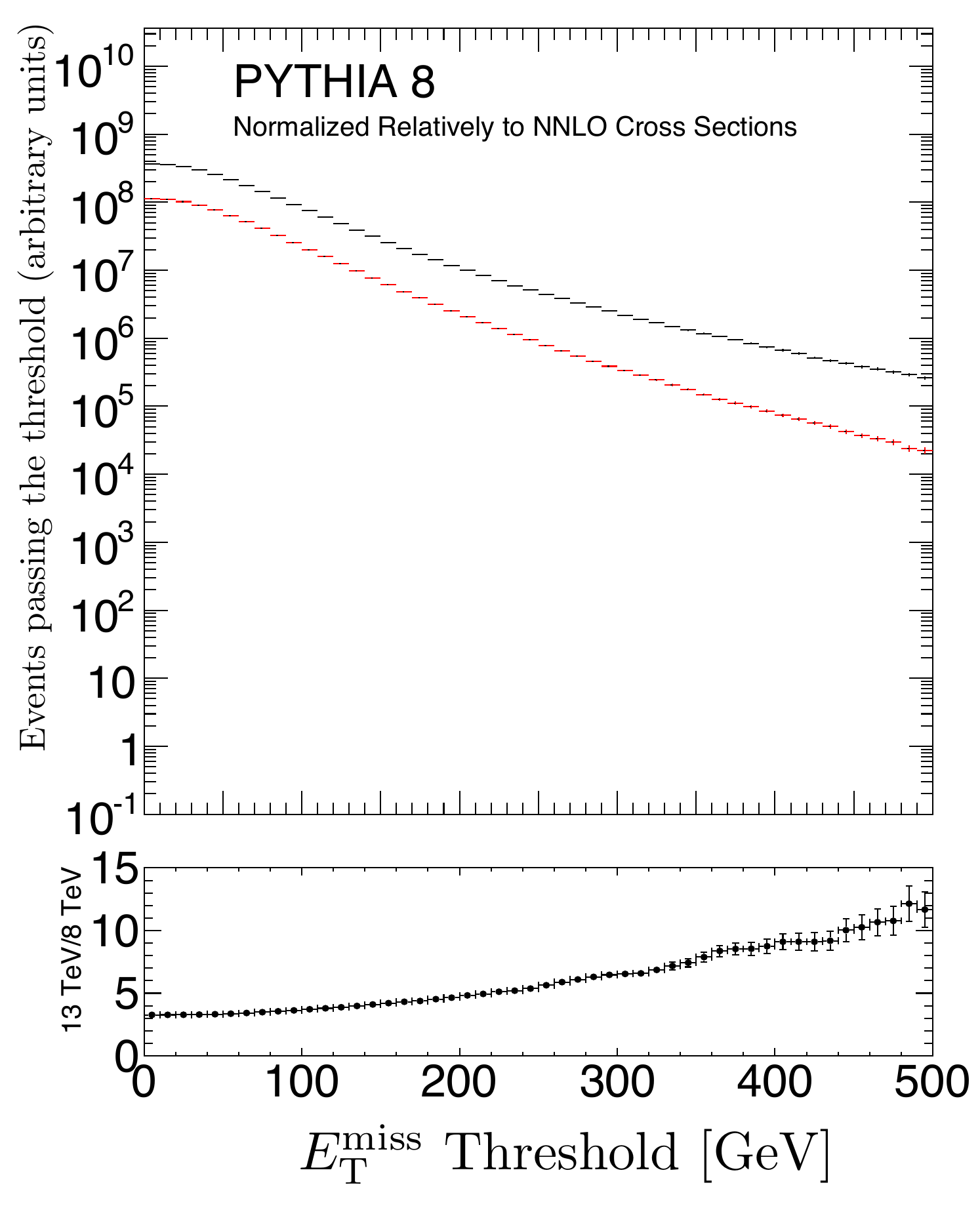}
\end{center}
\caption{Left: The stop pair production cross section at $\sqrt{s}=8$ TeV and $\sqrt{s}=13$ TeV as a function of the stop mass compared with important SM background processes $t\bar{t}$ and $t\bar{t}+Z(\rightarrow\nu\bar{\nu})$.  The SUSY cross section is calculated at NLO+NLL~\cite{Borschensky:2014cia,Kramer:2012bx} while the $t\bar{t}$ cross section has NNLO+NNLL accuracy using {\sc top++2.0}~\cite{Czakon:2011xx} with the PDF4LHC prescription~\cite{PDF4LHC}. The $t\bar{t}+Z$ cross section is computed at NLO from Ref.~\cite{Alwall:2014hca}. Right: the $t\bar{t}$ cross section as a function of the particle-level $E_\text{T}^\text{miss}$.}
\label{fig:stopcrosssection}
\end{figure}			
		
In addition to the stop mass, the other relevant mass scale is the neutralino mass, which sets how much phase space is available to the stop decay products for a fixed stop mass.  Fig.~\ref{fig:averagembffchi} shows the average invariant mass of the stop decay products for the decay $\tilde{t}\rightarrow b ff'\tilde{\chi}_1^0$.  Events are generated with a four-body phase-space and then re-weighted via

\begin{align}
\frac{1}{(m_{bff'}^2-m_t^2)^2+m_t^2\Gamma_t^2}\frac{1}{(m_{ff'}^2-m_W^2)^2+m_W^2\Gamma_W^2},
\end{align}			

\noindent where $m_t=175$ GeV, $\Gamma_t=1.3$ GeV, $m_W=80$ GeV and $\Gamma_t=2.5$ GeV.  The three stripes correspond to the on-shell ($m_\text{stop}>m_\text{top}+m_\text{LSP}$), three-body ($m_W+m_\text{LSP}<m_\text{stop}<m_\text{top}+m_\text{LSP}$), and four-body ($m_\text{stop}<m_W+m_\text{LSP}$) regions of parameter space.  Away from the on-shell region, the decay through a virtual top quark competes with the loop suppressed flavor changing neutral current process $\tilde{t}\rightarrow c\tilde{\chi}_1^0$ (could also be $\tilde{t}\rightarrow u\tilde{\chi}_1^0$).  Up until the CDF Tevatron Run II searches in Ref.~\cite{Aaltonen:2011rr,Aaltonen:2011na}, this was the only decay channel used for searching for flavor neutral stop decays.  The LEP experiments ruled out stops in this decay channel with $m_\text{stop}\lesssim 100$ GeV~\cite{Abbiendi:1999yz,Achard:2003ge,Abdallah:2003xe,Heister:2002hp} and the Tevatron experiments excluded these models for $m_\text{stop}\lesssim m_\text{t}$ and $m_\text{LSP}\lesssim m_\text{stop}-40$ GeV~\cite{Abazov:2008rc,Aaltonen:2012tq}.  At the LHC, the most powerful search strategies for this topology involve dedicated charm-jet tagging techniques~\cite{ATL-PHYS-PUB-2015-001} and the associated production of stops with initial state radiation (ISR) jets~\cite{Aad:2014nra}, excluding $m_\text{stop}$ up to $300$ GeV.  Traditionally, the $\tilde{t}\rightarrow c\tilde{\chi}_1^0$ decay mode was only considered in the four-body region of Fig.~\ref{fig:averagembffchi}.  However, the tradeoff between the two processes can be relevant all the way until the boundary of the on-shell top decay~\cite{Grober:2015fia}.  The off-shell regions of parameter space are briefly discussed in subsequent sections, but the remainder of Part~\ref{part:susy} will be focused on the on-shell regime, which gives rise to the $t\bar{t}+E_\text{T}^\text{miss}$ signature\footnote{The transition regions have finite width, so care is required when considering models in those regions where the phenomenology is rapidly changing.  The boundary region is discussed in the context of the search results in Sec.~\ref{chapter:results}.} from the simplified model shown in Fig.~\ref{fig:feynmanstopsimplifiedmodel}.
					
\begin{figure}[h!]
\begin{center}
\includegraphics[width=0.85\textwidth]{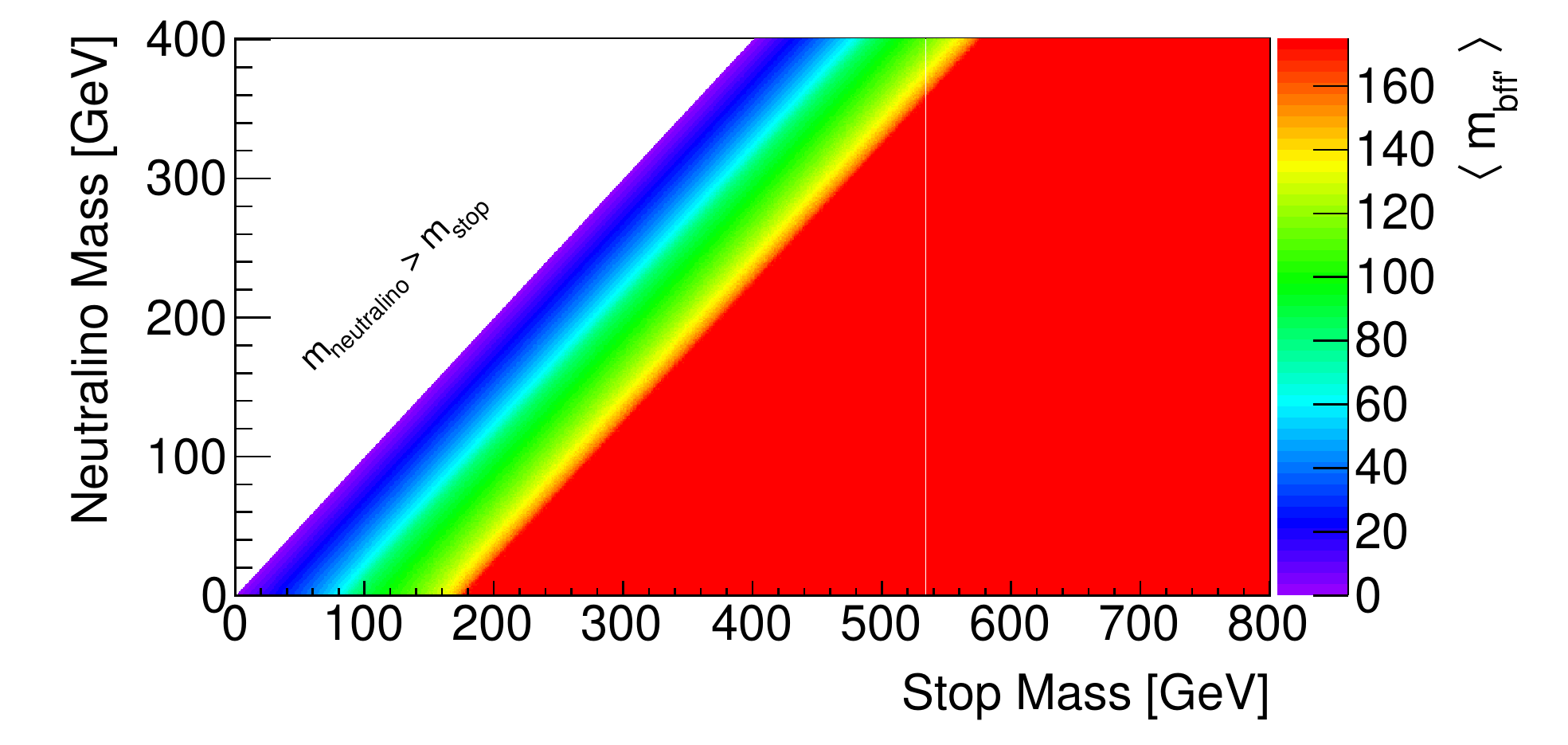}
\end{center}
\caption{The average invariant mass of the $b$-quark and SM fermions from the decay $\tilde{t}\rightarrow b ff'\tilde{\chi}_0^1$ as a function of the stop mass and neutralino mass. }
\label{fig:averagembffchi}
\end{figure}

\begin{figure}[h!]
\begin{center}
\includegraphics[width=0.5\textwidth]{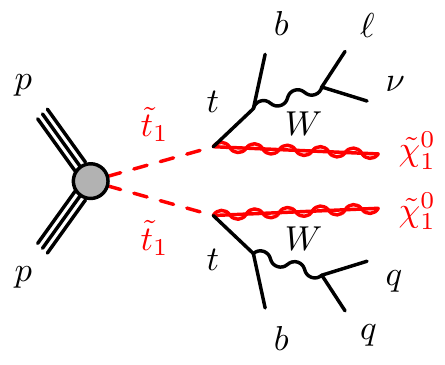}
\end{center}
\caption{A leading order Feynman diagram for the simplified stop model considered in Part~\ref{part:susy}.  The pair production of stops is followed by the subsequent decay to top quarks and neutralinos.  The top quark decays nearly 100\% of the time into a $W$ boson and a $b$-quark.  One of the $W$ bosons decays leptonically and the other decays hadronically (into quarks).}
\label{fig:feynmanstopsimplifiedmodel}
\end{figure}			

The kinematic properties of a stop decay event are determined by the momentum of the top quark and the neutralino, as illustrated by Fig.~\ref{fig:excll}.  Due to a steeply falling PDF, high mass stops are produced nearly at rest in the lab frame and so the magnitude of the top quark and neutralino momentum are nearly same in this frame.  For a given stop mass $M$ and neutralino mass $m$, this momentum is given by

\begin{align}
\label{eq:quartic}
p(M,m)=\sqrt{\frac{\left(M^2-(m_\text{top}-m)^2\right)\left(M^2-(m_\text{top}+m)^2\right)}{4M^2}}.
\end{align}

\noindent Figure~\ref{fig:acceptance} shows the top quark momentum as a function of the stop mass and neutralino mass using Eq.~\ref{eq:quartic}.  Over a large region of parameter space, the relative acceptance (using the top momentum as a proxy) is relatively constant and near $100\%$.  For $m_\text{LSP}\sim \frac{1}{2}m_\text{stop}+100$ GeV, there is a sharp transition where the acceptance drops to zero at the kinematic boundary $m_\text{stop}=m_\text{top}+m_\text{LSP}$.  The distributions of kinematic variables will be significantly different in this transition region compared with the `bulk' and therefore multiple event selections are required to maintain sensitivity across the full parameter space.  Changes in acceptance are combined with the falling cross section in Fig.~\ref{fig:stopcrosssection}.  Since the level curves of Fig.~\ref{fig:stopcrosssection} represent lines of constant signal yield, the sensitivity to stop models plotted in this plane should have the same form for a fixed signal region.

 \begin{figure}[h!]
 \begin{center}
  \includegraphics[width=0.5\textwidth]{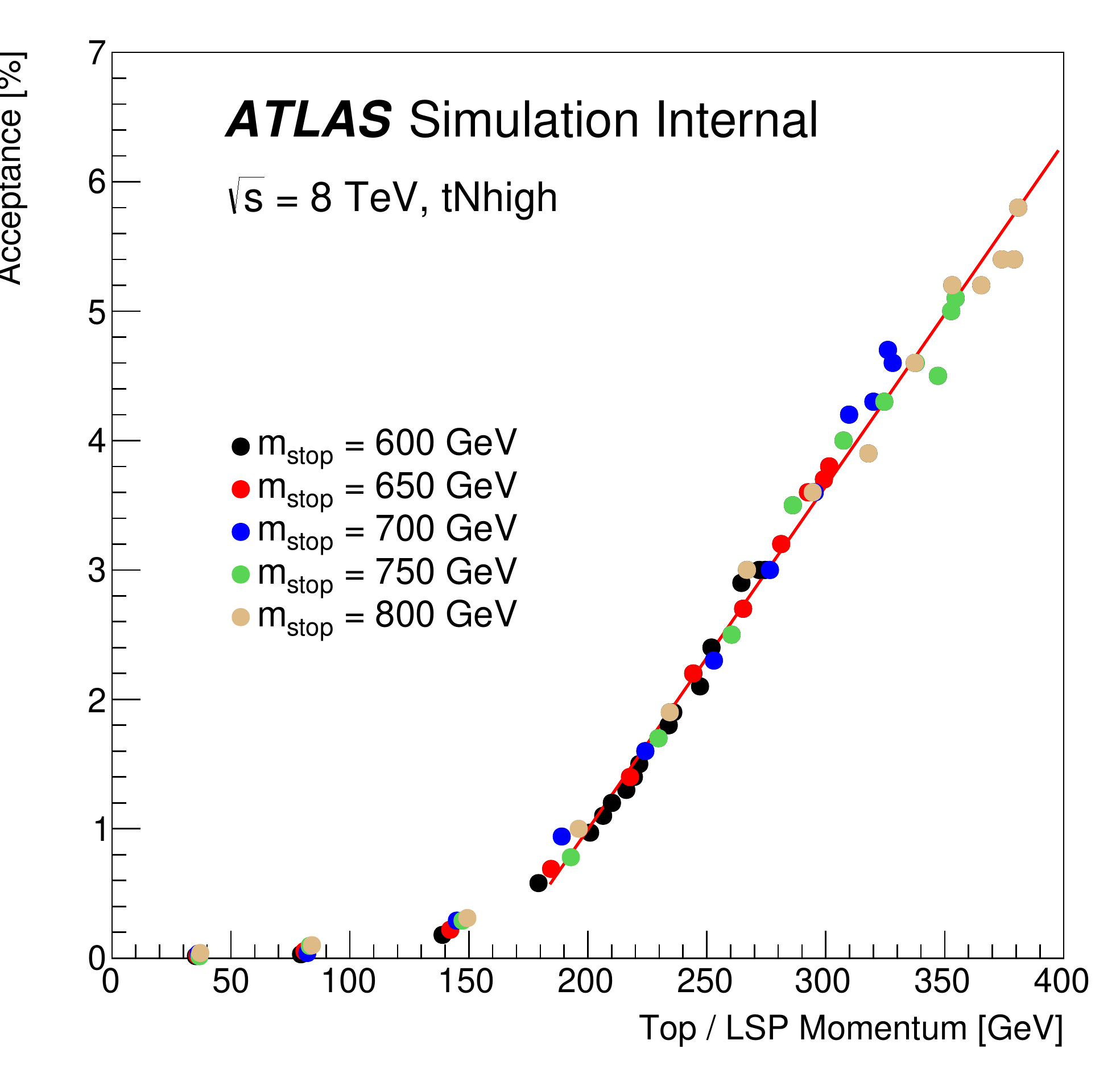}
 \caption{For the kinematically tightest event selection from the Run 1 stop search in the one-lepton channel (see Sec.~\ref{chapter:susy:signalregions}), the acceptance is plotted as a function of the top quark momentum given by Eq.~\ref{eq:quartic}.   The acceptance is defined as the fraction of simulated signal events that pass a particle-level version of the analysis (no detector simulation).  The ratio of the particle-level acceptance to the acceptance using the full detector simulation are all within a few percent of one in the relevant region and so can be safely ignored.  In the region beyond $600$ GeV, a the acceptance is well described by a straight line $\epsilon = 0.027\frac{\%}{\text{GeV}}\times p  -4.3\%$.  Every point with the same color corresponds to a model with the same stop mass.  The spread in the top/neutralino (LSP) momentum is due to the spread in the neutralino masses.}
 \label{fig:excll}
  \end{center}
 \end{figure}

\begin{figure}[h!]
\begin{center}
\includegraphics[width=0.85\textwidth]{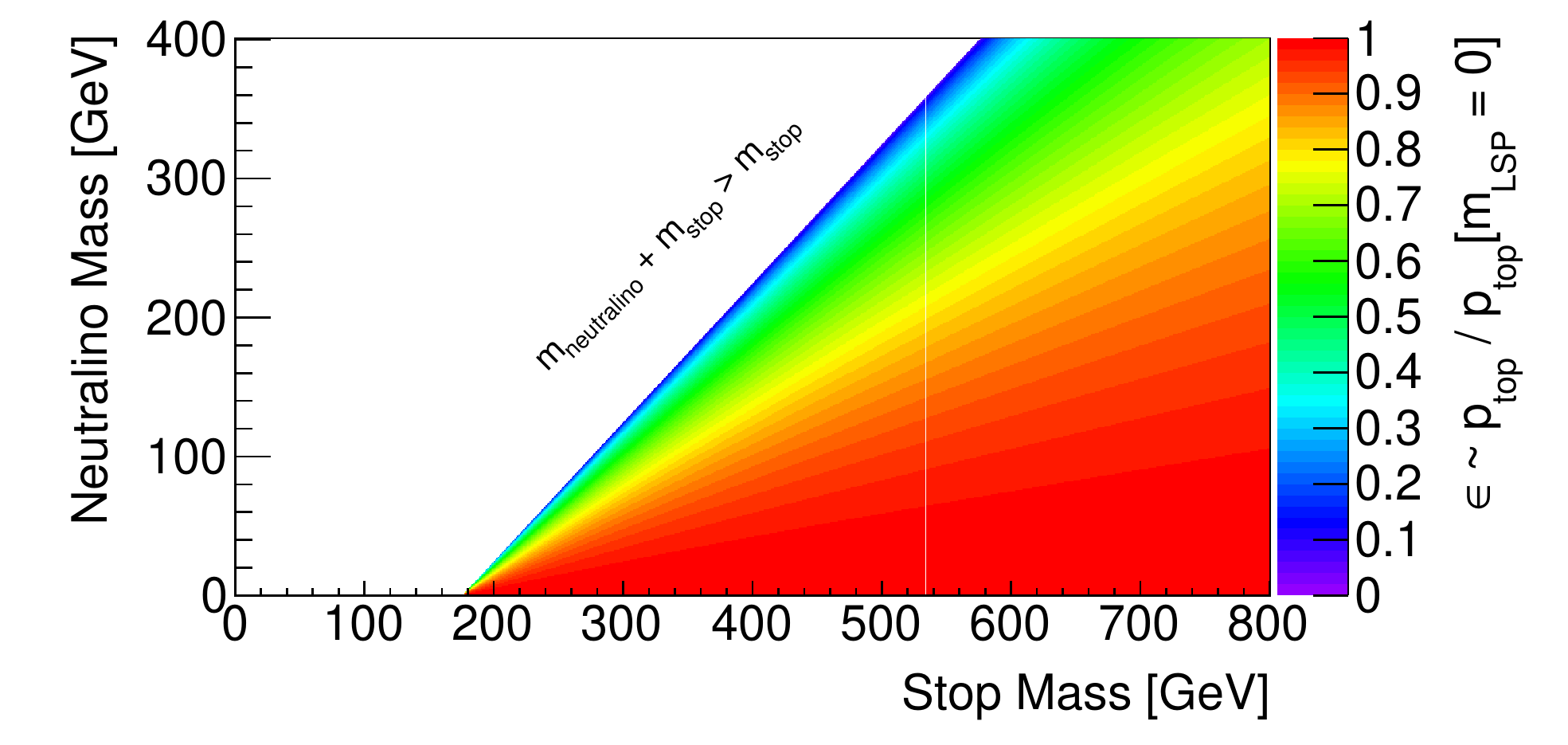}
\end{center}
\caption{The top (and neutralino) momentum given by Eq.~\ref{eq:quartic} as a function of the stop mass and neutralino mass.  The momentum is normalized to one at $m_\text{LSP}=0$ for a direct comparison of the acceptance different stop masses.}
\label{fig:acceptance}
\end{figure}

\begin{figure}[h!]
\begin{center}
\includegraphics[width=0.85\textwidth]{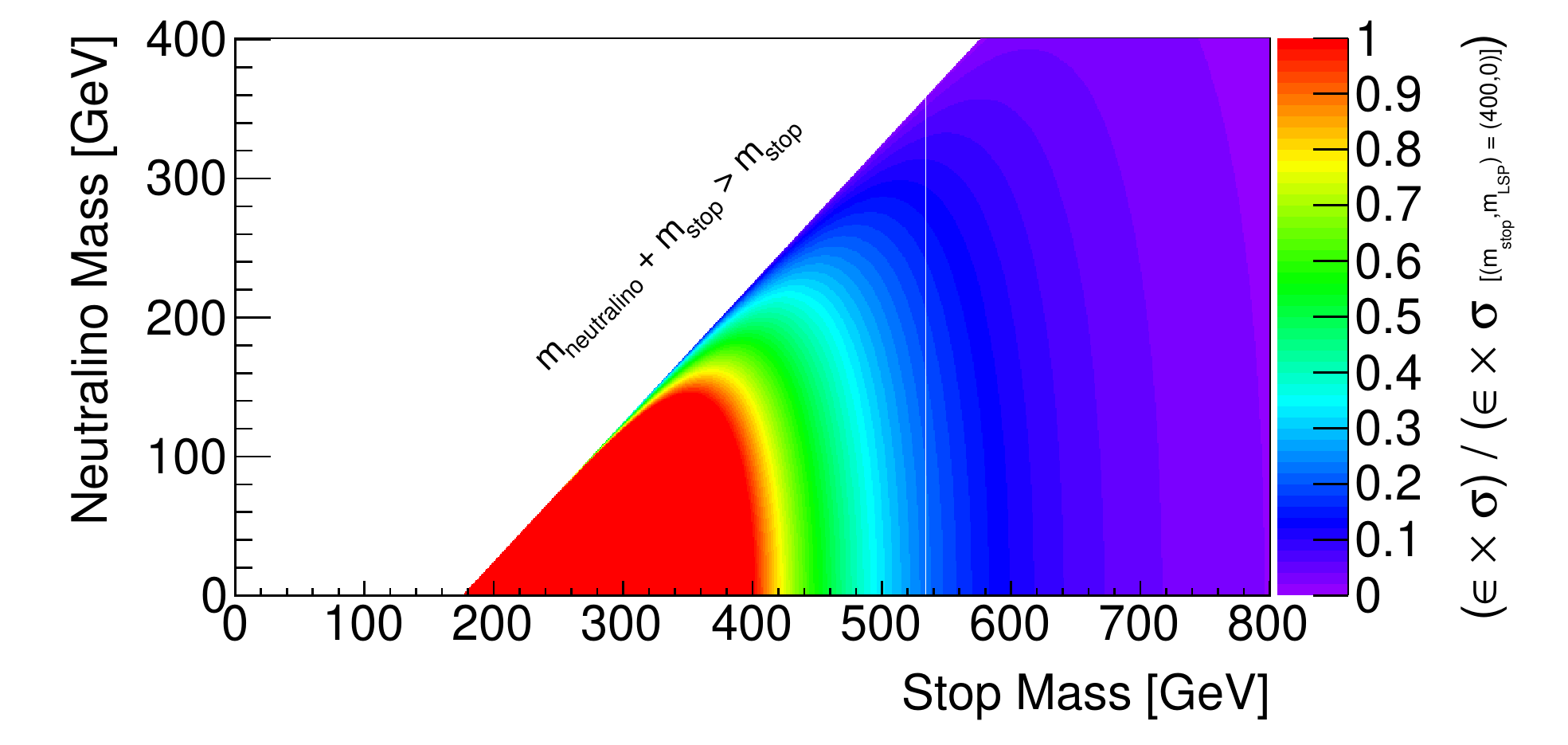}
\end{center}
\caption{A combination of the acceptance from Fig.~\ref{fig:acceptance} and the cross section from Fig.~\ref{fig:stopcrosssection}.  The $z$-axis is normalized to unity at $(m_\text{stop},m_\text{LSP})=(400,0)$.}
\label{fig:crosstimesacceptance}
\end{figure}
		
There is one other parameter needed to determine the kinematic properties of the stop decay products: the effective coupling to left- and right-handed top quarks $\theta_\text{eff}$ from Eq.~\ref{eq:effectivemixing}.  Before quantifying the impact of this effective coupling, Fig.~\ref{spinschematicstop} illustrates how it can impact the phenomenology.   For simplicity, suppose that all decays happen along a line.  The $b$-quark mass and the neutrino masses are negligibly small and so they always have left-handed helicity ($=$ chirality for massless particles).  Focusing on the left diagram of Fig.~\ref{spinschematicstop}, if the top quark is produced with a positive helicity (blue), then the fixed helically of the $b$-quark requires the $W$ to be longitudinally polarized.  In this case, there is no preferred momentum orientation (along the line) for the $W$ decay products.  However, if the top quark is produced with negative helicity, then only one configuration of $W$ boson decay products is allowed: the charged lepton must be going to the right in the $W$ boson rest frame.  This means that in the lab (stop) frame, the charged lepton will tend to have a softer momentum spectrum than the neutrino.   Following the same logic in the right diagram of Fig.~\ref{spinschematicstop} shows that for a (helicity) right-handed top, the charged lepton tends to be anti-parallel to the $W$ boson momentum but parallel to the top quark momentum.    The result is a slightly harder charged lepton momentum spectrum in the lab frame relative to the neutrino.  There are two other diagrams for all possible orientations of the momenta, but they are mirror images of the ones in Fig.~\ref{spinschematicstop} and give the same conclusions.  The overall conclusion is that (helicity) left-handed top quarks result in a softer charged lepton momentum spectrum while (helicity) right-handed top quarks result in a harder charged lepton momentum spectrum.  

\begin{figure}[h!]
\centering

\begin{tikzpicture}[line width=1.5 pt, scale=1.3, framed,background rectangle/.style={ultra thick, rounded corners=5pt, draw}]
	\everymath{\displaystyle}
	\node[] at (0,0) {$\tilde{t}$};
	\node[] at (-1,0) {$t$};
	\node[] at (1,0) {$\tilde{\chi}$};
	\node[] at (0.5,0) {$\rightarrow$};
	\node[] at (-0.5,0) {$\leftarrow$};
	
	\node[] at (-1,0.3) {{\color{blue}$\Leftarrow$}};

	\node[] at (-2,-1) {$W$};
	\node[] at (0,-1) {$b$};
	\node[] at (-0.5,-1) {$\rightarrow$};
	\node[] at (-1.5,-1) {$\leftarrow$};
	\node[] at (0,-1+0.3) {{\color{blue}$\Leftarrow$}};
	\node[] at (-2,-1+0.3) {{\color{blue}$0$}};

	\node[] at (-3,-2) {$l^+$};
	\node[] at (-1,-2) {$\nu$};
	\node[] at (-2.5,-2) {$\leftarrow$};
	\node[] at (-1.5,-2) {$\rightarrow$};
	\node[] at (-1,-2+0.3) {{\color{blue}$\Leftarrow$}};
	\node[] at (-3,-2+0.3) {{\color{blue}$\Rightarrow$}};

	\node[] at (-1,-3) {$l^+$};
	\node[] at (-3,-3) {$\nu$};
	\node[] at (-2.5,-3) {$\leftarrow$};
	\node[] at (-1.5,-3) {$\rightarrow$};
	\node[] at (-3,-3+0.3) {{\color{blue}$\Rightarrow$}};
	\node[] at (-1,-3+0.3) {{\color{blue}$\Leftarrow$}};
 \begin{scope}[shift={(5,0)}]
 
 \node[] at (0,0) {$\tilde{t}$};
	\node[] at (-1,0) {$t$};
	\node[] at (1,0) {$\tilde{\chi}$};
	\node[] at (0.5,0) {$\rightarrow$};
	\node[] at (-0.5,0) {$\leftarrow$};
	
	\node[] at (-1,0.3) {{\color{blue}$\Leftarrow$}};

	\node[] at (-2,-1) {$b$};
	\node[] at (0,-1) {$W$};
	\node[] at (-0.5,-1) {$\rightarrow$};
	\node[] at (-1.5,-1) {$\leftarrow$};
	\node[] at (-2,-1+0.3) {{\color{blue}$\Rightarrow$}};
	\node[] at (0,-1+0.3) {{\color{blue}$\Longleftarrow$}};

	\node[] at (-1,-2) {$l^+$};
	\node[] at (1,-2) {$\nu$};
	\node[] at (-0.5,-2) {$\leftarrow$};
	\node[] at (0.5,-2) {$\rightarrow$};
	\node[] at (1,-2+0.3) {{\color{blue}$\Leftarrow$}};
	\node[] at (-1,-2+0.3) {{\color{blue}$\Leftarrow$}};
 
 \end{scope}

	\end{tikzpicture}

\vspace{3mm}
	
\begin{tikzpicture}[line width=1.5 pt, scale=1.3, framed,background rectangle/.style={ultra thick, rounded corners=5pt, draw}]
	\everymath{\displaystyle}
	\node[] at (0,0) {$\tilde{t}$};
	\node[] at (-1,0) {$t$};
	\node[] at (1,0) {$\tilde{\chi}$};
	\node[] at (0.5,0) {$\rightarrow$};
	\node[] at (-0.5,0) {$\leftarrow$};
	
	\node[] at (-1,0.3) {{\color{red}$\Rightarrow$}};

	\node[] at (-2,-1) {$W$};
	\node[] at (0,-1) {$b$};
	\node[] at (-0.5,-1) {$\rightarrow$};
	\node[] at (-1.5,-1) {$\leftarrow$};
	\node[] at (0,-1+0.3) {{\color{red}$\Leftarrow$}};
	\node[] at (-2,-1+0.3) {{\color{red}$\Longrightarrow$}};

	\node[] at (-1,-3) {$l^+$};
	\node[] at (-3,-3) {$\nu$};
	\node[] at (-2.5,-3) {$\leftarrow$};
	\node[] at (-1.5,-3) {$\rightarrow$};
	\node[] at (-3,-3+0.3) {{\color{red}$\Rightarrow$}};
	\node[] at (-1,-3+0.3) {{\color{red}$\Rightarrow$}};
 \begin{scope}[shift={(5,0)}]
 
 \node[] at (0,0) {$\tilde{t}$};
	\node[] at (-1,0) {$t$};
	\node[] at (1,0) {$\tilde{\chi}$};
	\node[] at (0.5,0) {$\rightarrow$};
	\node[] at (-0.5,0) {$\leftarrow$};
	
	\node[] at (-1,0.3) {{\color{red}$\Rightarrow$}};

	\node[] at (-2,-1) {$b$};
	\node[] at (0,-1) {$W$};
	\node[] at (-0.5,-1) {$\rightarrow$};
	\node[] at (-1.5,-1) {$\leftarrow$};
	\node[] at (-2,-1+0.3) {{\color{red}$\Rightarrow$}};
	\node[] at (0,-1+0.3) {{\color{red}$0$}};

	\node[] at (-1,-2) {$l^+$};
	\node[] at (1,-2) {$\nu$};
	\node[] at (-0.5,-2) {$\leftarrow$};
	\node[] at (0.5,-2) {$\rightarrow$};
	\node[] at (1,-2+0.3) {{\color{red}$\Leftarrow$}};
	\node[] at (-1,-2+0.3) {{\color{red}$\Rightarrow$}};

	\node[] at (1,-3) {$l^+$};
	\node[] at (-1,-3) {$\nu$};
	\node[] at (-0.5,-3) {$\leftarrow$};
	\node[] at (0.5,-3) {$\rightarrow$};
	\node[] at (-1,-3+0.3) {{\color{red}$\Rightarrow$}};
	\node[] at (1,-3+0.3) {{\color{red}$\Leftarrow$}};
 
 \end{scope}

	\end{tikzpicture}

\caption{Diagrams indicating the various spin configurations of the stop decay products.  Single black arrow indicate the direction of the momentum in the rest frame of the particle above the origin of the arrows and double-lined colored arrows indicate the production of the spin along the momentum direction.  There are two other possible collinear diagrams per panel where the top begins moving to the right, but the conclusions are the same as for these two.}
\label{spinschematicstop}
\end{figure}

Chiral left-handed top quarks tend to result in helicity left-handed top quarks and vice versa~\cite{Belanger:2012tm}.  For a bino LSP, chiral right-handed stop\footnote{The stop is spin 0, so this is short-hand for the superpartner of the chiral right-handed top quark.  The helicity is with respect to the lab frame; for massive particles, one can always find a frame in which the sign of the helicity is reversed.} result in chiral right-handed top quarks and vice versa since $U(1)$ connects particles of like-chirality.  In contrast, since the higgs coupling is like the mass terms in the Lagrangian which couples left to right chiral states, for a higgsino LSP, chiral right-handed stops result in chiral left-handed top quarks.  The impact on the distribution of the lepton kinematics can be quantified using the effective mixing angle from Eq.~\ref{eq:effectivemixing}~\cite{Perelstein:2008zt}: 
		
\begin{align}
\label{leptondirection}
\frac{dN}{d\cos\theta_l} \propto E_{\tilde{\chi}_1^0}^\text{top frame}+2\sin(\theta_\text{eff})m_{\tilde{\chi}_1^0}+p_{\tilde{\chi}_1^0}^\text{top frame}\cos(2\theta_\text{eff})\cos\theta_l,
\end{align}		
		
\noindent where $\theta_l$ is the angle between the lepton and the neutralino (from the same stop decay) in the top quark rest frame, $E_{\tilde{\chi}_1^0}^\text{top frame}$ is the energy of the neutralino in the top quark rest frame and $p_{\tilde{\chi}_1^0}^\text{top frame}$ is the momentum of the neutralino in this frame.  Equation~\ref{leptondirection} shows that the behavior of the lepton angle in the top quark rest frame is determined by the quantity $\cos(2\theta_\text{eff})$.  Figure~\ref{fig:effectiveonstop} uses Eq.~\ref{eq:effectivemixing} to show how $\cos(2\theta_\text{eff})$ depends on the stop mixing angle $\theta_{\tilde{t}}$ for a pure bino LSP ($N_{11}=1$, $N_{1i}=0$, $i=2,3,4$) and a pure higgsino LSP ($N_{14}=1$, $N_{1i}=0$, $i=1,2,3$).  As expected from the previous description\footnote{Note that the graphs in this plot are $-1$ times the ones appearing in Ref.~\cite{Perelstein:2008zt}.  See the text for the explanation why the form given here agrees with the expectation of the coupling structure.}; for a purely chiral stop (i.e. $\tilde{t}_1=\tilde{t}_L$ or $\tilde{t}_1=\tilde{t}_R$), the top chirality will be the same as the stop for a bino LSP and the opposite for a higgsino LSP.
	
\begin{figure}[h!]
\begin{center}
\includegraphics[width=0.4\textwidth]{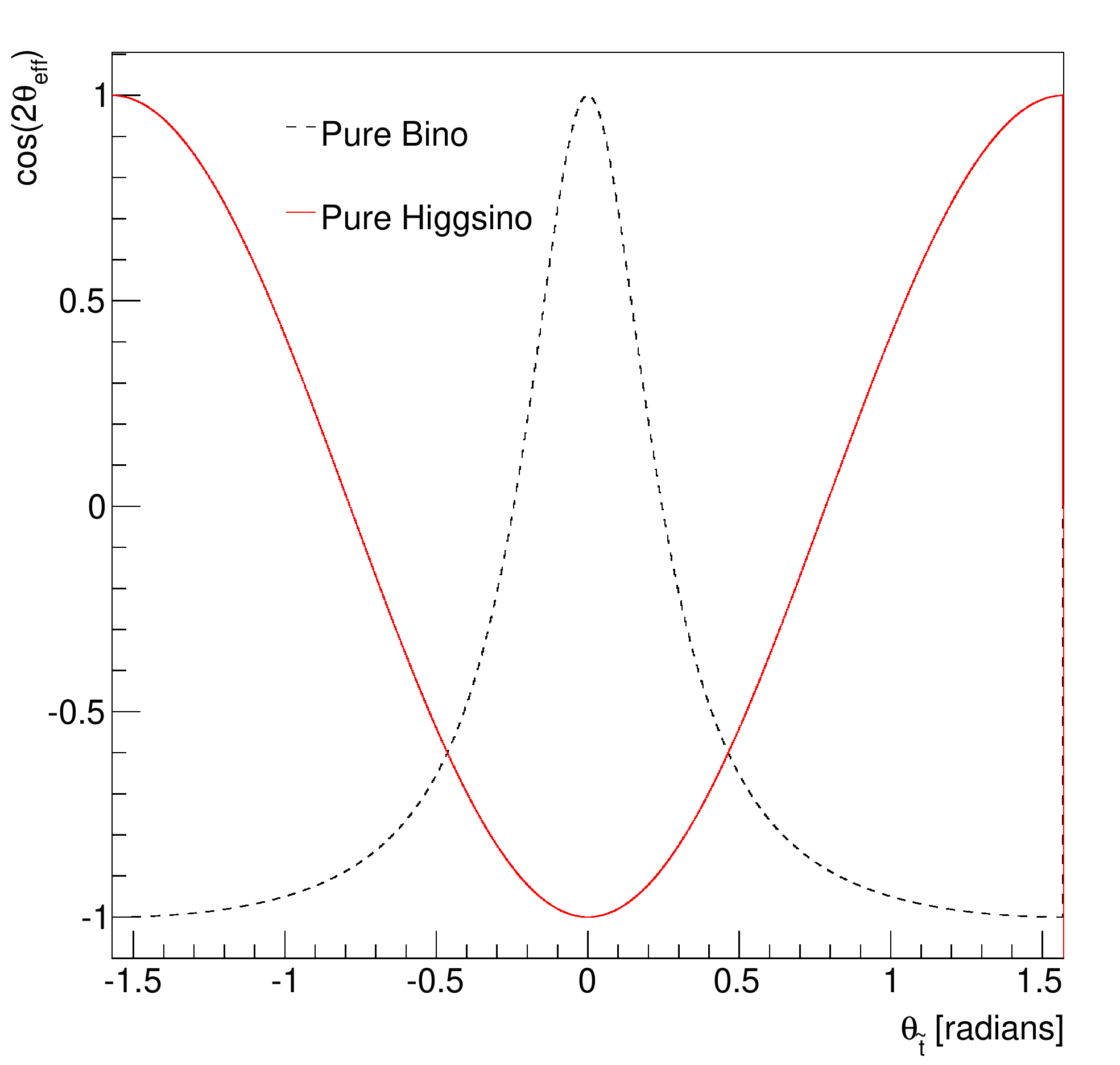}
\end{center}
\caption{The dependence of $\cos(\theta_\text{eff})$ on the stop mixing angle for a pure bino (black dashed) and a pure higgsino (red) LSP.}
\label{fig:effectiveonstop}
\end{figure}
	
The impact of the effective mixing angle $\theta_\text{eff}$ on the lepton $p_\text{T}$ spectrum is shown in Fig.~\ref{fig:effectivemixing}.  As expected from the simple description around Fig.~\ref{spinschematicstop}, the charged lepton $p_\text{T}$ spectrum is harder in the case of mostly right-handed top quarks ($\theta_\text{eff}=\pi/2$) than for mostly left-handed top quarks $(\theta_\text{eff}=0)$.  One way to quantify the impact of the change in the $p_\text{T}$ spectrum is to evaluate the efficiency for a fixed lepton $p_\text{T}$ requirement as a function of stop mass.  Figure~\ref{fig:polarizationonplane} shows the ratio of efficiencies between the mostly right- and mostly left-handed top quark configurations for a $p_\text{T}>25$ GeV threshold on the charged lepton momentum\footnote{The top quark polarization impacts the other decay products as well, but the effect is largest for the lepton $p_\text{T}$ because it is further down the decay chain compared to the $b$-quark and additional sources of jets can mitigate the impact from the hadronically decaying $W$ boson.}.  Over most of the $(m_\text{stop},m_\text{LSP})$ plane, the change in acceptance is about $25\%$.  Near the $m_\text{stop}\sim m_\text{LSP}$ diagonal, the impact of the polarization is less because the coefficient of the $\cos(2\theta_\text{eff})$ term in Eq.~\ref{leptondirection} is suppressed by the reduced phase space.  In the limit  $m_\text{stop}\rightarrow m_\text{LSP}$, $dN/d\cos\theta$ is constant, independent of $\theta_\text{eff}$.  Figure~\ref{fig:reductioninlimit} combines information from Fig.~\ref{fig:polarizationonplane} with the stop cross section in Fig.~\ref{fig:stopcrosssection} to estimate the reduction in the expected sensitivity for the extreme values of $\theta_\text{eff}$.  For a stop with $\theta_\text{eff}=\pi/2$ and $m_\text{stop}=500$ GeV, the number of predicted events is comparable to a stop with $\theta_\text{eff}=0$ and $m_\text{stop}\sim 480$ GeV, resulting in a predicted loss in sensitivity of about $20$ GeV.  The actual impact in the limit will be discussed in Sec.~\ref{8TeVresults} with the results.
	
\begin{figure}[h!]
\begin{center}
\includegraphics[width=0.5\textwidth]{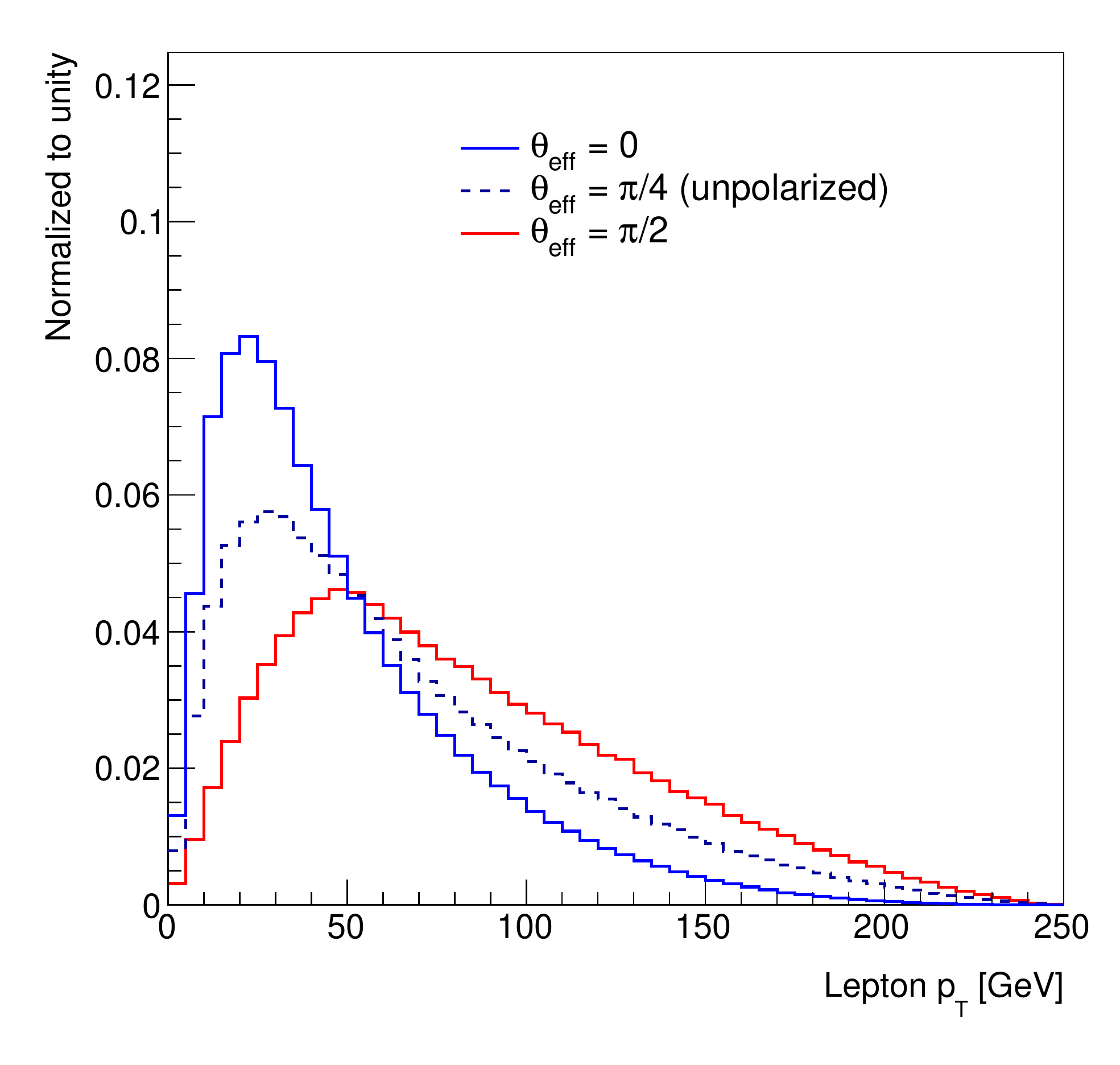}
\end{center}
\caption{The charged lepton $p_\text{T}$ spectrum in stop decays for three values of $\theta_\text{eff}$ for $(m_\text{stop},m_\text{LSP})=(500,0)$.}
\label{fig:effectivemixing}
\end{figure}	

\begin{figure}[h!]
\begin{center}
\includegraphics[width=0.85\textwidth]{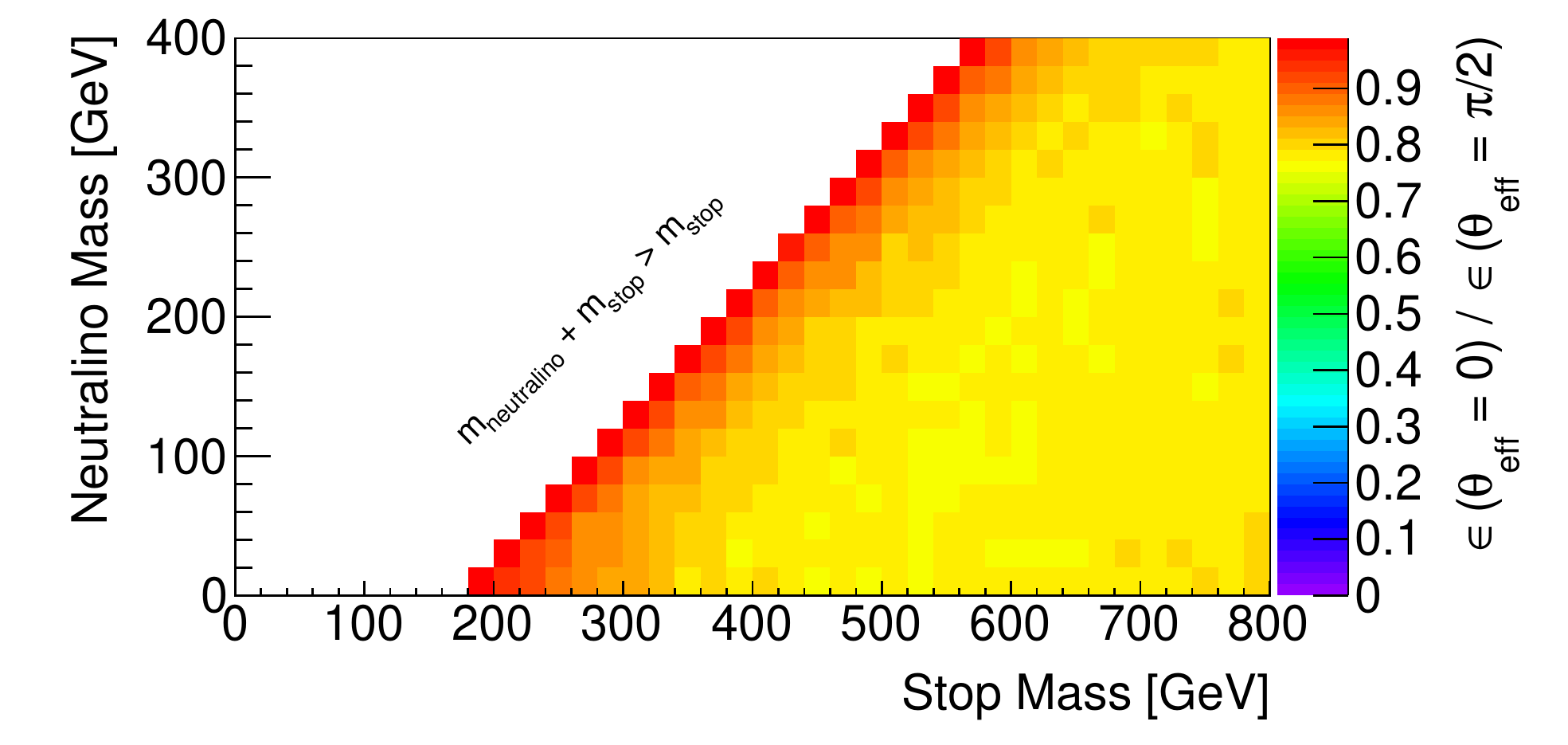}
\end{center}
\caption{The ratio of the efficiency for a $p_\text{T}^\text{charged lepton}>25$ GeV requirement between the mostly left-handed top quark ($\theta_\text{eff}=0$) and the mostly right-handed top quark ($\theta_\text{eff}=\pi/2$).}
\label{fig:polarizationonplane}
\end{figure}	

\begin{figure}[h!]
\begin{center}
\includegraphics[width=0.5\textwidth]{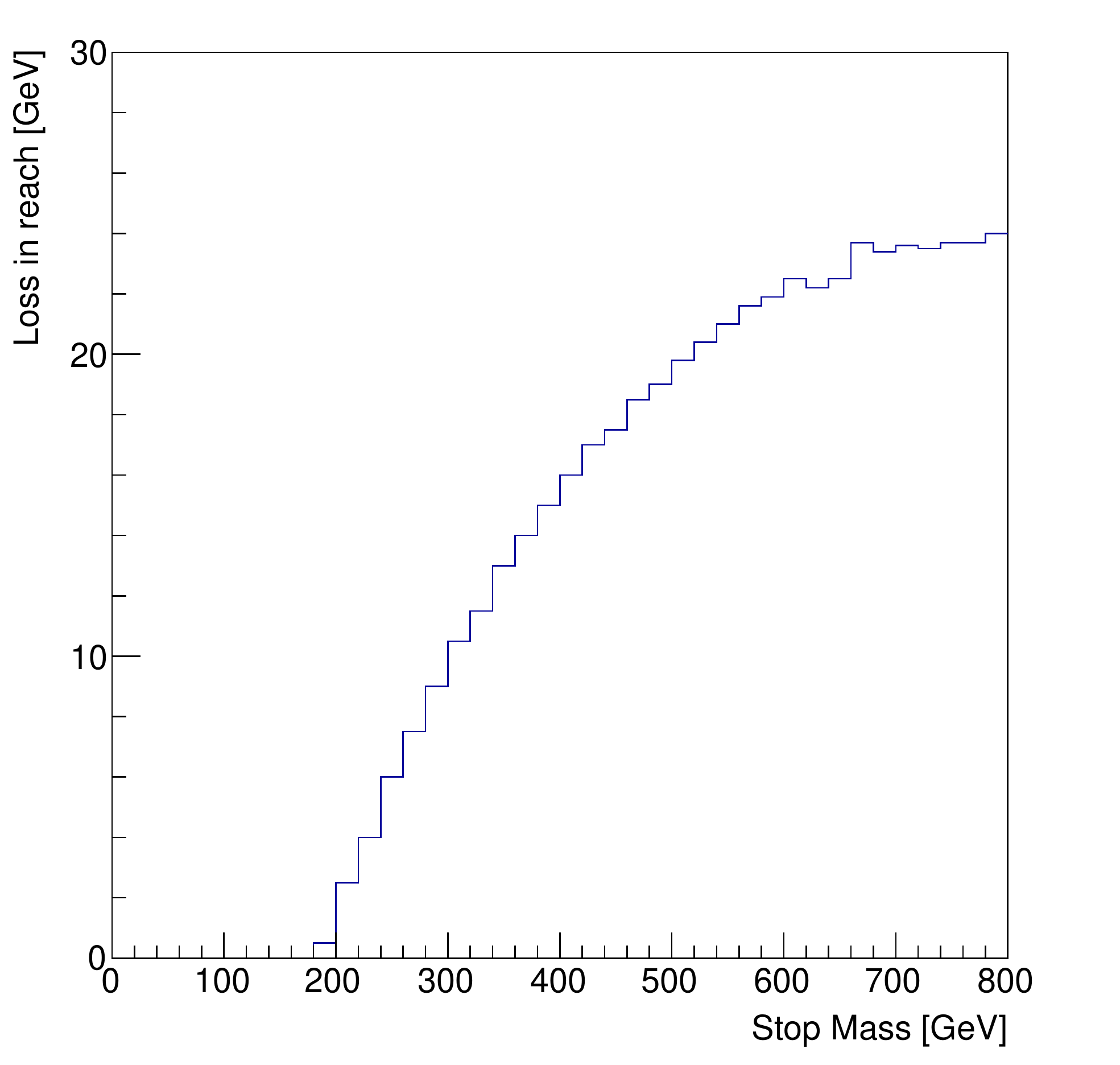}
\end{center}
\caption{For a fixed stop mass $m_{\text{stop},R}$ (LSP is massless), plotted is $m_{\text{stop},R}-m_{\text{stop},L}$ for $\sigma(m_{\text{stop},L})\times\epsilon_L=\sigma(m_{\text{stop},R})\times\epsilon_R$.  The value $m_{\text{stop},R}$ is the stop mass for $\theta_\text{eff}=\pi/2$ and $m_{\text{stop},L}$ is the stop mass for  $\theta_\text{eff}=0$, so the plot shows how much the stop mass needs to be reduced to have the same predicted number of events with a most left-handed top quark configuration compared to a given mostly right-handed top quark configuration.}
\label{fig:reductioninlimit}
\end{figure}	

\clearpage

For high mass $m_\text{stop}\gtrsim 200$ GeV and $m_\text{LSP}$ sufficiently far away the diagonal $m_\text{stop}=m_\text{top}+m_\text{LSP}$, the cross section $\times$ acceptance map in Fig.~\ref{fig:crosstimesacceptance} coupled with the top polarization completely categorizes the properties of the signal necessary for the search.  For these models, the strategy is to develop event selections estimated to have a high purity and a high yield of stop events for a particular {\it benchmark model}.  The reach of each selection will then be determined by the discussion above and the choice of benchmark models is set by the goal to cover a wide range of the ($m_\text{stop},m_\text{LSP})$ parameter space.  The stop signal is characterized by a large $\vec{p}_\text{T}^\text{miss}$ from the neutralinos, one isolated charged lepton and at least four high $p_\text{T}$ jets resulting from the tree-level top quarks shown in Fig.~\ref{fig:feynmanstopsimplifiedmodel}.  Two of these jets are expected to originate from $b$-quarks.  The construction of discriminating variable and their use in event selections are described in Chapters~\ref{chapter:susy:variables} and~\ref{chapter:susy:signalregions}, respectively.

When $m_\text{LSP}$ is near $m_\text{stop}-m_\text{top}$, the neutralinos and top quarks have very little momentum in the stop rest frame and so the signal $E_\text{T}^\text{miss}$ and lepton/jet momentum spectra start to approach the distributions of the $t\bar{t}$ background.  For relatively low $m_\text{stop}$, the signal cross section is sufficiently large so that one can take advantage of subtle differences in the shapes of various kinematic distributions.  Section~\ref{sec:shapefitsetup} describes how the single event selection paradigm can be modified to incorporate shape information.  Additional strategies are possible in this regime, and are likely needed for the sensitivity to cross the $m_\text{LSP}=m_\text{stop}-m_\text{top}$ limit.  These include requiring additional high $p_\text{T}$ (ISR) jets to boost the neutralino momentum~\cite{Hagiwara:2013tva,An:2015uwa,Macaluso:2015wja,Cheng:2016mcw} and forward jets in a vector-boson-fusion (VBF) topology~\cite{Dutta:2013gga}.

For $m_\text{stop}\lesssim 200$ GeV, the stop cross section is so high that top quark properties measurements are sensitive to the presence of a light stop.   Recent studies have suggested that the $t\bar{t}$ cross section~\cite{atlasxs,theoryxs} and spin-correlation measurements between the leptons from top quark decays~\cite{Aad:2014mfk} are sensitive to stops.  Exploiting precision measurements of the $t\bar{t}$ cross-section makes use of the current NNLO+NNLL accuracy that reduces the theoretical uncertainty in $\sigma_{t\bar{t}}$ to about 5\%, which is sensitive to the $\mathcal{O}(10\%)$ contribution of a degenerate stop.  The angular distribution between the two leptons from top quarks encodes information about the production mechanism and correlations between the top quark spins.  Like top quarks, stops are also produced mostly via gluon-gluon fusion at the LHC.  However, because stops are scalar particles, there is no direct correlation between the spins of the resulting top quarks.  Both ATLAS and CMS have performed measurements to constrain light stop models using the total cross section~\cite{atlasxs,Aad:2015pfx,Khachatryan:2016mqs} and spin correlations~\cite{Aad:2014mfk}, ruling out stop models with $m_\text{stop}$ between $150$ GeV and $190$ GeV.  A non-negligible ($\sim30\%$) contribution to the sensitivity for the spin correlation measurement is due to the constraint from the $t\bar{t}$ cross-section.  

Despite these innovative efforts to constrain degenerate $m_\text{stop}\sim m_\text{top}$ case, there is an important challenge with this regime that can obscure the results.  Measurements which exploit the cross section could be effected by a bias in the top quark mass measurement due the presence of a light stop.  In particular, since $\sigma_{t\bar{t}}$ increases with decreasing top quark mass, a negative shift in the measured top quark mass would increase the predicted $t\bar{t}$ cross-section and could hide the additional contribution to the measured cross-section from direct stop pair production.  Figure~\ref{fig:mjjj} illustrates the invariant mass distribution of the three jets associated to the hadronically decaying (off-shell) top quark in $pp\rightarrow t\bar{t}$ and $pp\rightarrow \tilde{t}_1\tilde{t}_1^*$ where the other (off-shell) top quark decays into $W^\pm\rightarrow l^\pm\nu$.  The invariant mass distribution is sensitive to the top quark mass and thus can be used to extract the mass from data.  When $m_\text{stop}$ is just below $m_\text{top}$, the distribution of $m_{jjj}$ is biased toward lower values.  This is also true even when $m_\text{stop}$ is just above $m_\text{top}$ because the top quark Breit-Wigner is skewed to lower values ($m_\text{top}$ must be less than $m_\text{stop}$).  The presence of these stop events could bias the top quark mass measurement to low values, if the {\it calibration curve} is derived using only $t\bar{t}$ simulation (right plot of Fig.~\ref{fig:mjjj}).  In particular, the measured values of the top quark mass and $t\bar{t}$ cross section are given by

\begin{align}
\label{eq:solve}
m_t^\text{measured} &= \frac{\langle m_{jjj}\rangle_{\tilde{t}\tilde{t}}  \times \sigma_{\tilde{t}\tilde{t}}(m_\tone)\times \epsilon+ \langle m_{jjj}\rangle_{t\bar{t}}(m_t) \times \sigma_{t\bar{t}}(m_t) }{ c_1(\sigma_{\tilde{t}\tilde{t}}(m_\tone) \times \epsilon + \sigma_{t\bar{t}}(m_t))}-\frac{c_0}{c_1}\notag\\
\sigma_\text{$t\bar{t}$}^\text{measured} &= \sigma_{\tilde{t}\tilde{t}}(m_\tone )\times \epsilon + \sigma_{t\bar{t}}(m_t),
\end{align}
	
\noindent where $\epsilon$ is the ratio of the SUSY acceptance to the $t\bar{t}$ acceptance and $c_0,c_1$ are the slope and intercept from the calibration curve in Fig.~\ref{fig:mjjj}, respectively.   For example, a stop with $m_{\tilde{t}} \sim 170$ GeV that decays via an off-shell top quark together with a true top quark mass of about $175$ GeV would cause
a bias in the top quark mass that makes it compatible with the measurements with Run 1 of the LHC (LHC8). As a consequence, the predicted $t\bar{t}$ cross-section would be over-estimated by about $16$ pb which in turn would make it much harder to find the stop with a cross-section of about $43$ pb (which is further reduced to about $60$\% since the acceptance is lower than for $t\bar{t}$). 
The cross-section over-estimation increases with the true top quark mass, while the compatibility of the measured top quark mass with the LHC8 decreases when going beyond about $175$ GeV.  Figure~\ref{fig:summary} and Table~\ref{tab:mass_shift} summarize how the change in the measured mass could hide such a {\it sneaky stop}\footnote{The analysis in this section has been published in Ref.~\cite{Eifert:2014kea}.}.  The impact for $m_\text{stop}>m_\text{top}$ is greatly reduced because the top quarks have nearly the same mass distribution as SM $t\bar{t}$ production.
	
\begin{figure}[h!]
\begin{center}
\includegraphics[width=0.5\textwidth]{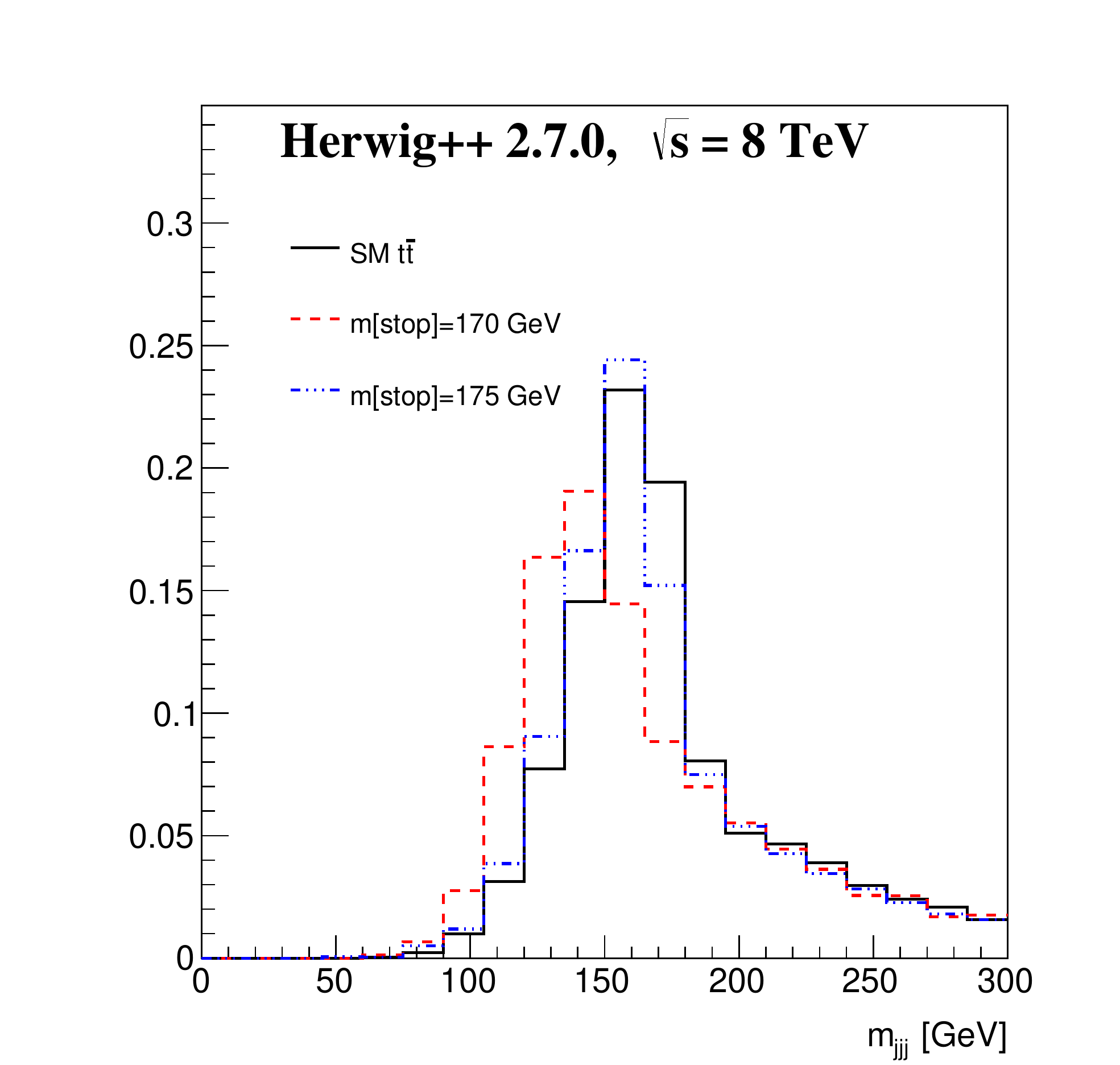}\includegraphics[width=0.5\textwidth]{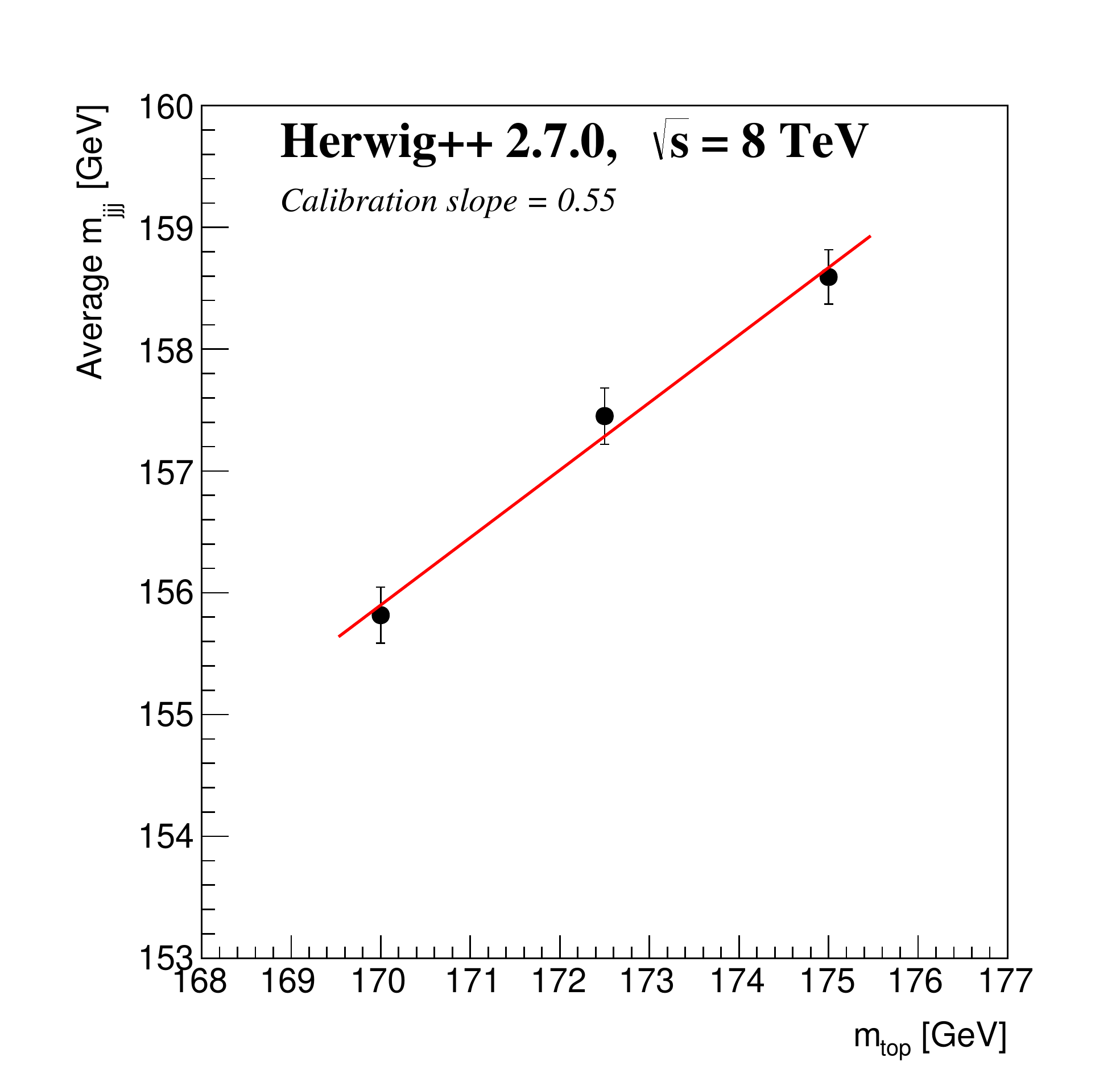}
\end{center}
\caption{Left: Unit normalized distributions of the $m_{jjj}$ variable for $t\bar{t}$ with $m_t=172.5$ GeV, and for $\tilde{t}$ pair production with a two-body $\tilde{t}\rightarrow tN$ decay with $m_{\tilde{t}}=175$ GeV (and $m_t=172.5$ GeV), and a three-body decay $\tilde{t}\rightarrow bWN$ for $m_{\tilde{t}}=170$ GeV.  The neutralino is assumed massless.  Jets are assigned to the hadronically decaying top quark by minimizing $\chi^2=(m_{j_1j_2b_1}-m_{b_2l\nu})^2/(20\text{ GeV})^2+(m_{j_1j_2}-m_W)^2/(10\text{ GeV})^2$ for $\{j_i\}$ the set of jets not identified as originating from a $b$-quark.  Even though the neutrino $p_z$ is unmeasured, it can be inferred by solving $m_{l\nu}=m_W$.  The simulations are performed using {\sc Herwig++ 2.7}~\cite{Bellm:2013hwb,Bahr:2008pv} and analyzed using the {\sc Rivet 1.8.2} framework~\cite{rivet} with {\sc Fastjet 3.0.6}~\cite{Cacciari:2011ma} for clustering anti-$k_t$ jets with $R=0.4$~\cite{Cacciari:2008gp}.  Right:  Calibration curve that relates the measured value $\langle m_{jjj}\rangle$ to the (MC) top quark mass, $m_{top}$ in $t\bar{t}$ events.  See Ref.~\cite{Eifert:2014kea} for more detail.}
\label{fig:mjjj}
\end{figure}	

\begin{figure}
\begin{center}
\includegraphics[width=0.8\textwidth]{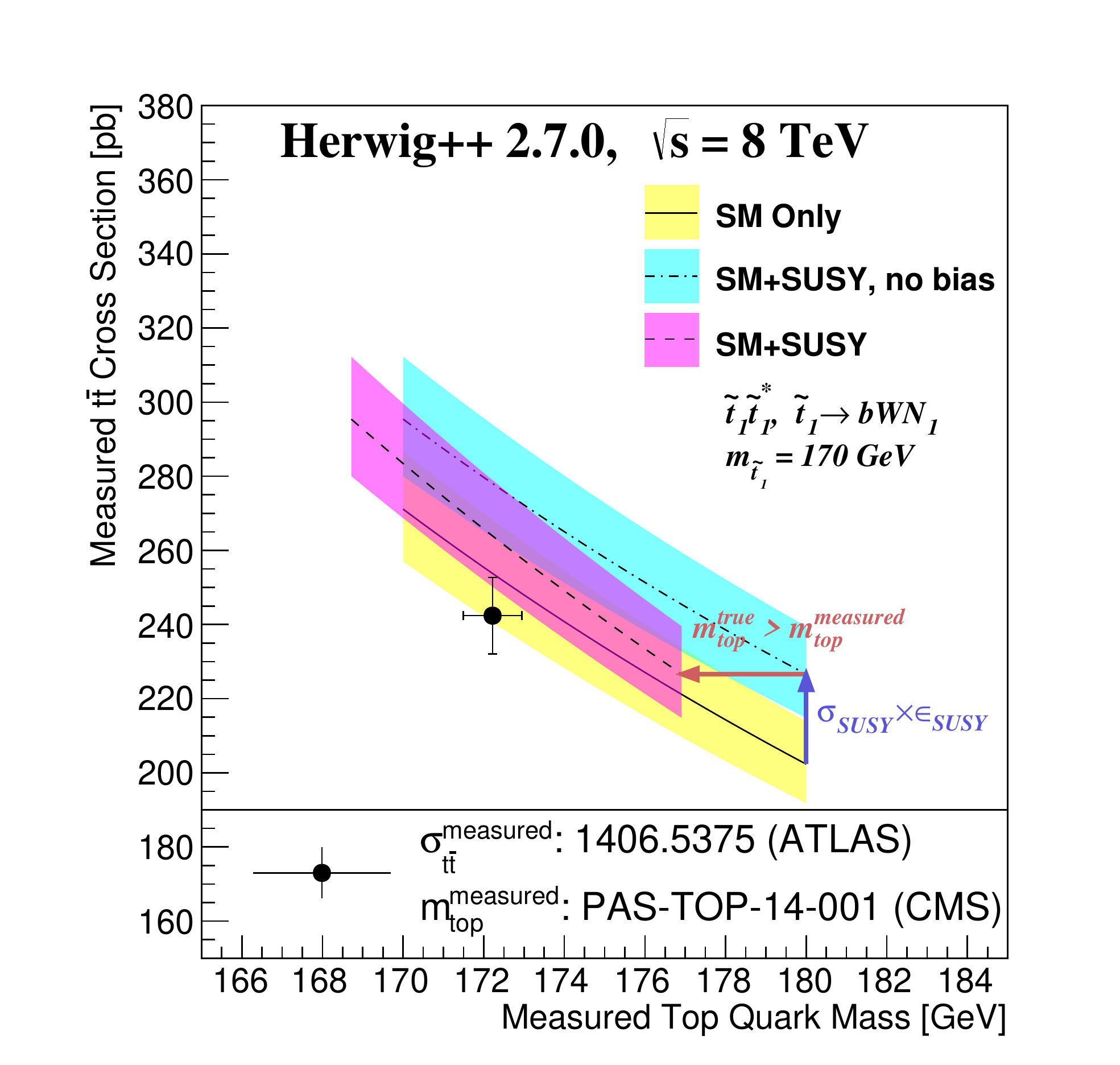}
\end{center}
\caption{
Summary of the effects leading to the sneaky stop scenario: the shifts in the measured $t\bar{t}$ cross-section and measured top quark mass.  The solid line corresponds to an unbiased measurement of the $t\bar{t}$ cross-section as a function of the top quark mass.  The dot-dashed line is what would be measured in the presence of a $\tilde{t}\rightarrow bWN$ with $m_{\tilde{t}}=170$ GeV for an unbiased top quark mass measurement.  However, under the SM+SUSY hypothesis the top quark mass measurement would be {\it biased} which translates into what would actually be observed shown in the dashed line.  For all three lines, the band reflects the $\sim 5-6\%$ theory uncertainty on the cross-section.  For comparison, the measured top quark mass and $t\bar{t}$ cross-section are shown from recent CMS~\cite{cmsmass} and ATLAS~\cite{atlasxs} results.  
}
\label{fig:summary}
\end{figure}	
	
\begin{table}
  \centering
\noindent\adjustbox{max width=\textwidth}{
{\small
\begin{tabular}{| c |c|c |c |c| c| c| c| c| c|c|}
\hline
$m_t^\text{true}$ & \multicolumn{2}{c|}{$m_t^\text{measured}$} & \multicolumn{2}{c|}{True $\sigma_{t\bar{t}}(m_t^\text{true})$}  & \multicolumn{2}{c|}{True $\sigma_{t\bar{t}}(m_t^\text{measured})$}&\multicolumn{2}{c|}{True $\sigma_{\tilde{t}\tilde{t}}$}& \multicolumn{2}{c|}{Measured $\sigma_{t\bar{t}}$} \\
\hline
	& LHC8 & Tevatron & LHC8 & Tevatron & LHC8 & Tevatron & LHC8 & Tevatron &LHC8 & Tevatron  \\
  \hline
  \hline
  170 & 168.6 & 169.0 & 271.1  & 8.0 & 279.0 & 8.1& 42.6 & 0.87  & 295.4&8.5\\
  172.5 & 170.8 & 171.3  & 251.7 & 7.3& 264.4&7.6& 42.6 & 0.87&276.0 &7.8\\
  175 & 172.9 & 173.5  & 233.8 &6.8 & 249.7&7.2& 42.6 & 0.87&258.1 &7.3\\
  \hline
\end{tabular}}}
\caption{
Bias in the measured top quark mass and $t\bar{t}$ cross-section due to the presence of a light stop ($m_{\tilde{t}} = 170$ GeV) that decays via the three-body process.
All masses are in GeV and all cross-sections are in pb.  The measured top quark mass is biased low from the true mass which results in the true cross-section at the measured top mass,  true $\sigma_{t\bar{t}}(m_t^\text{measured})$ to be higher than the true cross-section at the true mass, true $\sigma_{t\bar{t}}(m_t^\text{true})$. The former quantity is what would be predicted under the SM-only hypothesis in the presence of the 170 GeV stop.  The measured $\sigma_{t\bar{t}}$ is the sum of true $\sigma_{t\bar{t}}(m_t^\text{true})$ and true $\sigma_{\tilde{t}\tilde{t}}$, corrected for the lower acceptance for the three-body decay.}
\label{tab:mass_shift}
\end{table}	
	
\begin{figure}
\begin{center}
\includegraphics[width=0.5\textwidth]{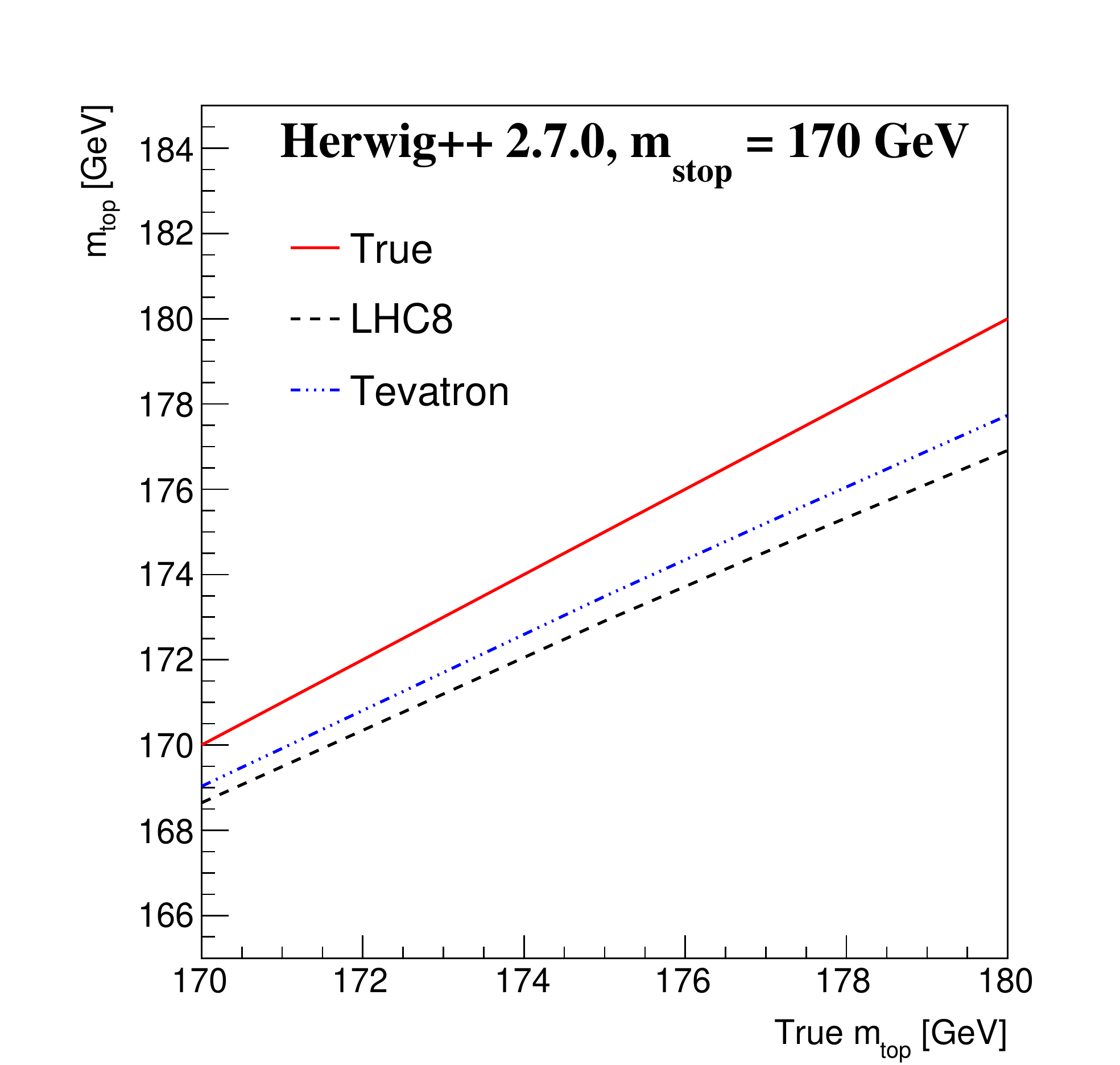}
\end{center}
\caption{The measured top quark mass as a function of the true top quark mass. The bias in the measurement arises from the presence of a light $\tilde{t}$ with $m_{\tilde{t}}=170$ GeV and decaying via the three-body process (left) or with $m_{\tilde{t}}=175$ GeV and decaying via the two-body process.   
}
\label{fig:mass_shift}
\end{figure}	
	
Due to the relatively lower stop cross section, the shift in the measured top quark mass is predicted to be smaller at the Tevatron compared to the LHC.  Figure~\ref{fig:mass_shift} shows the size of the shift as a function of the true top quark mass between the two colliders.  Interestingly, there is a small tension between measured top quark mass values between the Tevatron and LHC experiments that is in the correct direction predicted by a light stop.  However, all of the most precise measurements use a calibration scheme involving the {\it Monte Carlo} mass, which is related to a well-defined QFT top quark mass only within ambiguities of $\mathcal{O}{(\Lambda_\text{QCD})}$ and the relation may depend on $\sqrt{s}$; see e.g. ref.~\cite{Moch:2014lka}.  Since a wide range of simulation schemes where various MC mass definitions are used, this tension is not a significant indication of deviations from the SM.

The mass measurement based on $\langle m_{jjj}\rangle$ is highly simplified from the state-of-the-art.  However, prompted by the above argument, a detailed study using the most precise ATLAS top quark mass measurement technique~\cite{Aad:2015nba} indicates that the impact on the stop limits could be as much as $5$ GeV~\cite{Aad:2015pfx}.  As the LHC accumulates more data at $\sqrt{s}=13$ TeV and systematic uncertainties are reduced, cross section and other top quark properties should continue to be explited to ensure that no stop is hiding around $m_\text{stop}\lesssim m_\text{top}$.

The remainder of Part~\ref{part:susy} will focus on the direct search for a light stop in the one lepton + four jets + missing momentum final state using the {\it control region method}, described in Sec.~\ref{sec:CRmethod}.
	
	\clearpage
	
\section{The Control Region Method}
	\label{sec:CRmethod}

	In order to identify stop events among an overwhelming SM background, key variables are identified for which the probability distribution significantly differs between signal and background.  Figure~\ref{fig:CRschematic} illustrates how these variables are used to estimate, validate, and test the background predictions.  Each background process is separately estimated, but for the sake of simplicity, suppose that there is one type of SM background and one powerful variable $V$.  Examples of $V$ appear in Sec.~\ref{chapter:susy:variables}, but typically $V$ is associated with an energy scale in the event and the likelihood $p_S(V)/p_B(V)$ monotonically increases as a function of $V$.  A {\it signal region} (SR) is an interval of $V$ predicted to have low background and high signal yield.  As the likelihood is often monotonically increasing, these regions usually take the form $[v_0,\infty)$ for some fixed $v_0$.  The goal is compare the number of predicted signal events to the number of predicted background events in the SR.  If the Poisson fluctuations in the background are (much) larger than the predicted signal yield, then this is a hopeless exercise.  For this reason, the signal region is usually defined by $v_0\gg 1$ (with the appropriate units, often GeV) where $p_S(V)/p_B(V)\gtrsim 1$.  The shapes $p_S(V)$ and $p_B(V)$ are obtained from MC simulation.  The number of predicted events in the signal region\footnote{The background events in the signal region are also used for the final result, but have little influence due to the small total yield.  See Sec.~\ref{sec:susy:stats} for details.} is then given by $N_B\int_\text{SR} p_S(V)dV$ and $N_S\int_\text{SR} p_S(V)dV$ for the background and signal, respectively.  The factor $N_S$ is given by $\mathcal{L}_\text{int}\times\sigma\times\epsilon$, where $\mathcal{L}_\text{int}$ is the integrated luminosity, $\sigma$ is the cross section calculated for the signal, and $\epsilon$ is the efficiency of all event selections prior to the selection on $V$ (also estimated from simulation).  In contrast, $N_B$ is defined such that $N_B\int_\text{CR} p_B(V)dV$ agrees with the observed data in a {\it control region} (CR), which is an interval of $V$ where the signal is expected to be negligible compared with the background.  By normalizing the background in the control region, the predicted number of events at low values of $V$ is `correct' (see Sec.~\ref{drawbacks} for caveats) and the only uncertainty is due to extrapolating this prediction to the SR using $p_B(V)$.  The control region should be kinematically close to the SR in order to reduce any uncertainty from this extrapolation, but should also be loose enough so that the total number of events in the control region allows for a relatively precise measurement of $N_B$.  Often the requirement for higher event yields in the CR results in a significant gap between the CR and the SR.  Part of the region in between where $p_S(V)/p_B(V)$ is still small can be used to validate the CR prediction, albeit with a significant uncertainty.  Such a region is called a {\it validation region}.
	
	In practice, many variables are combined to form the signal and control regions and there are a variety of background processes.  Section~\ref{sec:singlebin} presents an overview of of the signal regions defined by a single set of kinematic requirements ({\it single bin} SR) and Sec.~\ref{sec:shapefitsetup} describes the setup when there are multiple SR bins used at the same time.  The control region method is a powerful tool for exploring regions of phase space that are not well constrained by previous measurements.  However, it does have some limitations (Sec.~\ref{drawbacks}) which are important to understand in the exploration of the TeV scale and beyond.
		
\begin{figure}[h!]
\begin{center}
\includegraphics[width=0.99\textwidth]{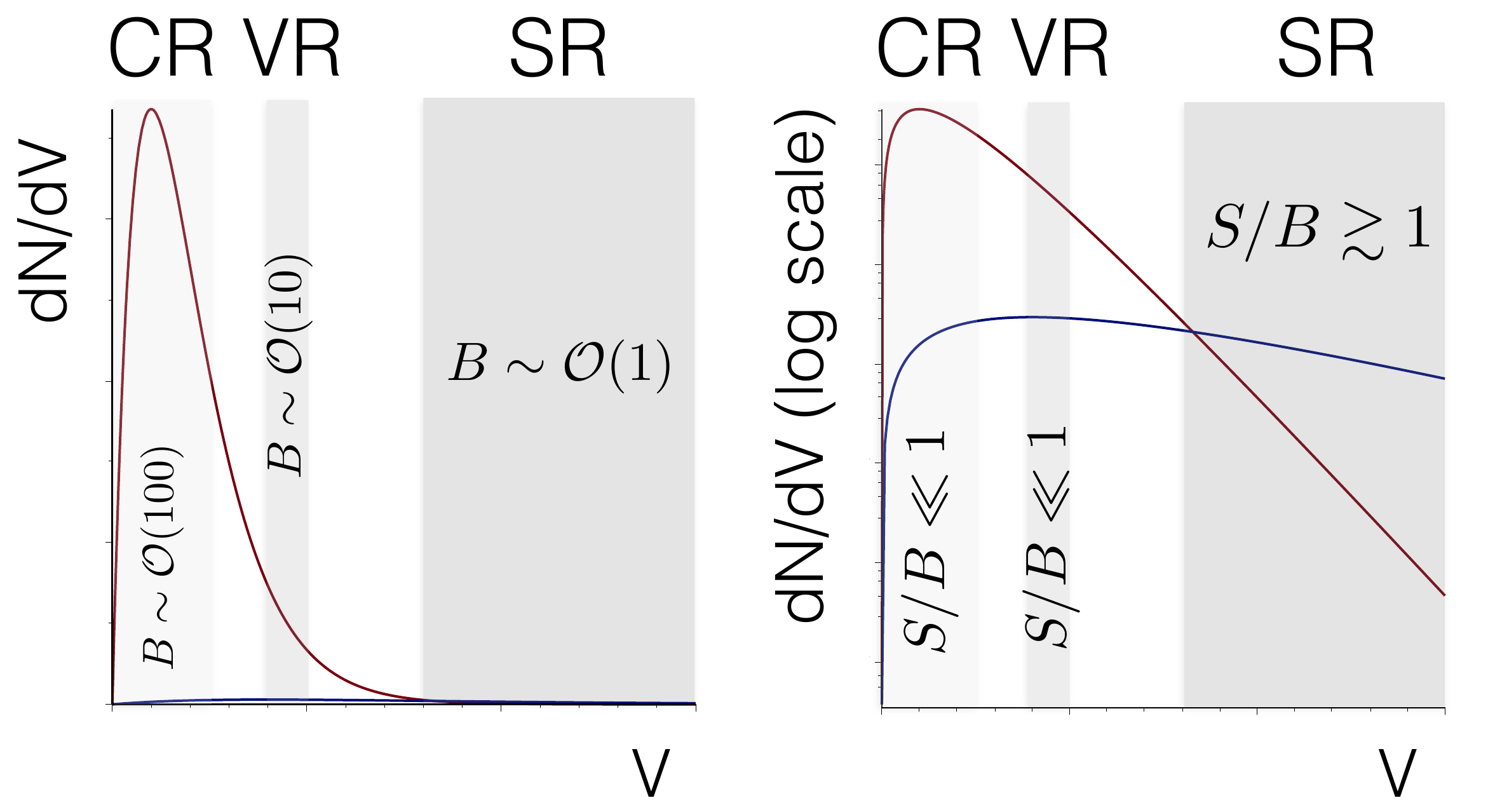}
 \caption{A schematic diagram to illustrate the control region method.  The red distribution represents $N_Bp_B(V)$ and the blue is $N_Sp_S(V)$ (see the text for details).  The left and right distributions are identical, only with a logarithmic scale for the vertical axis in the right plot. Control, validation, and signal regions are denoted by CR, VR, and SR, respectively.  The symbol $S$ is shorthand for $N_Sp_S(V)$ and $B$ represents $N_Bp_V(B)$.}
 \label{fig:CRschematic}
  \end{center}
\end{figure}		
		
\clearpage	
		
\subsection{Single Bin Signal Regions}
\label{sec:singlebin}			
			
In the most basic and widely used form of the control region method, there is one SR and multiple control regions that constrain various background processes.  All but one of the signal regions for Part~\ref{part:susy} have this structure.   Figure~\ref{fig:CRschematic} illustrates the setup using the early $\sqrt{s}=13$ TeV analysis as an example.  A variety of kinematic variables are used to define a signal region.  Two of these variables, in this case called $m_\text{T}$ (Sec.~\ref{sec:transmass}) and $am_\text{T2}$ (Sec.~\ref{sec:mt2}), are changed to form control regions that are disjoint from to the SR and to each other.  In addition to the kinematic requirements, the number of $b$-jets is a powerful tool for building control regions.  Requirements on other kinematic variables may also be loosened in order to increase the CR statistics, but the background composition is determined by a few key variables.  All of these control regions are described in more detail in Chapter~\ref{chapter:background}.  In Fig.~\ref{fig:CRschematic}, there are four control regions (TCR, WCR, STCR, TZCR) and in between these regions and the SR are three validation regions (TVR, WVR, and WVR-tail).  	Not all control regions have validation regions and some regions (e.g. WCR) can have multiple validation regions to probe different aspects of the extrapolation from the CR to the SR.
			
\begin{figure}[h!]
\begin{center}
\includegraphics[width=0.5\textwidth]{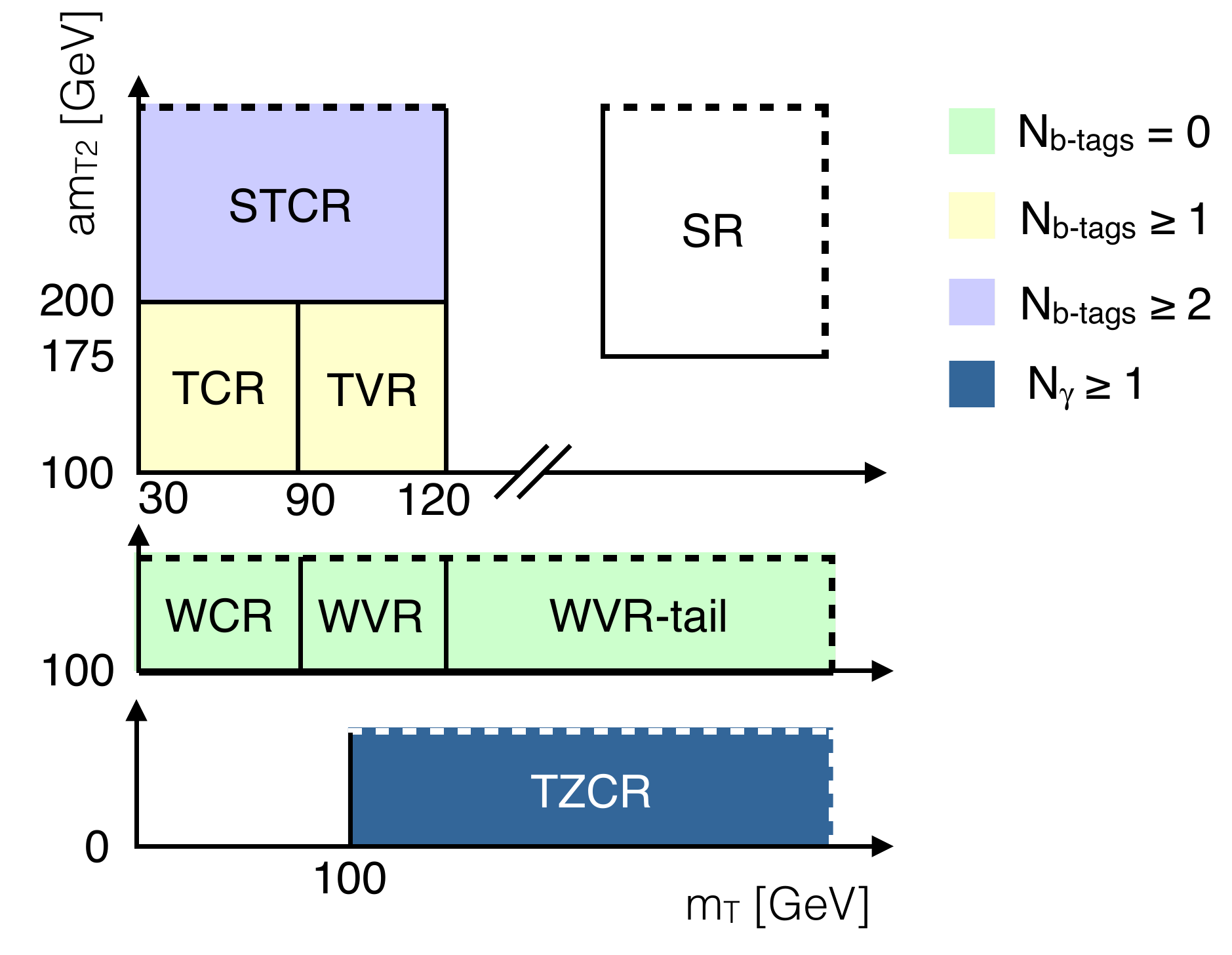}
 \caption{The single bin control and validation region setup for the $\sqrt{s}=13$ TeV analysis.  All regions share a common selection and then are further distinguished by requirements on $m_\text{T}$, $am_\text{T2}$ and the number of $b$-jets.  The prefixes stand for ST = single top, T = $t\bar{t}$, W = $W$+jets, and TZ = $t\bar{t}+Z$.  The exact definitions of the regions are given in Sec.~\ref{chapter:background}.}
 \label{fig:CRschematic}
  \end{center}
\end{figure}			
		
\clearpage			
			
\subsection{Multibin (Shape Fit) Signal Region}
\label{sec:shapefitsetup}
	
One way to increase the sensitivity of a signal region is to split it into multiple bins that have different $p_s/p_b$.  Multiple bins increase the sensitivity because it provides a finer scale for the likelihood and effectively gives a higher weight to events where the likelihood is higher.  Figure~\ref{fig:shapefittoy} quantifies this statement with an example; compared with the one-bin setup, the two-bin setup has a lower probability for rejecting the SM when there is SUSY for a fixed probability to reject the SM when it is in fact true.  The gain is bigger when the difference in the likelihoods between bins is bigger.

\begin{figure}[h!]
\begin{center}
\includegraphics[width=0.33\textwidth]{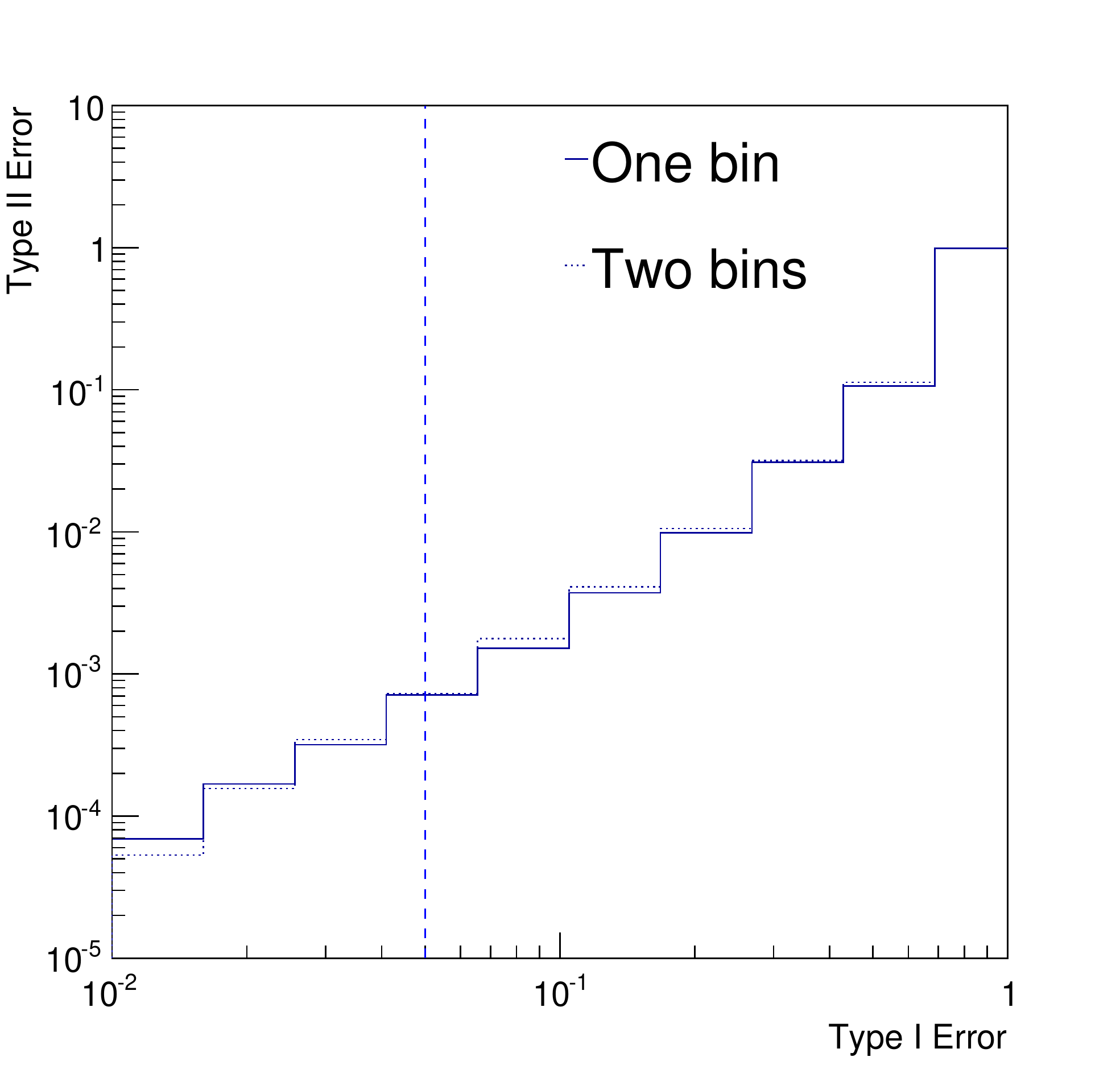}\includegraphics[width=0.33\textwidth]{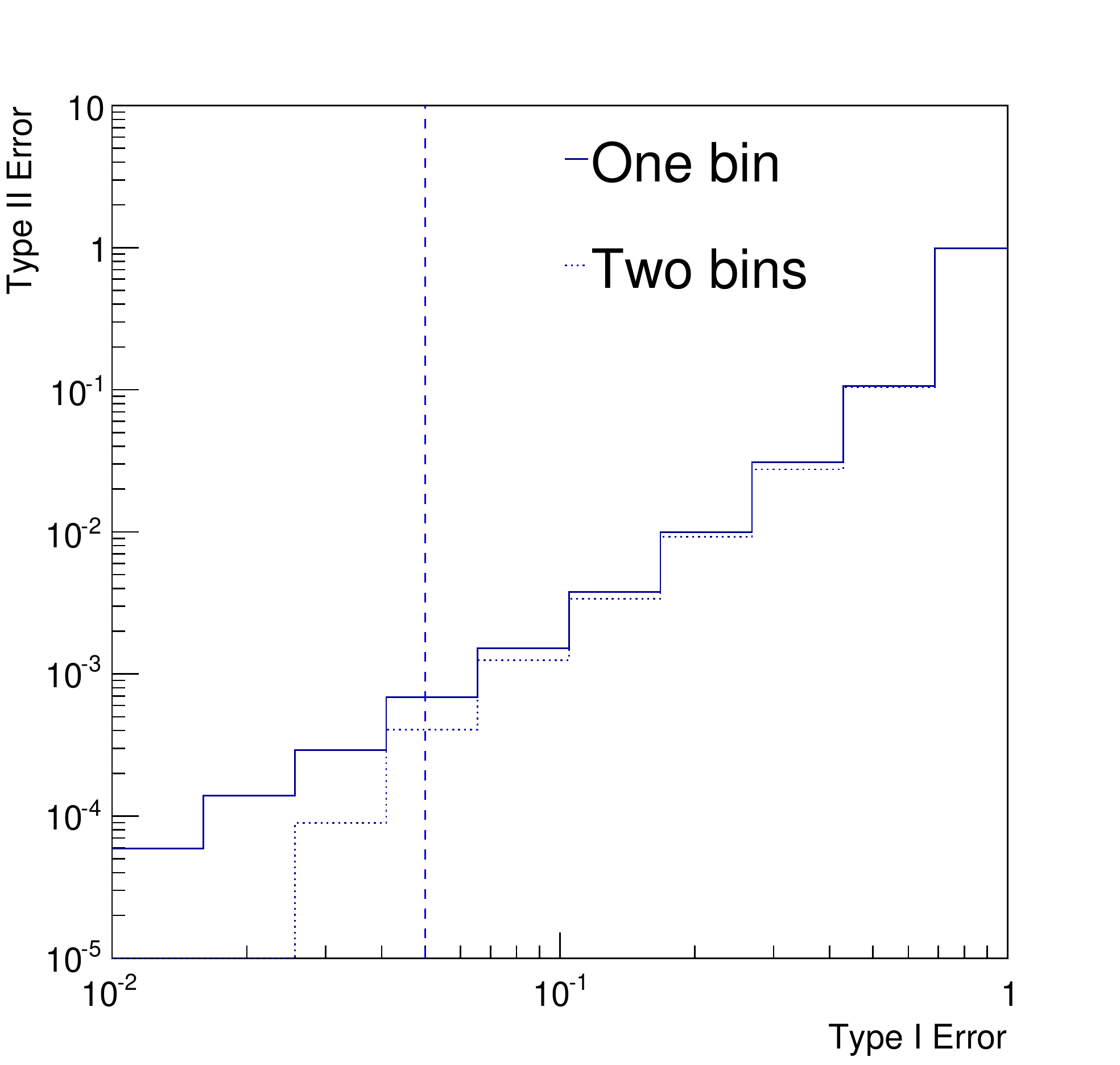}\includegraphics[width=0.33\textwidth]{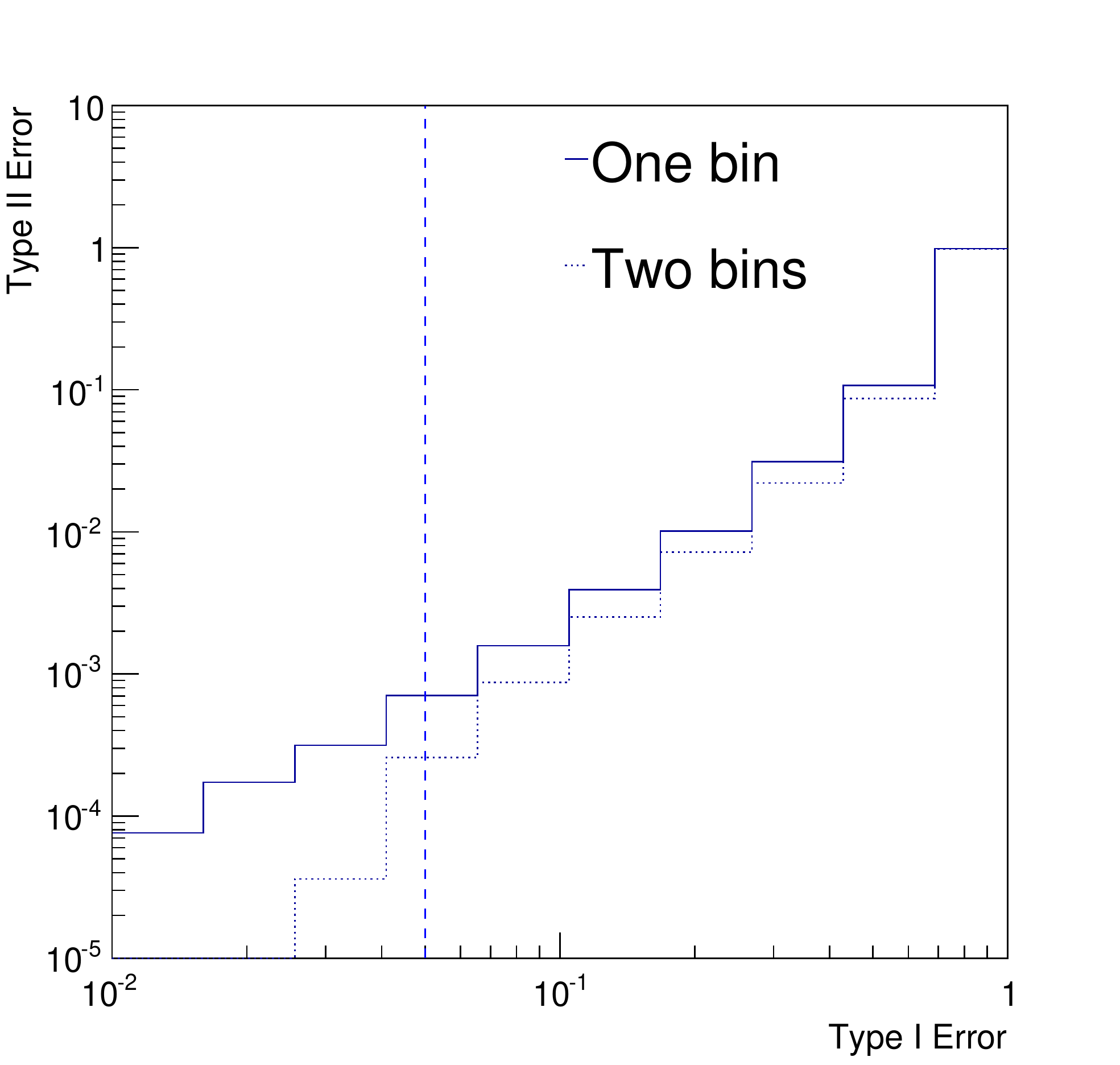}
 \caption{The tradeoff between Type II (do not reject SM when SUSY is true) and Type I errors (reject SM when SM is true) for three scenarios.  In all cases, the number of background events is $200$ in one bin and $100$ in a second bin.  The total number of signal events is $30$.  In the left plot, both bins have $10\%$ signal; in the middle plot the fraction is twice as high in the first bin; in the right plot, the fraction is four times as high in the first bin compared with the second bin.  The errors are computed by scanning threshold requirements on the (log) likelihood ratio distribution.  For the two-bin case, the likelihood ratio is a product of the individual bin likelihood ratios.  Only statistical uncertainties are included in the likelihood.  The vertical dashed line is at $5\%$. }
 \label{fig:shapefittoy}
  \end{center}
\end{figure}

For more intuition, consider a two-bin setup with mean background event yields of $B_1$ and $B_2$ and signal yields $S_1$ and $S_2$.  The optimal test procedure is based on the likelihood ratio $p_{s+b}(x)/p_b$ (see Sec.~\ref{sec:susy:stats}).  Suppose that $S_1=0$.  Ideally, one would remove the first bin, as it contains no useful discriminating information for the signal.  If the two bins are lumped together, the first bin dilutes the power of the two bins together.  However, if the two bins are split and the likelihood is a product over the two bins, then the first bin automatically does not contribute ($p_{s_1+b_1}(x_1)/p_{b_1}(x_1)=1$).  This is an extreme case, but it illustrates the main point.  In principle, the optimal procedure is to weight every event by its log likelihood ratio (i.e. put each event in its own bin\footnote{To see that these are equivalent, consider a case where there are only two possible values of $p_s/p_b$.  Label the bins $1$ and $2$ and then the log likelihood for the two bins is (up to a constant) $x_0\log(1+s_1/b_1)+x_1\log(1+s_2/b_2)$, where $x_i$ is the number of observed events, $s_i$ is the mean number of signal events, and $b_i$ is the mean number of background events in bin $i$.  Instead, suppose each event in bin $i$ is weighted by $\log(1+s_i/b_i)$ so that the total number of `measured events' is $x=x_0\log(1+s_1/b_1)+x_1\log(1+s_2/b_2)$.  Then, the log likelihood is (up to a constant) $x\log(1+s/b)$ for $s=s_1+s_2$ and $b=b_1+b_2$.  Since $\log(1+s/b)$ is a constant across bins, the likelihood with the weighted setup is a monotonic function of the binned likelihood.  Therefore, they result in the same statistical power for a fixed signal model.  However, this result may not (exactly) hold with a different test statistic.}), but this makes it difficult to validate the modeling of the weights and so the focus here is on a small number of bins.  Binning is not used (yet) for the SR setup introduced in Sec.~\ref{sec:singlebin} because of the explicit model dependence through $p_s$.  

The shape fit region still uses the control region method, but in a more integrated way than for the single bin regions.  Control regions and signal regions are simply bins of a multibin SR where the $p_s/p_b$ is very low in the CR-like bins and high in the SR-like bins.  This setup is illustrated in Fig.~\ref{fig:shapefit}.  The selection (Sec.~\ref{sec:shapefitregion}) and fit procedure (Sec.~\ref{sec:susy:stats}) are described in later sections.  Just like Fig.~\ref{fig:CRschematic}, there are two key kinematic variables in addition to $b$-tagging information to control the purity of various background processes.   The expected $p_s/p_b$ is about $20\%$ in some of the most signal-sensitive bins (upper-right of Fig.~\ref{fig:CRschematic}) and about $10\%$ in others.  In the CR-like regions, the likelihood is less than $1\%$.  A global normalization of the various data-driven backgrounds is possible but puts a stringent constraint on the modeling as a function of the variables defining the bins in Fig.~\ref{fig:shapefit}.  Since the background composition changes most strongly with the $m_\text{T}$ variable and not $E_\text{T}^\text{miss}$, each column in Fig.~\ref{fig:shapefit} is separately normalized using the CR method.  This also mitigates any potential mis-modeling of the $E_\text{T}^\text{miss}$ trigger turn-on for $E_\text{T}^\text{miss}\sim 100$ GeV.  By construction, the fitted background prediction is nearly identical to the data yield in the CR-like bins.  The implications the differences between data and simulation in the SR-like regions have for light stops are discussed in Sec.~\ref{8TeVresults}.
	
\begin{figure}[h!]
\begin{center}
\includegraphics[width=0.7\textwidth]{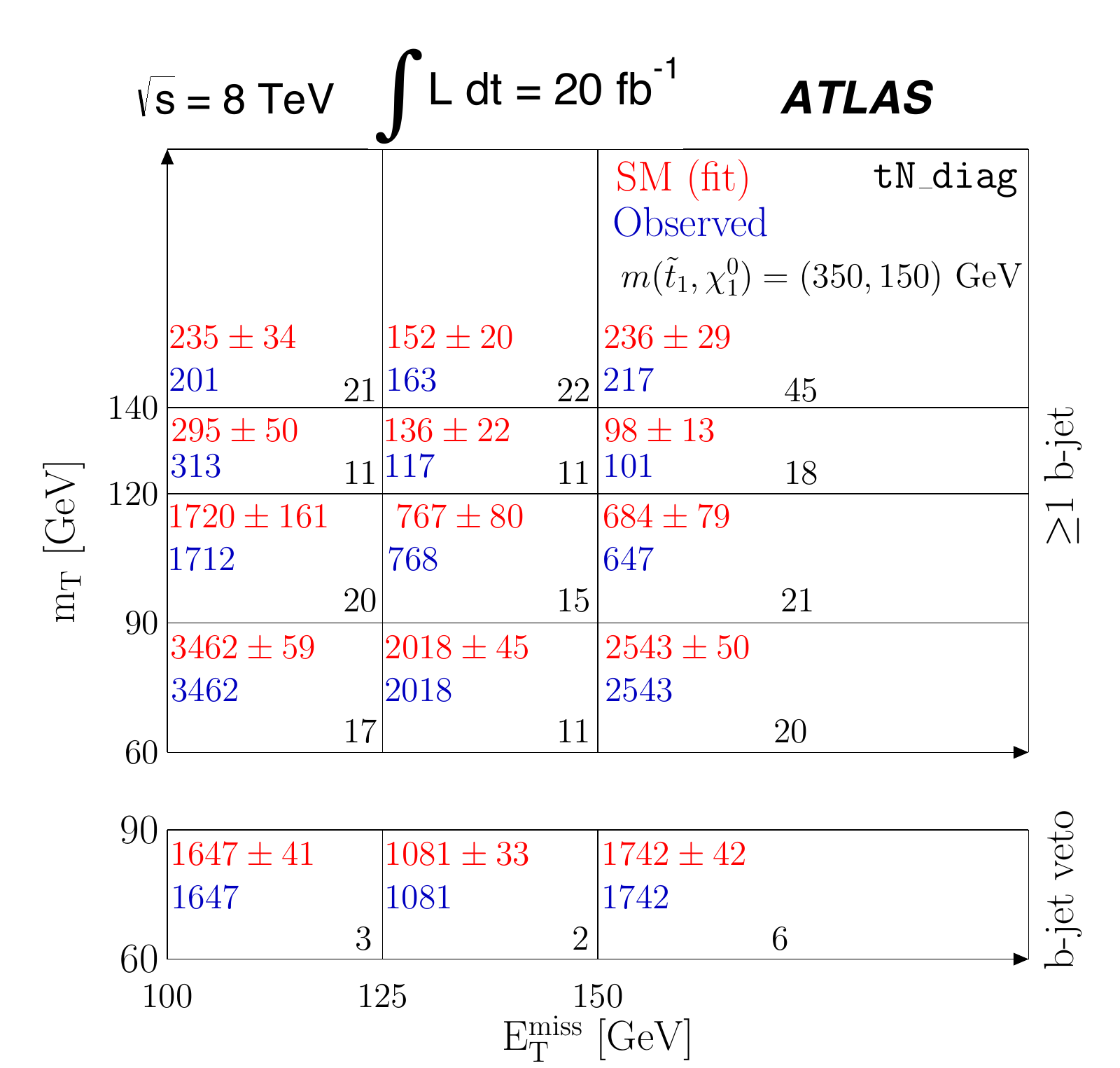}
 \caption{The various bins of the shape fit signal region tN\_diag from the $\sqrt{s}=8$ TeV analysis.  All the bins share a common selection and are distinguished by requirements on $m_\text{T}$, $E_\text{T}^\text{miss}$ and the number of $b$-jets.  The red numbers are the SM prediction, the blue numbers are the observed events, and the black numbers are the predicted signal yield.  The top right bin is the most signal sensitive bins, the lowest row is the most sensitive to $W$+jets and the row just above that is the most sensitive to the $t\bar{t}$ normalization.  These last two regions behave similarly to the control-regions from the one bin regions and as such the fitted background yield is nearly identical to the observed data.  The selection is described in Sec.~\ref{sec:shapefitregion} and the implications for light stops are discussed in Sec.~\ref{8TeVresults}.}
 \label{fig:shapefit}
  \end{center}
\end{figure}				
	
\clearpage	
			
\subsection{Drawbacks of the CR Method and Alternatives}
\label{drawbacks}	
		
The main disadvantage of the control region method is the assumption that the shape $p_B(V)$ from Sec.~\ref{sec:CRmethod} is known.  Differential distributions are usually known with less precision than the total cross-section and tails of distributions are known to be sensitive to higher order effects (and in some cases, non-perturbative modeling).  Therefore, a thorough investigation of potential sources of systematic bias in extrapolating from the control region to the signal region is presented in Sec.~\ref{chapter:uncertainites}.		
		
Another, more subtle disadvantage of the standard control region method is that it can be very sensitive to statistical fluctuations when the number of events in the control region is small.  If there is an under-fluctuation in the data, then the predicted background yield in the signal region will be too small.  This is partially accounted for in the statistical uncertainty from the data in the control region, but the {\it central value} will be biased.  Figure~\ref{fig:5} illustrates the source of bias by showing the probability of obtaining $2\sigma$ evidence for SUSY when there is only background.  The probability should be $5\%$.  It is not exactly so even when the number of events in the control region is infinite because the number of observed events in the signal region can only take discrete values.  The most striking feature of Fig.~\ref{fig:5} is that the probability for a $2\sigma$ excess is almost a factor of two higher for $N[\text{control region}]=10$ compared with $N[\text{control region}]\rightarrow\infty$.  Figure~\ref{fig:6} is another way to view the problem, but in the case where there is SUSY.  Suppose there would be a $3\sigma$ excess if the true expected number of events in the signal region were known (i.e. an infinite number of events in the control region).  Fig.~\ref{fig:6} shows the probability for this value to drop to less than $3\sigma$ when the number of events in the control region is finite.  For example, the probability for a $3\sigma$ to drop to a $2\sigma$ is about $10\%$ when the true expected number of events is $10$ in both the signal and control regions.  This has important implications for discovery as the threshold for `evidence' is usually set at $3\sigma$, while $2\sigma$ fluctuations are largely ignored.  These biases are mostly mitigated when $N[\text{control region}]\gtrsim\mathcal{O}(100)$, as is mostly true for the regions constructed in Sec.~\ref{chapter:background}.

\begin{figure}[h!]
\centering
\includegraphics[width=0.8\textwidth]{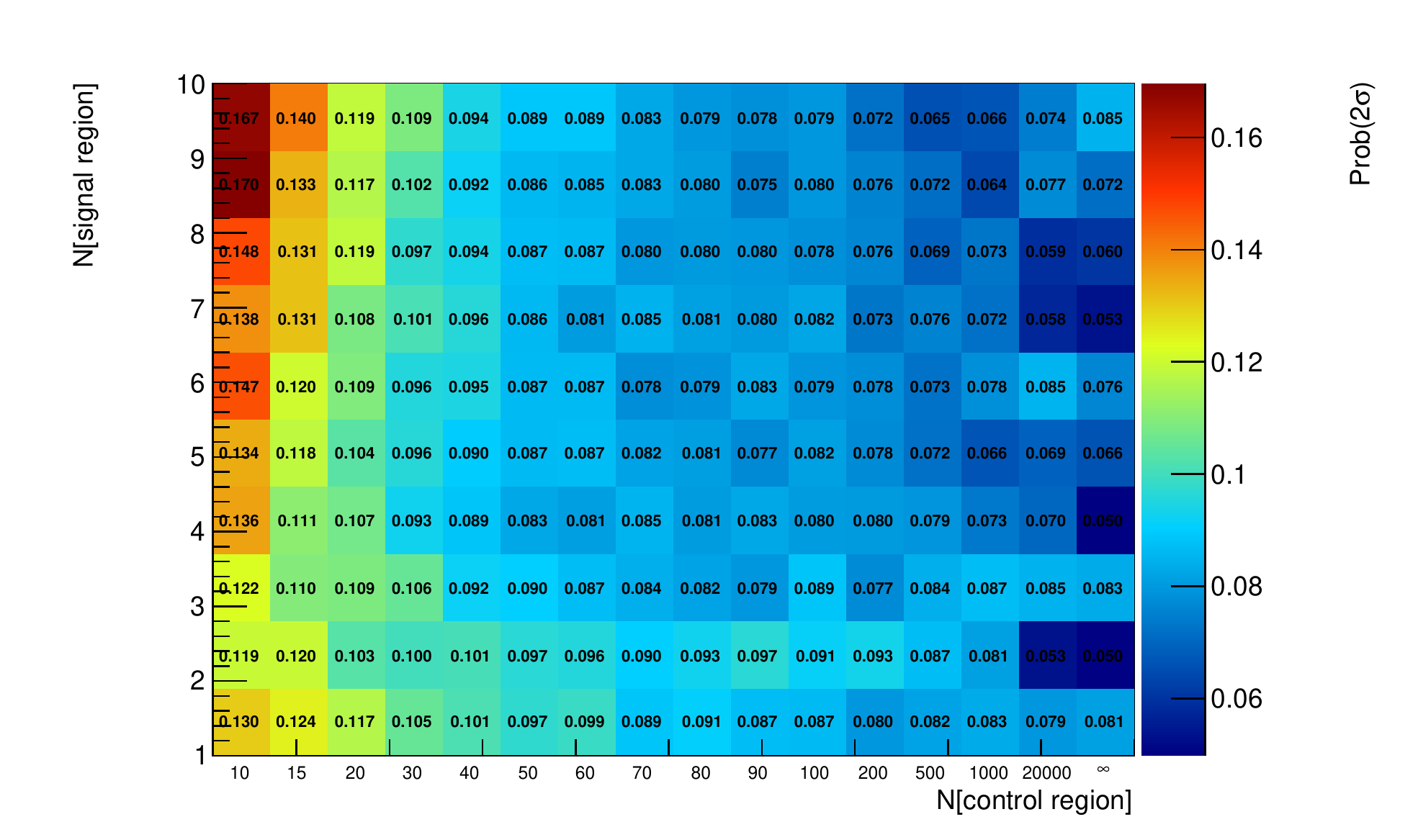}
\caption{The probability for a $2\sigma$ excess when there is only background as a function of the number of expected events in the control region and in the signal region using the standard control region method.  A $2\sigma$ excess is defined as a case when the probability for the observed number of events in the signal region to exceed the number of predicted events in the signal region to be less than $5\%$.  Only statistical uncertainties are included in this calculation.}
\label{fig:5}
\end{figure}

\begin{figure}[h!]
\centering
\includegraphics[width=0.33\textwidth]{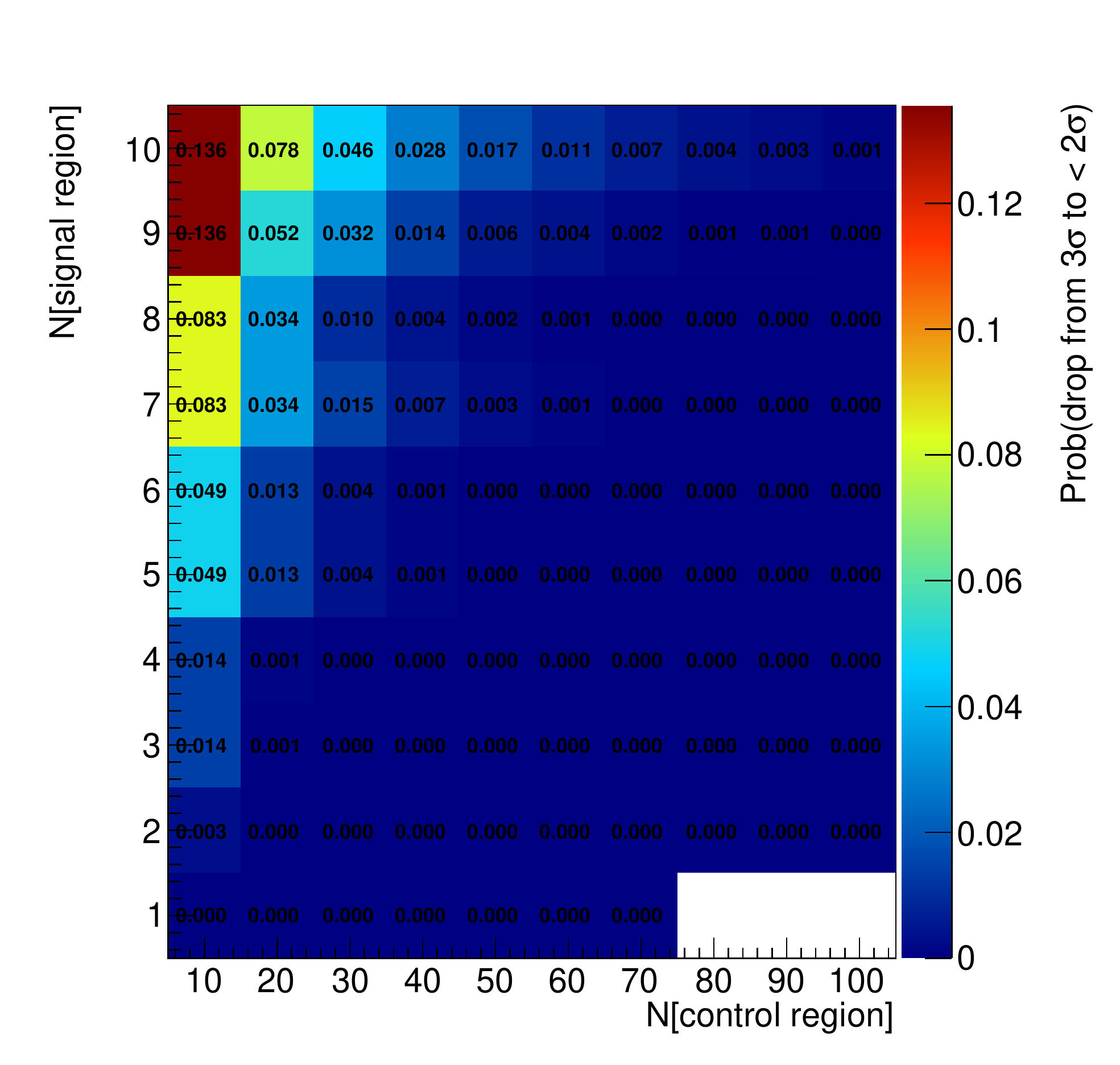}\includegraphics[width=0.33\textwidth]{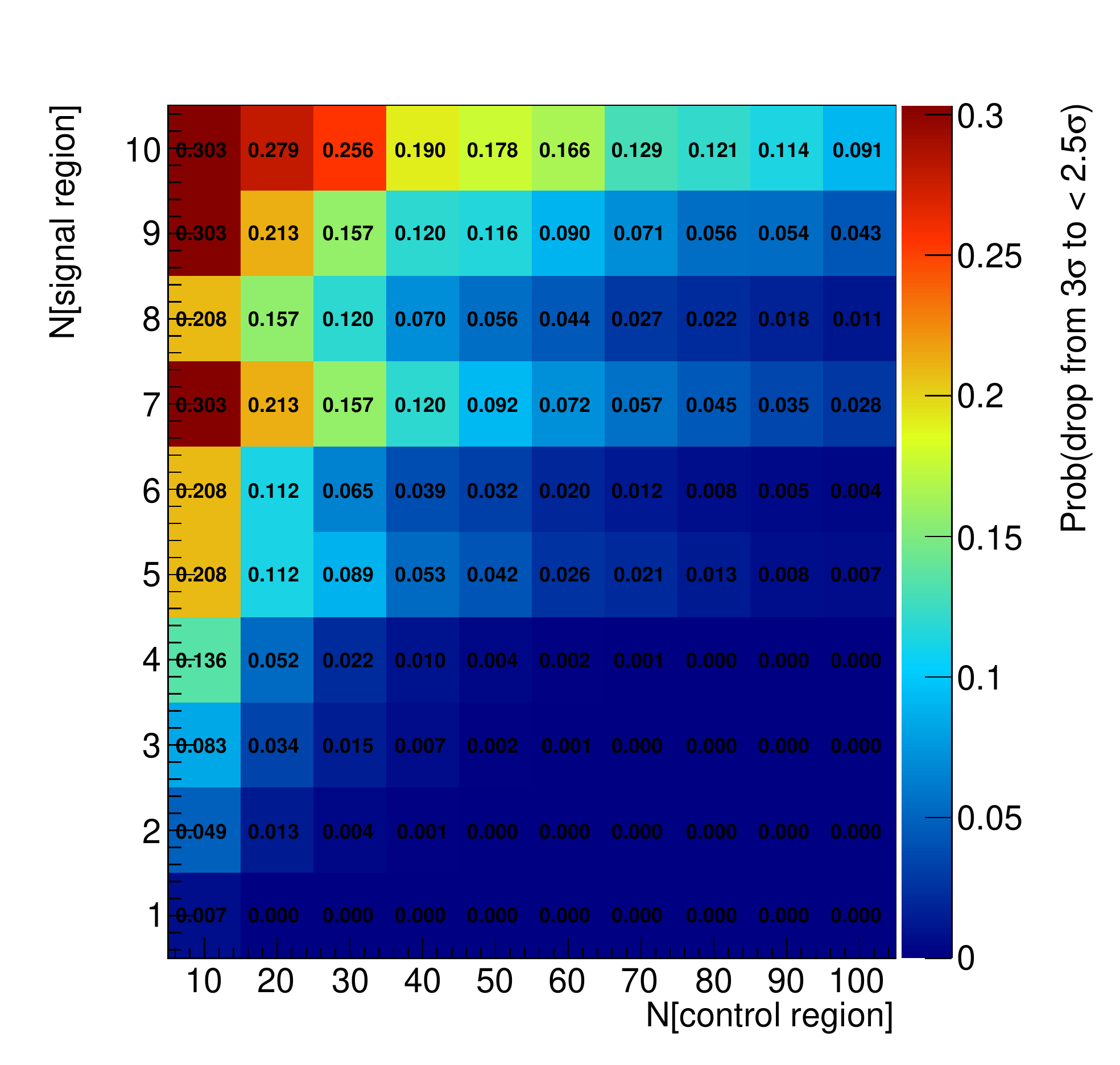}\includegraphics[width=0.33\textwidth]{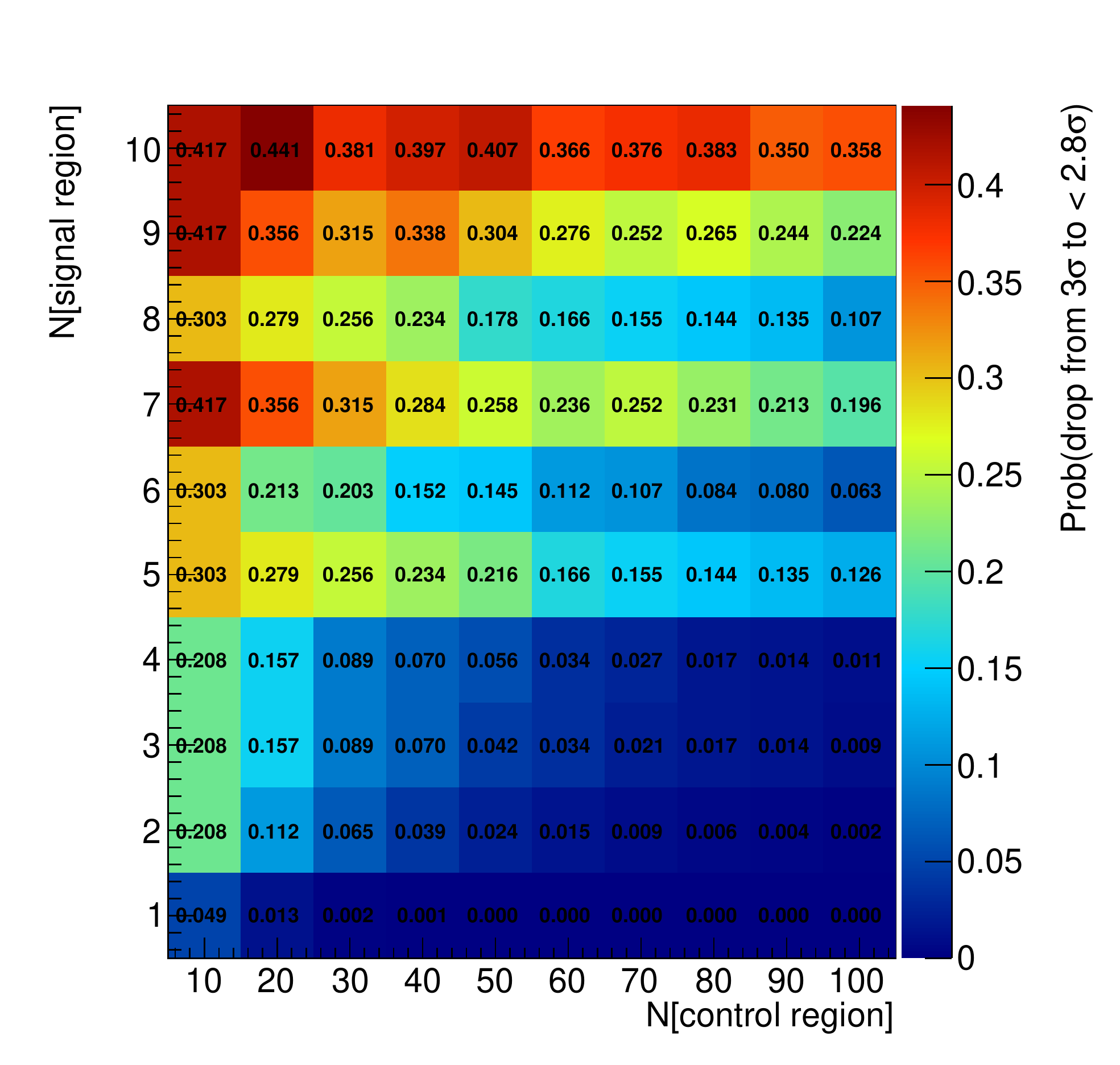}
\caption{The probability for a $3\sigma$ excess (with $\infty$ events in the control region) to be measured as a $<3\sigma$ excess.}
\label{fig:6}
\end{figure}

One simple modification of the standard control region method is to adapt it to a Bayesian framework.  The main problem arises because the standard method puts too much emphasis on the observed data in the CR.  A Bayesian approach would be:

\begin{align}
p(N_B|N_\text{CR}^\text{observed})\propto\text{Poisson}(N_\text{CR}^\text{observed}|N_B)p(N_B),
\end{align}

\noindent where $p(N_B)$ is a prior distribution.  One reasonable prior is the Gamma distribution, which is conjugate for the Poisson.  The posterior mean for a Gamma prior with parameters $\alpha$ and $\beta$ is

\begin{align}
\langle N_B|N_\text{CR}^\text{observed}\rangle=\left(\frac{1}{1+\beta}\right)N_\text{CR}^\text{observed}+\left(\frac{\beta}{1+\beta}\right)\frac{\alpha}{\beta},
\end{align}

\noindent which is a linear superposition of the prior mean $(\alpha/\beta)$ and the observed number of events from one observation.  The parameter $\beta$ plays the role of the number of effective events `observed' prior to seeing any data.  If there are auxiliary measurements that can be used to constrain $\alpha/\beta$, then the number of events in such a region could be used to set $\beta$.  Another possibility is to use the estimated systematic uncertainty on the number of events from the raw simulation.  For example, a systematic uncertainty of $30\%$ would correspond to an auxiliary measurement of $\sim 10$ events, since $1/\sqrt{10}\sim 30\%$.  Therefore, one could pick $\beta=10$ and then $\alpha=\beta\times N_B^\text{MC}$.  This procedure is illustrated in Fig.~\ref{fig:7}.  Compared to Fig.~\ref{fig:5}, the probability for observing an excess when there is no SUSY (often called Type 1 error) is significantly reduced. 

\begin{figure}[h!]
\centering
\includegraphics[width=0.8\textwidth]{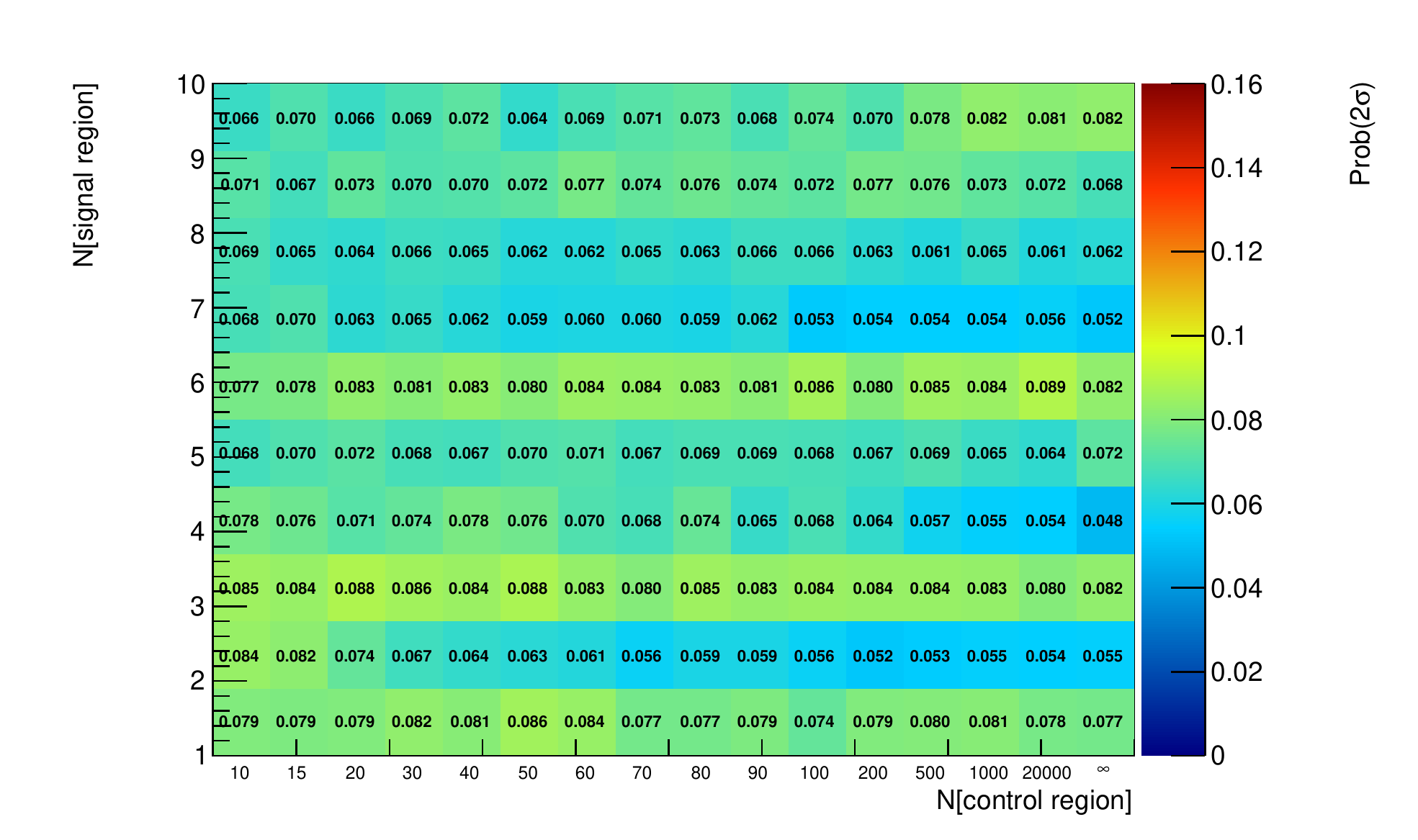}
\caption{The probability for a $2\sigma$ excess under the null (only known particles) hypothesis as a function of the number of expected events in the control region and in the signal region using the Bayesian control region method described in the text.  A $2\sigma$ excess is defined as a case when the probability for the observed number of events in the signal region to exceed the number of predicted events in the signal region to be less than $5\%$.  A systematic uncertainty of $30\%$ is assumed when choosing $\beta$.}
\label{fig:7}
\end{figure}

The Bayesian framework also allows a natural method for incorporating systematic uncertainties into the control region method.  The usual procedure for estimating the systematic uncertainty is to consider alternative plausible simulations and compare the differences in the predicted events in the signal region using the various models.  Usually, one simulation is taken as nominal and the differences with respect to the other models are taken as Gaussian uncertainties on the expected number of events in the signal region.  One way to avoid this ad-hoc approach is to use a hierarchical Bayes model in which there is a latent variable $z$ that describes which simulation is the best description of the data.  There could be many acceptable models and the final prediction is estimated by integrating out $z$.  The posterior variance would also give a sense of the uncertainty (in fact, any measure of spread based off of the posterior could be used).  This model is illustrated graphically in Fig.~\ref{fig:8}.  Additionally, it may be possible to combine this with an empirical Bayes approach in which many searches are simultaneously used to provide a prior for $z$.  If no other information is available, a discrete uniform random variable on $\{1,...,n\}$ could be used.  Figure~\ref{fig:9} illustrates the posterior under such a scheme with three plausible simulations, using the uniform prior for $z$.  As expected, the distribution is slightly broader when the sample variations are included.

\begin{figure}[h!]
\centering
\includegraphics[width=0.5\textwidth]{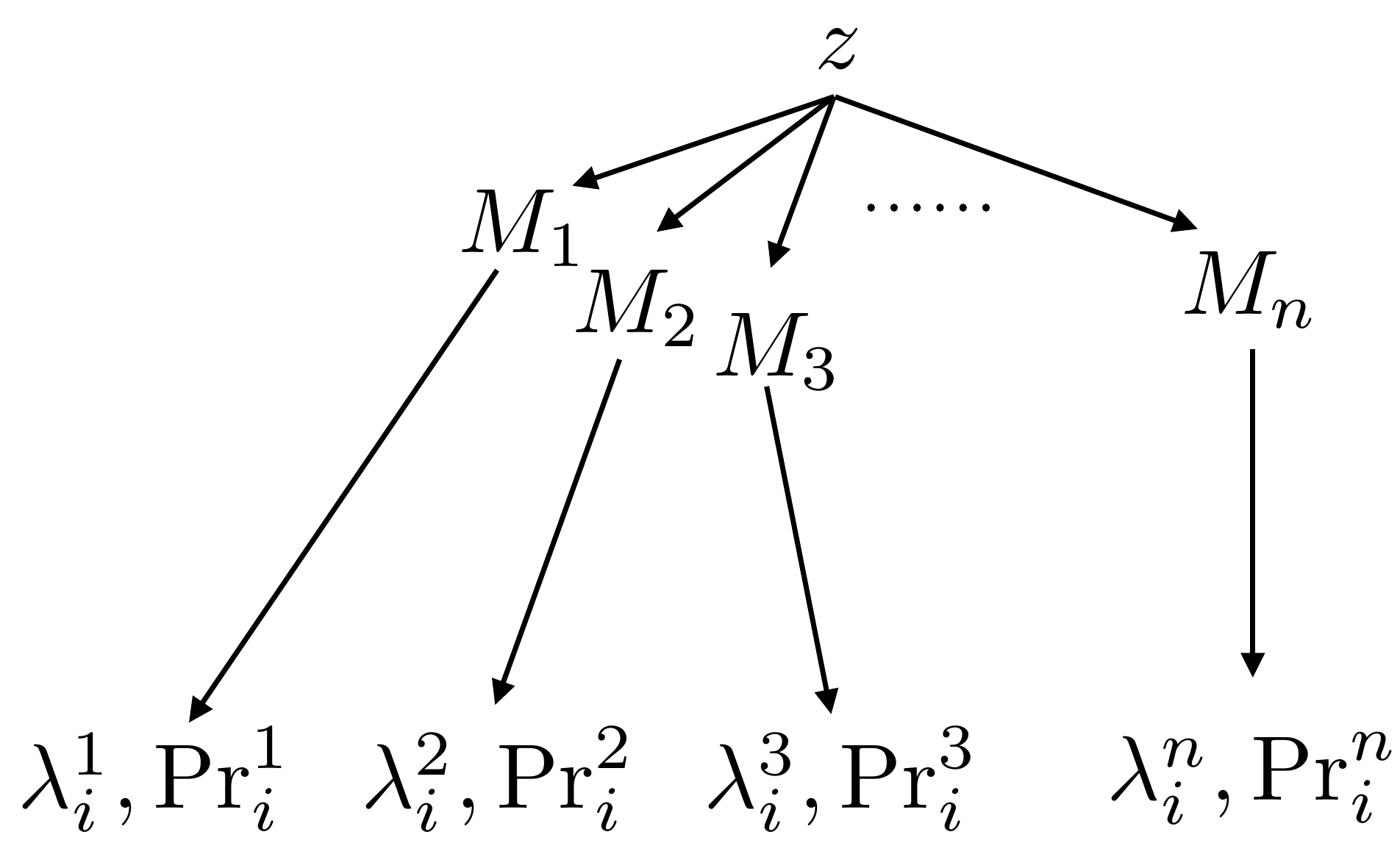}
\caption{A graphical representation of the hierarchical Bayes model described in the text.  For model $M_j$, the predicted background yield ($N_B^\text{MC}$) is denoted $\lambda_i^j$.}
\label{fig:8}
\end{figure}

\begin{figure}[h!]
\centering
\includegraphics[width=0.5\textwidth]{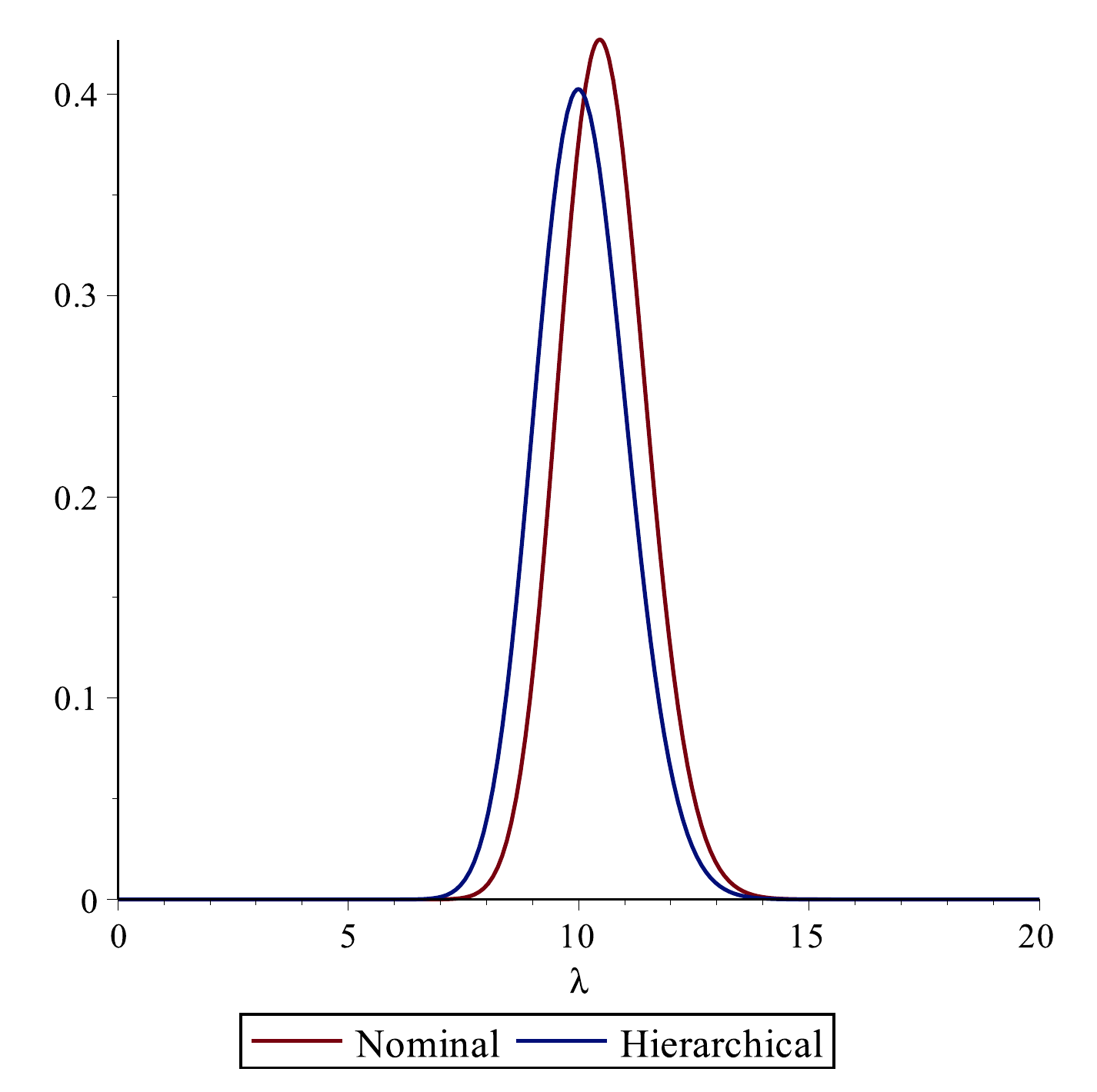}
\caption{Sample posteriors using the hierarchical Bayes model described in the text.  The nominal simulation predicts $(N_\text{CR}^\text{raw},N_\text{SR}^\text{raw})=(10,5)$ and the two alternative models have $(N_\text{CR}^\text{raw},N_\text{SR}^\text{raw})=(10.5,5.5)$ and $(N_\text{CR}^\text{raw},N_\text{SR}^\text{raw})=(9.5,4.5)$.  The number of observed events in the control region is 11 and the systematic uncertainty, used to construct $\beta$ is 30\%.}
\label{fig:9}
\end{figure}

One could additionally generalize the above procedures when there are multiple control regions or when a control region has multiple bins.  In that case, the simulation will be properly down-weighted to account for the new information from the data.   These Bayesian methods are promising ways to incorporate prior information from simulations and auxiliary measurements, but require more investigation before they can be properly incorporated into an analysis.

\clearpage		
		
		\section{Dataset and Monte Carlo Samples}
		\label{sec:datasetandMC}
		
The $\sqrt{s}=8$ TeV data from Run 1 were collected between March and December 2012 resulting in an integrated luminosity of $20.3$ fb{}$^{-1}$ and the $\sqrt{s}=13$ TeV data from the early part of Run 2 where collected between October and December 2015 leading to $\int L dt = 3.2$ fb${}^{-1}$.  These data were recorded using a combination of single lepton and $E_\text{T}^\text{miss}$ triggers.  Figure~\ref{fig:susytriggers} shows the efficiency for various trigger algorithms.   Isolated lepton triggers require $p_\text{T}>24$ GeV in addition to particle identification and isolation criteria.  The efficiency drops at high stop mass due to the isolation failing when the jets from the same boosted top quark as the lepton are close-by.  Inclusive lepton triggers have a higher $p_\text{T}$ threshold of $p_\text{T}>60,36$ GeV for electrons and muons, respectively.  The $E_\text{T}^\text{miss}$ trigger is fully efficient for offline $E_\text{T}^\text{miss}\gtrsim 150$-$200$ GeV.  Boosted $W$ and top quark jets can fire the single large-radius jet trigger, though the efficiency is not competitive with the other triggers as it is not fully efficient until $p_\text{T}\gtrsim400$ GeV.  

\begin{figure}[h!]
\begin{center}
\includegraphics[width=0.45\textwidth]{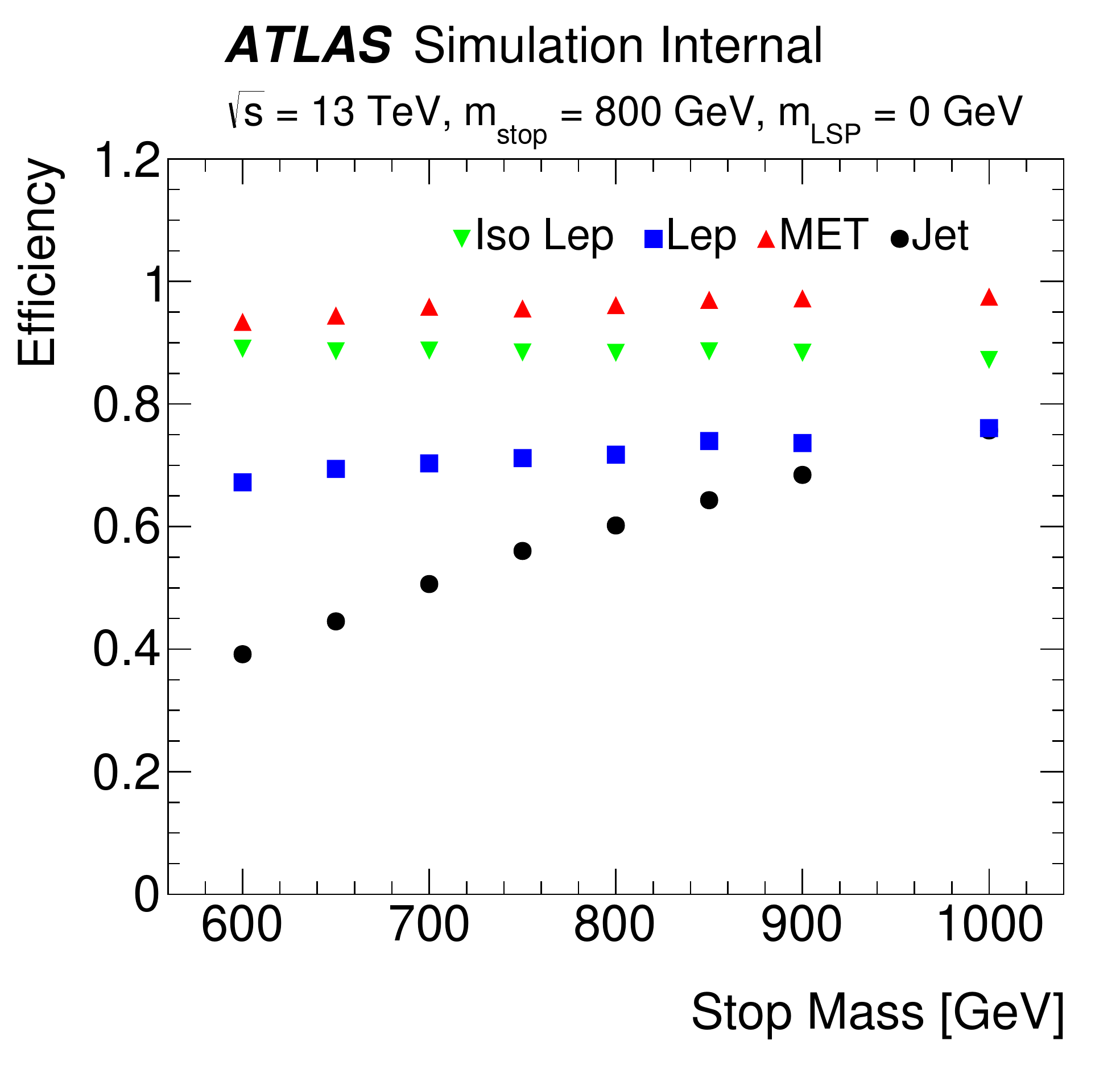}\includegraphics[width=0.45\textwidth]{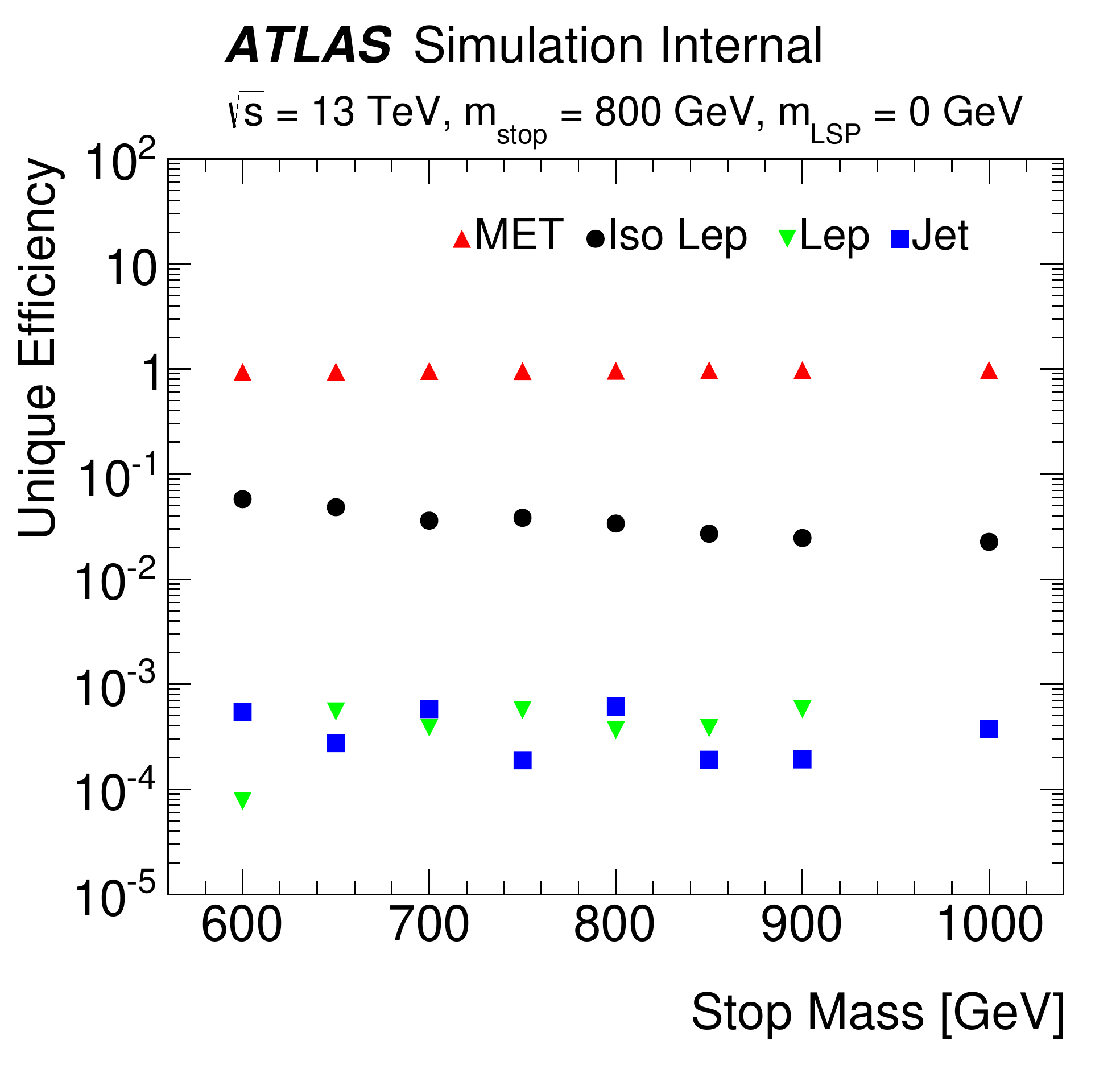}
 \caption{Left: the trigger efficiency for various algorithms as a function of the stop mass. Right: The additional efficiency provided by various triggers beyond the efficiency already provided by the triggers to the left of it in the legend. }
 \label{fig:susytriggers}
  \end{center}
\end{figure}

For $m_\text{stop}\gtrsim 500$ GeV and a massless LSP, the $E_\text{T}^\text{miss}$ trigger is the most efficient, recording $\gtrsim 95\%$ of events.  The isolated single lepton triggers are also highly efficient, but only add about $\lesssim 2\%$ beyond the $E_\text{T}^\text{miss}$ trigger for high mass stops.  For lower stop masses, the single lepton triggers are an essential complement to the $E_\text{T}^\text{miss}$ trigger.  As such, the early $\sqrt{s}=13$ TeV analysis uses only the $E_\text{T}^\text{miss}$ trigger, but the full $\sqrt{s}=8$ TeV analysis uses events that pass the $E_\text{T}^\text{miss}$ trigger or either of the isolated or inclusive single lepton triggers.

The diverse background processes relevant for the stop search require a complete set of simulated SM samples.  Several matrix element (ME) generators are combined with parton shower (PS) generators.  
Signal SUSY samples are generated at leading order (LO) with 
MG5\_aMC~v2~\cite{Alwall:2014hca} ({\sc Herwig++}) at $\sqrt{s}=13$ ($\sqrt{s}=8$) TeV.  All signal samples generated with MG5\_aMC are interfaced with {\sc Pythia} 
8.186~\cite{Sjostrand:2007gs}. The nominal stop mixing angle is given by $\theta_{\tilde{t}}=1$ ($\tilde{t}_1$ is mostly $\tilde{t}_R$) and $N_{11}=1$ (pure bino LSP).  This corresponds to an effective mixing angle of $\theta_\text{eff}\approx 1.4$.  Background samples use one of four setups:

\begin{itemize}
\item MG5\_aMC or MadGraph~\cite{Alwall:2011uj} interfaced with {\sc Pythia} or {\sc Herwig++} using the CKKW-L~\cite{Catani:2001cc,Lonnblad:2001iq} or the MC@NLO method for matching a LO or next-to-leading-order (NLO) ME to the PS, respectively.
\item {\sc Powheg-Box}~\cite{Alioli:2010xd,Re:2010bp,Frixione:2007nw,Frederix:2012dh,Alioli:2009je} interfaced to {\sc Pythia} 6~\cite{Sjostrand:2006za}, {\sc Herwig}+{\sc Jimmy}, or {\sc Herwig++} using the {\sc Powheg} method~\cite{Nason:2004rx,Frixione:2007vw} to match the NLO ME to the PS.
\item {\sc Sherpa}~\cite{Gleisberg:2008ta} using Comix~\cite{Gleisberg:2008fv} (LO+extra partons) and OpenLoops~\cite{Cascioli:2011va} (NLO) ME generators interfaced with the {\sc Sherpa} parton shower~\cite{Schumann:2007mg}.  Leading order samples with extra partons use the CKKW matching scheme.
\item {\sc AcerMC} 3.8~\cite{Kersevan:2004yg} interfaced with {\sc Pythia 6} for fragmentation.
\end{itemize}

\noindent The samples produced with MG5\_aMC and {\sc Powheg-Box} at $\sqrt{s}=13$ TeV use {\sc EvtGen} v1.2.0~\cite{EvtGen} for the modelling of $b$-hadron decays.   Similarly, the generators {\sc TAUOLA}~\cite{Jadach:1993hs} and {\sc PHOTOS}~\cite{Golonka:2005pn} are used to model $\tau$-lepton decays and QED radiative corrections.  The simulation setup is summarized in Table~\ref{tab:mc_samples1} and more details can be found in Ref.~\cite{ATL-PHYS-PUB-2016-004,ATL-PHYS-PUB-2016-003,ATL-PHYS-PUB-2016-005,ATL-PHYS-PUB-2016-002} for $t\bar{t}$ and single top, $W/Z$+jets, dibosons, and $t\bar{t}+W/Z$, respectively.   Due to the sub-optimal $h_\text{damp}=\infty$ setting for the $t\bar{t}$ sample at $\sqrt{s}=8$ TeV, a $p_\text{T}^{t\bar{t}}$ re-weighting is performed based on the dedicated early Run 1 measurement of this quantity~\cite{Aad:2014zka}.  Additional samples aside from those shown in Table~\ref{tab:mc_samples1} are used to assess theoretical modeling uncertainties and will be discussed in Section~\ref{chapter:uncertainites}.  Such samples are generated using one of the four setups listed above.

As in previous chapters, pileup is simulated by overlaying {\sc Pythia} 8 minimum bias events on the samples listed above.    The particle-level simulations are processed using either a full detector simulation~\cite{Aad:2010ah} based on {\sc Geant }4~\cite{Agostinelli:2002hh} or a fast simulation~\cite{ATLAS:1300517} with a parameterized calorimeter response and {\sc Geant} 4 for all other processes.  There is no significant difference between these setups for the event selections considered in Part~\ref{part:susy}.

\begin{table}[h!]
\centering
\noindent\adjustbox{max width=\textwidth}{
\begin{tabular}{| c | ccccc |}
\hline
Process & ME Generator & ME  & Fragmentation & UE  & Cross-section\\
 &  &  PDF &  & Tune & Order\\
\hline
$t\bar{t}$ & {\sc Powheg-Box} & CT10& {\sc Pythia} 6 & P2012 ({\color{blue}2011C}) &NNLO+NNLL~\cite{Czakon:2013goa,Czakon:2012pz,Czakon:2012zr,Baernreuther:2012ws,Cacciari:2011hy,Czakon:2011xx} \\ 
Single top& {\sc Powheg-Box} & CT10 & {\sc Pythia} 6 & P2012 ({\color{blue}2011C}) &NNLO+NNLL~\cite{Kidonakis:2011wy,Kidonakis:2010ux,Kidonakis:2010tc} \\ 
{\color{blue}Single top ($t$-chan.)}& {\sc \color{blue}AcerMC} & {\color{blue}CTEQ6L1} & {\color{blue}{\sc Pythia} 6} & {\color{blue}P2011C} &{\color{blue}NNLO+NNLL}~\cite{Kidonakis:2011wy} \\ 
$W/Z$+jets & {\sc Sherpa} 2.1.1 ({\color{blue} 1.4.1}) & CT10 & {\sc Sherpa} & Default & NNLO~\cite{Catani:2009sm}\\ 
Diboson & {\sc Sherpa} 2.1.1 ({\color{blue} 1.4.1}) & CT10 & {\sc Sherpa} & Default & NLO~\cite{Campbell:1999ah,Campbell:2011bn}\\ 
$t\bar{t}+W/Z$ & {\sc MG5\_aMC}~v2 & NNPDF2.3 & {\sc Pythia} 8 & A14 & NLO~\cite{Alwall:2014hca}\\ 
 &{\color{blue}MadGraph 5} & {\color{blue}CTEQ6L1} & {\color{blue}{\sc Pythia} 6} & {\color{blue}AUET2B} & {\color{blue}NLO}~\cite{Campbell:2012dh,Garzelli:2012bn}\\ 
$t\bar{t}+\gamma$& {\sc MG5\_aMC}~v2 & CTEQ6L1 & {\sc Pythia} 8 & A14 & NLO~\cite{Alwall:2014hca}\\ 
& {\color{blue}{\sc MadGraph 5} }& {\color{blue}CTEQ6L1} & {\color{blue}{\sc Pythia} 6} & {\color{blue}AUET2B} & {\color{blue}NLO}~\cite{Melnikov:2011ta}\\ 
 SUSY Signal &{\sc MG5\_aMC}~v2 & NNPDF2.3 & {\sc Pythia} 8& A14 & NLO+NLL~\cite{Borschensky:2014cia}\\ 
 &  {\color{blue}{\sc Herwig++}} & {\color{blue}CTEQ6L1} & {\color{blue}{\sc Herwig++}}  & {\color{blue}EE3} & {\color{blue}NLO+NLL}~\cite{Kramer:2012bx}\\ 
\hline
\end{tabular}}
\caption{Overview of the nominal simulated samples.  The last row indicates the order and reference for the inclusive cross-section to which the (lower order) MC simulations are normalized. The blue indicates a setup at $\sqrt{s}=8$ that differs from the one used at $\sqrt{s}=13$ TeV. More information about the $t\bar{t}+\gamma$ generation can be found in Sec.~\ref{sec:ttz:datadriven}.}
\label{tab:mc_samples1}
\end{table}

\clearpage

 \chapter{Object and Variable Definitions}
\label{chapter:susy:variables}
	
		The main difference between the stop search and a more inclusive search for squarks and gluinos is the particular stop pair production event topology resulting from high $p_\text{T}$ top quarks.  In the one-lepton channel, all reconstructable high $p_\text{T}$ objects are utilized: ($b$-tagged) jets, electrons, muons, photons, hadronically decaying $\tau$ leptons, and $E_\text{T}^\text{miss}$.  These objects are combined to form discriminating variables designed specifically for $t\bar{t}+E_\text{T}^\text{miss}$.  Section~\ref{sec:objects} provides an overview of the object reconstruction, including the procedures for resolving ambiguities in object labeling (overlap removal).  A detailed description of the discriminating variables used in the signal region optimization (Sec.~\ref{chapter:susy:signalregions}) follows in Sec.~\ref{sec:discriminating}.
		
		\clearpage
	
		\section{Object Selection}
		\label{sec:objects}
		
Many of the objects used by the stop search were already introduced in Part~\ref{parti} and Part~\ref{part:qpj}.  This section concisely describes each object and highlights the differences between the $\sqrt{s}=8$ analysis and the early Run 2 $\sqrt{s}=13$ TeV search.  The general strategy is to devise two sets of objects labeled {\it baseline} and {\it signal}, where the former are a subset of the latter.  Baseline objects are use in the ambiguity solving (overlap removal) and for vetoing events with a second reconstructed electron or muon.  Signal objects are used as inputs to the discriminating variables described in Sec.~\ref{sec:discriminating} and the final event selections.  Table~\ref{tab:objects} presents an overview of the object definitions, with references to more detailed documentation.  A quantitative comparison of the resolutions and reconstruction efficiencies is presented in Chapter~\ref{chapter:uncertainites}.  The paragraphs below briefly summarize the selections and notable changes between Runs 1 and 2.

The inputs to jet clustering at $\sqrt{s}=8$ TeV are calorimeter cell clusters with the LCW calibration while at $\sqrt{s}=13$ TeV, clusters directly at the electromagnetic scale are used for jet finding.  The EM-scale jet energy resolution is worse than LCW, but at $\sqrt{s}=13$ TeV, the global sequential calibration~\cite{ATLAS-CONF-2015-002} (EM+JES) reduces these differences.  An anti-$k_t$ radius parameter of $R=0.4$ is used for default jet clustering.  Large-radius jets based on re-clustering (see Sec.~\ref{sec:ReclusteredJetMass}) are used to identify boosted hadronically decaying top quarks and $W$ bosons for high mass stops.  These objects are discussed in more detail in Sec.~\ref{topmassreco}.  Due to the high energy nature of the target signal, the analysis is robust against the impact of pileup.  Nonetheless, at $\sqrt{s}=13$ TeV, the jet-vertex-tagger (JVT)~\cite{ATLAS-CONF-2014-018} based on the tracks associated to the jet used for $p_\text{T}^\text{jet}<50$ GeV to suppress spurious jets\footnote{For $p_\text{T}\sim 20$ GeV, about half the pileup jets are from random combinations of pileup interactions (stochastic pileup) while the other half are genuine quark and gluon jets (QCD pileup jets).  The fraction of QCD pileup jets increase with $p_\text{T}$ and the fraction of stochastic pileup jets increase with the number of pileup interactions.} from pileup interactions.  The JVT is configured for a $92\%$ hard-scatter jet efficiency, which corresponds to a pilep jet efficiency of about $1\%$.  Fake jets may also be generated by non-$pp$ collision processes such as calorimeter noise and beam-induced interactions with the imperfect beampipe vacuum.  Such jets are readily identified by various quality criteria such as the fraction of energy in the electromagnetic calorimeter compared with the fraction of the jet energy accounted for in reconstructed tracks~\cite{Aad:2014bia,ATLAS-CONF-2015-029}.  Since jets are used for many aspects of the event reconstruction, entire events are vetoed if any of these bad jets are identified.  The resulting efficiency for hard-scatter events is higher than $99\%$.  
Tracks and secondary/tertiary vertices associated with jets are also used to classify jets as resulting from $b$-quarks ($b$-tagged jets) using the MV1 (MV2c20) algorithm with 70\% (77\%) efficiency in simulated $t\bar{t}$ events~\cite{ATL-PHYS-PUB-2015-022,ATL-PHYS-PUB-2015-039,Aad:2015ydr} at $\sqrt{s}=8$ ($13$) TeV.   The `c20' in the Run 2 version indicates that the background composition in the algorithm training had 20\% charm-jets and $80\%$ light-flavor jets.  The $70\%$ MV1 working point has a light-quark jets rejection (=1/efficiency) of about 140 and a charm-quark jet rejection of about $5$.  For the same $b$-quark jet efficiency, the MV2c20 algorithm improves the light-quark jet rejection by about a factor of $4$ and the charm-quark jet rejection by about 50\%.  Part of this improvement came from the addition of the new pixel layer (IBL) and part from algorithmic improvements.  Scale factors are applied in the simulation to correct for differences in the efficiency between data and simulation.  The choice of the $b$-tagging working point was optimized for the search, as discussed in Sec.~\ref{chapter:susy:signalregions}.  

Jets and their associated tracks are additionally used to identify hadronically decaying $\tau$ leptons.  The distribution of energy inside the jet is combined with tracking information to form a multivariate classifier separately for $\tau$ leptons decaying into one ($\sim 85\%$) or three charged pions ($\sim 15\%$)~\cite{Aad:2014rga,ATL-PHYS-PUB-2015-025,ATL-PHYS-PUB-2015-045}.  At $\sqrt{s}=8$ TeV, a very loose working point was optimized that has a $3\%$ efficiency for $t\bar{t}$ events without a hadronically decaying $\tau$ and $31\%$ ($37\%$) efficiency for $t\bar{t}$ events with a one- (three-) prong $\tau$ decay.  This corresponds to a tight veto of $97\%$ in events without a hadronically decaying $\tau$.  At $\sqrt{s}=13$ TeV, tau jets are identified using the Loose identification algorithm~\cite{ATL-PHYS-PUB-2015-025,ATL-PHYS-PUB-2015-045} which has a 60\% and 50\% efficiency for reconstructing one- and three-prong $\tau$ decays, respectively.  In both Run 1 and Run 2, candidate hadronically decaying $\tau$ leptons are required to have no more than three tracks and if there three tracks, the sum of the track electric charges must be $\pm 1$.  All signal region event selections in Part~\ref{part:susy} require exactly one signal electron or muon; the $\tau$ candidate must have opposite electric charge to these leptons, unless there are two tracks and the net charge is zero.  Two track $\tau$ candidates are only permitted in the Run 1 analysis.  There is no distinction between baseline and signal reconstructed hadronically decaying $\tau$ leptons.

One of the most important background processes to the stop search is the pair production of top quarks resulting in two charged leptons, where one is not identified as such.  Therefore, it is advantageous to have the basline lepton definition be as inclusive as possible to efficiently veto such events.  The $p_\text{T}$ threshold for baseline leptons is $\leq 10$ GeV for both Run 1 and Run 2 and only loose quality criteria are imposed on the various track and electromagnetic calorimeter shower properties~\cite{ATLAS-CONF-2014-032,ATL-PHYS-PUB-2015-041,Aad:2014rra,Aad:2016jkr}. For the same efficiency, the Run 2 electron identification has a $\sim40\%$ larger background rejection due to the multivariate combination of reconstructed electron properties~\cite{ATLAS-CONF-2014-032}. Most high $p_\text{T}$ muons are measured by both the inner detector (ID) and the muon spectrometer (MS) resulting in combined muon candidates (CB).  Muons beyond the ID acceptance are selected using muons reconstructed only with the MS (stand-alone, SA) and the efficiency is recovered for $|\eta|<0.1$, where the MS is only partially instrumented due to calorimeter and ID services, by using ID tracks matched to either calorimeter energy deposits consistent with a minimum ionizing particle (calo-tagged, CT) or a track segment in the MS (segment-tagged, ST).  The electrons or muons from stops are predicted to be significantly harder than in background processes and so the signal leptons are required to have $p_\text{T}>25$ GeV.  This high threshold is also useful for suppressing QCD multijet backgrounds and is required for the $\sqrt{s}=8$ TeV analysis for the lepton trigger to be fully efficient. In addition, signal leptons must pass several quality criteria on their transverse ($d_0$) and longitudinal ($z_0$) impact parameters (IP).  In particular, $|d_0|<0.2$ mm and $z_0<1$ mm for muons at $\sqrt{s}=8$ TeV and $d_0/\sigma_{d_0} <5$ ($3$) and $z_0\sin(\theta) < 0.5$ mm for electrons (muons) at $\sqrt{s}=13$ TeV.  Leptons from $W$ boson decays are generally well-separated from other objects in the event, so other processes can be suppressed by imposing isolation criteria. These criteria are based on the scalar sum of the $p_\text{T}$ from tracks (excluding the electron or muon track) within a cone around the lepton.  When the top quarks from stop decays are produced with sufficient boost, the leptons are naturally close to the $b$-jet from the same top decay.  This is illustrated by Fig.~\ref{fig:deltaRlb}, which shows the joint distribution of the $\Delta(\ell,b)$ and stop mass.  To maintain efficiency for high stop masses, the isolation cone scales with the inverse of the lepton $p_\text{T}$ (see Chapter~\ref{cha:bosonjets}). 
\begin{figure}[h!]
\begin{center}
\includegraphics[width=0.5\textwidth]{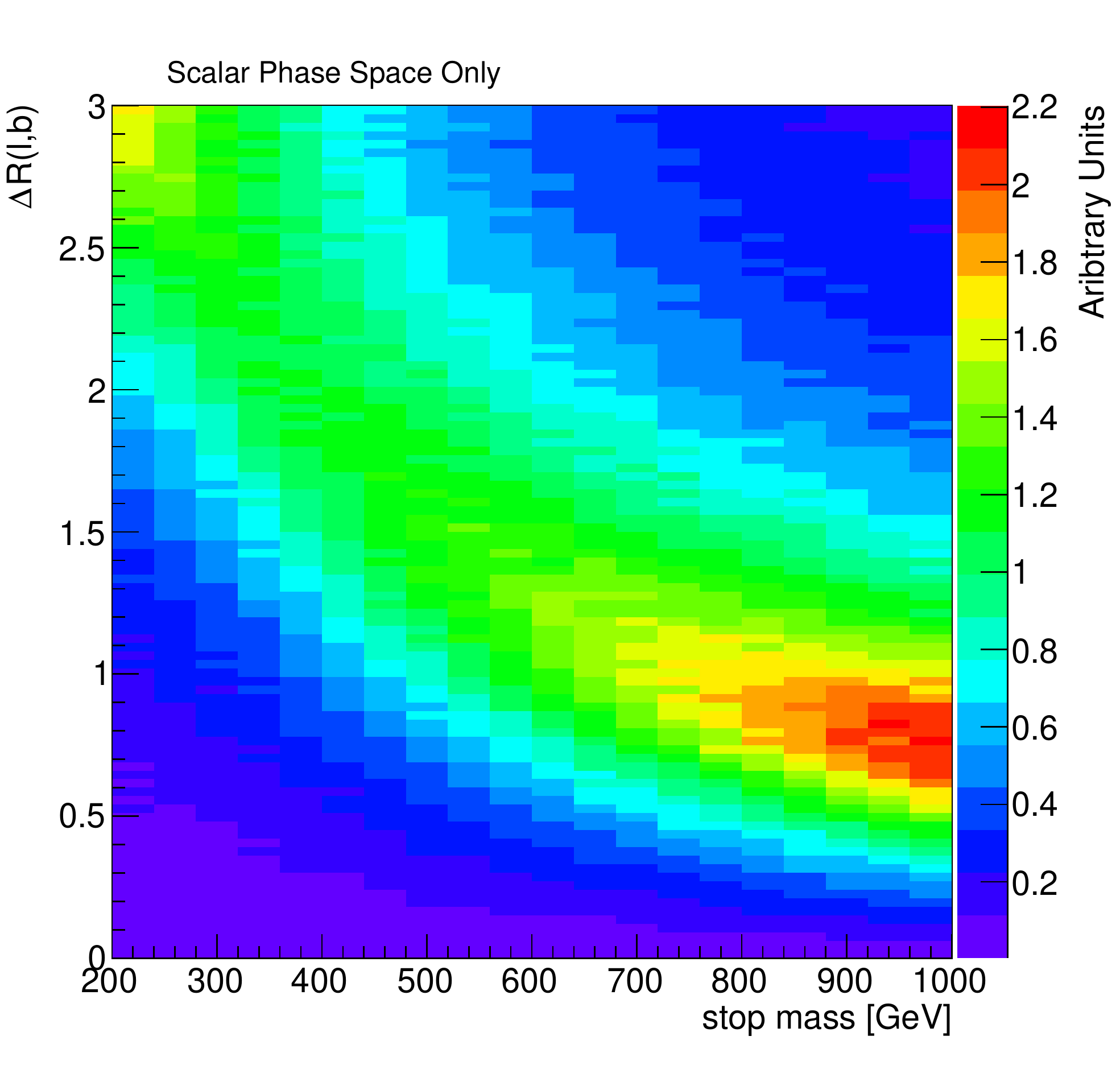}
 \caption{The joint distribution of the $\Delta R$ between the lepton and $b$-quark from the same top quark decay for $\tilde{t}\rightarrow t\tilde{\chi}^0$.  At $m_\text{stop}=500$ GeV, about $2\%$ ($10\%$) of events have $\Delta R<0.4$ ($0.8$) and at $m_\text{stop}=800$ GeV, this fraction increases to $\sim5\%$ ($25\%$).}
 \label{fig:deltaRlb}
  \end{center}
\end{figure}

Explicitly reconstructed isolated photons are only considered if their $p_\text{T}\gtrsim m_Z$ in order to mimic high $p_\text{T}$ $Z$-bosons as described in Sec.~\ref{sec:ttz:datadriven}. The actual threshold is chosen to be as high as possible while maintaining a sufficient event yield.  If not explicitly reconstructed, isolated photons will be labeled as jets.  At $\sqrt{s}=13$ TeV, the threshold was slightly raised with respect to the Run 1 search due to the use of the photon trigger that is nearly $100\%$ efficient\footnote{The single lepton triggers are never fully efficient - see Sec.~\ref{sec:datasetandMC}.} for $p_\text{T}^\gamma=125$ GeV.  There are many sources of high $p_\text{T}$ photons inside jets that are not useful for the $t\bar{t}+Z$ background estimation.  The Run 1 analysis did not impose any explicit isolation requirements, but photons do participate in the overlap removal (see below).  At $\sqrt{s}=13$ TeV, in addition to the overlap removal, an explicit isolation is imposed on the calorimeter energy inside a $\Delta R$ cone around the photon.  As with hadronically decaying $\tau$ leptons, there is no difference between signal and baseline photons.

All of the objects described above are used to form the $\vec{p}_\text{T}^\text{miss}$.  Jets, electrons, and muons that enter the calculation use the dedicated calibrations for those objects.  Hadronically decaying $\tau$ leptons enter in the calculation as electrons or jets without the dedicated $\tau$ calibration.  At $\sqrt{s}=8$ TeV, calibrated photons with $p_\text{T}>10$ GeV explicitly enter the $\vec{p}_\text{T}^\text{miss}$ calculation.  For Run 2, only the high $p_\text{T}$ photons for the $t\bar{t}+\gamma$ control region described in Sec.~\ref{sec:ttz:datadriven} enter the calculation as photons; otherwise photons are part of the jet collection.  The main difference between Runs 1 and 2 is that the former uses a calorimeter-based soft-term~\cite{Aad:2012re,ATLAS-CONF-2013-082} for the unclustered energy while the latter uses a track-based soft-term~\cite{ATL-PHYS-PUB-2015-027,ATL-PHYS-PUB-2015-023}.  Information about soft neutral particles is lost in the track-based soft term, but tracks are largely insensitive to pileup.  For the early Run 2 conditions, the insensitivity to pileup is the dominant effect and the $E_\text{T}^\text{miss}$ with a track-based soft-term has a $\sim 10\%$ better resolution than the calorimeter-based soft-term.

\vspace{10mm}

\begin{table}[h!]
  \centering
  \noindent\adjustbox{max width=\textwidth}{
  \vspace{3mm}
  \begin{tabular}{|c|c|c|c|c|c|}
	\hline
Object & $p_\text{T}>$ [GeV] & $|\eta|<$ & Identification & Isolation & Other\\
\hline
\hline
Baseline Jets & 20 & 2.8 & Looser${}^{*}$~\cite{Aad:2014bia} & -- & LCW\\
& {\color{blue}20}&{\color{blue}none} &{\color{blue}BadLoose}${}^{}$~\cite{ATLAS-CONF-2015-029}  & -- & {\color{blue}EM+GSC}\\
Signal Jets & 25 & 2.5 & same as baseline & -- & \\
&{\color{blue}25}&{\color{blue}2.5}&{\color{blue}JVT@92\%}~\cite{ATLAS-CONF-2014-018}&--&\\
$b$-tagged Jets & 25 & 2.5 & MV1@70\%~\cite{Aad:2015ydr} & -- & \\
& {\color{blue}25}& {\color{blue}2.5} & {\color{blue} MV2c20@77\%}~\cite{ATL-PHYS-PUB-2015-022,ATL-PHYS-PUB-2015-039} & -- & \\
\hline
Baseline Electron &10&2.47&Loose~\cite{ATLAS-CONF-2014-032}&none& \\
 &{\color{blue}7}&{\color{blue}2.47}&{\color{blue}VeryLooseLH}${}^{}$~\cite{ATL-PHYS-PUB-2015-041}&{\color{blue}none}& \\
Signal Electron &25&2.47&Tight~\cite{ATLAS-CONF-2014-032}&$p_\text{T}^{\Delta R<0.2}/p_\text{T}^e<0.1$& \\
 &{\color{blue}25}&{\color{blue}2.47}&LooseLH~\cite{ATL-PHYS-PUB-2015-041}+IP&{\color{blue}$p_\text{T}^{\Delta R<\max\{10\text{ GeV}/p_\text{T}^e,0.2\}}$@99\%}& \\
\hline
Baseline Muon &10&2.4&CB+ST~\cite{Aad:2014rra}&none& \\
&{\color{blue}6}&{\color{blue}2.7}&{\color{blue}Loose CB+ST+CT+SA}~\cite{Aad:2016jkr}&{\color{blue}none}& \\
Signal Muon &25&2.4&IP (see the text)&$p_\text{T}^{\Delta R<0.2}<1.8$ GeV& \\
 &{\color{blue}25}&{\color{blue}2.7}&{\color{blue}IP (see the text)}&{\color{blue}$p_\text{T}^{\Delta R<\max\{10\text{ GeV}/p_\text{T}^\mu,0.3\}}$@99\%}& \\
\hline
Photons & 125 & 2.37 & Tight~\cite{ATLAS-CONF-2012-123,Aad:2010sp} & none (see OR) & \\
&{\color{blue}100}&{\color{blue}2.37}&{\color{blue}Tight}~\cite{ATLAS-CONF-2012-123,Aad:2010sp}&{\color{blue}$E_\text{T}^{\Delta R<0.4}/ < 0.022\text{ }p_\text{T}^\gamma + 2.45/\text{GeV}$}&\\
Hadronic $\tau$ &15&2.47&Jet BDT@31\%-37\%${}^{}$~\cite{ATLAS-CONF-2011-152,ATLAS-CONF-2012-142}&none& \\
&{\color{blue}20}&{\color{blue}2.5}&{\color{blue}Loose}~\cite{ATL-PHYS-PUB-2015-025,ATL-PHYS-PUB-2015-045}&{\color{blue}none}&\\
$E_\text{T}^\text{miss}$ &--&--&--&--& CST~\cite{ATLAS-CONF-2013-082} \\
&--&--&--&--&TST~\cite{ATL-PHYS-PUB-2015-027}\\
\hline
  \end{tabular}}
  \caption{An overview of the objects definitions used for the stop search.  The color blue indicates the criteria at $\sqrt{s}=13$ TeV whereas black is for $\sqrt{s}=8$ TeV.  See the text for details.} 
   \label{tab:objects}
\end{table}

\clearpage

Baseline and signal jets and leptons considered for further use must pass an object ambiguity resolving algorithm.  These algorithms are tailored for the stop search and optimized using particle-level labels in simulation to maintain a low mis-classification rate in order to increase the one lepton signal and reduce the significant two-lepton background.  Tables~\ref{tab:OR8} and~\ref{tab:OR13} summarize these overlap removal (OR) procedures at $\sqrt{s}=8$ TeV and $\sqrt{s}=13$ TeV, respectively.  For example, if an electron and a jet overlap with $\Delta R<0.2$ and the jet is not $b$-tagged, then the object is interpreted as an electron and the overlapping jet is removed from the list of jets.  Overlap between leptons and $b$-jets is treated differently than inclusive jets because semi-leptonic $B$ decays can naturally produce overlapping leptons.  At high stop mass, signal leptons can also naturally be close to jets.  For this reason, at $\sqrt{s}=13$ TeV, a $p_\text{T}$-dependent $\Delta R$ cone is used to remove leptons in favor of jets.  Another change from Run 1 to Run 2 is the muon / non-$b$ jet overlap condition which only applies if the jet has less than three associated tracks with $p_\text{T}>500$ MeV or $p_\text{T}^\text{muon}/p_\text{T}^\text{jet} > 0.7$.  The track requirement removes jets that are seeded by muon radiation and the $p_\text{T}$ asymmetry requirement identifies jets that are unassociated with the muon, which is unlike the case for electrons that deposit most of their energy in the calorimeter and thus to the close-by jet.  Due to the use of calo-tagged muons at $\sqrt{s}=13$ TeV, there can be some overlap between electrons and muons near $|\eta|\approx 0$.  In nearly all cases, the simulation predicts that these objects are due to true electrons.  The overlap between electrons and photons has a significant improvment in the photon purity in the $t\bar{t}+\gamma$ control region (Sec.~\ref{sec:ttz:datadriven}).

\begin{table}[h!]
  \centering
  \noindent\adjustbox{max width=\textwidth}{
  \vspace{3mm}
  \begin{tabular}{|cccccc|}
	\hline
    Object 1     & e & $\mu$ & $l$  & $\gamma$ & $\tau$        \\
    Object 2  & $j$ & $j$ & $j$ & $j$ & $l$\\
    \hline
    $\Delta R<$  & 0.2 & 0.4 & 0.4 & 0.2 & 0.2 \\
    \hline
    Condition  & $j$ not $b$-tagged &  & -- & -- &  --\\
    \hline
    Resolution  & e & $\mu$ & $j$ & $\gamma{}^{*}$ & $l$ \\
\hline
  \end{tabular}}
  \caption{A summary of the procedure to resolve ambiguous object labels at $\sqrt{s}=8$ TeV.  The first two rows list the overlapping objects: electrons (e), muons ($\mu$), electron or muon ($l$), jets ($j$), photons ($\gamma$), and hadronically decaying $\tau$ leptons ($\tau$).  The third and fourth rows give the overlap condition and the last row lists which label is given to the ambiguous object.  The procedure is applied from left to right.  ($*$) Jets are removed only in the $t\bar{t}+\gamma$ validation region (see Sec.~\ref{sec:ttz:datadriven}).}
  \label{tab:OR8}
\end{table}

\begin{table}[h!]
  \centering
  \noindent\adjustbox{max width=\textwidth}{
  \vspace{3mm}
  \begin{tabular}{|cccccccc|}
	\hline
    Object 1 & e     & e & $\mu$ & $l$  & $\gamma$& $\gamma$ & $\tau$        \\
    Object 2 & $\mu$ & $j$ & $j$ & $j$ & $j$ & e & e\\
    \hline
    $\Delta R<$ & 0.1 & 0.2 & 0.2 &$\min\left(0.4,0.04+\frac{10}{p_\text{T}^l}\right)$ & 0.2 & 0.1 & 0.1 \\
    \hline
    Condition & CT $\mu$ & $j$ not $b$-tagged & $j$ not $b$-tagged and & -- & -- & -- & --\\
    & & & $n_\text{track}^j<3$ or $\frac{p_\text{T}^\mu}{p_\text{T}^j }> 0.7$ &&&&\\
    \hline
    Resolution & e & e & $\mu$ & $j$ & $\gamma$ & e & e \\
\hline
  \end{tabular}}
  \caption{A summary of the procedure to resolve ambiguous object labels at $\sqrt{s}=13$ TeV.  The first two rows list the overlapping objects: electrons (e), muons ($\mu$), electron or muon ($l$), jets ($j$), photons ($\gamma$), and hadronically decaying $\tau$ leptons ($\tau$).  The third and fourth rows give the overlap condition and the last row lists which label is given to the ambiguous object.  The procedure is applied from left to right.}
  \label{tab:OR13}
\end{table}

		\clearpage	
			
		\section{Discriminating Variables}
		\label{sec:discriminating}
		
		The key to a powerful and robust search is the use of a relatively small number\footnote{An alternative paradigm is to process all available information using sophisticated machine learning techniques (see Sec.~\ref{sec:HEPML}).  This can be a powerful approach, but for {\it tail searches} such as this one, by construction there is little data available near the signal region to thoroughly validate such methods.   Therefore, robust and powerful methods grounded in physical intuition are preferred. } of highly discriminating variables.  There are two strategies when developing variables.  One possibility is to target particular aspects of the background that are absent in the signal ({\it veto}).  A second tactic focuses on properties of the signal that are absent in the background ({\it tag}).  This section explores a series of veto and tag variables, many of which are specifically designed for the stop search and used here for the first time.  One of the key themes in the development of the variables is a focus to use {\it tailored variables} when possible.  Many simple variables such as $H_\text{T}=\sum_i p_\text{T, jet $i$}$ and $m_\text{eff}=E_\text{T}^\text{miss}+H_\text{T}+p_\text{T}^\text{lepton}$ are generically useful for signatures with high multiplicity final states involving multiple missing particles.  However, the price of simplicity is sub-optimality in particular situations such as the $t\bar{t}+E_\text{T}^\text{miss}$ topology.  One of the most powerful variables is the transverse mass ($m_\text{T}$), which has been mentioned at several points in earlier chapters.  Section~\ref{sec:transmass} describes $m_\text{T}$ in detail in order to demonstrate exactly how and why it is useful for the stop search.  A generalization of $m_\text{T}$ to cases when there are multiple missing particles is the $m_\text{T2}$ family of observables (Sec.~\ref{sec:mt2}).  Background events can be reconstructed with large $m_\text{T}$ and $m_\text{T2}$ when jets are sufficiently mis-measured.  Section~\ref{sec:significancevariables} described how resolution information can be incorporated into kinematic variables to suppress these events.  Another veto variable that can be combined with kinematic information is hadronic tau identification (Sec.~\ref{tauid}).  A significant fraction of $t\bar{t}$ contain a hadronically decaying $\tau$ to pass harsh requirements on $m_\text{T}$.  The section ends with a brief discussion of boosted top quark and $W$ boson tagging techniques (Sec.~\ref{topmassreco}).  Some of the techniques already described in Chapter~\ref{cha:bosonjets} are directly applicable to the stop search.
	
			\clearpage
		
			\subsection{Transverse Mass Variables}
			\label{sec:transmass}

			One of the most striking characteristics of top quarks and the targeted particles in theories beyond the SM is their large mass.  If all particles from these heavy particle decays could be reconstructed and unambiguously identified, the invariant mass would be a powerful variable.  However, due to neutrinos and neutralinos, a significant fraction of the resonance mass goes into undetected energy.   At a hadron collider, the $\sqrt{\hat{s}}$ is unknown and therefore only the sum of the transverse momentum of the undetected particles can be inferred.  Transverse mass variables are modifications of the usual invariant mass to cases where there is at least one undetected particle and the total longitudinal momentum is unknown.  Even though longitudinal information is missing, the transverse mass of the decay products of massive particles tends to be higher than for background processes.  This fact was first used to discover the $W$ boson  at CERN by the UA1~\cite{Arnison:1983rp} and UA2~\cite{Banner:1983jy} collaborations.  The left plot of Fig.~\ref{fig:susymtdist0} shows the transverse mass spectrum for the first six $W$ boson candidate events. 
						
\begin{figure}[h!]
\begin{center}
\includegraphics[width=0.5\textwidth]{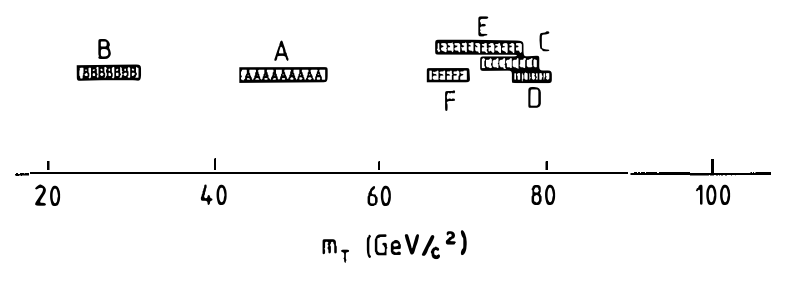}\includegraphics[width=0.5\textwidth]{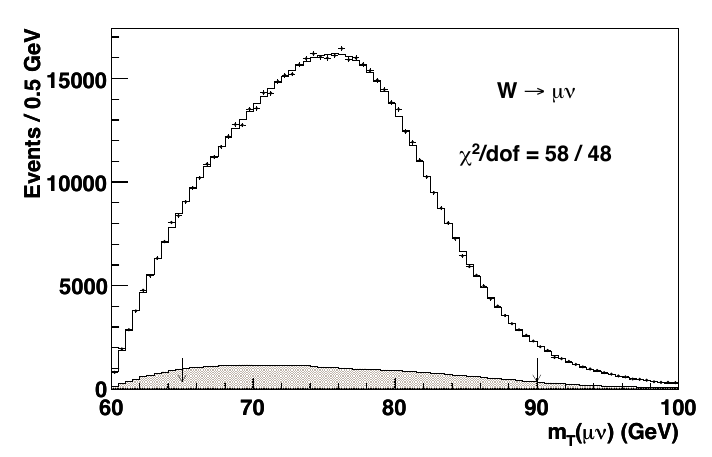}
 \caption{Left: The transverse mass distribution for the discovery of the $W$ boson by the UA1 collaboration~\cite{Arnison:1983rp}. Right:  the transverse mass distribution used by the CDF collaboration for one of the most precise determinations of the $W$ boson mass~\cite{Aaltonen:2012bp}.}
 \label{fig:susymtdist0}
  \end{center}
\end{figure}		

 The most important feature of transverse mass variables is that they are bounded by the resonance mass and tend to have a probability distribution that is concentrated near the kinematic endpoint.  These  feature allowed the UA1 and UA2 collaborations to estimate the mass of the $W$ boson and is still used today for the most precise determination of the $W$ boson mass by the Tevatron collaborations~\cite{Abazov:2012bv,Aaltonen:2012bp}.  The right plot of Fig.~\ref{fig:susymtdist0} shows an example from CDF~\cite{Aaltonen:2012bp} with $\mathcal{O}(10^6)$ events.  This bounded property of the transverse mass variables helps to suppress the $W$ boson and top quark backgrounds that have transverse mass well above other backgrounds, but well below the scale of new physics.  Section~\ref{sec:mT} contains a detailed description of the transverse mass in topologies with one missing particle.
	
	\begin{figure}[h!]
\begin{center}
\includegraphics[width=0.85\textwidth]{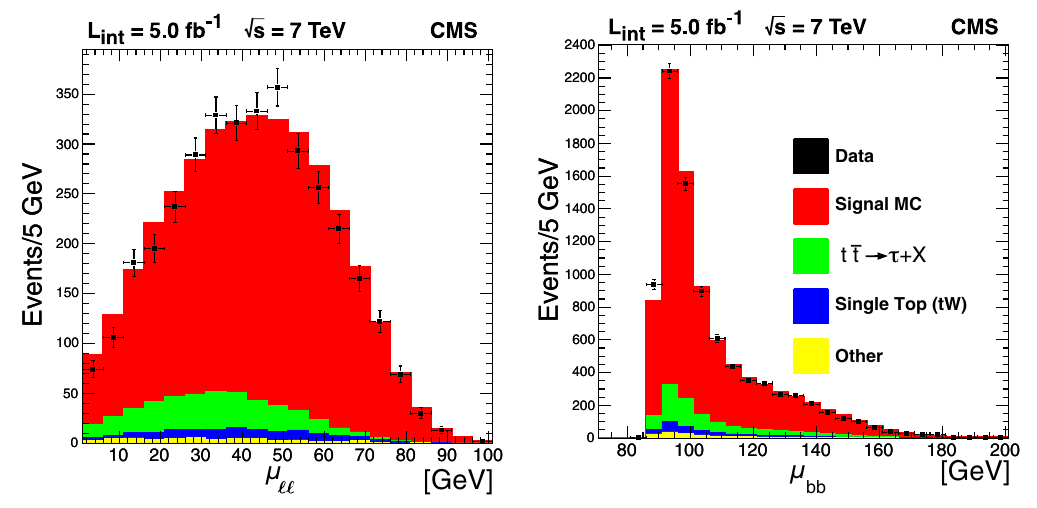}
 \caption{The distribution of $m_\text{T2}$ in dilepton $t\bar{t}$ events using the leptons ($b$-tagged jets) as the visible particles in the left (right) plot.  These distributions are used to measure the top quark, $W$ boson, and neutrino masses.  Published in Ref.~\cite{Chatrchyan:2013boa}.}
 \label{fig:susymtdist0mt2}
  \end{center}
\end{figure}	

When a decay chain has multiple missing particles, there is an ambiguity in the assignment of momentum to each particle because only the sum of their transverse momentum is measured by $\vec{p}_\text{T}^\text{miss}$.  However, in this case there are still ways to bound the parent particle mass by computing the minimum mass consistent with the observed decay products.  This is basis for a generalization of the transverse mass called $m_\text{T2}$ and is described in detail in Sec.~\ref{sec:mt2}.  Like the simple transverse mass described above, $m_\text{T2}$ will have the property that for $t\bar{t}$ events it is relativel large, but bounded well below the scale for signal.  Figure~\ref{fig:susymtdist0mt2} shows two $m_\text{T2}$ variables in dileptonic $t\bar{t}$ events.  The left (right) plot of Fig.~\ref{fig:susymtdist0mt2} is bounded by $m_W$ ($m_\text{top}$), a fact which is used to measure these masses just as the simple transverse mass is used to measure $m_W$ in inclusive $W$+jets events.
			
				\clearpage
			
				\subsubsection{Transverse Mass}
				\label{sec:mT}
				
				Consider two particles with four-momentum $p^\mu=(p_x,p_y,p_z,E)$ and $q^\mu=(q_x,q_y,q_z,F)$.   Define $\tilde{p}^\mu=(p_x,p_y,0,\sqrt{E^2-p_z^2})$ and $\tilde{q}_\mu=(q_x,q_y,0,\sqrt{F^2-q_z^2})$.  The {\it transverse mass} is defined by $m_\text{T}^2=(\tilde{p}^\mu+\tilde{q}^\mu)(\tilde{p}_\mu+\tilde{q}_\mu)$.  By construction, the transverse mass is invariant under longitudinal boosts.  If the particles have masses $m_p$ and $m_q$ and are the decay products of a two-body resonance decay with mass $M$ then,
				
				\begin{align}\nonumber
				M^2&=m_p^2+m_q^2+2(EF-p_zq_z-p_\text{T}q_\text{T})\\\nonumber
				&=m_p^2+m_q^2+2\left(E_\text{T}F_\text{T}\cosh(\Delta y)-p_\text{T}\cdot q_\text{T}\right)\\\nonumber
				&\geq m_p^2+m_q^2+2\left(E_\text{T}F_\text{T}-p_\text{T}\cdot q_\text{T}\right)\\\nonumber
				&=\left(\sqrt{E^2-p_z^2}+\sqrt{F^2-q_z^2}\right)^2-(p_x+q_x)^2-(p_y+q_y)^2\\				&=m_\text{T},\label{eq:fullmt}
				\end{align}
			
				\noindent where $E_\text{T}^2=m_p^2+p_\text{T}^2$ and $\Delta y$ is the difference in rapidity of the two particles.  The second line follows from trigonometry identities and tedious arithmetic to show that $p_z=E_\text{T}\sinh(y)$ ($p_z=p_\text{T}\sinh(\eta))$ and so $E=E_\text{T}\cosh(y)$ ($|p|=p_\text{T}\cosh(\eta)$).  Suppose that the resonance is a $W$ boson decaying into $W\rightarrow e\nu$.  Since $m_e,m_\nu\ll m_W$, $E\approx |\vec{p}|$ and $F\approx |\vec{q}|$ is an excellent approximation.   Therefore, $m_\text{T}^2\approx 2p_{T,e}p_\text{T,$\nu$}(1-\cos\theta_{e\nu})$, where $\theta_{e\nu}$ is the angle between $\vec{p}_\text{T,2}$ and $\vec{p}_\text{T,$\nu$}$.  Since the neutrino is not directly detected, the definition of the transverse mass used in practice is
				
		\begin{align}
		\label{eq:mtdefinition}
		m_\text{T}^2=2E_\text{T}^\text{miss}p_\text{T}^\text{lepton}\left(1-\cos\Delta\phi\left(\vec{E}_\text{T}^\text{miss},\vec{p}_\text{T}^\text{lepton}\right)\right).
		\end{align}
				
\noindent In the $W$ boson rest frame, $p_\text{T,e}=p_\text{T,$\nu$}=\frac{m_w}{2}|\sin\theta|$, where $\theta$ is the angle of the electron-neutrino axis from the $z$-axis.  With this formulation, it is clear that $m_\text{T}\leq m_W$ with equality only if $\theta=\pm \pi/2$.  Since the differential volume element in spherical coordinates is $dV=\rho^2\sin\theta d\rho d\theta d\phi=\rho^2d\rho d(\cos\theta)d\phi$, an isotropic distribution for the decay\footnote{The scalar decay is used for illustration; in reality the $W$ boson is a spin-1 particle.  In the $W$ boson rest frame, $dN/d\cos\theta\propto (1\pm\cos\theta)^2$ for transversely polarized bosons (with spin $\pm 1$) and $dN/d\cos\theta\propto \sin^2\theta$ for longitudinally polarized bosons (for the spin axis along $z$).  These correspond to $f_{\cos\theta}(x)=3(1\pm x)^2/8$ and $f_{\cos\theta}(x)=3(1-x^2)/4$, respectively.  $W$ bosons produced in top decays are mostly longitudinal whereas inclusive $W$ boson production results in mostly transversely polarized bosons.  See Appendix~\ref{sec:app:wpolarization}.} in the $W$ rest frame results in the following probability distribution for $m_\text{T}$:

				\begin{align}\nonumber
				f_\text{$m_\text{T}$}(m_\text{T})&=\sum_{\theta \geq \pi/2,\theta < \pi/2}\Bigg|\frac{\partial}{\partial m_\text{T}}m_\text{T}^{-1}(m_\text{T})\Bigg|f_{\cos\theta}(m_\text{T}^{-1}(m_\text{T}))\\\nonumber
				&=\sum_{\theta \geq \pi/2,\theta < \pi/2}\frac{1}{m_W}\frac{(m_\text{T}/m_W)}{\sqrt{1-(m_\text{T}/m_W)^2}}\times \frac{1}{2}\\
				&=\frac{1}{m_W}\frac{(m_\text{T}/m_W)}{\sqrt{1-(m_\text{T}/m_W)^2}}\label{eq:mtdist}
				\end{align}
				
				\noindent where $m_\text{T}(x)=m_W\sqrt{1-x^2}$ so $m_\text{T}^{-1}(y)=\pm\sqrt{1-(y/m_W)^2}$, one for each branch: $\theta \geq \pi/2,\theta < \pi/2$.  The probability distribution $f_{\cos\theta}=\frac{1}{2}$ because $\cos\theta$ is uniform on $[-1,1]$.  The sum in Eq.~\ref{eq:mtdist} is over the two values of $\cos\theta$ that result in the same $m_\text{T}$ value.  The most striking feature of Eq.~\ref{eq:mtdist} as a function is that it is monotonically increasing with its maximum at $m_\text{T}=m_W$.  Since the term in the absolute value in the first line of Eq.~\ref{eq:mtdist} is the Jacobian of the variable transformation, this peak is called the {\it Jacobian peak}.  The left plot of Fig.~\ref{fig:susymtdist} shows the distribution of the transverse mass for a $W$ boson produced at rest.  By construction, the analytical formula derived in Eq.~\ref{eq:mtdist} is identical to the red filled histogram for simulated (scalar) $W\rightarrow e\nu$ events.  For comparison, additional distributions are shown in Fig.~\ref{fig:susymtdist} for cases in which there are additional neutrinos in the event.  In these cases, Eq.~\ref{eq:mtdefinition} is used as the definition of the transverse mass, where the $E_\text{T}^\text{miss}$ includes all non-reconstructed particle momenta.  For leptonic $\tau$ decays $W\rightarrow \tau\nu_\tau, \tau\rightarrow e\nu_e\nu_\tau$, the endpoint of the $m_\text{T}$ distribution is still $m_W$, but the probability distribution is mostly concentrated at low values of the transverse mass.  Since events with only one true lepton-neutrino pair have an $m_\text{T}\leq m_W$, most of the events with $m_\text{T}>m_W$ originate from events with two true leptons, one of which is out of acceptance (lost) or reconstructed as a jet or is a hadronically decaying $\tau$ lepton (mis-id).  When the second lepton is lost, its momentum is part of the $E_\text{T}^\text{miss}$ while if it is mis-identified, its momentum is not directly\footnote{All the visible momenta are used to construct the $E_\text{T}^\text{miss}$, so these mis-identified leptons will contribute indirectly.} part of the $E_\text{T}^\text{miss}$.   Dilepton events are simulated by independently generating two $W$ bosons.  When these bosons are produced at rest, the momentum of the lost lepton cancels with its neutrino pair and so the $m_\text{T}$ spectrum is identical to the single $W\rightarrow e\nu$ case.  However, if the second lepton is mis-identified, then the $m_\text{T}$ distribution can exceed $m_W$.  The kinematic maximum is achieved when the decay axis of both $e$-$\nu$ pairs are parallel and in the transverse plane.  In that case, $m_\text{T}^\text{max} = \sqrt{2}m_W\approx $ 113 GeV.  Figure~\ref{fig:susymtdistspin1} shows the distribution of $m_\text{T}$ in simulated events with a full spin-1 $W$ boson.  All cases have the Jacobian peak, but the tail of the distribution toward zero depends on the spin.
							
\begin{figure}[h!]
\begin{center}
\includegraphics[width=0.5\textwidth]{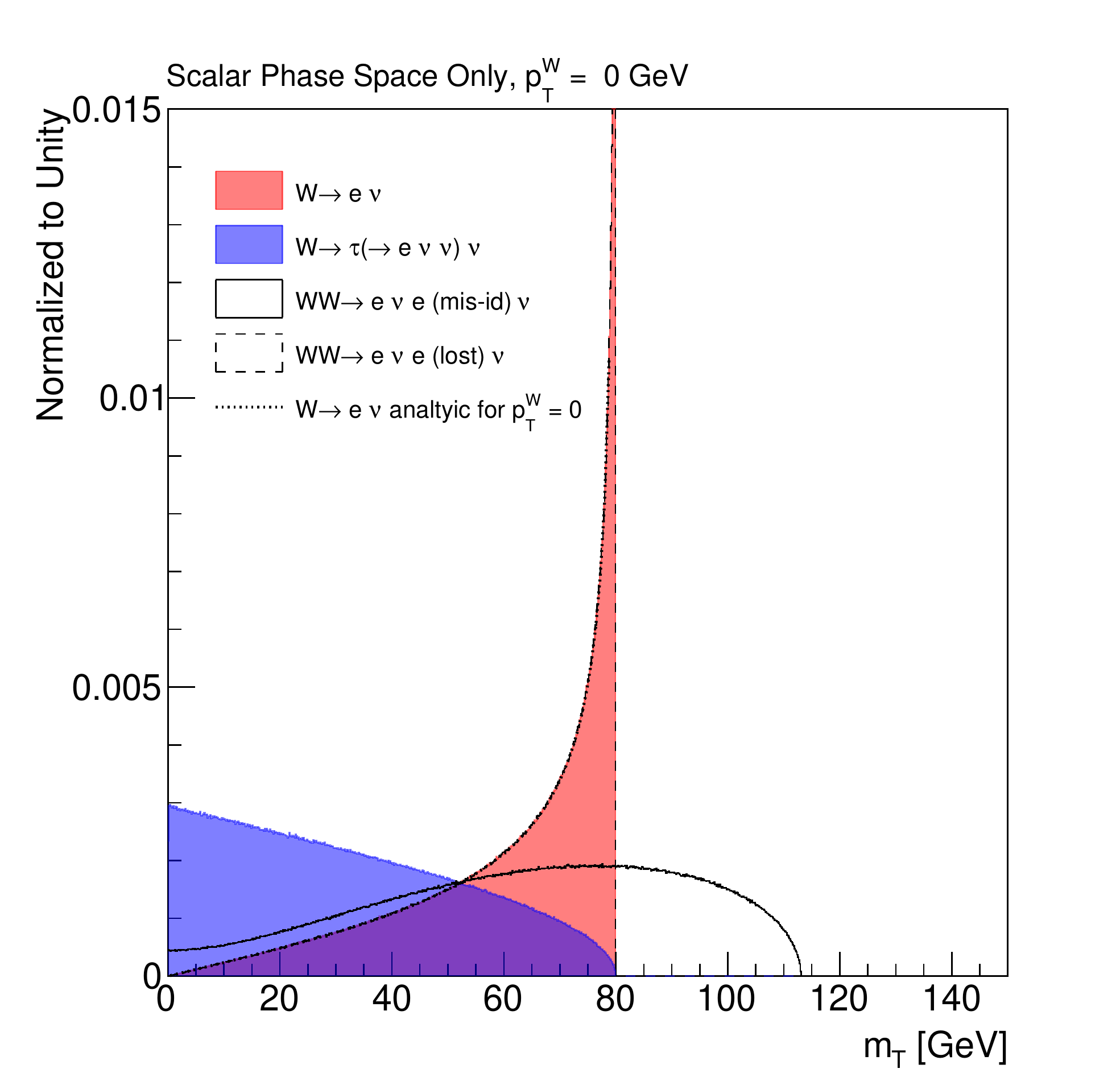}\includegraphics[width=0.5\textwidth]{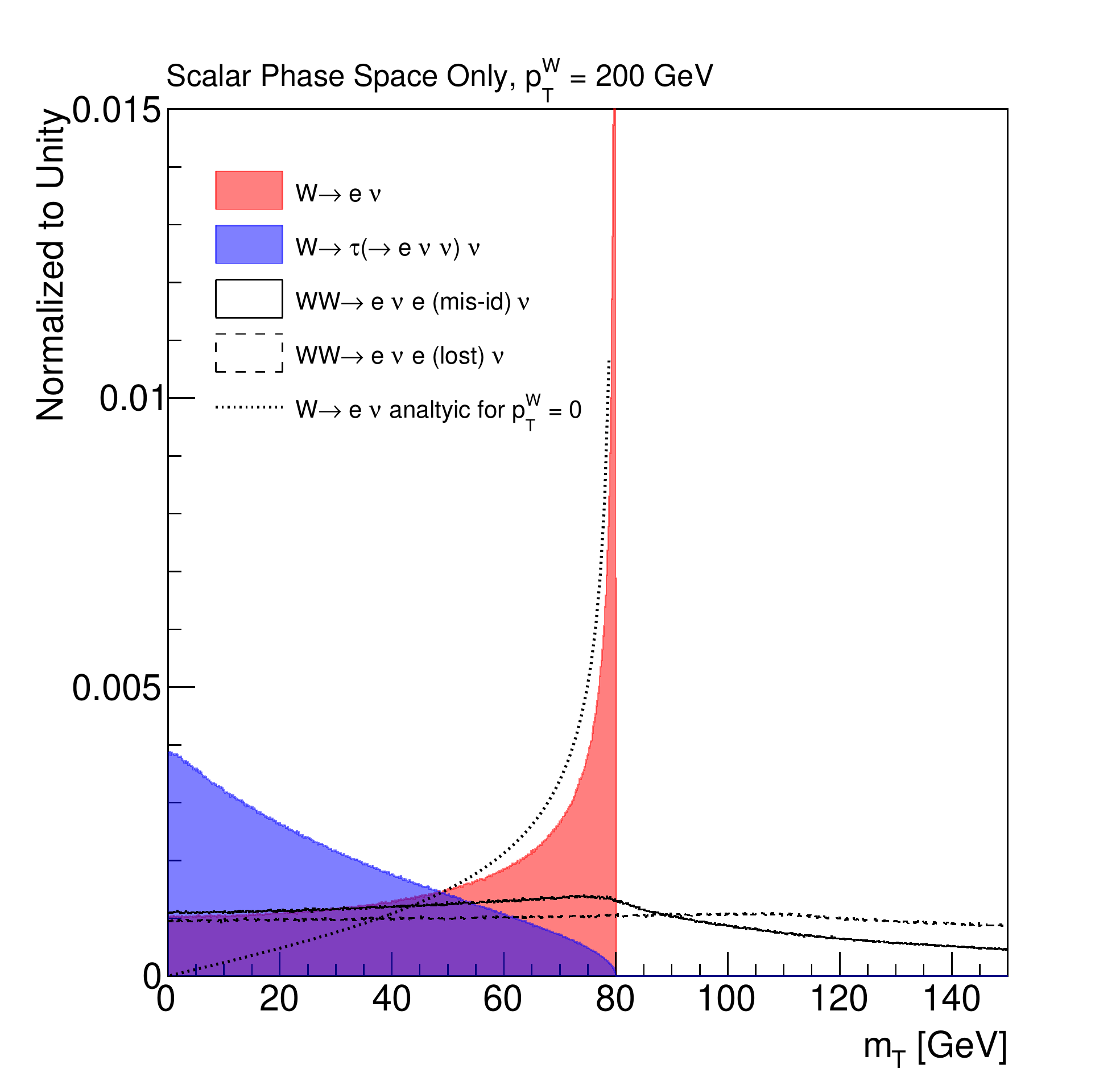}
 \caption{The distribution of the transverse mass for the isotropic decay of a $W$ boson for $p_\text{T}^W=0$ GeV on the left and $p_\text{T}^W=200$ GeV on the right. The red histogram shows the $m_\text{T}$ distribution for the decay $W\rightarrow e\nu$ and the blue histogram shows the distribution for $W\rightarrow \tau\nu$ where the $\tau$ decays to an electron and neutrinos.  The black histograms correspond to the pair production of $W\rightarrow e\nu$ where the second lepton is either included (lost) or not included (mis-id) in the $E_\text{T}^\text{miss}$.  The dotted line is the analytic formula derived in the text.}
 \label{fig:susymtdist}
  \end{center}
\end{figure}

\begin{figure}[h!]
\begin{center}
\includegraphics[width=0.5\textwidth]{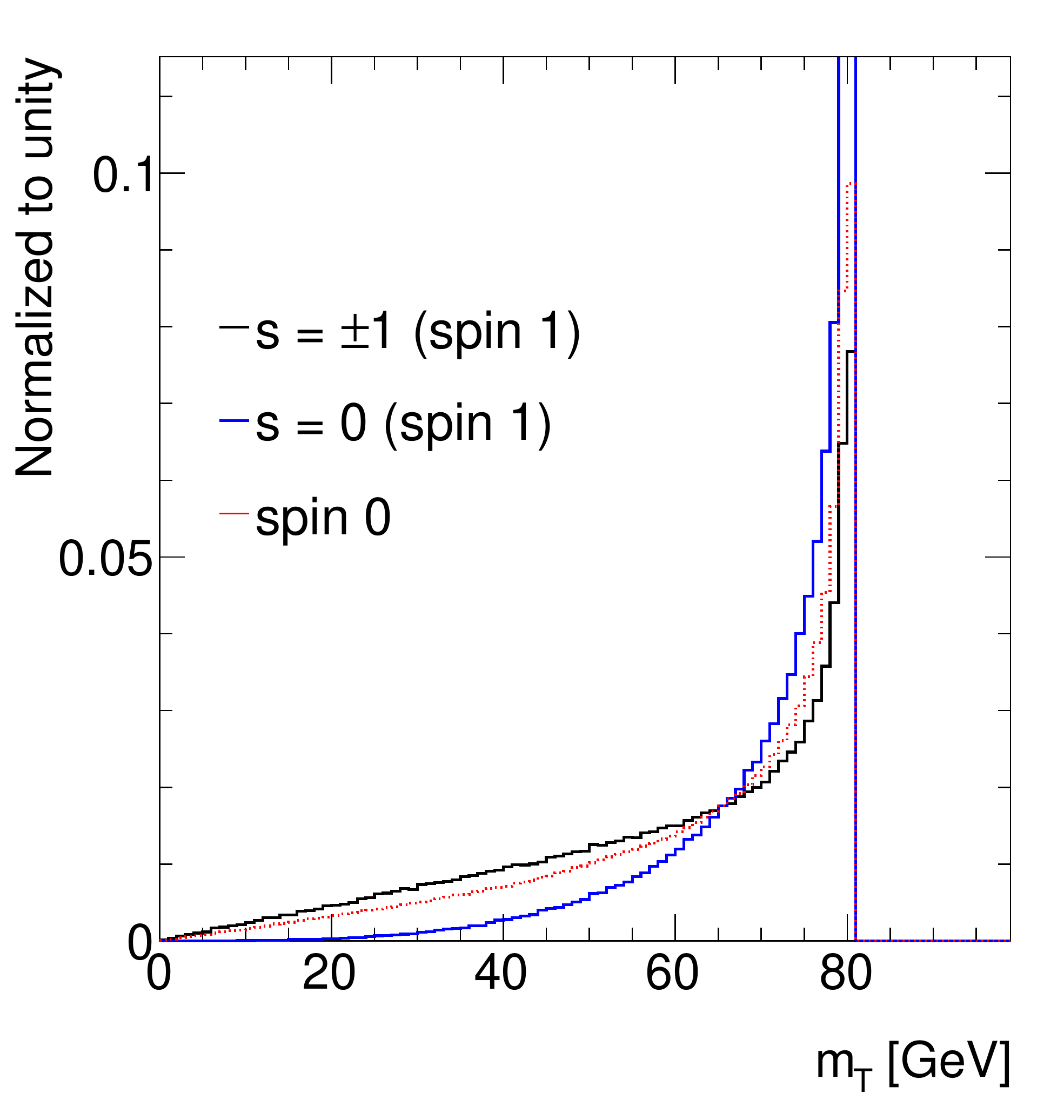}
 \caption{The transverse mass distribution for a scalar $W$ decay (spin 0) and a vector $W$ decay (spin 1).  Each distribution is generated by MG5\_aMC 2.1.1.  A Higgs boson with $m_H=m_W$ is used for the spin 0 line, which is identical to the lines from Fig.~\ref{fig:susymtdist} using a phase-space only generator and the analytic formula.}
 \label{fig:susymtdistspin1}
  \end{center}
\end{figure}

While the $m_\text{T}$ distribution is invariant under longitudinal boosts, it is not invariant under transverse boosts.  To illustrate the impact of a transverse boost, begin with a configuration close to the kinematic limit: $p^\mu=(\epsilon,m_W/2,0,m_W/2)+\mathcal{O}(\epsilon^2/m_W)$ and  $q^\mu=(-\epsilon,-m_W/2,0,m_W/2)+\mathcal{O}(\epsilon^2/m_W)$.  Under a boost along the $x$-axis with magnitude $\beta$, $p_x\rightarrow \gamma\epsilon+\gamma\beta\frac{m_W}{2}$. Therefore,

				\begin{align}\nonumber
				m_\text{T}^2&=(p_T+q_T)^2-(p_x+q_x)^2-(p_y+q_y)^2\\\nonumber
				&=m_W+\epsilon\gamma\beta m_W+\mathcal{O}\left(\frac{\epsilon^2}{m_W},\beta^2m_W^2\right)\\
				&=m_W^2\left(1-\frac{\epsilon\beta}{m_W}+\mathcal{O}\left(\frac{\epsilon^2}{m_W^2},\beta^2\right)\right).\label{eq:mt:lorentz}
				\end{align}
				
				\noindent The interesting properties of Eq.~\ref{eq:mt:lorentz} are that there is no impact of a boost if the maximum value is already achieved $(\epsilon=0)$ and if $\epsilon \neq 0$, a transverse boost {\it reduces} the transverse mass.  As the $W$ boson becomes more boosted, the momentum of the electron and neutrino increase, but the angle between them decreases.  The later happens faster than the former, which flattens out the $m_\text{T}$ distribution at high $p_\text{T}$.  The right plot of Fig.~\ref{fig:susymtdist} shows the $m_\text{T}$ distribution for the same configurations as the left plot, but now with $p_\text{T}^W=200$ GeV.  The kinematic maximum for $W\rightarrow e\nu$ is still $m_\text{T}\leq m_W$, but the distribution has filled in at lower values of the transverse mass.   For the pair production of $W$ bosons, both have the same $p_\text{T}$ but an arbitrary direction.  Since the angle between the $W$ bosons can be large, the transverse mass can significantly exceed $m_W$.

Figure~\ref{fig:susymtdist} showed that if there are anomalous contributions to the $E_\text{T}^\text{miss}$, the transverse mass can exceed its natural kinematic maximum.  Another way for $m_\text{T}>m_W$ in events with one reconstructed lepton is if there are additional genuine contributions to the $E_\text{T}^\text{miss}$.  For example, the high energy undetected neutralinos in stop events can push $m_\text{T}$ well beyond $m_W$.  Figure~\ref{fig:susymtdist2bb} shows the $m_\text{T}$ distribution for simulated signal $\tilde{t}\rightarrow t\tilde{\chi}^0$ events and dilepton $t\bar{t}$ events where one lepton is lost or mis-identified.  The stops are produced at rest, resulting in a kinematic maximum of $m_\text{T}\leq \sqrt{2}m_{\tilde{t}}$, which occurs when the neutralino-top axes are aligned and in the transverse plane.  The left and right plots of Fig.~\ref{fig:susymtdist2bb} differ in the source of the top quark boost.  In the left plot, the invariant mass of the $t\bar{t}$ system is zero, but its $p_\text{T}$ is significant.  Conversely, in the right plot of Fig.~\ref{fig:susymtdist2bb}, the $t\bar{t}$ system is produced at rest, but with significant invariant mass.  In both cases, the typical top quark $p_\text{T}$ is half of the $p_\text{T}$ or mass scale.  However, the $m_\text{T}$ distribution is significantly different between the $p_\text{T}^{t\bar{t}}$ and $m_{t\bar{t}}$ schemes.   The $m_{t\bar{t}}$ case is most similar to Fig.~\ref{fig:susymtdist} where the $W$ bosons can have a significant angle between them, while in the $p_\text{T}^{t\bar{t}}$ case, the two $W$ bosons are spatially close.  The larger opening angle results in a larger $m_\text{T}$ value.  In practice, both $p_\text{T}^{t\bar{t}}$ and  $m_{t\bar{t}}$ will be nonzero, but the former is more important for the stop search.  This is because a four-jet event selection requires dilepton events to be produced with additional jets, as discussed in more detail in Sec.~\ref{chapter:background}.

\begin{figure}[h!]
\begin{center}
\includegraphics[width=0.5\textwidth]{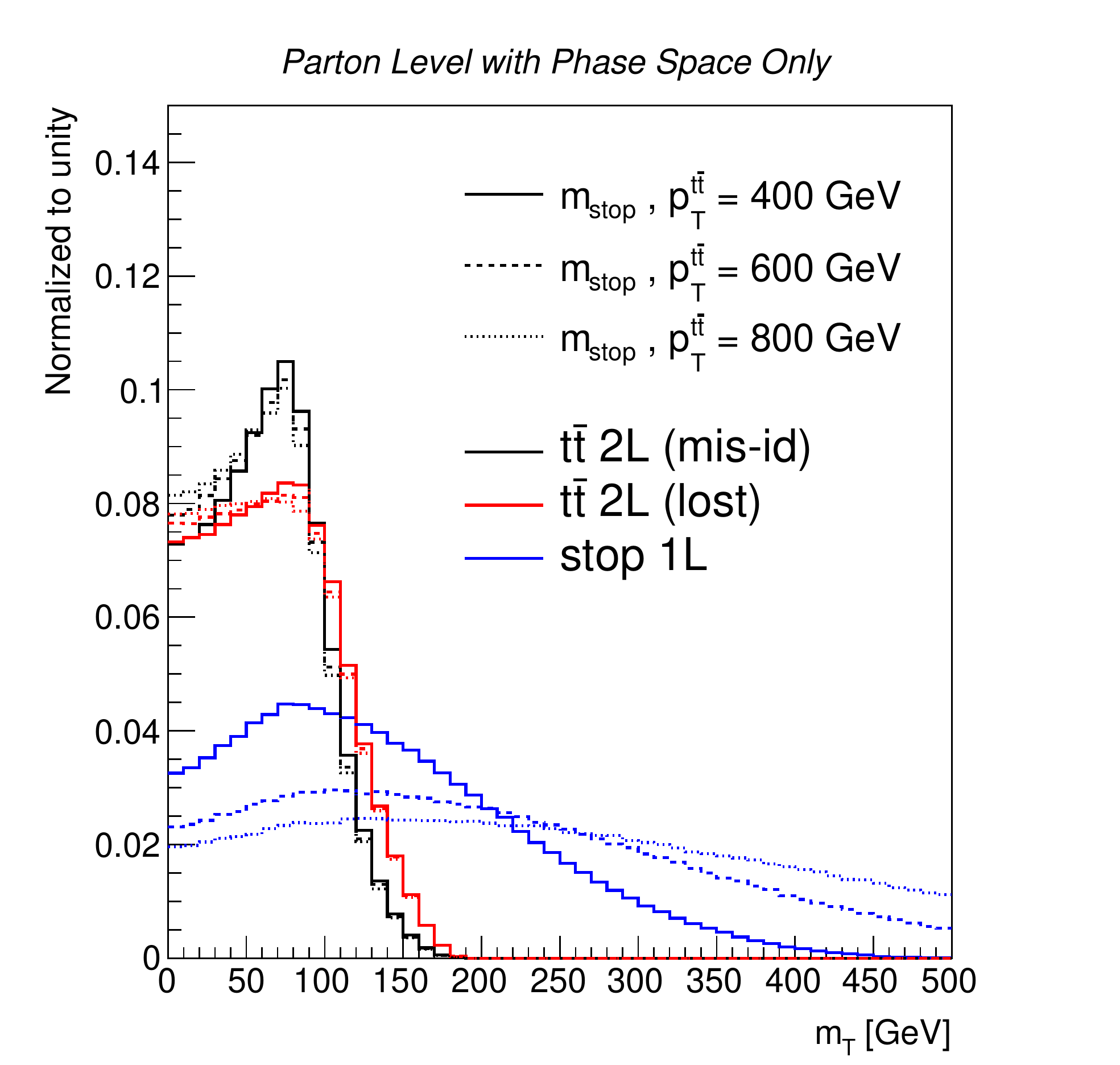}\includegraphics[width=0.5\textwidth]{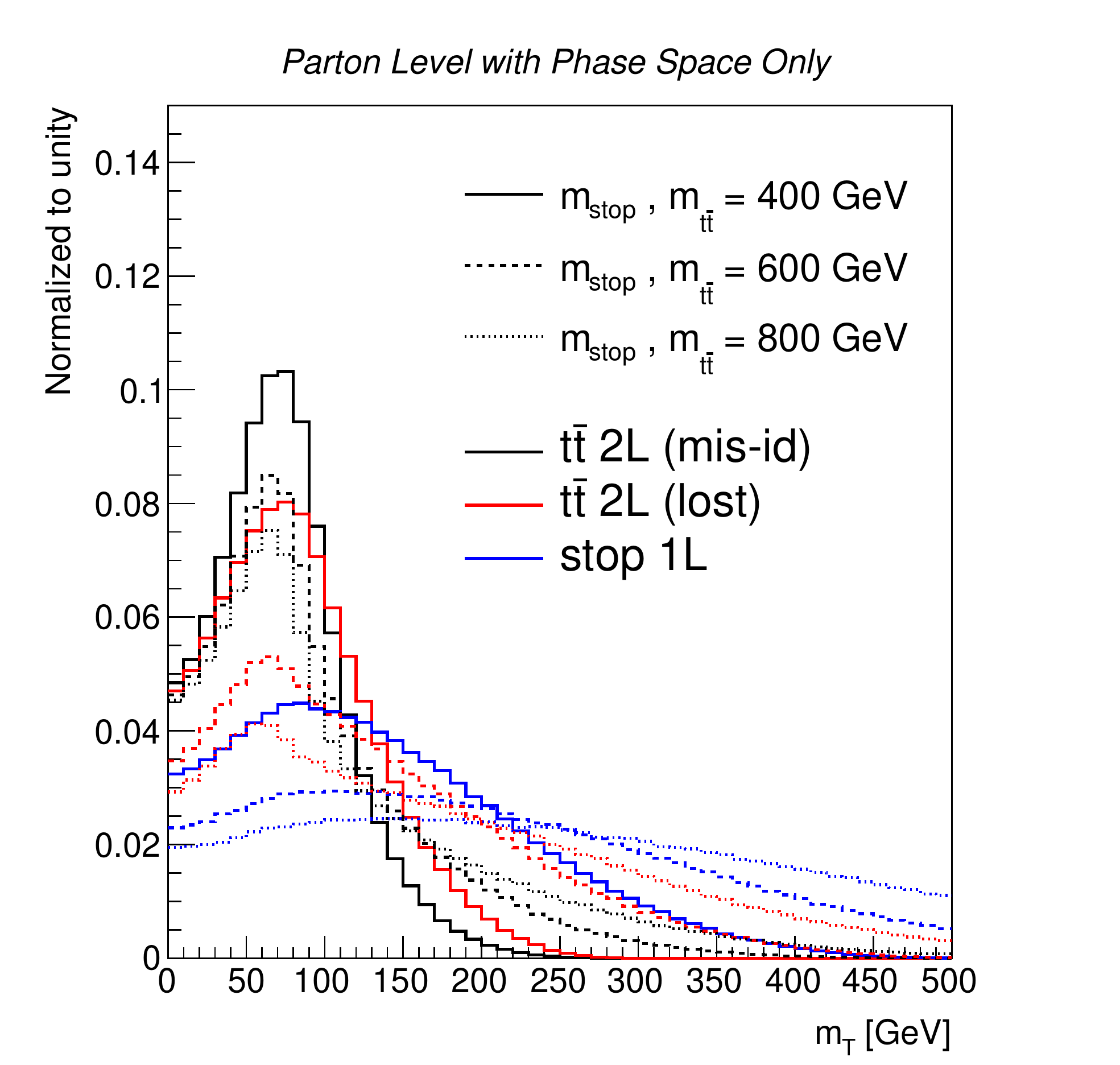}
 \caption{The distribution of the transverse mass for signal $\tilde{t}\rightarrow t\tilde{\chi}^0$ events and  dilepton $t\bar{t}$ events where one lepton is lost or mis-identified.  The signal histograms are identical in the left and right plots.  In the left plot, the $p_\text{T}$ of the $t\bar{t}$ system is boosted in an arbitrary direction with a fixed $p_\text{T}$.  In the right plot, the $t\bar{t}$ pair is produced at rest, but with a large $m_\text{$t\bar{t}$}$ so that the top quarks have significant boost.}
 \label{fig:susymtdist2bb}
  \end{center}
\end{figure}	

The right plot of Fig.~\ref{fig:susymtdist2bb} shows that there is a strong correlation between the $E_\text{T}^\text{miss}$ and $m_\text{T}$ in events where the two top quarks are independent.  Figure~\ref{fig:susymtdist3} quantifies the correlation for the various event types discussed above.  The correlation increases with mass for stop events and $t\bar{t}$ events with increasing $m_{t\bar{t}}$, reaching about 60\%.  In contrast, when there is a relationship between the direction of the two top quarks, as is the case when the entire $t\bar{t}$ system is boosted, there is little correlation between $E_\text{T}^\text{miss}$ and $m_\text{T}$.  These differences in correlation show that the $m_\text{T}$ can add useful information beyond what is already contained in the $E_\text{T}^\text{miss}$.  Note that the correlation is 100\% for a single $W\rightarrow e\nu$ produced at rest because both the $m_\text{T}$ and $p_\text{T,$\nu$}\sim\cos(\theta)$.
				
\begin{figure}[h!]
\begin{center}
\includegraphics[width=0.85\textwidth]{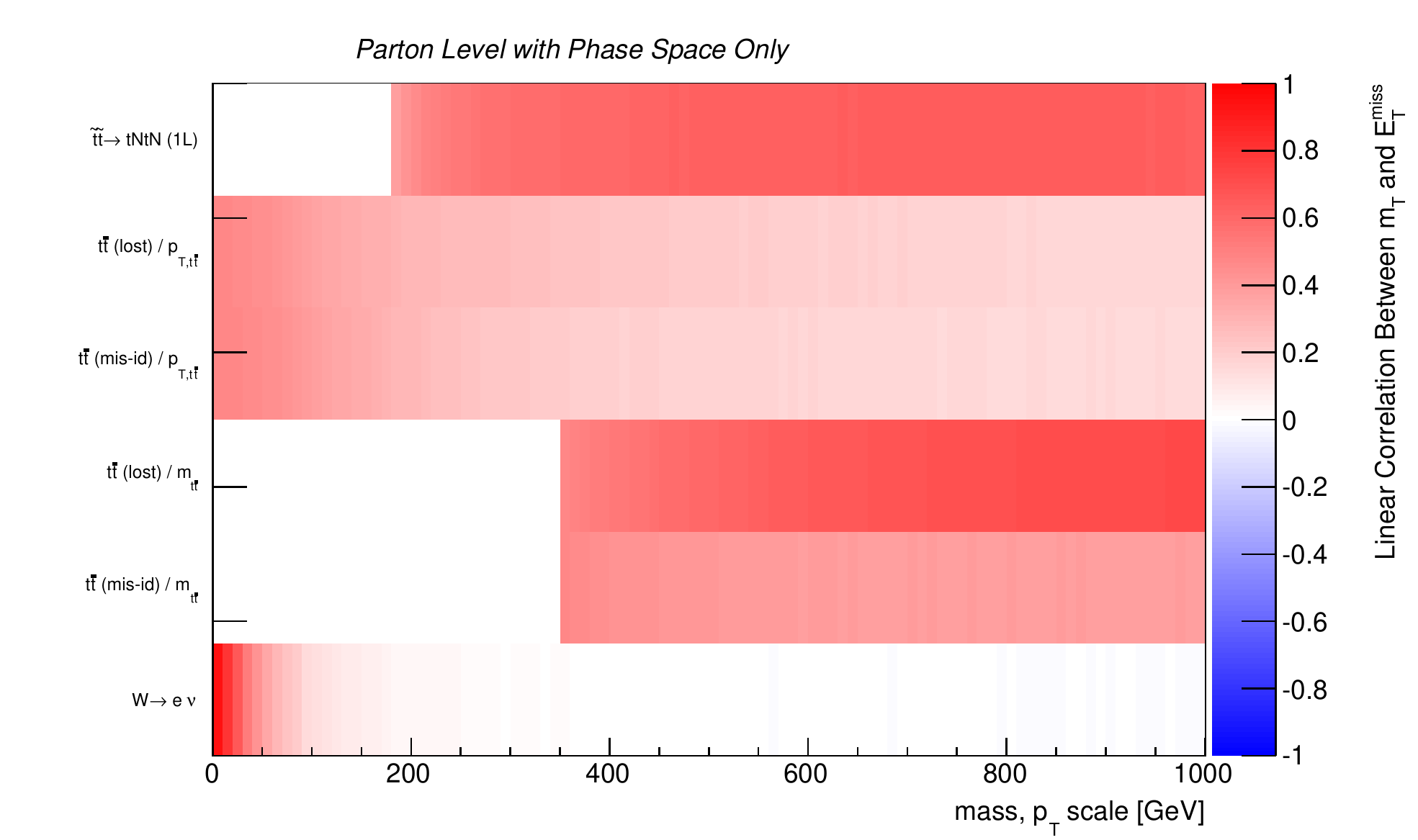}
 \caption{The linear correlation between the $E_\text{T}^\text{miss}$ and $m_\text{T}$ as a function of the $p_\text{T}$ or mass scale for various processes. The $p_\text{T}^{t\bar{t}}$ and $m_{t\bar{t}}$ are setup in the same way as for Fig.~\ref{fig:susymtdist2}.  The correlation is set to zero for unphysical parameter values such as $m_{t\bar{t}}<2m_t$.}
 \label{fig:susymtdist3}
  \end{center}
\end{figure}

For all the reasons described above, the transverse mass is a powerful variable at suppressing the single lepton $t\bar{t}$ and $W$+jets backgrounds. Figure~\ref{fig:susymtdatamc} shows the transverse mass distribution in the early Run 2 data for a selection of events enriched in the pair production of top quarks.  In particular, events are required to have exactly one reconstructed signal lepton with no additional baseline leptons, four jets with $p_\text{T}>25$ GeV and at least one $b$-tagged jet.  The resulting events are predicted to have a $t\bar{t}$ purity of about 80\%.  As expected, most of the $t\bar{t}$ events have one real lepton, due to the second lepton veto.  Therefore, there is a clear cutoff near the $W$ boson mass.  However, the right plot of Fig.~\ref{fig:susymtdatamc} shows that at high values of $m_\text{T}$, events with two real leptons dominate, split between events with a second real electron or muon and events with a hadronically decaying $\tau$.   The single lepton background at high $m_\text{T}$ is negligible.  Figure~\ref{fig:susymtdatamc2} shows the analogous plots for events enriched in the single production of a $W$ boson in association with jets.  The event selection for Fig.~\ref{fig:susymtdatamc2} differs from the one used for Fig.~\ref{fig:susymtdatamc} only by exchanging the $b$-jet requirement for a $b$-jet veto.  The Jacobian peak is clearly present in both the log and linear scale plots of Fig.~\ref{fig:susymtdatamc2}.  For $m_\text{T}>m_W$, there is an enhancement of diboson events with a second real lepton, but this is not as significant as for the $t\bar{t}$ case.  This is because the diboson to $W$+jets cross section ratio\footnote{The ratio is even smaller when including the leptonic branching ratio for the second boson.} is $\mathcal{O}(1\%)$~\cite{Aad:2014qxa,Aad:2014mda} while the dilepton branching ratio is about 25\% of the single lepton $t\bar{t}$ branching ratio.  Since the $W$+jets events with $m_\text{T}\gg m_W$ are well beyond the parton-level kinematic maximum, the shape of the distribution in the high $m_\text{T}$ tail is determined mostly by resolution effects.  Methods for incorporating resolution information into kinematic variables will be described in Sec.~\ref{sec:significancevariables}.
			
\begin{figure}[h!]
\begin{center}
\includegraphics[width=0.5\textwidth]{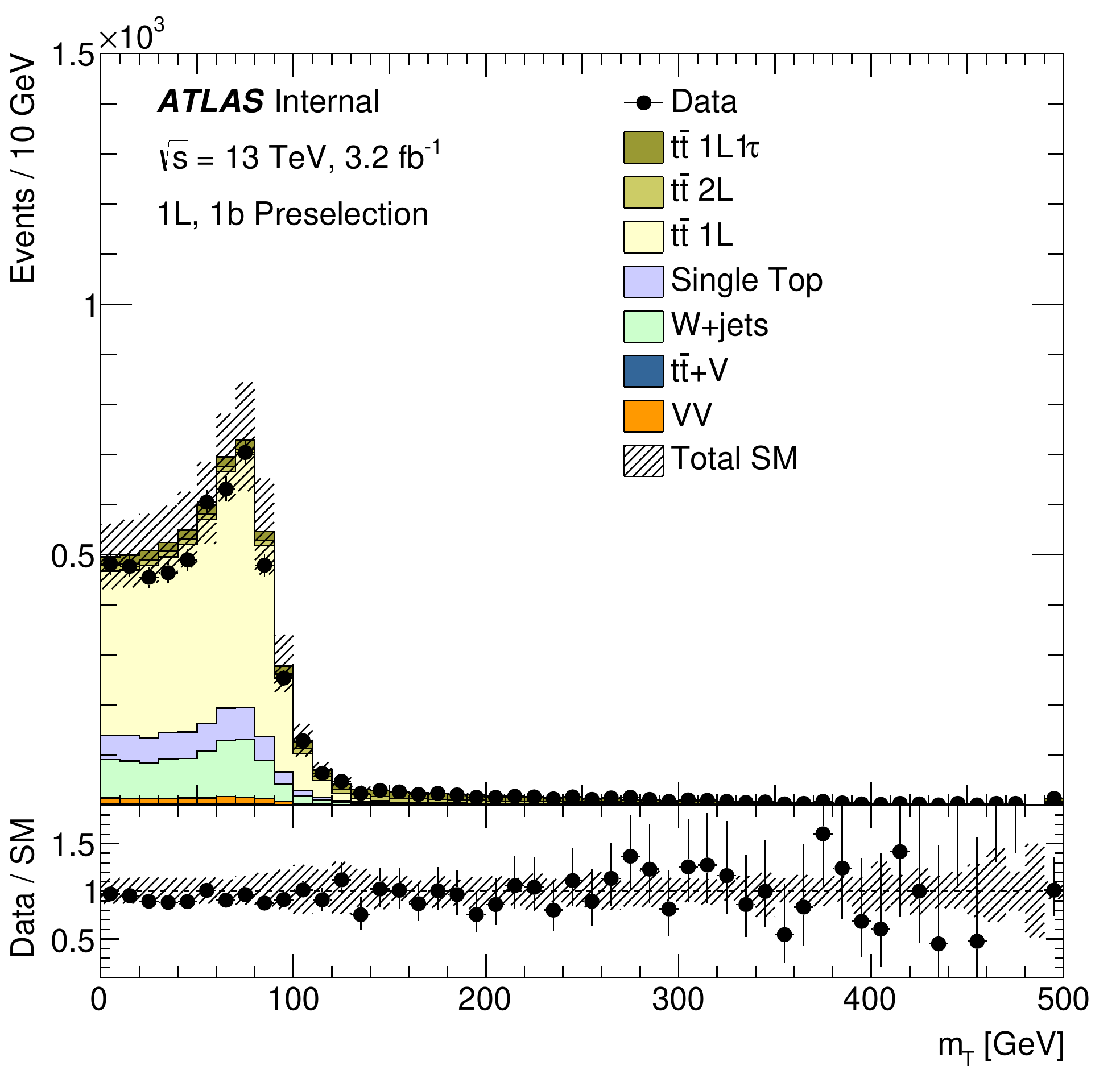}\includegraphics[width=0.5\textwidth]{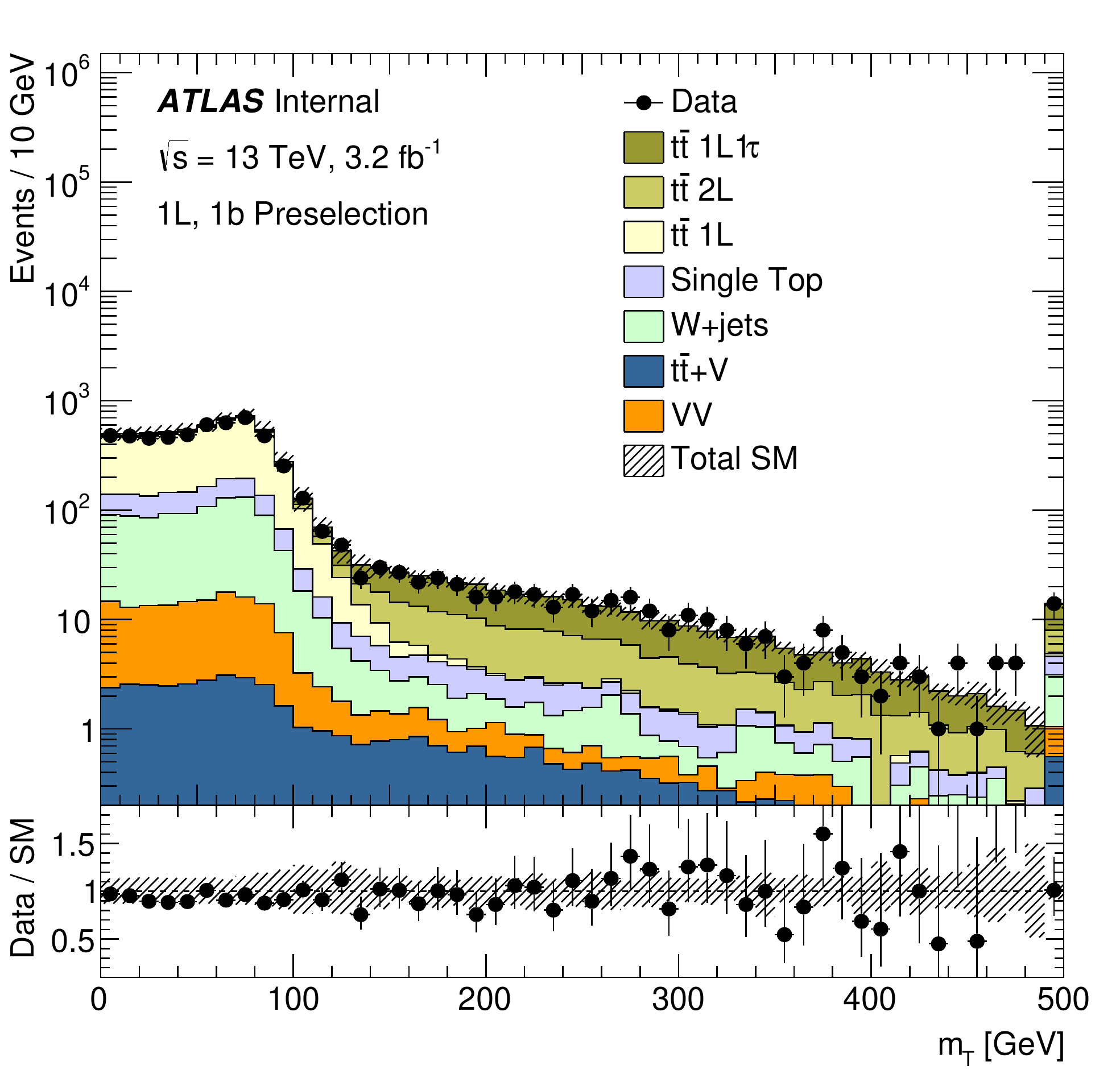}
 \caption{A comparison of data and simulation using a loose selection requiring exactly one signal lepton, four jets with $p_\text{T}>25$ GeV and at least one $b$-tagged jet.  The left and right plots differ only in the scaling of the vertical axis. The uncertainty band includes jet energy scale and resolution uncertainties (see Sec.~\ref{chapter:uncertainites}).}
 \label{fig:susymtdatamc}
  \end{center}
\end{figure}			

\begin{figure}[h!]
\begin{center}
\includegraphics[width=0.5\textwidth]{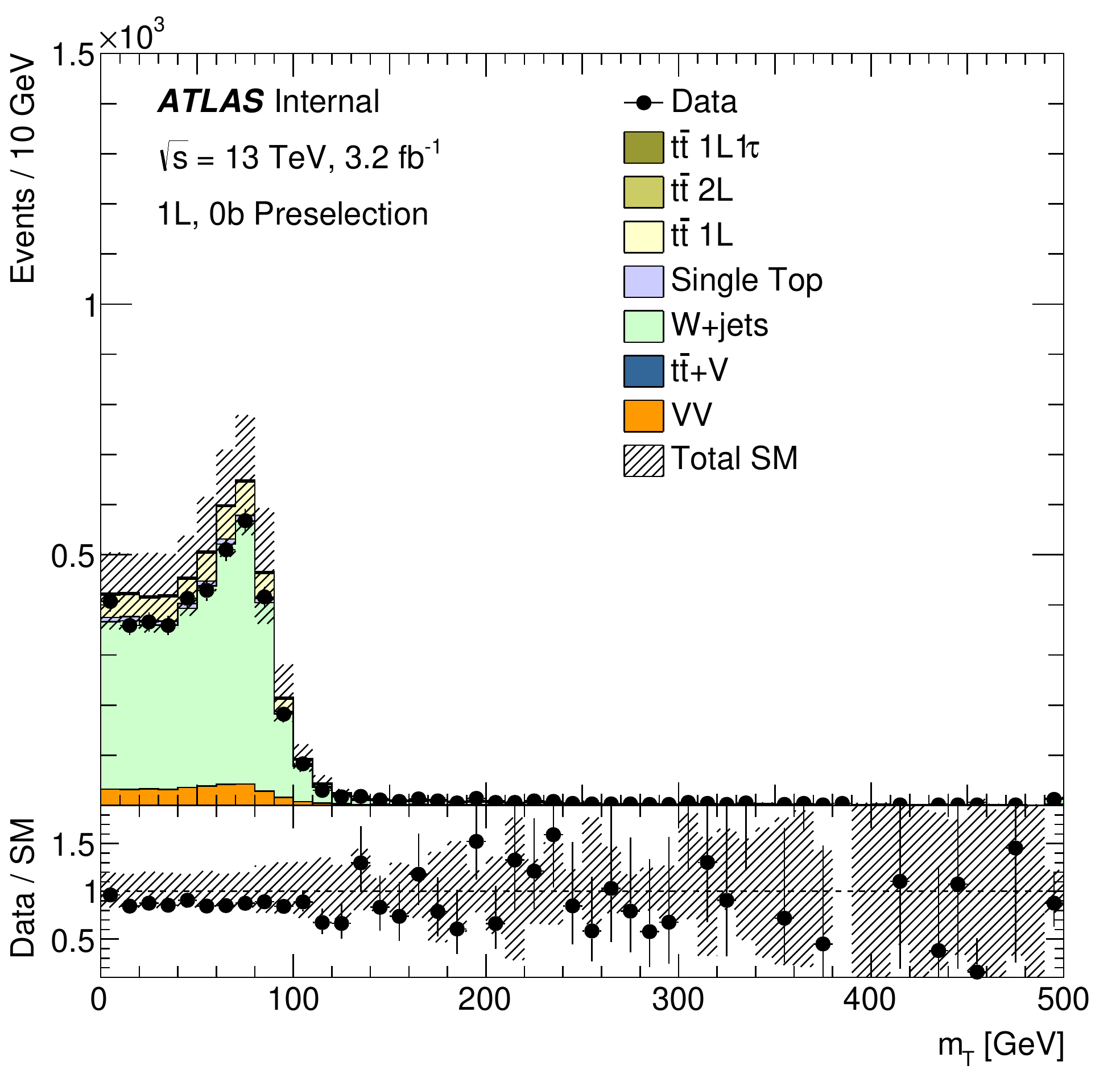}\includegraphics[width=0.5\textwidth]{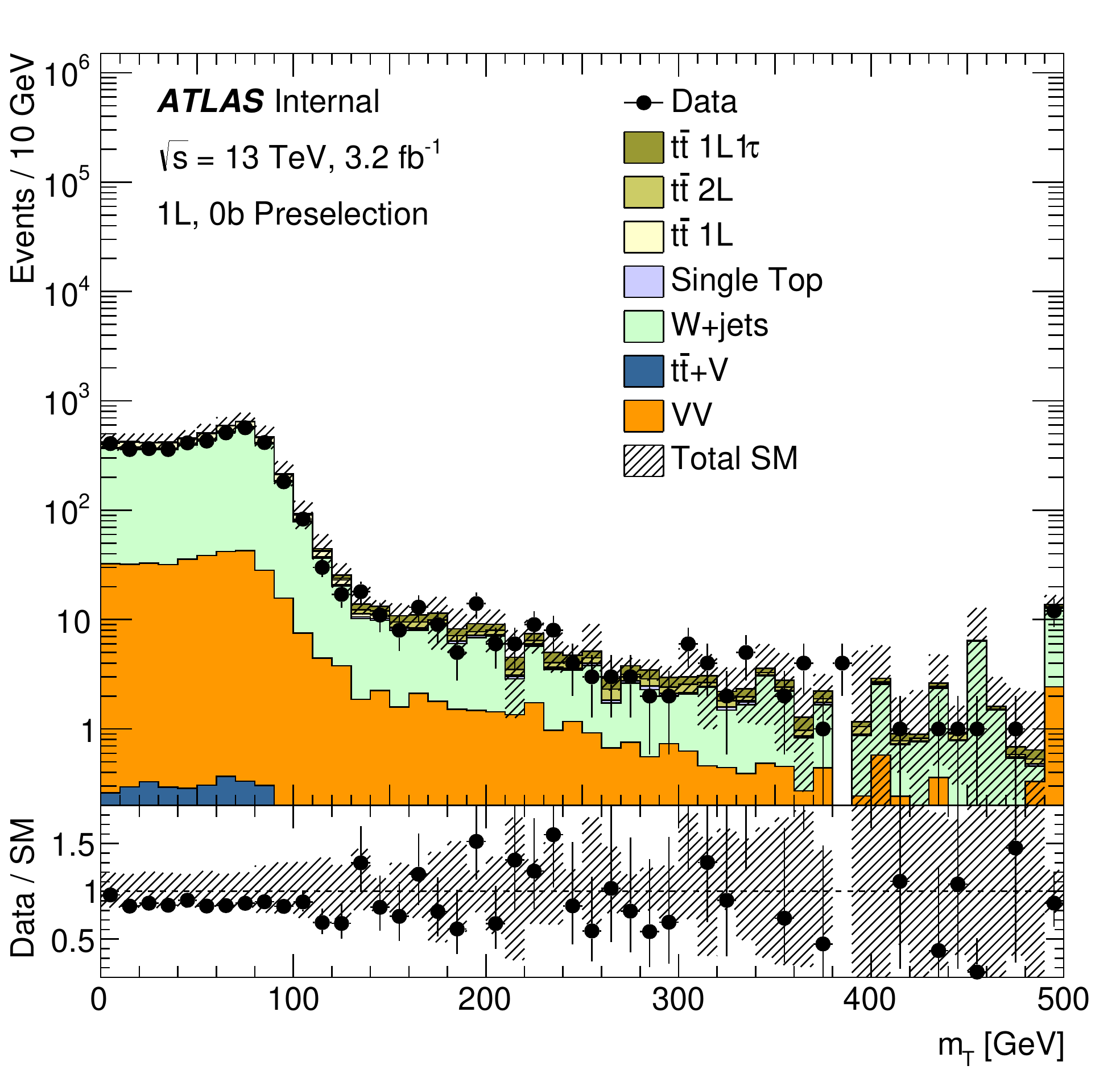}
 \caption{A comparison of data and simulation using a loose selection requiring exactly one signal lepton, four jets with $p_\text{T}>25$ GeV and at exactly no $b$-tagged jets.  The left and right plots differ only in the scaling of the vertical axis.  The uncertainty band includes jet energy scale and resolution uncertainties (see Sec.~\ref{chapter:uncertainites}).}
 \label{fig:susymtdatamc2}
  \end{center}
\end{figure}	
	
	The transverse mass constructed above assumes that the mass of the missing particle is known a priori.  This is a valid assumption when the missing object is a SM particle such as the neutrino.  However, in general, the full $m_\text{T}$ as defined by Eq.~\ref{eq:fullmt} (and not Eq.~\ref{eq:mtdefinition}) implicitly requires the input of a missing particle mass, otherwise $F^2-q_z^2=q_x^2+q_y^2+m_\text{test}^2$ cannot be computed from only transverse quantities.  This will be most relevant for generalizing the transverse mass in Sec.~\ref{sec:mt2}, but one can already see the consequences of a non-trivial test mass for the case of a single $W$ boson.  Figure~\ref{fig:susymtdist_testmass} shows the impact of choosing $m_\text{test}\neq m_\text{true}$.  The minimum value of $m_\text{T}$ is no longer zero - the transverse mass is bounded below by $m_\text{test}$.  This happens when the transverse momentum of the decay products is zero.  The most important change is that there is no longer a $p_\text{T}$-independent kinematic limit.  In the context of $W\rightarrow e\nu$, the generalized transverse mass can be written
	
	\begin{align}
	\label{eq:generalizedmt_reduced}
	m_\text{T}^2=m_\text{test}^2+2\left(p_\text{T,l} \sqrt{m_\text{test}^2+p_\text{T,$\nu$}^2}- p_\text{T,l} p_\text{T,$\nu$}\cos(\theta_{e\nu})\right).
	\end{align}
			
\noindent Eq.~\ref{eq:generalizedmt_reduced} is maximized when $\theta_{e\nu}=\pi$ and the electron and neutrino momentum are in the transverse plane.  This was clear earlier in the context of a $W$ boson at rest, but for a boosted $W$ boson the implication is that the boost must be collinear with the electron-neutrino axis.  A boost along the electron-neutrino axis will enhance the momentum of the electron relative to the neutrino or vice versa since in the $W$ rest frame, the two are back-to-back.  Since the neutrino momentum is added in quadrature with the test mass, the maximum value of $m_\text{T}$ is achieved when the boost is parallel to the electron direction.  If the boost has magnitude $\beta$:
	
	\begin{align}
	p_\text{T,l}\rightarrow &\gamma p_\text{T,l}+\beta\gamma p_\text{T,l}=\gamma p_\text{T,l}(1+\beta)=\frac{m_W}{2}\sqrt{\frac{1+\beta}{1-\beta}},
	\end{align}
			
	\noindent where without loss of generality, one can take the electron momentum to be aligned with the $x$-axis.  A similar calculation shows that $p_\text{T,$\nu$}\rightarrow \frac{m_W}{2}\sqrt{\frac{1-\beta}{1+\beta}}$.	  If $m_\text{test}=0$, these two factors exactly cancel and the endpoint is invariant under transverse boosts, as was observed earlier.  However, when $m_\text{test}>0$, the two boost factors do not cancel and therefore the endpoint scales with the boost.  This is a general feature of the transverse mass whenever $m_\text{test}\neq m_\text{true}$ and will be investigated further in Sec.~\ref{sec:mt2}.
				
\begin{figure}[h!]
\begin{center}
\includegraphics[width=0.5\textwidth]{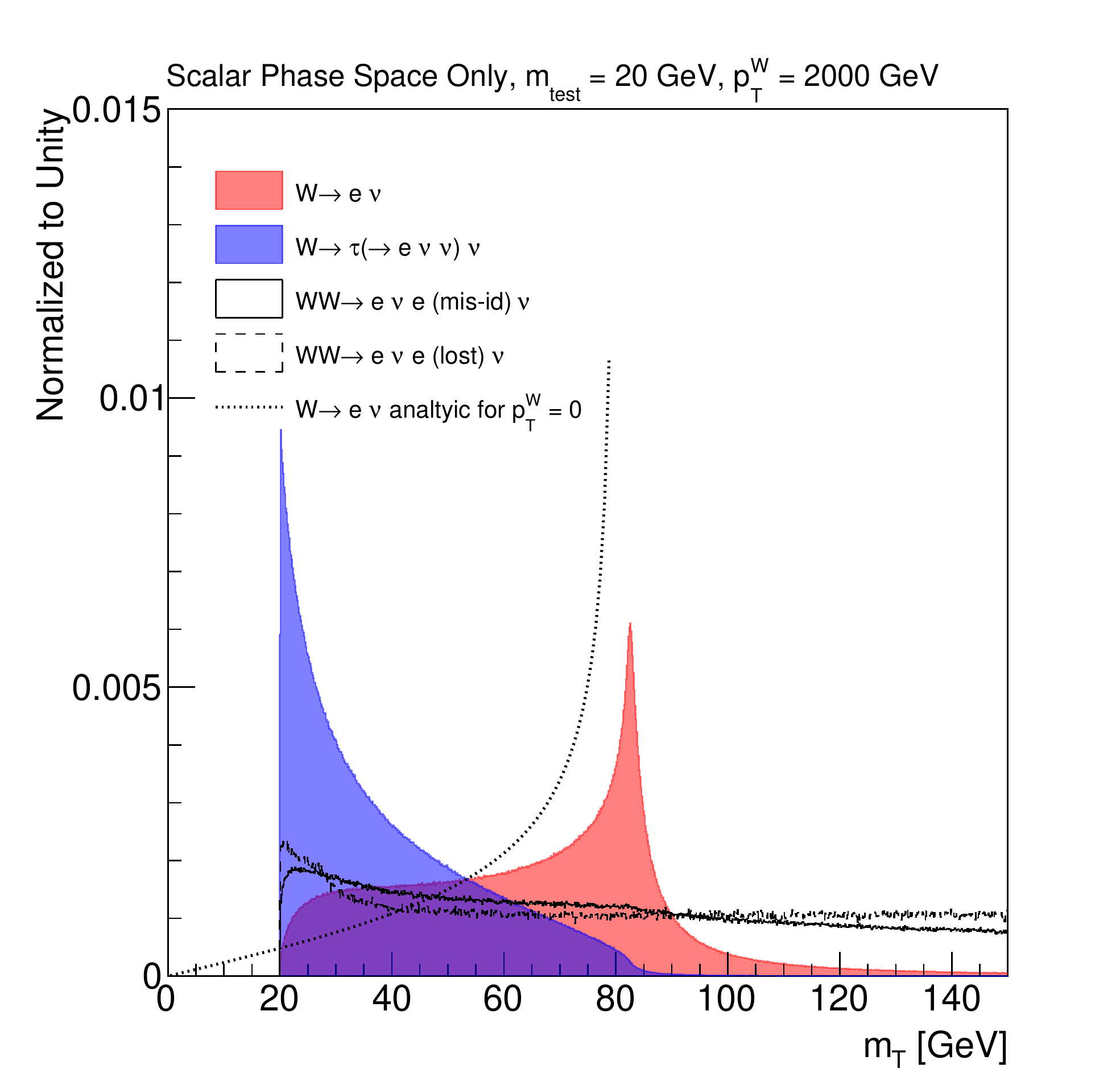}\includegraphics[width=0.5\textwidth]{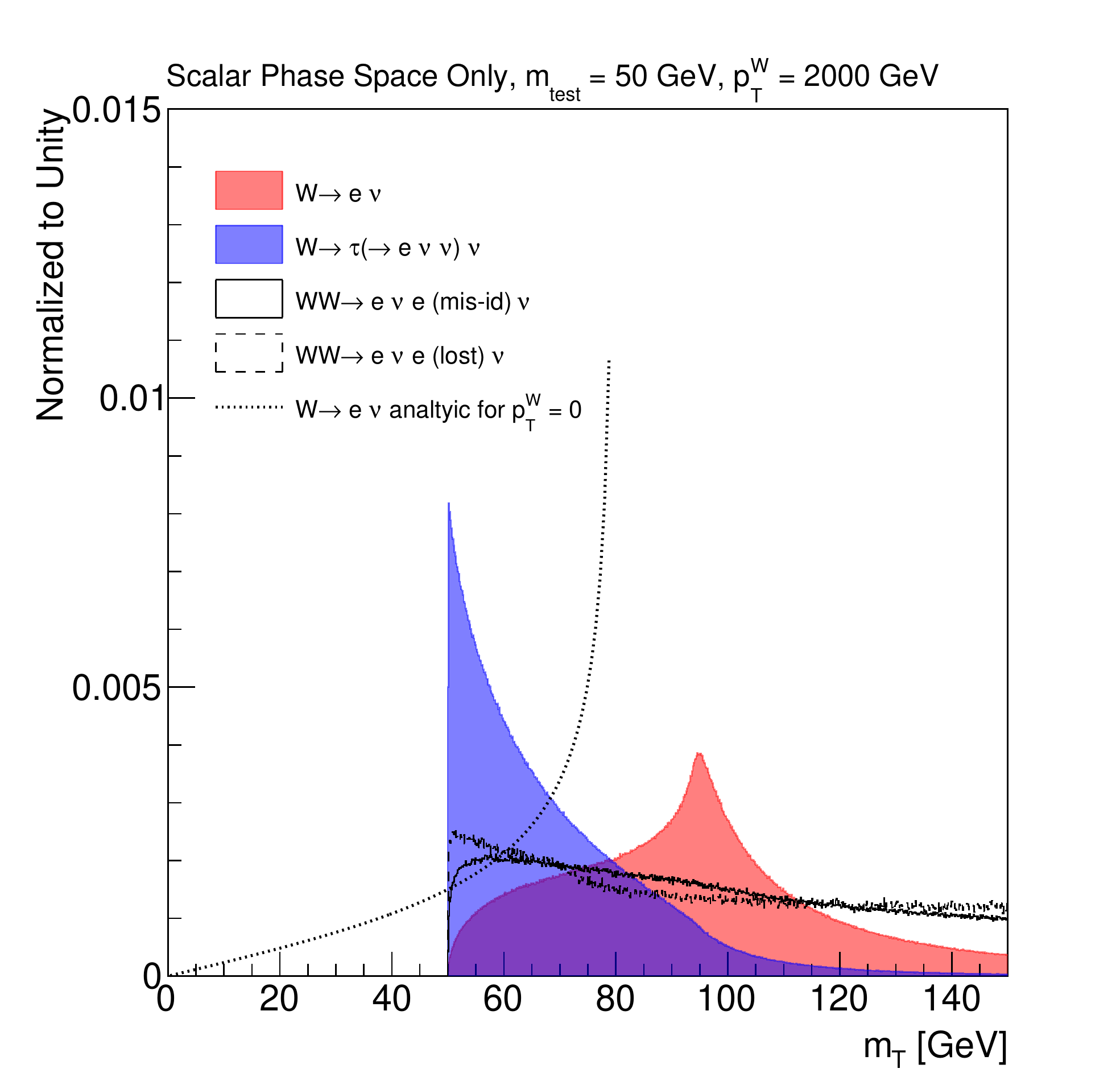}
 \caption{Identical $m_\text{T}$ distributions as in Fig.~\ref{fig:susymtdist}, except the neutrino mass parameter in the transverse mass calculation is set to 20 GeV (left) and 50 GeV (right)}
 \label{fig:susymtdist_testmass}
  \end{center}
\end{figure}			
			
				\clearpage
				
				\subsubsection{Multiple Missing Particles: $m_\text{T2}$}
				\label{sec:mt2}

				Sec.~\ref{sec:mT} demonstrated that a simple threshold requirement $m_\text{T}\gg m_W$ is an effective procedure for suppressing the single lepton $t\bar{t}$ and $W$+jets backgrounds while maintaining high stop signal efficiency.  One of the dominant residual backgrounds is $t\bar{t}$ with two real leptons where the second lepton is out of acceptance, fails the particle identification, or is a hadronically decaying $\tau$.  This section describes an extension of the transverse mass to cases where there are multiple missing particles in order to reduce the two lepton backgrounds with multiple neutrinos.
			
		Figure~\ref{fig:mt2setup} shows the generic setup: two particles $P$ with the same mass $m_P$ are pair produced and decay to visible particles $V_1,V_2$ and undetected particles $C$.  The momenta of $V_1$ and $V_2$ are measured in the detector but only the sum of the transverse momenta of the $C$ particles are inferred from momentum conservation. 	Define the variable $m_\text{T2}$\footnote{Pronounced `M-Tee-Two' and in the literature is also referred to as the `stransverse mass' or the `Cambridge $m_\text{T2}$'.}~\cite{Lester:1999tx,Barr:2003rg} as
		
		\begin{align}
		\label{eq:mt2eq}
		m_\text{T2}(m_{C_{1,\text{test}}},m_{C_{2,\text{test}}})=\min_{\vec{p}_\text{T,test}^{\hspace{1mm}C_1}+\vec{p}_\text{T,test}^{\hspace{1mm}C_2}=\vec{E}_\text{T}^\text{miss}}\max_{i}\left\{m_\text{T}\left(\vec{p}_\text{T}^{\hspace{1mm}V_i},\vec{p}_\text{T,test}^{\hspace{1mm}C_i},m_{V_i},m_{C_{i,\text{test}}}\right)\right\},
		\end{align}
		
		\noindent where $m_\text{T}(\vec{p},\vec{q},m_\text{vis},m_\text{test})$ is the generalized transverse mass introduced in Eq.~\ref{eq:fullmt} and $m_\text{test}$ is the assumed mass for the invisible particle $C_i$.  The {\it test} in $p_\text{T,test}^{C_i}$ and $m_{C_{i,\text{test}}}$ is to distinguish the dummy variable in the minimization from the true and unknown values of $p_\text{T}^{C_i}$ and $m_{C_{i}}$. In a form similar to Eq.~\ref{eq:generalizedmt_reduced} but including mass effects for the visible particle, the generalized transverse mass takes the form:
		
	\begin{align}
	\label{eq:generalizedmt}
	m_\text{T}^2(\vec{p},\vec{q},m_\text{vis},m_\text{test})=m_\text{vis}^2+m_\text{test}^2+2\left(\sqrt{m_\text{vis}^2+p^2}\sqrt{m_\text{test}^2+q^2}-\vec{p}_\cdot\vec{q}\right).
	\end{align}
			
\begin{figure}[h!]
\begin{center}
\begin{tikzpicture}[line width=1.5 pt, scale=1.3]
			\draw (-2,0.5) -- (0,0.5);
			\draw (-2,-0.5) -- (0,-0.5);
			\draw[dashed] (0,0.5) -- (2,0.5);
			\draw[dashed] (0,-0.5) -- (2,-0.5);			
			\draw (0,0.5) -- (1.4,1.7);
			\draw (0,-0.5) -- (1.4,-1.7);
			\node at (-2.2,0.5) {$P_1$};
			\node at (-2.2,-0.5) {$P_2$};
			\node at (2.2,0.5) {$C_1$};
			\node at (2.2,-0.5) {$C_2$};
			\node at (1.6,-1.8) {$V_1$};
			\node at (1.6,1.8) {$V_2$};	
			\node at (1.7,0) {$\vec{E}_\text{T}^\text{miss}$};	
			\draw[dotted, black!30] (2,0) ellipse (1.1 and 1.1);	
		\end{tikzpicture}
 \caption{A schematic diagram showing the generic $m_\text{T2}$ setup.}
 \label{fig:mt2setup}
 \end{center}
\end{figure}				
	
		\noindent Like the simple transverse mass, $m_\text{T2}$ is invariant under longitudinal boosts but not transverse boosts.  When the masses $m_{C_{1,\text{test}}},m_{C_{2,\text{test}}}$ are chosen correctly, $m_\text{T2}\leq m_P$.  This is because $m_\text{T2}\leq \max_{i}\left\{m_\text{T}\left(\vec{p}_\text{T}^{\hspace{1mm}V_i},\vec{p}_\text{T}^{\hspace{1mm}C_i},m_{V_i},m_{C_i}\right)\right\}\leq m_P$, as shown in Sec.~\ref{sec:mT}.  Figure~\ref{fig:susymtdist_testmass} demonstrates the power of $m_\text{T2}$ in events with two $W$ bosons where one $W$ decays into a hadronically decaying $\tau$ and the other $W$ boson decays into an electron or muon ($\ell$).  Multiple neutrinos contribute to the missing momentum so the $m_\text{T}$ of the $\ell$ and the $\vec{p}_\text{T}^\text{miss}$ can exceed the $m_W$ bound as long as the bosons are not produced at rest.  In contrast, the $m_\text{T2}$ using $V_1=\ell$, $V_2=\tau$, $m_{C_1}=m_{C_2}=0$ is kinematically bound by $m_W$ (neglecting $m_\tau$).  Unlike the $m_\text{T}$ case, the $m_\text{T2}$ endpoint is not as saturated due to the minimization in the definition.
			
\begin{figure}[h!]
\begin{center}
\includegraphics[width=0.45\textwidth]{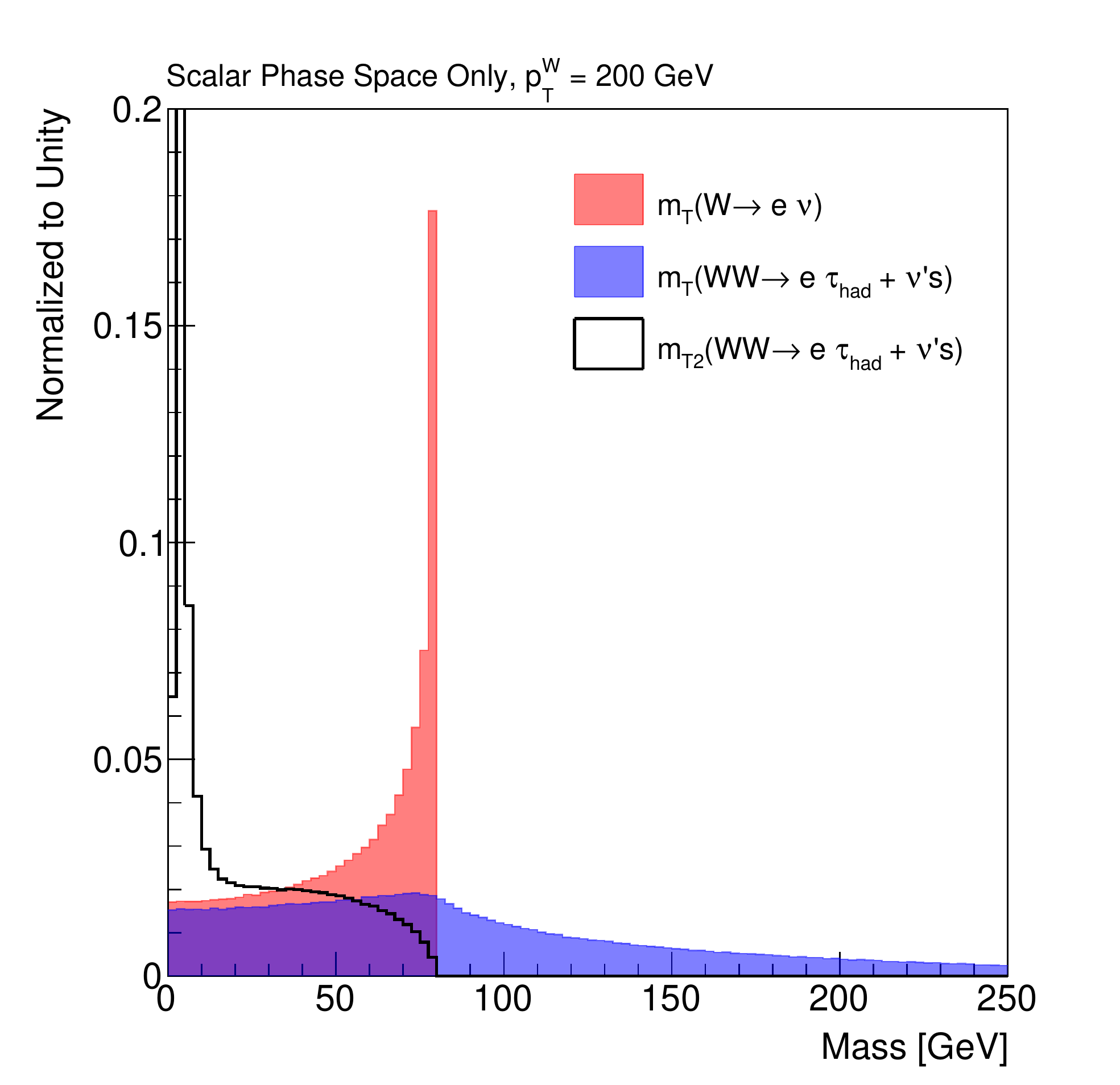}
 \caption{A comparison of $m_\text{T}$ and $m_\text{T2}$ for events with an identified hadronically decaying $\tau$.}
 \label{fig:susymtdist_testmass}
  \end{center}
\end{figure}			
	
	The visible objects $V_1$ and $V_2$ from Fig.~\ref{fig:mt2setup} are often {\it composite systems of particles}.  For example, $V_i$ might be the combination of a $b$-jet and a lepton.  In such cases, it is often true that $m_{V_1}\neq m_{V_2}$.  When the composite systems are not even made of the same types of constituent particles, the presumed lost children may not be the same and in general $m_{C_1}\neq m_{C_2}$.  Just like the generalized $m_\text{T}$ from Sec.~\ref{sec:mT}, the general $m_\text{T2}$ variable is bounded from below by $\max\{m_{C_1}+m_{V_1},m_{C_2}+m_{V_2}\}$.  This is readily calculated by taking the derivative of each $m_\text{T}$ branch with respect to the assigned missing particle momentum $p_\text{$\alpha$,test}^{C_i}$ for $\alpha\in\{x,y\}$, shown in Eq.~\ref{eq:minmt2}.  The second implication (2) in Eq.~\ref{eq:minmt2} is from summing and squaring the first implication (1) and the third implication (3) is the result of simplifying after inserting the second implication (2) back into the first one (1).  

	\begin{align}\nonumber
	\label{eq:minmt2}
	\frac{1}{2}&\frac{\partial }{\partial p_\text{$\alpha$,test}^{C_i}}m_\text{T}^2(\vec{p}_\text{T}^{V_i},\vec{p}_\text{T,test}^{C_i},m_{V_i},m_{C_i,\text{test}})=\frac{ p_\text{$\alpha$,test}^{C_i}\sqrt{m_{V_i}^2+(p_\text{T}^{V_i})^2}}{\sqrt{m_{C_i}^2+(p_\text{T,test}^{C_i})^2}}-p_\text{$\alpha$}^{V_i}\\\nonumber
	&\myeqa p_\text{$\alpha$,test}^{C_i}=p_\text{$\alpha$}^{V_i}\frac{\sqrt{m_{C_i}^2+(p_\text{T,test}^{C_i})^2}}{ \sqrt{m_{V_i}^2+(p_\text{T}^{V_i})^2}}\myeqb p_\text{T,test}^{C_i} = \frac{m_{C_i}}{m_{V_i}}	p_\text{T}^{V_i}\\
	&\myeqc p_\text{$\alpha$,test}^{C_i} = \frac{m_{C_i}}{m_{V_i}}	p_\text{$\alpha$}^{V_i} \myeqd m_\text{T,min}=m_{V_i}+m_{C_i}.
	\end{align}

	In addition to the flexibility to choose the particles composing $V_1$ and $V_2$ as well as the test masses $m_{C_i}$, there is an extensive literature on variations of $m_\text{T2}$ and related variables that aim to solve the same multiple-missing-particle problem.  See Ref.~\cite{Barr:2011xt,Barr:2010zj} for a extensive reviews of the existing methods.  One important variation is the {\it perpendicular} $m_\text{T2}$ {\it variables}.  Unlike the simple transverse mass, $m_\text{T2}$ generally does not have a closed-form solution to the minimization in Eq.~\ref{eq:generalizedmt}.  Numerical techniques for computing $m_\text{T2}$ are described in Sec.~\ref{sec:numericalmethodsmt2}, but first Sec.~\ref{mt2perp} documents the perpendicular $m_\text{T2}$ which does have a closed-form solution.
				
				\paragraph{Perpendicular $m_\text{T2}$} \mbox{}\\
				\label{mt2perp}
				
	Define the upstream transverse momentum as $\vec{p}_\text{T}^\text{up}=-\vec{p}_\text{T}^{V_1}-\vec{p}_\text{T}^{V_2}-\vec{p}_\text{T}^\text{miss}$, i.e. all transverse momentum aside from the momenta from $V_1$ and $V_2$.  The perpendicular momenta $\vec{p}_\text{T,$\perp$}=\vec{p}_\text{T}-(\hat{p}^\text{up}\cdot \vec{p}_\text{T})\hat{p}^\text{up}$, where $\hat{p}^\text{up}=\vec{p}_\text{T}^\text{up}/|\vec{p}_\text{T}^\text{up}|$.  In a topology like the one used to make Fig.~\ref{fig:susymtdist_testmass} but with $t\bar{t}$ production (Fig.~\ref{fig:susymtdist_testmass1}), the upstream momentum includes the $b$-jets directly from the top quark decay and any ISR radiated prior to the $t\bar{t}$ production.   At the end of Sec.~\ref{sec:mT}, Fig.~\ref{fig:susymtdist_testmass} showed that when the test mass $m_{C_i}$ is not equal to the mass of the missing object, the presence of nonzero upstream momentum makes the kinematic maximum $\vec{p}_\text{T}^\text{up}$-dependent.   By constructing a $m_\text{T2}$ variable with only perpendicular momenta, $m_\text{T2,$\perp$}$~\cite{Konar:2009wn}, the $\vec{p}_\text{T}^\text{up}$-independence of the kinematic maximum is restored because $\vec{p}_\text{T,$\perp$}^\text{up}=\vec{0}$ by construction.  
				
In addition to its $\vec{p}_\text{T}^\text{up}$-independence, $m_\text{T2,$\perp$}$ is useful because it has an analytic formula for the event-by-event\footnote{In the general case, even though there is no analytic formula for the event-by-event quantity, there are general formulae for the kinematic maxima - see Ref.~\cite{Burns:2008va}.  In general, this endpoint depends on $p_\text{T}^\text{up}$.} value.  To illustrate how this works, consider a special but important case where $m_{C_i}=m_{V_i}=0$.  Since all perpendicular momenta lie along a line:
				
	\begin{align}
	\label{eq:mt2perp}
	m^2_\text{T,$\perp$}(\vec{p}_\text{T,$\perp$}^{V_i},\vec{p}_\text{T,$\perp$,test}^{C_i})=\left\{\begin{matrix}4p_\text{T,$\perp$}^{V_i}p_\text{T,$\perp$,test}^{C_i}& \text{$\vec{p}_\text{T,$\perp$}^{V_i}$ is anti-parallel to $\vec{p}_\text{T,$\perp$,test}^{C_i}$} \cr  0 &\text{else} \end{matrix}\right. .
	\end{align}
	
\noindent The calculation of $m_\text{T2}$ is particularly simple because it is now a one-dimensional optimization problem.  Without loss of generality, suppose that $p_\text{T,$\perp$}^{V_1}\geq p_\text{T,$\perp$}^{V_2}$.  To ease the notation, let $x$ be the signed projected test momentum in the minimization, $x=\vec{p}^{C_1}_\text{T,$\perp$,test}\cdot \hat{p}_\text{T,$\perp$}^{V_1}$ where $\hat{p}_\text{T,$\perp$}^{V_1}=\vec{p}_\text{T,$\perp$}^{V_1}/p_\text{T,$\perp$}^{V_1}$ is a unit vector pointing in the direction of the projected momentum for $V_1$.   Analogously, define $p,q,y$ and $\epsilon$ as signed scalars representing the momenta of $V_1,V_2,C_1$ and $E_\text{T,$\perp$}$, i.e. $p=p^{V_1}_\text{T,$\perp$}$, $q=\vec{p}^{V_2}_\text{T,$\perp$}\cdot \hat{p}_\text{T,$\perp$}^{V_1}$, $y=\vec{p}^{C_2}_\text{T,$\perp$,test}\cdot \hat{p}_\text{T,$\perp$}^{V_1}$, and $\epsilon=\vec{E}^\text{miss}_\text{T,$\perp$}\cdot \hat{p}_\text{T,$\perp$}^{V_1}$.  By momentum conservation, $p+q+\epsilon=0$ and $x+y=\epsilon$.  Therefore, $y=-p-q-x$. With this notation, Eq.~\ref{eq:mt2perp} can be re-written as $(m_\text{T,$\perp$}^{V_1})^2=4|px|(1-\Theta(x))$ and $(m_\text{T,$\perp$}^{V_2})^2=4|qy|(1-\Theta(y))$ where $\Theta(x)$ is the Heavyside step function.  The two branches of Eq.~\ref{eq:mt2perp} are illustrated in the upper diagrams of Fig.~\ref{fig:susymtdist_testmass3}.  In the left diagram of Fig.~\ref{fig:susymtdist_testmass3}, the two visible perpendicular momenta are on opposite sides of the upstream momentum so $pq<0$.  Since $|p|>|q|$, $\epsilon < 0$.  When $x>0$, $m_\text{T,$\perp$}^{V_1}=0$ and for $x<0$, $(m_\text{T,$\perp$}^{V_1})^2=4p|x|$.  Similarly, $m_\text{T,$\perp$}^{V_2}=0$ when $y>0$, which occurs when $x<-|p|+|q|$.  For $x>0$, $m_\text{T,$\perp$}^{V_2}=m_\text{T,$\perp$}^{V_1}=\max_i\{m_\text{T,$\perp$}^{V_i}\}=0$, and therefore $m_\text{T2,$\perp$}=0$.  This is illustrated graphically in the lower left graph in Fig.~\ref{fig:susymtdist_testmass3}.   

The second possibility is that $pq>0$, as in the upper right diagram of Fig.~\ref{fig:susymtdist_testmass3}.  The curve for $(m_\text{T,$\perp$}^{V_1})^2$ in the lower right graph of Fig.~\ref{fig:susymtdist_testmass3} is unchanged from the first case.  However, now $y<0$ is required for $m_\text{T,$\perp$}^{V_2}$ to be nonzero.  This occurs when $x>-|p+q|$.  When $x$ is large, $\max_i\{m_\text{T,$\perp$}^{V_i}\}=m_\text{T,$\perp$}^{V_2}$ and when $x$ is much less than zero, $\max_i\{m_\text{T,$\perp$}^{V_i}\}=m_\text{T,$\perp$}^{V_1}$.  As illustrated by lower right graph of Fig.~\ref{fig:susymtdist_testmass3}, the global minimum occurs when $m_\text{T,$\perp$}^{V_1}=m_\text{T,$\perp$}^{V_2}$, which implies $px=qy=q(-p-q-x)$, or $x=-q$.  Substituting $x=-q$ in the formula for $m_\text{T,$\perp$}^{V_1}$ then gives $m_\text{T2,$\perp$}^2=4pq$.   A formula that covers both of the above cases is $m_\text{T2,$\perp$}^2=2A_\text{T,$\perp$}$ where $A_\text{T,$\perp$}=(|pq|-pq)=(p_\text{T,$\perp$}^{V_1}p_\text{T,$\perp$}^{V_1}-\vec{p}_\text{T,$\perp$}^{V_1}\cdot \vec{p}_\text{T,$\perp$}^{V_2})$.  A straightforward extension of the above argument to the case where the test masses are not zero, but are both equal to the same value $m_C$ gives~\cite{Konar:2009wn} $m_\text{T2,$\perp$}=\sqrt{\frac{1}{2}A_\text{T,$\perp$}}+\sqrt{\frac{1}{2}A_\text{T,$\perp$}+m_{C}^2}$.  Additionally including nonzero visible particle masses has the form\footnote{Note that this formula also holds in the limit that the upstream momentum is zero.  In that case, all $\perp$ quantities are replaced by the regular momenta.}~\cite{Lester:2007fq,Cho:2007dh}:

\begin{align}
m_\text{T2,$\perp$}^2=m_C^2+A_\text{T,$\perp$}+\sqrt{\left(1+\frac{4m_C^2}{2A_\text{T,$\perp$}-m_{V_1}^2-m_{V_2}^2}\right)\left(A_\text{T,$\perp$}^2-m_{V_1}^2m_{V_2}^2\right)},
\end{align}

\noindent where the generalized $A_\text{T,$\perp$}=\frac{1}{2}(E_\text{T,$\perp$}^{V_1}E_\text{T,$\perp$}^{V_1}-\vec{p}_\text{T,$\perp$}^{V_1}\cdot \vec{p}_\text{T,$\perp$}^{V_2})$ for $(E_\text{T,$\perp$}^{V_i})^2=m_{V_i}^2+(p_\text{T,$\perp$}^{V_i})^2$.  The full formula, including the possibility for $m_{C_1}\neq m_{C_2}$ has also been computed~\cite{Konar:2009qr}:

\begin{align}\nonumber
\label{eq:mt2general}
m_\text{T2,$\perp$}^2=\Sigma_C&+A_\text{T,$\perp$}+\frac{\Delta_C\Delta_V}{A_\text{T,$\perp$}-\Sigma_V}\\
&\pm\sqrt{\left(1+\frac{2\Sigma_C}{A_\text{T,$\perp$}-\Sigma_V}+\left(\frac{\Delta_C}{A_\text{T,$\perp$}-\Sigma_V}\right)^2\right)\left(A_\text{T,$\perp$}^2-m_{V_1}^2m_{V_2}^2\right)},
\end{align}

\noindent where $\Sigma_C=\frac{1}{2}(m_{C_1}^2+m_{C_2}^2)$, $\Delta_C=\frac{1}{2}(m_{C_2}^2-m_{C_1}^2)$ and similarly for $\Sigma_V, \Delta_V$ and $m_{V_i}$.

In all cases, the event-by-event formula for $m_\text{T2,$\perp$}$ depends on the particle momenta only though the quantity $A_\text{T,$\perp$}$.  When the visible particle masses are small compared to their momenta, this means that the distribution of $m_\text{T2,$\perp$}$ will have a $\delta$-function corresponding to the case $A_\text{T,$\perp$}=0$ that occurs whenever the visible particle momenta are on opposite sides of the upstream momentum.  Furthermore, the additional projection compared with the  original $m_\text{T2}$ means that $m_\text{T2,$\perp$}$ tends to be lower for background and signal events.  A full comparison of the discriminating power of $m_\text{T2,$\perp$}$ with $m_\text{T2}$ is described in Sec.~\ref{sec:mt2tmva}.

\begin{figure}[h!]
\begin{center}
\includegraphics[width=0.8\textwidth]{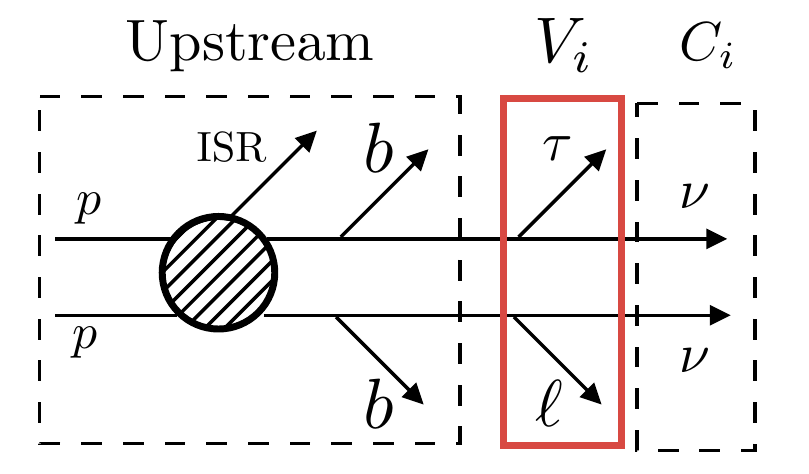}
 \caption{An illustration of the setup for the projected $m_\text{T2}$ variable. The upstream momentum is the sum of all particles not associated with the visible particles $V_i$ and child particles $C_i$.  In this $t\bar{t}$ topology, the upstream momentum is due to the $b$-jets directly from the top quark decay and any FSR produced before the $t\bar{t}$ prouction.}
 \label{fig:susymtdist_testmass1}
  \end{center}
\end{figure}
				
\begin{figure}[h!]
\begin{center}
\includegraphics[width=0.95\textwidth]{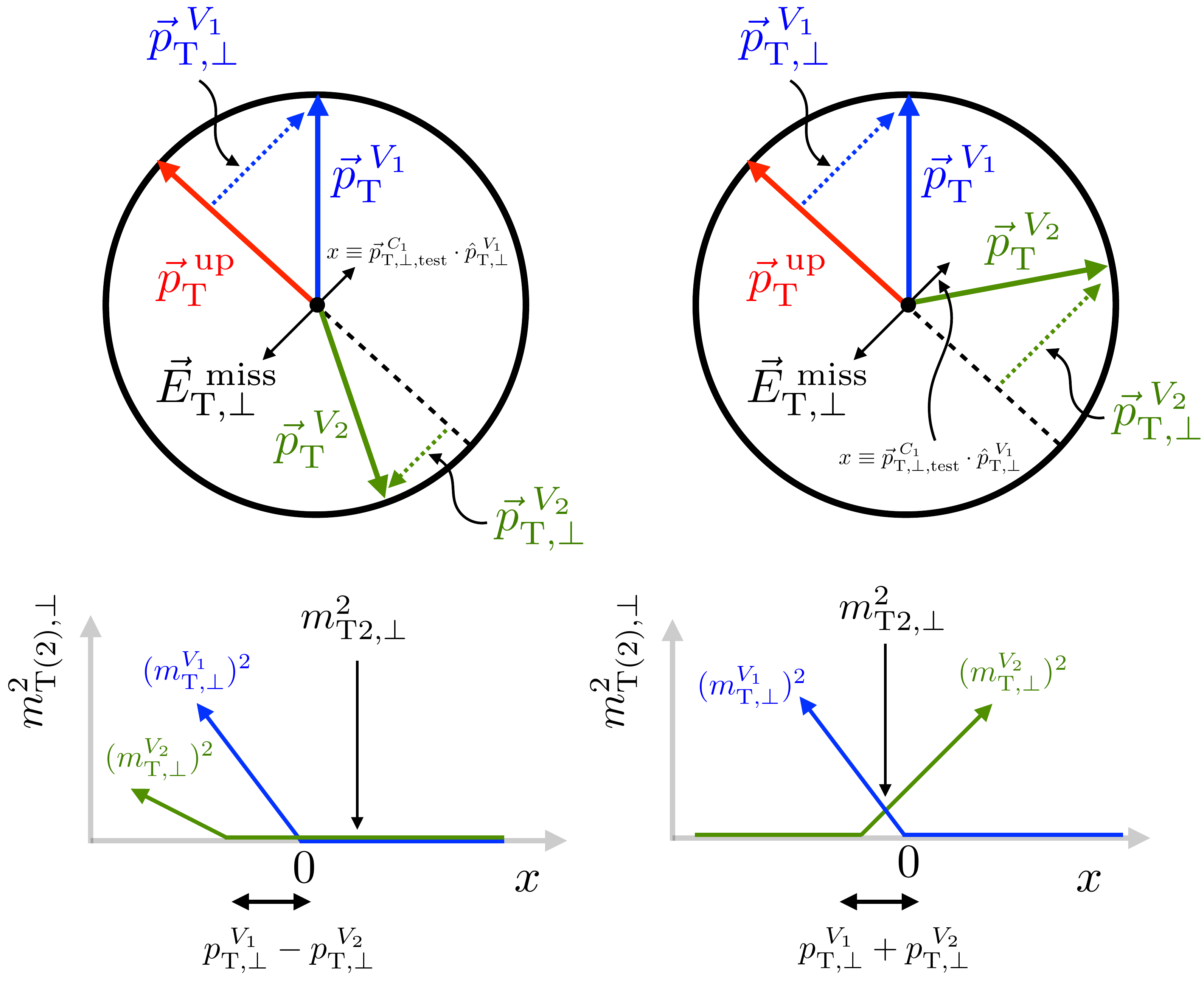}
 \caption{Diagrams illustrating the two possible orientations of projected transverse momenta (top) and the corresponding $m_\text{T,$\perp$}$ graphs for both branches of the decay (bottom). The circles represent a transverse cross-section of the detector - the beam is into and out of the page.  In the left configuration, the visible transverse momenta are on opposite sides of the upstream momentum and the resulting $m_\text{T2,$\perp$}$ value is equal to zero.  In the right diagrams, the visible momenta are on the same side of the upstream momentum which allows for a finite $m_\text{T2,$\perp$}$ value.}
 \label{fig:susymtdist_testmass3}
  \end{center}
\end{figure}				
				
				\clearpage
				
				\paragraph{Numerical Methods} \mbox{}\\
				\label{sec:numericalmethodsmt2}
			
Aside from special cases such as the perpendicular $m_\text{T2}$ in the previous section, there is no general analytic formula for the event-by-event value of $m_\text{T2}$.  The level sets of the $m_\text{T}^2$ curves in Eq.~\ref{eq:mt2eq} are conic sections in $x=\vec{p}_\text{$x$,test}^{C_i}\cdot\hat{x}$ and $y=\vec{p}_\text{$y$,test}^{C_i}\cdot\hat{y}$:

\begin{align}
\label{ellipse}
(m_{V_i}^2+(p_y^{V_i})^2)x^2+(m_{V_i}^2+(p_x^{V_i})^2)y^2-m^2p_x^{V_i}x-m^2p_y^{V_i}y-2p_x^{V_i}p_y^{V_i}xy-M=0,
\end{align}

\noindent where $m^2=m_\text{T}^2-m_{C_i}^2+m_{V_i}^2$ and $M=\frac{1}{4}m^4-(E_\text{T}^\text{$V_1$})^2(m_{C_i}^2)$.  The coefficients $A,B,C$ of the $x^2$, $xy$ and $y^2$ terms in Eq.~\ref{ellipse} satisfy $B-4AC\leq 0$:

\begin{align}
4(p_x^{V_i})^2(p_y^{V_i})^2-4(m_{V_i}^2+(p_y^{V_i})^2)(m_{V_i}^2+(p_x^{V_i})^2) = -4m_{V_i}^2(m_{V_i}^2+(p_\text{T}^{V_i})^2).
\end{align}

\noindent Therefore, the level sets are ellipses as long as $m_{V_i}>0$ and parabolas otherwise.  Writing Eq.~\ref{ellipse} as $f_i(x,y)=0$, it is possible to plot both curves with one set of coordinates using the conservation of momentum constraint, $f_1(x,y)=0$ and $f_2(E_x^\text{miss}-x,E_y^\text{miss}-y)=0$.  A vertical slice in $(x,m_\text{T})$ space that goes through the value of $m_\text{T2}$ looks like either the left or right graph in Fig.~\ref{fig:susymtdist_testmass1c}.  In the left graph of Fig.~\ref{fig:susymtdist_testmass1c}, the minimum over $\max\{m_\text{T}\}$ occurs at the intersection of the two $m_\text{T}$ curves (balanced) while in the right graph, the minimum of one $m_\text{T}$ curve is above the other curve and is thus equal to $m_\text{T2}$ (unbalanced).  Without loss of generality, assume that $m_{V_1}+m_{C_1}>m_{V_2}+m_{C_2}$.  From Eq.~\ref{eq:minmt2}, the minimum of $m_\text{T}^{V_1}$ occurs when $(x,y) = \frac{m_{C_1}}{m_{V_1}}	\vec{p}_\text{T}^{V_i}$.  Therefore, the condition for the unbalanced case is $m_{V_1}+m_{C_1}>m_\text{T}(\vec{p}_\text{T}^{V_2},\vec{E}_\text{T}^\text{miss}-\frac{m_{C_1}}{m_{V_1}}	\vec{p}_\text{T}^{V_1},m_{V_2},m_{C_2})$.  This case can only occur if $m_{V_1}+m_{C_1}\neq m_{V_2}+m_{C_2}$.  Figure~\ref{fig:susymtdist_testmass1d} demonstrates the probability for the unbalanced case as a function of $m_{C_1}$ and $m_{C_2}$ in $t\bar{t}$ events.  The fraction of unbalanced events is increased by choosing $V_1$ and $V_2$ with significantly different masses ($m_{bl}$ and $m_b$).  The larger the difference between the $m_\text{T}$ minima, the higher the probability for the unbalanced case.

\begin{figure}[h!]
\begin{center}
\includegraphics[width=0.85\textwidth]{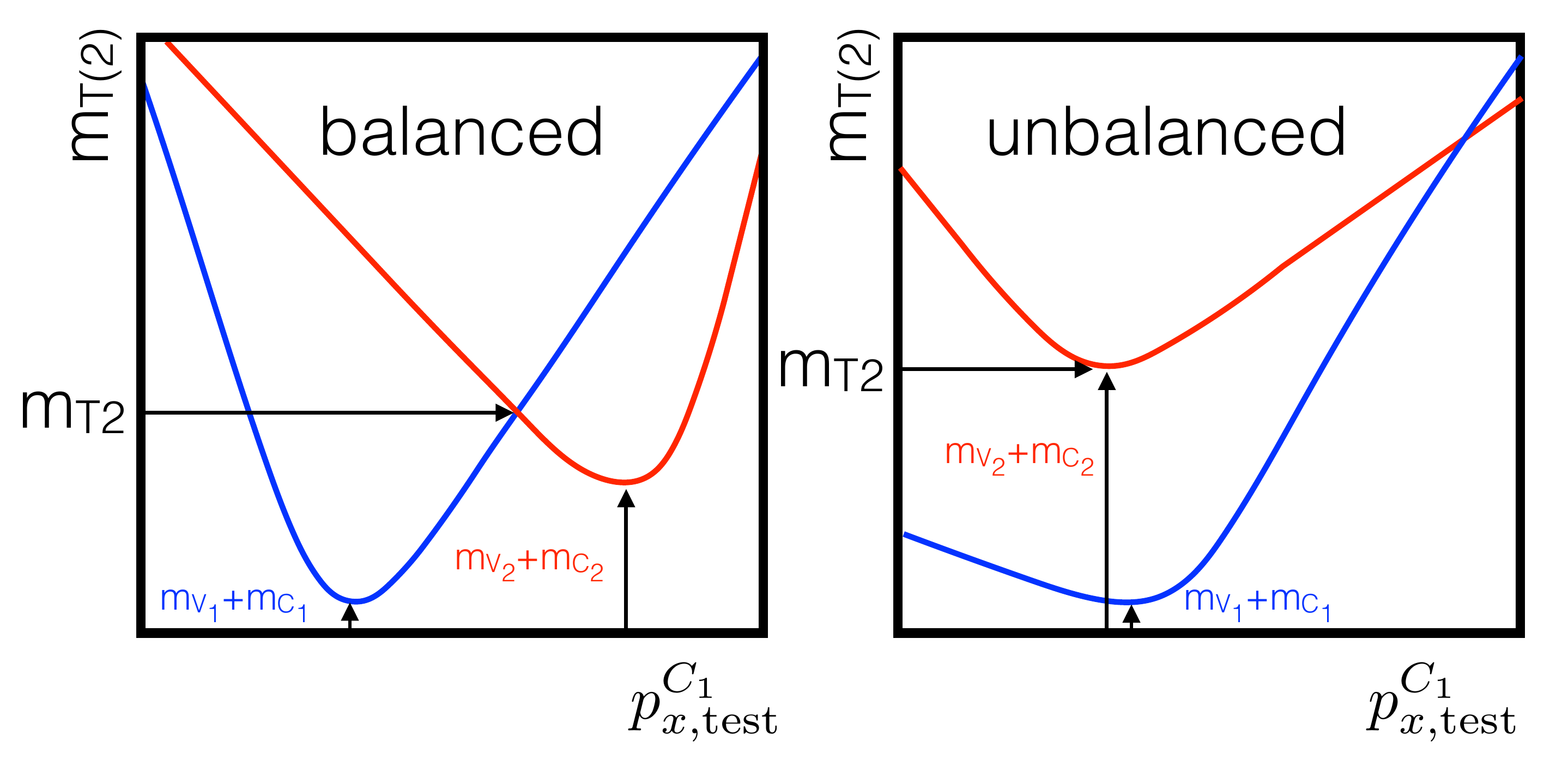}
 \caption{Graphs illustrating the two possible configurations leading to the value of $m_\text{T2}$ at the intersection of the two $m_\text{T}$ conic sections (left) or at the minimum of one of the sections, if it is above the other curve (right).}
 \label{fig:susymtdist_testmass1c}
  \end{center}
\end{figure}	

\begin{figure}[h!]
\begin{center}
\includegraphics[width=0.5\textwidth]{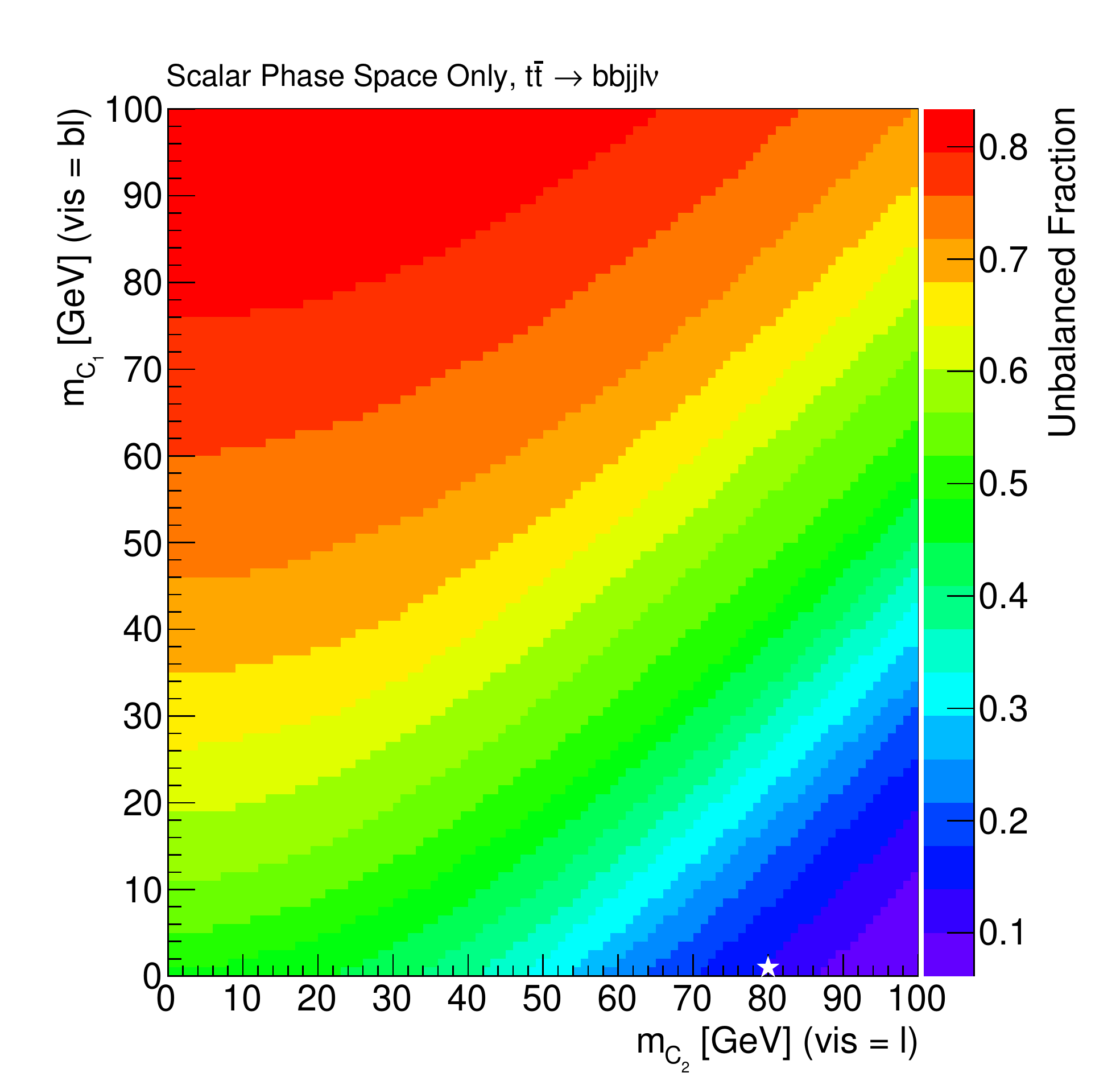}
 \caption{The fraction of events for which the $m_\text{T2}$ value is $m_{V_1}+m_{C_1}$ (unbalanced case) for $t\bar{t}$ events as a function of $m_{C_1}$ and $m_{C_2}$.  The visible object $V_1$ is the four-vector sum of the $b$-quark and lepton from the same top quark decay and $V_2$ is the $b$-quark from the other top quark decay.  A white star indicates the choice for the $am_\text{T2}$ variable described in Sec.~\ref{mt2forstop}.}
 \label{fig:susymtdist_testmass1d}
  \end{center}
\end{figure}	
			
In the balanced case, the value of $m_\text{T2}$ is determined by computing the minimum value along the intersection of the $m_\text{T}$ surfaces, as illustrated by Fig.~\ref{fig:susymtdist_testmass1e}.  Equivalently, the value of $m_\text{T2}$ is equal to the point at which the ellipses from the $m_\text{T}^2$ level curves are tangent.  This second condition, illustrated in Fig.~\ref{fig:susymtdist_testmass1f} is used for quickly and accurately computing $m_\text{T2}$.  Rewriting $f_1(x,y)=0$ and $f_2(E_x^\text{miss}-x,E_y^\text{miss}-y)=0$ with $m_\text{T}^{V_1}=m_\text{T}^{V_2}=m_\text{T}$:

\begin{align}\nonumber
\label{quadraticmt2}
a_{y2}y^2+(a_{xy}x+a_y(m_\text{T}^2))y+(a_{x2}x^2+a_x(m_\text{T}^2)x+a_0(m_\text{T}^4))&=0\\
b_{y2}y^2+(b_{xy}x+b_y(m_\text{T}^2))y+(b_{x2}x^2+b_x(m_\text{T}^2)x+b_0(m_\text{T}^4))&=0,
\end{align}

\noindent where the coefficients $a_i$ and $b_i$ are given by Eq.~\ref{ellipse}.  Solving for $y$ gives

\begin{align}
y=\tilde{a}_1(x,m_\text{T}^2)\pm\sqrt{\tilde{a}_2(x^2,m_\text{T}^4)+\tilde{a}_3(x^2,m_\text{T}^4)},
\end{align}

\noindent where the coefficients $\tilde{a}$ are the usual solution to the quadratic equation from Eq.~\ref{quadraticmt2}.  A similar expression holds for the second branch but with $a\leftrightarrow b$.  Where the two ellipses intersect, the values of $y$ will be the same:

\begin{align}
\tilde{a}_1(x,m_\text{T}^2)\pm\sqrt{\tilde{a}_2(x^2,m_\text{T}^4)+\tilde{a}_3(x^2,m_\text{T}^4)}=\tilde{b}_1(x,m_\text{T}^2)\pm\sqrt{\tilde{b}_2(x^2,m_\text{T}^4)+\tilde{b}_3(x^2,m_\text{T}^4)}.
\end{align}

\noindent Rearranging to remove the radicals:

\begin{align}
\label{quarticmt2}
\frac{1}{4}(c_1(x^2,m_\text{T}^4)-c_2(x^2,m_\text{T}^4))^2=\tilde{a}_2(x^2,m_\text{T}^4)+\tilde{a}_3(x^2,m_\text{T}^4)\tilde{b}_2(x^2,m_\text{T}^4)+\tilde{b}_3(x^2,m_\text{T}^4),
\end{align}

\noindent where $c_1(x^2,m_\text{T}^4)=(\tilde{a}_1(x,m_\text{T}^2)-\tilde{b}_1(x,m_\text{T}^2))^2$ and $c_2(x^2,m_\text{T}^4)=\tilde{a}_2(x^2,m_\text{T}^4)+\tilde{a}_3(x^2,m_\text{T}^4)+\tilde{b}_2(x^2,m_\text{T}^4)+\tilde{b}_3(x^2,m_\text{T}^4)$.  Equation~\ref{quarticmt2} is a quartic equation of $x$ for a fixed $m_\text{T}$.  From Fig.~\ref{fig:susymtdist_testmass1f}, when $m_\text{T}<m_\text{T2}$, there are no intersection points and so Eq.~\ref{quarticmt2} will have no real roots.  In contrast, when $m_\text{T}>m_\text{T2}$, Fig.~\ref{fig:susymtdist_testmass1f} shows that there are two points of intersection for the ellipses and so Eq.~\ref{quarticmt2} will have two real roots.  At exactly the value $m_\text{T}=m_\text{T2}$, Eq.~\ref{quarticmt2} will have one repeated root.  A polynomial has a repeated root if and only if its determinant is zero.   The discriminant of a quartic polynomial is a sixth degree polynomial in the coefficients of the polynomial.  However, the coefficients of Eq.~\ref{quarticmt2} are quadratic functions of $m_\text{T}^2$ and therefore solving for $m_\text{T2}$ using the discriminant requires finding the roots of a $12^\text{th}$ degree polynomial.  There is no general analytic formula for such a high degree polynomial so the roots must be computed numerically.  Even thought his is a well-defined procedure, it is not the usual way $m_\text{T2}$ is computed because it would be relatively slow and possibly numerically unstable. 

\begin{figure}[h!]
\begin{center}
\includegraphics[width=0.99\textwidth]{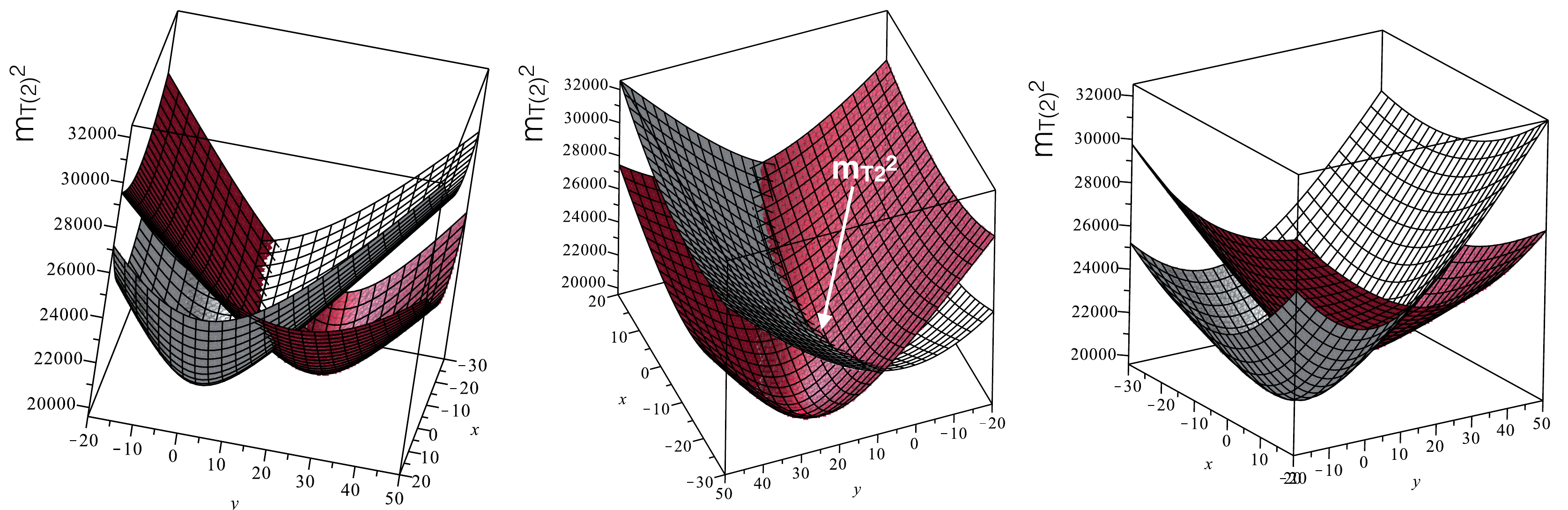}
 \caption{Three views of the $m_\text{T}^2$ surfaces from the same event, rotated so illustrate the structure of the intersection.  The value of $m_\text{T2}^2$ is indicated in the middle graph by a white arrow.}
 \label{fig:susymtdist_testmass1e}
  \end{center}
\end{figure}	

\begin{figure}[h!]
\begin{center}
\includegraphics[width=0.5\textwidth]{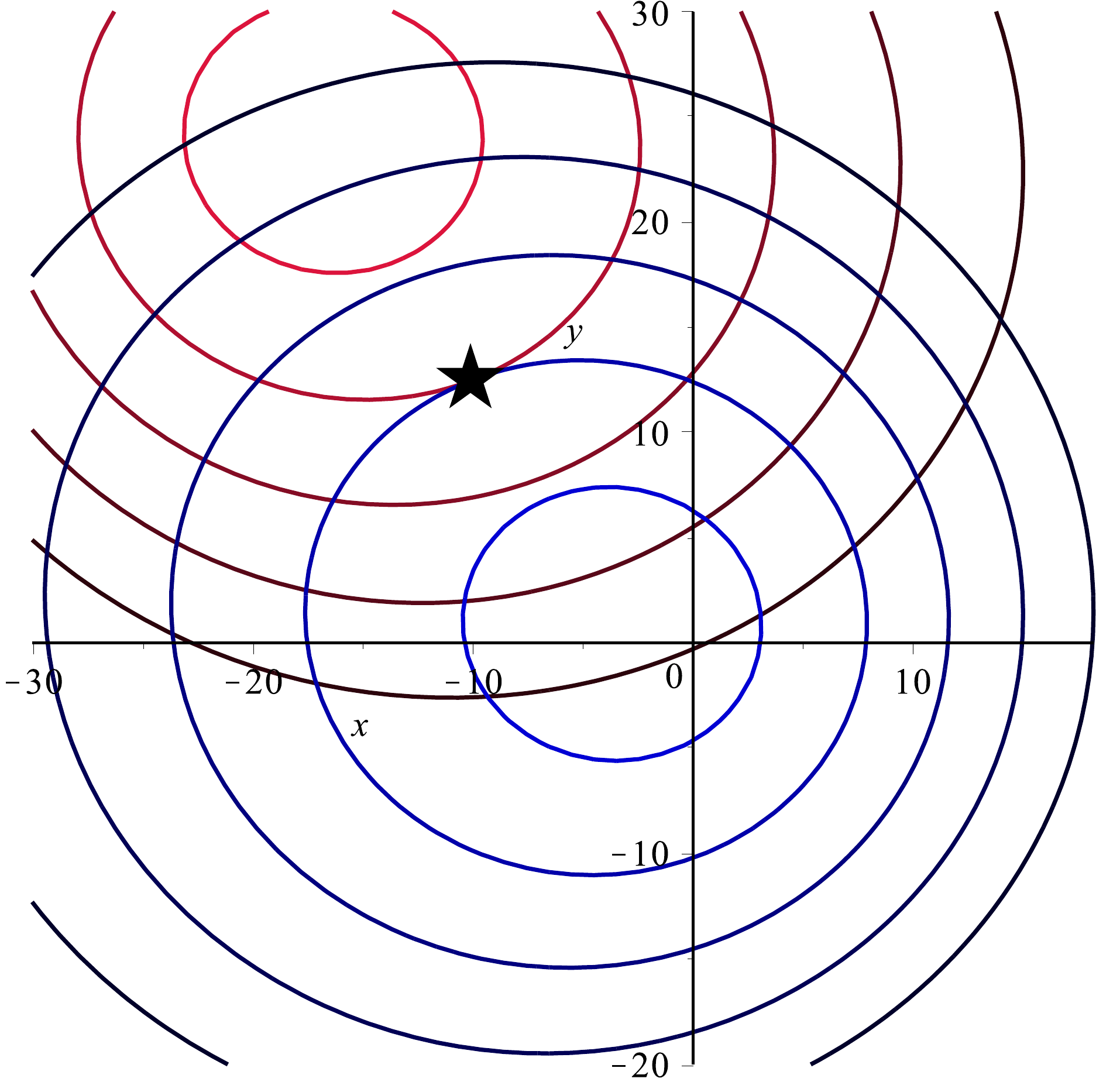}
 \caption{Level surfaces of the graphs from Fig.~\ref{fig:susymtdist_testmass1e}.  As noted above, these curves are ellipses and the point at which they are tangent (indicated by a black star) corresponds to the value of $m_\text{T2}$.}
 \label{fig:susymtdist_testmass1f}
  \end{center}
\end{figure}	

The state-of-the-art numerical calculators for $m_\text{T2}$ are based on the observation that it is quick and easy to check if two ellipses intersect.  Then, $m_\text{T2}$ is computed by iteratively bisecting an interval known to contain the point at which the ellipses are tangent.  Bisection techniques are powerful because they achieve a precision of $n$ decimal places with only $\log_2(10^n)$ bisections.  The first bisection method~\cite{Cheng:2008hk} used the above observation that the number of real roots of Eq.~\ref{quarticmt2} differs if $m_\text{T}$ is above or below $m_\text{T2}$.  A fast way to check the number of real roots of a polynomial is to use the {\it Sturm sequence}, which is based on a few evaluations of a simple series of five polynomials (the original one, its derivative, and various divisors).  An even faster method is based on the observation by C. Lester that it is easier and more robust to check if the area of two ellipses overlap than to check if their boundaries intersect~\cite{Lester:2014yga}.  The numerical procedure for the quick evaluation of overlapping conic sections is from Ref.~\cite{Etayo2006324}.  In addition to the speed of evaluation, this new procedure is more robust compared with the Sturm sequence method because it removes the need for special cases when e.g. one of the visible particles is massless and the $m_\text{T}$ level sets are parabolas instead of ellipses.  Furthermore, before the availability of the calculator from Ref.~\cite{Lester:2014yga}, there was no specialty $m_\text{T2}$ calculator for the case $m_{C_1}\neq m_{C_2}$ (Ref.~\cite{Cheng:2008hk} is only programed for the symmetric case).  Therefore, at $\sqrt{s}=8$ TeV $m_\text{T2}$ was evaluated using a generic function minimizer (Migrad - see Sec.~\ref{sec:optimizationprocedure}) initialized with $x=\vec{p}_x^\text{miss}\cdot\hat{x}/2$ and $y=\vec{p}_y^\text{miss}\cdot\hat{y}/2$.  Each $m_\text{T}$ surface is smooth, so the numerical minimization is robust except near the intersection of the surfaces where some instability is caused by the discontinuity in the first derivative.  The $\sqrt{s}=13$ TeV analysis uses the dedicated calculator from Ref.~\cite{Lester:2014yga} that is orders of magnitude faster than the generic approach for the same precision.  Figure~\ref{fig:susymtdist_testmass1g} shows that the bisection approach is just as good as the analytic formula in the case $p_\text{T}^\text{up}=0$.

\vspace{10mm}

\begin{figure}[h!]
\begin{center}
\includegraphics[width=0.95\textwidth]{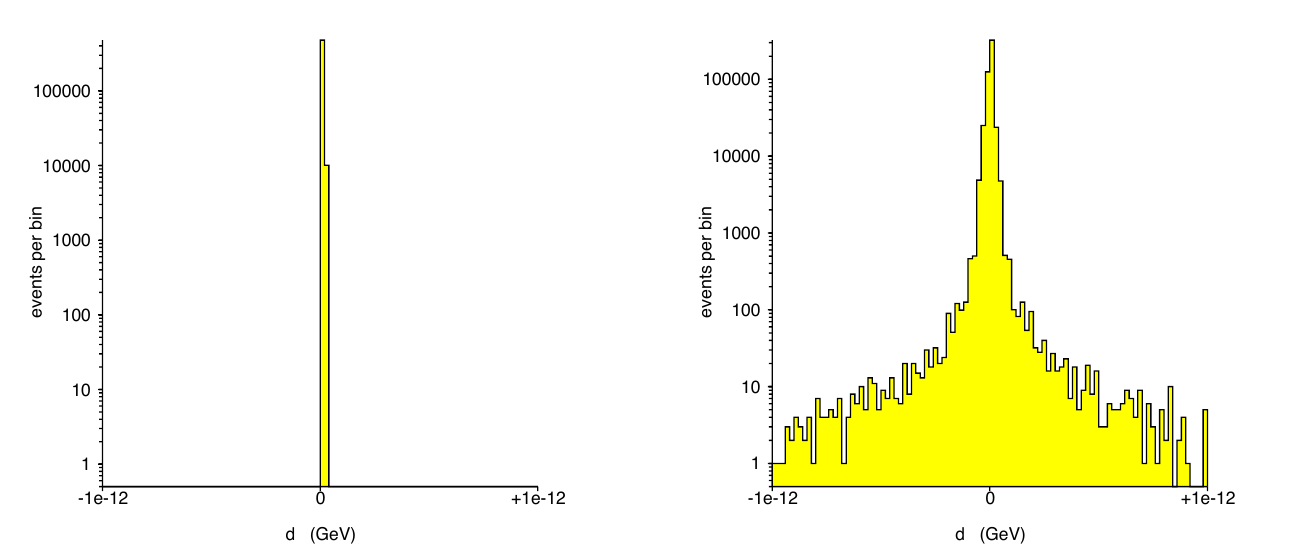}
 \caption{A demonstration of the numerical accuracy of the $m_\text{T2}$ calculator based on the overlap of conic sections instead of the intersection of their boundaries.  Plotted is the difference $d$ between the numerical calculation and the analytic formula using the projected $m_\text{T}$ from Sec.~\ref{mt2perp} in $t\bar{t}$ events for the variable $am_\text{T2}$ described in Sec.~\ref{mt2forstop}.  The scale of $m_\text{T}\sim\mathcal{O}(100$ GeV$)$.  See Ref.~\cite{Lester:2014yga} for more detail.}
 \label{fig:susymtdist_testmass1g}
  \end{center}
\end{figure}	
				\clearpage
				
				\paragraph{Tailoring $m_\text{T2}$ for the stop search}\mbox{}\\
				\label{mt2forstop}
				
				With a high multiplicity final state, there are many choices for $V_i$ and $m_{C_i}$ in constructing an $m_\text{T2}$ variable for the stop search.  This section describes two particular $m_\text{T2}$ variables that are designed to suppress dilepton $t\bar{t}$ events.  After requiring $m_\text{T}>m_W$, Fig.~\ref{fig:susymtdatamc} showed that the majority of surviving $t\bar{t}$ events have a second lepton that is either lost, mis-identified, or is a hadronically decaying $\tau$ lepton.  First, consider the case in which the second lepton is undetected so that $\vec{p}_\text{T}^\text{miss}\approx \vec{p}_\text{T}^{\nu_1}+\vec{p}_\text{T}^{\nu_2}+\vec{p}_\text{T}^\text{lost $\ell$}$.  One could construct an $m_\text{T2}$ variable using the $b$-jets as the $V_i$ and then grouping everything downstream of the $b$-quarks in the top decay chain into $\vec{p}_\text{T}^{C_1}+\vec{p}_\text{T}^{C_2}$.  However, additional information is available by using asymmetric objects~\cite{Barr:2009jv,Konar:2009qr} for the $V_i$.  Following an idea in Ref.~\cite{Bai:2012gs}, the {\it asymmetric} $m_\text{T2}$ ($am_\text{T2}$) is formed by letting $V_1$ be the $b$-jet from one top quark decay and setting $V_2$ to be the four-vector sum of the $b$-jet and lepton from the other top quark decay.  As illustrated in Fig.~\ref{fig:susymtdist2}, this means that the missing particle for the top branch is an entire $W$ boson and on the bottom branch, only a neutrino.  Therefore, $m_{C_1}=m_W$ and $m_{C_2}=m_\nu\approx 0$.  With these choices, $am_\text{T2}\leq m_\text{top}$ for the background depicted in Fig.~\ref{fig:susymtdist2}.
				
\begin{figure}[h!]
\begin{center}
\begin{tikzpicture}[line width=1.5 pt, scale=1.3]
			\draw (0,0.5) -- (1,0.5);
			\draw (0,-0.5) -- (1,-0.5);
			\draw (1,-0.5) -- (1.5,-1.5);
			\draw (1,0.5) -- (1.5,1.5);
			\draw (2,0.5) -- (2.5,1.5);
			\draw (2,-0.5) -- (2.5,-1.5);
			\draw (2,-0.5) -- (3,-0.5);
			\draw (2,0.5) -- (3,0.5);
			\draw[dashed, black!30] (3,1.2) ellipse (1 and 1.1);
			\draw[dashed, black!30] (3.3,-0.5) ellipse (0.5 and 0.5);
			\node[draw,circle] at (-0.25,0.5) {$t$};
			\node[draw,circle] at (-0.25,-0.5) {$t$};
			\node[draw,circle] at (3.25,0.5) {$\nu$};
			\node[draw,circle] at (3.25,-0.5) {$\nu$};
			\node[draw,circle] at (2.65,1.7) {$l$};
			\node[draw,circle] at (2.65,-1.7) {$l$};
			\node[draw,circle] at (1.65,1.7) {$b$};
			\node[draw,circle] at (1.65,-1.7) {$b$};
			\draw[snake=snake] (1,0.5) -- (2,0.5); 
			\draw[snake=snake] (1,-0.5) -- (2,-0.5); 
		\end{tikzpicture}
 \caption{A schematic diagram of dileptonic $t\bar{t}$ decay where one of the charged leptons is lost.  Lost particles are circled with a dashed line.  For the $am_\text{T2}$ variable, the visible particle on the top (bottom) branch is the $b$-jet (sum of the $b$-jet and lepton).  The missing particle in the top (bottom) branch is a $W$ boson (neutrino).}
 \label{fig:susymtdist2}
 \end{center}
\end{figure}				
	
An important practical complication for constructing $am_\text{T2}$ is the selection of the $b$-jets and the pairing of the lepton with the $b$-jet from the same branch.  The signal region event selections described in Sec.~\ref{chapter:susy:signalregions} only require one explicitly identified $b$-tagged jet.  Section~\ref{sec:mt2tmva} will explore two methods for choosing the two $b$-tagged jet: the two jets with the highest $b$-tag discriminant weights or the two highest $p_\text{T}$ jets.  As the $b$-quarks appear higher in the top quark decay chain than the other tree-level quarks, it is expected that they have a higher $p_\text{T}$ on average.  Section~\ref{sec:mt2tmva} also considers two algorithms for matching $b$-jets with the lepton: take the closest in $\Delta R$ or compute both possibilities and set $am_\text{T2}$ to be the minimum of the two $m_\text{T2}$ values.  Figure~\ref{fig:amt2parton} shows the scalar parton-level distribution of $am_\text{T2}$ for dileptonic $t\bar{t}$ and stop events.  By construction, $am_\text{T2}<m_\text{top}$ when the second lepton is lost.  In contrast, when the second lepton is measured but not identified as a lepton, there is a small tail of events with $am_\text{T2}>m_\text{top}$.  This is also true when the second lepton is lost, but the $b$-jet / lepton pairing is performed with $\Delta R$.  For stop events, $am_\text{T2}$ can greatly exceed $m_\text{top}$.  The peak around $150$ GeV is due to the unbalanced case in which $am_\text{T2}=m_{bl}$. 

\begin{figure}[h!]
\begin{center}
\includegraphics[width=0.49\textwidth]{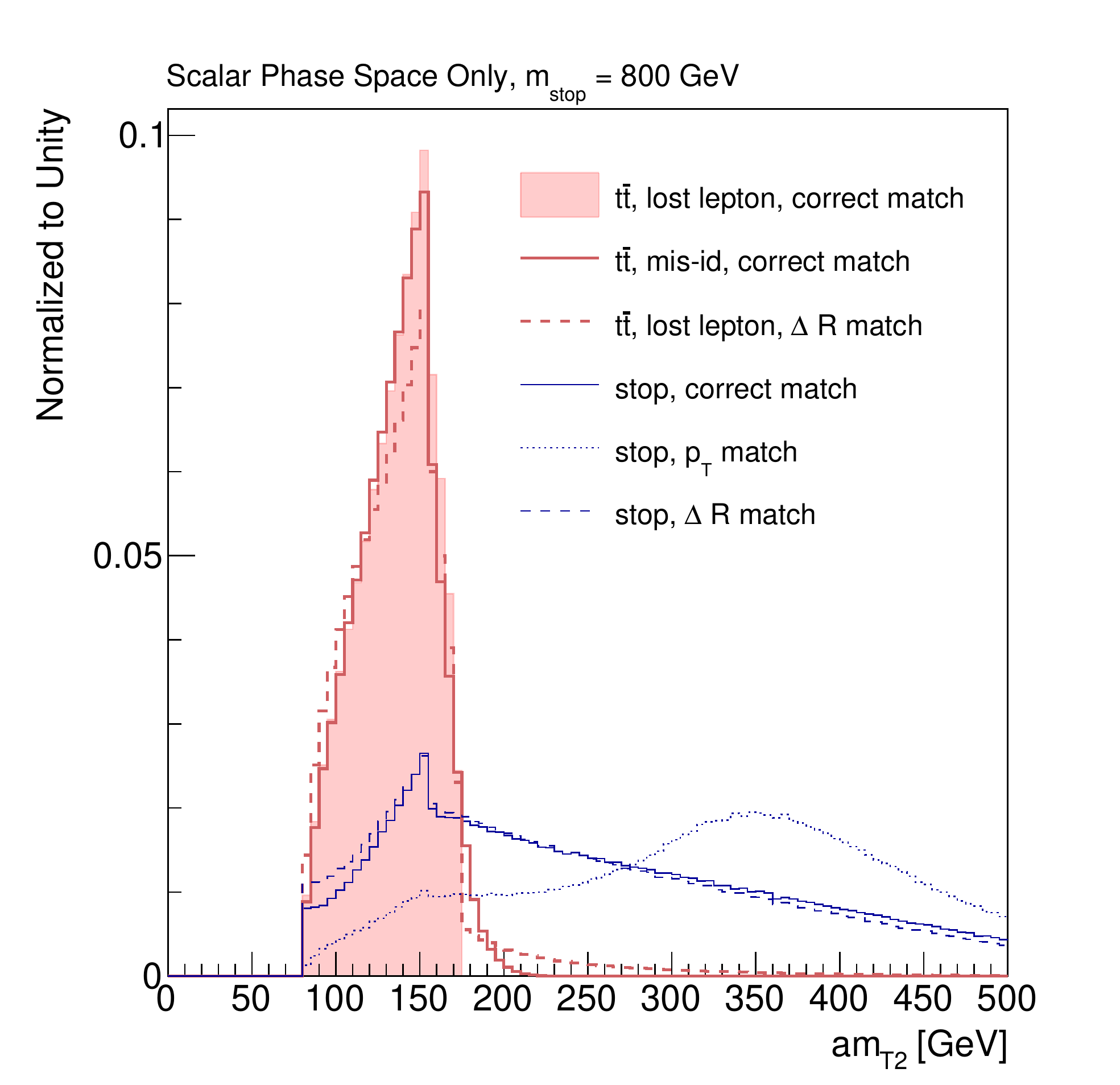}
 \caption{The distribution of $am_\text{T2}$ for dileptonic $t\bar{t}$ and stop events with various configurations as indicated in the legend.  The $p_\text{T}$ spectrum of the top quarks in $t\bar{t}$ events is chosen to be identical to the distribution for stop events.}
 \label{fig:amt2parton}
  \end{center}
\end{figure}

The distribution of $am_\text{T2}$ after a relatively loose event selection is shown in Fig.~\ref{fig:amt2data}.  All three $t\bar{t}$ components are significantly reduced for $am_\text{T2}>m_\text{top}$.  Interestingly, the other backgrounds have relatively uniform distributions over the plotted range; this observation is revisited in Sec.~\ref{singletop} to isolate single top events.
	
\begin{figure}[h!]
\begin{center}
\includegraphics[width=0.45\textwidth]{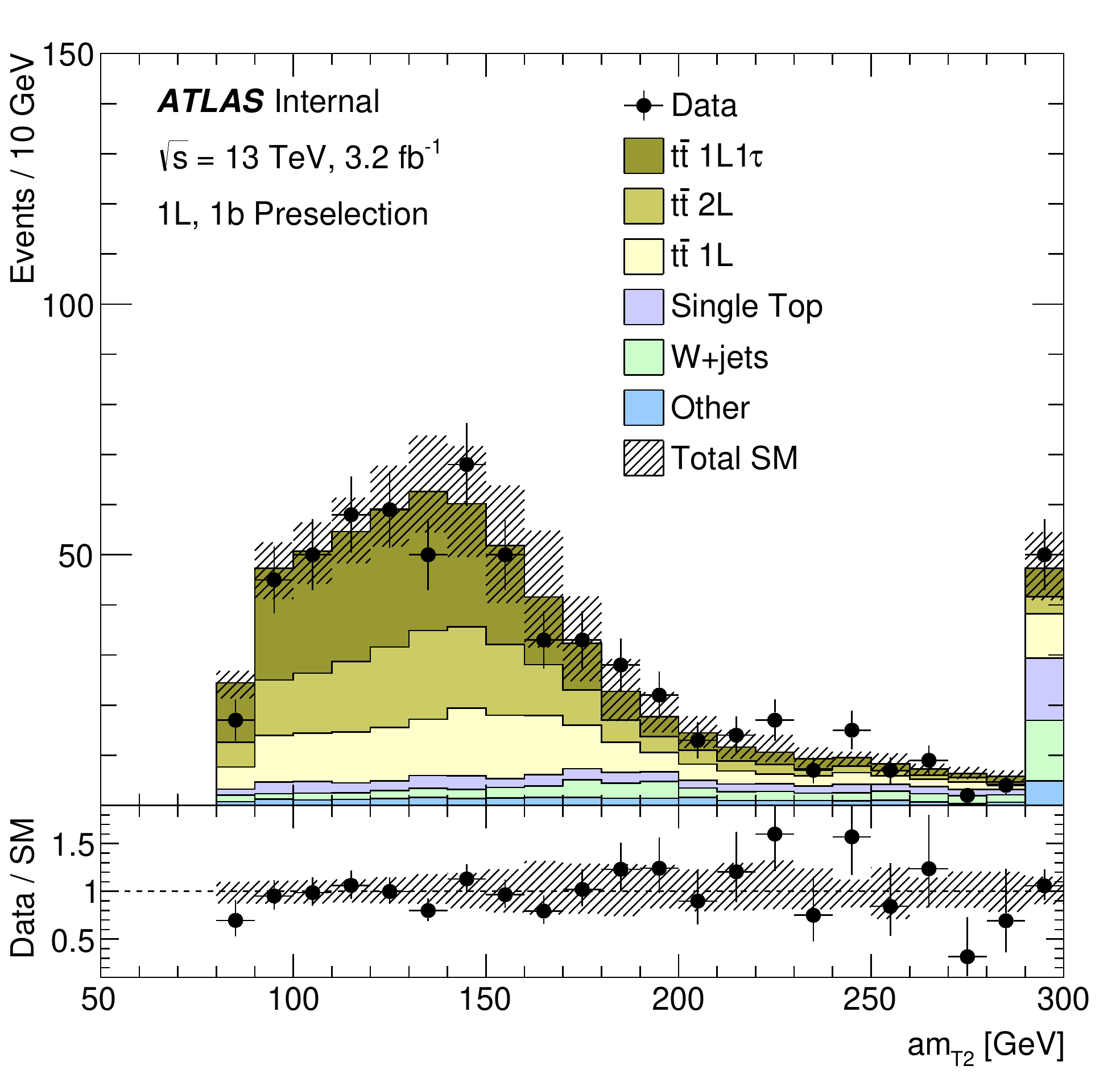}
 \caption{A comparison of data and simulation using a loose selection requiring exactly one signal lepton, four jets with $p_\text{T}>25$ GeV, at least one $b$-tagged jet, and $m_\text{T}>100$ GeV to enrich the dilepton contribution.  The $b$-tag weight is used to select the two $b$-jet candidates and the minimum value over both $b$-lepton pairings is used to resolve the matching ambiguity.  The uncertainty band includes jet energy scale and resolution uncertainties (see Sec.~\ref{chapter:uncertainites}).}
 \label{fig:amt2data}
  \end{center}
\end{figure}	
	
To target the case of dileptonic $t\bar{t}$, a second $m_\text{T2}$ variable called $m_\text{T2}^\tau$ is constructed based on the topology illustrated in Fig.~\ref{fig:susymtdist2tau}.  The visible particle on one branch is an identified hadronic $\tau$ and on the other branch is the reconstructed electron or muon.  For the lower branch, a single neutrino is the lost particle so $m_{C_2}=m_\nu$.  If the full hadronic $\tau$ were stable, than there would be a single neutrino on the upper branch.  However, the $\tau$ decays into a $\nu_\tau$ in addition to hadrons so the mass of the missing object in the upper branch is not strictly $m_\nu$.  However, when $m_{C_1}=0$, $m_\text{T2}$ is still bounded by $m_W$ and is therefore used in the construction of $m_\text{T2}^\tau$.  Figure~\ref{fig:mt2taumass} demonstrates that the $m_W$ bound is still preserved with $m_{C_1}=0$, but the kinematic maximum is not as saturated when the full $\tau$ energy is not measured. 

The only combinatorial challenge for $m_\text{T2}^\tau$ is the selection of the $\tau$ candidate.  Section~\ref{sec:mt2tmva} considers two possibilities: using the third highest $p_\text{T}$ jet or the highest $p_\text{T}$ jet that is not one of the two jets with the highest $b$-tagging weight.  The possibility of using an explicit $\tau$ candidate is investigated in Sec.~\ref{tauid}.

\begin{figure}[h!]
\begin{center}
		\begin{tikzpicture}[line width=1.5 pt, scale=1.3]
			\draw (0,0.5) -- (1,0.5);
			\draw (0,-0.5) -- (1,-0.5);
			\draw (1,-0.5) -- (1.5,-1.5);
			\draw (1,0.5) -- (1.5,1.5);
			\draw (2,0.5) -- (2.5,1.5);
			\draw (2,-0.5) -- (2.5,-1.5);
			\draw (2,-0.5) -- (3,-0.5);
			\draw (2,0.5) -- (2.55,0.5);
			\draw (3.05,0.5) -- (3.5,0.5);
			\draw (2.9,0.69) -- (3.5,1.5);
			\draw[dashed, black!30] (3.1,1.7) ellipse (1 and 0.5);
			\draw[dashed, black!30] (3.3,-0.5) ellipse (0.5 and 0.5);
			\node[draw,circle] at (-0.25,0.5) {$t$};
			\node[draw,circle] at (-0.25,-0.5) {$t$};
			\node[draw,circle] at (2.8,0.5) {$\tau$};
			\node[draw,circle] at (3.85,0.5) {$\mathrm{jet}$};
			\node[draw,circle] at (3.25,-0.5) {$\nu$};
			\node[draw,circle] at (2.65,1.7) {$\nu$};
			\node[draw,circle] at (3.6,1.7) {$\nu$};
			\node[draw,circle] at (2.65,-1.7) {$l$};
			\node[draw,circle] at (1.65,1.7) {$b$};
			\node[draw,circle] at (1.65,-1.7) {$b$};
			\draw[snake=snake] (1,0.5) -- (2,0.5); 
			\draw[snake=snake] (1,-0.5) -- (2,-0.5); 
		\end{tikzpicture}
 \caption{A dileptonic $t\bar{t}$ event with one hadronically decaying $\tau$.  Lost particles are circled with a dashed line.  For the $m_\text{T2}^\tau$ variable, the visible particle on the top (bottom) branch is the hadronic $\tau$ (reconstructed lepton).  The missing particle in the top (bottom) branch is the sum of $\tau$ neutrinos ($e$ or $\mu$ neutrino).}
 \label{fig:susymtdist2tau}
 \end{center}
\end{figure}	

\vspace{-10mm}

\begin{figure}[h!]
\begin{center}
\includegraphics[width=0.42\textwidth]{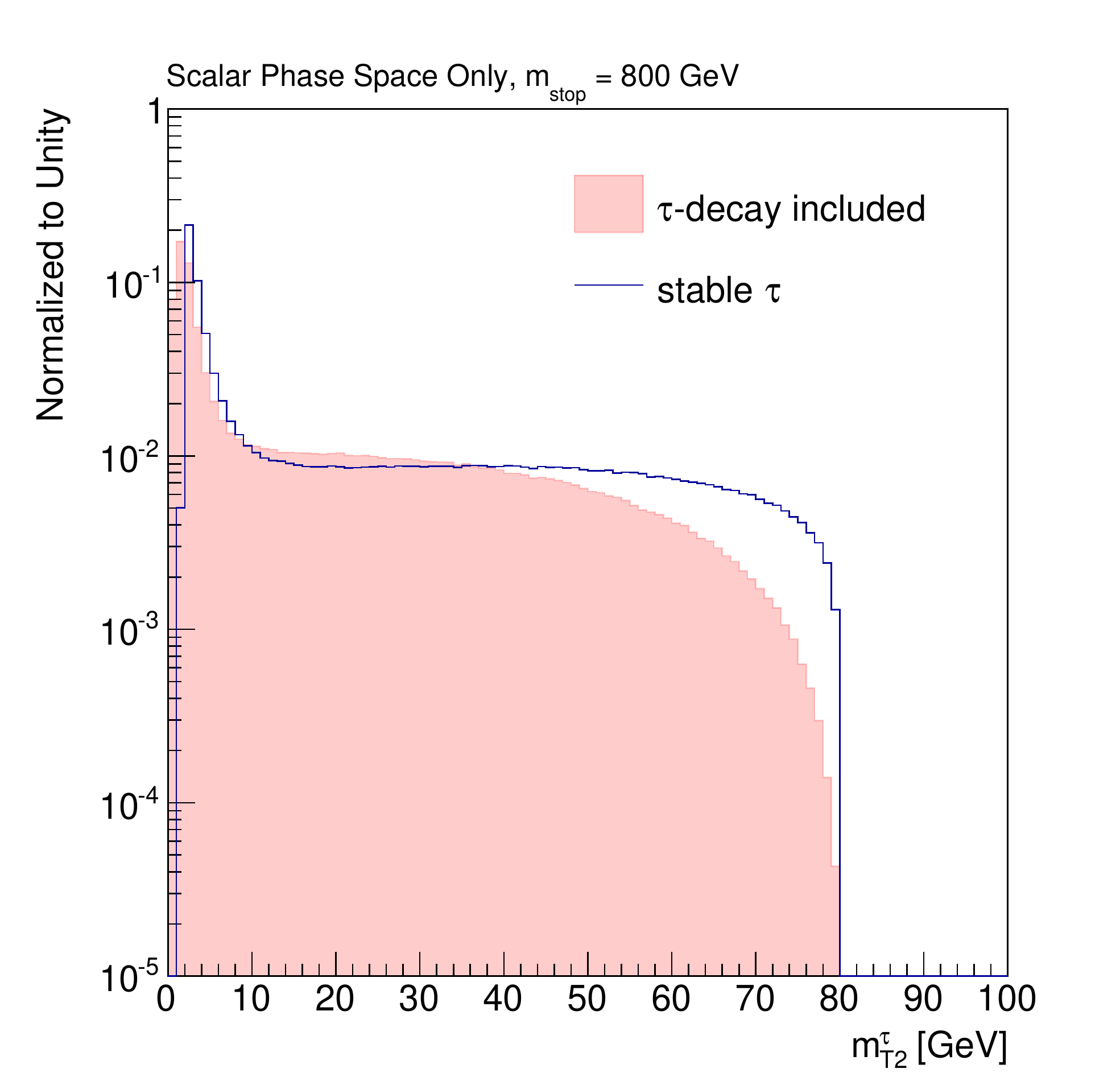}
 \caption{The distribution of $m_\text{T2}^\tau$ in dilepton $t\bar{t}$ events with top quark $p_\text{T}$ spectra that match that of a $800$ GeV stop and $m_\text{LSP}=0$.  For the filled histogram, the $\tau$ is decayed using a scalar three-body phase space to mimic $\tau\rightarrow \nu_\tau\pi^0\pi^\pm$.  The `pions' are added to form the visible $\tau$ and the $\nu_\tau$ is added to the $\vec{p}_\text{T}^\text{miss}$.}
 \label{fig:mt2taumass}
  \end{center}
\end{figure}	

\clearpage

				\paragraph{Comparisons between transverse mass variables }\mbox{}\\
				\label{sec:mt2tmva}
			
			Figure~\ref{fig:mtvariables} presents an overview of transverse mass variables constructed to identify and suppress dileptonic $t\bar{t}$ events.  The dashed lines indicate which subsystem of the $t\bar{t}$ decay is targeted by the variables with the same color code and the particles composing the $V_i$ are circled.  Section~\ref{mt2forstop} introduced the $am_\text{T2}$ and $m_\text{T2}^\tau$ variables.  The variable $m_{bl}$ is the $b$-jet / lepton invariant mass and is identical to $am_\text{T2}$ in the unbalanced case.  One new $m_\text{T2}$ variable, simply denoted $m_\text{T2}$ in Fig.~\ref{fig:mtvariables}, uses two $b$-tagged jets as the visible particles, adds the reconstructed lepton to the $\vec{p}_\text{T}^\text{miss}$ and then $m_{C_1}=m_{C_2}=m_W$.  The {\it contransverse mass}~\cite{Tovey:2008ui}, denoted $m_\text{CT}$, is similar in spirit to $m_\text{T2}$ and is defined by
			
			\begin{align}
			\label{eq:mct}
			m_\text{CT}^2=m_{V_1}^2+m_{V_2}^2+2\left(\sqrt{m_{V_1}^2+(p_\text{T}^{V_1})^2}\sqrt{m_{V_2}^2+(p_\text{T}^{V_2})^2}+\vec{p}_\text{T}^{V_1}\cdot\vec{p}_\text{T}^{V_2}\right).
			\end{align}
			
			\noindent Equation~\ref{eq:mct} is similar to the equation for $m_\text{T}^2$, but with the missing particle replaced with the second visible particle and with the sign of the last term flipped (compare with Eq.~\ref{eq:generalizedmt}).  The contransverse mass is constructed so that its kinematic maximum is invariant under equal and opposite boosts of the particles $V_1$ and $V_2$ in analogy to the invariance of the transverse mass under coherent boosts of $V_1$ and $V_2$ in the same direction\footnote{As with the transverse mass, the contransverse mass is not invariant under these {\it contra-linear} boosts event-by-event, but the endpoint of the $m_\text{CT}$ distribution is invariant.}.  The endpoint of the $m_\text{CT}$ distribution is a known combination of the parent, visible, and invisible particle masses, which makes it useful for discriminating signal events from background events.  One advantage of $m_\text{CT}$ over $m_\text{T2}$ is that Eq.~\ref{eq:mct} is simple, without requiring any optimization.  In the context of the stop search, $m_\text{CT}$ is constructed with two $b$-jet candidates as the visible particles.
			
\begin{figure}[h!]
\begin{center}

	\begin{tikzpicture}[scale=1.5] 
	\draw (1.8,0.5) -- (3.2,0.5);
	\draw (-0.5,0.5) -- (1.3,0.5);
	\draw (-0.5,-0.5) -- (1.3,-0.5);
	\draw (1.8,-0.5) -- (3.2,-0.5);
	\draw (0,0.5) -- (1,1);
	\draw (2,0.5) -- (3,1);
	\draw (0,-0.5) -- (1,-1);
	\draw (2,-0.5) -- (3,-1);
	\draw [green!70!black] (2.1,1) ellipse (1.5cm and 0.2cm);
	\draw [green!70!black] (2.1,-1) ellipse (1.5cm and 0.2cm);
	\draw [blue] (1.2,-1) ellipse (0.3cm and 0.3cm);
	\draw [blue] (1.2,1) ellipse (0.3cm and 0.3cm);
	\draw [red] (3.2,1) ellipse (0.3cm and 0.3cm);
	\draw [red] (3.2,-1) ellipse (0.3cm and 0.3cm);
	\draw[dotted,red,line width=2pt] (1.4,-1.5) -- (1.4,1.5);
	\draw[dotted,red,line width=2pt] (3.9,-1.5) -- (3.9,1.5);
	\draw[dotted,red,line width=2pt] (1.4,-1.5) -- (3.9,-1.5);
	\draw[dotted,red,line width=2pt] (1.4,1.5) -- (3.9,1.5);
	\draw[dotted,blue,line width=2pt] (-1.1,-1.5) -- (-1.1,1.5);
	\draw[dotted,blue,line width=2pt] (1.9,-1.5) -- (1.9,1.5);
	\draw[dotted,blue,line width=2pt] (-1.1,-1.5) -- (1.9,-1.5);
	\draw[dotted,blue,line width=2pt] (-1.1,1.5) -- (1.9,1.5);
	\draw[dotted,green!70!black,line width=2pt] (-1.17,-1.6) -- (-1.17,1.6);
	\draw[dotted,green!70!black,line width=2pt] (3.97,-1.6) -- (3.97,1.6);
	\draw[dotted,green!70!black,line width=2pt] (-1.17,-1.6) -- (3.97,-1.6);
	\draw[dotted,green!70!black,line width=2pt] (-1.17,1.6) -- (3.97,1.6);
		\node at (-0.8,0.5) {$t$};
	\node at (-0.8,-0.5) {$\bar{t}$};
	\node at (1.6,0.5) {$W^+$};
	\node at (1.6,-0.5) {$W^-$};
	\node at (3.5,0.5) {$\nu$};
	\node at (3.5,-0.5) {$\bar{\nu}$};
	\node at (1.2,1) {$b$};
	\node at (1.2,-1) {$\bar{b}$};
	\node at (3.2,1) {$l^+$};
	\node at (3.2,-1) {$l^-$};
	\node at (-0.2,1.1) {\normalsize\color{green!70!black} $am_\text{T2}, m_{bl}$};
	\node at (.5,0) {\normalsize \color{blue} $m_\text{T2}, m_\text{CT}$};
	\node at (2.8,0) {\normalsize \color{red} $m_\text{T2}^{\tau}$};
	\end{tikzpicture}
				
 \caption{A diagram of dileptonic the $t\bar{t}$ decay chain and kinematic variables targeting various components of the chain.  The dotted lines highlight which aspects of the top quark decay are involved with the construction of the variables with the same color.  Particle inputs to each variable are circled with the same colors.}
 \label{fig:mtvariables}
 \end{center}
\end{figure}				
		
		The above variables and their variations are quantitatively compared in terms of their ability to separate $t\bar{t}$ events from stop events.  Each variable has several variations, including the $b$-jet and $\tau$-jet identification algorithms for $(a)m_\text{T2}^{(\tau)}$, the $b$-jet / lepton matching scheme for $am_\text{T2}$ and $m_{bl}$, and the projection perpendicular to $\vec{p}_\text{T}^\text{up}$ (or not).  The $m_\text{CT,$\perp$}$~\cite{Matchev:2009ad} is defined analogously to the $m_\text{T2,$\perp$}$ variables.   Another method to reduce the $p_\text{T}^\text{up}$ dependence of $m_\text{CT}$ is to apply a {\it boost-correction}~\cite{Polesello:2009rn} based on the kinematic properties of the visible particles and the $\vec{p}_\text{T}^\text{miss}$. A metric to quantify the separation power is given by the overlap integral~\cite{babar,Hocker:2007ht}:
		
	\begin{align}
	\label{eq:separation}
	\langle S^2\rangle=\frac{1}{2}\int \frac{(f_S(x)-f_B(x))^2}{f_S(x)+f_B(x)} dx,
	\end{align}
	
	\noindent where $f_S$ and $f_B$ are the probability distribution functions for a random variable $X$ with signal and background processes, respectively\footnote{Note the similarity of the overlap integral with the $\chi^2$-divergence.  In fact, the overlap integral is an $f$-divergence with $f(u)=(u-1)^2/(u+1)$ (for the $\chi^2$-divergence, $f(u)=(u-1)^2$).  Despite this, the overlap integral has not been applied to e.g. signal processing outside of high energy physics.}.  Figure~\ref{fig:susymtdist2example} shows example distributions of $m_\text{T}$, $E_\text{T}^\text{miss}$, $am_\text{T2}$, and $am_\text{T2,$\perp$}$ for the $t\bar{t}$ background and a stop signal with $(m_\text{stop},m_\text{LSP})=(500,1)$.  As expected, the $m_\text{T}$ peak in $t\bar{t}$ events is around $m_W$ with a long tail due to resolution effects and the dilepton contribution.  The $am_\text{T2}$ distribution has an endpoint near $m_\text{top}$ for $t\bar{t}$ events while the $am_\text{T,$\perp$}$ distribution is concentrated at low values for both the background and signal.  As a result, the separation $\langle S^2\rangle$ is significantly worse for $am_\text{T2,$\perp$}$ compared with $am_\text{T2}$.
				
\begin{figure}[h!]
\begin{center}
\includegraphics[width=0.9\textwidth]{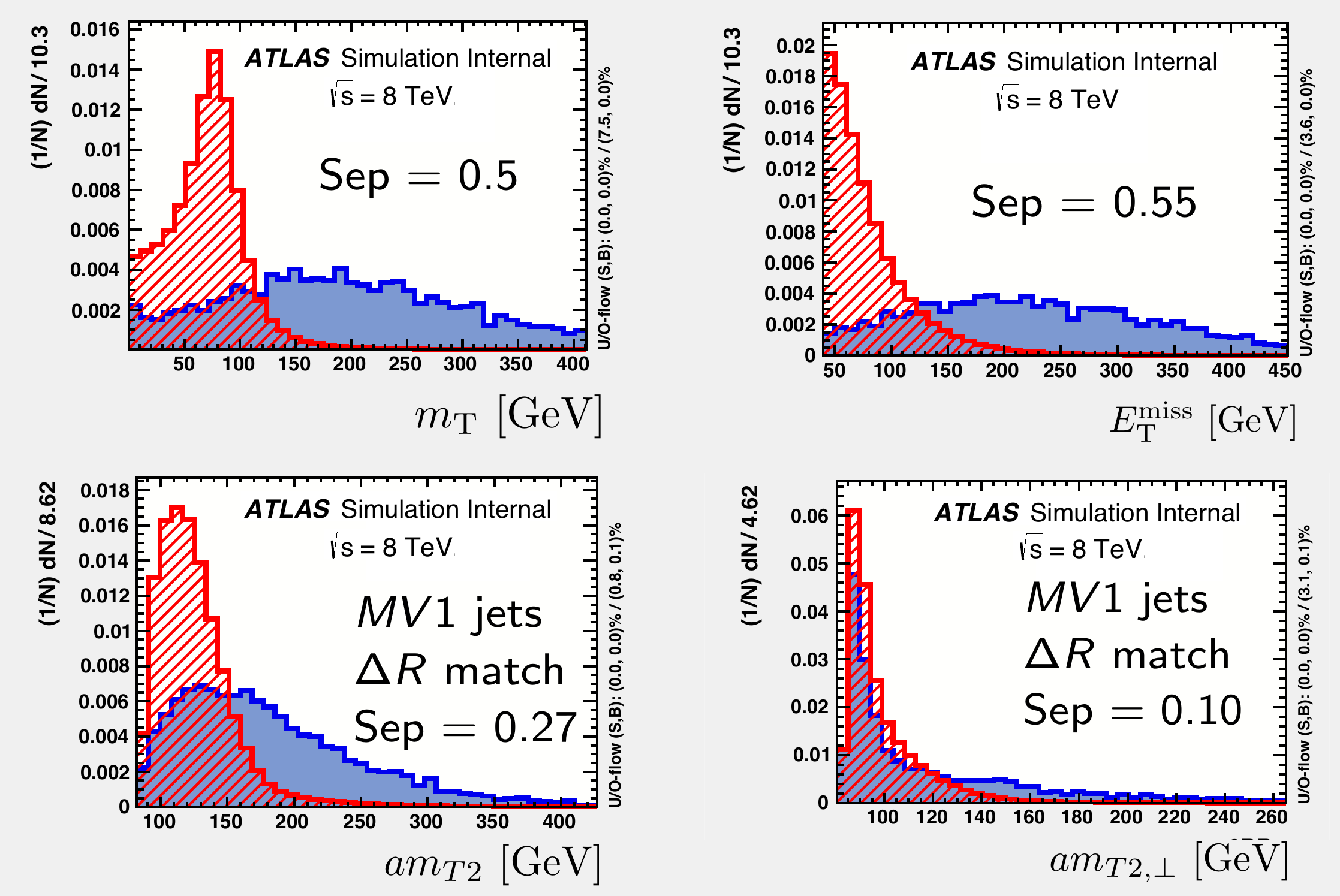}
\caption{The distributions of $m_\text{T}$ (top left), $E_\text{T}^\text{miss}$ (top right), $am_\text{T2}$ (bottom left), and $am_\text{T2,$\perp$}$ (bottom right) for $t\bar{t}$ and stop events with $(m_\text{stop},m_\text{LSP})=(500,1)$.  The separation ({\it sep}) is defined in Eq.~\ref{eq:separation}.  The $b$-tagging weight (MV1) is used to select the $b$-jets for $am_\text{T2}$ and a $\Delta R$ scheme matches the $b$-jet with the lepton.}
\label{fig:susymtdist2example}
\end{center}
\end{figure}	

Table~\ref{tab:mc_tablecomparison} summarizes the separation power for all the variables mentioned above.  In addition to the separation, the table also provides the correlation with the $E_\text{T}^\text{miss}$ and $m_\text{T}$.  These are two known powerful variables, so a low correlation is an important metric for deciding on the usefulness of a new technique.  The unprojected $am_\text{T2}$ and $m_\text{T2}^\tau$ variables have the highest separation power amongst the possible variables.  Additionally, $am_\text{T2}$ has only a modest correlation with $m_\text{T}$ and $E_\text{T}^\text{miss}$.  The variants have similar performance; for the sake of reducing the parameter space for later optimization, the ($B,\text{min}$) for $am_\text{T2}$ and the $B$ setting for $m_\text{T2}^\tau$ are chosen as default\footnote{The `$B$' method also has a significantly higher accuracy: the jet with the highest $b$-tag weight is nearly twice as likely as the highest $p_\text{T}$ jet to be matched to a particle-level $b$-jet.}.  Further comparisons between the transverse mass and other variables are described in Sec.~\ref{chapter:susy:signalregions}.

\begin{table}[h!]
\centering
\begin{tabular}{|c|c|c|c|c|}
\hline
Variable & Variant & Separation & Corr. $E_\text{T}^\text{miss}$ & Corr. $m_\text{T}$ \\
\hline 
\hline

$E_\text{T}^\text{miss}$ & -- & 0.55 & 100\% & 49\% \\
$m_\text{T}$ & -- & 0.5 & 49\% & 100\% \\
$m_\text{CT}$ & P & 0.19 &  53\%& 1\% \\
 & P, $\perp$ & 0.13 & 40\% & 1\% \\
  & P, BC & 0.23 &  60\%& 3\% \\
 & B &  0.06 & 29\% &6\%  \\
 & B,$\perp$ & 0.06 & 23\% &4\%  \\
  & B, BC & 0.13 & 39\% &9\%  \\  
$m_\text{T2}$ &  P & 0.24  &63\%  &2\%  \\
 &  B &0.16  & 47\% &8\%  \\
 &  P, $\perp$ & 0.13 & 40\% &1\%  \\
&  B, $\perp$ & 0.06 & 23\% &4\%  \\
$am_\text{T2}$ &  P, $\min$& 0.33 &65\%  & 24\%  \\
 &  P, $\Delta R$& 0.31 & 61\%  & 24\%  \\
&  B, $\min$ & 0.28 & 53\% &30\%  \\
&  B, $\Delta R$ & 0.27 &  51\%  &29\% \\
&   P, $\perp$, $\Delta R$ & 0.15  & 40\%  & 14\%  \\
&   B, $\perp$, $\Delta R$ & 0.10 & 25\% &18\%  \\
&   B, $\perp$, $\min$ & 0.08 & 26\% & 14\% \\
$m_\text{T2}^\tau$ &P  & 0.36 & 54\% &66\%    \\
 &  B&  0.40& 63\% & 70\% \\
  & P, $\perp$ &0.15  &19\% &34\%\\
   &  B, $\perp$& 0.14 & 24\% &31\%  \\
$m_\text{bl}$ &  B, $\min$ &0.02  & 0\% &15\%  \\
&  P, $\min$& 0.04 & 2\% &4\% \\
 &  B, $\Delta R$& 0.02 &3\% & 11\% \\
 &  P, $\Delta R$& 0.03 &4\%&7\% \\
\hline
\end{tabular}
\caption{The separation power and correlation with $E_\text{T}^\text{miss}$ and $m_\text{T}$ for a variety of variables described in the text.  The background is $t\bar{t}$ and the signal is a stop model with $(m_\text{stop},m_\text{LSP})=(500,1)$.  The variant $P$ means that $p_\text{T}$ is used to pick the $b$-jets while $B$ means that the $b$-tagging weight is used.  The symbol $\perp$ denotes the perpendicular variant of the variable in the first column.  A boost correction is indicated by the letters BC.  The symbols $\Delta R$ and min represent the scheme for addressing the matching ambiguity between the lepton and the $b$-jet by using the closest pair or considering the minimum of both possible pairings.}

\label{tab:mc_tablecomparison}
\end{table}

			\clearpage
			
				\paragraph{Additional Considerations}\mbox{}\\
				\label{sec:mt2tmva}

		This section briefly describes a few aspects of $m_\text{T2}$ that are slightly out of, or beyond the scope of the rest of the chapter.  In particular,
		
		\begin{itemize}
		\item The variable $am_\text{T2}$ is also useful for stop decays other than $\tilde{t}\rightarrow t\tilde{\chi}^0$.  For example, when $m_\text{stop}<m_\text{top}$, the stop can undergo a three-body $\tilde{t}\rightarrow bW\tilde{\chi}^0$ decay that has systematically {\it lower} $am_\text{T2}$ values than the dileptonic $t\bar{t}$ background for which $am_\text{T2}\sim m_\text{top}$.  This is illustrated in the left plot of Fig.~\ref{fig:susymtdist2threebody}.  The unbalanced case is set by $\max\{m_{bl}+m_\nu,m_b+m_W\}$, which is $m_{bl}\lesssim \sqrt{m_\text{top}^2-m_W^2}$ for $t\bar{t}$.  In contrast, $m_{bl}<m_W$ for the signal and so $am_\text{T2}$ tends to be close to the $m_W$ lower bound.  Therefore, an {\it upper threshold} on $am_\text{T2}$ is a powerful discriminant for targeting three-body stop decays.
		
		Another possibility is the flavor-changing decay $\tilde{t}\rightarrow b\tilde{\chi}^\pm$ (bC).  Without a resonant top quark, the $bl$ invariant mass for the bC decay is significantly higher than the $t\bar{t}$ case resulting in a larger separation between signal and background relative to the flavor-neutral decay.  This is illustrated in the right plot of Fig.~\ref{fig:susymtdist2threebody}.  The $am_\text{T2}$ variable has successfully improved the sensitivity to both three-body and bC stop decays in the {\tt 3body} and {\tt bCx} signal regions of Ref.~\cite{Aad:2014kra}.
		
\begin{figure}[h!]
\begin{center}
\includegraphics[width=0.45\textwidth]{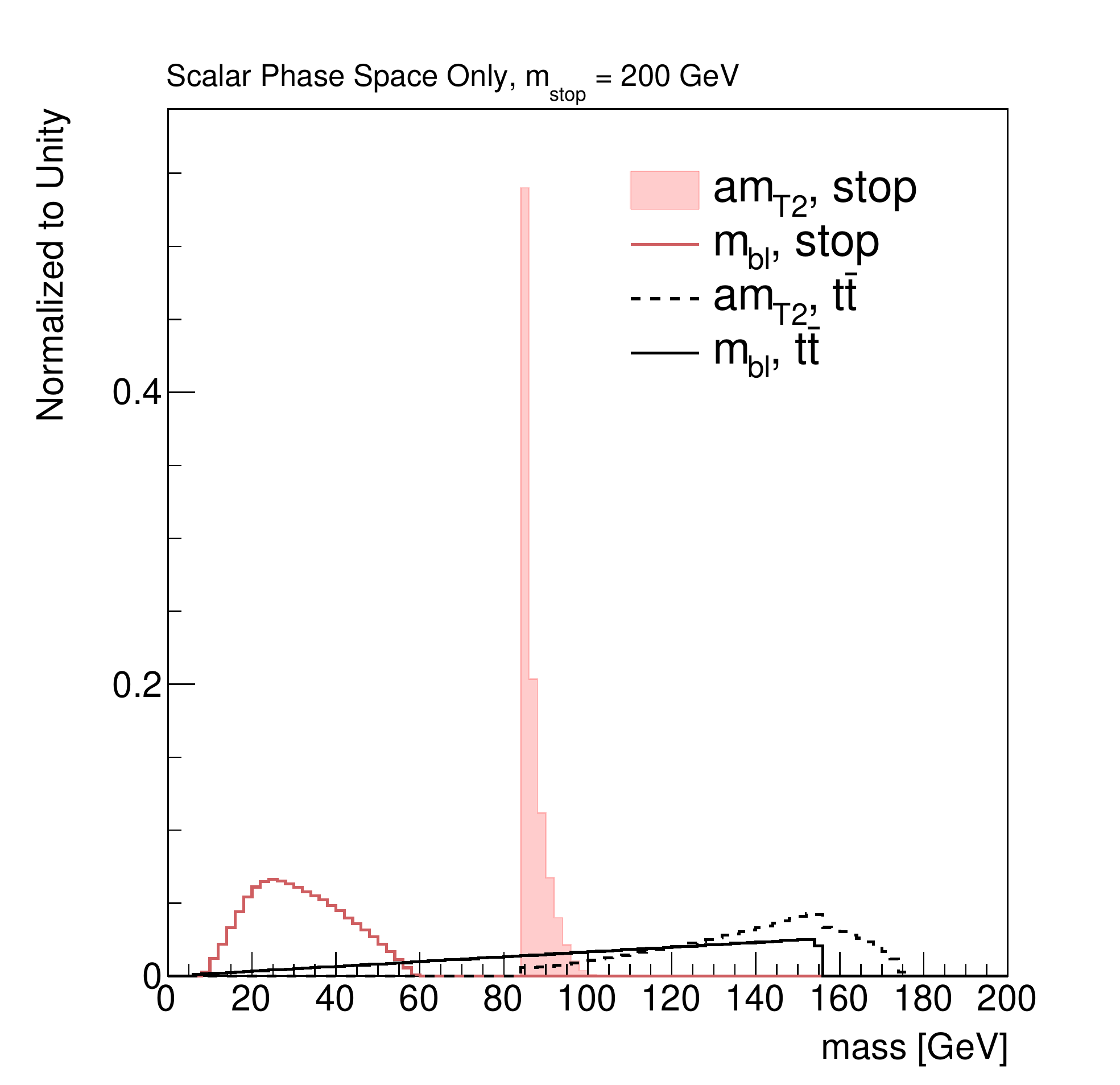}\includegraphics[width=0.45\textwidth]{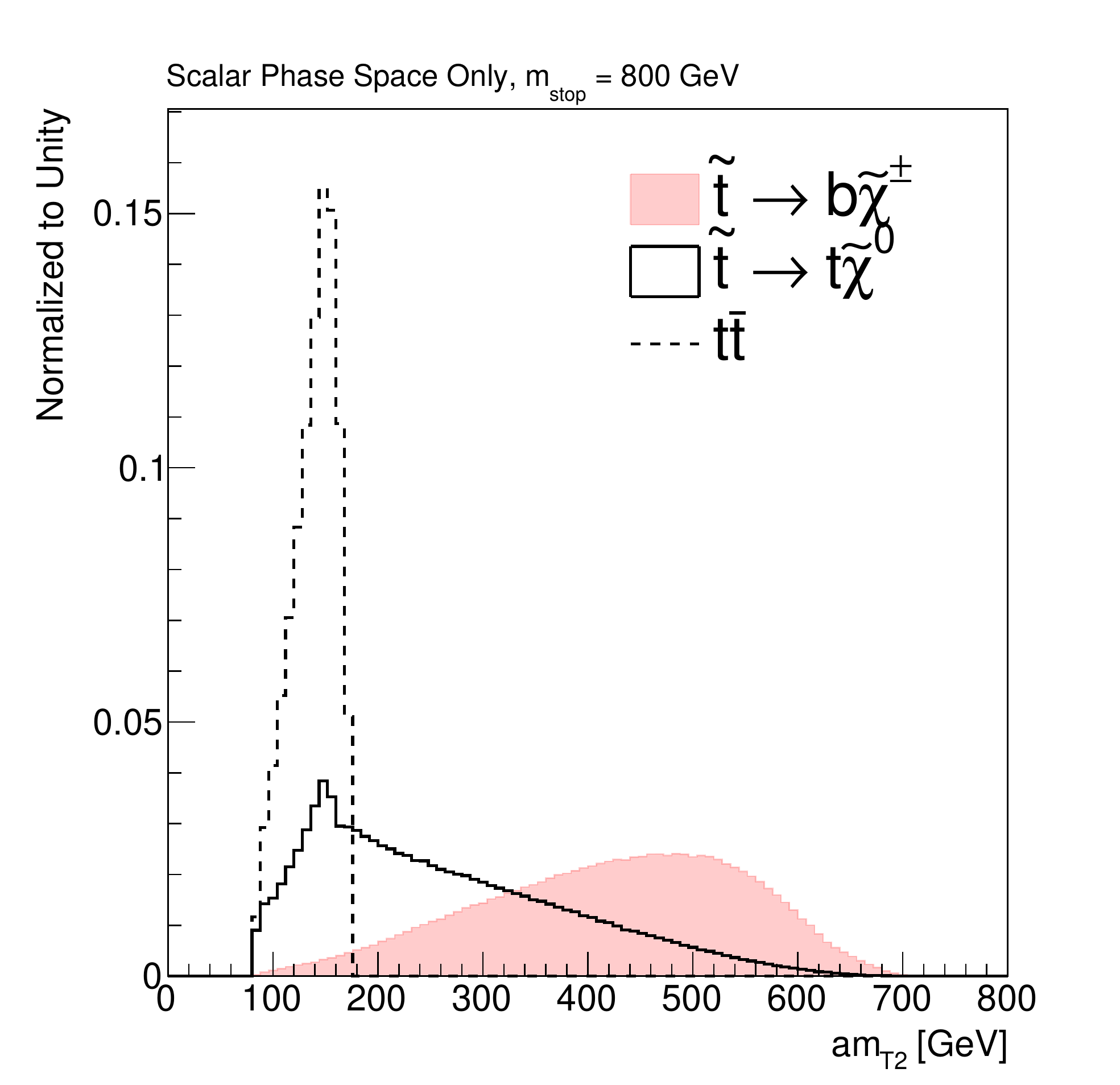}
\caption{The distribution of $am_\text{T2}$ for dileptonic $t\bar{t}$ where one lepton is lost and a three-body stop decay (left) with $(m_\text{stop},m_\text{LSP})=(200,100)$ GeV and for a flavor-changing $b+\tilde{\chi}^\pm$ decay and flavor-neutral two-body decay $t+\tilde{\chi}^0$ (right) with $(m_\text{stop},m_\text{chargino},m_\text{LSP})=(800,300,150)$ GeV.}
\label{fig:susymtdist2threebody}
\end{center}
\end{figure}		
		
		\item One can always improve or create $m_\text{T2}$ variables by incorporating more information.  Section~\ref{sec:significancevariables} will describe how to include resolution information and Sec.~\ref{tauid} will combine $m_\text{T2}$ with an explicit hadronically decaying $\tau$ reconstruction.  In a similar spirit, events with a second low quality lepton that does not pass the baseline criteria can be coupled with kinematic information via $m_\text{T2}$ to create a high-efficiency veto for dilepton events.  Often, low quality leptons are not well modeled, but large uncertainties are suppressed when the veto efficiency is close to unity.  Figure~\ref{fig:susymtdist2mt2lost} gives a concrete example where an $m_\text{T2}$ variable is formed from the selected lepton and the next highest $p_\text{T}$ muon.  All events in Fig.~\ref{fig:susymtdist2mt2lost} pass the second lepton veto, so the muon used in the $m_\text{T2}$ calculation does not pass the requirements to be baseline.  However, if the muon were truly from a $W$ decay, $m_\text{T2}^\text{lost}\leq m_W$.  Even in cases where there is not a lost muon, the scale of $m_\text{T2}^\text{lost}$ is significantly less than that of the signal.  Unfortunately, there are a significant fraction of signal events with $m_\text{T2}^\text{lost}\approx 0$, but it still may be useful to veto events in the first bin of Fig.~\ref{fig:susymtdist2mt2lost}.  It may also be possible to improve the performance by adding slightly more quality criteria to the muon definition (but still below the baseline requirements).
		
\begin{figure}[h!]
\begin{center}
\includegraphics[width=0.45\textwidth]{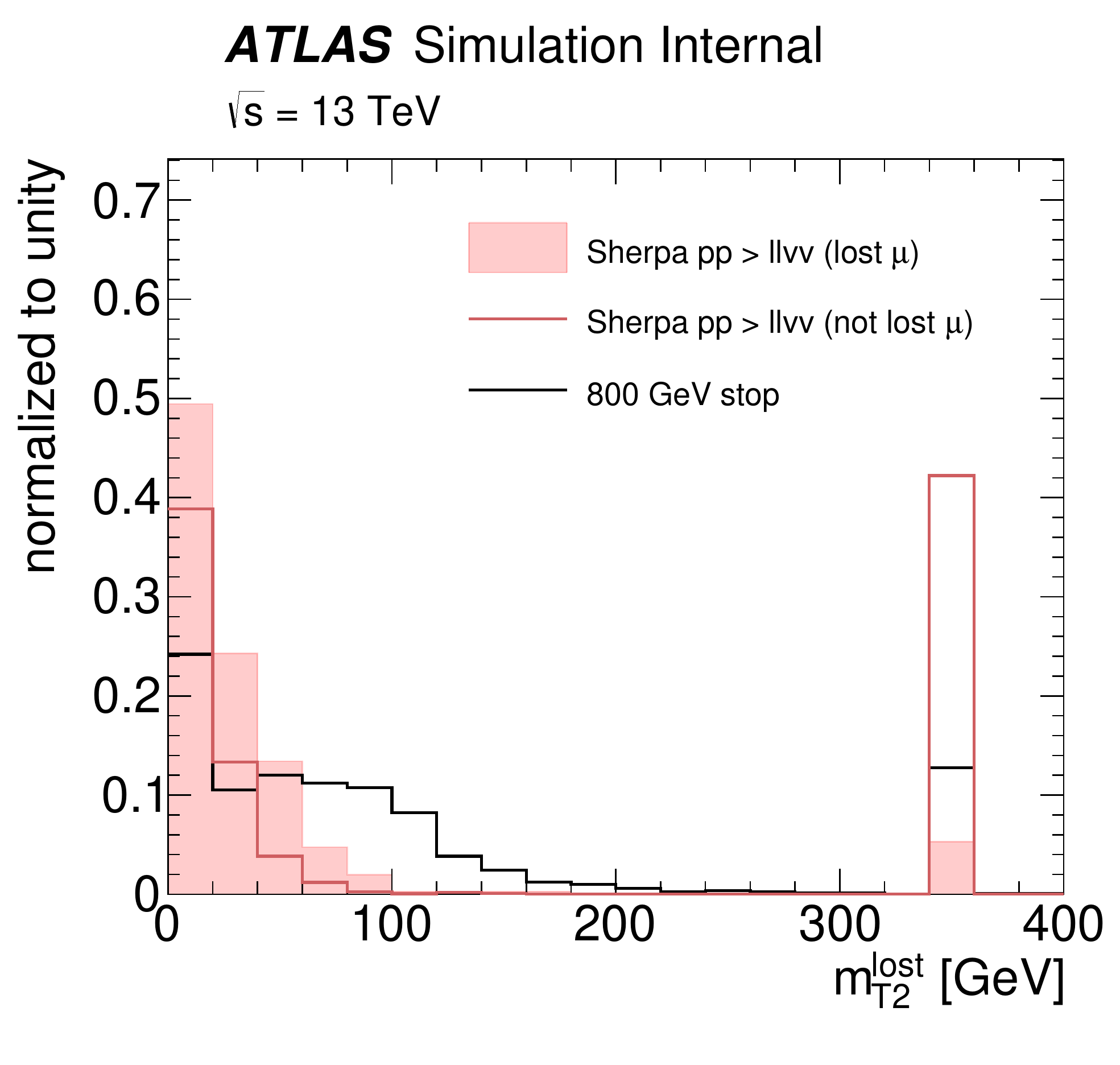}
\caption{The distribution of $m_\text{T2}$ in diboson ($ZZ/WW\rightarrow l^+l^-\nu\bar{\nu}$) and stop events using the selected lepton as one visible particle and the next highest $p_\text{T}$ muon for the second visible particle. For the stop model, $(m_\text{stop},m_\text{LSP})=(800,1)$ GeV.}
\label{fig:susymtdist2mt2lost}
\end{center}
\end{figure}		
		
		\item While all of the applications so far have been focused on vetoing the background, $m_\text{T2}$ variables can also be used to directly tag the signal.  The decay $\tilde{t}\rightarrow t\tilde{\chi}^0$ has the form of Fig.~\ref{fig:mt2setup} with top quarks for $V_i$ and neutralinos for $C_i$.  Therefore, one could construct an $m_\text{T2}^{\tilde{t}}$ with $m_{C_i}=m_{\tilde{\chi}^0}$.  The endpoint of such a variable would be $m_\text{T2}^{\tilde{t}}\leq m_{\tilde{t}}$ in stop events.  If signal events saturate this bound and background events are relegated to lower values, than this variable could be useful.  The left plot of Fig.~\ref{fig:mt2stop1} gives a concrete example of an $m_\text{T2}^{\tilde{t}}$ variable where one visible particle is the large-radius jet hadronic top quark candidate (see Sec.~\ref{topmassreco}) and the $b$-lepton pair is the other visible particle.  By construction, $m_\text{T2}^{\tilde{t}}\leq m_{\tilde{t}}$ and the average $m_\text{T2}^{\tilde{t}}$ is significantly larger for the signal than for the irreducible $t\bar{t}+Z(\rightarrow\nu\bar{\nu})$ background.  The right plot of Fig.~\ref{fig:mt2stop1} shows that $am_\text{T2}$ has a similar separation, suggesting that the two variables may be related.  The correlation between $am_\text{T2}$ and $m_\text{T2}^{\tilde{t}}$ is shown in Fig.~\ref{fig:mt2stop2} for both the background and the signal.  There is a strong correlation in the signal, but little correlation in the background.  One exception is at low $m_\text{T2}$ values where the unbalanced case can result in both variables giving the same value.  The $m_\text{T2}^{\tilde{t}}$ is promising tool for selecting stop events and suppressing events with a similar event topology; it will be interesting to expand upon these studies in the future.
		
		\end{itemize}				
				
\begin{figure}[h!]
\begin{center}
\includegraphics[width=0.45\textwidth]{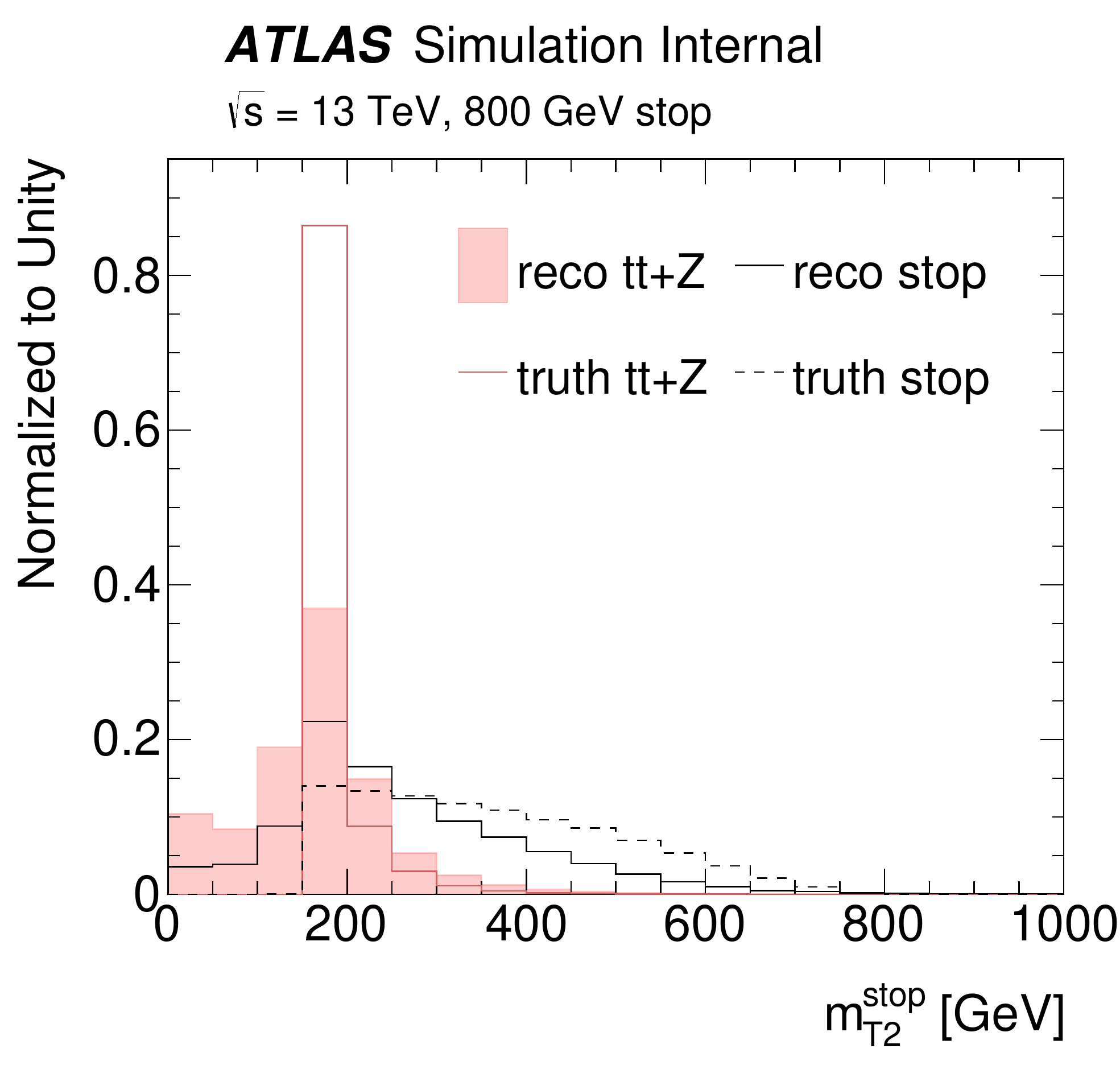}\includegraphics[width=0.45\textwidth]{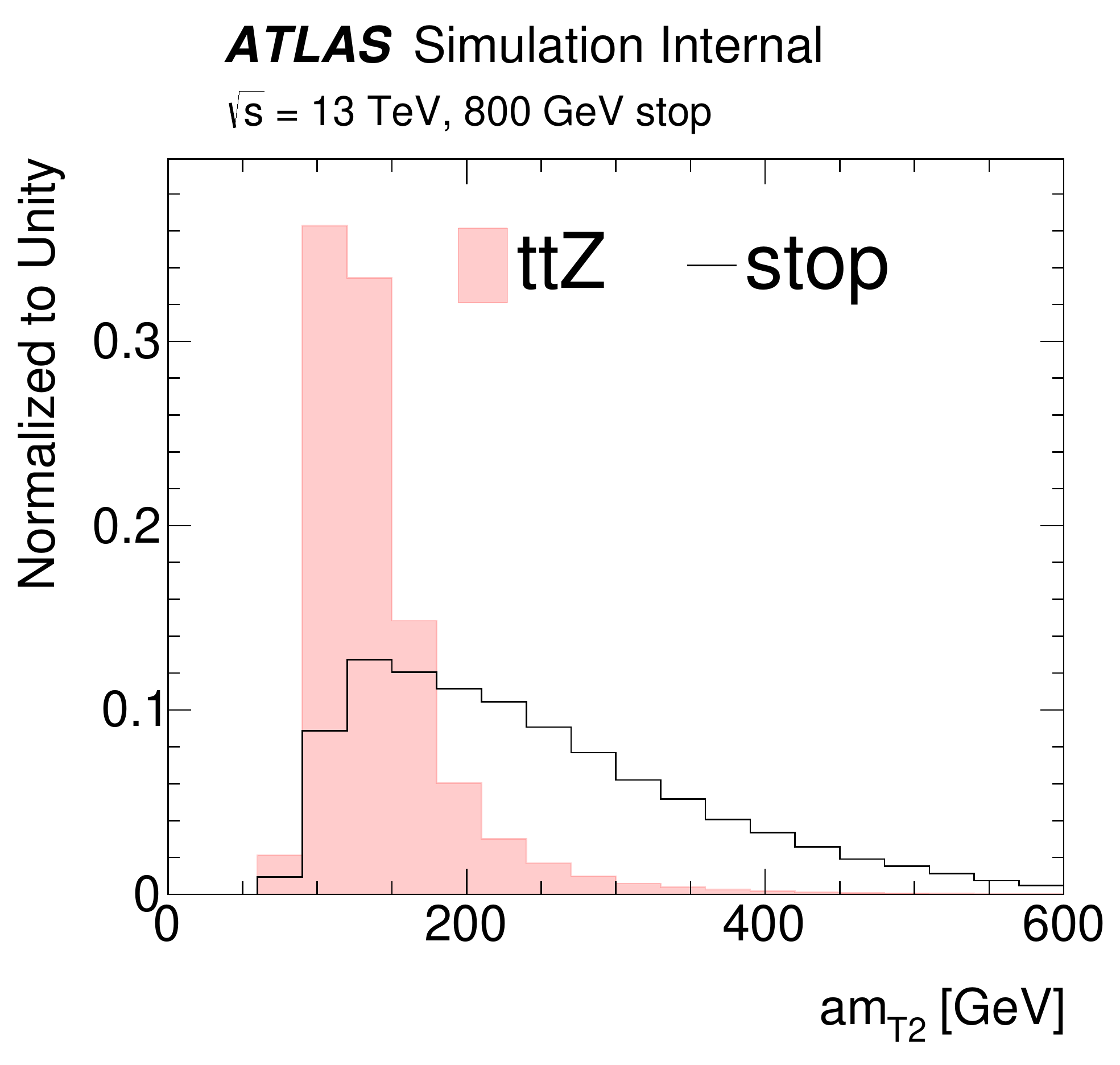}
\caption{The distribution of $m_\text{T2}^{\tilde{t}}$ (left) and $am_\text{T2}$ (right) for $t\bar{t}+Z(\rightarrow\nu\bar{\nu})$ and $\tilde{t}\tilde{t}$ events with $(m_\text{stop},m_\text{LSP})=(800,1)$ GeV.  The left plot also includes the distributions using particle-level top quarks for the visible particles.}
\label{fig:mt2stop1}
\end{center}
\end{figure}

\begin{figure}[h!]
\begin{center}
\includegraphics[width=0.45\textwidth]{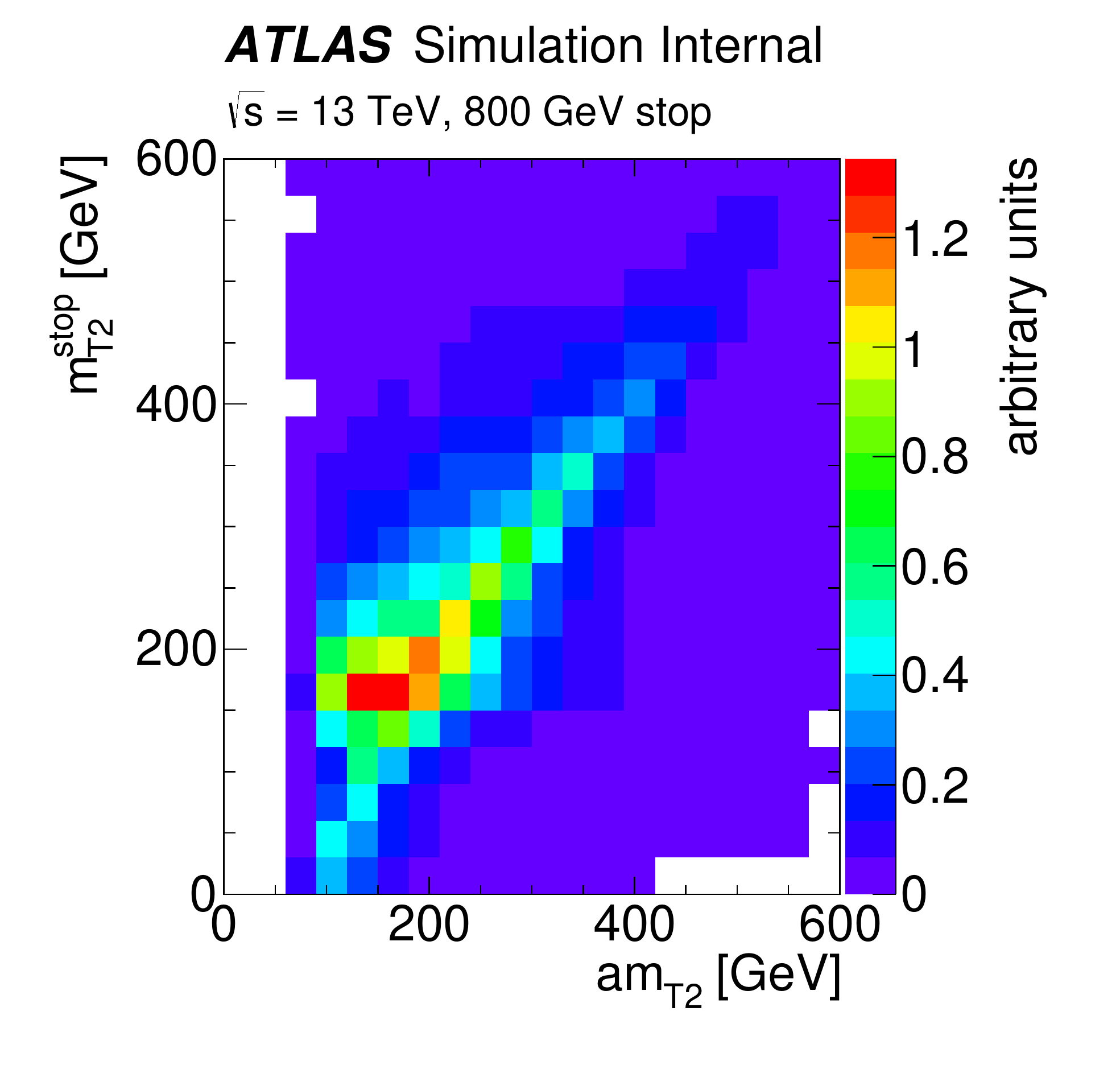}\includegraphics[width=0.45\textwidth]{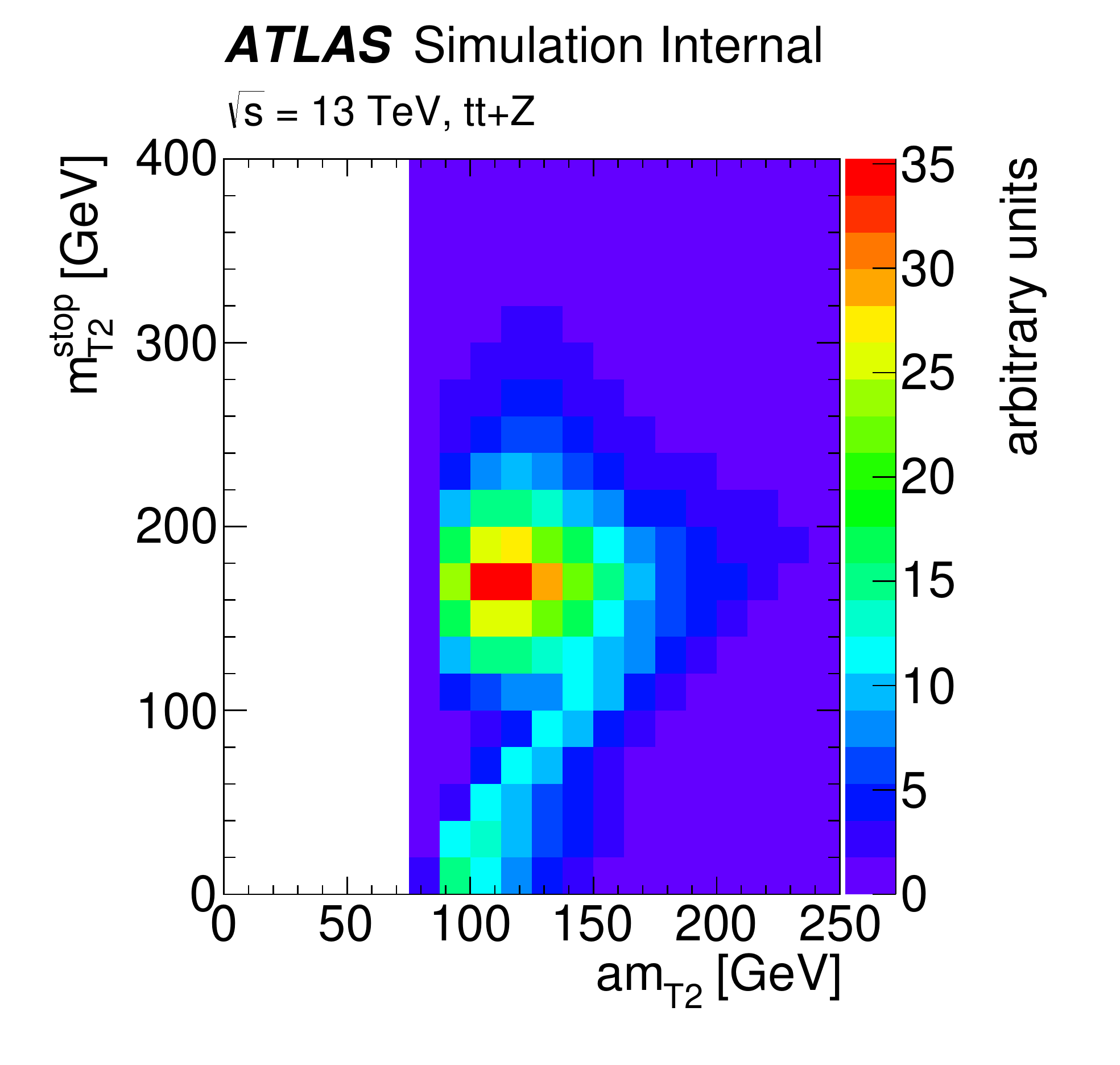}
\caption{The joint distribution of $m_\text{T2}^{\tilde{t}}$ and $am_\text{T2}$ for stop events (left) and $t\bar{t}+Z(\rightarrow\nu\bar{\nu})$ events (right).}
\label{fig:mt2stop2}
\end{center}
\end{figure}

Additional properties and uses of $m_\text{T2}$ are discussed in subsequent chapters.
				
				\clearpage
					
			\subsection{Significance Variables}
				\label{sec:significancevariables}
				
				Event-by-event and object-by-object resolutions can be estimated from simulation and auxiliary measurements.  These resolutions are often the source of background events with apparent signal-like kinematic properties.  For example, Fig.~\ref{fig:susy:fakemet} shows a schematic diagram illustrating a dijet event where one of the two jets has a significant mis-measurement in the direction transverse to the jet axis.  As a result of jet angular resolution, an event with no real missing momentum from neutrinos or other weakly interacting particles can have a large apparent missing momentum.  Jet-by-jet kinematic covariance matrices could be used to identify such events and rule out the mis-measurement as insignificant.  The uncertainty on the jets and other objects can be used for all kinematic variables, in addition to the $E_\text{T}^\text{miss}$.  However, this information is mostly unused in the construction of discriminating variables at the LHC.   This section describes simple procedures for optimally combining kinematic variables with estimates of their resolution to form {\it significance variables}.  
				
\begin{figure}[h!]
\begin{center}
\includegraphics[width=0.45\textwidth]{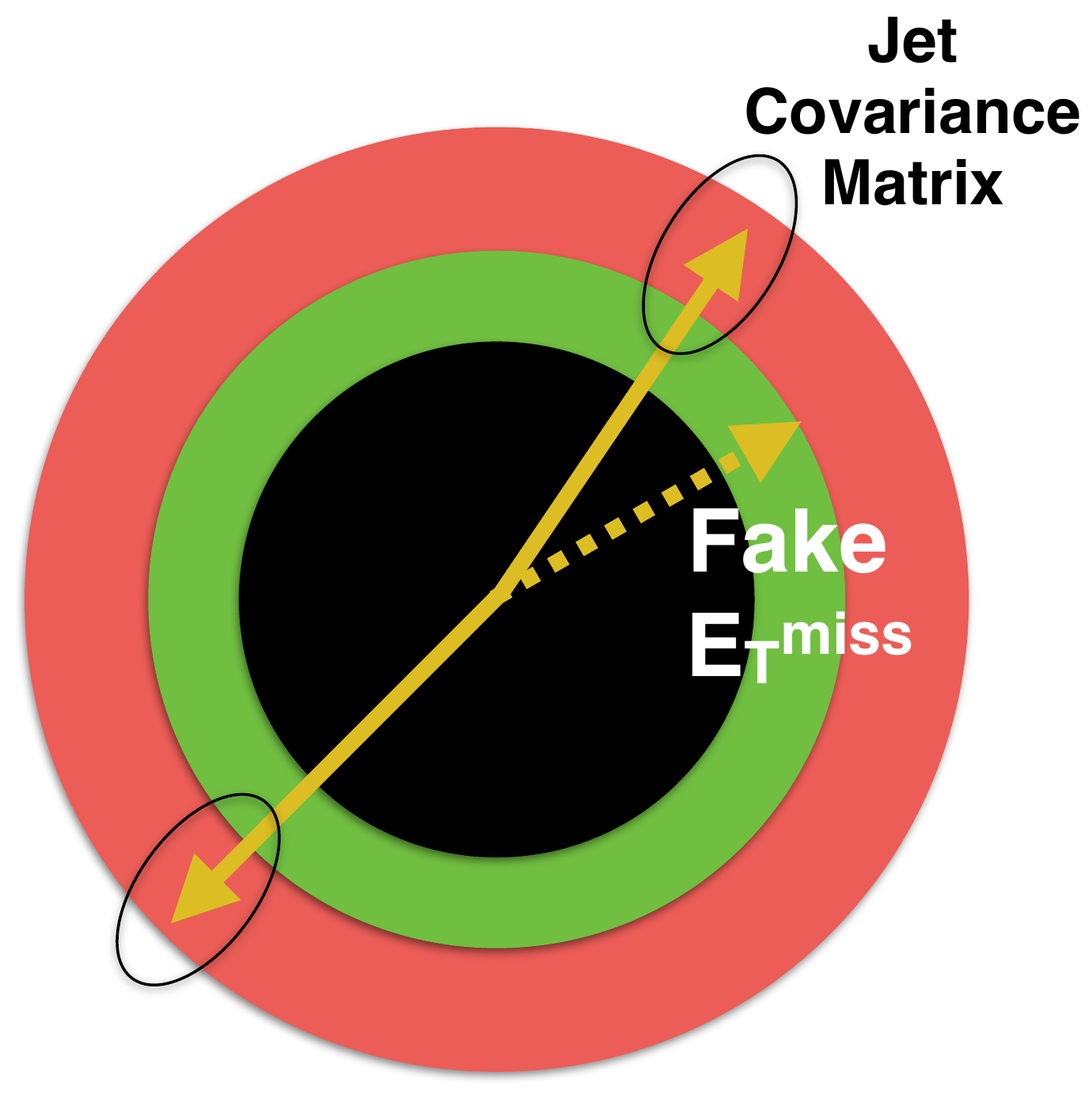}
 \caption{The green and red annuli represent the ATLAS calorimeters and the yellow arrows represent the measured jet directions.  The direction of the right jet is mis-measured leading to an apparent $E_\text{T}^\text{miss}$ represented by the dashed arrow.  Such events could be identified if the jet kinematic covariance matrices show that the jets are statistically consistent with being back-to-back.}
 \label{fig:susy:fakemet}
  \end{center}
\end{figure}		
			
	To concretely illustrate the idea\footnote{The ideas presented in this section are published in Ref.~\cite{Nachman:2013bia} and include input from C. Lester.}, consider a kinematic variable $m$ for a particular process which has a kinematic maximum $M$ in the absence of detector resolution.  For example, $m$ could be transverse momentum or the actual mass of some system of particles.   The usual procedure for using $m$ is to place a threshood $m_\text{threshold}$ and then to count the number of events for which $m>m_\text{threshold}$.  If this number significantly exceeds expectation, then one has evidence for new physics.  However, one can do better than this by including more information such as event-by-event resolutions and the mass scale $M$.  To construct an example, consider three variants of the variable $m$:

\begin{itemize}
	\item $m^{\text{truth}}$:  the value of $m$ for a single realization of a given experiment in the absence of detector resolution.
	\item $m^{\text{measured}}$: the measured value of $m$ for a single realization of a given experiment.
	\item $m^{\text{re-measured}}$: a random variable with probability distribution function given by the posterior distribution for the outcome of an experiment.  This variable only makes sense in the context of conditioning on a measured value from a single experiment $m^{\text{measured}}$.  In special cases, $m^{\text{measured}}$ may be the mean or median of the distribution of $m^{\text{re-measured}}$, but this is not true in general due to asymmetric detector responses and a falling truth-level spectrum. 
\end{itemize}

\noindent One quantity that captures resolution and scale information is the probability $P_M$ that the measured value $m^{\text{measured}}$ for a fixed event would exceed the scale $M$ if re-measured\footnote{A related quantity, is $Q_M=\Pr(m^\text{truth}> M | m^\text{measured})$.  One can show that $P_M$ and $Q_M$ do not induce the same ordering on events and therefore one may be better than the other for a particular application.}.  Symbolically, this is 

\begin{align}
\label{defofP}
P_M=\Pr(m^{\text{re-measured}}>M|m^{\text{measured}}),
\end{align}

\noindent The probability $P_M$ will depend on the probability distribution $p(m^{\text{(re-)measured}}|m^{\text{true}})$, known as the resolution function.  For many applications, the resolution function is well-approximated by a Gaussian centered at the true value with standard deviation $\sigma_m$.  If the true distribution is only slowly varying in a neighborhood $m^\text{measured}\pm \sigma_m$ such that $p(m^\text{true})$ is approximately constant, the value of $P_M$ can be computed analytically as shown in Eq.~\ref{IntroduceX}.

\begin{align}\nonumber
\label{IntroduceX}
P_M&=\int_M^\infty p(m^{\text{re-measured}}|m^\text{measured})dm^{\text{re-measured}}\\\nonumber
&=\int_M^\infty\int_{-\infty}^\infty p(m^{\text{re-measured}}|m^\text{measured},m^\text{true})p(m^\text{true}|m^\text{measured})|dm^{\text{re-measured}}dm^\text{true}\\\nonumber
&\propto\int_M^\infty\int_{-\infty}^\infty p(m^{\text{re-measured}}|m^\text{true})p(m^\text{measured}|m^\text{true})|dm^{\text{re-measured}}dm^\text{true}\\\nonumber
&\propto \int_M^\infty \exp\left(\frac{-(m^{\text{re-measured}}-m^{\text{observed}})^2}{4\sigma_m^2}\right)dm^{\text{re-measured}}\\
&=\frac{1}{2}\left(1+\text{erf}\left(\frac{m^{\text{observed}}-M}{2\sigma_m}\right)\right),
\end{align}

\noindent The second line in Eq.~\ref{IntroduceX} is the law of total probability, the third line is from the fact that $m^\text{re-measured}$ is independent of $m^\text{measured}$ given $m^\text{true}$, Bayes theorem, and the approximation that $p(m^\text{true})$ is approximately constant near $m^\text{measured}$.  The fourth line in Eq.~\ref{IntroduceX} is from completing the square and integrating out $m^\text{true}$.  Since the $\mathrm{erf}$ function is monotonic and smooth, the complete behavior of $P_M$ is determined by the quantity:

\begin{align}
X_M\equiv \frac{m^{\text{observed}}-M}{\sigma_m}.
\end{align}

The only current use of a variable like $X_M$ is the ``${E}_\text{T}^\text{miss}$ significance''.  First constructed at D\O~\cite{BruceKnuteson:1999pna}, the $E_\text{T}^\text{miss}$ significance in its most complete form usually refers to the log of a likelihood ratio:

\begin{align}
\label{METsig}
\log\left(\frac{p({E}_\text{T}^\text{miss}={E}_\text{T}^\text{miss,measured})}{p({E}_\text{T}^\text{miss}=0)}\right).
\end{align}

\noindent The purpose of ${E}_\text{T}^\text{miss}$ significance is to differentiate events with real missing energy from invisible particles like neutrinos from those without (see Fig.~\ref{fig:susy:fakemet}), and it is constructed from the resolution functions of all the objects used to construct the ${E}_\text{T}^\text{miss}$ itself.  For Gaussian resolutions, the ${E}_\text{T}^\text{miss}$ significance is a monotonic function of  $\left({E}_\text{T}^\text{miss}\right)^2/2\sigma_{{E}_\text{T}^\text{miss}}^2$.  The resolutions are well approximated by $\sigma_{{E}_\text{T}^\text{miss}}\propto \sqrt{H_T}$, the scalar sum of the visible $p_T$ in the event~\cite{Aad:2012re,Chatrchyan:2011tn}.  Therefore, an approximate ${E}_\text{T}^\text{miss}$ significance may be written as a monotonic function of $({E}_\text{T}^\text{miss})^2/{H_T}$.  The most widely used choice is ${E}_\text{T}^\text{miss}/\sqrt{H_T}$.  Note that the approximate ${E}_\text{T}^\text{miss}$ significance is a realization of $X_M$ in which $M=0$, the resolution function is Gaussian, and $\sigma\propto\sqrt{H_T}$.

Even though ${E}_\text{T}^\text{miss}/\sqrt{H_T}$ and ${E}_\text{T}^\text{miss}$ and
are correlated, one can gain statistical power by considering
${E}_\text{T}^\text{miss}/\sqrt{H_T}$ in addition to or instead of
${E}_\text{T}^\text{miss}$ itself.  This has been shown in numerous analyses spanning a wide range of physics processes including Standard Model measurements and searches for SUSY.   In addition to studying the general properties of significance variables, the next sections explore the potential gains from building significance variables for other kinematic variables.			

\subsubsection{Constructing Significance Variables}

The optimal method for using event-by-event and object-by-object resolutions is to combine them in multidimensional likelihood with the kinematic variables themselves.  By the Neyman-Pearson lemma~\cite{Neyman289}, a threshold requirement on the likelihood is no worse than any other possibility, and thus optimal.  However, it is often not possible or highly non-trivial to compute the likelihood combination.  The ${E}_\text{T}^\text{miss}$ significance example motivated the formation of a particular combination of the kinematic variable and its associated resolution into a single quantity.  This quantity is equivalent to the significance variable $X_M$, which may contain all of the relevant discriminatory information.   Ideally, it is a general trend that most of the relevant resolution information can be condensed into a single simple $X_M$-like variable.  Fortunately, this will be true under certain conditions -- principally those in which the signal and backgrounds are associated with different mass or energy scales.  Before showing specific examples, it is important to note that while the $X_M$ significance variables may capture most of the relevant resolution information, they may not always be optimal for every kinematic variable.   Any case in which resolutions are significantly non-Gaussian may require, for optimality, the use of a significance variable based on the full likelihood ratio.  Nonetheless, $X_M$ is simple to compute and contains information that is currently unused by most analyses.

\subsubsection{Examples of Optimal Significance Variables}
\label{sec:gumbel}

To begin, consider a simple model in which the variable $m$ has a delta function distribution, $(1/N)dm_i/dN=\delta(m-M_i)$, where $i\in\{s,b\}$ (signal/background).  For example, suppose that $m=m_\text{T}$ in a search with a resonance decaying into a lepton and a neutrino.   Due to the Jacobian peak, most of the probability for $m$ is near $M_i$, and so this simple model captures some important aspects of the analysis.   Let the resolution functions of $m$ be Gaussian with width $\sigma$.  Then, the joint probability distribution of $m$ and $\sigma$ is given by

\begin{align}
\label{gaus}
p_i(m,\sigma)=g(\sigma)\frac{1}{\sqrt{2\pi\sigma^2}}\exp\left(-\frac{(m-M_i)^2}{2\sigma^2}\right),
\end{align}

\noindent where $g(\sigma)$ is the distribution of $\sigma$.  The optimal use of $m$ and $\sigma$ is to place a threshold on the ratio $p_s(m,\sigma)/p_b(m,\sigma)$.  Dividing the probably functions from Eq.~\ref{gaus} and monotonically transforming them results in the following optimal significance variable

\begin{align}
\label{sigvars:gaussianopt}
V_\text{opt}^\text{(Gaussian)} = \frac{m-(M_s+M_b)/2}{\sigma^2}.
\end{align}

\noindent This significance variable is similar to $X_M$ (with $M=(M_s+M_b)/2$) and only differs in the use of the variance instead of the standard deviation of the resolution in the denominator.  The simple Gaussian example shows that while simple and intuitive, $X_M$ may not always be optimal.  However, as long as $g(\sigma)$ is not too broad, the difference between $\sigma$ and $\sigma^2$ in the denominator should be minimal, since linearizations about a characteristic scale $\sigma_0$ will give similar results:

$$\frac{x-M}{\sigma}=\frac{\sigma_0}{2}\frac{x-M}{\sigma^2}+\text{constant}+\mathcal{O}((\sigma-\sigma_0)^2)$$

Now, consider a variant of the previous example with an asymmetric resolution function defined by the Gumbel distribution~\cite{gumbel}:

\begin{align}
p_i(m)=\frac{1}{\beta}\exp\left(\frac{m-M_i}{\beta}\right)\exp\left(-\exp\left(\frac{m-M_i}{\beta}\right)\right).
\end{align}

\noindent The Gumbel is chosen because its Taylor series is the same as a Gaussian with parameters $\mu$ and $\sigma^2$ up to the third order term in $\frac{m-M_i}{\beta}$ with the identification $\sigma=\frac{e}{\sqrt{2\pi}}\beta$ and $\mu=M_i$.  Figure~\ref{fig:susy:gumbel} overlays a Gaussian on top of the Gumbel distribution with this identification scheme.  The two distributions have the same core, but the tail of the Gumbel distribution is heavier on the left than the right, which represents the generic case in which events are more likely to have smeared from lower values due to falling priors.  Taking the logarithm of the likelihood ratio results in the following optimal significance variable:

\begin{align}
V_\text{opt}^\text{(Gumbel)} =\exp\left(\frac{m-M_b}{\beta}\right)-\exp\left(\frac{m-M_s}{\beta}\right)+\frac{M_b-M_s}{\beta}. \label{eq:somethinggoes}
\end{align}

\begin{figure}[h!]
\begin{center}
\includegraphics[width=0.5\textwidth]{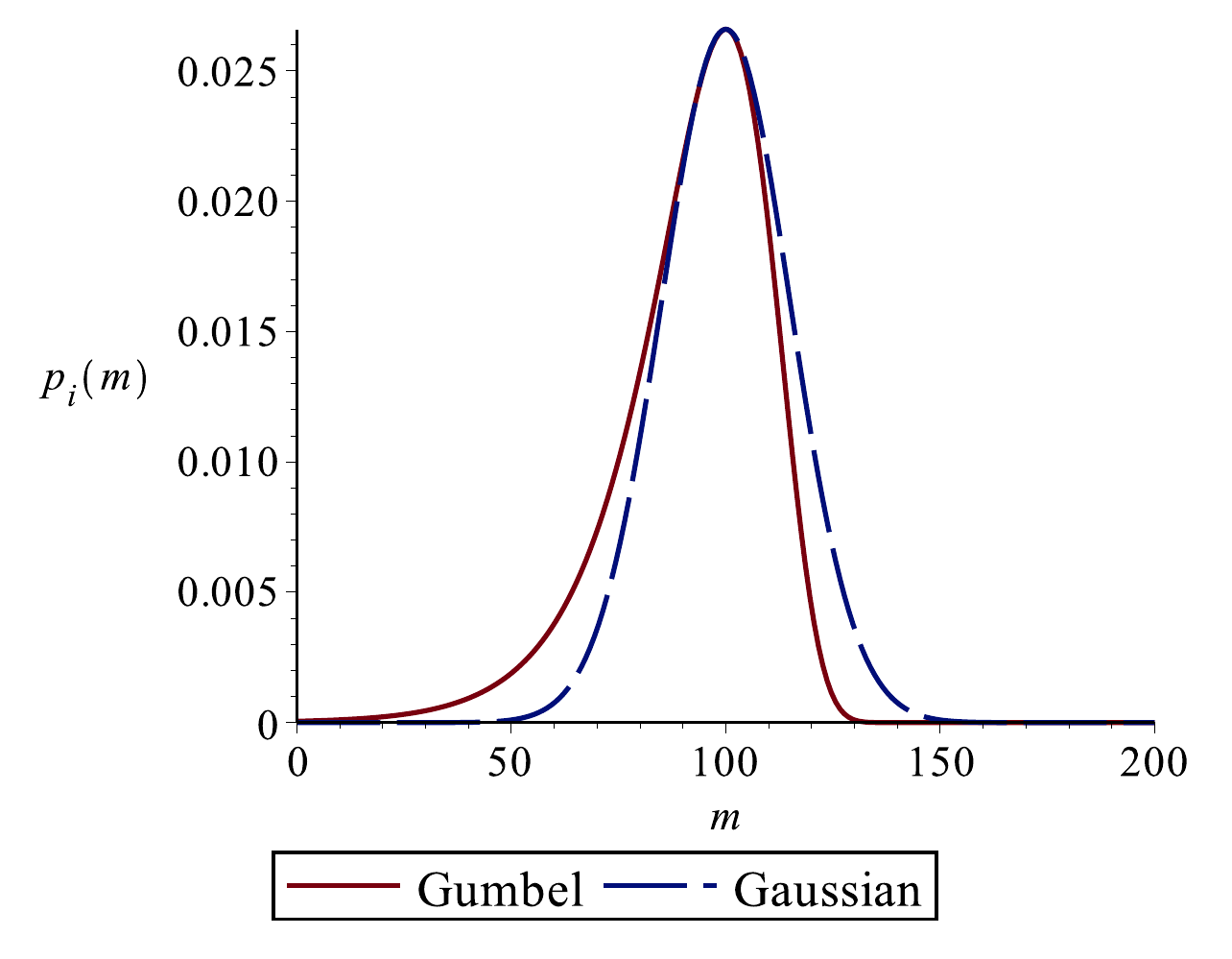}
 \caption{The Gumbel distribution and an equivalent Gaussian distribution.}
 \label{fig:susy:gumbel}
  \end{center}
\end{figure}

\noindent Lines of constant $V_\text{opt}^\text{(Gumbel)}$ are shown in Fig.~\ref{fig:susy:gumbel}.  When $m$ is small compared to $M_s$ and $M_b$, the first two terms in Eq.~\ref{eq:somethinggoes} are highly suppressed relative to the constant third term and so the lines of constant $V_\text{opt}^\text{(Gumbel)}$ are horizontal lines in Fig.~\ref{fig:susy:gumbel} on the left.  This region is uninteresting as usually the region of interest is $m>M_b$ since $M_b$ is often less than $M_s$.  For $M_b < m < M_s$, the first term in Eq.~\ref{eq:somethinggoes} dominates so lines of constant  $p_s/p_b$ are well approximated by lines of constant $X_M$ with $M=M_b$. These are found in Fig.~\ref{fig:susy:gumbel} as straight lines radiating from $(m=M_b, \beta=0)$.  Finally, in the region in which $m>M_s$ and $\beta$ is small compared $M_s-M_b$, both exponentials are large and so the dominant part of (\ref{eq:somethinggoes}) can be re-written as

\begin{align}
\exp\left(\frac{m-\bar{M}}{\beta}\right) \sinh\left(\frac{M_s-M_b}{\beta} \right),
\end{align}

\noindent where $\bar{M}$ is the average of $M_s$ and $M_b$.  Since the $\sinh$ term is relatively smaller and slowly varying, lines of constant likelihood ration are this just limes of constant $X_M$ with $M=\bar{M}$, which are once again straight lines in Fig.~\ref{fig:susy:gumbel}.  This simple example shows that an optimal use of $m, \sigma_m$, and $M$ is well approximated by a threshold requirement on $X_M$ even when the resolution function is realistically asymmetric.

\begin{figure}
\begin{center}
\includegraphics[width=0.45\textwidth]{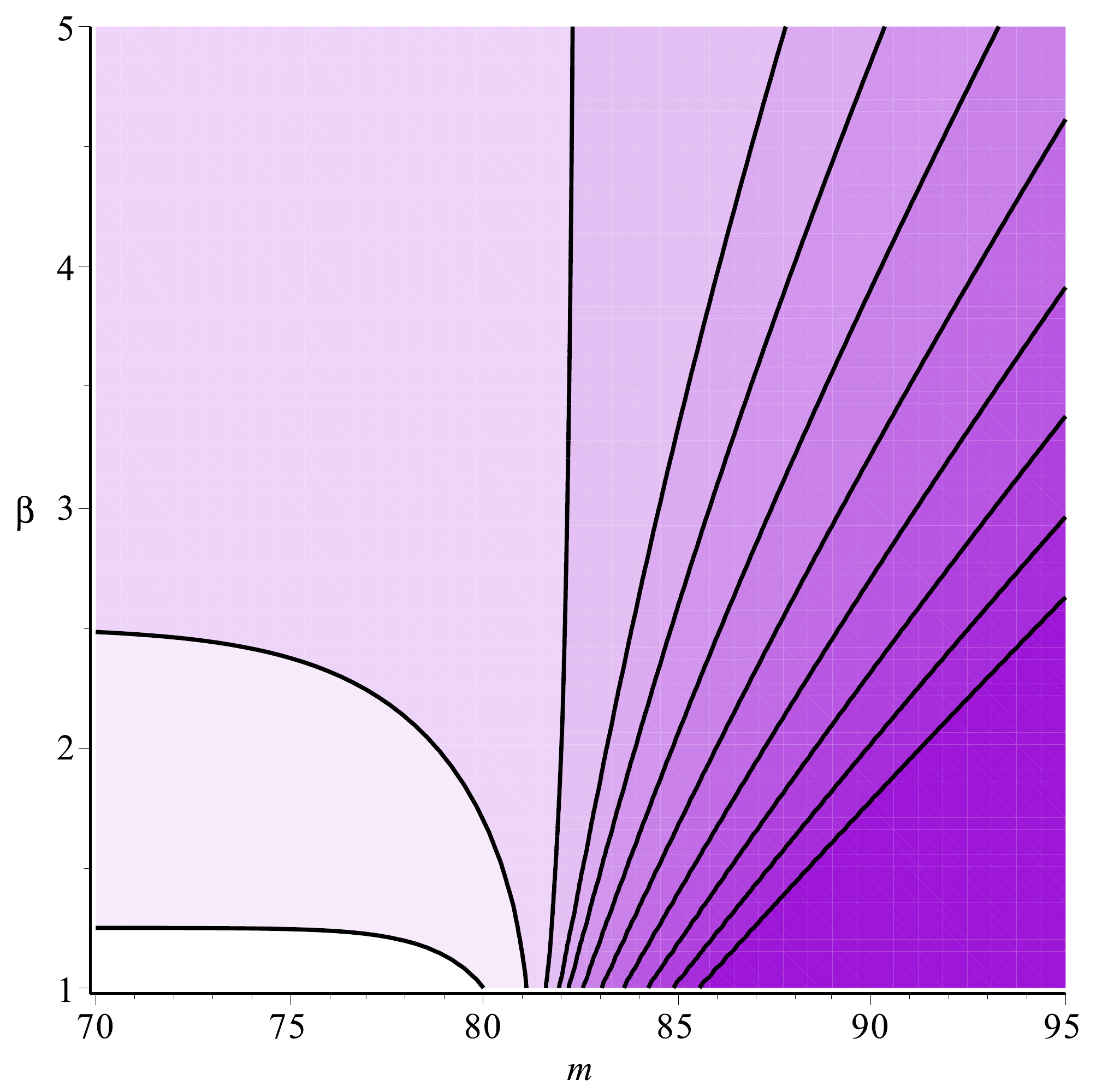}
\end{center}
\caption{Contours of constant $V_\text{opt}^\text{(Gumbel)}$ in the $(m,\beta)$ plane for $M_b=80$ and $M_s=85$.  The contours are drawn at powers of 2 from $-2^3,...,-2,0,2,...2^8$.}
\label{Gumbel}
\end{figure}

\subsubsection{Choosing the Separation Scale $M$}

The construction in Sec.~\ref{sec:gumbel} shows that $M$ can play a dynamic role in the definition of $X_M$.  First of all, note that an analysis that uses a threshold requirement on $X_M$ is truly a generalization of an analysis that uses a threshold on $m$.  Let $c$ be the threshold for the latter analysis, such that signal-like events are those with $m>c$.  Then, the events chosen by $X_c>0$ will be identical to those chosen with $m>c$ and therefore there is always a choice of $M$ that reduces the significance variable-based analysis to one based only on the kinematic variables themselves.  In particular, an optimal analysis based on $X_M$ can be no worse than one based on $m$ alone and will likely be better since $X_M$ incorporates more information and has an additional degree of freedom $(M)$. 

The interpretation of $M$ as the scale of Standard Model physics does not require that it be fixed ahead of time, since detector resolutions can distort the {\it reconstructed} scale away from the {\it true} scale.   It is often the case that the distribution of $\sigma$ itself is independent of the underlying process and thus not useful for distinguishing signal and background.  Another way to visualize how $M$ mixes with $\sigma$ to add discriminating power on top of $m$ is to consider the ordering of events induced by $X_M$ versus $m$.   For example, suppose that there are only two events with $m$ values $m_1,m_2$ and resolutions $\sigma_1$ and $\sigma_2$.  The quantity which controls the ordering of $X_M$ is $\Delta \equiv (m_2\sigma_1 -  m_1\sigma_2)/(\sigma_1-\sigma_2)$.   When $\Delta<0$ or infinite in magnitude, then $X_M^{1}>X_M^{2}$ for all $M$.  However, if $\Delta>0$, then there is a critical $M^*$ such that for $M<M^*$, $X_M^{1}>X_M^{2}$ for $M>M^*$, $X_M^{1}<X_M^{2}$.  The value of $M^*$ is $\Delta$.  For $N>2$, the situation is more complicated, but the result is the same; different values of $M$ can rearrange the distribution of events based on $X_M$ from the distribution based on $M$.   One can generalize the plots in Figure~\ref{Mdependance} for $N>2$.  Note that the distribution of points of intersection with the $M$ axis forms the observed distribution of $m$.

		\begin{figure}
         	\begin{center}
	\includegraphics{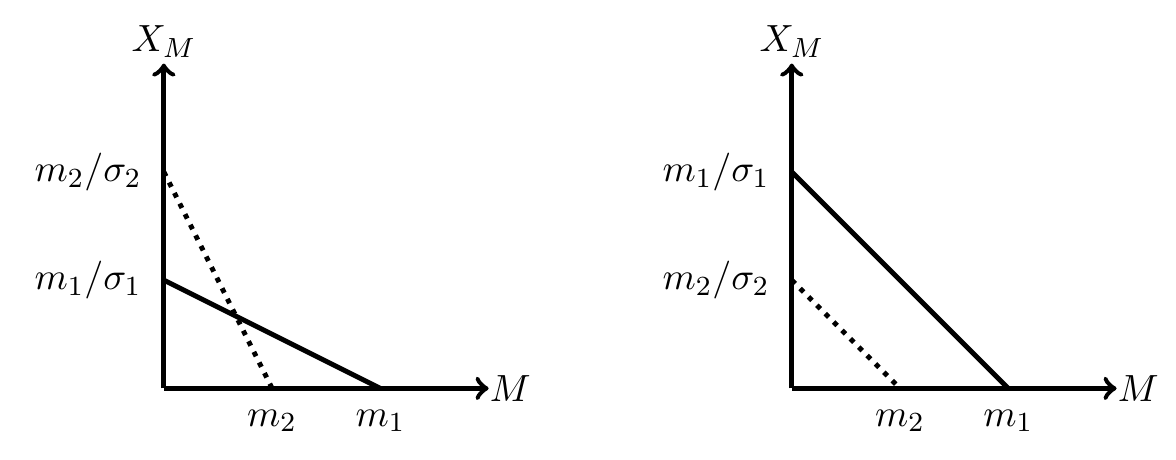}
		\end{center}
		\caption{The dependance of $X_M$ on $M$ for two events with $\Delta\equiv (m_2\sigma_1 -  m_1\sigma_2)/(\sigma_1-\sigma_2) > 0$ in the left plot and $\Delta\in [-\infty,0)\cup\{\infty\}$ in the right plot.}
		\label{Mdependance}
		\end{figure}

Consider a kinematic variable $m$ with zero resolution maximum $\tilde{m}$.   Intuitively, it may seem that $M$ should be equal to or greater than $\tilde{m}$. However, the value of $M$ which maximizes the significance $\hat{s}(c)\equiv s/\sqrt{b}$, for $c$ a threshold value on $X_m$, could be less than $\tilde{m}$.  If $\sigma$ is constant over all events, $X_M$ induces the same ordering on events as $m$ and so any value of $M$ maximizes $\hat{s}$.  As an example, recall the model in Eq.~\ref{gaus}.  If the distribution of $\sigma_m$ is also a delta function, then $X_M$ and $m$ will give the same significance.  Therefore, take a simple extension:

\begin{align}
g(\sigma) =p\delta(\sigma-\sigma_1)+(1-p)\delta(\sigma-\sigma_2),
\end{align}

\noindent where $\sigma_i$ are two fixed values of $\sigma$ and $p\in[0,1]$. With this simple model, one can easily compute the distributions of $m$, $X_M$ and $\hat{s}$, as seen in Figure~\ref{mstar} for $\tilde{m}=80$ for the background, $\tilde{m}=90$ for the signal, $p=1/2$ and $\rho$ is the signal efficiency, defined by $\rho(c)=\int_c^\infty\mathrm{d}x f(x)$ for $f(x)$ the signal probability density function and $c$ a cut value.  Furthermore, $\sigma_1=5$ and $\sigma_2=10$.  In this setup, there is an $M<\tilde{m}$ which outperforms the significance at $M=\tilde{m}$.  This is seen clearly in the second plot of the figure in which the low value of $M$ can allow for $X_M$ to distinguish between low and high resolution events for the signal.  In the limit as $\tilde{m}-M > \sigma$, $X_M$ will be able to distinguish the low and high resolution events, thus increasing $\hat{s}$.  For $\tilde{m}-M \gg \sigma$, the efficacy of $X_M$ approaches the constant resolution case and so one cannot gain more by decreasing $M$.  

\begin{figure}
\begin{center}
\includegraphics{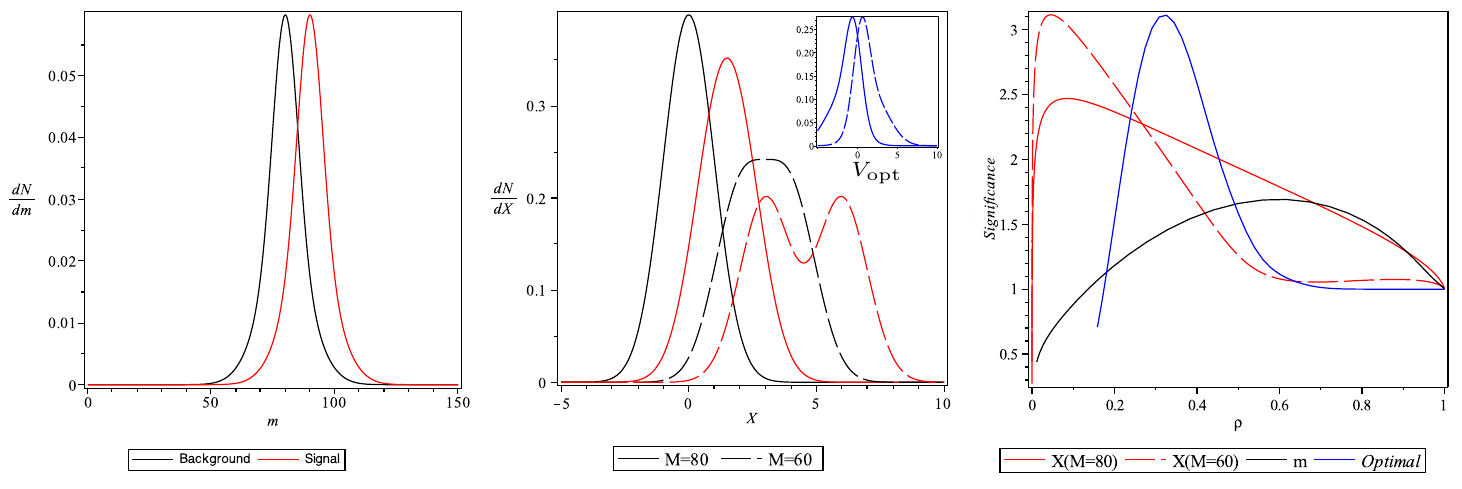}
\end{center}
\caption{These plots illustrate the distributions of $m$, $X_M$ and $\hat{s}$ for a simple model in which $m$ is always `on shell' at 80 for the background and 90 for the signal.  The resolutions can take one of two values with probability $1/2$, independent of the physics process.  The optimal ($V_{\text{opt}}$) variable is based on the likelihood ratio detailed in Sec.~\ref{sec:gumbel} and Eq.~\ref{sigvars:gaussianopt}.}
\label{mstar}
\end{figure}

Before proceeding, here are two further remarks about the above toy model.  First, note that due to the simplicity, one can actually derive the optimal variable, described in Sec.~\ref{sec:gumbel} as $V_{\text{opt}}=(m-\bar{M})/\sigma^2$, where $\bar{M}$ is the average of the signal and background true $m$ values.  The distribution and significance of $V_{\text{opt}}$ are shown alongside $X_M$ in Fig.~\ref{mstar}.  Once can see that while $X_M$ does no better than the optimal variable, for the appropriate choice of $M$ it can have essentially the same maximum significance.  A second remark is that since the distribution of $\sigma$ for signal and background is identical, the resolution alone cannot distinguish signal and background. Thus, the improved performance of $X_M$ over $m$ is due entirely to the event-by-event combination of $m$, $M$, and $\sigma$ to capture resolution and kinematic properties of the reconstructed objects.

\clearpage

\subsubsection{Empirical Examples}
\label{simulate}

This section contains a few illustrative examples of significance variables using realistic physics processes but simplistic models for the detector resolution.  Events are generated using \textsc{Pythia} 8.170~\cite{Sjostrand:2007gs,Sjostrand:2006za} to cover three canonical searches that exploit endpoints in kinematic distributions.   The resolution of the missing momentum is modeled as $\sigma_{E_\text{x,y}^\text{miss}}=0.5\sqrt{\sum E_T}$, where $\sum E_T$ is the sum of all visible momentum and follows the measured spectra in dijets~\cite{Aad:2012re}. The distributions of $X_M$ are computed by numerically propagating the underlying kinematic uncertainties.

A first example is the $W'\rightarrow \mu\nu$ search using the transverse mass of the muon and the neutrino as the main discriminant.   In this search, the $W$ mass is a natural choice for $M$ in constructing $X_M$, where $m=m_\text{T}$.  The $W'$ boson is created with a mass of 100 GeV\footnote{Excluded by~\cite{Aad:2012dm, Chatrchyan:2013lga}, useful here for illustration only.} and the SM CKM matrix.   The distributions of $m_\text{T}$, $X_M$ and $\hat{s}$ are shown in Fig.~\ref{W}.  The various rows of Fig.~\ref{W} demonstrate the affect of the $W$ width on the efficacy of $X_M$.  For a vary narrow resonance background, $X_M$ is much better than $m_T$, but as the width becomes large, the advantage decreases.  

\begin{figure}[h!]
\begin{center}
\includegraphics[scale=1.0]{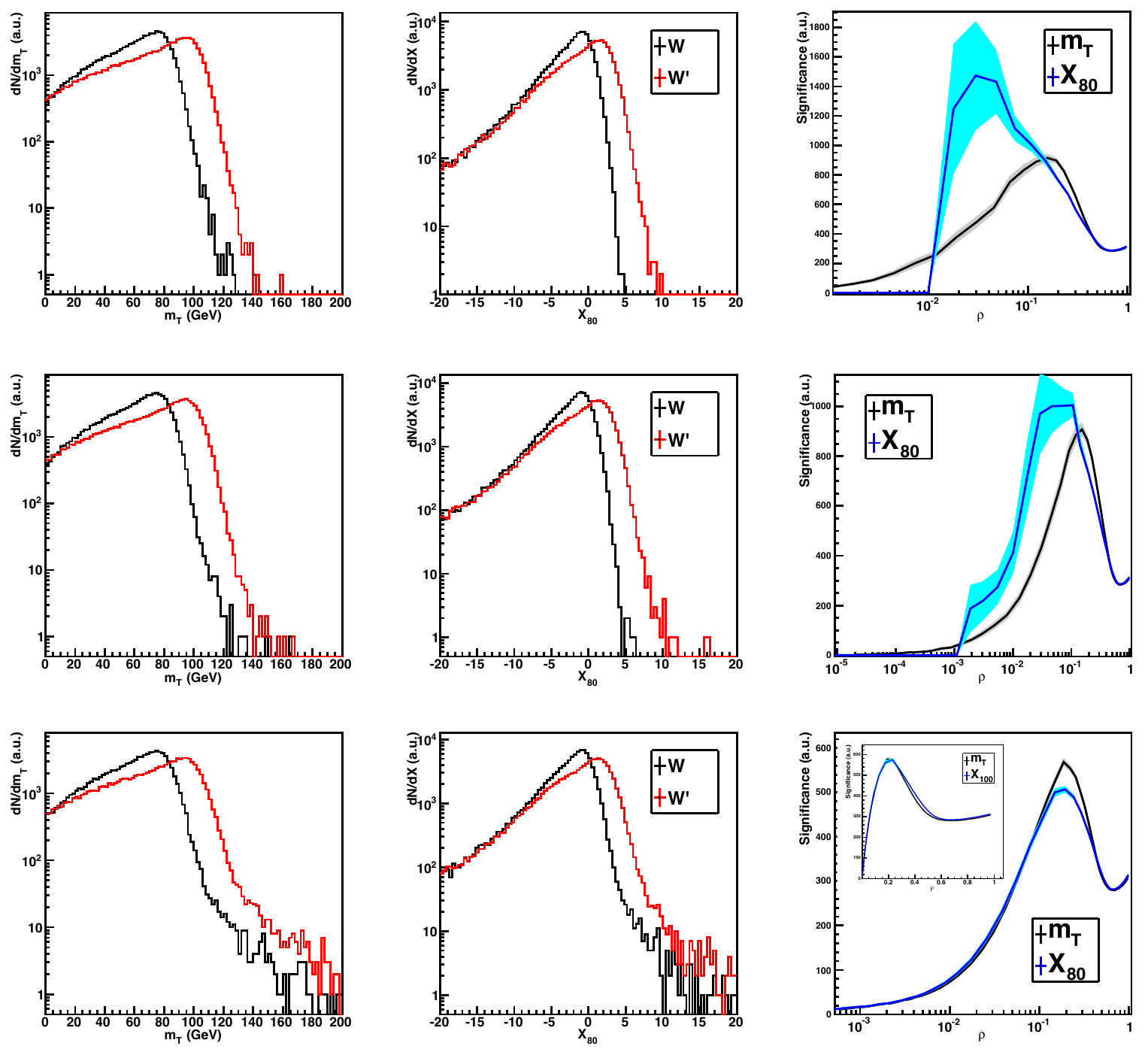}
\end{center}
\caption{In each row, the left plot compares the transverse mass distribution for a Standard Model W and a W' with mass 100 GeV.  The middle plot is the corresponding distributions of $X_M$ with $M=80$ GeV. The right plot shows the rejection $s\sqrt{b}$ as a function of the signal efficiency, in arbitrary units.  The bands show the statistical uncertainty due to limited Monte Carlo statistics.  The top row has a boson mass width of 0, the middle has a width of 20\% of the natural width, and the bottom row has the full width of about $2$ GeV~\cite{pdg}.  We can see that for this fixed value of $M$, the performance of $X_M$ is better than $m_T$ for a narrow width and then worse at higher width.  By construction, $X_M$ cannot be worse than $m_T$ and thus the optimal $M$ in the last row must be different than 80.  The inset plot shows $X_M$ for $M=100$, for which the performance of $X_M$ and $m_T$ is the same.}
\label{W}
\end{figure}

Another possible use of the $m_\text{T}$ significance is in the standard $H\rightarrow\tau\tau$ search (measurement)~\cite{Aad:2012mea,Chatrchyan:2012vp} where the di-tau system is the `visible particle' in the calculation of the transverse mass.  In the dilepton channel, the dominant background is $Z$ boson production and so the natural value for $M$ is $90$ GeV.  Figure~\ref{Higgs} shows the distributions of $m_\text{T}$, $X_M$, and $\hat{s}$ for a 125 GeV Higgs.  The optimal value of $M$ was found to be less than $90$, as indicated in the diagram.  The $\hat{s}$ figure shows that there can be a significant improvement from $X_M$ over $m_\text{T}$ by about 20\%.

\begin{figure}[h!]
\begin{center}
\includegraphics[scale=0.8]{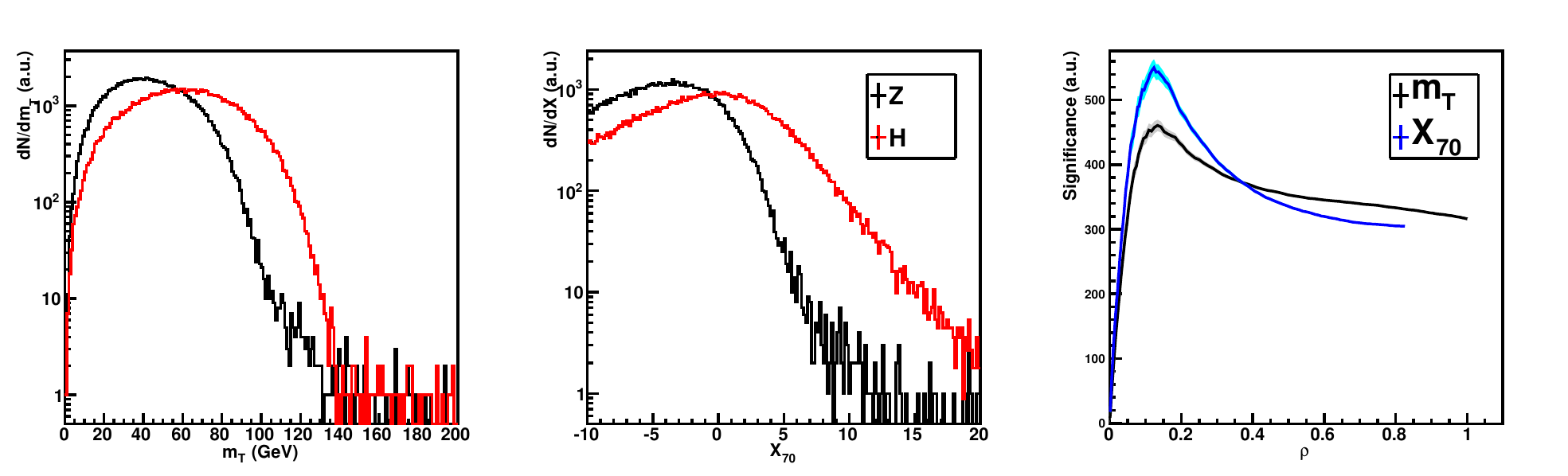}
\end{center}
\caption{The left plot is the $m_T$ distribution for dileptonic $Z\rightarrow\tau\tau$ and $H\rightarrow \tau\tau$  for a 125 GeV Higgs.  The middle plot is the corresponding $X_M$ curve with M=70 and the right plot is the rejection versus efficiency relationship.}
\label{Higgs}
\end{figure}

A third illustrative example is the pair production of stops with $\tilde{t}\rightarrow t+\mathrm{LSP}$ in the dilepton channel using $m_\text{T2}$.  With the leptons as the visible particles in the definition of $m_{T2}$, this system once again has the feature that the resolution is mostly due to the missing momentum vector.   With $t\bar{t}$ as the dominant background, the natural scale is $M=80$ GeV.   The $m_{T2}$ distribution, $m_\text{T2}$ significance, and $\hat{s}$ are shown in Fig.~\ref{Stop} for a compressed scenario of $m_{stop}=350$ GeV and $m_{LSP}=170$ GeV.  The use of $X_M$ improves the significance by about 30\% over $m_\text{T2}$ alone.

\begin{figure}[h!]
\begin{center}
\includegraphics[scale=0.8]{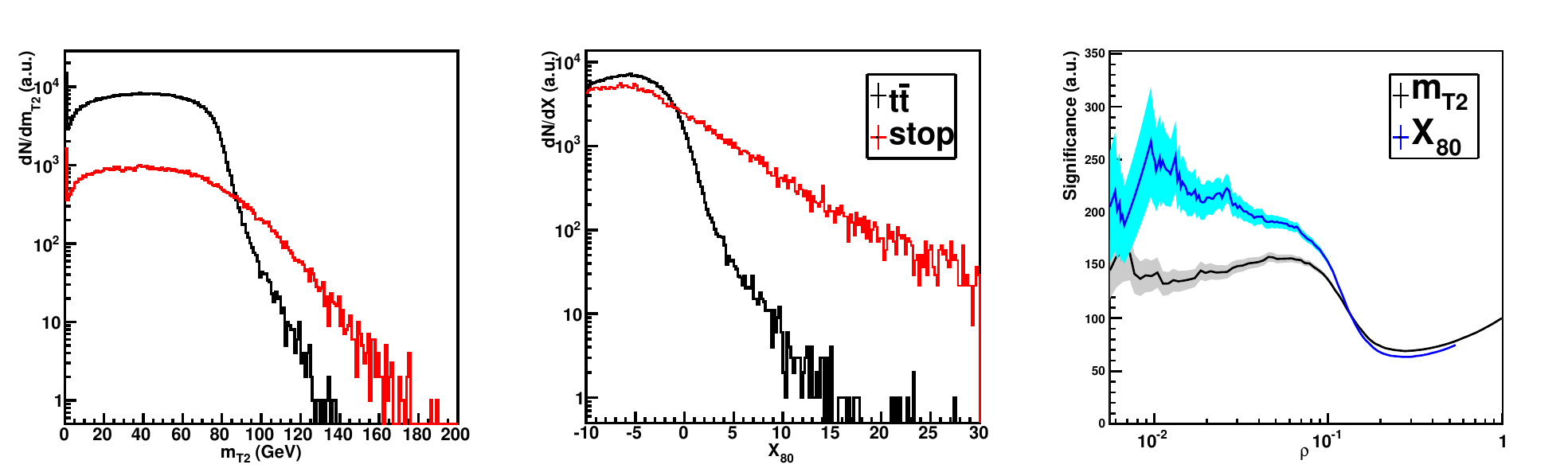}
\end{center}
\caption{The left plot is the $m_{T2}$ distribution for for dileptonic $t\bar{t}$ and $\tilde{t}\rightarrow t+\mathrm{LSP}$ for a 350 GeV stop and 170 GeV LSP.  The middle plot is the corresponding $X_M$ curve with M=80 and the right plot is the rejection versus efficiency relationship.}
\label{Stop}
\end{figure}

\clearpage

\subsubsection{Examples with Full Detector Simulation}

The examples in Sec.~\ref{simulate} show that the additional information from $X_M$ can be useful for improving the significance of bounded kinematic variables.  This section re-focuses on the stop search in the one lepton channel which makes use of several variables with kinematic edges (see Sec.~\ref{sec:transmass}).  Section~\ref{approxsig} begins with the construction of approximate significance variables that continue to utilize the relationship $\sigma_{E_\text{T}^\text{miss}}\propto \sqrt{H_T}$.  A more complex set of variables based on known jet-by-jet resolutions are built in Sec.~\ref{htsigmiss}. 

\paragraph{Approximate $m_\text{T}$ Significance}\mbox{}\\
\label{approxsig}

An approximate $m_\text{T}$ significance is defined as $X_{m_\text{T}}=(m_\text{T}-M)/\sigma$, where $\sigma$ is constructed from $\sqrt{H_\text{T}}$ and $E_\text{T}^\text{miss}$ by linearly propagating uncertainties assuming no angular resolution.  In the massless approximation, $m_\text{T}^2=2E_\text{T}^\text{miss}p_\text{T}^{\ell}(1-cos(\theta))$ where $p_\text{T}^{\ell}$ is the transverse momentum of the lepton and $\theta$ is the angle between $\vec{p}_\text{T}^{\ell}$ and $\vec{p}_\text{T}^\text{miss}$.  Linear error propagation results in the following formula for $\sigma$:

\begin{align}
\label{eq:sigmamtsig}
\sigma\propto (p_\text{T}^{\ell})^2(1-cos(\theta))^2\sigma_{E_\text{T}^\text{miss}}^2+(\text{term proportional to $\theta$ resolution}).
\end{align}

\noindent Neglecting the $\theta$ resolution and modeling  $\sigma_{E_\text{T}^\text{miss}}^2 \propto H_\text{T}$ results in $\sigma\propto m_\text{T}\sqrt{H_\text{T}}/E_\text{T}^\text{miss}$.  Figure~\ref{mtsiginsimulation} compares the distributions of $m_\text{T}$ and $X_{m_\text{T}}$ (with $M=100$ GeV) in simulations of $t\bar{t}$ and stop pair production. The $m_\text{T}$ significance distribution for the background falls off below $0$ while the peak in the signal is greater than zero.  A quantitative comparison of the performance between $m_\text{T}$ and $X_{m_\text{T}}$ is in Fig.~\ref{mtsignificancesig}.  An approximate statistical significance is given by $s/\sqrt{b}$, where $s$ and $b$ are the signal and background yield after a threshold requirement on $m_\text{T}$ or $X_{m_\text{T}}$.  The statistical significance of  $X_{m_\text{T}}$ is nowhere worse than $m_\text{T}$, even without a thorough optimization of $M$. At the peak of the statistical significance, around a signal efficiency of $\sim 1/3$, there is a $\sim 10\%$ improvement when incorporating the resolution information.  The joint distribution of $m_\text{T}$ and $X_{m_\text{T}}$ in Fig.~\ref{joitmtmtsig} shows that there is a strong relationship between these two variables, as might be expected from the simple form of $\sigma$ in Eq.~\ref{eq:sigmamtsig}.  However, there is still a significant spread, which leads to the improvement in Fig.~\ref{mtsignificancesig}.

\begin{figure}[h!]
\begin{center}
\includegraphics[width=0.99\textwidth]{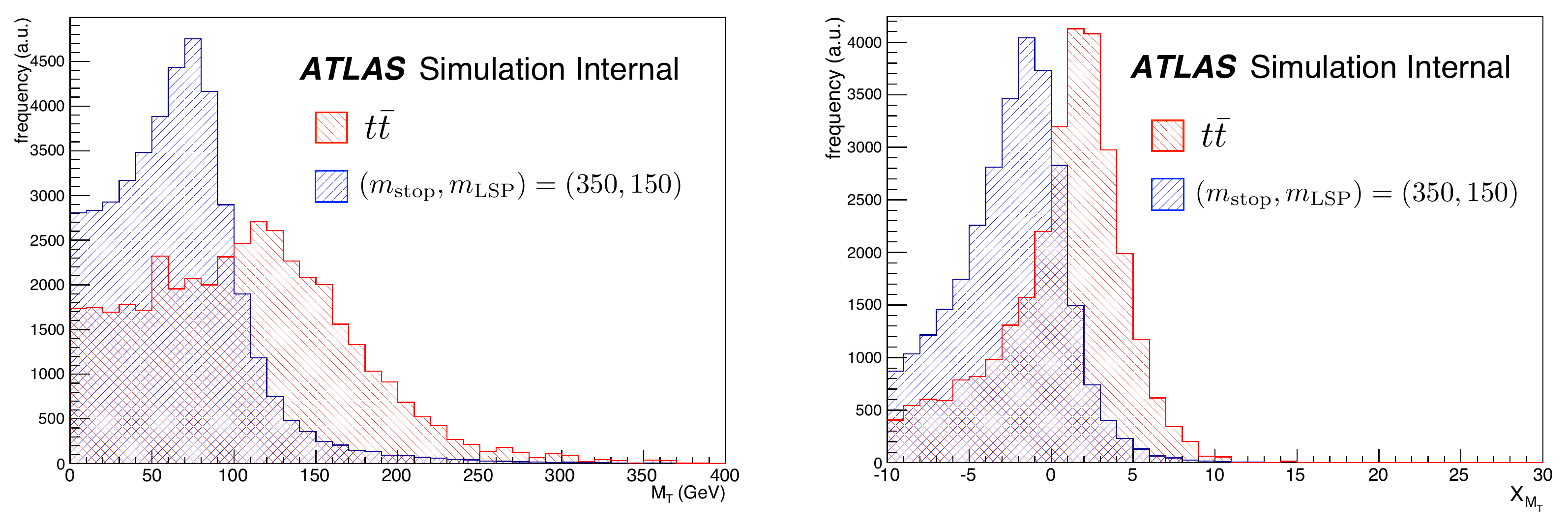}
\end{center}
\caption{The distributions of $m_\text{T}$ (left) and  $X_{m_\text{T}}$ (with $M=100$ GeV) (right) for $t\bar{t}$ and stop pair production.}
\label{mtsiginsimulation}
\end{figure}

\begin{figure}[h!]
\begin{center}
\includegraphics[width=1\textwidth]{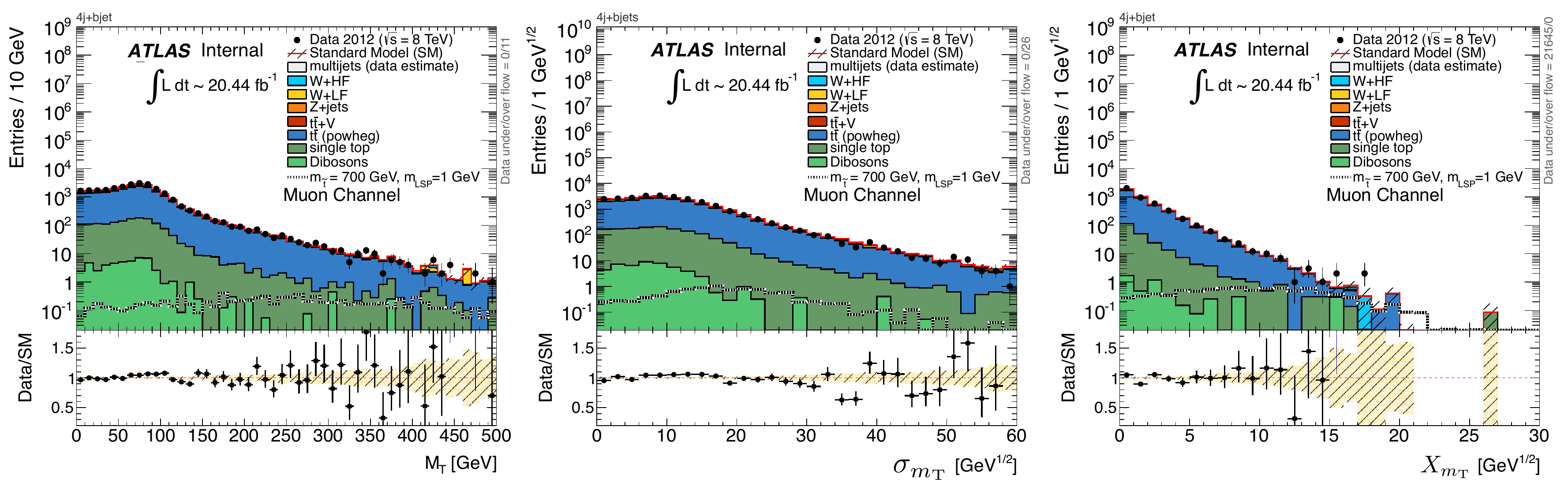}
\end{center}
\caption{A comparison of $m_\text{T}$ (left), $\sigma$ (middle), and $X_{m_\text{T}}$ (right) in data and simulation at $\sqrt{s}=8$ TeV.}
\label{mtsigdata}
\end{figure}

\begin{figure}[h!]
\begin{center}
\includegraphics[width=0.5\textwidth]{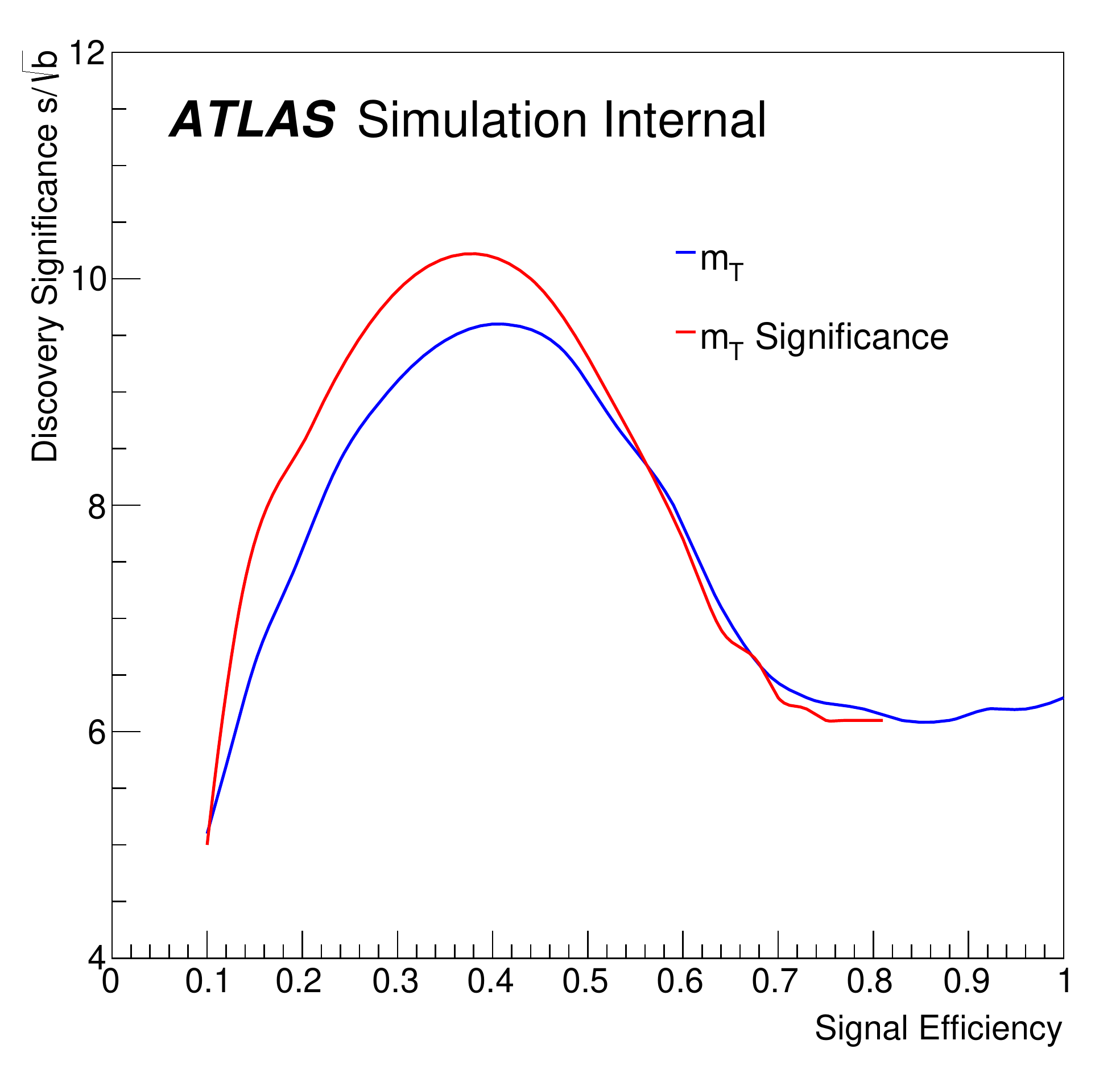}
\end{center}
\caption{A comparison of the statistical significance after a threshold requirement on $m_\text{T}$ and the (approximate) $m_\text{T}$ significance, $X_{m_\text{T}}$.  The background is $t\bar{t}$ and the signal is a stop model with ($m_\text{stop},m_\text{LSP})=(350,150)$ GeV. This statistical significance is the `discovery significance' because it quantifies the number of standard deviations a signal would be above the background-only noise.}
\label{mtsignificancesig}
\end{figure}

\begin{figure}[h!]
\begin{center}
\includegraphics[width=0.5\textwidth]{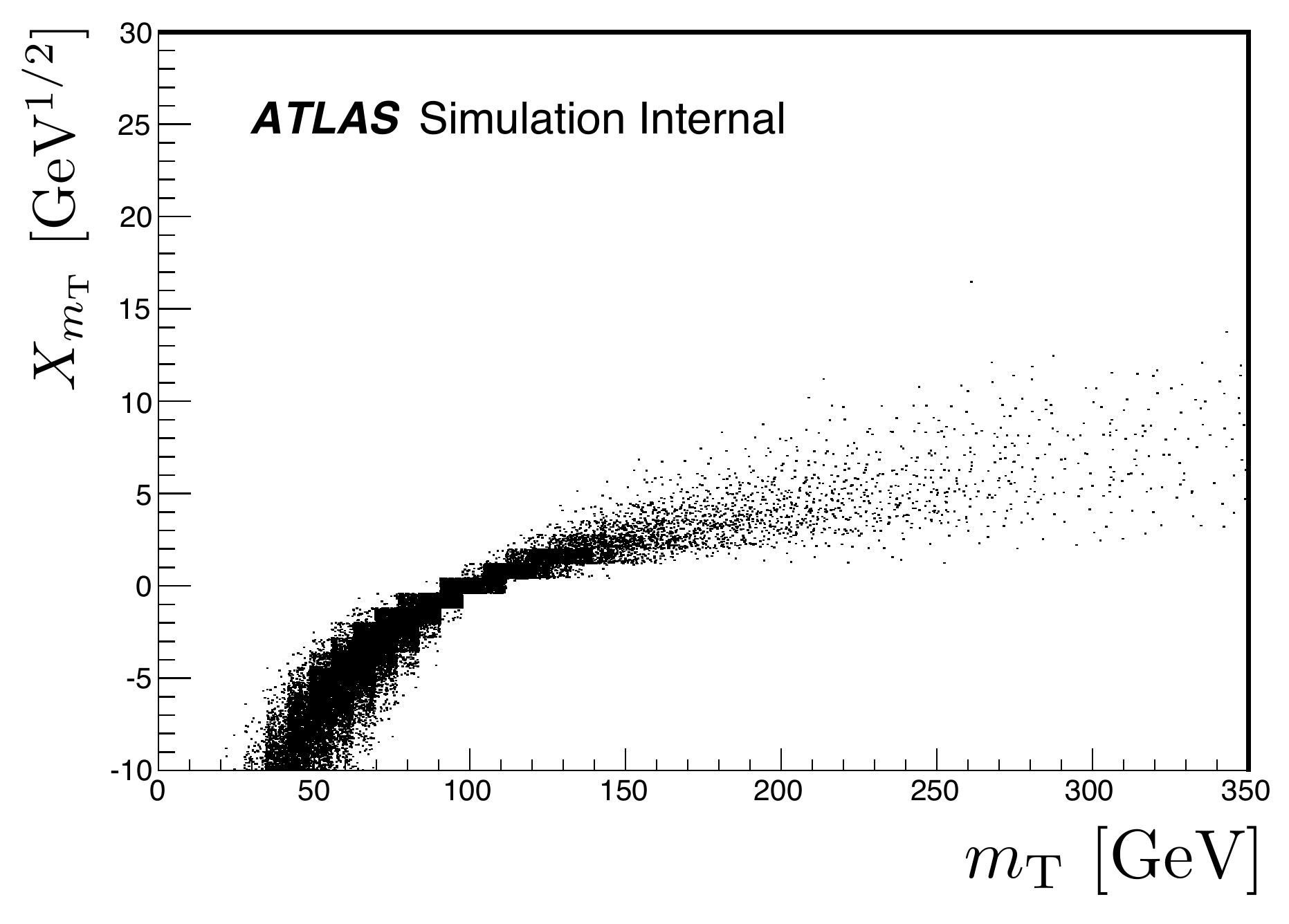}
\end{center}
\caption{The joint distribution of $m_\text{T}$ and $X_{m_\text{T}}$ with $M=100$.  When $m_\text{T}=100$, $X_{m_\text{T}}=0$ by construction.}
\label{joitmtmtsig}
\end{figure}

An important test for any new variable is the ability of the simulation to model the data.  Figure~\ref{mtsigdata} compares the numerator and denominator of $X_{m_\text{T}}$ as well as the significance variable itself.  All three quantities are well-modeled, suggesting that it is ready for use.  However, the $m_\text{T}$ significance constructed in this section is rather simple - a more sophisticated approach to significance variables that will be used for signal region optimization is discussed in the next section.

\clearpage

\paragraph{$H_\text{T,sig}^\text{miss}$ Significance} \mbox{}\\
\label{htsigmiss}

The $p_\text{T}$ and $\eta$ dependence of the jet resolutions are well-understood in simulation and have been well-measured in data (see e.g. Ref.~\cite{ATLAS-CONF-2015-037}).  Parameterizations of the resolutions can be used to calculate resolutions for kinematic quantities that depend on jets event-by-event.  Consider a quantity similar to the $E_\text{T}^\text{miss}$ called the $H_\text{T}^\text{miss}$:

\begin{align}
\label{htmiss}
H_\text{T}^\text{miss}=\Bigg|\sum_\text{jets $j$} \vec{p}_\text{T,j}+\vec{p}_\text{T}^{\ell}\Bigg|,
\end{align}

\noindent where the sum runs over all signal jets and the momentum of the lepton $\vec{p}_\text{T}^{\ell}$.  The symbol $H$ is used instead of $E$ to indicate that only the {\it hard-objects} are used to construct $H_\text{T}^\text{miss}$, whereas $E_\text{T}^\text{miss}$ also includes energy not associated with signal jets and leptons.  As expected, there is a strong correlation between the two definitions for high $E_\text{T}^\text{miss}$ when the contribution from these softer energy sources is small.  Figure~\ref{methtmiss} shows the distribution of $E_\text{T}^\text{miss}$ conditioned on $H_\text{T}^\text{miss}$ for $t\bar{t}$ events with $E_\text{T}^\text{miss}>100$ GeV.  For $H_\text{T}^\text{miss}\gtrsim 100$ GeV, there is a strong correlation with well over $50\%$ of $E_\text{T}^\text{miss}$ values within $15$-$30$ GeV of the $H_\text{T}^\text{miss}$.  The advantage of Eq.~\ref{htmiss} is that the resolutions of the jets are known parametrically and so the resolution $\sigma_{H_\text{T}^\text{miss}}$ can be computed as

\begin{align}
\label{htmisssig}
\sigma_{H_\text{T}^\text{miss}}^2=\frac{1}{N}\sum_{i=1}^N \left(\sum_\text{jets $j$} \Sigma_i^j\vec{p}_\text{T,j}+\vec{p}_\text{T}^{\ell}\right)^2-\left(\frac{1}{N}\sum_{i=1}^N\Bigg|\sum_\text{jets $j$} \Sigma_i^j\vec{p}_\text{T,j}+\vec{p}_\text{T}^{\ell}\Bigg|\right)^2,
\end{align}

\noindent where $\Sigma_i^j$ is a diagonal two-by-two matrix with entries $1+z_i^j$, for $z_i^j\sim\mathcal{N}(0,\sigma(p_\text{T,$j$},\eta_j))$.  To suppress the fluctuations in the calculation of $\sigma_{H_\text{T}^\text{miss}}^2$, $N$ in Eq.~\ref{htmisssig} is chosen to be $1000$.  As expected, there is a strong relationship between the approximate $E_\text{T}^\text{miss}$ resolution and the resolution computed with Eq.~\ref{htmisssig} (Fig.~\ref{corhtmisshtsig}).  By incorporating more local information about the resolution, $\sigma_{H_\text{T}^\text{miss}}$ should be a better approximation to the full significance.  The $H_\text{T}^\text{miss}$ significance, $H_\text{T,sig}^\text{miss}=(H_\text{T}^\text{miss}-M)/\sigma_{H_\text{T}^\text{miss}}$.

\begin{figure}[h!]
\begin{center}
\includegraphics[width=0.9\textwidth]{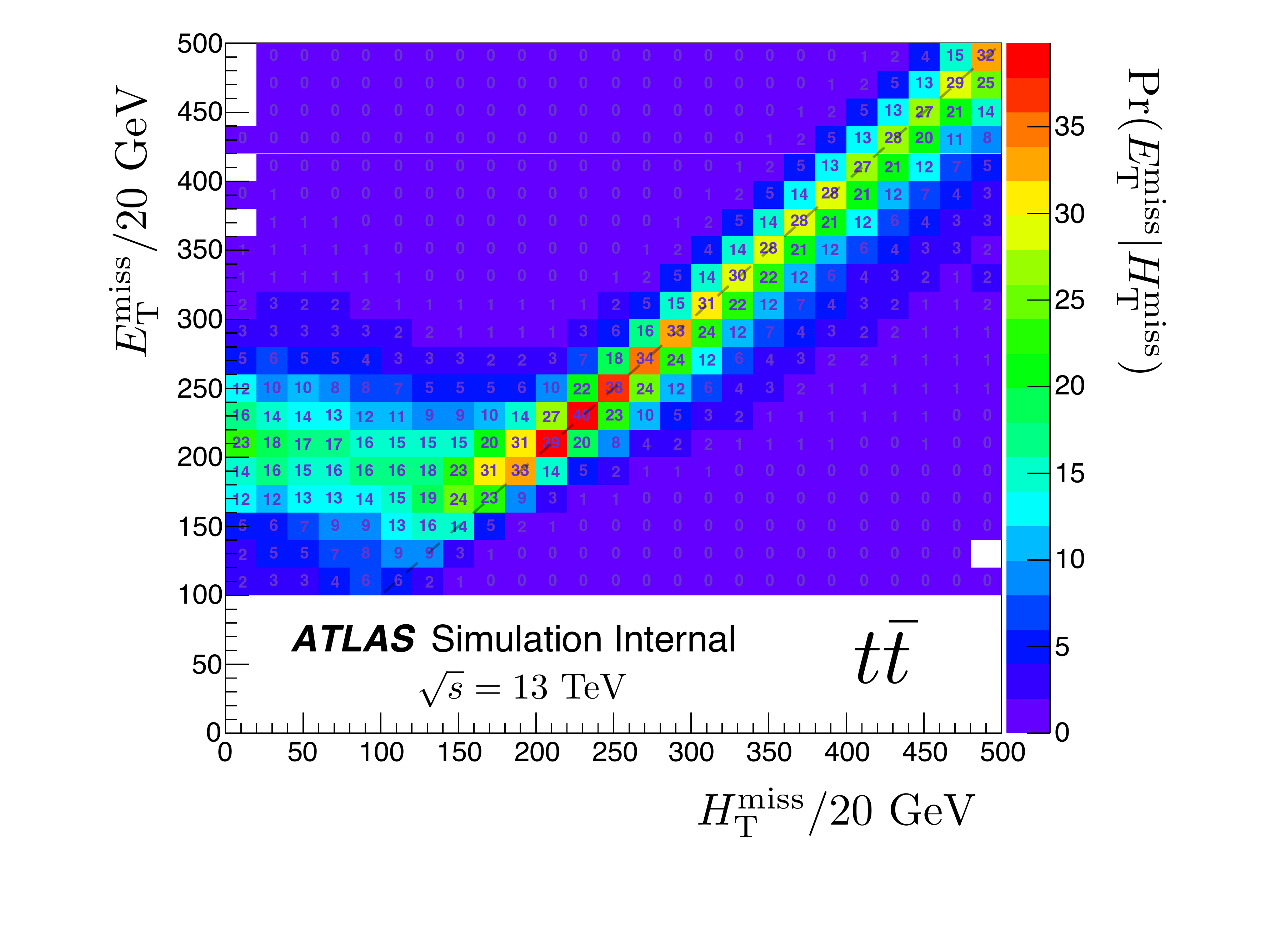}
\end{center}
\caption{The distribution of $E_\text{T}^\text{miss}$ conditioned on $H_\text{T}^\text{miss}$ in bins of $20$ GeV for each variable in $t\bar{t}$ events.  All events have $E_\text{T}^\text{miss}>100$ GeV.}
\label{methtmiss}
\end{figure}

\begin{figure}[h!]
\begin{center}
\includegraphics[width=0.5\textwidth]{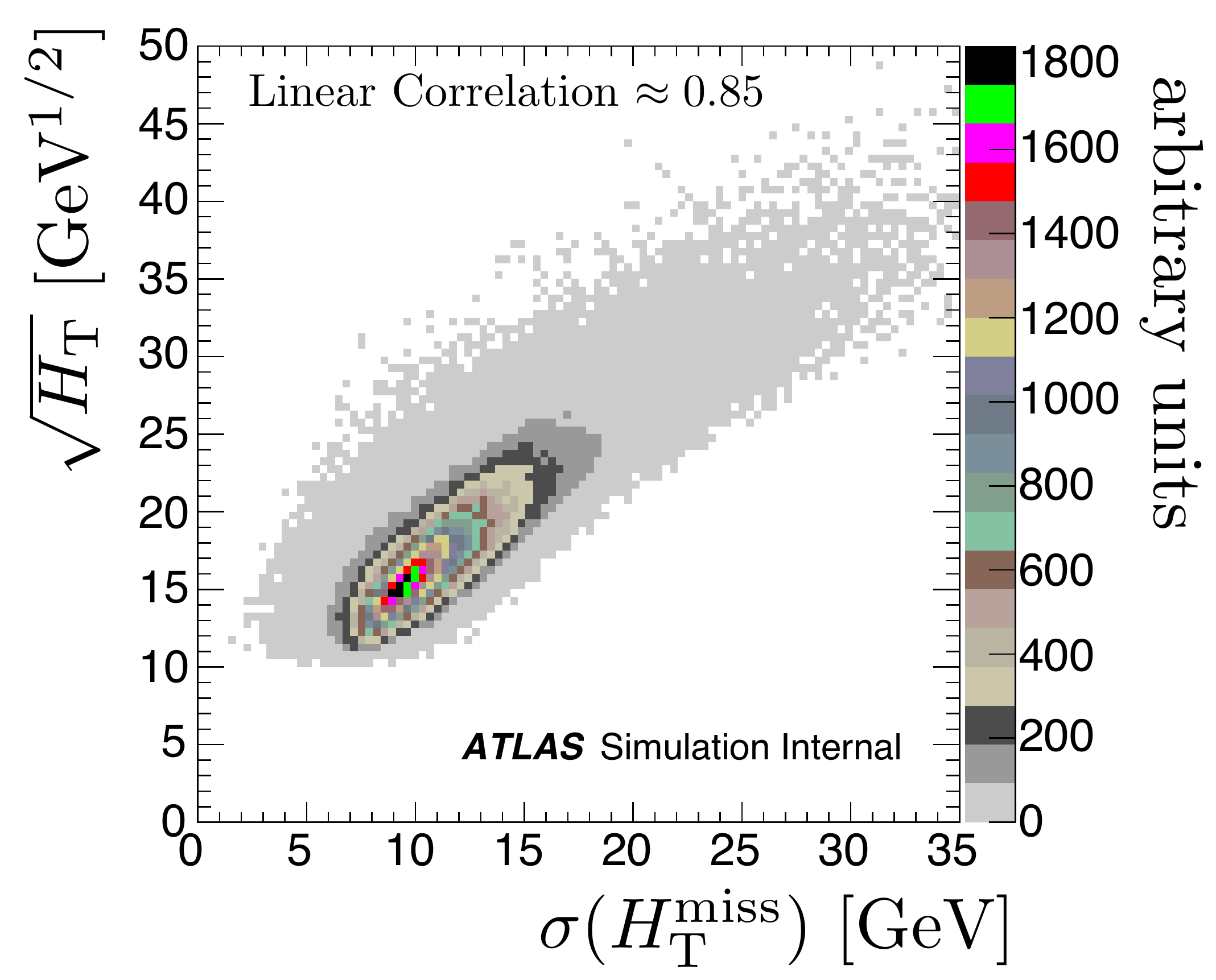}
\end{center}
\caption{The joint distribution of the approximate $E_\text{T}^\text{miss}$ resolution, $\sqrt{H_\text{T}}$ and the $H_\text{T}^\text{miss}$ resolution computed via Eq.~\ref{htmisssig}.}
\label{corhtmisshtsig}
\end{figure}

\clearpage

A quantitative comparison of $H_\text{T,sig}^\text{miss}$ with the traditional $E_\text{T}^\text{miss}/\sqrt{H_\text{T}}$ using the separation power from Eq.~\ref{eq:separation} is shown in Fig.~\ref{comparehtsigetsig}.  The separation is largest for $M=100$ GeV, with a $\sim 15\%$ improvement over $E_\text{T}^\text{miss}/\sqrt{H_\text{T}}$.  One of the disadvantages of $E_\text{T}^\text{miss}/\sqrt{H_\text{T}}$ is that it is strongly correlated with $E_\text{T}^\text{miss}$.  As a result of neglecting the soft energy and due to the scale shift $M$, $H_\text{T,sig}^\text{miss}$ can be less correlated with $E_\text{T}^\text{miss}$.  This intuition is quantified in Fig.~\ref{correlationhtsig} which shows that the linear correlation is smaller for all considered values of $M$.  Table~\ref{tablehtsig} summarizes the information from Fig.~\ref{comparehtsigetsig} and Fig.~\ref{correlationhtsig} and shows that $H_\text{T,sig}^\text{miss}(M=100)$ is strictly better than $E_\text{T}^\text{miss}/\sqrt{H_\text{T}}$ in the important metrics considered here and is therefore chosen as baseline for optimizations studies in later chapters.  Furthermore, Fig.~\ref{htsigmissdata} indicates that this $H_\text{T,sig}^\text{miss}$ (the $M=100$ GeV is henceforth dropped) is well-modeled by the simulation\footnote{In order to reduce data/MC differences in the measured jet resolutions, {\it the same} (simulation) resolution parameterizations are used for data and simulation.}. 

\vspace{5mm}

\begin{figure}[h!]
\begin{center}
\includegraphics[width=0.45\textwidth]{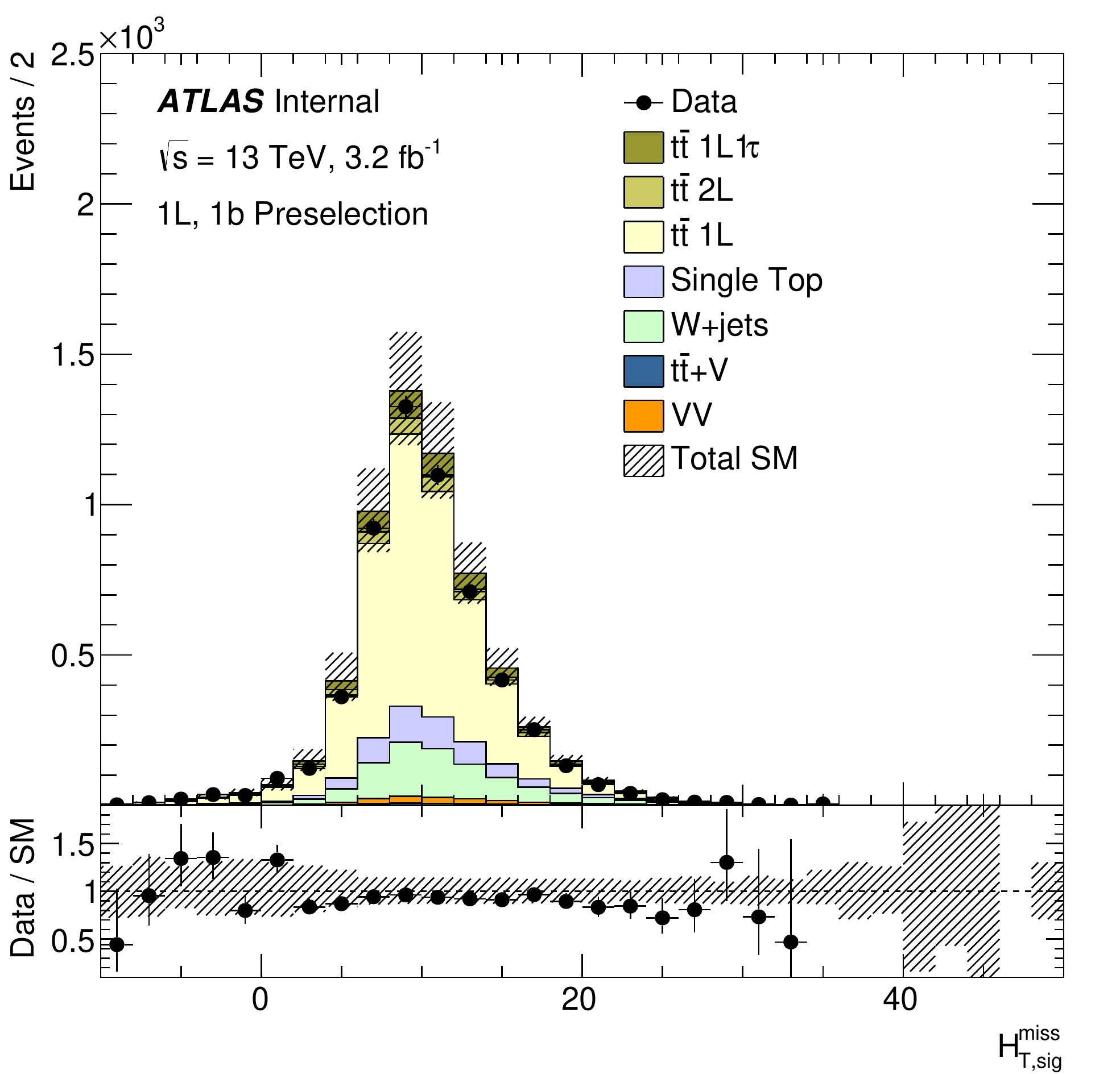}\includegraphics[width=0.45\textwidth]{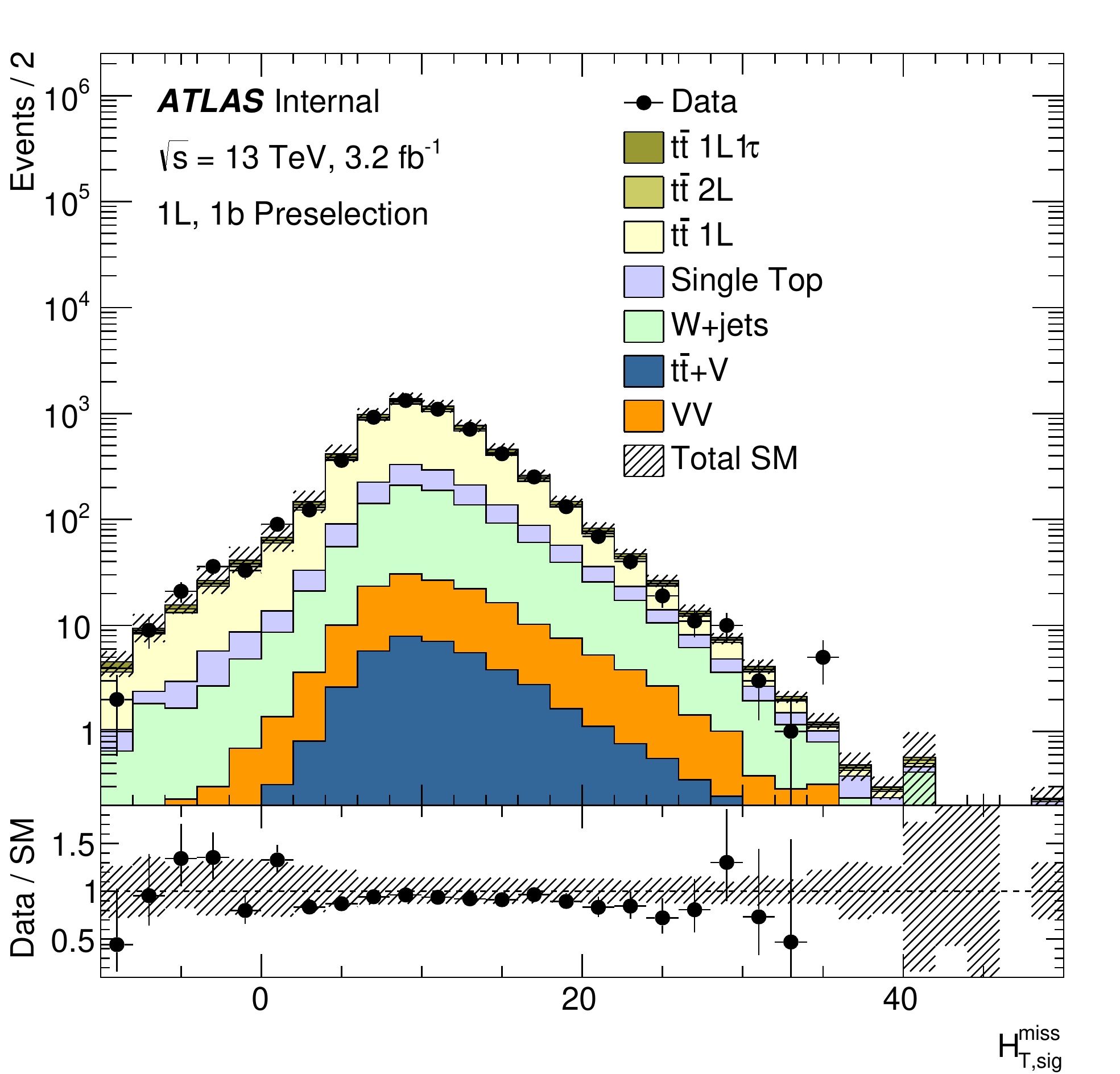}
\end{center}
\caption{A comparison of data and simulation using a loose selection requiring exactly one signal lepton, four jets with $p_\text{T}>25$ GeV and at least one $b$-tagged jet.  The left and right plots differ only in the scaling of the vertical axis. The uncertainty band includes jet energy scale and resolution uncertainties (see Sec.~\ref{chapter:uncertainites}).}
\label{htsigmissdata}
\end{figure}

\begin{figure}[h!]
\begin{center}
\includegraphics[width=0.9\textwidth]{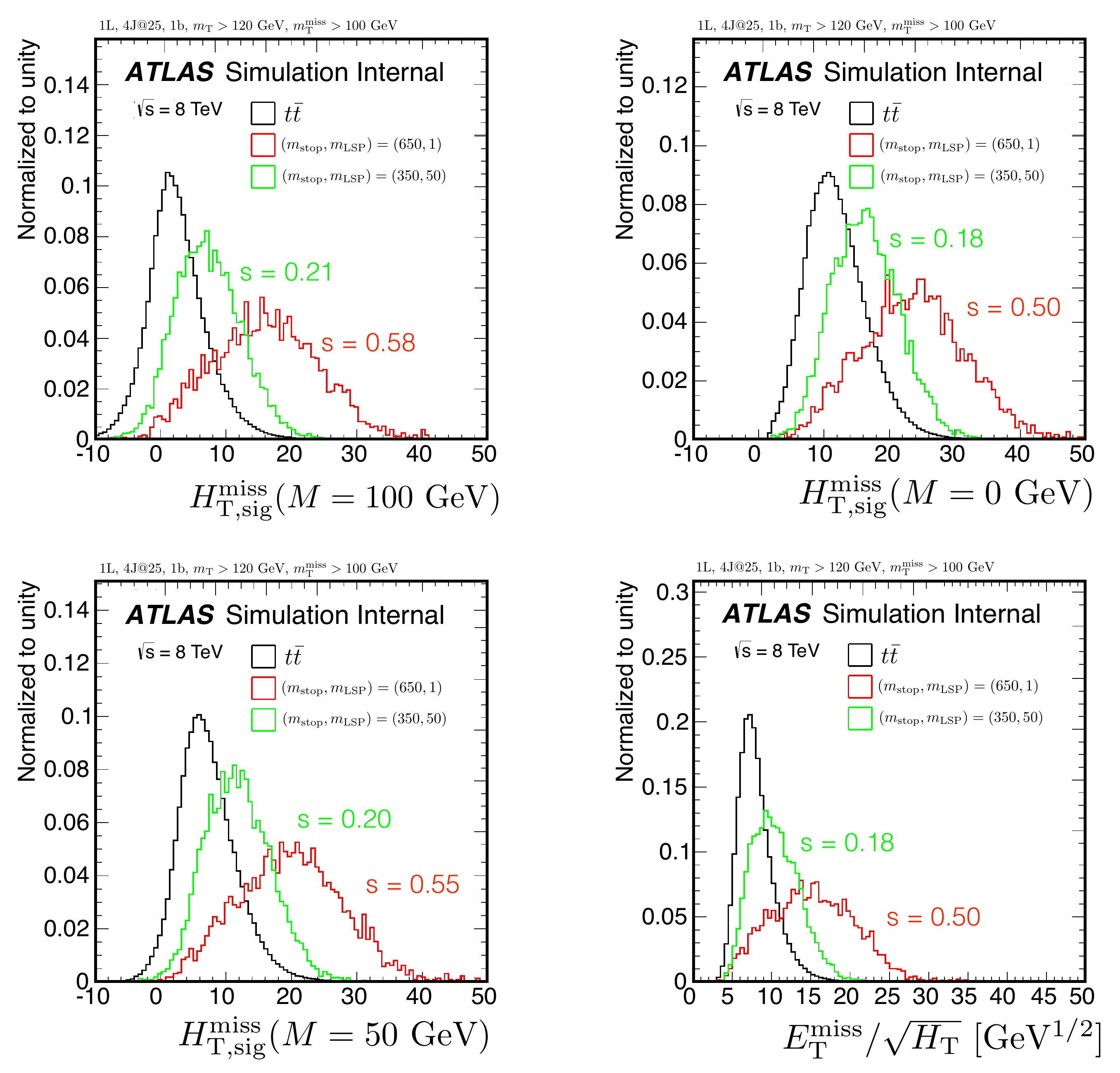}
\end{center}
\caption{The distributions of $H_\text{T,sig}^\text{miss}$ with $M=50$ GeV (bottom left), $M=100$ GeV (top left), and $M=0$ GeV (top right) along with the distribution of $E_\text{T}^\text{miss}/\sqrt{H_\text{T}}$ (bottom right) for $t\bar{t}$ and stop events.}
\label{comparehtsigetsig}
\end{figure}

\begin{figure}[h!]
\begin{center}
\includegraphics[width=0.9\textwidth]{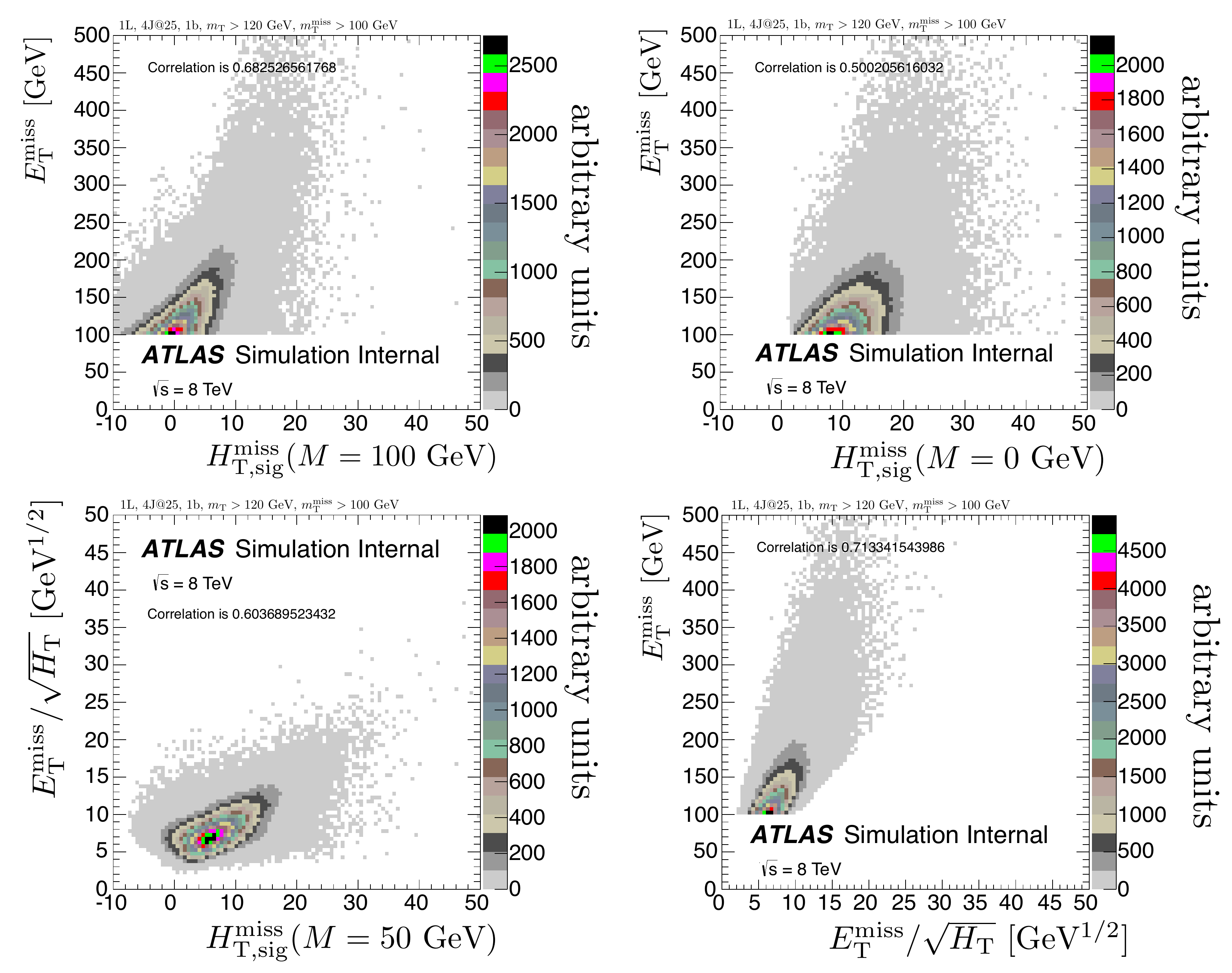}
\end{center}
\caption{The joint distribution of $H_\text{T,sig}^\text{miss}$ or $E_\text{T}^\text{miss}/\sqrt{H_\text{T}}$ with $E_\text{T}^\text{miss}$.  The linear correlation is indicated in each plot.}
\label{correlationhtsig}
\end{figure}

\begin{table}[h!]
\centering
\begin{tabular}{|c|c|c|c|}
\hline
\multirow{2}{*}{Variable} & Separation & Separation & Correlation \\
 & (650,1) & (350,50)& with $E_\text{T}^\text{miss}$ \\
\hline 
\hline

 $E_\text{T}^\text{miss}$&0.59&0.22&1.00\\
  $E_\text{T}^\text{miss}/\sqrt{H_\text{T}}$&0.50&0.18&0.71\\
   $H_\text{T,sig}^\text{miss}(M=100)$&0.58&0.21&0.68\\
    $H_\text{T,sig}^\text{miss}(M=0)$&0.50&0.18&0.50\\

\hline
\end{tabular}
\caption{A summary of the separation power and correlation with $E_\text{T}^\text{miss}$ for $H_\text{T,sig}^\text{miss}$ and $E_\text{T}^\text{miss}/\sqrt{H_\text{T}}$ based on Fig.~\ref{comparehtsigetsig} and Fig.~\ref{correlationhtsig}.}
\label{tablehtsig}
\end{table}

\clearpage

\subsection{Tau veto}
\label{tauid}

Dilepton $t\bar{t}$ events where one of the two leptons is a $\tau$ that decays hadronically is a major background to the search because the extra neutrinos allow events to evade $m_\text{T}$ and $E_\text{T}^\text{miss}$ thresholds and the hadronic activity contributes an extra jet to meet $n_\text{jet}$ requirements.  Section~\ref{sec:objects} introduced the explicit hadronically decaying $\tau$ reconstruction algorithms used at both $\sqrt{s}=8$ and $\sqrt{s}=13$ TeV.  This section explores how to create a powerful $\tau$ veto while maintaining a nearly $100\%$ efficiently for events without a hadronically decaying $\tau$.  Figure~\ref{fig:tau_id_type} shows that most hadronically decaying $\tau$ leptons are reconstructed as a signal jet.  Hadronically decaying $\tau$ leptons with $|\eta|<2.5$ are not reconstructed as a signal jet about $15\%$ of the time due to the $p_\text{T}>25$ GeV threshold.  Even if a $\tau$ lepton has $p_\text{T}>25$ GeV, a significant fraction of its energy can be lost to unmeasured neutrinos.  To begin, the next paragraph describes important properties of the $\sqrt{s}=13$ TeV $\tau$ reconstruction efficiency. 
\begin{figure}[h!]
\centering
\includegraphics[width=0.49\textwidth]{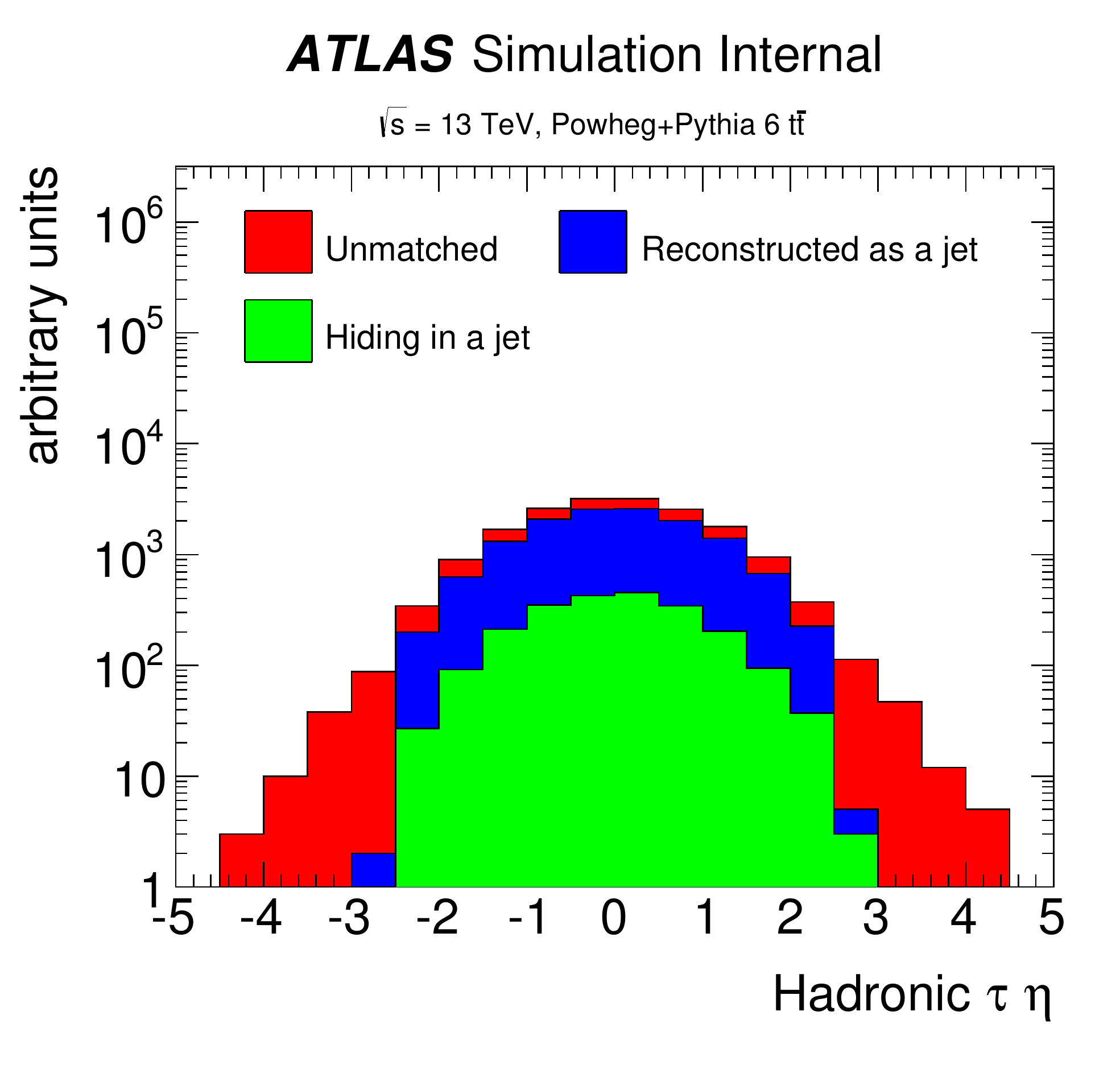}
\includegraphics[width=0.49\textwidth]{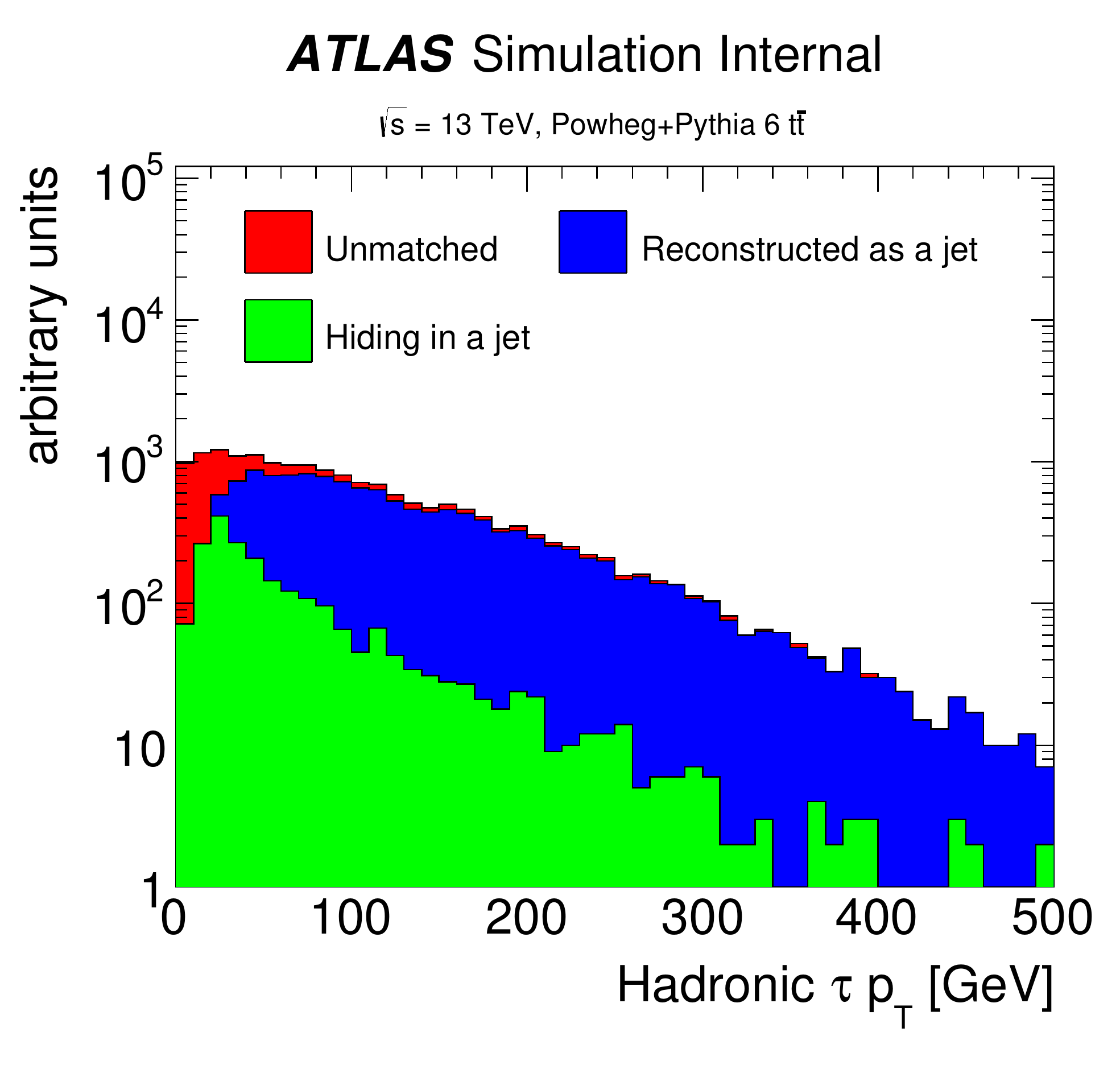}
\caption{The pseudorapidity (left) and $p_\text{T}$ (right) distributions of hadronically decaying $\tau$ leptons from $W$ decays in $t\bar{t}$ events decomposed by how the $\tau$ is reconstructed (if at all).  A $\tau$ lepton in the simulation is {\it matched} to a jet if $\Delta R<0.4$.  If the hadronic decay products of the $\tau$ constitute $\leq 90\%$ of the jet $p_\text{T}$, the $\tau$ is labeled as {\it hiding in a jet}.  Events are required to have exactly one signal lepton, at least four signal jets, at least one $b$-tagged jet, $E_\text{T}^\text{miss}>200$ GeV, $m_\text{T}>150$ GeV, and at least one large-radius jet with $p_\text{T}>150$ GeV (see Sec.~\ref{topmassreco}).}
\label{fig:tau_id_type}
\end{figure}

Three efficiency working points are optimized by combining track and calorimeter information such as the mass of the four-vector sum of tracks in the core of the jet~\cite{ATL-PHYS-PUB-2015-025,ATL-PHYS-PUB-2015-045}.  The points are called {\it loose}, {\it medium}, and {\it tight} and correspond to efficiencies for one- (three-)prong decays of about $60\%$ ($50\%$), $55\%$ ($40\%$), and $45\%$ ($30\%$), respectively.  Figure~\ref{fig:tau_cat3} shows the probability for a hadronically decaying $\tau$ lepton from a $W$ boson decay to be reconstructed and pass additional criteria.  With the same event selection as Fig.~\ref{fig:tau_id_type}, about $77\%$ of $\tau$ leptons are reconstructed as signal jets.  Since the jets used for the dedicated $\tau$ reconstruction have a threshold lower than for signal jets ($20$ GeV versus $25$ GeV), the efficiency to reconstruct a $\tau$ as a `reco $\tau$' is slightly higher than for signal jets by about $5\%$.  The number of tracks inside a jet is a powerful $\tau$ discriminant because it is $p_\text{T}$-independent for $\tau$ jets and increases with $p_\text{T}$ for quark and gluon jets (see Chapter~\ref{cha:multiplicity}).  The third bin of Fig.~\ref{fig:tau_cat3} shows the efficiency for reconstructing $\tau$ leptons as signal jets with less than five tracks.  This simple $\tau$ identification scheme has a similar efficiency to the dedicated $\tau$ reconstruction with exactly one or three tracks.  Tracks for the dedicated algorithm are only chosen from the jet core, $\Delta R<0.2$.  Additionally requiring that the reco tau has opposite electric charge to the signal lepton reduces the efficiency by a few percent and a $p_\text{T}>20$ GeV threshold further lowers the efficiency relatively by about $10\%$.  The overall efficiency of also applying the loose, medium, or tight identification criteria is about $38\%$, $34\%$, or $28\%$, respectively. 

\begin{figure}[h!]
\centering
\includegraphics[width=0.5\textwidth]{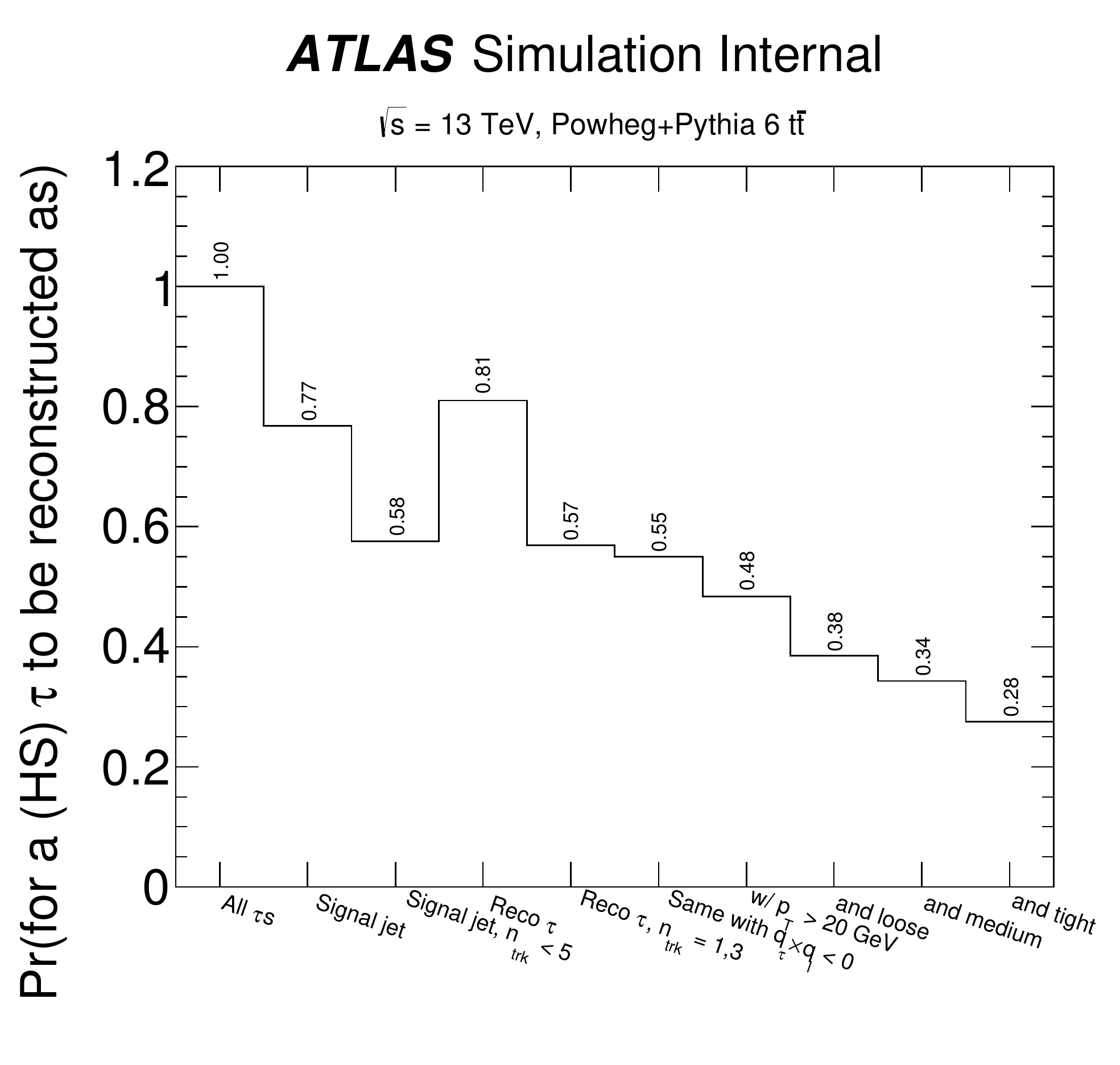}
\caption{The frequency for a hadronically decaying $\tau$ lepton to be reconstructed as one of a variety of objects, described by the labels on the horizontal axis. 
}
\label{fig:tau_cat3}
\end{figure}

The reconstruction efficiency in Fig.~\ref{fig:tau_cat3} does not include an important efficiency from combinatorics.   A reconstructed object may match to a generator-level $\tau$ lepton with high efficiency, but if an event has many such objects, then the ambiguity lowers the efficiency for reconstructing the correct $\tau$ momentum.  The left plot of Fig.~\ref{fig:nontau_cat} shows the probability that various objects are matched to the generator-level hadronically decaying $\tau$ lepton.  The Run 1 scheme that uses the leading non $b$-tagged jet to form $m_\text{T2}^\tau$ (see Sec.~\ref{mt2forstop}) has a low ($20\%$) efficiency for correctly selecting the $\tau$ lepton, while the $n_\text{track}$-based taggers have a much higher ($40$-$50\%$) efficiency.  Additionally, the right plot of Fig.~\ref{fig:nontau_cat} shows that the Run 1 scheme often chooses the wrong object as the leptonic $\tau$, while the dedicated $\tau$ reconstruction algorithms almost never picks the wrong object. In other words, even though the leading non $b$-tagged jet and the leading reco $\tau$ with a tight identification have similar efficiencies, the former is not the $\tau$ about $80\%$ of the time while the latter is not the $\tau$ $<1\%$ of the time.  A similar trend is true for events without a $\tau$ lepton fro a $W$ boson decay.  Figure~\ref{fig:nontau_cat2} shows the probability that a particular object is selected as a hadronic $\tau$ candidate when there is no particle-level $\tau$ lepton in the event.  The Run 1 scheme was chosen so that every event has a $m_\text{T}^\tau$ value; therefore it has a $100\%$ probability of picking an object in Fig.~\ref{fig:nontau_cat2}.  In contrast, loose, medium, or tight reco $\tau$ algorithms only have candidates in $7\%$, $6\%$, $4\%$ of events, respectively.

\begin{figure}[h!]
\centering
\includegraphics[width=0.5\textwidth]{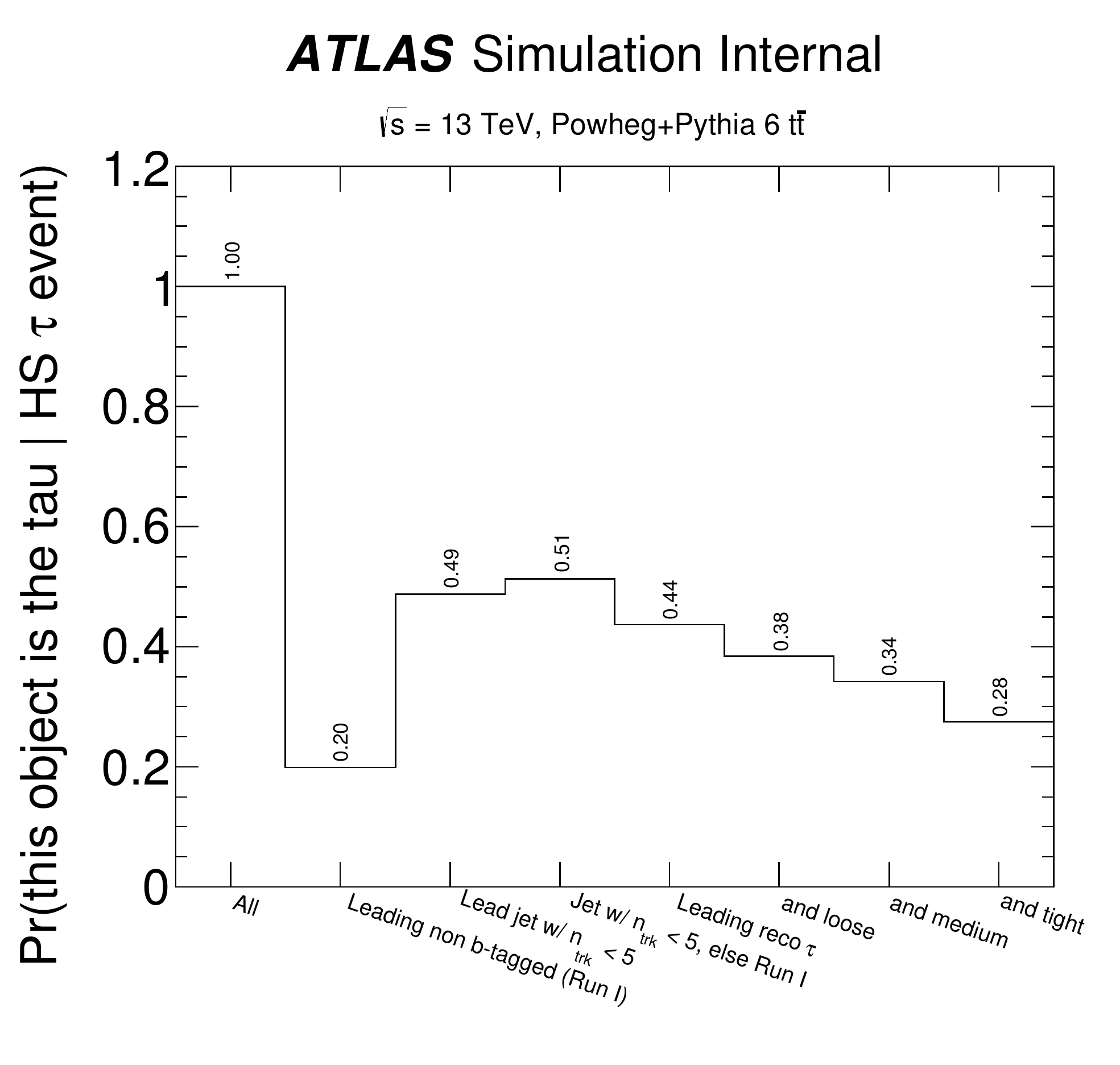}\includegraphics[width=0.5\textwidth]{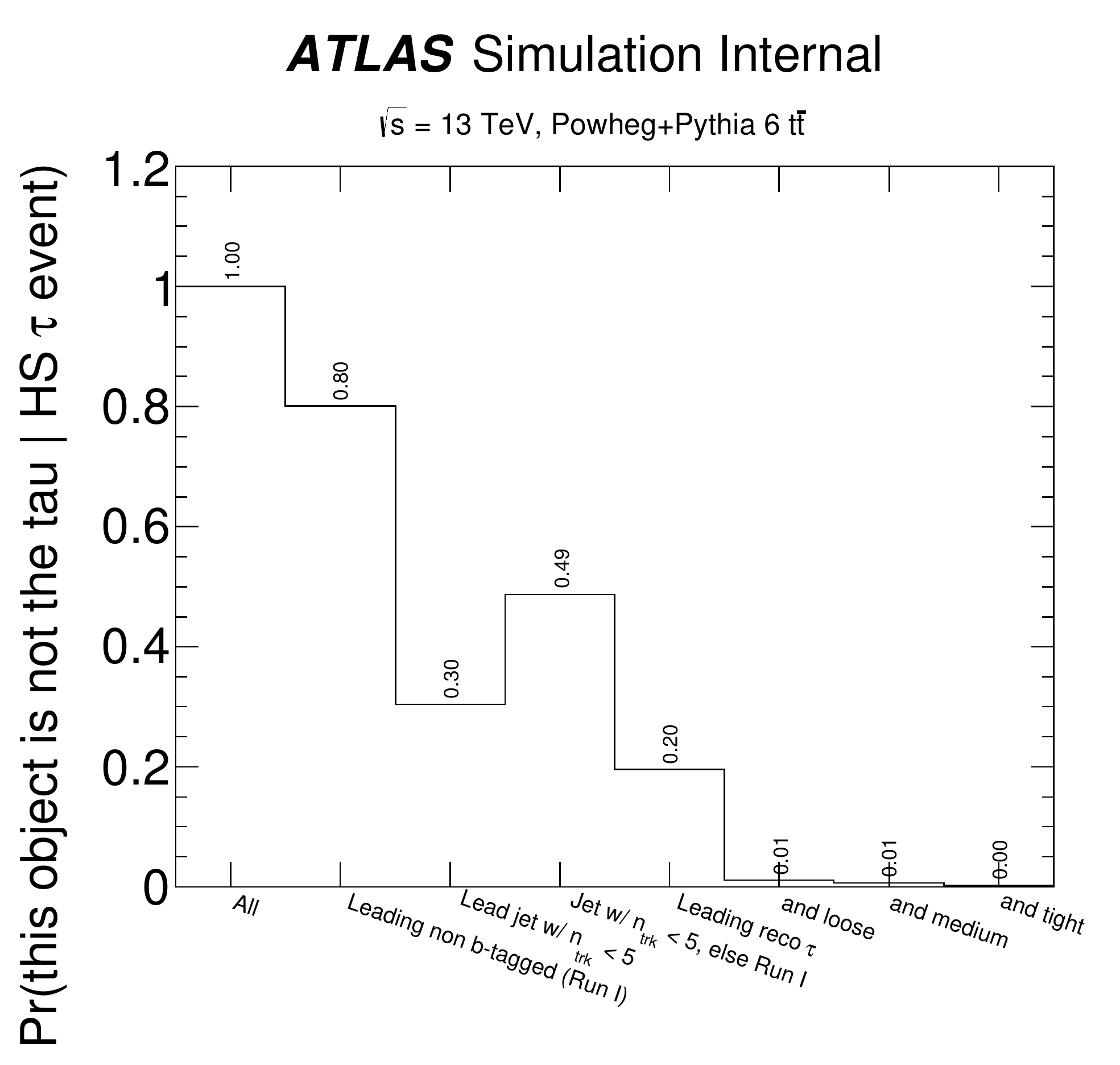}
\caption{Given an event has a hadronically decaying $\tau$, the frequency that a given $\tau$ identification technique selects the $\tau$ correctly (left) or incorrectly (right).
}
\label{fig:nontau_cat}
\end{figure}

\begin{figure}[h!]
\centering
\includegraphics[width=0.5\textwidth]{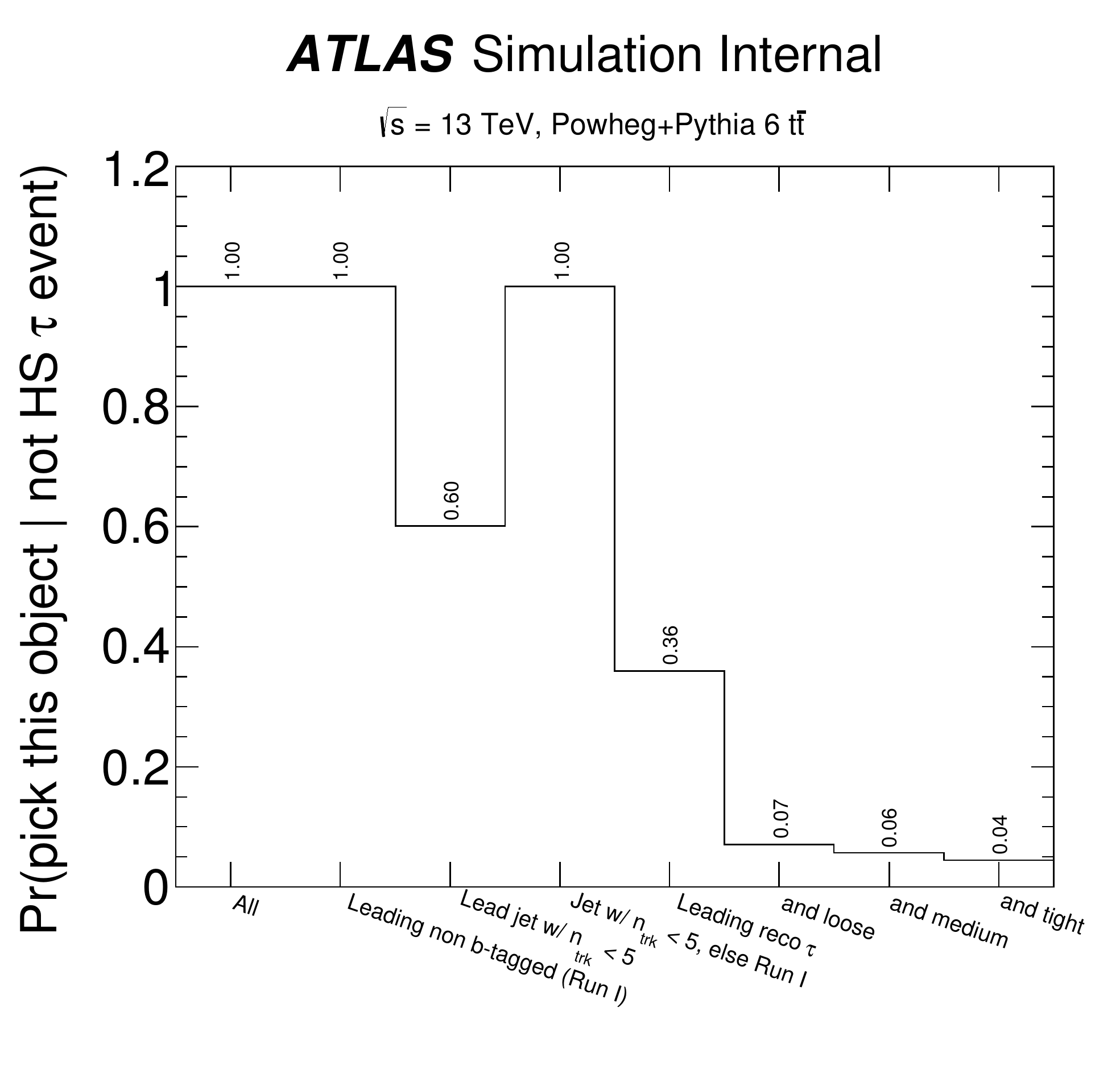}
\caption{Given an event does not have a hadronically decaying $\tau$, the frequency that a given $\tau$ identification algorithm produces a reconstructed $\tau$ candidate. 
}
\label{fig:nontau_cat2}
\end{figure}

The two main points from Fig.~\ref{fig:tau_cat3},~\ref{fig:nontau_cat}, and~\ref{fig:nontau_cat2} are that the `$\tau$' used for $m_\text{T2}^\tau$ does not usually correspond to a particle-level $\tau$ and the highest veto efficiency in stop events (no actual $\tau$) using one of the dedicated reco $\tau$ algorithms is $93\%$-$96\%$ (rightmost bins of Fig.~\ref{fig:nontau_cat2}).  One solution can improve both of these statistics: combining kinematic information from $m_\text{T2}^\tau$ with identification information from the reco $\tau$ algorithms.  A new $m_\text{T2}^\tau(\text{ID})$ variable is formed by using a reco $\tau$ with a particular identification algorithm (ID) as the visible particle for $m_\text{T2}$.  By construction, this variable can only be calculated a small fraction of the time.  Instead of vetoing events if a reco $\tau$ exists, events are only vetoed if $m_\text{T2}^\tau(\text{ID})\geq X$.  When $X=0$, then the veto has the $93\%$-$96\%$ efficiency quotes above for stop events and a $62\%$-$72\%$ efficiency for background $\tau$ events (Fig.~\ref{fig:tau_cat3}).  However, as $X\rightarrow\infty$, the veto is $100\%$ efficient for signal events.  The goal is to optimize $X$ and ID to achieve a $\sim 99\%$ efficiency for signal events and the best possible rejection of $\tau$ events.

Figure~\ref{fig:tau_money} summarizes the efficiencies for all combinations of of $\tau$ identification algorithms and $m_\text{T2}^\tau$.  The $\Delta R$ between the $\tau$ candidate and the leading large-radius jet also provides useful information for rejecting events with a hadronically decaying $\tau$ lepton.  In single lepton $t\bar{t}$ (and stop) events, the (fake) reco $\tau$ is usually within the large-radius jet, while in events with a $\tau$, there can be a large separation between the $\tau$ and the jet\footnote{In particular when leptons are part of the jet clustering, the large-radius including these leptons will tend to be harder than one from hadronically decaying $\tau$ leptons due to the lost energy in neutrinos.}.  For all the combinations in Fig.~\ref{fig:tau_money}, the threshold requirement on $\Delta R$ and $m_\text{T2}^\tau$ are optimized (where possible) so that the signal efficiency is $99\%$.  The best combination is for a loose $\tau$ identification and a threshold requirement on $m_\text{T2}^\tau\gtrsim m_W\approx 80$ GeV.  This combination is used for the $\sqrt{s}=13$ GeV signal region optimization described in Chapter~\ref{chapter:susy:signalregions}.

\begin{figure}[h!]
\centering
\includegraphics[width=\textwidth]{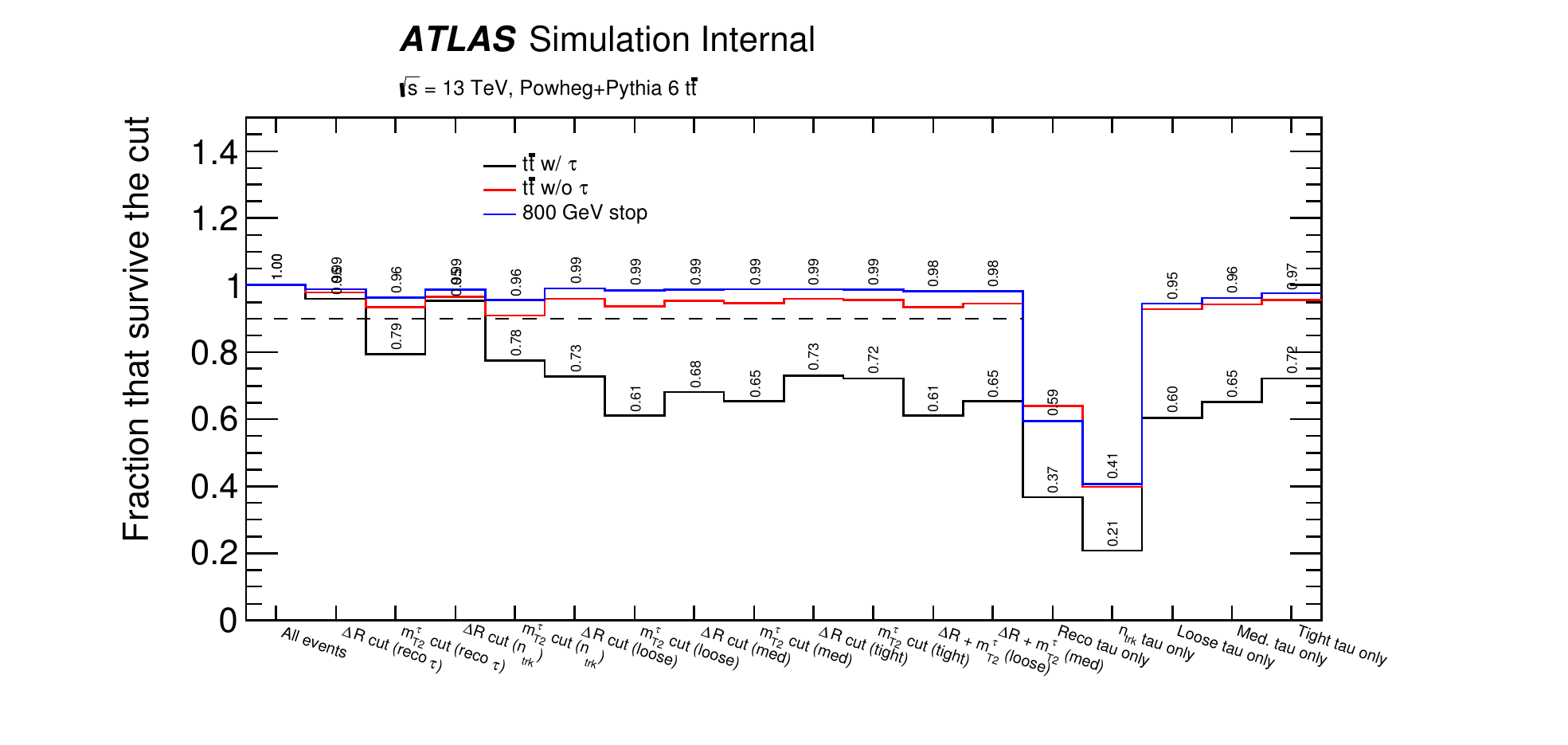}
\caption{A summary figure with all the considered hadronic $\tau$ vetoes based on the above considerations.  The (black) red line shows the efficiency for $t\bar{t}$ events with(out) a hadronically decaying $\tau$ lepton.  The signal $(m_\text{stop},m_\text{LSP})=(800,1)$ GeV is has mostly semi-leptonic $t\bar{t}$.  The $\Delta R$ is between the hadronically decaying $\tau$ candidate and the leading large-radius jet.  In addition to the loose, medium, and tight reco $\tau$ working points, a simple $n_\text{track}=n_\text{trk}<5$ identification scheme is part of the comparison.  The dashed line is at an efficiency of $90\%$.
}
\label{fig:tau_money}
\end{figure}

In addition to the improvement in the signal efficiency of the modified $m_\text{T2}^\tau$ veto, the new variable is significantly less correlated with $m_\text{T}$ (see Table~\ref{tab:mc_tablecomparison}).  The correlation is reduced by over a factor of $10$ for the $t\bar{t}$ background and by a factor of about two in the signal.  Figure~\ref{fig:mt2tau1L1tau} shows that the $m_\text{T2}^\tau$ distribution using a loose reco $\tau$ as one of the visible particles is relatively well-modeled and as expected, $m_\text{T2}^\tau\lesssim m_W$ for the background.  For the signal, $m_\text{T2}^\tau$ often significantly exceeds $m_W$, with only a small peak at $m_\text{T2}^\tau=0$ corresponding to the unbalanced case\footnote{It is stated in Sec.~\ref{sec:numericalmethodsmt2} that the unbalanced case can only occur when $m_{C_1}+m_{V_1}\neq m_{C_2}+m_{V_2}$.  However, when the $m_{V_i}=m_{C_i}=0$, the $m_\text{T}^2$ surfaces allow for the minimum value to be reached even if $E_\text{T}^\text{miss}>0$. This is related to Fig.~\ref{fig:susymtdist_testmass3} and is described in Ref.~\cite{Lester:2011nj} in detail.}.   

\begin{figure}[h!]
\begin{center}
\includegraphics[width=0.5\textwidth]{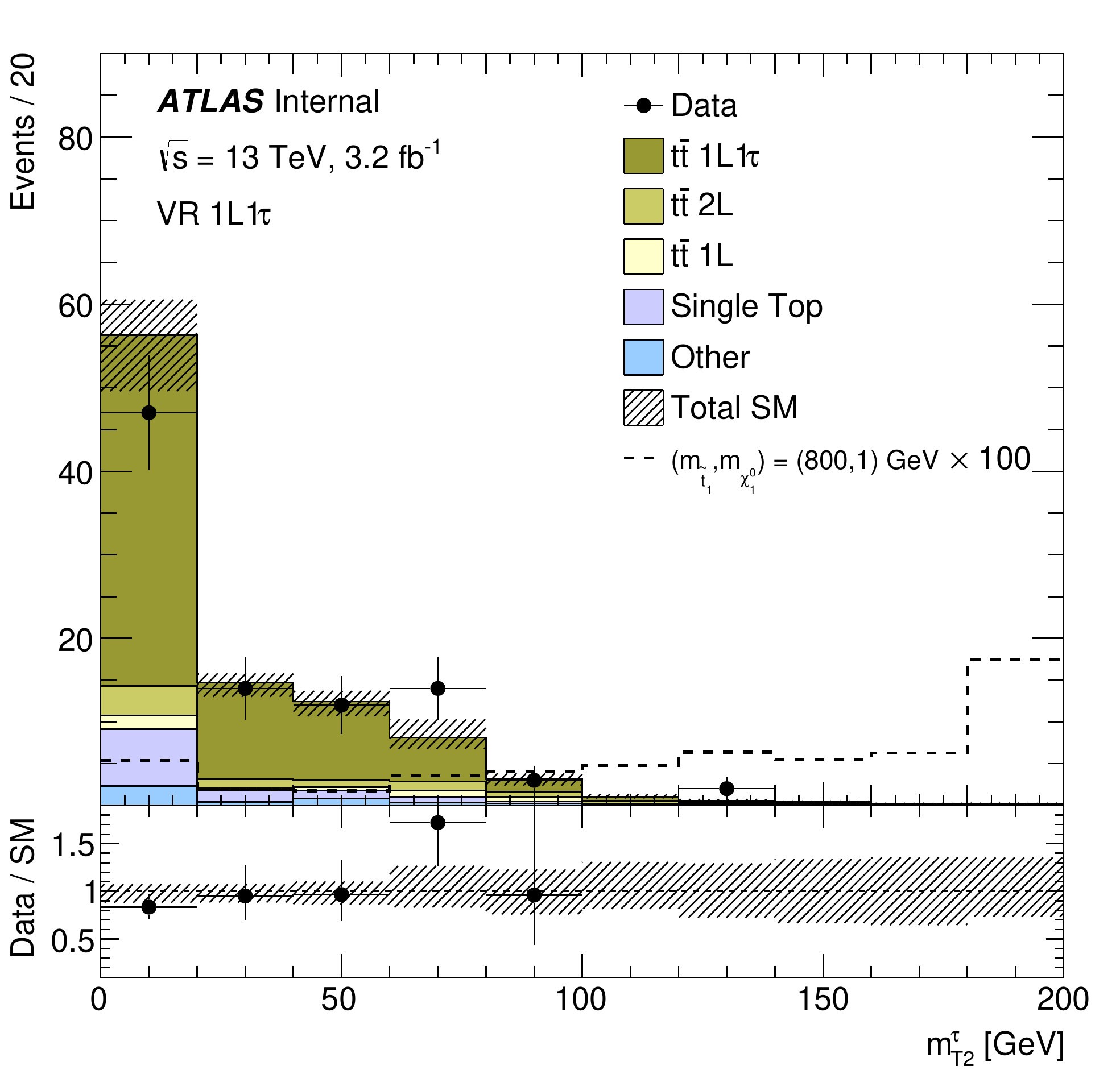}
 \caption{The distribution of $m_\text{T2}^\tau$ with an event selection enriched in dilepton $t\bar{t}$ events with a hadronically decaying $\tau$.  Events are required to have at least one reconstructed reco $\tau$ passing the loose identification.  In addition, events must have at least four jets with $p_\text{T}>80,50,40,25$ GeV, $E_\text{T}^\text{miss}>200$ GeV and at least one $b$-tagged jet.  In order to suppress semi-leptonic $t\bar{t}$ events, $m_\text{T}>100$ GeV.  See Sec.~\ref{sec:1l1tau} for details.}
 \label{fig:mt2tau1L1tau}
  \end{center}
\end{figure}	

\clearpage

\subsection{Hadronic Top Mass Reconstruction}
\label{topmassreco}

Must of the focus in the previous chapters is centered on identifying leptonically decaying top quarks for suppressing the dilepton $t\bar{t}$ background.  Another possibility is to target {\it hadronically} decaying top quarks that are present in the mostly semileptonic $t\bar{t}$ signal, but absent in the dileptonic $t\bar{t}$ background.  Hadronically decaying top quarks produced with a small or moderate boost often result in three daughter jets\footnote{This is an ill-defined notion, especially since the top quark is not colorless.  See Sec.~\ref{sec:colorflow:wcandidateselection} for detail.  In this context, the statement about the number of daughter jets is used heuristically and not quantitatively.}.  The first top-tagging technique for the stop search was introduced in the $\sqrt{s}=7$ GeV analysis~\cite{Aad:2012xqa}.  A relatively unoptimized simple combination of jets tries to capture a hadronically decaying $W$ boson matched with another jet to give the full top quark decay:

\begin{enumerate}
\item Let $j_1$ and $j_2$ be the two jets with $m_{j_1j_2}>60$ GeV closest in $\Delta R$ ($W$ candidate).  If no such jets exist, set $m_\text{had top} = 0$.
\item Take the signal jet $j_3$ with $m_{j_1j_2j_3}>130$ GeV closest in $\Delta R$ to the diejt system ($j_1+j_2$).  If no such jet exist, set $m_\text{had top} = 0$.
\item Define $m_\text{had top}=m_{j_1j_2j_3}$.
\end{enumerate}

\noindent In signal events with a hadronically decaying top quark, it is expected that $m_\text{had top}\sim m_\text{top}$.  The early $\sqrt{s}=8$ TeV signal regions use a loose top-quark tag defined by $0<m_\text{had top}<m_\text{top}+30$ GeV $=205$ GeV.  Two main challenges with this tag are that (a) the top quark and $W$ daughter jets need not be close when the top quark is produced with a small boost and (b) the signal efficiency for even the loose tag  $0<m_\text{had top}<205$ GeV is only $\sim 50\%$-$60\%$ in the signal.  In addition, there is more useful information in the event that can be used to improve the hadronic top quark mass candidate such as $b$-tag weights and jet resolutions.  This additional information could be used to select the jets and dynamically vary the thresholds in the algorithm that should depend on the jet resolutions.  Therefore, for the search based on the entire $\sqrt{s}=8$ TeV dataset, a new hadronic top quark identification technique is developed.  The new variable exists for all events ($100\%$ efficiency for $m_\text{top,$\chi^2$}>0$) and is given by:

\begin{enumerate}
\item Let $b_1,b_2$ be the signal jets with the highest $b$-tagging weight (consistent with the choice used for $m_\text{T2}$ in Sec.~\ref{mt2forstop}).
\item Consider all pairs $j_1,j_2$ of signal jets that are not $b_1,b_2$.
\item Compute $\sigma_{m_{j_1j_2}}^2=m_{j_1j_2}^2(r_1^2+r_2^2)$ and $\sigma_{m_{j_1j_2j_3}}^2=m_{j_1j_2j_3}^2(r_1^2+r_2^2+r_3^3)$, where $r_i$ is the fractional energy uncertainty of jet $i$ (same as for $H_\text{T,sig}^\text{miss}$ in Sec.~\ref{htsigmiss}).
\item Select $j_1,j_2$ and $i$ to minimize the following:
\begin{align}
\label{topmasschi2}
\chi^2 = \frac{(m_{j_1j_2j_3}-m_\text{top})^2}{\sigma^2_{m_{j_1j_2j_3}}}+\frac{(m_{j_1j_2}m_W)^2}{\sigma^2_{m_{j_1j_2}}}.
\end{align}
\end{enumerate}

\noindent Figure~\ref{fig:susy:mhadtop} compares the `simple' and $\chi^2$-based approaches for the dileptonic $t\bar{t}$ background and a stop model with  $(m_\text{stop},m_\text{LSP})=(600,250)$ GeV.  The top quark mass peak in the signal is sharper for the signal, but due to Eq.~\ref{topmasschi2}, the background also has a peak around $m_\text{top}$.  The separation power (Eq.~\ref{eq:separation}) is about $0.035$ for the simple definition and $0.05$ for $\chi^2$-definition.  In addition, the signal ($t\bar{t}$) efficiency for a $m<200$ GeV threshold increases from about $60\%$ ($47\%$) for the simple definition to about $69\%$ ($54\%$) for the $\chi^2$-definition.

When $m_\text{stop}\gtrsim 700$ GeV so that $p_\text{T}^t\gtrsim 350$ GeV, the top quark hadronic decay products are sufficiently collimated that a single large-radius jet can capture most of the energy.  For the stop search at $\sqrt{s}=13$ GeV that targets such high mass stops, re-clustered trimmed jets are used to form a hadronic top quark mass from $m_\text{jet}$.  Chapter~\ref{cha:bosonjets} describes boosted top quarks, large-radius jets, and re-clustering in detail.  One of the benefits of re-clustering is that the jet algorithm parameters can be easily optimized for each signal region.  Section~\ref{chapter:susy:signalregions} describes the parameter optimization in the case of the stop search.  Large-radius jets are formed with signal small-radius jet inputs after the overlap removal.  Figure~\ref{fig:susy:reclusteredjetmass} shows the modeling of the jet mass distribution in a selection enriched in semi-leptonic $t\bar{t}$ events.  There are peaks at the $W$ and top quark masses and most of the large-radius jets near $m_W$ have two constituent small-radius jets while most of the jets near $m_\text{top}$ have at least three small-radius jet constituents.  

\begin{figure}[h!]
\centering
\includegraphics[width=0.45\textwidth]{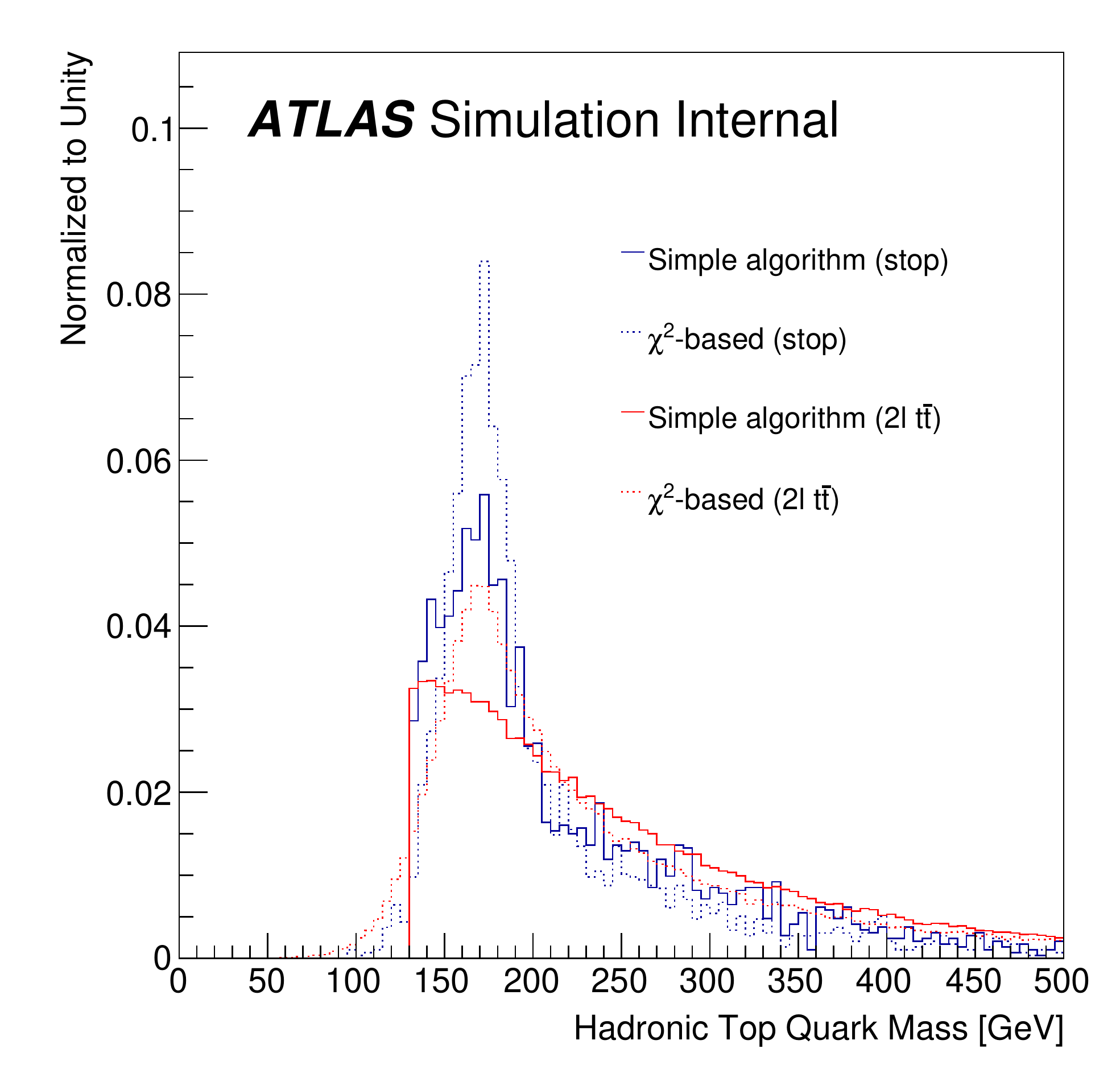}
\caption{A comparison of the two resolved hadronic top quark mass tagging techniques described in the text.  The stop model is $(m_\text{stop},m_\text{LSP})=(600,250)$ GeV.
}
\label{fig:susy:mhadtop}
\end{figure}

\begin{figure}[h!]
\centering
\includegraphics[width=0.45\textwidth]{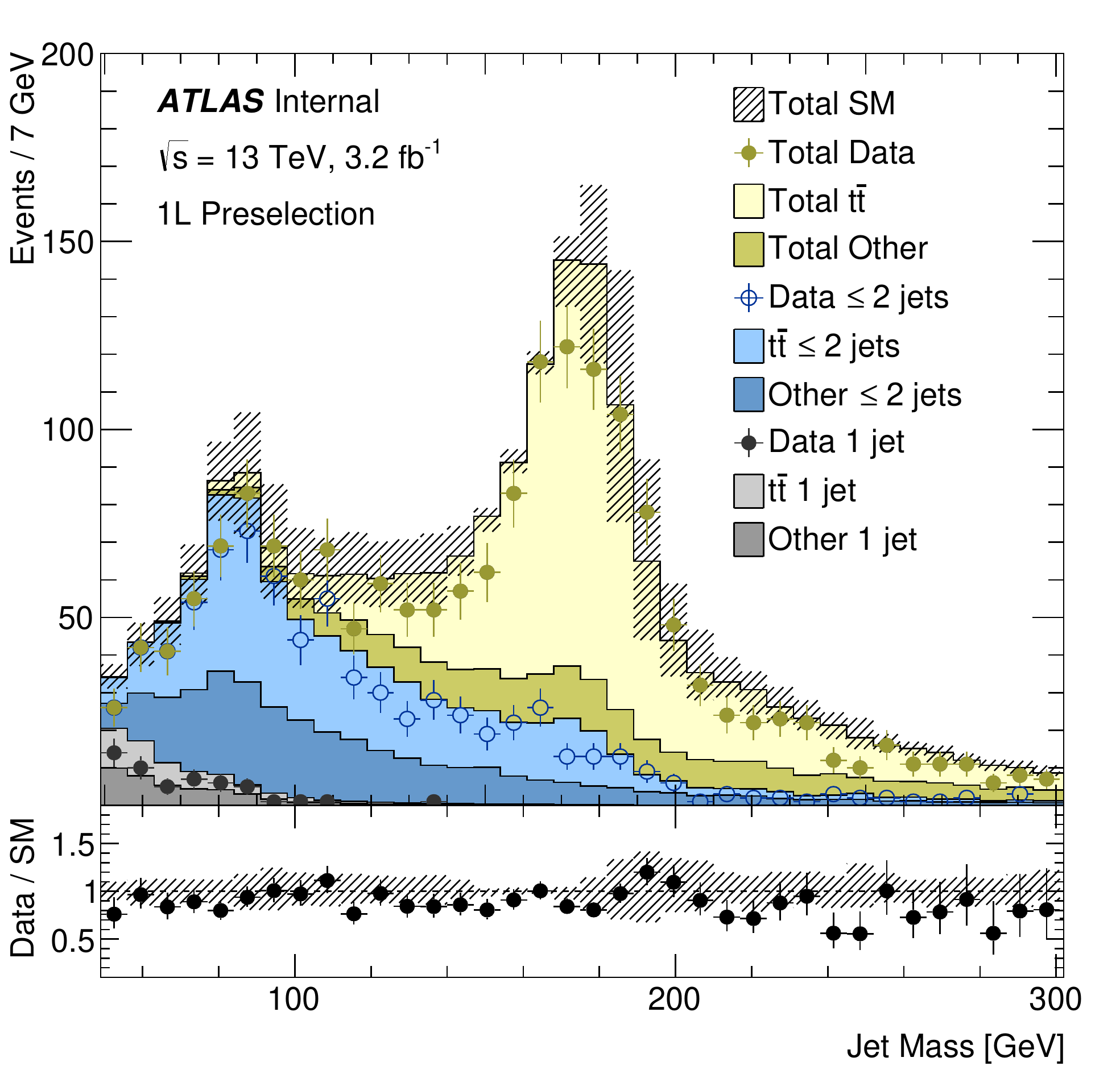}
\caption{The large-radius jet mass in a one-lepton $t\bar{t}$ selection requiring $\Delta\phi(\text{jet}_i,\vec{p}_\text{T}^\text{miss})>0.4$, at least four jets with $p_\text{T}>80,50,40,40$ GeV, $E_\text{T}^\text{miss}>200$ GeV, $m_\text{T}>30$ GeV, at least one $b$-jet, at least one large-radius jet with $R=1.2$ with $p_\text{T}>200$ GeV and $m_\text{jet}>50$ GeV.  Leptons are not included in the re-clustering and small-radius jets are dropped (trimmed) if they have $p_\text{T}<5\%$ of the jet $p_\text{T}$.  The jet mass distribution is decomposed by constituent multiplicity.
}
\label{fig:susy:reclusteredjetmass}
\end{figure}
 	\chapter{Signal Regions}
	\label{chapter:susy:signalregions}
	
	A series of increasingly complex event selections starting with the early $\sqrt{s}=8$ TeV data and covering the beginning of the $\sqrt{s}=13$ TeV data in Run 2 are sensitive to a broad region of the light stop parameter space.  Figure~\ref{fig:susy:exclusion:overview} presents an overview of the expected limits in the $m_\text{stop},m_\text{LSP}$ mass plane.  This chapter covers the optimization of seven signal regions from three datasets.  With $13$ fb${}^{-1}$ of $\sqrt{s}=8$ data, SR1, SR2, and SR3 extend the $\sqrt{s}=7$ TeV limits at intermediate masses, to higher neutralino masses, and to higher stop masses respectively.  With the full Run 1 $\sqrt{s}=8$ TeV dataset (20.3 fb${}^{-1}$) three additional signal regions further extend the limits in all three directions.  The tN\_diag signal region pushes the limit at low stop mass toward the challenging {\it diagonal} in the $m_\text{stop},m_\text{LSP}$ mass plane where $m_\text{stop}\approx m_\text{LSP}+m_\text{top}$.  Kinematically tighter regions tN\_med an tN\_high are analogues to SR2 and SR3 and extend the sensitivity at high stop and neutralino masses.  Even though the dataset at $\sqrt{s}=13$ TeV is significantly smaller than at $\sqrt{s}=8$ TeV (only 3.2 fb${}^{-1}$), the significant increase in the stop cross section coupled with new techniques allows the early Run 2 dataset to further expand the sensitivity to nearly $m_\text{stop}=800$ GeV.

	Each signal region is individually optimized starting from a loose event selection (Sec.~\ref{sec:preselection}) using procedures described in~\ref{sec:optimizationprocedure}.  There are two classes of signal regions: single-bin and multi-bin regions.  The single-bin regions are documented in Sec.~~\ref{sec:onebinregionoptimization} and the tN\_diag multi-bin region optimization and final event selection is described in Sec.~\ref{sec:shapefitregion}.

\begin{figure}[h!]
\begin{center}
\includegraphics[width=0.99\textwidth]{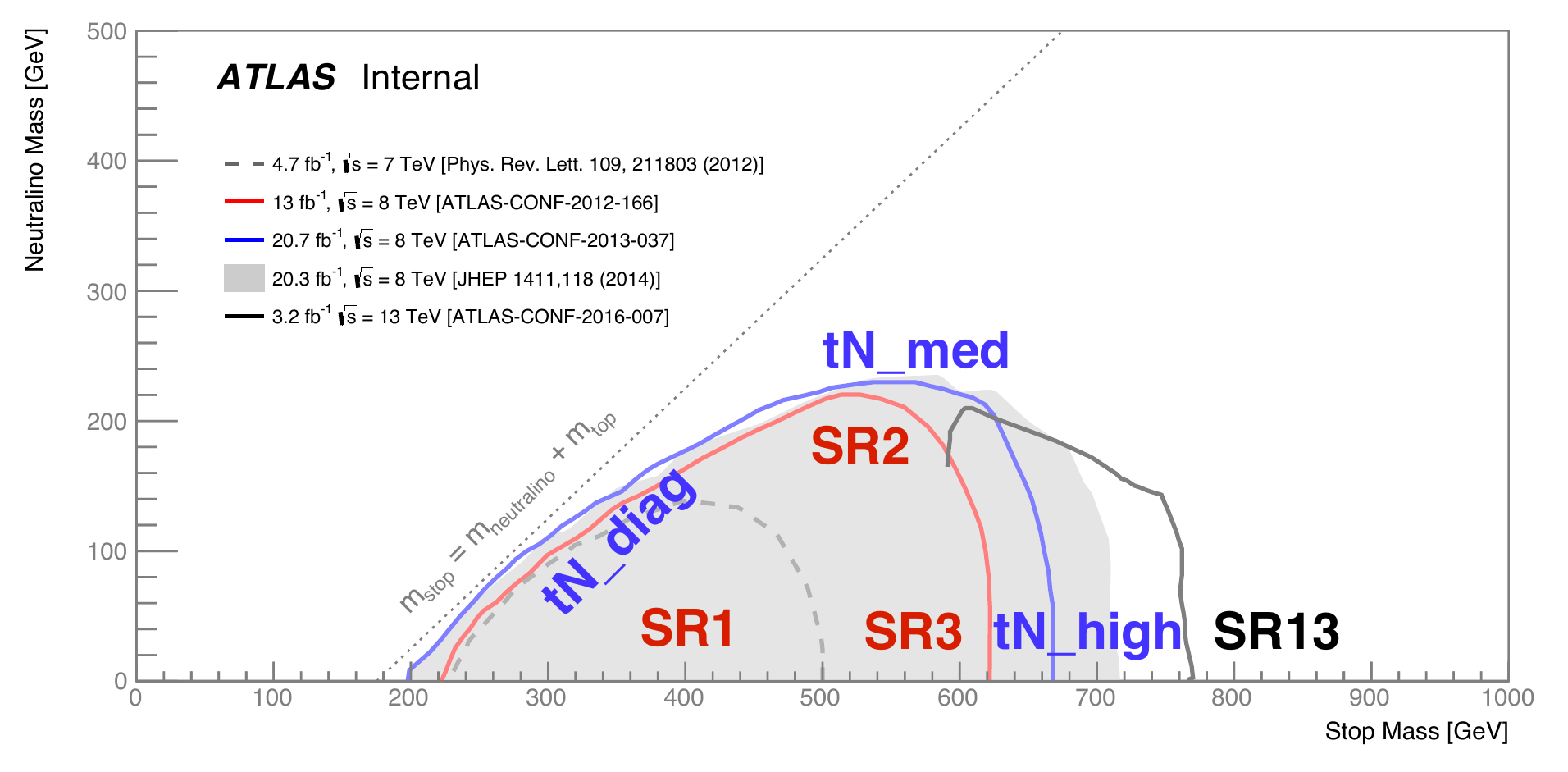}
 \caption{An overview of the various signal regions described in Chapter~\ref{chapter:susy:signalregions}.  The horizontal axis is the stop mass and the vertical axis is the neutralino mass; each point in this plane corresponds to a simplified stop model.  Model cross-sections are set by the stop mass and decrease from left to right.  The dashed line corresponds to the kinematic boundary above which stop decay to an on-shell top quark is forbidden.  The various lines and shaded region are the expected exclusion limits using the statistical procedures documented in Sec.~\ref{sec:susy:stats}.  Seven signal regions are optimized to ensure a broad sensitivity across the plane.  Names of the signal regions are placed in the locations of parameter space where they add the most unique sensitivity. }
 \label{fig:susy:exclusion:overview}
  \end{center}
\end{figure}		
	
		\clearpage
	
		\section{Preselection}
		\label{sec:preselection}
		
		The starting point for the signal region optimization is a loose event selection ({\it preselection}) with many of the irrelevant backgrounds already suppressed.  This preselection includes the trigger and isolated lepton requirements in addition to the second lepton veto.  Events are further required to have at least four signal jets, at least one $b$-tagged jet, $E_\text{T}^\text{miss}>100$ GeV, and $m_\text{T}>30$ GeV.   After the preselection, the $Z$+jets and QCD multijets backgrounds are negligible (see Chapter~\ref{chapter:background}) and $t\bar{t}$ events dominate.   Additional intermediate preselections are used to reduce the gap in phase space to the potential signal regions.  For example, at $\sqrt{s}=13$ TeV, the preselection used for the SR13 optimization additionally required $E_\text{T}^\text{miss}>150$ GeV, $m_\text{T}>100$ GeV, and $|\Delta\phi(\text{jet}_i,\vec{p}_\text{T}^\text{miss})|>0.4$ for $i=1,2$ (to suppress mis-measured $E_\text{T}^\text{miss}$).  Figure~\ref{fig:SR13pt1} shows the jet $p_\text{T}$ spectra after this preselection.  The pair production of top quarks is the dominant process and due to the relatively high $E_\text{T}^\text{miss}$ requirement, the leading jets have a hard $p_\text{T}$ spectrum.  Additional distributions with the SR13 preselection appear in Sec.~\ref{sec:onebinregionoptimization}.

		All signal region optimizations are performed prior to observing the data in signal-like regions of phase space.  In order to avoid looking at data with signal sensitive event selections while still monitoring the data in looser event selections, the optimization is performed {\it blinded}.  Prior to the finalization of the SR definitions, all data (and simulation) passing the preselection, $E_\text{T}^\text{miss}>200$ GeV, and $m_\text{T}>140$ ($150$) GeV at $\sqrt{s}=8$ ($13$) TeV are removed from monitoring plots.  Blinding does not effect simulation-only studies such as the optimization described in Sec.~\ref{sec:optimizationprocedure}.  None of the plots shown in subsequent sections have the blinding applied as all SR are now fixed.

\begin{figure}[h!]
  \centering
\includegraphics[width=0.5\textwidth]{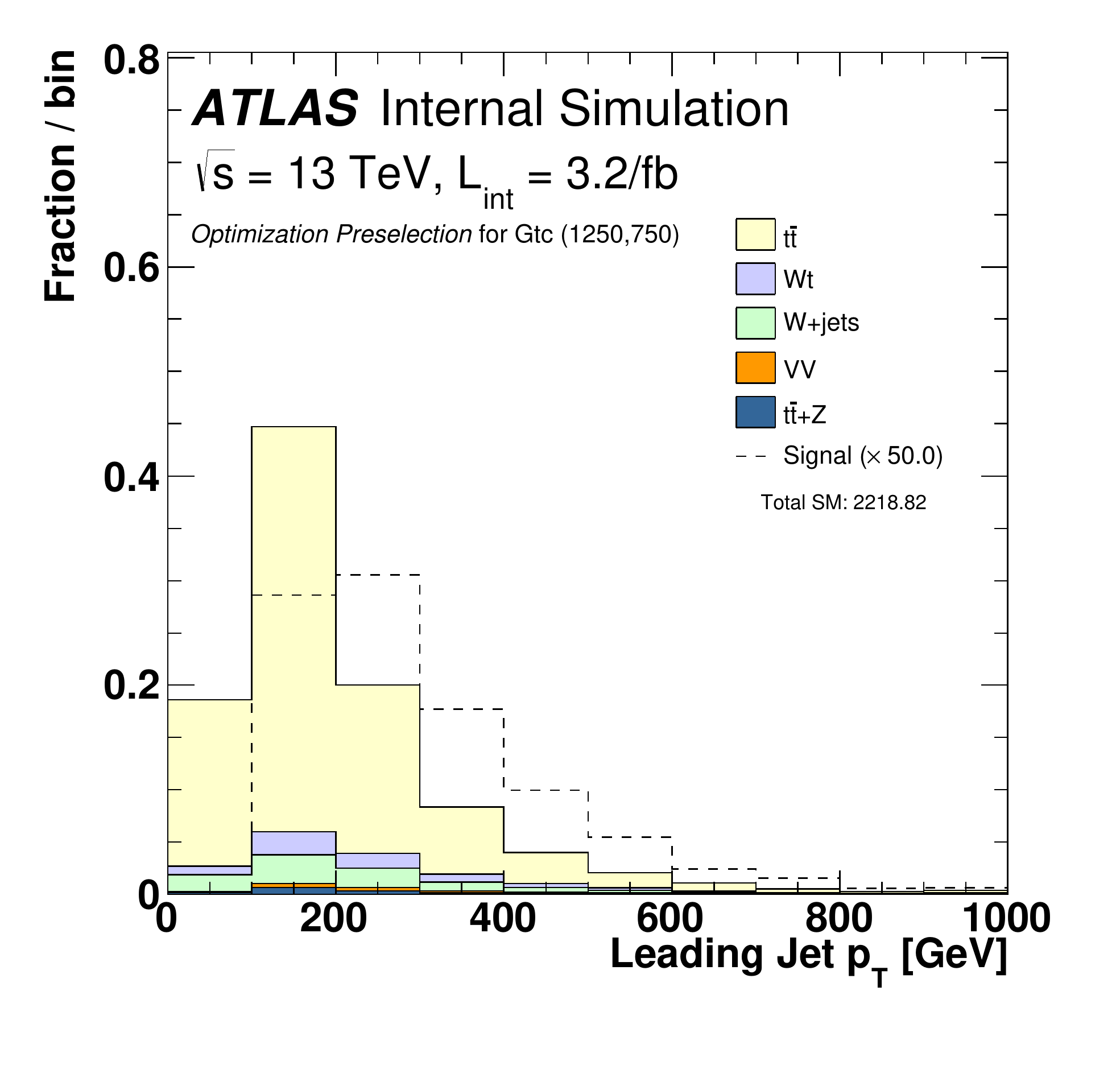}\includegraphics[width=0.5\textwidth]{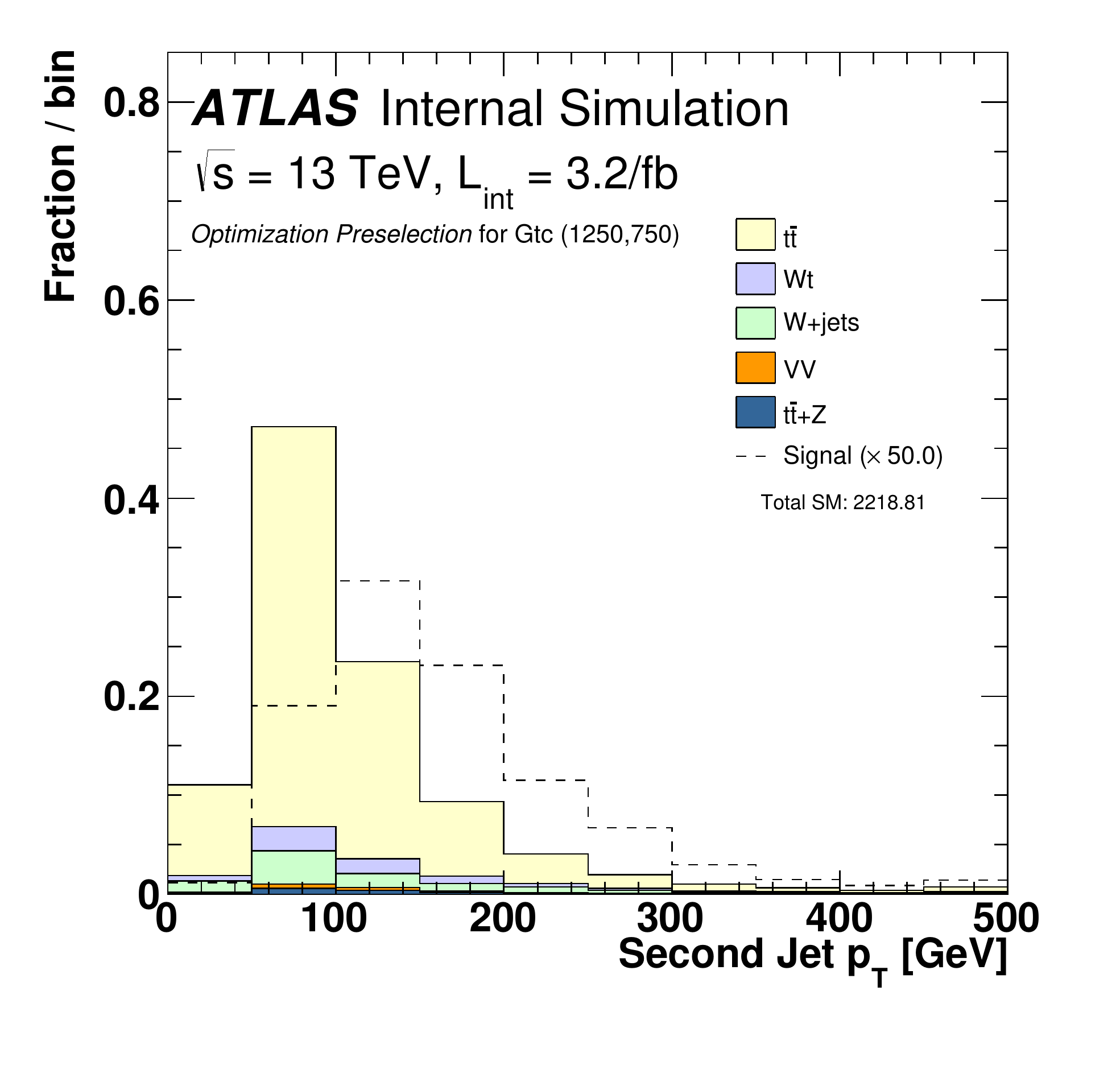}\\
\includegraphics[width=0.5\textwidth]{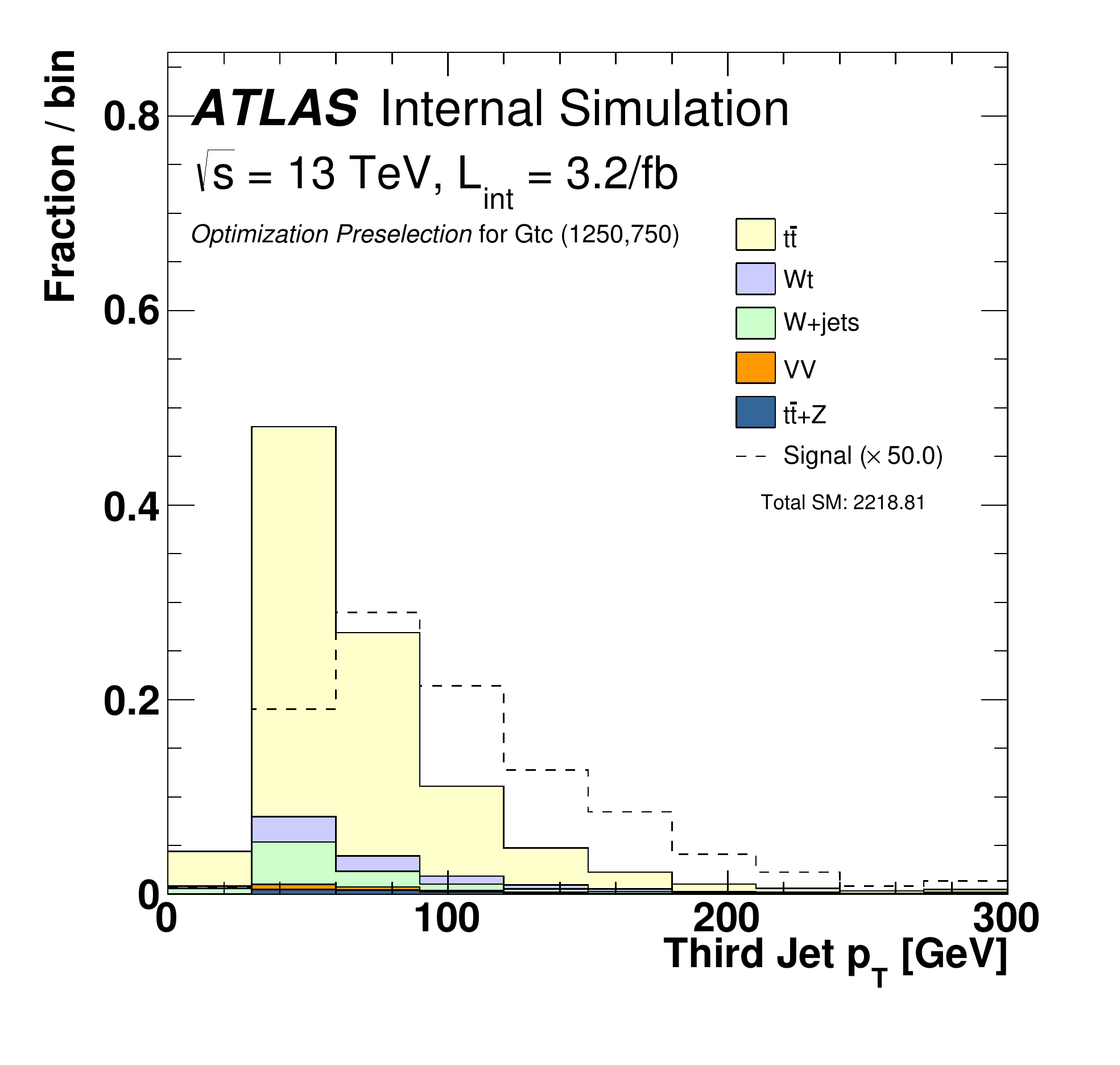}\includegraphics[width=0.5\textwidth]{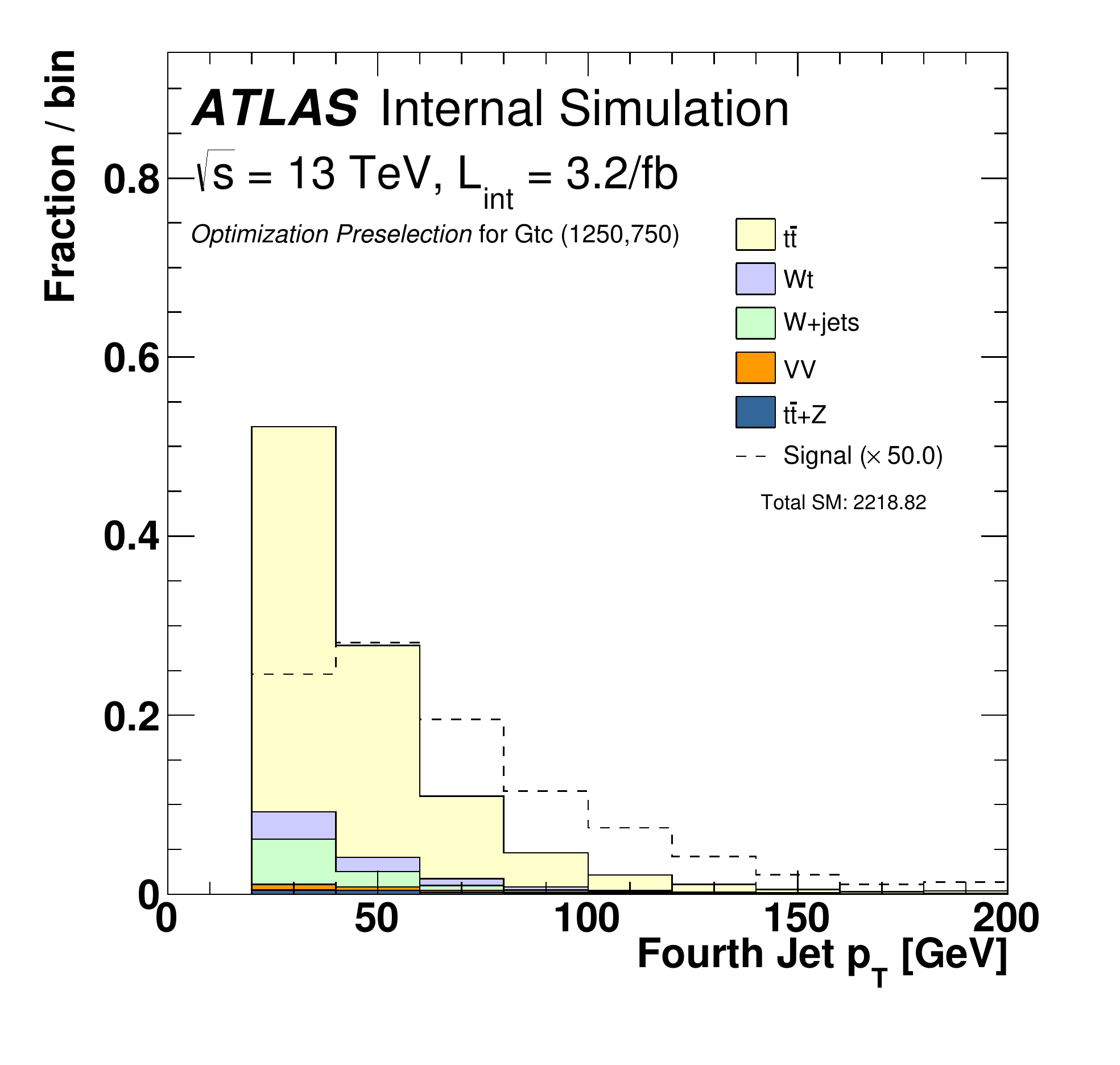}
  \caption{The distribution of the leading (top left), second leading (top right), third leading (bottom left) and fourth leading (bottom right) jet $p_\text{T}$ spectra after the preselection at $\sqrt{s}=13$ TeV.  All plots are normalized to unity.  Note that the horizontal axes have been chosen per distribution so that the distributions fill out the entire plot.}
  \label{fig:SR13pt1}
\end{figure}		
		
		\clearpage
		
		\section{Optimization Procedure}
		\label{sec:optimizationprocedure}
		
		In principle, the best event selection using only threshold requirements on a series of $N$ variables is obtained by scanning the entire $\sim \mathbb{R}^N$ space and computing the test statistic for each point, taking the selection that optimizes this statistic.  This is never possible in practice for the following reasons:
		
		\begin{description}
		\item[Number of combinations] If each dimension is discretized into $\sim10$ intervals and there are $\sim 10$ variables, then the number of combinations is $\text{\#intervals}^\text{\#variables}\sim 10^{10}$.  Ten intervals is already coarse; if instead there are $\sim 50$ intervals, then this number is $\sim 10^{16}$.  Large regions of parameter space can be eliminated based on simple criteria (e.g. no signal events remain), but this is still an unfeasible number of combinations to check.
		\item[Test Statistic Evaluation] The full test statistic (described in Sec.~\ref{sec:susy:stats}) for a given selection takes $\mathcal{O}(10 \text{ seconds})$ to compute.  With $10^{10}$ configurations and $1000$ full time batch nodes, the brute force optimization would require $\sim 3$ years.
		\end{description}
		
		\noindent For optimizing SR1, SR2, and SR3, the two challenges above where addressed by (1) only considering a small number of combinations and (2) using a simplified version of the test statistic.   About $10,000$ total combinations of threshold requirements on $am_\text{T2},m_\text{T2}^\tau,E_\text{T}^\text{miss},m_\text{T},E_\text{T}^\text{miss}/\sqrt{H_\text{T}},p_\text{T}^\text{jet 1}$, hadronic top mass, and the isolated track veto where studied using a brute force approach.  The simplified metric for (2) is $s/\sqrt{b+(0.25\times b)^2}$, for $s$ signal events and $b$ background events.  This formula is a comparison of $s$ signal events to the background uncertainty that is the sum in quadrature of a $\sqrt{b}$ Poisson uncertainty with a $25\%$ background systematic uncertainty.   If all yields could be treated as Gaussian, then this {\it significance} value would be the $Z$-score with $Z>2$ corresponding to a $2\sigma$ sensitivity.  This simple optimization procedure was able to quickly converge on signal regions that could extend beyond the $\sqrt{s}=7$ TeV performance, even after adjusting for the increase in luminosity.   For example, at $(m_\text{stop},m_\text{LSP})=(500,0)$, the significance was increased by a factor of $\sim 3$.  Because the procedure was so simple, each point in the coarse $(m_\text{stop},m_\text{LSP})$ plane was separately optimized.  Small ad-hoc adjustments of the optimized thresholds resulted in three distinct signal regions with broad sensitivity across the parameter space.
		
		A more sophisticated approach was used for the analysis of the full $\sqrt{s}=8$ TeV dataset which is a closer approximation to the optimal configuration described above.  For a given benchmark model (one for each target signal region), a set of selections is chosen that minimizes the background composition for a fixed number of predicted signal events.   This is repeated for a scan in the number of signal events between $5$ and $10$.  For each signal efficiency, the final test statistic is computed and the point with the best value is selected.  If the minimization step works successfully, then the selected signal region will be {\it globally optimal} because all (reasonable) test statistics will be improving\footnote{For a $p$-value, this means decreasing and for a significance, this means increasing.  At $\sqrt{s}=8$ TeV, the metric is the $\text{CL}_s$ value (see Sec.~\ref{sec:susy:stats}) and at $\sqrt{s}=13$ TeV, it is the `discovery significance' described in the caption of Table~\ref{sr2selection_b}.  With no evidence for SUSY, the final result is the exclusion limit based on the $\text{CL}_s$ in all cases.} for decreasing background yield for a fixed signal efficiency.  The minimization is performed using the Nelder-Mead simplex method~\cite{Nelder01011965} as implemented in the {\sc Minuit}~\cite{James:1975dr} package.  A simplex-based routine is used instead of the more standard Davidon-Fletcher-Powell switching method~\cite{doi:10.1137/0801001,Fletcher01081963,Fletcher01011970} (called {\sc Migrad} in {\sc Minuit}) because the number of simulated events passing a selection is discrete so methods based on derivatives do not perform well.  The objective function is:
		
		\begin{align}
		f(b,s,S)=\left\{\begin{matrix}-s/b & s> S \cr -s/b+g(s-S) & s\leq S\end{matrix}\right.,
		\end{align}
		
		\noindent where $g(x)$ is a penalty function that forces $s$ to be close to the target signal yield $S$.  The optimization of the $\sqrt{s}=8$ TeV data analysis used an exponential function $g(x)=\alpha\exp(\beta(x)-1)$, which makes $f$ continuous at $s=S$ and takes advantage of a large derivative to force $s\rightarrow S$.  The values $\alpha=\beta=1$ worked well.  One disadvantage of the exponential function is that $f$ is not smooth at $s=S$.  Therefore, the optimization at $\sqrt{s}=13$ TeV used the alternative function $g(x)=\gamma x^2$, which lead to better convergence because $\partial_xg(x)|_{x=0}=0$.  Values of $\gamma\sim 5-10$ resulted in relatively stable performance.  Figure~\ref{fig:susy:exclusion:overview} illustrates the above procedure.
	
\begin{figure}[h!]
\begin{center}
\includegraphics[width=0.4\textwidth]{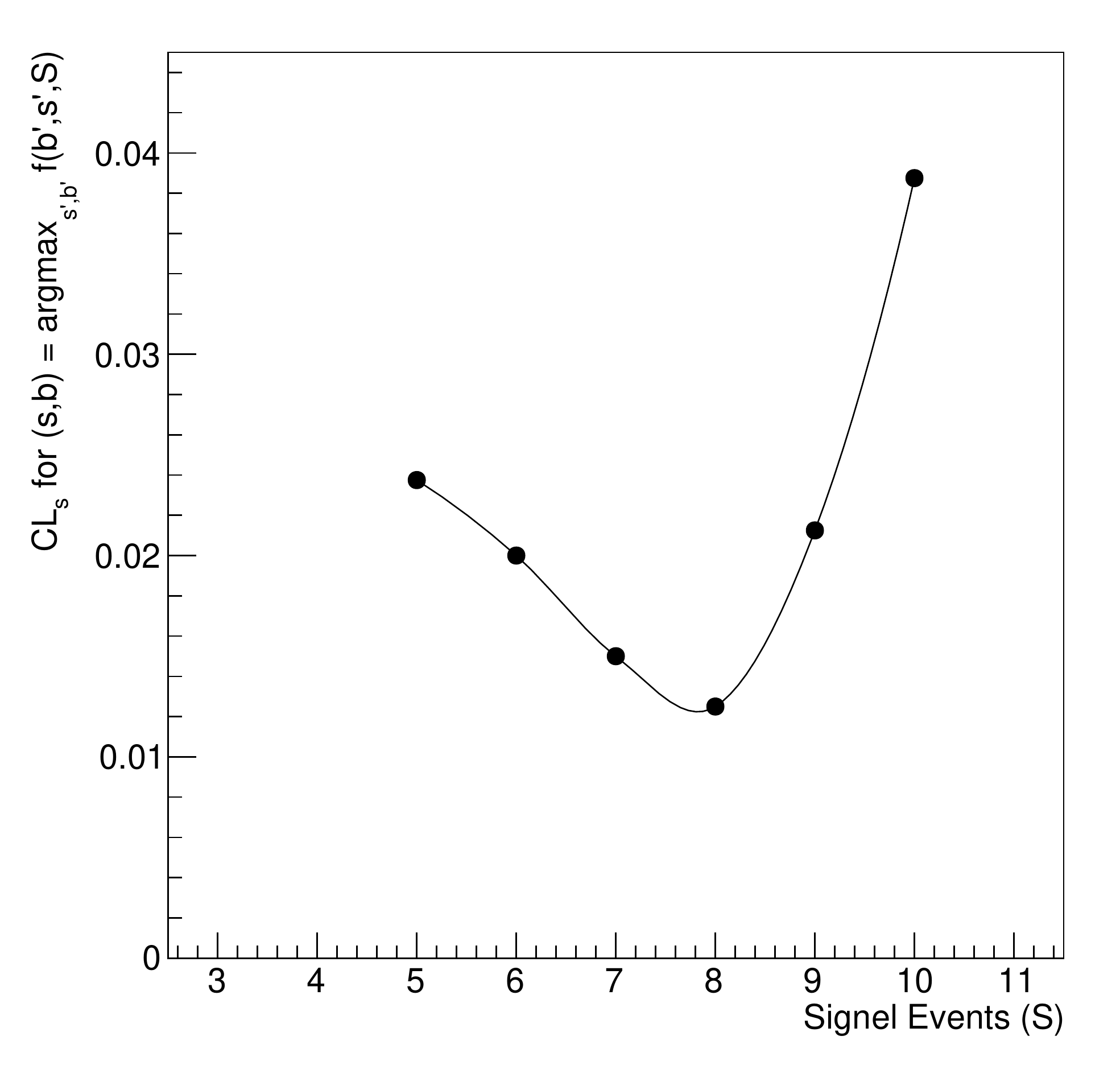}
 \caption{An example scan in the number of signal events.  For each fixed $S$, the background yield is minimized and the $\text{CL}_s$ value is computed (see Sec.~\ref{sec:susy:stats} for details).}
 \label{fig:susy:exclusion:overview}
  \end{center}
\end{figure}		
	
		One additional component of the algorithm is the use of background weights.  Background estimates that are at least partially data-driven are of higher quality than those constructed only from simulation.  Therefore, it is desirable to have a background composition that is enriched in backgrounds that can be well-estimated and suppressed in backgrounds that cannot be predicted using some insight from the data.  To achieve this in the minimization, the background yield $b$ is constructed as a weighted sum over all of the background components: $b=\sum \omega_i b_i$, where $b_i$ is the predicted background for process $i\in\{t\bar{t},t\bar{t}+V,W+\text{jets},\text{single top},\text{dibosons}\}$ and $\omega_i$ is a fixed weight.  By increasing the weight of one background relative to another, the algorithm can be steered toward suppressing a target background process.  For the $\sqrt{s}=8$ TeV analysis, $\omega_{t\bar{t}}=1, \omega_{t\bar{t}+V}=6, \omega_{W+jets}=3, \omega_\text{single top}=6$, and $\omega_\text{dibosons}=6$.  Even though there is a dedicated $W$+jets CR for the $\sqrt{s}=8$ TeV analysis, most of the $W$+jets events in the SR are produced in association with heavy flavor jets, while the events in the control region are mostly light flavor; this is the reason for $\omega_{W+jets}=3$.  With data driven estimates for $t\bar{t}+V$ and single top as part of the $\sqrt{s}=13$ TeV analysis, non-unity $\omega$ factors are no longer necessary\footnote{The diboson background is already negligible without imposing $\omega_\text{dibosons}>1$.}.  
		
		\clearpage
		
		\section{Single Bin Regions}
		\label{sec:onebinregionoptimization}
		
	Six single bin regions are constructed for a broad coverage of sensitivity in the $(m_\text{stop},m_\text{LSP})$ mass plane using the procedures described in the previous section.  Table~\ref{tab:benchmarks} provides an overview of the regions of parameter space targeted by each signal region.  There are three phenomenologically distinct regions.  At low stop mass, the signal cross section is relatively high ($\mathcal{O}(1\%)$ of $t\bar{t}$), but many kinematic distributions do not significantly differ from the dominate $t\bar{t}$ background (SR1).  In contrast, at high stop mass, the cross section is very low, but many kinematic distributions are significantly different between the signal and the $t\bar{t}$ background (SR3, tNhigh, SR13).  When additionally the LSP mass is large the difference between signal and background is reduced and requires a dedicated event selection to maintain sensitivity to this region of parameter space (SR2, tNmed).  In order to increase the sensitivity for discovering SUSY with the early $\sqrt{s}=13$ TeV data, a gluino mediated stop (GMS) model with a nearly degenerate stop and LSP was used for the optimization.  As described in Sec.~\ref{relatedmodels}, such a model also results in $t\bar{t}+E_\text{T}^\text{miss}$ when the stop decay products are too soft to reconstruct.  The gluino model $(m_{\tilde{g}},m_{\tilde{t}},m_\text{LSP})=(1250,750,745)$ GeV was chosen to be kinematically equivalent (see Sec.~\ref{sec:equiv} for details) to a model with stop pair production at $(m_{\tilde{t}},m_\text{LSP})=(800,0)$ GeV.   The only significant difference between the GMS and the direct stop models is that for a fixed mass, the former has a cross section that is about a factor of 50 more than the latter due to the additional spin and color states for the gluino.
	
Table~\ref{tab:signalregionselections} shows the defining selections for each region.  A complete description of each variable can be found in Chapter~\ref{sec:discriminating}.  Horizontal lines in Table~\ref{tab:signalregionselections} group variables with a similar purpose. Even though the event selections where constructed in a mostly automated fashion, it is useful to examine their anatomy to understand why each value was chosen. {\it A well-motivated event selection is a robust event selection}.  The logic for the various event selections is similar amongst the six event selections; for brevity SR13 is used an example.  Figure~\ref{fig:SR13ETmiss} shows the $E_\text{T}^\text{miss}$ distribution using the $\sqrt{s}=13$ TeV preselection and after the full SR13 event selection before the $E_\text{T}^\text{miss}$ requirement.  The LSPs can carry significant momentum and therefore the $E_\text{T}^\text{miss}$ is one of the most powerful variables in any search for $R$-parity conserving SUSY.  The pair of gluinos (or equivalently, a pair of $800$ GeV stops) for the model in Fig.~\ref{fig:SR13ETmiss} will be produced nearly at rest in the lab frame and therefore $E_\text{T}^\text{miss}\lesssim 2\times m_\text{stop}/2$.  The factor of two is from the two LSPs and the factor of $1/2$ is from the split in the stop mass between the top quark and the LSP boost.  As indicated by Fig.~\ref{fig:SR13ETmiss}, most of the stop events have significantly less $E_\text{T}^\text{miss}$ than this bound because the orientation of the two LSPs relative to each other is random and the neutrino from the top quark decay can further reduce the total missing momentum when it has a large momentum component anti-parallel to the LSP directions.   The optimal threshold value for SR13 is $E_\text{T}^\text{miss}\gtrsim 350$ GeV.  Beyond that value, the reduction in signal outweighs the reduction in background.  Note that the peak of the $E_\text{T}^\text{miss}$ distribution is higher after the event selection than it is with only the preselection.  This is due in part to the $H_\text{T,sig}^\text{miss}$ requirement.		
		
\begin{table}[h!]
\begin{center}
\begin{tabular}{|c|cc|cc|}
\hline  
Region & $L_\text{int}$ [fb${}^{-1}$]& $\sqrt{s}$ [GeV]& Stop Mass [GeV]& LSP Mass [GeV]\\
\hline
\hline   
SR1 & 13 & 8 & 250 & 50\\
SR2 & 13 & 8 & 500 & 200\\
SR3 & 13 & 8 & 650 & 50\\
\hline
tNmed & 20.3 & 8 & 550 & 200\\
tNhigh & 20.3& 8 & 650 & 1\\
\hline
SR13 & 3.2 & 13 & 800 & 1\\
\hline  
\end{tabular}
\end{center}
\caption{Benchmark signal models used for optimizing the single bin regions.  The regions SR1-3 were optimized using the entire grid and so the chosen models are representative of the regions of parameter space that the three regions target.  SR13 was optimized using a GMS model, but the equivalent stop model is the one in the table above.}
  \label{tab:benchmarks}
\end{table}

   A similar set of plots for the $m_\text{T}$ are in Fig.~\ref{fig:SR13mT}.  The peak of the $m_\text{T}$ distribution for the signal is lower than the $E_\text{T}^\text{miss}$ because the $m_\text{T}$ is essentially the geometric average of the $E_\text{T}^\text{miss}$ and the lepton $p_\text{T}$.  As it is further down the decay chain than the neutralinos, the lepton $p_\text{T}$ is expected to be softer and therefore brings down the geometric average.  Despite this, the $m_\text{T}$ is still one of the most powerful variables, with a separation\footnote{Using the same heuristic metric as introduced in Sec.~\ref{sec:mt2tmva}.} of about 14\% in the signal region -- the same as $E_\text{T}^\text{mis}$ (at preselection, the separation is 35\% for $m_\text{T}$ and about 55\% for $E_\text{T}^\text{miss}$).  While all of the SM backgrounds are suppressed at high values of $m_\text{T}$, $W$+jets events are reduced the most because they have no additional source of missing momentum to surpass the $m_\text{T}\lesssim m_W$ edge.

\begin{table}[h!]
\begin{center}
\noindent\adjustbox{max width=\textwidth}{
\begin{tabular}{|c|ccccc|cccc|c|}
\hline  
Variable & &SR1 & SR2 &SR3&  && tNmed & tNhigh && SR13 \\
\hline
\hline
Jet $p_\text{T} >$ [GeV] &&\multicolumn{3}{c}{$80,60,40,25$ }&&& $80,60,40,25$&$100,80,40,25$ && $100,80,50,25$  \\
$|\Delta\phi(\text{jet}_1,\vec{p}_\text{T}^\text{miss})|>$ && 0.8 & -- & 0.8 &&& -- & -- && 0.4 \\
$|\Delta\phi(\text{jet}_2,\vec{p}_\text{T}^\text{miss})|>$ && 0.8 & 0.8 & 0.8 &&& 0.8 & --&& 0.4 \\
\hline
$E_\text{T}^\text{miss}>$ [GeV] && 150 & 200 & 225& && 300 & 320 && 350 \\
$H_\text{T,sig}^\text{miss}>$ &&  \multicolumn{3}{c}{-- } &&& 12.5 & 12.5 && 20\\
\hline
$m_\text{T}>$ [GeV] && 140 $(*)$ & 140 & 180 &&& 140 & 200 & &200 \\
$am_\text{T2}>$ [GeV] && -- & 170 & 200 &&& 170 & 170 & &175 \\
$m_\text{T2}^\tau>$ [GeV] && -- & -- & 120 &&& -- & 120 & &80 \\
\hline
$m_\text{top,$\chi^2$}\in $ [GeV] &&  \multicolumn{3}{c}{[130,205] } &&& [130,195] & [130,250] && [140,$\infty$] \\
$\Delta R(b,l)<$ &&  \multicolumn{3}{c}{-- }  &&& --& 3 && 2.5\\ 
$\Delta\phi(\vec{p}_\text{T}^\text{miss},2^\text{nd}\text{ Large $R$ jet})>$&&  \multicolumn{3}{c}{-- }  &&&-- & -- && 1\\ 
\hline  
\end{tabular}}
\end{center}
\caption{A summary of the six single bin signal region event selections.  Dashed lines indicate that there is no requirement on the given variable.  Even though the same symbol might be used for all six regions, some of the variables have a different meaning across columns.  For example, the jets used in the first five columns have the LCW calibration while the jets in the last column are at the EM scale but with the GS calibration applied. Furthermore, the choice of the $\tau$ candidate differs between the first five columns and the last column.  The $m_\text{top,$\chi^2$}$ variable is an explicit tri-jet mass reconstruction in the first five columns (using jet resolution information in columns 4 and 5) and the large radius jet mass in the last column (see Sec.~\ref{topmassreco}).  $(*)$ There is also an upper threshold of $250$ GeV for this loose selection.}
  \label{tab:signalregionselections}
\end{table}
		
\begin{figure}[htbp]
  \centering
  \includegraphics[width=0.5\textwidth]{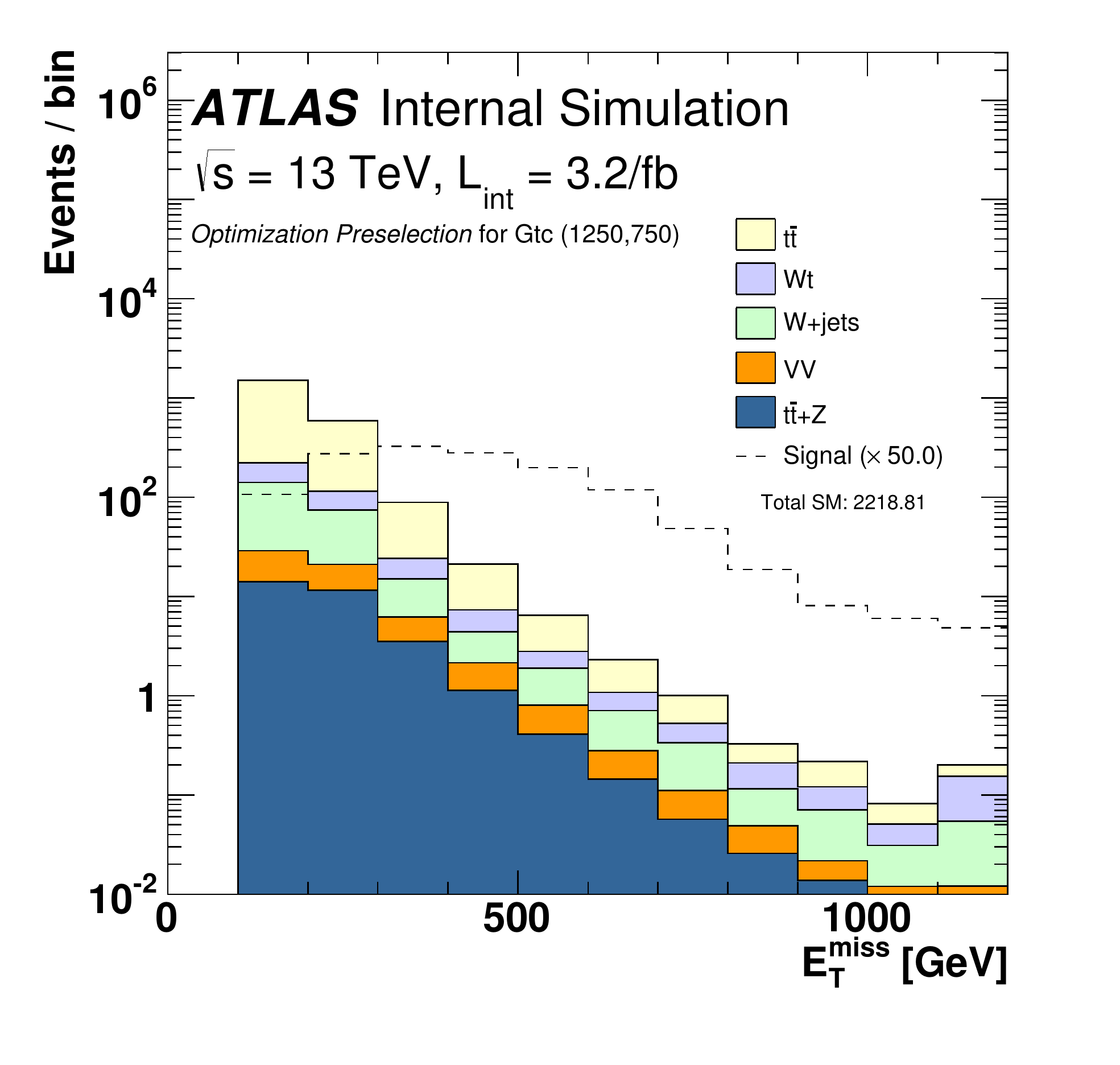}\includegraphics[width=0.5\textwidth]{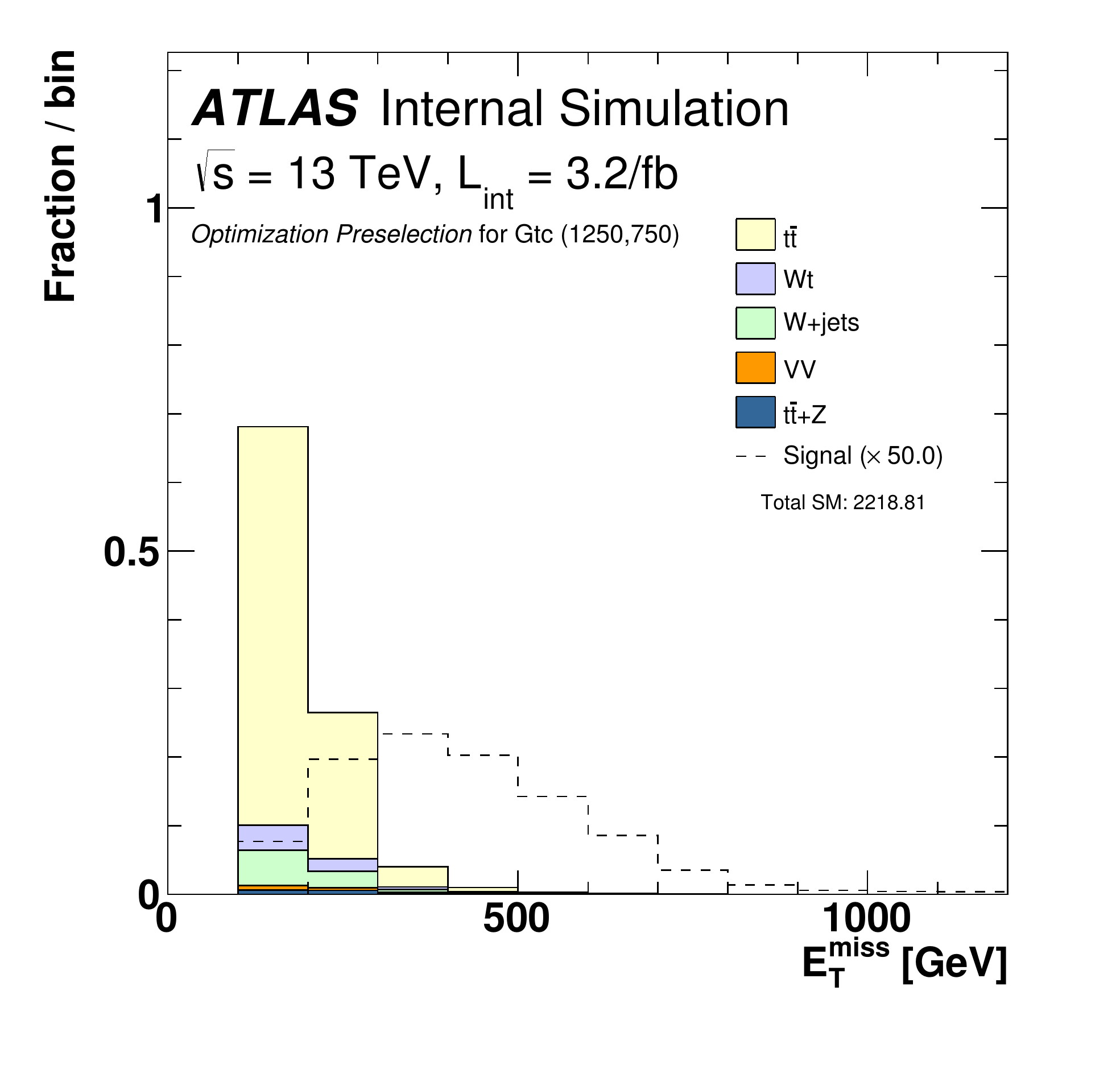}\\
\includegraphics[width=0.5\textwidth]{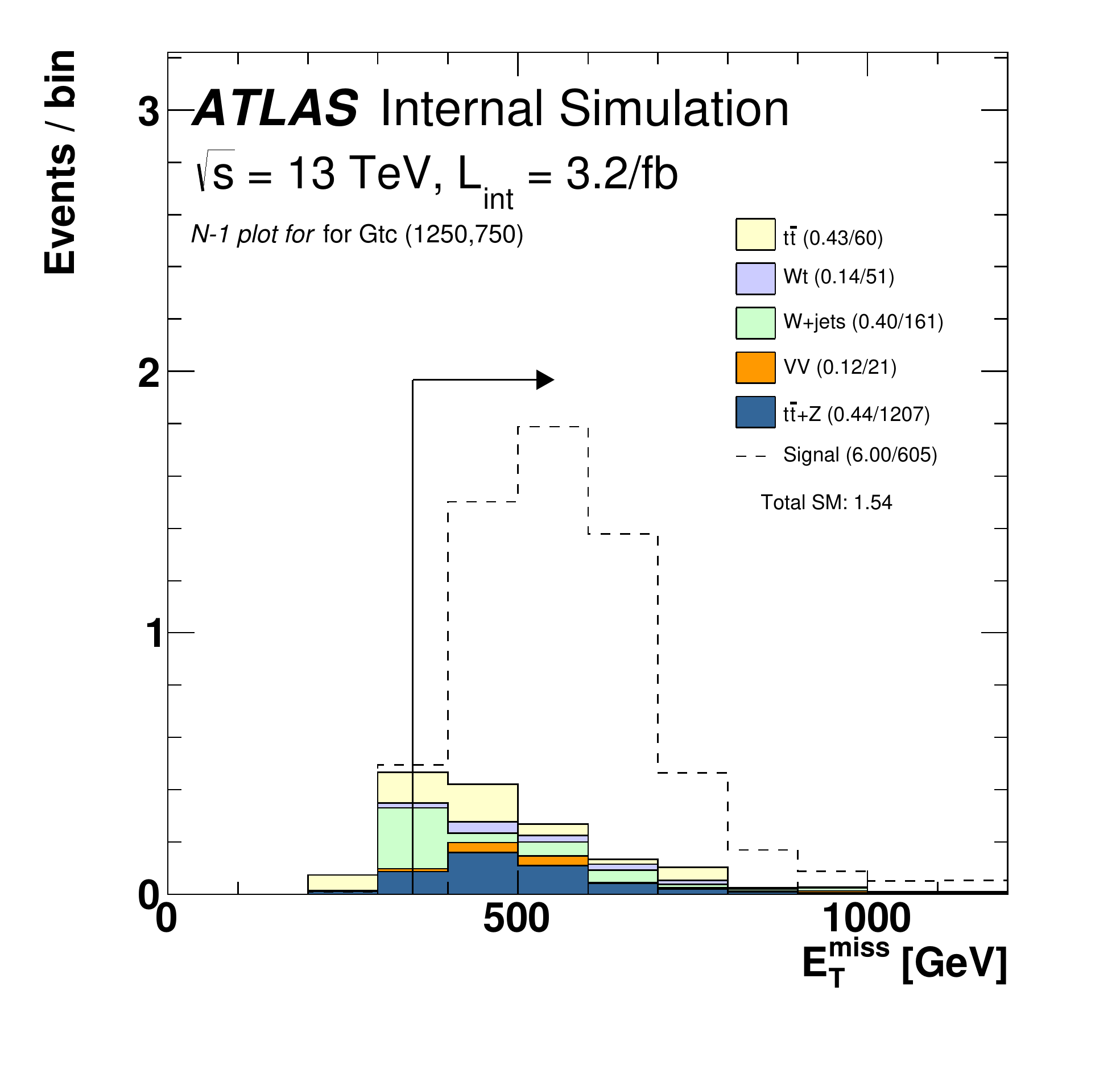}\includegraphics[width=0.5\textwidth]{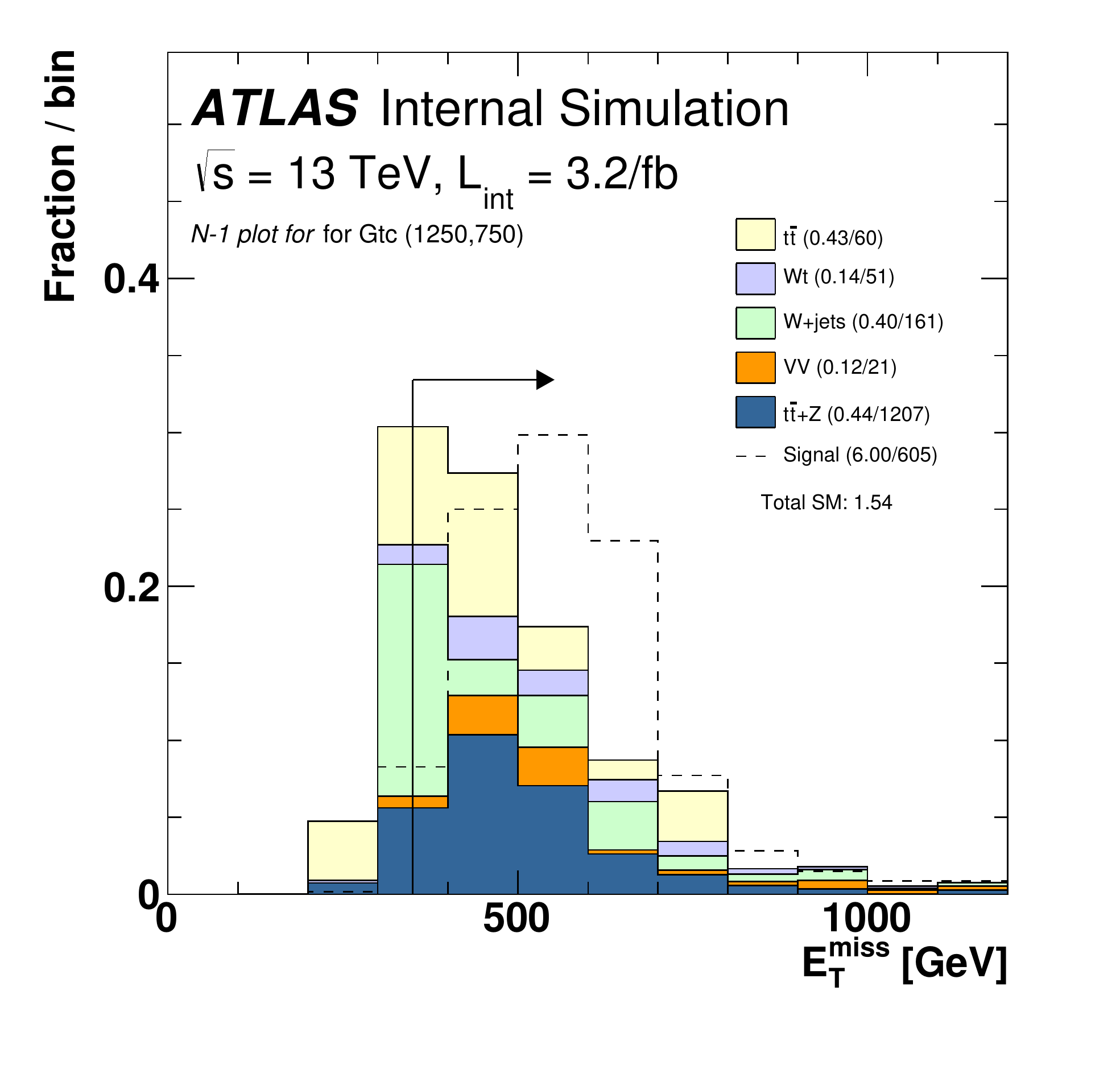}
  \caption{The distribution of the $E_\text{T}^\text{miss}$ for the event preselection (top) and after all SR13 requirements except $E_\text{T}^\text{miss}$ (bottom).  Figures with all signal region requirement but the one displayed are called $N$-$1$ plots.  Both the total background and signal yields are normalized to unity in the right plots.  An arrow indicates the signal region requirement.  The first number in parenthesis is the expected yield without applying any normalization factors and the second number is the raw event count in simulation (an indication of the statistical uncertainty).  The $Wt$ component includes all single to processes, but is dominated by the single production of a top quark in association with a $W$ boson.}
  \label{fig:SR13ETmiss}
\end{figure}		

\begin{figure}[htbp]
  \centering
  \includegraphics[width=0.5\textwidth]{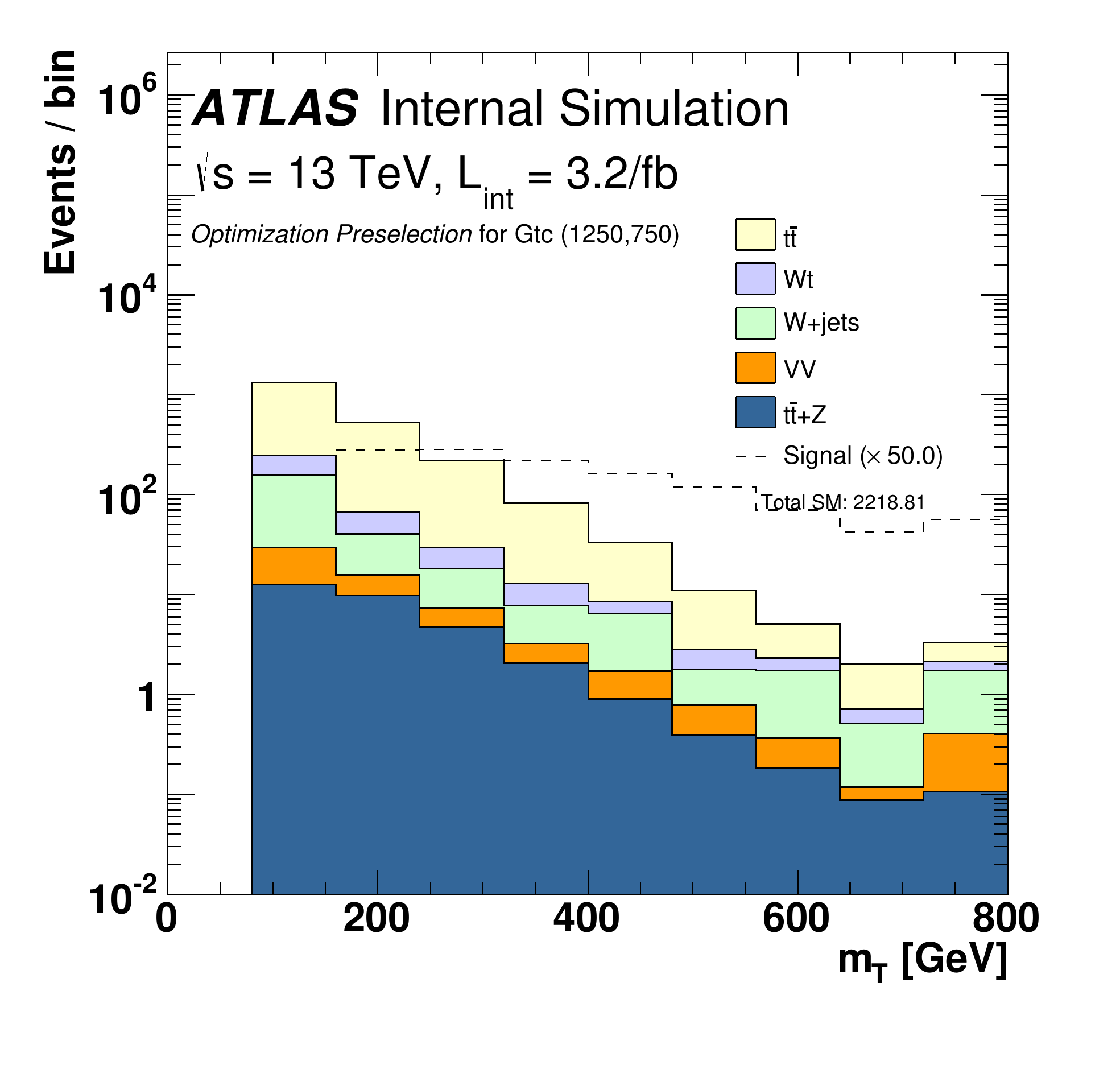}\includegraphics[width=0.5\textwidth]{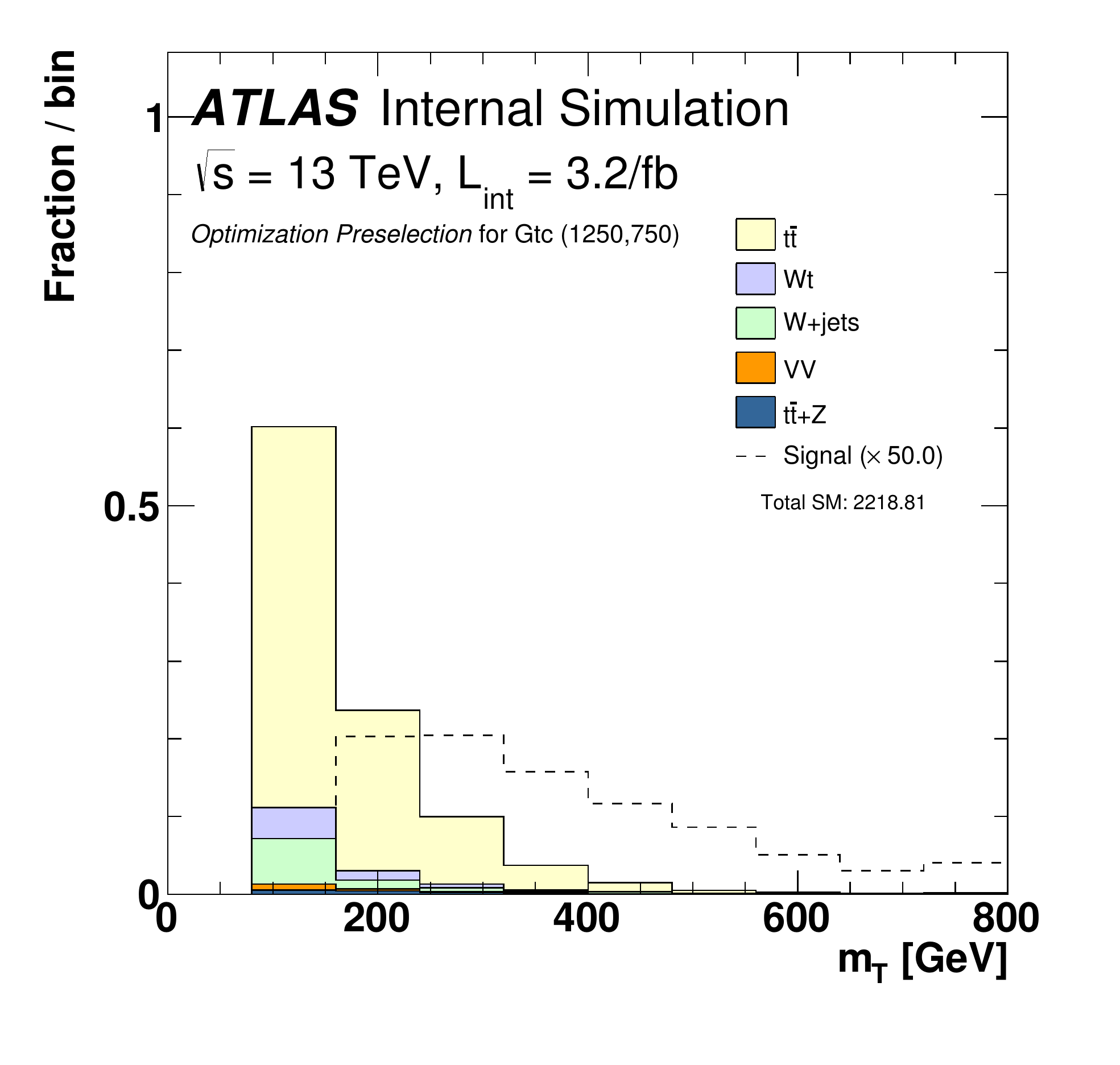}\\
\includegraphics[width=0.5\textwidth]{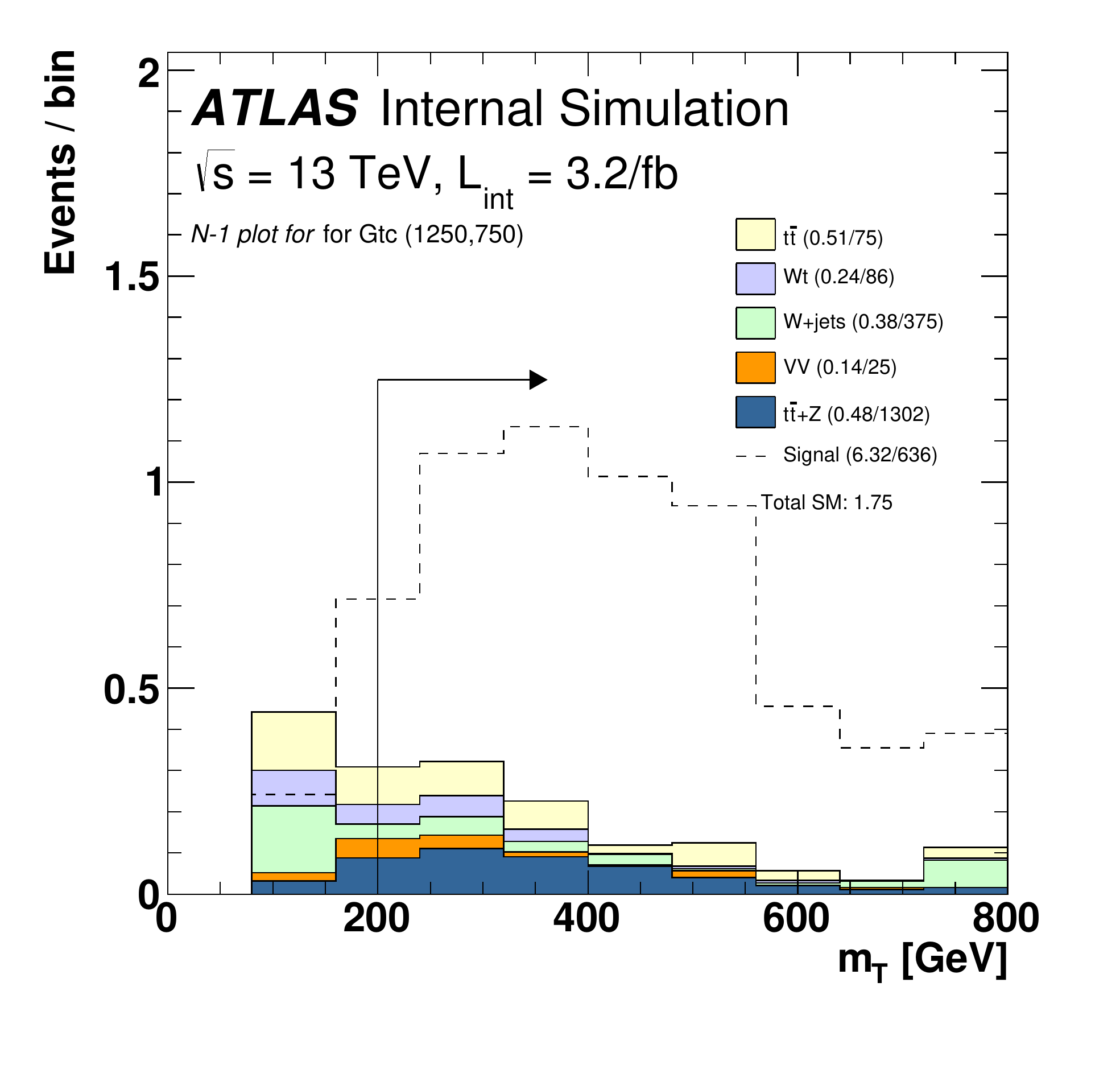}\includegraphics[width=0.5\textwidth]{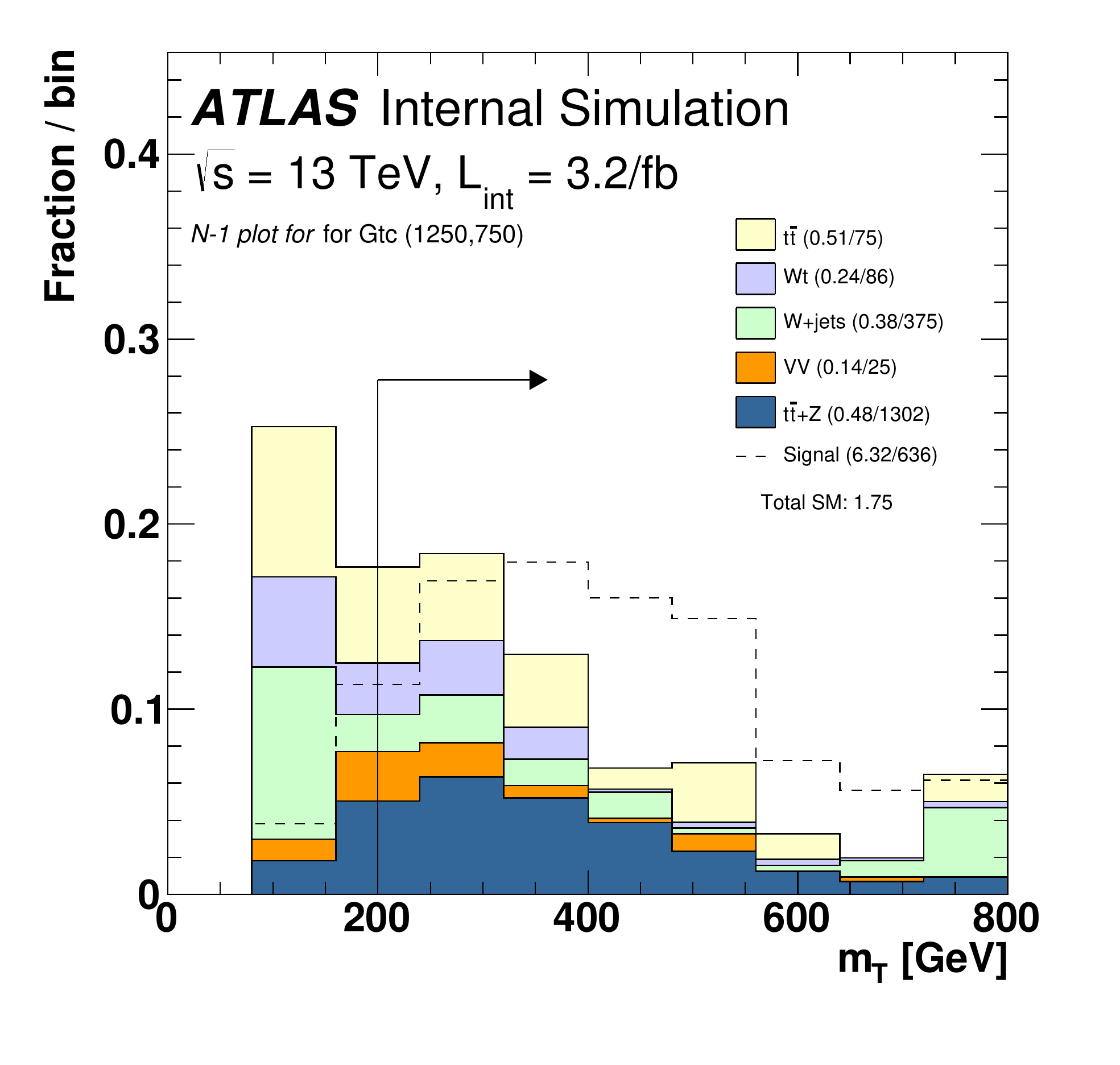}
  \caption{The same as Fig.~\ref{fig:SR13ETmiss}, but for $m_\text{T}$ instead of $E_\text{T}^\text{miss}$.}
  \label{fig:SR13mT}
\end{figure}	

The $am_\text{T2}$ distribution is shown in Fig.~\ref{fig:SR13amT2}.  At preselection, $am_{T2}$ has a similar separation between signal and background as $m_\text{T}$ ($\sim35\%$), but after the rest of the event selection, it offers the most discriminating power with a separation of about $\sim 20\%$.  In all signal regions that use $am_\text{T2}$, the threshold value is around $m_\text{top}\sim 175$ GeV.  The large drop in the $t\bar{t}$ distribution at this point is visible in all of the plots in Fig.~\ref{fig:SR13amT2}.  In contrast, single top events tend to be {\it above} $m_\text{top}$.  This will motivate a data-driven technique to estimate this background in Sec.~\ref{singletop:datadriven}.  

\begin{figure}[htbp]
  \centering
  \includegraphics[width=0.45\textwidth]{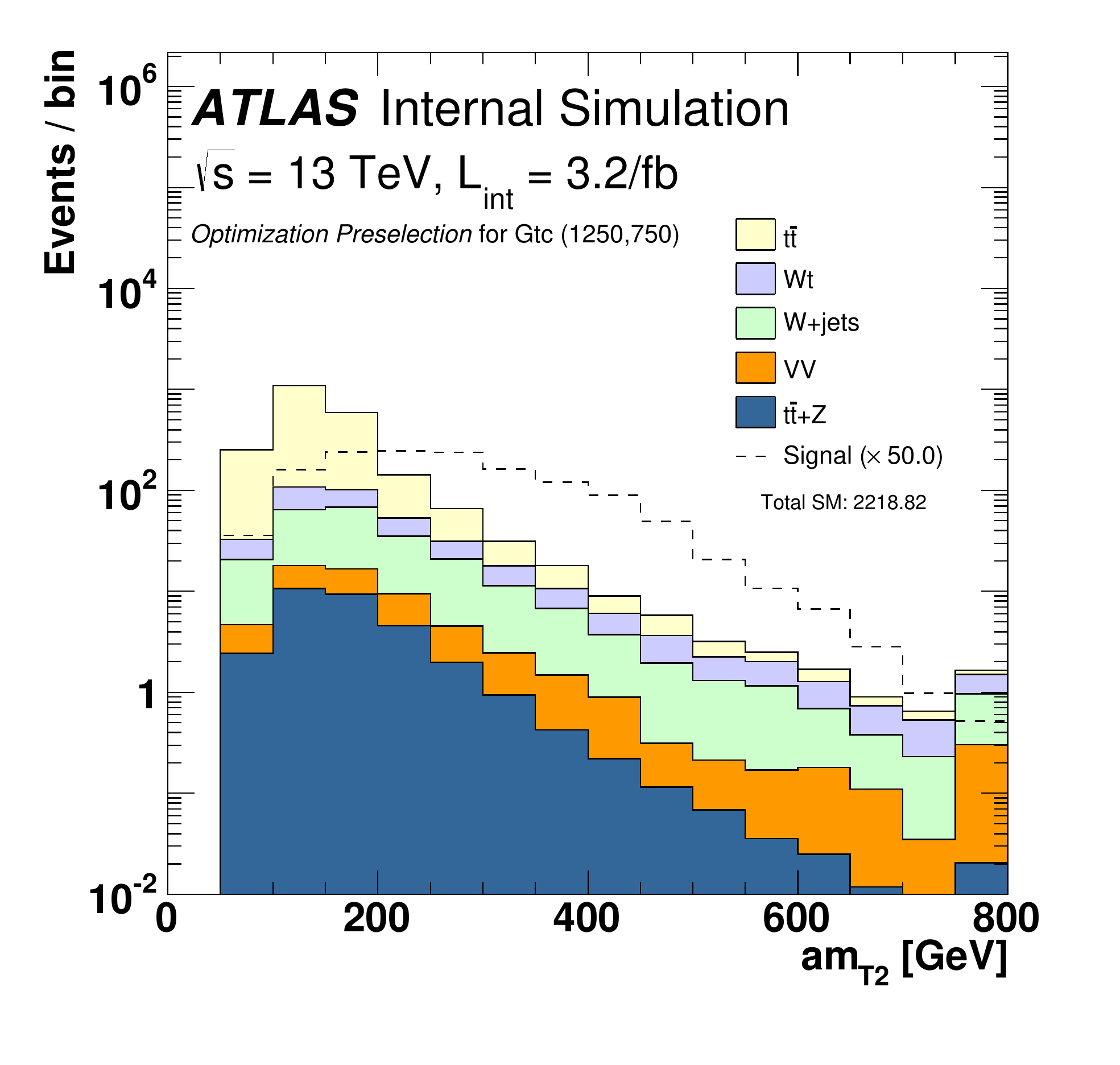}\includegraphics[width=0.45\textwidth]{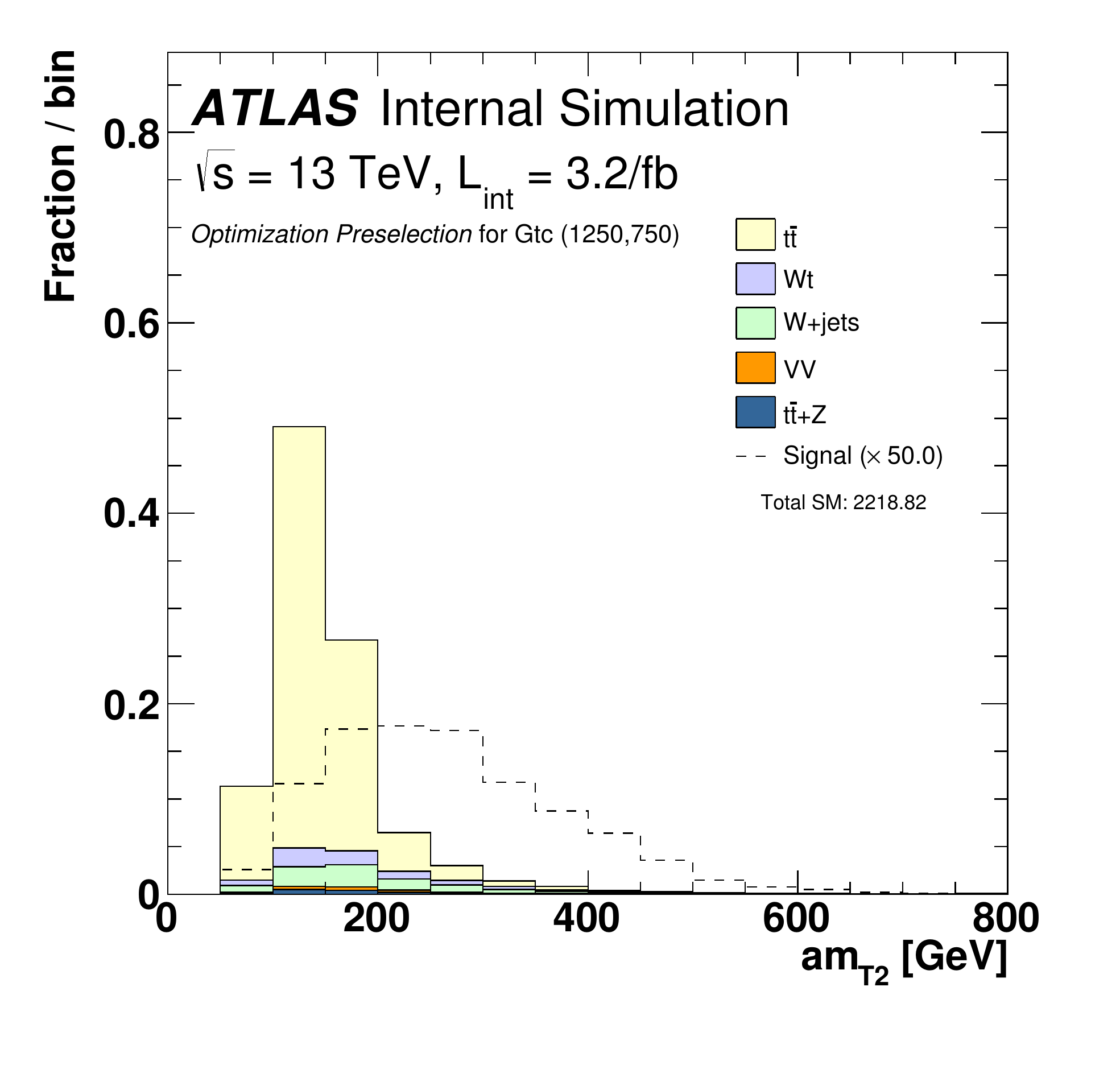}\\
\includegraphics[width=0.45\textwidth]{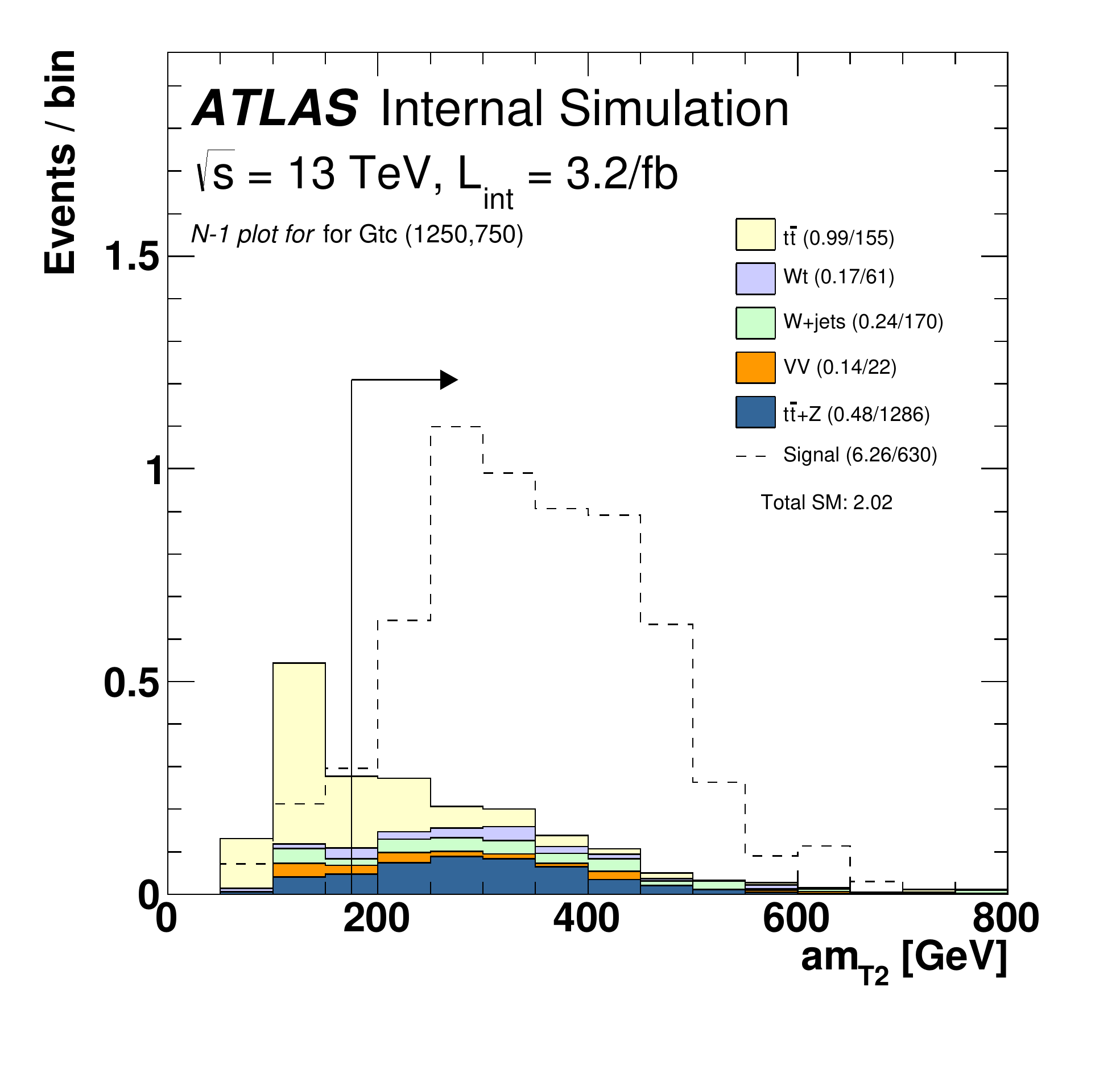}\includegraphics[width=0.45\textwidth]{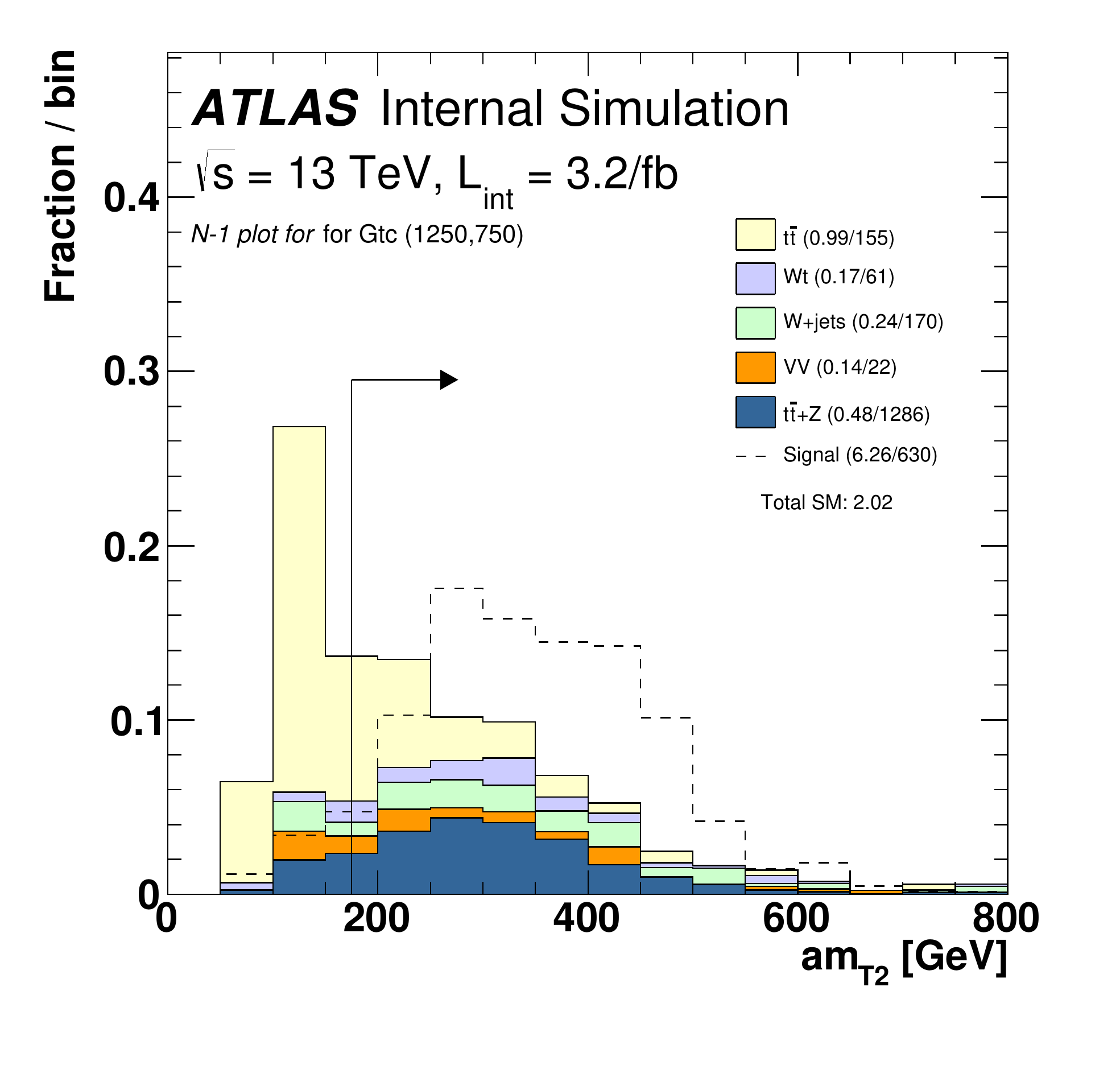}
  \caption{The same as Fig.~\ref{fig:SR13ETmiss}, but for $am_\text{T2}$ instead of $E_\text{T}^\text{miss}$.}
  \label{fig:SR13amT2}
\end{figure}

A technique proposed in the literature to further suppress $t\bar{t}$ is a $\chi^2$ compatibility test with the two-lepton $t\bar{t}$ background hypothesis~\cite{Graesser:2012qy}. For a missing leptonically decaying $W$ boson $W_2$ and one neutrino from the identified leptonically decaying $W$ boson $\nu_1$, there are $8$ total unknowns: $p_{W_2}^\mu$ and $p_{\nu_1}^\mu$.  Imposing $p_{W_2}^2=m_W^2, p_{\nu_1}^2=0, p_{W_2x}+p_{\nu_1x}=p_x^\text{miss}$ and $p_{W_2y}+p_{\nu_1y}=p_y^\text{miss}$ reduces this to only four unknowns.  One can choose these unknowns to be $p_{W_2x}, p_{W_2y},p_{W_2z}$, and $p_{\nu_1z}$.  A $\chi^2$ variable $S$ is then the sum $(p_{W_1}^2-m_W^2)^2/a_W^4+\sum_{i=1}^2(p_{t_i}^2-m_t^2)^2/a_t^4+(4m_t^2-(\sum_i p_i)^2)^2/a_\text{SM}$, where $p_{t_i}$ is the sum of $p_{W_i}$ and the four-vector of a $b$-tagged jet and the $a_x$ are resolution parameters (see Ref.~\cite{Graesser:2012qy}).  The {\it topness} variable is then given by $\log (\text{min} S)$.  The left plot of Fig.~\ref{fig:SR13topness} shows a double-peak structure that separates the dilepton background with low values of $S$ with the signal that has higher values of $S$.  However, after the $am_\text{T2}$ requirement (and other selections), shown in the right plot of Fig.~\ref{fig:SR13topness}, there is little additional discriminating power from topness.

\begin{figure}[htbp]
  \centering
\includegraphics[width=0.45\textwidth]{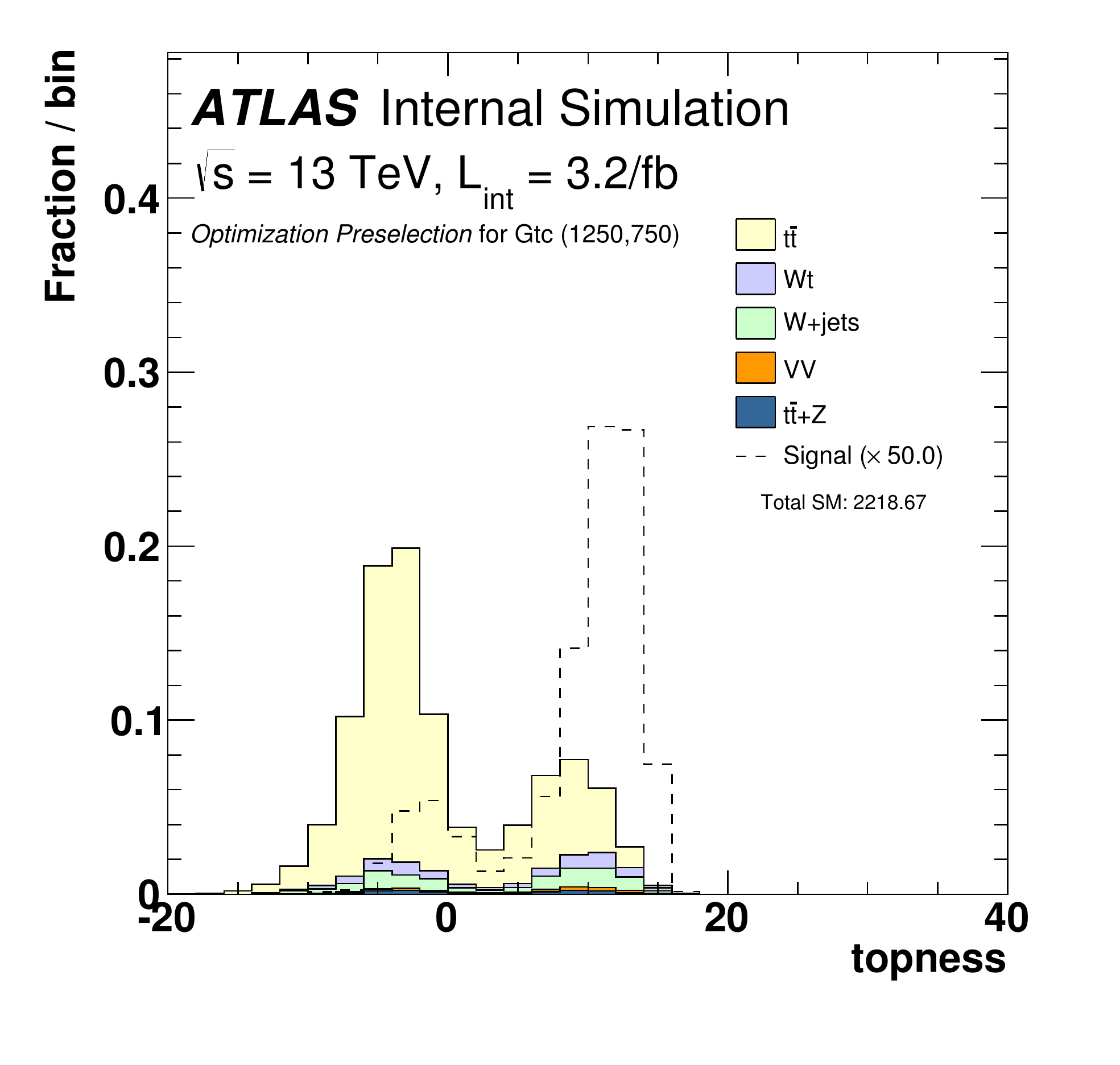}\includegraphics[width=0.45\textwidth]{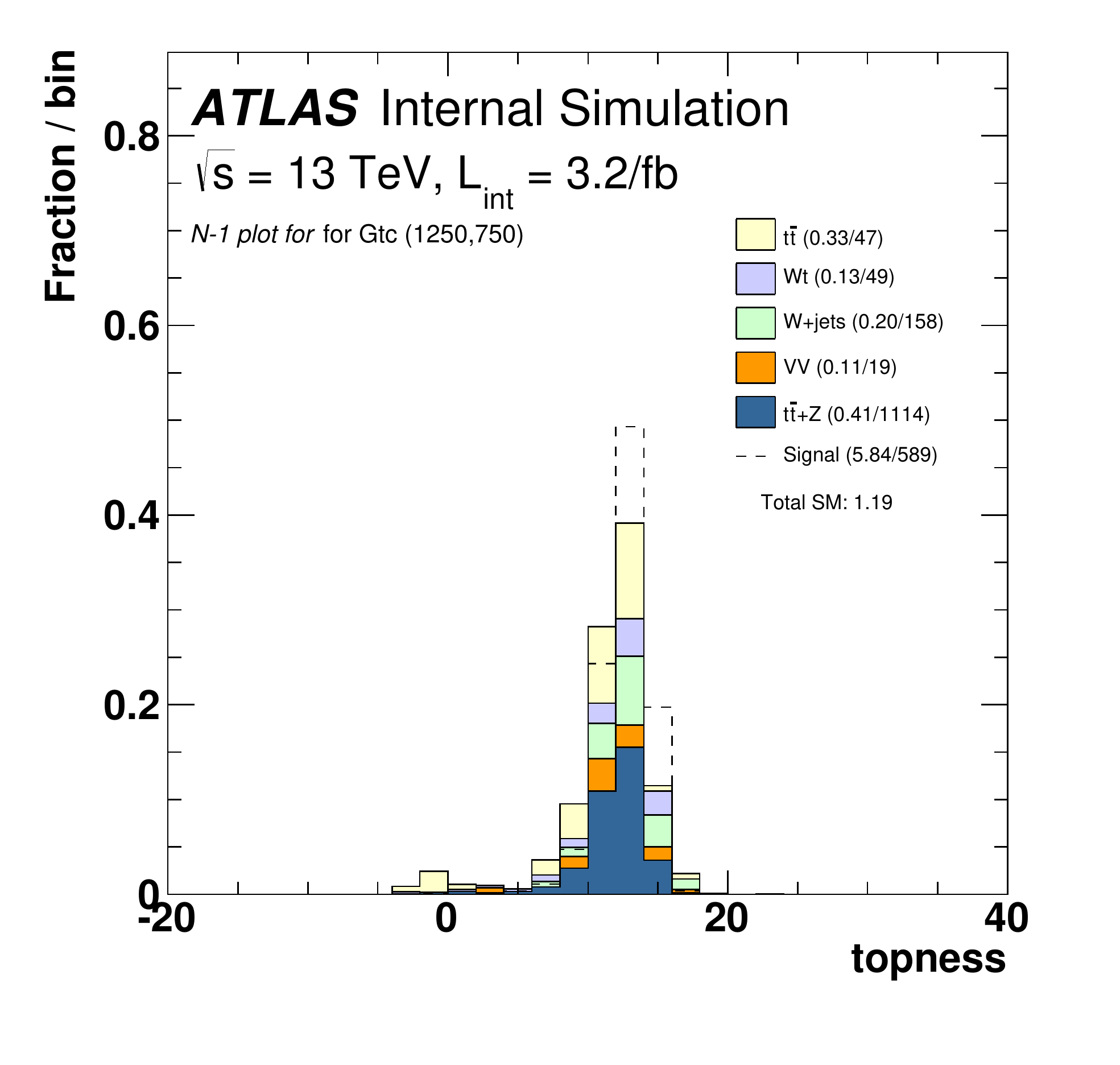}
  \caption{The distribution of the $m_\text{T}$ for the event preselection (top) and after all SR13 requirements except $m_\text{T}$ (bottom).  Both the total background and signal yields are normalized to unity in the right plots.  An arrow indicates the signal region requirement.}
  \label{fig:SR13topness}
\end{figure}

The $H_\text{T,sig}^\text{miss}$ has a significant correlation with the $E_\text{T}^\text{miss}$, but Fig.~\ref{fig:SR13htsig} shows that it is still has significant separation power after preselection and in the signal region.  Most events with the preselection have $H_\text{T,sig}^\text{miss}>0$ because $H_\text{T}^\text{miss}\sim E_\text{T}^\text{miss}$ and $H_\text{T,sig}^\text{miss}>0$ implies $H_\text{T}^\text{miss}>100$ GeV.  For the same reason, the peak of the $H_\text{T,sig}^\text{miss}$ distribution in the background shifts toward higher values in the SR due to the higher $E_\text{T}^\text{miss}$ requirement.

\begin{figure}[htbp]
  \centering
  \includegraphics[width=0.5\textwidth]{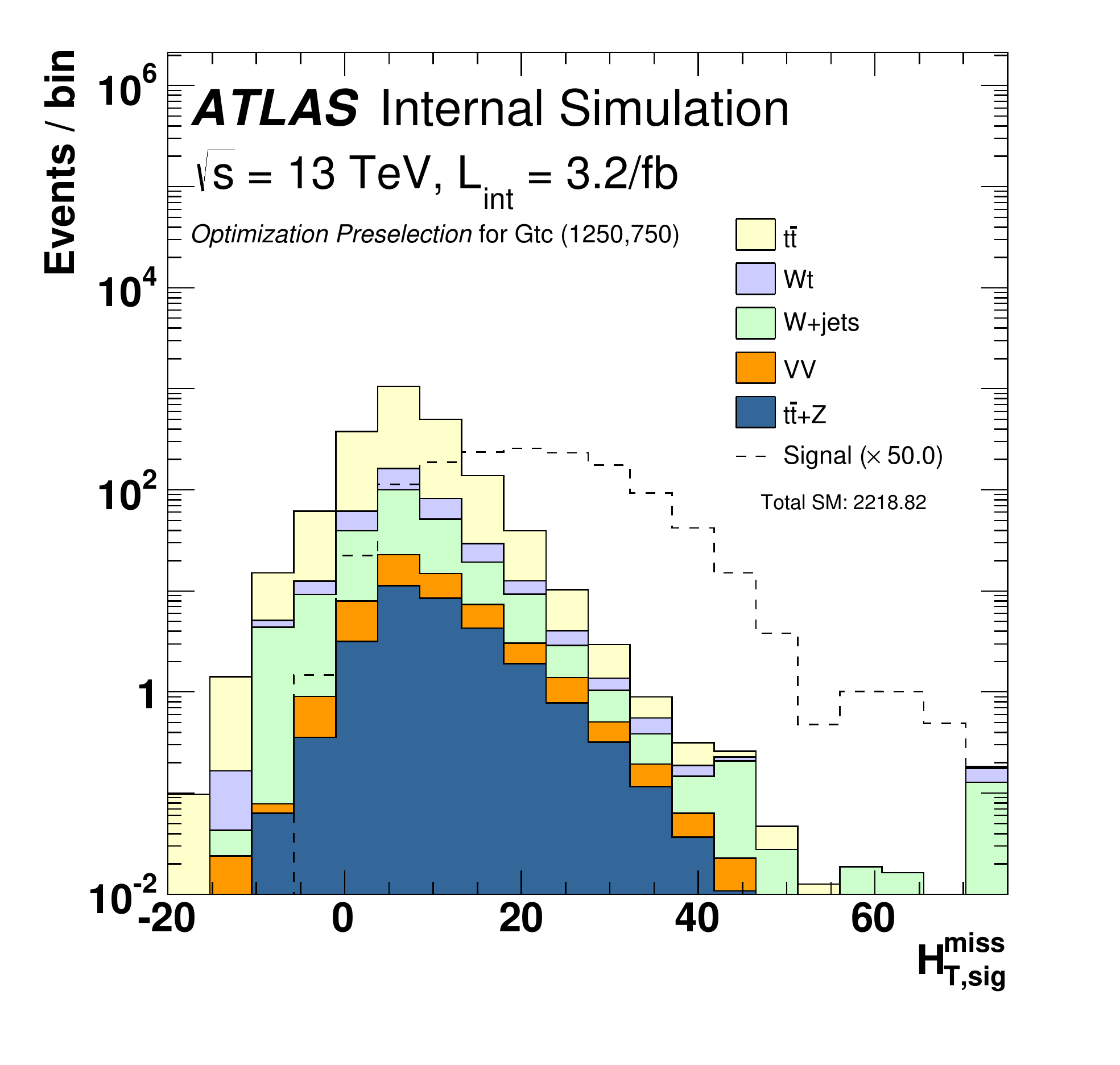}\includegraphics[width=0.5\textwidth]{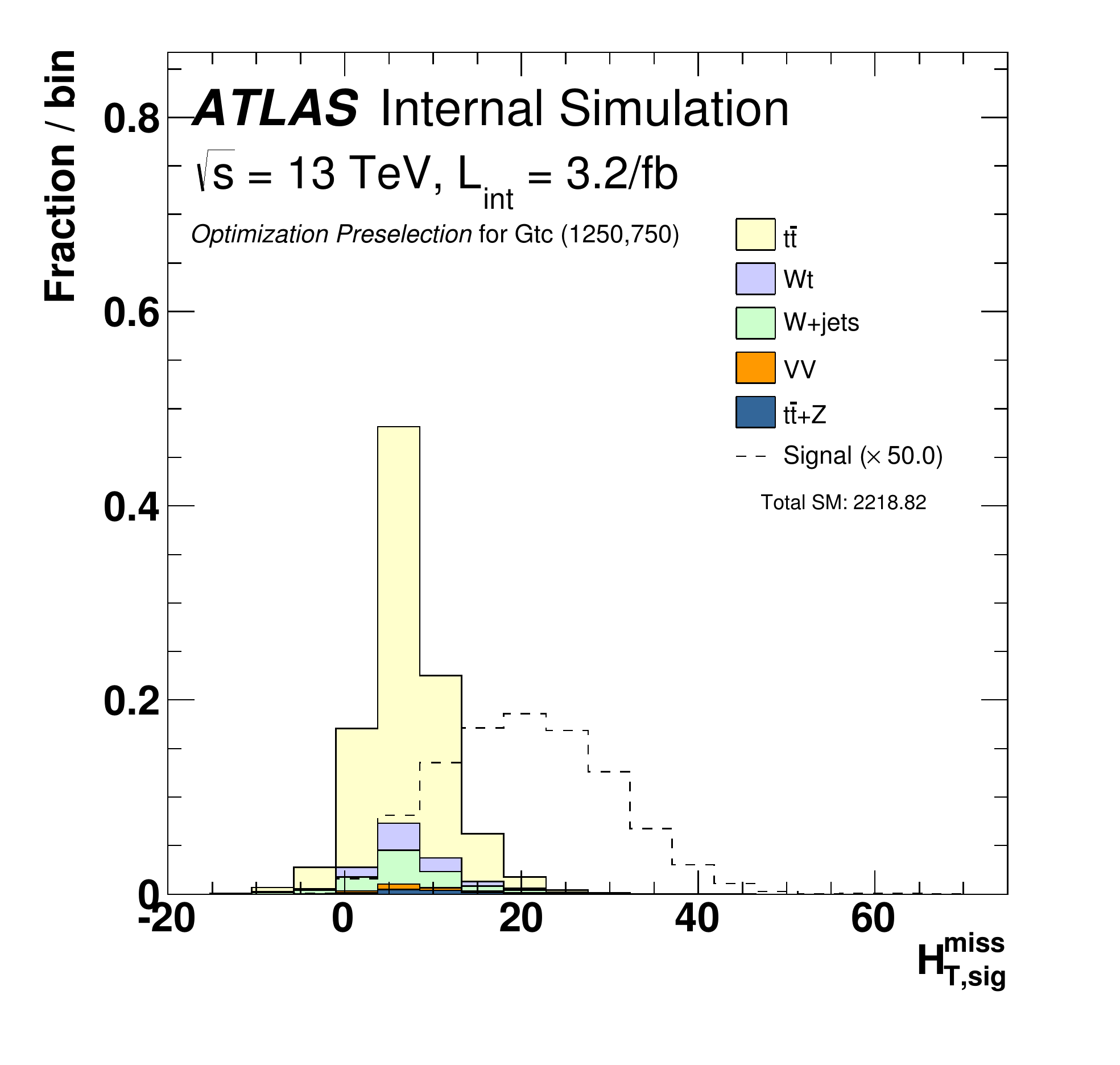}\\
\includegraphics[width=0.5\textwidth]{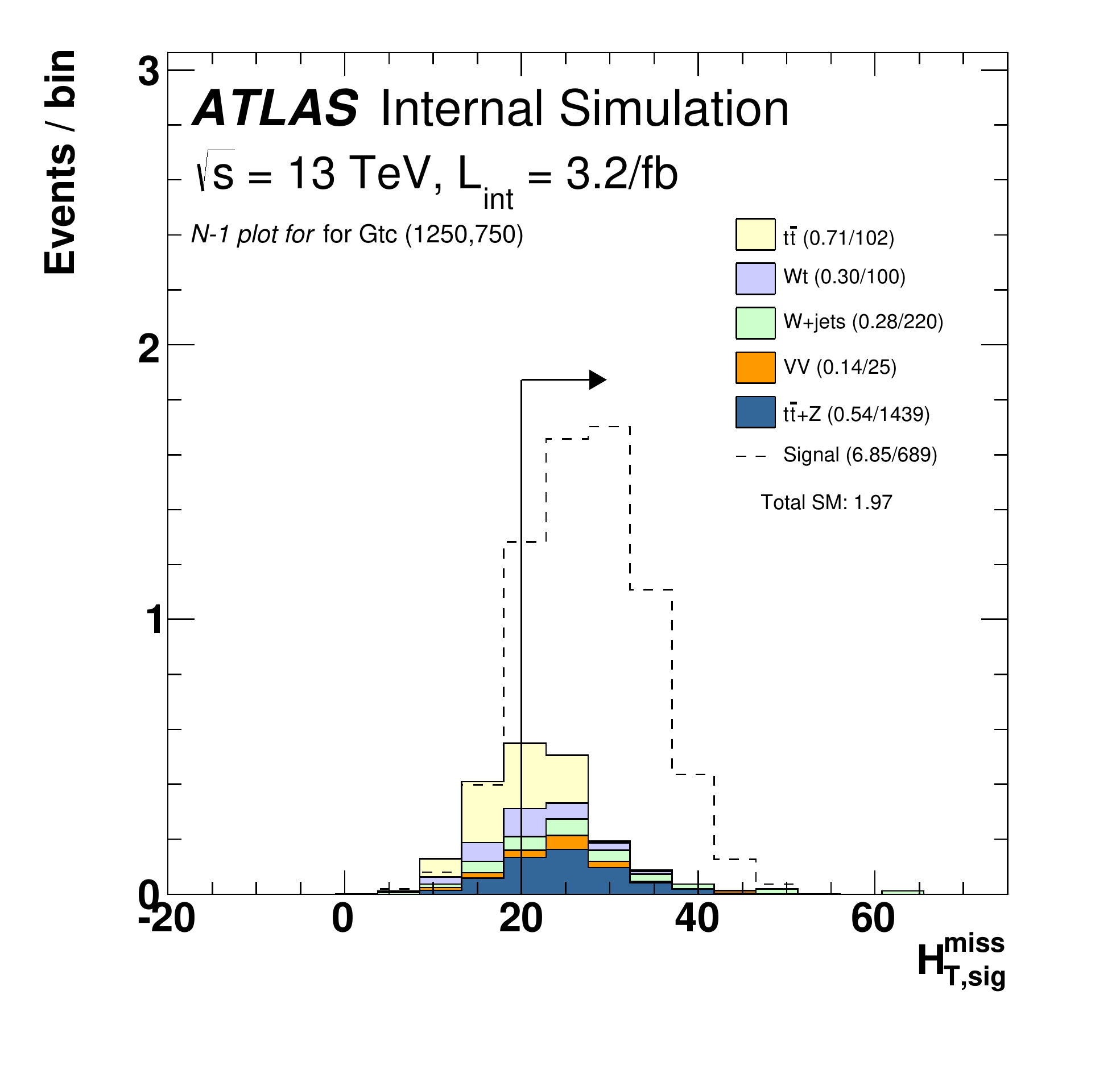}\includegraphics[width=0.5\textwidth]{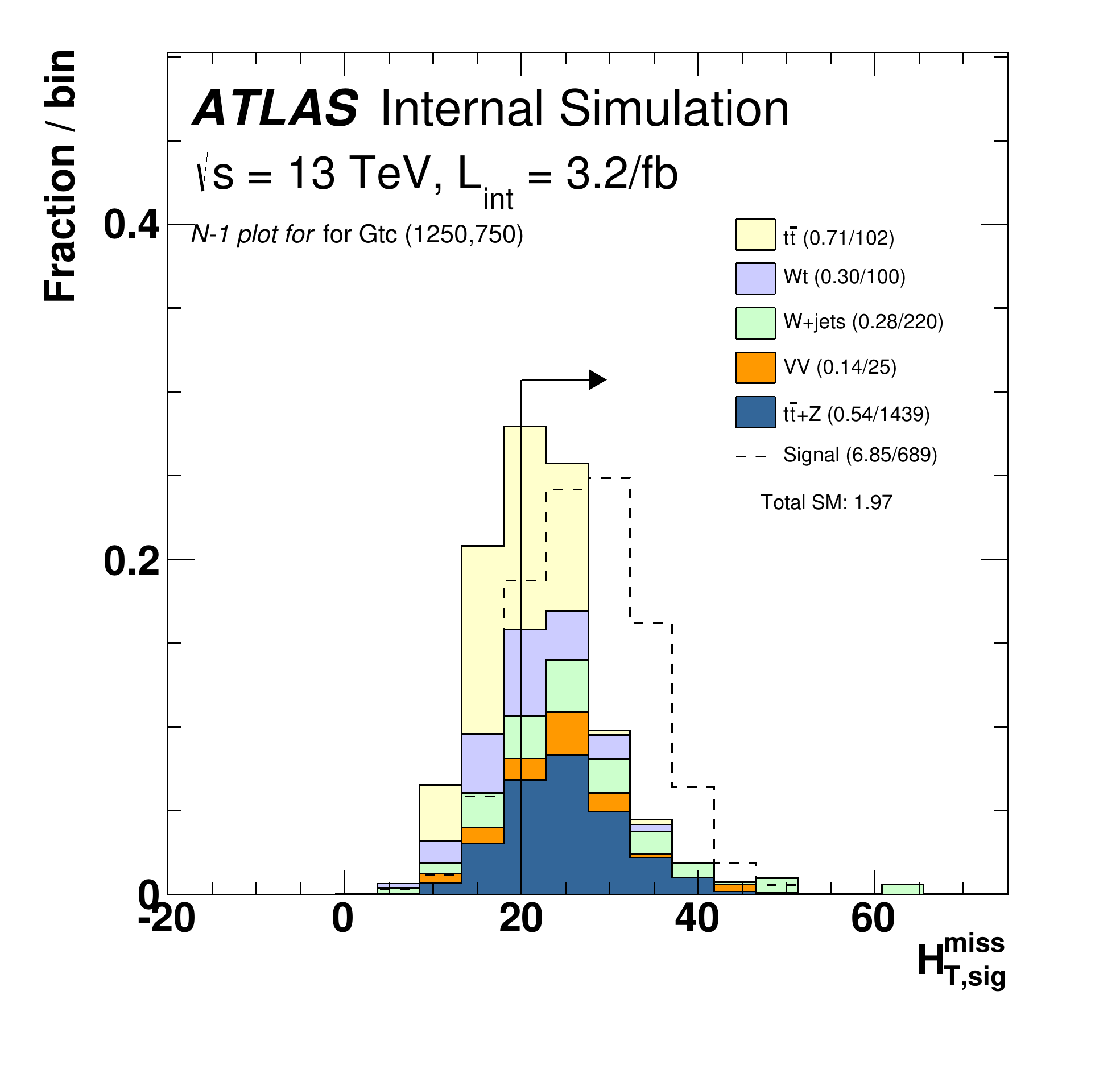}
  \caption{The same as Fig.~\ref{fig:SR13ETmiss}, but for $H_\text{T,sig}^\text{miss}$ instead of $E_\text{T}^\text{miss}$.}
  \label{fig:SR13htsig}
\end{figure}

Transverse mass variables are designed for cases where particles are lost or mis-identified; in contrast the next two variables target explicit top reconstruction.   The large-radius jet mass is shown in Fig.~\ref{fig:SR13mass}.   As discussed in Chapter~\ref{cha:bosonjets}, when the top quark has sufficient boost in the lab frame, its decay products can be captured by a single large-radius jet.  For $m_\text{stop}\sim 800$ GeV and a massless neutralino, $p_\text{T}^\text{top}\sim m_\text{stop}/2\sim 400$ GeV.  Re-clustered jets are used for this purpose, which allow for the large radius to be optimized per selection. Several radii were studied and the optimal value was found to be $R = 1.2$, which is consistent with the naive expectation that $R \sim 2m_t/p_\text{T}$.  All signal jets with $p_\text{T}>25$ GeV enter the re-clustering procedure and those with $p_\text{T}<5\%\times p_\text{T}^\text{large-radius}$ are trimmed away.  Re-clustering also allows for testing the inclusion (or not) of leptons in the re-clustering procedure. It was found that the sensitivity is higher when leptons are explicitly excluded from the re-clustering: both signal and background yields increase, but background increases more than the signal.  In particular, the signal yield for the SR13 benchmark increases by about 20\%, but the background increases by about 30\%, with the biggest increase from $t\bar{t}$.  Note that this exclusion of leptons is effectively an overlap removal procedure between large-radius jets and leptons. This overlap is trivial for re-clustered jets with a moderate boost as in this search; for large-radius jets clustered directly from calorimeter-cell clusters, the overlap is non-trivial due to significant energy deposits by electrons in the calorimeter.  The SUSY signal shows a clear top quark mass peak in the left plot of Fig.~\ref{fig:SR13mass} while the mostly dileptonic $t\bar{t}$ background has no resonant mass peak.  In the signal region, the separation is reduced due to the correlation with other variables, but the top quark mass peak for the signal and $t\bar{t}+V$ are still separated from the dileptonic $t\bar{t}$, which is concentrated at lower values of the jet mass.

\begin{figure}[htbp]
  \centering
\includegraphics[width=0.5\textwidth]{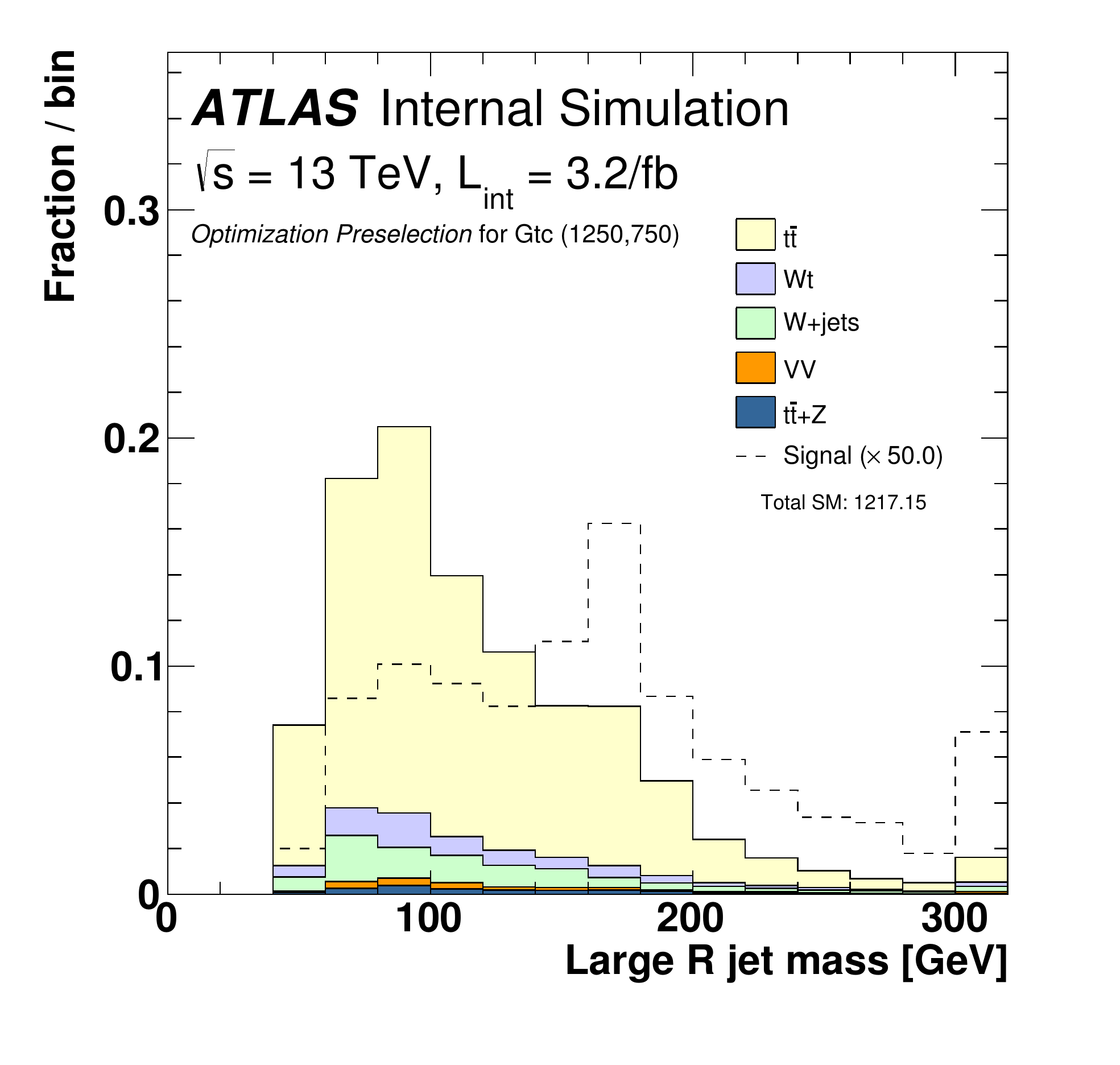}\includegraphics[width=0.5\textwidth]{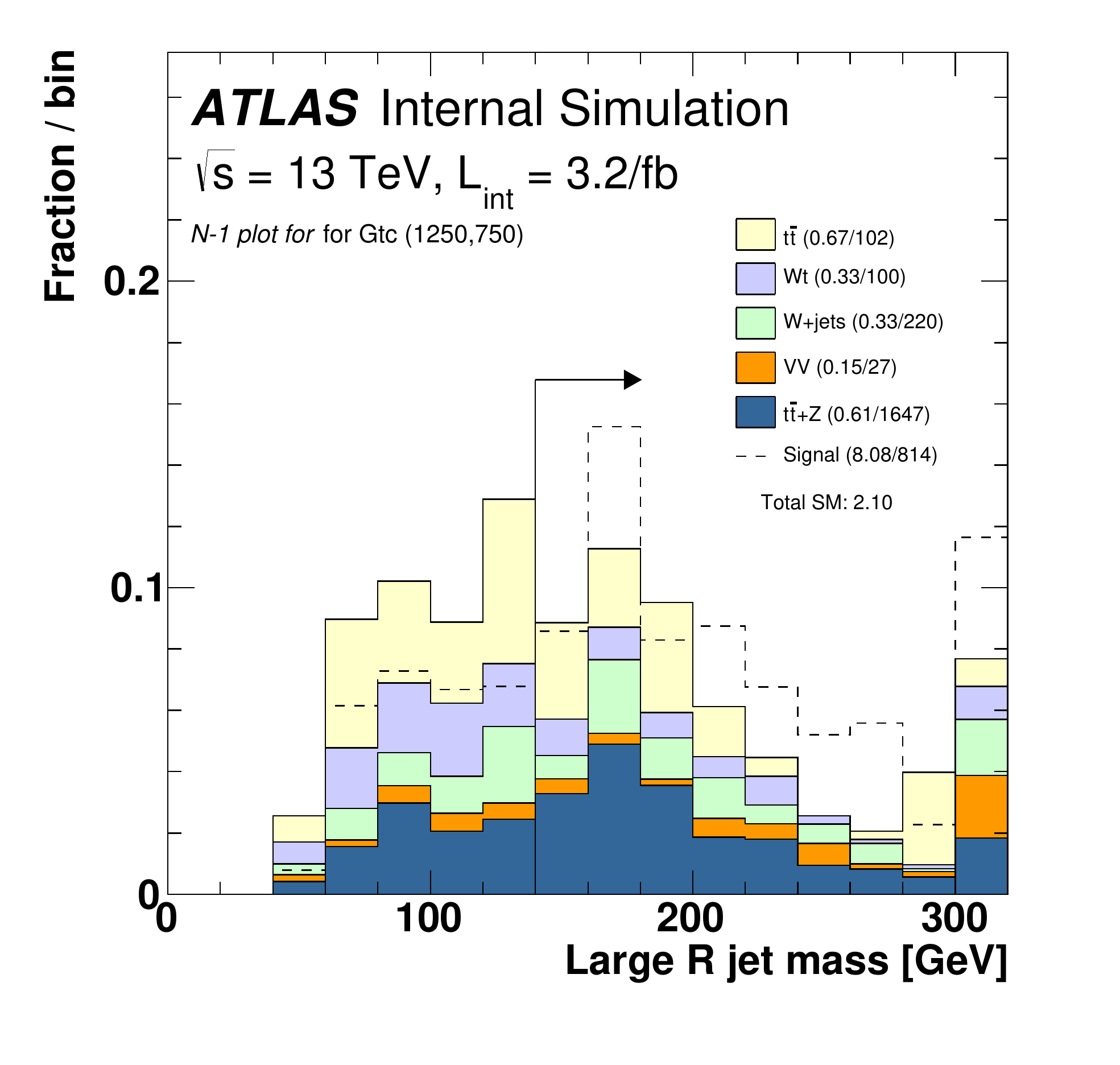}
  \caption{The distribution of the large-radius $R=1.2$ jet mass for the event preselection (top) and after all SR13 requirements except the jet mass (bottom).  Both the total background and signal yields are normalized to unity in the right plots.  An arrow indicates the signal region requirement.}
  \label{fig:SR13mass}
\end{figure}

\begin{figure}[htbp]
  \centering
\includegraphics[width=0.5\textwidth]{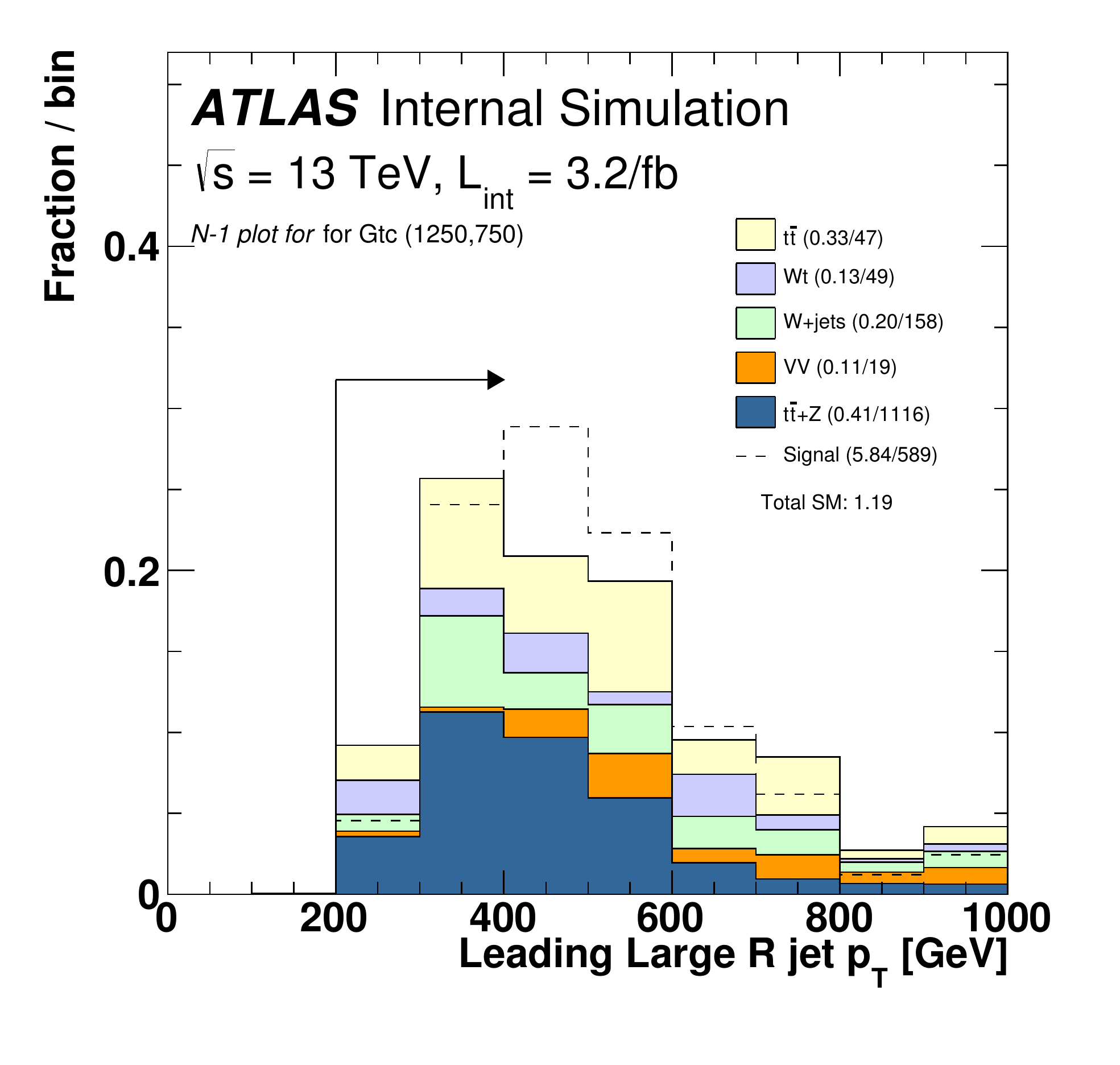}
  \caption{The distribution of the large-radius $R=1.2$ jet $p_\text{T}$ after all SR13 event selections except the large-radius jet $p_\text{T}$. For large-radius jets with $p_\text{T}>500$ GeV, the trimming actively removes low $p_\text{T}$ signal jets constituents.}
  \label{fig:SR13mass}
\end{figure}

The leading large-radius jet is generally back-to-back in $\phi$ with the $\vec{p}_\text{T}^\text{miss}$.  However, the subleading large-radius jet (if it exists) tends to be aligned with the $\vec{p}_\text{T}^\text{miss}$ in dileptonic $t\bar{t}$ events and back-to-back in signal events.  This is because the $E_\text{T}^\text{miss}$ from the neutrinos in $t\bar{t}$ events are generally close to at least the sub-leading large-radius jet formed in part by the lost or mis-identified second lepton.  In contrast, in signal events, both the hadronic and leptonic top quark candidates are recoiling from the $\vec{p}_\text{T}^\text{miss}$ from the neutralinos.  Figure~\ref{fig:SR13dphi} illustrates these properties of both the leading and sub-leading large-radius jets.  About 25\% of background events have a second signal large-radius jet ($p_\text{T} > 150$ GeV, $m_\text{jet}>50$ GeV, and $|\eta|<2.5$) while only about $10\%$ of signal events have such a jet.

\begin{figure}[htbp]
  \centering
\includegraphics[width=0.5\textwidth]{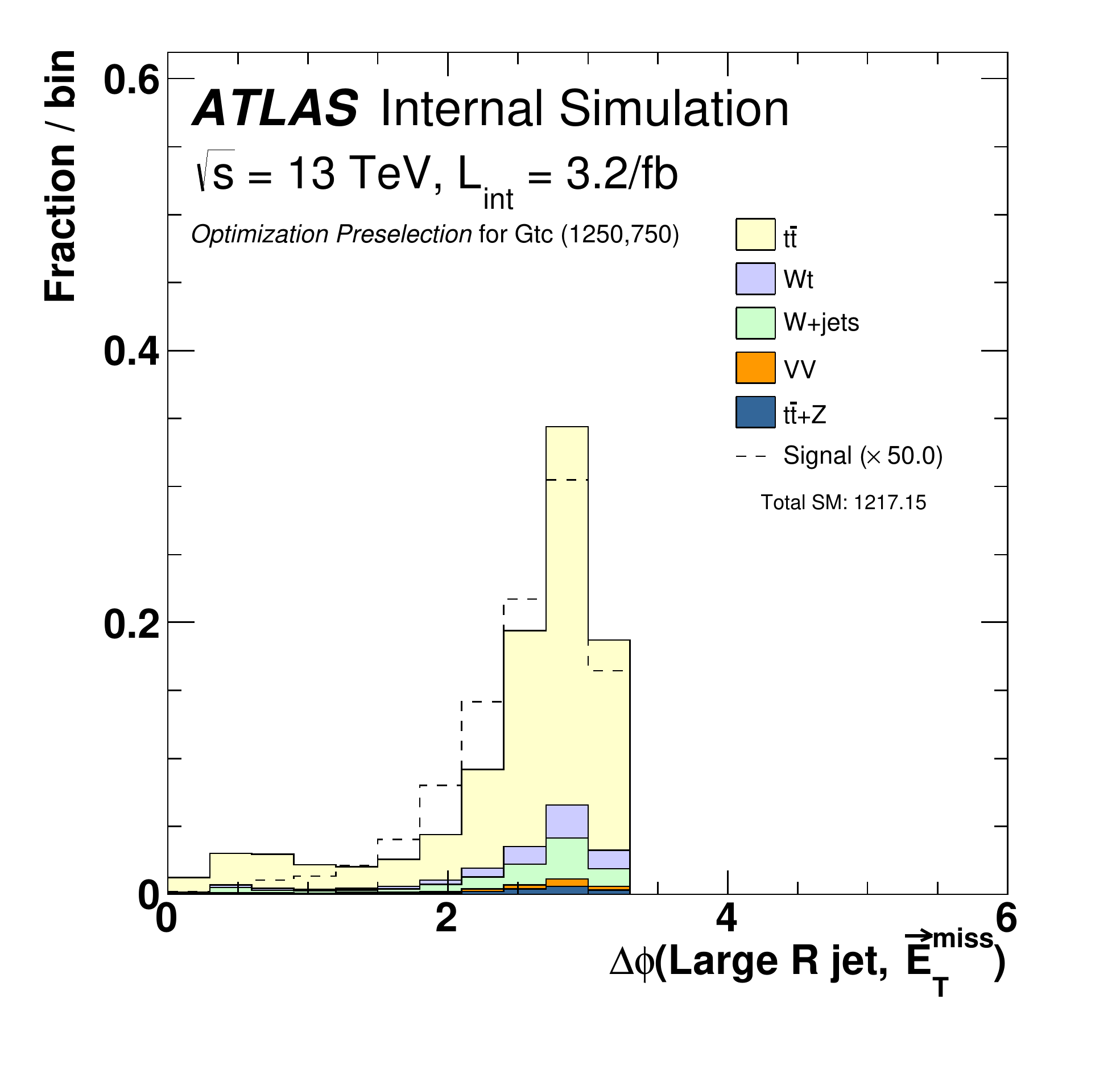}\includegraphics[width=0.5\textwidth]{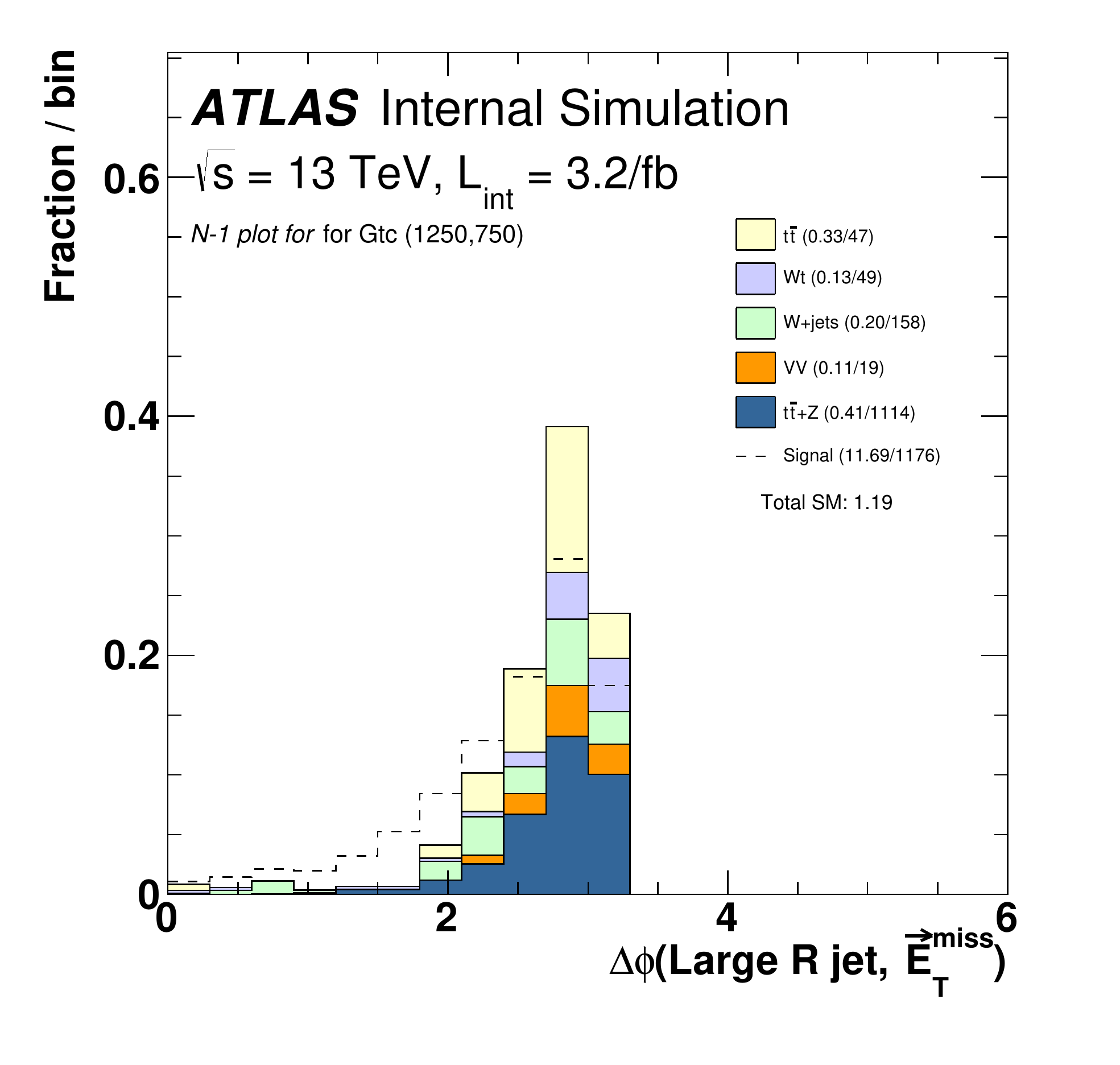}\\
\includegraphics[width=0.5\textwidth]{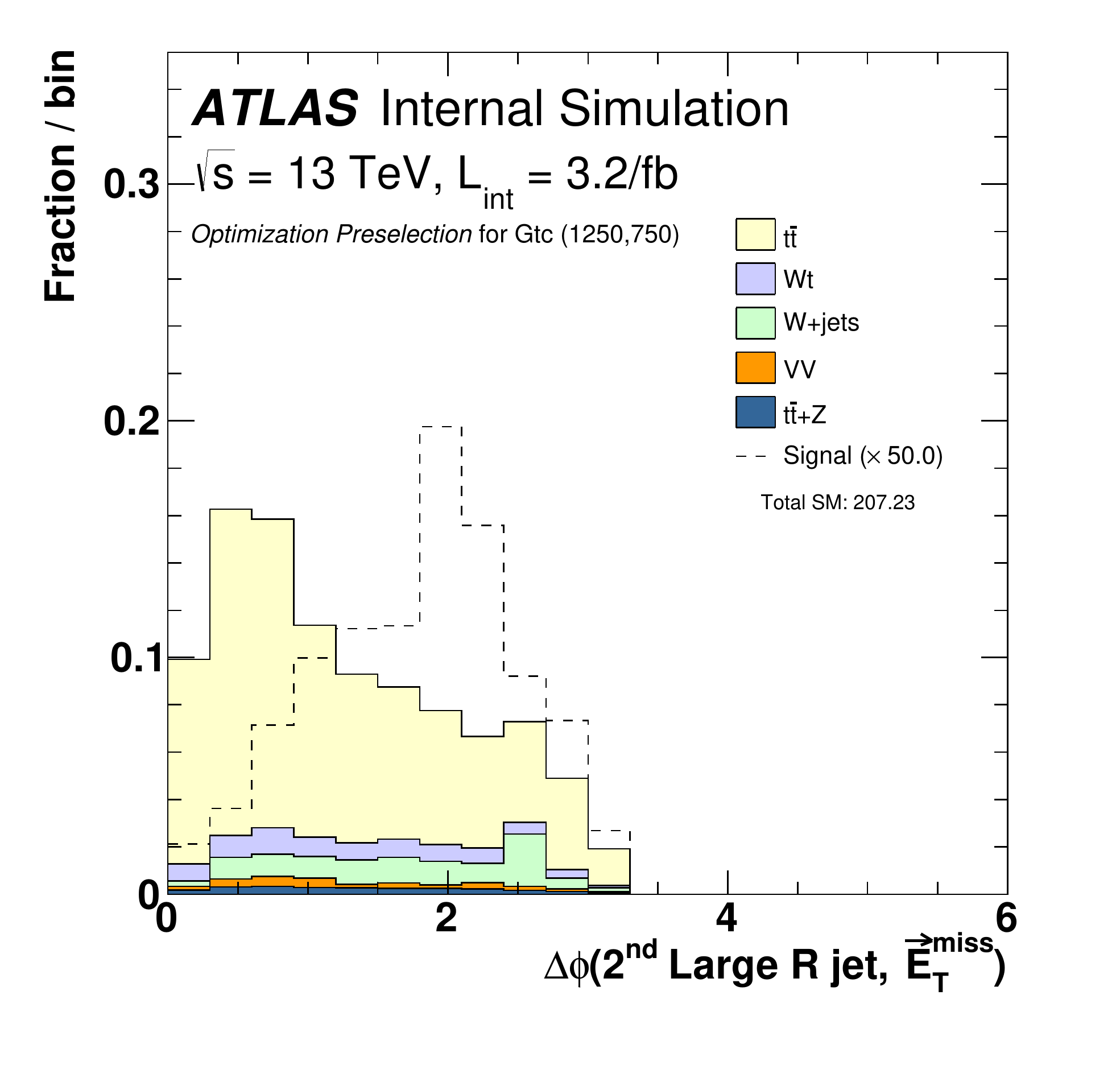}\includegraphics[width=0.5\textwidth]{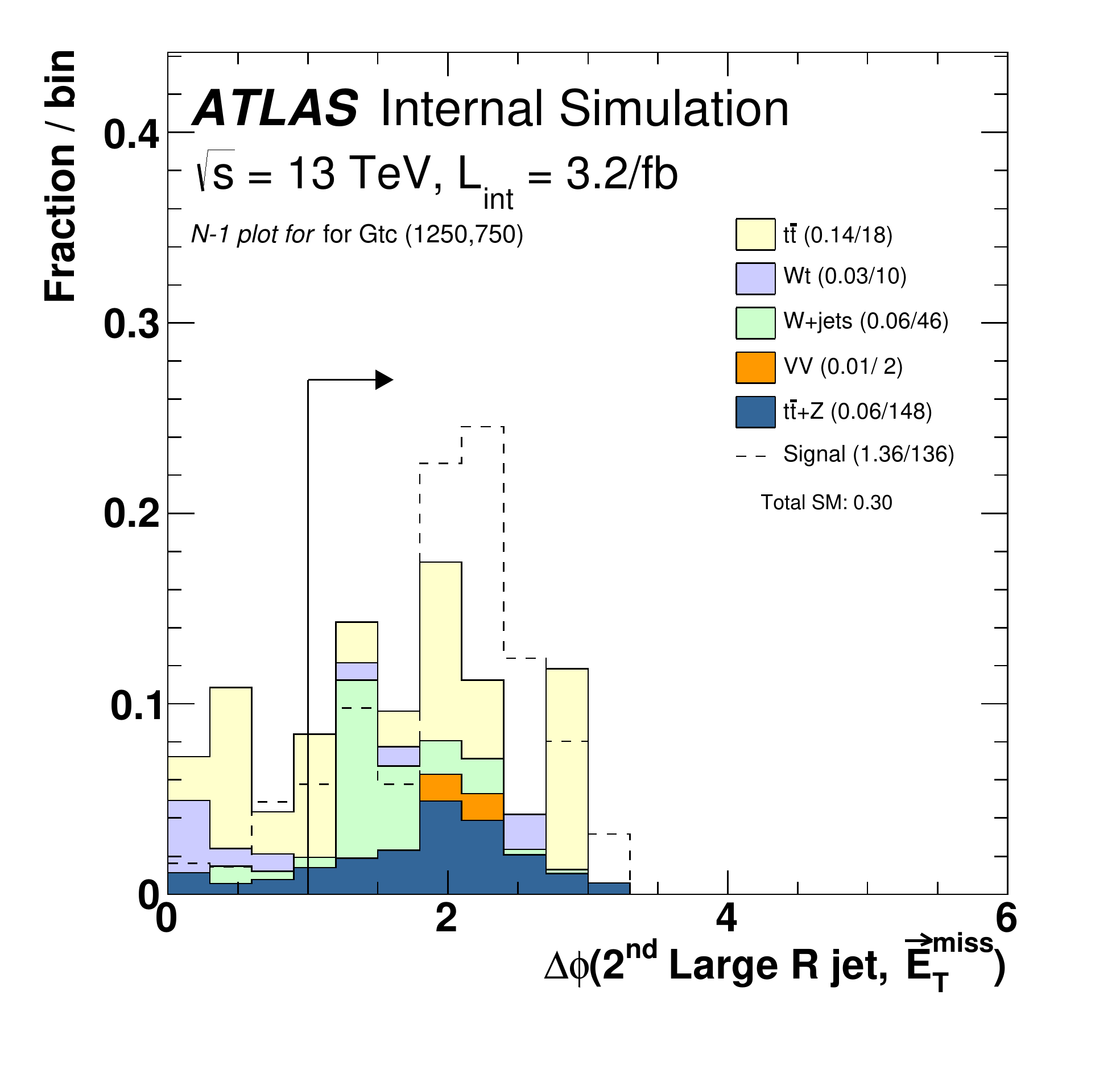}
  \caption{Top: the $\Delta\phi$ between the leading large radius jet and the $p_\text{T}^\text{miss}$.  Bottom: the same as the top, but for the subleading large radius jet if one exists.  Large radius jets are required to have $p_\text{T}>150$ GeV, $m>50$ GeV, and $|\eta|<2.5$.}
  \label{fig:SR13dphi}
\end{figure}

In addition to tagging the hadronically decaying top quark, it is possible to identify the boosted leptonically decaying top quark from the proximity of the lepton with a $b$-jet due to the same $2m/p_\text{T}$ scaling from above. Figure~\ref{fig:SR13dRbl} shows the distribution of $\Delta R(b,\ell)$ using the highest $p_\text{T}$ $b$-jet.   This distance tends to be $\lesssim 1$ for the signal, but there is a heavy tail due to combinatorics.  The $t\bar{t}$ background also has leptonically decaying boosted top quarks, but with less boost than for the signal.  The mass of the $b$-jet and lepton pair also contains information about the top quark mass, but $m_{b\ell}$ can be naturally large for the background: for the correct pairing it has the same distribution as the signal ($m_{b\ell}$ is a Lorentz invariant) and for the incorrect $b\ell$ pairing, it is naturally large due to the signifcant distance between the $b$ and the $\ell$.

\begin{figure}[htbp]
  \centering
\includegraphics[width=0.5\textwidth]{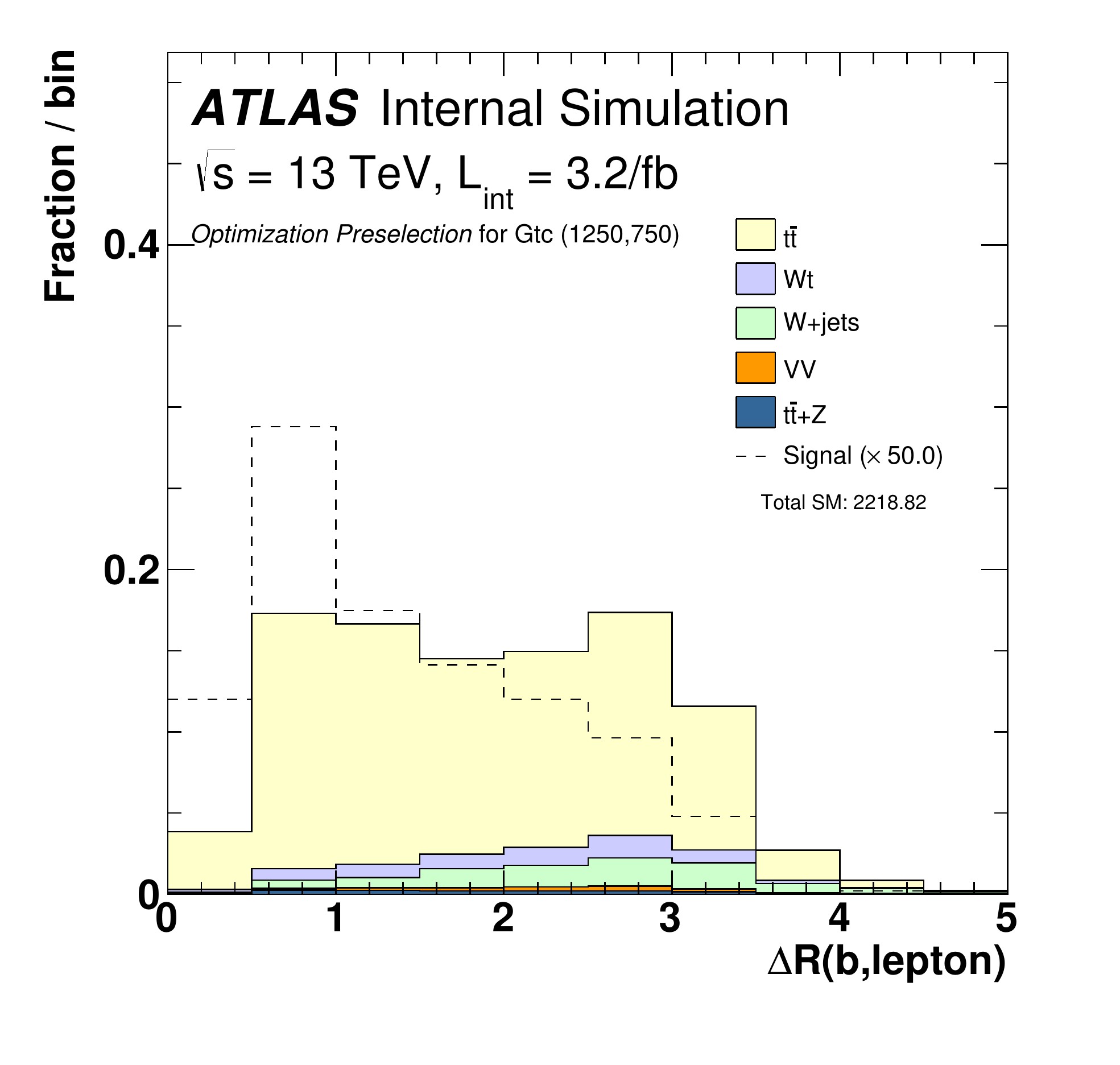}\includegraphics[width=0.5\textwidth]{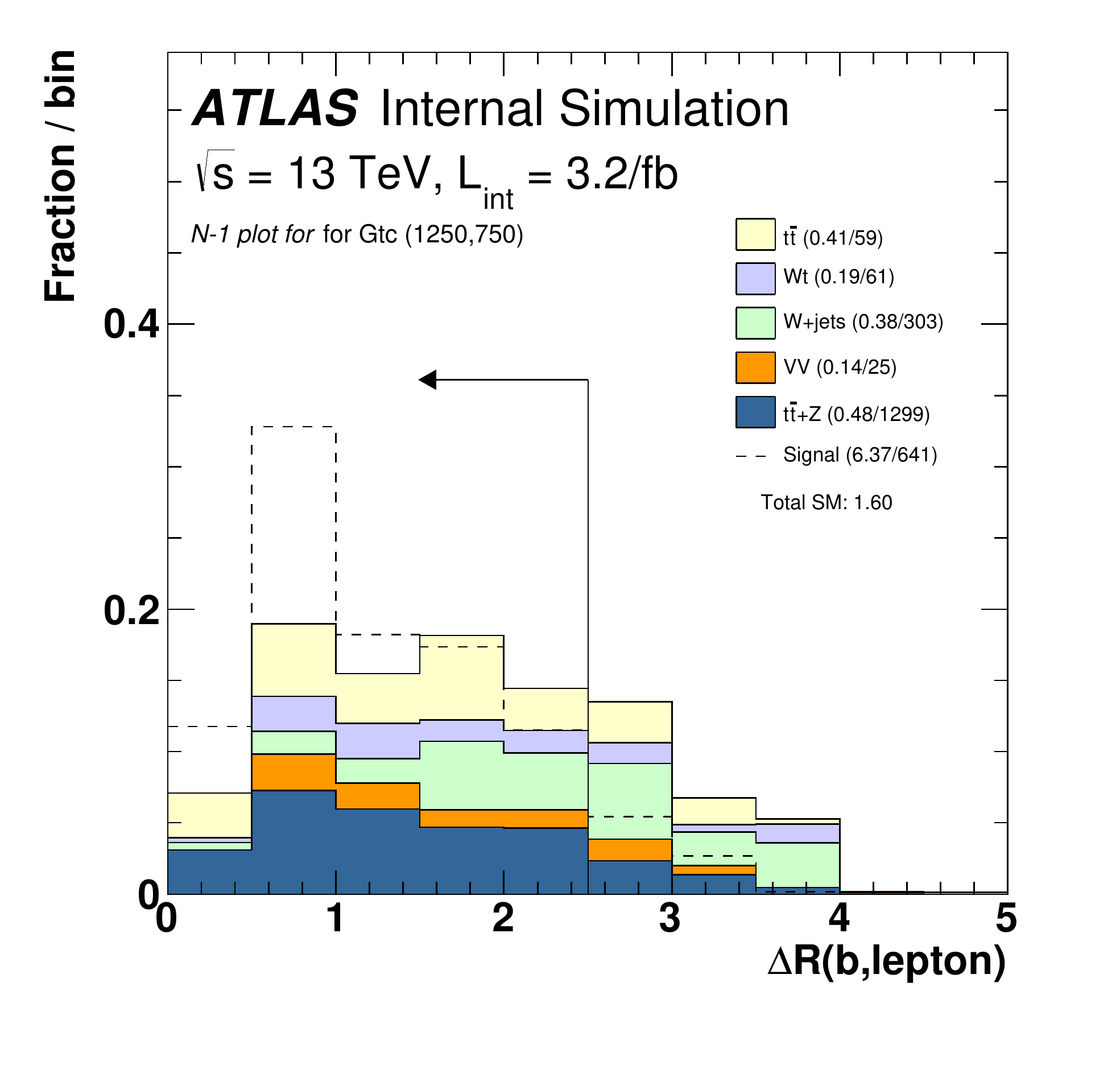}
  \caption{Same as Fig.~\ref{fig:SR13mass}, but with the $\Delta R$ between the highest $p_\text{T}$ $b$-jet and the selected lepton instead of the jet mass.}
  \label{fig:SR13dRbl}
\end{figure}

The distribution of $m_\text{T2}^\tau$ is shown in Fig.~\ref{fig:SR13mt2tau}.  As expected, the background distribution has an endpoint near $m_W$ while the signal is shifted toward much higher values.  There are come geometric orientations (the {\it unbalanced configuration} - see Sec.~\ref{sec:numericalmethodsmt2}) such that $m_\text{T2}^\tau$ is exactly zero.  This results in a finite loss in acceptance for any positive threshold on $m_\text{T2}^\tau$ and is the source of the non-negligible signal yield in the first bin of the histograms in Fig.~\ref{fig:SR13mt2tau}.  Even though the $W$+jets and $t\bar{t}+V$ processes are a significant contribution to the SR, they are largely absent from Fig.~\ref{fig:SR13mt2tau} because they do not usually have a second lepton, whereas $t\bar{t}$, $Wt$, and $VV$ events can use a hadronically decaying $\tau$ to exceed the $m_\text{T}$ threshold.   The number of $t\bar{t}$, $Wt$, and $VV$ events that are removed by the $m_\text{T2}^\tau$ requirement (i.e. present in Fig.~\ref{fig:SR13mt2tau}), $\sim(0.4,0.2, 0.1)$ are comparable to the total yield of these backgrounds with the full signal region selection, $\sim(0.3,0.1, 0.1)$ (see e.g. Fig.~\ref{fig:SR13mass}).

\begin{figure}[htbp]
  \centering
\includegraphics[width=0.5\textwidth]{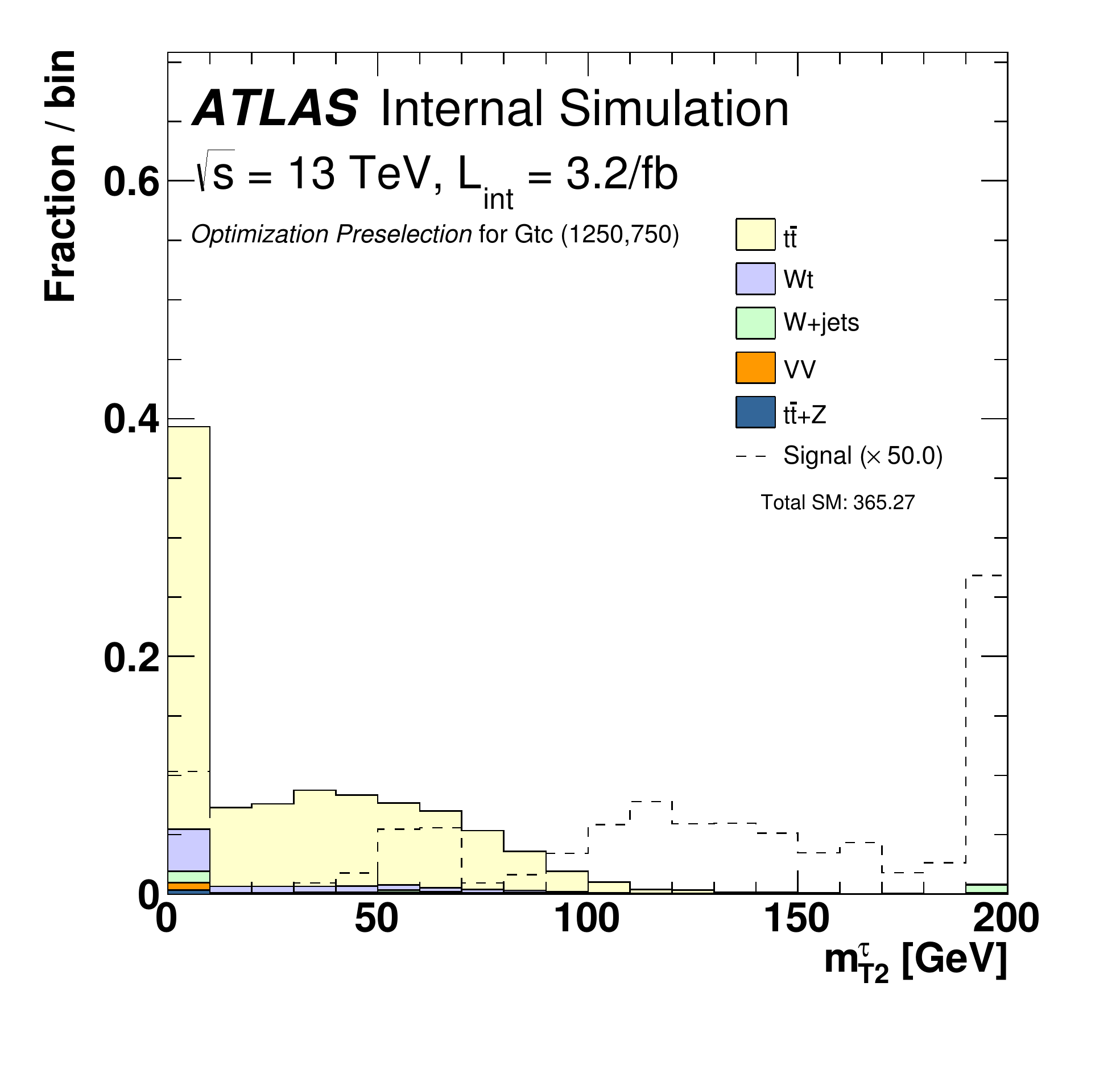}\includegraphics[width=0.5\textwidth]{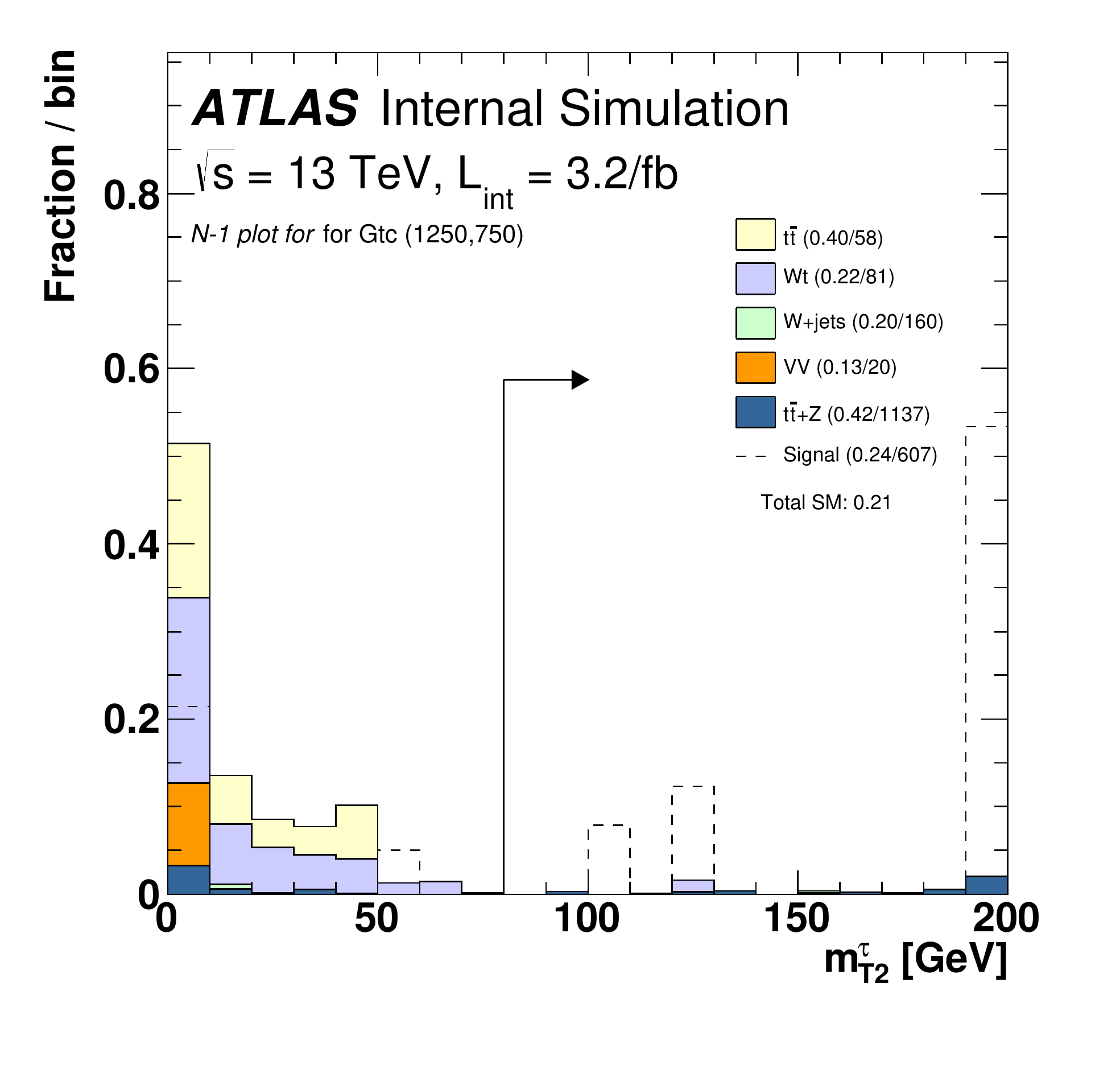}
  \caption{The normalized distribution of the $m_\text{T2}^\tau$ for the event preselection (left) and after all SR13 requirements except $m_\text{T}$ (right). An arrow indicates the signal region requirement in the right plot.}
  \label{fig:SR13mt2tau}
\end{figure}	

Figure~\ref{fig:SR13nbjets} shows the $b$-jet multiplicity at preselection and in the signal region.  The $b$-tagging algorithm is configured to be $77\%$ efficient for inclusive $t\bar{t}$ events.  Table~\ref{sr2selection_b} shows the yields in SR13 for a variety of $b$-tagging working points, ranging from $60\%$ to $85\%$ efficiency.  There is not a strong dependence on the working point for efficiencies below $77\%$, but the $85\%$ efficiency is clearly worse than the others as proportionally more background events pass the $\geq 1$ $b$-tagged jet requirement.  This is largely due to the significant increase in the $W$+jets background.  Figure~\ref{fig:tNhightbtags} shows the flavor breakdown of the $W$+jets $b$-tagged jet multiplicity for tNhigh, which is similar to SR13.  Between the $70\%$ and $80\%$ working points, there is a large increase in the charm-jet contribution to the one $b$-tagged jet bin.

Signal events in Fig.~\ref{fig:SR13nbjets} are significantly more likely to have a second reconstructed $b$-jet compared to background events.  However, requiring at least two $b$-tagged jets would remove too many signal events for such a threshold to be useful.  In the future, one could exploit the asymmetry in the $n_\text{$b$-jets}$ distribution by using event weights or dividing up the single bin SR into (at least two) bins.  Figure~\ref{fig:usingbjetweights} illustrates the improvement one might expect from using event-weights.  The optimal weights are signal-model dependent, but generic nearly optimal weights can be derived for a broad class of models.

\begin{table}[h!] 
\vspace{5mm}
\centering
\begin{tabular}{|ccccc|}
\hline
Process           & 60\%   & 70\% & 77\% &  85\%      \\
\hline
\ttbar\ 1L &0.0  &0.0 &0.0 &0.0 \\
\ttbar\ 1L1$\tau$ &0.1  &0.1&0.2&0.2\\
\ttbar\ 2L &0.1&0.1&0.1&0.1 \\
\ttbar\ total & 0.2&0.2&0.3&0.3\\
Single Top & 0.1&0.1&0.1& 0.1\\
$W$+jets & 0.1&0.1&0.2& 0.5\\
Dibosons &0.1&0.1&0.1& 0.2\\
$t\bar{t}+V$ & 0.3&0.4&0.4&0.4 \\
\hline
Total SM & 0.8 &1.0&1.1&1.6\\
Gtc ($1250,750$) & 5.0&5.1&5.4&5.7\\
Discovery $\sigma$ (30\% syst.) & 3.1 & 3.0 & 3.0 & 2.7 \\
\hline
\end{tabular}
\caption{Yields for the SR13 defined in Table~\ref{tab:signalregionselections} but with various $b$-tagging working points.  The last row is an approximate significance using the RooStat~\cite{Moneta:2010pm} {\tt NumberCountingUtils} routine {\tt BinomialExpZ}.  A $p$-value is computed with a likelihood given by the product of a Poisson term for the statistical uncertainty and a Poisson term for the systematic uncertainty treated as a statistical uncertainty from an auxiliary measurement, i.e. a Poisson with mean $\tau=1/(30\%)^2$ so that the fractional uncertainty of the auxiliary measurement is $1/\sqrt{\tau}=30\%$.  The conversion from $p$-value to $\sigma$ is given by $\sigma = \Phi^{-1}(1-p)$, for $\Phi$ the Gaussian cumulative distribution function.}
\label{sr2selection_b}
\end{table}

\begin{figure}[htbp]
  \centering
  \includegraphics[width=0.5\textwidth]{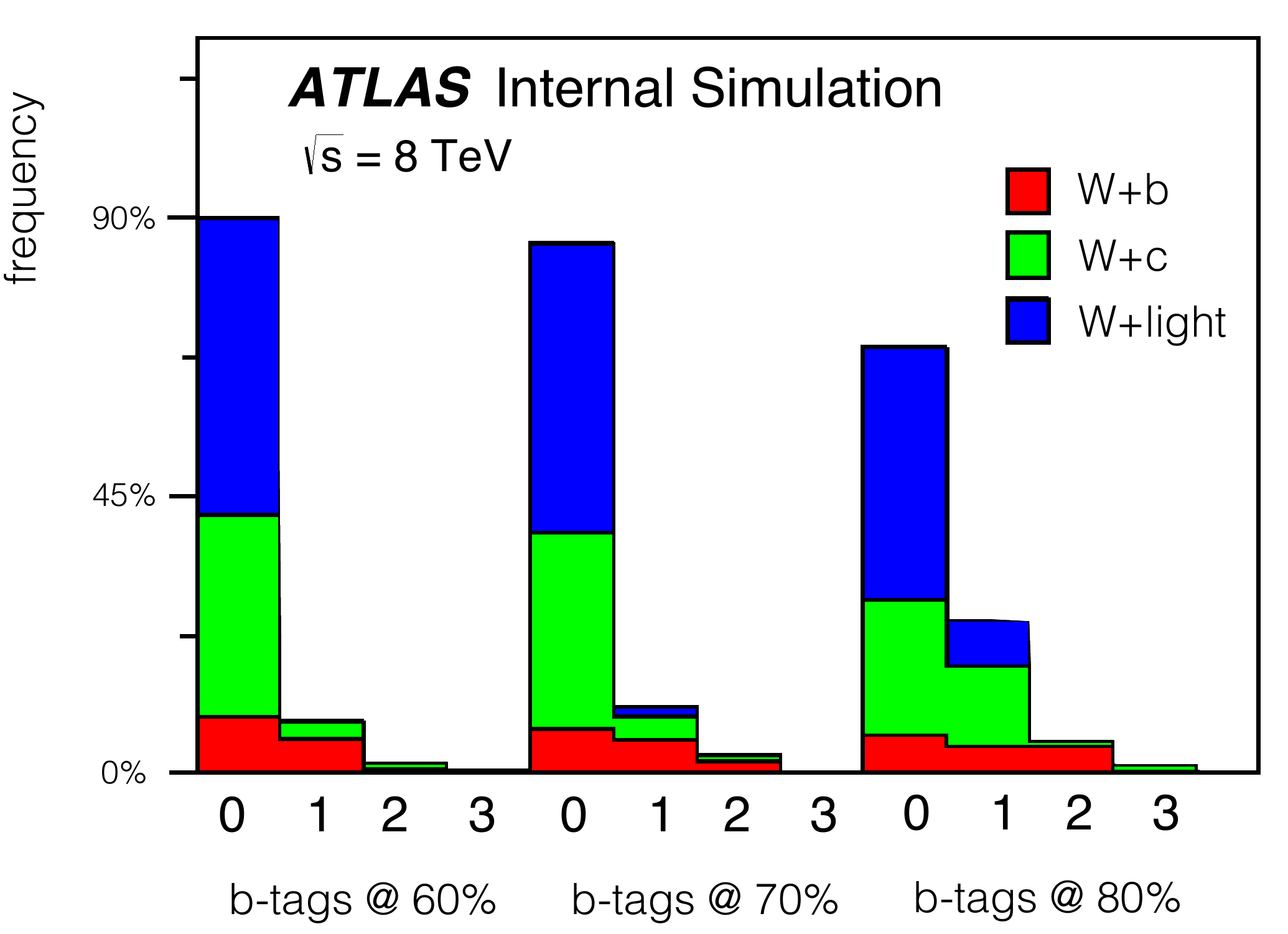}
  \caption{The $b$-tagged jet multiplicity for $W$+jets events in tNhigh broken down by jet flavor.  The defining efficiency of the $b$-tagging working point is evaluated in inclusive $t\bar{t}$ events.}
  \label{fig:tNhightbtags}
\end{figure}

\begin{figure}[htbp]
  \centering
  \includegraphics[width=0.5\textwidth]{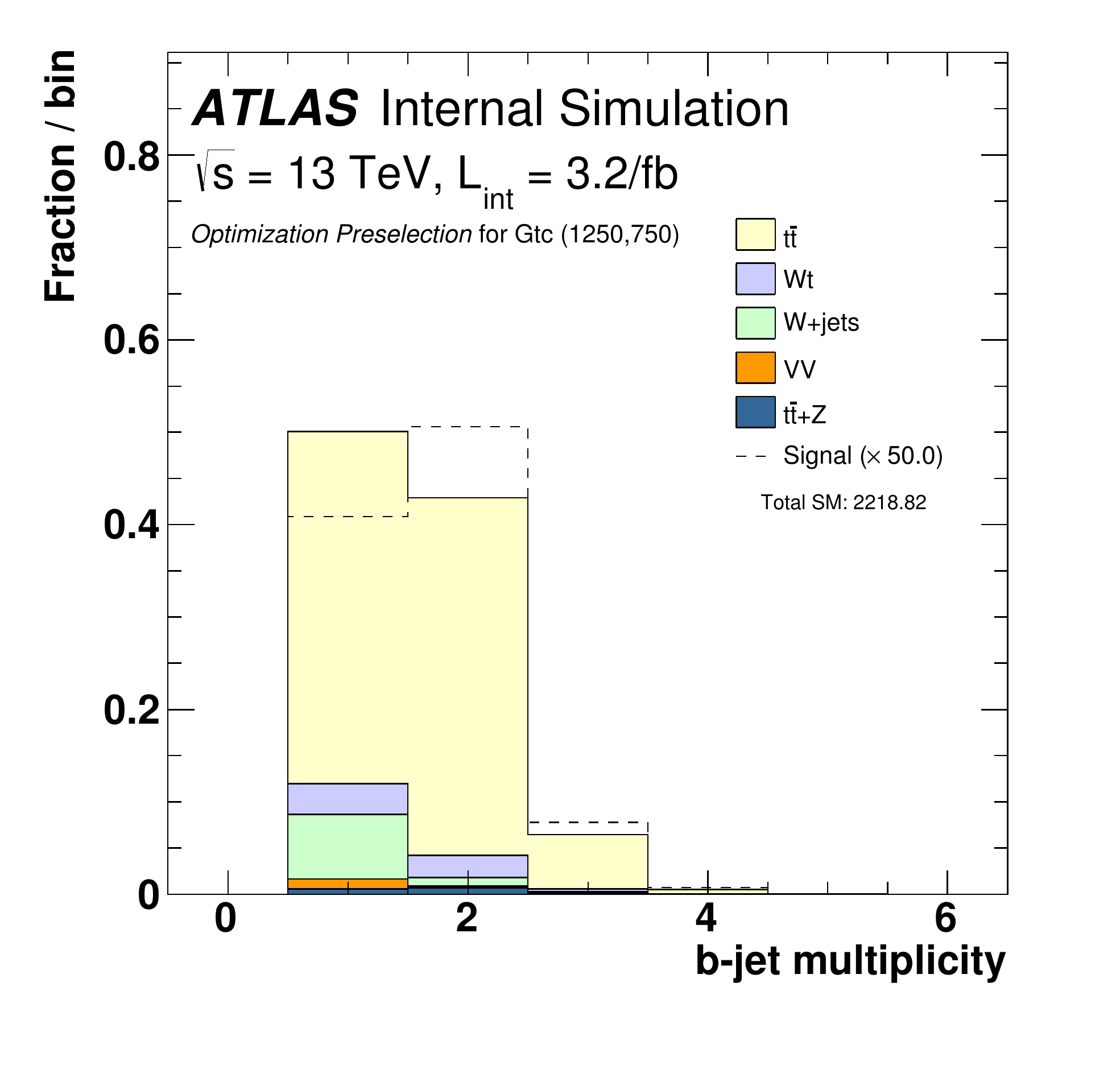}\includegraphics[width=0.5\textwidth]{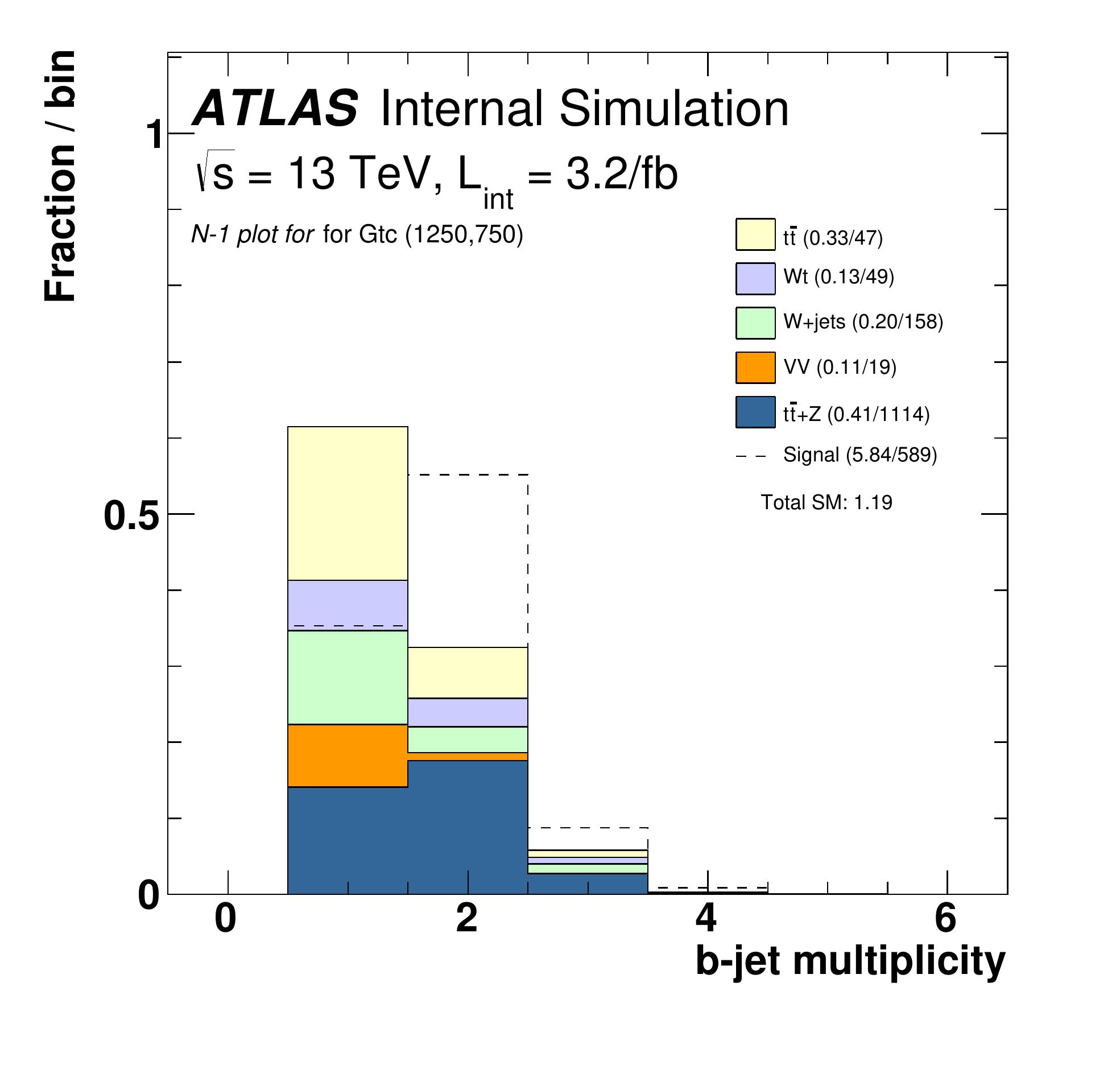}
  \caption{The distribution of the number of $b$-tagged jets at preselection (left) and in SR13 (right).}
  \label{fig:SR13nbjets}
\end{figure}

\begin{figure}[htbp]
  \centering
  \includegraphics[width=0.5\textwidth]{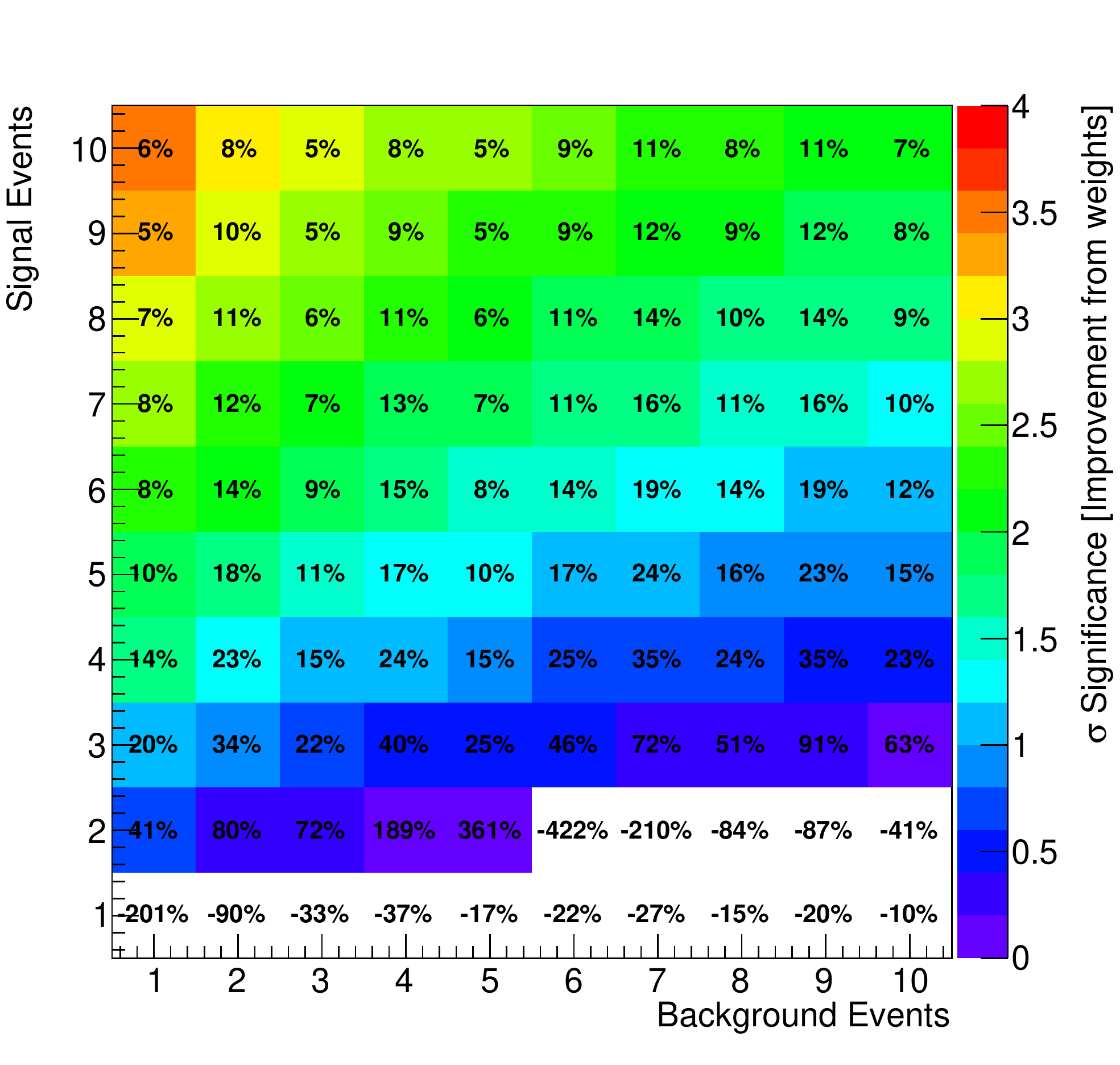}
  \caption{The median significance = $\Phi^{-1}(1-\text{CL}_s)$ (see Sec.~\ref{sec:susy:stats}) as a function of the number of background and signal events, assuming zero systematic uncertainty.  The numbers show the gain in significance when weighting each event by $2$ if there are $\geq 2$ $b$-jets.  With this simple setup, $2$ is nearly optimal, though it depends on $s$ and $b$ (see Sec.~\ref{sec:shapefitsetup}).  The relative frequencies of $b$-tagged jets is taken from Fig.~\ref{fig:SR13nbjets}.}
  \label{fig:usingbjetweights}
\end{figure}

		\clearpage
		
		\section{Compressed Signal Region}	
	\label{sec:shapefitregion}		

	Adding multiple bins to a signal region significantly increases the complexity of the optimization procedure.  To simplify the approach, the tN\_diag signal region uses SR1 as a base.  Two variables are loosened from their requirements in SR1 to define the bins of tN\_diag; the tightest bin roughly corresponds to the SR1 event selection.  Figure~\ref{fig:shapefit_matrix} shows the distribution of various kinematic variables with a one-lepton, four jets at $p_\text{T} > 80, 60, 40, 25$ GeV and $E_\text{T}^\text{miss}>100$ GeV preselection for $t\bar{t}$ and $\tilde{t}_1\tilde{t}_1$ with $(m_\text{stop},m_\text{LSP})=(250,50)$.  It is clear that the $m_\text{T}$ is the most discriminating variable given this preselection and is therefore chosen to define the shape fit.  The $E_\text{T}^\text{miss}$ is also a useful discriminating variable, which is partially hidden from Fig.~\ref{fig:shapefit_matrix} due to the $E_\text{T}^\text{miss}>100$ GeV requirement.  Figure~\ref{fig:shapefit_matrix2} shows the $E_\text{T}^\text{miss}, am_\text{T2}$ and $m_\text{had}^\text{top}$ after an upper requirement on $m_\text{T}$.  The signal and background distributions are nearly the same, but the the $E_\text{T}^\text{miss}$ shows the most difference and is therefore used as a second defining variable of the multibin signal region.  Even though the likelihood ratio may not significantly change as a function of $E_\text{T}^\text{miss}$, the modeling, in particular for the trigger, may depend on $E_\text{T}^\text{miss}$.  Therefore, the background normalization parameters and key systematic uncertainty nuisance parameters are assigned per $E_\text{T}^\text{miss}$ bin.

\begin{figure}[h!]
  \centering
\includegraphics[width=0.5\textwidth]{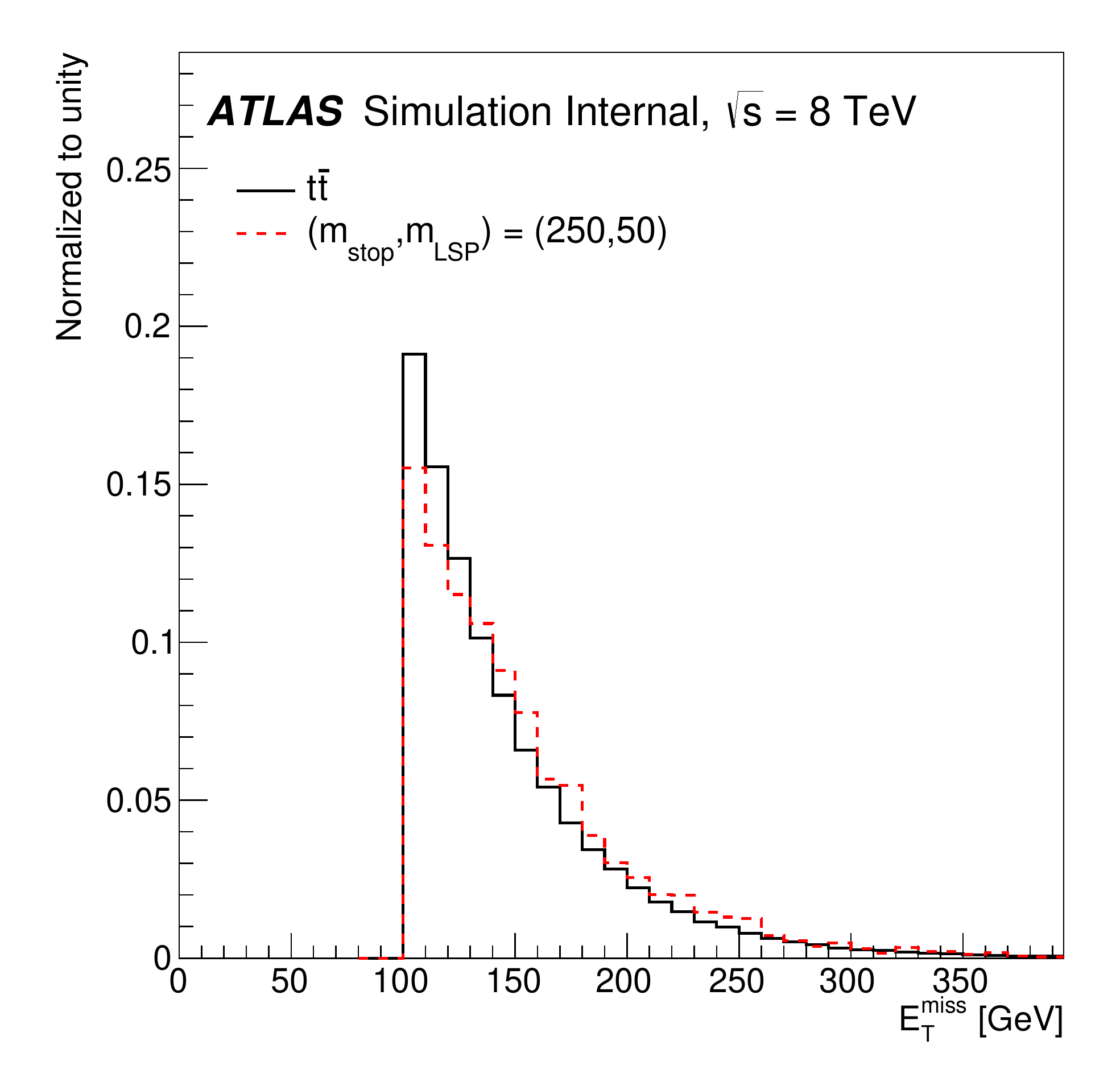}\includegraphics[width=0.5\textwidth]{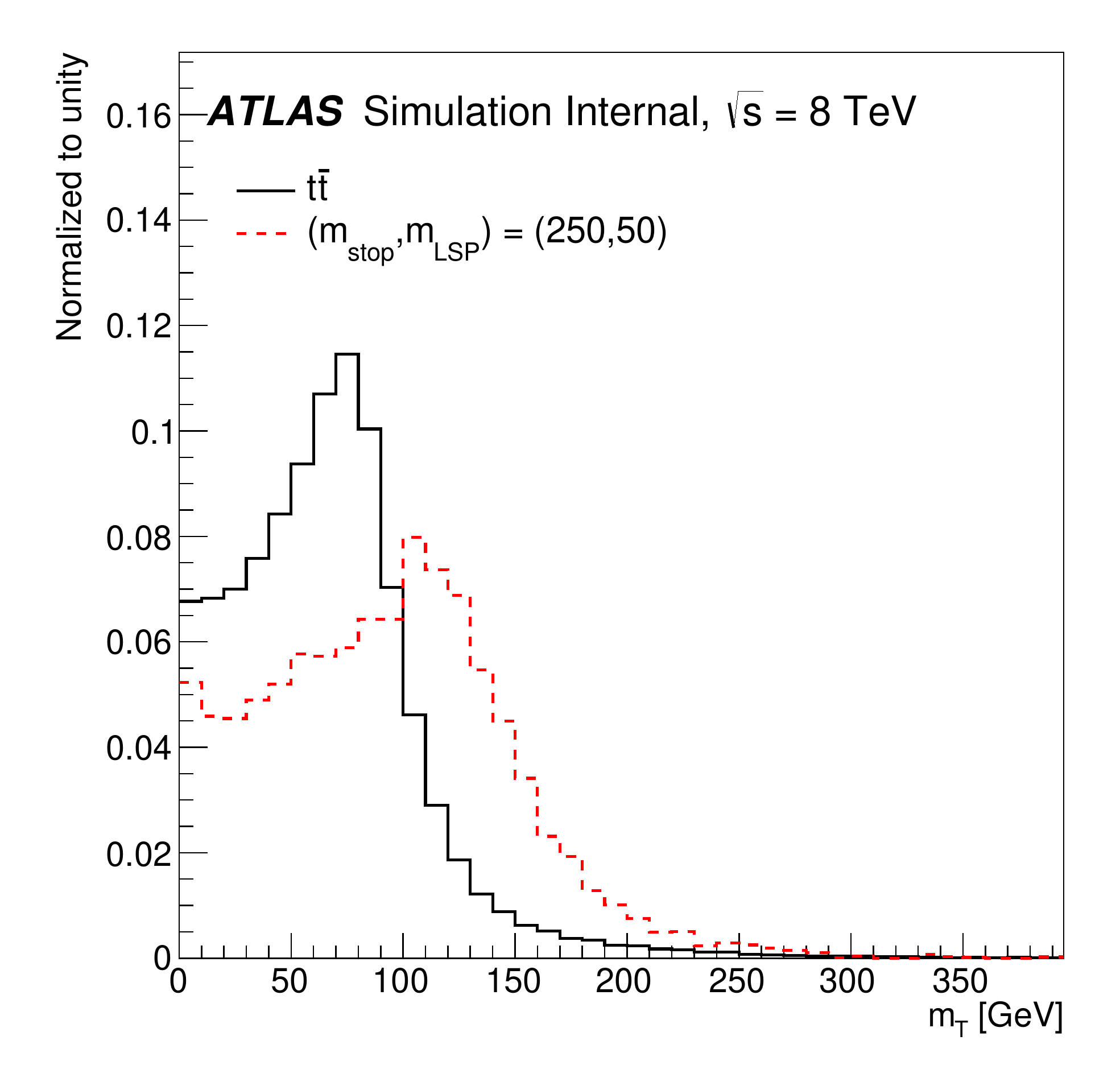}\\
\includegraphics[width=0.5\textwidth]{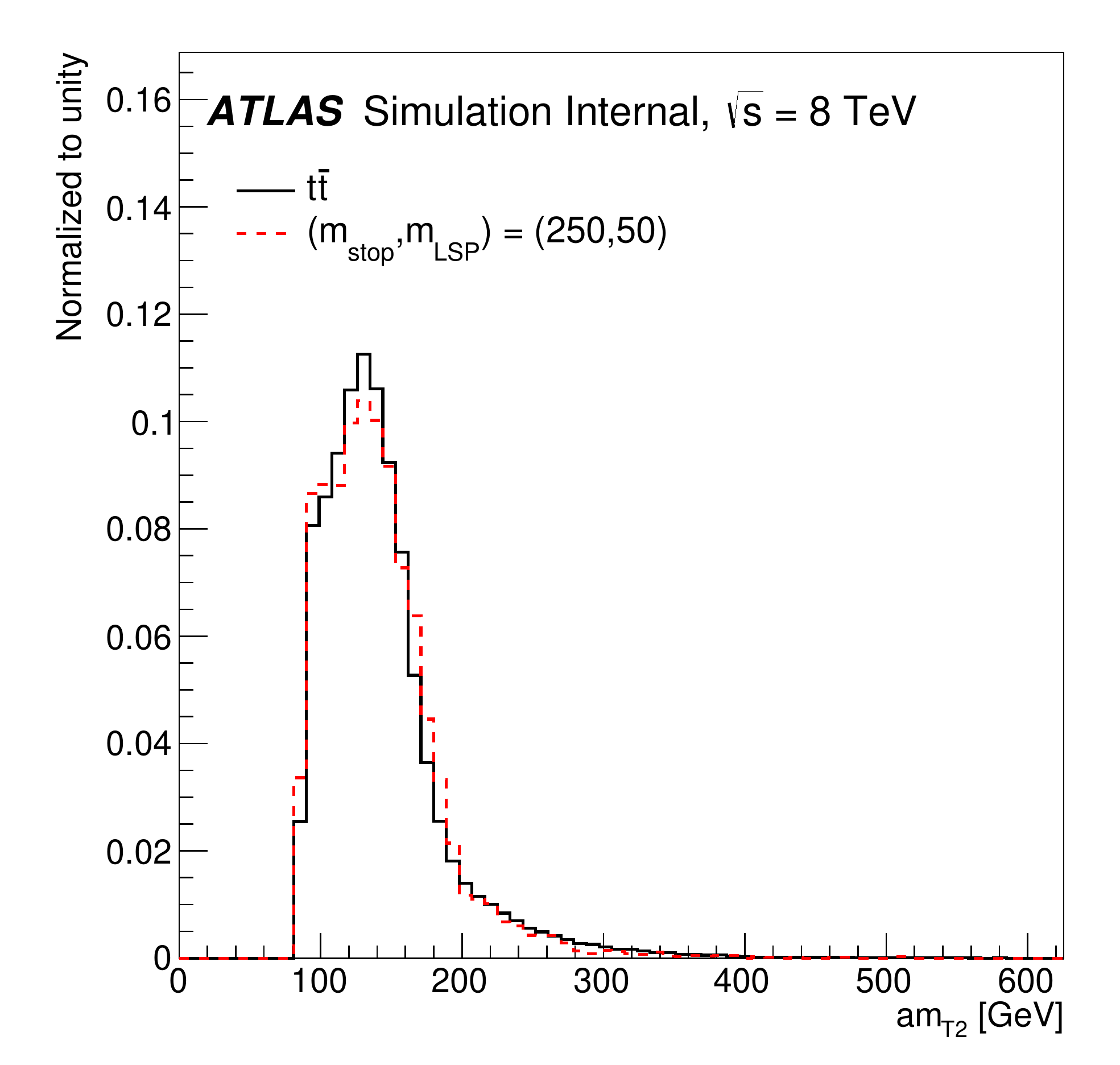}\includegraphics[width=0.5\textwidth]{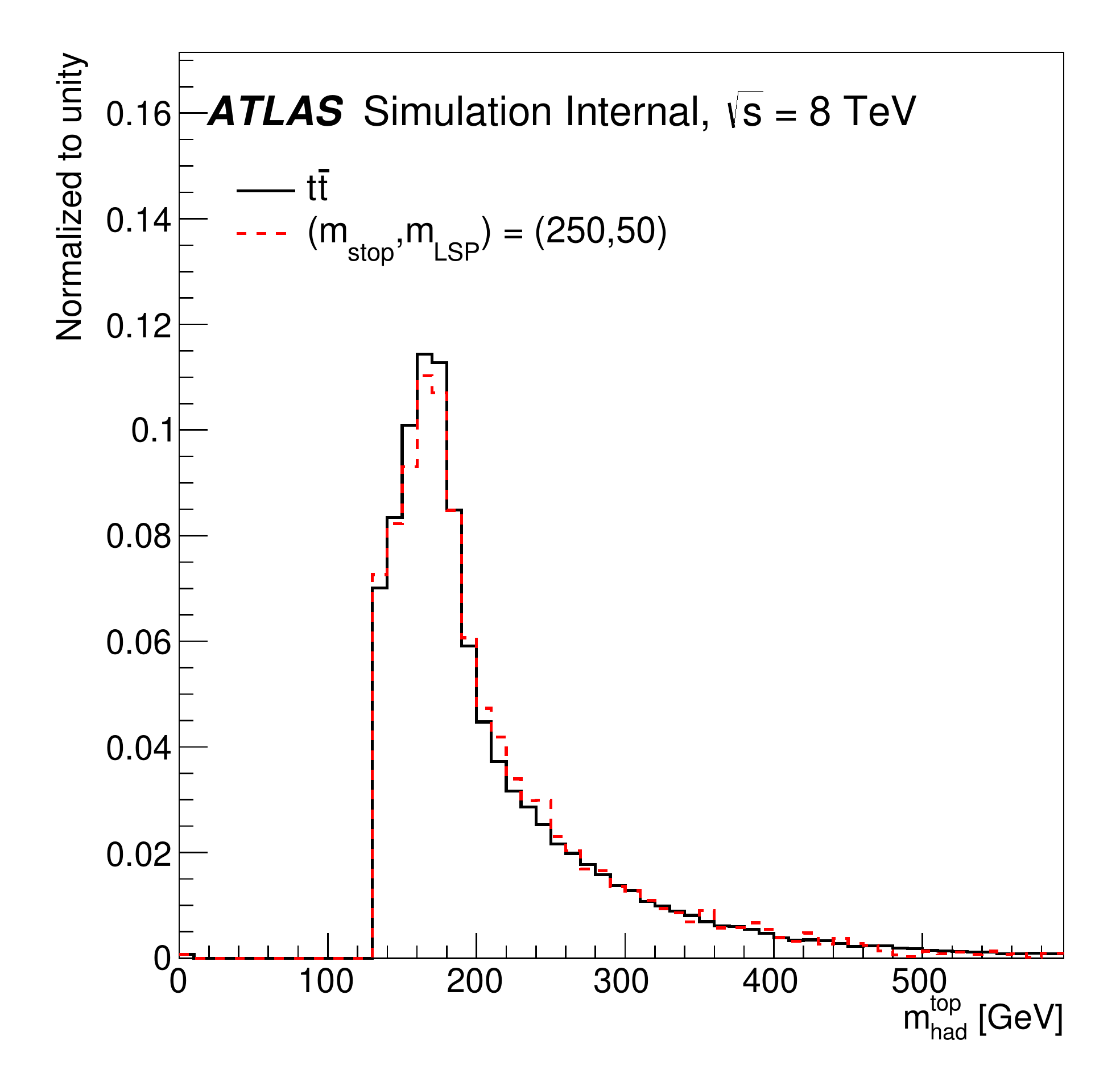}
  \caption{The distribution of $E_\text{T}^\text{miss}, m_\text{T}, am_\text{T2}$, and $m_\text{had}^\text{top}$ with a a one-lepton, four jets at $p_\text{T} > 80, 60, 40, 25$ GeV and $E_\text{T}^\text{miss}>100$ GeV preselection.}
  \label{fig:shapefit_matrix}
\end{figure}		

\begin{figure}[h!]
  \centering
\includegraphics[width=0.33\textwidth]{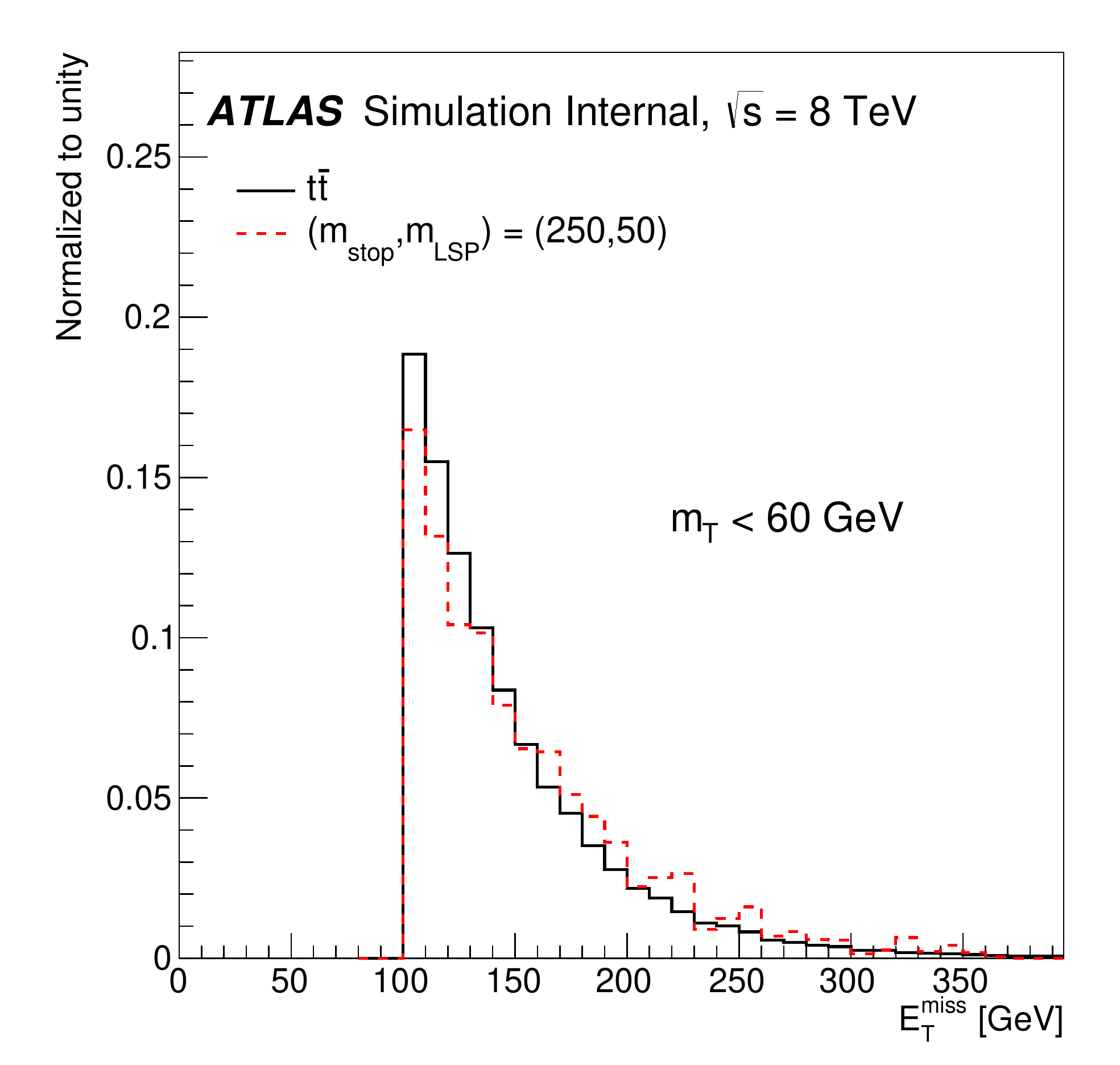}
\includegraphics[width=0.33\textwidth]{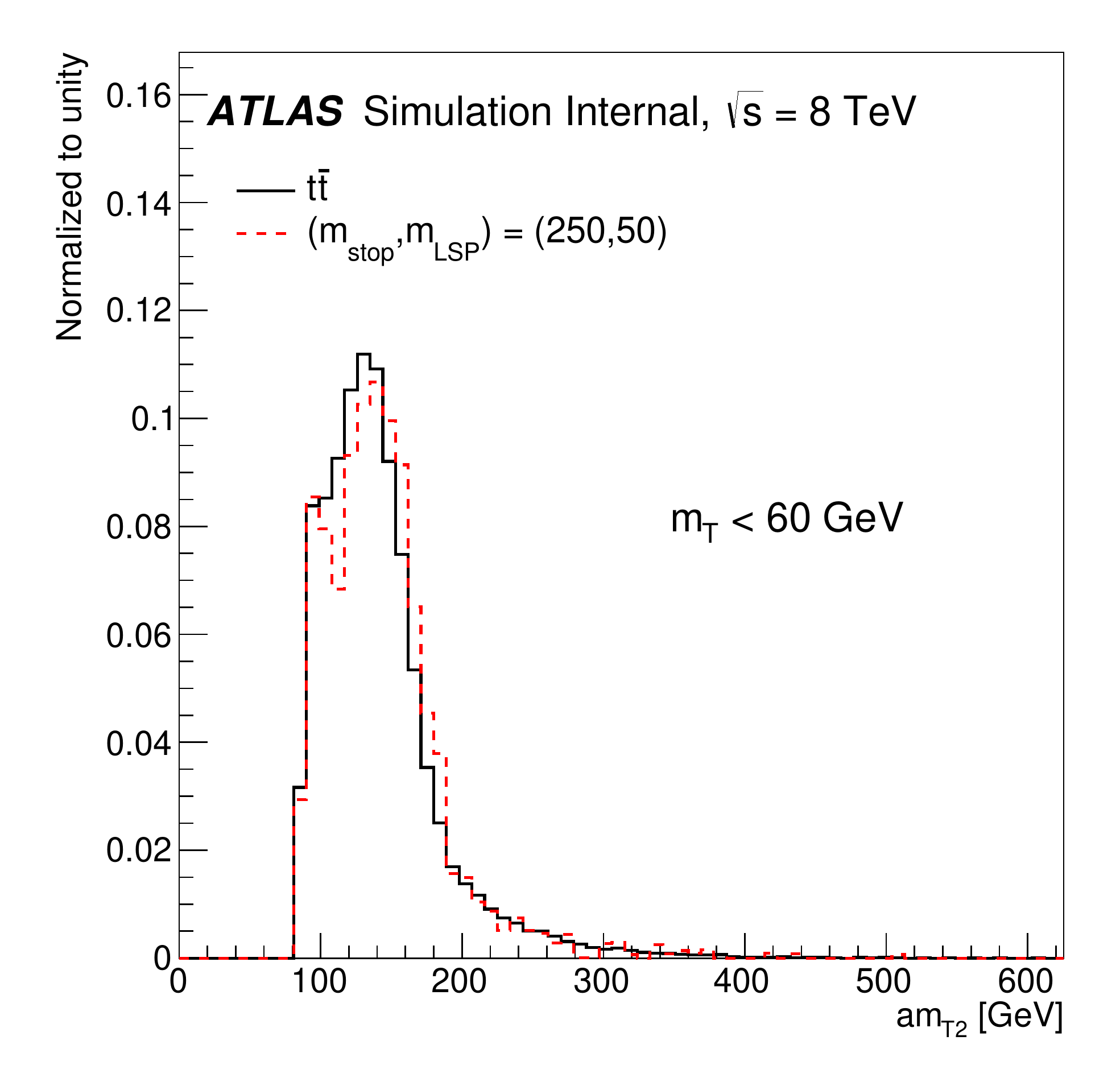}\includegraphics[width=0.33\textwidth]{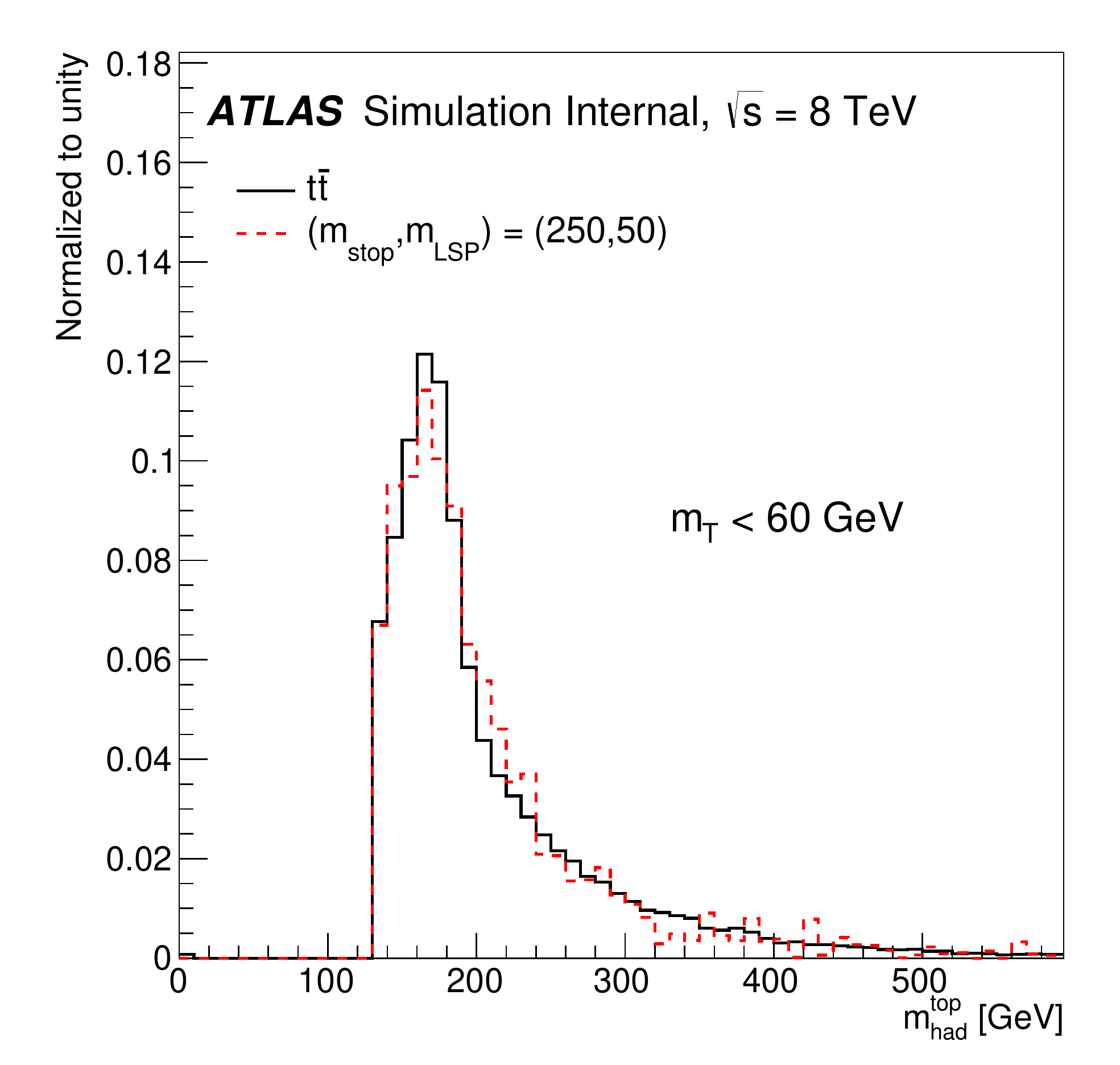}
  \caption{The distribution of $E_\text{T}^\text{miss}, am_\text{T2}$, and $m_\text{had}^\text{top}$ after the preselection from Fig.~\ref{fig:shapefit_matrix} with the additional requirement $m_\text{T}<50$ GeV.}
  \label{fig:shapefit_matrix2}
\end{figure}

The binning of the shape fit signal region is shown in Table~\ref{tab:fitmatrix}.  There are four bins of $m_\text{T}$ and three bins of $E_\text{T}^\text{miss}$.  Control regions are integrated into the signal region at low $m_\text{T}$ and with an inverted $b$-tag requirement, as indicated by the last column of Table~\ref{tab:fitmatrix}.  The upper boundaries for the tightest $m_\text{T}$ and $E_\text{T}^\text{miss}$ bins were optimized using a coarse scan in $E_\text{T}^\text{miss}\in (150\text{ GeV},175\text{ GeV})$ and $m_\text{T}\in(140\text{ GeV},160\text{ GeV})$.  Additionally, the $E_\text{T}^\text{miss}$ significance requirement is scanned in the range $(3,5,8)$ GeV${}^{1/2}$. For each combination of thresholds, pseudo-data from the simulation are fit using the procedure described in Sec.~\ref{sec:susy:stats}.  This fit incorporates the most important experimental systematic uncertainties, including the jet energy scale and jet energy resolution uncertainties with one nuisance parameter each per $E_\text{T}^\text{miss}$ bin. To demonstrate the power of the shape fit, the full multibin approach is compared with a single-bin region using only three bins from the $E_\text{T}^\text{miss}$ $>$ 150 GeV column: one for a SR, one for a $t\bar{t}$ CR, and one for a $W$+jets CR.  This region should have comparable sensitivity to SR1.  Table~\ref{tab:shapefit_res22} shows the results of the scan.  The most striking feature of Table~\ref{tab:shapefit_res22} is that the multibin setup improves upon the single-bin regions by nearly an order of magnitude for all parameter values.  The single bin regions are not able to exclude the benchmark model chosen for the optimization, but it is within reach of the multibin SR.  Due to the per $E_\text{T}^\text{miss}$ bin normalization/nuisance parameters, the level of profiling is minimal\footnote{In the usual CR method approach, the fit is setup so that each CR has significantly more events than the SR so that in the full fit, the background normalization is simply set by the ratio of data to MC in the CR.  However, the various bins of the shape fit region have a significant background yield and so many bins can play a role in normalizing the background.  There are $7$ total free parameters in the fit (one $t\bar{t}$ and one $W$+jets parameter per $E_\text{T}^\text{miss}$ bin and one overall signal normalization and $15$ total bins.  Breaking up the normalization parameters into $E_\text{T}^\text{miss}$ bins significantly reduces the amount of profiling from an over-constrained fit.  See Sec.~\ref{CR-onlyFit} for more details.}.  Therefore, the gain in sensitivity is mostly from the additional bins with various signal-to-background ratios, as desired.  A high $E_\text{T}^\text{miss}$ threshold of $8$ GeV${}^{1/2}$ is worse than the lower values in the scan, though there is not much difference between $3$ and $5$ GeV${}^{1/2}$.  Therefore, the SR1 requirement of $8$ GeV${}^{1/2}$ is loosened to $5$ GeV${}^{1/2}$ for tN\_diag.  More generally, the fit seems to perform best when the bin with the tightest selection has a relatively high signal yield (i.e. is relatively loose).  For this reason, the $m_\text{T}=140$ GeV and $E_\text{T}^\text{miss}=150$ GeV thresholds are chosen for the tightest bin.  The yields for this signal region are shown in Fig.~\ref{fig:shapefit_matrix2}\footnote{The background yields are {\it post-fit} - see Sec.~\ref{sec:susy:stats} for details of the fit.}.  For $m_\text{T}<120$ GeV, the bins have $\mathcal{O}(1000)$ events and for $m_\text{T}>120$ GeV, the bins have $\mathcal{O}(100)$ events.  The signal-to-background ratio ranges between $10$-$20\%$ in the tightest bins of the SR.

\begin{table}[h!]
\vspace{5mm}
\begin{center}
\begin{tabular}{|c | c | c | c|}
\hline
$m_\text{T}$ window [GeV] & $E_\text{T}^\text{miss}$ window [GeV] & $b$-tags & Comment\\
\hline  
\hline
  $60<m_\text{T}<90$  & 100 $<$ $E_\text{T}^\text{miss}$ $<$ 125& $=0$& $W$+jets enriched \\
   $60<m_\text{T}<90$ & 100 $<$ $E_\text{T}^\text{miss}$ $<$ 125 & $>1$& $t\bar{t}$ enriched\\
  $90<m_\text{T}<120$ & 100 $<$ $E_\text{T}^\text{miss}$ $<$ 125& $>1$& \\
  $120<m_\text{T}<140$ & 100 $<$ $E_\text{T}^\text{miss}$ $<$ 125&  $>1$&\\
   $m_\text{T}>140$  & 100 $<$ $E_\text{T}^\text{miss}$ $<$ 125 & $>1$&\\
\hline 
   $60<m_\text{T}<90$ & 125 $<$ $E_\text{T}^\text{miss}$ $<$ 150& $=0$& $W$+jets enriched \\
  $60<m_\text{T}<90$ & 125 $<$ $E_\text{T}^\text{miss}$ $<$ 150 & $>1$&$t\bar{t}$ enriched\\
  $90<m_\text{T}<120$ & 125 $<$ $E_\text{T}^\text{miss}$ $<$ 150 & $>1$&\\
  $120<m_\text{T}<140$ & 125 $<$ $E_\text{T}^\text{miss}$ $<$ 150& $>1$& \\
  $m_\text{T}>140$ & 125 $<$ $E_\text{T}^\text{miss}$ $<$ 150 & $>1$&\\
\hline 
   $60<m_\text{T}<90$ & $E_\text{T}^\text{miss}$ $>$ 150&   $=0$& $W$+jets enriched\\
   $60<m_\text{T}<90$ & $E_\text{T}^\text{miss}$ $>$ 150 &  $>1$&$t\bar{t}$ enriched\\
   $90<m_\text{T}<120$ & $E_\text{T}^\text{miss}$ $>$ 150 & $>1$&\\
   $120<m_\text{T}<140$ & $E_\text{T}^\text{miss}$ $>$ 150&   $>1$&\\
   $m_\text{T}>140$ & $E_\text{T}^\text{miss}$ $>$ 150  & $>1$&\\
\hline  
\end{tabular}
\end{center}
\caption{The definition of tN\_diag.  In addition to the variables shown above, the signal region is defined by a four-jet requiremenet with $p_\text{T}>80,60,40,25$ GeV, $m_\text{top}^\text{had}\in[130,205]$ GeV, $E_\text{T}^\text{miss}/\sqrt{H_\text{T}}>5$ GeV${}^{1/2}$, and $\Delta\phi(\text{jet}_i,\vec{p}_\text{T}^\text{miss}) > 0.8$ for $i=1$ and $2$.}
\label{tab:fitmatrix}
\end{table}

\begin{table}[h!]
\begin{center}
\noindent\adjustbox{max width=\textwidth}{
\begin{tabular}{c|c|c|c|c|c}
\hline  
 $E_\text{T}^\text{miss}/\sqrt{H_\text{T}}$ [GeV${}^{1/2}$] & $m_\text{T}$ [GeV] & $E_\text{T}^\text{miss}$ [GeV] & $\text{CL}_s$ multibin & $\text{CL}_s$ single
bin & Notes
\\
\hline   
 5 & 140 & 150 & 0.0151 & 0.146 & \\
 5 & 140 & 175 & 0.0179 & 0.301 & 0.0012*\\
 5 & 160 & 150 & 0.0156 & 0.221 & \\
 5 & 160 & 175 & 0.0177 & 0.451 & \\
 3 & 140 & 150 & 0.0152 & 0.145 & \\
 3 & 140 & 175 & 0.0176 & 0.301 & \\
 3 & 160 & 150 & 0.0162 & 0.224 & \\
 3 & 160 & 175 & 0.0168 & 0.451 & \\
 8 & 140 & 150 & 0.036 & 0.149 & \\
 8 & 140 & 175 & NaN & 0.301 & Fit Failed \\
8 & 160 & 150 & 0.0218 & 0.26 &  \\
 8 & 160 & 175 & NaN & 0.454 & Fit Failed\\
\hline  
\end{tabular}}
\end{center}
\caption{\CLs\ values (see Sec.~\ref{sec:susy:stats}) computed with the shape-fit (multibin) and single bin setups described in the text.  These values are approximately $p$-values for a hypothesis test (smaller values are better).   The $E_\text{T}^\text{miss}$ and $m_\text{T}$ values are the thresholds for the tightest region of the shape fit and define the single bin region.  The $*$ denotes the \CLs\ value for the shape fit without any systematic uncertainties.  As expected, systematic uncertainties have a big impact on the sensitivity.  In two cases (marked `Fit Failed') the multibin fit did not converge.}
  \label{tab:shapefit_res22}
\end{table}

\begin{figure}[h!]
  \centering
\includegraphics[width=0.65\textwidth]{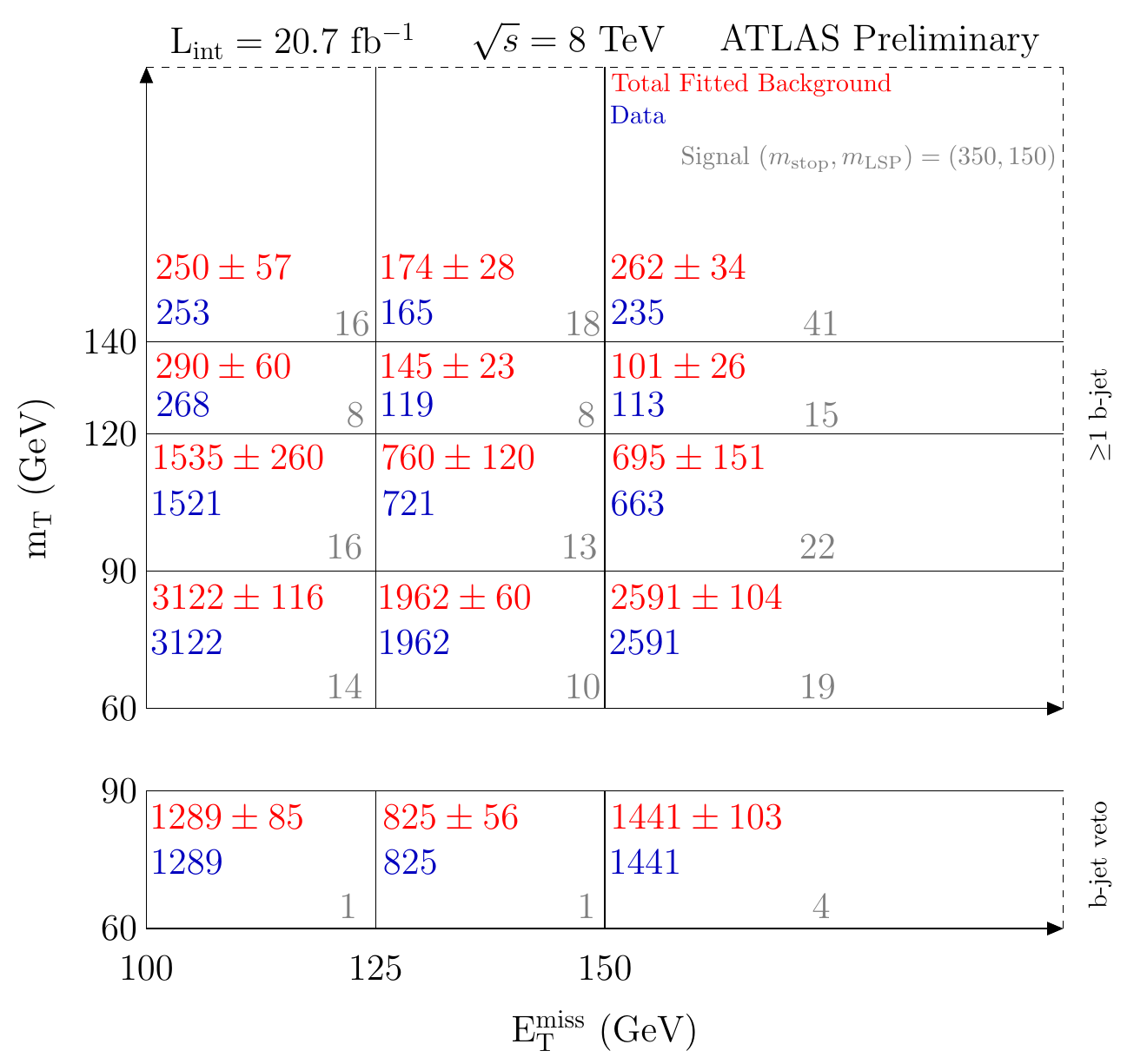}
  \caption{The yields for the preliminary version of tN\_diag.  An analogous table for the final version appears in Fig.~\ref{fig:shapefit}.  All background yields are after the post-fit - see Sec.~\ref{sec:susy:stats} for details.}
  \label{fig:shapefit_matrix2}
\end{figure}

The tN\_shape region described above was released as a preliminary result in the summer of 2013~\cite{ATLAS-CONF-2013-037}.  The additional event selection beyond the $m_\text{T}$ and $E_\text{T}^\text{miss}$ requirements was optimized slightly more for the final result published in Ref.~\cite{Aad:2014kra} and shown in Fig.~\ref{fig:shapefit} from Sec.~\ref{sec:shapefitsetup}.  In particular, the tight $\tau$ veto from Sec.~\ref{sec:objects} effectively removes about $10\%$ of the $t\bar{t}$ background with only a negligible impact on the signal.  Furthermore, the jet $p_\text{T}$ requirements are loosened to $p_\text{T}>60,60,40,25$ GeV.  The thresholds above have the same values as SR2 and SR3, whose benchmark models have much harder $p_\text{T}$ spectra than the tN\_diag benchmark model.  One other small change is the addition of a $\Delta R(b,\ell)<2.5$ requirement.  The top quarks from the tN\_diag benchmark model are not so boosted such that one of the $b$-tagged jets is always near the lepton, but this requirement is useful for suppressing dilepton $t\bar{t}$ where the two top quarks are back-to-back.   These and other modifications were studied using a similar setup to the one described above and also included approximate theoretical modeling systematic uncertainties for the $t\bar{t}$ and $W$+jets processes.  Between the two selections, the signal-to-background ratio increased from about $15\%$ in the tightest bin of the shape fit to about $20\%$ (for the same integrated luminosity).  Table~\ref{tab:signalregionselections_diag} summarizes the final tN\_diag event selection.

\begin{table}[h!]
\begin{center}
\noindent\adjustbox{max width=\textwidth}{
\begin{tabular}{|c|c|c|c|}
\hline  
Variable & tN\_diag (preliminary) & tN\_diag (final) & Comment\\
\hline
\hline
Jet $p_\text{T}>$ [GeV] & $80,60,40,25$ & $60,60,40,25$ &\\
 $\Delta\phi(\text{jet}_i,\vec{p}_\text{T}^\text{miss})>$ & 0.8 & 0.8 & $i=1,2$  \\
  $E_\text{T}^\text{miss}/\sqrt{H_\text{T}} > $ GeV${}^{1/2}$ & 5 & 5 & \\
 $E_\text{T}^\text{miss}>$ [GeV] & $100$ & $100$ & $3$ bins of the shape fit\\
 $m_\text{T}>$ [GeV] & $60$ & $60$ & $4$ bins of the shape fit\\
 $m_\text{had}^\text{top}$ & $\in [130,205]$ GeV&$\in [130,205]$ GeV&\\
 $\Delta R(b,\ell)$ & -- & $<2.5$ & \\
  $\tau$-veto & -- & tight & \\
\hline  
\end{tabular}}
\end{center}
\caption{A summary of the multibin shape fit region, tN\_diag.  This region was released with a preliminary selection in Ref.~\cite{ATLAS-CONF-2013-037} with a small change for the final result in Ref.~\cite{Aad:2014kra}.  Dashed lines indicate that there is no requirement on the given variable.  }
  \label{tab:signalregionselections_diag}
\end{table}

 	\chapter{Background Estimation}
	\label{chapter:background}
	
	The strategy for estimating the SM background in each signal region is to use the control region method to predict the yield for each distinct subprocess.  Control regions are constructed {\it for each signal region} to be as close as possible to the signal region phase space while maintaining a high yield and purity of the target background process.  Processes which cannot be normalized using data-driven techniques are subdominant and are estimated using simulation.  Figure~\ref{fig:background_comp} shows the background composition in each of the signal regions described in Chapter~\ref{chapter:susy:signalregions}.
	
\begin{figure}[h!]
\begin{center}
\includegraphics[width=0.5\textwidth]{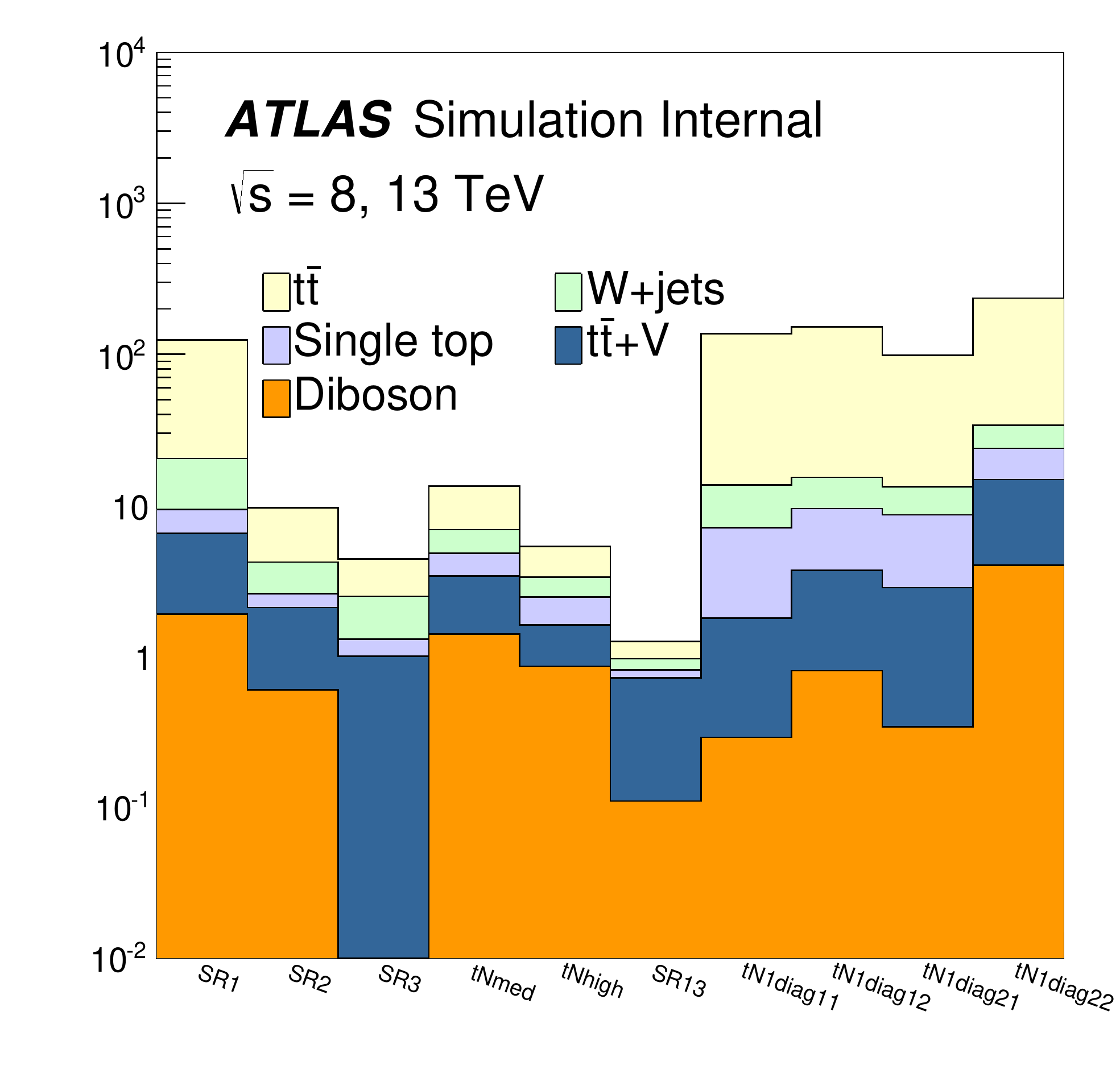}\includegraphics[width=0.5\textwidth]{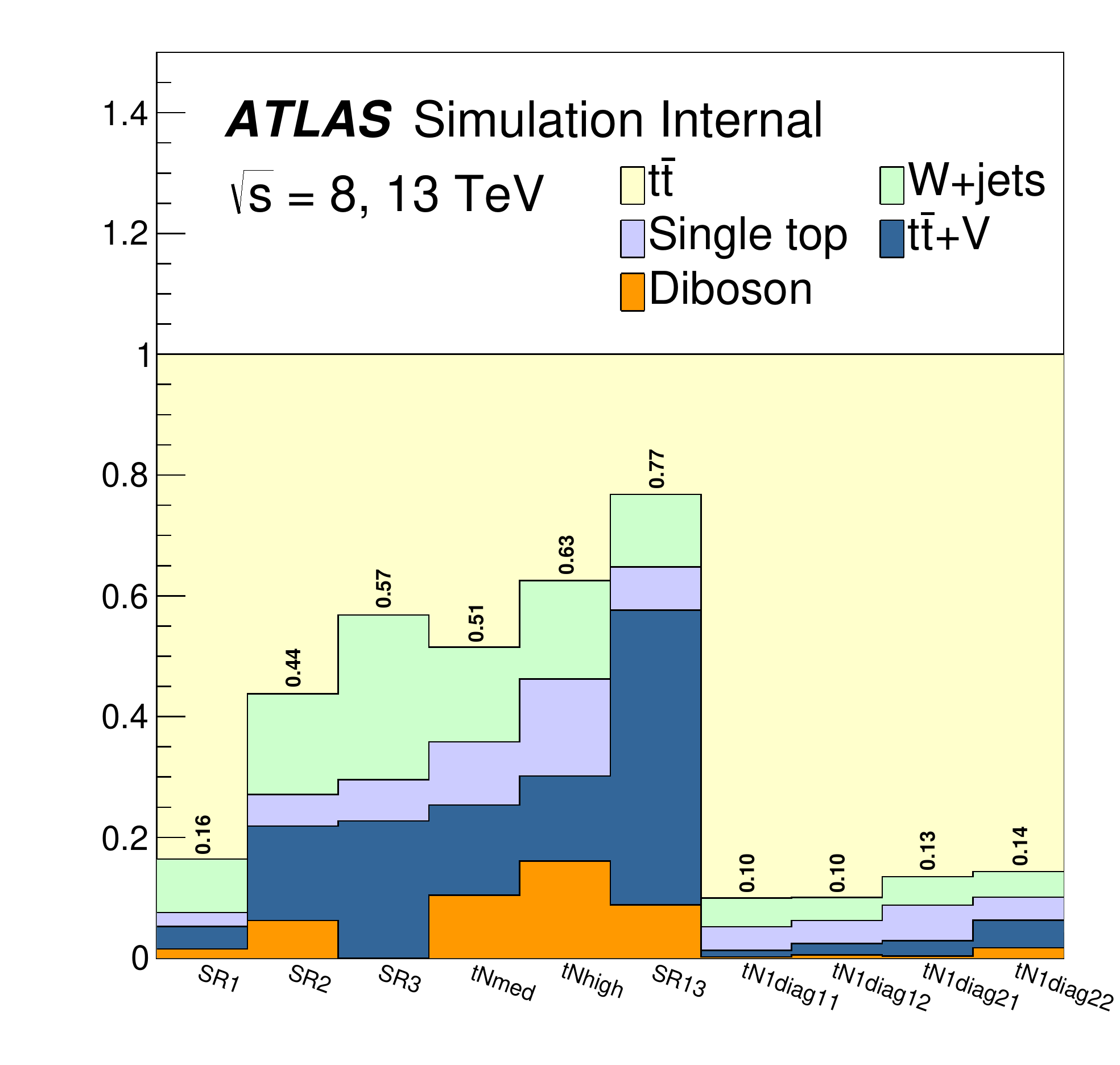}
 \caption{Left: The yields in each signal region broken down by background type after the CR-only fit setup described in Sec.~\ref{sec:susy:stats}.  Right: yields are normalized to unity to show the fractional background composition.  The numbers in the right plot show the fraction of non-$t\bar{t}$ background.  tNdiagxy is the $(x+1)^\text{th}$ $E_\text{T}^\text{miss}$ bin and $(y+2)^\text{th}$ $m_\text{T}$ bin of the shape fit region.  }
 \label{fig:background_comp}
  \end{center}
\end{figure}	
	
	Due to its large cross section and phenomenological similarity to the signal, top quark pair production is one of the most important background processes in all signal regions (see Fig.~\ref{fig:background_comp}).  As described in Sec.~\ref{sec:mT}, $m_\text{T}$ is a powerful tool for suppressing single lepton $t\bar{t}$ and therefore the dominant $t\bar{t}$ background in the signal regions has two real leptons, where one is not identified as a lepton or is a hadronically decaying $\tau$.  Section~\ref{ttbarCR} documents the construction of $t\bar{t}$ control regions.  As a result of the many tools available for reducing the yield of dilepton $t\bar{t}$ events, the remaining background composition in the signal regions is diverse.  In addition to $t\bar{t}$, the production of $W$ bosons in association with many jets (Sec.~\ref{wjets}), the single production of top quarks (Sec.~\ref{singletop}), and the associated production of top quark pairs with a $Z$ boson (Sec.~\ref{ttv}) are also significant contributions to the SM background.  As shown in Fig.~\ref{fig:background_comp} these backgrounds are increasingly relevant for higher target stop masses.  All of the background processes mentioned thus far are integrated into the control region method for a data-driven estimate of the normalization; for single top and $t\bar{t}+Z$, this is the first time data-driven techniques are used in the $t\bar{t}+E_\text{T}^\text{miss}$ topology.  The next most important background is the production of multiple electroweak bosons in association with jets.  Section~\ref{dibosons} describes the modeling of this background, which uses a simulation-based approach.  
	
	Other SM processes are negligibly small, due to a small cross-section or a low acceptance.  The latter category includes the single production of $Z$ bosons in association with jets and QCD multijets.  Both of these processes require significant fake $E_\text{T}^\text{miss}$, and fake leptons, or both.  The exception is $Z(\rightarrow \tau_\text{had}\tau_\text{lep})$+jets, which has the same final state as $W(\rightarrow l\nu)$+jets.  However, the former is suppressed by a factor of at least $\mathcal{BR}(Z\rightarrow \tau\tau)\times \mathcal{BR}(\tau_\text{lep})\times\mathcal{BR}(\tau_\text{had})\times \sigma_\text{$Z$+jets}/(\alpha_s\times\sigma_\text{$W$+jets}\times \mathcal{BR}(W\rightarrow l\nu))\sim 2\%$ with respect to $W\rightarrow l\nu$; the ratio $\sigma_\text{$Z$+jets}/\sigma_\text{$W$+jets}\sim 0.1$ and the factor of $\alpha_s$ is due to the fact that hadronically decaying $\tau$ would be reconstructed as a jet and thus reduce the number of needed quark and gluon jets.  A further suppression results from the required mis-measurement of $m_\text{T}$, which is naturally low for $\tau$ events relative to direct $W\rightarrow \mu/e+\nu$ events (see Sec.~\ref{sec:mT}) and thus needs to smear further to pass the high $m_\text{T}$ threshold.  The other possible $Z$+jets decays and estimates for their suppression factors are summarized in Table~\ref{zjetsyields}.  In all cases, since $W$+jets is already a small background, the $Z$+jets is negligible and is henceforth ignored.  Generic QCD multijet production has a large cross-section compared with $W$+jets, but requires both a fake lepton and fake $E_\text{T}^\text{miss}$.  The estimates for the regions described in sections~\ref{sec:ColorFlowEventSelection} and~\ref{sec:samples} showed that this background is already subdominant for a an inclusive one-lepton $t\bar{t}$ event selection and Ref.~\cite{ATLAS-CONF-2014-058} shows that it is negligible at high $E_\text{T}^\text{miss}$.  Therefore, multijets are ignored for the remainder of the chapter.
	
\begin{table}[h!]
\centering
\noindent\adjustbox{max width=\textwidth}{
\begin{tabular}{ccccc}
$Z$ Decay Mode & Lepton & Additional Jets & $E_\text{T}^\text{miss}$ &  Fraction of $W\rightarrow l\nu$   \\
\hline 
\hline
{\color{blue}$\nu\bar{\nu}$} & {\color{red}fake} &	{\color{orange}$4$}	& {\color{olive}correct} &$ 10\%\times {\color{red}\epsilon_f}\times{\color{blue}20\%}/30\% \lesssim 0.1\%$ \\ {\color{blue}$e^+e^-/\mu^+\mu^-$} & {\color{red}one lost} &	{\color{orange}$4$}	& {\color{olive}fake} &$10\%\times{\color{olive}\rho_f} \times  {\color{red}\epsilon_l}\times{\color{blue}6.6\%}/30\% \ll 0.1\%$ \\ {\color{blue}$\tau_\text{lep}\tau_\text{lep}$} & {\color{red}one lost} &	{\color{orange}$4$}	& {\color{olive}mis-measured}& $10\%\times{\color{olive}\rho_m} \times{\color{red}\epsilon_l}\times{\color{blue}3.3\%\times 35\%^2}/30\% \ll 0.1\%$\\
{\color{blue}$\tau_\text{lep}\tau_\text{had}$} & {\color{red}correct} &	{\color{orange}$3$}& {\color{olive}mis-measured}&$10\%\times{\color{olive}\rho_m}\times{\color{blue}3.3\%\times 35\%\times 65\%}/(30\%\times{\color{orange}\alpha_s}) \lesssim 2\%$	 \\ {\color{blue}$\tau_\text{had}\tau_\text{had}$} & {\color{red}fake} &	{\color{orange}$2$}	& {\color{olive}mis-measured}& $10\%\times{\color{olive}\rho_m} \times {\color{red}\epsilon_f}\times{\color{blue}3.3\%\times 65\%^2}/(30\%\times{\color{orange}\alpha_s^2}) \lesssim 0.1\%$\\
{\color{blue}$q\bar{q}$} & {\color{red}fake} &	{\color{orange}$2$}& {\color{olive}fake}&	$10\%\times{\color{olive}\rho_f} \times {\color{red}\epsilon_f}\times{\color{blue}70\%}/(30\%\times{\color{orange}\alpha_s^2})\ll 0.1\%$ \\
\hline
\hline
\end{tabular}}
\caption{Estimates for the yield of $Z$+jets events relative to the $W\rightarrow l\nu$ yield.  The number of additional jets only contributes to the last column when it differs from four, which is the necessary number of extra jets already needed by $W$+jets events to pass the event selection.  If a tau decays leptonically, it is considered to be reconstructed as a jet, reducing the number of extra jets required.  The rate of fake or non-prompt leptons $\epsilon_f\lesssim 1\%$~\cite{ATLAS-CONF-2014-058}.  The probability for events with no real $E_\text{T}^\text{miss}$ to be mis-reconstructed as events with large $E_\text{T}^\text{miss}$, $\rho_f$, is negligible because the resolution scales as $\sqrt{\sum E_\text{T}} \text{ GeV}^{1/2}\sim 20$ GeV.  Events with tau decays naturally have $E_\text{T}^\text{miss}$, but this needs to be severally mis-measured (with rate $\rho_m$) to pass at least the $m_\text{T}$ requirements.  }
\label{zjetsyields}
\end{table}

		The chapter ends with an overview in Sec.~\ref{overview} with all of the control region definitions and signal region yields.  In addition,  approximate scale factors for the data-driven background estimates are calculated as a function of key discriminating variables.  Uncertainties associated with the background estimates are presented in Chapter~\ref{chapter:uncertainites}.

		\clearpage
	
		\section{Top Quark Pair Production}
		\label{ttbarCR}
		
		Top quark pair production in the lepton+jets final state has the same signature at leading order as the targeted signal: one lepton, missing momentum (from the neutrino), four jets (two $b$-jets).  All of the single-bin signal regions use a strict $m_\text{T}$ requirement that effectively eliminates the single lepton background, replacing it with dilepton $t\bar{t}$ processes.   This technique is used in reverse to estimate the $t\bar{t}$ background in the signal region: the MC is normalized in a low $m_\text{T}$ window enriched in single lepton $t\bar{t}$ events.  Events that pass all signal requirements except have low $m_\text{T}$ are kinematically similar to the signal region events, but have a small predicted signal contamination and a high single lepton $t\bar{t}$ event yield and purity.  The disadvantage of the low $m_\text{T}$ method is the required extrapolation over lepton multiplicity from the control region to the signal region.  The cross section and event kinematics of a $t\bar{t}$ event are determined by $m_{t\bar{t}}$ and $p_\text{T,$t\bar{t}$}$.  Therefore, to reduce the theoretical systematic uncertainties on the extrapolation from one lepton events in the CR to two lepton events in the signal region, it is important to ensure that $m_{t\bar{t}}$ and $p_\text{T,$t\bar{t}$}$ are as similar as possible in the CR and SR.  For a fixed $m_{t\bar{t}}$ and $p_\text{T,$t\bar{t}$}$, the leading jets in one lepton $t\bar{t}$ events should have approximately the same distribution as the corresponding jets in two lepton $t\bar{t}$ events.  However, the third and fourth jets in one lepton $t\bar{t}$ events are already present at leading order (from $W$ boson decays) unlike in dilepton $t\bar{t}$ events.  Section~\ref{sec:njetsttbarTCR} explores the modeling of these subleading jets using an explicit dilepton event selection.   With the same fixed top quark kinematic properties, the $E_\text{T}^\text{miss}$ will be softer in one lepton $t\bar{t}$ events because there is only one neutrino from $W$ boson decays\footnote{How much softer depends on the reconstruction of the second lepton in dilepton events. If the second lepton is out of acceptance, then the difference between single lepton and dilepton events is larger than for events where the second lepton is within acceptance, but not identified as a lepton.  At high $E_\text{T}^\text{miss}$, the former is largely irrelevant by construction since a low $p_\text{T}$ lepton will not contribute significantly to the $E_\text{T}^\text{miss}$.  A high $p_\text{T}$ lepton that is too far forward to reconstruct as such that is also not reconstructed as a jet could could contribute significant additional $E_\text{T}^\text{miss}$, but this is suppressed because the $|\eta|$ distribution is falling from $0$.}.   This suggests a lower $E_\text{T}^\text{miss}$ threshold is appropriate for the $t\bar{t}$ control region compared with the signal region.  A lower $E_\text{T}^\text{miss}$ requirement is also useful because it can improve the $t\bar{t}$ purity.  Figure~\ref{fig:TCRopt} shows how the $t\bar{t}$ purity of the tNhigh control region depends on the $E_\text{T}^\text{miss}$ threshold.  A value of 70\% was chosen for the tNHigh TCR in order to increase the total event yield and also $t\bar{t}$ event purity.  Similar studies for all of the signal regions produced customized $t\bar{t}$ control regions that are summarized in Table~\ref{tab:TCRs}.  The $H_\text{T,sig}^\text{miss}$ is varied by a similar amount as the $E_\text{T}^\text{miss}$ when relevant because both variables scale the same way with additional real missing momentum.  Additional kinematic requirements are relaxed for the tighter signal regions where the total $t\bar{t}$ event yield is too low from simply using the low $m_\text{T}$ window.  Due to the Jacobian peak (see Sec.~\ref{sec:mT}), most of the single lepton events have $m_\text{T}\sim m_W$ and so the lower bound of the $m_\text{T}$ window is set greater than zero to reduce non $t\bar{t}$ backgrounds.  The upper edge of the $m_\text{T}$ window is chosen to allow a gap between the control region and signal region for validation purposes, described in Sec.~\ref{CR-onlyFit} in more detail.

\begin{figure}[h!]
\begin{center}
\includegraphics[width=0.95\textwidth]{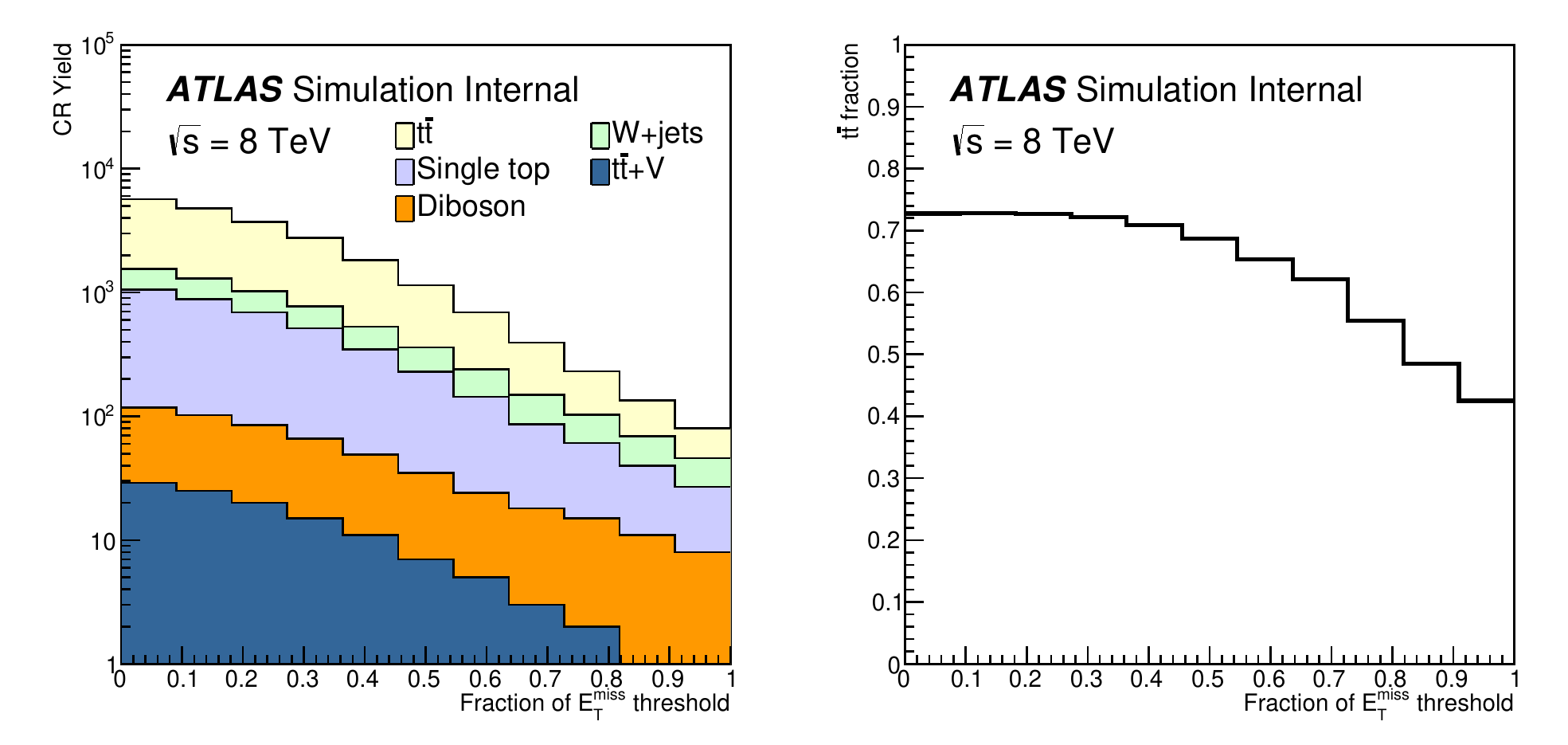}
 \caption{Optimization of the $t\bar{t}$ CR associated with tNhigh.  The left plot shows the control region composition for various fractional $E_\text{T}^\text{miss}$ thresholds after changing the $m_\text{T}$ window and removing the $m_\text{T}$ requirements.  The signal region requirement on $E_\text{T}^\text{miss}$ is $320$ GeV.  The $H_\text{T,sig}^\text{miss}$ threshold is varied coherently with the $E_\text{T}^\text{miss}$ requirement.}
 \label{fig:TCRopt}
  \end{center}
\end{figure}		
		
\begin{table}[h!]
\begin{center}
\noindent\adjustbox{max width=\textwidth}{
\begin{tabular}{|l |cc|cc|cc|cc|cc|cc|cc|cccc}
\hline
   Requirement & SR1 & TCR1 & SR2 & TCR2 & SR3 & TCR3 & tNmed & TCRmed & tNhigh & TCRhigh & SR13 & TCR13 \\
   \hline
     \hline
     $m_\text{T}$ [GeV] & $[140,250]$ & $[60,90]$ & $>140$ & [60,90] & $>180$ & $[60,90]$ & $>140$ & $[60,90]$ & $>200$ & $[60,90]$ & $200$ & $[30,90]$ \\
     $E_\text{T}^\text{miss}$ [GeV] & -- & -- & -- & -- & $>225$ & $>220$&--&--&$>320$&$>225$& $>350$ & $>250$\\
     $am_\text{T2}$ [GeV] & -- & -- & $>170$ & $>120$ & $>200$ & $>170$ & $>170$& $>120$ & $>170$&$>80$ & $>175$ & $[100,200]$ \\
     $m_\text{T2}^\tau$ [GeV] & -- & -- & -- & -- & $>120$ & $>0$ & -- & -- & $>120$ &$>0$ & -- & --\\ 
     $H_\text{T,sig}^\text{miss} $& -- & -- & -- & -- & -- & -- & --& --& $>12.5$ & $>8.8$ & $>20$ & $>15$ \\
     $\Delta R(b,l)$& -- & -- & -- & -- & -- & -- & --& --& -- & -- & $<2.5$ & $<\infty$ \\
     \hline
     \hline
    Total Yield & 125 &1661 & 9.6& 169 & 4.3 & 195  &13.0&159 &5.0&359& 1.3 &102\\
    $t\bar{t}$ Purity & 83\% &82\% &56\%& 66\% & 44\%& 57\% &50\%&79\%&39\%&80\%& 25\%& 88\% \\
    \hline
\end{tabular}}
\caption{The definition of the $t\bar{t}$ control regions for each signal region presented in Chapter~\ref{chapter:susy:signalregions}.  Only the requirements that differ from the corresponding signal region are indicated in the table, with a `--' if there is no change between the signal and control region.  The lower two rows show the total background yield and the fraction of $t\bar{t}$ events in both the signal and control region using the CR-only fit, described in Sec.~\ref{sec:susy:stats}.  The upper $am_\text{T2}$ requirement in TCR13 is to ensure orthogonality from STCR13, described in Sec.~\ref{singletop:datadriven}.}
  \label{tab:TCRs}
\end{center}
\end{table}		
		
The predicted signal yield in all of the $t\bar{t}$ control regions is negligible for regions of $(m_\text{stop},m_\text{LSP})$ near the corresponding benchmark models.  This is illustrating explicitly for SR13 in Fig.~\ref{fig:TCRsignalcontamination}, where the benchmark model is $(m_\text{stop},m_\text{LSP})=(800,0)$.  The predicted yield for $m_\text{stop}\sim 800$ GeV is less than $0.2$ events whereas the entire SM prediction for this region is about $100$ events (see Table~\ref{tab:TCRs}).

\begin{figure}[h!]
\begin{center}
\includegraphics[width=0.5\textwidth]{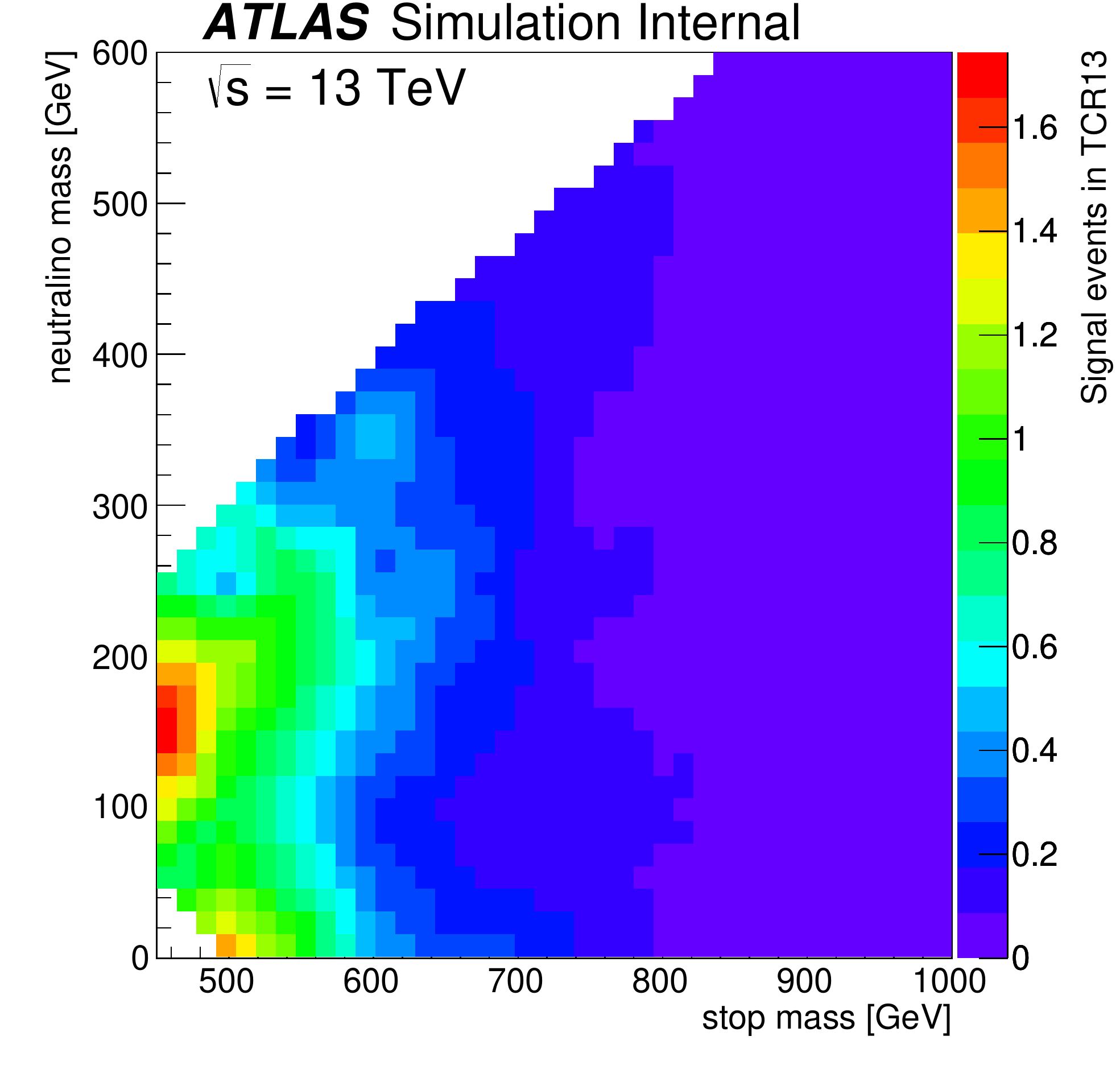}
 \caption{The number of signal events in TCR13 as a function of the stop mass and neutralino mass.  The benchmark model for SR2 has a stop mass of $800$ GeV and a neutralino mass of $0$ GeV.  The number of fitted SM events in TCR13 is $102$.}
 \label{fig:TCRsignalcontamination}
  \end{center}
\end{figure}

		\clearpage
		
		\subsection{Modeling Dilepton Events}
		\label{sec:njetsttbarTCR}
		
		As the dilepton $t\bar{t}$ background is estimated using a mostly single lepton $t\bar{t}$ event selection at low $m_\text{T}$, it is critical to validate the modeling of the jet-related variables in the high $m_\text{T}$ tail.  The leading order matrix element for dilepton events only has two out-going quarks and so at least two extra jets must originate from somewhere else in the simulation.  For a hadronic origin of the jets, the two possibilities are the real emission from the NLO matrix element and extra radiation at leading logarithm from the parton shower.  Another possibility is that the extra jets are mis-identified leptons, which nearly always occurs for hadronically decaying taus with sufficient $p_\text{T}$.  This section presents two event selections for probing the modeling of the extra jets in dilepton $t\bar{t}$ events by explicitly requiring a second lepton.  The events in Sec.~\ref{sec:1l1tau} are required to have a reconstructed tau candidate in addition to an electron or muon to form a $1L1\tau$ validation region (VR) and the events in Sec.~\ref{sec:2L} have an explicitly reconstructed electron-muon pair.  Note that these validation regions could be used for a direct estimation of the $t\bar{t}$ background yield in the signal region via the control region method in place of the low $m_\text{T}$ region.  However, due to the much lower cross-section for dilepton $t\bar{t}$ events, it is likely that the selection would need to be significantly looser than the one-lepton region and thus a larger phase-space extrapolation is required.
				
\begin{figure}[h!]
\begin{center}
\begin{tikzpicture}[line width=1.5 pt, scale=1.]
	
	\draw[gluon,color=black] (-1,1)--(0,0);
	\draw[gluon,color=black] (-1,-1)--(0,0);	
	\draw[gluon,color=black] (0,0)--(1,0);
	\draw[gluon] (-0.5,0.5)--(0.5,1.5);
	\draw[color=black](1,0)--(2,1);
	\draw[color=black](1,0)--(2,-1);
	\draw(2,1)--(3,2);
	\draw(2,-1)--(3,-2);
	\draw[vector,color=black](2,1)--(3,0.75);
	\draw[vector,color=black](2,-1)--(3,-0.75);	
	\draw[color=black](3,0.75)--(4,1.5);
	\draw[color=black](3,0.75)--(4,0.5);		
	\draw[color=black](3,-0.75)--(4,-1.5);
	\draw[color=black](3,-0.75)--(4,-0.5);			
					\node at (4.2,0.5) {\large $\mu$};
	\node at (4.2,1.5) {\large $\nu$};	
	\node at (4.2,-0.5) {\large $e$};
	\node at (4.2,-1.5) {\large $\nu$};
	\node at (4.1,-2.) {\large $b(\rightarrow\text{jet})$};	
	\node at (4.1,2.) {\large $b(\rightarrow\text{jet})$};
	\node at (0.5,1.8) {\large $g(\rightarrow\text{jet})$};
	\node at (3.5,0) {\large $g(\rightarrow\text{jet})$};
	\draw[gluon] (1.5,-0.5)--(2.5,0);	
	
	 \begin{scope}[shift={(7,0)}]
	\draw[gluon,color=black] (-1,1)--(0,0);
	\draw[gluon,color=black] (-1,-1)--(0,0);	
	\draw[gluon,color=black] (0,0)--(1,0);
	\draw[gluon] (-0.5,0.5)--(0.5,1.5);
	\draw[color=black](1,0)--(2,1);
	\draw[color=black](1,0)--(2,-1);
	\draw(2,1)--(3,2);
	\draw(2,-1)--(3,-2);
	\draw[vector,color=black](2,1)--(3,0.75);
	\draw[vector,color=black](2,-1)--(3,-0.75);	
	\draw[color=black](3,0.75)--(4,1.5);
	\draw[color=black](3,0.75)--(4,0.5);		
	\draw[color=black](3,-0.75)--(4,-1.5);
	\draw(3,-0.75)--(4,-0.5);			
					\node at (4.2,0.5) {\large $\mu$};
	\node at (4.2,1.5) {\large $\nu$};	
	\node at (5.1,-0.5) {\large $\tau(\rightarrow\text{jet})$};
	\node at (4.2,-1.5) {\large $\nu$};
	\node at (4.1,-2.) {\large $b(\rightarrow\text{jet})$};	
	\node at (4.1,2.) {\large $b(\rightarrow\text{jet})$};
	\node at (0.5,1.8) {\large $g(\rightarrow\text{jet})$}; 
	 \end{scope}
	
 \end{tikzpicture}
\end{center}
\caption{Feynman diagrams illustrating dilepton $t\bar{t}$ events passing the four-jet selection.  In the left diagram, one of the electron or muon is not identified or reconstructed as a jet.  In the right diagram, the tau is reconstructed as a jet.}
\label{fig:feynmansingletop}
\end{figure}				
				
		\clearpage		
				
		\subsubsection{Tau Validation Region}
		\label{sec:1l1tau}
		
		About half of the dilepton background has one hadronically decaying $\tau$ ($1L1\tau$).  Many of the signal regions have hadronic $\tau$ vetos designed to reject such events, but $\tau$ reconstruction is not as clean as electron or muon identification and therefore many hadronic $\tau$ events still pass the full event selection.  Events that would have been rejected due to the hadronic $\tau$ veto can be used to study the modeling of jets beyond those produced from the tree-level $t\bar{t}$ system.  A $\tau$ validation region is therefore constructed with the event selection shown in Table~\ref{tab:1L1tauVR}.  After requiring one reconstructed hadronically decaying $\tau$ in addition to the preselection, the validation region is still dominated by one lepton $t\bar{t}$ events with a fake $\tau$.  The one lepton component is suppressed by requiring $m_\text{T}>100$ GeV, as illustrated in the left plot of Fig.~\ref{fig:1L1tauMT}.   In total, there are about $100$ SM events predicted in the validation region with over $70\%$ $1L1\tau$ purity.
			
\begin{table}[h!]
\begin{center}
\noindent\adjustbox{max width=\textwidth}{
\begin{tabular}{|c|c|}
\hline
   Requirement & Value  \\
   \hline
     \hline
     Preselection & Exactly one lepton \\
     Reconstructed $\tau$ & $>0$\\
     $n_\text{jets}$ & $\geq 4$\\
$1^\text{st}$ jet $p_\text{T}$ [GeV] & $>80$ \\
$2^\text{nd}$ jet $p_\text{T}$ [GeV] & $>50$ \\
$3^\text{rd}$ jet $p_\text{T}$ [GeV] & $>40$ \\
$4^\text{th}$ jet $p_\text{T}$ [GeV] & $>25$ \\
$m_\text{T}$ [GeV] & $>100$ \\
$E_\text{T}^\text{miss}$ [GeV] & $>200$ \\
$n_\text{$b$-jets}$ & $>0$ \\
        \hline
   \end{tabular}}
\caption{An event selection requiring a hadronic $\tau$ candidate to study the modeling of $1L1\tau$ events. }
  \label{tab:1L1tauVR}
\end{center}
\end{table}

\begin{figure}[h!]
\begin{center}
\includegraphics[width=0.5\textwidth]{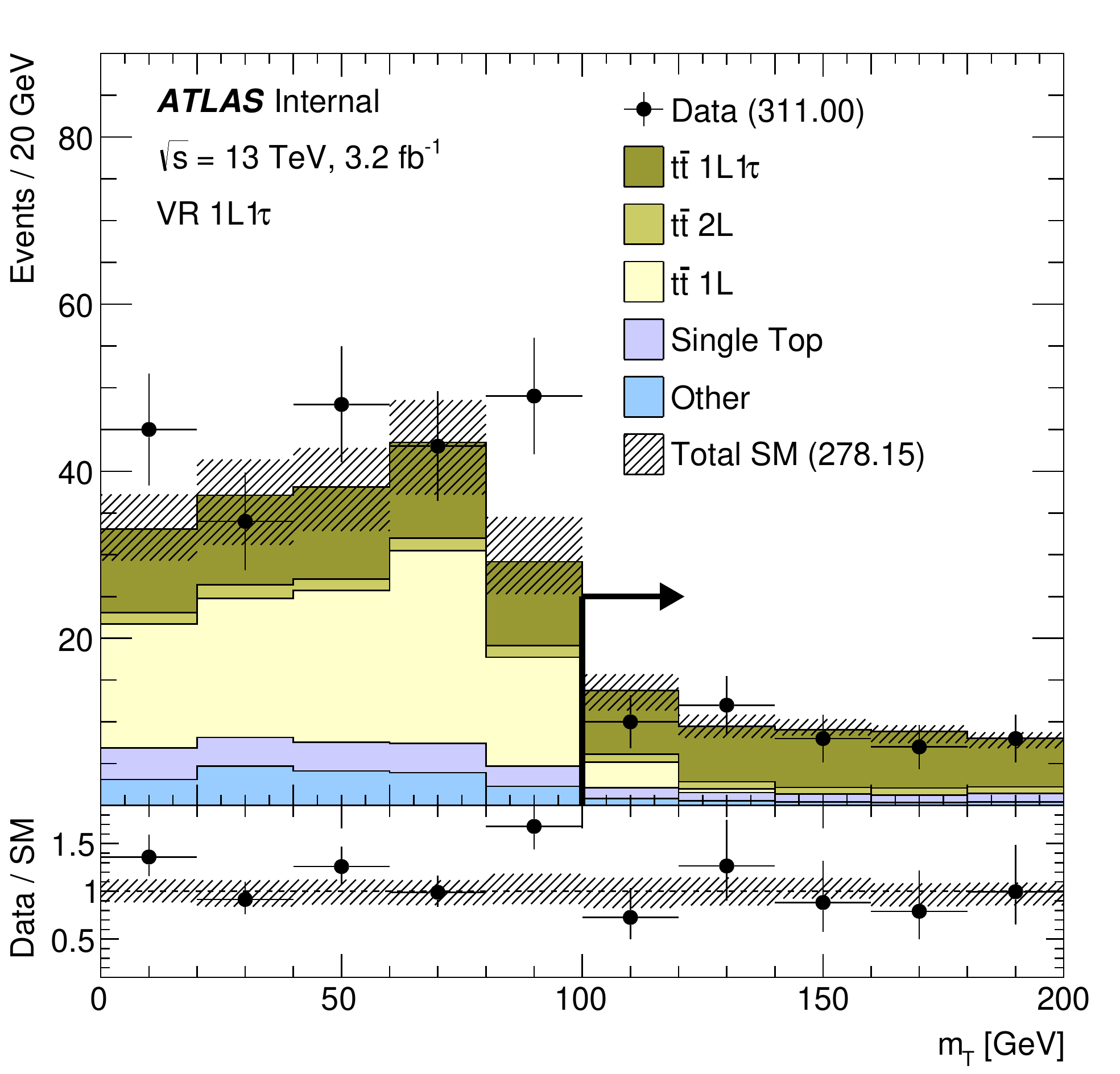}\includegraphics[width=0.5\textwidth]{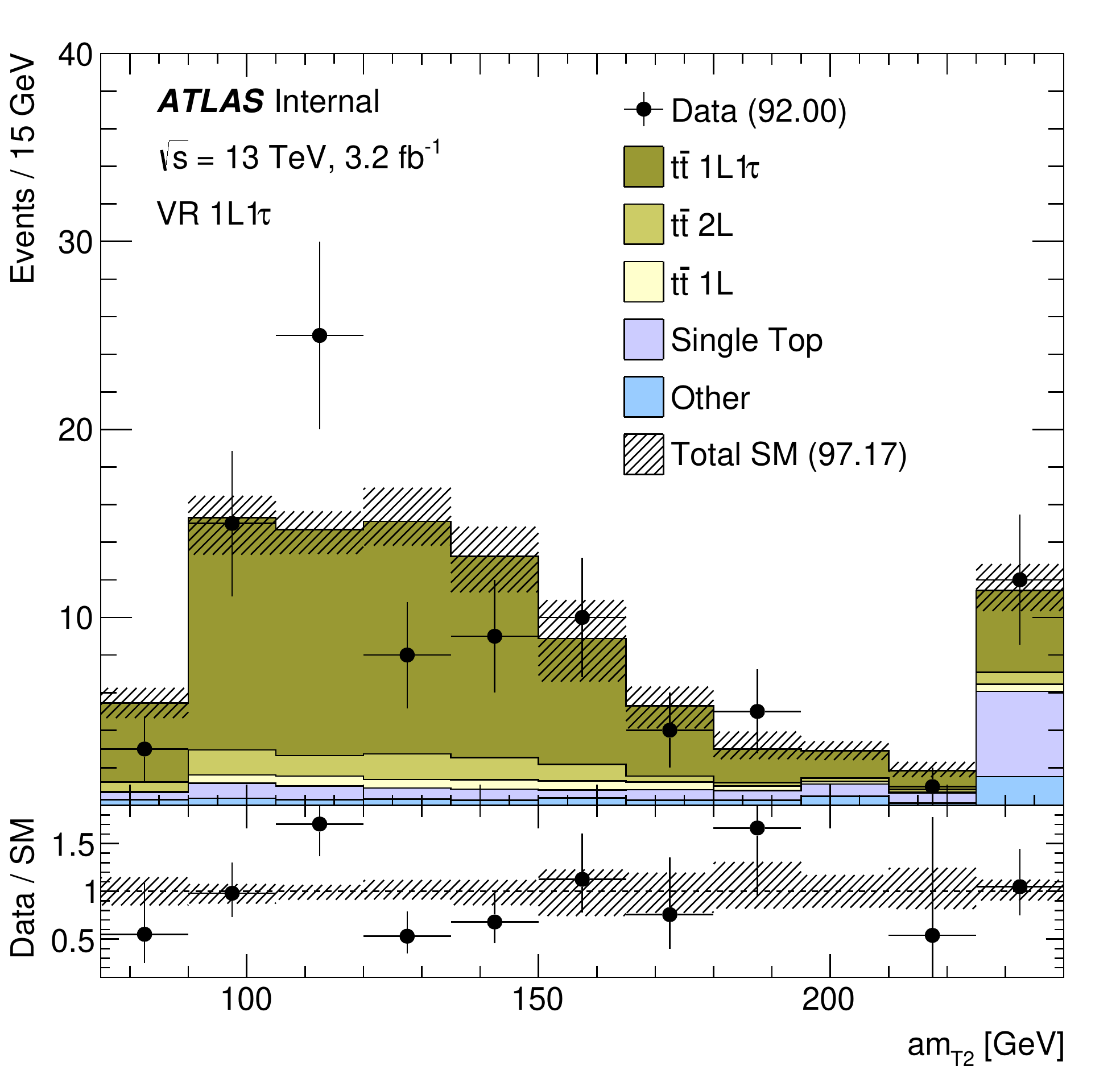}
 \caption{Left: the $m_\text{T}$ distribution in the $1L1\tau$ validation with all requirements except the $m_\text{T}$ threshold, which is indicated by an arrow.  Right: the $am_\text{T2}$ distribution in the $1L1\tau$ validation region.  Jet energy scale and resolution uncertainties are included in the error band. The last bin contains overflow.}
 \label{fig:1L1tauMT}
  \end{center}
\end{figure}

The jet multiplicity in the $\tau$ validation region is shown in Fig.~\ref{fig:1L1tauVRnjets}, beginning at four jets as required by all signal region selections.  Frequently, one of the four jets is the hadronically decaying $\tau$ itself as there is no $\tau$-jet overlap removal.  For this reason, it is slightly `easier' for a $1L1\tau$ event to pass the event selection compared with a dilepton $t\bar{t}$ event with only electrons or muons.  Formally, the {\sc Powheg-Box}+{\sc Pythia 6} simulation is NLO accurate to the fourth jet (assuming one of the four is the hadronically decaying $\tau$) and only leading logarithmically accurate for $n_\text{jets}>4$.  However, the agreement is significantly better than naively expected because the parton shower has been extensively tuned to collider data.  While the $\chi^2/\text{NDF}<1$, there is a small slope in the data/MC ratio.  Inclusive measurements of the `extra' jets in $t\bar{t}$ events find a slope in the opposite direction~\cite{ATLAS-CONF-2015-065}, suggesting that the potential trend in Fig.~\ref{fig:1L1tauVRnjets} is possibly insignificant. 

\begin{figure}[h!]
\begin{center}
\includegraphics[width=0.5\textwidth]{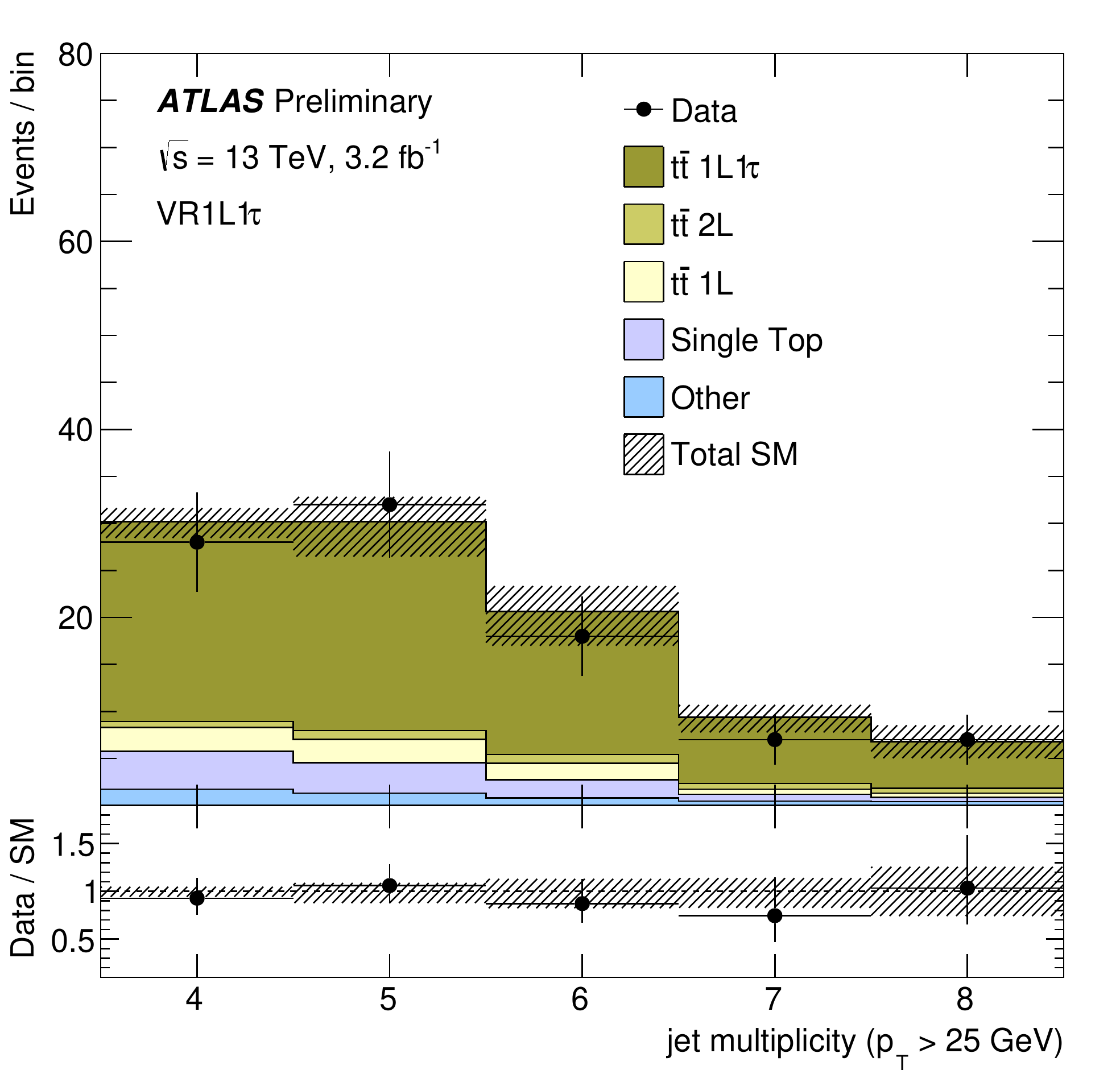}
 \caption{The number of signal jets in the $1L1\tau$ VR.  Jet energy scale and resolution uncertainties are included in the error band. The last bin contains overflow.}
 \label{fig:1L1tauVRnjets}
  \end{center}
\end{figure}

The subleading jets are examined in more detail in Fig.~\ref{fig:VR1l1taujetpT}.   There is no unambiguous way to select jets that are not produced from the leading order $t\bar{t}$ decay, but one useful proxy is to consider non $b$-tagged jets.  Figure~\ref{fig:VR1l1taujetpT} shows the $p_\text{T}$ distribution of the leading non $b$-tagged jets in the $1L1\tau$ validation region.  The data/MC ratio does not provide any significant evidence for a mis-modeling of these $p_\text{T}$ spectra.

\begin{figure}[h!]
\begin{center}
\includegraphics[width=0.5\textwidth]{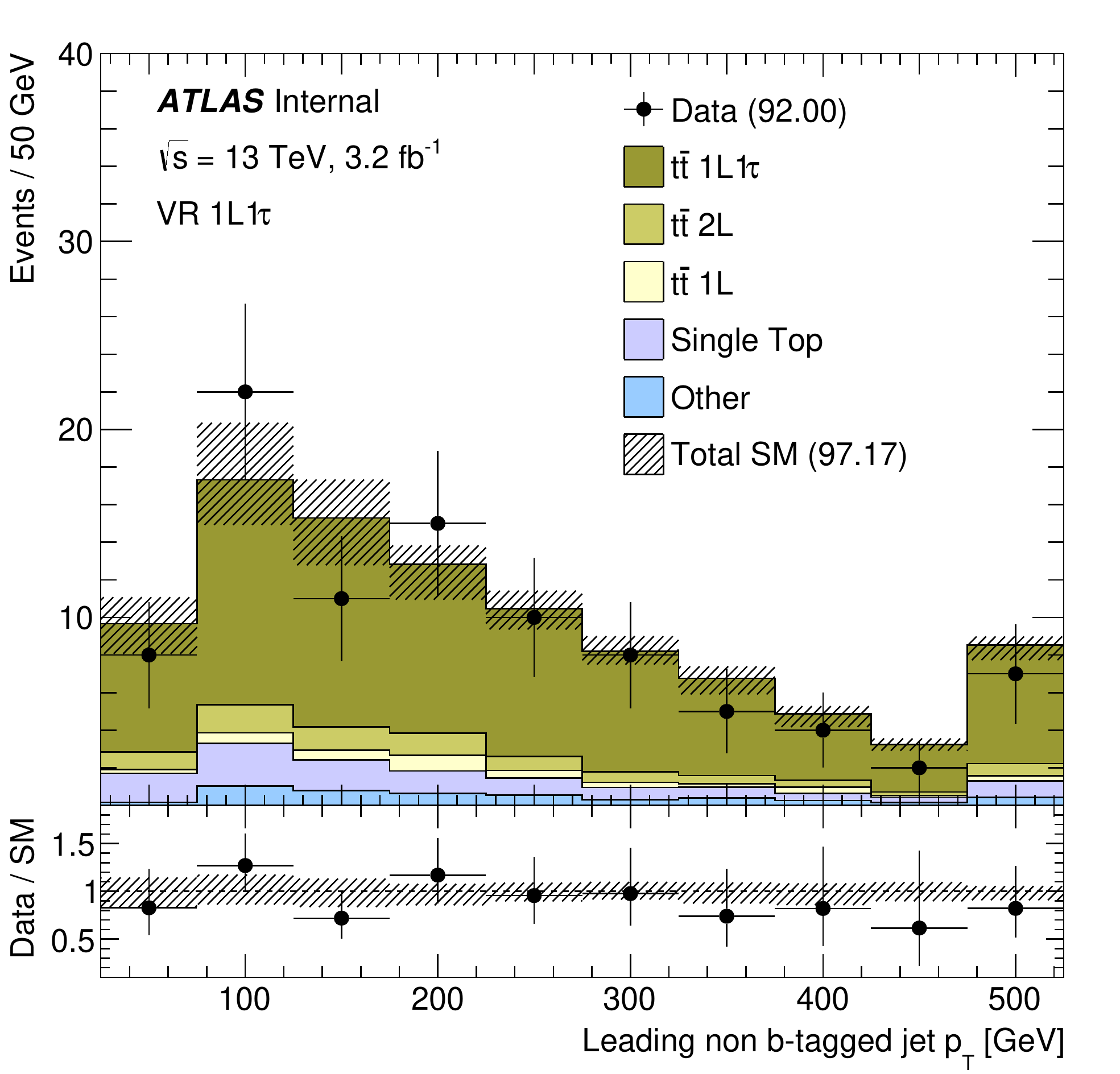}\includegraphics[width=0.5\textwidth]{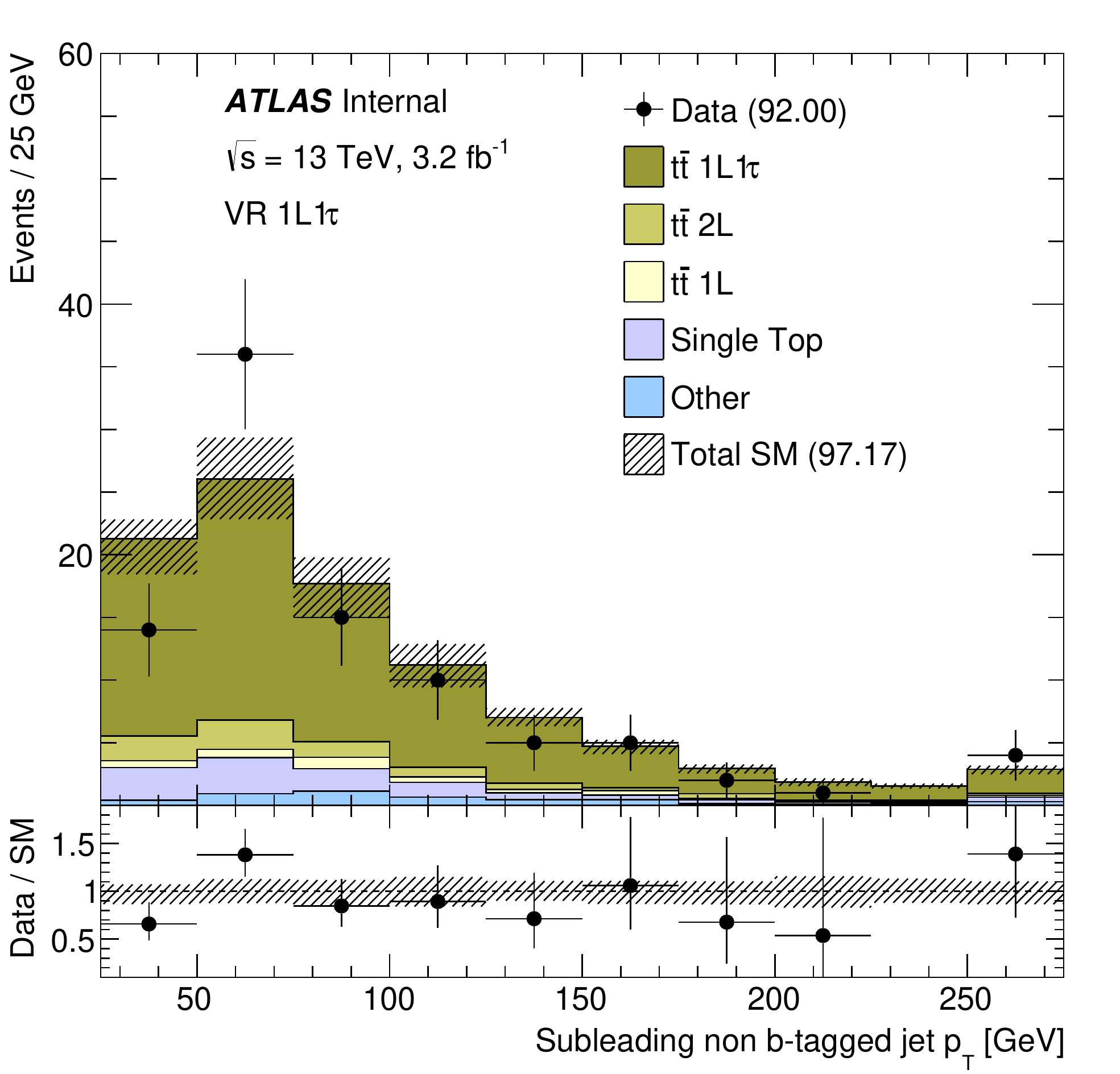}
 \caption{The leading (left) and sub-leading (right) non $b$-tagged jets in the 1L1$\tau$ validation region.  When there are more than two $b$-tagged jets in the event, the plotted jets are those after the leading two $b$-tagged jets, ordered by $p_\text{T}$. Jet energy scale and resolution uncertainties are included in the error band. The last bin contains overflow.}   \label{fig:VR1l1taujetpT}
  \end{center}
\end{figure}		
		
		\clearpage
		
		\subsubsection{Dilepton Validation Region}
		\label{sec:2L}

	The other half of the dilepton $t\bar{t}$ events that pass the event selection have a second electron or muon that is not reconstructed as such.  To study the modeling of the jets in this case, a minimal two lepton event selection is constructed and is summarized in Table~\ref{tab:2LVR}.  Low $p_\text{T}$ unprescaled dilepton triggers are available.  However, these triggers do not add a significant number of events for the region of interest $E_\text{T}^\text{miss}>200$ GeV where the $E_\text{T}^\text{miss}$ trigger is fully efficient.  An opposite flavor $e\mu$ selection is chosen to suppress $Z$+jets events\footnote{A selection with $ee$ or $\mu\mu$ would be possible with an additional requirement on $m_{ll}$ to be away from $m_Z$.}.  The predicted yield in the resulting validation region is about $400$ events with over $>85\%$ $t\bar{t}$ 2L purity. 
\begin{table}[h!]
\begin{center}
\noindent\adjustbox{max width=\textwidth}{
\begin{tabular}{|c|c|}
\hline
   Requirement & Value  \\
   \hline
     \hline
     Trigger & $E_\text{T}^\text{miss}$ \\
     $n_e$ & $=1$\\
      $n_\text{$\mu$}$ & $=1$\\
      $q_e\times q_\mu$ & $< 0$ \\
     $n_\text{jets}$ & $\geq 4$\\
$E_\text{T}^\text{miss}$ [GeV] & $>200$ \\
$n_\text{$b$-jets}$ & $>0$ \\
        \hline
   \end{tabular}}
\caption{An event selection requiring two reconstructed signal leptons.  The variable $q_l$ denotes the charge of lepton $l$.}
  \label{tab:2LVR}
\end{center}
\end{table}

There is no unique way to define the $m_\text{T}$ and $am_\text{T2}$ variables in two-lepton events, but a way to probe the case where the second lepton is reconstructed but mis-identified is shown in Fig.~\ref{fig:mt2L}.  The lepton with the higher $p_\text{T}$ is treated as the signal lepton and the second lepton is ignored.  Even though it is not explicitly identified, this second lepton still contributes to the $E_\text{T}^\text{miss}$ calculation.  The second neutrino allows events to exceed the $m_\text{T}=m_W$ and the population of the tail is determined by $p_\text{T}^{t\bar{t}}$ and $m_{t\bar{t}}$ as in Sec.~\ref{sec:mT}.  In contrast, the $am_\text{T2}$ distribution is mostly contained\footnote{The main topology motivating the $am_\text{T2}$ variable is when the second lepton is lost (not part of the $E_\text{T}^\text{miss}$, but Sec.~\ref{sec:mt2} shows that it is also useful when the lepton is only mis-identified.} within $am_\text{T2}\lesssim m_\text{top}\sim 175$ GeV which is the reason this variable is powerful at suppressing the dilepton $t\bar{t}$ background.

Figure~\ref{fig:njets2L} shows the jet multiplicity in the $2L$ validation region and is the analogue to Fig.~\ref{fig:1L1tauVRnjets} from the $1L1\tau$ region.  All other plots in this section require $n_\text{jets}\geq 4$, but the modeling of the third jet is already interesting because only two jets are expected from the ME in dilepton events.  There is a small slope in the data/MC ratio for $n_\text{jets}>4$, but it is well within the systematic uncertainty from the jet energy scale and resolution even though it is in the same direction as dedicated studies~\cite{ATLAS-CONF-2015-065}.  The modeling of the momentum of the `extra' jets is shown in Fig.~\ref{fig:jetpT2L}.  As in the $1L1\tau$ case, there is no unique way to identify such jets, but the leading non-$b$ tagged jets are a good proxy.  There is no significant evidence for mis-modeling the $p_\text{T}$ spectra.

\begin{figure}[h!]
\begin{center}
\includegraphics[width=0.5\textwidth]{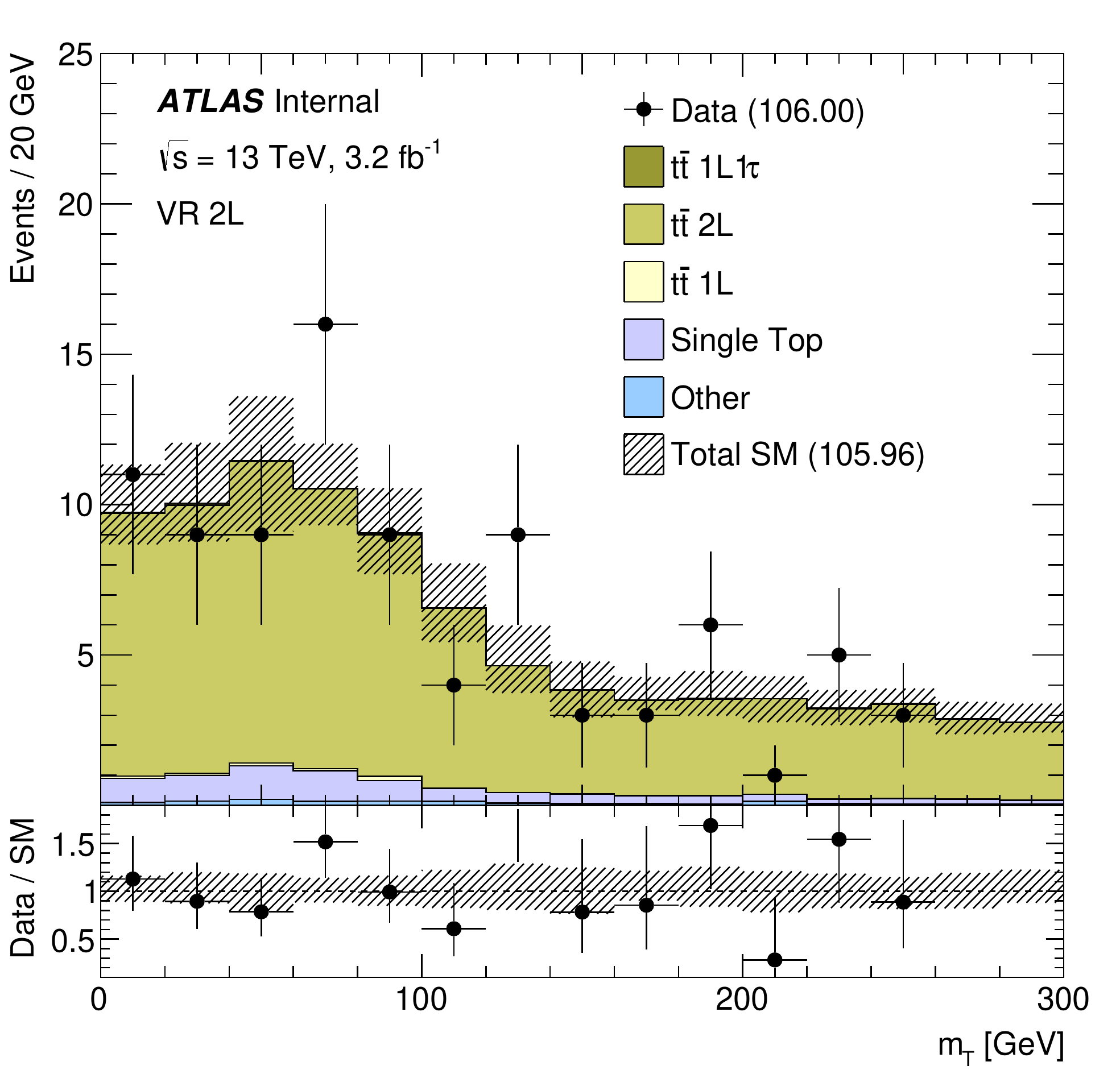}\includegraphics[width=0.5\textwidth]{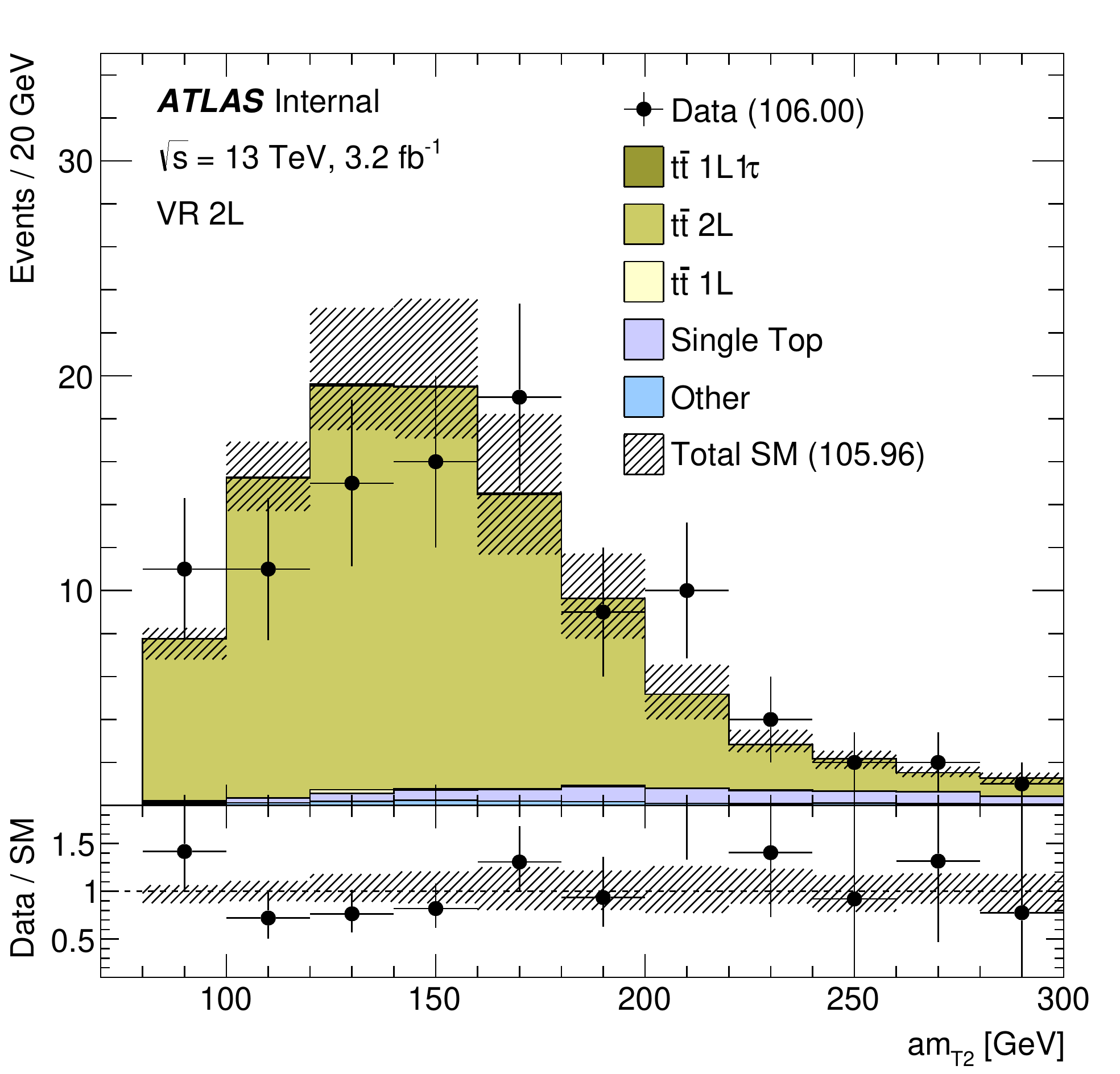}
 \caption{Left: the $m_\text{T}$ distribution in the $2L$ validation region where the softer lepton is treated as measured but not reconstructed.  Right: the $am_\text{T2}$ distribution with the same lepton treatment as the left plot. Jet energy scale and resolution uncertainties are included in the error band. The last bin contains overflow.}
 \label{fig:mt2L}
  \end{center}
\end{figure}

\begin{figure}[h!]
\begin{center}
\includegraphics[width=0.5\textwidth]{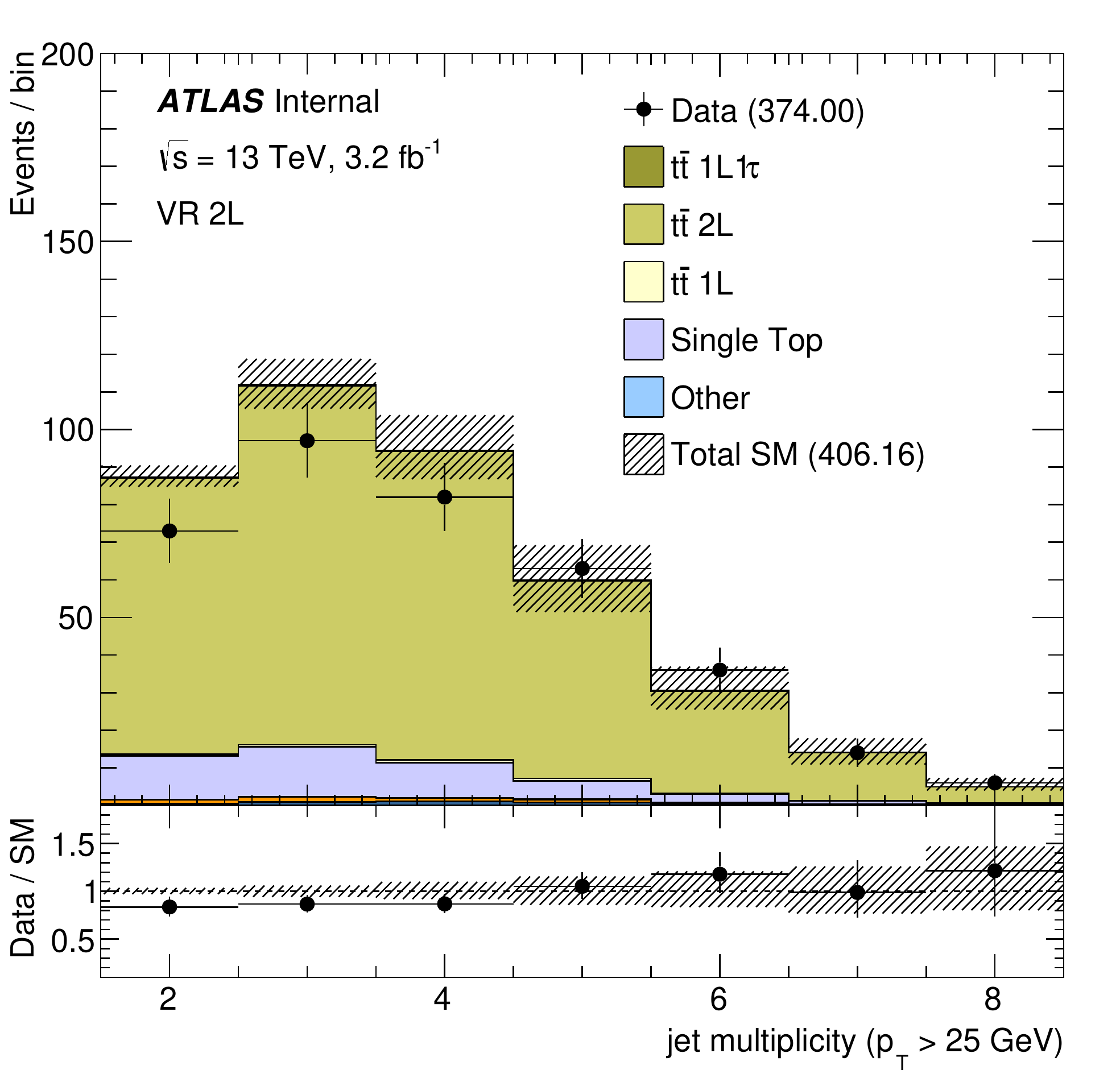}
 \caption{The jet multiplicity in the $2L$ VR. Jet energy scale and resolution uncertainties are included in the error band. The last bin contains overflow.}
 \label{fig:njets2L}
  \end{center}
\end{figure}
		
\begin{figure}[h!]
\begin{center}
\includegraphics[width=0.5\textwidth]{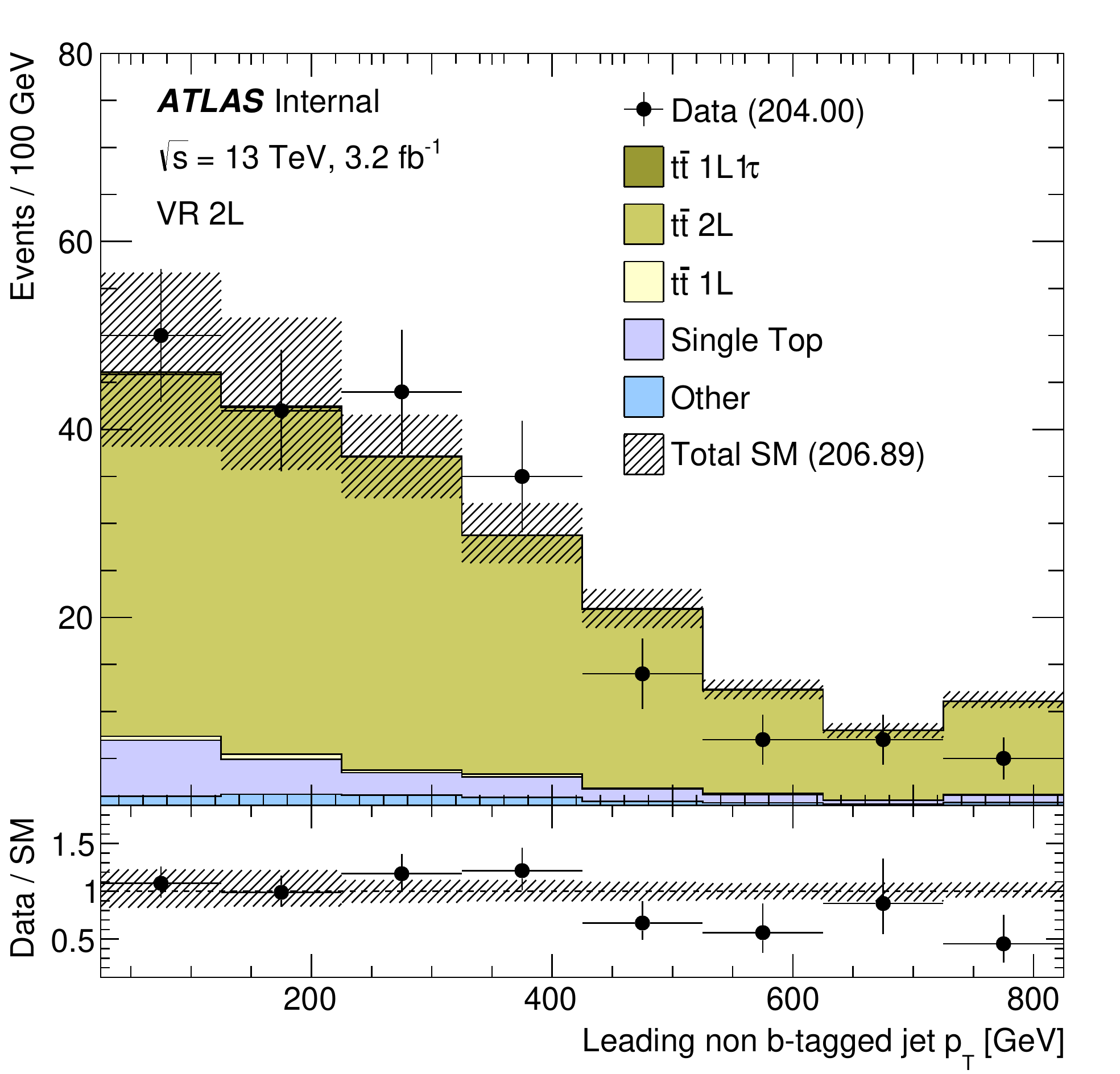}\includegraphics[width=0.5\textwidth]{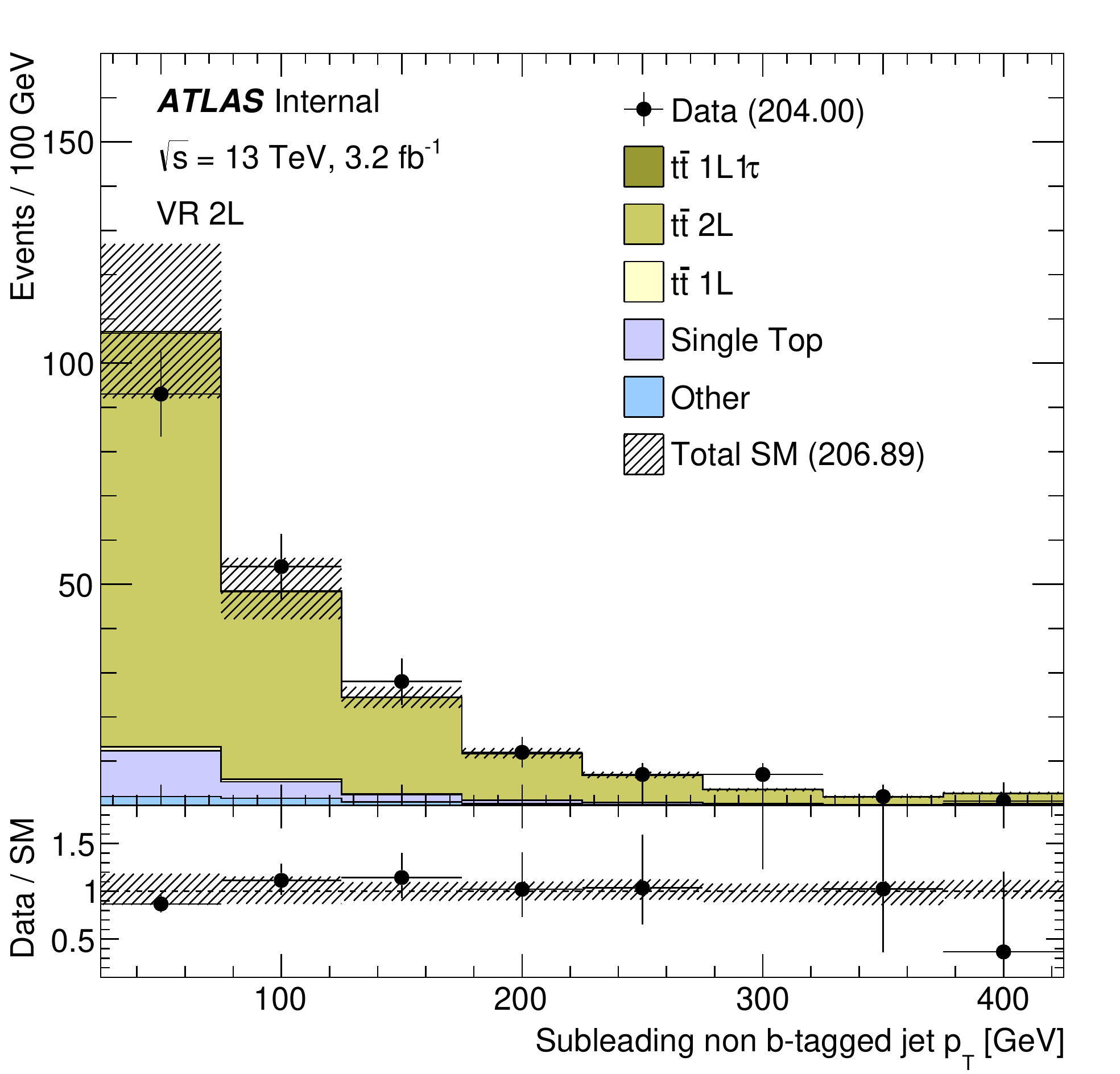}
 \caption{The leading (left) and sub-leading (right) non $b$-tagged jets in the 2L validation region.  When there are more than two $b$-tagged jets in the event, the plotted jets are those after the leading two $b$-tagged jets, ordered by $p_\text{T}$. Jet energy scale and resolution uncertainties are included in the error band. The last bin contains overflow.}
 \label{fig:jetpT2L}
  \end{center}
\end{figure}		
		
		\clearpage
		
		\section{$W$+jets}
		\label{wjets}

		The inclusive $W$+jets cross-section is about $500$ times higher than the inclusive $t\bar{t}$ cross-section.  Accounting for the production of extra jets (four for $W$+jets and $2$ for dilepton $t\bar{t}$) and the leptonic branching ratios, this factor reduces to approximately $500\times \alpha^4\times 30\%/(\alpha^2\times 10\%)\sim 15$.  However, unlike for $t\bar{t}$ events, there is no possibility for a second lepton to allow $W$+jets events to naturally exceed a stringent $m_\text{T}$ threshold.  Therefore, the $W$+jets events that pass the signal region event selections must have significant mis-measurement of the $\vec{p}_\text{T}^\text{miss}$.  Section~\ref{sec:mttail} investigates the modeling of $W$+jets events in the $m_\text{T}$ tail.  Control regions for $W$+jets are constructed analogously to the $t\bar{t}$ ones in Sec.~\ref{ttbarCR} with one additional modification.  In order to suppress $t\bar{t}$ events in the $W$+jets control region, events are required to have exactly no $b$-tagged jets.  The $b$-jet veto is a powerful tool for removing $t\bar{t}$ events while maintaining a high yield and purity of $W$+jets events, as demonstrated in Fig.~\ref{fig:wjetsnb}.  For a $\sim 70\%$ $b$-tag working point, one expects that $t\bar{t}$ events fall in the $2$ $b$-tag bin $\sim 0.7^2\sim50\%$ of the time, in the $1$ $b$-tag bin $\sim 2\times 0.3\times 0.7\sim 40\%$ of the time, and in the zero $b$-tag bin $\sim 10\%$ of the time.  However, one significant drawback of this method is that most of the $W$+jets events in the signal region are associated with heavy flavor jets (e.g. $W+b\bar{b}$ and $W+c$) while the ones in the control region are nearly all from light flavor jets.  The kinematic properties of the flavor extrapolation are discussed in Sec.~\ref{wjets:flavorextrap} and the associated systematic uncertainties are documented in Sec.~\ref{sec:susy:wjetsuncert}.
		
All of the $W$+jets control regions are recorded in Table~\ref{tab:wjetscr}, analogously to Table~\ref{tab:TCRs} for the $t\bar{t}$ control regions.  The regions are nearly identical to the $t\bar{t}$ control regions with only the $b$-jet multiplicity inverted.  One exception is WCR13, which has no upper bound on $am_\text{T2}$ as this is not needed to enforce orthogonality with STCR13, as described in Sec.~\ref{singletop:datadriven}.  The $W$+jets background is under $20\%$ of the total SM in all signal regions except SR3 where it is predicted to be less than $30\%$.   Conversely, the signal contamination in the $W$+jets control regions is negligible as the stops receive a similar suppression to $t\bar{t}$ and already have a small cross section.
		
\begin{figure}[h!]
\begin{center}
\includegraphics[width=0.5\textwidth]{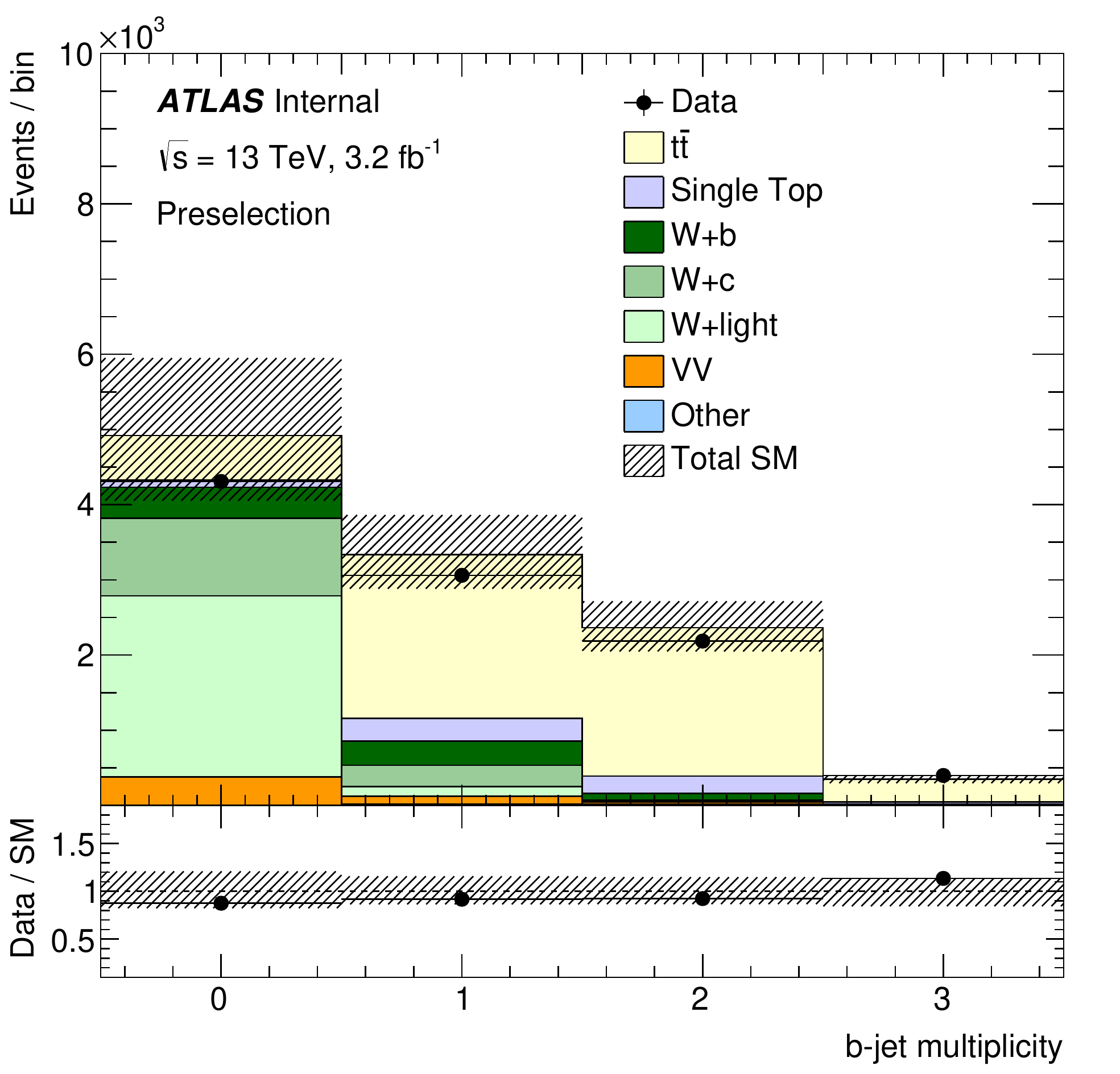}
 \caption{The distribution of the number of $b$-jets with the inclusive preselection.}
 \label{fig:wjetsnb}
  \end{center}
\end{figure}		
		
\begin{table}[h!]
\begin{center}
\noindent\adjustbox{max width=\textwidth}{
\begin{tabular}{|l |cc|cc|cc|cc|cc|cc|cc|cccc}
\hline
   Requirement & SR1 & WCR1 & SR2 & WCR2 & SR3 & WCR3 & tNmed & WCRmed & tNhigh & WCRhigh & SR213 & WCR213 \\
   \hline
     \hline
     $m_\text{T}$ [GeV] & $[140,250]$ & $[60,90]$ & $>140$ & [60,90] & $>180$ & $[60,90]$ & $>140$ & $[60,90]$ & $>200$ & $[60,90]$ & $200$ & $[30,90]$ \\
     $n_\text{$b$-jets}$ & $>0$ & {\color{red}$<1$} & $>0$ & {\color{red}$<1$} & $>0$ & {\color{red}$<1$} & $>0$ & {\color{red}$<1$} & $>0$ & {\color{red}$<1$} & $>0$ & {\color{red}$<1$} \\
     $E_\text{T}^\text{miss}$ [GeV] & -- & -- & -- & -- & $>225$ & $>220$&--&--&$>320$&$>225$& $>350$ & $>250$\\
     $am_\text{T2}$ [GeV] & -- & -- & $>170$ & $>120$ & $>200$ & $>170$ & $>170$& $>120$ & $>170$&$>80$ & $>175$ & {\color{red}$>100$} \\
     $m_\text{T2}^\tau$ [GeV] & -- & -- & -- & -- & $>120$ & $>0$ & -- & -- & $>120$ &$>0$ & -- & --\\ 
     $H_\text{T,sig}^\text{miss} $& -- & -- & -- & -- & -- & -- & --& --& $>12.5$ & $>8.8$ & $>20$ & $>15$ \\
     $\Delta R(b,l)$& -- & -- & -- & -- & -- & -- & --& --& -- & -- & $<2.5$ & $<\infty$ \\
     \hline
     \hline
    Total Yield         & 125 & 897\%& 9.6 &147  & 4.3 &  169   &13.0&161 &5.0&482& 1.3 &135\\
    $t\bar{t}$ Purity &   9\%    & 62\% &17\%&76\%  & 28\%& 79\%  &16\%&68\%&18\%&69\%&12\% & 71\%\\
    \hline
\end{tabular}}
\caption{The definition of the $W$ control regions for each signal region presented in Chapter~\ref{chapter:susy:signalregions}.  Only the requirements that differ from the corresponding signal region are indicated in the table, with a `--' if there is no change between the signal and control region.  All changes highlighted in red are different from the definition of the corresponding $t\bar{t}$ control region.  The lower two rows show the total background yield and the fraction of $W$+jets events in both the signal and control region using the CR-only fit, described in Sec.~\ref{sec:susy:stats}.}
  \label{tab:wjetscr}
\end{center}
\end{table}			

\clearpage		
		
		\subsection{Extrapolating in $b$-jet Multiplicity}
		\label{wjets:flavorextrap}
		
		The extrapolation from the $W$+jets control region to the signal region can be decomposed into two components: first a kinematic extrapolation across $m_\text{T}$ and then a flavor extrapolation from $0$ $b$-jets to $>0$ $b$-jets.  This section demonstrates that the kinematic extrapolation is similar for the $b$-veto and $b$-tag selection.  It is not possible to isolate a pure sample of $W$+jets events in the data with at least one $b$-tagged jet due to the contamination from $t\bar{t}$, so $W$+jets simulation is used for this study.  The top left and right plots in Fig.~\ref{fig:wjetsextrapolation} build upon the preselection with additional jet requirements to be kinematically similar to SR13: $p_\text{T}>100,80,50,25$ GeV.  There are significantly more events in the $b$-veto region than in the $b$-tag region, but the shape (in simulation) of the $m_\text{T}$ distribution is nearly the same, as seen in the top middle plot of Fig.~\ref{fig:wjetsextrapolation}.  There seems to be a small systematic feature just beyond $m_W$ that could be due to the difference in light jet and $b$-jet energy resolutions, which could shift the location of the Jacobian edge.  The ratio of the factors used to extrapolate from low $m_\text{T}$ to high $m_\text{T}$ (transfer factors, or TF) are largely independent of the SR $m_\text{T}$ requirement, as illustrated by the bottom middle plot of Fig.~\ref{fig:wjetsextrapolation}.  This is not strictly necessary for the method to work, but the similarity in transfer factors does make the extrapolation more robust.
		
\begin{figure}[h!]
\begin{center}
\includegraphics[width=0.95\textwidth]{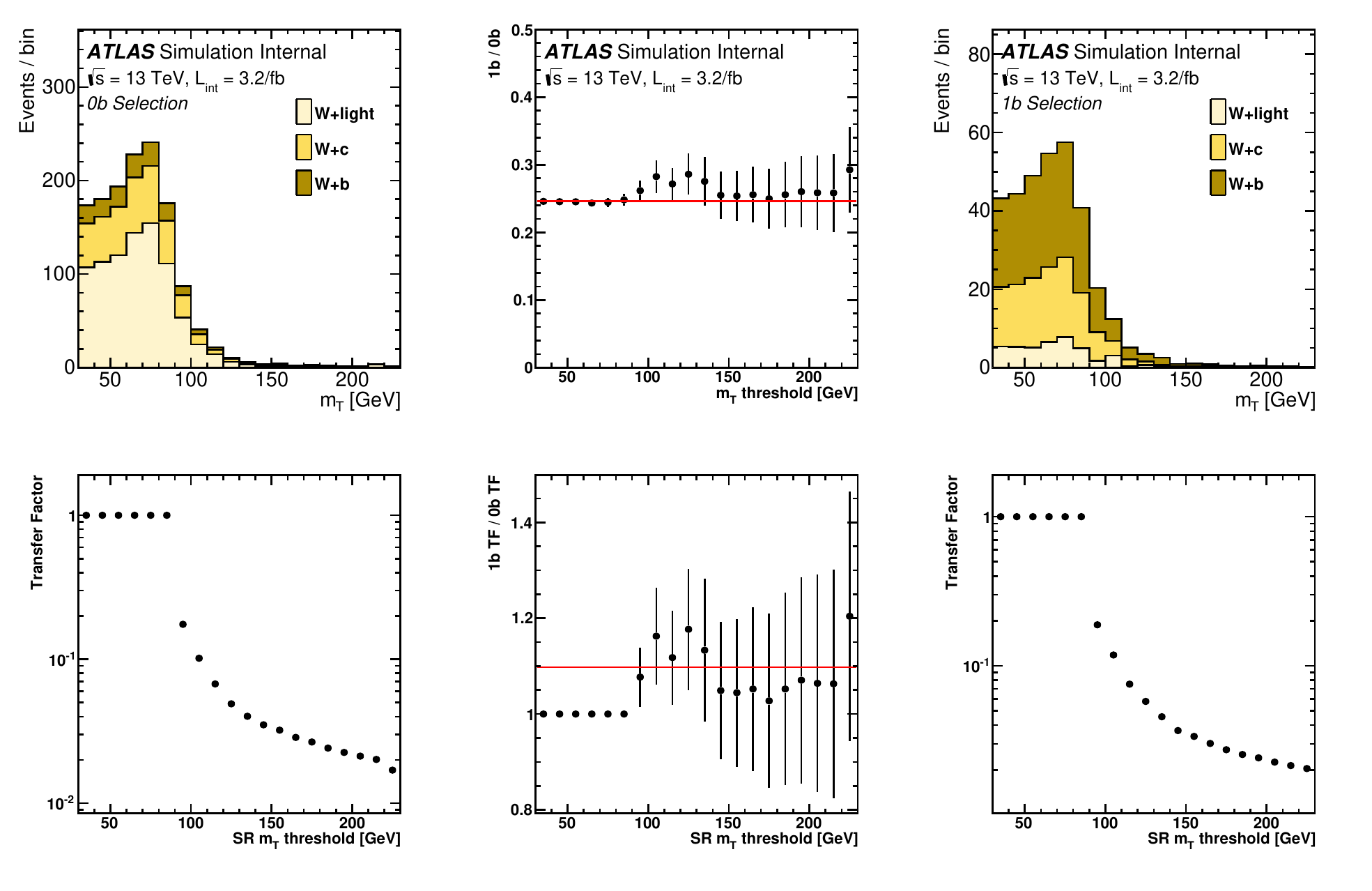}
 \caption{An illustration of the extrapolation from the $W$+jets control region to the signal region decomposed into two steps.  The top left plot is the $m_\text{T}$ distribution in a preselection with a $b$-jet veto.  The bottom left plot shows the transfer factor (TF) that relates the yield in the region $30$ GeV $<m_\text{T}<90$ GeV region to various regions with $m_\text{T}$ greater than the value on the $x$-axis.  The right column of plots are analogous to the first column, but with a $b$-jet requirement instead of a veto.  The middle column plots are ratios of the left and right columns.  The error bars represent statistical uncertainty.}
 \label{fig:wjetsextrapolation}
  \end{center}
\end{figure}			
		
		\subsection{Modeling the $m_\text{T}$ Tail}
		\label{sec:mttail}
		
		While $t\bar{t}$ events enter the signal region mostly through mis-identification of leptons, $W$+jets events pass the signal region event selection mostly through resolution smearing.  Therefore, it is crucial to validate the modeling of the high $m_\text{T}$ tail for $W$+jets events.  Since the signal is largely suppressed by a $b$-jet veto, there is little concern for signal contamination for nearly all of the $0$ $b$-tag phase space.  The kinematic region $90$ GeV $<m_\text{T}<120$ GeV is used to form a $W$+jets validation region in analogy to the $t\bar{t}$ validation regions and is discussed in Sec.~\ref{validationregions}.  The region $m_\text{T}>120$ GeV is investigated in this section, using SR13 as an example.  To be as kinematically close to the signal region as possible, events are required to have four jets with $p_\text{T}>100, 80, 50, 25$ GeV in addition to the preselection that includes $E_\text{T}^\text{miss}>200$ GeV.  Figures~\ref{fig:wvrtail1} and~\ref{fig:wvrtail2} show various kinematic distributions in the resulting WVR-tail validation region which has about $100$ events with approximately $60\%$ $W$+jets purity.  Even though events in the WVR-tail have significantly mis-measured $\vec{p}_\text{T}^\text{miss}$, the MC is a good model within the statistical uncertainites.

\begin{figure}[h!]
\begin{center}
\includegraphics[width=0.5\textwidth]{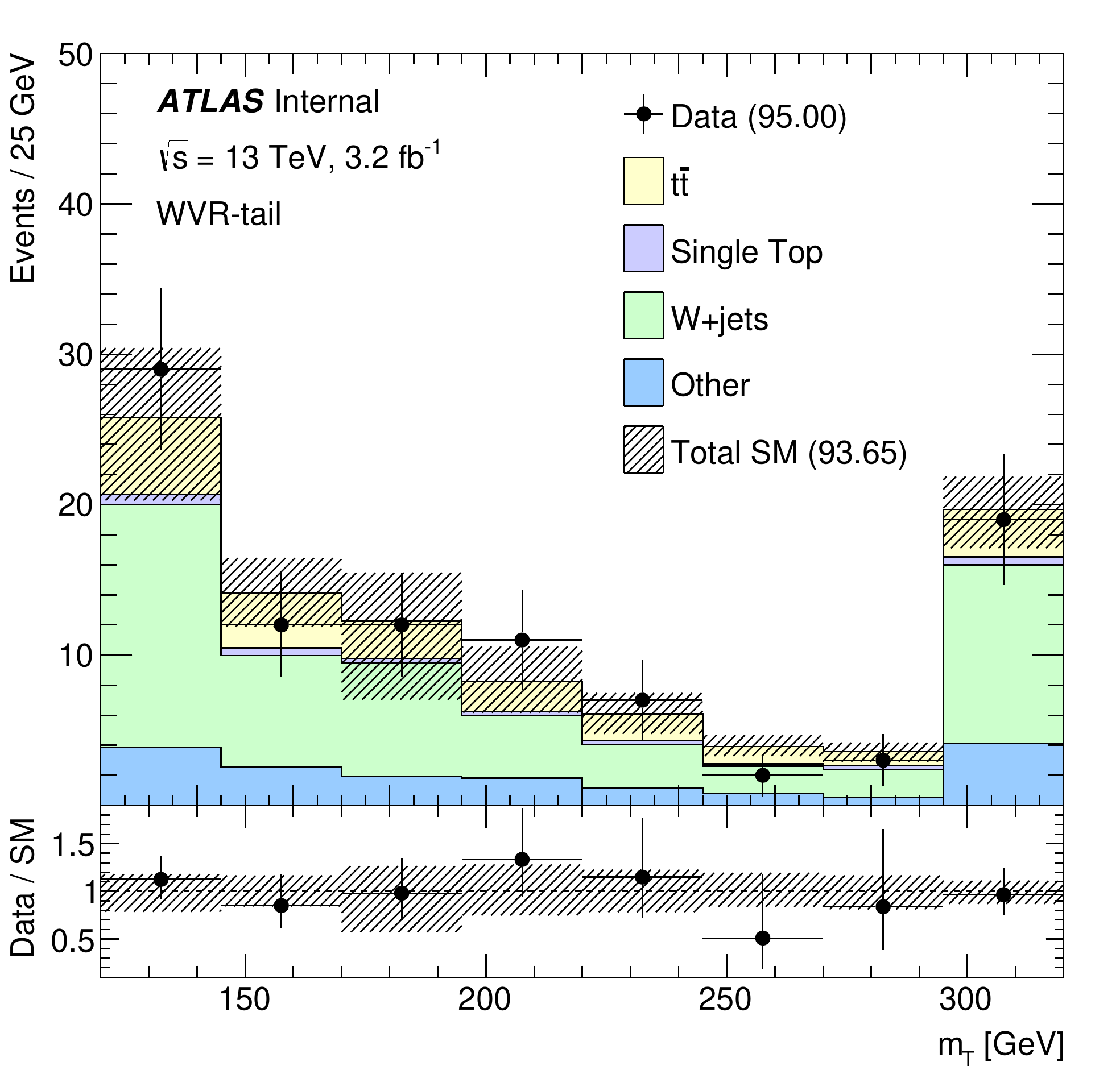}\includegraphics[width=0.5\textwidth]{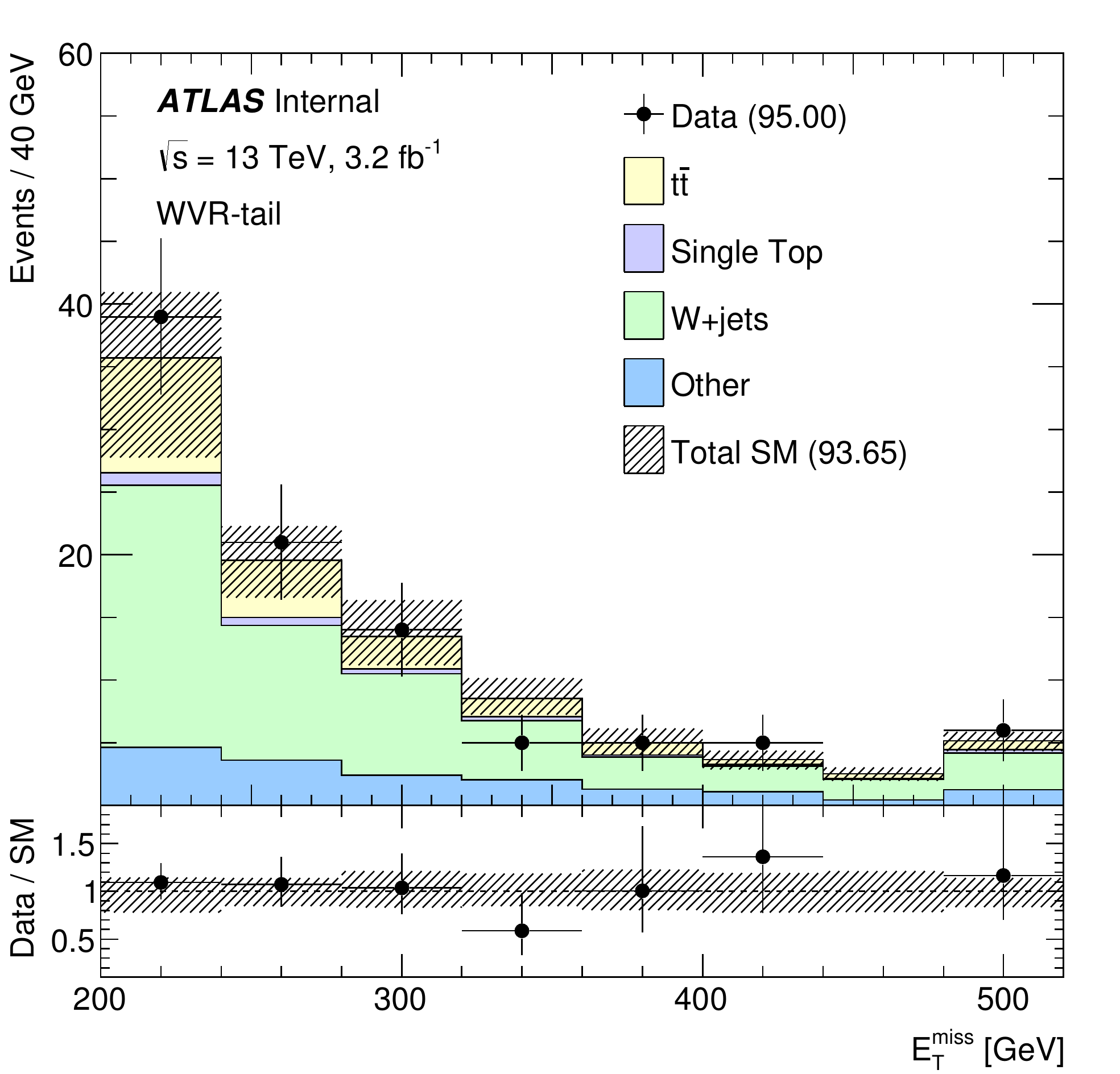}
\includegraphics[width=0.5\textwidth]{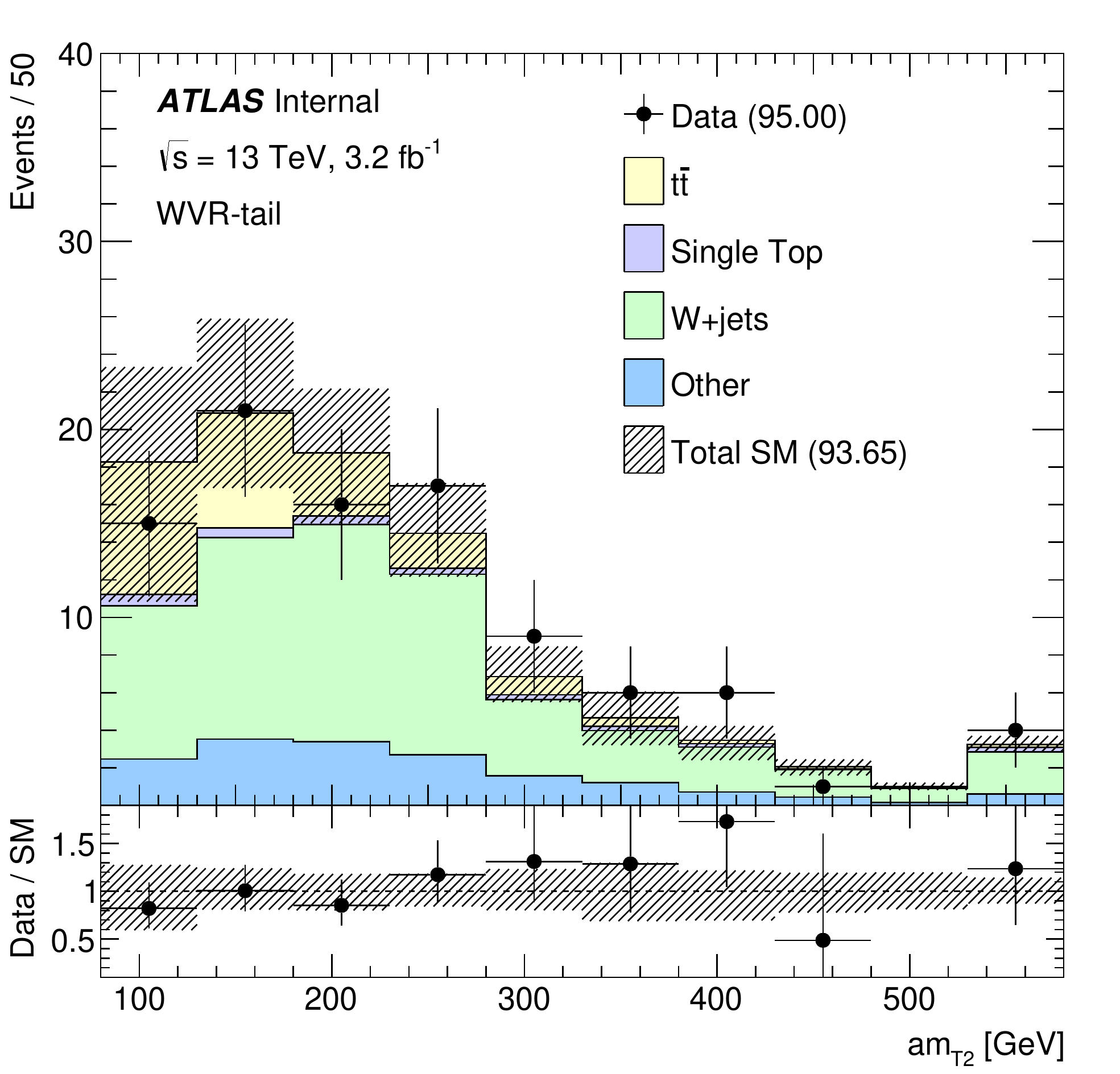}\includegraphics[width=0.5\textwidth]{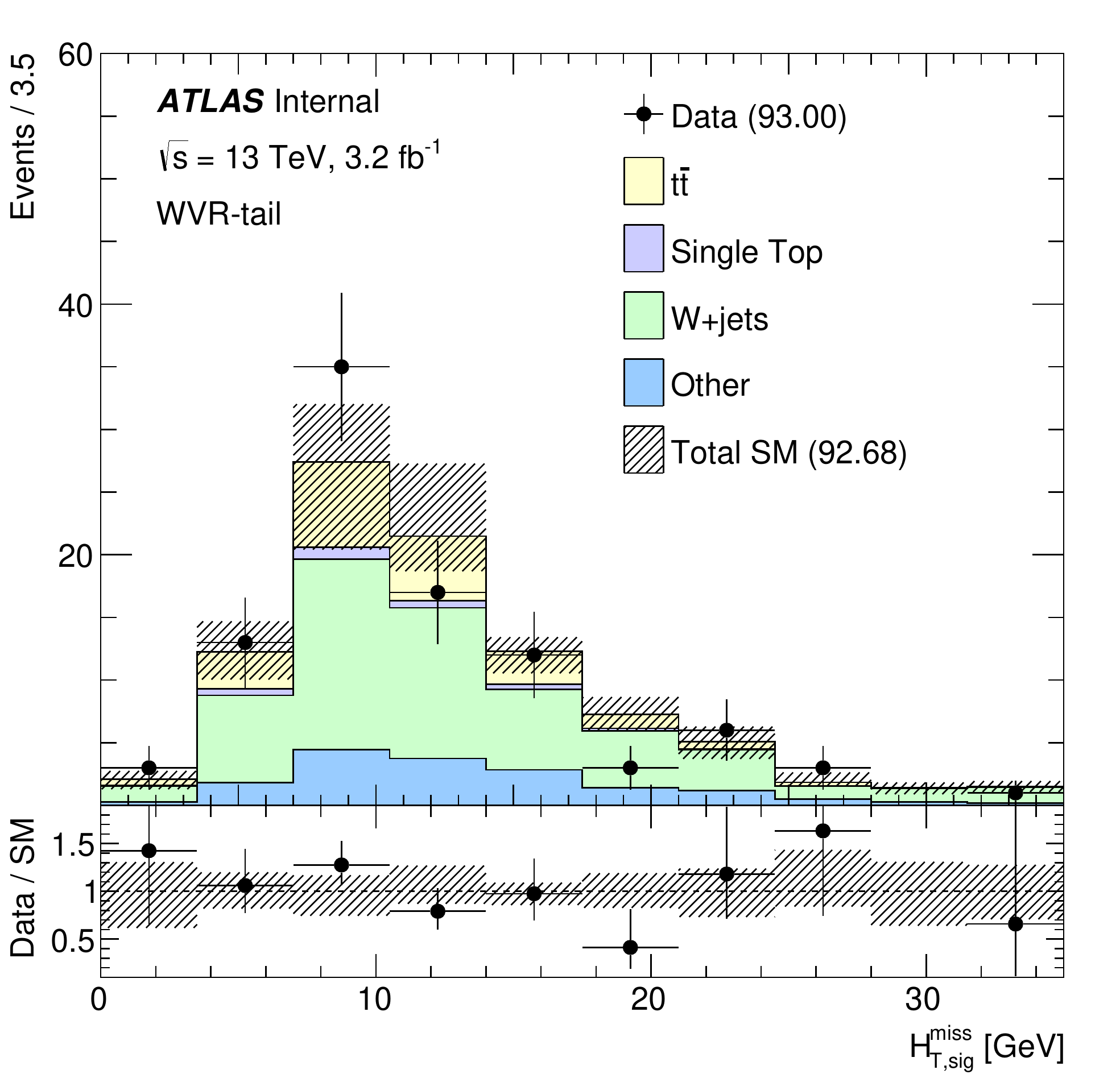}
 \caption{The $m_\text{T}$ (top left), $E_\text{T}^\text{miss}$ (top right), $am_\text{T2}$ (bottom left), and $H_\text{T,sig}^\text{miss}$ (bottom right) distributions in the WVR-tail validation region.}
 \label{fig:wvrtail1}
  \end{center}
\end{figure}		

\begin{figure}[h!]
\begin{center}
\includegraphics[width=0.5\textwidth]{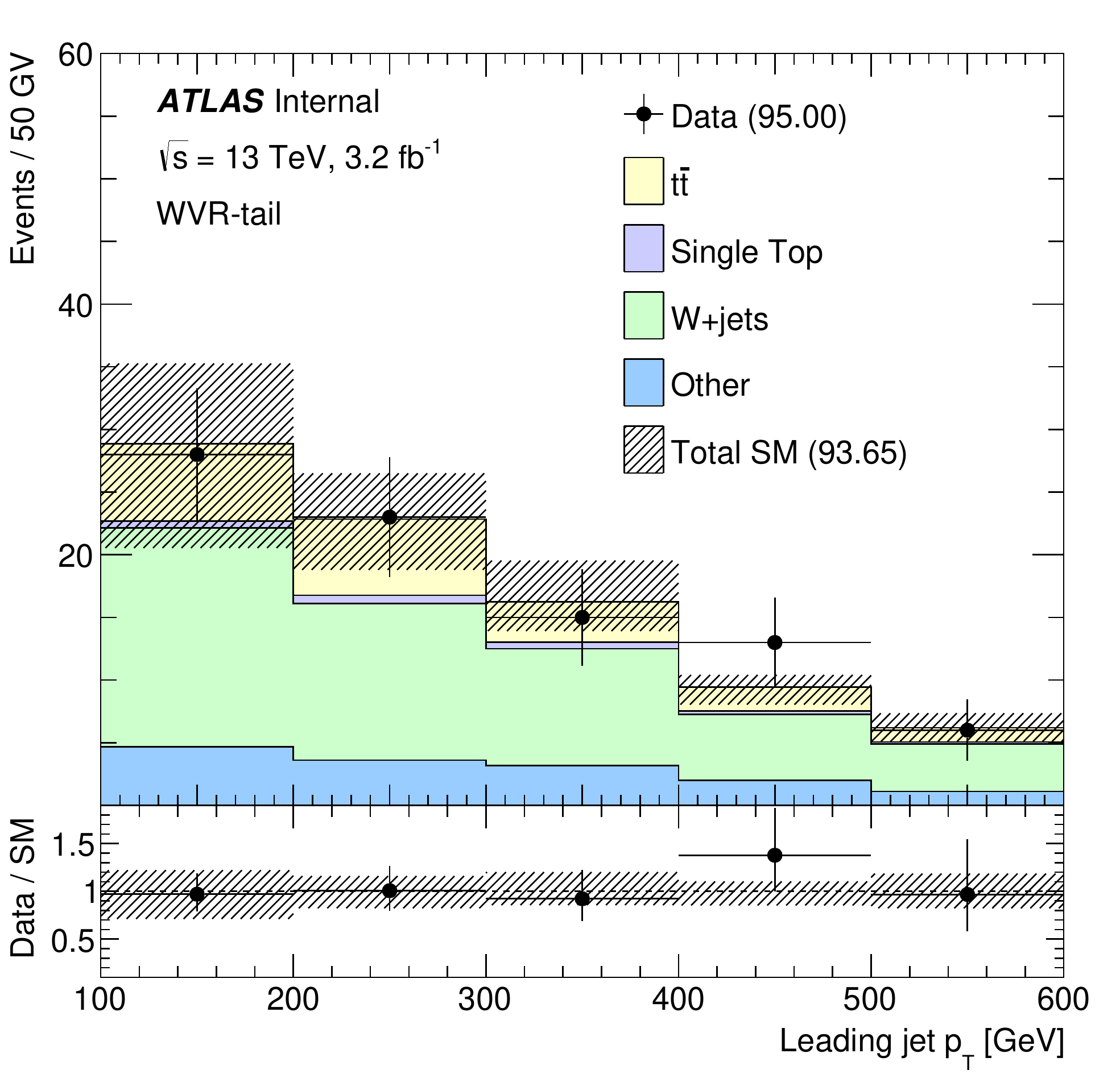}\includegraphics[width=0.5\textwidth]{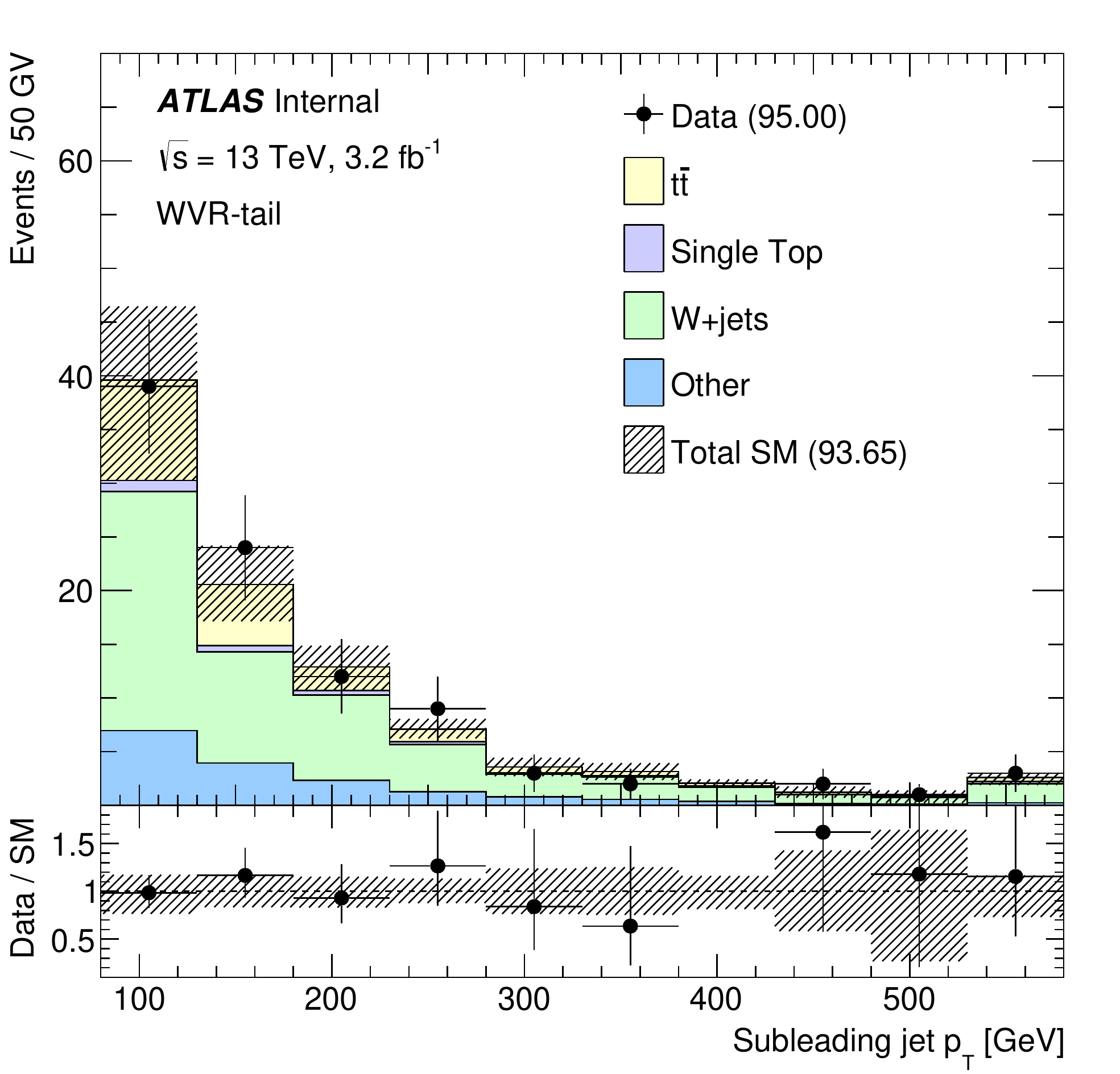}
\includegraphics[width=0.5\textwidth]{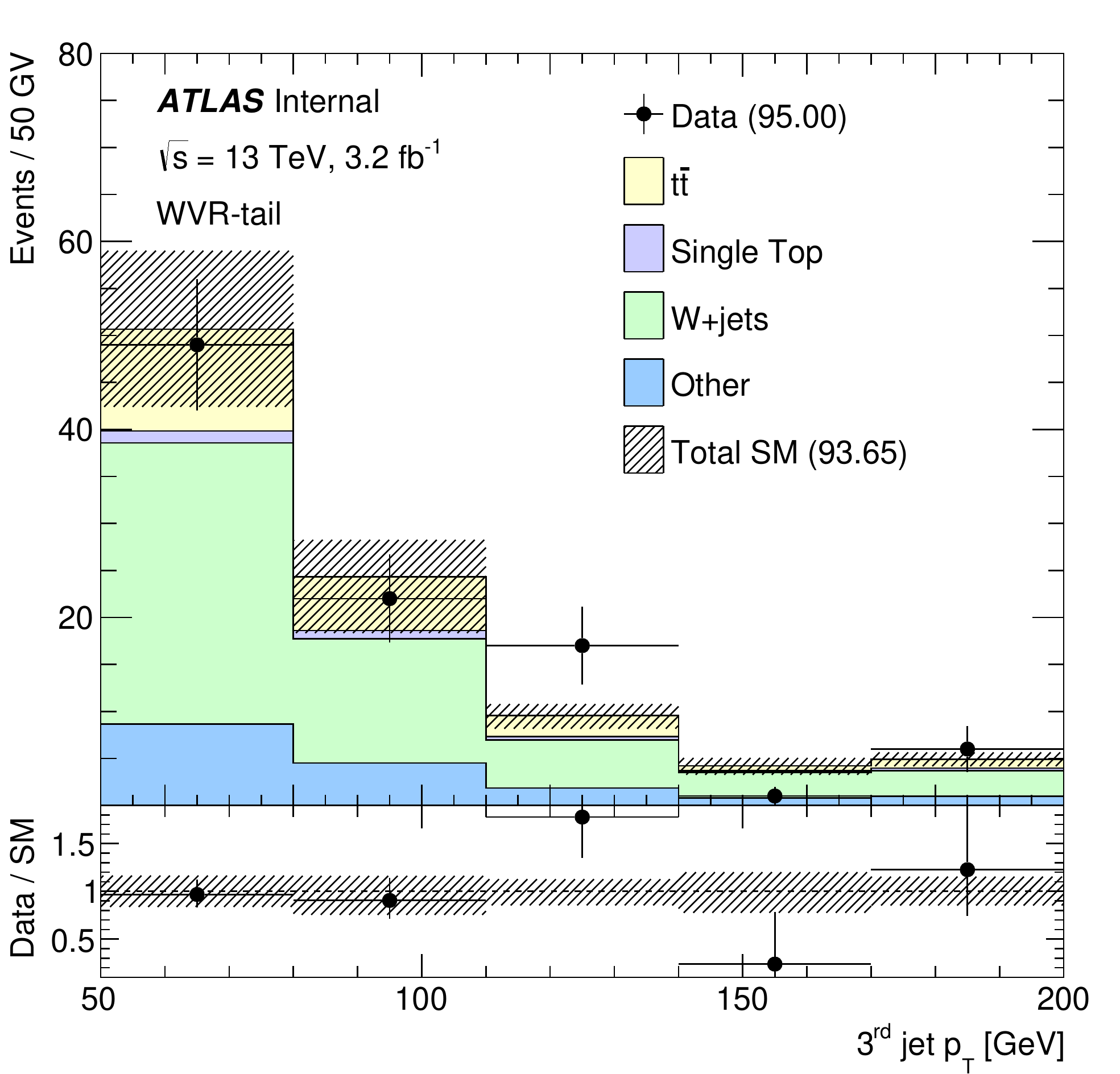}\includegraphics[width=0.5\textwidth]{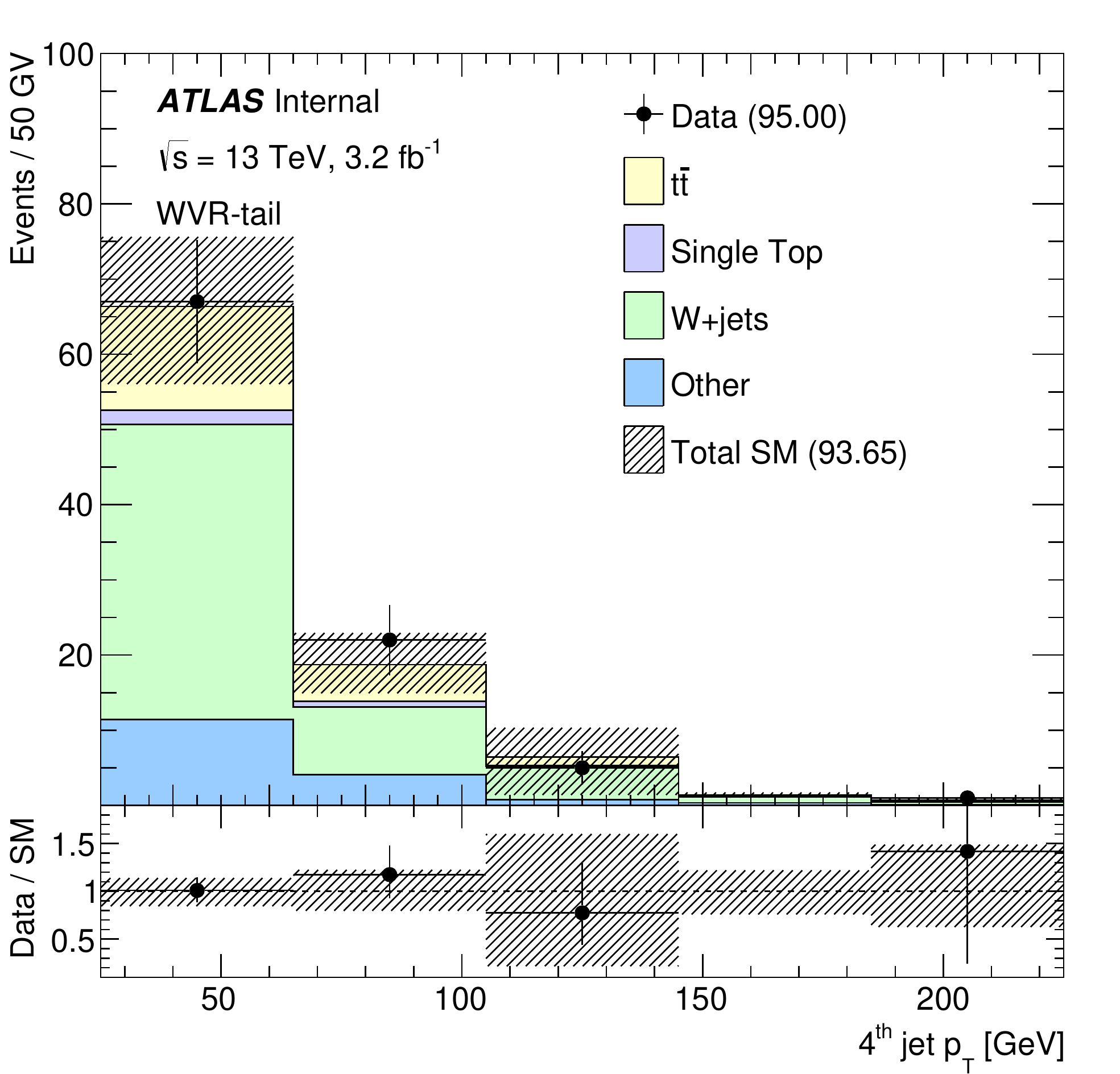}
 \caption{The leading (top left), subleading (top right), third leading (bottom left) and fourth leading (bottom right) $p_\text{T}$ distributions in the WVR-tail validation region.}
 \label{fig:wvrtail2}
  \end{center}
\end{figure}	
		
		\clearpage		
		
		\section{Single Top}
		\label{singletop}
		
			Even though the required $\sqrt{\hat{s}}$ is lower, the cross-section for the single production of top quarks is lower than for $t\bar{t}$ due to the involvement of electroweak couplings and/or $b$-quarks in the proton.  Inclusively, the dominant single top production mechanism is the $t$-channel exchange of a $W$ boson, illustrated by the middle diagram of Fig.~\ref{fig:feynmansingletop}.  However, with only one out-going $W$ boson (from the top quark) and a small number of out-going matrix-element quarks and gluons (two if the $W$ decays leptonically in Fig.~\ref{fig:feynmansingletop}), the $t$-channel single top process is effectively suppressed by $m_\text{T}$ and mild jet requirements.  In contrast, the $Wt$-channel process has two $W$ bosons, and similarly to top quark pair production, the doubly leptonically decaying $W$ boson final state can effectively circumvent an $m_\text{T}$ threshold.  As with $t\bar{t}$, nearly all of the $Wt$ events predicted to pass the SR requirements have two real leptons and so much of the discussion from Sec.~\ref{ttbarCR} related to validating the modeling of extra jets directly applies to the single top process.
			
\begin{figure}[h!]
\begin{center}
\begin{tikzpicture}[line width=1.5 pt, scale=1.]
	
	\draw (-1,1)--(0,0);
	\draw (-1,-1)--(0,0);	
	\draw[vector] (0,0)--(1,0);
	\draw (1,0)--(2,1);
	\draw(1,0)--(2,-1);
	\node at (2.3, 1.) {\large $t$};
	\node at (2.3, -1.) {\large $b$};		
	\node at (-1.2, 1.) {\large $q$};
	\node at (-1.2, -1.) {\large $q'$};	
	\node at (0.5,0.5) {\large $W$};	
		
	 \begin{scope}[shift={(4.8,0)}]
	\draw (-1,1)--(0,1);
	\draw (-1,-1)--(0,-1);	
	\draw[vector] (0,1)--(0,-1);
	\draw (0,1)--(1,1);
	\draw(0,-1)--(1,-1);
	\node at (1.3, 1.) {\large $t$};
	\node at (1.3, -1.) {\large $q'$};		
	\node at (-1.2, 1.) {\large $b$};
	\node at (-1.2, -1.) {\large $q$};	
	\node at (0.5,0.0) {\large $W$};
  	\end{scope}
	
	 \begin{scope}[shift={(8.6,0)}]	
	\draw[gluon2] (-1,1)--(0,1);
	\draw (-1,-1)--(0,-1);	
	\draw (0,1)--(0,-1);
	\draw (0,1)--(1,1);
	\draw[vector] (0,-1)--(1,-1);
	\node at (1.3, 1.) {\large $t$};
	\node at (1.3, -1.) {\large $W$};		
	\node at (-1.2, 1.) {\large $g$};
	\node at (-1.2, -1.) {\large $b$};	
	  	\end{scope}	
	
 \end{tikzpicture}
\end{center}
\caption{Feynman diagrams for the $s$-, $t$-, and $Wt$-channels on the left, middle, and right, respectively.  Even though these are leading order in $\alpha_s$, these are not necessarily the dominant diagrams because the $b$-quark PDF is highly suppressed due to the $b$-quark mass.}
\label{fig:feynmansingletop}
\end{figure}			
			
Several variables described in Sec.~\ref{sec:discriminating} are designed to select events with a resonant hadronically decaying $W$ boson.   Single top events with two leptonically decaying $W$ bosons would fail this requirement, except the invariant mass of one of the leptons and one of the $b$-quarks is naturally much larger than the corresponding observable in top quark pair production.  Figure~\ref{fig:singletopWmb} shows the invariant mass of the non-resonant $W$ boson and $b$ quark from $Wt$ events.    By construction, $m(bW)>m(W)$, but approximately 50\% of events have $m(bW)>m_\text{top}\sim 175$ GeV.  Requirements that select relatively high $m(bl)$ such as the large-radius jet mass and $am_\text{T2}$ will have higher efficiency for $Wt$ events compared with top quark pair production.   As a result, $Wt$ is predicted to be a non-negligible background in many of the signal regions.
		
\begin{figure}[h!]
\begin{center}
\includegraphics[width=0.5\textwidth]{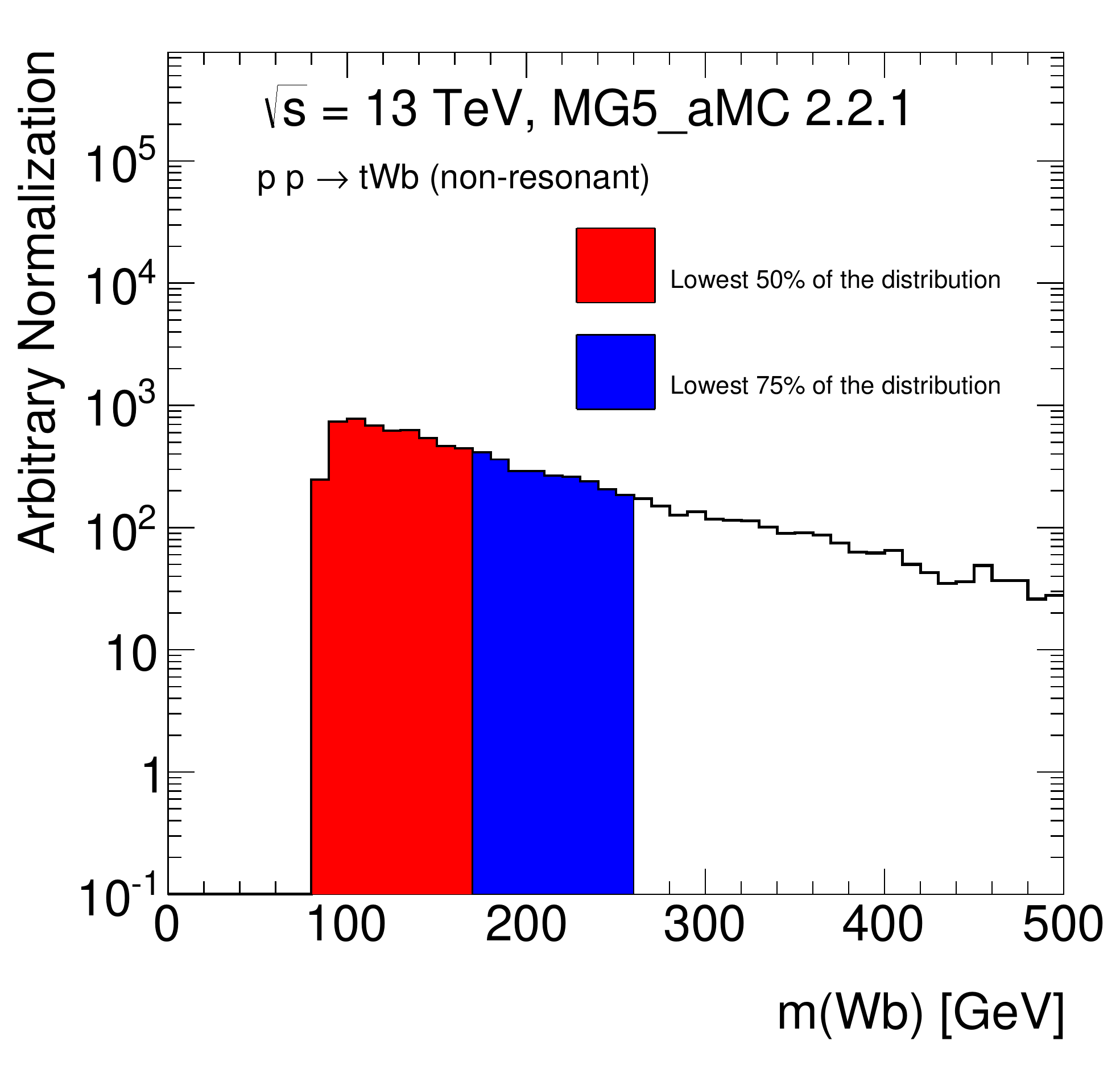}
 \caption{The distribution of the invariant mass of the $W$ boson and $b$-quark from the non-top quark side.  Diagrams with doubly resonant top quarks are explicitly removed.  About $50\%$ of the distribution has $m(Wb)>m_\text{top}\sim 175$ GeV.}
 \label{fig:singletopWmb}
  \end{center}
\end{figure}

The $Wt$-channel Feynman diagram in Fig.~\ref{fig:feynmansingletop} only contains one out-going $b$-quark from the matrix element.  However, {\sc Powheg-Box} predicts that inclusively $40\%$ of the events have a second out-going $b$-quark in the NLO ME using the 5-flavor scheme in which $b$-quarks are treated as constituents of the proton.  Figure~\ref{fig:singletopWmb} shows that this fraction increases with $E_\text{T}^\text{miss}$.  After the full event preselection, nearly all $Wt$ events have a second $b$-jet at particle-level. 
		
\begin{figure}[h!]
\begin{center}
\includegraphics[width=0.45\textwidth]{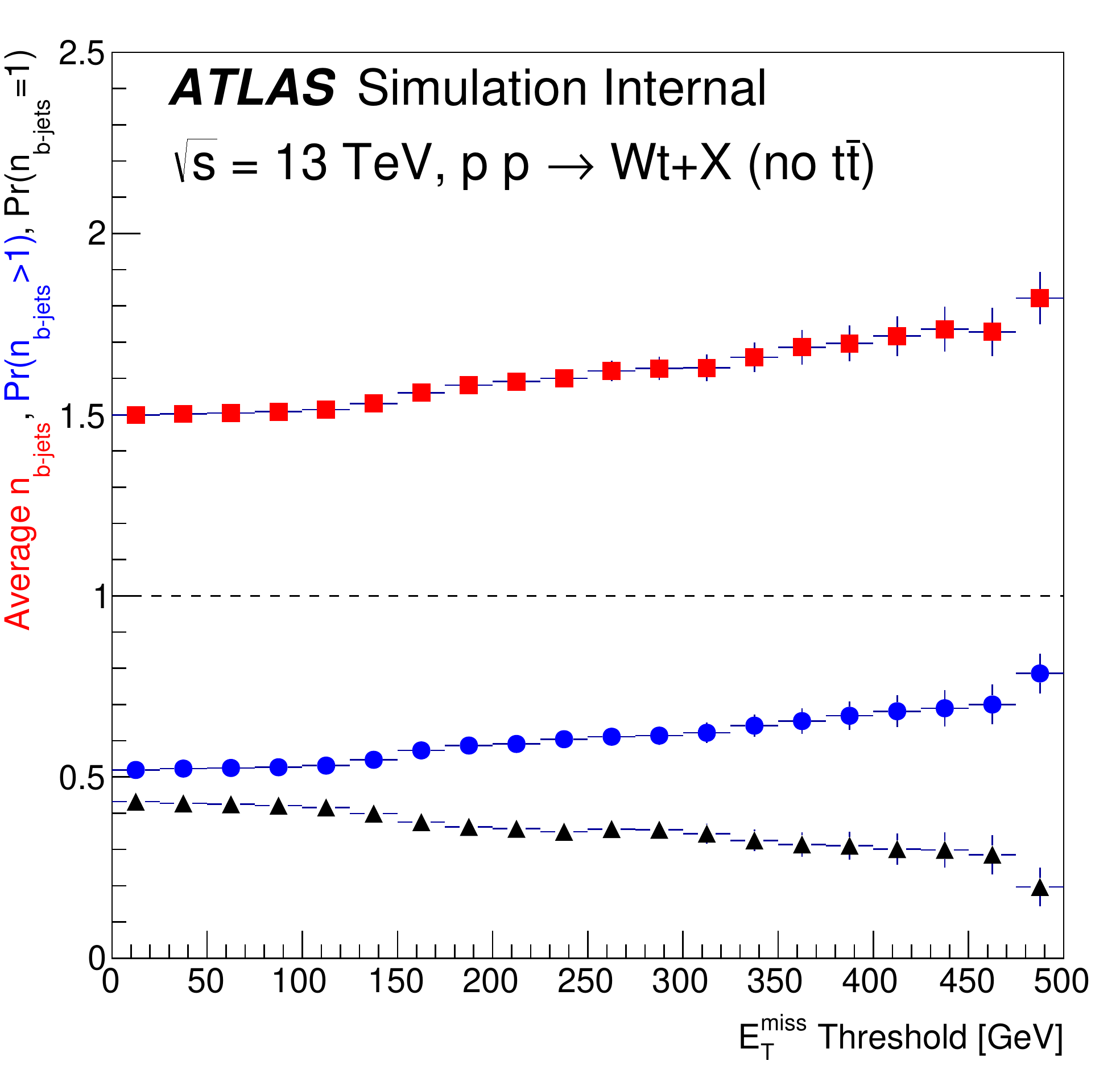}
 \caption{The average number of $b$-jets, the probability for at least and for exactly one $b$-jet at particle-level from {\sc Powheg-Box}+{\sc Pythia 6}.}
  \label{fig:singletopWmb}
  \end{center}
\end{figure}
		
One key challenge with the $Wt$ process is the interference between $Wt$ simulated at NLO and LO $t\bar{t}$.   Representative beyond leading order $Wt$ Feynman diagrams with a second $b$-quark in the final state are shown in Fig.~\ref{fig:singletopWmb2}.  There are some diagrams that overlap with $t\bar{t}$ when a $Wb$ pair go on-shell.  This is further discussed in Sec.~\ref{sec:singletopuncerts} in the context of systematic uncertainties, but is an important motivation for constraining aspects of this process with data.  Another motivation is that unlike $t\bar{t}$, the $Wt$ process has only recently been observed~\cite{Aad:2015eto,Chatrchyan:2014tua} and has essentially no constraints on the modeling of its kinematic properties.  Section~\ref{singletop:datadriven} describes a single-top CR used for the first time in a $t\bar{t}+E_\text{T}^\text{miss}$ search in the early $\sqrt{s}+13$ TeV search.  For the signal regions at $\sqrt{s}=8$ TeV, the single top background is predicted directly from simulation.

\begin{figure}[h!]
\begin{center}
\includegraphics[width=0.4\textwidth]{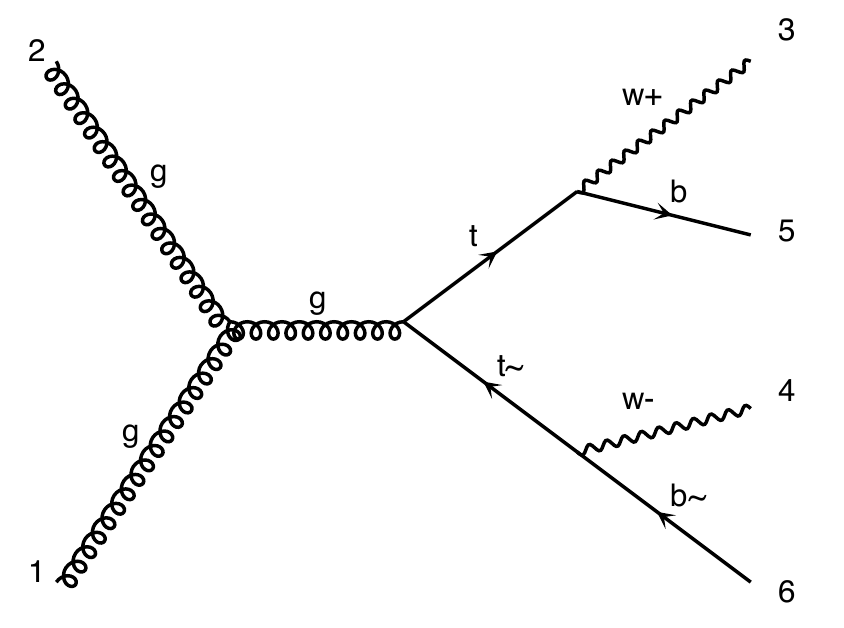}\hspace{5mm}\includegraphics[width=0.4\textwidth]{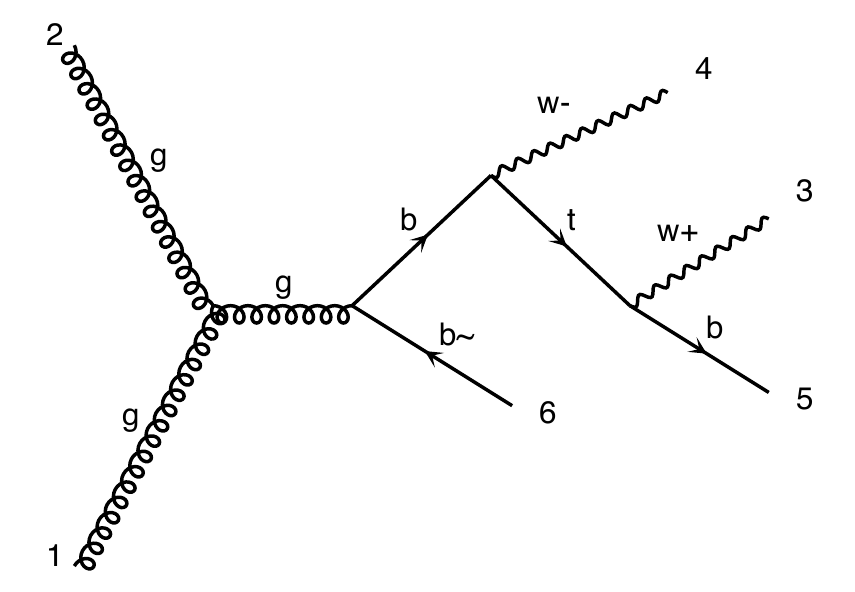}\\
\includegraphics[width=0.4\textwidth]{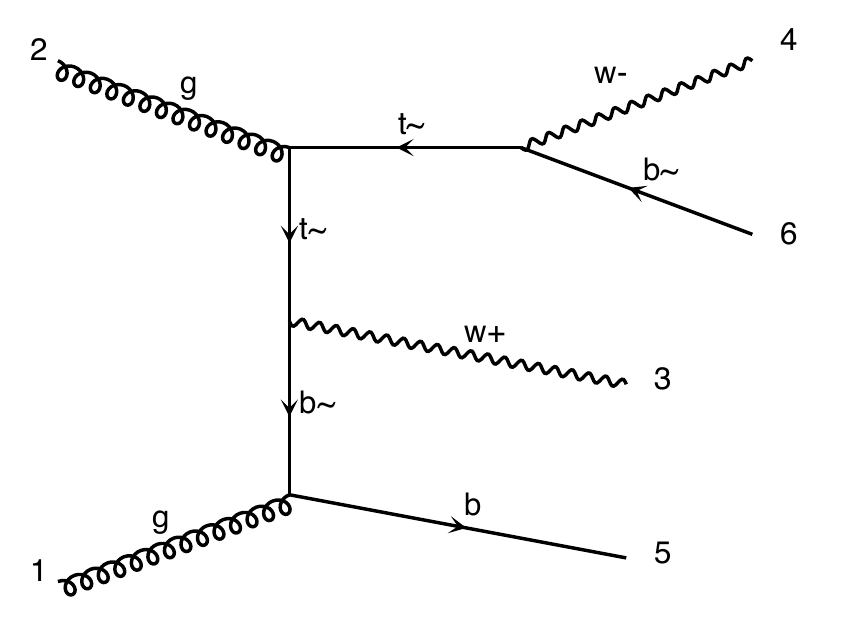}\hspace{5mm}\includegraphics[width=0.4\textwidth]{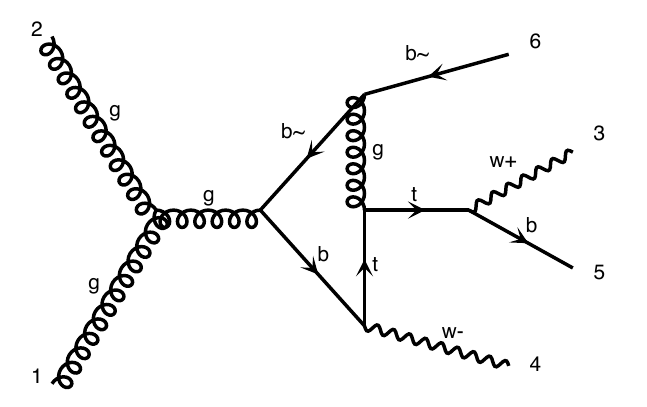}
 \caption{Next-to-leading-order (up to $\alpha_s^4\alpha_w^2$) Feynman diagrams from {\sc MG5\_aMC} that have the same outgoing particles as $t\bar{t}$ at tree level.  The top left diagram overlaps with $t\bar{t}$ when the two intermediate top quarks go on-shell.  The top right diagram has no top quarks at all and the bottom diagrams contain top quarks but do not interfere with top quark pair production even when the intermediate top quark(s) go on-shell.}
 \label{fig:singletopWmb2}
  \end{center}
\end{figure}

			\subsection{A Data-driven Approach}
			\label{singletop:datadriven}
			
			The only difference between $Wt$ and $t\bar{t}$ when there are two out-going $b$-quarks in the ME is the presence of one non-resonant $Wb$ pair.  Therefore, variables aimed at reconstructing the hadronic or leptonically decaying top quarks can (partially) separate $Wt$ from $t\bar{t}$.  One particularly powerful variable for this task is $am_\text{T2}$, for which $t\bar{t}$ events are significantly reduced beyond $am_\text{T2}\gtrsim m_\text{top}$.  As mentioned in the previous section, many $Wt$ events have a second $b$-jet at particle-level.  Requiring two $b$-jets is crucial for obtaining a high $Wt$ purity because of the contamination from $W$+jets events which are also not bounded by $am_\text{T2}\lesssim m_\text{top}$.  Additional $t\bar{t}$ suppression is possible in the two $b$-jet selection when the $\Delta R$ between $b$-jets is required to be relatively large.  This is because one way for one-lepton $t\bar{t}$ events to exceed the $am_\text{T2}$ endpoint is for a charm jet from the hadronically decaying $W$ boson to be $b$-tagged with higher $b$-tagging weight than a second $b$-jet from the top quark decay.   This is illustrated in Fig.~\ref{fig:amt2charm} at parton level.  For a given choice of $b$-jet, $am_\text{T2}\sim max(80,m(bl))$.  To account for combinatorics, the selected $am_\text{T2}$ is minimized over both pairings of $b$-jets; therefore $am_\text{T2}\sim min(m(b_1l),m(b_2l))$.  When the $b$-jet entering the $am_\text{T2}$ calculation is on the same side as the lepton, then $m(bl)$ is bounded from kinematics by $\sqrt{m_\text{top}^2-m_W^2}\sim 155$ GeV independent of the top quark $p_\text{T}$ so this will generally be smaller than the invariant mass of the lepton and the charm-jet which increases with $p_\text{T}^\text{top}$.  However, if both the (true) $b$-jet and the charm-jet are from the opposite top quark from the lepton, the minimum can be much larger than $m_\text{top}$.  
			
\begin{figure}[h!]
\begin{center}
\includegraphics[width=0.45\textwidth]{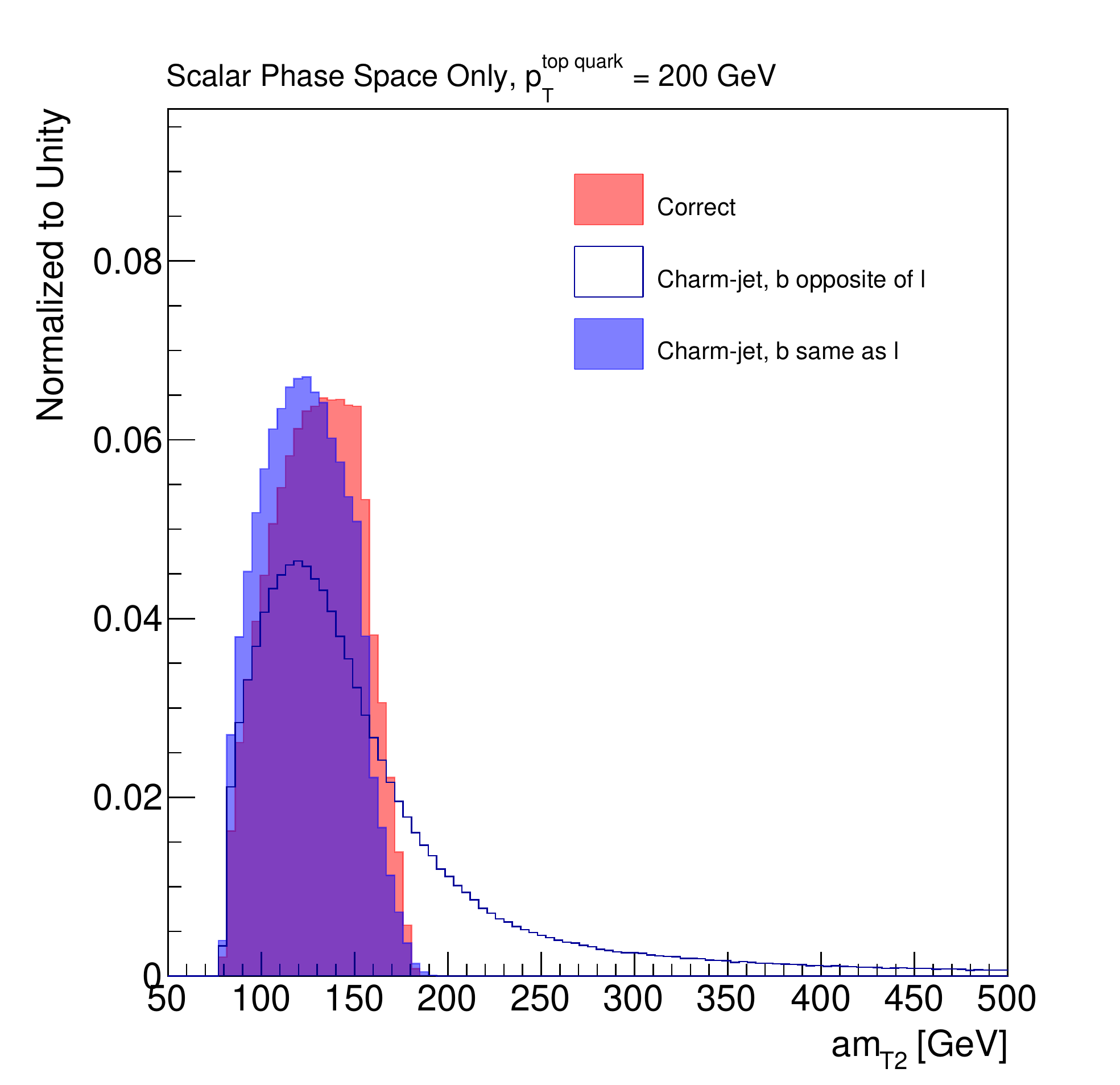}
 \caption{The value of $am_\text{T2}$ from a simple scalar parton-level simulation.  In the correct assignment, the two $b$-quarks from the top decay are used to calculate $am_\text{T2}$.  When the charm quark from the $W$ decay is used in place of one of the $b$-quarks, then the second $b$-quark used in the calculation can either be from the same top quark as the lepton or the opposite top quark. In all cases, the plotted value is the minimum over both pairings of the lepton and `$b$-jets'.}
 \label{fig:amt2charm}
  \end{center}
\end{figure}

Table~\ref{tab:stcr} summarizes the event selection for the single top control region.  The $m_\text{T}$ window is larger than for the $t\bar{t}$ control region in order to increase statistics; there are not enough events for a validation region at high $m_\text{T}$ which is what the region beyond $m_\text{T}=90$ GeV is used for in the $t\bar{t}$ case.  A few of the other requirements are also loosened from the $t\bar{t}$ case in order to increase statistics.   With about $80$ predicted events in the single top control region, the $Wt$ purity is about $40\%$.  About $3\%$ of the single top events are due to other single top processes, dominated by $t$-channel production.  

\begin{table}[h!]
\begin{center}
\noindent\adjustbox{max width=\textwidth}{
\begin{tabular}{|l |ccc|}
\hline
   Requirement & SR13 & TCR & STCR \\
   \hline
     \hline
     $m_\text{T}$ [GeV] & $>200$ & $[30,90]$ & $[30,120]$ \\
     $n_\text{$b$-jets}$ & $>0$ & >0 & $>1$  \\
     $E_\text{T}^\text{miss}$ [GeV] & $>350$ & $>250$ &  $>200$ \\
     $H_\text{T,sig}^\text{miss} $&$>20$ & $>15$ & $>5$\\
     $am_\text{T2} $& $>175$ & $[100,200]$ & $>200$\\ 
     Large $R$ jet mass [GeV] & $>140$ & $>140$ & --\\
     $\Delta R(b,l)$& $<1.5$ & -- & -- \\
     $\Delta R(b,b)$& -- & -- &  $<1.2$\\
     \hline
     \hline
    $t\bar{t}$ $1L$  &  0.0  &  72.0& 30.8\\
    $t\bar{t}$ (other)  &   0.3  &12.4  & 4.7\\
    $Wt$ &    0.1  &3.8 & 29.9\\
    Single top (other) & 0.0    & 1.6&1.0\\
    $W$+jets &   0.2  & 7.9 &11.9\\
    $VV$ &   0.1   & 1.8 &2.5\\
    $t\bar{t}+V$ & 0.4     &0.7  &1.2\\
    \hline
    \hline
    $Wt$ purity & 12\%& 4\%&36\% \\
     SM (pre-fit) &  1.1  & 100.3 & 82.0\\
      Data  &  Sec.~\ref{chapter:results}    &102  &71\\
    \hline
\end{tabular}}
\caption{The definition of the single top control region for SR13 compared with the definitions of SR13 and the corresponding $t\bar{t}$ control region.  Only the requirements that differ from the corresponding signal region are indicated in the table, with a `--' if there is no requirement on the given variable.  The lower panel indicates pre-fit predictions from the simulation compared with the data.}
  \label{tab:stcr}
\end{center}
\end{table}
		
Figure~\ref{fig:stcramt2} shows the $am_\text{T2}$ distribution in the $Wt$ control region with all selections applied aside from the $am_\text{T2}$ requirement. There may be a small slope in the data/MC ratio for $am_\text{T2}\lesssim 200$ GeV, but in the control region there is no significant evidence for mis-modeling.  The bottom ratio panel in Fig.~\ref{fig:stcramt2} shows that the $Wt$ purity increases by nearly two orders of magnitude between $\sim 150$ GeV and $\sim 300$~GeV.  The $b$-jet related selections are illustrated in Fig.~\ref{fig:stcrbjets}.  The lower ratio in the $n_\text{$b$-jets}$ clearly shows the significant improvement in the $Wt$ purity by explicitly requiring a second $b$-jet.  As motivated earlier, the $\Delta R(b_1,b_2)$ distribution peaks at low values for $t\bar{t}$ and the purity significantly increases for $\Delta R\gtrsim 1$.  The overall MC prediction is a slightly above the measured data so the $Wt$ contribution in the final results is scaled down from the control region method.
		
\begin{figure}[h!]
\begin{center}
\includegraphics[width=0.45\textwidth]{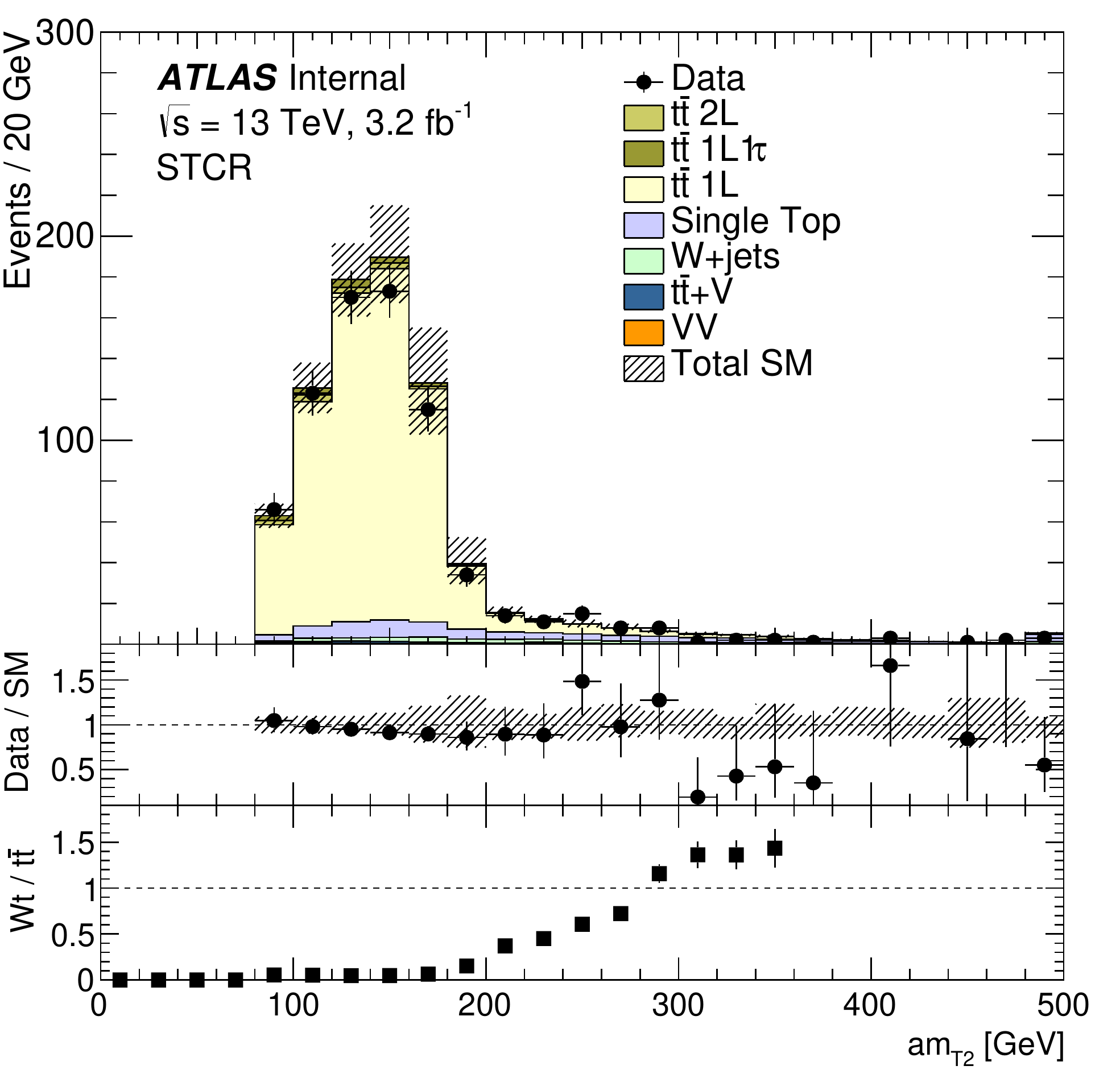}\includegraphics[width=0.45\textwidth]{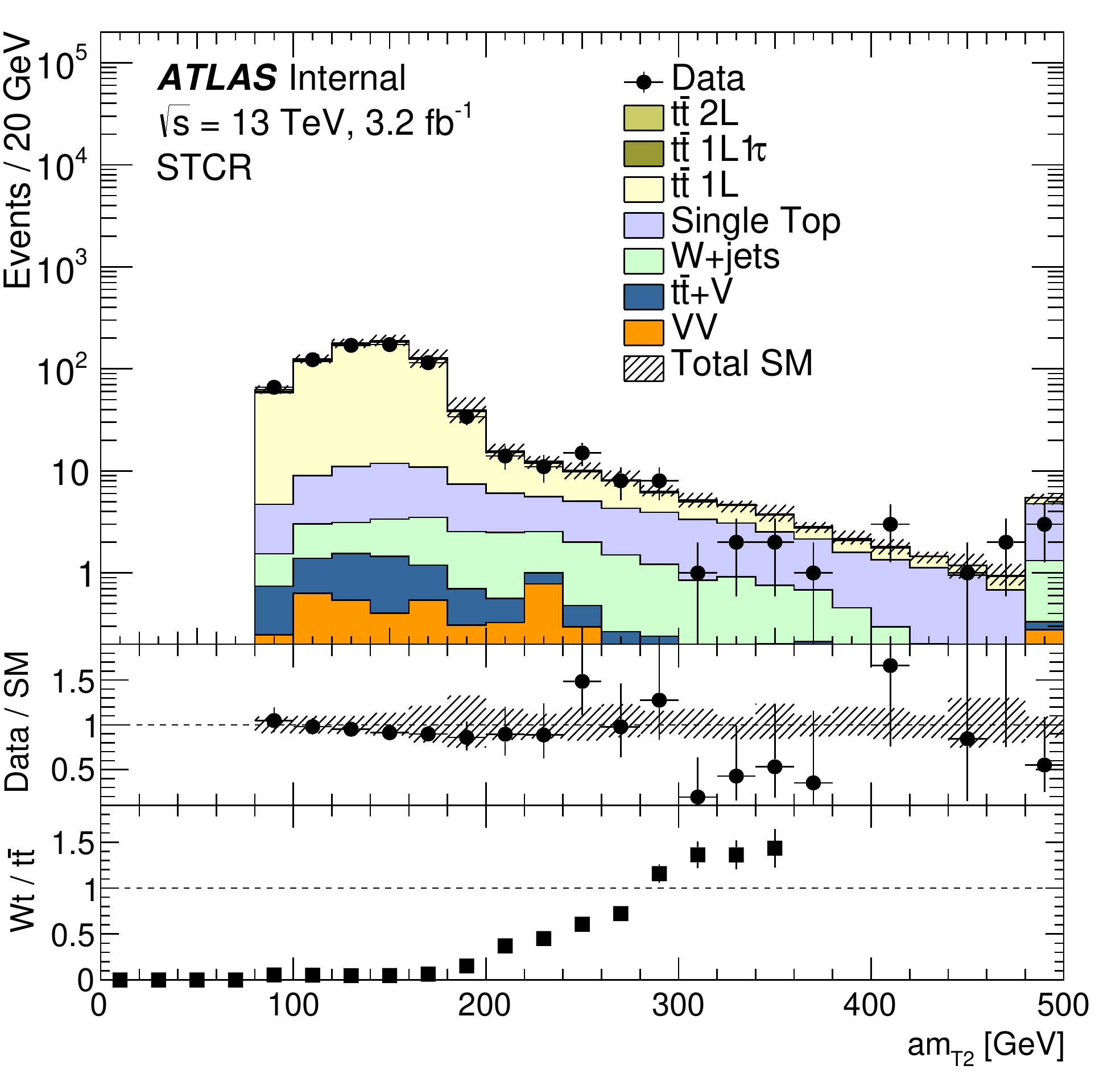}
 \caption{The distribution of $am_\text{T2}$ in the single top control region with all selections applied except the $am_\text{T2}$ requirement.  The left and right plots are identical except for the difference in scale.}
 \label{fig:stcramt2}
  \end{center}
\end{figure}			

\begin{figure}[h!]
\begin{center}
\includegraphics[width=0.45\textwidth]{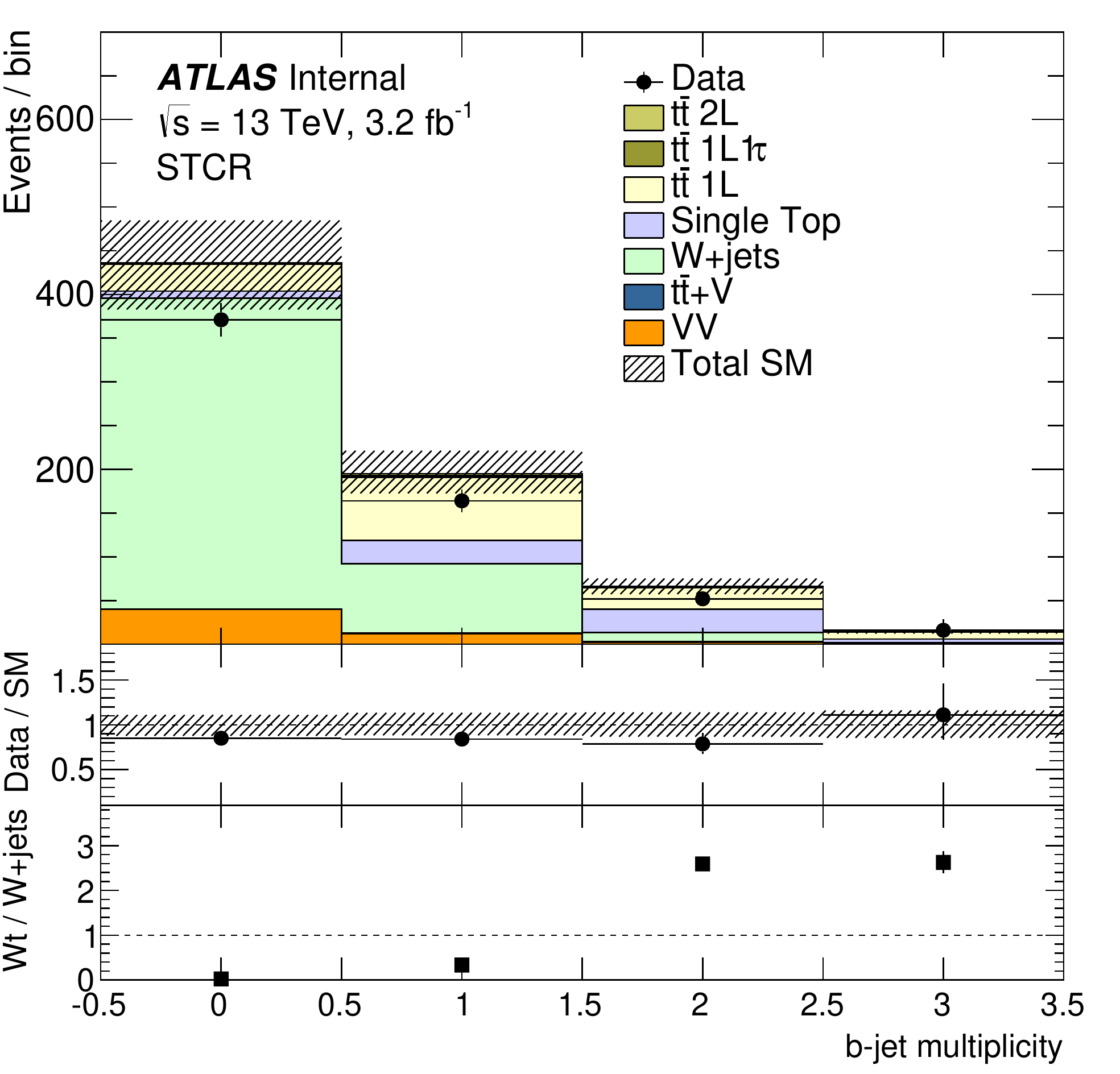}\includegraphics[width=0.45\textwidth]{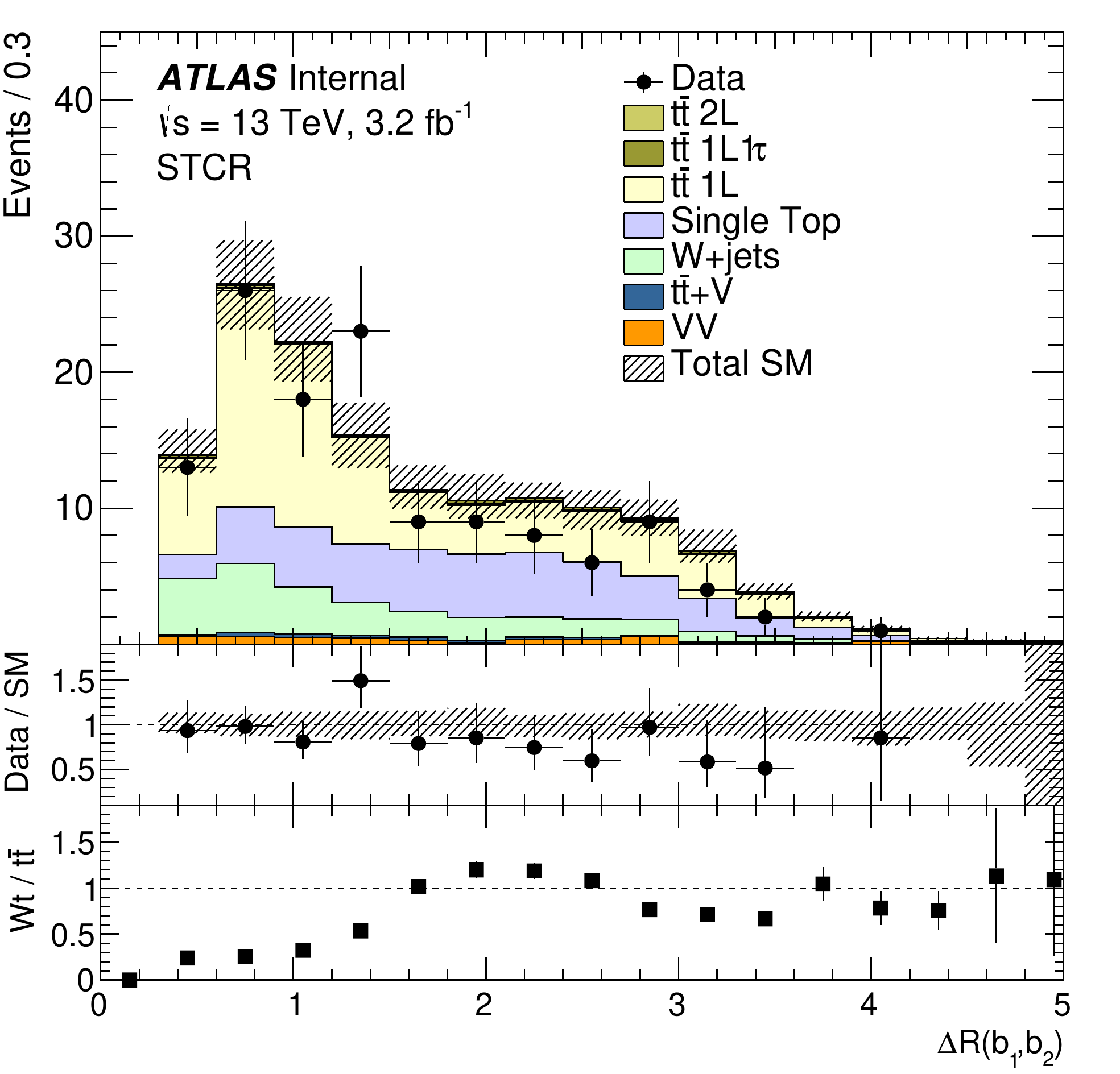}
 \caption{The distribution of $n_\text{$b$-jets}$ and $\Delta R(b_1,b_2)$ in the single top control region with all selections applied except the ones shown in the plots.}
 \label{fig:stcrbjets}
  \end{center}
\end{figure}	
		
		\clearpage
		
		\section{Top Quark Pair Production with a $Z$ Boson}
		\label{ttv}
		
			 The production of a $Z$ boson in association with a top quark pair that decays into neutrinos is an irreducible background for the stop search with $\tilde{t}\rightarrow t+\tilde{\chi}^0$.  Figure~\ref{fig:feynmanttz} shows two representative Feynman diagrams where the phenomenological similarity between the two processes is highlighted with dashed circles around the missing momentum from neutralinios/neutrinos.  The cross-section for $t\bar{t}+Z(\rightarrow\nu\nu)$ is suppressed with respect to generic top quark pair production by $\sim \alpha_w\times\text{BR}(Z\rightarrow\nu\nu)\times P\sim 0.001P$ where $P \sim (m_{t\bar{t}}/m_{t\bar{t}Z})^2\sim\mathcal{O}(0.1)$ phase space factor (see Fig.~\ref{fig:stopcrosssection}).  Therefore, this process is only a significant background for the high stop mass search where the signal cross-section is comparably small.  In this regime, the $t\bar{t}+Z$ is a dominant background even though its cross-section is only known with a $30\%$ uncertainty from dedicated measurements~\cite{ATLAS-CONF-2016-003,Aad:2015eua,CMS-PAS-TOP-16-009,Khachatryan:2015sha}.  The early $\sqrt{s}=8$ TeV analysis used a simulation-only method to estimate the $t\bar{t}+Z$ background and the full $\sqrt{s}=8$ TeV began using a data-driven technique to validate this estimate.  The data-driven estimate became fully integrated into the control region method for the default estimate in the early $\sqrt{s}=13$ TeV analysis.  Section~\ref{ttz:simulation} briefly describes the simulation-only estimate and Sec.~\ref{sec:ttz:datadriven} details the data-driven technique using photons.
	
	\vspace{10mm}
	
\begin{figure}[h!]
\begin{center}
\begin{tikzpicture}[line width=1.5 pt, scale=1.]
	
	\draw (-1,1)--(0,0);
	\draw (-1,-1)--(0,0);	
	\draw[gluon2] (0,0)--(1,0);
	\draw[dashed,color=red] (1,0)--(2,1);
	\draw[dashed,color=red] (1,0)--(2,-1);
	\draw[color=black] (2,1)--(3,1.5);
	\draw[color=red] (2,1)--(3,0.5);
	\draw[vector,color=red] (2,1)--(3,0.5);
	\draw[color=black] (2,-1)--(3,-1.5);
	\draw[color=red] (2,-1)--(3,-0.5);
	\draw[vector,color=red] (2,-1)--(3,-0.5);
			\node at (1.5, 1.) {\large $\tilde{t}$};
	\node at (1.5, -1.) {\large $\tilde{t}$};
	\node at (3.4, 1.5) {\large $t$};
	\node at (3.4, -1.5) {\large $t$};
	\node at (3.4, 0.5) {\large $\tilde{\chi}^0$};
	\node at (3.4, -0.5) {\large $\tilde{\chi}^0$};
	\draw[dashed] (3.2,0) circle (1);	
	
	\draw[implies-implies,double equal sign distance] (5.,0) -- (5.5,0);
	
	 \begin{scope}[shift={(7.5,0)}]

	\draw[gluon2] (-1,1)--(0,1);
	\draw[gluon2] (-1,-1)--(0,-1);	
	\draw (0,1)--(0,-1);
	\draw[color=black] (0,1)--(2,1.5);
	\draw[color=black] (0,-1)--(2,-1.5);
	\draw[vector,color=black] (0,0)--(1,0);
	\draw[color=black] (1,0)--(2,0.5);
	\draw[color=black] (1,0)--(2,-0.5);		
	\node at (2.3, 0.5) {\large$\nu$};
	\node at (2.3, -0.5) {\large$\nu$};
	\node at (0.5, 0.5) {\large$Z$};
	\draw[dashed] (2.2,0) circle (1);	
	\node at (2.3, 1.5) {\large $t$};
	\node at (2.3, -1.5) {\large $t$};
  	\end{scope}
	
 \end{tikzpicture}
\end{center}
\caption{Feynman diagrams for stop pair production and decay (left) and top quark pair production in association with a $Z$ boson that decays to neutrinos (right).  The dashed circles show the dominant contribution to the $E_\text{T}^\text{miss}$.  There are many other leading order Feynman diagrams for the $t\bar{t}+Z$ process, which are described in Sec.~\ref{sec:ttz:datadriven}.}
\label{fig:feynmanttz}
\end{figure}	
		
		\clearpage
		
			\subsection{Estimation from Simulation}
			\label{ttz:simulation}
			
		MadGraph 5 (MG5\_aMC) is used to simulate $t\bar{t}+Z$ events at $\sqrt{s}=8$ ($13$) TeV.  The top row of Fig.~\ref{fig:ttzMCMETmt} shows the distribution of the $E_\text{T}^\text{miss}$ and $m_\text{T}$ for $t\bar{t}+Z$, $t\bar{t}$ and signal events.  The stop mass sets a natural scale for these kinematic variables, which are steeply falling for the SM processes.  However, the bottom row of Fig.~\ref{fig:ttzMCMETmt} illustrates the challenge with $t\bar{t}+Z$: it does not need a second lepton to pass the $m_\text{T}$ requirement and therefore has a hadronically decaying top quark\footnote{In contrast to $t\bar{t}$ events without a $Z$ boson that require a second lepton in order for $m_\text{T}>m_W$.}.  As a result, the $m_\text{jet}$ distribution is nearly the same as for signal.  Furthermore, the $am_\text{T2}$ distribution is generally harder than the $t\bar{t}$ background due to the extra energy from the $Z$ boson.

\begin{figure}[h!]
\begin{center}
\includegraphics[width=0.35\textwidth]{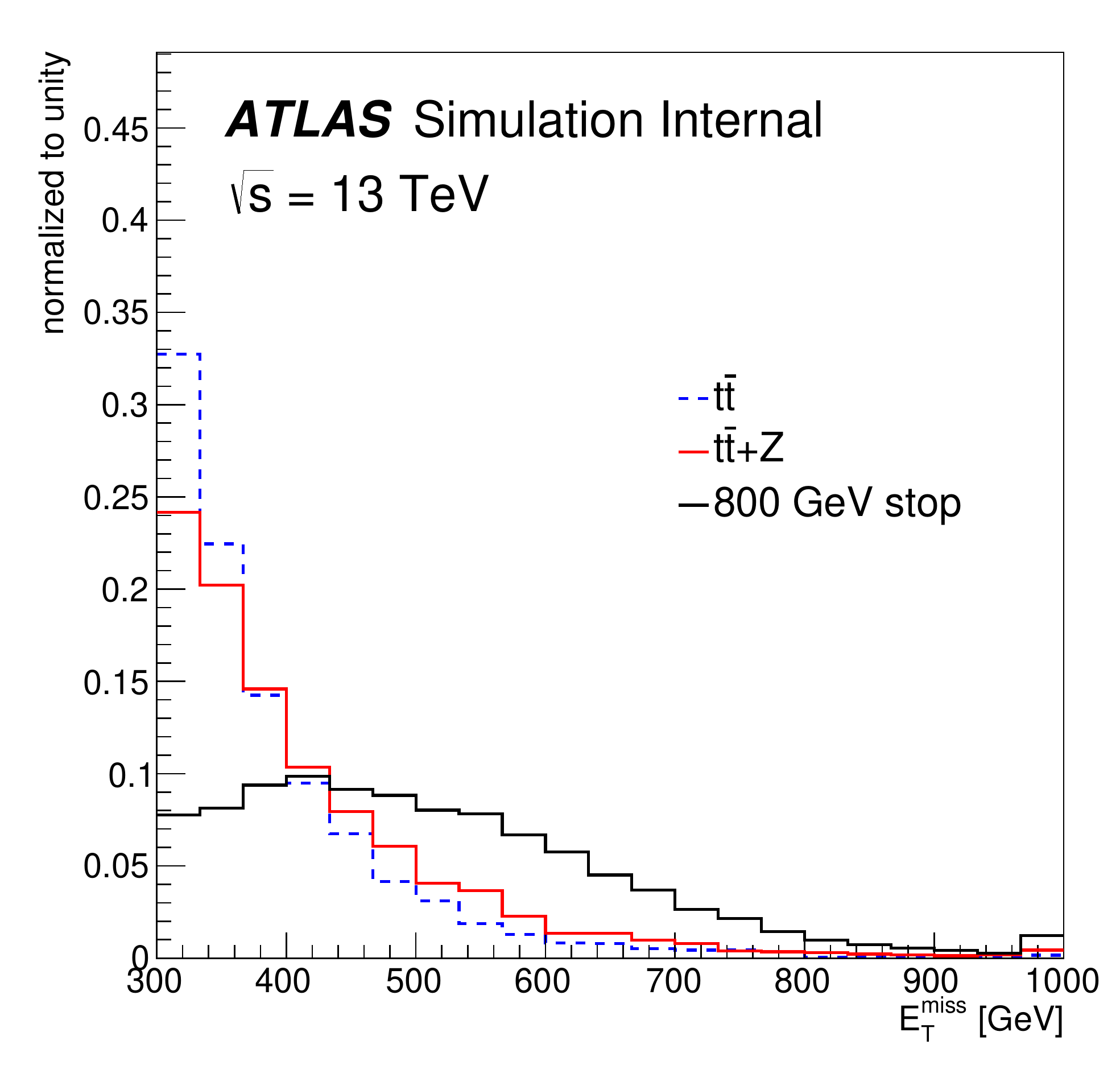}\includegraphics[width=0.35\textwidth]{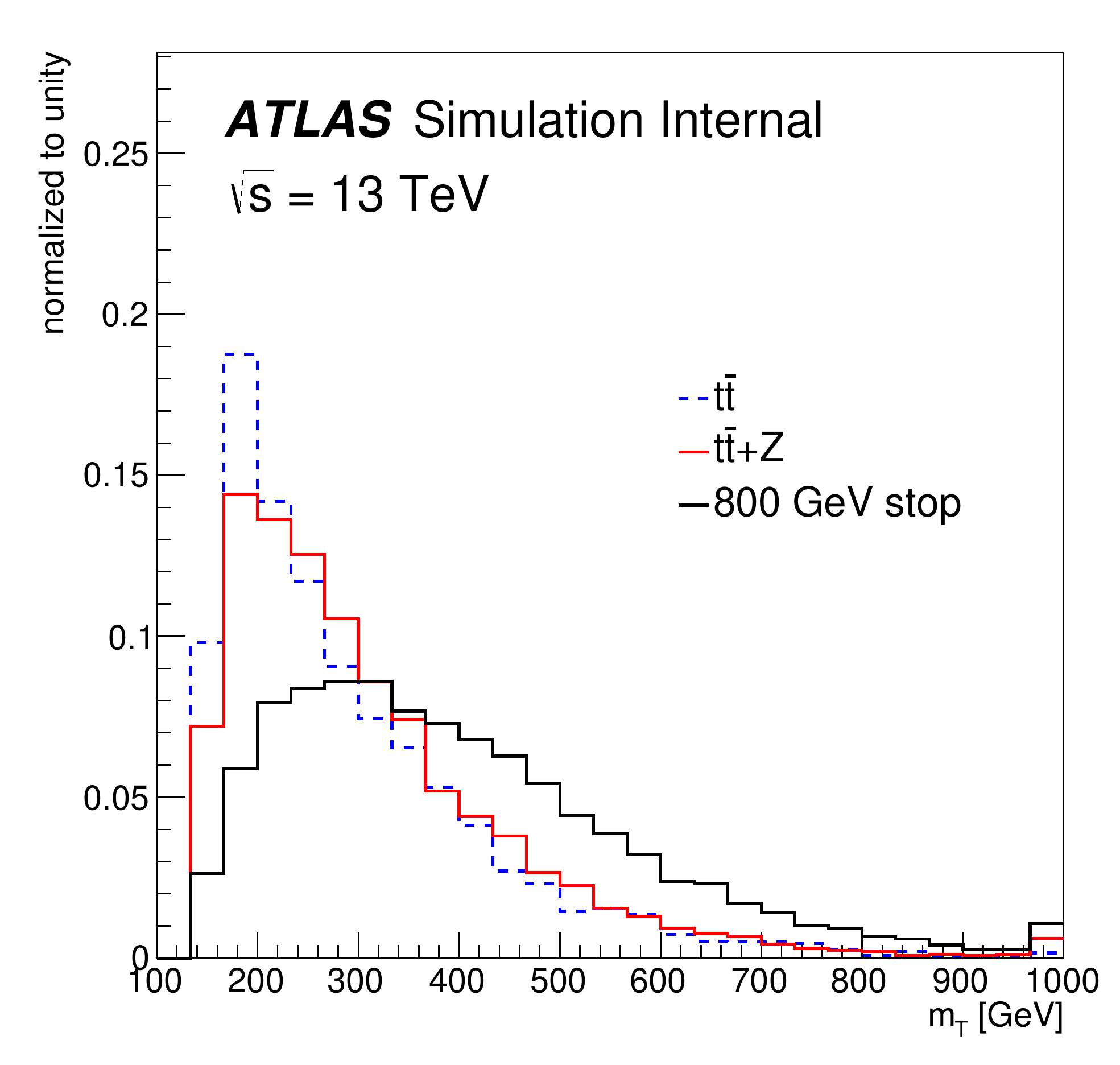}\\
\includegraphics[width=0.35\textwidth]{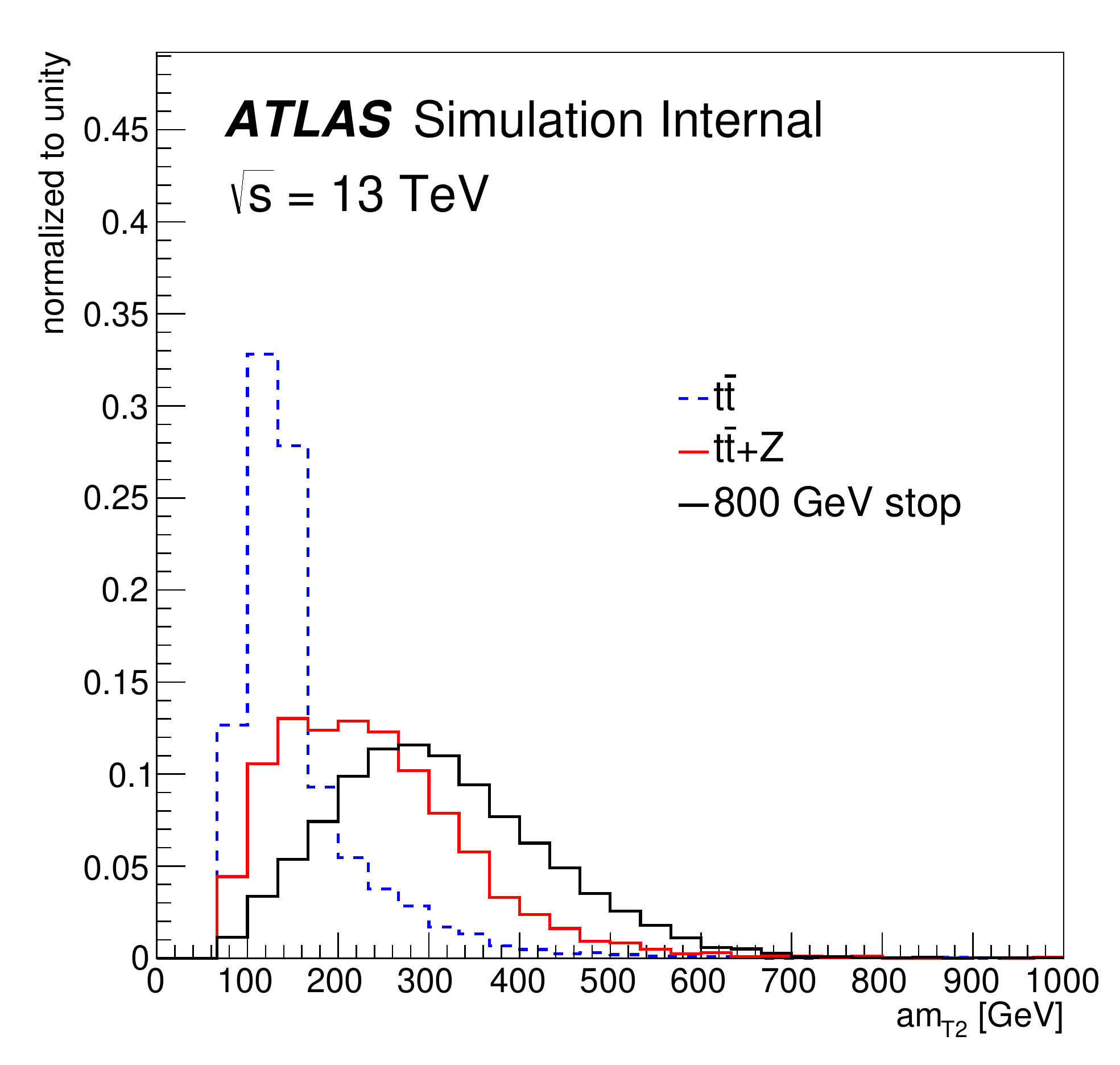}\includegraphics[width=0.35\textwidth]{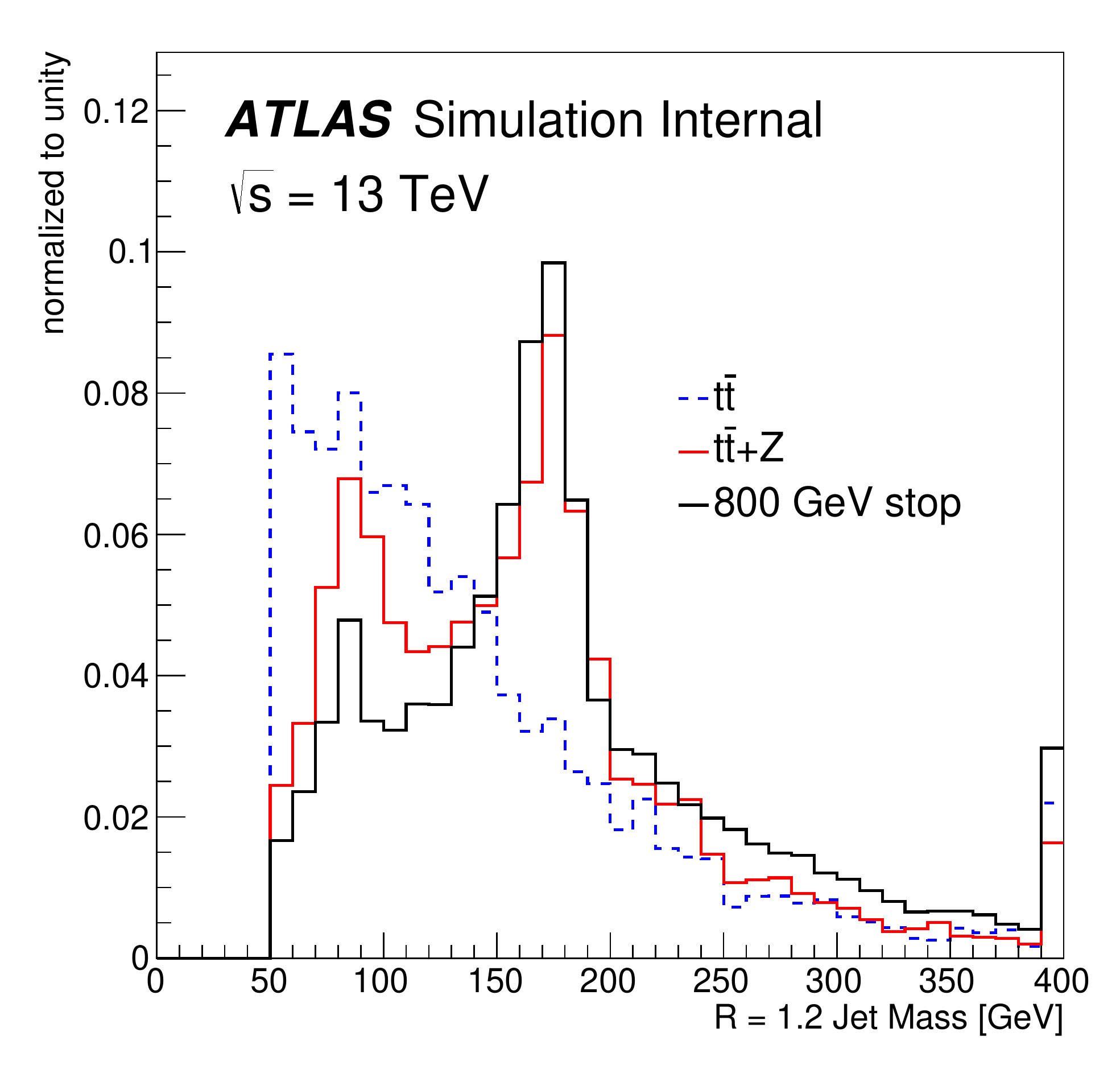}
 \caption{Distributions of (clockwise) $E_\text{T}^\text{miss}$, $m_\text{T}$, $m_\text{jet}$, and $am_\text{T2}$.  The stop model has $(m_\text{stop},m_\text{LSP})=(800,0)$.  Events are required to have four jets with $p_\text{T}>50,50,25,25$ GeV, $E_\text{T}^\text{miss}>300$ GeV, $m_\text{T}>120$ GeV and at least one $R=1.2$ large radius re-clustered jet with $p_\text{T}>300$ GeV.  The last bin contains overflow.}
 \label{fig:ttzMCMETmt}
  \end{center}
\end{figure}	

			\clearpage
		
			\subsection{A Data-driven Method with Photons}
			\label{sec:ttz:datadriven}
			
			\subsubsection{Motivation and Overview}
				
			Since the process $t\bar{t}+Z(\rightarrow\nu\bar{\nu})$ is a significant and irreducible background, in particular for signal regions targeting high mass stops, it is desirable to constrain the normalization using data-driven techniques.  However, it is not possible to isolate a pure sample of $t\bar{t}+Z$ events with sufficient statistics to set a useful constraint on the yield in the signal region.  Leptonic $Z$ decays allow for a pure sample, but the cross-section times branching ratio ($\sim10\%$) is too small - similar to the pair production of 800 GeV stops (see Fig.~\ref{fig:stopcrosssection}).  One possibility is to use a similar process: $t\bar{t}+\gamma$.   Using photons to constrain $Z$ production is a standard technique for estimating inclusive $Z(\rightarrow\nu\bar{\nu})$+jets background processes~\cite{Aaltonen:2009fd,daCosta:2011qk,Khachatryan:2011tk,Collaboration:2011ida,CMS:2009wxa} that has been extensively studied theoretically~\cite{Bern:2011pa,Bern:2012vx,Ask:2011xf} and experimentally~\cite{Khachatryan:2015ira}, but has never before been studied or applied to $t\bar{t}+Z$.  The main benefit of using photons is that they can be directly identified with high purity with no loss due to a small branching ratio to leptons.   Even though the $t\bar{t}+\gamma$ process itself was only recently observed~\cite{Aaltonen:2011sp,Aad:2015uwa,CMS-PAS-TOP-13-011}, the studies in this section show that a selection with high purity and sufficient statistics at high $p_\text{T}$ can be constructed.  Since the Feynman diagrams for $t\bar{t}+Z$ and $t\bar{t}+\gamma$ are nearly identical, the theoretical uncertainty in extrapolating from $t\bar{t}+\gamma$ to $t\bar{t}+Z$ is expected to be small.  Section~\ref{relating} quantifies this similarity at leading order.
			
\begin{figure}[h!]
\begin{center}
\begin{tikzpicture}[line width=1.5 pt, scale=1.]
	
	\draw (-1,1)--(0,0);
	\draw (-1,-1)--(0,0);	
	\draw[gluon2] (0,0)--(1,0);
	\draw[color=black] (1,0)--(2,1);
	\draw[color=black] (1,0)--(2,-1);
	\draw[vector,color=black] (1.5,-0.5)--(2,0);		
	\node at (3.2, 0.) {\large$Z(\rightarrow\nu\bar{\nu})$};
	\node at (2.3, 1.3) {\large $\bar{t}$};
	\node at (2.3, -1.3) {\large $t$};
	
	\draw[implies-implies,double equal sign distance] (5.,0) -- (5.5,0);
	
	 \begin{scope}[shift={(7.5,0)}]

	\draw (-1,1)--(0,0);
	\draw (-1,-1)--(0,0);	
	\draw[gluon2] (0,0)--(1,0);
	\draw[color=black] (1,0)--(2,1);
	\draw[color=black] (1,0)--(2,-1);
	\draw[vector,color=black] (1.5,-0.5)--(2,0);		
	\node at (2.3, 0.) {\large$\gamma$};
	\node at (2.3, 1.3) {\large $\bar{t}$};
	\node at (2.3, -1.3) {\large $t$};
  	\end{scope}
	
 \end{tikzpicture}
\end{center}
\caption{Leading order Feynman diagrams for $t\bar{t}+Z$ (left) and $t\bar{t}+\gamma$ (right) with the boson produced in the final state.  In addition to these diagrams, there are two additional sets of FSR diagrams with gluons in the initial state: one with the same setup as above and one with a $t$-channel exchange of top quarks fusing into the boson (see Fig.~\ref{fig:feynmanttz}). }
\label{fig:feynmanFSR}
\end{figure}			
			
			\subsubsection{Relating $t\bar{t}+\gamma$ to $t\bar{t}+Z$ at Leading Order}
			\label{relating}
			
			The properties of $t\bar{t}+Z$ and $t\bar{t}+\gamma$ matrix elements are similar: the sets of Feynman diagrams are nearly identical.  In addition to the final state radiation diagrams in Fig.~\ref{fig:feynmanFSR}, the other leading order diagrams with the boson radiated in the initial state are shown in Fig.~\ref{fig:feynmanISR}.  Since the gluon is not charged under the electroweak force, only $q\bar{q}$ initial states contribute to the ISR diagrams while both gluon-gluon and $q\bar{q}$ diagrams contribute at leading order to the FSR diagrams.  The only diagrams which are different between $t\bar{t}+Z$ and $t\bar{t}+\gamma$ are the ones that begin at NLO (such as Fig.~\ref{fig:feynman2}) due to the coupling of neutrinos to $Z$ bosons that does not exist for photons.

\begin{figure}[h!]
\begin{center}
\begin{tikzpicture}[line width=1.5 pt, scale=1.]
	
	\draw (-1,1)--(0,0);
	\draw (-1,-1)--(0,0);	
	\draw[gluon2] (0,0)--(1,0);
	\draw[color=black] (1,0)--(2,1);
	\draw[color=black] (1,0)--(2,-1);
	\draw[vector,color=black] (-0.5,0.5)--(0,1);		
	\node at (0.2, 1) {\large$Z$};
	\node at (2.3, 1.3) {\large $\bar{t}$};
	\node at (2.3, -1.3) {\large $t$};
	
	 \begin{scope}[shift={(6.5,0)}]

	\draw (-1,1)--(0,0);
	\draw (-1,-1)--(0,0);	
	\draw[gluon2] (0,0)--(1,0);
	\draw[color=black] (1,0)--(2,1);
	\draw[color=black] (1,0)--(2,-1);
	\draw[vector,color=black] (-0.5,0.5)--(0,1);		
	\node at (0.2, 1) {\large$\gamma$};
	\node at (2.3, 1.3) {\large $\bar{t}$};
	\node at (2.3, -1.3) {\large $t$};
  	\end{scope}
	
 \end{tikzpicture}
\end{center}
\caption{Leading order Feynman diagrams for $t\bar{t}+Z$ (left) and $t\bar{t}+\gamma$ (right).}
\label{fig:feynmanISR}
\end{figure}

\begin{figure}[h!]

\begin{center}
\begin{tikzpicture}[line width=1.5 pt, scale=1.]
	
	\draw (-1,1)--(0,0);
	\draw (-1,-1)--(0,0);	
	\draw[vector] (0,0)--(1,0);
	\draw (1,0)--(2,1);
	\draw (1,0)--(2,-1);
	\draw (2,1)--(2,-1);
	\draw[vector] (2,1)--(3,2);
	\draw (3,2)--(4,3);
	\draw (3,2)--(4,1);
	\draw[vector] (2,-1)--(3,-2);
	\node at (-1.2,1) {$u$};
	\node at (-1.2,-1) {$\bar{u}$};
	\node at (0.4,0.4) {$Z$};
	\node at (1.4,0.9) {$\nu$};
	\node at (1.4,-0.9) {$\nu$};
	\node at (2.3,0) {$\nu$};
	\node at (3.3,-2) {$Z$};
	\node at (2.2,1.8) {$Z$};
	\node at (4.2,1.1) {$t$};
	\node at (4.2,3) {$t$};
	
	 \begin{scope}[shift={(7.,0)}]

	\draw (-1,1)--(0,0);
	\draw (-1,-1)--(0,0);	
	\draw[gluon2] (0,0)--(1,0);
	\draw (1,0)--(2,1);
	\draw (1,0)--(2,-1);
	\draw (2,1)--(2,-1);
	\draw[gluon2] (2,1)--(3,2);
	\draw (3,2)--(4,3);
	\draw (3,2)--(4,1);
	\draw[vector] (2,-1)--(3,-2);
	\node at (-1.2,1) {$u$};
	\node at (-1.2,-1) {$\bar{u}$};
	\node at (0.4,0.4) {$g$};
	\node at (1.4,0.9) {$q$};
	\node at (1.4,-0.9) {$q$};
	\node at (2.3,0) {$q$};
	\node at (3.3,-2) {$Z$};
	\node at (4.2,1.1) {$t$};
	\node at (4.2,3) {$t$};
	\node at (2.,1.8) {$g$};
		
  	\end{scope}
	
 \end{tikzpicture}
\end{center}
\caption{A lowest order ($\alpha_w^5$) diagram that exists for $t\bar{t}+Z$ that has no $t\bar{t}+\gamma$ version (left) and the strong production ($\alpha_w\alpha_s^4$) analogue (right). The left diagram is suppressed with respect to the right one by $(\alpha_w/\alpha_s)^4\sim 10^{-4}$.}
\label{fig:feynman2}
\end{figure}

Even though the set of Feynman diagrams are basically identical for $t\bar{t}+Z$ and $t\bar{t}+\gamma$, the relative contributions are different because the $Z$ boson couples stronger to down-type quarks and the photon couples stronger to up-type quarks.   In particular, the photon couples with strength $eQ_q$ which is $Q_q=2/3$ for up-type quarks (including the top quark) and $Q_q=1/3$ for down-type quarks.  The $Z$ boson coupling is different for left- and right-handed fermions with $e(T^3-\sin^2\theta_WQ_q)/(cos\theta_W\sin\theta_W)$, where $T^3$ is weak isospin.  For up-type quarks, the first term is $(\frac{1}{2}-\frac{2}{3}\sin^2\theta_W)$ for left-handed quarks and $-\frac{2}{3}\sin^2\theta_W$ for right-handed quarks.  Likewise, for down-type quarks, the $Z$ boson coupling strength is proportional to $(-\frac{1}{2}+\frac{1}{3}\sin^2\theta_W)$ for left-handed quarks and $\frac{1}{3}\sin^2\theta_W$ for right-handed quarks.   The $t\bar{t}+Z$ to $t\bar{t}+\gamma$ cross-section ratio for a fixed quark type is given by

\begin{align}
R_q=\frac{\sigma_{t\bar{t}+Z}^\text{via $q$}}{\sigma_{t\bar{t}+\gamma}^\text{via $q$}} =\frac{\frac{1}{2}\left(\sigma_{t\bar{t}+Z}^\text{via $q_L$}+\sigma_{t\bar{t}+Z}^\text{via $q_R$}\right)}{\sigma_{t\bar{t}+\gamma}^\text{via $q$}}=\frac{\left(\frac{1}{2}-|Q_q|\sin^2\theta_W\right)^2+(Q_q\sin^2\theta_W)^2}{2Q_q^2\cos\theta_W^2\sin\theta_W^2},
\end{align}

\noindent where the factor of two in the denominator is from averaging over the initial state spins.  For $\cos\theta_W=m_W/m_Z$ with $m_W\approx 80.385$ GeV and $m_Z\approx 91.1876$ GeV, $R_u\approx 0.945$ and $R_d\approx 4.851$.  If the bosons were only produced via the FSR process, then the total cross-section ratio is expected to be $R_u$ since the top quark is up-type.  In contrast, if the bosons were only produced via the ISR processes, then there would be a tradeoff between $R_u$ and $R_d$ due to the mixture of $u\bar{u}$ and $d\bar{d}$ initial states.  As the valence up quark PDF dominates at high $p_\text{T}$, asymptotically the initial state cross-section ratio should also approach $R_u$.  The behavior in the cross-section ratio for the ISR processes is similar to the inclusive $V$+jets~\cite{Ask:2011xf} case.  However, the FSR processes dominate for $t\bar{t}+V$ ($\gtrsim 80\%$) for $p_\text{T}^V\gtrsim 100$ GeV.

Even if the couplings were identical between $t\bar{t}+Z$ and $t\bar{t}+\gamma$, there would still be a significant difference in the cross-sections due to the boson masses.  The biggest impact of the large $Z$ boson mass is at low $p_\text{T}^V$.   Generically, electroweak radiation receives double Sudakov logarithm enhancements~\cite{Peskin:1995ev}:

\begin{align}
\label{eq:brems}
d\sigma(p\rightarrow p'+\gamma/Z)\approx d\sigma(p\rightarrow p')\times\frac{\alpha}{\pi}\log\left(\frac{-q^2}{\mu^2}\right)\log\left(-\frac{-q^2}{m^2}\right),
\end{align}

\noindent where the first logarithm is due to a soft singularity that is cutoff by the detectability scale $\mu$ for photons and the boson mass for $t\bar{t}+Z$ and the second logarithm is the collinear divergence that is cutoff by the mass of the emitting particle\footnote{See Ref.~\cite{Hook:2014rka} for a nice discussion on electroweak radiation in the ultra high $p_\text{T}\gtrsim$ 10 TeV regime.}.  Since $\mu\ll m_Z$, there is a large enhancement of photon bremsstrahlung at low $p_\text{T}$.  This is further complicated by the fact that this enhancement occurs for all electrically charged particles in the final state, independent of their relationship to the hardscatter process (see Sec.~\ref{sec:MEP}).  Fortunately, the region where $t\bar{t}+Z$ is relevant corresponds to high $p_\text{T}^V$ where differences due to the $Z$ boson mass are less important.  Figure~\ref{fig:syst:ttv:corptmet} shows the conditional distribution of the $Z$ $p_\text{T}$ given the particle-level $E_\text{T}^\text{miss}$.  For a selection requiring $E_\text{T}^\text{miss}>300$ GeV, 2/3 of the $Z$ bosons have $p_\text{T}>300$ GeV and about 90\% have $p_\text{T}>200$ GeV.

\begin{figure}[h!]
\begin{center}
\includegraphics[width=0.5\textwidth]{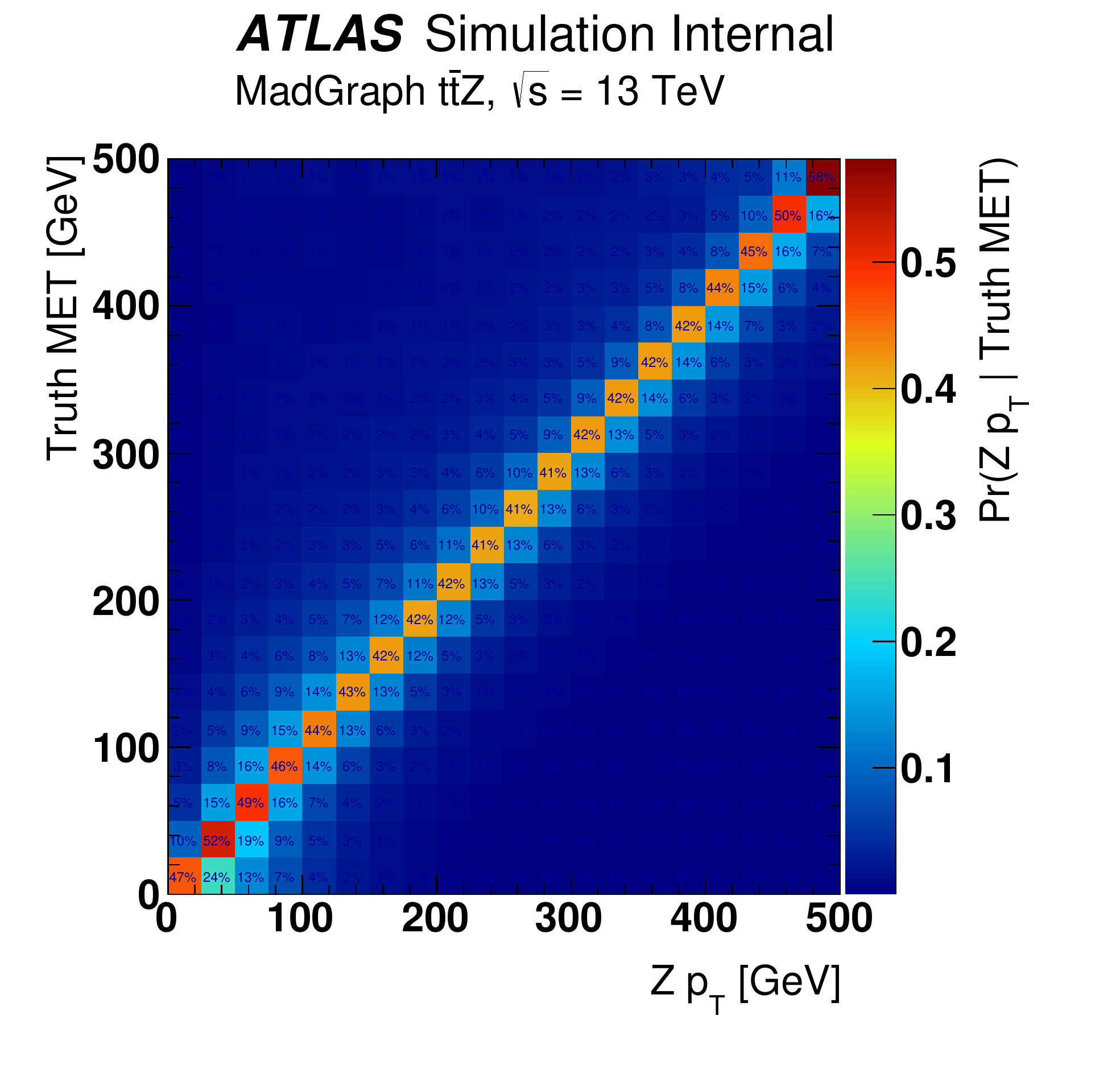}
\caption{The distribution of the $Z$ $p_\text{T}$ given the particle level $E_\text{T}^\text{miss}$ in $t\bar{t}+Z(\rightarrow\nu\bar{\nu})$ events.  All neutrinos contribute to the particle level $E_\text{T}^\text{miss}$.}
\label{fig:syst:ttv:corptmet}
\end{center}
\end{figure}

Figure~\ref{fig:ttzratio} shows the cross-section ratio of $t\bar{t}+Z$ to $t\bar{t}+\gamma$ for parton-level calculations of the ISR processes and the FSR processes with gluon-gluon initial states.  At low $p_\text{T}^V\ll m_Z$, the ratio is very small due to the large enhancement for photons that is cutoff for $Z$ bosons (see Eq.~\ref{eq:brems}).  The kinematic differences between $t\bar{t}+Z$ and $t\bar{t}+\gamma$ are mostly eliminated as $p_\text{T}^V\gg m_Z$.  As expected, the ratio for the ISR processes (red) lies between $R_u$ and $R_d$ and is closer to $R_u$ due to the valence up quarks (as $p_\text{T}\rightarrow \sqrt{s}, R\rightarrow R_u)$.  This ratio slightly decreases with $p_\text{T}^V$ as the fraction of up quarks increases with $\sqrt{\hat{s}}$.  However, this is a small effect, illustrated by Fig.~\ref{fig:ttzpdffracs}.  Over three orders of magnitude in $p_\text{T}^V$, the relative contribution from $u\bar{u}$ increases by only 5-10\%.   The most puzzling aspect of Fig.~\ref{fig:ttzratio} in the context of the above discussion is the blue line.  Since the top quark is an up-type quark, the above argument suggests that the blue line should be $R_u\sim 1$, about half of what is observed.  The reason is subtle and is a new feature of the $t\bar{t}+Z$ that is not relevant for generic $Z$+jets.  Due to its mass, the $Z$ boson has three polarization states while the photon only has two (transverse) states.  The bosons produced in inclusive $Z$+jets are mostly transverse and so the additional polarization state is irrelevant.  However, just as $W$ bosons from $t\bar{t}$ are mostly longitudinally polarized, the FSR $Z$ bosons have a significant ($\sim 50\%$) longitudinal polarization.  When only the transverse polarizations are considered\footnote{This is accomplished by observing all longitudinal helicity states {\tt DATA (NHEL(I,   2),I=1,5) /-1,-1,-1,-1, 0/} in {\tt matrix1.f} and then skipping these states (2 in this case) in the loop {\tt DO I=1,NCOMB} so that {\tt TS(I)} remains zero.  Thank you to Michael Peskin for the idea and Valentin Hirschi for the assistance in implementation. }, the ratio is indeed close to $R_u$, as shown by the green line ratio in Fig.~\ref{fig:ttzratio}.  

Figure~\ref{fig:ttzratiopaperr} shows the cross-section ratio for all sub-processes using the simulation setup described in Sec.~\ref{sec:datasetandMC}.  The next section describes the identification of photons in the simulation in more detail, which in part accounts for some differences (larger photon contribution) between Fig.~\ref{fig:ttzratiopaperr} and the expectations from Fig.~\ref{fig:ttzratio}.  A lower ratio is expected at $\sqrt{s}=8$ TeV compared with $\sqrt{s}=13$ TeV in part because a fixed $p_\text{T}$ samples a lower momentum fraction at $\sqrt{s}=13$ TeV which moves the ISR process ratio away from $R_u$.
	
\begin{figure}[h!]
\begin{center}
\includegraphics[width=0.55\textwidth]{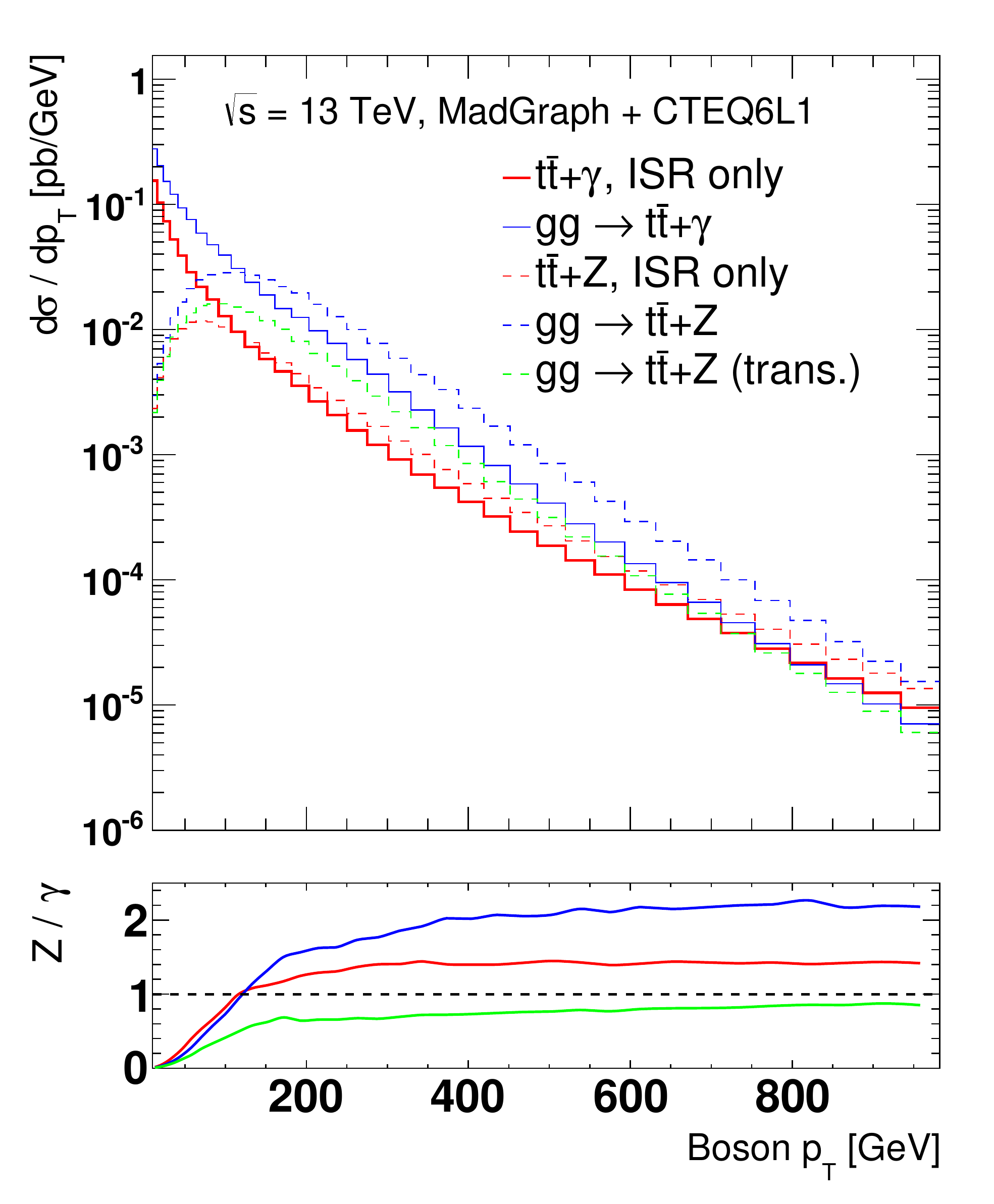}
 \caption{The cross section ratio for $t\bar{t}+Z$ and $t\bar{t}+\gamma$ for various sub-processes described in the text.  Unlike the simulation setups described in Sec.~\ref{sec:datasetandMC}, the $t\bar{t}+Z$ and $t\bar{t}+\gamma$ use exactly the same setup: {\sc MG5\_aMC} 2.1.1 with PDF set {\sc CTEQ6L1}.  No extra partons are generated in the ME and photon radiation from top decay products is not included.  The ISR only processes are generated with the syntax {\tt generate p p $>$ t t}{\fontfamily{ptm}\selectfont  \textasciitilde} {\tt a / t t}{\fontfamily{ptm}\selectfont  \textasciitilde}.}
 \label{fig:ttzratio}
  \end{center}
\end{figure}	

\begin{figure}[h!]
\begin{center}
\includegraphics[width=0.55\textwidth]{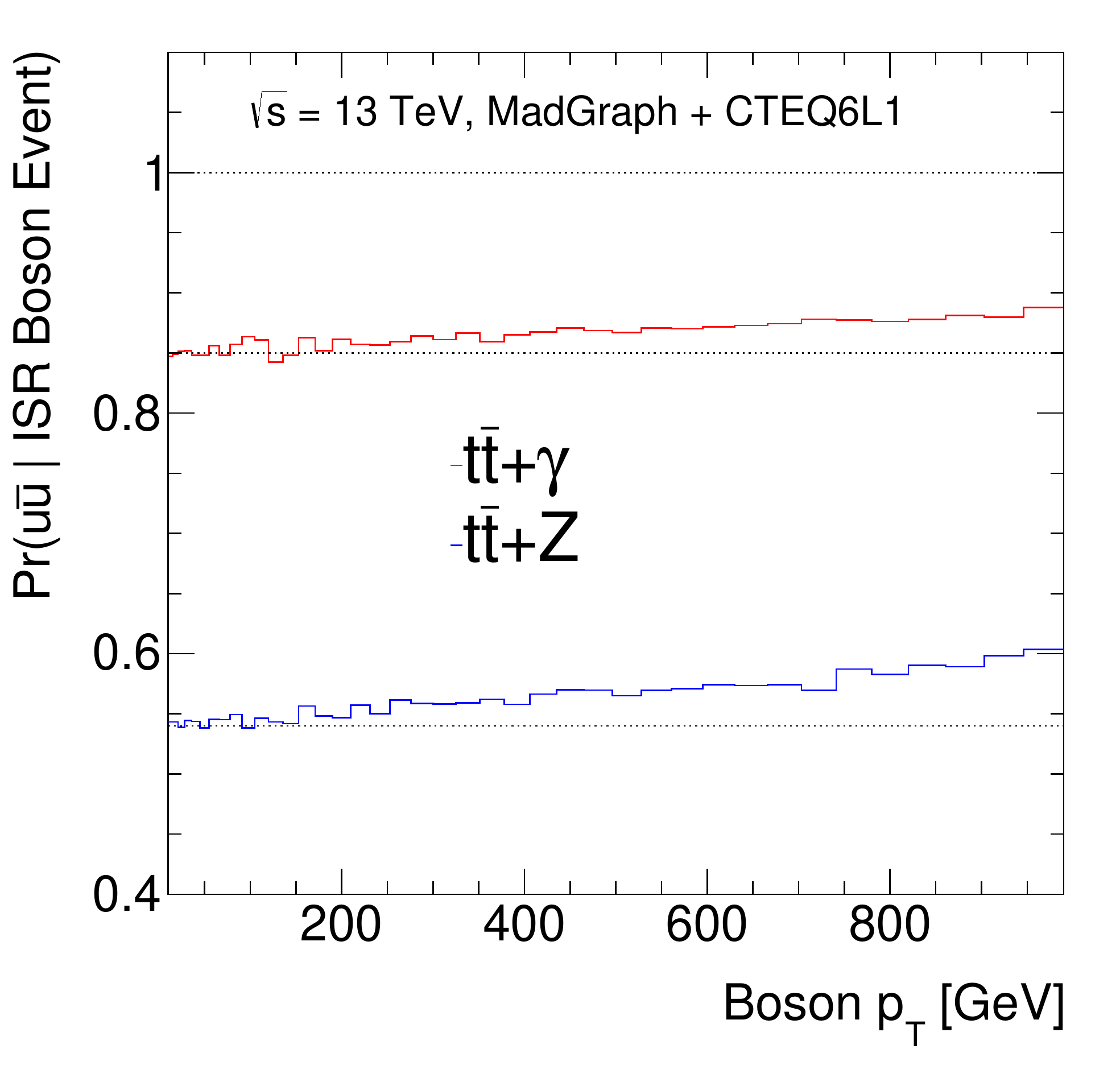}
 \caption{Given a $t\bar{t}+V$ ISR event, this is the fraction of events originating from a $u\bar{u}$ initial state.  The dotted lines show the $p_\text{T}=0$ fractions.}
 \label{fig:ttzpdffracs}
  \end{center}
\end{figure}

\begin{figure}[h!]
\begin{center}
\includegraphics[width=0.45\textwidth]{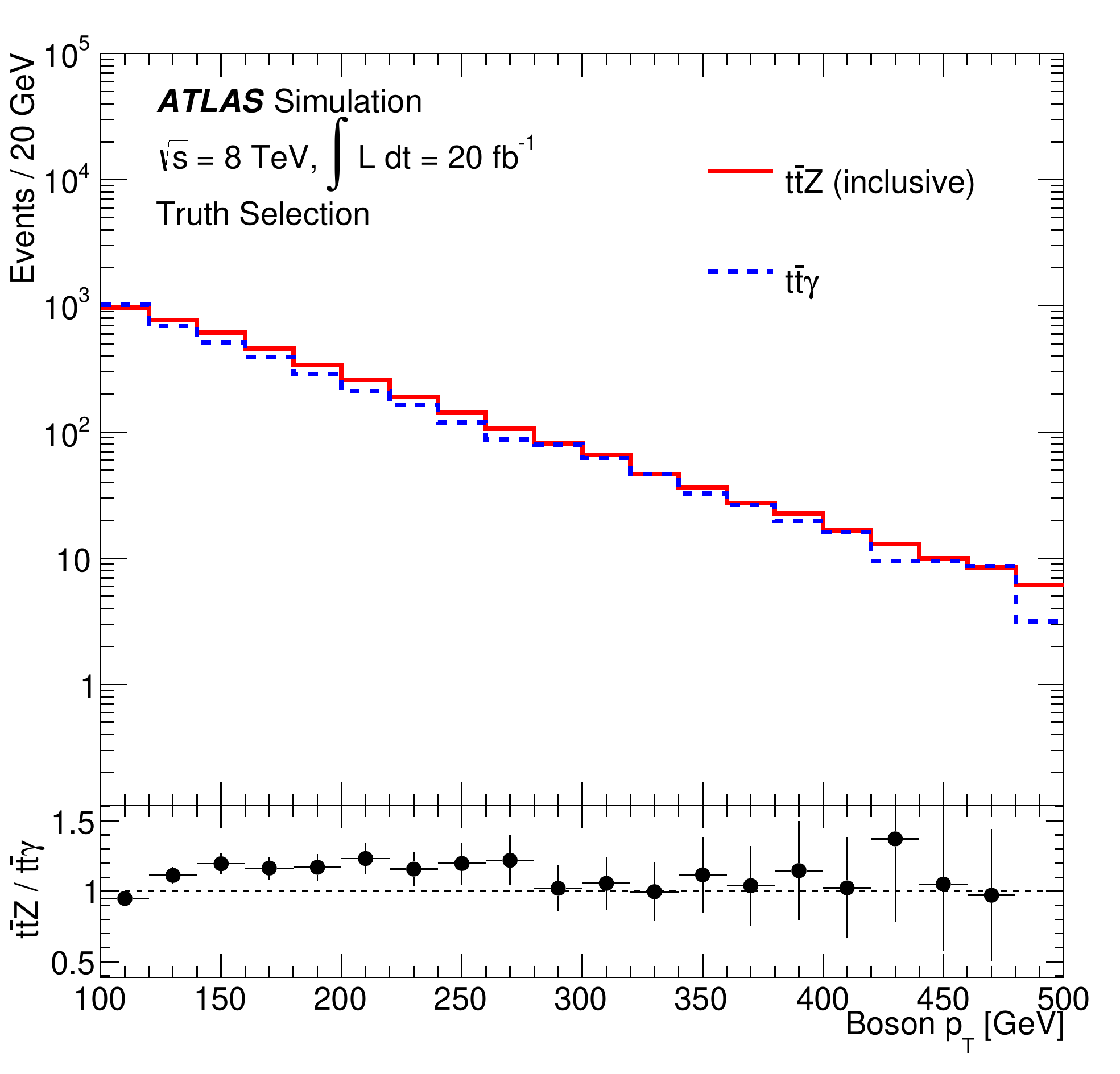}\includegraphics[width=0.45\textwidth]{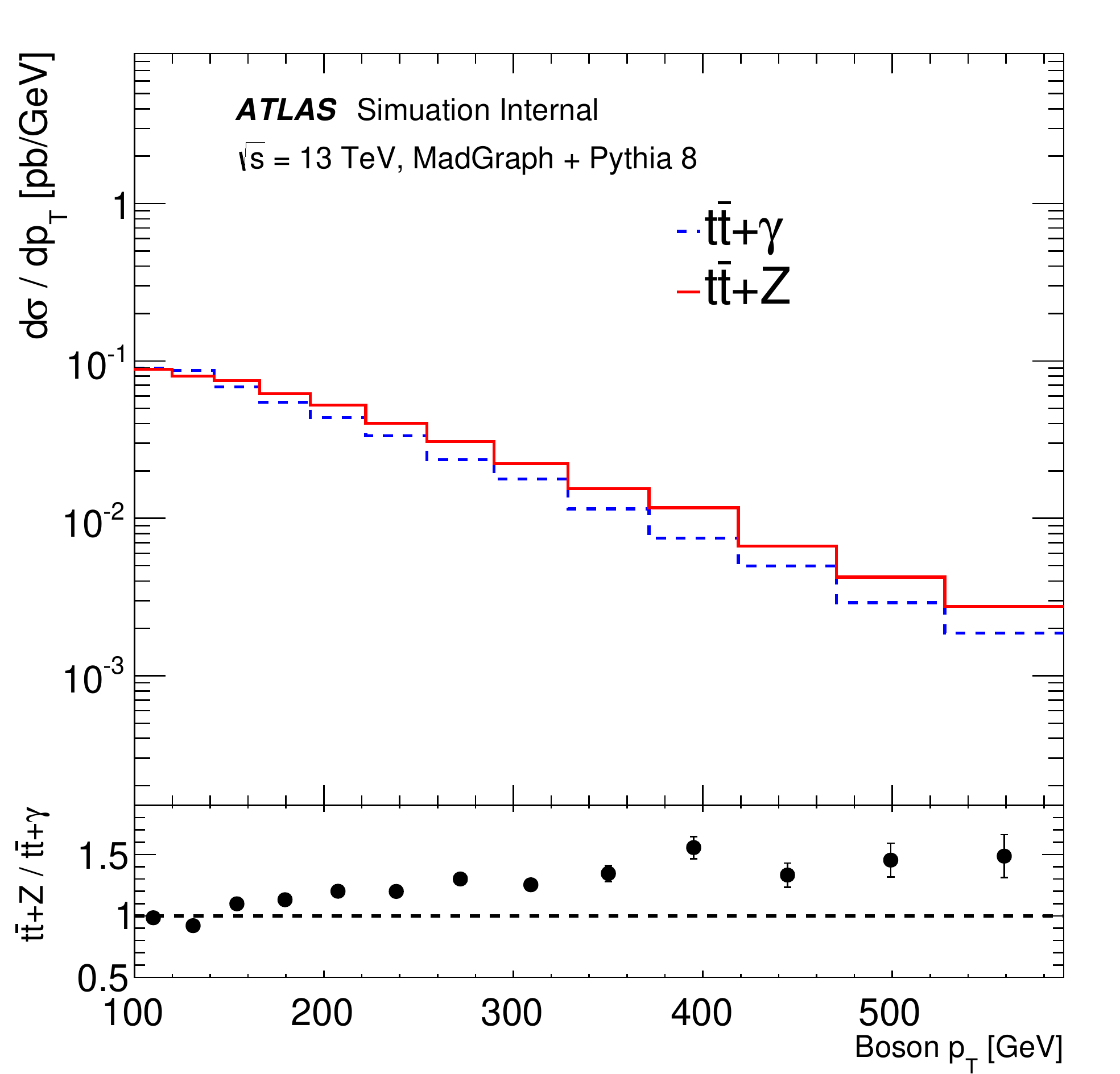}
 \caption{The full cross-section ratio between $t\bar{t}+Z$ and $t\bar{t}+\gamma$ at 8 TeV (left) and 13 TeV (right).}
 \label{fig:ttzratiopaperr}
  \end{center}
\end{figure}

For high $p_\text{T}$ bosons, Fig.~\ref{fig:ttzratiopaperr} shows that the cross-section ratio is nearly unity.  The power of the photon method is that the $Z(\rightarrow e^+e^-/\mu^+\mu^-)$ branching ratio is about $6\%$ so the usable cross-section for the photon process is about 20 times larger than the $Z$ process.  As will be described in Sec.~\ref{sec:ttbargammaeventselection}, photons can be identified and reconstructed with high efficiency and purity.  However, there is a finite acceptance for photon reconstruction while the neutrinos from $t\bar{t}+Z(\rightarrow\nu\bar{\nu})$ can be anywhere in the detector.  Especially at high $p_\text{T}$ where the photons and $Z$ bosons are mostly central, this is a subdominant effect to the others discussed above.   For example, about $2.2\%$ of photons have $|\eta|>2.5$ at $p_\text{T}>100$ GeV, $1.6\%$ for $p_\text{T}>200$ GeV and about $1.2\%$ for $p_\text{T}>300$ GeV.

				\clearpage
				
				\subsubsection{Simulation and Matrix Element Photons}
				\label{sec:MEP}
				
	The discussion in Sec.~\ref{relating} was focused on photons originating directly from the hard scatter process.  However, there are two significant sources of additional photons at particle-level: radiation from charged particles from the top quark decays and the decays of neutral pions, $\pi^0\rightarrow \gamma\gamma$.  Figure~\ref{fig:radiatedphotondiagrams} shows representative diagrams from {\sc MadGraph} when the photon is radiated from one of the charged decay products from the top quark.  As noted in Sec.~\ref{relating}, $Z$ Bremsstrahlung is highly suppressed compared to photon radiation and so these photons are not directly useful for constraining the $t\bar{t}+Z$ cross-section.  The fraction of photons from the charged decay products of the top quark decreases with photon $p_\text{T}$.  Figure~\ref{fig:bremratio} shows the fraction of photons produced from the charged top quark decay products ($b,W^\pm,l$) as a function of the photon $p_\text{T}$.  In agreement with the fractions reported by Ref.~\cite{Melnikov:2011ta}, photons from the charged top quark decay products dominate until about $p_\text{T}\gtrsim 60$ GeV and this fraction decreases to reach about $25$-$30\%$ by $p_\text{T}\gtrsim 100$ GeV.  
		
\begin{figure}[h!]
\begin{center}
\includegraphics[width=0.33\textwidth]{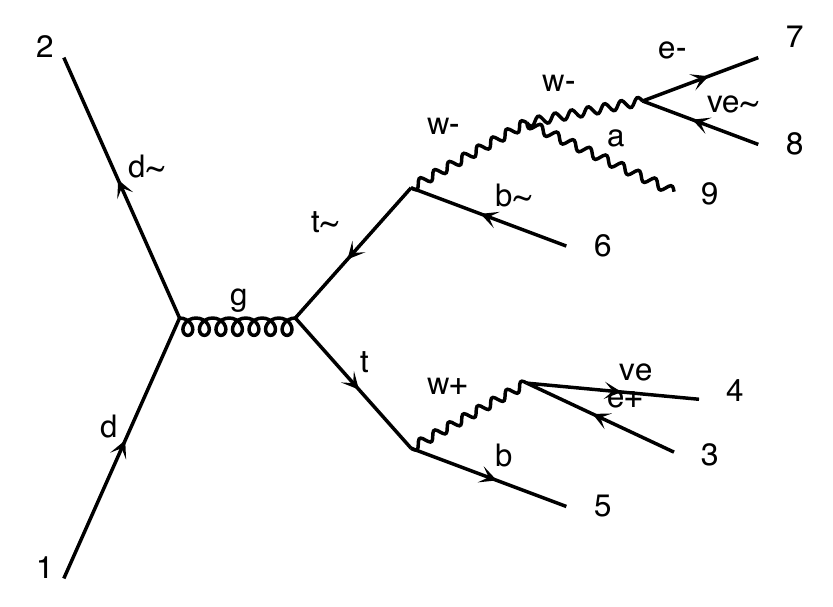}\includegraphics[width=0.33\textwidth]{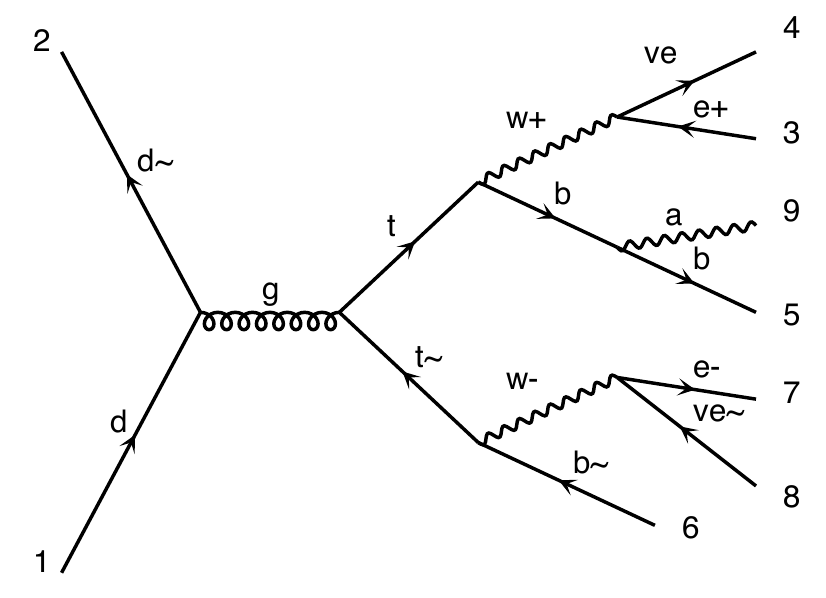}\includegraphics[width=0.33\textwidth]{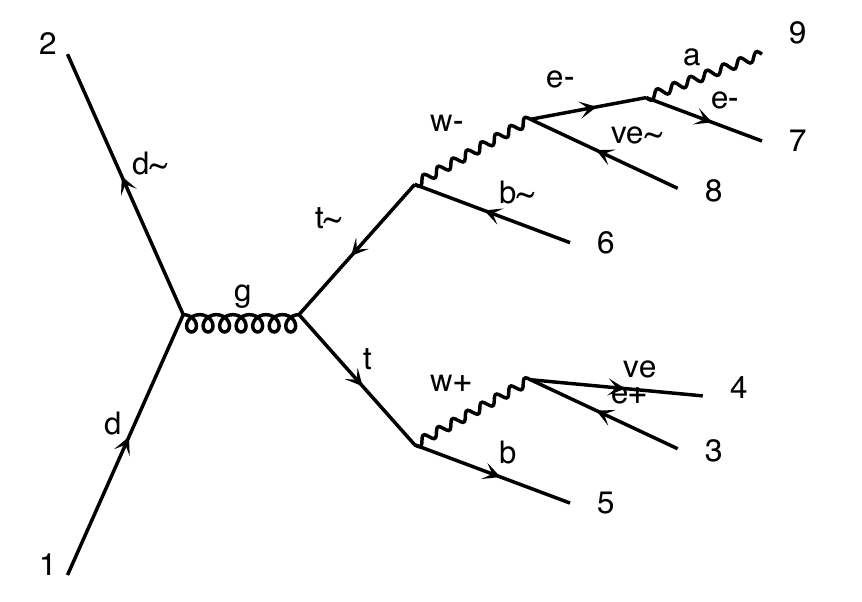}
 \caption{Representative diagrams from {\sc MadGraph} where the photon originates from one of the charged decay products of the top quark.}
 \label{fig:radiatedphotondiagrams}
  \end{center}
\end{figure}		
		
\begin{figure}[h!]
\begin{center}
\includegraphics[width=0.5\textwidth]{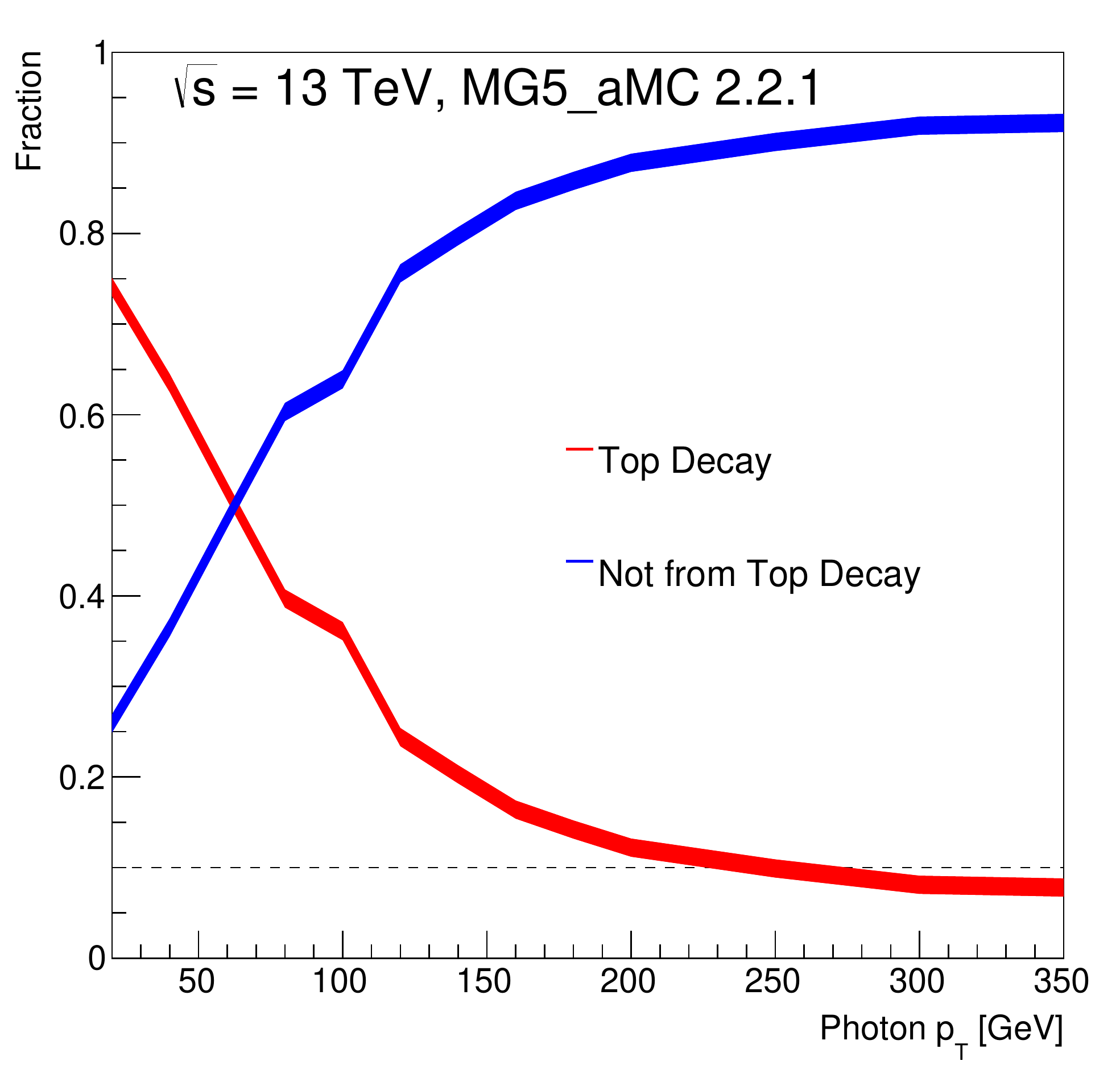}
 \caption{The fraction of photons radiated from the charged top quark decay products (Top Decay) versus the fraction of photons directly from the hard-scatter (Not from Top Decay) as a function of the photon $p_\text{T}$.  Events are generated with {\sc MG5\_aMC} at leading order with the full process {\tt generate p p $>$ t t}{\fontfamily{ptm}\selectfont  \textasciitilde} {\tt $>$ l+ vl b b}{\fontfamily{ptm}\selectfont  \textasciitilde} {\tt l- vl}{\fontfamily{ptm}\selectfont  \textasciitilde} {\tt a}.  Photons are categorized as originating from the charged decay products of the top quark by parsing the ancestry stored in the LHE output of {\sc MG5\_aMC}.  The band is the MC statistical uncertainty.}
 \label{fig:bremratio}
  \end{center}
\end{figure}					
		
	High $p_\text{T}$ photons are also abundantly produced from neutral pions.  A majority of the hadrons from quark and gluon fragmentation are pions and about $1/3$ of the pions are $\pi^0$.  When one of the $\pi^0$ inside a jet carries a large momentum fraction of the initiating quark or gluon, the photons from the $\pi^0$ decay can have significant $p_\text{T}$.  Most of these photons can be separated from the hard-scatter photons because they are non-isolated from the remaining hadronic activity inside the jet.  However, on occasion a real photon from a $\pi^0$ will be reconstructed as an isolated photon, which will artificially decrease the $t\bar{t}+Z$ to $t\bar{t}+\gamma$ cross-section ratio.  Experimental tools for suppressing these photons are described in Sec.~\ref{sec:ttbargammaeventselection}.  The remainder of this section focuses on the labeling of photons in the simulation and the removal of overlap between different generators that cover the same regions of phase space.
					
	Photons are generated at nearly every stage of event simulation.  The MadGraph (or MG5\_aMC) matrix elements include $t\bar{t}+\gamma$ (but not the {\sc Powheg-Box} matrix elements), {\sc Pythia} and {\sc Photos} ($\sqrt{s}=8$ TeV only) add photons as ISR and FSR during fragmentation, and photons can be generated by {\sc Geant4} during the interactions of particles with the detector.	Furthermore, photons generated at one stage can be removed at another stage.  For example, there is a small probability that photons from the ME can be converted to fermion pairs in {\sc Pythia} and photon conversions in the detector are common.  It is therefore crucial to specify a hierarchy in order to avoid double-counting of photons.  The highest preference is given to photons that originate from the ME generator.  Therefore, ISR photons from {\sc Pythia} or {\sc Photos} must be removed as they cover the same region of phase space.  Table~\ref{fig:ISRphotonpythia} shows an example event where {\sc Pythia} adds a high $p_\text{T}$ ISR photon to a $t\bar{t}$ event that needs to be removed as it is covered by the ME $t\bar{t}+\gamma$ sample.  The composition of high $p_\text{T}$ photons in the {\sc Powheg-Box} $t\bar{t}$ sample is shown in Figure~\ref{fig:ttgammacomposition}.  Figure~\ref{fig:ttgammalabeling} illustrates how the labeling is performed.  Most of the photons in particle level events with at least one photon with $p_\text{T}>80$ GeV are from (asymmetric) neutral hadron decays.  Only $3\%$ originate from ISR and only $15\%$ are radiated off of the top quark or its immediate decay products.  These events are the ones that need to be removed.  

\begin{figure}[h!]
\begin{center}
\begingroup
    \fontsize{8pt}{8pt}\selectfont
\begin{Verbatim}[commandchars=\\\{\},codes={\catcode`$=3\catcode`^=7\catcode`_=8}]
 --------  PYTHIA Event Listing  (complete event)  ---------------------------------------------------------
 
    no        id   name            status     mothers   daughters     colours      px        py
     0        90   (system)           -11     0     0     0     0     0     0      0.000      0.000
     1      2212   (p+)               -12     0     0   307     0     0     0      0.000      0.000
     2      2212   (p+)               -12     0     0   308     0     0     0      0.000      0.000
     3        21   (g)                -21     7     0     5     6   101   102      0.000      0.000
     4        21   (g)                -21     8     8     5     6   102   103      0.000      0.000
     5         6   (t)                -22     3     4     9     9   101     0     -4.567    -88.578
     6        -6   (tbar)             -22     3     4    10    10     0   103      4.567     88.578
     7         2   (u)                -41    12    12    11     3   101     0      0.000     -0.000
    11         2   (u)                -43     7     0    16    16   102     0    175.819    -37.109
    16         2   (u)                -44    11    11    22    22   102     0    184.181    -42.930
    22         2   (u)                -44    16    16    30    30   102     0    183.933    -42.817
    30         2   (u)                -52    22    22    38    38   102     0    182.014    -42.370
    38         2   (u)                -44    30    30    64    64   102     0    182.669    -39.669
    64         2   (u)                -44    38    38   104   105   102     0    182.666    -39.696
   104         2   (u)                -51    64     0   128   128   125     0    181.192    -38.581
   105        21   (g)                -51    64     0   122   122   102   125      1.494     -1.236
   128         2   (u)                -52   104   104   143   144   125     0    167.269    -35.617
   143         2   (u)                -51   128     0   177   177   136     0    159.955    -34.472
   177         2   (u)                -44   143   143   232   233   136     0    160.057    -34.206
   232         2   (u)                -51   177     0   265   266   136     0     25.298     -5.486
  \color{red} 233        22   (gamma)            -51   177     0   350   350     0     0    134.759    -28.720
  \color{red} 350        22   gamma               62   233   233     0     0     0     0    134.760    -28.362 \color{black}
\end{Verbatim}      
\endgroup    
\end{center}
 \caption{An example (abridged) event record from {\sc Pythia} showering a $t\bar{t}$ event in which a high $p_\text{T}$ photon is added as ISR (red).}
 \label{fig:ISRphotonpythia}
\end{figure}

\begin{figure}[h!]
\begin{center}
\includegraphics[width=0.5\textwidth]{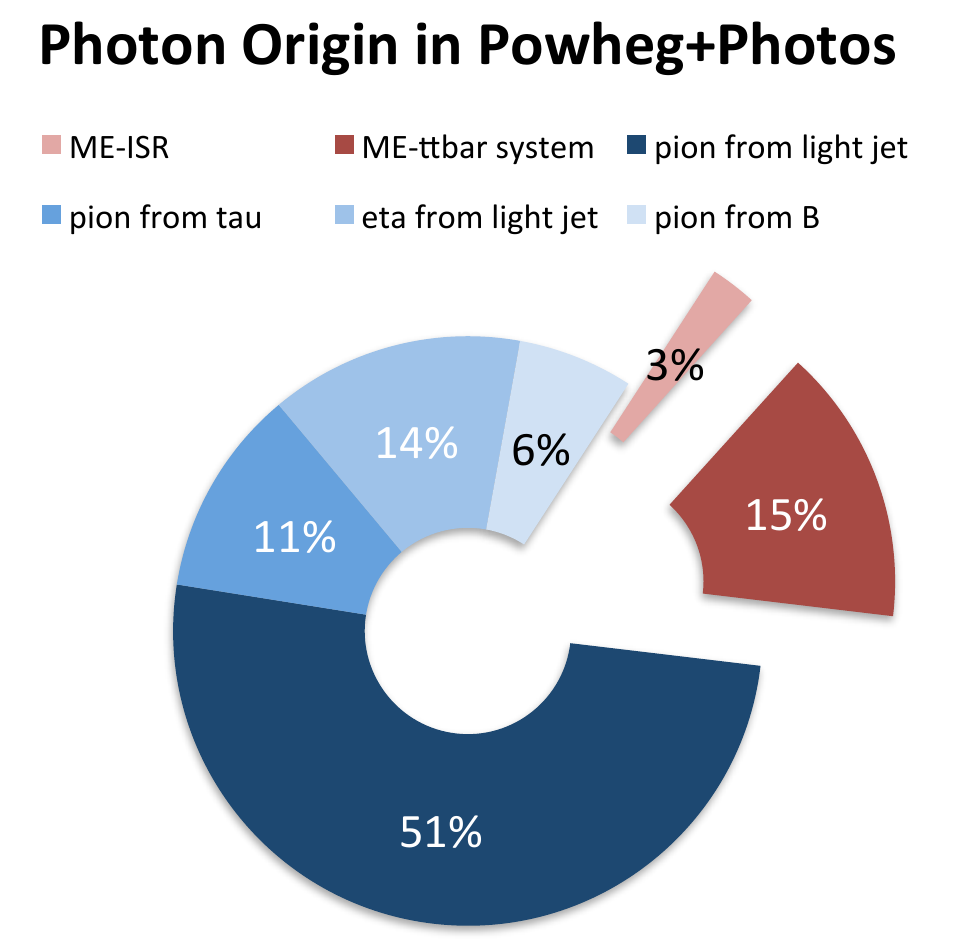}
 \caption{The composition of the leading photon origin in {\sc Powheg-Box}+{\sc Pythia}~6+{\sc Photos} $t\bar{t}$ events with at least one photon at particle-level with $p_\text{T}>80$ GeV.  The ME-$t\bar{t}$ system category includes photons radiated from the decay products of the top quarks.}
 \label{fig:ttgammacomposition}
  \end{center}
\end{figure}

\begin{figure}[h!]
\begin{center}
\begin{tikzpicture}[line width=1.5 pt, scale=1.5]
			\node at (0,0.5) {$\pi^0$ from light jet};
			\node at (0,0.0) {22 ($\gamma$) 1 557};
			\draw[<-] (0,-0.15) -- (0,-0.35);
			\node at (0,-0.5) {111 ($\pi^0$) 2 432};
			\draw[<-] (0,-0.65) -- (0,-0.85);
			\node at (0,-1) {-213 2 245};
			\draw[<-] (0,-1.15) -- (0,-1.35);
			\node at (0,-1.5) {92 2 244};
\begin{scope}[shift={(2,0)}]
			\node at (0,0.5) {$\pi^0$ from $\tau$};
			\node at (0,0.0) {22 ($\gamma$) 1 687};
			\draw[<-] (0,-0.15) -- (0,-0.35);
			\node at (0,-0.5) {111 ($\pi^0$) 2 686};
			\draw[<-] (0,-0.65) -- (0,-0.85);
			\node at (0,-1) {-213 2 684};
			\draw[<-] (0,-1.15) -- (0,-1.35);
			\node at (0,-1.5) {15 ($\tau$) 2 22};
			\draw[<-] (0,-1.65) -- (0,-1.85);
			\node at (0,-2) {-24 ($W^-$) 2 25};		
			\draw[<-] (0,-2.15) -- (0,-2.35);
			\node at (0,-2.5) {-24 ($W^-$) 2 19};		
			\draw[<-] (0,-2.65) -- (0,-2.85);
			\node at (0,-3) {-2 ($\bar{u}$) 3 3};	
			\draw[<-] (0,-3.15) -- (0,-3.35);
			\node at (0,-3.5) {2212 ($p$) 3 1};										
 \end{scope}
 \begin{scope}[shift={(4,0)}]
			\node at (0,0.5) {ME-$t\bar{t}$ system};
 			\node at (0,0.0) {22 ($\gamma$) 1 10002};
			\draw[<-] (0,-0.15) -- (0,-0.35);
			\node at (0,-0.5) {-24 ($W^-$) 2 25};
			\draw[<-] (0,-0.65) -- (0,-0.85);
			\node at (0,-1) {-24 ($W^-$) 2 19};
			\draw[<-] (0,-1.15) -- (0,-1.35);
			\node at (0,-1.5) {-1 ($\bar{d}$) 3 3};
			\draw[<-] (0,-1.65) -- (0,-1.85);
			\node at (0,-2) {2212 ($p$) 3 1};		
\end{scope}	
\begin{scope}[shift={(6,0)}]
			\node at (0,0.5) {ME-ISR};
			\node at (0,0.0) {22 ($\gamma$) 1 22};
			\draw[<-] (0,-0.15) -- (0,-0.35);
			\node at (0,-0.5) {1 ($d$) 3 4};
			\draw[<-] (0,-0.65) -- (0,-0.85);
			\node at (0,-1) {2212 ($p$) 3 2};									
 \end{scope}
\end{tikzpicture}
\end{center}
 \caption{Example particle ancestries for the categories shown in Fig.~\ref{fig:ttgammacomposition}.  Photons are identified with their PDG ID (22) and then the MC event ancestry is parsed to find the origin of the photon. Each line contains four items: PDG ID, particle name, HepMC status code~\cite{Dobbs:2001ck}, and location in the event record.  Not all numbers are used for the location in the event record, but higher numbers do correspond to later in the simulation.}
 \label{fig:ttgammalabeling}
\end{figure}		

Table~\ref{ttgammacompositiontable} shows the photon labeling hierarchy and the relative fractions of the various categories in the $t\bar{t}+\gamma$ validation region that is described in Sec.~\ref{sec:ttbargammaeventselection}.  After the basic event selection, most events are from the dedicated $t\bar{t}+\gamma$ sample with a matrix element photon with $p_\text{T}>80$ GeV.  The $\sim 18\%$ of events from {\sc Powheg-Box} with a MEP with $p_\text{T}>80$ GeV should be removed to avoid double-counting.  The next biggest category of events have a particle-level photon with $p_\text{T}>80$ GeV that originates from somewhere other than the ME.  The dedicated $t\bar{t}+\gamma$ sample is generated with a $p_\text{T}>15$ GeV photon filter in the ME and a $p_\text{T}>80$ GeV photon filter after {\sc Pythia}.  While most events have a ME photon with $p_\text{T}>80$ GeV, about $2\%$ of events pass the {\sc Pythia} filter due to a pion decay.  These events should be removed in favor of the higher order corrections in the {\sc Powheg-Box} sample.  The remaining small fraction of events have no particle-level photon with $p_\text{T}>80$ GeV and are the result of fake photons either from electrons or jets that have a large electromagnetic fraction.

\begin{table}[h!]
\centering
\begin{tabular}{ccc}
Condition & Label & VR Fraction   \\
\hline 
\hline
if $\exists$ MEP with $p_T>80$ GeV, & $t\bar{t}+\gamma$&	$\sim 80\%$	 \\
else if $\exists$ a truth $\gamma$ with $p_T>80$ & $t\bar{t}$+shower $\gamma$ (with $t\bar{t}$) & $\sim 15\%$\\
else if $\Delta R($reco $\gamma$,truth e)$<0.1$  & $t\bar{t}$ + fake ($e\mapsto\gamma$ fake) &$\sim 2\%$\\
else & $t\bar{t}$+fake (jet $\mapsto\gamma$ fake) &$\sim 3\%$\\
\hline
\hline
\end{tabular}
\caption{The composition of photon origins in the $t\bar{t}+\gamma$ and $t\bar{t}$ samples at $\sqrt{s}=8$~TeV.  The order of photon precedence is described in the first column.  Except for the first row, all other rows are labeled $t\bar{t}$ and not $t\bar{t}+\gamma$.  The last column quantifies the fraction of the various categories in the $t\bar{t}+\gamma$ validation region described in Sec.~\ref{sec:ttbargammaeventselection}}
\label{ttgammacompositiontable}
\end{table}				
				
				\clearpage
				
				\subsubsection{Higher Order QCD Corrections}
				\label{sec:kfactorttgamma}

Next-to-leading-order QCD corrections\footnote{This section benefited from many useful conversations with Till Eifert, Javier Montejo Berlingen, Josh McFayden, Stefan Hoche, and Lance Dixon.} for the $t\bar{t} Z$ and $t\bar{t} \gamma$ processes have been calculated and are non-negligible.   Table~\ref{tab:many} summarizes the values of the $k$-factors from the literature.   Only the {\sc MG5\_aMC} collaboration has reported $k$-factors for both processes using the same setup, which is desirable to minimize theoretical uncertainties.  Scale and PDF uncertainties in the individual $k$-factors are also not small - about $20\%$ when reported.  Some of the calculations include stable top quarks while others allow for radiation from the top quark decay products.  The $\sqrt{s}=8$ TeV version of the analysis used a $k$-factor of 1.9 for the $t\bar{t}+\gamma$ validation region following the procedure of Ref.~\cite{ATLAS-CONF-2013-080} based on the calculation in Ref.~\cite{Melnikov:2011ta}.  However, a careful investigation of this $k$-factor reveals that it is likely over-estimated because it is based off of a result using a fixed order calculation with jet requirements and thus artificially increases the NLO cross-section via a higher acceptance.  One other difference is that the top decay is correctly treated in the calculation for the case when the $k$-factor was $1.9$.  However, the authors state that this is likely not the cause of the higher $k$-factor, and when considering the fraction of events with high $p_\text{T}$ photons from the top decay products (see Sec.~\ref{sec:MEP}), this cannot account for the difference with the {\sc MG5\_aMC} calculation.  The inclusive $k$-factor from Ref.~\cite{Melnikov:2011ta} is closer to $1.5$, which agrees with the inclusive $k$-factor from the {\sc MG5\_aMC} collaboration.  In the control region method, only the cross-section {\it ratio} between $t\bar{t}+\gamma$ and $t\bar{t}+Z$ is relevant (working only at high $p_\text{T}^V$).  Since the diagrams are basically identical, one may expect that the QCD corrections are nearly the same for the two processes.  This is supported by the $k$-factor ratio calculation in Ref.~\cite{Alwall:2014hca} shown in the second row of Table~\ref{tab:many} and is further examined in this section.  Additionally, this section explores the $p_\text{T}$ dependence of the $k$-factor ratio.  It is not known from the calculations presented in Table~\ref{tab:many} if there is a significant $p_\text{T}$ dependence to the $k$-factor or the ratio (the external studies only go to $p_\text{T}$ $<200$ GeV) of $k$-factors between $t\bar{t}+Z$ and $t\bar{t}+\gamma$. 

Tables~\ref{tab:ttz:sherpaxs} and~\ref{tab:ttz:madgraphFOxs} show the result of fixed-order calculations of the LO and NLO $t\bar{t}\gamma$ and $t\bar{t} Z$ cross sections as a function of the boson $p_\text{T}$ threshold for {\sc Sherpa}+{\sc Openloops} (\ref{tab:ttz:sherpaxs}) and {\sc MG5\_aMC} (\ref{tab:ttz:madgraphFOxs}) at $\sqrt{s}=13$ TeV.  In both cases, an isolation for the photon of $\Delta R=0.4$ is used implemented by the Frixione cone with $n=2$ and $\epsilon=0.025$~\cite{Frixione:1998jh}.  The two calculations give similar results and show that the $k$-factor ratio is consistent with unity and independent of $p_\text{T}$ within 10\% over the range $100$ GeV $<p_\text{T}^V< 600$ GeV.  One reason\footnote{This idea is due to Stefan Hoche.} it might decrease is that at low boson $p_\text{T}$, the dominant contributions are gluon-gluon fusion where the boson comes from a top quark line, whereas at high boson $p_\text{T}$, the quark-quark annihilation dominates where the Z and $\gamma$ come from ISR and thus the $k$-factor decreases and tends toward the Z+jets/$\gamma$+jets k-factor ratio, which is $\sim 90\%$~\cite{Bern:2012vx}.  Based on these calculations, a $k$-factor ratio of $1$ is used for the extrapolation from the $t\bar{t}+\gamma$ CR to the $t\bar{t}+Z$ in the SRs.  Uncertainties associated with this choice are described in Sec.~\ref{sec:susy:ttzuncert}.

\vspace{5mm}

\begin{table}[h!]
\centering
\label{tab:many}
  \centering
\noindent\adjustbox{max width=\textwidth}{
\begin{tabular}{c|ccccccc}
Reference & $\sigma^\text{LO}_{t\bar{t}\gamma}$ & $\sigma^\text{NLO}_{t\bar{t}\gamma}$ &$k_{t\bar{t}\gamma}$ & $\sigma^\text{LO}_{t\bar{t}Z}$ & $\sigma^\text{NLO}_{t\bar{t}Z}$ & $k_{t\bar{t}Z}$ & $k_{t\bar{t}Z}/k_{t\bar{t}\gamma}$ \\
\hline
~\cite{Melnikov:2011ta}                             &       1.96${}^{+0.64}_{-0.45}$    &  2.93${}^{+0.42}_{-0.39}$           &      1.49                &       &         &                  &                \\
~\cite{Alwall:2014hca}                              &    1.203(1)${}^{+29.6}_{-21.3}$         &   1.744(5)${}^{+9.8}_{-11.0}$                     & 1.45     &        0.5273(41)${}^{+30.5}_{-21.8}$         &   0.7598(26)${}^{+9.7}_{-11.1}$                   &     1.44     &   0.99   \\
~\cite{Lazopoulos:2008de} &          &           &                  &          0.808      &      1.09           &         1.35${}^{+0.25}_{-0.25}$    &    \\
~\cite{Kardos:2011na} &          &           &                  &          0.808      &      1.121(2)           &         1.39    &    \\
~\cite{Rontsch:2014cca}(1) &          &           &                  &          0.1035(1)     &  0.1370(3)             &      1.32    &    \\
~\cite{Rontsch:2014cca}(2) &          &           &                  &          0.00379(0)     &  0.00516(1)             &      1.36    &    \\
~\cite{Rontsch:2014cca}(3) &          &           &                  &          0.00325(0)     &  0.00480(1)             &      1.48    &    \\
~\cite{Garzelli:2012bn} &          &           &                  &          0.1539(1)     &          0.2057(2)     &    1.34${}^{+0.22}_{-0.27}$      &    \\
\hline            
\end{tabular}}
\caption{NLO QCD corrections to the $t\bar{t}+\gamma$ and $t\bar{t}+Z$ cross-sections.  Ref.~\cite{Melnikov:2011ta} is for 14 TeV and has a second $k$-factor given for a second selection that has a harder jet requirement.  Since the calculation is fixed-order, this artificially increases the $k$-factor to the 1.9 value that was used for the 8 TeV analysis.  The $k$-factor in Ref.~\cite{Garzelli:2012bn} is for 8 TeV.  Both Ref.~\cite{Lazopoulos:2008de}  and Ref.~\cite{Kardos:2011na} show the differential (in $Z$ $p_\text{T}$) k-factor up to 200 GeV, which appears to be relatively flat in that range.  The value ~\cite{Rontsch:2014cca}(1) is for the zero-width approximation while~\cite{Rontsch:2014cca}(2) is for a narrow-width approximation and uses the MSTW08 PDF set for both LO and NLO.  The third value~\cite{Rontsch:2014cca}(3) is for a narrow-width approximation and mixes CTEQ6L1 at LO with CT10 at NLO.  Electroweak corrections have also been reported in Ref.~\cite{Frixione:2015zaa}.}
\label{tab:many}
\end{table}

\begin{table}[h!]
\centering
\label{tab:ttz:sherpaxs}
  \centering
\noindent\adjustbox{max width=\textwidth}{
\begin{tabular}{c|ccccccc}
$p_\text{T,cut}^\text{boson}$ {[}GeV{]} & $\sigma^\text{LO}_{t\bar{t}\gamma}$ & $\sigma^\text{NLO}_{t\bar{t}\gamma}$ &$k_{t\bar{t}\gamma}$ & $\sigma^\text{LO}_{t\bar{t}Z}$ & $\sigma^\text{NLO}_{t\bar{t}Z}$ & $k_{t\bar{t}Z}$ & $k_{t\bar{t}Z}/k_{t\bar{t}\gamma}$ \\
\hline
100                              & 0.2002(4)           & 0.329(2)             & 1.62                      & 0.2330(3)      & 0.367(1)        & 1.59                 & 0.98               \\
200                              & 0.0479(1)           & 0.0784(5)            & 1.62                      & 0.0812(1)      & 0.1278(8)       & 1.58                 & 0.97               \\
300                              & 0.01428(3)          & 0.0227(2)           & 1.58                      & 0.02768(5)     & 0.04244(2)      & 1.53                 & 0.97               \\
400                              & 0.00489(1)          & 0.00775(6)           & 1.59                      & 0.01002(2)     & 0.01512(8)      & 1.51                 & 0.95               \\
500                              & 0.001872(6)         & 0.00291(3)           & 1.57                      & 0.003917(9)    & 0.00583(4)      & 1.49                 & 0.95               \\
600                              & 0.000791(3)         & 0.00121(2)           & 1.55                      & 0.001654(5)    & 0.00240(2)      & 1.45                 & 0.93  \\
\hline            
\end{tabular}}
\caption{LO and NLO cross-sections for $t\bar{t}\gamma$ and $t\bar{t}Z$ as a function of the boson $p_\text{T}$ threshold computed with {\sc Sherpa+OpenLoops} by Stefan Hoche. All cross-sections are in pb.  The numbers in parentheses are the statistical uncertainties.   The PDF is CT14.  A scale of $H_\text{T}=\sum p_\text{T}$ over all final state objects is used (the difference between the scalar sum of $p_\text{T}$ and $m_\text{T}$ was found to be negligible in this range).}
\label{tab:ttz:sherpaxs}
\end{table}

\begin{table}[h!]
\centering
\label{tab:ttz:madgraphFOxs}
  \centering
\noindent\adjustbox{max width=\textwidth}{
\begin{tabular}{c|ccccccc}
$p_\text{T,cut}^\text{boson}$ {[}GeV{]} & $\sigma^\text{LO}_{t\bar{t}\gamma}$ & $\sigma^\text{NLO}_{t\bar{t}\gamma}$ &$k_{t\bar{t}\gamma}$ & $\sigma^\text{LO}_{t\bar{t}Z}$ & $\sigma^\text{NLO}_{t\bar{t}Z}$ & $k_{t\bar{t}Z}$ & $k_{t\bar{t}Z}/k_{t\bar{t}\gamma}$ \\
\hline  
100                              &     0.2634(8)      &  0.3842(3)${}^{+12.9\%}_{-13.4\%}$            &     1.46                 &    0.3122(10)   &    0.4209(2)     &    1.35              &        0.92        \\
200                              &       0.06305(2)   &     0.08864(6)${}^{+13.2\%}_{-13.8\%}$        &   1.41                    & 0.1077(3)     &    0.1433(9)    &       1.33          &     0.95           \\
300                              &       0.01842(5)   &      0.02608(2)${}^{+13.6\%}_{-14.1\%}$       &     1.42                 &   0.03587(1)   &    0.04760(4)  &     1.33             &         0.94      \\
400                              &        0.00615(2)   &      0.008737(7)${}^{+15.5\%}_{-15.0\%}$      &     1.42                 &  0.01274(4)    &    0.01673(2)   &      1.31            &    0.92          \\
500                              &       0.002305(7)   &      0.003234(2)${}^{+17.0\%}_{-15.8\%}$     &   1.40                  &   0.00489(2)&               0.00643(7)        &      1.31     &0.94    \\
600                              &         0.000947(3)  &    0.001342(10)${}^{+14.4\%}_{-15.4\%}$        &      1.42                 &     0.002032(8)    &  0.00258(2) &     1.27            & 0.89  \\
\hline            
\end{tabular}}
\caption{LO and NLO cross-sections for $t\bar{t}\gamma$ and $t\bar{t}Z$ as a function of the boson $p_\text{T}$ threshold computed with {\sc MG5\_aMC}.  A custom fortran filter is used to isolate $t\bar{t}+Z$ events with a fixed $Z$ boson threshold.   All cross-sections are in pb.  The uncertainties on the NLO cross section are from variations of the factorization and renormalization scale.  The numbers in parentheses are the statistical uncertainties.  The PDF is NNPDF2.3 LO for the LO calculations and NNPDF2.3NLO for the NLO calculations.  The for both LO and NLO scale is half the scalar sum of the transverse mass of all out-going partons (default for NLO and scale option 3 for LO~\cite{Hirschi:2015iia}).  The impact of adding a PS was found to be small ($\lesssim 10\%$).}
\label{tab:ttz:madgraphFOxs}
\end{table}

\clearpage

As stated earlier, given a choice of the $k$-factor ratio, the actual $k$-factors themselves do not impact the prediction.  Nonetheless, it is useful to make an informed choice for the $k$-factor in order to directly compare the simulation with the data in the CR.  The $k$-factor used by the $\sqrt{s}=13$ TeV analysis is $1.33$, which is based off of the leading order $t\bar{t}+Z$ cross-section from the ATLAS generation and the NLO cross-section from the {\sc MG5\_aMC} collaboration~\cite{Alwall:2014hca}.  Note that the $k$-factor directly from the {\sc MG5\_aMC} collaboration is about $10\%$ larger because their leading order calculation used a different PDF set (NNPDF2.3LO versus MSTW2008nlo68cl) and top quark mass (172.5 GeV versus 173.2 GeV used by aMC).

In order to justify the use of unity for the cross-section ratio, the leading order simulation for $t\bar{t}+\gamma$ must be as similar as possible to that for $t\bar{t}+Z$.  The simulation for both processes are based on {\sc MG5\_aMC} interfaced with {\sc Pythia 8}, but there are some significant differences.  In particular, the $t\bar{t}+\gamma$ ($t\bar{t}+Z$) sample uses the CTEQ6L1 (NNPDF2.3) PDF set, a fixed (variable) factorization and renormalization scale of $2\times m_\text{top}$ (transverse mass), and no extra partons (up to two extra partons) are generated in the matrix element.  Using the ATLAS simulation framework, small $t\bar{t}+Z$ samples were generated with variations to study the impact of these settings.  Changing the PDF from CTEQ6L1 fro NNPDF2.3 resulted in a 12\% higher cross-section.  The cross section is reduced by 2\% when no additional partons are considered in the calculation and by 5\% when adopting the fixed scale choice of the $t\bar{t}\gamma$ simulation. The combination of the three effects yields a 4\% difference in cross section from the choice of generator settings. The $t\bar{t}\gamma$ cross section is increased by 4\% to account for these known differences.  

\clearpage

		\subsubsection{Event Selection}
		\label{sec:ttbargammaeventselection}
		
		In order for the $t\bar{t}+\gamma$ process to be as kinematically close as possible to the $t\bar{t}+Z(\rightarrow\nu\bar{\nu})$ process, the $\gamma$ is added\footnote{A more pragmatic adjective would be `remove' instead of 'add' since the photon is already part of the $E_\text{T}^\text{miss}$ calculation as a visible object.} to the $E_\text{T}^\text{miss}$, mimicking the lost neutrinos.  The sum of $\vec{p}_\text{T}^\text{miss}$ and $\vec{p}_\text{T}^\gamma$ will be denoted $\tilde{p}_\text{T}^\text{miss}$.  This new variable is then used to construct $\tilde{E}_\text{T}^\text{miss}$ and $\tilde{m}_\text{T}$ with the standard definitions, replacing $\vec{p}_\text{T}^\text{miss}$ with $\tilde{p}_\text{T}^\text{miss}$.  Table~\ref{tab:ttgselections} shows the event selections used for the $\sqrt{s}=8$ TeV $t\bar{t}+\gamma$ validation region (VR8) and the $\sqrt{s}=13$ TeV $t\bar{t}+\gamma$ control region (CR13).  The jet $p_\text{T}$ requirements are chosen to match the the signal regions.  The upper $E_\text{T}^\text{miss}$ requirement for CR13 ensures orthogonality with the $t\bar{t}$ CR.  Single lepton and $E_\text{T}^\text{miss}$ triggers are used to collect the data for the VR and a dedicated high $p_\text{T}$ photon trigger ($p_\text{T}>120$ GeV) is used for the CR in order to increase the available statistics.  
								
\begin{table}[h!]
  \centering
\noindent\adjustbox{max width=\textwidth}{
    \setlength{\tabcolsep}{1.0pc}
    \begin{tabular}{lcc}
      \hline
      Requirement           & VR8 & CR13 \\
      \hline
      At least four jets with $p_\text{T}$ [GeV]$>  $ & $80,60,40,25$  & $120,80,50,25$ \\[0.15cm]
      At least one signal photon with $p_\text{T}>$ [GeV] & 100 &  125  \\[0.15cm]
      $\tilde{E}_\text{T}^\text{miss}$ [GeV]    $>$               &120 & 120\\[0.15cm]   
            $\tilde{m}_\text{T}$ [GeV]   $>$                 &110 &  110\\[0.15cm]
      $\tilde{H}_\text{T,sig}^\text{miss}$   $>$                 &-- &  5\\[0.15cm]
      $E_\text{T}^\text{miss}$ [GeV]    $<$               &-- & 200\\[0.15cm] 
      \hline
    \end{tabular}}
    \caption{The requirements for the $t\bar{t}+\gamma$ VR ($\sqrt{s}=8$ TeV) and CR ($\sqrt{s}=13$ TeV).  In both regions, exactly one signal lepton is required with no other baseline leptons.  Furthermore the event selections require at least one $b$-tagged jet.  The tilde variables include the photon in the $\vec{p}_\text{T}^\text{miss}$ as described in the text.
    }
    \label{tab:ttgselections}
\end{table}

The predicted composition both the VR and CR are summarized in Table~\ref{tab:ttgammayields}.  In addition to the changes in cross-section and integrated luminosity between the two energies, the main difference between the regions is the jet $p_\text{T}$ requirements in the CR that are kinematically tighter in order to be close to SR13.  In addition, there is a photon-electron overlap removal at $\sqrt{s}=13$ TeV\footnote{This idea is due to J. Montejo Berlingen.} that additionally helps to reduce the $t\bar{t}$ contamination.  Both regions have a very high $t\bar{t}+\gamma$ purity, with about $75\%$ in the VR and $92\%$ in the VR.  

\begin{table}[h!]
\begin{center}
\begin{tabular}{l |cc}
\hline
   process & VR8 &  CR13 \\
   \hline
         $t\bar{t}$ + $\gamma$  & $75.2 \pm 1.6$  & $29.2\pm 1.4$ \\
      $t\bar{t}$ & $27.0 \pm 1.2$  & $1.6\pm 0.3$ \\
            Other & $1.7 \pm 0.5$  &$0.9\pm 0.2$ \\
        total SM & $103.9 \pm 2.1$ & $31.6\pm 1.5$  \\
        \hline
     data & 104 &  45 \\     
     \hline
\end{tabular}
\caption{Expected and observed event yields in the $t\bar{t}$+$\gamma$ validation/control regions.  All MC numbers are normalized to 20.3 fb${}^{-1}$ for the VR at $\sqrt{s}=8$ TeV and $3.32$ fb${}^{-1}$  for the CRs at $\sqrt{s}=13$ TeV.   The $t\bar{t}$ sample at $\sqrt{s}=8$ TeV is  reweighted according to the standard procedure described in Section~\ref{ttbarCR}.  The displayed uncertainties are due to limited statistics.}
  \label{tab:ttgammayields}
\end{center}
\end{table}	
		
The remainder of this section shows key kinematic distributions in the VR and CR.  Figure~\ref{fig:ttgamma8TeVmet} compares the $E_\text{T}^\text{miss}$ distribution in the VR with the $\tilde{E}_\text{T}^\text{miss}$ distribution.  Nearly all events have $E_\text{T}^\text{miss}\lesssim 200$ GeV, as the $\tilde{E}_\text{T}^\text{miss}$ is dominated by the photon momentum, as also expected for $t\bar{t}+Z(\rightarrow\nu\bar{\nu})$ events (see Fig.~\ref{fig:syst:ttv:corptmet}).  Analogous plots for $m_\text{T}$ and $\tilde{m}_\text{T}$ are shown in Fig.~\ref{fig:ttgamma8TeVmt}.  Overall, the simulation agrees well with the data within the large statistical uncertainties, though this is partly coincidental due\footnote{Additionally, the overlap removal described at the end of Sec.~\ref{sec:MEP} is not applied, which would further reduce the total SM by removing approximately $5$ $t\bar{t}$ events. } to the large $k$-factor (see Sec.~\ref{sec:kfactorttgamma}).  Supporting plots for the $\sqrt{s}=13$ TeV CR are in Fig.~\ref{fig:ttzphotptCR} and Fig.~\ref{fig:ttzMETCR}.  The photon $p_\text{T}>125$ GeV by construction and has a broad spectrum.  Most photons are central, with most photons contained in $|\eta|\lesssim 1$.  The $\tilde{E}_\text{T}^\text{miss}$ and $\tilde{m}_\text{T}^\text{miss}$ distributions in Fig.~\ref{fig:ttzMETCR} are similar to the corresponding $\sqrt{s}=8$ TeV ones.  There is no significant evidence for mis-modeling any of the kinematic distributions, though the statistical precision is limited.

\begin{figure}[h!]
\begin{center}
\includegraphics[width=0.5\textwidth]{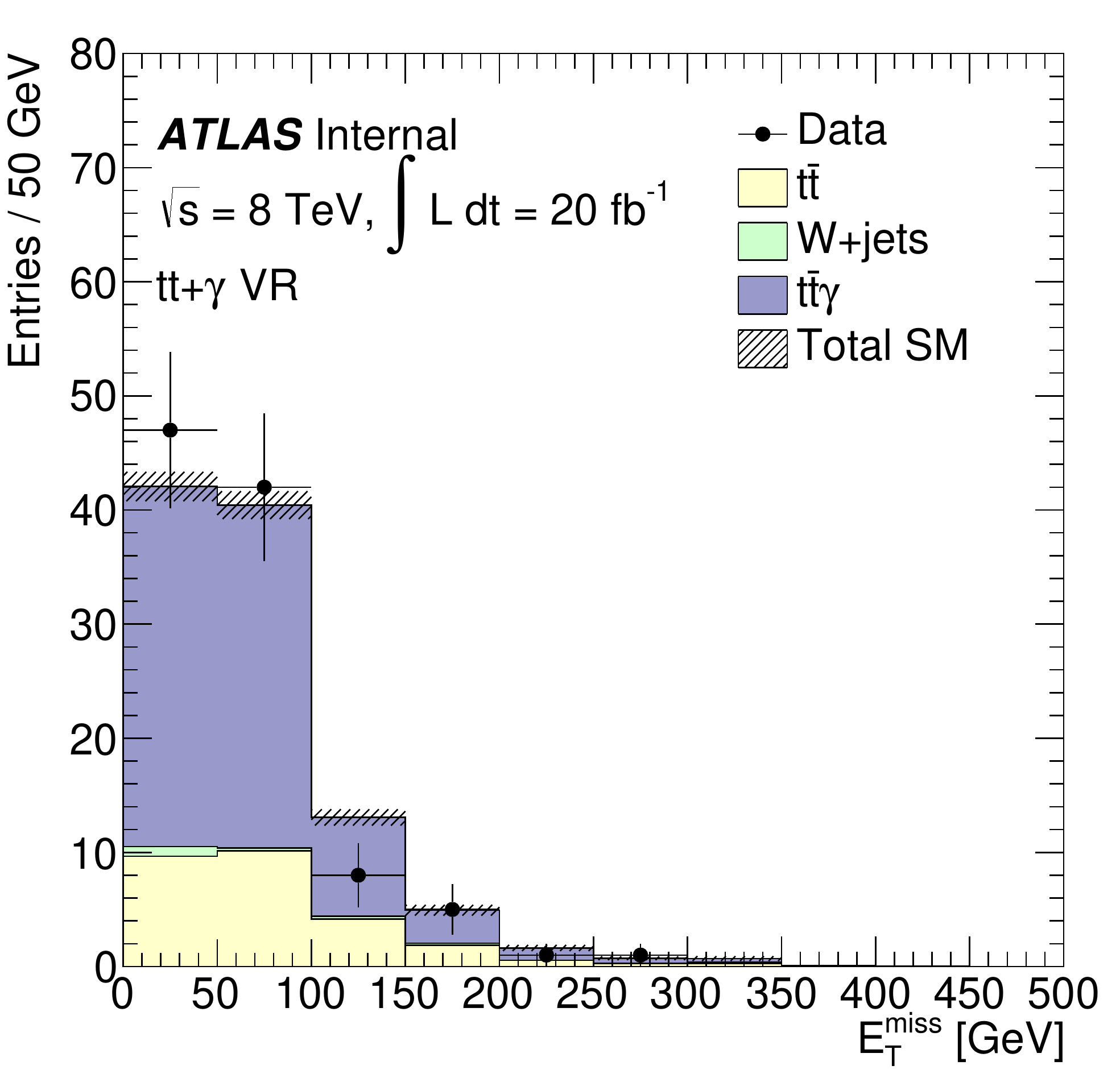}\includegraphics[width=0.5\textwidth]{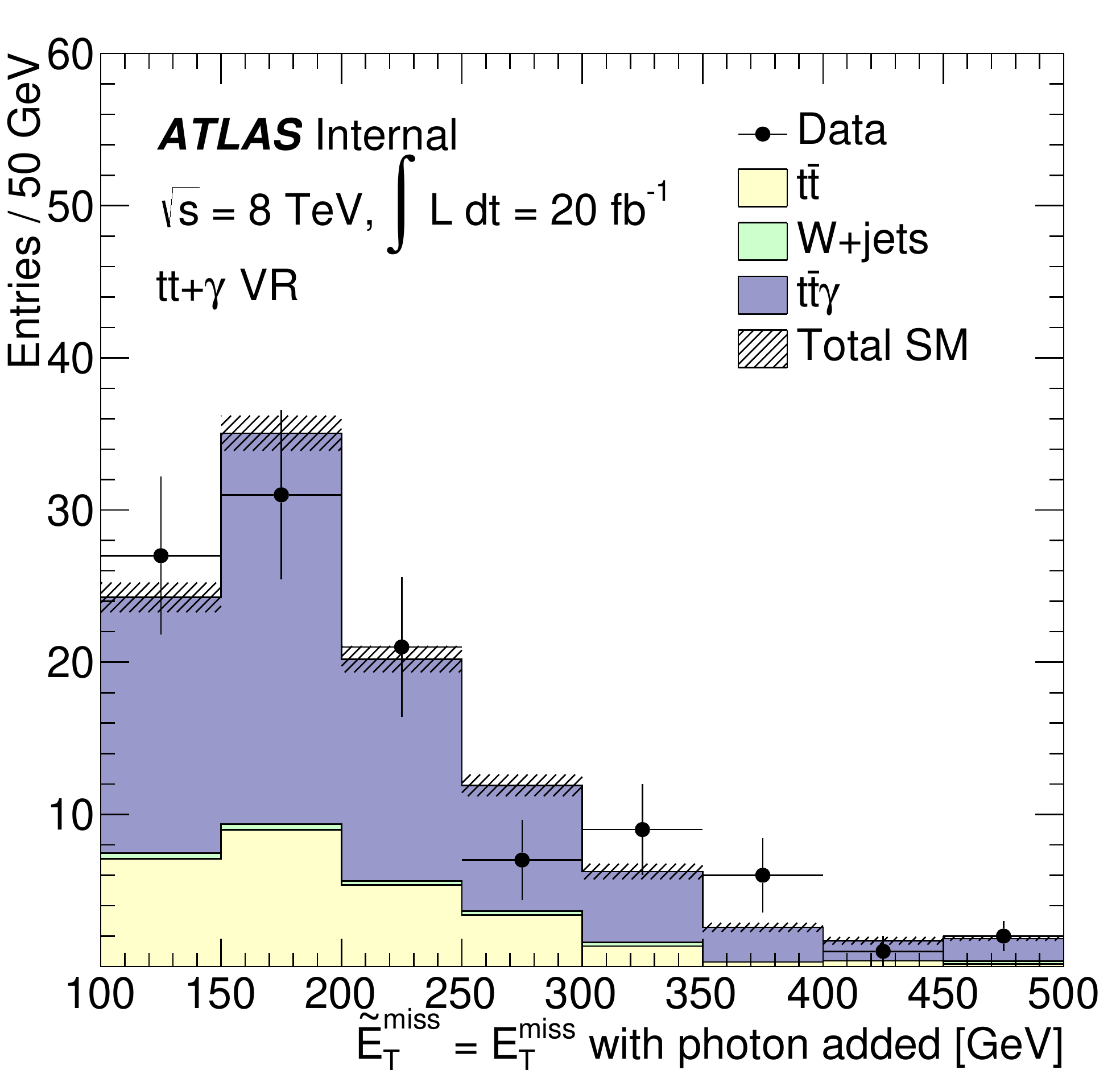}
 \caption{The distribution of $E_\text{T}^\text{miss}$ (left) and $\tilde{E}_\text{T}^\text{miss}$ (right) constructed from $\vec{p}_\text{T}^\text{miss}$ and $\vec{p}_\text{T}^\gamma$.  Both distributions use data and simulation in the VR at $\sqrt{s}=8$ TeV.  Only statistical uncertainties are included in the error bars and bands.}
 \label{fig:ttgamma8TeVmet}
  \end{center}
\end{figure}		
		
\begin{figure}[h!]
\begin{center}
\includegraphics[width=0.5\textwidth]{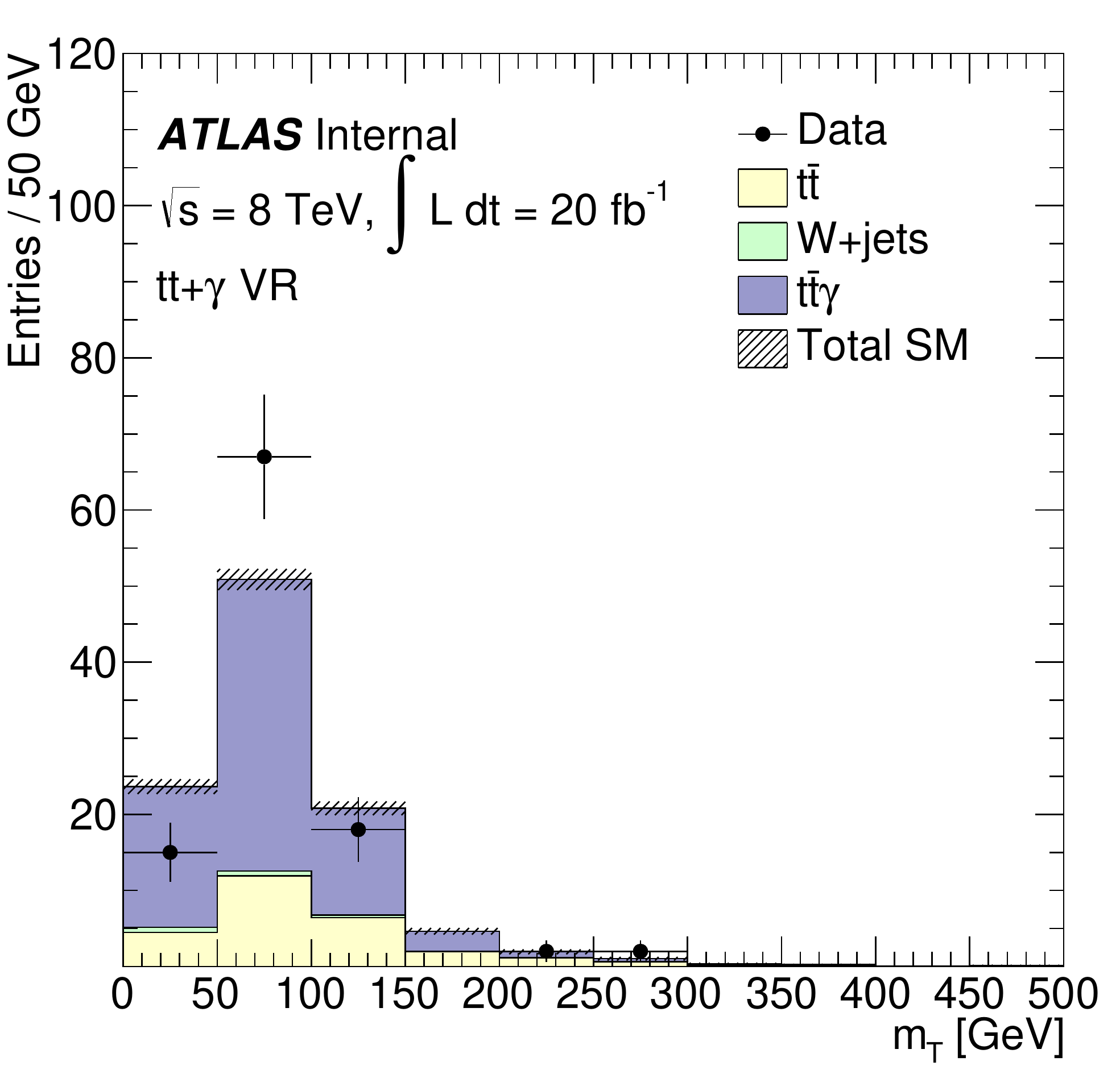}\includegraphics[width=0.5\textwidth]{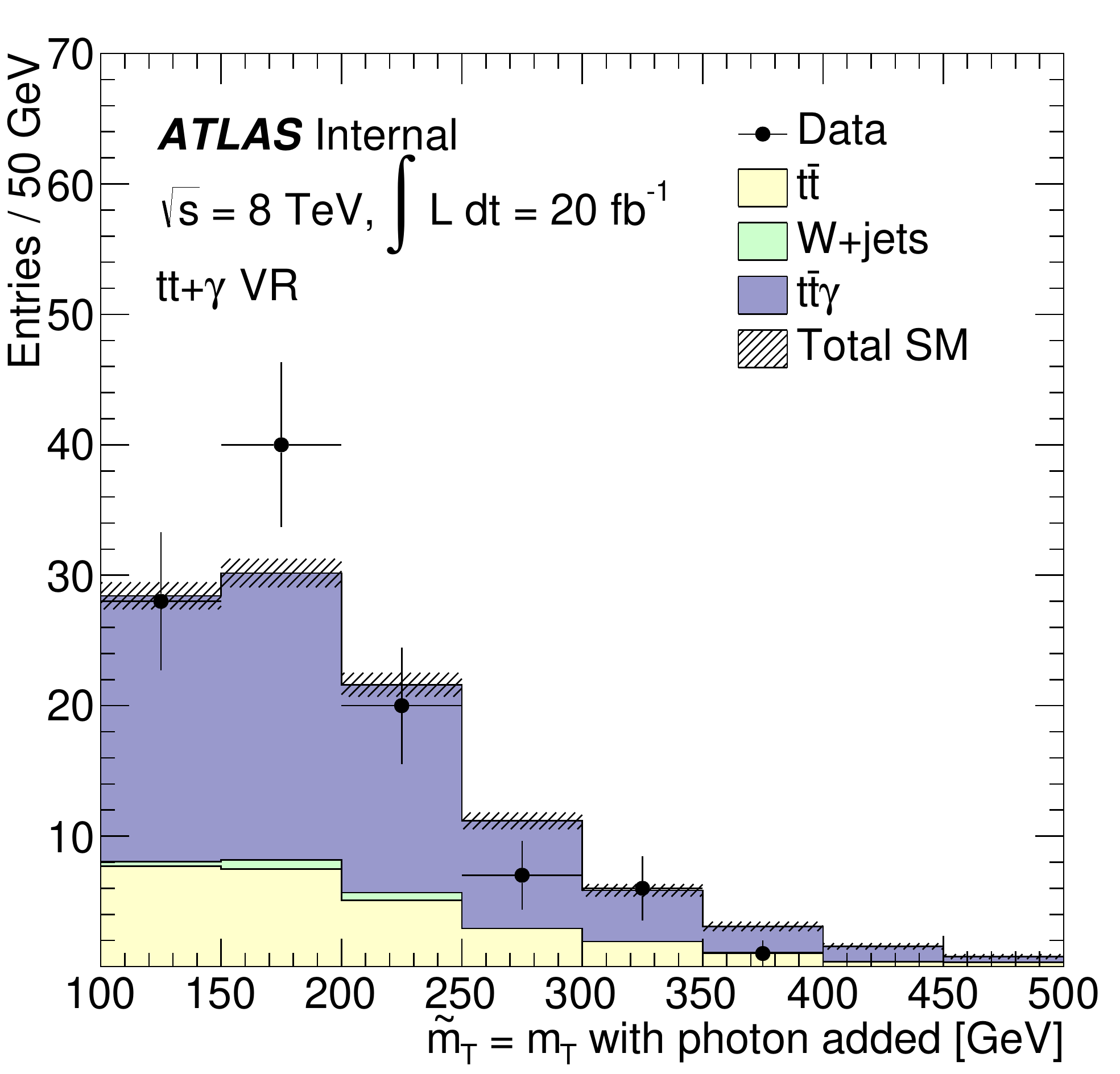}
 \caption{The distribution of $m_\text{T}^\text{miss}$ (left) and $\tilde{m}_\text{T}^\text{miss}$ (right) constructed from $\vec{p}_\text{T}^\text{miss}$ and $\vec{p}_\text{T}^\gamma$.  Both distributions use data and simulation in the VR at $\sqrt{s}=8$ TeV. Only statistical uncertainties are included in the error bars and bands.}
 \label{fig:ttgamma8TeVmt}
  \end{center}
\end{figure}				

\begin{figure}[h!]
\begin{center}
\includegraphics[width=0.5\textwidth]{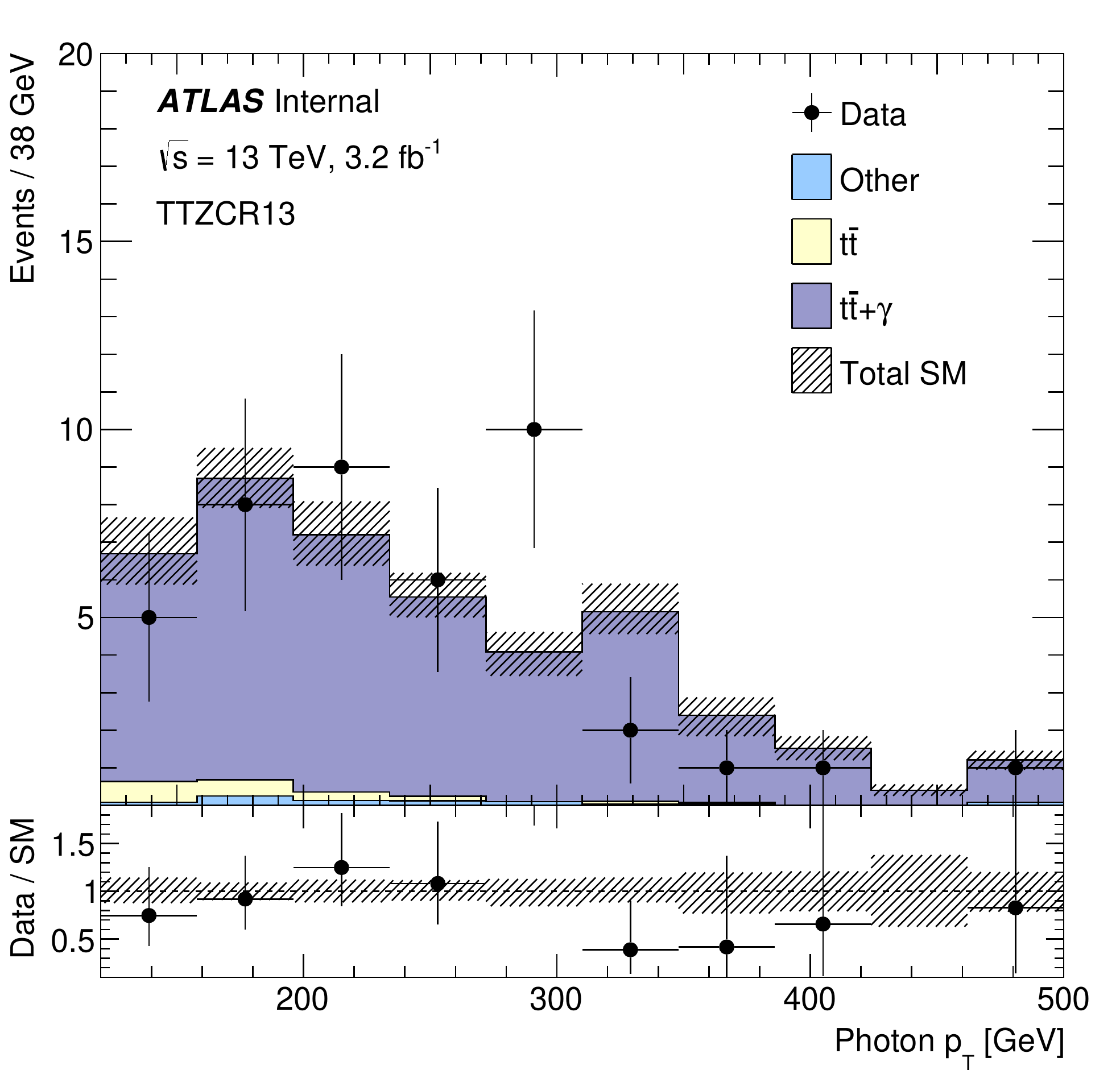}\includegraphics[width=0.5\textwidth]{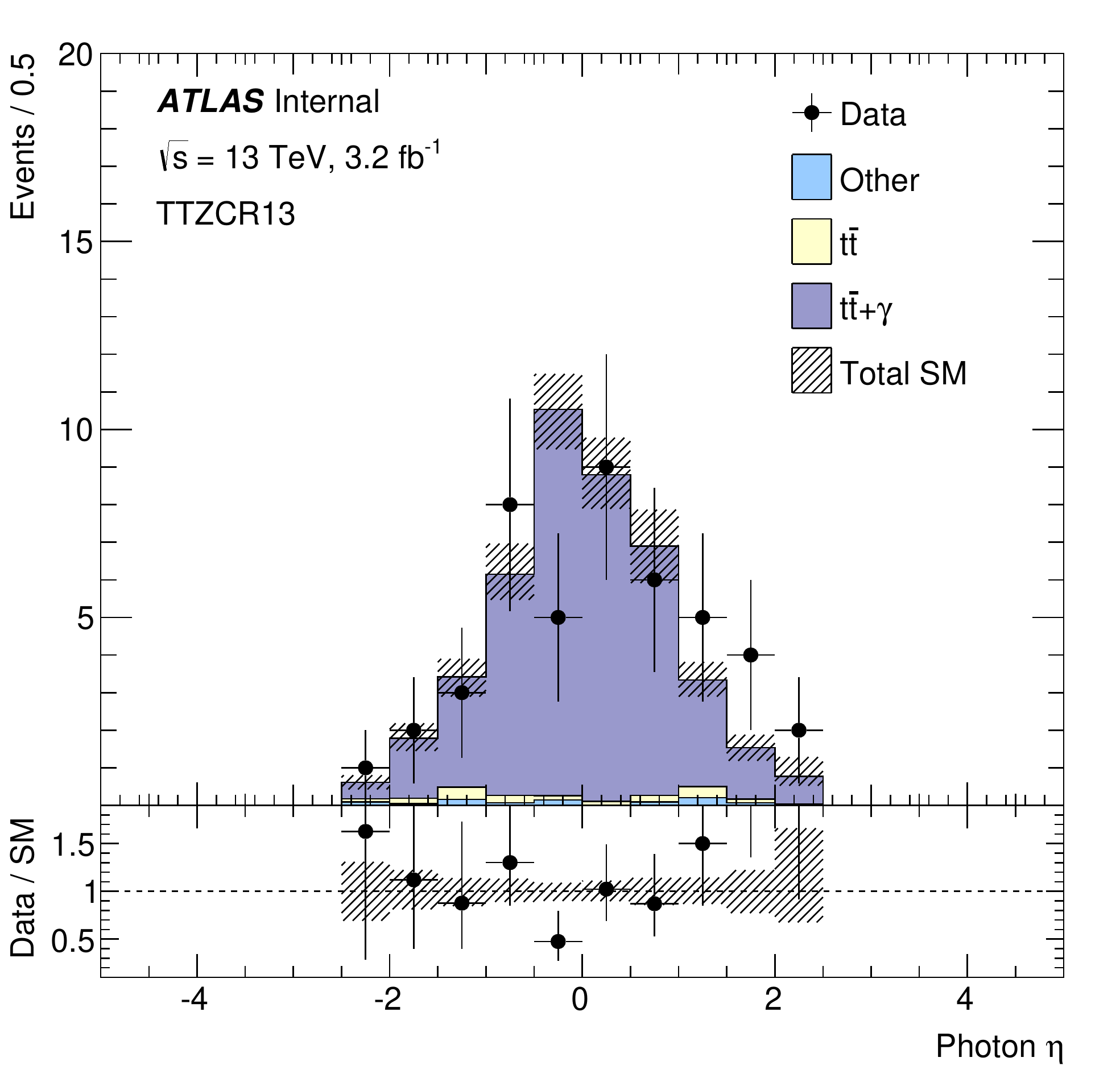}
 \caption{The distribution of photon $p_\text{T}$ and photon $\eta$ in the $t\bar{t}+\gamma$ CR at $\sqrt{s}+13$ TeV.  Jet energy scale and resolution uncertainties in addition to statistical uncertainties are included in the error bars and bands.  A normalization factor of $1.42$ is applied.  The last bin includes overflow.}
 \label{fig:ttzphotptCR}
  \end{center}
\end{figure}	

\begin{figure}[h!]
\begin{center}
\includegraphics[width=0.5\textwidth]{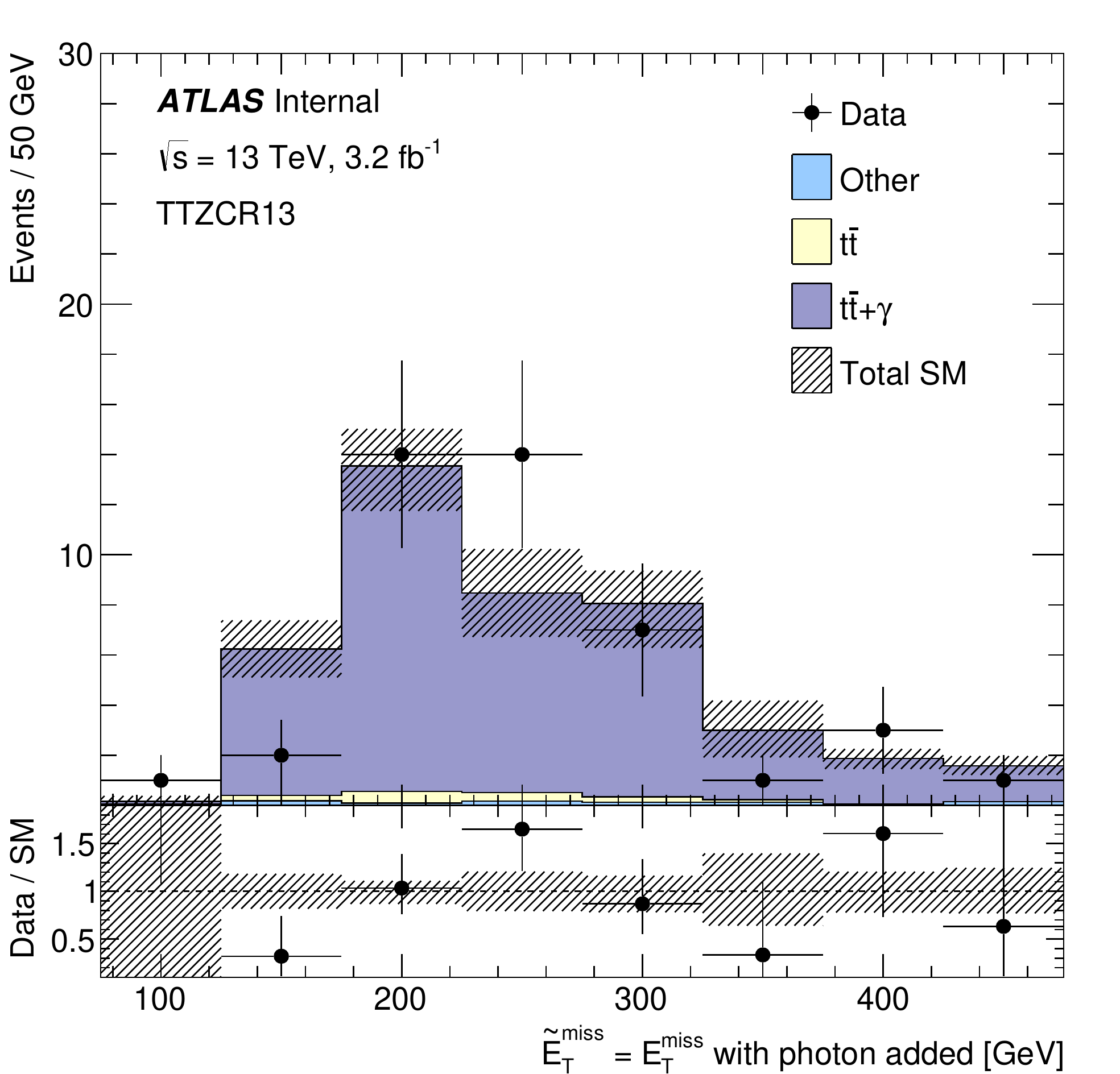}\includegraphics[width=0.5\textwidth]{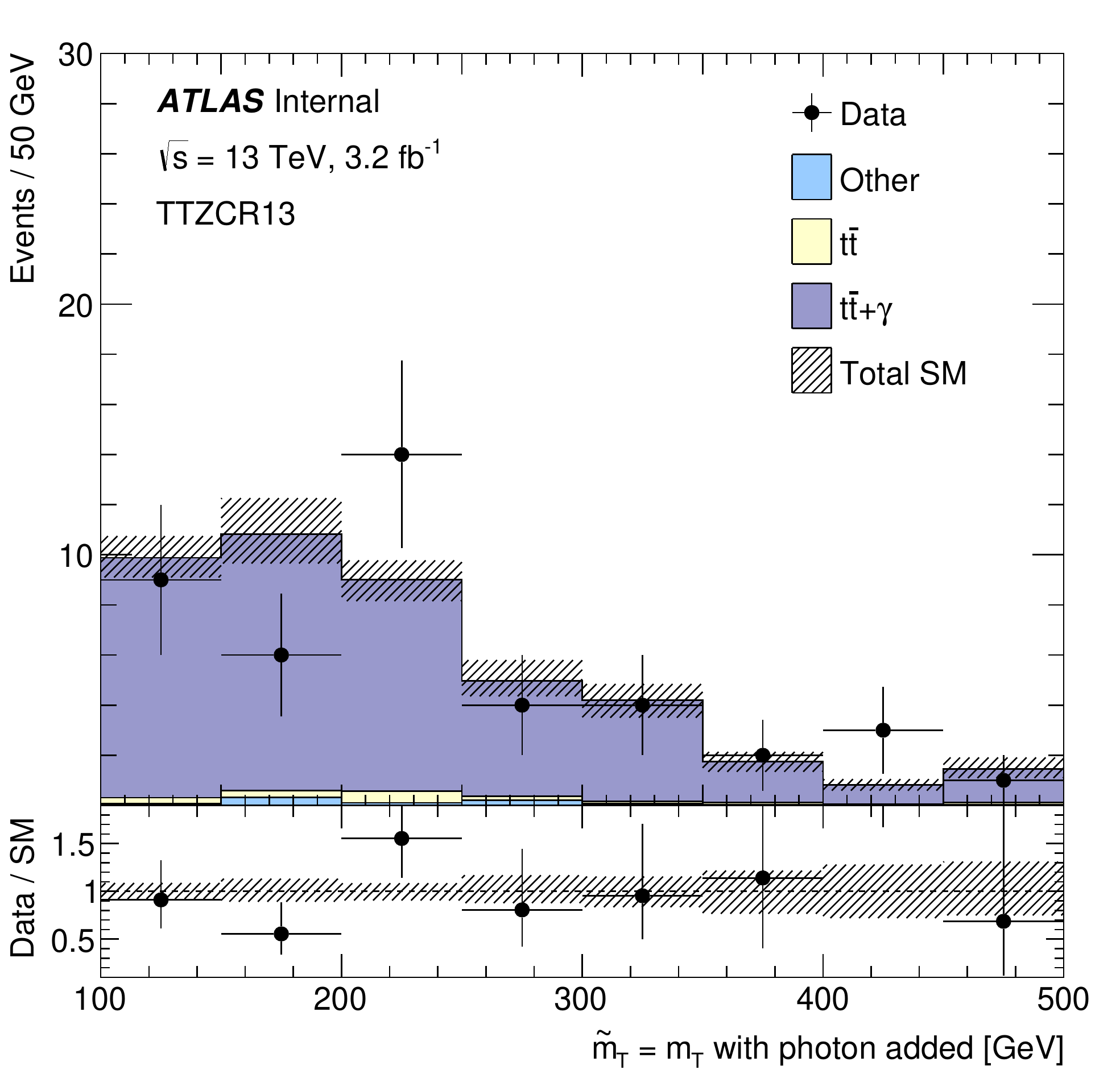}
 \caption{The distribution of $\tilde{E}_\text{T}^\text{miss}$ and $\tilde{m}_\text{T}$ built from $\tilde{p}_\text{T}^\text{miss}=\vec{p}_\text{T}^\text{miss}+\vec{p}_\text{T}^\gamma$ in the $t\bar{t}+\gamma$ CR at $\sqrt{s}+13$ TeV.  Uncertainties are the same as Fig.~\ref{fig:ttzphotptCR}.  A normalization factor of $1.42$ is applied.  The last bin includes overflow. }
 \label{fig:ttzMETCR}
  \end{center}
\end{figure}

		\clearpage
		
		\section{Dibosons}
		\label{dibosons}
		
		The cross section for the double production of electroweak bosons is suppressed by $3$-$4$ orders of magnitude with respect to the inclusive $W$+jets cross section.  However, unlike generic $W$+jets, there are some diboson processes which can have a second lepton, possibly with additional neutrinos, that when not identified as such can allow these events to exceed the $m_\text{T}$ threshold.  Figure~\ref{fig:dibosonprocesses} shows the number of diboson events predicted in simulation after a basic preselection.  Due to the large hadronic branching ratio, the semileptonic $WW$ and $WZ$ processes dominate at low $m_\text{T}$.  However, with only one leptonically decaying $W$ boson, the semileptonic processes are highly suppressed for $m_\text{T}\gtrsim m_W$, after which the dominate processes have multiple leptons/neutrinos.  The $WZ\rightarrow l\nu\nu\nu$ process can naturally have large $m_\text{T}$, but its cross-section is slightly lower than the $WW\rightarrow ll\nu\nu$ process.   The $ZZ\rightarrow ll\nu\nu$ component of the $VV\rightarrow ll\nu\nu$ is subdominant to the $WW$ part and the split is similar to dilepton $t\bar{t}$: roughly half of the dilepton diboson events have a hadronically decaying $\tau$.  Diboson events are a sub-dominant contribution to all signal regions and are estimated using the {\sc Sherpa} event generator.
		
\begin{figure}[h!]
\begin{center}
\includegraphics[width=0.5\textwidth]{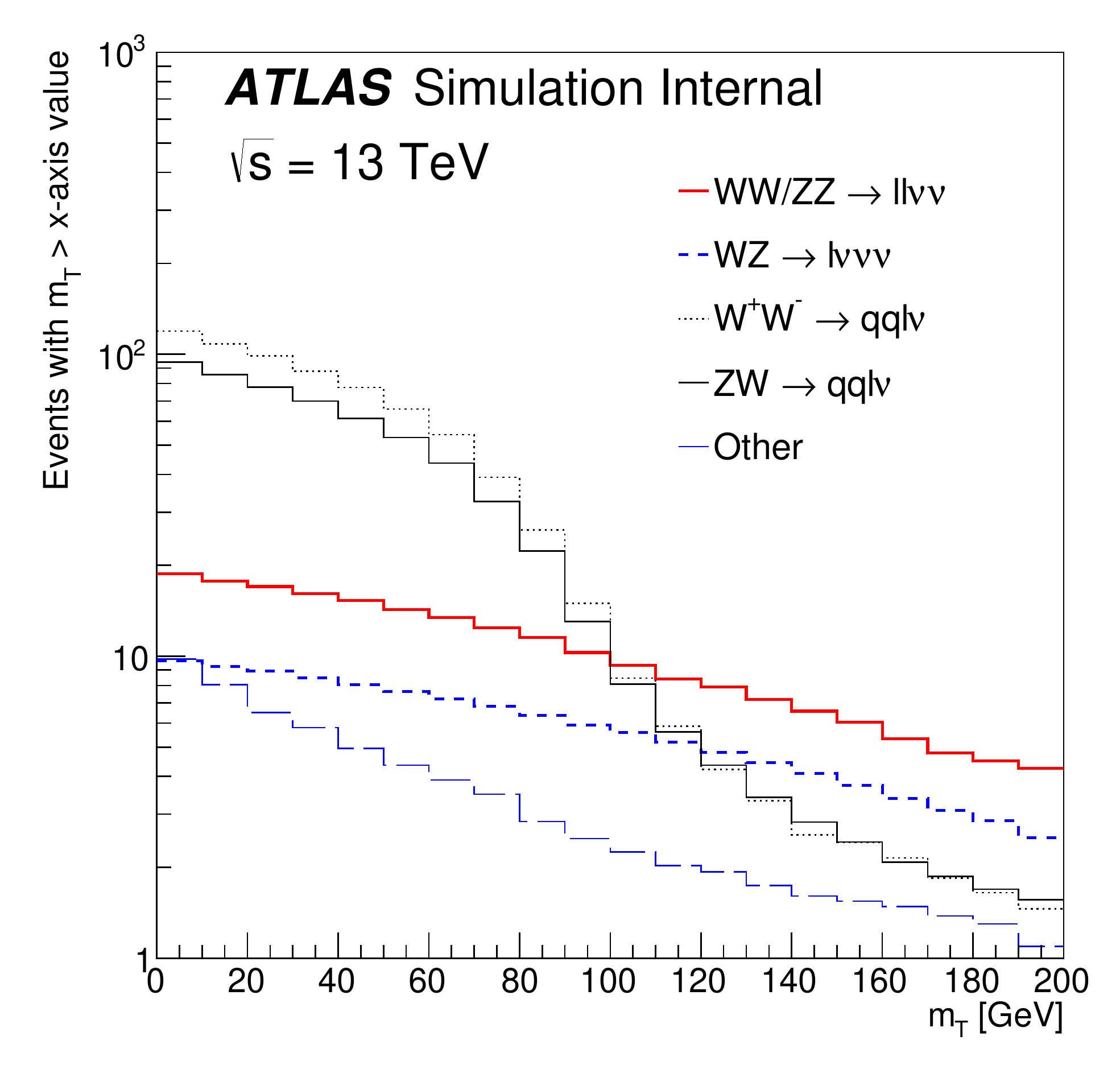}
 \caption{The $m_\text{T}$ distribution of various diboson subprocesses with {\sc Sherpa 2.1}.}
 \label{fig:dibosonprocesses}
  \end{center}
\end{figure}

		\clearpage

		\section{Overview}
		\label{overview}
		
This section explores the overall agreement between the data and simulation using all of the background estimation techniques described earlier in this chapter.  The final background estimate is performed using the simultaneous fit with systematic uncertainties as nuisance parameters as described in Sec.~\ref{sec:susy:stats}.  However, it is possible to estimate the per process scale factors by solving the following system of equations:

	\begin{align}\nonumber
	\label{scalefactors}
	N^\text{data}_\text{WR}&=\mu_W N^\text{W+jets}_\text{WR}+\mu_\text{$t\bar{t}$} N^\text{$t\bar{t}$}_\text{WR}+\mu_\text{$Wt$} N^\text{$Wt$}_\text{WR}\\
	N^\text{data}_\text{TR}&=\mu_W N^\text{W+jets}_\text{TR}+\mu_\text{$t\bar{t}$} N^\text{$t\bar{t}$}_\text{TR}+\mu_\text{$Wt$} N^\text{$Wt$}_\text{TCR}\\\nonumber
	N^\text{data}_\text{StR}&=\mu_W N^\text{W+jets}_\text{StR}+\mu_\text{$t\bar{t}$} N^\text{$t\bar{t}$}_\text{StR}+\mu_\text{$Wt$} N^\text{$Wt$}_\text{StR},	
	\end{align}	
	
\noindent where $\mu_x$ is the normalization factor (NF) for process $x$ and $N_y^z$ is the number of simulated or measured events of type $z$ in the $y$ event selection.  The data $N_y^\text{data}$ must be corrected for the simulation-based estimates, $N_y^\text{data}=	N_y^\text{data,observed}-N_y^\text{$VV$}-N_y^\text{$t\bar{t}$+V}$.  The $t\bar{t}+V$ yield in the $t\bar{t}$, $W$+jets and single top control regions is negligible so the data-driven estimate with the photon mostly decouples from the rest of Eq.~\ref{scalefactors}.  As a system of three equations with three unknowns ($\mu_\text{$t\bar{t}$}, \mu_W$, and $\mu_\text{$Wt$}$), there is a unique solution.  Looser versions of the control regions described in Sec.~\ref{ttbarCR},~\ref{wjets}, and~\ref{singletop:datadriven} are used in order to study the dependence of the normalization factors on key even kinematic properties.   In addition to the preselection, events are required to have the leading four jets with $p_\text{T}>50,50,50,25$ GeV, $E_\text{T}^\text{miss}>150$ GeV, and $30$ GeV $<m_\text{T}<90$ GeV.  The $t\bar{t}$ enriched region additionally requires $n_\text{$b$-jets}>0$ and $am_\text{T2}<200$ GeV or $n_\text{$b$-jets}=1$ (orthogonality to the single top region), the $W$+jets enriched region requires $n_\text{$b$-jets}=0$, and the single top enriched region requires $n_\text{$b$-jets}>1$ and $am_\text{T2}>200$ GeV. Figures~\ref{fig:scalefactorsjetpt},~\ref{fig:scalefactorshtsig},~\ref{fig:scalefactorsmet}, and~\ref{fig:scalefactorsmass} show the distributions of the leading jet $p_\text{T}$, $H_\text{T,sig}^\text{miss}$, $E_\text{T}^\text{miss}$, and the leading large-radius ($R=1.2$) jet mass in each of the $t\bar{t}$, $W$+jets, and single top enriched event selections.  The $t\bar{t}$ enriched region has about $6000$ events, of which about $80\%$ are predicted to be $t\bar{t}$ events; the $W$+jets enriched region has about $4000$ events, of which which about $75\%$ are predicted to be $W$+jets events, and the single top enriched region has about $300$ events, of which about $25\%$ are predicted to be single top events.

\begin{figure}[h!]
\begin{center}
\includegraphics[width=0.33\textwidth]{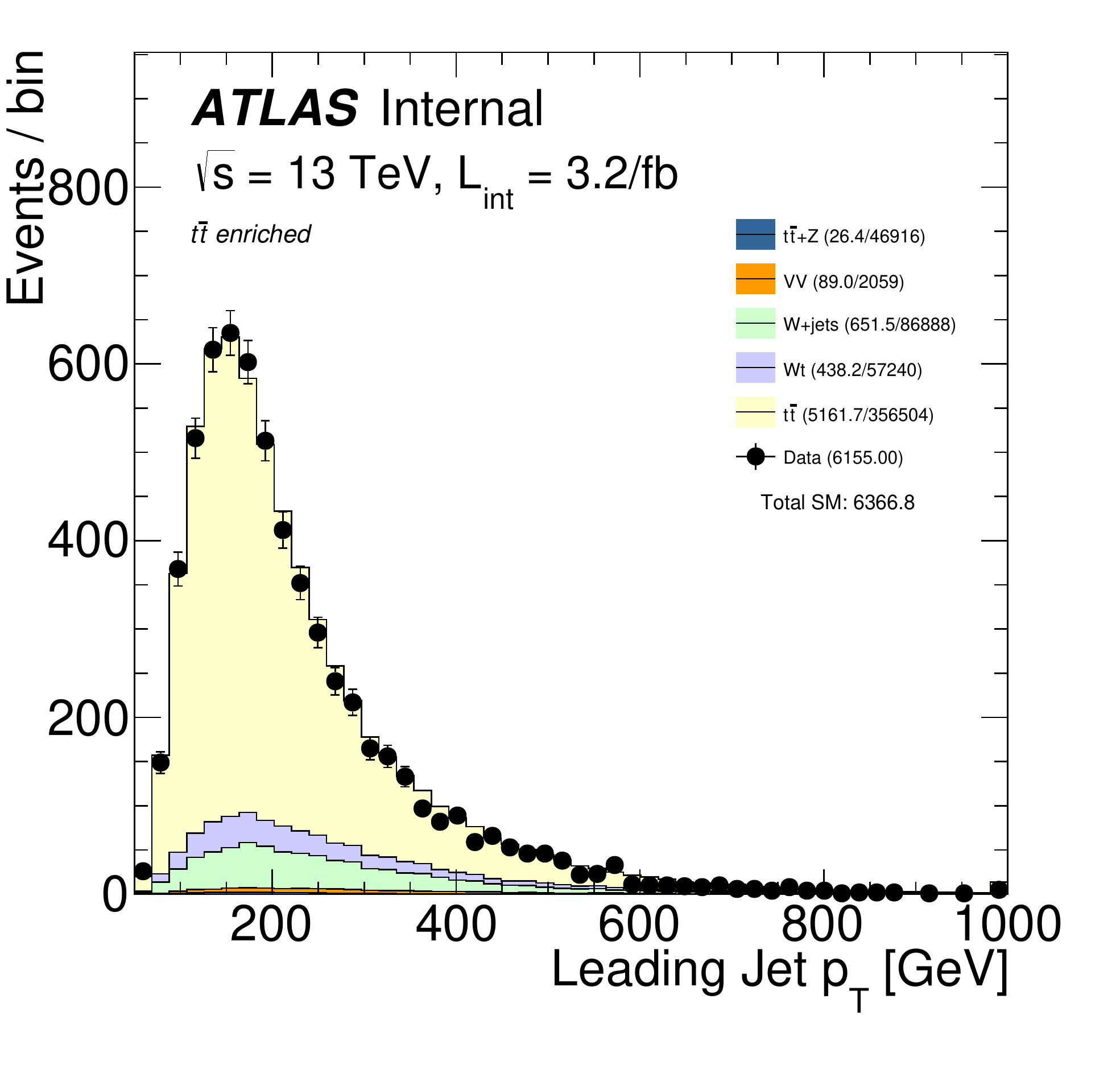}\includegraphics[width=0.33\textwidth]{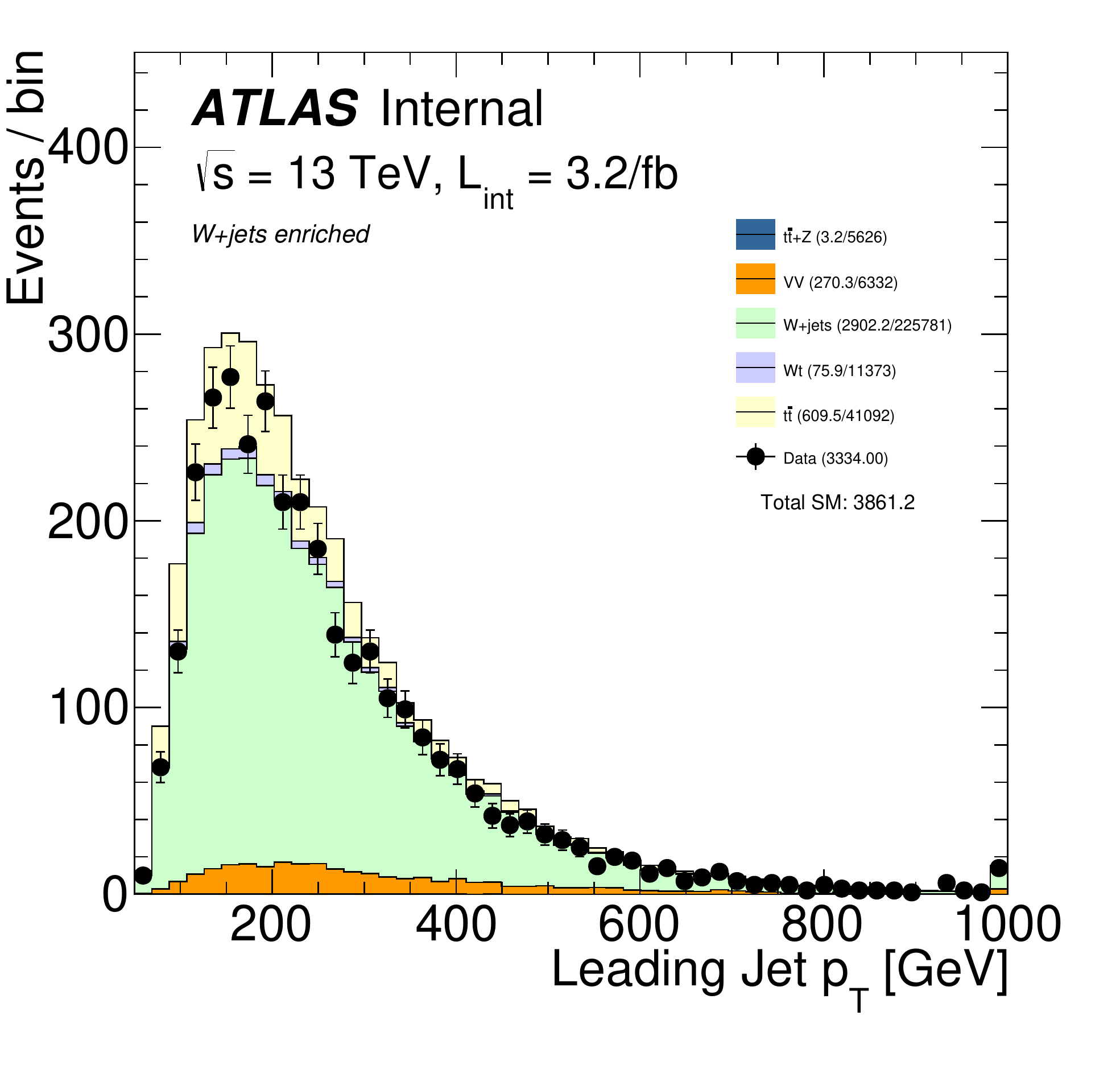}\includegraphics[width=0.33\textwidth]{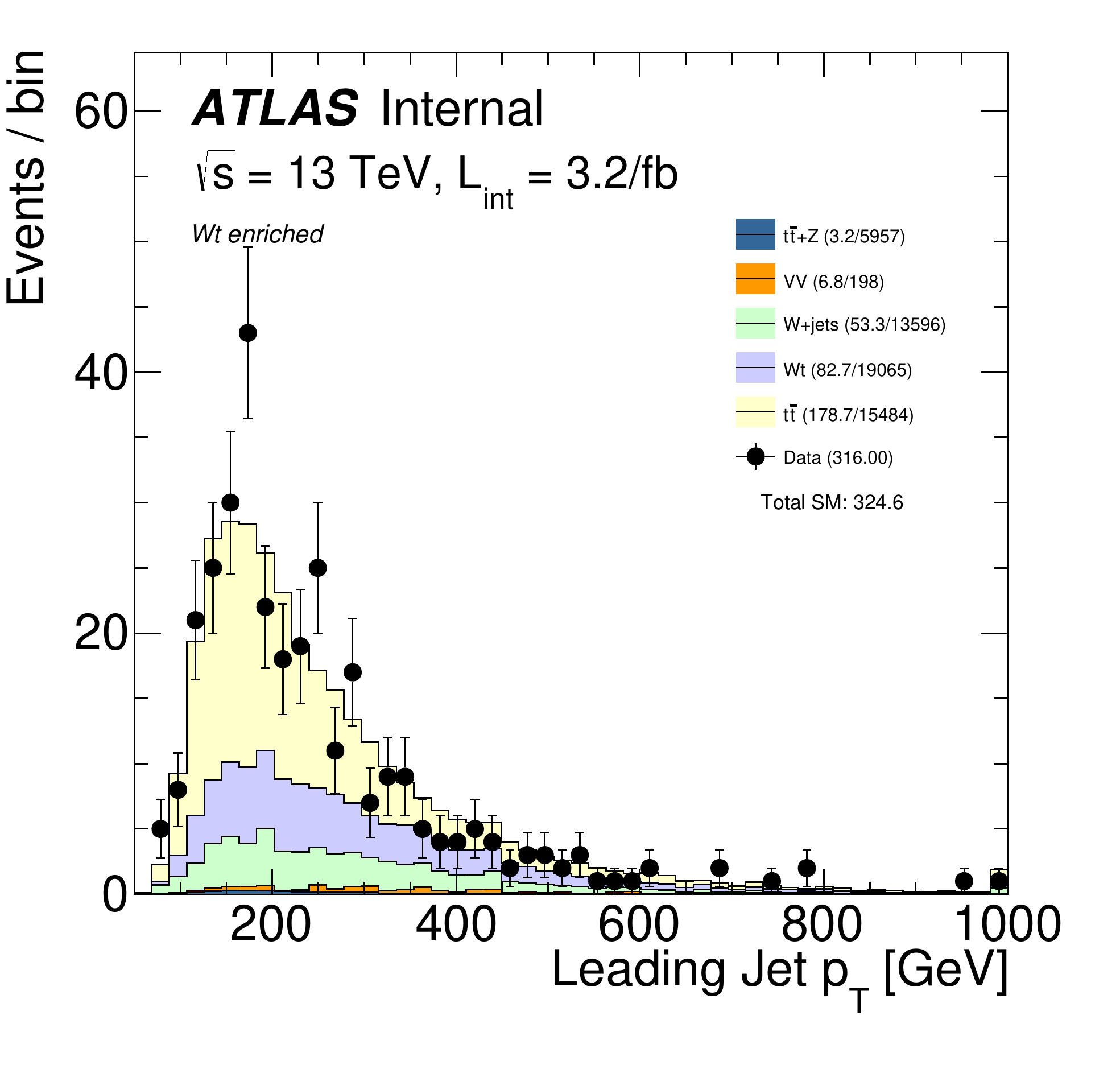}
 \caption{The distribution of the leading jet $p_\text{T}$ in the $t\bar{t}$ enriched region (left), the $W$+jets enriched region (middle), and the single top enriched region (right).  See the text for the event selections.  The first number in parenthesis after the process in the legend is the estimated event yield and the second number is the number of raw MC events used to make that prediction.}
 \label{fig:scalefactorsjetpt}
  \end{center}
\end{figure}

\begin{figure}[h!]
\begin{center}
\includegraphics[width=0.33\textwidth]{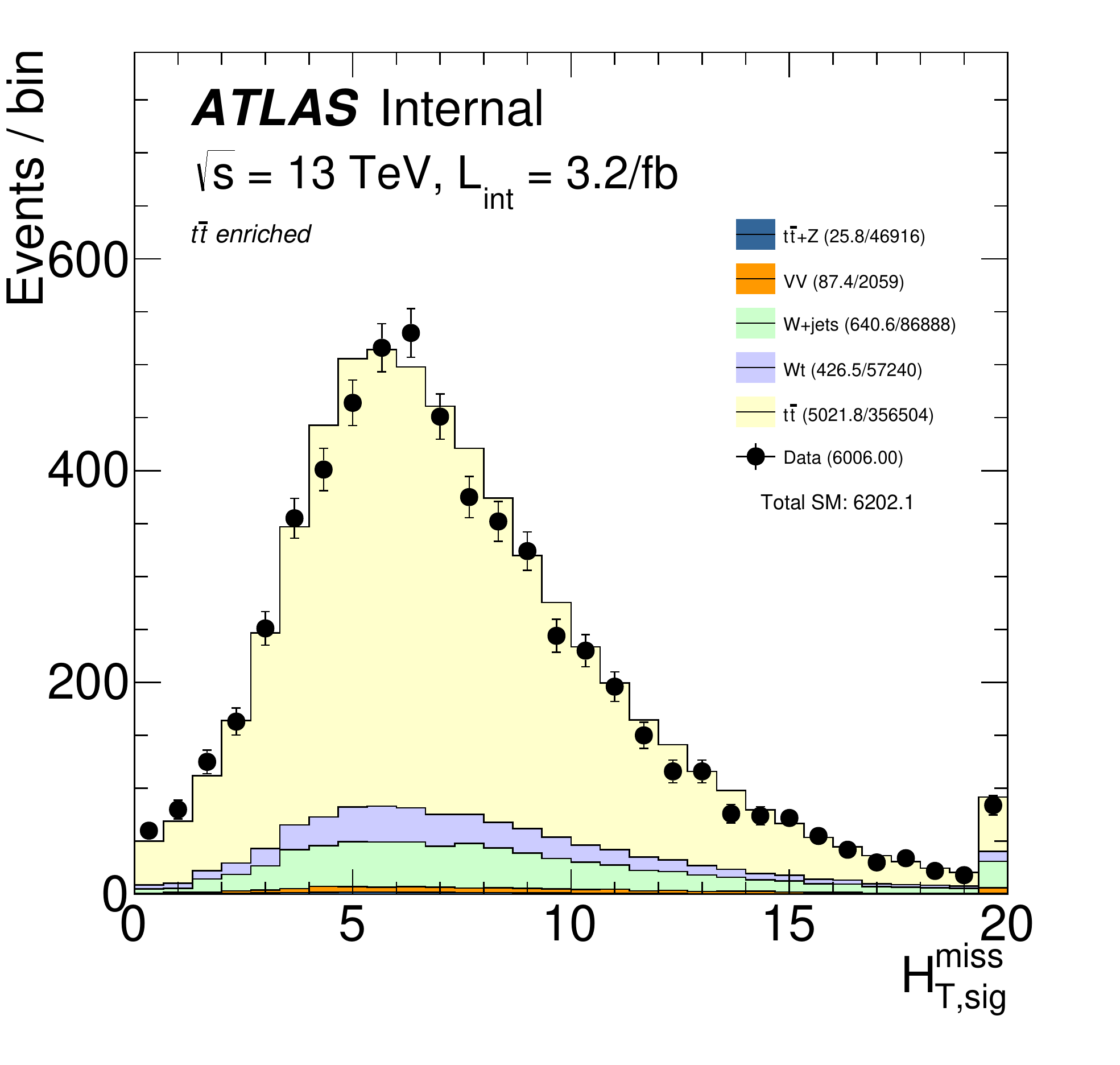}\includegraphics[width=0.33\textwidth]{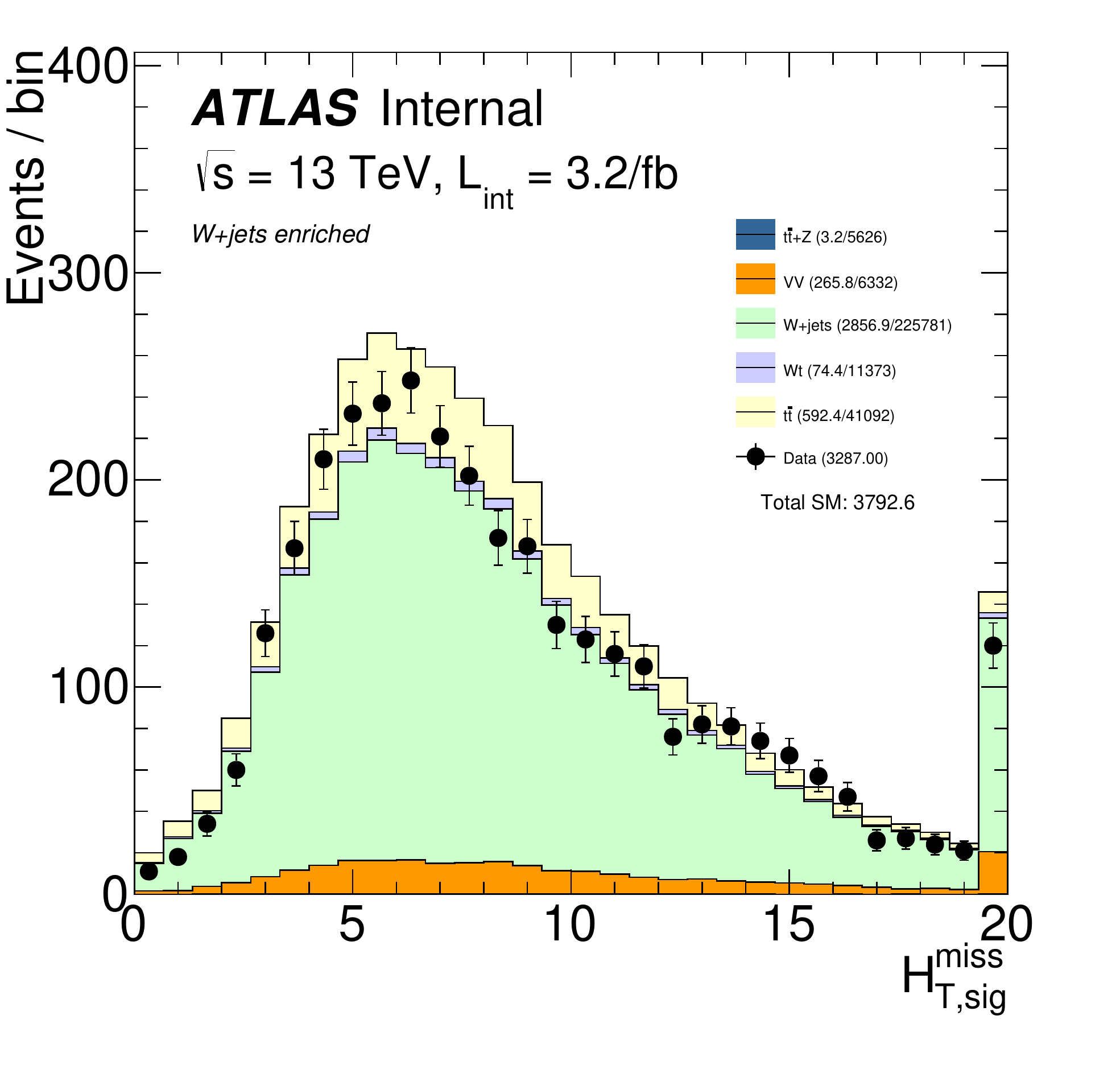}\includegraphics[width=0.33\textwidth]{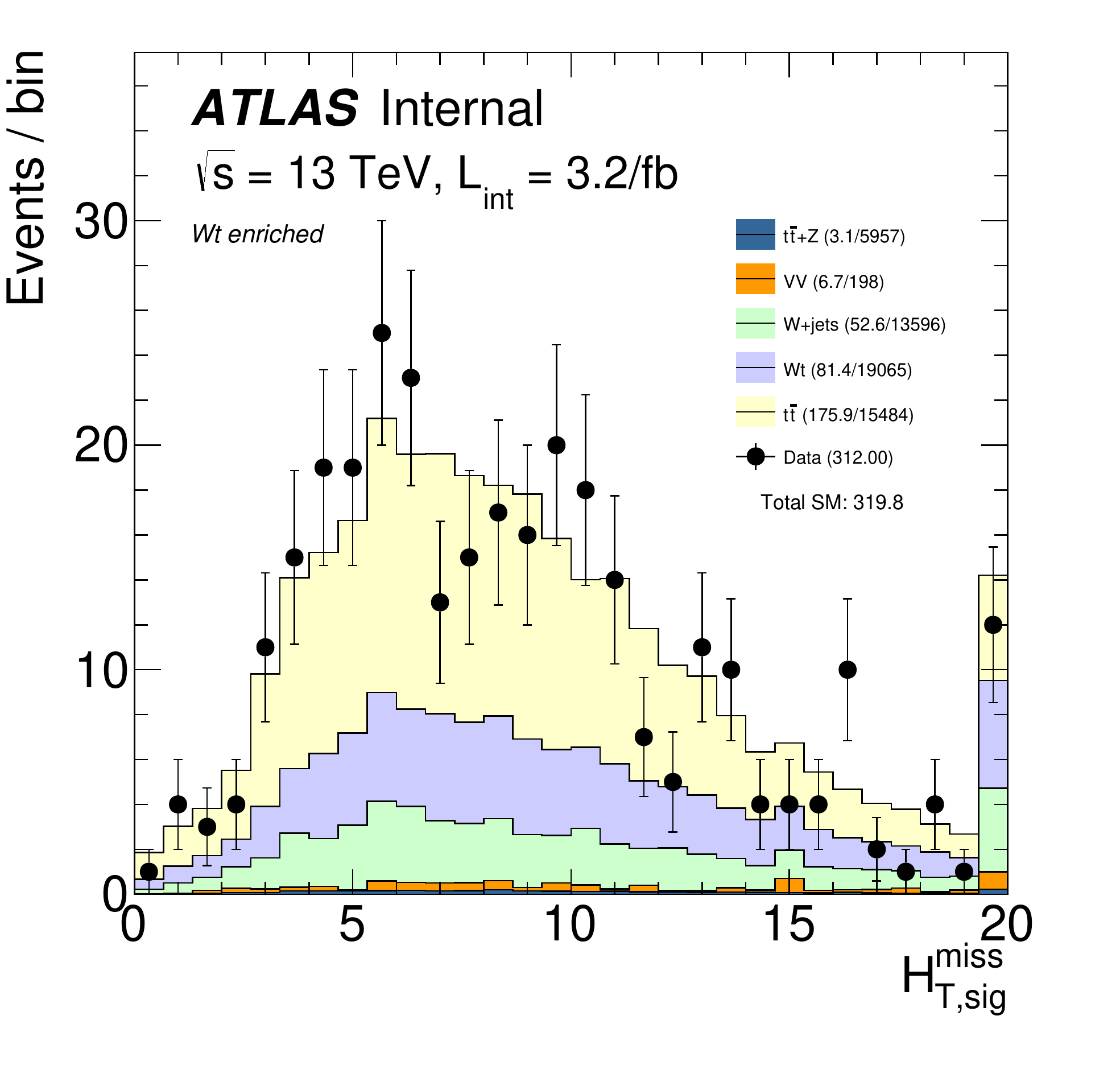}
 \caption{The distribution of $H_\text{T,sig}^\text{miss}$ in the $t\bar{t}$ enriched region (left), the $W$+jets enriched region (middle), and the single top enriched region (right).  See the text for the event selections.  The first number in parenthesis after the process in the legend is the estimated event yield and the second number is the number of raw MC events used to make that prediction.}
 \label{fig:scalefactorshtsig}
  \end{center}
\end{figure}

\begin{figure}[h!]
\begin{center}
\includegraphics[width=0.33\textwidth]{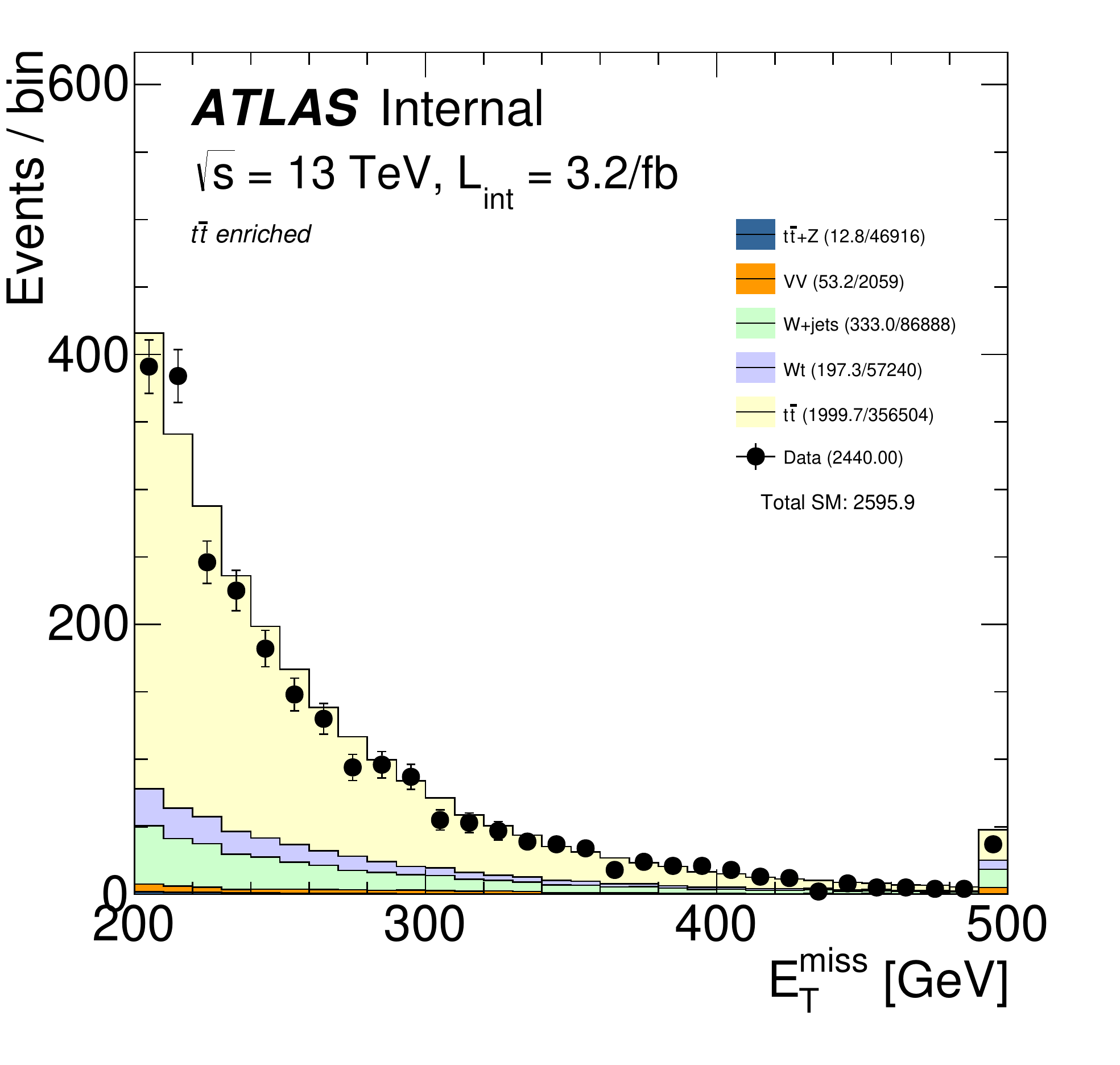}\includegraphics[width=0.33\textwidth]{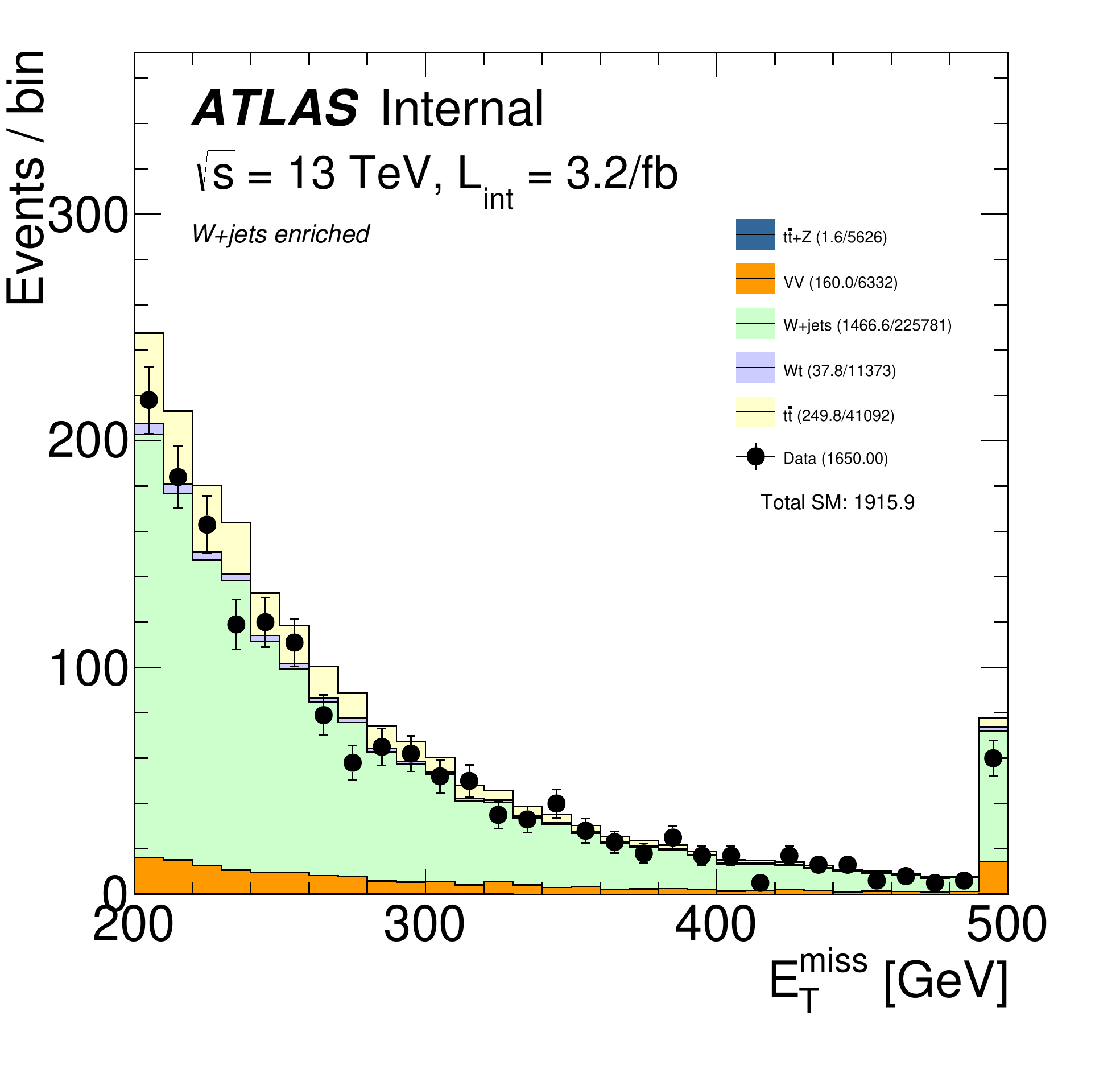}\includegraphics[width=0.33\textwidth]{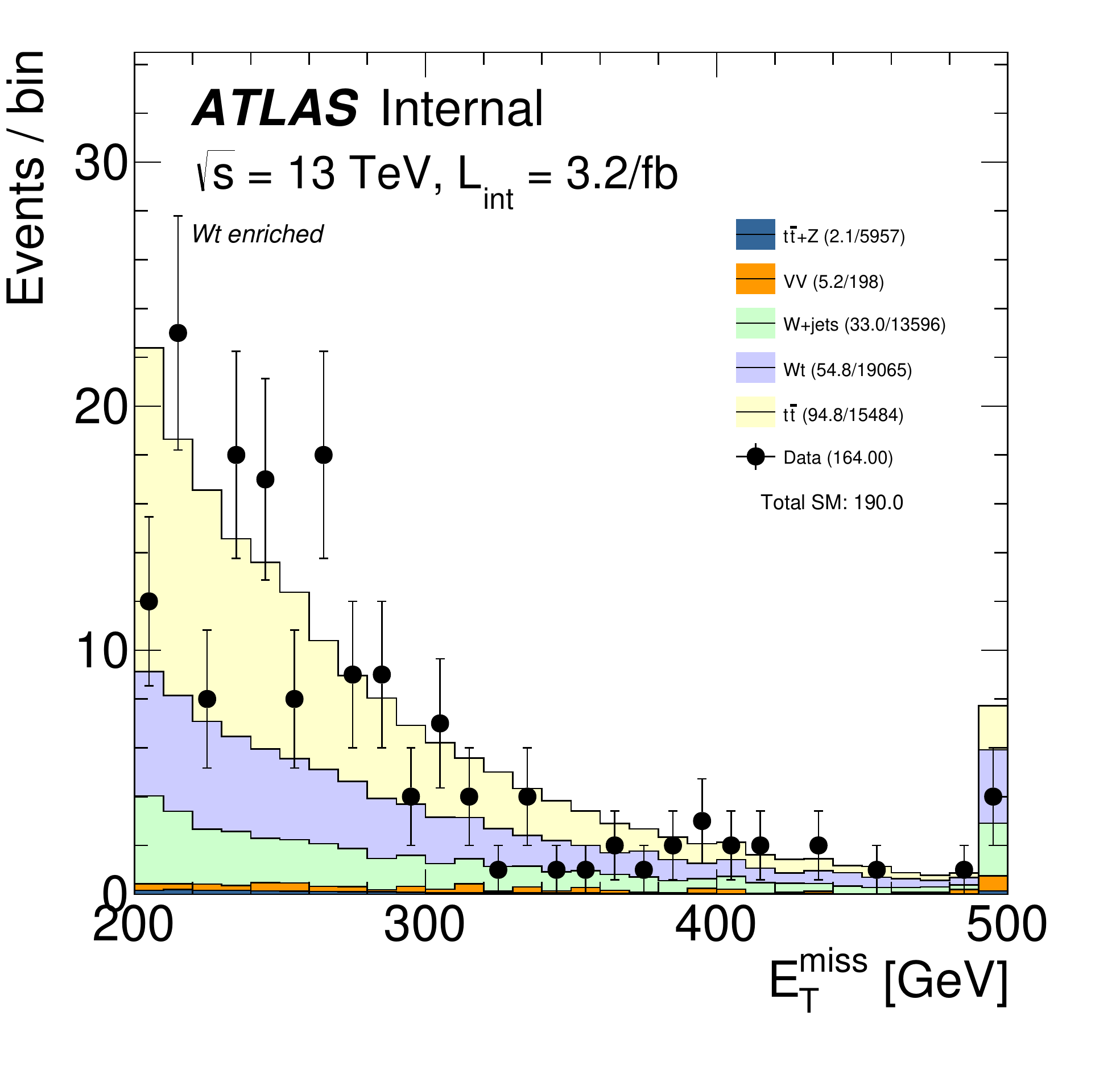}
 \caption{The distribution of $E_\text{T}^\text{miss}$ in the $t\bar{t}$ enriched region (left), the $W$+jets enriched region (middle), and the single top enriched region (right).  See the text for the event selections.  The first number in parenthesis after the process in the legend is the estimated event yield and the second number is the number of raw MC events used to make that prediction}
 \label{fig:scalefactorsmet}
  \end{center}
\end{figure}

\begin{figure}[h!]
\begin{center}
\includegraphics[width=0.33\textwidth]{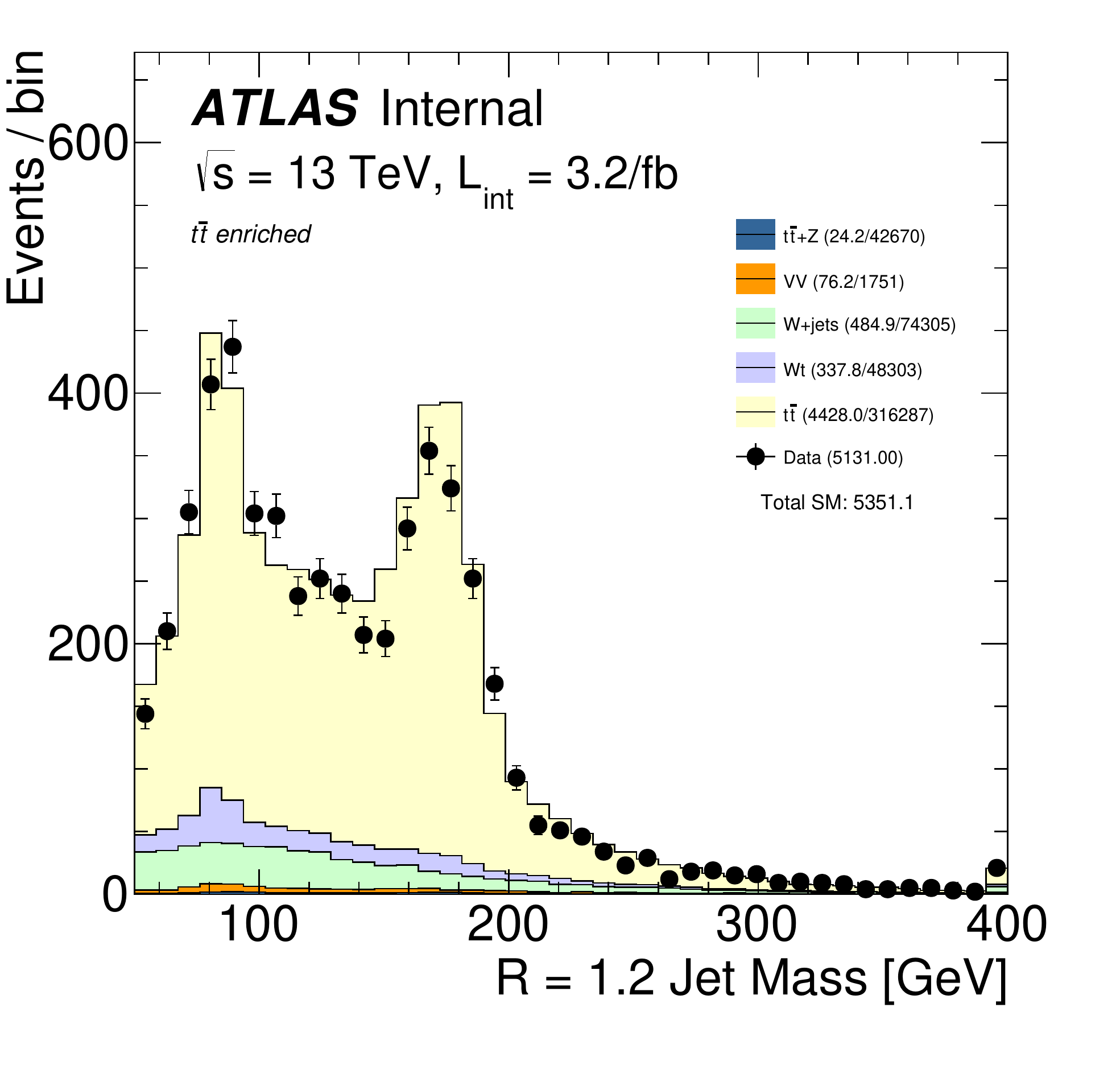}\includegraphics[width=0.33\textwidth]{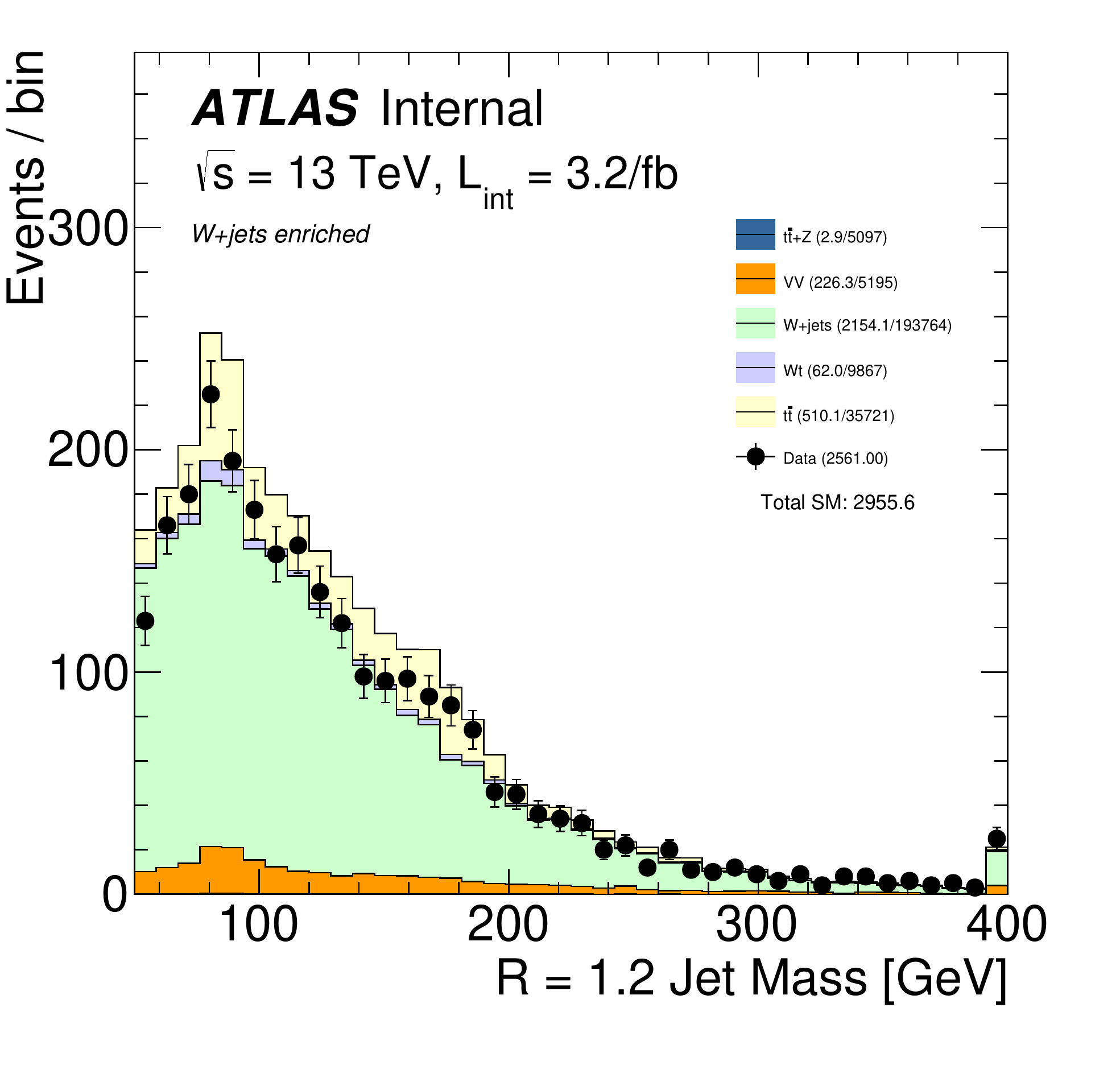}\includegraphics[width=0.33\textwidth]{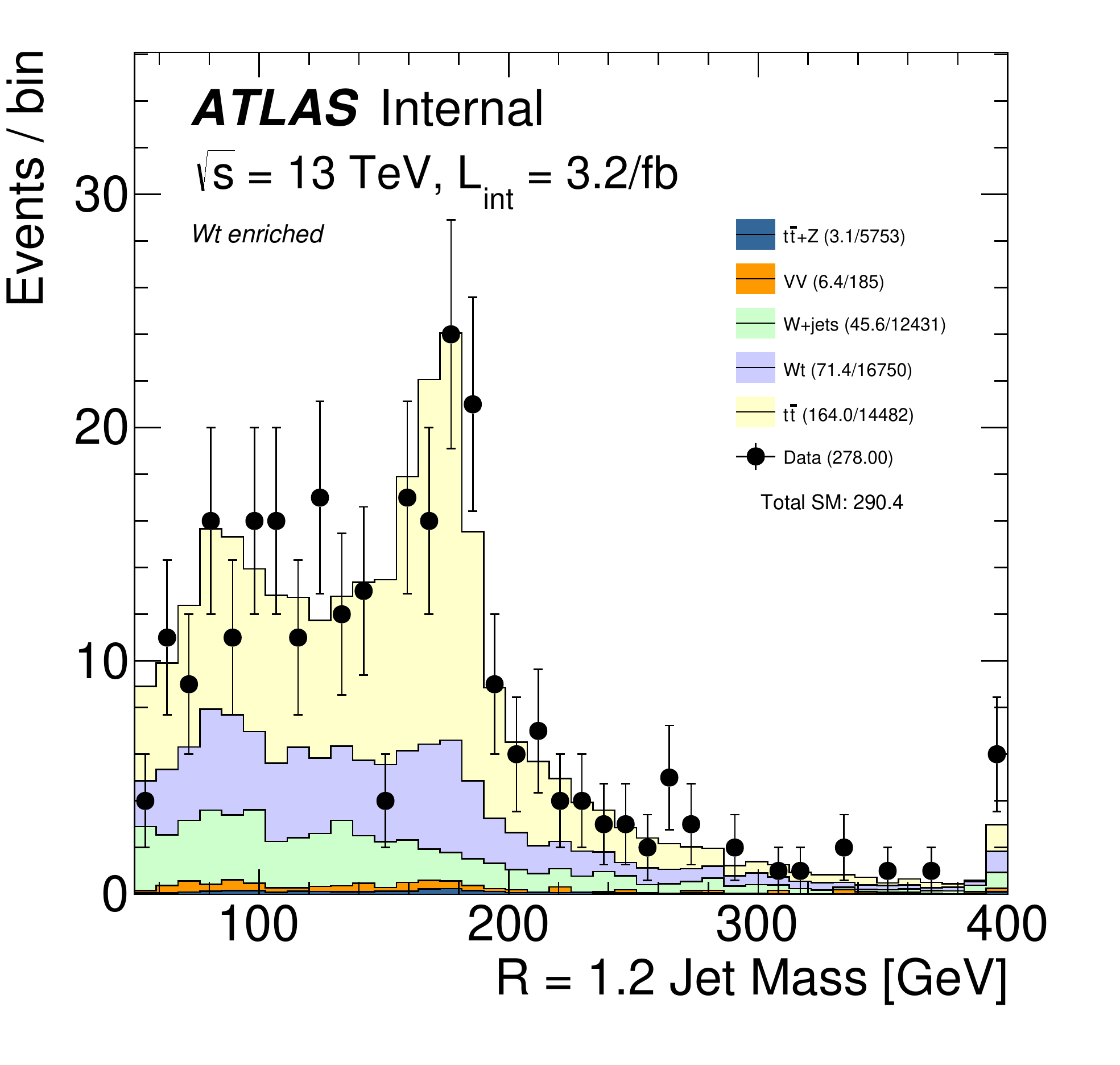}
 \caption{The distribution of the leading large-radius ($R=1.2$) jet mass in the $t\bar{t}$ enriched region (left), the $W$+jets enriched region (middle), and the single top enriched region (right).  See the text for the event selections.  The first number in parenthesis after the process in the legend is the estimated event yield and the second number is the number of raw MC events used to make that prediction}
 \label{fig:scalefactorsmass}
  \end{center}
\end{figure}
		
Each of the kinematic variables in Fig.~\ref{fig:scalefactorsjetpt},~\ref{fig:scalefactorshtsig},~\ref{fig:scalefactorsmet}, and~\ref{fig:scalefactorsmass} are scanned to compute the dependence of the normalization factors $\mu$ on the variables.  Figure~\ref{fig:scalefactor_depend_etmiss_htsig} shows the dependence on $E_\text{T}^\text{miss}$ and $H_\text{T,sig}^\text{miss}$.  The $t\bar{t}$ and $W$+jets normalization factors are relatively constant as a function of these variables while there is a decreasing trend for the single top normalization factors, albeit with significant statistical uncertainties due to the low yield and purity in the single top enriched region.  Similar plots are shown in Fig.~\ref{fig:scalefactor_depend_pt_mass} for the leading jet $p_\text{T}$ and the leading large-radius ($R=1.2$) jet mass.  The $t\bar{t}$ and $W$+jets normalization factors are relatively constant as a function of the jet mass, but there is a significant decrease in the $t\bar{t}$ normalization factor as a function of the leading jet $p_\text{T}$.  This is likely related to the well-known mis-modeling of the top quark $p_\text{T}$ (see e.g. Ref.~\cite{Aad:2014zka}).  All of the control regions described in Sec.~\ref{ttbarCR},~\ref{wjets}, and~\ref{singletop:datadriven} have the same jet $p_\text{T}$ requirements as the signal region in order to remain largely insensitive to this mis-modeling.  			
		
\begin{figure}[h!]
\begin{center}
\includegraphics[width=0.5\textwidth]{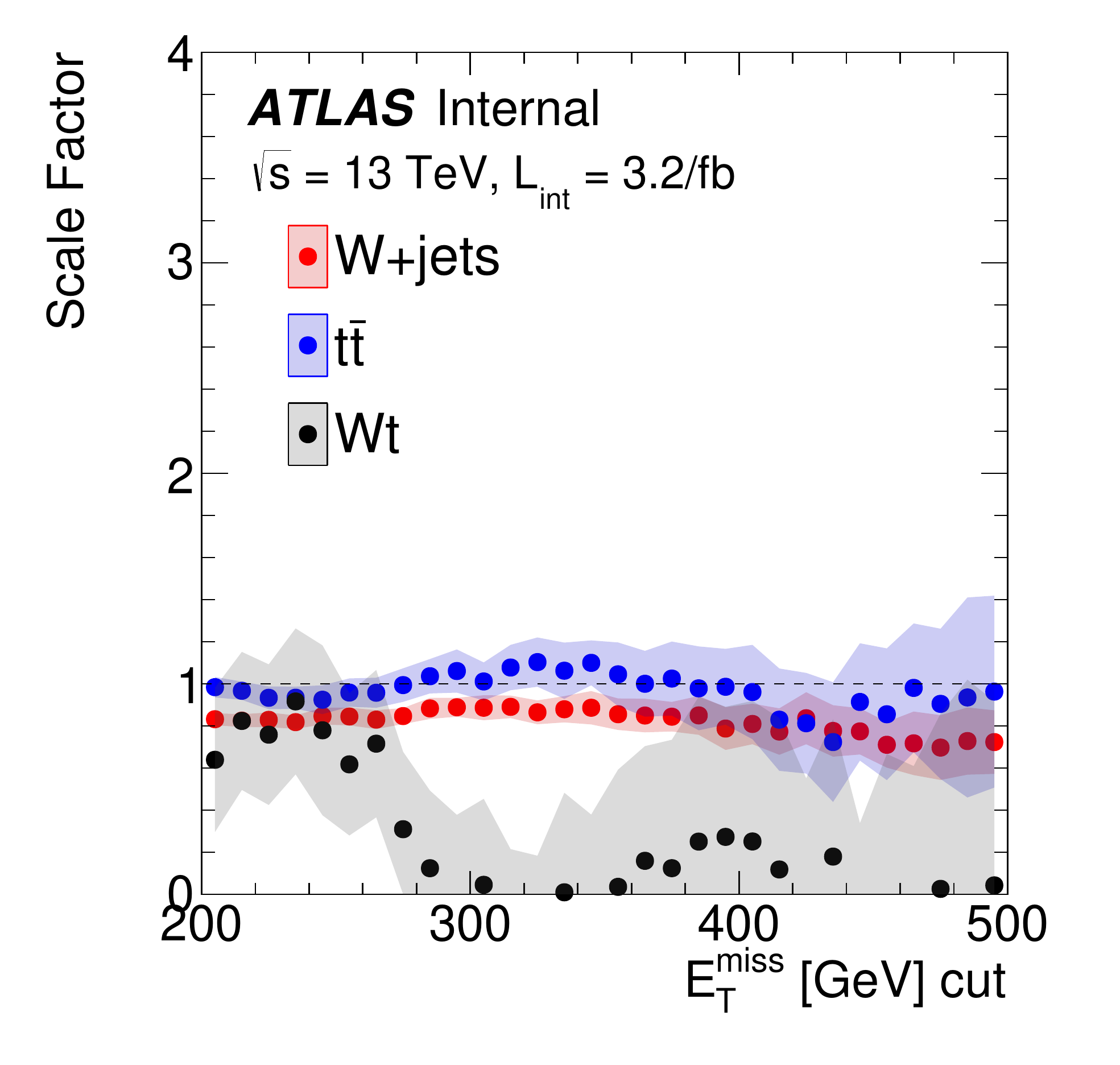}\includegraphics[width=0.5\textwidth]{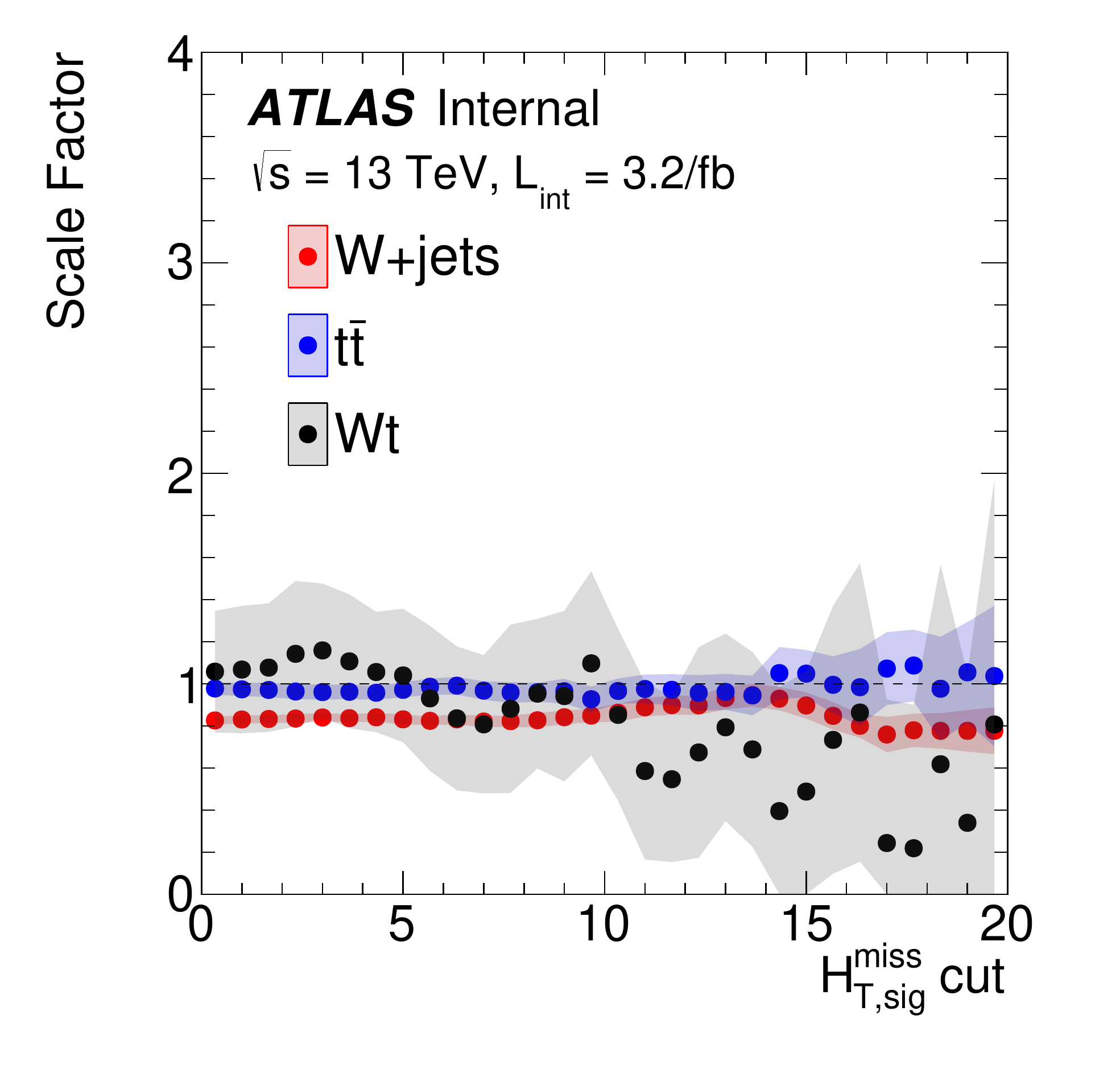}
 \caption{The dependence of the normalization factors on $E_\text{T}^\text{miss}$ (left) and $H_\text{T,sig}^\text{miss}$ (right).  The error band represents the statistical uncertainty derived from bootstrapping the data in the enriched regions and resolving Eq.~\ref{scalefactors}.}
 \label{fig:scalefactor_depend_etmiss_htsig}
  \end{center}
\end{figure}

\begin{figure}[h!]
\begin{center}
\includegraphics[width=0.5\textwidth]{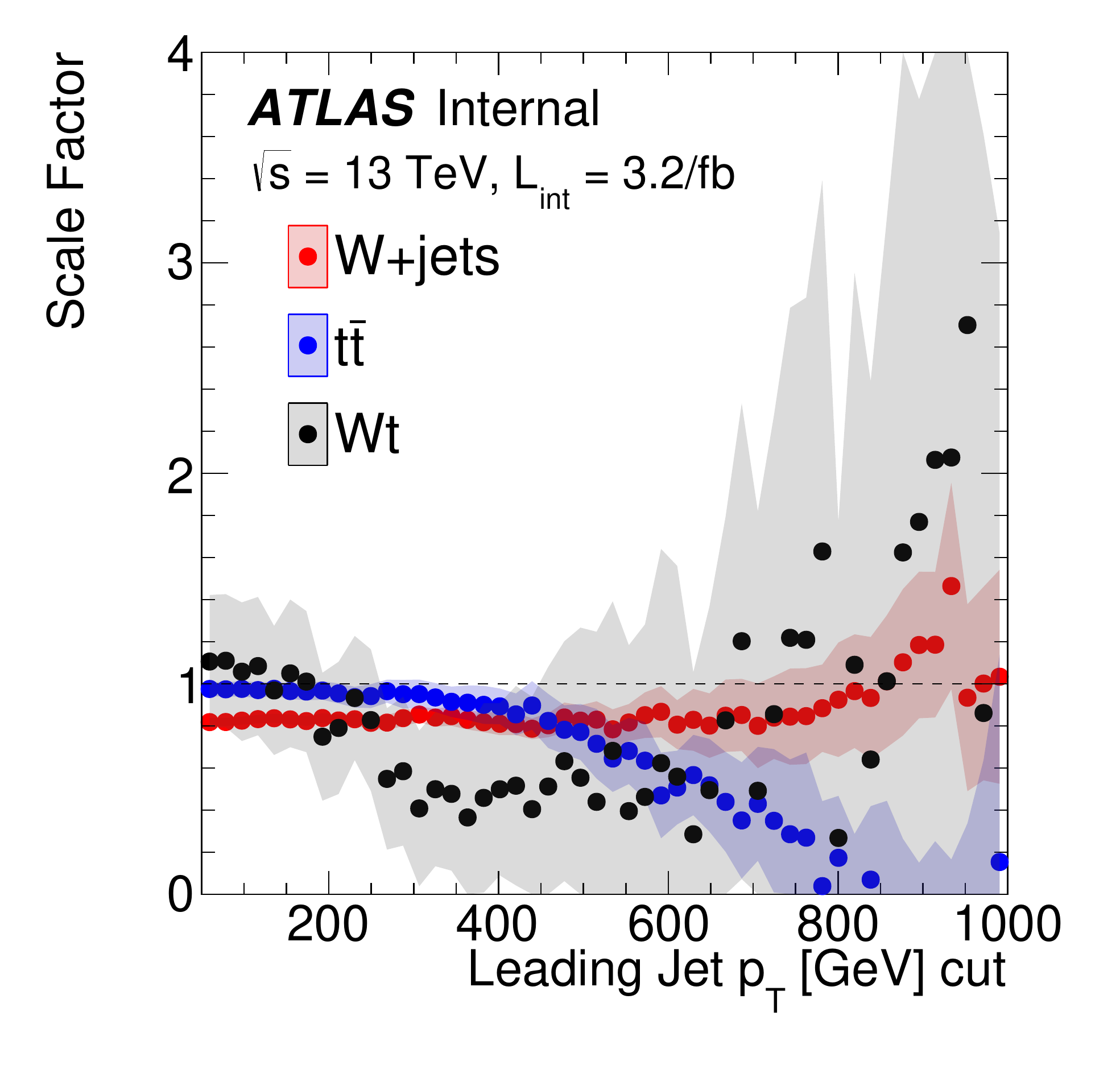}\includegraphics[width=0.5\textwidth]{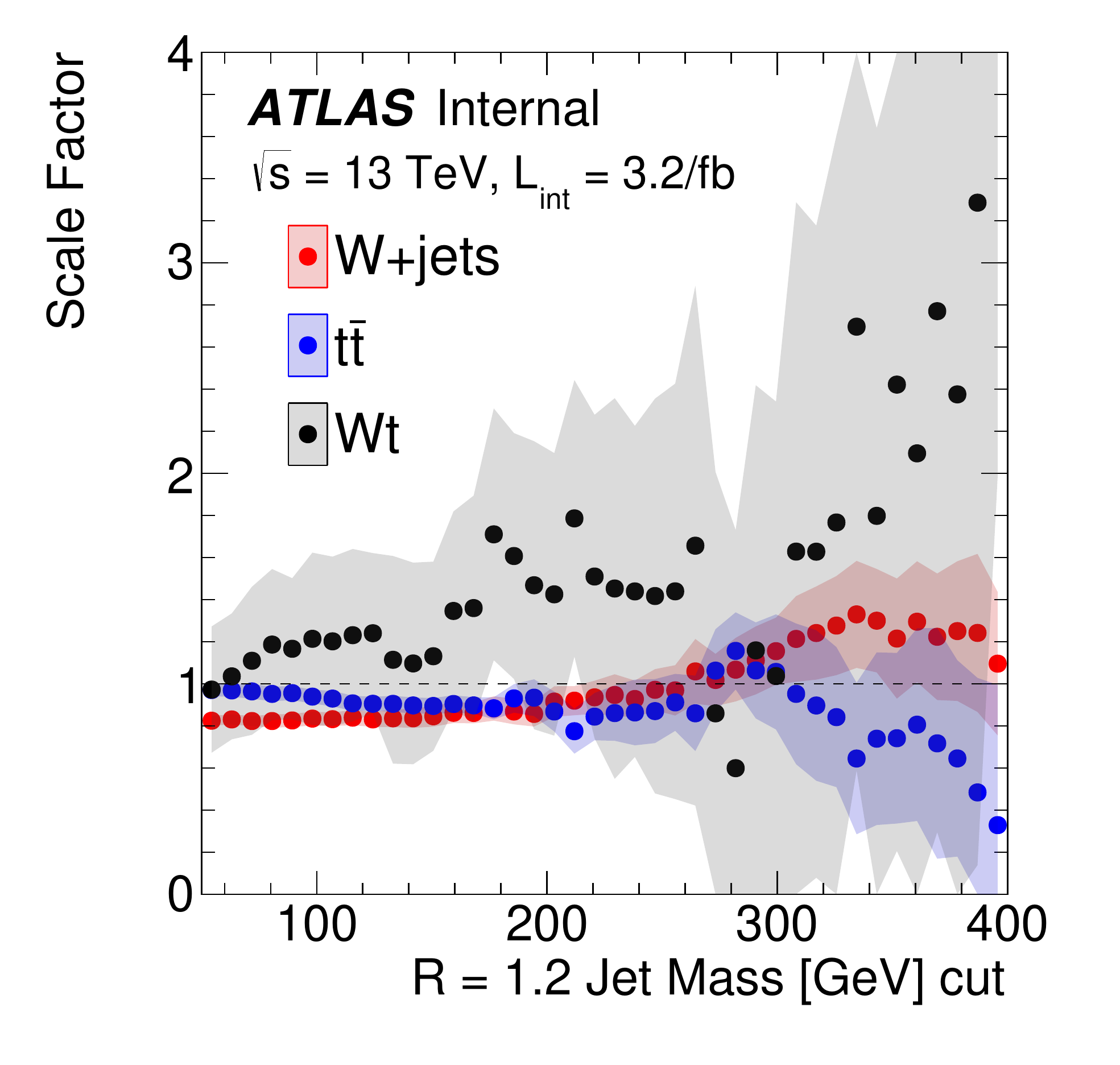}
 \caption{The dependence of the normalization factors on the leading jet $p_\text{T}$ (left) and the leading large radius ($R=1.2$) jet mass (right).  The error band represents the statistical uncertainty derived from bootstrapping the data in the enriched regions and resolving Eq.~\ref{scalefactors}.}
 \label{fig:scalefactor_depend_pt_mass}
  \end{center}
\end{figure}

Both Fig.~\ref{fig:scalefactor_depend_etmiss_htsig} and~\ref{fig:scalefactor_depend_pt_mass}	 show that the uncertainty on the single top normalization factor is much larger than the uncertainties for the $t\bar{t}$ and $W$+jets factors.  This is due in part to the low event yield in the single top enriched region but also to the contamination of single top events in the $t\bar{t}$ enriched region and $t\bar{t}$ events in the single top enriched region.  The top row of Fig.~\ref{fig:scalefactorcorrelations} shows the statistical correlations between the various normalization factors when using the enriched samples described earlier without any further requirements.  The $t\bar{t}$ and $W$+jets normalization factors are largely uncorrelated, but the $t\bar{t}$ and single top factors are nearly 100\% anti-correlated.  Fixing the number of single top events in the single top enriched region, the middle and lower panels of Fig.~\ref{fig:scalefactorcorrelations} demonstrate the impact of reducing contamination in the $t\bar{t}$ and single top enriched regions.  In the middle panel, the $t\bar{t}$ contribution to the single top enriched region is set to zero.  This reduces the correlation between the $t\bar{t}$ and single top normalization factors and the overall single top normalization uncertainty by nearly a factor of two.  However, the lower panel shows that the correlation between the normalization factors is not the only relevant quantity.  When the single top contamination in the $t\bar{t}$ enriched region is set to zero, the single top and $t\bar{t}$ normalization factors are nearly uncorrelated but the uncertainty in the single top factor is only reduced by about $15\%$.  The additional $t\bar{t}$ reduction in the single top control region by the $\Delta R(b,b)$ requirement (Sec.~\ref{singletop:datadriven}) is therefore a well-motivated technique for reducing the single top normalization factor uncertainty.

\begin{figure}[h!]
\begin{center}
\includegraphics[width=0.95\textwidth]{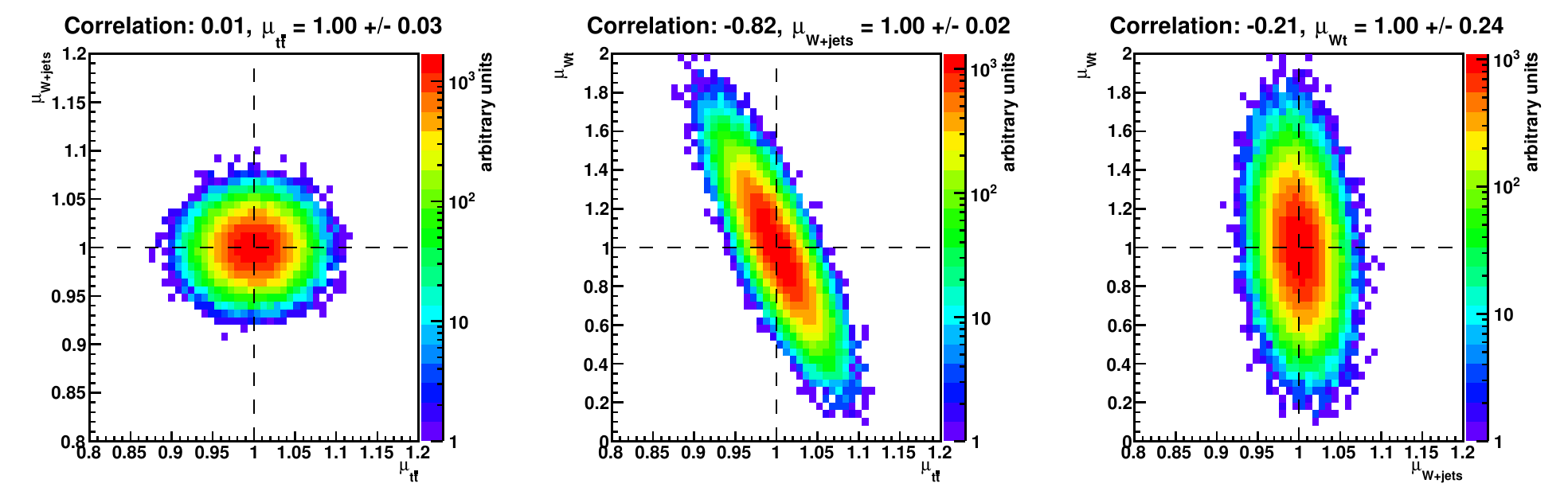}\\
\includegraphics[width=0.95\textwidth]{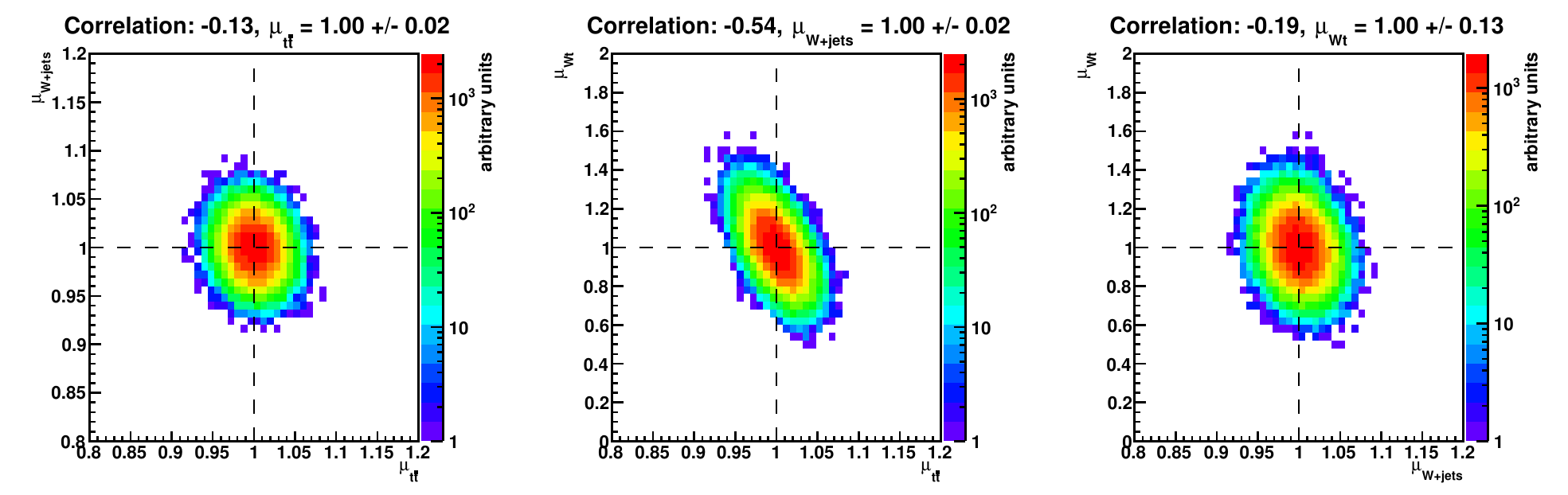}
\includegraphics[width=0.95\textwidth]{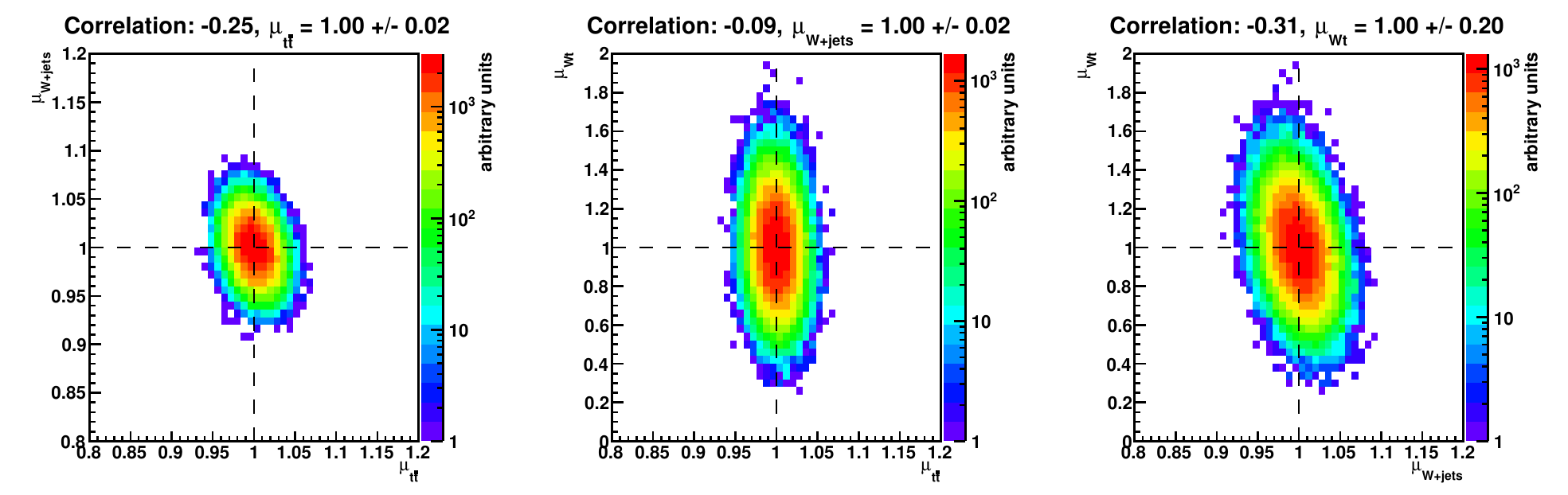}
 \caption{Statistical correlations between the various normalization factors determined from bootstrapping.  The data yields are set to the prediction so that the scale factors are centered at $\mu=1$.  In the middle row, the $t\bar{t}$ contribution to the single top enriched region is set to zero and in the bottom row the single top contribution to the $t\bar{t}$ enriched region is set to zero.}
 \label{fig:scalefactorcorrelations}
  \end{center}
\end{figure}

The results from the full control region fit used to extract the normalization factors from the control regions described earlier in this chapter are described in Chapter~\ref{chapter:results}.  First, Chapter~\ref{chapter:uncertainites} quantifies the accuracy of the transfer factors and MC background estimates with a full assessment of systematic uncertainties. 
 	\chapter{Systematic Uncertainties}
	\label{chapter:uncertainites}		
	
	The background estimation procedures described in Sec.~\ref{chapter:background} are only useful if the precision and accuracy are known.  The precision is set by the various sources of statistical uncertainty, including the MC statistical uncertainty and the data statistical uncertainty in both the control and signal regions.   The MC statistical uncertainty can be reduced by running larger simulations and the uncertainty from the finite control region statistics can be reduced by loosening the selection, at the cost of a larger extrapolation to the signal region.  This chapter describes a variety of techniques that are used to estimate potential sources of systematic bias impacting the accuracy of the background estimates.  For signal regions with harsh selections, the systematic uncertainties are largely subdominant to the uncertainty from the data statistical uncertainty.  This is illustrated quantitatively in Sec.~\ref{fig:tradeoffstatsyst} with a simple one-bin region.  As long as the systematic uncertainty is below the data (Poisson) statistical uncertainty, there is little impact on the sensitivity.   However, for the looser signal regions selections (in particular for the shape fits), systematic uncertainties can have a significant impact on the sensitivity to stops.
	
\begin{figure}[h!]
\begin{center}
\includegraphics[width=0.45\textwidth]{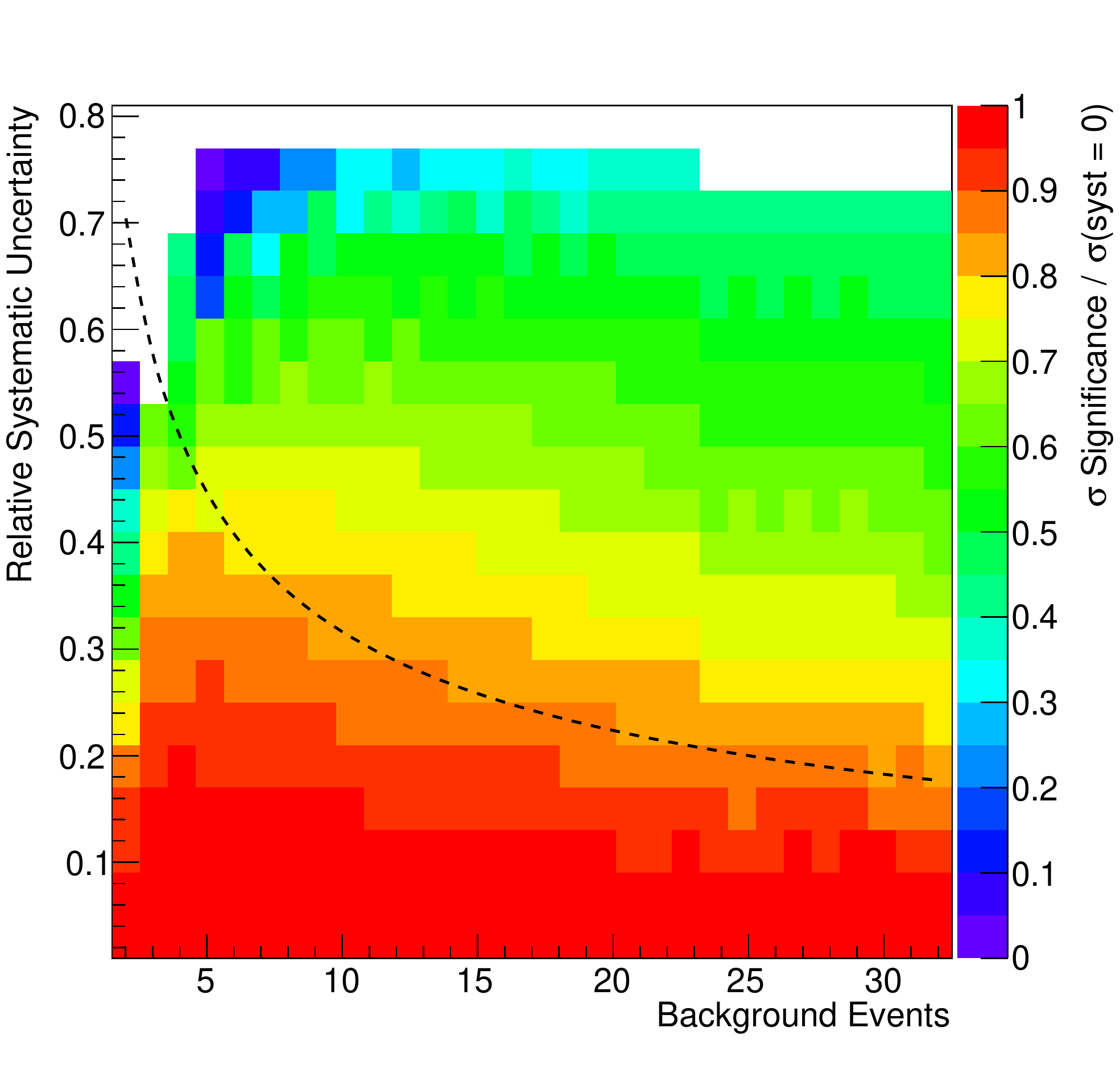}\hspace{10mm}\includegraphics[width=0.45\textwidth]{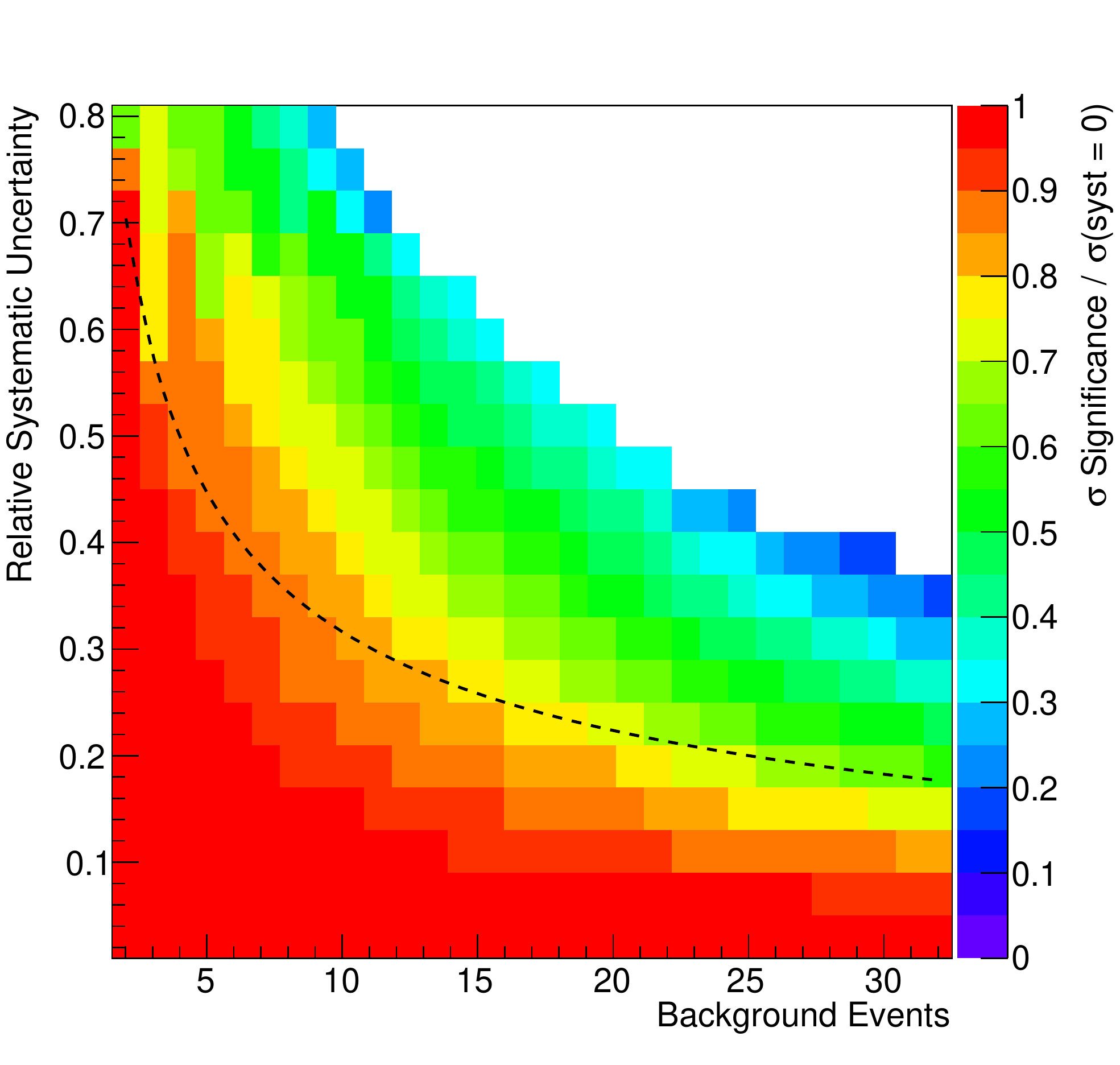}
 \caption{The normalized significance as a function of the number of background events and the fractional systematic uncertainty for a one-bin fit with log-normal uncertainties and no profiling.  The `significance' is computed as $\Phi^{-1}(1-\text{CL}_s)$ for $\Phi$ the cumulative distribution function of the normal distribution.  The number of signal events is set to the number of background events (10 events) in the left (right) plot and each column is normalized to be one in the first row.  The dashed line represents the data statistical uncertainty: $1/\sqrt{\text{Background Events}}$.}
 \label{fig:tradeoffstatsyst}
  \end{center}
\end{figure}	
	
	There are two sources of potential bias: experimental systematic uncertainties and theoretical modeling uncertainties.  The experimental uncertainties are related to the reconstructed object efficiencies and resolutions.  They are constrained in dedicated auxiliary studies and then the impact on this search is estimated by varying some aspect of the simulation, such as the event weight or per-object kinematic quantities.  Generally these uncertainties are constructed to cover differences between data and simulation in the auxiliary studies.  In contrast, the theoretical modeling uncertainties are largely unconstrained by data since they are used to estimate the background predictions in extreme kinematic regimes.  These regions of phase space by construction have little or no data and therefore the modeling uncertainties are constructed to cover all reasonable variations in the simulation.
	
	This chapter is organized as follows.  The experimental systematic uncertainties and their impact on the signal region predictions are described in Sec.~\ref{experimentaluncerts}.  Section~\ref{SUSY:modelinguncerts} documents the procedures for the theoretical modeling uncertainties, including several that are unique to this analysis.  An overview of all the uncertainties for the various stages of the search is presented in Sec.~\ref{sec:susy:summary:uncerts}.
	
	\clearpage
	
	\section{Experimental Systematic Uncertainties}
	\label{experimentaluncerts}
	
Each of the reconstructed objects have an associated uncertainty on their energy scale, energy resolution, and reconstruction efficiency.  In addition, there is an uncertainty on the modeling of various global properties such as pileup and the instantaneous luminosity.   Even though a diverse set of reconstructed objects is used to construct the signal regions, the total measured energy scale and energy resolution are dominated by jets.  Furthermore, due to the complexity and the lack of a conceptually and experimentally clean resonance constraint for jets, the uncertainty on the jet energy scale and resolution are significantly larger than for other reconstructed objects.  A comparison of the per-object systematic uncertainties is presented in Table~\ref{tab:objectcomparison}.   The jet energy resolution decreases with $p_\text{T}$ because of the Poisson nature of the calorimeter energy resolution while the $p_\text{T}$ resolution of track-based objects increases with $p_\text{T}$.  The uncertainties on jet properties decrease with $p_\text{T}$ due in part to the sub-dominance of pileup and other effects.  In contrast, the uncertainties on electrons, photons, muons, and taus increases with $p_\text{T}$ due to the limited availability of resonance decays in data.  The resolution and uncertainty on the $E_\text{T}^\text{miss}$ soft term can be significant, but the presence of neutrinos and real missing momentum renders the soft term largely irrelevant even for the more inclusive event selections presented in this analysis.  The impact of the per-object and per-event uncertainties on the analysis are detailed in the following sections.  Sections~\ref{sec:JESuncert} and~\ref{sec:susy:jer} cover the jet energy scale and resolution uncertainties.  Uncertainties associated with $b$-tagging and the $E_\text{T}^\text{miss}$ are described in sections~\ref{sec:susy:btag} and~\ref{sec:susy:met}.  All other (minor) uncertainties are summarized in Sec.~\ref{susy:exp:other}.  An overview of the impact of the leading experimental systematic uncertainties is shown in Fig.~\ref{fig:experimentalsystoverview}, using SR13 as an example.  Normalization factors are extracted using the same technique described in Sec.~\ref{overview}, expanded to a fourth equation (the $t\bar{t}+\gamma$ CR) and a fourth unknown $(\mu_{t\bar{t}+Z})$.  The $t\bar{t}+Z$ and $t\bar{t}+\gamma$ normalization factors are set equal by construction and the diboson contribution is determined from simulation and subtracted from the data before solving the equations.  The resulting normalization factors are the same as the simultaneous fit described in Sec.~\ref{sec:susy:stats}.  After multiplying each component by its respective normalization factor, the impact on the total background prediction is about $10\%$, indicated by the difference between the last and penultimate rows in Fig.~\ref{fig:experimentalsystoverview}.  The statistical uncertainty from the control regions results in a $10\%$ systematic uncertainty on the total background prediction (error band on the {\it CR stats} row).  In general, the uncertainties are reduced using the control region constraint (black versus red error bands) and the dominant systematic uncertainties are from the jet energy scale and jet energy resolution uncertainty.

 \begin{table}[h!]
\begin{center}
\noindent\adjustbox{max width=\textwidth}{
\begin{tabular}{| c |cc|ccc|}
\hline
Object  & $\sigma$ & $\epsilon$ & $\mu$ uncertainty & $\sigma$ uncertainty & $\epsilon$ uncertainty \\
     \hline
    \hline
 Jets & {\color{red}20\%}/{\color{blue}8\%}~\cite{ATLAS-CONF-2015-037} &  {\color{red}92\%}/{\color{blue}100\%}~\cite{Aad:2015ina,ATLAS-CONF-2015-029} & {\color{red}4\%}/{\color{blue}1\%}~\cite{ATLAS-CONF-2015-037} & {\color{red}2\%}/{\color{blue}0.5\%}~\cite{ATLAS-CONF-2015-037} &  {\color{red}2\%}/{\color{blue}0\%}~\cite{Aad:2015ina}  \\  $b$-jets~\cite{Aad:2015ydr} & -- & {\color{red}60\%}/{\color{blue}80\%} & --&  -- & {\color{red}5\%}/{\color{blue}3\%}\\
 Electrons & {\color{red}0.025\%}/{\color{blue}0.012\%}~\cite{Aad:2014nim} &  {\color{red}90\%}/{\color{blue}97\%}~\cite{ATLAS-CONF-2014-032} & {\color{red}0.1\%}/{\color{blue}0.3\%}~\cite{Aad:2014nim} & {\color{red}5\%}/{\color{blue}17\%}~\cite{Aad:2014nim} & {\color{red}3\%}/{\color{blue}0.5\%}~\cite{ATLAS-CONF-2014-032} \\
 Photons & {\color{red}0.02\%}/{\color{blue}0.01\%}~\cite{Aad:2014nim} &  {\color{red}70\%}/{\color{blue}93\%}~\cite{ATLAS-CONF-2012-123}&{\color{red}0.2\%}/{\color{blue}0.2\%}~\cite{Aad:2014nim} & {\color{red}5\%}/{\color{blue}20\%} ~\cite{Aad:2014nim}& {\color{red}3.5\%}/{\color{blue}0.5\%}~\cite{ATLAS-CONF-2012-123}\\
  Muons~\cite{Aad:2014rra} &  {\color{red}1.5\%}/{\color{blue}2\%} & {\color{red}99\%}/{\color{blue}99\%}& {\color{red}0.06\%}/{\color{blue}0.04\%} & {\color{red}4\%}/{\color{blue}6\%} & {\color{red}0.1\%}/{\color{blue}0.2\%}\\
 Taus~\cite{ATL-PHYS-PUB-2015-045} & {\color{red}22\%}/{\color{blue}6\%} & {\color{red}75\%}/{\color{blue}80\%} & {\color{red}4\%}/{\color{blue}4\%} & -- & {\color{red}8\%}/{\color{blue}5\%}\\ 
   $E_\text{T}^\text{miss}$~\cite{ATL-PHYS-PUB-2015-023}&{\color{red}130\%}/{\color{blue}85\%} & -- & {\color{red}10\%}/{\color{blue}20\%} & {\color{red}20\%}/{\color{blue}40\%} & --\\ 
       \hline
\end{tabular}}
\caption{Performance metrics for the various reconstructed objects and their systematic uncertainties at {\color{red}$p_\text{T}=25$ GeV} (upper number) and {\color{blue}$p_\text{T}=100$ GeV} (lower number).  The symbol $\sigma$ denotes the energy resolution (width / mean), $\mu$ the energy scale, and $\epsilon$ the reconstruction efficiency.  Many of the object reconstruction algorithms changed between Run 1 and Run 2; the selection in this table mixes the defaults between the two runs to give an idea of the overall performance. The jet reconstruction efficiency is based on the {\it loose} quality criteria~\cite{ATLAS-CONF-2015-029} (and refs. therein) which is over 99.9\% efficient and a jet vertex tagger (JVT) threshold of $0.59$, which corresponds to a pileup jet efficiency of just over $1\%$~\cite{Aad:2015ina}.  The $b$-tagging efficiency is determined using the $70\%$ working point of the MV1 algorithm~\cite{Aad:2015ydr}.  Combined (CB) + segment tagged (ST) muons~\cite{Aad:2014rra} are considered without any further quality criteria and the scale and resolution (uncertainties) are on $m_{\mu\mu}$ from various resonance decays.  Electrons are reconstructed with the {\it loose} criteria.  Unconverted {\it tight} photons~\cite{ATLAS-CONF-2012-123} are used for photon reconstruction properties.  All numbers refer to central objects only (except for $E_\text{T}^\text{miss}$, which uses objects across $\eta$).  Medium one-prong tau reconstruction at $\sqrt{s}=13$ TeV is used for illustration~\cite{ATL-PHYS-PUB-2015-045}.  The last line indicates the properties of the component of the track-based soft term parallel to $p_\text{T}^\text{hard}$~\cite{ATL-PHYS-PUB-2015-023} (and refs. therein). }
  \label{tab:objectcomparison}
\end{center}
\end{table}

\begin{figure}[h!]
\begin{center}
\includegraphics[width=0.9\textwidth]{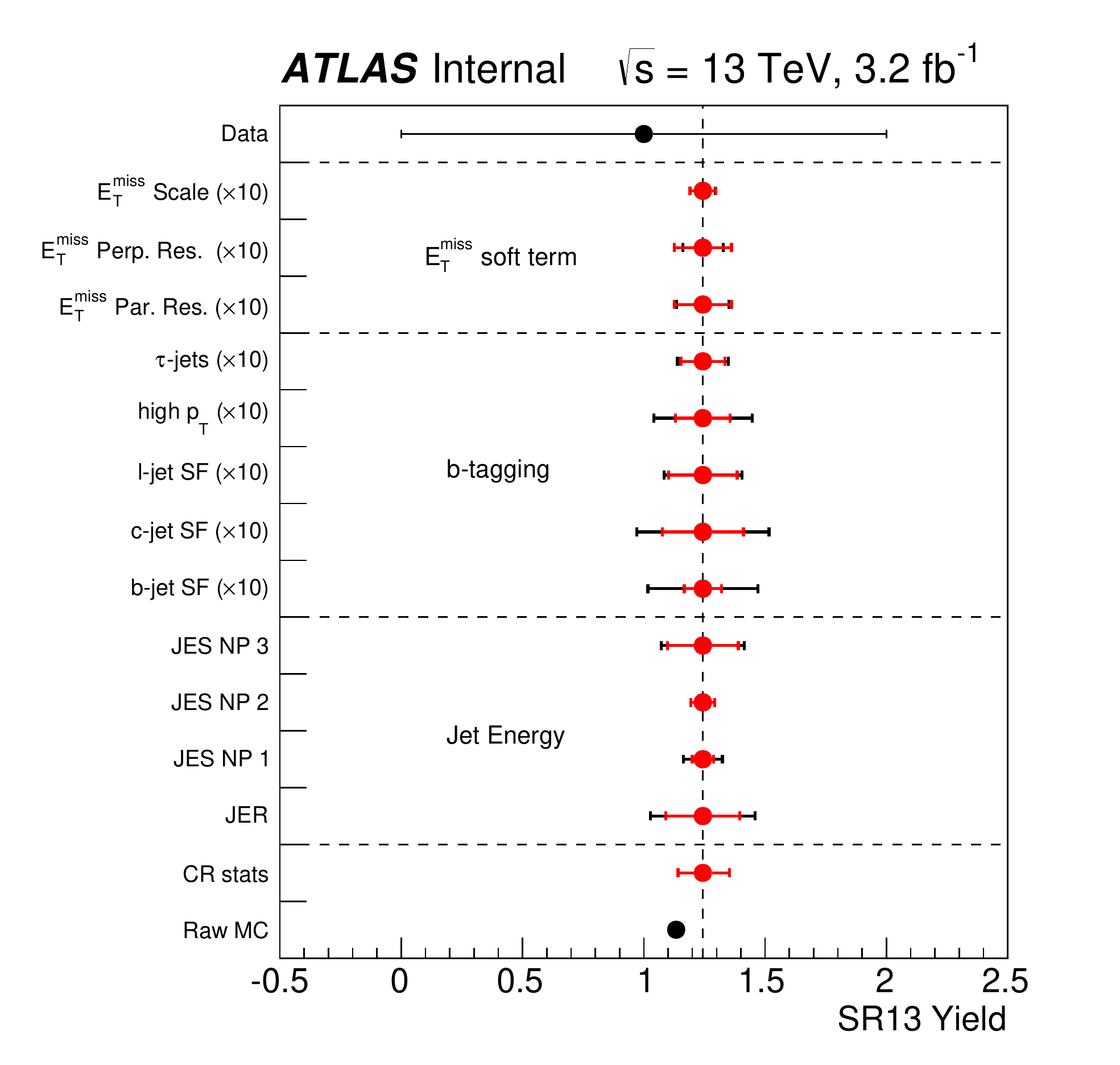}
 \caption{An overview of the impact of the experimental uncertainties on the predicted yield in SR13.  The Raw MC prediction does not include any data constraints from the control regions.  The vertical dashed line indicates the nominal SM prediction after normalizing the simulation to the data in each of the four control regions described in Chapter~\ref{chapter:background}.  The black error bands show the size of the uncertainties before the control region constraint and the red bands show the error after using the control region method.  Sections~\ref{sec:JESuncert},~\ref{sec:susy:jer},~\ref{sec:susy:btag}, and~\ref{sec:susy:met} describe how the various uncertainties are applied.  To make them visible on the plot, the $b$-tagging and $E_\text{T}^\text{miss}$ soft term uncertainties are multiplied by $10$.  The data and CR stats error bands represent the symmetric $68\%$ inter-quantile range centered about the median determined from bootstrapping the data.}
 \label{fig:experimentalsystoverview}
  \end{center}
\end{figure}	

\clearpage
	
	\subsection{Jet Energy Scale}
	\label{sec:JESuncert}
	
	The jet energy scale (JES) and its uncertainty at $\sqrt{s}=8$ TeV are described in Sec.~\ref{sec:calojet}.  A similar procedure is used to calibrate jets and determine the systematic uncertainty at $\sqrt{s}=13$ TeV~\cite{ATL-PHYS-PUB-2015-015}.   For Run 1, the default jet calibration is LCW+JES while in Run 2, the local cluster weighting is not used by default for $R=0.4$ jets.  At the beginning of Run 1, the EM+JES scheme had significantly larger uncertainties than the LCW+JES scheme at low jet $p_\text{T}$ due to the dependence on the quark/gluon composition~\cite{Aad:2014bia}.  However, with the global sequential calibration (GSC) add-on to the calibration procedure, the flavor dependence is significantly reduced (in part because of $n_\text{track}$ - see Chapter~\ref{cha:multiplicity}) and therefore the uncertainties for the EM+JES scheme in Run 2 are only slightly larger than those with the LCW+JES scheme~\cite{ATLAS-CONF-2015-037}.  The full JES uncertainty has many components that could each be included in the simultaneous background fit as an independent nuisance parameter.   The $\mathcal{O}(10)$ parameters introduced in Sec.~\ref{sec:calojet} is already a reduction from the full $\mathcal{O}(100)$ parameters due to all the auxiliary in-situ measurements~\cite{ATL-PHYS-PUB-2015-014}.  In most of the parameter space probed by the stop search, the fit is not sensitive to the intricate details of the JES uncertainty correlations.  Therefore, the Run 1 single-bin regions use a single nuisance parameter to capture the total jet energy scale uncertainty.   This uncertainty is largely independent of $\eta$ and is about $1\%$ for $100$ GeV $<p_\text{T}<1$ TeV.   Below $100$ GeV, the uncertainty grows to about $4\%$ at $25$ GeV and above $1$ TeV, there is an increase in the uncertainty to about $3\%$ due to a change in the uncertainty method (lack of statistics for an in-situ constraint).   The early $\sqrt{s}=13$ TeV analysis uses three nuisance parameters, which capture most of the relevant correlations~\cite{ATL-PHYS-PUB-2015-014} while the shape fit region at $\sqrt{s}=8$ TeV uses the same $17$-parameter setup as in in Sec.~\ref{sec:calojet} because the fit is over-constrained (more bins than normalization parameters) leading to the potential for a reduction in the uncertainty from the fit ({\it profiling}) and the increased importance of the systematic uncertainties for the more inclusive selection.  The impact of the JES uncertainty on the shape fit is revisited at the end of this section.
		
	The JES uncertainty impacts the analysis directly through an uncertainty in the acceptance and indirectly by changing high level variables that depend on  jet $p_\text{T}$.  Figure~\ref{fig:susy:jes:njets} shows the jet multiplicity after the preselection at $\sqrt{s}=13$ TeV for $t\bar{t}$ events.  The JES nuisance parameter with the largst variation is shown for illustration.  There is a $\pm 10\%$ impact on the total number of events in the four-jet bin, which is the JES-induced uncertainty on the acceptance.  The residual impact on the shape of the $n_\text{jets}$ distribution (middle plot) is less pronounced.   A similar trend is observed for the leading jet $p_\text{T}$ in Fig.~\ref{fig:susy:jes:jetpt}, where the JES uncertainty is at the percent-level for jets beyond the peak of the distribution at $\sim 200$ GeV.  The JES uncertainty induces a jet mass scale uncertainty for large-radius reclustered jets.  Figure~\ref{fig:susy:reclusteredjetmass} showed the reclustered jet mass in data after the preselection.  The error band is dominated by the JES uncertainty, but seems strangely asymmetric.  A similar trend is observed in the jet mass distribution in Fig.~\ref{fig:wwmass}.  This is a general trend for resonance peaks and the reason is illustrated by Fig.~\ref{fig:susy:jes:jetmass}.  The middle plot of Fig.~\ref{fig:susy:jes:jetmass} looks as expected for nearly symmetric variations in the JES: the peak position and width are decreased when the JES is reduced and vice versa when the JES is increased.  Small bumps in the ratio of the right plot of Fig.~\ref{fig:susy:jes:jetmass} clearly illustrate the importance of these changes around the resonance peak.  The reason for the asymmetric uncertainty band in earlier figures is because of the change in acceptance in addition to the change in the shape.  When the JES is shifted down, there are also fewer events that pass the selection.  Therefore, the red distribution in the middle plot is scaled down in the left plot of Fig.~\ref{fig:susy:jes:jetmass} and coincidentally is on top of the nominal histogram, which makes the overall uncertainty seem small.  In contrast, when the JES is shifted up, there are more jets that pass the event selection and so the blue histogram in the middle plot is shifted up in the left plot of Fig.~\ref{fig:susy:jes:jetmass}, leading to the large uncertainty only on the right side of the peak.
	
\begin{figure}[h!]
\begin{center}
\includegraphics[width=0.95\textwidth]{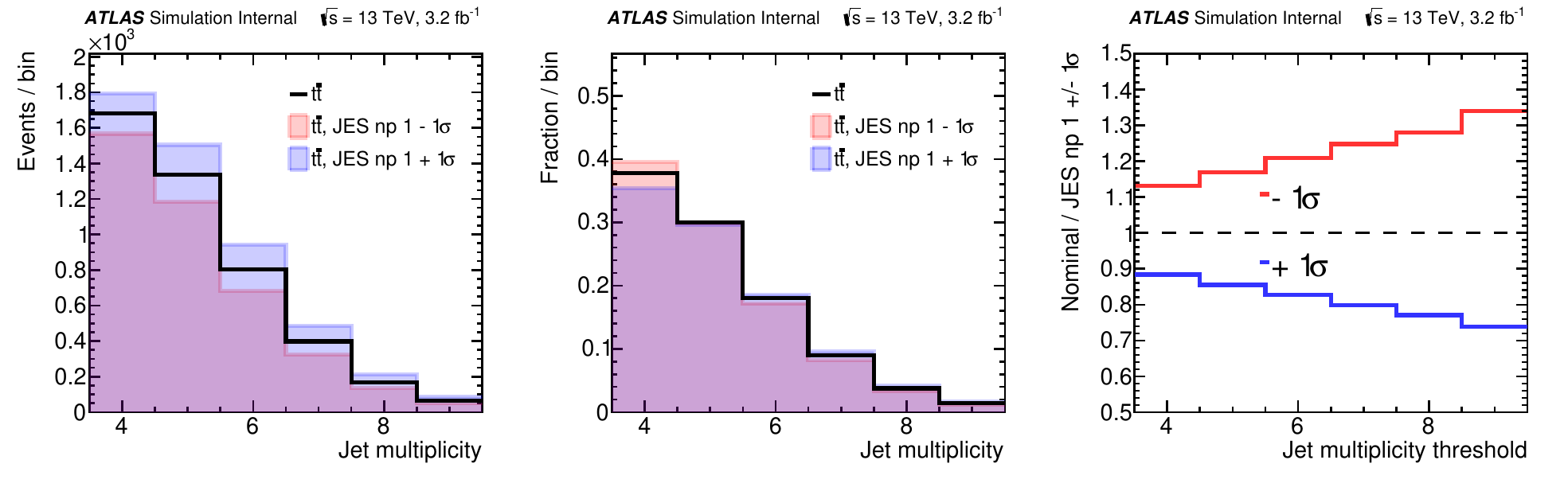}
 \caption{The signal jet multiplicity in $t\bar{t}$ events after the preselection before (left) and after (middle) normalizing the distributions to unity.  The ratio of the JES $\pm 1\sigma$ to the nominal in the left plot is shown in the right plot.}
 \label{fig:susy:jes:njets}
  \end{center}
\end{figure}

\begin{figure}[h!]
\begin{center}
\includegraphics[width=0.95\textwidth]{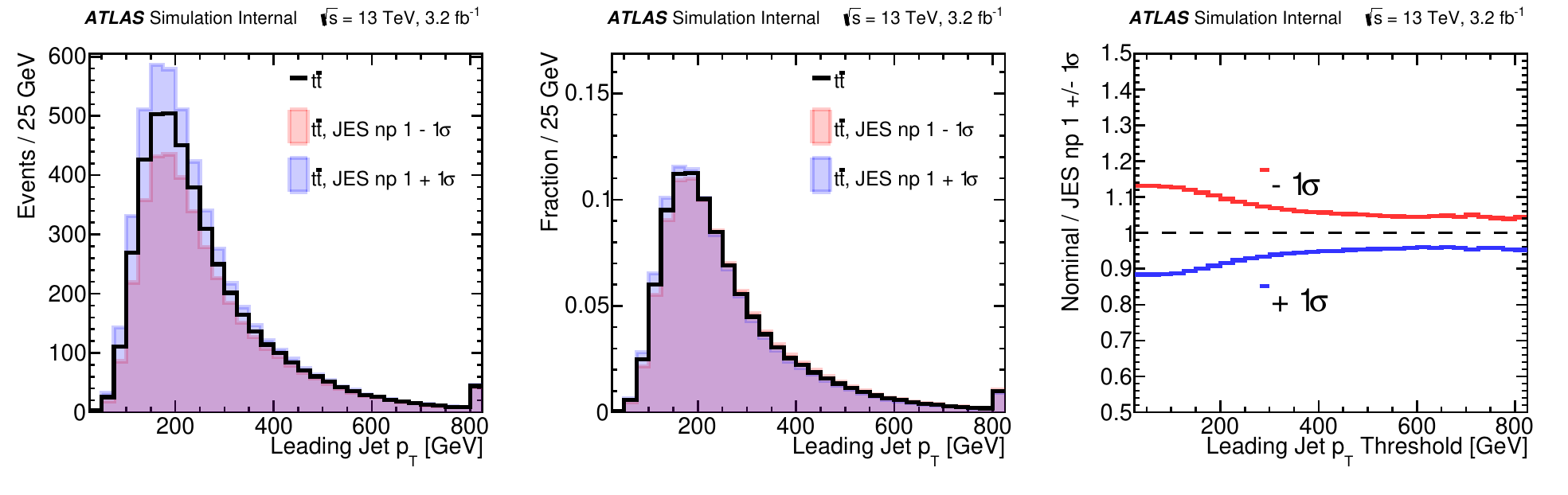}
 \caption{The leading jet $p_\text{T}$ in $t\bar{t}$ events after the preselection before (left) and after (middle) normalizing the distributions to unity.  The ratio of the JES $\pm 1\sigma$ to the nominal in the left plot is shown in the right plot.}
 \label{fig:susy:jes:jetpt}
  \end{center}
\end{figure}

\begin{figure}[h!]
\begin{center}
\includegraphics[width=0.95\textwidth]{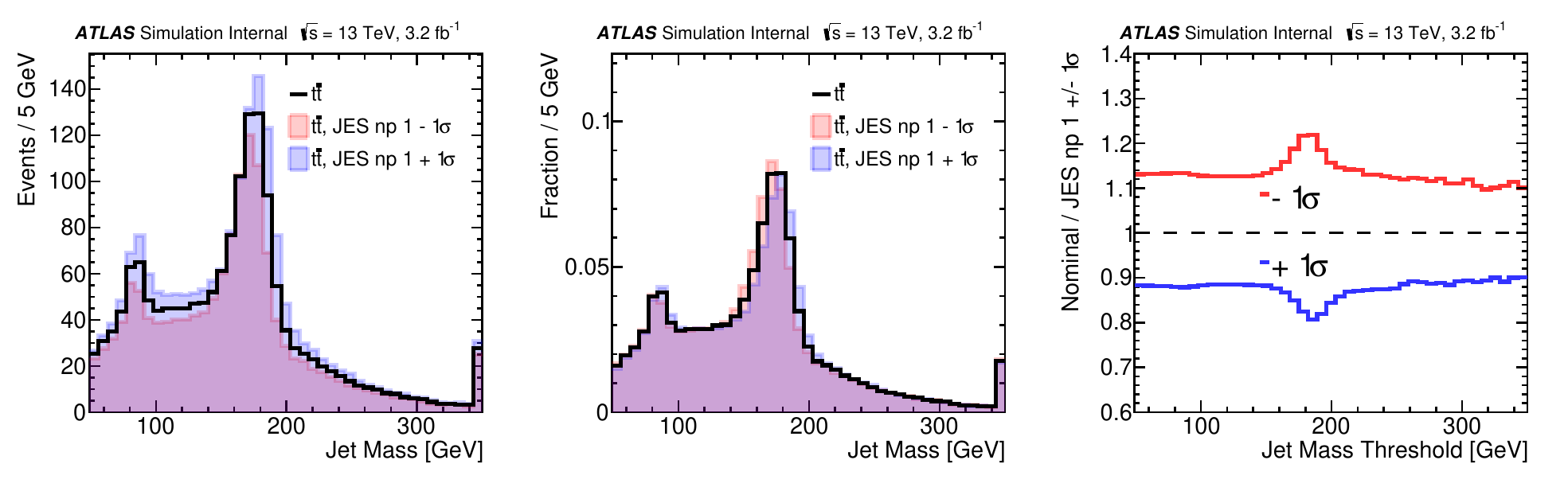}
 \caption{The large-radius $(R=1.2)$ jet mass in $t\bar{t}$ events after the preselection before (left) and after (middle) normalizing the distributions to unity.  The ratio of the JES $\pm 1\sigma$ to the nominal in the left plot is shown in the right plot.}
 \label{fig:susy:jes:jetmass}
  \end{center}
\end{figure}
	
Figure~\ref{fig:experimentalsystoverview} in the previous section showed that a significant fraction of the JES uncertainty cancels from the control region method because the shifts have a similar impact on acceptance in the CR and SR.  This cancellation is demonstrated for the shape fit region in Fig.~\ref{fig:susy:jes:shapefit}.  This normalization reduces the uncertainties from $20$-$30\%$ to $1$-$5\%$ for most bins.  However, even after the reduction, the JES uncertainty is comparable or larger than the data statistical uncertainty.  Therefore, the fit is expected to be sensitive to the JES uncertainty and in particular can constrain it in the analysis phase space.  More details on this profiling are discussed in Sec.~\ref{CR-onlyFit}.

\begin{figure}[h!]
\begin{center}
\includegraphics[width=0.5\textwidth]{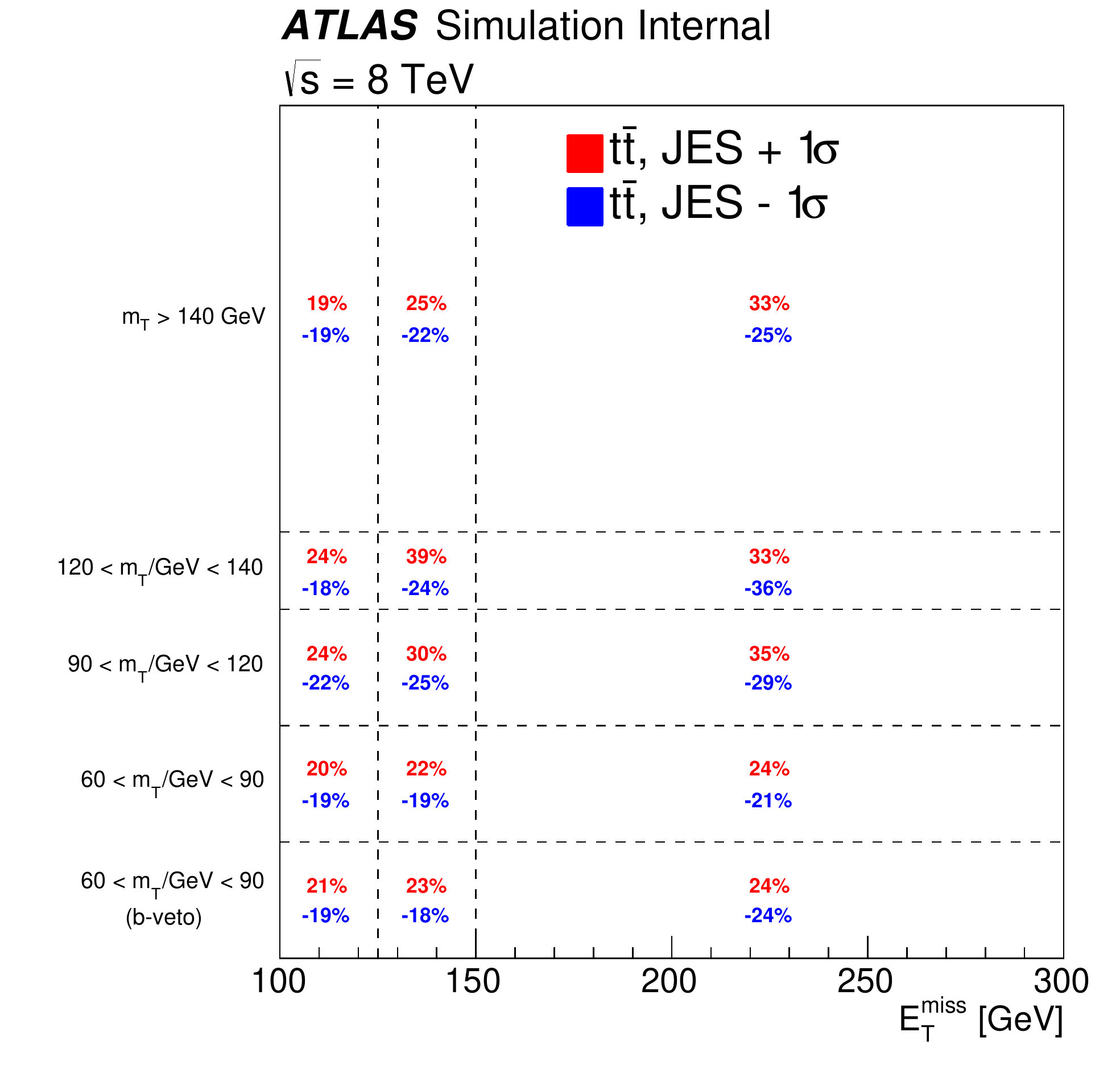}\includegraphics[width=0.5\textwidth]{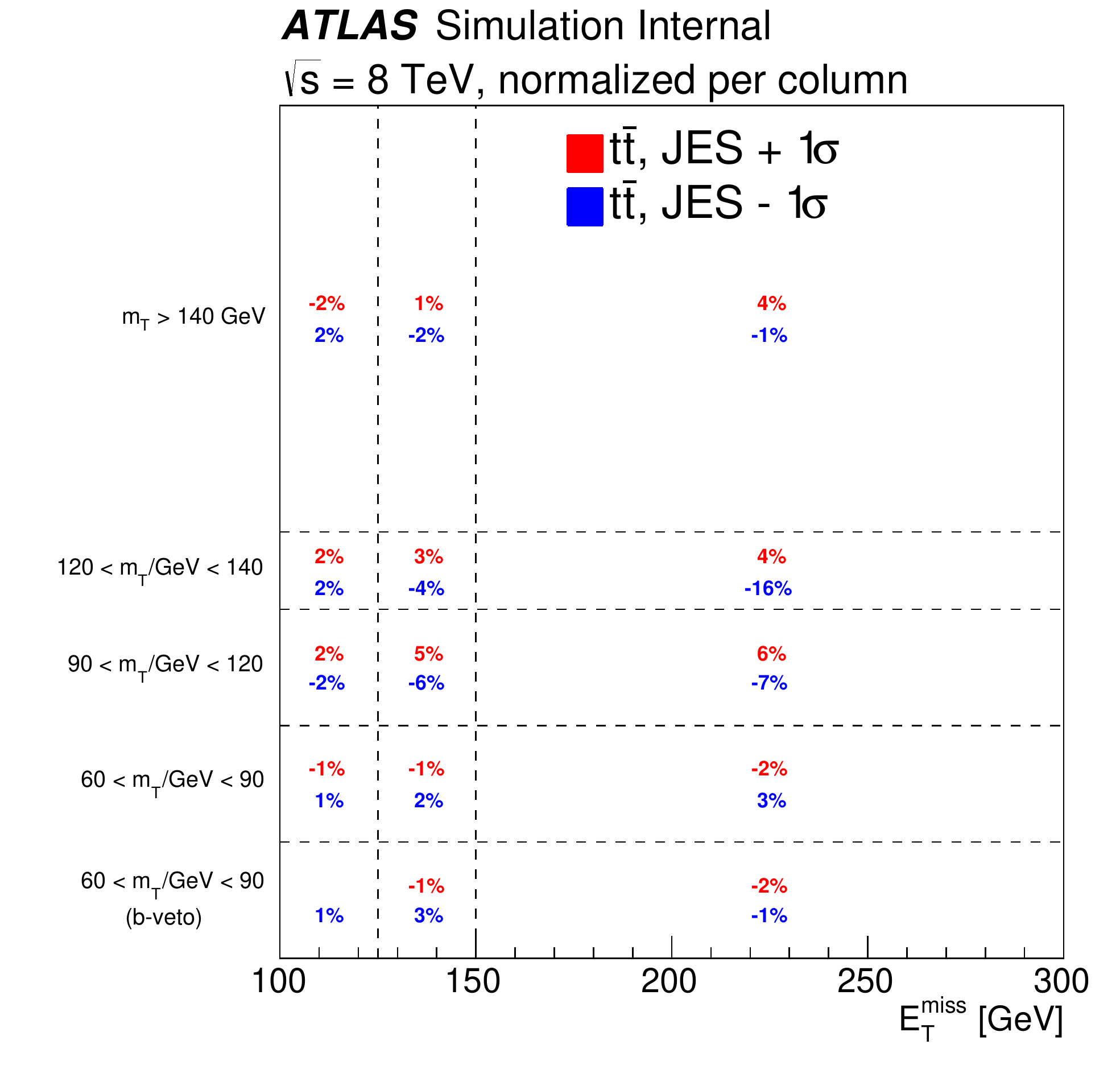}
 \caption{The size of the total JES uncertainty on each bin of the tNshape signal region before (left) and after (right) normalizing the total yields in each $E_\text{T}^\text{miss}$ bin.}
 \label{fig:susy:jes:shapefit}
  \end{center}
\end{figure}	
	
	\clearpage
	
	\subsection{Jet Energy Resolution}
	\label{sec:susy:jer}
	
Similarly to the jet energy scale, the jet energy resolution (JER) uncertainty was introduced already in Sec.~\ref{sec:calojet}.  Due to the complexity in reducing the JER (see Sec.~\ref{sec:JMR:resmethod}), the uncertainty is evaluated simply by inflating the resolution and then symmetrizing the effect on the analysis.  In addition to changes in acceptance, increasing the JER generally broadens peaks (jet mass) and softens edges ($m_\text{T}$ and $m_\text{T2}$).  Figure~\ref{fig:susy:jer:mass} illustrates the broadening of the top quark mass peak in the jet mass spectrum and Fig.~\ref{fig:susy:jer:mt} shows the softening of the $m_\text{T}$ edge when the jet energy resolution is increased.  For high $m_\text{T}$, the increase in acceptance from an increase in the JER by $1\sigma$ is about $15$-$20\%$.

\begin{figure}[h!]
\begin{center}
\includegraphics[width=0.95\textwidth]{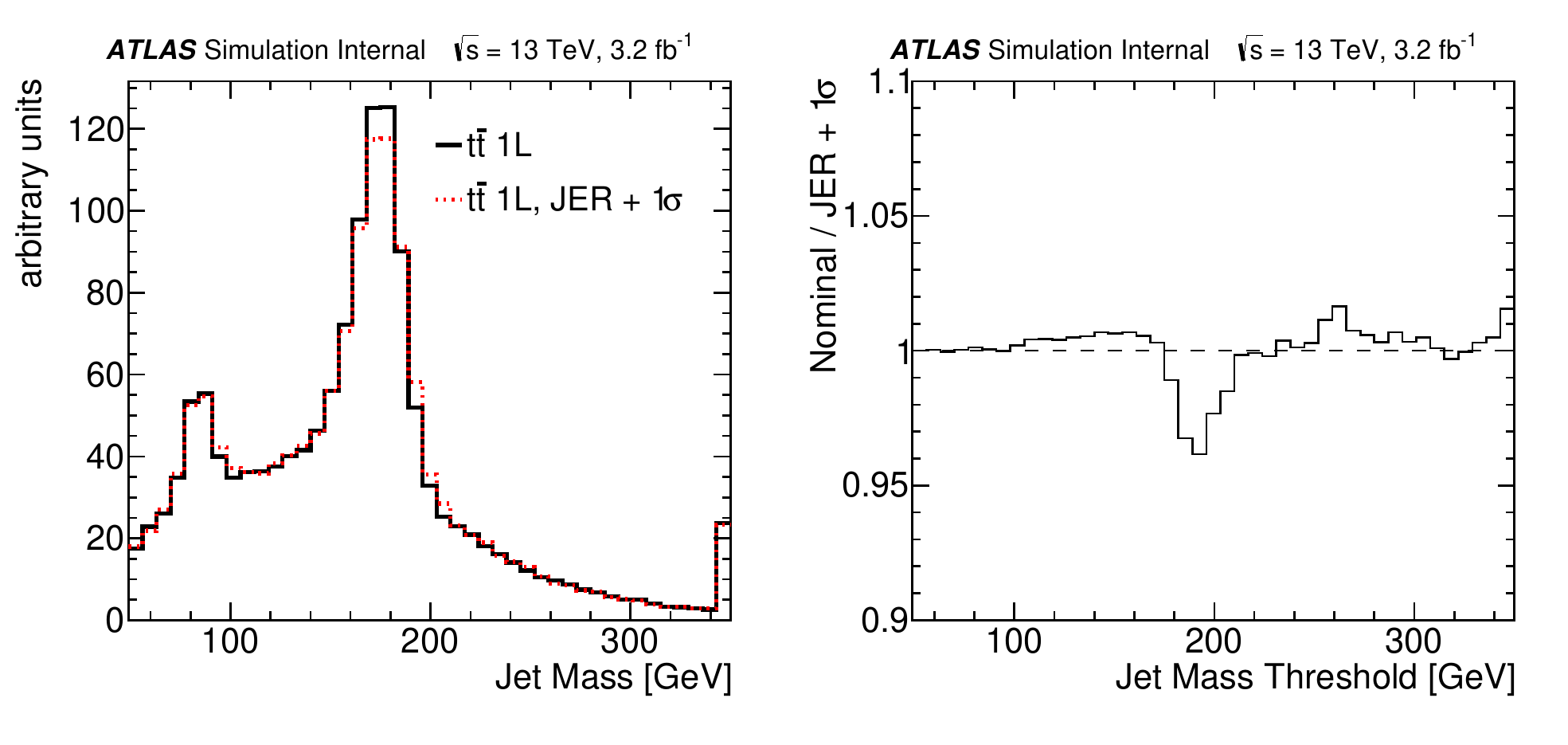}
 \caption{Left: The large-radius $(R=1.2)$ jet mass distribution in $t\bar{t}$ events after the preselection with the nominal JER and the JER inflated within its $1\sigma$ uncertainty.  Right: the ratio of the two histograms in the left plot.}
 \label{fig:susy:jer:mass}
  \end{center}
\end{figure}	
	
\begin{figure}[h!]
\begin{center}
\includegraphics[width=0.95\textwidth]{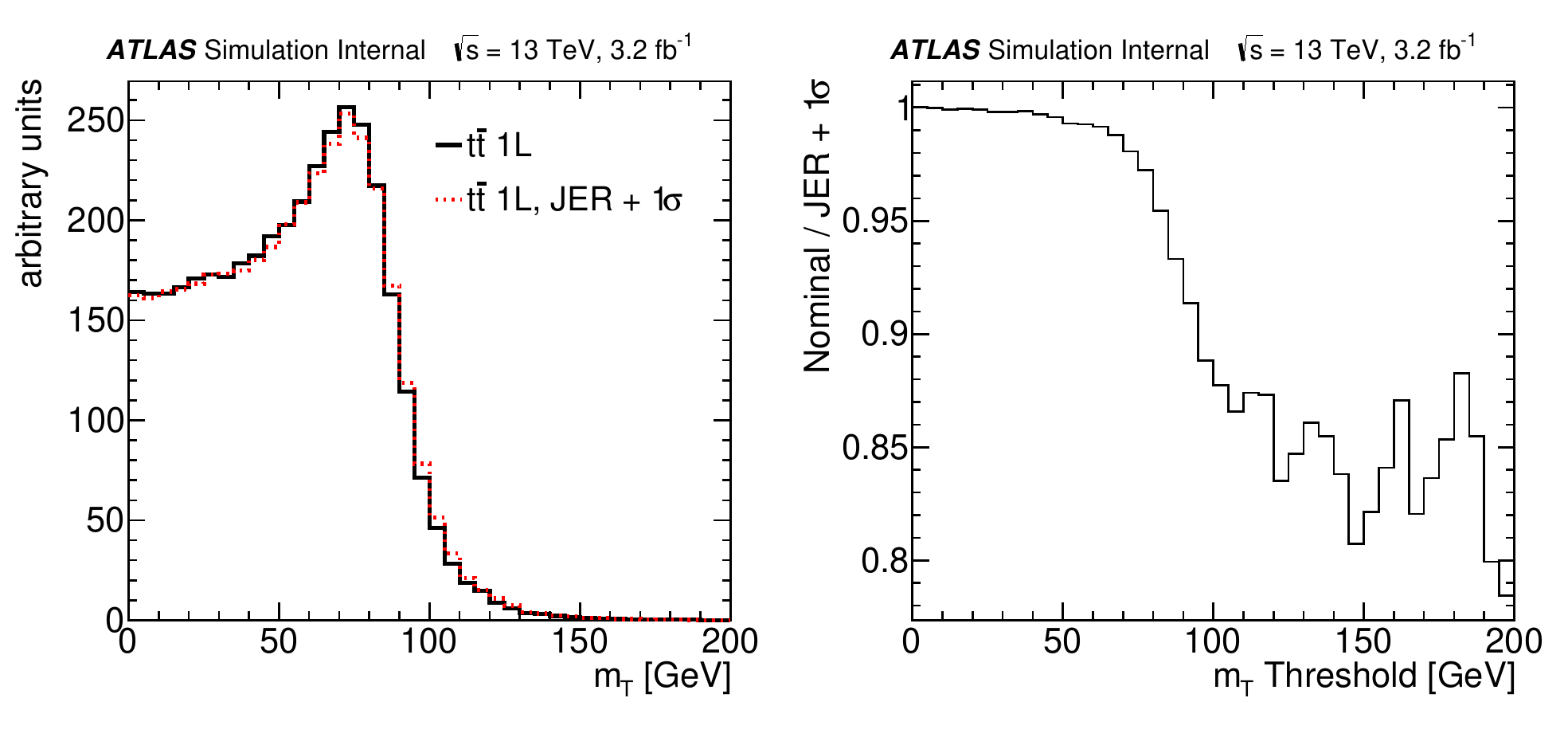}
 \caption{Left: The $m_\text{T}$ distribution in $t\bar{t}$ events after the preselection with the nominal JER and the JER inflated within its $1\sigma$ uncertainty.  Right: the ratio of the two histograms in the left plot.}
 \label{fig:susy:jer:mt}
  \end{center}
\end{figure}	

For all one-bin signal regions, a single JER nuisance parameter is used in the fit.  To allow for more flexibility in the shape fit, there is one JER nuisance parameter per $E_\text{T}^\text{miss}$ bin.  Fig.~\ref{fig:susy:jer:shapefit} shows the change in each bin after increasing the JER by $1\sigma$ before and after normalizing per column.  The reduction in the uncertainty is not as large as for the JES uncertainty because of the qualitatively different impact of JER in the CR-like regions and SR-like bins.
	
\begin{figure}[h!]
\begin{center}
\includegraphics[width=0.5\textwidth]{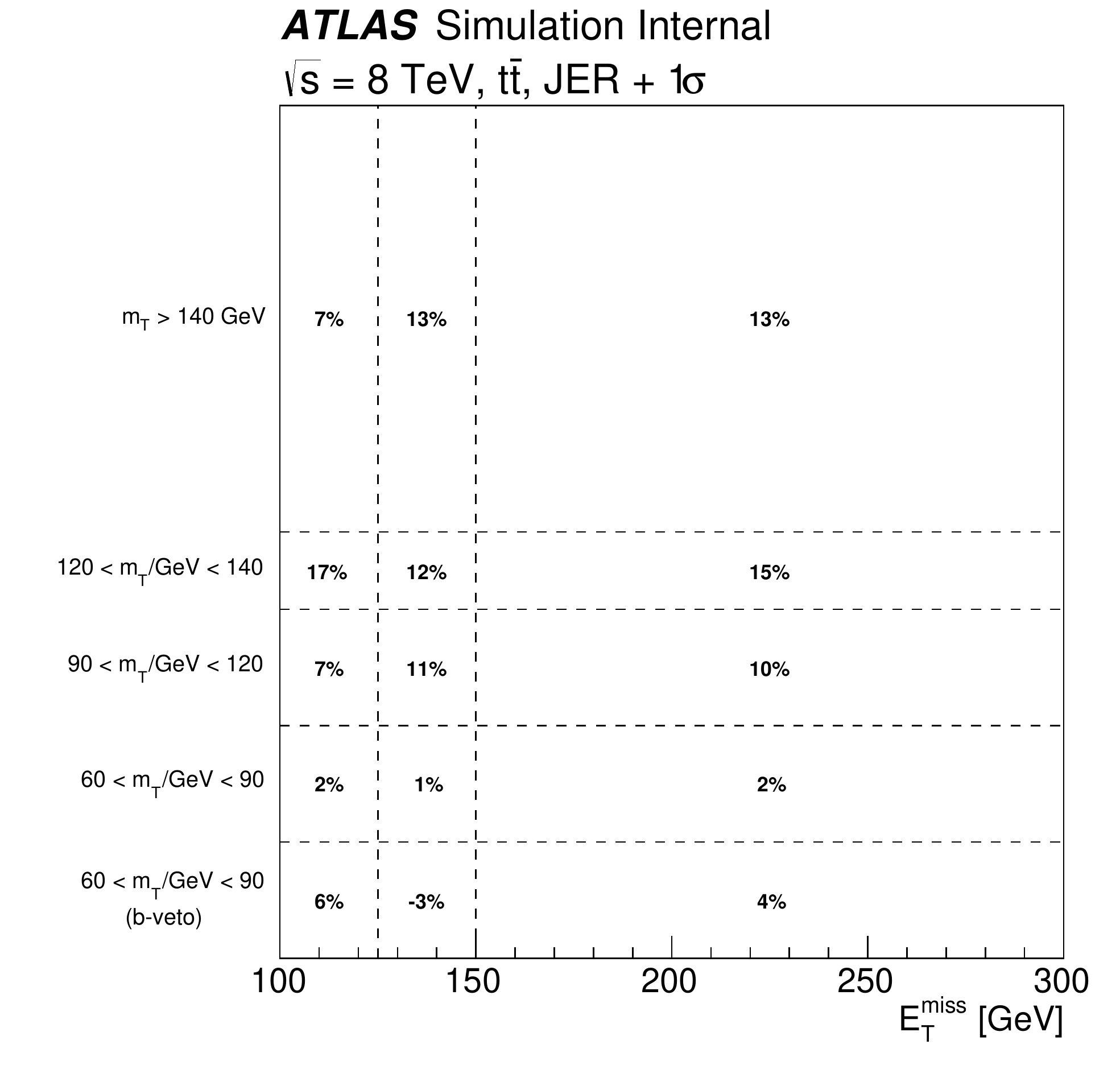}\includegraphics[width=0.5\textwidth]{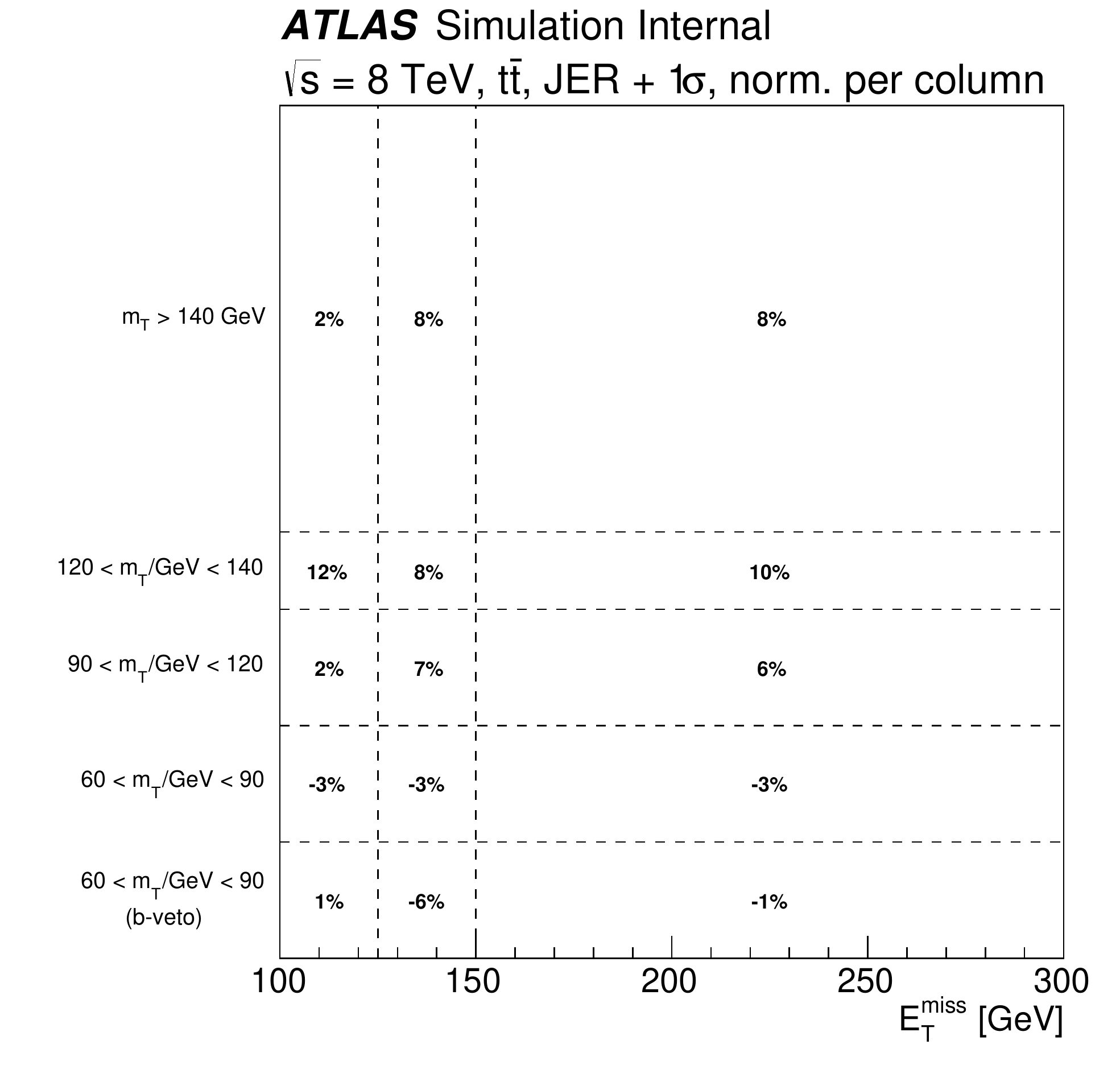}
 \caption{The size of the total JER uncertainty on each bin of the tNshape signal region before (left) and after (right) normalizing the total yields in each $E_\text{T}^\text{miss}$ bin.}
 \label{fig:susy:jer:shapefit}
  \end{center}
\end{figure}	

	\clearpage	
	
	\subsection{$b$-tagging}
	\label{sec:susy:btag}
	
	Jet flavor tagging is used in two ways: directly in the event selection by requiring at least one $b$-tagged jet and indirectly by specifying which jets are used in higher level variable calculations such as $am_\text{T2}$.  Uncertainties related to the second use are highly suppressed with respect to the first.  In order for a bias in the $b$-tagging to impact variable calculations, the relative ordering of jet $b$-tagging weights needs to be permuted.  In contrast, an overall shift in the $b$-tagging weights\footnote{Practically, instead of shifting the weights, the uncertainty is estimated by varying the actual $b$-tagging efficiency.  This is accomplished by applying event weights.} changes the acceptance but leaves the ordering unchanged.  The multi-binned $b$-tagging setup from Sec.~\ref{sec:objs} and its uncertainty correctly account for permutations, but since this is a subleading effect for the stop search, only the overall changes in acceptance are considered for $b$-tagging uncertainties.
	
	The uncertainty on the $b$-tagging efficiency is estimated by comparing data and simulation in auxiliary measurements as described in Sec.~\ref{sec:systs}.  As with the JES uncertainty, the $b$-tagging efficiency uncertainty has many components.  The high mass stop search is not sensitive to the intricate correlation between the many nuisance parameters and therefore a reduced set is used.  Figure~\ref{fig:susybtagginguncert} shows the uncertainty on the number of $b$-tagged jets by varying the $b$-jet\footnote{This is an unfortunate but standard nomenclature: $b$-jets are jets originating from $b$-quarks, in contrast to $b$-tagged jets (often also called just $b$-jets), which are any jet that is tagged with a $b$-tagging algorithm.} efficiency scale factors within their uncertainties.  The JES uncertainty results in a much larger overall difference in acceptance, but has little impact on the normalized $b$-jet multiplicity distribution.  A comparison of the various $b$-tagging efficiency scale factor uncertainties at $\sqrt{s}=13$ TeV is shown in Fig.~\ref{fig:susybtagginguncert2}.  For low $b$-tagged jet multiplicity, the uncertainty on the $b$-jet efficiency is the largest source of uncertainty.  Since charm and light jets make a significant contribution to higher $b$-tagged jet multiplicities (only two $b$-quarks expected in $t\bar{t}$ at tree-level), the uncertainties in their $b$-tagging efficiencies are also important at higher multiplicities. 
	
\begin{figure}[h!]
\begin{center}
\includegraphics[width=0.95\textwidth]{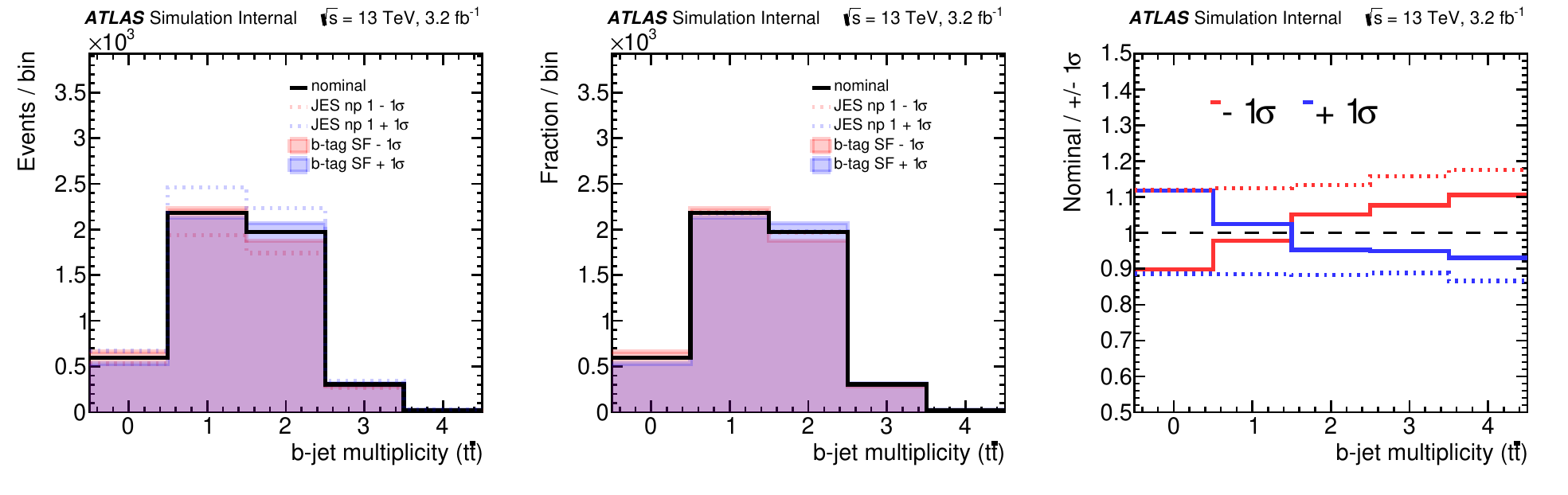}
 \caption{The $b$-jet multiplicity in $t\bar{t}$ events after the preselection before (left) and after (middle) normalizing the distributions to unity.  The ratio of the $\pm 1\sigma$ variations to the nominal in the left plot is shown in the right plot.  The dashed lines in the right plot correspond to the JES uncertainty while the solid lines are the $b$-jet efficiency scale factor uncertainties.}
 \label{fig:susybtagginguncert}
  \end{center}
\end{figure}	

\begin{figure}[h!]
\begin{center}
\includegraphics[width=0.5\textwidth]{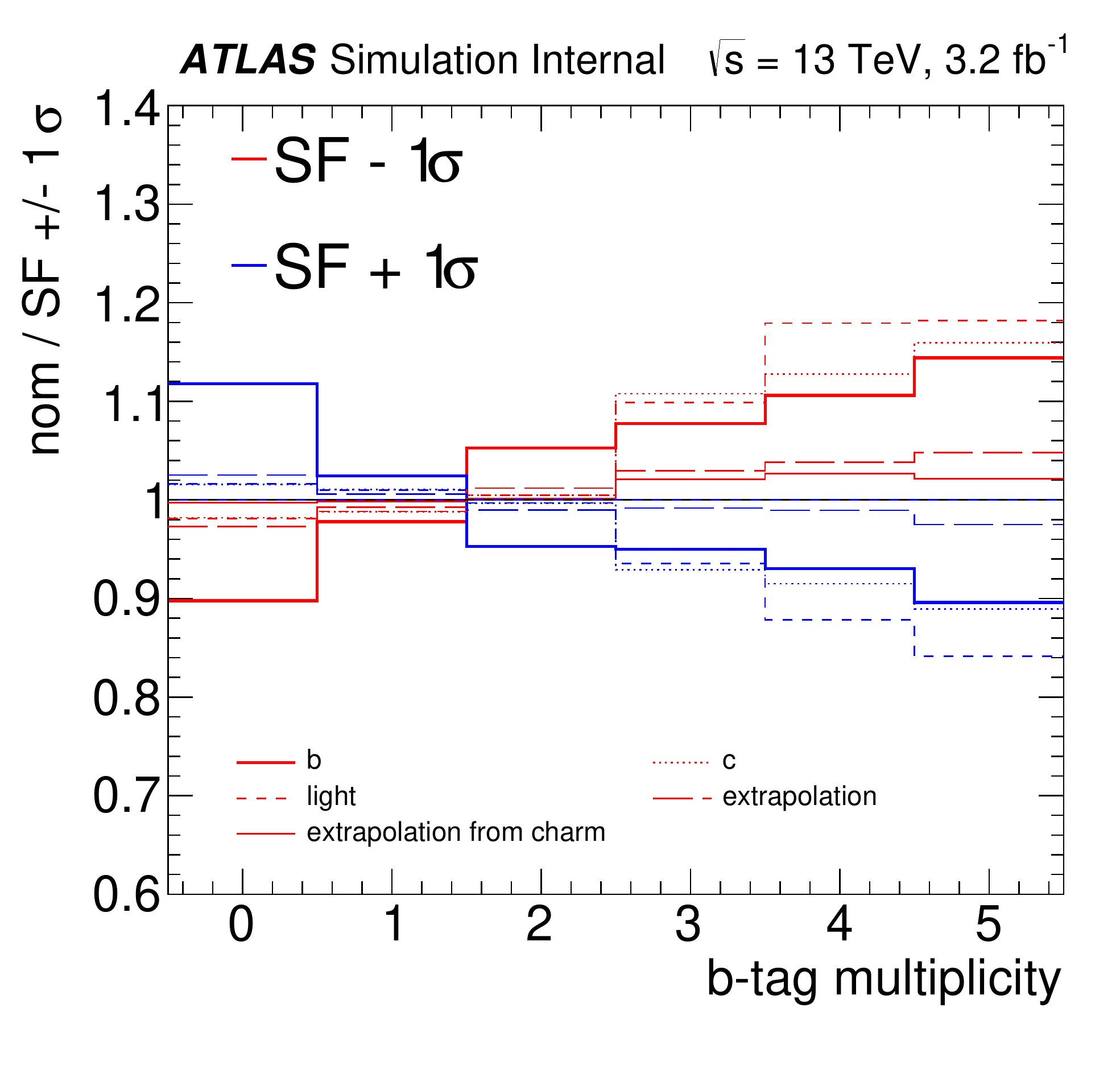} \caption{The relative $b$-tagging efficiency scale factor uncertainty for various components of the uncertainty as a function of the $b$-tagged jet multiplicity.  The `extrapolation from charm' uncertainty is for $\tau$ jets that are $b$-tagged and the `extrapolation' uncertainty is for extending the in-situ constraints to high $p_\text{T}$ where the data statistics are limited.}
 \label{fig:susybtagginguncert2}
  \end{center}
\end{figure}
	
	\clearpage
	
	\subsection{Missing Transverse Momentum}
	\label{sec:susy:met}
	
	The missing transverse momentum is a key input to many of the discriminating variables used in the stop search.  As it is composed of all objects, the uncertainty in each object is coherently propagated to arrive at an uncertainty on the $\vec{p}_\text{T}^\text{miss}$ that is correctly correlated with the input object uncertainties.  The one component that is not accounted for this way is the momentum not associated with any other reconstructed object (soft term).  Uncertainties on the soft term at are estimated using auxiliary studies with $Z\rightarrow \mu\mu$ events in simulation ($\sqrt{s}=13$ TeV~\cite{ATL-PHYS-PUB-2015-023}) and with additional comparisons in data ($\sqrt{s}=8$ TeV~\cite{ATLAS-CONF-2013-082}).  These uncertainties are parameterized based on $\vec{p}_\text{T}^\text{hard}$, which is the sum of all hard objects, including neutrinos (in simulation).  An uncertainty is estimated on the scale of the $E_\text{T}^\text{miss}$ soft term parallel to $\vec{p}_\text{T}^\text{hard}$ and on the resolution parallel and perpendicular to $\vec{p}_\text{T}^\text{hard}$.  Figure~\ref{fig:susyuncertaintyetmiss} illustrates the size of the soft term scale uncertainty at $\sqrt{s}=13$ TeV.  For the high $E_\text{T}^\text{miss}$ probed by the stop search, the impact of the soft term scale uncertainty is small compared with the impact of the JES uncertainty on the $E_\text{T}^\text{miss}$.  The uncertainties on the soft term resolution are comparably small.
	
\begin{figure}[h!]
\begin{center}
\includegraphics[width=0.95\textwidth]{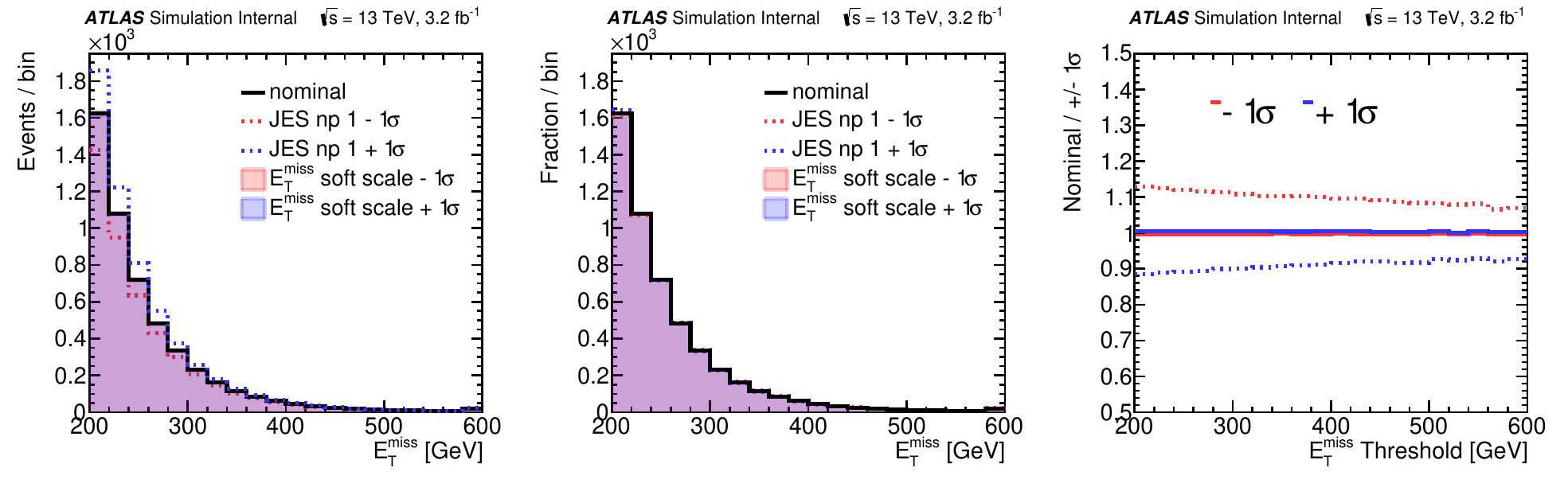} \caption{$E_\text{T}^\text{miss}$ in $t\bar{t}$ events after the preselection before (left) and after (middle) normalizing the distributions to unity.  The ratio of the $\pm 1\sigma$ variations to the nominal in the left plot is shown in the right plot.  The dashed lines in the right plot correspond to the JES uncertainty while the solid lines are the $E_\text{T}^\text{miss}$ soft term scale uncertainties.}
 \label{fig:susyuncertaintyetmiss}
  \end{center}
\end{figure}	

	\clearpage
	
	\subsection{Other}	
	\label{susy:exp:other}		
	
	In addition to the $E_\text{T}^\text{miss}$ soft term systematic uncertainties, there are a series of subdominant uncertainties related to all reconstructed objects as well as general event properties.  The per object uncertainties were summarized in Table~\ref{tab:objectcomparison}.  The electron and muon related energy scale, resolution and reconstruction efficiency uncertainties are all precisely known from resonance decays.  Scale factors from these auxiliary measurements are applied to correct the simulation and the uncertainty on the corrections is a source of systematic uncertainty.  One purpose of the efficiency scale factors is to correct for any mis-modeling of the isolation criteria.  Hadronic tau decays are also constrained from resonance decays ($Z$ boson), but are inherently less clean than for electrons and muons and so the resolutions and uncertainties are generally larger.  However, reconstructed taus are only used as a (highly efficient) veto in this analysis.  Similarly, there are uncertainties associated with mis-modeling in the trigger, but these are suppressed by working in a regime where the trigger is nearly $100\%$ efficient.  The one exception is for the shape fit, where the lowest $E_\text{T}^\text{miss}$ bin is in the trigger turn on region below $E_\text{T}^\text{miss}=200$ GeV.  This could in principle result in significant systematic uncertainties, but because of the control region method, most of the mis-modeling is absorbed into the normalization of the background at low $m_\text{T}$. 
	
	There are also uncertainties on more global properties such as the amount of pileup and the integrated luminosity.  All of the techniques designed for the stop search have some corrections or other protection from the effects of pileup and therefore the impact on a mis-modeling of the pileup spectrum is expected to have a small effect.  This is quantified by reweighting events so that the average number of interactions per bunch crossing varies by $10\%$ ($\sqrt{s}=8$ TeV) or $15\%$ ($\sqrt{s}=13$ TeV).  The luminosity is precisely calibrated and measured using techniques described in Ref.~\cite{Aad:2013ucp}.  The uncertainty in the luminosity for the analysis of $13$ fb${}^{-1}$ of $\sqrt{s}=8$ TeV is 3.6\%, of the full $20.3$ fb${}^{-1}$ of $\sqrt{s}=8$ TeV is 2.8\%, and of the first 3.2 fb${}^{-1}$ at $\sqrt{s}=13$ TeV is $5\%$.  Even though it is has an experimental origin, the luminosity uncertainty is mostly relevant for the theory uncertainties as various background components are normalized by the $\sigma\times \int \mathcal{L} dt$.  A detailed description of the theory modeling uncertainties is in the next section, Sec.~\ref{SUSY:modelinguncerts}.

	\clearpage

	\section{Theoretical Modeling Uncertainties}
	\label{SUSY:modelinguncerts}
	
	In addition to the experimental uncertainties that impact all events, each SM process has an associated uncertainty because the background estimation is performed separately for all the processes.  For the background processes that are estimated using the control region method, the theory modeling uncertainty is associated with the extrapolation from the control region to the signal region.  All other backgrounds have an additional uncertainty on the inclusive cross-section.  Unlike the experimental uncertainties, the theory modeling uncertainties are not usually determined from the difference between data and simulation in auxiliary measurements.  By construction, the regions of phase space probed by the search have little or no overlap with previous measurements and so the uncertainties are derived entirely from a complete set of reasonable variations in the simulation.  There is no unique way to compute these uncertainties.  One way to build a reasonable set of uncertainties for a given process is to decompose the total uncertainty into categories that probe different aspects of the simulation.    Such a decomposition might look like the one in Table~\ref{systematictheoryuncerts}.  This decomposition provides a quantitative procedure that probes nearly\footnote{For example, varying the factorization and renormalization scales changes both the inclusive and differential cross section.  However, the inclusive cross-section is usually known with much higher precision than the differential one.} independent sources of uncertainty arising from fixed order calculations and phenomenological models describing non-perturbative effects.  Individual processes may have additional sources of theoretical modeling uncertainties.   Sections~\ref{sec:susy:syst:ttbar},~\ref{sec:singletopuncerts},~\ref{sec:susy:ttzuncert},~\ref{sec:susy:wjetsuncert}, and~\ref{sec:susy:VVuncert} describe the application of Table~\ref{systematictheoryuncerts} as well as any additional uncertainties for the $t\bar{t}$, single top, $t\bar{t}+V$, $W$+jets, and dibosons processes, respectively.
	
	  \begin{table}[h!]
\centering
\label{my-label}
\noindent\adjustbox{max width=\textwidth}{
\begin{tabular}{|c|c|}
\hline
Source & Procedure \\
\hline
Inclusive cross-section & Uncertainty of the most precise calculation\\
Parton momentum &PDF uncertainty/compare PDF sets,\\
&Vary factorization scale $\mu_f$\\
Differential cross-section & Vary the renormalization scale $\mu_r$\\
Merging Scheme (NLO)  & Compare {\sc Powheg} and {\sc MC@NLO} \\
Matching Scheme (LO) & Vary the {\sc CKKW} or {\sc MLM} parameters\\
Fragmentation Model & Compare {\sc Pythia} and {\sc Herwig} \\
`Extra' Radiation (ISR/FSR/MPI) & Vary PS tune, vary $h_\text{damp}$ ({\sc Powheg-Box}) \\
\hline
\end{tabular}}
\caption{A decomposition of theory modeling uncertainties into several categories. }
\label{systematictheoryuncerts}
\end{table}	
	
	Even though Table~\ref{systematictheoryuncerts} describes a clear decomposition for evaluating the theory modeling uncertainties, there is no unique way to ascribe a `$1\sigma$ uncertainty' for each source.  The general prescription is to take the difference in the predicted yield between the nominal sample $N$ and a variation $V$ and compute $\sigma=|N-V|$.  When a particular process is normalized in a control region, only the difference in the yield after normalizing both samples in the control region is used for the uncertainty (transfer factor).  When the procedure calls for a comparison between two samples $V_1$ and $V_2$, neither of which is the nominal, the general strategy is to take $|V_1-V_2|$ when they are both MC samples with nominal settings and $\frac{1}{2}|V_1-V_2|$ when $V_1$ is an `up' variation and $V_2$ is a `down' variation of some simulation parameter.  Furthermore, simulations are computationally expensive and need to be sufficiently large to make MC statistical uncertainties negligible.  Therefore, most of the theoretical modeling uncertainties are evaluated at particle-level using particle-level event selections analogous to the detector-level ones.  This may still not be sufficient to populate the extreme kinematic tails distributions near the signal regions and therefore an additional strategy is to compare two samples with a looser event selection and then extrapolate the difference to tighter selections.

		\clearpage
		
		\subsection{Top Quark Pair Production}
		\label{sec:susy:syst:ttbar}
		
The uncertainty on the $t\bar{t}$ extrapolation from the control region to the signal region closely follows the prescription from Table~\ref{systematictheoryuncerts}.  Table~\ref{systematictheoryuncerts_ttbar} summarizes the specific procedure for $t\bar{t}$ at both $\sqrt{s}=8$ TeV and $\sqrt{s}=13$ TeV.  The most important uncertainties are those related to the number and spectra of the `extra' jets that are not from the ME. The extrapolation in $m_\text{T}$ changes the $t\bar{t}$ composition from a mostly one-lepton topology with ME-induced jets to a mostly two-lepton topology with extra jets from ISR/FSR.  An uncertainty on the fragmentation model is estimated by comparing {\sc Pythia} 6 and {\sc Herwig}(++), fixing {\sc Powheg-Box} as the ME generator.  This results in a $\sim 10\%$ uncertainty in the extrapolation from the CR to the SR.  In addition, the amount of radiation within and around jets is varied using dedicated parton shower tune variations.  At $\sqrt{s}=8$ TeV, {\sc AcerMC} +{\sc Pythia} 6 with the AUET2B tune~\cite{ATL-PHYS-PUB-2011-009} was modified based on the measurement of radiation gaps in dilepton $t\bar{t}$ events at $\sqrt{s}=7$ TeV~\cite{ATLAS:2012al}.  {\sc Pythia} 6 parameters related to the value of $\alpha_s$ used in generating ISR and FSR are varied to bracket the measurement~\cite{ATL-PHYS-PUB-2014-005}.  At $\sqrt{s}=13$ TeV, the {\sc Pythia 6} Perugia2012 tune variations radHi and radLo, which vary the shower $\alpha_s$, are used in conjunction with simultaneous variations of the factorization and renormalization scales as well as $h_\text{damp}$.  The combination (tune, $\mu_f$, $\mu_r$, $h_\text{damp}$)=(radLo/Hi, $\times 2/0.5$, $\times 2/0.5$, $m_\text{top}/2m_\text{top}$) is based on the $\sqrt{s}=7$ TeV gap fraction measurement as well as other $t\bar{t}$ properties measurements~\cite{ATL-PHYS-PUB-2015-002}.  Related to the amount of radiation in the event is the interface between the NLO matrix element and the parton shower.  At $\sqrt{s}=8$ TeV, {\sc MC@NLO}+{\sc Herwig} (NLO) and Alpgen+{\sc Pythia} 6 (LO) with MLM matching were studied, but found to be significantly worse models of the data in inclusive event selections and therefore were not considered for the final uncertainty.  For the early Run 2 analysis, the {\sc MC@NLO} and {\sc Powheg} methods are compared, fixing {\sc Herwig++} for fragmentation.  Figure~\ref{fig:ttbarsystsusy} shows a comparison between the nominal $t\bar{t}$ sample and MG5\_aMC+{\sc Herwig++} using a particle-level selection at $\sqrt{s}=13$ TeV.  After updating $h_\text{damp}\rightarrow m_\text{top}$ and swapping {\sc Herwig} with {\sc Herwig++} ($\sqrt{s}=8\rightarrow 13$ TeV), both of these simulations are reasonable models of the data inclusively and the fact that they predict similar event yields gives confidence in the extrapolation from the CR to the SR.  The PDF4LHC procedure~\cite{PDF4LHC} is used to estimate the PDF uncertainty on the acceptance at $\sqrt{s}=8$ TeV.  As expected, this uncertainty is subdominant to others ($\mathcal{O}(1\%)$) as PDF variations only slightly change the energy and rapidity distributions.  These uncertainties were ignored for the early $\sqrt{s}=13$ TeV analysis.  The total uncertainty on the extrapolation from the TCR to the SR is in the range $15\%$-$25\%$.
		
 \begin{table}[h!]
\centering
\label{my-label}
\noindent\adjustbox{max width=\textwidth}{
\begin{tabular}{|c|cc|}
\hline
Source & Procedure ($\sqrt{8}$ TeV) & Procedure ($\sqrt{13}$ TeV) \\
\hline
Inclusive cross-section & N/A (CR method) & N/A (CR method)\\
Parton momentum & PDF4LHC&Ignored\\
Differential cross-section & $\mu_f,\mu_r$ by $\times\frac{1}{2}$ and $\times 2$& (see last row)\\
Merging / Matching &{\sc MC@NLO}*, {\sc Alpgen}*&P+{\sc Herwig++} v. M+{\sc Herwig++} \\
Fragmentation Model &P+{\sc Pythia} 6 v. P+{\sc Herwig} & P+{\sc Pythia} 6 v. P+{\sc Herwig++}  \\
Amount of `Extra' Radiation  & AcerMC variations&$\mu_f,\mu_r$, {\sc Pythia} 6 tune, $h_\text{damp}$  \\
\hline
\end{tabular}}
\caption{A summary of the theoretical modeling uncertainties for $t\bar{t}$ at $\sqrt{s}=8$ TeV and $\sqrt{s}=13$ TeV.  The * indicates that this uncertainty was studied but not applied because of worse agreement between the alternative generators and data in an inclusive event selection. P stands for {\sc Powheg-Box} and M stands for MG5\_aMC.}
\label{systematictheoryuncerts_ttbar}
\end{table}

\begin{figure}[h!]
\begin{center}
\includegraphics[width=0.5\textwidth]{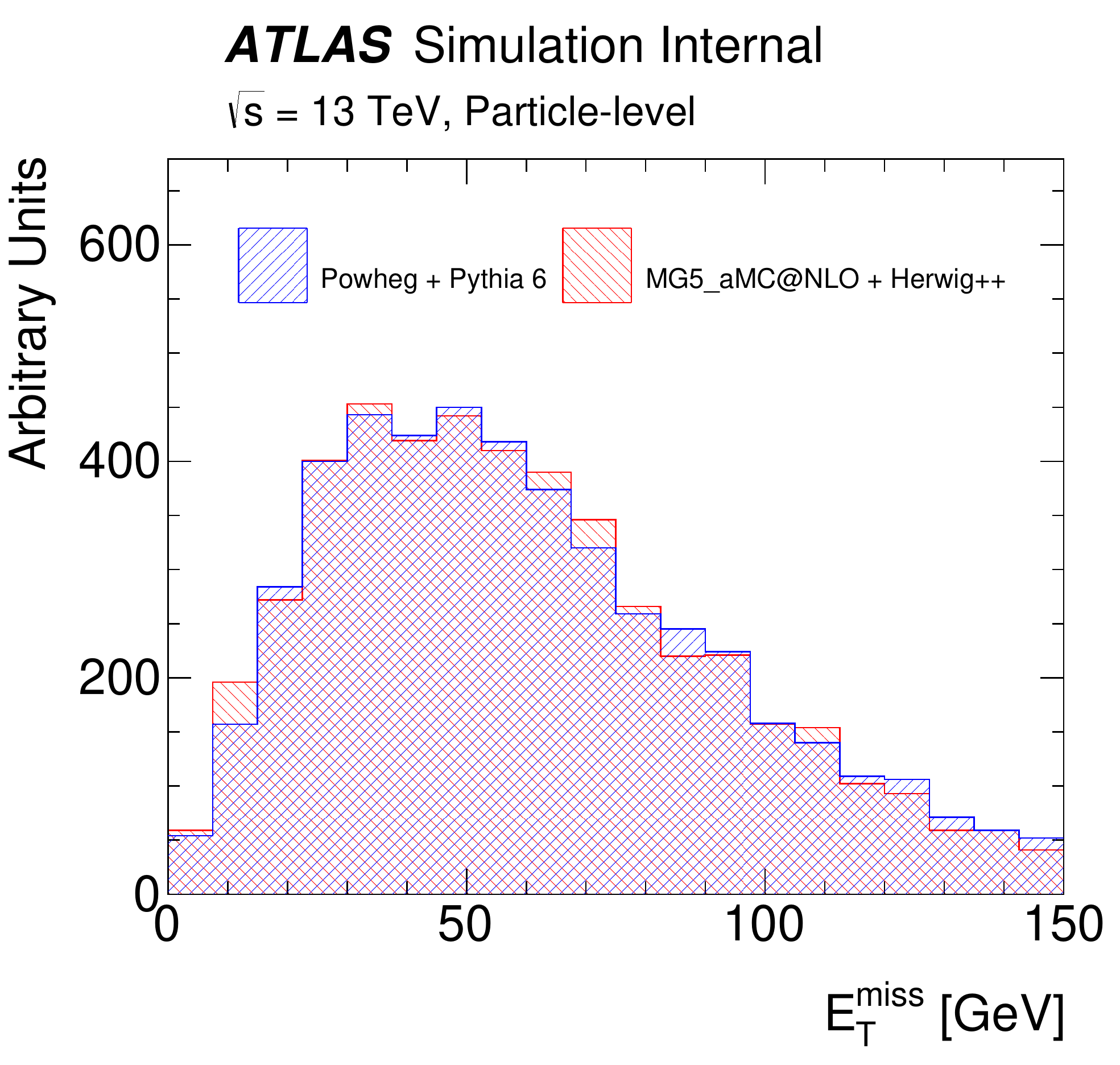}\includegraphics[width=0.5\textwidth]{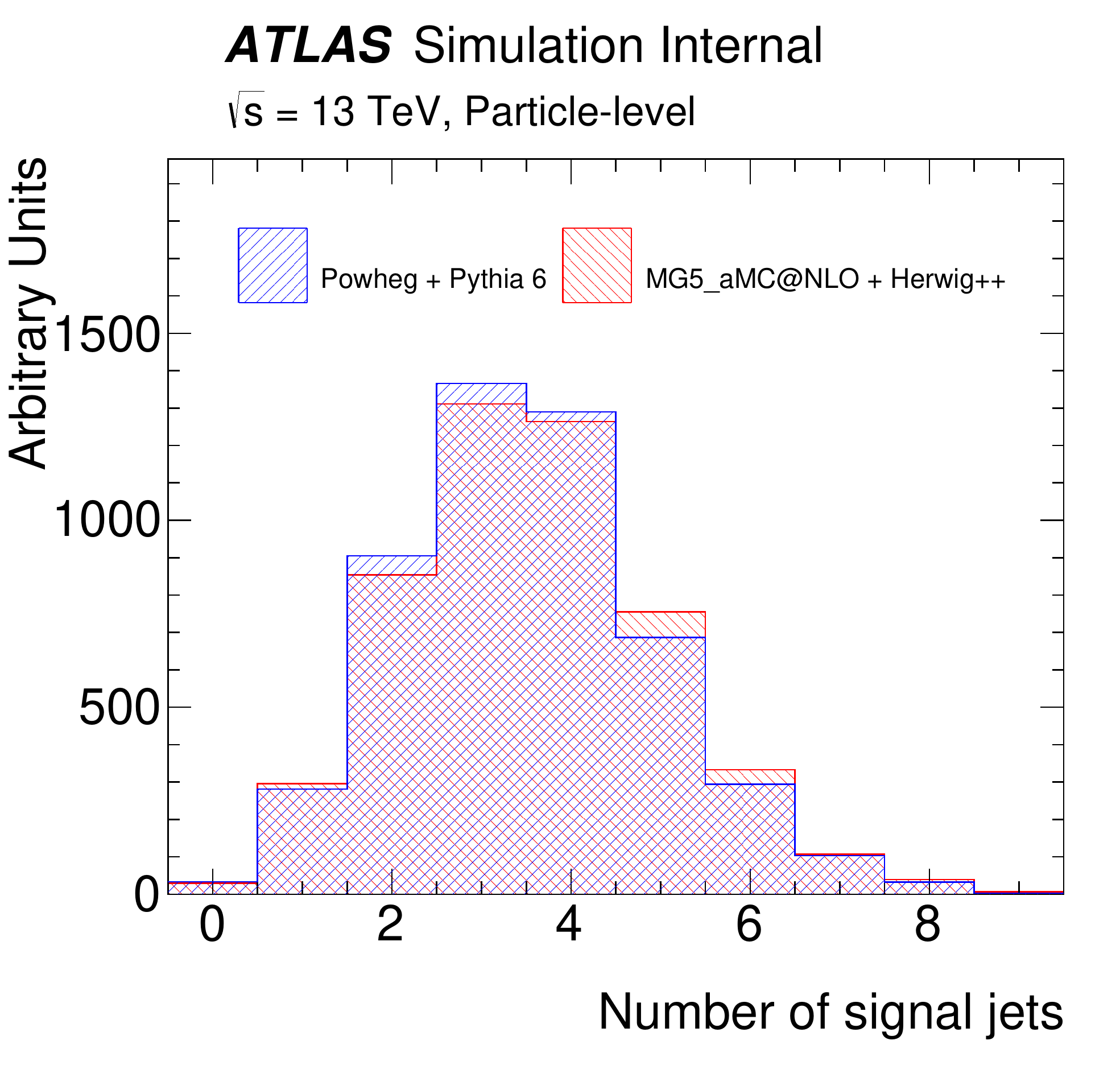}
 \caption{The particle-level $E_\text{T}^\text{miss}$ (left) and the number of particle-level jets (right) for {\sc Powheg}+{\sc Pythia} 8 and MG5\_aMC+{\sc Herwig++}.}
 \label{fig:ttbarsystsusy}
  \end{center}
\end{figure}			
		
\clearpage		
		
		\subsection{Single Top}
		\label{sec:singletopuncerts}
		
	As described in Sec.~\ref{singletop}, the dominant single top process is the $Wt$-channel.  Systematic uncertainties are estimated for this process and the other single top channels in a similar manner as for $t\bar{t}$, summarized in Table~\ref{systematictheoryuncerts}.  The main difference is that without a control region at $\sqrt{s}=8$, the search is sensitive to the inclusive cross section uncertainty.  Furthermore, without any region enriched in single top events, there is no way to conclude that one model is better than another and so the {\sc MC@NLO}+{\sc Herwig} that was not used for $t\bar{t}$ at $\sqrt{s}=8$ TeV is used to set an uncertainty on the merging scheme for single top.  This is a $\sim 10\%$ uncertainty on the yield in the signal region.  The dominant uncertainty for single top is due to the modeling of the interference between single top and $t\bar{t}$ ($30\%$), as described below.
		
 \begin{table}[h!]
\centering
\label{my-label}
\noindent\adjustbox{max width=\textwidth}{
\begin{tabular}{|c|cc|}
\hline
Source & Procedure ($\sqrt{8}$ TeV) & Procedure ($\sqrt{13}$ TeV) \\
\hline
Inclusive cross-section & 6.8\% & N/A (CR method)\\
Parton momentum & PDF4LHC&Ignored\\
Differential cross-section & Ignored& (see penultimate row)\\
Merging / Matching &{\sc MC@NLO}& Ignored \\
Fragmentation Model &P+{\sc Pythia} 6 v. P+{\sc Herwig} & P+{\sc Pythia} 6 v. P+{\sc Herwig++}  \\
Amount of `Extra' Radiation  & AcerMC variations&$\mu_f,\mu_r$, {\sc Pythia} 6 tune  \\
Interference with $t\bar{t}$ &DR/DS*, AcerMC & MG5\_aMC \\
\hline
\end{tabular}}
\caption{A summary of the theoretical modeling uncertainties for single top at $\sqrt{s}=8$ TeV and $\sqrt{s}=13$ TeV.  The * indicates that this uncertainty was studied but not applied (see the text). P stands for {\sc Powheg-Box}.}
\label{systematictheoryuncerts}
\end{table}		

At NLO accuracy, there is a non-trivial interference between leading order $t\bar{t}$ and NLO $Wt$ with one real emission.  For example, the reaction $gg\rightarrow \bar{t}t^*\rightarrow \bar{t} bW^{+}$ contributes at LO to $t\bar{t}$ and at NLO to single top.  When $m(bW^{+})\sim m_t$ this interference is large.  The interference is treated by removing a contribution from the $Wt$ simulation using either the Diagram Removal (DR) or Diagram Subtraction (DS)~\cite{Frixione:2008yi} schemes and the difference can be an estimate of the uncertainty.  Both schemes result in unphysical results that are either not gauge-invariant (DR) or not intended for exclusive event selections (DS).   When the interference between NLO single top and LO $t\bar{t}$ is small, the two interference schemes are comparable~\cite{Frixione:2008yi,Re:2010bp,White:2009yt}.  However, the stop search selects single top events with relatively high purity and such events have kinematic properties that result in a non-trivial interference with LO $t\bar{t}$.  Using separated $t\bar{t}$ and $Wt$ processes simulated at NLO may not be meaningful and the difference between the DR and DS interference removal schemes may not give an accurate estimate of the uncertainty on the modeling of the composite process.  

Specific examples at particle level with $E_\text{T}^\text{miss}> 200$ GeV are shown in Fig.~\ref{fig:Wt:problems} and Fig.~\ref{fig:Wt:problems2} for a single lepton and a dilepton selection, respectively.  The two-lepton events are a useful complement to the single lepton ones because most $Wt$ events in the SR have two leptonically decaying $W$ bosons.  To enhance the interference, events are required to have at least four jets with $p_\text{T}>25$ GeV and at least two such jets must have originated from $b$-quarks.  For both the single and double lepton selections, there is a significant and increasing difference between the single top simulations with the DR and DS schemes for the $E_\text{T}^\text{miss}$ (Fig.~\ref{fig:Wt:problems}) and the leading jet $p_\text{T}$ distributions (Fig.~\ref{fig:Wt:problems2}).  In both cases, the DR scheme results in both a harder  $E_\text{T}^\text{miss}$ and leading jet $p_\text{T}$ spectrum. For this selection the $Wt$ contribution is a small fraction (5--10\%) of the total top quark contribution.   However, in the $E_\text{T}^\text{miss}$ tail for the dilepton events, the $\gtrsim 50\%$ difference between the single top DR and DS results in $15$-$20\%$ differences in the combined $t\bar{t}+Wt$ simulation when comparing the two interference schemes.  Even higher purities occur for tighter selections where the difference between the two setups can approach $100\%$ and $\gtrsim 50\%$ overall uncertainties.

At $\sqrt{s}=8$ TeV, the interference in the signal regions was further studied by looking at a LO sample generated by AcerMC~\cite{Kersevan:2004yg} with the inclusive $2\rightarrow 6$ reaction $pp\rightarrow W^+W^-b\bar{b}$ ($WWbb$) that includes the double resonant $t\bar{t}$ production, the single resonant $Wt$ production in association with a $b$-quark and the non-resonant diboson production in association with jets (see Fig.~\ref{fig:feynman}).  The rest of this section presents an analogous study using MG5\_aMC interfaced with {\tt Pythia} 8 at $\sqrt{s}=13$ TeV.  Figures~\ref{fig:Wt:problemsb} and~\ref{fig:Wt:problems2b} include this inclusive $WWbb$ sample in comparison with the {\sc Powheg-Box} simulation using the DR scheme.  There are significant differences in both the $E_\text{T}^\text{miss}$ and leading jet $p_\text{T}$ distributions.  This is due in part to the fact that the $t\bar{t}$ component of the $WWbb$ sample is LO and there are significant NLO corrections.  Further studies of the full ME comparison will benefit from a multileg $WWbb$ simulation, which requires a non-trivial merging setup, and ultimately a full NLO $WWbb$ simulation interfaced with a parton shower.  Both the Sherpa+OpenLoops~\cite{Cascioli:2013wga} and MG5\_aMC~\cite{Frederix:2013gra} collaborations have calculated fixed order inclusive calculations, but there is currently no general matrix element at NLO matched to a parton shower.  It is not even currently possible with the existing frameworks to compute $WWbb$ with extra partons in the matrix element because of the non-trivial overlap in the phase space between the $b$-quarks at NLO for the single top process and quarks and gluons from the parton shower.  The fact that the stop search is so sensitive to the interference between the processes and that a single top control region can be constructed with relatively high purity suggests that the data can be used to directly constrain the existing and future models of higher order interference.

\begin{figure}
\centering
\includegraphics[width=0.33\textwidth]{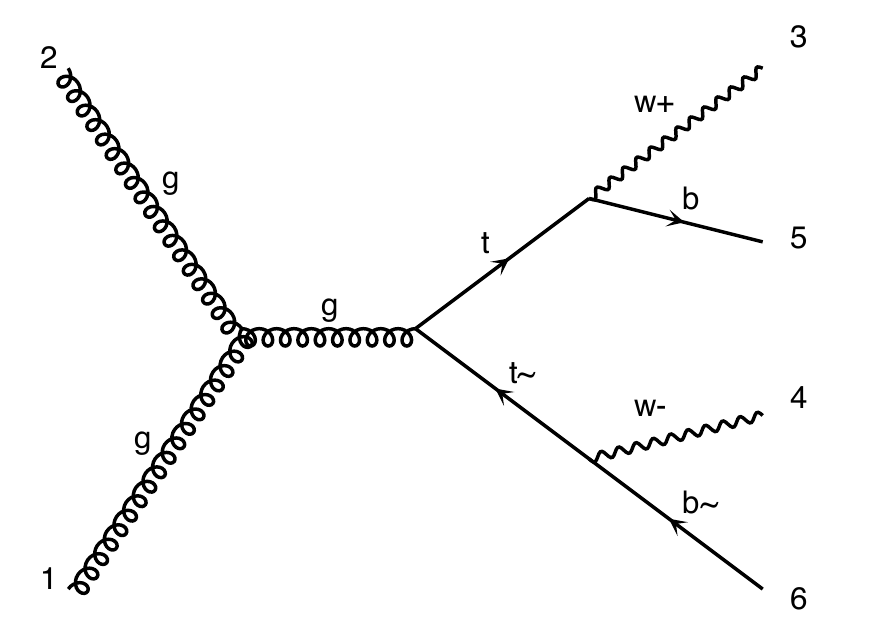}\includegraphics[width=0.33\textwidth]{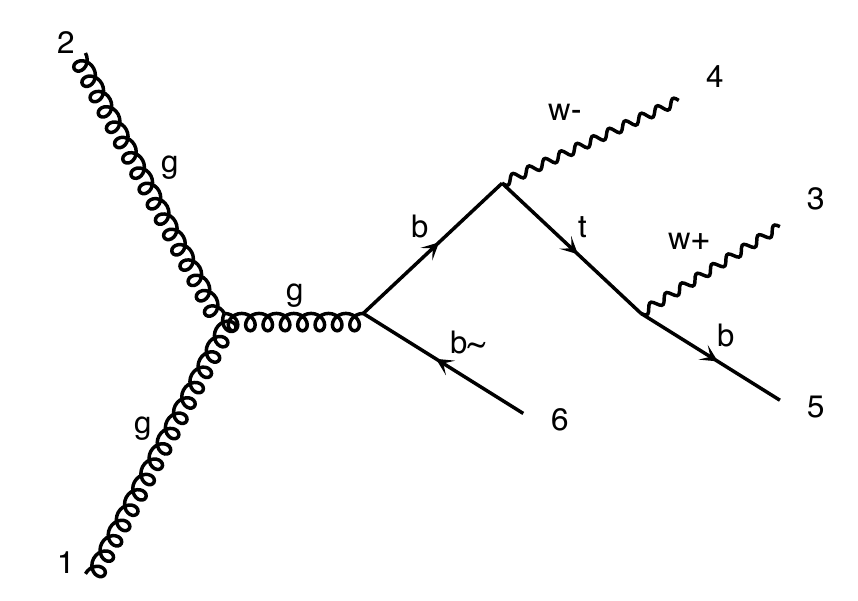}\includegraphics[width=0.33\textwidth]{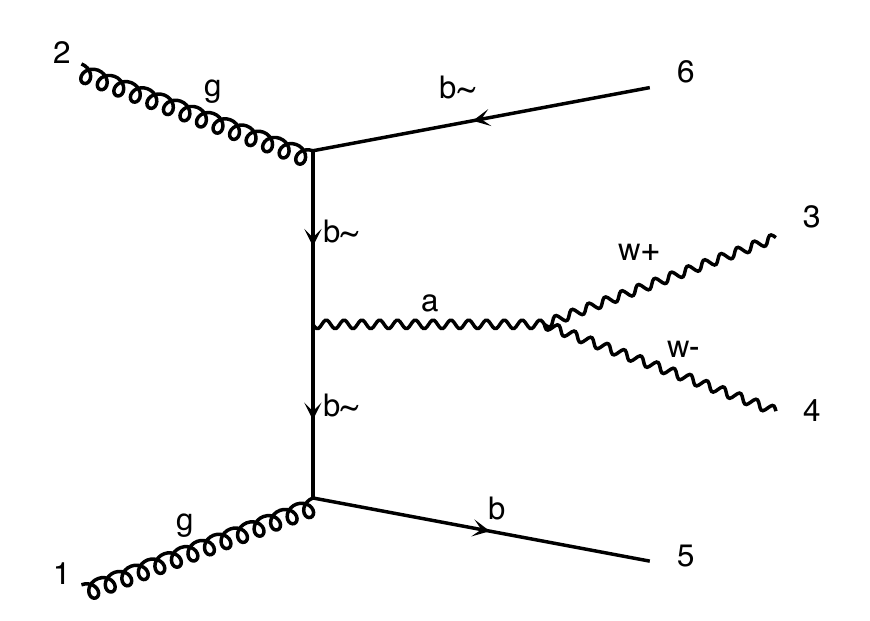}
\caption{The inclusive $WWbb$ process includes Feynman diagrams with doubly resonant (left), singly resonant (middle) and non-resonant top quark contributions. Feynman diagrams from MG5\_aMC.}
\label{fig:feynman}
\end{figure}

Despite the disadvantages of comparing the leading order simulation with the NLO setup, it may produce a conservative uncertainty and therefore is used as the baseline method\footnote{This is not fully satisfactory and will be an important topic of future study. Fortunately, the MC community has made significant progress very recently - see Ref.~\cite{Jezo:2016ujg}.}. The uncertainty at $\sqrt{s}=8$ and $13$ TeV from comparing the $WWbb$ samples with the NLO $t\bar{t}+Wt$ (DR scheme) simulations using {\sc Powheg-Box} result in $\sim 30\%$ uncertainties on the modeling of the interference.

\begin{figure}
\centering
\includegraphics[width=0.45\textwidth]{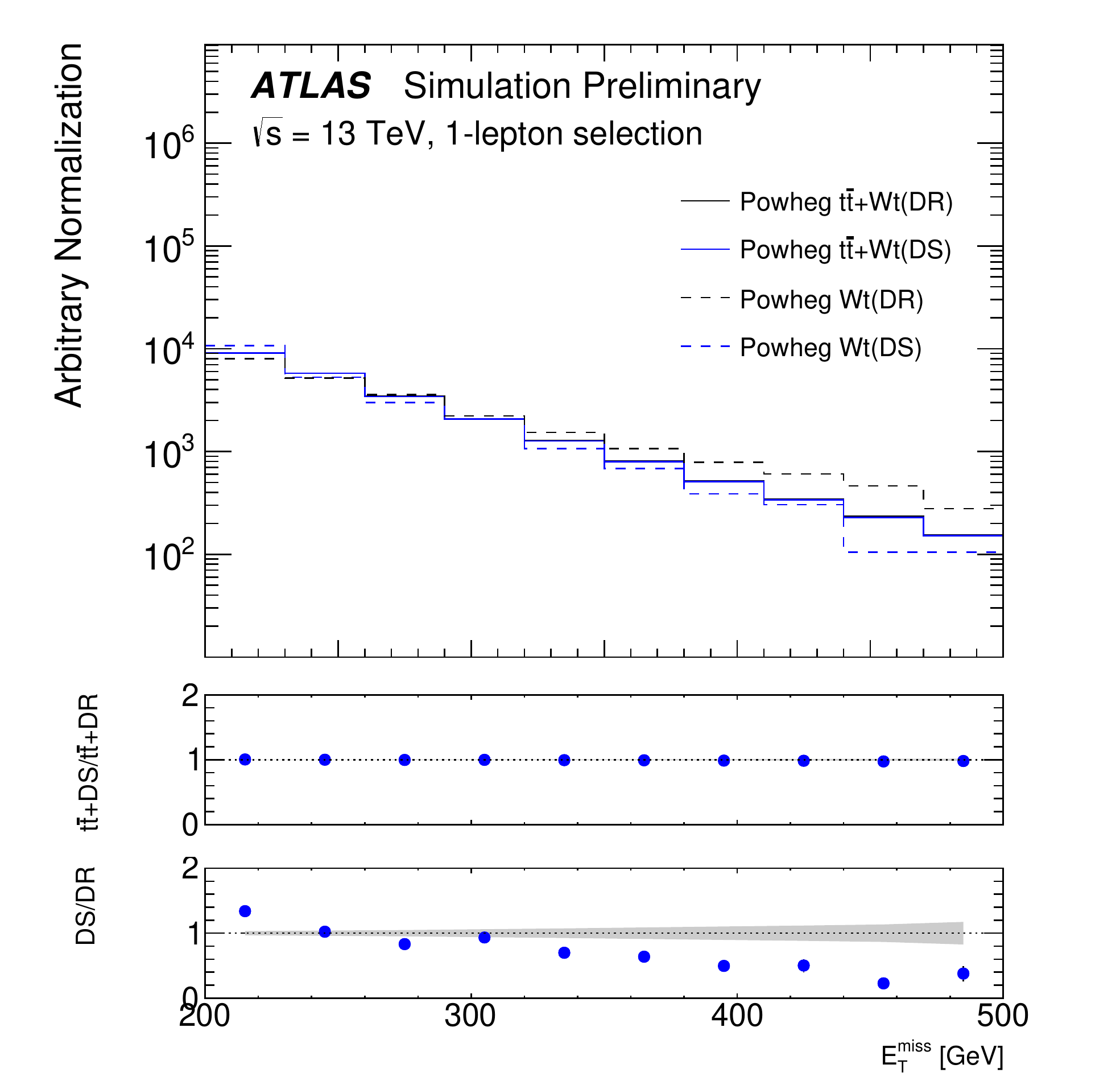}
\includegraphics[width=0.45\textwidth]{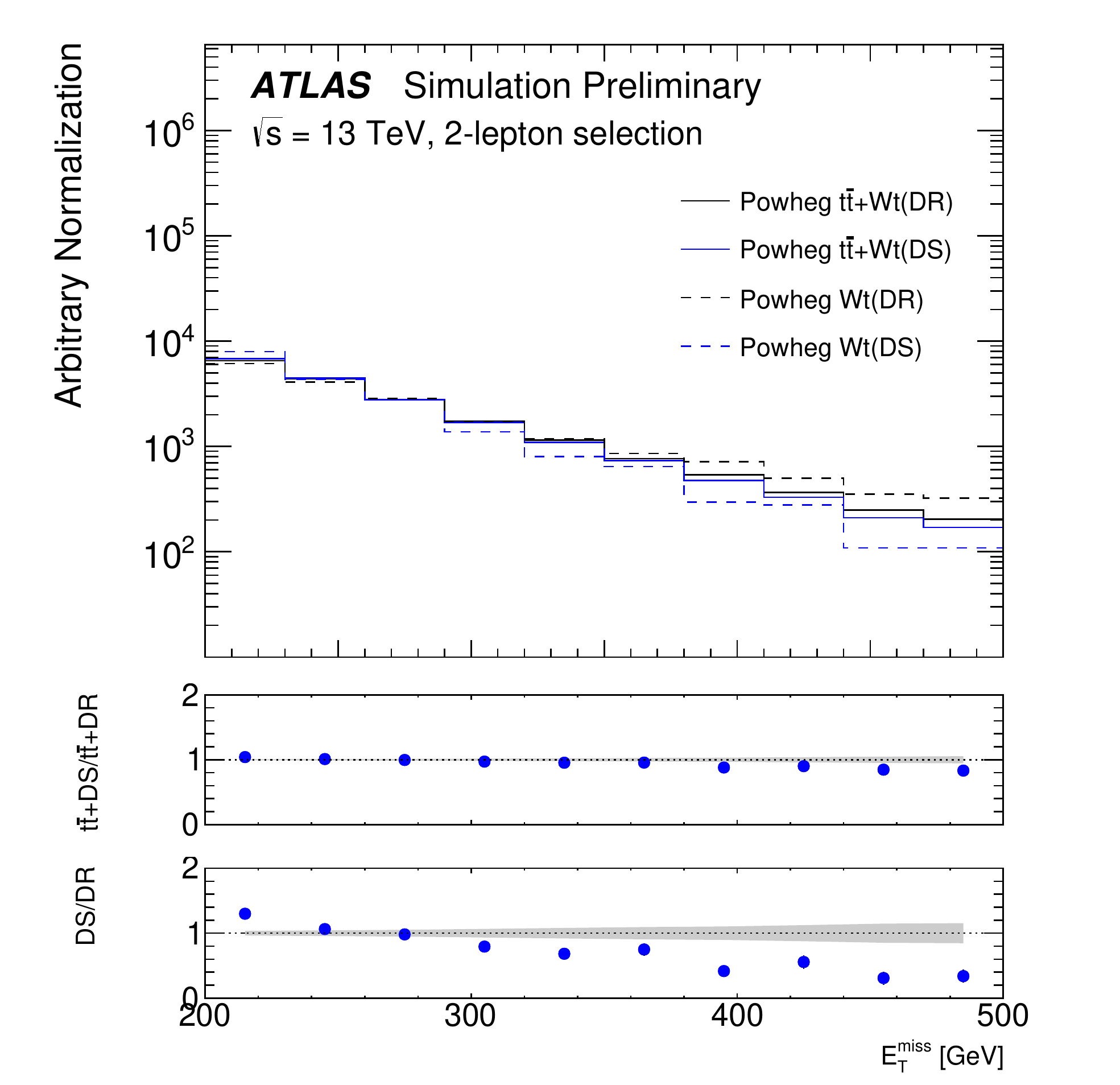}
\caption{The particle-level $E_\text{T}^\text{miss}$ for events passing a one lepton (a) and a two lepton (b) selection.  Both selections require at least four jets with $p_\text{T}>25$ GeV, of which two must have originated from $b$-quarks. All distributions are normalized to have the same integral in the above range.  The gray band in the ratio is the statistical uncertainty from the simulation using the DR scheme and the uncertainty on the markers is from the simulation used in the numerator of the ratio.  Most of these uncertainties are smaller than the markers.}
\label{fig:Wt:problems}
\end{figure}

\begin{figure}
\centering
\includegraphics[width=0.45\textwidth]{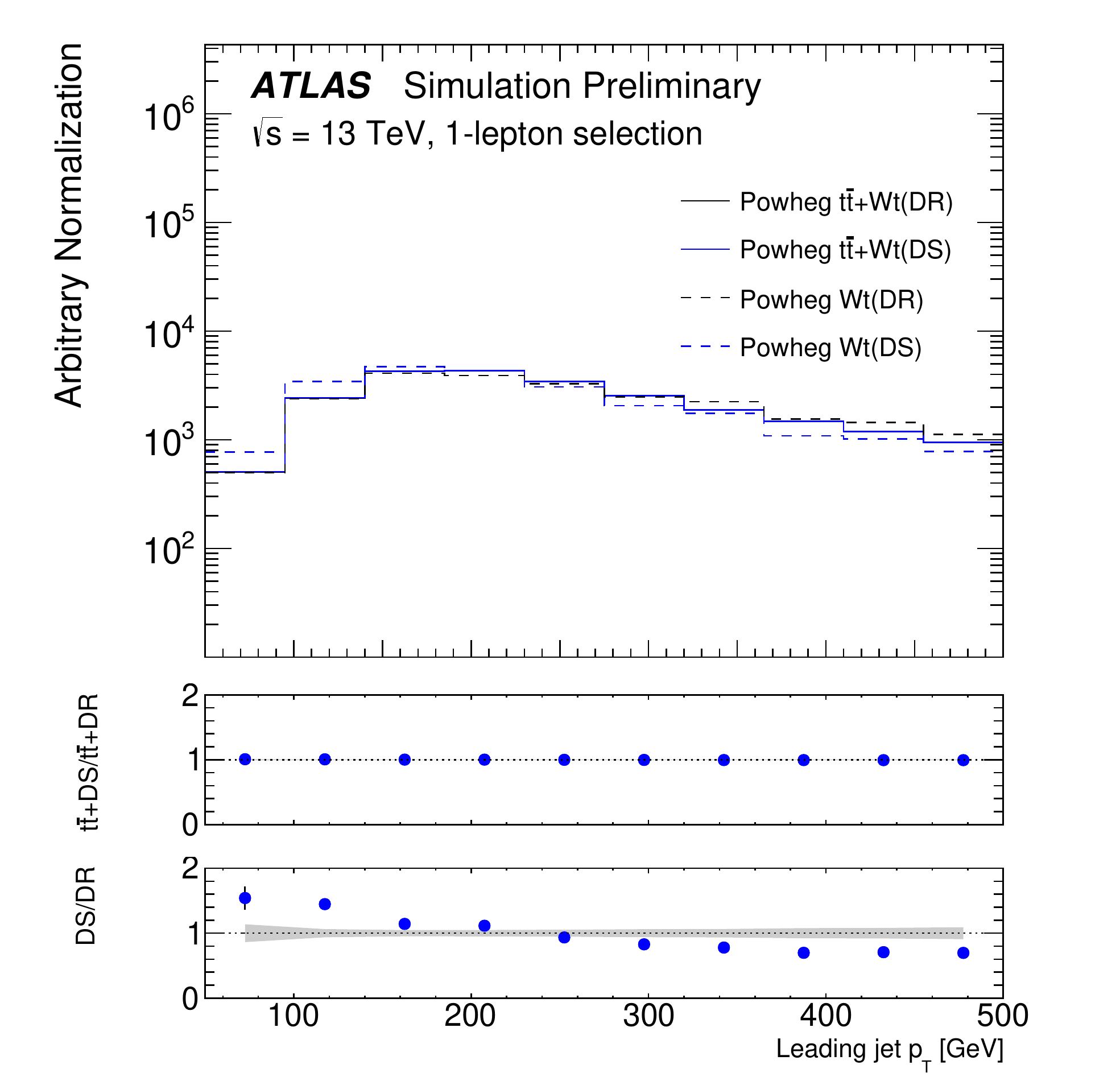}
\includegraphics[width=0.45\textwidth]{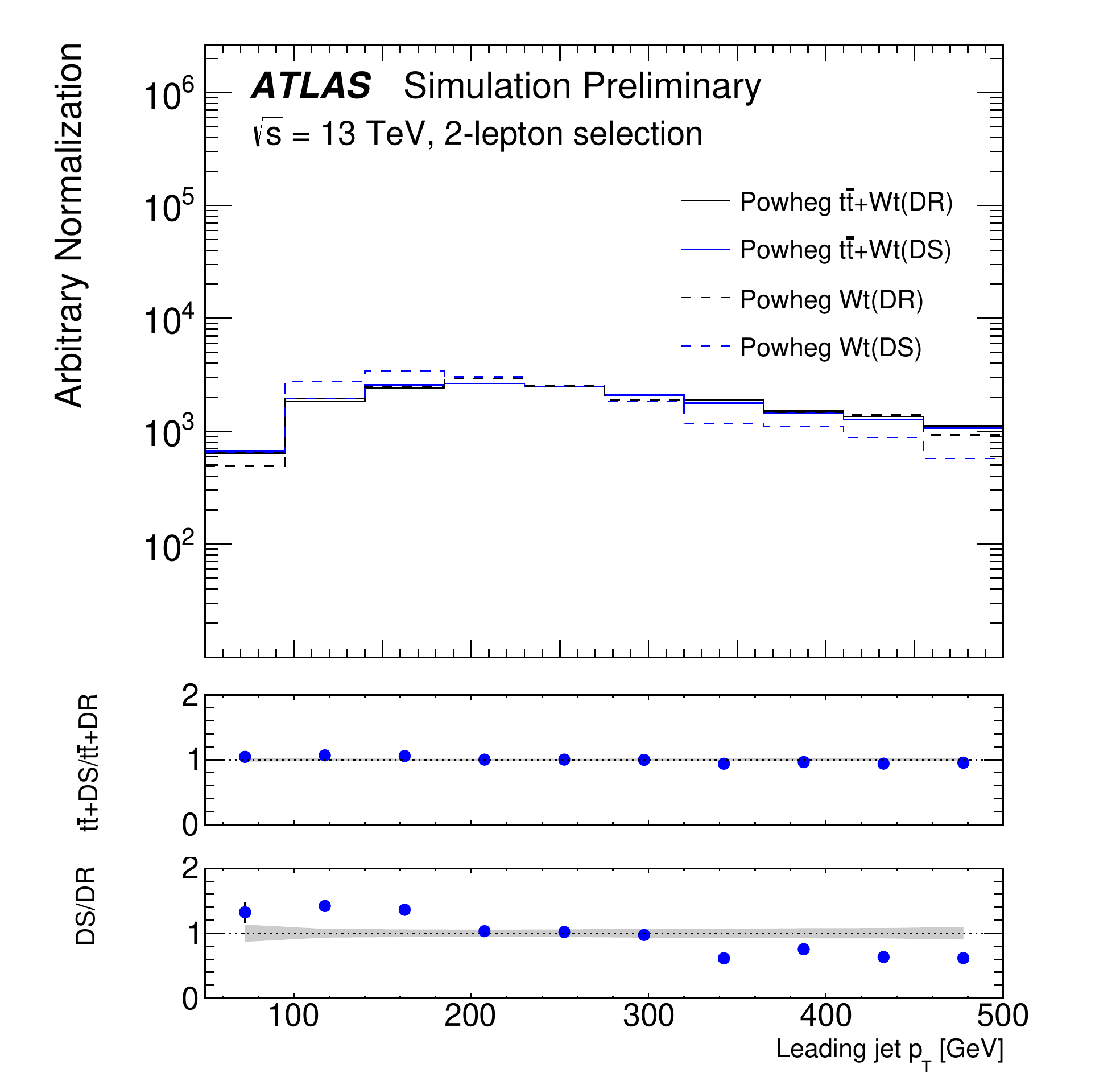}
\caption{Same as Fig.~\ref{fig:Wt:problems} but for the leading jet $p_\text{T}$ instead of the $E_\text{T}^\text{miss}$.}
\label{fig:Wt:problems2}
\end{figure}

\begin{figure}
\centering
\includegraphics[width=0.45\textwidth]{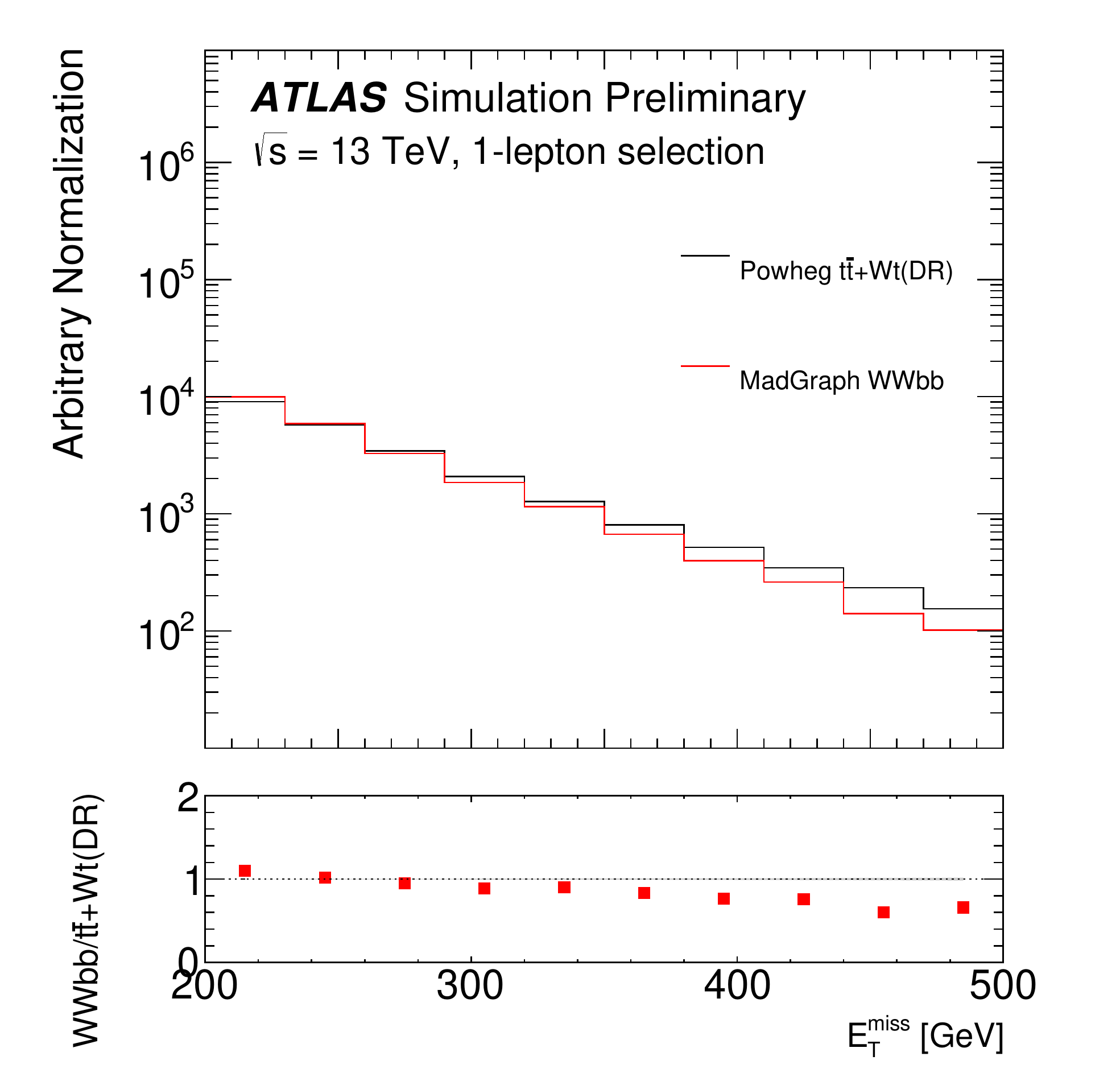}
\includegraphics[width=0.45\textwidth]{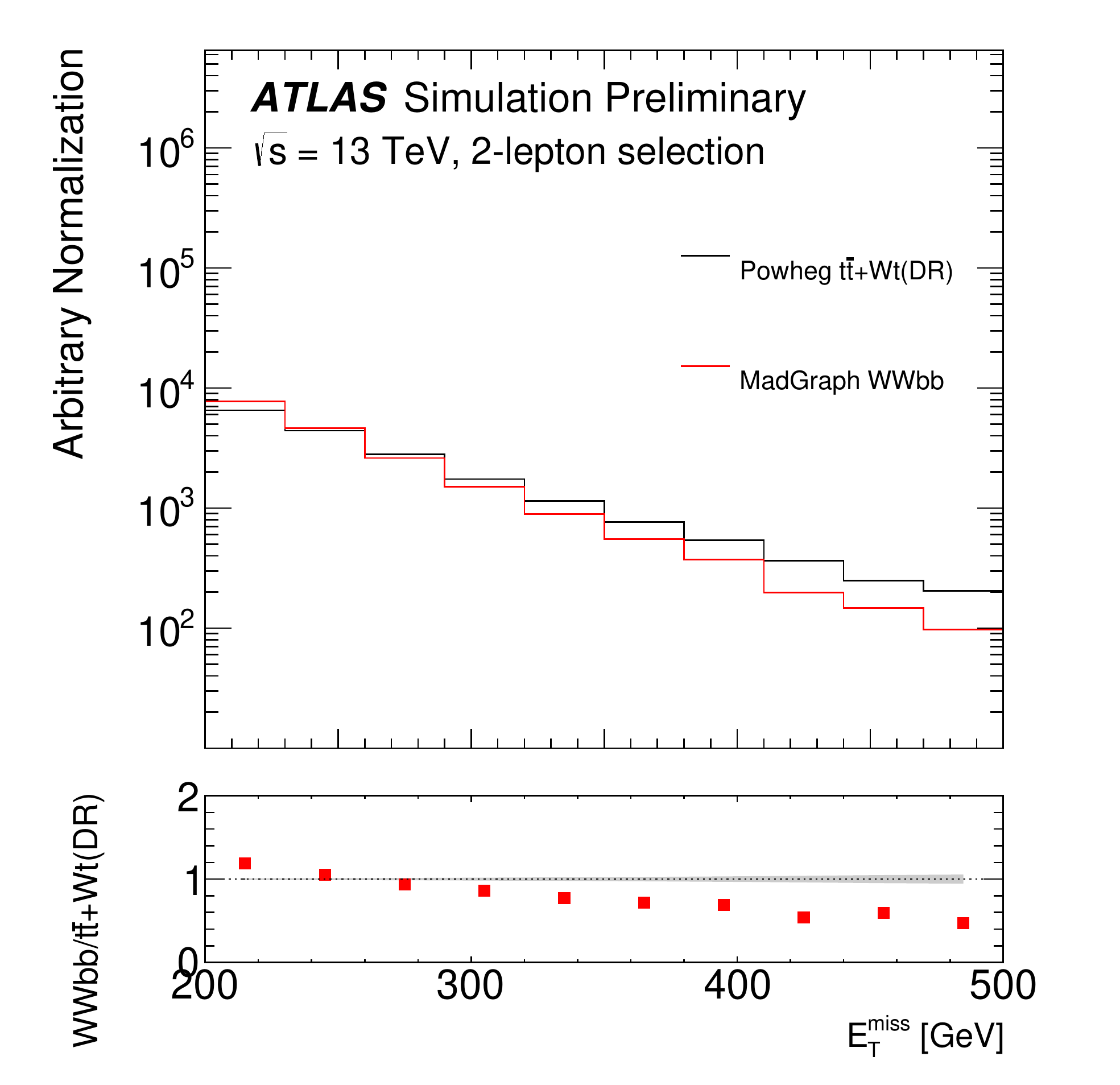}
\caption{Same as Fig.~\ref{fig:Wt:problems}, but comparing the NLO calcluation with interference removal with an inclusive $WWbb$ sample generated with MG5\_aMC.}
\label{fig:Wt:problemsb}
\end{figure}

\begin{figure}
\centering
\includegraphics[width=0.45\textwidth]{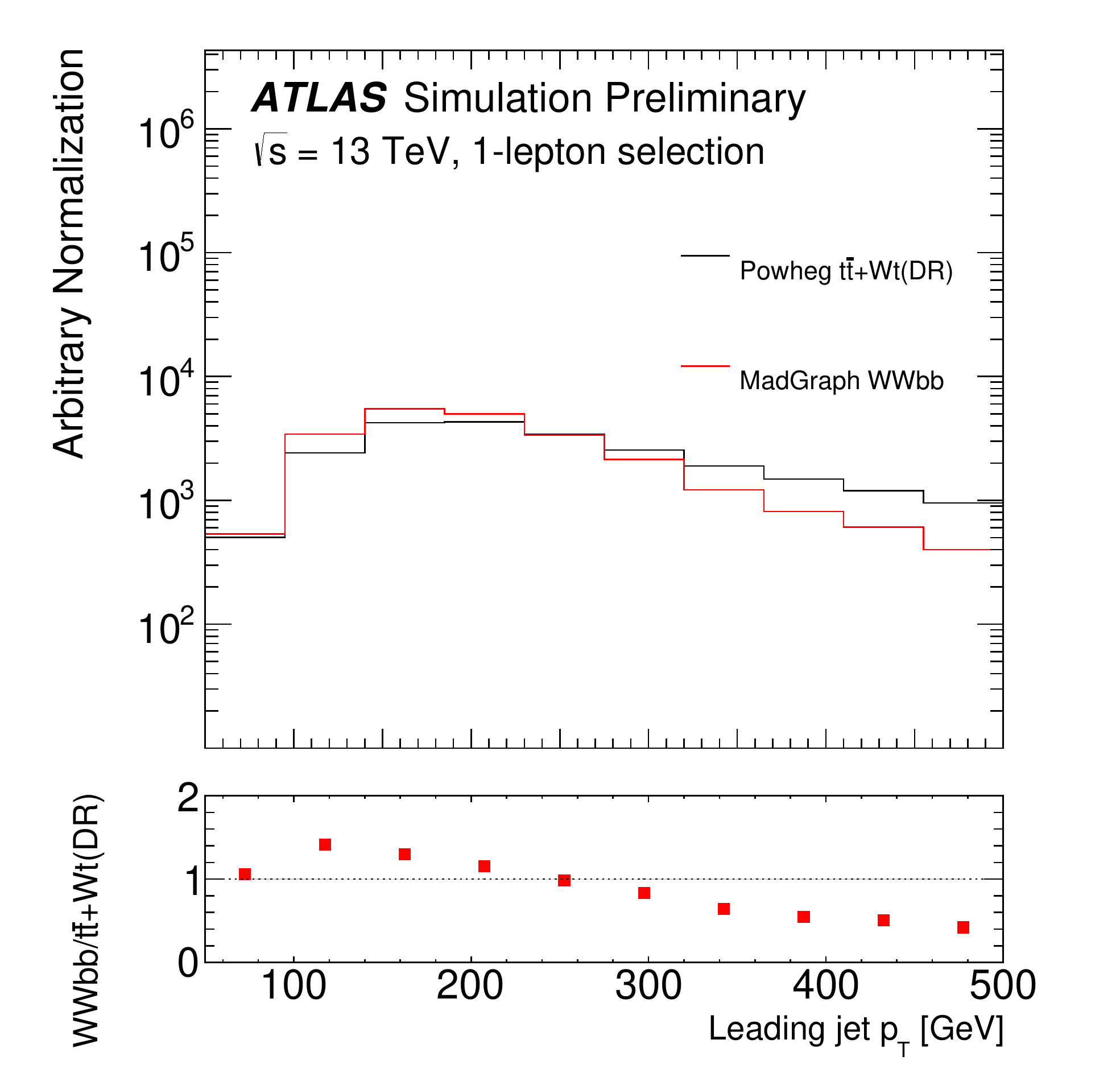}
\includegraphics[width=0.45\textwidth]{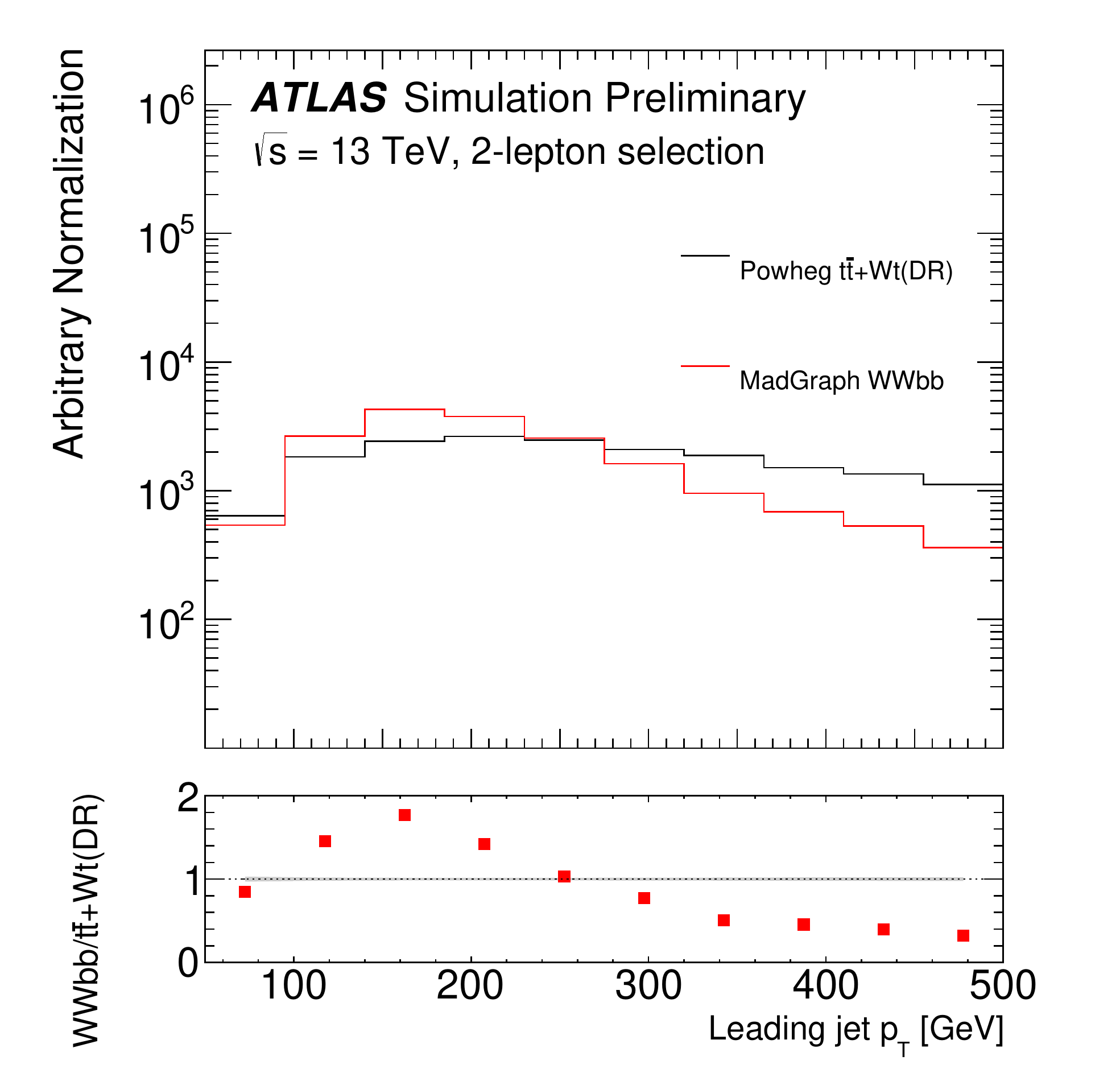}
\caption{Same as Fig.~\ref{fig:Wt:problemsb} but for the leading jet $p_\text{T}$ instead of the $E_\text{T}^\text{miss}$.}
\label{fig:Wt:problems2b}
\end{figure}
	
\clearpage		
		
		\subsection{Top Quark Pair Production with a $Z$ Boson}
		\label{sec:susy:ttzuncert}
		
	Unlike the $t\bar{t}$ and $Wt$ backgrounds, the $t\bar{t}+Z(\rightarrow\nu\bar{\nu})$ background does not need a second lepton in order to pass the event selection.  Therefore, the four jets used in all signal regions already exist at tree-level and so the sensitivity to the modeling of additional radiation is significantly reduced.	Table~\ref{systematictheoryuncertsttv} summarizes the procedure for both the simulation-based approach at $\sqrt{s}=8$ TeV and the data-driven approach at $\sqrt{s}=13$ TeV.  In the simulation-based case, the uncertainty is directly evaluated on the predicted yield in the signal region (Sec.~\ref{ttzsec:simulationbased}) while the data-driven uncertainties are on the relative yield between the CR and SR and between $t\bar{t}+Z$ and $t\bar{t}+\gamma$ (Sec.~\ref{ttzdatadriven}).
		
 \begin{table}[h!]
\centering
\label{my-label}
\noindent\adjustbox{max width=\textwidth}{
\begin{tabular}{|c|cc|}
\hline
Source & Procedure ($\sqrt{8}$ TeV) & Procedure ($\sqrt{13}$ TeV) \\
\hline
Inclusive cross-section & 22\% & N/A (CR method)\\
Parton momentum & PDF4LHC&CT14 and NNPDF3.0\\
Differential cross-section & $\mu_f,\mu_r$& $\mu_f,\mu_r$\\
Merging / Matching &Matching scale, $n_\text{partons}$& {\sc Sherpa} and MG5\_aMC (both MEPS)\\
Fragmentation Model &Ignored & Ignored  \\
Amount of `Extra' Radiation  & ISR/FSR variations& Ignored  \\
\hline
\end{tabular}}
\caption{A summary of the theoretical modeling uncertainties for $t\bar{t}+Z$ at $\sqrt{s}=8$ TeV and $\sqrt{s}=13$ TeV.}
\label{systematictheoryuncertsttv}
\end{table}			
		
		\subsubsection{Simulation-based}	
		\label{ttzsec:simulationbased}
		
		Without a CR, there is an overall cross-section uncertainty of $22\%$~\cite{Campbell:2012dh,Garzelli:2012bn}.  The default $t\bar{t}+V$ simulation at $\sqrt{s}=8$ TeV was produced by {\sc MadGraph} 5 with up to two extra partons in the matrix element.  To assess the impact of the extra partons, an additional sample was generated with only one extra parton in the matrix element.  In principle, the nominal two-parton setup should be more accurate, but with no data to constrain the modeling, the difference between the two samples is taken as a crude and likely conservative uncertainty.  A similar probe of the extra radiation is from varying the MLM matching scale that connects {\sc MadGraph} 5 and {\sc Pythia} 6.  The uncertainty is estimated by changing the scale ({\tt xqcut}) between $15$ and $25$ GeV ($20$ GeV is nominal).  Additionally, the amount of ISR and FSR can be varied using the renormalization scale of the $\alpha_s$ used for both processes coherently in {\sc MadGraph} 5 and {\sc Pythia} 6.   The ISR variation scales {\tt alpsfact} ({\sc MadGraph} 5) and {\tt PARP(64)} ({\sc Pythia} 6) by factors of $2$ and $4$, respectively.   FSR from {\sc Pythia}~6 is varied by changing the value of $\Lambda_\text{QCD}$ used in the running $\alpha_s$ ({\tt PARP(72)}) from $0.2635$ GeV to $0.7905$ GeV ($0.527$ GeV is nominal).  In addition, the infrared cutoff for FSR ({\tt PARJ(82)})  is simultaneously varied between $0.5$ GeV and $1.66$ GeV ($0.83$ GeV is nominal). Figure~\ref{fig:syst:ttv:scalevariations} illustrates the differences in the distributions of two key kinematic variables when using one or two extra partons in the matrix element and Fig.~\ref{fig:syst:ttv:scalevariations2} summarizes the slopes from fitting the ratios of all the above variations for $am_\text{T2}$, $E_\text{T}^\text{miss}$ and $m_\text{T}$.  The corresponding uncertainties are largely statistically consistent with zero, except for the conservative finite partons slope.  A combination of an uncertainty at preselection and an extrapolation into a single-region like selection for all three variables results in a $\sim 20\%$ uncertainty on the $t\bar{t}+V$ yield in addition to the inclusive cross-section uncertainty.

\begin{figure}[h!]
\begin{center}
\includegraphics[width=0.85\textwidth]{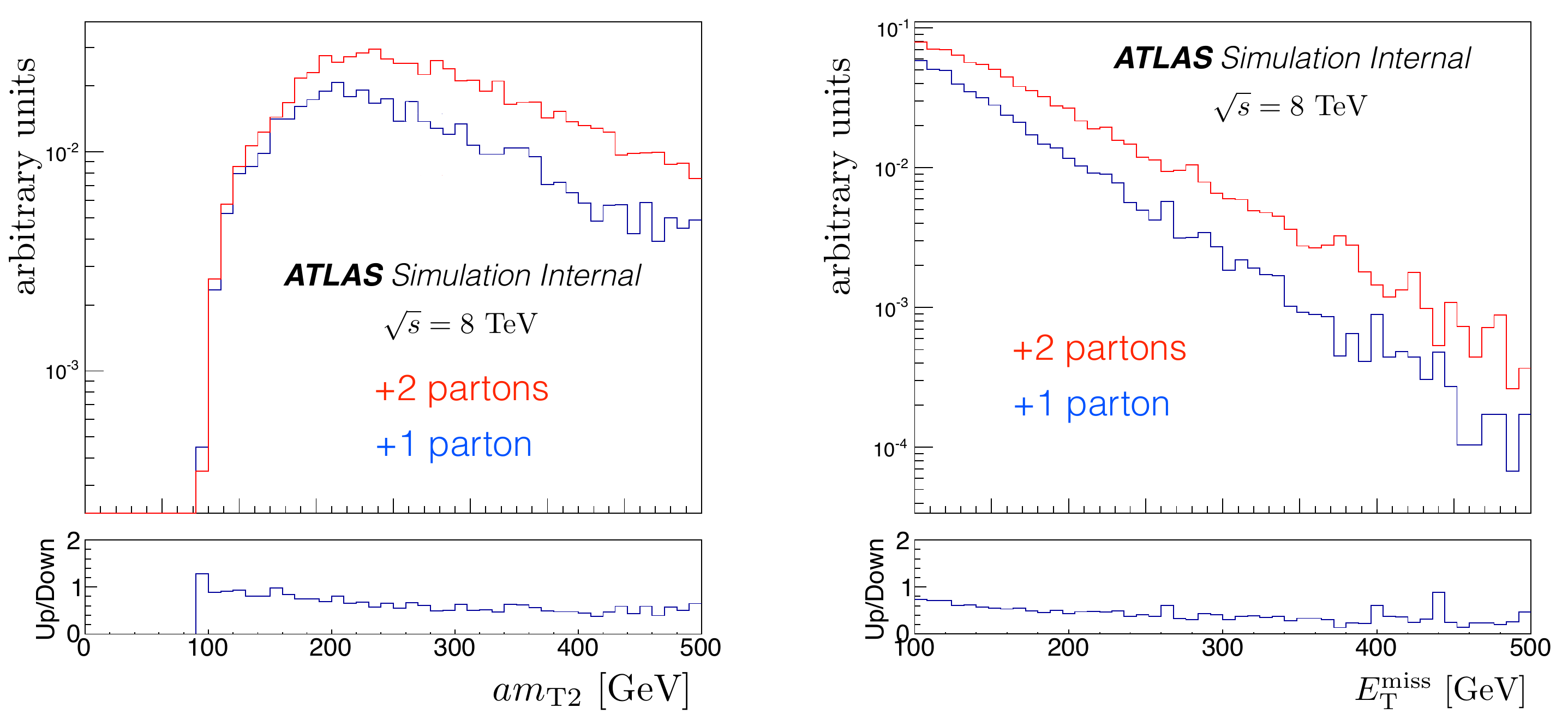}
\caption{The $am_\text{T2}$ (left) and $E_\text{T}^\text{miss}$ (right) distributions when using one or two extra partons in the ME for $t\bar{t}+Z$ with {\sc MadGraph} 5 + {\sc Pythia} 6.}
\label{fig:syst:ttv:scalevariations}
\end{center}
\end{figure}
		
\begin{figure}[h!]
\begin{center}
\includegraphics[width=0.45\textwidth]{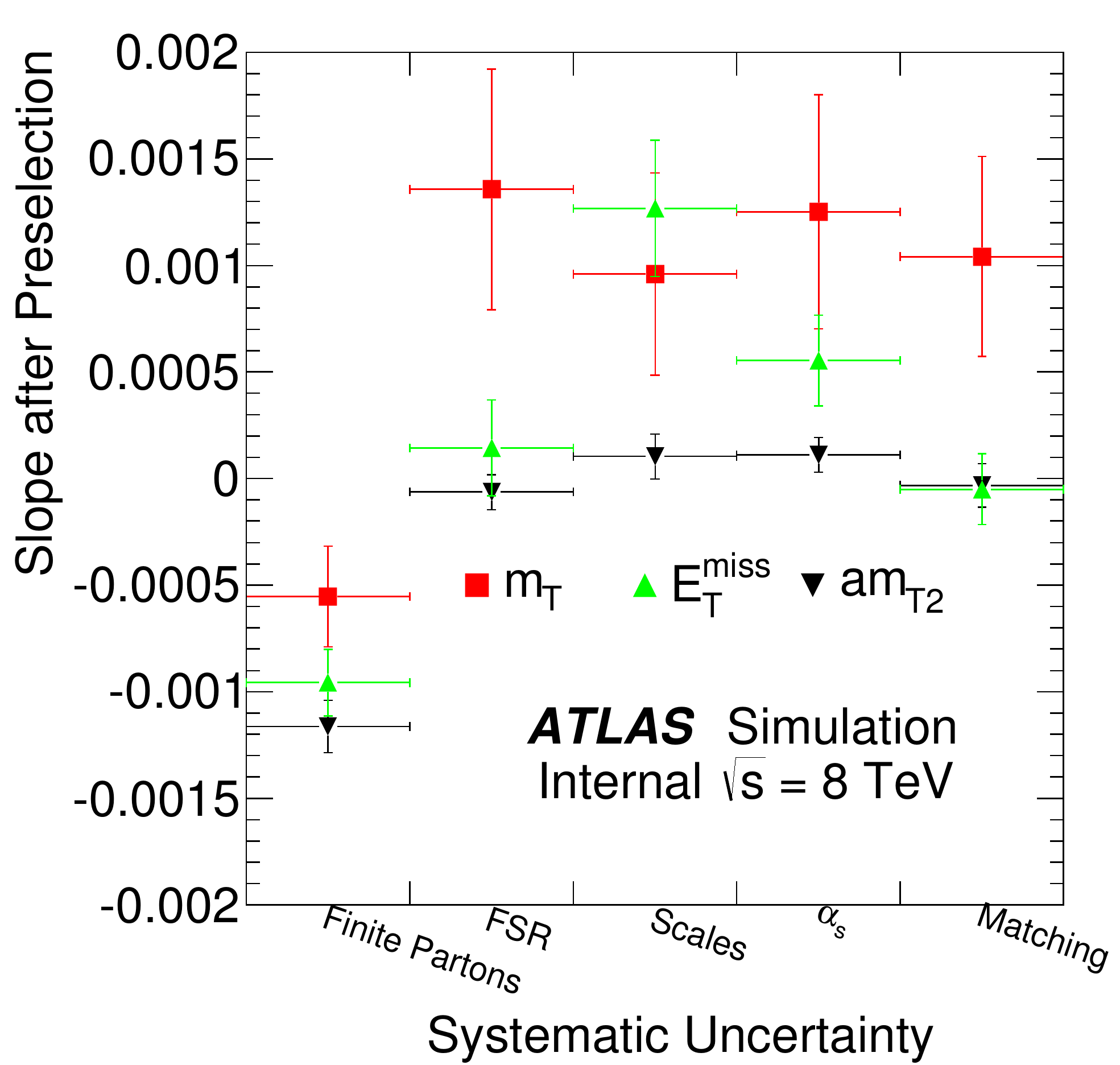}
\caption{The slope parameter from a linear fit to the ratio of the variations listed on the horizontal axis.  For `finite partons', the comparison is between one and two extra partons in the ME; for `FSR', {\sc Pythia} 6 parameters are varied (see the text), for `Scales', the $\mu_f$ and $\mu_r$ are simultaneously doubled and halved; for $\alpha_s$, ISR parameters in {\sc MadGraph} 5 and {\sc Pythia} 6 are varied (see the text); for `Matching', the MLM matching scale is varied.  The error bars are due to MC statistical uncertainty.  The units of the slope parameter are $1/\text{GeV}$.}
\label{fig:syst:ttv:scalevariations2}
\end{center}
\end{figure}

		\subsubsection{Data-driven}
		\label{ttzdatadriven}
		
At $\sqrt{s}=13$ TeV, the nominal background estimation method for $t\bar{t} + V$ uses a $t\bar{t}+\gamma$ control region, as described in Sec.~\ref{sec:ttz:datadriven}.  The total systematic uncertainty on the transfer factor from the $t\bar{t}+\gamma$ CR to $t\bar{t}+Z$ in the SR is $12\%$ and consists of\footnote{All of the calculations in this section are based on fixed order results without a parton shower.  In the case of MG5\_aMC, it was checked that the addition of {\sc Pythia} 8 does not have a significant impact on the reported cross-section differences.}:

\begin{enumerate}
\item A $10\%$ systematic due to coherent factorization and renormalization scale variations as a function of boson $p_\text{T}$ for the LO samples.  This value is based on Fig.~\ref{fig:syst:ttv:scalevariations}, which shows that the $10\%$ in the double ratio $(\mu^{t\bar{t}+\gamma}_\text{up}/\mu^{t\bar{t}+\gamma}_\text{down})/(\mu^{t\bar{t}+Z}_\text{up}/\mu^{t\bar{t}+Z}_\text{down})$ is relatively independent of the $E_\text{T}^\text{miss}$\ for $E_\text{T}^\text{miss} \gtrsim300$ GeV.  
\item A 5\% systematic due to variation of the $k$-factor ratio resulting from scale variations.  There is no uncertainty in the absolute cross-section (and thus $k$-factor) because of the CR normalization, but there is an uncertainty in the difference in the higher order corrections between the $t\bar{t}+\gamma$ and $t\bar{t}+Z$ processes.  Figure~\ref{fig:syst:ttv:kfactor} shows the $k$-factor ratio as a function of boson $p_\text{T}$ for various NLO matrix element generator, PDF, and scale choices\footnote{Thank you to Stefan Hoche for useful discussions about these uncertainties and for providing the {\sc Sherpa}+{\sc OpenLoops} numbers.}.   Fixing {\sc Sherpa}+{\sc OpenLoops} as the matrix element generator and either NNPDF3.0 or CT14 as the PDF set, the variation in the $k$-factor ratio (up versus down triangles in Fig.~\ref{fig:syst:ttv:kfactor}) is about $5\%$ when the scale is varied by a factor of two.  The default scale is the sum of the transverse mass $m_\text{T}^2=m^2+p_\text{T}^2$ of all out-going particles. Note that these scale variations are the LO to NLO coherent scale variations so partially related with the first bullet but not fully correlated.
\item A 5\% systematic to cover the differences in $k$-factor ratios between Sherpa/OpenLoops and Madgraph/aMC@NLO as shown in Fig.~\ref{fig:syst:ttv:kfactor}.  One difference between the setups is the electroweak parameter scheme, i.e. which parameters are taken as input and which are calculated to a fixed order internally.  
\item $\sim 1$-$2$\% to cover the differences in $k$-factor ratios between different PDF sets for a fixed matrix element generator and scale choice as in Fig.~\ref{fig:syst:ttv:kfactor}.
\end{enumerate}

\begin{figure}[h!]
\begin{center}
\includegraphics[width=0.6\textwidth]{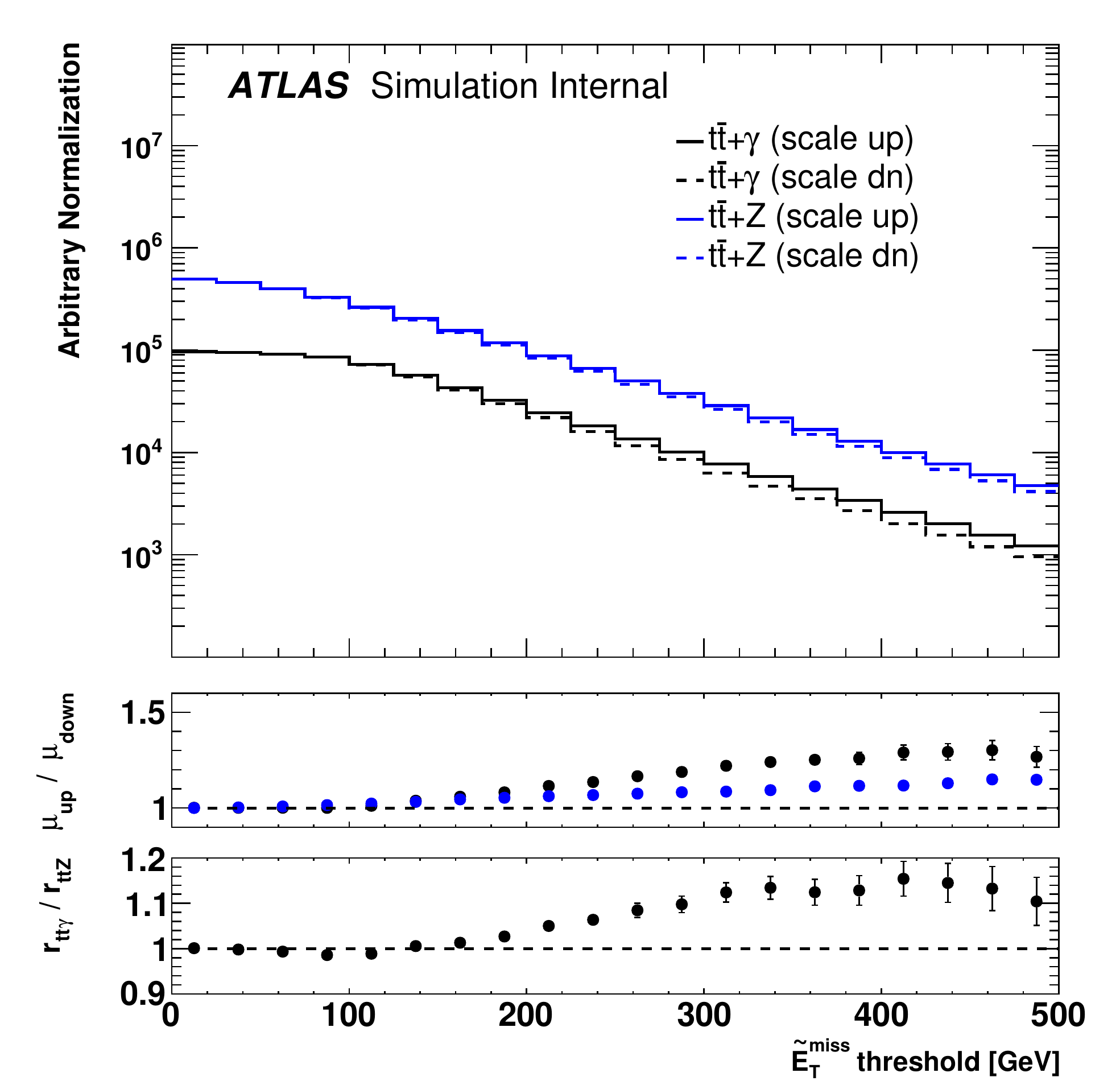}
\caption{Variations of the leading order renormalization and factorization MadGraph scales coherently by factors of two and one-half.   The default scale is the transverse mass (see Table 13 in Ref.~\cite{Hirschi:2015iia}). The $t\bar{t}+\gamma$ sample has a ME cut of 100 GeV.  The $\tilde{E}_\text{T}^\text{miss}$ is the regular $E_\text{T}^\text{miss}$ in the $t\bar{t}+V$ case and the magnitude of the vector sum of the missing momentum and the photon momentum. The relative normalization of the $t\bar{t}+V$ and $t\bar{t}+\gamma$ samples is arbitrary.  The upper ratio compares the scale up with the scale down for the $t\bar{t}+V$ and $t\bar{t}+\gamma$ separately.  The lower panel is the ratio of the ratios: $r=\mu_\text{up}/\mu_\text{down}$.}
\label{fig:syst:ttv:scalevariations}
\end{center}
\end{figure}

\begin{figure}[h!]
\begin{center}
\includegraphics[width=0.5\textwidth]{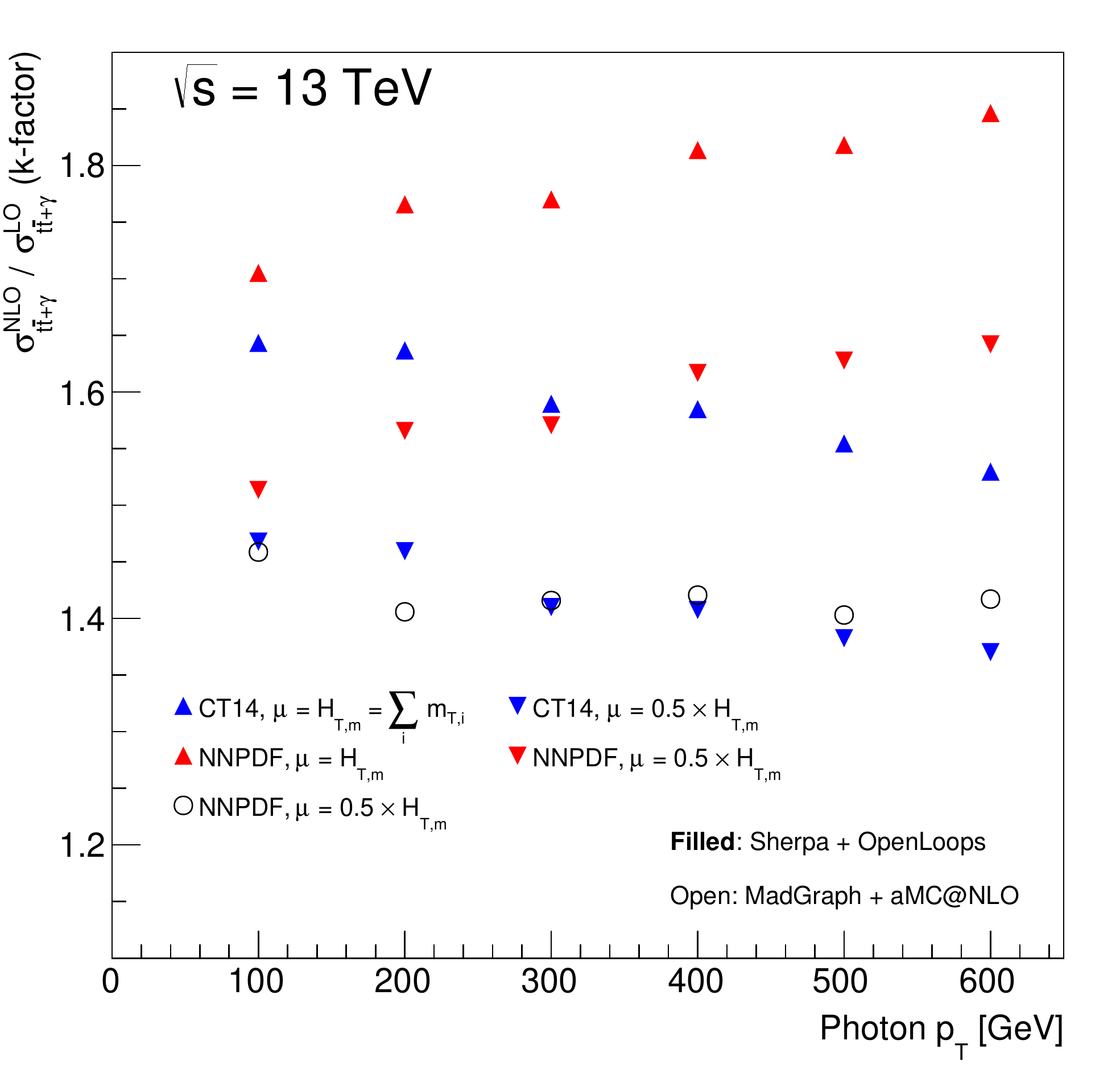}\includegraphics[width=0.5\textwidth]{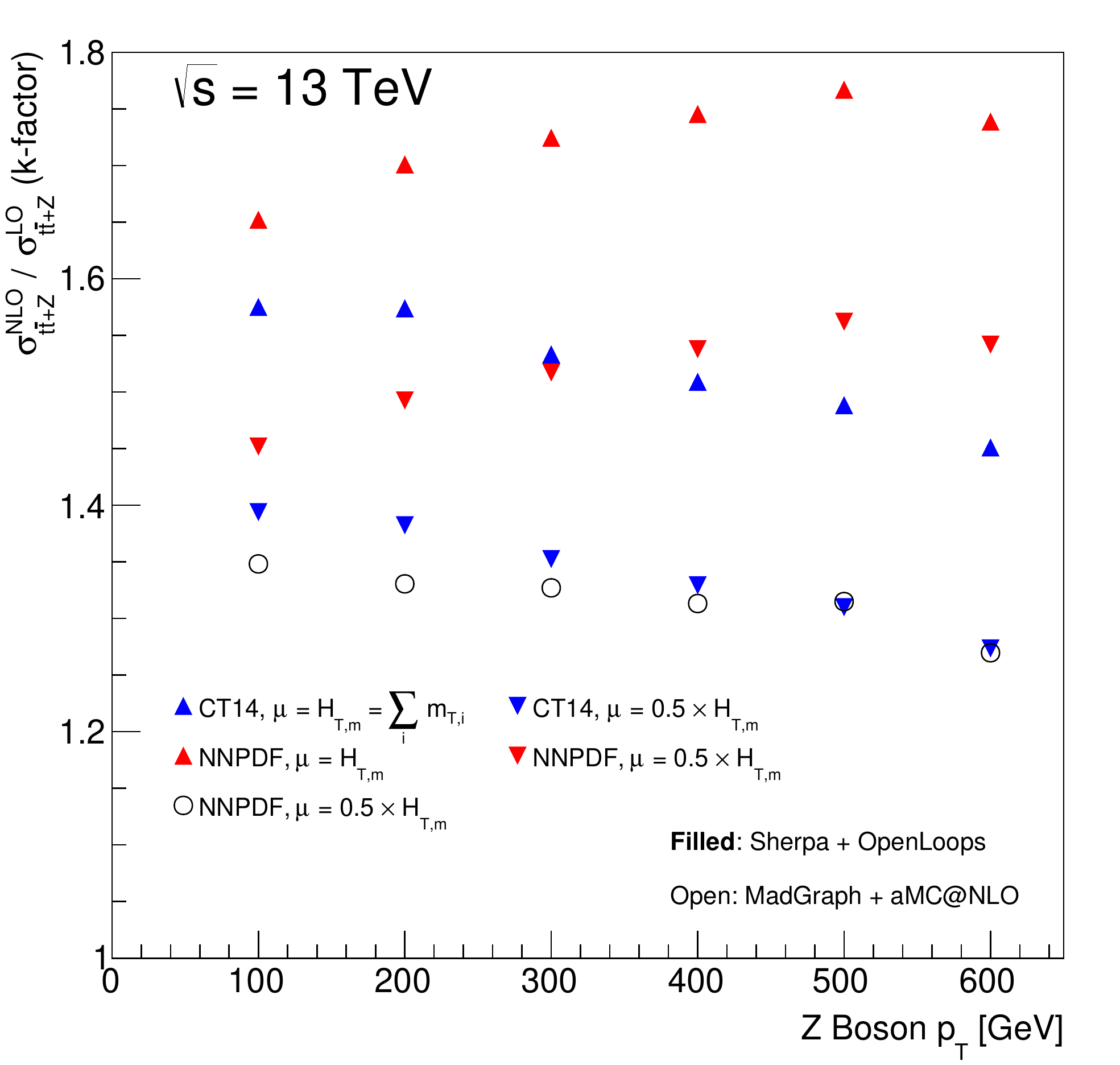}
\caption{Left: The $k$-factor for $t\bar{t}+\gamma$ (left) and $t\bar{t}+Z$ (right) as a function of the boson $p_\text{T}$ for {\sc Sherpa}+{\sc OpenLoops} and MG5\_aMC.  The PDF and scale choice are given in the legend.}
\label{fig:syst:ttv:kfactor}
\end{center}
\end{figure}

\begin{figure}[h!]
\begin{center}
\includegraphics[width=0.5\textwidth]{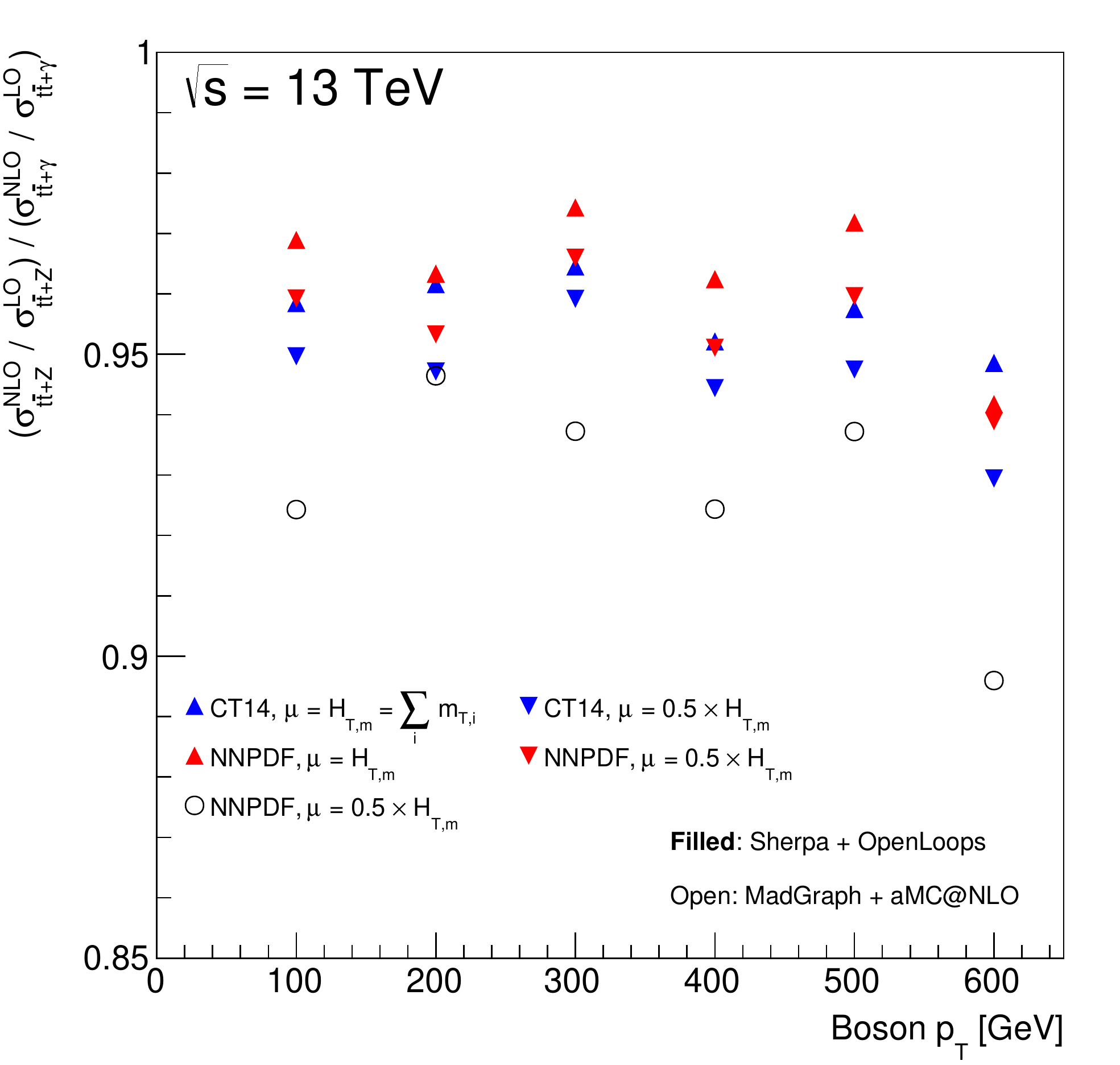}
\caption{The $k$-factor ratio between $t\bar{t}+\gamma$ and $t\bar{t}+Z$ as a function of the boson $p_\text{T}$ for {\sc Sherpa}+{
\sc OpenLoops} and MG5\_aMC.  The PDF and scale choice are given in the legend.}
\label{fig:syst:ttv:kfactor}
\end{center}
\end{figure}

		\clearpage
		
		\subsection{W+jets}
		\label{sec:susy:wjetsuncert}
		
Unlike all previous samples, both the diboson and $W$+jets backgrounds are estimated using {\sc Sherpa} as the nominal MC generator.  Both of these processes require many extra hard jets to pass the event selection and only {\sc MadGraph}, {\sc Sherpa}, and {\sc Alpgen} have this capability.  Both {\sc Sherpa} and MG5\_aMC can model extra jets at NLO and the {\sc Sherpa}+{\sc OpenLoops} setup is used at $\sqrt{s}=13$ TeV for two extra partons at NLO and four partons at LO.  Table~\ref{systematictheoryuncertswjets} summarizes the systematic uncertainties for $W$+jets at both $\sqrt{s}=8$ and $13$ TeV. At both enerties, one of the main systematic uncertainties is from scale variations to probe shape differences that could change the extrapolation from the CR to the SR.  The modeling of the extra radiation is probed at $\sqrt{s}=8$ TeV by varying the number of partons in the matrix element (see Sec.~\ref{ttzsec:simulationbased}).  At $\sqrt{s}=13$ TeV, the extra radiation is varied by changing the {\sc Sherpa} resummation scale {\tt QSF} as well as comparing the {\sc Sherpa} sample with a simulation from MG5\_aMC+{\sc Pythia 8} with up to four extra partons in the matrix element.  This comparison simultaneously varies the ME setup and the parton shower\footnote{Ideally, these variations would be decomposed, but it is not possible to vary only the PS within the {\sc Sherpa} framework.}.

In addition to the standard comparisons listed above, there is an additional source of uncertainty due to the extrapolation from a mostly no $b$-jet region in the $W$+jets CR (with a $b$-tag veto) to a mostly $b$- and $c$-jet selection in the SR.  Section~\ref{wjets:flavorextrap} showed that there is little flavor dependence on the $m_\text{T}$ extrapolation, but nonetheless it is important to estimate the extrapolation in flavor on the yield in the SR.  A $25\%$ uncertainty from the ATLAS $W+bb$ cross-section measurement~\cite{Aad:2013vka} is combined with a $\sim 15\%$ uncertainty from {\sc Alpgen} 2.14~\cite{Mangano:2001xp} parameter variations\footnote{{\sc Alpgen} is combined with {\sc Herwig} and the factorization, renormalization, matching scales are varied.  Additionally, the minimum $\Delta R$ between and minimum $p_\text{T}$ of partons are varied.  Strictly speaking these uncertainties only apply for {\sc Alpgen}, but because the jets beyond the leading two in {\sc Sherpa} are also at LO, these uncertainties may be a useful proxy for the {\sc Sherpa} sample as well.} on the extrapolation from two jets to the four jets as required by all signal regions.  This prescription is certainly conservative, as it includes a total cross-section uncertainty that should be canceled by the control region normalization.

 \begin{table}[h!]
\centering
\label{my-label}
\noindent\adjustbox{max width=\textwidth}{
\begin{tabular}{|c|cc|}
\hline
Source & Procedure ($\sqrt{8}$ TeV) & Procedure ($\sqrt{13}$ TeV) \\
\hline
Inclusive cross-section & N/A (CR method) & N/A (CR method)\\
Parton momentum & PDF4LHC&Ignored\\
Differential cross-section & $\mu_f,\mu_r$& $\mu_f,\mu_r$\\
Merging / Matching &$n_\text{partons}$& Matching Scale; MG5\_aMC\\
Fragmentation Model &Ignored & {\sc Sherpa} and MG5\_aMC+{\sc Pythia} 8  \\
Amount of `Extra' Radiation  & Ignored& Resummation scale (QSF)  \\
$W$+HF & $28\%$ to $W+bb$& $28\%$ to $W+bb$  \\
\hline
\end{tabular}}
\caption{A summary of the theoretical modeling uncertainties for $W$+jets at $\sqrt{s}=8$ TeV and $\sqrt{s}=13$ TeV.}
\label{systematictheoryuncertswjets}
\end{table}

		\subsection{Dibosons}
		\label{sec:susy:VVuncert}
		
	The uncertainties for dibosons at $\sqrt{s}=8$ TeV are similar to the analogous $W$+jets ones from Sec.~\ref{sec:susy:wjetsuncert}.  Due to the sub-dominance of dibosons for the $\sqrt{s}=13$ TeV SR, a crude and likely conservative approach compares the {\sc Sherpa} sample with up to three extra partons in the ME to a {\sc Powheg-Box}+{\sc Pythia} 8 sample with no extra partons in the matrix element.  This is a simultaneous variation of the matrix element calculation and the parton shower.  Figure~\ref{fig:dibosons} shows that the leading jet $p_\text{T}$ spectrum is similar for the two generators, but the number of jets and the $p_\text{T}$ spectrum of the subleading jets are significantly different.  The extra jet activity is likely significantly underestimated by the {\sc Powheg-Box} sample, but serves as a crude and likely conservative approach. Table~\ref{systematictheoryuncertsVV} summarizes the diboson uncertainties.

 \begin{table}[h!]
\centering
\label{my-label}
\noindent\adjustbox{max width=\textwidth}{
\begin{tabular}{|c|cc|}
\hline
Source & Procedure ($\sqrt{8}$ TeV) & Procedure ($\sqrt{13}$ TeV) \\
\hline
Inclusive cross-section & $5$-$7\%$ & $6\%$\\
Parton momentum & PDF4LHC&Ignored\\
Differential cross-section & $\mu_f,\mu_r$& Ignored\\
Merging / Matching &Matching scale, $n_\text{partons}$& {\sc Powheg-Box}\\
Fragmentation Model &Ignored & {\sc Sherpa} and {\sc Pythia} 8  \\
Amount of `Extra' Radiation  & Ignored& Ignored  \\
\hline
\end{tabular}}
\caption{Theoretical modeling uncertainties for dibosons.}
\label{systematictheoryuncertsVV}
\end{table}	

\begin{figure}[h!]
\begin{center}
\includegraphics[width=0.5\textwidth]{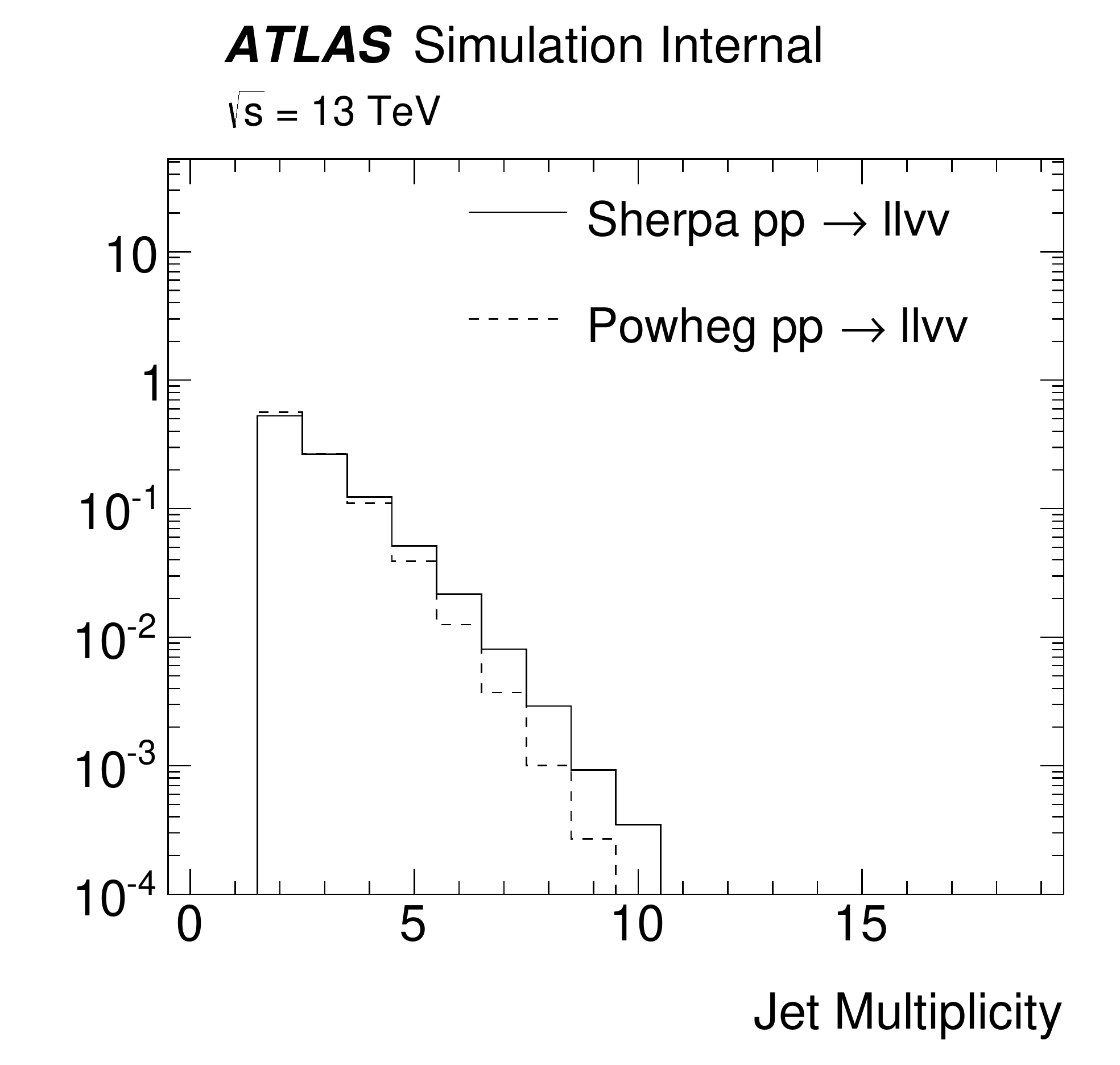}\includegraphics[width=0.5\textwidth]{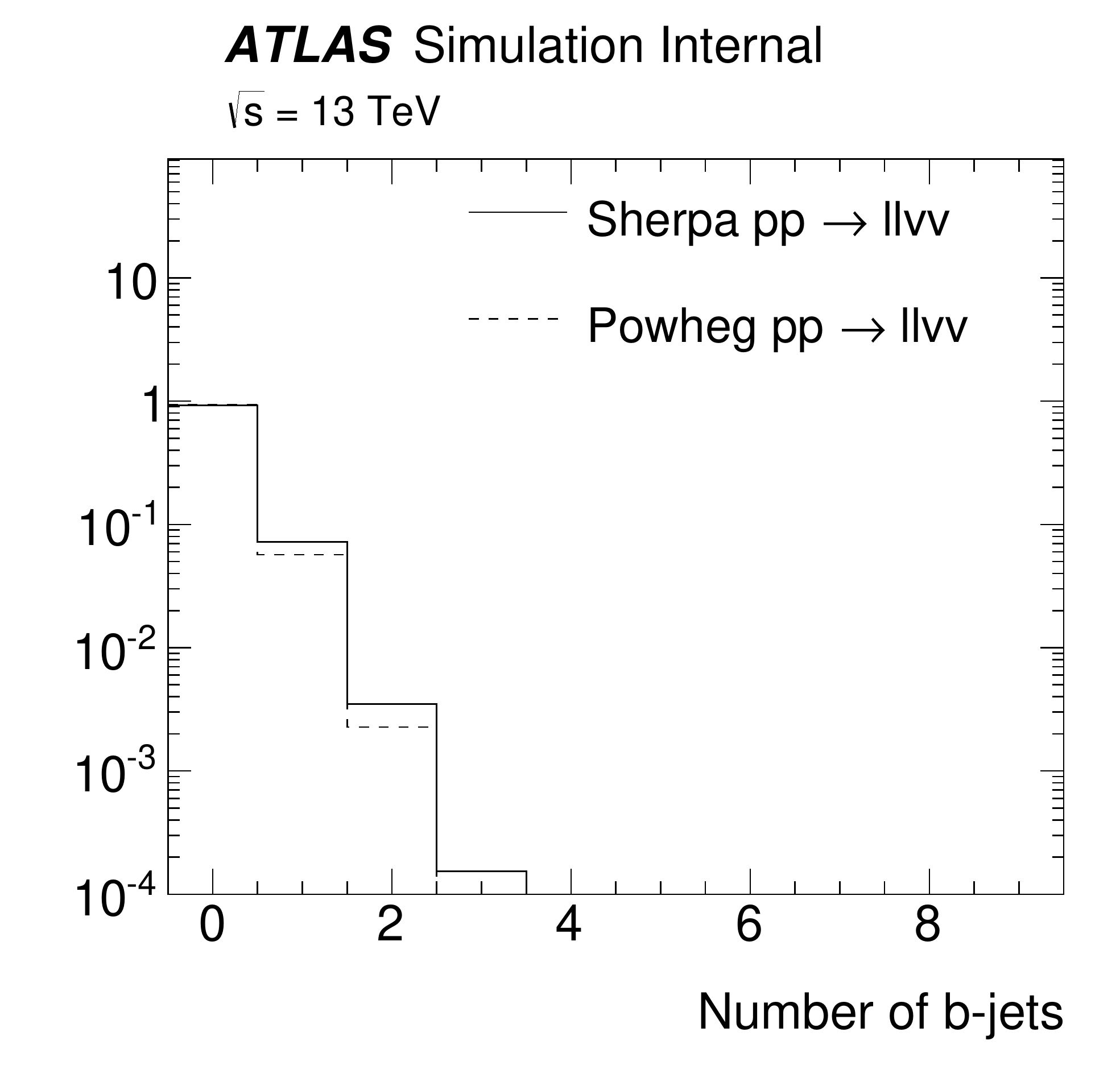}\\
\includegraphics[width=0.5\textwidth]{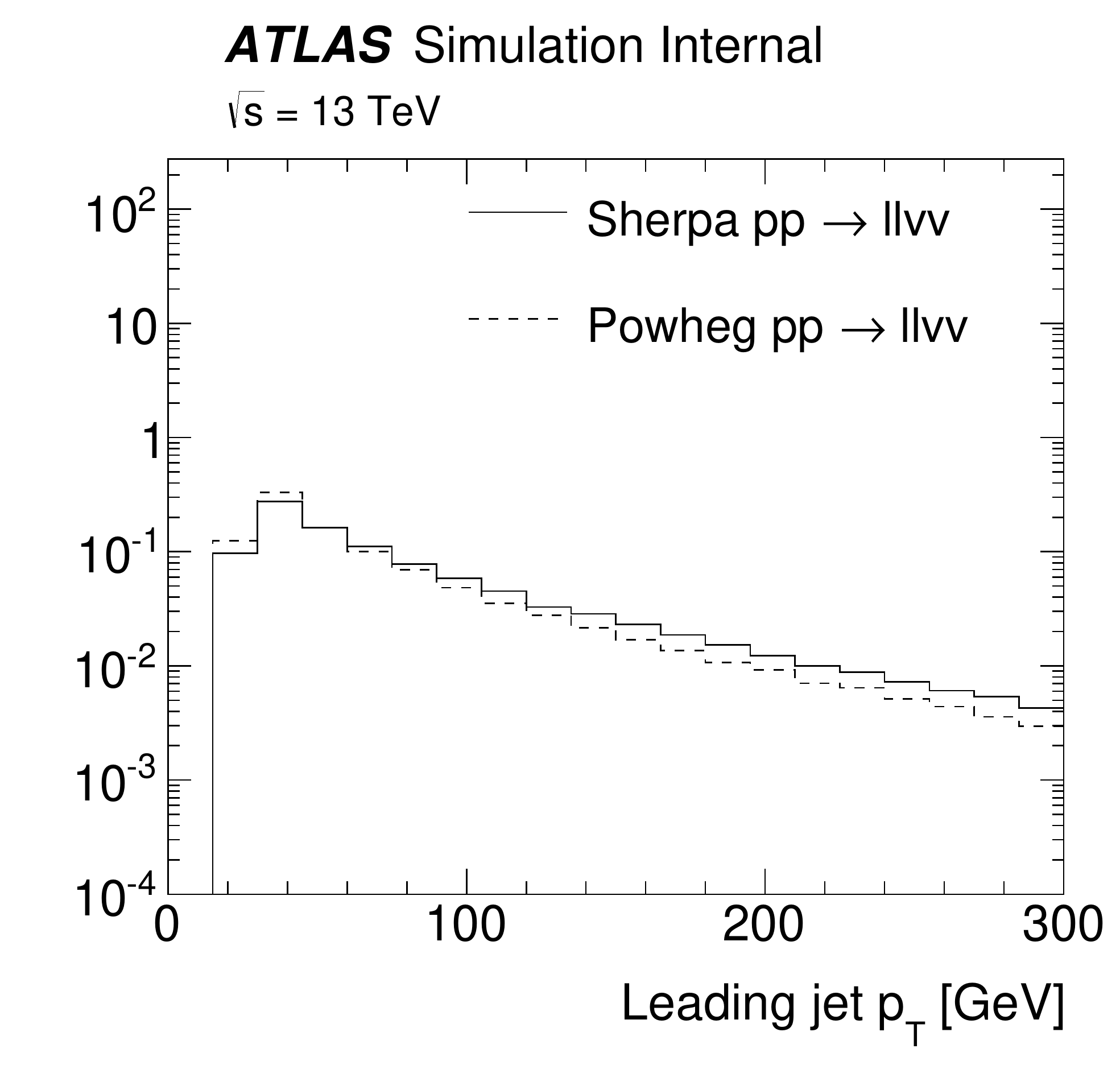}\includegraphics[width=0.5\textwidth]{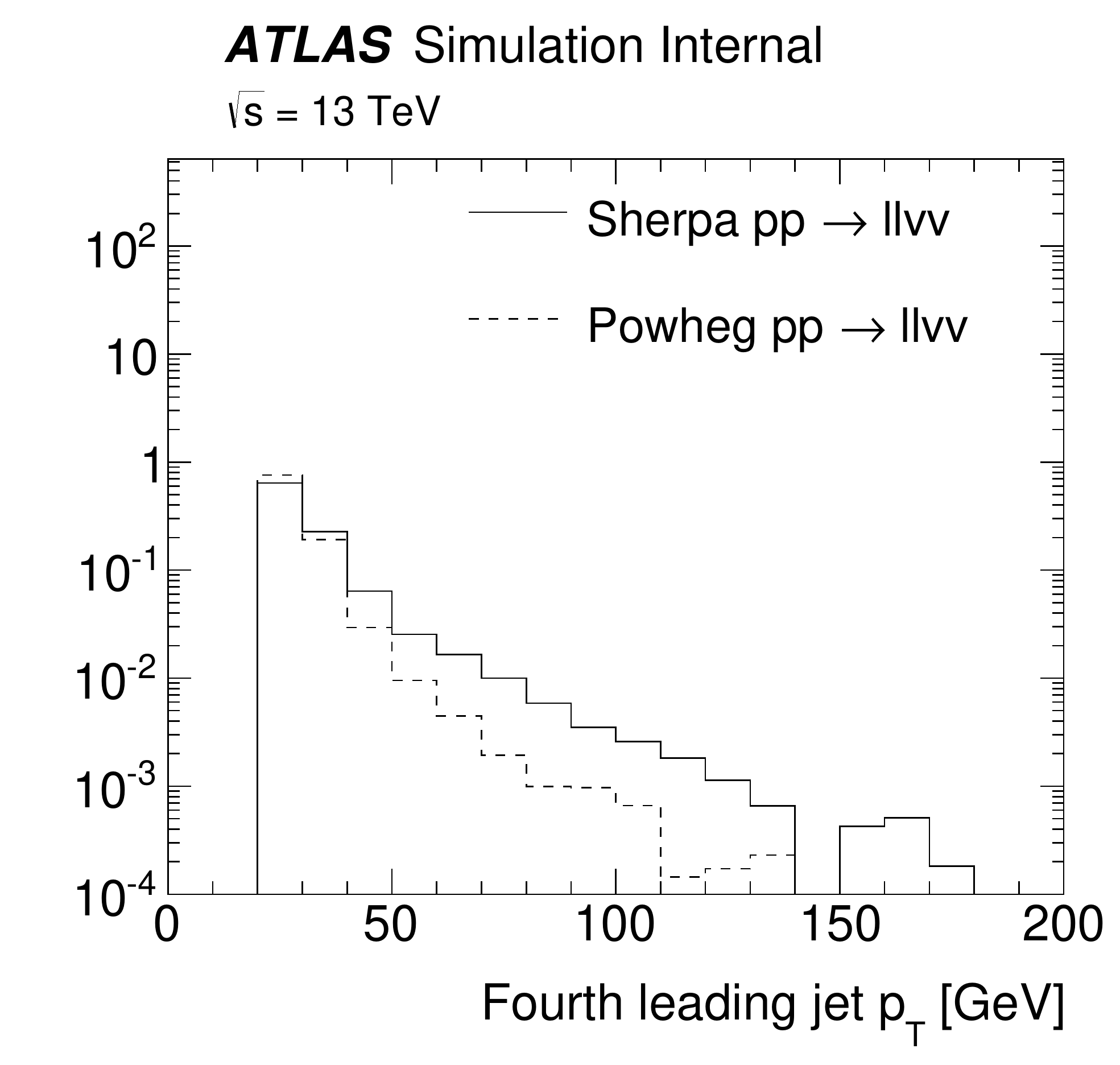}
 \caption{Various comparisons of jet related quantities between {\sc Sherpa} and {\sc Powheg-Box}+{\sc Pythia} 8 for the process $ll\nu\nu$. There is a $\sim 50\%$ difference between the predicted yields at preselection and an additional $\sim 20\%$ from extrapolating between the preselection and the SR.}
 \label{fig:dibosons}
  \end{center}
\end{figure}	
		
	\clearpage	
		
	\section{Summary}
	\label{sec:susy:summary:uncerts}

Table~\ref{tab:systematicsummary} presents a summary of the uncertainties for the signal regions with the full $\sqrt{s}=8$ TeV data and the early $\sqrt{s}=13$ TeV data.  These uncertainties are shown after the control region method is applied, so the any coherent uncertainties between the CR and SR are eliminated.  The data statistical uncertainty dominates over the systematic uncertainties for the single bin regions, while the systematic uncertainty is much larger than the statistical uncertainty for the more inclusive shape fit signal region.  The JES and JER uncertainties are the largest experimental uncertainties.  For the single bin regions that have higher $m_\text{T}$ thresholds and are thus more sensitive to the resolution tail beyond $m_\text{T}=m_W$, the JER is a bigger uncertainty than the JES.  At $\sqrt{s}=8$ TeV, there are only three components to the $b$-tagging efficiency uncertainty, while at $\sqrt{s}=13$ TeV, two additional components describe various extrapolation uncertainties (see Sec.~\ref{sec:susy:btag}).  The $b$-jet component of the $b$-tagging efficiency uncertainty is about $2\%$ in all regions and the other components are only relevant for the single bin regions.  This is due in part to the presence of $c$-jets which can allow events to exceed stringent $am_\text{T2}$ requirements.  The luminosity uncertainty is small because most of the backgrounds are normalized using control regions; this is especially true for tN13 (only dibosons are directly from MC) and the shape fit region for which the backgrounds other than $t\bar{t}$ are small. This also explains why the uncertainty from single top and other backgrounds is small for the shape fit region.  For example, the interference between $Wt$ and $t\bar{t}$ results in a $30\%$ uncertainty for the $\sqrt{s}=8$ TeV analyses, but due to the relatively small fraction of single top events in the SR, the total impact of this uncertainty is only a few percent at most.  Inclusive cross-section uncertainties are only relevant when the background process is not normalized in a control region.  The total systematic uncertainty is between $10\%$ and about $20\%$ in all regions.

 \begin{table}[h!]
\begin{center}
\noindent\adjustbox{max width=\textwidth}{
\begin{tabular}{|c|c|cccc|c|c|c|}
\hline
Type & Source & tN11 & tN12 & tN21 & tN22 & tNmed & tNhigh & tN13\\
     \hline
    \hline
\multirow{10}{*}{Experimental} & JES (leading) & 13\% & 10\% & 12\% & 12\% & 8\% & 6\% &6\% \\
& JES (subleading) & 9\% & 7\% & 8\% & 8\% & N/A & N/A & 3\% \\
& JER & 9\% & 9\% & 8\% & 6\% & 11\% & 11\% & 13\% \\
& $b$-tagging ($b$-jets) & 2\% & 2\% & 2\% &2\% & 2\% & 2\% & 2\% \\
& $b$-tagging ($c$-jets) & -- & -- & -- & -- & 1\% & 2\% & 2\% \\
& $b$-tagging (light-jets) & -- & -- & -- & -- & 1\% & 1\% & 1\% \\
& $b$-tagging (other) & N/A & N/A & N/A & N/A & N/A & N/A & 2\% \\
& $E_\text{T}^\text{miss}$ soft-scale & 5\% & 3\% & 5\% & 2\%& 1\% & 2\% & 1\% \\
& $E_\text{T}^\text{miss}$ soft-resolution & 1\% & -- & -- & -- & 1\% & 2\% & 1\% \\
& Other Experimental & 2\% & 2\% & 2\% & 1\% & 1\% & 1\% & 1\% \\
& Luminosity & -- & -- & -- & -- & 1\% & 1\% & -- \\
       \hline
\multirow{13}{*}{Theoretical} & $t\bar{t}$ Fragmentation  & 3\% & 2\% & 1\% & -- & 1\% & 3\% & 2\% \\    
& $t\bar{t}$ Extra Radiation  & 6\% & 2\% & 1\% & 5\% & 2\% & 4\% & 5\% \\     
& $t\bar{t}$ Hard-scatter  & 1\% & 1\% & 2\% & 3\% & 1\% & 1\% & 2\% \\  
& $Wt$ Cross-section  & -- & -- & -- & -- & 1\% & -- & N/A \\    
& $Wt$ Fragmentation  & -- & -- & -- & -- & -- & -- & 1\% \\    
& $Wt$ Extra Radiation  & -- & -- & -- & -- & -- & -- & -- \\     
& $Wt$ Hard-scatter  & -- & -- & 1\% & -- & 1\% & 1\% & -- \\    
& $Wt$/$t\bar{t}$ Interference & 1\% & -- & 2\% & 1\% & 2\% & 3\% & 3\% \\  
& $W$+jets Modeling & 1\% & 1\% & 1\% & -- & 1\% & 2\% & 3\% \\  
& $W$+HF  & 1\% & -- & -- & -- & 2\% & 2\% & 5\% \\      
& $t\bar{t}+V$ Modeling & -- & -- & -- & 1\% & 3\% & 3\% & 6\% \\  
& $t\bar{t}+V$ Cross-section & -- & -- & -- & 1\% & 3\% & 3\% & N/A \\   
& $VV$ Total & -- & -- & -- & -- & 3\% & 2\% & 5\% \\   
    \hline
    \hline
    \multicolumn{2}{|c|}{Total Systematic Uncertainty} &16\%- & 13\% & 13\% & 12\% & 17\% & 19\% & 21\% \\ 
    \multicolumn{2}{|c|}{Data Statistical Uncertainty} & 9\% & 8\% & 10\% & 7\% & 28\% & 45\% & 90\% \\ 
   \hline    
\end{tabular}}
\caption{A summary of the uncertainty in the total signal region yield after the background-only fit from the control region method (see Sec.~\ref{sec:susy:stats}).  If an uncertainty is less than $1\%$, it is marked with `--' while uncertainties that are not applicable are labeled N/A.  PDF uncertainties are included in the HS uncertainty.  tNxy is the $(x+1)^\text{th}$ $E_\text{T}^\text{miss}$ bin and $(y+2)^\text{th}$ $m_\text{T}$ bin of the shape fit.  When there is more than one JES nuisance parameter, there are at least three, but only the two biggest ones are shown here.  Due to correlations in the uncertainties after the fit, the total systematic uncertainty is not the sum in quadrature of the individual uncertainties.}
  \label{tab:systematicsummary}
\end{center}
\end{table}
 	\chapter{Search Results}	
	\label{chapter:results}
			
		Unfortunately, despite extensive efforts to search for stops with the $\sqrt{s}=8$ and early $\sqrt{s}=13$ TeV datasets, there is no significant evidence for a deviation from the Standard Model.  Section~\ref{sec:susy:stats} describes the statistical framework used to quantify the compatibility with the SM and to set limits on models of SUSY.  The statistical fit from Sec.~\ref{sec:susy:stats} is exercised in Sec.~\ref{CR-onlyFit} using only the control regions and the fidelity of the predictions are tested in validation regions in Sec.~\ref{validationregions}.  Limits on stop models are described in Sec.~\ref{susy:limits} for each of the signal regions, including the evolution of sensitivity with more data and technique improvements.  The chapter and Part~\ref{part:susy} ends in Sec.~\ref{epi} with a broad overview of all ATLAS and CMS Run 1 SUSY searches
			
		\clearpage	
			
		\section{Statistical Methods}
		\label{sec:susy:stats}
	
		Formally, the statistical analysis of the search results is a hypothesis test with the null hypothesis $H_1=\text{\bf SM only}$ and the alternative hypothesis $H_0=\text{\bf SM$+$stop}$.  A given signal model is excluded if the corresponding null hypothesis is rejected\footnote{The setup is different when optimizing the sensitivity of the test to {\it discover} SUSY; in that case the null hypothesis is the SM only case.  The focus of Chapter~\ref{chapter:results} will be on the exclusion of signal models given the lack of a significant excess in any signal region.}.  By the Neyman-Pearson lemma~\cite{Neyman289}, for a fixed upper bound on the probability of rejecting the null hypothesis when it is true (type 1 error), the likelihood ratio test minimizes the probability of not rejecting the null when the alternative is true (type II error) i.e. maximizes the probability of rejecting the SM-only hypothesis when there is SUSY.  The likelihood function is given by
		
		\begin{align}\nonumber
		\label{susy:likelihood}
		L(\nu,\vec{\theta},\vec{\mu})&:=p(\vec{n},\vec{\theta}_0|\nu,\vec{\mu},\vec{s}(\vec{\theta}),\vec{b}(\vec{\theta}))=\prod_{i=1}^{n_\text{bins}} \Pr(n_i|\nu,\vec{\mu},\vec{s}(\vec{\theta}),\vec{b}(\vec{\theta})) \times p(\vec{\theta}|\vec{\theta}_0)\\
		&=\prod_{i=1}^{n_\text{bins}} \frac{\left(\nu s_i(\vec{\theta})+\sum_{k=1}^{n_\text{backs}} \mu_k b_{ki}(\vec{\theta})\right)^{n_i}}{n_i!}e^{-\left(\nu s_i(\vec{\theta})+\sum_{k=1}^{n_\text{backs}} \mu_k b_{ki}(\vec{\theta})\right)}\times p(\vec{\theta}|\vec{\theta}_0),
		\end{align}
		
		\noindent where there are $n_\text{bins}$ total SR and CR bins with MC predictions for $s_i$ signal events and $b_{ki}$ background events of the $k^\text{th}$ SM background process in bin $i$.  The values $\mu_k$ are the normalization factors.  For the processes without a data-driven background estimate, $\mu_k$ is fixed to unity.  The values $\theta_i$ are all of the nuisance parameters associated with each systematic uncertainty; the input values of these uncertainties are given by $\vec{\theta}_0$.  The last term $p(\vec{\theta}|\vec{\theta}_0)$ is the constraint on the nuisance parameters.  The nuisance parameters are constructed so that they are mostly independent and therefore $p(\vec{\theta}|\vec{\theta}_0)$ factorizes for each parameter $\theta_i$\footnote{Note that even though the input nuisance parameters are indepenent, the output $\theta_i$ can be correlated given the data.}.  For all theoretical modeling uncertainties and all systematic uncertainties for the shape fit region, $\theta_i|\theta_{0,i}$ follows a standard normal distribution.  The experimental systematic uncertainties in the single-bin regions and the dedicated signal model uncertainties in all regions are modeled with a standard log-normal distribution, $\theta_i|\theta_{0,i}\sim\exp(\mathcal{N}(0,1))$.  The impact of the nuisance parameters on the yield $y$ (such as $b_{ki}$ or $s_i$) is given by $y = y_0(1+\sum_i\theta_i(H(\theta_i)\sigma_i^++(1-H(\theta_i))\sigma_i^-))$, where $\sigma_i$ is the fractional uncertainty on the yield for systematic uncertainty source $i$ and $H$ is the {\it Heaviside step function}.  The purpose of $H$ is to allow the impact of `up' $(\sigma^+)$ and `down' $(\sigma^-)$ shifts of the nuisance parameter to asymmetric effects on the yield.  When $\sigma_i^+=\sigma_i^-=\sigma_i$, the contribution to the yield is simply $\theta_i\sigma_i$ as $H(x)+(1-H(x))=1$.  The {\it parameter of interest} in Eq.~\ref{susy:likelihood} is $\nu$, which is $1$ under $H_0$ and $\nu=0$ under $H_1$.  The {\it test statistic} used to perform the hypothesis test is the log of the {\it profile likelihood ratio}:
		
		\begin{align}
		\label{teststat}
		t = -2\ln\left(\frac{\max_{\vec{\mu},\vec{\theta}}L(1,\vec{\mu},\vec{\theta})}{\max_{\nu',\vec{\mu}',\vec{\theta}'}L(\nu',\vec{\mu}',\vec{\theta}')}\right).
		\end{align}
	
		\noindent Note that since $t$ is not a monotonic transformation of the likelihood ratio, its type II error is not optimal even in the absence of nuisance parameters\footnote{Interestingly, even though the log profile likelihood ratio is standard for the LHC collaborations, the Tevatron collaborations used the log ratio of the profile likelihoods, which is optimal in the absence of nuisance parameters.}.  However, the value of $\nu$ that maximizes $L(\nu,\vec{\mu},\vec{\theta})$ will be close to zero (no evidence for SUSY) and therefore $t$ is close to optimal.  Near-optimality is also true in general in the asymptotic limit of large event yields due to a result by A. Wald~\cite{1990256,Cowan:2010js}.  Profiling refers to the maximization of the likelihood over the nuisance parameters in Eq.~\ref{teststat}.  The maximized (`fitted') values of the $\theta_i$ and their post-fit uncertainty can deviate from zero and unity, respectively.  When this is significant, the relevant nuisance parameter is said to be `profiled'.  Profiling is revisited in Sec.~\ref{CR-onlyFit}.  The calculation of $t$ for the search results presented in Sec.~\ref{susy:limits} are implemented using HistFitter~\cite{Baak:2014wma} based on RooStats~\cite{Moneta:2010pm}, RooFit~\cite{Verkerke:2003ir}, and ROOT~\cite{Brun:1997pa} through HistFactory~\cite{Cranmer:1456844}.
	
		In addition to the likelihood in Eq.~\ref{susy:likelihood}, a useful related quantity is the CR-only likelihood that is identical to Eq.~\ref{susy:likelihood}, but with $\nu=0$ and the signal regions removed from the product:
		
		\begin{align}
		\label{susy:likelihood-cronly}
		L_\text{CR-only}(\vec{\theta},\vec{\mu})&=\prod_{i=1}^{n_\text{CR bins}} \Pr(n_i|\vec{\mu},\vec{b}(\vec{\theta})) \times p(\vec{\theta}|\vec{\theta}_0).
		\end{align}		
	
		\noindent The CR-only fit referred to in several places in earlier sections is simply $(\vec{\mu},\vec{\theta})=\text{argmax}_{\vec{\mu}',\vec{\theta}'}L_\text{CR-only}(\vec{\theta}',\vec{\mu}')$.  Since $\vec{\mu}$ is not directly constrained by a PDF in the likelihood, when the number of CR bins is equal to the number of normalization factors, the CR fit simply returns values of $\mu_i$ that solve the system of equations (or a subset/superset if there are fewer/more data-driven regions) in Eq.~\ref{scalefactors} and $\theta_i=0$.
	
		The distribution of the test-statistic $t$ can be estimated numerically by sampling from the distributions of the input stochastic variables or with asymptotic formulae~\cite{Cowan:2010js}.  Due to its computational simplicity and accuracy, the asymptotic approximation is used as default and a few signal models are checked with the full numeric approach.  The formula is based on the observation that in the asymptotic regime, the log likelihood approaches a (non-central) chi-square distribution~\cite{1990256}, for which the $p$-value can be readily computed:

		\begin{align}
		\label{eq:asimov}
		\text{$p$-value}=1-\Phi^{-1}\left(\sqrt{t_\text{asymptotic}}-\frac{1-\nu}{\sigma_\text{asymptotic}}\right),
		\end{align}
		
		\noindent where $\Phi$ is the standard normal cumulative distribution function and $t_\text{asymptotic}$ is the value of $t$ when the number of events in each bin is $\nu s_i(\vec{\theta}_0)+\sum_{k=1}^{n_\text{backs}} \hat{\mu}_k b_{ki}(\vec{\theta}_0)$ and $\sigma_\text{asymptotic}^2=(1-\nu)^2/t_\text{asymptotic}$.  The parameter $\hat{\mu}$ is the value of $\mu$ that maximizes $L$ given $\nu$ and $\vec{\theta}_0$; with these values, $\max_{\nu',\vec{\mu}',\vec{\theta}'}L(\nu',\vec{\mu}',\vec{\theta}')=L(\nu,\hat{\vec{\mu}},\vec{\theta}_0)$.  Amazingly, even though the asymptotic convergence is only $\mathcal{O}(1/\sqrt{N})$, the approximation in Eq.~\ref{eq:asimov} well-approximates the full value even when the number of events $N$ is $\gtrsim 10$.  Figure~\ref{fig:susy:clsasympotitc} illustrates the convergence for various values of $N$ using one signal region with one background process and one systematic uncertainty.  When there are $3$ background events and $2$ signal events, the asymptotic formula overestimates the $p$-value at the measured $t$-value by $15$-$20\%$, while when there are $12$ background events and $5$ signal events, the $p$-value under $H_1$ is only off by $3\%$ and the $p$-value under $H_0$ is over-estimated by about $10\%$.
				
\begin{figure}[h!]
\begin{center}
\includegraphics[width=0.5\textwidth]{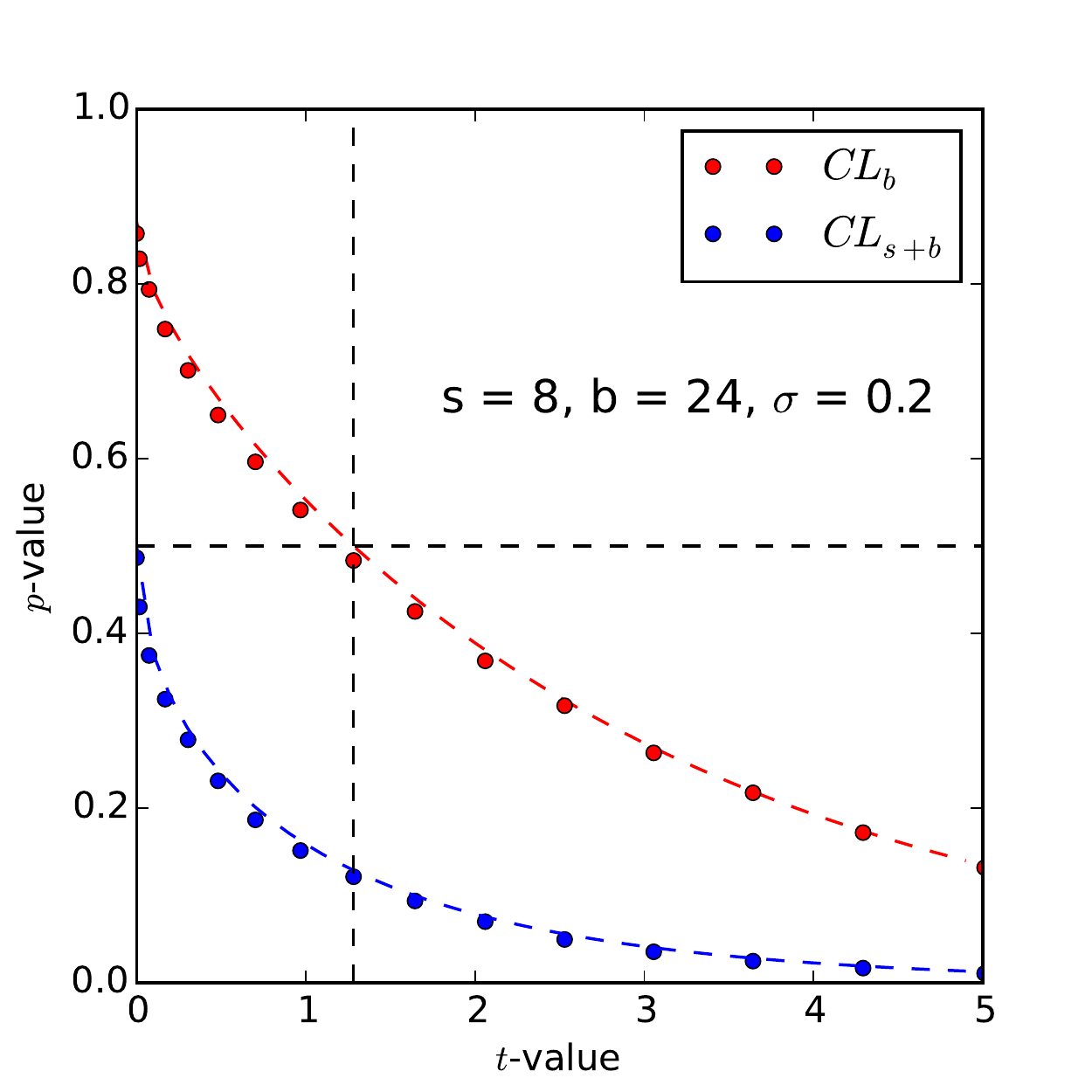}\includegraphics[width=0.5\textwidth]{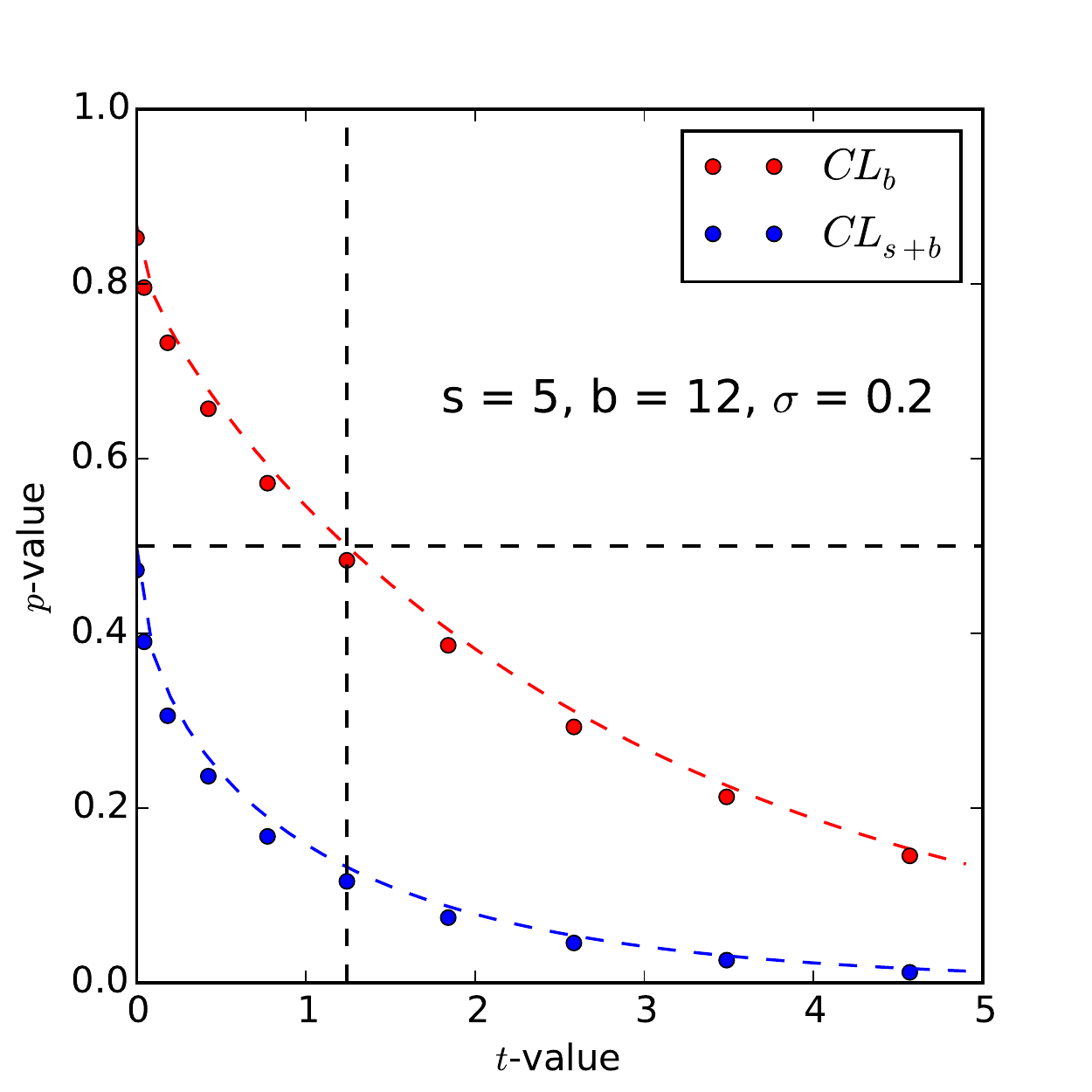}\\
\includegraphics[width=0.5\textwidth]{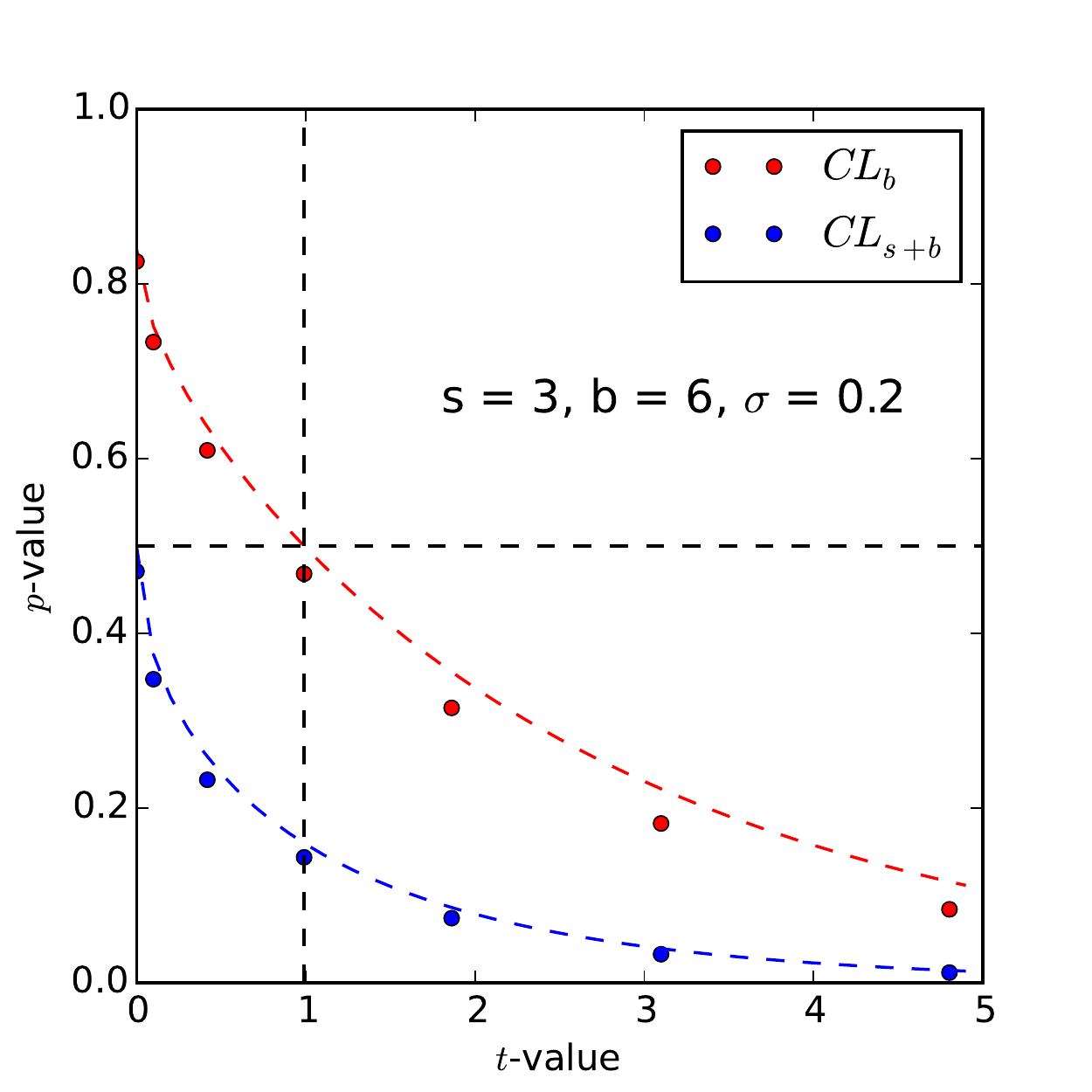}\includegraphics[width=0.5\textwidth]{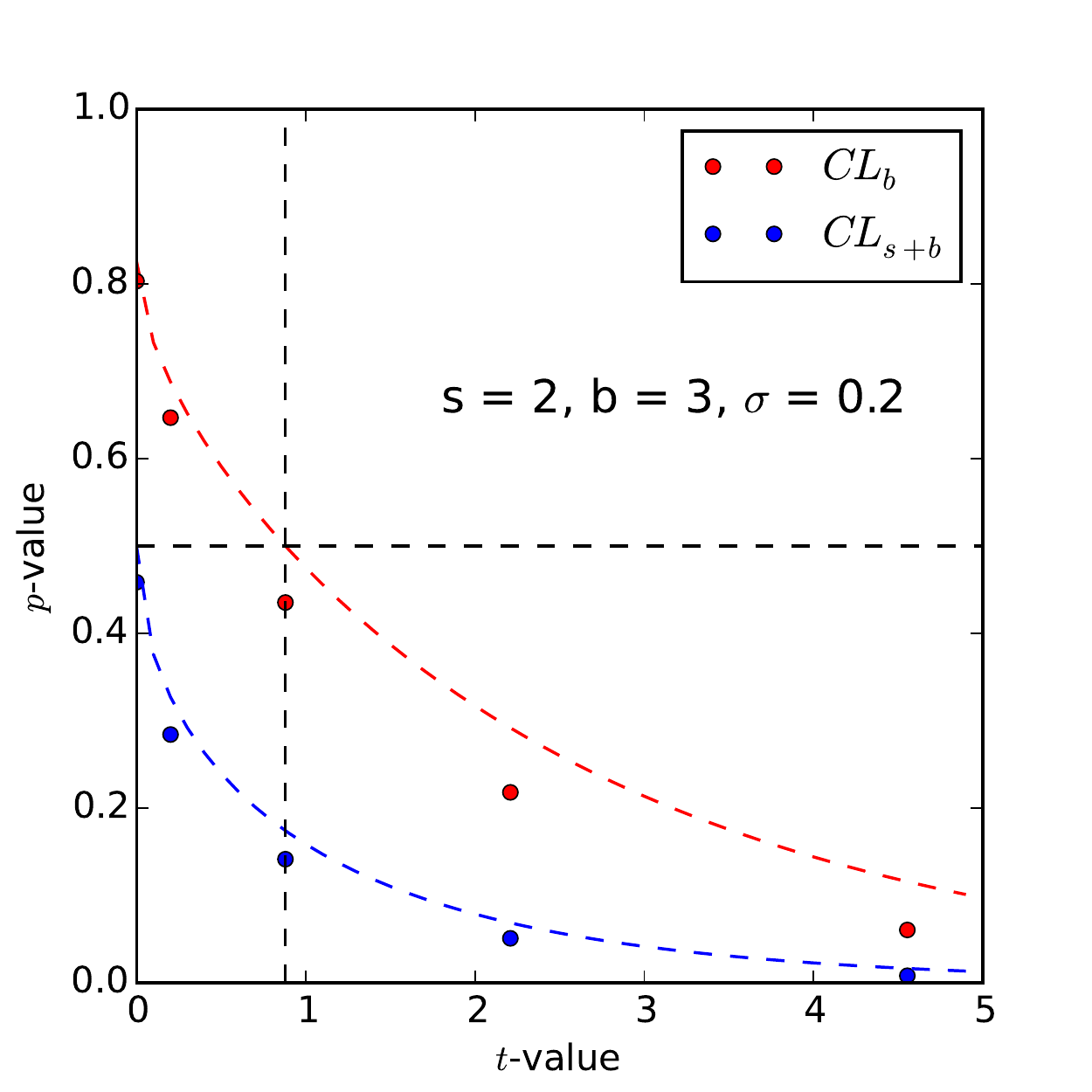}
 \caption{A comparison of the $p$-value under $H_1$ ($\text{CL}_b$) and under $H_0$ ($\text{CL}_{s+b}$) computed using numeric methods (10,000 toys) and the asymptotic formula given in Eq.~\ref{eq:asimov}. The vertical dashed line indicates the observed value of $t$ and the dashed line is at the $p$-value corresponding to the observed $t$-value using the asymptotic formula.}
 \label{fig:susy:clsasympotitc}
  \end{center}
\end{figure}				
				
	A given signal model is excluded if the $p$-value from the test described above is sufficiently small.  One undesirable feature of the $p$-value under $H_0$ which is a general property of two-model hypothesis testing is that the value can be small even if the data are inconsistent with both the SM and SM-only hypotheses.   For example, suppose that for a one-bin signal region there are $M\gg 1$ predicted background events and $\epsilon \ll M$ predicted signal events.   The $p$-value under $H_0$ when there are $N\ll M$ observed events will be small, but the $p$-value under $H_1$ will also be small.  This is a general feature when comparing two models that do not partition the space of all possible models and in particular when the null and alternative hypotheses are similar.  In high energy physics, the standard solution to this problem is to regulate the $p$-value under the null hypothesis ($\text{CL}_{s+b}$) by dividing by the $p$-value under the alternative hypothesis ($\text{CL}_{b}$) to form the $\text{CL}_s=\text{CL}_{s+b}/\text{CL}_{b}$ value~\cite{Junk:1999kv,Read:2002hq}.   This new quantity has the property that it will be large in the example described above, i.e. when both the null and alternative hypothesis are inconsistent with the data.  The community standard is to treat $\text{CL}_s$ {\it as if it were a proper $p$-value} by declaring a model excluded when $\text{CL}_s<0.05$.  However, it should be noted that the $\text{CL}_s$ is not a $p$-value and is not unique.  Any function $f(x)$ that has the property $\lim_{x\rightarrow 0}f(x)=0$ will be able to regulate the $\text{CL}_{s+b}$ by $\text{CL}_{s+b}/f(\text{CL}_{b})$.  One simple function is
	
	\begin{align}
	f_r(x) = \left\{\begin{matrix}x & x \leq r\cr 1 & x > r\end{matrix}\right.,
	\end{align}
	
	 \noindent where $0\leq r\leq 1$ is a fixed value.  One natural choice is $r=0.5$.  When $\text{CL}_{b}$ is small, this {\it regulated} $\text{CL}_{s}$ is enlarged just like the usual $\text{CL}_{s}$\footnote{The regulated $\text{CL}_s$ is similar to the idea of power-constrained limits in Ref.~\cite{Cowan:2011an}.  Without a proper loss function for Type 1 errors under the background-only hypothesis, there is no unique way to regulate the $p$-value.  Thank you K. Cranmer for pointing out this interesting paper.}.  However, when $\text{CL}_{b}>0.5$, a regime where presumably there is no need for the correction, the {\it power} ($=1-\Pr(\text{type II error})$) is strictly larger for the regulated $\text{CL}_{s}$ than for the usual $\text{CL}_{s}$.   Figure~\ref{regulatedCLs} demonstrates the increased power of the regulated $\text{CL}_{s}$.  For a fixed background yield in the left plot of Fig.~\ref{regulatedCLs} shows that the power of the regulated $\text{CL}_{s}$ interpolates between the baseline $\text{CL}_{s}$ definition and the $\text{CL}_{s+b}$, which is a proper $p$-value and has maximal power by the Neyman-Pearson lemma (with the caveats discussed above).  The right plot of Fig.~\ref{regulatedCLs} shows the minimum number of signal events that are needed to exclude a model given the number of estimated background events.  The regulated $CL_{s}$ requires about 15\% fewer signal events than the baseline $CL_{s}$ procedure.  More sophisticated choices for $f$ are possible to increase the power in the low $\text{CL}_{b}$ regime and still build in protection from the undesirable properties of $\text{CL}_{s+b}$.  Despite the promise of the regulated $\text{CL}_{s}$, the community standard is the baseline $\text{CL}_{s}$ and therefore it is important to use the same definition when comparing results with other analysis.  Thus, the baseline $\text{CL}_{s}$ is used for all subsequent results.
	
\begin{figure}[h!]
\begin{center}
\includegraphics[width=0.5\textwidth]{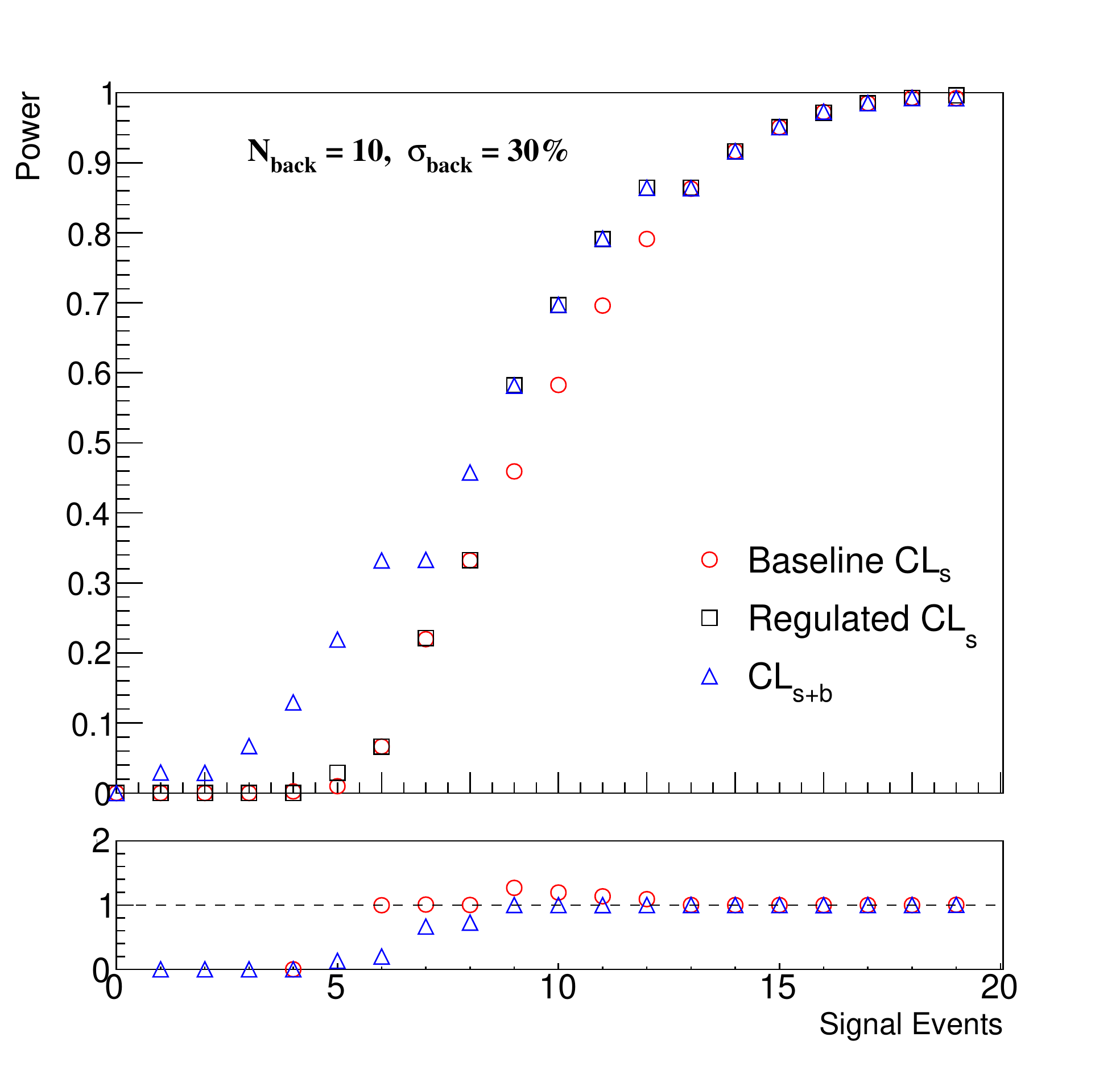}\includegraphics[width=0.5\textwidth]{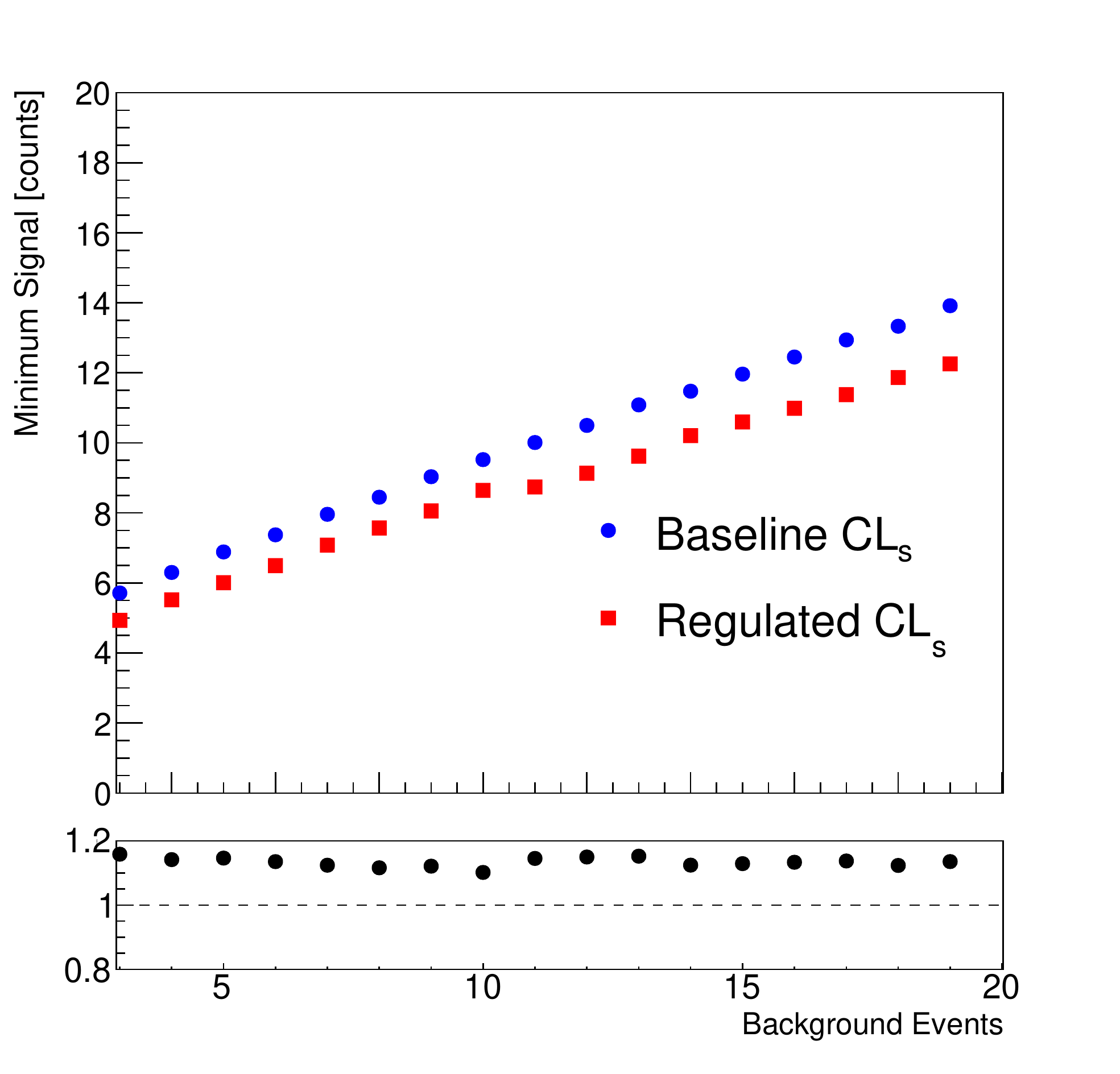}
\end{center}
\caption{Left: The power for three exclusion techniques (see the text for details) as a function of the number of signal events, for a fixed number of background events.  Right: The minimum number of signal events that are needed to exclude a model given the number of estimated background events.  A log-normal constraint is used to model the background uncertainty and $100,000$ toys are used for each $p$-value calculation. }
\label{regulatedCLs}
\end{figure}

	\clearpage		
		
		\section{CR-only Fit}
		\label{CR-onlyFit}
		
	Before describing the compatibility of the predicted yields with the observations in the signal regions, this section documents the outcome of the CR-only fit from maximizing Eq.~\ref{susy:likelihood-cronly}.  Figure~\ref{fig:susy:cronly:muvalues} shows the normalization factors $\mu_i$ for each signal region.  Except for the tN1shape regions, each signal region has dedicated $t\bar{t}$ and $W$+jets control regions that participate in the fit.  The $\sqrt{s}=13$ signal region additionally has the signal top and $t\bar{t}+\gamma$ control regions to constrain $\mu_\text{single top}$ and $\mu_\text{$t\bar{t}+Z$}$.  For each point, the outer error bar is the total uncertainty from the fit, including the impact of systematic uncertainties.  The inner error bar represents the control region statistical uncertainty and is determined by bootstrapping the data in the control regions and resolving the system of equations in Eq.~\ref{scalefactors}.  The single bin regions have the same number of control region bins as normalization parameters and so the central value from these fits are the same.  All of the control regions have $\mathcal{O}(100)$ events and so the statistical uncertainty is $\mathcal{O}(10\%)$ and is the dominant uncertainty for most regions.  The normalization factor uncertainties for the single top and $t\bar{t}+Z$ processes for SR13 are significantly larger than the corresponding $t\bar{t}$ and $W$+jets factor uncertainties due to the small yield ($t\bar{t}+Z$) and purity (single top) in the control regions.  The shape fit regions have one normalization parameter per $E_\text{T}^\text{miss}$ bin, as motivated in Sec.~\ref{sec:shapefitregion}.  Each $E_\text{T}^\text{miss}$ bin has four $m_\text{T}$ bins, so the CR-only fit (which is actually the full likelihood in Eq.~\ref{susy:likelihood} only with $\nu=0$) is already over-constrained.  For this reason, the system of equations method does not apply and so there are no inner error bars in the last three points in Fig.~\ref{fig:susy:cronly:muvalues}.  Across all bins, the $W$+jets scale factors are less than unity.  This is comparable to the $15$-$20\%$ over-estimation of the $W$+jets process by {\sc Sherpa} in the inclusive phase space probed by the ATLAS cross-section measurement reported in Ref.~\cite{Aad:2014qxa}.  The $t\bar{t}$ normalization factors are approximately consistent with unity, though there is a slight trend for $\lesssim 10\%$ upward corrections.  The single top normalization factor is much less than one, but its uncertainty is too large to make conclusions.  Despite the large uncertainty on the $t\bar{t}+Z$ normalization factor, it is significantly greater than one.  The inclusive $t\bar{t}+Z$ cross-section measurement at $\sqrt{s}=13$ TeV also observes an excess, but there is not enough events yet to determine if it is significant~\cite{ATLAS-CONF-2016-003}; the $\sqrt{s}=8$ TeV measurement does not see the same excess, though the statistical uncertainty is comparably large~\cite{Aad:2015eua}.  If the same (lower) $k$-factor is used for $\sqrt{s}=8$ TeV $t\bar{t}+\gamma$ validation region (see Sec.~\ref{sec:ttz:datadriven}) as for the $t\bar{t}+\gamma$ CR at $\sqrt{s}=13$ TeV, the data suggests a normalization factor that is also $\sim1.5$.
		
\begin{figure}[h!]
\begin{center}
\includegraphics[width=0.5\textwidth]{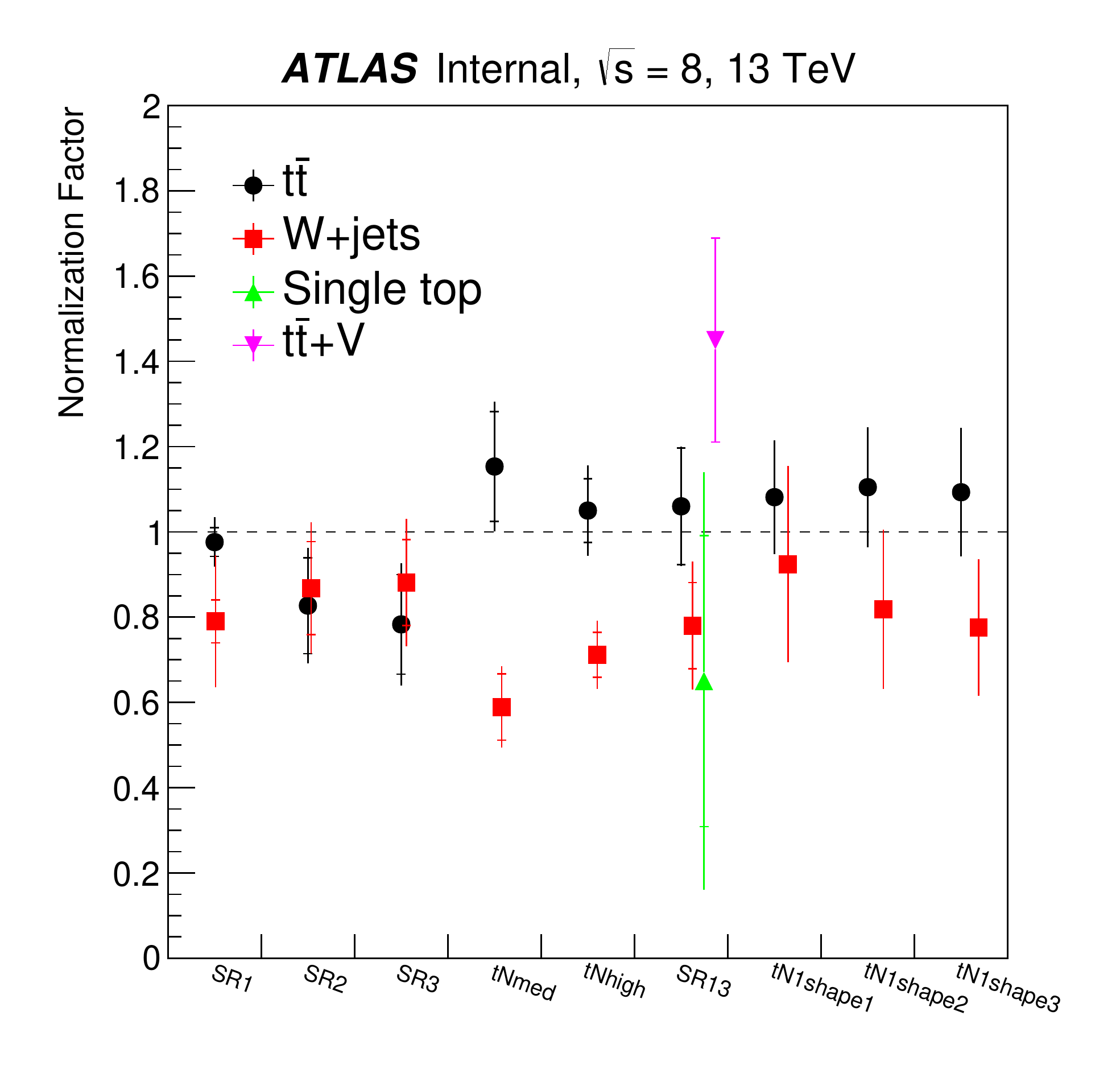}
 \caption{The $\mu$ values from the CR-only fit for all single bin signal regions.  There is one normalization factor per $E_\text{T}^\text{miss}$ bin for the tN1shape fit.  The outer error bars indicate the total post-fit uncertainty while the inner error bars show the statistical uncertainty only (see the text for details).}
 \label{fig:susy:cronly:muvalues}
  \end{center}
\end{figure}	
		
Another important aspect of the fit to investigate before showing the full results is the level of nuisance parameter profiling.   In the signal regions with equal numbers of control regions and normalization factors, the CR-only fit will not profile the nuisance parameters by construction.   However, once the signal regions are included and in general for the shape fit signal region, the fit is over-constrained and so the nuisance parameters can change from their initial values.  Figure~\ref{fig:susy:jesjerprofiling} shows the impact of a {\it background-only} fit using the control and signal regions associated with tNmed, tNhigh, and tN1shape for JES and JER nuisance parameters.  The background-only fit is identical to the CR-only fit, but including the data and simulation in the signal region (i.e. maximize Eq.~\ref{susy:likelihood} with $\nu=0$).   In the absence of profiling, the mean is zero and the standard deviation is unity.   As expected, since the number of events in the single bin regions is small compared to the number of events in the control regions, there is essentially no profiling of the jet energy related nuisance parameters.  In contrast, there is significant profiling of the JES and JER nuisance parameters for the tN1shape fit.   The six reduced nuisance parameters from the in-situ measurements (NP1-6) are ordered from the biggest to smallest impact on the jet energy scale.  Therefore, the largest profiling occurs for the first NPs (at the $60\%$ level) whereas there is little sensitivity and thus little profiling for NP5 and NP6.  Due to their size, there is also significant profiling for the nuisance parameters associated with the $\eta$ inter-calibration (increases with $|\eta|$), the jet flavor~\cite{ATLAS-CONF-2011-053,ATLAS-CONF-2013-002}, the technical closure\footnote{There was a small change in the simulation from the calibration sample to the applied sample.}, and pileup\footnote{This includes an uncertainty on the $\mu$, NPV, and $p_\text{T}$ dependence of the pileup corrections and an uncertainty on the modeling of the median pileup density $\rho$.  See Ref.~\cite{areasATLAS} for more detail.} (see Ref.~\cite{Aad:2014bia} for more details).  Interestingly, the largest change in the central value of a nuisance parameter is for the $b$-jet energy scale, which is reduced by about $25\%$.  The last JES nuisance parameter is associated with the high $p_\text{T}$ JES derived from the single-hadron response.  As this is only relevant for jets with $p_\text{T}\gtrsim 1$ TeV, the corresponding nuisance parameter is not profiled.   The last three points in Fig~\ref{fig:susy:jesjerprofiling} show the profiling of the JER uncertainty nuisance parameters, one per $E_\text{T}^\text{miss}$ slice.  Due to its significant impact on the $m_\text{T}$ shape, this parameter is significantly profiled in all three $E_\text{T}^\text{imss}$ regions.

\begin{figure}[h!]
\begin{center}
\includegraphics[width=0.99\textwidth]{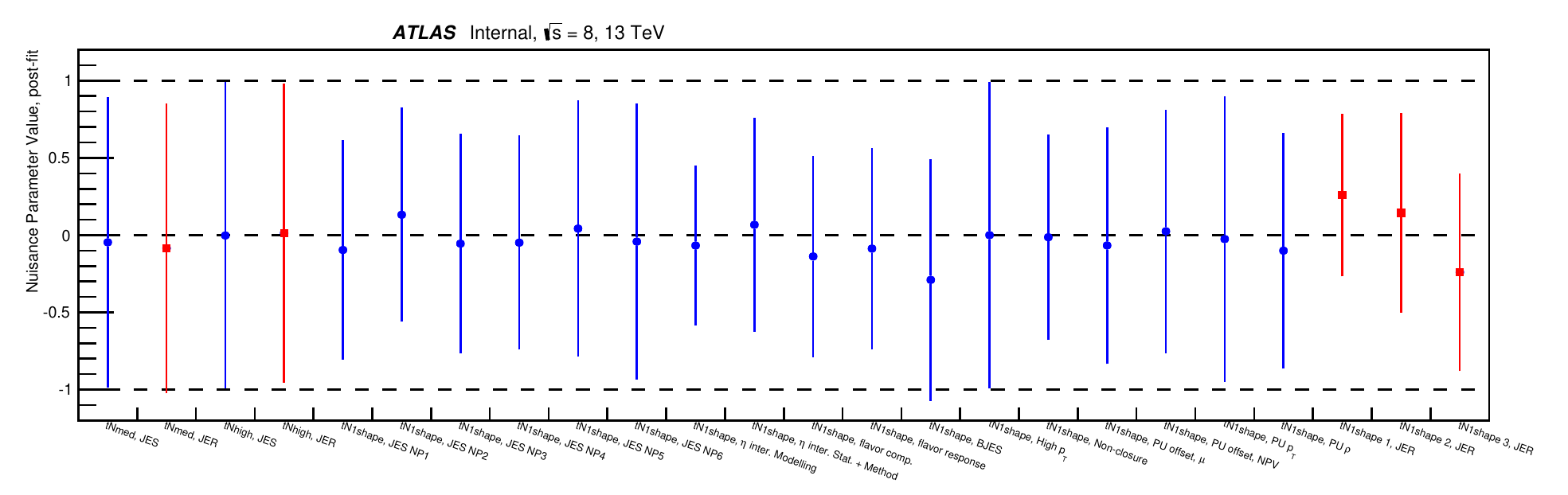}
 \caption{The post-fit JES and JER nuisance parameters for the background-only all-bins fit for tNmed, tNhigh, and tN1shape.  Red (blue) lines indicate the JER (JES) parameters.  Single bin (shape fit) regions are on the left (right).}
 \label{fig:susy:jesjerprofiling}
  \end{center}
\end{figure}		
			
		\clearpage	
			
	\section{Validation Regions}
	\label{validationregions}
		
		Data and predictions in the signal regions using the CR-only fit are shown in Fig.~\ref{fig:susy:pullplot}.  In addition to the signal regions, Fig.~\ref{fig:susy:pullplot} also shows comparisons for a set of {\it validation regions} that are kinematically between the control regions and signal regions.  The validation regions have the same selections as the corresponding control regions, but instead of $60$ GeV $<m_\text{T}<90$ GeV, they require $90$ GeV $<m_\text{T}<120$ GeV.  Both $t\bar{t}$ and $W$+jets validation regions are associated with tNmed, tNhigh, and tN13.  Overall, there is excellent agreement between the predictions and the data; the $\chi^2/\text{NDF}\sim 0.4$ with a $p$-value of about $98\%$\footnote{See Sec.~\ref{epi} for a discussion about the possible over-estimation of uncertainties.}.
		
\begin{figure}[h!]
\begin{center}
\includegraphics[width=0.65\textwidth]{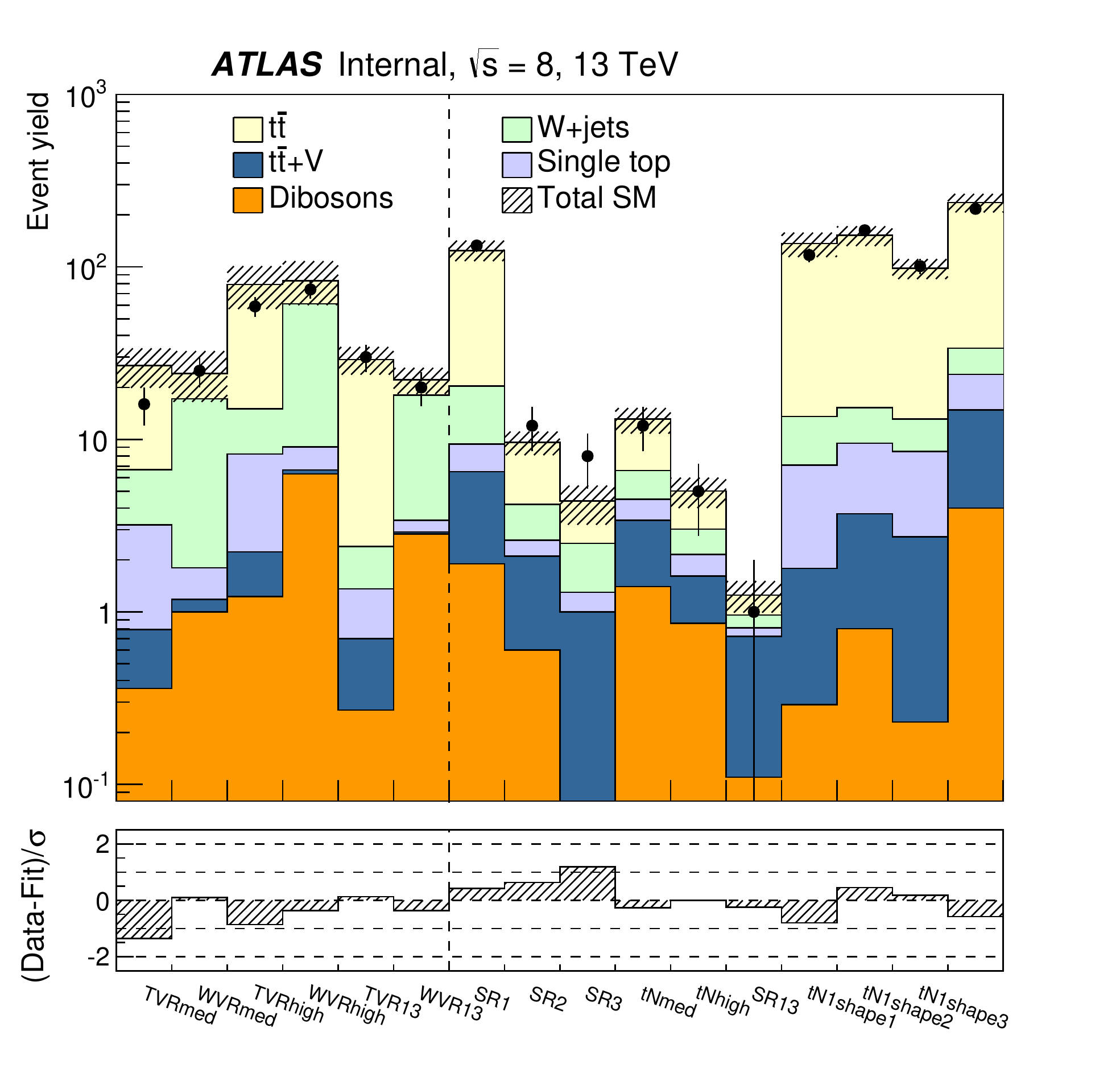}
 \caption{A comparison of the data and simulation in the validation and signal regions using the CR-only background fit.  The error bar in the ratio panel shows the {\it pull}, defined as the difference between the data and the prediction, divided by the uncertainty.  In this case, the uncertainty is the sum in quadrature of the data statistical uncertainty and the total background uncertainty.}
 \label{fig:susy:pullplot}
  \end{center}
\end{figure}		
		
		\clearpage	
			
		\section{Exclusion Limits}
		\label{susy:limits}
	
		In the absence of a significant excess, limits are set on simplified models with $\tilde{t}\rightarrow t\tilde{\chi}^0$.  The following sections will show a series of contours, similar to the schematic one shown in Fig.~\ref{fig:susy:limitexplain}.  The horizontal axis will be the stop mass, which sets the cross-section (see Fig.~\ref{sec:targetpheno}) and the vertical axis will be the neutralino mass or the mass difference between the neutralino mass and the stop mass, quantities which are responsible for determining how much phase space is available for the stop decay products.  A black dashed line indicates the {\it expected exclusion limit}, which is determined by computing the median $\text{CL}_s$ assuming that the data follow a Poisson distribution with mean value given by the SM prediction.  The $1\sigma$ systematic uncertainty is represented by a yellow band around the dashed line.  The exclusion limit from the observed data is represented by a solid red line and the theoretical cross-section uncertainty on the signal is represented by red dashed lines around the solid line.  Only the total cross-section uncertainty on the signal is included in the red dashed lines; all other uncertainties on the signal model are included in the yellow band.  Due to computing and storage limitations, a grid of models with approximately $50$ GeV spacing in $m_\text{stop}$ and $25$ GeV in $m_\text{LSP}$ is used to estimate the full contour.  A bilinear interpolation between grid points is performed using the significance, $\sigma=\Phi^{-1}(1-\text{CL}_s)$.
				
\begin{figure}[h!]
\begin{center}
\includegraphics[width=0.95\textwidth]{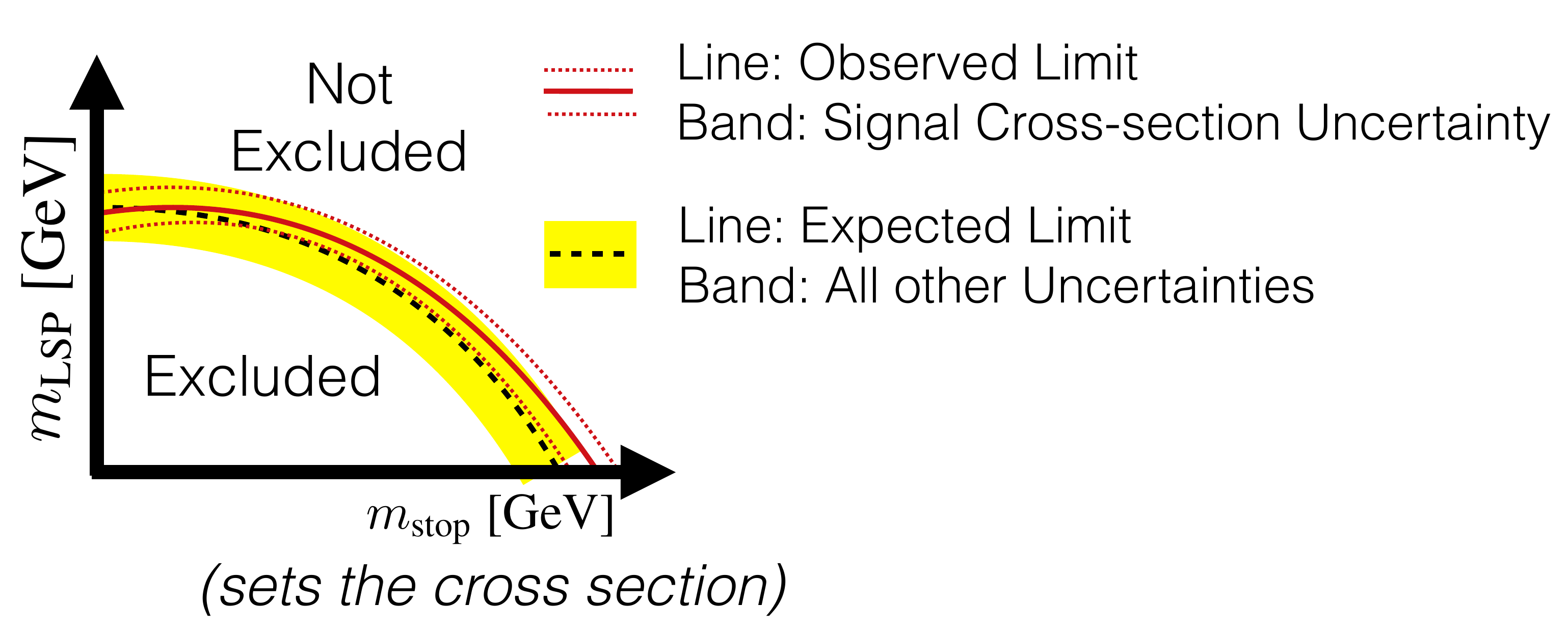}
 \caption{A schematic diagram demonstrating how the exclusion limits are presented.  See the text for details.}
 \label{fig:susy:limitexplain}
  \end{center}
\end{figure}	
		
\clearpage		
		
\subsection{Early $\sqrt{s}=8$ TeV Results}	
\label{results:early8TeV}

Figure~\ref{fig:susy:limitHCP} shows the exclusion contour in the $(m_\text{stop},m_\text{LSP})$ plane after collecting $13$~fb${}^{-1}$ of data at $\sqrt{s}=8$ TeV.   The three signal regions SR1-3 are combined using the mapping shown in Fig.~\ref{fig:susy:limitHCP2} based on the lowest expected $\text{CL}_s$ value.  The limits extend significantly beyond the full $\sqrt{s}=7$ TeV Run sensitivity~\cite{Aad:2012xqa}, pushing the limit up to about $m_\text{stop}=625$ GeV for a massless LSP.   There are three factors that led to the improved limit.  First, the total integrated luminosity at $\sqrt{s}=7$ TeV was only $4.7$ fb${}^{-1}$ resulting in a factor of $2.8$ more events with the $13$ fb${}^{-1}$ at $\sqrt{s}=8$ TeV.  Second, the increase in the center-of-mass energy increased the cross-section for stops by about a factor of $2$ for $m_\text{stop}\sim 500$ GeV.  The stop cross section is a factor of $4.6$ lower for $m_\text{stop}=625$ GeV ($\sqrt{s}=8$ TeV limit) than at $m_\text{stop}=500$ GeV ($\sqrt{s}=7$ TeV limit).  Therefore, there would be $20\%$ more $m_\text{stop}=625$ GeV stop events at $\sqrt{s}=8$ TeV than $m_\text{stop}=500$ GeV events at $\sqrt{s}=7$ TeV.  However, the number of background events also increases with $\sqrt{s}$.  The dominant $t\bar{t}$ cross section increases by at least $50\%$ between $\sqrt{s}=7$ and $\sqrt{s}=8$ TeV (see Fig.~\ref{fig:stopcrosssection}).  To achieve sensitivity to $m_\text{stop}=625$ GeV, a kinematically tighter event selection is required.  The third factor that led to the improvement in the early $\sqrt{s}=8$ TeV analysis is the addition of the $m_\text{T2}$ variables that allowed for a harsher event selection with a higher background rejection than with the toolkit from the $\sqrt{s}=7$ TeV analysis.   These tools also helped improve the the sensitivity for high LSP masses (SR2), where the maximum height of the contour at $\sqrt{s}=8$ TeV nearly doubled with respect to the $\sqrt{s}=8$ TeV analysis.  Note that the new tools mostly helped to improve the the kinematically tight signal regions (SR2, SR3); there was essentially no improvement at low stop mass (SR1).

\begin{figure}[h!]
\begin{center}
\includegraphics[width=0.65\textwidth]{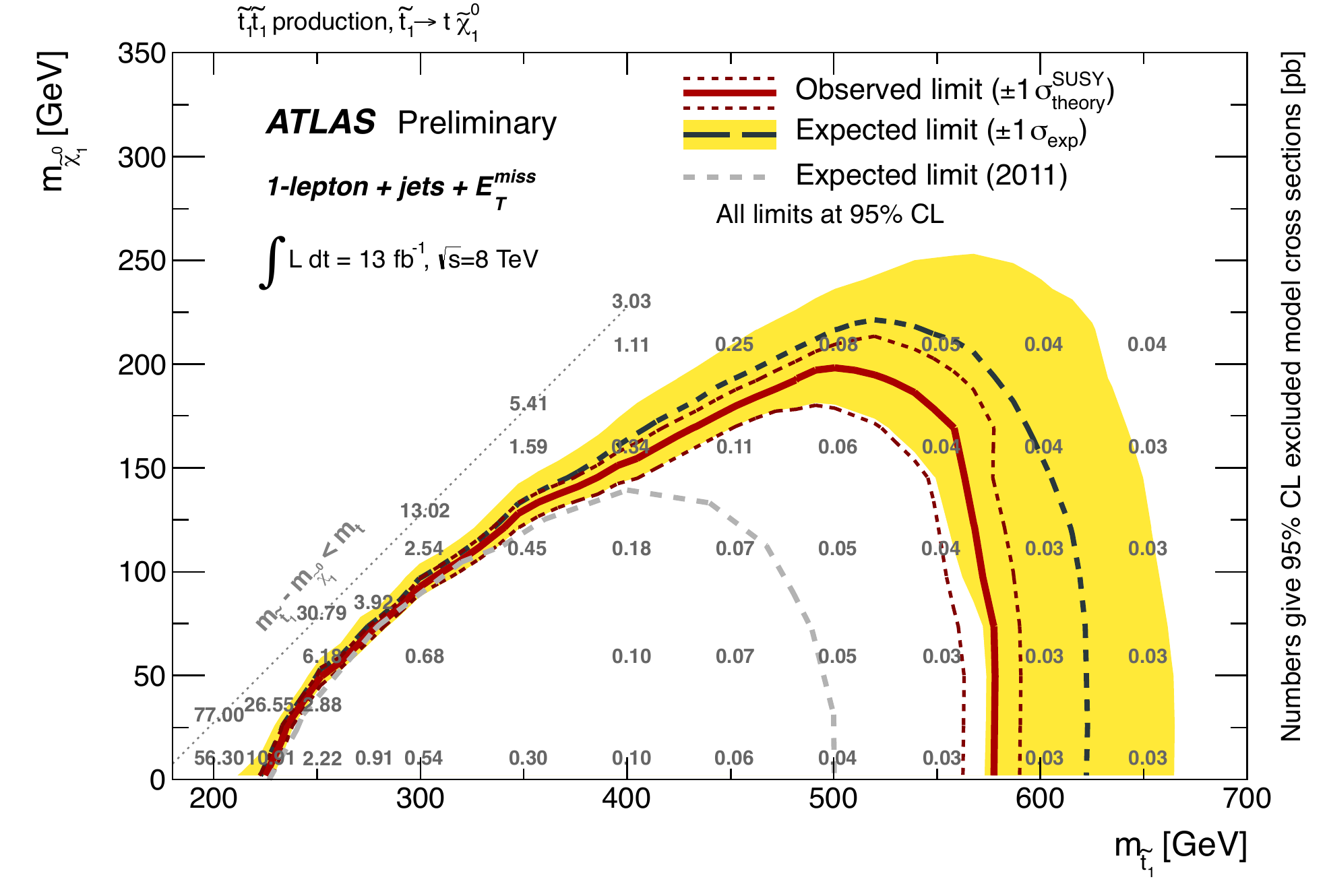} \caption{The exclusion contour of simplified stop models using the early $\sqrt{s}=8$ TeV data.  The observed limit is computed using the SR with the best expected sensitivity (lowest expected $\text{CL}_s$) for the given model as shown in Fig.~\ref{fig:susy:limitHCP2}.  For comparison, the exclusion limits with the full $\sqrt{s}=7$ TeV Run are overlaid with a gray dashed line~\cite{Aad:2012xqa}.  The numerical values at each signal mass point show the smallest cross-section that would be excluded for a model with exactly the same acceptance.  This is computed by scanning the signal cross-section and re-running the fit.}
 \label{fig:susy:limitHCP}
  \end{center}
\end{figure}

\begin{figure}[h!]
\begin{center}
\includegraphics[width=0.65\textwidth]{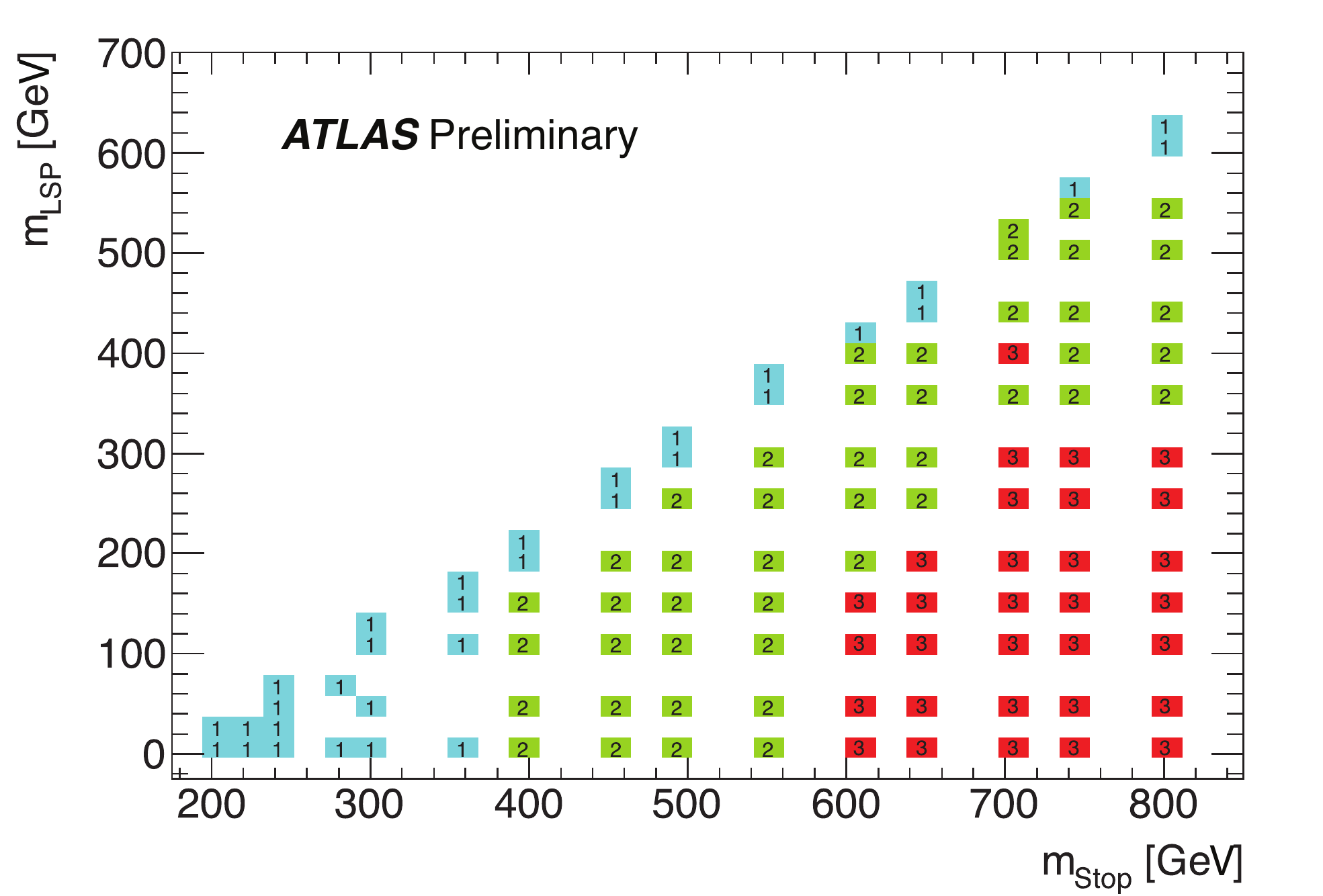} \caption{The signal region used for each mass point in the $(m_\text{stop},m_\text{LSP})$ mass plane to compute the expected and observed limits in Fig.~\ref{fig:susy:limitHCP}.}
 \label{fig:susy:limitHCP2}
  \end{center}
\end{figure}

\clearpage

\subsection{Full $\sqrt{s}=8$ TeV Results}		
\label{8TeVresults}

New techniques and more integrated luminosity further improved the limits from the partial to the full $\sqrt{s}=8$ TeV dataset.  The updated exclusion limits are presented in Fig.~\ref{fig:susy:exclusion:SRmapping8TeV}.  For a massless neutralino, the high mass limit extends to about $m_\text{stop}=675$~GeV, a $50$~GeV improvement over the expected limit from Sec.~\ref{results:early8TeV} and evidence that the small excess in SR3 from the partial dataset is a statistical fluctuation.   For the same integrated luminosity, there would be about $57\%$ fewer stop events with $m_\text{stop}=675$ than for $m_\text{stop}=625$.  Accounting for the difference in dataset size, there would be about $10\%$ fewer stops at the limit with the full dataset compared to the number of stops at the (expected) limit for the partial dataset if the acceptance was constant.  Figure~\ref{fig:susy:exclusion:acceptance} shows the acceptance for the tNmed and tNhigh signal regions.  Near $m_\text{stop}=625$ GeV, the acceptance is about $5\%$ for tNhigh and increases by about $20\%$ when $m_\text{stop}$ is increased by $50$ GeV.  Therefore, the number of signal events near the exclusion limit with the full dataset would be slightly higher than the number of of signal events near the partial dataset exclusion for a fixed event selection.  However, the number of background events increases by about $55\%$.  This means that tNhigh is able to reject about $55\%$ more background events than SR3 while only reducing the signal by about $10\%$.

\begin{figure}[h!]
\begin{center}
\includegraphics[width=0.64\textwidth]{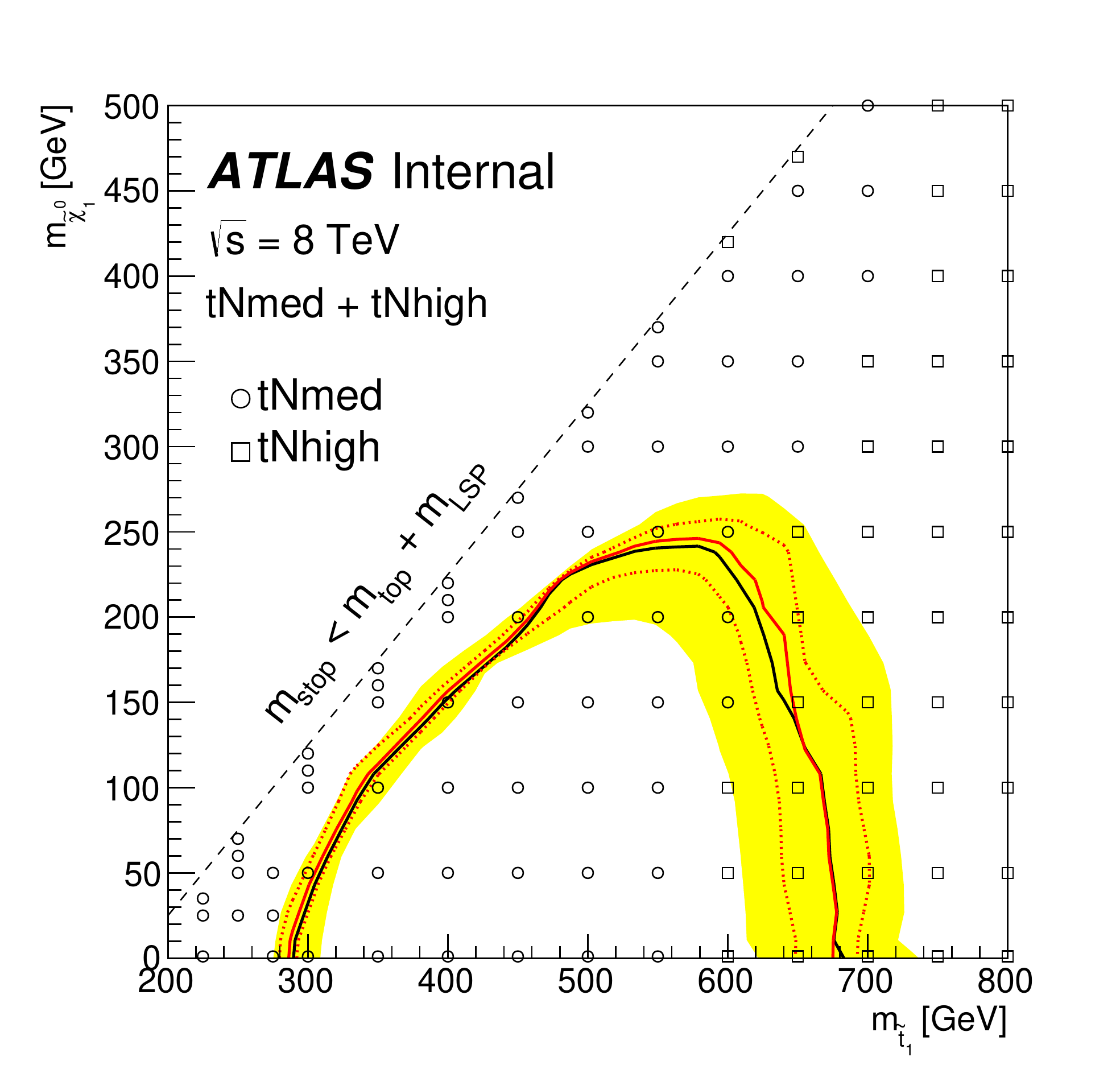}
 \caption{The exclusion contour and SR mapping of simplified stop models using the full $\sqrt{s}=8$ TeV dataset based on the lowest expected $\text{CL}_s$ value. By construction, tNmed is the most sensitive for intermediate $m_\text{stop}$ and large $m_\text{LSP}$ and tNhigh is the most sensitive at $m_\text{stop}$.}
 \label{fig:susy:exclusion:SRmapping8TeV}
  \end{center}
\end{figure}

\begin{figure}[h!]
\begin{center}
\includegraphics[width=0.57\textwidth]{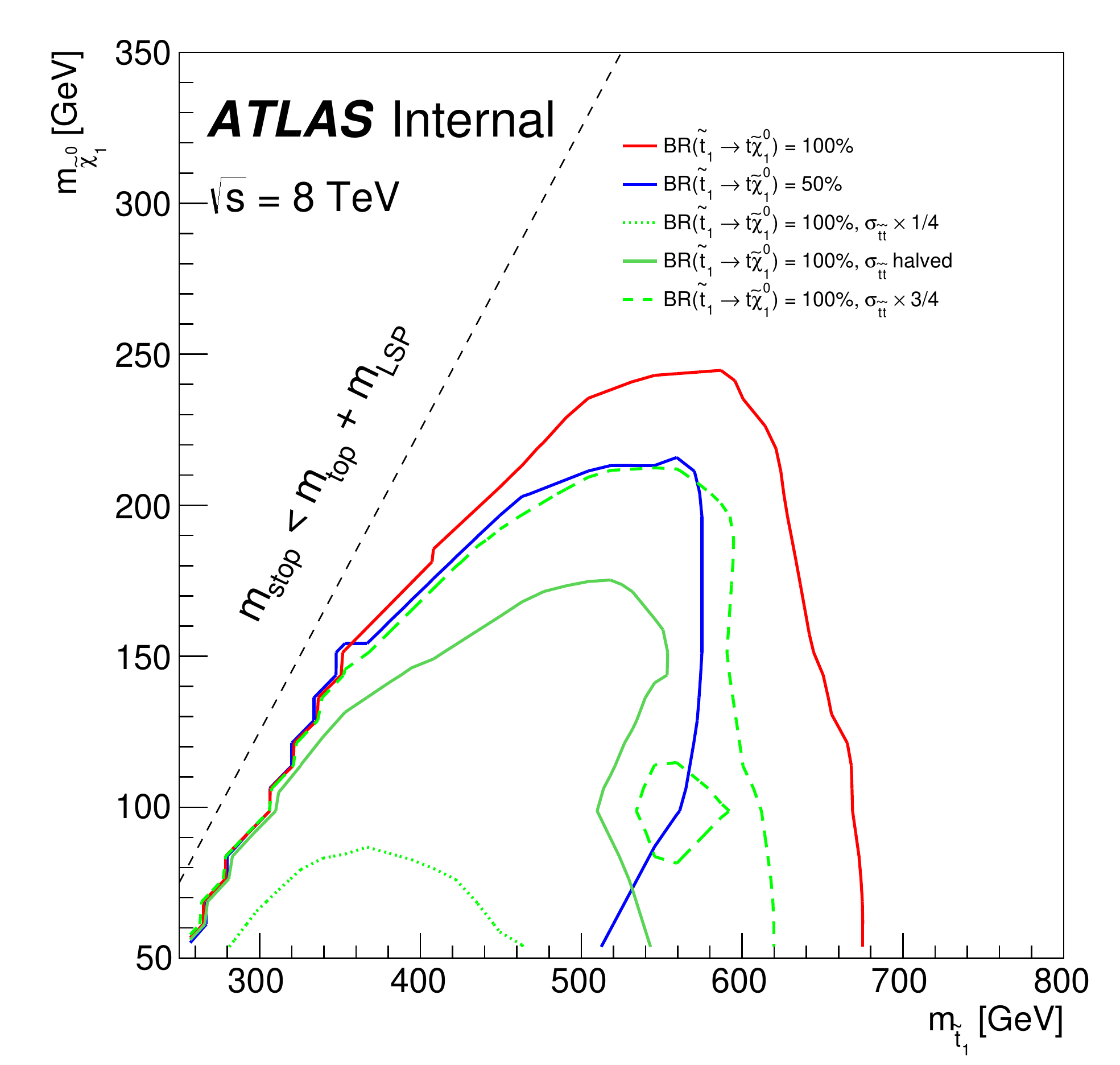}
 \caption{The observed limits using the best-expected map from Fig.~\ref{fig:susy:exclusion:SRmapping8TeV} for two different branching ratio assumptions ($\mathcal{BR}(\tilde{t}\rightarrow t\tilde{\chi}_1^0)+\mathcal{BR}(\tilde{t}\rightarrow b\tilde{\chi}_1^\pm)=1$).  The green lines are computed by comparing the observed cross-section limits from Fig.~\ref{fig:susy:exclusion:SRmapping8TeV} to $25\%$, $50\%$, or $75\%$ of the predicted cross-section.  To make a smooth contour, the significance is set to $2\sigma_\text{excluded}/\sigma_\text{predicted}$.  The wavy line on the left-hand side is due to the lack of signal models above the limit.}
 \label{fig:susy:exclusion:overviewtN12_mix}
  \end{center}
\end{figure}

\begin{figure}[h!]
\begin{center}
\includegraphics[width=0.5\textwidth]{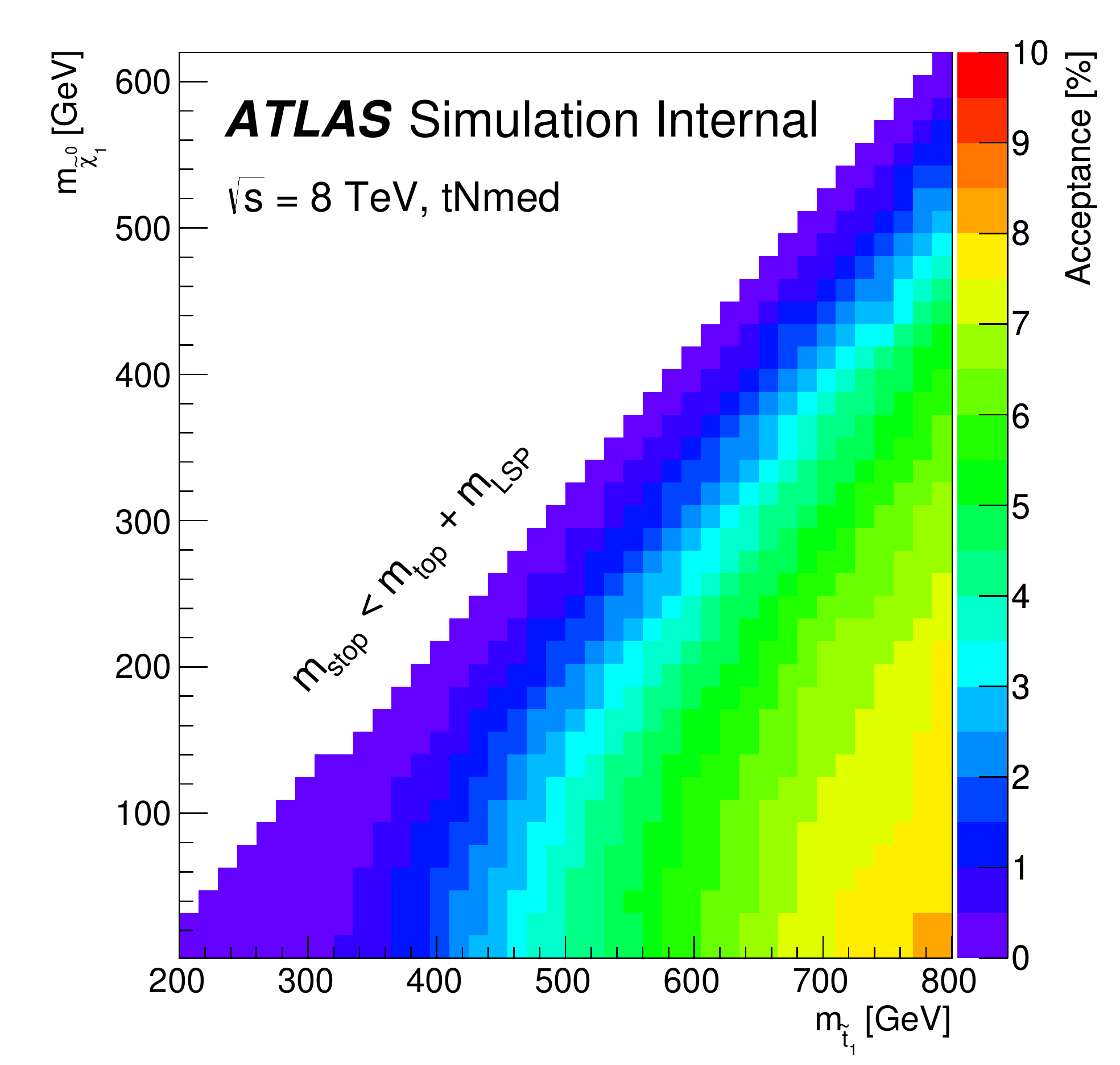}\includegraphics[width=0.5\textwidth]{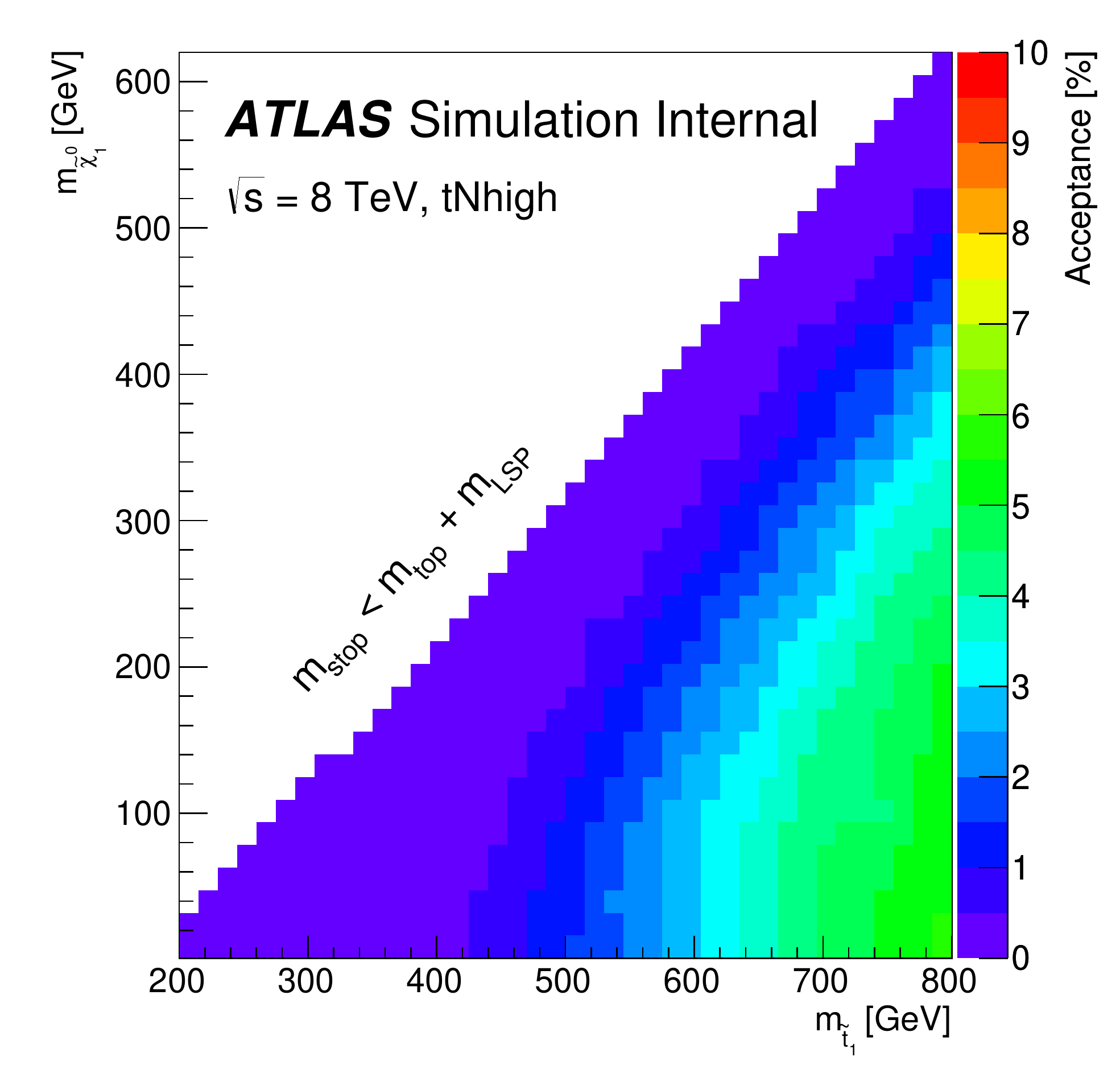}
\includegraphics[width=0.5\textwidth]{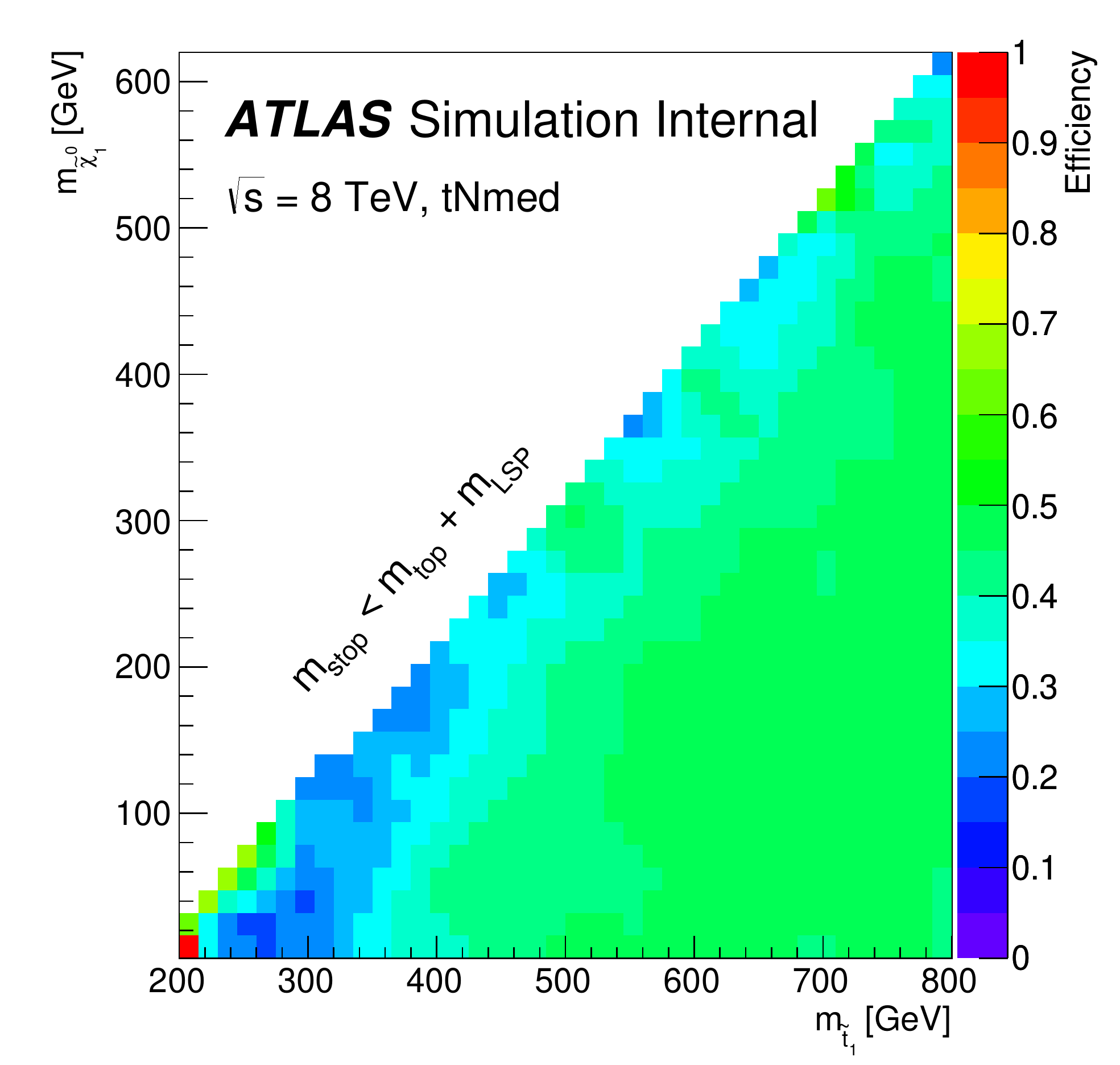}\includegraphics[width=0.5\textwidth]{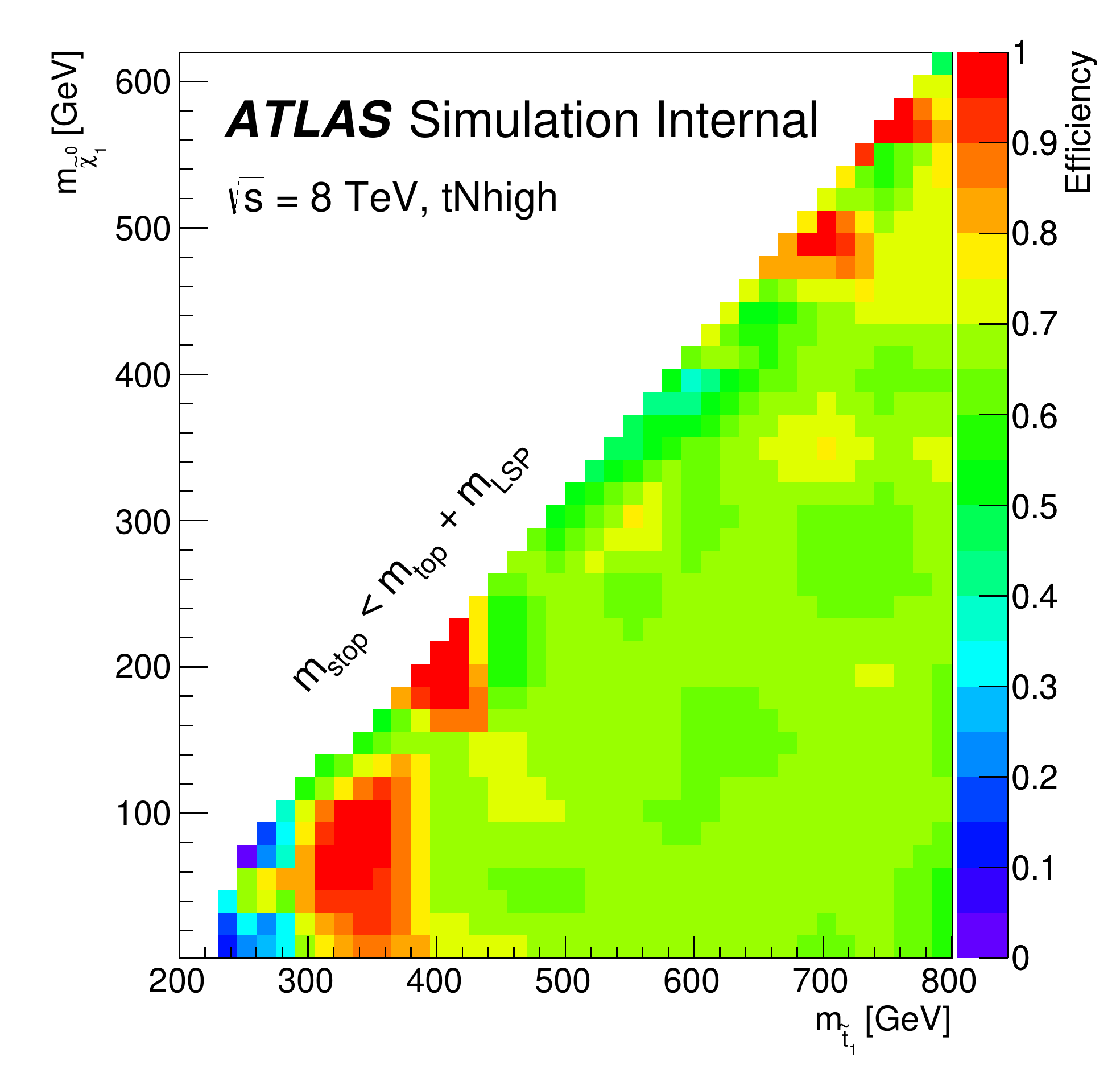}
 \caption{Top: the acceptance for the tNmed (left) and tNhigh (right) event selections as a function of $m_\text{stop}$ and $m_\text{LSP}$ using a {\it particle-level} event selection analogous to the ones described in Sec.~\ref{chapter:susy:signalregions}. The particle-level objects are similar to the ones used throughout Part~\ref{part:qpj} and are detailed in the appendix of Ref.~\cite{Aad:2014qaa}.  Bottom: the ratio of the acceptance using detector-level objects to the acceptance using particle-level objects.  This ratio is a correction for detector-effects in the event selection and is mostly uniform in the sensitive regions of parameter space.  The acceptance in the upper plots is defined using particle-level objects in order to facilitate comparisons with other models for which a detector-simulation is not available.  }
 \label{fig:susy:exclusion:acceptance}
  \end{center}
\end{figure}		
		
For low stop masses, a significant increase in the sensitivity is from the multi-bin signal region (tNshape).  Figures~\ref{fig:susy:exclusion:overviewtN11} and~\ref{fig:susy:exclusion:overviewtN12} highlight this challenging region of parameter space.  The smallest mass gap $m_\text{stop}-m_\text{top}-m_\text{LSP}$ that is excluded is about $12$-$14$ GeV for a stop mass near $250$ GeV.  This is a significant improvement of about $12$ GeV over previous limits.  Over this range, the top quark $p_\text{T}$ drops by nearly a factor of two based on Eq.~\ref{eq:quartic}.  For higher stop masses, the limit weakens as the cross-section is too small for the inclusive tNshape event selection to have any sensitivity.  The sensitivity also decreases for lower stop masses as the signal is less distinguished from the background.  The limit at low stop mass will not improve with more data unless the dominant systematic uncertainties can be reduced, additional variables are identified with a larger variation in $s/b$, and/or modeling uncertainties are sufficiently small to employ new techniques based on ISR or ME jets mentioned in Sec.~\ref{sec:targetpheno}.
		
All of the limits presented thus far assume the stop is mostly the partner of the right-handed top quark.  As discussed in Sec.~\ref{chapter:susy:analysisstrategy}, it is expected that the limits are slightly weaker for a mostly left-handed stop due in part to the softer lepton $p_\text{T}$ spectrum.  The stop mass limit for mostly right-handed stops and $m_\text{LSP}=50$ GeV is about $50$ GeV higher than for mostly left-handed stops.		
		
\begin{figure}[h!]
\begin{center}
\includegraphics[width=0.45\textwidth]{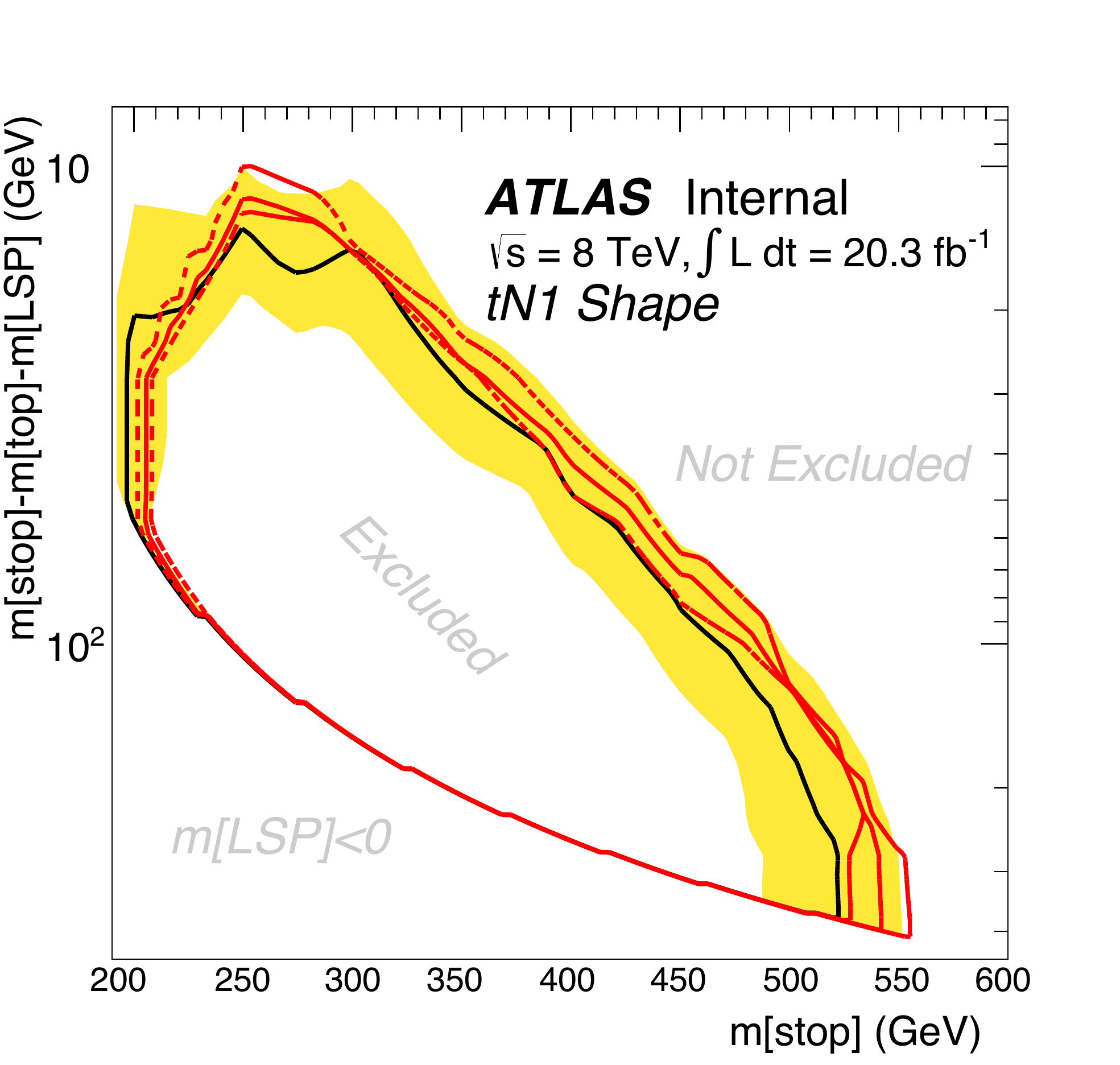}\includegraphics[width=0.45\textwidth]{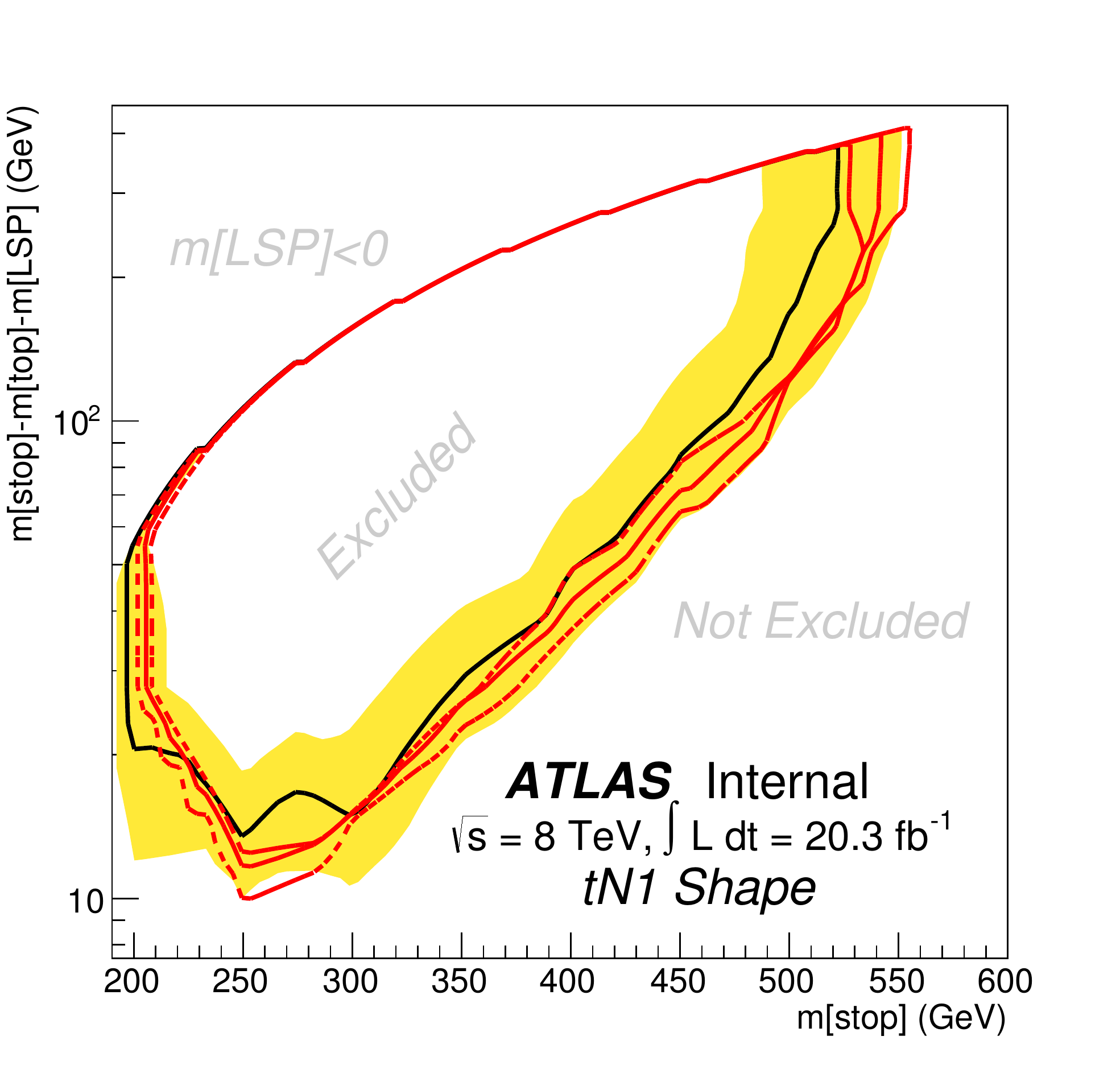}
 \caption{Excluded regions in the $(m_\text{stop},m_\text{stop}-m_\text{top}-m_\text{LSP})$ plane.  The plots are identical aside from a vertical inversion.  The region inside the bounded contour is excluded using the $\text{CL}_s<0.05$ criteria (N.B. this is {\bf not} the $95\%$ confidence level).  In the region below (above) the contour in the left (right) plot, the LSP is a Tachyon ($m<0$) and in the region above (below) the contour, models are allowed by the data.}
 \label{fig:susy:exclusion:overviewtN11}
  \end{center}
\end{figure}	

\begin{figure}[h!]
\begin{center}
\includegraphics[width=0.45\textwidth]{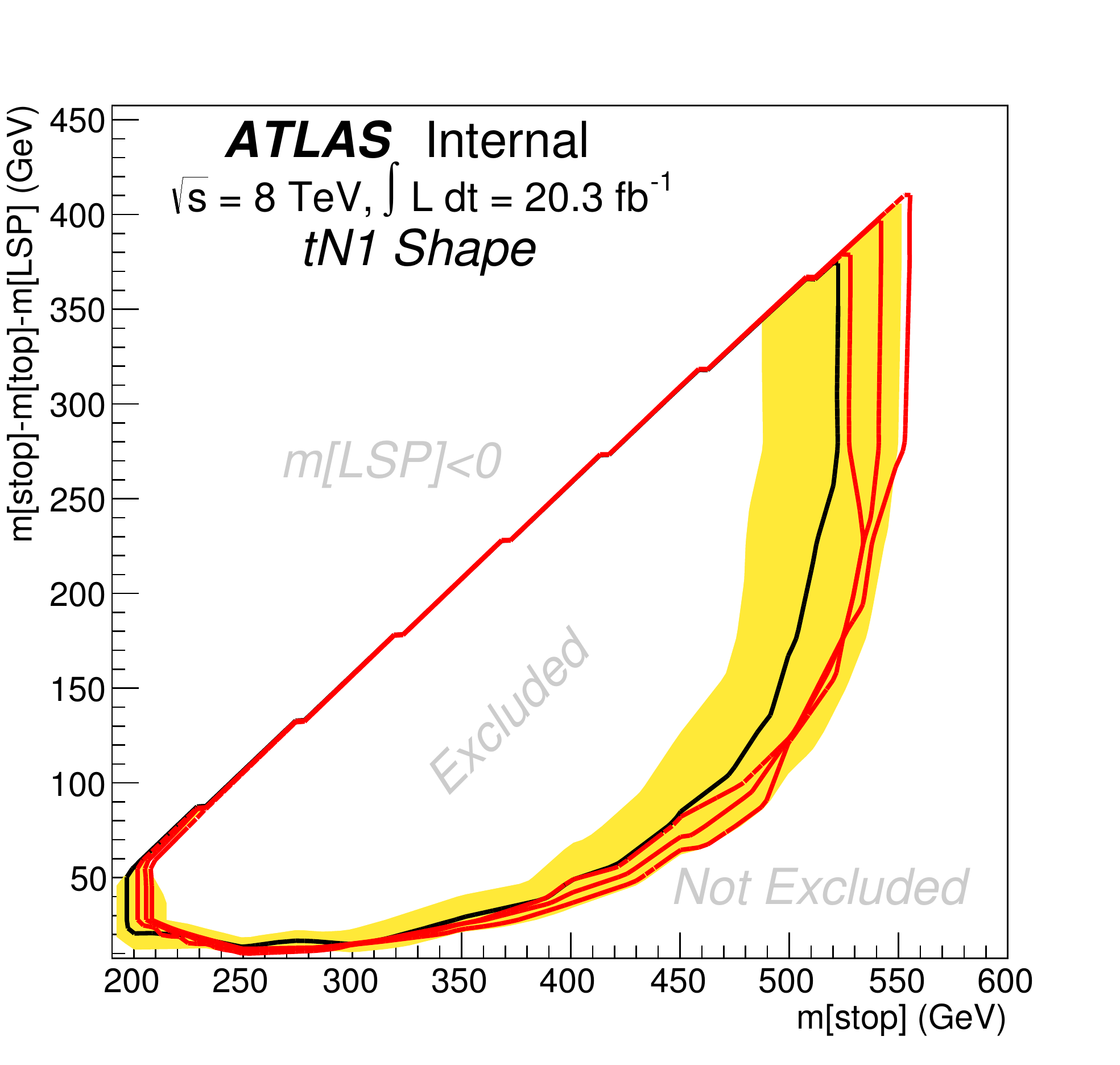}\includegraphics[width=0.45\textwidth]{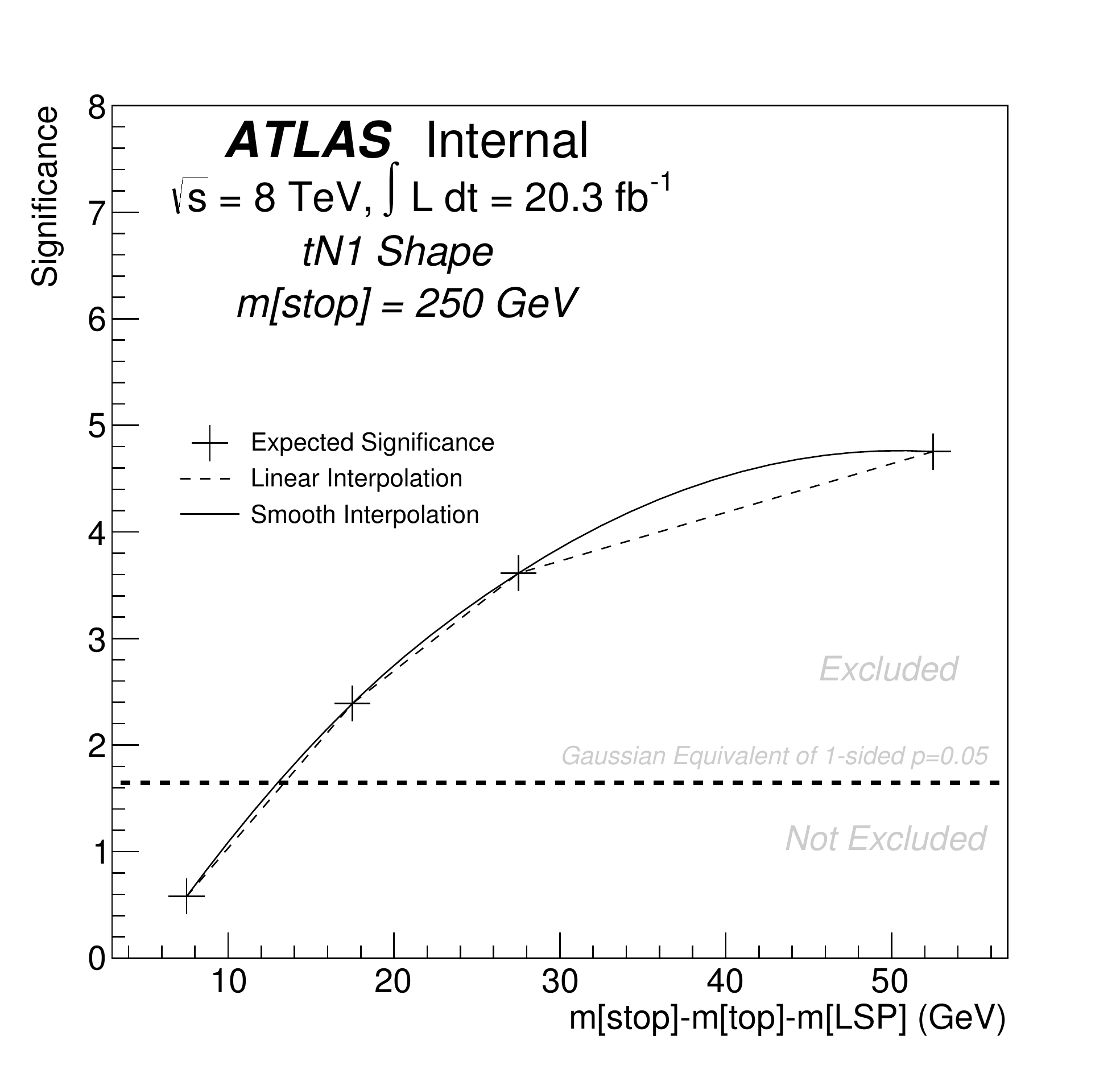}
 \caption{Left: the excluded region of the $(m_\text{stop},m_\text{stop}-m_\text{top}-m_\text{LSP})$ mass plane (linear-scale version of Fig.~\ref{fig:susy:exclusion:overviewtN11}).  Right: a one-dimensional projection of the significance $\sigma=\Phi^{-1}(1-\text{CL}_s)$ as a function of the mass gap $m_\text{stop}-m_\text{top}-m_\text{LSP})$ for $m_\text{stop}=250$ GeV.  Crosses indicate grid points and the solid/dashed lines interpolate between points.  The horizontal dashed line is the exclusion threshold of $\Phi^{-1}(0.95)$.}
 \label{fig:susy:exclusion:overviewtN12}
  \end{center}
\end{figure}		

Even though the event selections presented in this section were optimized using simplified stop models with a $100\%$ branching ratio $\tilde{t}\rightarrow t\tilde{\chi}_1^0$, the signal regions are sensitive to many extensions of the SM.  The next section will discuss the sensitivity to other models that predict $t\bar{t}+E_\text{T}^\text{miss}$ topologies.  To close this section, consider a slightly less-simplified scenario in which $\mathcal{BR}(\tilde{t}\rightarrow t\tilde{\chi}_1^0)<100\%$.  Figure~\ref{fig:susy:exclusion:overviewtN12_mix} shows the observed exclusion limits for $\mathcal{BR}(\tilde{t}\rightarrow t\tilde{\chi}_1^0)=50\%$, with the other $50\%$ of the time the stop decays via the flavor-changing decay $\tilde{t}\rightarrow b\tilde{\chi}_1^\pm$ followed by $\tilde{\chi}_1^\pm\rightarrow W^\pm \tilde{\chi}_1^0$.  A two-dimensional mass plane is no longer sufficient to uniquely specify such a model.  To reduce the parameter space, Fig.~\ref{fig:susy:exclusion:overviewtN12_mix} assumes $m_{\tilde{\chi}^\pm_1}=2m_{\tilde{\chi}^0_1}$, motivated by gaugino universality (see Chapter~\ref{sec:theory}).   For high neutralino mass, the limit is only reduced by about $50$ GeV whereas at low neutralino mass, the limit is reduced by over $150$ GeV.  At low stop mass, there is little impact from the branching ratio reduction.  All of the exclusion at high stop mass is due to tNmed and not tNhigh.  As one might expect, the tighter event selections reduce the breadth of sensitivity.   Interestingly, the exclusion limit for $\mathcal{BR}(\tilde{t}\rightarrow t\tilde{\chi}_1^0)=50\%$ is significantly better than would be expected if tNmed and tNshape where only sensitive to events where both stops decayed via $t\tilde{\chi}_1^0$ ($\sigma\mapsto 25\%\sigma$).  The limit is actually similar to the case where the signal regions are not sensitive to events where both stops decay via $b\tilde{\chi}_1^\pm$ ($\sigma\mapsto 75\%\sigma$).  
\clearpage		
		
\subsubsection{Recasting Stop Limits}		
\label{stoprecast}	
	
As introduced in Sec.~\ref{relatedmodels}, there are several extensions of the SM that predict new particles resulting in $t\bar{t}+E_\text{T}^\text{miss}$ topologies.  Table~\ref{tab:models} categorizes models based on their spin and color charge, which determine the cross-section.  As neither the spin nor the color are measured, the cross-section is a sum over states and therefore the cross-section increases with spin and the dimension of the color representation.  Scalar leptoquarks have the same cross-section has stops\footnote{With small differences in acceptance that depend on the stop mixing.} while vector leptoquarks have a significantly higher cross-section.  Figure~\ref{fig:crosssectionsmodels} illustrates the cross-section differences as a function of new particle mass\footnote{Thanks to Marat Freytsis for providing the UFO model for the vector lepto-quark, which was used through MG5\_aMC 2.1.1 to compute the cross-sections. }.  The remainder of this section focuses on GMS, but the methods could be applied to any of the models in Table~\ref{tab:models}.
	
\begin{table}[h!]
  \centering
\noindent\adjustbox{max width=\textwidth}{
\begin{tabular}{|c|c|c|c|c|}
\hline
Name & Spin & Color Charge & Electric Charge & Relative Cross-section\\
\hline
-- & 0 & {\bf 1} & $2/3$& Tiny\\
Stop & 0 & {\bf 3} & $2/3$& Small\\
Scalar Leptoquark & 0 & {\bf 3} & $2/3$& Small\\
-- & 0 & {\bf 8} & $2/3$& Medium\\
-- & $1/2$ & {\bf 1} & $2/3$ & Tiny\\
$T$ (or $T'$) & $1/2$ & {\bf 3} & $2/3$ & Medium\\
GMS & $1/2$ & {\bf 8} & $2/3$& Large\\
-- & 1 & {\bf 1} & $2/3$ & Tiny\\
Vector Leptoquark & 1 & {\bf 3} & $2/3$ & Medium-Large\\
-- & 1 & {\bf 8} & $2/3$ & Large\\
\hline   
\end{tabular}}
\caption{Example models that result in $t\bar{t}+E_\text{T}^\text{miss}$ categorized by their spin, color charge, electric charge (has to be $2/3$ if the missing particles are neutral), and the relative cross-section.  The acronym GMS stands for gluino mediated stop. The particle $T$ exists in a variety of models and could decay via $T\rightarrow tA_0$ for a new weakly interacting particle $A_0$ (see Ref.~\cite{Aad:2011wc} and the references therein), or it could be a vector-like quark decaying via $T'\rightarrow tZ(\rightarrow\nu\bar{\nu})$, which also has the $t\bar{t}+E_\text{T}^\text{miss}$ topology.  A `--' indicates that there is no standard model with these properties.}
\label{tab:models}
\end{table}	
	
 \begin{figure}[h!]
 \begin{center}
 \includegraphics[width=0.8\textwidth]{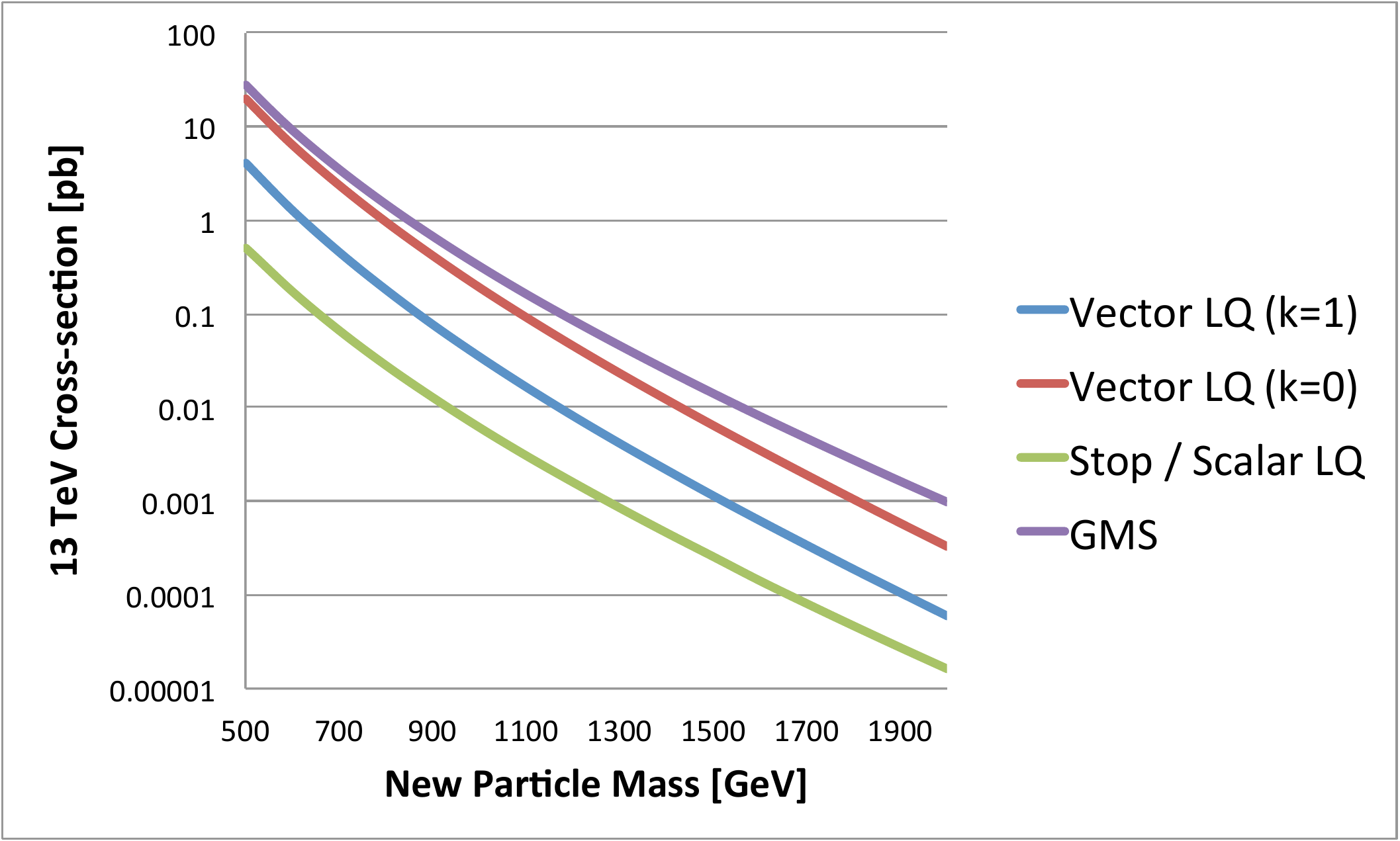}
 \caption{The cross-section for various models highlighted in Table~\ref{tab:models}.  The parameter $k$ for the vector lepto-quarks corresponds to the $\kappa$ model parameter in Ref.~\cite{Freytsis:2015qca}. }
 \label{fig:crosssectionsmodels}
  \end{center}
 \end{figure}

As introduced in Sec.~\ref{relatedmodels}, gluino mediated stops (GMS) with nearly mass degenerate stops and neutralinos have a similar signature to direct stop production.  However, the gluino pair production cross section is much larger than the cross section for direct pair produced stops.  For example, at $\sqrt{s}=8$ TeV, the stop pair production cross-section for $m_\text{stop}=800$ GeV is about $0.002$ pb whereas the cross-section for stops produced from the decay of $1$ TeV pair produced gluinos is about $0.02$ pb~\cite{Kramer:2012bx}.  Therefore, mass limits in the GMS model will be higher than those for direct stop production.   At $\sqrt{s}=13$ TeV, the effective cross section for the gluino mediated process is twice the direct stop pair production cross section, which is exploited by the early $\sqrt{s}=13$ TeV search to be sensitive to discover new particles earlier than expected.  The results of that search are presented in Sec.~\ref{sec:13TeV}.  This section describes how the the limits on GMS models can be extracted indirectly from the stop limits discussed in Sec.~\ref{8TeVresults}.  

General GMS models are well-motivated by naturalness (see Sec.~\ref{relatedmodels}).  Both ATLAS~\cite{Aad:2013wta,Aad:2014lra,Aad:2014pda,Aad:2014wea,Aad:2015mia} and CMS~\cite{Chatrchyan:2014lfa,Chatrchyan:2013iqa,Chatrchyan:2013fea,CMS-PAS-SUS-14-011,CMS-PAS-SUS-13-016,CMS-PAS-SUS-13-008} have searched extensively for generic GMS models, excluding spectra with large mass splittings up to $m_{\tilde{g}}\lesssim1.5$ TeV.  Compressed mass spectra are generally more difficult to identify than spectra with large splittings, but are still well-motivated by e.g. dark matter.  Gluino pair production with four high energy top or bottom quarks leaves a striking signature in a detector.  However, if any of the mass splittings are compressed, the power of traditional techniques may deteriorate.  Figure~\ref{fig:masshierarchy3} shows the possible GMS mass  hierarchies, highlighting the presence of direct stop-like and direct sbottom-like signatures.   Searches for direct stop/sbottom pair production can be recast as searches for GMS in order to extend the sensitivity.   This section will show that compressed GMS limits at $\sqrt{s}=8$ TeV can be extended by at least $225$ GeV for a $1.1$ TeV stop.  Before describing the GMS limits, the general notion of {\it equivalence} for model reinterpretation is described in Sec.~\ref{sec:equiv}.

\begin{figure}[h!]

\begin{center}
\begin{tikzpicture}[line width=1.5 pt, scale=1.3]
	
	\node at (-0.5, 4.5) {$m_{\tilde{g}}$};
	\node at (-0.5, 3) {$m_{\tilde{t}}$};
	\node at (-0.5, 1.5) {$m_{\tilde{\chi}_1^\pm}$};
	\node at (-0.5, 0) {$m_{\tilde{\chi}_1^0}$};	

	\draw[dotted,color=gray!90!white] (0.,4.5)--(8.5,4.5);		
	\draw (0.5,4.5)--(1.,4.5);	
	\draw (1.5,4.5)--(2.,4.5);	
	\draw (2.5,4.5)--(3.,4.5);	
	\draw (3.5,4.5)--(4.,4.5);	
	\draw (4.5,4.5)--(5.,4.5);	
	\draw (5.5,4.5)--(6.,4.5);	
	\draw (6.5,4.5)--(7.,4.5);	
	\draw (7.5,4.5)--(8.,4.5);		
	
	\draw[dotted,color=gray!90!white] (0.,3)--(4.1,3);
	\draw[dotted,color=gray!90!white] (4.1,3)--(4.4,4);
	\draw[dotted,color=gray!90!white] (4.4,4)--(8.5,4);		
	\draw (0.5,3)--(1.,3);	
	\draw (1.5,3)--(2.,3);	
	\draw (2.5,3)--(3.,3);	
	\draw (3.5,3)--(4.,3);	
	\draw (4.5,4)--(5.,4);	
	\draw (5.5,4)--(6.,4);	
	\draw (6.5,4)--(7.,4);	
	\draw (7.5,4)--(8.,4);	
	
	\draw[dotted,color=gray!90!white] (0.,1.5)--(1.1,1.5);	
	\draw[dotted,color=gray!90!white] (1.1,1.5)--(1.4,2.5);
	\draw[dotted,color=gray!90!white] (1.5,2.5)--(2.1,2.5);
	\draw[dotted,color=gray!90!white] (2.1,2.5)--(2.4,1.5);
	\draw[dotted,color=gray!90!white] (2.4,1.5)--(3.1,1.5);
	\draw[dotted,color=gray!90!white] (3.1,1.5)--(3.4,2.5);
	\draw[dotted,color=gray!90!white] (3.4,2.5)--(4.1,2.5);
	\draw[dotted,color=gray!90!white] (4.1,2.5)--(4.4,1.5);
	\draw[dotted,color=gray!90!white] (4.4,1.5)--(5.1,1.5);
	\draw[dotted,color=gray!90!white] (5.1,1.5)--(5.4,3.5);
	\draw[dotted,color=gray!90!white] (5.4,3.5)--(6.1,3.5);
	\draw[dotted,color=gray!90!white] (6.1,3.5)--(6.4,1.5);
	\draw[dotted,color=gray!90!white] (6.4,1.5)--(7.1,1.5);
	\draw[dotted,color=gray!90!white] (7.1,1.5)--(7.4,3.5);
	\draw[dotted,color=gray!90!white] (7.4,3.5)--(8.5,3.5);		
	\draw (0.5,1.5)--(1.,1.5);	
	\draw (1.5,2.5)--(2.,2.5);	
	\draw (2.5,1.5)--(3.,1.5);	
	\draw (3.5,2.5)--(4.,2.5);	
	\draw (4.5,1.5)--(5.,1.5);	
	\draw (5.5,3.5)--(6.,3.5);	
	\draw (6.5,1.5)--(7.,1.5);	
	\draw (7.5,3.5)--(8.,3.5);	

	\draw[dotted,color=gray!90!white] (0.,0)--(2.1,0);
	\draw[dotted,color=gray!90!white] (2.1,0)--(2.4,1);
	\draw[dotted,color=gray!90!white] (2.4,1)--(3.1,1);
	\draw[dotted,color=gray!90!white] (3.1,1)--(3.4,2);	
	\draw[dotted,color=gray!90!white] (3.4,2)--(4.1,2);	
	\draw[dotted,color=gray!90!white] (4.1,2)--(4.4,0);
	\draw[dotted,color=gray!90!white] (4.4,0)--(6.1,0);
	\draw[dotted,color=gray!90!white] (6.1,0)--(6.4,1);	
	\draw[dotted,color=gray!90!white] (6.4,1)--(7.1,1);	
	\draw[dotted,color=gray!90!white] (7.1,1)--(7.4,3);
	\draw[dotted,color=gray!90!white] (7.4,3)--(8.5,3);		
	\draw (0.5,0)--(1.,0);	
	\draw (1.5,0)--(2.,0);	
	\draw (2.5,1.)--(3.,1.);	
	\draw (3.5,2.)--(4.,2);	
	\draw (4.5,0)--(5.,0);	
	\draw (5.5,0)--(6.,0);	
	\draw (6.5,1)--(7.,1);	
	\draw (7.5,3)--(8.,3);	

	\draw[->] (0.75,4.4)--(0.75,3.1);	
	\node at (0.5, 3.8) {$t$};

	\draw[->] (1.75,4.4)--(1.75,3.1);	
	\node at (1.5, 3.8) {$t$};

	\draw[->] (2.75,4.4)--(2.75,3.1);	
	\node at (2.5, 3.8) {$t$};

	\draw[->] (3.75,4.4)--(3.75,3.1);	
	\node at (3.5, 3.8) {$t$};

	\draw[->] (0.75,2.9)--(0.75,1.6);	
	\node at (0.5, 2.3) {$b$};

	\draw[->] (2.75,2.9)--(2.75,1.6);	
	\node at (2.5, 2.3) {$b$};

	\draw[->] (4.75,3.9)--(4.75,1.6);	
	\node at (4.5, 2.8) {$b$};

	\draw[->] (6.75,3.9)--(6.75,1.6);	
	\node at (6.5, 2.8) {$b$};

	\draw[->] (0.75,1.4)--(0.75,0.1);	
	\node at (0.5,0.8) {$W$};

	\draw[->] (1.75,2.4)--(1.75,0.1);	
	\node at (1.5,1.4) {$W$};

	\draw[->] (4.75,1.4)--(4.75,0.1);	
	\node at (4.5,0.8) {$W$};

	\draw[->] (5.75,3.4)--(5.75,0.1);	
	\node at (5.5,1.8) {$W$};

	\node at (0.75, -0.5) {(a)};
	\node at (1.75, -0.5) {(b)};	
	\node at (2.75, -0.5) {(c)};
	\node at (3.75, -0.5) {(d)};	
	\node at (4.75, -0.5) {(e)};	
	\node at (5.75, -0.5) {(f)};		
	\node at (6.75, -0.5) {(g)};		
	\node at (7.75, -0.5) {(h)};		
	
 \end{tikzpicture}
\end{center}
\caption{All possible mass hierarchies for $\tilde{g}\rightarrow t\tilde{t}$, $\tilde{t}\rightarrow b\chi_1^\pm$.  The arrows indicate which, if any, high energy particles are produced in the cascade decay.  Decay (d) is phenomenologically the same as the compressed GMS and (g) is equivalent to the gluino mediated sbottom. The other possibilities are better covered by dedicated searches than direct stop/sbottom production.  The gluino-mediated sbottom is similar, but with $\tilde{t}\leftrightarrow\tilde{b}$ and $t\leftrightarrow b$. }
\label{fig:masshierarchy3}
\end{figure}

\clearpage

\paragraph{Equivalent Models} \mbox{}\\
\label{sec:equiv}

The two ingredients needed to set experimental limits on a SUSY model are the acceptance and the cross-section.  The acceptance is the predicted fraction of SUSY events that pass the experimental event selection and the cross-section is the rate of production for SUSY events\footnote{As with Fig.~\ref{fig:susy:exclusion:acceptance}, there is often a distinction between the particle-level acceptance (often simply called acceptance) and the detector-level acceptance.}.  Let $M$ be a particular SUSY model and define $M_s=(\epsilon,\sigma)$, where $\epsilon$ is the acceptance of the model $M$ under an experimental selection $s$ and $\sigma$ is the cross section for $M$.  Note that the cross section does not depend on $s$.   Two distinct SUSY models $M$ and $M'$ are defined to be {\it equivalent} under the experimental event selection $s$ if $M_s=M_s'$.  If two models $M$ and $M'$ are equivalent under the experimental selection $s$, then one is excluded by $s$ if and only if the other is also excluded.  In SUSY simplified models, the cross section and acceptance depend on only a few key parameters.  For direct stop production ($\mathcal{M}_{\tilde{t}}$), the cross section $\sigma$ for models in $\mathcal{M}_{\tilde{t}}$ depend only on the stop mass, $m_{\tilde{t}}$ and the acceptance under a given experimental selection depends on both\footnote{As noted in Sec.~\ref{sec:targetpheno}, the acceptance also depends on the top polarization.  Top quarks in the GMS model originate directly from the scalar gluino and thus are unpolarized (the stops are produced on-shell in these models).  This will have a small impact on the exclusion which is ignored in the following.} the stop mass and the neutralino mass $m_{\tilde{\chi}^0_1}$.  For the GMS production ($\mathcal{M}_{\tilde{g}}$), the cross section is set by the gluino mass $m_{\tilde{g}}$ and the acceptance depends on all three masses: $m_{\tilde{g}},m_{\tilde{t}},$ and $m_{\tilde{\chi}^0_1}$.  Since models in $\mathcal{M}_{\tilde{t}}$ and $\mathcal{M}_{\tilde{g}}$ both need the stop and neutralino masses as input, for clarity, $m_{\tilde{t}_1}^{\tilde{t}}$ will denote the stop mass in a model $M_{\tilde{t}}\in \mathcal{M}_{\tilde{t}}$ and $m_{\tilde{t}_1}^{\tilde{g}}$ represents the stop mass in a model $M_{\tilde{g}}\in \mathcal{M}_{\tilde{g}}$ (and analogously for the neutralino). The next sections describe a procedure for associating to every model $M_{\tilde{g}}\in \mathcal{M}_{\tilde{g}}$, an equivalent model $M_{\tilde{t}}\in \mathcal{M}_{\tilde{t}}$.  Experimental limits on $M_{\tilde{t}}$ can then be used to place limits on $M_{\tilde{g}}$.

\clearpage

\paragraph{Limits on Compressed Gluino Mediated Stop Production} \mbox{}\\
\label{sec:limits}

The lost sensitivity to compressed $\tilde{g}\rightarrow t\tilde{t}$ from direct gluino searches with multi-top quark, multi-$b$ quark, or multi-lepton final states can be recovered by direct stop searches. There are only subtle differences between the models due to the fact that the gluino is a fermionic color octet, instead of a scalar triplet like the stop, so there will be small changes in angular distributions and radiation patterns between jets.  However, most analysis techniques are not sensitive to these effects.  One non-negligible difference is the electric charge, as stops can have the same charge when from gluinos (as it is a Majorana particle), but must be oppositely charged for direct stop production.  For this reason, same-sign lepton searches can retain sensitivity even when the decay chains are compressed.  However, the results below indicate that the one- and zero-lepton searches are more powerful, due to the much larger branching ratio.

Given an experimental selection $s$, for a particular model $M_{\tilde{t}}\in \mathcal{M}_{\tilde{t}}$, the goal is to find an equivalent model $M_{\tilde{g}}\in \mathcal{M}_{\tilde{g}}$.  The first step in finding an equivalent model is to match the cross sections $\sigma(M_{\tilde{g}}) = \sigma(M_{\tilde{t}})$.  There is a one-to-one correspondence between $m_{\tilde{g}}$ and $\sigma(M_{\tilde{g}})$ and between $m_{\tilde{t}}^{\tilde{t}}$ and $\sigma(M_{\tilde{t}})$.  The numerical relationships can be found in Ref.~\cite{Kramer:2012bx}.  Therefore, given $m_{\tilde{t}}^{\tilde{t}}$, there is a unique $m_{\tilde{g}}$ such that 

\begin{align}
\sigma(M_{\tilde{t}}) =\sigma(m_{\tilde{t}}^{\tilde{t}})=\sigma(m_{\tilde{g}})= \sigma(M_{\tilde{g}}).
\end{align}

\noindent The second step is to find $m_{\tilde{\chi}^0}^{\tilde{g}}$ (chosen to be nearly identical to $m_{\tilde{t}}^{\tilde{g}}$) and $m_{\tilde{\chi}^0}^{\tilde{t}}$ such that the acceptances under $s$ are the same for $M_{\tilde{t}}$ and $M_{\tilde{g}}$.   As described in Sec.~\ref{sec:targetpheno}, this can be accomplished by choosing $m_{\tilde{\chi}^0}^{\tilde{g}}$ and $m_{\tilde{\chi}^0}^{\tilde{t}}$ such that the final state objects have the same top quark momentum spectrum $p(M,m)$ given in Eq.~\ref{eq:quartic}, where $(M,m)=(m_{\tilde{t}}^{\tilde{t}},m_{\tilde{\chi}^0}^{\tilde{t}})$ for $\mathcal{M}_{\tilde{t}}$ and $(M,m)=(m_{\tilde{g}}, m_{\tilde{\chi}^0}^{\tilde{g}})$ for $\mathcal{M}_{\tilde{g}}$. Given $m_{\tilde{t}}^{\tilde{t}},m_{\tilde{\chi}^0}^{\tilde{t}}$ and determining $m_{\tilde{g}}$ by the equality of the cross sections between $M_{\tilde{t}}$ and $M_{\tilde{g}}$, $m_{\tilde{t}}^{\tilde{g}}$ is chosen by solving $p(m_{\tilde{t}}^{\tilde{t}},m_{\tilde{\chi}^0}^{\tilde{t}})=p(m_{\tilde{g}},m_{\tilde{t}}^{\tilde{g}})$.  The solution to this equation is quartic in $m_{\tilde{t}_1}^{\tilde{g}}$, so in general there can be up to four real solutions.  Fortunately, two solutions are negative (or imaginary) and of the two possible positive solutions, only one is smaller than $m_{\tilde{g}}$ and thus there is at most one physical solution.  Various kinematic distributions for one particular set of equivalent models are shown in Fig.~\ref{fig:kins}.  The model ${M}_{\tilde{t}}$ is specified by a 700 GeV stop mass and a massless neutralino and the equivalent model ${M}_{\tilde{g}}$ has a 1.1 TeV gluino and a $\sim 650$ GeV stop/neutralino.  By construction, all of the kinematic distributions are nearly identical between these two models and as a result, any selection $s$ based on kinematic variables should have the same acceptance.  The difference between the two models is quantified in Fig.~\ref{fig:kins2}, which shows that the approximation equating these models is valid only when $\delta\equiv|m_{\tilde{t}}^{\tilde{g}}-m_{\tilde{\chi}^0}^{\tilde{g}}|$ is sufficiently small.  For $\delta\lesssim 15$ GeV for this representative model, the correction to the efficiency is a few percent and grows to about ten percent when $\delta\sim 10 $ GeV.  The number of jets with a particular transverse momentum increases when $\delta$ increases because the charm quarks can produce measurable jets when there is enough phase space.  In contrast, the magnitude of the missing transverse momentum decreases because the charm quarks take energy away from the neutralinos.  For this reason, the product of efficiencies for jet variables and missing momentum variables is much less dependent on $\delta$.  Note that if the four-body decay of the stop, $\tilde{t}\rightarrow b f f' \tilde{\chi}^0$, for fermions $f$ and $f'$, dominated over the two-body decay $\tilde{t}\rightarrow c \tilde{\chi}^0$, the dependence on $\delta$ is also reduced because there are more objects that need to share the sparse phase space.

A set of model equivalences between direct stop and gluino mediated compressed stop models are summarized in Table~\ref{tab:translate}.  These models are chosen because they are at the edge of the high mass exclusion limit from Sec.~\ref{8TeVresults}.

\begin{table}[h]
\begin{center}
\begin{tabular}{|c|c|c|c|c|}
\hline
$m_{\tilde{t}}^{\tilde{t}}$ [GeV] & $m_{\tilde{\chi}_1^0}^{\tilde{t}}$ [GeV] & $\sigma(m_{\tilde{t}}^{\tilde{t}})$ [pb] & $m_{\tilde{g}}^{\tilde{g}}$ [GeV] & $m_{\tilde{t}}^{\tilde{g}}$ [GeV]  \\
\hline
675             & 100                         & 0.011                               & 1090                                            & 670                                             \\
625             & 220                         & 0.018                               & 1030                                            & 690                                             \\
600             & 240                         & 0.025                               & 995                                             & 680                                             \\
550             & 240                         & 0.045                               & 930                                             & 660      \\\hline                                    
\end{tabular}
\caption{A set of direct stop models that are equivalent to GMS models.  The third and fourth columns are from Ref.~\cite{Kramer:2012bx}.} \label{tab:translate}
\end{center}
\end{table}

 \begin{figure}[h!]
 \begin{center}
 \includegraphics[width=0.45\textwidth]{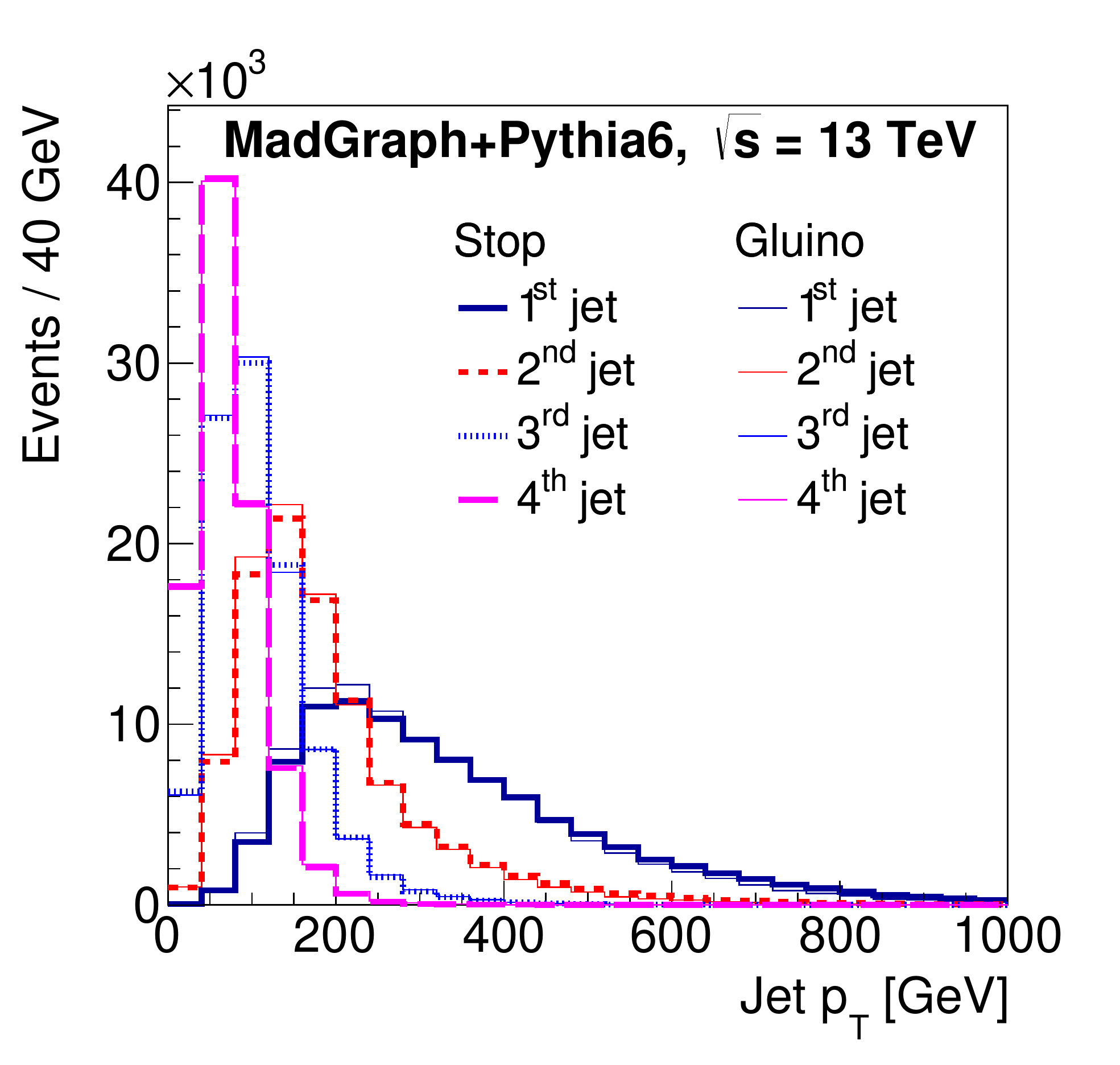}\\\includegraphics[width=0.45\textwidth]{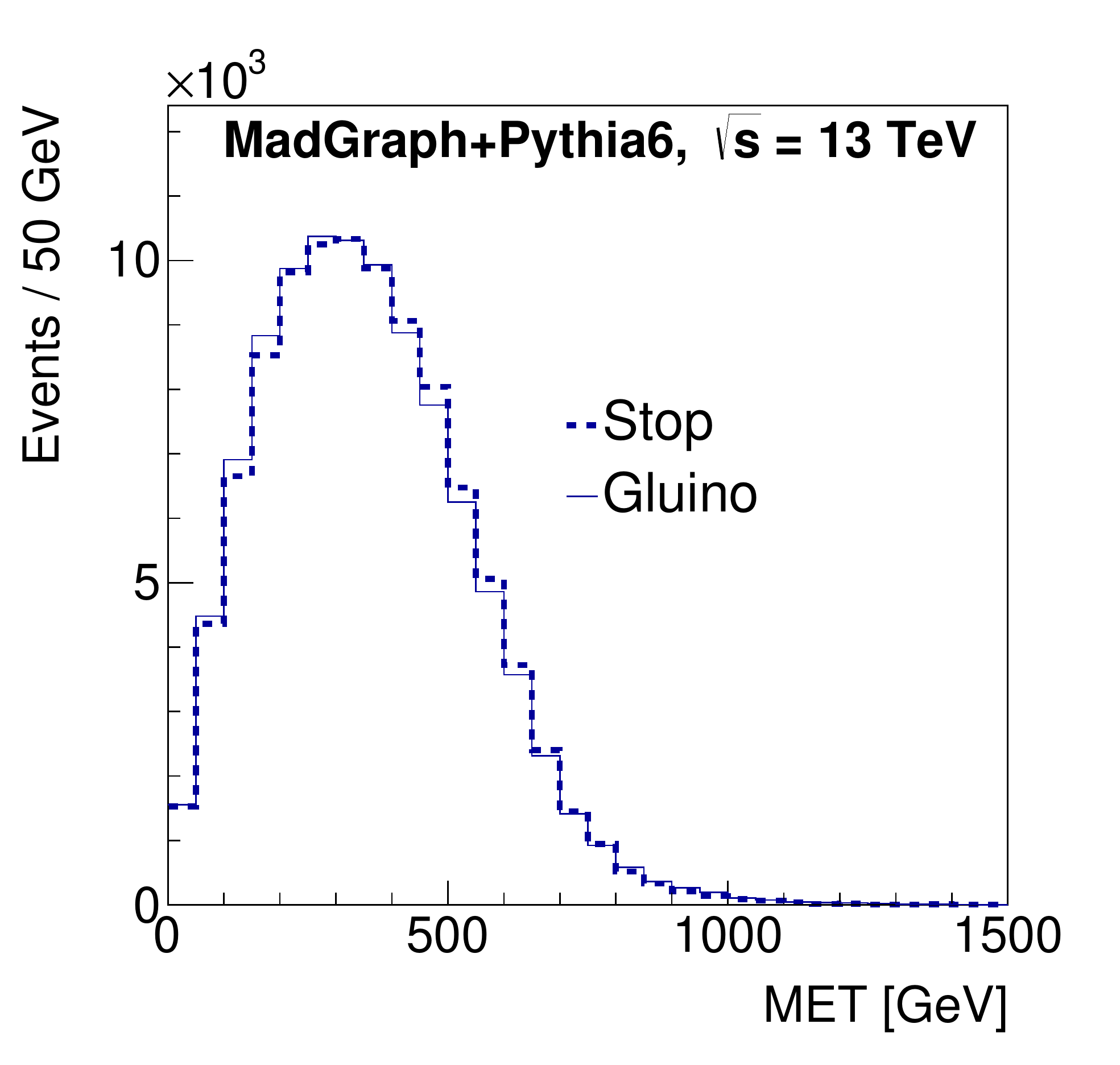} \includegraphics[width=0.45\textwidth]{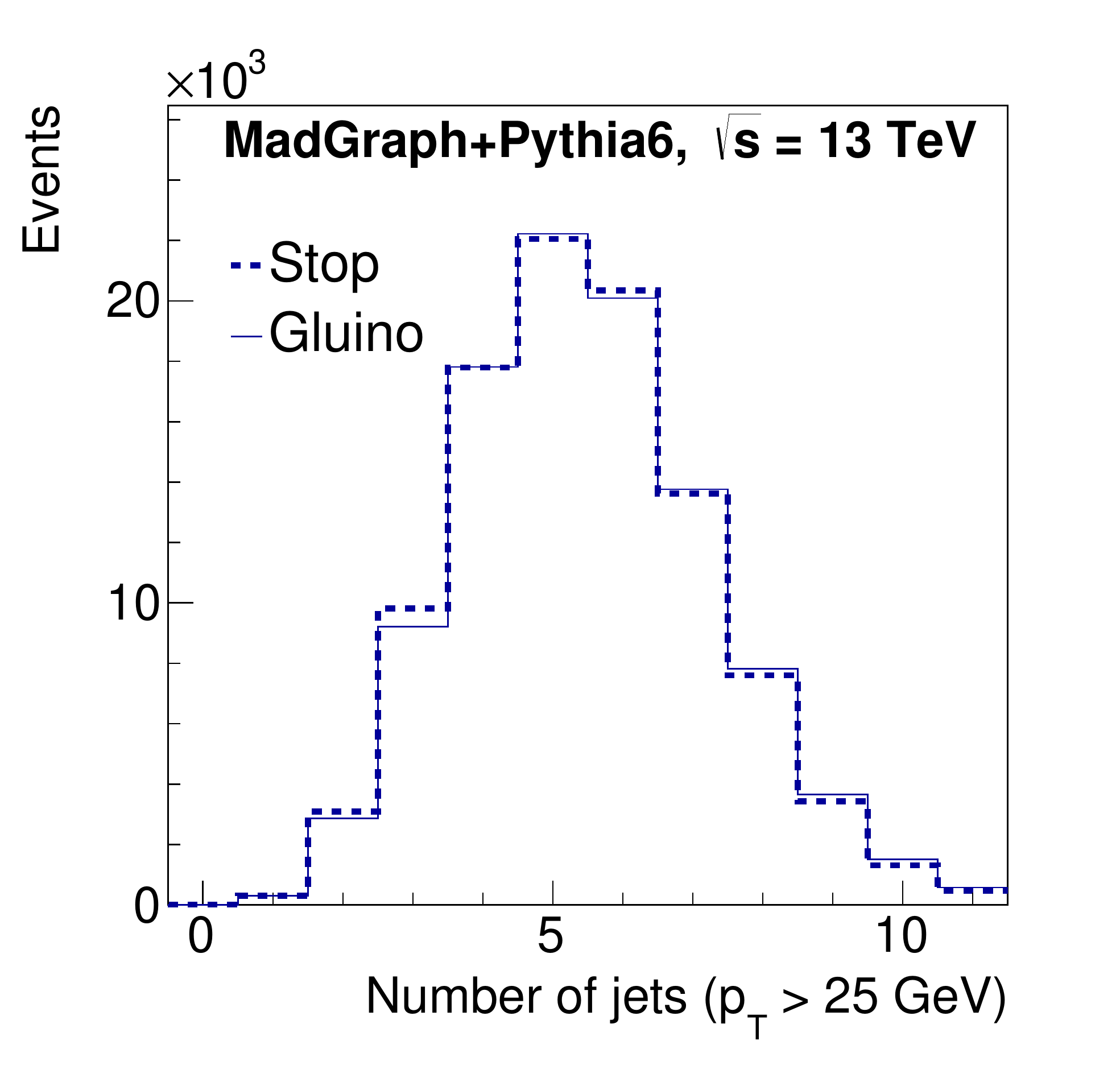}
 \caption{Kinematic distributions for two distinct, but equivalent SUSY models.  The model labeled {\it stop} is direct stop pair production with a 700 GeV stop mass and a massless neutralino.  The model labeled {\it gluino} is a gluino mediated compressed stop model with a 1.1 TeV gluino, a 652 GeV stop and a 650 GeV neutralino.  The generation is performed with MadGraph5\_aMC@NLO version 5.2.1.1~\cite{Alwall:2014hca} for the matrix element and Pythia 6.428~\cite{Sjostrand:2006za} for the parton shower and hadronization.  The 2 GeV difference between the stop mass and neutralino mass in the gluino mediated stop model is due to a 1.5 GeV charm mass in Pythia (for $\tilde{t}\rightarrow c\tilde{\chi}^0$).  A detector simulation is modeled with Delphes v3.1.2~\cite{deFavereau:2013fsa}.  Jets are clustered with the anti-$k_t$ algorithm~\cite{Cacciari:2008gp} with $R=0.4$ using the fastjet program~\cite{Cacciari:2011ma}.  The top plot shows the distribution of the leading four jet $p_T$, the bottom left plot shows the magnitude of the missing transverse momentum and the bottom right plot shows the number of jets with $p_T>25$ GeV.}
 \label{fig:kins}
  \end{center}
 \end{figure}

 \begin{figure}[h!]
 \begin{center}
 \includegraphics[width=0.8\textwidth]{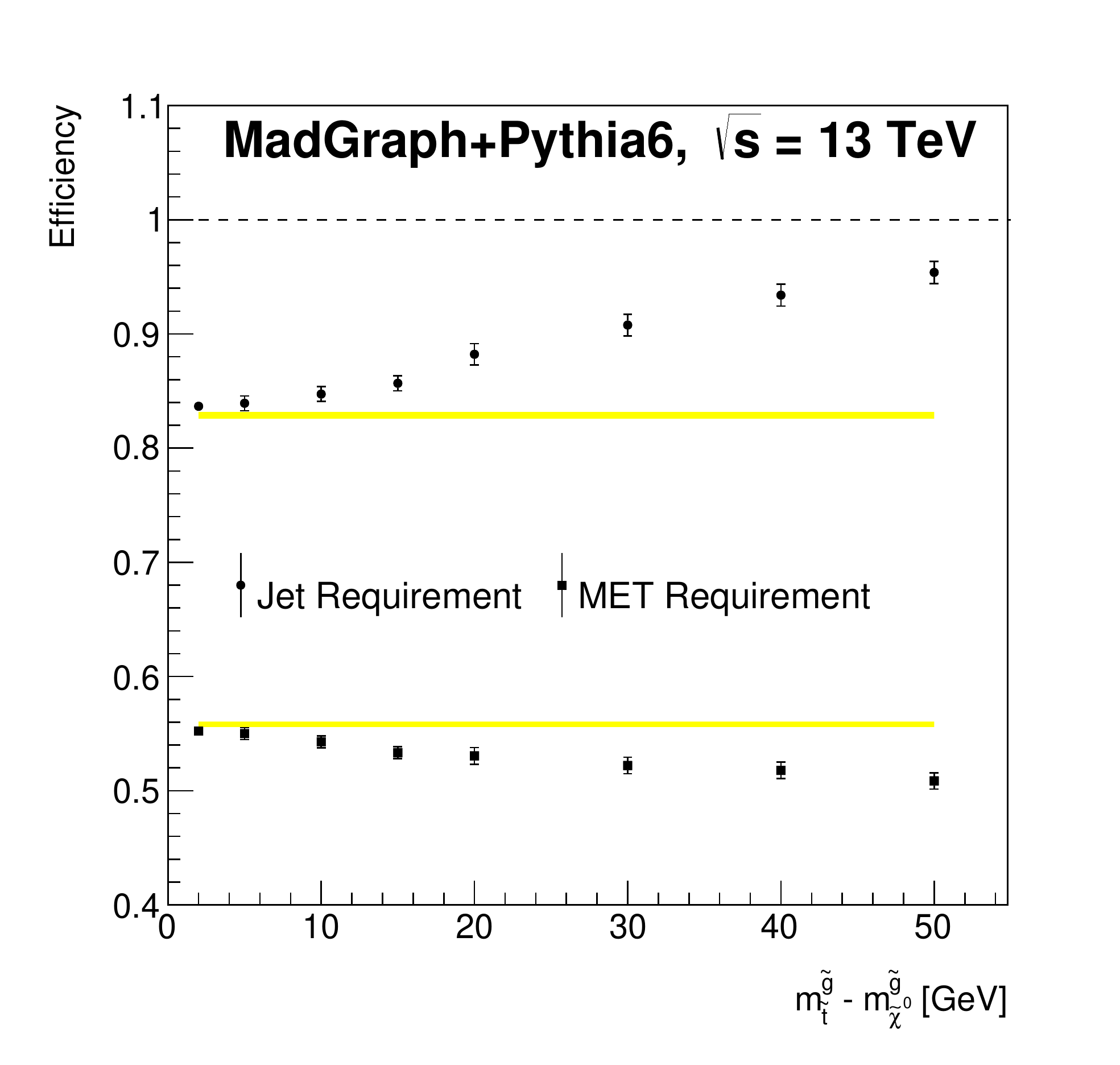}
 \caption{The efficiency of a 315 GeV threshold on the magnitude of the missing transverse momentum (square markers) and the efficiency of a four jet requirement with transverse momentum thresholds (75, 65, 40, 25) GeV (circles).  The markers indicate the efficiency for the compressed gluino model with $m_{\tilde{g}}^{\tilde{g}}=1.1$ TeV and $m_{\tilde{t}}^{\tilde{g}}=650$ GeV.  The yellow band is the efficiency for the equivalent stop model with $m_{\tilde{t}}^{\tilde{t}}=700$ GeV and a massless neutralino.  The band and the error bars represent statistical uncertainties from finite simulated datasets.  The generation is performed with MadGraph5\_aMC@NLO version 5.2.1.1~\cite{Alwall:2014hca} for the matrix element and Pythia 6.428~\cite{Sjostrand:2006za} for the parton shower and hadronization.   A detector simulation is modeled with Delphes v3.1.2~\cite{deFavereau:2013fsa}.  Jets are clustered with the anti-$k_t$ algorithm~\cite{Cacciari:2008gp} with $R=0.4$ using the {\sc FastJet} program~\cite{Cacciari:2011ma}.}
 \label{fig:kins2}
  \end{center}
 \end{figure}
 
 \clearpage
 \newpage

For every direct stop production model, the procedure above assigns an equivalent gluino model.  However, there are gluino models that do not have an equivalent stop pair production model.  At a fixed gluino mass $m_{\tilde{g}}$, let $m_{\tilde{t}}^{\tilde{t}}$ be the stop mass such that $\sigma(m_{\tilde{t}}^{\tilde{t}})=\sigma(m_{\tilde{g}})$.  Stop masses $m_{\tilde{t}}^{\tilde{g}}$ with $p(m_{\tilde{g}},m_{\tilde{t}}^{\tilde{g}}) > p(m_{\tilde{t}}^{\tilde{t}},0)$ have no equivalent stop pair production model.  For example, at $\sqrt{s}=8$ TeV, $m_{\tilde{t}}^{\tilde{t}}=700$ GeV and $m_{\tilde{g}}=1.1$ TeV have the same cross section, but clearly the gluino model with $m_{\tilde{t}}^{\tilde{g}}=0$ has no equivalent direct stop pair production model since the available momentum in the gluino model exceeds the direct stop mass.  However, this leads to an artificial truncation of gluino models that can be excluded by direct stop searches.  Acceptance generally increases with the top/neturalino momentum for a fixed cross section.  Therefore, if the point $(m_{\tilde{g}},m_{\tilde{t}}^{\tilde{g}})$ is excluded by a particular search, then all models specified by $(m_{\tilde{g}},x)$ with $x<m_{\tilde{t}}^{\tilde{g}}$ will also be excluded.

One can take this argument further to extrapolate to a region of phase space applicable to gluino searches, but forbidden to direct stop searches.  Consider a direct stop model with $m_{\tilde{t}}^{\tilde{t}}$ just beyond the exclusion limit.  The equivalent gluino model with mass $m_{\tilde{g}}$ will correspondingly not be excluded.  However, since the acceptance increases in decreasing $m_{\tilde{t}}^{\tilde{g}}$, there may be a model with gluino mass $m_{\tilde{g}}$ that is excluded, but has no equivalent direct stop model.  One way to estimate the excluded region is to fit the acceptance curve from Fig.~\ref{fig:excll} and predict the acceptance of a particular gluino model. For large values of $p$, the acceptance should be roughly linear in $p$ as the missing momentum in the event is linear in $p$.  Therefore, a linear fit for $p>200$ GeV is shown in Fig.~\ref{fig:excll} for extrapolating the acceptance to higher values of $p$.  Values of $(m_{\tilde{g}},m_{\tilde{t}}^{\tilde{g}})$ can be declared excluded if $\mathcal{L}_\text{int}\times\sigma(m_{\tilde{g}})\times \epsilon(p(m_{\tilde{g}},m_{\tilde{t}}^{\tilde{g}}))\times \kappa > n_\text{excluded}$, where $\kappa$ is the efficiency from Fig.~\ref{fig:susy:exclusion:acceptance} (roughly independent of stop/LSP mass) and $n_\text{excluded}$ is the model-independent limit on the number of BSM events from scanning over the number of BSM events in the SR and then re-running the exclusion fit.  For tNmed/tNhigh, the observed (expected) $n_\text{excluded}$ is $8.5/6.0$ ($9.2/6.0$).

One can do even better than naively recasting limits based on $n_\text{excluded}$ by tightening thresholds on the key variables (e.g. $E_\text{T}^\text{miss}$, $m_\text{T}$, and $am_\text{T2}$), but this change would require a careful assessment of the change in the background yield which is beyond the scope of this section.  
\clearpage

\paragraph{Derived Limits} \mbox{}\\

Re-casted direct stop limits are shown Fig.~\ref{fig:excl} alongside existing limits from the ATLAS same-sign search~\cite{Aad:2014pda} and the inclusive one lepton\footnote{A similar search exists in the zero lepton final state, with slightly weaker limits~\cite{Aad:2014wea} } search~\cite{Aad:2015mia}.   The same-sign limits are optimistic because the selection in Ref.~\cite{Aad:2014pda} requires a third hard jet, which is not part of the leading order description of the final state.  Estimates based on calculations with MG5\_aMC version 2.1.1~\cite{Alwall:2014hca} indicate that the fraction of the time an additional jet from initial or final state radiation has enough $p_T$ to pass the jet selection is roughly 40\%.  This agrees well with the three jet selection efficiency published in auxiliary material Table 64~\cite{aux} of the ATLAS search for a model with a large stop mass for which kinematically the soft $c$-quark jets will not pass the hard jet $p_\text{T}$ threshold.  As the mass splitting between the stop and the neturalino goes to zero, the reduction in the limit for the highest mass splitting reduces by $\lesssim100$~GeV (not shown).  The inclusive one lepton search is based on generic variables such as $E_\text{T}^\text{miss}$, $m_\text{T}$, effective mass, etc. and is not optimized for the $t\bar{t}+E_\text{T}^\text{miss}$ final state (the limits may even degrade as $m_{\tilde{t}}\rightarrow m_{\tilde{\chi}^0}$).  The improvement over these existing analyses for the reinterpreted direct search are shown in shaded blue in Fig.~\ref{fig:excl}.  The darkest blue is from the strict re-interpretation based on the strategy leading up to Table~\ref{tab:translate}.  The light blue area below the dark blue area is assumed excluded because the signal efficiency increases for the larger mass splitting.  The light blue area to the right of the dashed line is from interpolating and extrapolating the efficiency and comparing to the $n_\text{excluded}$.  For a 1.1 TeV gluino, the inclusive one lepton limit is extended vertically by about 225 GeV.
 \begin{figure}[h!]
 \begin{center}
 \includegraphics[width=0.9\textwidth]{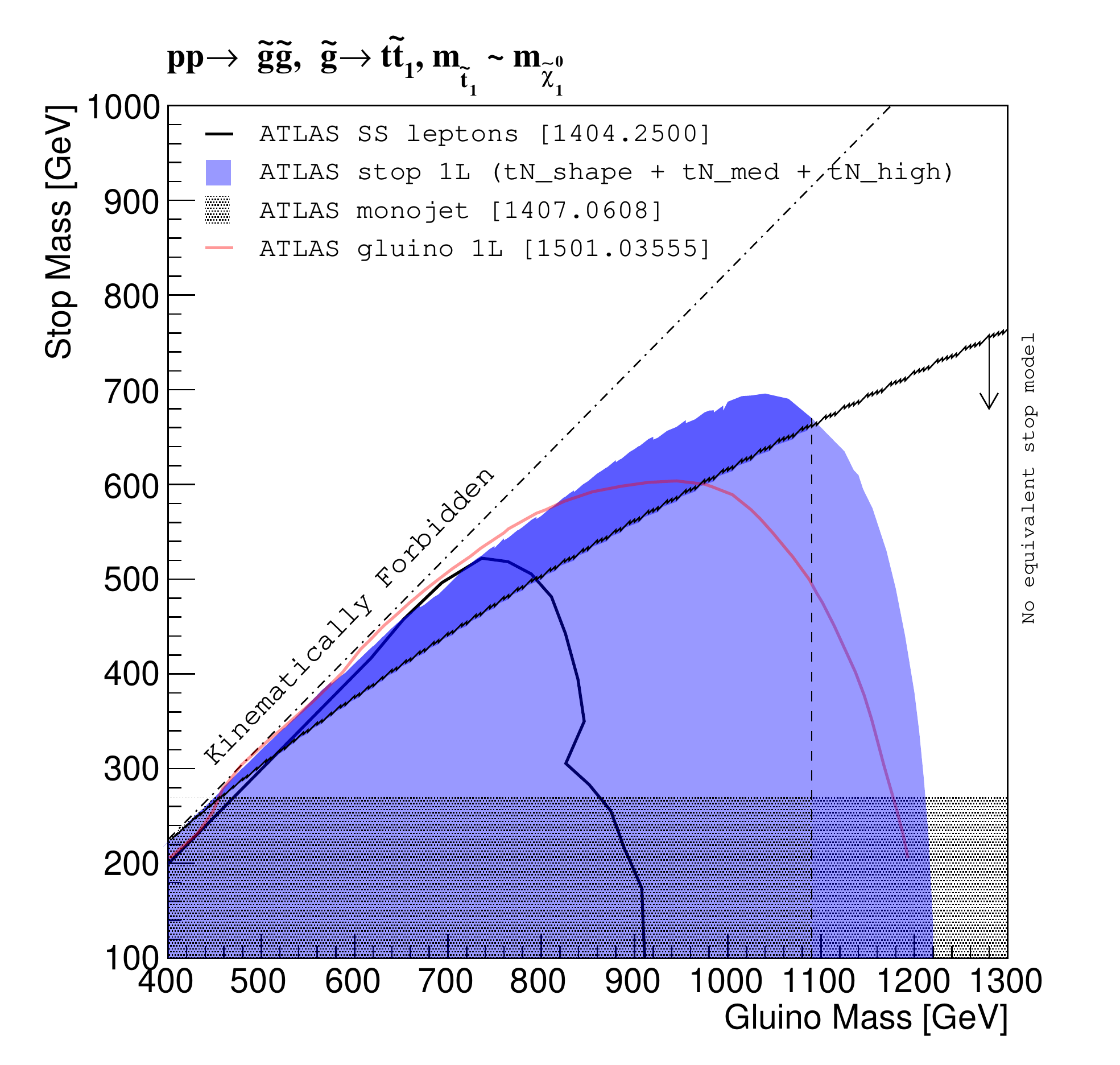}
 \caption{A comparison of existing limits and the re-interpretation of the direct stop search limits at $\sqrt{s}=8$ TeV (see Sec.~\ref{8TeVresults}).  The expected limits (based on the $\text{CL}_{s}$ procedure~\cite{Read:2002hq}) are used to control for statistical fluctuations in the observations.   The blue shaded region is the re-interpretation of the direct stop search.  Above the marked diagonal line, every gluino model has an equivalent stop model.  Below this line, there is no equivalent stop model and the exclusion limits are estimated by extrapolating the signal region acceptance as described in the text. The hatched region is from the $\sqrt{s}=8$ TeV ATLAS search for compressed direct stop production via an ISR monojet~\cite{Aad:2014nra}.  The red line is from the ATLAS inclusive one lepton search~\cite{Aad:2015mia} (Fig. 18a) and the black line is from the ATLAS same-sign lepton search~\cite{Aad:2014pda}.  }
 \label{fig:excl}
  \end{center}
 \end{figure}

\clearpage
\newpage

\paragraph{Transitioning to $\sqrt{s}=13$ TeV} \mbox{}\\
\label{sec:conc}

The sensitivity of the direct stop search to GMS models is a strong motivation for performing the stop search with the early $\sqrt{s}=13$ TeV data.  Table~\ref{tab:13TeV} summarizes the relative increase in cross-sections for direct stop models and GMS models from $\sqrt{s}=8$ to $13$ TeV.  Larger masses generally have a larger increase in cross-section because they are probing a smaller momentum fraction of the proton.   At the edge of the Run 1 sensitivity, the expected increase in the yield of stops from GMS is twice the corresponding yield for directly produced stops. 

\vspace{5mm}

\begin{table}[h!]
  \centering
\noindent\adjustbox{max width=\textwidth}{
\begin{tabular}{|c|c|c|c|c|c|c|}
\hline
$m_{\tilde{t}}^{\tilde{t}}$ & $\sigma^\text{8 TeV}(m_{\tilde{t}})$ & $m_{\tilde{g}}$ & $\sigma^\text{13 TeV}(m_{\tilde{t}}^{\tilde{t}})$ &  $\sigma^\text{13 TeV}(m_{\tilde{g}})$ & $\sigma(m_{\tilde{t}}^{\tilde{t}})$ 13 TeV/8 TeV&  $\sigma(m_{\tilde{g}})$ 13 TeV/8 TeV \\
\hline
600             & 0.03                                      & 1000                        & 0.2                                   & 0.3                                               & 7.0                  & 13.6        \\
700             & 0.008                                     & 1125                        & 0.07                                  & 0.1                                               & 8.3                  & 17.1        \\
800             & 0.003                                     & 1250                        & 0.03                                 & 0.06                                             & 9.8                  & 21.7    \\
\hline   
\end{tabular}}
\caption{The expected increase in yields for the direct stop search and the re-interpreted gluino search from $\sqrt{s}=8$ to $13$ TeV.  The first column is the stop mass in GeV, the second column is the stop cross section at $\sqrt{s}=8$ TeV in pb from Ref.~\cite{Kramer:2012bx}.  The third column is in pb and also uses Ref.~\cite{Kramer:2012bx} to solve $\sigma(m_{\tilde{t}}^{\tilde{t}})=\sigma(m_{\tilde{g}})$. The fourth and fifth columns give the cross sections for stop and gluino production at $\sqrt{s}=13$ TeV from Ref.~\cite{Borschensky:2014cia}.  The last two columns give the ratio of the increase in yields for direct stop and GMS production, respectively. }
\label{tab:13TeV}
\end{table}

\vspace{5mm}

All possibilities for natural SUSY should be targeted, including those with compressed scenarios.  If there is a light enough gluino to mediate, more territory for light stops and sbottoms will be accessible to the direct searches with the early data.  As discussed in Sec.~\ref{chapter:susy:signalregions}, GMS models were used as benchmarks for optimizing the $\sqrt{s}=13$ TeV analysis.  The results of that search are presented in the next section.

\clearpage		
		
\subsection{Early $\sqrt{s}=13$ TeV Results}			
\label{sec:13TeV}	
		
Figure~\ref{fig:susy:exclusion:overviewSR13} shows the exclusion limits from SR13 using the $3.2$ fb${}^{-1}$ from the 2015 $\sqrt{s}=13$ TeV dataset.  The signal region was optimized with the GMS benchmark model $(m_{\tilde{g}},m_{\tilde{t}},m_{\tilde{\chi}^0})=(1250,750,745)$, which is just on the edge of the exclusion limit in the right plot of Fig.~\ref{fig:susy:exclusion:overviewSR13}.   For a stop mass of $m_{\tilde{t}}\approx 650$ GeV, the GMS limit is extended over $400$ GeV in gluino mass.  Part of this gain is in a genuinely new region of parameter space that does not have an equivalent stop model.  A $m_{\tilde{g}}\sim 1.45$ TeV has the same cross-section at $\sqrt{s}=13$ TeV as $m_{\tilde{t}}\sim 850$ GeV; therefore the highest stop mass in the GMS model that corresponds to a physical direct stop model is about $920$ GeV.  The highest gluino mass for $m_{\tilde{t}}\approx 650$ GeV that has a physical direct stop model equivalent is about $1.2$ TeV.

By construction, SR13 is sensitive to $t\bar{t}+E_\text{T}^\text{miss}$ topologies and therefore it can be used to set limits on direct stop models as well as the target GMS models.  The left plot of Fig.~\ref{fig:susy:exclusion:overviewSR13} shows that the observed limit improves by over $75$ GeV for a massless LSP, albeit with a large uncertainty.  It is likely that with a dedicated optimization, the limits would be even stronger, though the limit statistics of the 2015 dataset is prohibitive for a much stronger direct stop limit.  With the full Run 2 dataset, it may be possible to discover or rule out $m_{\tilde{t}}\lesssim 1 $ TeV, the benchmark for naturalness. 
		
\begin{figure}[h!]
\begin{center}
\includegraphics[width=0.5\textwidth]{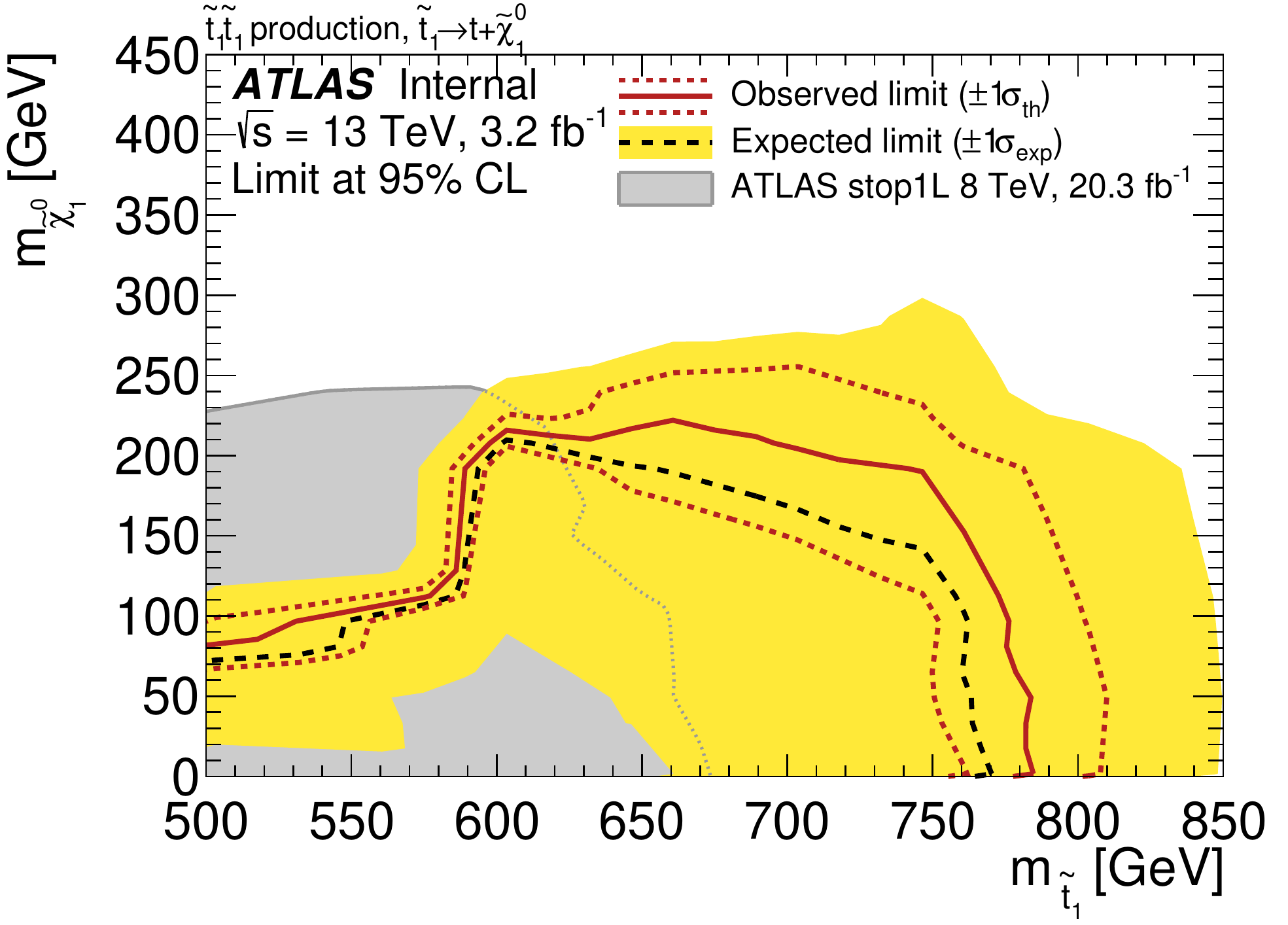}\includegraphics[width=0.5\textwidth]{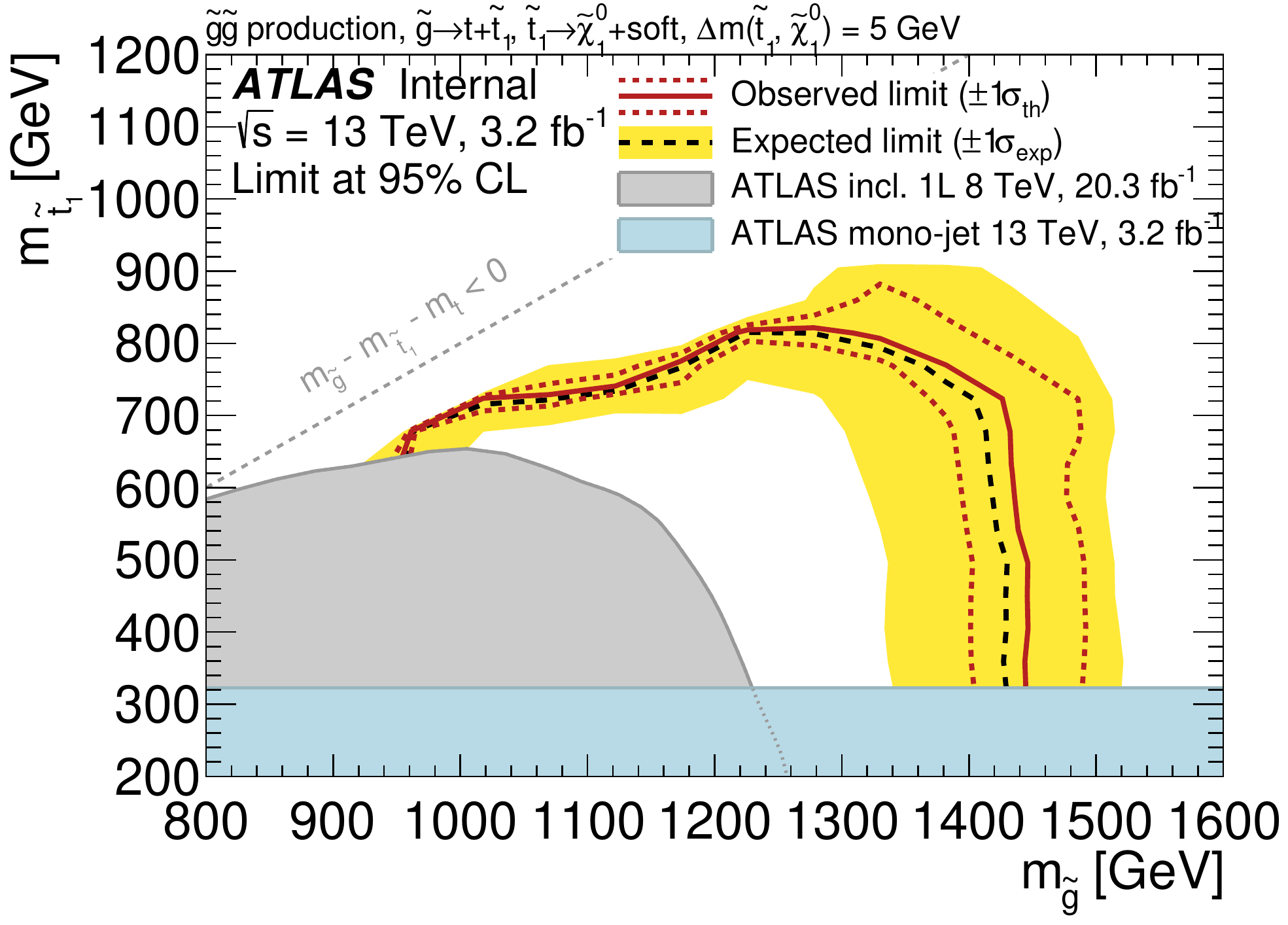}
 \caption{The exclusion limits using SR13 at $\sqrt{s}=13$ TeV for direct stop pair production (left) and GMS with $m_{\tilde{t}}-m_{\tilde{\chi}^0}=5$ GeV (right).  The blue filled area in the right plot is from the early Run 2 ATLAS mono-jet search~\cite{Aaboud:2016tnv}.}
 \label{fig:susy:exclusion:overviewSR13}
  \end{center}
\end{figure}		
		
		\clearpage	
		\section{The LHC Run I SUSY Epilogue}
		\label{epi}

In addition to the lack of evidence for stops, the Run 1 (and early Run 2) data do not support the existence of weak-scale SUSY in general.  Both ATLAS and CMS have conducted extensive searches for SUSY in a multitude of final states, with various numbers of jets, leptons, and photons.  The kinematic reach of the detectors have been exploited in order to be sensitive to high mass particles, which may be produced with a low cross section.  However, with the large number of searches ($\mathcal{O}(100)$ between ATLAS and CMS), some low $p$-value results are expected due to statistical fluctuations.  This section presents\footnote{The analysis presented here is published in Ref.~\cite{Nachman:2014hsa} and includes input from T. Rudelius.} a meta-analysis of the $\sqrt{s}=8$ TeV ATLAS and CMS SUSY searches, studying the distribution of $p$-values associated with the SM-only hypothesis.

\subsection{Constructing the Dataset}

Even though the 8 TeV dataset was collected in 2012, both ATLAS and CMS continue(d) to analyze the data.  This section presents data from all analyses prior to an arbitrarily cutoff at the SUSY 2014 conference (July 20, 2014).  This includes 17 ATLAS papers~\cite{Aad:2014nra,Aad:2014mra,Aad:2014lra,Aad:2014kra,Aad:2014yka,Aad:2014bva,Aad:2014wea,Aad:2014iza,Aad:2014pda,Aad:2014vma,Aad:2014mha,Aad:2014qaa,Aad:2014nua,Aad:2013gva,Aad:2013yna,Aad:2013ija,Aad:2013wta} and 12 CMS papers~\cite{Khachatryan:2014qwa,Khachatryan:2014doa,Chatrchyan:2014aea,Chatrchyan:2014lfa,Chatrchyan:2013mya,Chatrchyan:2013fea,Chatrchyan:2013iqa,Chatrchyan:2013xna,Chatrchyan:2013xsw,Chatrchyan:2013wxa,Chatrchyan:2013lya,Chatrchyan:2012paa}.  The difficulty in assembling the dataset is to understand the correlations between measurements.  The general strategy is to categorize the various searches by their selections on jets, leptons, and photons.  Two analyses which have non-overlapping requirements in the number and properties of these objects are treated as uncorrelated.  For the data, this is an excellent assumption and only breaks down in the rare case that the data in one signal region is used for the background estimate of another signal region.  If two signal regions are such that one is a subset of the other, then a decorrelation procedure is attempted in order to produce two orthogonal regions.  If the yields are $x\pm \sigma_x$ and $y\pm \sigma_y$ with $x<y$, then the decorrelated regions have yields $x\pm \sigma_x$ and $(y-x)\pm \sqrt{\sigma_y^2-\sigma_x^2}$.  In all other cases, it is not possible with the information given to determine the correlations and the signal regions in question are simply not used.  In general, if there are two analyses with an unknown correlation, the one with more signal regions is preferred unless the one with fewer regions already has orthogonal selections.  The regions to be included where selected before looking at any $p$-values in order to minimize potential biases.  Tables~\ref{tab:cms} and~\ref{tab:atlas} give some summary information about the dataset construction given the general guidelines from above.  In total, there are 124 ATLAS regions and 325 CMS regions.

\vspace{10mm}

\begin{table}[h!]
  \centering
\noindent\adjustbox{max width=\textwidth}{
\begin{tabular}{ c c c }
	arXiv reference & Category & Note \\
	\hline
  1303.2985 & Multijets &  \begin{tabular}{@{}c@{}} Regions orthogonal; drop those with \\ $H_T>800$ GeV due to overlap with 1402.4770\end{tabular}    \\
  \hline
  1402.4770 & Multijets &   \begin{tabular}{@{}c@{}} Regions orthogonal; drop those with \\ $H_T\in[500,800]$ GeV due to overlap with 1303.2985\end{tabular}    \\
    \hline
  1305.2390 & Multijets &  Unknown correlation with 1303.2985 and 1402.4770: remove\\
    \hline
  1311.4937 & One Lepton &  \begin{tabular}{@{}c@{}} Regions orthogonal; use the LS method \\ for uncertainties when given a choice\end{tabular} \\
    \hline
  1308.1586 & One Lepton &  \begin{tabular}{@{}c@{}} Unknown correlation with 1311.4937. \\ Prefer 1311.4937 as its regions are orthogonal\end{tabular}   \\
    \hline
  1212.6194 & Same sign leptons & SR6 $\subseteq $ SR3 $\subseteq $ SR4 $\subseteq $ SR1 $\subseteq $ SR0.  Drop other regions.\\
    \hline
  1311.6736 & Same sign leptons & \begin{tabular}{@{}c@{}} Drop regions with $\geq 2$ b-jets due to overlap with 1212.6194. \\    Arbitrarily pick the low $p_T$ region \end{tabular}  \\
    \hline
  1306.6643 & Multileptons &  \begin{tabular}{@{}c@{}} Unknown overlap with 1404.5801.  \\ Use 1404.5801 as it has more regions. \end{tabular} 
   \\
     \hline
  1404.5801 & Multileptons & Regions orthogonal.\\
    \hline
  1405.3886 & Multileptons & Use the two lepton OS regions only.\\
    \hline
  1405.7570 & Multileptons &  \begin{tabular}{@{}c@{}} Use the two lepton OS regions only.  \\ Use signal sensitive regions (as described in the text)\end{tabular} 
   \\
     \hline
  1312.3310 & Diphoton & Regions orthogonal.\\
  \hline
\end{tabular}
}
\caption{An overview of the signal regions used in the meta-analysis from 8 TeV CMS searches.}
\label{tab:cms}
\end{table}

\begin{table}[h!]
\centering
\noindent\adjustbox{max width=\textwidth}{
\begin{tabular}{ c c c }
	arXiv reference & Category & Note \\
	\hline
1308.1841  & Multijets &\begin{tabular}{@{}c@{}} 8j80$x$b $\subseteq$ 8j50$x$b, $x\in\{0,1,2\}$. Unknown \\ 
 correlations between $M_J^\Sigma$ regions and others, drop
\end{tabular}  \\
  \hline
1308.2631  & Multijets & SRA $m_{CT}(350)$ $\subseteq$  SRA $m_{CT}(300)$ $\subseteq$ $\cdots$ $\subseteq$   SRA $m_{CT}(150)$  \\
  \hline
1407.0608 & Multijets & M3 $\subseteq$ M2 $\subseteq$ M1; C2 $\subseteq$ C1\\
  \hline
1405.7875 & Multijets &\begin{tabular}{@{}c@{}}  2jt $\subseteq$ 2jm $\subseteq$ 2jl.  2jW $\cap$ 3j unknown, drop 2jW. \\ 
 6jt $\subseteq$ 6jm $\subseteq$ 5j and 6jl (5j $\cap$ 6jl $=\{\}$ once 6jm is removed).\\
 Drop all other regions due to unknown correlations.
\end{tabular} 
\\
  \hline
1406.1122 & Multijets & \begin{tabular}{@{}c@{}} SRA2 $\subseteq$ SRA1; SRA4 $\subseteq$ SRA3.  Drop SRB.  \\  SRC3 $\subseteq$ SRC2 $\subseteq$ SRC1
\end{tabular} \\
  \hline
1407.0600 & Multijets &\begin{tabular}{@{}c@{}} SR-0l-7j-C $\subseteq$ B $\subseteq $ A; Drop 4j regions due to 4j $\cap$ 7j = ?
\end{tabular} \\
  \hline
1407.0583 & One Lepton & \begin{tabular}{@{}c@{}} Unknown correlations between shape fit regions, \\ consider only tN\_diag (signal sensitive regions).\\
tN\_high $\subseteq$ tN\_med $\subseteq$ tightest tN\_diag region.\\
bCb\_high $\subseteq$ bCb\_med1.  Unknown relation \\
between bCa\_low and bCa\_med, drop low.  Unknown\\
correlation between bCd, tNbC\_mix and other regions, drop
\end{tabular} \\
  \hline
1407.0603 &At Least One $\tau$ & \begin{tabular}{@{}c@{}} 1$\tau$Tight $\subseteq$ 1$\tau$Loose, 2$\tau$ GMSB $\subseteq$ 2$\tau$ nGM $\subseteq$ 2$\tau$ Incl.\\
Unknown overlap between 2$\tau$ bRPV and 2$\tau$ GM,\\
drop bRPV.  $\tau$+l bGM $\subseteq$ $\tau$+l mSUGRA.  Unknown overlap\\
 between $\tau$+l GMSB and bRPV, drop bRPV
\end{tabular}  
 \\
   \hline
1407.0350 & At least two $\tau$s & \begin{tabular}{@{}c@{}} C1C1 $\cap$ C1N2 = ?, drop C1C1 \\ DS-lowMass $\cap$ DS-highMass = ?, drop lowMass
  \end{tabular}  \\
    \hline
1403.4853 & Two OS Leptons&   \begin{tabular}{@{}c@{}} unknown correlation of L90,120 with 1403.5294, drop \\ unknown correlation between L110-100, drop L100 \\ H160 orthogonal, drop MVA region\end{tabular}      \\
  \hline
1403.5294 & Two OS Leptons &  \begin{tabular}{@{}c@{}} Jet veto regions orthogonal to other searches, drop Zjets  \\ $m_{T2}^{150}(x)\subseteq m_{T2}^{120}(x) \subseteq WWc(x) \subseteq m_{T2}^{90}(x)$, $x\in\{SF,DF\}$ \\
Overlap of $WWb(x)$ with $m_{T2}^{90}(x)$ unknown, drop
 \end{tabular} \\
   \hline
1404.2500 & Same Sign Leptons &   \begin{tabular}{@{}c@{}}  Regions orthogonal.  Drop SR3Llow/high \\ due to unknown overlap with 1402.7029\end{tabular}     \\
  \hline
1403.5222 & Multileptons & SR$x$b $\subseteq $ SR$x$a, $x\in\{2,3\}$ \\
  \hline
1402.7029 & Three Leptons &  Regions orthogonal except SR2$\tau$a $\cap$ SR2$\tau$b = ?, drop b
\\
  \hline
1405.5086 & $\geq 4$ Leptons &  SR$x$noZb $\subseteq $ SR$x$noZa, $x\in\{0,1,2\}$  \\
  \hline
 1310.3675 & Disappearing Tracks & Region inclusion by increasing $p_T$ cut\\
   \hline
1310.6584 & Out-of-time & For the muon veto, inclusion by jet $p_T$\\
   \hline
\end{tabular}
}
\caption{An overview of the signal regions used in the meta-analysis from 8 TeV ATLAS searches.  Note that OS = opposite sign.  The stop search results are part of the `One Lepton' category.}
\label{tab:atlas}
\end{table}

\clearpage
\newpage

\subsection{Statistical Analysis}

Once the ATLAS and CMS datasets are constructed, the expected and observed distributions of $p$-values are computed for both a Gaussian and a lognormal distribution of the expected number of counts (the number of counts itself is assumed to be Poisson).  A $p$-value was assigned to each data point according to
\begin{equation}
\mbox{p-value} = \int_0^\infty{\phi(\lambda|\mu,\sigma) P_{\geq n}(\lambda)d\lambda}.
\label{pvalueeq}
\end{equation}
Here, $P_{\geq n}$ is the probability of observing $n$ or more counts given a Poisson distribution with parameter $\lambda$,
\begin{equation}
P_{\geq n}(\lambda) = \sum_{k=n}^\infty \frac{e^{-\lambda} \lambda^k}{k!} =1- \sum_{k=0}^{n-1} \frac{e^{-\lambda} \lambda^k}{k!}.
\end{equation}

\noindent In addition to analyzing the excesses, one can also study the deficits in the SUSY search regions by replacing $P_{\geq n}$  in Eq.~\ref{pvalueeq} with $P_{\leq n}$: the probability of observing $n$ or less counts given a Poisson distribution with parameter $\lambda$.
The function $\phi(\lambda|\mu,\sigma)$ is the probability distribution function of the specified random variable with mean $\mu$ and standard deviation $\sigma$.  These parameters are the expected value for the number of counts ($\mu$) and the uncertainty on that value ($\sigma$).  For the Gaussian distribution,
\begin{equation}
\phi(\lambda|\mu,\sigma) = \frac{1}{N\sigma\sqrt{2\pi}} e^{-(\lambda-\mu)^2/2\sigma^2},
\end{equation}
where $N$ is a normalization constant correcting for the fact that $\lambda$ cannot be negative, and so the negative part of the distribution must be cut off.  For the lognormal distribution, whose support is $\mathbb{R}_+$, no such normalization constant is required,
\begin{equation}
\phi(\lambda|\mu,\sigma) = \frac{1}{\lambda\tilde{\sigma}\sqrt{2\pi}} e^{-(\ln{\lambda}-\tilde{\mu})^2/2\tilde{\sigma}^2},
\end{equation}
with $\tilde{\mu} := \ln{\mu^2/\sqrt{\mu^2+\sigma^2}}$, $\tilde{\sigma} := \sqrt{\ln{1 + \sigma^2/\mu^2}}$ defined so that the lognormal distribution is precisely the distribution of $Y = e^X$ for a Gaussian random variable $X$ with mean $\tilde{\mu}$ and variance $\tilde{\sigma}^2$.

One might expect the distribution of $p$-values defined in this way to be uniformly distributed on the interval $[0,1]$ under the null hypothesis, in accordance with the usual interpretation of $p$-values as the probability of observing a more significant result in precisely $p\times100\%$ of studies.  However, this intuitive understanding is only correct when the distribution is continuous~\cite{Hartung}, not in the case of Poisson distribution considered here.  As a result, the first step of of the analysis is to compute the {\it expected distribution} of $p$-values under the null hypothesis and then compare this with the observed distribution of $p$-values.  The expected distribution of p-values is determined by summing up the probability that each particular trial would fall into one of ten bins, $(\frac{i}{10},\frac{i+1}{10}], i = 0,...,9$,
\begin{equation}
\mbox{Pr}\left(\frac{i}{10} < \mbox{p-value} \leq \frac{i+1}{10} \right) = \int_0^\infty{d\lambda f_i(\lambda) \phi(\lambda|\mu,\sigma)},
\end{equation}
where
\begin{equation}
f_i(\lambda) = \sum_{m=0}^\infty\left[ \mbox{Pr}(X = m) \times \left\{ \begin{array}{lc}
1 & \mbox{ if } \mbox{Pr}(X \geq m) \in (\frac{i}{10},\frac{i+1}{10}]  \\
0 & \mbox{ otherwise } \end{array}   \right\} \right].
\label{expecteddisteq}
\end{equation}
Here, $X \sim$ Poisson$(\lambda)$ is the random variable measuring the number of counts, and the $\geq$ in Eq.~\ref{expecteddisteq} is replaced by a $\leq$ when computing deficits below rather than excesses above the expected signal.

Some of the studied signal regions had $0$ expected events.  There is no lognormal distribution with a mean of 0, so these regions had to be discarded in performing the lognormal analysis.  Fortunately, this only applied to seven of the CMS signal regions and none of the ATLAS ones.  However, a fairly sizable fraction had an expected mean that was very close to zero.  For these trials, it is reasonable to suspect that neither a Gaussian with a cutoff imposed at $0$ nor a lognormal will provide a good approximation to the true error distribution.  As a check, the analysis was repeated after removing all data points with $\mu-2 \sigma < 0$ ($\approx 10\%$ for ATLAS, $30\%$ for CMS) .  The results of this second analysis did not differ qualitatively from the first, indicating that the results of the original analysis are not significantly affected by the statistical modeling of these data points.

Note that the both the log-normal and Gaussian distributions are simple approximations to complicated likelihood functions (see e.g. Sec.~\ref{sec:susy:stats}); however, they should capture the essential features of the distributions and the difference between the two approaches will give a sense of the robustness of the procedure.

\subsection{Results and Discussion}

The results of the combined ATLAS and CMS analysis are shown in Figures~\ref{figcomb1} and the results of statistical tests are presented in Tables~\ref{tablecomb1}-\ref{tablecomb2}. There is a lack of deficits with $p<0.1$ at a level of $3.23\sigma$ and a lack of deficits with $p<0.3$ at a level of $3.15\sigma$ in the Gaussian case and $4.10\sigma$ in the lognormal case.  This trend is also observed separately in both the ATLAS and CMS datasets.  The observed $p$-value distribution is significantly different from the expected one, but the difference is not concentrated at low $p$-values. It is interesting to note that the distributions observed here are somewhat different from those observed in the $\sqrt{s}=7$~TeV version of this study~\cite{NachmanRudelius}.  That analysis also revealed a deficit of $p$-values in the tails of the distribution, but there were significantly fewer $p$-values $<0.1$, indicating a possible overestimation of the mean background as well as the uncertainty.  Here, there is actually a slight (statistically insignificant) surplus of $p$-value excesses $<0.1$ in the Gaussian case, but a clear lack of $p$-value deficits $<0.1$ in both the Gaussian and lognormal cases.  
The results presented here indicate:

\begin{enumerate}
\item The uncertainties are not well-modeled by Gaussian or lognormal distributions.
\item SM predictions have an inherent bias not captured by systematic uncertainties.
\item There is a contribution of SUSY or another model of BSM that causes the observed distribution of $p$-values to deviate from the expected one.
\end{enumerate}

\noindent The present analysis cannot distinguish between these three possibilities.  At the least, the differences indicate that the true uncertainty distributions are not well described by Gaussian or lognormal distributions with the reported means and uncertainties\footnote{This is hopefully a strong motivation for making additional statistical details about search results public.  There is also a sociological aspect of SUSY searches related to (2); in particular, people often `worry less' about deficits than excess so they recieve less scrutiny.}.  It will be interesting to see how this picture changes with the Run 2 dataset; hopefully the new physics is not so subtle that a meta-analysis is required to identify it.  The analysis presented in Part~\ref{part:susy} has significantly reduced the parameter space of natural SUSY.  In doing so, new discriminating variables and background estimation techniques have been developed that will continue to be useful for probing the high energy nature of the SM and beyond.

\vspace{10mm}

\begin{figure}[h!]
\begin{center}
\includegraphics[width=0.5\textwidth]{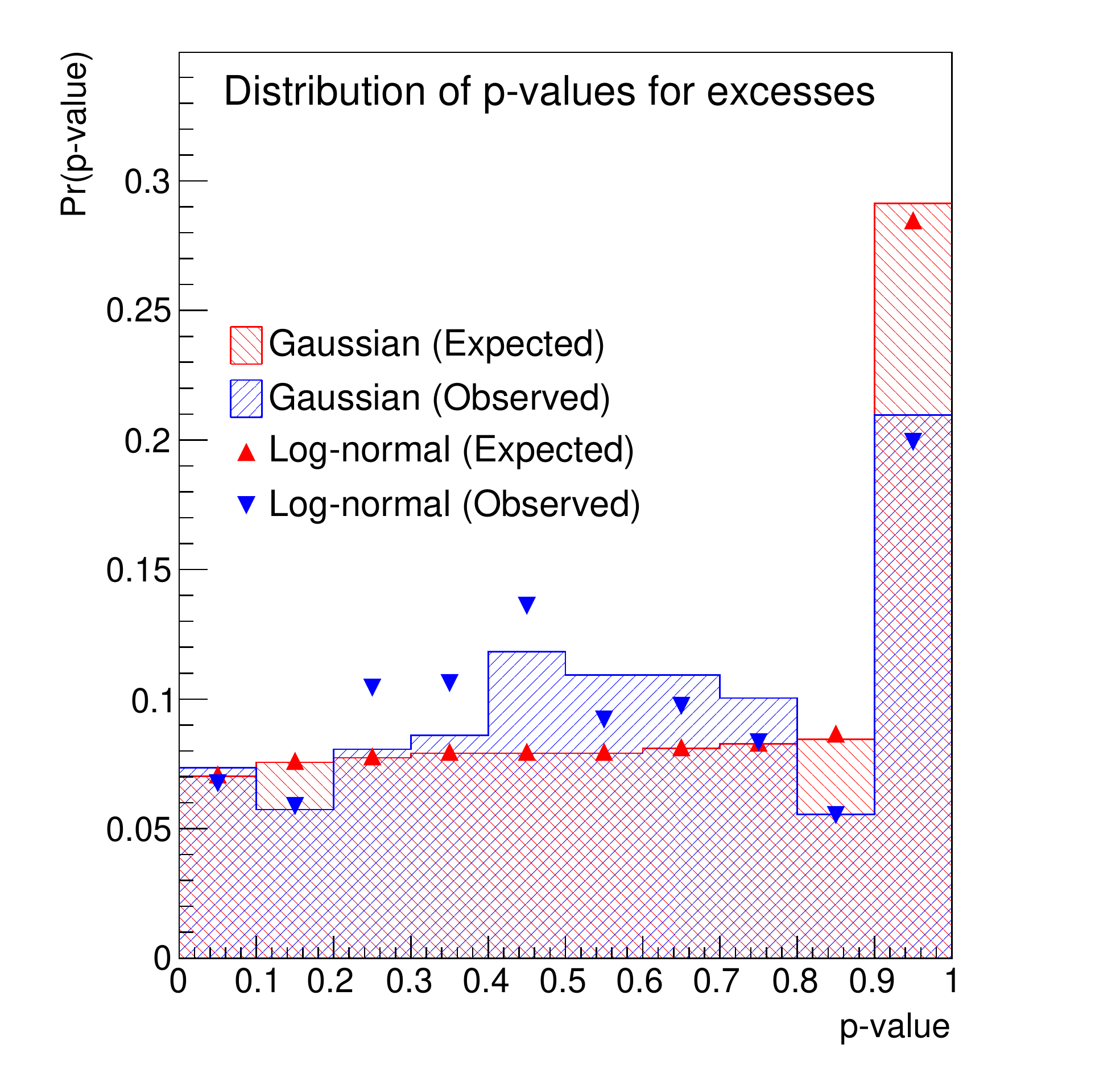}\includegraphics[width=0.5\textwidth]{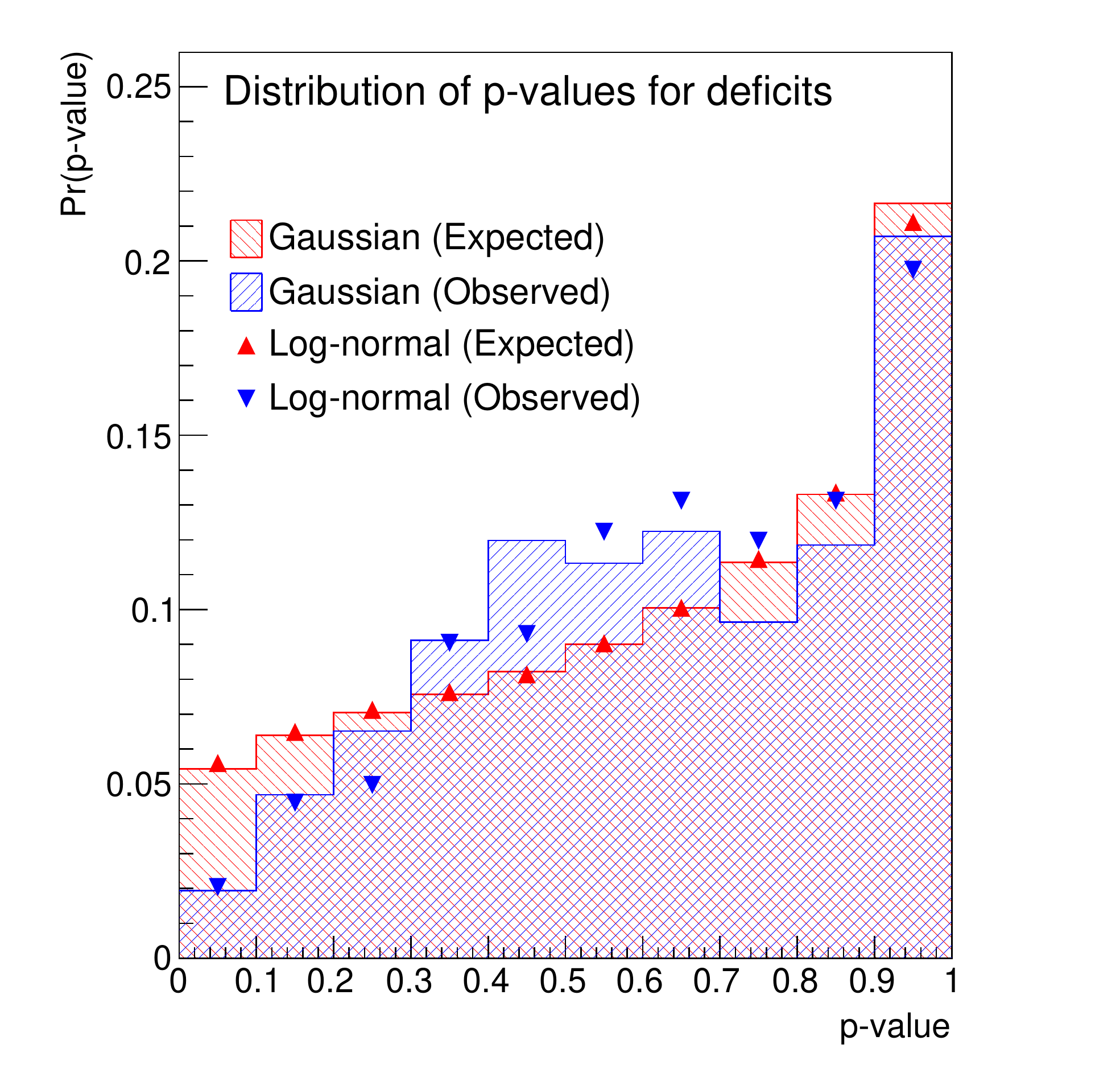}
\end{center}
\caption{The distribution of $p$-values for excesses (left) and deficits (right) for both Gaussian and log-normal uncertainty distributions.  For a continuous probability distribution, one expects the distribution of $p$-values to be uniform on $[0,1]$.}
\label{figcomb1}
\end{figure}

\clearpage

\vspace{10mm}

\begin{table}[h!]
\centering
\noindent\adjustbox{max width=\textwidth}{
\begin{tabular}{||c|c|c|c|c|c||} \hline
\multirow{2}{*}{Quantity} & \multirow{2}{*}{Dist. under $H_0$ ($T$)}  &\multicolumn{2}{|c|}{Test statistic ($t$)} &\multicolumn{2}{|c||}{Pr($|T|>t$)} \\\cline{3-6}
 &&Gaussian &LN&Gaussian& LN\\ \hline
Trials with $p<0.1$& N(0,1) & 0.23 & $-0.27$  &  0.82 & 0.79 \\ \hline
Trials with $p<0.3$&N(0,1)& $-0.57$  & 0.31 & 0.57 & 0.75  \\ \hline
Trials with $p<0.2$ or $p>0.8$ &N(0,1)& $-5.28$ & $-5.77$ & $\ll 0.001$ & $\ll 0.001$ \\ \hline
Expected vs. observed dist. & $\chi^2_9$&36.18 & 45.74  & $\ll 0.001$ & $\ll 0.001$ \\ \hline
\end{tabular}}
\caption{Results for statistical hypothesis tests on combined ATLAS and CMS excesses, under the assumptions of Gaussian and lognormal error distributions.}
\label{tablecomb1}
\end{table}

\vspace{10mm}

\begin{table}[h!]
\centering
\noindent\adjustbox{max width=\textwidth}{
\begin{tabular}{||c|c|c|c|c|c||} \hline
\multirow{2}{*}{Quantity} & \multirow{2}{*}{Dist. under $H_0$ ($T$)}  &\multicolumn{2}{|c|}{Test statistic ($t$)} &\multicolumn{2}{|c||}{Pr($|T|>t$)} \\\cline{3-6}
 &&Gaussian &LN&Gaussian& LN\\ \hline
Trials with $p<0.1$& N(0,1) &$ -3.23$ & $-3.23$ & 0.001 & 0.001  \\ \hline
Trials with $p<0.3$&N(0,1)& $-3.15$ & $-4.10$ & 0.002& $\ll 0.001$  \\ \hline
Trials with $p<0.2$ or $p>0.8$ &N(0,1)&$-3.24$& $-3.04$ & 0.001 & 0.002 \\ \hline
Expected vs. observed dist. & $\chi^2_9$& 29.04 &  27.11  &  0.0006 &  0.001 \\ \hline
\end{tabular}
}
\caption{Results for statistical hypothesis tests on combined ATLAS and CMS deficits, under the assumptions of Gaussian and lognormal error distributions.}
\label{tablecomb2}
\end{table}

\chapter{Conclusions and Future Outlook}

Undoubtedly, particle physics will embark on many grand adventures in the near future.  Our experimental and theoretical tools allow us to probe the SM to unprecedented precision and we have begun a full expedition of the unexplored TeV landscape.  The SM does not predict any undiscovered particles or forces, but they must be there.  Something new is expected, but anything new will be a surprise.  Supersymmetric models with a light stop still remain some of the most tantalizing theories.  There are many extensions of the search presented in Part~\ref{part:susy} that will push the sensitivity to higher mass scales, more complex decay chains, and more compressed mass spectra.  These extensions will benefit from and extend the techniques presented here to identify signal-like events and to suppress and estimate background processes.  At the same time, all the searches from Sec.~\ref{epi} (and more) have extensively mapped out the tails of kinematic distributions in the $\sqrt{s}=8$ TeV and the early $\sqrt{s}=13$ TeV data.  Now, we know a lot about where there is nothing, but there is a lot we can learn about where there is something\footnote{Think about how we know the ocean floor with 5~km precision~\cite{ocean} while the surface of Mars has been mapped with 100~m precision~\cite{nasa}.  There are likely no Martians amongst high energy quarks and gluons, but there is a lot of rich structure at the bottom of the ocean.}.  The phenomenology of a multi-TeV jet is mostly governed by a single number\footnote{At LHC energies, we will also be able to probe electroweak radiation in a regime with reduced phase space suppression for $W$ and $Z$ emission during jet formation.}: $\alpha_s$. Yet, there are qualitatively different physics processes that occur on all scales spanning $\Lambda_\text{QCD}\sim 1$ GeV all the way to the energy of the initiating quark or gluon.  With jet substructure techniques, we can probe jet formation by studying the quantum properties of jets.  Pushing this frontier beyond what is presented in Part~\ref{part:qpj} will require both experimental and theoretical advances\footnote{Significant progress on track reconstruction inside high $p_\text{T}$ jets between Run 1 and Run 2 will already boost sensitivity in the future~\cite{Aad:2014yva}.}.  Track reconstruction inside jets will play an increasingly important role in reconstructing jet substructure and will allow us to push boson and top quark tagging to the multi-TeV regime.  The LHC has performed exceedingly well and the ATLAS collaboration, as a team, has shown that we can harness our detector to measure extreme energies with great precision.  With more data and new ideas, together we will uncover the next clue in Nature's captivating mystery.  

\vspace{10mm}

\begin{flushright}
Benjamin Philip Nachman\\
Geneva, Switzerland, July 2016
\end{flushright}

    \clearpage
    \appendix   
    
\chapter{Radiation Damage}
\label{raddamage}

As the closest subdetector to the interaction point, the ATLAS pixel detector will be exposed to an extreme amount of radiation over its lifetime ($\gtrsim 10^{15}$ $n_\text{eq}/\text{cm}^2$).  The modules composing the detector are designed to be radiation tolerant, but their performance will degrade over time.  It is therefore critical to model the impact of radiation damage for accurate simulation of tracking in the future.  Including a radiation damage model is especially relevant for the high luminosity upgrade of the LHC; the instantaneous and integrated luminosity will significantly exceed current values, but simulations of the upgraded inner detector (ITK)~\cite{ATLAS:1502664} do not include the effects of radiation damage.  This section briefly documents a digitization model\footnote{This work is built on previous studies by many people and benefited from direct technical input from M. Benoit, M. Bomben, C. Bertsche, and R. Carney.} designed for the ATLAS software system that includes the impact of radiation damage.  For a detailed account of the impact of radiation damage to silicon sensors, see Ref.~\cite{MichaelMoll} and Sec. 5 in Ref.~\cite{Aad:2008zz}.  The model described here includes two impacts of bulk defects: modifications to the electric field inside the sensors and charge trapping.  Energy deposition in the silicon is modeled with {\sc Geant4} and then various effects illustrated in Fig.~\ref{fig:app:raddamge1} are accounted for during {\it digitization}: the modeling of the detection and readout of energy deposited by charged particles.  A minimum ionizing particle (MIP) is a charged particle that has momentum corresponding to the minimum average energy loss per distance ($\langle dE/dx\rangle $) in a given material.  Since the stopping power increases only logarithmically for several decades in momentum beyond the minimum, the definition of a MIP is extended to include momenta up to the point where radiative losses become important.  For example, muons between about $1$ GeV and $1$ TeV in silicon are MIPs.   When a MIP traverses silicon, it generates electron-hole pairs.  The energy required to generate such pairs is about 3.6 eV (depends mildly on temperature).  This leads to about $80$ electron-hole pairs deposited per micron.  The digitization model converts the energy deposited by {\sc Geant4} into discrete charge clumps which are propagated to the electrode.  The number of collected charges is then converted into a discrete {\it time over threshold} (TOT) value, which is exactly the same output of a real pixel module.  Due to time constraints, it is not possible for each charge clump to represent a fundamental charge (electron or hole).  The implications of this clumping are described at the end of this section.

\begin{figure}[h!]
\centering
\includegraphics[width=0.95\textwidth]{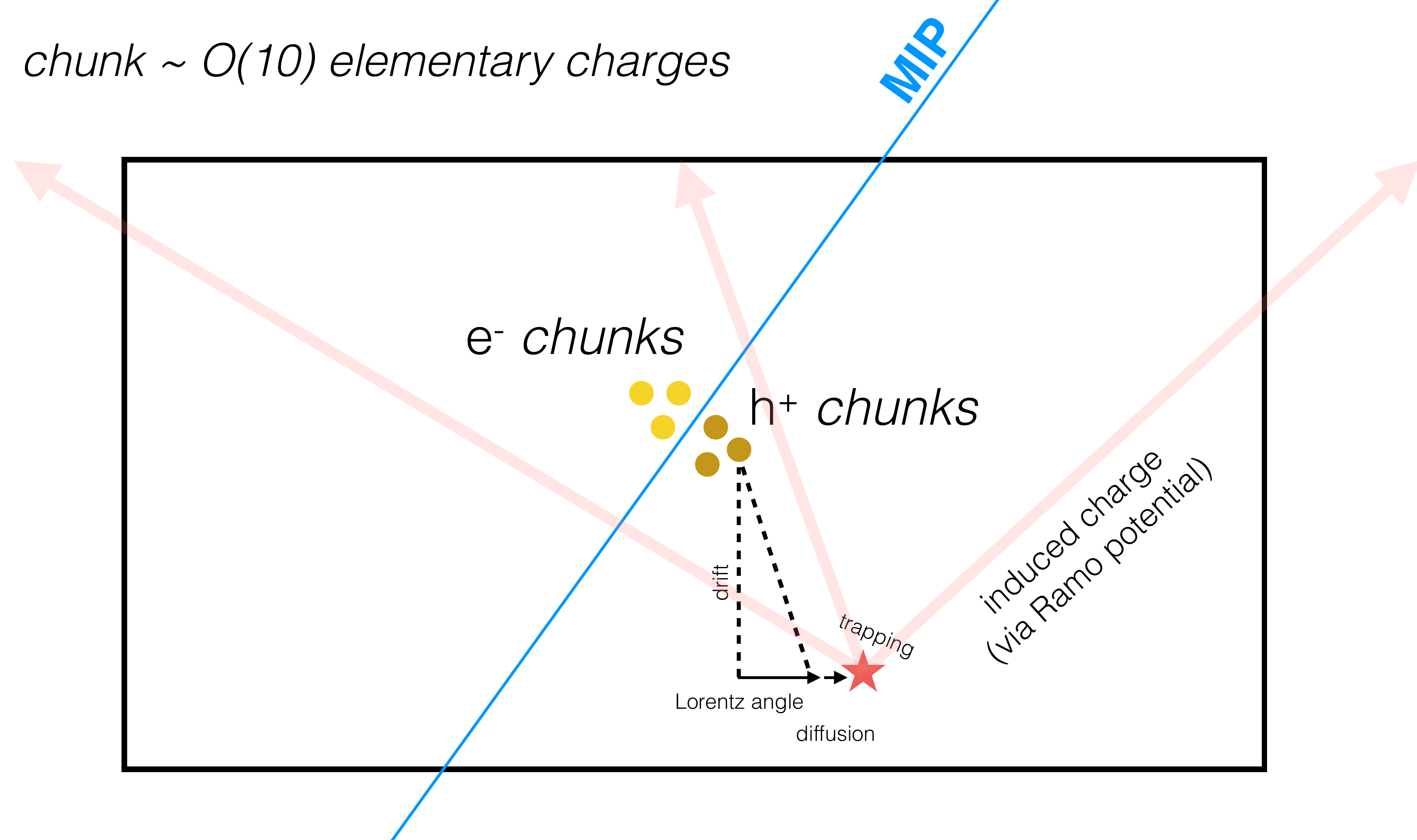}
\caption{A schematic diagram illustrating the components of the digitizer model described in this section.  While included in the model, thermal diffusion is a generally small effect and is not discussed in the following section.}
\label{fig:app:raddamge1}
\end{figure}

After generating clumps of charge, several process are simulated as the clumps propagate to the electrode.  The bias voltage applied to the sensor generates a large electric field that causes the electrons (holes) to drift toward (away from) the collecting electrode.  The velocity of this motion is determined by the charge carrier mobility $\mu$ via $v=\mu E$.  The mobility has a small $E$-field and temperature dependence:

\begin{align}
\label{eq:mobility}
\mu^{p}(E)=\frac{v_s^p/E_c^p}{\left(1+\left(\frac{E}{E_c^p}\right)^{\beta^p}\right)^{1/{\beta^p}}},
\end{align}

\noindent where $p$ stands for electron or hole.  The values for the saturation velocity $v_s$, critical $E$-field $E_c$ and temperature exponent $\beta$ can be found in Table~\ref{eq:constants_rad}.  In addition to the electric field from the bias voltage, there is a magnetic field generated by the solenoid surrounding the inner detector.  One effect from this field is that it modifies the mobility so that the average velocity of charge carriers follows the {\it Hall mobility}, which is the drift mobility (Eq.~\ref{eq:mobility}) multiplied by the Hall factor $r$ found in Table~\ref{eq:constants_rad}.  Another impact of this field is that the charges do not travel parallel to the electric field: they travel at an angle called the {\it Lorentz angle}.  This angle is only relevant in the direction perpendicular to the beam and is approximately $\tan\theta\sim 0.2$ in the inner detector barrel. 

\begin{table}[h!]
\centering
\begin{tabular}{|c|c|c|}
  \hline
   quantity & electrons & holes		\\
   \hline	
  $v_s$ ($\mu$m/ns) & $116\times (T/273\text{ K})^{-0.87}$ & $88\times (T/273\text{ K})^{-0.52}$ \\
  $E_c$ (kV/cm) & $6.0 \times (T/273\text{ K})^{1.55}$ & $15 \times  (T/273\text{ K})^{1.68}$ \\
  $\beta$ & $1.0 \times (T/273\text{ K})^{0.66} $& $1.1\times (T/273\text{ K})^{0.17} $\\
  $r$ &$1.13+8\times 10^{-4}\times (T/\text{K}-273)$ & $0.72-5\times 10^{-4}\times (T/\text{K}-273)$\\
  \hline  
\end{tabular}
\caption{Physical constants describing the mobility of charge carriers in silicon.  The first three rows are reformatted from Ref.~\cite{mobility} and the Hall scale factor is from Ref.~\cite{hall1}.}
\label{eq:constants_rad}
\end{table}

The time for a charge chunk to reach the electrode is estimated by integrating the mobility:

\begin{align}
\label{eq:drifttime}
t_\text{electrode}=\int_{z_\text{initial}}^{z_\text{final}} \frac{dz}{\mu_p(z)E(z)},
\end{align}

\noindent where $z_\text{final}$ is the depth of the electrode (200 $\mu$m for the ATLAS IBL) for electrons and $0$ for holes (the velocity is negative).  Figure~\ref{fig:app:raddamge2} shows the average electric field as a function of depth in an IBL sensor with and without radiation damage.  There are several models for simulating radiation damage effects on the electric field - for this study, the Chiochia model~\cite{Chiochia1,Chiochia2} was simulated using a TCAD model\footnote{Input E-fields and Ramo potential maps are from M. Bomben.}.  The main effect of irradiation is that the field strength in the center of the sensor is reduced while the field strength near the edges increases.  Using these fields as input, Fig.~\ref{fig:app:raddamge3} shows the projected drift time from Eq.~\ref{eq:drifttime}.

\begin{figure}[h!]
\centering
\includegraphics[width=0.45\textwidth]{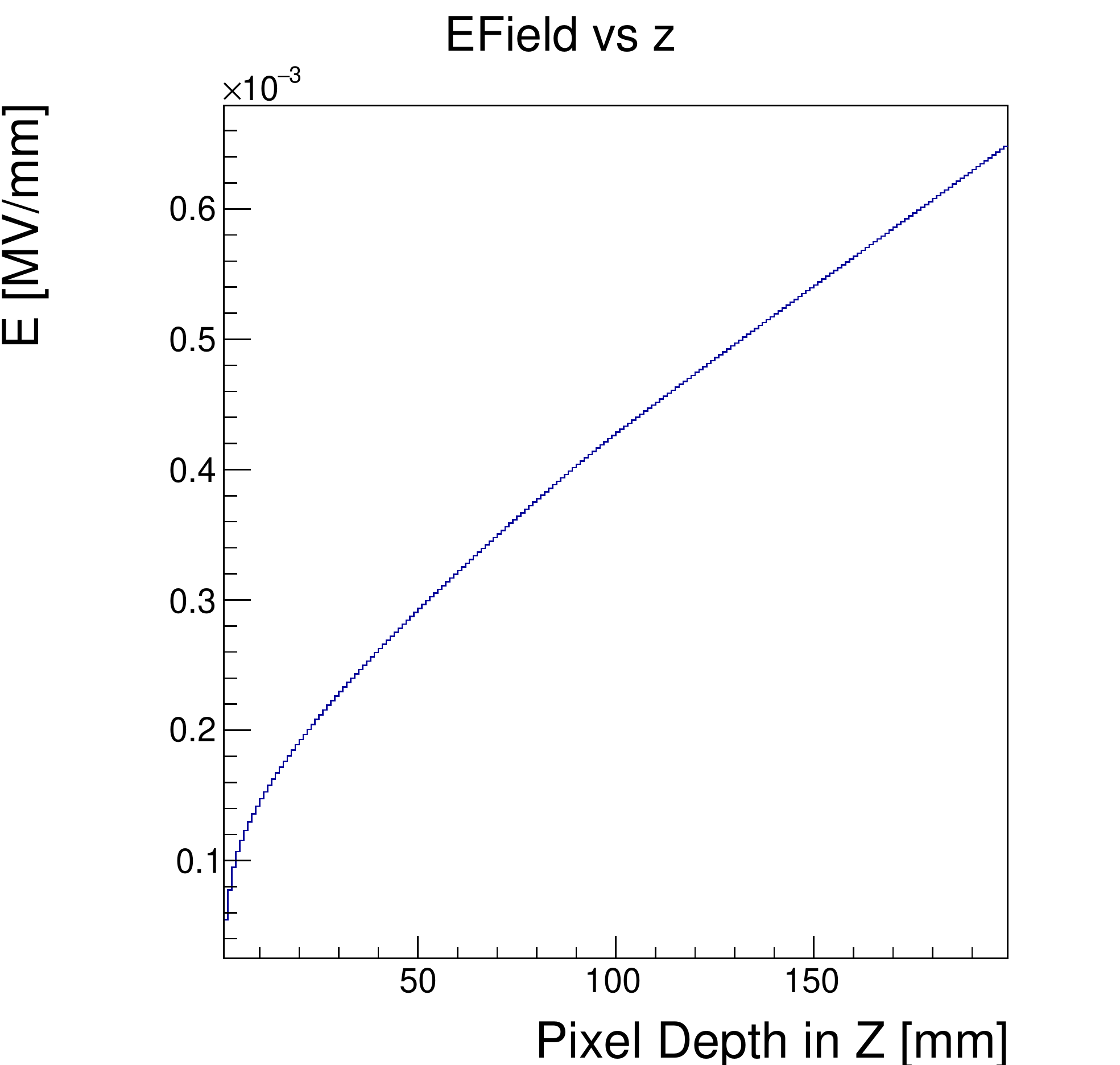}\includegraphics[width=0.45\textwidth]{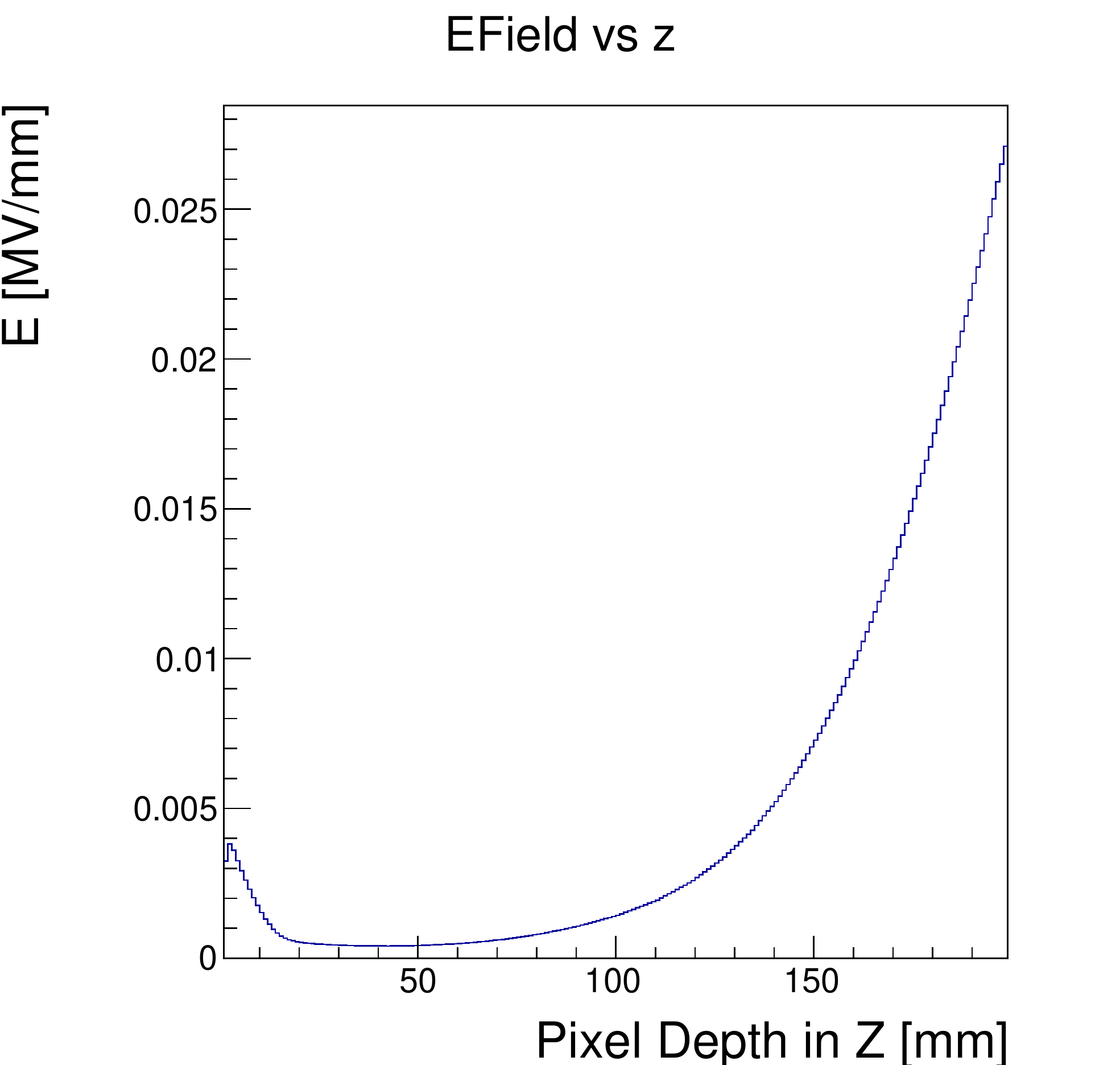}
\caption{The average (over $x$ and $y$) electric field as a function of the depth ($z$) inside an unirradiated planar $200$ $\mu m$ deep planar sensor with a bias voltage of $80$ V ($1000$ V) for an unirradiated ($5\times 10^{15}$ $n_\text{eq}/\text{cm}^2$) sensor on the left (right).}
\label{fig:app:raddamge2}
\end{figure}

\begin{figure}[h!]
\centering
\includegraphics[width=0.45\textwidth]{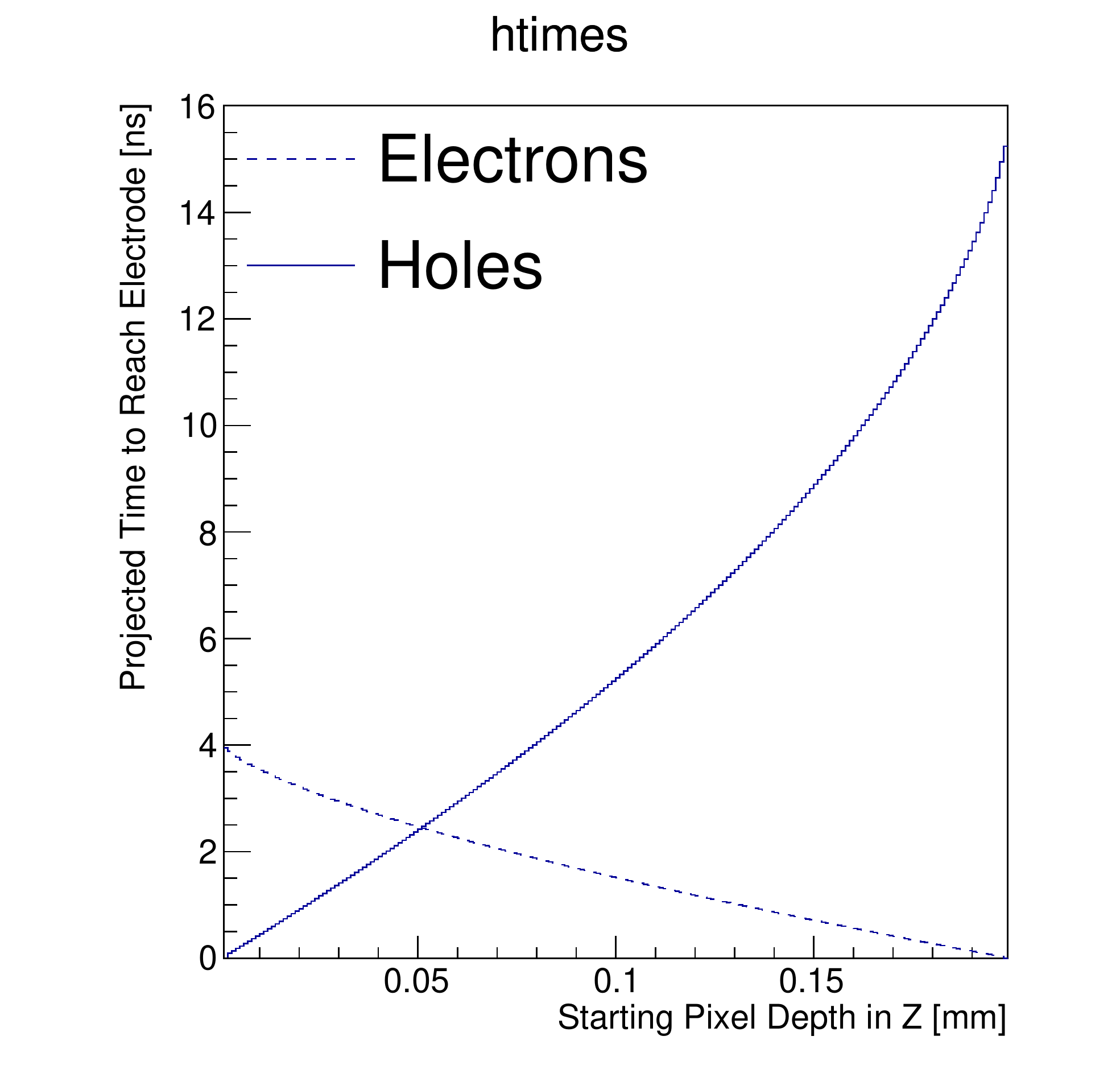}\includegraphics[width=0.45\textwidth]{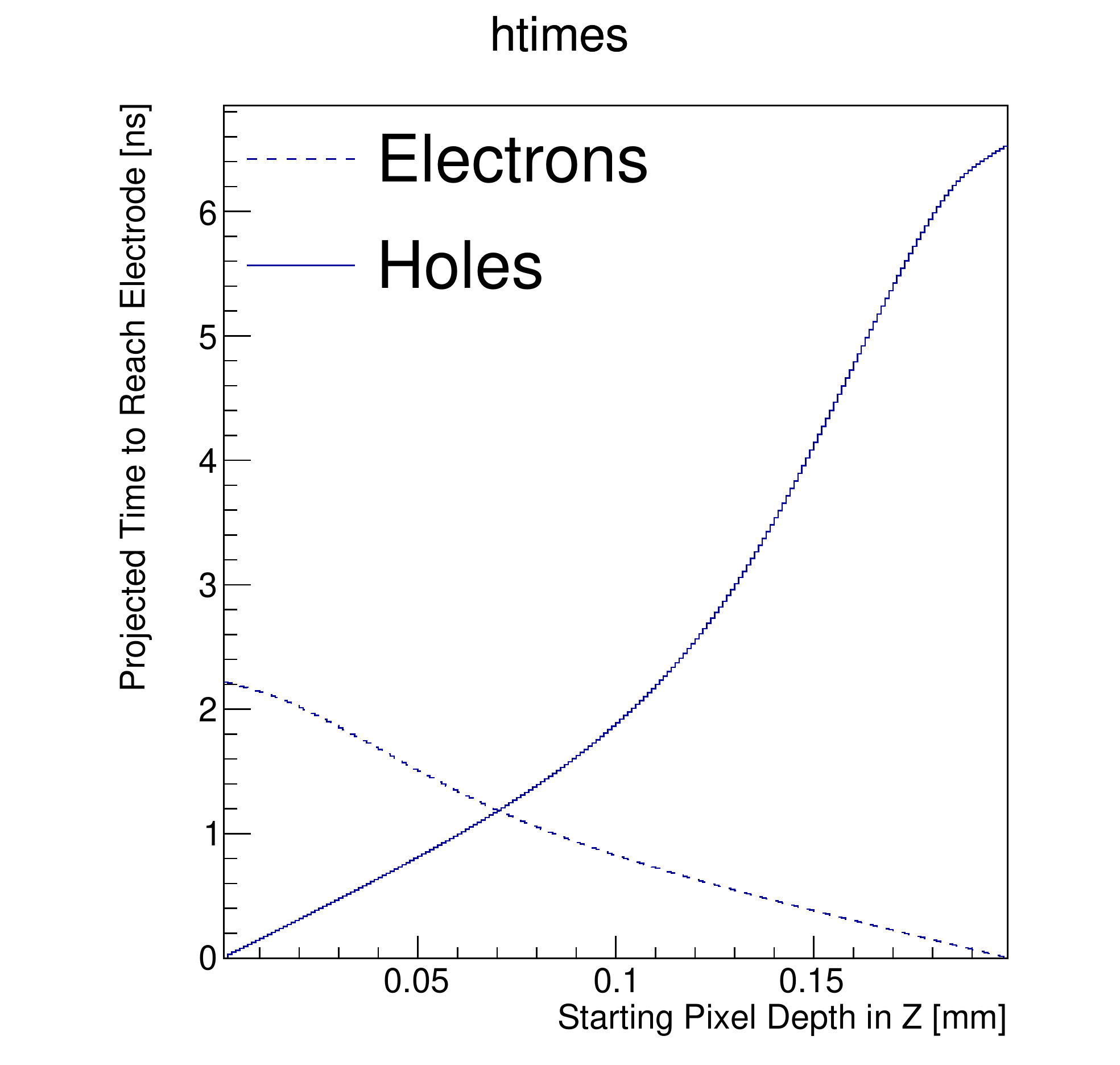}
\caption{The time for an electron or hole to drift to the top (collecting electrode) or bottom of the sensor as a function of the depth ($z$) using the averaged $E$ fields shown in Fig.~\ref{fig:app:raddamge2} for a planar sensor with a bias voltage of $80$ V ($1000$ V) for an unirradiated ($5\times 10^{15}$ $n_\text{eq}/\text{cm}^2$) sensor on the left (right).  }
\label{fig:app:raddamge3}
\end{figure}

As a result of irradiation, defects form in the silicon and are sites for charge trapping.  In the simulation, charge chunks are declared trapped if the projected time to reach the electrode from Fig.~\ref{fig:app:raddamge3} exceeds a random trapping time $t$ that is exponentially distributed with mean value $1/(\kappa\Theta)$, where $\Theta$ is the fluence.  The constant $\kappa$ (called $\beta$ in the literature) has been measured at the 2001 CERN test beam and is approximately $\kappa=3\times 10^{-16}$ cm${}^2$/ns~\cite{trapping2}.  Charge trapping reduces the collected signal and thus degrades track reconstruction efficiency.

However, not all the trapped charge is lost.  Charge is induced on the electrode as soon as the electrons or holes start to move.  The amount of induced charge can be readily calculated using the {\it Ramo potential} from the Shockley-Ramo theorem~\cite{Shockley,Ramo}.  This theorem states that the amount of induced charge is the particle charge multiplied by the difference in the Ramo potential from its starting and ending (trapped) location.  The Ramo potential for a particular electrode is calculated by calculating the electrostatic potential by holding the given electrode at unit voltage and setting all other electrodes to have zero potential.  For example, for a an infinite parallel plate capacitor, the field is constant in between the plates, so the Ramo potential is linear (starting at 1 and decreasing to zero).  Figure~\ref{fig:app:raddamge5} shows the Ramo potential for parallel plate capacitors that have various widths\footnote{Example inspired by Ref.~\cite{pixeldetectors}.}.  As the width decreases, the area over which the charge is collected becomes increasingly small, i.e. the Ramo potential is increasingly peaked at zero relative to the rest of the sensor.  This trend is also illustrated in two dimensions in Fig.~\ref{fig:app:raddamge6}.  Note that since the Ramo potential extends beyond the extend of the sensor, charge is also induced in neighboring pixels.

\begin{figure}[h!]
\centering
\includegraphics[width=0.5\textwidth]{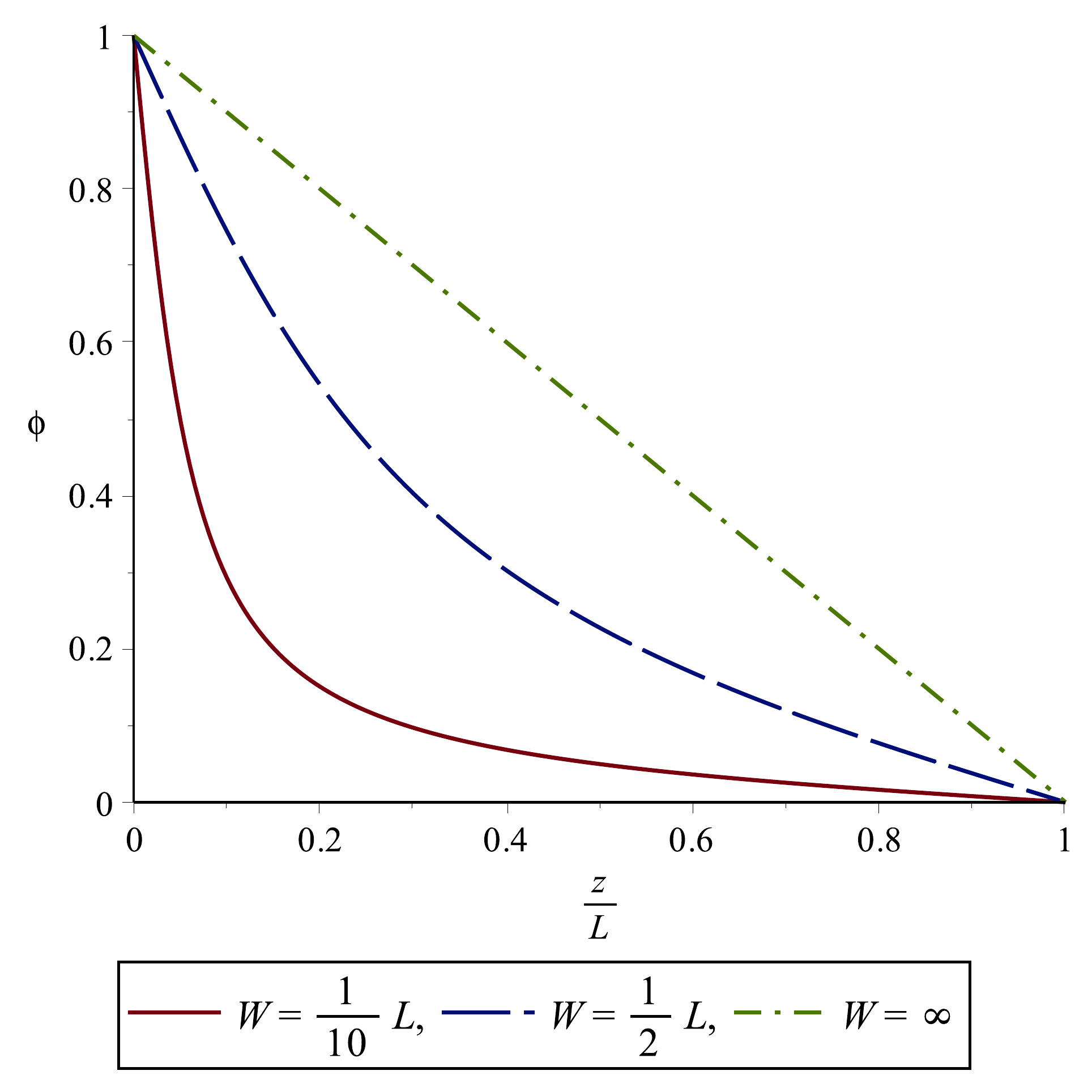}
\caption{The Ramo potential at $x=y=0$ (centered on the collecting electrode $\epsilon$) as a function of the distance $z$ away from $\epsilon$ for three sizes of $\epsilon$: $W=L/10, W=L/2,$ and $W=\infty$, where $L$ is the sensor thickness.}
\label{fig:app:raddamge5}
\end{figure}

\begin{figure}[h!]
\centering
\includegraphics[width=0.33\textwidth]{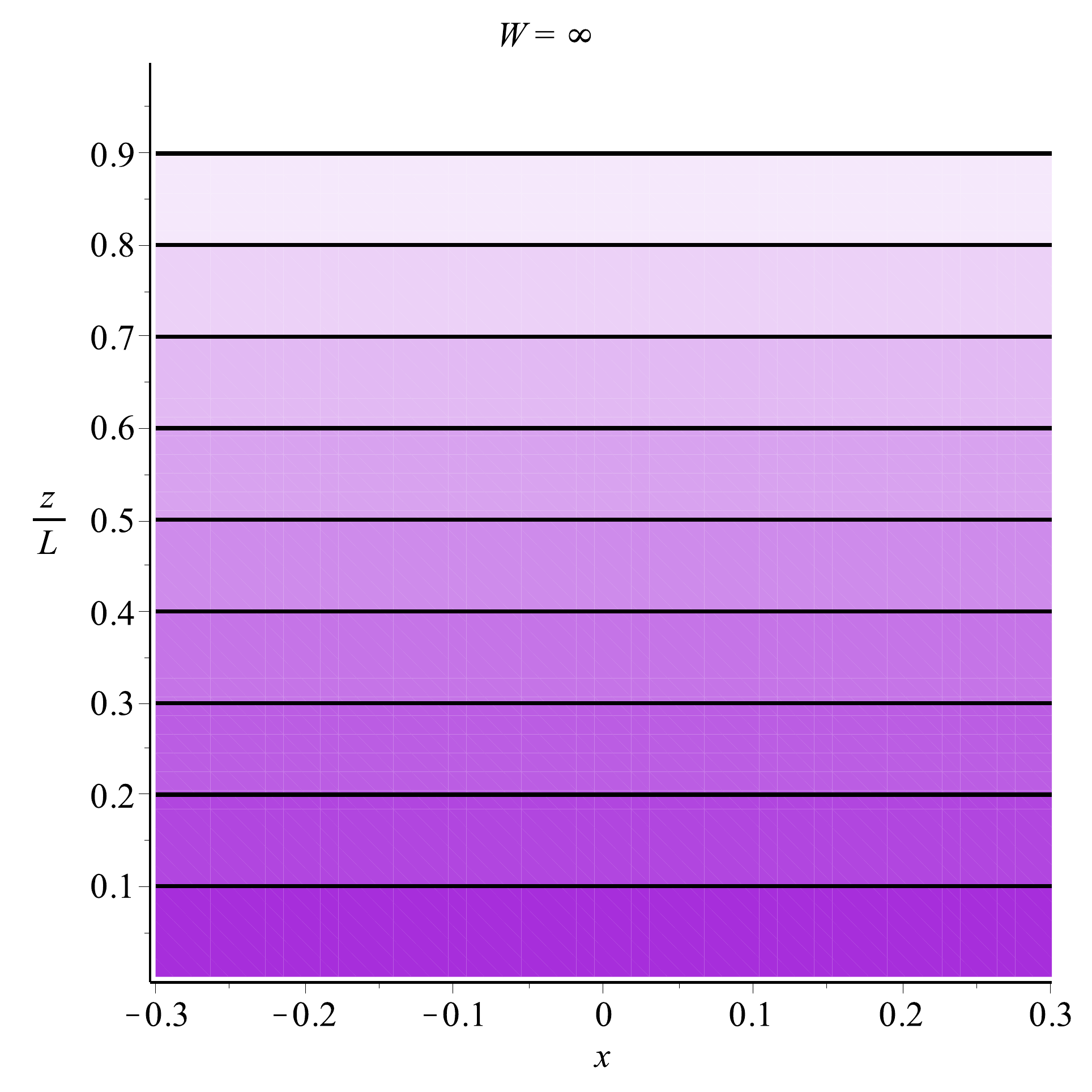}\includegraphics[width=0.33\textwidth]{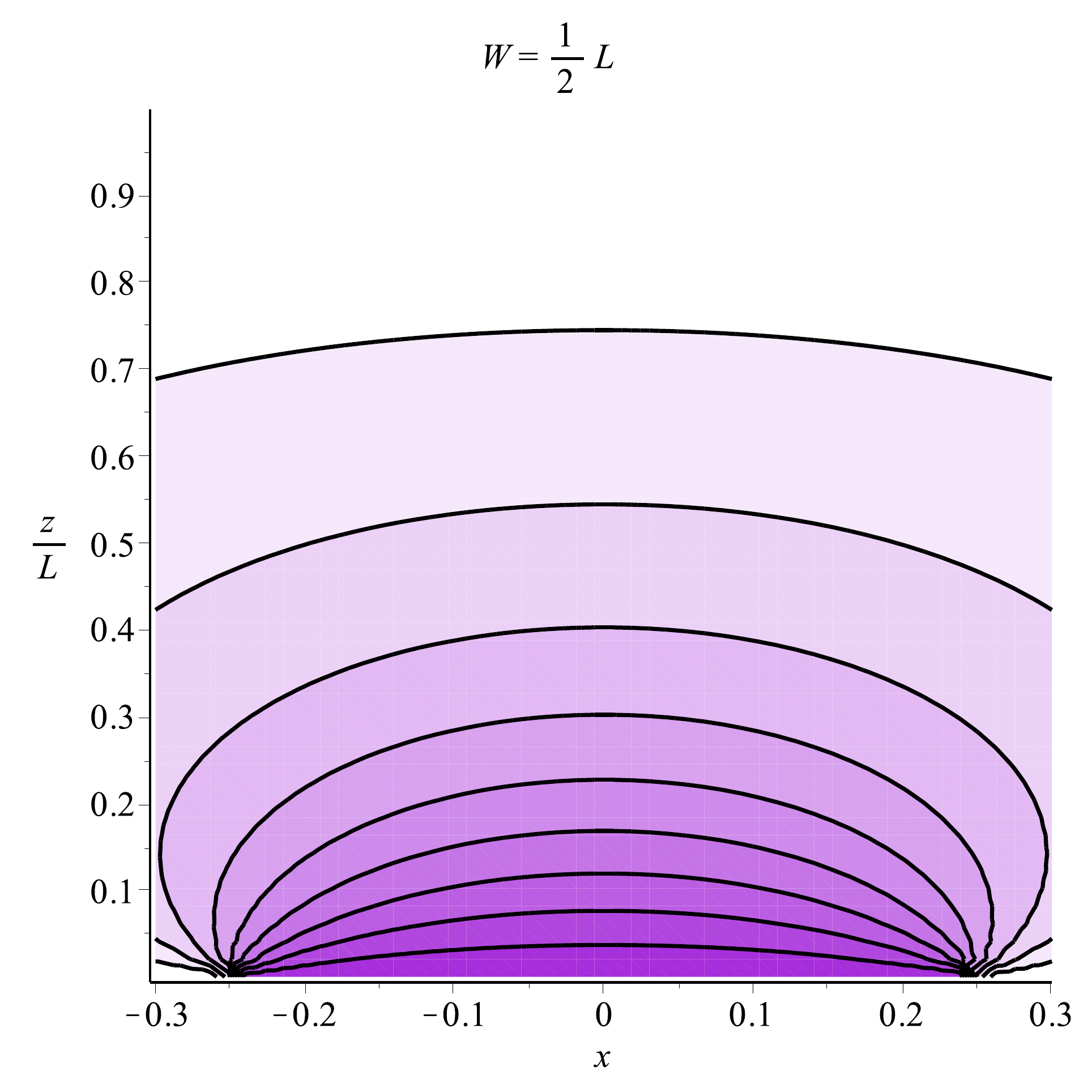}\includegraphics[width=0.33\textwidth]{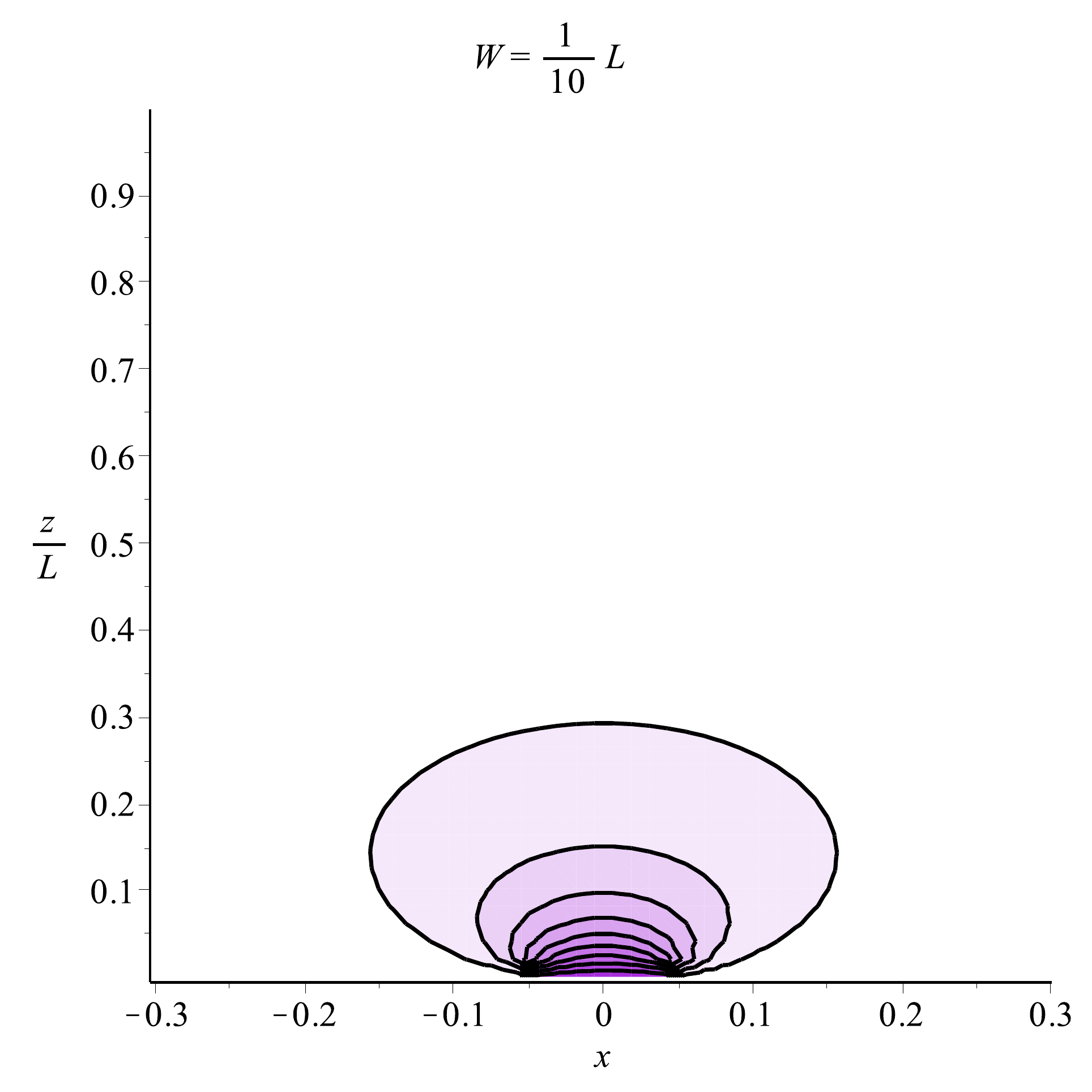}
\caption{The Ramo potential at $y=0$ (centered on the collecting electrode $\epsilon$) as a function of the distance $z$ away from $\epsilon$ for three sizes of $\epsilon$: $W=L/10, W=L/2,$ and $W=\infty$, where $L$ is the sensor thickness.}
\label{fig:app:raddamge6}
\end{figure}

By construction, without charge trapping, the total induced charge on the primary electrode must be the total charge.  Similarly, the charge induced on the neighboring electrodes must be zero without trapping.  However, while the induced charge on the primary electrode increases monotonically with time, the charge induced on the neighboring electrodes increases when the electron or hole is far away and then decreases once it is close enough.  To understand this, consider a point unit charge that is a distance $z$ away from an infinite plate that has been cut into strips, where each strip is grounded.  The surface charge density is peaked at zero and the peak increases the closer the charge is to the surface.  The left plot in Fig.~\ref{fig:app:raddamge7} shows a transverse slice of the surface charge density.  A comparison of the various colored lines in this plot shows how the field moves over the neighboring electrode.  The middle plot in Fig.~\ref{fig:app:raddamge7} is the integral of the charge density on the electrode neighboring the primary one to the right and the induced charge calculated with the Ramo potential is shown in the right plot.

\begin{figure}[h!]
\centering
\includegraphics[width=0.33\textwidth]{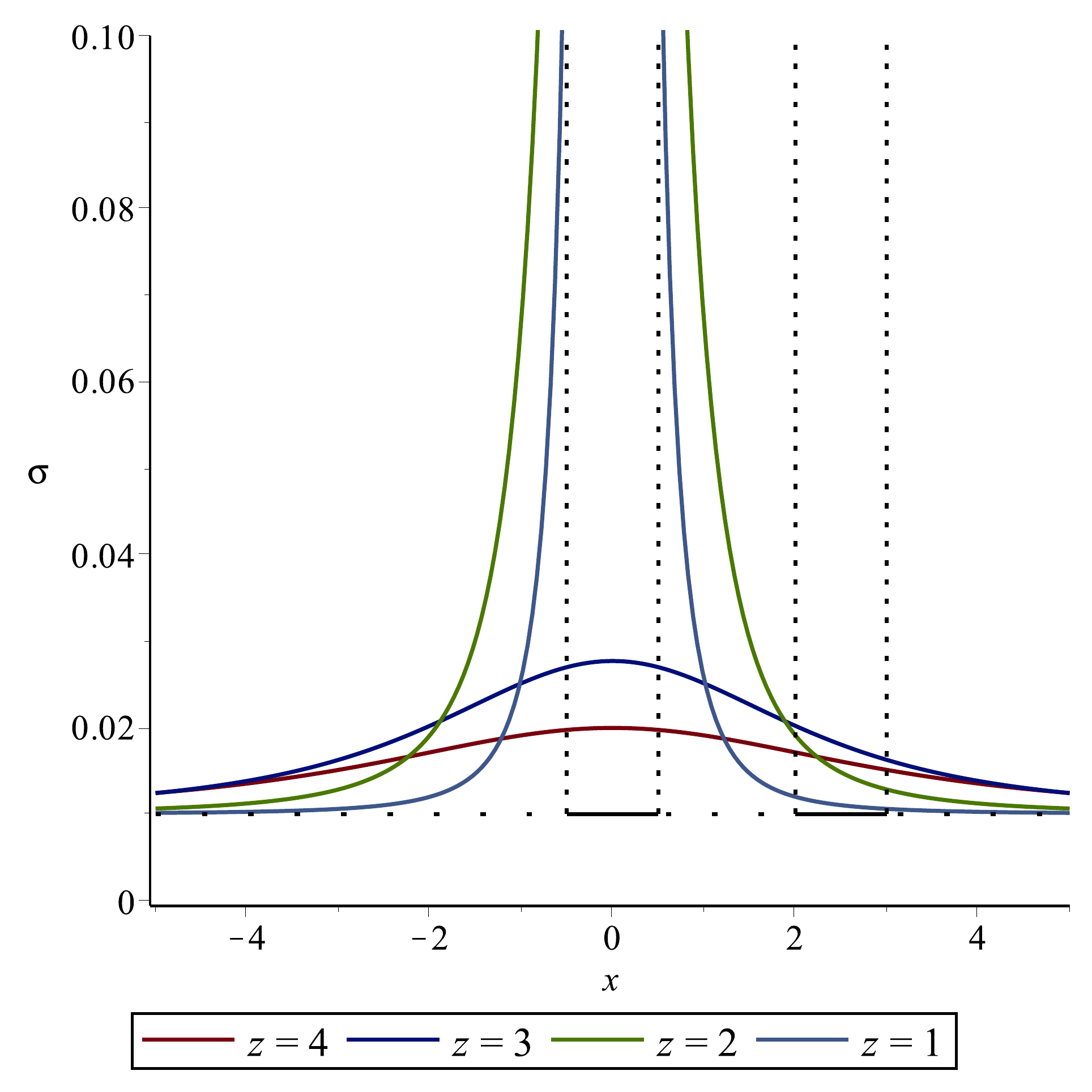}\includegraphics[width=0.33\textwidth]{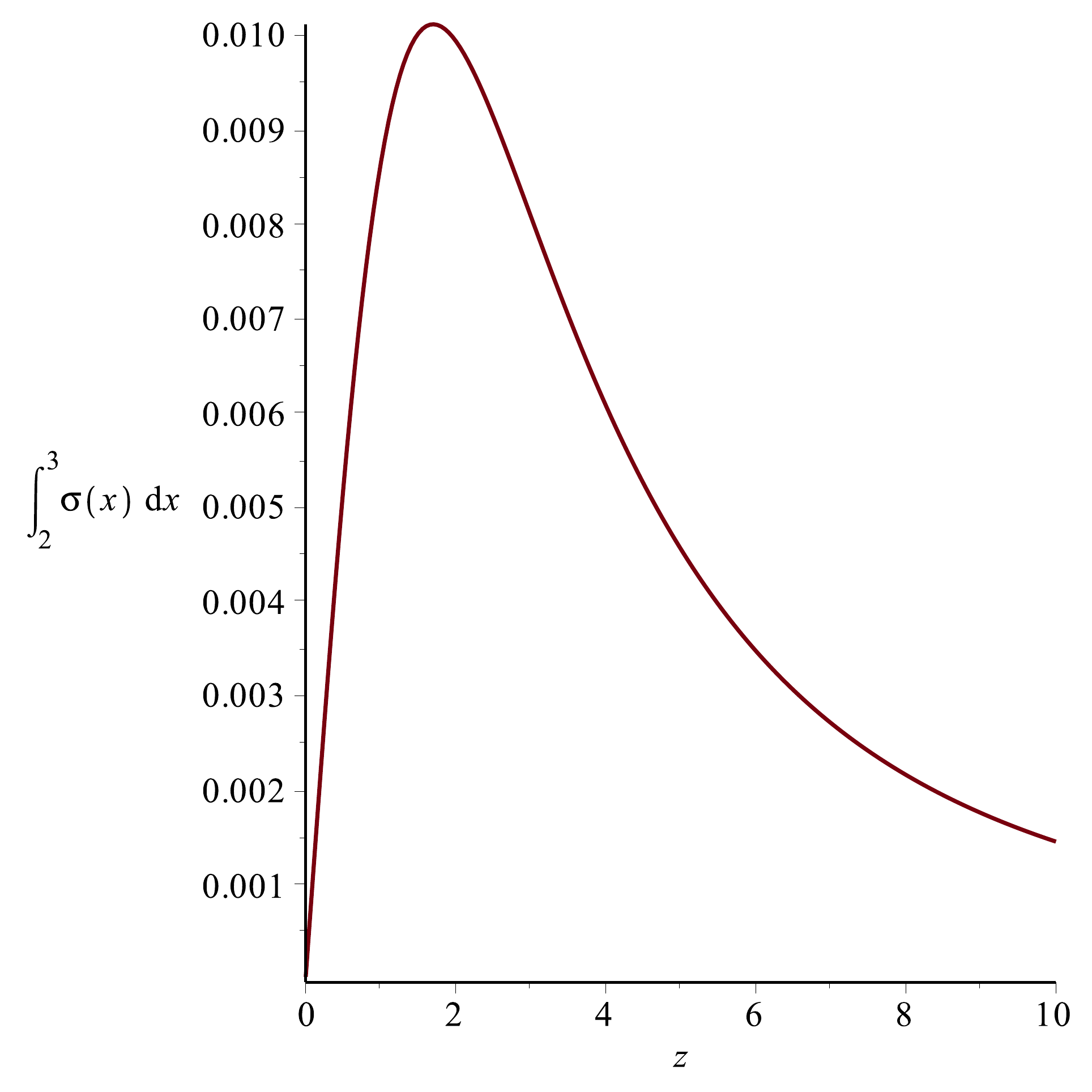}\includegraphics[width=0.33\textwidth]{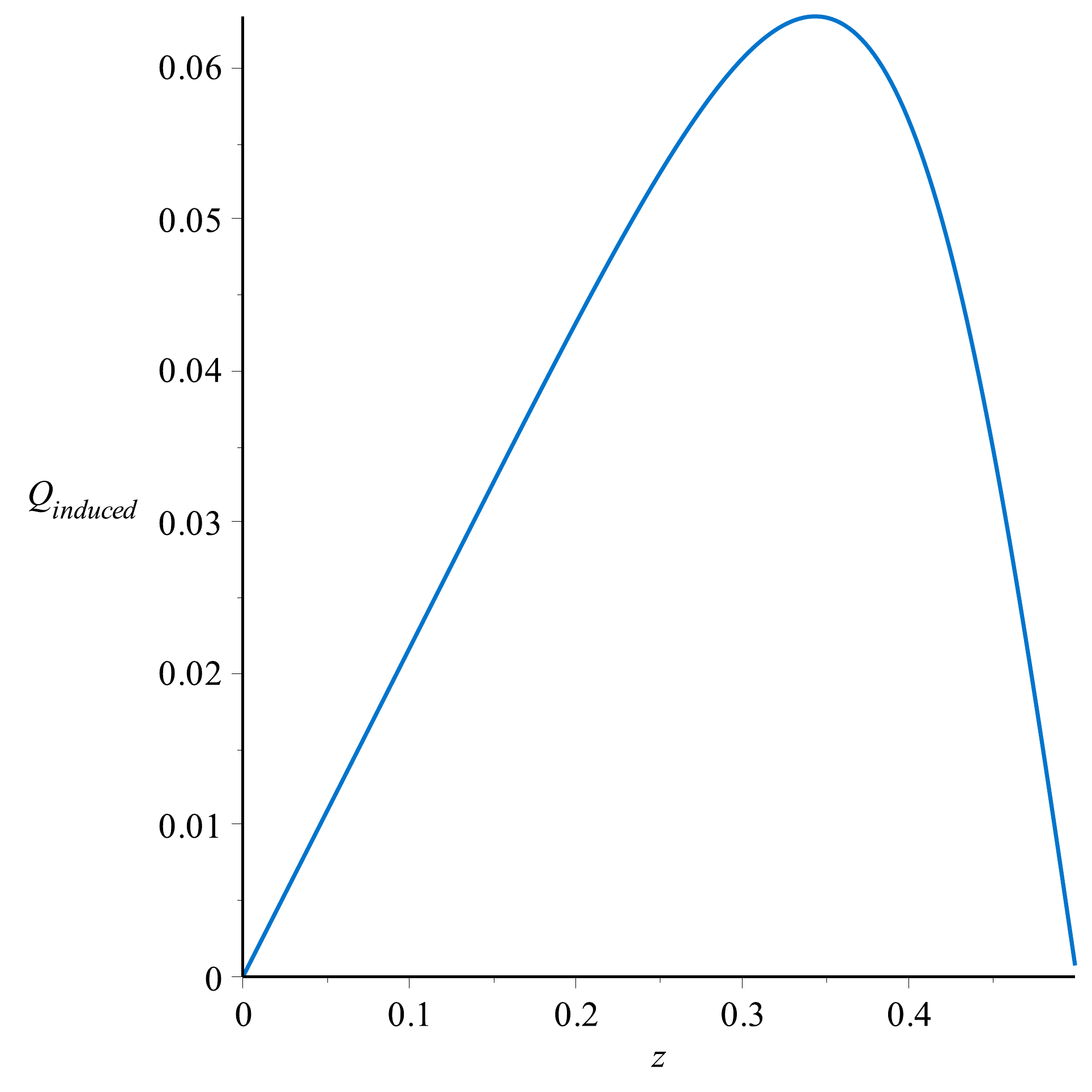}
\caption{A cross-section of the surface charge density from a point charge a distance $z$ away from a series of grounded strips.  The `primary electrode' is the strip centered at $0$ (extending to $\pm \infty$ in the direction into and out of the page) and the `neighboring electrode' is the one shown between $2$ and $3$.  The middle plot shows the total charge on the neighboring electrode as a function of the distance the charge is from the origin and the plot on the right shows the fraction of the charge that is induced on the neighbor using the Ramo potential.}
\label{fig:app:raddamge7}
\end{figure}

Figures~\ref{fig:app:raddamge5} and~\ref{fig:app:raddamge4} show the final depth for electrons and holes that start at some depth and are trapped after a time $t$ shown on the vertical axis.  The induced charge is computed as the difference in the Ramo potential between the initial and final depths\footnote{The Ramo potential depends only on geometry and not on the fluence~\cite{ramorad}.}.  The induced charge based on the unirradiated $E$-field is shown in Fig.~\ref{fig:app:raddamge9}.  In practice, the time to the trap is always $\infty$ in this case, but Fig.~\ref{fig:app:raddamge9} provides a technical closure of the setup.  As expected, the charge induced on the primary electrode reaches $100\%$ as the trapping time goes to infinity.  The asymmetry with respect to the center of the detector is due in part to the difference in mobilities between electrons and holes.  Similarly, the charge induced on the neighboring electrodes goes to zero as time goes to infinity.  The induced charge is much larger for the electrode that is only $50$ $\mu$m away (short direction) compared with the one that is $200$ $\mu$m away (long direction).

\begin{figure}[h!]
\centering
\includegraphics[width=0.45\textwidth]{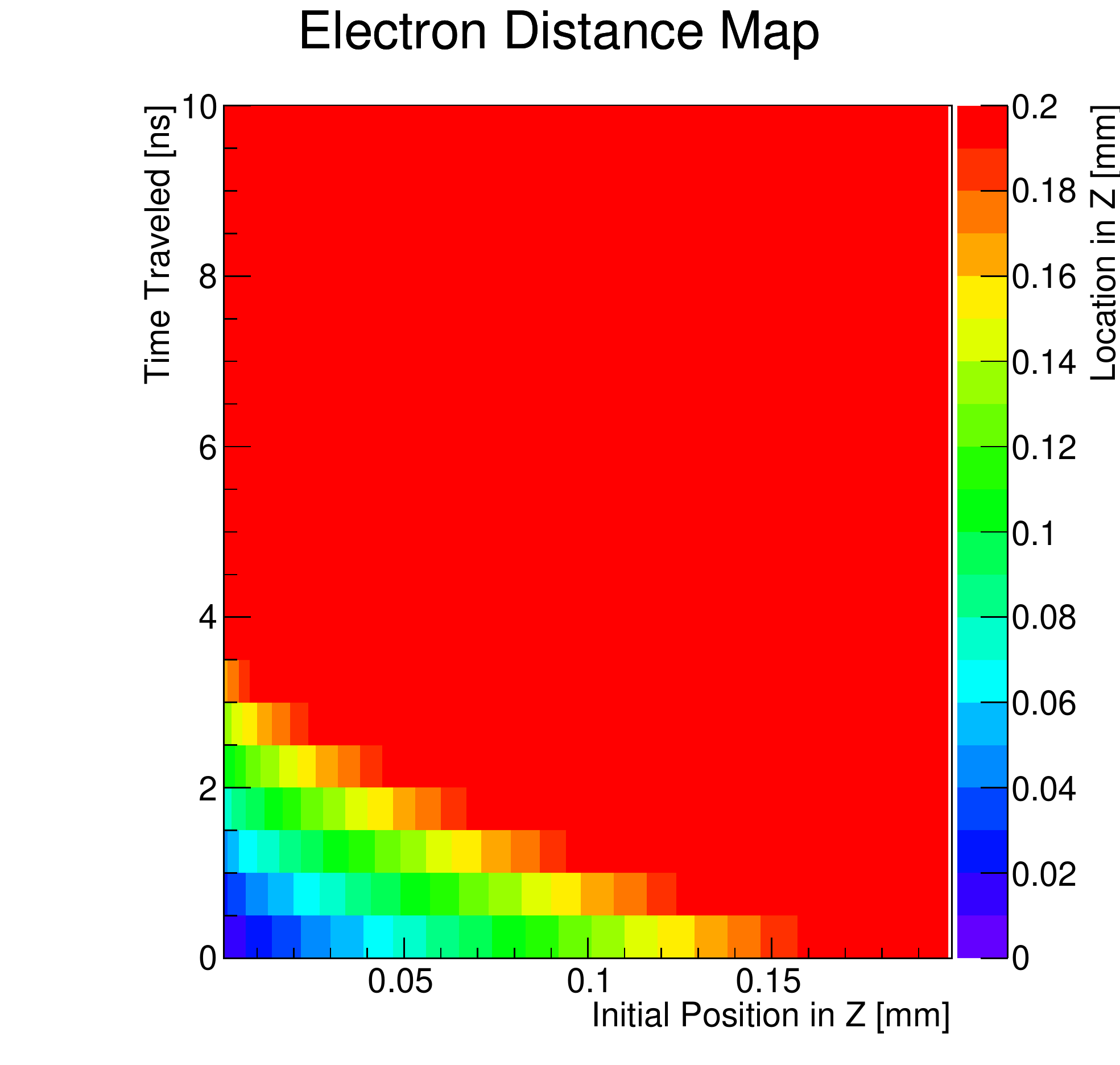}\includegraphics[width=0.45\textwidth]{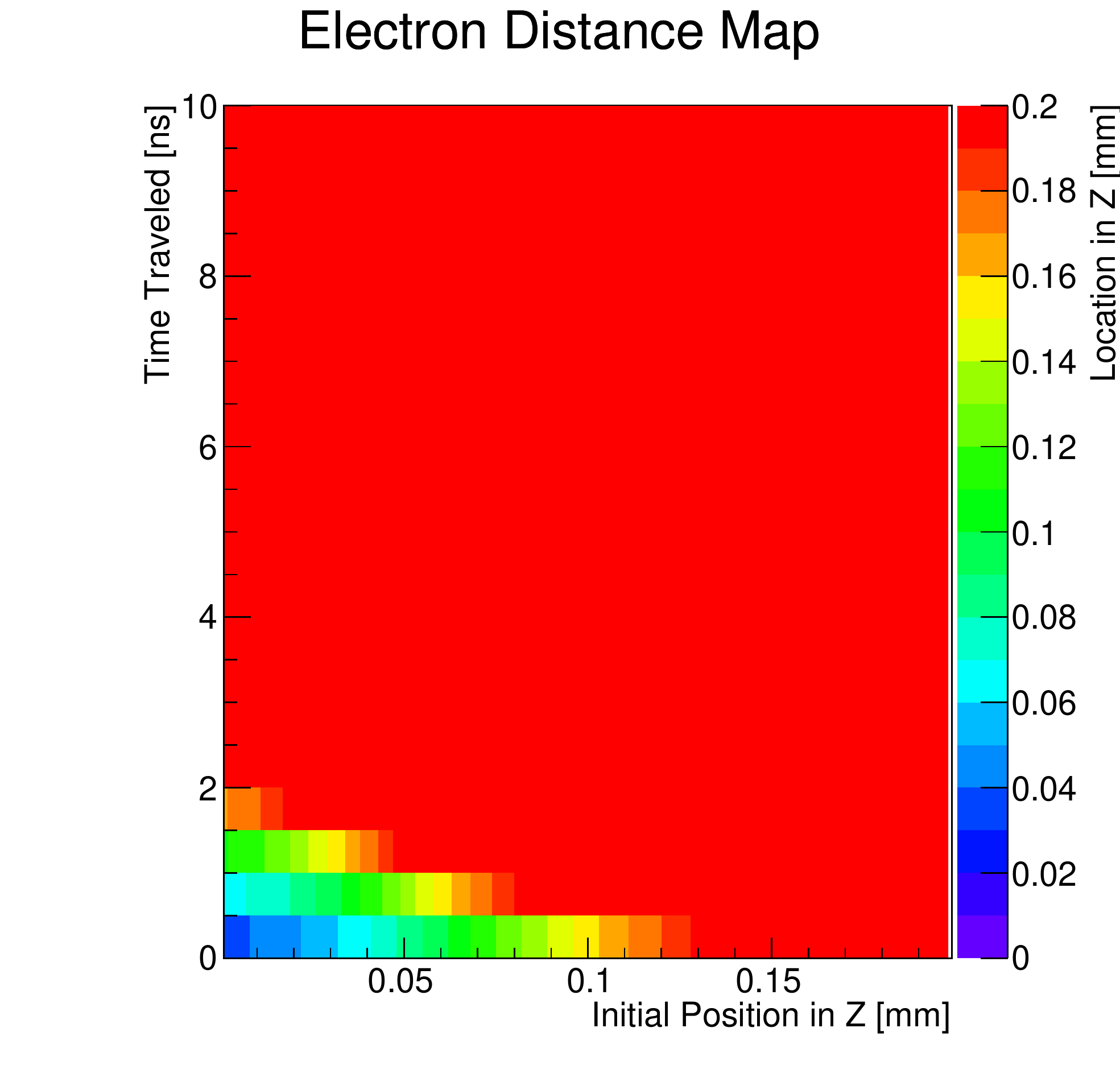}
\caption{The final location (based only on drift) for electrons starting at a depth $z$ and traveling a time given by the vertical axis. The coordinate $z$ is measured with respect to the back-side (away from the collecting electrode) of the sensor using the averaged $E$ fields shown in Fig.~\ref{fig:app:raddamge2} for a planar sensor with a bias voltage of $80$ V ($1000$ V) for an unirradiated ($5\times 10^{15}$ $n_\text{eq}/\text{cm}^2$) sensor on the left (right).}
\label{fig:app:raddamge5}
\end{figure}

\begin{figure}[h!]
\centering
\includegraphics[width=0.45\textwidth]{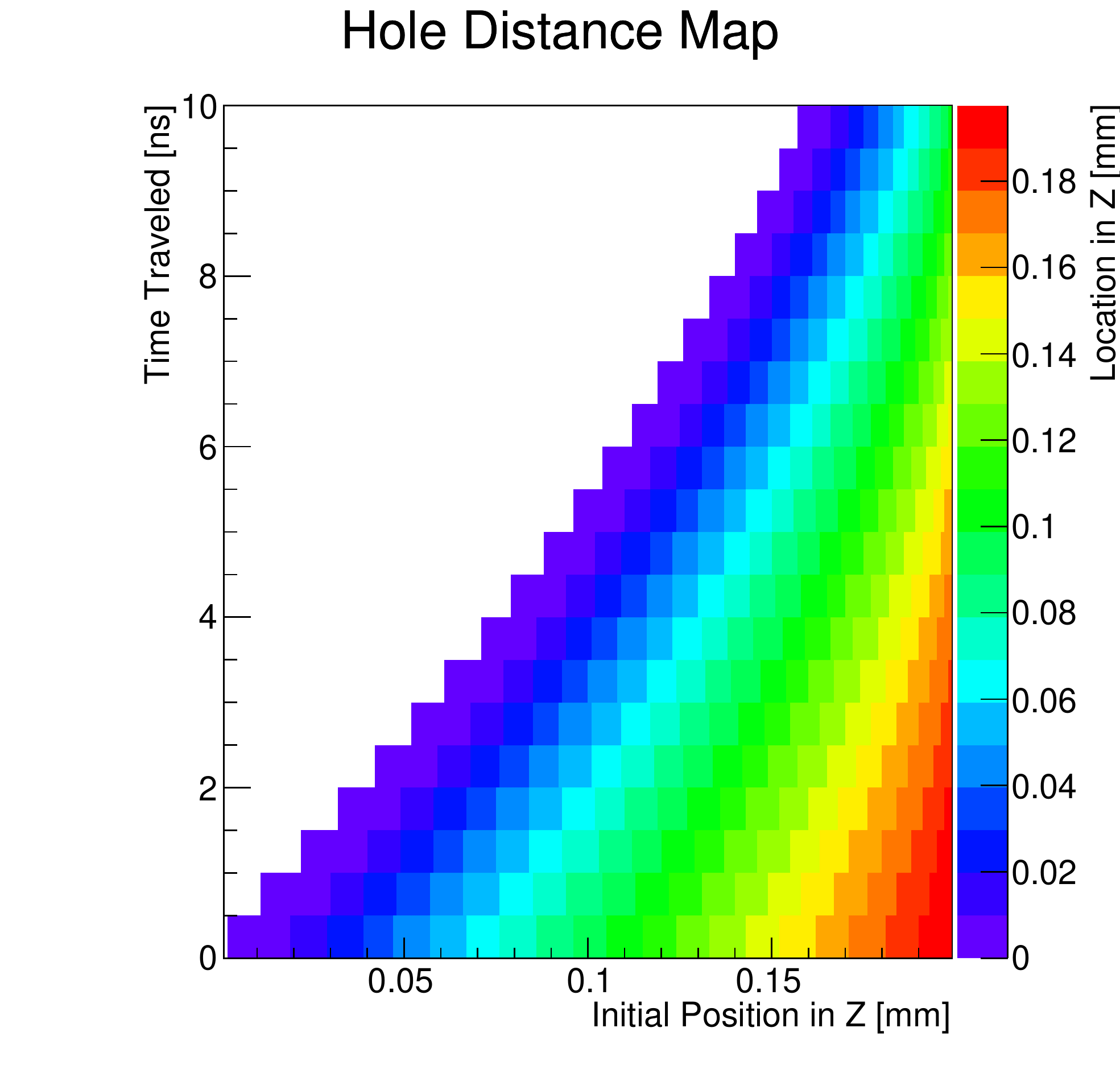}\includegraphics[width=0.45\textwidth]{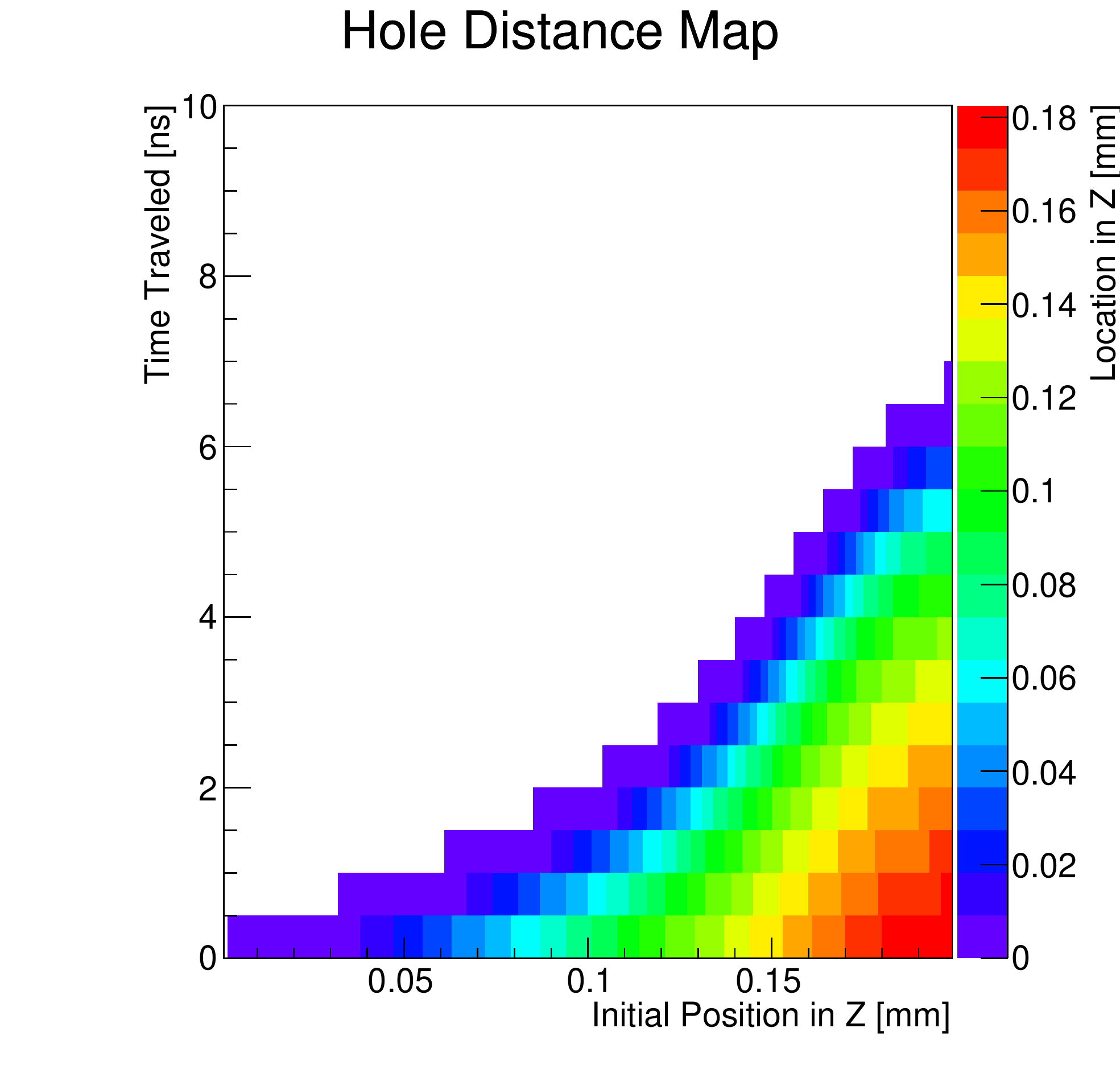}
\caption{The final location (based only on drift) for holes starting at a depth $z$ and traveling a time given by the vertical axis. The coordinate $z$ is measured with respect to the back-side (away from the collecting electrode) of the sensor using the averaged $E$ fields shown in Fig.~\ref{fig:app:raddamge2} for a planar sensor with a bias voltage of $80$ V ($1000$ V) for an unirradiated ($5\times 10^{15}$ $n_\text{eq}/\text{cm}^2$) sensor on the left (right).}
\label{fig:app:raddamge4}
\end{figure}

\begin{figure}[h!]
\centering
\includegraphics[width=0.33\textwidth]{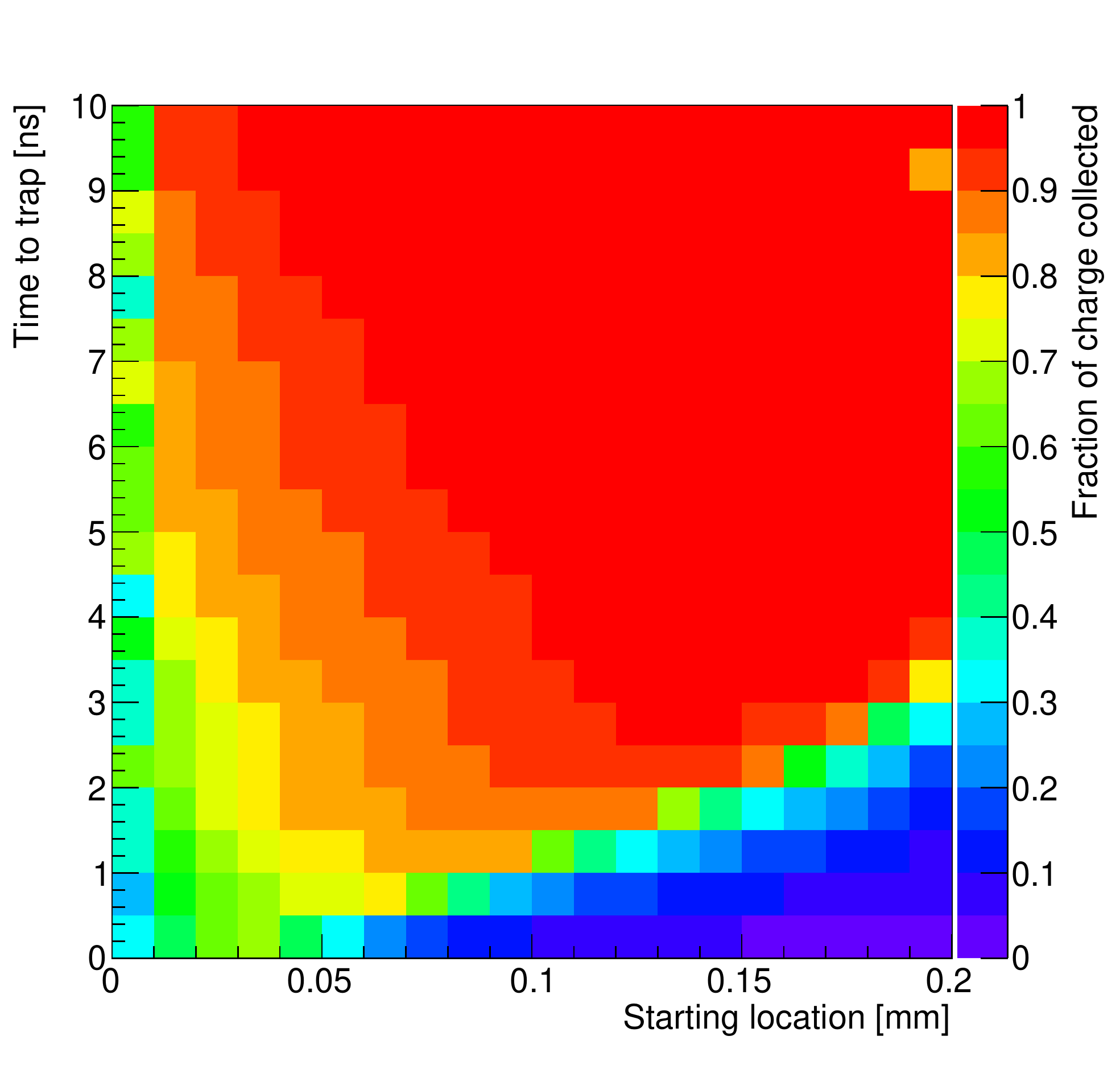}\includegraphics[width=0.33\textwidth]{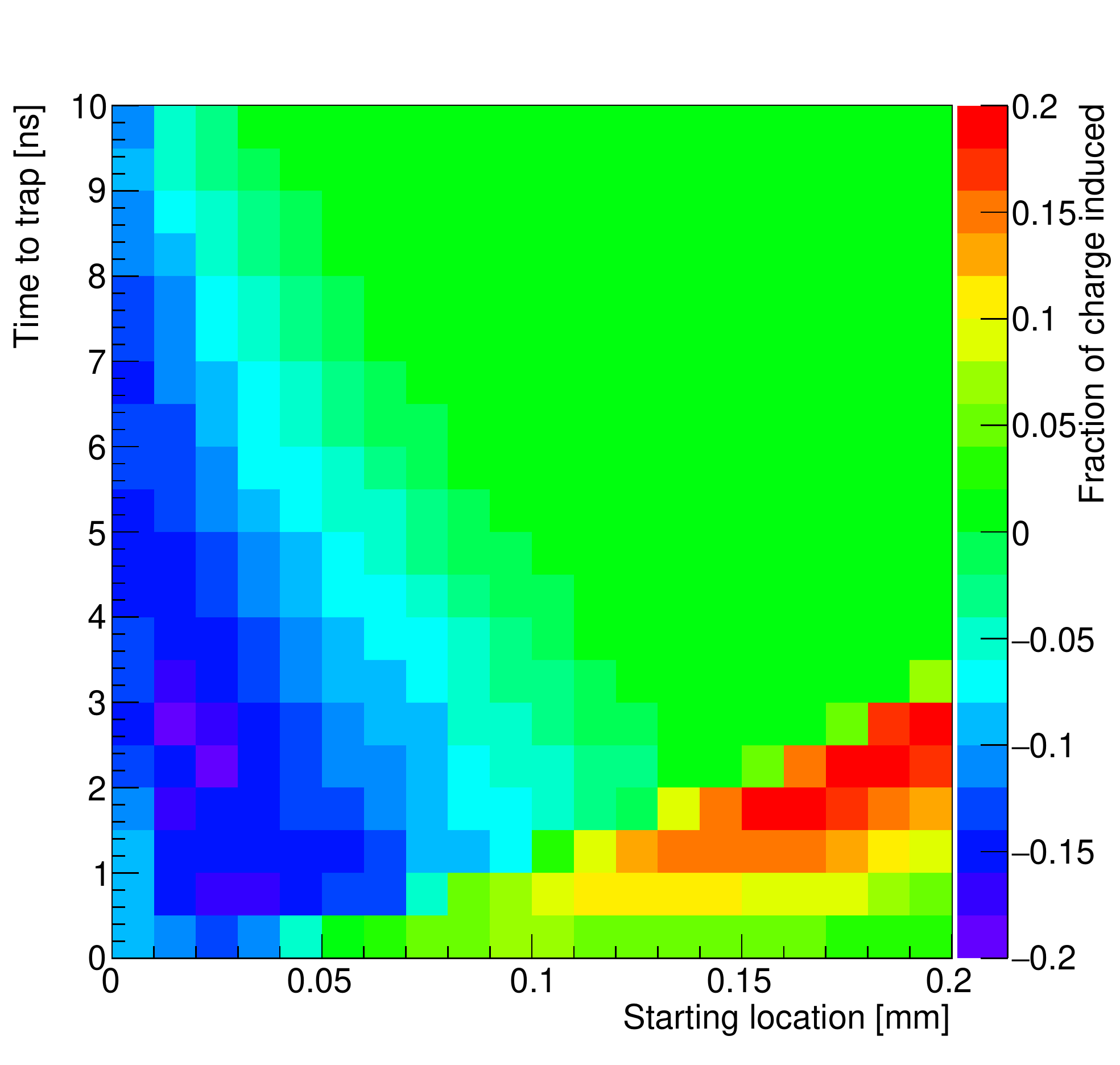}
\includegraphics[width=0.33\textwidth]{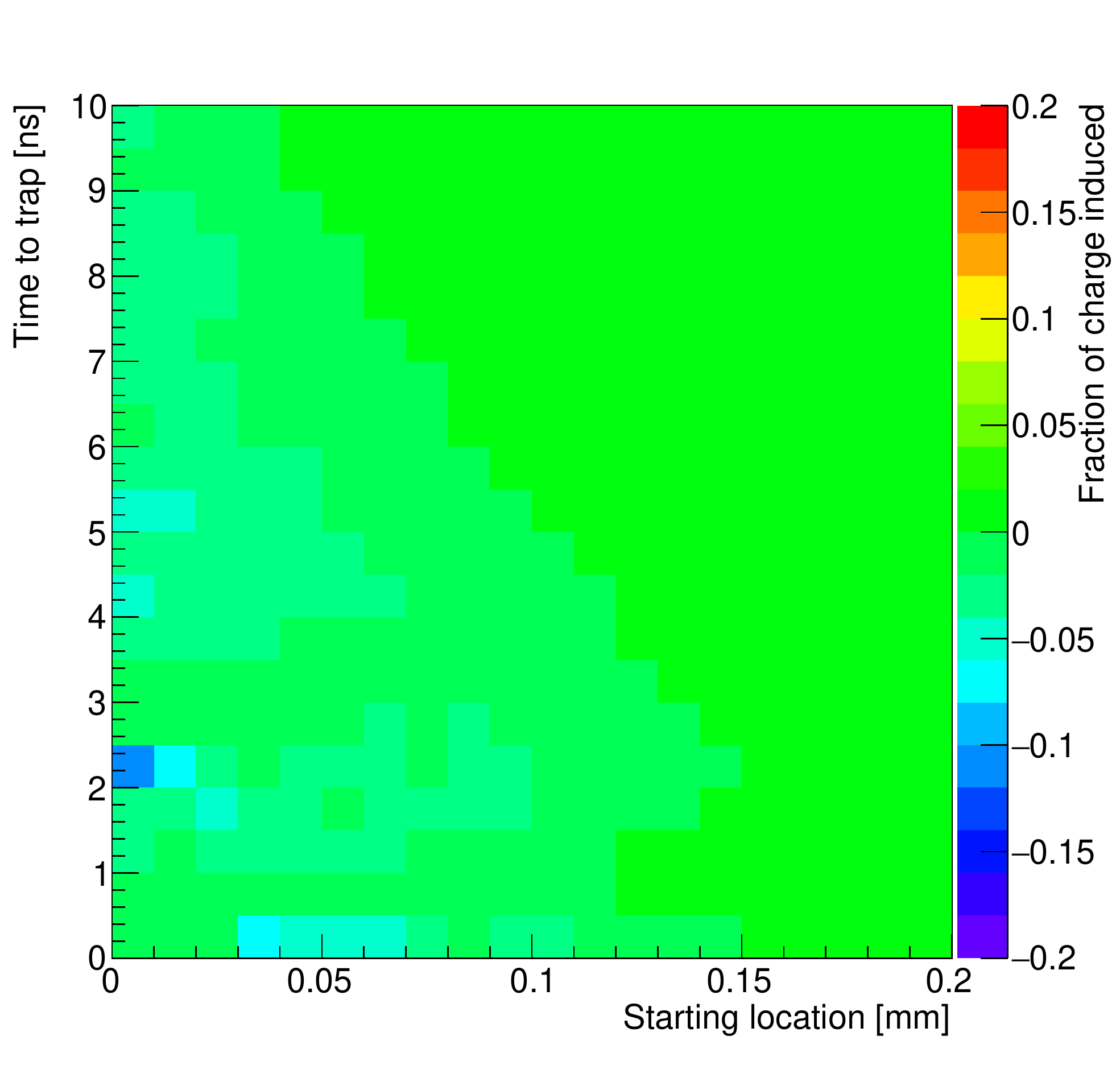}
\caption{The charge induced on the primary (left) and neighboring electrodes (short direction in the middle and long direction on the right) computed using the Ramo potential.}
\label{fig:app:raddamge9}
\end{figure}

One last effect that can play an important role in modeling the collected charge is related to charge chunking.  Representing many fundamental charges as one multi-charged chunk does not change the average charge collected, but does impact the charge collection resolution\footnote{Thank you to M. Garcia-Sciveres for pointing this out.}.  Figure~\ref{fig:app:raddamge8} quantifies how the resolution increases as more and more fundamental charges are combined together into one chunk.  This effect can be corrected by using a method inspired by the forward-folding method from Sec.~\ref{sec:JMR:resmethod}.  In particular, if $X$ is a random variable with mean $\mu$ and standard deviation $\sigma$, then $Y=\mu+\kappa(X-\mu)$ will have mean $\mu$ and standard deviation $\kappa\sigma$.  The resolution for chunks is larger than for fundamental charges, so $\kappa \leq 1$ (i.e. unsmearing is required).  In this case, $\kappa=1/\sqrt{n}$, where $n$ is the number of fundamental charges that one chunk represents. The average value is also known: for a charge $Q$, the average charge that will be collected is $e^{-t_\text{electrode}/t_\text{trapping}}Q$.  Therefore, the corrected charge is

\begin{align}
Q\times\delta_\text{collected}\mapsto Q\left[e^{-t_\text{electrode}/t_\text{trapping}}+\kappa\left(\delta_\text{collected}-e^{-t_\text{electrode}/t_\text{trapping}}\right)\right],
\end{align}

\noindent where $Q$ is the charge per chunk and $\delta_\text{collected}$ is one if the charge is collected and zero if it is trapped.  Note that even if a charge is trapped, it will still contribute to the collected charge.

\begin{figure}[h!]
\centering
\includegraphics[width=0.33\textwidth]{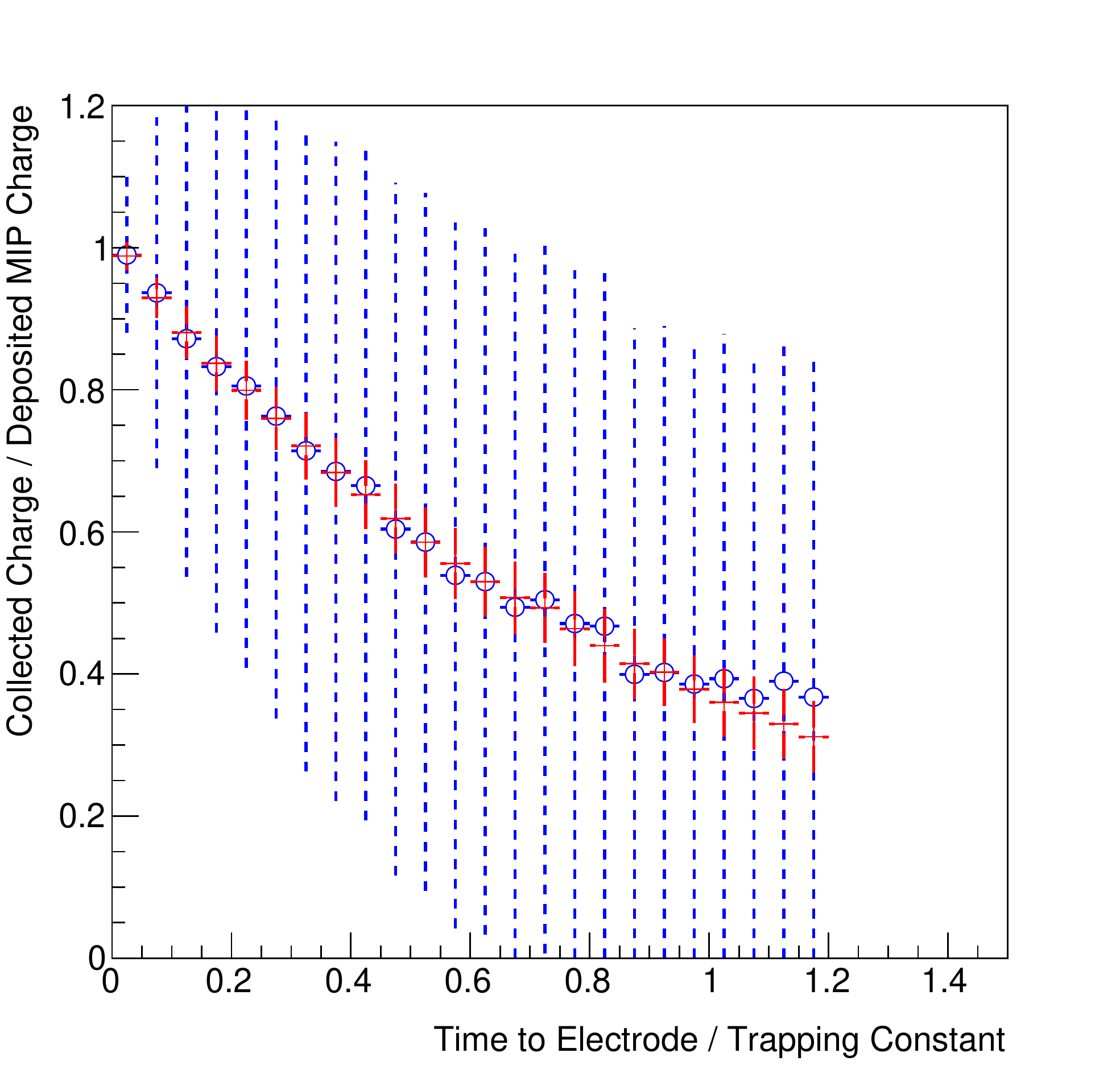}\includegraphics[width=0.33\textwidth]{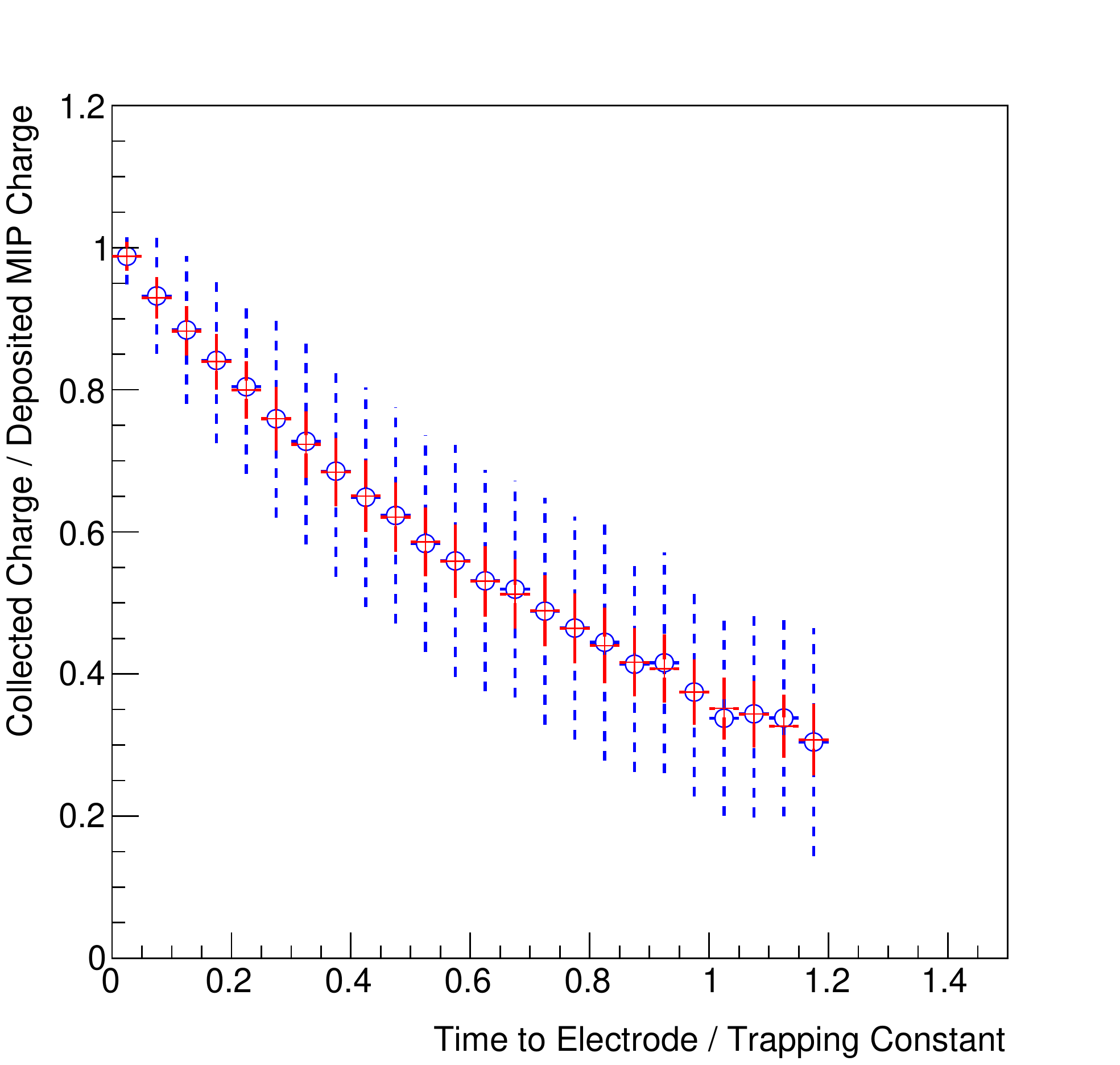}\includegraphics[width=0.33\textwidth]{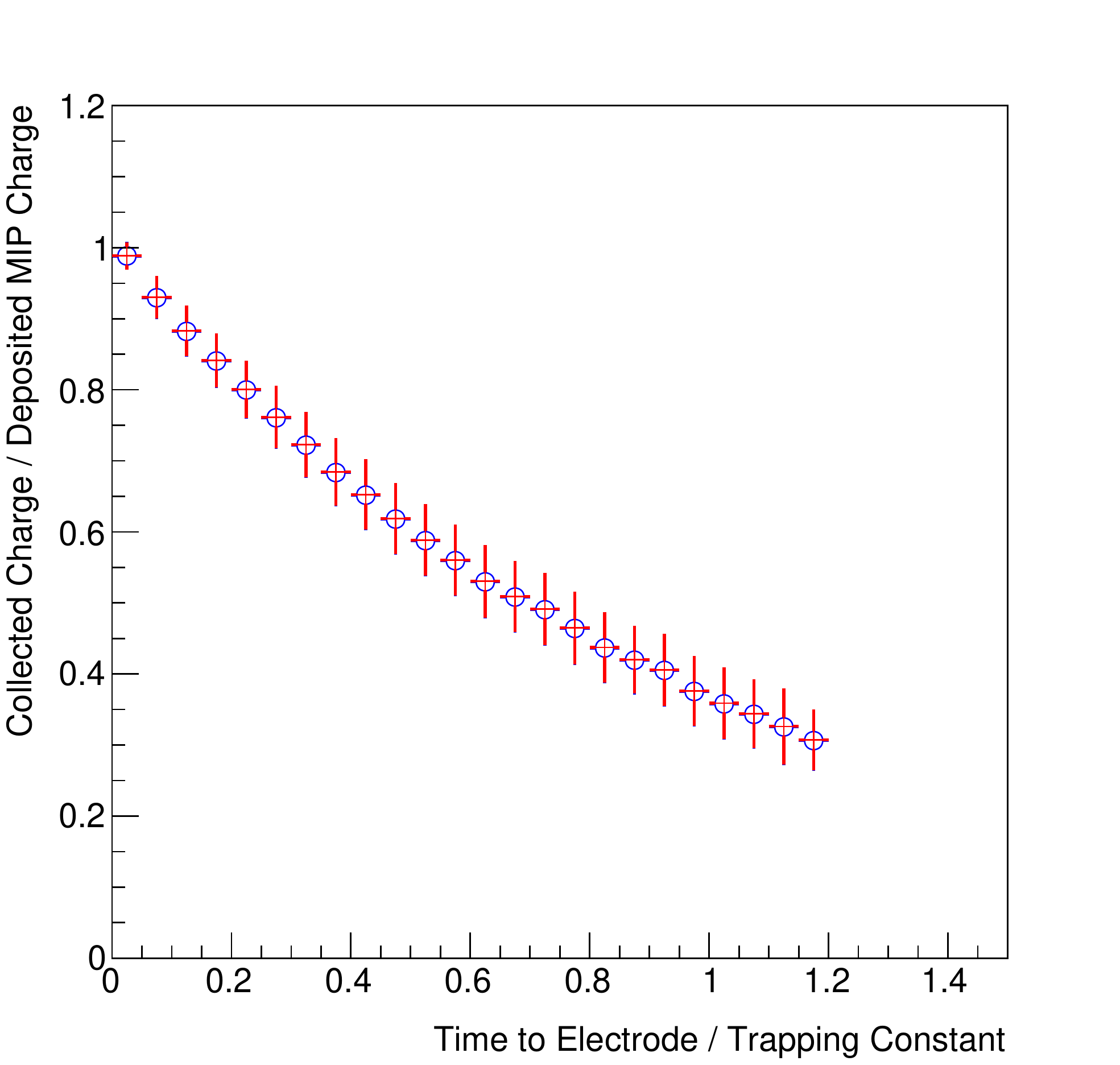}
\caption{An illustration of the impact of representing multiple fundamental charges by a single multi-charged chunk.  Displayed is the fraction of collected charge as a function of the exponential trapping time constant.  $N$ chunks are propagated to the electrode and some fraction $f$ of them reach the electrode before a random exponential time.  If a chunk reaches the electrode, a charge of $Q$ is recorded.  The points show $\langle fQN\rangle$ and the error bars are the standard deviation.  Red corresponds to $Q=1$ and blue corresponds to $Q=100$ (left), $Q=10$ (middle), and $Q=1$ (right). }
\label{fig:app:raddamge8}
\end{figure}

In order to validate the radiation damage model, modules with and without a significant radiation dose\footnote{Irradiated at Ljubljana with neutrons to $5\times 10^{15}$ $n_\text{eq}$/cm${}^{2}$.} are tested with a dedicated testbeam at the SLAC End Station A\footnote{Thank you to M. Benoit for helping acquiring the samples and to Su Dong for an extensive amount of time in the lab/testbeam for preparations and operations.  Thank you also to M. McCulloch and R. Carney for help with the setup and operations.}.  A kicker magnet extracts a 5 Hz electron beam that is incident on a copper target and focused to produce an $11$ GeV electron beam.  A {\it telescope} of six planes with Mimosa26~\cite{mimosa} sensors allow for $\mathcal{O}(\mu\text{m})$ precision tracking.  Three of these planes are on either side of a gap for the Device Under Test (DUT), which in this case is an irradiated or unirradiated IBL-like planar FEI4 module.  Figure~\ref{fig:app:raddamge9} shows the setup inside the SLAC beamline.  

\begin{figure}[h!]
\centering
\includegraphics[width=0.45\textwidth]{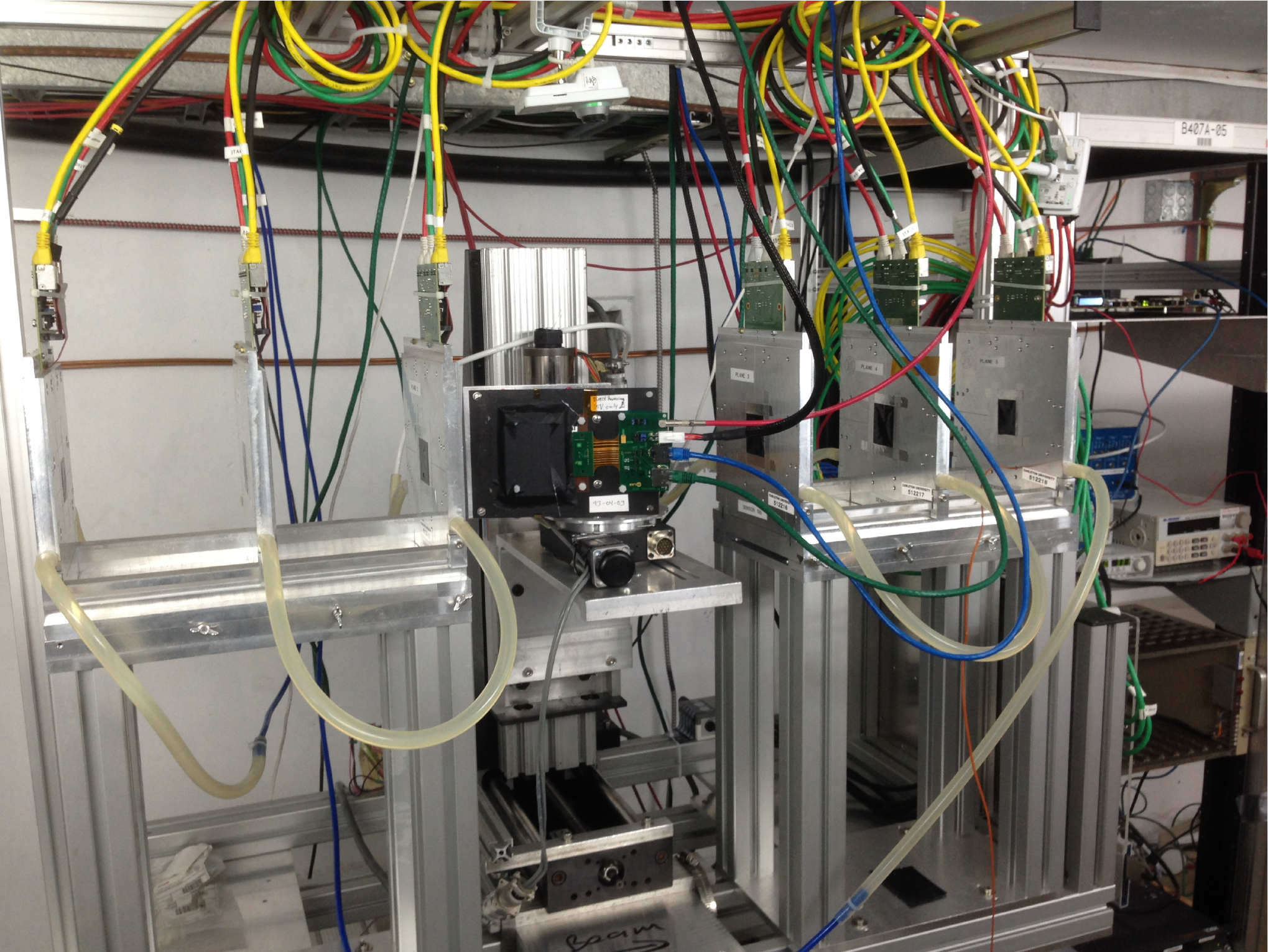}\hspace{2mm}\includegraphics[width=0.45\textwidth]{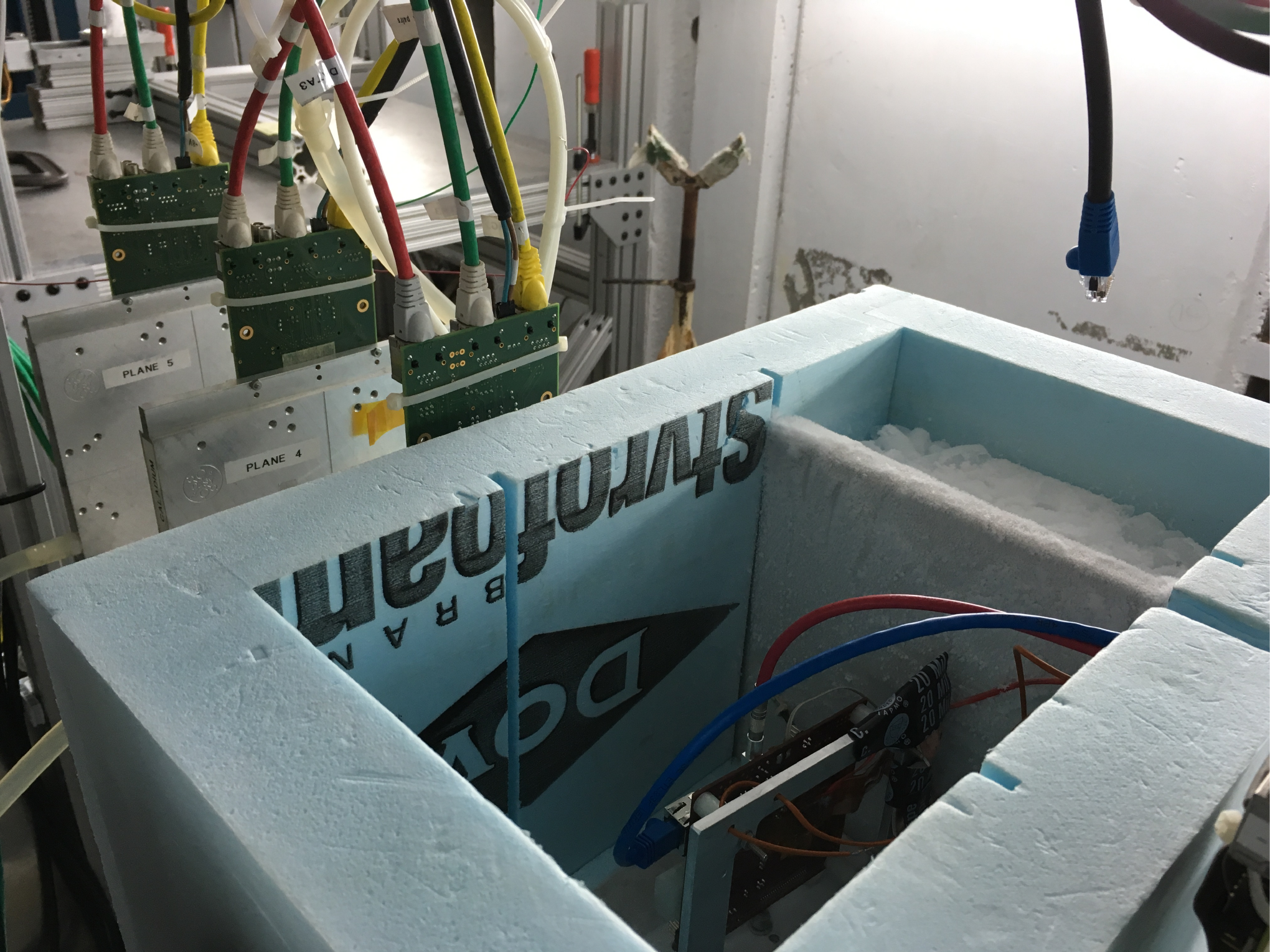}
\caption{Pictures of the testbeam setup. The left picture shows the six telescope planes, three on each side of the DUT.  The unirradiated module does not need to be cooled and sits on a stage that can be automatically raised, lowered, and tilted.  The beam goes from left to right, piercing the center of the black squares shown in the middle of each telescope plane.  The blue, red, and black cables from the DUT are for the data, low voltage, and high voltage, respectively.  The right picture shows a close-up of the irradiated module inside the box used to keep it cool (with dry ice).  The orange cables are temperature sensors. }
\label{fig:app:raddamge9}
\end{figure}

One of the most striking features of the irradiated sensors is the predicted electric field, as shown in Fig.~\ref{fig:app:raddamge2}.  To expose this dependence, the modules are rotated and tilted so that instead of the electrons traversing the $200$ $\mu$m depth of the sensor, they pass through the $50$ $\mu$m edge.  Figure~\ref{fig:app:raddamge10} illustrates this configuration: electron-hole pairs from pixels near the beginning or end of the cluster will probe the (large) field closest to and furthest away from the collecting electrode, while those passing through the center of the cluster will see the lower electric field in the middle of the sensor.  The angle corresponds to a cluster length of 15 pixels.  Representative event displays from sensors with and without irradiation are shown in Fig.~\ref{fig:app:raddamge11}.  As expected, particles form long streaks in the short pixel direction ($Y$ in these rotated coordinates).  Figure~\ref{fig:app:raddamge12} shows how the TOT distribution depends on the position inside one of the long clusters.  For the unirradiated module, the charge distribution is nearly independent of the position inside the cluster, as expected.  This also seems to be true for clusters of length 10 inside the irradiated module.  Even though the tilt angle is the same for both modules, the irradiated sensor is not fully depleted, so the clusters do not reach the full length.  Part of the degradation to the electric field in the middle of the sensor is compensated by the induced charge from the Ramo potential.  Further studies using more sophisticated clustering algorithms that can account for gaps in the clusters (from pixels below threshold) may reveal a structure that will be useful for tuning the simulation.

\begin{figure}[h!]
\centering
\includegraphics[width=0.6\textwidth]{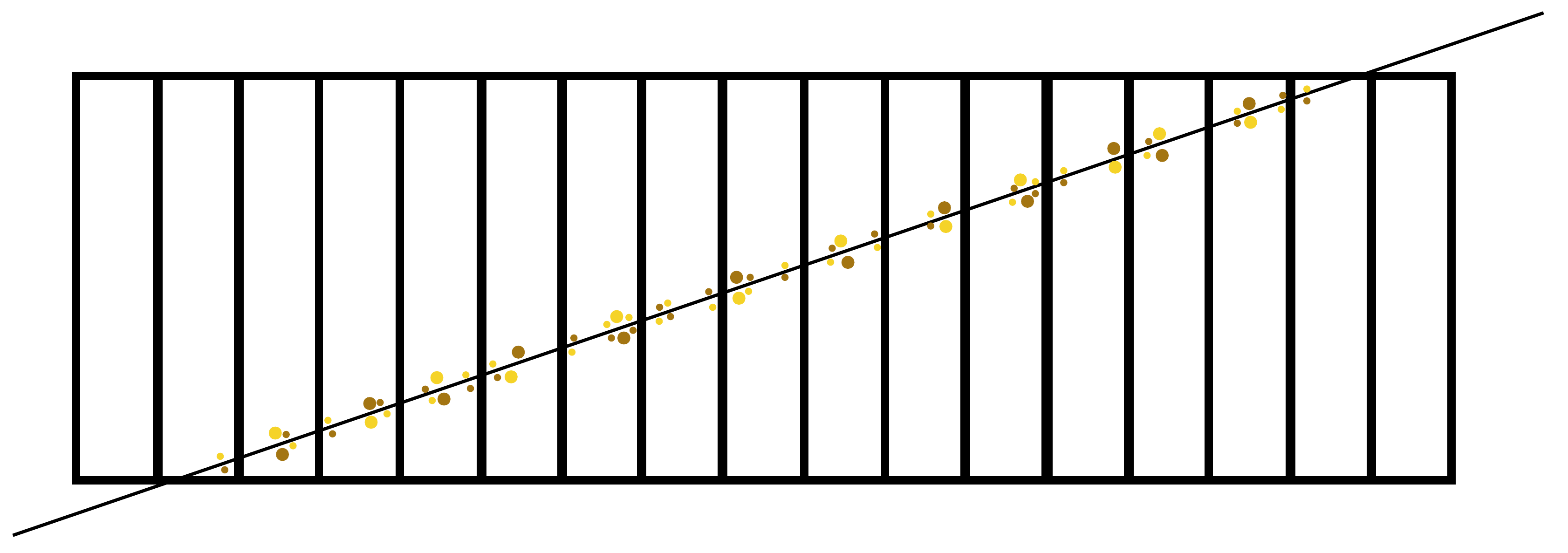}
\caption{A illustration of the tilted sensor configuration.  A charged particle travels from the bottom left to the top right and traverses about $50$ $\mu$m of silicon in each sensor.  The tilt angle is chosen so that the particle will traverse about 15 pixels, probing different depths along its path.}
\label{fig:app:raddamge10}
\end{figure}

\begin{figure}[h!]
\centering
\includegraphics[width=0.45\textwidth]{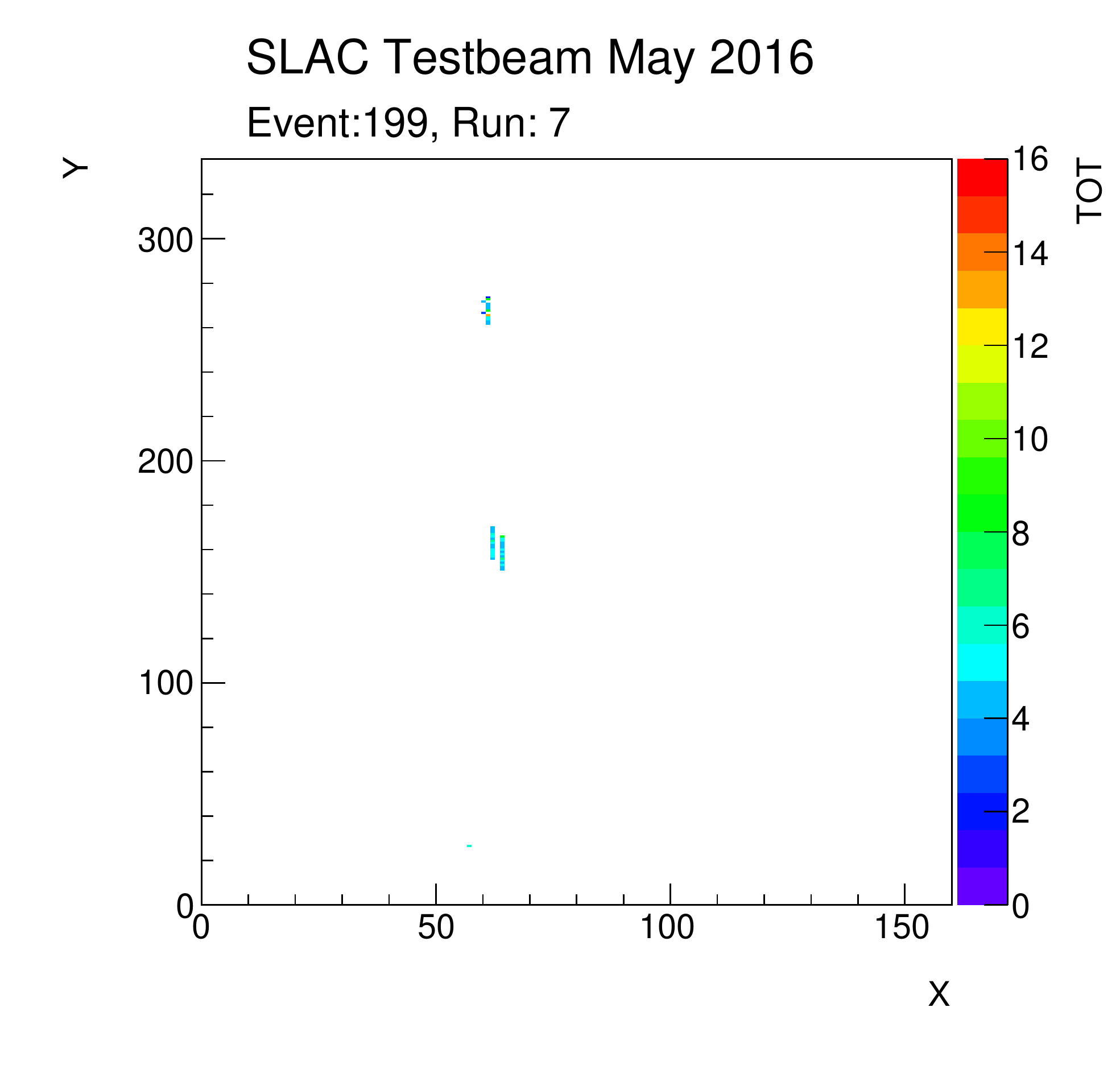}\includegraphics[width=0.45\textwidth]{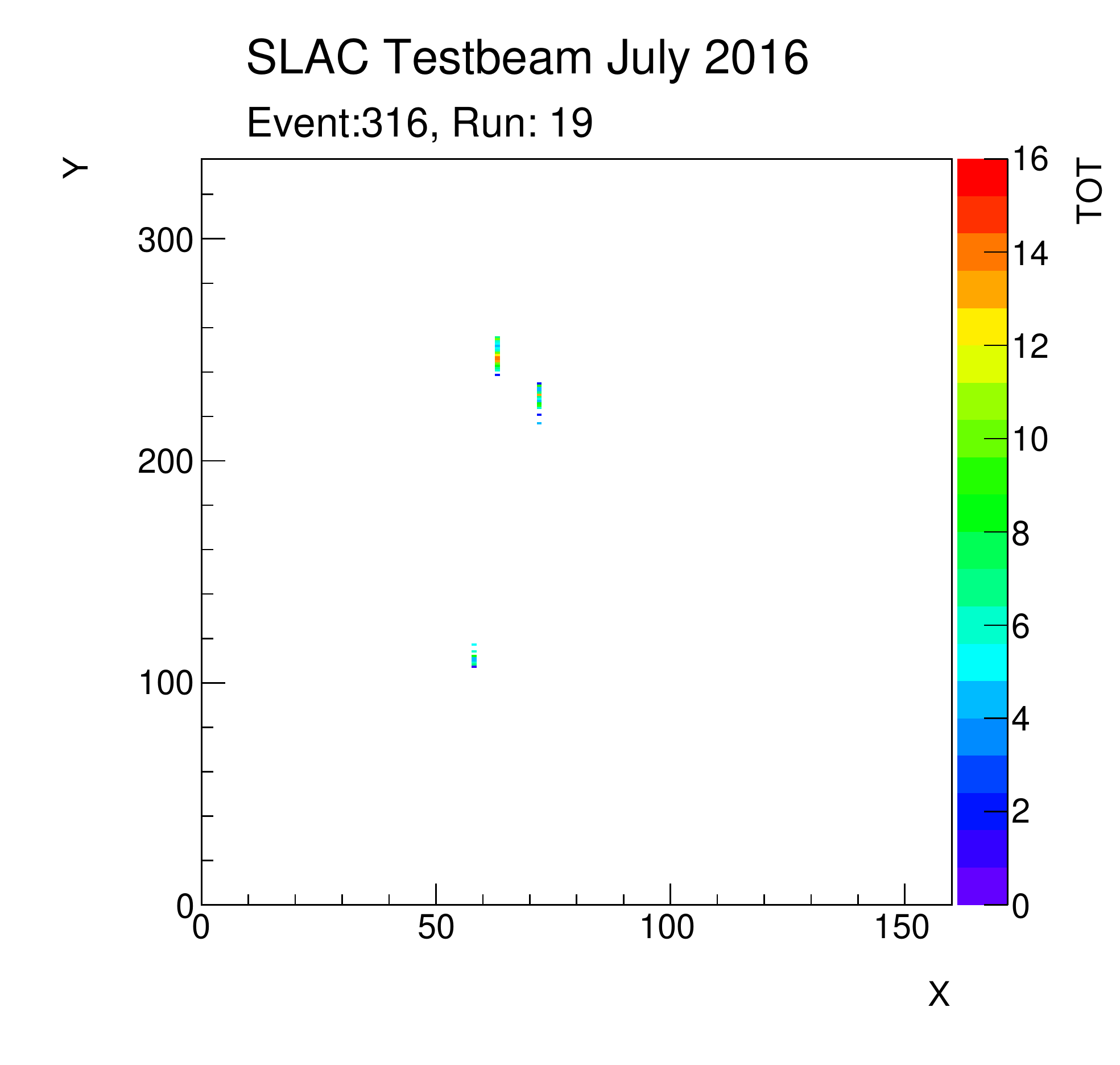}
\caption{Representative event displays from the unirradiated (left) and the irradiated (right) sensors.  In both cases, there are three clusters.  The sensors are tilted (Fig.~\ref{fig:app:raddamge10}) so that one expects about 15 pixel clusters per particle. The unirradiated sensor uses a 2000 electron threshold and 8 TOT is tuned to 11000 collected electrons, while the unirradiated sensor uses a 1500 electron threshold with 8000 electrons corresponding to 5 TOT.  The unirradiated sensor is biased with 60 V while the irradiated sensor is biased by 1.1 kV and cooled to about $-35^{\circ c}$.}
\label{fig:app:raddamge11}
\end{figure}

\begin{figure}[h!]
\centering
\includegraphics[width=0.45\textwidth]{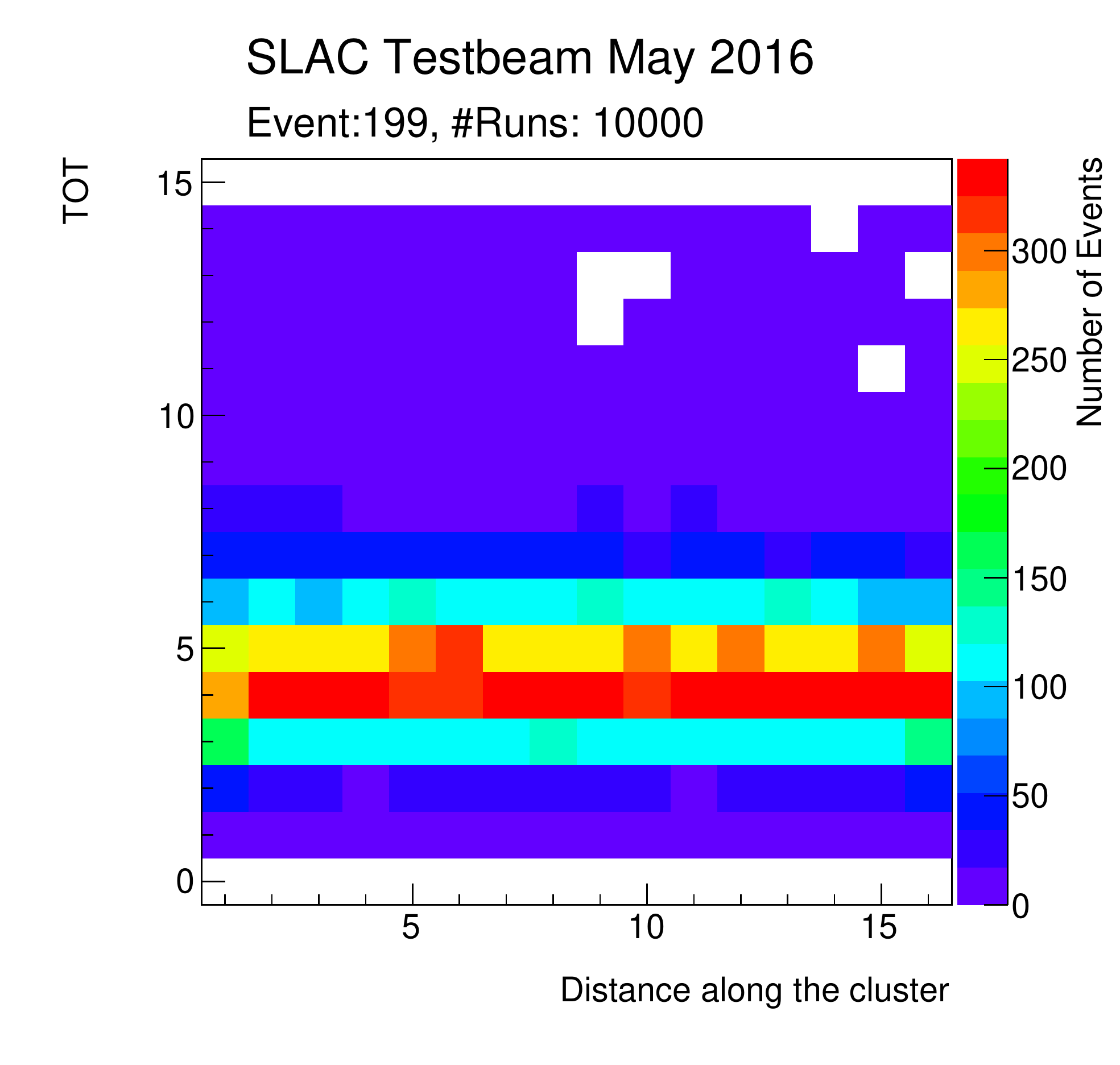}\includegraphics[width=0.45\textwidth]{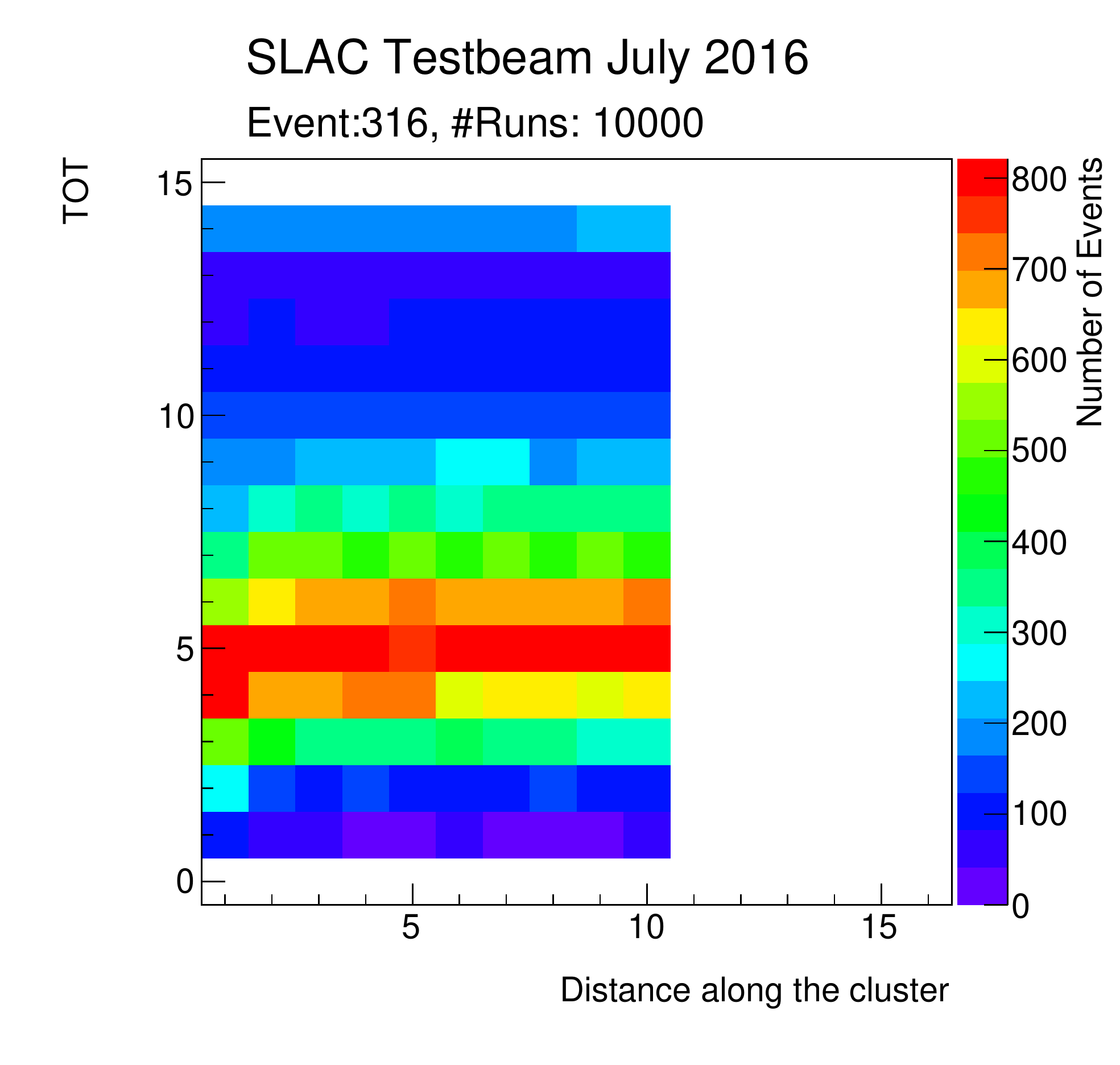}
\caption{The TOT distribution as a function of depth inside clusters of length exactly 15 (left) or 10 (right) for the unirradiated and irradiated sensors from Fig.~\ref{fig:app:raddamge11}.  The tilt angle is the same for both modules, but the pixel size distribution was peaked at lower values (10 versus 15) for the irradiated sensor, indicating that it is not fully depleted (even at 1.1 kV). }
\label{fig:app:raddamge12}
\end{figure}

\clearpage

\chapter{Boson Polarizations}

\clearpage

\section{Angular Distributions in $W$ and $Z$ Boson Decays}
\label{sec:app:wzdecay}

Without loss of generality, assume that the $W$ or $Z$ boson is moving in the $+z$ direction with momentum $p_V^\mu=(0,0,p_z,\sqrt{p_z^2+m_V^2})$.  The massive gauge bosons have three polarization states corresponding to a spin that is anti-aligned (transverse, spin $-1$), aligned (transverse, spin $+1$), or orthogonal (longitudinal, spin $0$) to the boson momentum.  The three corresponding polarization vectors are $\epsilon_{-1}^\mu=\frac{1}{\sqrt{2}}(0,1,-i,0),\epsilon_{+1}=-\frac{1}{\sqrt{2}}(0,1,i,0)$, and $\epsilon_0=\frac{1}{m_V}(p_z,0,0,\sqrt{p_z^2+m_V^2})$.  The weak charged and neutral currents have the form

\begin{align}
\label{eq:bosondecay}
j^\mu = \bar{u}(f)\frac{1}{2}\gamma^\mu(c_V^f-c_A^f\gamma^5)v(\bar{f}'),
\end{align}

\noindent where the $W$ boson only couples to left-handed fermions and right-handed anti-fermions, whereas the $Z$ boson couples to both left- and right-handed fermions, but with unequal couplings $c$.  The factors $\bar{u}$ and $v$ are the spinors for the out-going fermion $f$ and the out-going anti-fermion $\bar{f}'$.   For the $W$ decay, $c_A=c_V=1$ (i.e. the parenthetical term in Eq.~\ref{eq:bosondecay} is a pure projection operator) and for the $Z$ decay, $c_A=\frac{1}{2}$ for up-type quarks (u,c,t) and $-\frac{1}{2}$ for down-type quarks (d,s,b) while $c_V =\frac{1}{2}-2\times\frac{2}{3}\times \sin^2\theta_W\approx 0.19$ for up-type quarks and $\frac{1}{2}+2\times\frac{1}{3}\times \sin^2\theta_W\approx -0.35$ for down-type quarks.  In terms of pure left- and right-handed out-going fermions, one can write $j^\mu=c_{L/R} \bar{u}(f)\gamma^\mu v(\bar{f}')$, where the couplings $c_{L/R}$ can be extracted from Eq.~\ref{eq:bosondecay} using projection operators and are given by $c_L\approx -0.35$ for up-type quarks, $c_L\approx -0.42$ for down-type quarks, $c_R\approx -0.15$ for up-type quarks, and $c_R\approx 0.08$ for down-type quarks.  The matrix element is $M^2\propto |\epsilon_\mu j^\mu|^2$, where the proportionality constant is a coupling factor for the weak vertices multiplied by the number of colors $N_C$.  Using the setup shown in Fig.~\ref{fig:spinor} and working in the boson rest frame, the momentum of the fermions are $p_f^\mu=\frac{m_V}{2}(\sin\theta,0,\cos\theta,1)$ and $p_{\bar{f}'}^\mu=\frac{m_V}{2}(-\sin\theta,0,-\cos\theta,1)$ (ignoring the fermion mass).  With these momenta, the leading order matrix element is given by (see e.g. the polarized $e^+e^-\rightarrow \mu^+\mu^-$ calculations from Ref.~\cite{Peskin:1995ev}):

\begin{align}
|M_{-1}^W|^2&=\frac{3g_2^2m_W^2V^2}{4}(1+\cos\theta)^2\\
|M_{0}^W|^2&=\frac{3g_2^2m_W^2V^2}{2} \sin^2\theta\\
|M_{+1}^W|^2&=\frac{3g_2^2m_W^2V^2}{4}(1-\cos\theta)^2,
\end{align}

\noindent where $V$ is an element of the CKM matrix.  For $Z$ bosons,

\begin{align}
|M_{-1}^Z|^2_\text{up}&=\frac{3g_2^2m_Z^2}{2\cos^2\theta_W}\left[c^2_\text{L,up}(1+\cos\theta)^2+c^2_\text{R,up}(1-\cos\theta)^2\right]\\
|M_{0}^Z|^2_\text{up}&=\frac{3g_2^2m_Z^2}{\cos^2\theta_W}\left[c^2_\text{L,up}+c^2_\text{R,up}\right]\sin^2\theta\\
|M_{+1}^Z|^2&=\frac{3g_2^2m_Z^2}{2\cos^2\theta_W}\left[c^2_\text{L,up}(1-\cos\theta)^2+c^2_\text{R,up}(1+\cos\theta)^2\right],
\end{align}

\noindent and the equivalent formula for down-type quarks but with up $\leftrightarrow$ down.

\begin{figure}[h!]
\begin{center}
\begin{tikzpicture}[line width=1.5 pt, scale=1.3]
	\draw[->] (-2,0) -- (2,0);
	\draw[->] (0,0) -- (2,1);
	\draw[->] (0,0) -- (-2,-1);
	\node at (0.8,0.2) {$\theta$};
	\node at (2.2,1.1) {$f$};
	\node at (2.4,0) {$+z$};
	\node at (-2.2,-1.1) {$\bar{f}'$};
 \end{tikzpicture}
 \end{center}
 \caption{A diagram illustrating the setup for the calculation described in the text.  The boson spin is along the $z$-axis.}
 \label{fig:spinor}
 \end{figure}
 
 \clearpage
 
 \section{Polarization of $W$ Bosons}
\label{sec:app:wpolarization}

A calculation similar to Appendix~\ref{sec:app:wzdecay} can be used to determine the fractions of transverse and longitudinally polarized $W$ bosons from various production modes.  To begin, consider $W$ bosons produced from top quark pair production.  Consider a top quark decay $t\rightarrow W^+b$ from the top quark rest frame with the top spin aligned along the $+z$ axis.  For illustration, suppose that the $W^+$ and $b$ momenta are (anti-)parallel to the $+z$ axis and that the top quark spin is $+\frac{1}{2}$.  Neglecting the $b$-quark mass, there are only two possibilities: (a) the $b$ is moving in the $-z$ direction with spin $+\frac{1}{2}$ (left-handed) and by conservation of angular momentum, the $W$ boson is longitudinally polarized and (b) the $b$-quark is moving in the $+z$ direction with spin $-\frac{1}{2}$ (left-handed) and by conservation of angular momentum, the $W$ boson has spin $+1$.  The matrix element is given by

\begin{align}
\label{eq:polarizationtop}
M=g_2m_W\bar{u}(b)\epsilon_\mu^*(W^+)\gamma^\mu\frac{1}{2}(1-\gamma^5)u(t)
&=g_2m_Wu^\dag(b)\gamma^0\epsilon_\mu^*(W^+)\gamma^\mu\frac{1}{2}(1-\gamma^5)u(t).
\end{align}

\noindent In the chiral basis (the one used by Ref.~\cite{Peskin:1995ev}), 

\begin{align}
{\scriptsize
\gamma^\mu=\left[\begin{pmatrix}0&0&1&0\cr 0&0&0&1\cr1&0&0&0\cr0&1&0&0\end{pmatrix},\begin{pmatrix}0&0&0&1\cr 0&0&1&0\cr0&-1&0&0\cr-1&0&0&0\end{pmatrix},\begin{pmatrix}0&0&0&-i\cr 0&0&i&0\cr0&i&0&0\cr-i&0&0&0\end{pmatrix},\begin{pmatrix}0&0&1&0\cr 0&0&0&-1\cr-1&0&0&0\cr0&1&0&0\end{pmatrix}\right]},
\end{align}

\noindent and

\begin{align}
u(p)=\frac{1}{2}\begin{pmatrix}(\mathbb{I}_2-\hat{p}\cdot\sigma)\xi\cr (\mathbb{I}_2+\hat{p}\cdot\sigma)\xi\end{pmatrix},
\end{align}

\noindent where $\mathcal{I}_2$ is the $2\times 2$ identity matrix and $\xi$ is a two-component spinor.  The top quark is at rest and has $\xi=(1,0)^T$ (spin up) so $u(t)=\sqrt{2m_t}(1,0,0,0)^T$.  When the $b$-quark is spin up, it is moving in the $-z$ direction so $\hat{p}\cdot\sigma = -\sigma^3$.  Therefore, $u(t)=\sqrt{2E_b}(1,0,0,0)^T$.  In contrast, when the $b$-quark is spin down ($\xi=(0,1)^T$), it is moving in the $+z$ direction and so $u(t)=\sqrt{2E_b}(0,1,0,0)^T$.  The longitudinal $W^+$ polarization vector is $\frac{1}{m_W}(p_W,0,0,E_W)$ and the transverse (spin -1) $W^+$ polarization vector is $\frac{1}{\sqrt{2}}(0,1,-i,0)$.  Putting all of these pieces together with Eq.~\ref{eq:polarizationtop} produces the following results (dropping constants appearing in both terms):

{\small
\begin{align}
M_0&\propto\frac{1}{m_W} \begin{pmatrix}0&0&1&0\end{pmatrix}\begin{pmatrix}0 & 0 & E_W-p_W& 0\cr 0 & 0 & 0 & E_W+p_W\cr E_W+p_W & 0 & 0 & 0\cr 0 & E_W-p_W & 0 & 0\end{pmatrix}\begin{pmatrix}1\cr 0\cr 0\cr 0\end{pmatrix}\\
&=\frac{(E_W+p_W)}{m_W}=\frac{m_t}{m_W},
\end{align}
}
\noindent where the last equality holds because $p_W=p_b$ in the top quark rest frame by conservation of momentum and by conservation of energy, $m_t=E_b+E_W=p_b+E_W=p_W+E_W$ (ignoring the $b$-quark mass).  Likewise,

{\small
\begin{align}
M_-&\propto \frac{1}{\sqrt{2}}\begin{pmatrix}0&0&0&1\end{pmatrix}\begin{pmatrix}0 & 0 & 0& 0\cr 0 & 0 & 2 &0\cr 0& 0 & 0 & 0\cr -2 & 0& 0 & 0\end{pmatrix}\begin{pmatrix}1\cr 0\cr 0\cr 0\end{pmatrix}\\
&=\frac{2}{\sqrt{2}}
\end{align}
}

\noindent Therefore, the ratio of the number of longitudinally polarized $W^+$ bosons to transversely polarized $W^+$ bosons is

\begin{align}
\frac{|M_0|^2}{|M_-|^2}=\frac{m_t^2}{2m_W^2}\approx 2.3,
\end{align}

\noindent so there are {\it more} longitudinally polarized $W$ bosons from top quark decays relative to transversely polarized $W$ bosons.  The above observation is true even if the $bW$ axis is not aligned with the top quark spin axis and has been computed at NNLO in QCD to be $0.21\pm 0.05$~\cite{Czarnecki:2010gb}.  In contrast, $W$ bosons produced via inclusive $W$+jets processes, are predominately produced with a transverse polarization.  This is nicely explained, along with studies of higher order QCD effects, in Ref.~\cite{Bern:2011ie}.

 \clearpage

\chapter{Additional Statistical Considerations}
\label{additionastats}

\clearpage

 \section{Uncertainty Ellipses}
\label{sec:app:uncertaintyellipse}

In one dimension, for a random variable $X\sim\mathcal{N}(\mu,\sigma^2)$, $(X-\mu)/\sigma\sim \mathcal{N}(0,1)$ and so an interval centered at the mean that contains $p$-percent of the probability distribution of $X$ is given by $\mu \pm Z\sigma$, where $p=\frac{1}{\sqrt{2\pi}}\int_{-Z}^Zdx \exp(-x^2/2)$.  An equivalent way to arrive at the same interval that generalizes to higher dimensions is to note that $(X-\mu)^2/\sigma^2\sim \chi^2_1$, a chi-squared distribution with one-degree of freedom.  Then, the same interval can be constructed as $\mu \pm \sqrt{C}\sigma$, where $p=\frac{1}{\sqrt{2\pi}}\int_0^C dxx^{-1/2}\exp(-x/2)$.  Now, suppose that $\vec{X}\sim\mathcal{N}(\vec{\mu},\Sigma)$ where $\mu$ is an $n$-dimensional vector and $\Sigma$ is the $n\times n$ covariance matrix (symmetric, positive semi-definite).  Then, $(\vec{X}-\vec{\mu})^\text{T}\Sigma^{-1}(\vec{X}-\vec{\mu})\sim\chi^2_n$, a chi-squared distribution with $n$ degrees of freedom.  When $n=1$, this reduced to the one-dimensional case above.  An ellipsoid centered about the mean which contains $p$-percent of the probability distribution distribution is then given implicitly by $(\vec{x}-\vec{\mu})^\text{T}\Sigma^{-1}(\vec{x}-\vec{\mu}) \leq C$,
where $p=\frac{1}{2^{n/2}\Gamma(n/2)}\int_0^C dxx^{n/2-1}\exp(-x/2)$.  In two dimensions $(X,Y)\sim\mathcal{N}((\mu_x,\mu_y),\Sigma)$, this is an ellipse.  One can write 

\begin{align}
\Sigma=\begin{pmatrix}\sigma_x^2 & \rho\sigma_x\sigma_y \cr \rho\sigma_x\sigma_y & \sigma_y^2\end{pmatrix},
\end{align}

\noindent where $\rho$ is the correlation between $X$ and $Y$.  The matrix $\Sigma$ is diagonalizable such that after a suitable rotation of $X$ and $Y$,

\begin{align}
\Sigma=\begin{pmatrix}\lambda_+ & 0 \cr 0& \lambda_-\end{pmatrix},
\end{align}

\noindent where $\lambda_\pm$ are the eigenvalues of $\Sigma$ and are found by solving $\text{Det}(\Sigma-I\lambda)=0$:

\begin{align}
\lambda_{\pm}=\frac{1}{2}(\sigma_x^2+\sigma_y^2)\pm \frac{1}{2}\sqrt{(\sigma_x^2+\sigma_y^2)^2-4(1-\rho^2)\sigma_x^2\sigma_y^2}.
\end{align}

\noindent Let $X'$ and $Y'$ be the centered and rotated versions of $X$ and $Y$.  In these transformed coordinates, the uncertainty ellipse is given by

\begin{align}
\frac{(x')^2}{C\lambda_+}+\frac{(y')^2}{C\lambda_-} \leq 1,
\end{align}

\noindent which is the standard form of an ellipse with radii $\sqrt{C\lambda_\pm}$.  The tilt of the ellipse with respect to the original coordinates can be computed from the orientation of the eigenvectors $v_\pm$ of $\Sigma$.  For example, $\Sigma v_+=\lambda_+ v_+$ gives the condition

\begin{align}
(v_+)_x=\left(\frac{\rho\sigma_x\sigma_y}{\lambda_+-\sigma_x^2}\right)(v_+)_y.
\end{align}

\noindent Therefore, the rotation angle counter close-wise from the $x$-axis is $\theta=\text{tan}^{-1}[(\lambda_+-\sigma_x^2)/(\rho\sigma_x\sigma_y)]$.  As expected, as $\rho\rightarrow 0$, $\lambda_+\rightarrow\sigma_x^2, \lambda_-\rightarrow\sigma_y^2$ and $\theta\rightarrow0$.  Figure~\ref{fig:ellipse} shows the general form of the uncertainty ellipse.

\begin{figure}[h!]
\begin{center}
\includegraphics[width=0.5\textwidth]{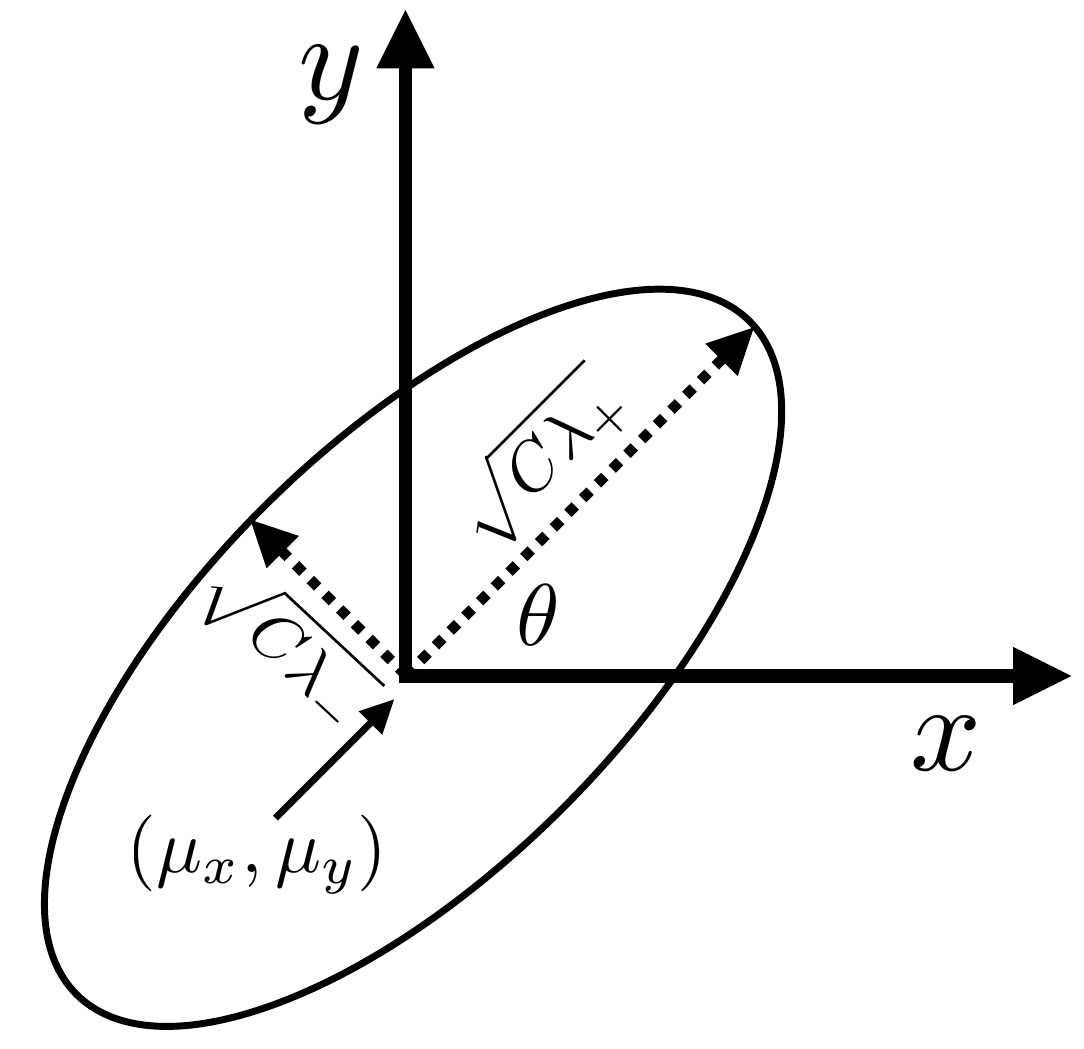}\\
\caption{A schematic diagram of an uncertainty ellipse for a bivariate normal distribution.  See the text for details. }
\label{fig:ellipse}
\end{center}
\end{figure}

\clearpage

\section{Wrapped Gaussian}
\label{sec:wrapped}

The EM algorithm (Sec.~\ref{sec:EMalgorithm}) depends on the event topology.  For instance, if a Gaussian density is used to model $\phi$, then, in the E step, a particle with $\phi_i$ near $2\pi$ will be deemed far from a cluster with location $\phi_j$ near $0$.  To avoid this undesirable behavior and enforce the equivalence of the angles $0$ and $2\pi$, $\phi$ is associated with a {\it wrapped Gaussian density} and $y$ with a standard Gaussian density:

\begin{align}
\label{eq:wrap}
\Phi(y,\phi | \mu_\phi,\mu_y,\sigma^2)=\Phi_y(y|\mu_y,\sigma^2)\frac{1}{\sqrt{2\pi\sigma^2}}\sum_{I=-\infty}^\infty\exp\left[\frac{-(\phi-\mu_\phi(I))^2}{2\sigma^2}\right],
\end{align}

\noindent where $\Phi_y$ is a normal distribution and $\mu_\phi(I)=\mu_\phi+2\pi I$.  In order to approximate the sum in Eq.~(\ref{eq:wrap}), only the leading contribution is retained by choosing $\mu_\phi(I^*)$ for $I^*= \text{argmin}_{I'}|\phi-\mu_\phi+2\pi I'|$.  Other contributions are exponentially suppressed and this part recovers continuity near $0$ and $2\pi$, as illustrated in Fig.~\ref{fig:wrappedgaussian}.

\begin{figure}[h!]
\vspace{1cm}
\begin{center}
\begin{tabular}{cc}
\begin{overpic}[width=0.43\textwidth]{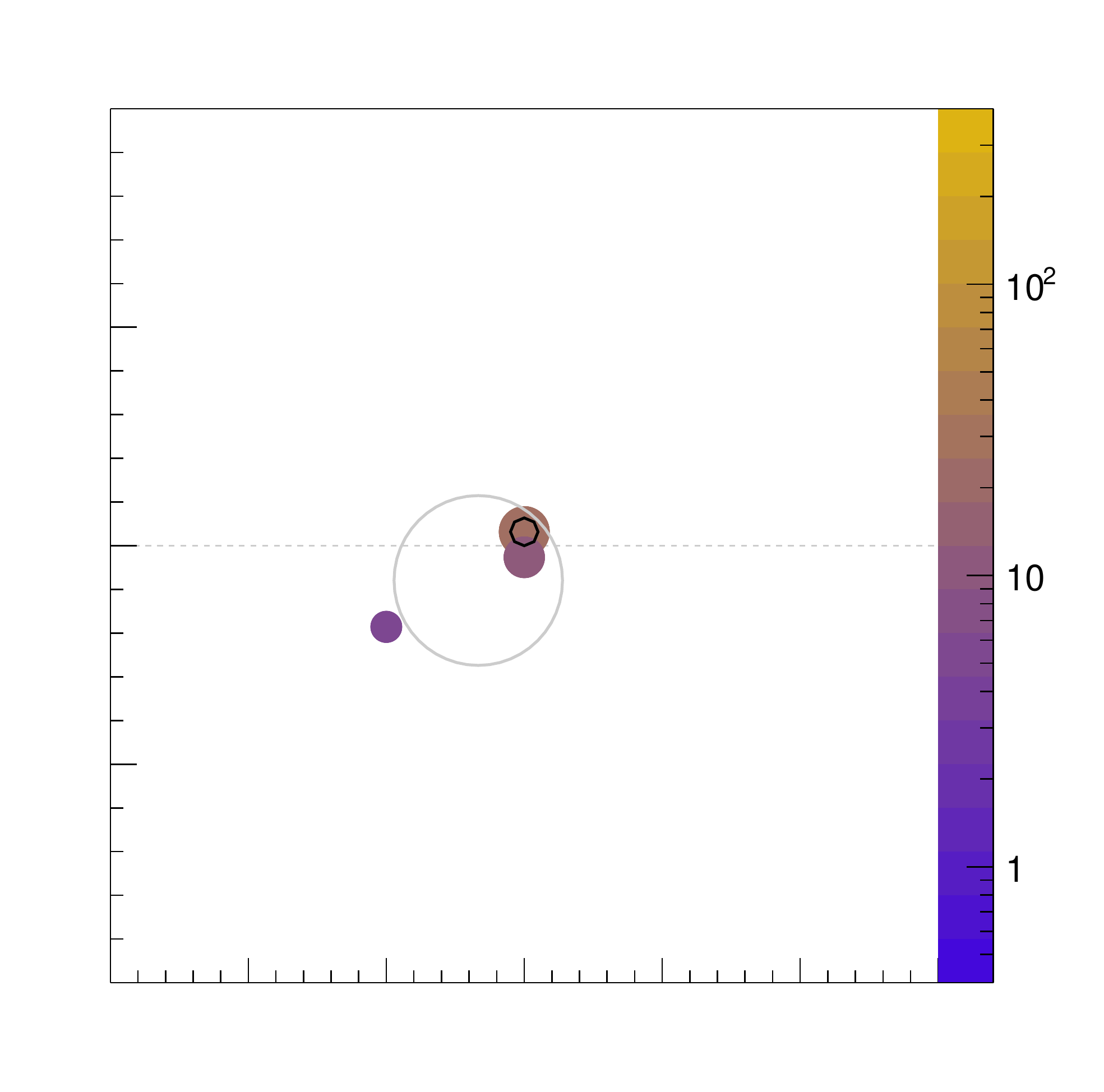}
\put(11, 91){  \small Naive Gaussian Density}
\put(72.2, 50.8){ \sffamily \tiny \textcolor[rgb]{0.8, 0.8, 0.8}{Internal}}
\put(70, 46.2){\bfseries \sffamily \tiny \textcolor[rgb]{0.8, 0.8, 0.8}{boundary}}
\put(24, -4){ \small Pseudorapidity ($\eta$)}
\put(-4, 10){\rotatebox{90}{ \small Rotated Azimuthal Angle
    ($\phi$)}}
\put(95, 37){\rotatebox{90}{ \small $p_T \text{ [GeV]}$}}

\put(5, 8){\bfseries \small \sffamily $\pi$}
\put(4, 28.3){\bfseries \small \sffamily $\frac{3\pi}{2}$}
\put(4, 48.3){\bfseries \small \sffamily $2\pi$}
\put(4, 68){\bfseries \small \sffamily $\frac{5\pi}{2}$}
\put(4, 88){\bfseries \small \sffamily $3\pi$}

\put(4,  4){\bfseries \small \sffamily $-3$}
\put(17, 4){\bfseries \small \sffamily $-2$}
\put(29.6, 4){\bfseries \small \sffamily $-1$}
\put(46.2, 4){\bfseries \small \sffamily $0$}
\put(58.8, 4){\bfseries \small \sffamily $1$}
\put(71.3, 4){\bfseries \small \sffamily $2$}
\put(83.8, 4){\bfseries \small \sffamily $3$}
\end{overpic} &
\begin{overpic}[width=0.43\textwidth]{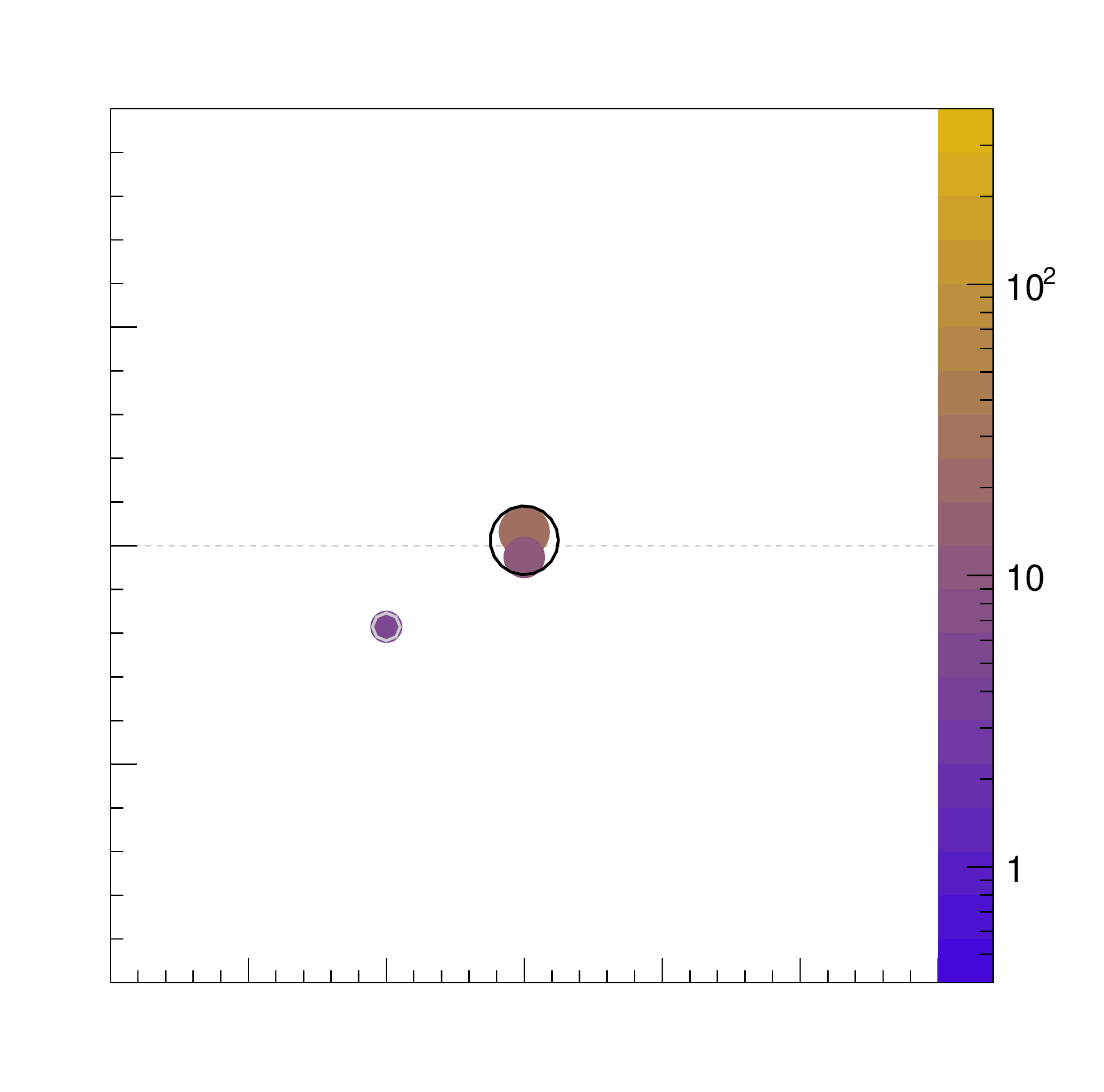}
\put(11, 91){  \small Wrapped Approximation}
\put(72.2, 50.8){\bfseries \sffamily \tiny \textcolor[rgb]{0.8, 0.8, 0.8}{Internal}}
\put(70, 46.2){\bfseries \sffamily \tiny \textcolor[rgb]{0.8, 0.8, 0.8}{boundary}}
\put(24, -4){ \small Pseudorapidity ($\eta$)}
\put(-4, 10){\rotatebox{90}{ \small Rotated Azimuthal Angle
    ($\phi$)}}
\put(95, 37){\rotatebox{90}{ \small $p_T \text{ [GeV]}$}}

\put(5, 8){\bfseries \small \sffamily $\pi$}
\put(4, 28.3){\bfseries \small \sffamily $\frac{3\pi}{2}$}
\put(4, 48.3){\bfseries \small \sffamily $2\pi$}
\put(4, 68){\bfseries \small \sffamily $\frac{5\pi}{2}$}
\put(4, 88){\bfseries \small \sffamily $3\pi$}

\put(4,  4){\bfseries \small \sffamily $-3$}
\put(17, 4){\bfseries \small \sffamily $-2$}
\put(29.6, 4){\bfseries \small \sffamily $-1$}
\put(46.2, 4){\bfseries \small \sffamily $0$}
\put(58.8, 4){\bfseries \small \sffamily $1$}
\put(71.3, 4){\bfseries \small \sffamily $2$}
\put(83.8, 4){\bfseries \small \sffamily $3$}
\end{overpic} \\
\end{tabular}
\end{center}
\caption{A three-particle event display illustrating the results of fuzzy jet clustering using a Gaussian density for $\phi$ (left) and a wrapped Gaussian density approximation for $\phi$ (right).  Figure from C. Stansbury.}  
\label{fig:wrappedgaussian}
\end{figure}

\clearpage
\newpage

\section{The EM algorithm}
\label{sec:emalgo}

This appendix contains two derivations: the modified EM algorithm updates in Eq.~(\ref{eq:emupdates}) and the proof that the modified EM algorithm generically improves the original modified log likelihood Eq.~(\ref{eq:mm2}) with every iteration.  Recall the expected modified complete log likelihood (mmCLL) from Eq.~(\ref{eq:cll}):

\begin{align*}
\sum_{i=1}^n\sum_{j=1}^kp_{Ti}^\alpha\left(q_{ij}\log\Phi(\vec{\rho}_i;\vec{\mu}_j,\Sigma_j)+q_{ij}\log\pi_j\right).
\end{align*}

\noindent Viewing the mCLL as a function of $\vec{\mu},\Sigma$ and $\pi$ for fixed $\lambda$ and $\vec{\rho}$ we can maximize.  For $\pi$, we optimize

\begin{align*}
\sum_{i=1}^n\sum_{j=1}^kp_{Ti}^\alpha\left(q_{ij}\log\pi_j\right)+\lambda\left(\sum_{j=1}^k \pi_j-1\right),
\end{align*}

\noindent where the last term is needed so that the optimal $\pi^*$ is a probability.  The derivative of this expression with respect to $\pi_j$ is

\begin{align*}
\pi_j=-\frac{1}{\lambda}\sum_{i=1}^np_{Ti}^\alpha q_{ij},
\end{align*}

\noindent and then summing the equation over $j$ and using $\sum_{j=1}^kq_{ij}=1$ and the constraint equation $\sum_{j=1}^k\pi_j=1$, we find that

\begin{align*}
\pi_j^*=\frac{1}{\sum_{i=1}^np_{Ti}^\alpha}\sum_{i=1}^np_{Ti}^\alpha q_{ij}
\end{align*}

\noindent The updates for $\vec{\mu}$ and $\Sigma$ follow from the standard derivation (by similarly taking derivatives of the mCLL with respect to components of these multi-dimensional objects) by noting that the only difference is that $q_{ij}\mapsto q_{ij}p_{Ti}^\alpha$ and there are no Lagrange multipliers needed unlike for $\pi_j^*$.

Finally, we prove the claim that the modified EM algorithm described in the body of the text monotonically improves the modified log likelihood in Eq.~(\ref{eq:mm2}).  First, we note that we can rewrite the (log) likelihood as

\begin{align*}
p_T^\alpha \log p(\rho|\theta) &= p_T^\alpha \log\left(\sum_{\lambda\in \{1,2,...,k\}} p(\rho,\lambda;\theta)\right)\\
&=p_T^\alpha \log\left(\sum_{\lambda\in \{1,2,...,k\}} \frac{q(\lambda)p(\rho,\lambda;\theta)}{q(\lambda)}\right)\\
&=p_T^\alpha \log \mathbb{E}_q\left[\frac{p(\rho,\lambda;\theta)}{q(\lambda)}\right]\\
&\geq  \mathbb{E}_q\left[p_T^\alpha\log\left(\frac{p(\rho,\lambda;\theta)}{q(\lambda)}\right)\right]\equiv \mathcal{L}(q,\theta),
\end{align*}

\noindent where the inequality in the last line follows from Jensen's inequality.  Now, we are ready to prove the claim that $p_T^\alpha p(\rho|\theta^{(t}))$ improves monotonically with $t$, the index for the iteration of the EM algorithm.  First, note that

\begin{align*}
\mathcal{L}(q,\theta)&= \mathbb{E}_q\left[p_T^\alpha\log\left(\frac{p(\rho,\lambda;\theta)}{q(\lambda)}\right)\right]\\
&= \mathbb{E}_q\left[p_T^\alpha\log\left(p(\rho,\lambda;\theta)\right)\right]- \mathbb{E}_q\left[p_T^\alpha\log\left(q(\lambda)\right)\right],
\end{align*}

\noindent where the first term is the mCLL and the second term has no $\theta$ dependance and so maximize $\mathcal{L}(q,\theta)$ over $\theta$ is equivalent to maximize the mCLL over $\theta$.  Therefore, $\mathcal{L}(q^{(t+1)},\theta^{(t)})\leq \mathcal{L}(q^{(t+1)},\theta^{(t+1)})$.  By the inequality above, $\mathcal{L}(q^{(t+1)},\theta^{(t+1)})\leq p_T^\alpha p(\rho|\theta^{(t+1)}) $.  The E step can be recast as choosing

\begin{align*}
q^{(t+1)}(\lambda_i=j)=q_{ij}(\theta^{(t)})=\mathbb{E}_{\theta^{(t)}}[q_{ij}]=p(\lambda|\rho,\theta^{(t)}).
\end{align*}

\noindent This enforces:

\begin{align*}
\mathcal{L}(p(\lambda|\rho,\theta^{(t)}),\theta^{(t)})&=\mathbb{E}_{p(\lambda|\rho,\theta^{(t)})}\left[p_T^\alpha\log\left(\frac{p(\rho,\lambda;\theta^{(t)})}{p(\lambda|\rho,\theta^{(t)})}\right)\right]\\
&=\mathbb{E}_{p(\lambda|\rho,\theta^{(t)})}\left[p_T^\alpha\log\left(p(\rho;\theta^{(t)})\right)\right]\\
&=p_T^\alpha\log\left(p(\rho;\theta^{(t)})\right)
\end{align*}

\noindent Putting this together with the bounds from the M step, we arrive at the desired result: $p_T^\alpha p(\rho|\theta^{(t)})\leq p_T^\alpha p(\rho|\theta^{(t+1)})$, i.e., every step of the modified EM algorithm improves or leaves the same the original likelihood.

\clearpage
\newpage

\section{A Leading Order Description of Fuzzy Jet $\sigma$}
\label{sec:leadingordersigma}

Section~\ref{sec:tagging} demonstrated that the fuzzy jet $\sigma$ is correlated with $\rho=m/p_\text{T}$.  One can build some intuition for this relationship by considering a leading order QCD calculation of $\sigma$.  Consider an isolated quark jet with energy $E$ which radiates a gluon with angle $\theta\ll 1$ from the jet axis and with energy fraction $z\ll1$.  Without loss of generality, suppose the quark is moving in the $\phi=0$ direction and the splitting happens in the $\phi=\pi/2$ direction so that the four vector of the quark is $q^\mu=E(1-z)(1,0,0,1)$, and the gluon four-vector is $g^\mu=Ez(1,\theta,0,1)$, to leading order.  To this order, the jet mass is simply $m=Ez\theta^2$.  What is $\sigma$?  Consider $k=1$ and something like the event-jet applied so that we can treat this jet in isolation from other hadronic activity in the event.  Since $k=1$, the soft memberships are all one, i.e., $q_{i1}=1$ and there is only one step of the EM algorithm.  The anti-$k_t$ jet has $(y,\phi)$ coordinates $(0,\theta)$, which could be used for the seed, but since $k=1$, the seed is not used.  The quark has coordinates $(0,0)$, and the gluon has coordinates $(0,\theta)$.  One can compute the fuzzy jet coordinates in the (single) M step:

\begin{align}
\mu_y&=0\\
\mu_\phi&=\frac{0\times E(1-z)+\theta\times Ez}{E(1-z)+Ez}=z\theta\\
\sigma^2&=\frac{(0-z\theta)^2\times E(1-z)+(\theta-z\theta)^2\times Ez}{2(E(1-z)+Ez)}\\
&=z\theta^2+\mathcal{O}(\theta^2z^2).
\end{align}

\noindent Therefore, to leading order and $k=1$, the learned $\sigma$ is the jet mass.  For $k=2$, there are enough degrees of freedom to resolve the substructure of the hard splitting and so the relationship between the jet mass and $\sigma$ breaks down.

\clearpage
\newpage

\section{Controlling Jet Multiplicity with $p_T$}
\label{sec:pt_multiplicity}

In contrast to most uses of hierarchical-agglomerative clustering algorithms, the number of fuzzy jets is fixed before clustering
begins. Whereas a single traditional jet can reasonably be considered to
correspond to a parton in appropriate cases, mGMM jets should not be,
as several mGMM jets can together express structure of what would be
one or several jets according to another algorithm.  The choice of the number of jets used in mGMM jet clustering therefore
controls the expressive power of the algorithm to look at the event
structure. In practice, choosing too many jets does not greatly affect
the value of the leading learned $\sigma$ variable, because the
additional jets learn finer features of the event structure. On the
other hand, choosing too few jets is often problematic as can be seen
in Figure~\ref{fig:pt_cut_ed} - the fuzzy jets need to grow in order to cover the full energy distribution in the event.  Using anti-$k_t$ jets as seeds for fuzzy jets has the feature that the number of fuzzy jets change dynamically with the complexity of the event.  The algorithm is not very sensitive to the exact locations of the anti-$k_t$ jets - studies
which randomly perturbed the initial jet locations inside a disc of
radius $1.0$ found that $\sigma$ was robust to such
fluctuations, even on an event by event basis.   However, the $p_T$ threshold for the seed anti-$k_t$ jets can have a significant impact on the fuzzy jets as this alters the number of seeds.  The $p_T$ threshold for the anti-$k_t$ seeds is typically lower than the $p_T$ threshold one would use to consider anti-$k_t$ jets alone because the fuzzy jets algorithm needs enough seeds to populate the low energy regions of the detector.  One way of mitigating the impact of the $p_T$ cut on the fuzzy jet clustering is to introduce an {\it event jet}, described in Section~\ref{fuzzypileup}.

\begin{figure}[h!]
\vspace{1cm}
\begin{center}
\begin{tabular}{cc}
\begin{overpic}[width=0.43\textwidth]{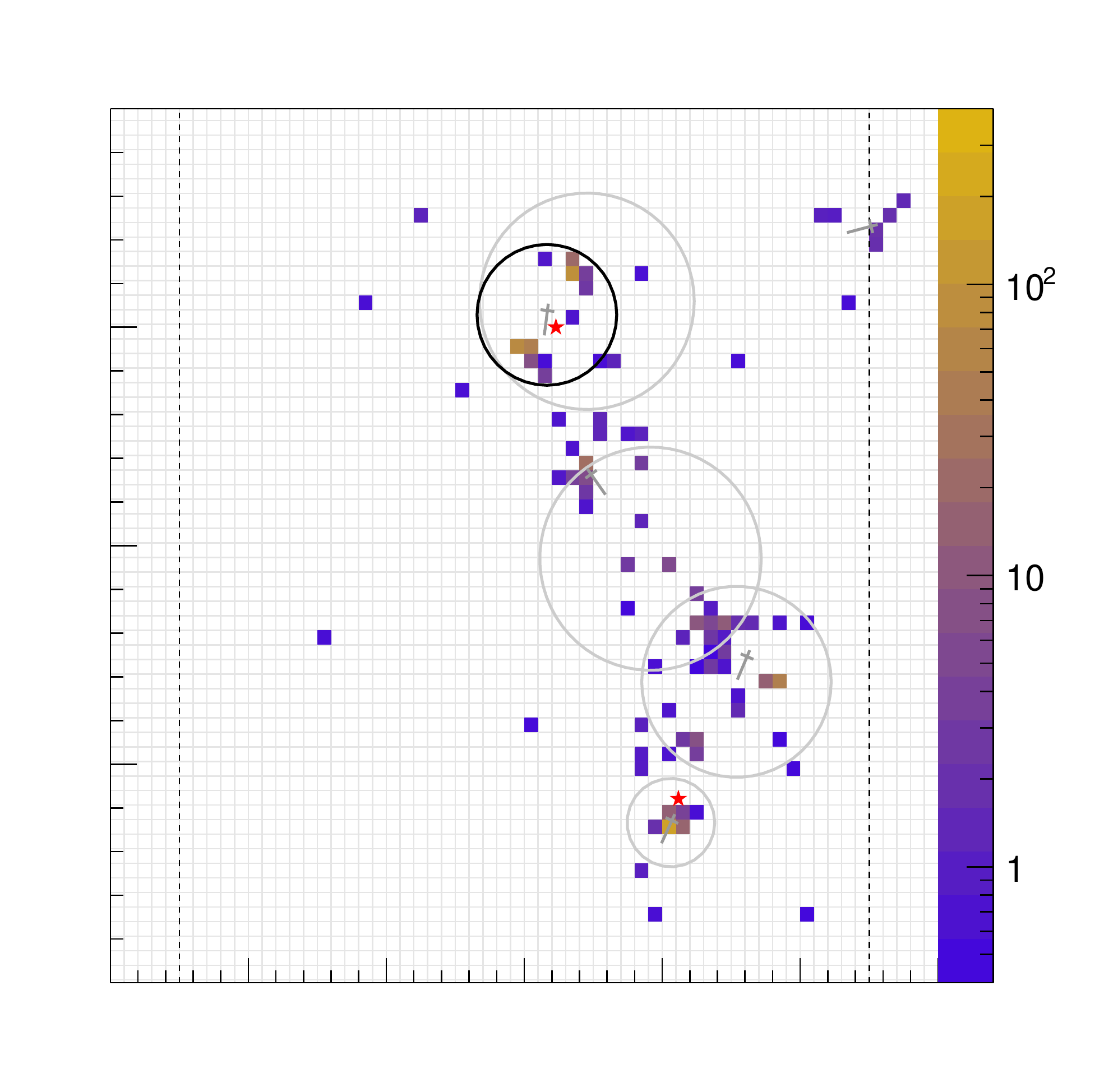}
\put(12, 98){  \Large {\sc Pythia 8}}
\put(12, 91){  $\sqrt{s} = 8 \text{ TeV}$}
\put(62, 91){  \large $Z'
    \rightarrow \text{t}\bar{\text{t}}$}
\put(62, 98){  \large $n_{\text{PU}} = 0$}
\put(14, 14){  \large $p_T^{\text{cut}} = 5 \text{ GeV}$}

\put(24, -4){ \small Pseudorapidity ($\eta$)}
\put(-4, 10){\rotatebox{90}{ \small Rotated Azimuthal Angle
    ($\phi$)}}
\put(95, 27){\rotatebox{90}{ \small Tower $p_T \text{ [GeV]}$}}

\put(5, 8){\bfseries \small \sffamily $0$}
\put(4, 28.3){\bfseries \small \sffamily $\frac{\pi}{2}$}
\put(5, 48.3){\bfseries \small \sffamily $\pi$}
\put(4, 68){\bfseries \small \sffamily $\frac{3\pi}{2}$}
\put(4, 88){\bfseries \small \sffamily $2\pi$}

\put(4,  4){\bfseries \small \sffamily $-3$}
\put(17, 4){\bfseries \small \sffamily $-2$}
\put(29.6, 4){\bfseries \small \sffamily $-1$}
\put(46.2, 4){\bfseries \small \sffamily $0$}
\put(58.8, 4){\bfseries \small \sffamily $1$}
\put(71.3, 4){\bfseries \small \sffamily $2$}
\put(83.8, 4){\bfseries \small \sffamily $3$}
\end{overpic} &
\begin{overpic}[width=0.43\textwidth]{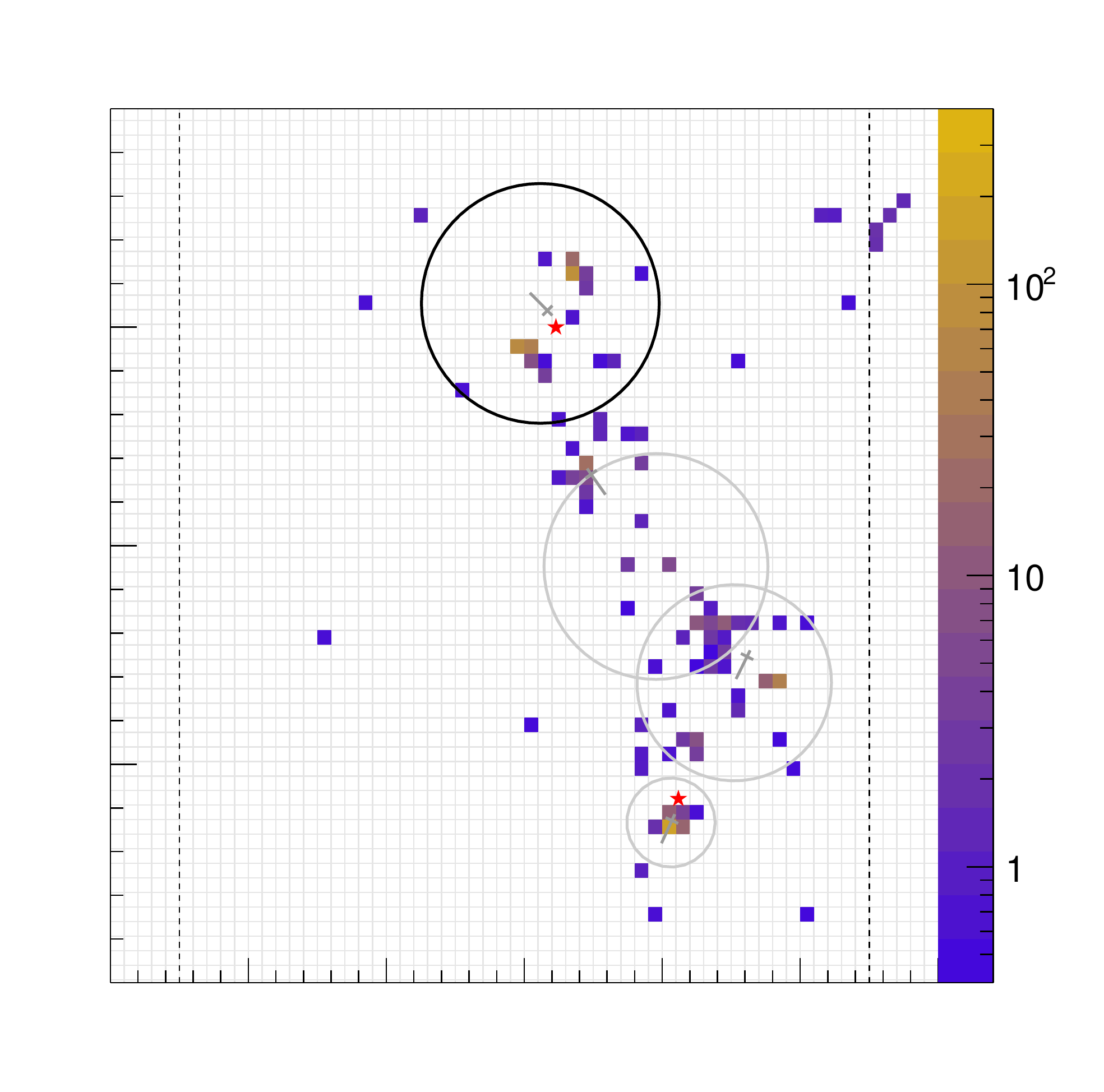}
\put(12, 98){  \Large {\sc Pythia 8}}
\put(12, 91){  $\sqrt{s} = 8 \text{ TeV}$}
\put(62, 91){  \large $Z'
    \rightarrow \text{t}\bar{\text{t}}$}
\put(62, 98){  \large $n_{\text{PU}} = 0$}
\put(14, 14){  \large $p_T^{\text{cut}} = 50 \text{
    GeV}$}

\put(24, -4){ \small Pseudorapidity ($\eta$)}
\put(-4, 10){\rotatebox{90}{ \small Rotated Azimuthal Angle
    ($\phi$)}}
\put(95, 27){\rotatebox{90}{ \small Tower $p_T \text{ [GeV]}$}}

\put(5, 8){\bfseries \small \sffamily $0$}
\put(4, 28.3){\bfseries \small \sffamily $\frac{\pi}{2}$}
\put(5, 48.3){\bfseries \small \sffamily $\pi$}
\put(4, 68){\bfseries \small \sffamily $\frac{3\pi}{2}$}
\put(4, 88){\bfseries \small \sffamily $2\pi$}

\put(4,  4){\bfseries \small \sffamily $-3$}
\put(17, 4){\bfseries \small \sffamily $-2$}
\put(29.6, 4){\bfseries \small \sffamily $-1$}
\put(46.2, 4){\bfseries \small \sffamily $0$}
\put(58.8, 4){\bfseries \small \sffamily $1$}
\put(71.3, 4){\bfseries \small \sffamily $2$}
\put(83.8, 4){\bfseries \small \sffamily $3$}
\end{overpic} \\
\end{tabular}
\end{center}
\caption{Changing the choice of the $p_T$ cut used to select seeds can
  make a vast difference in the values of the constructed variables,
  like $\sigma$. In this event, clustered on the left with a cut of $5
  \text{ GeV}$ resulting in five jets, and on the right with a cut of $50
  \text{ GeV}$ resulting four jets. Fewer degrees of freedom in
  the four jet case means a much larger learned value for the $\sigma$ variable.  Figures from C. Stansbury.}
\label{fig:pt_cut_ed}
\end{figure}

\clearpage

\section{Computation of Significance Variables}
\label{compute}

This section briefly describes how to numerically compute the significance variables introduced in Sec.~\ref{sec:significancevariables}.  If Gaussian approximations to the input object resolution functions are valid and the observable is sufficiently close to a linear combination of the input object kinematic quantities, then an annalytic approximation using linear error propagation should be sufficient.  However, to capture non-Gaussian attributes or important non-linear behavior of the kinematic variable, numeric propagation may be necessary.  In particular, if $m$ is a mass-like variable with a restriction $m>0$, the resolution function will necessarily be non-Gaussian near $m=0$.  In such cases, one can estimate how many random draws are necessary to accurately compute $\sigma_m$.   If $s^2$ is the sample variance, then the variance of the sample variance is given by Eq.~\ref{varofsamplevar}, where $\kappa$ is the excess kurtosis~\cite{VarSampleVar}.  

\begin{align}
\label{varofsamplevar}
\text{Var}[s^2]=\sigma^4\left(\frac{2}{n-1}+\frac{\kappa}{n}\right).
\end{align}

\noindent For an absolute uncertainty on the standard deviation $f$ and an $\mathcal{O}(1)$ standard deviation, one needs

\begin{align}
n=\frac{2+\kappa+f^2+\sqrt{4+4\kappa+4f^2+\kappa^2-2f^2\kappa+f^4}}{2f^2}.
\end{align}

\noindent For $f\ll 1$ and an order $1$ or smaller $\kappa$ (this is zero for a Gaussian),

\begin{align}
n\approx\frac{2+\kappa+\sqrt{4+4\kappa+\kappa^2}}{2f^2}\sim\frac{3}{f^2}.
\end{align}

\noindent For example, one needs $n\approx 300$ for an accuracy of $0.1$ GeV.   

\section{The Non-closure of Numerical Inversion}
\label{numericalinversion}

The jet calibration procedures of ATLAS~\cite{Aad:2011he} and CMS~\cite{Chatrchyan:2011ds} involve several steps to correct for pileup, the non-linear detector response, the $\eta$-dependence of the jet response, flavor-dependence of the jet response, and residual data/simulation differences in the jet response.  The simulation-based corrections to correct for the calorimeter non-linearities in $p_T$ and $\eta$ are accounted for using {\it numerical inversion}.  Let $X$ will be a random variable representing the particle-jet $p_\text{T}$ and $Y$ will be a random variable representing the reconstructed jet $p_\text{T}$.  Define\footnote{Capital letters represent random variables and lower case letters represent realizations of those random variables, i.e. $X=x$ means the random variable $X$ takes on the (non-random) value $x$.}

\begin{align}
f(x)&=\mathbb{E}[Y|X=x]\\
R(x) &= \mathbb{E}\left[\frac{Y}{x}\middle| X=x\right] = \frac{f(x)}{x}. 
\end{align}

\noindent Often, the normal approximation is valid: $Y|X=x\sim \mathcal{N}(f(x),\sigma(x))$, where this notation means `Y given $X=x$ is normally distributed with mean $f(x)$ and standard deviation $\sigma(x)$.'  The function $R(x)$ is called the {\it response function}.  Formally, numerical inversion is the following procedure:

\begin{enumerate}
\item Compute $f(x)$, $R(x)$.  
\item Let $\tilde{R}(Y) = R(f^{-1}(Y))$.
\item Apply a jet-by-jet correction: $Y\mapsto Y/\tilde{R}(Y)$.
\end{enumerate}

\noindent The intuition for the second step is that $f^{-1}(Y)$ is an estimate for $x$ and then $R(f^{-1}(Y))$ is an estimate for the response at the estimate of $x$ that gives rise to $Y$.  Note that $\mathbb{E}[X|Y]$ is not useful instead of $f^{-1}(Y)$ because the former depends on $p(x)$, the underlying $p_\text{T}$ spectrum, whereas $f$ (and thus $f^{-1}$) to do not depend on $p(x)$, by construction.   

In principle, a biased jet calibration is usable, even beneficial if the resolution can be made small.  However, for a variety of reasons, it is desirable for the calibration procedure to {\it closes}:

\begin{align}
\mathbb{E}\left[\frac{Y}{\tilde{R}(Y)x}\middle| X=x\right] = 1.
\end{align}

\noindent The random variable $Y/\tilde{R}(Y)=f^{-1}(Y)$.  To see this, let $\tilde{x}=f^{-1}(Y)$, Then,

\begin{align}
\frac{Y}{\tilde{R}(Y)}=\frac{f(\tilde{x})}{R(\tilde{x})}=\tilde{x}=f^{-1}(Y).
\end{align}

\noindent Now, suppose that $Y|X=x\sim\mathcal{N}(f(x),\sigma(x))$.  One can calculate the non-closure for a given function $f$.  First, a lemma:

\vspace{4mm}

\noindent {\it Lemma.} Suppose that $X\sim \mathcal{N}(\mu,\sigma)$.  Then, $f(X)\sim\mathcal{N}(\mu',\sigma')$ if and only if $f(x)$ is linear in $x$.

\vspace{4mm}

\noindent The proof is in Sec.~\ref{sec:lemma}.  Now a corollary for numerical inversion:

\vspace{4mm}

\noindent {\it Corollary.}  Suppose that $Y|X=x\sim\mathcal{N}(f(x),\sigma(x))$.  Then, the calibrated jet $p_\text{T}$ response $Y/\tilde{R}(Y)|X=x$ is normally distributed if and only if $f$ is linear in $x$.

\vspace{4mm}

\noindent This corollary is surprising because the ATLAS response function is non-linear and therefore numerical inversion {\it spoils} normality.  However, for Run 1 conditions with moderate pileup, this is a small effect.  Figure~\ref{fig:atlas:res2} shows the {\it theoretical} non-closure as a function of $x$.  The most relevant curve is the red one, which closely models the ATLAS response function.  The method clearly does not close, but the amount of non-closure is already less than 0.5\% at 20 GeV (and decreases with $x$). 

\vspace{4mm}

\begin{figure}[h!]
\begin{center}
\includegraphics[width=0.6\textwidth]{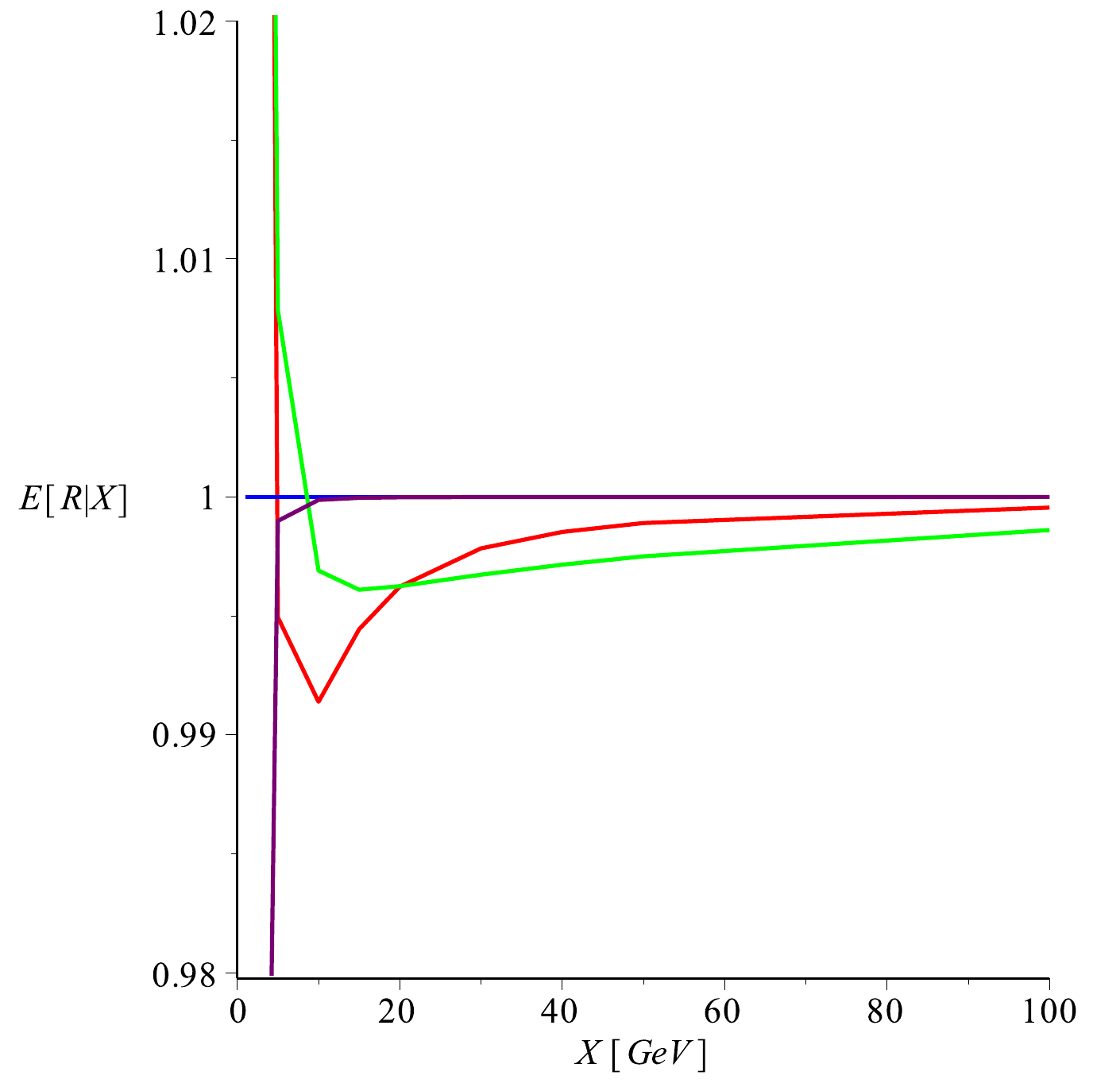}
\end{center}
\caption{The calculated non-closure for numerical inversion assuming $Y|X=x\sim\mathcal{N}(f(x),\sigma(x))$, where $\sigma(x)/x=1/\sqrt{x}$ estimated from Ref.~\cite{Aad:2011he} and $f(x)$ has several possibilities.  The blue line is {\color{blue}$f(x)=ax+b$} (which closes identically), red is {\color{red}$f(x)=a+b\log(x)$} (derived by estimating the curve from the lowest $|\eta|$ bin in Ref.~\cite{Aad:2011he} - $a=0.5$ and $b=0.09$), purple is {\color{purple}$f(x)=\sqrt{x}$}, and green is {\color{green}$f(x)=ax^2+bx+c$ with $b=.73, b=0.002, c=0.0$}.}
\label{fig:atlas:res2}
\end{figure}

\clearpage

\section{Gaussian Invariance Lemma}
\label{sec:lemma}

{\it Lemma.} Suppose that $X\sim \mathcal{N}(\mu,\sigma)$.  Then, $f(X)\sim\mathcal{N}(\mu',\sigma')$ if and only if $f(x)$ is linear in $x$.

\begin{proof}
The converse is trivial.  For the other direction, suppose that $f(X)\sim\mathcal{N}(\mu',\sigma')$.  Let $Y=(X-\mu)/\sigma$ and define 

\begin{align}
g(y)=\frac{f(\sigma y+\mu)-\mu'}{\sigma'}.
\end{align}

\noindent so that $Y$ and $g(Y)$ both have a standard normal distribution.  First, note that for $Z=g(Y)$, the following relation holds amongst the probability distributions for $Z$ and $Y$:

\begin{align}
f_Z(z)=f_Y(g^{-1}(z))\frac{\partial g^{-1}(z)}{\partial z}.
\end{align}

\noindent In particular, since the normal probability distribution is never non-positive, $g$ has to be monotonic (the derivative term can never be zero).  Then, we can write for any $c$:

\begin{align}
\nonumber
\Phi(c)=\Pr(Z<c)&=\Pr(g(Y)<c)\\
&=\Pr(Y<g^{-1}(c))=\Phi(g^{-1}(c)),
\end{align}

\noindent where the second line holds because $g$ preserves ordering.  Since the normal distribution cumulative distribution function is invertible, we then have that $g(c)=c$.  Inserting the definition of $g$ then gives us the final result:

\begin{align}
f(x)=\frac{\sigma'}{\sigma} (x-\mu)+\mu'
\end{align}

\end{proof}

\chapter{Changes Since Submission}
\label{sec:changes}

\begin{itemize}
\item {\bf September 11, 2016}: First submission to arXiv.  No changes to content (only restructuring of tex files).
\end{itemize}
 
\bibliographystyle{JHEP-2}	
\bibliography{NachmanThesis_v3-stripped.bib}{}

\end{document}